\def\sketchesmode{1}    
\def\seriousmode{0}

\def\maincontribstyle{0}  
      
\def\parainTOC{0}      
  
\def\doublecol{1}

\documentclass[twocolumn, aps, rmp, amsmath, amssymb, nofootinbib, superscriptaddress, floatfix, eqsecnum]{revtex4-2}

\usepackage[pdftex]{graphicx}
\usepackage[usenames, dvipsnames, svgnames, table]{xcolor}
\usepackage{mathrsfs}

\usepackage{hyperref}
\hypersetup{colorlinks = true, linkcolor = [rgb]{0.19411,0.51882,0.667058}, urlcolor = [rgb]{0.125490,0.29542,0.1647058}, citecolor = [rgb]{0.75882,0.37411,0.14117}}

\usepackage{url}
\usepackage{makeidx}
\usepackage{nomencl}
\usepackage{colortbl}
\usepackage[section]{placeins}
\usepackage{qcircuit}
\usepackage{array}
\usepackage{amsmath}
\usepackage[noend]{algpseudocode}
\usepackage{mdframed}
\usepackage{type1cm}
\usepackage{lettrine}
\usepackage{afterpage}
\usepackage[english]{babel}
\usepackage{lmodern}
\usepackage{multirow}
\usepackage[margin=0pt, font=small, labelfont=bf, labelsep=endash, justification=centerlast, labelsep=colon]{caption}
\usepackage{microtype}
\usepackage{booktabs}
\usepackage{fancyhdr}
\usepackage[T1]{fontenc}
\newcommand{\bra}[1]{\langle#1|}
\newcommand{\ket}[1]{|#1\rangle}
\newcommand{\braket}[1]{\langle#1\rangle}

\newcommand\gobbleone[1]{}

\newcommand{\smallaffiliation}[1]{\affiliation{\small #1}}
\newcommand{\boldauthor}[1]{\author{\textbf{#1}}}

\newcommand{\captionspacealg}[0]{\vspace{-1.5em}}
\newcommand{\captionspacefig}[0]{\vspace{-0.8em}}
\newcommand{\captionspacetab}[0]{\vspace{-0.6em}}

\if 1\seriousmode
	\newcommand{\latinquote}[1]{}
\else
	\newcommand{\latinquote}[1]{\textit{--- #1}\index{Latin}}
\fi

\newcommand{\sectionby}[1]{}

\newcommand{\famousquote}[2]{\noindent\textit{``#1''} --- #2.\index{Quotes}\index{#2}}

\newcommand{\dropcap}[2]{{#1}{#2}}

\newcounter{regtablecntr}
\newcounter{algcntr}
\newcommand{\startnormtable}{
	\renewcommand{\tablename}{Table}
	\setcounter{algcntr}{\value{table}}
	\setcounter{table}{\value{regtablecntr}}
}

\newcommand{\startalgtable}{
	\renewcommand{\tablename}{Algorithm}
	\setcounter{regtablecntr}{\value{table}}
	\setcounter{table}{\value{algcntr}}
}

\if 0\sketchesmode
	\newcommand{\sketch}[1]{}
\fi
\if 1\sketchesmode
	\newcommand{\sketch}[1]{\begin{center}\includegraphics[clip=true, width=0.4\textwidth]{#1}\end{center}\index{Artwork}}
\fi

\addto\captionsenglish{}

\renewcommand{\tablename}{Algorithm}

\if 1\doublecol
	\newmdtheoremenv[innertopmargin=3pt, innerbottommargin=3pt, nobreak]{definition}{Def}
	\newmdtheoremenv[innertopmargin=3pt, innerbottommargin=3pt, nobreak]{postulate}{Post}
\else
	\newmdtheoremenv[innertopmargin=3pt, innerbottommargin=3pt, nobreak]{definition}{Definition}
	\newmdtheoremenv[innertopmargin=3pt, innerbottommargin=3pt, nobreak]{postulate}{Postulate}
\fi

\renewcommand{\cite}{\citep}

\makeatletter
\def\BState{\State\hskip-\ALG@thistlm}
\makeatother

\graphicspath{{./Figures/}}

\afterpage{\clearpage}

\makeatletter
\let\UrlSpecialsOld\UrlSpecials
\def\UrlSpecials{\UrlSpecialsOld\do\/{\Url@slash}\do\_{\Url@underscore}}%
\def\Url@slash{\@ifnextchar/{\kern-.11em\mathchar47\kern-.2em}%
    {\kern-.0em\mathchar47\kern-.08em\penalty\UrlBigBreakPenalty}}
\def\Url@underscore{\nfss@text{\leavevmode \kern.06em\vbox{\hrule\@width.3em}}}
\makeatother

\hypersetup{
	pdfauthor = {Peter P. Rohde},
	pdftitle = {The Quantum Internet},
	pdfsubject = {Quantum Internet},
	pdfkeywords = {Quantum, Computing, Information, Networks, Internet, Cryptography}
}

\makenomenclature
\makeindex

\usepackage{hyperref}

\begin{document}

	\title{ The Quantum Internet (Technical Version) \\ ~~ \large -- the Second Quantum Revolution }

%
% Peter P. Rohde
%

\if 1\maincontribstyle
\boldauthor{\newline\newline {\footnotesize Main author:}\newline Peter P. Rohde}
\else
\boldauthor{Peter P. Rohde}
\fi
\email[To whom correspondence should be addressed: ]{dr.rohde@gmail.com}
\homepage{http://www.peterrohde.org}
\smallaffiliation{Centre for Quantum Software and Information, Faculty of Engineering and Information Technology, University of Technology Sydney, Sydney, NSW, Australia}
\smallaffiliation{Hearne Institute for Theoretical Physics, Louisiana State University, Baton Rouge, United States}

%
% Zixin Huang
%

\if 1\maincontribstyle
\boldauthor{\newline\newline {\footnotesize Contributing authors:}\newline Zixin Huang}
\else
\boldauthor{Zixin Huang}
\fi
\email{zixin.huang@mq.edu.au}
\smallaffiliation{School of Mathematical and Physical Sciences, Macquarie University, NSW 2109, Australia}
\smallaffiliation{Centre for Quantum Software and Information, Faculty of Engineering and Information Technology, University of Technology Sydney, Sydney, NSW, Australia}

\boldauthor{Yingkai Ouyang}
\smallaffiliation{School of Mathematical and Physical Sciences, University of Sheffield, Sheffield, S3 7RH, United Kingdom}

% He-Liang Huang

%

\boldauthor{He-Liang Huang}
%\email{quanhhl@mail.ustc.edu.cn}
% \smallaffiliation{Hefei National Laboratory for Physical Sciences at Microscale and Department of Modern Physics, University of Science \& Technology of China, Hefei, China}
% \smallaffiliation{CAS Centre for Excellence and Synergetic Innovation Centre in Quantum Information and Quantum Physics, University of Science \& Technology of China, Hefei, China}
% \smallaffiliation{CAS-Alibaba Quantum Computing Laboratory, Shanghai, China}
% \smallaffiliation{Henan Key Laboratory of Quantum Information \& Cryptography, Zhengzhou Information Science \& Technology Institute, Zhengzhou, China}
\smallaffiliation{Henan Key Laboratory of Quantum Information and Cryptography, Zhengzhou, Henan 450000, China}

%
% Zu-En Su
%

\boldauthor{Zu-En Su}
%\email{zuensu@gmail.com}
\smallaffiliation{Hefei National Laboratory for Physical Sciences at Microscale and Department of Modern Physics, University of Science \& Technology of China, Hefei, China}
\smallaffiliation{CAS Centre for Excellence and Synergetic Innovation Centre in Quantum Information \& Quantum Physics, University of Science \& Technology of China, Hefei, China}
\smallaffiliation{CAS-Alibaba Quantum Computing Laboratory, Shanghai, China}

%
% Simon Devitt
%

\boldauthor{Simon Devitt}
\smallaffiliation{Centre for Quantum Software and Information, Faculty of Engineering and Information Technology, University of Technology Sydney, Sydney, NSW, Australia}
\smallaffiliation{InstituteQ, Aalto University, 02150 Espoo, Finland.}

%
% Rohit Ramakrishnan
%

\boldauthor{Rohit Ramakrishnan}
%\email{rohit.k.ramakrishnan@gmail.com}
\smallaffiliation{Indian Institute of Science, Bangalore, India}

%
% Atul Mantri
%

\boldauthor{Atul Mantri}
%\email{atul_mantri@mymail.sutd.edu.sg}
\smallaffiliation{Singapore University of Technology \& Design, Singapore}
\smallaffiliation{Centre for Quantum Technologies, National University of Singapore, Singapore}

%
% Si-Hui Tan
%

\boldauthor{Si-Hui Tan}
% \email{sihui.tan@gmail.com}
\smallaffiliation{Horizon Quantum, Ireland, 24 Fitzwilliam Place, Dublin 2, D02 T296, Ireland}

%
% Nana Liu
%

\boldauthor{Nana Liu}

\smallaffiliation{Institute of Natural Sciences, Shanghai Jiao Tong University, Shanghai 200240, China}
\smallaffiliation{School of Mathematical Sciences, Shanghai Jiao Tong University, Shanghai 200240, China}
\smallaffiliation{Ministry of Education Key Laboratory in Scientific and Engineering Computing, Shanghai Jiao Tong University, Shanghai 200240, China}
\smallaffiliation{Shanghai Artificial Intelligence Laboratory, Shanghai, China}
\smallaffiliation{University of Michigan-Shanghai Jiao Tong University Joint Institute, Shanghai 200240, China.}

%
% Scott Harrison
%

\boldauthor{Scott Harrison}
%\email{harrison@dipf.de}
\smallaffiliation{Leibniz Institute for Research \& Information in Education, Frankfurt am Main, Germany}

%
% Chandrashekar Radhakrishnan
%

\boldauthor{Chandrashekar Radhakrishnan}
%\email{chandrashekar10@gmail.com}
\smallaffiliation{Department of Computer Science and Engineering, NYU Shanghai, 567 West Yangsi Road, Shanghai, 200124, China.}

\boldauthor{Gavin K. Brennen}
\smallaffiliation{School of Mathematical and Physical Sciences, Macquarie University, NSW 2109, Australia}
\smallaffiliation{Centre of Excellence in Engineered Quantum Systems, Macquarie University, New South Wales, Australia}

\boldauthor{Ben~Q.~Baragiola}
\smallaffiliation{Centre for Quantum Computation and Communication Technology,
School of Science, RMIT University, VIC 3000, Australia}

\boldauthor{Jonathan P. Dowling}
%\email{jdowling@lsu.edu}
\smallaffiliation{Hearne Institute for Theoretical Physics and Department of Physics \& Astronomy, Louisiana State University, Baton Rouge, United States}

%
% Tim Byrnes
%

\boldauthor{Tim Byrnes}
%\email{tim.byrnes@nyu.edu}
\smallaffiliation{New York University Shanghai; NYU-ECNU Institute of Physics at NYU Shanghai, 567 West Yangsi Road, Shanghai, 200124, China.}
\smallaffiliation{State Key Laboratory of Precision Spectroscopy, School of Physical and Material Sciences, East China Normal University, Shanghai 200062, China}
\smallaffiliation{Center for Quantum and Topological Systems (CQTS), NYUAD Research Institute, New York University Abu Dhabi, UAE.}
\smallaffiliation{Department of Physics, New York University, New York, NY 10003, USA}

%
% William J. Munro

%

\boldauthor{William J. Munro}
%\email{bill.munro@me.com}
\smallaffiliation{Okinawa Institute of Science and Technology Graduate University, Onna-son, Okinawa 904-0495, Japan}
 \smallaffiliation{National Institute of Informatics, 2-1-2 Hitotsubashi, Chiyoda-ku, Tokyo 101-8430, Japan}

\date{\today}
% \fi

\frenchspacing

%
% Abstract
%

\begin{abstract}
\newpage

The desire to share and unite remote digital assets motivated the development of the classical internet, the enabler of the entire 21st century economy and our modern way of life. As we enter the quantum era, it is to be expected there will be a similar demand for networking quantum assets, motivating a \textit{global quantum internet} for bringing together the world's quantum resources, leveraging off their exponential trajectory in capability. We present models for quantum networking, how they might be applied in the future, and the implications they will have. Socially, economically, politically and geostrategically, the upcoming era of quantum supremacy will be as significant for the 21st century as the transistor was for the 20th.
\\[4pt]
The inherently different scaling in the computational power of quantum computers fundamentally changes the dynamics of how they will operate in the future. Given their high expected initial cost, a client/server model for outsourcing computation will be essential for enabling the accessibility and proliferation of this technology, and ensuring its economic viability. We therefore anticipate the emergence of \textit{cloud quantum computing}, a model for outsourcing quantum computations to the network. We argue that economic efficiency will mandate that all future quantum computers be united into a single global \textit{virtual quantum computer}, offering exponentially more power to all network participants than if they were to keep their resources to themselves. This model for the allocation of computational resources is uniquely quantum, with no classical analogue, completely altering the economic landscape for the future of computation.
\\[4pt]
Given the sensitivity of much of the data to which future quantum computers are going to foreseeably be applied, protocols for encrypted quantum computation will be essential -- the outsourcing of computations that neither an eavesdropper nor even the server performing the computation can spy upon. This will enable new models for the commercialisation and proliferation of quantum technologies, unlike any existing models for classical computing.
\\[4pt]
% We argue that the quantum internet will not create a system of winners and losers, but will rather be of benefit to all of humanity, to an even greater extent than the classical internet. It is therefore essential that it be imminently established and pursued.
% 
We argue that the quantum internet has the potential to benefit all of humanity, surpassing even the impact of the classical internet. Its development and implementation should be prioritised as a critical global endeavor to ensure these benefits are realised.
\\[4pt]
While this work is only an early step in a rapidly developing field, still in its infancy, the central concepts we present will be highly relevant to future developments, laying the groundwork for this blossoming field.
\\[4pt]
We present both original ideas, as well as an extensive review of relevant and related background material. 
The work is divided into technical sections (requiring only a basic knowledge of the notation of quantum mechanics), for those interested in mathematical details, as well as extensive, entirely non-technical sections for those seeking a more general understanding.
\\[4pt]
We target this work very broadly at quantum and classical computer scientists, classical computer systems, software and network engineers, physicists, economists, artists, musicians, and those just generally curious about the future of quantum technologies and what they might bring to humanity.
\\[4pt]
\latinquote{Carpe futurum. Ad astra per alas fideles. Scientia potentia est. Benedictus benedicat.}

%\if 1\sketchesmode
%	\includegraphics[clip=true, width=0.55\linewidth]{sketch_cover}
%\fi

\clearpage
\begin{center}
\large 
\vspace*{100pt}

~~~~~~~~~~In memory of Prof Jonathan P. Dowling.\\
~~~~~~~~~He referred to this project as ``the doorstop'' 
\footnote{In the last chapter of ``Schr{\"o}dinger’s Web: Race to Build the Quantum Internet'', \cite{dowling2020schrodinger}}. 
\end{center}
\clearpage
\end{abstract}

\maketitle

% \latinquote{Dum spiro, spero.}

%
% Statement of Contributions
%

\textbf{Statement of contributions}
\footnote{Note: apart from a few sections, most of this project has been written before 2021.}

\begin{itemize}
	\item Peter P. Rohde conceived, directed and edited this project, developed the original ideas, and prepared most of the manuscript.
	\item Zixin Huang prepared Secs.~\ref{sec:intro_to_QO}, \ref{sec:quant_inf_th}, \ref{sec:telescopy}, \ref{sec:CV_QKD}, \ref{sec:adiabatic_QC}, \ref{sec:CV_QC}, \ref{sec:essay_future_QKD}, \ref{sec:attacks_QKD}, \ref{sec:shallow_circs} \& \ref{sec:NISQ} and contributed to the editting of this manuscript.
	\item He-Liang Huang \& Zu-en Su prepared Part.~\ref{part:SotA}.
	\item Simon Devitt prepared Sec.~\ref{sec:sneakernet}.
	\item Rohit Ramakrishnan prepared Sec.~\ref{sec:interfer_switches}.
	\item Yingkai Ouyang, Si-Hui Tan \& Atul Mantri contributed to Sec.~\ref{sec:homo_blind}.
	\item Nana Liu prepared Secs.~\ref{sec:verification} \& \ref{sec:quantum_mind}.
	\item Scott Harrison prepared Secs.~\ref{sec:econ_prop}, \ref{sec:policy} \& \ref{sec:game_theory}.
	\item Darryl Vietch contributed to Part.~\ref{part:class_net}.
	\item Chandrashekar Radhakrishnan prepared Secs.~\ref{sec:superconducting_qubits} \& \ref{sec:artificial_atoms}.
	\item Time Byrnes prepared Secs.~\ref{sec:clock_sync} \& \ref{sec:quant_space_race}.
	\item Gavin Brennen prepared Sec.~\ref{subsec:BS_PoW}.
	\item Ben Q. Baragiola prepared and edited sections of Parts.~\ref{part:quant_net} and ~\ref{part:protocols}.
	\item Jonathan P. Dowling contributed to Sec.~\ref{sec:quant_space_race} and to the editing.
	\item William J. Munro prepared Sec.~\ref{sec:rep_net}. 
	\item All authors played active roles in the discussions that developed this work. 
\end{itemize}

\clearpage

\input{Sections/table_of_contents}

\sketch{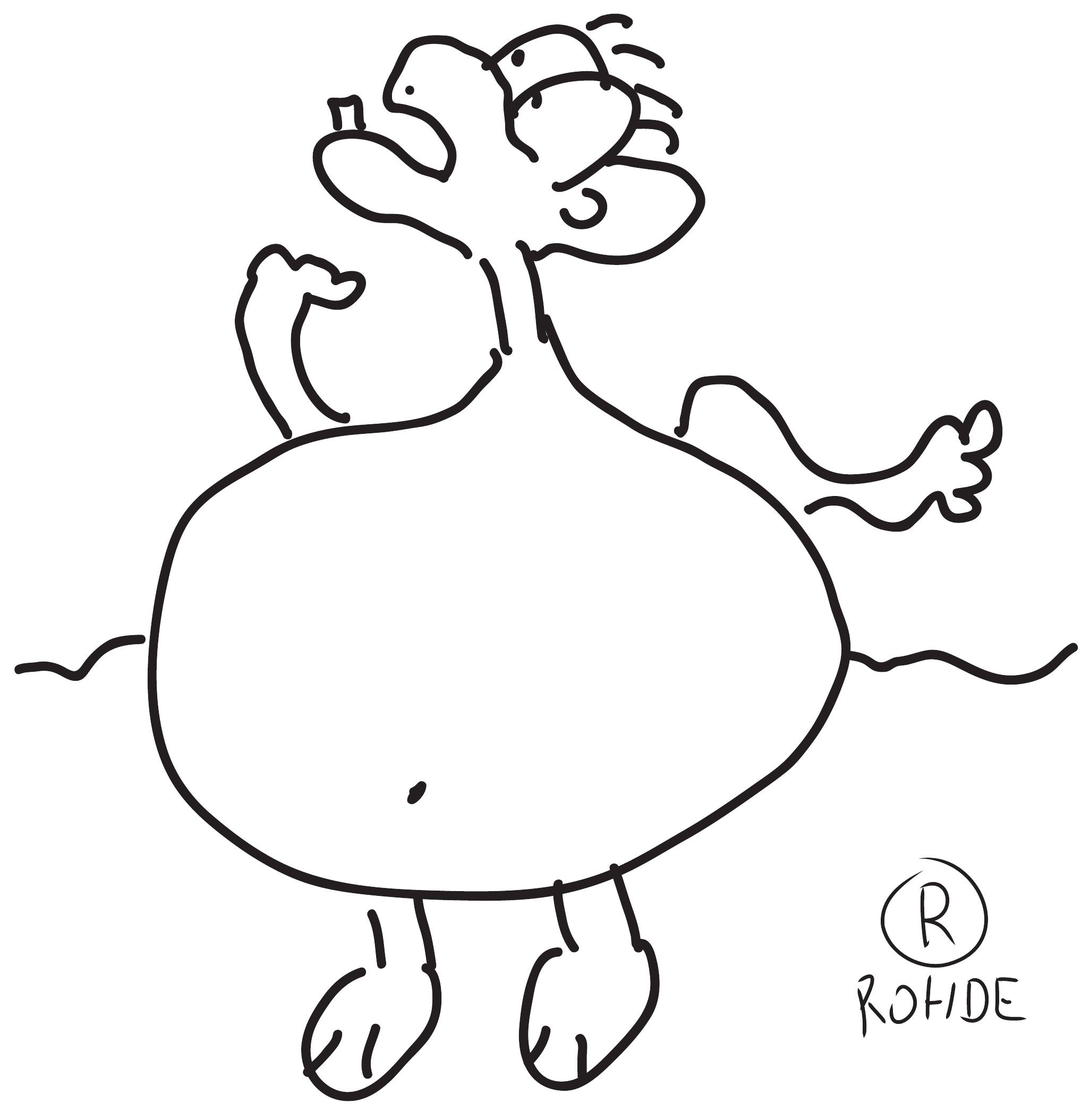}

\clearpage
%
% Introduction
%

\part{Introduction}\label{part:introduction}\index{Introduction}

\famousquote{Any sufficiently advanced technology is indistinguishable from magic.}{Arthur C. Clarke}
\newline

\famousquote{There may be babblers, wholly ignorant of mathematics, who dare to condemn my hypothesis, upon the authority of some part of the Bible twisted to suit their purpose. I value them not, and scorn their unfounded judgement.}{Nicolaus Copernicus}\index{Grandad}\index{Mansplaining}
\newline

\famousquote{Imagination is more important than knowledge.}{Albert Einstein}
\newline

%
% Foreword
%

\section{Foreword}\label{sec:foreword}

\dropcap{Q}{uantum} technologies\index{Quantum technologies} are not just of interest to quantum physicists, but will have transformative effects across countless areas -- the next technological revolution. For this reason, this work is directed at a general audience of not only preexisting quantum computer scientists, but also classical computer scientists, physicists, economists, artists, musicians, and computer, software and network engineers. More broadly, we hope this work will be of interest to those who recognise the future significance of quantum technologies, and the implications (or even just curiosities) that globally networking them might have -- the creation of the global quantum internet \cite{bib:van2014quantum, bib:kimble2008quantum}. We expect the answer to that question will look very different to what emerged from the classical internet.

A basic understanding of quantum mechanics \cite{bib:Sakurai94}, quantum optics \cite{bib:GerryKnight05}, quantum computing and quantum information theory \cite{bib:NielsenChuang00}\footnote{Throughout this manuscript we use the Nielsen \& Chuang convention for the pronunciation of `zed' \cite{bib:NielsenChuang00}.\index{Zed}}, classical networking \cite{bib:TanenbaumNet}, and computer algorithms \cite{bib:RivestAlgBook} are helpful, but not essential, to following our discussion. Some mathematical sections require a basic understanding of the mathematical notation of quantum mechanics. Although the reader without this background ought to be able to nonetheless follow the broader arguments. To bring readers from a mathematical but non-quantum background up to scratch, in Part.~\ref{part:mathematical_foundations} we present introductory tutorials on quantum mechanics and quantum optics, covering the essential mathematics necessary for following this book.

The entirely technically disinterested or mathematically incompetent reader may refer to just Parts~\ref{part:introduction}, \ref{part:essays} \& \ref{part:the_end} -- essentially brief non-technical, highly speculative essays about the motivation, applications and implications of the future quantum internet.

This work is partially a review of existing knowledge relevant to quantum networking, and partially original ideas, to a large extent based on the adaptation of classical networking concepts and quantum information theory to the context of quantum networking. A reader with an existing background in these areas could calmly skip the respective review sections.

Our goal is to present a broadly accessible technical and non-technical overview of how we foresee quantum technologies to operate in the era of quantum globalisation, and the exciting possibilities and emergent phenomena that will evolve from it.

We don't shy away from making bold predictions about the future of the quantum internet, how it will manifest itself, and what its implications will be for humanity and for science. Inevitably, some of our predictions will turn out to be accurate, whilst others will completely miss the mark entirely. We have no fear of controversy. How accurate our vision will be will have to be seen, but the most important goal in presenting grandiose predictions is to inspire new research directions, encourage future work, and stimulate lively and rigorous scientific debate about future technology. If we succeed at achieving these things, yet every last one of our predictions turn out to be completely and utterly wrong, we will consider this work a resounding success. Our goal, first and foremost, is to inspire future science.

\if 0\seriousmode
The theme music for this work may be found at \url{http://soundcloud.com/peter-rohde/wir-sind-ein-volk} \copyright\index{Theme music}\footnote{We use the copyright (\copyright) and trademark (\texttrademark) symbols liberally throughout this text when referring to things where we believe commercial opportunities may exist in the future. The authors do not own copyrights or trademarks to anything in this text.}.
\fi

\latinquote{Liber magnus est.}

%
% Introduction
%

\section{Introduction} \label{sec:introduction}

\famousquote{Nothing in life is to be feared, it is only to be understood. Now is the time to understand more, so that we may fear less.}{Marie Curie}
\newline

\dropcap{T}{he} internet is one of the key technological achievements of the 20th century, an enabling factor in every aspect of our everyday use of modern technology. While digital computing was the definitive technology of the 20th century, quantum technologies will be for the 21st \cite{bib:NielsenChuang00, bib:Bennett00}. 

Perhaps the most exciting prospect in the quantum age is the development of quantum computers\index{Quantum computing}. Richard Feynman \cite{bib:Feynman85} was the first to ask the question \textit{``If quantum systems are so exponentially complex that we are unable to simulate them on our classical computers, can those same quantum systems be exploited in a controlled way to exponentially outperform our classical computers?''}\index{Richard Feynman}. Subsequently, the Deutsch-Jozsa algorithm \cite{bib:DeutschJozsa92}\index{Deutsch-Jozsa algorithm} demonstrated for the first time that algorithms can run on a quantum computer, exponentially outperforming any classical algorithm. Since then, an enormous amount of research has been dedicated to finding new quantum algorithms, and the search has indeed been a very fruitful one\footnote{See the Quantum Algorithm Zoo\index{Quantum Algorithm Zoo} for a comprehensive summary of the current state of knowledge on quantum algorithms (\url{http://math.nist.gov/quantum/zoo/}).}, with many important applications having been found, including, amongst many others\index{Quantum algorithms}:

\begin{itemize}
	\item Searching unstructured databases (Sec.~\ref{sec:quantum_search}):\index{Grover's algorithm}
		\begin{itemize}
		\item Grover's algorithm \cite{bib:Grover96}.
		\item Quadratic speedup.
		\end{itemize}
	\item Satisfiability \& optimisation problems\footnote{A satisfiability problem is one where we search a function's input space for a solution(s) satisfying a given output constraint. The hardest such problems, like the archetypal 3-\textsc{SAT}\index{3-SAT problem} problem, are \textbf{NP}-complete.}\index{NP \& NP-complete}\index{Optimisation}\index{Satisfiability problems} (Sec.~\ref{sec:optimising_the_world}):
		\begin{itemize}
			\item Grover's algorithm.
			\item Quadratic speedup.
			\item Includes solving \textbf{NP}-complete problems, and brute-force cracking of private encryption keys.\index{Satisfiability problems}
			\item Many optimisation problems are \textbf{NP}-complete or can be approximated in \textbf{NP}-complete.\index{Optimisation}
			\end{itemize}
	\item Period finding and integer factorisation\index{Shor's algorithm} (Sec.~\ref{sec:shors_alg}):
		\begin{itemize}
		\item Shor's algorithm \cite{bib:ShorFactor}.
		\item Exponential speedup.
		\item This compromises both Rivest, Shamir \& Adleman (RSA) and elliptic-curve public-key cryptography \cite{bib:RSA}\index{RSA encryption}\index{Elliptic-curve cryptography}\index{Public-key cryptography}, the most widely used cryptographic protocols on the internet today.
		\item This problem is believed to be \textbf{NP}-intermediate -- an \textbf{NP} problem that lies outside \textbf{P} (and is therefore classically hard), but which is not \textbf{NP}-complete (the `hardest' of the \textbf{NP} problems).
		\end{itemize}
	\item Simulation of quantum systems\index{Quantum simulation} (Sec.~\ref{sec:quantum_sim_alg}):
		\begin{itemize}
			\item Lloyd's algorithm \cite{bib:lloyd1996universal}
			\item Exponential speedup.
			\item This includes simulation of: molecular and atomic interactions in the study of quantum chemistry or nuclear physics; interactions between drug molecules and organic molecules for drug design; genetic interactions for the study of genetics and genetic medicine; nanoscale semiconductor physics for integrated circuit design; and much more.
			\end{itemize}
	\item Simulation of quantum field theories\index{Simulating quantum field theories}:
		\begin{itemize}
		 \item Jordan-Lee-Preskill algorithm \cite{bib:JLP, bib:RohdeWavelet15}
		 \item Exponential speedup.
		 \item A key area of fundamental physics research.
		 \end{itemize}
	\item Topological data analysis (Sec.~\ref{sec:TDA}):\index{Topological!Data analysis}
		\begin{itemize}
		\item Lloyd's algorithm \cite{bib:lloyd2016quantum}.
		\item Exponential speedup.
		\item Broad applications including: social media network analysis\index{Social media network analysis}; consumer behaviour\index{Consumer behaviour}; behavioural dynamics\index{Behavioural dynamics}; neuroscience\index{Neuroscience}; and higher-dimensional signal and image processing\index{Signal \& image processing}.
		\end{itemize}
	\item Solving linear systems of equations:\index{Linear systems}
		\begin{itemize}
		\item Algorithms by \cite{bib:harrow2009quantum, bib:BerryLinear}.
		\item Exponential speedup.
		\item Widespread applications in linear algebra and calculus.
		\end{itemize}
	\item Quantum machine learning (Sec.~\ref{sec:quantum_mind}):\index{Quantum machine learning}
		\begin{itemize}
		\item Lloyd's algorithm \cite{bib:lloyd2013quantum}.
		\item This includes putting an end to humanity \latinquote{Ne obliviscaris.}
		\end{itemize}
\end{itemize}
An elementary technical overview of some of these archetypal algorithms is presented in Sec.~\ref{sec:quantum_algs}.

It's likely we haven't yet begun to fully recognise the capabilities of quantum computers, and the full plethora of applications they may have in the future. We stand at the beginning of the emergence of an entirely new type of technology.

In addition to many practical applications, the onset of quantum computing carries with it deep philosophical implications. Specifically, the Extended Church-Turing (ECT)\index{Extended Church-Turing (ECT) thesis} thesis hypothesises that any physically realisable system can be \textit{efficiently}\footnote{The term `efficient' is one coined by the computer scientist to mean that a problem can be solved in time at most polynomial in the size of the problem.\index{Computational!Efficiency}} simulated by a universal Turing machine\index{Turing machines} (i.e classical computer). The believed exponential complexity of quantum systems inclines quantum computer scientists to believe that the ECT thesis is therefore false \cite{bib:Deutsch85}\footnote{We have discovered a truly marvellous proof of this, which this footnote is too narrow to contain.}. The demonstration of large-scale quantum computers, while unable to prove or disprove the ECT thesis\footnote{When one talks about `scalability' or the `ECT thesis', we are talking about asymptotic relationships. Clearly no finite-sized experiment can prove asymptotic scaling with certainty. But with a sufficiently large quantum computer at our disposal, demonstrating exponentially more computational power than its classical sibling, we might be reasonably satisfied in convincing ourselves about the nature of the scaling of different computational models.}, could at least provide some convincing evidence against the ECT conjecture.

\begin{figure}[!htbp]
\if 1\doublecol
	\includegraphics[clip=true, width=0.475\textwidth]{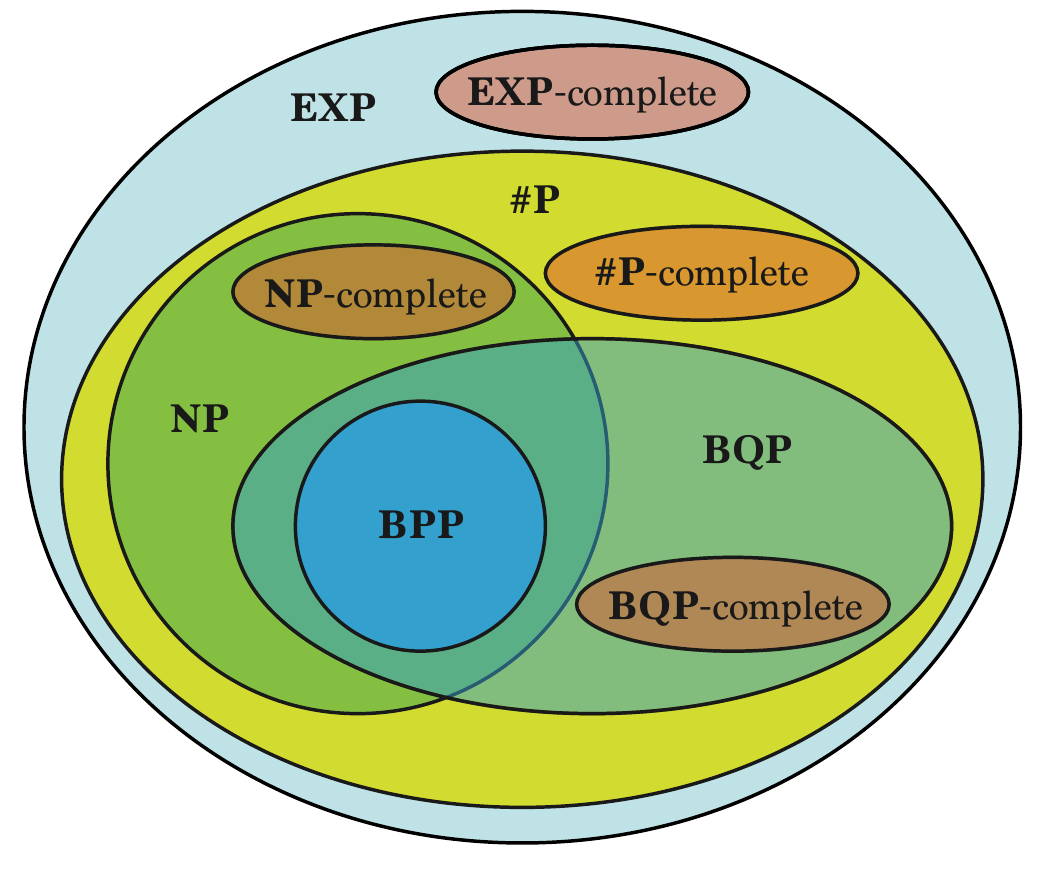}
\else
	\includegraphics[clip=true, width=0.6\textwidth]{complexity_classes}
\fi
	\index{Complexity classes} \index{P}\index{BPP} \index{NP \& NP-complete} \index{BQP} \index{\#P} \index{EXP}
	\vspace{-1.2em} \captionspacefig \caption{Believed relationships between the complexity classes most relevant to quantum computing. \textbf{BPP} is the class of polynomial-time probabilistic classical algorithms. \textbf{NP} is the class of problems verifiable in polynomial time using classical algorithms. \textbf{NP}-complete are the subset of \textbf{NP} problems polynomial-time reducible\index{Reducibility} to any other problem in \textbf{NP}, similarly for other `complete'\index{Complete problems} problems. \textbf{BQP} is the class of probabilistic algorithms solvable in polynomial time on universal quantum computers. $\#\mathbf{P}$ is the set of counting problems, that count satisfying solutions to \textbf{P} problems (\textbf{P} is the same as \textbf{BPP}, but deterministic, rather than probabilisitic). \textbf{EXP} is the class of all algorithms that require exponential time. Note that it is actually unproven whether \mbox{$\mathbf{P}=\mathbf{BPP}$} or \mbox{$\mathbf{P}\subset\mathbf{BPP}$}. There are examples where the best known \textbf{BPP}\index{BPP} algorithms outperform the best known \textbf{P} algorithms, which could arise because the two classes are inequivalent, or that we simply haven't tried hard enough to find the best deterministic algorithms. Furthermore, while it is known that \mbox{$\mathbf{P}\subseteq\mathbf{NP}$}, it is not known whether \mbox{$\mathbf{BPP}\subseteq\mathbf{NP}$}. For the sake of illustration in our Venn diagram we have taken the view that it is. \textbf{BPP} is regarded as the class of problems efficiently solvable on universal Turing machines (i.e classical computers), whereas \textbf{BQP} is that efficiently solvable on universal quantum computers. The computational superiority of quantum computers is based on the (strongly believed, yet unproven) assumption that \mbox{$\mathbf{BPP}\subset\mathbf{BQP}$}.} \label{fig:complexity_classes}
\end{figure}

From a computational complexity\index{Computational!Complexity} theorist's perspective, it is strongly believed that the complexity classes of problems efficiently solvable on classical computers (\textbf{P} \& \textbf{BPP}\index{P}\index{BPP}) and quantum computers (\textbf{BQP}\index{BQP}) are distinct. Specifically, it is believed that \mbox{$\mathbf{BPP}\subset\mathbf{BQP}$}. If this conjecture is correct, it implies the existence of quantum algorithms super-polynomially faster than the best classical ones, and that the ECT thesis is not correct. More specifically, Fig.~\ref{fig:complexity_classes} illustrates the believed relationships between some of the most important complexity classes relevant to quantum computing.

Aside from quantum computing, quantum cryptography\index{Quantum cryptography} holds the promise of uncrackable cryptographic protocols, guaranteed not by the assumed complexity of solving certain mathematical problems like integer factorisation or brute-force searching, but by the laws of quantum mechanics. That is, provided our understanding of quantum mechanics is correct, quantum cryptographic protocols exist, which cannot be cracked, irrespective of the computational resources of an adversary.

Already we are beginning to see elementary realisations of essential quantum technologies such as quantum computing, cryptography, and metrology. As these technologies become increasingly viable and more ubiquitous, the demand for networking them and sharing quantum resources between them will become a pressing issue. Most notably, quantum cryptography and \textit{cloud quantum computing}\index{Cloud quantum computing} will be pivotal in the proliferation of quantum technology, which necessarily requires reliable quantum communications channels.

The first demonstrations of digital computer networks were nothing more than simple two-party, point-to-point (P2P) communication\index{P2P topology}. However, the internet we have today extends far beyond this, allowing essentially arbitrary worldwide networking across completely ad hoc networks comprising many different mediums, with any number of parties, in an entirely plug-and-play\index{Plug-and-play} and decentralised fashion. Similarly, elementary demonstrations of quantum communication\index{Quantum communication} have been performed across a small number of parties, and much work has been done on analysing quantum channel capacities\index{Quantum channels!Capacity} in this context \cite{bib:wilde2013quantum,khatri2020principles}. But, as with digital computing, demand for a future \textit{quantum internet} is foreseeable, enabling the arbitrary communication of quantum resources, between any number of parties, over ad hoc networks.

The digital internet may be considered a technology stack, such as TCP/IP (Transmission Control Protocol/Internet Protocol)\index{Transmission Control Protocol/Internet Protocol (TCP/IP)}, comprising different levels of abstraction of digital information \cite{bib:TanenbaumNet}. At the lowest level we have raw digital data we wish to communicate across a physical medium. Above this, we decompose the data into packets. The packets are transmitted over a network, and TCP is responsible for routing the packets to their destination, and guaranteeing data integrity and Quality of Service (\textsc{QoS})\index{Quality of service (QoS)}. Finally, the packets received by the recipient are combined and the raw data reconstructed.

The TCP layer remains largely transparent to the end-user, enabling virtual software interfaces to remote digital assets that behave as though they were local. This allows high-level services such as the File Transfer Protocol (FTP)\index{File Transfer Protocol (FTP)}, the worldwide web, video and audio streaming, and outsourced computation on supercomputers, as though everything was taking place locally, with the end-user oblivious to the underlying networking protocols, which have been abstracted away. To the user, YouTube videos or Spotify tracks behave as though they were held as local copies. And FTP or DropBox allow storage on a distant data-centre to be mounted as though it were a local volume. We foresee a demand for these same criteria in the quantum era.

In the context of a quantum internet, packets of data will instead be quantum states, and the transmission control protocol is responsible for guiding them to their destination and ensuring quality control.

Here we present a treatment for such Quantum Transmission Control Protocols (QTCPs)\index{Quantum Transmission Control Protocol (QTCP)} as a theoretical foundation for a future quantum internet. We consider how such ad hoc networks may be described mathematically, how to quantify network performance, and present a QTCP stack for operating it. While the goals of QTCP are similar as for classical TCP, there are major conceptual differences between the classical and quantum internets, owing to the unique properties of quantum states with no classical analogue.

Our treatment of quantum networks will be optics-heavy, based on the reasonable assumption that communications channels will almost certainly be optical, albeit with many possible choices of optical states and mediums. However, this does not preclude non-optical systems from representing quantum information that is not in transit, and we consider such `hybrid' architectures\index{Hybrid!Architectures} in detail, as well as the interfacing between optical and non-optical systems. Indeed, it is almost certain that future large-scale quantum computers will not be all-optical, necessitating interfacing different physical architectures. We accommodate for this requirement in the design of the QTCP.

Shared quantum entanglement\index{Quantum entanglement} is a primitive resource with direct applications in countless protocols. This warrants special treatment of quantum networks, which do not implement a full QTCP network stack, but instead specialise in just this one task -- entanglement distribution. We will see that such a specialised network will already be immensely useful for a broad range of applications, and its simplicity brings with it many inherent advantages.

The quantum internet will enable advances in the large-scale deployment of quantum technologies. Most notably, in the context of quantum computing it will allow initially very expensive technology to be economically viable and broadly accessible via the outsourcing of computations from consumers who can't afford quantum computers, to well-resourced hosts who can -- \textit{cloud quantum computing}\index{Cloud quantum computing}.

With the addition of recent advances in homomorphic encryption and blind quantum computing\index{Encrypted quantum computation}, such cloud quantum computing can be performed securely, guaranteeing privacy of both data and algorithms, secure even against the host performing the computation. This opens up entirely new economic models and applications for the licensing of compute time on future quantum computers in the cloud.

The unique behaviour of quantum computing, in terms of the super-classical scaling in its computational power, brings with it many important economic and strategic considerations that are extremely important to give attention to in the post-classical world.

But quantum technologies extend far beyond computation. Many other exciting applications for controlled quantum systems exist, with new ones frequently emerging. Thus, the quantum internet will find utility beyond cloud quantum computing, enabling the global exchange of quantum resources and assets. This could include the networking of elementary quantum resources such as state preparation, entanglement sharing, teleportation and quantum measurements, or scale all the way up to massively distributed quantum computation or a global quantum cryptography network.\index{Quantum assets}

It is hard to foresee the future trajectory of quantum technology, much as no one foresaw the advances digital technology has made over the last half century. But it is certain that as the internet transformed digital technology, the quantum internet will define the future of quantum technologies.

\latinquote{Vincit qui se vincit.}

\sketch{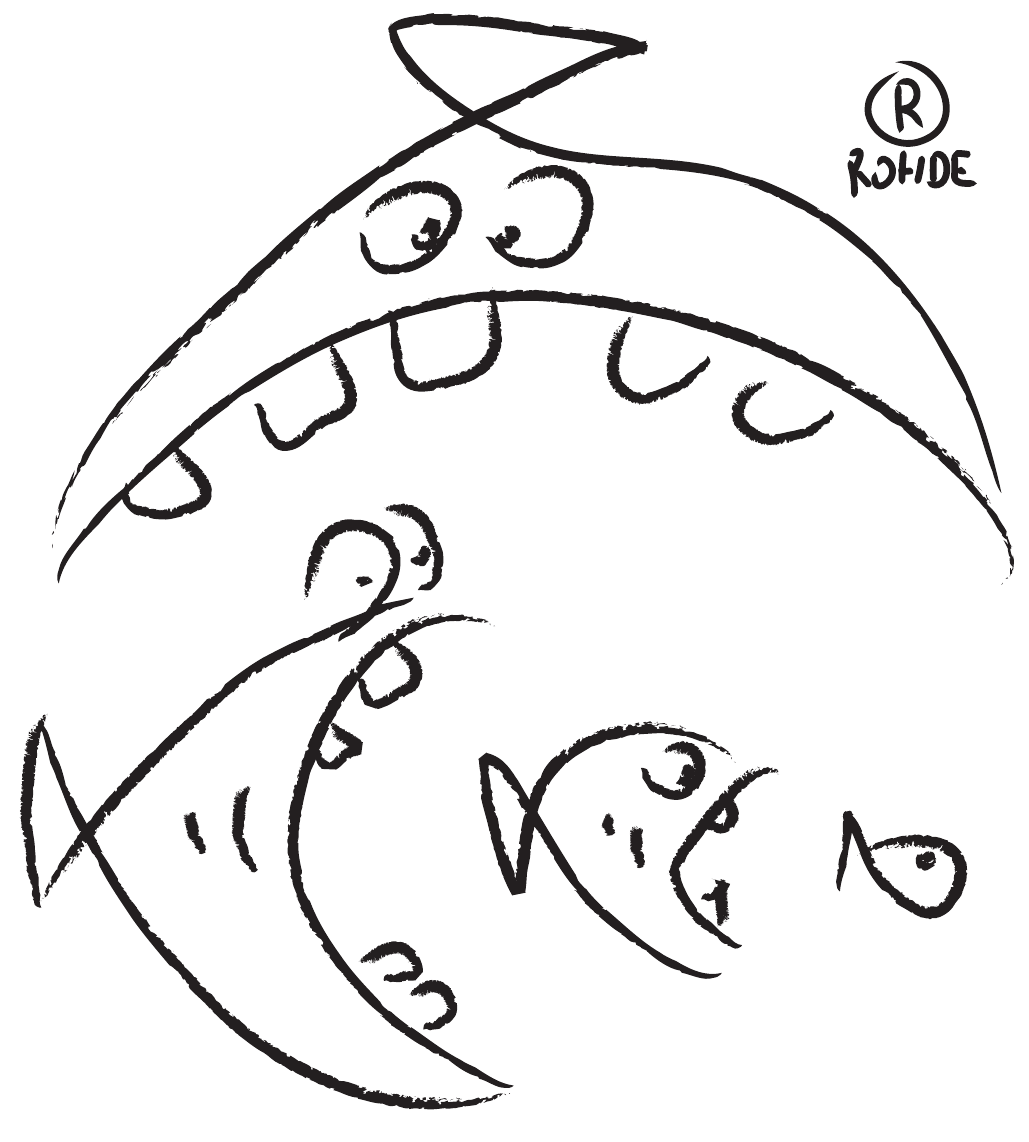}

\clearpage
% Mathematical foundations
%
\part{Mathematical foundations}\label{part:mathematical_foundations}\index{Mathematical foundations}

\famousquote{I come in peace. I didn't bring artillery. But I'm pleading with you, with tears in my eyes: If you fuck with me, I'll kill you all.}{General James ``Mad Dog'' Mattis}
\newline

\dropcap{T}{o} make the upcoming discourse accessible to those from a non-quantum background, we will briefly level the playing-field upfront by presenting an overview of the key concepts and mathematical formulations for quantum mechanics and quantum optics, both relied upon heavily throughout this work. All of this mathematical formulation fits within the confines of familiar basic undergraduate linear algebra\index{Linear algebra}\index{Quantum mechanics}\index{Quantum optics}.

This part ought provide the mathematical background necessary to follow the upcoming discussions. Those already familiar with the laws and formulation of quantum mechanics and quantum optics can calmly skip this part, without risk of offending the authors.

However, rather than just present the mathematical formulations and expect you to reverently accept the maths as the truth, the light, and the way, we will try as best we can to inject some intuition and meaningful interpretation of the mathematics into our tutorial, so that hopefully the reader is able to walk away with not just the ability to follow basic quantum mechanical derivations, but also grasp the intuition of the underlying physical reality.
%
% Introduction to quantum mechanics
%

\section{Introduction to quantum mechanics}\index{Quantum mechanics}\label{sec:intro_to_QM}

\famousquote{I think I can safely say that nobody understands quantum mechanics.}{Richard Feynman}
\newline

\dropcap{T}{he} laws of quantum mechanics are inherently different to those of classical mechanics in virtually every respect, and bring with them their own unique mathematical notation for:
\begin{itemize}
	\item The representation of quantum states.
	\item The evolution of quantum systems.
	\item The measurement of quantum states.
\end{itemize}

\noindent We review all of these formalisms in the upcoming sections, and present the necessary mathematical framework to enable following the derivations presented here.

%
% Features of quantum mechanics
%

\subsection{Features of quantum mechanics}\index{Quantum mechanics!Features}

Classical and quantum physics are fundamentally different in a number of extremely important ways. These differences are what allow quantum mechanics to enable the entire new field of quantum technologies, most notably quantum computing, which offers substantial computational gains over traditional classical computing. The same applies to the novel and groundbreaking promises of the other key quantum technologies, such as quantum cryptography, quantum teleportation, and quantum-enhanced metrology. All of these gains are achieved by exploiting these unique new characteristics offered by quantum systems, that are physically impossible in a classical context -- all technologies must obey the laws of physics, and therefore new laws of physics necessarily imply new technological possibilities.

The key conceptual differences between classical and quantum mechanics are:
\begin{itemize}
	\item Superposition\index{Superpositions}: the ability of quantum systems to be in multiple states simultaneously (simultaneously, not probabilistically!). This arises since quantum states can always be interpreted as waves, which are inherently delocalised and non-particle-like.
	\item Quantum amplitudes: quantum states are not governed by classical probabilities, but by \textit{quantum amplitudes}, which are complex numbers rather than real numbers.\index{Quantum amplitudes}
	\item Entanglement: a type of non-local quantum correlation, stronger than any possible classical correlation.\index{Entanglement}
	\item Unitary evolution: the free evolution of quantum systems is restricted to being described by unitary operators. This implies that all quantum evolution is perfectly time-reversible\index{Time-reversal}, since all unitary operators are invertible.\index{Evolution}
	\item Measurement collapse: the measurement of a quantum system changes its state in general, \textit{collapsing} it to the state corresponding to the measurement outcome\footnote{An open and intuitively confusing conundrum in quantum theory is to resolve the question of what exactly it is that constitutes `measurement' -- the \textit{measurement paradox}\index{Measurement!Paradox}. Given that all quantum evolution is supposed to be governed by unitary dynamics, whereas measurements are not, at what point in physical reality does an operation transition from one to the other? This is an extremely important open question, particularly from the perspective of quantum foundations and the philosophy of physics. This has led to the emergence of countless theories for the interpretation of quantum mechanics\index{Interpretation of quantum mechanics}, which is an essentially untestable field of research, more akin to fundamentalist religious movements than physical science. However, the resolution is largely unimportant in a practical technological context, where it suffices to define `measurement' as being `when quantum information becomes classical information', e.g once we are able to read it off a measurement device's LCD screen!}.\index{Measurement!Collapse}
	\item Non-determinism: quantum systems exhibit genuine non-determinism, associated with the randomness of the measurement collapse\index{Measurement!Collapse} process\footnote{When we say `random' we \textit{really} mean random, not `quasi-random but actually deterministic'\index{Quasi-randomness}. This true randomness is another of the unique and surprising features of quantum mechanics. Conversely to classical physics, quantum mechanics is inherently and provably non-deterministic! Specifically, it has been proven theoretically \cite{bohm1966proposed} that so-called `hidden variable theories'\index{Hidden variable theories} are incompatible with quantum physics, which rules out any form of quasi-randomness and mandates true randomness.}.\index{Non-determinism of quantum mechanics}
	\item Exponential complexity: the number of unique parameters required to fully describe a quantum system grows exponentially with its `size' (e.g the number of constituent subsystems, such as quantum particles or qubits). This is in contrast to classical systems whose state complexity grows only linearly with system size, since it's always possible to describe constituent subsystems independently (there is no notion of entanglement). Therefore the number of parameters required to characterise classical systems accumulates additively, whereas for quantum systems it grows multiplicatively.\index{Exponential complexity}
	\item Wave-particle duality: all quantum objects can simultaneously be considered to be particles and waves, where the wave-like and particle-like characteristics express themselves differently depending on the application.\index{Wave-particle duality}
	\item Uncertainty principle: unlike classical systems, where we can essentially know everything about them to arbitrary accuracy, quantum systems necessarily exhibit uncertainties. Heisenberg's uncertainty principle places strict physical bounds on the amount of information that can be known about quantum systems, exhibiting tradeoffs between how much we can know about different parameters simultaneously. The best-known example of this is the position-momentum uncertainty principle, which tells us that the more information we know about an object's position, the less we can simultaneously know about its momentum, and vice versa. It is not possible even in principle to have perfect information about both (this is a physical limitation, not an engineering one!).\index{Uncertainty principle}
\end{itemize}

All of these phenomena will become more clear and intuitive as we introduce the basic relevant mathematical constructions in the upcoming sections.

%
% Quantum States
%

\subsection{Quantum states}\index{Quantum states}

A quantum state is a mathematical description of a quantum object, fully characterising its quantum mechanical properties, much like classical attributes like momentum and position characterise the state of classical objects. This captures both the wave- and particle-like properties of quantum objects, given the quantum principle of \textit{wave-particle duality}\index{Wave-particle duality}. The key mathematical difference is that quantum objects exhibit unique, non-classical properties like superposition and entanglement. This requires introducing new constructs for their mathematical representation.

%
% State Vectors
%

\subsubsection{State vectors}\index{State vectors}

The most fundamental mathematical object in representing quantum systems is the \textit{state vector}, a vector representation for the quantum wave-function\index{Wave-functions}.

A state vector resides in an $N$-dimensional complex Hilbert space\index{Hilbert space}, \mbox{$\mathcal{H}\in\mathbb{C}^N$}, where $N$ denotes the number of basis states\index{Basis states} supporting the system. This support could be, for example, the available electron energy levels in an atom, or the two orthogonal polarisations available to a photon in an optical mode.

These vector representations for quantum states can quickly become unwieldy, since their dimensionality grows exponentially with the number of constituent subsystems. For example, a single qubit resides in a 2-dimensional Hilbert space, \mbox{$\mathcal{H}\in\mathbb{C}^2$}, whereas an $n$-qubit system resides in a $2^n$-dimensional one, \mbox{$\mathcal{H}\in\mathbb{C}^{2^n}$}. This is in contrast to $n$ classical bits, which require just $n$ parameters to collectively describe. Note that the difference in scaling between the vector representations of quantum systems and classical ones is exponential in nature. It is from this simple observation that the first intuitive inkling arises that perhaps this inherent astronomical complexity can be exploited for our benefit! \latinquote{Ad astra!}

Mathematically, we can employ the following alternate state vector representations, where the latter is referred to as \textit{Dirac notation} (with $\ket{n}$ denoting the basis states),
\begin{align}\label{eq:gen_state_vec}
	\ket\psi &= \begin{pmatrix}
	\alpha_1\\
	\alpha_2\\
	\vdots\\
	\alpha_N
\end{pmatrix}\nonumber\\
	&= \sum_{n=1}^N \alpha_n \ket{n}.
\end{align}
Here \mbox{$\alpha_n\in\mathbb{C}$} are referred to as the \textit{amplitudes} of the state, which for normalisation satisfy,
\begin{align}\label{eq:state_norm_cond}\index{Normalisation}
\sum_n |\alpha_n|^2 = 1.
\end{align}
Contrast this with the mathematical definition for the normalisation of classical probability distributions,
\begin{align}\index{Normalisation}
\sum_n p_n = 1.
\end{align}
The mathematical objects denoted as $\ket{\cdot}$ are commonly referred to as \textit{kets}\index{Kets}.

In this book we will largely be dealing with qubits\index{Qubits}, the simplest unit of quantum information, which are supported by exactly two basis states, denoted $\ket{0}$ and $\ket{1}$. A qubit can therefore be expressed as,
\begin{align}
\ket\psi &= \begin{pmatrix}
	\alpha \\
	\beta
\end{pmatrix}\nonumber\\
&= \alpha\ket{0} + \beta\ket{1}.
\end{align}
These are referred to as the \textit{logical basis states}\index{Logical basis states}, which are treated abstractly, detached from a particular physical realisation (e.g by associating them with electron energy levels, photon polarisation, or indeed \textit{any} two-level quantum system), much in the same way that classical bits are treated abstractly as a unit of information, detached from physical reality.

Also mathematically useful is the \textit{dual} of a state vector\index{Dual vector}, which is simply the complex conjugate transpose ($\dag$) of the vector, capturing the same information about the state in conjugate form,
\begin{align}
\bra\psi &= \begin{pmatrix}
	\alpha \\
	\beta
\end{pmatrix}^\dag\nonumber\\
&=\begin{pmatrix}
	\alpha^*, \beta^*
\end{pmatrix}\nonumber\\
&= \alpha^*\bra{0} + \beta^*\bra{1}.
\end{align}
The objects $\bra{\cdot}$ are referred to as \textit{bras}\index{Bras}.

% State overlap

\paragraph{State overlap}\index{State!Overlap}

For two generic state vectors, supported by the same basis,
\begin{align}
	\ket\psi = \sum_n \alpha_n \ket{n},\nonumber\\
	\ket\phi = \sum_n \beta_n \ket{n},
\end{align}
the \textit{overlap}\index{State!Overlap} between them is defined as,
\begin{align}
\braket{\psi|\phi} = \sum_n \alpha_n\beta_n^*,
\end{align}
which is mathematically equivalent to their vector dot product\index{Dot product}. The magnitude of the overlap of two quantum states lies in the range,
\begin{align}
	0\leq|\braket{\psi|\phi}|\leq 1,
\end{align}
where \mbox{$|\braket{\psi|\phi}|=0$} indicates that the two states were orthogonal, and \mbox{$|\braket{\psi|\phi}|=1$} implies they are identical (up to an irrelevant global phase\index{Global phase}\footnote{In quantum mechanics, all quantum states are physically invariant under the introduction of a \textit{global} phase-factor, i.e \mbox{$\ket\psi\to e^{i\theta}\ket\psi$}. Thus, global phases may always be safely ignored without changing the underlying physical reality. This most certainly does not apply to \textit{local} phases, which are a part of physical reality, and critically affect interference dynamics.}). The object $\braket{\cdot|\cdot}$ is referred to as a \textit{braket}\footnote{We never quite appreciated the corny sense of humour of the author of the terms \textit{bra}, \textit{ket}, and \textit{braket}. But somehow it stuck.}\index{Braket}. The normalisation condition from Eq.~(\ref{eq:state_norm_cond}) can therefore equivalently be written,
\begin{align}\index{Normalisation}
\braket{\psi|\psi} = 1.	
\end{align}
Owing to the cyclic property of the trace operator\index{Trace operator}, it follows that the trace,
\begin{align}
\mathrm{tr}(\ket\psi\bra\phi)=\braket{\psi|\phi},	
\end{align}
is a state overlap. This quantity has the useful intuitive interpretation as the \textit{distinguishability}\index{Distinguishability} between quantum states, which plays an important role in measurement theory and the theory of quantum information. It is also widely used to characterise the quality of quantum experiments, as a way of comparing quantum states prepared in the lab with the ideal states obtained from the theory.

Because basis states are orthonormal\index{Orthonormality}, this implies that for basis states $\ket{m}$ and $\ket{n}$ we have the trivial delta function overlap identity,
\begin{align}
	\braket{m|n}=\delta_{m,n}.
\end{align}

%
% Composite Systems
%

\subsubsection{Composite systems}\index{Composite systems}

The joint state vector of a multi-partite system is represented using a tensor\index{Tensor product} (or Kronecker) product Hilbert space, which has dimensionality equal to the product of their individual dimensions\footnote{This is in contrast to the dimensionality of composite classical systems, which grow additively.}, hence yielding the dreaded (cherished!) exponential growth in the complexity of quantum systems,
\begin{align}
\mathcal{H}_{A,B} = \mathcal{H}_A\otimes \mathcal{H}_B,	
\end{align}
where,
\begin{align}
\mathcal{H}_A &\in\mathbb{C}^M\,\,(\mathrm{system}\,A),\nonumber\\
\mathcal{H}_B &\in\mathbb{C}^N\,\,(\mathrm{system}\,B),\nonumber\\
\mathcal{H}_{A,B} &\in\mathbb{C}^{MN}\,\,(\mathrm{composite\,\,system}).	
\end{align}

The state vector of two \textit{independent} subsystems is simply given by their tensor product\footnote{Often the $\otimes$ symbol is omitted for brevity and left implicit.},
\begin{align}
\ket\psi_{A,B} &= \ket\psi_A \otimes \ket\phi_B\nonumber\\
&=\sum_{m,n} \alpha_m\beta_n\ket{m}_A\otimes\ket{n}_B.
\end{align}
This is a so-called \textit{separable state}\index{Separable states}, since the state vector can be factorised into independent terms characterising each of the subsystems -- each subsystem has its own complete and independent physical description in isolation.

%
% Entangled states
%

\subsubsection{Entangled states}\index{Entanglement}

More generally, we may also have states of the form,
\begin{align}
\ket\psi_{A,B} =\sum_{m,n} \lambda_{m,n}\ket{m}_A\otimes\ket{n}_B,
\end{align}
where in general $\lambda_{m,n}$ may not be separable as,
\begin{align}
\lambda_{m,n}=\alpha_m\beta_n.
\end{align}
In this case no complete physical description exists for either subsystem in isolation -- they may only be described collectively. Now the state is said to be \textit{entangled}\index{Entanglement}, a uniquely quantum class of physical states with no classical analogue. The simplest and best-known example is the \textit{Bell state}\index{Bell!States} on two qubits,
\begin{align}\label{eq:bell_state_vec_def}
\ket\Phi^+_{A,B} &= \frac{1}{\sqrt{2}}\begin{pmatrix}
  1\\
  0\\
  0\\
  1
\end{pmatrix}\nonumber\\
&= \frac{1}{\sqrt{2}}(\ket{0}_A\ket{0}_B+\ket{1}_A\ket{1}_B).
\end{align}
It's obvious upon inspection that the Bell state cannot be written in separable form,
\begin{align}
	\ket\Phi^+_{A,B} \neq \ket\psi_A \otimes \ket\phi_B,
\end{align}
and is hence entangled.

Entanglement is one of the key unique features of quantum physics, which underpins the operation of most quantum information processing protocols, including quantum computing in particular, where extremely complex many-qubit entanglement is present and exploited for computational advantage. This immediately hints that in future quantum information systems, where information will be encoded into entangled states in general, information will be highly non-local\index{Non-locality}, in contrast to most standard methods for representing classical information.

%
% Density Operators
%

\subsubsection{Density operators}\index{Density operators}

The state vector formalism presented above applies to \textit{pure states}\index{Pure states}, i.e ones which exhibit perfect quantum coherence\index{Coherence} (i.e superposition), and no classical randomness. But realistic physical systems combine both classical probabilities \textit{and} quantum superposition amplitudes, so-called \textit{mixed states}\index{Mixed states}. To capture both of these features simultaneously we employ \textit{density operators}. For an $N$-dimensional Hilbert space, the density operator, $\hat\rho$, is an \mbox{$N\times N$} complex Hermitian matrix\index{Hermitian matrices} of the form,
\begin{align}
	\hat\rho = \hat\rho^\dag,
\end{align}
which satisfies,
\begin{align}
\mathrm{tr}(\hat\rho)=\sum_i \hat\rho_{i,i} = 1,	
\end{align}
for normalisation.\index{Normalisation}

Thus, in the special case of a qubit, density operators are \mbox{$2\times 2$} complex matrices,
\begin{align}\label{eq:2x2density}
\hat\rho = \begin{pmatrix}
  a & b \\
  b^* & 1-a
\end{pmatrix},
\end{align}
where diagonal elements are necessarily real, and off-diagonal ones may be complex in general.

The intuitive physical interpretation of density matrices is as follows:
\begin{itemize}
\item Rows and columns enumerate the basis states of the Hilbert space.
\item Diagonal elements represent the classical probabilities\index{Classical probabilities} of the respective basis states.
\item Off-diagonal elements represent \textit{coherences}\index{Coherences} between pairs of basis states, i.e whether they exist as a classical probability distribution (when the coherence is zero) or as a quantum superposition (when non-zero). The $\hat\rho_{m,n}$ matrix element (for \mbox{$m\neq n$}) represents the degree of coherence between basis states $\ket{m}$ and $\ket{n}$.
\end{itemize}

Therefore, referring back to the qubit example in Eq.~(\ref{eq:2x2density}), $a$ and \mbox{$1-a$} can be regarded as the \textit{classical} probability amplitudes associated with being in the $\ket{0}$ or $\ket{1}$ basis state, while $b$ characterises to what extent it is a classical mixture of these two states (for \mbox{$b=0$}) or a coherent superposition of them (for \mbox{$|b|=\sqrt{a(1-a)}$}).

The density operator captures \textit{all} information (both classical and quantum) that can be known about a physical system under the standard laws of quantum physics, and is therefore the most general representation for quantum states, and a very powerful representation to employ.

For a pure state (a qubit in this example), the density operator takes the form of a vector outer product\index{Outer product},
\begin{align}
\hat\rho &= \ket\psi\bra\psi\nonumber\\
&= \begin{pmatrix}
  |\alpha|^2 & \alpha\beta^* \nonumber\\
  \alpha^*\beta & |\beta|^2
\end{pmatrix}.
\end{align}

A classical mixture of quantum states $\hat\rho_i$ with probability distribution $p_i$ takes the form,
\begin{align}\label{eq:rho_sum_interp}
	\hat\rho = \sum_i p_i \hat\rho_i,
\end{align}
where the probabilities are normalised such that,
\begin{align}
	\sum_i p_i = 1.
\end{align}
Eq.~(\ref{eq:rho_sum_interp}) has the elegant interpretation that the state $\hat\rho$ is in quantum state $\hat\rho_i$ with classical probability $p_i$.

% Purity

\paragraph{Purity}

An important measure on density operators is their \textit{purity}\index{Purity}, which quantifies to what extent the system is coherent (i.e \textit{not} classically probabilistic),
\begin{align}
\mathcal{P} = \mathrm{tr}(\hat\rho^2),
\end{align}
where,
\begin{align}
\frac{1}{N}\leq \mathcal{P}\leq 1,	
\end{align}
and \mbox{$\mathcal{P}=1$} only for pure states.

%
% Reduced States
%

\subsubsection{Reduced states}\index{Reduced states}

Suppose there exists a bipartite system, to which we only have access to one subsystem. Since in general quantum states are not always separable, we now necessarily have an incomplete physical description of what we have access to. This physical description is obtained by taking the \textit{partial trace}\index{Partial trace} of the joint system, tracing out the subsystem to which we don't have access,
\begin{align}
\hat\rho_A = \mathrm{tr}_B(\hat\rho_{A,B}).	
\end{align}
$\hat\rho_A$ is referred to as the \textit{reduced state} of subsystem $A$, and captures all that can be known about  $A$ in isolation of $B$.

In general, if $A$ and $B$ are entangled, the reduced state of either will provide an incomplete physical description, since it ignores the entanglement between them, capturing only local properties of the state. For this reason, attempting to reconstruct a joint density operator from its reduced density operators, in general does not faithfully reproduce the original joint density operator, unless the two subsystems were perfectly separable to begin with. That is, for,
\begin{align}
	\hat\rho_A &= \mathrm{tr}_B(\hat\rho_{A,B}),\nonumber\\
	\hat\rho_B &= \mathrm{tr}_A(\hat\rho_{A,B}),
\end{align}
in general,
\begin{align}
	\hat\rho_{A,B} \neq \hat\rho_A \otimes \hat\rho_B.
\end{align}

The partial trace is a linear operator defined to satisfy the following properties\index{Partial trace!Properties},
\begin{align}
\mathrm{tr}_B(\hat\rho_A\otimes\hat\rho_B) &= \mathrm{tr}(\hat\rho_B) \cdot \hat\rho_A,\nonumber\\
\mathrm{tr}_B(\ket{i}_A\bra{j}_A\otimes\ket{k}_B\bra{l}_B) &= \braket{k|l}\cdot\ket{i}_A\bra{j}_A,\nonumber\\
\mathrm{tr}_B(\hat\rho+\hat\sigma) &= \mathrm{tr}_B(\hat\rho) + \mathrm{tr}_B(\hat\sigma).
\end{align}

A useful property of the reduced state of a larger pure state is that its purity\index{Purity} characterises how entangled that subsystem is with the peripheral system. For example, a joint system which is separable and pure exhibits reduced states which are also pure. But the reduced state of an entangled state, such as that shown in Eq.~(\ref{eq:bell_state_vec_def}), is necessarily mixed. Thus, the purity $\mathcal{P}$ of the reduced states of a joint system may be used as a metric for quantifying the degree of entanglement in the system. This provides an extremely important link between entanglement and decoherence -- the loss of coherence in an observed quantum system can always be attributed to it becoming entangled to the external environment, to which we don't have access.

%
% Evolution
%

\subsection{Evolution}\index{Evolution}

Thus far we have only provided a description for the state of quantum systems. What about their evolution? We can't execute any algorithm without evolution! Evolution in quantum mechanics has two forms, discrete-time and continuous-time.

% Discrete-time evolution

\subsubsection{Discrete-time evolution}\index{Evolution!Discrete-time}

In quantum mechanics the discrete-time evolution of an $N$-dimensional state is given by arbitrary \mbox{$N\times N$}-dimensional unitary transformations\index{Unitary!Operators}, $\mathrm{U}(N)$, satisfying,
\begin{align}
\hat{U}^\dag\hat{U}=\hat\openone.	
\end{align}

Intuitively, this corresponds to the class of higher-dimensional complex rotations that preserve the orthonormality of quantum states -- i.e just a change of basis in the higher-dimensional Hilbert space. In principle, any unitary operator is physically allowed, although most are challenging to artificially engineer, motivating quantum engineers to typically decompose quantum protocols into \textit{universal gate sets}\index{Universal gate sets} -- a restricted but accessible set of operations that are practical to implement, yet sufficient to constructively build up whatever complex transformations we need.

The evolution of quantum state vectors under unitary transformation is simply given by matrix multiplication,
\begin{align}
\ket{\psi'} = \hat{U}\ket\psi,	
\end{align}
or for density operators by conjugation,
\begin{align}
\hat\rho' = \hat{U}\hat\rho\hat{U}^\dag.	
\end{align}

% Continuous-time evolution

\subsubsection{Continuous-time evolution}\index{Evolution!Continuous-time}

Sometimes an alternate continuous-time description is employed. Here the system is governed by a \textit{Hamiltonian} operator\index{Hamiltonians}, $\hat{H}$, the direct quantum analogue of a classical Hamiltonian, which simply relates to unitary evolution via exponentiation,
\begin{align}
	\hat{U} = e^{-i\hat{H}t},
\end{align}
over evolution time $t$.

% Schroedinger vs Heisenberg pictures

\subsubsection{Schr{\" o}dinger vs Heisenberg pictures}

The above formalism describes quantum evolution by acting the unitary operator on the quantum state directly,
\begin{align}
\ket{\psi'} = \hat{U}\ket\psi.
\end{align}
This is known as the \textit{Schr{\" o}dinger picture}\index{Schr{\" o}dinger!Picture}. An alternate, but entirely mathematically equivalent approach is the \textit{Heisenberg picture}\index{Heisenberg!Picture}, whereby we evolve the operators acting upon a state, rather than the state itself. Now the operator evolution is described by conjugating the operators with the unitary operator describing the evolution. That is, representing the quantum state in terms of some operator $\hat{o}$, the operator evolution proceeds as,
\begin{align}
	\hat{o}' = \hat{U}\hat{o}\hat{U}^\dag.
\end{align}

For example, representing photonic states using creation operators $\hat{a}^\dag$ (introduced in Sec.~\ref{sec:creation_ann_ops}), the evolved creation operator is simply given by $\hat{U}\hat{a}^\dag\hat{U}^\dag$. In quantum optics in particular, where creation operator representations are highly convenient, this is a very useful formulation for describing quantum evolution.

%
% Measurement
%

\subsection{Measurement}\index{Measurement}

The final ingredient in our toolbox of allowed quantum operations is measurement. After all, what use is a quantum protocol if we can't read out the answer?

So-called \textit{projective measurements} in quantum mechanics are described by sets of \textit{measurement projectors}\index{Measurement!Projectors},
\begin{align}
\hat{M}_i = \ket{m_i}\bra{m_i},	
\end{align}
where,
\begin{align}
\sum_i \hat{M}_i = \hat\openone,
\end{align}
for normalisation. Each projector in the set corresponds to a possible allowed measurement outcome. After measuring a given outcome the state is disturbed and collapses\index{Measurement!Collapse} (i.e is \textit{projected}) into being in the state corresponding to the measurement outcome.

If the set of projectors corresponds to the eigenbasis of some unitary operator $\hat{U}$, we can say that the measurement is performed in the $\hat{U}$-basis.

Importantly -- and this is a key feature of quantum mechanics -- we cannot control what the measurement outcome is. It occurs randomly according to some classical probability distribution, given by the overlaps between the measurement projectors and the measured quantum state.

The set of possible projected states following measurement is given by,
\begin{align}\label{eq:meas_proj_def}
\ket{\psi_i} &= \frac{\hat{M}_i \ket\psi}{\sqrt{p_i}},\nonumber\\
\hat\rho_i &= \frac{\hat{M}_i\hat\rho\hat{M}_i^\dag}{p_i},
\end{align}
where those outcomes occur with probabilities,\index{Measurement!Probabilities}
\begin{align}\label{eq:meas_prob_def}
p_i &= \bra\psi \hat{M}_i^\dag \hat{M}_i\ket\psi,\nonumber\\
p_i &= \mathrm{tr}(\hat{M}_i\hat\rho\hat{M}_i^\dag).
\end{align}

Taking the state vector representation from Eq.~(\ref{eq:gen_state_vec}), performing a measurement in the basis in which the vector is expressed (i.e measurement projectors \mbox{$\ket{n}\bra{n}$}), the measurement probabilities are simply given by the absolute squares of the respective amplitudes,
\begin{align}
p_i = |\alpha_i|^2.	
\end{align}

More general classes of measurements are allowed, given by measurement operators $\hat{M}_i$ comprised of linear combinations of projectors. The same rules above for the action of these measurement operators apply, Eqs.~(\ref{eq:meas_proj_def}, \ref{eq:meas_prob_def}).

The described formalism in terms of measurement projectors gives us `singe-shot' information about a measurement -- what is the probability of a given single measurement outcome, and what is the state of the system following this measurement? Alternately, we might be interested not in single-shot events, but statistical measurement information, such as the average measurement outcome in the limit of a large number of repeated independent trials.

We define a measurement \textit{observable}\index{Observables} as an operator $\hat{O}$ acting on a state-space, with eigenvalues $O_i$ over some range of $i$. The eigenvalues $O_i$ denote the allowed values of the measurement outcome for the associated eigenstate, which corresponds to the projected post-measurement state. The statistical expectation value\index{Expectation values} of these measurement outcomes for a given state $\ket\psi$ may now be obtained as the expectation value of the observable,
\begin{align}
	\braket{\hat{O}} = \braket{\psi|\hat{O}|\psi},
\end{align}
and their variance\index{Variance} by,
\begin{align}
	\braket{\hat{O}^2} - \braket{\hat{O}}^2.
\end{align}
Both of these are in the limit of an infinite number of independent trials of the measurement performed on identical copies of the state.

% Example derivation

\subsection{Example derivation}

To illustrate the basic techniques presented above, we now present a derivation of the operation of a very simple quantum protocol comprising state preparation, evolution, and measurement.

Let us begin by preparing a single-qubit state into the logical basis state,
\begin{align}
\ket\phi = \ket{0},
\end{align}
where we have employed Dirac notation for simplicity.

Next we will apply the so-called Hadamard gate\index{Hadamard!Gate}, given by the \mbox{$2\times 2$} unitary matrix,
\begin{align}
\hat{H} = \frac{1}{\sqrt{2}}\begin{pmatrix}
  1 & 1 \\
  1 & -1
\end{pmatrix}.
\end{align}

Applying this gate to our input state yields,
\begin{align}
	\ket\psi &= \hat{H}\ket{0}\nonumber\\
	&= \frac{1}{\sqrt{2}}\begin{pmatrix}
  1 & 1 \\
  1 & -1
\end{pmatrix}\begin{pmatrix}
1\\
0	
\end{pmatrix}\nonumber\\
&=\frac{1}{\sqrt{2}}\begin{pmatrix}
1\\
1	
\end{pmatrix}\nonumber\\
&= \frac{1}{\sqrt{2}}(\ket{0}+\ket{1}),
\end{align}
which is an equal superposition of the two logical basis states.

Applying our measurement identities, we obtain measurement probabilities,
\begin{align}
p_0 = |\braket{0|\psi}|^2 = \frac{1}{2},\nonumber\\
p_1 = |\braket{1|\psi}|^2 = \frac{1}{2},
\end{align}
and respective post-measurement projected states,
\begin{align}
\ket{\psi_0} &= \sqrt{2}\ket{0}\braket{0|\psi} = \ket{0},\nonumber\\
\ket{\psi_1} &= \sqrt{2}\ket{1}\braket{1|\psi} = \ket{1}.
\end{align}

Thus, we see that with 50:50 probability the measurement randomly collapses the measured state onto the $\ket{0}$ or $\ket{1}$ states.
\section{Introduction to quantum optics}\index{Quantum optics}\label{sec:intro_to_QO}

\famousquote{All the fifty years of conscious brooding have brought me no closer to answer the question, `What are light quanta?'. Of course today every rascal thinks he knows the answer, but he is deluding himself.}{Albert Einstein}
\newline

\dropcap{M}{uch} as the classical internet is heavily dependent upon optics to mediate communication, any future quantum internet will almost inevitably rely heavily on quantum optics to mediate quantum communication. It is therefore worth introducing the language and basic set of tools employed by quantum opticians.

%
% Discrete-variables
%

\subsection{Discrete-variables}\index{Discrete-variables}

Like most things in the quantum world, light is discretised into fundamental, indivisible particles called \textit{photons}\index{Photons} -- light quanta\index{Quanta}. This yields the most common representation for quantum states of light, referred to as the \textit{discrete-variables} (DV) picture\index{Discrete-variables}.

% Photon-number states

\subsubsection{Photon-number states}

The most basic optical quantum state comprising photons is the \textit{photon-number state}\index{Photon-number!States} or \textit{Fock state}. These states form a discrete basis, labelled by the integer number of photons in the state,
\begin{align}
\ket{n},\,n\in\mathbb{Z}^+.	
\end{align}
The special case of $\ket{0}$ is referred to as the \textit{vacuum state}\index{Vacuum state}, since it contains no photons.

The Fock states can be thought of as the energy levels in a quantum harmonic oscillator\index{Harmonic oscillators}, and the energy\index{Photons!Energy} of the $n$-photon Fock state is therefore given by,
\begin{align}
E_n = \left(n+\frac{1}{2}\right)\hbar\omega,
\end{align}
where $\omega$ is the optical frequency\index{Optical!Frequency} (in radians per second).

Any optical state in a single mode can be represented in the photon-number basis, which generalises to the multi-mode case in the obvious way, using a basis of composite photon-number states,
\begin{align}
	\ket{\vec{n}} = \ket{n_1}\otimes\ket{n_2}\otimes\cdots\otimes\ket{n_m},
\end{align}
for $m$ optical modes.

% Measurement

\subsubsection{Measurement}

A measurement in the photon-number basis is represented using photon-number projectors\index{Photon-number!Projectors},
\begin{align}
\hat\Pi_n = \ket{n}\bra{n},
\end{align}
which obey the usual completeness relation for measurement operators,
\begin{align}
\sum_{n=0}^\infty \hat\Pi_n &= \hat\openone.	
\end{align}

% Creation & annihilation operators

\subsubsection{Creation \& annihilation operators}\label{sec:creation_ann_ops}

In many cases it's convenient to represent states using photonic \textit{creation} ($\hat{a}^\dag$) and \textit{annihilation} ($\hat{a}$) operators\index{Creation operators}\index{Annihilation operators}. These (non-commuting) operators have the effect of acting upon a photon-number state and incrementing or decrementing its photon-number. Note that these operators are not unitary, and therefore do not represent legitimate quantum evolutions on their own. However, they may be employed to construct unitary operators representing physically legitimate evolution processes.

These operators satisfy the following basic algebraic properties:
\begin{align}
\hat{a}^\dag\ket{n} &= \sqrt{n+1}\ket{n+1},\nonumber\\
\hat{a}\ket{n} &= \sqrt{n}\ket{n-1},\nonumber\\
\hat{a}\ket{0} &= 0.
\end{align}
From this, it follows that an arbitrary Fock state can be expressed in terms of creation operators as,
\begin{align}
\ket{n} = \frac{1}{\sqrt{n!}}(\hat{a}^\dag)^n\ket{0}.
\end{align}
The creation and annihilation operators obey the commutation relation\index{Commutation relations},
\begin{align}
	[\hat{a},\hat{a}^\dag] &= 1,
\end{align}
where,
\begin{align}
[\hat{A},\hat{B}]=\hat{A}\hat{B}-\hat{B}\hat{A},
\end{align}
is the \textit{commutator}\index{Commutators} of two operators (operators that commute obviously necessarily exhibit \mbox{$[\hat{A},\hat{B}]=0$}). For multiple modes ($i$ and $j$) the different creation operators commute, as do the respective annihilation operators,
\begin{align}
[\hat{a}^\dag_i,\hat{a}^\dag_j] &= 0,\nonumber\\
[\hat{a}_i,\hat{a}_j] &= 0.
\end{align}
The physical interpretation of this property is that because photons are bosons, they are exchange-symmetric, and thus the physical state is invariant under swapping them.

% Photon-number operators

\subsubsection{Photon-number operators}\index{Photon-number!Operators}

A particularly useful and ubiquitous operator is the photon-number operator, defined as,
\begin{align}
\hat{n}=\hat{a}^\dag\hat{a},
\end{align}
which satisfies the eigenvalue relation\index{Eigenvalue equations} with photon-number states,
\begin{align}
\hat{n}\ket{n} = n\ket{n}.	
\end{align}
Treating it as a measurement observable\index{Observables}, the expectation value\index{Expectation values} of the photon-number operator, $\braket{\hat{n}}$, gives us the average measured photon-number when a state is measured in the photon-number basis, yielding the statistical average of the photon-number measurement outcome.

%
% Continuous-variables
%

\subsection{Continuous-variables}\index{Continuous-variables}

A completely alternate, but entirely equivalent formalism for the representation of quantum optical states is in the \textit{continuous-variables} (CV) picture. Here we no longer think in terms of a discretised photon-number basis, but in terms of a continuous basis in the complex plane, referred to as \textit{phase-space}\index{Phase-space}, completely analogous to the familiar classical phase-space. This alternate representation is introduced purely as a matter of convenience, since some states (typically referred to as CV states), are particularly well-suited to elegant representation in this picture, as are some types of evolution.

% Position & momentum operators

\subsubsection{Position \& momentum operators}

In the CV picture, instead of representing states using photonic creation operators, we express them in terms of the \textit{position} ($\hat x$)\index{Position operator} and \textit{momentum} ($\hat p$)\index{Momentum!Operator} operators, which are related to the creation and annihilation operators via,
\begin{align}
\hat x &=    \sqrt{\frac{\hbar}{2 \omega}}(\hat a + \hat a^\dag), \nonumber \\
\hat p &= -i \sqrt{\frac{\hbar  \omega}{2}}(\hat a - \hat a^\dag), 
\end{align}
where $\omega$ is the optical frequency\index{Optical!Frequency}. These operators obey the commutation relation\index{Commutation relations},
\begin{align}
[\hat x, \hat p] = i \hbar,
\end{align}
and satisfy the eigenvalue relations with the position and momentum eigenstates,
\begin{align}
\hat{x}\ket{x} &= x\ket{x},\nonumber\\
\hat{p}\ket{p} &= p\ket{p}.	
\end{align}
These eigenstates satisfy the completeness relations,
\begin{align}
\int_{-\infty}^\infty \ket{x}\bra{x}\,dx &= \hat\openone,\nonumber\\
\int_{-\infty}^\infty \ket{p}\bra{p}\,dp &	= \hat\openone.
\end{align}

The position and momentum operators represent the \textit{quadratures}\index{Quadratures} of a mode, and correspond to the real and imaginary components of a harmonic oscillator's\index{Harmonic oscillators} amplitude.

% Position & momentum representations

\subsubsection{Position \& momentum representations}\index{Position representation}\index{Momentum!Representation}

The \textit{position representation} of an optical state corresponds to expressing a state vector $\ket{\psi}$ in the position basis\index{Position representation},
\begin{align}
\ket{\psi} = \int \psi(x)\ket{x}\, dx.
\end{align}

Here the wave-function\index{Wave-functions} in the position representation\index{Position representation} is defined as,
\begin{align}
\psi(x) = \braket{x|\psi}.
\end{align}
The state can similarly be expressed via a \textit{momentum representation}\index{Momentum!Representation},
\begin{align}
	\ket\psi = \int {\tilde\psi}(p) \ket{p}\, dp,
\end{align}s
with wave-function\index{Wave-functions},
\begin{align}
	{\tilde\psi}(p) = \braket{p|\psi},\nonumber\\
\end{align}

The quadrature eigenstates\index{Quadratures!Eigenstates} are mutually related to one another by a Fourier transform\index{Fourier transform},
\begin{align}
\ket{x} &= \frac{1}{\sqrt{\pi}} \int_{-\infty} ^\infty e^{-2 i x p} \ket{p} \,dp,\nonumber \\
\ket{p} &= \frac{1}{\sqrt{\pi}} \int_{-\infty} ^\infty e^{2 i x p} \ket{x} \,dx.
\end{align}

% Coherent states

\subsubsection{Coherent states}\index{Coherent states}

A particularly ubiquitous state that emerges in CV representations is the \textit{coherent state}\index{Coherent states}, which closely approximates classical laser light\index{Lasers},
\begin{align}\index{Coherent states}
\ket{\alpha} = e^{-\frac{|\alpha|^2}{2}} \sum_{n=0}^\infty \frac{\alpha^n}{\sqrt{n!}} \ket{n},
\end{align}
where \mbox{$\alpha\in\mathbb{C}$} is the complex coherent amplitude. Alternately, coherent states can be defined as the eigenstates of the annihilation operator,
\begin{align}
\hat{a}\ket{\alpha} = \alpha\ket{\alpha}.
\end{align}
Writing the coherent amplitude as,
\begin{align}
	\alpha=re^{i\theta},
\end{align}
$r$ can be interpreted as the strength of the classical field, and $\theta$ as its local phase.

Note that the coherent state has indeterminate photon-number (and hence energy) -- it is in fact a coherent superposition of \textit{all} photon-numbers (except when \mbox{$\alpha=0$}). However, the \textit{mean} photon-number\index{Coherent states!Mean photon-number} is a useful quantity, given by,
\begin{align}
\bar{n} = \braket{\hat{n}} = |\alpha|^2.
\end{align}

% Phase-space representations

\subsubsection{Phase-space representations}\index{Phase-space}

We can graphically represent phase-space on the complex plane, where the two axes denote expectation values of the $\hat{x}$ and $\hat{p}$ operators. Unlike classical phase-space, in quantum phase-space states are `blobs'\index{Blobs} rather than points. The variance of these blobs represents the measurement uncertainty in a given direction, which is strictly non-zero in the quantum world, owing to the Heisenberg uncertainty principle\index{Uncertainty principle}.

\begin{figure}[!htbp]
\includegraphics[clip=true, width=0.4\textwidth]{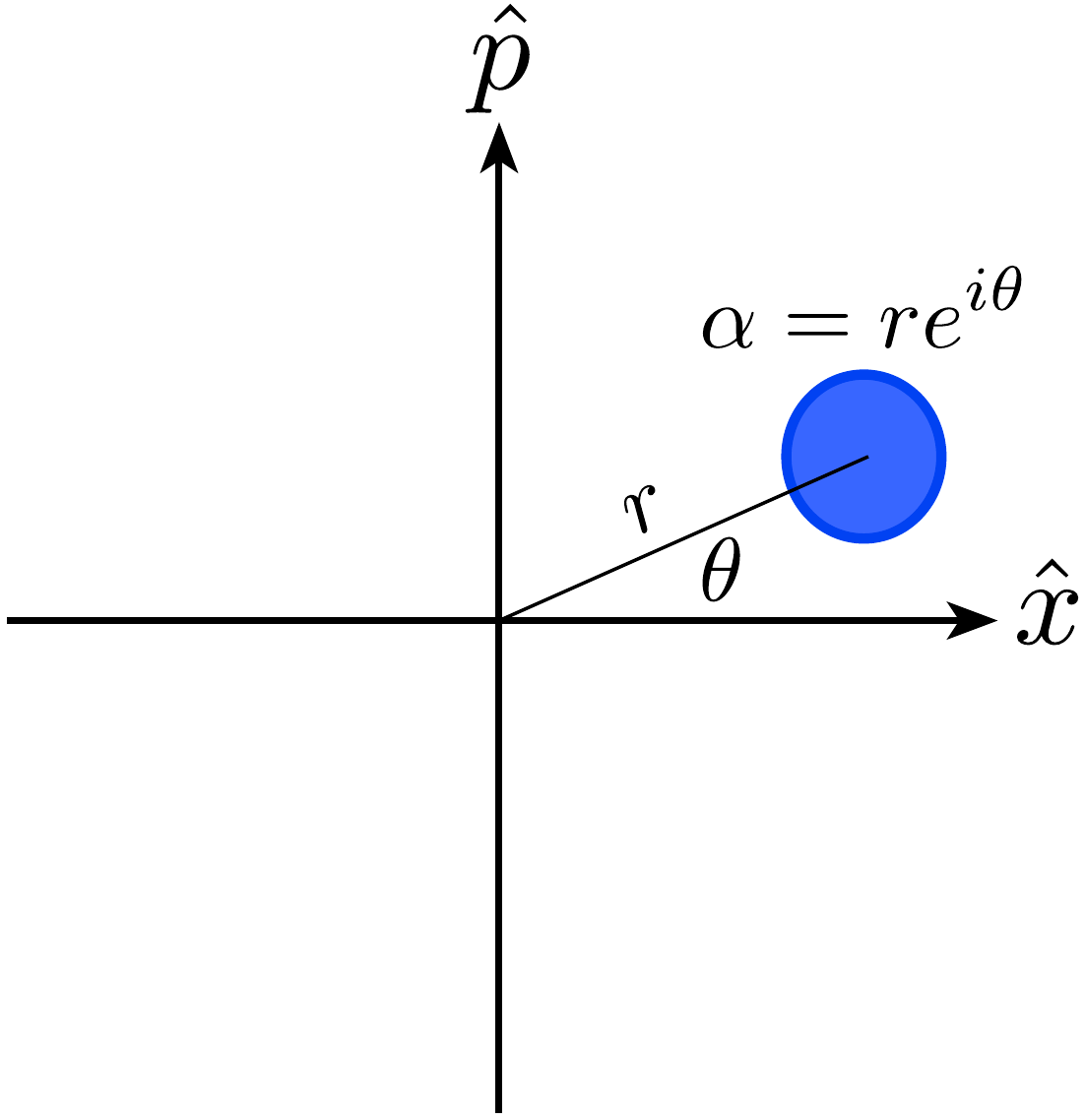}
\captionspacefig \caption{Phase-space representation for a coherent state of complex amplitude \mbox{$\alpha=re^{i\theta}$}, where $r$ can be regarded as the field strength, and $\theta$ its phase.}\label{fig:phase_space}
\end{figure}

The phase-space representation for a coherent state is a circular blob with mean $\alpha$ in the complex plane\index{Complex plane}, shown in Fig.~\ref{fig:phase_space}. Its variance is indicative of the quantum uncertainty\index{Uncertainty} of the field in the two quadrature directions. A coherent state is a \textit{minimum uncertainty state}\index{Minimum uncertainty state}, where the product of the variances in the two orthogonal directions is minimised. For other states it will be greater in general, but strictly not less.

Beyond coherent states, represented by circular blobs, all manner of other geometries are available, allowing states such as \textit{squeezed states}\index{Squeezed states} (another useful type of CV state), which are rather complex to work with in the DV picture, to be elegantly graphically represented.

In phase-space, the evolution of a state under the accumulation of optical phase\index{Phase-space!Evolution} simply manifests itself as a rotation of the complex plane around the origin over time. Thus, the coherent state blob simply revolves around the origin. On the other hand, photon-number states manifest themselves as concentric rings in phase-space. Since this geometry is circular-symmetric, its phase-space representation is invariant under phase evolution.
%
% Introduction to quantum information theory
%
% \clearpage

\section{Introduction to quantum information theory}\index{Quantum information theory} \label{sec:quant_inf_th}

\famousquote{When you find yourself in a room surrounded by your enemies you tell yourself, `I am not locked in here with you, you are locked in here with me'. This is the kind of mindset you should have if you want to succeed in life. Get rid of that victim mentality.}{Bruce Lee}

\subsection{Probability, information \& classical correlation measures}

% The fundamental unit in Shannon information is the \textit{bit}. 

``Information theory begins with the observation that there is a fundamental link between probabilities and information'' \cite{barnett2009quantum}. 

Suppose that there are two parties trying to communicate, conventionally named Alice and Bob. Alice sends information using an alphabet $\{x_j \}$, and each letter occurs with probability $p_j$. If some letter occur less likely than others, then Bob will be more surprised receiving those, and one might naturally expect that the information content of those events are higher.

This suggests a way to measure the \textit{information content}\index{Information content} of an outcome, or how surprised Bob is,
\begin{align}
i(x_j) \equiv \log\left(\frac{1}{p(x_j)}  \right) =  -\log\left(p(x_j)  \right)
\end{align}

We can model Alice's source as a random variable $X$, whose outcomes are $\{ x_j\}$ each occurring with probability $p_j$. Then the expected information content of the source, or \textit{Shannon entropy}\index{Shannon entropy} associated with the $X$ is defined as,

\begin{align}\index{Shannon entropy}
H(X) = -\sum_i p_i\log_2(p_i).
\end{align}

Entropy plays a central role in information theory. The intuitive interpretation is that it quantifies an experimenter's uncertainty about $X$ before measuring it, and his expected information gain is $H(X)$ bits upon learning the outcome. The information $H(X)$ is zero if and only if one of the probabilities $p(x)$ is unity, with the others being zero. In this case the value of $X$ is already known and so there is no information to be gained from observing it.  

 If there are two random variables $X$ and $Y$, the joint entropy of the two is simply given by,
\begin{align}\index{Joint Shannon entropy}
H(X,Y) =  -\sum_{x,y} p_{x,y}\log_2(p_{x,y}).
\end{align}

Note that the joint entropy $H(X,Y)= H(X)+H(Y)$ if and only if $X$ and $Y$ are independent, i.e. the occurrence of one does not change the probability of the other.  

If the two distributions are not independent, i.e. knowing something about $X$ reveals information on $Y$, the two variables are said to be \textit{correlated}. Suppose Alice possesses the random variable $X$ and Bob has the random variable $Y$, the \textit{mutual information} specifies the number of bits in common between the two distributions. Equivalently, this represents the maximum number of bits that one party can learn about the other just by inspecting their own information.

For two classical distributions, the classical mutual information is given by,
\begin{align}\index{Mutual information}
I(X;Y) = H(X) + H(Y) - H(X,Y),
\label{eq:classical_mutual_info}
\end{align}
a measure of the correlation between the events $X$ and $Y$. The quantity in Eq.~\eqref{eq:classical_mutual_info} is important because it upper bounds the amount of information Alice and Bob can reliably communicate. For a discrete random variable with $d$ possible outcomes, the maximum mutual information between $X$ and $Y$ is $I(X,Y)=\log_2 d$ bits.

Another quantity of interest is the \textit{conditional entropy}, which describes the entropy of $Y$ should $X$ be known. The entropy of $Y$ conditioned on $X$ is given by,
\begin{align}
H(X|Y) &= H(X,Y)- H(Y) \nonumber\\
       &=- \sum_{x,y} p(x,y) \log_2 \frac{p(x,y)}{p(x)}.
\end{align}
Classically the conditional entropy is always positive, but as we will see later, this is not the case if quantum states are involved.

\begin{figure}[!htbp]
	\includegraphics[clip=true, width=0.475\textwidth]{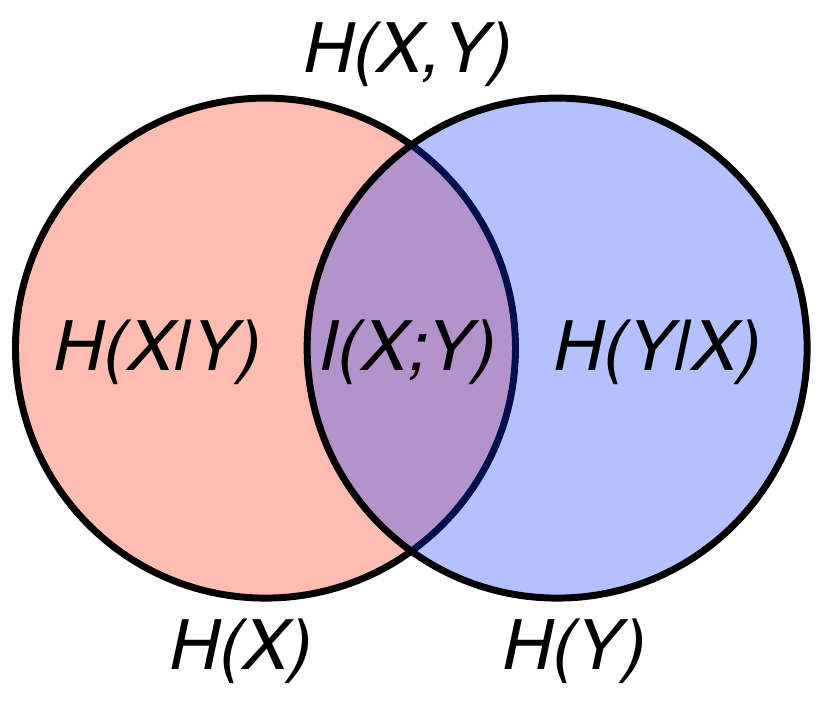}
	\captionspacefig \caption{Venn diagram showing the relationship between some elementary information theoretic entropy relations.} \label{fig:mutual_info}
\end{figure}

The relationship between the quantities mentioned above is graphically summarised in Fig.~\ref{fig:mutual_info}.

\subsection{Quantum correlation measures}

The von Neuman entropy\index{von Neuman entropy} \cite{bib:bengtsson2017geometry} for quantum density operators, $S(\hat\rho)$, is defined analogously, replacing probabilities with density operator eigenvalues,
\begin{align}\index{von Neuman entropy}
S(\hat\rho) &= - \sum_x \lambda_x \log_2 (\lambda_x) \nonumber \\
&= -\mathrm{tr}(\hat\rho\,\log \,\hat\rho),
\end{align}
where $\{\lambda\}$ is the eigenvalue spectrum of $\hat\rho$. This modification is logically justified, as the eigenvalues can be interpreted directly as a purely classical probability distribution of orthogonal states when the density operator is transformed into a basis with no coherences between basis states (i.e a diagonal basis or spectral decomposition\index{Spectral decomposition}). In that case, the Shannon and von Neuman entropies essentially have identical physical interpretations.

There are a few properties of the von Neumann entropy which are often usefuL:
\begin{enumerate}
	\item For pure states $\hat \rho = \ket{\psi}\bra{\psi}$,
		\begin{align}
			S(\hat\rho) = 0.
		\end{align}
	\item The von Neumann entropy is invariant under change of basis transformations (i.e conjugation by a unitary operator), 
			\begin{align}
			S(U\hat \rho U^\dagger) = S(\hat \rho)
			\end{align}
	\item For a state living in a Hilbert space of dimension $d$,
		\begin{align}
			S(\hat\rho) \leq \log_2 d,
		\end{align}
	with equality being achieved when the quantum state is maximally mixed.
	\item For a given bipartite system $AB$,
			\begin{align}
			S(\hat \rho_{AB}) \leq S(\hat \rho_A) + S(\hat \rho_B),
			\end{align}
			where equality holds if $\hat\rho_{AB}= \hat\rho_A \otimes \hat\rho_B$.
\end{enumerate}

Analogously, the quantum mutual information\index{Quantum mutual information} for bipartite state $\hat\rho_{A,B}$ is defined as,
\begin{align}
I(A;B)_{\hat\rho} = S(\hat\rho_A) + S(\hat\rho_B) - S(\hat\rho_{A,B}),
\end{align}
using the von Neuman entropy.

The mutual information between two quantum states is invariant under local unitary transformation,
\begin{align}
I(A;B)_{\hat\rho} = I(\hat{U}_A\hat\rho_A \hat{U}_A^\dag; \hat{U}_B\hat\rho_B \hat{U}_B^\dag),
\end{align}
since the eigenvalue spectrum of a density operator is invariant under unitary transformations. Therefore, the mutual information represents the maximum amount of information Bob can learn about Alice's state under \textit{any} local operations.

Now, in sharp contrast with the classical case, the conditional entropy for quantum states,
\begin{align} \label{cond_quant_ent}
H(A| B)_{\hat \rho} = S(\hat \rho_{AB}) - S(\hat \rho_B),
\end{align}
can be negative! If $\hat\rho_{AB}$ is a maximally entangled state, then $S(\hat \rho_{AB}) =0$, and $H(A|B) =- S(\hat \rho_{AB}) <0$. Intuitively, this can be interpreted as the fact that bipartite maximally entangled states can be more strongly correlated than classically possible, and that knowing the state of $A$ reveals an `unnatural' amount of information about the state of $B$.

\begin{figure}[!htbp]
\includegraphics[clip=true, width=0.3\textwidth]{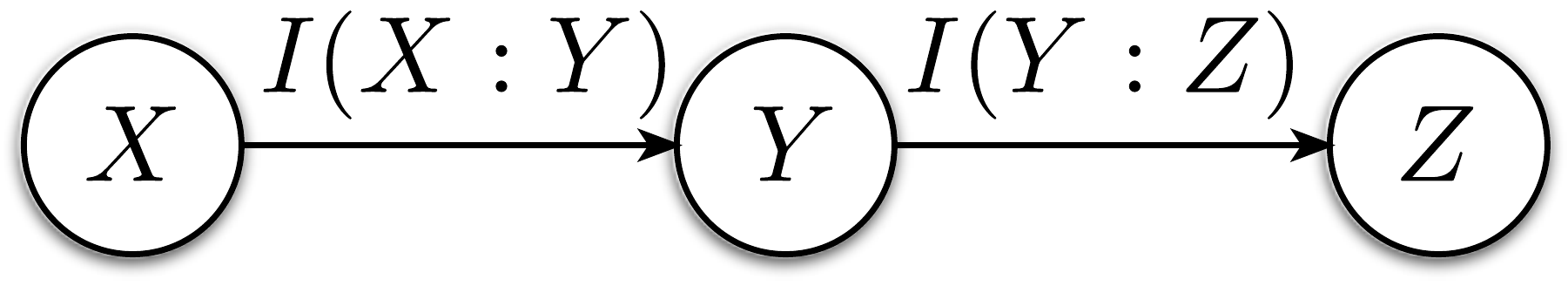}
\captionspacefig \caption{\label{fig:data_proc_ineq}A sequence of events \mbox{$X\to Y\to Z$}. The data processing inequality states that the mutual information from beginning to end is upper-bounded by the mutual information between neighbouring stages, as per Eq.~(\ref{eq:data_proc_ineq}).}	
\end{figure}

A quantum process cannot increase the mutual information between two parties. This yields the \textit{data processing inequality}\index{Data processing inequality} that, for a sequence of channels \mbox{$X\to Y\to Z$},
\begin{align}\index{Data processing inequality}\label{eq:data_proc_ineq}
I(X:Z)&\leq I(X:Y), \nonumber \\
I(X:Z)&\leq I(Y:Z),
\end{align}
with equality if and only if the channel not specified in the identity on the right hand side (\mbox{$Y\to Z$} or \mbox{$X\to Y$} respectively) is unitary, i.e one of the links in the chain perfectly preserves information content. The progression is shown in Fig.~\ref{fig:data_proc_ineq}.

The mutual information is defined as being between a particular known pair of states. Of course, in a quantum channel we have the flexibility to manipulate the input state and the measurement at the output. From this, we can then define the \textit{classical information of the channel}\index{Classical information of the channel}, $I_c(\hat\rho_{A,B})$ --- the maximum of the mutual information, optimised over \textit{all} possible measurement settings. Suppose $\hat\rho_{A,B}$ is shared between Alice and Bob, from which they would like to extract maximal correlations (i.e joint information). They measure using local measurement bases $\{\Lambda_A^x\}$ and $\{\Lambda_B^x\}$, yielding random variables $A$ and $B$. The quantum mutual information of the channel $\mathcal{E}$ is defined as,
\begin{align} \label{eq:channel_quant_mutual}
I_c(\mathcal{E}) = \max_{\rho,\{\Lambda_A^x\},\{\Lambda_B^x\}} I(A;B)_\rho. 
\end{align}

Eq.~\eqref{eq:channel_quant_mutual} quantifies the maximum amount of correlation that can be established between input and output states that can be achieved across the channel. This dictates the physical upper bound on the achievable bitrate (classical communication) or bandwidth of the channel.

All of the correlations discussed above concern classical information. Whilst classical correlation measures have a clear operational meaning, quantum channel capacities less so --- these quantities can be interpreted as how well the channel preserves the quantum state.

That the conditional entropy for quantum states can be negative is significant. In fact, it has its own name and notation, the \textit{coherent information},
\begin{align}
I(A\rangle B) = H(B)_{\hat\rho} - H(A,B)_{\hat\rho},
\end{align}
For a maximally entangled state, the coherent information is equal to one. It measures to what extent we know less about a subsystem compared to the whole.

%
% Channel Capacity
%

\subsection{Channel capacity} \label{sec:channel_cap} \index{Channels!Capacity}

The measures considered until now have quantified the preservation of quantum states. Alternately, one might consider information theoretic measures\index{Information theoretic!Measures}, which quantify the number of bits/qubits transmitted by a link, i.e the number of bits/qubits in common before and after the channel. This is an extremely powerful tool as it upper bounds the amount of information the receiver can extract from the transmitter under \textit{any} measurement scheme, very useful in a cryptographic context, where we want security to be attack-independent\index{Information-theoretic!Security} (Sec.~\ref{sec:comp_vs_inf_th_sec}).

Suppose $\hat\rho_{A,B}$ is shared between Alice and Bob, from which they would like to extract maximal correlations (i.e joint information). They measure using local measurement bases $\{\Lambda_A^x\}$ and $\{\Lambda_B^x\}$, yielding random variables $A$ and $B$. Recall from Sec.~\ref{sec:quant_inf_th} that the quantum mutual information of the channel is defined as,

\begin{align} \label{eq:quant_mut}
I_c(\hat\rho_{A,B}) = \max_{\rho,\{\Lambda_A^x\},\{\Lambda_B^x\}} I(A;B).
\end{align}

The intuitive interpretation is that this is the maximum mutual information between input and output states that can be achieved across the channel. This effectively places a physical upper bound on the achievable bitrate or bandwidth of the channel.

\noindent We are then led to the definition of the \textit{classical channel capacity} of a quantum channel $\mathcal{E}$,
\begin{align}\index{Classical!Channels!Capacity}
\mathcal{C}(\mathcal{E}) = \lim_{n \rightarrow \infty} \frac{I_c(\mathcal{E}^{\otimes n})}{n},
\end{align}
which can be interpreted as the maximum bitrate of the channel, per use of the channel, in the limit of an infinite number of uses, where entangled between multiple applications is allowed in general.

 For a noisy channel, if the input contains product states as well as separable measurements, the classical capacity is equal to Eq.~(\ref{eq:quant_mut}), that is, the single-use and the asymptotic classical capacities are equal.

If we also allow joint measurement (but not entanglement between input states), the classical channel capacity can be expressed as,
\begin{align}
\chi(\mathcal{E}) = \max_{p_i,\hat\rho_i} S(\hat\rho) - \sum_i p_i S(\hat\rho_i), \label{eq:holevoinfo}
\end{align}
known as the Holevo bound\index{Holevo bound}\index{Holevo bound}.

However, if entangled input states are allowed in conjunction with joint measurement setting, the Holevo bound no longer holds, and the asymptotic classical channel capacity,
\begin{align}
\chi(\mathcal{E}) < \mathcal{C(E)}.
\end{align}

It has been shown that using entangled input states in conjunction with joint measurements can increase the amount of classical information which can be transmitted over a noisy quantum channel. In this case, 
\begin{align}
C_\text{ent} = \lim _{n\to \infty} \frac{1}{n}\chi(\mathcal{E}^{\otimes n}).
\label{eq:ent_ent}
\end{align}

The private classical capacity\index{Private channel capacity} $\mathcal{P(E)}$ of a quantum channel $\mathcal{E}$ describes the maximum rate at which the channel is able to send classical information through the channel reliably and privately, that is, the trusted parties do not want the environment to have access to their classical correlations.

Here Alice prepares the quantum state $\hat\rho_x^A$ according to the probability distribution $\{p_X(x)\}$. The expected density operator of the ensemble is of the form,
\begin{align}
\hat\rho^{XA} \equiv \sum_x p_X(x)\ket{x}\bra{x}  \otimes p_x^A.
\end{align}

The difference between the amount of classical correlation that she can establish with Bob, minus the correlation obtainable by Eve is given by,
\begin{align}
I(X;B)_\rho - I(X;E)_\rho,
\end{align}
which leads to the definition of the \textit{private information of a quantum channel}, where Alice maximises over all of her inputs,
\begin{align}
I_\mathcal{P} = \max_{\rho^{XA}} \{I(X;B)_\rho - I(X;E)_\rho \}.
\end{align}

The private classical capacity is defined as,
\begin{align}
\mathcal{P(E)} = \lim_{n\rightarrow \infty} \frac{1}{n} I_\mathcal{P}(\mathcal{E}^{\otimes n })
\end{align}

The last capacity in regards to classical communication over quantum channel is the \textit{entanglement-assisted classical capacity}\index{Entanglement-assisted classical capacity}, $\mathcal{C}_E(\mathcal{E})$. This quantity measures the classical information that can be transmitted through the channel, if Alice and Bob possess shared entanglement prior to the transmission,
\begin{align}
\mathcal{C}_E(\mathcal{E}) = \max_{p_i,\hat\rho_i} I(A:B).
\end{align}

The difference between $\mathcal{C(E)}$ and $\mathcal{C}_E(\mathcal{E}) $ is that $\mathcal{C}_E(\mathcal{E}) $ is equal to the maximised quantum mutual information (i.e additivity holds), and is equal to the single-use version of $\mathcal{C}_E(\mathcal{E})$. This implies that shared entanglement does not change the additivity of quantum mutual information.

\subsection{Quantum channel capacity}

The quantum coherent information exhibits much of the same mathematical structure as the classical mutual information. And analogously,
\begin{align}\index{Quantum channels!Capacity}
\mathcal{I}_Q(\mathcal{E}) = \max_{\hat\rho} I_\text{coh}(\hat\rho,\mathcal{E}).
\label{eq:mutual_info_quantum_single}
\end{align}

The \textit{quantum channel capacity}\index{Quantum channel capacity} is then analogously defined as,
\begin{align}\label{eq:quantcohinfo}
Q(\mathcal{E}) = \lim_{n \rightarrow \infty} \frac{\mathcal{I}_Q(\mathcal{E}^{\otimes n})}{n}.
\end{align}

The most important distinction between quantum mutual information and quantum coherent information is that the mutual information is always additive, but the quantum coherent information is not.

Analytic solutions to Eq.~(\ref{eq:quantcohinfo}) are notoriously difficult to calculate. However, once net dephasing or depolarisation rates have been calculated across a route, the \textit{single-use} quantities can be calculated, which can serve as a lower bound to the ultimately achievable rates, if a number representing \textit{cost}, rather than an analytic solution, is all we need.

A summary of the different measure of classical and quantum capacities are given in Tab.~\ref{tab:capacities}.

\startnormtable
\begin{table*}[!htbp]
\begin{tabular}{ |c | c | c| c| } \hline
Capacity & Type of information & Correlation measure &Capacity formula  \\ \hline
Classical &Classical information &Holevo information &Eq.~(\ref{eq:holevoinfo})       \\
Private & Classical information & Private information  & See \cite{bib:PhysRevLett.103.120501} \\
Entanglement-assisted classical & Classical information & Quantum mutual information & See \cite{bib:PhysRevLett.83.3081} \\
Quantum &Quantum information &Quantum coherent Information &Eq.~(\ref{eq:quantcohinfo}) \\ \hline
\end{tabular}
\captionspacetab
\caption{\label{tab:capacities} Measure of classical and quantum channel capacities. Taken from \cite{bib:8242350}.}
\end{table*}
\startalgtable

We now discuss the quantum capacity for the erasure channel and the amplitude damping channel, both of which the quantum capacities can be calculated.

The erasure channel acts on a state as follows is,
\begin{align}
\hat\rho \rightarrow (1-\eta) \hat\rho + \eta\ket{e}\bra{e}
\end{align}
where $\eta$ is the probability of losing the state and $\ket{e}\bra{e}$ is orthogonal to $\rho$. Given a system of dimension $d_A$, For $\eta \in [1/2,1]$, the quantum capacity is zero. For $\eta \in [0,1/2]$, the quantum capacity is $(1-2\eta)\log d_A.$

For an amplitude damping channel with damping parameter $\gamma$, the Kraus operators can be written as,
\begin{align}
A_0 &= \ket{0}\bra{0} + \sqrt{1-\gamma}\ket{1}\bra{1},\nonumber\\
A_1 &= \sqrt{\gamma}\ket{1}\bra{1}.
\end{align}

For $\gamma \in [0,1/2]$,the quantum capacity is equal to,
\begin{align}
\max_{p \in [0,1]} h_2 [(1-\gamma)p] - h_2[\gamma_p],
\end{align}
where $h_2$ represents the binary entropy function. For $\gamma \in [1/2,1]$, the quantum capacity is equal to zero.

Two of the most surprising discoveries of quantum Shannon theory is the \textit{superadditivity}\index{Superadditivity} of coherent information for the depolarising channel, and \textit{superactivation} of quantum capacity, where two channels whose individual capacities are zero can be combined to make a channel with non-zero capacity.

\subsection{Superadditivity of coherent information}

For qubits, the depolarising channel transmits its input with probability $(1-p)$ and replaces it with the maximally mixed state $\hat\openone/2$ with probability $p$,
\begin{align}
\rho \rightarrow (1-p) \rho + p \frac{\openone}{2}.
\end{align}
It is known that its coherent information is strictly superadditive when $p$ is large.

\begin{align}
I_\mathcal{Q}(\mathcal{E}) < 5 I_\mathcal{Q}(\mathcal{E}^{\otimes 5}).
\end{align}

It can be shown that the state which maximises the coherent information is the Bell state $\frac{1}{\sqrt2}(\ket{00}_{AA_1} + \ket{11}_{AA_1} )$, and the one-use coherent information is,
\begin{align} \label{eq:qcohdepol}
I_\mathcal{Q} = 1+ \left( 1-\frac{3p}{4} \right )\log\left[1-\frac{3p}{4}\right] + \frac{3p}{4}\log\left[\frac{p}{4}\right].
\end{align}

At $p = 0.25$, Eq.~\eqref{eq:qcohdepol} is equal to zero, but if we calculate the coherent information for a five-qubit repetition code,
\begin{align}
\frac{1}{\sqrt2}&(\ket{00}_{AA_1} + \ket{11}_{AA_1} ) \rightarrow \nonumber \\ 
\frac{1}{\sqrt2}&(\ket{000000}_{AA_1A_2 A_3 A_4 A_5} + \ket{111111}_{AA_1A_2 A_3 A_4 A_5}),
\end{align}
at $p$ just above $0.25$, the quantity of interest is positive \cite{bib:PhysRevA.57.830}. This demonstrates that the quantum capacity of the depolarising channel is superadditive. Despite the fact that the channel looks simple, the best strategy for achieving the quantum capacity of the depolarising channel remains very poorly understood \cite{bib:wilde2013quantum}.

\subsection{Superactivation}\index{Superactivation}

Suppose that Alice is connected to Bob by two quantum channel $\mathcal{E}_1,\mathcal{E}_2$ both with zero quantum capacity. Intuitively one would expect that Alice would not be able to transmit quantum information over the channel $\mathcal{E}_1\otimes \mathcal{E}_2$. However, examples of two zero-capacity channels are known to have non-zero joint quantum capacity. The phenomenon is known as \textit{superactivation}.

One of the channels is a 50\% erasure channel (which we know to have zero quantum capacity from the non-cloning theorem), and the other is known as an entanglement-binding channel \cite{bib:horodecki2001separability}.

This phenomenon has important implications for quantum information transmission. It implies that if there are other seemingly useless channels available, one could transmit more quantum information than if the channels were used alone. It also implies that the coherent information is strongly non-additive, and there is much to be understood about reliable communication rates over quantum channels.

\latinquote{Alis volat prorilis.}

\sketch{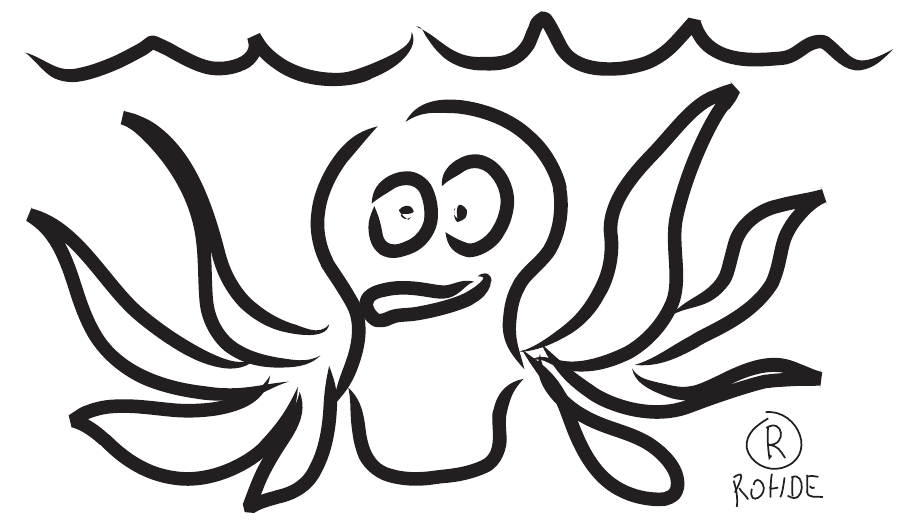}

\clearpage
% %
% Classical networks
%

\part{Classical networks}\label{part:class_net}\index{Classical networks}

%
% Classical Networking Protocols
%

%\sectionby{Darryl Veitch \& Peter Rohde}

\famousquote{Mobile phones should be left to the kids. They're the only ones who can operate them.}{Anthony Hincks}
\newline

\dropcap{T}{o} set the context for our upcoming treatment of quantum networks, we begin by discussing \textit{classical} networks, and some of the key protocols behind their operation.

\section{Classical networking protocols} \label{sec:classical_nets} \index{Classical networking protocols}

\dropcap{T}{here} have been numerous approaches employed in the past for sharing communications links between multiple users\index{Shared communication channels}. This includes:
\begin{itemize}
	\item Channel-switching: an entire communications channel is designated for exclusive use by a given user. \index{Channel-switched networks}
	\item Packet-switching: data is divided into packets, which are routed independently by the network, being reconstructed by the recipient once all packets have been received.\index{Packet!Switching}
	\item Time- or frequency-multiplexing: each user is designated a particular frequency spectrum or series of time-slots exclusively for their use. \index{Time-multiplexing}\index{Frequency!Multiplexing}
	\item Code Division Multiple Access (CDMA): all users can broadcast over a channel simultaneously, and the construction of the coding technique enables demultiplexing of the distinct signals, despite their interference with one another.\index{Code Division Multiple Access (CDMA)}
	\item Ethernet: all users are free to broadcast\index{Broadcast networks} over a shared channel at will, and \textit{collision detection}\index{Collisions!Detection} identifies when packets interfere, after which they are discarded and rebroadcast following a random waiting period, repeating until success.\index{Ethernet}
\end{itemize}

Nowadays packet-switched networks have become the norm in most digital networks, as they facilitate far greater efficiency in the use of network bandwidth, and are more easily scaled to greater numbers of users in a dynamic and ad hoc manner. It is foreseeable the same trend will continue with quantum technologies, especially given their initial high cost, where maximising network utility is paramount.

In this work we will focus on packet-switched networks when we later introduce our quantum networking protocols. However, with sufficient flexibility in the design of our upcoming quantum protocols, packet-switched networks can easily be made to effectively implement channel-switched, or time-/frequency-multiplexed communication.

%
% TCP/IP
%

\subsection{TCP/IP} \index{Transmission Control Protocol/Internet Protocol (TCP/IP)}

The present-day internet is built on top of a protocol stack comprising primarily the Internet Protocol (IP), User Datagram Protocol (UDP), and Transmission Control Protocol (TCP). Most commonly, these are simply referred to as TCP/IP. These define a stack of different layers of abstraction for communicating data packets between nodes in a network, determining their routing, and enforcing any quality of service requirements.

%
% Internet Protocol
%

\subsubsection{Internet Protocol} \index{Internet Protocol (IP)}

IP is the standard protocol employed in the internet for P2P communication of data packets. It is a low-level protocol that encapsulates digital data into packets containing a header field, which specifies routing information, most notably the IP addresses of the source and destination. IP does not enforce any kind of quality control, which is instead delegated to higher-level protocols like TCP (Sec.~\ref{sec:TCP}), a higher-level of abstraction built on top of IP (Sec.~\ref{sec:TCP}).

Multiple packets with the same source and destination needn't follow the same route -- the routing is determined dynamically in realtime by routers, based on network characteristics such as load or latency. Thus, packets belonging to the same underlying data may arrive out of order, or some may go missing altogether. IP does not address these issues, instead engaging in only `best-effort' delivery. 

In IP there is no central authority with knowledge of the state of the entire network, which tells routers in the network how to best route packets. Thus, IP must be complemented with up-to-date routing tables, held by routers/nodes in the network, which make routing decisions on a per-packet basis. This is achieved using gateway protocols, discussed next.

%
% User Datagram Protocol
%

\subsubsection{User Datagram Protocol} \index{User Datagram Protocol (UDP)}

The UDP is a simple protocol built on top of IP, based on a `send-and-forget' principle for sending data packets. That is, there is no quality of service guarantee, and no notifications are provided to the sender as to whether packets successfully reached their destination. However, a checksum (hash) forms a part of the packet headers to enable error detection by the recipient. The lack of quality control bypasses the associated latency, making it particularly useful in time-critical applications, where the late arrival of a packet is useless and therefore needn't be retransmitted.

UDP is connectionless, meaning that no designated connection is established between hosts. Instead data is simply transmitted and then forgotten about. The receiver may not even be operational on the network, in which case the packets are lost without notice.

Key examples for the use of UDP are realtime audio and video transmission. If a packet associated with a frame in a video link is delayed and arrives several frames late, it is useless, since it is associated strictly with a previous frame in the video that has already passed. Quality control, in the form of contacting the sender to request a retransmission, would therefore achieve nothing. This applies similarly to live audio streaming, such as voice over IP (VoIP)\index{Voice over IP (VoIP)}, where the late arrival of a packet cannot possibly be correctly inserted into the audio playback and might as well be discarded.

Therefore, UDP prioritises latency over reliability, and is best suited to time-critical applications where quality of service is not relevant.

%
% Transmission Control Protocol (TCP)
%

\subsubsection{Transmission Control Protocol} \label{sec:TCP} \index{Transmission Control Protocol/Internet Protocol (TCP/IP)}

TCP differs from UDP in that it intrinsically supports quality control. The protocol is able to determine whether a packet successfully reached its destination, and if not, retransmit it as often as necessary to guarantee packet delivery. A checksum is also included in packet headers to enable error detection. This quality control has made TCP the dominant protocol employed in the present-day internet, where, in most scenarios, we wish to guarantee that data has been correctly delivered -- if an email is missing random segments of its text, users will become irate very quickly!

TCP is connection-oriented, meaning that a handshaking protocol establishes a dedicated bidirectional channel between two hosts. It also enforces packet reordering, to counter out-of-order packet arrival.

However, the enforced quality control and handshaking protocols incur a network performance overhead that UDP does not, since handshaking protocols consume bandwidth. Thus, TCP should not be used instead of UDP if there are no quality of service requirements.

%
% Ethernet
%

\subsection{Ethernet} \index{Ethernet}

\begin{figure}[!htbp]
	\includegraphics[clip=true, width=0.475\textwidth]{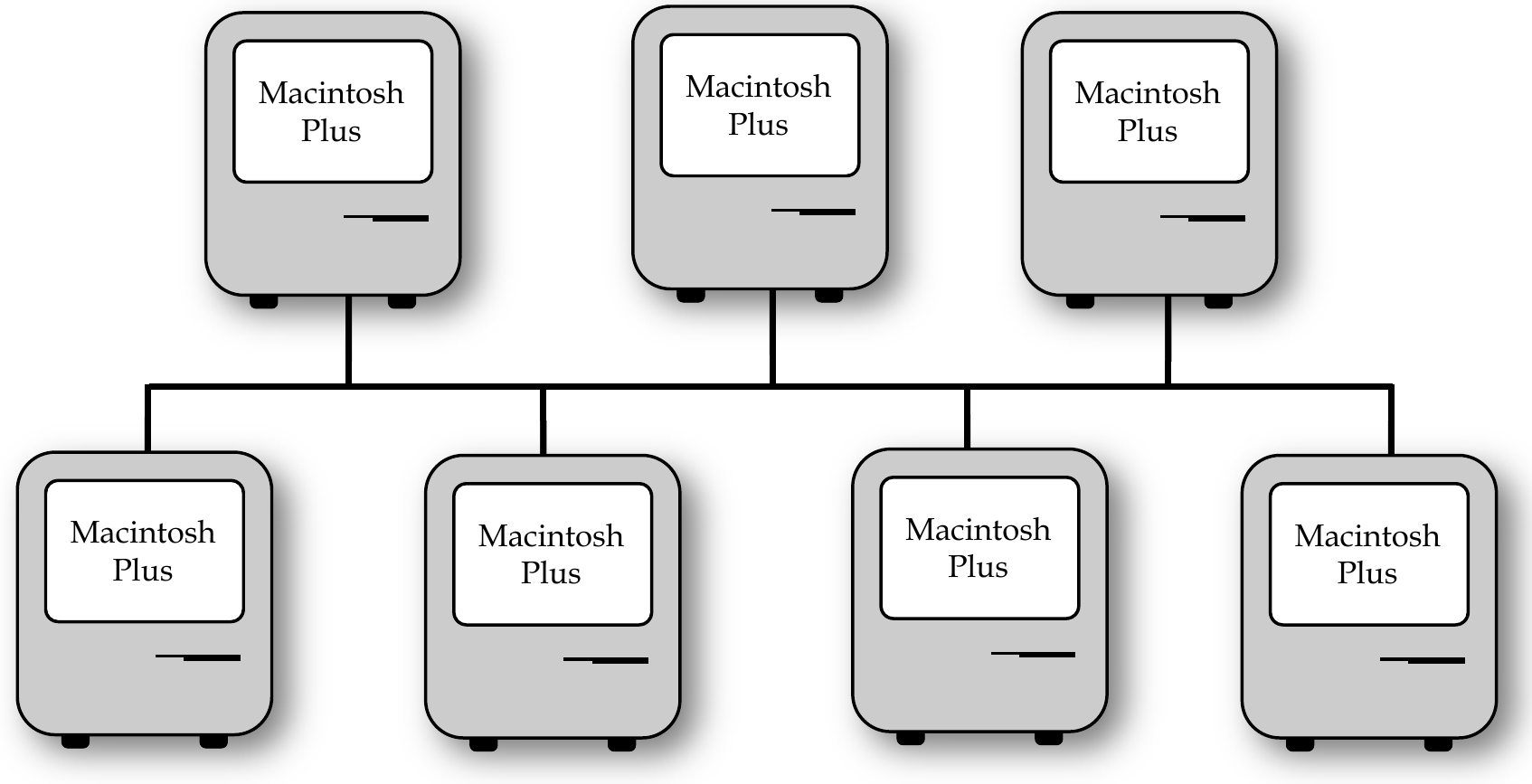}
	\captionspacefig \caption{The topology of an Ethernet network, whereby all users share a common channel, which they can broadcast to at their leisure. If data packets collide, it is detected via the packets' checksums, and the corrupted packets may be re-broadcast after a random `backoff' waiting period, repeating this process until packet delivery is successful. Obviously the chance of collisions occurring increases with the number of connected users, thus network performance is inversely related to the number of nodes.} \label{fig:ethernet}
\end{figure}

Ethernet is a networking protocol based on `broadcasting' on a shared network. This model is particularly suited to local area networks (LANs), where all users share a single communications channel rather than dedicated P2P links, as shown in Fig.~\ref{fig:ethernet}.

In the Ethernet protocol, every user is free to broadcast data onto the shared channel as they please -- all users transmit to, and receive from a single shared channel. However, clearly sometimes packet `collisions' will occur\index{Collisions!Handling}, resulting in packet corruption. To overcome this, Ethernet packets contain a checksum that can be used to verify upon arrival whether a packet has been corrupted by a collision. If a collision is detected, the respective users are able to re-broadcast, following a randomly chosen waiting period (known as `backoff')\index{Backoff}. Collisions therefore reduce network performance, and it follows that network bandwidth decreases with the number of users competing for bandwidth\footnote{Think of that awkward dinner table conversation, where two people start talking simultaneously (Peter \& Jon). It's immediately obvious to them both that they are interfering with one another, and if they were to just talk over one another (packet collision), no one would understand either of them. So, they both awkwardly pause, before starting to speak again. In a \textit{really} awkward conversation, they will both start again simultaneously, after which there will be an even longer awkward pause before recommencing. Eventually, this self-regulating system will resolve itself probabilistically, with a sole victor controlling the airwaves, commanding the attention of the listeners. Provided that all dinner guests adhere to social etiquette and backoff appropriately, with repeated conversations, all guests will statistically receive an equitable share of attention, albeit with some wastage of conversation time owing to the periods of silence. Clearly, the proportion of the time wasted due to collisions will scale up with the number of guests, limiting the protocol to not-too-large tables (or very quiet guests).}.

From this protocol, any given packet will eventually be successfully transmitted uncorrupted, collision-free, albeit with uncertain timing that grows with the number of competing users. For this reason, the Ethernet protocol is not ideal for time-critical applications requiring hard guarantees on network latency.

The beauty of this approach is that only a single channel is required for connecting all users. No dedicated P2P connections are required. As the number of users increases, the complexity of the network topology does not -- requiring only the addition of a node to the existing backbone. For small LANs this is clearly reasonably functional. However, as the size of networks increases, the rate at which packet collisions occur increases, resulting in a reduction in network bandwidth. Thus, the Ethernet protocol is ideally suited to small LANs, but is clearly not viable at a global level, where network competition is astronomical and the overhead from backoff would reduce network performance to a standstill, wasting most of the bandwidth.

Another elegant feature of the Ethernet protocol is that bandwidth allocation is self-regulating, with bandwidth fairly and equitably allocated between users, not prioritising any user over another. This applies even in completely ad hoc networks, with users joining and leaving the network willy nilly. Provided all users are correctly and honestly implementing the \textsc{Broadcast and Backoff} protocol, network bandwidth is allocated evenly amongst users, and no mediating, overriding central authority is needed to oversee network resource allocation. This allows Ethernet networks to be truly `plug-and-play'.

%
% Gateway Protocols & Routing Tables
%

\subsection{Gateway protocols \& routing tables} \label{sec:gateway} \index{Gateway protocols} \index{Routing!Tables}

In the absence of a central mediating authority, routing decisions must be made by individual nodes in the network, upon receipt of packets. For routers to make sensible routing decisions, they must have some idea of the overall structure and state of the network. This is achieved using gateway protocols, which communicate information about the state of the network on a nearest-neighbour basis. There are various gateway protocols in use, with the Exterior Gateway Protocol (EGP)\index{Exterior Gateway Protocol (EGP)} and Border Gateway Protocol (BGP)\index{Border Gateway Protocol (BGP)} being very common.

We let every node in the network have a routing table, initially empty, that will ultimately be populated with information on how to best route incoming packets further along the route to their destination.

To mitigate the need for a central authority, nodes engage in only nearest neighbour communication, sharing their routing tables with one another, to query about the distance metrics (Sec.~\ref{sec:costs}) associated with routes to different destinations. This communication is taking place regularly, and as nodes' routing tables become populated, updating in real-time, they will (hopefully) reach a steady-state. From these tables, single-source shortest path algorithms (Sec.~\ref{sec:single_source_sp}) can be applied by nodes to construct a complete picture of costs to every point in the network. Such a nearest neighbour algorithm is effectively a distributed breadth-first-search\index{Breadth-first-search (BFS) algorithm} algorithm (Sec.~\ref{sec:path_exp}).

%
% Network Hierarchies
%

\subsection{Network hierarchies} \index{Network!Hierarchies}

The disadvantage of Ethernet's \textsc{Broadcast and Backoff} principle is that packets are often wasted -- whenever a collision occurs. Because there is no mediating central authority, packet collisions are a certainty in a heavily-utilised shared network, each time resulting in packet loss, and an associated reduction in usable network bandwidth.

To the other extreme, we could have dedicated P2P channels between every pair of users. Then there would be guaranteed no packet collisions, and therefore maximum bandwidth efficiency, but the network would be extremely costly, and plug-and-play extremely challenging.

To address this dilemma, the topology and subdivision of networks need to be carefully designed. If we consider a large organisation, for example, potentially networking thousands of desktop PCs, the bandwidth wastage associated with packet collisions could grind the entire network to a halt, were all thousands of PCs to be communicating large amounts of data simultaneously. However, if a hierarchy of subnetworks could be implemented, rather than a single monolithic network, efficiency could be improved drastically.

\begin{figure}[!htbp]
\if 1\doublecol
	\includegraphics[clip=true, width=0.475\textwidth]{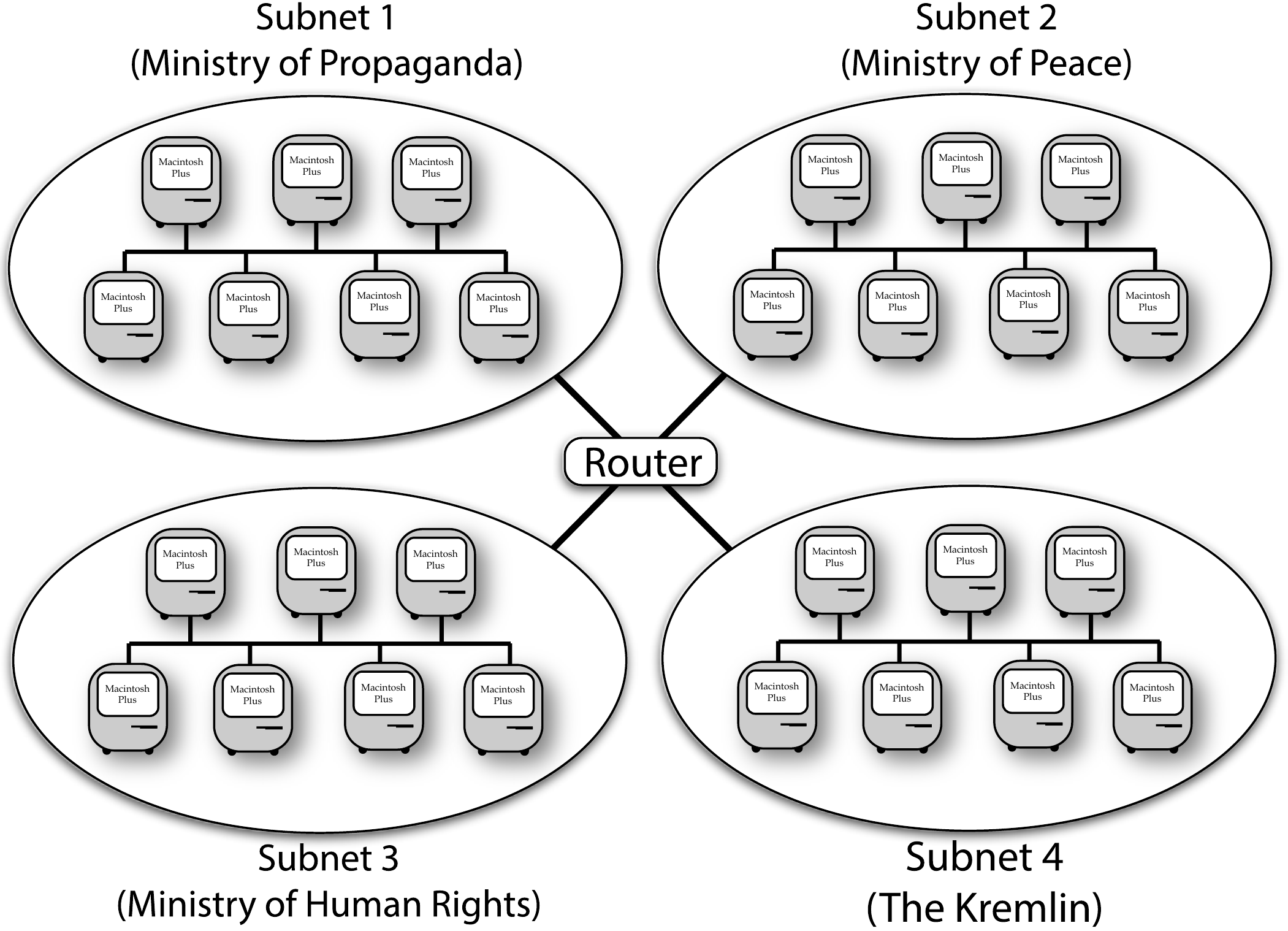}
\else
	\includegraphics[clip=true, width=0.6\textwidth]{network_hierarchy}
\fi
	\captionspacefig \caption{Simple example of a network with two levels in its hierarchy. At the lowest level are 4 different subnets belonging to different departments within an organisation, each of which implements Ethernet networking. Above this, the subnets connect together in a star network via a central router. By subdividing the network hierarchy as such, if most traffic emanating within a subnet stays within that same subnet, collisions between packets on different subnets are avoided, thereby improving the efficiency of the subnets' Ethernet implementations.} \label{fig:net_hierarchy}
\end{figure}

Suppose our hypothetical organisation had several different departments, and users had a tendency to communicate primarily with other users in the same department. By defining distinct departmental subnets, which individually implement Ethernet, but interconnect with one another using an alternate routing framework, we can easily see that many unnecessary packet collisions may be entirely avoided. That is, why broadcast data to users who we know don't want it? A simple example of this is shown in Fig.~\ref{fig:net_hierarchy}.

Extending upon this simple intuitive example, enormous amounts of research and development have been invested into the design of network hierarchies, and how to optimise their efficiency. A pressing consideration in the design of network protocol stacks is therefore to accommodate for multiple routing protocols, and enabling their inter-compatibility.

\latinquote{Te futueo et caballum tuum.}

%
% Mathematical Representation of Networks
%

\section{Mathematical representation of networks}

\dropcap{W}{e} now turn our attention to defining a mathematical construction for the representation of (quantum and/or classical) networks, that we will subsequently rely on heavily in our framework for quantum networks. This encompasses representing networks as graphs, representing the cost of communications within the network, and how to optimise network routing to minimise costs.  These notions will be essential in our treatment of quantum networks.

%
% Graph-Theoretic Representation
%

\subsection{Graph-theoretic representation} \index{Network!Graphs}

We consider a classical network to be a weighted, directed graph,
\begin{align}\index{Network!Graphs}\index{Graphs}
	G=(V,E),
\end{align}
where vertices represent \textit{nodes} (\mbox{$v\in V$}) in the network, and the weighted edges represent communication \textit{links} (\mbox{$e\in E$}) between neighbouring nodes.

A node could be, for example, data storage, a classical computer implementing a computation, a router that switches the connections between incoming and outgoing links, or an end-user -- anything that communicates with the network, sender or receiver. A link on the other hand is any arbitrary means of communication between nodes, such as optical fibre, satellite, radio, electrical, smoke signals, tin cans connected by a taut piece of string, or well-trained carrier pigeon. In the protocols to be described here, it is completely irrelevant what the specific mediums for communication are. Rather what matters are \textit{costs} and \textit{attributes}, quantifying the relative performance of different links.

A key feature of the global internet is redundancy. In a packet-switched environment, sending identical packets twice might each follow entirely different routes to their common destination. Node-to-node redundancy is easily accommodated for in the graph-theoretic model by allowing multiple distinct edges between nodes. It is extremely important to accommodate multiple edges in network graphs, since redundant routes provide a direct means by which to load-balance a route. So, for example, a hub in Australia might connect to a sister hub in New Zealand using both a fibre-optic undersea cable, and simultaneously via a satellite uplink. If the faster of the two connections is running out of capacity, a proportion of the packets can simply be switched to the other link, thereby balancing the load. For this reason we abstain from using an adjacency matrix representation for network graphs, as they do not accommodate redundancy.

%
% Cost Vector Analysis
%

\subsection{Cost vector analysis} \label{sec:costs} \index{Cost vector analysis}\index{Attributes}

The edge weights in $G$ represent the \textit{costs} ($\vec c$) and \textit{attributes} ($\vec a$) associated with using that link. In general these needn't be single numbers, but would rather be sets or data-structures, representing different types of costs and attributes of links, of which there may be many. These could include, for example, latency, bandwidth, dollar cost, and quality measures.

The distinction between costs and attributes, is that costs may be expressed in terms of units which may be interpreted as distances metrics in a Euclidean sense, obeying the following requirements:

\begin{definition}[Network cost metrics] \label{def:metric} Cost metrics satisfy the properties:\index{Network!Cost metrics}
	\begin{itemize}
    	\item Identity operations: If a channel performs nothing, its associated cost is zero, \mbox{$c(\hat\openone) = 0$}.
    	\item Triangle inequality: \\ $c(v_1\to v_2\to v_3) \leq c(v_1\to v_2) + c(v_2\to v_3)$, \\ across all paths \mbox{$v_1 \to v_2 \to v_3$}. In the case of strict equality under addition we refer to the cost as a \textit{strictly additive cost}.
    	\item Positivity: \mbox{$c\geq 0$}. This ensures that shortest-path algorithms will function correctly. It is also congruent with the intuitive expectation that data traversing a communications channel is not somehow better off than if it hadn't traversed that channel at all.
	\end{itemize}
\end{definition}
Attributes, on the other hand do not have a distance interpretation, and may have arbitrary structure. A detailed discussion on the relationship between costs and attributes is presented in Sec.~\ref{sec:c_vs_a}.

The reason we demand costs have a distance interpretation is so that graph-theoretic pathfinding algorithms (Sec.~\ref{sec:shortest_path}) are applicable, allowing us to build upon the vast pre-existing understanding of graph theory. Ideally we would like equality in costs' triangle inequality, which yields an exact cost. But often this isn't possible and we are satisfied with the inequality, which simply dictates an upper bound on cost.

A detailed discussion of some of the major costs and attributes that realistic quantum networks will be subject to is presented in Sec.~\ref{sec:quantum_meas_cost}.

A \textit{route}\index{Routes} between two nodes, Alice ($A$) and Bob ($B$), of the network, $G$, is an acyclic subgraph connecting those nodes, \mbox{$R_{A\to B}\subseteq G$}. In general ad hoc networks there will typically be multiple paths between two nodes \mbox{$A\to B$}. For a particular cost metric, the cost of an entire route is simply the sum of the costs of each of the constituent links,
\begin{definition}[Route costs]
The net cost of a route \mbox{$A\to B$}, using cost metric $c(A\to B)$, traversing nodes $v_i$, is,
\begin{align}\index{Route costs}
c(R_{A\to B}) = \sum_{i=1}^{|R_{A\to B}|-1} c(v_i \to v_{i+1}),
\end{align}
where $v_i$ is the $i$th node in the route $R_{A\to B}$.
\end{definition}

\begin{figure}[!htbp]
\includegraphics[clip=true, width=0.325\textwidth]{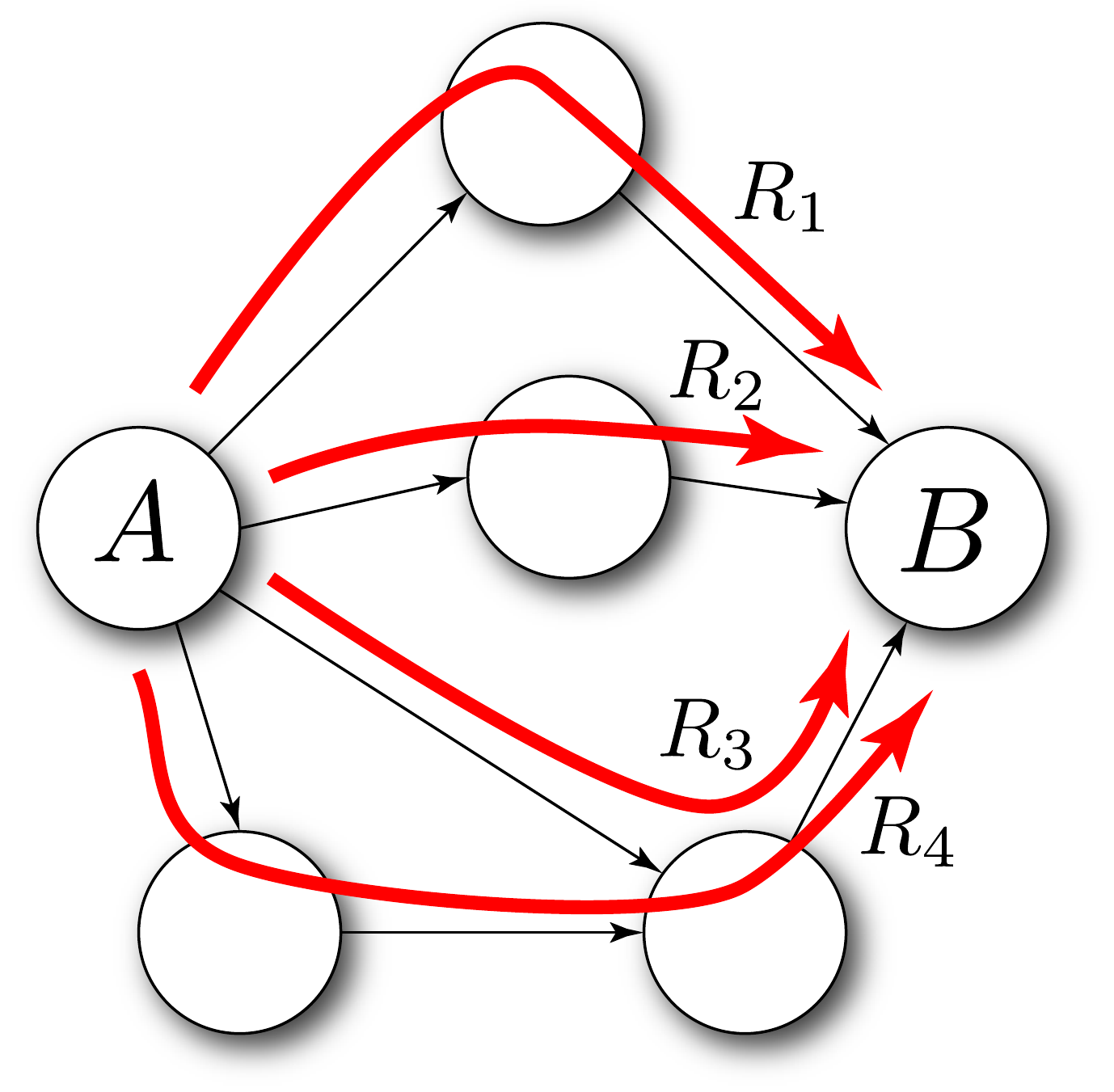}
\captionspacefig \caption{Example of a simple network with multiple routes \mbox{$A\to B$}. Note that $R_3$ and $R_4$ are competing with one another for use of the last link, which the routing strategy, $\mathcal{S}$, will need to resolve if multiple simultaneous transmissions are taking place.} \label{fig:example_routes}
\end{figure}

\begin{figure}[!htbp]
\includegraphics[clip=true, width=0.3\textwidth]{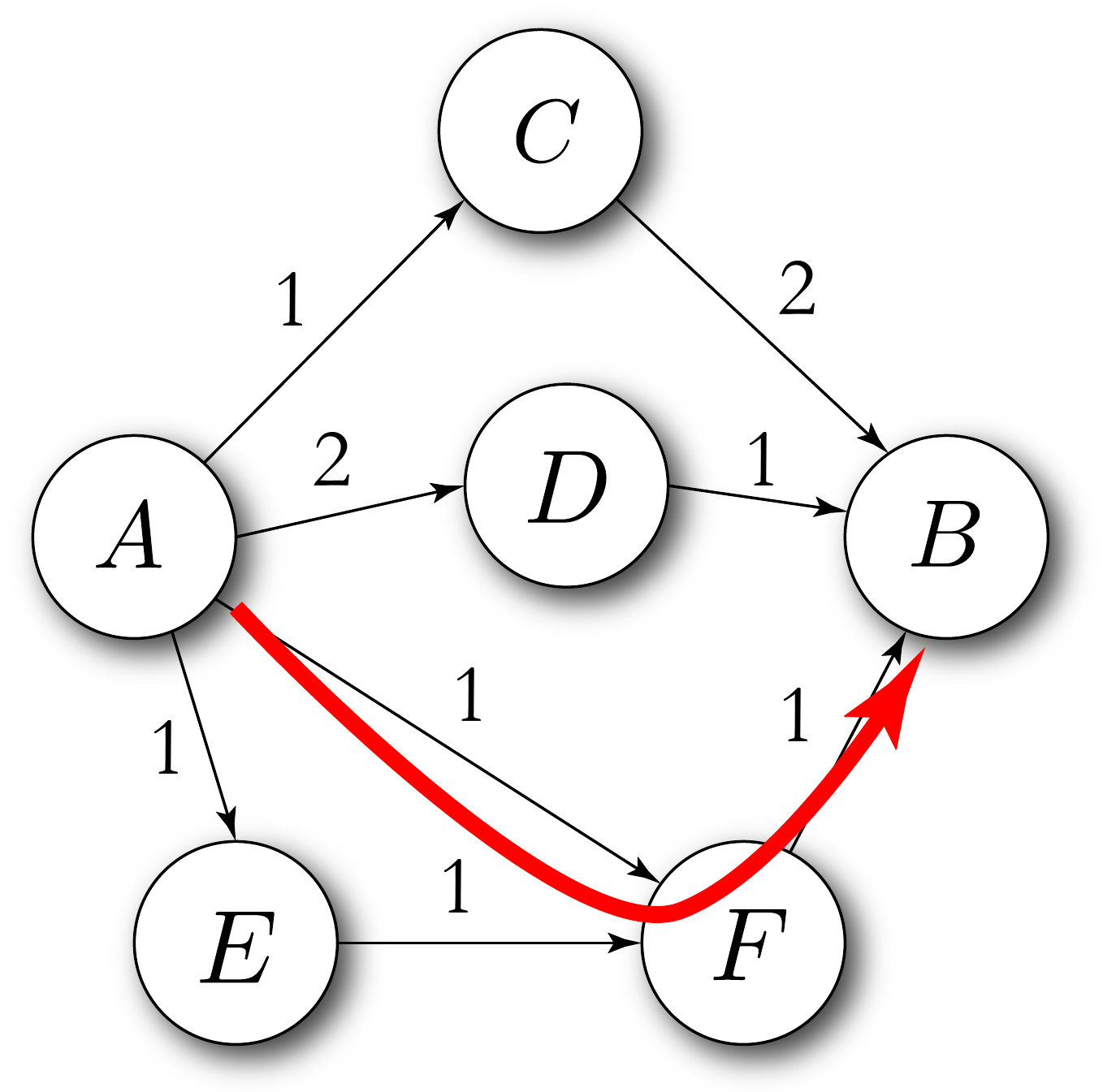}
\captionspacefig \caption{The same network graph from Fig.~\ref{fig:example_routes}, with links weighted by some arbitrary cost metric. Applying a shortest-path algorithm yields the optimal route between Alice and Bob to be \mbox{$A\to F\to B$}, which incurs a net cost of \mbox{$c=2$}, as opposed to all other routes, which incur a net cost of \mbox{$c=3$}.} \label{fig:simp_route_opt}
\end{figure}

Fig.~\ref{fig:example_routes} illustrates a simple example network with all of its available routes, \mbox{$R_{A\to B} \subseteq G$}. Fig.~\ref{fig:simp_route_opt} illustrates the optimal path for \mbox{$A\to B$} based on edge weights.

In a given network, it is unlikely that only a single cost metric or attribute will be of interest when determining optimal routings. There may be a tradeoff between different measures. For example, for time-critical applications the cost of a route might be considered a combination of both dollar cost and latency -- a satellite has very low latency but is extremely expensive, while a carrier pigeon is slow but cheap (and prohibited by PETA). What is the best tradeoff between the two?

To accommodate this, we allow the \textit{net cost} of a route to be defined as an arbitrary function of other primitive cost metrics and attributes of the route,
\begin{definition}[Net routing cost]\index{Net routing cost}
The net cost of a route \mbox{$A\to B$} is given by,
\begin{align} \label{eq:net_cost_R} \index{Net routing cost}
c_\mathrm{net}(R) = f_\mathrm{cost}(\vec{c}(R),\vec{a}(R)),
\end{align}
where $c_\mathrm{net}$ is a single numeric value representing the net cost as calculated from an arbitrary cost function, $f_\mathrm{cost}$, of the vector of associated costs and attributes.
\end{definition}
Note that the net routing cost needn't be a metric, as the cost function could be arbitrary. The net cost can be thought of as a ranking for routes, but not necessarily as a metric that accumulates across routes, since it already captures all these accumulations.

Eq.~(\ref{eq:net_cost_R}) gives us the net cost of a given route. For multiple users we would like to simultaneously optimise the cost across all users of the network. Thus we define the routing cost for the entire network to be,
\begin{definition}[Network routing cost]
	The net routing cost of all costs, over all active routes $\vec{R}$ is,
\begin{align} \label{eq:c_total}
c_\mathrm{total}(\vec{R}_{\vec{A}\to \vec{B}}) = \sum_{r \in {\vec R}_{\vec{A}\to \vec{B}}} c_\mathrm{net}(r),
\end{align}
where $\vec{R}_{\vec{A}\to \vec{B}}$ is a set of active routes connecting each pair \mbox{$A_i\to B_i~\forall ~ i$}.
\end{definition}

%
% Flow Networks
%

\subsection{Flow networks} \label{sec:flow_networks} \index{Flow networks}

On a shared network with many users utilising the network simultaneously, it may be the case that the preferred goal for the network is to maximise \textit{flow} -- the total amount of information that can be transmitted per unit time, i.e the net utilisation of the network's resources, summed over all users. In this case we can build on the existing theory of \textit{flow networks} \cite{goldberg1989network}, which characterise the load of links within the network.

A flow network is easily obtained from the network graph by associating a `capacity' attribute with each link and defining the graph weighted by the capacities as the flow network, preserving the underlying structure of the network graph.

When a route within the graph is utilised, we decrement the capacities of each link in that route, generating the so-called \textit{residual network}, which will now take the place of the original network in subsequent calculations. This process effectively tallies the links' utilisation, and when the tally hits zero, the link can no longer be used for any new routes. This forms a basic building block for more complex flow network algorithms.

There are many variations on flow networks. The simplest case is of a single user transmitting multiple packets simultaneously to a recipient. Depending on link capacities, different packets may need to follow different routes through the network, if network performance is to be maximised. Alternately, it may not be possible to send the desired number of packets simultaneously if the network capacity saturates.

The more complex (and realistic) scenario is of multiple users each transmitting from distinct starting nodes to distinct recipient nodes across a shared network. This is known as a \textit{multi-commodity flow network} \cite{ahuja1995network} and is likely to be the dominant class of networks in real-world networking applications.

%
% Routing Strategies
%

\subsection{Routing strategies} \label{sec:route_strats} \index{Routing!Strategies}

A \textit{strategy}, $\mathcal{S}$, is simply an algorithm that chooses a route, based on the starting and finishing nodes of a communication, and also updates the vectors of costs and attributes within the network associated with the utilisation of that route,
\begin{definition}[Routing strategies]
A routing strategy is defined by,
	\begin{align}
\mathcal{S}(i,j,\vec{c},\vec{a}) &\to \{k,{\vec{c}}~',{\vec{a}}~'\}, \nonumber \\
i,j &\in V, \nonumber\\
k &\in \{R_{v_i\to v_j}\},
\end{align}
where $\mathcal{S}$ denotes the strategy, $k$ is a route, $i$ and $j$ are the source and destination nodes of the route, and $\vec{c}$ and $\vec{a}$ are vectors of associated costs and attributes.
\end{definition}
The goal of the strategy $\mathcal{S}$ is to minimise a chosen cost measure.

No particular route through a network is going to have infinite capacity, and therefore we cannot typically always reemploy the same most cost-effective route for all data. Particularly in multi-user networks, as routes are employed for communicating quantum states, their cost metrics may change according to load, or other external influences. Alternately, some routes may come into and out of operation. For example, a satellite requiring line-of-sight communication may oscillate in and out of sight, thereby periodically enabling and disabling respective network routes. For this reason, it is important that strategies accommodate dynamic changes in the network. This is easily accounted for by letting the edge weights in our network graph be a function of time, $G_t$, which are updated via the application of a strategy, which may also be time-dependent,
\begin{definition}[Time-dependent routing strategies]
A time-dependent strategy, $\mathcal{S}_t$, updates the network graph, $G_t$, at each time-step $t$,
\begin{align} \label{eq:S_G}
G_{t+1} = \mathcal{S}_t(G_t).
\end{align}
$S_t$ could be any \textbf{BPP} algorithm, deterministic or probabilistic.
\end{definition}
For example, the network might have bandwidth restrictions on some links, in which case if more than a certain amount of data is transmitted through a link, it is no longer available for use until previous transmissions have completed. Or, based on market dynamics, the dollar cost of utilising a link may change with its demand.

This type of cost minimisation approach to routing is analogous to \textit{distance-vector routing protocols}\index{Distance-vector routing protocols} in classical networking theory.

A detailed exposition of routing strategies is provided in Sec.~\ref{sec:strategies}.

%
% Strategy Optimisation
%

\subsection{Strategy optimisation} \label{sec:strat_opt} \index{Strategy!Optimisation} 

Clearly the goal when choosing routing strategies is to minimise the total cost, Eq.~(\ref{eq:net_cost_R}). That is, solving the optimisation problem,
\begin{definition}[Strategy optimisation]
The optimisation of strategies with a network comprising net costs $c_\mathrm{total}$ is given by,
\begin{align}
c_\mathrm{min} &= \underset{\mathcal{S}}{\mathrm{min}}(c_\mathrm{total}), \nonumber \\
\mathcal{S}_\mathrm{opt} &= \underset{\mathcal{S}}{\mathrm{argmin}} (c_\mathrm{total}).
\end{align}
\end{definition}

Choosing optimal strategies is a challenging problem, potentially requiring complex, computationally inefficient optimisation techniques. Strategy optimisation is an example of resource allocation, whose optimal solutions are often notoriously difficult to solve exactly, residing in complexity classes like \textbf{NP}-complete\index{NP \& NP-complete} (or worse!). In general, the number of possible routes through a graph will grow exponentially with the number of vertices. Thus, explicitly enumerating each possible route is generally prohibitive for large networks, unless some known structure provides `shortcuts' to optimisation. Having said this, Dijkstra's shortest path algorithm (discussed in Sec.~\ref{sec:shortest_path}) is the perfect counterexample, demonstrating that although an exponential number of routes may exist between two points, an optimal one can be found in \textbf{P}.

%
% Ad hoc Operation vs. Central Authorities
%

\subsubsection{Ad hoc operation vs. central authorities} \index{Central mediating authority} \index{Ad hoc networks}

When considering strategy optimisation, the first question to ask is `Who performs the optimisation, and who has access to what information?'.

In terms of who performs the optimisation, the two main options are that either each node is responsible for optimising the routes of packets passing through it (\textsc{Individual} algorithms), or there is a reliable and trusted central mediating authority\index{Central mediating authority} who oversees network operation and performs all strategy decision-making (\textsc{Central} algorithms).

In the case of \textsc{Individual} algorithms, the required knowledge of the state of the network could be obtained using network exploration algorithms (Sec.~\ref{sec:path_exp}) or gateway protocols (Sec.~\ref{sec:gateway}). 

On the other hand, for \textsc{Central} algorithms, either network exploration could be employed, or alternately the network policy could require nodes to notify the central authority upon joining or leaving the network. The former introduces an overhead in classical networking resource usage, since network exploration must be performed routinely to keep the ledger of nodes up-to-date. The latter, on the other hand, avoids this, but introduces a point of failure, in that all network participants must be reliable in notifying the central authority as required by the network policy. Failure to do so could result in invalid or suboptimal strategies.

%
% Local vs. Global Optimisation
%

\subsubsection{Local vs. global optimisation} \index{Local optimisation}\index{Global optimisation}

There are two general approaches one might consider when choosing strategies -- \textit{local optimisation} (\textsc{Local}) and \textit{global optimisation} (\textsc{Global}). \textsc{Local} simply takes each state to be communicated, one-by-one, and allows it to individually choose an optimal routing strategy based on the state of the network at that moment. \textsc{Global} is far more sophisticated and simultaneously optimises the sum of the routing costs, Eq.~(\ref{eq:c_total}), of all currently in-demand routes.

To implement \textsc{Local} optimisation, either \textsc{Individual} or \textsc{Central} algorithms may be employed. On the other hand, \textsc{Global} optimisation necessarily requires a \textsc{Central} algorithm, since it requires knowledge of the entire state of the network, which is collectively optimised.

Since \textsc{Global} represents the class of all algorithms that take all network costs by all packets into consideration, it must clearly perform at least as well as \textsc{Local}, which only takes into consideration the costs of a given packet. But we expect \textsc{Global} to perform better than \textsc{Local} in general, owing to the additional information it takes into consideration. We express this as \mbox{\textsc{Local}$\subset$\textsc{Global}}. However, \textsc{Global} requires solving a complex, simultaneous optimisation problem, which is likely to be computationally hard, whereas \textsc{Local} can be efficiently solved using multiple independent applications of, for example, an efficient shortest-path algorithm (so-called \textsc{Greedy} algorithms), discussed in Sec.~\ref{sec:shortest_path}.

A further stumbling block for \textsc{Global} is that it requires some central authority, responsible for the global decision-making, to have complete, real-time knowledge of the state of the entire network. This may be plausible for small LANs, but would clearly be completely implausible for the internet as a whole. So it is to be expected that different layers and subnets in the network hierarchy will employ entirely different strategy optimisation protocols. This is certainly reminiscent of the structure of the present-day internet.

Roughly speaking, we might intuitively guess that at lower levels in the network hierarchy, responsible for smaller subnets, there will be a tendency towards the adoption of \textsc{Global} strategies, as full knowledge of the state of the subnet is readily obtained and maintained. However, as we move to the highest levels of the network hierarchy (e.g routing of data across international or intercontinental boundaries), we might expect more laissez-faire (i.e \textsc{Greedy}) strategies to be adopted, since the prospects of enforcing a central authority with full knowledge of the state of the internet, who is also trusted by all nations to fairly and impartially allocate network resources and mediate traffic, is highly questionable.

We will not aim to comprehensively characterise the computational complexity of \textsc{Global} strategies. However, in Sec.~\ref{sec:strategies} we will present some elementary analyses of several toy models for realistic strategies. Some such strategies are efficient although not optimal, but nonetheless satisfy certain criteria we might expect.

Future developments in the optimisation techniques required for \textsc{Global} strategies may improve network performance, leaving our techniques qualitatively unchanged.

When employing \textsc{Local}, on the other hand, things are often far simpler. If we are optimising over a cost metric satisfying the distance interpretation, we may simply employ a shortest-path algorithm to find optimal routes through the network.

If one were to become even more sophisticated, one might even envisage treating network resource allocation in a game theoretic context, which we won't even begin to delve into here.

%
% Message- vs. Packet-Level Routing
%

\subsection{Message- vs. packet-level routing}

In Eq.~(\ref{eq:S_G}) we defined the action of a strategy, $\mathcal{S}$, on a network, $G$. However, we were intentionally ambiguous in our introduction of the time-dependence, given by $t$. This is to allow us to consider changes at one of two different time-scales: the packet level, or the message level. The \textit{message} is the entire data stream transmitted from Alice to Bob, whereas the \textit{packet} is a small block of data taken from the message, where each packet may be independently routed.

When defining the action of strategies, we could do so at either of these time-scales. We could choose routes in their entirety, from start to finish, at the beginning of the message transmission, under the assumption that the costs in the network will be constant over that duration and no one will misbehave. We refer to such strategies as \textit{message-level strategies}. Alternately, and perhaps more realistically in many scenarios, the costs and attributes of a network could be highly dynamic and readily change within the transmission time-window. In that case, we will employ \textit{packet-level strategies}, which reevaluate the strategy independently for each packet and for each of their hops between nodes.

In our future discussions on routing strategies, context will make it clear when we are referring to packet- or message-level strategies.

\latinquote{Nil sine magno labore.}

%
% Network Topologies
%

\section{Network topologies} \index{Network!Topologies}\label{sec:network_topologies}

\dropcap{A}{s} quantum (or classical) networks inherently reside on graphs, it is important to introduce some of the key graph structures of relevance to networking and some of their properties of relevance to quantum networking protocols.

Let the graph $G$ representing the network be,
\begin{align}\index{Graphs}
G=(V,E),	
\end{align}
with vertices $V$ and edges $E$. In principle a network could be characterised by any connected graph whatsoever. However, there are certain structures and patterns that emerge very frequently and deserve special attention.

It is paramount that QTCP protocols have the capacity to deal with the diverse network topologies that are likely to present themselves in the future real-world quantum internet. Some of the graph-theoretic algorithms that we rely on in our QTCP protocol (Sec.~\ref{sec:graph_theory}) are computationally efficient for \textit{arbitrary} graph topologies, even more so for certain classes of graphs exhibiting particular structure, such as tree graphs or complete graphs. Others, however, are computationally inefficient in general, but may have efficient approximation algorithms for some or all classes of topologies.

We will now review some of the graph structures most likely to arise in quantum networks, learning from the structures that have become ubiquitous in classical networking. Understanding the basic mathematical properties of these different network topologies is extremely important to take into consideration when designing future quantum networks, since they strongly impact important features such as construction cost of the network infrastructure, routing cost vector analysis (Sec.~\ref{sec:quantum_meas_cost}), likelihood of successful routing, and transmission time.

A summary of the basic mathematical characteristics of the topologies presented is shown in Table.~\ref{tab:net_top_sum}, specifically showing the number of edges and vertices, the \textit{diameter} of the topologies (i.e the distance between extremal points in the network).

\startnormtable
\begin{table*}[!htbp]
	\begin{tabular}{|c|c|c|c|}
		\hline
  		Topology & Vertices ($|V|$) & Edges ($|E|$) & Diameter ($d$) \\
      	\hline
      	\hline
      	Point-to-point & $2$ & $1$ & $1$ \\
      	Linear & $|V|$ & \mbox{$O(|V|)$} & \mbox{$|V|$} \\
      	Complete & $|V|$ & \mbox{$O(|V|^2)$} & $1$ \\
      	Lattice & $mn$ & \mbox{$O(mn)$} & \mbox{$O(m+n)$} \\
      	Tree & $|V|$ & \mbox{$O(|V|)$} & \mbox{$O(\log |V|)$} \\
      	Percolation & \mbox{$p_\mathrm{vertex}\cdot |V|$} & \mbox{$p_\mathrm{edge}\cdot |E|$} & variable \\
      	Random & \mbox{$p_\mathrm{vertex}\cdot |V|$} & \mbox{$p_\mathrm{edge}\cdot O(|V|^2)$} & variable \\
      	Scale-free & $|V|$ & $|E|$ & \mbox{$O(\log\log |V|)$}\\
      	\hline
	\end{tabular}
	\captionspacetab \caption{Summary of the mathematical characteristics of different network topologies.} \label{tab:net_top_sum}
\end{table*}
\startalgtable

%
% Point-To-Point
%

\subsection{Point-to-point} \index{Point-to-point (P2P)!Topologies}\label{sec:P2P_topol}

The most trivial network topology, which also acts as the elementary primitive from which our other topologies will be constructed is a simple dedicated point-to-point (P2P) connection between two parties, where the sender and recipient of a packet reside on neighbouring nodes.

Such P2P connections may be reserved exclusively for the two connected neighbouring nodes. In this instance, the packets' \textsc{Routing Queue}s trivially specify just the recipient. Alternately, the P2P link may be an intermediate step between more distant sender/recipient pairs.

In the case whereby the P2P connection is reserved exclusively for a particular sender/recipient pair, the link has the property that there is no competition between multiple users sharing the channel, and the QTCP stack needn't concern itself with dynamic routing strategies\footnote{Assuming the P2P channel has sufficient capacity to meet demand and exhibits better cost metrics than other potential redundant, indirect routes.}. This significantly simplifies network scheduling algorithms (Sec.~\ref{sec:strategies}), and a \textsc{First-Come First-Served} (i.e chronologically ordered FIFO queue) strategy may be employed. Furthermore, packet collisions cannot occur, thereby improving network efficiency.

In the case whereby the P2P connection is not reserved for exclusive use between a single sender/recipient pair, but shared between different competing routes in the network, the importance of network routing strategies manifests itself. Now competition for access to the channel will reduce network efficiency, scaling inversely against the number of network participants, and the priorities and costs of packets must be tallied for the purpose of implementing routing strategies.

%
% Linear
%

\subsection{Linear}\index{Linear topologies}\label{sec:linear_topol}

\begin{figure}[!htbp]
\includegraphics[clip=true, width=0.4\textwidth]{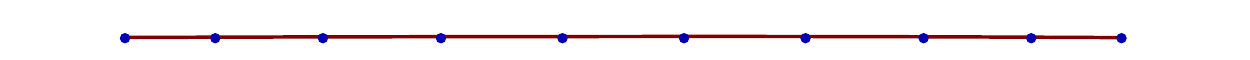}
\caption{A simple linear graph topology with \mbox{$|V|=10$} vertices.}\label{fig:linear_graph}
\end{figure}

A linear graph topology, shown in Fig.~\ref{fig:linear_graph}, has very simple properties. The number of edges simply scales as,
\begin{align}
|E|=|V|-1,	
\end{align}
and the graph diameter is simply the number of vertices,
\begin{align}
d = |V|.	
\end{align}
There are limited routing considerations for such a topology since there is always exactly one route between two points, although buffering issues may still arise under congestion.

Because there is no path redundancy, linear graphs are vulnerable to node failures, since the deletion of a since node makes disconnects the network.

%
% Complete
%

\subsection{Complete}\label{sec:complete_topol} \index{Complete topologies}

The complete graph, denoted $K_{|V|}$, is a $|V|$-vertex graph where every vertex has an undirected link to every other. From a networking point of view, this can be regarded as the extremity of exclusive-use P2P networking, whereby every node has a direct link with every other. Thus, any sender can directly communicate with any receiver, via a dedicated direct channel, with no need to utilise any indirect routes. This topology has the favourable property that although any node can communicate with any other, by exclusively utilising direct P2P links we achieve several benefits:
\begin{itemize}
\item Packet collisions can be mitigated entirely, thereby maximising network efficiency.
\item Competition for the use of links can be eliminated, minimising congestion and the need for buffering (i.e quantum memory).
\item Network costs can typically be minimised, as every route only traverses a single link, and there will be no accumulation of costs.
\item The network has maximal route redundancy, making it the most tolerant against link failures\footnote{To disconnect a given node $v$ from the network, all $|v|$ links emanating from it must be broken, otherwise redundant routes to the remainder of the network will exist.}.
\item A trivial \textsc{First-Come First-Served} routing strategy can be employed, eliminating the need for any dynamic or computationally complex strategies.
\item If the network allows indirect routes to be established, the maximal redundancy of the topology also maximises the ability for routing strategies to engage in load-balancing across routes.
\item In the special case of a symmetric complete graph, whereby all edge weights are approximately equal, the shortest path between any two nodes is trivially the P2P link between them, and no complex scheduling algorithms are required.
\end{itemize}
However, these highly desirable benefits come at the expense of requiring the most elaborate and expensive network, with maximal interconnectedness.

This type of topology could arise in, for example, international-scale networks, where links of very high bandwidth (and value) between nations or continents need to be maximally utilised, which would be undermined by sparse, shared network topologies. Additionally, in this instance route redundancy will be highly valued, as the isolation of one continent from another would be catastrophic to the functioning of the global network.

\begin{figure}[!htbp]
\includegraphics[clip=true, width=0.35\textwidth]{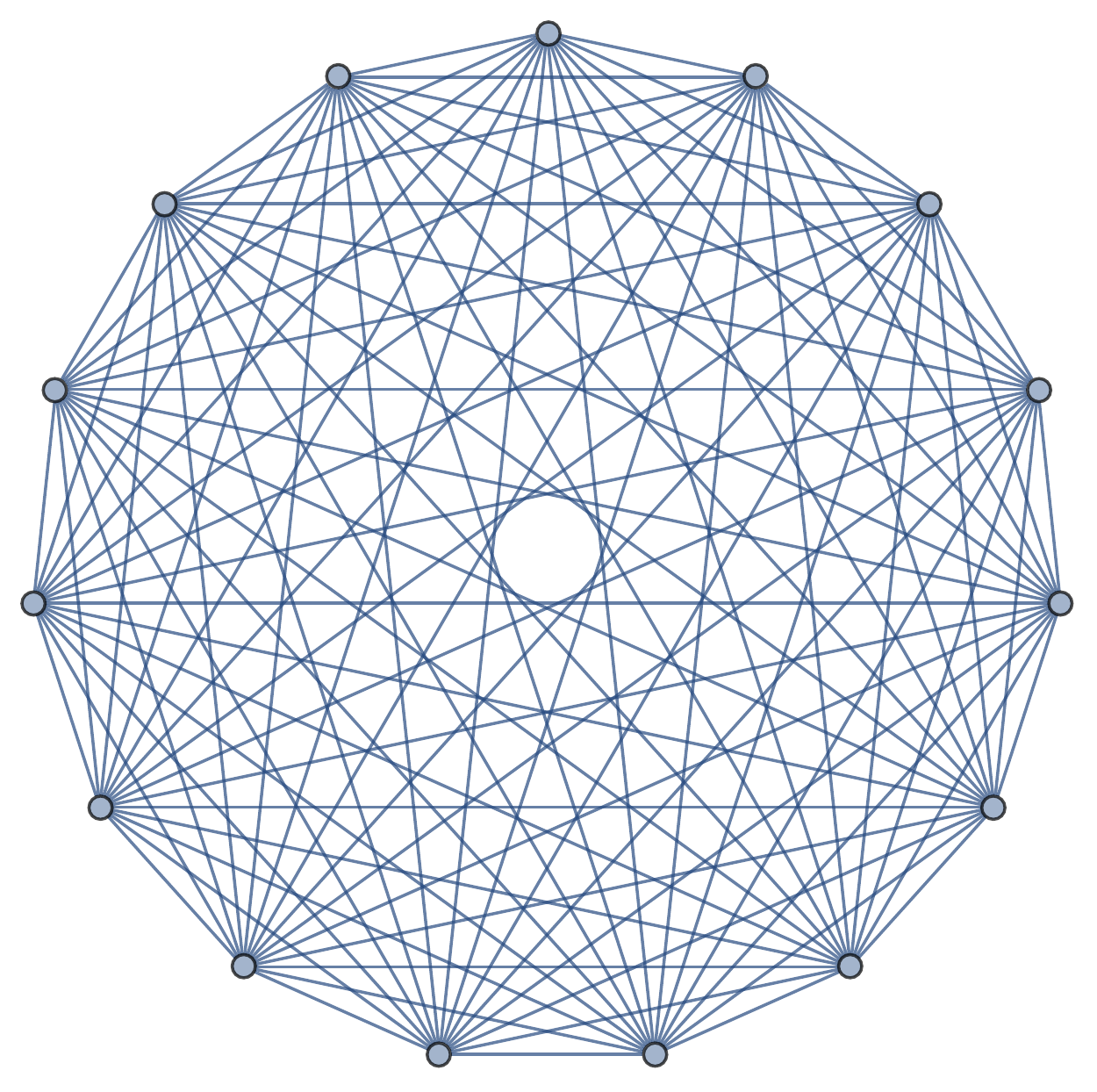}
\captionspacefig \caption{The 15-vertex complete graph, $K_{15}$. Every vertex has an edge to every other, with a total of 105 edges.} \label{fig:complete_graph}
\end{figure}

Fig.~\ref{fig:complete_graph} illustrates the $K_{15}$ graph. The number of edges scales as,
\begin{align}
	|E|=O(n^2).
\end{align}
Clearly route-finding is trivial, since there is always a direct P2P link from sender to receiver, with no possibility of collisions with other packets, requiring $O(1)$ search time (assuming all users are communicating only via their direct links with one another, which may not strictly be the case when costs are factored into strategies).

%
% Lattice
%

\subsection{Lattice}\label{sec:lattice_topol} \index{Lattice!Topologies}

\begin{figure}[!htbp]
\includegraphics[clip=true, width=0.35\textwidth]{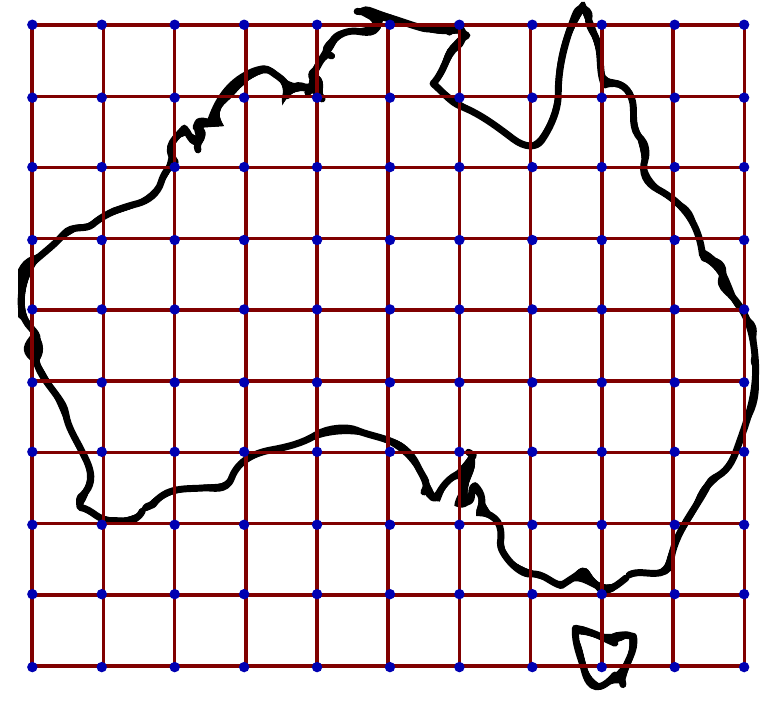}
\captionspacefig \caption{A \mbox{$10\times 11$} square lattice graph, and how it might represent a network topology with geographically associated costs. Notice that Hobart has no internet connection (why even include Tasmania at all?).} \label{fig:lattice}
\end{figure}

A lattice graph is simply an \mbox{$n\times m$} lattice of vertices (of any geometry, e.g squares), connecting each vertex to its immediate geometric neighbours. The number of edges scales obviously as,
\begin{align}
	|E|=O(mn).
\end{align}
This type of graph is useful when link costs are measured in terms of Euclidean distances, and nodes have nearest neighbour links, as per Fig.~\ref{fig:lattice}.

A slightly distorted lattice graph, in which vertices have been dragged around geometrically to match, for example, cities within a country, closely resembles the topology of the network. Similarly, if the nodes represent houses in the street layout of a highly regular city like Manhattan, a lattice may be a good approximation.

In the case of a balanced lattice, in which all edges are of equal weight, the cost of a route is the sum of the number of steps in the vertical and horizontal directions, also known as the Manhattan or $L_1$ distance,
\begin{align}
L_1 = |x_\mathrm{start} - x_\mathrm{finish}| + |y_\mathrm{start} - y_\mathrm{finish}|.
\end{align}
In this case, route finding is simplified, since \textit{all} routes, which strictly traverse in one direction vertically and one direction horizontally, are optimal and of equal distance. Thus, the diameter (maximum number of hops between any two points) on the network is,
\begin{align}
	d=O(m+n).
\end{align}

%
% Tree
%

\subsection{Tree}\label{sec:tree_topol} \label{sec:tree_graph} \index{Tree topologies}

A tree is a graph containing no cycles, only \textit{branches}\index{Branches}. There are many uses for tree graphs, but one property is of particular convenience in many applications: because the graph is acyclic, there is always exactly one path from any vertex to any other. This mitigates the need for shortest-path algorithms designed for general graphs, and simplifies route-finding algorithms (to be discussed in Sec.~\ref{sec:path_exp}). However, this brings with it the drawback that the topology is most vulnerable to link failures, since the removal of any link from the tree will separate it into a multipartite graph\index{Multipartite graphs}, making communication between the disjoint subgraphs (which are also trees) impossible, as there are no redundant routes. In a sense, tree graphs can be considered the polar opposites of complete graphs.

Trees are specified entirely by \textit{branching parameters} ($b_i$) -- the number of child nodes emanating from a given node, $i$. In general, branching parameters may be distinct for each node, although often trees with symmetries in their branching structures are considered, such as the balanced trees discussed in Sec.~\ref{sec:bal_tree}. A node terminates a branch if its branching parameter is zero (i.e it has no children).

The \textit{depth} ($d$) of a tree is the maximum number of steps from the root node to a terminating node with no children. The depth scales between \mbox{$d=O(|V|)$}, for the trivial linear tree (\mbox{$b_i=1$}), and \mbox{$d=O(\log |V|)$} for non-trivial branching parameters (\mbox{$b_i\neq 1$}).

The worst-case number of edges that must be traversed to reach any vertex from any other is,
\begin{align}
	O(\log|V|),
\end{align}
known as the \textit{diameter} of the graph\index{Diameter}, which implies that accumulated cost metrics scale similarly. Trees are the most frugal graphs in their number of edges, which are fixed at,
\begin{align}
	|E|=|V|-1,
\end{align}
irrespective of the branching parameters, since because the graph is strictly acyclic, every addition of an edge requires the addition of exactly a single vertex. This makes tree graphs the cheapest to construct in terms of physical resource usage.

%
% Balanced Tree
%

\subsubsection{Balanced tree} \label{sec:bal_tree} \index{Balanced!Tree topologies}

A balanced tree is a tree with a regular, self-similar structure, in which every node at a given depth is the parent of the same number of sub-nodes, all separated by the same edge weights. That is, the network has a hierarchical structure, subdividing into identically structured subnetworks. Such a network is characterised by just two parameters -- the branching parameter, $b$, and the depth, $d$. Some examples of balanced trees with different $b$ and $d$ are shown in Fig.~\ref{fig:tree_example}.

\if 1\doublecol
	\begin{figure}[!htbp]
	\includegraphics[clip=true, width=0.325\textwidth]{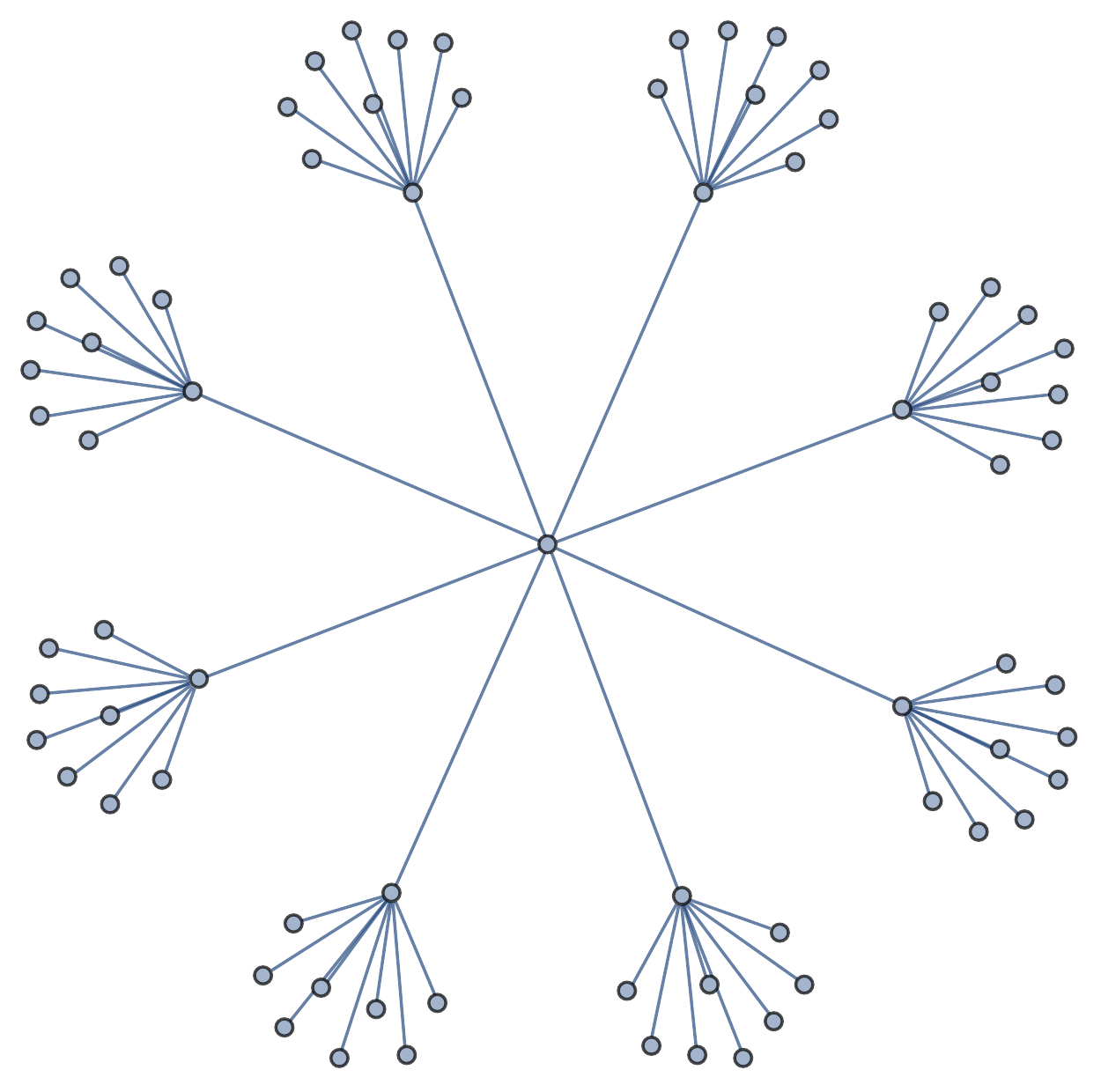}\\
	\includegraphics[clip=true, width=0.325\textwidth]{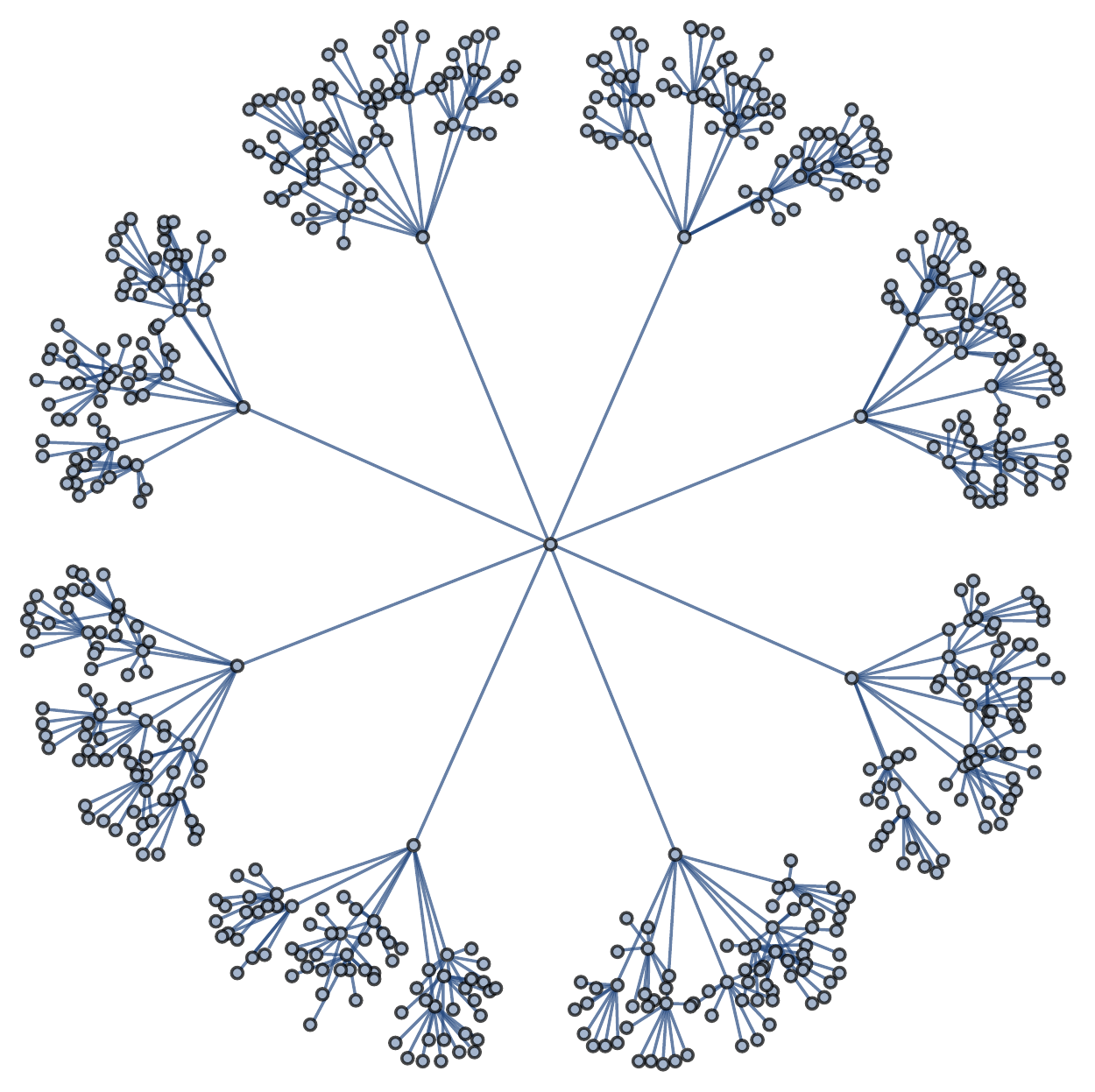}\\
	\includegraphics[clip=true, width=0.325\textwidth]{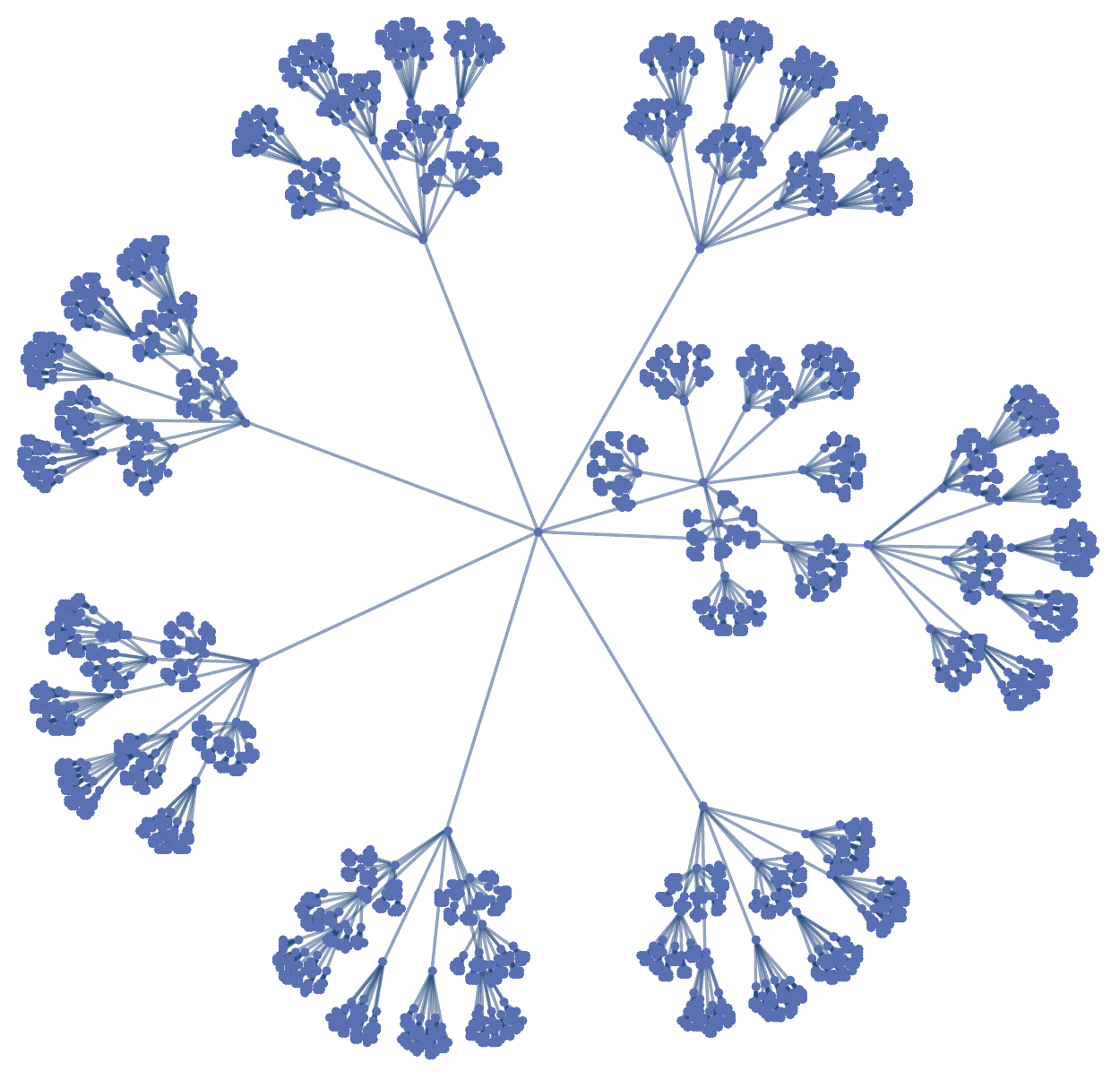}
	\captionspacefig \caption{Balanced tree graphs with branching factor \mbox{$b=8$}, and depths \mbox{$d=3,4,5$}. Despite having no redundant paths, the hierarchical structure of balanced trees somewhat resembles that of real-world networks, which are typically decomposed into a pyramid scheme of progressively smaller subnetworks.} \label{fig:tree_example}
	\end{figure}
\else
	\begin{figure*}[!htbp]
	\includegraphics[clip=true, width=0.325\textwidth]{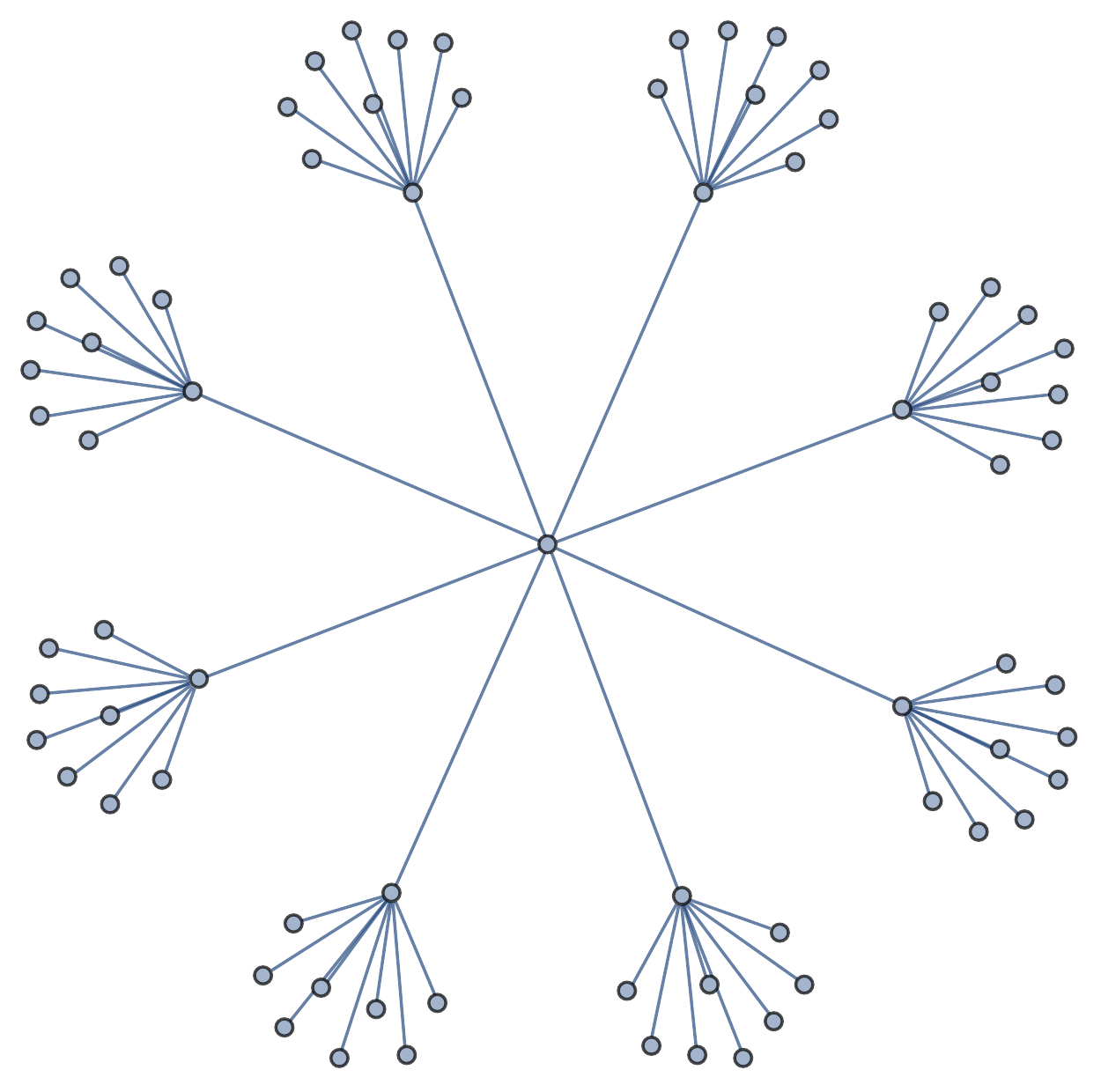}
	\includegraphics[clip=true, width=0.325\textwidth]{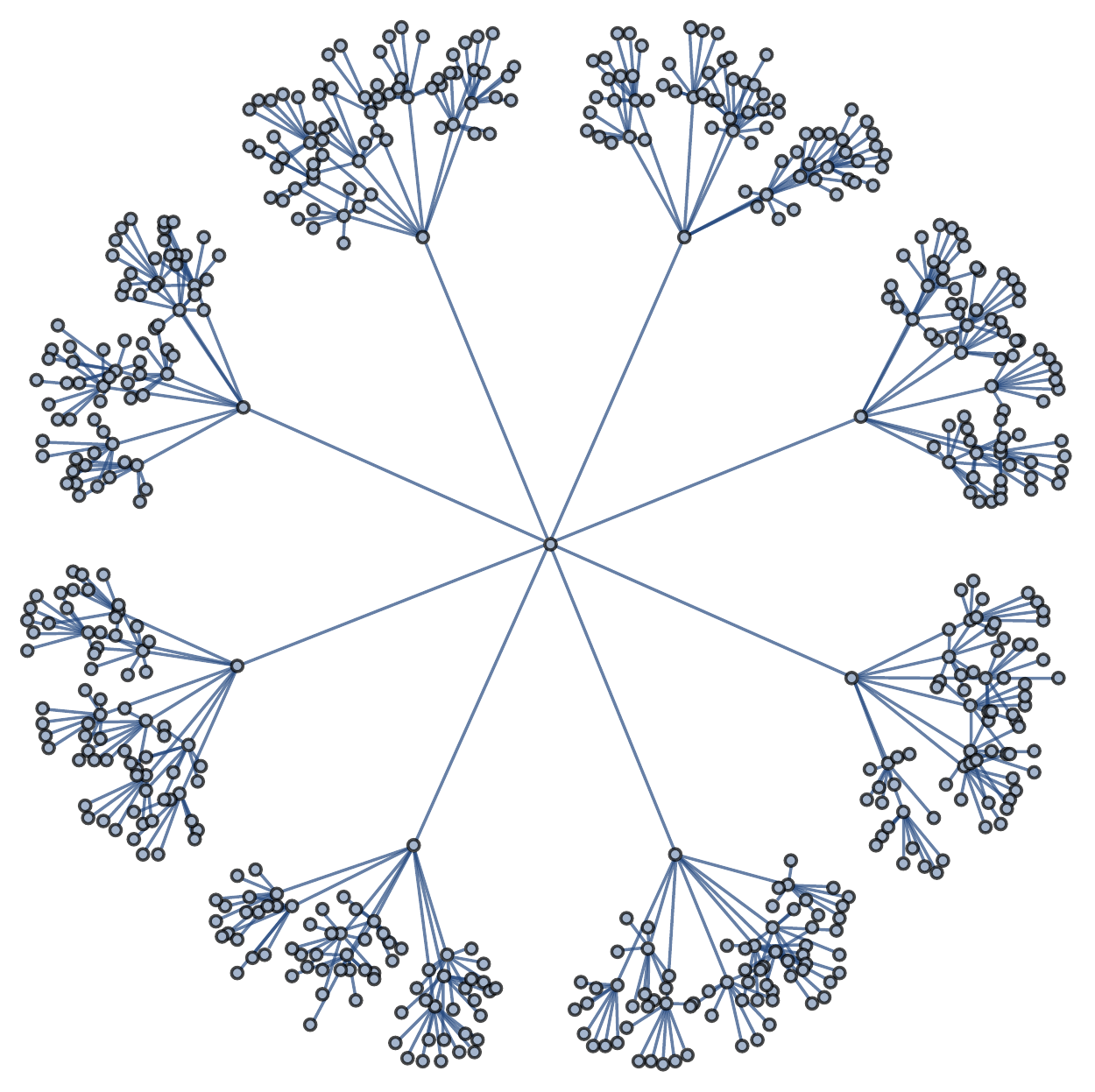}
	\includegraphics[clip=true, width=0.325\textwidth]{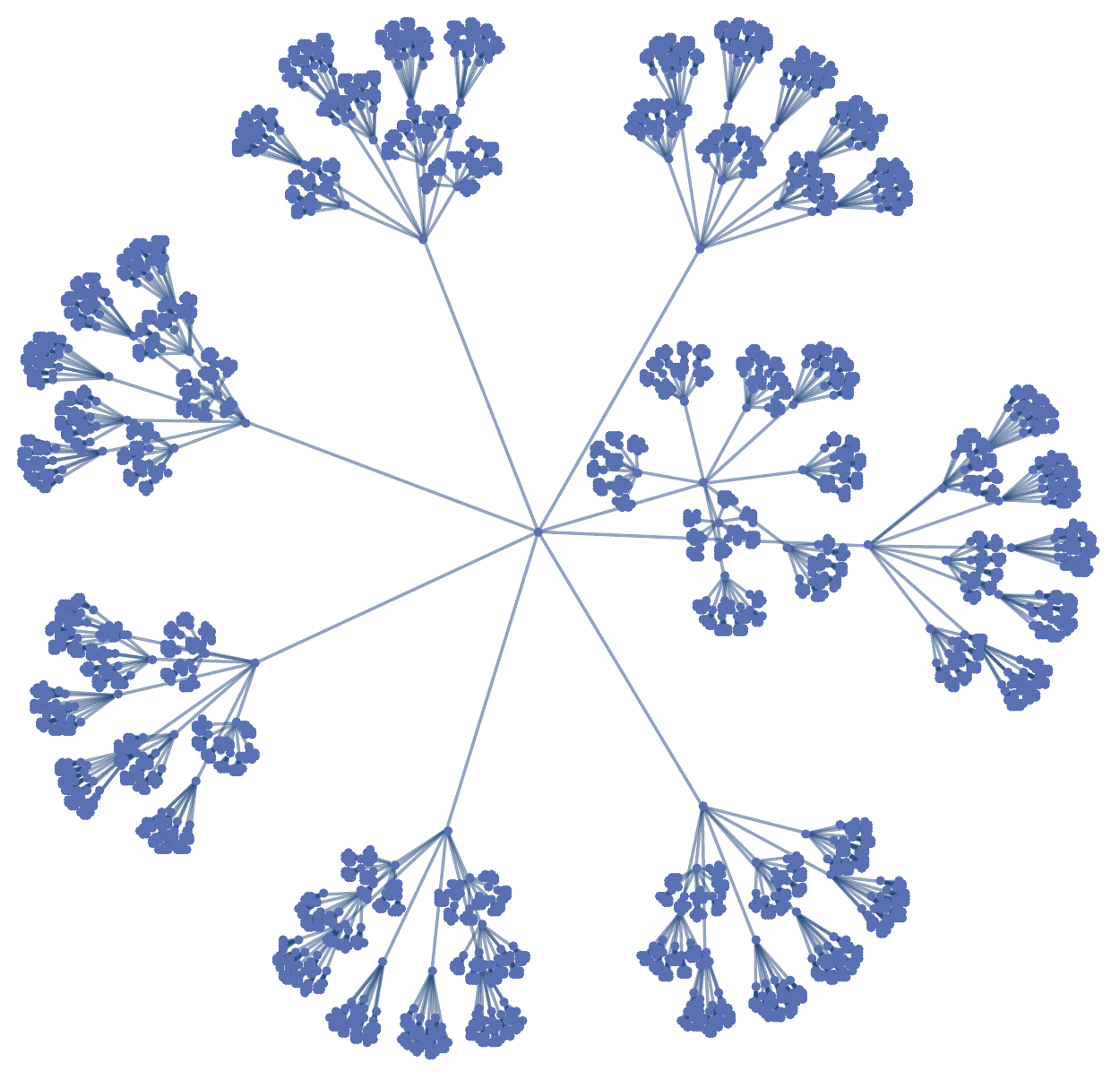}
	\captionspacefig \caption{Balanced tree graphs with branching factor \mbox{$b=8$}, and depths \mbox{$d=3,4,5$}. Despite having no redundant paths, the hierarchical structure of balanced trees somewhat resembles that of real-world networks, which are typically decomposed into a pyramid scheme of progressively smaller subnetworks.} \label{fig:tree_example}
	\end{figure*}
\fi

This type of structure is (approximately) natural in many realistic scenarios. Consider for example a network containing a hierarchy of clusters of nodes representing a LAN, followed by a neighbouring internet router, followed by a city-wide router, followed by a country-wide router. In such a case, this type of general structure is typical (although more realistically one might expect the branching parameter to vary with depth).

A special case is when \mbox{$d=1$}, which we refer to as a \textit{star} graph. This might arise naturally when a series of subnets are connected together via a central router (e.g Fig.~\ref{fig:net_hierarchy}), with no further hierarchy in the network.

%
% Random Tree
%

\subsubsection{Random tree} \index{Random!Tree topologies}

While balanced trees accurately capture the hierarchical nature of realistic networks, they are somewhat contrived in their perfect symmetry. The subnetworks in a given network are not likely to actually all be identical. Random trees are perhaps more realistic, in that their tree structure captures the hierarchical nature of real-world networks, and also their highly ad hoc nature.

To construct a random tree we simply randomly choose a branching parameter, according to some arbitrary distribution, for every node. When a node has \mbox{$b_i=0$}, it terminates the lineage. Some examples of random trees are shown in Fig.~\ref{fig:random_tree}.

\if 1\doublecol
	\begin{figure}[!htbp]
	\includegraphics[clip=true, width=0.475\textwidth]{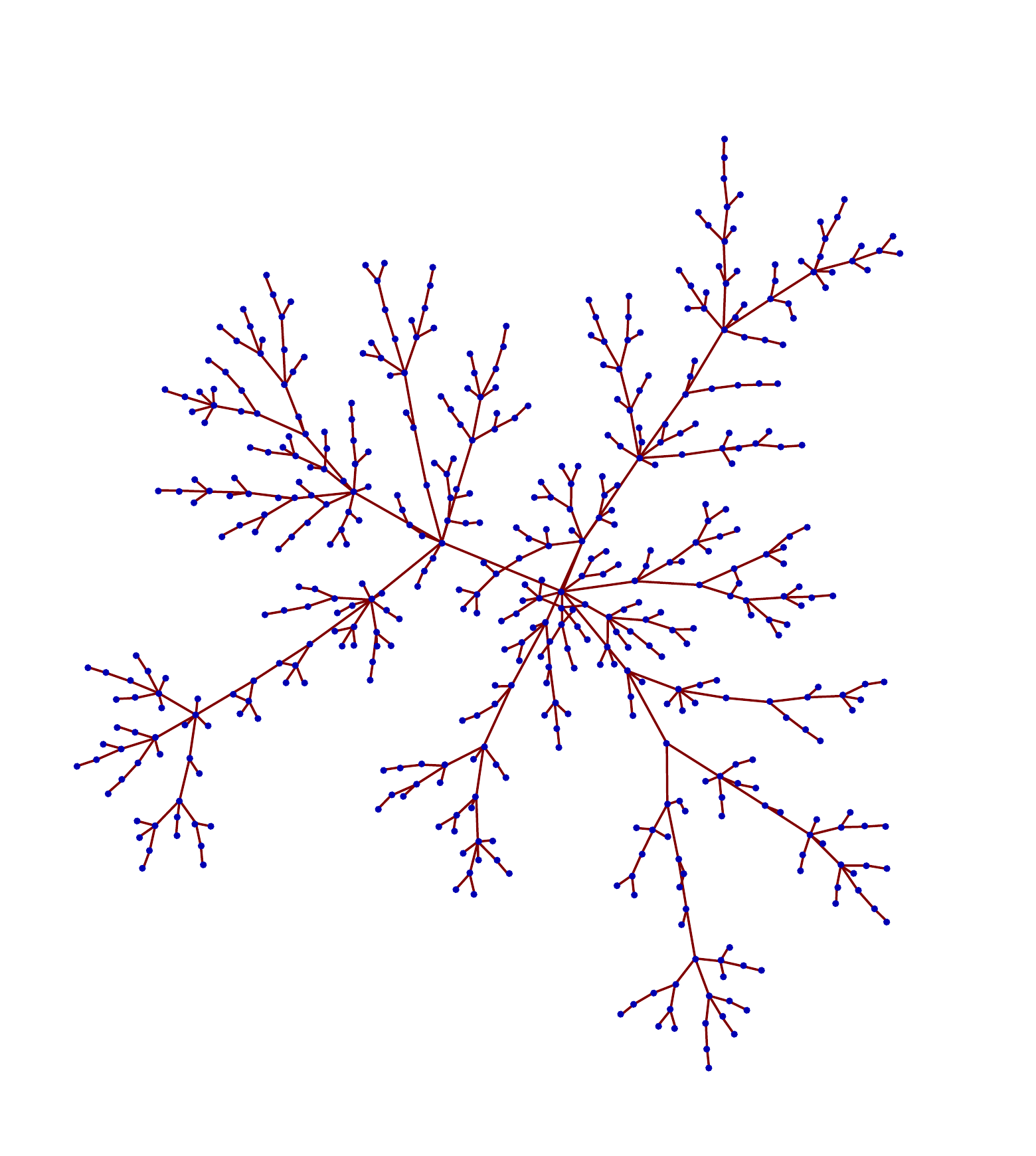}\\
	\includegraphics[clip=true, width=0.475\textwidth]{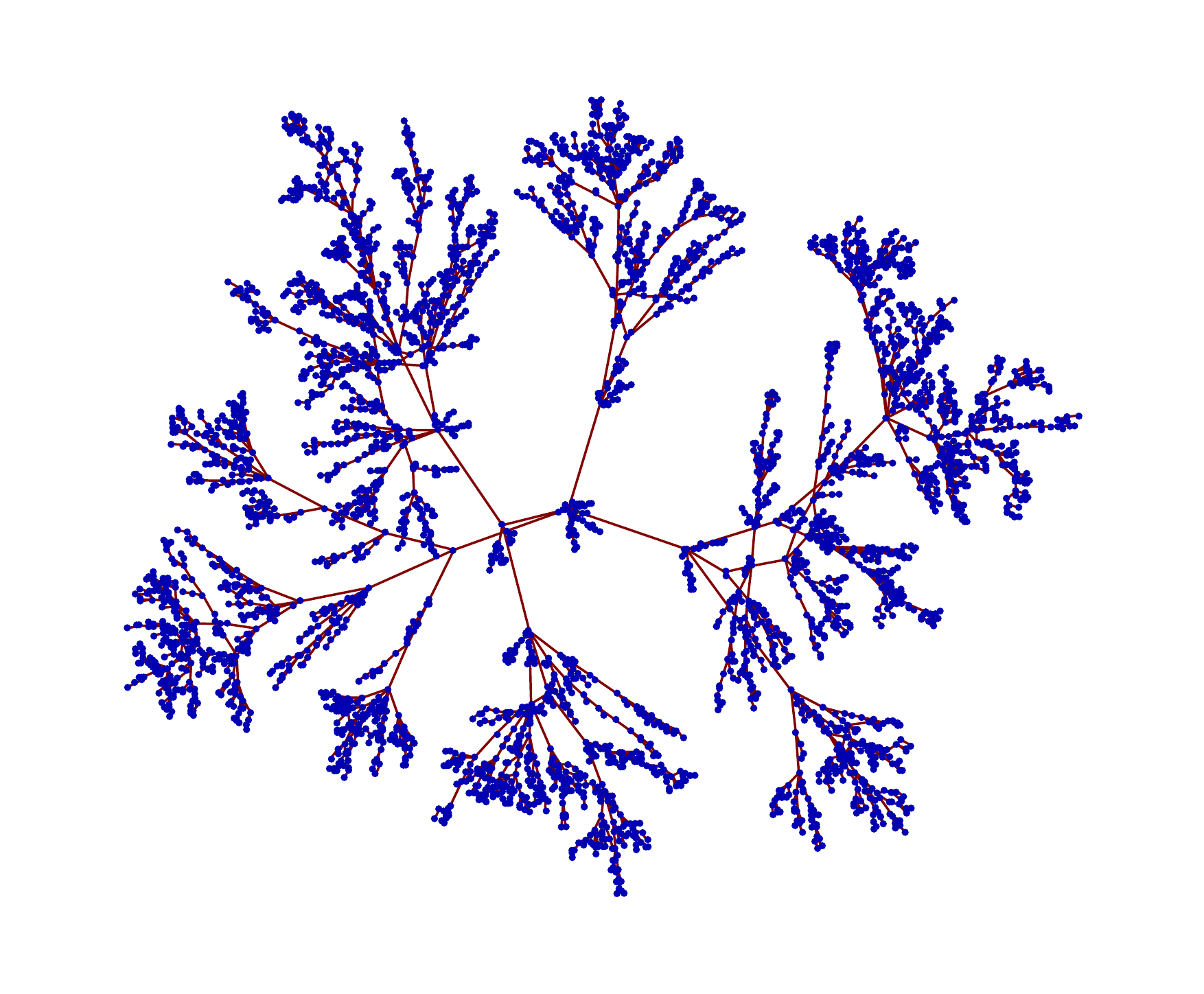}
	\captionspacefig \caption{Random trees with different randomised branching parameters (higher $b$ at the bottom). When a node has zero branches, it terminates the branch. This type of graph topology qualitatively captures the hierarchical, yet ad hoc qualities of many real-world networks, and may act as a useful test model for simulations.} \label{fig:random_tree}
	\end{figure}
\else
	\begin{figure*}[!htbp]
	\includegraphics[clip=true, width=0.475\textwidth]{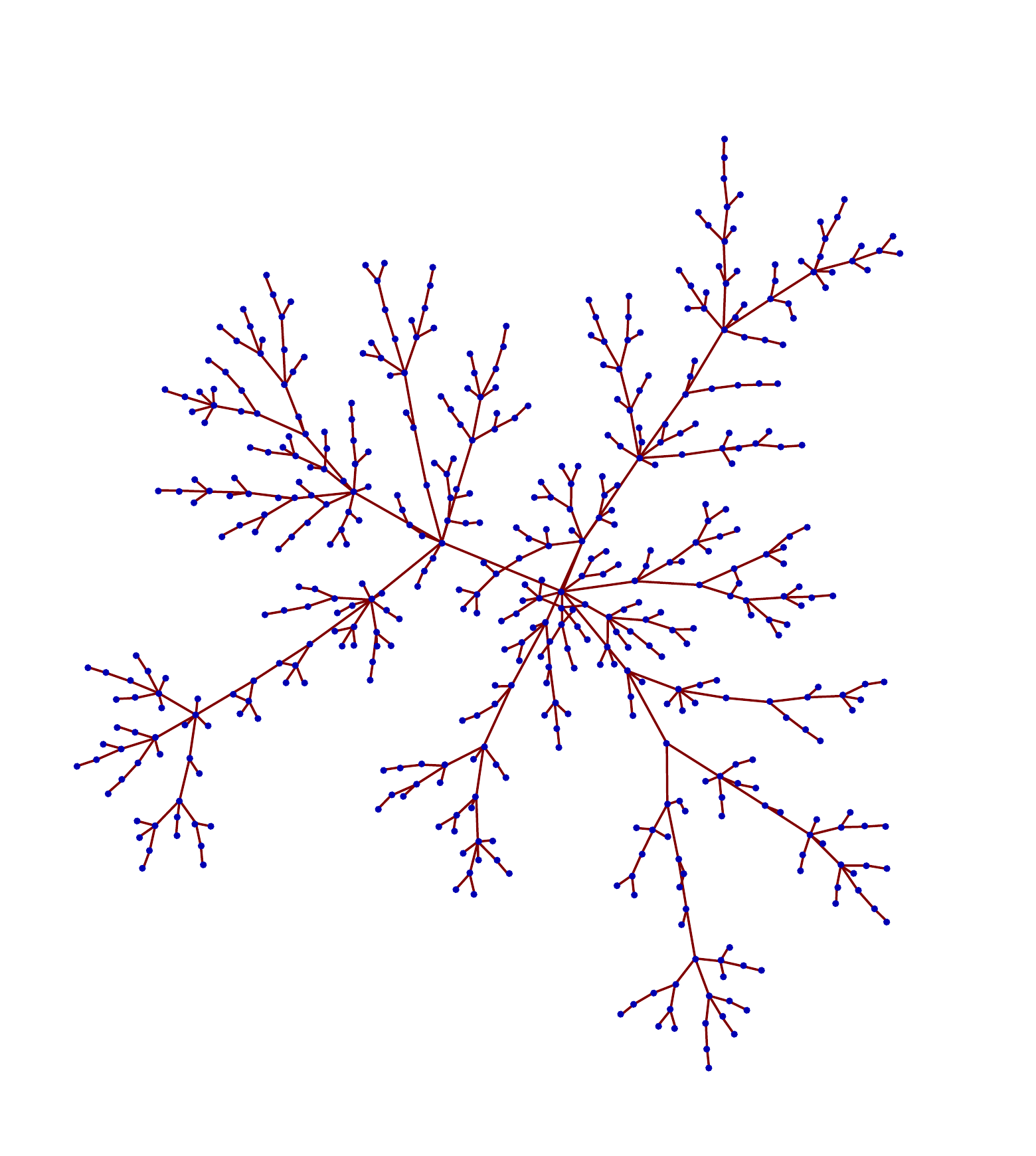}
	\includegraphics[clip=true, width=0.475\textwidth]{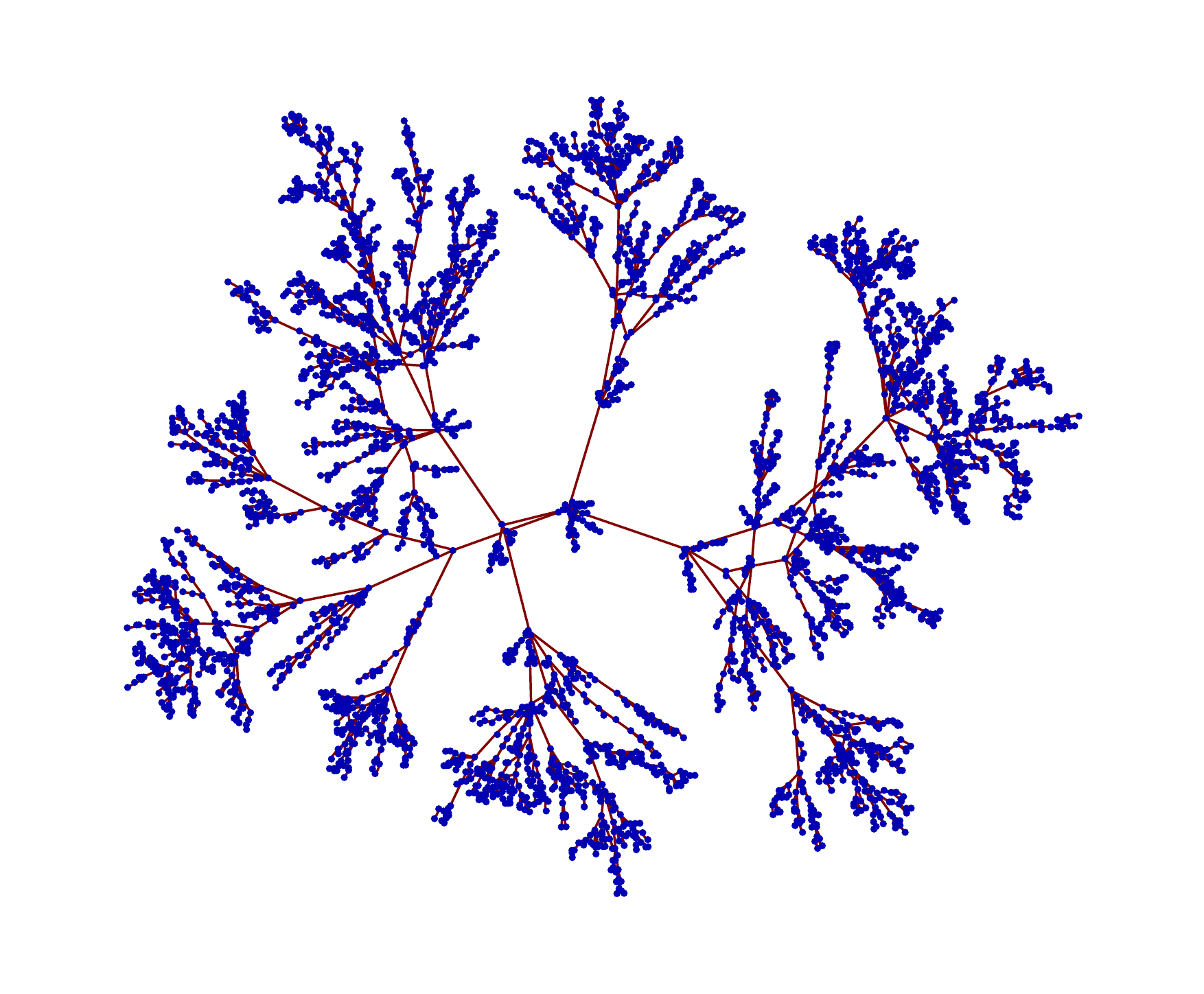}
	\captionspacefig \caption{Random trees with different randomised branching parameters (higher $b$ on the right). When a node has zero branches, it terminates the branch. This type of graph topology qualitatively captures the hierarchical, yet ad hoc qualities of many real-world networks, and may act as a useful test model for simulations.} \label{fig:random_tree}
	\end{figure*}
\fi

%
% Minimum Spanning Tree
%

\subsubsection{Minimum spanning tree} \label{sec:graph_MST} \index{Minimum spanning tree}

\begin{figure}[!htbp]
\includegraphics[clip=true, width=0.4\textwidth]{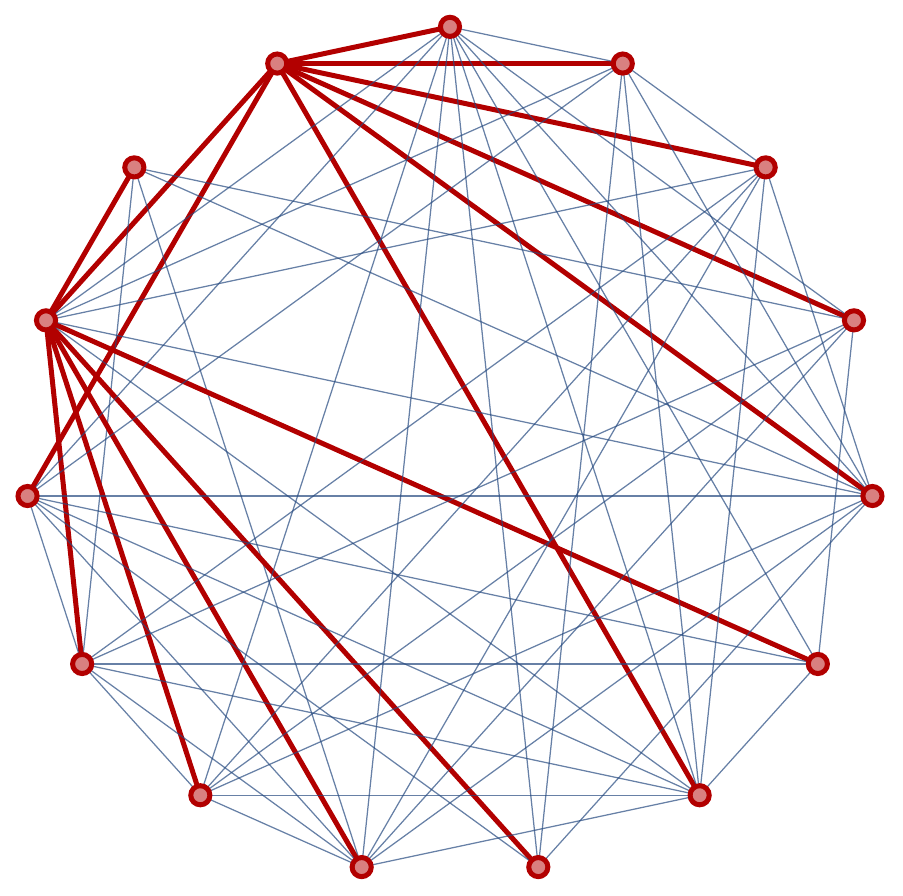}
\captionspacefig \caption{A random graph (blue) with an MST highlighted (red).} \label{fig:mst}
\end{figure}

A \textit{spanning tree}\index{Spanning tree} $S$, of a graph $G$, is a tree subgraph \mbox{$S\subset G$}, containing every vertex of $G$. The \textit{weight} of a spanning tree is the sum of all its constituent edge weights. Thus, the \textit{minimum spanning tree} (MST) is a spanning tree that minimises net weight. An example is shown in Fig.~\ref{fig:mst}. See Sec.~\ref{sec:min_tree} for a discussion on MST algorithms.

The calculation of MSTs is most likely to come into consideration when actually performing the initial construction of networks, where we wish to connect all nodes in the network, but using the most frugal possible physical resources. MSTs serve this purpose, and since they are trees, inherit all the same properties of tree networks.

In general, the MST of a graph is not unique, and there may be an arbitrarily large number of completely differently structured MSTs all with the same minimum weight.

%
% Percolation
%

\subsection{Percolation}\index{Percolation!Topologies}\label{sec:perc_topol}

A variation on any graph is to instead have a randomised implementation of it, whereby each of the possible edges or vertices occur with some probability, $p_\mathrm{edge}$ or $p_\mathrm{vertex}$, otherwise deleted. These are referred to as \textit{edge percolation} and \textit{site percolation} graphs respectively.

For any given graph, its associated percolation graph has average vertex and edge counts,
\begin{align}
|E|_\mathrm{av} &= p_\mathrm{edge}\cdot |E|,\nonumber\\
|V|_\mathrm{av} &= p_\mathrm{vertex}\cdot |V|.
\end{align}

Adjusting $p_\mathrm{edge/vertex}$ allows us to tune between the desired graph $G$ (when $p_\mathrm{edge/vertex}=1$) and the completely disconnected graph (when $p_\mathrm{edge/vertex}=0$).

This model is very useful in real-world applications, allowing unreliable channels/nodes to be incorporated into our network model. The analysis of such percolation networks is invaluable for understanding the robustness of such networks to channel and node failures.

Note that percolation graphs might be disjoint with sufficient defects, in which case the respective network becomes unreliable. Specifically, with sufficiently low $p_\mathrm{edge/vertex}$, `islands' may form in the network topology -- small segregated networks, which are unable to interface with the remainder of the network.

For asymptotically large percolation graphs, \textit{percolation theory}\index{Percolation!Theory} \cite{LI20211} provides thresholds for $p_\mathrm{edge/vertex}$ such that routes across the network exist in asymptotic limits.

Fig.~\ref{fig:perc_graph} illustrates several square lattice graphs with different percolation probabilities, and how the larger network segregates into smaller disconnected islands as failure rates increase.

\if 1\doublecol
	\begin{figure}[!htbp]
	\includegraphics[clip=true, width=0.35\textwidth]{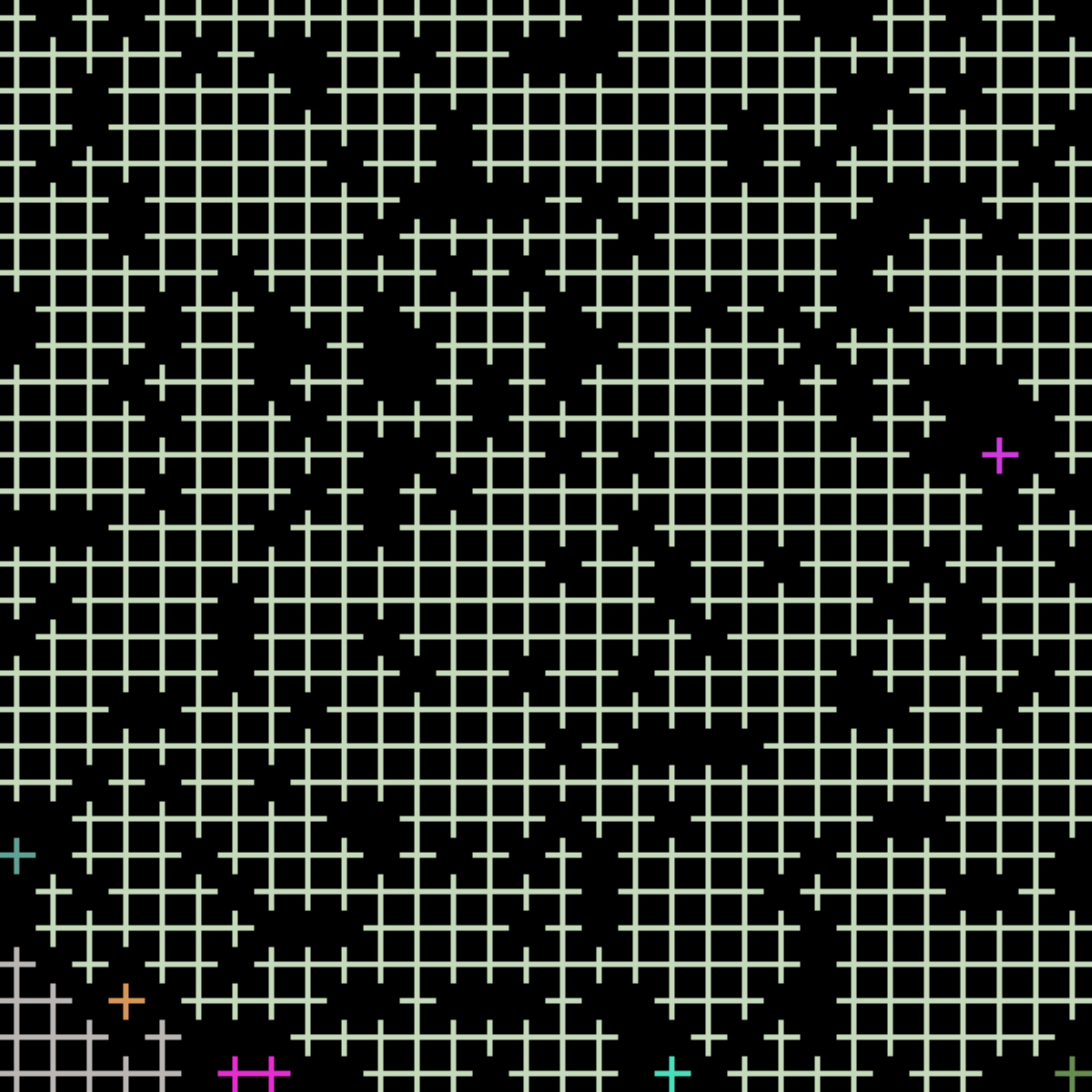}\\
	\includegraphics[clip=true, width=0.35\textwidth]{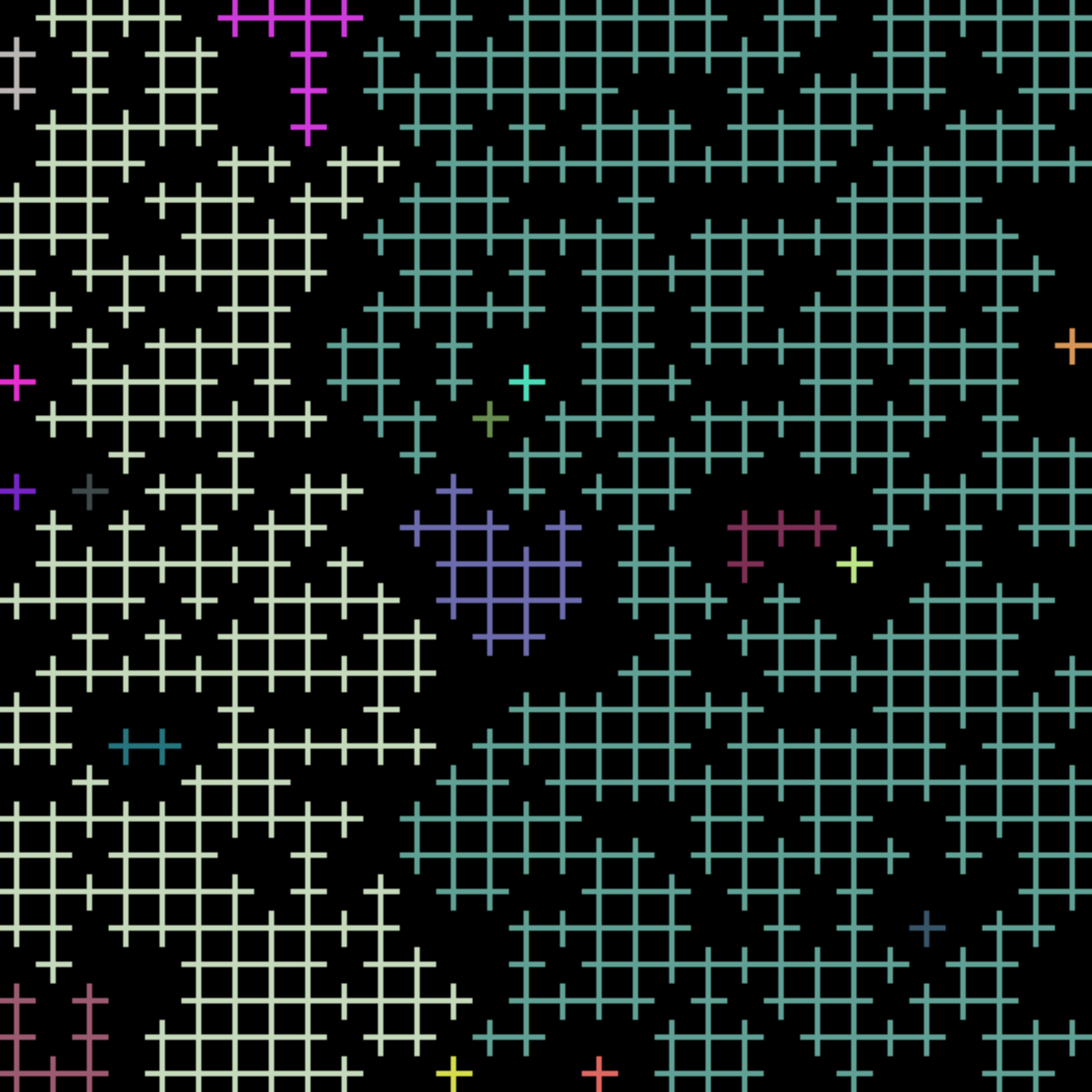}\\
	\includegraphics[clip=true, width=0.35\textwidth]{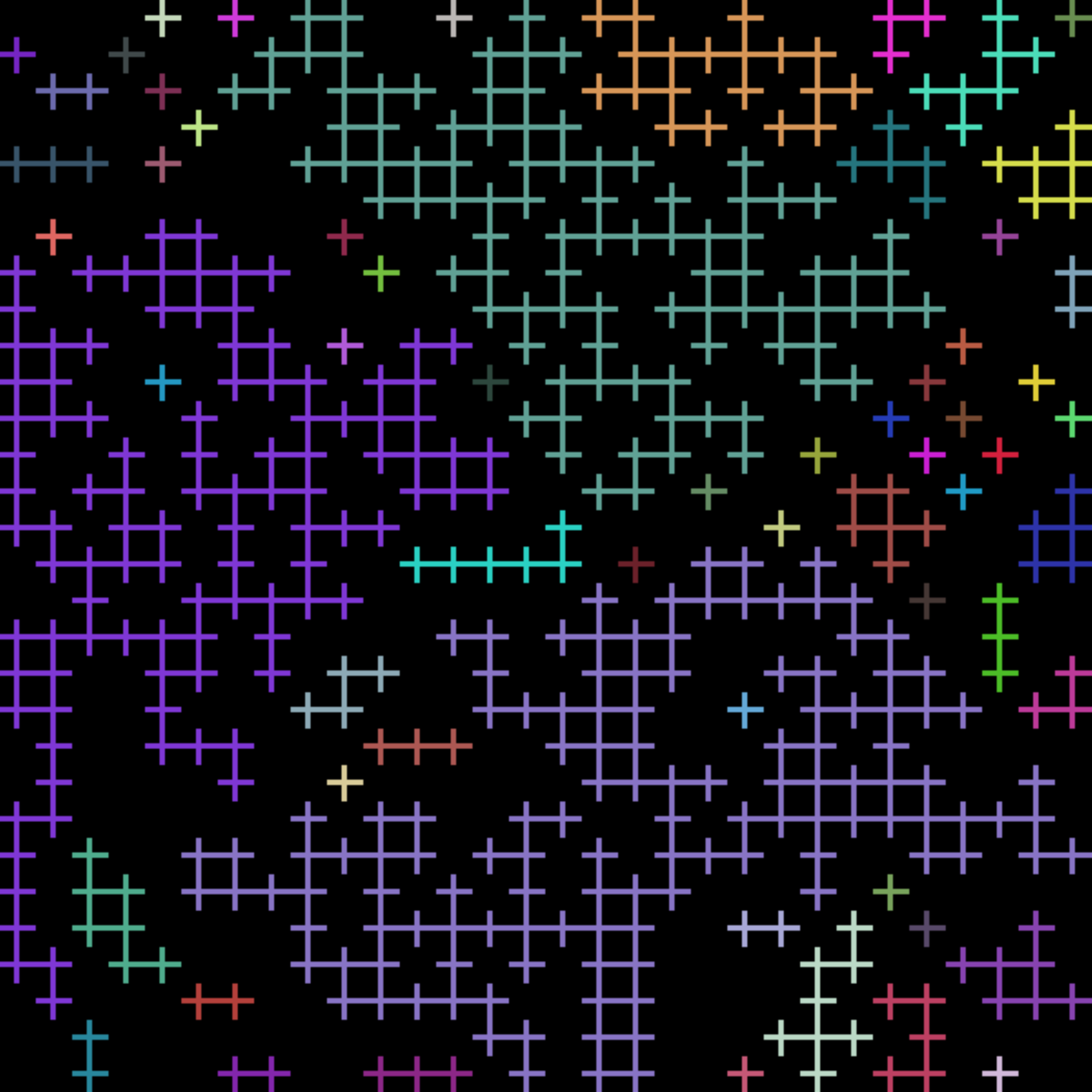}
	\captionspacefig \caption{A square lattice graph subject to different percolation rates (node defects). As the failure rate increases (top to bottom), the larger network segregates into a multipartite graph of smaller disjoint islands (denoted by colour).} \label{fig:perc_graph}\index{Percolation!Topologies}
	\end{figure}
\else
	\begin{figure*}[!htbp]
	\includegraphics[clip=true, width=0.325\textwidth]{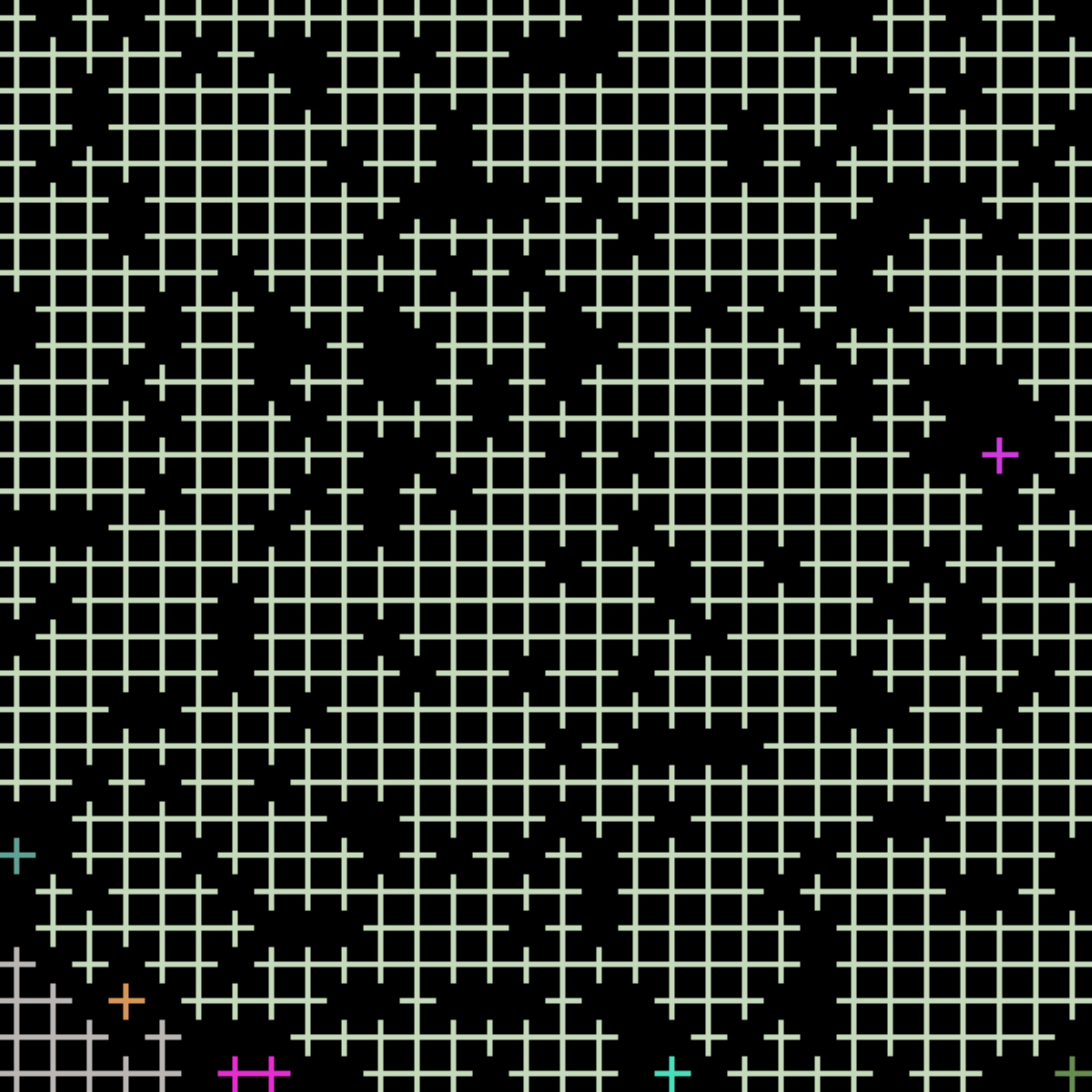}
	\includegraphics[clip=true, width=0.325\textwidth]{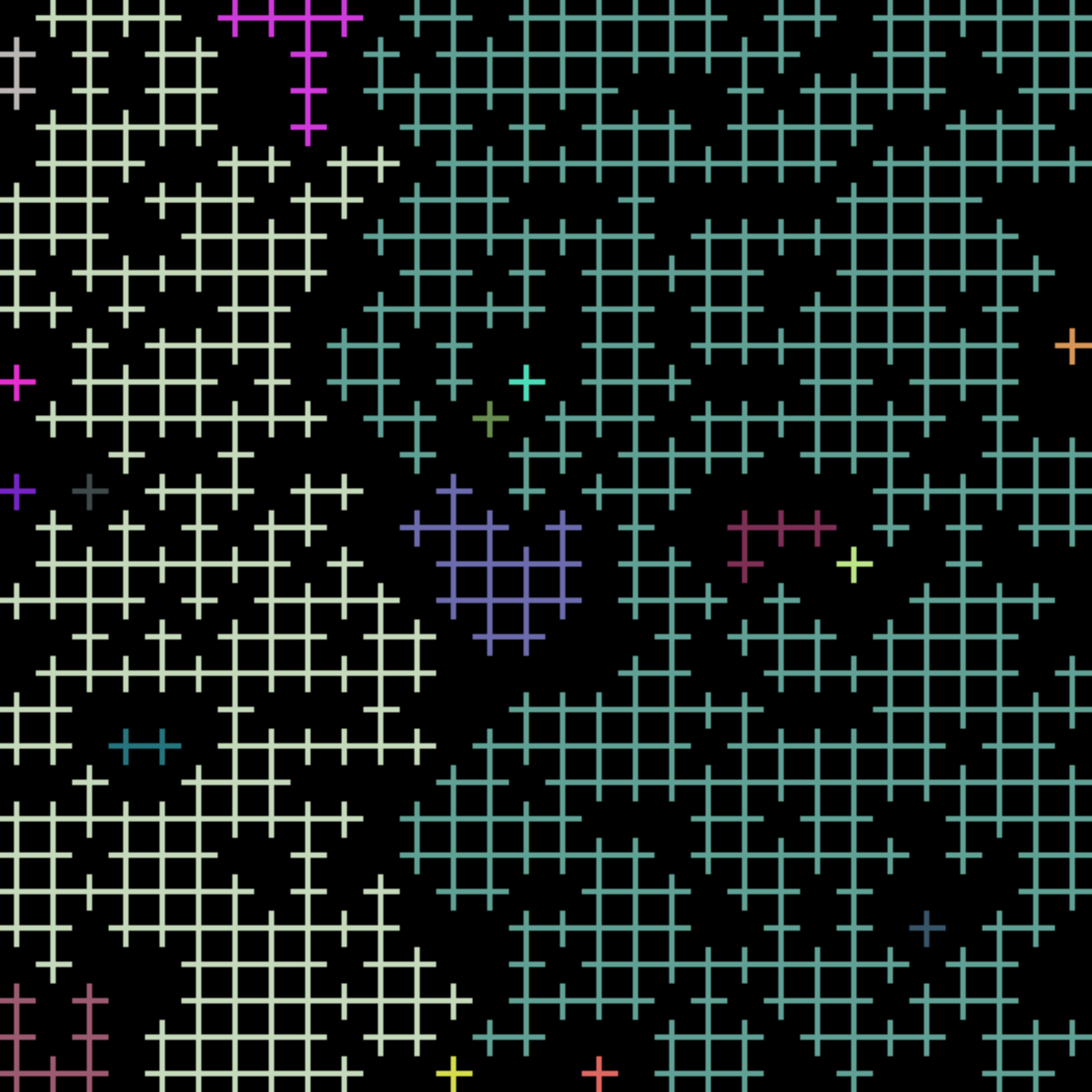}
	\includegraphics[clip=true, width=0.325\textwidth]{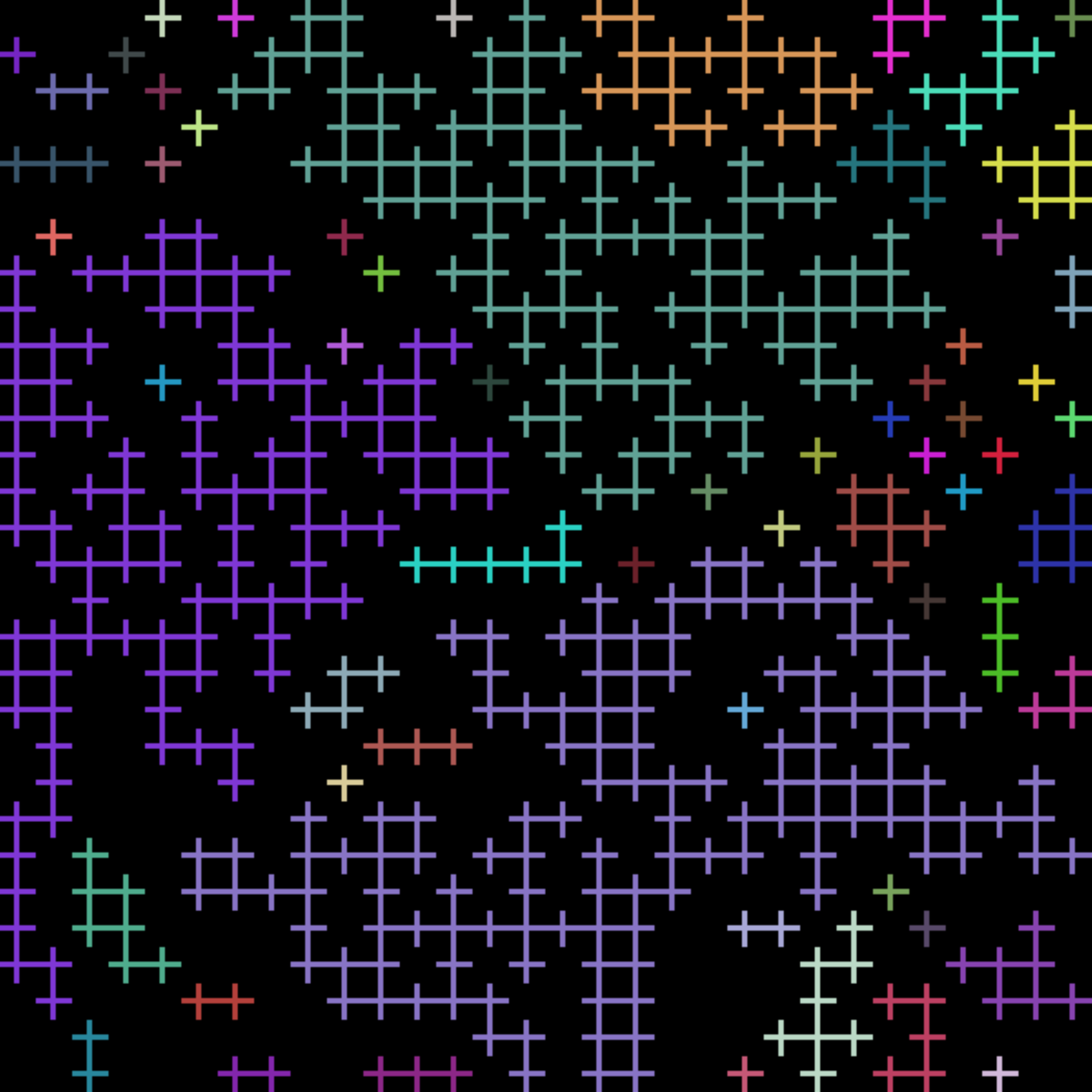}
	\captionspacefig \caption{A square lattice graph subject to different percolation rates (node defects). As the failure rate increases (left to right), the larger network segregates into a multipartite graph of smaller disjoint islands (denoted by colour).} \label{fig:perc_graph}\index{Percolation!Topologies}
	\end{figure*}
\fi

%
% Random
%

\subsection{Random}\label{sec:rand_topol}\index{Random!Topologies}

We refer to a random graph as being one in which edges between each pair of vertices occur with some probability $p_\mathrm{edge}$. No vertices are removed from the network, although some may have order \mbox{$|v|=0$}, i.e \mbox{$p_\mathrm{vertex}=1$}. This can be thought of as the edge percolation graph of the complete graph $K_{|V|}$.

The average number of edges in such a network scales as,
\begin{align}
|E|_\mathrm{av} = p_\mathrm{edge}\cdot O(|V|^2).	
\end{align}

Some examples are shown in Fig.~\ref{fig:random_graph}.

\if 1\doublecol
	\begin{figure}[!htbp]
	\includegraphics[clip=true, width=0.325\textwidth]{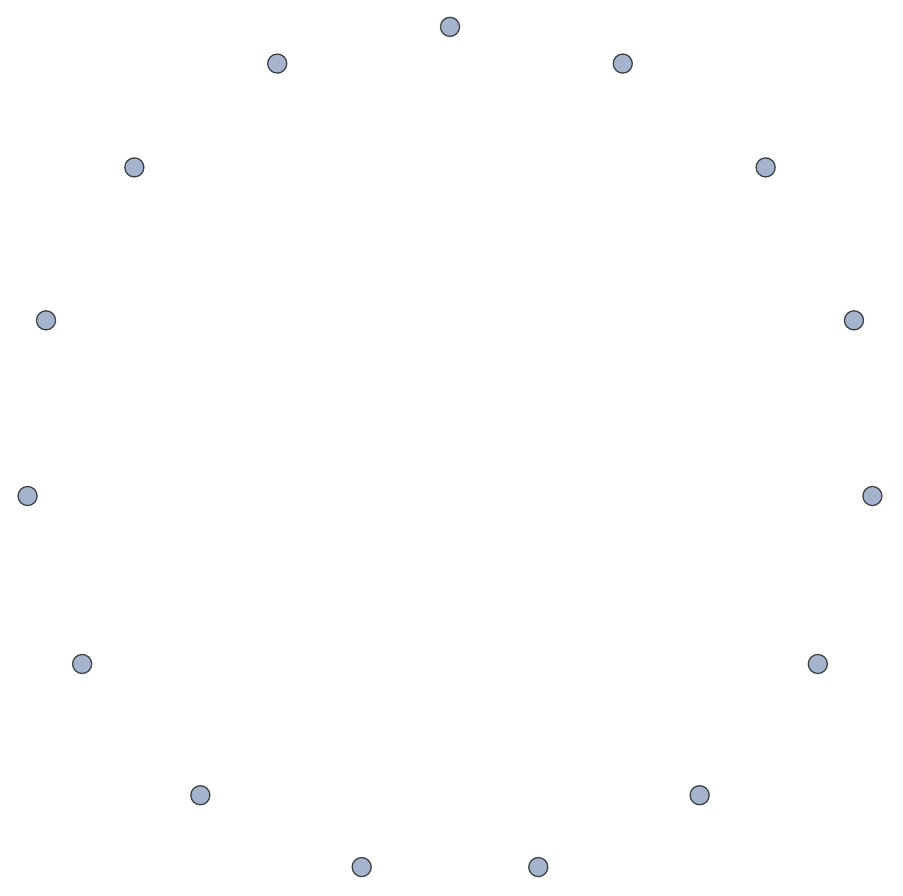}\\
	\includegraphics[clip=true, width=0.325\textwidth]{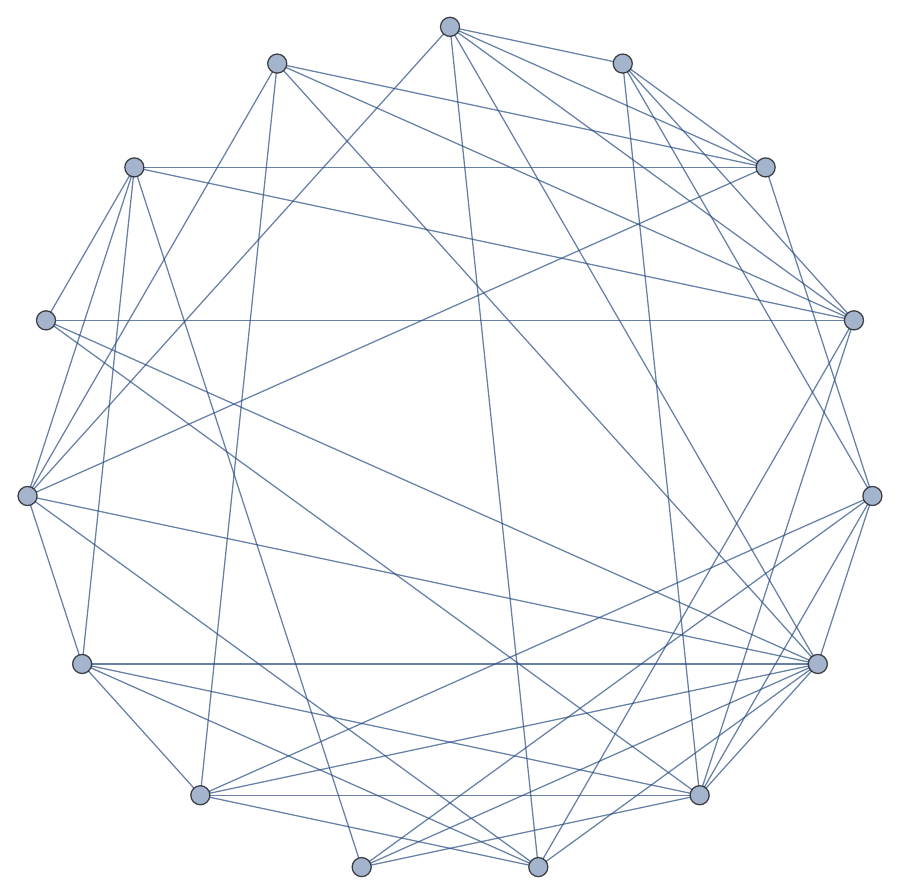}\\
	\includegraphics[clip=true, width=0.325\textwidth]{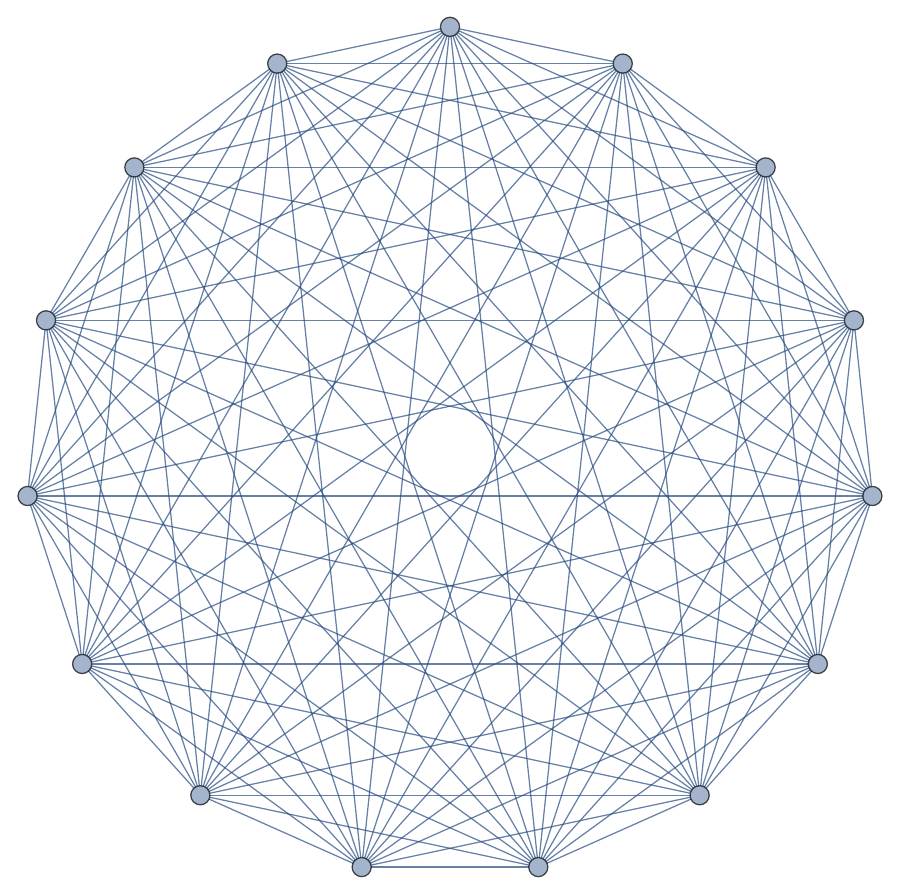}
	\captionspacefig \caption{The 15-vertex random graph. This is the same as $K_{15}$ in Fig.~\ref{fig:complete_graph}, but where edges are present with probabilities \mbox{$p_\mathrm{edge}=0,0.5,1$} (top to bottom).} \label{fig:random_graph}\index{Random!Topologies}
	\end{figure}
\else
	\begin{figure*}[!htbp]
	\includegraphics[clip=true, width=0.325\textwidth]{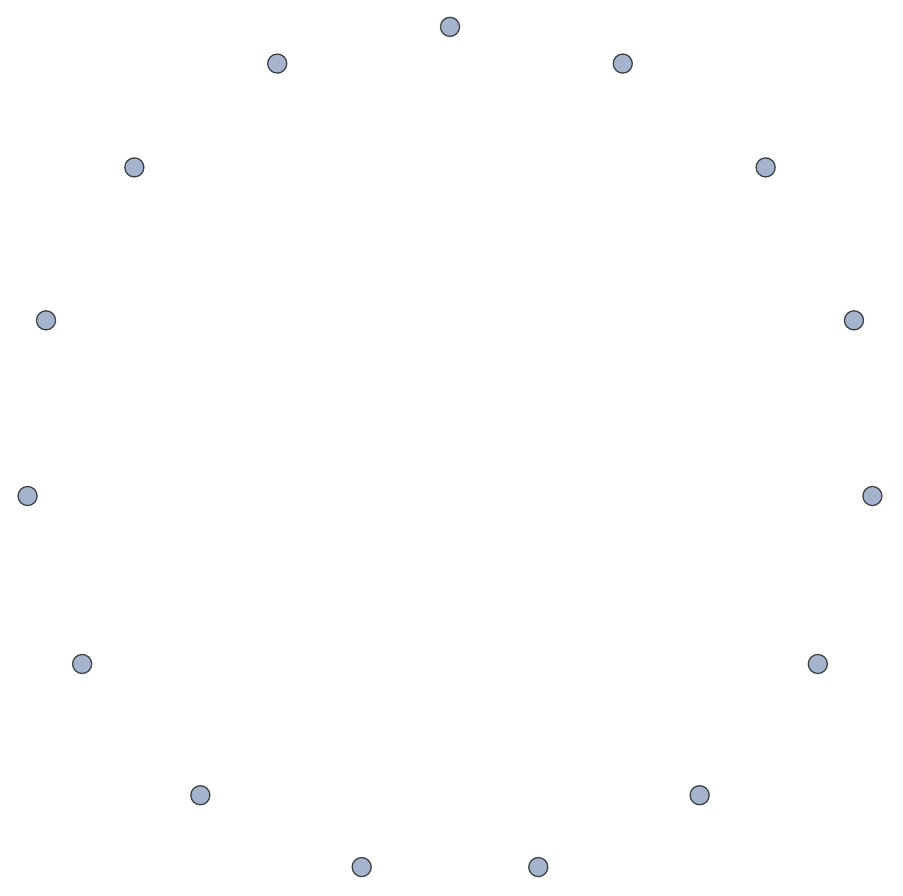}
	\includegraphics[clip=true, width=0.325\textwidth]{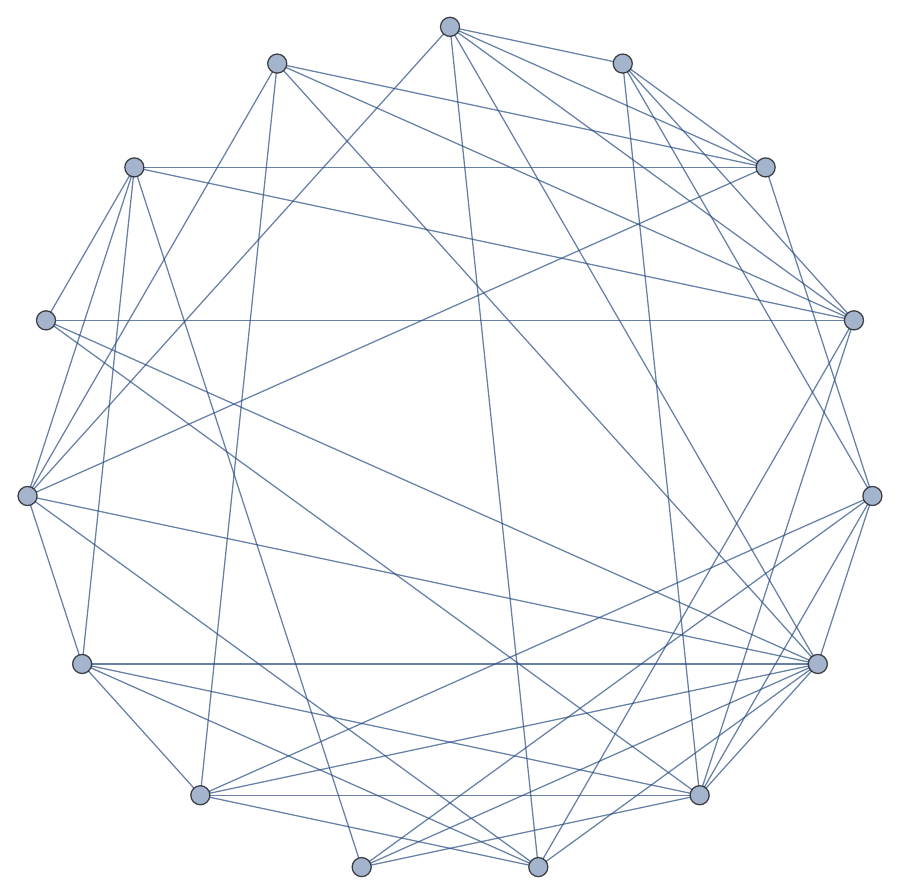}
	\includegraphics[clip=true, width=0.325\textwidth]{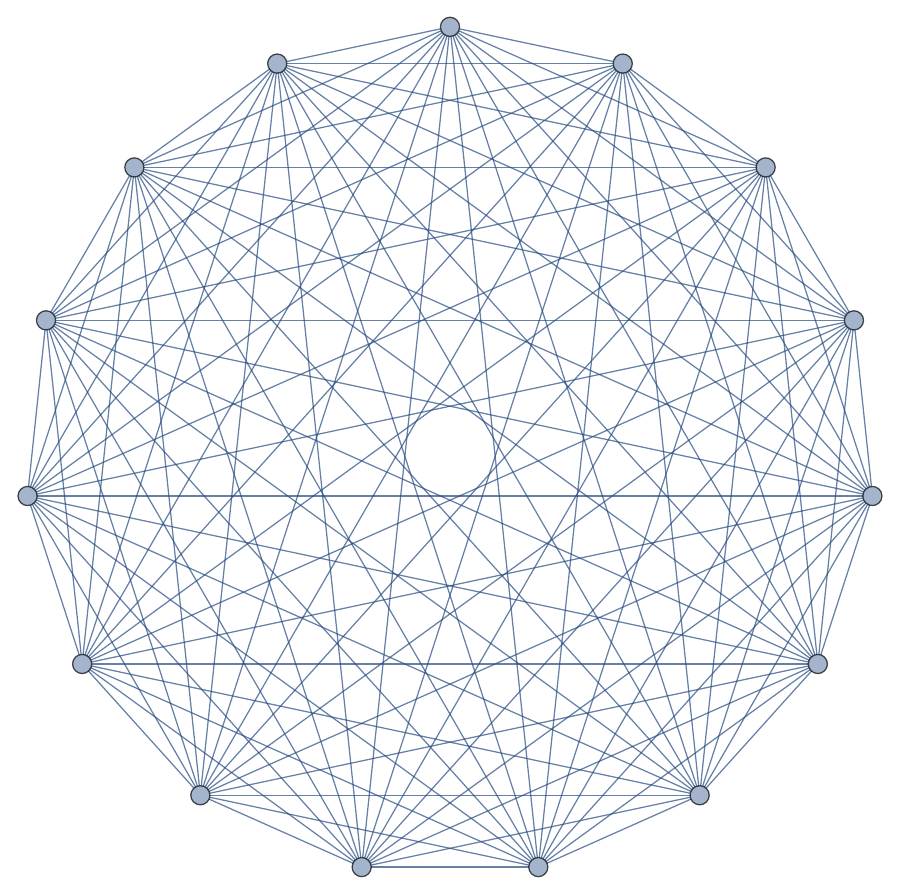}
	\captionspacefig \caption{The 15-vertex random graph. This is the same as $K_{15}$ in Fig.~\ref{fig:complete_graph}, but where edges are present with probabilities \mbox{$p_\mathrm{edge}=0,0.5,1$} (left to right).} \label{fig:random_graph}\index{Random!Topologies}
	\end{figure*}
\fi

%
% Hybrid
%

\subsection{Hybrid} \index{Hybrid!Topologies}

Real networks are highly unlikely to fit the exact form factor of any of the classes of graphs presented above. Rather, a truly global internet is inevitably going to comprise many subnetworks, each structured completely independently of one another, with little consistency or large-scale planning between them. Who thinks about the broader structure of the global internet when setting up their office network?

For example, at the global scale, it is entirely plausible that the internet might take on a random tree-like structure. But when we get down to a lower level, the tree structure vanishes and is replaced by all manner of different network topologies, run and maintained by different organisations in their own distinct ways.

Furthermore, the real-world internet is not simply a hierarchy of different types of well-known graph structures. Rather, it takes the form of `glued' graphs, whereby networks running over different mediums, or via different operators, each exhibit their own independent graph topologies, meeting at interconnect points that join the different networks. Typically this yields redundancy in the routes between different nodes, ushering in the need for combinatorial optimisation techniques when allocating network resources.

This hybrid network topology is the norm today in our classical internet, and it is entirely foreseeable that a similar trend will emerge in the future quantum internet as quantum technologies become more mainstream, their networking less well structured, and competing, redundant links are in place.

%
% Scale-Free Networks
%

\subsection{Scale-free networks}\index{Scale-free networks}\label{sec:scale_free_networks}

\begin{figure}[!htbp]
\includegraphics[clip=true, width=0.475\textwidth]{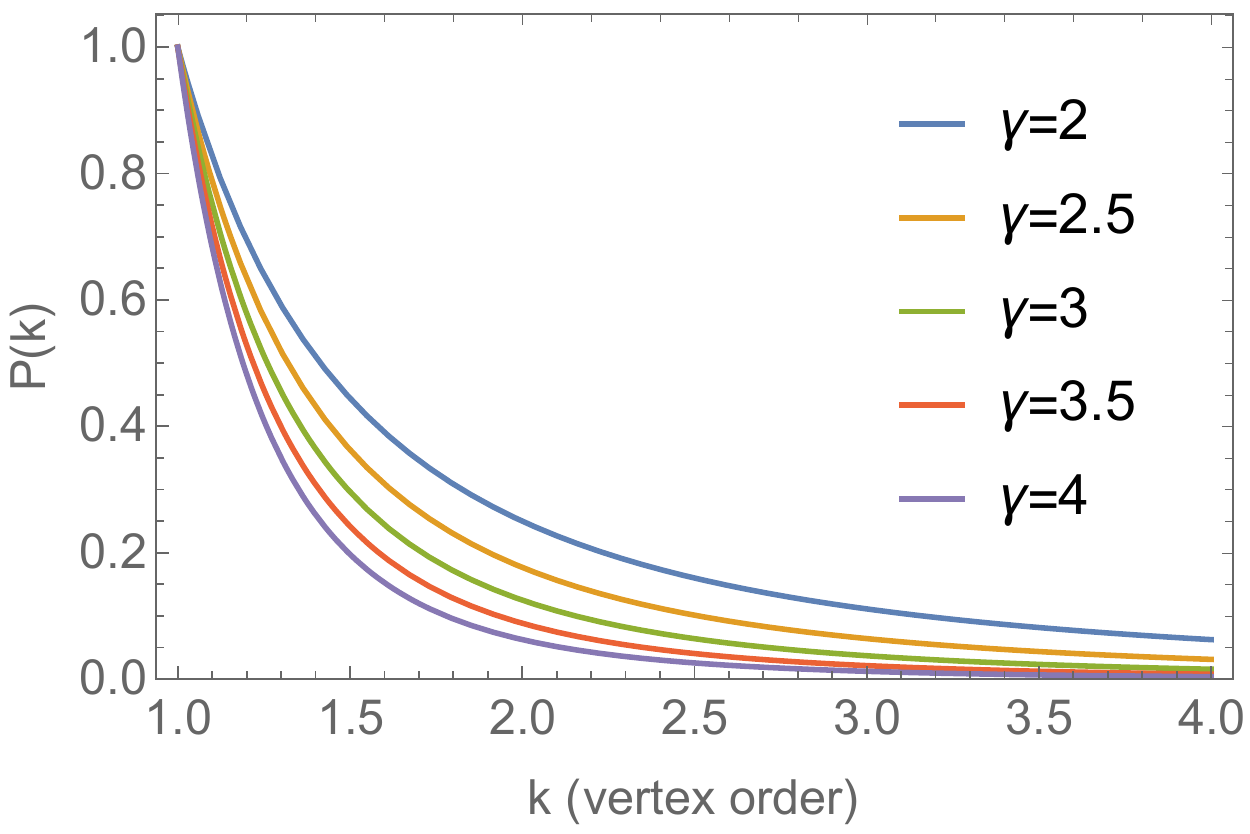}
\captionspacefig \caption{Examples of the power law, characteristic of the vertex order distribution in scale-free networks.}\label{fig:power_law}\index{Scale-free networks}\index{Power!Law}	
\end{figure}

Scale-free networks are not defined as obeying a specific topological structure, but rather as following a particular statistical distribution in the connectedness of their nodes. Specifically, the probability distribution function\index{Probability distribution function} for the order of vertices (degree distribution\index{Degree distribution}) roughly follows a Pareto distribution or power law\index{Power!Law},
\begin{align}\label{eq:pareto_dist}
	P(k) \sim k^{-\gamma},
\end{align}
where $P(k)$ is the probability of a vertex having order $k$ (up to normalisation), and \mbox{$\gamma>1$}. Most commonly \mbox{$2\leq\gamma\leq 3$}. Fig.~\ref{fig:power_law} illustrates this scaling behaviour for different power coefficients.

This distribution is observed empirically in many real-world networks and sociological structures, and as such is more an observation about the typical behaviour of naturally occurring and evolving human-made networks than an explicit definition for their construction. However, owing to this particular statistical behaviour, and the underlying causations for their Pareto distribution, much has been researched and is known about the properties of scale-free networks.

Scale-free networks arise naturally in systems exhibiting \textit{preferential attachment}\index{Preferential attachment}, i.e when a new node is added to the system it preferentially attaches to nodes that are already more highly connected. This yields a so-called \textit{fitness model}\index{Fitness!Model}.

According to the Bianconi-Barab{\'a}si fitness model\index{Bianconi-Barab{\'a}si fitness model} \cite{bib:BBfitness, bib:BAfitness}, let \mbox{$\eta_i>0$} be the \textit{fitness factor}\index{Fitness!Factors} of node $i$, which follow a distribution $\rho(\eta)$, a characteristic of the network. Then the fitness parameters\index{Fitness!Parameters} are defined to be normalised such that,
\begin{align}
\Pi_i = \frac{\eta_i k_i}{\sum_j \eta_j k_j}.	
\end{align}
Upon adding a new node of degree $m$ to the network, the temporal dynamics will satisfy,
\begin{align}
\frac{\partial P(k_i)}{\partial t}= m\Pi_i.	
\end{align}
The probability distribution can then be shown to have solution,
\begin{align}
P(k) \approx \int \rho(\eta)\frac{C}{\eta}\left(\frac{m}{k}\right)^\frac{C}{\eta+1}\,d\eta,
\end{align}
where,
\begin{align}
	C &= \int \frac{\rho(\eta)\cdot\eta}{1-\beta(\eta)}\,d\eta,\nonumber\\
	\beta(\eta) &= \frac{\eta}{C},
\end{align}
which is a linear combination of power law relationships, as required for the definition for scale-free.

Intuitively, why would we expect computer networks (classical or quantum) to be scale-free? To answer this, we simply must establish whether the preferential attachment property will hold. In computer networks there are several reasons why we might expect this to be the case:
\begin{itemize}
	\item Distance\index{Cost vector analysis}: connecting to more highly-connected nodes reduces (on average) the number of channels data packets must traverse to reach their destination, making them `cheaper' in terms of their cost vector analysis (Sec.~\ref{sec:quantum_meas_cost}).
	\item Availability\index{Availability}: larger nodes are more likely to have unused network sockets\index{Sockets} available for use by new nodes, and they are likely to be more readily accessible. For example, one is more likely to be able to successfully sign up for an internet connection with a major national ISP than a small, local upstart player.
	\item Economies of scale\index{Economies of scale}: the dollar cost per connection of a larger node is likely to be less than for a smaller one, owing to economies of scale. For example, the cost per FLOP of a large-scale supercomputer is far less than for a desktop PC, and Google's data-centres experience lower cost-per-bandwidth on their internet connections than home-users connecting via their ISPs.
\end{itemize}

\begin{figure}[!htbp]
\includegraphics[clip=true,width=0.4\textwidth]{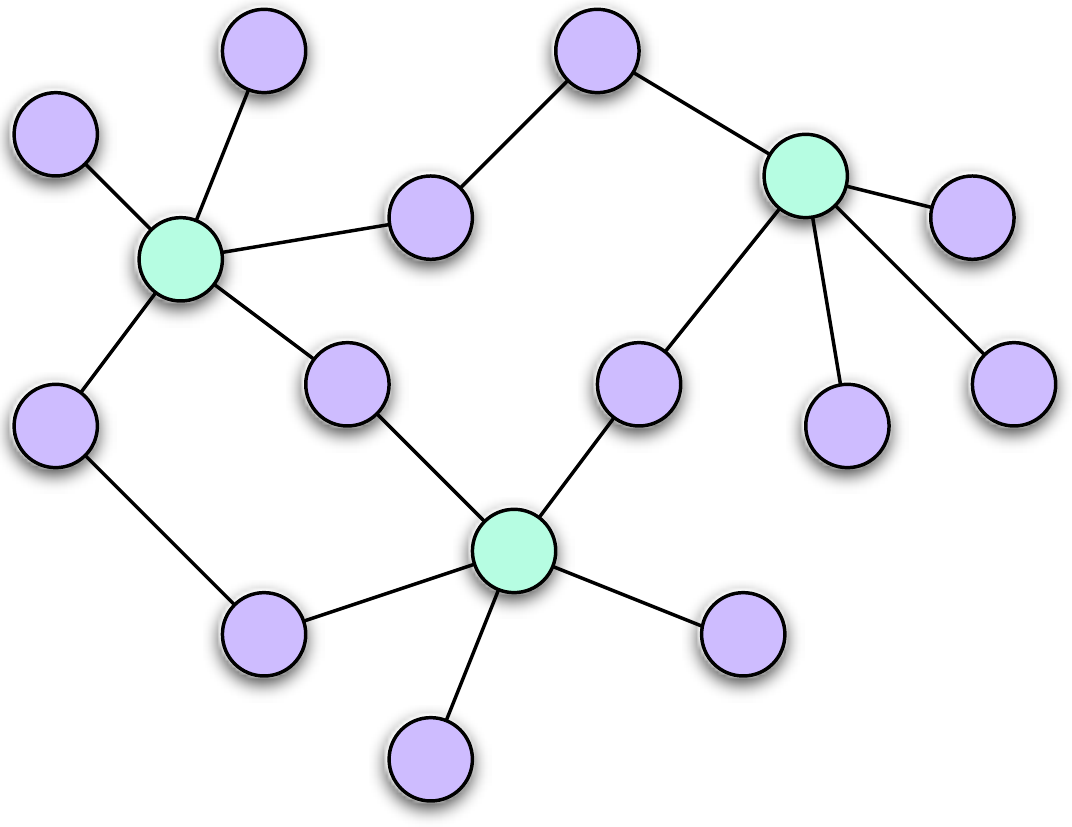}
\captionspacefig \caption{Example of a scale-free network. The graph exhibits a highly hierarchical structure, with a small number of `hub' nodes that are highly connected (green, with vertex order 5), that connect downward to a larger number of less connected nodes (purple, with vertex order 1-2).}\label{fig:scale_free_net}
\end{figure}

One notable characteristic of scale-free networks is their hierarchical structure, with a small number of very highly-connected `hubs'\index{Hubs} at the top of the food chain, which quickly connect onto smaller hubs, and so on down the food chain with decreasing connectivity, as shown in Fig.~\ref{fig:scale_free_net}.

A feature of scale-free networks of especial interest in the context of computer networks is their robustness against node failure. Suppose we constructed a percolated (Sec.~\ref{sec:perc_topol}) instance of a typical scale-free network. Such a network is highly robust against \textit{random} node/edge deletions compared to many other graph constructions, in the sense that a relatively large number of failures must occur to disconnect the graph. This makes the scale-free network characteristic a particularly attractive one from the perspective of the failure-tolerance of a network. It should be noted that, on the other hand, a scale-free network is highly vulnerable to \textit{targeted} node/edge deletions, that specifically target the highly-connected nodes. A targeted attack against major hubs could disconnect the network with relatively few successful attacks. This brings with it important geostrategic\index{Geostrategic politics} considerations when constructing network infrastructure.

Scale-free networks typically exhibit extremely small diameter\index{Graphs!Diameter} (average distance between nodes), scaling as \cite{bib:PhysRevLett.90.058701},
\begin{align}
	d = O(\log \log |V|).
\end{align}
That is, they exhibit (exponentially) smaller diameter than tree graphs (which already exhibit only logarithmic depth). Thus, expanding the network (in a manner consistent with the model) by adding a moderate number of new nodes effectively leaves graph diameter unchanged -- the graph diameter is virtually a constant under modest evolution.

%
% The Internet Web-Graph
%

\subsection{The internet web-graph} \index{Internet web-graph}

Of course, all the topological structures described until now are in-principle constructs. Of most relevance is the \textit{internet web-graph}, the graph of the actual internet (or some other real-world network).

Fig.~\ref{fig:webgraph} illustrates some example web-graphs, constructed from subsets of data from the actual internet. The combination of random, densely and sparsely connected, and tree structures, and its clear hierarchy are all evident. This encourages our intuition of the different types of structures present in realistic networks.

\if 1\doublecol
	\begin{figure}[!htbp]
	\includegraphics[clip=true, width=0.475\textwidth]{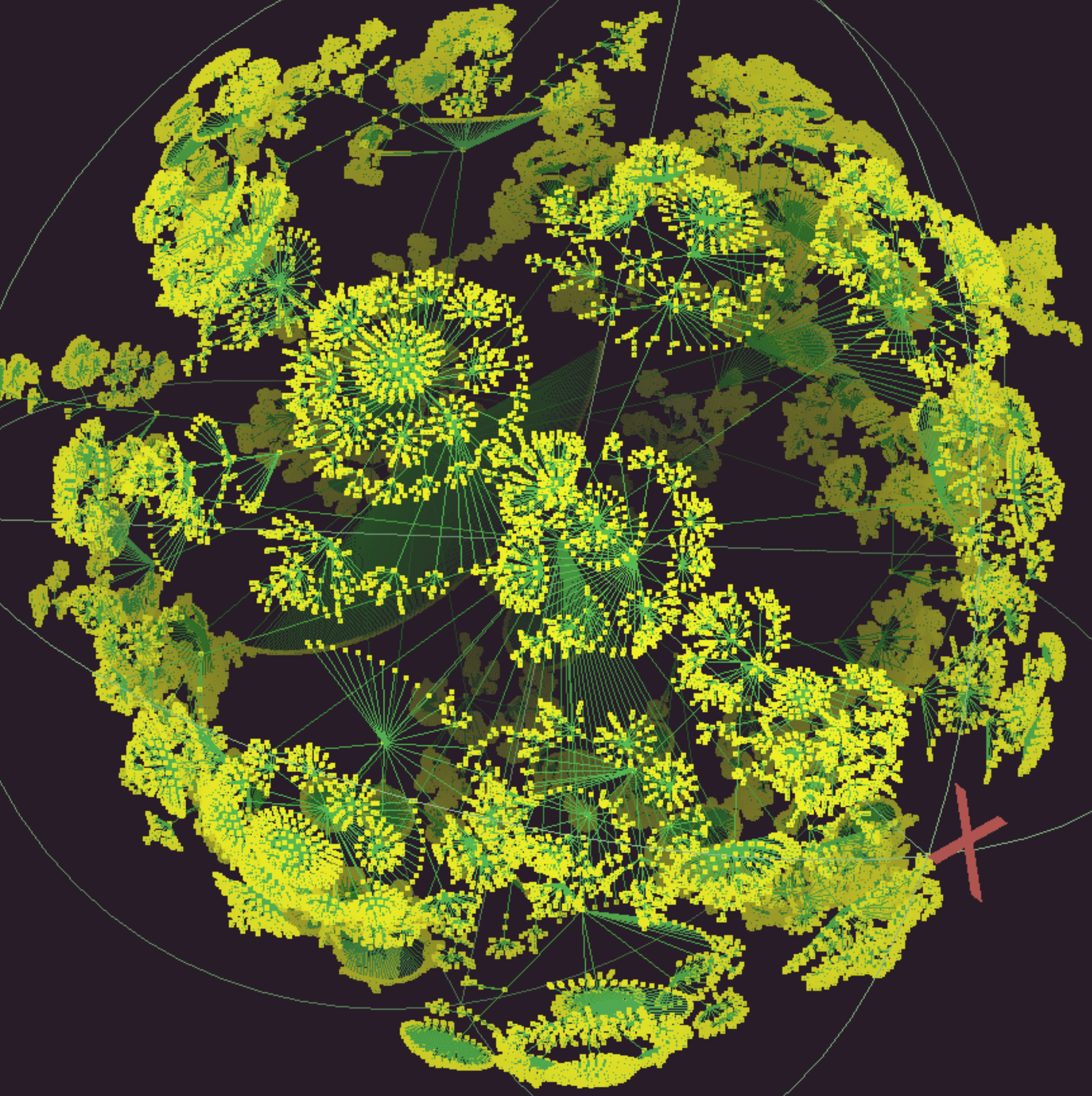}\\
	\includegraphics[clip=true, width=0.475\textwidth]{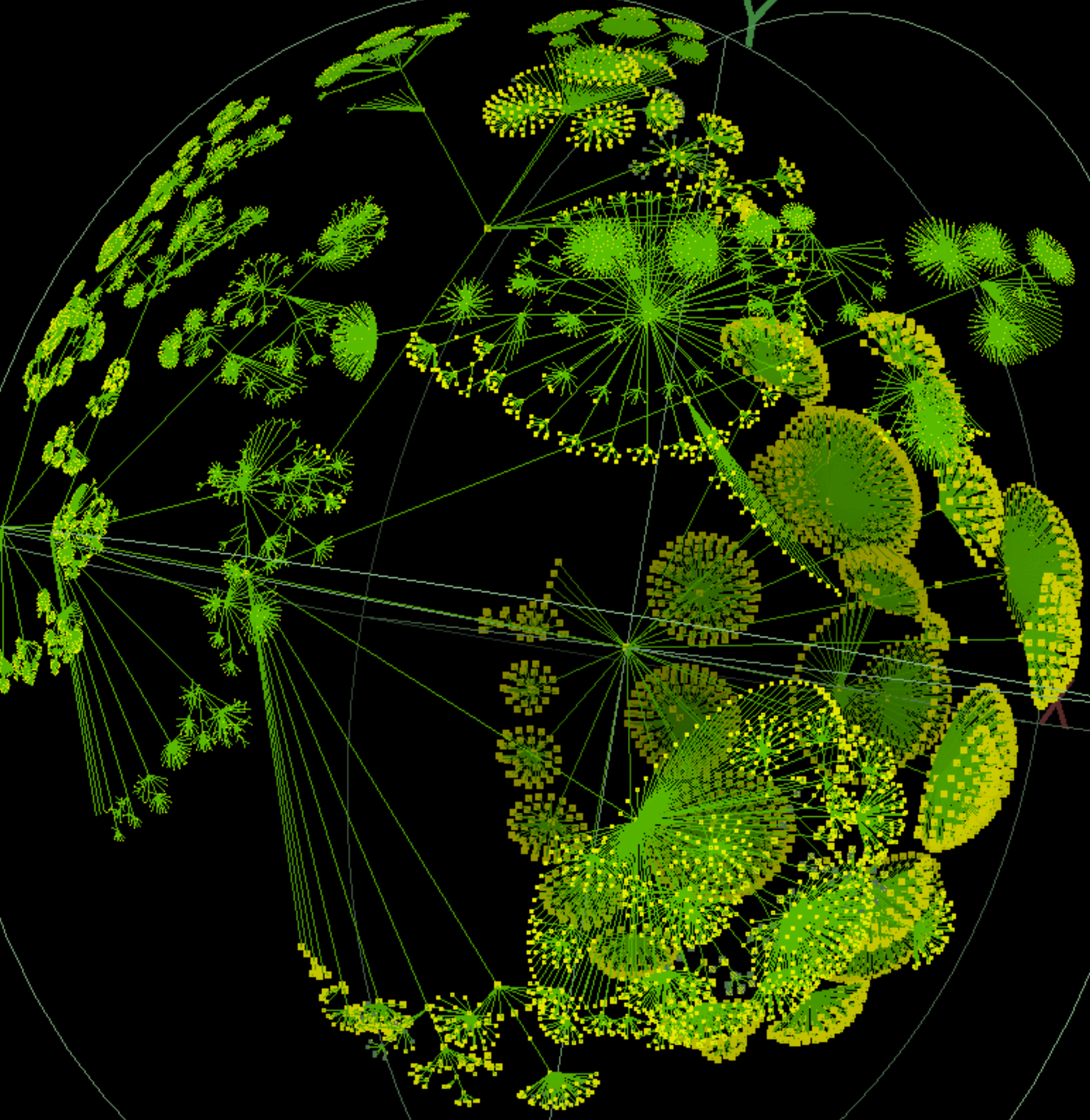}
	\captionspacefig \caption{Examples of real-world web-graphs of the internet, capturing their high-level random tree-like structure. Graphics attributed to the Center for Applied Internet Data Analysis (CAIDA), \url{http://www.caida.org}.}\index{Internet web-graph} \label{fig:webgraph}
	\end{figure}
\else
	\begin{figure*}[!htbp]
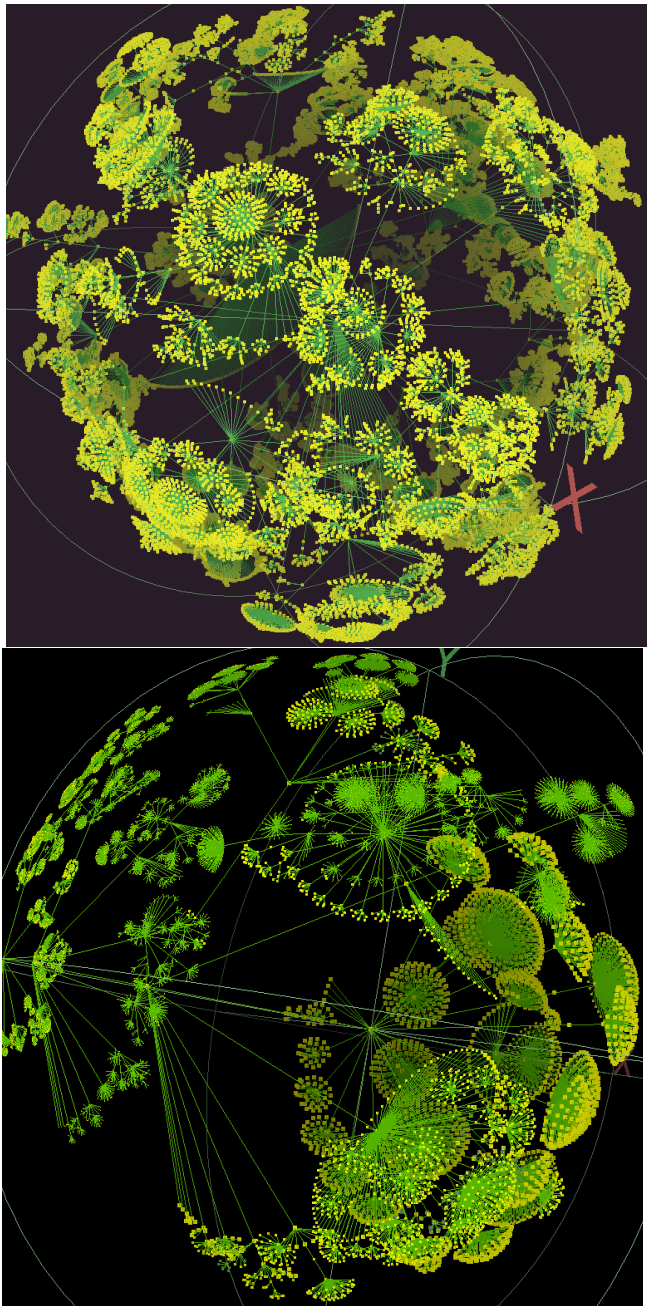

	\includegraphics[clip=true, width=0.481\textwidth]{webgraph_1}
	\includegraphics[clip=true, width=0.475\textwidth]{webgraph_2}
	\captionspacefig \caption{Examples of real-world web-graphs of the internet, capturing their high-level random tree-like structure. Graphics attributed to the Center for Applied Internet Data Analysis (CAIDA), \url{http://www.caida.org}.}\index{Internet web-graph} \label{fig:webgraph}
	\end{figure*}
\fi

The internet web-graph has been observed to be a scale-free network (Sec.~\ref{sec:scale_free_networks}), observing its power law distribution in node connectivity, as per Eq.~(\ref{eq:pareto_dist}).

%
% Network Robustness
%

\subsection{Network robustness}\index{Network!Robustness}

A key feature of any network topology is its robustness against node or channel failures. This is important from the perspective of naturally occurring hardware faults, and also from a geostrategic perspective, where adversaries may be launching attacks against the network. In general, there are two main contributing factors to network robustness:
\begin{itemize}
	\item Redundancy\index{Redundancy}: the number of redundant paths between two points in a graph stipulates how many backups there are to finding a route to a destination in the advent of one route failing.
	\item Diameter\index{Diameter}: the chance of a data packet encountering a faulty node/channel increases with the number of hops required to the reach its destination. Graphs with smaller diameter are hence less vulnerable.
\end{itemize}

The extreme case of network robustness is the complete graph, $K_n$, which has P2P links between every pair of nodes. Therefore, if a single channel fails, there are \textit{always} alternate paths taking us between nodes. On the opposing extreme are tree graphs, which contain no redundancy whatsoever, and just a single failure will disconnect the network, making certain routes impossible. Scale-free networks sit in the intermediate zone, but are relatively robust against the failure of random nodes/links, but are vulnerable to conspiratorial failures, which target the elite, highly connected hub-nodes\footnote{The 1\%.}.

Fig.~\ref{fig:graph_deletions} illustrates some examples of the robustness of these two extreme cases to link and node failure.

\if 1\doublecol
	\begin{figure}[!htbp]
	\includegraphics[clip=true, width=0.475\textwidth]{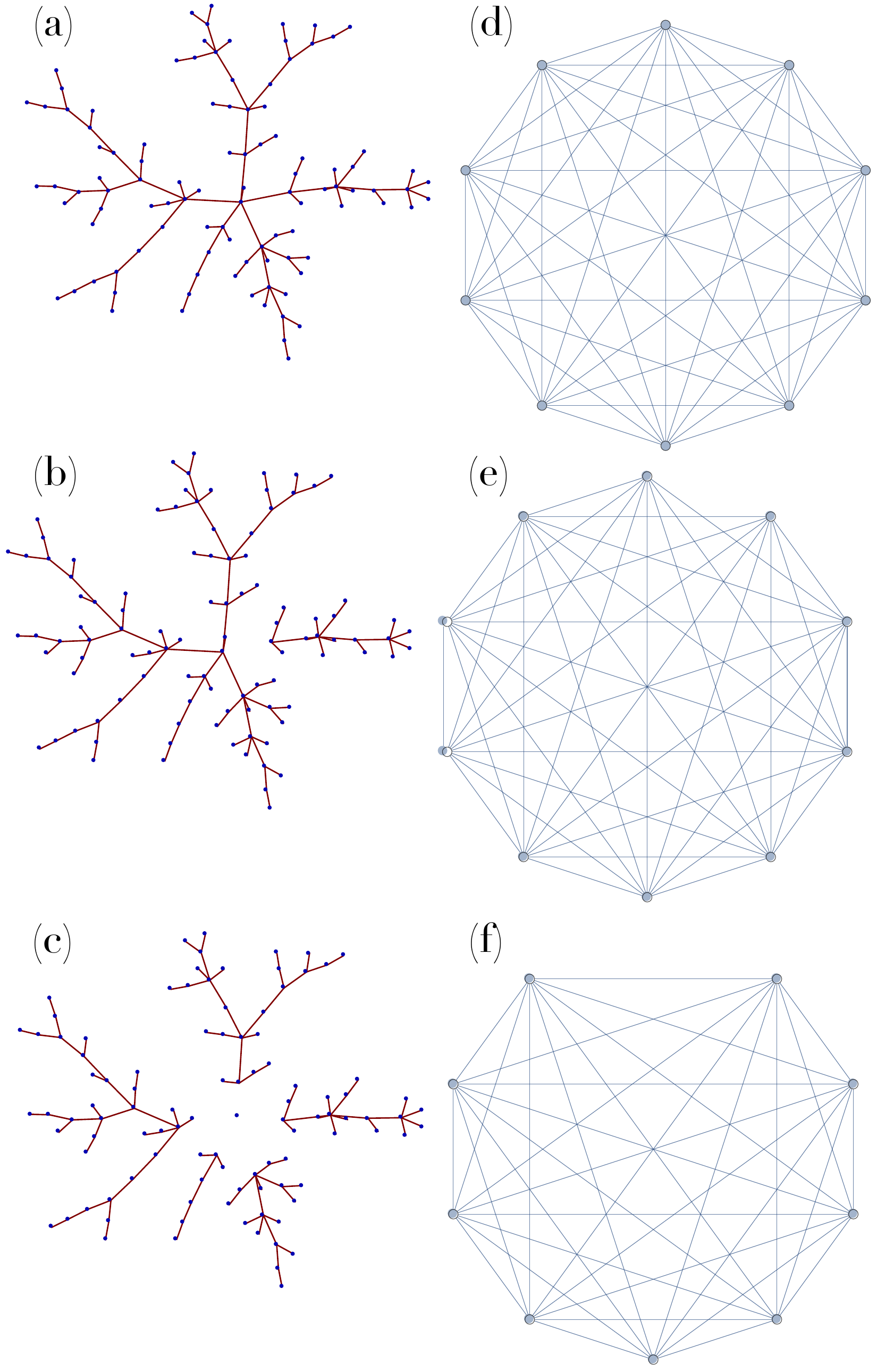}
	\captionspacefig \caption{Robustness of network topologies to node and link deletion. Examples of a tree graph (a) and a complete graph, $K_n$, with \mbox{$n=10$} (d). (b,e) The same graphs subject to a single link failure. The failure disconnects the tree graph into a bipartite graph (b), whereas the complete graph's connectivity is unhindered as alternate routes exist between all nodes (e). A single node failure disconnects the tree graph into a $|v|$-partite graph (c), where $|v|$ is the order of the vertex at which failure occurs. The complete graph, on the other hand, is simply reduced to a $K_{n-1}$ graph, with no loss of connectivity (f). Thus, tree graphs are the most vulnerable network topologies to node/link failures, whereas complete graphs are the most robust.}\label{fig:graph_deletions}
	\end{figure}
\else
	\begin{figure*}[!htbp]
	\includegraphics[clip=true, width=\textwidth]{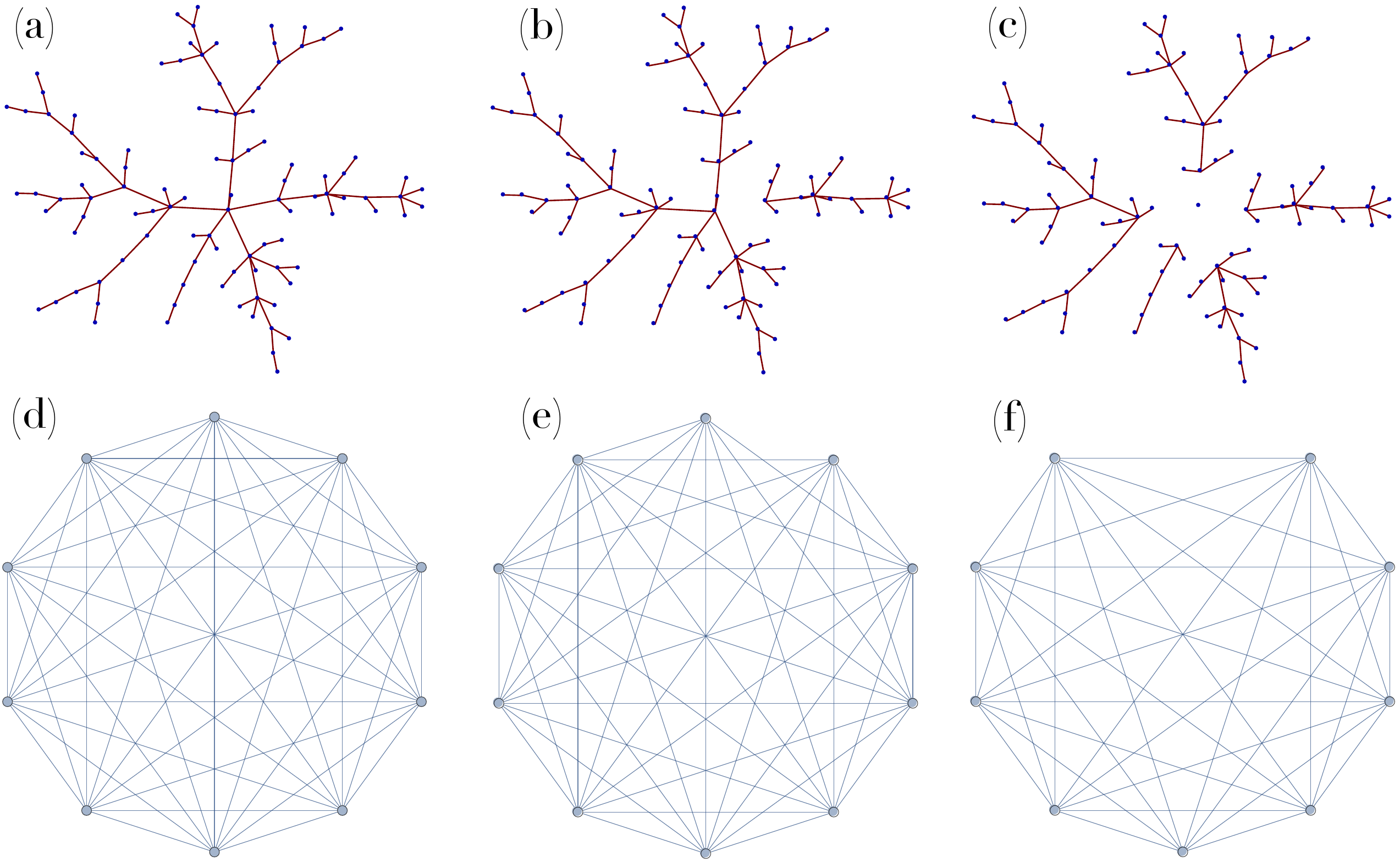}
	\captionspacefig \caption{Robustness of network topologies to node and link deletion. Examples of a tree graph (a) and a complete graph, $K_n$, with \mbox{$n=10$} (d). (b,e) The same graphs subject to a single link failure. The failure disconnects the tree graph into a bipartite graph (b), whereas the complete graph's connectivity is unhindered as alternate routes exist between all nodes (e). A single node failure disconnects the tree graph into a $|v|$-partite graph (c), where $|v|$ is the order of the vertex at which failure occurs. The complete graph, on the other hand, is simply reduced to a $K_{n-1}$ graph, with no loss of connectivity (f). Thus, tree graphs are the most vulnerable network topologies to node/link failures, whereas complete graphs are the most robust.}\label{fig:graph_deletions}
	\end{figure*}
\fi

\latinquote{Vitam regit fortuna, non sapientia.}

%
% Network Algorithms
%

\section{Network algorithms} \label{sec:graph_theory} \index{Network!Algorithms}

\dropcap{H}{aving} introduced some of the more relevant graph structures, we now introduce some of the key graph-theoretic algorithms of direct relevance to networking theory \cite{bib:RivestAlgBook}. In graph theory, many fundamental problems are believed to be computationally hard to solve, often \textbf{NP}-complete\index{NP \& NP-complete}. However, there are several important graph algorithms that are (very) classically efficient to solve, and which are of great utility to us as network architects.

We will focus heavily on combinatorial optimisation techniques, where the goal is to allocate network resources so as to optimise some cost metric. This includes both single- and multi-user algorithms, the latter being the far more relevant ones in the context of shared networks like the internet.

In Tab.~\ref{tab:net_alg_sum} we summarise the upcoming discussion on important network algorithms, and their associated complexities.

\startnormtable
\begin{table*}[!htbp]
	\begin{tabular}{|c|c|c|c|}
		\hline
  		\rowcolor{Dandelion} Algorithm & Description & Complexity class & Scaling \\
  		\hline
  		\hline
  		\rowcolor{LimeGreen} Breadth-first-search & Explore all vertices in a graph & \textbf{P} & $O(|V|+|E|)$ \\
  		\hline
  		\rowcolor{LimeGreen} Depth-first-search & (same as above) & \textbf{P} & $O(|V|+|E|)$ \\
		\hline
  		\rowcolor{LimeGreen} Shortest-path (Dijkstra) & Find the shortest route between two nodes & \textbf{P} & $O(|E|+|V|\log|V|)$ \\
  		\rowcolor{LimeGreen} & in a directed graph & &  \\
  		  		\hline
		\rowcolor{LimeGreen} Shortest-path (\textit{A*}) & (same as above) & \textbf{P} & (varying) \\
  		  		\hline
		\rowcolor{LimeGreen} Single-source shortest path & Find the shortest paths from a given node to & \textbf{P} & $O(|V|\cdot |E|)$\\
  		\rowcolor{LimeGreen} & \textit{all} other nodes & & \\
 		\hline
		\rowcolor{LimeGreen} Minimum spanning tree & Find a spanning tree of a graph that minimises & \textbf{P} & $O(|E|\log |V|)$ \\
  		\rowcolor{LimeGreen} & the total of the edge weights & & \\
  		\hline
  		\rowcolor{LimeGreen} Minimum-cost flow & Minimise total costs in a network &  \textbf{P} & $O(|V|\log |V|(|E|$\\
  		\rowcolor{LimeGreen} (Orlin) & with a specified amount of flow& & $+|V|\log |V|))$ \\
  		\hline
  		\rowcolor{LimeGreen} Maximum flow & Maximise flow in a network, regardless of costs & \textbf{P} & $O(|E|\cdot c_\mathrm{max})$ \\
  		\rowcolor{LimeGreen} (Ford-Fulkerson) & & & \\
  		\hline
  		\rowcolor{Lavender} Multi-commodity flow & Same as maximum flow, but generalised to & \textbf{NP}-complete & $O(\mathrm{exp}(|V|))$ \\
  		\rowcolor{Lavender} & arbitrary numbers of users & (exactly), & \\
  		\rowcolor{Lavender} & & \textbf{P} (approximation & \\
  		\rowcolor{Lavender} & & using heuristics) & \\
  		\hline
  		\rowcolor{Lavender} Vehicle routing problem & Generalises the shortest-path algorithm to multiple & \textbf{NP}-complete & $O(\mathrm{exp}(|V|))$ \\
  		\rowcolor{Lavender} & users, with distinct sources and destinations & & \\ 
  		\hline
  		\rowcolor{Lavender} Vehicle rescheduling problem & Same as above but with dynamically changing costs & \textbf{NP}-complete & $O(\mathrm{exp}(|V|))$ \\
    	\hline
	\end{tabular}
	\captionspacetab \caption{Summary of some important network algorithms and their complexities. The \textbf{NP}-complete algorithms are not believed to have efficient classical algorithms, and their exact scaling is not well understood.} \label{tab:net_alg_sum} \index{Network!Algorithms}\index{Breadth-first-search (BFS) algorithm}\index{Depth-first-search (DFS) algorithm}\index{Shortest-path algorithm}\index{Single-source shortest path algorithm}\index{Minimum spanning tree!Algorithm}\index{Minimum-cost flow algorithm}\index{Maximum flow algorithm}\index{Multi-commodity flow algorithm}\index{Vehicle routing problem}\index{Vehicle rescheduling problem}
\end{table*}
\startalgtable

%
% Network Exploration & Pathfinding
%

\subsection{Network exploration \& pathfinding} \label{sec:path_exp} \index{Network!Exploration}\index{Pathfinding}

Here the goal is to systematically explore every vertex in an unknown graph exactly once, so as to reconstruct the entire network graph, or to find a target node with unknown location (which can obviously be achieved if the former can be). The two main approaches are \textit{breadth-first-search} (BFS) and \textit{depth-first-search} (DFS) algorithms\index{Breadth-first-search (BFS) algorithm} \index{Depth-first-search (DFS) algorithm}. In both cases we begin at a starting (root) node, from which we wish to explore the entire graph by only following edges to nearest neighbours one at a time.

In BFS we proceed from the root node to visit every one of its neighbours. Having done so, and created a list of those neighbours, we proceed onto the neighbours of the neighbours, and so on, until every vertex in the graph has been visited, or the target node found.

In DFS, on the other hand, we begin by following a single arbitrary path until we reach a dead-end, at which point we backtrack until we reach a branch leading to a vertex we hadn't previously visited.

Examples of these two algorithms are shown in Fig.~\ref{fig:BFS_DFS}.

\if 1\doublecol
	\begin{figure}[!htbp]
	\includegraphics[clip=true, width=0.3\textwidth]{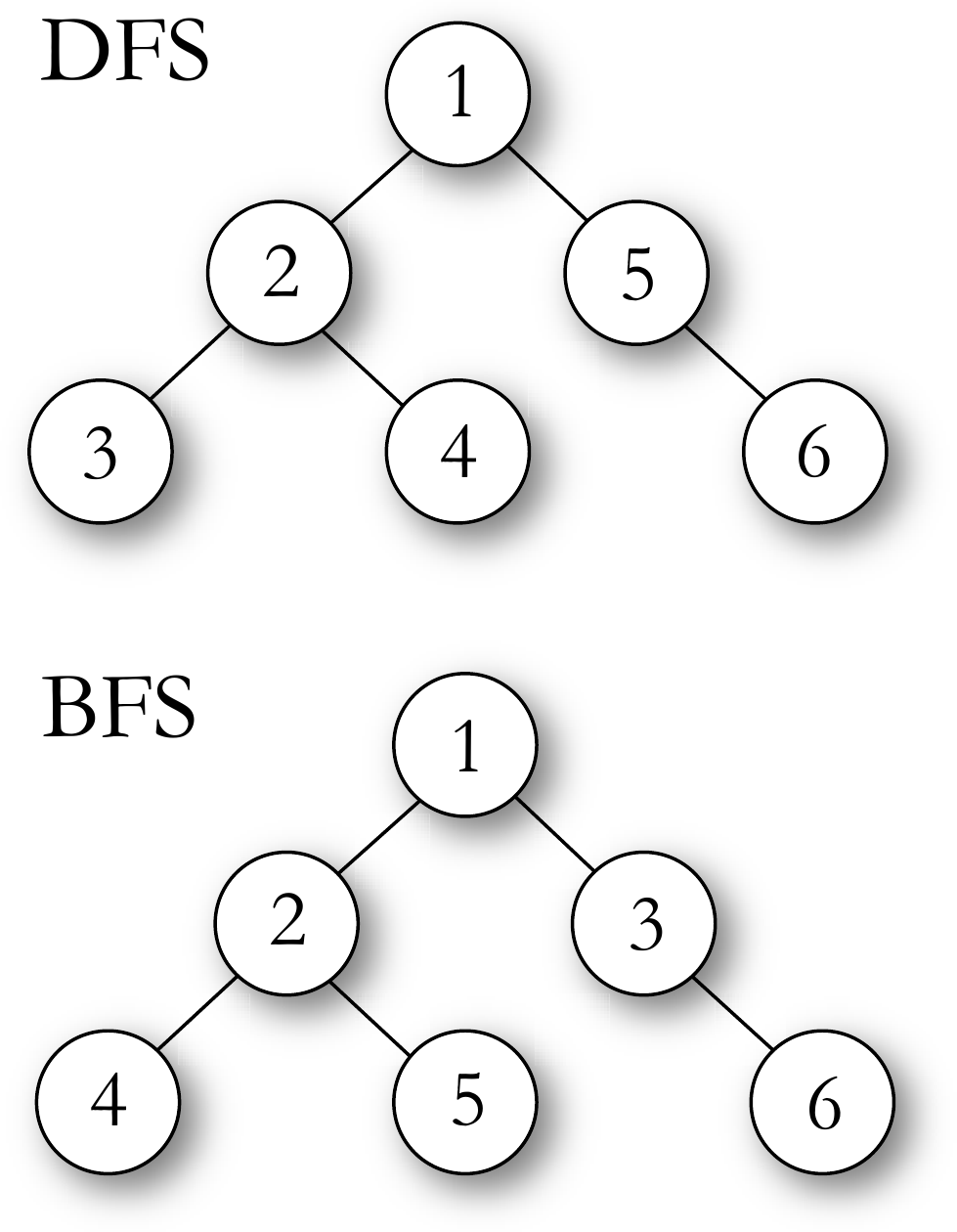}
	\captionspacefig \caption{Comparison of the order in which vertices are explored, using the breadth-first-search (BFS) and depth-first-search (DFS) algorithms, where vertex 1 is the root vertex.} \label{fig:BFS_DFS}
	\end{figure}
\else
	\begin{figure*}[!htbp]
	\includegraphics[clip=true, width=0.65\textwidth]{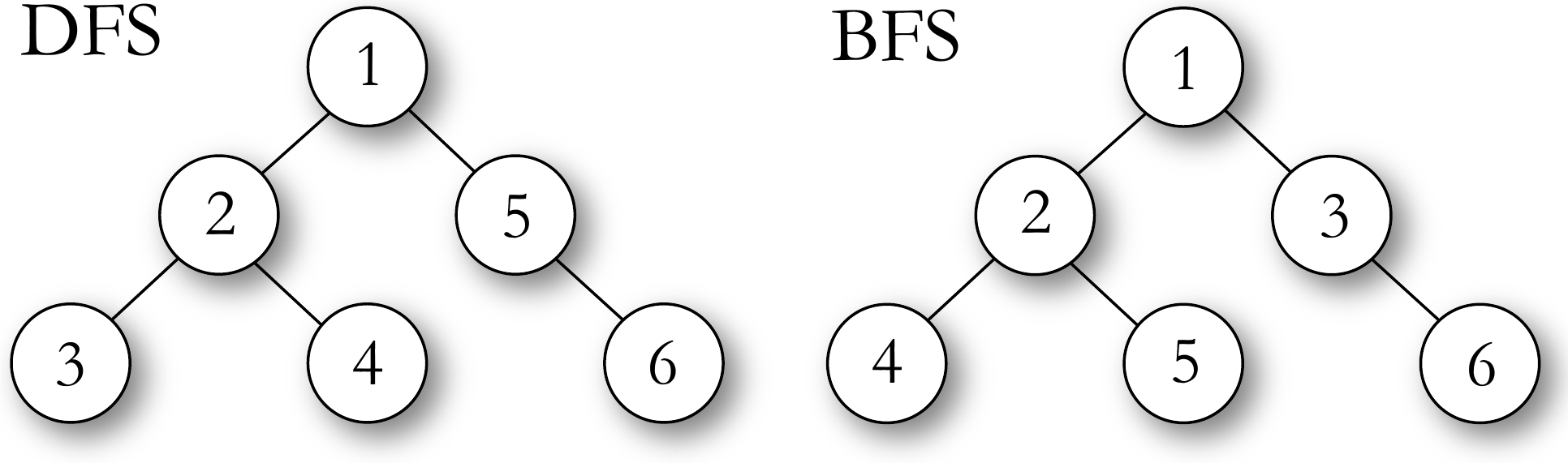}
	\captionspacefig \caption{Comparison of the order in which vertices are explored, using the breadth-first-search (BFS) and depth-first-search (DFS) algorithms, where vertex 1 is the root vertex.} \label{fig:BFS_DFS}
	\end{figure*}
\fi

Both BFS and DFS guarantee visiting every vertex in a connected graph, and do so using only nearest neighbour transitions. Such algorithms are therefore very useful for network discovery.

The BFS algorithm is particularly applicable to pathfinding in ad hoc networks. Consider the situation where there is no central authority with full knowledge of the network, overseeing network operation. Rather, everyone needs to figure things out for themselves by only interrogating their neighbours, to whom they have direct connections. This directly leads to a BFS algorithm, where a node speaks to each of its neighbours in turn, who subsequently do the same thing, yielding a recursive algorithm. This can be naturally parallelised, as each node can be interrogating its neighbours independently, thereby implementing a distributed BFS algorithm. Note that, when searching for a target node, while the BFS algorithm obviously finds the target using the smallest number of hops (i.e a lowest-order route), it needn't necessarily find the route with the lowest cost (which is distinct from the number of hops in general). Shortest-path algorithms require \textit{a priori} knowledge of the full network graph, discussed in Sec.~\ref{sec:shortest_path}.

Both BFS and DFS exhibit runtime	,
\begin{align}
	O(|V|+|E|),
\end{align}
where $|V|$ and $|E|$ are the number of vertices and edges respectively. Thus, these graph exploration algorithms reside in the complexity class \textbf{P}, and are classically efficient.

%
% Shortest-Path
%

\subsection{Shortest-path} \label{sec:shortest_path} \index{Shortest-path algorithm}

In graph theory, the shortest-path problem is that of finding a subgraph of a given graph $G$, connecting two vertices, \mbox{$A\to B \subset G$}, such that the sum of its edge weights is minimised. In the context of our application to route-finding, this amounts to finding a route that minimises cost.

The first proven shortest-path algorithm was invented by and named after Dijkstra \cite{bib:Dijkstra59}, which requires runtime,
\begin{align}
	O(|V|^2),
\end{align}
which has since been improved to,
\begin{align}
	O(|E|+|V|\log |V|),	
\end{align}
by \cite{bib:FredmanLawrence84} using min-priority queues\index{Min-priority queues} implemented via Fibonacci heaps\index{Fibonacci heaps}. Thus, the shortest-path algorithm resides in \textbf{P}\index{P} -- one of the relatively few, and highly valuable optimisation problems that is classically efficient. Subsequently, a number of improvements and variations on Dijkstra's algorithm have been proposed, most notably the $A^*$ algorithm \cite{bib:Astar}\index{Shortest-path algorithm}, which has found widespread modern use, using a heuristic approach to improve performance over Dijkstra. An alternate implementation of Dijkstra using

Formally, let $\vec{R}$ be the set of all routes \mbox{$A\to B$}. Then,
\begin{align}
c_\mathrm{opt} = \min_{r\in R} \left(\sum_{i\in r} c_\mathrm{net}(i) \right),
\end{align}
where \mbox{$i\in r$} denotes the $i$th edge in the route $r$. Intuitively, the (in general) exponential number of possible paths through a graph might lead one to believe the above optimisation problem is a computationally inefficient one (such as \textbf{NP}-complete, or worse). However, perhaps surprisingly, Dijkstra's algorithm cleverly manages to reduce this to a polynomial-time problem. A sketch of the algorithm is provided in Alg.~\ref{alg:dijkstra}, which needn't be understood by the reader desperate to read further.

\begin{table}[!htbp]
\begin{mdframed}[innertopmargin=3pt, innerbottommargin=3pt, nobreak]
\texttt{
function DijkstraShortestPath($G$,$A$,$B$):
\begin{enumerate}
	\item currentNode = $A$
    \item tentativeDistances[$A$] = 0
    \item tentativeDistances[others] = $\infty$
    \item nodesVisited[$A$] = True
    \item nodesVisited[others] = False
    \item loopStart:
    \item neighbours = currentNode.neighbourhood
    \item nodesVisited[neighbours] = True
    \item for(n$\in$neighbours) \{
    \setlength{\itemindent}{0.2in}
    \item newTentativeDist = \\ min(tentativeDistances[currentNode] \\+ edgeWeight[currentNode,n],\\
        tentativeDistances[n])
    \item nodesVisited[currentNode] = True
    \setlength{\itemindent}{0in}
    \item \}
    \item if(nodesVisited[$B$] = True) \{
    \setlength{\itemindent}{0.2in}
	\item return(tentativeDistances[$B$])
	\item $\Box$
    \setlength{\itemindent}{0in}
    \item \}
	\item currentNode = \\
	tentativeDistances[unvisitedNodes].
	nodeWithSmallest()
	\item goto(loopStart)
    \item $\Box$
\end{enumerate}}
\end{mdframed}
\captionspacealg \caption{Dijkstra's original shortest-path algorithm for finding the lowest weight path through a graph, $G$, between two vertices, $A$ (source) and $B$ (destination). The algorithm has $O(|E|+|V|\log|V|)$ runtime (in \textbf{P}). An example application of this algorithm is shown in Fig.~\ref{fig:dijkstra_eg}.} \label{alg:dijkstra}\index{Shortest-path algorithm}
\end{table}

\begin{figure}[!htbp]
\if 1\doublecol
\includegraphics[clip=true, width=0.475\textwidth]{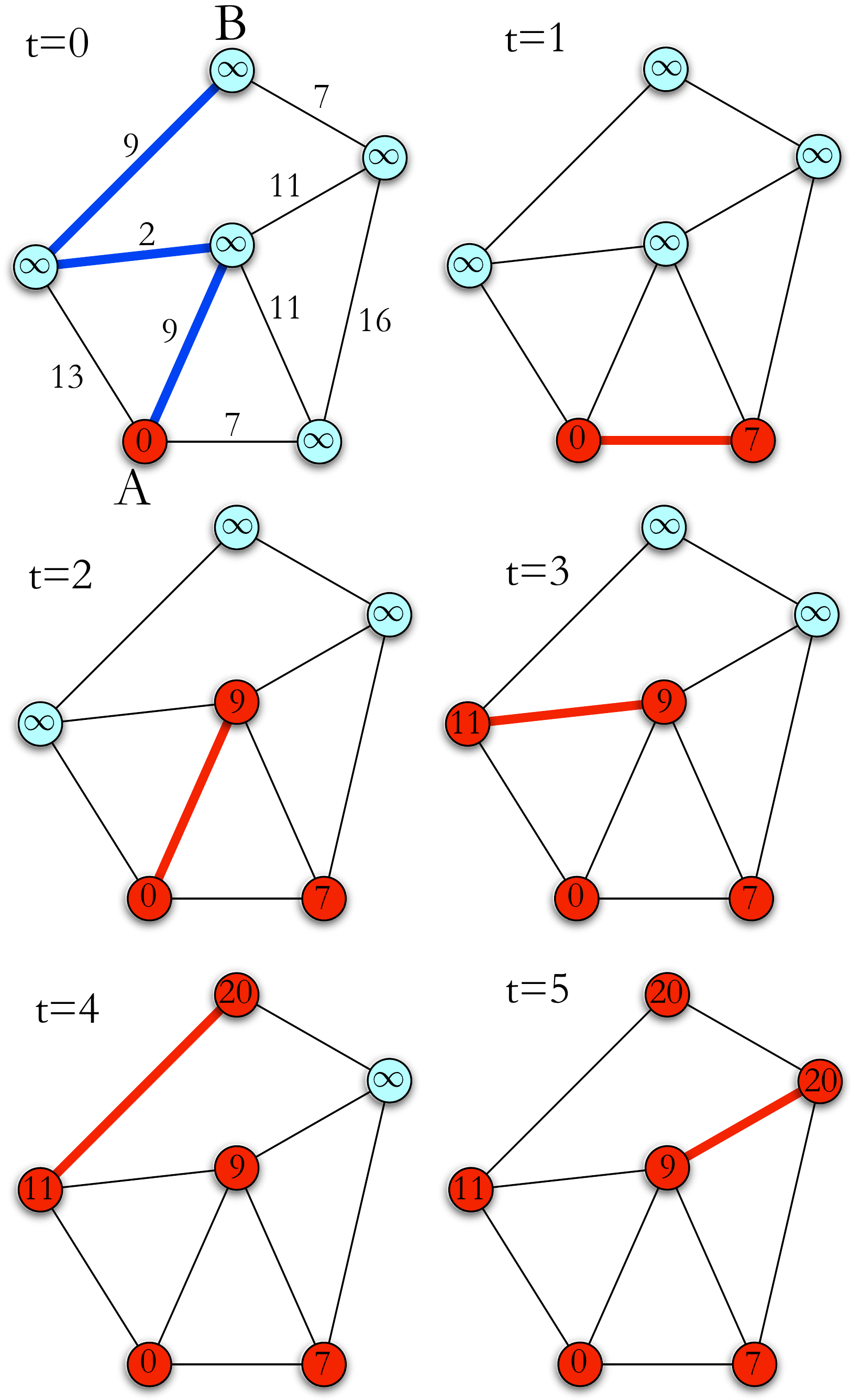}
\else
\includegraphics[clip=true, width=0.7\textwidth]{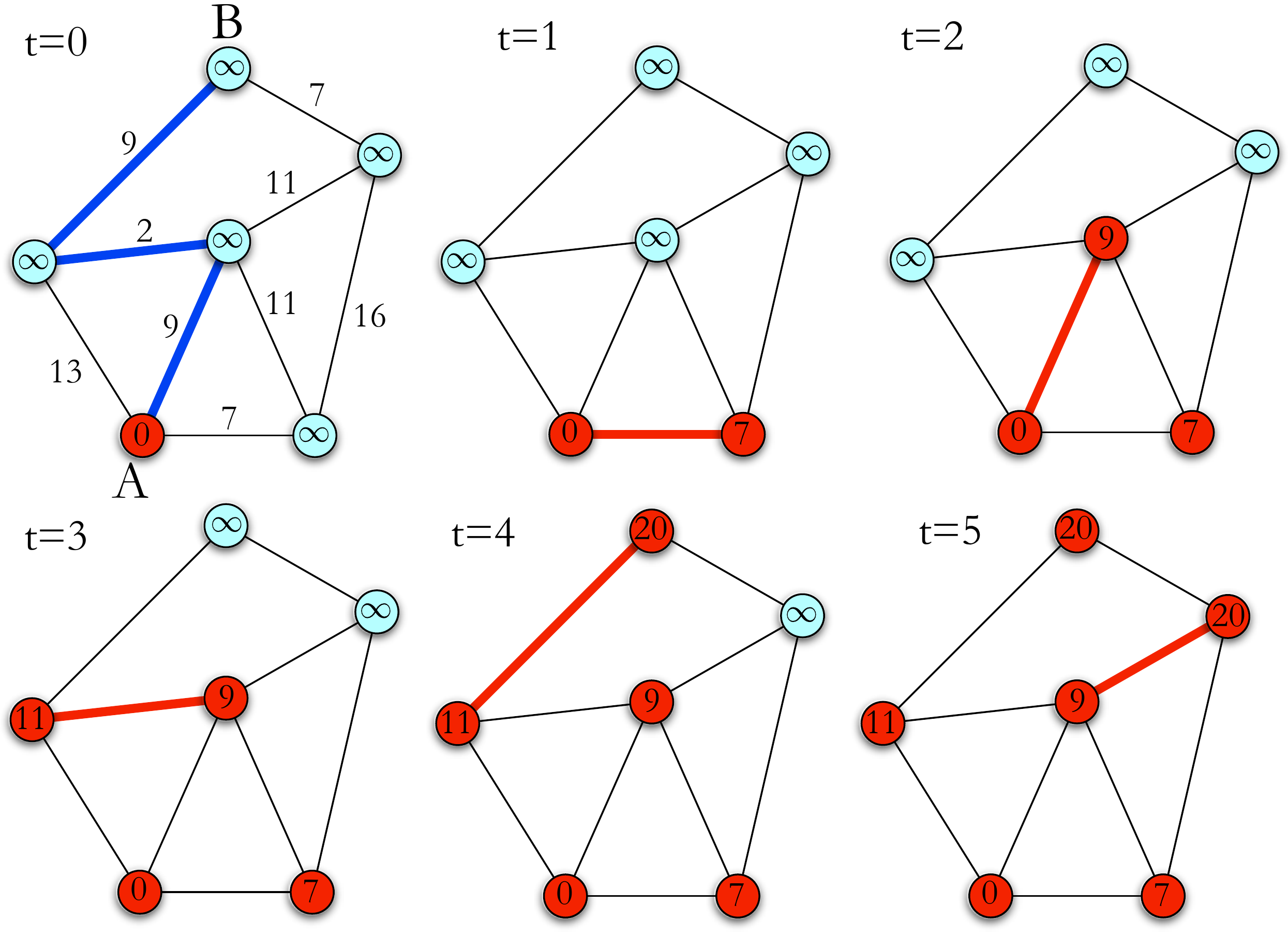}
\fi
\captionspacefig \caption{Example of Dijkstra's shortest-path algorithm for finding the shortest route from $A$ to $B$ (blue route), yielding a shortest route of distance 20 using 5 algorithmic steps. Initially (\mbox{$t=0$}) all nodes are designated a tentative distance of $\infty$ and marked as unvisited (blue vertices), except the starting node $A$, which is designated distance $0$ and visited (red vertices). At each step we choose the lowest cost path to a previously unexplored node (red edges). This updates the tentative distance of the newly-visited neighbour, which is now marked as visited. This is iterated until all nodes have been visited. Each node is visited exactly once.}\label{fig:dijkstra_eg}\index{Shortest-path algorithm}
\end{figure}

Fig.~\ref{fig:simp_route_opt} illustrates a directed, edge-weighted graph. A shortest-path algorithm applied between vertices $A$ and $B$ would return \mbox{$R_\mathrm{shortest} = A\to F\to B$} as the minimum-cost route.

When introducing network graphs earlier, we insisted upon all costs being associated with edges rather than vertices, and presented a trivial means by which to convert vertex costs to edge costs in Fig.~\ref{fig:remove_nodes}. This adamance arose because the presently described shortest-path algorithms operate purely in terms of edge weights, not vertex weights. But the mapping we presented from the latter to the former obviates this issue.

This is the motivating factor behind representing network graphs purely in terms of edge weights (Sec.~\ref{sec:quant_proc_in}), thereby enabling compatibility with shortest-path algorithms.

For the purposes of the QTCP protocol, we are interested in the case of directed graphs (recall that in terms of cost metrics, undirected graphs can be converted to directed graphs by replacing undirected edges with a pair of identical edges in opposite directions).

Shortest-path techniques find widespread application in many areas. Computer networks are an obvious candidate, since networks are inherently graph-theoretic by nature.

To implement the shortest-path algorithms discussed above, the party performing the calculation requires knowledge of the full network graph. In an ad hoc network, where users might be added to or removed from the network arbitrarily, this isn't necessarily the case.

One solution is for a central authority to be responsible for maintaining a ledger of all network participants and their connectivity, which users are required to notify upon joining or leaving the network. The central authority may then apply shortest-path calculations, which may be queried by users. However, a disruption in connection to the central authority, or failure of nodes to notify the central authority upon joining or leaving the network, introduces a point of failure into the operation of the protocol.

Another approach, which does not require a reliable central authority, is for users to implement network exploration algorithms each time they wish to perform a shortest-path calculation. This facilitates truly ad hoc networking, but incurs the cost overhead associated with nodes frequently implementing network exploration. However, network exploration is a purely classical algorithm, which may run entirely over the classical network, and therefore incurs no cost in quantum resources.

With this approach, a new node can join the network, without having to know anything about the topology of the network. Similarly, upon leaving the network, it needn't notify anyone, since a future interrogation by a neighbour will be detected as a non-existent node. The BFS is therefore highly suited to ad hoc operation. In fact, present-day internet gateway protocols (Sec.~\ref{sec:gateway}) essentially implement a distributed version of BFS.

%
% Constrained Shortest-Path
%

\subsection{Constrained shortest-path}\label{sec:const_short_path}\index{Constrained shortest-path algorithm}

In some scenarios we may wish to find a shortest-path through a graph, subject to some constraints. In general, the addition of constraints can make such algorithms far more computationally complex, undermining the efficiency of Dijkstra's algorithm. However, in some circumstances such constraints can easily be incorporated, without undermining the performance of the algorithm.

In particular, if there are constraints on the relationships between nearest-neighbours in the graph, this can be incorporated by pre-processing the graph via deletions of edges that violate the constraints. Then the usual shortest-path algorithm may be applied to the reduced graph. The rather trivial algorithm is shown in Alg.~\ref{alg:const_short}, with a simple example shown in Fig.~\ref{fig:constrained_shortest_path}.

\begin{table}[!htbp]
\begin{mdframed}[innertopmargin=3pt, innerbottommargin=3pt, nobreak]
\texttt{
function ConstrainedShortestPath($G$,$A$,$B$,$C$):
\begin{enumerate}
	\item For the set of nearest-neighbour constraints $C$, generate the set of edges $E(C)$ that violate the constraints.
	\item $G'=G-E(C)$
	\item route = DijkstraShortestPath($G'$,$A$,$B$)
	\item return(route)
    \item $\Box$
\end{enumerate}}
\end{mdframed}
\captionspacealg \caption{Efficient algorithm for finding a constrained shortest-path, where the constraints are in terms of nearest-neighbour relationships.} \label{alg:const_short}\index{Constrained shortest-path algorithm}
\end{table}

\if 1\doublecol
\begin{figure}[!htbp]
\includegraphics[clip=true, width=0.35\textwidth]{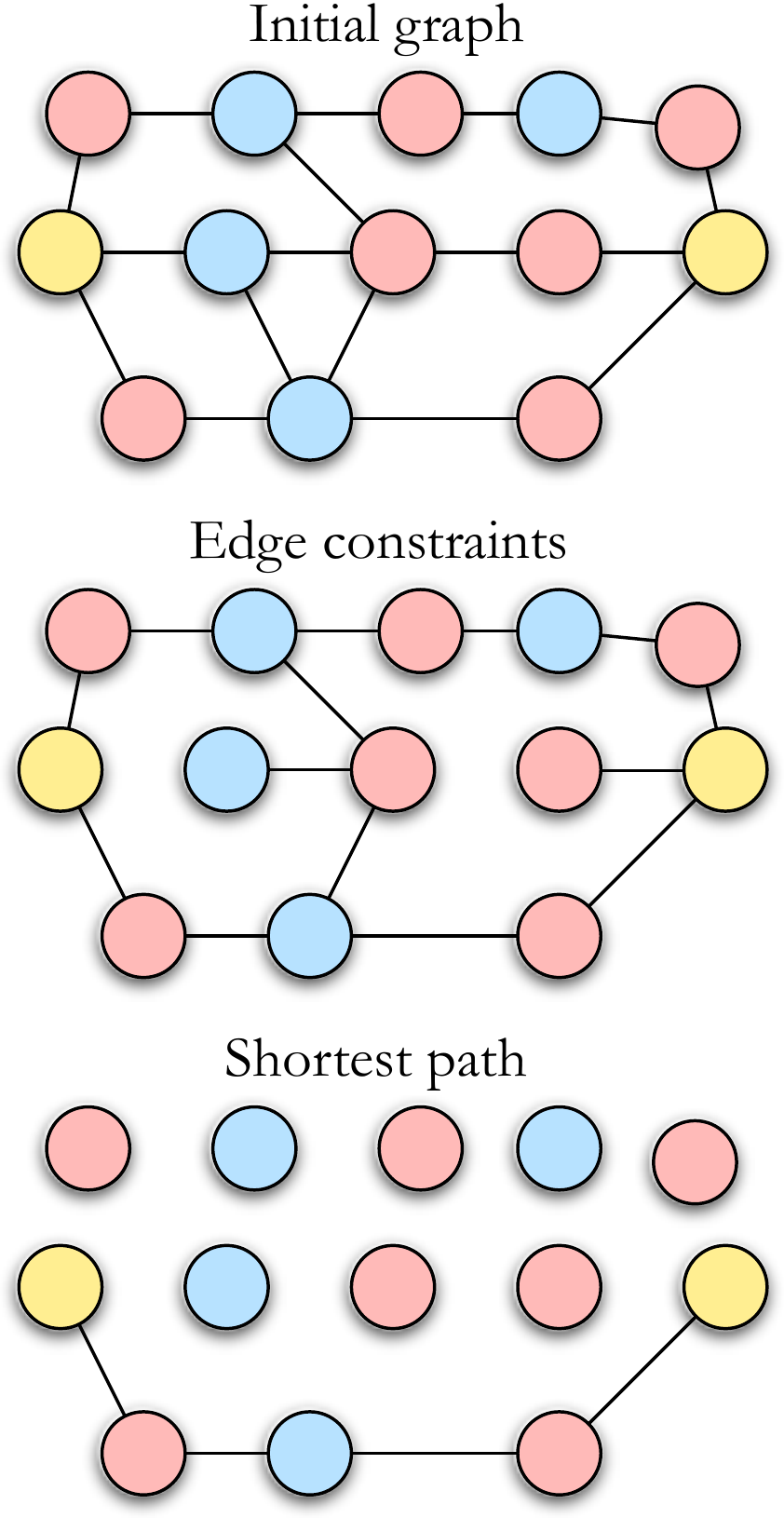}
\captionspacefig \caption{Example of a constrained shortest-path algorithm. Beginning with the initial graph we eliminate edges that do not satisfy imposed nearest-neighbour constraints. This is performed as a pre-processing stage. Then a usual shortest-path algorithm is applied to the reduced graph, yielding the optimal route subject to the required constraints.}\label{fig:constrained_shortest_path}
\end{figure}
\else
\begin{figure*}[!htbp]
\includegraphics[clip=true, width=\textwidth]{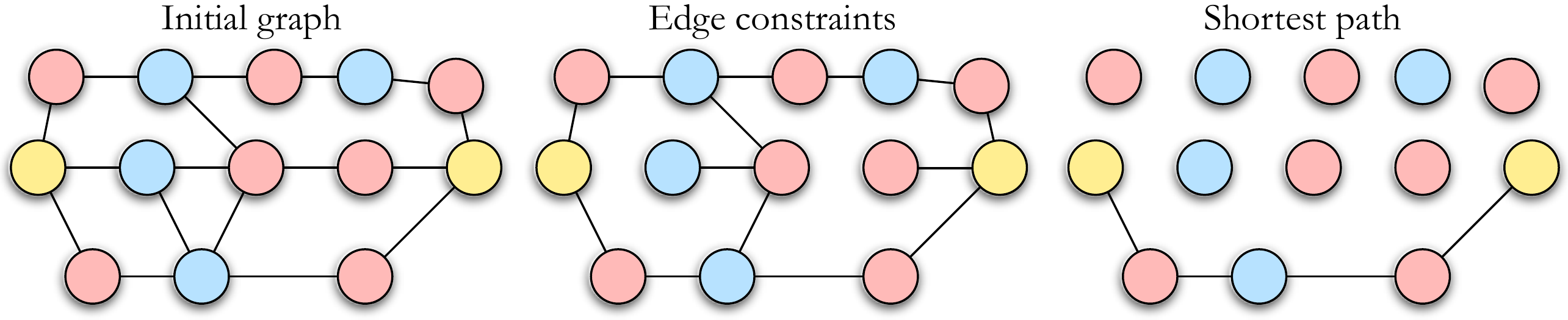}
\captionspacefig \caption{Example of a constrained shortest-path algorithm. Beginning with the initial graph we eliminate edges that do not satisfy imposed nearest-neighbour constraints. This is performed as a pre-processing stage. Then a usual shortest-path algorithm is applied to the reduced graph, yielding the optimal route subject to the required constraints.}\label{fig:constrained_shortest_path}
\end{figure*}
\fi

A notable example of where this technique finds applicability is in entanglement swapping networks, discussed later in Secs.~\ref{sec:swapping} \& \ref{sec:rep_net}. Consider the network shown in Fig.~\ref{fig:racetime}. Here the entanglement swapping network comprises nodes alternating between two different functionalities: Bell pair preparation, and entanglement swapping. Representing these as colours, this effectively introduces the edge constraint that edges between nearest neighbours of the same colour ought to be removed. Additionally, the end-points of the network must neighbour Bell pair sources, not entanglement swappers. This enforces the additional constraint of removing edges between end-points with the wrong coloured neighbours. Having applies these edge deletions enforcing these constraints, the shortest-path algorithm will now find the constrained optimal route. Thus, this example of an entanglement swapping network is isomorphic to the example presented in Fig.~\ref{fig:constrained_shortest_path}.

%
% Single-Source Shortest-Path Algorithm
%

\subsection{Single-source shortest-path} \label{sec:single_source_sp} \index{Single-source shortest path algorithm}

The shortest-path algorithm by Dijkstra presented above finds the shortest route between two specified nodes in a network. However, when employing \textsc{Individual} routing strategies, where there is no central mediation of the network, each node desires an up-to-date routing table, showing the best route to take to any other point in the network. Then, upon receiving packets with particular destinations, rather than repeatedly applying Dijkstra's algorithm, we can simply look up the destination on the node's local routing table.

Single-source shortest-path algorithms address this problem by calculating the shortest paths from the current node to \textit{every} other node in the network topology.

The best-known algorithm for this problem is the Bellman-Ford (or Bellman-Ford-Moore)\index{Bellman-Ford-Moore algorithm} algorithm \cite{bellman1958routing,ford1956network,moore1959shortest}, which requires worst-case runtime of,
\begin{align}
	O(|V|\cdot |E|).
\end{align}
Clearly this is more complex than Dijkstra's algorithm for finding a particular shortest-path. But it is more efficient than using brute-force to find the shortest-path between every pair of nodes in the network via $O(|V|^2)$ repeated applications of Dijkstra.

%
% Minimum Spanning Tree
%

\subsection{Minimum spanning tree} \label{sec:min_tree} \index{Minimum spanning tree!Algorithm}

MST algorithms find an MST\footnote{There may be multiple distinct MSTs for a given graph.} of some arbitrary graph. Like the shortest-path problem, it has a polynomial-time, deterministic algorithm (i.e it resides in \textbf{P}\index{P}). The first MST algorithm \cite{bib:Boruvka26} required,
\begin{align}
	O(|E|\log |V|),
\end{align}
runtime. Numerous variations have since been proposed, with little change to the underlying scaling.

Because MST algorithms are efficient, they play a very useful role in the design of real-world network topologies, where resource minimisation is crucial.

Fig.~\ref{fig:mst} shows an example of a graph with its MST.

%
% Minimum-Cost Flow
%

\subsection{Minimum-cost flow} \label{sec:min_cost_flow_prob} \index{Minimum-cost flow algorithm}

The \textit{minimum-cost flow problem} \cite{goldberg1989network} is that of minimising costs through a network for a specified amount of flow (i.e total bandwidth or throughput), which acts as a constraint on the problem. The definition of `cost' in this context is compatible with our earlier definition of cost metrics (Def.~\ref{def:metric}).

This problem can be efficiently solved using linear programming. Specifically, cost metrics along links in series are given by linear combinations of individual link costs. If, in addition, we let our net cost function be linear in the constituent costs then the net cost will also be linear in all the edge weights. This lends itself directly to optimisation via linear programming techniques. Algorithms for linear programming, such as the \textit{simplex} algorithm, have polynomial-time solutions (i.e reside in \textbf{P}\index{P}), and a plethora of software libraries are available for implementing them numerically.

One polynomial-time algorithm, by \cite{orlin1997polynomial}, for solving this problem does so in, 
\begin{align}
	O(|V|\log |V|(|E|+|V|\log |V|)),
\end{align}
time.

%
% Maximum Flow
%

\subsection{Maximum flow} \label{sec:max_flow_prob} \index{Maximum flow algorithm}

The \textit{maximum flow problem} \cite{goldberg1989network} is the seemingly simple goal of -- as the name suggests -- maximising network flow, without consideration for any of the other cost metrics or attributes associated with the network. This type of problem is relevant when brute bandwidth is the dominant goal.

This problem can be tackled using a number of techniques. In some circumstances, linear programming techniques can be employed. The best-known algorithm is the Ford-Fulkerson algorithm\index{Maximum flow algorithm} \cite{ford1956maximal}, which finds a solution in,
\begin{align}
	O(|E|\cdot c_\mathrm{max}),
\end{align}
runtime, where $|E|$ is the number of links in the network and $c_\mathrm{max}$ is the maximum cost present in the network. The algorithm behaves pathologically in some conditions, which can easily be overcome in the context we present here. Using Ford-Fulkerson as a starting point, numerous other more sophisticated algorithms have been developed.

%
% Multi-Commodity Flow
%

\subsection{Multi-commodity flow} \label{sec:multi_comm_flow} \index{Multi-commodity flow algorithm}

The \textit{multi-commodity flow problem} \cite{ahuja1995network} generalises the previous algorithms to be applicable to multi-user networks. The generalisation is that there may be a number of distinct senders, residing on different nodes, each transmitting to distinct recipients, residing on different nodes. This is the most realistic scenario we are likely to encounter in a real-world quantum internet, where networks will inevitably be shared by many users, residing at different nodes.

Unfortunately the computational complexity of solving this problem is much harder than the previous algorithms in general. Specifically, solving the problem exactly is \textbf{NP}-complete\index{NP \& NP-complete} in general. However, in specific circumstances it can be approached using linear programming or polynomial-time approximation schemes.

%
% Vehicle Routing Problem
%

\subsection{Vehicle routing problem} \label{sec:VRP} \index{Vehicle routing problem}

The vehicle routing problem (VRP) is a multi-user generalisation of the shortest-path problem, where the goal is to minimise total network cost (i.e the sum of all individual users' costs) when there are multiple users sharing the network, each with distinct sources and destinations.

\begin{figure}[!hbtp]
\includegraphics[clip=true, width=0.475\textwidth]{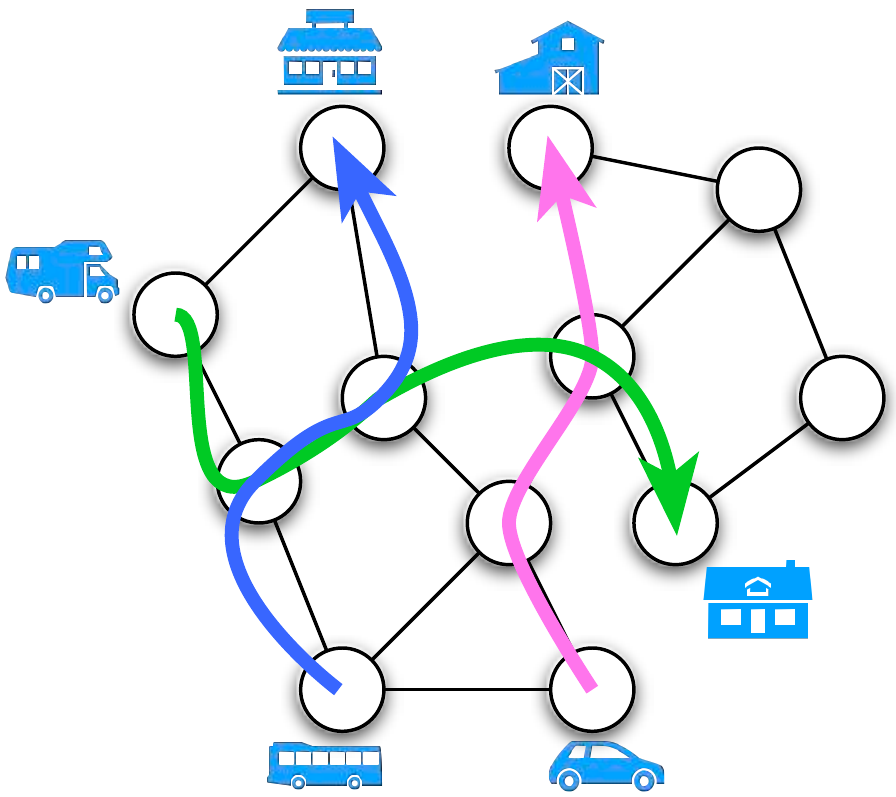}
	\captionspacefig \caption{Example of the vehicle routing problem, the multi-user generalisation of the shortest-path problem, where the goal is to minimise the total cost across all users. Intuitively, because whenever individual shortest-paths intersect one must trial different prioritisations, the combinatorics of this grow exponentially with the number of competing users.}\label{fig:VRS}
\end{figure}

Unlike the polynomial-time shortest-path algorithm, exactly solving the VRP is \textbf{NP}-hard\index{NP \& NP-complete} in general. However, heuristic methods can find approximate suboptimal solutions far more efficiently, and there is a multitude of software packages available for doing so.

The VRP has found widespread use in, for example, the routing of transport networks for delivery companies or public transportation networks (hence the name), and many commercial companies exist, which perform these kinds of optimisations on behalf of transport providers to enhance their efficiency.

It is obvious that this algorithm is directly applicable to multi-user communications networks, which are conceptually identical to transport networks, albeit a bit faster. 

A multitude of variations on the VRP exist, accommodating for different types of constraints (or additional flexibilities) in the operation of the network.

%
% Vehicle Rescheduling Problem
%

\subsection{Vehicle rescheduling problem} \label{sec:VRSP} \index{Vehicle rescheduling problem}

The vehicle rescheduling problem (VRSP) generalises the VRP to the case where properties of the network undergo changes dynamically within the course of transmissions over the network. To use the analogy of transport networks, this could entail, for example, a truck breaking down en route to its destination, requiring real-time rescheduling of the other vehicles.

Solving the VRSP exactly is \textbf{NP}-complete\index{NP \& NP-complete} in general, but as with the VRP, heuristic methods can often be applied, which efficiently find approximate solutions.

In the context of communications over networks, the VRSP has obvious applicability -- a quantum internet is likely going to be largely ad hoc in nature, with users coming and going, and many non-deterministic points of failure, requiring ongoing updating of routing decisions if resource allocation is to remain as efficient as possible.

%
% Improving Network Algorithms Using Quantum Computers
%

\subsection{Improving network algorithms using quantum computers} \index{Network!Algorithms!on quantum computers}

Given that we are directing this work at the upcoming quantum era, where quantum computing will become a reality, it is pertinent to ask whether quantum computers might improve the aforementioned network algorithms, some of which are computationally hard problems. Most notably, several of the discussed algorithms are \textbf{NP}-complete\index{NP \& NP-complete} in general, a complexity class strongly believed to be exponentially complex on classical computers. Can quantum computers help us out here, and improve network resource allocation? Can quantum computers help themselves?

While it is not believed that quantum computers can efficiently solve such \textbf{NP}-complete\index{NP \& NP-complete} problems, it is known that they can offer a quadratic speedup using Grover's unstructured search algorithm. Specifically, \textbf{NP}-complete\index{NP \& NP-complete} problems can be treated as satisfiability problems, where we are searching for an input to a classical algorithm that yields a particular output.

To gain a quantum advantage, we treat the classical algorithm as an oracle whose input configurations form an unstructured search space. Then, Grover's algorithm can perform a search over the space of input configurations to find a satisfying solution, with quadratically enhanced runtime.

While a quadratic improvement is far short of the exponential improvement we might hope for, Grover's algorithm is known to be optimal for the unstructured search problem \cite{grover1996fast}. Nonetheless, despite only yielding a quadratic improvement, a quadratic speedup may already be sufficient to significantly improve network resource allocation.

\sketch{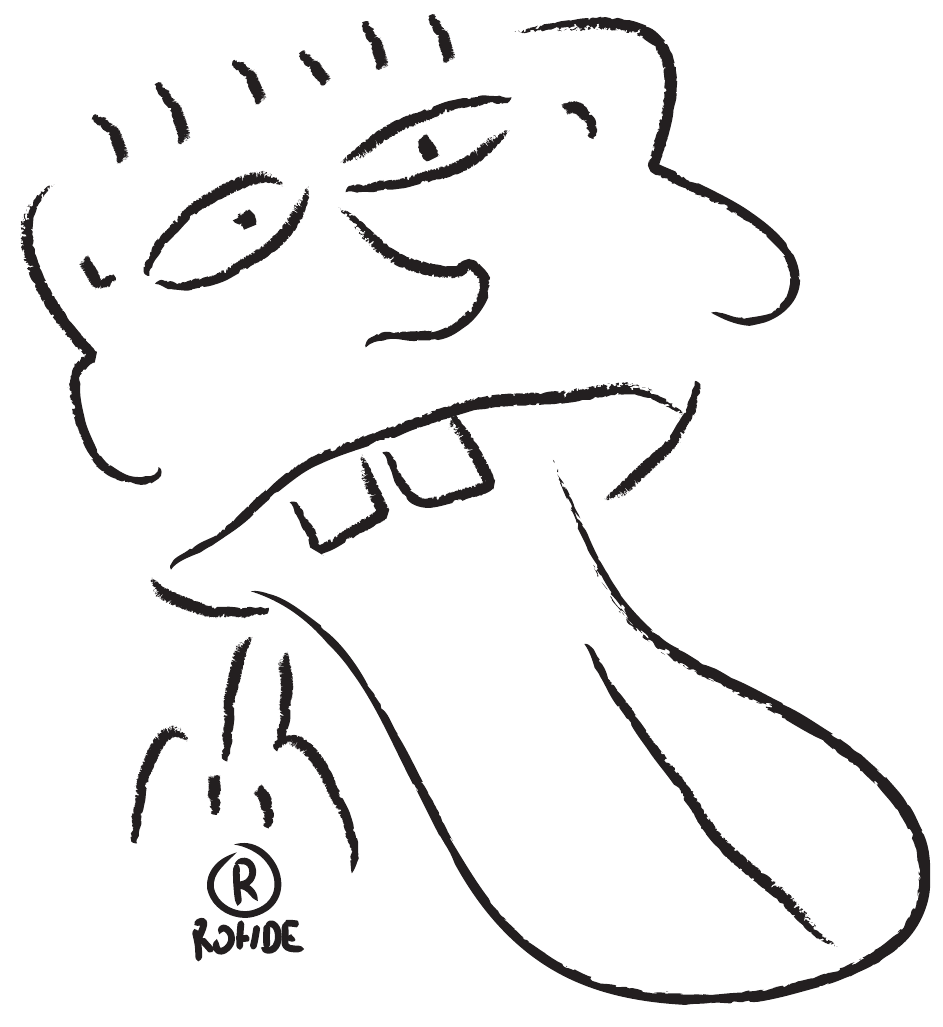}

\clearpage
%
% Quantum networks
%

\part{Quantum networks}\label{part:quant_net}\index{Quantum networks}

%
% Quantum Networks
%

\famousquote{Advances in the technology of telecommunications have proved an unambiguous threat to totalitarian regimes everywhere.}{Rupert Murdoch}
\newline

\dropcap{W}{e} have reviewed some of the key aspects of classical networks, including the real-world implementation of classical networking via the TCP/IP protocol stack, and the essential mathematical foundations for networking theory, including cost vector analysis and routing strategies.

Let us now lay the foundations and some of the key motivations and assumptions we will make in our upcoming discourse on quantum networks, and lay out some of the key differences between classical networks and future quantum ones.

Quantum networks comprise all the same ingredients as classical networks, but with some very important non-classical additions. Nodes can additionally implement quantum computations, quantum-to-classical interfaces (i.e measurements), quantum-to-quantum interfaces (i.e switching data between different physical systems), quantum memories, or any quantum process in general. Many of these are not allowed by the laws of classical physics.

The cost vectors associated with links could include measures that are uniquely quantum, such as fidelity, purity or entanglement measures, none of which are applicable to classical digital data.

As in the classical case, our goal is to find routing strategies that optimise a chosen cost measure. But in the quantum context costs will be constructed entirely differently owing to the quantum nature of the information being communicated.

We envisage a network with a set of senders and receivers, all residing on a time-dependent network graph as before. Senders have sets of quantum states they wish to communicate. For each state they must choose appropriate strategies, such that the overall cost is optimised, for some appropriate cost measure. Compared to classical resources, equivalent quantum resources are costly and must be used efficiently and frugally. Indeed, the no-cloning theorem\index{No-cloning theorem} imposes the constraint that arbitrary unknown states cannot be replicated at all! This makes resource allocation strategies of utmost importance in the quantum world.

Routing strategies will not always guarantee that packets have immediate access to network bandwidth the moment they demand it. One needs to think about the others too! Inevitably, in shared networks there will sometimes be competition and congestion, forcing some users to wait their turn. For this reason, many quantum networks will require at least some nodes (the ones liable to competition) to have access to quantum memories, such that quantum packets can be buffered for a sufficient duration that they can wait their turn on the shared network resources for which there is high competition. The required lifetime of a quantum memory will then be related to overall network congestion. Of course, quantum memories induce unwanted quantum processes of their own, which need to be factored into cost calculations.

Given that classical networking is decades more advanced than quantum networking, and extremely cheap and reliable in comparison, we will assume that classical resources `come for free', and only quantum resources are of practical interest in terms of their cost. That is, classical communication and computation is a free resource available to mediate the operation of the quantum network. We therefore envisage a \textit{dual network}\index{Dual network} with two complementary networks operating in parallel and in tandem -- the quantum network for communicating quantum data, and a topologically identical classical network operating side-by-side and synchronised with the quantum network, overseeing and mediating the quantum network.

Data packets traversing the network will comprise both quantum and classical fields, which will be separated to utilise the appropriate network, but synchronised such that they arrive at their destination as a single package of joint quantum and classical information to be at the disposal of the recipient.

The motivation for the dual network is to ensure that classical and quantum data that jointly represent packets remain synchronised and subject to the same QoS issues, such as packet collisions and network congestion.

We envisage quantum networks to extend beyond just client/server quantum computation, to include the free trade of any quantum asset. This includes state preparation, measurement, computation, randomness, entanglement, and information. Much like the classical internet, by allowing quantum assets to be exchanged, we can maximise utility, improve economy of scale, and enable new models for commercialisation.\index{Quantum assets}

May the games begin.

\latinquote{Gladiator in arena consilium capit.}

%
% Quantum Channels
%

\section{Quantum channels} \label{sec:quant_chan} \index{Quantum channels}

\dropcap{L}{ike} classical channels, quantum channels are inevitably subject to errors. These errors could be an intrinsic part of the system, or induced by interaction with the external environment. The \textit{quantum process} formalism provides an elegant mathematical description for all physically realistic error mechanisms \cite{bib:NielsenChuang00, bib:Gilchrist05}. Here we review the quantum process formalism and how it applies to quantum networks. This paves the way for the quantum notion of costs and attributes.

%
% Quantum Processes
%

\subsection{Quantum processes}\label{sec:quantum_processes}\index{Quantum processes}

To quantify the operation of nodes and links within our network, we must characterise the evolution they impose upon quantum states passing through them. Quantum processes, also known as \textit{trace-preserving, completely positive maps} (CP-maps) are able to capture all the physical processes relevant to quantum networking, such as: unitary evolution; decoherence; measurement; quantum memory; state preparation; switching; and, indeed entire quantum computations. And they are able to capture physical processes in any degree of freedom, most commonly in the qubit basis, but also, for photons, in the spatio-temporal, photon-number, phase-space, or polarisation degrees of freedom.

Quantum processes are most easily represented using \textit{Kraus operators}\index{Kraus operators}, $\{\hat{K}_i\}$,
\begin{align} \label{eq:kraus_rep}
\mathcal{E}(\hat\rho) = \sum_i \hat{K}_i \hat\rho \hat{K}_i^\dag,
\end{align}
where,
\begin{align}
\sum_i \hat{K}_i^\dag \hat{K}_i = \hat\openone,
\end{align}
for normalisation. Here $\mathcal{E}$ is a super-operator, denoting the action of the process on state $\hat\rho$. This is also referred to as the \textit{operator-sum representation}\index{Kraus operators}. This representation has the elegant interpretation as the probabilistic application of each of the Kraus operators $\hat{K}_i$, with probability,
\begin{align}
p_i = \mathrm{tr}(\hat{K}_i \hat\rho \hat{K}_i^\dag).
\end{align}
In the ideal case, the two types of evolution of interest are unitary evolution, in which case there is only one Kraus operator, \mbox{$\hat{K}_1=\hat{U}$}, and projective measurement, where there is again only one Kraus operator, \mbox{$\hat{K}_1=\ket{m}\bra{m}$}, for measurement outcome $m$.

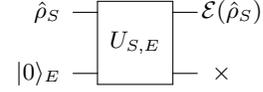
\begin{figure}[!htbp]
\begin{align}
\Qcircuit @C=1em @R=1.6em {
    \lstick{\hat\rho_S} & \multigate{1}{{U}_{S,E}} & \qw & \,\,\,\,\,\mathcal{E}(\hat\rho_S) \\
    \lstick{\ket{0}_E} & \ghost{\hat{U}_{S,E}} & \qw & \times \\
} \nonumber
\end{align}
\captionspacefig \caption{Model for the quantum process formalism, as a system state $\hat\rho_S$ undergoing joint unitary evolution with an environment state $\ket{0}_E$, which is subsequently traced out, yielding an arbitrary quantum process $\mathcal{E}(\hat\rho_S)$ acting on the primary system. For the most general possible class of quantum processes to be enabled, the dimension of the environment Hilbert space must grows quadratically with the dimension of the primary Hilbert space.\index{Hilbert space}} \label{fig:q_proc}
\end{figure}

Mathematically, quantum processes are equivalent to a state jointly undergoing unitary evolution with an external environment that is not observed (i.e traced out),
\begin{align} \label{eq:proc_environment}
\mathcal{E}(\hat\rho_S) = \mathrm{tr}_E (\hat{U}_{S,E} [\hat\rho_S\otimes \ket{0}_E\bra{0}_E] \hat{U}^\dag_{S,E}),
\end{align}
where $S$ denotes the primary system to which the process is applied, and $E$ is an auxiliary environment system, as shown in Fig.~\ref{fig:q_proc}.

We will require that all our states are normalised,
\begin{align}
\mathrm{tr}(\hat\rho) = 1,
\end{align}
and that our processes are \textit{trace preserving}. That is, they preserve normalisation,
\begin{align}
\mathrm{tr}[\mathcal{E}(\hat\rho)] = 1.
\end{align}

Multiple consecutive processes may be composed using the notation,
\begin{align}
\mathcal{E}_n(\dots \mathcal{E}_2(\mathcal{E}_1(\hat\rho)))=(\mathcal{E}_n \circ \dots \circ \mathcal{E}_2\circ\mathcal{E}_1)(\hat\rho).
\end{align}
In general, processes do not commute, i.e \mbox{$\mathcal{E}_1\circ \mathcal{E}_2 \neq \mathcal{E}_2\circ \mathcal{E}_1$}. Unless unitary, quantum processes are irreversible, meaning that errors accumulate and cannot be overcome without the overhead of some form of quantum error correction (QEC) \cite{bib:Shor95, bib:CalderbankShor96, bib:NielsenChuang00}. The linearity of Eq.~(\ref{eq:kraus_rep}) implies that quantum processes are also linear\index{Linearity},
\begin{align}
	\mathcal{E}(\hat\rho_1+\hat\rho_2) = \mathcal{E}(\hat\rho_1)+\mathcal{E}(\hat\rho_2).
\end{align}

The only limitation faced by the quantum process formalism is that it is described over discrete-time only. To consider continuous-time evolution, \textit{master equations}\index{Master equations} can be used. These represent the continuous-time evolution of a quantum state as a differential equation in time, combining a usual Hamiltonian term as well as decoherence terms,
\begin{align}
\frac{d\hat\rho}{dt} = -\frac{i}{\hbar}[\hat{H},\hat\rho] + \sum_j (2\hat{L}_j\hat\rho\hat{L}_j^\dag - \{\hat{L}_j^\dag\hat{L}_j,\hat\rho\}),
\end{align}
where $\hat{H}$ is the Hamiltonian of the isolated system undergoing coherent evolution, and $\hat{L}_j$ are the \textit{Lindblat operators}\index{Lindblat operators}, capturing the incoherent component of the dynamics (i.e environmental couplings). Here $[\cdot,\cdot]$ and $\{\cdot,\cdot\}$ are the commutator and anti-commutator respectively.

In this work we will only make use of discrete-time quantum processes, since they naturally correspond to the evolution of states between discrete points within a network -- we are typically only interested in the process undergone by a state from one end of a link to another, not the continuous-time dynamics of what takes place within them.

%
% Quantum Process Matrices
%

\subsection{Quantum process matrices} \index{Quantum process matrices}

In general, the Kraus operator representation for quantum processes is not unique -- there may be multiple choices of Kraus operators that implement identical physical processes. But if the representation is not unique, how do we compare different quantum processes? To address this, it is common to choose a `standard' basis for representing quantum processes, such that they may be consistently and fairly compared. This requires choosing a basis which is complete for operations on the Hilbert space acted upon by the process.

For example, for a single qubit, the Pauli operators\index{Pauli!Operators} -- $\hat\sigma_1$ (identity, $\hat\openone$), $\hat\sigma_2$ (bit-flip, $\hat{X}$), $\hat\sigma_3$ (bit-phase-flip, $\hat{Y}$), and $\hat\sigma_4$ (phase-flip, $\hat{Z}$) -- are complete for single-qubit operations ($\mathbb{C}_2$). Therefore by decomposing our Kraus operators into linear combinations of these basis operators we have a standardised representation for single-qubit processes. Formally, for one qubit,
\begin{align} \label{eq:process_matrix}
\mathcal{E}(\hat\rho) = \sum_{i,j=1}^4 \chi_{i,j} \hat{\sigma}_i\hat\rho\,\hat{\sigma}_j^\dag.
\end{align}

The Hermitian matrix $\chi$ is known as the \textit{process matrix}\index{Process matrices}, from which many other metrics of interest may be directly computed (some of which are discussed in Sec.~\ref{sec:quantum_meas_cost}).

Process matrices share many algebraic properties and interpretations in common with density matrices. The diagonal elements can be regarded as the amplitudes associated with applying each of the four Pauli operators, all of which are non-negative, while the off-diagonal elements represent the coherences between them, i.e whether the operations on the diagonal are being applied probabilistically or coherently. For example, a process that simply randomly applies Pauli operators would have a diagonal process matrix in the Pauli basis. But off-diagonal elements would be indicative of applying coherent superpositions of the operators. Like density matrices, the dimensionality of process matrices grows exponentially with the number of qubits in the system being characterised, and for exactly the same conceptual reasons.

For the process to be trace preserving we require,
\begin{align}
\mathrm{tr}(\chi) = 1.
\end{align}
We will typically enforce this constraint on our processes. $\mathrm{tr}(\chi) < 1$ implies non-determinism, i.e the process sometimes fails. 

\begin{figure}[!htbp]
\includegraphics[clip=true, width=0.4\textwidth]{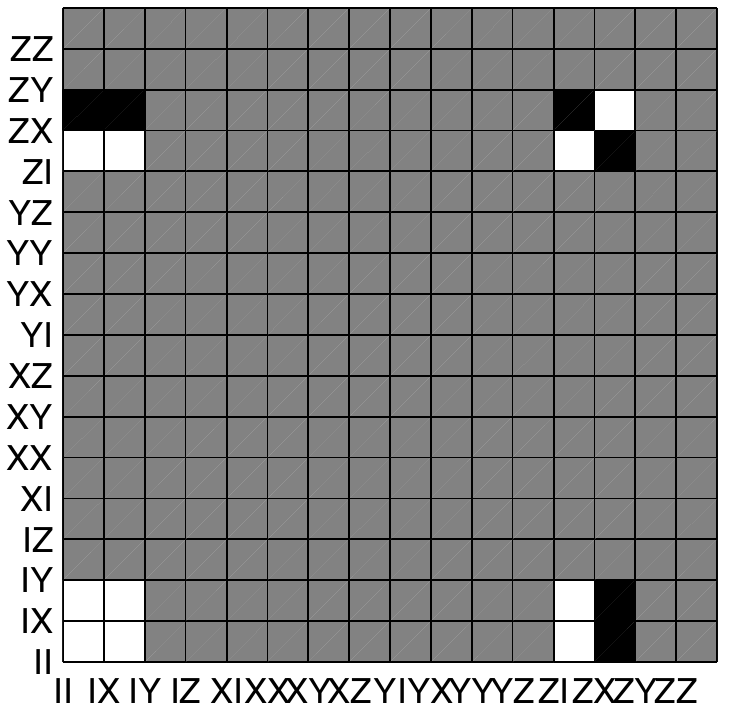}
\captionspacefig \caption{Process matrix for the CNOT gate, expressed in the Pauli basis. Colour coding: grey=0, white=$1/4$, black=$-1/4$.} \label{fig:CNOT_proc_matrix}
\end{figure}

As an illustrative example of the interpretation of process matrices, in Fig.~\ref{fig:CNOT_proc_matrix} we show the process matrix for the CNOT gate, represented in the Pauli basis. The CNOT operator can be expressed in the Pauli operator basis as,
\begin{align}
\hat{U}_\mathrm{CNOT} = \frac{1}{2}(\hat\openone\otimes \hat\openone + \hat\openone \otimes \hat{X} + \hat{Z}\otimes \hat\openone - \hat{Z}\otimes \hat{X}).
\end{align}
Then, some density operator evolved under the CNOT gate is simply \mbox{$\hat{U}_\mathrm{CNOT}\hat\rho \,\hat{U}_\mathrm{CNOT}^\dag$}. Expanding this out, we obtain a new state comprising 16 terms, each representing the action of some combination of Pauli operators from the left and from the right. The amplitudes of these terms exactly correspond to the 16 non-zero elements of the process matrix shown in Fig.~\ref{fig:CNOT_proc_matrix}.

%
% Quantum Processes in Quantum Networks
%

\subsection{Quantum processes in quantum networks} \label{sec:quant_proc_in} \index{Quantum processes}

\begin{figure}[!htbp]
\includegraphics[clip=true, width=0.3\textwidth]{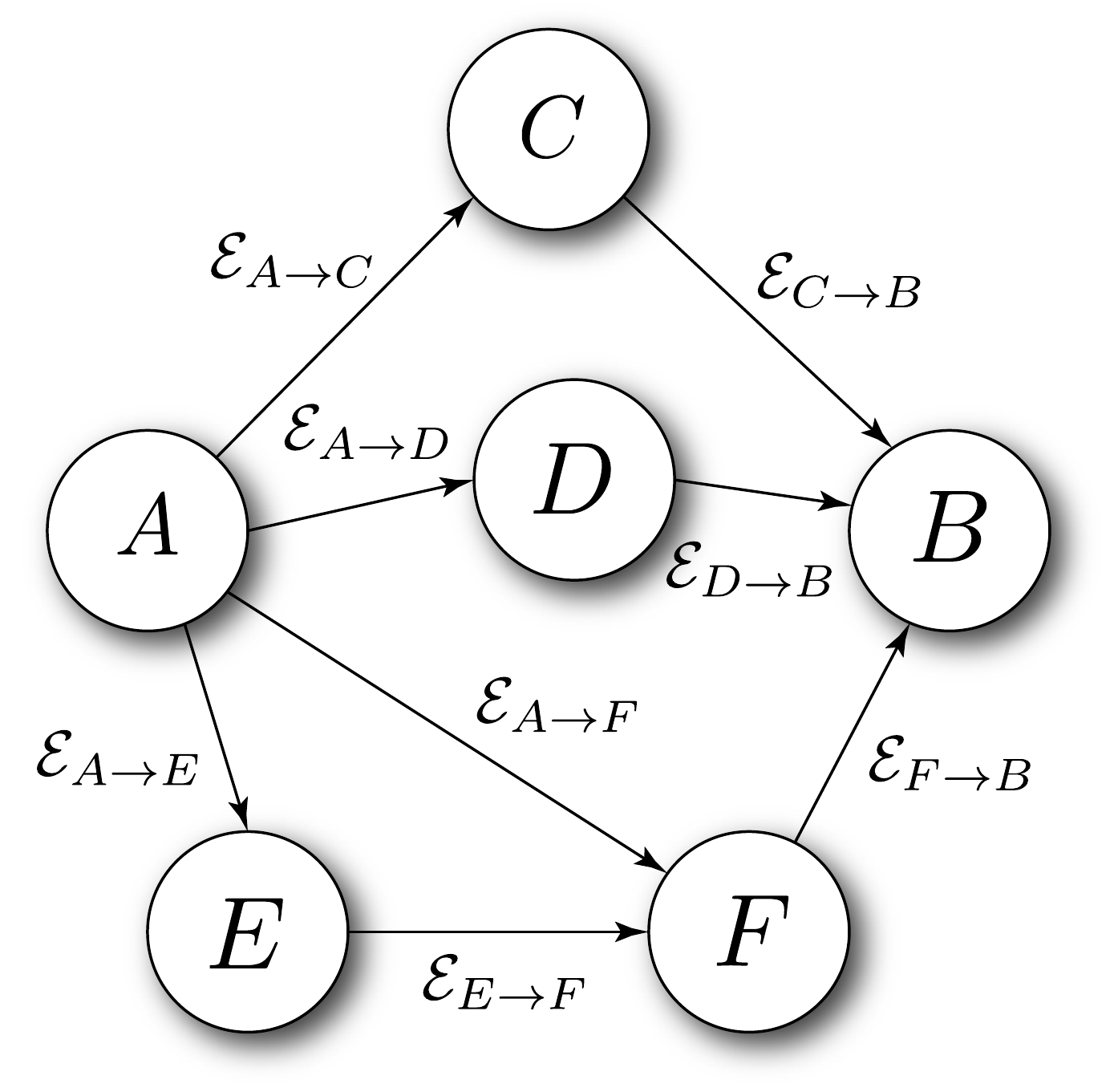}
\captionspacefig \caption{The network from Fig.~\ref{fig:example_routes}, with the quantum processes associated with each link. The net process associated with a route is given by the composition of each of the processes over the length of the route. For example, the route \mbox{$R_1=A\to C\to B$} induces the process \mbox{$\mathcal{E}_{R_1} = \mathcal{E}_{C\to B} \circ \mathcal{E}_{A\to C}$}.} \label{fig:example_proc_graph}
\end{figure}

Letting $v_i$ represent the $i$th node within a route $R$, the process associated with communication from that node to the next is $\mathcal{E}_{v_i\to v_{i+1}}$. For the same network used previously, Fig.~\ref{fig:example_proc_graph} shows the quantum processes associated with the links in the network. The cumulative process associated with an entire route is therefore,
\begin{align}
\mathcal{E}_R = \mathcal{E}_{{v_{|R|-1}}\to v_{|R|}} \circ \dots \circ \mathcal{E}_{v_2\to v_3} \circ \mathcal{E}_{v_1\to v_2},
\end{align}
where $|R|$ is the number of nodes in $R$, and to simplify notation, all $v_i$ are implicitly over the route $R$.

\begin{figure}[!htbp]
\includegraphics[clip=true, width=0.35\textwidth]{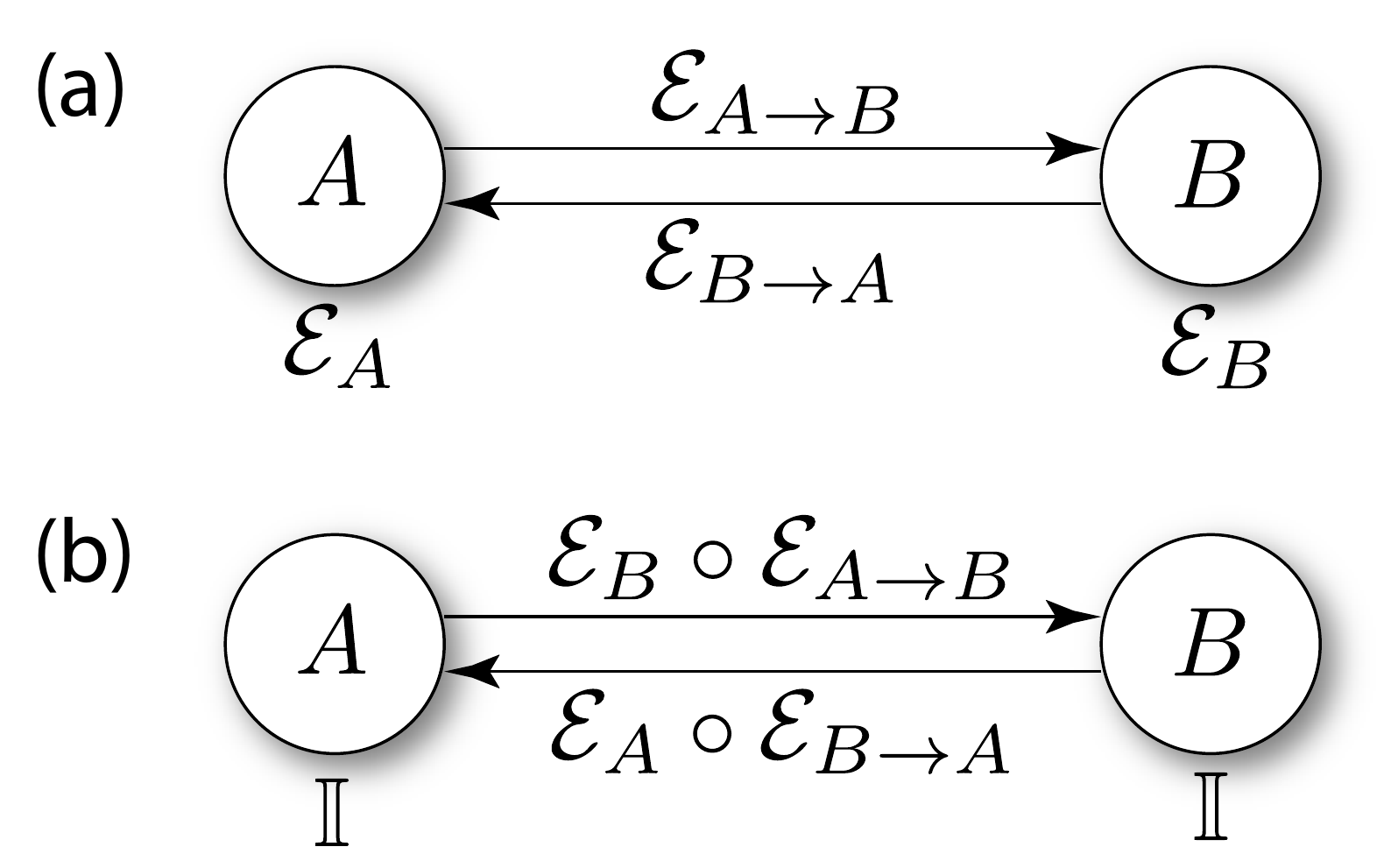}
\captionspacefig \caption{Removing node processes from network graphs on a trivial network with two nodes, $A$ and $B$. Each node is associated with a quantum process ($\mathcal{E}_A$ and $\mathcal{E}_B$). Similarly, each link is associated with a process ($\mathcal{E}_{A\to B}$ and $\mathcal{E}_{B\to A}$). (a) Representation where the node and link processes are shown explicitly. (b) The node processes are replaced with identity operations by replacing each link process with the composition of the link process and its target node process. Equivalently, the cost of each node process is added to the cost of \textit{every} incoming link and then eliminated. The same may be applied for attributes rather than costs. This procedure requires that all links be directed. If undirected links are present, they may simply be replaced by two directed links, one in each direction, implementing identical quantum processes each way.} \label{fig:remove_nodes}
\end{figure}

In general, both nodes and links in a quantum network may implement quantum processes. However, for the purposes of compatibility with the graph-theoretic algorithms described in Sec.~\ref{sec:graph_theory}, we will eliminate node processes by merging them into link processes, such that the processes in the network are described entirely by links. This reduction procedure is straightforward, shown in Fig.~\ref{fig:remove_nodes}.

%
% Characterising Quantum States & Channels
%

\subsection{Characterising quantum states \& channels} \label{sec:QPT}

Given a link implementing some arbitrary quantum process, it is essential that it can be experimentally determined such that network performance may be characterised. For example, if an optical channel is lossy, what is the loss rate? This is crucial when attempting to choose routing strategies that optimise certain cost metrics.

Treating a link or node as an unknown black box, \textit{quantum process tomography} (QPT) \cite{bib:ChuangNielsen97} is a technique that may be applied to fully characterise the quantum process it implements, reproducing its complete process matrix. QPT has become a standard procedure, demonstrated in numerous architectures, most notably in optics \cite{bib:OBrien04, bib:RohdeGateChar05}.

QPT works in general for processes in any degree of freedom, e.g the qubit degree of freedom. However, it is important to note that full QPT requires statistics across the entire basis over which measurements are defined, which typically grows exponentially with the size of the system. For example, the number of measurement bases required to perform full QPT on $n$ qubits grows exponentially with $n$.

However, often full process characterisation is not necessary. Instead, knowing particular metrics of interest may suffice. Some of the more noteworthy such metrics will be discussed in Sec.~\ref{sec:quantum_meas_cost}. In this instance, much work has been done in the field of \textit{compressed sensing} or \textit{compressed quantum process tomography} \cite{gross2010quantum}, in which some process metrics of interest can be experimentally determined using far fewer physical resources (with efficient scaling!) than via a full reconstruction of the process matrix using QPT. As a most trivial example, if the loss associated with a fibre-optic channel is the metric of interest, this can be much more easily determined than by performing full QPT.

On the other hand, however, most quantum channels are designed to accommodate systems with very limited Hilbert space dimensionality per clock-cycle -- e.g a fibre-optic link might transmit just one photon at a time -- in which case there is no exponentiality to be terribly concerned about (QPT of a single-photon channel is trivial).

Importantly, it is often the case that the quantum process associated with a channel will remain constant over time. The efficiency of a length of fibre, for example, does not change. In this instance, characterising the channel need only be performed once in advance, without requiring ongoing dynamic updating. On the other hand, when communicating with satellites in low Earth orbit it is to be expected that the properties of links will be highly dynamic.

We will now explain QPT in the archetypal context of single-qubit channels, which logically generalises to multiple qubits, and can similarly be generalised to non-qubit systems also.

%
% Quantum State Tomography
%

\subsubsection{Quantum state tomography} \index{Quantum state tomography (QST)}

The first stage in QPT is \textit{quantum state tomography} (QST), where the goal is to reconstruct and unknown density matrix via measurements upon multiple copies of the state. QST is based upon the simple observation that the completeness relation\index{Completeness relation} for an arbitrary state can be expressed,
\begin{align}
\hat\rho = \sum_i \mathrm{tr}(\hat{E}_i\hat\rho)\cdot\hat{E}_i,
\end{align}
where $\{\hat{E}_i\}$ forms a complete basis for the Hilbert space of $\hat\rho$. For a single qubit this decomposition is most often performed in the Pauli basis, 
\begin{align}
\hat\rho &= \mathrm{tr}(\hat\rho)\cdot\hat\openone + \mathrm{tr}(\hat{X}\hat\rho)\cdot\hat{X} + \mathrm{tr}(\hat{Y}\hat\rho)\cdot\hat{Y} +\mathrm{tr}(\hat{Z}\hat\rho)\cdot\hat{Z} \nonumber\\
	&= \sum_{i=1}^4 \mathrm{tr}(\hat{\sigma}_i\hat\rho)\cdot\hat{\sigma}_i,
\end{align}
where $\sigma_i$ denote the four Pauli operators. Of course, \mbox{$\mathrm{tr}(\hat{E}\hat\rho) = P(\hat{E}|\hat\rho)$} is just the expectation value of the measurement operator $\hat{E}$ when measuring $\hat\rho$. Thus, measuring the expectation values in each of the four Pauli bases reconstructs $\hat\rho$.

This generalises straightforwardly to multi-qubit systems, where we measure all combinations of tensor products of the Pauli operators, the number of which grows exponentially with the number of qubits $n$, as $4^n$. This introduces scalability issues for systems comprising a large number of qubits.

In the case of optical systems, entirely alternate, but equivalent, approaches may be used, such a probing the Wigner function directly using homodyne detection\index{Homodyne detection}.

%
% Quantum Process Tomography
%

\subsubsection{Quantum process tomography} \index{Quantum process tomography (QPT)}

Now to perform QPT we apply the unknown process to a complete basis of input states $\{\hat\rho_i\}$, and perform QST on the output state for each. This yields,
\begin{align}
\mathcal{E}(\hat\rho_j) = \sum_{i} c_{i,j} \hat\rho_i,
\end{align}
where the sum runs over the basis of states. From QST, all the coefficients $c_{i,j}$ may be determined. Next we define the following decomposition for each of the terms in the sum of Eq.~(\ref{eq:process_matrix}),
\begin{align}
\hat{E}_m \hat\rho_j \hat{E}_n^\dag = \sum_k B^{m,n}_{j,k} \hat\rho_k,
\end{align}
where $B$ defines a decomposition in the chosen basis, not dependent on any measurement results. Then we can write,
\begin{align}
\mathcal{E}(\hat\rho_j) &= \sum_{m,n} \chi_{m,n} \hat{E}_m\hat\rho_j\hat{E}_n^\dag \nonumber \\
&= \sum_{m,n} \sum_k \chi_{m,n} B^{m,n}_{j,k} \hat\rho_k.
\end{align}
Because $\hat\rho_k$ form a linearly independent basis, we can write the decomposition,
\begin{align}
c_{j,k} = \sum_{m,n} \chi_{m,n} B_{j,k}^{m,n},
\end{align}
for all \mbox{$j,k$}. From this, standard linear algebra techniques allow an inversion to obtain,
\begin{align}
\chi_{m,n} = \sum_{j,k} (B_{j,k}^{m,n})^{-1} c_{j,k},
\end{align}
thereby obtaining the full process matrix $\chi$, in the chosen basis.

\latinquote{Bulla crustulum.}

%
% Optical Encoding of Quantum Information
%

\section{Optical encoding of quantum information} \label{sec:opt_enc_of_qi} \index{Optical!Encoding of quantum information}

\dropcap{W}{hile} all-optical quantum computing is an unlikely architecture for future scalable quantum computers, it is all but inevitable that optics will play a central role in quantum communications networks. Foremost, this is because photons are `flying' by their very nature and can very easily be transmitted across large distances -- it's quite challenging to transmit a superconducting circuit containing information from Australia to Mozambique in the blink of an eye! Additionally, optical states are, in many cases, relatively easy to prepare, manipulate and measure, and can also be readily interfaced with other physical quantum systems (Sec.~\ref{sec:opt_inter}), allowing the transfer of quantum information from optical communications systems to some other architecture better suited to a given task.

Optical systems are very versatile, allowing quantum information to be optically encoded in a number of ways -- into single photons, many photons, or even an indeterminate number of photons, and in both discrete or continuous degrees of freedom. Different types of encodings may have very different properties in terms of the errors they are susceptible to (Sec.~\ref{sec:errors_in_nets}).

When dealing with single photons, information can be encoded in a number of ways. Most obviously, it can be encoded into the polarisation basis, allowing one qubit of information per photon (i.e horizontal and vertical polarisation represent the logical $\ket{0}$ and $\ket{1}$ states). Or it could be directly encoded into the photon-number basis. However, other degrees of freedom, such as the spectral/temporal degrees of freedom could be employed, encoding information into time- or frequency-bins, with potentially far more levels than a simple polarisation qubit \cite{bib:RohdeInfCap13}. Next we discuss some primary methods for optical encoding of quantum information.

%
% Single Photons
%

\subsection{Single photons} \label{sec:single_phot_enc} \index{Single-photons!Encoding}

A very attractive feature of single photons is that they undergo very little decoherence, even over large distances -- dephasing (Sec.~\ref{sec:dephasing_error}\index{Dephasing}) in the polarisation degree of freedom, for example, is negligible in free-space. They are, however, very susceptible to loss, and protocols relying on many single-photon states suffer exponential decay in their success rates as the number of photons is increased (Sec.~\ref{sec:eff_err}).

We can encode a single qubit into a single photon in the polarisation basis using the horizontal and vertical polarisation degrees of freedom. Equivalently, one can employ `dual rail' encoding, whereby a single photon is placed into a superposition across two spatial modes. Finally, one can use time-bin encoding, whereby discrete windows of time represent logical basis states when occupied by a photon. This leads to the equivalent representations for logical qubits ($L$),
\begin{align} \label{eq:single_photon_enc}\index{Polarisation!Encoding} \index{Dual-rail encoding}
	\index{Time-bin encoding}
\ket{\psi}_\mathrm{qubit} &\equiv \alpha\ket{0}_L + \beta\ket{1}_L, \nonumber \\
\ket{\psi}_\mathrm{pol} &\equiv \alpha\ket{H} + \beta\ket{V}, \nonumber \\
\ket{\psi}_\mathrm{dual} &\equiv \alpha\ket{0,1} + \beta\ket{1,0}, \nonumber \\
\ket{\psi}_\mathrm{temporal} &\equiv \alpha\ket{0_t,1_{t+\tau}} + \beta\ket{1_t,0_{t+\tau}},
\end{align}
shown graphically in Fig.~\ref{fig:photonic_qubit_encodings}.

\if 1\doublecol
	\begin{figure}[!htbp]
	\includegraphics[clip=true, width=0.15\textwidth]{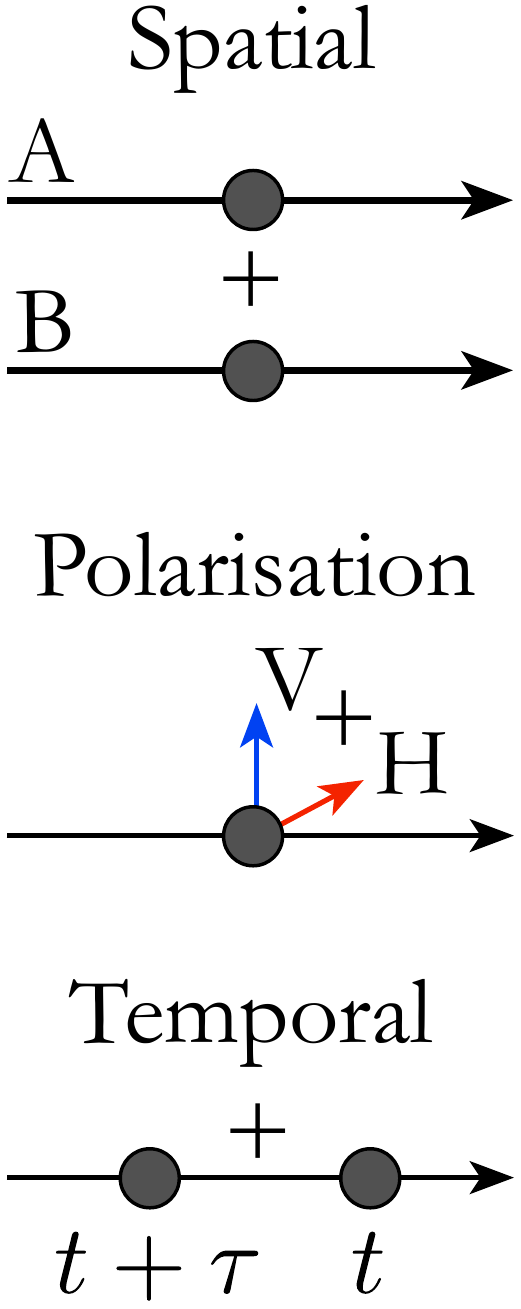}
	\captionspacefig \caption{Three approaches to encoding a single qubit using a single photon, via a superposition across two spatial ($A$ and $B$), polarisation ($V$ and $H$) or temporal ($t$ and \mbox{$t+\tau$}) modes.} \label{fig:photonic_qubit_encodings}
	\end{figure}
\else
	\begin{figure*}[!htbp]
	\includegraphics[clip=true, width=0.5\textwidth]{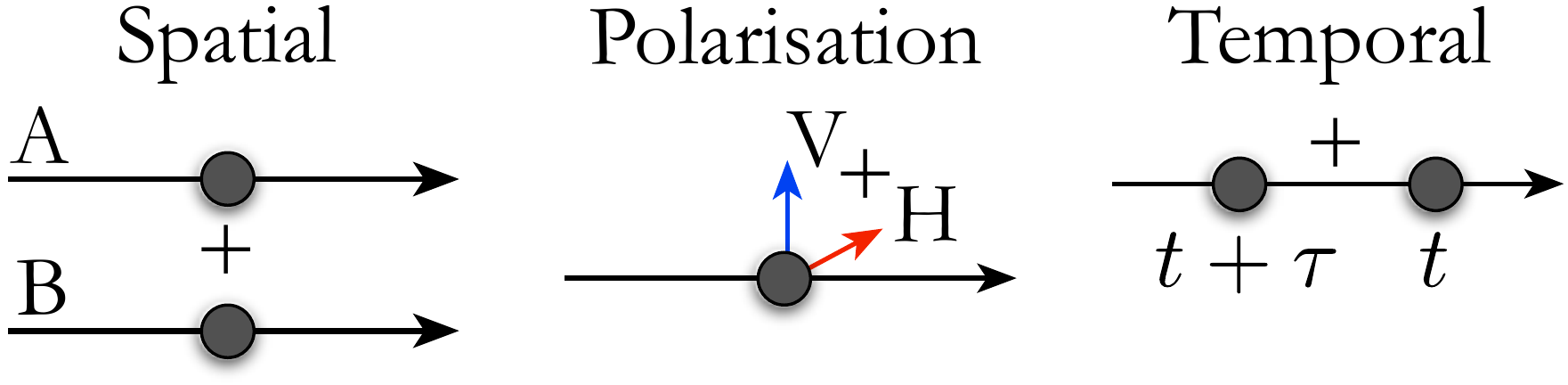}
	\captionspacefig \caption{Three approaches to encoding a single qubit using a single photon, via a superposition across two spatial ($A$ and $B$), polarisation ($V$ and $H$) or temporal ($t$ and \mbox{$t+\tau$}) modes.} \label{fig:photonic_qubit_encodings}
	\end{figure*}
\fi

\begin{figure}[!htbp]
\includegraphics[clip=true, width=0.3\textwidth]{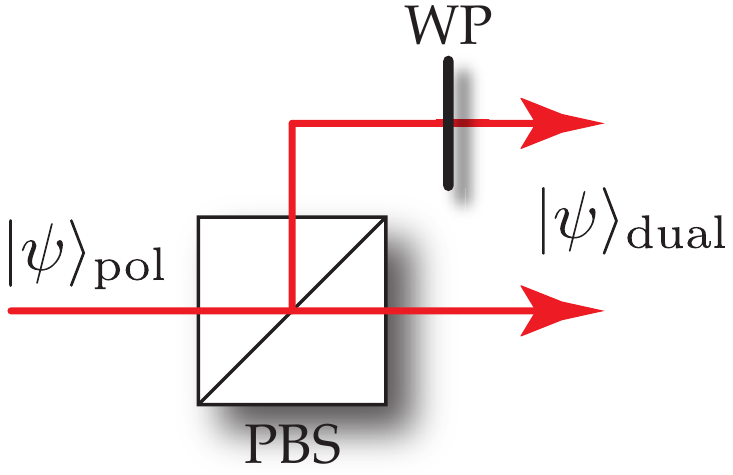}
\captionspacefig \caption{Conversion from single-photon polarisation encoding to dual-rail encoding, using a polarising beamsplitter (PBS) and wave-plate (WP). The PBS separates the polarisation components into two distinct spatial modes. The WP then rotates the polarisation of one of the spatial modes such that it has the same polarisation as the other. Conversion from dual-rail to polarisation encoding is just the reverse of this procedure.} \label{fig:pol_to_dual_conv}\index{Polarising beamsplitters}
\end{figure}

Conversion between polarisation and dual-rail encoding is straightforward and deterministic using standard optical components, as described in Fig.~\ref{fig:pol_to_dual_conv}.

Note that polarisation encoding requires a single spatial mode per qubit, whereas dual-rail encoding requires two. Polarisation encoding brings with it the advantage that arbitrary single-qubit operations may be implemented using wave-plates, which maintain coherence between the basis states extraordinarily well. In dual-rail encoding, on the other hand, single-qubit operations are implemented using beamsplitter operations between the two spatial modes, which must be interferometrically stable, since consecutive single-qubit operations yields Mach-Zehnder (MZ) interference \cite{bib:Zehnder1, bib:Zehnder2}\index{Mach-Zehnder (MZ) interference}, to be discussed in detail in Sec.~\ref{sec:MZ_inter}.

Single-photon encodings are extremely important, as they form the basis for universal linear optics quantum computing (Sec.~\ref{sec:KLM_univ}), \textsc{BosonSampling} (Sec.~\ref{sec:boson_sampling}) and quantum walks (Sec.~\ref{sec:QW}). They are also the simplest optical states for representing single qubits.

%
% Photon-number
%

\subsection{Photon-number} \index{Photon-number!Encoding}

Of course, the photon-number degree of freedom needn't be limited to 0 or 1 photons. By fully exploiting the photon-number degree of freedom, we can encode a qudit\footnote{A $d$-level system, as opposed to a qubit's two levels.}\index{Qudits} of arbitrary dimension into a single optical mode,
\begin{align} \label{eq:number_qudit}
\ket\psi_\mathrm{qudit} \equiv \sum_{n=0}^\infty \alpha_n \ket{n}.
\end{align}
This may give the impression that a single optical mode has infinite information capacity. Needless to say, this sounds too good to be true, and it is. Loss decoheres photon-number-encoded states exponentially with photon-number, since for large photon-number the probability of a number state retaining its photon-number exponentially asymptotes to zero. So although in principle we can encode an $\infty$-level qudit, the moment any non-zero loss is introduced, this exponential dependence destroys the state (Sec.~\ref{sec:eff_err}).

While photon-number encoding can be useful for communications purposes, it is not very practical for quantum information processing tasks, since operations between basis states are not energy preserving, with each basis state having energy \mbox{$E=n\hbar\omega$}, where $\omega$ is frequency, and $\hbar$ is Planck's constant. Thus, qudit operations would need to be active processes.

%
% Spatio-Temporal Qudit Encoding
%

\subsection{Spatio-temporal} \label{sec:spatio_temporal} \index{Spatio-temporal!Encoding}

Completely independent of the photon-number degree of freedom are the spatio-temporal degrees of freedom, which encode the spatial, temporal, or spectral structure of photons. In the temporal domain, for example, we could define the temporal structure of a single photon as,
\begin{align}
\ket\psi_\mathrm{temporal} = \int_{-\infty}^\infty \psi(t) \hat{a}^\dag(t)\,dt\,\ket{0},
\end{align}
where $\hat{a}^\dag(t)$ is the time-specific photonic creation operator, and $\psi(t)$ is the temporal distribution function \cite{bib:RohdeFreqTemp05}. Equivalently, one could take the Fourier transform\index{Fourier transform} of the temporal distribution function and represent the same state in the frequency basis,
\begin{align}
\tilde\psi(\omega) = \mathcal{FT}(\psi(t)).	
\end{align}
Likewise, one could employ a similar representation in the transverse spatial degrees of freedom, with spatial distribution function $\psi(x,y)$.

Alternately, we can define \textit{mode operators} \cite{bib:RohdeMauererSilberhorn07,baragiola2012photons}\index{Mode operators}, which are mathematically equivalent to creation operators, but create photons with a specific temporal envelope,
\begin{align}
\hat{A}^\dag_\psi &= \int_{-\infty}^\infty \psi(t) \hat{a}^\dag(t)\,dt, \nonumber \\
\ket\psi_\mathrm{temporal} &= \hat{A}^\dag_\psi \ket{0}.
\end{align}
Mode operators commute, inheriting this property directly from photonic creation operators,
\begin{align}
\left[\hat{A}^\dag_{\psi_1},\hat{A}^\dag_{\psi_2}\right]=0.
\end{align}

Now by defining an orthonormal basis of temporal distribution functions, $\{\xi_i\}$, such that,
\begin{align} \label{eq:spec_orth_def}
\bra{0} \hat{A}_{\xi_i} \hat{A}^\dag_{\xi_j}\ket{0} = \delta_{i,j},
\end{align}
we can encode a qudit of arbitrary dimension into the spatio-temporal degrees of freedom,
\begin{align}
\ket\psi_\mathrm{qudit} \equiv \sum_{i=0}^\infty \alpha_i \hat{A}^\dag_{\xi_i} \ket{0}.
\end{align}

This encoding allows a qudit of arbitrary dimension to be encoded into a single spatial mode. Again, however, summing to infinity is somewhat fanciful, given any physically realistic spatio-temporal error model, such as an imperfect frequency response in the channel, e.g a bandpass response of an optical fibre or photo-detector.

\begin{figure}[!htbp]
	\includegraphics[clip=true, width=0.475\textwidth]{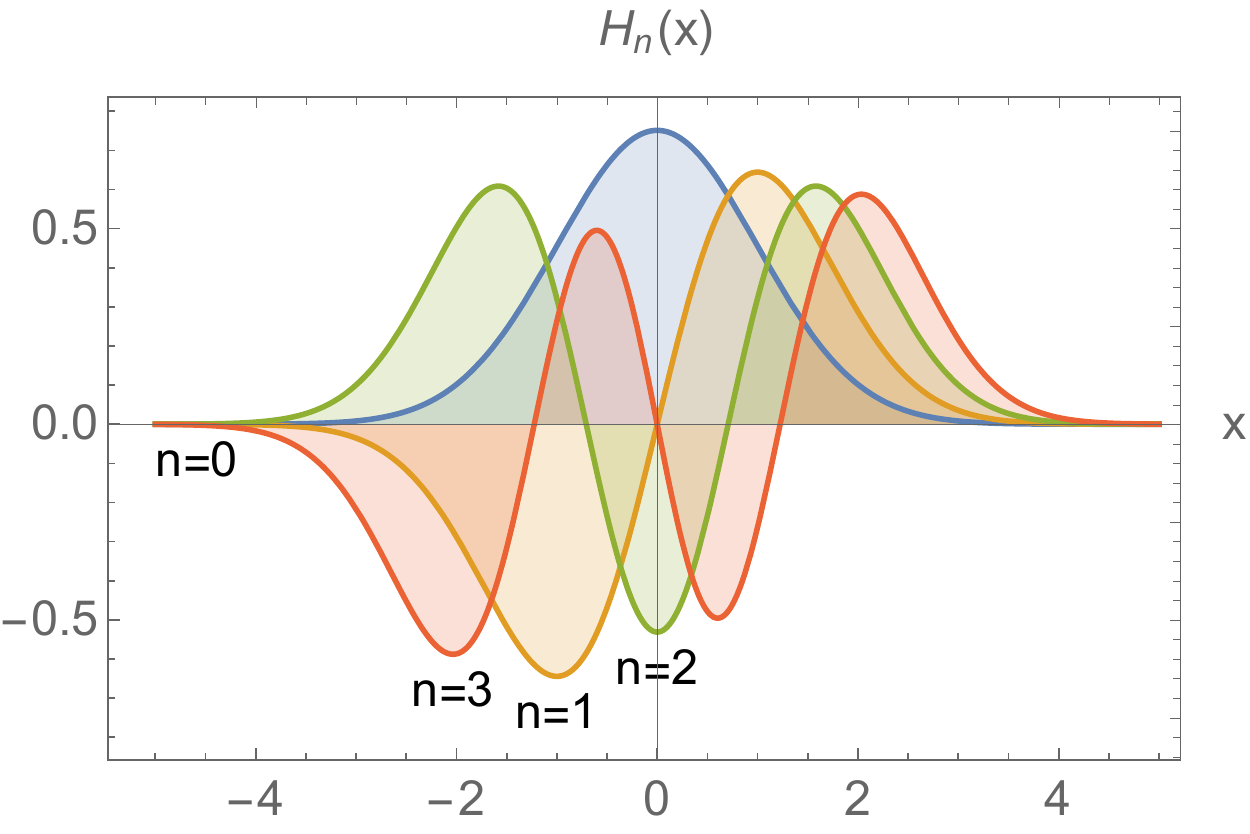}
	\captionspacefig \caption{The family of Hermite functions, $H_n(x)$, modulated by Gaussian envelopes, are an example of an orthonormal basis of continuous functions that could be utilised in the spatio-temporal encoding of qudits into photonic wave-packets. Shown here are \mbox{$n=0,1,2,3$}.}\index{Hermite!Functions}\label{fig:hermite_basis}
\end{figure}

The spectral basis functions could take the form of any orthonormal basis of complex functions, such as wavelets 
\cite{chui2016introduction,brennen2015multiscale}\index{Wavelets}, Hermite functions\index{Hermite!Functions} (shown in Fig.~\ref{fig:hermite_basis}), or well-separated functions with finite support.

\subsubsection{Transverse electro-magnetic modes}\index{Transverse electro-magnetic (TEM) modes}

Of particular interest are transverse electro-magnetic (TEM) modes, which are the two-dimensional eigenfunctions of the electro-magnetic field in the transverse direction of a light field. The are obtained by imposing boundary conditions on the electro-magnetic field, depending on the medium the geometry of the medium (e.g optical fibre, free-space, wave-guides).

Specifically, for cylindrical and rectangular boundary conditions, the intensity profile of the TEM modes are given by,
\begin{widetext}
\begin{align*}
	I_{p,l}^{\mathrm{cyl}}(\rho,\varphi) &= I_0 \rho^l L_p^l(\rho)^2 \cos(l\varphi)e^{-\rho},\nonumber\\
	I_{m,n}^{\mathrm{rect}}(x,y,z) &= I_0\left(\frac{\omega_0}{\omega}\right)^2 \left[H_m\left(\frac{\sqrt{2}x}{\omega}\right)e^{-\frac{x^2}{\omega^2}}\right]^2 \left[H_n\left(\frac{\sqrt{2}y}{\omega}\right)e^{-\frac{y^2}{\omega^2}}\right]^2,
\end{align*}
\end{widetext}
where,
\begin{align}
	H_n(x) &= n! \sum_{m=0}^{\lfloor \frac{n}{2}\rfloor} \frac{(-1)^m}{m!(n-2m)!} (2x)^{n-2m},\nonumber\\
	L_n(x) &= \sum_{k=0}^n \binom{n}{k} \frac{(-1)^k}{k!}x^k,
\end{align}
are the Hermite and Laguerre polynomials\index{Hermite!Polynomials}\index{Laguerre polynomials} respectively. These examples are shown in Fig.~\ref{fig:TEM_modes}.

\if 1\doublecol
	\begin{figure}[!htbp]
	\includegraphics[clip=true, width=0.475\textwidth]{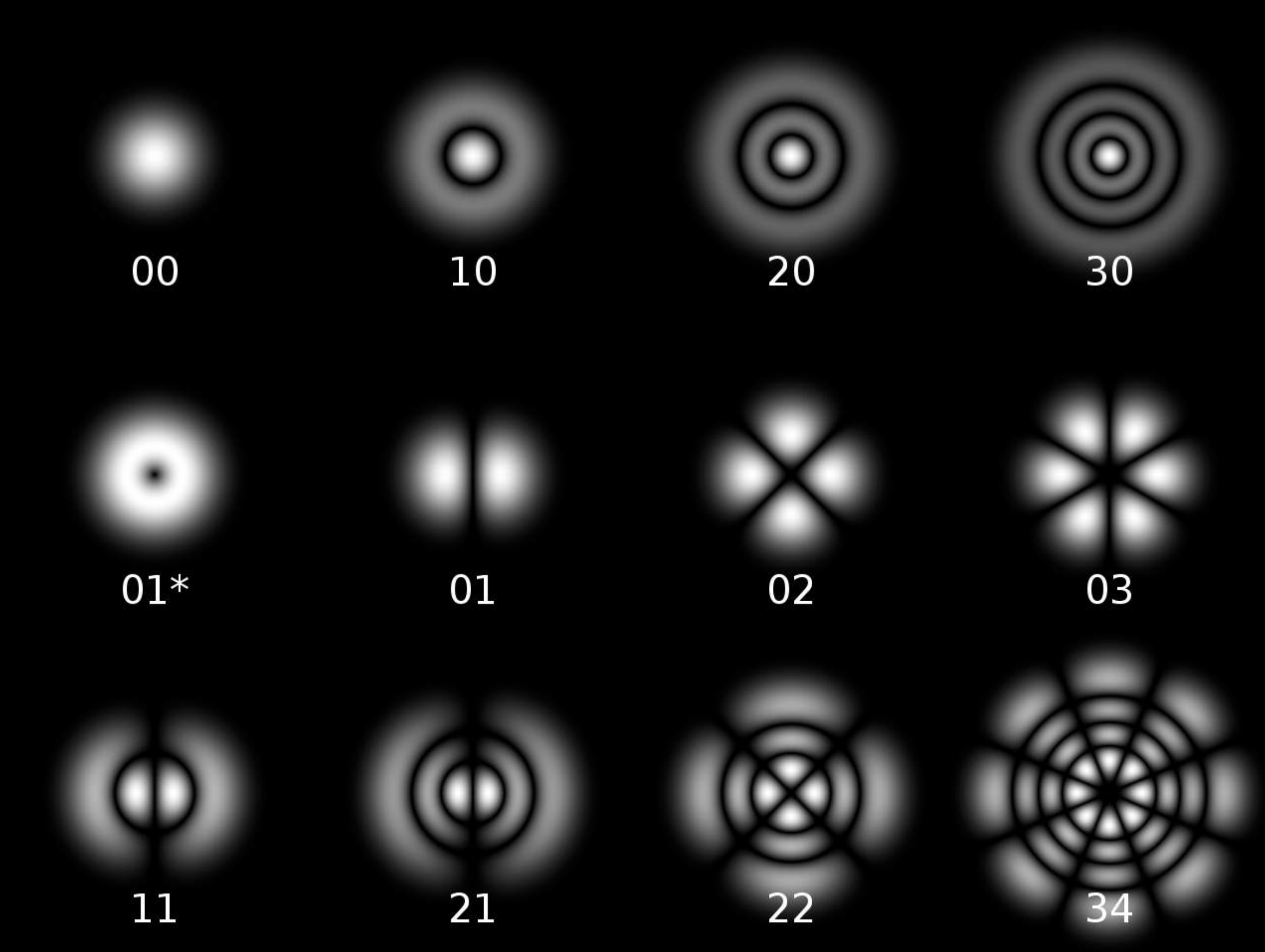}\\
	\includegraphics[clip=true, width=0.475\textwidth]{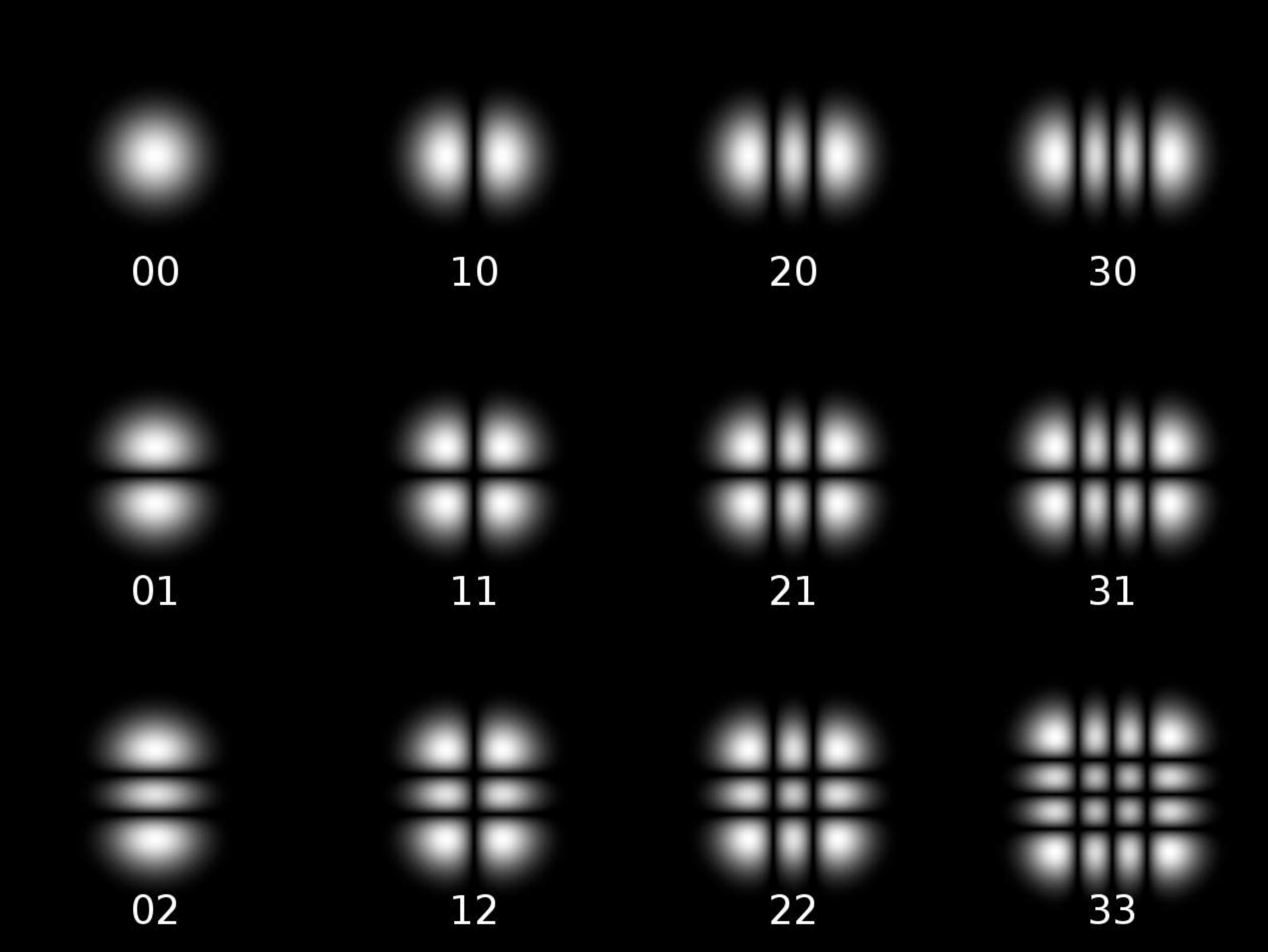}
	\captionspacefig \caption{TEM$_{m,n}$ modes for cylindrical (top) and rectangular (bottom) mode profiles.}\index{Transverse electro-magnetic (TEM) modes}\label{fig:TEM_modes}
	\end{figure}
\else
	\begin{figure*}[!htbp]
	\includegraphics[clip=true, width=0.49\textwidth]{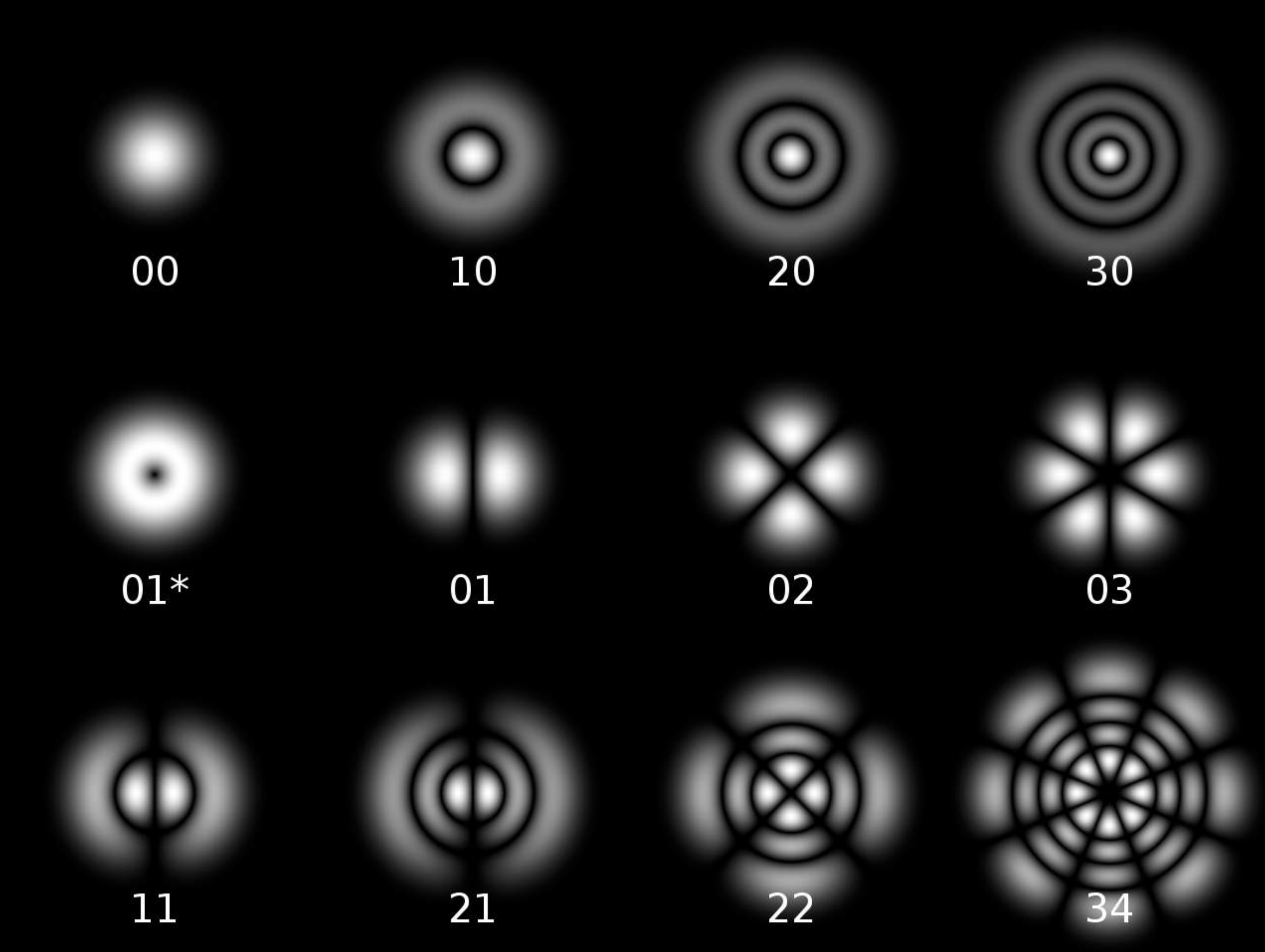}
	\includegraphics[clip=true, width=0.49\textwidth]{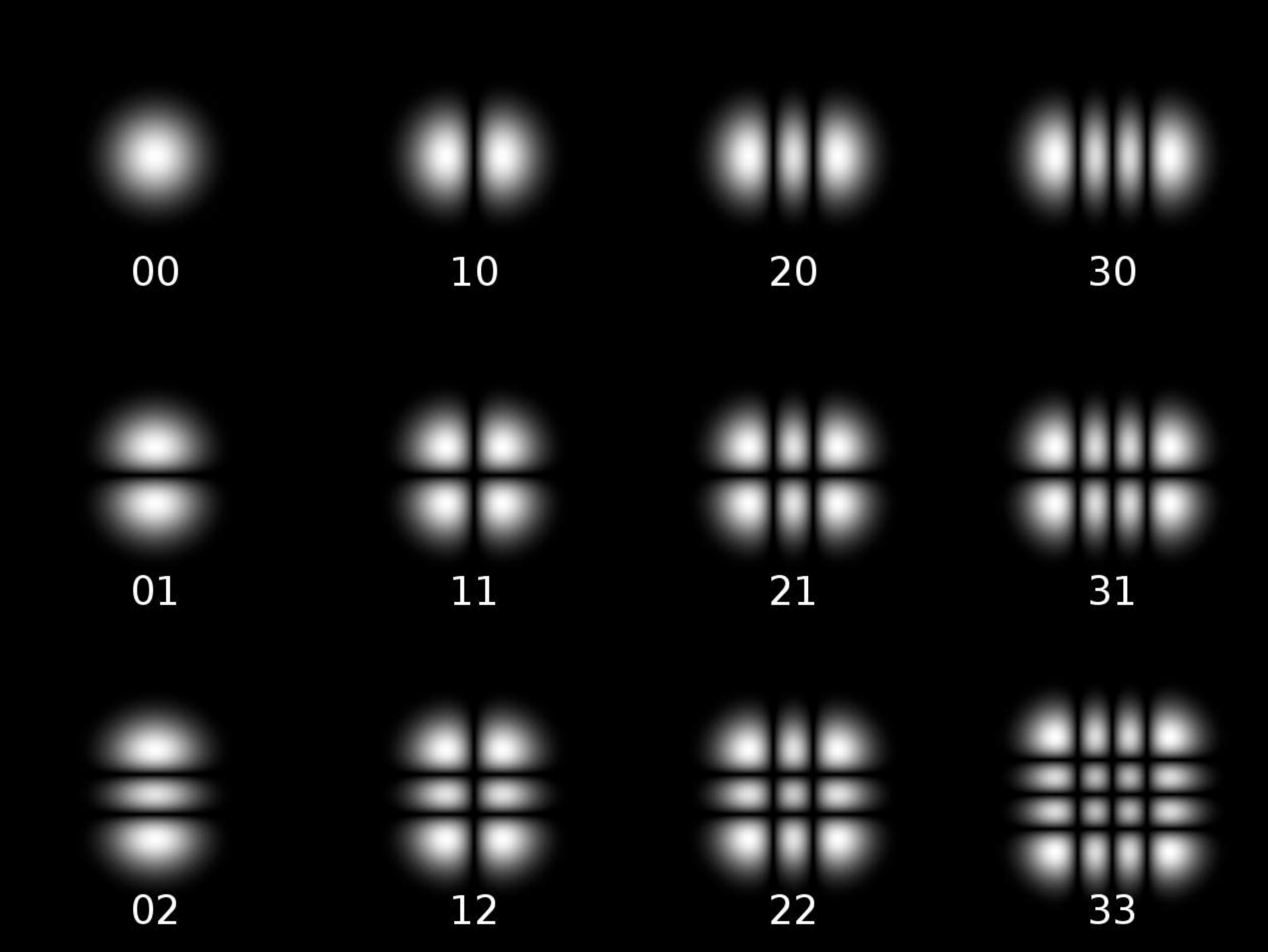}
	\captionspacefig \caption{TEM$_{m,n}$ modes for cylindrical (left) and rectangular (right) mode profiles.}\index{Transverse electro-magnetic (TEM) modes}\label{fig:TEM_modes}
	\end{figure*}
\fi

The TEM modes are discrete, and denoted TEM$_{mn}$, where \mbox{$m,n\in \mathbb{Z}_+$}. TEM modes can be prepared and manipulated using holograms\index{Holograms} in the form of phase-masks\index{Phase!Masks}. 

%
% Time-Bins
%

\subsubsection{Time-bins} \label{sec:time_bin} \index{Time-bin encoding}

\begin{figure}[!htbp]
\includegraphics[clip=true, width=0.25\textwidth]{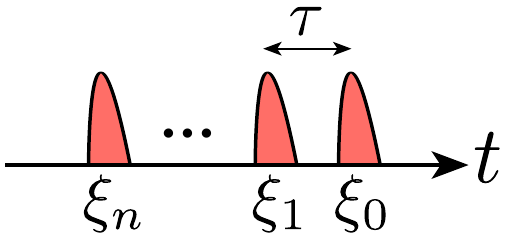}
\captionspacefig \caption{Time-bin encoding on an $n$-level quantum system. Time-bin are temporally localised with envelopes $\xi_j$, orthogonal to all others, different only via temporal displacements. The temporal separation between each consecutive time-bin is given by $\tau$.}\label{fig:time_bin_encoding}	
\end{figure}

In time-bin encoding we define our basis of modes (whether it be qubits or higher-dimensional qudits\index{Qudits}) as distinct, non-overlapping time-bins, which are localised wave-packets in the temporal degree of freedom, each separated from the next by some fixed interval $\tau$. This can be considered a special case of spatio-temporal encoding\index{Spatio-temporal!Encoding}, where the basis mode functions satisfy the relation,
\begin{align}
\xi_{j}(t) = \xi_0(t-j\tau),
\end{align}
as well as the usual orthonormality constraints. Here $\tau$ is sufficiently large, and $\xi_i(t)$ sufficiently temporally localised, that the temporal modes are orthogonal as per Eq.~(\ref{eq:spec_orth_def}). The encoding is shown graphically in Fig.~\ref{fig:time_bin_encoding}.

Time-bin encoding arises naturally in architectures where the photon source driving the system is operating at a high repetition rate\index{Repetition rate}, $R$, in which case \mbox{$\tau=1/R$}. Architectures for optical quantum computing have been described \cite{bib:RohdeLoop15, bib:RohdeUnivLoop15}, and experimentally demonstrated \cite{he2017time}, based entirely on time-bin encoding.

These schemes can be very resource efficient, since a single source operating at high repetition rate can replace an entire bank of distinct sources that would ordinarily be required in spatial architectures. Similarly, a single time-resolved detector, with resolution at least $\tau$, can replace a bank of detectors operating in parallel. And only a single spatial mode is required to store an arbitrary number of qubits/qudits, so long as it is long enough to support the entire pulse-train --- at least $2n\tau$ for $n$ qubits.

In the schemes of \cite{bib:RohdeLoop15, bib:RohdeUnivLoop15}, entire optical quantum computing protocols can be efficiently constructed using only a single source, a single detector, two delay-lines, and three dynamically-controlled beamsplitters, irrespective of the size of the computation, an enormous resource saving compared to traditional spatial encodings. Furthermore, in these schemes, there is only a single point of interference, greatly simplifying optical interferometric alignment, which would ordinarily require simultaneously aligning a large number of optical elements, as many as $O(m^2)$ elements for an $m$-mode network \cite{bib:Reck94}.

%
% Thermal State Encoding
%

\subsection{Thermal states} \index{Thermal!State encoding}\label{sec:thermal_states}

In some quantum protocols, although the inner workings may be quantum mechanical in nature, the inputs and outputs needn't capture any quantum coherence -- sometimes \textit{classical} information is sufficient for communications. As discussed above, coherent states are the archetypal example of this, and this is in fact the norm in present-day classical fibre-optic communication, where coherent states prepared via laser diodes\index{Lasers!Diodes} are employed.

Another, and even simpler option, is thermal states. These are obtained by fully dephasing a coherent state, retaining the amplitude distribution, while nullifying all the coherence terms,
\begin{align}
\hat\rho_\mathrm{thermal}(\alpha) = e^{-|\alpha|^2} \sum_{n=0}^\infty \frac{|\alpha|^2}{n!}\ket{n}\bra{n}.
\end{align}

Thermal states can encode classical information into their amplitudes, polarisations, or time-bins, as before. The advantage of this type of encoding is that thermal states are trivial to prepare and measure (e.g ordinary blackbody radiation\index{Blackbody radiation} emits thermal states -- this is how a normal incandescent lightbulb\index{Incandescent lightbulb} emits light). However, they are purely classical states, do not undergo interference with one another, and are therefore useless for, for example, entangling qubits via which-path erasure\index{Which-path erasure}, any other type of coherent interferometric process, or for representing coherent quantum information such as qubits.

%
% Phase-Space
%

\subsection{Phase-space} \label{sec:exotic} \index{Phase!Space!Encoding}

When encoding information optically, we needn't restrict ourselves to photon-number states (discrete variables\index{Discrete variables}). We also have a lot of flexibility to encode information in phase-space using continuous-variable (CV) states, where phase and amplitude relations encode quantum information \cite{bib:CahillGlauber69}. In this formalism, rather than expressing states in terms of photonic creation operators, $\hat{a}^\dag$, we represent them using phase-space position ($\hat{x}$) and momentum ($\hat{p}$) operators.

In phase-space, the most common method for visualising optical states is in terms of quasi-probability functions\index{Quasi-probability functions}\footnote{The term `quasi-probability' arises because in some regimes (for example, strictly non-negative $P$-functions), the function has a true probabilistic interpretation. However this interpretation breaks down for any negativity in the $P(\alpha)$, since negative probabilities have no meaningful classical interpretation.}, of which there are a multitude. The best known quasi-probability representations are:
\begin{itemize}\index{P-function}\index{Q-function}\index{Wigner function}
\item $P$-function: represents a state as a quasi-mixture of coherent states. When the $P$-function is strictly non-negative, it can be interpreted as a perfect classical mixture of coherent states. However, with any negativity this classical interpretation breaks down, hence `quasi'-probability. In general, the $P$-function representation for a state is not unique.
\begin{align}
\hat\rho = \int\!\!\!\int P(\alpha) \ket{\alpha}\bra{\alpha} d^2\alpha.
\end{align}
\item $Q$-function: represents a state in terms of its overlap with the complete set of all coherent states, which form an over-complete basis.
\begin{align}
Q(\alpha) = \frac{1}{\pi} \braket{\alpha|\hat\rho|\alpha}.
\end{align}
\item Wigner function: also has a quasi-probability interpretation, and negativity is qualitatively associated with `quantumness'. The Wigner function of a state is unique, and isomorphic to the density operator, making it perhaps the most useful phase-space representation for quantum states of light.
\begin{align}
W(x,p) = \int e^{ips/\hbar} \left\langle x-\frac{s}{2}\right| \hat\rho \left|x+\frac{s}{2}\right\rangle ds.
\end{align}
\end{itemize}
These representations, whilst entirely equivalent to a photon-number basis representation, are far easier to work with for many types of states. Most notably, Gaussian states are conveniently represented and manipulated using phase-space representations.

%
% Coherent States
%

\subsubsection{Coherent states} \label{sec:coherent_state_enc} \index{Coherent states!Encoding}

As the most trivial CV encoding of quantum information, consider coherent states. These are particularly useful since they are pure states, with well defined coherence relationships, and are closely approximated by laser sources, and therefore readily available in the lab.

\begin{figure}[!htbp]
\includegraphics[clip=true, width=0.4\textwidth]{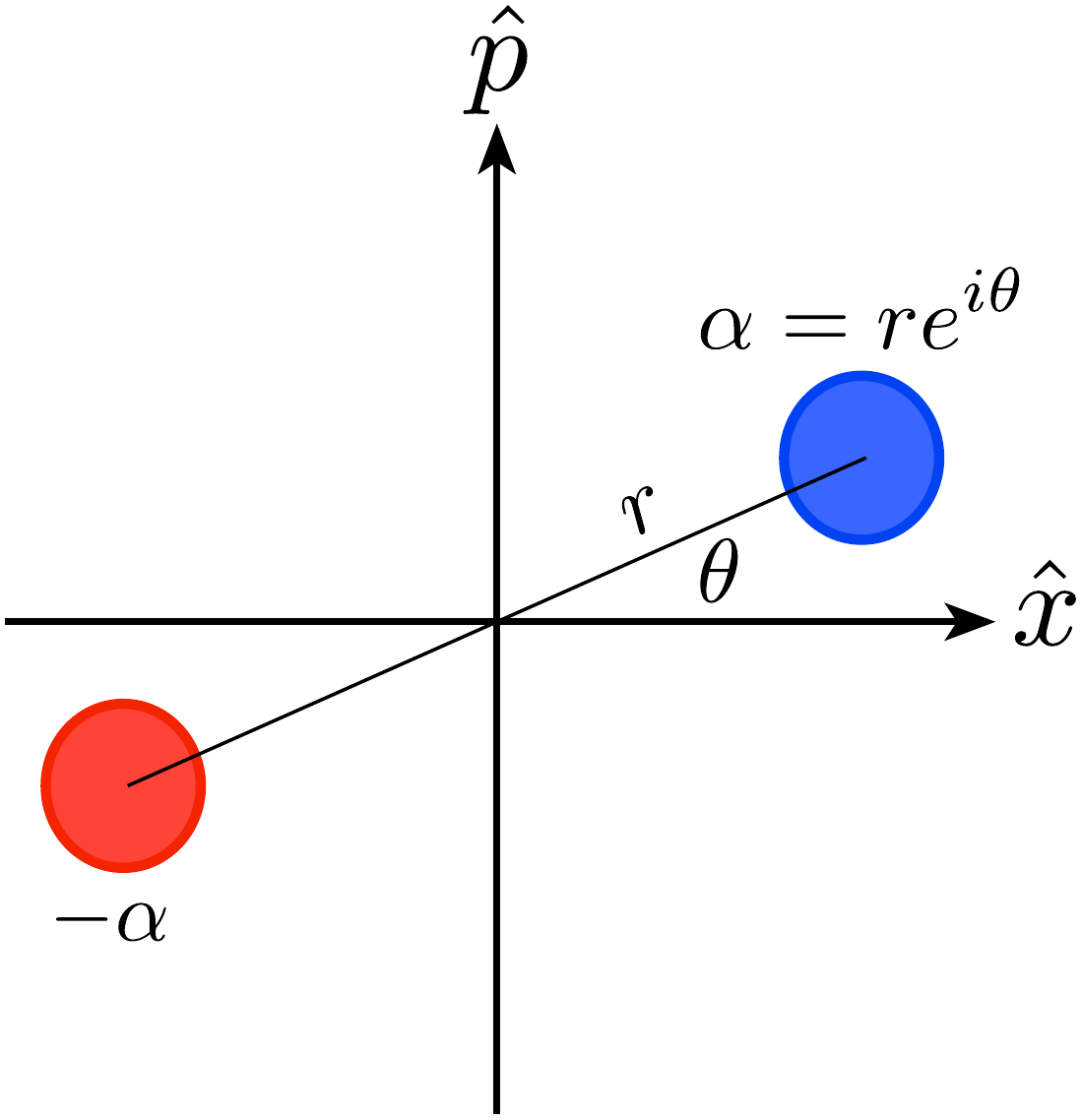}
\captionspacefig \caption{Phase-space representation of a coherent state with complex amplitudes $\pm\alpha$. For sufficiently large amplitude one can approximate orthogonality, enabling a qubit encoding.} \label{fig:coherent_state_encoding}	
\end{figure}

A coherent state, $\ket\alpha$, is parameterised by a single complex parameter, $\alpha$, given by a phase and amplitude,
\begin{align}\index{Coherent states}
\ket{\alpha} = e^{-\frac{|\alpha|^2}{2}} \sum_{n=0}^\infty \frac{\alpha^n}{\sqrt{n!}} \ket{n}.
\end{align}
Fig.~\ref{fig:coherent_state_encoding} illustrates the phase-space representation for two approximately orthogonal coherent state basis states.

By manipulating these parameters, information can be encoded into coherent states. We could, for example, define two coherent states of opposite phase to represent qubit basis states,
\begin{align}
\ket{0}_L &\equiv \ket{\alpha}, \nonumber \\
\ket{1}_L &\equiv \ket{-\alpha}.
\end{align}
Note, however, that this representation for qubits is only approximate, since coherent states are not orthogonal,
\begin{align}\label{eq:coh_st_overlap}
\braket{\alpha|\beta} &= e^{-\frac{1}{2}(|\alpha|^2+|\beta|^2-2\alpha^*\beta)}\nonumber\\
&\neq \delta(\alpha-\beta),	
\end{align}
thus the two logical basis states are not perfectly orthogonal,
\begin{align}
\braket{-\alpha|\alpha} = e^{-2|\alpha|^2},
\end{align}
which is non-zero for any finite $\alpha$, whereas for ideal qubits we require \mbox{$\braket{0|1} = 0$}. However, for large $\alpha$, $\ket{\pm\alpha}$ closely approximate orthogonality, allowing them to be used as qubits.

This representation for qubits using coherent states is easily generalised to qudits by considering coherent states orbiting the origin of phase-space at equal angular intervals of \mbox{$2\pi/d$}, for a $d$-level qudit. The $k$th qudit basis state is then,
\begin{align}
\ket{k}_d = \ket{\alpha e^{ik/d}},
\end{align}
for \mbox{$k=0,\dots,d-1$}, where again the basis states are non-orthogonal, but closely approximate orthogonality for large $\alpha$. The qudit value $k$ can easily be manipulated using simple phase-shift operators\index{Phase!Shifts},
\begin{align}
\hat\Phi(\phi) = e^{i\phi\hat{n}},	
\end{align}
where \mbox{$\hat{n}=\hat{a}^\dag\hat{a}$} is the photon-number operator. These phases are trivially implemented in the laboratory as wavelength-scale modulations in optical path length (i.e a thin piece of glass or other transmissive material with a different refractive index).

Note that despite being pure states, with well-defined coherence, coherent states are considered classical, as they are unable to encode quantum information. That is, the coherence relationships cannot be exploited for the encoding of qubits or qudits.

Coherent states are useful in that they are easy to prepare using modern lasers, including laser diodes, and by turning up the amplitude can be transmitted over long distances, with loss not affecting quantum coherence, only the amplitude (Sec.~\ref{sec:eff_err}).

Coherent state encoding can be regarded as encoding via the displacement operator,
\begin{align}\label{eq:disp_op}
\hat{D}(\alpha) = \exp \left[\alpha\hat{a}^\dag - \alpha^*\hat{a}\right],
\end{align}
which implements translations in phase-space via the addition of coherent amplitude to a state. Coherent states are simply obtained as displaced vacuum states,
\begin{align}
\hat{D}(\alpha)\ket{0} = \ket{\alpha}.
\end{align}

%
% Cat States
%

\subsubsection{Cat states} \label{sec:cat_enc} \index{Cat states!Encoding}\index{Cat states}

Another type of CV state, which can in fact encode quantum information, is superpositions of coherent states (colloquially known as `cat' states), with the encoding \cite{bib:JeongRalph05},
\begin{align}\label{eq:cat_state_enc}
\ket{0}_L &\equiv \mathcal{N}_+ (\ket{\alpha}+\ket{-\alpha}) \nonumber \\
&= 2\mathcal{N}_+ e^{-\frac{|\alpha|^2}{2}} \sum_{n=0}^\infty \frac{\alpha^{2n}}{\sqrt{(2n)!}} \ket{2n}, \nonumber\\
&= \ket{\mathrm{cat}_+(\alpha)},\nonumber \\
\ket{1}_L &\equiv \mathcal{N}_- (\ket{\alpha}-\ket{-\alpha}) \nonumber \\
&= 2\mathcal{N}_- e^{-\frac{|\alpha|^2}{2}} \sum_{n=0}^\infty \frac{\alpha^{2n+1}}{\sqrt{(2n+1)!}} \ket{2n+1},\nonumber\\
&= \ket{\mathrm{cat}_-(\alpha)},
\end{align}
where the normalisation factors are,
\begin{align}
\mathcal{N}_\pm = \frac{1}{\sqrt{2(1\pm e^{-2|\alpha|^2})}},
\end{align}
which arise due to the non-orthogonality of coherent states, Eq.~(\ref{eq:coh_st_overlap}). These two basis states contain strictly even or odd photon-number terms respectively (i.e they have well-defined parity\index{Parity}), implying that, unlike coherent states, they are always orthogonal, regardless of amplitude,
\begin{align}
\braket{\mathrm{cat}_+(\alpha)|\mathrm{cat}_-(\alpha)} = 0 \,\,\forall\,\alpha,
\end{align}
making them directly appropriate for qubit encoding, even for weak coherent amplitudes.

Unfortunately, cat states are notoriously difficult to prepare, and extremely sensitive to loss (Sec.~\ref{sec:eff_err}) and dephasing (Sec.~\ref{sec:dephasing_error}). This arises because loss of a single photon flips the parity of the state to an orthogonal one, meaning that as $\alpha$ increases, the state is exponentially more susceptible to decohering into a mixture of the logical basis states. However, modulo these difficulties, with a resource of cat states at one's disposal, universal quantum computation may be realised using post-selected linear optics \cite{bib:JeongRalph05, bib:Gilchrist04}.

Generalizations of cat codes using more than two coherent states in superposition~\cite{zaki2103cats} can be used to increase resilience to photon-loss errors at the expense of reduced resilience to dephasing-type errors. These codes and others like them exhibit discrete rotational symmetry in phase space; their properties and a scheme to perform quantum computation can be found in Ref.~\cite{Grimsmo2020rotation}.

%
% Squeezed States
%

\subsubsection{Squeezed states}\index{Squeezed states}\label{sec:squeezed_enc}

In the same way that information can be encoded using the displacement operator via coherent states, we can encode information via the single-mode squeezing operator\index{Squeezing!Operators},
\begin{align}\label{eq:sq_op}
\hat{S}(\xi) = \exp\left[ \frac{1}{2}(\xi^*\hat{a}^2 - \xi{\hat{a}^{\dag 2}})\right],
\end{align}
where \mbox{$\xi  = r e^{i \varphi}\in \mathbb{C}$}, $r$ is known as the squeezing parameter\index{Squeezing!Parameter}, which will determine the magnitude of the squeezing, and \mbox{$\varphi \in [0, 2\pi]$} denotes the axis along which the squeezing is taking place.

\begin{figure}[!htbp]
\includegraphics[clip=true, width=0.3\textwidth]{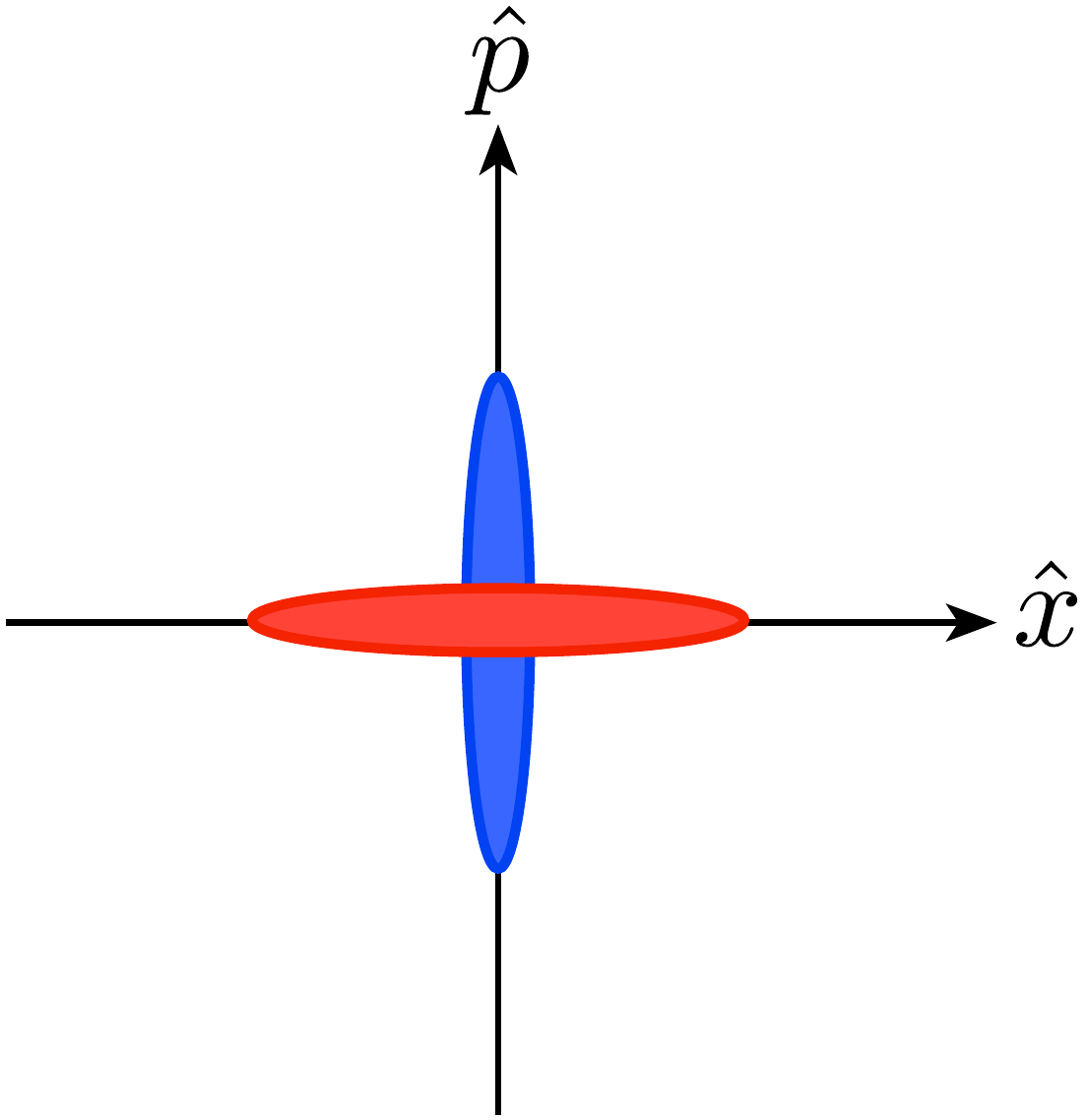}
\captionspacefig \caption{Phase-space representation of two squeezed vacuum states, squeezed in the $\hat{x}$ (blue) and $\hat{p}$ (red) directions. In the limit of large squeezing these states become approximately orthogonal and may therefore approximate the encoding of a qubit.} \label{fig:squeezed_state_encoding}
\end{figure}

Graphically, in terms of their phase-space visualisation, squeezing implements dilations about a given axis. Strongly squeezing the vacuum state along the $\hat{x}$ or $\hat{p}$ directions yields two states that are approximately orthogonal for large squeezing amplitudes, as shown in Fig.~\ref{fig:squeezed_state_encoding}. Thus, with sufficient squeezing they can be used as a basis for \textit{approximating} a qubit. This encoding can be exploited for full universal quantum computing, to be discussed in Sec.~\ref{sec:CV_QC}.

%
% GKP States
%

\subsubsection{Gottesman-Kitaev-Preskill states}\index{GKP states}\label{sec:gkp_enc}

More exotic states can be used to encode a logical qubit. An important example are Gottesman-Kitaev-Preskill (GKP) states~\cite{gottesman2001encoding,brady2024GKPreview}, because they have a logical gate set that can be implemented deterministically using linear optics (beam splitters and phase shifters), require only standard homodyne detection~\cite{baragiola2010allgaussian}, and are extremely good for protection against loss~\cite{noh2019capacity}. The logical states are best described as a comb of position (or momentum) eigenstates,
	\begin{align}
		\ket{0}_L & \equiv \sum_{n = -\infty}^\infty \ket{x = 2n\sqrt{\pi} } \\
		\ket{1}_L & \equiv \sum_{n = -\infty}^\infty \ket{x = (2n+1)\sqrt{\pi} }. 
	\end{align}
Thus they can be though of as a superposition of highly squeezed states. This periodicity gives GKP states a translationally-invariant grid structure in phase space, too~\cite{mensen2021phase}. The GKP states above are unphysical and require infinite squeezing; finite-energy approximations used in practice and for analysis still have good resilience to photon loss and can be used for fault-tolerant universal quantum computation.

%
% Non-Optical Encoding
%

\subsection{Non-optical encoding}\index{Non-optical encodings}

In a non-optical context, the elementary unit of quantum information -- the qubit -- can be naturally encoded into any system with a natural or engineered two-level structure. This actually encompasses a broad range of possibilities, including, amongst many others:
\begin{itemize}
\item \index{2-level systems}Two-level atoms: let two distinct electron energy levels, with long lifetimes, represent the two logical basis states.
\item \index{$\lambda$-configuration systems}$\lambda$-configuration atoms: atoms with two degenerate ground states, which encode the logical qubit, and an additional excited state, which may be transitioned to upon excitation from only one of the ground states. Relaxation from the excited state enables optical coupling via the emitted photon.
\item \index{Quantum dots}Quantum dots: are essentially artificial atoms, which can be engineered with custom band-structures, allowing two- or higher-level qudits to be easily fabricated.
\item \index{Nitrogen-vacancy (NV) centres}Nitrogen-vacancy (NV) centres: are a type of point defect in diamond, which has a very well defined energy level structure that may be utilised to represent qubits.
\item \index{Atomic!Ensembles}Atomic ensembles: encode quantum information similarly to a single atom, except that the excitation is a \textit{collective} one, in superposition across all the atoms in the ensemble.
\item \index{Superconductors!Rings}Superconducting rings: a superposition of current flow direction in a superconducting ring represents the two logical basis states.
\item \index{Trapped ions}Trapped ions: qubits are encoded into stable electronic states of electromagnetically trapped ions.
\end{itemize}

Clearly the non-optical elements in a quantum network must somehow interface with optical states, such that communication is facilitated. This is discussed later in Sec.~\ref{sec:opt_inter}. \label{sec:optical_encodings}

\latinquote{Audaces fortuna juvat.}

%
% Errors in Quantum Networks
%

\section{Errors in quantum networks} \label{sec:errors_in_nets} \index{Errors in quantum networks}

\dropcap{A}{s} with classical data, quantum data is susceptible to corruption during transmission. However, in addition to all the usual classical error models, quantum information is subject to further uniquely quantum errors. These errors can be represented using the quantum process formalism and fully characterised using QPT (Sec.~\ref{sec:QPT}). We now briefly discuss several of the dominant errors arising in quantum systems, paying especial attention to error models acting on qubits and optical states, as these are the most relevant in a quantum networking context.

%
% Known Unitaries
%

\subsection{Known unitaries} \index{Unitary!Errors}

The most trivial error mechanism is when a (potentially multi-qubit) unitary channel (e.g an identity channel for the purposes of quantum memory) actually implements some unitary transformation, $\hat{U}$, that is not that which is desired. However, the unitary is constant, not varying from trial to trial, and is known, which can be easily determined by performing QPT on the channel. For example, an optical fibre might induce a polarisation rotation on transmitted photons, but the fibre isn't changing and neither is the rotation. If consistently implementing the same known unitary then reversing it is straightforward in most architectures, by applying $\hat{U}^\dag$, since $\hat{U}^\dag\hat{U}=\hat\openone$.

%
% Unknown Unitaries
%

\subsection{Unknown imperfect unitaries} \index{Unitary!Errors}

Alternate to known unitaries, the unitary operation implemented by a node/channel may deviate from that which is desired, in an unknown manner, thereby implementing a slightly different operation than that which we intended to engineer. Specifically, the effective unitary can be represented as the ideal unitary, augmented by some deviation matrix,
\begin{align}
	\hat{U}_\mathrm{effective} = \hat{U}_\mathrm{ideal} + \hat{\Delta}_\mathrm{error},
\end{align}
where the matrix elements of $\hat{\Delta}_\mathrm{error}$ (which is not unitary in general) are unknown, but hopefully small. Since the unknown deviation matrix needn't be constant, it will be a function of random variables, evaluated independently for each trial of the process. Furthermore, since the deviation matrix may vary from trial to trial, QPT cannot be employed to characterise it, unlike unitaries with fixed errors.

%
% Loss
%

\subsection{Loss} \label{sec:eff_err} \index{Loss!Channel}

Given that quantum communication links will typically be optical, the dominant error mechanism is likely to be loss. We let the \textit{efficiency}, $\eta$, of an optical quantum process be the probability that a given photon entering the channel leaves the channel in the desired mode, or probability \mbox{$1-\eta$} of being lost. In the case of information encoded into single-photon states, e.g using the polarisation degree of freedom, $\eta$ corresponds exactly to the success probability of the communication.

When implementing protocols employing post-selection upon detecting all photons, the protocol will be non-deterministic, where loss dictates the protocol's success probability. Specifically, with $n$ photons, each with efficiency $\eta$, the net post-selection success probability\index{Post-selection success probability} of the entire device is,
\begin{align}
	P=\eta^n.
\end{align}
This implies an exponential number of trials\index{Trials},
\begin{align}
	N = \frac{1}{P} = \frac{1}{\eta^n},
\end{align}
is required in post-selected protocols. Clearly this exponential scaling is of concern, requiring demanding efficiencies in future large-scale implementations.

% Model

\subsubsection{Model}\index{Loss model}

\begin{figure}[!htbp]
	\includegraphics[clip=true, width=0.3\textwidth]{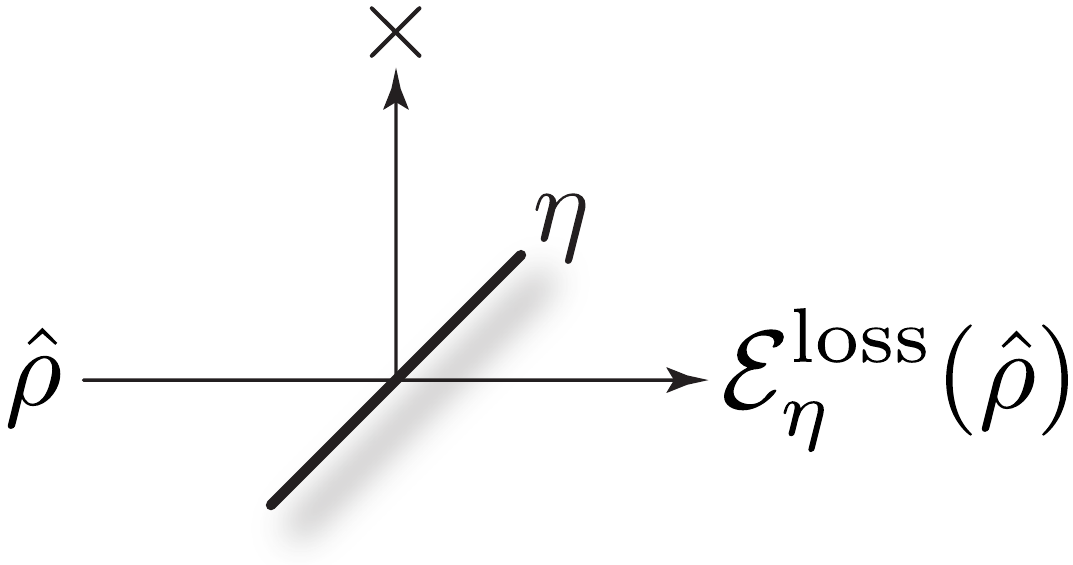}
	\captionspacefig \caption{Model for the loss channel. The input state, $\hat\rho$, passes through a beamsplitter of transmissivity $\eta$, and the reflected mode discarded, yielding the lossy output state \mbox{$\mathcal{E}^\mathrm{loss}_\eta(\hat\rho)$}.} \label{fig:loss_model} \index{Loss!Channel}
\end{figure}

Formally, let $\mathcal{E}^\mathrm{loss}_\eta$ be the loss channel with efficiency $\eta$. The channel acting on an initially pure single-photon state, $\ket{1}$, can be modelled as a beamsplitter with transmissivity $\eta$ acting on the state, where the reflected mode is traced out, shown in Fig.~\ref{fig:loss_model}. This yields the quantum process,
\begin{align}
\mathcal{E}^\mathrm{loss}_\eta(\hat\rho) = \mathrm{tr}_B[\hat{U}_\mathrm{BS}(\hat\rho_A\otimes\ket{0}_B\bra{0}_B)\hat{U}_\mathrm{BS}^\dag],
\end{align}
where $\hat{U}_\mathrm{BS}$ is the beamsplitter operation.

Consecutive loss channels act multiplicatively (in the net efficiency) and commutatively,
\begin{align}
\label{eq:LO_unitary_map}
\mathcal{E}_{\eta_1}^\mathrm{loss} \circ \mathcal{E}_{\eta_2}^\mathrm{loss} = \mathcal{E}_{\eta_2}^\mathrm{loss} \circ \mathcal{E}_{\eta_1}^\mathrm{loss} = \mathcal{E}_{\eta_1 \eta_2}^\mathrm{loss}.
\end{align}

% Linear optics networks

\subsubsection{Linear optics networks}\index{Linear optics networks}

\begin{figure}[!htbp]
	\includegraphics[clip=true, width=0.475\textwidth]{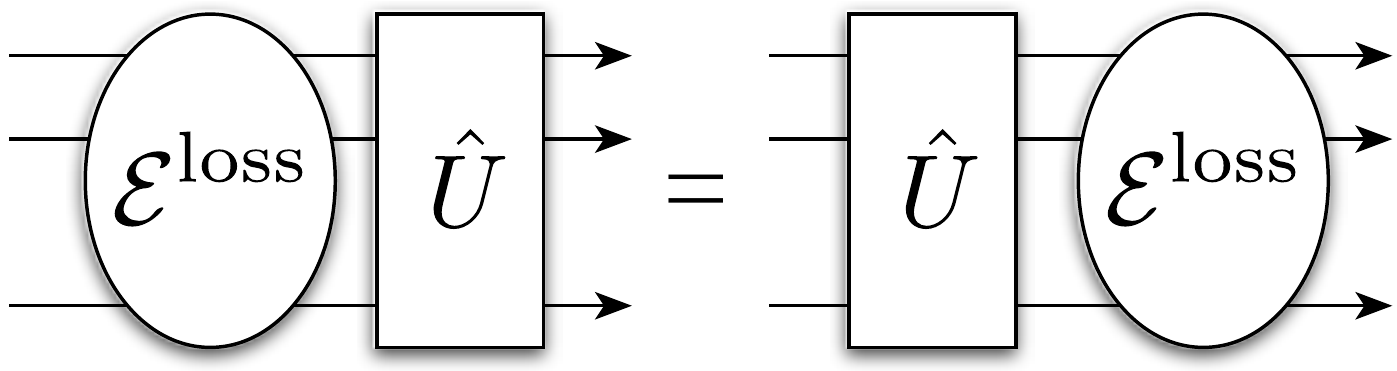}
	\captionspacefig \caption{Commutation of a uniform loss channel (i.e identical efficiency on each mode), $\mathcal{E}^\mathrm{loss}$, through a passive linear optics network, $\hat{U}$.} \label{fig:loss_LO_commutation} \index{Loss!Commutation}
\end{figure}

In the special case of linear optics circuits, loss channels have the elegant property that, provided the loss rate is uniform across all modes, they can be commuted through the circuit to the front or back \cite{oszmaniec2018classical,wiebe2012quantum}\index{Loss!Commutation}. Specifically,
\begin{align}
(\mathcal{E}_{\eta}^\mathrm{loss})^{\otimes m} \circ \mathcal{E}_U = \mathcal{E}_U \circ (\mathcal{E}_{\eta}^\mathrm{loss})^{\otimes m},
\end{align}
where $\mathcal{E}_U$ is a unitary linear optics process, implementing a photon-number-preserving map of the form of Eq.~(\ref{eq:LO_unitary_map}). This is represented by the circuit diagram shown in Fig.~\ref{fig:loss_LO_commutation}. This simplifies the treatment of distinct system inefficiencies (such as source, network and detector inefficiencies) by allowing us to commute them to the beginning or end of the circuit and combine them together into a single net efficiency. In many scenarios, this allows the different system inefficiencies to be dealt with via post-selection.

% Single-photon encoding

\subsubsection{Single-photon encoding}\index{Single-photons!Encoding}

In the case of the vacuum and single-photon states, which is most common to qubit encodings, we obtain,
\begin{align}
	\mathcal{E}^\mathrm{loss}_\eta(\ket{0}\bra{0}) &= \ket{0}\bra{0}, \nonumber \\
\mathcal{E}^\mathrm{loss}_\eta(\ket{1}\bra{1}) &= (1-\eta)\ket{0}\bra{0} + \eta\ket{1}\bra{1}.
\end{align}
This dynamic is of the same form as amplitude damping (Sec.~\ref{sec:amp_damp}).

% Polarisation & dual-rail encoding

\subsubsection{Polarisation \& dual-rail encoding}\index{Polarisation!Encoding}\index{Dual-rail encoding}

This process would apply equivalently to both horizontal and vertical polarisations. Therefore, via linearity, the loss channel acting on a polarisation-encoded qubit (Sec.~\ref{sec:single_phot_enc}) yields,
\begin{align}
\mathcal{E}^\mathrm{loss}_\eta(\ket\psi_\mathrm{pol}\bra\psi_\mathrm{pol}) = (1-\eta) \ket{0}\bra{0} + \eta\ket\psi_\mathrm{pol}\bra\psi_\mathrm{pol}.
\end{align}

The same applies in the context of dual-rail encoding. Note that while this transformation mixes the state in the photon-number degree of freedom, it preserves coherence between the horizontal and vertical single-photon components. Thus, upon successful post-selection, the state is projected back onto the desired qubit state.

% Photon-number encoding

\subsubsection{Photon-number encoding}\index{Photon-number!Encoding}

In the general case of an $n$-photon Fock state, we obtain,
\begin{align}
	\mathcal{E}^\mathrm{loss}_\eta(\ket{n}\bra{n}) = \sum_{i=0}^n \binom{n}{i} \eta^i(1-\eta)^{n-i} \ket{i}\bra{i}.
\end{align}

In the case of higher order photon-number encoding of qudits\index{Qudits}, as per Eq.~(\ref{eq:number_qudit}), the probability of an $n$-photon basis state being maintained scales as $\eta^n$. That is, if the highest photon-number term in our qudit is $n$, that component has an exponentially low probability of being preserved through the loss channel. For this fundamental reason, photon-number encoding does not enable infinite-dimensional qudits to be encoded.

% Coherent state encoding

\subsubsection{Coherent state encoding}\index{Coherent state!Encoding}

Coherent states are the one example of states, which are in a sense robust against loss, since a lossy coherent state is another coherent state with lower amplitude, but without any loss in coherence,
\begin{align}
\mathcal{E}^\mathrm{loss}_\eta(\ket\alpha\bra\alpha) = \ket{\eta\alpha}\bra{\eta\alpha}.
\end{align}
This arises because coherent states are eigenstates of the photonic annihilation operator, \mbox{$\hat{a}\ket{\alpha}=\alpha\ket{\alpha}$}.

However, although coherence is maintained under the loss channel, the process is irreversible, since noise-free amplitude amplification is not possible in general \cite{serafini2023quantum}. Thermal states exhibit the same property, that a loss channel simply yields another thermal state with reduced amplitude, although these exhibit no coherence.

% Cat state encoding

\subsubsection{Cat state encoding}

To the contrary, while cat states (Sec.~\ref{sec:cat_enc}) are simple superpositions of coherent states, they are extremely sensitive to loss. This is because cat states have well-defined photon-number parity (strictly even or odd photon-number), and therefore the loss of just a single photon will flip a cat state to an orthogonal one. Since the probability of photon loss occurring increases exponentially with photon-number, large amplitude cat states are exponentially sensitive to loss channels.

\begin{figure}[!htbp]
	\includegraphics[clip=true, width=0.475\textwidth]{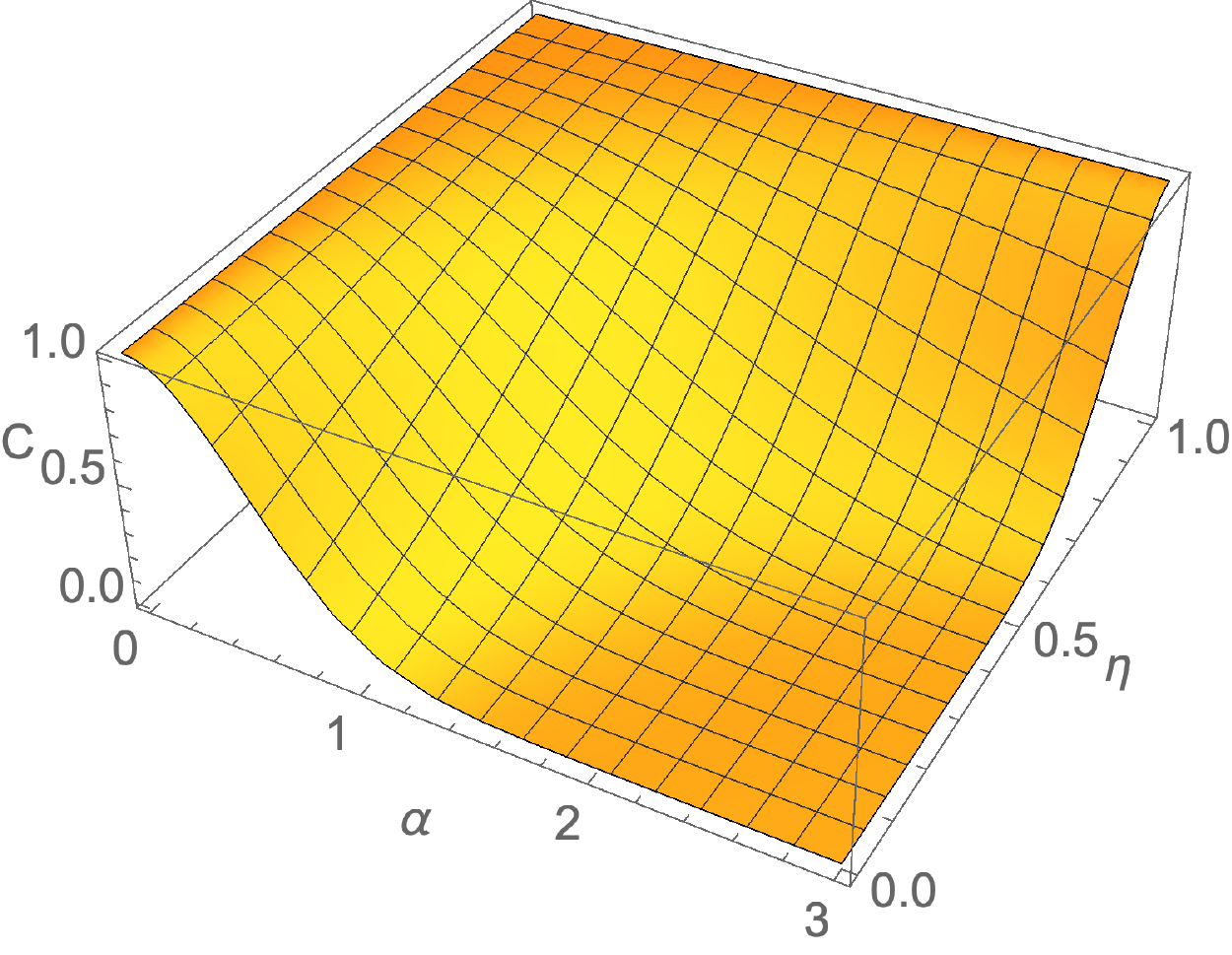}
	\caption{Coherence factor in a cat state subject to decoherence via loss, as a function of efficiency $\eta$, and coherent amplitude $\alpha$.}\label{fig:cat_decoherence}
\end{figure}

Consider a cat state (either even or odd parity), as per Eq.~(\ref{eq:cat_state_enc}). Subjecting this to a loss channel yields the mixed state,
\begin{align}
&\hat\rho_\mathrm{loss} = \mathcal{E}_\eta^\mathrm{loss}(\ket{\mathrm{cat}_\pm(\alpha)}\bra{\mathrm{cat}_\pm(\alpha)}) \nonumber\\
&= {\mathcal{N}_\pm}^2 [\mathcal{E}_\eta^\mathrm{loss}(\ket{\alpha}\bra{\alpha}) + \mathcal{E}_\eta^\mathrm{loss}(\ket{-\alpha}\bra{-\alpha}) \nonumber\\
&\pm \mathcal{E}_\eta^\mathrm{loss}(\ket{-\alpha}\bra{\alpha}) \pm \mathcal{E}_\eta^\mathrm{loss}(\ket{\alpha}\bra{-\alpha})]\nonumber\\
&= {\mathcal{N}_\pm}^2[\ket{\eta\alpha}\bra{\eta\alpha} + \ket{-\eta\alpha}\bra{-\eta\alpha}\nonumber\\
&\pm e^{-2|(1-\eta)\alpha|^2}(\ket{\eta\alpha}\bra{-\eta\alpha}+\ket{-\eta\alpha}\bra{\eta\alpha})].
\end{align}
Now it's evident that the off-diagonal terms (i.e those capturing coherence) accumulate a factor of,
\begin{align}
C=e^{-2|(1-\eta)\alpha|^2},
\end{align}
which rapidly asymptotes to zero for large $\alpha$ with any \mbox{$\eta<1$}, as shown in Fig.~\ref{fig:cat_decoherence}.

% NOON states

\subsubsection{NOON states}\index{NOON states}

Similarly, NOON states (Sec.~\ref{sec:NOON}) undergo complete wave-function collapse if just a single photon is lost to the environment. This is because any single photon reveals complete information about the location of the remaining \mbox{$N-1$} photons. Therefore, a NOON state subject to loss of a single photon decoheres to the mixture,
\begin{align}
\hat\rho = \frac{1}{2}(\ket{N-1,0}\bra{N-1,0} + \ket{0,N-1}\bra{0,N-1}),
\end{align}
which contains no entanglement.

\begin{figure}[!htbp]
	\includegraphics[clip=true, width=0.475\textwidth]{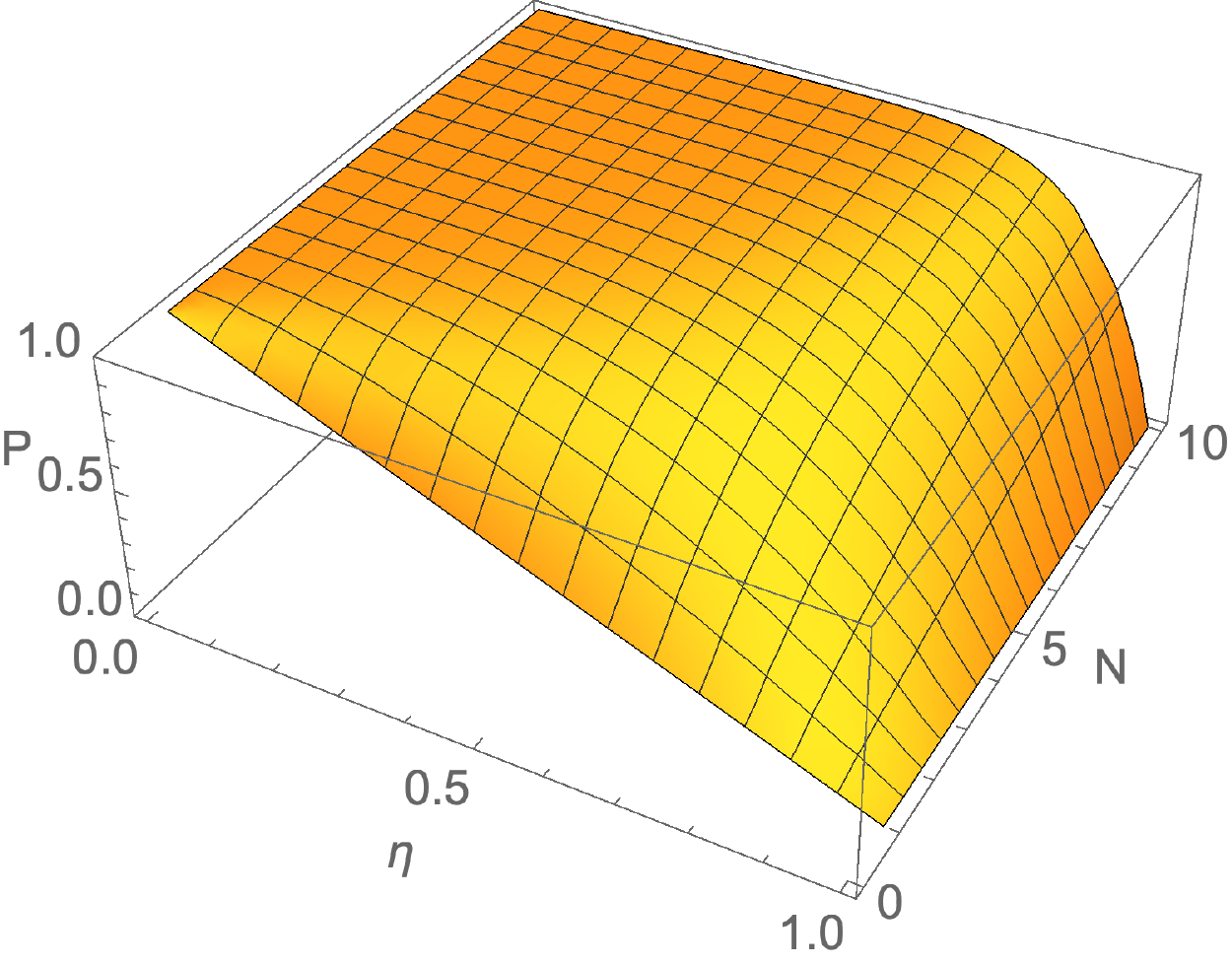}
	\caption{Probability of an $N$-photon NOON state completely collapsing and losing all entanglement, under a loss channel with efficiency $\eta$.}\label{fig:NOON_loss}
\end{figure}

Because there are $N$ photons in total, the probability of wave-function collapse grows exponentially with photon-number. Specifically, the probability of completely collapsing and losing all entanglement is,
\begin{align}
	P=1-\eta^N,
\end{align}
as shown in Fig.~\ref{fig:NOON_loss}.

% Scaling

\subsubsection{Scaling}

The scaling of loss over distance $d$ varies depending on the medium through which the light traverses. We will consider two dominant mediums, most relevant to future quantum networking:
\begin{itemize}
	\item Optical fibre\index{Optical fibres}: mode geometry is well-preserved, but optical medium is intrinsically lossy.
	\item Free-space\index{Free-space}: mode geometry is subject to dispersion\index{Dispersion}, but the medium is either lossless (in vacuum), or very low-loss (in atmosphere).
\end{itemize}

When propagating through fibre (or atmosphere, or some other lossy medium) net efficiency scales inverse exponentially as,
\begin{align}
	\eta = O(e^{-\alpha d}),
\end{align}
where $\alpha$ is a characteristic of the medium\footnote{With present-day fibre technology, this characteristic decay rate is on the order of \mbox{$\alpha = \frac{1}{22\mathrm{km}}$}.}.

\begin{figure}[!htbp]
	\includegraphics[clip=true, width=0.475\textwidth]{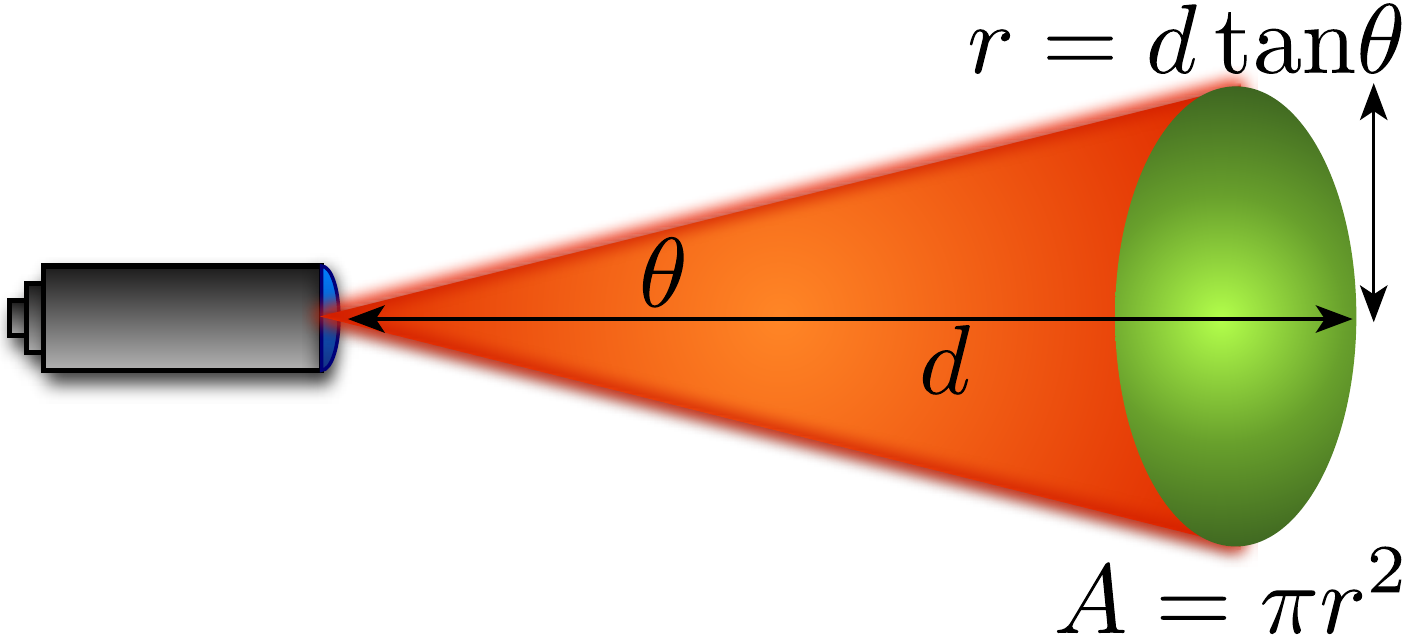}
	\captionspacefig \caption{Spot-size of an optical beam grows quadratically with distance from the source. If the beam is then collected via a camera with aperture area $A'$, any component of $A$ falling outside of $A'$ is effectively lost, yielding an effective loss channel with efficiency \mbox{$\eta = \frac{A'}{A} = \frac{A'}{\pi d^2 \tan^2\theta} = O\left(\frac{1}{d^2}\right)$}.} \label{fig:free_space_disp}
\end{figure}

In free-space on the other hand, where the medium of propagation is vacuum, which is effectively lossless, the effective loss rate is not determined by the medium, but rather by the fact that the spot-size\index{Spot-size} of an optical state is subject to dispersion\index{Dispersion} and grows only quadratically with distance, as shown in Fig.~\ref{fig:free_space_disp}. Then when the light is detected, if the spot-size is greater than the detector aperture, the undetected component effectively translates to loss. Thus through free-space the effective efficiency scales as,
\begin{align}
	\eta = O\left(\frac{1}{d^2}\right),
\end{align}
which is far more favourable than the exponential scaling inherent to lossy mediums. This provides space-based quantum networks with an inherent competitive advantage compared to any form of ground-based network.

Through atmospheric channels we will have both dispersion and distance-dependent loss, yielding an effective loss rate given by the product of the two effects,
\begin{align}
	\eta = O\left(\frac{e^{-\alpha d}}{d^2}\right).
\end{align}

%
% Dephasing
%

\subsection{Dephasing} \label{sec:dephasing_error} \index{Dephasing!Channel}

The dephasing error model describes the deterioration of quantum coherence in a state. It does not change the actual amplitudes of the components in the superposition, but rather reduces the state to a mixture of those components. Thus, dephasing can be thought of as destroying quantum information (coherence), while retaining classical information (probability amplitudes). 

% Qubits

\subsubsection{Qubits}

In terms of qubits, dephasing is most commonly represented using the Kraus representation\index{Kraus representation},
\begin{align} \label{eq:dephasing_channel}
\mathcal{E}_p^\mathrm{dephasing}(\hat\rho) = p\cdot\hat\rho + (1-p)\cdot \hat{Z}\hat\rho\hat{Z},
\end{align}
where $\hat\rho$ is the state of a single qubit, and $\hat{Z}$ is the Pauli phase-flip operator\footnote{Bit-flip\index{Bit-flip!Channel} and bit-phase-flip\index{Bit-phase-flip channel} channels may be represented similarly by replacing $\hat{Z}$ with $\hat{X}$ or $\hat{Y}$ respectively, although these don't arise as naturally as dephasing in many physical contexts.}. Intuitively this tells us that the dephasing channel creates a mixture of an input state with its phase-flipped self.

An alternate interpretation for the dephasing channel is that it is equivalent to the outside environment measuring $\hat\rho$ in the logical ($\hat{Z}$) basis, but unknown to us, thereby projecting the state onto one basis state or another, yielding a mixture of the two.

Dephasing acting on $\hat\rho$ can be very elegantly visualised as simply nullifying the off-diagonal matrix elements, i.e eliminating coherence terms. Dephasing is a ubiquitous error mechanism and affects all current quantum computing architectures.

Consecutive dephasing channels accumulate into another dephasing channel,
\begin{align} \label{eq:multi_deph}
\mathcal{E}_{p_1}^\mathrm{dephasing} \circ \mathcal{E}_{p_2}^\mathrm{dephasing} = \mathcal{E}_{p'}^\mathrm{dephasing},
\end{align}
where,
\begin{align}
	p' = p_1 p_2 + (1-p_1)(1-p_2),
\end{align}
i.e the probability that an even number of phase-flips have occurred.

As a simple example, consider the \mbox{$p=1/2$} dephasing channel acting on the \mbox{$\ket{+} = \frac{1}{\sqrt{2}}(\ket{0}+\ket{1})$} state. Then we have,
\begin{align}
\mathcal{E}^\mathrm{dephasing}_{1/2}(\ket{+}\bra{+}) &= \frac{1}{2} (\ket{+}\bra{+} + \hat{Z}\ket{+}\bra{+}\hat{Z}) \nonumber \\
&= \frac{1}{2} (\ket{+}\bra{+} + \ket{-}\bra{-}) \nonumber \\
&= \frac{1}{2} (\ket{0}\bra{0} + \ket{1}\bra{1}) \nonumber \\
&= \frac{\mathbb{\hat\openone}}{2},
\end{align}
is the completely mixed state. That is, the state has completely decohered. Note, however, that this complete decoherence depended on the choice of input state. A computational basis state, on the other hand, would be left unchanged by this channel,
\begin{align}
\mathcal{E}^\mathrm{dephasing}_{1/2}(\ket{0}\bra{0}) &= \frac{1}{2} (\ket{0}\bra{0} + \hat{Z}\ket{0}\bra{0}\hat{Z}) \nonumber \\
&= \ket{0}\bra{0}, \nonumber \\
\mathcal{E}^\mathrm{dephasing}_{1/2}(\ket{1}\bra{1}) &= \frac{1}{2} (\ket{1}\bra{1} + \hat{Z}\ket{1}\bra{1}\hat{Z}) \nonumber \\
&= \ket{1}\bra{1}.
\end{align}

Note that the probability of no dephasing occurring over multiple dephasing channels in series is given by the product of the respective probabilities for the individual channels.

% T_2-times

\subsubsection{$T_2$-times}\index{T$_2$-time}

\begin{figure}[!htbp]
	\includegraphics[clip=true, width=0.475\textwidth]{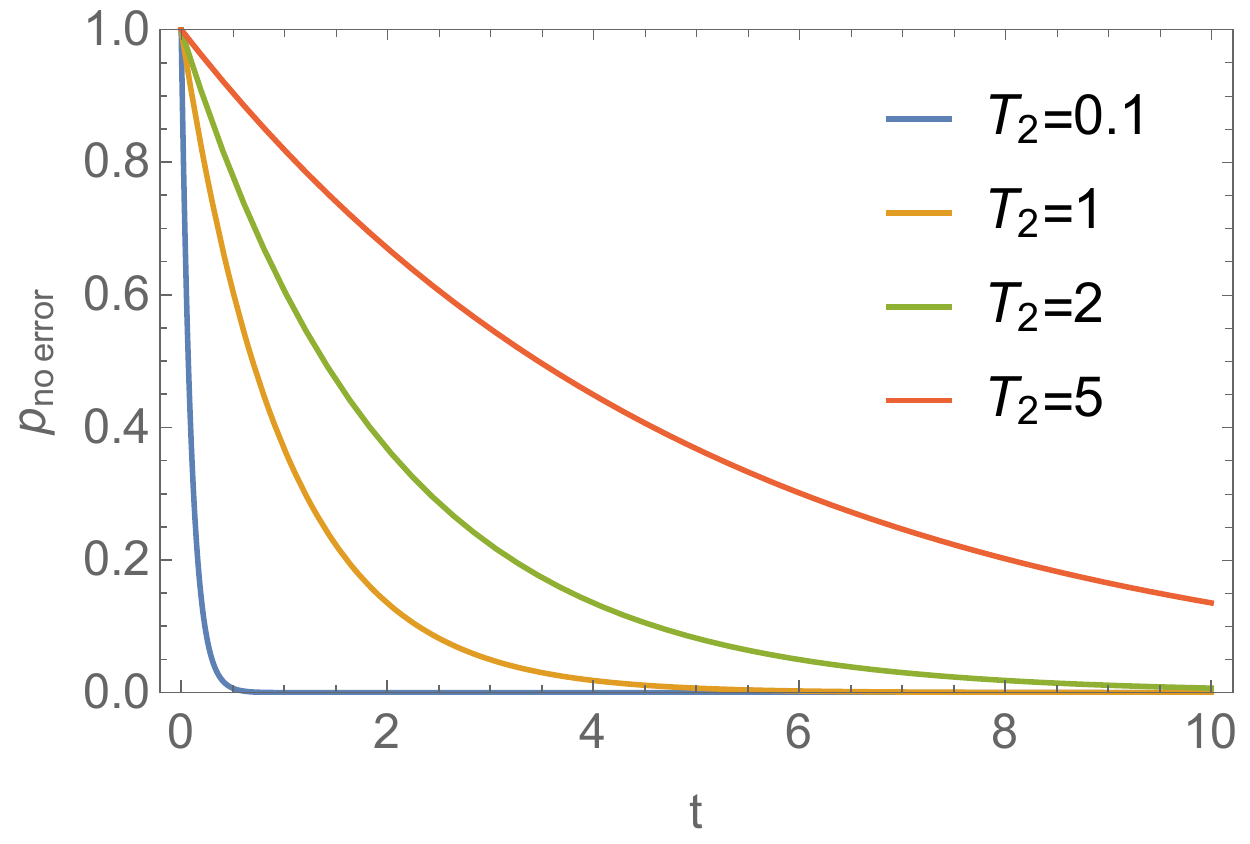}
	\captionspacefig \caption{Dephasing under continuous-time evolution, with characteristic decay rate given by the $T_2$-time. $p_\mathrm{no\,error}$ is the probability that the state is left unchanged, whereas \mbox{$1-p_\mathrm{no\,error}$} is the probability that it is replaced with the completely dephased state.}\label{fig:T2_time}
\end{figure}

A qubit dephasing channel is often quoted in terms of its $T_2$-time, a characteristic time for dephasing to occur under continuous time-evolution. Specifically, the probability of no dephasing occurring scales as,
\begin{align}
p_\mathrm{no\,error} = e^{-t/T_2},	
\end{align}
yielding a the equivalent dephasing channel,
\begin{align} \label{eq:T2_time}
\mathcal{E}_t^\mathrm{dephasing}(\hat\rho) = e^{-t/T_2} \hat\rho + \frac{1}{2}(1-e^{-t/T_2})(\hat\rho + \hat{Z}\hat\rho\hat{Z}),
\end{align}
as shown in Fig.~\ref{fig:T2_time}. For a qubit density matrix,
\begin{align}
	\hat\rho = \begin{pmatrix}
 \alpha & \gamma \\
 \gamma^* & \beta
\end{pmatrix},
\end{align}
this is equivalent to adding a factor of $e^{-t/T_2}$ to the two off-diagonal (coherence) elements,
\begin{align}
	\hat\rho_t = \begin{pmatrix}
 \alpha & e^{-t/T_2}\gamma \\
 e^{-t/T_2}\gamma^* & \beta
\end{pmatrix}.
\end{align}

Note that Eq.~\eqref{eq:T2_time} is parameterised into an ideal term ($\hat\rho$) and a completely dephased term ($(\hat\rho+\hat{Z}\hat\rho\hat{Z})/2$), and thus multiple rounds of this channel yields an equivalent channel where the probability associated with the former term accumulates multiplicatively,
\begin{align}
e^{-t'/T_2} = \prod_i e^{-t_i/T_2}	
\end{align}
or in logarithmic form additively, making it applicable to cost vector analysis (Sec.~\ref{sec:cost_as_dist}),
\begin{align}
t' = \sum_i t_i,	
\end{align}
providing a direct mechanism for calculating the effective dephasing rate across multiple subsequent sections in a qubit network route. The same can easily be seen to apply to depolarising (Sec.~\ref{sec:depolarising_channel}), amplitude damping (Sec.~\ref{sec:amp_damp} and loss channels (Sec.~\ref{sec:eff_err}).

% Optical states

\subsubsection{Optical states}

The notion of dephasing can be easily generalised to non-qubit states of light, i.e with photon-number \mbox{$n>1$}. In general, dephasing has the property of mapping a superposition of basis states to a mixture of the same basis states, whilst preserving amplitudes. Thus, for perfect dephasing,
\begin{align}
\mathcal{E}^\mathrm{dephasing}\left(\sum_i \alpha_i\ket{i} \cdot \sum_j \alpha_j^*\bra{j} \right) \to \sum_i |\alpha_i|^2 \ket{i}\bra{i},
\end{align}
for some arbitrary basis enumerated by $i$ and $j$. As an example, this process decoheres coherent states into thermal states. For partial dephasing, we can express the channel as creating a mixture over the input state with different phase rotations applied,
\begin{align} \label{eq:deph_int}
\mathcal{E}_{\phi}^\mathrm{dephasing}(\hat\rho) = \int_{0}^{2\pi} \phi(\omega) \hat{\Phi}(\omega)\hat\rho\,\hat{\Phi}(\omega)^\dag\,d\omega,
\end{align}
where $\hat{\Phi}(\omega)$ is a phase-shift operator with phase $\omega$, obeying \mbox{$\hat\Phi(\omega)^\dag = \hat\Phi(-\omega)$}, and $\phi(\omega)$ is a normalised probability density function characterising the distribution of phase-shifts. In the case of optical states, the phase-shift operators take the form,
\begin{align}\index{Phase!Shifts}
\hat\Phi(\omega) = e^{-i\omega\hat{n}},
\end{align}
in the photon-number basis, where $\hat{n}=\hat{a}^\dag\hat{a}$ is the photon-number operator\index{Photon-number!Operators}, satisfying \mbox{$\hat{n}\ket{n}=n\ket{n}$}. With no dephasing, \mbox{$\phi(\omega)=\delta(\omega)$} and $\mathcal{E}$ reduces to the identity channel. Otherwise, the off-diagonal (coherence) terms in the density operator begin to cancel out, leaving the diagonal (amplitude) terms unchanged. Thus, a perfect dephasing channel acting on a coherent state yields a thermal state of equal amplitude.

From this definition it can be seen that susceptibility to dephasing increases with photon-number, since the number operator adds a multiplicative factor to the acquired phase-shift,
\begin{align}
\hat\Phi(\omega) \ket{n} = e^{-i\omega n}\ket{n}.
\end{align}
For number states not in superposition, this corresponds to a simple unimportant global phase, since number states are phase-invariant. However, in superposition this adds relative phases, thereby destroying coherences upon applying the integral from Eq.~(\ref{eq:deph_int}).

%
% Depolarisation
%

\subsection{Depolarisation} \index{Depolarising channel} \label{sec:depolarising_channel}

Depolarisation is a noise model more general than dephasing, that probabilistically replaces a state with the completely mixed state (regardless of the input state). That is, with some probability we lose \textit{all} quantum \textit{and} classical information, i.e both coherences and probability amplitudes. Note that the dephasing channel introduced above only destroys quantum coherence, whilst preserving amplitudes. Formally, the depolarising channel can be expressed as,
\begin{align} \label{eq:depolarizing_channel}
\mathcal{E}^\mathrm{depolarising}_p(\hat\rho) = p \cdot \hat\rho + (1-p)\cdot \frac{\mathbb{\hat\openone}}{\mathrm{dim}(\hat\rho)},
\end{align}
where $\mathbb{\hat\openone}/\mathrm{dim}(\hat\rho)$ is the completely mixed state in the $d$-dimensional Hilbert space.

When acting on qubits, the depolarising channel can equivalently be represented as the action of each of the four Pauli matrices with equal probability, since,
\begin{align}
\frac{\hat\openone}{2} = \frac{1}{4}(\hat\rho + \hat{X}\hat\rho\hat{X} + \hat{Y}\hat\rho\,\hat{Y} + \hat{Z}\hat\rho\hat{Z}).
\end{align}
Thus, both dephasing and depolarisation are examples of Pauli error models.

In the qubit basis (i.e not including loss, for example), the Pauli matrices form a complete basis for quantum operations. Thus, the depolarising channel is the most general qubit error model, since it effectively applies all four Pauli error channels. For this reason, when evaluating fault-tolerance thresholds for QEC codes, thresholds are typically quoted in terms of the depolarising error rate.

Like the dephasing and loss channels, the error probability of multiple channels in series accumulates multiplicatively,
\begin{align}
\mathcal{E}_{p_1}^\mathrm{depolarising} \circ \mathcal{E}_{p_2}^\mathrm{depolarising} = \mathcal{E}_{p_1 p_2}^\mathrm{depolarising}.
\end{align}

%
% Amplitude Damping
%

\subsection{Amplitude damping} \index{Amplitude damping!Channel} \label{sec:amp_damp}

An error not so much relevant to optics, but which arises very naturally in some other systems, such as atomic systems or quantum dots, is amplitude damping, also referred to as a \textit{relaxation channel}. Here the process models the relaxation of a higher energy level, $\ket{1}$, to a lower energy one, $\ket{0}$. The $\ket{0}$ state is assumed to be the ground state and cannot relax any further, but the $\ket{1}$ state can spontaneously relax to the ground state. After complete amplitude damping, any input state will be left in the ground state $\ket{0}$. This model can be thought of as energy dissipating from the qubit system and being measured by the environment, leading to a type of decoherence whereby the input state is probabilistically replaced by the ground state.

The amplitude damping channel is easily represented in the quantum process formalism using two Kraus operators,
\begin{align}
\hat{K}_1 &= \ket{0}\bra{0} + \sqrt\eta\ket{1}\bra{1}, \nonumber \\
\hat{K}_2 &= \sqrt{1-\eta}\ket{0}\bra{1}, 
\end{align}
where \mbox{$0\leq\eta\leq 1$} quantifies the degree of damping (\mbox{$\eta=0$} represents complete damping, and \mbox{$\eta=1$} represents the identity channel).

The physical intuition is clear upon inspection of the structure of the projectors in the Kraus operators, with $\hat{K}_2$ representing relaxation from the excited state to the ground state, with probability \mbox{$1-\eta$}.

In the specific context of optics, the loss channel (Sec.~\ref{sec:eff_err})\index{Loss!Channel} is the equivalent of amplitude damping.

% T_1-times

\subsubsection{$T_1$-times}\index{T$_1$-time}

The degree of amplitude damping is often quoted in terms of a channel's $T_1$-time, characterising the expected time for the excited state to undergo spontaneous emission and relax to the ground state. Using this parameterisation we can express the amplitude damping channel as,
\begin{align}
\mathcal{E}_t^\mathrm{relax}(\hat\rho) = e^{-t/T_1}\hat\rho + (1-e^{-t/T_1})\ket{0}\bra{0},
\end{align}
for which the output state is of the form,
\begin{align}
\hat\rho_t = \begin{pmatrix}
1 - (1-\alpha) e^{-t/T_1} & \gamma e^{-t/T_1} \\
\gamma^* e^{-t/T_1} & \beta e^{-t/T_1}
\end{pmatrix}.
\end{align}

%
% Mode-Mismatch
%

\subsection{Mode-mismatch} \label{sec:MM_error} \index{Mode-mismatch}

Mode-mismatch is an error model unique to optical implementations. For perfect interference to take place between two optical modes, which is necessary to entangle them or perform ideal `which-path erasure'\footnote{Which-path erasure is the phenomenon whereby a beamsplitter interaction between two modes makes processes associated with those two modes indistinguishable, thereby projecting them into a superposition state of both possibilities. This is most commonly used to entangle distinct photon-emitting systems. This is discussed in detail in Sec.~\ref{sec:hybrid}.}\index{Which-path erasure}, the photons in those modes must be perfectly indistinguishable, i.e they must exhibit identical spatio-temporal structure \cite{bib:RohdeMauererSilberhorn07}\index{Spatio-temporal!Structure of photons} and must be pure states.

This phenomenon arises very naturally whenever optical path-lengths are not perfectly aligned, or there is imperfect spatial mode-overlap between optical modes interfering at beamsplitters. Furthermore, even if optical networks are perfect, photon distinguishability\index{Photon distinguishability} may arise during state preparation, since no two photon sources are absolutely identical -- engineering photon sources is a precise business and no two are ever exactly alike.

In real-world experiments, the most common form of mode-mismatch is temporal mode-mismatch, whereby the timing of different photons are not perfectly synchronised, yielding temporal distinguishability, thereby reduced quantum interference. This type of error is easily introduced via mismatched path lengths in an experiment, or incorrectly accounted for changes in refractive index. This is easily represented mathematically via translations in the temporal distribution functions (Sec.~\ref{sec:spatio_temporal}) of photons,
\begin{align} \label{eq:mode_mismatch_shift}
\psi(t) \to \psi(t-\Delta_t),
\end{align}
for temporal mismatch $\Delta_t$. Of course, this logically generalises to other degrees of freedom, such as spatial mode-mismatch, in which case a translation of the following form would take place,
\begin{align}
\psi(x,y) \to \psi(x-\Delta_x,y-\Delta_y),
\end{align}
where $x$ and $y$ are the two transverse spatial dimensions perpendicular to the direction of propagation.

The Hong-Ou-Mandel (HOM) \cite{bib:HOM87}\index{Hong-Ou-Mandel (HOM) interference} \textit{visibility} is a direct measure of the indistinguishability of two photons based on their interference fringes. Specifically, interference fringes are reduced as the photons become more distinguishable. Once completely distinguishable, they obey classical statistics.

Let us consider this in detail. Consider the two-mode, two-photon state,
\begin{align}
\ket{\psi_\mathrm{in}} = \hat{A}^\dag_{\psi_1} \hat{B}^\dag_{\psi_2} \ket{0},
\end{align}
where $\hat{A}^\dag$ and $\hat{B}^\dag$ denote the mode operators for two spatial modes, with respective temporal distribution functions $\psi_1$ and $\psi_2$. Evolving this though a 50:50 (Hadamard) beamsplitter yields,
\begin{align}
\ket{\psi_\mathrm{out}} &= \hat{U} \ket{\psi_\mathrm{in}} \\
&= \frac{1}{2} \left[\hat{A}^\dag_{\psi_1}+\hat{B}^\dag_{\psi_1}\right]\left[\hat{A}^\dag_{\psi_2}-\hat{B}^\dag_{\psi_2}\right] \ket{0} \nonumber \\
&= \frac{1}{2} \left[\hat{A}^\dag_{\psi_1}\hat{A}^\dag_{\psi_2} - \hat{A}^\dag_{\psi_1}\hat{B}^\dag_{\psi_2} + \hat{A}^\dag_{\psi_2}\hat{B}^\dag_{\psi_1} - \hat{B}^\dag_{\psi_1}\hat{B}^\dag_{\psi_2}\right] \ket{0} \nonumber.
\end{align}
Post-selecting upon detecting a coincidence event (i.e one photon per mode), the conditional state is projected onto,
\begin{align}
\ket{\psi_\mathrm{cond}} = \frac{1}{2} \left[\hat{A}^\dag_{\psi_1}\hat{B}^\dag_{\psi_2} - \hat{A}^\dag_{\psi_2}\hat{B}^\dag_{\psi_1}\right] \ket{0}.
\end{align}
The probability of this coincidence event occurring is then given by the normalisation of the residual state,
\begin{align}
P_\mathrm{coincidence} &= \left| \braket{\psi_\mathrm{cond}|\psi_\mathrm{cond}} \right|^2 \nonumber \\
&= \frac{1}{2} - \frac{1}{2} \left| \int^\infty_{-\infty} \psi_1(t)\psi_2^*(t)\,dt\right|^2.
\end{align}

\begin{figure}[!htbp]
\includegraphics[clip=true, width=0.475\textwidth]{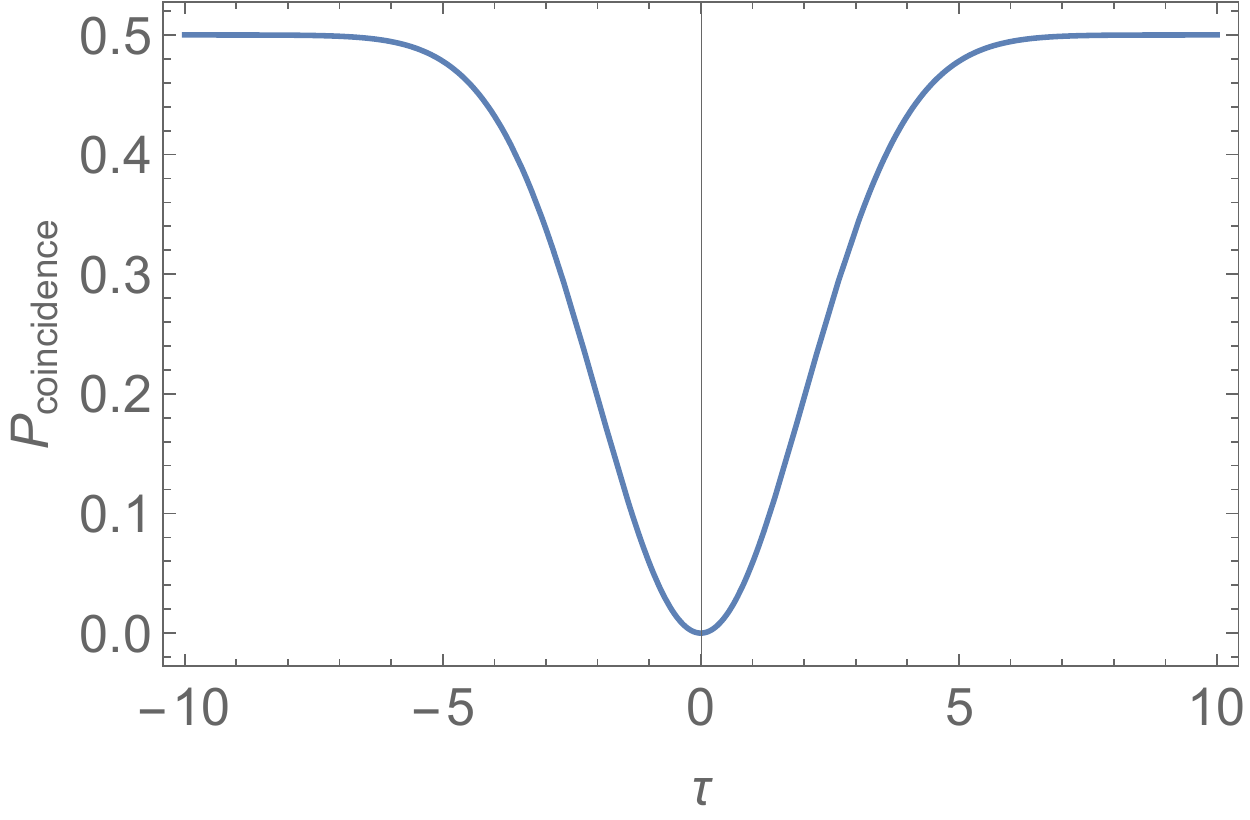}
\captionspacefig \caption{Hong-Ou-Mandel dip for two photons with normal Gaussian temporal distribution functions, and temporal offset $\tau$ between them. $\tau$ effectively characterises the degree of photon distinguishability, where \mbox{$\tau=0$} represents complete indistinguishability (quantum statistics), and \mbox{$\tau\to\pm\infty$} represents complete distinguishability (classical statistics). Thus, performing this experiment and measuring $P_\mathrm{coincidence}$ can be used to characterise the degree of photon distinguishability.} \label{fig:HOM_dip}\index{Hong-Ou-Mandel (HOM) interference}
\end{figure}

Now if we let both input photons have identical temporal structure, $\psi$, but with a time-delay $\tau$ between them, this reduces to,
\begin{align}
P_\mathrm{coincidence} = \frac{1}{2} - \frac{1}{2} \left| \int^\infty_{-\infty} \psi(t)\psi^*(t-\tau)\,dt\right|^2.
\end{align}
It is clear upon inspection that when \mbox{$\tau=0$}, the coincidence probability \mbox{$P_\mathrm{coincidence}=0$}, and we observe perfect photon bunching at the output (quantum statistics). On the other hand, as \mbox{$\tau\to\pm\infty$}, the photons become completely distinguishable, and we reduce to classical statistics, whereby \mbox{$P_\mathrm{coincidence}=1/2$}. In the intermediate regime, there will be a monotonic tradeoff between distinguishability (determined by $|\tau|$) and the coincidence probability. As an example, if we let the temporal distribution function be a normal Gaussian distribution,
\begin{align}
\psi(t) = \frac{1}{\sqrt[4]{2\pi}}e^{-\frac{t^2}{4}},
\end{align}
then,
\begin{align}
P_\mathrm{coincidence} = \frac{1}{2} - \frac{1}{2} e^{-\frac{\tau^2}{8}},
\end{align}
which is shown in Fig.~\ref{fig:HOM_dip}. Thus, experimentally measuring $P_\mathrm{coincidence}$ directly determines the degree of photon distinguishability.

In the above representation of mode-mismatch as a temporal or spatial translation, the process is entirely coherent, and could in principle be reversed if the translation were known (which might easily be established using tomographic characterisation techniques). Of course, such translations could occur incoherently also. In particular, `time-jitter' is where this process occurs incoherently, and the photons are subject to probabilistic temporal displacements. In this instance, a pure single-photon state would evolve into a mixture of states subject to different displacements. Since the mode-mismatch is now probabilisitic, it is not reversible. The state of a single photon subject to time-jitter would be of the form,
\begin{align}\index{Time-jitter}
\hat\rho_\mathrm{jitter} = \int_{-\infty}^\infty p_\mathrm{jitter}(\Delta_t) \ket{\psi-\Delta_t}\bra{\psi-\Delta_t}d\Delta_t,
\end{align}
where $p_\mathrm{jitter}(\Delta_t)$ characterises the classical probability distribution of the temporal displacement. Time-jitter is particularly natural in heralded spontaneous parametric down-conversion (SPDC) sources (Sec.~\ref{sec:single_phot_src}), where imprecision in the measurement time of the heralding mode projects that temporal uncertainty onto the heralded state. For this reason, much time is being invested into engineering SPDC sources with separable output photons, such that pathological behaviour of the detection of the heralding photon does not project the heralded photon onto a mixed state. Time-jitter is a major consideration in all present-day single-photon source technologies.

When considering mode-mismatch, there are two general regimes for how it manifests itself in an optical system. The first is when the interference taking place is between distinct, independent photons, i.e HOM interference (or its equivalent generalisations to higher-photon-number systems). The second is when multiple paths followed by a given photon interfere it with itself, i.e Mach-Zehnder (MZ)\index{Mach-Zehnder (MZ) interference} interference. The former only requires mode-matching on the scale of the photons' wave-packets, whereas the latter requires interferometric stability on the order of the photons' wavelength, a far more demanding requirement. This is discussed in greater detail in Sec.~\ref{sec:opt_stab}.

Mode-mismatch has been studied extensively in the context of linear optics quantum computing (LOQC), introduced in Sec.~\ref{sec:KLM_univ}. In particular, it was shown that in the cluster state formalism (Sec.~\ref{sec:CSQC}), mode-mismatch in a fusion gate is equivalent to a dephasing error model, where the dephasing rate is related to the degree of photon distinguishability (i.e visibility) \cite{bib:RohdeRalph06}. More generally, the operation of entangling gates \cite{bib:RohdeFreqTemp05, bib:RohdeGateChar05, bib:RohdeOptPhot05, bib:RohdeTimeRes11} and \textsc{BosonSampling} (Sec.~\ref{sec:boson_sampling}) \cite{bib:RohdeArbSpec15, bib:RohdeArbLow12} have been considered, and explicit error models derived.

%
% Dispersion
%

\subsection{Dispersion} \label{sec:dispersion}\index{Dispersion}

Dispersion is the phenomenon of frequency-dependent velocity of light in a given medium. These effects can be very diverse, but can always be expressed in the mode-operator representation using an appropriate transformation in the temporal or spectral wave-function,
\begin{align}
f_\mathrm{disp}: \,\tilde\psi(\omega)\to\tilde\psi(\omega)'.
\end{align}

%
% Spectral Filtering
%

\subsection{Spectral filtering} \label{sec:spectral_filt} \index{Spectral filtering}

In Sec.~\ref{sec:eff_err} we discussed the loss channel, whereby with some fixed probability photons are lost to the environment. In reality, this process is often not uniform, but frequency-dependent, resulting in spectral filtering effects. For example, optical fibres are typically designed to operate with a particular optical frequency in mind, and will attenuate frequencies outside a given range, implementing, for example, low-pass, high-pass or band-pass spectral filtering.

\begin{figure}[!htbp]
	\includegraphics[clip=true, width=0.25\textwidth]{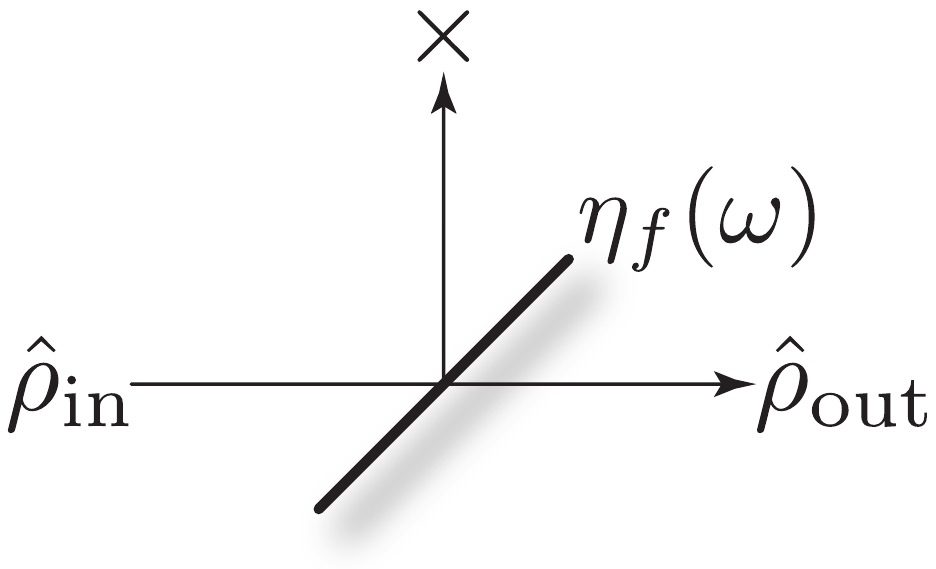}
	\captionspacefig \caption{Model for the spectral filtering channel. The input state, $\hat\rho_\mathrm{in}$, passes through a beamsplitter of frequency-dependent transmissivity $\eta_f(\omega)$, and the reflected mode discarded, yielding the lossy output state \mbox{$\hat\rho_\mathrm{out} = \mathcal{E}^\mathrm{filter}_{\eta_f}(\hat\rho_\mathrm{in})$}.} \label{fig:spectral_filter_model} \index{Spectral filtering}
\end{figure}

Because spectral filtering can be regarded as frequency-dependent loss, it can be modelled in the same way as per the loss channel, but using a frequency-dependent beamsplitter with transmissivity $\eta_f(\omega)$, which models the frequency response of the channel. The model is shown in Fig.~\ref{fig:spectral_filter_model}.

This channel has the effect of modulating the spectral distribution function of a photonic mode operator $\hat{A}_\psi^\dag$ to $\hat{A}_{\psi'}^\dag$, where,
\begin{align}
\psi'(\omega) = \sqrt{\eta_f(\omega)}\psi(\omega).	
\end{align}
Note that unless \mbox{$\eta_f(\omega)=1\,\forall\,\,\psi(\omega)\neq 0$}, the new distribution function $\psi'(\omega)$ will not be normalised, where the normalisation reflects the loss probability,
\begin{align}
p_\mathrm{loss} = 1 - \int_{-\infty}^\infty \eta_f(\omega)|\psi(\omega)|^2\,d\omega.
\end{align}

%
% Phase-Space
%

\subsection{Phase-space} \index{Phase!Space!Errors}

In Sec.~\ref{sec:non_lin_opt} we introduce the displacement and squeezing operations, two non-linear operations which are important ingredients in CV quantum information processing schemes. Of course, such processes are subject to errors.

In the case of the displacement operation, which is implemented by mixing a state with a coherent state on a beamsplitter, errors in the amplitude of the coherent state or in the beamsplitter reflectivity will introduce an offset in the displacement amplitude. Thus, instead of implementing $\hat{D}(\alpha)$, we might over- or under-displace the state, implementing,
\begin{align}
	\hat{D}(\Delta)\hat{D}(\alpha)\propto \hat{D}(\alpha+\Delta),
\end{align}
for some error $\Delta$.

In the case of the squeezing operation, we might similarly have uncertainty in the squeezing parameter, thus implementing $\hat{S}(\xi+\Delta)$ instead of $\hat{S}(\xi)$.

\latinquote{Igne natura renovatur integra.}

%
% Cost Vector Analysis
%

\section{Quantum cost vector analysis} \label{sec:quantum_meas_cost}
\index{Costs}\index{Attributes}\index{Cost vector analysis}

\dropcap{A}{s} with the classical case in Sec.~\ref{sec:costs}, there will be costs associated with the links and nodes in a network -- nothing is free! In the quantum case, all the usual classical costs are valid, but there are some very important additions of far greater relevance to most quantum applications. Classical digital data is discretised, resulting in data transmission highly robust against noise. In a quantum setting this is necessarily not the case, as the coefficients in quantum superpositions are continuous, meaning that errors accumulate during transmission and states will inevitably deteriorate, unlike digital states. This requires a rethinking of appropriate cost metrics.

%
% Costs
%

\subsection{Costs}

We now briefly introduce some of the key measures for quantifying the quality of quantum communications links, and how they may be expressed as metrics with meaningful operational interpretations. Many of the measures typically employed for characterising quantum systems are not true metrics (i.e costs), but in many cases can be converted to metrics, or used meaningfully as attributes instead.

%
% Efficiency
%

\subsubsection{Efficiency} \index{Loss!Channel}

The efficiency measure introduced previously is multiplicative, so for consecutive lossy channels the net efficiency is,
\begin{align}
\eta_\mathrm{net}=\prod_i \eta_i,
\end{align}
where $\eta_i$ is the efficiency of the $i$th channel. Intuitively, this is simply telling us that if a photon passes through a channel with success probability $\eta_1$, followed by another with $\eta_2$, the total success probability is \mbox{$\eta_1\eta_2$}.

When employing single-photon encoding of qubits (e.g using the polarisation degree of freedom), there are three basis states of interest: a single photon horizontally polarised ($\ket{H}$); a single photon vertically polarised ($\ket{V}$); and, the vacuum state ($\ket{0}$). The effect of the loss channel on this type of state is to map $\ket{H}$ and $\ket{V}$ to $\ket{\mathrm{0}}$ with probability \mbox{$1-\eta$}, while doing nothing to $\ket{\mathrm{0}}$. Note that because the loss process affects both logical basis states ($\ket{H}$ and $\ket{V}$) identically, its action is invariant under unitary operations in the logical (i.e polarisation) basis space.

%
% Spectral filtering
%

\subsubsection{Spectral filtering} \index{Loss!Channel}\index{Spectral filtering}

Because spectral filtering can be regarded as a frequency-dependent loss channel, its associated cost can be treated in the same manner, except that rather than keeping track of a single efficiency $\eta$, we track a frequency response function $\eta_f(\omega)$, with the same multiplicative property (on a per-frequency basis),
\begin{align}
	\eta_f^\mathrm{(net)}(\omega)=\prod_i \eta_f^{(i)}(\omega).
\end{align}

If we are keeping track of the frequency response, the usual efficiency metric can be made redundant and absorbed into the frequency response function as a uniform response,
\begin{align}
	\eta_f(\omega)=\eta\,\,\forall\,\omega.
\end{align}

%
% Decoherence
%

\subsubsection{Decoherence} \index{Dephasing!Channel} \index{Depolarising channel} \index{Decoherence}

The dephasing and depolarising channels, given by Eqs.~(\ref{eq:dephasing_channel} \& \ref{eq:depolarizing_channel}), also behave multiplicatively. If $p_i$ is the probability that the state passing through the $i$th channel in series does not undergo the error process, then the probability of the state passing though the entire series without error is simply,
\begin{align}
p_\mathrm{net}=\prod_i p_i,
\end{align}
exhibiting the same multiplicative behaviour as the loss channel. The same observation applies to any of the other Pauli error channels.

%
% Mode-Mismatch
%

\subsubsection{Mode-mismatch} \index{Mode-mismatch}

In Sec.~\ref{sec:MM_error} we introduced a simple model for temporal mode-mismatch as a displacement in the temporal wave-function of photons propagating through a channel. Clearly, such a process is cumulative -- a temporal displacement of $\Delta_1$ followed by another of $\Delta_2$ yields a net displacement of \mbox{$\Delta_1+\Delta_2$}. Thus, for a chain of such channels we simply accumulate a net temporal displacement of,
\begin{align}
\Delta_\mathrm{net} = \sum_i \Delta_i.
\end{align}

For an incoherent mode-mismatching process, such as time-jitter, an upper bound on the accumulated mismatch may be obtained by summing the maximum temporal displacements at each step.

%
% Distance Measures
%

\subsubsection{Distance measures} \label{sec:fid_metric} \index{Distance measures}

The fidelity of two states directly quantifies how close they are to one another in a geometric sense, i.e on the Bloch sphere, or, in the context of a state passing through a quantum channel, a measure of how well the state is preserved.

The fidelity\index{Fidelity} between two states is defined as,
\begin{align}
\mathcal{F}(\hat\rho_1,\hat\rho_2) = \mathrm{tr}\left(\sqrt{\hat\rho_1^{1/2}\cdot\hat\rho_2\cdot\hat\rho_1^{1/2}}\right),
\end{align}
where,
\begin{align}
& \mathcal{F}(\hat\rho_1,\hat\rho_2) = \mathcal{F}(\hat\rho_2,\hat\rho_1), \nonumber \\
& 0\leq \mathcal{F}(\hat\rho_1,\hat\rho_2) \leq 1.
\end{align}
\mbox{$\mathcal{F}(\hat\rho_1,\hat\rho_2)=1$} iff the states are equal, and \mbox{$\mathcal{F}(\hat\rho_1,\hat\rho_2)=0$} iff they are orthogonal.
In the case where one of the states is a pure state, this simplifies to,
\begin{align}
\mathcal{F}(\hat\rho_1,\ket{\psi_2}) = \bra{\psi_2}\hat\rho_1\ket{\psi_2},
\end{align}
and when both states are pure to simply,
\begin{align}
\mathcal{F}(\ket{\psi_1},\ket{\psi_2}) = |\braket{\psi_1 | \psi_2}|^2.
\end{align}

The fidelity is invariant under a common unitary applied to both states,
\begin{align}
\mathcal{F}(\hat\rho_1,\hat\rho_2) = \mathcal{F}(\hat{U}\hat\rho_1 \hat{U}^\dag,\hat{U} \hat\rho_2\,\hat{U}^\dag).
\end{align}

We define the fidelity of two processes, the process fidelity \index{Fidelity} \cite{bib:Gilchrist05}, to be the fidelity between two identical copies of a state that have been evolved under each of those processes, minimised over all possible states. That is, it provides a lower bound on the fidelity between identical states evolved under the two processes. In the context of networking, where quality must be guaranteed, this definition is more appropriate than, say, the average case fidelity. Specifically,
\begin{align}
\mathcal{F}(\mathcal{E}_1,\mathcal{E}_2) = \mathrm{tr}\left( \sqrt{\chi_1^{1/2}\cdot\chi_2\cdot\chi_1^{1/2}}\right),
\end{align}
where $\chi_1$ and $\chi_2$ are the process matrices\index{Process matrices} for $\mathcal{E}_1$ and $\mathcal{E}_2$.

The fidelity of two processes is invariant under a common unitary applied to both channels before or after the process. Specifically,
\begin{align}
\mathcal{F}(\mathcal{E}_1,\mathcal{E}_2) &= \mathcal{F}(\mathcal{E}_U\circ\mathcal{E}_1,\mathcal{E}_U\circ\mathcal{E}_2) \nonumber \\
&= \mathcal{F}(\mathcal{E}_1\circ \mathcal{E}_U,\mathcal{E}_2\circ \mathcal{E}_U),
\end{align}
where $\mathcal{E}_U$ is an arbitrary unitary process.

In the special case of an identity channel, $\hat\openone$, which is of special interest in many communications scenarios, we employ the shorthand,
\begin{align}
\mathcal{F}(\mathcal{E}) = \mathcal{F}(\mathcal{E},\hat\openone) = \min_{\hat\rho} \left[\mathcal{F}(\hat\rho,\mathcal{E}(\hat\rho))\right].
\end{align}
By definition \mbox{$\mathcal{F}(\mathcal{E})=1$} iff \mbox{$\mathcal{E}=\hat\openone$}.

A lower bound on the process fidelity of multiple processes in series is multiplicative,
\begin{align}
\mathcal{F}(\mathcal{E}_2\circ\mathcal{E}_1,\mathcal{E}_3) &\geq \mathcal{F}(\mathcal{E}_2,\mathcal{E}_3)\cdot\mathcal{F}(\mathcal{E}_1,\mathcal{E}_3), \nonumber \\
\mathcal{F}(\mathcal{E}_2\circ\mathcal{E}_1) &\geq \mathcal{F}(\mathcal{E}_2)\cdot\mathcal{F}(\mathcal{E}_1).
\end{align}

Generalising to a sequence of $n$ processes in series yields,
\begin{align}
\mathcal{F}(\mathcal{E}_n\circ\dots\circ\mathcal{E}_1) \geq \prod_{i=1}^n \mathcal{F}(\mathcal{E}_i).
\end{align}

An alternate measure for the distance between two quantum states is the trace-norm distance\index{Trace-norm distance}, defined as,
\begin{align}
D(\hat\rho_1,\hat\rho_2) &= \frac{1}{2}\|\hat\rho_1 - \hat\rho_2\|_1 \nonumber\\
&= \frac{1}{2}\sum_i |\lambda_i|,
\end{align}
where $\lambda_i$ are the eigenvalues of \mbox{$\hat\rho_1-\hat\rho_2$}. Like the fidelity, the trace-norm distance is invariant under unitary transformation. Furthermore, it is contractive under the action of quantum processes,
\begin{align}
D(\mathcal{E}(\hat\rho_1),\mathcal{E}(\hat\rho_2)) \leq D(\hat\rho_1,\hat\rho_2).
\end{align}
The trace-norm distance relates to the fidelity according to the following bounds,
\begin{align}
1-F(\hat\rho_1,\hat\rho_2) \leq D(\hat\rho_1,\hat\rho_2) \leq \sqrt{1-F(\hat\rho_1,\hat\rho_2)^2}.
\end{align}

%
% Purity
%

\subsubsection{Purity} \index{Purity}

The purity of a state that was initially pure quantifies how well quantum coherence was maintained during evolution, equivalently how well superpositions are maintained. The purity is defined as,
\begin{align}
\mathcal{P}(\hat\rho) = \mathrm{tr}(\hat\rho^2),
\end{align}
where,
\begin{align}
\frac{1}{\mathrm{dim}(\hat\rho)} \leq \mathcal{P}(\hat\rho) \leq 1.
\end{align}
We have \mbox{$\mathcal{P}(\hat\rho) = 1$} iff \mbox{$\hat\rho=\ket{\psi}\bra{\psi}$} is a pure state, and \mbox{$\mathcal{P}(\hat\rho)=1/\mathrm{dim}(\hat\rho)$} iff \mbox{$\hat\rho=\hat\openone/\mathrm{dim}(\hat\rho)$} is the maximally mixed state.

The purity is invariant under unitary operations,
\begin{align}
\mathcal{P}(\hat\rho) = \mathcal{P}(\hat{U}\hat\rho\,\hat{U}^\dag).
\end{align}

The purity of a process is defined analogously to the fidelity of a process,
\begin{align}
\mathcal{P}(\mathcal{E}) = \mathrm{tr}(\chi^2),
\end{align}
and as with the fidelity, a lower bound on the purity of multiple processes in series is multiplicative,
\begin{align}
\mathcal{P}(\mathcal{E}_2\circ\mathcal{E}_1)
\geq \mathcal{P}(\mathcal{E}_2)\cdot\mathcal{P}(\mathcal{E}_1).
\end{align}
If the channel implements a unitary operation then necessarily \mbox{$\mathcal{P}(\mathcal{E})=1$}.

Like the process fidelity, the purity of a quantum process is invariant under unitary operations,
\begin{align}
\mathcal{P}(\mathcal{E}) &= \mathcal{P}(\mathcal{E}_U\circ\mathcal{E}) \nonumber \\
&= \mathcal{P}(\mathcal{E}\circ\mathcal{E}_U).
\end{align}

Generalising to a sequence of $n$ processes in series yields,
\begin{align}
\mathcal{P}(\mathcal{E}_n\circ\dots\circ\mathcal{E}_1) \geq \prod_{i=1}^n \mathcal{P}(\mathcal{E}_i).
\end{align}

%
% Entanglement
%

\subsubsection{Entanglement} \label{sec:ent_meas} \index{Entanglement!Measures}

When distributing entanglement between separate nodes, metrics quantifying bipartite entanglement are relevant. For pure bipartite states $\ket{\psi}_{A,B}$, the purity of one of the reduced subsystems directly quantifies the degree of entanglement between them,
\begin{align}
\mathcal{M}(\ket{\psi}_{A,B})) &= \mathcal{P}(\mathrm{tr}_A(\ket{\psi}_{A,B})) \nonumber \\
&= \mathcal{P}(\mathrm{tr}_B(\ket{\psi}_{A,B})),
\end{align}
The entanglement between two systems in invariant under local unitaries,
\begin{align}
\mathcal{M}(\ket{\psi}_{A,B}) = \mathcal{M}([\hat{U}_A\otimes \hat{U}_B]\ket{\psi}_{A,B}).
\end{align}

%
% Phase-Space
%

\subsubsection{Phase-space} \index{Phase!Space!Errors}

Displacements in phase-space accumulate additively, up to a phase-factor. Specifically, the composition of two displacements is given by,
\begin{align}
\hat{D}(\alpha)\hat{D}(\beta) = e^{\frac{1}{2}(\alpha\beta^*-\alpha^*\beta)}\hat{D}(\alpha+\beta).
\end{align}
Thus, the composition of a chain of unwanted or uncertain displacements yields, up to phase, a displacement with amplitude given by the sum of the individual displacement amplitudes.

Similarly, from the definition of the squeezing operator,
\begin{align}\label{eq:sq_op}
\hat{S}(\xi) = \exp\left[ \frac{1}{2}(\xi^*\hat{a}^2 - \xi{\hat{a}^{\dag 2}})\right],
\end{align}
it is evident that squeezing accumulates additively as well,
\begin{align}
\hat{S}(\xi_1)\hat{S}(\xi_2) = \hat{S}(
\xi_1+\xi_2).	
\end{align}

%
% Latency
%

\subsubsection{Latency} \label{sec:latency_metric} \index{Latency}

Aside from the actual information content of a transmitted quantum state, the latency associated with its transmission is a key consideration in many time-critical applications.

By defining the latency of a link/node as the time between receipt of a quantum state and its retransmission, the total latency of a route is simply the sum of all the individual node and link latencies across the route,
\begin{align}
\mathcal{L}(R) = \sum_{i\in R} \mathcal{L}_i,
\end{align}
where $\mathcal{L}_i$ is the latency associated with the $i$th link in route $R$.

%
% Dollars
%

\subsubsection{Dollars} \label{sec:dollars} \index{Dollar cost}

Not to be overlooked is the actual dollar cost of communicating information. It is unlikely that Alice and Bob outright own the entire infrastructure of particular routes. Rather, different links and nodes are likely to be owned by different operators (particularly in ad hoc networks), who are most likely going to charge users for bandwidth in their network (quantum networks won't be cheap). Clearly dollar costs are additive over the links and nodes within routes,
\begin{align}
\mathcal{C}(R) = \sum_{i\in R} \mathcal{C}_i,
\end{align}
where $\mathcal{C}_i$ is the dollar cost of utilising the $i$th link in route $R$.

%
% Costs as Distance Metrics
%

\subsection{Costs as distance metrics} \label{sec:cost_as_dist} \index{Cost distance metrics}

Def.~\ref{def:metric} defines the properties of a cost metric in the classical context. We now wish to consider this in the quantum context, such that we are empowered to ask questions like ``what is the the total cost across a network route?'' or ``which route minimises a cost between two parties?'', where \textit{cost} now refers to some metric relevant to quantum state distribution, such as accumulated decoherence or loss.

If we consider a lossy photonic channel for example, efficiencies ($\eta$) are multiplicative -- for a route \mbox{$v_1\to v_2\to v_3$}, the net efficiency is given by the product of the individual efficiencies, \begin{align}
\eta_{v_1\to v_2 \to v_3} = \eta_{v_1\to v_2} \eta_{v_2\to v_3}.
\end{align}
This is multiplicative rather than additive, clearly not satisfying our definition for a cost metric. However, multiplicative metrics\index{Multiplicative metrics} such as this can easily be made additive\index{Additive metrics} by shifting to a logarithmic scale, since
\begin{align}\index{Logarithmic scale}
\log(\eta_{v_1\to v_2\to v_3}) = \log(\eta_{v_1\to v_2}) + \log(\eta_{v_2\to v_3}),
\end{align}
which now has a legitimate interpretation as a distance. The same applies to, for example, frequency response functions, which are equivalent to frequency-dependent loss.

In general, for a series of links \mbox{$v_1\to v_2 \to \dots \to v_n$} characterised by multiplicative measure $m$, the equivalent cost metric is,
\begin{align} \label{eq:dist_log}\index{Logarithmic distance}
c_{v_1\to v_2 \to \dots \to v_n} = -\sum_{i=1}^{n-1} \log (m_{v_i\to v_{i+1}}).
\end{align}
We have assumed that \mbox{$0\leq m \leq 1$}, where \mbox{$m=0$} represents complete failure, and \mbox{$m=1$} represents ideal operation.

With these properties, the costs in our graph have an elegant interpretation. In the case of perfect operation, \mbox{$m=1$}, the cost is \mbox{$c=0$}, creating an ideal direct link between neighbouring nodes at no cost. On the other hand for complete failure, \mbox{$m=0$}, the cost metric is \mbox{$c=\infty$}, effectively removing the link from the network and prohibiting pathfinding algorithms from following that route altogether.

Such a logarithmic scale is particularly convenient when a cost metric over links accumulates on a per physical distance basis, in which case the cost metric is simply the physical length of the link multiplied by the metric per unit distance. For example, if a fibre channel implements loss at 3dB/km, the loss over 10km is 10$\times$3dB.

Note that lower bounds on fidelity, purity, efficiency and dephasing are all multiplicative on a scale of 0 to 1, and thus their logarithms may be regarded as cost metrics. Spatio-temporal mode-mismatch, latency, dollar cost and displacements are clearly automatically metrics as they are additive.

A dephasing channel can be easily converted to a distance metric as follows. First we reparameterise the dephasing channel into,
\begin{align}
\mathcal{E}(\hat\rho) &= p\hat\rho + (1-p)\hat{Z}\hat\rho\hat{Z}\nonumber\\
&= (2p-1)\hat\rho + (1-p)(\hat{Z}\hat\rho\hat{Z} + \hat\rho).	
\end{align}
Now \mbox{$2p-1$} is the probability that the state is not dephased and \mbox{$1-p$} is the probability that the state is replaced with the completely dephased state. Therefore the probability of multiple applications of the channel (\mbox{$\mathcal{E}_n\circ\dots\circ\mathcal{E}_1$}) not dephasing the state scales multiplicatively as\footnote{With this parameterisation, in the limit of many applications of the dephasing channel, an input state asymptotes to the completely dephased state, \mbox{$\lim_{n\to\infty} \mathcal{E}^n(\hat\rho) = \frac{1}{2}(\hat{Z}\hat\rho\hat{Z}+\hat\rho)$}. Thus, \mbox{$2p-1$} can be regarded as a discretised parameterisation of a system's $T_2$-time\index{T$_2$-time}.},
\begin{align}
p_\mathrm{no\, error} = \prod_{i}(2p_i-1),
\end{align}
which is additive in a logarithmic scale as before,
\begin{align}
\log(p_\mathrm{no\, error}) = \sum_i \log(2p_i-1),
\end{align}
which acts as a distance metric. This approach can similarly be applied to other Pauli channels.

In the case of mutual information and channel capacity, it makes most sense to consider the number of bits that are lost by a channel, rather than the number communicated, since then we have a measure with quasi-metric properties. Specifically, let the number of bits lost by a channel be the difference between the number of bits in the input state and the channel capacity,
\begin{align}
B_\mathrm{lost}(\mathcal{E},\hat\rho) = S(\hat\rho) - \mathcal{C}(\mathcal{E}).
\end{align}
Then there are two cases to consider -- upper and lower bounds on accumulated lost bits.

The best-case scenario is that subsequent channels lose the same bits, giving us a lower bound on the number of lost bits as the maximum number of bits lost by the constituent channels,
\begin{align}
B_\mathrm{lower}(\mathcal{E}_2\circ\mathcal{E}_1,\hat\rho) = \mathrm{max}[B_\mathrm{lost}(\mathcal{E}_1,\hat\rho), B_\mathrm{lost}(\mathcal{E}_2,\hat\rho)].
\end{align}
Alternately, each subsequent channel could lose a different set of bits, in which case the number of lost bits accumulates additively,
\begin{align}
B_\mathrm{upper}(\mathcal{E}_2\circ\mathcal{E}_1,\hat\rho) = B_\mathrm{lost}(\mathcal{E}_1,\hat\rho) + B_\mathrm{lost}(\mathcal{E}_2,\hat\rho). 
\end{align}
Then, the number of actual bits lost is bounded from above and below as,
\begin{align}
B_\mathrm{lower} \leq B_\mathrm{lost} \leq B_\mathrm{upper}.	
\end{align}

%In the case of mutual information, which is not a metric, one can use it to define the \textit{variation of information} metric,
%\begin{align}\index{Variation of information}
%d_\mathrm{VI}(X,Y) &= H(X,Y) - I(X;Y), \nonumber \\
%d_\mathrm{VI}(\hat\rho_A,\hat\rho_B) &= S(\hat\rho_A,\hat\rho_B) - I(\hat\rho_A;\hat\rho_B),
%\end{align}
%or for a channel,
%\begin{align}
%d_\mathrm{VI}(\hat\rho,\mathcal{E}) &= S(\hat\rho,\mathcal{E}(\hat\rho)) - I(\hat\rho;\mathcal{E}(\hat\rho)),
%\end{align}
%which obeys the metric properties.

%
% Non-Trivial Node Operations
%

\subsection{Non-trivial node operations}

Thus far we have considered how to accumulate cost metrics across routes through a network, where each link is subject to some quantum process obeying our notion of a cost metric. But what happens when the links are interspersed with nodes that may be doing more than just simple switching?

A more general scenario to consider is where the nodes are not restricted to routing, but can additionally implement arbitrary unitary operations. This substantially broadens the class of networks under consideration, to encompass nodes capable of doing everything from straightforward routing to entire quantum computations.

All of the examples for cost metrics we introduced in Sec.~\ref{sec:quantum_meas_cost} have the property that they are invariant under unitary operations. Therefore the costs along a route may simply be accumulated as before, summing up the edge weights, without needing any special treatment for node operations, provided they are unitary. For non-unitary node processes, we can merge them into their neighbouring link processes as before (see Fig.~\ref{fig:remove_nodes}).

\subsection{Negative cost vectors}\index{Negative cost vectors}

When we initially introduced cost vector analysis in the classical context (Sec.~\ref{sec:costs}) we insisted that costs be positive by definition. However, in the quantum scenario we will loosen this demand since negative costs arise quite naturally in the context of operations that \textit{improve} quantum data. Specifically this arises naturally when nodes implement operations such as entanglement purification or quantum error correction, to be discussed in detail in Sec.~\ref{sec:QOS_chap}. In that case making what would otherwise be a routing detour can yield net benefit, and so the cost vector analysis must take these negative costs into consideration and give them the favourable treatment they deserve.

\latinquote{Flectere si nequeo superos, acheronta movebo.}

% \clearpage
%
% Quantum Transmission Control Protocol (QTCP)
%

\section{Quantum Transmission Control Protocol (QTCP)} \index{Quantum Transmission Control Protocol (QTCP)}\label{sec:QTCP}

\dropcap{I}{n} the classical world, TCP is employed for data transmission and routing. Next we present a simple toy model for a proposed quantum equivalent -- the Quantum Transmission Control Protocol (QTCP). The stack is described in detail in Sec.~\ref{sec:prot_stack}.

We emphasise that this toy model is not intended to be a proposal suited to immediate implementation, solving all the problems of quantum communication in the most effective way. Rather, we simply aim to construct a sketch of the data structures and algorithms that might form a basis for future, more well-considered real-world implementations. Alternately, it is plausible that a QTCP protocol may never reach the light of day at all, instead being made redundant by networks built entirely on entanglement distribution, discussed in detail in Secs.~\ref{sec:rep_net} \& \ref{sec:ent_ultimate}. The answer to this question is difficult to foresee, largely depending on the future requirements of quantum communication protocols.

The goal of QTCP is to abstract away the low-level physical operation of a quantum network to create a virtual interface between Alice and Bob, allowing direct access to data as if it were held locally, in much the same way that high-level services like classical FTP facilitate interaction with remote data as though it were a local asset, blind to the intermediate networking.

The design goals of our elementary toy model are simply to capture the quintessential feature requirements of real-world protocols, and a sketch for their implementation. The designs we present should not be interpreted as a final proposal, but merely as laying a foundation of ideas to build upon.

We consider the scenario where Alice (or a set of Alices) is in possession of some quantum state, which she wishes to communicate to Bob (Bobs), with the aim of optimising some arbitrary cost measure. Bob is no guru and doesn't want to concern himself with how the state was communicated from Alice to himself -- his only concern is that he receives it and that it satisfies quality constraints he and Alice have agreed upon.
 
QTCP is the joint software/hardware stack that facilitates these objectives. QTCP begins by logically separating different levels of network functionality into distinct layers of abstraction. This includes primarily:
\begin{itemize}
	\item Encapsulation of data into packets of quantum information (Secs.~\ref{sec:data_message_layer} \& \ref{sec:packet_layer}).
	\item Cost vector analysis (Sec.~\ref{sec:costs}).
	\item Routing decisions (Secs.~\ref{sec:intro_strat} \& \ref{sec:strategies}).
	\item Reconstruction of communicated quantum states upon receipt (Sec.~\ref{sec:reconstruction_layer}).
	\item Enforcement of quality of service requirements and error correction (Sec.~\ref{sec:reconstruction_layer}).
	\item Providing a high-level virtual interface between end-users, which abstracts away low-level operations (Sec.~\ref{sec:services_apps}).
\end{itemize}

All the while, Alice and Bob, as end-users, ought to be as blind as possible to the lower-level layers, instead only directly interfacing with the highest layer of abstraction, that which provides the virtual interface between end-users.

\latinquote{O derint dum metuant.}

%
% QTCP Protocol Stack
%

\subsection{QTCP protocol stack} \label{sec:prot_stack} \index{QTCP protocol stack}

As with classical networking, our protocols for quantum networks will be separated into distinct layers, each performing a specific set of tasks with different levels of abstraction.

\begin{figure}[!htbp]\index{QTCP protocol stack}
\if 1\doublecol
\includegraphics[clip=true, width=0.475\textwidth]{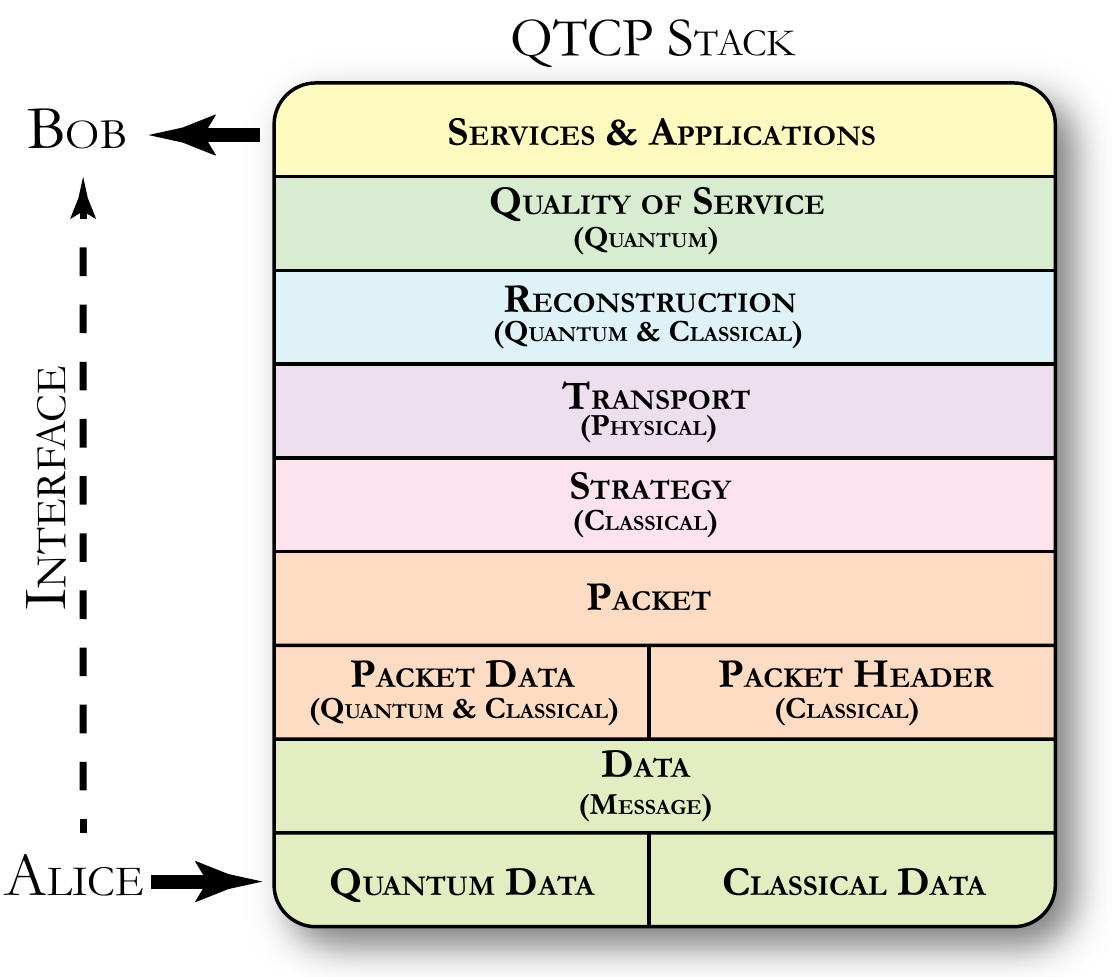}
\else
\includegraphics[clip=true, width=0.6\textwidth]{stack}
\fi
\captionspacefig \caption{Protocol stack for QTCP. The protocol stack mediates communication of quantum data from Alice to Bob using the shown layers of abstraction. The end goal is to provide Bob a virtual interface to Alice's transmitted data, while remaining oblivious to the underlying protocol.} \label{fig:stack}\index{QTCP protocol stack}
\end{figure}

The structure of the protocol stack for QTCP is shown in Fig.~\ref{fig:stack}. In summary, the layers in the protocol stack are, beginning from the lowest level:
\begin{itemize}
\item \textsc{Data (Message)}\index{Data (message) layer}: Raw data (`payload') Alice wishes to transmit to Bob. Comprises both \textsc{Quantum Data}\index{Quantum data layer} and \textsc{Classical Data}\index{Classical data layer}.
\item \textsc{Packet}\index{Packet!Layer}: Decomposition of \textsc{Data} into blocks (\textsc{Packet Data}\index{Packet!Data layer}), and associated classical \textsc{Packet Headers}\index{Packet!Header layer} containing metadata (e.g routing information).
\item \textsc{Strategy}\index{Strategy!Layer}: Construct \textsc{Packet} routing strategies based on a cost optimisation algorithm.
\item \textsc{Transport}\index{Transport layer}: Physical routing of \textsc{Packets} to arrive at their destination, based upon metadata contained in \textsc{Packet Header}. Perform collision detection during transit.
\item \textsc{Reconstruction}\index{Reconstruction layer}: Reconstruct \textsc{Data} from received \textsc{Packets}.
\item \textsc{Quality of Service (QoS)}\index{Quality of service (QoS)!Layer}: Apply QEC and determine whether \textsc{QoS} requirements have been satisfied.
\item \textsc{Services \& Applications}\index{Services \& applications layer}: High-level interface to \textsc{Data} presented to Bob's services and applications. The interface abstracts away lower levels of the protocol stack, presenting Bob with only \textsc{Data} and its associated metadata.
\end{itemize}

Next we describe the operation of these layers in detail.

%
% Data (Message)
%

\subsubsection{Data (Message)} \index{Data (message) layer} \label{sec:data_message_layer}

At the lowest level of the protocol we have the raw \textsc{Data} Alice wishes to communicate to Bob. \textsc{Data} is allowed to comprise both \textsc{Quantum Data} and \textsc{Classical Data} components, and may contain one or the other, or both.

%
% Quantum Data
%

\paragraph{Quantum data} \index{Quantum data layer}

\textsc{Quantum Data} is allowed to be an arbitrary quantum state. It could be a pure or mixed state, of arbitrary (but predetermined) dimension, or even a subsystem of a larger external state (i.e entangled with another system). We stress that it needn't be expressed using a conventional qubit representation, in the way digital data is necessarily represented using bits. Keep in mind that the quantum internet isn't just there to communicate qubit data streams. Rather, it is intended to act as generally as possible, such that essentially arbitrary \textit{quantum assets}\index{Quantum assets} can be exchanged. These needn't be restricted to any particular type of encoding, such as those discussed in Sec.~\ref{sec:opt_enc_of_qi}. For example, in addition to something `standard' like polarisation-encoded qubits in single photons, one network user might like to share an exotic CV state of light, like a cat state, with his mate whose cat died. Indeed, multiple types of state encoding might be encapsulated within a single \textsc{Packet}. The QTCP acts only as an abstract interface for quantum networking, but is completely blind as to what the underlying data in the network is. QTCP is only concerned with getting that state from Alice to Bob.

%
% Classical Data
%

\paragraph{Classical data} \index{Classical data layer}

\textsc{Classical Data} is a purely classical state with no coherence (i.e a diagonal density matrix), which may be represented as a classical bit-string. We very intentionally segregate the \textsc{Classical} and \textsc{Quantum} components of \textsc{Data}, since the classical network is expected to be cheaper and more reliable than the quantum network operating in parallel to it. The \textsc{Classical Data} could, for example, provide nodes with classical instructions on what quantum computations to perform on the \textsc{Quantum Data}.

%
% Packet
%

\subsubsection{Packet} \label{sec:packet_layer} \index{Packet!Layer}

\textsc{Data} is transmitted as \textsc{Packets}, much in the same way as conventional TCP. The \textsc{Data} is decomposed into three components: \textsc{Quantum Data}, \textsc{Classical Data}, and \textsc{Packet Header}.

We can express the state of an entire \textsc{Packet} as,
\begin{align}
\hat\rho_\mathrm{packet}(i) = \hat\rho_\mathrm{quantum}(i) \oplus \hat\rho_\mathrm{classical}(i) \oplus \hat\rho_\mathrm{header}(i),
\end{align}
where $i$ denotes the $i$th packet. Here $\hat\rho_\mathrm{quantum}$ ($\hat\rho_\mathrm{classical}$) is a block of \textsc{Quantum Block Size} qubits (\textsc{Classical Block Size} bits) taken from the user's \textsc{Quantum Data} (\textsc{Classical Data}), while $\hat\rho_\mathrm{header}$ is the \textsc{Packet's} classical \textsc{Packet Header}. As discussed earlier, since $\hat\rho_\mathrm{quantum}$ is quantum, and $\hat\rho_\mathrm{classical}$ and $\hat\rho_\mathrm{header}$ are classical, they needn't be transmitted together over the same quantum network. Instead all \textsc{Classical Data} and \textsc{Packet Header} could be transmitted over a classical network operating in parallel to and synchronised with the quantum network, which carries the \textsc{Quantum Data}.

Note that while one can always measure classical data without disturbance, this is not the case with quantum data, where measurements cause wave-function collapse. Thus, while Alice is always able to know $\hat\rho_\mathrm{classical}$ and $\hat\rho_\mathrm{header}$, she may or may not know $\hat\rho_\mathrm{quantum}$. Clearly if she prepared the state herself, she would (hopefully) know what she was doing. But in general, quantum networks could be used for far less trivial networking, where Alice is, for example, an intermediary in a distributed quantum computation. In this instance, Alice is unlikely to know what her \textsc{Quantum Data} is.

%
% Packet Data
%

\paragraph{Packet data} \index{Packet!Data layer}

Comprises blocks of both \textsc{Quantum Data} and \textsc{Classical Data}, of sizes \textsc{Quantum Block Size} and \textsc{Classical Block Size} respectively. \textsc{Packet Data} requires that \textsc{Data} be decomposed into distinct subsystems which are independently transmitted by QTCP. For easy of exposition, we will restrict ourselves to the case where the \textsc{Data} is encoded into a stream of qubits (\textsc{Quantum Data}) and bits (\textsc{Classical Data}). But of course other encodings could be used. Also bear in mind that any quantum (classical) information can be encoded into qubits (bits), and once represented as such, the decomposition of \textsc{Data} into \textsc{Packets} arises very naturally and intuitively. The \textsc{Packet's} \textsc{Quantum Data} is what is transmitted via the quantum channels, while the \textsc{Classical Data} is communicated via classical channels.

%
% Packet Header
%

\paragraph{Packet header} \label{sec:packet_header} \index{Packet!Header layer}

\textsc{Packet Header} is purely classical and needn't be transmitted over the costly quantum network, instead being transmitted over a complementary classical network, running in parallel to, and synchronised with the quantum network. \textsc{Packet Header} contains no information content from \textsc{Data}, instead comprising only metadata relevant to the higher levels of the protocol stack. In particular, \textsc{Packet Header} contains the following fields:
\begin{itemize}
    \item \textsc{Header Size}\index{Header size}: The number of bits in the \textsc{Packet Header}.
    \item \textsc{Message ID}\index{Message ID}: A unique identifier for the complete message to which this \textsc{Packet} belongs. This field mitigates ambiguity as to which \textsc{Message} this packet belongs when performing \textsc{Reconstruction}.
    \item \textsc{Lifetime}\index{Lifetime}: How long the \textsc{Packet} has been in existence for, i.e since it was initially sent by the \textsc{Sender}. This is used by strategies to prevent collisions.
    \item \textsc{Sender}\index{Sender}: A unique node identifier for the sender (Alice).
    \item \textsc{Recipient}\index{Recipient}: A unique node identifier for the recipient (Bob).
    \item \textsc{Order}\index{Order}: To which block taken from \textsc{Data} does this \textsc{Packet Data} belong? This is extremely important in networks where \textsc{Packets} may arrive out of order. The \textsc{Order} field forms the basis for the \textsc{Reconstruction} layer.
    \item \textsc{Quantum Block Size}\index{Quantum block size}: The number of qubits contained in the \textsc{Quantum} component of \textsc{Packet Data}. This is important for the \textsc{Reconstruction} layer.
    \item \textsc{Classical Block Size}\index{Classical block size}: The number of bits contained in the \textsc{Classical} component of \textsc{Packet Data}, also important for the \textsc{Reconstruction} layer.
    \item \textsc{Routing Queue}\index{Routing!Queues}: A first in, first out (FIFO) queue of node identifiers, tracing out the entire route for the \textsc{Packet} to follow, in chronological order from the next node to visit all the way to the \textsc{Recipient}.
    \item \textsc{Costs}\index{Costs}: A tuple characterising all the accumulated costs of the \textsc{Packet} at the current stage in the route. These are treated as accumulators that are incremented appropriately after each step, since costs are additive.
    \item \textsc{Attributes}\index{Attributes}: A tuple characterising all the non-\textsc{Cost} properties associated with the \textsc{Packet}. Examples include: the \textsc{Priority} of routing a \textsc{Packet} to its destination; suggesting a preferred routing \textsc{Strategy}; or, indicating whether or not a \textsc{Resend Until Success} protocol may be applied to this \textsc{Packet}.
    \item \textsc{Padding}\index{Padding}: Null data to pad the joint \textsc{Classical Packet Data} and \textsc{Packet Header} fields to be of the same length as the \textsc{Quantum Packet Data}. This ensures that the components of the \textsc{Packet} traversing the quantum and classical channels remain in perfect tandem -- bit for qubit. In Sec.~\ref{sec:transport} we show that this facilitates collision detection without the need to measure quantum states.
    \item \textsc{Checksum}\index{Checksums}: A regular checksum of the entire \textsc{Classical} component of the \textsc{Packet}, including both \textsc{Classical Packet Data} and \textsc{Packet Header}. This also forms a part of the collision detection protocol.
\end{itemize}

One might question why \textsc{Packet Headers} tally accumulated costs when we ought to already know all the costs, since these were employed by the algorithm for choosing strategies in the first case. In the ideal case where all strategies are determined \textit{a priori} and are implemented as intended, this is certainly valid. However, for generality we retain this option since more realistic networks may require dynamically updating strategies during the course of propagation, in which case dynamically tallying costs is appropriate.

%
% Strategy
%

\subsubsection{Strategy} \label{sec:intro_strat} \index{Strategy!Layer}

Based on the \textsc{Packet Headers} of all users sharing the network, choose routing strategies to optimise cost metrics. The notion of strategies is introduced in Sec.~\ref{sec:strat_opt}, and a detailed discussion of example strategies is presented in Sec.~\ref{sec:strategies}.

Once routings have been determined for all \textsc{Packets}, the \textsc{Routing Queues} in their \textsc{Packet Headers} are initialised accordingly by pushing the sequence of node identifiers tracing out the desired route.

In the case of dynamic, time-dependent strategies, which can be updated within the duration of transmissions, the \textsc{Routing Queues} may need to be updated. A change in a \textsc{Packet's} route simply requires flushing the queue and pushing new node identifiers for each of the nodes in the new route, in chronological order.

The \textsc{Strategy} layer is responsible for evaluating the net cost function $f_\mathrm{cost}$ from Eq.~(\ref{eq:net_cost_R}), which accounts for both \textsc{Costs} and \textsc{Attributes} to calculate a single effective cost measure that may be employed in routing decisions.

%
% Transport
%

\subsubsection{Transport} \label{sec:transport} \index{Transport layer}

The \textsc{Transport} layer is responsible for actual routing at the physical level, making direct decisions as to what to do with a \textsc{Packet} at each step, based upon the metadata contained in \textsc{Packet Header}, most notably the \textsc{Routing Queue}, which specifies the full route a \textsc{Packet} is destined to follow. It is also responsible for keeping track of costs that accumulate over their route.

Additionally, the \textsc{Transport} layer is responsible for collision detection, whereby multiple packets being transmitted simultaneously over a network interfere with one another, corrupting the data. In classical networking, collision detection is straightforward using checksums. But the usual classical approach breaks down in the quantum setting. In Sec.~\ref{sec:collision} we discuss in detail collision detection in QTCP.

The pseudo-code algorithm implemented by the \textsc{Transport} layer, including collision detection, is shown in Alg.~\ref{alg:transport_alg}.

\startalgtable
\begin{table}[!htbp]
\begin{mdframed}[innertopmargin=3pt, innerbottommargin=3pt, nobreak]
\texttt{
function Transport(Packet):
\begin{enumerate}
    \item nextNode = Packet.RoutingQueue.Pop()
    \item Packet.PhysicallySendTo(nextNode)
    \item Packet.WaitUntilArrivesAt(nextNode)
    \item checksum = Hash(Packet.Header + Packet.ClassicalData)
    \item if(checksum $\neq$ Packet.Header.Checksum) \{
    \setlength{\itemindent}{0.2in}
    \item Packet.Sender.Notify(\textsc{Failure})
    \item Packet.Recipient.Notify(\textsc{Failure})
    \item Packet.Discard()
    \item $\Box$
        \setlength{\itemindent}{0in}
\item \}
    \item Packet.Costs += IncomingLink.Costs
    \item Packet.Attributes.Update()
    \item if(Packet.RoutingQueue.Length = 0) \{
    \setlength{\itemindent}{0.2in}
    \item Return(Packet)
        \setlength{\itemindent}{0in}
    \item \}
    \setlength{\itemindent}{0in}
    \item $\Box$
\end{enumerate}}
\end{mdframed}
\captionspacealg \caption{Algorithm implemented by the \textsc{Transport} layer of QTCP for each \textsc{Packet}. The \texttt{Attributes.Update()} function is left undefined. This is where arbitrary \textsc{Attribute} dynamics may take place.} \label{alg:transport_alg}
\end{table}

%
% Reconstruction
%

\subsubsection{Reconstruction} \index{Reconstruction layer} \label{sec:reconstruction_layer}

The \textsc{Reconstruction} layer only serves one purpose -- to chronologically reorder the received \textsc{Packets} based on the \textsc{Order} field in their \textsc{Packet Headers}. This stage is only performed by Bob -- the final recipient -- and not at any intermediate stage. In general this will require Bob to have a quantum memory, able to hold all \textsc{Packet Data} for a sufficient duration as to enable an arbitrary permutation of \textsc{Packets} to be applied, reproducing the correct chronological order. The algorithm for this is shown in Alg.~\ref{alg:reconstruction}.

\begin{table}[!htbp]
\begin{mdframed}[innertopmargin=3pt, innerbottommargin=3pt, nobreak]
\texttt{
function Reconstruction(Packets):
\begin{enumerate}
    \item Packets.WaitUntilAllReceived()
    \item message =\\ 
    Packets.SortByOrderAscending().data
    \item Packets.Receiver.Notify(message)
     \item $\Box$
\end{enumerate}}
\end{mdframed}
\captionspacealg \caption{The goal of the \textsc{Reconstruction} layer, is to take a collection of received \textsc{Packets} and reassemble them into the \textsc{Message}.} \label{alg:reconstruction}
\end{table}

%
% Quality of Service (QoS)
%

\subsubsection{Quality of service} \label{sec:QOS} \index{Quality of service (QoS)!Layer}

In classical networking theory, error detection and correction is an important element of networking protocols. Communication links may be unreliable, or subject to external noise, which users must be able to detect so as to guarantee the quality of their data.

Classically, error detection is typically performed using checksums (hash functions), which generate a short digest of a packet's data that can be recalculated upon arrival to verify integrity. The checksum can be included in the header component of each packet, allowing the remainder of the protocol to remain unchanged.

In the quantum context the elegant notion of checksums is complicated by the fact that calculating a hash function of a quantum state would necessarily entangle the state with the output hash. This would have the undesired effect of causing measurement of the checksum to collapse the quantum state of the data, thereby altering it in an uncontrollable way.

As an alternative to checksums, we could borrow the notion of quantum error correction (QEC)\index{Quantum error correction (QEC)} and fault-tolerance\index{Fault-tolerance} from quantum computing theory. Here we encode a quantum state into a (polynomially) larger Hilbert space. \textit{Syndrome measurements}\index{Syndromes!Measurements} on some of the states in this larger space allow us to both detect and correct universal error models, such as depolarisation or dephasing, provided that error rates are within the fault-tolerance threshold of the code being employed.

Thus, enforcing QoS in fundamentally different for quantum data packets than for classical ones, requiring entirely different techniques. Given how large a field this has become in its own right, we dedicate Sec.~\ref{sec:QOS_chap} entirely to protocols suited to the implementation of quantum QoS requirements in quantum networks.

%
% Services & Applications
%

\subsubsection{Services \& applications} \index{Services \& applications layer} \label{sec:services_apps}

Having communicated all the \textsc{Packet Data} from Alice to Bob, performed \textsc{Reconstruction}, and applied \textsc{QoS} protocols, Bob ought to have $\hat\rho_\mathrm{data}$ to a good approximation. The quality of Bob's received state can be inferred directly from the \textsc{Costs} vector contained in the \textsc{Packet Headers}. The final state and its associated quality metrics (\textsc{Costs} and QEC outcomes) may then be provided to Bob as a software interface for end use.

\latinquote{Divide et impera.}

%
% Collision Handling & Classical Errors
%

\subsection{Collision handling \& classical errors} \label{sec:collision} \index{Collisions!Handling} \index{Classical errors}

In classical networking, protocols such as Ethernet allow users to simply broadcast data at their leisure and rely on \textit{collision detection} to detect when the broadcasts of multiple users have interfered, signalling that both users ought to retransmit following backoff, to minimise the chances of another collision occurring. While this \textsc{Resend Until Success} approach has certainly proven to be effective in classical networking, in a quantum setting the rules of the game are entirely different.

First, collision detection necessarily requires measuring a communications channel to test whether data has been corrupted. Classical networks typically do this by transmitting a checksum with the data, which is recalculated upon arrival for comparison. This raises the obvious problem that quantum measurements are destructive, which means that testing the integrity of our data destroys it in the process -- the last feature we'd like our network to exhibit! However, to overcome this, in Sec.~\ref{sec:transport} we describe a protocol based on the dual classical/quantum network that allows collision detection without measuring quantum states.

Second, collision detection is not always even allowed at all. If one of Alice's packets was entangled with another (i.e she was communicating an entangled system, where the different subsystems resided in different packets), she would not be able to simply retransmit an identical copy of the corrupted packet, since the entanglement with the other system would have been lost and there are no local operations she can do to recover it.

Alternately, Alice might be a part of a distributed computation, where she didn't prepare the data in the first place. In this instance, the no-cloning theorem\index{No-cloning theorem}\footnote{The no-cloning theorem states that it is impossible to take a completely general, unknown quantum state and prepare an identical copy of it. That is, there is no quantum process\index{Quantum processes} implementing the map \mbox{$\mathcal{E}(\hat\rho\otimes\ket{0}\bra{0}) \to \hat\rho\otimes\hat\rho$} for unknown $\hat\rho$. Note that this result applies when $\hat\rho$ is arbitrary and unknown, but in general needn't apply if restricted to a particular basis, subject to other constraints, or we have \textit{a priori} knowledge of the state. \textit{Proof}:\index{No-cloning theorem!Proof} To see this, consider the unitary evolution \mbox{$\hat{U}\ket\psi\ket{e} = e^{i\alpha(\psi,e)}\ket\psi\ket\psi$}, for some normalised ancillary state $\ket{e}$, where $\alpha$ denotes an irrelevant global phase factor. We have \mbox{$\braket{\psi|\phi}\braket{e|e} = \bra\psi\bra{e}\hat{U}^\dag\hat{U}\ket{\phi}\ket{e} = e^{-i[\alpha(\psi,e)-\alpha(\phi,e)]} \braket{\psi|\phi}^2$}. Since $\ket{e}$ is normalised, this implies \mbox{$|\braket{\psi|\phi}| = |\braket{\psi|\phi}|^2$}, which can only be true if \mbox{$ |\braket{\psi|\phi}|=0$} or \mbox{$ |\braket{\psi|\phi}|=1$}, which does not hold in general for arbitrary states $\ket\psi$ and $\ket\phi$. However, clearly if $\ket\psi$ and $\ket\phi$ are restricted to being equal or orthogonal then this constraint can be satisfied with an appropriate choice of $\hat{U}$, since this corresponds to classical cloning\index{Classical cloning}. This proof applies to unitary evolution. What about generic quantum processes? In that case we merely need to assume that the ancillary state $\ket{e}$ acts on a joint primary/environment system, and the same argument applies.} implies that she cannot, in general, learn what the quantum state was, and therefore would be unable to make a second transmission attempt.

From Alice's point of view, \textsc{Resend Until Success} would clearly work if she was preparing a known state, separable from the other packets. However, collisions on the network caused by her reckless resending would likely corrupt the communications between other parties, leaving them rather ticked off at her.

There are therefore two main approaches to dealing with collisions. First, central planning of all routing could be employed, precisely scheduling all routes \textit{a priori} so as to entirely eliminate any possibility for collisions. Second, if all users in the network were communicating data where packet loss could be tolerated, they could all mutually agree to use the \textsc{Resend Until Success} protocol. This would not require a central authority, and be highly desirable for ad hoc networks. It is important to stress, however, that the latter requires unanimity amongst network participants to function, and the restriction to known, separable states is a major limitation that would prohibit many important uses for quantum networks, such as distributed quantum computation. However, both these approaches are entirely valid in their appropriate context.

In classical TCP all components of data packets are classical and are kept together throughout every stage of transmission. In the quantum case we will instead have a mixture of both quantum and classical data. As mentioned, we will assume that classical communication and computation resources `come for free' (or are at least cheap compared to quantum resources), so there will be a clear disambiguation as to what data is quantum or classical within packets.

As discussed, quantum collision detection is complicated by the fact that measuring quantum data to determine whether it has been corrupted disturbs the quantum state. We address this problem by taking advantage of the duality of the quantum/classical network, discussed in Part.~\ref{part:quant_net}. Because the quantum and classical components of the \textsc{Packets} are synchronised and of equal length (thanks to the \textsc{Padding} field of the \textsc{Packet Header}), and because the same applies to all other \textsc{Packets} on the network, a collision in the quantum data necessarily implies a collision in the classical data, and vice versa. Therefore, by applying regular classical collision detection techniques based on checksums (recall \textsc{Packet Header} contains a \textsc{Checksum} field), we can infer collisions in quantum data without actually measuring it. We refer to this as \textit{indirect collision detection}\index{Indirect collision detection}. This guarantees us the ability to detect when a collision has occurred, in which case both quantum \textit{and} classical data are corrupted, or has not occurred, in which case both quantum and classical data are uncorrupted and the quantum data remains unmeasured. Collision detection is incorporated into the pseudo-code implementation of the \textsc{Transport} layer shown in Alg.~\ref{alg:transport_alg}.

This algorithm could, depending upon implementation, be executed locally on the node currently hosting the \textsc{Packet}, or it could be delegated to a central authority, but with the overhead of additional classical communication.

In instances where corruption or loss of packets cannot be tolerated, a more proactive approach may be applied. To preempt the risk of packet collision, one could introduce \textit{probe packets}\index{Probe packets} -- packets containing only classical data, that query a route ahead to negotiate channel usage for the following proper packet, thereby avoiding collisions. Of course, if an upcoming node is unable to guarantee channel capacity immediately, the packet may need to be stored in quantum memory until the channel is available. Thus, it is important to accommodate for this by ensuring that quantum memory is available in nodes preceding links/nodes where collisions are not guaranteed to be mitigated immediately. Quantum memory will be discussed in Sec.~\ref{sec:memory}.

\latinquote{Esse est percipi.}

% \input{Sections/multi_packet_operations}

% \latinquote{Capax infiniti.}

%
% Extensibility of QTCP
%

\subsection{Extensibility of QTCP}\index{Extensibility of QTCP} \label{sec:c_vs_a}

In Sec.~\ref{sec:costs} we introduced the notion of the \textit{costs} and \textit{attributes} of links in a classical network, and in Sec.~\ref{sec:quantum_meas_cost} generalised these notions to the quantum case. In Sec.~\ref{sec:packet_header} we described the header format for quantum data packets.

The \textsc{Costs} and \textsc{Attributes} fields within the \textsc{Header} are very powerful data structures, implemented as ordered sets of arbitrary dimension, comprising arbitrary data fields. The intention here is to allow QTCP to be extensible into the future, with the flexible addition of new data structures into the protocol. These can be custom designed to, in conjunction with appropriate routing strategies and cost functions, influence the operation of QTCP completely arbitrarily, and easily implement entirely different quantum networking paradigms than presented here.

\textsc{Costs} naturally capture characteristics of the network that accumulate additively along routes, whereas \textsc{Attributes} capture any other characteristics that aren't additive. A network needn't have both costs \textit{and} attributes. It may have one or the other, or both, but not neither, since there must be some measure by which to judge routes, even if via a very trivial measure.

% \comment{To do!}

\latinquote{Cedere nescio.}

% \input{Sections/scalability_of_QTCP}

% %----------

%
% Routing Strategies
%

\section{Routing strategies} \label{sec:strategies} \index{Routing!Strategies}

\dropcap{I}{n} Sec.~\ref{sec:costs} we introduced the notion of network costs, strategies for allocating network resources in Sec.~\ref{sec:route_strats}, and a general formalism for optimising strategies so as to minimise costs in Sec.~\ref{sec:strat_opt}. In this section we present some meaningful example strategies and associated pseudo-code fragments, illustrating the implementation of various aspects of strategies of practical real-world interest.

%
% Single User
%

\subsection{Single user} \label{sec:single_user_shortest} \index{Single user strategies}

Let us begin our discussion of strategies by considering the simplest case of just a single user on the network. Consider the graph shown in Fig.~\ref{fig:simp_route_opt}. This is the same example used earlier, but now the edges have been weighted by some arbitrary cost metric. There are four routes from $A$ to $B$. All have cost \mbox{$c=3$} except the route indicated by the red arrow, which has cost \mbox{$c=2$}. Clearly the latter is optimal in terms of cost minimisation, and any shortest-path algorithm applied between $A$ and $B$ will accurately come to that conclusion. Thus, single-user networks are very trivial to optimise, and there is no distinction between \textsc{Local} and \textsc{Global} strategies.

The very trivial algorithm for this route finding is shown in Alg.~\ref{alg:single_user}, where the \texttt{ShortestPath()} function could be any of the existing, well-known shortest path algorithms (Sec.~\ref{sec:shortest_path}).
\begin{table}[!htbp]
\begin{mdframed}[innertopmargin=3pt, innerbottommargin=3pt, nobreak]
\texttt{ 
function Strategy.SingleUser(Packets):
\begin{enumerate}
    \item for(packet$\in$Packets) \{
        \setlength{\itemindent}{.2in}
                \item currentNode = packet.RoutingQueue.Pop()
        \item shortestRoute = \\
        ShortestPath(currentNode,packet.Recipient)
        \item packet.RoutingQueue.Flush()
        \item packet.RoutingQueue.Push(shortestRoute)
    \setlength{\itemindent}{0in}
    \item \}
    \item $\Box$
\end{enumerate}}
\end{mdframed}
\captionspacealg \caption{For a single user, a simple shortest-path algorithm necessarily finds the optimal route, as there is no potential for packet collisions or competition for network resources.} \label{alg:single_user}
\end{table}

%
% Multiple Users
%

\subsection{Multiple users} \label{sec:two_user} \index{Multiple user strategies}

\begin{figure}[!htbp]
\includegraphics[clip=true, width=0.475\textwidth]{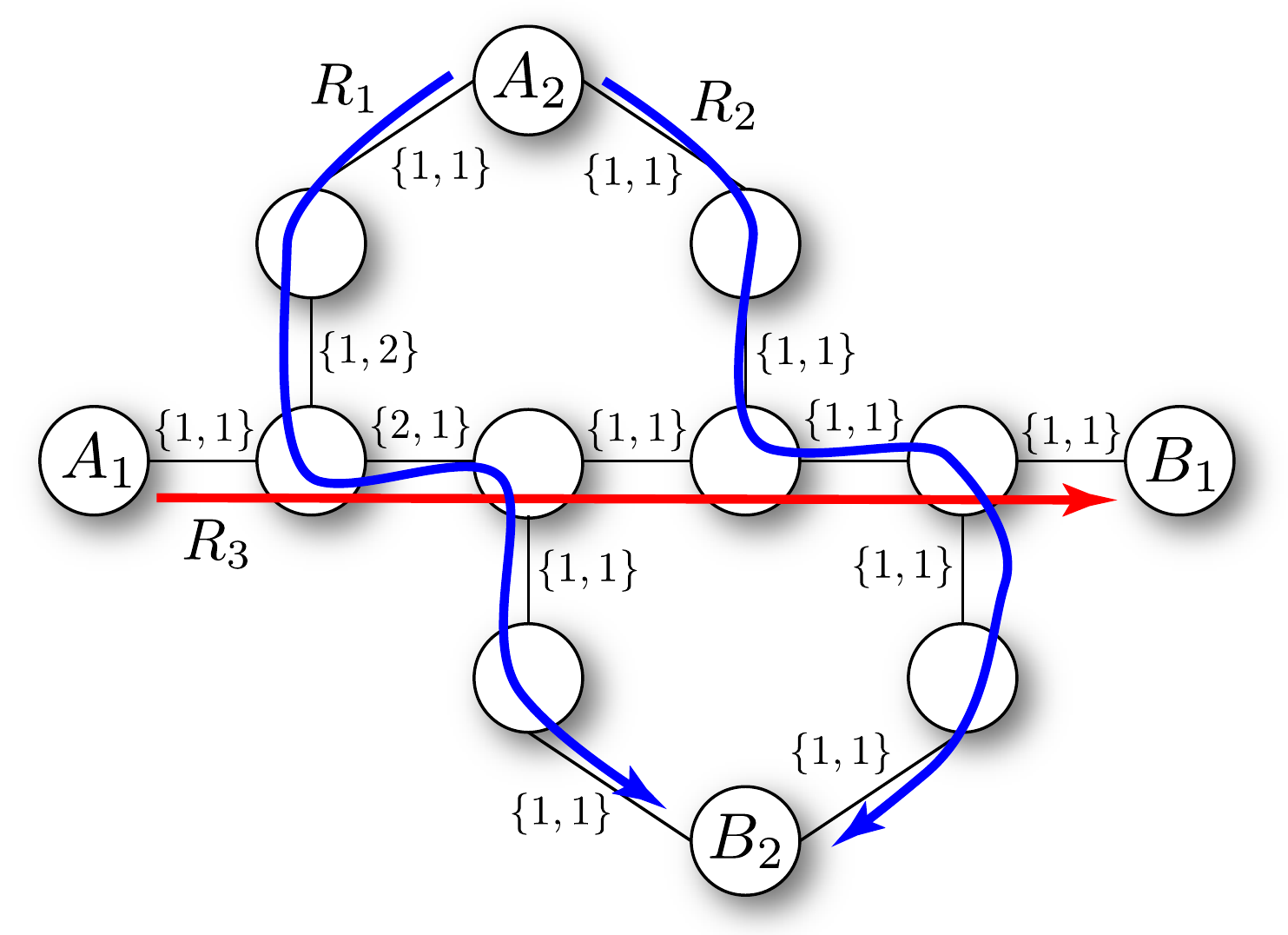}
\captionspacefig \caption{A simple network with two competing pairs of senders and receivers, \mbox{$A_1\to B_1$} and \mbox{$A_2\to B_2$}. Edges are labelled by \mbox{$\{b,d\}$}, where $b$ is the bandwidth attribute of the link (i.e number of states that can be communicated simultaneously), and $d$ is the cost metric associated with the link, e.g loss in dB. (blue line) $R_1$ and $R_2$ are two routes from $A_2\to B_2$. Either of these routes could be declared optimal, depending on the choice of cost function. For a trivial additive cost function, $R_2$ would be declared optimal. (red lines) $R_3$ is the optimal route from \mbox{$A_1\to B_1$}.} \label{fig:conflict}
\end{figure}

Next consider the more complex network shown in Fig.~\ref{fig:conflict}. We consider two sender/receiver pairs, \mbox{$A_1\to B_1$} and \mbox{$A_2\to B_2$}. The available routes connecting both pairs overlap, creating competition for network resources.

Let us assume there are just two properties of interest when deciding strategies -- cost in dollars (which may differ for different links), and availability (i.e how many states can the channel handle at once). Let $c_1$ be the dollar cost, and \mbox{$a_1$} be the amount of available channel capacity. Our network is very primitive and each channel can only accommodate one state at a time. Thus, we let \mbox{$a_1=1$} for all links, except for the one common to both $R_1$ and $R_3$, \mbox{($R_1\cap R_3$)}, which we invest more heavily into, since both routes are going to be wanting to use this link.

To define our net cost measure, we combine $c_1$ and $a_1$ according to,
\begin{align}
\mathcal{S} : f_\mathrm{net}(\vec{c}) = \left\{
\begin{array}{l l}
c_1, & \quad \mathrm{if}~ a_1>0 \\
\infty, & \quad \mathrm{if}~~ a_1=0 \\
\end{array} \right..
\end{align}
That is, provided bandwidth is available, the link will have the dollar cost $c_1$. If no bandwidth is available, the cost is infinite, thereby removing the respective link from the graph.

Next the cost metrics are updated by the strategy $\mathcal{S}$ following each communication. In this instance this simply decrements the bandwidth attribute for the links that were utilised,
\begin{align}
\mathcal{S} : a_1 \to a_1-1.
\end{align}

Suppose the strategy optimises the \mbox{$A_1\to B_1$} route first, yielding $R_3$, before moving onto the \mbox{$A_2\to D_2$} route. In this case, the reduction of the bandwidth attribute signals that the cheapest route $R_2$ is no longer available to be utilised simultaneously to $R_3$, and must therefore wait its turn on the following clock-cycle. Alternately, the strategy could employ $R_1$ for \mbox{$A_2\to B_2$}, in which case their common link with capacity for two states would eliminate the competition between the two communications, allowing both to take place simultaneously. Thus, there is a tradeoff: for \mbox{$A_2\to B_2$}, we could achieve a net cost of \mbox{$c(A_2\to B_2)=5$}, requiring 2 clock-cycles; or we could achieve simultaneous communication at the expense of increasing cost to \mbox{$c(A_2\to B_2)=6$}. This indicates that when choosing strategies, we must carefully define its goals.

Suppose net cost, rather than clock-cycles, was the key measure of interest. Then choosing the routes $R_1$ and $R_3$ would be the optimal choice. An optimal \textsc{Global} optimisation would recognise this. However, a \textsc{Local} optimisation, based on choosing shortest-paths one-by-by for each sender/receiver pair, may or may not choose the optimal routes, depending on the order in which the decisions were made.

Suppose the \mbox{$A_2\to B_2$} route were optimised first. We would choose $R_2$. Then there would be a traffic jam on the \mbox{$A_1\to B_1$} route, and it would necessarily have to wait its turn. In a time-critical application, where waiting is intolerable, this effectively renders the network useless to the first sender/receiver pair.

If, however, the \mbox{$A_1\to B_1$} were optimised first, choosing $R_3$, then $R_2$ would be prohibited once the bandwidth attributes were updated, and the second best option, $R_1$, would be chosen. Now both communications could take place simultaneously. So we see that the outcomes of \textsc{Local} optimisations needn't always be consistent or unique. Rather, they can be highly dependent upon circumstantial issues, such as the arbitrary order in which routes are chosen for optimisation.

Generalising this to any number of users is a straightforward extension to the route optimisation problem, incurring a higher computational overhead due to the increased optimisation complexity.

In the upcoming sections we discuss multi-user strategies in more detail. None of these are true \textsc{Global} strategies, but nonetheless address some of the problems facing \textsc{Local} strategies mentioned above.

Truly \textsc{Global} strategies could employ, for example, the vehicle routing problem (Sec.~\ref{sec:VRP}) or vehicle rescheduling problem (Sec.~\ref{sec:VRSP}) algorithms. However, both of these are \textbf{NP}-hard\index{NP \& NP-complete} in general. Thus, the approximation heuristics to be discussed in the following sections are highly applicable.

%
% Round Robin
%

\subsection{Round robin} \label{sec:round_robin} \index{Round-Robin!Strategies}

Perhaps the simplest and most elegant multi-user scheduling strategy is to borrow from the idea of time-division multiplexing for preemptive multitasking employed by UNIX operating systems. Here we simply put all live packets in a list, and go through the list, one-by-one, giving each packet an equal time-share of network resources, independent of costs. The algorithm for this is shown in Alg.~\ref{alg:round_robin}.

\begin{table}[!htbp]
\begin{mdframed}[innertopmargin=3pt, innerbottommargin=3pt, nobreak]
\texttt{
function Strategy.RoundRobin(Packets):
\begin{enumerate}
    \item for(packet$\in$Packets) \{
        \setlength{\itemindent}{.2in}
                \item currentNode = packet.RoutingQueue.Pop()
        \item shortestRoute = \\
        ShortestPath(currentNode,packet.Recipient)
        \item packet.RoutingQueue.Flush()
        \item packet.RoutingQueue.Push(shortestRoute)
    \setlength{\itemindent}{0in} \}
    \item $\Box$
\end{enumerate}}
\end{mdframed}
\captionspacealg \caption{In the \textsc{Round Robin} strategy we simply iterate through the list of active packets, with no regard for any metrics, or conflicts between them. Rather, we strive for perfect time-sharing equality, and every packet entirely ignores the actions of all other packets, performing a completely selfish optimisation strategy.} \label{alg:round_robin}
\end{table}
The \textsc{Round Robin} strategy can be considered base skeleton code for more sophistic algorithms to build upon, simply by reordering the packet queue.

While such a protocol clearly ensures scheduling that gives all packets attention, it is the perfect example of an algorithm subject to the resource allocation imbalance discussed in Sec.~\ref{sec:two_user}. Specifically, the routes being followed by some packets may systemically receive more favourable treatment than others, based on the arbitrary ordering of the list of packets. Also, equal timesharing fails to accommodate for the fact that some routes are inherently more costly than others and deserve a greater share of network resources.

%
% Data Priority
%

\subsection{Data priority} \label{sec:data_priority} \index{Data priority strategies}

Are all men created equal? No. Some packets may inherently be more important than others, and ought to receive priority when allocating network resources. A simple variation on the \textsc{Round Robin} strategy is to, before iterating through the list of packets, order them according to a \textsc{Priority} attribute. Thus, when applying a shortest-path algorithm, it is deemed most important to minimise the costs of the more important packets first.

This is trivially achieved by taking the existing \textsc{Round Robin} strategy, and first ordering the packet list by their priority attributes, i.e by inserting a new line 1, \mbox{\texttt{Packets.SortByPriority()}}.

%
% Randomisation
%

\subsection{Randomisation} \label{sec:random} \index{Randomised strategies}

The imbalance issue facing the \textsc{Round Robin} strategy (Sec.~\ref{sec:round_robin}) may be most trivially addressed using randomisation of the strategy, such that routes are optimised in an order chosen randomly each time. This would allow the different sender/receiver pairs to have equal access to network resources, when averaged over many network uses.

This is also a straightforward variation of the \textsc{Round Robin} strategy, achieved by first randomising the list of packets before the other stages, i.e insert a new line 1, \mbox{\texttt{Packets.RandomizeOrder()}}.

%
% Cost Priority
%

\subsection{Cost priority} \label{sec:cost_priority} \index{Cost priority strategies}

The \textsc{Random} strategy overcomes one key problem facing any \textsc{Local} optimisation strategy. But it is nonetheless merely a mild variation on the \textsc{Round Robin} strategy, guaranteeing equal time-share to each sender/receiver pair. But does equal time-sharing actually represent the best allocation of resources?

It isn't just the order in which routes are chosen, which creates imbalance between users. The costs and attributes of the routes themselves is inevitably biased more in favour of some users than others. To accommodate this we now introduce the \textsc{Cost Priority} strategy. Here, rather than prioritising packets on a random basis, or according to a fixed, predetermined priority attribute, we do so according to their net accumulated cost. Those who have accumulated the highest cost will subsequently be treated with highest priority. This strategy effectively introduces a negative feedback loop into resource allocation, creating a self-regulating (and hopefully stable!) time-multiplexed packet-switched network. The pseudo-code for the \textsc{Cost Priority} strategy is shown in Alg.~\ref{alg:cost_prior_alg}.

\begin{table}[!htbp]
\begin{mdframed}[innertopmargin=3pt, innerbottommargin=3pt, nobreak]
\texttt{
function Strategy.CostPriority(Packets):
\begin{enumerate}
    \item packetsAndCosts = []
    \item for(packet$\in$Packets) \{
        \setlength{\itemindent}{.2in}
        \item cost = costFunction(packet)
        \item packetsAndCosts.Append([packet,cost])
    \setlength{\itemindent}{0in}
    \item     \}
    \item sorted = \\
        SortByCostDescending(packetsAndCosts)
    \item for(packet$\in$sorted) \{
        \setlength{\itemindent}{.2in}
        \item currentNode = packet.RoutingQueue.Pop()
        \item shortestRoute = \\
        ShortestPath(currentNode,packet.Recipient)
        \item packet.RoutingQueue.Flush()
        \item packet.RoutingQueue.Push(shortestRoute)
    \setlength{\itemindent}{0in}
    \item \}
    \item $\Box$
\end{enumerate}}
\end{mdframed}
\captionspacealg \caption{The \textsc{Cost Priority} strategy scheduling algorithm that gives highest routing priority to \textsc{Packets} with the highest accumulated cost (i.e which have suffered the most). The as-yet undefined \texttt{costFunction()}, which refers to $f_\mathrm{cost}$ from Eq.~(\ref{eq:net_cost_R}), is where the details of the priority decisions are made, which could be entirely arbitrary. In this example, the shortest route is recalculated at each step, based on the expectation that network metrics are dynamic.} \label{alg:cost_prior_alg}
\end{table}

This is an example of a \textsc{Greedy} optimisation algorithm, which attempts to optimise routing by always optimising the most desperate packets first, in descending order down to the least. It is well-known that \textsc{Greedy} algorithms often do not find global optima. Nonetheless, this approach improves on the previous multi-user protocols.

Let us consider a simple example scenario. Imagine we begin with an ordinary network graph, with edges weighted by costs and attributes. For generality, we will additionally assume the available network resources are very dynamic and unpredictable. The costs associated with links are at the whim of market forces we do not understand (do we ever?). And, for the sake of example, and to make matters worse, the links have been very unreliable lately, and are routinely dropping in and out -- `blackouts'. This effectively rules out \textit{a priori} route optimisation, requiring something dynamic.

There are many users on the network, with many active packets at any give time, but because of the constant oscillations in network resources, some packets have received second-class treatment, and through neglect accumulated an unfair share of state degradation. This simple toy model is, at least qualitatively, something that could arise quite naturally in networks with constrained or unreliable resources.

Let us define an example \textsc{Cost Priority} strategy using the following:
\begin{itemize}
\item \textsc{Latency} cost: How long has the packet has been in transit for? This is actually a very general cost metric, since any other cost metric measured in units per time will be directly proportional to this metric. That is, loss, fidelity, purity, efficiency, and so on, all mirror this metric when expressed on a decibel scale. Of course, the same strategy could have easily been applied to any other cost metric.
\item \textsc{Blackout} attribute: Is our unreliable link actually working right now? A given link will have probability $p_\mathrm{op}$ of being operational at any given time, chosen independently for each link at each clock-cycle. The \texttt{Attributes.Update()} function from Alg.~\ref{alg:transport_alg} is responsible for implementing this.
\item \mbox{\textsc{Cost Function}} ($f_\mathrm{cost}$, \texttt{costFunction()} in Alg.~\ref{alg:cost_prior_alg}): The strategy must make sensible decisions based upon only the above two parameters. Because of the previously mentioned generality of the \textsc{Latency} metric, we would like the net cost to directly reflect this metric, but only of course, if the link is operational. If it is not, then that link must be ruled out entirely by assigning it an infinite cost. Thus, we simply choose,
\begin{align}
\mathcal{S} : f_\mathrm{cost}(c,a) = \left\{
\begin{array}{l l}
c, & \quad \mathrm{if}~ a=\mathrm{\texttt{True}} \\
\infty, & \quad \mathrm{if}~ a=\mathrm{\texttt{False}} \\
\end{array} \right..
\end{align}
Note that different packets could be associated with different net cost functions, $f_\mathrm{net}$, to accommodate for the different QoS requirements of different users and messages.
\end{itemize}

In other words, the net cost is taken directly from the underlying cost metric, and modulated by an attribute, yielding a net cost for each packet, which is used to determine which packets receive priority.

This provides us with a simple illustration of how costs and attributes can compliment one another to yield meaningful strategies, that improve network performance over na\"ive, but well-intentioned, time-sharing approaches.

%
% All or Nothing
%

\subsection{All or nothing} \label{sec:all_or_nothing} \index{All or nothing strategies}

In some cases, end-user applications may have strict QoS constraints associated with any data they receive. For example, in a time-critical enterprise, say high-frequency trading, receiving information a millisecond too late is worthless, and it would be best to discard the out of date information to free up bandwidth for the next round of information. Alternately, if the fidelity of a state is required to strictly fall within a fault-tolerance threshold, it will be useless if the threshold is violated. In such a context, the \textsc{Strategy} will apply hard boundaries on QoS metrics, discarding anything violating it, after which some other \textsc{Strategy} is applied. The algorithm is summarised in Alg.~\ref{alg:all_or_nothing}.

\begin{table}[!htbp]
\begin{mdframed}[innertopmargin=3pt, innerbottommargin=3pt, nobreak]
\texttt{
function Strategy.AllOrNothing(Packets, threshold):
\begin{enumerate}
    \item for(packet$\in$Packets) \{
        \setlength{\itemindent}{.2in}
        \item cost = packet.costFunction()
        \item if(cost $\geq$ threshold) \{
        \setlength{\itemindent}{.4in}
            \item packet.Sender.Notify(\textsc{Failure})
            \item packet.Recipient.Notify(\textsc{Failure})
            \item packet.Discard()
                    \setlength{\itemindent}{.2in}
            \item \}
        \setlength{\itemindent}{0in}
    \item \}
    \item Strategy.SomeOtherStrategy(Packets)
    \item $\Box$
\end{enumerate}}
\end{mdframed}
\captionspacealg \caption{The \textsc{All or Nothing} strategy. If the net cost of a packet exceeds a certain \texttt{threshold}, it is discarded outright, and the sender and recipient notified.} \label{alg:all_or_nothing}
\end{table}

%
% Optimal Flow
%

\subsection{Optimal flow} \index{Optimal flow strategies}

In Sec.~\ref{sec:flow_networks} we introduced flow networks as a means for analysing networks where maximising network flow (throughput) is the primary objective. Formulating our quantum networks in this manner is extremely convenient since, combined with our existing definitions for cost metrics and attributes, we can easily exploit a plethora of known results from flow network theory.

As an example of how load allocation might be applied in a simple network, consider again the network shown in Fig.~\ref{fig:simp_route_opt}, where the edge weights are regular cost metrics (not capacities). Alice wishes to send two packets to Bob, simultaneously if possible. Clearly she would transmit her first packet over the \mbox{$A\to F\to B$} route, since this has lowest cost. But let us assume that every link has a maximum capacity of one packet per unit time. In this case Alice will be unable to send her second packet via the same route and must instead resort to using \mbox{$A\to C \to B$} or \mbox{$A\to D\to B$}. The optimisation is straightforward in this instance. However, in general these types of optimisations are somewhat more involved.

These scenarios are handled by flow network optimisation algorithms, of which there are many. We discuss a few of the most relevant ones for our purposes in Sec.~\ref{sec:graph_theory}. Note that these algorithms are \textsc{Global} optimisation algorithms, requiring complete knowledge of the status of the entire network to perform the optimisation.

The routing strategy is very straightforward, shown in Alg.~\ref{alg:opt_flow}, since the \textsc{Global} flow-optimisation algorithm completely specifies the entire configuration of routes through the network.

\begin{table}[!htbp]
\begin{mdframed}[innertopmargin=3pt, innerbottommargin=3pt, nobreak]
\texttt{
function Strategy.OptimalFlow(Packets):
\begin{enumerate}
    \item routes = Packets.OptimalFlowRoutes()
    \item for(packet$\in$Packets) \{
        \setlength{\itemindent}{.2in}
                \item packet.RoutingQueue.Flush()
                \item packet.RoutingQueue.Push(routes[packet])
                \setlength{\itemindent}{0in} 
    \item \}
    \item $\Box$
\end{enumerate}}
\end{mdframed}
\captionspacealg \caption{A generic optimal flow routing strategy. \textsc{Packets} is the array of all packets that ought to be transmitted simultaneously, which are collectively optimised using some flow optimisation algorithm before undergoing transport.} \label{alg:opt_flow}
\end{table}

\latinquote{Disce quasi semper victurus vive quasi cras moriturus.}

%
% Interconnecting & Interfacing Quantum Networks
%

\section{Interconnecting \& interfacing quantum networks} \label{sec:inter} \index{Interfacing!Quantum networks}

\dropcap{A}{ny} global-scale network will inevitably comprise participants choosing to go about things their own way. The physical architecture and medium may vary from one subnetwork to the next, as may the QTCP policies they adopt. The key then is to construct efficient \textit{interconnects} between different levels of the network hierarchy, each of which may subscribe to their own QTCP policies and cross between different physical mediums. Note that the QTCP protocol presented here does not enforce any particular networking policies, but rather provides a high-level framework that can be customised essentially arbitrarily.

For example, the cost metrics and attributes employed at the intercontinental level would most certainly be very different to those in a small LAN. A small LAN might be running applications whereby they can easily reproduce packets and thereby tolerate packet loss. But for a warehouse-scale commercial quantum computing enterprise, responsible for performing one stage of a distributed quantum computation, the loss of a single packet could be extremely costly, requiring the entire computation to be performed completely from scratch due to no-cloning\index{No-cloning theorem} and no-measurement limitations, something that may not come cheaply.

Such interconnects will typically comprise a combination of:
\begin{itemize}
\item Packet switching\index{Packet!Switching}: such that packets can be arbitrarily switched between the different levels of the network hierarchy.
\item Physical interface: interconnect may be switching between different media. Such physical interfaces have costs associated with them. For example, coupling between free-space and fibre is typically very lossy. Sec.~\ref{sec:opt_inter} discusses optical interfacing with matter qubits, and Sec.~\ref{sec:hybrid} discusses hybrid architectures, where optics mediates entanglement generation between matter qubits.
\item Quantum memory\index{Quantum memory}: such that data can be buffered while it awaits its turn at being switched between networks, as different networks may have different loads and operate at different clock-rates. This is discussed in Sec.~\ref{sec:memory}.
\item Packet format conversion\index{Packet!Format conversion}: different levels of the network hierarchy may be employing entirely different cost metrics, attributes, and cost functions, requiring packet headers to be reformatted upon switching between networks.
\end{itemize}

The packet switching and quantum memory are implemented as quantum processes at nodes, using the usual quantum process formalism. The physical interface between different mediums, if there is one, could be very diverse, encompassing many types of physical systems, but can always be characterised using the quantum process formalism. Packet headers, which contain all formatting, cost, and routing information are represented entirely classically and communicated entirely by the classical network. Thus, this operation also takes place at nodes, but no quantum processes are taking place.

%
% Optical Interfacing
%

\subsection{Optical interfacing} \label{sec:opt_inter} \index{Optical!Interfacing}

Unless the entire pipeline of quantum operations through the course of a protocol is all-optical, there will be a need to exchange information between physical systems, for example via light-matter interactions \cite{bib:Cohen-Tannoudji92}. We will now discuss optical interfacing with some of the significant types of matter systems, such that their intercommunication can be optically mediated over the network.

%
% Two-Level Systems
%

\subsubsection{Two-level systems} \index{2-level systems}

The archetypal interface is that between a photonic qubit in the \mbox{$\{\ket{0},\ket{1}\}$} photon-number basis, and a two-level matter qubit\index{Matter qubits} in the $\ket{g}$ (ground) and $\ket{e}$ (excited) state basis. The logical qubit is defined as,
\begin{align}
	\ket{0}_L &\equiv \ket{g}, \nonumber \\
	\ket{1}_L &\equiv \ket{e}.
\end{align}.
Examples include atoms in cavities\index{Atoms in cavities}, NV centres\index{Nitrogen-vacancy (NV) centres}, and engineered quantum dots\index{Quantum dots}.

\begin{figure}[!htbp]
\includegraphics[clip=true, width=0.3\textwidth]{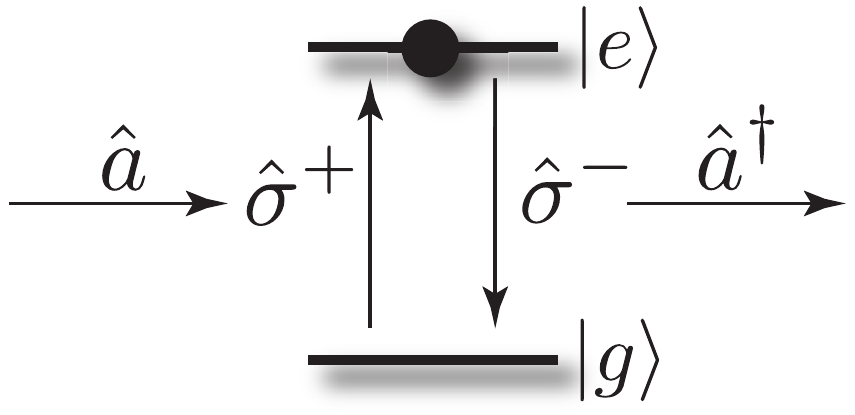}
\captionspacefig \caption{Light-matter interfacing between a single-photon state ($\hat{a}$, $\hat{a}^\dag$) and a two-level matter qubit ($\ket{g}$, $\ket{e}$). The absorption (emission) of a photon is associated with the excitation (relaxation) of the matter qubit ($\hat\sigma^\pm$).} \label{fig:opt_int}
\end{figure}

In the case of a photon interacting with a two-level matter qubit, the interface can be expressed via the Jaynes-Cummings\index{Jaynes-Cummings Hamiltonian} interaction Hamiltonian of the form,
\begin{align} \label{eq:two_level_hamil}
\hat{H}_\mathrm{int} = \hbar \chi (\hat{a}\,\hat\sigma^+ + \hat{a}^\dag\hat\sigma^-),
\end{align}
where $\hat{a}$ ($\hat{a}^\dag$) is the photonic annihilation (creation) operator, $\hat\sigma^\pm$ are the Pauli spin-flip operators, and $\chi$ is the interaction strength\index{Interaction!Strength}. The interpretation of this Hamiltonian is very clear upon inspection -- the annihilation (creation) of a photon is associated with the excitation (relaxation) of the two-level matter system, thereby directly coherently exchanging quantum information between the two systems, as shown in Fig.~\ref{fig:opt_int}.

%
% Lambda-Configuration Systems
%

\subsubsection{$\lambda$-configuration systems} \index{$\lambda$-configuration systems}

\begin{figure}[!htbp]
\includegraphics[clip=true, width=0.225\textwidth]{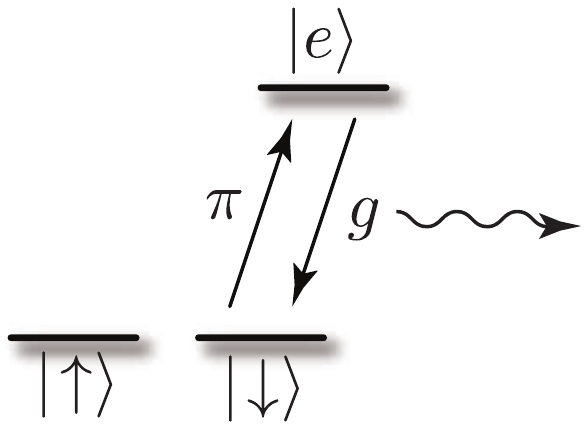}
\captionspacefig \caption{Light-matter interfacing between an optical mode and a $\lambda$-configuration system. The two degenerate ground states represent the logical qubit (\mbox{$\ket{0}_L\equiv\ket{\!\uparrow}$}, \mbox{$\ket{1}_L\equiv\ket{\!\downarrow}$}), only one of which may undergo transition to the excited state $\ket{e}$. Upon pumping the \mbox{$\ket{\!\downarrow}\to\ket{e}$} transition with a $\pi$-pulse, a relaxation back to the ground state maps the logical qubit value to photon-number.} \label{fig:lambda_atom}
\end{figure}

Alternately, one can easily optically interface with a $\lambda$-configuration system, as shown in Fig.~\ref{fig:lambda_atom}. Here there are two degenerate ground states representing the logical qubit basis states (\mbox{$\ket{0}_L\equiv\ket{\!\uparrow}$}, \mbox{$\ket{1}_L\equiv\ket{\!\downarrow}$}), one of which may undergo a transition to an excited state, $\ket{e}$. By pumping the system to the excited state and waiting for a coherent relaxation, the emitted photon may be used to couple the qubit state of the $\lambda$-configuration to an optical mode, mapping the qubit value of the matter qubit to a photon-number representation.

%
% Atomic Ensembles
%

\subsubsection{Atomic ensembles} \label{sec:atomic_ens} \index{Atomic!Ensembles}

In addition to single atoms with well-defined electronic structure, atomic ensembles \cite{bib:DLCZ, bib:Chou05} can be used, whereby the absorption of a photon creates a \textit{collective excitation}\index{Collective excitations} -- a superposition of a single excitation across all the atoms in the ensemble. Specifically, excitations are represented using collective excitation operators,
\begin{align}\index{Collective excitations!Operator}
\hat{S}^\dag = \frac{1}{\sqrt{N}}\sum_{i=1}^N \hat{S}_i^\dag,
\end{align}
where,
\begin{align}
\hat{S}_i^\dag=\ket{e}_i\bra{g}_i,
\end{align}
is the excitation operator for the $i$th atom in the ensemble, $\ket{g}_i$ and $\ket{e}_i$ are the ground and excited states for the $i$th particle, and there are $N$ atoms. The state of a single collective excitation is then given by,
\begin{align}
\ket{\psi_\mathrm{collective}} = \hat{S}^\dag \ket{g}^{\otimes N}.	
\end{align}

Atomic ensembles are essentially well-engineered clouds of atomic gasses, trapped in a glass container, coupled to an optical mode. Atomic ensembles have been demonstrated with extremely long coherence lifetimes ($T_2$-times on the order of milliseconds\index{T$_2$-time}), operating at room temperatures (a very attractive feature on its own). They exhibit \textit{collective enhancement}\index{Collective enhancement} in their coupling to the optical mode -- the optical coupling strength is amplified by a factor quadratic in $N$ compared to single-atom optical coupling, mitigating the need for a cavity.

The collective excitations exhibit the same general mathematical structure as single-atom excitations -- the absorption (emission) of a single photon is associated with a single collective excitation (relaxation), albeit with the favourable collective enhancement in the coupling strength.

\begin{figure}[!htbp]
\includegraphics[clip=true, width=0.225\textwidth]{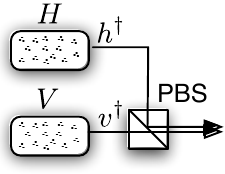}
\captionspacefig \caption{Coupling a polarisation-encoded photonic qubit to a pair of atomic ensembles, each of which corresponds to one of the qubit's logical basis states ($\ket{0}$ or $\ket{1}$). The horizontal and vertical components of the photonic qubit are spatially separated using a PBS, which subsequently independently interface with distinct atomic ensembles via collective excitation.} \label{fig:atomic_ensemble_qubit}
\end{figure}

To couple with a polarisation-encoded photonic qubit, a PBS can be employed to spatially separate the horizontal and vertical modes, each of which couples to a separate atomic ensemble, which jointly represent the logical qubit, as shown in Fig.~\ref{fig:atomic_ensemble_qubit}.

Atomic ensembles have been proposed as quantum memories, given their long coherence lifetimes. Additionally, a protocol for universal cluster state quantum computation (Sec.~\ref{sec:CSQC}) based upon atomic ensemble qubits has been described \cite{bib:RohdeAtEns10}.

Essentially, the long coherence lifetimes of collective excitations owes to the fact that the excitation is effectively encoded as a W-state (Sec.~\ref{sec:W_state_prep})\index{W-states}, an equal superposition of a single excitation across many ($N$) atoms, of the form,
\begin{align}
\ket{\psi_W^{(N)}} = \frac{1}{\sqrt{N}}(&\ket{e,g,g,\dots}\nonumber \\
+&\ket{g,e,g,\dots}\nonumber \\
+&\ket{g,g,e,\dots}\nonumber \\
+&\dots\nonumber \\
+&\ket{g,g,\dots,e}).
\end{align}
W-states are favourable from a decoherence perspective as tracing out a single particle has minimal impact on the coherence of the residual state, which preserves most entanglement, with this robustness growing with the number of particles. This is in stark contrast to GHZ states, which completely decohere under the loss of just a single particle.

Specifically, if $\ket{\psi_W^{(N)}}$ is the $N$-particle W-state (collective excitation), tracing out a single particle yields,
\begin{align}
\hat\rho_\mathrm{tr} &= \mathrm{tr}_1(\ket{\psi_W^{(N)}}\bra{\psi_W^{(N)}}) \\
&= \left(1-\frac{1}{N}\right)\ket{\psi_W^{(N-1)}}\bra{\psi_W^{(N-1)}} + \frac{1}{N}(\ket{g}\bra{g})^{\otimes (N-1)}, \nonumber
\end{align}
which for \mbox{$N\gg 1$} approaches the pure state $\ket{\psi_W^{(N-1)}}$, i.e a W-state with one fewer particles.

%
% Superconducting Qubits
%

\subsubsection{Superconducting qubits}\label{sec:superconducting_qubits}\index{Superconductors!Qubits}\index{Quantum transducers}\index{Microwave qubits}

\sectionby{Chandrashekar Radhakrishnan}\index{Chandrashekar Radhakrishnan}

In the context of superconducting qubits (Sec.~\ref{sec:artificial_atoms}), the energy difference between the energy levels being utilised to encode the qubit is extremely small. Therefore photons coupled to these transitions sit in the microwave regime, whose wavelength lies in the range \mbox{$\lambda\sim100\mu$m-1m}.
x
%To build a scalable quantum computer, multiple qubits are required, with the ability to interact with one another. Hence, it is essential to find ways to transfer quantum information between distinct superconducting qubits. When doing so, it is essential that quantum coherence be preserved. This implies we need to find a different kind of qubit, which can carry information from one place to another, but interacts only weakly with the other qubits and the environment. Photons are the ideal candidate for this (Sec.~\ref{sec:opt_enc_of_qi}). Photonic qubits for carrying information should have frequencies equal to the energy difference between qubit levels. For a superconducting qubit, the energy difference is very small, hence the frequency of optical qubits is in microwave regime, .

Information transfer between distinct superconducting qubits is achieved using a resonator, which acts as a quantum data bus\index{Quantum data bus}. A simple resonator\index{Resonators} is an LC circuit\index{LC!Circuit}, which can support only one frequency mode, but a waveguide resonator can support multiple modes. In general, the transmission line circuits used in non-linear quantum electric circuits\index{Non-linear!Quantum electric circuits} are in the form of coplanar waveguides\index{Waveguides}. These waveguides are engineered to handle a particular set of frequencies, and produce transmission lines with tuneable frequency. Tuneable resonators\index{Tuneable resonators} are very important in quantum optics, and are useful in implementing controllable coupling between different quantum elements, and also in shaping photon wave-packets. 

\begin{figure}[!htbp]
\if 1\doublecol
	\includegraphics[clip=true, width=0.475\textwidth]{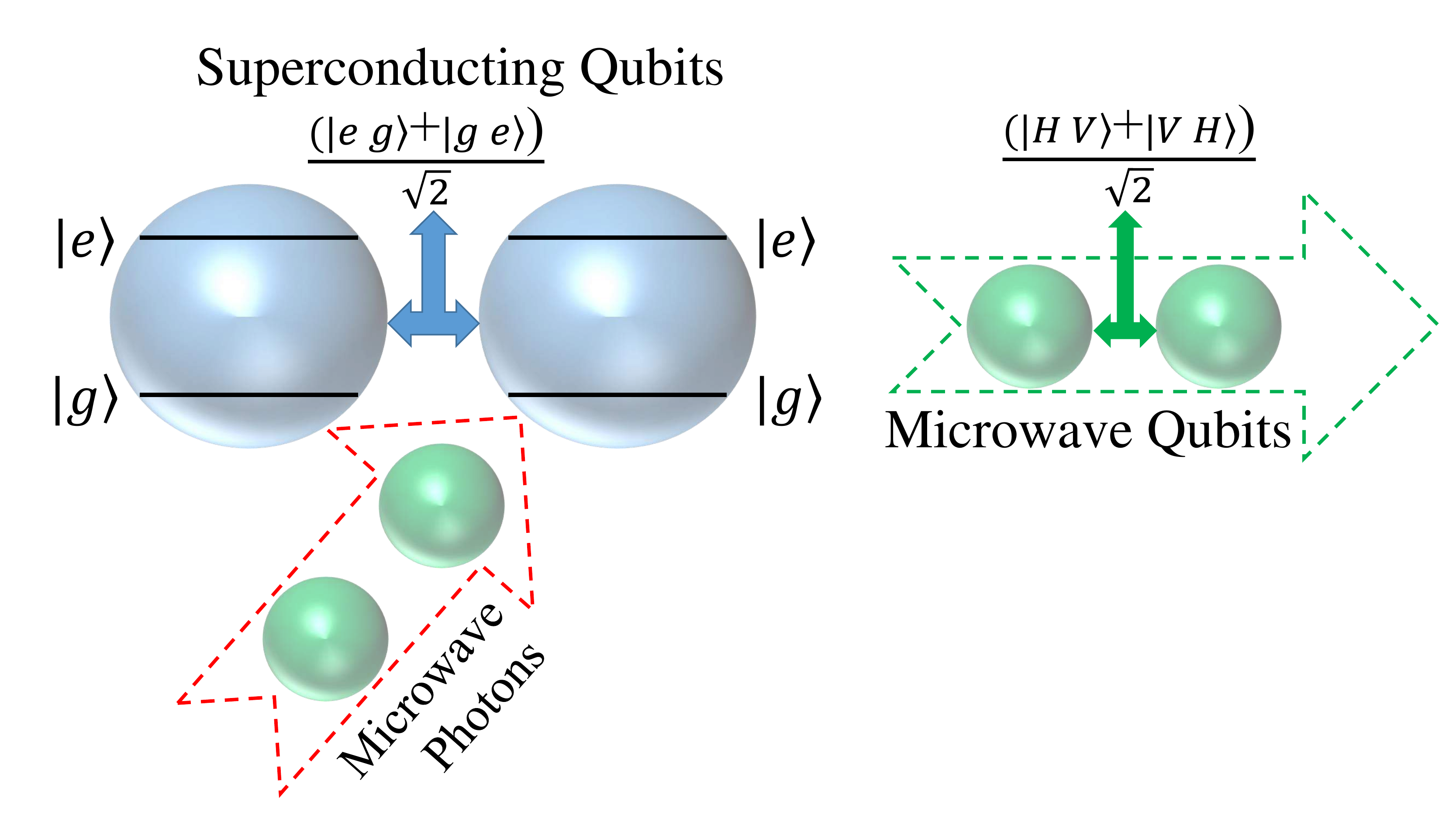}
\else
\includegraphics[clip=true, width=0.7\textwidth]{microwave_qubits}
\fi
\captionspacefig \caption{Schematic sketch of the interaction between superconducting qubits and microwave qubits. The superconducting qubits are in the entangled state \mbox{$\frac{1}{\sqrt{2}}(\ket{e,g} + \ket{g,e})$}, where $\ket{g}$ and $\ket{e}$ are the ground and excited states of the superconducting qubits. These are subjected to interact with the flying qubits, the microwave qubit (red dashed arrow). The photons become entangled and the output state of the photons are \mbox{$\frac{1}{\sqrt{2}}(\ket{H,V} + \ket{V,H})$}, where $\ket{H}$ and $\ket{V}$ are the horizontal and vertical polarisation modes respectively.}\label{fig:microwave_qubits}\index{Microwave qubits}
\end{figure}

In cavity quantum electrodynamics (QED)\index{Quantum electrodynamics} the interaction of a natural atom with an optical photon in the visible wavelength regime is considered. Similarly, the interaction between quantum non-linear electrical circuits and  microwave photons are investigated in circuit QED. The coupling-strength between a natural atom and visible light photon is fixed, where atoms couple weakly with photons \cite{bib:raimond2001manipulating}. Meanwhile, the coupling-strength between a superconducting qubit and a microwave can be manipulated by engineering the parameters of the qubit and resonator, yielding strong and ultra-strong coupling\index{Strong coupling}\index{Ultra-strong coupling} between qubits and photons \cite{bib:wallraff2004strong} (see Fig.~\ref{fig:microwave_qubits}). Furthermore, the coupling between an atom and a photon can be tuned dynamically during the course of an experiment. Several quantum optics components such as mirrors, beamsplitters, circulators and switches can also be designed based on quantum electric circuits.

%\subsubsection{Quantum transducers}

Due to scalability requirements, superconducting qubits are the most widely used qubit implementation used today. The energy-level spacing in superconducting qubits lie in the microwave regime. Hence to control and transfer information from a superconducting qubit one might need to use microwave photons. In principle, we can use microwave photons to transfer information from one node to another, but such a transmission process is extremely lossy. Also, such processes have very demanding technical requirements, like the design of specialised Niobium waveguides\index{Niobium waveguides}, maintained at extremely low temperatures. Hence it isn't feasible to use microwave photons for the long-distance transfer of quantum information. Meanwhile, it is well known that photons in the visible spectrum can be transmitted easily using optical fibres, with favourable efficiency.

\begin{figure}[!htbp]
\if 1\doublecol
\includegraphics[clip=true, width=0.475\textwidth]{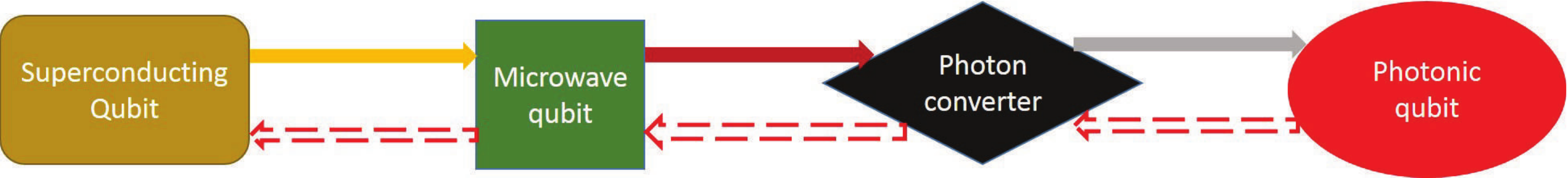}
\else
\includegraphics[clip=true, width=\textwidth]{quantum_transducer}
\fi
\captionspacefig \caption{Block diagram for the quantum transducer. The quantum computer comprising superconducting qubits couples with the microwave qubit. The microwave qubit and the photonic qubit are coupled by a three-level system in which the energy difference between the levels correspond to both microwave and visible optical frequencies. The thick arrows denote the forward process, required to convert the microwave to optical frequencies, and the dashed arrows represent the reverse process, required at the receiving end to convert the optical frequency to a microwave frequency.}\label{fig:quantum_transducer}
\end{figure}

To convert microwave photons to optical photons we can use a quantum transducer. A sketch of a typical design for a quantum transducer\index{Quantum transducers} is shown in Fig.~\ref{fig:quantum_transducer} through a flowchart diagram. The quantum computer is made up of superconducting qubits, which feed quantum information to microwave qubits. The microwave qubits are then interfaced to optical qubits in the visible spectrum through a 3-level quantum system\index{3-level systems}, which can couple at both microwave and optical frequencies. These steps should be reversible. Hence it should be possible to convert the optical qubits back into microwave qubits at the receiving end. 

There are several proposals for quantum transducers in existence and they can be classified into two major classes:
\begin{itemize}
	\item Opto-mechanical \cite{bib:rabl2010quantum, bib:barzanjeh2011entangling, bib:bochmann2013nanomechanical, bib:didier2014quantum, bib:schuetz2015universal, bib:shumeiko2016quantum, bib:stannigel2010optomechanical}.
	\item Spin-ensembles \cite{bib:imamouglu2009cavity, bib:blum2015interfacing}.
\end{itemize}

\begin{figure}[!htbp]
\if 1\doublecol
\includegraphics[clip=true, width=0.475\textwidth]{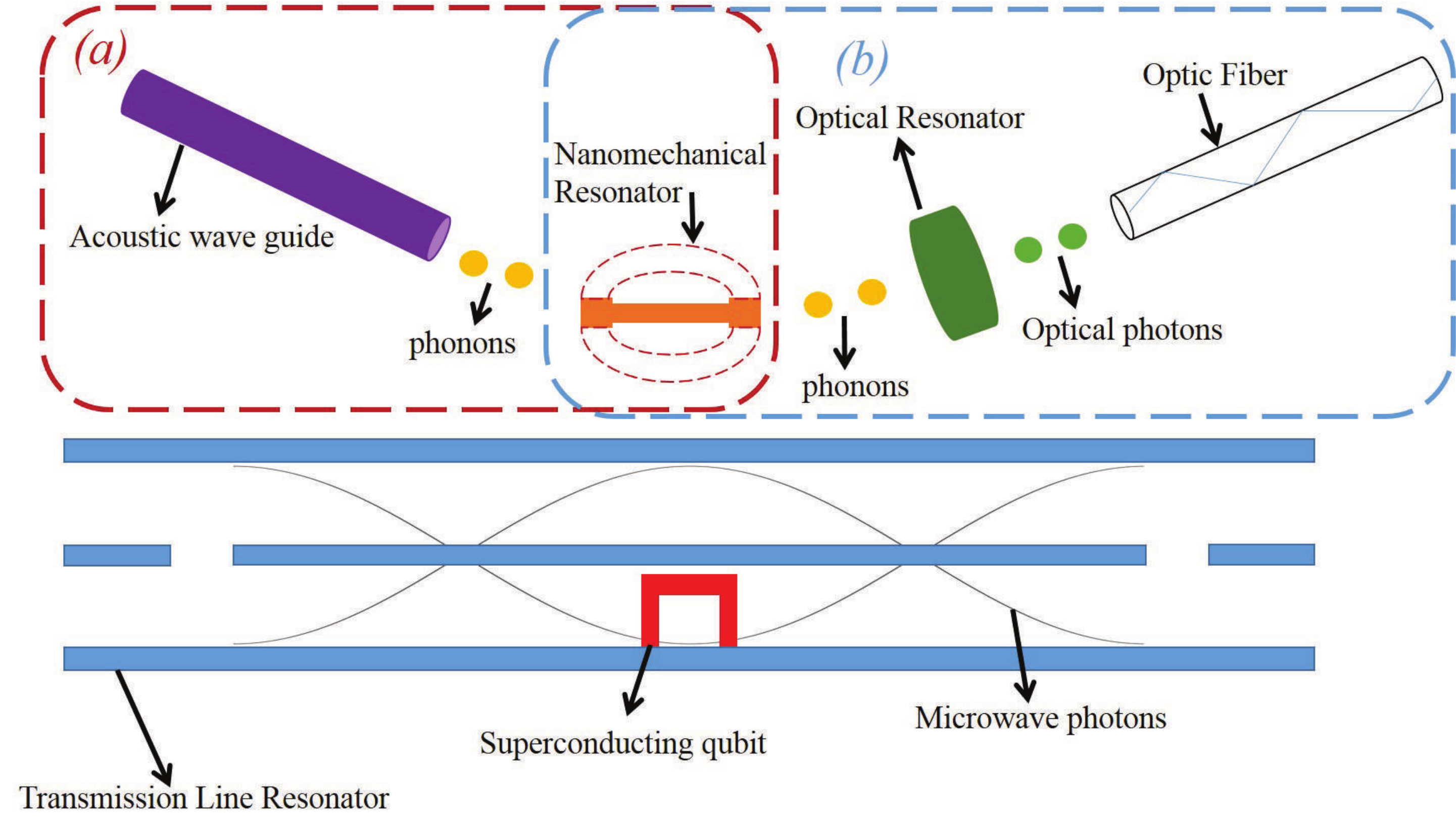}
\else
\includegraphics[clip=true, width=\textwidth]{transducer_scheme_1}
\fi
\captionspacefig \caption{Scheme for an opto-mechanics-based quantum transducer. There are two possible ways of converting microwaves. The final steps for the process in which acoustic modes are transported using waveguides is enclosed by a brown coloured box with dashed outlines labeled (a). Similarly the blue coloured box labeled (b) represents the scheme where phonons are converted to optical photons which are then transmitted via optical fibres. The initial few steps consisting of the superconducting qubit, transmission line resonator and microwaves are common to both processes.}\label{fig:opto_mechanics_transducer}\index{Opto-mechanical quantum transducer}
\end{figure}

The opto-mechanical quantum transducer\index{Opto-mechanical quantum transducer} (see Fig.~\ref{fig:opto_mechanics_transducer}), as the name suggests, combines optical components with a nano-mechanical resonator and converts the microwave photon\index{Phonons} into a phonon (acoustic) mode. The acoustic mode is then transmitted via waveguides\index{Waveguides}. Since we need to fabricate waveguides with very high precision to transmit phonon modes, this scheme is not suitable for communicating between two distant quantum computers. A coupling of the phonon mode to the optical photon mode in the visible region was suggested to enable long-distance transfer of quantum information.

The Hamiltonian of an opto-mechanical quantum transducer, which converts a microwave photon to optical photon using an intermediate nano-mechanical resonator\index{Nano-mechanical resonator} reads,
\begin{align}
\hat{H} &= \hbar \omega_{1} \, \hat{a}_{1}^{\dag} \hat{a}_{1} + \hbar \omega_{2} \, \hat{a}_{2}^{\dag} \hat{a}_{2} \nonumber\\
&+ \hbar \Omega \, \hat{b}^{\dag} \hat{b} + \hbar \, g \, (\hat{b}+\hat{b}^{\dag}) (\hat{a}_{2}^{\dag} \hat{a}_{1} + \hat{a}_{1}^{\dag} \hat{a}_{2}),
\end{align}
where $\omega_{1}$ ($\omega_{2}$) is the frequency of the microwave (optical) photon, and $\Omega$ is phonon frequency. The operators $\hat{a}_{1}$ ($\hat{a}_{1}^{\dag}$) and $\hat{a}_{2}$ ($\hat{a}_{2}^{\dag}$) denote the annihilation (creation) operators corresponding to the microwave and optical photons respectively. Meanwhile, $\hat{b}^{\dag}$ ($\hat{b}$) denotes the phonon creation (annihilation) operator corresponding to the phonons. The factor $g$ is the coupling-strength between the microwave, phonon and photon modes. This design for a quantum transducer is widely preferred, since the optical photons in the visible spectrum can be transmitted over long distances using fibre optics. But the scheme requires two intermediate conversions, each of which reduces overall efficiency.

\begin{figure}[!htbp]
\if 1\doublecol
\includegraphics[clip=true, width=0.475\textwidth]{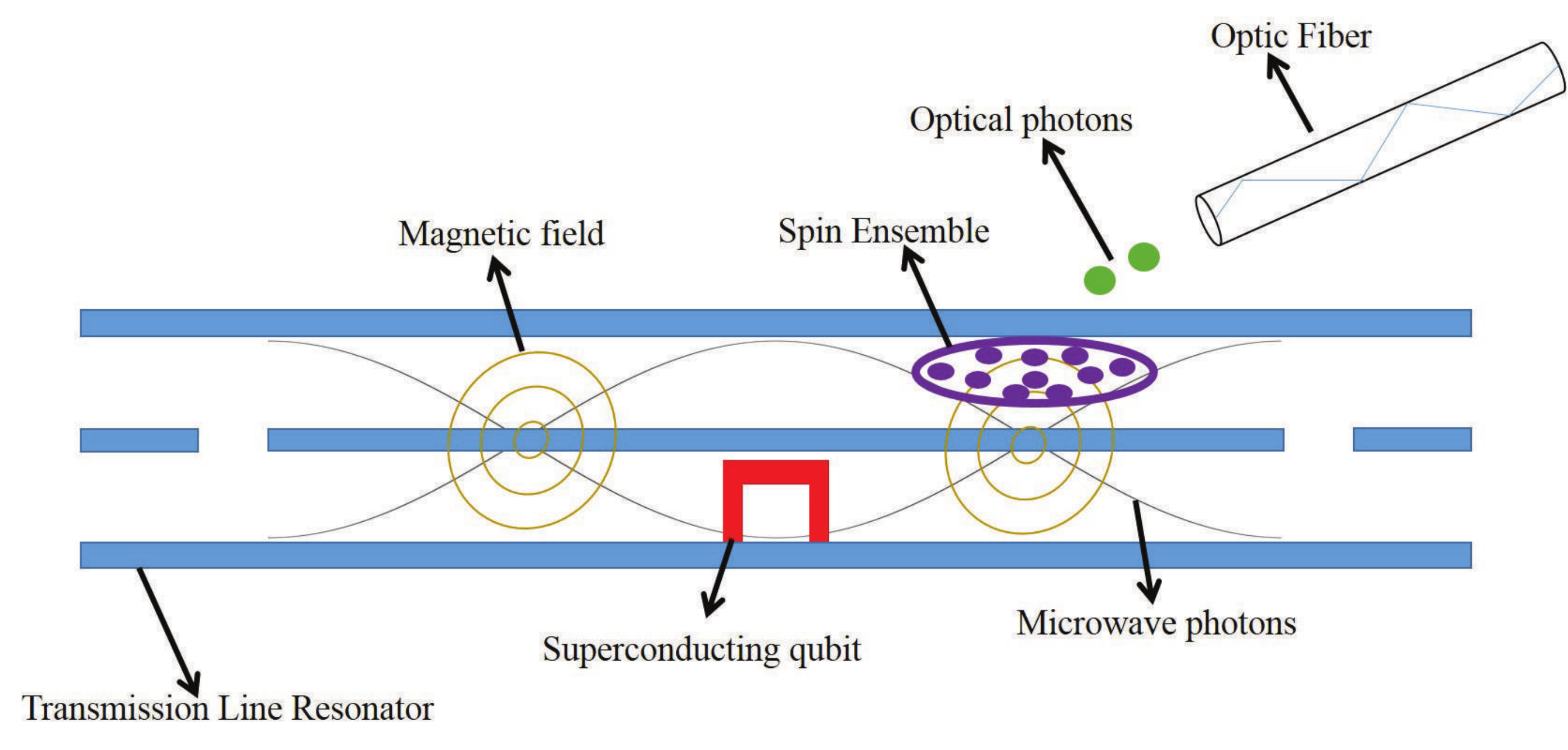}
\else
\includegraphics[clip=true, width=0.9\textwidth]{transducer_scheme_2}
\fi
\captionspacefig \caption{The spin-ensemble-based quantum transducer. Microwave photons are converted to optical photons via a spin-ensemble.}\label{fig:spin_ens_transducer}\index{Spin-ensemble quantum transducer}
\end{figure}

The spin-ensemble-based quantum transducer\index{Spin-ensemble quantum transducer} (see Fig.~\ref{fig:spin_ens_transducer}) is an alternative to the opto-mechanical quantum transducer. In this scheme an ensemble of spins interact with microwave qubits via magnetic dipole coupling\index{Magnetic dipole coupling}, while the superconducting qubits interact via electric dipole coupling\index{Electric dipole coupling} with the microwave coupling. The Hamiltonian for such a system is,
\begin{align}
\hat{H} = \hat{H}_{\mathrm{mw}} + \hat{H}_{\mathrm{spin}} + \hat{H}_{\mathrm{opt}},
\end{align}
where,
\begin{align}
\hat{H}_{\mathrm{mw}} &= \hbar \omega_\mathrm{sq} \hat\sigma_\mathrm{sq}^{\dag} \hat\sigma_\mathrm{sq} + \hbar \omega_{\mu} \hat{a}^{\dag} \hat{a} + \hbar g_{\mu} (\hat{a}^{\dag} \hat\sigma_\mathrm{sq} + \hat{a} \hat\sigma_\mathrm{sq}^{\dag}),\nonumber \\
\hat{H}_{\mathrm{spin}} &= \hbar g_\mathrm{s} (\hat\sigma_\mathrm{ba}^{\dag} \hat\sigma_\mathrm{ba} + \hat\sigma_\mathrm{bs}^{\dag} \hat\sigma_\mathrm{bs}), \nonumber\\
\hat{H}_{\mathrm{opt}} &= \hbar g_\mathrm{ab} (\hat\sigma_\mathrm{ba}^{\dag} \hat{c} + \hat{c}^{\dag} \hat\sigma_\mathrm{ba}).
\end{align}

The factor $\omega_\mathrm{sq}$ is the frequency of the superconducting qubit, and $\hat\sigma_\mathrm{sq}^{\dag}$ and $\hat\sigma_\mathrm{sq}$ are the raising and lowering operators\index{Raising operators}\index{Lowering operators} corresponding to the superconducting qubits. The frequency of the microwaves is given by $\omega_{\mu}$, and the $\hat{a}^{\dag}$ and $\hat{a}$ are the creation and annihilation operators for the microwave photons. The factor $g_{s}$ is the coupling strength between the various levels of the spin, and $\hat\sigma_\mathrm{ba}$ and $\hat\sigma_\mathrm{bs}$ are the spin operators corresponding to the transition between the level $a$, $b$ and $s$. Finally, the coupling strength of the spin interaction with the photon is denoted by $g_\mathrm{ab}$, and $\hat{c}^{\dag}$ ($\hat{c}$) is the creation (annihilation) operator corresponding for the photon. Again this is a two-step process, which is in addition beset with the problem of inhomogenous line broadening\index{Inhomogenous line broadening}. The design of an experimental, high-fidelity quantum transducer is still an open and ongoing challenge in the field of quantum technology.

\latinquote{Vir sapit qui pauca loquitur.}

%
% Optical Routers
%

\section{Optical routers} \index{Optical!Routers}

\dropcap{P}{erhaps} the most fundamental building block in any network is routers, devices which switch data packets between multiple inputs and outputs so as to relay them to a destination. Indeed, in many real-world networks, many nodes will purely implement routing, and nothing more elaborate such as computations or other end-user protocols, to be discussed in Part.~\ref{part:protocols}.

We now discuss the implementation of optical routers, beginning with the simplest two-port switch, upon which we build to construct more general and powerful routers.

There are many parameters of interest characterising the operation of optical routes. We will introduce the terminology convention that:
\begin{itemize}
	\item \textit{Ports}\index{Ports}: number of input and output optical modes in a device.
	\item \textit{Channels}\index{Channels}: number of simultaneous communications streams running in parallel through the device.
	\item \textit{Optical depth}\index{Optical!Depth}: number of primitive optical elements/devices an optical path traverses through the course of its trajectory from input to output.
	\item \textit{Directionality}\index{Directionality}: whether information is transferred in one (unidirectional) or two (bidirectional) directions.
	\item \textit{Switching time}\index{Switching!Time}: time for the switch to be reconfigured from one state to another.
	\item \textit{Delay time}\index{Delay time}: time taken by signal to reach the output line of the switch from input.
	\item \textit{Throughput}\index{Throughput}: maximum data rate that can flow through the switch.
	\item \textit{Switching energy}\index{Switching!Energy}: energy input required for activating and deactivating the switch.
	\item \textit{Power dissipation}\index{Power!Dissipation}: power dissipated during the process of switching.
	\item \textit{Insertion loss}\index{Insertion loss}: loss in signal power when the switch is connected.
	\item \textit{Crosstalk}\index{Crosstalk}: coupling to other optical modes.
\end{itemize}

A summary of the routing devices we consider, and their associated resource requirements, is provided in Tab.~\ref{tab:router_summary}.

Of course, real-world routers will not only switch optical paths, but also implement some (probably undesired) quantum processes across those paths, such as a loss channel or temporal mode-mismatch. Thus, proper analysis of optical router performance in quantum networks requires treating them as legitimate nodes in the network graph, with associated costs and attributes, as per the QTCP framework.

\startnormtable
\begin{table*}[!htbp]
	\begin{tabular}{|c|c|c|}
		\hline
  		Device & Resource requirements & Optical depth \\
  		\hline
  		\hline
  		Two-channel two-port switch & \mbox{$N_\mathrm{bs}=2$}, \mbox{$N_\mathrm{ps}=1$} & \mbox{$d=1$} \\
  		Linear $n$-port multiplexer & \mbox{$N_\mathrm{s}=n-1$} & \mbox{$1\leq d\leq n-1$} \\
  		Pyramid $n$-port multiplexer & \mbox{$N_\mathrm{s}=n-1$} & \mbox{$d=\log_2 n$} \\
    	Single-channel multi-port switch (linear) & \mbox{$N_\mathrm{s}=2n-3$} & \mbox{$2\leq d\leq 2n-3$} \\
  		Single-channel multi-port switch (pyramid) & \mbox{$N_\mathrm{s}=2n-3$} & \mbox{$d=2\,\log_2 n-1$} \\
  		Multi-channel multi-port switch & \mbox{$N_\mathrm{s} = \left\lceil \frac{n^2}{2}\right\rceil - n + 1$} & \mbox{$\left\lceil \frac{n}{2} \right\rceil \leq d\leq n-1$} \\
  		Crossbar switch & \mbox{$N_\mathrm{s}=n^2$} & \mbox{$1\leq d\leq 2n-1$}\\
    	\hline
	\end{tabular}
	\captionspacetab \caption{Summary of different primitives for constructing optical routers. $n$, $N_\mathrm{bs}$, $N_\mathrm{ps}$ and $N_\mathrm{s}$ are the number of input/output ports, beamsplitters, phase-shifters, and two-port switches respectively. $d$ is the optical depth (in units of number of two-port switches). Since all of the multi-port devices are constructed from two-port switches, in all cases \mbox{$N_\mathrm{bs} = 2 N_\mathrm{s}$} and \mbox{$N_\mathrm{ps} = N_\mathrm{s}$}.} \label{tab:router_summary} \index{Optical!Routers}\index{Optical!Depth}\index{Optical!Routers!Resource requirements}
\end{table*}
\startalgtable

%
% Mechanical Switches
%

\subsection{Mechanical switches}\index{Mechanical switches}

Most obviously, optical switching could be performed mechanically, by physically displacing fibre endpoints, directing them towards different routes\footnote{Remember, the telephone network used to be mechanically routed by human switchboard operators, manually routing point-to-point connections!}. Such switches have found use in other areas, but are not particularly appropriate for quantum information processing applications, as they are extremely slow compared to electro- or acousto-optic technologies. Certainly, mechanical switching would not be applicable to optical fast-feedforward, such as that required by optical quantum computing, on the order of nanoseconds.

A second disadvantage of mechanical switches is that the introduction of moving parts into quantum optics protocols makes optical stabilisation extremely challenging. The mechanical control required to preserve wavelength-level coherence, for example, is effectively ruled out by moving mechanical parts.

%
% Interferometric Routers
%

\subsection{Interferometric switches} \index{Interferometric!Switches}\label{sec:interfer_switches}

\sectionby{Rohit Ramakrishnan}\index{Rohit Ramakrishnan}

Interferometric routers are based on the principle that the evolution implemented by interferometers are in general highly dependent on the phase relationships within them. This reduces the seemingly uphill task of high-speed, dynamic switching between modes to the problem of implementing dynamically-controllable phases. Thankfully there are a number of techniques for implementing such phase-switching. We will discuss these phase-modulation techniques, before moving onto combining them into more complex routing systems.

A phase modulator\index{Phase!Modulators} is a classically-controlled device that lets us tune the local phase accumulated by an optical path, ideally over the full range of $\{0,2\pi\}$. These may be implemented in several ways:

%
% Electro-Optic Modulators
%

\subsubsection{Electro-optic modulators} \index{Electro-optic!Modulators}

Electro-optics modulators (EOMs) are based on anisotropic materials\index{Anisotropic materials}, in which the refractive index\index{Refractive index} changes according to an applied electric field. There are two primary variations on this:
\begin{itemize}
	\item Pockel's effect\index{Pockel's!Effect}: a linear electro-optic effect, where the refractive index\index{Refractive index} change is proportional to the applied electric field.
	\item Kerr's effect\index{Kerr's!Effect}: a quadratic electro-optic effect, where the refractive index change is proportional to the square of the applied electric field.
\end{itemize}

These changes in refractive index are typically small, such that the effects are significant over propagation distances larger than the light's wavelength. For example, in a material where the refractive index increases by $10^{-4}$, an optical wave propagating a distance of $10^{-4}$ wavelengths will acquire a phase-shift of $2\pi$.

The refractive index of an electro-optic medium\index{Electro-optic!Medium} is a function $n(E)$ of the applied electric field $E$. This function varies only slightly with $E$, such that using a Taylor series expansion\index{Taylor series} about \mbox{$E=0$} we obtain,
\begin{align}
n(E) = n+a_1E + \frac{1}{2}a_2E^2+\dots.
\end{align}

In a Pockel's medium\index{Pockel's!Medium} this relation becomes (after approximating and simplifying),
\begin{align}
n(E) = n-\frac{1}{2}\chi n^3 E,
\end{align}
where $\chi$ is called the Pockel's coefficient\index{Pockel's!Coefficient} or linear electro-optic coefficient. Typical values of $\chi$ lie in the range $10^{-12}-10^{-10}$mV$^{-1}$. The most common crystals used as the medium for Pockel's cells are NH$_4$H$_2$PO$_4$ (ADP), KH$_2$PO$_4$ (KDP), LiNbO$_3$, LiTaO$_3$, and CdTe.

In a centrosymmetric material or Kerr's medium\index{Kerr's!Medium} this relation becomes (again after approximating and simplifying),
\begin{align}
n(E) = n-\frac{1}{2}\xi n^3 E^2,
\end{align}
where $\xi$ is the Kerr's coefficient\index{Kerr's!Coefficient} or the quadratic electro-optic coefficient. Typical values of $\xi$ lies in the range $10^{-18}-10^{-14}$m$^2$V$^{-2}$.

\begin{figure}[!htbp]
\includegraphics[clip=true, width=0.475\textwidth]{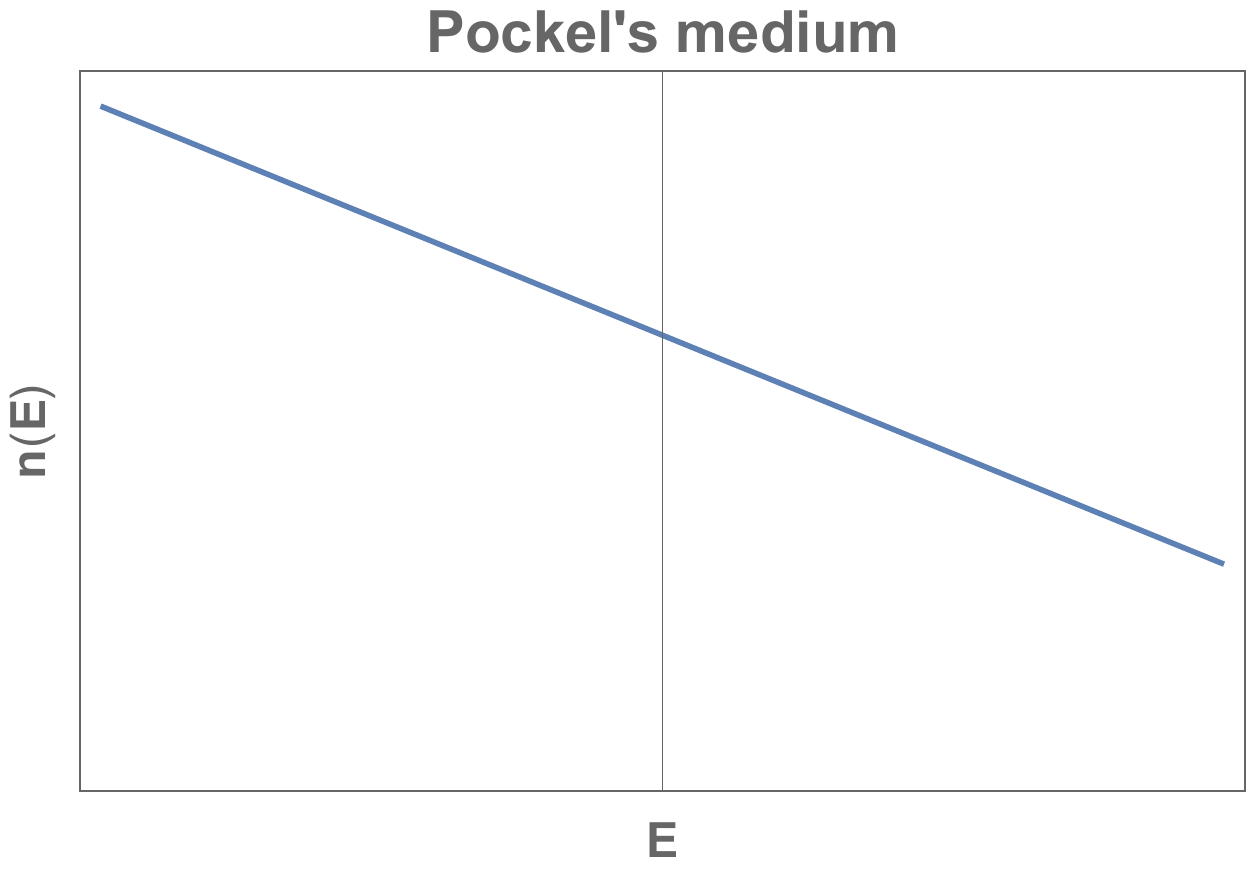} \if 1\doublecol \\ \fi
\includegraphics[clip=true, width=0.475\textwidth]{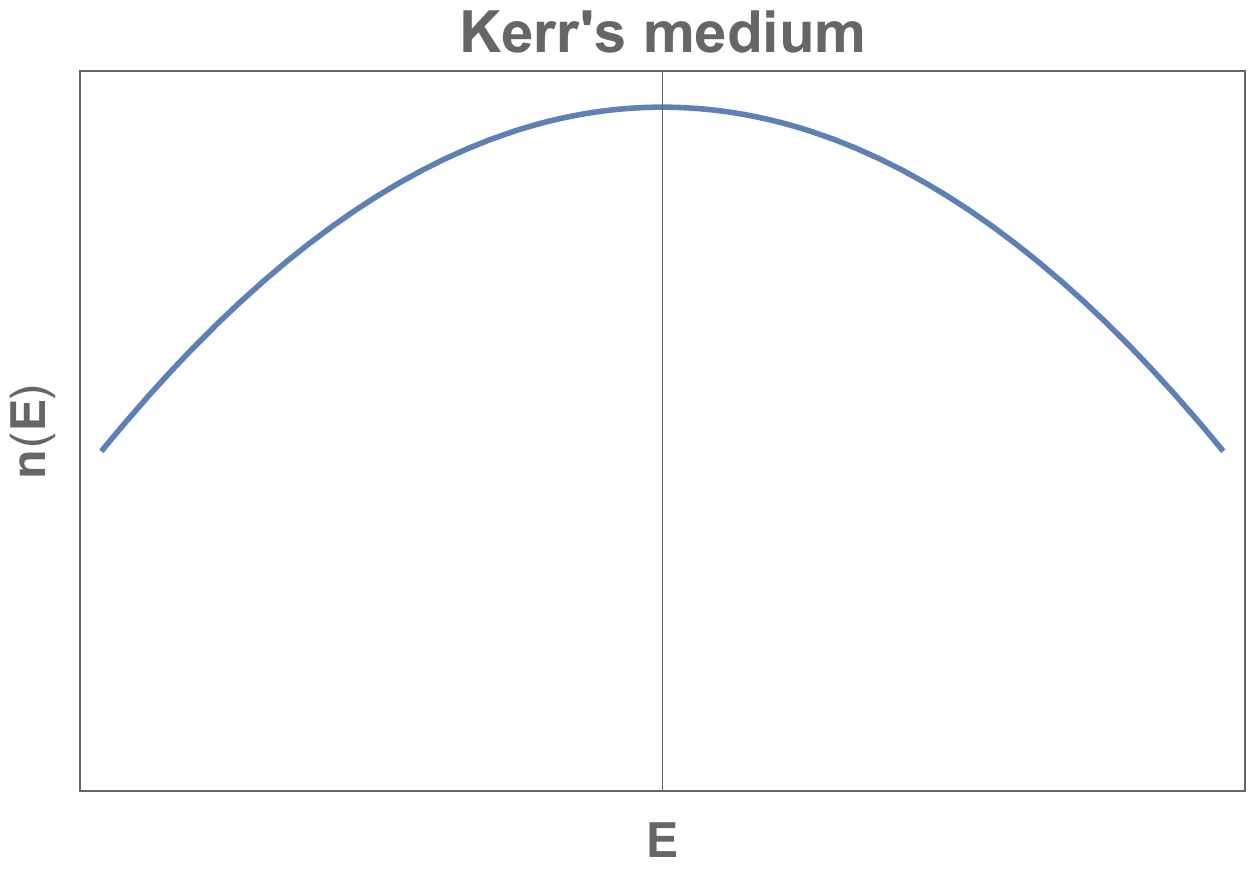}
\captionspacefig \caption{Dependence of refractive index on electric field in: Pockel's medium\index{Pockel's!Medium}, exhibiting linear electric field dependence; Kerr's medium\index{Kerr's!Medium}, exhibiting quadratic electric field dependence. Graphs express qualitative behaviour only, hence no numbers are provided.}\label{fig:EOM_ref_index}
\end{figure}

The refractive index profiles as a function of applied electric field strength for Kerr's and Pockel's mediums are shown in Fig.~\ref{fig:EOM_ref_index}.

Light transmitted through a transparent plate with controllable refractive index undergoes a controllable phase-shift. This plate can be used as an optical phase modulator.

Consider light traversing a Pockets cell of length $L$ to which an electric field $E$ is applied. The phase-shift undergone is given by,
\begin{align}
\phi \approx \phi_0 - \pi\frac{\chi n^3 E L}{\lambda_0},
\end{align}
where,
\begin{align}
\phi_0 = \frac{2\pi nL}{\lambda_0}.
\end{align}

If the electric field generated by applying a voltage $V$ across the faces of the cell of dimension $d$ is,
\begin{align}
	E=\frac{V}{d},
\end{align}
then,
\begin{align}
	\phi=\phi_0-\pi \frac{V}{V_\pi},
\end{align}
where,
\begin{align}
	V_\pi=\frac{d\lambda_0}{L\chi n^3},
\end{align}
is the half-wave voltage\index{Half-wave voltage}, the voltage at which the phase-shift changes by $\pi$.

The electric field is applied either perpendicular (transverse modulators\index{Transverse modulators}) or parallel (longitudinal modulators\index{Longitudinal modulators}) to the direction of the propagation light. The value of the electro-optic coefficient $\chi$ depends on the directions of propagation and the applied field. The speed of operation is limited by the capacitive effects and the transit time of the signal through the material. 

State of the art electro-optic modulators are integrated optic devices based on LiNbO$_3$, in which materials like titanium are used to increase the refractive index. The typical operation speed is above 100GHz. Light signals can be coupled in and out using optical fibres.

%
% Acousto-Optic Modulators
%

\subsubsection{Acousto-optic modulators} \index{Acousto-optic modulators}

\begin{figure}[!htbp]
\includegraphics[clip=true, width=0.35\textwidth]{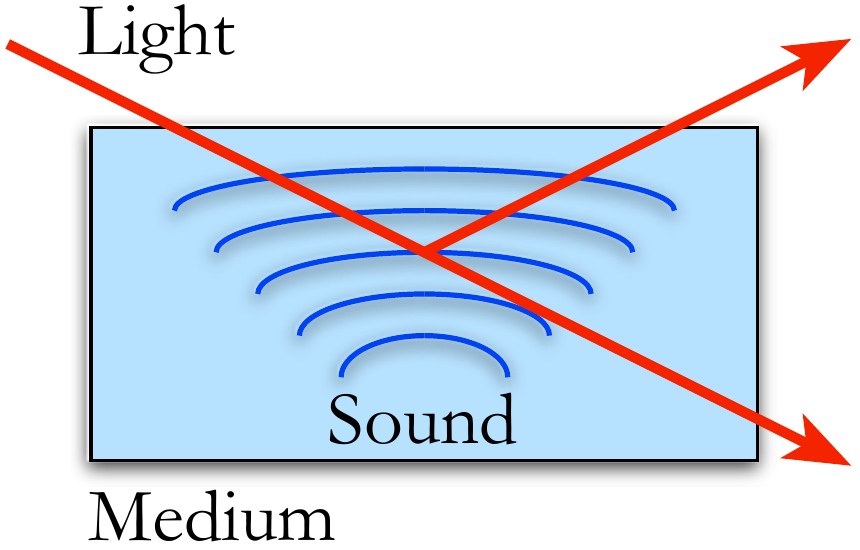}
\captionspacefig \caption{Acousto-optic modulators as a classically-controlled optical switch. The light signal is refracted depending on the applied sound wave.}\label{fig:AOM}
\end{figure}

Sound, or acoustic waves\index{Acoustic waves}, are vibrations that travel through a medium with a velocity characteristic of the medium. This can create perturbations in the refractive index of the optical medium, thus modifying the velocity of light passing through the medium. Thus sound can be used to modify the effect of the medium on light. That is, sound can control the direction of propagation of light. This acousto-optic effect is used to make a variety of devices like optical modulators, switches, deflectors\index{Deflectors}, filters\index{Filters}, isolators\index{Isolators}, frequency shifters\index{Frequency!Shifters} and spectrum analysers\index{Spectrum analysers}. This is shown in Fig.~\ref{fig:AOM}.

\begin{figure}[!htbp]
\includegraphics[clip=true, width=0.25\textwidth]{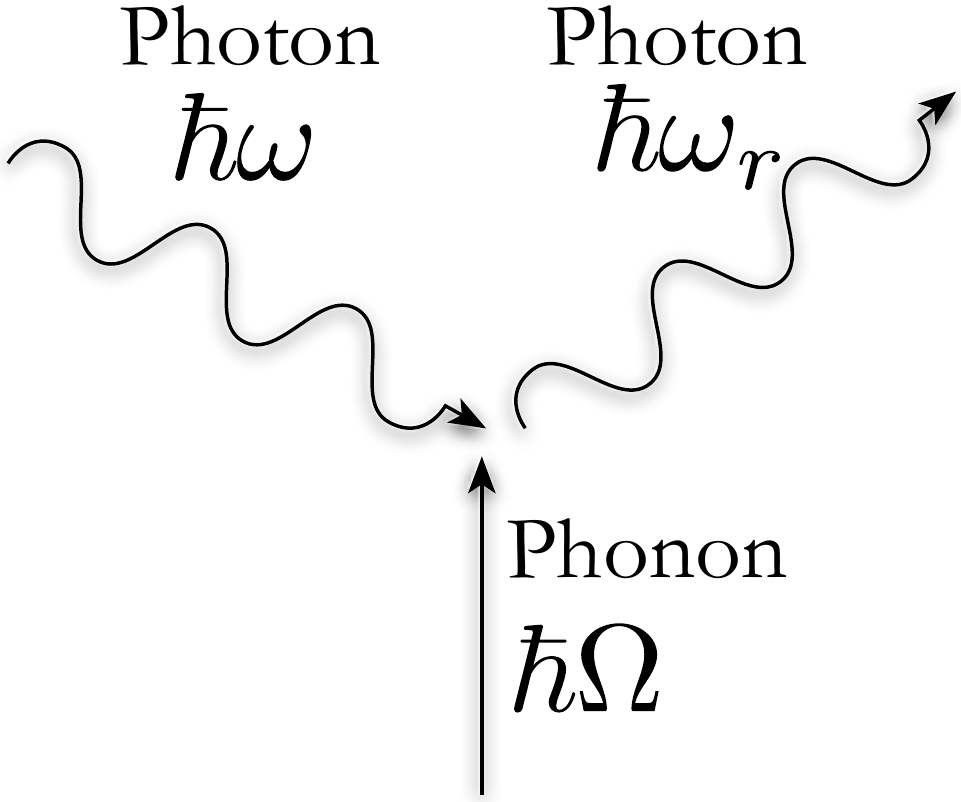}
\captionspacefig \caption{Energy diagram for an acousto-optic modulator, based on Eq.~(\ref{eq:AOM_energy}). Energy and momentum must be conserved from the incident photon of energy $\hbar\omega$ and phonon of energy $\hbar\Omega$, yielding a scattered photon of energy $\hbar\omega_r$.}\label{fig:AOM_energy_diagram}
\end{figure}

According to quantum theory, a light wave of angular frequency $\omega$ and wave-vector $k$ is a stream of photons each with energy $\hbar\omega$ and angular momentum $\hbar k$. Additionally, acoustic waves with frequency $\Omega$ and wave-vector $q$ are a stream of phonons each with energy $\hbar\Omega$ and momentum $\hbar q$. When light and sound interact, a photon combines with a phonon to generate a new photon with energy and wave-vector subject to energy and momentum conservation laws\index{Conservation!Energy}\index{Conservation!Momentum}, 
\begin{align}\label{eq:AOM_energy}
\hbar\omega_r &= \hbar\omega + \hbar\Omega,\nonumber\\
\hbar k_r &= \hbar k + \hbar q.
\end{align}
The associated energy conservation diagram is shown in Fig.~\ref{fig:AOM_energy_diagram}.

Since the intensity of the reflected light is proportional to the intensity of the sound (provided the intensity of sound is low), the intensity of reflected light can be varied proportionally by using an electrically controlled acoustic transducer. This device can be used as a linear modulator of light.

When the acoustic power increases beyond a certain threshold level, total reflection of light occurs whereby the modulator behaves as an optical switch. By switching the sound on and off, the reflected light can be turned on and off, yielding an acoustically-controlled switch.

%
% Magneto-Optic Modulators
%

\subsubsection{Magneto-optic modulators} \index{Magneto-optic modulators}

In the presence of a static magnetic field, certain materials act as polarisation rotators, known as the Faraday effect\index{Faraday effect}. The angle of rotation is proportional to distance and the rotary power\index{Rotary power} $\rho$ (angle per unit length), which is proportional to the component $B$ of the magnetic flux density in the direction of wave propagation,
\begin{align}
	\rho=VB,
\end{align}
where $V$ is known as the Verdet constant\index{Verdet constant}, which is a function of wavelength $\lambda_0$.
 
Examples of materials that exhibit the Faraday effect include glass, Yttrium-iron-garnet (YIG), Terbium-gallium-garnet (TGG) and Terbium-aluminium-garnet (TbAlG).

A simple form of magneto-optic modulator comprises a parallel-sided disk of material placed in a small coil. An alternating current in the coil provides a magnetic field normal to the plane of the disk. The material becomes magnetised in this direction and light propagating through the disk undergoes a polarisation rotation about its plane of polarisation. The modulation of the angle of the plane of polarisation induced by the alternating current may be converted to amplitude modulation\index{Amplitude modulation} by subsequently passing the beam through a polariser. 

%
% Two-Channel Two-Port Switches
%

\subsection{Two-channel two-port switches} \index{Two-channel two-port switches}

\begin{figure}[!htbp]
\includegraphics[clip=true, width=0.475\textwidth]{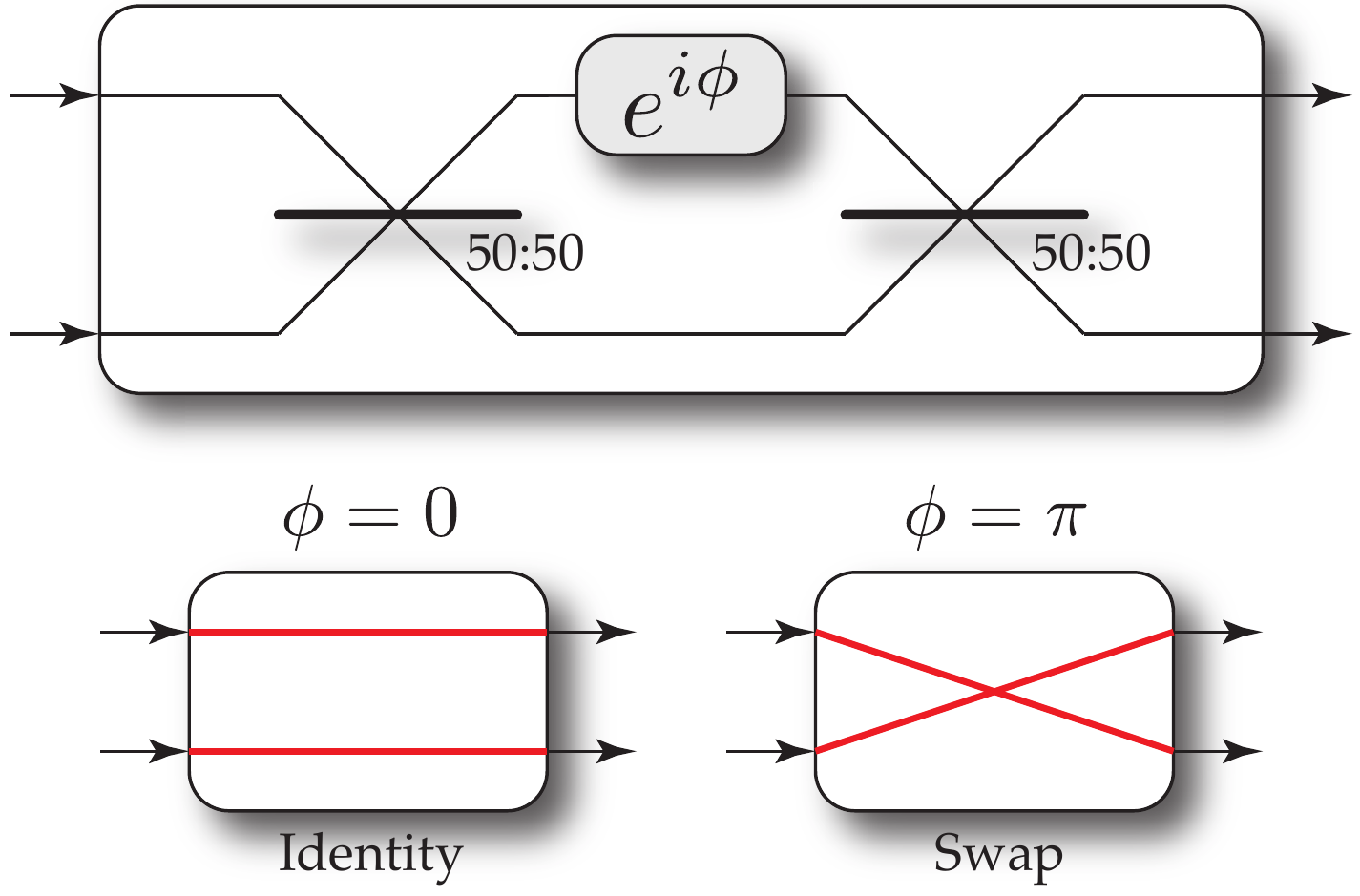}
\captionspacefig \caption{(top) A two-channel two-port switch has two inputs and two outputs, implementing either an identity or swap operation between them. This may be constructed using a Mach-Zehnder interferometer with a variable, classically-controlled phase-shift, $e^{i\phi}$, in one of the arms, which may be implemented using an acousto-optic or electro-optic modulator (AOM or EOM). The phase-shift is allowed to be either \mbox{$\phi=0$} for an identity channel (bottom left) or \mbox{$\phi=\pi$} for a swap operation (bottom right). Because the switch is based on MZ interference, this technique only applies to optical states which undergo MZ interference. The total resource requirements are two 50:50 beamsplitters and a single phase-shifter.} \label{fig:two_channel_two_port_switch} \index{Two-channel two-port switches}\index{Mach-Zehnder (MZ) interference}
\end{figure}

The elementary primitive switch from which more complicated routers may be constructed is the two-channel two-port switch. This switch may be constructed from a Mach-Zehnder interferometer\index{Mach-Zehnder (MZ) interference}, with a classically-controlled phase-shifter in one arm. By switching the phase to either \mbox{$\phi=0$} or \mbox{$\phi=\pi$}, the MZ may be tuned to implement either an identity or swap operation respectively. This is shown in Fig.~\ref{fig:two_channel_two_port_switch}.

In the upcoming diagrams we present, arrows are used to indicate the time-ordering of the flow of data. However, it should be noted that a MZ interferometer is reversible and therefore bidirectional, and so too are all of the more complex routers based upon them.

Because the two-port switch is based upon MZ interference, it will only function for optical states subject to such MZ interference. Thus, single-photons and coherent states are applicable, whereas thermal states, for example, are not.

%
% Multiplexers & Demultiplexers
%

\subsection{Multiplexers \& demultiplexers} \index{Multiplexers}\index{Demultiplexers}

From the two-port switch, which implements a controlled permutation of two optical modes, we can construct multi-port multiplexers and demultiplexers, which controllably route a single input port to one of $n$ multiple output ports, or vice versa.

\begin{figure}[!htbp]
\includegraphics[clip=true, width=0.425\textwidth]{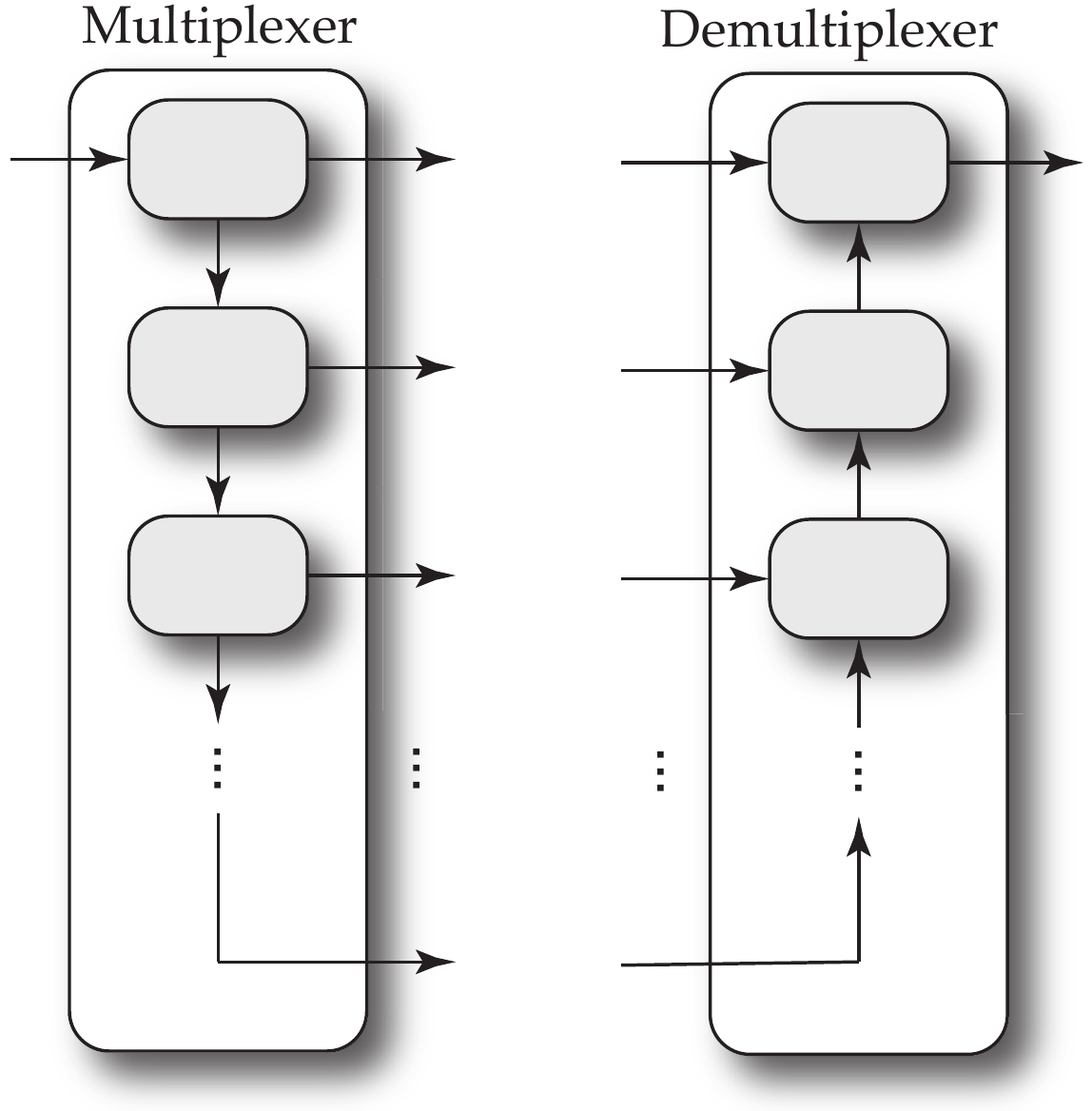}
\captionspacefig \caption{Linear multiplexers (left) and demultiplexers (right) may be constructed from a linear chain of two-port switches (grey boxes), cascading into one another. These switch a single optical channel between $n$ ports. The total resource requirements are \mbox{$n-1$} two-port switches. The optical depth ranges from $1$ (for the first port) to \mbox{$n-1$} (for the final port).} \label{fig:linear_multiplexer} \index{Linear multiplexers \& demultiplexers}
\end{figure}

\begin{figure}[!htbp]
\includegraphics[clip=true, width=0.35\textwidth]{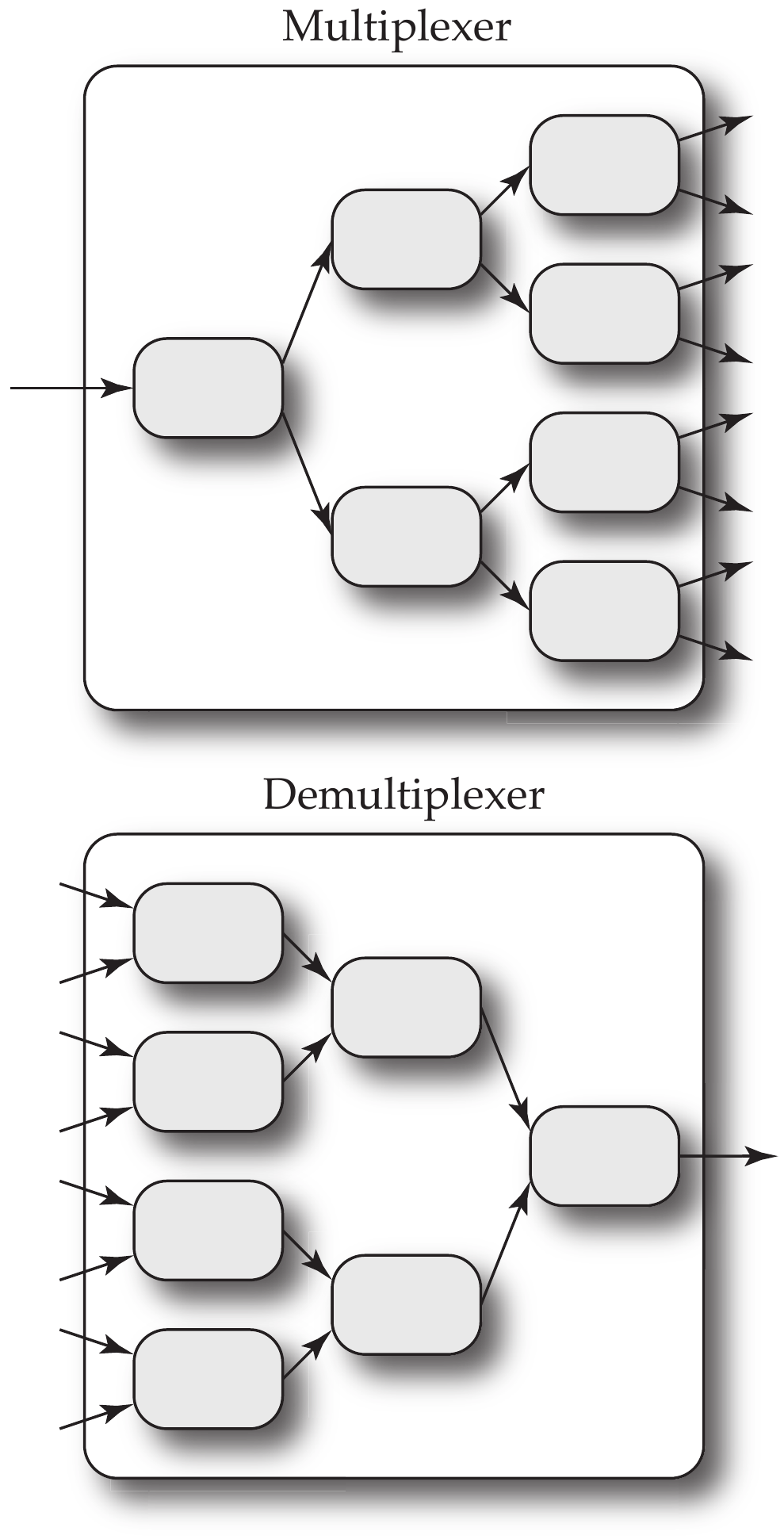}
\captionspacefig \caption{Pyramid multiplexers (top) and demultiplexers (bottom) decompose the multiplexing into a binary tree-structure of two-port switches (grey boxes), shown here for the case of \mbox{$n=8$} ports. For $n$ ports, all optical paths observe an optical depth of \mbox{$d=\log_2(n)$} two-port switches, of which there are \mbox{$n-1$} in total.} \label{fig:pyramid_multiplexer} \index{Pyramid multiplexers \& demultiplexers}
\end{figure}

There are two main architectures that may be employed for implementing such multiplexers/demultiplexers. The first is to use a linear cascade of two-port switches, shown in Fig.~\ref{fig:linear_multiplexer}\index{Linear multiplexers \& demultiplexers}. The second is to use a pyramid cascade, shown in Fig.~\ref{fig:pyramid_multiplexer}\index{Pyramid multiplexers \& demultiplexers}. Both layouts require,
\begin{align}
N_\mathrm{s} = n-1,
\end{align}
two-port switches to implement. However, they differ in one important respect. In the linear multiplexer, different routes experience different optical depth\index{Optical!Depth}, ranging from \mbox{$d=1$} (for the first port) to \mbox{$d=n-1$} (for the final port). This will lead to asymmetry in accumulated errors. In the pyramid multiplexer, on the other hand, all optical paths have the same optical depth, \mbox{$d=\log_2 n$}, yielding completely symmetric operation.

The differing optical depths of linear and pyramid multiplexers lend themselves naturally to different applications. Suppose that in a network a single input-to-output route through a multiplexer is used far more often than the others. In that case, utilising a linear multiplexer will minimise average optical depth since that route can be designated to the first output port, which has an optical depth of only \mbox{$d=1$}. On the other hand, in a very balanced network, in which all optical routes are used roughly uniformly, the average case logarithmic optical depth of the pyramid multiplexer outperforms the average case linear optical depth of the linear multiplexer.

Note that the logarithmic optical depth of the pyramid configuration grows less quickly than the linear average optical depth of the linear configuration. Thus, on average, optical paths pass through fewer optical elements in the pyramid configuration, reducing average accumulated error rates when using noisy optical elements. This, in conjunction with the pyramid's perfect symmetry, makes the pyramid multiplexer configuration generally most favourable.

%
% Single-Channel Multi-Port Switches
%

\subsection{Single-channel multi-port switches} \index{Single-channel multi-port switches}

\begin{figure}[!htbp]
\includegraphics[clip=true, width=0.375\textwidth]{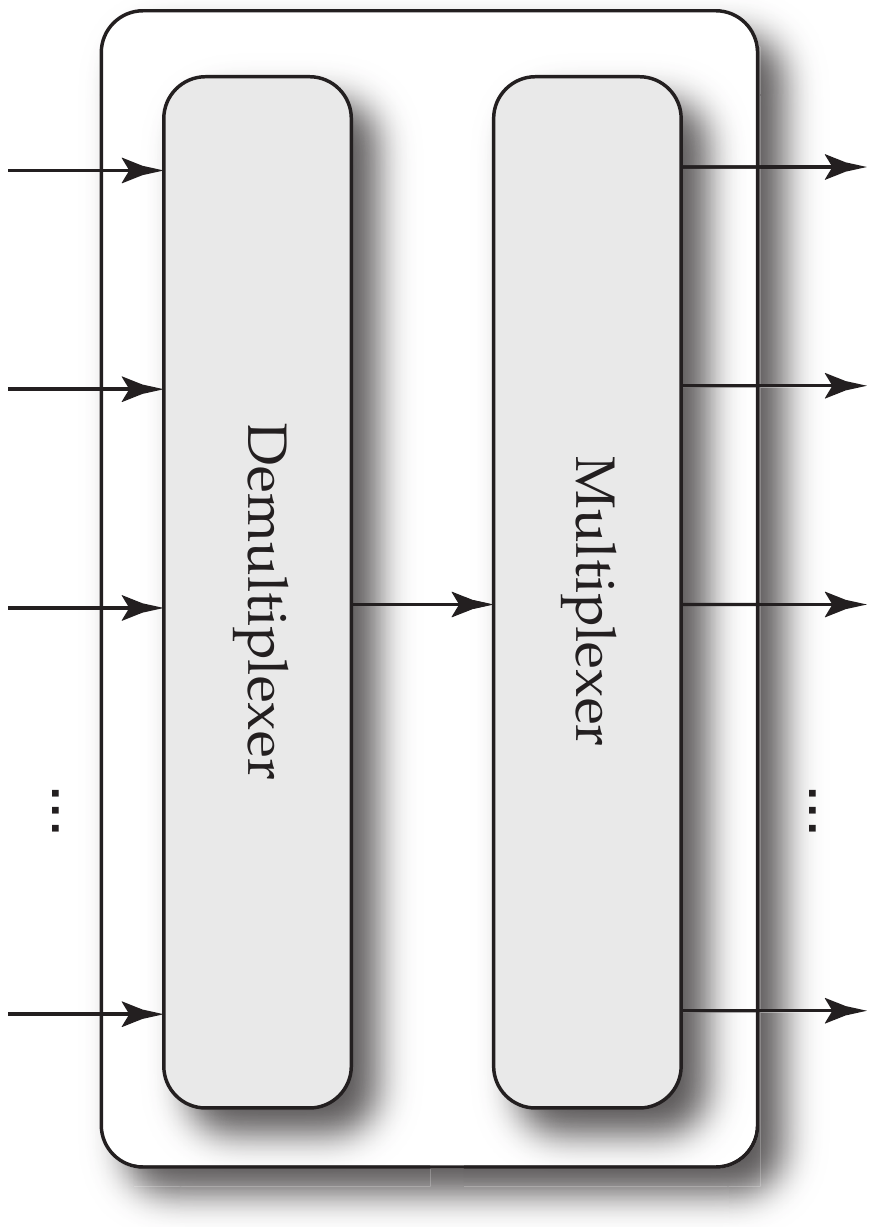}
\captionspacefig \caption{A single-channel multi-port switch may be constructed by demultiplexing the $n$ input ports to a single port, routing the desired input channel to that port, before multiplexing it back out to the desired output port. This allows an arbitrary input to be routed to an arbitrary output, but only one channel at a time. This requires \mbox{$2n-3$} two-port switches in total.} \label{fig:single_channel_multi_port_switch} \index{Single-channel multi-port switches}	
\end{figure}

The multiplexers and demultiplexers route between one port and $n$ ports. In the more general and useful case, we wish to route between $n$ inputs and $n$ outputs. If we only require one active channel at a given time, such a router may be trivially constructed from an $n$-port multiplexer connected to and $n$-port demultiplexer, as shown in Fig.~\ref{fig:single_channel_multi_port_switch}. Here, the demultiplexer chooses one of the input modes to route to its single output, which then feeds into the multiplexer to fan it out to the desired output. The multiplexers/demultiplexers could be implemented using either of the aforementioned layouts, yielding a total resource count of,
\begin{align}
	N_\mathrm{s} = 2n-3,
\end{align}
two-port switches\footnote{Note that the multiplexer and demultiplexer each require \mbox{$2(n-1)$} two-port switches, but one of the central ones adjoining the multiplexer and demultiplexer is redundant and may be eliminated, reducing the number of two-port switches to \mbox{$2n-3$}.}.

%
% Multi-Channel Multi-Port Switches
%

\subsection{Multi-channel multi-port switches} \index{Multi-channel multi-port switches}

The single-channel multi-port switch enables switching between an arbitrary number of input/output ports, but suffers that it can only route a single channel at a time. The most general scenario to consider is multi-channel multi-port switching, which implements an arbitrary permutation between $n$ inputs and $n$ outputs. That is, all $n$ ports may be routing active channels, enabling simultaneous routing of multiple data-flows.

\begin{figure}[!htbp]
\includegraphics[clip=true, width=0.475\textwidth]{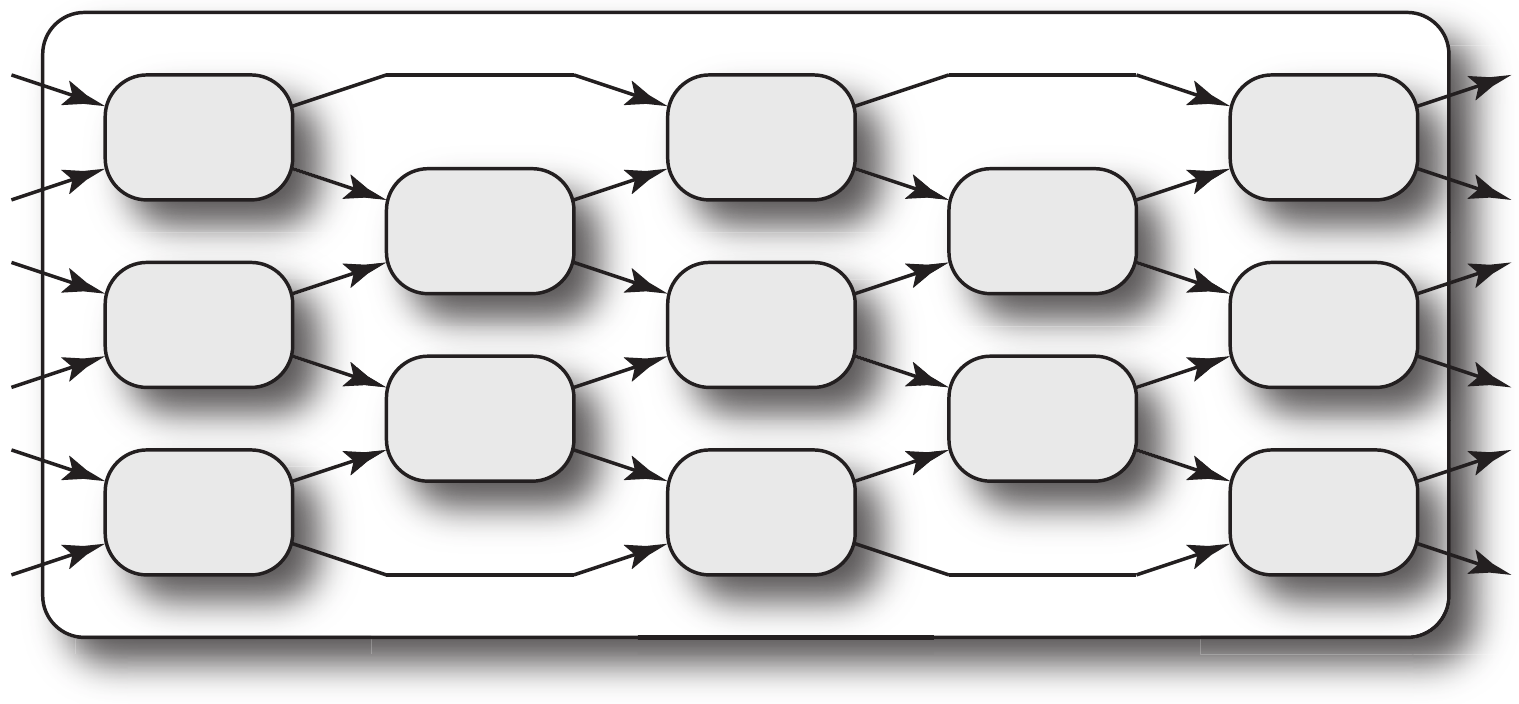}
\captionspacefig \caption{A completely general multi-channel multi-port switch may be constructed using a staggered grid of two-port switches (grey boxes), shown here for \mbox{$n=6$} ports. This allows the implementation of an arbitrary permutation between input and output ports, enabling all $n$ channels to be simultaneously utilised and routed across distinct input-to-output routes. This requires \mbox{$\left\lceil \frac{n^2}{2}\right\rceil - n + 1$} two-port switches in total. Optical depth is approximately equal across all input-to-output paths.} \label{fig:multi_channel_multi_port_switch} \index{Multi-channel multi-port switches}	
\end{figure}

Such a switch may be constructed from a staggered, rectangular lattice of two-port switches, as shown in Fig.~\ref{fig:multi_channel_multi_port_switch}. It is easy to see upon inspection that optical paths exist between every input/output pair of ports. The total resource count for this device is,
\begin{align}
N_\mathrm{s} = \left\lceil \frac{n^2}{2}\right\rceil - n + 1,
\end{align}
two-port switches.

The operation implemented by this device can therefore be expressed as,
\begin{align}
	\begin{pmatrix}
  		\hat{b}^\dag_1 \\
  		\hat{b}^\dag_2 \\
  		\vdots \\
  		\hat{b}^\dag_m
\end{pmatrix}=\hat\sigma \cdot \begin{pmatrix}
  		\hat{a}^\dag_1 \\
  		\hat{a}^\dag_2 \\
  		\vdots \\
  		\hat{a}^\dag_m
\end{pmatrix},
\end{align}
where \mbox{$\hat\sigma\in S_m$} is an arbitrary element of the symmetric group (i.e a permutation matrix), and $\hat{a}_i^\dag$ ($\hat{b}_i^\dag$) are the input (output) photonic creation operators.

Note that this decomposition is more favourable than the completely general Reck \textit{et al.} decomposition presented in Fig.~\ref{fig:LO_archs}(a), since the circuit is balanced, with (almost!) identical optical depths across all input-to-output paths.

%
% Crossbar Switches
%

\subsection{Crossbar switches}\index{Crossbar switches}

\begin{figure}[!htbp]
\includegraphics[clip=true, width=0.4\textwidth]{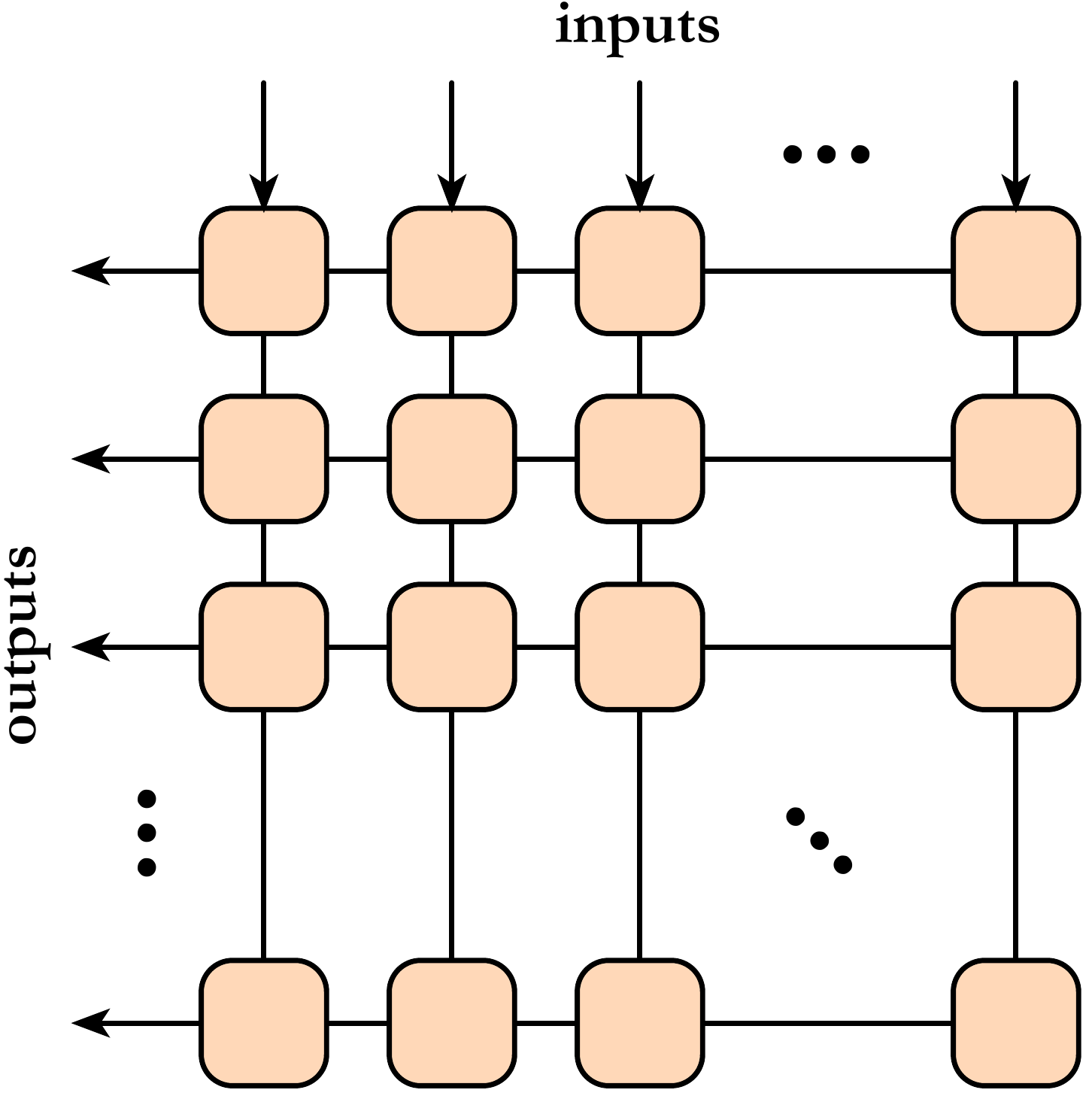}
\captionspacefig \caption{Crossbar architecture for multi-port switching. Each orange box represents a \mbox{$2\times 2$} switch, of any physical implementation. The switching sequence of the constituent two-port switches is defined by a binary \mbox{$n\times n$} permutation matrix, whose elements determine whether a given two-port switch flips modes or doesn't.} \label{fig:crossbar_switch}	
\end{figure}

\begin{figure}[!htbp]
\includegraphics[clip=true, width=0.4\textwidth]{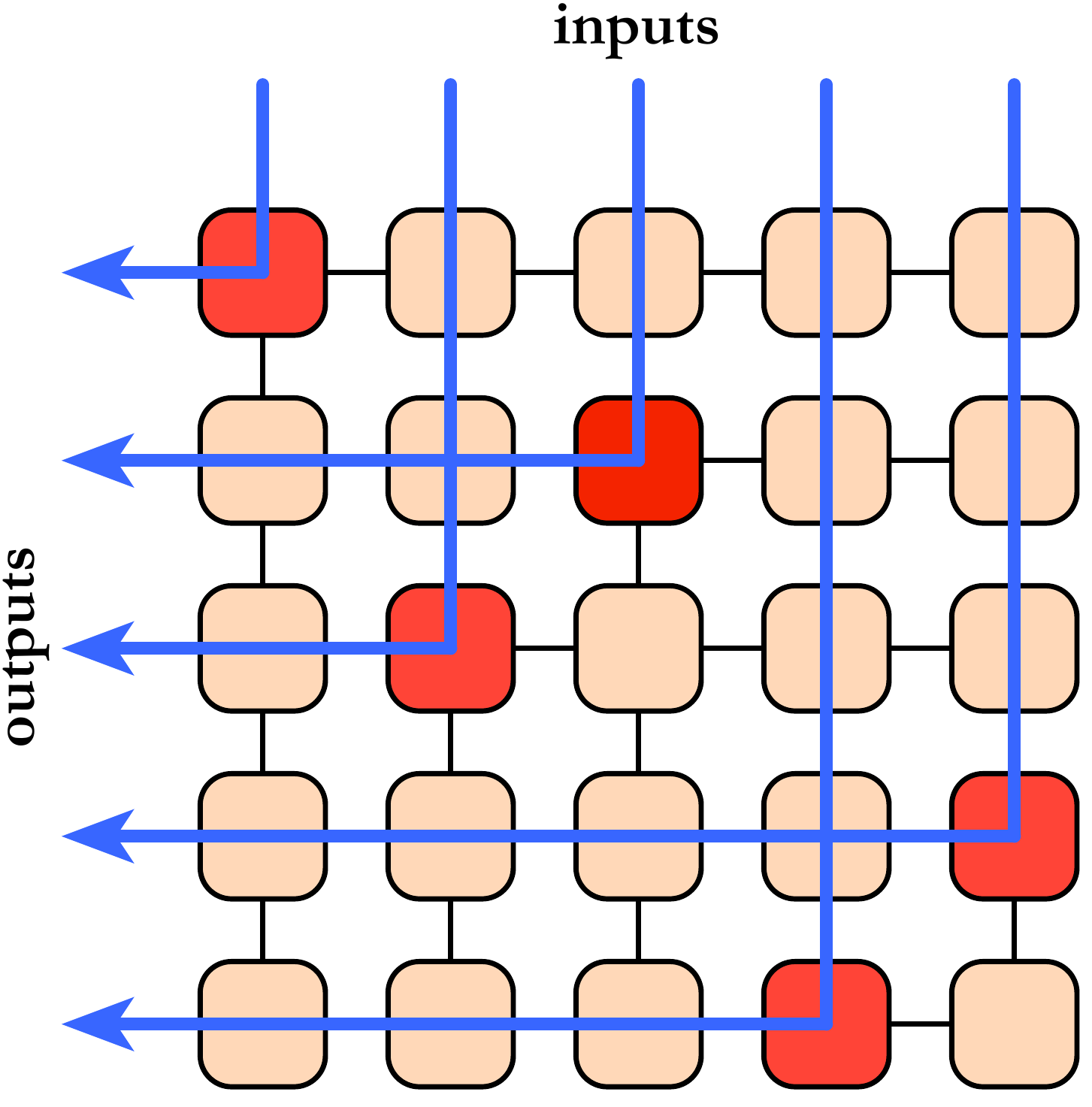}
\captionspacefig \caption{Example switching configuration for a \mbox{$5\times 5$} crossbar switch. The switch colours represent whether the respective \mbox{$2\times 2$} switch is set to flip (red) the modes or not (orange).} \label{fig:crossbar_example}	
\end{figure}

A general multi-port switching architecture, that gained popularity in the early days of channel-switched telecommunications networks\index{Channel-switched networks}, is the crossbar architecture, whereby $n$ inputs are mapped to $n$ outputs via a binary permutation matrix, which controls a lattice of \mbox{$2\times 2$} switches. The general layout of the architecture is shown in Fig.~\ref{fig:crossbar_switch}, and an example of a routing sequence corresponding to a particular binary control matrix is shown in Fig.~\ref{fig:crossbar_example}.

Clearly the scheme requires $n^2$ two-port switches to implement arbitrary \mbox{$n\times n$} mode permutations. The main disadvantage of this scheme is that in general different paths within a given permutation experience differing optical depths, ranging from 1 (best case) to \mbox{$2n-1$} (worst case).

\latinquote{Semper inops quicumque cupit.}

%
% Optical Stability In Quantum Networks
%

\section{Optical stability in quantum networks} \label{sec:opt_stab} \index{Optical!Stability}

\dropcap{G}{iven} that communications links in quantum networks are expected to be optical, an issue of central importance is optical stability when signals from remote sources interfere or interact with local quantum states. For example, in an entanglement swapping protocol (Sec.~\ref{sec:swapping}) forming a part of a quantum repeater network (Sec.~\ref{sec:rep_net}), if the entangling operation between the remotely prepared qubits suffers errors, so too will the prepared distributed entangled state.

If we consider the simplest scenario of employing a PBS (Sec.~\ref{sec:bell_proj}) to implement the entangling operation in the polarisation degree of freedom, photon distinguishability in the form of mode-mismatch (Sec.~\ref{sec:MM_error}) will undermine quantum interference, thereby reducing the entangling power of the gate. Similar observations apply to many other protocols involving entangling measurements, or multi-photon interference more generally.

In present-day laboratories, mode-mismatch and photon-distinguishability can be controlled with exceptionally high fidelity. However, in the networking context this is likely to not be so easy, since perfectly aligning states emanating over long-distance communications channels, which we do not have exquisite control over in a well-controlled laboratory setting, is going to be a somewhat unpredictable and time-varying technological challenge.

Such processes are likely to arise in a multitude of ways, including, but certainly not limited to:
\begin{itemize}
	\item Optical fibre: slight variations in temperatures induce refractive index changes, or changes in physical dimension, resulting in temporal displacements of optical wave-packets.
	\item Satellite: precise knowledge of the distance to a rapidly moving target, at the scale of photon wave-packets, is an extremely daunting prospect.
	\item Free-space (including via satellite): unpredictable temperature and pressure fluctuations in the atmosphere cause unpredictable variations in the speed of light.
\end{itemize}

For these inevitable reasons, it is important to understand the susceptibility of different network protocols to optical stability. 

There are two dominant forms of photonic interference that must be considered, each with quite distinct behaviours under the influence of optical instability. These are:
\begin{itemize}	
	\item Hong-Ou-Mandel (HOM) interference (Sec.~\ref{sec:HOM_inter}):  interference between two distinct photons at a beamsplitter.
	\item Mach-Zehnder (MZ) interference (Sec.~\ref{sec:MZ_inter}): self-interference of a single-photon traversing multiple paths in superposition within an interferometer.
\end{itemize}

%
% Photon Wave-Packets
%

\subsection{Photon wave-packets} \index{Wave-packets}

Before describing optical interference in detail, we must first formalise a definition for the optical wave-packets we will be dealing with. We will assume wave-packets with Gaussian temporal envelope of width $\sigma$ (the coherence length\index{Coherence!Length}), frequency-shifted by some carrier frequency $\omega_0$ (the wavelength\index{Wavelength}).

The temporal distribution function is then,
\begin{align} \label{eq:wavepacket_modulated}
\psi(t) = \sqrt[4]{\frac{2}{\sigma\pi}}e^{-\frac{t^2}{\sigma}-i\omega_0t},
\end{align}
with associated mode-operator $\hat{A}^\dag_\psi$ (Sec.~\ref{sec:spatio_temporal}). This wave-packet is normalised such that,
\begin{align}
|\braket{0| \hat{A}_\psi \hat{A}_\psi^\dag |0}|^2 = \int_{-\infty}^\infty |\psi(t)|^2 \, dt = 1.
\end{align}

\begin{figure}[!htbp]
	\includegraphics[clip=true, width=0.475\textwidth]{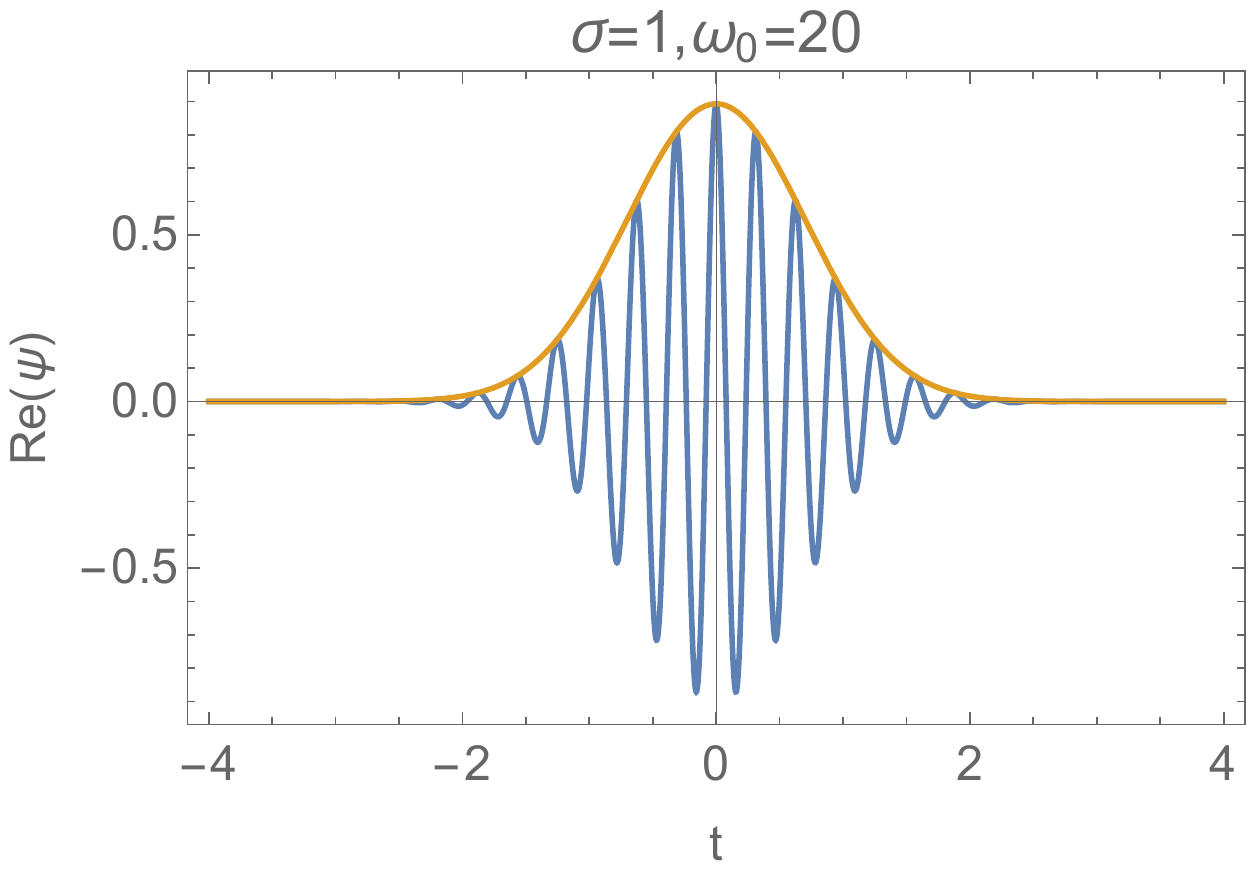} \\
	\captionspacefig \caption{A photonic wave-packet of the form of Eq.~(\ref{eq:wavepacket_modulated}), with Gaussian temporal envelope of width $\sigma$ (orange), shifted by a carrier frequency $\omega_0$ (blue).} \label{fig:HOM_vs_MZ}
\end{figure}

Of course the temporal envelope needn't be Gaussian in general, and could take any other form, subject to normalisation. In Fig.~\ref{fig:HOM_vs_MZ} we illustrate the two main features of this representation: the temporal envelope, and the underlying carrier frequency that it modulates.

In real-world scenarios we are likely to encounter carrier frequencies sufficiently large that oscillations at the carrier frequency level are far more rapid than that of the temporal envelope. For this simple reason, it is to be expected that interference dependent only on $\sigma$ will be far more robust against temporal instability than interference dependent on $\omega_0$.

%
% Mach-Zehnder Interference
%

\subsection{Mach-Zehnder interference} \index{Mach-Zehnder (MZ) interference} \label{sec:MZ_inter}

\begin{figure}[!htbp]
	\includegraphics[clip=true, width=0.475\textwidth]{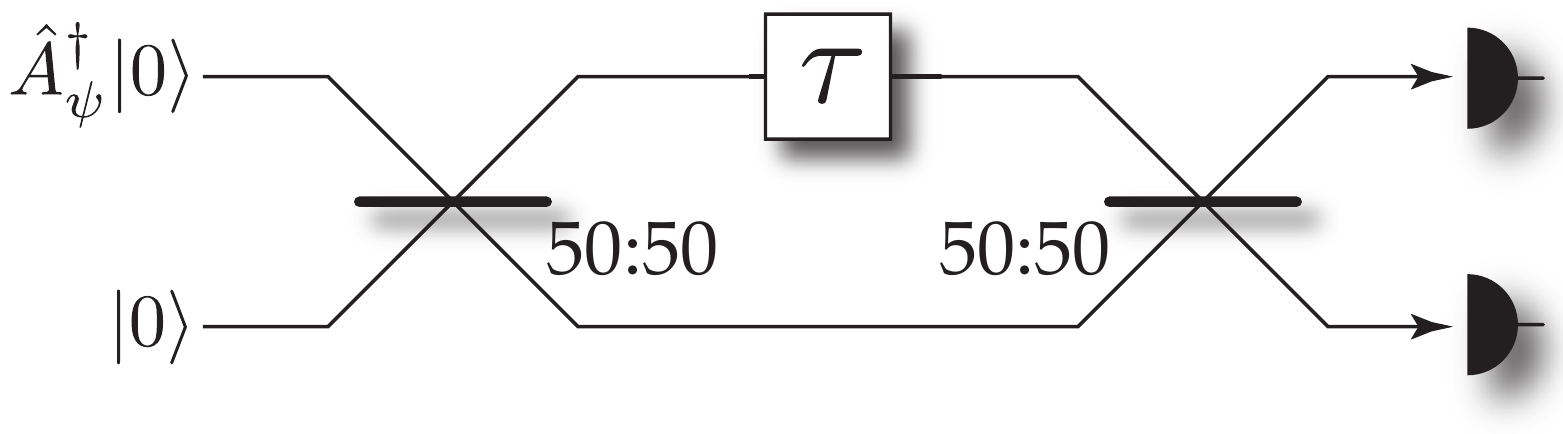} \\
	\includegraphics[clip=true, width=0.475\textwidth]{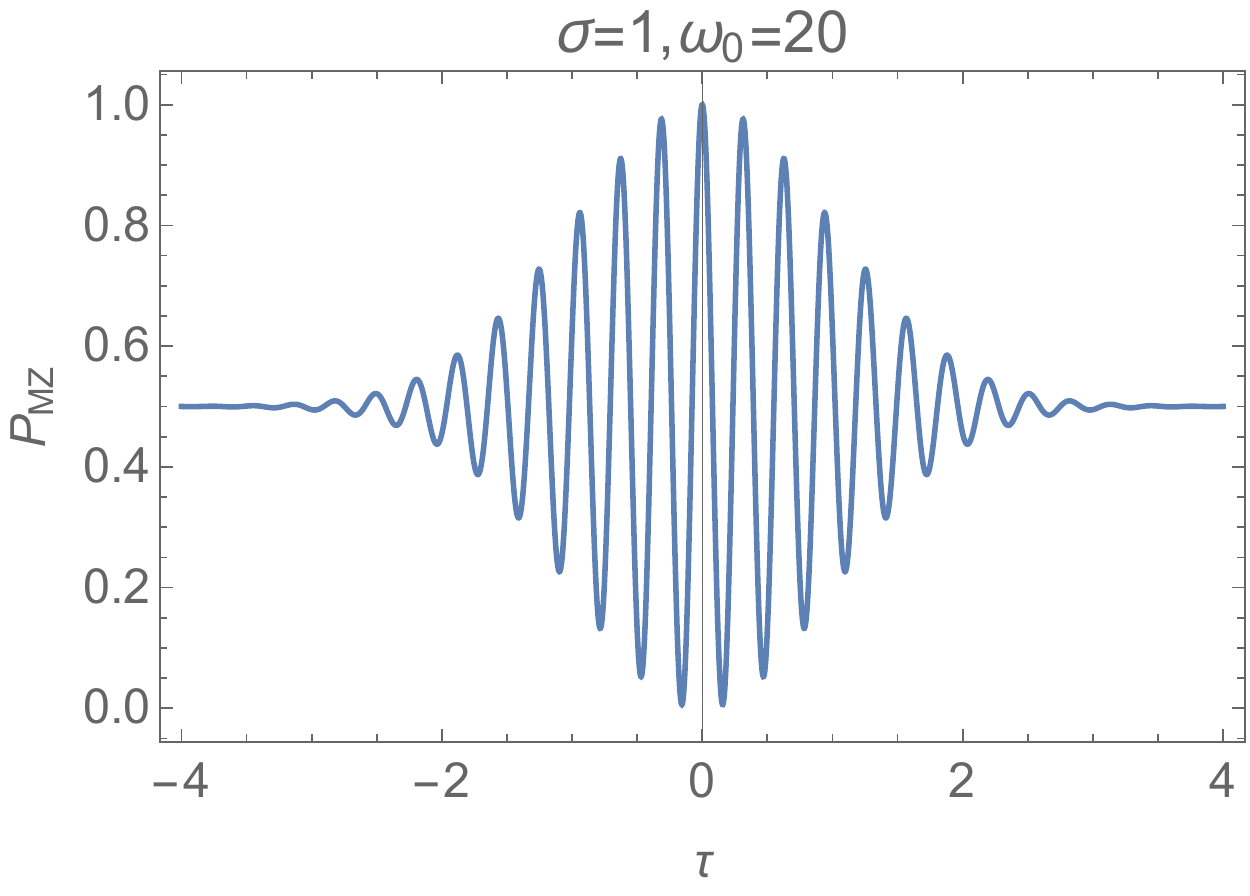}
	\captionspacefig \caption{Mach-Zehnder self-interference of a single photon. (top) Layout of the interferometer. The photon is subject to a time-delay $\tau$ within only the upper arm of a balanced interferometer, comprising two 50:50 beamsplitters. (bottom) Interference fringe $P_\mathrm{MZ}(\tau)$ as a function of the time-delay. The interference is sensitive at the scale of the photon wavelength, (blue) in Fig.~\ref{fig:HOM_vs_MZ}.} \label{fig:MZ_inter}
\end{figure}

Mach-Zehnder (MZ) interference is the interference of a photon or coherent state with itself in a two-mode interferometer constructed from two 50:50 beamsplitters in series, as shown in Fig.~\ref{fig:MZ_inter}(top). This is MZ interference in its simplest form, which can of course be generalised to more complex networks involving self-interference across multiple optical paths.

Within the interferometer is a time-delay, $\tau$, which acts as a temporal mismatch between the two optical paths. 

Let us calculate explicitly the evolution of a single-photon through this device, beginning with a photon described by mode operator $\hat{A}^\dag_\psi$\index{Mode operators} (Sec.~\ref{sec:spatio_temporal}), with the temporal distribution function from Eq.~(\ref{eq:wavepacket_modulated}). We have,
\begin{align}
	\ket{\psi_\mathrm{in}} &= \hat{A}^\dag_\psi \ket{0,0} \nonumber \\
	&\underset{\mathrm{BS}}{\to} \frac{1}{\sqrt{2}} [\hat{A}^\dag_\psi + \hat{B}^\dag_\psi] \ket{0,0} \nonumber \\
	&\underset{\tau}{\to} \frac{1}{\sqrt{2}} [\hat{A}^\dag_{\psi-\tau} + \hat{B}^\dag_\psi] \ket{0,0} \nonumber \\
	&\underset{\mathrm{BS}}{\to} \frac{1}{2} [\hat{A}^\dag_{\psi-\tau} + \hat{B}^\dag_{\psi-\tau} + \hat{A}^\dag_\psi - \hat{B}^\dag_\psi] \ket{0,0} \nonumber \\
	&\underset{\mathrm{PS}}{\to} \frac{1}{2} [\hat{A}^\dag_{\psi-\tau} + \hat{A}^\dag_\psi] \ket{0,0} \nonumber \\
	&= \frac{1}{2} \int_{-\infty}^\infty [\psi(t) + \psi(t-\tau)] \hat{a}^\dag(t)\,dt,
\end{align}
where BS denotes the evolution implemented by a 50:50 beamsplitter, and PS denotes post-selecting upon detecting a single-photon in the first output mode.

We now characterise the operation of the device in terms of the probability of detecting the photon in the first output mode,
\begin{align}
P_\mathrm{MZ}(\tau) &= \frac{1}{4} \int_{-\infty}^\infty |\psi(t) + \psi(t-\tau)|^2 \,dt \nonumber \\
&= \frac{1}{2} \left[ 1 + e^{-\frac{\tau^2}{2\sigma}}\cos(\omega_0\tau) \right].
\end{align}
These dynamics are shown in Fig.~\ref{fig:MZ_inter}(bottom). There are two key features in the behaviour of $P_\mathrm{MZ}(\tau)$. First, there is a slowly varying Gaussian term. Second, the Gaussian term modulates a rapidly oscillating sinusoidial term associated with the carrier frequency. This implies that $\tau$ on the order of the photon's wavelength dominates the measurement dynamics, making it extremely sensitive to temporal instability.

%
% Hong-Ou-Mandel Interference
%

\subsection{Hong-Ou-Mandel interference} \index{Hong-Ou-Mandel (HOM) interference} \label{sec:HOM_inter}

\begin{figure}[!htbp]
	\includegraphics[clip=true, width=0.325\textwidth]{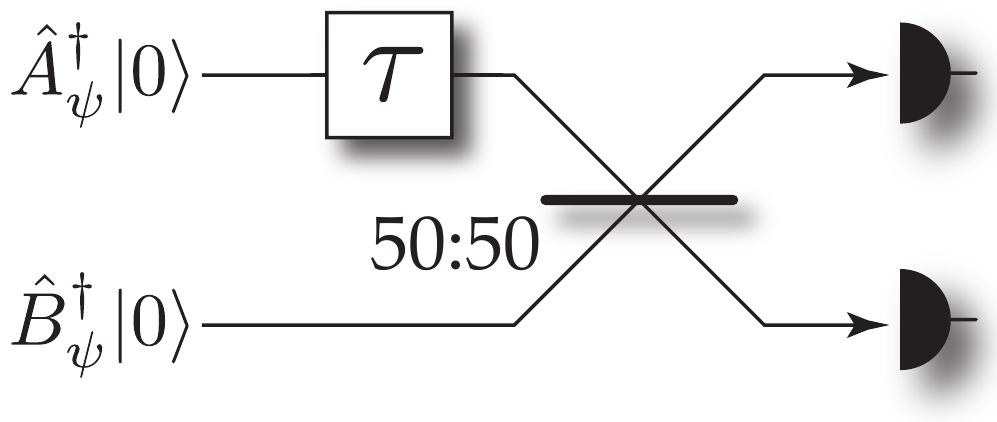} \\
	\includegraphics[clip=true, width=0.475\textwidth]{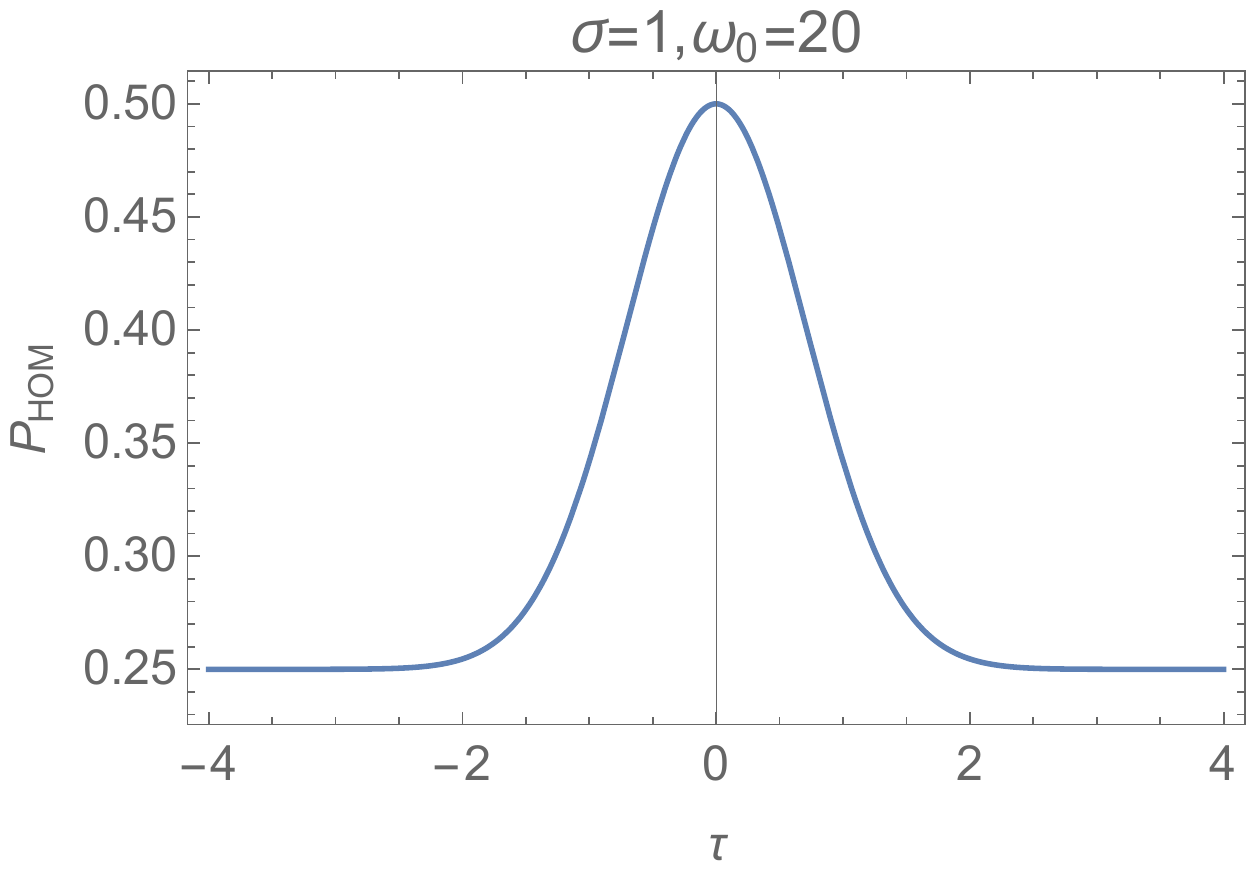}
	\captionspacefig \caption{Hong-Ou-Mandel interference between two independent photons, $A$ and $B$. (top) Layout of the interferometer. Two photons given by mode operators $\hat{A}^\dag$ and $\hat{B}^\dag$\index{Mode operators} (Sec.~\ref{sec:spatio_temporal}), both with temporal distribution function $\psi(t)$, interfere at a 50:50 beamsplitter, where mode $A$ is first subject to a time-delay $\tau$. (bottom) Interference fringe $P_\mathrm{HOM}(\tau)$ as a function of the time-delay. The fringe is only sensitive at the scale of the wave-packet envelope, (orange) in Fig.~\ref{fig:HOM_vs_MZ}.} \label{fig:HOM_inter}
\end{figure}

In Hong-Ou-Mandel (HOM) interference, there is no self-interference as per MZ, but rather interference between two independent but indistinguishable photons. The interference takes place at a single 50:50 beamsplitter, with a temporal delay in one input mode modelling temporal instability. The model is shown in Fig.~\ref{fig:HOM_inter}(top).

Performing the same evaluation of the evolution of the system as before, we obtain,
\begin{align}
	\ket{\psi_\mathrm{in}} &= \hat{A}^\dag_\psi \hat{B}^\dag_\psi \ket{0,0} \nonumber \\
	&\underset{\tau}{\to} \hat{A}^\dag_{\psi-\tau} \hat{B}^\dag_\psi \ket{0,0} \nonumber \\
	&\underset{\mathrm{BS}}{\to} \frac{1}{2} [\hat{A}^\dag_{\psi-\tau} + \hat{B}^\dag_{\psi-\tau}] [\hat{A}^\dag_\psi - \hat{B}^\dag_\psi] \ket{0,0} \nonumber \\
	&\underset{\mathrm{PS}}{\to} \frac{1}{2} \hat{A}^\dag_\psi \hat{A}^\dag_{\psi-\tau} \ket{0,0} \nonumber \\
	&= \frac{1}{2} \int_{-\infty}^\infty \int_{-\infty}^\infty \psi(t)\psi(t'-\tau)\hat{a}^\dag(t)\hat{a}^\dag(t')\,dt\,dt'\ket{0,0}.
\end{align}

We then characterise the operation of the device in terms of the probability of detecting both photons in the first output mode (photon bunching),
\begin{align}
	P_\mathrm{HOM}(\tau) &= \frac{1}{4} \left[1 + \left|\int_{-\infty}^\infty \psi(t)\psi(t-\tau)^*\,dt\right|^2 \right] \nonumber \\
	&= \frac{1}{4}\left[ 1 + e^{-\frac{\tau^2}{\sigma}} \right].
\end{align}
These dynamics are shown in Fig.~\ref{fig:HOM_inter}(bottom). Now, unlike MZ interference, we observe no dependence on the carrier frequency and its associated rapidly oscillating terms. Rather, operation depends only on the temporal envelope, which exists over a far larger time-scale.

Importantly, unlike MZ interference, HOM interference is not applicable to coherent states, which do not entangle or enter into superposition at beamsplitters. The photon bunching effect is unique to single-photons.

The intuition behind the HOM-dip phenomenon is as follows. We know that for identical, indistinguishable photons, an input photon-pair evolves as,
\begin{align}
\ket{1,1} \underset{\mathrm{BS}}{\to} \frac{1}{\sqrt{2}}(\ket{2,0}-\ket{0,2}),
\end{align}
yielding perfect photon-bunching. This bunching effect arises from quantum mechanical interference between the photons. Next imagine that the two photons arrived a long time apart from one another, so long that their wave-packets do not overlap at all. In that instance, the photons do not `see' one another and no quantum interference takes place. Instead, rather than a two-photon quantum interference experiment, we effectively have two independent instances of single-photon experiments, given by,
\begin{align}
\ket{1,0} &\underset{\mathrm{BS}}{\to} \frac{1}{\sqrt{2}}(\ket{1,0}+\ket{0,1}), \nonumber \\	
\ket{0,1} &\underset{\mathrm{BS}}{\to} \frac{1}{\sqrt{2}}(\ket{1,0}-\ket{0,1}).
\end{align}
Note that each of these independent instances obeys the classical statistics of a 50/50 distribution. Combining the two instances using classical probability theory, we now observe a 50\% chance of measuring a coincidence, as opposed to the 0\% chance for true HOM interference.

%
% HOM vs MZ Interference
%

\subsection{HOM vs MZ interference} \index{Mach-Zehnder (MZ) interference}\index{Hong-Ou-Mandel (HOM) interference}\index{Hong-Ou-Mandel (HOM) interference!vs Mach-Zehnder (MZ) interference}

Let us now examine the implications of these different types of interference. The key observation was that MZ is far more sensitive to temporal mismatch than HOM, the former at the scale of the photons' wavelength, the latter at the scale of their temporal envelope, which is far larger.

This leads to the immediate conclusion that network protocols relying on HOM interference will be far more robust against temporal instability than those relying on MZ interference. Realistically, it is to be expected that the latter might be impossibly challenging in many contexts, as wavelength-scale stabilisation over long distances seems implausible.

In Tab.~\ref{table:summary_inter} we summarise the network protocols discussed in Part.~\ref{part:protocols} in terms of the types of interference they rely upon. This creates a picture of which are more realistic from a near-term engineering perspective.

\startnormtable
\begin{table}[!htbp]
	\begin{tabular}{|c|c|}
		\hline
  		\rowcolor{Dandelion} Protocol & Interference type \\
  		\hline
  		\hline
  		\rowcolor{LimeGreen} Cluster state measurement & None \\
   		\rowcolor{LimeGreen} Quantum anonymous broadcasting & None \\
  		\rowcolor{LimeGreen} QKD (BB84, E91) & None \\
  		\rowcolor{LimeGreen} Quantum memory & None \\
  		\rowcolor{LimeGreen} Quantum process tomography & None \\
  		\rowcolor{LimeGreen} Quantum state tomography & None \\
  		\rowcolor{LimeGreen} Random number generation & None \\
  		\rowcolor{LimeGreen} Separable measurements & None \\
  		\rowcolor{LimeGreen} Separable state preparation & None \\
  		\rowcolor{LimeGreen} Optical interfacing & None \\
  		\hline
  		\rowcolor{Apricot} Cluster state preparation (fusion gates) & HOM \\
  		\rowcolor{Apricot} Entanglement purification & HOM \\
  		\rowcolor{Apricot} Entanglement swapping & HOM \\ 
  		\rowcolor{Apricot} Matter qubit entangling operations & HOM \\
  		\rowcolor{Apricot} Partial Bell state measurements & HOM \\
   		\rowcolor{Apricot} Quantum gate teleportation & HOM \\
  		\rowcolor{Apricot} Quantum state teleportation & HOM \\
  		\rowcolor{Apricot} Superdense coding & HOM \\
  		\hline
  		\rowcolor{Lavender} \textsc{BosonSampling} & MZ \\
  		\rowcolor{Lavender} General linear optics networks & MZ \\
  		\rowcolor{Lavender} Universal LOQC (KLM) & MZ \\
  		\rowcolor{Lavender} Quantum metrology & MZ \\
  		\rowcolor{Lavender} Quantum walks & MZ \\
    	\hline
	\end{tabular}
	\captionspacetab \caption{Summary of the major quantum network protocols presented in Part.~\ref{part:protocols}, and their required type of interference.}\index{Interferometric!Requirements}\label{table:summary_inter}
\end{table}

%
% Optical Stabilisation
%

%\subsection{Optical stabilisation} \index{Optical!Stabilisation}

%\comment{To do!}

%\comment{Discussion of both static and dynamic stabilisation}

\latinquote{Dum inter homines sumus, colamus humanitatem.}

\sketch{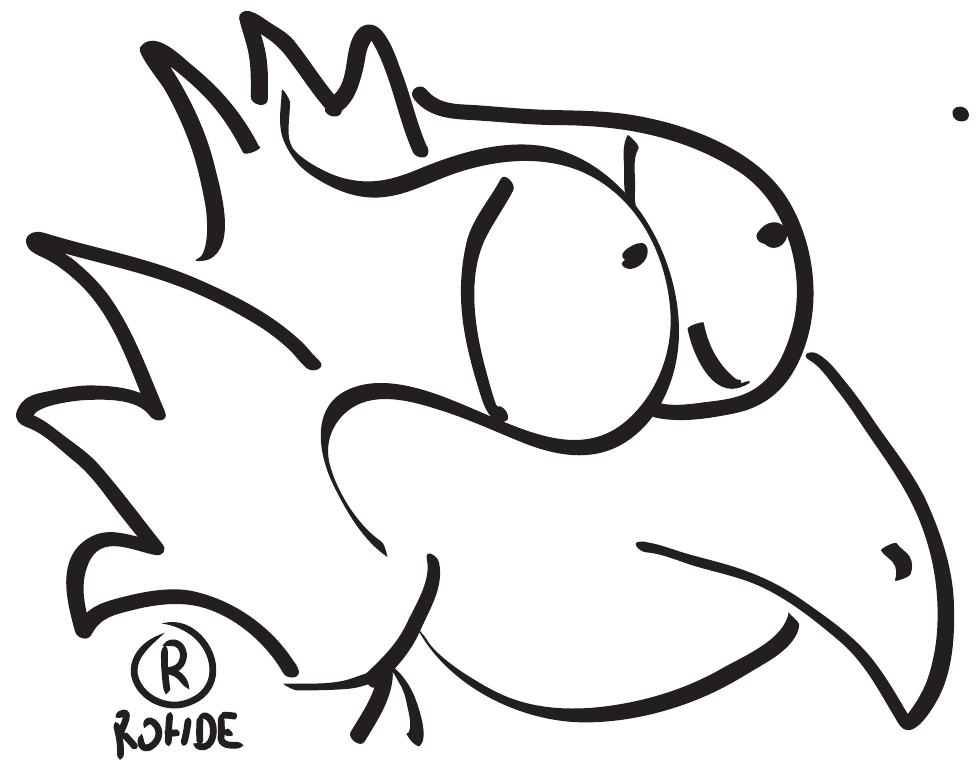}

% %
% % Protocols for the quantum internet
% %

\clearpage

\part{Protocols for the quantum internet}\label{part:protocols}\index{Protocols}
% 
%
% Protocols for the Quantum Internet
%

\dropcap{T}{here} are countless applications for the long-distance communication and processing of quantum data. We will outline some of the most notable examples. Broadly, we will begin with discussion of \textit{low-level protocols} that form the primitives upon which other protocols are built. We will then progressively move towards \textit{high-level protocols}, culminating with full \textit{cloud quantum computing}.

Much of the recent experimental progress in quantum technology has been in the area of low-level protocols, although demonstrations of higher-level protocols are rapidly accelerating.

We keep in mind that although throughout this presentation we have been very quantum computing-centric, quantum computing is not the \textit{only} quantum resource worth communicating. In the same way that \textit{digital assets} encompass a broad range of digital systems and information, any aspect of a quantum system -- from a state, to an operation, storage, to a measurement, or anything else -- could be treated as a \textit{quantum asset}\index{Quantum assets}, which, for generality, we would like our quantum networks to be able to handle.

At the lowest physical level, quantum protocols have in common that they involve state preparation, evolution, and measurement as the fundamental primitives upon which more complex protocols are constructed. We consider these primitive resources in detail, before building upon them to consider some of the major elementary quantum protocols that implement tasks of practical interest. All of those discussed here have been subject to extensive experimental investigation and demonstration, which will be summarised in Part.~\ref{part:SotA}. We treat full quantum computation separately in Secs.~\ref{sec:models_QC} \& \ref{sec:archs_QC}, as this is such an involved topic in its own right.

We will employ circuit model diagrams when describing some protocols. The unfamiliar reader may refer to Sec.~\ref{sec:circuit_model} for a very brief introduction to quantum circuits.

Throughout this section the material will be optics-heavy and does not include discussion of some purely non-optical architectures based on the reasonable assumption that networked quantum protocols will be optically mediated.

%
% State Preparation
%

\section{State preparation} \index{State preparation}

\dropcap{T}{he} first step in any quantum protocol involves the preparation of some kind of quantum state. Some quantum states are easy and cheap to prepare. Others are complex and costly. Thus, the most fundamental quantum asset that a quantum network must handle is the preparation and communication of quantum states.

A state prepared by Bob and sent to Alice might be prepared in isolation, or it might be entangled with a much larger system held by Charlie that Alice does not have full access to. In that case, it would be impossible for Alice to prepare the state on her own, unless she were to first establish a relationship with Charlie. Alternately, maybe Alice just isn't very well-resourced and can't do much on her own. The ability to let someone else prepare her desired quantum states for her would be highly appreciated.

Given the emphasis on quantum optics in quantum networking, it should be noted that optical quantum state engineering has broad applications but can be very challenging in general. Single-photon state engineering, for example, finds ubiquitous applications in quantum information processing protocols, and has become commonplace. Most notably, linear optics quantum computing (Sec.~\ref{sec:KLM_univ}), and some quantum metrology protocols (Sec.~\ref{sec:metrology}) rely on single-photon state preparation. `Push-button' (i.e on-demand) single-photon sources would be a prized asset to many undergraduate experimentalists, were they able to afford them. With access to the quantum internet, they could purchase single photons from another better-resourced lab, with QoS constraints guaranteed by QTCP.

The QTCP protocol is ideally suited to facilitating this kind of transaction. With the use of efficiency and purity cost metrics, QoS guarantees could be established for the efficiency and purity of a licensed single-photon source. In the case of single photons, the dephasing metric is irrelevant, since photon-number states are phase-invariant. This is an elegant example of the value of the versatility of having the QTCP protocol track multiple cost metrics for quantum packets, since different metrics will be of relevance to different messages. Were the message a coherent state, $\ket\alpha$, on the other hand, dephasing would be of utmost importance, whereas loss would be less critical, as lossy coherent states remain as coherent states and retain their coherence.

We see that even the most basic primitive in quantum technologies -- state preparation -- already brings with it much to take into consideration when designing quantum networks. However, the QTCP protocols we described earlier are versatile enough to be capable of mediating their distribution across quantum networks, whilst providing QoS guarantees.

%
% Coherent States
%

\subsection{Coherent states} \label{sec:coherent_states} \index{Coherent states!Preparation}

Coherent states (Sec.~\ref{sec:coherent_state_enc}), although not strictly \textit{quantum} states, nonetheless find broad applications in quantum protocols, for example as the pump for SPDC\index{Spontaneous parametric down-conversion (SPDC)} sources (Sec.~\ref{sec:single_phot_src}), or as a phase-reference\index{Phase!Reference} for homodyne detection (Sec.~\ref{sec:homodyne})\index{Homodyne detection}. Coherent states are rather trivial to prepare, as they are closely approximated by laser sources. Despite this, high quality lasers can nonetheless become very expensive, large, and inaccessible to the not-so-well-resourced end-user. It is not uncommon for laser sources in contemporary labs to be valued in the \$100k's.

%
% Single-Photons
%

\subsection{Single photons} \label{sec:single_phot_src} \index{Single-photons!Preparation}

Single-photon sources (Sec.~\ref{sec:single_phot_enc}) \cite{bib:Oxborrow05} are of particular interest, as a foundational building block in many optical quantum information processing applications, such as linear optics quantum computing (Sec.~\ref{sec:KLM_univ}) and quantum key distribution (Sec.~\ref{sec:QKD}).

The most common approach to preparing single-photon states is via heralded SPDC\index{Spontaneous parametric down-conversion (SPDC)} \cite{bib:URen03, bib:URen05}, whereby a coherent pump source is down-converted into two-mode photon-pairs via a second-order non-linear crystal with interaction Hamiltonian of the form,
\begin{align}
\hat{H}_\mathrm{SPDC} = \xi(\hat{a}_p\hat{a}_s^\dag\hat{a}_i^\dag + \hat{a}_p^\dag\hat{a}_s\hat{a}_i),
\end{align}
where $\xi$ is the interaction strength, and $\hat{a}_p$, $\hat{a}_s$ and $\hat{a}_i$ are the photonic annihilation operators for the pump (input), and \textit{signal} and \textit{idler} (output) modes respectively. This has the clear intuitive interpretation as the coherent exchange of photon-pairs in the output modes with photons in the coherent pump.

\begin{figure}[!htbp]
\includegraphics[clip=true, width=0.45\textwidth]{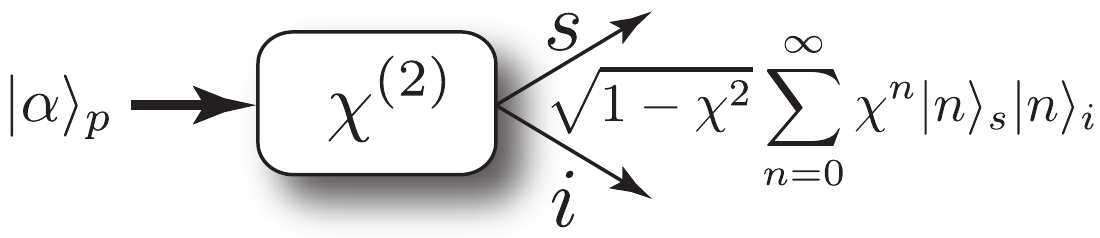}
\captionspacefig \caption{Layout of an SPDC single-photon source. A second-order non-linear crystal is pumped with a coherent state (i.e laser source), yielding a two-mode output state with perfect photon-number correlation between the two modes. Then, post-selecting upon detecting a single photon in one mode in principle guarantees a single photon in the other.} \label{fig:SPDC_source}
\end{figure}

Specifically, a two-mode SPDC state takes the form,
\begin{align}
\ket\psi_\mathrm{SPDC} = \sqrt{1-\chi^2} \sum_{n=0}^\infty \chi^n \ket{n}_s\ket{n}_i,
\end{align}
where $\chi$ is the squeezing parameter, a function of the pump power and properties of the crystal. The layout is shown in Fig.~\ref{fig:SPDC_source}.

Applying the single-photon projector, \mbox{$\ket{1}\bra{1}$}, to the first mode yields the single-photon state in the other, up to normalisation, which reflects the inherent non-determinism. The preparation success probability is derived from the amplitude of the \mbox{$n=1$} term as,
\begin{align} \label{eq:SPDC_p_prep}
P_\mathrm{prep}=\chi^2(1-\chi^2),
\end{align}
assuming ideal photo-detection. Thus, the perfect photon-number correlation in an SPDC state enables heralded preparation of states with exactly one photon in principle.

\begin{figure}[!htbp]
\includegraphics[clip=true, width=0.38\textwidth]{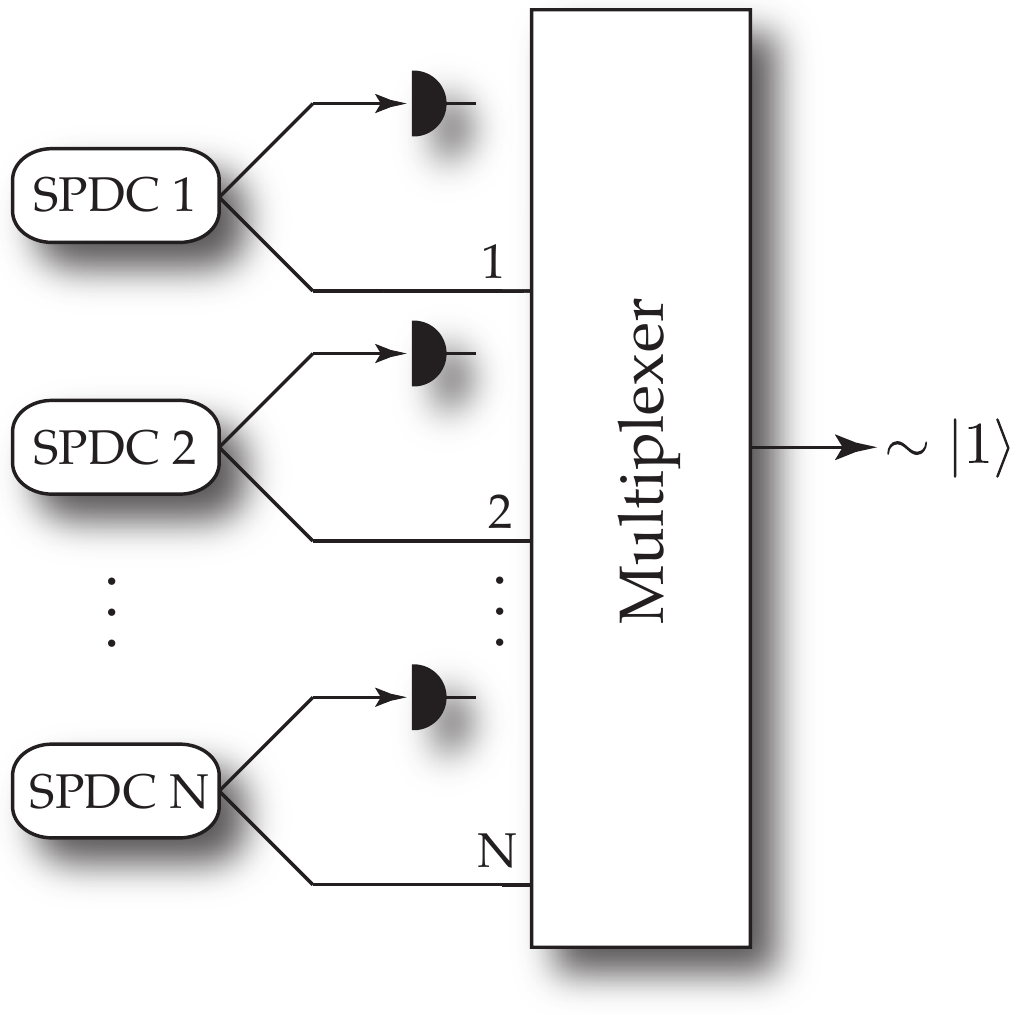}\index{Multiplexed!Single-photon sources}
\captionspacefig \caption{Quasi-deterministic single-photon state preparation using $N$-fold multiplexing of heralded SPDC sources (or any other non-deterministic, but heralded source). All $N$ SPDC sources are triggered simultaneously. The heralding detectors feedforward to the multiplexer, which routes a successfully heralded single-photon state (if there is one) to the output mode. With a sufficiently large bank of sources in parallel, the probability of successfully preparing a single-photon state approaches unity.} \label{fig:SPDC_multiplexing_arch}
\end{figure}

Transitioning from heralded state preparation to quasi-deterministic state preparation may then be achieved by operating a bank of such sources in parallel, and multiplexing their outputs, such that when all sources are triggered simultaneously, if any one succeeds, the respective single photon is routed to the desired output mode, as shown in Fig.~\ref{fig:SPDC_multiplexing_arch}.

\begin{figure}[!htbp]
\includegraphics[clip=true, width=0.475\textwidth]{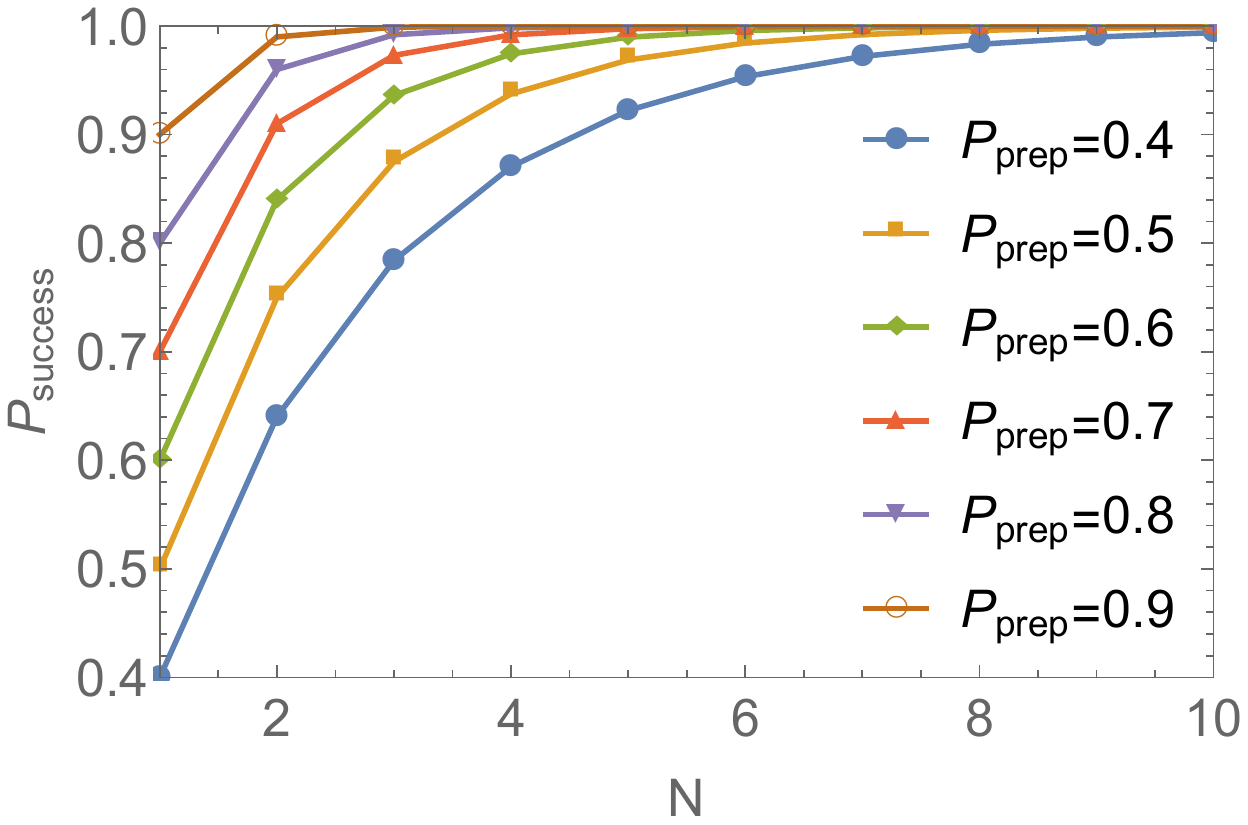}
\captionspacefig \caption{Single-photon state preparation probability, $P_\mathrm{success}$, using $N$-fold multiplexing, where the individual heralded sources have heralding probability $P_\mathrm{prep}$. $P_\mathrm{success}$ always exponentially asymptotes to unity with increasing $N$, for any \mbox{$P_\mathrm{prep}>0$}.} \label{fig:SPDC_multiplexing_plot}
\end{figure}

The success probability of the multiplexed source exponentially asymptotes to unity as the number of in-parallel sources increases,
\begin{align} \label{eq:SPDC_multiplex}
P_\mathrm{success} = 1 - (1-P_\mathrm{prep})^N,
\end{align}
where there are $N$ sources in parallel. This relationship is shown in Fig.~\ref{fig:SPDC_multiplexing_plot} for sources with varying heralding probabilities. This principle could also obviously be applied to any other type of non-deterministic, but heralded source.

\begin{figure}[!htbp]
\includegraphics[clip=true, width=0.4\textwidth]{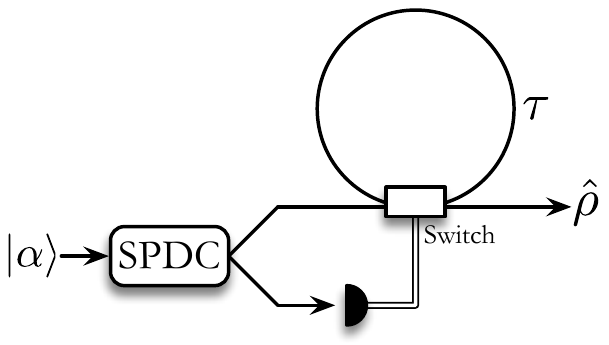}
\captionspacefig \caption{Multiplexed single-photon state preparation in the temporal domain. An SPDC source operating at high repetition rate, with time-bin separation $\tau$, enters a fibre-loop with an in/out coupling switch classically controlled by the heralding outcomes. The fibre-loop acts as a quantum memory, keeping the most recent successfully heralded time-bin in memory until the procedure terminates. The output is a pulse-train where the last time-bin closely approximates a single-photon.} \label{fig:SPDC_time_multiplexing}
\end{figure}

The multiplexing approach needn't be restricted to the spatial domain, but could also be equivalently implemented in the temporal domain \cite{bib:RohdeLoopMulti15}, as shown in Fig.~\ref{fig:SPDC_time_multiplexing}.

Of course, operating a large bank of sources in parallel, along with the associated multiplexing, which requires nanosecond-scale fast-feedforward, is experimentally costly (both in physical size, complexity and dollars), making outsourcing of this technology potentially highly desirable.

This description is purely in the photon-number basis. However, as discussed in Sec.~\ref{sec:spatio_temporal}, photons also have spatio-temporal characteristics. This strongly affects state preparation when using heralded SPDC, particularly state purity, and much effort has been invested into engineering the spectral structure of SPDC states so as to maximise purity and indistinguishability \cite{bib:Aichele02, bib:Branning00}. Specifically, we wish to engineer the photon-pairs to be spectrally separable, such that the heralded photon remains spectrally pure even if the heralding photon was measured with undesirable spectral characteristics (e.g finite resolution).

SPDC is relatively cheap, and widely used, but nonetheless might be out of reach for many end-users, particularly when the previously discussed multiplexing techniques are employed to boost heralding efficiencies. It is quickly being challenged by rival technologies in cutting-edge labs, such as quantum dot sources, which have deterministic, push-button potential \cite{bib:Santori01, bib:Kiraz04}. Techniques based on cavity quantum electrodynamics (QED) \cite{bib:Brattke01} and molecular fluorescence \cite{bib:Brunel99} have also been demonstrated. However, such sources are very much in their developmental stages, and relatively expensive.

Generally speaking, a push-button photon source could be constructed from any two-level system, comprising a ground state, $\ket{g}$, and an excited state, $\ket{e}$, with short lifetime, whereby relaxation via the \mbox{$\ket{e}\to\ket{g}$} transition emits a photon. Then, pumping the system to excite it to the $\ket{e}$ state, and waiting for spontaneous decay yields a single photon.

%
% NOON States
%

\subsection{NOON states} \label{sec:NOON} \index{NOON states!Preparation}

So-called NOON states, path-number entangled two-mode states of the form,
\begin{align}
\ket\psi_\mathrm{NOON} = \frac{1}{\sqrt{2}}(\ket{N,0}+\ket{0,N}),
\end{align}
may be exploited to perform Heisenberg-limited quantum metrology\index{Quantum metrology} (Sec.~\ref{sec:metrology}), allowing extremely precise phase measurement with large photon-number $N$ \cite{bib:Dowling08}. 

The extreme sensitivity of NOON states owes to the $N$-fold enhancement in phase-dependence associated with the $N$-photon component of the superposition. Specifically, if a phase $\phi$ is present in only the second mode, the state evolves to,
\begin{align}
e^{i\phi\hat{n}} \ket\psi_\mathrm{NOON} = \frac{1}{\sqrt{2}}(\ket{N,0}+e^{i\phi N}\ket{0,N}),
\end{align}
where \mbox{$\hat{n}=\hat{a}^\dag\hat{a}$} is the photon-number operator associated with the second mode\index{Photon-number!Operators}. The phase enhancement arises because \mbox{$\hat{n}\ket{N}=N\ket{N}$}. Then, a simple interferometric procedure is able to extract this enhanced phase-dependence as an observable.

However, these states are notoriously difficult and technologically challenging to prepare, and can only be prepared non-deterministically using linear optics \cite{bib:Cable07, bib:PhysRevA.65.030101, bib:PhysRevA.76.063808}. If a remote server had the capacity to prepare such states, they would be in high demand across the globe. Hindering this, NOON states are very fragile creatures. First, they exhibit exponentially increased susceptibility to loss -- loss of just a single photon completely decoheres the state, rendering it useless for metrological purposes. Second, the large photon number, $N$, amplifies unwanted dephasing by a factor of $N$, as discussed in Sec.~\ref{sec:dephasing_error}---the phase sensitivity of the state works against it in this case! These considerations can be readily accommodated for in the QTCP protocol by tracking dephasing and loss metrics of the packets encapsulating the NOON states.

%
% Cluster States
%

\subsection{Cluster states} \index{Cluster states!Preparation}

In addition to the simple single- or two-mode states discussed above, an entire universal quantum computation can be performed using the measurement-based model for quantum computation (explained in detail in Sec.~\ref{sec:CSQC}). An entangled resource, called a \emph{cluster state}, is prepared beforehand, and a particular calculation is then ``carved out'' of the cluster state by performing local measurements in different bases. Parts of the computation may require adapting measurement bases to previous outcomes and performing local feedforward operations, but no further entanglement generation is required.

The beauty of measurement-based quantum computation with cluster states is that there is a natural separation between state preparation and computation, with the preparation stage being far more technologically challenging than the computation stage. Thus, Alice might ask better-resourced Bob to prepare a cluster state and send it to her, at which point she implements the computation herself using only simple single-qubit (local) measurements and feedforward. 

Since cluster states are a computation-based optical resource, their specific structure is determined by the way in which information is encoded. For information encoded in single photons, non-deterministic gates are required to produce a so-called \emph{optical cluster state}, and photo-detection is required to perform computations. A detailed discussion of optical cluster state preparation is presented in Sec.~\ref{sec:CS_LO} with a focus on protocols employing non-deterministic gates. For information encoded into the position and momentum quadratures of the electric field, the associated \emph{CV cluster state}~\cite{menicucci2006universal,walsh2021streamlined} is a multi-mode Gaussian state that can be prepared by sending single-mode squeezed states through a large multi-mode interferometer. The computation is then performed using homodyne detection~\cite{asavanant2021logicalgates}. Other types of encoding lead to other cluster states with their own features and measurements. Notably, optical GKP cluster states can be more challenging to make than others, but they can be made fault tolerant~\cite{fukui2018analog,bourassa2021blueprint,pantaleoni2021hidden,walshe2024totl} and require only homodyne detection to operate.

%
% Greenberger-Horne-Zeilinger States
%

\subsection{Greenberger-Horne-Zeilinger states} \index{Greenberger-Horne-Zeilinger (GHZ) states}\label{sec:GHZ_states}

Another class of states is Greenberger-Horne-Zeilinger (GHZ) states \cite{bib:GHZ89}, which are maximally-entangled states across an arbitrary number of qubits, $n$, of the form,
\begin{align}
\ket\psi_\mathrm{GHZ}^{(n)} = \frac{1}{\sqrt{2}}(\ket{0}^{\otimes n} + \ket{1}^{\otimes n}).
\end{align}

GHZ states are useful for various quantum information processing applications, including quantum anonymous broadcasting (Sec.~\ref{sec:anon_broad}). These states are particularly susceptible to loss, since the loss of a single qubit completely decoheres the state into a perfect mixture of the \mbox{$\ket{0}^{\otimes (n-1)}$} and \mbox{$\ket{1}^{\otimes (n-1)}$} states, with complete loss of entanglement and coherence,
\begin{align}
\hat\rho_\mathrm{GHZ}^\mathrm{loss} &= \mathrm{tr}_1(\ket\psi_\mathrm{GHZ}^{(n)}\bra\psi_\mathrm{GHZ}^{(n)})\nonumber \\
&= \frac{1}{2}(\ket{0}^{\otimes (n-1)}\bra{0}^{\otimes (n-1)}+\ket{1}^{\otimes (n-1)}\bra{1}^{\otimes (n-1)}).
\end{align}

\begin{figure}[!htbp]
	\includegraphics[clip=true, width=0.475\textwidth]{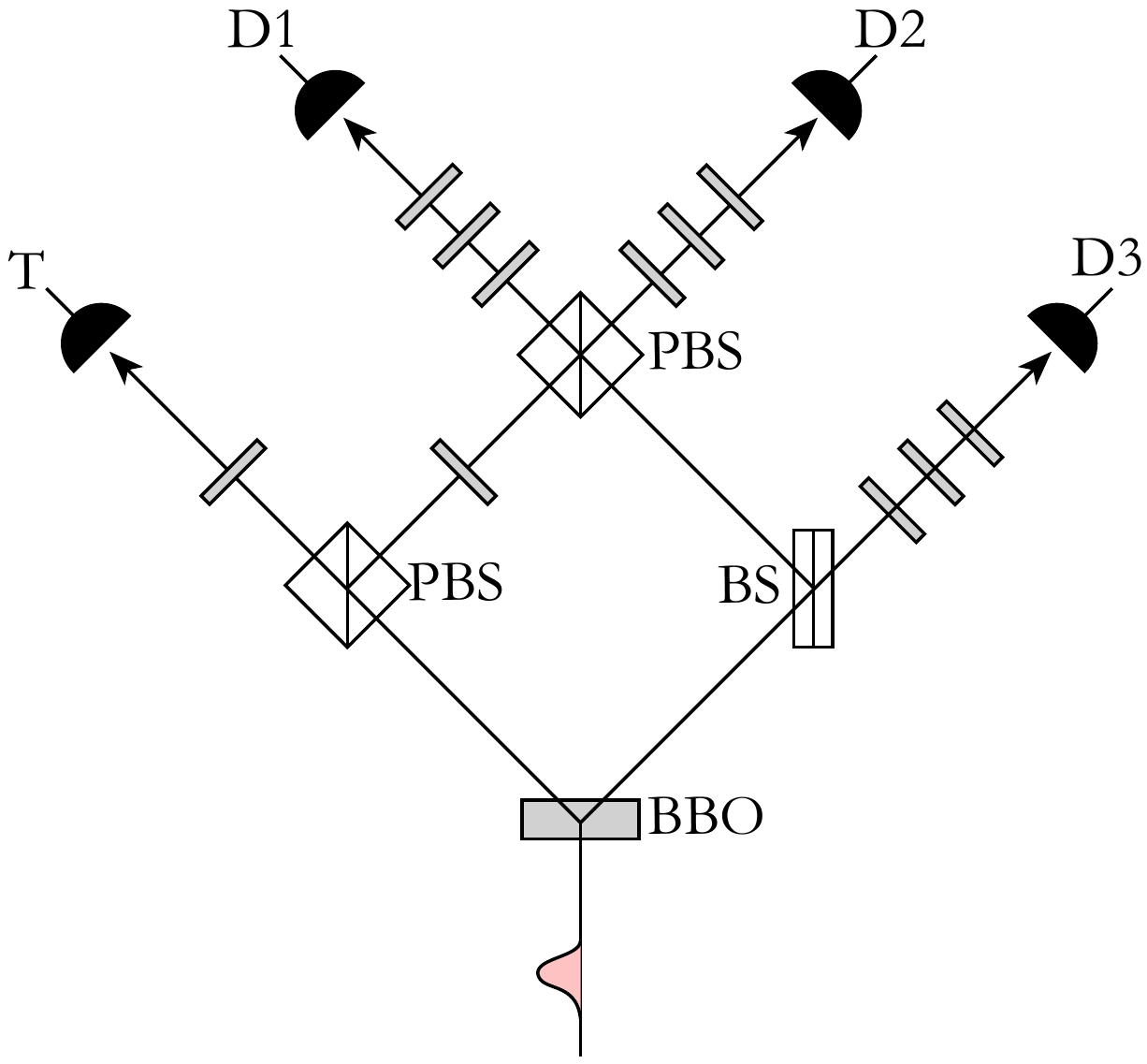}
	\captionspacefig \caption{Linear optics circuit for the non-deterministic preparation of 3-qubit polarisation-encoded GHZ states \cite{pan2000experimental}. A BBO SPDC source is pumped into the double excitation regime. Following evolution through the linear optics network, measurement of a photon in the tigger mode ($T$) heralds the preparation of a GHZ state across modes $D1$, $D2$ and $D3$.}\label{fig:GHZ_LO_prep}
\end{figure}

A simple linear optics circuit for the preparation of 3-qubit polarisation-encoded GHZ states is shown in Fig.~\ref{fig:GHZ_LO_prep}.

%
% W-states
%

\subsection{W-states}\index{W-states}\label{sec:W_state_prep}

W-states are a class of non-maximally-entangled states across an arbitrary number of qubits/modes. They are given by the equal superposition of a single excitation/`1' across all $n$ qubits. Thus, there are $n$ terms in the superposition, of the form,
\begin{align}
\ket\psi_\mathrm{W}^{(n)} &= \frac{1}{\sqrt{n}} \sum_{i=1}^n \hat{a}_i^\dag \ket{0}^{\otimes n}\nonumber\\
&= \frac{1}{\sqrt{n}}(\ket{1,0,0,0,\dots} + \ket{0,1,0,0,\dots}\nonumber\\
&+ \ket{0,0,1,0,\dots} + \ket{0,0,0,1,\dots} + \dots).
\end{align}

W-states are a class of states distinct from GHZ states (for any \mbox{$n\geq 3$}). Unlike GHZ states, W-states preserve some entanglement when a single qubit is traced out of the system, giving them a degree of robustness against qubit loss. This property is discussed in further detail in Secs.~\ref{sec:atomic_ens} \& \ref{sec:error_averaging}, where this property is exploited for error correction purposes.

\begin{figure}[!htbp]
\includegraphics[clip=true, width=0.475\textwidth]{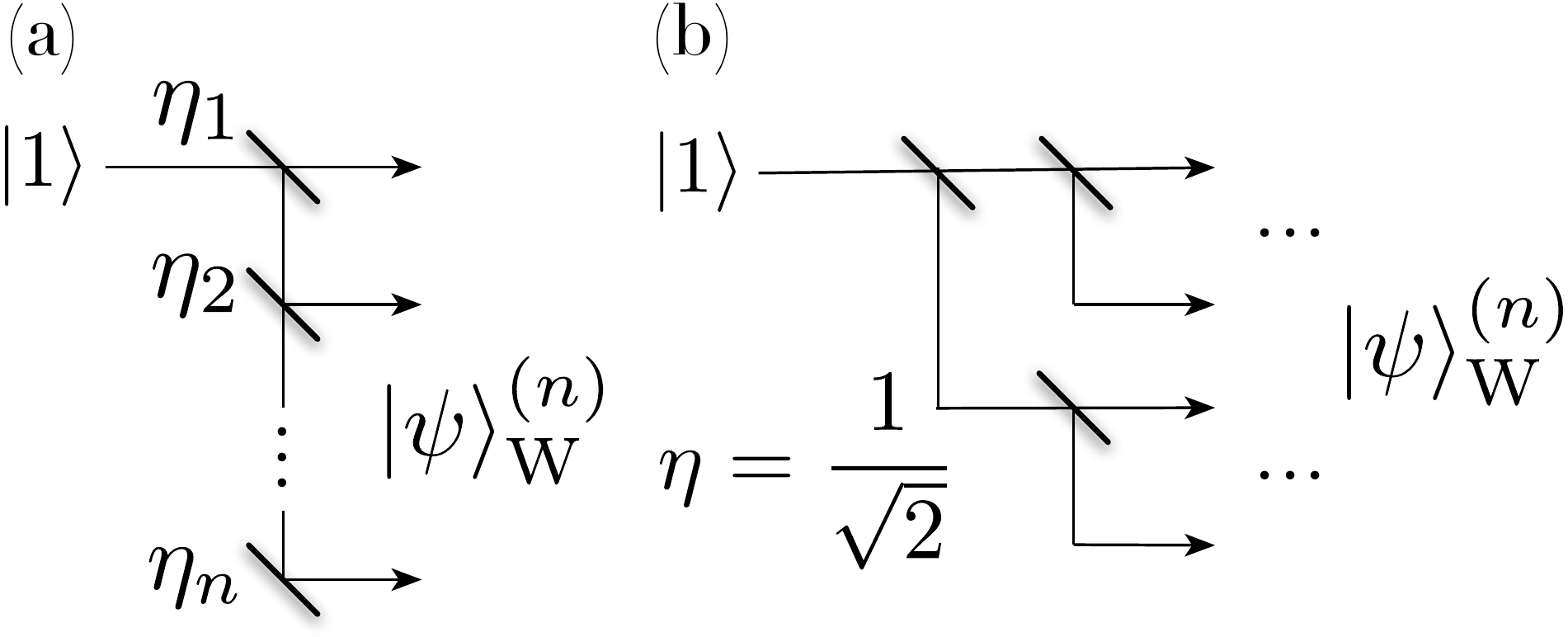}
\captionspacefig \caption{Preparation of an $n$-mode W-state using a single-photon source and a (a) linear cascade of beamsplitters, (b) a pyramid network of beamsplitters. The latter has the advantage that paths are balanced, in the sense that all routes pass through the same number of optical elements. Additionally the pyramid network has only \mbox{$O(\log n)$} depth, compared to the worst-case \mbox{$O(n)$} depth of the linear cascade. This is a relevant consideration when optical elements are lossy or induce errors.}\label{fig:W_state_LO_prep}	
\end{figure}

Using linear optics, W-states are amongst the most trivial entangled states to prepare, using a simple fanout operation. An $n$-mode W-state can be prepared directly using $n$ beamsplitters in a linear cascade, as shown in Fig.~\ref{fig:W_state_LO_prep}(a), where the beamsplitter reflectivities are chosen so as to create the uniform superposition, given by the pattern,
\begin{align}
	{\eta_1}^2 &= 1 - \frac{1}{n},\nonumber\\
	{\eta_1}^2 {\eta_2}^2 &= \frac{1}{n},\nonumber\\
	{\eta_1}^2 {\eta_3}^2 (1-{\eta_2}^2)  &= \frac{1}{n},\nonumber\\
	{\eta_1}^2 {\eta_4}^2(1-{\eta_2}^2) (1-{\eta_3}^2)  &= \frac{1}{n},\nonumber\\
	{\eta_1}^2 {\eta_i}^2 \prod_{j=2}^{i-1} (1-{\eta_j}^2) &= \frac{1}{n},\nonumber\\
	{\eta_n}^2 &= 1.
\end{align}
Alternately, they can be made from a pyramid fanout network of \mbox{$\eta=1/\sqrt{2}$} beamsplitters, as shown in Fig.~\ref{fig:W_state_LO_prep}(b).

More generally, an entire basis of W-states\footnote{We refer to a `basis' of W-states as being a set of states that are all equivalent to a standard W-state up to local phases.} is accessible using linear optics networks described by unitaries where all matrix elements have amplitude $1/\sqrt{n}$, differing only via local phases. This includes the generalised Hadamard transform\index{Hadamard!Transform}, and quantum Fourier transform\index{Quantum Fourier transform}.

%
% Bell States
%

\subsection{Bell states} \label{sec:bell_state_res} \index{Bell!States!Preparation}

Bell states, also known as Einstein-Podolsky-Rosen (EPR) pairs\index{Bell!States} \cite{bib:EPR35}, which are maximally-entangled 2-qubit states, are particularly useful for many applications, including quantum teleportation (Sec.~\ref{sec:teleport}), cluster state preparation (Sec.~\ref{sec:CSQC}), and entanglement swapping (Sec.~\ref{sec:swapping}). Bell states are the special case of 2-qubit cluster states, or equivalently, 2-qubit GHZ states,
\begin{align}
	\ket{\Phi^+} = \ket\psi_\mathrm{GHZ}^{(2)}.
\end{align}

Bell pairs may be directly prepared as the two-mode output from a type-II\footnote{In type-II SPDC the photon-pair is polarisation-entangled, directly preparing a Bell pair in the polarisation basis. In type-I SPDC both photons have the same polarisation, yielding only photon-number correlation, but no polarisation entanglement.} SPDC source, or using non-deterministic linear optics from single-photon sources.

There are four Bell states\footnote{$\ket{\Psi^-}$ is also referred to as a \textit{singlet} state, and $\ket{\Psi^+}$ as a \textit{triplet} state.}, defined as, 
\begin{align} \label{eq:bell_basis}
\ket{\Phi^{\pm}} &= \frac{1}{\sqrt{2}} (\ket{0}_A\ket{0}_B \pm \ket{1}_A\ket{1}_B), \nonumber \\
\ket{\Psi^{\pm}} &= \frac{1}{\sqrt{2}} (\ket{0}_A\ket{1}_B \pm \ket{1}_A\ket{0}_B),
\end{align}
which are locally equivalent to one another via the application of Pauli operators, and may therefore be transformed to one another without classical or quantum communication between the two parties. Specifically,
\begin{align}
\ket{\Phi^+} = \hat{Z}\ket{\Phi^-} = \hat{X}\ket{\Psi^+} = \hat{X}\hat{Z}\ket{\Psi^-},
\end{align}
where $\hat{X}$ and $\hat{Z}$ could apply to either qubit, up to global phase. 

In Sec.~\ref{sec:ent_ultimate} we present the case that these states are so useful on their own that one might be justified in building entire quantum networks based purely upon the distribution of Bell pairs. This is the basis for \textit{quantum repeater networks}, which will be discussed in Sec.~\ref{sec:rep_net}.

%
% Cat States
%

\subsection{Cat states} \index{Cat states!Preparation}

Cat states (Sec.~\ref{sec:cat_enc}) -- superpositions of coherent states -- are extremely difficult to prepare, most easily via non-linear processes. However, they are very useful for optical quantum computation and for the study of macroscopic quantum systems, when using large coherent amplitudes. Because of the difficulty of their preparation, the ability to outsource it would be very valuable.

However, it must be cautioned that cat states decohere very readily, since their well-defined parity implies decoherence upon loss of just a single photon. This makes QoS considerations particularly pertinent. Note that detection of a lost photon signals a shift in parity but keeps the structure of the superposition preserved.

Generalizations of cat states featuring more than two coherent states in superposition around a ring, also called compass states~\cite{Zurek2001subplanck}, are also useful for quantum computing and sensing~\cite{Grimsmo2020rotation}, but they too suffer from increased sensitivity to loss. 

%
% Squeezed States
%

\subsection{Squeezed states} \label{sec:squeezed_prep} \index{Squeezed states!Preparation}

Of particular interest to metrology\index{Quantum metrology} and CV quantum computing\index{Continuous-variables!Quantum computation} in particular, are squeezed states, states which have been longitudinally distorted in phase-space. In the metrological context, squeezed states enable sub-shot-noise\index{Shot-noise} limited metrology \cite{demkowicz2015quantum}, thereby outperforming many classical protocols, using, for example, coherent states.

Mathematically, squeezing is represented using the squeezing operator introduced in Eq.~(\ref{eq:sq_op}). Experimentally, such states are prepared using non-linear crystals. It is intuitively obvious that linear optics alone cannot prepare such states, owing to the non-linear terms in the definition of the operator, which do not preserve photon-number, and therefore cannot be passive.

Of particular interest are squeezed coherent states, $\hat{S}(\xi) \ket{\alpha}$, which are minimum uncertainty states, saturating the Heisenberg uncertainty relation. A special case of this is squeezed vacuum states, $\hat{S}(\xi) \ket{0}$, which are even-parity\index{Parity} states (i.e containing strictly even photon-number terms). This implies that, like cat states, they are very vulnerable to decoherence for the same reason.

%
% GKP States
%

\subsection{GKP states} 

Gottesman-Kitaev-Preskill (GKP) states (Sec.~\ref{sec:gkp_enc}) exhibit periodic structure in phase space, which can be leveraged to encode quantum information for computation and communication~\cite{menicucci2014fault,fukui2021GKPcomm,brady2024GKPreview}. Additionally, they are attractive for metrology, because their periodic structure makes them simultaneously sensitive to small shifts in position and momentum~\cite{terhal2017sensor,valahu2024sensing}. 

Experimentally, optical GKP states can be produced the using the same machinery as single photons from SPDC but at a larger scale. A multimode Gaussian state is produced using squeezing and interferometers, and then photon-number-resolving detectors are used to count photons on all modes but one~\cite{furusawa2024opticalGKPprep, XanaduNature}. Specific outcomes herald approximate GKP states, whose quality can be improved using methods such as breeding~\cite{glancy2010breeding,takase2024breeding}. Other methods to prepare GKP states include coupling to matter qubits~\cite{traviglione2002GKPprep,motes2017encoding} and transduction to optics from the other systems, such as microwave cavities~\cite{yale2020makingGKP} and the mechanical motion of a trapped ion~\cite{fluhmann2019makingGKP}, where they are regularly produced.  

GKP states can be represented as periodic superpositions of coherent states or squeezed states. Thus they inherit sensitivity to loss and dephasing, although these can in some cases be mitigated with error correction strategies, which underpins their value for quantum information tasks.

%
% More Exotic States of Light
%

%\subsection{More exotic states of light}

%As discussed in Sec.~\ref{sec:exotic}, many other far more elaborate states are very difficult to prepare, and it would be highly desirable if they could be obtained/purchased over a quantum network. This includes certain CV states, which have utility in alternate models for quantum computing \cite{bib:Menicucci06, Ralph, Lund}.

%
% Matter Qubits
%

\subsection{Matter qubits} \index{Matter qubits}

Optical quantum states can also be used to mediate entanglement and assist in the state preparation for systems comprising matter qubits, using which-path erasure techniques (Sec.~\ref{sec:hybrid} \& Fig.~\ref{fig:barrett_kok}) or light-matter interactions (Sec.~\ref{sec:opt_inter} \& Fig.~\ref{fig:opt_int}). These techniques are very versatile, and apply to many different matter qubit systems, such as two-level atoms, trapped ions, nitrogen-vacancy centres, quantum dots, Rydberg atoms, and atomic ensembles.

This is useful when matter-based architectures are more scalable or technologically simpler than all-optical architectures (Sec.~\ref{sec:KLM_univ}), and particularly for quantum memory (Sec.~\ref{sec:memory}), when using matter qubits with long lifetimes.

%
% Measurement
%

\section{Measurement} \index{Measurement}

\dropcap{A}{s} a last (and possibility also intermediate) step in any quantum protocol is the measurement of quantum states. State measurement is, in the most general context, essentially state preparation in reverse, and brings with it many of the same challenges.

Different detection schemes bring with them their own (potentially substantial) costs and technological challenges. State of the art 
%micro-pillar photo-detectors\index{Micro-pillar photo-detectors} \cite{???}, at the time of writing, 
number-resolved detectors cost many thousands of dollars to buy and maintain and require sophisticated laboratory setup including large and expensive dilution fridges for cooling. Clearly this type of infrastructure is inaccessible to many players, and borrowing or licensing access to such equipment over a quantum network would pave the way for broader accessibility to state of the art technology.

Each type of state being measured, in combination with the nature of the detection scheme, comes with its own limitations. Specifications of interest include dead-time, speed (relevant when implementing feedforward operations), efficiency, and spatio-spectral filtering characteristics.

Optimization over these specifications presents significant technological challenges, which are costly to overcome, necessitating outsourcing over the quantum internet to become economically viable on a large scale. However, the QTCP protocol is able to accommodate error metrics and attributes covering all the above error models, enabling reliable, predictable QoS for outsourced quantum measurement.

With the ability to perform measurements over a complete basis for the respective system, QST, and consequently QPT (Sec.~\ref{sec:QPT}), can also be outsourced, as both these protocols are built entirely upon determining measurement expectation values in some known basis.

%
% Photo-detection
%

\subsection{Photo-detection} \label{sec:photo_detection} \index{Photo-detection}

Perhaps the most useful, and ubiquitous, type of optical state measurement is photo-detection, where we would like to count photon-number. Broadly, there are two main classes of photo-detectors -- \textit{number-resolved}\index{Number-resolved photo-detectors} and \textit{non-number-resolved}\index{Non-number-resolved photo-detectors} (or `bucket' detectors). These behave exactly as the names suggest, with the former typically being more expensive and technologically demanding than the latter.

%
% Mathematical Representation
%

\subsubsection{Mathematical representation}

A general photo-detector can be modelled as a POVM\index{POVM},
\begin{align}
\hat\Pi_m = \sum_{n=0}^\infty P(m|n) \ket{n}\bra{n},	
\end{align}
where $P(m|n)$ is the conditional probability of measuring $m$ photons given $n$ incident photons. The POVM\index{POVM} is fully characterised by the conditional probabilities, which must be inferred from the specifics of the architecture. Alternately, a quantum process\index{Quantum processes} formalism can be constructed as,
\begin{align}
\mathcal{E}_m(\hat\rho) = \sum_{n=0}^\infty P(m|n) \hat{E}_n\hat\rho\hat{E}_n^\dag,	
\end{align}
where \mbox{$\hat{E}_n=\hat{E}_n^\dag=\ket{n}\bra{n}$} are the Kraus operators\index{Kraus operators}.

These mathematical representations very conveniently reduce the characterisation and representation of photo-detectors to calculating the matrix of conditional probabilities, $P(m|n)$. This readily allows various experimental effects and imperfections to be accommodated.

%
% Avalanche Photo-Diodes (APDs)
%

\subsubsection{Avalanche photo-diodes}\index{Avalanche photo-diodes (APDs)}

The most common form of photo-detection is using avalanche photo-diodes (APDs), which are cheap but non-number-resolving. Here, a single photon excites an electron into the conduction band at a semiconductor junction, enabling a detectable current flow. However, a single excitation triggers an `avalanche' of further excitation making the magnitude of the detected current essentially unrelated to exact photon number.

%
% Superconducting Photo-Detectors}
%

\subsubsection{Superconducting photo-detectors}\index{Superconductors!Photo-detectors}

\begin{figure}[!htbp]
\includegraphics[clip=true, width=0.25\textwidth]{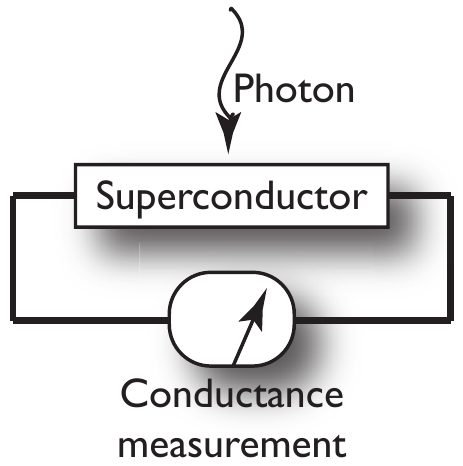}
\captionspacefig \caption{A superconducting photo-detector. A superconductor is held at just below its critical temperature. The absorption of a photon is sufficient to heat it above the critical temperature, yielding a detectable change in conductance across the device.} \label{fig:super_det}
\end{figure}

Recently, superconducting detectors have been adopted, as they have the potential for number resolution. Here a superconductor is kept just below its critical temperature, and the absorption of a photon is enough to heat the superconductor above the critical temperature, creating a detectable change in resistance across the superconductor. This is shown in Fig.~\ref{fig:super_det}. Despite their superior performance, however, superconducting detectors are very expensive (for obvious reasons) and only accessible to well-resourced labs. However, unlike APDs they can be photon-number-resolving.

Two leading types of superconducting photo-detector are transition-edge sensors (TESs)~\cite{Irwin2005TES}, which operate by measuring changes in resistance near a superconducting transition, and superconducting single-photon nanowire detectors (SNSPDs)~\cite{zadeh2021SNSPDreview}, which operate by detecting local disruptions of superconductivity in nanowires. Each has its own merits and drawbacks. TESs are slower, more expensive, and require more cooling than SNSPDs, but TESs have higher efficiency, lower dark-count rates, and they allow for number-resolved detection.

%
% Quantum Dot Photo-Detectors
%

\subsubsection{Quantum-dot photo-detectors}\index{Quantum dots!Photo-detectors}

Quantum dots are confined charge carriers (electrons and holes), often in a semiconductor material. Quantum-dot detectors~\cite{kim2021quantumdotdetectors} rely on the fact that photons with more energy than the bandgap of the quantum dot excite charge carriers, which produces a measureable photocurrent. Quantum-dot detectors can be sensitive to a broad spectrum of light, can operate at room temperature, and can have high efficiency. However, they suffer from relatively high dark counts and timing jitter compared to their superconducting counterparts, and they can be unstable over time due to material changes. Also, a quantum dot is a saturable absorber, so the detectors are not number-resolving without multiplexing.

%\subsubsection{Other photo-detector}\index{Photo-diodes}
%
%Other photo-detectors are available

%
% Experimental issues
%

\subsubsection{Experimental issues}

The key parameter of interest in a photo-detector, in addition to whether or not it is number-resolving, is its efficiency\index{Detector!Efficiency}, $\eta$ -- the probability that a given incident photon will trigger the detector. For most applications, the goal is to maximise $\eta$. As one might expect, there is a direct tradeoff between $\eta$ and cost, with very high-efficiency detectors often economically out of reach for many experimentalists. Also of interest is the `dark-count' rate -- the rate at which the detector falsely clicks in the absence of photons. However, this is often ignored as modern detectors typically exhibit very low dark-count-rates.

Mathematically, the measurement operators for inefficient number-resolved detection are,
\begin{align}\index{Number-resolved photo-detectors}
\hat\Pi_n = \eta^{n} \sum_{m=n}^\infty \binom{m}{n} (1-\eta)^{m-n} \ket{m}\bra{m},
\end{align}
for measurement outcome $n$, in the photon-number basis. And for non-number-resolved detection,
\begin{align}\index{Non-number-resolved photo-detectors}
&\hat\Pi_0 = \sum_{m=0}^\infty (1-\eta)^{m} \ket{m}\bra{m}, \nonumber \\
&\hat\Pi_{>0} = \hat\openone - \hat\Pi_0.
\end{align}
Thus, inefficiency results in projection onto the wrong photon-number, making measurement outcomes incorrect.

In addition to their operation in the photon-number basis, photo-detectors exhibit spatio-temporal characteristics, which affect their operation in quantum information processing protocols. For example, imperfect spectral response can undermine photonic interference, affecting which-path erasure protocols, such as Bell state projection (Sec.~\ref{sec:bell_proj}). However, in many cases this can be improved upon using spectral filtering or time-gating techniques, also at the expense of experimental complexity and resource overhead.

Furthermore, photo-detectors are subject to `dead-time'\index{Dead-time}, which renders them inactive for a finite recovery period following a detection event. This is of especial importance in time-bin-encoded schemes (Sec.~\ref{sec:time_bin}), where detectors must resolve photons over very short timescales on the order of nanoseconds. Dead-time can be modelled as a time-dependent efficiency of the form,
\begin{align}
\eta(t) = \left\{\begin{array}{l l}
 0, & t<\tau_\mathrm{dt} \\
 \eta_\mathrm{ss}, & t\geq\tau_\mathrm{dt} \\
\end{array}\right.,
\end{align}
where $t$ is time, $\tau_\mathrm{dt}$ is the detector's dead time\index{Dead time}, and $\eta_\mathrm{ss}$ is the detector's steady-state efficiency (i.e when not dead).

Photo-detectors of all types are inevitably subject to `dark-counts'\index{Dark-counts}, whereby thermal noise\index{Thermal!Noise}, either within the detector or coupled from the noisy external environment, triggers non-existent detection events. The distribution follows exactly that of the thermal state photon-number distribution (Sec.~\ref{sec:thermal_states}). Thus, the probability of $n$ dark-counts occurring is,
\begin{align} \index{Thermal!Distribution}
p_\mathrm{dc}(n) = e^{-|\alpha|^2} \frac{|\alpha|^2}{n!},
\end{align}
where $\alpha$ is a parameterisation of the temperature of the environmental noise. Fortunately, modern detector technology is able to keep dark-count rates very low for some technologies, making this far less of an issue than the aforementioned ones.
%, loss being the dominant.

Finally, all photo-detection techniques are subject to some degree of `time-jitter'\index{Time-jitter} -- noise in the detector's reported time of detection. This can be extremely important in the context of temporal mode-matching, where post-selection upon detection events in an extremely narrow time-window effectively enforces temporal indistinguishability.

%
% Multiplexed Photo-Detection
%

\subsection{Multiplexed photo-detection}\index{Multiplexed!Photo-detection}

Number-resolved detectors are the more challenging ones to experimentally realise. However, using multiplexing techniques\index{Multiplexed!Photo-detection}, non-number-resolved detectors can be used to closely approximate number-resolution \cite{bib:Fitch03, bib:Banaszek03, bib:Achilles04, bib:RohdeCompDet07}, at the expense of an (efficient) overhead in the complexity of the experiment, which comes at a cost. 

\begin{figure}[!htbp]
\includegraphics[clip=true, width=0.4\textwidth]{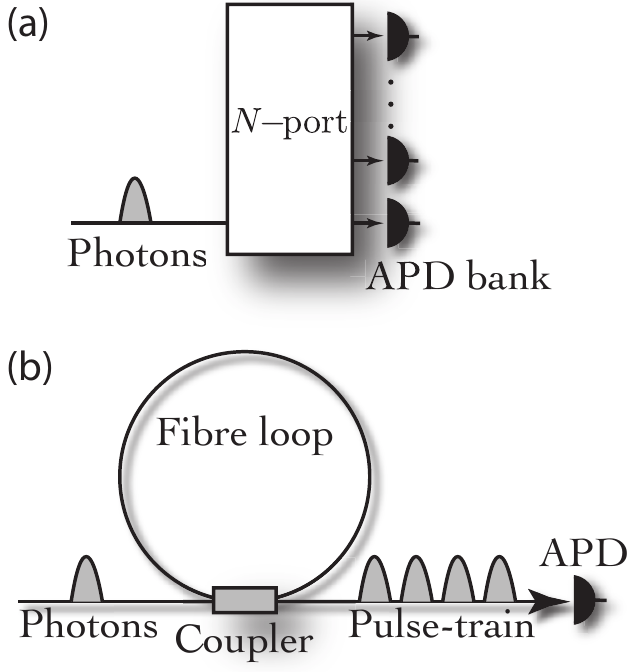}
\captionspacefig \caption{Multiplexed number-resolved photo-detection using non-number-resolved photo-detectors. The principle is to spread out (as uniformly as possible) a multi-photon state across a large number of modes, sufficiently large that it is unlikely that more than one photon will be present in any given mode. Then, the sum of the number of clicks at each mode closely approximates the incident photon-number. (a) In the spatial domain. (b) In the temporal domain. The advantage of employing the temporally multiplexed architecture is that only a single detector is required, unlike the multiple independent detectors required in the spatially multiplexed scheme. However this requires that the dead-time\index{Dead-time} of the detector is less than the round-trip time of the fibre loop. An alternate, but conceptually equivalent approach is to spatially disperse the optical field across a charge-coupled device (CCD)\index{Charge-coupled devices (CCDs)}, much like that found in a regular digital camera, except with single-photon resolution per-pixel. This achieves an effectively very large number of optical modes.} \label{fig:det_mult}
\end{figure}

Specifically, there is a direct tradeoff between the confidence in photon-number outcomes and experimental overhead. The idea behind this is simple. We spread out an $n$-photon state evenly across a large number of modes, $m$, and detect each one independently using a non-number-resolved photo-detector. If \mbox{$m\gg n$}, it is unlikely that more than a single photon will reach any given detector. Thus, by summing the total number of clicks across all detectors, we closely approximate the true photon-number. This multiplexing can be performed in the spatial- or temporal-domains, shown in Fig.~\ref{fig:det_mult}, and has been a widely employed technique in laboratories without access to expensive number-resolved detectors.

\begin{figure}[!htbp]
	\includegraphics[clip=true, width=0.475\textwidth]{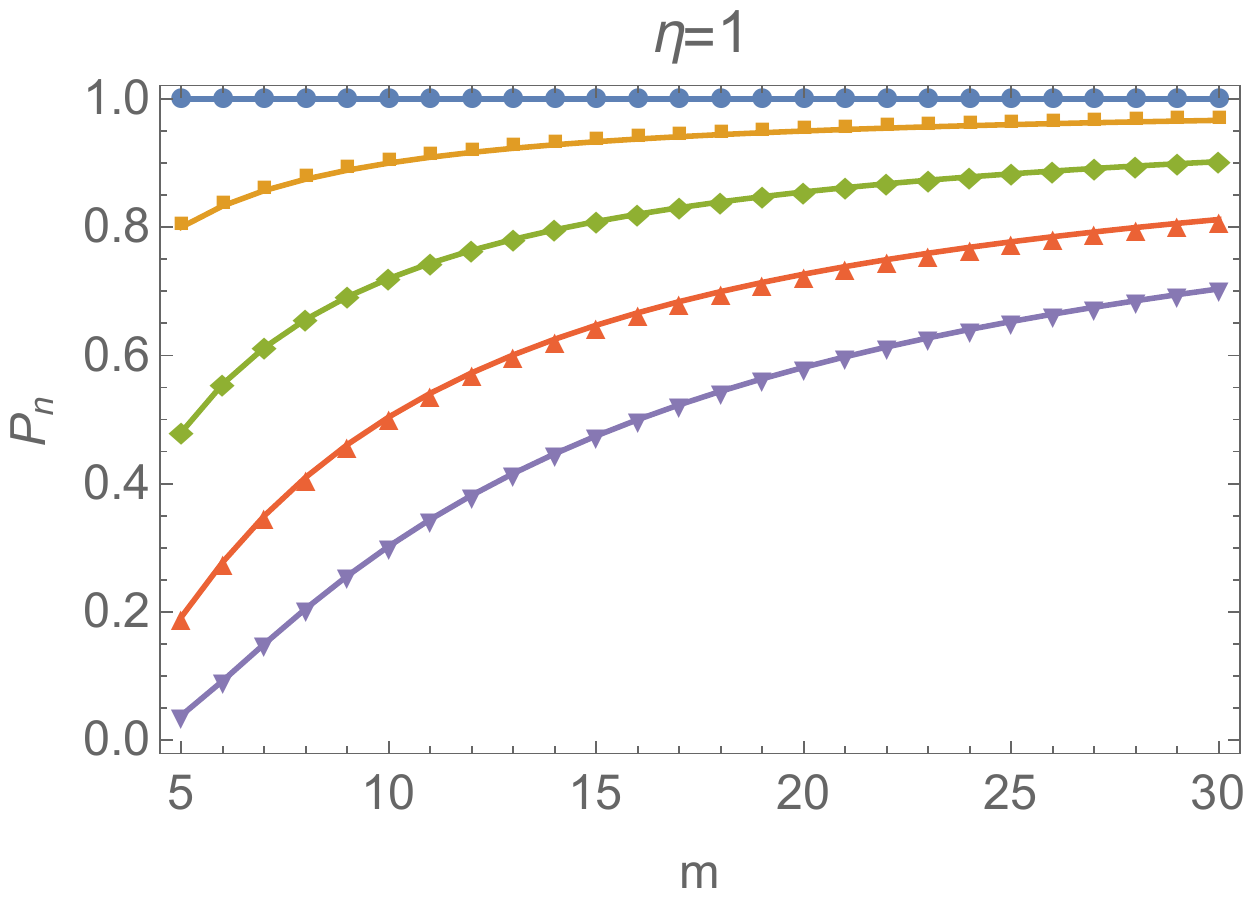}
	\captionspacefig \caption{Probability of a balanced multiplexed detector inferring the correct number of photons with $n$ incident photons across $m$ modes. The detectors have perfect efficiency, \mbox{$\eta=1$}, and all other errors are ignored. The number of incident photons ranges from \mbox{$n=1$} (top) to \mbox{$n=5$} (bottom). All curves asymptotically approach \mbox{$P_n=1$} for large $m$.}\label{fig:multiplexed_pd}
\end{figure}

Mathematically, we are interested in the probability \mbox{$P(n_\mathrm{meas}=n_\mathrm{inc})$} that the measured number of photons ($n_\mathrm{meas}$) matches the actual number of incident photons ($n_\mathrm{inc}$). The structure of this expression will vary enormously depending on the details of the architecture (e.g multi-port interferometer vs fibre-loop). However, \cite{bib:RohdeCompDet07} presented a very general mathematical formalism applicable to all architectural variants. The simplest case to consider is the multi-port interferometer, owing to its perfect symmetry. The probability is simply the probability that no output mode from the multi-port contain multiple photons. A quick calculation yields,
\begin{align}
	P(n_\mathrm{meas}=n_\mathrm{inc}) = \frac{\eta^n m!}{m^n(m-n)!},
\end{align}
for efficiency $\eta$, not accounting for other lesser errors such as dark-counts. This is shown in Fig.~\ref{fig:multiplexed_pd}. For perfect efficiency, \mbox{$\eta=1$}, this probability always approaches unity in the limit of a large number of modes,
\begin{align}
\lim_{m\to\infty}P(n_\mathrm{meas}=n_\mathrm{inc})=1.
\end{align}

%
% Homodyne detection
%

\subsection{Homodyne detection} \label{sec:homodyne} \index{Homodyne detection}

\begin{figure}[!htbp]
\includegraphics[clip=true, width=0.3\textwidth]{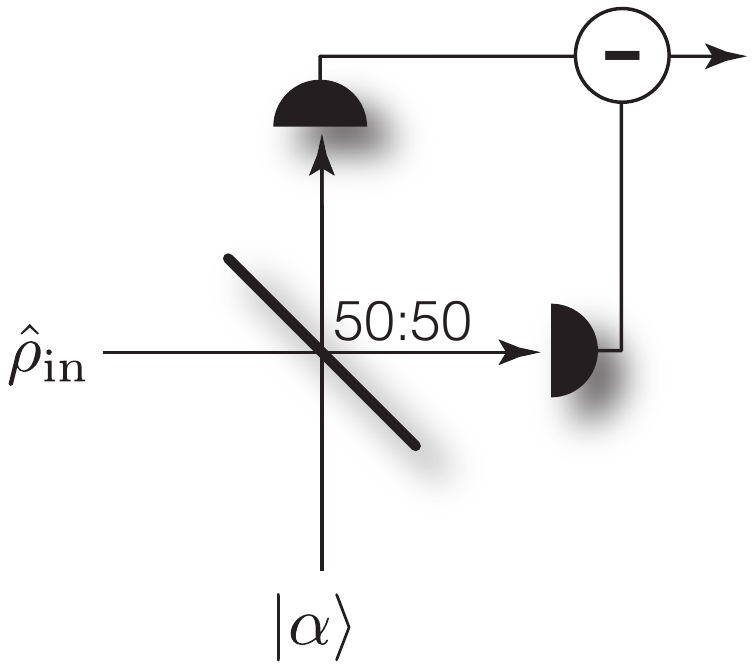}
\captionspacefig \caption{Homodyne detection of an unknown optical state $\hat\rho_\mathrm{in}$, by mixing it with a reference coherent state $\ket\alpha$ on a 50:50 beamsplitter, and taking the difference in the photo-detection rates at the output modes. By sweeping through the phase and amplitude of $\ket\alpha$, we can directly sample the Wigner function of $\hat\rho_\mathrm{in}$, allowing its full reconstruction.} \label{fig:homodyne}
\end{figure}

Homodyne detection is designed to measure the electric field rather than intensity, which direct photo-detection measures. A homodyne detector mixes the signal state on a beam splitter with a coherent state, called a local oscillator, which acts as a phase reference\index{Phase!Reference}. Then, photo-detecting both output modes and taking the difference in the photon count rates yields a photocurrent that is proportional to the electric field in the signal mode (Fig.~\ref{fig:homodyne}). This effectively allows us to observe `beating'\index{Beating} effects between the signal and reference probe. The operation of homodyne measurement can be visualised in phase-space, where it can be regarded as integrating along an infinite line with arbitrary rotation determined by the phase-reference. While conceptually straightforward, preparing the reference beam requires a coherent source, which can become costly (Sec.~\ref{sec:coherent_states}).

By changing the phase of the local oscillator, homodyne detection can directly sample position ($\hat x$), momentum ($\hat p$), or any linear combination of the two. Sweeping across local oscillator phases allows one to directly all of phase space, allowing the Wigner function\index{Wigner function} -- which is isomorphic to the density operator -- of an unknown state to be fully reconstructed.

%This measurement technique is typically applied to CV states rather than photon-number states.

%
% Bell State & Parity Measurements
%

\subsection{Bell state \& parity measurements} \label{sec:bell_proj} \index{Bell!Measurements}

For the purposes of which-path erasure, essential for optical cluster state (Sec.~\ref{sec:CSQC}) preparation and quantum teleportation (Sec.~\ref{sec:teleport}), Bell state measurements [i.e projections onto the Bell basis given in Eq.~(\ref{eq:bell_basis})], or equivalently parity measurements, are important.

To realise this, there are two primary options. The first is to use a CNOT gate, for example an LOQC gate (Sec.~\ref{sec:KLM_univ}). The second is to perform a \textit{partial} Bell state projection using a polarising beamsplitter (PBS) -- a beamsplitter which completely transmits vertical polarisation, and completely reflects horizontal polarisation \cite{bib:BraunsteinMann95}.

\begin{figure}[!htbp]
\includegraphics[clip=true, width=0.375\textwidth]{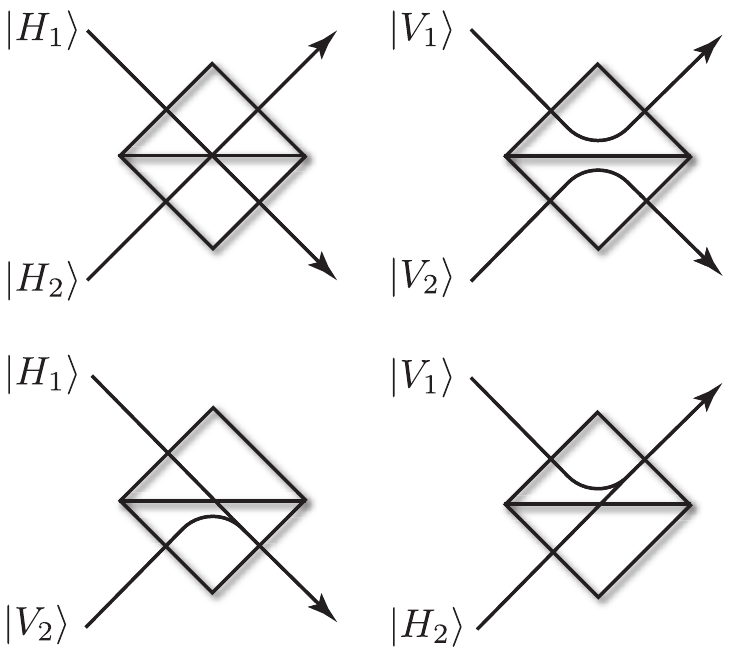}
\captionspacefig \caption{Partial Bell state projection using a polarising beamsplitter (PBS). The PBS completely transmits horizontally polarised light, whilst completely reflecting vertically polarised light. Shown are the four possible two-photon input states, and the respective trajectories followed by the photons. To complete the partial Bell projection we measure the output modes in the diagonal basis, \mbox{$\ket{\pm} = \frac{1}{\sqrt{2}}(\ket{H}\pm\ket{V})$}, such that horizontally and vertically polarised photons cannot be distinguished. If the input state was $\ket{H,H}$ or $\ket{V,V}$, we would measure one photon in each output mode (both transmitted or both reflected). Since the detectors cannot distinguish $\ket{H}$ from $\ket{V}$, this effectively projects us onto the coherent superposition of both possibilities (`which-path erasure'), implementing the measurement projector \mbox{$\hat\Pi_\mathrm{Bell}^\pm = \ket{H,H}\bra{H,H}\pm\ket{V,V}\bra{V,V}$}. If, on the other hand, we measure two photons at one output mode, we know with certainty what the polarisations of both incident photons were and we probabilistically implement one of the projectors \mbox{$\hat\Pi_
\mathrm{HV}=\ket{H,V}\bra{H,V}$} or \mbox{$\hat\Pi_
\mathrm{VH}=\ket{V,H}\bra{V,H}$}, effectively performing polarisation-resolved detection upon both modes, which equates to a $\hat{Z}$ measurement on the logical qubits. The practical outcome of this is that, when using a PBS to prepare cluster states, with probability $1/2$ we are able to successfully fuse two smaller cluster states together into a larger one, and with probability $1/2$ we fail to do so, instead removing two qubits from the clusters.} \label{fig:partial_bell}
\end{figure}

In the Heisenberg picture, the transformation of the photonic creation operators implemented by a PBS is,
\begin{align}\index{Polarising beamsplitters}
\hat{h}_1^\dag &\to \hat{h}_2^\dag, \nonumber \\
\hat{h}_2^\dag &\to \hat{h}_1^\dag, \nonumber \\
\hat{v}_1^\dag &\to \hat{v}_1^\dag, \nonumber \\
\hat{v}_2^\dag &\to \hat{v}_2^\dag,
\end{align}
where $\hat{h}_i^\dag$ ($\hat{v}_i^\dag$) are the horizontal (vertical) creation operators for the $i$th mode. The measurement projectors implemented by the PBS, when both modes are measured in the diagonal ($+/-$) basis\footnote{By measuring in the diagonal basis we erase information about whether photons were horizontally or vertically polarised, thereby projecting onto the coherent subspace of both possibilities. Such a diagonal basis measurement may be implemented using a wave-plate to perform a Hadamard polarisation rotation, followed by another PBS, separating the horizontal and vertical components, which are then independently measured via regular photo-detection.}, are then,
\begin{align}
	\hat\Pi_\mathrm{Bell}^+ &= \ket{H,H}\bra{H,H}+\ket{V,V}\bra{V,V}, \nonumber \\
	\hat\Pi_\mathrm{Bell}^- &= \ket{H,H}\bra{H,H}-\ket{V,V}\bra{V,V}, \nonumber \\
	\hat\Pi_
	\mathrm{HV} &= \ket{H,V}\bra{H,V}, \nonumber \\
\hat\Pi_
	\mathrm{VH} &= \ket{V,H}\bra{V,H},
\end{align}
where the former two represent successful projection onto the Bell basis, and the latter two represent failures, effectively measuring both modes in the $H/V$ basis. This approach is described in Fig.~\ref{fig:partial_bell}.

Technically, $\hat\Pi^\pm_\mathrm{Bell}$ are not Bell measurements, but rather projections onto the even parity subspace. A true Bell projection would implement $\ket{\Phi^\pm}\bra{\Phi^\pm}$. However, in an optical context the two terms are often used interchangeably, since they exhibit effectively the same behaviour, given that the detection process is destructive.

Bell projections using CNOT gates can be implemented with arbitrarily high success probability in principle. However, in most scenarios of interest (such as cluster state preparation and entanglement purification) Bell projection using a PBS succeeds with probability of $1/2$, since a PBS is only able to uniquely distinguish two of the four Bell states. To its advantage, such `partial' Bell measurements only require high HOM visibility, avoiding the need for the interferometric stability inherent internally within LOQC CNOT gates.

While partial Bell state projection using a PBS is relatively straightforward, LOQC CNOT gates (which are very desirable owing to their near-determinism) are very technologically challenging, with drastic resource overheads, particularly for high success probability. Thus, outsourcing them to the cloud may be very economically efficient.

\subsubsection{CV equivalent: EPR measurement}

The CV equivalent to the Bell projection above is projection onto an EPR state, which is a two-mode state that is perfectly correlated in one quadrature (and perfectly anti-correlated in another). An example is the canonical EPR state
	\begin{align}
		\ket{\text{EPR}} = \frac{1}{\sqrt{2\pi}} \int dx \, \ket{x,x} = \frac{1}{\sqrt{2\pi}}  \int dp \, \ket{p,-p}.
	\end{align}
EPR states are the infinite-squeezing limit of two-mode squeezed states. As such, EPR states can be created by mixing an (infinitely squeezed) $\hat{x}$-eigenstate and an $\hat{p}$-eigenstate on a beam splitter. 

An EPR \emph{measurement} is projection onto displaced EPR states 
	\begin{align}
		\ket{\text{EPR}(m_1,m_2) } = \hat{D} \big( \tfrac{1}{\sqrt{2}} (m_1 + i m_2) \big) \ket{\text{EPR}},
	\end{align}
described by the measurement projectors 
 \begin{align}
 	\hat{\Pi}(m_1, m_2) 
	= \ket{\text{EPR}(m_1,m_2) } \bra{\text{EPR}(m_1,m_2) } 
\end{align}
where $m_1$ and $m_2$ are real numbers corresponding to homodyne measurement outcomes, and $\hat{D}$ is a displacement operator~\cite{walshe2020gateteleportation}.

 Measurement-based quantum computing with CV cluster states is based on EPR measurements, which are realized by a beamsplitter and homodyne detection on each of the two modes. This measurement model has the additional benefit that locally adjusting the quadrature bases can realize effective gates before the measurement~\cite{walshe2020gateteleportation,walshe2023equivalent}. Unlike the Bell measurements above, which are designed for single-photon encodings, the EPR measurements for CV encodings are deterministic --- \emph{i.e.} there is no probability of failure, although one is typically required to perform a local, outcome-dependent feedforward operation. An example of this arises in CV quantum teleportation, which is the gadget at the heart of CV cluster state computation.

%
% Matter Qubits
%

\subsection{Matter qubits} \index{Matter qubits}

Many non-optical systems can be indirectly measured by first entangling optical states with the matter qubits and then measuring the optical state. Because of the entanglement, projective measurement on the optical state teleports the measurement onto the matter qubit.

In Fig.~\ref{fig:barrett_kok} we illustrate a scheme for entangling two $\lambda$-configuration atoms using which-path erasure. Consider just one of these qubits in isolation. If a $\pi$-pulse is applied to the atom, the $\ket{\!\downarrow}$ state is excited to the $\ket{e}$ state, after which, upon relaxation, it emits a photon. Thus, upon measurement, the presence or absence of a photon directly indicates whether the qubit was in the $\ket{\!\uparrow}$ or $\ket{\!\downarrow}$ state.

The attractive feature of this is that although the matter qubit is stationary, its indirect measurement via optical coupling may be performed over arbitrary distances across the optical network, allowing the measurement stage to be outsourced. This includes entangling measurements, useful for, for example, cluster state preparation (Sec.~\ref{sec:CSQC}).

%
% Quantum non-demolition measurement
%

\subsection{Quantum non-demolition measurement}\index{Quantum non-demolition measurements (QND)}

When it comes to optical systems, measurement is typically \textit{destructive}\index{Destructive measurements}, i.e the photon (or other optical state) is destroyed in the process of measurement. An APD, for example, converts a photon to an electrical current and then the photon is gone. Can we measure the logical value of optical states without destroying them?

The answer is yes, via \textit{quantum non-demolition measurements} (QND). The central idea here is to entangle a system which mustn't be destroyed with an ancillary system that we can happily afford to lose. Because the two systems are correlated, a destructive measurement on the ancillary state yields an effective measurement on the primary system, but without destroying it.

The simplest example to illustrate this is by considering the measurement of a single qubit,
\begin{align}
\ket\psi = \alpha\ket{0} + \beta\ket{1}.	
\end{align}
We begin by introducing an additional ancillary qubit in the logical $\ket{0}$ state, which we maximally entangle to the primary system with a CNOT gate. This yields two redundantly encoded\index{Redundant encoding} qubits,
\begin{align}
\hat{\mathrm{CNOT}}\ket\psi \ket{0} = \alpha\ket{0}\ket{0} + \beta\ket{1}\ket{1},
\end{align}
characterised by the same amplitudes as the original qubit. The ancilla is then measured (destructively), which reads out the original state of the primary system to which it was correlated, but which has not been destroyed and may freely continue on its journey.

Unfortunately, the required CNOT gate is cumbersome in optical architectures. But it can be done, and in Sec.~\ref{sec:KLM_univ} we describe in detail how such gates can be constructed.

\begin{figure}[!htbp]
\begin{align}
\Qcircuit @C=1em @R=1.6em {
    \lstick{\ket\psi} & \ctrl{1} & \qw & \hat{M}\ket\psi \\
    \lstick{\ket{0}} & \targ & \qw & \meter \\
} \nonumber
\end{align}
\captionspacefig \caption{Quantum non-demolition measurement of a single qubit. $\hat{M}$ is the measurement operator applied during measurement to the ancillary state, which may be destructive. The primary qubit has now been projected onto the measurement outcome $\hat{M}$, but has not been destroyed.}\index{Quantum non-demolition measurements (QND)}\label{fig:QND_circ}
\end{figure}
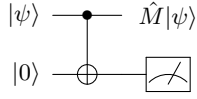

The quantum circuit implementation for QND is shown in Fig.~\ref{fig:QND_circ}.

%
% Weak Measurement
%

\subsection{Weak measurement}\index{Weak measurements}

The measurements considered until now have been \textit{projective} measurements. These completely collapse the wave-function of a state onto an eigenstate of the measurement operator, and also give us perfect information about the associated eigenvalue (i.e measurement outcome). But in some scenarios this might be too aggressive -- could we instead just extract \textit{some} information about the system, and only \textit{partially} collapse it. This is achieved using \textit{weak measurement}.

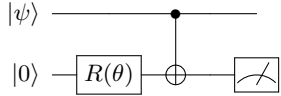
\begin{figure}[!htbp]
\begin{align}
\Qcircuit @C=1em @R=1.6em {
    \lstick{\ket\psi} & \qw & \ctrl{1} & \qw & \qw \\
    \lstick{\ket{0}} & \gate{R(\theta)} &  \targ & \qw & \meter \\
} \nonumber
\end{align}
\captionspacefig \caption{Quantum weak measurement. The choice of rotation on the ancillary qubit determines how much information about the state is extracted from the measurement, and also how much the state is disturbed by the measurement.}\label{fig:weak_meas}
\end{figure}

Consider the circuit shown in Fig.~\ref{fig:weak_meas}. This is identical to the previous circuit for QND, but with the addition of the arbitrary rotation on the ancilla qubit prior to the CNOT gate.

When \mbox{$\theta=0$} this circuit implements an ordinary projective measurement, projecting onto the eigenbasis of the measurement operator. We'll refer to this as a \textit{strong} measurement -- maximum information and maximum disturbance.
On the other hand, consider the case where $\theta = \frac{\pi}{2}$, giving \mbox{$\hat{R}(\theta)=\hat{H}$}. This transforms the ancillary $\ket{0}$ state into the \mbox{$\ket{+}=\frac{1}{\sqrt{2}}(\ket{0}+\ket{1})$} state. However, the $\ket{+}$ state is an eigenstate of the $\hat{X}$ operator that resides in the CNOT gate, and is therefore invariant under the action of the CNOT. Now the measurement outcome will be completely randomised and independent of $\ket\psi$, revealing no information about $\ket\psi$ whatsoever. But simultaneously, $\ket\psi$ will be unaffected by the measurement since it was not entangled with the ancilla, remaining separable. We have effectively implemented no measurement at all.

More generally, we can choose $\theta$ to be any value in between these two extremes. The strength of the measurement, how much information is extracted from it, and how much it disturbs the state $\ket\psi$, are functions of $\theta$. The rule is, the more information our measurement extracts, the more we disturb the state. For \textit{weak measurements}, when $\theta$ is near but not exactly equal to $\frac{\pi}{2}$, only a small amount of information is extracted and the state $\ket{\psi}$ is only mildly perturbed.

%
% Evolution
%

\section{Evolution}

\dropcap{T}{he} evolution of optical states represents an extremely broad category of quantum operations, including passive linear optics, post-selected linear optics, non-linear optics, and light-matter interactions. Clearly, the items in this list present technological challenges, inaccessible to many users.

The error models in the evolution of optical states are largely accounted for by those discussed in Sec.~\ref{sec:errors_in_nets}. 

%
% Linear Optics
%

\subsection{Linear optics} \label{sec:LO_ev_archs} \index{Linear optics!Evolution}

Linear optics networks \cite{bib:TanRohdeRev} implement unitary linear maps on the photonic creation operators, of the form,
\begin{align} \label{eq:LO_unitary_map}
\hat{U}\hat{a}_i^\dag \hat{U}^\dag \to \sum_{j=1}^m U_{i,j} \hat{a}^\dag_j,
\end{align}
where $\hat{a}^\dag_i$ is the photonic creation operator on the $i$th of the $m$ modes, and $U$ may be any $\mathrm{SU}(m)$ matrix. It was shown by \cite{bib:Reck94} that arbitrary transformations of this form may be decomposed into $O(m^2)$ linear optical elements (beamsplitters and phase-shifters), enabling efficient construction of arbitrary linear transformations.\index{Linear optics!Decompositions} Furthermore, the algorithm for determining the decomposition of such transformations has polynomial classical runtime (i.e residing in \textbf{P}\index{P}). Note that the original Reck \textit{et al.} decomposition is not unique, and various other topologies of optical elements also enable universality, each with their own implementational advantages and disadvantages.

Each individual beamsplitter in such a decomposition is described by an $\mathrm{SU}(2)$ matrix acting on two photonic creation operators, $\hat{a}^\dag$ and $\hat{b}^\dag$,
\begin{align}\index{Beamsplitters}
\begin{pmatrix}
\hat{a}^\dag_\mathrm{out} \\
\hat{b}^\dag_\mathrm{out}
\end{pmatrix}=\begin{pmatrix}
e^{i\phi_1} \sqrt{\eta} & e^{i\phi_2}\sqrt{1-\eta} \\
e^{-i\phi_2}\sqrt{1-\eta} & -e^{-i\phi_1}\sqrt{\eta}
\end{pmatrix}\begin{pmatrix}
\hat{a}^\dag_\mathrm{in} \\
\hat{b}^\dag_\mathrm{in}
\end{pmatrix},
\end{align}
where \mbox{$0\leq\eta\leq1$} is the reflectivity, and \mbox{$0\leq\phi_1,\phi_2\leq 2\pi$} determine the phase relationships.

When operating in the polarisation basis, wave-plates\index{Wave-plates} enable the same transformation as beamsplitters do in dual-rail encoding. The phase-shifters implement the unitary operation,\index{Phase!Shifts}
\begin{align}
\hat\Phi(\phi)=e^{-i\phi\hat{n}},
\end{align}
or equivalently,
\begin{align}
\hat{U}\hat{a}^\dag\hat{U}^\dag \to e^{-i\phi}\hat{a}^\dag,
\end{align}
for phase-shift $\phi$.

The linear optical transformations above are all \emph{passive}, meaning that they do not change the energy of input states. An \emph{active} linear transformation is generated by the displacement operator, which translates the Wigner function by some arbitrary amplitude in phase-space, whilst preserving all other features of the phase-space representation. This is described by the unitary operator,\index{Displacement operator}
\begin{align}
\hat{D}(\alpha) = \exp \left[\alpha\hat{a}^\dag - \alpha^*\hat{a}\right],
\end{align}
where $\alpha$ is the displacement amplitude. This transformation is implemented by mixing a state on a low-reflectivity beamsplitter with a coherent state of  arbitrary complex amplitude, which determines the displacement amplitude. 
Displacement operators introduce a classical offset to annihilation or creation operators,
	\begin{align}
		\hat{D}^\dagger(\alpha) \hat{a}^\dagger \hat{D}(\alpha) = \hat{a}^\dagger + \alpha^*,
	\end{align}
but do not couple them or introduce nonlinearities. In the special case of a displacement operator acting on the vacuum state, we obtain a coherent state of equal amplitude, \mbox{$\hat{D}(\alpha)\ket{0}=\ket\alpha$}. Note that this definition of ``linear optics'' is not universal and may only include passive transformations without displacements.

These linear optics evolutions are most commonly implemented using either:
\begin{itemize}
\item Bulk optics: discrete optical elements are arranged on an optical table.\index{Bulk optics}
\item Time-bin architectures: time-bin encoded qubits (Sec.~\ref{sec:time_bin}) evolve through delay lines and interfere at a single central optical component.\index{Time-bin encoding}\index{Fibre-loops}
\item Integrated waveguides: all passive components are etched into a chip. When optical modes are brought physically close together, evanescent coupling\index{Evanescent coupling} allows photons to coherently hop between modes. This gives rise to evolution described by the coupled oscillator Hamiltonian\index{Coupled oscillator Hamiltonian}, given in Eq.~(\ref{eq:coupled_osc_ham}), where the coupling coefficients are dictated by the proximity and geometry of the waveguides.\index{Waveguides}
\end{itemize}
Displacements are implemented by introducing coherent state (lasers) into some modes. These three main contenders are illustrated in Fig.~\ref{fig:LO_archs}.

\if 1\doublecol
	\begin{figure}[!htbp]
	\includegraphics[trim={0cm 0cm 0cm 0cm},clip=true, width=0.475\textwidth]{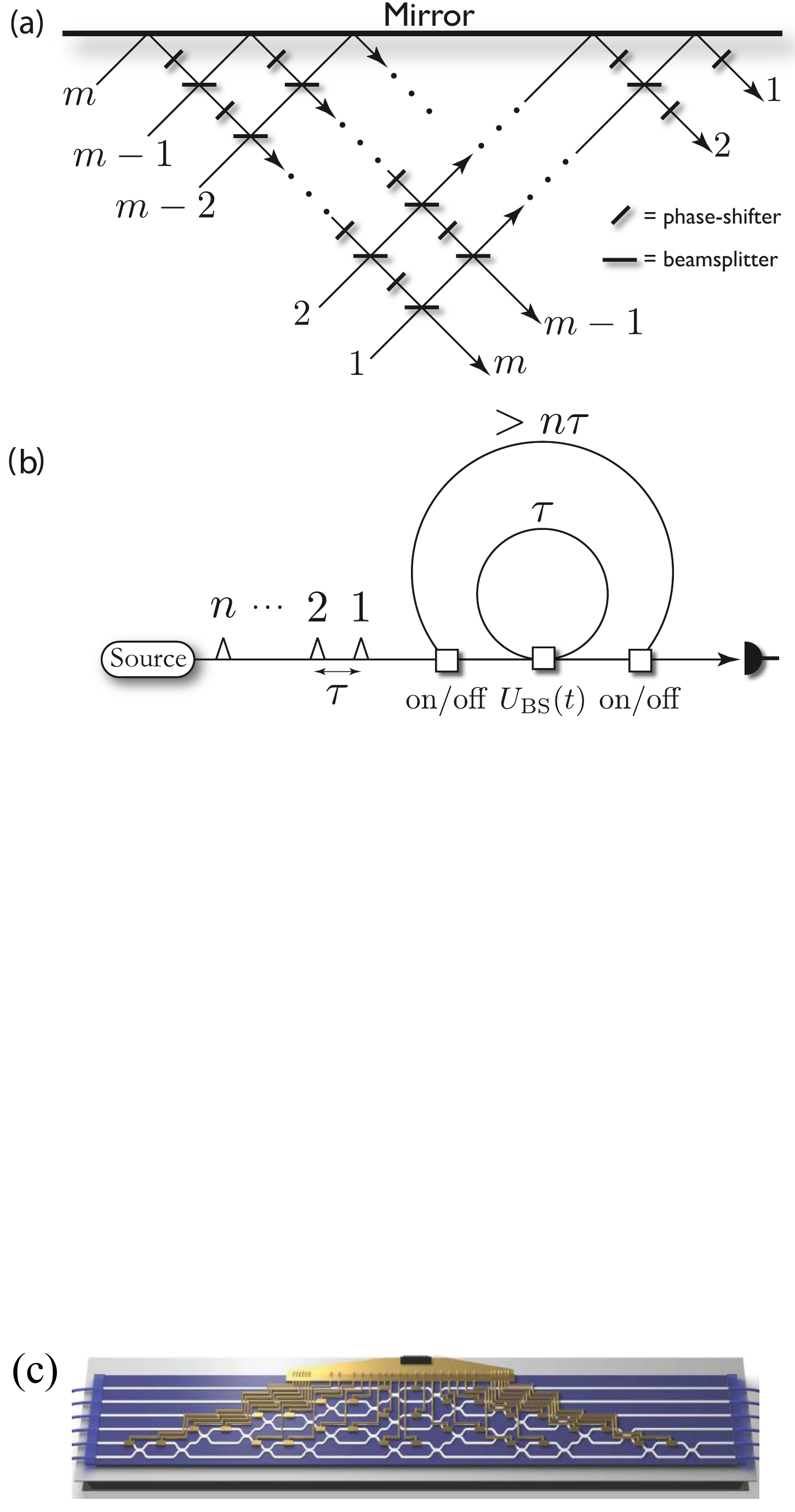}
	\captionspacefig \caption{The three primary approaches for implementing linear optics transformations. (a) Bulk optics\index{Bulk optics}, where each optical mode is spatially encoded, and the linear transformation is decomposed into a discrete array of beamsplitters and phase-shifters, an appropriate choice of which enables arbitrary linear optics transformations to be implemented. (b) Time-bin encoding\index{Time-bin encoding}\index{Fibre-loops}, where each optical mode is designated a distinct time-bin. Fibre-loops meeting at dynamically reconfigurable beamsplitters enable arbitrary linear transformations to be implemented. 
	(c) Integrated wave-guide chips\index{Waveguides}, where evanescent coupling between neighbouring wave-guides within a chip facilitates interference between modes\cite{bib:PeruzzoQW}.
	 (c) Integrated wave-guide chip with electrically controllable phase-shifters, implementing a programmable, universal \mbox{$6\times 6$} linear optics network (graphic courtesy of Jeremy O'Brien \cite{bib:carolan2015universal}).} \label{fig:LO_archs}
	\end{figure}
\else
	\begin{figure*}[!htbp]
	\captionspacefig \caption{The three primary approaches for implementing linear optics transformations. (a) Bulk optics\index{Bulk optics}, where each optical mode is spatially encoded, and the linear transformation is decomposed into a discrete array of beamsplitters and phase-shifters, an appropriate choice of which enables arbitrary linear optics transformations to be implemented. (b) Time-bin encoding\index{Time-bin encoding}\index{Fibre-loops}, where each optical mode is designated a distinct time-bin. Fibre-loops meeting at dynamically reconfigurable beamsplitters enable arbitrary linear transformations to be implemented. 
	% (c) Integrated wave-guide chips\index{Waveguides}, where evanescent coupling between neighbouring wave-guides within a chip facilitates interference between modes
	 % (graphic courtesy of Alberto Peruzzo \cite{bib:PeruzzoQW}). 
	 (c) Integrated wave-guide chip with electrically controllable phase-shifters, implementing a programmable, universal \mbox{$6\times 6$} linear optics network (graphic courtesy of Jeremy O'Brien \cite{bib:carolan2015universal}).} \label{fig:LO_archs}
	\end{figure*}
\fi

%
% Non-Linear Optics
%

\subsection{Non-linear optics} \label{sec:non_lin_opt} \index{Non-linear!Optics}

In addition to the linear transformations described above, various active, non-linear interactions are also of interest to optical quantum information processing. The most prominent of these is squeezing, discussed in Sec.~\ref{sec:squeezed_prep}, is described by the unitary operator\index{Squeezing!Operators},
\begin{align}
\hat{S}(\xi) = \exp \left[ \frac{1}{2}(\xi^*\hat{a}^2 - \xi{\hat{a}^{\dag 2}})\right],
\end{align}
where $\xi$ is the squeezing parameter, which has the effect of applying a dilation of some arbitrary factor along a particular axis in phase-space.
Together, displacements, phase shiters, and squeezing operators enable arbitrary translations, rotations, and dilations in phase space\index{Phase!Space}. These operations form the basis for CV quantum computing schemes, to be discussed in more detail in Sec.~\ref{sec:CV_QC}. More general non-linear unitary operators, such as a cubic phase gate~\cite{budinger2024phasegate}, provide access to all possible unitary transformations when supplemented with linear optics, squeezing, and displacements. However, such unitaries are not native to optical systems and can be quite tricky to engineer.

%
% Non-Optical Systems
%

\subsection{Non-optical systems}

There are countless non-optical systems applicable to quantum information processing applications. For example, quantum computing schemes have been described using:
\begin{itemize}
	\item Two-level\index{2-level atoms}.
	\item $\lambda$-configuration atoms\index{$\lambda$-configuration systems}.
	\item Superconducting rings\index{Superconductors!Rings}.
	\item Ion traps\index{Ion traps}.
	\item Atomic ensembles\index{Atomic!Ensembles}.
	\item Countless more\ldots
\end{itemize}

From a networking perspective, we are not terribly interested in the inner workings of all these schemes, as we are reasonably confident that optics will be mediating networking, even if other aspects of the protocol are non-optical. Thus, we will not go into great detail about the evolution of non-optical systems. Instead, for our purposes, the relevant issue is interfacing between optical and non-optical systems, such that networking protocols between them may be implemented. Optical interfacing is discussed in detail in Sec.~\ref{sec:opt_inter}.

%
% Quantum Memory
%

\section{Quantum memory} \label{sec:memory} \index{Quantum memory}

% \comment{Discuss atomic ensembles, two-level systems, lambda systems, 3-level systems with two ground states (better since no decay).}

\dropcap{A}{} final building block, that will be essential in many networks, is quantum memory, which simply delays a packet by some fixed amount of time, ideally implementing an identity channel ($\hat\openone$) in the non-temporal degrees of freedom. This will be required when, for example, quantum data packets reach a network bottleneck, and face one of two options: wait, or be discarded. As discussed earlier, discarding quantum packets is often a highly undesirable enterprise, as they often cannot be easily recreated, most notably when entangled with other systems. Quantum memory is an essential component in quantum repeater networks, to be discussed in Sec.~\ref{sec:rep_net}.

%
% Network Graph Representation
%

\subsection{Network graph representation}\index{Network!Graphs}

\begin{figure}[!htbp]
\includegraphics[clip=true, width=0.4\textwidth]{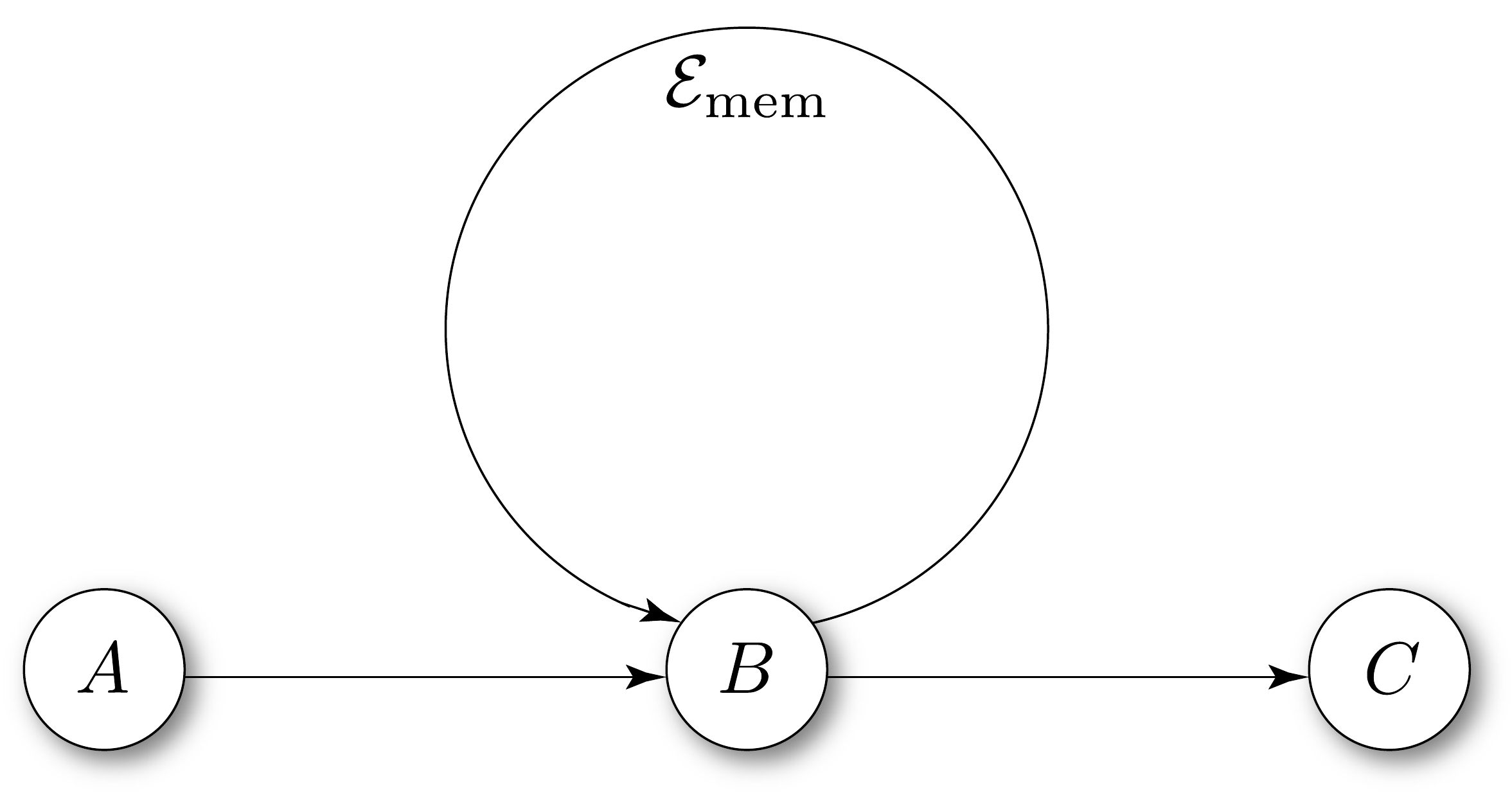}
\captionspacefig \caption{Simple model for a quantum memory via a self-loop that passes through a memory process, $\mathcal{E}_\mathrm{mem}$. Ideally, $\mathcal{E}_\mathrm{mem}$ does not affect any of the costs or attributes of states passing through the link, except for the \textsc{Latency} cost, which is incremented according to the duration of the memory.} \label{fig:memory}
\end{figure}

Quantum memory is modelled in our network graph representation as per Fig.~\ref{fig:memory}, via a self-loop implementing a process that delays packets. Ideally, the associated process should implement the identity operation in all degrees of freedom, except the temporal one, affecting only the \textsc{Lifetime} metric of the packets passing through it, incrementing it by the duration of the quantum memory.

Note that this is not directly compatible with conventional shortest-path algorithms, which ignore self-loops. One approach is to modify our strategy optimisation algorithms to accommodate self-loops. Alternately, we could construct a `virtual' graph, obtained by adding additional nodes to the network, with connections determined by `unravelling' the self-loops. For example, in Fig.~\ref{fig:memory}, we could eliminate the self-loop, and instead replace $B$ with multiple redundant nodes in series between $A$ and $C$, each associated with their own latency cost.

%
% Physical Implementation
%

\subsection{Physical implementation}

\begin{figure}[!htbp]
\includegraphics[clip=true, width=0.3\textwidth]{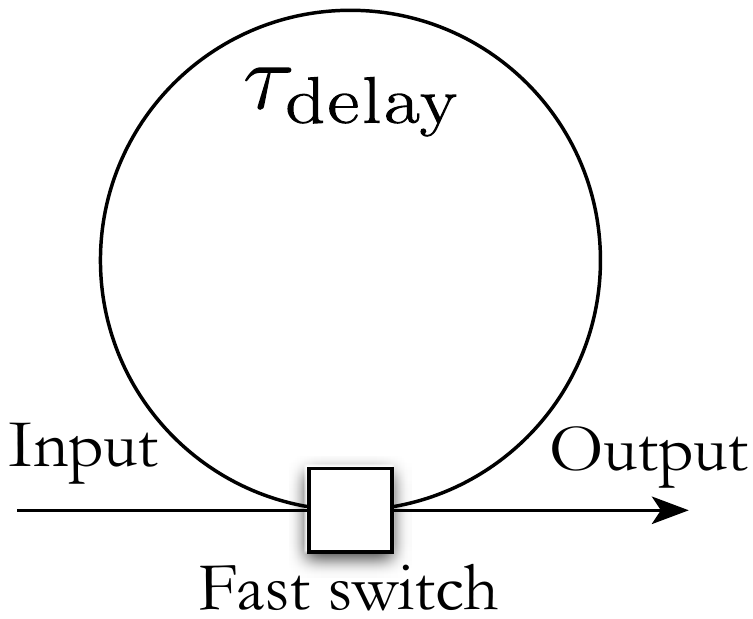}
\captionspacefig \caption{Simple architecture for a controllable optical quantum memory using an optical delay line. A length of optical fibre (or free-space), with roundtrip time $\tau_\mathrm{delay}$, couples to a fast controllable switch, with switching time \mbox{$\tau_\mathrm{switch}\ll\tau_\mathrm{delay}$}. The switch allows the optical state to be coupled into the loop, maintained there for an arbitrary number of roundtrips, and then coupled out. The storage time of the memory is restricted to being integer multiples of the roundtrip time, \mbox{$\tau_\mathrm{storage}=n\cdot\tau_\mathrm{delay}$}, where \mbox{$n\in\mathbb{Z}_+$} is the number of roundtrips. If the efficiency of the fibre-loop is $\eta$, the effective efficiency of the quantum memory after readout is \mbox{$\eta_\mathrm{eff}=\eta^n$}.}\label{fig:delay_line_mem}	
\end{figure}

At the physical level, there are two main approaches we could use to put optical states into memory. The first is simply to employ optical delay lines\index{Optical delay lines} (shown in Fig.~\ref{fig:delay_line_mem}), either in free-space or in fibre. The second is to interface the state with a non-optical system with a long coherence lifetime, which holds the information content until needed before being out-coupled. This can be achieved using, for example, the light-matter interfacing techniques discussed in Sec.~\ref{sec:opt_inter}.

The former is experimentally straightforward, but plagued by loss, and is only suitable over short timescales, on the order of nanoseconds. The latter is more experimentally challenging, but can achieve longer storage times, limited by the lifetime ($T_1$- and $T_2$-times)\index{T$_2$-time}\index{T$_1$-time} of the non-optical system. For some physical systems, this can be very high, on the order of milliseconds for atomic ensemble qubits \cite{bib:Duan01, bib:Duan02, bib:LauratKimble07}, for example, which is typically adequate for the purposes of waiting out network bottlenecks.

%
% Error correction
%

\subsection{Error correction}\index{Quantum error correction (QEC)}

Since quantum memories are subject to errors that accumulate with time, long-life quantum memories will necessarily require error correction mechanisms to preserve qubit states held within them.

To achieve this, standard QEC codes (Sec.~\ref{sec:QOS}) can be employed to encode a number of logical qubits into a larger number of physical qubits, which are held in quantum memory and undergo active error correction. Any of the previously discussed QEC techniques are applicable to this.

\begin{figure}[!htpb]
	\if 1\doublecol
		\includegraphics[clip=true, width=0.475\textwidth]{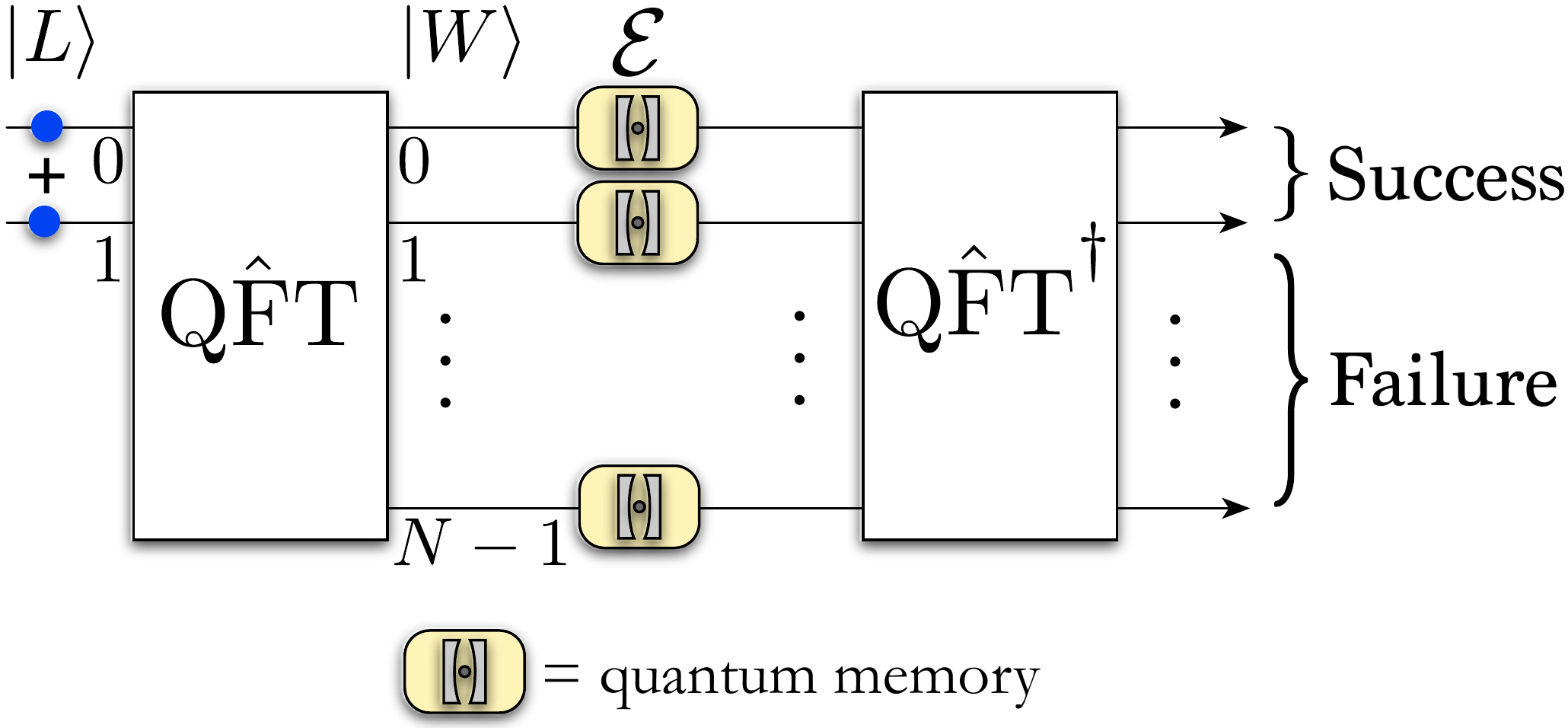} 
	\else
		\includegraphics[clip=true, width=0.7\textwidth]{W_state_memory}
	\fi
	\captionspacefig \caption{Quantum memory based on W-state encoding. Rails represent optical modes, and the input logical qubit state, $\ket{L}$, is represented using dual-rail encoding. The QFT fanout operation maps each of the two input basis states to one of two orthogonal $N$-qubit W-states, $\ket{W}$, differing only by their phase relationships. These are then fed into a bank of $N$ quantum memories. To readout the memories we convert back to optical encoding and decode using the inverse QFT operation. Success of the protocol is defined as there being exactly one photon between the first two `success' output modes. If a photon leaks into any of the other `failure' output modes we discard the state and assume it was erroneous. Thus error detection is heralded via the presence or absence of a photon in the `failure' modes.}\label{fig:W_state_memory}
\end{figure}

An alternate approach is to use W-state encoding\index{W-states!Encoding} to implement unitary error averaging (Sec.~\ref{sec:error_averaging})\index{Unitary!Error averaging}, where the unitaries are single-qubit channels. The protocol is shown in Fig.~\ref{fig:W_state_memory}. We begin by using a QFT fanout\index{Fanout} operation to encode a dual-rail encoded qubit\index{Dual-rail encoding} across $N$ optical modes. Each optical mode then feeds into a solid-state qubit (e.g a two-level atom), which ideally implement an identity channel but are inevitably subject to usual error processes such as dephasing and amplitude damping (characterised by $T_1$- and $T_2$-times\index{T$_2$-time}\index{T$_1$-time}). Finally, the inverse fanout operation is applied and success of the protocol is defined as there being exactly one photon between the first two (`success') output modes. If a photon is detected in any of the other (`failure') modes the state is presumed erroneous and discarded.

For a logical qubit of the form,
\begin{align}
\ket\psi_L &= \alpha\ket{0}+\beta\ket{1},\nonumber\\
\alpha &= \cos \left(\frac{\theta}{2}\right),\nonumber\\
\beta &= e^{i\phi}\sin \left(\frac{\theta}{2}\right),
\end{align}
the heralding success probability, and associated fidelity of the heralded events is given by,
\begin{align}
P_H &=1-e^{-t/T_1} + e^{-t/T_1}(e^{-t/T_2} + \frac{2}{N}(1-e^{-t/T_2})),\nonumber\\ 
F_H &= e^{-t/T_1}\frac{e^{-t/T_2} + (1-e^{-t/T_2})\left(\frac{2+\sin^2\theta}{2N}\right)}{e^{-t/T_2}+\frac{2}{N}(1-e^{-t/T_2})}.	
\end{align}

\begin{figure}[!htbp]
\if 1\doublecol
	\includegraphics[clip=true, width=0.475\textwidth]{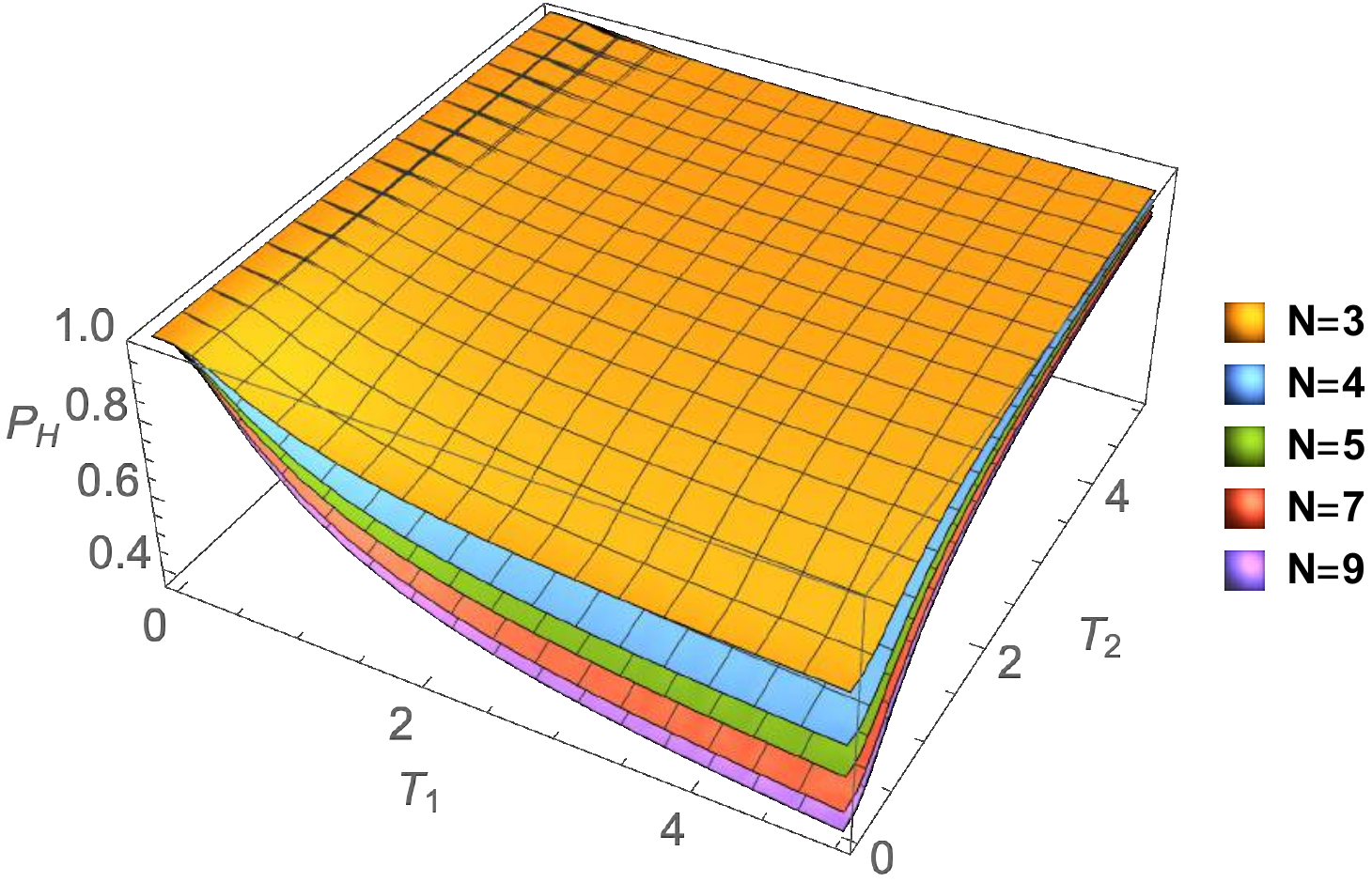}\\
	\includegraphics[clip=true, width=0.475\textwidth]{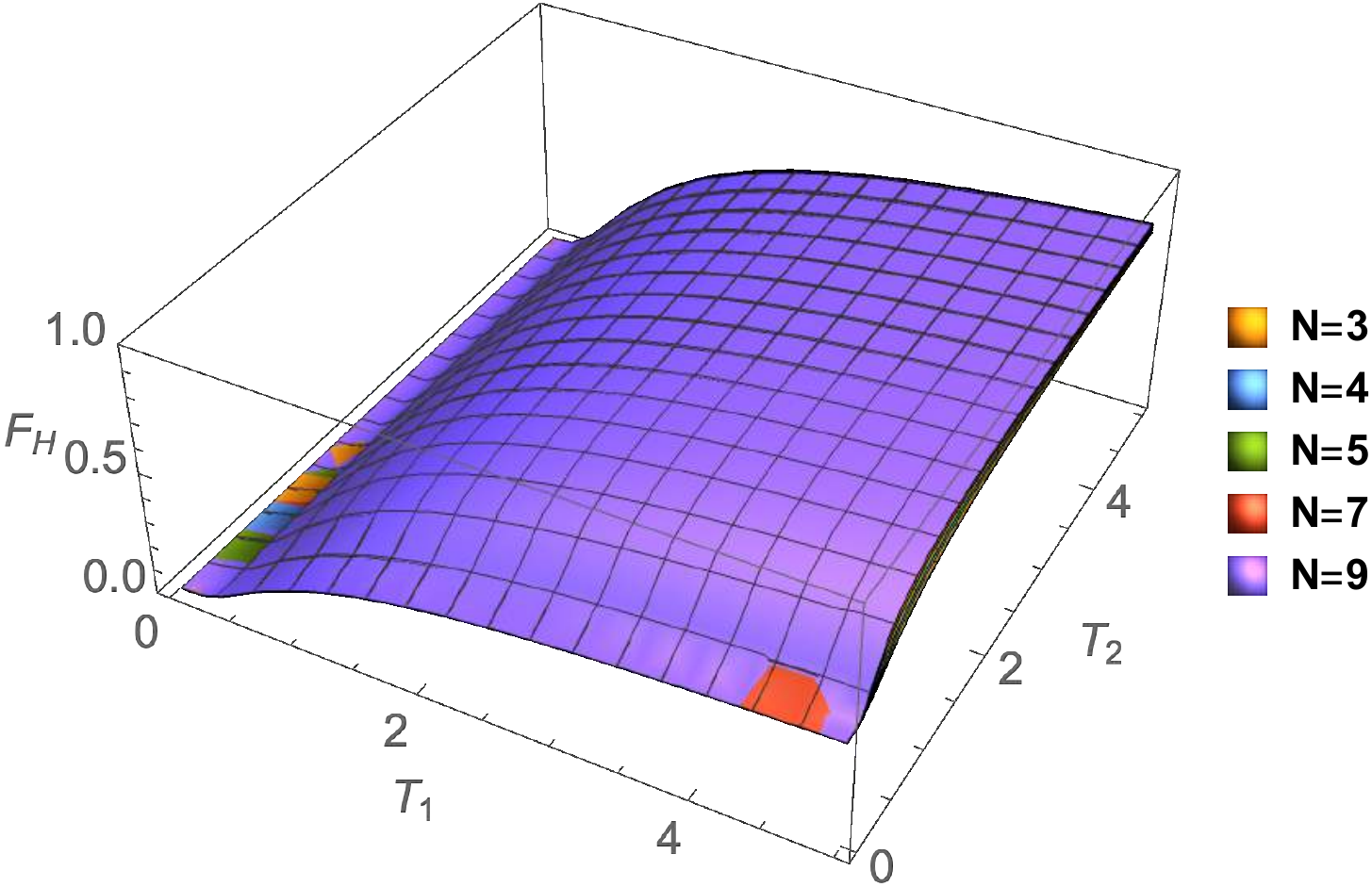}
\else
	\includegraphics[clip=true, width=0.475\textwidth]{W_state_QEC_prob}
	\includegraphics[clip=true, width=0.475\textwidth]{W_state_QEC_fidelity}
\fi
\captionspacefig \caption{Heralding probability ($P_H$) and associated error-corrected fidelity ($F_H$) for a single photonic qubit stored in a W-state quantum error correction circuit with decoherence times $T_1$ (amplitude damping)\index{Amplitude damping} and $T_2$ (dephasing)\index{Dephasing}\index{T$_1$-time}\index{T$_2$-time}. We have chosen \mbox{$\theta=0$} (representing worst-case behaviour), storage time \mbox{$t=1$}, and several levels of encoding $N$.}\label{fig:W_state_QEC_fidelity}
\end{figure}

Fig.~\ref{fig:W_state_QEC_fidelity} illustrates the fidelity of an error-corrected logical qubit using W-state encoding.

%
% High-Level Protocols
%

\section{High-level protocols} \index{High-level protocols}

\dropcap{B}{uilding} upon the aforementioned primitive protocols for quantum networking, we can construct a plethora of higher-level protocols that implement more powerful end-user applications. These high-level protocols are ubiquitous in quantum information processing and form building blocks for even more powerful architectures, such as full cloud quantum computing, to be discussed in Sec.~\ref{sec:cloud}.

%
% Random Number Generation
%

\subsection{Random number generation} \index{Random!Number generation}

\famousquote{God does not play dice!}{Albert Einstein}
\newline

Perhaps the simplest quantum information processing task is that of perfect random number generation. True random numbers have widespread applications in cryptography, Monte-Carlo simulations\index{Monte-Carlo simulations}, and any type of randomised (e.g \textbf{BPP}\index{BPP}) algorithm.

Classical random number generators are actually deterministic, following the laws of classical physics, but so difficult to predict that we accept them to be as good as random. But for some applications this isn't enough, and we must make sure that no correlations of any type exist between different random numbers, or between the random numbers and their environment.

\begin{figure}[!htpb]
	\includegraphics[clip=true, width=0.4\textwidth]{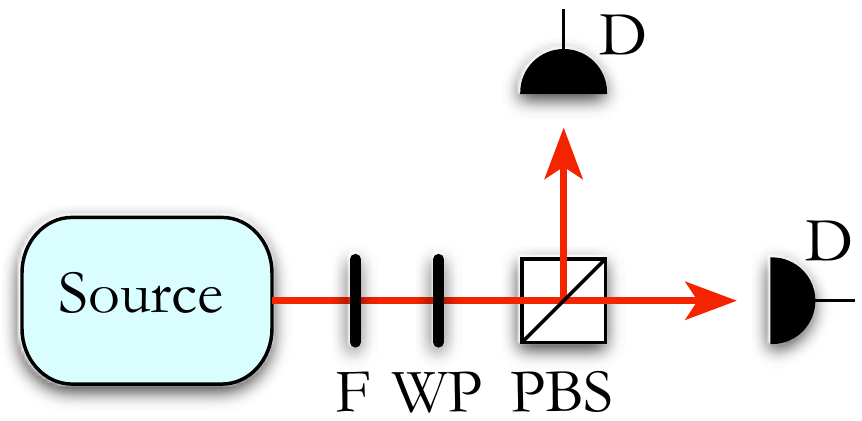}
	\captionspacefig \caption{A simple optical implementation for a quantum binary random number generator. The source prepares single photons, which pass through a polarisation filter (\textit{F}) to retain only horizontally polarised photons, $\ket{H}$. A wave-plate (\textit{WP}) then rotates the polarisation into the diagonal basis, yielding a stream of $\ket{+}$ polarised photons. These are then separated into horizontal and vertical components via a polarising beamsplitter (\textit{PBS}), each of which are independently photo-detected (\textit{D}). If correctly implemented, the two detectors click one at a time, each with 50\% probability, and the random stream is inherently non-deterministic, guaranteed by the non-determinism of quantum mechanics. Random numbers are produced at a rate of 1 bit per photo-detection.}\label{fig:random_number_generation}
\end{figure}

This can be achieved in many different ways quantum mechanically. Ultimately, they are all based on the Heisenberg uncertainty principle, that certain quantum mechanical measurements yield uncertainty. The procedure is shown in Alg.~\ref{alg:random_number}, and a simple optical implementation is shown in Fig.~\ref{fig:random_number_generation}.

\startalgtable
\begin{table}[!htbp]
\begin{mdframed}[innertopmargin=3pt, innerbottommargin=3pt, nobreak]
\texttt{
function RandomBit():
\begin{enumerate}
    \item Prepare the equal superposition state,
    \begin{align}
    \ket\psi_\mathrm{in} = \ket{+} = \frac{1}{\sqrt{2}}(\ket{0}+\ket{1}).
    \end{align}
    \item Most commonly this is in the polarisation basis,
    \begin{align}
    \ket{H}&\equiv\ket{0}, \nonumber \\
    \ket{V}&\equiv\ket{1}.
    \end{align}
    \item Measure the state in the logical basis, with measurement projectors,
    \begin{align}
    \hat\Pi_0 &= \ket{0}\bra{0}, \nonumber \\
    \hat\Pi_1 &= \ket{1}\bra{1}.
    \end{align}
    \item The measurement outcomes occur with probabilities,
    \begin{align}
    P_0&=|\braket{0|+}|^2 = \frac{1}{2}, \nonumber \\
    P_1&=|\braket{1|+}|^2 = \frac{1}{2},
    \end{align}
    following a uniform, random, binary distribution.
    \item Repeat for as many random bits as are required.
    \item $\Box$
\end{enumerate}}
\end{mdframed}
\captionspacealg \caption{Procedure for the generation of random bit-strings. Assuming the device is perfectly implementing this procedure, we will measure a perfect random 50/50 distribution between the two measurement outcomes. Note that the procedure requires no quantum interference, and no entanglement. Only single-qubit state preparation and measurement are required. Thus, a single-photon source, wave-plate, polarisation filter, and photo-detector are sufficient for its realisation. Favourably, if the detector is inefficient it simply reduces the bit-rate, but does not compromise the randomness of the distribution. The scheme is very robust, in the sense that there are no temporal synchronisation or mode-matching requirements.} \label{alg:random_number}
\end{table}

The cynics amongst us might question the non-determinism\index{Non-determinism of quantum mechanics} of the laws of Nature, and ask whether quantum random numbers really are truly random (in the sense of non-determinism), or whether they also are just too hard to predict that we treat them as effectively random. The answer to this is that it has been proven that quantum mechanics is inconsistent with `hidden variable theories' \cite{bohm1966proposed} \index{Hidden variable theories}, i.e that there is an underlying, but inaccessible determinism in the world, which is guiding quantum measurements in a completely deterministic manner. This disproof effectively validates the notion of quantum mechanical perfect random number generation.

Consider the scenario where a client needs a stream of true random numbers for use in her Monte-Carlo simulation algorithm or as a secret-key for her email encryption. She has limited quantum resources herself, so she outsources it to her better-equipped mate. Depending on her own resource limitations and potential security considerations, her friend could either: (1) implement the full protocol described above, providing her with a classical random bitstream; or (2) only take care of photon generation, providing her with a perpetual source of high-quality photons for her to measure herself using a simple photo-detector. (1) and (2) would both be suitable if the intention was to apply the source of randomness to a Monte-Carlo simulation. But in a cryptographic scenario, where the randomness is being used for key generation, clearly Alice could not outsource the measurement stage without revealing her secret-key. In this instance, Bob can act as the provider of photons, while Alice does the measurements herself so as to keep her random bit-string secret.

For cryptographic purposes, there are far more stringent constraints placed upon our random number generator than for use in say a Monte-Carlo computer simulation -- cryptographic random number generation\index{Cryptographic!Random number generation}. In this context, the demands placed upon the amount of bias or correlations in the random number stream are very stringent. An enormous amount of research has been invested into this topic, and sophisticated statistical tests have been developed for establishing crypto-worthiness of a random number stream. The fact that quantum random numbers, via the inherent non-determinism of quantum mechanics, do not obey any hidden variable theory, implies there is intrinsically no underlying `seed'\index{Random!Seed} to the random number stream, which reveals the entire deterministic sequence. This is unlike any classical generator, where there always is such a seed, but it is simply taken to be too hard to determine.

This scenario is an obvious example of where a UDP-like \textsc{Send-and-forget} protocol may be viable. Unlike most other applications, Bob is broadcasting a stream of identical, pure quantum states, that are not entangled with any peripheral system, and are easily replicated, with no correlations between distinct photons. Thus, if any particular photon fails to reach Alice, it matters not, as she can simply await the next one emanating from Bob's bombardment of photons (the `shotgun' approach\index{Shotgun!Protocol}). There are no QoS requirements.

%
% Entanglement Purification
%

\subsection{Entanglement purification} \label{sec:ent_purif} \index{Entanglement!Purification}

Entangled states, most notably Bell pairs (Sec.~\ref{sec:bell_state_res}), play a central role in many quantum technologies. These maximally entangled states are easily represented optically using polarisation encoding of single photons, and can be non-deterministically prepared directly using SPDC (Sec.~\ref{sec:single_phot_src}), or post-selected linear optics.

Bell pairs are the basis for building cluster states (Sec.~\ref{sec:CSQC}), some quantum cryptography protocols (Sec.~\ref{sec:QKD}), and quantum teleportation (Sec.~\ref{sec:teleport}), to name just a few applications. Therefore distributing entangled states with the highest entanglement metrics is extremely important. In short, entanglement can be considered a valuable quantum resource (discussed in detail in Sec.~\ref{sec:ent_ultimate}), upon which many other protocols may be built.

Suppose Alice and Bob share an entangled pair. Quantum mechanics, specifically the very definition of entanglement itself, prohibits local operations performed by Alice and Bob from increasing the level of entanglement. However, if Alice and Bob share multiple pairs, they can perform an operation known as \textit{entanglement purification} or \textit{entanglement distillation}, whereby two lower-fidelity entangled pairs are consumed and projected onto a single entangled pair with higher fidelity \cite{bib:PRA_53_2046, bib:PRA_54_3824, bib:Deutsch96}. Such protocols will be extremely useful in protocols where achieving the highest possible degree of entanglement is paramount, for example when error thresholds must be achieved for the purpose of error-correction and fault-tolerance \cite{bib:NielsenChuang00}.

Taking two polarisation-encoded photonic Bell pairs, say $\ket{\Psi^+}$, and subjecting them to a dephasing error model (Sec.~\ref{sec:dephasing_error})\index{Dephasing!Channel} yields a mixed state of the form,
\begin{align}
\hat\rho_\mathrm{in} = F\ket{\Psi^+}\bra{\Psi^+} + (1-F)\ket{\Psi^-}\bra{\Psi^-},
\end{align}
where $F$ is the entanglement fidelity, which is a function of the dephasing rate. Note that $\ket{\Psi^+}$ and $\ket{\Psi^-}$ are related by local Pauli phase-flip operations ($\hat{Z}$) applied to either qubit,
\begin{align} \label{eq:psi_minus}
\ket{\Psi^-} = \hat{Z}_A \ket{\Psi^+} = \hat{Z}_B \ket{\Psi^+}.
\end{align}

\begin{figure}[!htbp]
\includegraphics[clip=true, width=0.45\textwidth]{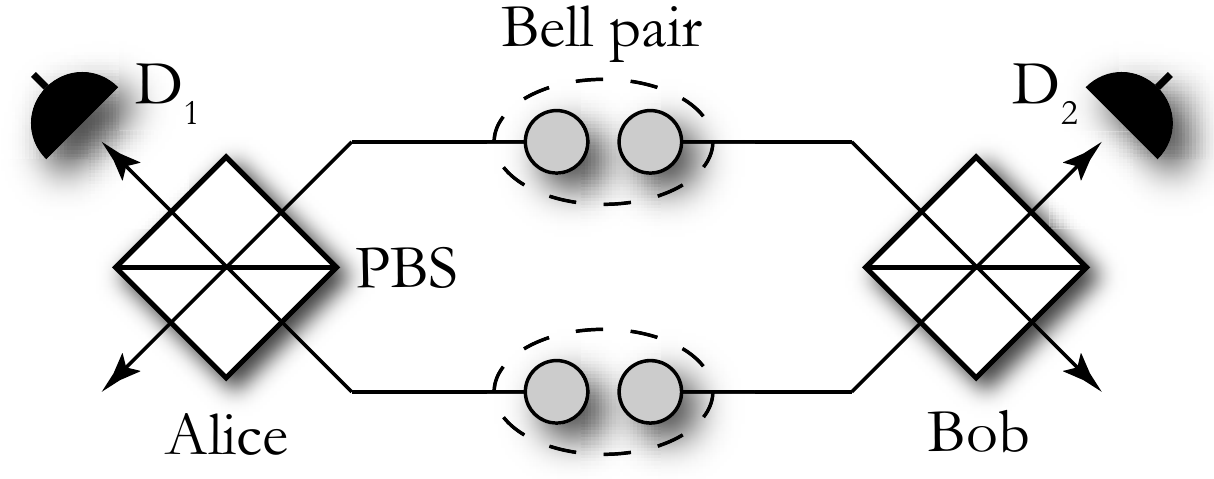}
\captionspacefig \caption{Elementary entanglement purification using linear optics. Two Bell pairs are distributed between Alice and Bob, each of which has been subject to a dephasing error model. Alice and Bob perform Bell measurements on their two qubits using a PBS and polarisation-resolved photo-detection ($D_1$ and $D_2$). Upon successful Bell state projection (Bell measurements are necessarily non-deterministic using linear optics), Alice and Bob will share a single Bell pair with higher fidelity than the two input pairs.} \label{fig:ent_purif_prot}
\end{figure}

A linear optics entanglement purification protocol can be simply implemented using two polarising beamsplitters PBSs \cite{bib:Pan01, bib:Pan03}. Alice uses one PBS to interfere the photons from her side of each of the photon pairs, measuring one output only, which implements a non-deterministic, partial Bell state projection. Bob does the same on his side. What's left is one photon in Alice's hands and one in Bob's. When successful, they will now be sharing a single entangled pair of higher Bell state fidelity than the two starting states. The protocol is shown in Fig.~\ref{fig:ent_purif_prot}.

Note that when using PBSs to perform the Bell projections, the protocol is necessarily non-deterministic, since PBSs are only able to distinguish two of the four Bell states. Thus, each PBS has a success probability of $1/2$. And there are two PBSs per instance of the protocol, therefore the net success probability is $1/4$. When concatenated, $n$ applications of the protocol thus has an exponentially low success probability of $1/4^n$. This could be overcome using deterministic CNOT gates (Sec.~\ref{sec:KLM_univ}), but these are challenging using linear optics.

Furthermore, the protocol consumes two Bell pairs upon each trial, only one quarter of which are successful. Thus, on average, 8 Bell pairs are consumed for every purified Bell pair prepared, and the expected number of Bell pairs required to perform $n$ iterations of entanglement purification grows exponentially as $8^n$.

\begin{figure}[!htbp]
\includegraphics[clip=true, width=0.475\textwidth]{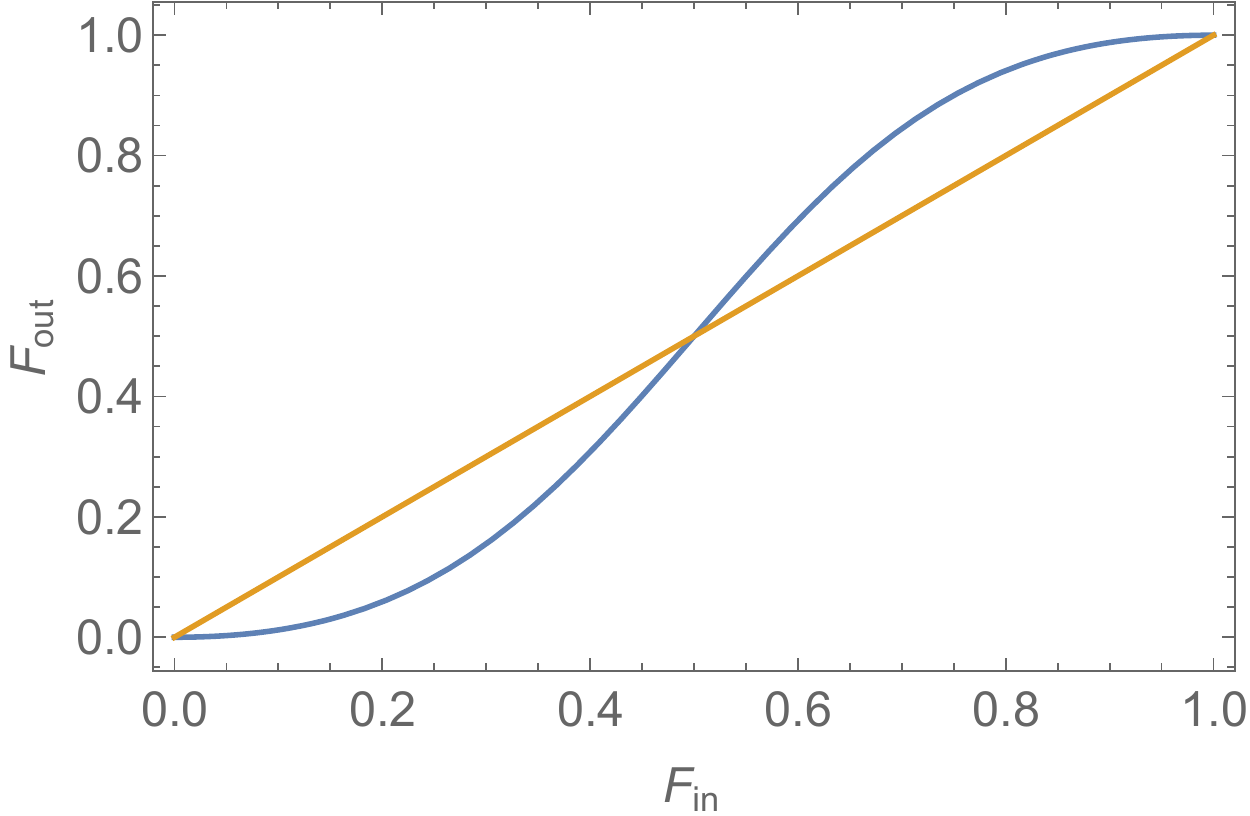}
\captionspacefig \caption{Entanglement purification of two polarisation-encoded, photonic Bell pairs. $F_\mathrm{in}$ ($F_\mathrm{out}$) are the input (output) fidelities of the Bell pairs. The protocol consumes two Bell pairs for every one purified pair. The straight line represents the break-even point in terms of state fidelity, above which the protocol enhances fidelity, and below which reduces it. This places a strict bound on the fidelity of Bell pairs reaching the purifier. This equates to route cost, if measured by the fidelity metric, stipulating network performance requirements. This threshold requirement presents an example of where an \textsc{All or Nothing} strategy might be appropriate.} \label{fig:ent_purif}
\end{figure}

Specifically, the relationship between the input ($F_\mathrm{in}$) and output ($F_\mathrm{out}$) fidelities of the protocol is,
\begin{align}
F_\mathrm{out} = \frac{{F_\mathrm{in}}^2}{{F_\mathrm{in}}^2 + (1-F_\mathrm{in})^2}.
\end{align}
This input/output relationship is shown in Fig.~\ref{fig:ent_purif}.

Note that there is a break-even point, above which the protocol strictly increases fidelity, and below which strictly decreases it. This occurs at \mbox{$F_\mathrm{in}=1/2$}. Provided pairs can be communicated above this fidelity threshold, bootstrapped application of the protocol could be employed to boost entanglement fidelity asymptotically close to unity (but with exponential resource overhead, since each operation non-deterministically consumes two pairs to produce one). But below this threshold it is impossible to recover any more entanglement than we started with. This provides an example of an application where the protocol being implemented dictates strict requirements on network cost metrics. Specifically, assuming perfect Bell pairs to begin with, the routes by which they are communicated must strictly ensure entanglement fidelities of at least \mbox{$F=1/2$} upon reaching their destination. Here, a type of \textsc{All or Nothing} networking strategy would be applicable -- if the fidelity requirement is not met, the state cannot be purified and might as well be thrown away to make way for other traffic.

A theoretical analysis of this protocol has been performed, accounting for mode-mismatch (Sec.~\ref{sec:MM_error}) in the protocol \cite{bib:RohdeOptEntPur06}, where it was found that mode-mismatch shifts the break-even point upwards, and lowers the maximum value of $F_\mathrm{out}$ -- with more mode-mismatch, a higher starting fidelity is required to break even, and we achieve a lower, sub-unity output fidelity. In this case, a cost function that combines the dephasing and mode-mismatch metrics of the network will be required.

Importantly, this protocol is based on partial Bell state measurement, and therefore does not require interferometric stability, only high HOM visibility, thus making stabilisation comparatively easy over long distances.

Entanglement purification can also be performed using physical encodings other than single photons. For example, this has been demonstrated using Gaussian CV quantum states \cite{bib:Duan00}.

%
% Quantum State Teleportation
%

\subsection{Quantum state teleportation} \label{sec:teleport} \index{Quantum state teleportation}

Quantum state teleportation \cite{bib:PRL_70_1895} is an essential ingredient in many higher-level protocols. It forms the basis of cluster state quantum computing (Sec.~\ref{sec:CSQC}), some QEC codes, the KLM linear optics quantum computing scheme (Sec.~\ref{sec:KLM_univ}), and can act as a mediator for long-range transmission of quantum states, amongst others.

In the standard teleportation protocol, Alice begins with a single qubit,
\begin{align}
\ket\phi = \alpha\ket{0} +\beta\ket{1},
\end{align}
which she would like to teleport to Bob. Importantly, no quantum communication between the two is allowed, since obviously this would make the problem trivial. However, classical communication is allowed (and turns out to be necessary), and furthermore they share an entangled Bell pair as a resource. Thus, Alice begins with two qubits, and Bob begins with one -- his half of the entangled pair onto which Alice's state ought to be teleported. The initial state is therefore,
\begin{align}
\ket\psi_\mathrm{in} &= \ket{\phi}_{A_1} \ket{\Psi^+}_{A_2,B} \\
&= \frac{1}{\sqrt{2}} (\alpha\ket{0}_{A_1}+\beta\ket{1}_{A_1}) (\ket{0}_{A_2}\ket{1}_B + \ket{1}_{A_2}\ket{0}_B). \nonumber
\end{align}

The first step of the protocol is for Alice to perform a 2-qubit entangling measurement on her two qubits, projecting onto the Bell basis, Eq.~(\ref{eq:bell_basis}). She obtains one of four measurement outcomes. For illustration, suppose she measures the $\ket{\Psi^+}$ outcome. Then the projected state is,
\begin{align}
\ket\psi_\mathrm{proj}^{\Psi^+} &= \bra{\Psi^+}_{A_1,A_2} \ket\psi_\mathrm{in} \nonumber \\
&= \frac{1}{\sqrt{2}} \bra{\Psi^+}_{A_1,A_2}\ket\psi_{A_1}(\ket{0}_{A_2}\ket{1}_B + \ket{1}_{A_2}\ket{0}_B) \nonumber \\
&= \frac{1}{2} (\bra{0}_{A_1}\bra{1}_{A_2} + \bra{1}_{A_1}\bra{0}_{A_2}) \nonumber \\
&\cdot (\alpha\ket{0}_{A_1}+\beta\ket{1}_{A_1})(\ket{0}_{A_2}\ket{1}_B + \ket{1}_{A_2}\ket{0}_B) \nonumber \\
&= \frac{1}{2} (\alpha \ket{0}_B + \beta \ket{1}_B)\nonumber \\
&= \frac{1}{2} \ket\phi_B,
\end{align}
which is Alice's initial state. For all four possible Bell measurement outcomes we have,
\begin{align}
\ket\psi_\mathrm{proj}^{\Psi^+} &= \frac{1}{2} (\alpha \ket{0}_B + \beta \ket{1}_B) \nonumber \\
&= \frac{1}{2} \ket\phi_B, \nonumber \\
\ket\psi_\mathrm{proj}^{\Psi^-} &= \frac{1}{2} (\alpha \ket{0}_B - \beta \ket{1}_B) \nonumber \\
&= \frac{1}{2} \hat{Z}\ket\phi_B, \nonumber \\
\ket\psi_\mathrm{proj}^{\Phi^+} &= \frac{1}{2} (\alpha \ket{1}_B + \beta \ket{0}_B) \nonumber \\
&= \frac{1}{2} \hat{X} \ket\phi_B, \nonumber \\
\ket\psi_\mathrm{proj}^{\Phi^-} &= \frac{1}{2} (\alpha \ket{1}_B - \beta \ket{0}_B) \nonumber \\
&= \frac{1}{2} \hat{X}\hat{Z}\ket\phi_B,
\end{align}
which are all locally equivalent to $\ket\phi$ under Pauli gates, and can be corrected by Bob, given communication of the classical Bell measurement outcome provided by Alice. The full protocol is described in Alg.~\ref{alg:state_teleport}.

\begin{table}[!htbp]
\begin{mdframed}[innertopmargin=3pt, innerbottommargin=3pt, nobreak]
\texttt{
function StateTeleportation($\ket\phi_{A_1}$, $\ket{\Phi^+}_{A_2,B}$):
\begin{enumerate}
    \item Alice prepares the state $\ket\phi_{A_1}$, which she would like to teleport to Bob.
    \item Alice and Bob share the Bell pair $\ket{\Phi^+}_{A_2,B}$.
    \item Alice performs a Bell state projection between qubits $A_1$ and $A_2$.
    \item Alice communicates the classical measurement outcome to Bob - one of four outcomes.
    \item Bob applies an appropriate local correction to his qubit - some combination of the Pauli operators $\hat{X}$ and $\hat{Z}$ - according to the classical measurement outcome:
    \begin{align}
    \ket{\Psi^+}\bra{\Psi^+} &\to \hat\openone, \nonumber \\
    \ket{\Psi^-}\bra{\Psi^-} &\to \hat{Z}, \nonumber \\
    \ket{\Phi^+}\bra{\Phi^+} &\to \hat{X}, \nonumber \\
    \ket{\Phi^-}\bra{\Phi^-} &\to \hat{Z}\hat{X}.
    \end{align}
    \item Bob is left with the state $\ket\phi_B$.
    \item $\Box$
\end{enumerate}}
\begin{align}
\Qcircuit @C=1em @R=1.6em {
    \lstick{\ket\phi} & \qw & \multimeasureD{1}{\mathrm{Bell}} & \cw  & \control \cw \\
    \lstick{} & \qw & \ghost{\mathrm{Bell}} & \control \cw & \cwx \\
    \lstick{} & \qw & \qw & \gate{X} \cwx & \gate{Z} \cwx & \qw & \qw & \ket\phi
    \inputgroupv{2}{3}{.8em}{.8em}{\ket{\Phi^+}}
} \nonumber
\end{align}
\end{mdframed}
\captionspacealg \caption{Quantum state teleportation of a single qubit.} \label{alg:state_teleport}
\end{table}

In general, the protocol is deterministic, although using PBSs to perform partial Bell measurements, the success probability is at most $1/2$.

The question now is what error metrics apply and how do they accumulate in the teleportation protocol. The answer is straightforward -- the final teleported qubit accumulates all local Pauli errors (e.g dephasing or depolarisation) associated with Alice's input state as well as any that acted upon the shared Bell pair. That is, the errors get teleported along with the state being teleported, plus any errors on the Bell pair.

In the case of loss, loss of either of Alice's qubits will immediately be detected when she performs her Bell measurement. Thus, loss becomes a located error, and the knowledge of the error allows the associated packet to be discarded, and the sender and recipient notified. On the other hand, loss of Bob's qubit will behave no differently than loss acting on an ordinary qubit channel.

Thus, in terms of Pauli errors, no special treatment is required by the QTCP protocol -- it is almost as if the teleportation protocol weren't there. And in terms of loss, the Bell state projection diagnoses lost qubits, allowing appropriate action to be taken, which is actually better than if the error were undiagnosed. These are often referred as \textit{located}\index{Located errors} and \textit{unlocated}\index{Unlocated errors} errors.

The total resources required to teleport a single-qubit state are:
\begin{enumerate}
\item The qubit to be teleported.
\item A shared Bell pair.
\item A 2-qubit entangling measurement in the Bell basis.
\item The transmission of two classical bits.
\item Two classically-controlled Pauli gates for correction.
\end{enumerate}

This is more costly than sending the qubit directly over a quantum channel, but may be the only approach if a direct link is not available. In the context of an internet where entanglement distribution is treated as the fundamental resource (Sec.~\ref{sec:ent_ultimate}), state teleportation is the natural approach for communicating quantum states, since no quantum communication of any kind is required once the two parties have a shared Bell pair between them.

The important feature of this protocol to note is that there is no direct quantum communication between Alice and Bob, only a classical communications channel. Rather, the Bell pair mediates the transfer of quantum information, despite there being no direct quantum channel between Alice and Bob.

Relying on teleportation rather than direct quantum communication makes frugal use of quantum channels, since there is no need for direct quantum routes between every pair of nodes in the network. Instead, each node need only have a direct one-way quantum link with the central authority responsible for entanglement distribution, thereby significantly reducing the complexity of the topology of the quantum network.

The Bell state measurement can be implemented either using a CNOT gate, or as a non-deterministic partial Bell state measurement using a PBS (Sec.~\ref{sec:bell_proj}), both of which are non-deterministic using purely linear optics.

The above describes quantum state teleportation at the level of single qubits. However, when dealing with more general QTCP packets, which may have multi-qubit payloads, we may wish to teleport an entire packet\index{Packet!Teleportation}. This is implemented as a simple extension of the above procedure -- we simply implement $n$ multiple independent teleportation protocols to all of the packet's $n$ constituent qubits. Via linearly, although the teleportation protocols are being applied independently to each qubit, the net packet teleportation operation will preserve their joint state, including entanglement between them. Note, however, that if the qubit state teleportation protocols are individually non-deterministic with success probability $p_\mathrm{teleport}$, the net success probability for the teleportation of the entire packet scales inverse exponentially with $n$, as ${p_\mathrm{teleport}}^n$.

%
% Open-destinations
%

\subsubsection{Open-destinations}\index{Open-destination quantum state teleportation}\label{sec:open_dest_qst}

In the standard quantum state teleportation protocol there is a single sender and a single receiver. A generalisation of the protocol is \textit{open-destination quantum state teleportation}, whereby there is still just one sender, but any number of potential recipients. At the time of transmission, the sender does not specify the recipient, but rather wishes to `broadcast'\index{Broadcasting} the state to \textit{all} recipients, such that any \textit{one} of them can subsequently read out the state. Note that only a single recipient may actually perform the readout, since multiple readouts would violate the no-cloning theorem\index{No-cloning theorem}. However, the key new feature introduced by this variant of the protocol is that the final choice of which recipient performs the readout needn't be known in advance, but can be decided at an arbitrary later stage, well after the sender has completed their side of the protocol.

\if 1\doublecol
	\begin{figure}[!htbp]
	\includegraphics[clip=true, width=0.475\textwidth]{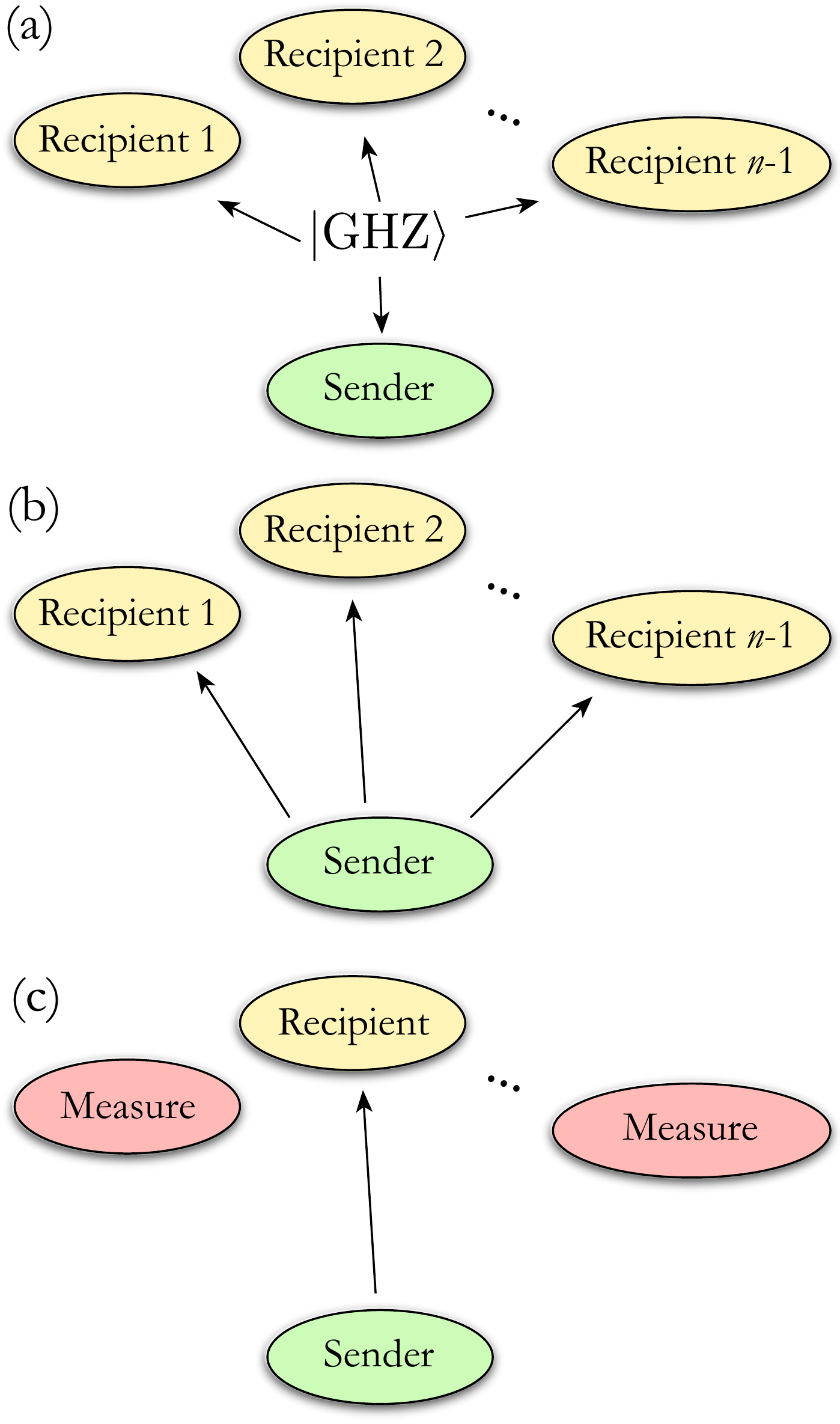}
	\captionspacefig \caption{Open-destination quantum state teleportation. (a) GHZ state distribution from a central server between the sender and all potential recipients. (b) Sender performs quantum state teleportation, resulting in the state being teleported to all recipients in redundantly-encoded\index{Redundant encoding} form. (c) All non-receivers measure out their qubits in the $\hat{X}$-basis, resulting in the teleported state arriving at the destination of just the chosen true receiver.}\index{Open-destination quantum state teleportation}\label{fig:open_dest_teleport}
	\end{figure}
\else
	\begin{figure*}[!htbp]
	\includegraphics[clip=true, width=\textwidth]{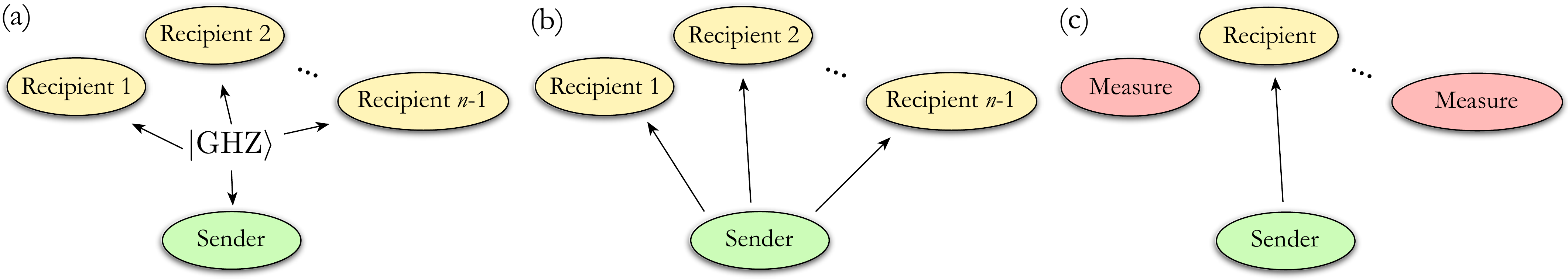}
	\captionspacefig \caption{Open-destination quantum state teleportation. (a) GHZ state distribution from a central server between the sender and all potential recipients. (b) Sender performs quantum state teleportation, resulting in the state being teleported to all recipients in redundantly-encoded\index{Redundant encoding} form. (c) All non-receivers measure out their qubits in the $\hat{X}$-basis, resulting in the teleported state arriving at the destination of just the chosen true receiver.}\index{Open-destination quantum state teleportation}\label{fig:open_dest_teleport}
	\end{figure*}
\fi

To implement this protocol, outlined in Fig.~\ref{fig:open_dest_teleport}, rather than first distributing a Bell pair between sender and recipient, we distribute an $n$-party GHZ state between the sender and the \mbox{$n-1$} recipients,
\begin{align}
\ket{\psi_\mathrm{GHZ}^{(n)}} = \frac{1}{\sqrt{2}}(\ket{0}^{\otimes n} + \ket{1}^{\otimes n}).
\end{align}

Next, note that performing an $\hat{X}$-basis measurement on one qubit in a GHZ state reduces it to an \mbox{$n-1$}-qubit GHZ state, up to a potential local $\hat{Z}$-correction,
\begin{align}
	\braket{+|\psi_\mathrm{GHZ}^{(n)}} &= \ket{\psi_\mathrm{GHZ}^{(n-1)}},\nonumber\\
	\braket{-|\psi_\mathrm{GHZ}^{(n)}} &= \hat{Z}\ket{\psi_\mathrm{GHZ}^{(n-1)}}.
\end{align}
Performing this contraction repeatedly ultimately reduces us all the way down to a single Bell pair, since,
\begin{align}
	\ket{\psi_\mathrm{GHZ}^{(2)}} = \ket{\Phi^+}.
\end{align}

Thus, to teleport from sender to recipient, all non-recipients simply measure their qubits in the $\hat{X}$-basis, and publicly report their classical measurement outcomes for the purposes of performing local corrections. What is left is a Bell pair between sender and recipient. This Bell pair may then be employed in the conventional teleportation protocol to perform the teleportation.

The key observation now is that the dynamics of this entire system must be invariant under the time-ordering of the $\hat{X}$-basis measurements. Thus, whether they are performed at the beginning of the protocol (thereby directly reducing us to standard 2-party teleportation), or deferred until later, makes no difference to the final state obtained by the recipient. This is the basis for the ability of the protocol to broadcast the teleported state, without first specifying the intended recipient.

The full protocol is described in Alg.~\ref{alg:open_dest_qst}. This protocol was successfully experimentally demonstrated photonically using 5 qubits \cite{bib:PanOpenDest}, with high teleported state fidelities.

\begin{table}[!htbp]
\begin{mdframed}[innertopmargin=3pt, innerbottommargin=3pt, nobreak]
\texttt{
function OpenDestTeleportation($\ket\phi_A$, $\ket{\psi_\mathrm{GHZ}^{(n)}}$):
\begin{enumerate}
    \item An $n$-qubit GHZ state is distributed between the sender and \mbox{$n-1$} potential recipients,
    \begin{align}
    	\ket{\psi_\mathrm{GHZ}^{(n)}} = \frac{1}{\sqrt{2}}(\ket{0}^{\otimes n} + \ket{1}^{\otimes n}).
    \end{align}
    \item The sender completes their side of the protocol as per the ordinary 2-party quantum state teleportation protocol.
    \item The sender's classical measurement outcome is broadcast to all potential recipients.
    \item Wait until the potential recipients have decided between themselves (classically) who shall be the recipient.
    \item All non-recipients measure their qubits in the $\hat{X}$-basis and publicly announce their measurement outcomes.
    \item The recipient calculates the parity of the announced measurement outcomes, which determines whether they apply a local $\hat{Z}$-correction.
    \item The recipient completes their side of the protocol as per the usual quantum state teleportation protocol, obtaining $\ket\phi$.
    \item $\Box$
\end{enumerate}}
\end{mdframed}
\captionspacealg \caption{Open-destination quantum state teleportation of a single qubit from a single sender to \mbox{$n-1$} potential recipients.}\index{Open-destination quantum state teleportation}\label{alg:open_dest_qst}
\end{table}

\subsubsection{CV Teleportation} \label{sec:CVteleportation} \index{CV teleportation}

The continuous-variable analogue to the above teleportation protocol is more general in the sense that \emph{any} state can be teleported, not just a state encoded in the subspace spanned by vacuum and a single photon. In CV teleportation, Alice holds an arbitrary pure state that she would like to transmit to Bob,
	\begin{align}
		\ket{\phi} = \int dx \, \psi(x) \ket{x}
	\end{align}
here described by the position wave function $\psi(x) = \braket{ \psi | x } $. To make this possible, Alice also holds one half of an EPR state. The other half is held by Bob, who may be spacelike separated from Alice. The initial state is given by
	\begin{align}
	    \ket\psi_\mathrm{in} &= \ket{\phi}_{A_1} \otimes \ket{\text{EPR}}_{A_2,B}.
	\end{align}
Just as above, Bob and Alice are not able to perform any quantum communication or joint quantum measurements, but classical communication is allowed. 

To teleport the state, Alice performs an EPR measurement of the two modes she holds by mixing them on a beam splitter and using homodyne detectors set to measure position on one mode and momentum on the other. This yields two real-number outcomes, $m_1$ and $m_2$, and projects the input state into 
	\begin{align}
	    \ket{\psi}^{m_1,m_2}_\mathrm{proj} &= \bra{\text{EPR} (m_1,m_2)}_{A_1,A_2} \ket{\psi}_\mathrm{in}.
	\end{align}
Using CV identities~\cite{walshe2020gateteleportation}, we can show that this is equal to  
	\begin{align}
	    \ket{\psi}^{m_1,m_2}_\mathrm{proj} = \hat{D}^\dagger \big( \tfrac{1}{\sqrt{2}} (m_1 + i m_2) \big) \ket{\phi}_B
	\end{align}
where $\hat{D}$ is a displacement operator. Alice then sends her measurement outcomes to Bob using their classical communication channel, which he can use to undo the displacement and fully recover the teleported state $\ket{\phi}$.

In practice, the EPR measurement can be performed with high fidelity, but the shared EPR state inevitably contains noise, even if it is pure. That's because EPR states are infinite-energy limits of two-mode squeezed states. The teleportation protocol is corrupted by the finite squeezing, with better squeezing teleporting more faithfully. CV teleportation was first demonstrated in 1998~\cite{furusawa1998teleportation} and is now commonplace.

%
% Quantum Gate Teleportation
%

\subsection{Quantum gate teleportation} \label{sec:teleport_gate} \index{Quantum gate teleportation}

Using quantum \textit{state} teleportation as a primitive building block, quantum \textit{gate} teleportation may be implemented \cite{bib:GottesmanChuang99}. Here rather than teleporting a quantum state from one physical system to another, we teleport the action of a quantum gate onto a physical system (archetypically a maximally entangling 2-qubit gate, such as a CNOT gate).

The general outline of the derivation of the protocol for teleporting a CNOT gate onto a 2-qubit state is shown in Alg.~\ref{alg:gate_teleport}.

\begin{table}[!htbp]
\begin{mdframed}[innertopmargin=3pt, innerbottommargin=3pt, nobreak]
\texttt{
function GateTeleportation($\ket\psi_A\ket\phi_B$):\
\begin{enumerate}
\item We wish to apply a CNOT gate to \mbox{$\ket\psi_{A}\ket\phi_{B}$}.
\item Introduce two additional qubits, $C$ and $D$.
\item Teleport states \mbox{$\ket\psi_{A}\to\ket\psi_C$}, \mbox{$\ket\psi_{B}\to\ket\psi_D$}.
\item Apply \mbox{$\hat{\mathrm{CNOT}} \ket \psi_C \ket\phi_D$}.
\begin{align}
\Qcircuit @C=1em @R=1.6em {
\lstick{\ket\psi} & \qw & \multimeasureD{1}{\mathrm{Bell}} & \cw  & \control \cw \\
\lstick{} & \qw & \ghost{\mathrm{Bell}} & \control \cw & \cwx \\
\lstick{} & \qw & \qw & \gate{X} \cwx & \gate{Z} \cwx & \ctrl{1} & \qw & \qw & \qw \inputgroupv{2}{3}{.8em}{.8em}{\ket{\Phi^+}} \\
\lstick{} & \qw & \qw & \gate{X} & \gate{Z} & \targ & \qw & \qw & \qw \inputgroupv{4}{5}{.8em}{.8em}{\ket{\Phi^+}} \\
\lstick{} & \qw & \multimeasureD{1}{\mathrm{Bell}} & \control \cw \cwx & \cwx \\
\lstick{\ket\phi} & \qw & \ghost{\mathrm{Bell}} & \cw  & \control \cw \cwx
} \nonumber
\end{align}
\item The CNOT is a Clifford gate and can therefore be commuted to the front of the Pauli operators to yield a CNOT followed by some different configuration of Pauli operators.
\item The CNOT now acts jointly upon the Bell pairs that were acting as a resource for the state teleportation, independent of \mbox{$\ket\psi_{A}\ket\phi_{B}$}.
\item Group the CNOT gate and Bell pairs together, and treat them as a 4-qubit resource state preparation stage, which does not depend on \mbox{$\ket\psi_{A}\ket\phi_{B}$}. 
\item Prepare the 4-qubit resource state, \mbox{$\ket\chi=\hat{\mathrm{CNOT}}_{2,3}\ket{\Psi^+}_{1,2}\ket{\Psi^+}_{3,4}$}, offline in advance.
\item If the CNOT is non-deterministic, employ \textsc{Repeat Until Success} to prepare $\ket\chi$.
\item The output state is \mbox{$\hat{\mathrm{CNOT}}_{C,D} \ket\psi_{C}\ket\phi_{D}$}.
\item $\Box$
\end{enumerate}}
\begin{align}
\Qcircuit @C=1em @R=1.6em {
\lstick{\ket\psi} & \qw & \multimeasureD{1}{\mathrm{Bell}} & \cw & \cw & \control \cw \\
\lstick{} & \qw & \ghost{\mathrm{Bell}} & \cw & \cw & \cw \cwx & \control \cw \\
\lstick{} & \qw & \qw & \gate{X} & \gate{Z} & \qw \cwx & \gate{Z} \cwx & \qw & \qw & \qw \\
\lstick{} & \qw & \qw & \gate{X} \cwx & \qw \cwx & \gate{X} \cwx & \gate{Z} \cwx & \qw & \qw & \qw \\
\lstick{} & \qw & \multimeasureD{1}{\mathrm{Bell}} & \control \cw \cwx & \cwx \inputgroupv{2}{5}{0.8em}{4.1em}{\ket{\chi}} \\
\lstick{\ket\phi} & \qw & \ghost{\mathrm{Bell}} & \cw  & \control \cw \cwx
} \nonumber
\end{align}
\end{mdframed}
\captionspacealg \caption{Teleporting a CNOT gate onto a 2-qubit state.} \label{alg:gate_teleport}
\end{table}

Most notably, gate teleportation is useful when attempting to apply 2-qubit entangling operations using non-deterministic gates, in which case gate teleportation allows the non-deterministic elements to be performed offline as a resource state preparation stage, overcoming the non-determinism during the gate application stage.

Specifically, when a CNOT gate acting directly upon two qubits fails, it corrupts those qubits, whereas if it fails during a state preparation stage, it can simply be reattempted until a success occurs, without corrupting the target qubits. A concatenated version of the gate teleportation protocol forms the basis for constructing near-deterministic entangling gates in linear optics, to be explained in detail in Sec.~\ref{sec:KLM_univ}.

Quantum gate teleportation effectively reduces the problem of implementing CNOT gates to:
\begin{enumerate}
\item Offline preparation of highly-entangled 4-qubit resource states. This needn't be deterministic, since the resource state does not depend on the state to which the CNOT gate ought to be applied.
\item Two Bell measurements.
\item Some configuration of local Pauli operators, dependent upon the Bell measurement outcomes.
\end{enumerate}
Importantly, like quantum state teleportation, there is no need for a quantum communications channel between the two parties holding the qubits to which the gate is applied -- classical communication is sufficient.

The gate teleportation idea is conceptually interesting as it converts the problem of `gate application' to that of `state preparation'\footnote{The resource state is prepared from two Bell pairs and a single CNOT gate, which is locally equivalent to a 4-qubit GHZ state. In the absence of a direct source of Bell pairs, they can be prepared using separable single-qubit states and a CNOT gate. Thus, the full resource state may be prepared from separable single qubits via three CNOT gates.}, by commuting all the entangling operations to the beginning of the protocol. Cluster state quantum computing (Sec.~\ref{sec:CSQC}) is actually the extremity of this logic, whereby an entire quantum computation is transformed into a sequence of state and gate teleportations. One may interpret this to mean that teleportation is a universal resource for quantum computation \cite{bib:GottesmanChuang99}.

The resource states required for gate teleportation are highly-entangled 4-qubit states, which are challenging to prepare, especially in the optical context. Thus, as with cluster states, if the preparation of these resource states were to be outsourced to a specialised provider, they could be in high demand.

Note that this technique works for the CNOT gate because it is a Clifford gate\index{Clifford gates} (i.e it commutes with the classically-controlled Pauli gates to yield a different combination of classically-controlled Pauli gates). Thus, this technique does not automatically apply to \textit{any} 2-qubit gate.

%
% Entanglement Swapping
%

\subsection{Entanglement swapping} \label{sec:swapping} \index{Entanglement!Swapping}

\begin{figure}[!htbp]
\includegraphics[clip=true, width=0.475\textwidth]{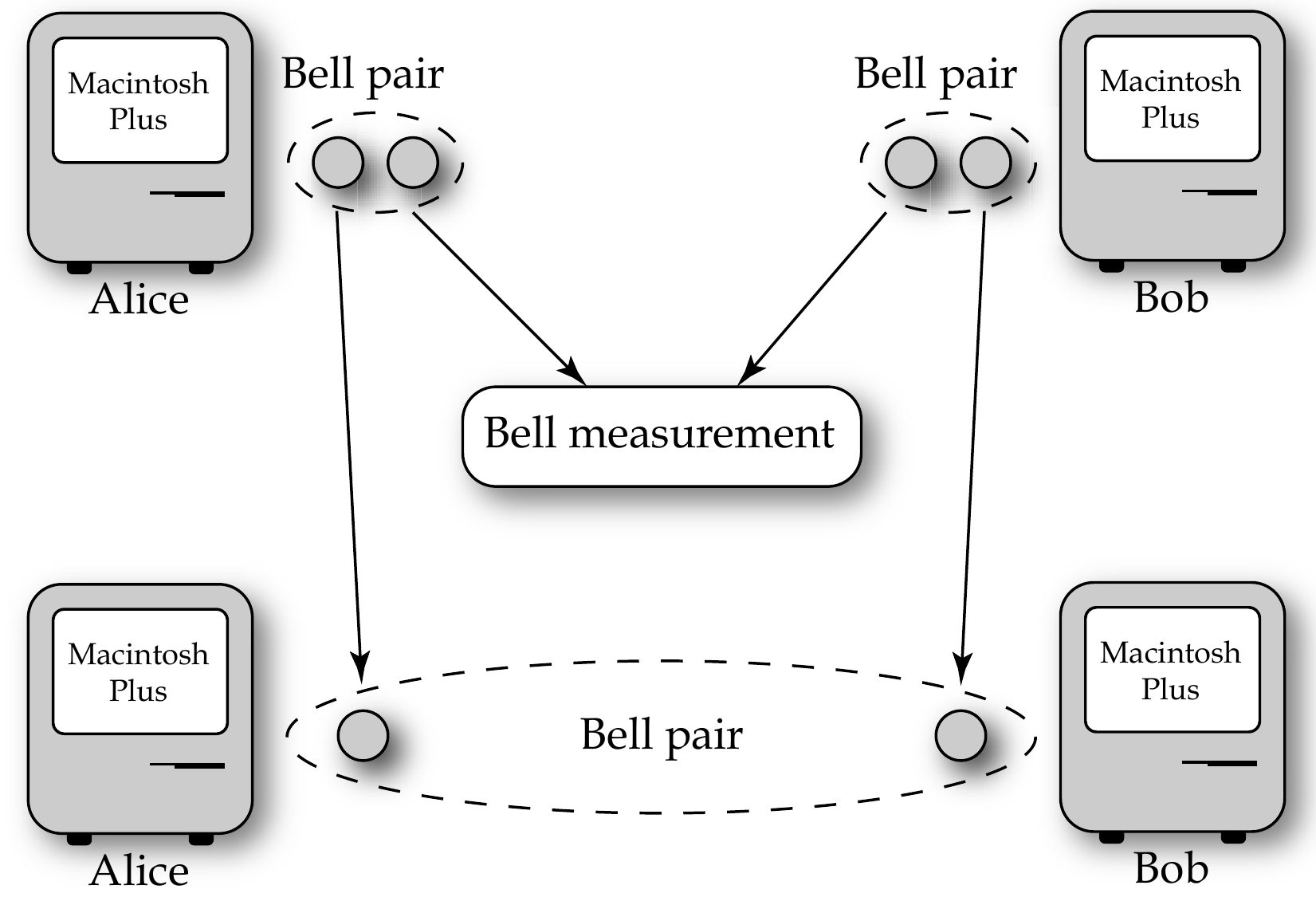}
\captionspacefig \caption{Entanglement swapping between two nodes. Each node initially holds a Bell pair (dotted ellipses) comprising two qubits (grey circles). One qubit from each pair is sent to the repeater between them, which measures them in the Bell basis. After local unitary corrections, the two nodes share an entangled pair.} \label{fig:ent_swap}
\end{figure}

The obvious approach to sending a qubit from Alice to Bob is to send a qubit from Alice to Bob (duh!). However, over long distances this may accrue impractical error rates, particularly losses. The other alternative is to employ the quantum state teleportation protocol (Sec.~\ref{sec:teleport}) to teleport the state between the two parties. However, this requires that Alice and Bob first share an entangled Bell pair, which must itself be distributed across the same distances. Entanglement swapping \cite{bib:PhysRevLett.75.4337} is the process of taking two Bell pairs, one held by each party, and swapping the entanglement between them such that the two parties share an entangled state. This procedure can be bootstrapped to progressively swap the entanglement over longer and longer distances, yielding \textit{quantum repeater networks} (Sec.~\ref{sec:rep_net}). The procedure for this protocol is shown in Alg.~\ref{alg:ent_swap} and Fig.~\ref{fig:ent_swap}

\begin{table}[!htbp]
\begin{mdframed}[innertopmargin=3pt, innerbottommargin=3pt, nobreak]
\texttt{
function EntanglementSwapping($\ket{\Phi^+}^{\otimes 2}$):
\begin{enumerate}
    \item Alice locally prepares the Bell pair,
    \begin{align}
    \ket{\Phi^+}_{A_1,A_2}.
    \end{align}
    \item Bob locally prepares the Bell pair,
    \begin{align}
    \ket{\Phi^+}_{B_1,B_2}.
    \end{align}
    \item The net initial state is,
    \begin{align}
    \ket\psi_\mathrm{in} = \ket{\Phi^+}_{A_1,A_2} \ket{\Phi^+}_{B_1,B_2}.
    \end{align}
    \item Alice sends qubit $A_1$ to third-party Eve.
    \item Bob sends qubit $B_1$ to third-party Eve.
    \item Eve performs a Bell projection between $A_1$ and $B_1$, yielding,
    \begin{align}
    \bra{\Phi^+}_{A_1,B_1} \ket\psi_\mathrm{in} = \ket{\Phi^+}_{A_2,B_2}.
    \end{align}
    \item In the case of the other Bell projection outcomes ($\bra{\Phi^-}_{A_1,B_1}$, $\bra{\Psi^+}_{A_1,B_1}$ or $\bra{\Psi^-}_{A_1,B_1}$), local corrections (Pauli operators) are made by Alice and/or Bob, as dictated by classical communication from Eve,
    \begin{align}
    \bra{\Phi^+}_{A_1,B_1} \ket\psi_\mathrm{in} &= \ket{\Phi^+}_{A_2,B_2}, \nonumber \\
    \bra{\Phi^-}_{A_1,B_1} \ket\psi_\mathrm{in} &= \hat{Z}_{B_2} \ket{\Phi^+}_{A_2,B_2}, \nonumber \\
    \bra{\Psi^+}_{A_1,B_1} \ket\psi_\mathrm{in} &= \hat{X}_{B_2} \ket{\Phi^+}_{A_2,B_2}, \nonumber \\
    \bra{\Psi^-}_{A_1,B_1} \ket\psi_\mathrm{in} &= \hat{X}_{B_2} \hat{Z}_{B_2} \ket{\Phi^+}_{A_2,B_2}.
    \end{align}
    \item Alice and Bob now possess a joint Bell pair between qubits $A_2$ and $B_2$,
    \begin{align}
    \ket\psi_\mathrm{out} = \ket{\Phi^+}_{A_2,B_2}.
    \end{align}
    \item $\Box$ \\
\end{enumerate}}
\begin{align}
\Qcircuit @C=1em @R=1.6em {
    \lstick{} & \qw & \qw & \qw & \qw & \qw & \qw \\
    \lstick{} & \qw & \multimeasureD{1}{\mathrm{Bell}} & \cw  & \control \cw
    \inputgroupv{1}{2}{.8em}{.8em}{\ket{\Phi^+}} \\
    \lstick{} & \qw & \ghost{\mathrm{Bell}} & \control \cw & \cwx \\
    \lstick{} & \qw & \qw & \gate{X} \cwx & \gate{Z} \cwx & \qw & \qw
    \inputgroupv{3}{4}{.8em}{.8em}{\ket{\Phi^+}}
} \nonumber
\end{align}
\end{mdframed}
\captionspacealg \caption{Entanglement swapping protocol between two parties. Two Bell pairs held locally by two users, \mbox{$\ket{\Phi^+}_{A_1,A_2}\ket{\Phi^+}_{B_1,B_2}$}, are converted to a single Bell pair shared between the users, $\ket{\Phi^+}_{A_2,B_2}$.} \label{alg:ent_swap}
\end{table}

In a sense, entanglement swapping can be regarded as `indirect' entanglement distribution, whereby entanglement is created between two distant parties who do not directly exchange any quantum information.

Alternately, note that the entanglement swapping is structurally almost identical to two instances of quantum state teleportation side-by-side. This is not a coincidence, and entanglement swapping can indeed be thought of as Bell pair state teleportation.

Now if instead of Alice and Bob we have a long chain of these operations in series, then the entanglement can be swapped across the entire length of the chain, enabling the preparation of end-to-end entangled pairs, which can be employed for state teleportation.

The advantage to this approach is that the range of each repeater can be much smaller than the entire length of the channel, easing constraints imposed by errors, notably loss. Furthermore, the entanglement swapping needn't be actually performed in any chronologically linear sequence. The operations could be arbitrarily ordered, since the measurements are independent and commute. Thus, if some segments are detected as failing (e.g qubits are lost), just those segments can be performed again without requiring the entire protocol to start from scratch, unlike the na{\" i}ve direct communication technique. This \textsc{Divide and Conquer} approach can drastically improve performance of the network in terms of channel capacity, improving the exponential dependence of loss on distance. 

The protocol is conceptually very similar to teleportation, where instead of teleporting a qubit state, we are teleporting entanglement. Because of this similarity, it inherits similar error propagation characteristics as for teleportation discussed previously. That is, errors acting on the qubits upon which the Bell measurements are performed are effectively teleported onto the remaining qubits. Then, entanglement purification can be implemented as a higher-level layer on top of the repeaters, enabling high-fidelity entanglement distribution.

Each Bell measurement can be implemented non-deterministically using a PBS, mitigating the need for interferometric stability, as before, but therefore introducing non-determinism into the protocol.

%
% Quantum Cryptography
%

\subsection{Quantum cryptography}\index{Quantum key distribution (QKD)}\index{Quantum cryptography}\label{sec:QKD_prot}

One of the most widely demonstrated class of quantum protocols is the cryptographic ones. Most importantly, these protocols allow, at least in principle, provably secure communication between two parties, immune to any attack. Because these protocols are so important and thoroughly researched, we dedicate Part.~\ref{part:quant_crypto} entirely to this topic.

%
% Superdense Coding
%

\subsection{Superdense coding}\label{sec:superdense}\index{Superdense coding}

\textit{Superdense coding} is a hybrid quantum/classical communications protocol for increasing classical bit-rates between two parties, who share entanglement as a resource.

Suppose Alice wishes to send classical information to Bob over a quantum channel. The HSW Theorem\index{HSW theorem} \cite{bib:holevo1998capacity, bib:schumacher1997sending} tells us that Alice can send information to Bob at a maximum rate of one bit per qubit. However, if Alice and Bob share Bell pairs, superdense coding allows information to be transmitted at a maximum rate of two bits per qubit.

Let Alice and Bob begin with the shared Bell state,
\begin{align}
    \ket{B_{00}} = \frac{1}{\sqrt{2}}\left(\ket{0}_{A}\ket{0}_{B}+\ket{1}_{A}\ket{1}_{B}\right),
\end{align}
where the first qubit, $A$, belongs to Alice and the second qubit, $B$, belongs to Bob. This entangled pair is provided to them by a third-party entanglement server. The protocol exploits the fact that all four Bell states are locally-equivalent, and can be transformed into one another using operations performed only by Alice. Specifically, the four Bell states can be prepared from $\ket{B_{00}}$ via the local operations,
\begin{align}
    \ket{B_{00}} &= (\hat\openone\otimes\hat\openone)\ket{B_{00}} \nonumber \\
    &= \frac{1}{\sqrt{2}}\left(\ket{0}\ket{0}+\ket{1}\ket{1}\right), \nonumber \\
    \ket{B_{01}} &= (\hat{Z}\otimes\hat\openone)\ket{B_{00}} \nonumber \\
    &= \frac{1}{\sqrt{2}}\left(\ket{0}\ket{0}-\ket{1}\ket{1}\right), \nonumber \\
    \ket{B_{10}} &= (\hat{X}\otimes\hat\openone)\ket{B_{00}} \nonumber \\
    &= \frac{1}{\sqrt{2}}\left(\ket{0}\ket{1}+\ket{1}\ket{0}\right), \nonumber \\
    \ket{B_{11}} &= (\hat{Z}\hat{X}\otimes\hat\openone)\ket{B_{00}} \nonumber \\
    &= \frac{1}{\sqrt{2}}\left(\ket{0}\ket{1}-\ket{1}\ket{0}\right).
\end{align}

Suppose Alice wishes to send Bob the two-bit string $x\in\{0,1\}^2$. She applies local operations on her qubit to transform the shared Bell state into the Bell state $\ket{B_{x}}$. There are four such states, therefore this encodes two classical bits of information. She then sends her qubit to Bob, who already holds the other half of the entangled pair. Now by measuring in the Bell basis Bob can determine which two-bit string Alice encoded. The algorithm is described in Alg.~\ref{alg:superdense}.

\begin{table}[!htbp]
\begin{mdframed}[innertopmargin=3pt, innerbottommargin=3pt, nobreak]
\texttt{
function SuperdenseCoding($\ket{B_{00}}$, $x$):
\begin{enumerate}
\item Alice and Bob share the Bell pair $\ket{B_{00}}$.
\item Alice encodes the two-bit string \mbox{$x\in\{0,1\}^2$} into a choice of the four possible Bell pairs.
\item Alice prepares the respective Bell pair using operations local to only her half of the shared state ($\hat\openone$, $\hat{X}$, $\hat{Z}$ or \mbox{$\hat{Z}\hat{X}$}).
\item Alice sends her qubit to Bob.
\item Bob measures in the Bell basis, with four possible measurement outcomes.
\item The measurement outcome corresponds to the bit-string $x$.
\item $\Box$
\end{enumerate}
\begin{align}
\Qcircuit @C=1em @R=1.6em {
\lstick{x_0} & \cw & \cw & \control \cw \\
\lstick{x_1} & \cw & \control \cw \\
\lstick{} & \qw & \gate{X} \cwx & \gate{Z} \cwx[-2] & \qw & \multimeasureD{1}{\mathrm{Bell}} & \cw & \rstick{x_0} \\
\lstick{} & \qw & \qw & \qw & \qw & \ghost{\mathrm{Bell}} & \cw & \rstick{x_1}
\inputgroupv{3}{4}{.8em}{.8em}{\ket{B_{00}}\quad} \\
} \nonumber
\end{align}
}
\end{mdframed}
\captionspacealg \caption{Superdense coding protocol for communicating two classical bits via transmission of a single qubit. The protocol requires the two parties share a Bell pair as a resource, provided by a third-party.} \label{alg:superdense}
\end{table}

Note that the protocol in a sense `cheats', since it assumes a resource of Bell pairs between Alice and Bob, which doesn't come for free. However, in an environment where both parties have access to the same entanglement server or repeater network (Sec.~\ref{sec:rep_net}), in addition to their own direct line of quantum communication, they can utilise this protocol to double classical communication rates from one bit per qubit to two.

However, this doubling in communication rate requires using quantum infrastructure, which, at least for the foreseeable future, will come at a greater cost than our present-day commodified classical hardware. It may therefore be the case that the technological effort of implementing this protocol outweighs the gain, or that for the same effort other classical bandwidth-increasing technologies could be employed.

Alternately, in a future quantum world where such technologies are cheap off-the-shelf commodities, as with our current classical ones, why not double our classical network bandwidths if we can?

%
% Quantum Metrology
%

\subsection{Quantum metrology} \label{sec:metrology} \index{Quantum metrology}

The goal of quantum metrology is to estimate an unknown phase\index{Phase!Estimation} with the greatest degree of precision. This finds many applications, perhaps most notably the recent gravity wave measurement protocols \cite{???}. The shot-noise limit (SNL)\index{Shot-noise!Limit} represents the maximum achievable precision using classical states, whereas the Heisenberg limit (HL)\index{Heisenberg!Limit} is the best that can be achieved using quantum resources. The goal of quantum metrology is to beat the SNL, ideally saturating the HL.

Achieving the SNL is easily done using a Mach-Zehnder interferometer\index{Mach-Zehnder (MZ) interference} (Sec.~\ref{sec:MZ_inter}) fed with coherent states (Sec.~\ref{sec:coherent_states}), which are not true quantum states. Referring to Fig.~\ref{fig:MZ_inter}, if a coherent state is inputted into one arm of the interferometer, with no phase-shift (\mbox{$\tau=0$}) all the coherent amplitude would exit the corresponding output port. If on the other hand there were a $\pi$ phase-shift, all the amplitude would exit the other output port. For intermediate $\tau$ there will be varying degrees of coherent amplitude distributed between the two outputs. Thus, the relative amplitude exiting the two output ports acts as a signature for the internal phase-shift $\tau$.

Improving upon this, HL metrology can be achieved using NOON states (Sec.~\ref{sec:NOON}) \cite{bib:Dowling08}. 
% An alternate recent proposal (known as the MORDOR protocol, after the authors), employs only single-photon states (Sec.~\ref{sec:single_phot_src}) and passive linear optics, which, although not saturating the HL, significantly beats the SNL \cite{bib:MORDOR15, bib:MORDOR2}. This was recently experimentally demonstrated by \cite{???}. Squeezed states\index{Squeezed states} have also been shown to beat the SNL.
% 
NOON states in particular are difficult to prepare, as they cannot be deterministically prepared using linear optics, and no current source natively prepares them directly. Thus, outsourcing these state preparation stages could be of great value to end-users of metrology, were there a specialised server dedicated to this task.\index{NOON states}

%
% Quantum State & Process Tomography
%

\subsection{Quantum state \& process tomography} \index{Quantum state tomography (QST)} \index{Quantum process tomography (QPT)}

In Sec.~\ref{sec:QPT} we introduced QST and QPT, as procedures by which to experimentally reconstruct unknown density matrices or process matrices respectively. It is conceivable that these tomographic procedures might want to be performed over a quantum network in a distributed fashion.

Consider the case where a node joins an existing ad hoc network. Before thinking about routing its packets through the network, it must understand the network's relevant cost metrics. Suppose that metric is one that is calculated directly from a channel's process matrix. Then, to characterise the channels in the network connecting the node to its new nearest neighbours, it could apply distributed QPT, whereby the new node is responsible for preparing the complete basis of input states required for QPT, which are transmitted to the chosen neighbour across the respective channel, after which the recipient performs all the necessary measurements in the required bases. Purely classical communication is obviously required to communicate measurement settings and outcomes.

In this simple example scenario it is immediately clear that QPT of new links in a network is perfectly suited to distributed implementation. In fact, having a node attempt to characterise a channel from start to finish could be entirely unrealistic if the channel ran over long distances -- the owner of the node would never be able to reach the other end of the channel in time for the photons' arrival! This necessitates a cooperative protocol.

%
% Quantum Clock Synchronisation
%

\subsection{Quantum clock synchronisation} \index{Quantum clock synchronisation}\label{sec:clock_sync}

\sectionby{Tim Byrnes}\index{Tim Byrnes}

Clock synchronisation is a fundamental task with widespread applications, ranging from navigation, telecommunications, and financial transactions, to the internet as a whole, and many scientific applications. Of these, the global positioning system (GPS)\index{Global positioning system (GPS)} has become a day-to-day necessity for much of humanity, having been increasingly incorporated into smartphones and other commodity devices.

The GPS system famously relies upon very precise clock synchronisation to perform its task through a process of quadrangulation\index{Quadrangulation} from a constellation of several satellites\index{Constellation network}. Due to the high speed of light, we require highly synchronised clocks, accurate to the nanosecond level. This allows positioning to be performed to the level of meters, a level of precision required for many routine applications. GPS satellites have atomic clocks that are stable to one part in $10^{13}$, so that active correction can maintain this level of accuracy.

The great success of the GPS system has created a further demand for increasingly precise navigation. For example, autonomous vehicles would immediately benefit from a more precise navigation system.

In principle, technology for more stable clocks already exists, with atomic clocks\index{Atomic!Clocks} exceeding stabilities of those on satellites being routinely produced, and optical atomic clocks now reaching stabilities of one part in $10^{18}$ \cite{bib:ludlow2015optical}. An outstanding question is then how to synchronise these clocks given their remarkable stabilities. 

\begin{figure*}[!htbp]
\includegraphics[clip=true, width=0.7\textwidth]{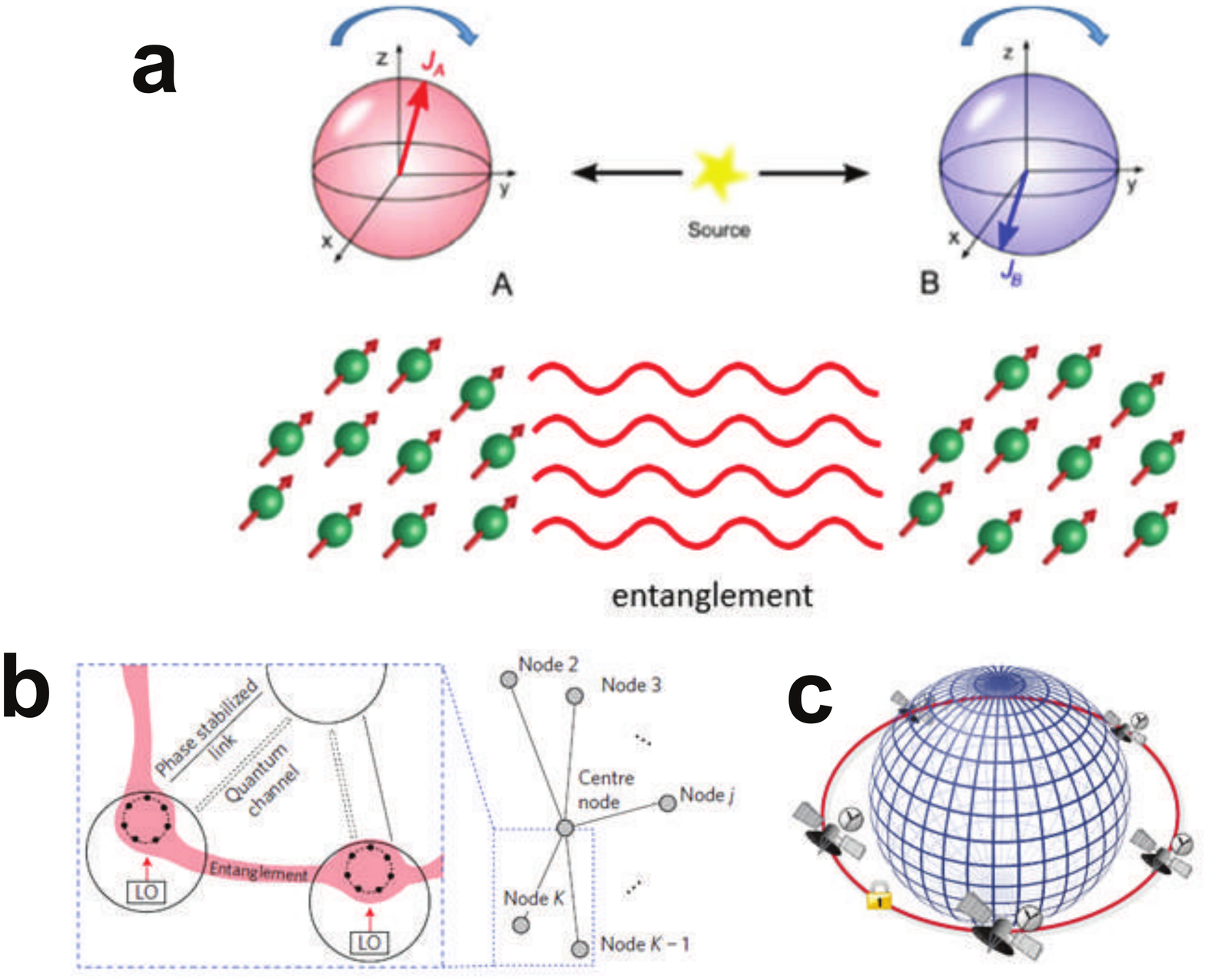}
\captionspacefig 
\caption{
% \comment{Better figure, this is shit}
Quantum clock synchronisation schemes. (top) The proposal by Jozsa \& Dowling \textit{et al.} where a singlet state\index{Bell!States} is shared and measured in an ensemble of qubits \cite{bib:jozsa00}. (bottom) The proposal by Lukin \& Ye \textit{et al.} where a GHZ state\index{Greenberger-Horne-Zeilinger (GHZ) states} is distributed between satellites to measure the frequency drift at the Heisenberg limit \cite{bib:komar14}.}
\label{fig:space_3}
\end{figure*}

Previous works have examined the problem of clock synchronisation in space. In the proposal of Jozsa \textit{et al.}, many copies of shared entanglement in a singlet state\index{Bell!States} is first distributed and stored on the clock states of an atomic clock \cite{bib:jozsa00}. The measurement is then performed by one party, which collapses the states simultaneously across all parties, and the time evolution of the states begins.

Classical information is exchanged between them, which reveals the time elapsed since the measurement, which can be used to synchronise the clocks. While the original protocol only allowed clock synchronisation between two parties, similar ideas were used to extend this to the multiparty context \cite{bib:krvco2002quantum, bib:ben2011optimized, bib:ren2012clock}.

In a more recent proposal, a shared GHZ state\index{Greenberger-Horne-Zeilinger (GHZ) states} is prepared across all nodes in the quantum network, which allows for quantum metrologically enhanced detection\index{Quantum metrologically enhanced detection} of the clock signal drift at the Heisenberg limit\index{Heisenberg!Limit} \cite{bib:komar14}. The use of shared resources acts to improve the overall precision, allowing for an optimal scheme for the qubit resources that are employed. Several other proposals have also been made, which are quantum versions of Eddington's slow clock transport protocol\index{Eddington's slow clock transport protocol} where the qubit keeps time of the transmission \cite{bib:chuang2000quantum, bib:tavakoli2015quantum}. 

Experimentally, there have been several demonstrations of the protocol, albeit at relatively short distances. Continuous time-bin\index{Time-bin encoding} entangled photons were used as the entanglement resource to obtain a time-correlation between a distance of 3km \cite{bib:valencia2004distant}, and another technique based on Hong-Ou-Mandel interferometry\index{Hong-Ou-Mandel (HOM) interference} was performed across a 4km fibre link \cite{bib:quan2016demonstration}. Several other demonstrations based on nuclear magnetic resonance (NMR)\index{Nuclear magnetic resonance (NMR)} \cite{bib:zhang2004nuclear, bib:kong2017implementation} have also been performed. 

There are however several outstanding problems with the quantum clock synchronisation scheme as presented above. In the scheme of Jozsa \textit{et al.}, if one starts in a perfect singlet state\index{Bell!States}, the scheme works as intended, but if one instead starts with the state, 
\begin{align}
\frac{1}{\sqrt{2}} ( \ket{0}_A \ket{1}_B - e^{i \delta} \ket{1}_A \ket{0}_B ),
\end{align}
(a Bell pair augmented by a local phase) one obtains an offset to the synchronisations between the two parties. In practice, such a phase could arise from decoherence induced noise, or differences in the basis conventions chosen by the two parties. Thus, in practice entanglement purification\index{Entanglement!Purification} would be required to produce a singlet state with \mbox{$\delta=0$} prior to executing the protocol. However, it was argued that to perform the entanglement purification quantum circuit correctly, the timing of the quantum gates would need to be controlled, which requires synchronised clocks \cite{bib:preskill2000quantum} --  this renders synchronisation impossible. It has previously been shown that such a phase cannot be eliminated using asynchronous entanglement purification\index{Entanglement!Purification} \cite{bib:yurtsever02}, and hence the protocol remains incomplete in the general case where imperfect singlet pairs\index{Bell!States} are shared.

A quantum protocol for clock synchronisation based on shared entanglement is given in Alg.~\ref{alg:clock_sync}.

\begin{table}[!htbp]
\begin{mdframed}[innertopmargin=3pt, innerbottommargin=3pt, nobreak]
\texttt{
function ClockSync($\ket{\Psi^-}$):
\begin{enumerate}
\item Distribute a Bell state between Alice and Bob,
\begin{align}
\ket{\psi_0} &= \frac{1}{\sqrt{2}}(\ket{0}_A\ket{1}_B - \ket{1}_A\ket{0}_B)\nonumber\\
	&= \frac{1}{\sqrt{2}}(\ket{+}_A\ket{-}_B - \ket{-}_A\ket{+}_B).
\end{align}
\item The joint system is freely-evolving under the Hamiltonian,
\begin{align}
\hat{H} = \hbar\omega(\hat{Z}_A + \hat{Z}_B).	
\end{align}
\item Over time $t_\mathrm{free}$ this evolves to,
\begin{align}
	\ket{\psi_1} &= e^{-i\frac{\hat{H}}{\hbar} t_\mathrm{free}}\frac{1}{\sqrt{2}}(\ket{+}_A\ket{-}_B - \ket{-}_A\ket{+}_B)\nonumber\\
	&= e^{-i\omega t_\mathrm{free}}\frac{1}{\sqrt{2}}(\ket{+}_A\ket{-}_B - \ket{-}_A\ket{+}_B),
\end{align}
yielding only an irrelevant global phase.
\item Alice measures her qubit in the $\hat{X}$ basis ($\ket\pm\bra\pm$), and classically announces the time at which she performed the measurement, $t_A$, and her measurement outcome, \mbox{$m_A=\pm$}.
\item The system evolves for time $t_\mathrm{ev}$,
\begin{align}
	\ket{\psi_{+}} &= e^{-i\hat{Z} t_\mathrm{ev}}\ket{-}_B\nonumber\\
	&\propto \ket{0}_B-e^{-i\omega t_\mathrm{ev}}\ket{1}_B,\nonumber\\
	\ket{\psi_{-}} &= e^{-i\hat{Z} t_\mathrm{ev}}\ket{+}_B\nonumber\\
	&\propto \ket{0}_B+e^{-i\omega t_\mathrm{ev}}\ket{1}_B.
\end{align}
\item Bob measures his qubit in the $\hat{X}$ basis, with measurement probabilities,
\begin{align}
	m_+ &\propto \sin^2(\omega t_\mathrm{ev}),\nonumber\\
	m_- &\propto \cos^2(\omega t_\mathrm{ev}),
\end{align}
and infers $t_\mathrm{ev}$.
\item Bob sets his clock to,
\begin{align}
	t_B = t_A+t_\mathrm{ev}.
\end{align}
\item $\Box$
\end{enumerate}
}
\end{mdframed}
\captionspacealg \caption{Algorithm for performing quantum clock synchronisation between two parties using shared Bell pairs and classical communication.} \label{alg:clock_sync}
\end{table}

%
% Quantum-Enabled Telescopy
%

\subsection{Quantum-enabled telescopy}\index{Quantum-enabled telescopy}\label{sec:telescopy}

\sectionby{Zixin Huang}\index{Zixin Huang}

For direct imaging of an object the image resolution is constrained by the aperture size and the wavelength. For high resolution, we want to use large light collectors and small wavelengths. For astronomy, building telescopes larger than tens of metres is infeasible; since we cannot illuminate distant objects, all we can do is analyse the light that reaches us.

Image formation requires the light collected from an object to be focused. To do so,
different parts of the imaging system need to have a well-established phase relationship; the
lens is a prototypical example of this. Since it is physically unfeasible to build lenses or
mirrors larger than tens of metres in diameter, to achieve a large baseline, telescope arrays
are used: a telescope array is a set of separate telescopes, mirror segments, and radio
antennae that work together effectively as a single telescope to provide higher resolution.

Current large-baseline telescope arrays operate in the radio and microwave bands, where
the electric field of the received signal (amplitude as well as phase, as opposed to intensity
alone) can be measured directly. Therefore, the phase relationship between the signal
wavefronts arriving at the individual telescopes is known, and the signal is then post
processed into an image. If we move into the optical band, this task becomes very
difficult: even the fastest electronics cannot directly measure the oscillations of the electric field at optical frequencies.
So the collected signals are coherently processed for image reconstruction.

When we consider two neighbouring points on the object separated by a small angle, the minimum angular separation resolvable is,
\begin{align}
	\theta_\mathrm{min} = 1.22 \frac{\lambda}{D},
\end{align}
known as the Rayleigh criterion\index{Rayleigh criterion}, $D$ being the diameter of the aperture\index{Aperture}.

In principle one can build a telescope array with a synthetic aperture\index{Synthetic aperture} of arbitrary $D$.  But, if we feed the collected light into fibres and other optical elements, the losses and noise would destroy the signal if the distances are large. A quantum information protocol has been developed to side-step this problem: instead of losing the signal (starlight), we can have the loss act upon the pre-distirbuted entanglement instead.

The light arriving from the distant object is a thermal state, but the average photon number per mode is much less than 1, therefore higher order terms are negligible. The state that reaches the telescope is therefore approximated by,
\begin{align}
\ket{\psi_\mathrm{star}} = \frac{1}{\sqrt{2}}(\ket{0}_A\ket{1}_B + e^{i\phi}\ket{1}_A\ket{0}_B),
\end{align}
where $\phi$ is the relative phase-shift between the two telescopes, which depends on the difference in distance of propagation. If $\phi$ can be measured accurately, this can give a precise estimate on the location of the object,
\begin{align}
\phi = \frac{b \sin(\theta)}{\lambda},
\end{align}
where $\lambda$ is wavelength.

Often the light that arrives will be formed by a mixture of photons from different sources that emit incoherently, and different locations give rise to different phase-shifts $\phi$, resulting in a density matrix of the form,
% \begin{align}
% \hat\rho = \frac{1}{2} \begin{pmatrix}
%   0 & 0 & 0 & 0 \\
%   0 & 1 & \mathcal{V}^* & 0 \\
%   0 & \mathcal{V} & 1 & 0 \\
%   0 & 0 & 0 & 0
% \end{pmatrix},
% \end{align}
\begin{align}
\rho_\mathrm{star} = \left(1+\mathcal{V} \right)/2 \ket{\psi_+^\phi}\bra{\psi_+^\phi} + \left(1-\mathcal{V} \right)/2 \ket{\psi_-^\phi}\bra{\psi_-^\phi}
\end{align}
where $\ket{\psi_\pm^\phi} = \frac{1}{\sqrt{2}}(\ket{0}_A\ket{1}_B + e^{i\phi}\ket{1}_A\ket{0}_B)$ and
$\mathcal{V}$ is the visibility. The parameter $\phi \in  [0,2\pi)$ is related to the location of the sources, and 
$\mathcal{V} \in [0,1]$ is related to the angular size of the sources. Optimally estimating $\phi$ and $\gamma$ provides complete information of the source distribution

If we interfere the two modes $A$ and $B$ at a 50:50 beamsplitter, the photon will exit port 1 with probability,
\begin{align}
	p = \frac{1}{2} (1 + \mathrm{Re}[\mathcal{V}e^{-i\delta}]),
\end{align}
from which $\mathcal{V}$ can be determined by taking measurements while sweeping through $\delta$.

The problem with implementing the measurement is the difficulty of transporting the single photon state over long distances without incurring loss or additional phase-shifts.

Instead of sending a valuable quantum state directly over a noisy quantum channel, one can distribute entanglement between the two telescopes. The entangled state is known, the preparation can be repeated, and one can use an entanglement distillation protocol to eliminate losses.

The entangled state we want to distribute is,
\begin{align}
\ket{\psi_\mathrm{shared}} = \frac{1}{\sqrt{2}}(\ket{0}_A\ket{1}_B + e^{i\delta}\ket{1}_A\ket{0}_B),	
\end{align}
where $\delta$ is determined by a controllable phase, allowing completion of the protocol to determine $\phi$.

\begin{figure}[!htbp]
\includegraphics[clip=true, width=0.4\textwidth]{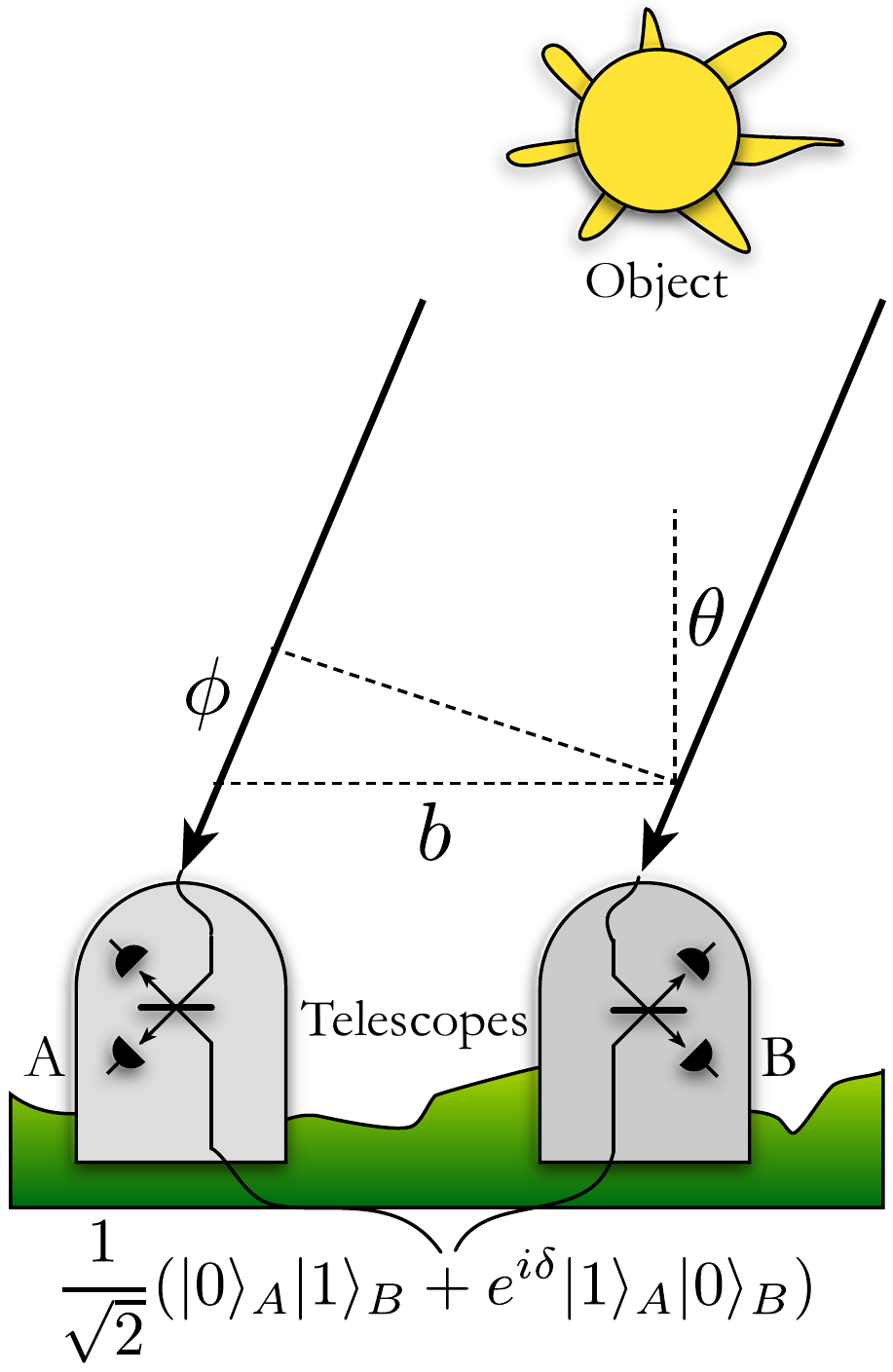}
\captionspacefig \caption{Architecture for quantum-enabled telescopy using two widely separated telescopes, which have shared Bell pairs. The basic idea is that the Bell pair mediates teleportation of one telescope's photon to the other, at which point an interferometric technique measures their phase-difference, thereby determining $\phi$. It is assumed the baseline separation is large, \mbox{$b\gg 1$}.}\label{fig:telescopy}\index{Quantum-enabled telescopy}	
\end{figure}

Now, we can use the entangled pair directly to measure the visibility, as in Fig.~\ref{fig:telescopy}. We post-select on the measurement results, considering the events where a single photon is observed at $A$ and $B$ simultaneously.
 The total probability of seeing a correlated photons, conditioned on having one click at each telescope is,
\begin{align}
	\frac{1}{2} (1 + \mathrm{Re}[\mathcal{V}e^{-i\delta}]),
\end{align}
and the total probability of seeing an anti-correlation is \cite{bib:PhysRevLett.109.070503},
\begin{align}
	\frac{1}{2} (1 - \mathrm{Re}[\mathcal{V}e^{-i\delta}]).
\end{align}

\begin{figure}[!htbp]
\includegraphics[clip=true, width=0.475\textwidth]{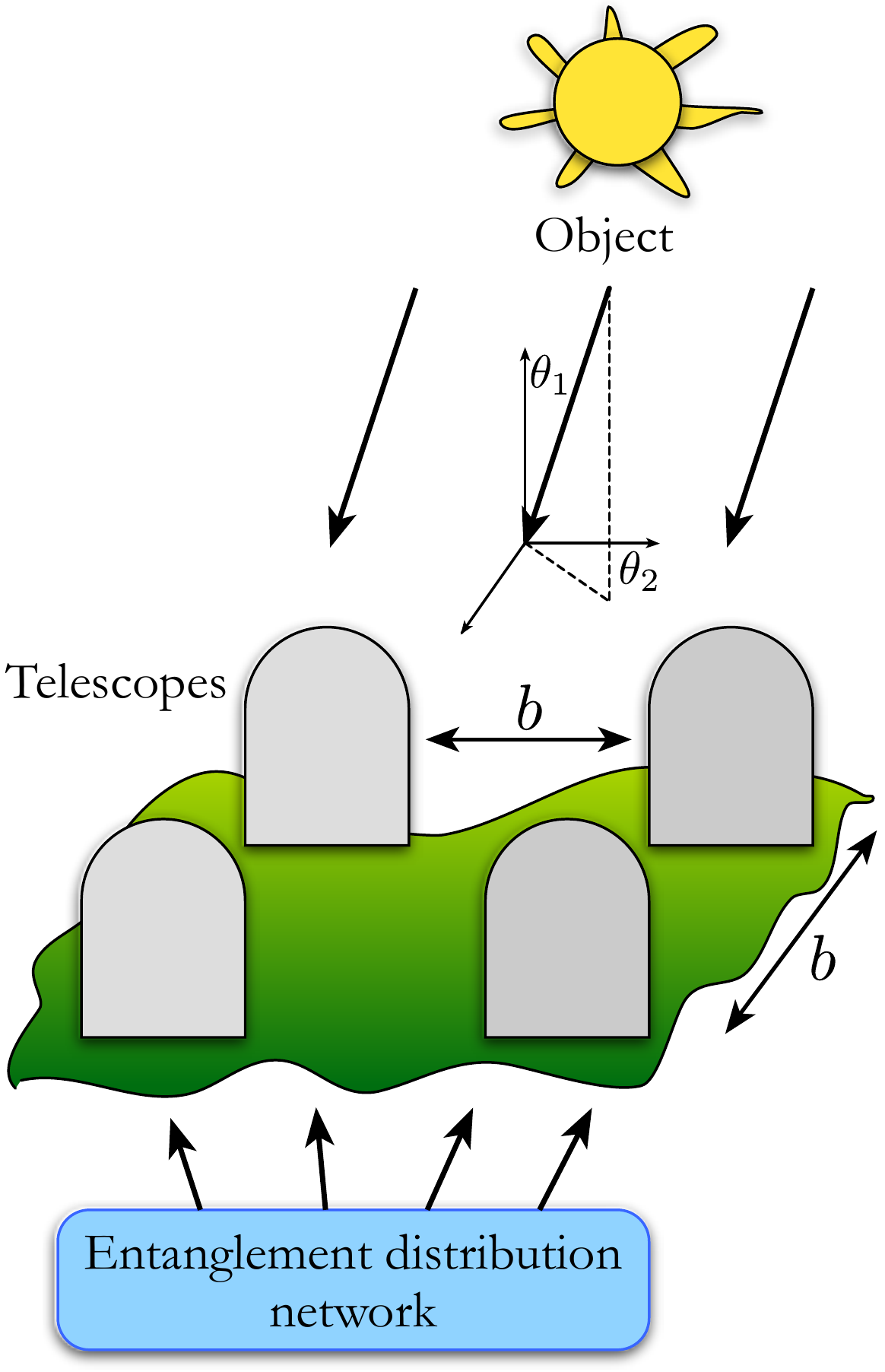}
	\captionspacefig \caption{An array of quantum-enabled telescopes, paired with one another via Bell pairs, distributed over a quantum network, where \mbox{$b\gg 1$}. With more telescopes in the array, signal strength can be increased, and if the array has a 2D grid topology, rather than a linear one, the angles of incident light fields can be measured along multiple axes.}\label{fig:telescope_array}\index{Telescope arrays}
\end{figure}

Thinking futuristically, in a future large-scale quantum internet, whereby Bell pairs are a readily available resource across the globe, quantum-enabled telescopy needn't be limited to pairs of telescopes, but could expand to become large-scale telescope arrays\index{Telescope arrays} comprising numerous telescopes, all sharing pairwise entanglement, distributed via the quantum internet (see Fig.~\ref{fig:telescope_array}). Using a 2D grid would enable $\theta$ to be measured along different axes, and the increased number of detectors would increase signal strength.

Large-baseine interferometry with quantum networks has been studied in the context that includes multi-mode entanglement distribution \cite{PhysRevLett.130.160801} and quantum memories \cite{khabiboulline2019optical,PhysRevA.100.022316} and quantum error correction \cite{PhysRevLett.129.210502}. With quantum memories, the amount of entanglement required can be significantly reduced \cite{khabiboulline2019optical,PhysRevA.100.022316}. The same protocol has also been considered in the continuous-variable formalism \cite{PhysRevA.109.052434,wang2023astronomical}.

\latinquote{Etiam capillus unus habet umbram.}

\sketch{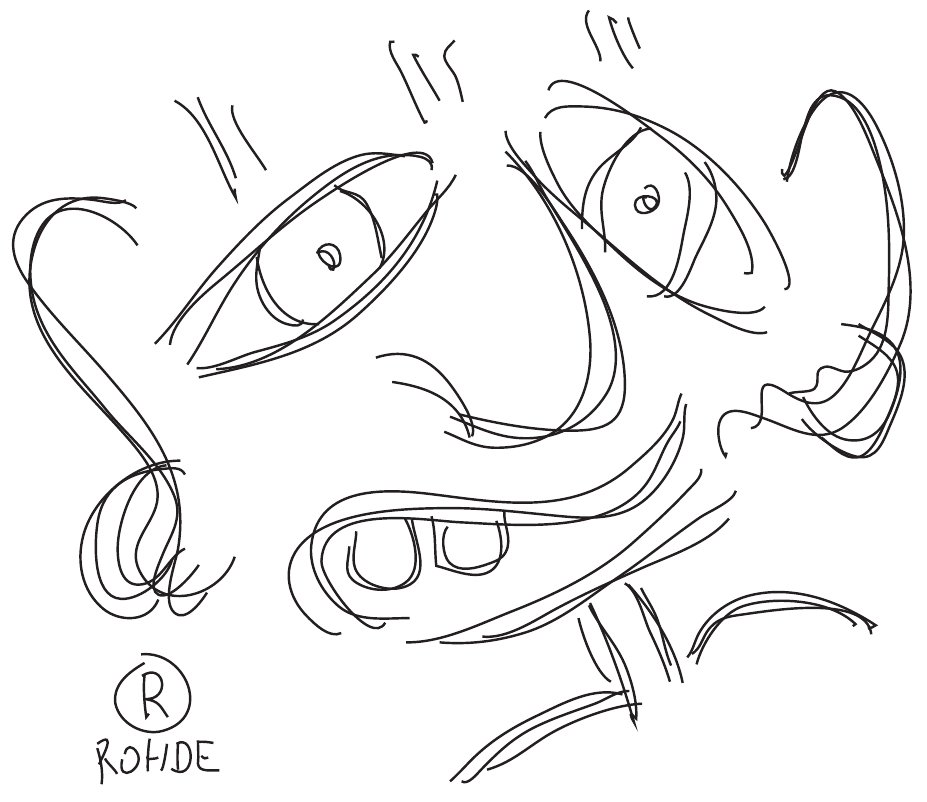}

\clearpage
% %
% Entanglement distribution
%

\part{Entanglement distribution}\label{part:ent_dist}\index{Entanglement!Distribution}

%
% Entanglement - The Ultimate Quantum Resource
%

\section{Entanglement -- The ultimate quantum resource} \label{sec:ent_ultimate} \index{Entanglement}

\dropcap{A}{s} we have seen, the diversity of quantum states that may be communicated, and protocols implemented over the quantum internet is extremely diverse, encompassing many different types of encodings and communications protocols.

Given this plethora of protocols and encodings, discussed in detail in Part.~\ref{part:protocols}, one might ask whether there is a single primitive resource that might be applicable to all, or at least most of these quantum protocols, thereby reducing the technological requirements of the nodes and quantum channels forming the network mediating them -- if networks were able to specialise in a very limited number of tasks, we might reasonably expect them to be better optimised and exhibit better performance than a `Jack of all trades, master of none' network!

It turns out that there is one particularly useful quantum resource, that finds applicability in many of these protocols -- \textit{distributed entanglement}\index{Entanglement!Distributed}, which comes in many flavours and varieties, some of which we discuss now.

%
% Bell States
%

\subsection{Bell states}\index{Bell!States}

Foremost, Bell pairs (Sec.~\ref{sec:bell_state_res}) -- the simplest entangled states -- are an utterly indispensable resource for countless quantum protocols. In brief, Bell pairs find applicability in, amongst many others, the following key protocols:
\begin{itemize}
\item Cluster states (Sec.~\ref{sec:CSQC})\index{Cluster states}: a Bell pair is also a 2-qubit cluster state, a supply of which can be employed in fusion strategies to prepare larger cluster states, enabling universal, distributed MBQC.
\item Quantum state teleportation (Sec.~\ref{sec:teleport})\index{Quantum state teleportation}: a shared Bell pair between Alice and Bob forms the elementary quantum resource upon which the state teleportation protocol is constructed.
\item QKD (Sec.~\ref{sec:QKD})\index{Quantum key distribution (QKD)}: the E91 QKD\index{E91 protocol} protocol is built upon a reliable stream of distributed Bell pairs, enabling private communication with perfect information theoretic security.
\item Modularised quantum computation (Sec.~\ref{sec:module})\index{Modularised quantum computation}: using Bell pairs, entanglement swapping (Sec.~\ref{sec:swapping})\index{Entanglement!Swapping} can be employed to fuse neighbouring, but potentially distant modules together using operations local to each module.
\item Superdense coding (Sec.~\ref{sec:superdense})\index{Superdense coding}: a shared Bell pair enables the communication of two classical bits of information via transmission of a single qubit, thereby doubling classical channel capacity\index{Channels!Capacity}.
\item Quantum-enabled telescopy (Sec.~\ref{sec:telescopy}): a shared Bell pair between two telescopes allows a photon received at one telescope to be teleported to the other, at which point interferometric techniques yield extremely sensitive phase information.
\end{itemize}

We see that Bell pairs form a ubiquitous resource, covering many of the most significant quantum protocols in quantum computation, distributed quantum computation, quantum state teleportation, and quantum cryptography.

%
% GHZ States
%

\subsection{GHZ states}\index{Greenberger-Horne-Zeilinger (GHZ) states}

Beyond Bell pairs, multi-qubit GHZ states\index{Greenberger-Horne-Zeilinger (GHZ) states} (Sec.~\ref{sec:GHZ_states}) (the direct generalisation of Bell pairs to $n$ qubits) are useful in a variety of settings.

For the purposes of quantum anonymous broadcasting\index{Quantum anonymous broadcasting} (Sec.~\ref{sec:anon_broad}), multi-party GHZ entanglement is the primitive resource upon which the cryptographic protocol is constructed. As with Bell pairs, GHZ states are a known state and infinitely reproducible. They can also be purified. Thus, GHZ entanglement distribution is another useful primitive, which future quantum hubs might specialise in preparing and distributing.

Additionally, quantum gate teleportation (Sec.~\ref{sec:teleport_gate})\index{Quantum gate teleportation} of a maximally-entangling 2-qubit gate (e.g a CNOT or CZ gate) is mediated via a shared 4-qubit GHZ state. In a distributed environment the sharing of such a state between two parties (2 qubits per party) enables implementation of a truly distributed 2-qubit entangling gate.

%
% Cluster States
%

\subsection{Cluster states}\index{Cluster states}

Finally, cluster states\index{Cluster states} (Sec.~\ref{sec:CSQC}) are a primitive resource for measurement-based quantum computation\index{Cluster states!Model for quantum computation}. Owing to their handy ability to fuse together to one another, forming larger clusters, the preparation and distribution of relatively small cluster states lends itself well to distributed implementation by specialised providers. Providers could distribute small cluster states, which are subsequently fused together using simple 2-qubit entangling operations to form desired topologies, held either locally or distributed in the cloud.

%
% Why Specialise in Entanglement Distribution?
%

\subsection{Why specialise in entanglement distribution?}

These observations warrant special treatment of entanglement distribution as a fundamental building block in the quantum era. One might envisage a quantum internet in which a central server(s), who specialises in only entangled state preparation and distribution, serves the sole role of pumping out Bell pairs or other entangled states across the quantum internet to whomever requests them, who subsequently use them for protocols such as those mentioned above. This could be in the form of a server transmitting over fibre networks, across free-space, or via a satellite in orbit, transmitting at an intercontinental level.

What's the advantage of this approach to quantum networking? Why specialise in entanglement distribution, rather than implementing more capable networks with the ability to perform arbitrary operations? There are numerous:
\begin{enumerate}
\item Dedicated servers can specialise in this one particular task, as can be the transmission infrastructure.
\item The entanglement servers are entirely passive, not involved interactively with clients.
\item The server needn't concern itself with the nitty-gritty of the protocols implemented by the end-user. It acts purely as a provider of a single resource, remaining uninvolved in their subsequent applications.
\item Because servers are providing a single standardised product, they can be commodified, enabling mass production of the hardware devices and the associated economy of scale. For example, mass production of simple ground-based Bell state relays, or the construction of a comprehensive globe-enveloping constellation of satellites, would inevitably improve economies of scale.
\item Unlike generic quantum states, Bell pairs, GHZ states and cluster states are known states that are infinitely reproducible, without having to worry about no-cloning\index{No-cloning theorem} limitations.
\item Photonic Bell pairs are easily prepared via type-II SPDC at very high repetition rates (\mbox{$\sim 100$MHz-1GHz}), enabling rapid state preparation.
\item Small entangled states like Bell pairs are relatively `cheap' to prepare, and can be readily manufactured using widely accessible, present-day technology that has already been well-demonstrated on Earth and in space.
\item QoS is a lesser issue in most scenarios. We can employ a \textsc{Send-and-forget}\index{Send-and-forget strategy} protocol for the distribution of entanglement (much like classical UDP\index{User Datagram Protocol (UDP)}) -- since every state is identical, we needn't be concerned about missing ones. Instead, we can simply wait for the next one (a \textsc{Repeat-until-success}\index{Repeat-until-success strategy} strategy), knowing it will be exactly the same. We also call this the \textsc{Shotgun}\index{Shotgun!State preparation} approach -- keep firing away until we hit something, and if we lose a few, who cares?
\item Rather than transmitting quantum states between distant parties directly, if we instead use state teleportation (Sec.~\ref{sec:teleport}) mediated by Bell states, the state to be transmitted will not be corrupted if the communications channel fails (e.g via loss). Instead we can wait for the next successfully transmitted Bell pair until we are ready to teleport the state, which then proceeds without directly utilising the quantum communications channel, accumulating its associated costs, or risking losing the state altogether should link failure occur. Only classical communication is required to complete the protocol, which can be regarded as error-free for all intents and purposes.
\item Entanglement purification may be employed by parties to improve the cost metrics associated with their shared entanglement, thereby partially overcoming the limitations imposed by the quantum communication channels.
\item If no direct link exists between server and clients, bootstrapped entanglement swapping can be employed to concatenate servers to create longer-distance `virtual' links. This is the basis for \textit{quantum repeater networks}, to be discussed next in Sec.~\ref{sec:rep_net}.
\end{enumerate}

%
% Why Not Distributed Entangling Measurements
%

\subsection{Why not distributed entangling measurements?}

In addition to entanglement distribution, entangling measurements, e.g Bell state projections (Sec.~\ref{sec:bell_proj}), may be used as a primitive for many protocols. This is effectively entangled state distribution in reverse\index{Distributed entangling measurements}, whereby two clients transmit states to a host, who performs a joint entangling measurement upon them. For example, in the modularised model for cluster state quantum computing\index{Cluster states!Model for quantum computation}, two adjacent but distant modules might transmit optical qubits to a satellite, which projects them into the Bell basis, thereby creating a link between the respective modules via entanglement swapping\index{Entanglement!Swapping}. This isn't as powerful as entangled state distribution, since it cannot be used for, for example, E91 QKD\index{E91 protocol}, but nonetheless remains a powerful primitive for many protocols.

So which ought our quantum hubs specialise in, entanglement distribution, entangling measurements, or both? For most practical purposes the former is far more powerful and robust. Let us take the example of fusing two remote cluster states together to form a larger, distributed virtual cluster. Imagine that their fusion operations are optically mediated by a satellite overhead. The options for satellite-mediated state fusion are (see Fig.~\ref{fig:sat_up_down}):
\begin{itemize}
	\item Downlink mode\index{Satellites!Downlink}: the satellite uses the downlink to distribute an entangled Bell pair between the two nodes. Each node performs a Bell projection between their half of the Bell pair and their respective qubit from their local cluster state, thereby swapping the entanglement and creating a link.
	\item Uplink mode\index{Satellites!Uplink}: each node takes an optical qubit from their cluster state (or entangles their cluster state qubit with an optical qubit), which is uplinked to the satellite. The satellite performs a Bell projection between the two received optical qubits, thereby implementing entanglement swapping between the two nodes, creating a link.
\end{itemize}

\if 1\doublecol
	\begin{figure}[!htbp]
		\includegraphics[clip=true, width=0.475\textwidth]{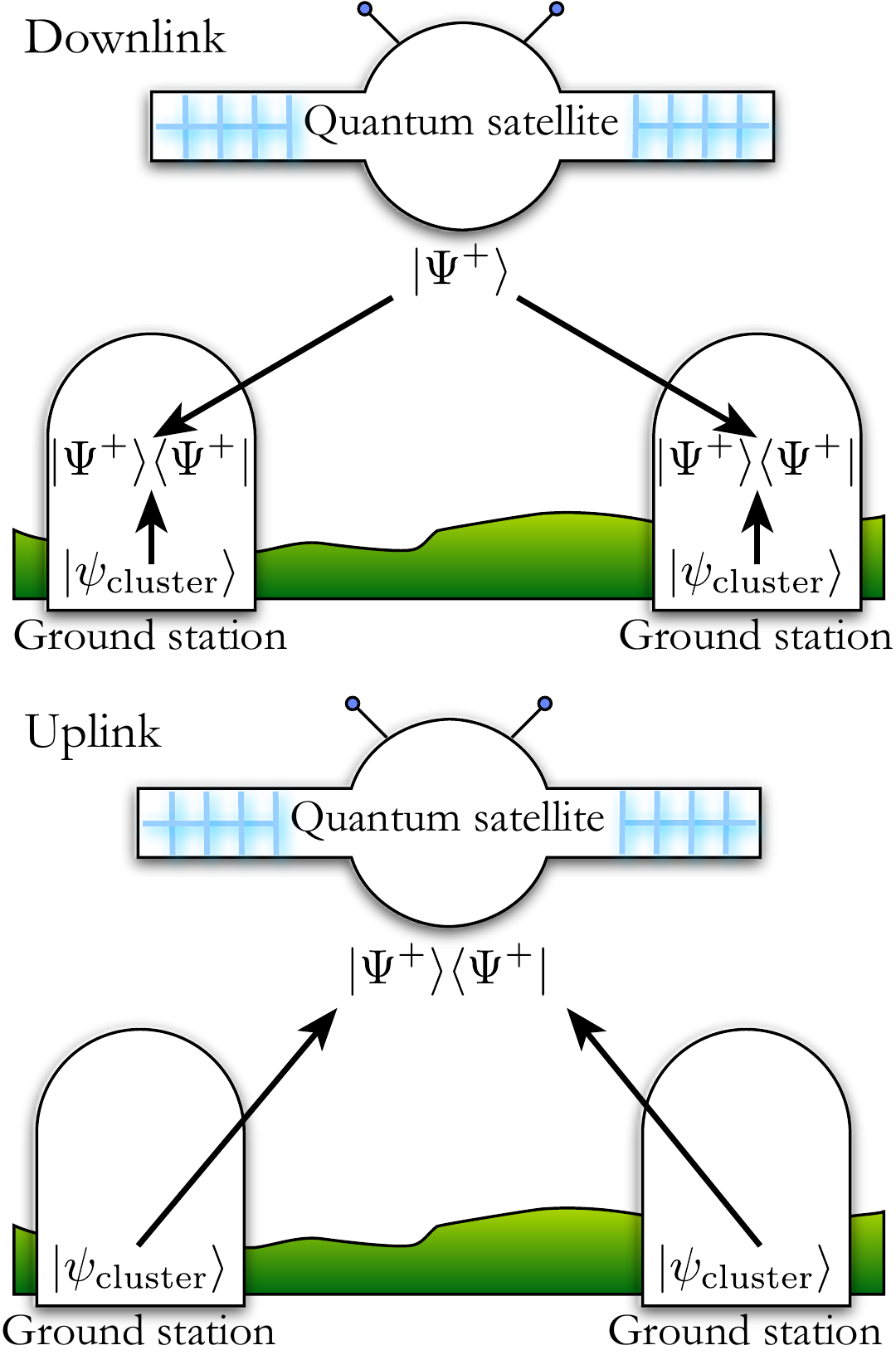}
		\captionspacefig \caption{Satellite-mediated cluster state fusion operations for creating a link between two cluster states held by distant ground nodes. (top) Via entanglement distribution over a downlink channel. (bottom) Via distributed Bell projection over an uplink channel. When performing a distributed Bell measurement it is essential that the optical qubits arrive synchronously at the entangling measurement device, denoted \mbox{$\ket{\Psi^+}\bra{\Psi^+}$} (e.g a PBS), which is technologically challenging to implement on satellite given the unpredictable nature of the atmospheric quantum channel, necessitating on-board quantum memories to synchronise the qubits. This is likely to make downlinks cheaper, faster and more efficient than uplinks.} \label{fig:sat_up_down}
	\end{figure}
\else
	\begin{figure*}[!htbp]
		\includegraphics[clip=true, width=\textwidth]{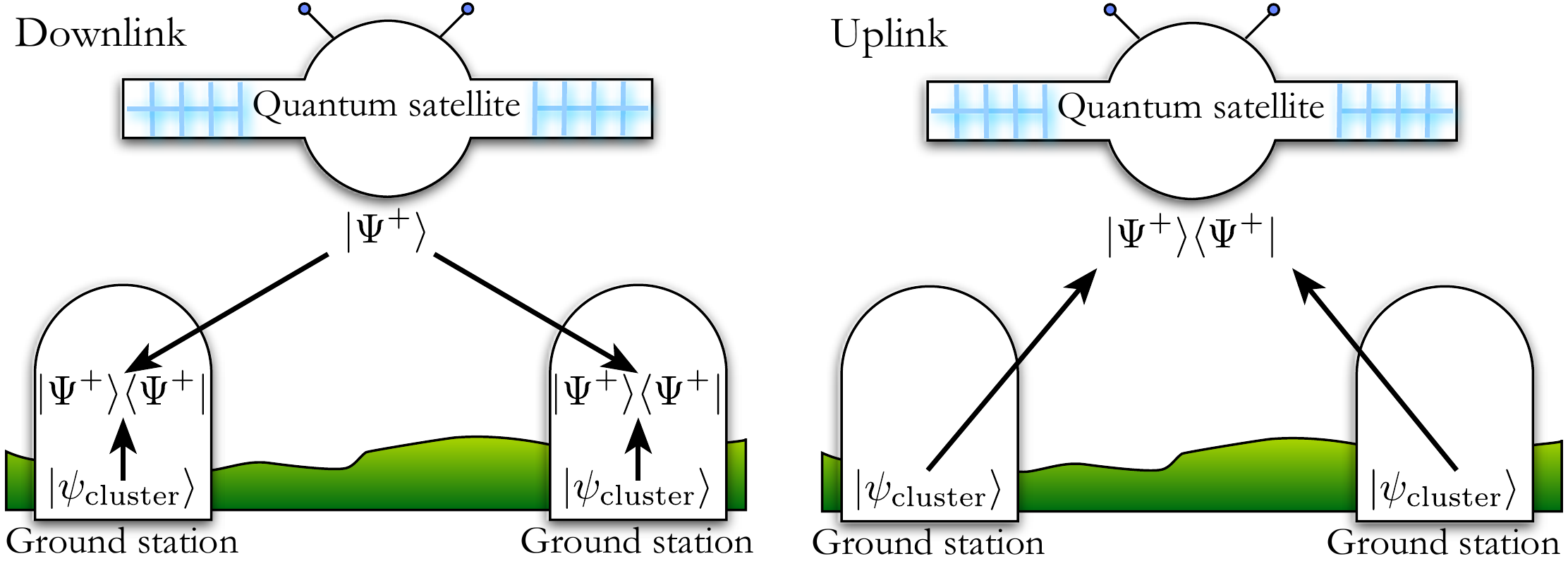}
		\captionspacefig \caption{Satellite-mediated cluster state fusion operations for creating a link between two cluster states held by distant ground nodes. (left) Via entanglement distribution over a downlink channel. (right) Via distributed Bell projection over an uplink channel. When performing a distributed Bell measurement it is essential that the optical qubits arrive synchronously at the entangling measurement device, denoted \mbox{$\ket{\Psi^+}\bra{\Psi^+}$} (e.g a PBS), which is technologically challenging to implement on satellite given the unpredictable nature of the atmospheric quantum channel, necessitating on-board quantum memories to synchronise the qubits. This is likely to make downlinks cheaper, faster and more efficient than uplinks.} \label{fig:sat_up_down}
	\end{figure*}
\fi

Mathematically, these two processes are almost identical in their operation, differing only in direction. However, the former has the key advantage that it requires no time-synchronisation operations on the server side, whereas the latter does, and satellite-based hardware is orders of magnitude more expensive than Earth-based hardware.

Both scenarios involve Bell projections. These entangling measurements require active synchronisation to ensure that the measured qubits arrive at the entangling measurement device (typically a PBS) simultaneously, so as to achieve high HOM-visibility\index{HOM-visibility}, requiring synchronisation on the order of the photons' coherence length\index{Coherence!Length}. This can be achieved either using a brute-force \textsc{Repeat-Until-Success} mode of operation (post-selecting on events where both qubits arrive within a required temporal window), or storing one qubit in quantum memory\index{Quantum memory} until the other arrives. However, post-selection is expensive, requiring a massive overhead in the number of trials, and quantum memory is technologically challenging to implement, more so in space.

In the former case, the time ordering of the Bell projections performed locally on the ground nodes is irrelevant. Although within each ground station the two qubits being projected must be synchronised, requiring quantum memories within ground stations.

On the other hand, in the latter case it is essential that both optical qubits arrive at the satellite's entangling measurement device simultaneously, which is extremely difficult to enforce when our quantum channels are tracking moving targets in low-Earth orbit and traversing a turbulent atmospheric channel in between. An on-satellite quantum memory would be extremely costly!

This yields several key advantages in favour of entanglement distribution as opposed to server-side joint entangling measurements:
\begin{enumerate}
	\item The challenging prospect of quantum memory may operate on Earth, far less onerous and expensive than incorporating this technology into a satellite in low-Earth orbit.
	\item Because the server is not storing any qubits in quantum memory, it does not suffer downtime associated with the periods between receiving the first photon and waiting for the second -- it can continue to spit out Bell pairs at maximum capacity.
	\item The satellite remains passive, implementing only the simplest of possible operations, reducing mass-production costs.
	\item The satellite does not require any interaction with its clients (classical or quantum).
	\item Because Bell pairs are known, infinitely reproducible states, the server can operate in a UDP-like\index{User Datagram Protocol (UDP)} mode and it is not problematic if any given pair was lost. In the reverse direction, loss of a qubit could compromise the entire peripheral state associated with it in the ground station.
	\item Entanglement purification can be employed to enhance the effective quality of the transmission channel.
\end{enumerate}

We therefore anticipate that distributed entangling operations are likely to be mediated via entanglement distribution rather than distributed entangling measurements in the future quantum internet.

These observations lead us to naturally conclude that a quantum network specialised to this one particular task -- entanglement distribution -- would already be immensely useful, and on its own enable many key applications.

\latinquote{Credo in unum deum.}

%
% Quantum Repeater Networks
%

\section{Quantum repeater networks} \label{sec:rep_net} \index{Quantum repeater networks}

\sectionby{William Munro}\index{William Munro}

\dropcap{I}{n} the previous section (Sec.~\ref{sec:ent_ultimate}) we concluded that quantum networks specialising purely in entanglement distribution (Bell pairs in the simplest case), would already be extremely capable in enabling many distributed quantum protocols. This motivates the development of protocols for entanglement distribution over noisy, long-distance quantum networks.

Any useful future quantum internet is going to require the communication of quantum information over arbitrarily long distances. While intercity communication might be implemented via point-to-point connections\index{Point-to-point (P2P)}, intracity and intercontinental communication will require extremely long distance links, well beyond the attenuation length of the optical fibres connecting them or the line-of-sight of satellites in orbit.

\begin{figure}[!htbp]
\includegraphics[clip=true, width=0.475\textwidth]{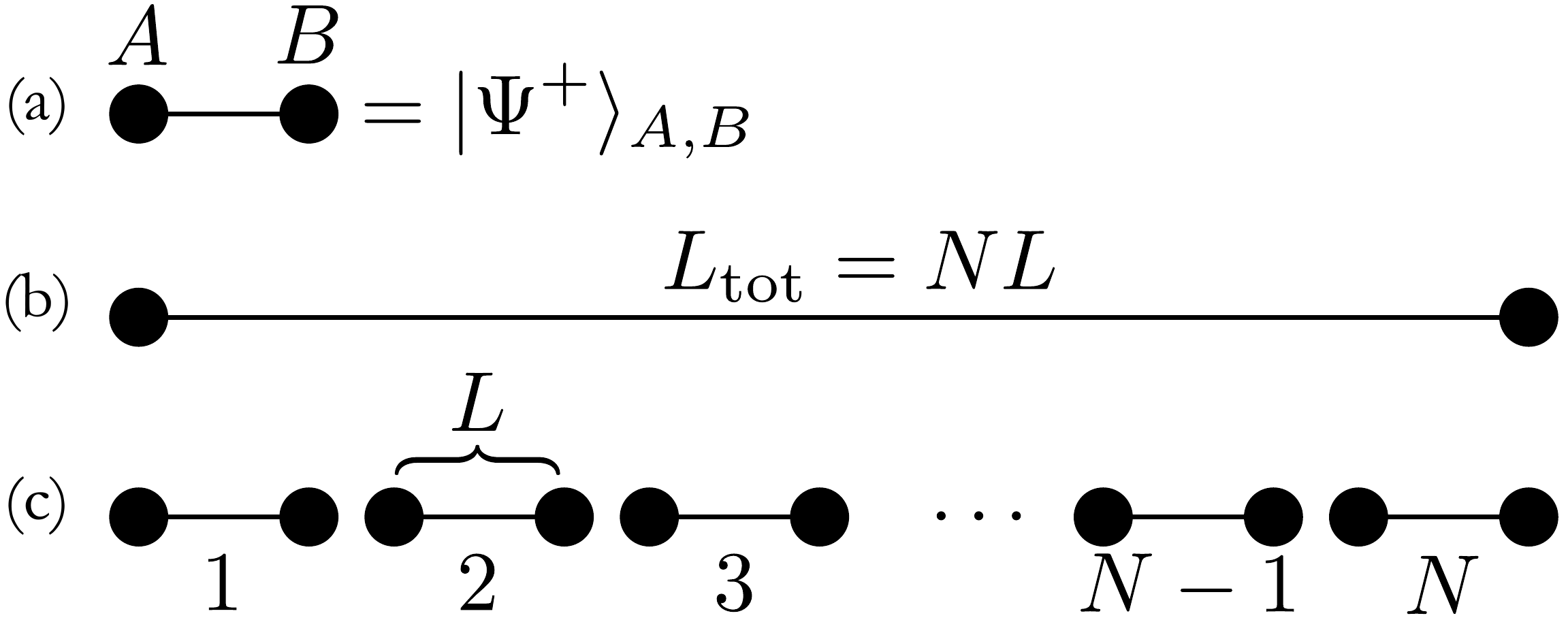}
\captionspacefig \caption{(a) Schematic representation of an entangled Bell pair $\ket{\Psi}_{A,B}$ shared between remote parties Alice and Bob. The two solid dots represent physical qubits, while the edge represents entanglement. (b) The link may be over a long distance $L_{\rm tot}$. (c) Due to channel losses the link may be broken into $N$ smaller segments of length $L$. The links for each of the smaller segments can be independently generated and combined to form the longer distance link.} 
\label{fig:repeaters_1}
\end{figure}

\textit{Quantum repeaters}\index{Quantum repeaters} \cite{bib:Gisin2007, bib:sangouard11, bib:WJM2015} are devices that allow high-quality entanglement to be shared between distant nodes, when no direct line of communication is available from a server to its two clients. This is achieved by dividing long-distance links into a finite number of segments interspersed with repeaters (see Fig.~\ref{fig:repeaters_1}).

 For example, a satellite in low Earth orbit (Sec.~\ref{sec:quant_space_race}) may be outside simultaneous line-of-sight\index{Line-of-sight} to two distinct ground stations\index{Ground stations}, owing simply to the curvature of the Earth\index{Earth curvature}. But this can be overcome by relaying a channel through several satellites in line-of-sight of one another. This is achieved using a bootstrapped entanglement swapping and purification protocol (Secs.~\ref{sec:swapping} \& \ref{sec:ent_purif}). Most commonly, this entanglement is in the form of Bell pairs\index{Bell!States}, which, as discussed previously in Sec.~\ref{sec:ent_ultimate}, form a ubiquitous resource for many essential quantum protocols. The actual physical encoding of the entangled states may vary, but is most commonly and archetypically in the form of polarisation-encoded\index{Polarisation!Encoding} single-photons or CV states\index{Continuous-variables!States}.
 
The links are now over much shorter distances and so can be generated with far higher probability. Then by stitching these together using entanglement swapping (Sec.~\ref{sec:swapping})\index{Entanglement!Swapping}, we can generate our required long-range entanglement link.

Beginning from this simple principle, the field of quantum repeater networks has grown enormously, leading to several generations of repeater designs, of ever increasing power and sophistication, and ever more challenging technological demands.

\subsection{First-generation repeaters}\index{First-generation!Repeaters}

The above description is very hand-wavy, and of course things are a little more complicated in practise. We now examine these ideas in a little more detail, starting with a simple linear chain of repeater stations. 

\begin{figure}[!htbp]
\includegraphics[clip=true, width=0.4\textwidth]{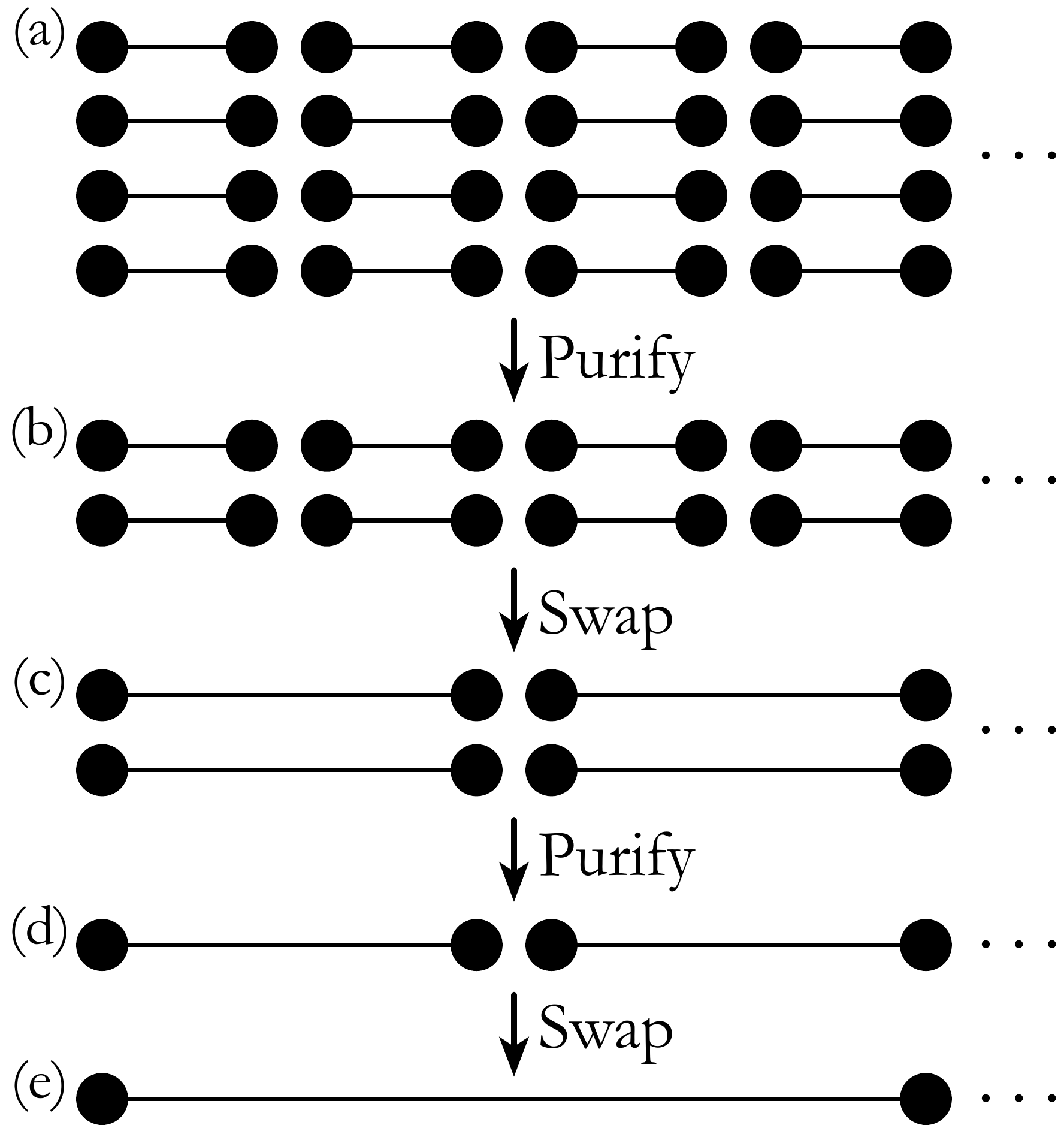}
\captionspacefig \caption{Basic operation of a (first-generation) quantum repeater network. (a) Preparation of multiple entangled links between adjacent repeater nodes. (b) They are then purified to create higher fidelity links. (c) Entanglement swapping between adjacent pairs creates links of twice the original length. (d) These new links are purified to create higher fidelity ones. (e) Entanglement swapping in creates a link four times the original size. This process continues as necessary to reach target distance and purity.} 
\label{fig:repeaters_2}
\end{figure}

In a quantum repeater network, there are three main operations required:
\begin{enumerate}
\item Entanglement distribution (Sec.~\ref{sec:reps_ent_dist}): to create entangled links between adjacent repeater nodes.\index{Entanglement!Distribution}
\item Entanglement purification (Sec.~\ref{sec:reps_ent_purif}): to improve the quality of entanglement between nodes\index{Entanglement!Purification}.
\item Entanglement swapping (Sec.~\ref{sec:reps_ent_swap}): to join adjacent entangled links together to form longer distance links\index{Entanglement!Swapping}.
\end{enumerate}
The basic operation of a repeater, as shown in Fig.~\ref{fig:repeaters_2}, works as follows:

We begin our preparation of a long-range entangled link by creating multiple entangled pairs between adjacent repeater nodes (the number will depend both on the quality of the pairs we initially generate and also the target quality we want our final pair to have). Once we have enough pairs established between two repeater nodes, we perform entanglement purification\index{Entanglement!Purification}, which converts multiple entangled links (pairs) into a fewer number with higher quality. 

These purification steps, shown in Fig.~\ref{fig:repeaters_2}(a-b), are performed on the links between all adjacent repeaters, increasing the quality of the links between those adjacent repeater stations to the required degree. Entanglement swapping, as shown in Fig.~\ref{fig:repeaters_2}(c), then creates links twice as long. The resulting entanglement links can then be used iteratively for further rounds of purification and swapping until one generates a high quality link between the desired points in the network. 

\subsubsection{Entanglement distribution}\index{Entanglement!Distribution}\label{sec:reps_ent_dist}

Probably the most important operation for any quantum repeater setup is entanglement distribution, the process of creating entanglement between two remote parties (Alice and Bob) connected by a quantum channel (generally an optical fibre or free-space link). This can be implemented in a number of ways \cite{bib:Bennett96, bib:enk98, bib:bennett93, bib:sangouard11, bib:childress06, bib:loock06, bib:munro08}, but can be broadly categorised into three basic schemas:
\begin{itemize}
\item Photon emission from quantum memories in the repeater nodes, followed by which-path erasure.\index{Which-path erasure}
\item Absorption of entangled photons by quantum memories.
\item Photon emission at one node and absorption at another.
\end{itemize}
By far, the emission based schemes are the most common, which we will concentrate on here. Such schemes operate by using an entangling operation -- \textit{which-path erasure} -- to entangle two quantum memories via photons to which they were coupled. Effectively the process teleports the action of an entangling gate\index{Quantum gate teleportation} (Sec.~\ref{sec:teleport_gate}) acting on the photons onto the quantum memories to which they were entangled.

We now describe such a which-path entangling operation in the context of 2-level quantum memories coupled to polarisation-encoded photons. A closely related scheme for preparing cluster states on $\lambda$-configuration systems for the purposes of quantum computation is discussed in Sec.~\ref{sec:hybrid}.

Ideally one wants to initially generate a maximally entangled state of the form \cite{bib:WJM2015},
\begin{align}
\ket{\Psi}=\frac{1}{\sqrt{2}} (\ket{g}\bra{H} + \ket{e}\bra{V}),
\end{align}
within the repeater node, where $\ket{g}$ and $\ket{e}$ are the two states (ground and excited) of the quantum memory, and $\ket{H}$ and $\ket{V}$ are the polarisation states of a single photon. 
\begin{figure}[!htbp]
\includegraphics[clip=true, width=0.3\textwidth]{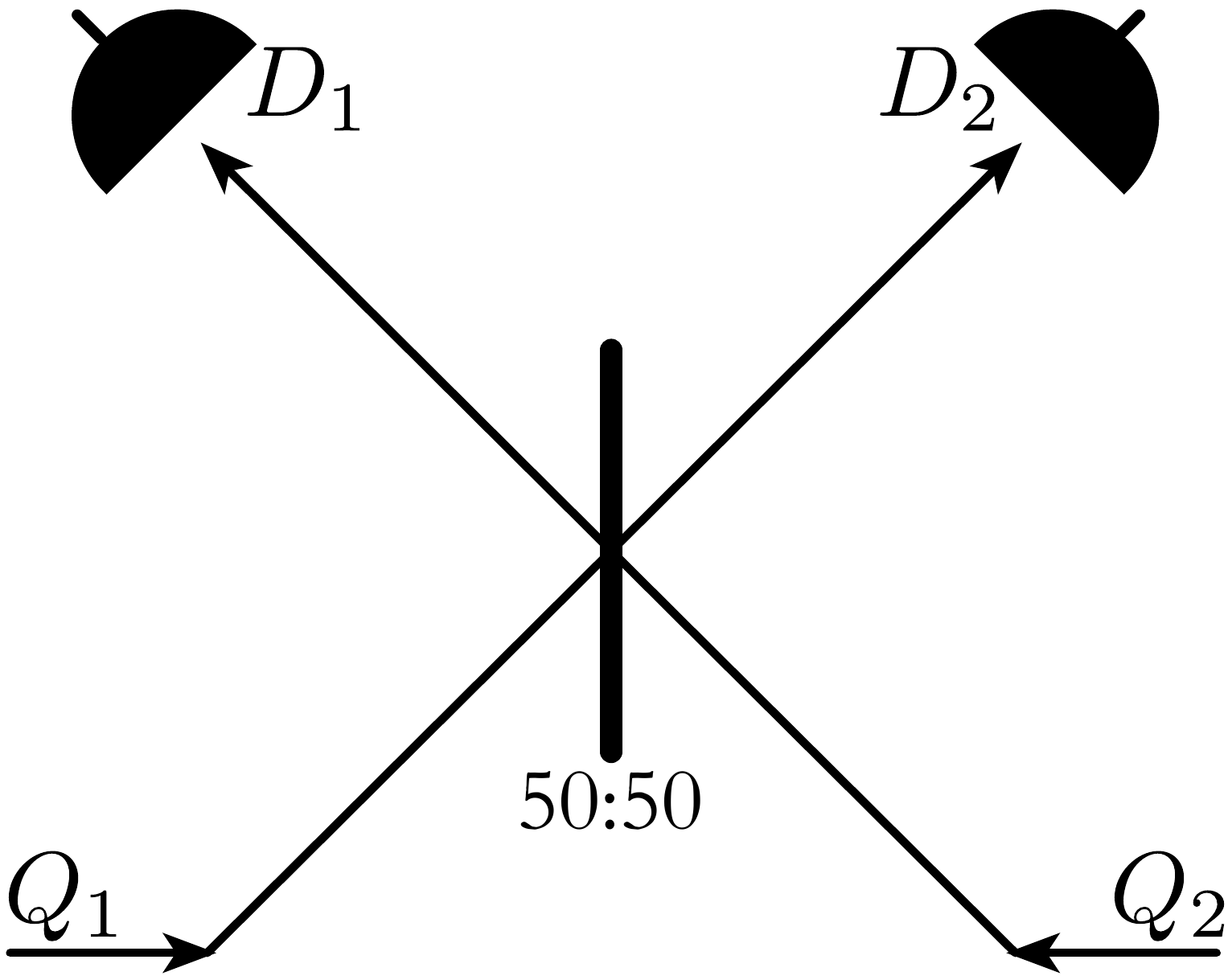}
\captionspacefig \caption{Entanglement distribution scheme based on quantum emitters and which-path erasure\index{Which-path erasure}. Each node emits a photon entangled with the quantum memories present within that nodes. The photons from the adjacent repeater nodes then interfere on a beamsplitter (or polarising beamsplitter)\index{Beamsplitters}\index{Polarising beamsplitters} which erases information about which path the photon took. The photons are then measured in an appropriate basis to project the quantum memories within the nodes onto an entangled state.} 
\label{fig:repeaters_3}
\end{figure} 
The photons from the two repeater nodes (Fig.~\ref{fig:repeaters_3}) are then transmitted to a beamsplitter (or PBS in this example), after which the state of the system is,
\begin{align}
\ket{\Psi} &= \frac{1}{2} \ket{g} \ket{g} \ket{H} \ket{H} +\frac{1}{2} \ket{e} \ket{e} \ket{V} \ket{V} \nonumber \\
&+\frac{1}{2} \ket{g} \ket{e} |\ket{HV} \ket{0} + \frac{1}{2} \ket{e} \ket{g} \ket{0}\ket{HV}. 
\end{align}

One immediately notices that the $\ket{g} \ket{e}$ and $\ket{e}\ket{g}$ contributions are associated with two photons in one of the PBS exit modes, the other being in the vacuum state. However, the $\ket{g}\ket{g}$ and $\ket{e}\ket{e}$ terms have one photon in each of the output modes. They are of opposite polarisation, but measuring those photons in the diagonal/anti-diagonal ($\hat{X}$) basis erases this `which-path' information\index{Which-path erasure} yielding an equal superposition of the two alternative histories -- an entangled Bell state\index{Bell!States} of the form,
\begin{align}
\ket{\Psi_\pm}=\frac{1}{\sqrt{2}} (\ket{g}\ket{g} \pm \ket{e}\ket{e}),
\end{align}
where the sign is given by the parity of the two photo-detection outcomes in the $\hat{X}$ basis. This entangled state is stored in the quantum memories between nodes. 

The scheme based on photon absorption by the quantum memories is effectively the time reversal of the emission-based scheme. Instead of using the beamsplitter to entangle the photons emitted from each memory, a source of entangled photon(s) is employed. Of course, the emission and absorption schemes can be used together in a hybrid architecture.

In any entanglement distribution scheme for quantum networks, the repeater nodes are spatially separated and one must consider channel losses, which are the dominant error source. Channel loss in this situation implies that we do not register a coincidence event between $D_1$ and $D_2$, which heralds the entanglement. Thus our entanglement distribution success probability is reduced. In fact, the heralded probability of success can be expressed as,
\begin{align}
p_\mathrm{ED}= \frac{1}{2} e^{-L/L_0} {p_\mathrm{det}}^2,
\end{align} 
where $L$ is the distance between the two repeater nodes with $L_0$ being the attenuation length of the channel, while $p_\mathrm{det}$ is the detector efficiency. Here we have ignored the source and coupling efficiencies. It is immediately obvious from this expression that the further the repeater nodes are apart, the lower the probability of success, on an exponentially decaying trajectory. The attenuation length of typical telecom optical fibre is approximately 22.5km and so the average time to generate a distributed entangled pair is,
\begin{align}
T_\mathrm{av} &\sim \frac{L}{ c \cdot p_\mathrm{ED}}\nonumber\\
&= \frac{2 L e^{L/L_0}}{ c \cdot {p_\mathrm{det}}^2}
\end{align} 
where $c$ is the speed of light in the channel. This grows exponentially against node separation and so places important constraints on the lifetime of the quantum memories. If we consider pure dephasing effects on our matter qubits, the state of our system can be represented by,
\begin{align}
\hat\rho(F) = F \ket{\Psi^+} \bra{\Psi^+}+(1-F) \ket{\Psi^-}\bra{\Psi^-},
\label{eq:rho_bell_fid}
\end{align} 
where $F$ is the fidelity of our entangled state given by,
\begin{align}
F=\frac{1+e^{-t/\tau_\mathrm{D}}}{2},
\end{align} 
with $t$ being the duration over which the entangled state is held in memory, while $\tau_\mathrm{D}$ is the coherence time of the memory. If one only requires a single Bell pair and no further operation are performed, then \mbox{$t=c/L$}. However in a more general setting where multiple pairs are required, the time will be $T_\mathrm{av}$ on average, which is inversely proportional to the probability of generating the entangled state. The quality of the prepared remote entangled state may therefore not be sufficient for the tasks it is required for due to these finite memory lifetimes or operational gate errors. One needs to be able to purify these entangled resources. 

\subsubsection{Entanglement purification}\index{Entanglement!Purification}\label{sec:reps_ent_purif}

\begin{figure}[!htbp]
\includegraphics[clip=true, width=0.35\textwidth]{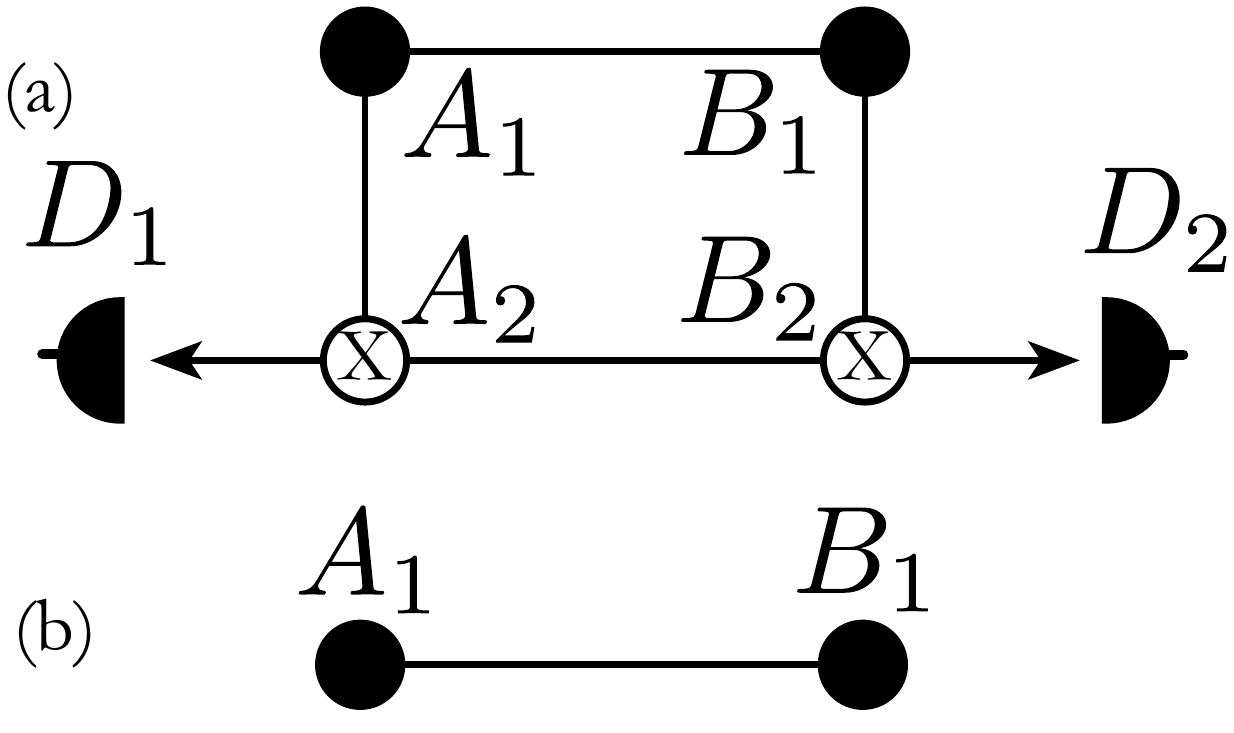}
\captionspacefig \caption{Entanglement purification: (a) The simplest purification scheme involving two pairs of shared remote entangled quantum memories (\mbox{$A_1-B_1$} and \mbox{$A_2-B_2$}). The purification operation begins with Alice performing a CNOT operation between memories $A_1$ and $B_1$. Similarly Bob performs a CNOT operation between his memories. Alice and Bob then measure qubits $A_2$ and $B_2$ in the computational ($0$, $1$) basis and share their results. They discard the resulting state if between them they measured odd parity ($0,1$ or $1,0$). They keep the state if they measured an even parity between them ($0,0$ or $1,1$) which should have higher fidelity. (b) Two qubits are removed, leaving a residual 2-qubit state between $A_1$ and $B_1$ with improved fidelity.} 
\label{fig:repeaters_4}
\end{figure}

The finite coherence-time of quantum memories and operational errors caused by quantum gates means some mechanism will be required to improve the fidelity of the distributed entangled state, especially if the spatial separation is large. This is generally achieved by entanglement purification \cite{bib:Bennett96, bib:Deutsch96, bib:dur98, bib:Pan01, bib:dur07, bib:Aschauer2004, bib:jiang09, bib:munro12, bib:Stephens2013} which  as it name implies purifies the entanglement to a higher value. The purification operation uses either an error detection code (probabilistic but heralded operations) \cite{bib:Bennett96, bib:Deutsch96, bib:dur98} or deterministic error correction codes \cite{bib:Aschauer2004, bib:jiang09, bib:munro12}. While the error correction codes purify in a deterministic way, they place tough constraints on both the required initial fidelity of entangled states and also the quality of the quantum gates implementing the purification \cite{bib:Aschauer2004}. Given this, we will focus on the simplest error detection code which requires only a pair of shared entangled quantum memories (as shown in Fig.~\ref{fig:repeaters_4}). This scheme is equivalent to the entanglement purification protocol described in Sec.~\ref{sec:ent_purif}, although the graphical notation is somewhat different.

In this simplest purification protocol, Alice and Bob share two pairs of entangled states of the form given by Eq.~(\ref{eq:rho_bell_fid}). These states are a mixture of only two Bell states. We begin our purification protocol by using local operations to transform $\hat\rho$ to,
\begin{align}\label{eq:rho_bell_fid_dash}
\hat\rho(F)=F \ket{\Psi^+}\bra{\Psi^+}+(1-F) \ket{\Phi^+}\bra{\Phi^+},
\end{align}

As shown in Fig.~\ref{fig:repeaters_4} we then apply a CNOT gate between Alice's two memories and Bob's two memories following by measuring $A_2, B_2$ in the computational basis. Upon measurement of even parity our resulting state $\hat\rho(F')$ has the form, but with new fidelity,
\begin{align}
	F'=\frac{F^2}{F^2+(1-F)^2}.
\end{align}

\begin{figure}[!htbp]
\includegraphics[clip=true, width=0.35\textwidth]{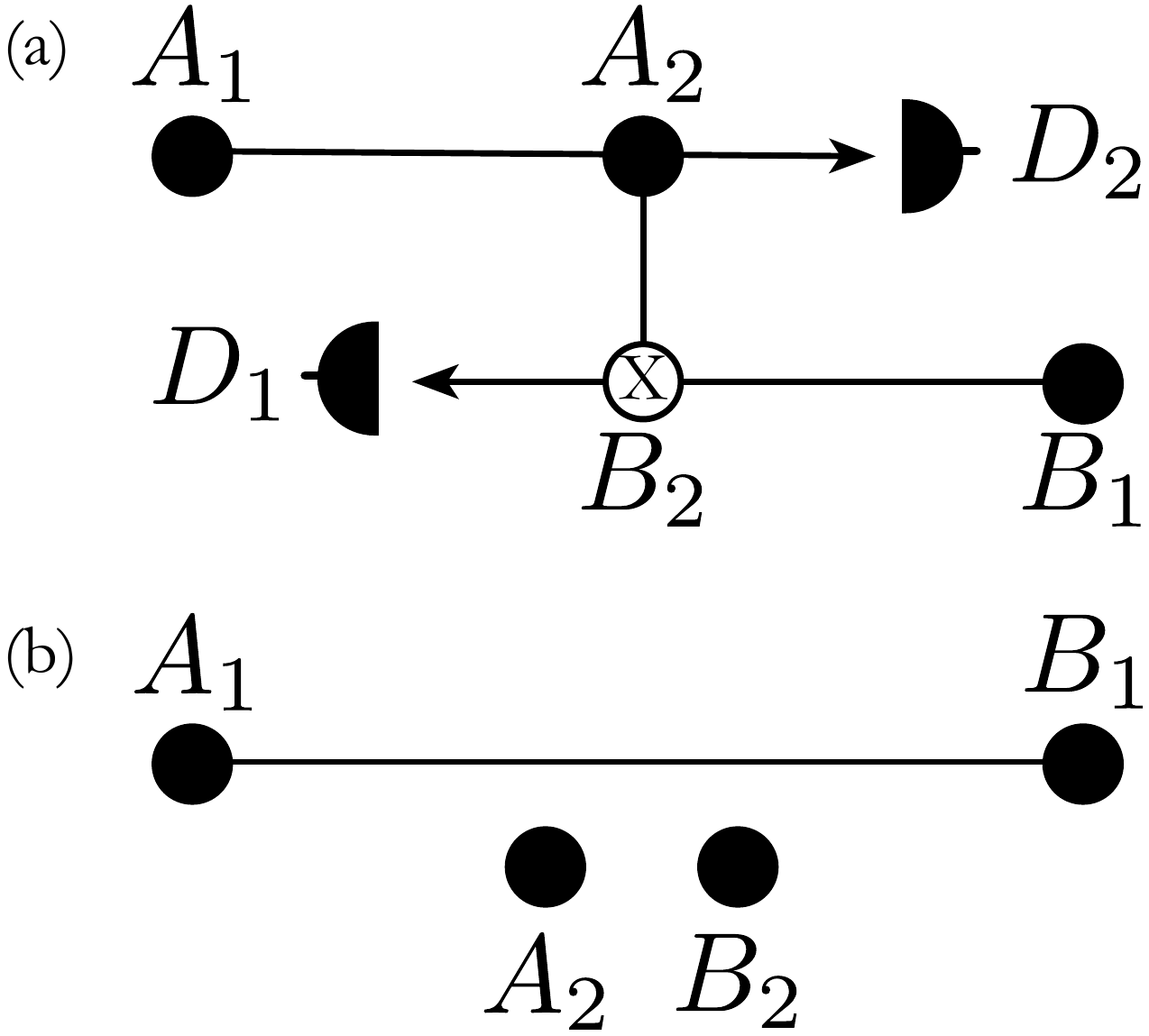}
\captionspacefig \caption{Plot of the increased fidelity and success probability for entanglement purification for a mixture of two Bell states with initial fidelity $F$. The dashed lines show how multiple pairs with an initial fidelity $F=0.7$ can be purified iteratively to a final fidelity above 0.95.} 
\label{fig:rep_purification}
\end{figure}

It is immediately obvious that our resulting state $\hat\rho(F')$ is more entangled than $\hat\rho(F)$ when \mbox{$F>1/2$} (see Fig.~\ref{fig:rep_purification}). In fact the degree of entanglement as measured by the concurrence\index{Concurrence} increases from,
\begin{align}
	C=2 F-1,
\end{align}
to,
\begin{align}
C' &=2 F'-1 \nonumber\\
&= \frac{2 F^2}{F^2+(1-F)^2}-1.
\end{align}

It is important to mention that the entanglement purification doesn't allow one to distribute a \textit{perfect} Bell state. Rather it \textit{asymptotically} approaches perfection (under ideal conditions) with repetition of the protocol.

The probability of obtaining the even parity outcome is,
\begin{align}
	p_\mathrm{even}=\frac{F^2+(1-F)^2}{2}.
\end{align}
Alternatively for the odd parity measurement results, which occur with probability,
\begin{align}
	p_\mathrm{odd}=F(1-F),
\end{align}
the resulting state is an equal mixture of $\ket{\Psi^+}$ and $\ket{\Phi^+}$ and is not entangled at all. In this case we must start again from scratch with the entanglement distribution.

So far we have discussed one round of entanglement purification but the protocol naturally works in a recursive way where two copies of a state with the same fidelity are used for the next purification round. Using this bootstrapped approach one can in principle generate a near unit fidelity entangled pair from a finite fidelity pair (provided initial input fidelity \mbox{$F>1/2$}). 

There are two common variants of these purification protocols: the Deutsch and D{\"u}r variants:
\begin{itemize}
\item \textit{Deutsch protocol} \cite{bib:Deutsch96}\index{Deutsch protocol}: This is an efficient purification protocol utilising Bell diagonal states that reaches a high fidelity in a few purification rounds. It is assumed that both entangled pairs have the same form. The purification protocol is the same at the one described above in Fig.~\ref{fig:repeaters_4}, but begins with Alice (Bob) applying $\pi/2$ $(-\pi/2)$ rotations about the $X$-axis on their qubits before the usual CNOT gates and measurements are performed. Two copies of the successfully purified pair can then be used in a recursive approach to purify either further. This in turns means multiple copies of the originally distributed states are required. We must have enough entangled pairs available to perform the multiple rounds of purification that are required, which grows exponentially with the number of purification rounds. 

\item \textit{D{\"u}r protocol} \cite{bib:dur98}\index{D{\"ur} protocol}: This uses the same core purification elements as shown in Fig.~\ref{fig:repeaters_4} but relaxes the traditional constraint that both Bell pairs must have the same fidelity. Instead we begin with two pairs of the same fidelity $F$, and perform the traditional purification. If successful we perform the next round of purification using the improved fidelity pair from the previous round and a fresh fidelity $F$ pair. In effect this new auxiliary pair is used to boost the fidelity of the original pair higher. This can continue until we reach a limiting fidelity dependent on the original $F$. This limiting fidelity may be above the desired resultant fidelity, at which point we can terminate the purification protocol. A significant difference between the Deutsch and D{\"u}r protocols is that the number of memories in the D{\"u}r situation is linear in the number of nesting levels.
\end{itemize}

It is critical in repeater protocols to also discuss how fast these purification protocols can be performed. Even with ideal gates one has to wait for the parity information to be shared between the repeater nodes. For nodes separated by a distance $L$, the communication time for a single trial is $L/c$. However, remembering that purification is probabilistic but heralded in nature our waiting time could be many multiples of $L/c$. This will have a dramatic effect on performance, especially if performed at many different stages in the network with increasing distances between nodes.

\subsubsection{Entanglement swapping}\label{sec:reps_ent_swap}\index{Entanglement!Swapping}

The entanglement distribution and purification scheme discussed previously allow one in principle to create high fidelity entangled states between adjacent repeaters nodes. The next task is to extend the range of our entangled states, and this occurs via simple entanglement swapping \cite{bib:BDCZ98, bib:Zukowski93, bib:goebel08, bib:Duan01}. This was described previously in Sec.~\ref{sec:swapping}, although the notation is modified.

\begin{figure}[!htbp]
\includegraphics[clip=true, width=0.4\textwidth]{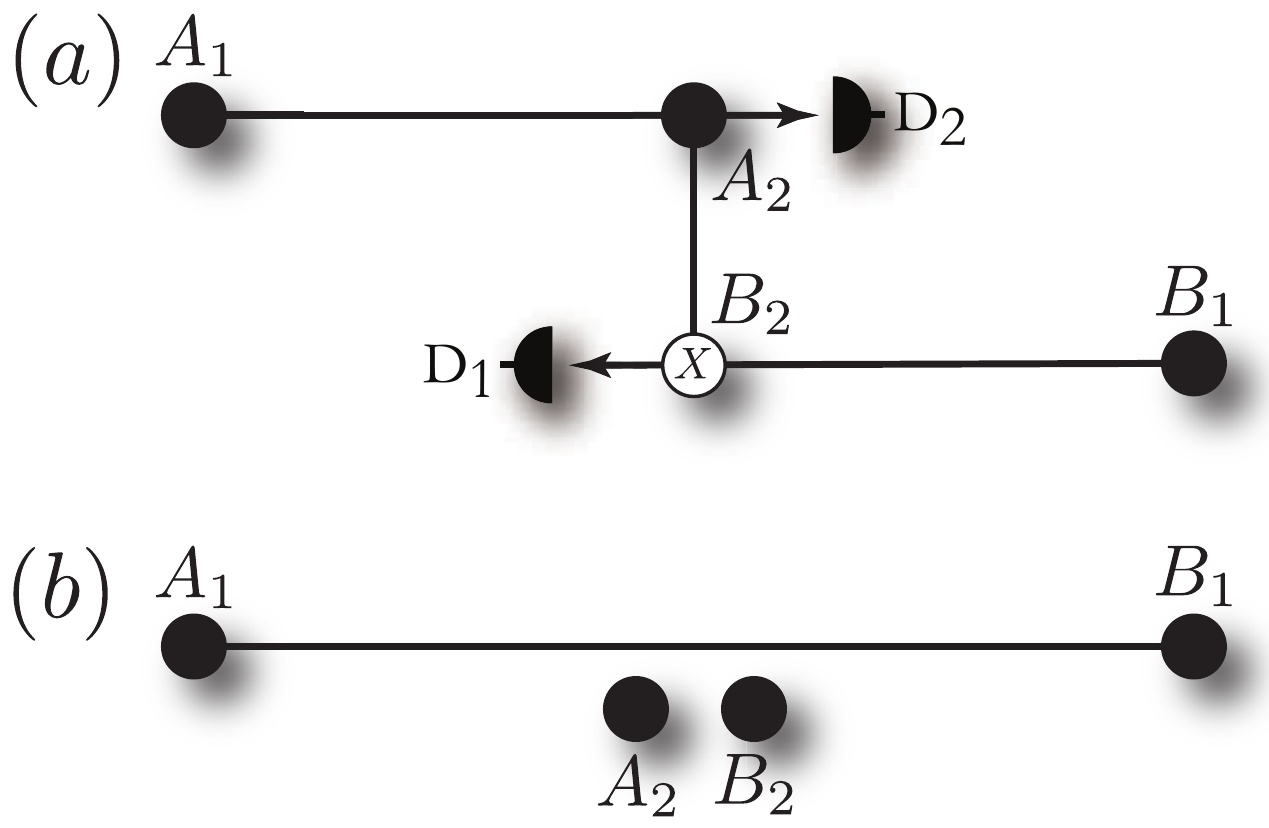}
\captionspacefig \caption{Entanglement swapping: (a) An entangled state is shared between Alice and a repeater station ($A_1-A_2$), and also between the repeater station and Bob ($B_2-B_1$). The entanglement operation begins by performing a Bell state measurement between $A_2$ and $B_2$ using a CNOT gate, and measurements at $D_1$, $D_2$.  The measurements indicate which Bell state we have projected our state $A_1$, $B_1$ onto. (b) The resultant entangled state between $A_1$, $B_1$ with the qubits $A_2$ and $B_2$ disentangled from it.}
\label{fig:repeaters_6}
\end{figure} 

Consider the situation where we have an entangled Bell pairs between nodes $A_1$ and $A_2$ and also between $B_2$ and $B_1$. The entanglement swapping operation involves a Bell state measurement between the qubits $A_2$ and $B_2$ as shown in Fig.~\ref{fig:repeaters_6}. After the Bell measurement we have the resultant state, 
\begin{align}
\hat\rho_{A_1,A_2} (F)\otimes \hat\rho_{B_2,B_1}(F)\rightarrow \hat\rho_{A_1,B_1} (F')
\end{align}
with,
\begin{align}
	F'=F^2+(1-F)^2,
\end{align}
where a local correction operation is performed on either $A_1$ or $B_1$ depending on the measurement outcome. It is clear that the longer range entangled state $\hat\rho_{F'}$ is less entangled that the states $\hat\rho_F$ used to generate it. In fact, to first order our fidelity drops from $F$ to $F^2$. This in turn means that we can not simply purify adjacent repeater pairs and swap them all to create the long range pairs. If we had $n$ links, our final fidelity from all the swapping would scale as $F^n$. For high fidelity end-to-end entangled links we need to follow the approach outlined in Fig.~\ref{fig:repeaters_2}. Finally, depending on how the Bell measurement is implemented, this process could be probabilistic (but heralded) or deterministic in nature. We assign the success probability as $p_\mathrm{ES}$.

\subsubsection{Performance}\index{Repeater!Performance}

We now have all the operations required for a repeater to create long-range entanglement. The natural question to ask is how well it performs.

There are several important points to initially consider here. The majority of the repeater operations are probabilistic in nature (entanglement distribution and purification fundamentally, and entanglement swapping dependent upon implementation). While these probabilistic operations may be heralded, classical signalling must be performed between involved nodes  to inform them of successes or failures. For entanglement distribution this time is just that associated with the signalling between adjacent nodes. However, purification and swapping are likely to require such signalling over the entire length of the network. This has a dramatic effect on the performance of the repeater network. The normalised rate for generating Bell pairs over a total distance $L_\mathrm{tot}$ is given by,
\begin{align}
R(n,k,L_\mathrm{tot})= \frac{1}{T_{n,k,L_\mathrm{tot}} M_{n,k}}
\label{eq:rep_net_resources}
\end{align}
where $T_{n,k,L_\mathrm{tot}}$ is the time to generate a Bell pair over the total distance using an $n$-nested repeater configuration with $k$ rounds of purification per nesting level. The distance between repeater nodes is given by,
\begin{align}
	L=\frac{L_\mathrm{tot}}{2^n},
\end{align}
meaning there are \mbox{$2^n-1$} intermediate repeater nodes with Alice and Bob at the endpoints. In Eq.~(\ref{eq:rep_net_resources}) we discount our rate by $M_{n,k}$, the total number of quantum memories used. The justification for this is that this provides a fairer comparison when different purification approaches are used. The Deutsch protocol for instance achieves its target fidelity much faster (fewer rounds) than the D{\"u}r protocol, but consumes far more resources in doing so.

Now it's straightforward, albeit tedious, to show that $T_{n,k,L_\mathrm{tot}}$ is given by \cite{bib:braztzik2013},
\begin{widetext}
\begin{align*}
 T_{n,k,L_\mathrm{tot}} &\sim \frac{3^n}{2^{n-1} p_\mathrm{ED}} \prod_{i=0}^{n-1} \left(\frac{3}{2}\right)^{k}  \frac{1}{P_\mathrm{ES}(n-i)  }\prod_{j=0}^{k-1} \frac{1}{p_\mathrm{P}(k-j,n-i)}  \nonumber \\
 &+\sum_{m=1}^n\left(\frac{3^{n-m}}{2^{n-1}}\right) \prod_{i=0}^{n-m}   \left(\frac{3}{2}\right)^{k} \frac{1}{P_\mathrm{ES}(n-i)}  \prod_{j=0}^{k-1}  \frac{1}{p_\mathrm{P}(k-j,n-i)}  \\
&+\sum_{m=1}^n {\sum_{q=0}^{k-1} \left(\frac{3^{n-m+q}}{2^{n- 2 m+q}}\right) \prod_{r=0}^{q}\frac{1}{p_\mathrm{P}(k-r,m)}} 
\prod_{i=0}^{n-m-1}   \left(\frac{3}{2}\right)^{k} \frac{1}{P_\mathrm{ES}(n-i)}  \prod_{j=0}^{k-1}  \frac{1}{p_\mathrm{P}(k-j,n-i)} \nonumber 
\end{align*}
\end{widetext}
where $p_\mathrm{ED}$ is the probability of successfully distributing entanglement between adjacent repeater nodes, while $p_\mathrm{P}(j,i)$ [$p_\mathrm{ES}(i)$] represents the purification [entanglement swapping] probability at the $i$th nesting level with $j$ rounds of purification. The factors of $3/2$ present in all entanglement distribution, purification and swapping operations is a multiplicative factor associated with the extra time required for the two pairs to be available for the various quantum operations \cite{bib:sangouard11}.

It can be easily seen from this formula that,
\begin{align}
	T_{n,k,L_\mathrm{tot}} \gg \frac{2 L_\mathrm{tot}}{c},
\end{align}
especially if probabilistic gates are included. Next the resources scale polynomially with,
\begin{align}
	M_{n,k} &\sim 2^{(k+1)n}\nonumber\\
	&= \left(\frac{L_\mathrm{tot}}{L}\right)^{k+1},
\end{align}
for the Deutsch protocol, which in turn implies it is efficient. However for long distances $L_\mathrm{tot}$, our normalised rate \mbox{$R(n,k,L_\mathrm{tot})\ll 1\mathrm{Hz}$}, especially when probabilistic CNOT gates and Bell state measurements are employed \cite{bib:jiang09, bib:munro10}.

\subsection{Second-generation repeaters \& error correction}\index{Second-generation repeaters}\index{Error correction}

The previous approach for entanglement distribution over long distances based on first-generation quantum repeaters has its performance heavily constrained by both the probabilistic nature of the various quantum operations and the associated classical communication time. We know that the classical communication in entanglement distribution is only between the adjacent nodes, whereas for the purification and swapping operations it can be very long-range, potentially over the entire network length. This is the fundamental reason why the time to create a pair is of order $O(L_\mathrm{tot}/c)$ or longer. This will not change significantly even if we have deterministic CNOT gates and Bell measurements as the entanglement purification protocols will remain probabilistic in nature (even though the swapping operations will be deterministic). We thus need to replace our usual entanglement purification protocols with a similar operation that is deterministic in nature \cite{bib:jiang09, bib:munro10}.

The typical entanglement purification protocols are a form of quantum error detection code \cite{bib:WJM2015, bib:devitt2013} (see Sec.~\ref{sec:QOS} for further discussion on quantum error detection and correction). Such codes herald whether an error has occurred or not, and in the situation considered above, detection of errors means one must discard the entangled pairs associated with the purification protocol. No errors means the purification protocol has worked.

Error correction codes which operate in a deterministic fashion can also detect errors and can be used in this fashion \cite{bib:jiang09, bib:munro10}. More critically, quantum error correction codes have the potential to correct some errors that have occurred, mitigating the need to completely discard states affected by errors. For normal error correction protocols used in quantum computations, we encode our physical qubits into logical qubits using the code, and then use syndrome measurements to determine where an error has potentially occurred.

\begin{figure*}[!htbp]
\includegraphics[clip=true, width=\textwidth]{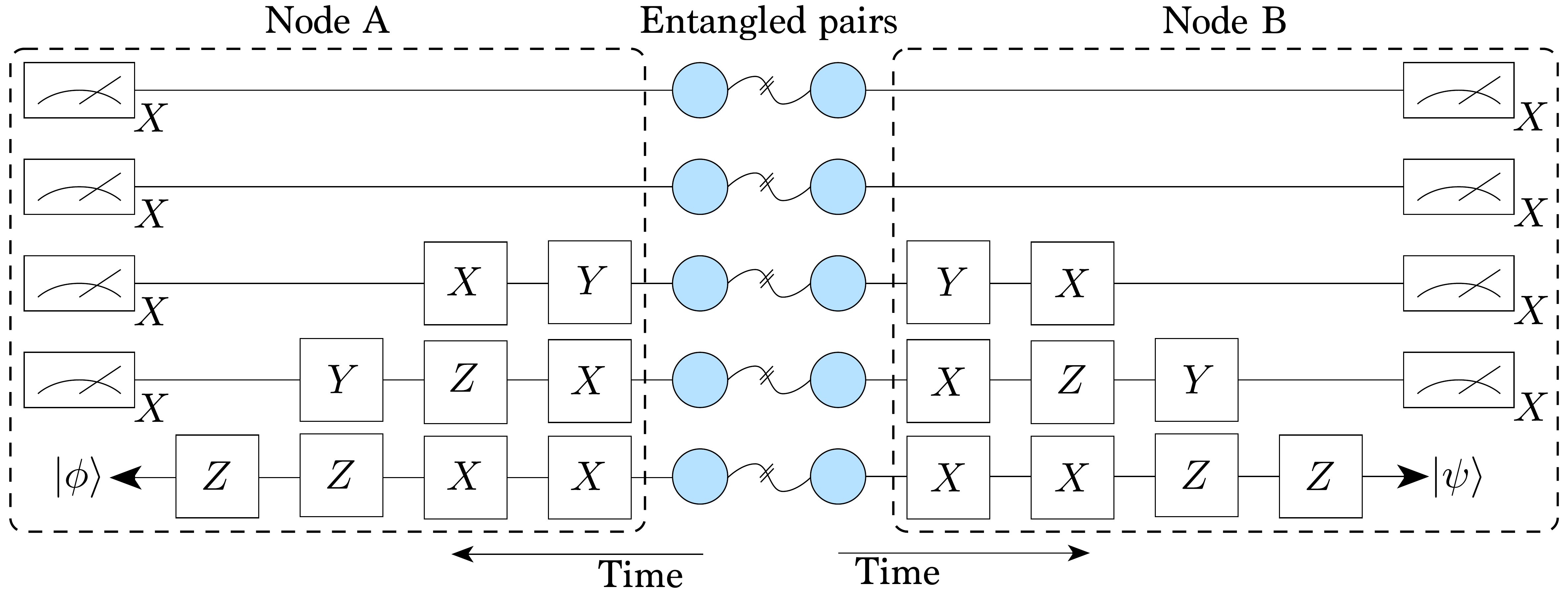}
\captionspacefig \caption{Purification circuit based on quantum error correction. The specific example shown is for the \mbox{[[5,1,3]]} code \cite{bib:Bennettr1996a, bib:Knill97}. We assume that entanglement distribution has allowed Alice and Bob to create 5 copies of their imperfect Bell pairs. The error correction circuit is executed independently between the two nodes. While we show the situation when the measurements at both sides are done directly on four pairs of entangled qubits (leaving us with one unencoded Bell pair), one can also use ancilla qubits to measure the appropriate syndromes. As soon as the measurements are complete both nodes' qubits are available for continued use as the error correction is deterministic and there are no failure events that need to be heralded. In this case the classical message between nodes just carries Alice's measurement results, allowing either node to interpret which Bell state was generated and for one of them to apply the bit-flip or phase-flip correction operation if needed to recover the desired Bell state. In many cases this correction is classically tracked in the Pauli frame, which keeps a record of whether $\hat{X}$ and/or $\hat{Z}$ corrections need to be performed at some stage \cite{bib:jiang09, bib:munro10}. Note that it's not necessary to measure out all but one of the qubits involved in the entangled links. Instead the logical qubit can be maintained by the use of ancilla qubits within that node with the syndrome being measured with the help of the ancilla qubits. Entanglement swapping could then be performed on the logical qubits enabling a much more error resilient system.
}
\label{fig:repeaters_7}
\end{figure*}

Quantum communication however is different in this case as we must assume we have generated a number of imperfect Bell pairs between the repeater nodes before we utilise the error correction schemes. The error correction protocol in this case operates by using the error correction encoding circuit on Alice's qubits and the decoding circuit on Bob's \cite{bib:Aschauer2004} as illustrated in Fig.~\ref{fig:repeaters_7} for the 5-qubit code\index{5-qubit code} \cite{bib:Bennettr1996a, bib:Knill97}.

It's important to state here that error correction-based purification is deterministic in nature (there is however a significant cost that must be paid -- the fidelity of the originally generated entanglement between adjacent nodes must be quite high) \cite{bib:jiang09, bib:Aschauer2004}. There are no measurement events that need to be discarded. Instead the measurement results only inform us of which particular imperfect Bell states we have and the correction operation required to return to the desired state. In effect the measurement is updating the Pauli reference frame\index{Pauli!Reference frame} \cite{bib:Knill2005}. This does not need to be executed immediately, and may be deferred until later. In turn this means once the measurements have been performed, we can immediately use the purified Bell state without having to wait for the classical signalling (at some stage the correction operation needs to be executed but this can be once the long distance entanglement has been generated). 

Mitigating having to wait for the measurement results to be sent and received in both the quantum error correction-based purification and entanglement swapping protocols has a profound effect on the rate of generating long-range entangled pairs. We still need to perform long-range classical messaging (potentially between end nodes), and thus it's immediately obvious that the preparation time can scale solely as,
\begin{align}
	T = \frac{2 L_\mathrm{tot}}{c},
\end{align}
which was the lower bound on the first-generation schemes \cite{bib:munro10}.

\begin{figure*}[!htbp]
\includegraphics[clip=true, width=\textwidth]{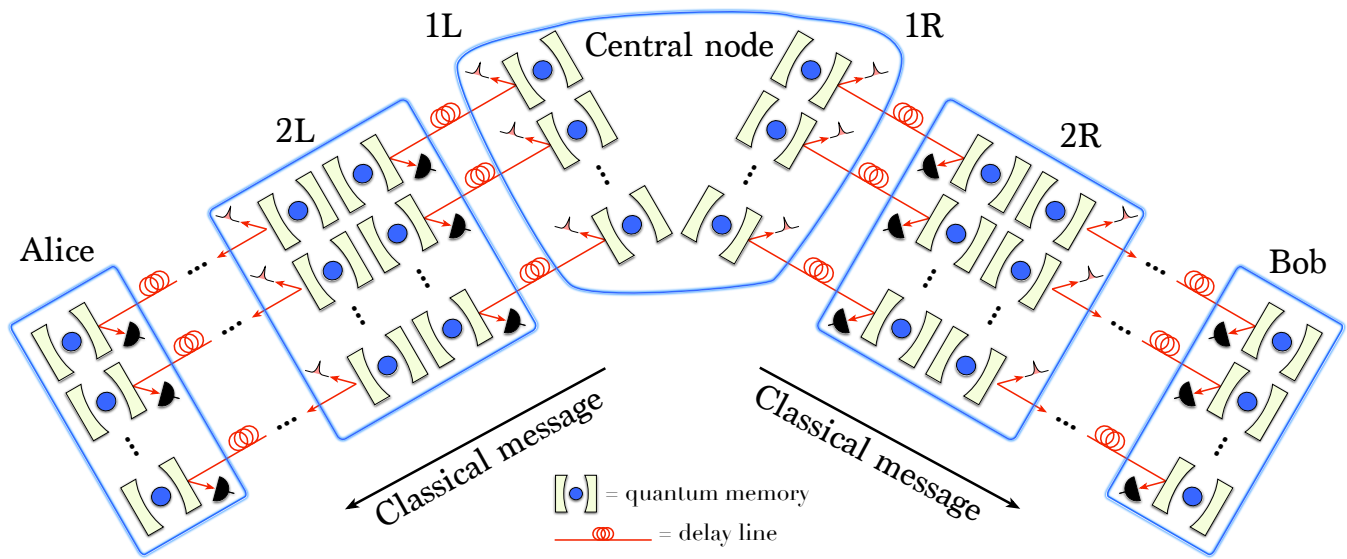}
\captionspacefig \caption{A butterfly design quantum repeater network protocol that reduces the requirements on all the quantum memory times to only that associated with the signalling time between adjacent repeater nodes \cite{bib:munro10}. Enough pairs must be generated between the node to ensure that we can use them in the error correction code in a single round trip time between adjacent nodes. The scheme relies on multiple entangled pairs being generated temporally, starting from the mid point of the network. The protocol begins with the central node creating links to both the left and right nearest neighbour nodes in sufficient number to allow an error correction code to be implemented. Once they are created the error correction circuits are applied to the links left and right of this central node (effectively creating encoded logical links). Entanglement swapping at the middle node is then applied  between these logical links, creating a logical link between the left and right adjacent nodes. The left and right nodes can then do the same to their next adjacent repeater nodes, error correcting as they go, until the desired end-to-end entangled link is achieved.}
\label{fig:repeaters_8}
\end{figure*}

Na{\" i}vely this seems to imply that the generation rate between end nodes cannot be faster than this. However, one can in fact do far better! This is shown in Fig.~\ref{fig:repeaters_8} (protocol described in caption). The key issue is that the generation rate depends on how long the adjacent nodes need to store part of an entangled state \cite{bib:jiang09, bib:munro10, bib:Muralidharan2016}. 

Fundamentally we know that the time to attempt to generate a single entangled Bell pair between two nodes is scaling as $L/c$ (where $L$ is the distance between those two nodes). With channel losses we need to make,
\begin{align}
m=\frac{\log_{10} (\varepsilon)}{\log_{10} (1-p_\mathrm{ED})} - \log_{10} \left(\frac{\varepsilon}{p_\mathrm{ED}}\right),
\end{align}
attempts to generate a single Bell pair with error probability $\varepsilon$. We can make these attempts simultaneously and not affect the generation time. Now by using a butterfly repeater design, as illustrated in Fig.~\ref{fig:repeaters_8}, one immediately notices that the qubits with the repeater nodes are only used for duration $\sim 2 L/c$. After this time those qubits have been freed up and are available to generate new entangled links. This means in turn that the time to generate the long-range entangled pair will scale as \mbox{$T\sim O(2L/c)$} (independent of the overall distance $L_\mathrm{tot}$) \cite{bib:jiang09, bib:munro10, bib:Muralidharan2016}. The exact resources used depends heavily on the error correcting code, but we know they in principe scale as \mbox{$M \sim O(\mathrm{polylog}(L_\mathrm{tot}))$} \cite{bib:Muralidharan2016}.

This is quite a dramatic decrease in both $T$ and $M$ compared to the first-generation. In fact one could expect the normalised rates to be on the order of kHz \cite{bib:munro10}. However this is a significant cost in terms of the quality of the original Bell pairs that must be prepared. In the first-generation schemes a fidelity just over 50\% was sufficient. However with the second-generation schemes using normal error correcting codes, it is likely this initial fidelity will have to be over 90\% \cite{bib:jiang09, bib:munro10}. 

\subsection{Third-generation repeaters}\index{Third-generation repeaters}

The use of error correcting codes significantly improves the performance of second-generation quantum repeaters compared to first-generation ones. The second-generation schemes are now limited by the communication time between adjacent repeater nodes to herald whether entanglement distribution was successful or not \cite{bib:munro10, bib:munro12}. The communication (both quantum and classical) is ultimately limited by the speed of light (either in fibre or over free-space). The natural question is whether we can improve performance even further.

The only remaining avenue at our disposal is to move from probabilistic to deterministic entanglement distribution. Remembering that we have losses in the channel, the only way to achieve deterministic entanglement distribution will be by transmitting encoded error-correctable states between repeaters. This means we must turn to loss-based error correction codes \cite{bib:ralph05, bib:munro12, bib:Fowler10, bib:ATL13, bib:MKLLJ14}.  

\subsubsection{Loss-tolerant codes}\index{Loss!Tolerance!Codes}

There are quite a number of error codes that can correct for loss events, but here for illustration we consider \textit{parity codes}\index{Parity!Codes} in their simplest form \cite{bib:ralph05, bib:munro12}. Other well-known approaches are based on cluster states\index{Cluster states}.

Consider a four photon state of the form,
\begin{align}
\ket{\Psi} &= \alpha \ket{0}_1 \ket{0}_2+\ket{1}_1 \ket{1}_2) \otimes (\ket{0}_3 \ket{0}_4+\ket{1}_3 \ket{1}_4) \nonumber \\
&+ \beta (\ket{0}_1 \ket{1}_2+\ket{1}_1 \ket{0}_2) \otimes (\ket{0}_3 \ket{1}_4+\ket{1}_3 \ket{0}_4),
\end{align}
where $\ket{0}$ and $\ket{1}$ represent orthogonal degrees of freedom (e.g polarisation). This state can be rewritten in the form,
\begin{align}\label{eq:third_gen_red_enc}
\ket{\Psi} = \alpha \ket{\Phi_{12}^+} \ket{\Phi_{34}^+}+\beta \ket{\Psi_{12}^+} \ket{\Psi_{34}^+},
\end{align}
and thus the state has been encoded into terms of a tensor product of two redundantly encoded Bell states. Now photon loss will remove one of these photons. As an example, let us consider what happens when photon 4 is lost. The resultant state can be represented by the density matrix,
\begin{align}
	\hat\rho= \ket{\zeta^+} \bra{\zeta^+} +\ket{\zeta^-}\bra{\zeta^-},
	\end{align}
where,
\begin{align}
\ket{\zeta^+} &=  \alpha \ket{\Phi_{12}^+} \ket{0}_3 + \beta  \ket{\Psi_{12}^+} \ket{1}_3, \nonumber \\
\ket{\zeta^-} &=  \alpha \ket{\Phi_{12}^+} \ket{1}_3 + \beta  \ket{\Psi_{12}^+} \ket{0}_3.
\end{align}

We immediately notice that \mbox{$\ket{\zeta^-}=\hat{X}_3 \ket{\zeta^+}$} and so by measuring the third photon in the $\hat{X}$ basis our state reduces to the pure state \mbox{$\alpha \ket{\Phi_{12}^+} \pm \beta \ket{\Psi_{12}^+}$}, where the $\pm$ sign is given by the $\hat{X}$ measurement outcome. This is then correctable using local operations. After the loss event of photon 4 and the measurement of the third photon our state thus becomes,
\begin{align}
\alpha \ket{\Phi_{12}^+} + \beta  \ket{\Psi_{12}^+},
\end{align}
which has exactly the same information in it as $\ket{\Psi}$ but without the redundant encoding.

It is now straightforward to re-encode back to our original state, Eq.~(\ref{eq:third_gen_red_enc}). We considered photon loss only on the fourth qubit. However the same principle applies for any lost photon. Unfortunately we can only tolerate the loss of a single photon using this encoding, so the loss rate must be small.

The above example illustrates how the smallest optical loss code work. The general code with \mbox{$n - 1$} redundancy can be written as \cite{bib:ralph05, bib:munro12},
\begin{align}
\ket{\Psi} = \alpha \ket{\Phi_{e}}_1 \ldots  \ket{\Phi_e}_n+\beta \ket{\Psi_{o}}_1 \ldots  \ket{\Psi_{o}}_n
\end{align}
where $\ket{\Phi_{e,o}}$ are the even and odd parity m photon states given by,
\begin{align}
\ket{\Phi_{e,o}} = \frac{1}{\sqrt{2}}(\ket{+}_1 \ldots  \ket{+}_m\pm \ket{-}_1 \ldots  \ket{-}_m),
\end{align}
with \mbox{$\ket{\pm}=\frac{1}{\sqrt{2}}(\ket{0}\pm \ket{1})$}. This redundancy-based parity code is composed of $n$ logical qubits each containing $m$ photons. For this code to correct loss errors we have two constraints,
\begin{itemize}
\item At least one logical qubit must arrive without photon loss.
\item Every logical qubit must have at least one photon arrive successfully.
\end{itemize}
If these constraints are met, the loss events during transmission between adjacent repeater nodes can be corrected. Of course, such codes can not correct more than fifty percent errors and so the distance between repeater nodes is limited. Remembering that the probability of a photon being successfully transmitted through a channel of length $L$ with attenuation length $L_0$ is given by \mbox{$p=e^{-L/L_0}$}, the maximum distance between repeater nodes is \mbox{$L/L_0\sim 0.69$} (which corresponds to approximately 17km in present-day commercial telecom fibre). This is much shorter than what we would typically consider for the first and second-generation schemes.

\subsubsection{Operation}

\begin{figure}[!htbp]
\includegraphics[clip=true, width=0.475\textwidth]{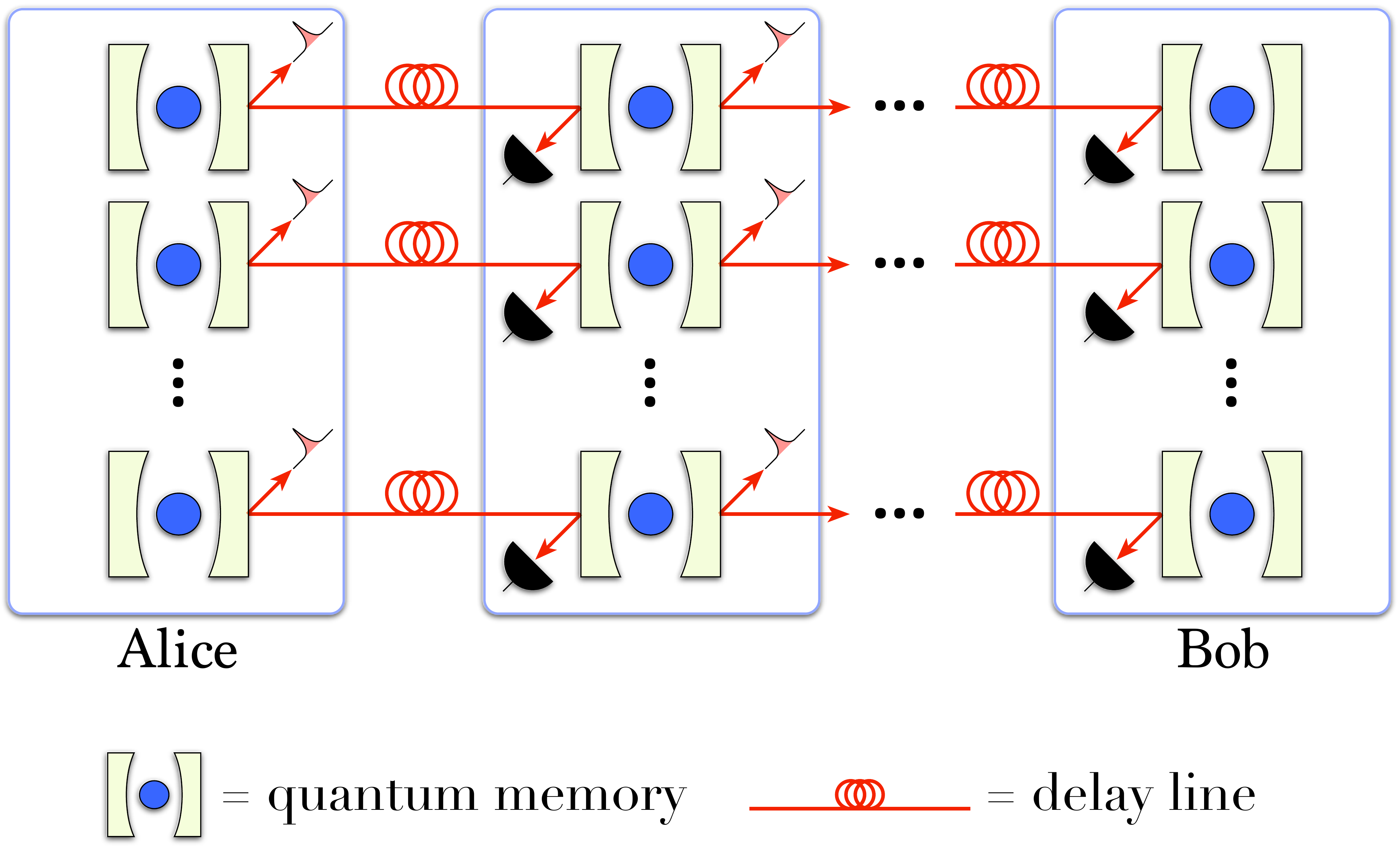}
\captionspacefig \caption{Transmission of a quantum signal using loss-based error correction codes in a quantum network.}\label{fig:repeaters_9}
\end{figure}

Let us now describe the operation of the third-generation repeater scheme  depicted in Fig.~\ref{fig:repeaters_9} in detail \cite{bib:munro12, bib:MKLLJ14}.

It begins at the left hand node by Alice encoding her message into a redundant parity code created on a series of matter qubits using local quantum gates within that repeater node.

The quantum state is then transferred/teleported via photons which are transmitted through a lossy channel to the adjacent repeater node. Here, two specific operations occur: first the information encoded on each photon is transferred to a matter qubit within that repeater node and then that photon is measured. The photon measurement is critical as it heralds which photons have been lost and allows us to measure the remaining qubit in that block in the $\hat{X}$ basis, which removes the damaged parity blocks from our encoded state, leaving our information intact. We can now add the full redundancy back into our encoded state in the matter qubits. The fully encoded state can then be transferred to photons and transmitted to the next repeater node where the same procedure occurs again. This continues until our state reaches the last repeater node where Bob is.

There is one immediate observation that can be made from this scheme. The matter qubits (quantum memories) within the local nodes are only used to encode and error correct the redundancy code as well as transmitting those quantum states as photons. Entanglement is not stored within the nodes while the photons are being sent to the adjacent repeater nodes. This in turn means the resources within that repeater node can be used immediately again (once the photons have been transmitted), and so the rate of communication is now limited by the time to perform the local operations within a node, rather than the round trip time between adjacent nodes. 

The focus so far has been only on loss-based errors but this code is fault-tolerant to general errors as well \cite{bib:MKLLJ14}. Furthermore, this redundancy code was only an illustrative example that photon loss in the channel can be corrected. Many other codes can be used in a similar fashion \cite{bib:munro12, bib:Fowler10, bib:MKLLJ14}. Finally the scheme we have presented in Fig.~\ref{fig:repeaters_9} transmits a quantum signal from Alice and Bob. It can however be adapted to use the butterfly design from Fig.~\ref{fig:repeaters_8} to create remote entanglement between Alice and Bob while maintaining the performance advantages our direct transmission scheme gave. 

\subsection{Resource scalings across repeater generations}\index{Resource!Scaling}

As can seen the various quantum repeater generations take quite different approaches as to how they distribute entanglement between Alice and Bob over a long distance \cite{bib:Muralidharan2016}. It is useful thus to summarise in Tab.~\ref{tab:rep_nets_scale} the performance of the various repeater approaches and their requirements.

\startnormtable
\begin{widetext}
\begin{center}
\begin{table}[!htbp]
\centering
\begin{tabular}{ccccc}
\hline
\multicolumn{1}{|l|}{\textbf{Repeater generation}} & \multicolumn{1}{l|}{\rm $T_\mathrm{av}$}   & \multicolumn{1}{l|}{\rm Resources consumed}    & \multicolumn{1}{l|}{\rm  $L_\mathrm{max}$}     & \multicolumn{1}{l|}{\rm  Local gate precision}     \\ \hline \hline
\multicolumn{1}{|l|}{First-generation}    & \multicolumn{1}{l|}{$O(L_\mathrm{tot}/c)$} & \multicolumn{1}{l|}{$O(\mathrm{poly}(L_\mathrm{tot}))$} & \multicolumn{1}{l|}{\rm arbitrary}  & \multicolumn{1}{l|}{\rm arbitrary}    \\ \hline
\multicolumn{1}{|l|}{Second-generation}   & \multicolumn{1}{l|}{$O(2 L/c)$}     & \multicolumn{1}{l|}{$O(\mathrm{polylog}(L_\mathrm{tot}))$} & \multicolumn{1}{l|}{\rm arbitrary}  & \multicolumn{1}{l|}{\rm high}   \\ \hline
\multicolumn{1}{|l|}{Third-generation}   & \multicolumn{1}{l|}{$O(t_\mathrm{local})$}     & \multicolumn{1}{l|}{$O(\mathrm{polylog}(L_\mathrm{tot}))$} & \multicolumn{1}{l|}{$L/L_0<0.69$}   & \multicolumn{1}{l|}{\rm fault tolerant levels}   \\
\hline
\end{tabular}
\captionspacetab \caption{Quantum repeater approaches and their expected performance scalings.  $T_\mathrm{av}$ corresponds to the time between which the protocol can be attempted (the time to generate a single Bell state is at least $L_\mathrm{tot}/c$) \cite{bib:Muralidharan2016}. The generation rate is $R\sim 1/T_\mathrm{av}$. Further given are the resources (quantum memories) required as well as the precision for the local gate operations within repeater nodes. $L_\mathrm{max}$ is the maximum spacing between repeater nodes. $L_\mathrm{tot}$ is the total distance between Alice and Bob while $L$ is the distance between adjacent repeater nodes. $t_\mathrm{local}$ is the time required to perform the local operations within the repeater node, while $L_0$ is the attenuation length of the channel/fibre.}
\label{tab:rep_nets_scale}
\end{table}
\end{center}
\end{widetext}
\startalgtable

The table clearly shows that the average time to generate the Bell pair between the end nodes of the repeater network decreases significantly as we move to higher generation quantum repeaters. In the first-generation our generation rate is $O(c/L_\mathrm{tot})$, which increases to $O(c/L)$ for the second-generation schemes, and finally to $O(1/t_\mathrm{local})$ for the third-generation ones. The difference here could be more than nine orders of magnitude. Next the number of quantum memories required decreases from $O(\mathrm{poly}(L_\mathrm{tot}))$ for the first-generation approach to $O(\mathrm{polylog}(L_\mathrm{tot}))$ for the higher ones. The higher generation schemes however come at quite a cost, with the requirement for fully fault-tolerant (or near fault-tolerant) quantum gates within nodes. In fact, it's likely that Alice and Bob will have multiple potential routes between themselves.

%
% The transition to quantum networks
%

\subsection{The transition to quantum networks}\index{Transition to quantum networks}

The previous quantum repeater networks we discussed have been simple point-to-point\index{Point-to-point (P2P)!Network} linear networks\index{Linear network}. While there may have been a number of ways to establish end-to-end entangled links between Alice and Bob, they knew they were connected via a simple, direct, linear chain.

Of course this is highly unrealistic. Alice and Bob are likely to be members of a complex quantum network, that supports multiple users simultaneously and offers multiple routes from a given source to a given destination (see Sec.~\ref{sec:network_topologies}). This leads to a number of interesting considerations going forward. 

\begin{itemize}
\item For large-scale networks the users may not know its exact network topology or even the best route between them. In fact there could be multiple paths between Alice and Bob. Probing the entire network to establish the best route would be slow and costly (in practice). Still, every node should have a unique identifier (quantum IP address\index{Quantum IP addresses}), uniquely indicating its location and resource availability in the network.
\item Most complex networks dynamically change in time as resources become congested or nodes break. This in turn means using a butterfly approach to create Alice and Bob's links is problematic as one does not know the middle point between them to start the entanglement creation process. If one has to determine the route in advance and restrict access to those parts of the network required to establish the entire links, congestion will quickly follow. The generation rate will be very slow. 
\item Finally, it's unlikely that repeater nodes will be equally spatially separated (making the first-generation repeater schemes extremely hard to use in this situation). 
\end{itemize}

The above issues lead us to a network model where Alice and Bob suspect there is a route between them but do not know the exact route, which is likely to be dynamic, i.e the availability of routes and their relative costs are liable to change. In such a case, if Alice wants to send a message to Bob, she uses her knowledge of Bob's rough location (from the quantum IP address) and her knowledge of the nodes close to her to send a message to a repeater node, who will have more knowledge of Bob's part of the network. This node can then forward the message to further nodes (who know even more about Bob's location) until it finally reaches Bob. The quantum IP address\index{Quantum IP addresses} is essential here as that identifier indicates to the repeater node who to forward to next. In principle as the message (or entanglement) is being established node-by-node, those repeater nodes who have already been used are free to work for tasks for other users. 

There is another interesting aspect of our general complex quantum networks. There are likely to be many paths between Alice and Bob which could be attempted in superposition fashion. This will not only increase the capacity between Alice and Bob but also its robustness.

%
% Repeater Synchronisation
%

\subsection{Repeater synchronisation}\index{Repeater!Synchronisation}\index{Race-time conditions}

\sectionby{Peter P. Rohde}

When employing non-deterministic entanglement sources in a quantum repeater network\index{Quantum repeater networks} there is no guarantee that pumping the source will actually yield an output entangled pair. Indeed this is extremely unlikely for sources such as SPDC. For this reason quantum memories will be required when performing entanglement swapping, so as to temporally synchronise the unpredictable arrival times of qubits.

However, future technologies may enable push-button entanglement sources\index{Push-button source}, in which case there is no ambiguity in the preparation times of pairs. In this instance quantum memories may be avoided entirely. Instead we can trigger all the sources at exactly the right times so as to ensure that at every joint measurement device in the repeater network the photons arrive simultaneously.

\begin{figure*}[!htbp]
\includegraphics[clip=true, width=0.8\textwidth]{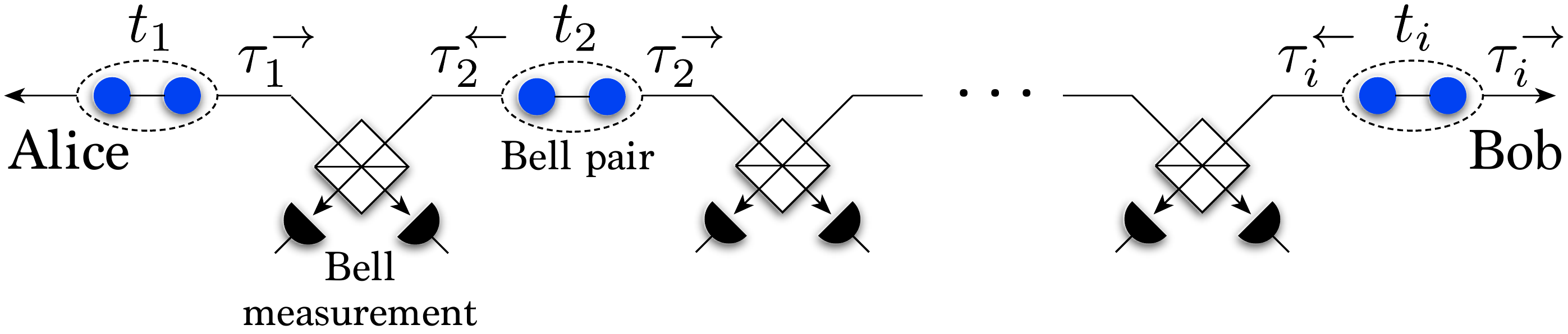}
\captionspacefig \caption{A linear quantum repeater network comprising Bell pair sources, and Bell measurements for entanglement swapping (there is no purification stage in this example). Upon success this yields a single Bell pair between the leftmost and rightmost photonic qubits. $t_i$ is the triggering time of the $i$th source, and $\tau_i^\leftarrow$ ($\tau_i^\rightarrow$) are the channel propagation times between the $i$th source and the entanglement swapper to its immediate left (right). With push-button sources\index{Push-button source}, if the $t_i$ are chosen appropriately, as per Eq.~(\ref{eq:repeater_trig_time_sol}), we can satisfy the race-time condition\index{Race-time conditions} of simultaneous arrival times of photons at Bell measurements, mitigating the need for quantum memories to synchronise them.}\label{fig:racetime}	
\end{figure*}

Consider a simple quantum repeater network, comprising a linear chain of alternating Bell pair sources and entanglement swappers, as shown in Fig.~\ref{fig:racetime}.

Let $t_i$ be the triggering time of the $i$th source, and $\tau_i^\leftarrow$ ($\tau_i^\rightarrow$) be the channel propagation times from the $i$th source to the entanglement swapper immediately to its left (right) in the chain. Imposing the race-time condition\index{Race-time conditions} that two photons arriving at a measurement device are simultaneous,
\begin{align}\label{eq:ent_sync_cond}
t_i + \tau_i^\rightarrow &= t_{i+1} + \tau_{i+1}^\leftarrow,	
\end{align}
where we set \mbox{$t_1=0$} as an arbitrary reference. This yields the linear system of equations,
\begin{align}
\hat{T}\cdot\vec{t} = \vec{\delta},
\end{align}
where,
\begin{align}
\hat{T} = \begin{pmatrix}
 -1 & 0 & 0 & 0 &\dots \\
 1 & -1 & 0 & 0 & \\
 0 & 1 & -1 & 0 & \\
 0 & 0 & 1 & -1 & \\
 \vdots & & & & \ddots
\end{pmatrix},
\end{align}
\begin{align}
\vec{t} = \begin{pmatrix}
t_2\\
t_3\\
t_4\\
t_5\\
\vdots	
\end{pmatrix},
\end{align}
\begin{align}
\vec\delta &= \begin{pmatrix}
\tau_{2}^\leftarrow - \tau_1^\rightarrow	 \\
\tau_{3}^\leftarrow - \tau_2^\rightarrow	 \\
\tau_{4}^\leftarrow - \tau_3^\rightarrow	 \\
\tau_{5}^\leftarrow - \tau_4^\rightarrow	 \\
\vdots
\end{pmatrix}.
\end{align}
Solving,
\begin{align}\label{eq:repeater_trig_time_sol}
	\vec{t} = \hat{T}^{-1}\cdot\vec\delta,
\end{align}
where,
\begin{align}
	\hat{T}^{-1} & = \begin{pmatrix}
 -1 & 0 & 0 & 0 &\dots \\
 -1 & -1 & 0 & 0 & \\
 -1 & -1 & -1 & 0 & \\
 -1 & -1 & -1 & -1 & \\
 \vdots & & & & \ddots
\end{pmatrix},
\end{align}
yields the required triggering times for all sources (relative to \mbox{$t_1=0$}) to ensure synchronisation of photon arrival times at all entanglement swappers.

The total execution time of the protocol (i.e time taken to successfully distribute an entangled pair) is given by,
\begin{align}
	t_\mathrm{exec} = \max_i(t_i +\max(\tau_i^\rightarrow,\tau_i^\leftarrow)) - \min_i(t_i
	),
\end{align}
just the difference between the latest photon arrival time and the earliest source triggering time.

\latinquote{Acta non verba.}

%
% The Irrelevance of Latency
%

\section{The irrelevance of latency}\index{Latency}\index{Irrelevance of latency}

\dropcap{E}{ntanglement} distribution\index{Entanglement!Distribution} can be executed in a highly varying manner of ways -- from transmitting optical qubits through space via satellites, to across land surfaces via optical fibre, to dumping solid-state qubits into cargo containers and shipping them via land or sea freight. These bring with them associated transmission latencies. The former two distribute entanglement at the speed of light with latencies on the order of microseconds, whereas the latter induces enormous latencies on the order of days or weeks.

At first glance it may appear that this renders the Sneakernet\texttrademark\index{Sneakernet} approach to entanglement distribution useless. Who wants to wait several weeks to communicate their qubits?

If these transmission methods were being utilised for direct transmission of quantum data, this would certainly be a major concern. However, we are not employing them to communicate unknown quantum data packets directly. Rather we are using them to distribute many instances of completely identical Bell states. This changes the impact of latency entirely. That is to say, we treat known entangled states as a \textit{resource} rather than as an actual unit of data, and provided we can store it (i.e we have a good quantum memory), whether it arrives sooner or later is not terribly important. More important is that we have a `buffer' of entangled states at hand to draw upon when needed.

If our goal is to transmit a quantum state between two parties, the obvious approach is to send the qubits directly over the quantum channel. Alternately, they could initially share Bell pairs then employ quantum state teleportation\index{Quantum state teleportation} to teleport the state between parties. In this case all that matters is that they hold a shared Bell pair in time for execution of the teleportation protocol. It could have been distributed between them at any point in the past, held in a quantum memory\index{Quantum memory} until needed. The latency is now determined entirely by the latency of the \textit{classical} channel, which communicates the associated local corrections required to complete the teleportation protocol. In most classical networks, communication rates are on the order of the speed of light, with very little latency.

We see that the latency associated with entanglement distribution does not affect the latency of quantum state transmission when implemented via teleportation. The quantum network could continually be sharing entangled pairs between parties in a UDP-like\index{User Datagram Protocol (UDP)} mode, who hold them in quantum memory. They ensure that Bell pairs are being distributed at a sufficient rate that parties have a buffer of entangled pairs sufficient to accommodate demand for future teleportations. This irrelevance of quantum latency is a uniquely quantum phenomena, not applicable to any classical protocols\footnote{One minor exception might be to treat randomness as a resource for randomised classical computation, i.e for application in \textbf{BPP} algorithms\index{BPP}. In that restricted instance the latency of our source of random bit-strings is also irrelevant since randomness is invariant under temporal displacement and can be buffered for future use.}.

Teleportation-based quantum communication is additionally favourable in that shared Bell pairs can be purified before being utilised, allowing errors accrued during quantum communication to be minimised, something not so straightforward (or impossible) when transmitting data qubits directly.

\latinquote{Ceterum censeo carthaginem delendam esse.}

\section{The quantum Sneakernet\texttrademark}\index{Sneakernet}\label{sec:sneakernet}

\sectionby{Simon Devitt}

\dropcap{R}{eliable}, fast, long-distance communication has always been a key driving force behind technological and economic progress. Whether couriers on foot or horseback, carrier pigeons\index{Carrier pigeons}, smoke signals\index{Smoke signals}, semaphores\index{Semaphores}, or electronic and optical signals, humanity has continually refined technology to make long-distance communications easier, faster, cheaper, and more reliable. We now live in a world where near light speed, global communications is universally available and for a significant fraction of the world's population, accessible with a device residing in our pockets. 

The development of a model for the quantum internet \cite{SD-Wehner:2018aa} has a distinct advantage over all other communication networks that have preceded it, namely the ability to leverage a treasure trove of information about how global communications systems are used, their inefficiencies, choke-points and vulnerabilities. The knowledge we have gained provides a unique design opportunity in the quantum realm, namely, 
\\
\\
\textit{If we had the ability to redesign an entirely new internet from scratch, how would we go about it? What would we do differently?}
\\
\\
The hardware required to transmit quantum information around the globe will coexist in parallel with the classical internet, since at the very least a fast classical network is required to communicate the classical information necessary to complete many quantum information processing protocols, such as quantum state teleportation or entanglement swapping. At a minimum, we can conclude that a global quantum network will be \textit{rate limited}\index{Rate limiting} by the classical network supporting it. However, given that the classical internet presently has the capacity to transmit information at approximately 70\% of the speed of light (in standard silica fibre optics), over global distances, at extremely high data-rates, means we will have to be industrious in establishing large-scale quantum networks before the classical side-channel becomes a potential rate-limiting factor of the quantum communications protocols themselves\footnote{We are referring here to the classical component of a quantum protocol such as teleportation or key exchange, not bottlenecks that may be encountered in {\em establishing} quantum entanglement through a technology such as repeaters because of a high amount of back-and-forth classical information exchange in the repeater network}.

There have traditionally been three major candidates for distributing quantum information over long-distance communications channel:
\begin{itemize}
\item Optic fibre transmission and quantum repeaters \cite{SD-Fowler:2010aa,SD-Sangouard:2011aa,SD-Munro:2012aa,SD-Azuma:2015aa}.
\item Direct free-space transmission \cite{SD-Ursin:2007aa,SD-Ma:2012aa}.
\item Quantum satellites \cite{SD-Tang:2016aa,SD-Takenaka:2017aa,SD-Yin:2017aa}. 
\end{itemize}

All of these techniques employ photons as the carrier of quantum information and have already been deployed in QKD protocols \cite{SD-Schmitt-Manderbach:2007aa, SD-Peev:2009aa, SD-Liao:2017aa}. However, there are several issues with these techniques that could make building a global quantum network challenging.

While QKD remains the foremost protocol deployed on current quantum communication hardware, a true global quantum internet must handle a vast array of potential applications, including linking fully error-corrected quantum computing systems \cite{SD-Devitt2011,SD-Fitzsimons:2017aa}. Consequently, we need to consider several metrics in potential hardware implementations. Four primary metrics can be identified:
\begin{itemize}
\item Speed: How many qubits can be transmitted per second?
\item Distance: How far can we transmit qubits, while maintaining the required speed?
\item Fidelity: How accurately can each qubit be transmitted, while maintaining the required speed and distance?
\item Cost: What is the monetary cost of building, deploying and maintaining the hardware underlying a quantum channel with given speed, distance, and fidelity?
\end{itemize}

We want to maximise the first three metrics as much as physically possible, while minimising the fourth. Classical networks provide an upper bound on the speed and distance of a quantum internet, and fidelity is ultimately bounded by the intended application: QKD provides a useful lower bound (as key distribution can be performed with low fidelity); connecting fully error-corrected quantum computers running large-scale quantum algorithms are a useful upper bound (requiring channel fidelities arbitrarily close to one).

In this chapter we discuss some of the practical demands that must be taken into account when constructing a quantum internet (in the context of QKD), and introduce a fourth type of model for global quantum networking, the quantum Sneakernet \cite{SD-Devitt:2016aa}. We will take a qualitative look at its structure and operation, and argue that there are many potential practical and economic merits of employing it as the backbone technology in a global network.

\subsection{Security}

\subsubsection{Long-term security}\index{Long-term security}

The problem is more pronounced with information that requires long-term security after its initial transmission. This may include national security secrets, trade or business secrets or personal information such as medical and banking records. The risk when transmitting this kind of information is that its value may not be related to how quickly it can be decrypted, but instead whether it can be decrypted at all. It is well known that intelligence operations (both by governments, private sector actors or nefarious organisations) routinely intercept encrypted traffic on classical networks and simply put it into long-term storage, hoping that, some day in the future, technology progresses to the point where the information can be decrypted and exploited. 

It is this kind of information that is particularly vulnerable to quantum computers. If data is encrypted using protocols that are susceptible to quantum attacks, eventually this information will become exposed. If secrecy needs to be maintained many years in the future (which is certainly the case for much national security, trade secrets, business secrets and personal information), this is a significant problem -- one that does not yet have a satisfactory solution beyond either preventing the encrypted information from being intercepted in the first place or crossing our fingers that quantum technology will either not be built or only be available forever to friends and allies. 

\subsubsection{Security through obfuscation}\index{Security through obfuscation}

Some techniques for key and message exchange attempt to hide from adversaries any evidence that such protocols are taking place. This may be as simple as giving a secret-key or message to a 20-something hipster\index{Hipsters} riding a bike, instead of a man in a well-tailored suit with dark sunglasses and a briefcase handcuffed to his wrist. Security through obfuscation is more focused on not arousing suspicion or misdirecting adversaries to use a false attack vector when attempting to compromise information exchange, rather than directly defending against a potential attack. 

\subsubsection{Security through segregated networks}\index{Network!Segregation}

Segregating classical networks from the public and/or potential adversaries is a commonly-employed security technique, simply because the speed and range of radio/optical signals is so large. Ensuring that a military or intelligence network is not connected to anything more widely accessible, in principle, enables a very high level of security, and the ability to have confidence that cryptographic keys can be sent without interception. 

These basic techniques are rarely implemented in isolation. Complex network security policies employ many combinations of strong encryption, segregated networks, obfuscation, and other techniques to minimise the chance of data being compromised. However, none of these techniques fully satisfy the assumptions of an information-theoretically secure cryptographic protocol, and they often rely heavily on the honesty and competence of human personnel (alternately \href{https://xkcd.com/538/}{https://xkcd.com/538/}) in order to be effective. 

\subsection{QKD vs classical key exchange}\index{Classical key exchange}

A 2016 assessment from the UK's cyber intelligence agency GCHQ\index{GCHQ} cites four concerns with QKD as an effective replacement for classical techniques \href{https://www.ncsc.gov.uk/whitepaper/quantum-key-distribution}{https://www.ncsc.gov.uk/whitepaper/quantum-key-distribution}:
\begin{itemize}
\item \textit{QKD protocols only address key exchange}: Ubiquitous on-demand modern services (such as verifying identities and data integrity, establishing network sessions, providing access control, and automatic software updates) rely more on authentication and integrity mechanisms such as digital signatures than on encryption.
\item \textit{Commercial QKD systems operate over relatively short distances, and BB84 (and other similar protocols) are inherently point-to-point protocols}: This means that QKD does not integrate easily with the classical internet, or with widespread mobile technologies, apps, and services.
\item \textit{Hardware is relatively expensive to purchase and maintain}: Unlike software, hardware cannot be patched remotely or cheaply when it degrades or vulnerabilities are discovered.
\item \textit{Real-world QKD systems are built from classical components (e.g sources, detectors, fibres, and ancillary classical network devices), any of which may prove a point of failure}: A number of attacks have been proposed and demonstrated on deployed QKD systems that subvert hardware components, enabling the recovery of secret keys without detection.
\item \textit{Denial-of-service (DoS) attacks\index{Denial-of-service (DoS) attacks} that disrupt transmissions are potentially easier with QKD than with classical internet or mobile network technologies}: Since QKD devices typically abort a key establishment session upon detecting tampering, it is difficult to recommend QKD in situations where DoS attacks are likely.
\end{itemize}

As we will discuss, Sneakernet-based system can address each of these concerns, or reduce their significance.

\subsection{Trusted couriers \& one-time-pads}\index{Trusted couriers}\index{One-time-pads}

Before detailing the specifics of a Sneakernet-based entanglement distribution network, let us first discuss the general framework the system leverages in the context of QKD. This is the principle of physically-trusted couriers \cite{SD-Merkle:1978:SCO:359460.359473} and one-time-pads \cite{SD-Shannon:1949aa}. The general technique, utilised routinely for extremely sensitive applications, does not transmit key material over any classical network, instead entrusting a physical courier with pre-loaded key material on a portable memory stick. 

The key material, generated at home base, is physically couriered to its destination and security is, in principle, achieved using a combination of obfuscation, deception and the trust of the courier. The couriers themselves can be heavily vetted, achieving the highest levels of security clearance within an organisation, and dummy data may be handed to couriers at random to determine whether it is appearing in places where it shouldn't. 

Today's technology makes the other requirements -- easy access to high capacity storage, and sufficient one-time-pad material for large-scale applications -- straightforward to achieve. Consequently, distribution of the key material itself is typically the hardest requirement to satisfy when relying on the information-theoretic security\index{Information-theoretic security} of one-time-pads.

A quantum Sneakernet is a technological solution to the problem of trusting the courier. We exploit the nature of quantum information to develop an analogue of the trusted-courier network, wherein human couriers are no longer a point of vulnerability. 

\subsection{Active quantum memory units}\index{Active quantum memory units}

The key technology underpinning a quantum Sneakernet is the fabrication and packaging of long-lived \textit{active} quantum memories \cite{SD-Terhal:2015aa}. Quantum memories are inherently fragile, owing to the extreme susceptibility of quantum information to decoherence induced by environmental interactions. Today's best physical qubits typically have lifetimes on the order of $\sim$1ms-1s. While there has been significant research into finding or engineering quantum systems with lifetimes on the order of minutes or hours \cite{SD-Zhong:2015aa,SD-Rancic:2017aa,SD-Astner:2018aa}, these systems are notoriously hard to control (i.e information read-in/out is challenging), because they have been purposefully engineered to be so isolated from the environment that control signals are effectively isolated too. The ability to build useful quantum technology requires increasing effective quantum memory times (e.g $T_1$- and $T_2$-times\index{T$_1$-times}\index{T$_2$-times}), such that they can be utilised over timescales appropriate for human manipulation. 

The problem of increasing effective lifetimes was solved with the formulation of active QEC protocols \cite{SD-Devitt:2013aa}, where a single \textit{logical} qubit\index{Logical qubits} is encoded into a finite number of \textit{physical} qubits\index{Physical qubits}. Continuous operations on the array of physical qubits extract information about physical errors, and, provided physical error rates are sufficiently low, the logically encoded information can be preserved over arbitrary timescales. Unlike specially engineered quantum systems that are isolated to achieve long lifetimes, encoded memories consist of intrinsically unstable physical qubits. However, encoded information in the system is maintained for arbitrary timescales via the active QEC. Active quantum memories form the foundational building blocks of a quantum Sneakernet network, allowing us to maintain encoded quantum information for extremely long durations (potentially even decades), sufficient for most important real-world applications.

The second major element in a Sneakernet network is \textit{portability}\index{Portability} -- the underlying qubit technology must be amenable to constructing active quantum memories that can be packaged and physically moved between geographic destinations. 

Considering the major, well-developed qubit technologies, we summarise their relative advantages and disadvantages. In many systems, \textit{infrastructure requirements} appear to be the most significant barrier to portability. These major technologies are:
\begin{itemize}
\item \textbf{Superconducting qubits}\index{Superconducting qubits}: require operational temperatures of $\sim$30mK to maintain operational coherence. Such temperatures require dilution refrigeration\index{Dilution refrigeration}, which is clumsy and undermines portability. 
\item \textbf{Ion traps}\index{Ion traps}: require an almost perfect vacuum environment, which again may prove to be problematic for physical portability. 
\item \textbf{Linear optics}\index{Linear optics}: avoids issues such as vacuums and extremely low temperatures, but gate non-determinism introduces -- currently -- impractical physical overheads. 
\item \textbf{Quantum dots}\index{Quantum dots}: similar to superconducting technology, dilution refrigeration is required.
\item \textbf{Phosphorus in Silicon}\index{Phosphorus in Silicon}: Also requires dilution refrigeration. 
\item \textbf{Colour centres (e.g. NV-diamond)}\index{Colour centres}\index{NV-diamond}: Neither vacuums nor dilution refrigeration are required. 
\item \textbf{Neutral Atoms}\index{Colour centres}\index{NV-diamond}: Requires vacuum environments, similar to ion-trap computers. Motional stability during transport may be of concern.
\end{itemize}

Colour centres, at least from the context of building large, portable qubit arrays, may be the most promising candidate. In colour centres, a single atomic defect is inserted into a crystal substrate. For example, a Nitrogen defect in a perfect diamond crystal of Carbon, providing a so-called \textit{spin vacuum substrate}\index{Spin vacuum substrates} \cite{SD-Aharonovich:2011aa}. Essentially, the crystal surrounding the defect provides the same isolation properties that an actual physical vacuum does in other technologies. Additionally, many colour centres operate at temperatures of only $\sim$4K. While still cold, 4K cryogenic technology is far simpler than the dilution refrigeration systems required to achieve a 30mK thermal environment. 4K cryogenic technology is so advanced that we are able to effectively launch them into space \cite{SD-Gehrz2007}. Thirdly, colour centres are generally \textit{optically accessible}\index{Optical accessibility}. An optically accessible colour centre has, in general, an electronic transition in the donor system that is resonant at optical frequencies. This allows for remote coupling of colour centres using optical photons, which are comparatively easy to route in and out of the crystal and surrounding 4K infrastructure. This is in contrast to superconducting qubits, which interact with microwave photons that are difficult to route \cite{SD-Kurpiers:2018aa}, or systems such as Phosphorus in Silicon that generally utilise direct dipole/dipole or exchange-based coupling to realise gates \cite{SD-Tosi:2017aa}. 

\subsection{Long-lived qubits}\index{Long-lived qubits}

Active quantum error-corrected devices are the cornerstone of scalable quantum computation and communication. There has been significant discussion surrounding so-called Noisy Intermediate-Scale Quantum (NISQ) applications\index{Noisy Intermediate-Scale Quantum (NISQ) technology} \cite{SD-Preskill2018quantumcomputingin}. That is, applications small enough to not necessarily require QEC. However, no NISQ application has yet been identified of scientific or commercial relevance that can be performed without employing resource-costly QEC protocols. QKD is a case in point. One of the major implementation-related drawbacks of current QKD technology arises from the physical errors that occur during qubit transmission, and the limited ranges and rates induced by them. 

Active QEC protocols are central to a Sneakernet communications architecture. A Quantum Memory Unit (QMU)\index{Quantum memory units} is ideally designed to house an array of optically coupled qubits that are encoded using QEC into a long-lived quantum memory. Without loss of generality, we will use a particular, well-known error correction technique to illustrate the basic operational principles of a QMU, and how they can be scaled up to the scale necessary for a quantum Sneakernet network.

\begin{figure*}[htbp!]
	\includegraphics[clip=true, width=0.9\textwidth]{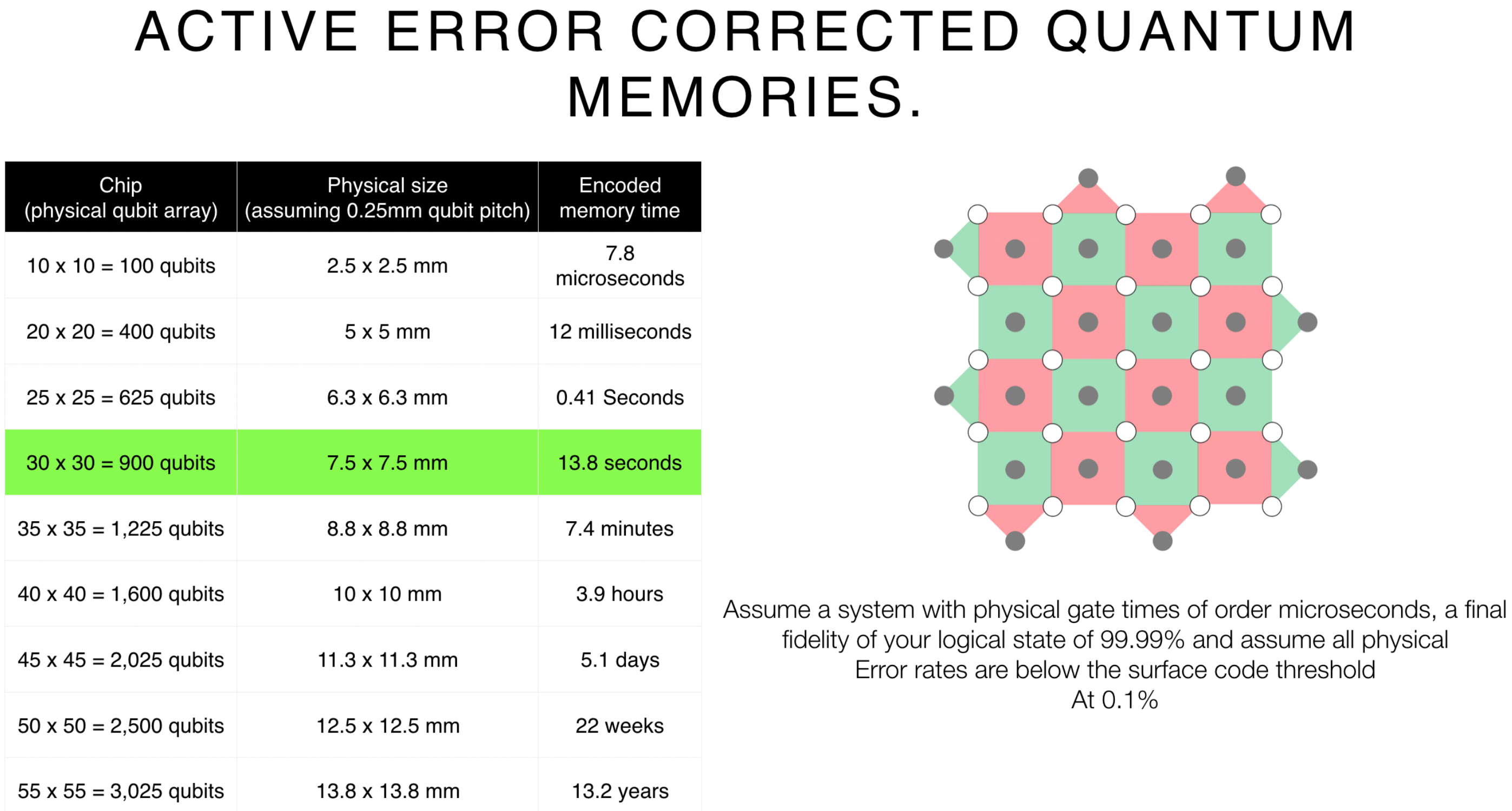}
	\caption{Expected resource overhead for a Quantum Memory Unit based on rotated planar codes \cite{SD-Horsman:2012aa}. A 2D nearest neighbour array of physical qubits can be used to encode a single {\em logical qubit} of quantum information. Provided the underlying hardware has a physical error rate associated with initialisation, measurement, single and two-qubit gates less than the fault-tolerant threshold of the surface code -- approximately 0.7\% -- increasing the size of the lattice will exponentially increase the effective coherence time of the stored information.  In the table we illustrate an example of a QMU built from a physical array of qubits that have a physical separation of 250$\mu$m, physical gate times of $t=1\mu$s and physical error rates of $p=0.1\%$. We specify the expected {\em logical} memory time at a {\em logical} fidelity of 99.99\% for a hardware with stochastic, balanced qubit noise \cite{SD-Devitt:2016aa}.} 
	\label{fig:array}
\end{figure*}

Surface code\index{Surface codes} QEC is a well-known technique for active QEC, that is being developed by several of the major hardware vendors in quantum computation \cite{SD-Fowler:2012aa}. The basic idea is that a square 2D array of interacting qubits is needed to encode a \textit{logical} piece of quantum information to extend its natural lifetime beyond the individual \textit{physical} lifetimes of the constituent physical qubits. The basic schematic is shown in Fig.~\ref{fig:array}. 

The physical array of qubits is arranged in a 2D nearest-neighbour connection geometry, wherein individual physical qubits can couple to immediate neighbours to the north, south, east, and west as shown in the right image in Fig.~\ref{fig:array}. 

The surface code has been extremely well studied over the past 15 years. We have extremely accurate simulations that detail the threshold behaviour of the code under a variety of different assumptions and we have quantified very well how the scaling of the code behaves as a function of physical error rate \cite{SD-Stace:2010aa, SD-Wang:2011aa, SD-Stephens:2014aa, SD-Nagayama:2017aa, SD-Tuckett:2018aa}. For this reason we are able to quantify expected \textit{logical} memory time, $T$, in terms of the physical operation time of the quantum gates in our hardware, $t$, the number of physical qubits in the QMU array, $N$, and the logical error rate we want to target, $P$. Through direct numerical simulation, we find \cite{SD-Devitt:2016aa},
\begin{align} \label{eq:scale}
T \approx 10t\sqrt{N} P(70p)^{-\sqrt{N}/4},
\end{align}
where $p$ is the physical error rate associated with each of the qubits within each QMU -- assumed to be at least below the fault-tolerant threshold of the planar code, $p_\mathrm{th} \approx 0.7\%$ \cite{SD-Fowler:2012aa}.

In the table illustrated in Fig.~\ref{fig:array} we detail the number of physical qubits in a QMU chip, the physical size of the chip-set under the assumption that physical qubits are physically separated by 250$\mu$m, and the amount of time we can maintain the coherence of a single piece of logically encoded quantum information. For this table we assume that physical gates in the QMU chip -- i.e. physical qubit initialisation, measurement, single qubit gates and two qubit gates are all, $t = 1\mu$s. Once the chip set contains approximately 900 physical qubits at a physical dimension of about 7.5mm$\times$7.5mm, we can expect to extend the natural coherence time of a piece of encoded information and reliably store it for approximately 14 seconds at a \textit{logical} error of $P= 10^{-4}$. The performance of the error correction scales exponentially. Consequently, if we roughly triple the size of the chip set, the effective quantum memory time increases to 13 years!

The point where logical information within the QMU can be reliably stored for longer than approximately 1 seconds is the boundary of what we define as \textit{macroscopic quantum memories}. A macroscopic quantum memory is something that can store information long enough to start physically moving the QMU. Clearly the longer you can reliably store the information in the QMU the further it can be physically moved, but a memory of at least 1 second is needed before you can start doing something interesting using Sneakernet principles.

It should be emphasised that the estimates in Fig.~\ref{fig:array} are based on QEC techniques that are unoptimised for a specific hardware chipset. Performing error correction decoding on a system with biased noise \cite{SD-Tuckett:2018aa} can change resource overheads and it is possible for other QEC coding techniques to be invented that is compatible with hardware engineering constraints that scales better than Eq.~(\ref{eq:scale}) for the rotated planar code \cite{SD-Bombin:2015aa,SD-Breuckmann:2017aa,SD-Fawzi:2018aa}. This could reduce the required physical qubit array sizes for a given QMU memory. However, the key message here is that a physical qubit array of on the order of 1cm$^2$, with physical gate times of order 1$\mu$s can be made into very long-lived quantum memories using the same architectural structures and assumptions of quantum computing systems that are already under development (we just require the added property of portability) \cite{SD-Jones:2012aa, SD-Gimeno-Segovia:2015aa, SD-Nemoto:2014aa, SD-Hill:2015aa, SD-Lekitsch:2017aa}. 

\begin{figure}[htbp!]
	\includegraphics[clip=true, width=0.475\textwidth]{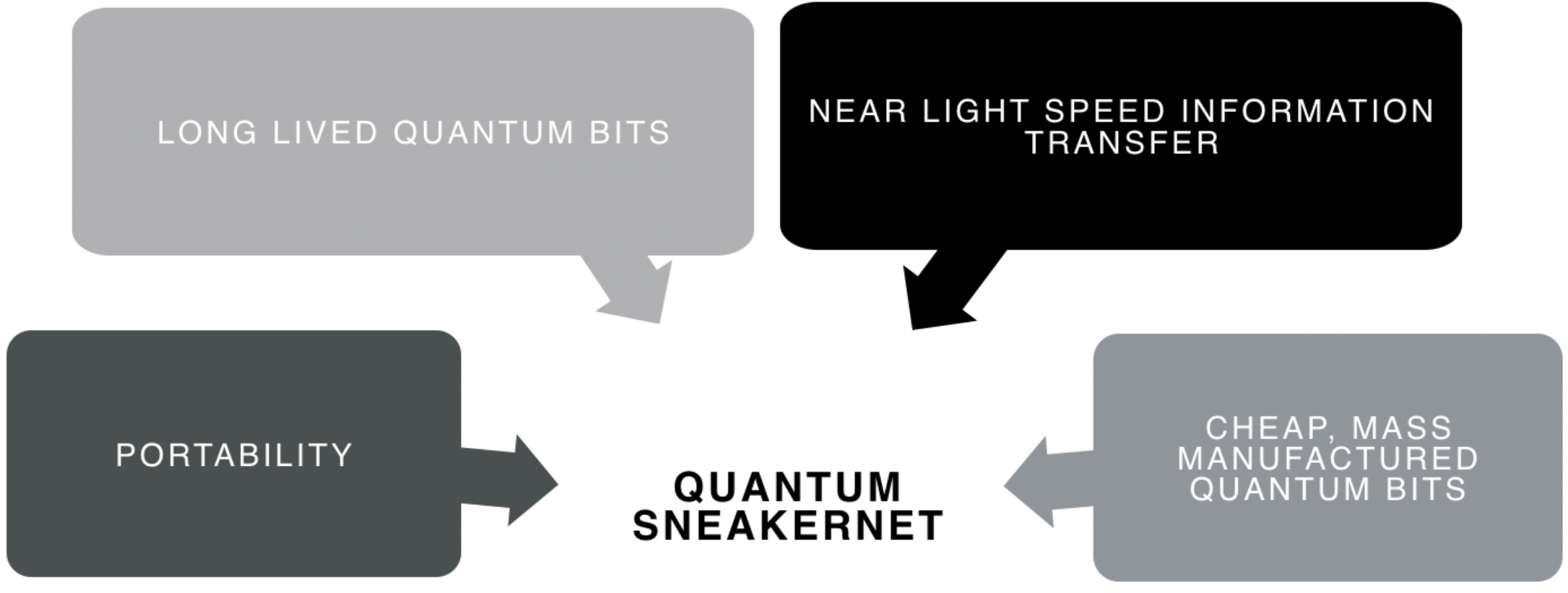}
	\caption{Key properties of the underlying physical hardware technology that allows for quantum Sneakernet operations.} \label{fig:goal}
\end{figure}

Illustrated in Fig.~\ref{fig:goal} are the four primary technological properties that makes Sneakernet based quantum communications a possible game changer for QKD and global quantum networking. While many are hunting for commercially relevant NISQ \textit{quantum algorithms}, quantum memory units and Sneakernet based communications may be the actual killer application for physical chipsets of the order of 1,000 qubits. However, this will only be of academic interest if systems cannot be be manufactured economically at scale.

\subsubsection{Cheap, mass-manufactured qubit chipsets}\index{Qubits!Chipsets}

Cost will always be a major issue for any quantum computing or communication/QKD system. While quantum technology opens up a plethora of possibilities in terms of computational power and communications flexibility, it cannot do so with simply a handful of qubits. Unlike the development of classical computation (where the competition for early computers was a room full of people with slide rules) or communications systems (where the state of the art was message exchange using carrier pigeon), quantum technology is competing with an extremely sophisticated and powerful classical infrastructure. 

Many have made the claim that building a quantum computing system or quantum communications network will be akin to other major scientific projects such as the Large Hadron Collider (LHC)\index{Large Hadron Collider} at CERN\index{CERN} or an array of LIGO\index{LIGO} gravitational wave detectors for astronomy. However, this analogy is not completely accurate. While LIGO and the LHC are extremely expensive scientific projects, in the case of the LHC there is only one and with LIGO there are only currently four completed operational units. Quantum computing and communications systems will hopefully be ubiquitous in the future. We already know better than to believe, as IBM chairman reputedly did,
\\
\\
\famousquote{I think there is a world market for about five computers.}{Thomas Watson}
\\
\\
Consequently, being able to mass manufacture physical qubits cheaply will be of paramount concern for any hardware developer in this space. And when we say \textit{cheaply}, we really mean it. Physical qubits in a quantum system are often analogised with physical transistors in a classical information processing system. As you can see in Fig.~\ref{fig:price} the cost of individual classical transistors is \textit{insanely} low. 
 
\begin{figure*}[htbp!]
	\includegraphics[clip=true, width=0.9\textwidth]{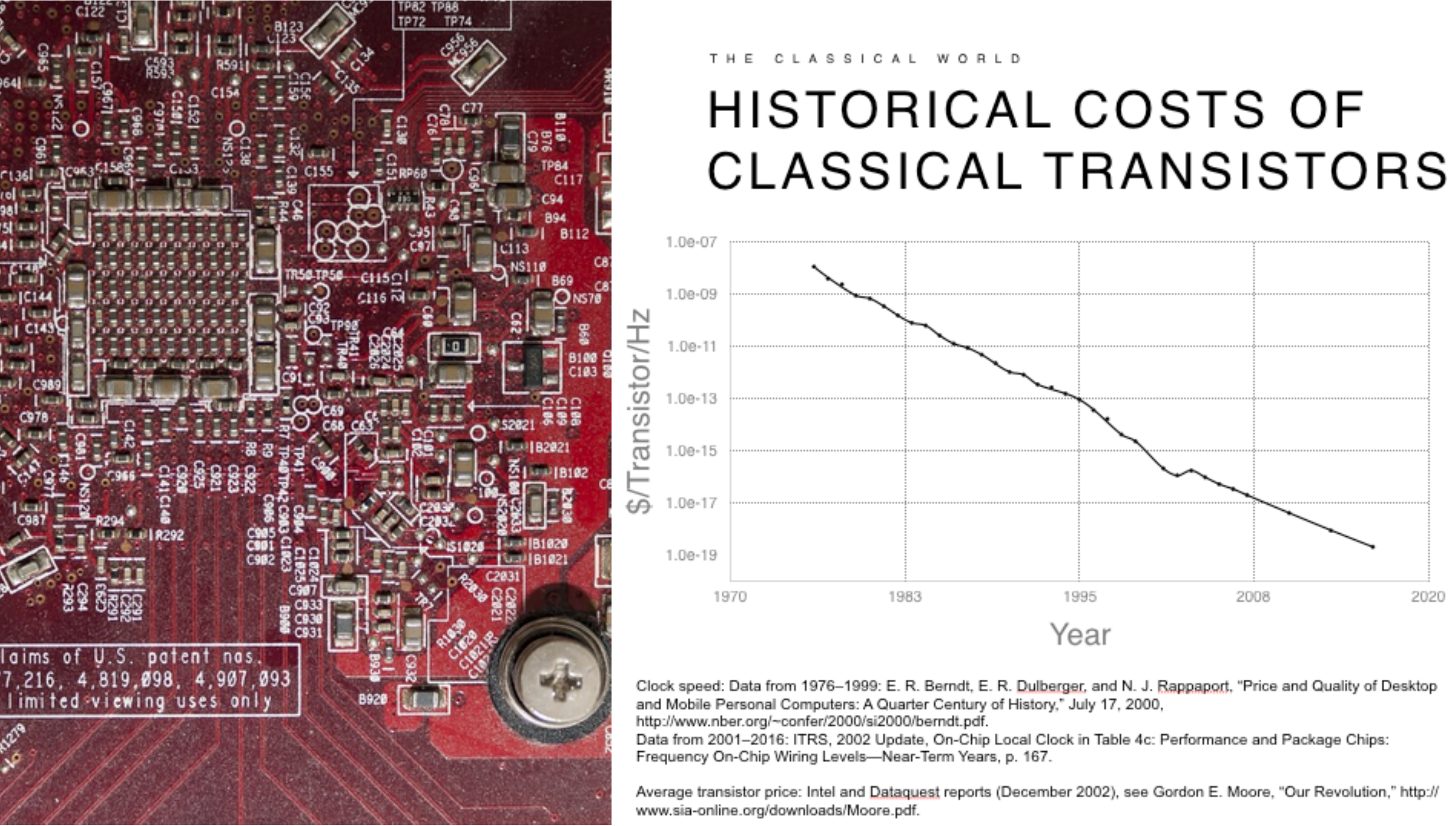}
	\caption{Exponential decrease of the cost of classical transistors over the past 50 years.  Plot was constructed using data from \cite{SD-Berndt2000}.} \label{fig:price}.
\end{figure*}

While it is certainly not expected that the price per qubit (PPQ) will reach these levels anytime soon, we need to keep Fig.~\ref{fig:price} in mind if we ultimately want to make quantum technology at the scale needed to take full advantage of the power of quantum computing and communications. 

This issue of PPQ is of critical concern for large-scale quantum computing systems. Shown in Fig.~\ref{fig:price} are costing out of a quantum computing system of sufficient size to perform Shor's factoring algorithm for large key sizes and for quantum simulation of nitrogenase\index{Nitrogenase} molecules for nitrogen fixation \cite{Gidney:2021,SD-Reiher:2017aa}. If an ultimate PPQ goal is not \$1 or lower, there may be little economic motivation for building more than a handful of quantum computers globally. It should be noted that PPQ needs to take into account the infrastructure and control systems as well as the physical qubit itself. For example, in superconducting systems, the required dilution refrigeration system is expensive -- around \$500,000. Additionally, the sample chamber for the dilution refrigerator is quite small in comparison to the `footprint' of a superconducting qubit. A commercial dilution refrigeration system can accommodate perhaps 100 such physical qubits. Consequently, at a minimum, the PPQ in this system would be \$5,000. Additionally Fig.~\ref{fig:price} does not include any R\&D expenditure or the costs associated with building the fabrication infrastructure to build the qubits themselves. 

\begin{figure*}[htbp!]
	\includegraphics[clip=true, width=0.9\textwidth]{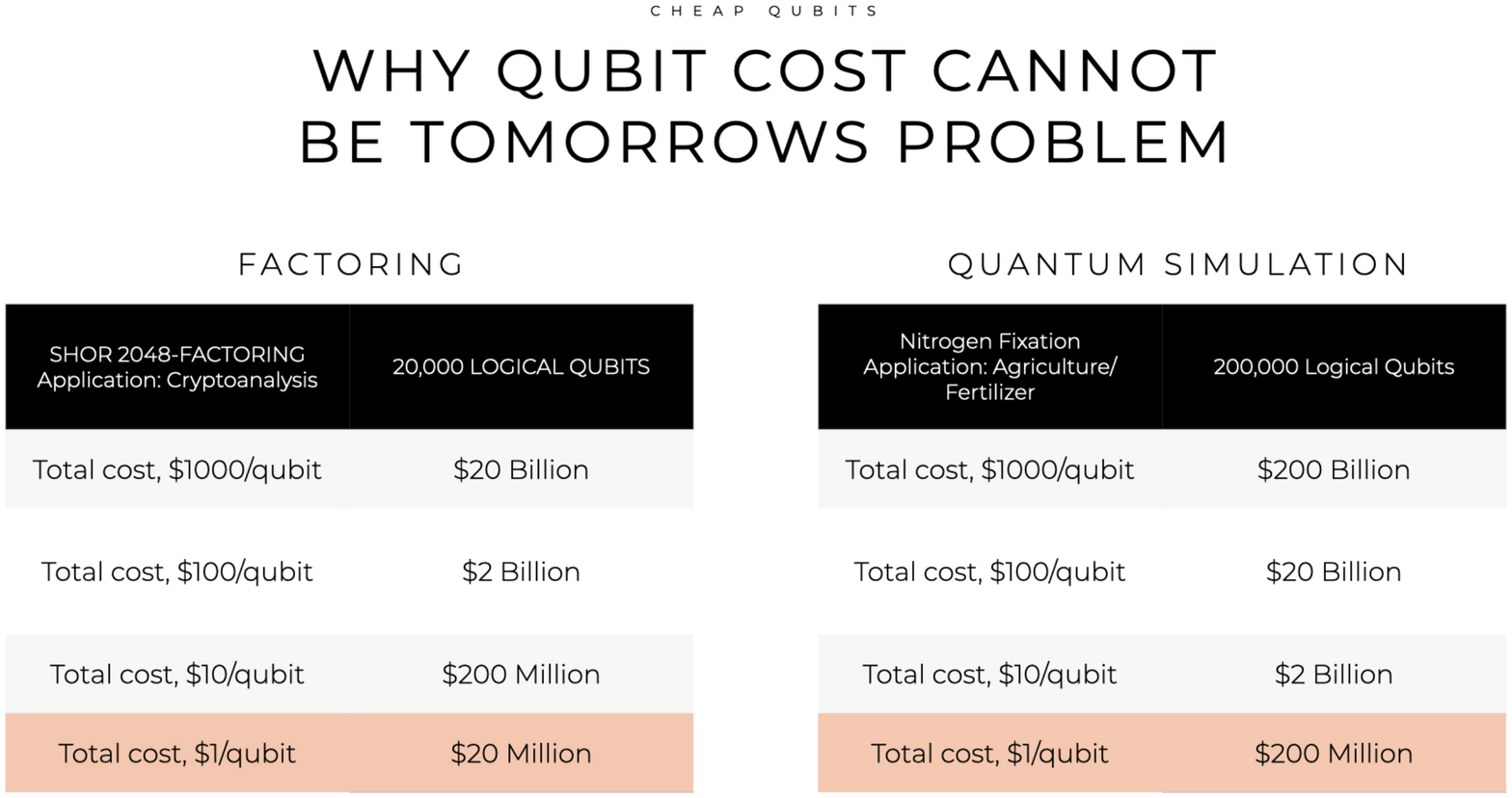}
	\caption{Examples of resource estimates for Shor's algorithm \cite{Gidney:2021} and quantum chemistry simulations for Nitrogen fixation \cite{SD-Reiher:2017aa}. Large numbers of physical qubits are needed for a fully error-corrected implementation. If the price per qubit for the actual hardware is not at least close to the \$1 mark, the price of the machine may outweigh any economic benefit to the calculation. These costs do not include R\&D or infrastructure costs associated with the computers development.} \label{fig:price}.
\end{figure*}

All current qubit architectures have associated considerations with regards to ultimate costs of manufacturing a large number of physical qubits and we do expect costs to drop, thanks to economies of scale and efficiency in manufacturing processes as various systems mature. However, we should always keep the \$1 PPQ in the back of our minds as system architectures are invented and experimentally developed. 

\subsection{Operating principles}

The basic operational primitive of the quantum Sneakernet system is an old technique for the long range distribution of information. As encapsulated in this wonderful quote from the former director of the University of Toronto Computing Services (UTCS):
\\
\\
\famousquote{Never underestimate the bandwidth of a station wagon full of tapes hurtling down the highway.}{Warren Jackson}
\\
\\
While essentially everyone today is familiar with the principle of classical Sneakernets, many people are not familiar with the name \href{https://en.wikipedia.org/wiki/Sneakernet}{https://en.wikipedia.org/wiki/Sneakernet}. Rather than sending information via radio waves or fibre optic cables, information it's loaded onto storage memory and physically transported from source to destination. The viability of Sneakernets in the classical world depends heavily on three practical considerations:
\begin{itemize}
\item Availability of cheap, high capacity classical memories.
\item Limited bandwidth and high costs associated with radio or optical fibre communications channels.
\item The need or lack of need for low latency communications.
\end{itemize}

The first practical consideration for Sneakernet communications has been solved. Classical memories have fallen in price and increased in capacity to an even greater degree than transistor price and density. In 2018, Toshiba released a single 3.5-inch hard disk drive with a capacity of 14TB available for \$550 USD %\href{https://www.amazon.com/Toshiba-14TB-SATA-7200RPM-Enterprise/dp/B07DHY61JP}{https://www.amazon.com/Toshiba-14TB-SATA-7200RPM-Enterprise/dp/B07DHY61JP}. 

Let's contrast this capacity with fibre optic connections. In 2016 the new FASTER\index{FASTER} fibre optic communications link was brought online between northern California and Japan. Costing \$300m to deploy, it has a total data capacity of 7.5Tbps/s across the Pacific %\href{https://www.nec.com/en/press/201408/global_20140811_01.html}{https://www.nec.com/en/press/201408/global_20140811_01.html}. Consequently, it takes two seconds, utilising the full design capacity of a \$300m dollar piece of telecommunications infrastructure, to transmit the data contained on a \$550 device that could fit into a handbag. This \$550 hard drive can be shipped via FedEx from northern California to Tokyo for about an extra \$150 and arrive about twenty hours later -- based on advertised prices from FedEx for door-to-door express shipping of a 1kg unit. Using that approach, we could achieve the same bandwidth as the FASTER network utilising approximately 36,000 hard drives shipped continuously back and forth. The total cost of this would be approximately \$20m dollars for the hard-drives and \$5m per day in commercial shipping costs (which could be made significantly cheaper by using more dedicated cargo transportation systems than FedEx, which is arguably the most expensive).

\begin{figure}[htbp!]
	\includegraphics[clip=true, width=0.475\textwidth]{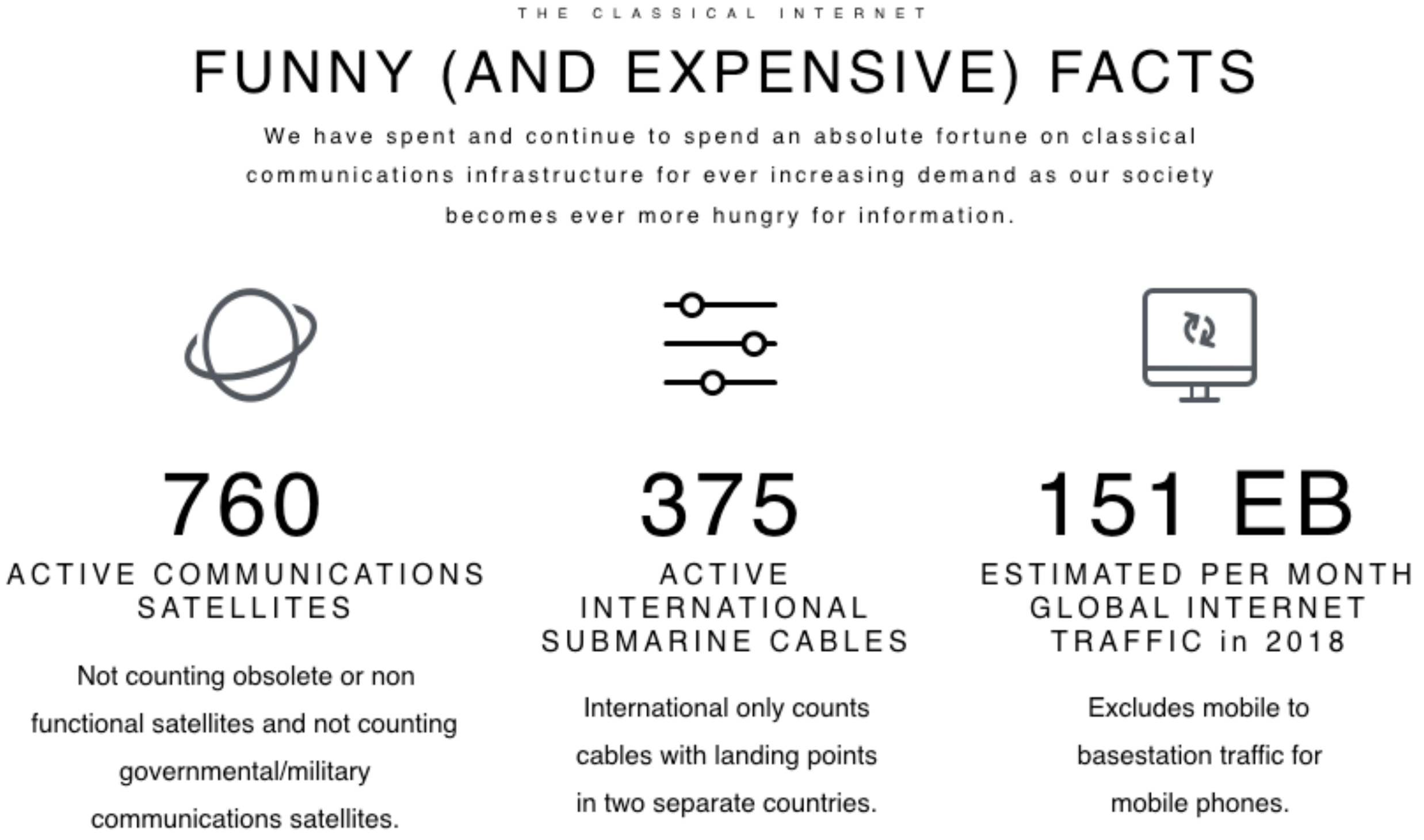}
	\caption{
    Satellite information from the Union of Concerned Scientists Satellite database
    \footnote{www.ucsusa.org \cite{sneakernet_web_1}}.
    Submarine optic cable data from Wikipedia
    \footnote{en.wikipedia.org \cite{sneakernet_web_2}}.
    Global internet traffic sourced from Statistica \footnote{www.statista.com \cite{sneakernet_web_3}} }
      \label{fig:classical}
\end{figure}

The capital cost alone associated with the FASTER network is 15 times more expensive than the capital cost of using hard drive-based Sneakernet. The operational costs of the hard drive system would be comparable if not lower than these fibre-optic links. So why do we spend so much time and money on capital and maintenance-intensive classical communications infrastructure such as submarine fibre or communications satellites (Fig.~\ref{fig:classical})?

The answer is the third point noted above -- \textit{information latency}\index{Latency}. When we transmit information from one side of the planet to another, we don't want to wait twenty hours for the data to arrive. A classical communications system that has that amount of information latency is unusable for most practical applications. Consequently, our classical infrastructure for communications is designed to operate with high bandwidth (capacity) at as close to the speed of light as physically possible. 

Once we move into the quantum regime, we can get the best of both worlds -- the infrastructure benefits of a Sneakernet-based communications link without the latency problem traditionally associated with that approach. 

We can consider a collection of individual QMUs in a single packaged device, forming basically a Quantum memory STICK (QuSTICK)\index{QuSTICK}. As we have described, a chip set array of physical qubits can maintain and store quantum information for time periods ranging from a day to many decades, depending on the number of physical qubits in the chip set QMU. In Fig.~\ref{fig:array}, a 1.6cm$^2$ chip-set has enough QEC hardware to protect the information for approximately 22 weeks. A QuSTICK consists of hundreds, potentially thousands, of these chip sets in a common cryogenic environment, packaged into a single device would be the quantum version of a USB portable memory stick. However, a classical memory stick and a QuSTICK transport very different things. A memory stick carries classical data; a QuSTICK carries quantum entanglement. This distinction is the key to the power of the quantum Sneakernet. 

\begin{figure}[htbp!]
	\includegraphics[clip=true, width=0.475\textwidth]{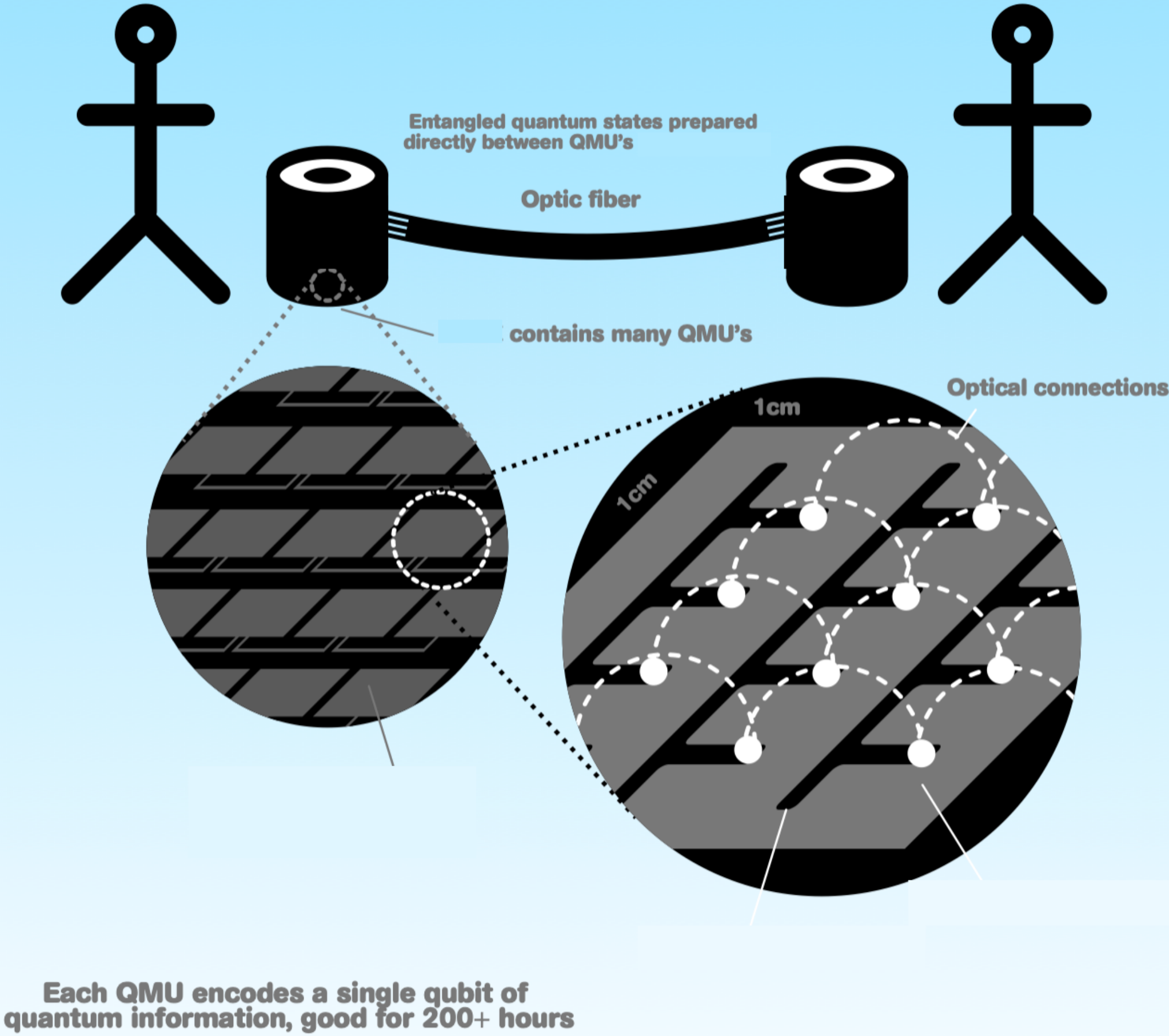}
	\caption{A QuSTICK is a portable device containing a collection of active QMU's.  The QuSTICK is to be portable and ideally with an optical interface to the physical qubits comprising each QMU.  In the case of colour centers, which in general have optical transitions available, interacting physical qubits within QMU's and between QMU's separated by a small amount of optic fibre should be nearly identical.  Hence two separate QuSTICK units can be coupled together, entangling respective QMU's in each unit into Bell states.} \label{fig:qustick1}
\end{figure}

Shown in Fig.~\ref{fig:qustick1} is the basic structure of the QuSTICK communications link. Two parties, Alice and Bob each have a QuSTICK. The device itself contains multiple QMUs. Each QMU is chip-set containing sufficient physical qubits to create a long-lived quantum memory of some defined timescale. Referencing back to Fig.~\ref{fig:array}, if the desired memory time is about a day (at a fidelity of $99.99\%$), each QMU chipset would consist of approximately 1,600 physical qubits. If we wanted each QMU to protect its respective quantum information for a year, each chip-set would contain approximately 2,100 physical qubits \footnote{Advances in error correction techniques such as quantum low density parity check codes have the potential to drastically reduce this overhead}. 

The packaged device would need to contain not only the qubit chipsets themselves, but also all infrastructure required to operate the machine. This may include cooling, laser or microwave control infrastructure, classical computing systems and power. This packaging will ultimately dictate the size of the device and how many individual QMUs could be placed in a single QuSTICK.

\subsection{Transporting quantum entanglement}

Entanglement is a unique property of quantum mechanics that has no classical analogue. Once referred to by Einstein\index{Albert Einstein} as `spooky action at a distance', entanglement is the ability for quantum particles to remain linked after they have been interacted together regardless of physical separation or physical obstruction. Unless decoherence occurs, there is no evidence that quantum entanglement can be disturbed, blocked or otherwise tampered with. According to the basic principles of quantum mechanics, if two quantum particles are entangled and isolated well enough from the outside world, they can be transported to the opposite sides of the observable universe, with all the stars, planets and black-holes between them, and the entangled correlations will remain undisturbed. It is this entangled state that will be transported by the QuSTICKS. Entanglement isn't information, but rather a quantum resource that can criss-cross the globe and be used as a consumable for quantum related communications protocols such as QKD. 

In Fig.~\ref{fig:array} we illustrated a simple two-party point-to-point connection that is possible with QuSTICKs. Two QuSTICK units, each containing a number of QMU chip sets are built. Each QMU is designed to maintain its integrity for some predetermined amount of time using active QEC protocols built into each QMU. Both QuSTICK units start out in the same room. If physical qubits within a QMU (and between QMUs) are optically connected to each other (if we use colour centres as the underlying hardware technology), it is no more difficult to interact qubits in two separate QuSTICKS than it is to interact qubits that exist in the same QMU. This allows us to create entangled states between QMUs in separate QuSTICKs by simply connecting them together with a suitable optical connection. 

The most basic entanglement protocol is to match up each individual QMU in QuSTICK-1 with a partner QMU in QuSTICK-2 and sequentially create an entangled Bell states between each pair of QMUs. Once this is done, the optic link physically connecting the two QuSTICKs can be removed and the internal error correction will preserve the quantum entanglement up to the time specified by the physical number of qubits inside each QMU.

\begin{figure}[htbp!]
	\includegraphics[clip=true, width=0.475\textwidth]{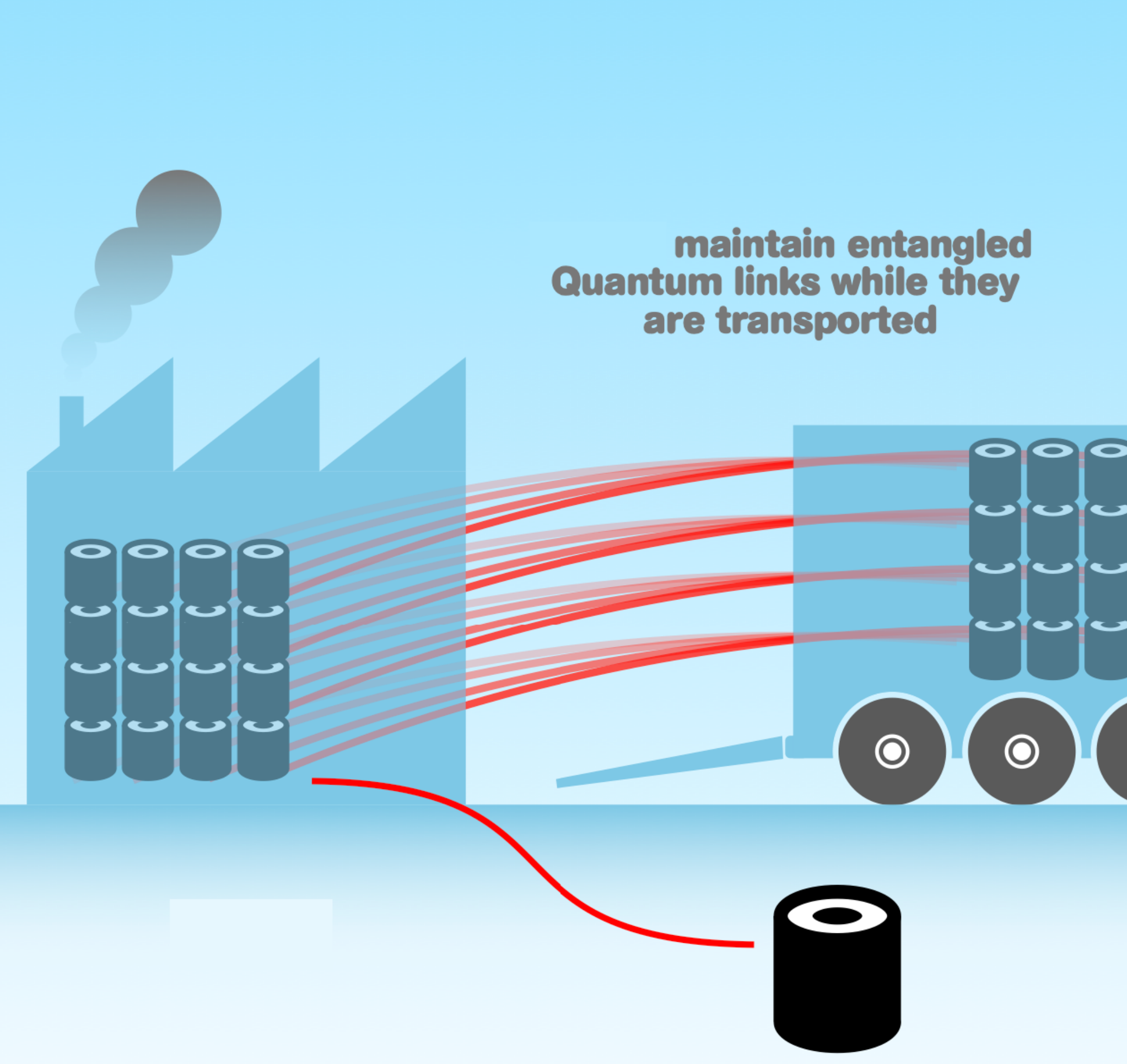}
	\caption{QuSTICK units can be manufactured and entangled locally.  Due to the portability of the units, they can be physically shipped.  This distributes the entanglement directly, without transmitting quantum information via free space or optic fibre over long distances.  The internal error correction of each QMU will maintain the coherence of the entangled states during the physical transport.} \label{fig:loading}
\end{figure}

Each QuSTICK unit has a finite number of QMUs, but there is no limit to the number of QuSTICKs that can be manufactured and deployed in the entire Sneakernet network. For each pair of QuSTICKs we execute this pairwise entangling protocol locally and then then load half of them into a transport vehicle. Shown in Fig.~\ref{fig:loading} is a scenario where a manufacturing factory is `home base' and we are preparing long-lived entangled states with a set of QuSTICKs that remain at `home' and partner sets that are physically moved to different locations. 

It is important to note that when the entanglement is initially prepared, we do not need to know the destination of the transported QuSTICKs or their eventual application. The application could be highly sensitive QKD distribution or it could be a very public scientific experiment. The only thing we care about is that the QMUs inside each QuSTICK have sufficient quantum memory time to get to their destination with their entangled state intact at the desired fidelity.

Once the entanglement is prepared, the entanglement links are \textit{automatically} preserved while the devices are being physically moved. The internal error correction protocols for each QMU ensures that errors produced by the physical qubits themselves or the act of physically moving the system are effectively corrected. The portability of the architecture is crucial to this. 

\begin{figure}[htbp!]
	\includegraphics[clip=true, width=0.475\textwidth]{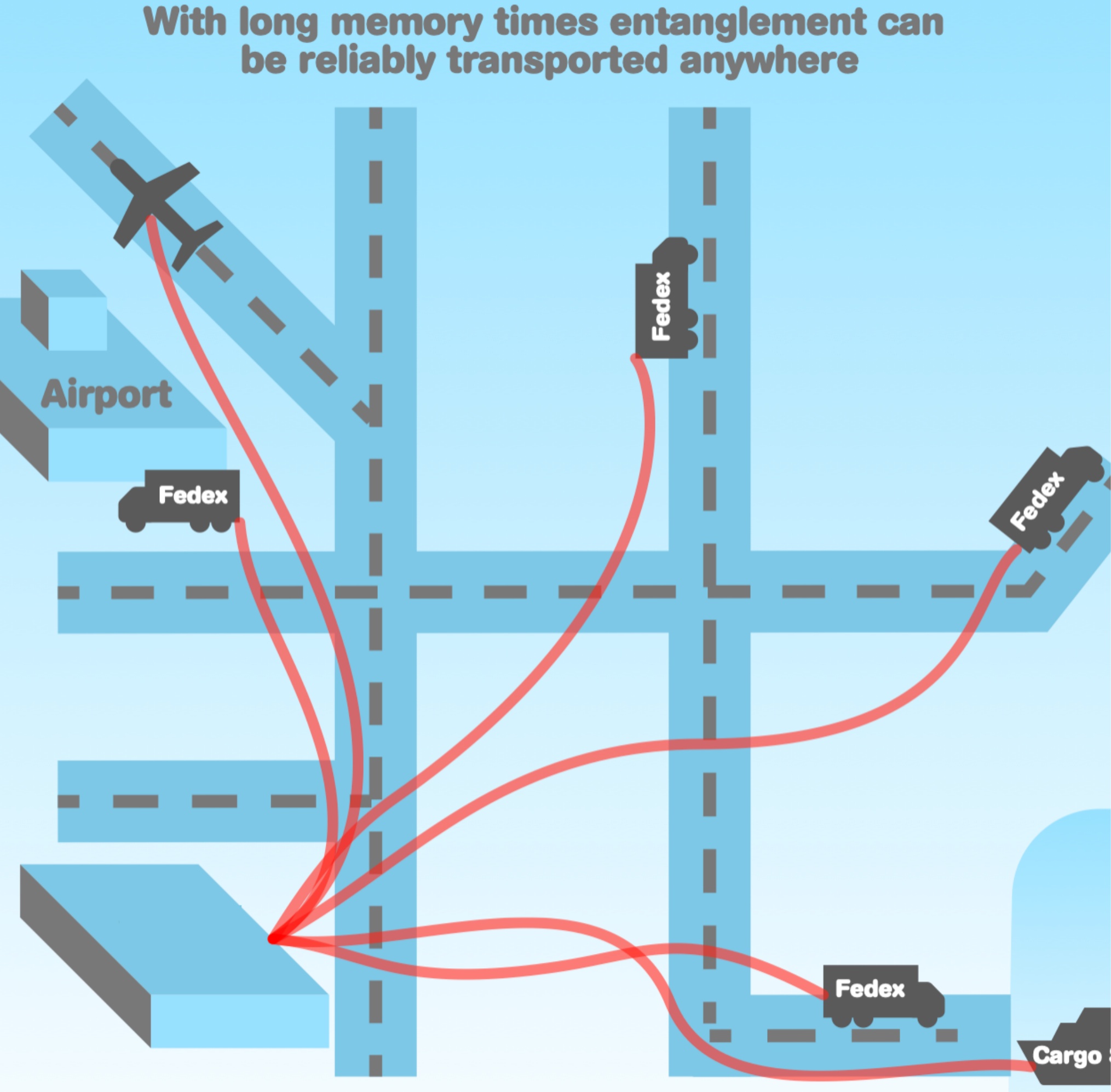}
	\caption{Entangled QMU units can be distributed anywhere that conventional shipping transport can go.  A distributed, high-fidelity quantum network over global distances can be established by simply putting subsets of QuSTICK units onto different transports and moving them to separate locations.} \label{fig:distribution}
\end{figure}

Not all QuSTICKS have to go to the same destination. Once the entanglement is prepared at home base, a subset of QuSTICKS may be transported to destination $A$, another subset to destination $B$, and so on. In fact, destination $A$ may be 5km up the road, while destination $B$ is half way around the world. Destination $C$ might use low-cost cargo shipping to receive their QuSTICKs, while destination $D$'s need might be for fewer units sooner, and so shipment by air is preferred (as shown in Fig.~\ref{fig:distribution}). The flexibility enabled by portable quantum memories allows for dynamic allocation of both QuSTICK resources and the use of whatever physical transport method is appropriate to support the final application. This is in contrast to the `one size fits all' approach used by infrastructure-intensive communications systems like satellites and repeaters. Such systems operate in exactly the same way regardless of the demands of the ultimate application.

\subsection{Utilising \& re-charging QuSTICKs}\index{QuSTICK!Re-charing}\index{QuSTICK!Utilisation}

We have described a system where a finite amount of entangled quantum states is maintained by a set of QuSTICKs for a predefined period of time. This period of time needs to be sufficient to entangle the QuSTICKs directly together at the source and physically transport them to the destination. This mechanism now mimics the behaviour of a quantum satellite or quantum repeater system, but with a crucial difference that ultimately makes the Sneakernet system superior from a practical standpoint:
\begin{itemize}
\item The sender and receiver now share a persistent, high fidelity entanglement link that \textit{does not have to be used immediately}. Quantum satellites and ballistic repeater designs generate entanglement using photons. They are not designed to `store' the entanglement for future use -- the entanglement needs to be consumed for some type of communications protocol immediately upon receipt.
\item The total amount of entanglement a destination shares with home base is dependent only on the number of QMUs and QuSTICKs used. If only 100 entangled states are needed for an application, a single QuSTICK (containing 100 QMU chips) will suffice. One million entangled states would require 10,000 QuSTICKS at 100 QMUs per stick. 
\item The entanglement the Sneakernet provides is \textit{by design} ultra-high-fidelity. All of the numbers in Fig.~\ref{fig:array} assumes that each QMU can maintain an encoded piece of quantum information for a given period of time \textit{to a fidelity of $99.99\%$}. Increasing the fidelity even further requires only a small increase in the size of the QMU. If a target application requires a fidelity of 99.999999999\% (say for direct connections between fully fault-tolerant quantum computing systems), simply use a few more physical qubits per QMU. i.e. a $T= 22$ week QMU at 99.99\% fidelity requires 2,500 physical qubits, the same $T=22$ week QMU at fidelity $99.999999999\%$ requires 5,400 physical qubits. That's just over double the physical size for 7 orders of magnitude increase in fidelity.
\item Sneakernet entanglement connections can reach global distances (Alice's QuSTICK in Sydney can be connected to Bob's QuSTICK in London using conventional transportation services).
\item The persistent entanglement shared can continue to move before it is used. Alice may not want to use this entanglement in Sydney, but instead transport it to central Australia. The entanglement link with London will continue to move with Alice as long as her QuSTICK provides sufficient error-correction for the additional time required.
\end{itemize}

\subsection{Entanglement swapping}\index{Entanglement swapping}

Another extremely beneficial aspect to the quantum Sneakernet design is the ability to perform entanglement swapping \cite{SD-Zukowski:1993aa}. What does this mean? 

In Fig.~\ref{fig:distribution} we assume that all QuSTICKs that will be physically transported are initially physically connected to partners sitting at their home base. This creates a star-network structure, where the home base acts as an endpoint node for all of the entanglement links that are physically distributed outwards. But suppose two distributed QuSTICK units want to share entanglement directly? Do they need to physically meet and have their optical links hooked up? No, instead they can ask for an entanglement swapping protocol to occur. 

Let us take three parties: Alice, Bob and Alan. Alan is located in Berlin, and, as in the previous example, Alice is in Sydney and Bob is in London. In total there are four QuSTICKs. Alan has initially manufactured all four, and we will call them ALICE, BOB and ALAN-(1,2). Before putting ALICE and BOB onto airplanes or ships, Alan entangles all of ALAN-1 QMUs to the QMUs in ALICE, and similarly between ALAN-2 and BOB. ALICE and BOB are then shipped off to Sydney and London respectively. Again, at this point, nobody has decided exactly how they are going to use this entanglement; but (depending on the amount of memory time provided by each QMU in each QuSTICK) they have a fixed time during which they can use it for anything. 

Once ALICE and BOB arrive, Alice and Bob only share entanglement with Alan directly. However, at some point, Alice and Bob decide that they want a direct entanglement connection with each other (for example, if they are intending to establish some cryptographic keys).

\begin{figure}[htbp!]
	\includegraphics[clip=true, width=0.475\textwidth]{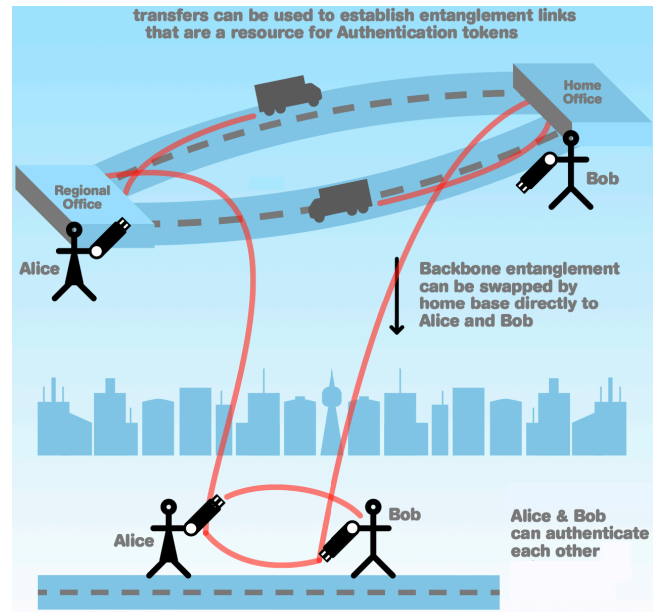}
	\caption{A backbone network of physically shipping a large volume of QuSTICKS between a home office and regional office can generate continuous Bell states between these two locations.  Alice and Bob can take QuSTICK units from the two respective offices and travel to a third location where they meet.  If an authentication protocol is needed -- because Alice and Bob have never met and need to know they have come from the same organisation -- the entanglement in each of their QuSTICKS can be SWAPPED by their respective offices such that they now share Bell states together.  These Bell states can then be used, authenticating that their quantum resource was given to them by the same source.} \label{fig:SWAPPING}
\end{figure}

Instead of physically moving again, Alice and Bob can request an entanglement swap from Alan. Alan will then connect ALAN-1 and ALAN-2 directly together (since he still has physical possession of both units in Berlin) and entangle each of their QMUs. If Alan then measures each of the QMUs in his two QuSTICKs in the $\hat{X}$-basis, the entanglement he initially shared with Alice and Bob is swapped to them. By performing this operation, Alice and Bob can \textit{directly} share two QuSTICKs of entangled QMUs without ever having to physically meet each other. They can then proceed to perform whatever communications protocol they wish. This swapping protocol is illustrated in Fig.~\ref{fig:SWAPPING} 

The above example highlights an important point. The physical distribution of a set of QuSTICKs creates an \textit{initial network topology} for their entanglement. However, this topology can be modified after the fact to create direct entanglement connections between parties that were never initially in the same distribution channel. The physical distribution network can be thought of as a network graph. Initially, each node is the physical location of a QuSTICK, and each edge connects the node where a QuSTICK began its journey to a node where it ended its journey. Entanglement swapping allows us to change the structure of the graph (i.e. change where edges are and are not) without having to move the QuSTICKs again. 
 
This kind of flexibility is simply not possible in classical networking design. It would be as if a classical data packet could be sent \textit{directly} between the Australia and Iceland despite there being no direct telecommunications link between those two countries. Classically, the data packet would have to first travel from Australia to Singapore, then perhaps from Singapore through routers in west Asia, the Middle East, central and northern Europe, the UK and finally to Iceland. At any point in its journey the packet might be intercepted, copied, lost, tampered with, inspected, corrupted, or accidentally routed to the completely wrong destination. 

With entanglement swapping using the Sneakernet network, we can define a network topology first, before it is used, and then later define direct entanglement connections between parties that were never in direct physical contact. This opens up a whole new world in network design theory. 

\subsection{Entanglement depletion \& network persistence}\index{Entanglement!Depletion}\index{Network persistence}

The use of the entanglement that is prepared between two parties to actually perform a given communications or computational protocol generally consists of the following broad steps: 
\begin{itemize}
\item The parties to the protocol first configure the entanglement network, through swapping, to match the requirements of the protocol.
\item Each party measures the \textit{logical state} of each QMU by physically measuring every physical qubit that QMU comprises. These measurements can occur in many different ways (technically referred to as the \textit{basis} in which the logic state is measured). Measuring a logic state in each QMU results in a yes/no answer. Choosing a different \textit{basis} to measure the QMU is akin to asking a different yes/no question. 
\item The parties in the protocol announce, publicly, over a classical communication system, what \textit{question} the QMUs are being asked. i.e. each party only announces the BASIS they chose to measure their respective QMUs in. They never reveal the ANSWER the QMUs gave. 
\item All the classical \textit{answers} between the QMUs are now completely correlated due to the initial quantum entanglement the parties shared. 
\item The entanglement carried by each QMU has now been destroyed or \textit{consumed} to perform the protocol and is no longer present to use again (although, of course, the QuSTICK itself isn't affected and can be re-entangled any number of times). 
\end{itemize}

What happens when the collection of QMUs within a QuSTICK has been depleted? The entanglement resource of a QuSTICK is analogous to a battery that carries a fixed amount of charge. A battery is a physical unit that can be used in many different ways and moved around to different devices; but it contains a resource that is finite. Once that finite resource has been exhausted, it must be replenished again from an appropriate source. In the case of a battery, it's recharged from via electrical outlet. In the case of a QuSTICK, the replenishment comes from another QuSTICK.

A fabrication factory may be the source of the physical QuSTICKs, but is not, in general, the source of entanglement. Returning to the battery analogy, the Tesla's\index{Tesla} Gigafactory\index{Gigafactory} may be the source the physical batteries, but it isn't the source for the battery \textit{energy} in your Tesla vehicle. 

Network connections between QuSTICKs can be established whenever and wherever QuSTICKs can be physically connected to each other. And, as discussed above, via swapping protocols, entanglement connections can be reconfigured over very long distances. While the system is certainly designed to be portable over global distances, we can also exploit the `small-world phenomenon' \cite{SD-Milgram1967} to construct long range entanglement with only relatively short physical movement of QuSTICKS. 

Once the entanglement in a QuSTICK is consumed and depleted, the owner or operator of the QuSTICK simply has to physically find another QuSTICK that has some entanglement `charge' left\index{Entanglement!Charge}. If Alice has no entanglement left, she could go find Charlie (who hasn't consumed his entanglement yet) and physically connect her QuSTICK to his and re-entangle all the QMUs. At this point, Alice has reentered the entanglement network. Once she has, she can use swapping to establish connections to another network member to perform more communications protocols. There is no need for her to reconnect to a central entanglement hub or a distant source that would require her to ship her QuSTICK to the other side of the planet and wait for it to return. 

Entanglement is a consumable resource, and may ultimately end up a marketable commodity in a future {\em entanglement based economy} \cite{SD-Devitt:2016aa}. But, unlike a battery, the entanglement `charge' is a free resource that comes as a consequence of two QuSTICKS being in the same room.  

The processes described above enable a persistent network to exist. Even though users will consume QuSTICK entanglement whenever they use their device to perform a quantum communications protocol, that networked entanglement can be replenished easily. Again, the nature and structure of a Sneakernet network allows us considerable flexibility in network topology. We can essentially design a two-stage embedded network: One stage in which the structure of the network is determined by the physical interaction of the users who have access to physical QuSTICKs, and a second stage in which quantum information flow is dictated by the structure of the underlying entanglement network. This is inconceivable in the classical world of information networking. We can only begin to imagine what remarkable things are possible with this kind of technology infrastructure. We can be certain that it will unlock capabilities well beyond the basic ultra-high security quantum communications applications that we are discussing here. 

\subsection{Quantum Sneakernet protocols}\index{Quantum Sneakernet!Protocols}

Many different communications protocols can be enacted using QuSTICK entanglement and the Sneakernet. The principal determining factor for what can be done with this technology is the number of units that can be physically manufactured and deployed. As production increases, and the total number of QuSTICKs in the world grows, new communications applications become possible. One of the major benefits of a Sneakernet design is that \textbf{the entire network is expandable}. We don't replace a QuSTICK every time we a new one is manufactured; we simply add it to the network to increase overall capacity. Generation one of the QuSTICK design should be compatible with future generations. New design iterations will primarily increase quantum data capacity (as is the case with classical memory chips), and a QuSTICK-based communications network will become more complex, higher bandwidth and more flexible as additional QuSTICKS are deployed. 

A central principle (violation of Bell's inequality\index{Bell's inequality} \cite{SD-Clauser:1969aa}) underlies each of these protocols, and the non-specialist reader may want to take a moment to review the fundamentals of quantum entanglement and its utility in secure communications throughout the rest of this book. These discussions will be largely qualitative as the main take away points is the flexibility that a Sneakernet based network allows for when realising detailed communications protocols that have been analysed extensively in other chapters. Most of this quantitative analysis for these protocols talk about the number of entangled bits (or eBits required). In our discussion, eBits are equivalent to a pair of QMUs within a pair of QuSTICKS that have been entangled locally and distributed physically. 

\subsubsection{Authentication protocols}\index{Authentication protocols}

An authentication protocol is generally a small data-exchange mechanism used to confirm that a user is who he says he or she is, and that the information that is to follow will not be coming from a third party or impostor. It is used extensively in classical communications for applications that range from the routine (when I visit \href{href://www.google.com}{http://www.google.com}, am I really talking to Google?) to communications that determine the survival of the human race (did the launch order received by an Ohio-class nuclear submarine really come from the White House?) In a Sneakernet based communications system, authentication tokens will be one of the first applications due to the comparatively small number of QMUs needed \cite{SD-Curty:2001aa}. 

The key question that the authentication process must answer is this: How do we ensure that messages are coming from the requisite trusted source? A quantum Sneakernet solution is similar to physical authentication tokens that some bank customers are given to access their accounts for high-value transactions. These classical authentication tokens utilise a variety of methods to ensure that the server (e.g. a bank website) and the client (the person with a bank account) can share a secret message that can be used to authenticate a log-in session. Some utilise static password tokens, some use synchronous dynamic tokens and some use a type of `call and response' method. There are even services that use a smart phone APP to verify the phone's SIM card and then generate a one-time key. All of these techniques are vulnerable to security flaws that present themselves if:
\begin{enumerate}
\item The implementation is imperfect.
\item The token is lost, stolen, or duplicated.
\item If some classical calculation that was assumed to be computationally difficult is compromised in some way. 
\end{enumerate}

A quantum Sneakernet solution uses the same basic principle as the classical authentication token -- the client and server share a physical object -- but, because our QuSTICK units are quantum in nature, we can use the laws of physics to ensure complete security of the authentication protocol. The underlying idea is that two parties, Alice and Bob, can violate a Bell inequality if and only if they share an entangled Bell state. That is, two QuSTICKs that are entangled will behave differently when queried than two QuSTICKs that aren't entangled.

To illustrate the protocol, we stipulate that our two parties share a finite number of Bell states between them, with one half of the Bell states contained within Alice's QuSTICK and the other half within Bob's QuSTICK. This shared Bell state will typically be prepared ahead of time in QMUs inside a QuSTICK system by a third party and then distributed to Alice and Bob. That is, Alice and Bob have received their respective halves of the entangled state from some other QuSTICK source of entanglement. The source could be a much larger backbone quantum network that is distributing long-range entanglement via the swap mechanism as described in the previous section. Alternatively, the physical QuSTICKS themselves could have been delivered to Alice and Bob by a third party or parties. However, as we will see below, this third party source of QuSTICKs or entanglement does not need to be trusted by either Alice or Bob.

If Alice and Bob share Bell states between each other, then they can perform a \textit{Bell violation test}. We will briefly summarise the basic test below. In a nutshell: The Bell parameter, $S$ is calculated by performing measurements over a finite set of shared Bell states. In classical theory, this parameter, $S$ must be $\leq 2$, while in quantum theory, $S=2\sqrt{2}$ \cite{SD-Clauser:1969aa}. Hence to confirm that Alice and Bob indeed shared a quantum state, we must determine if $S>2$.

Using a standard error analysis, for $M = 100$ shared Bell states between Alice and Bob, the standard error is approximately 0.4, implying that with 99\% confidence, $S > 2$. Hence 100 QMUs in each of two entangled QuSTICKs can be used to violate a Bell inequality with very high confidence, guaranteeing -- assuming that quantum mechanics is correctly describing the behaviour of the QMUs -- that Alice and Bob do actually share entangled states. Each QuSTICK is designed to distribute extremely high fidelity Bell states, and any possible hardware or implementation inaccuracy is addressed with the internal error correction of each QMU. As a result, the primary source of error in this example is sampling error -- i.e. by design, we don't waste resources to error correct the shared entangled state itself within the protocol and assume essentially pure shared quantum states within any communications protocol itself.

The portability of QuSTICK units is what drives many of the potential applications of this design. While 100 error-corrected QMUs is a large number compared to what is currently available, the system is designed to scale to this level almost immediately upon the first demonstration of a portable quantum memory unit. The long-lived nature of our QMUs and QuSTICKs -- and the ability to move them anywhere -- allows for deployment of these units across a wide range of platforms and environments. 

Whenever ultra-high security is needed, particularly for message authentication, the Sneakernet is particularly valuable. The fact that quantum entanglement is prepared locally, at home base, before any of these units are physically deployed in the field adds tremendous security benefits. Such benefits are not available with quantum protocols based on the transmission of the entangled pairs. Why? To put this a different way: It is very difficult to detect if an eavesdropper is trying to steal a photon from an optical fibre or a photon flying through the air. It's a lot easier to know if someone has hit you over the head and stolen your QuSTICK! \footnote{This is the rational used by the US  for the development of much smaller nuclear reactors for use in the Navy.  Smaller reactors require more highly enriched nuclear fuel which would normally causes proliferation concerns.  However, the US Navy argues that because they have 100\% control over the entire fuel and reactor system, they can ensure the security of much more highly enriched fuel.}

As quantum entanglement is not obstructed by any known physical process, the links provided by the QuSTICKS cannot be disrupted or destroyed unless someone has physical control of the QuSTICK units themselves -- or the internal error correction of the QMU fails for some reason. In the case where QMU units are assumed to be physically secure and no additional quantum security protocols are needed to pre-verify shared entanglement, like a Bell test, the QMUs can be used to create a shared authentication token by simply measuring out each unit in pre-defined bases.  i.e. not even the classical reconciliation channel is required. 

QuSTICK-based message authentication will likely be the initial application of a Sneakernet based quantum communications platform as it can occur with a small number of QMUs and hence physical qubits. However, the framework is adaptable. Initially, while QuSTICKS are sparse, we can run the Sneakernet system as simply a network for secure authentication. As we fabricate and deploy more QMUs, the authentication network grows until we hit a critical volume of devices, at which point they can be re-tasked to more complex quantum communications protocols, and the growth cycle starts again. Ultimately, as QuSTICK volume increases, we can run authentication protocols, QKD protocols, distributed quantum computing and communications and anything in between over a shared network that does not require segregated sub-networks for highly secure applications. 

\subsubsection{Key exchange}\index{Key exchange}

The next step after authentication protocols is key distribution using standard QKD protocols and additional QuSTICK units. 

The QKD literature is now very rich, so we won't go into as much detail on how QKD can be implemented and optimised. The basic process involves distributing a shared random bit-string that can be used to encrypt messages using strong, symmetric classical cryptographic protocols. In the most current, declassified standards, the U.S. National Institutes of Standards and Technology (NIST) states\footnote{\href{http://nvlpubs.nist.gov/nistpubs/FIPS/NIST.FIPS.197.pdf}{http://nvlpubs.nist.gov/nistpubs/FIPS/NIST.FIPS.197.pdf}}:
\\
\\
{\em ``The design and strength of all key lengths of the AES\index{AES} algorithm (i.e. 128, 192 and 256 bits) are sufficient to protect classified information up to the \textsc{SECRET} level. \textsc{TOP SECRET} information mandates use of either 192 or 256 key lengths. The implementation of AES in products intended to protect national security systems and/or information must be reviewed and certified by the NSA\index{NSA} prior to their acquisition and use".}
\\
\\
Hence, key exchange for strongly secure encryption requires key lengths that are roughly a factor of three higher than what would be needed for the authentication protocols described previously. Note that this factor of three does not account for error correction or privacy amplification protocols, which we will discuss below. Note also that the information exchange as a whole is not provably \textit{quantum} secure. That is, AES encryption has not yet been proven to be immune to attacks from quantum computers (although the consensus is that symmetric protocols are not vulnerable to quantum attack).

In the conventional QKD analysis, a significant amount of the `raw' key material is sacrificed for error correction and privacy amplification in order to compensate for hardware errors that occur naturally (even though the protocol assumes that ALL errors are induced by eavesdropping). In the case of QuSTICKS, naturally occurring hardware errors are corrected -- the QMUs in each QuSTICK are, by design, able to maintain ultra-high-fidelity Bell state entanglement. Additionally, as the entanglement is prepared locally, we are not required to correct for possible eavesdropping on the communications channel. To perform a man-in-the-middle attack on a Sneakernet network would require an eavesdropper to steal a physical unit, entangle it with her own unit and then return the original QuSTICK -- all while remaining undetected. As a result, we will need to sacrifice much less of the `raw' key in order to perform error correction and privacy amplification. 

The Sneakernet system handles the distribution of the entangled states between Alice and Bob in the same way as it handles distribution for authentication protocols -- requiring only a higher volume of QuSTICKS. As with the authentication protocols, the backbone network can be used to distribute and swap entanglement to Alice and Bob through a more complex networking structure if necessary. 

The key distribution system using Sneakernet principles has extraordinary flexibility and significant benefits over a more traditional quantum key distribution network. As noted previously, the portability of the QuSTICKs allow QKD nodes to be placed essentially anywhere that a QuSTICK can be physically transported. Optic fibre, satellites,  dedicated receiving stations, and other infrastructure-intensive hardware are not required. This allows for the distribution of QKD nodes to mobile platforms, field units or remote outposts. Additionally the entanglement network can be reconfigured as required. If QKD nodes need to be relocated from a mobile outpost in central Africa to northern Australia, simply move the QuSTICKs using trucks or planes to redistribute the quantum hardware. We do not need to rebuild optic fibre links, reposition or relaunch costly satellite systems or rebuild complex optical receiving stations. 

Key exchange in the Sneakernet environment essentially mimics the trusted courier model in which hard drives full of sensitive key material are physically transported around the world by trusted personnel. The security of classical keys are completely dependent on the reliability of these couriers. While vetting protocols used by the most skilled intelligence agencies worldwide are generally excellent, the possibility of key material being lost, stolen, sold, or surreptitiously copied is the most significant failure mode of these networks. 

In the Sneakernet system, \textit{entanglement is the only thing being transported}. The keys themselves are not generated until just before being used. This allows us to generate and purify keys milliseconds before they are used to encrypt a classical message, then destroy them immediately thereafter. This does not completely close the window in which a key can be intercepted or copied, but it reduces it from hours or days (the time needed for a trusted courier to transport a hard drive from home base to the field) to mere milliseconds. Additionally, the only way in which keys could be copied or otherwise intercepted is if an individual physically steals or otherwise takes control of the QuSTICK itself. If an adversary does compromise a QuSTICK unit, they would also need to compromise the classical authentication channel that is used for key reconciliation and the transmission of the encrypted message. At the same time, this adversary would have to ensure that whatever physical act was performed in order to gain physical control over the QuSTICK had not been detected (refer back to the note above about hitting someone over the head and stealing his or her briefcase). 

\subsubsection{One-time-pads}\index{One-time-pads}

With the Sneakernet system, generation and usage of one-time-pad material occurs in exactly the same way as for key exchange. The only difference is the physical volume of QuSTICKs necessary. For key exchange, we generate a small, fixed amount of key material that is then combined with strong, classical encryption techniques to generate a secure message. This opens up a possible security flaw because we have to trust the security of this classical encryption protocol (e.g. AES) at a time when quantum computers are on the verge of becoming practical. For highly sensitive information that needs to be secured for a long period of time, hoping that classical encryption techniques will remain secure for decades to come is an assumption that many may not wish to make. 

One-time-pads are the only encryption technique discovered that is completely secure when implemented correctly. However, as discussed previously, one-time-pad encryption requires a key that is exactly the same length as the message to be transmitted. The key material is hashed with the message and, provided that the key is only known by the sender and receiver, it is provably impossible for any eavesdropper to decrypt the message. 

At the extreme end of the scale, a high-definition video stream (4K at 24fps) requires transmitting approximately 35MBps of classical data. Hence, securing this kind of transmission using a one-time-pad would require at least 70 million QMUs for each second of video. This is clearly an extraordinarily large number of QuSTICKS. However, as we have noted, the network continues to grow as more physical QuSTICKS are manufactured. 

\begin{figure}[htbp!]
	\includegraphics[clip=true, width=0.475\textwidth]{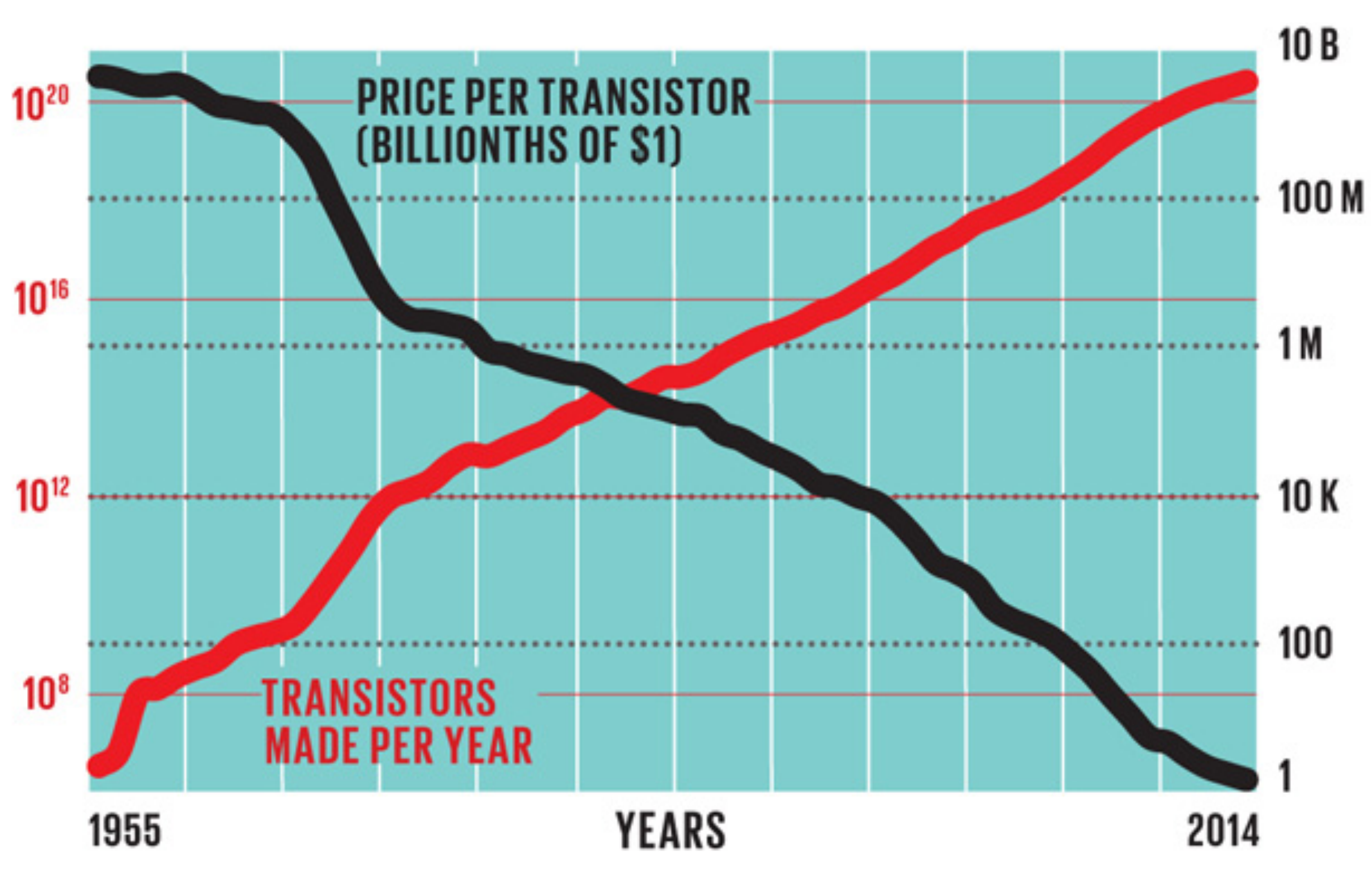}
	\caption{Historical number of transistors manufactured per year since their invention (red curve).  (source: VSLI research, \href{http://www.vsliresearch.com}{http://www.vsliresearch.com})} \label{fig:transistor}
\end{figure}

In Fig.~\ref{fig:transistor} we illustrate this assuming Moore's law scaling in the total number of QMUs fabricated (note that Moore's law actually quantifies the number of transistors that can be placed on a single chip; the numbers here are much, much larger). Fig.~\ref{fig:transistor} shows the total number of transistors manufactured per year, worldwide, starting with the first demonstration of the transistor in 1947, until now. 

\begin{figure}[htbp!]
	\includegraphics[clip=true, width=0.475\textwidth]{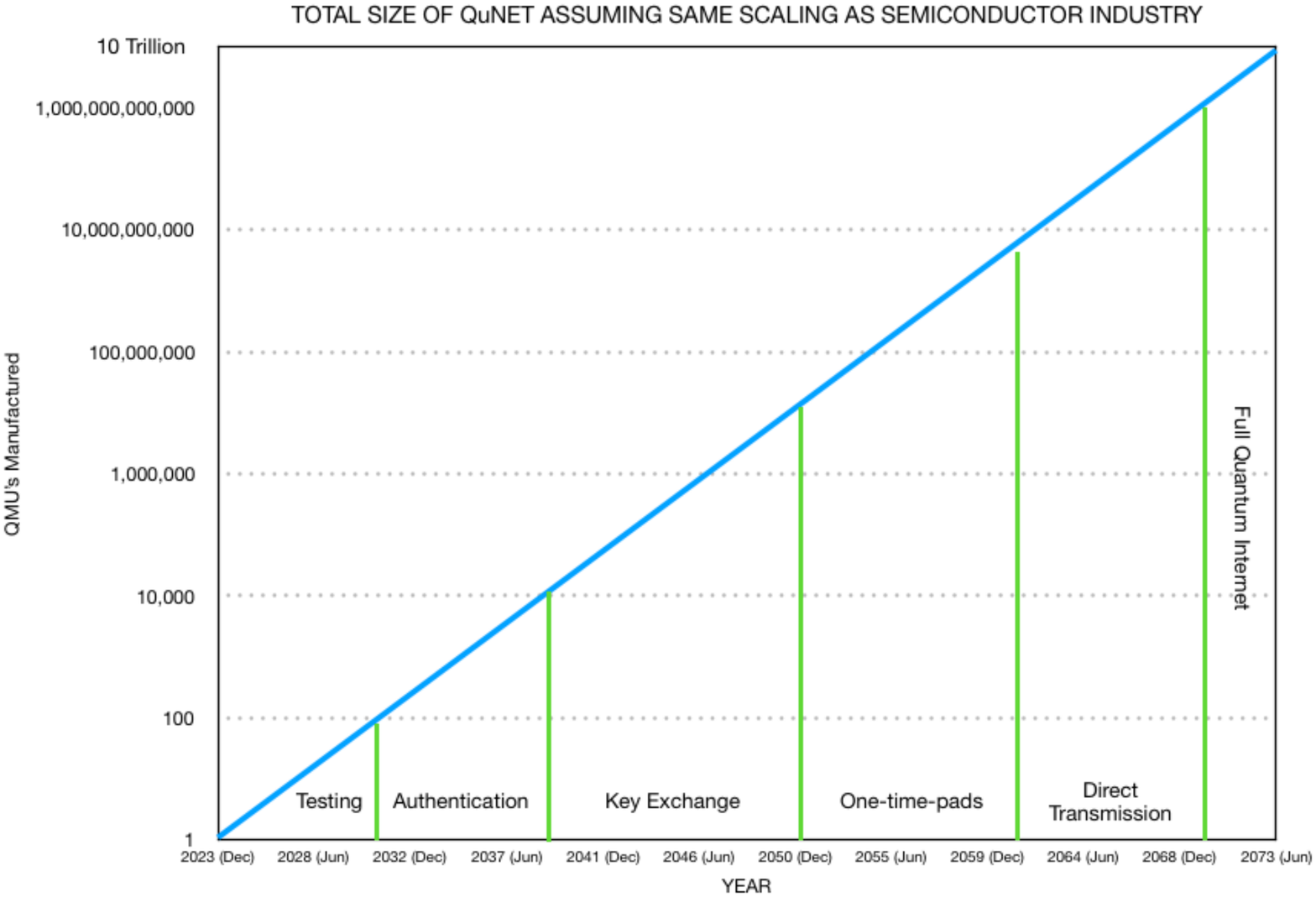}
	\caption{Assuming the same scaling as Figure \ref{fig:transistor} for the quantum world, the cumulative number of QMU's in existence for the next 50 years.  This plot assumes that the first QMU is demonstrated in 2023 and a Moore's law scaling kicks in at that point.  Illustrated are the boundaries when certain quantum protocols ``come online", from small scale testing all the way to a functional, global quantum internet.} \label{fig:QMU}
\end{figure}

Assuming an initial R\&D time frame of approximately 5 years for the first demonstration of a QMU, we further assume the same scaling in terms of the number of QMUs manufactured and illustrate the growth of the Sneakernet network though each stage: 1) a global authentication network, 2) a global key exchange network, and 3) a global one-time-pad encryption network. We also show a forth stage of the Sneakernet network: the ability to build a true quantum internet that connects large-scale, error-corrected quantum computing systems. Fig.~\ref{fig:QMU} illustrates network volume extrapolated using the same scaling as Fig.~\ref{fig:transistor} (in terms of the total number of QMUs manufactured after 2023) for the next 50 years.  This extrapolation is somewhat tongue-in-cheek as while we are confident that a Moore's law type of scaling will occur with qubit technology, we have no way of knowing if the scaling will be the same as the classical silicon industry or when that exponential trend will begin. 

However, if a similar trend line is followed, approximately ten trillion QMUs may be deployed worldwide by the latter half of this century. This is an astronomical figure for active quantum devices; and it would also be more than sufficient to construct multiple, large-scale quantum computing systems capable of executing any large-scale algorithm that has ever been developed. But keeping our discussion limited to the communications space, the Sneakernet system is, as we have said, designed to incorporate new QMUs as they become available, rather than replacing them. So the total size of the network is cumulative. 

How such a network will be used is less clear. As noted above, we can make the following arguments as to when a particular application `comes online':
\begin{itemize}
\item Authentication tokens can come online once the number of QMUs $> 10^2$. 
\item Key exchange can come online once the number of QMUs $> 10^4$.
\item One-time-pads can come online once the number of QMUs $> 5\times 10^7$. This assumes voice communications using the 600 bps NATO standard STANAG-4591 vocoding technique with a cumulative total of 10 hours of communication that needs to be encoded using QuSTICKS\footnote{\href{https://www.public.navy.mil/jtnc/APIs/API_1.4_20150226_VocoderService.pdf}{https://www.public.navy.mil}}.
\item Direct coupling of fully error-corrected quantum computers at a MHz logic-gate rate once the number of QMUs $> 10^{10}$. 
\item Multi-link, global quantum internet once the number of QMUs $> 10^{12}$.
\end{itemize}

Between 100 and 10,000 QMUs, the network can expand into a network consisting of 100 separate authentication links, with each link implemented at arbitrary distance scales. These separate links could be deployed between the same two parties (increasing the speed of a single link) or it could be spread out over multiple locations. How the network is configured and re-configured is completely at the discretion of whoever owns or controls the physical QuSTICKS. 

Once a total of 10,000 QMUs have been manufactured, the operator of the network may choose to re-task \textit{all} the QMUs then deployed to form a single link dedicated to key exchange for symmetric encryption protocols such as AES. Once this is achieved, the process starts again. At this point the network can become a hybrid of both key exchange and authentication, with resources deployed depending on need. The expansion process continues using these two protocols until one-time-pad quantum encryption becomes feasible for data transmission of significant volume (at approximately 50 Million QMUs). At each stage of network expansion, resources can be dynamically redesigned to support whatever protocols are both possible (given the total number of units in the field) and what is desired by whomever controls the physical QuSTICK units. 

Once the number of QMUs exceeds the number needed to connect together fully error-corrected quantum computing systems, the network will continue to expand to encapsulate all the sub-protocols, completing the transition to a multi-purpose quantum internet. 

Our extrapolations based upon the historical evolution of the classical semiconductor industry is extremely optimistic; but given the design of a particular underlying technology, once a single QMU chip-set can be fabricated at low enough cost, expansion of the network becomes purely a function of how quickly high volume manufacturing can be developed and how much PPQ can be reduced as that manufacturing infrastructure becomes more advanced. As with classical storage memory, requiring petabytes or exabytes of cheap, stable memories may have seemed completely unrealistic in the 1940's or 1950's but continual development of the technology has now made that that level of classical storage routine.

\subsection{Revisiting the GCHQ criticisms}\index{GCHQ}

Now that we have described the structure and operation of a Sneakernet based communications system, let us revisit some of the criticisms of the GCHQ (the UK's Government Communications Headquarters office) mentioned earlier in using QKD to secure classical communications channels. 
\\
\\
\textit{``QKD protocols address only the problem of agreeing on keys for encrypting data. Ubiquitous on-demand modern services, such as verifying identities and data integrity, establishing network sessions, providing access control, and automatic software updates, rely on authentication and integrity mechanisms (e.g. digital signatures) as well as encryption."}
\\
\\
As we have discussed, key distribution is only one protocol in a larger stack of applications of a Sneakernet network. The network can be used for authentication and session integrity by using shared entanglement between two parties with or without quantum validation via the violation of Bell inequalities; and the portability of the QuSTICK system allows for these authentication channels to be set up wherever physical access to transport is available.
\\
\\
\textit{``The two major functional limitations of commercial QKD systems are the relatively short effective range of transmission, and the fact that BB84 and similar proposals are fundamentally point-to-point protocols. This means that QKD does not integrate easily with the Internet or with the mobile technologies, apps and services that dominate public and business life today."}
\\
\\
This system is designed from the ground up to be functional over global distances. It doesn't need extensive infrastructure along the communications channel and can leverage classical transport mechanisms that are already available, efficient and ubiquitous. 

While we have illustrated much of the function of the Sneakernet system through point-to-point protocols\index{Point-to-point!Protocols} between two parties, the design is a fully-formed quantum network. Entanglement can be distributed and shared between an arbitrary number of parties and more complex protocols, such as secret sharing or distributed communication/computation, can be performed. In the context of QKD, we do not utilise the point-to-point protocol of BB84. Instead we base QKD protocols on the more advanced E91\index{E91 protocol} (which utilises two properties of entanglement, the perfect correlation of measurements made by Alice and Bob as well as the ability to detect eavesdropping by noting disturbances in the quality of that correlation). A Sneakernet's ability to distribute, share and maintain entanglement on global distances makes Ekert-91 practical for real-world cryptography for the first time.
\\
\\
{\em ``Hardware is expensive to obtain and maintain. Unlike software, hardware cannot be patched remotely or cheaply when it degrades or when vulnerabilities are discovered."}
\\
\\
The Sneakernet network is built from the QMU chip sets and the QuSTICK devices. These devices are ideally designed to be cheap and to be mass manufactured, with network capabilities increasing as more and more units are produced. One of the major benefits to this system is that we can easily replace/repair or augment QuSTICKS within the network to fix potential vulnerabilities in the future or to simply fix faulty units. Unlike infrastructure-intensive quantum communication systems (such as quantum repeater networks or satellites), we have trivial access to each physical device within the network, and they can be repaired or replaced as simply as a non-functional hard-drive in a classical Sneakernet communications channel. Patching hardware remotely is not required, as the hardware itself is easily portable. Hence, if repairs or patches are needed, QuSTICK units can be rotated in and out of the larger network and then immediately redeployed in the field without significantly affecting the performance of the network. Once large volumes of QuSTICKs are deployed in the network, regular servicing and repairs of individual units will be background noise to overall network performance. 
\\
\\
\textit{``Any real-world QKD system will be built from classical components, such as sources, detectors and fibres, and potentially ancillary classical network devices, any one of which may prove to be a weak link. A number of attacks have been proposed and demonstrated on deployed QKD systems that subvert one of more of these hardware components, enabling the secret shared key to be recovered without triggering an alarm."}
\\
\\
Sources, detectors and fibres are not components of the Sneakernet system. The integrity of the network rests with the functional integrity of individual QuSTICKs. As these units will be continuously moved, entangled locally, and then moved again, QuSTICKs can be tested and verified when the entanglement is initially prepared. Various unit testing can be performed between two QuSTICKs at home base to ensure quantum integrity of the system before they are ever deployed in the field. Compromised units can be removed from the network or returned for a complete rebuild and redeployment. Any units that become compromised in the field will be detected through the entanglement links that were initially prepared when QuSTICKs were present at home base. As no further entanglement operations are performed between QuSTICK units \textit{after} they have been locally entangled and verified points of failure (where security could be affected) are drastically reduced. 
\\
\\
\textit{``Denial of service (DoS)\index{Denial of service (DoS) attacks} attacks that interfere with the paths carrying the QKD transmissions also seem potentially easier with QKD than with contemporary Internet or mobile network technologies. Since QKD devices typically abort a key establishment session when they detect tampering, this makes it difficult to recommend QKD for contexts where DoS attacks are likely to be attempted."}
\\
\\
As we have mentioned, denial of service attacks require an actual physical QuSTICK to be compromised. For an effective denial of service attack to be launched against the entire network, an adversary would have to steal or physically compromise \textit{every} QuSTICK unit one of the parties possesses -- a much more difficult thing to do than simply cutting an optic fibre link or jamming the transmission of photons. If only a subset of QuSTICKS are stolen or otherwise compromised, network performance will decrease, but provided there are at least two uncompromised units somewhere, a viable entanglement connection will exist. 

\subsection{A vision for a Sneakernet based quantum internet}

A fully functional quantum internet (that is, a network with the ability to provide quantum communication channels between large-scale, error corrected quantum computers) requires the following:
\begin{itemize}
\item Transcontinental communication links spanning distances anywhere up to 10,000km.
\item A high speed network spanning those distances for $>$THz operational speeds.
\item End-to-end error rates of $10^{-10}$ or lower.
\end{itemize}

Two ideas for constructing such worldwide quantum communication networks have received extensive theoretical examination and experimental demonstration:
\begin{enumerate}
\item Quantum repeater systems.
\item Quantum-based satellite communications.
\end{enumerate}
However, both approaches run into significant practical limitations.

High-speed quantum repeater networks only exist for \textit{theoretical} transmission rates of about 1-10MHz \cite{SD-Fowler:2010aa} and require repeater stations every 20-50km \cite{SD-Munro:2012aa}. This upper limit to repeater station separation is necessitated by the loss rates of current optic fibre technology and it is unclear if that range will ever be significantly extended. An associated problem with the small separation distances of quantum repeater stations is the difficulty of deploying networks across oceans or otherwise inhospitable environments. Quantum repeaters are small-scale quantum computers, consisting of thousands of qubits and associated control infrastructure. This requirement currently precludes their deployment at high densities across the planet. 

Regarding satellite technology, there has been significant experimental progress, with entanglement-based satellite platforms deployed by the Chinese, as well as proof-of-principle payloads deployed by the Singaporeans, Japanese and Austrians. These platforms are not designed for general purpose quantum communications. They are built for QKD applications, which do not have the stringent constraints listed above. 

While fully error-corrected, high bandwidth, and low error-rate satellite systems have not received significant theoretical attention as yet, we can safely assume that such technology \textit{could} be built and deployed. However, deployment and maintenance costs, bandwidth sufficient for fully error-corrected communications channels, and the infrastructure associated with receiving stations represent very big hurdles to overcome when utilising space based platforms as backbones to a quantum internet. 

QuSTICK-based quantum networks can satisfy the constraints noted above as well as address the issues associated with quantum repeaters and satellite communication systems. Initial analysis shows that $>$THz transpacific networks can be realised in a systems of moderate physical speed and size, provided that the cost per physical qubit is low enough. Beyond the challenge of simply building a sufficient number of QuSTICKs (a challenge that is common for any hardware targeting the construction of large-scale quantum computing platforms), there is no additional hardware development work needed to realise a global network. The only additional infrastructure needed by a global QuSTICK network is traditional global shipping channels that already exist. 

Prototype quantum communication networks using Sneakernet principles and QuSTICKs will first be demonstrated at much shorter ranges and at slower communication rates. Scaling up to higher fidelity, higher speed, and longer ranges is conceptually straightforward. Additionally, communications channels do not influence infrastructure development. Building a quantum communications system between Tokyo and Osaka is no different to building a communications system from Japan to Australia. 

\begin{figure}[htbp!]
	\includegraphics[clip=true, width=\columnwidth]{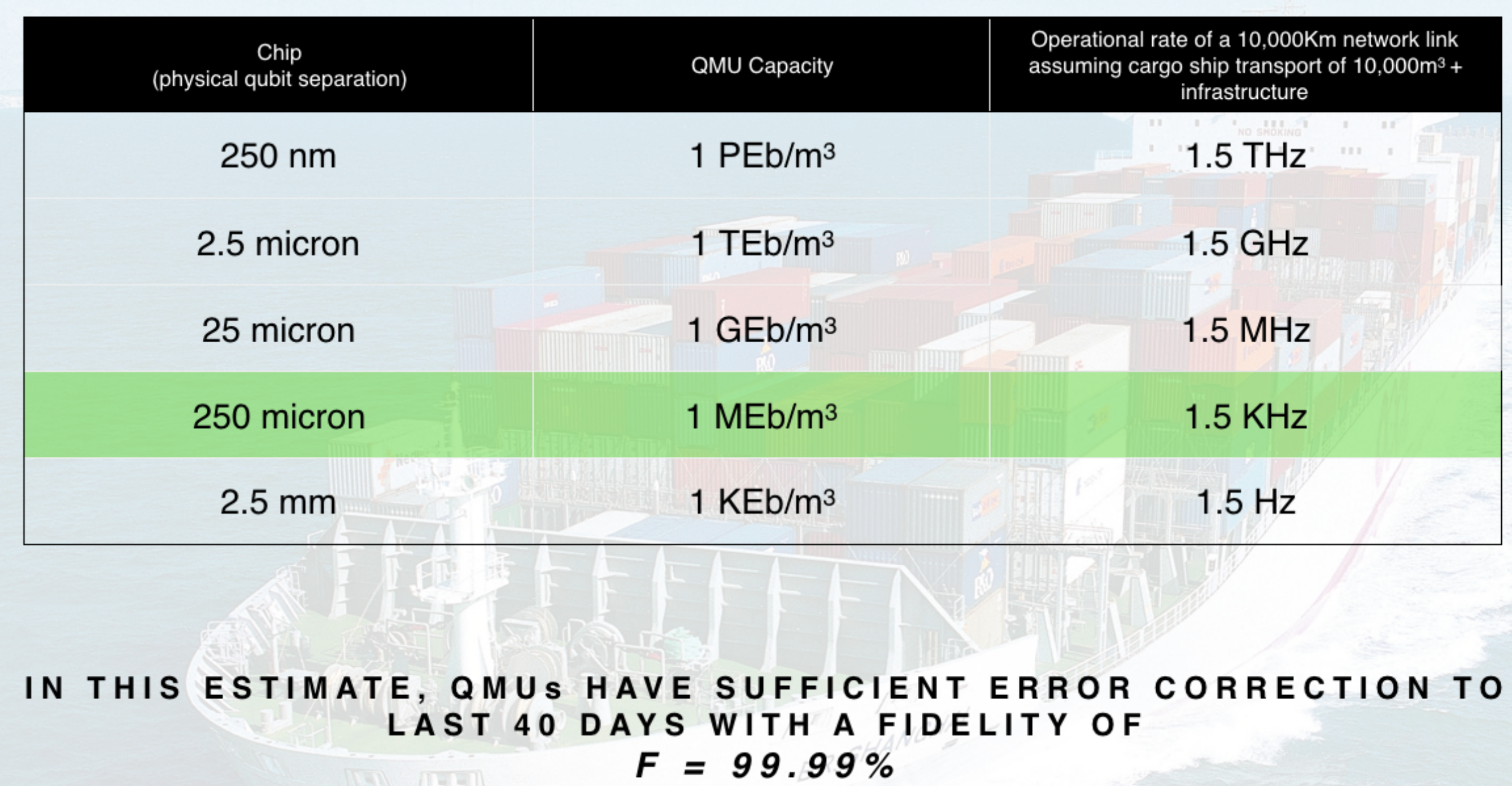}
	\caption{} \label{fig:link}
\end{figure}

Fig.~\ref{fig:link} illustrates the performance of a single link based on physically transporting QuSTICK units on cargo containers, over a distance of 10,000km with a link fidelity of 99.99\% as a function of the density of physical qubits inside each QMU. These are only back of the envelope calculations, but does illustrate that overall QMU density and number of units is the only impediment to achieving truely global link distances at ultra-high fidelity at bandwidths that approach what is currently possible in classical fibre optics communications. No other system proposed can scale to this level.
 
While many of the other core technologies for quantum communications (satellites, fibre, free-space) can be used to augment secondary and tertiary communication links appropriate to the applications needed, Sneakernet based communication networks appear to be the best approach to form the major high-bandwidth truck lines of the future. 

\latinquote{Veritas vos liberabit.}

\sketch{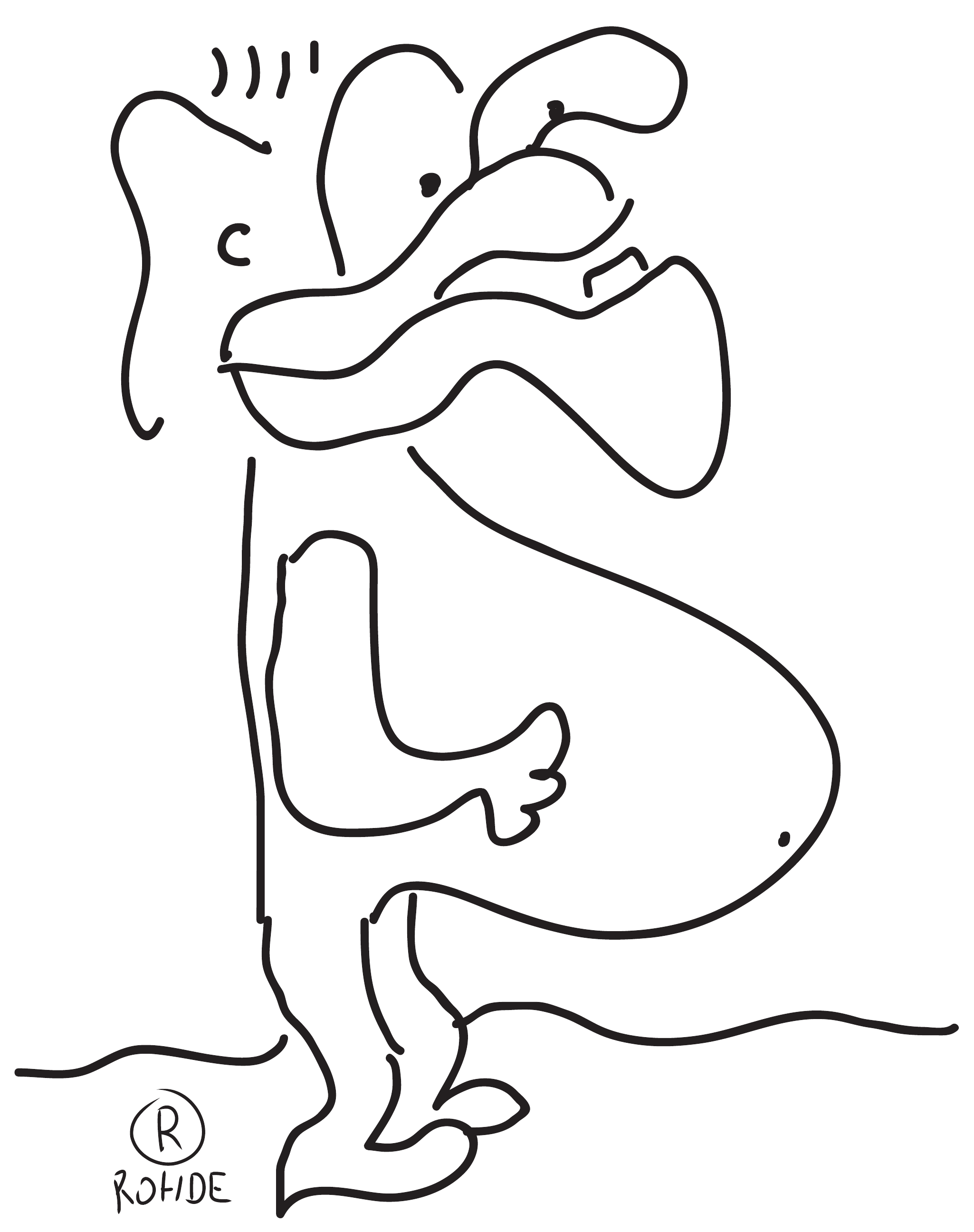}

\clearpage

% % Quantum cryptography
% %

\part{Quantum cryptography}\label{part:quant_crypto}\index{Quantum cryptography}

%
% CryptoWars (TM)
%

\famousquote{I would rather have questions that can't be answered than answers that can't be questioned.}{Richard Feynman}
\newline

\dropcap{U}{ndoubtedly}, quantum technologies will be most impactful (and disruptive!) in the area of information security, something of fundamental importance to us all on a daily basis, vital to the entire world economy. Quantum technologies will be important both in terms of breaking and maintaining security, with the former mandating interest in the latter.

In Sec.~\ref{sec:homo_blind} we discussed encrypted outsourced quantum computation as an important concept in future cloud quantum computing. In this section we will step back from full-fledged distributed quantum computation, instead focussing on more elementary protocols for simple secure communication or protocols.

Today, the ability to communicate secretly with others is completely taken for granted in all but a few nations and resides in every smartphone and desktop PC. Furthermore, the encryption technologies available to the average consumer are extremely strong, the same as those used by large organisations, including world governments.

%
% What is Security?
%

\section{What is security?}\index{Computational!Security}\index{Information-theoretic!Security}\label{sec:comp_vs_inf_th_sec}

\famousquote{The only true wisdom is knowing you know nothing.}{Socrates}
\\
\\
\famousquote{Ignorance more frequently begets confidence than does knowledge.}{Charles Darwin}
\\

\dropcap{B}{efore} describing any specific cryptographic protocols, let us define what is meant by `security' in a cryptographic context. We differentiate between \textit{information theoretic security}\index{Information-theoretic!Security}, as opposed to \textit{computational security}\index{Computational!Security}:

\begin{itemize}
	\item Information-theoretic security: the laws of quantum information bound the amount of information that can be extracted from a system, irrespective of measurement or computational operations. Thus, such security can be regarded as attack-independent, making no assumptions about our adversary's capabilities.
	\item Computational security: is based on the assumption that an adversary's computational resources are insufficient to perform cryptanalysis or brute-force cracking.
\end{itemize}
Clearly the former makes a far stronger statement about the security of a protocol than the latter.

Classical public- and private-key encryption protocols are typically based upon the assumption of computational security (e.g the computational complexity of performing integer factorisation in the case of RSA public-key encryption, or solving a complex satisfiability problem in the case of private-key encryption), whereas quantum encryption protocols are typically information theoretically secure (e.g the one-time pad using QKD).

%
% Classical Cryptography
%

\section{Classical cryptography}\index{Classical cryptography}

\dropcap{W}{e} begin with an introduction into \textit{classical} cryptography, so as to understand its limitations, which logically leads us into how quantum mechanics can assist in overcoming them. We only scratch the surface of this extremely well-researched field, reviewing some of the most important and widely used protocols. For a deeper understanding of classical cryptography we refer the interested reader to the excellent and comprehensive \cite{bib:Schneier96}.

%
% Private-Key Cryptography
%

\subsection{Private-key cryptography}\index{Private-key!Cryptography}

Private- (or symmetric-) key cryptography is perhaps the most basic (and useful) cryptographic primitive, enabling encryption of a channel between two parties who share a secret-key\index{Private-key} -- a random bit-string of length determined by the encryption algorithm. The same secret-key is employed for both encryption and decryption operations (hence `symmetric'), making it of utmost importance that it be retained secret.

Private-key cryptography has a long history, in fact going back to ancient times, enabling the secret sharing of diplomatic messages between emperors and empires, e.g the so-called \textit{Caesar cipher}\index{Caesar cipher}, a simple substitution cipher\index{Substitution!Cipher} based on shifting the letters of the alphabet. However it was a niche technology that very few utilised, since it had to be implemented by hand without computers or automation.

Today there are countless freely available private-key cryptographic protocols available online, and some have been standardised by standards institutes. Currently, the Advanced Encryption Standard (AES)\index{Advanced Encryption Standard (AES)} is a standard endorsed by the US government, replacing the earlier standardised Data Encryption Standard (DES)\index{Data Encryption Standard (DES)} whose mere 56-bit key-length is today considered insecure in light of present-day computing power. AES is a block cipher\index{Block cipher}, meaning that it divides data into small blocks of 128 bits, each of which are encrypted independently, and operates with key lengths of up to 256 bits (referred to as AES256), making it very robust against (even quantum) brute-force attacks (Sec.~\ref{sec:attacks_on_class}). The length of the plaintext and ciphertext is the same, meaning there is no bandwidth overhead when communicating encrypted data across a network.

%
% One-Time Pad Cipher
%

\subsection{One-time pad cipher}\index{One-time pad}

There is one and only one \textit{provably} secure (in the sense of information-theoretic security\index{Information-theoretic!Security} as opposed to computational security\index{Computational!Security}) encryption protocol -- the \textit{one-time pad}\index{One-time pad}. This protocol requires Alice and Bob to share a random secret-key as long as the message (plaintext\index{Plaintext}) being communicated between them. The two bit-strings undergo bit-wise XOR operations to form the ciphertext\index{Ciphertext}. Mathematically,\index{One-time pad}
\begin{align}
c = s \oplus k,
\end{align}
where $\oplus$ is the bitwise XOR operation (equivalently addition modulo 2), and $c$, $s$ and $k$ are the ciphertext, plaintext and key bit-strings respectively, all of which are of the same length,
\begin{align}
	|c|=|s|=|k|.
\end{align}

The security of this protocol is easy to see intuitively -- with an appropriate choice of key, \textit{any} plaintext of the same length could be inferred from \textit{any} ciphertext. This means that there is no possibility of performing any kind of frequency analysis\index{Frequency!Analysis}, as the ciphertext string has maximum entropy\index{Entropy} (inherited from the maximum entropy of the random key, and assuming a strong cryptographic random bit generator) and thus no correlations. Since every possible valid plaintext can be recovered using an appropriate key, a cracking algorithm is unable to find a unique plaintext matching the ciphertext, since all are equally valid decryptions.

Importantly, the secrecy of the one-time pad\index{One-time pad} strictly requires that a key never be reused. A fresh key must be generated for each message sent, otherwise trivial frequency analysis\index{Frequency!Analysis} techniques can be employed to compromise security. If the same key $k$ is used to encode two messages $s_1$ and $s_2$, yielding ciphertexts,
\begin{align}
c_1&=s_1\oplus k,\nonumber\\
c_2&=s_2\oplus k,
\end{align}
then we trivially obtain,
\begin{align}
c_1 \oplus c_2 &= (s_1 \oplus k) \oplus (s_2 \oplus k) \nonumber \\
&= (s_1 \oplus s_2) \oplus (k \oplus k) \nonumber \\
&= s_1 \oplus s_2,
\end{align}
which is independent of the key. Now a frequency analysis on the bitwise XOR of two plaintexts can be applied, without requiring any knowledge of the key whatsoever.

Needless to say, the requirement for keys of the same length as the plaintexts, which cannot be reused, raises the obvious criticism that now secret-key-sharing is as difficult as sharing a secret message in the first place. This reduces the problem of perfect secrecy of arbitrary messages to the secrecy of shared randomness. 

Although during the Cold War Soviet diplomats would literally carry briefcases between countries full of paper with random data for use in a one-time pad, it is clearly not suitable for everyday applications!

%
% Public-Key Cryptography
%

\subsection{Public-key cryptography}\index{Public-key cryptography}\label{sec:public_key_crypt}

While private-key cryptography solves the problem of end-to-end cryptography, it has one main downfall -- how does one share a private-key between two parties? After all, if we had the ability to secretly share keys between ourselves, wouldn't we just use that same method to directly communicate, bypassing the unnecessary cryptographic protocol?

Public- (or asymmetric-) key cryptography addresses this issue by replacing the private-key with two keys (known as a key-pair\index{Key-pair}), one used solely for \textit{encryption}, the other solely for \textit{decryption}. Importantly, these two keys are non-trivially related and cannot be efficiently computed from one another. To send a message to a friend I can send him my encryption (public) key that he is only able to use for preparing an encrypted message for me. No security is required when sharing the public-key since an eavesdropper can't use it for decryption. Finally, I am able to decrypt the message using my decryption (private) key, which I kept completely to myself and never shared with anyone.

RSA \cite{bib:RSA} was the first published public-key cryptographic protocol, and forms the backbone for most encryption used on the internet today. It achieves its security based on the (strongly held, but unproven) belief that factorising large integers into constituent primes is a computationally hard problem -- a so-called `trapdoor function'\index{One-way functions}. The algorithm is built upon number theory using modular arithmetic. Alg.~\ref{alg:RSA} described the RSA key-generation, encryption, and decryption protocols.

\begin{table}[!htbp]
\begin{mdframed}[innertopmargin=3pt, innerbottommargin=3pt, nobreak]
\texttt{
function RSAGenerateKey():
\begin{enumerate}
\item $p$ = randomPrime()
\item $q$ = randomPrime()
\item if($p=q$) goto 1
\item $n = pq$
\item $\lambda = \mathrm{lcm}(p-1,q-1)$
\item $e = \mathrm{coprime}(\lambda) \,\,\mathrm{s.t} \,\,e<\lambda$
\item $d = e^{-1}\,(\mathrm{mod} \, \lambda)$
\item publicKey = $\{n,e\}$
\item privateKey = $\{n,d\}$
\item keyPair = \{publicKey,privateKey\}
\item return(keyPair)
\item $\Box$	
\end{enumerate}
function RSAEncrypt(plaintext, publicKey):
\begin{enumerate}
\item $\mathtt{cipertext} = \mathtt{plaintext}^e \, (\mathrm{mod} \, n$)
\item return(ciphertext)
\item $\Box$
\end{enumerate}
function RSADecrypt(ciphertext, privateKey):
\begin{enumerate}
\item $\mathtt{plaintext} = \mathtt{ciphertext}^d \, (\mathrm{mod} \, n)$
\item return(plaintext)
\item $\Box$
\end{enumerate}
}
\end{mdframed}
\captionspacealg \caption{Number-theoretic algorithms based on modular arithmetic for RSA key generation, encryption and decryption.} \label{alg:RSA}
\end{table}

Since RSA, numerous other public-key cryptosystems have been developed, based on different choices of trapdoor function. Most notably, elliptic-curve cryptography\index{Elliptic-curve cryptography} has gained much attention. However, RSA remains the most widely used and well-studied public-key cipher.

To mitigate the need for constant one-on-one exchange of public-keys, many key servers\index{Key servers} exist around the globe, which maintain databases of people's public-keys. These servers are in a position of trust, vouching for the identities associated with their stored public-keys.

%
% Key Exchange Protocols
%

\subsection{Key exchange protocols}

A downside of RSA is that ciphertexts are in general much longer than plaintexts, unlike private-key protocols where the ciphertext is always the same length as the plaintext. For this reason it is typically not used to directly encrypt long messages, since the memory and bandwidth overheads would be undesirable. Instead, RSA is typically employed in conjunction with private-key cryptography in a key exchange protocol\index{Key exchange protocol}. Here, the public-key system communicates a private \textit{session key}\index{Session key} between parties, which is subsequently employed in a private-key protocol, without incurring the time and memory overhead that RSA does. In Sec.~\ref{sec:brute_force_attacks} we point out that quantum computers are not believed to be able to efficiently crack private-key protocols, but instead effectively reduce their key length by a factor of 1/2, which is easily counteracted with longer keys to restore security.

Numerous key exchange protocols have been formulated, the best known being the Diffie-Hellman\index{Diffie-Hellman protocol} protocol \cite{diffie2022new}, which has been widely adopted on the internet.

%
% Digital Signatures
%

\subsection{Digital signatures} \label{sec:dig_sig} \index{Digital signatures}

Rather than cryptographically ensuring the secrecy of messages, a user may wish to prove their identity when sending a message, such that the recipient can be certain it originated from who it says it does, and accurately conveys what they said. This is achieved using \textit{digital signatures}.

Digital signatures can be easily implemented using the RSA protocol\index{RSA encryption!Protocol}, by reversing the roles of the public and private-keys. Now the public-key can only be used for decrypting a message, and the private-key can only be used for encrypting it. As before, it is computationally hard to infer one from the other. The protocol for sending and verifying a digitally signed message is shown in Alg.~\ref{alg:dig_sig}.

\begin{table}[!htbp]
\begin{mdframed}[innertopmargin=3pt, innerbottommargin=3pt, nobreak]
\texttt{
function DigitalSignature(message,keyPair):
\begin{enumerate}
\item Alice prepares a short \textit{digest}\index{Digest} of her message using a cryptographic hash function\index{Hash!Functions}, such as SHA256\index{SHA256} (Sec.~\ref{sec:hashing}),\\
	\vspace{1mm} 
	$digest=SHA256(message)$
	\item Alice encrypts the digest using her private-key. This forms the `digital signature',\\
	\vspace{1mm} 
	$signature = RSAEncrypt(digest,privateKey)$
	\item Alice transmits the digital signature and original message to Bob,\\
	\vspace{1mm} 
	$signedMessage = \{message,signature\}$
	\item Bob hashes the received message,\\
	\vspace{1mm} 	
	$hash=SHA256(message)$
	\item Bob uses Alice's public-key to decrypt her digital signature,\\
	\vspace{1mm} 
	$decryptedHash =$\\
	$RSADecrypt(signature,publicKey)$
	\item Bob compares his calculated hash with Alice's decrypted hash for consistency. If the two hashes are identical, Bob concludes the message was authentic,\\
	\vspace{1mm} 
	$if(decryptedHash=hash)$\\
	$\,\,return(pass)$\\
	$else$\\
	$\,\,return(fail)$
	\item $\Box$
\end{enumerate}
}
\end{mdframed}
\captionspacealg \caption{Protocol for digitally signing a message using public-key cryptography (RSA) and a cryptographic hash function (SHA256).} \label{alg:dig_sig}
\end{table}

The key point from the security perspective is that the private-key cannot be efficiently inferred from the public-key. So although everyone has access to Alice's public-key, no one is able to counterfeit messages since they cannot create encrypted signatures without access to her private-key -- signatures can be easily verified but not created.

Because this protocol is implemented using ordinary RSA\index{RSA encryption!Protocol}, albeit with reversed roles for the key-pair, it shares the same security strengths and vulnerabilities as RSA public-key cryptography (Sec.~\ref{sec:attacks_on_class}).

Like RSA cryptography, key servers\index{Key servers} exist, maintaining databases of people's public-keys and their associated identities.

Because RSA-encrypted messages are long, digital signature protocols typically do not sign the full document directly. Instead they create a message digest\index{Message digests} of the document using a cryptographic hash function, which is signed using RSA. These hash functions have the property that they cannot for forged or manipulated, providing an accurate summary of a document, but with extremely low memory overhead (256 bits is typical). Hash functions are discussed next in Sec.~\ref{sec:hashing}.

%
% Hashing
%

\subsection{Hashing} \label{sec:hashing} \index{Hash!Functions}

Hash functions are functions that map a long bit-string of arbitrary length to a short, fixed-size bit-string with quasi-random behaviour,
\begin{align}
	f_\mathrm{hash}:\,\{0,1\}^n \to \{0,1\}^m,
\end{align}
for an $n$-bit input and $m$-bit output hash, where $n$ is variable and $m$ is fixed. They are an example of `one-way functions'\index{One-way functions} that are computationally easy to compute in the forward direction, but extremely hard to invert. That is, given a hash, it is computationally unviable to find input strings that map to that value.

Hash functions have broad applicability throughout computer science, but here we are most interested in \textit{cryptographic hash functions} for use in cryptography, which impose strong conditions on the difficulty of inversion and their quasi-random characteristics. Most notably, the desired characteristics of a cryptographic hash function include:
\begin{itemize}
	\item The distribution of hashes ought not exhibit any biases, following a uniform distribution with quasi-random\index{Quasi-randomness} behaviour.
	\item Changing a single bit in the input string ought to flip approximately half the bits of the hash on average.	
	\item It's computationally efficient to calculate a hash from an input.
	\item It's computationally complex to find an input that hashes to a given value (that is, they are one-way or trapdoor functions).\index{One-way functions}
	\item The hashes of two very similar inputs ought to yield hashes that a very different.
	\item Two different inputs are extremely unlikely to hash to the same value.
\end{itemize}

The standard cryptographic hash function with mainstream adoption is the 256-bit Secure Hashing Algorithm (SHA256\index{SHA256}), which generates 256-bit hashes. The algorithm is extremely efficient to implement digitally, and exhibits $O(n)$ runtime for input string length $n$.

Cryptographically, hash functions are useful for creating message digests\index{Message digests}, which act as a highly condensed checksum\index{Checksums} of a document that can be utilised in a digital signature (Sec.~\ref{sec:dig_sig}).

Note that because the function in general maps longer strings to shorter ones, there are necessarily \textit{collisions}\index{Hash!Collisions} -- multiple inputs for a given output. However, for strong cryptographic hash functions their behaviour is sufficiently random that two distinct messages will almost certainly yield completely different hashes (even if the messages are very similar), making it all but impossible for someone to make the claim that Alice said something she did not. This property is extremely important for the security of digital signatures.

%
% Attacks On Classical Cryptography
%

\section{Attacks on classical cryptography}\index{Cryptographic!Attacks}\label{sec:attacks_on_class}

\famousquote{While I thought that I was learning how to live, I have been learning how to die.}{Leonardo da Vinci}
\\

\dropcap{H}{aving} introduced the main classes of classical cryptographic protocols, we now turn our attention to their weaknesses and vulnerabilities, both against adversaries with classical or quantum computational resources.

%
% Classical Attacks
%

\subsection{Classical attacks}

All known classical attacks against any respected classical cryptosystem involve tremendous computational resources. After all, were this not the case the cryptosystem would be considered weak and would never have become widely adopted in the first place!

%
% Brute-Force
%

\subsubsection{Brute-force}\index{Brute-force!Attacks}\label{sec:brute_force_attack}

The most obvious approach to cracking a cryptosystem is to systematically try out all possible keys until we find one that correctly decodes the encrypted message. This is also the most na\"ive approach, and one which is computationally intractable for real-world key lengths. Specifically, for a key length of $k$ bits (\mbox{$k=256$} for AES256), there are $2^k$ possible keys to try, and on average we will wait for $2^{k-1}$ trials before choosing the right one. Clearly an average waiting time of $2^{255}$ is not plausible!

%
% Cryptanalysis
%

\subsubsection{Cryptanalysis}\index{Cryptanalysis}

Far better than waiting the age of the universe for the right key to turn up, is \textit{cryptanalysis}. Here we study patterns between input and output strings from a cipher utilising a particular key. There are many variations on this, but include techniques such as \cite{bib:Schneier96}:

\begin{itemize}
	\item Known plaintext attacks (KPA)\index{Known plaintext attack}: Through alternate means of espionage, the attacker is able to possess \textit{both} a ciphertext and its associated plaintext. Knowing both the input and output to the encryption algorithm may then reveal information about the key relating them. This technique was important to Alan Turing's successful cracking of the German Enigma\index{Enigma machines} encryption protocol during World War II.
	\item Chosen plaintext attack (CPA)\index{Chosen plaintext attacks}: The same as a KPA except that the adversary has the ability to choose what the known plaintext is, a more challenging prospect to orchestrate.
	\item Linear cryptanalysis\index{Linear cryptanalysis}: A technique for representing ciphers as linear systems, to which KPA are applied.
	\item Differential cryptanalysis\index{Differential cryptanalysis}: We analyse how changes in input bits propagate through the cipher to modulate output bits. Typically this type of technique operates as a CPA.
\end{itemize}
 
%
% Integer Factorisation
%
 
\subsubsection{Integer factorisation}\index{Integer factorisation}

In the case of RSA encryption, whose security derives from the believed computational hardness of factorising large integers, the most efficient known classical algorithm for integer factorisation is the general number field sieve (GNFS)\index{General number field sieve}, with time-complexity,
\begin{align} \label{eq:GNFS_scaling}
	O(\exp (O(1) (\log n)^{\frac{1}{3}} (\log\log n)^{\frac{2}{3}})),
\end{align}
which scales poorly for large $n$, keeping in mind that present-day implementations of RSA accommodate key lengths of up to 4,096 bits, as for example is implemented by the widely-used Pretty Good Privacy (PGP)\index{Pretty Good Privacy (PGP)} package.

%
% Quantum Attacks
%

\subsection{Quantum attacks}

Having established that classical attacks against strong classical cryptosystems are quite limited by their implausible computational requirements, what if our adversary now has quantum computational resources? Does this change the game?

%
% Brute-Force
%

\subsubsection{Brute-force}\index{Brute-force!Attacks}\label{sec:brute_force_attacks}

A brute-force attack by a quantum computer does not offer us the exponential improvement attacker Eve might hope for. However, we can gain a quadratic improvement by cleverly exploiting Grover's search algorithm (Sec.~\ref{sec:quantum_algs})\index{Grover's algorithm}.

To do this, we treat the brute-force cracking algorithm as a satisfiability problem, similar to how Grover's is employed to enhance \textbf{NP}-complete problems. Specifically, our oracle implements the code's decryption operation, taking as input a qubit string representing the key. After decoding the message with the key, the oracle runs an appropriate test on the decrypted message to determine whether it is a legitimate decoded message. For example, it could run an English language test -- a message decoded incorrectly with the wrong key will appear very random and almost certainly won't pass such a test. The oracle tags an element passing this test, which the Grover algorithm searches for, yielding the associated key.

Note that when performing a brute-force attack against a private encryption key\index{Private-key!Cryptography}, a quadratic speedup effectively halves the key length in terms of algorithmic runtime, since \mbox{$O(\sqrt{2^k}) = O(2^{k/2})$}. Thus, in the quantum era private-key lengths will need to be doubled to maintain an equivalent level of security against brute-force attacks.

This same technique of treating encryption as an oracle within a quantum search algorithm can be utilised to invert hash functions\index{Hash!Functions}. However, in this case there will necessarily be multiple solutions owing to collisions.

%
% Cryptanalysis
%

\subsubsection{Cryptanalysis}\index{Cryptanalysis}

In the case of private-key cryptosystems such as AES\index{Advanced Encryption Standard (AES)}, no quantum-enhanced cryptanalytic techniques have been described, which offer an exponential enhancement. Thus, modulo doubling key-lengths to counter a Grover attack, these cryptosystems are not regarded as being compromised by quantum computing.

%
% Integer Factorisation
%

\subsubsection{Integer factorisation}\index{Integer factorisation}

In the case of RSA public-key cryptography the attack is more direct -- with access to a scalable quantum computer, Shor's algorithm\index{Shor's algorithm} can be employed to efficiently factorise large integers, allowing private-keys to be retrieved from public-keys. Unlike the brute-force attacks, which yielded only a quadratic enhancement, Shor's algorithm is exponentially faster than the classical GNFS, requiring runtime of only,
\begin{align}
	O((\log n)^2(\log\log n)(\log\log\log n)).
\end{align}
Compare this with the classical case given in Eq.~(\ref{eq:GNFS_scaling}).

%
% Bitcoin & The Blockchain
%

\section{Bitcoin \& the Blockchain}\index{Blockchain}\index{Bitcoin}\label{sec:bitcoin_blockchain}

\dropcap{O}{ne} of the most exciting new cryptographic applications that has emerged in recent years is the Blockchain, a secure distributed ledger\index{Distributed ledger} for recording the execution of contracts and transactions. This has enabled cryptocurrencies\index{Cryptocurrencies}, most notably Bitcoin, to emerge as a secure digital alternative to conventional fiat currencies.

More recent developments, such as the Ethereum project\index{Ethereum}, develop the distributed ledger further to allow executable code to be committed to the Blockchain, opening the prospects for self-enforcement and -execution of completely arbitrary `smart contracts'\index{Smart contracts}, a potential game-changer for the operation of financial and derivative markets.

In the Blockchain protocol, the validity of contracts and transactions is recognised collectively by participants using an encrypted digital ledger\index{Ledger}. The ledger records the complete history of all Blockchain transactions, which are digitally signed (Sec.~\ref{sec:hashing}) by network participants using elliptic-curve public-key cryptography\index{Digital signatures}\index{Elliptic-curve cryptography}. A democratic process ensures that, provided a single user doesn't monopolise the network, recorded transactions are legitimate, recognised collectively and democratically. This is secured by network participants digitally signing off on transactions as they take place.

The Bitcoin protocol builds on top of the Blockchain to create a secure digital cryptocurrency. This requires the introduction of another sub-protocol, \textit{mining}\index{Bitcoin!Mining}, where units of currency (`coins'\index{Coins}) are created. The protocol cryptographically ensures that there is an upper-bound on the number of coins that can exist, thereby preventing forgery and an inflationary blowout in the money supply.

The mining process is based upon the computational hardness of inverting (double) SHA256 hashing\index{Hash!Functions}\index{SHA256}. A legitimate Bitcoin is defined by a string with a hash satisfying a thoughtfully chosen constraint, specifically one which hashes to a value within some range,
\begin{align}
	\epsilon_\mathrm{lower}\leq \mathrm{SHA256}(\mathrm{SHA256}(x_\mathrm{coin})) \leq \epsilon_\mathrm{upper}.
\end{align}
This is slightly weaker than inverting hash functions, but is nonetheless a task that can only be approached via brute-force hashing in the forward direction. This associates computational complexity with the mining process, and hence computational integrity of the money supply, whilst upper bounding the number of unique coins that can exist. This technique is known as `proof-of-work'\index{Proof-of-work}, for associating something of value with proof that certain amounts of computation were invested into achieving it\footnote{In future implementations of Blockchain protocols, the proof-of-work required for a given protocol can be arbitrarily manipulated to accommodate for technological advances in computational power, for example via the adoption of quantum computing. The amount of work required to satisfy the constraint grows as we narrow the range \mbox{$\epsilon_\mathrm{upper}-\epsilon_\mathrm{lower}$}, providing us with much leverage to manipulate the complexity of the proof-of-work, and hence the rate of growth in the money supply, equivalently the rate of inflation\index{Inflation!Rate}.}. This idea was originally borrowed from the Hashcash protocol\index{Hashcash}, where proof-of-work is employed to associate work (and hence monetary value) with sending emails so as to eliminate automated spamming bots.

The two key algorithms for Bitcoin and the Blockchain are therefore hashing and public-key digital signatures. Both of these are subject to enhanced quantum attacks.

Inverse hashing does not have any known quantum algorithm with exponential improvement, however using a Grover search\index{Grover's algorithm} one can achieve a quadratic speedup, using the same idea as for enhancing \textbf{NP}-complete problems by treating the hash function as a search oracle\index{Oracles}. This however does not pose a fundamental security concern as it will speed up the Bitcoin mining process, but does not circumvent the upper-bound on the number of coins that may be in existence. Already classical mining has pushed the Bitcoin money supply close to its asymptotic maximum and there is limited room for additional mining\footnote{Bitcoin mining has gained so much traction and become so competitive that desktop PCs have become uneconomical for mining. Instead miners are resorting to utilising specialised hardware in the form of CUDA cores\index{CUDA}, FPGAs\index{FPGA} and ASICs\index{ASIC} (or by secretly using the company supercomputer while the boss isn't looking).}.

Elliptic-curve public-key cryptography, like RSA, has a known efficient quantum attack via Shor's algorithm\index{Shor's algorithm}. In the context of implementing digital signatures this implies that an adversary could fraudulently sign off on illegitimate transactions, thereby committing falsified contracts to the Blockchain.

A detailed investigation into the vulnerability of the Blockchain to quantum attacks was performed by \cite{bib:TomamichelBlockchain}. However, it is near impossible to predict the future rate of growth in quantum computer technology and hence over what kind of timescale the Blockchain will be compromised. But it is certain that a full compromise is inevitable at some point in the future when scalable, universal quantum computing becomes a reality.

To address this security threat, quantum-resistant hashing and public-key cryptographic protocols will need to be developed. In the former case this can easily be achieved by increasing hash lengths so as to offset the quadratic enhancement offered by Grover's algorithm\index{Grover's algorithm}. In the latter case this will require post-quantum public-key cryptosystems (to be discussed in the next section, Sec.~\ref{sec:end_of_class_crypto}).

Evidently, the lifespan of exisiting Blockchain technologies is limited and in the quantum future post-quantum Blockchain algorithms will be required to ensure the survival of cryptocurrencies.

\subsection{Quantum PoW based on boson sampling}
\label{subsec:BS_PoW}

What would such a post-quantum blockchain look like? Aside from addressing the above mentioned security concerns by adopting post-quantum cryptography, which is classical technology, it would be desirable to find a way to use quantum technology to alleviate the enormous amount of energy spent solving proof-of-work puzzles. Indeed, in 2024 the Bitcoin network consumed more energy than the country of Argentina, and while the ASICs used for crypto-mining can be located near renewable energy sources, a scalable network with high volume transactions does not appear to be sustainable. One could instead consider a variant of proof-of-work consensus where the miners are equipped with quantum processors and they need to prove they invested quantum computational effort into solving puzzles. But what kind of problems should be chosen? Natural candidates are problems in {\bf NP} which are additionally efficient for quantum computers to solve but not for classical computers. If the difficulty of the problem could be tuned by just adding a few more qubits, then the energy cost would grow only weakly with network demand, and the economics would favour investing in higher quality quantum components over buying more ASICs. Such problems exist; factoring is one, but as far as we know they would require full scale quantum computers to solve and would be expensive to access, thus limiting adoption in a practical setting. 
\\
An alternative is to use a problem not in {\bf NP}. This is what is done by the authors in Ref.~\cite{singh2023proof} where the proof-of-work is based on solving the boson sampling problem. This is a problem that is likely exponentially hard for classical computers to solve but that can be efficiently solved using a quantum platform involving a linear optical interferometer with single photon input states and single photon measurements (see Sec. XXXIII). Such devices are not universal for quantum computing but are much cheaper and easier to build. Because the problem isn't in ${\bf NP}$ a direct validation of miners computational effort would be exponentially hard. Instead the consensus protocol needs to be altered so that nodes commit sample sets to the network first and then afterward an efficiently computed coarse grained distribution can be checked against the committed sample sets to determine which miners correctly solved the sampling problem. 
\\
Another, even more radical use of quantum technologies for distributed consensus could do away with blockchains altogether. This would be enabled using what are called one-shot signatures. The scenario here is Alice wishes to delegate her digital signing rights to another player Bob, one time and one time only. Classically it isn't possible since Bob could simply copy the secret-key she provided him and  sign multiple messages at once. Strikingly, the authors in Ref.~\cite{bib:RMAZ20} showed it is possible when Bob is equipped with a quantum computer. The essence of the protocol is Bob uses classical information from Alice to construct a quantum circuit which performs a quantum computation known as a coherent quantum Oracle call, where the Oracle performs hashing. His input state is a uniform superposition of all strings $\ket{x}$ on the input register which then becomes entangled with the output register in state $\ket{h(x)}$. He then measures the output register, say without outcome $y$, which collapses the input register to a superposition: $\ket{sk(y)}\propto\sum_{z|h(z)=y}\ket{z}$. This state constitutes the ``quantum secret key". Due to the no-cloning theorem, Bob cannot copy this state, and because he can't solve the inverse hashing problem, he also wouldn't have a way to prepare it again efficiently. However, given $\ket{sk(y)}$ he can perform a single Grover Oracle call to target all strings which share the first bit $m_1$ of a message $m$. Upon measuring the input register, he uses the output $z$  as the signature for $m_1$. The remaining bits of the message can be signed in a similar manner. Suprisingly, while Alice and Bob do need some rounds of classical communication, together with a verifier, no quantum communication is needed whatsoever.

%
% The End Of Classical Cryptography?
%

\section{The end of classical cryptography?} \label{sec:end_of_class_crypto}

\dropcap{T}{he} vulnerability of RSA to attacks by quantum computers raises the question whether this spells the end of classical cryptography and compromises the security of much of the present-day internet.

Thankfully, there are two saving graces. First of all, much research is being carried out into \textit{post-quantum classical cryptography}\index{Post-quantum classical cryptography}. That is, public-key cryptosystems based upon trapdoor functions\index{One-way functions} that reside outside of \textbf{BQP} and are therefore not efficiently attacked by quantum computers. One such line of research is to construct cryptosystems based upon \textbf{NP}-complete problems, such as the McEliece protocol\index{McEliece protocol} \cite{mceliece1978public}. Recall from Fig.~\ref{fig:complexity_classes} that \textbf{NP}-complete is strongly believed to reside completely outside of \textbf{BQP}. However, while many computer scientists might be comfortable with such a level of security, it is nonetheless based on the unproven conjecture that \textbf{NP}-complete and \textbf{BQP} do not intersect, i.e \mbox{$\mathbf{NP}\nsubseteq\mathbf{BQP}$}. What would be much more satisfying would be protocols demonstrating information-theoretic security\index{Information-theoretic!Security} rather than computational security\index{Computational!Security}. Here, quantum mechanics can help us -- \textit{quantum cryptography}.

%
% Quantum Cryptography
%

\section{Quantum cryptography}\index{Quantum cryptography}

\dropcap{A}{s} quantum physics can compromise some important aspects of classical cryptography, can it perhaps be similarly exploited to make new cryptosystems that are immune even to quantum adversaries? Thankfully the answer is yes\ldots at least some of the time.

%
% Quantum Key Distribution (QKD)
%

\subsection{Quantum key distribution} \label{sec:QKD} \index{Quantum key distribution (QKD)}

Aside from quantum computing, a central use for quantum technologies is in cryptography \cite{bib:Gisin02}. The demand for secure cryptography is now extremely important in the context of electronic commerce and general security of information transmission in the internet age. Electronic currencies such as Bitcoin\index{Bitcoin} depend on cryptographic protocols in order to secure the value of assets, assign ownership certificates\index{Ownership certificates}, and secure the currency against fraud. However, such protocols are based upon the computational complexity of certain mathematical problems (i.e computational security\index{Computational!Security}), and are not fundamentally secure in the presence of limitless computational resources, or quantum computers. Therefore, using quantum mechanical protocols based on physical principles (i.e information-theoretic security\index{Information-theoretic!Security}) rather than computational limitations, are favourable for future-proofing ourselves.

Quantum key distribution (QKD) protocols facilitate shared, secret randomness, where any intercept-resend\index{Intercept-resend attacks} (or man-in-the-middle) attack may be detected and rejected, guaranteed by the laws of quantum physics (specifically the Heisenberg uncertainty principle\index{Heisenberg!Uncertainty principle} and no-cloning theorem\index{No-cloning theorem}). This shared, secret randomness may subsequently be employed in a one-time pad cipher\index{One-time pad}, presenting us with true information-theoretic security\index{Information-theoretic!Security}.

The central notion to QKD protocols, in their numerous manifestations, is that measurement of quantum states invokes a wave-function collapse. When measuring a state in a basis for which that state is not an eigenstate, this necessarily changes the state. QKD relies on this simple result from quantum mechanics to reveal any eavesdropper performing an intercept-resend attack\index{Intercept-resend attacks} via the changes to transmitted quantum states that this would induce.

QKD is a relatively mature technology with already several commercial systems being available off-the-shelf\footnote{Examples of companies selling off-the-shelf QKD hardware include \href{http://www.magiqtech.com}{MagiQ} and \href{http://www.idquantique.com}{ID Quantique}.} and initial space-based implementations have been successfully demonstrated 
\cite{liao2017satellite}.

It's easy to see the utility of quantum networks in enabling commodity deployment of QKD -- users desire to communicate photons across long-range ad hoc networks, with low loss and dephasing. A global quantum internet would allow quantum cryptography to truly supersede classical cryptography, bypassing the vulnerabilities faced by classical cryptography in the era of quantum computing.

%
% BB84 Protocol
%

\subsubsection{BB84 protocol}\index{BB84 protocol}

The first described QKD scheme was the \textit{BB84} \cite{bib:BennetBrassard84}\index{BB84 protocol} protocol, which exploits the fact that states encoded in the $\hat{Z}$-basis but measured in the $\hat{X}$-basis (and vice versa) collapse randomly, yielding completely random measurement outcomes, whereas states measured in the same basis in which they were encoded always correctly communicate a single bit of information.

Implemented photonically, BB84 requires only the transmission of a sequence of single photons, polarisation-encoded\index{Polarisation!Encoding} with random data.

\begin{figure}[!htbp]
\includegraphics[clip=true, width=0.475\textwidth]{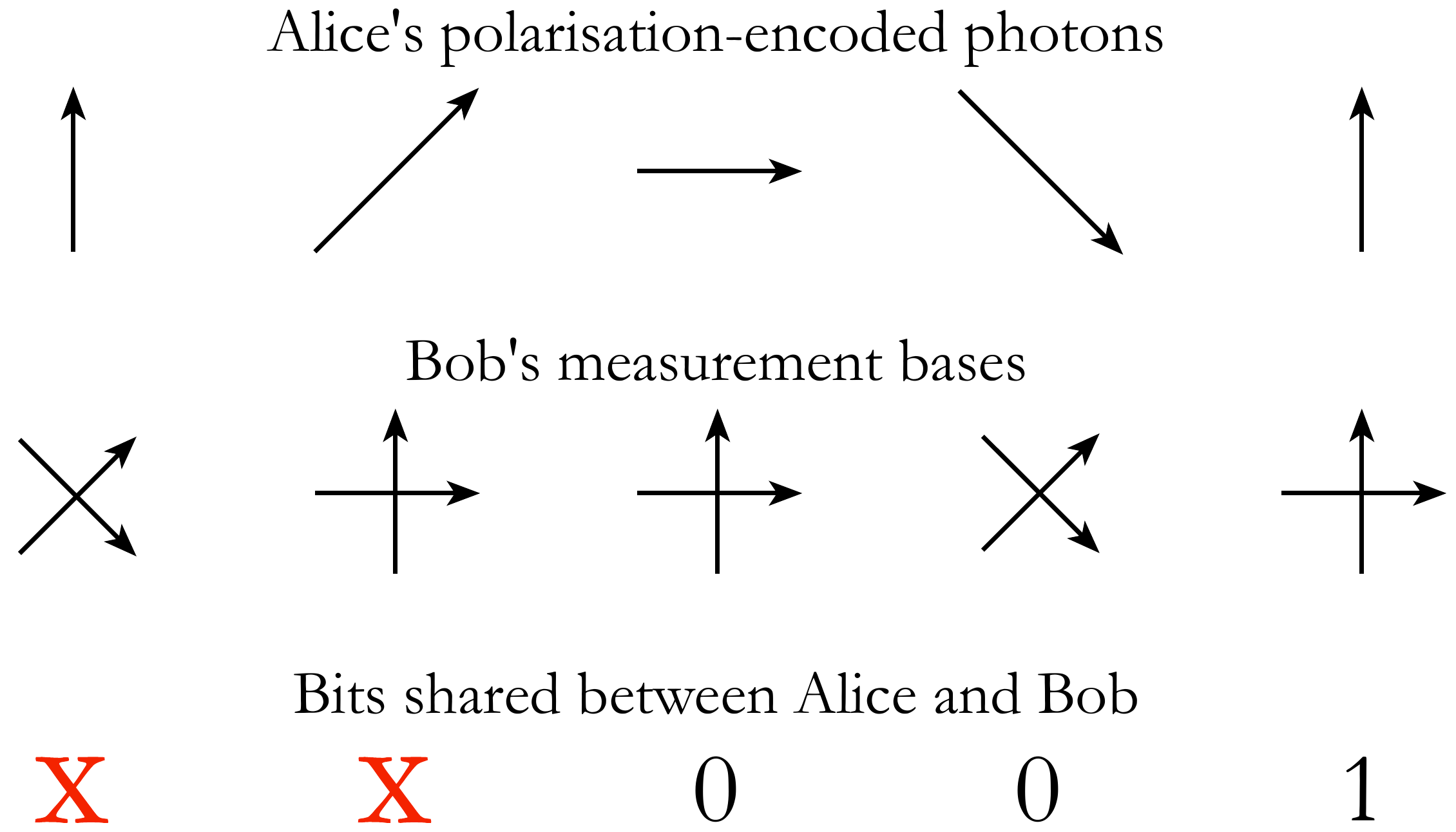}
\captionspacefig \caption{Example execution of the BB84 protocol for securely sharing random bit-strings between Alice and Bob as per Alg.~\ref{alg:bb84}. At the conclusion of the protocol, some bits are discarded (red `X'), with those remaining guaranteed to be secret between the two parties.} \label{fig:BB84_example}	
\end{figure}

The BB84 protocol is described in detail in Alg.~\ref{alg:bb84} in the context of polarisation-encoded photons, which is the most natural (but not only) setting for this protocol. An example evolution of the protocol is illustrated in Fig.~\ref{fig:BB84_example}.

\begin{table}[!htbp]
\begin{mdframed}[innertopmargin=3pt, innerbottommargin=3pt, nobreak]
\texttt{
function BB84():
\begin{enumerate}
\item Alice chooses a random bit, $0$ or $1$.
\item Alice randomly chooses a basis, $\hat{X}$ or $\hat{Z}$.
\item Depending on the choice of basis, she encodes her bit into the polarisation of a single photon as:
\begin{align}
\ket{0}_Z &\equiv \ket{H}, \nonumber \\
\ket{1}_Z &\equiv \ket{V},
\end{align}
or,
\begin{align}
\ket{0}_X &\equiv \frac{1}{\sqrt{2}}(\ket{H}+\ket{V}), \nonumber \\
\ket{1}_X &\equiv \frac{1}{\sqrt{2}}(\ket{H}-\ket{V}).
\end{align}
\item Encoding into the randomly chosen basis, she transmits the randomly chosen bit to Bob.
\item She does not announce the choice of bit or basis.
\item Bob measures the bit in a randomly chosen basis, $\hat{X}$ or $\hat{Z}$.
\item The above is repeated many times.
\item Upon receipt of all qubits, Alice (publicly) announces the basis used for encoding each bit sent.
\item Qubits where Bob measured in the opposite basis to which Alice encoded are discarded, as they will be decorrelated from Alice.
\item The remaining measurement outcomes are guaranteed to yield identical bits between Alice and Bob.
\item Remaining is roughly half as many bits as were sent, which are random, but guaranteed to be identical between Alice and Bob.
\item Alice and Bob sacrifice some of their bits by publicly communicating them to check for consistency. This rules out intercept-resend attacks.
\item Privacy amplification may be used to distill the partially compromised key into a shorter but more secret one.
\item $\Box$
\end{enumerate}}
\end{mdframed}
\captionspacealg \caption{BB84\index{BB84 protocol} QKD protocol using polarisation-encoded\index{Polarisation!Encoding} photons. Upon completion of the protocol, Alice and Bob share a random bit-string for use in a one-time pad cipher\index{One-time pad}, yielding perfect information-theoretic security.}\label{alg:bb84}
\end{table}

To understand the secrecy of the protocol as described in Alg.~\ref{alg:bb84}, suppose an eavesdropper, Eve, were to perform an intercept-resend\index{Intercept-resend attacks} attack on the channel between Alice and Bob. At that stage in the protocol Alice had not yet announced her choice of encoding bases, and Eve will not know the bases in which to measure states without randomly collapsing them onto values inconsistent with Alice's encoding. Thus, by sacrificing some of their shared bits, via openly communicating them to one another for comparison, such an attack will be detected with asymptotically high probability. Now Alice and Bob have great confidence that they have a shared, secret, random bit-string, which may subsequently be employed in a one-time pad\index{One-time pad} with perfect secrecy.

The BB84 protocol has no measurement timing, mode-matching or interferometric stability requirements, making it a very robust protocol, readily achievable with present-day photonics technology. The scheme has been adapted to physical architectures beyond just polarisation-encoded photons, such as CV encodings\index{Continuous-variables!Quantum key distribution (QKD)} (see Sec.~\ref{sec:CV_QKD}).

%
% Privacy amplification
%

\subsubsection{Privacy amplification}\index{Privacy amplification}

When Alice and Bob sacrifice and compare a randomly chosen subset of their key bits to detect eavesdroppers, they also need to accept the inescapable fact that their qubits propagated through imperfect channels and were subject to noise en route. This has the same effect as an eavesdropper -- it corrupts some of the bits -- and it's impossible to distinguish which took place, a noisy channel (which is ok) or an eavesdropper (which is not).

Because the channel was necessarily noisy, Alice and Bob \textit{must} tolerate some number of corrupted bits. But if the corruption came from Eve rather than the noisy channel they would effectively be tolerating her knowing some of the key. We don't want her to know \textit{any} of the key!

\textit{Privacy amplification} is a mathematical technique based on hashing algorithms for taking a shared key with a number of unknown compromised bits and distilling it to a shorter key of which Eve has almost zero knowledge.

Specifically, if we know that Eve has compromised $t$ of our $n$ shared random bits, privacy amplification allows us to distill a new key from the compromised one of approximately length \mbox{$n-t$} over which Eve knows almost nothing.

This is an information-theoretic security\index{Information-theoretic!Security} result, not a computational security\index{Computational!Security} one, thereby rescuing the perfect security of the BB84 QKD protocol.

%
% E91 Protocol
%

\subsubsection{E91 protocol}\index{E91 protocol}

E91 is slightly different to BB84. Here Alice and Bob share an entangled Bell pair provided by a central authority. Then both Alice \textit{and} Bob measure their qubits in random bases. As with BB84, after measuring all qubits, they compare their choices of random bases. When they coincide, they have a shared, random bit. When they don't, they discard their result. From here the remainder of the protocol is the same as for BB84. The protocol is summarised in Alg.~\ref{alg:e91}.

%\comment{Fix this up. Discuss using Bell violation to prove security.}

\begin{table}[!htbp]
\begin{mdframed}[innertopmargin=3pt, innerbottommargin=3pt, nobreak]
\texttt{
function E91($\ket{\Phi^+}$):
\begin{enumerate}
\item A central server shares a Bell pair between Alice and Bob,
\begin{align}
\ket{\Phi^+} = \frac{1}{\sqrt{2}}(\ket{0}_A\ket{0}_B+\ket{1}_A\ket{1}_B).
\end{align}
\item Alice randomly measures her qubit in either the $\hat{X}$ or $\hat{Z}$ basis.
\item Bob randomly measures his qubit in either the $\hat{X}$ or $\hat{Z}$ basis.
\item Alice and Bob share what their measurement bases were (classically and unencrypted).
\item When Alice and Bob's bases were consistent they store the measurement outcomes as a shared random bit.
\item Alice and Bob sacrifice some of their bits by publicly communicating them to check for consistency. This rules out intercept-resend attacks.
\item Privacy amplification may be used to distill the partially compromised key into a shorter but more secret one.
\item $\Box$
\end{enumerate}}
\end{mdframed}
\captionspacealg \caption{E91\index{E91 protocol} QKD protocol using polarisation-encoded\index{Polarisation!Encoding} photonic Bell pairs. Upon completion of the protocol, Alice and Bob share a random bit-string for use in a one-time pad cipher\index{One-time pad}.}\label{alg:e91}
\end{table}

Like BB84, E91 has no mode-matching\index{Mode-matching} or interferometric stability\index{Interferometric!Stability} requirements, and Alice and Bob both only require single-photon detection. Unlike BB84, however, E91 requires a central authority that is able to prepare entanglement on-demand as a resource.

An advantage of E91 over BB84 is that it does not require a direct quantum communications link between Alice and Bob. The protocol could be mediated from above by a Bell pair-producing satellite within line-of-sight of both Alice and Bob.

%
% Continuous-Variable Protocols
%

\subsubsection{Continuous-variable protocols}\index{Continuous-variables!Quantum key distribution (QKD)}\label{sec:CV_QKD}

\sectionby{Zixin Huang}\index{Zixin Huang}

Like quantum computing, QKD protocols may be adapted to the CV domain also. Alg.~\ref{alg:cv_qkd} describes a simple such scheme based on encoding in phase-space\index{Phase!Space}, where the basis states are coherent states of different amplitudes and phases. The goal is the same as BB84 -- to securely share a random bit-string for use in a one-time-pad\index{One-time pad}.

Note that this protocol provides information-theoretic security\index{Information-theoretic!Security}, as per BB84\index{BB84 protocol}, despite the fact that coherent states are non-orthogonal, forming an over-complete basis in phase-space\index{Phase!Space}.

Conceptually, the operation of the CV QKD protocol is virtually identical to photonic BB84, differing only in that now the different choices of encodings correspond to phase-space transformations. Like BB84, if Eve were to perform an intercept-resend attack\index{Intercept-resend attacks} she would probabilistically re-encode in the wrong quadrature\index{Quadratures}, thereby revealing herself to Alice and Bob, who could then terminate and start over. 

\begin{table}[!htbp]
\begin{mdframed}[innertopmargin=3pt, innerbottommargin=3pt, nobreak]
\texttt{
function CV\_QKD():
\begin{enumerate}
\item Alice chooses two Gaussian-distributed random numbers with mean zero,
\begin{align}
x_A &= \mathcal{N}(0,V_\mathrm{mod}),\nonumber \\
p_A &= \mathcal{N}(0,V_\mathrm{mod}),
\end{align}
where $V_\mathrm{mod}$ is the modulation variance\index{Modulation variance}.
\item Alice prepares the coherent state,
\begin{align}
\ket\alpha = \ket{x_A+ip_A}.	
\end{align}
\item Alice transmits $\ket\alpha$ to Bob.
\item Bob randomly measures either $\hat{x}$ or $\hat{p}$ using homodyne detection\index{Homodyne detection}.
\item Alice and Bob use classical communication to determine for which transmissions their preparation and measurement were consistent.
\item The remainder of the protocol proceeds as per BB84.
\item $\Box$
\end{enumerate}}
\end{mdframed}
\captionspacealg \caption{CV QKD protocol using coherent states, encoded in the quadrature basis.}\label{alg:cv_qkd}
\end{table}

%
% Security
%

\subsubsection{Security}\index{Security!of QKD}

Importantly, unlike classical cryptographic protocols, QKD makes no assumptions about the computational complexity of inverting encoding algorithms or trapdoor functions\index{One-way functions}. The protocols are information-theoretically secure\index{Information-theoretic!Security}, and therefore no physically realisable computer, even a quantum computer, can compromise them. Thus, usual cryptanalytic techniques, like linear and differential cryptanalysis \cite{bib:Schneier96}\index{Linear cryptanalysis}\index{Differential cryptanalysis}, or the ability to factor large numbers\index{Integer factorisation}, that are employed to attack other encryption protocols, do not compromise QKD.

However, this is not to say that QKD is actually perfectly secure in real-life. Recent history has demonstrated that this is certainly not the case, with many attacks against various quantum cryptographic protocols being described and successfully demonstrated. The reason for this schism between theory and experiment is that no experiment ever \textit{perfectly} mimics the theoretical proposal it is trying to implement. Laboratory components might be imprecise in an unfortunate way, opening up avenues for attack, or they might perform unwanted additional actions that leak information to Eve. The prospects for such so-called `side-channel attacks'\index{Side-channel attacks} must be carefully considered and satisfactorily addressed.

The best known attack against photonically implemented BB84 is the `photon-number splitting attack'\index{Photon-number-splitting attacks}. This attack targets implementations where Alice's photon source does not produce perfect single-photon states, but may have some amplitude of higher photon-number. Weak coherent states or SPDC states exhibit this property. The attack is very simple. Eve simply performs a man-in-the-middle attack\index{Intercept-resend attacks}, but not of an intercept-resend\index{Intercept-resend attacks} variety. Rather than intercepting the entire channel, she inserts a low reflectivity beamsplitter and measures only the reflected mode, the other following its desired trajectory to Bob. Now there is a chance that Eve can extract just one of the multiple photons in the signal, such that Bob still receives a photon. Eve holds the split-off signal in memory until the classical communication of encoding bases, at which point she measures all her split signals in the correct basis, thereby recovering the associated secret-key bit.

This trivial attack vector clearly demonstrates the importance of well-considered engineering decisions when physically implementing QKD. No piece of hardware is ever 100\% to specification!

\subsubsection{Public-key cryptography}

The BB84 protocol is used exclusively for private-key cryptography. For many applications (notably digital signatures and easy key exchange with unidentified parties), public-key cryptosystems would be highly desirable.

Are there any viable public-key quantum protocols that could fill the vacancy of the soon-to-be-compromised RSA? Unfortunately the answer is `not yet'. As appealing as it would be, and despite many highly intelligent people putting their minds to it, to-date no one has presented a viable public-key quantum cryptosystem.

This is problematic since when quantum computing becomes a reality it will immediately compromise the classical public-key cryptosystems we all rely on on a daily basis, and it would be highly desirable for a quantum replacement to be available to fill its shoes.

%
% Quantum Enigma machines
%

\subsection{Quantum Enigma machines}\index{Quantum Enigma machines}

While the BB84\index{BB84 protocol} and other related QKD\index{Quantum key distribution (QKD)} protocols are perfectly secure, they suffer the major drawback that because they are based upon the one-time pad\index{One-time pad}, the number of successfully communicated key-bits must equal the message length and cannot be reused (not even once). This means that for real-time or high-bandwidth applications, the quantum communications channel must have similarly high bandwidth to be applicable.

What would be more useful would be a quantum equivalent of private-key cryptography\index{Private-key!Cryptography}, whereby a short key can be
used to lock a much longer message.
 % reused over and over again for different messages, meaning that only a short, shared secret-key would need to be established once-off (or at least very infrequently).

\begin{figure}[!htbp]
\includegraphics[clip=true, width=0.475\textwidth]{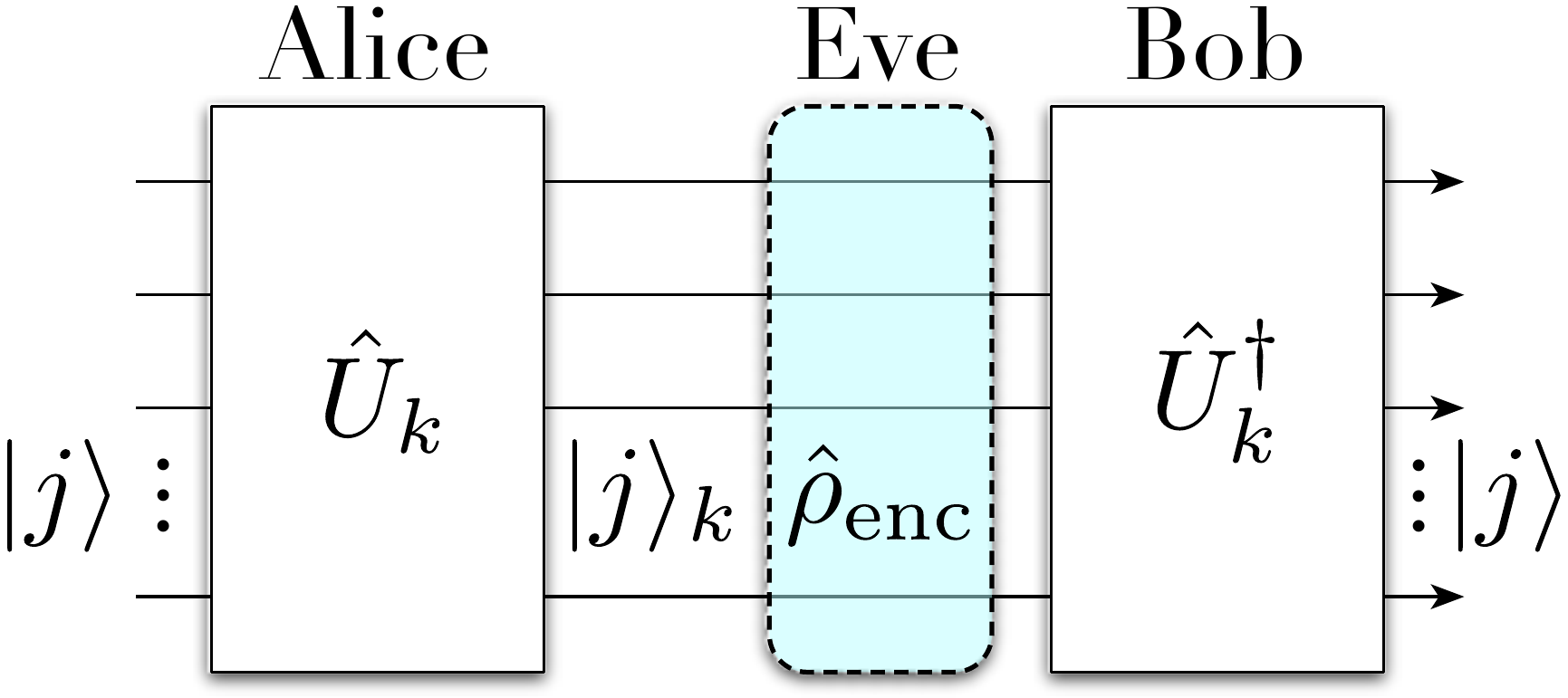}
\captionspacefig \caption{Schematic of the quantum Enigma machine protocol for quantum private-key cryptography. $\ket{j}$ is the message, $k$ is the key, and $\hat{U}_k$ is the encryption operation associated with key $k$, chosen randomly from the Haar measure\index{Haar measure}. The encrypted message is $\ket{j}_k$. An eavesdropper intercepting the channel without knowledge of the key observes the state $\hat\rho_\mathrm{enc}$.}\index{Quantum Enigma machines}\label{fig:enigma}    
\end{figure}

This led to the proposal for \textit{quantum Enigma machines}\footnote{The Enigma machine\index{Enigma machines} was the classical cryptosystem employed by the Germans during World War 2 for military and intelligence communication.} \cite{bib:LloydEnigma}, based on the phenomenon of quantum data locking \cite{DiVin}. The protocol is shown in Fig.~\ref{fig:enigma}.

Quantum data locking is a uniquely quantum phenomenon, and presents one of the most striking differences between classical and quantum information theory \cite{bib:guha2014enigma}.
Quantum data locking occurs when the accessible information about a classical message encoded into a quantum state decreases by a much larger amount when ``locked" with a much smaller key.

Here Alice and Bob share a short secret-key of length $K$, and wish to communicate a message $M$ of length $n$-bits. The secret-key is assumed to be unconditionally secure, and can be established via conventional QKD. First off, they agree upon a set of $K$ Haar-random unitaries (but it is sufficient to consider 2-designs) \index{Haar measure}, $\{\hat{U}_k\}$, associating one with each of the possible $K$ keys. The set of $\{\hat{U}_k\}$ unitaries need not be a secret. 

Alice then encodes her message into the state $\ket{j}$. Her encryption operation is to apply $\hat{U}_k$ to this state according to Alice and Bob's shared secret-key. 
\begin{align}
\ket{j}_k=\hat{U}_k\ket{j},
\end{align}
which Bob is easily able to decrypt using the inverse unitary,
\begin{align}
\hat{U}_k^\dag\ket{j}_k=\ket{j}.
\end{align}

To characterise the security of the scheme, we first note that in the absence of knowing the key or message, and assuming all $j$ and $k$ are equally likely, Eve observes the mixed state,
\begin{align}\label{eq:ghwoirsfjnk}
 \hat \rho_\text{enc}= \frac{1}{2^n} \sum_{j=1}^{2^n} \ket{j}_A\bra{j} \otimes \frac{1}{K}\sum_{k=1}^K U_k\ket{\psi_j}_E\bra{\psi_j}U_k^\dagger   .
\end{align}

The security can now be quantified in terms of the accessible information
\footnote{The accessible information is the maximum of the mutual information between Alice's input states and measurements performed on the encoded state by Eve.}\index{Accessible information} between this state and the plaintext state, which can be upper-bounded as \cite{lupo2015NJP,lupo2014pra},
\begin{align}
    I_c \leq n + \frac{1}{2^n}\max_{\ket\phi}\sum_{j,k} |\braket{ \phi|j}_k|^2 \log |\braket{ \phi|j}_k|^2.
\end{align}
It was shown that this quantity can be made arbitrarily small with key-size scaling as,
\begin{align}
m=O(\epsilon\log n),
\end{align}
\noindent for an $\epsilon$ that can be made arbitrarily small, given $\log K \sim \log{(1/\epsilon)}$, therefore the key size is exponentially smaller than the length of the message. Thus, we have an encryption scheme that requires very frugal requirements in key-size versus message length. It is further showed that such a scheme can be made robust against noise and loss.

Note that the security of the scheme is information-theoretic\index{Information-theoretic!Security}, and does not make any assumptions about computational security\index{Computational!Security} (except we require the assumption that Eve has finite-time quantum memory \cite{lupo2015Entropy} in order to preserve composable security required for QKD \cite{muller2009composability}). Thus, this represents a strong form of quantum private-key cryptography, requiring only very short keys compared to the message length.

Since then, quantum data locking has found its application for Boson Sampling \cite{huang2019boson}. 
However, the scheme is very challenging to implement on a large scale over long distances. Consider an optical implementation, where the message is photonically encoded. Now the scheme represents a complex, multi-mode, generalised Mach-Zehnder interferometer\index{Mach-Zehnder (MZ) interference}, meaning that the channel between Alice and Bob, who might be far apart, must be interferometrically stabilised\index{Interferometric!Stability} (Sec.~\ref{sec:opt_stab}) on the order of the photons' wavelength (hundreds of nanometers for optical frequencies), which is extremely challenging over long distances.

%
% Hybrid Quantum/Classical Cryptography
%

\subsection{Hybrid quantum/classical cryptography}\index{Hybrid!Quantum/classical cryptography}

As discussed, the RSA public-key cryptosystem is vulnerable to an efficient quantum attack, whereas private-key schemes like AES\index{Advanced Encryption Standard (AES)} are not (believed to be). Thus, combining QKD schemes with private-key classical schemes does not compromise security in the quantum era.

Why would we combine quantum and classical encryption techniques when QKD is already provably secure, whereas the classical schemes are not?

In the near future, as QKD schemes begin their rollout in space and on Earth, random bits from the QKD implementation will be very expensive and exhibit low bandwidth. Suppose we wanted to securely videoconference across the globe. For just a single user this would require megabits per second of shared random bits, which will quickly saturate the capacity of overhead quantum satellites.

Instead, let us use the QKD system to securely share just a 256-bit private session key\index{Session key} between two users. This is subsequently employed for AES256\index{Advanced Encryption Standard (AES)} encryption that operates entirely over the classical network, which we regard as extremely cheap and high-bandwidth. Importantly, unlike one-time pad implementations, this session key may be reused. Now we have a hybrid system which is not quantum-compromised, but which overcomes the cost and bandwidth issues associated with emerging QKD networks.

While such a hybrid scheme is not information-theoretically secure (AES is not proven to be quantum-safe), the computational security assumptions are far stronger than for say RSA, since there are no known efficient quantum attacks against strong private-key schemes.

%
% Quantum Anonymous Broadcasting
%

\subsection{Quantum anonymous broadcasting} \label{sec:anon_broad} \index{Quantum anonymous broadcasting}

The previously described protocols all focussed on preserving the secrecy of messages. Alternately, it may not be the message that is sensitive, but rather the identity of the person who says it. \textit{Anonymous broadcasting}\index{Quantum anonymous broadcasting} is a protocol for achieving this.

Consider the following scenario. A group of users share a classical broadcast channel that anyone is able to transmit to, and everybody is able to listen to unencrypted. But it is of importance that the identity of whoever broadcasts to the channel must be kept secret from all users. \cite{christandl2005quantum} described a scheme for achieving this quantum mechanically using shared GHZ states -- \textit{quantum anonymous broadcasting} (QAB).

Let there be a (trusted\footnote{Note that if the server is not trusted, he could easily conspire to reveal people's identities by distributing $\ket{+}^{\otimes n}$ states instead of GHZ states.}) server that distributes GHZ states (of arbitrary numbers of qubits) to a group of users, one qubit per user. This can be prepared as described in, for example, Sec.~\ref{sec:GHZ_states}. Now if every user measures in the \mbox{$\ket\pm=\frac{1}{\sqrt{2}}(\ket{0}+\ket{1})$} basis the joint \textit{parity} (i.e whether an even or odd number of $+$'s were measured) is guaranteed to be even. For example, all users might measure $\ket{+}\bra{+}$, or exactly 2, but never exactly 1 or 3.

On the other hand, if a $\hat{Z}$ gate were applied to any one qubit, this would flip the parity outcome. Note that a GHZ transforms according to,
\begin{align}
	\hat{Z}_i \frac{1}{\sqrt{2}}(\ket{0}^{\otimes n} + \ket{1}^{\otimes n}) \to \frac{1}{\sqrt{2}}(\ket{0}^{\otimes n} - \ket{1}^{\otimes n})\,\,\forall \, i,
\end{align}
for any qubit $i$. This invariance in the location of the $\hat{Z}$ gate is the basis for the anonymity of the protocol. If a user wishes to broadcast `0' he does nothing, whereas if he wishes to broadcast `1' he applies a $\hat{Z}$ gate to his local qubit.

Finally, all users measure their qubits in the $\pm$-basis and publicly (without encryption) broadcast their measurement outcomes. All users now see all other users' measurement outcomes and are able to calculate the collective parity of the measurements. Now if the parity is even, the speaker must have said `0', whereas if it is odd he must have said `1'. The protocol is shown in Fig.~\ref{fig:QAB} and described in detail in Alg.~\ref{alg:QAB}.

\if 1\doublecol
\begin{figure}[!htbp]
\includegraphics[clip=true, width=0.475\textwidth]{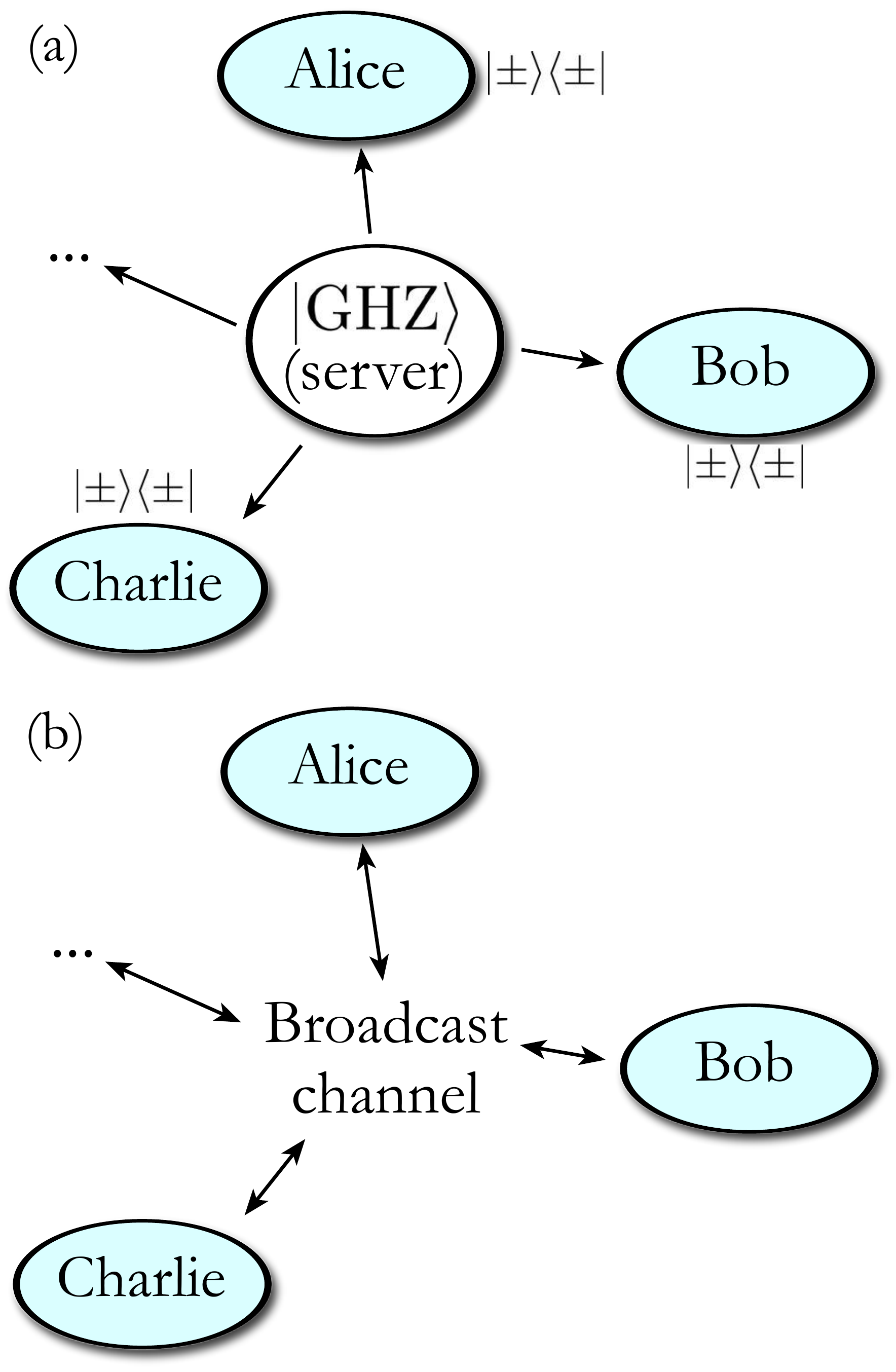}
\captionspacefig \caption{Protocol for quantum anonymous broadcasting. (a) A central trusted server prepares GHZ states and distributes them amongst a group of users, one qubit per user. All users measure in the $\pm$-basis. (b) All users classically broadcast their measurement outcomes yielding shared random parity. During broadcast, the broadcaster lies about his measurement outcome to flip the joint parity if he wishes to transmit `1', or tells the truth to transmit `0'. The joint parity encodes the message of the anonymous user, which all listeners are able to recover. Importantly, only one user may broadcast at a time, otherwise the recovered message will be given by the XOR of all the simultaneously broadcast messages.} \label{fig:QAB}
\end{figure}
\else
\begin{figure*}[!htbp]
\includegraphics[clip=true, width=0.8\textwidth]{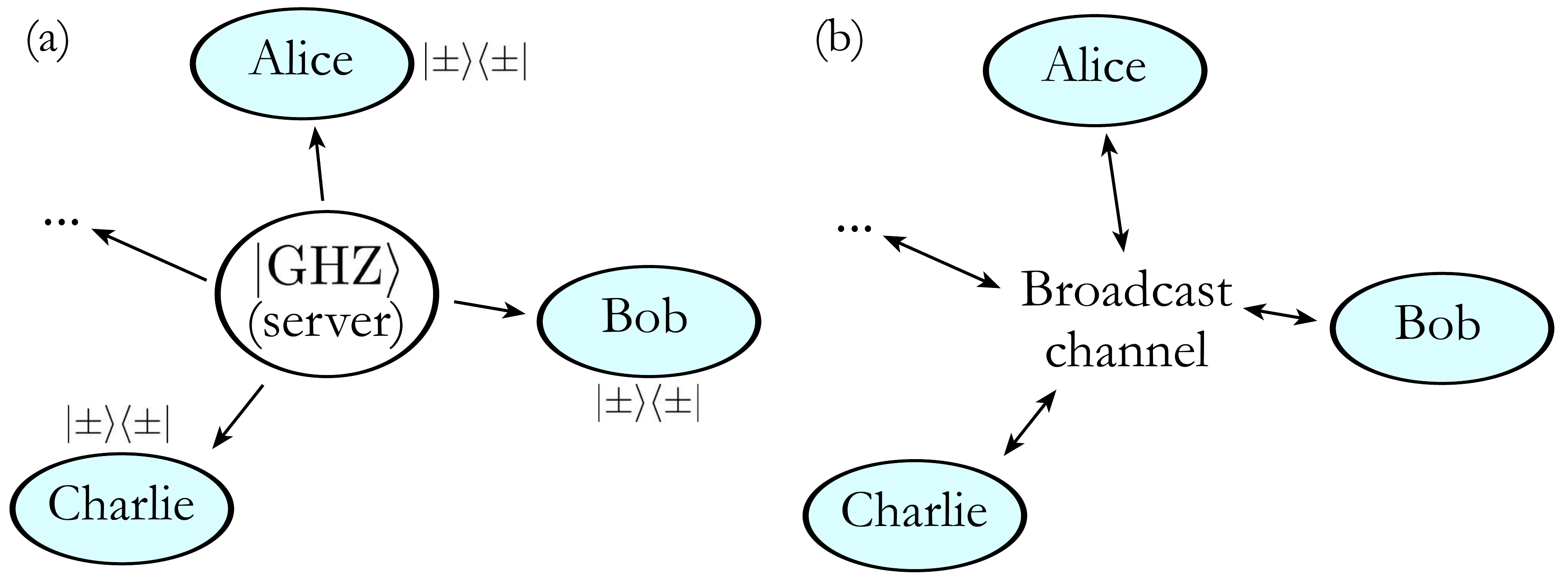}
\captionspacefig \caption{Protocol for quantum anonymous broadcasting. (a) A central trusted server prepares GHZ states and distributes them amongst a group of users, one qubit per user. All users measure in the $\pm$-basis. (b) All users classically broadcast their measurement outcomes yielding shared random parity. During broadcast, the broadcaster lies about his measurement outcome to flip the joint parity if he wishes to transmit `1', or tells the truth to transmit `0'. The joint parity encodes the message of the anonymous user, which all listeners are able to recover. Importantly, only one user may broadcast at a time, otherwise the recovered message will be given by the XOR of all the simultaneously broadcast messages.} \label{fig:QAB}
\end{figure*}
\fi

\begin{table}[!htbp]
\begin{mdframed}[innertopmargin=3pt, innerbottommargin=3pt, nobreak]
\texttt{
function QuantumAnonymousBroadcasting(message, speaker):
\begin{enumerate}
\item $\ket\psi = \frac{1}{\sqrt{2}}(\ket{0}^{\otimes n}+\ket{1}^{\otimes n})$
\item $\ket\psi \to (\hat{Z}_\mathrm{speaker})^\mathrm{message}\ket\psi$
\item for(i$\in$users) \{
	\setlength{\itemindent}{.2in}
\item outcome$_i$ = measureInXBasis($\ket\psi_\mathrm{i}$)
\setlength{\itemindent}{0in}
\item \}
\item parity = $\sum_i$ outcome$_i$\,(mod\,2)
\item return(parity)
\item $\Box$
\end{enumerate}
}
\end{mdframed}
\captionspacealg \caption{Protocol for quantum anonymous broadcasting. The GHZ state is distributed in advance, one qubit per user. The measurement outcomes are classically broadcast without encryption. The final parity of the classical measurements reflects the message bit without identifying the speaker.} \label{alg:QAB}
\end{table}

Note that the scheme can be slightly simplified by rather than the speaker applying the $\hat{Z}$ to his qubit, upon announcing his measurement outcome he instead simply lies about his outcome and flips it. This follows simply because a $\hat{Z}$ gate prior to a $\pm$ measurement bit-flips the classical measurement outcome, \mbox{$\hat{Z}\ket\pm=\ket\mp$}.

There are no constraints on time-ordering of the measurements, nor, much like BB84, are there any interferometric stability requirements (not including the GHZ preparation stage), making this protocol very experimentally practical and robust over long distances.

Because of the time invariance in the measurements, distribution and measurement of the GHZ states can be performed well in advance of the actual message broadcast. This allows us to treat `shared parity'\index{Shared parity} as a fundamental resource (Sec.~\ref{sec:ent_ultimate}) for the QAB cryptoprotocol.

Since the parity-sharing can be isolated from the broadcasting stage it is unimportant if the GHZ source is non-deterministic or the channels for distributing it lossy. We can instead simply repeat GHZ distribution over and over at high repetition rate, post-selecting upon measurement outcomes where all users signal that they successfully received and measured their photons.

For these reasons, this scheme lends itself readily to photonic implementation, provided a reliable GHZ preparation circuit. The scheme has since been ported to operate on distributed toric codes\index{Toric code} to facilitate error correction of the distributed GHZ states \cite{bib:MenicucciExpQAB}. The anonymous protocol has been generalised classically \cite{broadbent2007information} and experimentally implemented by a fully connected network of Bell pairs \cite{huang2022experimental}.

%
% Quantum Voting
%

\subsection{Quantum voting}\index{Quantum voting}

The security of voting systems, and the anonymity of their voters are pressing issues in the modern free-world, and there have been countless high-profile instances of voting systems being compromised nefariously.

Based on similar ideas to quantum anonymous broadcasting is quantum voting, whereby a group of parties can anonymously vote such that no party, including the tallyman, is able to learn any individual voter's vote, but at the conclusion all are able to see the collective outcome of the vote.

There are a multitude of different models for voting, and a number of quantum implementations for them have been described. The two most well-known are:
\begin{itemize}
	\item Binary voting\index{Binary voting}: whereby each party votes `yes' or `no'. 
	\item Anonymous surveys\index{Anonymous surveys}: whereby each party votes a number and we wish to determine the sum of all the votes.
\end{itemize}

\begin{figure}[!htbp]
\includegraphics[clip=true, width=0.35\textwidth]{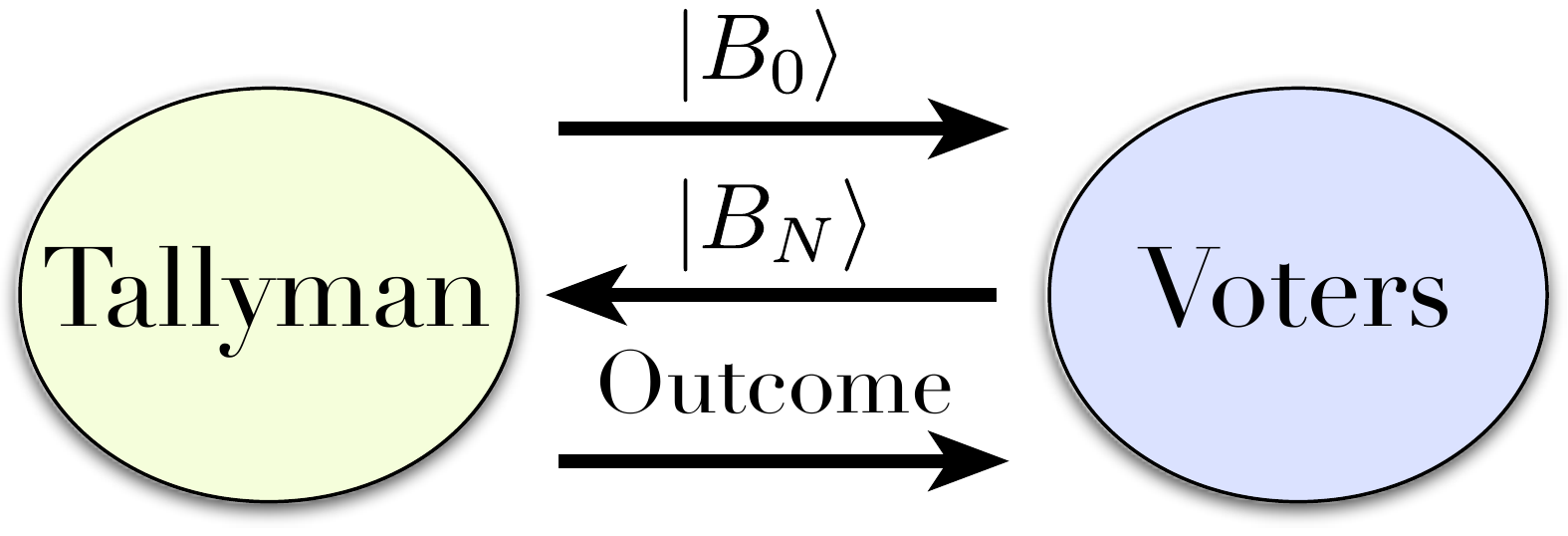}
\captionspacefig \caption{Protocol for quantum voting via anonymous surveying. The tallyman prepares the entangled state $\ket{B_0}$, half of which is shared with the voters. The voters each perform local phase-shift operations on their half of the state, as per Alg.~\ref{alg:quantum_voting}. The phase-transformed state is then returned to the tallyman, who is able to extract the phase and hence the cumulative vote.}\label{fig:quantum_voting}\index{Quantum voting}	
\end{figure}

Fig.~\ref{fig:quantum_voting} and Alg.~\ref{alg:quantum_voting} describe a quantum implementation for anonymous surveys, using the protocol described by \cite{bib:VaccaroVoting}.

Conceptually, the scheme is similar to quantum anonymous broadcasting in that it hides votes in phases within an entangled state, which are not accessible to individual parties, but are rather a global property of the state. The scheme relies on the preparation and distribution of a particular entangled state as a resource. Unfortunately this particular state is not one which is known how to be trivially prepared optically, and would therefore lend itself well to the outsourcing of preparation and distribution to a capable host via the quantum internet.

\begin{table}[!htbp]
\begin{mdframed}[innertopmargin=3pt, innerbottommargin=3pt, nobreak]
\texttt{
function QuantumVoting($\ket{B_0}$):
\begin{enumerate}
\item Prepare \textit{ballot state} for $N$ voters,
\begin{align}
\ket{B_0} = \frac{1}{\sqrt{N+1}}\sum_{n=0}^N \ket{N-n}_T\ket{n}_V,	
\end{align}
in the photon-number basis, where $T$ and $V$ represent the tallyman and the voters.
\item Each voter $j$ applies their vote operator,
\begin{align}
	\hat{v}_j = e^{i\pi\hat{n}_V \frac{\nu_j}{N+1}},
\end{align}
where $\hat{n}_V$ is the photon-number operator, and $\nu_j$ is the vote cast by $j$.
\item Following the $m$th vote,
\begin{align}
	\ket{B_m} = \frac{1}{\sqrt{N+1}}\sum_{n=0}^N e^{in\Delta_m} \ket{N-n}_T\ket{n}_V,
\end{align}
where,
\begin{align}
\Delta_m = \frac{2\pi}{N+1}\sum_{j=0}^m \nu_j,
\end{align}
is a scaled sum of the votes.
\item The state observed by the $T$ is,
\begin{align}
\mathrm{tr}_V(\ket{B_m}\bra{B_m}) = \frac{1}{N+1}\sum_{n=0}^N \ket{n}_T\bra{n}_T,	
\end{align}
which contains no voting information, similarly for $V$.
\item After all $N$ voters have voted, the voter state $V$ is transferred to the tallyman $T$.
\item Define the tally operator as,
\begin{align}
\hat{T} = \sum_{n=0}^N n\ket{T_n}\bra{T_n},	
\end{align}
where,
\begin{align}
\ket{T_n} = \frac{1}{\sqrt{N+1}}\sum_{j=0}^N e^{inj\frac{2\pi}{N+1}}\ket{N-j}\ket{j}.	
\end{align}
\item The tallyman finds the expectation value of the tally operator,
\begin{align}
r = \bra{B_N}\hat{T}\ket{B_N} = \sum_{j=0}^N \nu_j,
\end{align}
yielding the sum of all the votes.
\item return(r)
\item $\Box$
\end{enumerate}
}
\end{mdframed}
\captionspacealg \caption{Protocol for performing secure quantum anonymous surveying.} \label{alg:quantum_voting}\index{Quantum voting}\index{Ballot state}
\end{table}

%
% Quantum Secret Sharing
%

%\subsection{Quantum secret sharing}\index{Quantum secret sharing}

%\comment{To do}

%
% Attacks on QKD
%

\section{Attacks on quantum cryptography}\index{Attacks on quantum cryptography}\label{sec:attacks_QKD}

\sectionby{Zixin Huang}\index{Zixin Huang}

\dropcap{F}{or} a QKD system, information-theoretic security\index{Information-theoretic!Security} is achieved only when security against collective, coherent attacks is proven. We mustn't make assumptions about the limitations of our adversaries. For a more comprehensive review
on quantum cryptography, see Ref.~\cite{pirandola2019advances}.

Hacking attacks in this context exploit weaknesses in the physical implementation, rather than weakness of the theory -- so-called `side-channel attacks'\index{Side-channel attacks}. Some examples of weaknesses which allow zero-error attacks\index{Zero-error attacks} include:

\begin{itemize}
	\item Losses: Genuinely lossy channels or components are indistinguishable from ideal ones where some of the signal has been tapped off by Eve. Therefore, we must always assume the worst, that whatever is lost from our system is in the hands of Eve.
	\item Imperfect components: Our physical implementation might just not be operating strictly according to the theory.
	\item Correlations: Mutual information between signals in our system may leak information from the secure system to the environment, where Eve might be waiting patiently.
\end{itemize}

When considering the security of noisy channels, one must assume that all noise is due to manipulation by an eavesdropper -- the worst-case scenario. An attempted attack is considered successful if it can be proven that the eavesdropper can gain a non-negligible (i.e not exponentially small) amount of mutual information with the final secret-key established between Alice and Bob, without alerting them. Some of the best-known attacks follow:

\subsection{Hacking discrete-variable protocols}

\subsubsection{Beamsplitter \& photon-number-splitting attacks}\index{Beamsplitter attacks}\index{Photon-number-splitting attacks}

The security proofs for many discrete variable protocols assumes that Alice's signals consist of single photons. However, true single-photon soures are not yet widely available, and QKD systems often make use of strongly-attenuated laser pulses, and there is some probability of the source emitting multiple photons. This fact can be exploited 
by Eve, who employs the photon number splitting (PNS) attack \cite{bib:PhysRevLett.68.3121}: Eve can perform a non-demolition measurement to determine the number of photons 
in the signal, she steals the excess photons and sends the rest to bob. She stores these in her quantum memory and wait until the classical communication between Alice and Bob, hence finding out Alice's preparation basis. 

A beamsplitting (BS) attack translates the fact that any signal lost over a channel is acquired by Eve. Here, Eve induces losses in the communications channel by putting a beamsplitter outside Alice's device, then forwards the remaining photons to Bob. The BS attack does not modify the optical mode that Bob receives: it's therefore always possible for lossy channels, and does not introduce any errors.

 % The BS and PNS attacks were discovered as zero-error attacks against BB84 implemented with weak laser pulses.  

A method used to counter the PNS attack is the decoy-state\index{Decoy states} method \cite{bib:PhysRevLett.91.057901, bib:PhysRevLett.94.230504} . In the decoy-state protocol, Alice randomly replaces some of her signal states with multi-photon pulses from a decoy source. Eve cannot distinguish between decoy pulses from the encoding signals, and can only act identially on both. In the post-process stage, Alice public announces which states were the decoy pulses. The trusted parties can then characterise the action of the channel on the multiple-photon pulses and detect the presence of a PNS attack.

\subsubsection{Trojan horse and flash-back attacks}\index{Trojan horse attacks}

Another family of hacking that can be used against discrete-variable QKD is the Trojan horse attacks.
These attack involve Eve probing the settings of Alice and Bob by sending light into their devices and collecting the reflected signal. The first of this kind of attack actually came for free for the eavesdropper \cite{bib:RevModPhys.81.1301}: it was discovered that some photon-counters emit light when a photon is detected \cite{bib:kurtsiefer2001breakdown}. If the emitted light carries correlated information about which detector was triggered, it must be prevented from leaking outside the secure space and becoming accessible to Eve.

In general, Eve probes into the optical channel that Alice and Bob use to communicate. She send her own states into Alice's system, which will reflect off the same apparatus Alice uses used to encode her signal. Eve's states can be imprinted upon some information about the encoding used by Alice, when Eve measures them. She can then use the result of this measurement, combined with some operation on Alice's signals to make a best estimate of the quantum state that Alice sent to Bob, thus giving her some non-negligible mutual information with the key \cite{bib:PhysRevA.97.042335}.

\subsubsection{Detector attacks}\index{Detector!Attacks}

The faked-state attack\index{Faked-state attack} is based on the weak-laser implementation of BB84. Here, Eve manipulates Bob's detectors to force him to measure in the same basis. It exploits the fact that the detectors may have a dead-time\index{Dead-time}, and the eavesdropper can trigger the detector whenever she chooses. It follows that Bob's detection outcomes are controlled by the eavesdropper.

Eve can also go beyond detector blinding. She can send in a powerful laser pulse to optically damage components in the QKD system and permanently change its characteristics \cite{bib:jain2016attacks}. If the new characteristics then assist the eavesdropper in an attack without Alice or Bob being notified, the security of the QKD system would be severely compromised.

A more detailed discussion on attacks on physical implementation can be found in \cite{bib:jain2016attacks}.

\subsection{Attacks on continuous-variable protocols}

\subsubsection{Attacks on the local oscillator}

In order to measure Alice's signal states, Bob needs to carry out quadrature measurements, which
are defined with respect to Alice's local oscillator. Since it is difficult to maintain coherence 
between Alice's and Bob's local oscillators, often the Alice sends the local oscillators through the channel together with the signal states. This leaves open some side-channels which Eve can exploit. For example, she can reduce the error which the trusted parties would expect (e.g. of an intercept and resend attack), if she replaces the signal and local oscillators with squeezed states. Other attacks of the local oscillators include Eve exploiting the wavelength dependence of the beam splitters, shape of the local oscillators pulse, phase noises and the non-linearity photo-detectors.

\subsubsection{Saturation attacks on the detector}

The security proofs of CV-QKD assume a linear relationship between the quadratures of the state and the measurements, but in reality, homodyne detection have a finite range of linearity. The detectors can also be saturated. By causing Bob's measurement results to overlap with the saturation region, Eve can artificially reduce the trusted parties' error estimation.

Proposed countermeasures include using a Gaussian post-selection filter to ensure that the measurement results used for key generation fall within the linear region of the detector, and to use random attentuation of Bob's signal to monitor whether the measurements are linearly related to the inputs \cite{bib:qin2016pra}

\subsubsection{Trojan horse attacks}

Continuous variable protocols are also vulnerable to trojan horse attacks. By sending Trojan 
states into Alice's encoder, Eve can gain information on how Alice's states have been modulated. Active monitoring of incoming light is suggested as a countermeasure.

\subsection{Quantum digital signatures}\index{Quantum digital signature}\label{sec:q_digital_signatures}

\sectionby{Zixin Huang}\index{Zixin Huang}

% \comment{Zixin to do: Say something about one way functions and hardness, computational complexity etc}

As discussed previously, a classical digital signature has three main tasks. It ensures that the message
\begin{itemize}
\item was created by the claimed sender.
\item has not been altered.
\item is non-repudiable: the sender cannot deny having sent this message.
\end{itemize}
\noindent A digital signature is a vital tool for a huge range of modern applications, ranging from software distribution, financial transaction, emails, cryptocurrencies and voting etc \cite{pirandola2019advances}.

A quantum digital signature (QDS) scheme involves one sender and potentially many receivers. It consists of three stages, each consists of a corresponding algorithm.

\begin{enumerate}
\item Key generation: a private key is obtained by the sender, and public keys are delivered to the receivers.
\item Signature: the sender chooses a message $M$ and uses the private key to generate a signature, $\sigma = \sigma(M)$. She sends the pair $(M,\sigma(M))$ to the desired receiver(s)
\item Verification: a receiver receives the message and signature pair $(M,\sigma(M))$ and the public key, he/she checks whether to accept the message as having originated from the claimed sender.
\end{enumerate}

After the key generation phase, it is important that the actions of the involved parties are determined by this point. That is, they decide to accept or reject the signature without further classical or quantum communication.

Ensuring all the parties receive the same and correct quantum public key is a non-trivial task, given that quantum states need to be distributed. Instead of using the cryptographic primitives from classical digital signatures, when dealing with QDS schemes, one often aims to ensure
that the QDS scheme has the following properties:
\begin{enumerate}
\item Unforgeability: a dishonest party cannot send a message pretending to be someone else.
\item Transferability: if a receiver receives a signature, they should be confident that any other receiver should also accept the signature.
\item Non-repudiation: the sender cannot deny having sent the message.
\end{enumerate}

There are various protocols that fall under the category of QDS. The protocols we describe in this section deals with the task of signing a classical message using tools from quantum information theory.

\subsubsection{The (classical) Lamport one-time signature}

QDS schemes are inspired Lamport's one-time signatures \cite{lamport1979constructing}.  Here we rely on a classical one-way function $f$. Given $x$, it is easy to evaluate $f(x)$, but if $f(x)$ is given, it is hard to invert $f(x)$ to find $x$. The algorithm is as follows
\begin{enumerate}
\item Key generation: Alice chooses two random inputs $x_0$ and $x_1$, and evaluates $f(x_0)$, $f(x_1)$. She publically broacasts the pair $(0,f(x_0))$, $(1,f(x_1))$ whilst keeping $x_0, x_1$ secret. 
\item Signing: Alice sends the message $b$ along with her stored corresponding secret key $x_b$.
\item Verification: the receivers evaluate $f(x_b)$ and check if this agrees with the public key in order to choose to accept or reject.
\end{enumerate}

The security of such a scheme comes from the fact that $f(x)$ is classically difficult to invert, and therefore an adversary who has access to only the public key cannot find the secret key to forge a signature. However, any scheme that is based on computational-complexity arguments cannot offer information-theoretic security, and schemes based purely on the physics of quantum mechanics are much more desirable.

\subsubsection{The Gottesman-Chuang QDS}

In 2001, Gottesman and Chuang \cite{gottesman2001quantum} proposed the first QDS protocol. The idea is to use
the non-orthogonality of state to realize a ``quantum one-way function", where the difficulty to invert the function is based on the laws of physics rather than computational complexity.

If we have a quantum state $\ket{f(x)}$ based on the classical description $f(x)$, no one should be able to characterize $\ket{f(x)}$ with certainty, unless they already know $f(x)$. The function $f$ does not need to be difficult to invert, since the computational complexity  in the classical protocol is replaced by non-orthogonality.
    \begin{enumerate}
\item A function $f$ is chosen and is made public, it takes the input $x$ and generates $f(x)$ that describes the quantum state.
\item Key generation: for the private key, Alice chooses a pair of bit-string 
$\{x_0^i,x_1^i\}$ where $1\leq i \leq L$. The bit-string $x_0$ ($x_1$) will be used to sign the 0's (1's) in the message. $L$ is determined by the security level required.
\item For the public key, Alice generates multiple copies of $\{\ket{x_0^i},\ket{x_1^i}\}$. She distributes to each receiver the corresponding states. To check that the public keys are the same, each of the receivers interacts by performing SWAP tests. A SWAP test provides an affirmative answer without disturbing the states.
\item Signing: to sign a message, Alice chooses the message vale $b$, and sends $(b,x_b^i)$ to the receiver. This can be done classically.
\item Verification: the receiver uses $x_b^i$ and generates the corresponding quantum state
$\ket{f(x)_b^i}$ and checks whether this is consistent with the stored public key.
\end{enumerate}

Whilst the Gottesman-Chuang QDS protocol is intuitive, it is also highly impractical due to the experimental requirements. Firstly, the protocol requires for long-storage quantum memories, since the signature and the verification can be separated by long periods of time -- this requirement is one of the major bottle-necks for building a scalable quantum computer. Secondly, multiple copies of the public key are needed, and each receiving party need to perform the SWAP test, the latter involves more quantum communication and ancillary qubits

\subsubsection{A practical QDS protocol}

Since the development of the Gottesman-Chuang QDS protocol, the field has undergone major developments and lifted all the restricting requirements of the protocol.

The storage of the quantum state until verification renders the scheme impractical. This requirement can be removed by replacing the quantum public key with classical verification keys, which is now different for each receiver \cite{bib:Dunjko2014prl}.
After removing the quantum memory requirement, the other experimentally challenging task is ensuring that the same quantum public key was sent to different receivers. This step can be bypassed by adding an extra step, replacing the SWAP test with a "symmetrization" \cite{bib:Wallden2015pra}. This ensures that even if the states distributed by Alice are not identical, the classical verification keys shared by the receivers will be ``symmetric" \cite{pirandola2019advances}. 

We explain the symmetrization protocol now: Alice generates two sets of BB84 states and sends them to Bob and Charlie respectively. Bob (Charlie) either measures it or forwards it to Charlie (Bob), and he similarly measures the states he receives from Charlie (Bob). Depending on the outcome, Bob knows for certainty which state Alice did \textit{not} send. For example, if the measurement outcome is $\ket{0}$, he knows for sure Alice did not send $\ket{1}$. Bob stores the sequence of eliminated state, and whether it was received directly from Alice or Charlie. This classical information forms Bob’s and Charlie’s eliminated signatures, and will be the verification keys. 

Now we are equipped to describe a memory-less QDS protocol that can be realized with the same technology as QKD. The protocol works for three parties, but can be generalised. In this protocol, Alice signs the message, Bob first receives the message and needs to authenticate it, and Charlie receives the forwarded message from Bob, and verifies that the initial source was indeed Alice.

\begin{table}[!htbp]
\begin{mdframed}[innertopmargin=3pt, innerbottommargin=3pt, nobreak]
\texttt{
function QDS():
\begin{enumerate}
    \item Key generation: Alice performs QKD (up to the point where the raw keys are obtained) with Bob and Charlie separately, twice. Now, Alice has four bit-strings: $A_0^B, A_1^B, A_0^C, A_1^C$, Bob has two: $K_0^B, K_1^B$, and Charlie has $K_0^C, K_1^C$. The private key is the concatenation of the two corresponding strings, $(A_0^B||A_0^C, A_1^B||A_1^C)$.
    \item Error rates of the quantum channel are estimated at this stage.
    \item Bob and Charlie perform a symmetrization by exchanging secretly half of their strings via another QKD link. The new strings for Bob are $S_0^B, S_1^B$ and Charlie $S_0^C,S_1^C$. They are composed of half of the string initially sent to Bob and half of that to Charlie, but Alice does not know from which parts of Bob's and Charlie's strings they came from.
    \item The verification keys for Bob and Charlie are $(S_0^B,S_1^B)$ and 
          $(S_0^C,S_1^C)$.
    \item Signing: to sign a message $M$, Alice sends $(M, A_M^B||A_M^C)$ to Bob.
    \item Verification: Bob checks the mismatch rate between the signature received 
          $A_M^B||A_M^C$ and his stored verification key $S_M^B$. If the error is compatible with the channel noise, he accepts. Charlie receives a message with Alice's signature, but from Bob. He performs a similar check and may choose to accept the message as having originated from Alice. 
\item $\Box$
\end{enumerate}}
\end{mdframed}
\captionspacealg \caption{A quantum digital signature (QDS) protocol compatible with a QKD network.}\label{alg:qds}
\end{table}

% Quantum Crypto-Assets

%
% Quantum Crypto-Assets
%

\section{Quantum crypto-assets}\label{sec:quantum_crypto_assets}\index{Quantum crypto-assets}\index{Blockchain}

\dropcap{A}{side} from the value of outsourcing actual computations is the value of users' data itself. In the classical world we can generically refer to high-value data (especially in the context of digital tokens representing units of crypto-currencies\index{Crypto-currencies}) as \textit{crypto-assets}. Similarly, and especially given the nature of data likely to be subject to quantum treatment, we anticipate future \textit{quantum crypto-assets} -- high-value quantum states. Although the ways in which such assets are handled will be highly application-dependent, we comment on several specific use-cases that will foreseeably arise from observing recent developments in classical crypto-assets.

% Secure Quantum Data

\subsection{Secure quantum data}\label{sec:secure_quantum_date}\index{Secure quantum data}

Suppose Alice wishes to store quantum data offsite, for example in a repository for safekeeping, or within remote or decentralised data structures (such as a Blockchain\index{Blockchain}). Since her data is not held locally, there is the concern that an unauthorised third-party might simply steal it. How can she ensure its integrity, without maintaining any quantum data (assuming she can locally maintain classical data)?

This is a trivial problem to solve. Essentially we can think of this as a trivialised special case of blind quantum computing (Sec.~\ref{sec:blind_qc}), where the outsourced computation is just the identity operation. Employing the same ideas as the blind cluster state quantum computing protocol, Alice simply takes each qubit in her quantum data structure, and with equal probability (\mbox{$p=1/4$}) applies one of the four Pauli operators to it ($\hat\openone$, $\hat{X}$, $\hat{Y}$ or $\hat{Z}$). She of course keeps track of which ones were applied, which requires 2 classical bits per qubit.

From her perspective (or anyone she chooses to share the 2 classical bits with), she is always able to perfectly recover the hidden qubit, simply by applying the same Pauli operators a second time to invert them. However, from the perspective of a third-party without access to the 2 classical bits, they observe a perfect depolarising channel\index{Depolarising channel},
\begin{align}
	\mathcal{E}^\text{depolarising}_{0}(\hat\rho) = \frac{\hat\openone}{4},
\end{align}
yielding the completely mixed state, independent of the input state, from which no state information can be inferred. Thus, this approach confers \textit{perfect} information-theoretic security\index{Information-theoretic security}.

% Quantum Atomic Swaps

\subsection{Quantum atomic swaps}\label{sec:quantum_atomic_swaps}\index{Quantum atomic swaps}\index{Atomic swaps}

In a crypto-market we inherently wish to engage in the free exchange of different types of crypto-assets. In the context of Blockchain-based asset ledgers, we may wish to directly exchange tokens residing on entirely distinct Blockchains. For example, we may wish to trade a Bitcoin\index{Bitcoin} for an Ether\index{Ethereum} (Ethereum coin). Enabling this kind of exchange is vital for creating a fully functional crypto-market\index{Crypto-market} of arbitrarily interconvertible assets.

It's essential that such exchanges be performed with integrity, such that in a potentially anonymous transaction one party cannot run off with everything. The obvious solution here is to employ a trusted third-party to mediate the transaction, as is done in many real-world high-value exchanges. But this in undesirable for several reasons:
\begin{itemize}
\item Trust\index{Trust}: Both parties must have complete confidence in the integrity of the mediating third-party. This necessarily introduces risk, which manifests itself as an indirect transaction cost.
\item Monetary cost: A third-party is most likely to charge for the service they provide. Even if this margin is slim, in high-volume markets this becomes a consideration.
\item Latency\index{Latency}: Mediation introduces an additional layer of communication, with associated latency. Even in today's high-frequency markets, minute latency improvements yield major competitive advantage.
\item Resource overheads: Mediation imposes greater resource requirements, most notably communication or computational ones.
\item Regulatory\index{Regulations}: `Credible' mediators typically comply with regulatory and taxation agencies. Traders of an anarcho-capitalistic\index{Anarcho-capitalism} mindset would rather avoid such nonsense altogether, and establish a truly globalised free-market. \latinquote{Lux libertas. Lux tua nos ducat.}
\end{itemize}
This motivates the question, can we securely implement \textit{direct} exchanges, in the absence of any trusted mediating authority or regulatory oversight?

In classical Blockchain technology, \textit{atomic swaps} can be employed for this purpose. Such algorithms allow the direct exchange of crypto-assets, cryptographically enforced to guarantee one of two outcomes: a successful mutual exchange, or no exchange at all. There is cryptographically no possibility for a partial exchange to occur, in which one party ends up with both assets.

With quantum crypto-assets we can easily construct such \textit{quantum atomic swaps} by exploiting some well-known identities relating CNOT and SWAP gates. The first identity is that two consecutive CNOT gates cancel one another to yield an identity operation,
\begin{align}
\Qcircuit @C=1em @R=1.6em {
    \lstick{\ket\psi} & \ctrl{1} & \ctrl{1} & \qw & \ket\psi \\
    \lstick{\ket\phi} & \targ & \targ & \qw & \ket\phi
}\nonumber
\end{align}
Second, a sequence of three alternating CNOT gates in series yields a SWAP\index{SWAP gate} operation,
\begin{align}
\Qcircuit @C=1em @R=1.6em {
    \lstick{\ket\psi} & \ctrl{1} & \targ & \ctrl{1} & \qw & \ket\phi \\
    \lstick{\ket\phi} & \targ & \ctrl{-1} & \targ & \qw & \ket\psi
}\nonumber
\end{align}

Important is that in neither of the above two identities does partial implementation (i.e not executing one of the CNOTs) manifest itself as a one-way transaction, the key security consideration in the construction of a cryptographic atomic swap.

Note that these two decompositions for the identity and SWAP gates differ only via the presence of the central CNOT gate within the latter decomposition. By replacing it with a doubly-controlled CNOT gate, the additional control qubit effectively specifies whether or not a regular CNOT gate is applied between the first two qubits,
\begin{align}
\Qcircuit @C=1em @R=1.6em {
    \lstick{\ket\psi} & \ctrl{1} & \targ & \ctrl{1} & \qw \\
    \lstick{\ket\phi} & \targ & \ctrl{-1} & \targ & \qw \\
    \lstick{\ket{\text{execute}}} & \qw & \ctrl{-2} & \qw & \qw
}\nonumber
\end{align}

This ancillary `execute' qubit (restricted to $\ket{0}$ or $\ket{1}$, i.e effectively a classical bit), acts as a toggle between the two modes of operation, without changing the underlying physical circuit implementation -- it's now software-controlled, rather than hardware-controlled.

% Quantum Smart Contracts

\subsection{Quantum smart contracts}\index{Quantum smart contracts}

An atomic swap implementation with an `execute' control signal is particularly useful in an environment involving \textit{smart contracts}\index{Smart contracts} -- self-executing generalisations of conditional contracts, such as credit default swaps (CDSs) -- where the execution of an exchange depends upon an algorithmically-determined outcome. In this instance, the `execute' qubit will be held and controlled by the smart contract algorithm. The algorithm making this choice can essentially be arbitrary, enabling extremely sophisticated contractual arrangements and exotic derivative instruments to be implemented in a self-enforced manner, without reliance upon third parties.

In the most general case in the quantum era, not only will the crypto-assets under exchange be quantum in nature, but the smart contract algorithms controlling their execution could be too (in principle any \textbf{BQP}\index{\textbf{BQP}} algorithm). This paves the way towards generic \textit{quantum smart contracts}\index{Quantum smart contracts}, the direct quantum extension of existing classical smart contract techniques.

The implications of complimenting smart contracts with quantum algorithms (even if only trading classical crypto-assets) are potentially immense. Should quantum-enhanced algorithms facilitate improved pricing models for example, crypto-markets could benefit from improved market efficiency\index{Market efficiency}, more accurate price signals\index{Price signals} from futures markets\index{Futures markets} (i.e reduced risk spreads), enhanced allocation of capital, and greater investor confidence, with higher market volume and liquidity\index{Market volume}\index{Market liquidity}.

\latinquote{A mari usque ad mare.}

\sketch{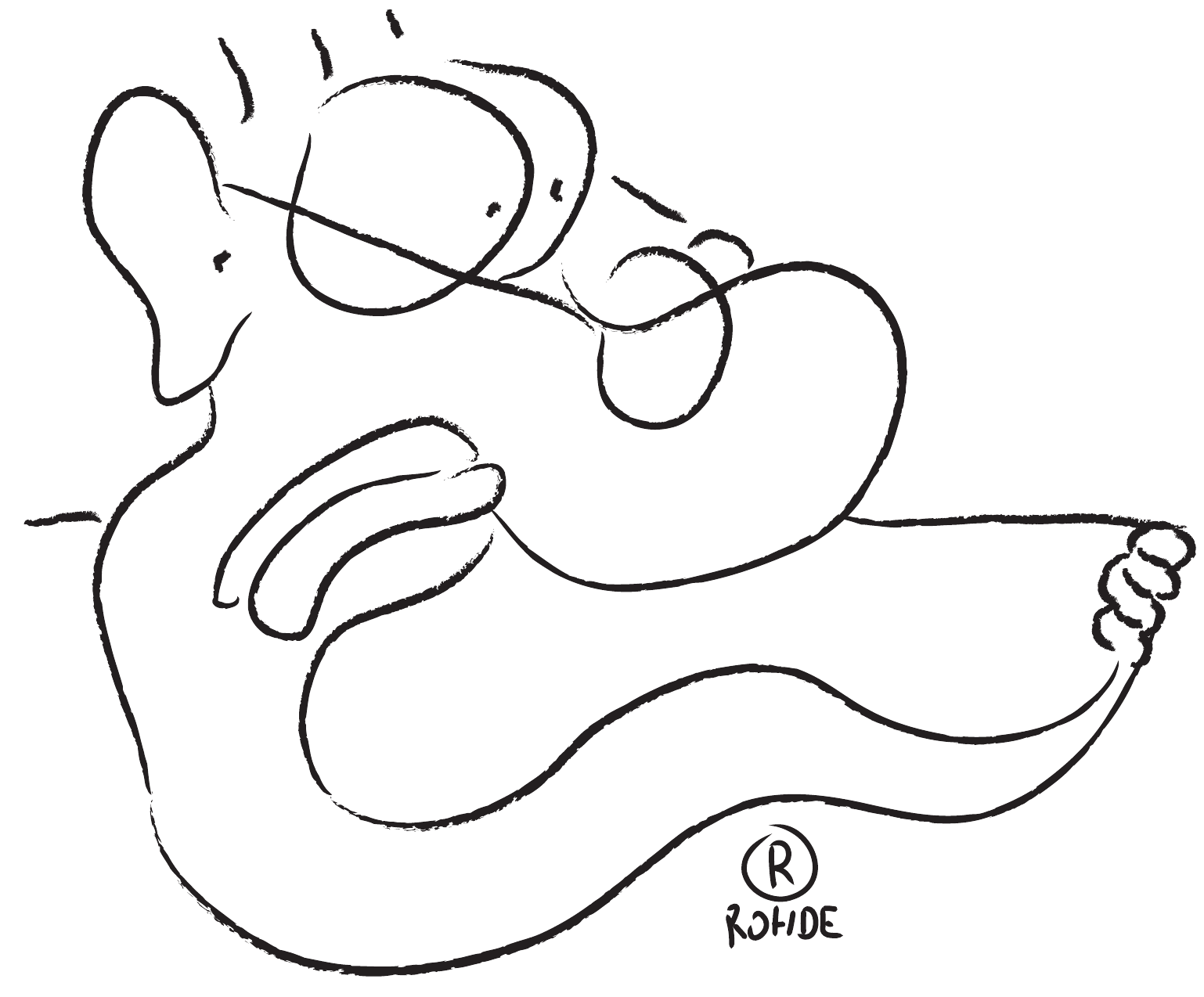}

\clearpage
Quantum computing

\part{Quantum computing}\label{part:QC}\index{Quantum computing}

\dropcap{S}{ince} quantum computing is perhaps the most exciting of the emerging quantum technologies, which we treat as the foremost application for the quantum internet, we now introduce quantum computing, covering models and physical implementations for realising it, and some of its well-known algorithmic applications.

%
% Models For Quantum Computation
%

\section{Models for quantum computation} \label{sec:models_QC} \index{Models for quantum computation}

\dropcap{W}{e} begin by reviewing the models for quantum computation that we will refer to throughout this work. There are various approaches to implementing and representing quantum computations. We now briefly introduce the ones most relevant to our discussions on networked quantum computation.

%
% Quantum Circuits
%

\subsection{Circuit model} \label{sec:circuit_model} \index{Circuit!Model}

The \textit{circuit model} is the conventional and most intuitive approach for expressing quantum algorithms, decomposing them into chronological sequences of elementary operations, comprising state preparation, single- and multi-qubit gates, measurement, and classical feedforward. We recommend referring to the introductory sections of \cite{bib:NielsenChuang00} for a far more comprehensive introduction to quantum circuits than is presented here. This model will be naturally intuitive to those familiar with classical circuit diagrams, albeit with some important differences, such as time-ordering.

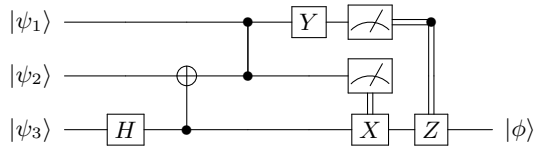
\begin{figure}[!htbp]
	\begin{align}
		\Qcircuit @C=.7em @R=.4em @! {
		\lstick{\ket{\psi_1}} & \qw & \qw & \ctrl{1} & \gate{Y} & \meter & \control \cw\\
		\lstick{\ket{\psi_2}} & \qw & \targ & \ctrl{-1} & \qw & \meter & \cwx\\
		\lstick{\ket{\psi_3}} & \gate{H} & \ctrl{-1} & \qw & \qw & \gate{X} \cwx & \gate{Z} \cwx & \rstick{\ket{\phi}} \qw
		} \nonumber
	\end{align}
	\captionspacefig \caption{Simple example of a quantum circuit on 3 qubits, comprising several single- and 2-qubit quantum gates and measurements. Rows represent qubits, and time flows from left-to-right.} \label{fig:eg_circuit}
\end{figure}

Fig.~\ref{fig:eg_circuit} illustrates a simple 3-qubit quantum circuit comprising all of these elements. The interpretation of this diagram is as follows:
\begin{itemize}
	\item Horizontal lines represent individual qubits.
	\item Time flows from left to right (feedback is not allowed in the typical formalism for this representation).
	\item The three input qubits are labelled on the far-left as $\ket{\psi_1}$, $\ket{\psi_2}$ and $\ket{\psi_3}$.
	\item Single-qubit gates are denoted as boxes containing the name of the associated unitary operation. Here, the examples are the Hadamard ($\hat{H}$), Pauli bit-flip ($\hat{X}$), Pauli bit-phase-flip ($\hat{Y}$), and Pauli phase-flip ($\hat{Z}$) gates\index{Pauli!Operators}\index{Hadamard!Gate},
	\begin{align}
		\hat{H} &= \frac{1}{\sqrt{2}}\begin{pmatrix}
		1 & 1 \\
		1 & -1
		\end{pmatrix},\nonumber \\
		\hat{X} &= \begin{pmatrix}
		0 & 1 \\
		1 & 0
		\end{pmatrix},\nonumber \\
		\hat{Y} &= \begin{pmatrix}
		0 & -i \\
		i & 0
		\end{pmatrix},\nonumber \\
		\hat{Z} &= \begin{pmatrix}
		1 & 0 \\
		0 & -1
		\end{pmatrix}.
	\end{align}
	\item 2-qubit gates are denoted by vertical lines between the respective qubits.
	\item The maximally-entangling 2-qubit controlled-NOT (CNOT) gate\index{Controlled-NOT (CNOT) gates} is denoted via a control ($\bullet$) and a target ($\oplus$),
	\begin{align}
		\hat{\mathrm{CNOT}}=\begin{pmatrix}
		1 & 0 & 0 & 0 \\
		0 & 1 & 0 & 0 \\
		0 & 0 & 0 & 1 \\
		0 & 0 & 1 & 0
		\end{pmatrix}.
	\end{align}
	This is the quantum equivalent of the classical XOR gate, flipping the target ($\hat{X}$) if the control is on.
	\item All quantum gates have the same number of input as output qubits. This is a necessary condition for the unitarity of quantum gates (\mbox{$\hat{U}^\dag \hat{U} = \hat\openone$}).
	\item The maximally-entangling 2-qubit controlled-phase (CZ)\index{Controlled-Z (CZ) gates} gate is denoted by two targets ($\bullet$) (the gate operates symmetrically on its two qubits),
	\begin{align}
		\hat{\mathrm{CZ}}=\begin{pmatrix}
		1 & 0 & 0 & 0 \\
		0 & 1 & 0 & 0 \\
		0 & 0 & 1 & 0 \\
		0 & 0 & 0 & -1
		\end{pmatrix},
	\end{align}
	applying a phase-gate ($\hat{Z}$) to the target if the control is on.
	\item The `meter' symbol represents a classical measurement in the Pauli $\hat{Z}$-basis (the computational or logical basis).
	\item Double lines represent classical feedforward of measurement outcomes, controlling a subsequent gate.
\end{itemize}

The circuit in Fig.~\ref{fig:eg_circuit} can be interpreted mathematically as implementing the following operation,
\begin{align}
	\ket\phi &= {\hat{Z}_3}^{m_1} \cdot {\hat{X}_3}^{m_2} \cdot \hat{M}_2 \cdot \hat{M}_1 \cdot \hat{Y}_1 \nonumber \\
	&\cdot \hat{\mathrm{CZ}}_{1,2} \cdot \hat{\mathrm{CNOT}}_{3,2} \cdot \hat{H}_3 \cdot \ket{\psi_1}\otimes\ket{\psi_2}\otimes\ket{\psi_3},
\end{align}
where $m_1$ and $m_2$ are the binary measurement outcomes of the two single-qubit $\hat{Z}$-basis measurements, $\hat{M}_1$ and $\hat{M}_2$.

Using the circuit model, arbitrary quantum computations can be elegantly and intuitively represented. To enable \textit{universal} quantum computation within this model, a \textit{universal gate set} must be available at our disposal. Most commonly, this is chosen to be the maximally-entangling 2-qubit CZ or CNOT operation, in addition to arbitrary single-qubit gates. Any quantum (i.e \textbf{BQP}) algorithm may be efficiently decomposed into a polynomial-depth circuit comprising elements from this universal gate set\index{Universal gate sets}. Note that the universal gate set is not unique, and there are many distinct sets. However, this set must contain at least one entangling operation acting on two or more qubits (such as a CZ or CNOT gate), and at least one non-Clifford gate\index{Clifford gates}\footnote{The Clifford group is that which commutes with the CNOT gate, such as the Pauli group.}.

%
% Cluster States
%

\subsection{Cluster states} \label{sec:CSQC} \index{Cluster states!Model for quantum computation}

The \textit{cluster state} model for quantum computation \cite{bib:Raussendorf01, bib:Raussendorf03, bib:Nielsen06} (also referred to as the \textit{one-way}, \textit{measurement-based}, or \textit{graph state} models for quantum computation) is an extremely powerful paradigm that warrants treatment of its own, owing to its significant distinction from the more familiar circuit model, and its applicability to distributed models for quantum computation, to be discussed in Sec.~\ref{sec:dist_QC}.

In the cluster state model, we begin by preparing a particular, highly-entangled state, called a \textit{cluster state} or \textit{graph state}. The state is associated with a graph $G$, comprising vertices, $V$, and edges, $E$,
\begin{align}
	G=(V,E),
\end{align}
of some topology, although rectangular lattice graphs are usually considered as they are sufficient for universal quantum computation\footnote{Note that the graph upon which a cluster state resides is not to be confused with the network graph. Rather it is just a convenient graphical representation for a class of multi-qubit states.}. That is, they act as a `substrate' for implementing arbitrary quantum computations.

In the graph, vertices represent qubits initialised into the,
\begin{align}
	\ket{+}=\frac{1}{\sqrt{2}}(\ket{0}+\ket{1}),
\end{align}
state, and edges represent the application of maximally entangling CZ gates between vertices,
\begin{align}
	\ket\psi_\mathrm{cluster} = \prod_{e\in E} \hat{\mathrm{CZ}}_e \cdot \bigotimes_{v\in V}\ket{+}_v.
\end{align}
Alternately, but equivalently, cluster states may be defined in the stabiliser formalism\index{Stabiliser!Formalism}. Specifically, a cluster state is defined to be the joint +1 eigenstate of all the stabilisers,
\begin{align} \label{eq:CS_stab} \index{Cluster states!Stabilisers}
	\hat{S}_v = \hat{X}_v \prod_{i\in n_v} \hat{Z}_i,
\end{align}
where there is one stabiliser $\hat{S}_v$ per vertex $v$, and $n_v$ is the set of vertices neighbouring $v$. The cluster state therefore satisfies,
\begin{align}
	\hat{S}_v\ket\psi_\mathrm{cluster} = \ket\psi_\mathrm{cluster}\,\forall\, v,
\end{align}
and the full set of stabilisers, $\hat{S}_v$, over all vertices $v$ is sufficient to fully characterise the cluster state, $\ket{\psi}_\mathrm{cluster}$, for a given graph topology.

\begin{figure}[!htbp]
	\includegraphics[clip=true, width=0.3\textwidth]{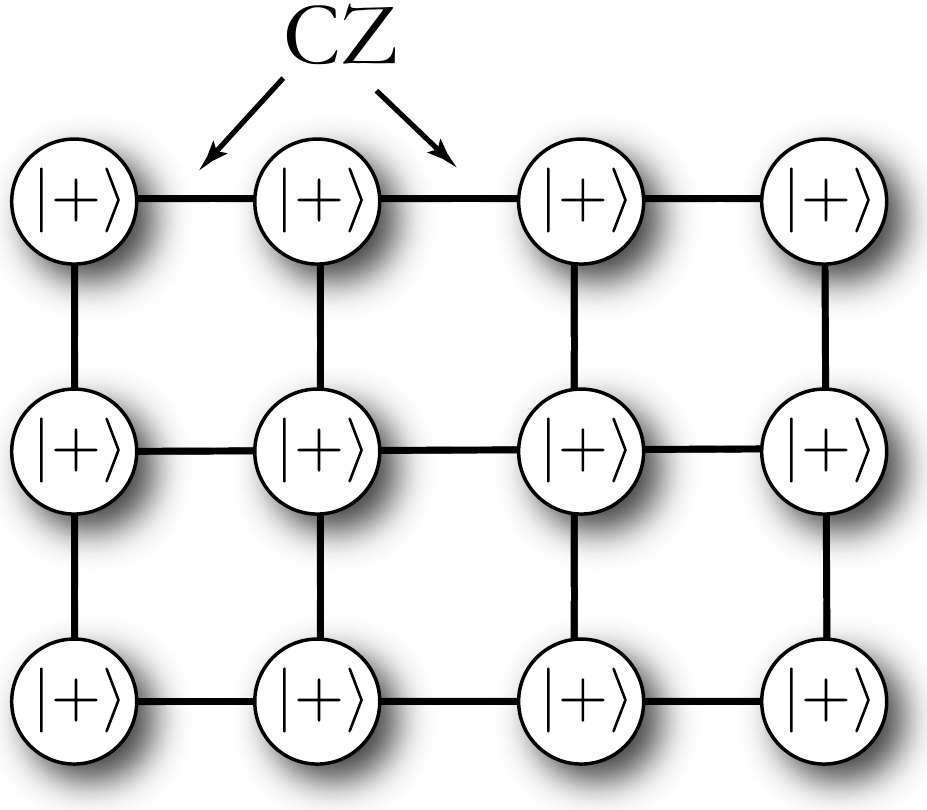}
	\captionspacefig \caption{Example of a \mbox{$4\times 3$} rectangular lattice cluster state. Each vertex in the graph represents a qubit initialised into \mbox{$\ket{+}=\frac{1}{\sqrt{2}}(\ket{0}+\ket{1})$}. Edges represent the application of CZ gates between qubits (CZ gates commute, so the order is unimportant). Of sufficient dimension, states of this topology enable universal measurement-based quantum computation, whereby computation proceeds purely via single-qubit measurements, and all entangling operations have been commuted to the state preparation stage. Because CZ gates commute, the preparation of cluster states is time-independent, and easily implemented in a distributed or parallelised manner. The time-ordering of the single-qubit measurements is dependent on the structure of the graph and the algorithm.} \label{fig:cluster_state}
\end{figure}

An example of a rectangular lattice cluster state is presented in Fig.~\ref{fig:cluster_state}. Cluster states are easily encoded optically using photonic polarisation encoding (Sec.~\ref{sec:single_phot_enc}), and therefore readily lend themselves to optical networking.

Having prepared this state, the computation is implemented purely via a well-orchestrated routine of single-qubit measurements. The order and basis in which they are performed (which depends on previous measurement outcomes in general -- i.e we require fast-feedforward) then stipulates the computation. In the context of distributed computation (Sec.~\ref{sec:dist_QC}), this requires classical communication between nodes.

\begin{figure}[!htbp]
	\includegraphics[clip=true, width=0.375\textwidth]{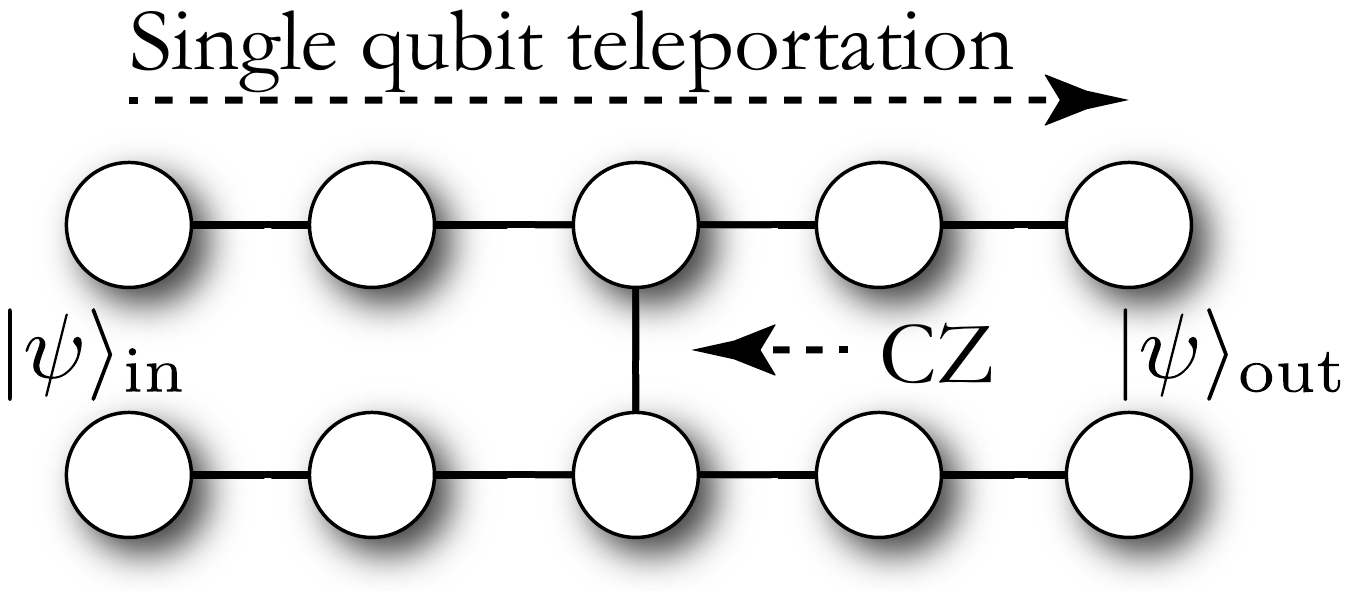}
	\captionspacefig \caption{Simple example of a cluster state that performs a computation comprising single-qubit operations and a CZ gate between two logical qubits. Let the two horizontal chains represent our two logical qubits. After inputting our input state from the left, we progressively measure out the cluster state qubits chronologically from left-to-right. Upon each single-qubit measurement, the choice of measurement basis teleports the action of a single-qubit gate. These accumulate sequentially. When we reach the point of measuring the two cluster state qubits joined with the vertical edge, the logical qubits accumulate the action of a CZ gate between them, since this is identically what that vertical edge physically corresponds to. Reaching the final two qubits, one from the upper rail and one from the lower, we obtain our two output logical qubits.} \label{fig:cluster_state_circuit}
\end{figure}

Mapping a circuit model computation to a cluster state topology can be most na{\" i}vely performed by taking a circuit acting on $n_\mathrm{qubits}$ qubits with depth $n_\mathrm{depth}$, preparing an \mbox{$n_\mathrm{qubits}\times n_\mathrm{depth}$} rectangular lattice cluster, and `etching' the circuit directly into the cluster state substrate. To perform this mapping we choose a universal gate set comprising CZ and single-qubit gates, retaining vertical edges where CZ gates ought to be present, eliminating the remaining vertical edges. Now we have a substrate that looks topologically very much like its equivalent circuit construction, and the computation proceeds chronologically in the same manner. The only conceptual distinction is that in the circuit model gates are directly applied chronologically to the set of qubits, whereas in the cluster state gate teleportation (Sec.~\ref{sec:teleport_gate}) is effectively implemented upon each measurement, with the action of gates accumulating as these teleportations are successively applied. A simple example of this notion is shown in Fig.~\ref{fig:cluster_state_circuit}.

The distinctive feature of this model is that all the entangling CZ gates are performed at the very beginning of the protocol, during the state preparation stage. The algorithm itself is purely measurement-based, requiring only single-qubit measurements (no entangling measurements).

\begin{figure}[!htbp]
	\begin{align}
		\Qcircuit @C=.7em @R=.4em @! {
		\lstick{\ket{\psi}} & \ctrl{1} & \gate{H} & \meter & & \\
		\lstick{\ket{+}} & \ctrl{0} & \qw & \gate{X} \cwx & \gate{H} & \rstick{\ket\psi} \qw \\
		} \nonumber
	\end{align}
	\captionspacefig \caption{The single-qubit teleporter, based upon a CZ gate, a single-qubit measurement, and classical feedforward.} \label{fig:single_qubit_teleporter} \index{Single-qubit teleporter}
\end{figure}
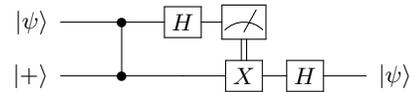

An alternate interpretation of the cluster state model is that it is a complicated network of state and gate teleportation protocols (Sec.~\ref{sec:teleport}). Specifically, a CZ gate with a $\ket{+}$ state as a resource, followed by measurement of one of the two qubits acts as a single-qubit teleporter, as shown in Fig.~\ref{fig:single_qubit_teleporter}\footnote{This is an alternative, but equivalent implementation for quantum state teleportation to that presented in Sec.~\ref{sec:teleport}.}. Thus, with a substrate state of CZ gates applied between $\ket{+}$ states, the single-qubit measurements progressively teleport the input state through the graph topology, at each stage accumulating the action of more gates, which are related to the choices of the previous single-qubit measurement bases, and the graph topology.

The cluster state formalism has proven very useful, enabling the development of models for linear optics quantum computing (Sec.~\ref{sec:KLM_univ}), orders of magnitude more efficient than the originally proposed protocol. It has been found that bonding strategies -- i.e the order in which smaller clusters are fused into larger ones when using non-deterministic gates -- plays a major role in resource overhead, and much work has been performed on efficient preparation strategies for various topologies 
\cite{bib:Nielsen04, bib:BarrettKok05, bib:BrowneRudolph05, bib:BenjaminEisert05, bib:Gross06, bib:RohdeBarrett07, bib:Kieling06, bib:KielingRudolphEisert06, bib:Kieling07, bib:Campbell07}.

These cluster states are highly valuable, given their computational power, and the ability to communicate them from Alice, who is able to prepare them, to Bob, who lacks the technology, would be a boon for Bob.

It would be most practical, economical, and resource efficient to have a single, well-equipped server with the ability to prepare such states, who does so on behalf of everyone else, and communicates the fresh cluster states to them over the quantum internet (for a price, perhaps).

Importantly, the preparation of cluster states is readily parallelised. All the entangling CZ operations commute, the order in which they are applied is irrelevant, and a rectangular lattice cluster is completely uniform. Thus, the graphs representing smaller cluster states may be easily `fused' together to form larger cluster states using, for example, CZ gates. Several other types of entangling gates can also be employed, such as polarising beamsplitters -- so-called \textit{fusion gates} \cite{bib:BrowneRudolph05}. This allows the preparation of cluster states to be performed in a `patchwork quilt'-like manner -- a number of nodes each prepare small lattice clusters, they are all put side-by-side, and stitched together using CZ gates. This type of distributed state preparation is a perfect application for in-parallel distributed quantum processing (Sec.~\ref{sec:dist_QC}).

Consider the scenario whereby Alice requests a large cluster state from Bob, but, while she was unable to prepare the cluster state herself, she has the technological ability to perform the measurement-based computation on the state (i.e simple single-qubit measurements). This would effectively bypass the need for secure quantum computation (Sec.~\ref{sec:homo_blind}) on Bob's hardware altogether, enabling computation with \textit{perfect} secrecy, since no foreign parties would be involved in the computation stage, and no secret data is communicated -- only the \textit{substrate} for the computation is communicated, which could be used for any purpose whatsoever. By commuting all the technologically challenging aspects of a quantum computation to the state preparation stage, we can effectively mitigate the need for blind quantum computing entirely, since the `hard work' has been done in advance by the host, and Alice gets to fulfil the computation on her own, completely bypassing poor old Bob, who was just dying to read Alice's secret love letters before processing them into Hallmark cards.

\begin{figure}[!htbp]
\if 1\doublecol
	\includegraphics[clip=true, width=0.475\textwidth]{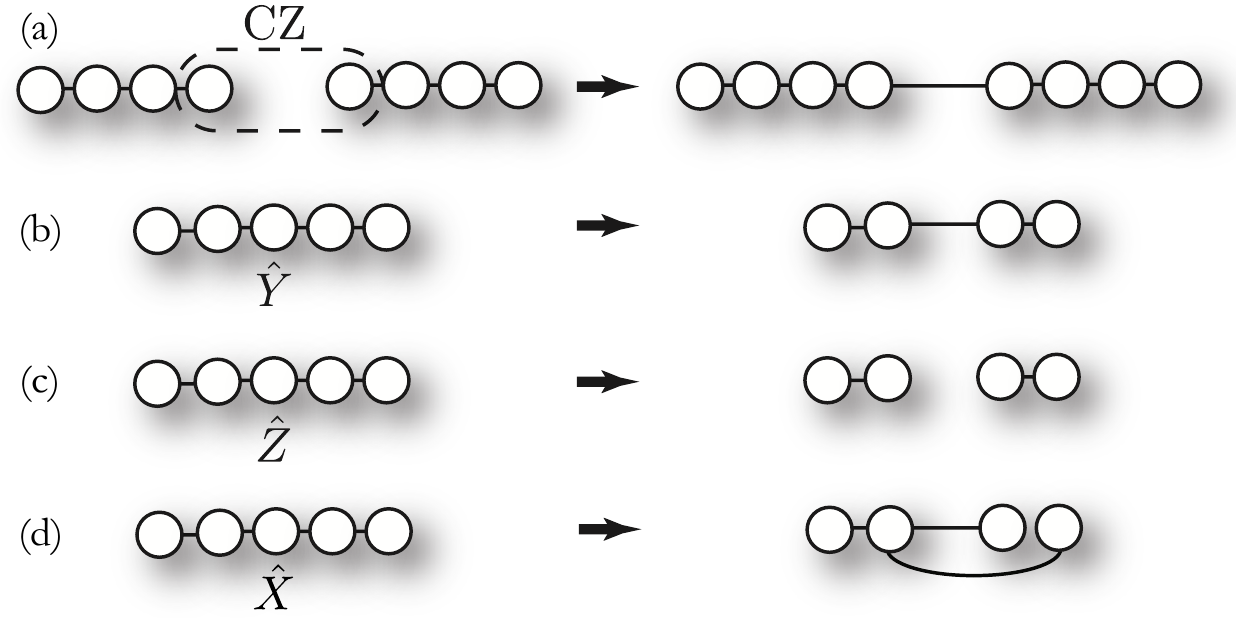} 
\else
	\includegraphics[clip=true, width=0.7\textwidth]{cluster_identities} 
\fi
	\index{Cluster states!Identities}
	\captionspacefig \caption{Several cluster state identities, demonstrated in the case of linear clusters. (a) a CZ gate between two qubits creates an edge between them in the graph. (b) Measurement of a qubit in the Pauli $\hat{Y}$ basis removes that qubit from the graph, whilst creating new edges between the neighbouring qubits. (c) Measurement of a qubit in the Pauli $\hat{Z}$ basis removes that qubit and any neighbouring edges. (d) Measurement of a qubit in the Pauli $\hat{X}$ basis removes that qubit, leaving one of its neighbours as a `dangling node'.} \label{fig:cluster_ident} 
\end{figure}

There are several cluster state identities we will utilise later, summarised in Fig.~\ref{fig:cluster_ident}.

When using non-deterministic gates (i.e ones that probabilistically fail) to prepare cluster states\index{Non-deterministic cluster state preparation}, there are approaches to nonetheless preparing ideal cluster states. There have been two main approaches that have become particularly well known. 

The first is to use the ideas of \textit{micro-clusters}\index{Micro-cluster states} and cluster state recycling\index{Cluster states!Recycling} to incrementally build up larger clusters, progressing as a random walk\index{Random!Walks}, which is biased in the direction of state growth. This approach is discussed in more detail in Secs.~\ref{sec:CS_LO} \& \ref{sec:module}.

The second approach is to borrow techniques from percolation theory\index{Percolation!Theory} to simply tolerate defects in a cluster state lattice by working around them \cite{brown2013defects}. Specifically, if the defect probability (i.e probability of a missing vertex or edge) is below some \textit{percolation threshold}\index{Percolation!Threshold}, \mbox{$p_\mathrm{defect}\leq \epsilon_\mathrm{threshold}$}, in the asymptotic limit we are guaranteed that routes exist through the lattice, enabling the required flow of information. This allows defective graphs to be employed for quantum computation.

%
% Adiabatic Quantum Computation
%

\subsection{Adiabatic quantum computation} \label{sec:adiabatic_QC} \index{Adiabatic!Quantum computation}

\sectionby{Zixin Huang}\index{Zixin Huang}

Adiabatic quantum computation (AQC) began as an approach for solving optimisation problems, but now has evolved into a universal alternative to the circuit model. AQC is based on an idea that is distinct from the circuit model \cite{bib:RevModPhys.90.015002}; in the circuit model, the computation is encoded into a series of gates, whereas in AQC the computation starts from an initial Hamiltonian whose ground state is easy to prepare, and evolves to a final state that encodes the solution to the computational problem.

The adiabatic theorem\index{Adiabatic!Theorem} forms the backbone of AQC, which states that for a system initially prepared in an eigenstate of a time-dependent Hamiltonian $\hat{H}(t)$, the is dictated by the Schr{\"o}dinger equation\index{Schr{\" o}dinger!Equation},
\begin{align}
i \frac{\partial \ket{\psi(t)}}{\partial t} = \hat{H}(t)\ket{\psi(t)},
\end{align}
\noindent which will keep approximately to the instantaneous ground state, if $\hat{H}(t)$ varies sufficiently slowly.

In the circuit model, the cost of the computation is equated with the number of gates. In AQC, the cost of adiabatic algorithms is defined to be the dimensionless quantity \cite{bib:aharonov2008adiabatic},
\begin{align}
\mathrm{cost} = t_f \cdot \mathrm{max}_s ||\hat{H}(s)||,
\end{align}
where $t_f$ is the algorithmic runtime, and $||\cdot||$ denotes the operator norm\index{Operator norm}, the largest singular value\index{Singular value} of the operator.
 
The worst-case runtime of an adiabatic algorithm scales as,
\begin{align}
t_f = O\left(\frac{1}{\Delta^{(3)}}\right), 
\end{align}
where $\Delta^{(3)}$ is the minimum eigenvalue gap between the ground state and the first excited state of the Hamiltonian \cite{bib:jansen2007bounds}. This is the reason why AQC has generated much interest -- it has a rich connection to the well-studied field of condensed matter physics\index{Condensed matter physics}.

The analysis for AQC involves bounding the eigenvalue gap of a complicated many-body Hamiltonian, which is a notoriously difficult problem. Nevertheless, a number of examples are known exhibiting a gap between the classical and quantum cases. 

\begin{figure}[!htbp]
	\includegraphics[clip=true, width=0.475\textwidth]{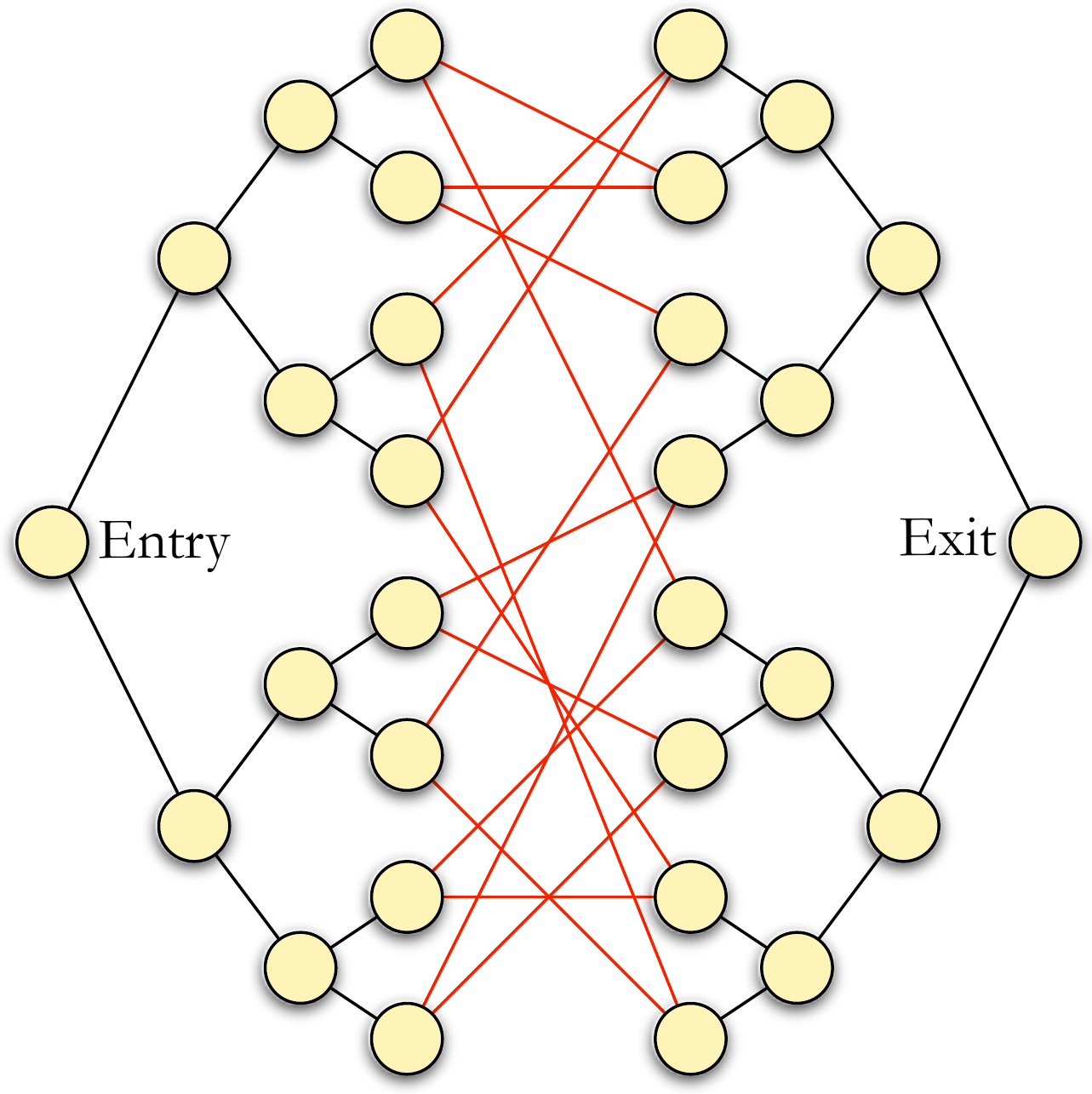}
	\captionspacefig \caption{A glued-trees graph of depth \mbox{$n=4$}. The `gluing' edges, chosen randomly such that each leaf vertex connects to exactly two of them, are shown in red.} \label{fig:glued_trees}\index{Glued-trees graph}
\end{figure}

Some of the well-known examples of AQC algorithms\index{Adiabatic!Algorithms} include:
\begin{itemize}
\item Adiabatic Grover algorithm\index{Adiabatic!Grover algorithm}: as its name suggests, this is the adiabatic version of Grover's search algorithm. The initial Hamiltonian is,
\begin{align}
	\hat{H}_0 = \hat\openone - \ket{\phi}\bra{\phi},
\end{align}
where $\ket{\phi}$ is the uniform superposition state $\ket{+}^{\otimes n}$. The final Hamiltonian is given by,
\begin{align}
\hat{H}_1 = \hat\openone - \ket{m}\bra{m},
\end{align}
where $\ket{m}$ is the marked item we are searching for \cite{bib:PhysRevA.65.042308}.
\item Adiabatic Deutsch-Jozsa algorithm\index{Adiabatic!Deutsch-Jozsa algorithm}\index{Deutsch-Jozsa algorithm}: given a binary function,
\begin{align}
f:\{0,1 \}^n \rightarrow \{0,1\},
\end{align}
which is either constant\index{Constant functions} or balanced\index{Balanced functions}, the Deutsch-Jozsa algorithm can determine which type of function it is, exhibiting quantum speedup. An adiabatic implementation \cite{bib:PhysRevLett.95.250503} is derived to match the speedup obtained in the original circuit model implementation.
\item Adiabatic glued-trees problem\index{Adiabatic!Glued-trees problem}: we consider two binary trees\index{Binary trees} of depth $n$. Each tree has \mbox{$2^{n+1}-1$} vertices, and the two trees are randomly glued together, as shown in Fig.~\ref{fig:glued_trees}. A leaf is chosen from the left, connected to a random leaf on the right, which is in turn glued to another leaf on the left and so on. Every leaf on is connected to two on the other side, creating a random cycle that alternates between the two trees.
 The computation problem is to start from the left root and find a path to the right root in the smallest possible number of steps, given that one can only see adjacent vertices. Classical algorithms require at least a sub-exponential number of queries, but there exists a polynomial-time quantum algorithm based on quantum walks \cite{bib:childs2003exponential}.
\item Adiabatic \textsc{PageRank} algorithm\index{Adiabatic!PageRank algorithm}: the \textsc{PageRank}\index{PageRank algorithm} vector is a crucial tool in data mining and information retrieval, employed heavily by the Google search engine. The goal of the algorithm is to rank the importance, impact or influence of some entity in a network (webpages on the internet in the case of a Google search). The algorithm achieves this by representing the network as a flow network\index{Flow networks}, whereby flow from one vertex to another represents a vote by the source vertex for the importance of the destination vertex. By finding a steady-state flow to each edge in the network, the net flow into vertices represents their cumulative importance, as determined collectively by participants in the network. However, the magnitude of votes cast by vertices is weighted by their own ranking. Therefore, the algorithm aims to filter out bogus gaming of the system via the creation of dummy nodes that cast votes en masse to rig the election. The \textsc{PageRank} approach to ranking webpages has been by far the most successful and robust such algorithm, and was perhaps the key development behind the global success of Google. The best-known classical algorithms for solving the \textsc{PageRank} problem are via: representing the flow network as a matrix equation, from which the solution is obtained via solving an eigenvalue equation; representing the flow network as a random walk, whereby walkers randomly follow paths influenced by the flows, which in the long-time limit yields walks approximating the solution. The current best-known quantum \textsc{PageRank} algorithm outperforms all currently known classical algorithms with polynomial or even exponential speedup \cite{bib:PhysRevLett.108.230506}, but better future classical algorithms have not been ruled out and are under active investigation.
\end{itemize}

The computational power of the circuit model and the adiabatic model for quantum computing are equivalent up to a polynomial overhead. In the circuit model, a set of gates is said to be universal if any unitary operation may be approximated to arbitrary accuracy by a quantum circuit involving only those gates. The analog of such a set of gates in AQC is to efficiently map any circuit to a Hamiltonian.

If we have in the circuit model an $n$-particle pure state $\ket{\psi}$, acted upon by unitary $\hat{U}$ with circuit depth\index{Circuit depth} $L$, a time-dependent Hamiltonian $\hat{H}(t)$ is universal if:
\begin{itemize}
\item The ground state of $\hat{H}(t_f)$ is equal to $\hat{U}\ket{\psi}$ with non-zero probability.
\item The number of particles in $\hat{H}(t_f)$ is polynomial in $n$ and $L$, and $t_f$ is also polynomial in $n$ and $L$.
\end{itemize}
It has been proven that AQC can efficiently simulate the entire circuit model \cite{bib:RevModPhys.90.015002}, which implies that it is itself universal for efficient quantum computation.

%
% Restricted Models for Quantum Computation
%

\subsection{Restricted models for quantum computation} \label{sec:restricted_models} \index{Restricted models for quantum computation}

In the near future we are unlikely to have devices with the full power and versatility of universal quantum computers. Instead, we will gradually evolve towards that challenging goal via many incremental, intermediate steps. Those steps will take us along a path of restricted quantum computers, that solve specific problems of relatively small size, and are not fault-tolerant. Probably we can expect them to contain on the order of hundreds of qubits in the coming few years, as of the time of writing this.

We dedicate Sec.~\ref{sec:NISQ} to discussing these so-called \textit{Noisy Intermediate-Scale Quantum} (NISQ) devices \cite{bib:preskill2018quantum}, speculating on what that might exactly entail for our quantum journey in the near future.

One thing is for certain -- full-fledged quantum computing will remain an extremely ambitious goal for some time, but we will learn a lot from traversing the path towards it, and hopefully uncover new restricted quantum applications along the way.

%
% Fault-Tolerance
%

\subsection{Fault-tolerance}\index{Fault-tolerance}\label{sec:fault_tolerance}

In Sec.~\ref{sec:QOS_chap} we discussed QoS\index{Quality of service (QoS)} in the context of quantum networks, where we wish to protect the quantum information being communicated via packets of quantum data. In particular, QEC\index{Quantum error correction (QEC)} allows us to detect and correct errors introduced into quantum data during transmission across noisy channels.

Much more broadly, in the context of an entire quantum computation we will want to achieve the same goal, except that our techniques will need to extend far beyond defending individual quantum states against errors during transmission, but defending an entire computation and all the information residing within it at every stage throughout its execution.

This is achieved by extending techniques from QEC to achieve \textit{fault-tolerant quantum computing}. The primary difficulty here is that a quantum computation is not a passive operation, but involves the successive application of a potentially enormous number of quantum gates, each of which subject to its own error processes, all of which must be mitigated for the computation to succeed.

Because a quantum computation is not a passive operation but highly active, fault-tolerance protocols are also active and it does not suffice to simply perform an encoding at the beginning and error correction at the end. Instead, error correction procedures must be applied repeatedly throughout execution, at each stage projecting the encoded computation onto an error-free state.

%As with conventional QEC, this introduces (potentially large, but efficient) overheads associated with encoding logical qubits into fault-tolerant codes. Similarly, there is the notion of \textit{fault-tolerance thresholds}\index{Fault-tolerance!Thresholds} -- thresholds on gate error rates that must be achieved if fault-tolerant execution is to be successful. These thresholds are typically depressingly low (well below 1\%) and are the primary reason humanity has not yet achieved scalable quantum computing.

The concept of Fault-tolerance in computation is not a new idea, it was first developed 
in relation to classical computing~\cite{bib:N55,bib:G83,bib:A87}.  However, in recent years the precise manufacturing 
of digital circuitry has made large scale error correction and fault-tolerant circuits largely unnecessary.

The basic principle of Fault-tolerance is that the circuits used for gate operations and 
error correction procedures should not cause errors to cascade.  Quantum gates not only covert errors, i.e. a Hadamard operation can convert an $X$-error into a $Z$-error and visa versa, but multi-qubit gates can also {\it copy} errors.  
Hence if quantum circuits are not designed carefully, a correctable number of {\it physical} errors could occur which are consequently copied so many times that they overwhelm the error-correction capabilities of the encoding scheme.  

Fault-tolerance, in the context of error correction, is a function of how circuits and protocols are implemented, not a function of the underlying physical hardware.  It is assumed that all single qubit gates can introduce single qubit errors at some probability, $p$, and it is assumed that all two-qubit gates will {\it copy} pre-existing errors that exist at the input and also has the possibility of introducing a two-qubit correlated error on the two-qubits, with probability, $p$.  

In some cases there are examples of higher order gates being defined as primitives, for example the three qubit Toffoli gate.  However, it should be noted that in almost all cases, the physical implementation of these multi-qubit gates occur through an implicit decomposition into single and two-qubit gates.  This is due to the fact that the vast majority of physical systems, the highest order coupling term in a system Hamiltonian is weight two.  Higher weight coupling terms, which would be required to enable native multi-qubit gates (i.e. weight three terms in the Hamiltonian would be needed to natively enact a Toffoli gate), simply do not arise in natural and easily controllable quantum systems.  

To determine how errors are copied by gate operations, and error operator $E$ is conjugated through the gate operation to create a new error operator, $E' = G^{\dagger}EG$, for some gate unitary, $G$.  A single qubit example is the transformation of $X$-errors to $Z$-errors and visa versa through a Hadamard gate, due to the identity $\hat{X} = \hat{H}\hat{Z}\hat{H}$ and $\hat{Z} = \hat{H}\hat{X}\hat{H}$.  

A two-qubit example is more involved as we need to check all combinations of error mappings on both qubits involved in the gate.  If $\hat{G} = \hat{\mathrm{CNOT}}$, we can examine how $X$- and $Z$-errors change via $G$,
\begin{equation}
\begin{aligned}
&\mathrm{CNOT} (I\otimes X) \mathrm{CNOT} = I \otimes X \\
&\mathrm{CNOT} (X\otimes I) \mathrm{CNOT} = X \otimes X \\
&\mathrm{CNOT} (I\otimes Z) \mathrm{CNOT} = Z \otimes Z \\
&\mathrm{CNOT} (Z\otimes I) \mathrm{CNOT} = Z \otimes I \\
\end{aligned}
\end{equation}
where the notation, $A \otimes B$, are error operators, $\{A,B\} \in \{I,X,Y,Z\}$, on qubits one and two of the gate and, 
$\hat{G} = \hat{G}^{\dagger} = \hat{\mathrm{CNOT}}$.

So, for a controlled-not operation, $X$-errors are copied from control qubit to target, and $Z$-errors are copied from target to control.  pre-existing $X$-errors on the target qubit or $Z$-errors on the control qubit are unchanged through the gate.  

The fact that quantum circuits can cause errors to be copied implies that if circuits are designed badly, errors can cascade during error correction protocols even when only one or two {\it physical} errors actually took place.  Considering that error correction codes have a finite correcting power, i.e. the Steane code will deterministically correct an arbitrary {\it single} qubit error, but if more than a single error occurs between correction cycles, logical errors are likely to be induced.  

Fault-tolerance is a discrete feature of a quantum circuit construction, either a construction is fault-tolerant or it is not. However, what is defined to be fault-tolerant can be a function of what type of error-correction code is used.  For example, for a single error correcting code, fault-tolerance is defined as:
\begin{enumerate}
\item A single error will cause \textbf{at most} one error in the output for each logical qubit block.
\end{enumerate}
However, if the quantum code employed is able to correct multiple errors, then the definition of fault-tolerance can be relaxed, i.e. if the code can correct three errors then circuits may be designed such that a single failure results in at most two errors in the output (which is then correctable).  In general, for a code correcting $t=\lfloor (d-1)/2 \rfloor$ errors, fault-tolerance requires that $\leq t$ errors during an operation does not result in $> t$ errors in the output for each logical qubit.    
 
\subsection{The threshold theorem}
The threshold theorem is a consequence 
of fault-tolerant circuit design and the ability to perform dynamical error correction.  
Rather than present a detailed derivation of the theorem for a variety of noise models, we will 
instead take a very simple case where we utilize a quantum code that can only correct for a 
single error, using a model that assumes uncorrelated, errors on individual qubits.  For 
more rigorous derivations of the theorem see~\cite{bib:AB97,bib:G97+}.  

Consider a quantum computer where each physical 
qubit experiences either an $X$ and/or $Z$ error independently 
with probability $p$, per gate operation.  
Furthermore, it is assumed that each logical gate operation and error 
correction circuit is designed fault-tolerantly and that a cycle of 
error correction is performed after each elementary {\em logical} gate operation.  If an error occurs 
during a logical gate operation, then the fault-tolerant constructions ensure this error will only propagate 
to at most one error in each block, after which a cycle of error correction will remove the error.  

Hence if the failure probability of un-encoded qubits per time step is $p$, then a single level 
of error correction will ensure that the logical step fails only when two (or more) errors occur.  Hence 
the failure rate of each logical operation, to leading order, is now $p^1_L = cp^2$, where $p^1_L$ is the 
failure rate (per logical gate operation) of a 1st level logical qubit and $c$ is the upper bound for the 
number of possible 2-error combinations 
which can occur at a physical level within the circuit consisting of the 
correction cycle $+$ gate operation $+$ 
correction cycle.  

We now repeat the process, re-encoding the computer 
such that a level-2 logical qubit is formed, using the same $[[n,k,d]]$ 
quantum code, from $n$, level-1 encoded 
qubits.  It is assumed that all error correcting procedures and gate operations at the 2nd level are 
self-similar to the level-1 operations (i.e. the circuit structures for the level-2 encoding are 
identical to the level-1 encoding).  Therefore, if the level-1 failure rate per logical time step is $p^1_L$, 
then by the same argument, the failure rate of a 2-level operation is given by,
$p^2_L = c(p^1_L)^2 = c^3p^4$.  This iterative procedure is then repeated (referred to as concatenation) 
up to the $k$th level, such that the logical failure rate, per time step, of a $k$-level encoded qubit is given by,
\begin{equation}
p^k_L = \frac{(cp)^{2^k}}{c}.
\label{eq:threshold}
\end{equation}   
Eq. \ref{eq:threshold} implies that for a finite {\em physical} error rate, $p$, per qubit, per time step, 
the failure rate of the $k$th-level encoded qubit can be made arbitrarily small by simply increasing $k$,  
dependent on $cp < 1$.  This inequality defines the threshold.  The physical error rate 
experienced by each qubit per time step must be $p_{th} < 1/c$ to ensure that multiple levels of 
error correction reduce the failure rate of logical components.

Hence, provided sufficient resources are available, an arbitrarily large quantum circuit can be 
successfully implemented, to arbitrary accuracy, once the physical error rate is below threshold.  Initial estimates at the threshold, which gave $p_{th} \approx 10^{-4}$~\cite{bib:K97,bib:AB97,bib:G97+} did not sufficiently model physical systems in an accurate way.  Recent results~\cite{bib:SFH07,bib:SDT07,bib:SBFRYSGF06,bib:MCTBCCC04,bib:BSO05} have been estimated for more realistic quantum processor architectures, showing significant differences in threshold when architectural considerations are taken into account.  The most promising thresholds that have been calculated for expected, circuit level noise, are based on surface codes~\cite{bib:WFSH09,bib:WFH11,bib:FMMC12,bib:S14}, with thresholds slightly less than 1\%.  This has now become the target for experimental groups as a large number of scalable systems architectures utilise the surface code as the underlying correction model~\cite{bib:Gimeno-Segovia:2015aa,bib:Hill:2015aa,Lekitsch:2017aa,Nemoto:2014aa,Jones:2012aa,bib:Mukai_2020}.

\latinquote{Aqua vitae.}

%
% Quantum Algorithms
%

\section{Quantum algorithms} \index{Quantum algorithms}\label{sec:quantum_algs}

\dropcap{T}{he} ultimate goal of quantum computing is to implement algorithms with a quantum speedup compared to classical algorithms. The degree of speedup achieved varies between algorithms, and it is important to note that not every classical algorithm exhibits any speedup when implemented quantum mechanically.

To provide context for the excitement of quantum computing and motivate interest in their development, we now summarise some of the key quantum algorithms that have been described exhibiting quantum speedup.

%
% Deutsch-Jozsa
%

\subsection{Deutsch-Jozsa} \index{Deutsch-Jozsa algorithm}

The first quantum algorithm demonstrating a provable improvement over the best classical algorithm was the Deutsch-Jozsa algorithm \cite{bib:DeutschJozsa92}. Unfortunately the algorithm solves a very contrived problem, designed for the purposes of demonstrating post-classicality rather than solving a problem of actual practical interest. Nonetheless, the algorithm is straightforward to explain and understand, making it a useful starting point in understanding quantum algorithms and the computational enhancement they may offer.

The algorithm relies on a `black box', referred to as an \textit{oracle}\index{Oracles}, which takes an input bit-string and outputs a single bit, evaluating the function $f(x)$ for the $n$-bit input bit-string $x$. In this contrived problem $f(x)$ is guaranteed to be either \textit{uniform}\index{Uniform functions} or \textit{balanced}\index{Balanced!Functions}. In the former case, the output to the oracle is always \mbox{$f(x)=0$} or always \mbox{$f(x)=1$}, but it doesn't matter which, they simply must always be the same. In the latter case, the output is \mbox{$f(x)=0$} for exactly half the inputs $x$, and \mbox{$f(x)=1$} for the other half of $x$, but the ordering of which inputs generate which outputs may be arbitrary. The goal of the algorithm is to determine whether $f(x)$ is uniform or balanced using the least number of queries to the oracle.

While it's clear that the dimensionality of the input state space is exponentially large, $2^n$, it is fairly obvious that a trivial \textbf{BPP}\index{BPP} algorithm exists for solving this problem with confidence exponentially asymptoting to unity against the number of oracle queries. We simply evaluate the oracle for randomly chosen inputs. If we measure any occurrences of measurement outcomes that are not all 0 or all 1 we know with certainty that the function must have been balanced. If on the other hand we measure all 0s or all 1s for more than half the input state space $x$, we know with certainty the function was uniform.

However, if the function were balanced, there is the possibility that it might conspire against us to fool us into thinking the function was uniform until we evaluate half plus one of the input states, requiring $O(2^n)$ oracle queries, although this will occur with exponentially low probability against the number of queries. Thus, the algorithm can be approximated with exponential asymptotic certainty in \textbf{BPP}\index{BPP}. But considering the \textit{worst} case\index{Worst case complexity} rather than the \textit{average} case\index{Average case complexity}, we may have to perform an exponential number of evaluations, $O(2^n)$, to know the answer with absolute certainty.

The Deutsch-Jozsa algorithm solves this rather specialised problem in the worst case using only a single quantum evaluation of the oracle.

The algorithm implementing the Deutsch-Jozsa protocol and its circuit diagram are shown in Alg.~\ref{alg:deutsch_jozsa}. The engine room of the algorithm is in the Hadamard transform\index{Hadamard!Transform}, $\hat{H}^{\otimes n}$, which prepares an equal superposition of all $2^n$ possible input bit-strings $x$, which are then evaluated in superposition by the oracle. To ensure unitarity, the oracle is defined to implement the transformation\footnote{Note that the seemingly more obvious choice of \mbox{$\hat{U}_f\ket{x}=\ket{f(x)}$} is not unitary. This trick of introducing an additional ancillary state to enable unitary construction of arbitrary functions is a common one in quantum algorithm design, as will be discussed further in Sec.~\ref{sec:oracles}.},
\begin{align}
	    \hat{U}_f \ket{x}\ket{y} &= \ket{x}\ket{y\oplus f(x)}.
\end{align}
That is, it flips bit $y$ if \mbox{$f(x)=1$} (equivalently addition modulo 2 or an XOR operation). An inverse Hadamard transform subsequently yields a measurement outcome with one of two possibilities:
\begin{itemize}
	\item The 0 and 1 terms outputted from the oracle interfere perfectly constructively, if the function was uniform.
	\item They interfere perfectly destructively, if the function was balanced.
\end{itemize}
Then, with a single-shot measurement of the inverse Hadamard transformed output from the oracle we establish whether $f(x)$ was balanced or uniform with certainty. This exhibits an exponential worst case speedup compared to a randomised classical sampling algorithm (which is classically optimal).

\begin{table}[!htbp]
\begin{mdframed}[innertopmargin=3pt, innerbottommargin=3pt, nobreak]
\texttt{
function DeutschJozsa(f,n):
\begin{enumerate}
    \item Prepare the \mbox{$n+1$}-bit state,
    \begin{align}
    \ket\psi_0 = \ket{0}^{\otimes n}\ket{1}.	
    \end{align}
    \item Apply the \mbox{$n+1$}-bit Hadamard transform,
    \begin{align}
    \ket\psi_1 &= \hat{H}^{\otimes(n+1)}\ket\psi_0 \nonumber \\
    &= \frac{1}{\sqrt{2^{n+1}}} \sum_{x=0}^{2^n-1}\ket{x}(\ket{0}-\ket{1}),	
    \end{align}
    where $x$ enumerates all $n$-bit binary bit-strings.
    \item Apply the unitary oracle, implementing the transformation,
    \begin{align}
    \hat{U}_f \ket{x}\ket{y} &= \ket{x}\ket{y\oplus f(x)},
    \end{align}
    where $\oplus$ denotes addition modulo 2, yielding,
    \begin{align}
    \ket\psi_2 = \hat{U}_f \ket\psi_1.	
    \end{align}
    \item Apply another Hadamard transform,
    \begin{align}
    \ket\psi_3 = \hat{H}^{\otimes n} \ket\psi_2.
    \end{align}
    \item The full evolution is thus given by,
    \begin{align}
    	\ket\psi_\mathrm{out} = (\hat{H}^{\otimes n}\otimes\hat\openone) \cdot \hat{U}_f \cdot \hat{H}^{\otimes (n+1)}\ket{0}^{\otimes n}\ket{1}.
    \end{align}
	\item Measure the first $n$ qubits to determine the probability of measurement outcome $\ket{0}^{\otimes n}$.
	\item This probability is given by,
	\begin{align}
	P_0 = \left| \frac{1}{2^n} \sum_{x=0}^{2^n-1} (-1)^{f(x)} \right|^2.	
	\end{align}
	\item Depending on whether $f(x)$ was uniform or balanced, the alternating sign terms in this sum interfere constructively or destructively, yielding \mbox{$P_0=1$} or \mbox{$P_0=0$} respectively.
	\item Thus, a single measurement outcome suffices to determine whether $f(x)$ was balanced or uniform.
	\item $\Box$
\end{enumerate}
\begin{align}
\Qcircuit @C=1em @R=1.6em {
    \lstick{\ket{0}^{\otimes n}} & \gate{{H}^{\otimes n}} & \multigate{1}{{U}_f} & \gate{{H}^{\otimes n}} & \meter \\
    \lstick{\ket{1}} & \gate{{H}} & \ghost{{U}_f} & \qw & \\
} \nonumber
\end{align}
}
\end{mdframed}
\captionspacealg \caption{Deutsch-Jozsa algorithm for evaluating whether the function $f(x)$ is balanced or uniform, exhibiting exponential worst case speedup compared to the best classical \textbf{BPP} algorithm\index{BPP}\index{Deutsch-Jozsa algorithm}.} \label{alg:deutsch_jozsa}
\end{table}

%
% Quantum Search
%

\subsection{Quantum search}\label{sec:quantum_search} \index{Grover's algorithm}

The problem of finding specific entries in unstructured data spaces is a ubiquitous one. This class of \textit{search algorithms} have amongst the broadest applicability of any class of algorithms. Computer scientists have invested excruciating man-hours\index{SJW}\footnote{Presently, most computer science research institutions are equal opportunity employers.} into organising and structuring data so as to minimise the resource overhead (in time and/or space) associated with extracting desired components. However, the methodology for achieving this, and the favourability of associated resource overheads, is highly dependent on the structure of the underlying data, or whether there even is any. To this end, numerous data structures and algorithms have been developed, accommodating for every mutation and variation of the posed problem imaginable. Often, there is a tradeoff between the overheads induced in time and memory, as well as in pre-processing and data structure maintenance requirements.

For example, \textit{hash tables}\index{Hash!Tables} enable theoretical $O(1)$ lookup times on data with a \textit{key-value pair}\index{Key-value pair} data structure. In a key-value pair each data entry (value) is tagged with a unique identifier (key) used for lookup. The value can observe any structure whatsoever, whereas the key is designed so as to minimise search times. When storing telephone numbers one might represent entries as key-value pairs, where the keys are people's names, and the values their respective telephone numbers. An efficient algorithm for mapping keys to physical memory addresses would imply efficient lookup of telephone numbers by name.

In the absence of a key-value representation one might simply store data in sorted form. However, this requires pre-sorting the entire data space, which may become costly for large data sets, and requires continual rearrangement whenever the data space is modified, making it computationally costly for mutable datasets.

For the end user, who wishes to find data elements, the worst-case data space is one with no order or underlying structure. Suppose we want to find whether a number exists in the telephone directory, but we don't know its associated name. In this instance, it can easily be seen that the best one can hope for, in terms of algorithmic runtime, is to simply look through the data space brute-force\index{Brute-force} until we find what we are looking for. It is clear that with an unstructured space of $N$ elements, this brute-force search algorithm requires on average $O(N)$ queries to find the desired entry. We call this the \textit{unstructured search problem}\index{Unstructured search problem}.

The brute-force classical algorithm, despite already being technically `efficient' (i.e $O(N)$ linear runtime), could nonetheless become unwieldy for very large datasets. Google doesn't want to exhaustively scan their entire collection of data-centres each time they want to lookup a database element. The quantum search algorithm, first presented by Grover \cite{bib:Grover96}, provides a solution to this problem using only $O(\sqrt{N})$ runtime (oracle queries), a quadratic enhancement. Whilst this falls far short of the exponential quantum enhancement one might have hoped for, which has also shown to be optimal \cite{bennett1997strengths, zalka1999grover}, it is nonetheless still extremely helpful for many purposes, given the broad applications for this algorithm.

We will formulate the quantum search algorithm as an oracular algorithm\index{Oracles}, where the oracle takes as input an $n$-bit string, and outputs 1 if the input matches the entry we are looking for, otherwise 0. This formulation of the problem makes the algorithm naturally suited to solving satisfiability problems (many of which are \textbf{NP}-complete and of great practical interest).

The Grover quantum search algorithm is shown explicitly in Alg.~\ref{alg:quant_search}.

\begin{table}[!htbp]
\begin{mdframed}[innertopmargin=3pt, innerbottommargin=3pt, nobreak]
\texttt{
function Grover(f,n):
\begin{enumerate}
	\item Using a Hadamard transform, prepare the $n$-qubit equal superposition of all $2^n$ logical basis states,
    \begin{align}
    \ket\varphi &= \hat{H}^{\otimes n}\ket{0}^{\otimes n}	 \nonumber \\
    &= \frac{1}{\sqrt{2^n}}\sum_{x = 0}^{2^n - 1}\ket{x},
    \end{align}
    where $x$ denotes a bit-string of length $n$.
	\item The oracle is defined as a unitary black-box, which tags a target element $T$ using a phase-flip,
	\begin{align}
		\hat{U}_T \ket{x} &= (-1)^{f(x)}\ket{x} \nonumber \\
		&= (\hat\openone - 2\ket{T}\bra{T})\ket{x},
	\end{align}
	where \mbox{$f(x)=\{0,1\}$} is the black-box function determining whether input $x$ is the target element $T$ (\mbox{$f(x)=1$}) or not (\mbox{$f(x)=0$}).
	\item The Grover diffusion operator\index{Grover diffusion operator} is defined to implement,
	\begin{align}
	\hat{U}_s = \hat\openone - 2\ket{T}\bra{T}.
	\end{align}
    \item repeat $O(N)$ times:
    \setlength{\itemindent}{.2in}
    \item $\ket{\varphi_{i+1}} = \hat{U}_s\cdot\hat{U}_T\ket{\varphi_i}$.
    \setlength{\itemindent}{0in}
    \item $\Box$
    %\item \comment{What about measurement stage}
\end{enumerate}
}
\end{mdframed}
\captionspacealg \caption{Grover's algorithm for performing a quantum search over an oracle, exhibiting a quadratic quantum enhancement.} \label{alg:quant_search}
\end{table}

%
% Oracles
%

\subsection{Oracles}\index{Oracles}\label{sec:oracles}

From Sec.~\ref{sec:quantum_search} we know that the quantum search algorithm requires a black box oracle that evaluates a classical function or database lookup query as a subroutine. How do we construct these oracles?

If we consider the case of solving a satisfiability problem, to find a satisfying input such that the function evaluates to \mbox{$f(x)=1$}, then it would na{\"i}vely appear that the required quantum operation needs to implement the transformation,
\begin{align}\label{eq:wrong_oracle}
	\hat{U}_f\ket{x}=\ket{f(x)}.
\end{align}
However, it's easy to see that in general such a transformation is non-unitary, and therefore cannot be implemented quantum mechanically.

To overcome this obstacle and enable unitary implementation, we introduce additional ancillary bits\index{Ancillary bits} to $f(x)$, and first write it as a reversible classical circuit\index{Reversible classical circuits}, which then always lends itself to a unitary quantum construction. Specifically, if instead of attempting to construct Eq.~(\ref{eq:wrong_oracle}) we  introduce some extra qubits and attempt to implement the transformation,
\begin{align}
\hat{U}_f\ket{x}\ket{0} 	= \ket{x}\ket{f(x)},
\end{align}
then it can easily be seen that this preserves orthonormality and may be implemented unitarily for any function $f(x)$, even if $f(x)$ is not an invertible function.

To provide the simplest possible example, consider the classical XOR gate, given by,
\begin{align}
	y_1 = x_1\oplus x_2.
\end{align}
This operation is obviously not reversible (and hence not unitary), since for a given output $y_1$, the inputs $x_1$ and $x_2$ are not unique. However, if we trivially modify the transformation to be,
\begin{align}
	y_1 &= x_1\oplus x_2,\nonumber\\
	y_2 &= x_2,
\end{align}
then a quick back-of-the-envelope calculation verifies that the transformation (simply a CNOT gate\index{CNOT gate}) is now reversible, unitary, and output $y_1$ encodes the desired function evaluation.

\begin{figure}[!htbp]
\includegraphics[clip=true,width=0.4\textwidth]{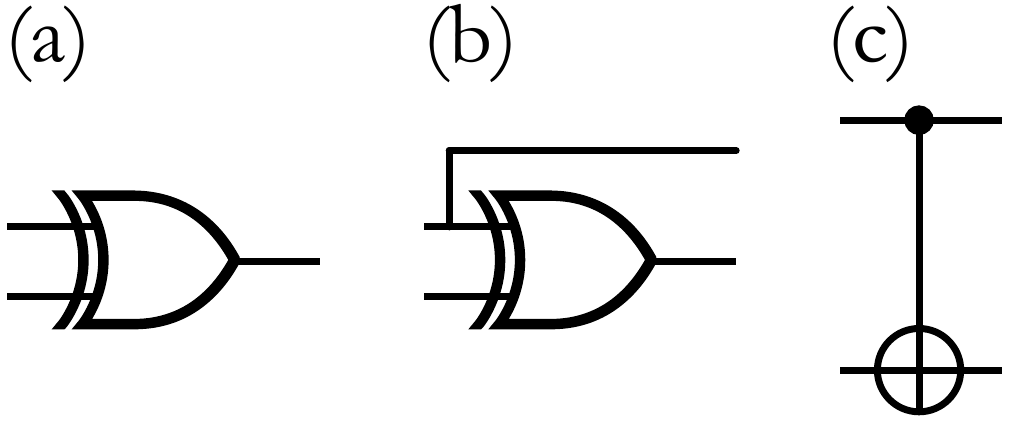}
\captionspacefig \caption{Simple example of the progression of turning a classical function into a quantum oracle. (a) A simple classical circuit, the XOR gate, which is not invertible, implementing 2-bit modulo 2 addition. (b) By introducing an ancillary bit we are able to make a reversible implementation of the same circuit. (c) With a reversible circuit construction, we can directly substitute the reversible gates for their unitary equivalents, yielding a quantum oracle implementation of the starting function.}\label{fig:oracle_circ}	
\end{figure}

Having re-expressed our desired oracle function as a reversible classical circuit, we now simply make direct substitutions of all the reversible classical gates with their logically equivalent quantum counterparts, yielding a quantum implementation of the function that is unitary and preserves quantum coherence. The progression for this example is shown in Fig.~\ref{fig:oracle_circ}, and the general procedure for oracle construction is shown in Alg.~\ref{alg:oracle_const}.

\begin{table}[!htbp]
\begin{mdframed}[innertopmargin=3pt, innerbottommargin=3pt, nobreak]
\texttt{
function BuildOracle(f):
\begin{enumerate}
	\item Let $f$ be an arbitrary, efficiently-computable function in \textbf{BPP}.
	\item Express $f$ as a classical logic circuit.
	\item Make ancillary bits available.
	\item Rewrite the classical circuit as a reversible classical circuit, exploiting the introduced ancillary bits as necessary.
	\item Make direct substitutions of all reversible classical gates with their quantum counterparts that implement the equivalent logical operations in qubit space.
	\item Return $\hat{U}_f$.
    \item $\Box$
\end{enumerate}
}
\end{mdframed}
\captionspacealg \caption{Outline of the general procedure for constructing quantum oracles from classical logic descriptions.} \label{alg:oracle_const}
\end{table}

%
% Quantum Fourier Transform
%

\subsection{Quantum Fourier transform}\index{Quantum Fourier transform}\label{sec:QFT_alg}

The quantum Fourier transform (QFT) is not an algorithm per se, but rather a unitary operator that finds widespread use in other quantum algorithms -- a quantum subroutine of sorts. The QFT operator is so ubiquitous as a component in other quantum algorithms, that it warrants special treatment.

The QFT matrix simply contains coefficients taken from the discrete Fourier transform (DFT)\index{Discrete Fourier transform}. Specifically, the \mbox{$N\times N$} QFT matrix has elements,
\begin{align}
\hat{\mathrm{QFT}}_{j,k} = \frac{1}{\sqrt{N}} \omega^{(j-1)(k-1)},	
\end{align}
where,
\begin{align}
\omega = e^{\frac{2\pi i}{N}},	
\end{align}
is a complex root of unity\index{Complex root of unity}.

In matrix form the $N$-dimensional QFT is therefore,
\begin{align}
\hat{\mathrm{QFT}}_N = \frac{1}{\sqrt{N}} \begin{pmatrix}
  1 & 1 & 1 & \dots & 1\\
  1 & \omega & \omega^2 & \dots & \omega^{N-1} \\
  1 & \omega^2 & \omega^4 & \dots & \omega^{2(N-1)} \\
  \vdots & \vdots & \vdots & \ddots & \vdots \\
  1 & \omega^{N-1} & \omega^{2(N-1)} &\dots &\omega^{(N-1)(N-1)}
\end{pmatrix},
\end{align}
which is symmetric, \mbox{$\hat{\mathrm{QFT}}=\hat{\mathrm{QFT}}^\top$}. Note that all matrix elements are phases with magnitude $1/\sqrt{N}$,
\begin{align}
	|\hat{\mathrm{QFT}}_{i,j}|=\frac{1}{\sqrt{N}}\,\,\forall\, i,j.
\end{align}

Equivalently, in terms of basis state transformations this implements the map,
\begin{align}
	\hat{\mathrm{QFT}} \ket{j} \to \frac{1}{\sqrt{N}}\sum_{k=0}^{N-1} \omega^{jk}\ket{k},
\end{align}
on the $N$ basis states.

Unlike the closely-related Hadamard transform\index{Hadamard!Transform}, $\hat{H}^{\otimes n}$, which shares many properties\footnote{Like the QFT, all matrix elements of the Hadamard transform have magnitude \mbox{$1/\sqrt{N}$}, differing only in their phases, which are all simple $\pm$.} with the QFT, the QFT is a highly entangling operation. The QFT of any dimension has an efficient circuit implementation, making it an important subroutine in many algorithms. Specifically, the $n$-qubit QFT (i.e a transformation on $2^n$ amplitudes) requires only $O(n^2)$ elementary gates (Hadamards and CZs). If an approximation of the QFT is sufficient for one's purposes, this can be further reduced to only \mbox{$O(n\log n)$}.

Note that while the circuit only requires $O(n^2)$ gates, it implements a Fourier transform on $2^n$ amplitudes, giving this subroutine an exponential quantum improvement over classical DFTs.

The circuit construction for the QFT is shown in Alg.~\ref{alg:QFT}.

\begin{table}[!htbp]
\begin{mdframed}[innertopmargin=3pt, innerbottommargin=3pt, nobreak]
\texttt{
function QFT($\ket{x}$):
\begin{enumerate}
\item Apply the recursively-defined circuit below to the $n$-qubit input state $\ket{x}$, where,
\begin{align}
\hat{R}_\theta=\begin{pmatrix}
1 & 0 \\
0 & e^{i\theta}	
\end{pmatrix},
\end{align}
is a generalised phase-flip gate.
\item The amplitudes in the $n$-qubit output state $\ket{y}$ are given by the quantum Fourier transform of the input amplitudes, implementing the transformation,
\begin{align}
	\hat{\mathrm{QFT}} \ket{j} \to \frac{1}{\sqrt{N}}\sum_{k=0}^{N-1} \omega^{jk}\ket{k},
\end{align}
on the logical basis states, or in matrix form,
\begin{align}
\frac{1}{\sqrt{N}} \begin{pmatrix}
  1 & 1 & 1 & \dots & 1\\
  1 & \omega & \omega^2 & \dots & \omega^{N-1} \\
  1 & \omega^2 & \omega^4 & \dots & \omega^{2(N-1)} \\
  \vdots & \vdots & \vdots & \ddots & \vdots \\
  1 & \omega^{N-1} & \omega^{2(N-1)} &\dots &\omega^{(N-1)(N-1)}
\end{pmatrix},
\end{align}
where there are,
\begin{align}
N=2^n,
\end{align}
logical basis states.
\item The quantum circuit requires $O(n^2)$ gates.
\item Return $\ket{y}$.
\item $\Box$
\end{enumerate}
\begin{center}
\if 1\doublecol
\includegraphics[clip=true, width=\textwidth]{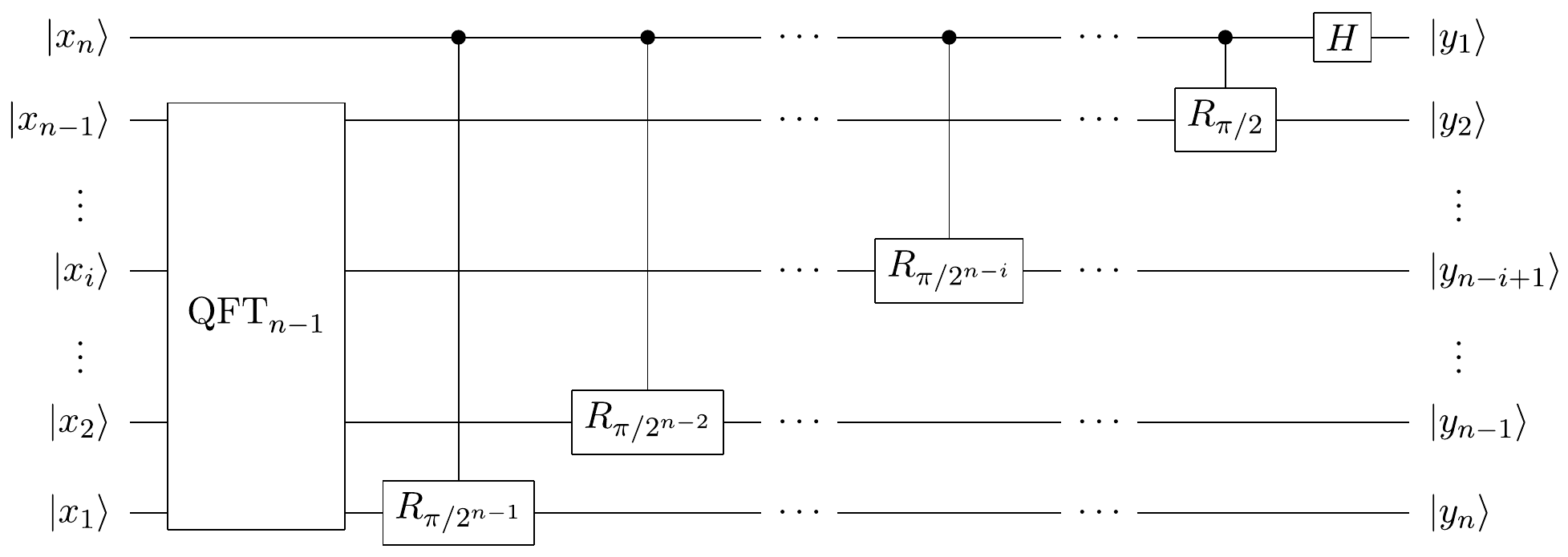}
\else
\includegraphics[clip=true, width=0.75\textwidth]{QFT}
\fi
\end{center}
}
\end{mdframed}
\captionspacealg \caption{The quantum Fourier transform (QFT) algorithm. The circuit is defined recursively, defining the $n$-qubit QFT in terms of an \mbox{$n-1$}-qubit QFT.} \label{alg:QFT}\index{Quantum Fourier transform}
\end{table}

%
% Phase-Estimation
%

\subsection{Phase-estimation}\index{Phase!Estimation!Algorithm}\label{sec:phase_est_alg}

The goal of the phase-estimation algorithm \cite{kitaev1995quantum, bib:NielsenChuang00} is to calculate the eigenvalue of a given unitary operator for a given eigenvector of that unitary. Since the eigenvalues of unitary operators are always phases of amplitude 1, we can write,
\begin{align}
	\hat{U}\ket\psi = e^{2\pi i\theta}\ket\psi,
\end{align}
where $\theta$ parameterises the phase of the eigenvalue and $\ket\psi$ is an eigenvector of $\hat{U}$.

The algorithm for this is baked from several ingredients. To implement phase-estimation with $n$ bits of precision, we require:
\begin{itemize}
	\item An $n$-qubit Hadamard transform, $\hat{H}^{\otimes n}$\index{Hadamard!Transform}.
	\item $n$ controlled-$\hat{U}$s\footnote{Any unitary has a coherently controlled-unitary equivalent, of the form $\hat{U}_\mathrm{controlled} = \ket{0}\bra{0}\otimes \hat\openone + \ket{1}\bra{1}\otimes\hat{U}$. These always have an efficient circuit construction, given $\hat{U}$.}\index{Controlled-unitaries}.
	\item An $n$-qubit inverse quantum Fourier transform\index{Quantum Fourier transform}, $\hat{\mathrm{QFT}}_n^\dag$ (Sec.~\ref{sec:QFT_alg}).
\end{itemize}

Alg.~\ref{alg:q_phase_est} shows the circuit implementation of the phase-estimation algorithm. The output is a binary representation of \mbox{$2^n\theta$}, where the precision is determined entirely by the number of qubits in the output register, which can be arbitrarily large in principle.

\begin{table}[!htbp]
\begin{mdframed}[innertopmargin=3pt, innerbottommargin=3pt, nobreak]
\texttt{
function PhaseEstimation($\hat{U}$, $\ket\psi$, $n$):
\begin{enumerate}
	\item For unitary $\hat{U}$ with eigenvector $\ket\psi$ and respective eigenvalue $e^{2\pi i\theta}$, apply the circuit below, on the input state \mbox{$\ket{0}^{\otimes n}\ket\psi$}.
	\item The register of qubits at the output to the QFT encodes $2^n\theta$ in binary representation, where $n$ determines the number of bits of precision in the output.
	\item $\Box$
\end{enumerate}
\begin{align}
\Qcircuit @C=1em @R=1.6em {
    \lstick{\ket{0}} & \gate{{H}} & \qw &\qw & \dots && \ctrl{3} & \multigate{2}{{QFT}^\dag} & \meter \\
    \lstick{\vdots} & \gate{{H}} & \qw &\qw & \dots & & \qw & \ghost{{QFT}^\dag} & \meter \\
    \lstick{\ket{0}} & \gate{{H}} & \ctrl{1} &\qw & \dots && \qw & \ghost{{QFT}^\dag} & \meter \\
    \lstick{\ket{\psi}} & \qw & \gate{{U}^{2^0}} &\qw & \dots & &\gate{{U}^{2^{n-1}}} & \qw & \qw \\
} \nonumber
\end{align}
}
\end{mdframed}
\captionspacealg \caption{Quantum phase-estimation algorithm.} \label{alg:q_phase_est}
\end{table}

The phase-estimation algorithm is not extremely useful as a standalone product, but acts as a necessary subroutine in many important algorithms, such as Shor's algorithm\index{Shor's algorithm} (Sec.~\ref{sec:shors_alg}) and topological data analysis\index{Topological!Data analysis} (Sec.~\ref{sec:TDA}).

%
% Quantum Simulation
%

\subsection{Quantum simulation} \index{Quantum chemistry}\index{Quantum simulation}\label{sec:quantum_sim_alg}

The field of quantum computation was originally inspired by Feynman's\index{Richard Feynman} observation that quantum systems cannot be efficiently classically simulated, and therefore maybe computers based on quantum principles could handle this problem. Indeed they can, as was shown by \cite{bib:lloyd1996universal}.

It requires little imagination to recognise that the applications for quantum simulation are enormous, given the multitude of quantum systems under active investigation by researchers across countless fields.

Consider a quantum system comprised of a global Hamiltonian, which may be decomposed into smaller local interaction terms,
\begin{align} \label{eq:Ham_sim_Ham}
\hat{H} = \sum_{i=1}^N \hat{H}_i	,
\end{align}
where each $\hat{H}_i$ acts on a subspace of dimension $m_i$ within the larger system. The evolution of the entire system is given by the unitary operator,
\begin{align}
\hat{U} = e^{i\hat{H}t}.
\end{align}
We wish to simulate this evolution.

From the Baker-Campbell-Hausdorff lemma\index{Baker-Campbell-Hausdorff lemma} we can approximate this as,
\begin{align}\label{eq:CBH_lemma}
	\hat{U} \approx \left(e^{i\hat{H}_1\frac{t}{n}}\dots e^{i\hat{H}_N\frac{t}{n}}\right)^n + O\left(\frac{t^2}{n}\right).
\end{align}
This representation effectively decomposes the global evolution into $nN$ discretised stages of,
\begin{align}
	\hat{U}_j = e^{i\hat{H}_j\frac{t}{n}},
\end{align}
each of which operates on an $m_i$-dimensional subspace, and may therefore be directly efficiently implemented as a unitary gate within the circuit model on a quantum computer. 

Clearly the error terms vanish in the limit of \mbox{$n\to\infty$}, whereby the simulation becomes exact. However, this requires an infinite number of gates via infinitesimal discretisation. We would rather approximate the solution using a finite number of gates. It follows from Eq.~(\ref{eq:CBH_lemma}) that for simulation accuracy $\delta$, the number of discrete steps scales as,
\begin{align}
n = O\left(\frac{t^2}{\delta}\right),
\end{align}
which scales efficiently with the duration of time being simulated and the accuracy of the simulation.

This approach is efficient and applies to any Hamiltonian which may be decomposed into local terms as per Eq.~(\ref{eq:Ham_sim_Ham}). However many other quantum simulation algorithms have since been described for simulating different types of quantum systems with different Hamiltonian structures \cite{bib:JLP, bib:RohdeWavelet15}. The algorithm is summarised in Alg.~\ref{alg:ham_sim}.

\begin{table}[!htbp]
\begin{mdframed}[innertopmargin=3pt, innerbottommargin=3pt, nobreak]
\texttt{
function HamiltonianSimulation($\hat{H}$, $t$, $\epsilon$):
\begin{enumerate}
	\item Hamiltonian to be simulated is of the form,
    \begin{align} \label{eq:Ham_sim_Ham}
\hat{H} = \sum_{i=1}^N \hat{H}_i	,
\end{align}
where $\hat{H}_i$ are local Hamiltonians operating on low-dimensional spaces, and $\hat{H}$ is the global Hamiltonian for the entire system.
\item For each $\hat{H}_i$ construct a quantum circuit implementing the unitary,
\begin{align}
	\hat{U}_j = e^{i\hat{H}_j\frac{t}{n}}.
\end{align}
\item The degree of discretisation, $n$, is chosen according to,
\begin{align}
n = O\left(\frac{t^2}{\delta}\right),
\end{align}
where $t$ is evolution time and $\delta$ is the accuracy.
\item Apply the circuits for simulating the local Hamiltonians according to the sequence,
\begin{align}
	\hat{U} = \left(\prod_{j=1}^N \hat{U}_j \right)^n.
\end{align}
\item The algorithm has runtime,
\begin{align}
	O\left(\frac{Nt^2}{\delta}\right).
\end{align}
\item $\Box$
\end{enumerate}
}
\end{mdframed}
\captionspacealg \caption{Quantum Hamiltonian simulation algorithm.} \label{alg:ham_sim}
\end{table}

%
% Integer Factorisation
%

\subsection{Integer factorisation} \index{Shor's algorithm}\index{Hidden subgroup problem}\index{Period-finding}
	\index{Integer factorisation}\label{sec:shors_alg}

By far the most influential quantum algorithm, and one of the first, is Shor's integer factorisation algorithm \cite{bib:ShorFactor}. The problem is simply to find \mbox{$x,y\in\mathbb{Z}^+$} given $z=xy$.
 
 While this algorithm is known to reside in \textbf{BQP} (since it has an efficient quantum algorithm), it is strongly believed not to be \textbf{BQP}-complete. Similarly, while it is known to reside in \textbf{NP} (since it can be efficiently classically verified using simple multiplication), it is strongly believed not to be \textbf{NP}-complete, thereby placing it in the `limbo zone' of \textbf{NP}-intermediate complexity.\index{NP \& NP-complete}

This problem is of immense interest to the field of cryptography, since finding private RSA keys (Sec.~\ref{sec:public_key_crypt}) computationally can be reduced to this problem. An adversary with access to an efficient factoring algorithm could completely compromise RSA cryptography. For this reason, Shor's algorithm is responsible for the large investments made into quantum computing by nation states, and their military and intelligence agencies!

Shor's algorithm works by first reducing integer factorisation to another problem, \textit{period finding}\index{Period-finding}. For the function,
\begin{align}
f(x)= a^x \,\mathrm{mod}\, N,	
\end{align}
find its period \mbox{$r\in\mathbb{Z}^+$}, the smallest integer such that,
\begin{align}
f(x)=f(x+r) \,\mathrm{mod}\, N.	
\end{align}
With the ability to solve the period finding problem, an efficient classical algorithm exists for transforming this solution to a solution for factoring.

The algorithm derives its power from a quantum Fourier transform subroutine (Sec.~\ref{sec:QFT_alg}), and has runtime,
\begin{align}
O((\log N)^2(\log\log N)(\log\log\log N)),	
\end{align}
making it quantum-efficient. By contrast, the best-known classical algorithm, the \textit{general number field sieve}\index{General number field sieve}, requires time,
\begin{align}
	O\left((e^{1.9(\log N)^{\frac{1}{3}}(\log\log N)^\frac{2}{3}}\right).
\end{align}

The algorithm is described in Alg.~\ref{alg:shor}.

\begin{table}[!htbp]
\begin{mdframed}[innertopmargin=3pt, innerbottommargin=3pt, nobreak]
\texttt{
function Shor(a,N):
\begin{enumerate}
	\item For function,
	\begin{align}
		f(x)= a^x \,\mathrm{mod}\, N,	
	\end{align}
	we wish to find the smallest integer $r$ such that,
	\begin{align}
		f(x)=f(x+r) \,\mathrm{mod} \,N.	
	\end{align}
	\item Run the circuit below.
	\item Efficient classical processing of the measured output registers reveals $r$ with high probability.
	\item Classically verify that,
	\begin{align}
		f(x)=f(x+r) \,\mathrm{mod} \,N.	
	\end{align}
	\item If it isn't then repeat the circuit to obtain more samples. 
    \item $\Box$
\end{enumerate}
\begin{center}
\if 1\doublecol
\includegraphics[clip=true, width=\textwidth]{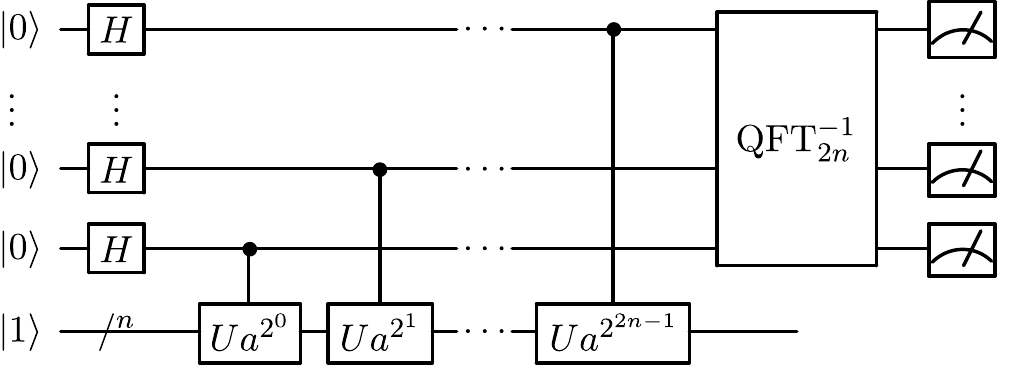}
\else
\includegraphics[clip=true, width=0.55\textwidth]{shor_circuit}
\fi
\end{center}
}
\end{mdframed}
\captionspacealg \caption{Shor's quantum algorithm for integer factorisation.} \label{alg:shor}
\end{table}

%
% Quantum Machine Learning
%

\subsection{Quantum machine learning} \index{Quantum machine learning}

With present-day classical computing power becoming ever more powerful, the desire to employ machine learning algorithms has become immense. Machine learning techniques have become indispensable in many fields. Quantum machine learning has emerged as an exciting application for quantum computing power, to enhance the power of machine learning algorithms. We devote the entirety of Sec.~\ref{sec:quantum_mind} to this topic.

%
% Topological Data Analysis
%

\subsection{Topological data analysis} \index{Topological!Data analysis}\label{sec:TDA}

The internet currently comprises extraordinary amounts of data, from which useful information must be extracted if this vast amount of data is to be utilised effectively. For example, firms like Google and Facebook must extract meaningful information from their databases of user behaviour in order to make appropriate advertising suggestions. This task is extremely valuable -- a small improvement in a recommendation engine\index{Recommendation engine}, for example, could be worth many millions of dollars in revenue.

Performing analyses like these is extremely computationally challenging when dealing with such enormous datasets as Facebook's user database or Google's website database, so-called \textit{big data analysis}\index{Big data analysis}.

One avenue for the analysis of large, complex datasets is via homology theory\index{Homology theory}, which yields \textit{topological data analysis} (TDA)\index{Topological!Data analysis}. In particular, the so-called \textit{Betti numbers}\index{Betti numbers} characterise the nature of interconnectedness within a dataset. Specifically, the $k$th Betti number, $\beta_k$, is the number of $k$-dimensional holes and voids in a dataset. For example, $\beta_0$, $\beta_1$ and $\beta_2$ represents the number of connected components, 1-dimensional holes, and 2-dimensional voids in a dataset respectively.

Calculating Betti numbers exactly is known to be \textbf{PSPACE}-complete\index{PSPACE}\footnote{\textbf{PSPACE} is the complexity class of problems requiring polynomial memory with unbounded runtime, a class that is not classically efficient.}, and the best-known classical approximation algorithm has exponential runtime,
\begin{align}
O\left(2^n \log \left(\frac{1}{\delta}\right)\right),
\end{align}
for accuracy $\delta$ on $n$ data-points.

Recently, improved quantum algorithms for approximating the Betti numbers have been presented \cite{bib:lloyd2016quantum, bib:PhysRevLett.113.130503}, with polynomial runtime of only,
\begin{align}
O\left(\frac{n^5}{\delta}\right).
\end{align}
An elementary photonic experimental demonstration of this algorithm has been performed using a small dataset \cite{huang2018demonstration}.

\begin{figure}[!htbp]
\if 1\doublecol
	\includegraphics[clip=true, width=0.475\textwidth]{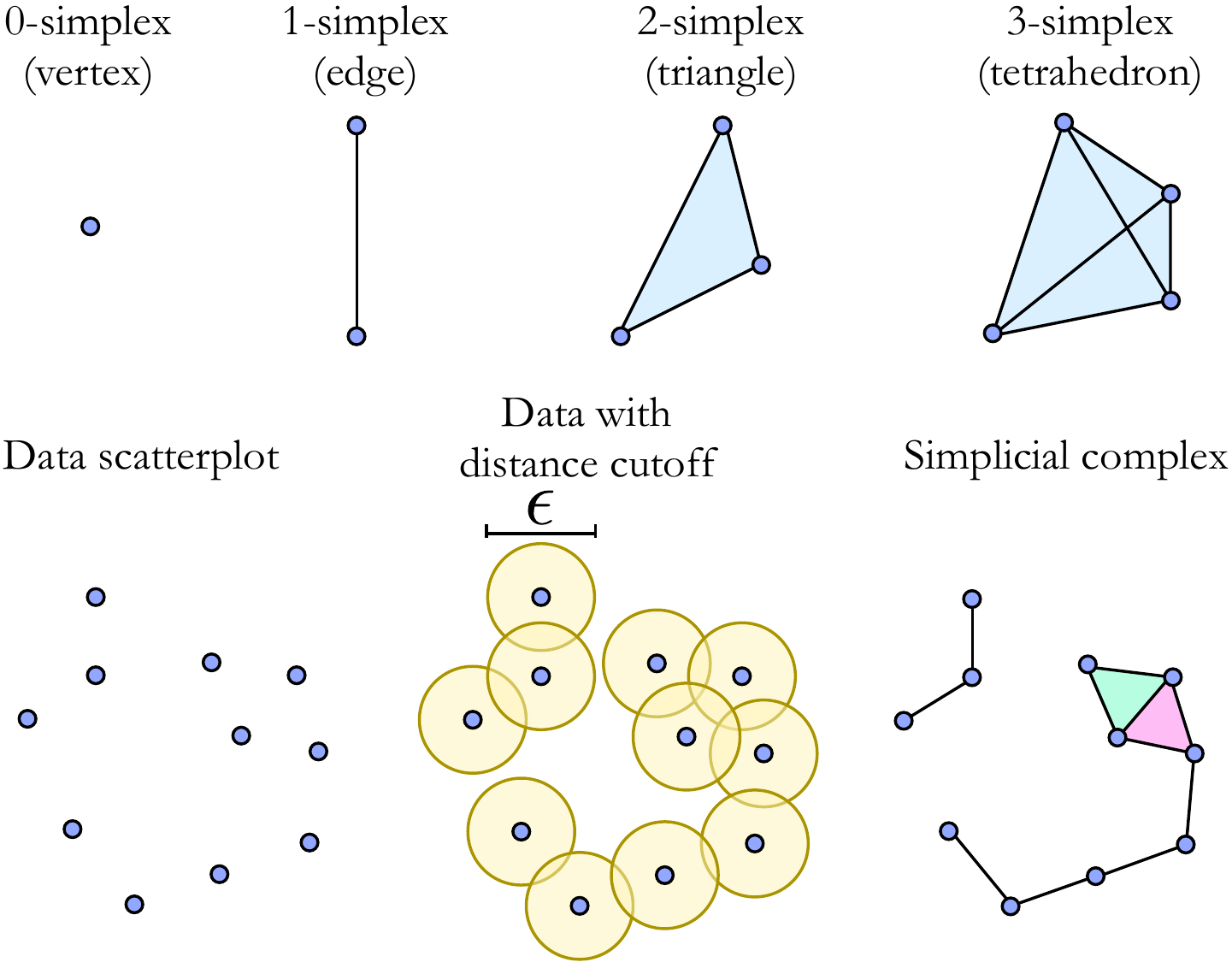}
\else
	\includegraphics[clip=true, width=0.6\textwidth]{TDA_simplices_complex}
\fi
\captionspacefig \caption{(top) $k$-simplices\index{Simplex} are constructed as fully connected subsets of the data of different dimensions $k$. (bottom) A distance cutoff is applied to create edges within a scatterplot of raw data, from which the simplicial complex\index{Simplicial complex} is constructed. The shown example of a simplicial complex contains two 2-simplices (coloured triangles), and 6 1-simplices (edges). The final simplicial complex is highly dependent on the choice of cutoff $\epsilon$. In general, as $\epsilon$ is increased the complex will contain more simplices of higher dimension, since vertices will have more immediate neighbours.} \label{fig:TDA_simplex}	
\end{figure}

Taking a dataset with well-defined distances between data-points\footnote{`Distance' could be any arbitrary metric\index{Distance metric} of any dimension.}, we begin by applying a distance cutoff $\epsilon$\index{Distance cutoff} to define connections between data-points. We define $k$-simplices\index{Simplex} within the dataset, which are fully connected subsets of \mbox{$k+1$} data-points, with \mbox{$k(k+1)/2$} undirected edges between them. The full set of simplices defines the dataset's \textit{simplicial complex}\index{Simplicial complex} for a given distance cutoff $\epsilon$. Construction of the so-called Vietoris-Rips simplicial complex is described in Fig.~\ref{fig:TDA_simplex}.

The first step in the algorithm is to construct the simplicial complex superposition state,
\begin{align}
\ket\psi_k^\epsilon = \frac{1}{\sqrt{|S_k^\epsilon|}} \sum_{s_k\in S_k^\epsilon} \ket{s_k},
\end{align}
where $s_k$ denotes a $k$-simplex from the simplicial complex $S_k^\epsilon$. This superposition can be constructed by employing a Grover search\index{Grover's algorithm} using a set-membership oracle function\index{Oracles},
\begin{align}
f_\epsilon(s_k) = \left\{ \begin{array}{l l}
 1, & \mathrm{if}\,\,s_k\in S_k^\epsilon \\
 0, & \mathrm{otherwise}
\end{array}\right.,
\end{align}
yielding quadratically enhanced efficiency in the simplicial complex state preparation.

From the superposition state, the uniform mixture of simplices state,
\begin{align}
\hat\rho_k^\epsilon = \frac{1}{|S_k^\epsilon|} \sum_{s_k\in S_k^\epsilon} \ket{s_k}\bra{s_k},
\end{align}
is easily prepared with the addition of CNOT gates and ancillary qubits\footnote{Using parallel CNOT gates one can transform an arbitrary superposition state into a redundantly encoded\index{Redundant encoding} equivalent, \mbox{$\hat{U}_\mathrm{CNOTs}\sum_i \lambda_i \ket{\psi_i}\ket{0} \to \sum_i \lambda_i \ket{\psi_i}\ket{\psi_i}$}, following which tracing out the redundant copy takes us to its uniform mixture, \mbox{$\hat\rho=\sum_i |\lambda_i|^2\ket{\psi_i}\bra{\psi_i}$}.}.

The quantum TDA algorithm then takes the simplicial complex mixed state and estimates the full set of Betti numbers by employing a phase-estimation algorithm\index{Phase!Estimation!Algorithm} (Sec.~\ref{sec:phase_est_alg}), which induces an exponential algorithmic runtime improvement. The full TDA algorithm is summarised in Alg.~\ref{alg:TDA}.

Performing the TDA across a range of $\epsilon$ yields a \textit{barcode}\index{Barcode} representation for the dataset's topology. Topological features which persist over large ranges of $\epsilon$ can then be regarded as robust features of the dataset, whereas ones which only persist over a small range of $\epsilon$ can be regarded as localised, non-persistent features, which might correspond to noise for example, and be filtered out prior to further analysis. The barcode representation thereby gives us an extremely robust characterisation of the topology of the data in a scale-dependent way.

As an example for how this type of technique might be applied, consider Facebook's\index{Facebook} user database. The distance metric might relate to how similar users' interests are, or how closely related they are via their friendship networks. Then examining the barcode representation of the data by scanning over $\epsilon$ would provide insight into the persistence and robustness of these relationships at different scales within the network. At different scales we could investigate topological relationships and features at the level of individuals, family or friendship networks, communities, common interest groups, between cities, across demographic characteristics, or between nations, for example.

\begin{table}[!htbp]
\begin{mdframed}[innertopmargin=3pt, innerbottommargin=3pt, nobreak]
\texttt{
function TDA(dataPoints):
\begin{enumerate}
	\item Implement a Grover search on the set of data-points with set-membership oracle function,
	\begin{align}
f_\epsilon(s_k) = \left\{ \begin{array}{l l}
 1, & \mathrm{if}\,\,s_k\in S_k^\epsilon \\
 0, & \mathrm{otherwise}
\end{array}\right.,
\end{align}
which prepares the simplicial complex superposition state,
\begin{align}
\ket\psi_k^\epsilon = \frac{1}{\sqrt{|S_k^\epsilon|}} \sum_{s_k\in S_k^\epsilon} \ket{s_k},
\end{align}
	\item Using ancillary qubits, CNOT gates and a trace-out operation, prepare the simplicial complex uniform mixed state,
	\begin{align}
\hat\rho_k^\epsilon = \frac{1}{|S_k^\epsilon|} \sum_{s_k\in S_k^\epsilon} \ket{s_k}\bra{s_k},
\end{align}
	\item Perform quantum phase-estimation on the simplicial complex mixed state.
	\item Classical post-processing reveals the Betti numbers.
	\item Algorithm has runtime,
\begin{align}
O\left(\frac{n^5}{\delta}\right),
\end{align}
for accuracy $\delta$.
\item $\Box$
\end{enumerate}
\begin{center}
\if 1\doublecol
	\includegraphics[clip=true, width=\textwidth]{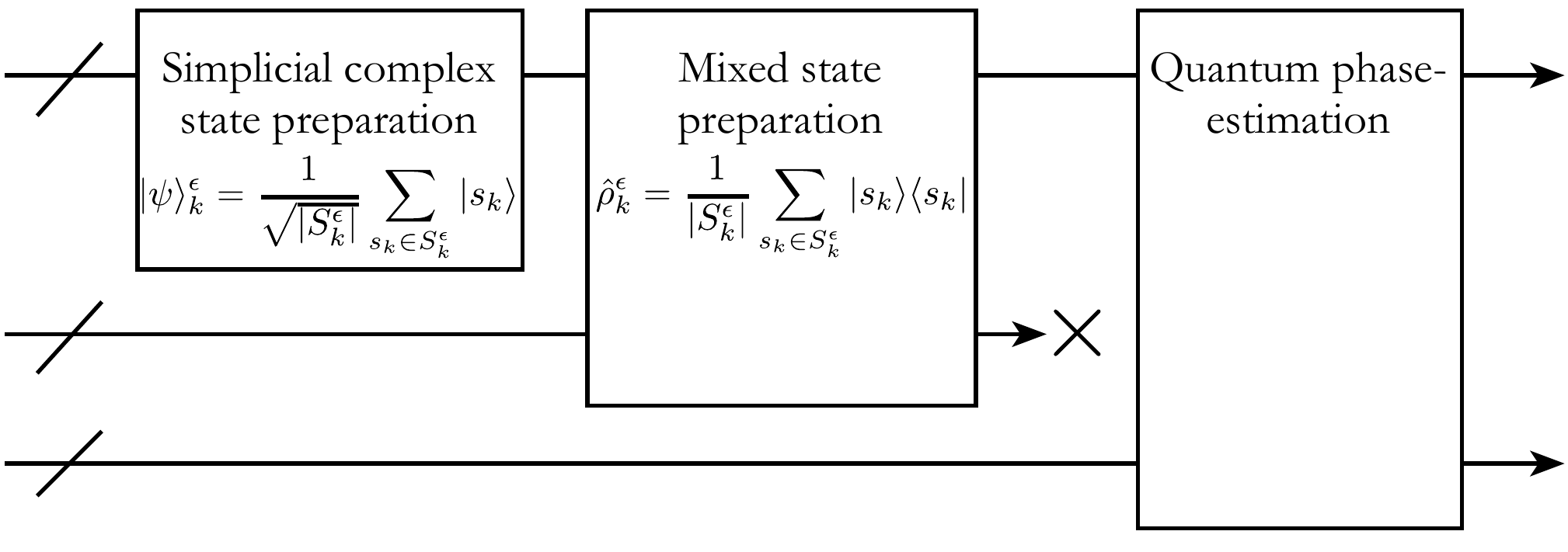}
\else
	\includegraphics[clip=true, width=0.6\textwidth]{TDA_circuit}
\fi
\end{center}
}
\end{mdframed}
\captionspacealg \caption{Quantum topological data analysis algorithm for calculating Betti numbers.} \label{alg:TDA}
\end{table}

%
% Sampling Problems
%

\subsection{Sampling problems}\index{Sampling problems}\label{sec:sampling_problems}

The algorithms described previously are examples of \textit{decision problems}\index{Decision problems}, whereby the computation answers a question, providing a well-defined output for a well-defined input. Another entirely different class of problems are the so-called \textit{sampling problems}, whereby the goal is to accurately reproduce samples taken from some probability distribution. By their nature, these algorithms are statistical and generally their output cannot be associated with the answer to a decision problem. Nonetheless, despite being an entirely different category of problems, they reside in distinct sampling complexity classes, some of which are classically efficient to simulate, others not.

\begin{figure}[!htbp]
\includegraphics[clip=true, width=0.3\textwidth]{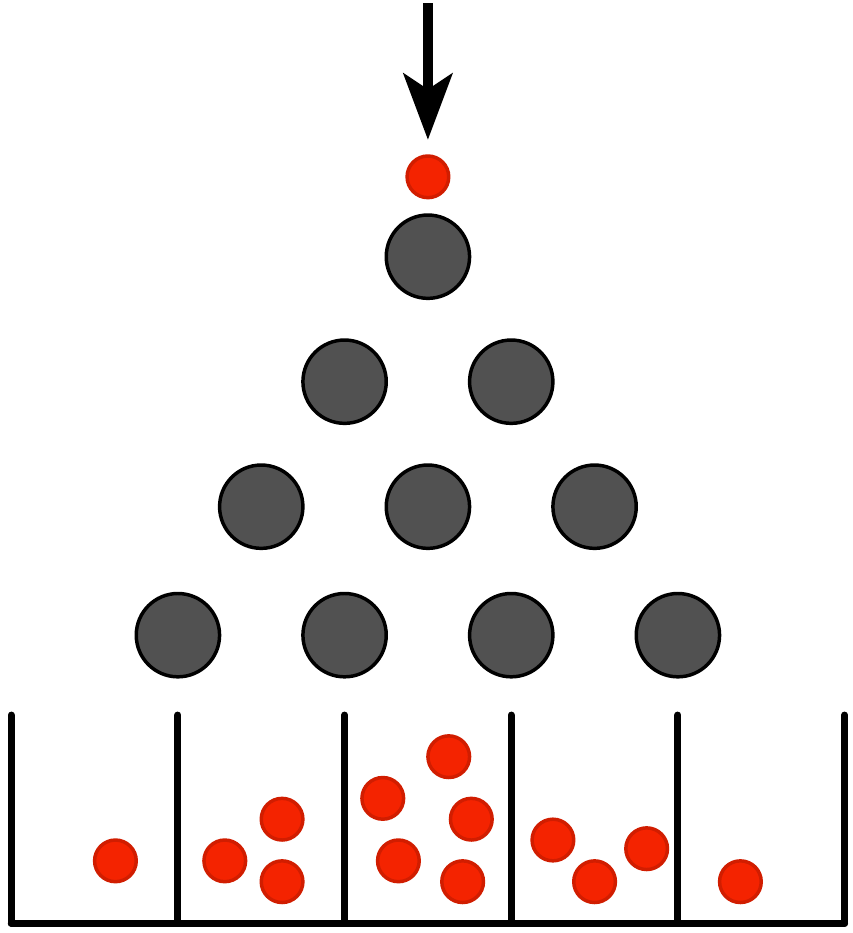}
\captionspacefig \caption{The Galton board yields a very simple classical sampling problem. Balls (red) are input at the top and allowed to fall freely through the pyramid network of pegs (grey), at which balls bounce to the left or right with 50\% probability. At the output the balls are collected into buckets, which populate according to a binomial distribution\index{Binomial distribution}. The computational problem is to produce statistically accurate samples from such a device, which classically is efficiently implemented by sampling the binomial distribution.}	\label{fig:galton_board}\index{Galton board}
\end{figure}

The simplest example of a classical sampling problem is the propagation of balls through a Galton board\index{Galton board}, as shown in Fig.~\ref{fig:galton_board}. This is a computationally easy problem, whose simulation requires only calculating a binomial distribution, which is classically straightforward.

An equivalent quantum sampling problem would be to replace the pegs in the Galton board with beamsplitters, and the balls with photons. Now we have an analogous problem, but defined in terms of photonic wave-function amplitudes rather than classical transition probabilities. When generalised to the multi-photon context, this problem turns into the so-called \textsc{BosonSampling}\index{Boson-sampling} problem, which is discussed in detail in Sec.~\ref{sec:boson_sampling}, a problem believed to not have an efficient classical simulation algorithm \cite{bib:AaronsonArkhipov10}.

In computationally hard sampling problems, e.g quantum ones, the goal is generally not to reconstruct the complete probability distribution. This is generally impossible in the quantum context, since the number of basis states we are sampling from is exponentially large, and therefore cannot be efficiently fully reconstructed. Thus the goal of \textit{sampling} is distinct from the goal of \textit{reconstruction}. In the case of sampling problems, we are satisfied with incomplete data, provided that we only have to take a polynomial number of samples for the sake of computational efficiency.

However, this limitation to having incomplete data makes the problem of verification\index{Verification} a conceptually challenging one. We are unable to simply compare our statistical results with that of a reliable reference, since our massively incomplete samples are almost certainly going to be entirely different ones, providing no benchmark for comparison. Therefore verification by comparison is effectively ruled out, requiring more elaborate verification techniques, something which has become a highly active area of research on its own. Addressing the verification problem is of course an important one -- when we reach the milestone of achieving `quantum supremacy'\index{Quantum supremacy}, we'd sure like to be able to convincingly prove to the world that we in fact did!

Countless other quantum sampling problems have been described. Most notably, the IQP (instantaneous quantum protocol)\index{Instantaneous quantum protocol (IQP)} sampling problem is a very simple prescription for a quantum algorithm that has been shown to likely be classically hard despite having very shallow circuit depth and all-commuting gates \cite{lund2017quantum}.

Although not as obviously algorithmically useful as decision problems, sampling problems have received much interest owing to their generally very simple construction. For example, \textsc{BosonSampling} requires only single-photon inputs, evolved via a passive linear optics\index{Linear optics} beamsplitter network, and measured using photo-detection\index{Photo-detection}. IQP sampling requires only single-qubit Hadamard gates\index{Hadamard!Gate} and generalised multi-qubit controlled-phase gates\index{Generalised controlled-phase gates}\footnote{Generalised CZ gates are simply gates diagonal in the logical basis, where the diagonal elements are arbitrary phases, \mbox{$\hat{U}_\mathrm{CZ} = \mathrm{diag}(e^{i\theta_1},\dots,e^{i\theta_n})$}.}, thereby sampling from the logical state,
\begin{align}
\ket{\psi_\mathrm{out}} = \hat{H}^{\otimes n} \cdot \hat{U}_\mathrm{CZs} \cdot \hat{H}^{\otimes n} \ket{0}^{\otimes n}.	
\end{align}
Because CZ gates commute, they can be performed in parallel, giving the IQP circuit construction very low circuit depth\index{Circuit!Depth}. The IQP protocol is shown in Alg.~\ref{alg:IQP_samp}.

\begin{table}[!htbp]
\begin{mdframed}[innertopmargin=3pt, innerbottommargin=3pt, nobreak]
\texttt{
function IQP\_sampling():
\begin{enumerate}
    \item Prepare the $n$-qubit state,
    \begin{align}
    \ket{\psi_\mathrm{in}} = \ket{0}^{\otimes n}.
    \end{align} 
    \item Apply the $n$-qubit Hadamard transform,
    \begin{align}
	\hat{H}^{\otimes n}.
	\end{align}
    \item Apply some choice of generalised CZ gates,
	\begin{align}
	\hat{U}_\mathrm{CZs} = \mathrm{diag}(e^{i\theta_1},\dots,e^{i\theta_n}).
	\end{align}
	\item Apply another $n$-qubit Hadamard transform,
    \begin{align}
	\hat{H}^{\otimes n}.
	\end{align}
	\item The output state is,
	\begin{align}
		\ket{\psi_\mathrm{out}} = \hat{H}^{\otimes n} \cdot \hat{U}_\mathrm{CZs} \cdot \hat{H}^{\otimes n} \ket{0}^{\otimes n}.	
	\end{align}
	\item Measure all qubits in the computational basis, yielding bit-string $\vec{x}$, which occurs with probability,
	\begin{align}
		P_{\vec{x}} = |\bra{\vec{x}} \hat{H}^{\otimes n} \cdot \mathrm{diag}(e^{i\theta_1},\dots,e^{i\theta_n}) \cdot \hat{H}^{\otimes n}  \ket{0}^{\otimes n}|^2.
	\end{align}
	\item Repeat protocol $O(\mathrm{poly}(n))$ times.
	\item $\Box$
\end{enumerate}
	\item \begin{align}
\Qcircuit @C=1em @R=1.6em {
    \lstick{\ket{0}} & \gate{{H}} & \multigate{2}{{U}_\mathrm{CZs}} & \gate{{H}} & \meter \\
    \lstick{\vdots} & \gate{{H}} & \ghost{{U}_\mathrm{CZs}} & \gate{{H}} & \meter \\
    \lstick{\ket{0}} & \gate{{H}} & \ghost{{U}_\mathrm{CZs}} & \gate{{H}} & \meter \\
} \nonumber
\end{align}
}
\end{mdframed}
\captionspacealg \caption{The IQP sampling problem, which is believed to be a classically hard problem.} \label{alg:IQP_samp}
\end{table}

These reduced resource requirements makes both analysis and physical construction of some sampling problems far simpler than a universal quantum computer, yet nonetheless they implement computationally hard problems. For these reasons, many researchers regard non-universal sampling problems as being likely candidates for the first demonstration of quantum supremacy\index{Quantum supremacy}.

%
% Shallow quantum circuits
%

\subsection{Shallow quantum circuits}\index{Shallow quantum circuits}\label{sec:shallow_circs}

\sectionby{Zixin Huang}\index{Zixin Huang}

As discussed in Sec.~\ref{sec:NISQ}, we have good reason to believe that quantum computers have super-classical capabilities. Recently, an unconditional proof of a quantum speed-up has been shown for a particular class of circuits.

It is incredibly difficult to compare quantum versus classical polynomial time computation, and currently we have only noisy gates in the lab, further complicating the goal of achieving quantum supremacy. Given $n$ qubits, circuit depth $d$, and error rate $\epsilon$, intuitively, we demand the condition,
\begin{align}
nd \ll \frac{1}{\epsilon},
\end{align}
to hold, in order for a quantum algorithm to run successfully. If $d$ is large, then noise dominates and our circuit is classically simulable. On the other hand, if we have large $n$ and small $d$ (constant depth circuits), we may observe a quantum speed up in the near future. Given these factors, we are motivated to examine `shallow circuits'. Now we will see that there is a distinction between constant depth circuits run using classical versus quantum algorithm.

The result is for the following hidden linear function (HLF) problem\index{Hidden linear function problem} \cite{bib:bravyi2018quantum}. Given an \mbox{$n\times n$} symmetric binary matrix\index{Symmetric binary matrices} $A$, we specify a quadratic form\index{Quadratic form},
\begin{align}
q(\vec x) = \vec x^T A \vec x \, (\mathrm{mod} \,4).
\end{align}
The goal is to find a binary vector $\vec z \in \{0,1\}^n$ such that,
\begin{align}
q(\vec x) = 2 \vec z^T \vec x,
\end{align}
for all $\vec x$ in the binary null-space of $A$.

The quantum algorithm that solves the 2D HLF problem is similar to the Bernstein-Vazirani problem\index{Bernstein-Vazirani problem} \cite{bib:bernstein1997quantum}. The algorithm, shown in Alg.~\ref{alg:shallow_circs}, has similar structure to the IQP sampling problem (Alg.~\ref{alg:IQP_samp}).

\begin{table}[!htbp]
\begin{mdframed}[innertopmargin=3pt, innerbottommargin=3pt, nobreak]
\texttt{
function ShallowQuantumCircuits():
\begin{enumerate}
    \item Prepare the $n$-qubit state,
    \begin{align}
    \ket{\psi_\mathrm{in}} = \ket{0}^{\otimes n}.
    \end{align}
    \item Apply the $n$-qubit Hadamard transform\index{Hadamard!Transform},
    \begin{align}
    \hat{H}^{\otimes n}.
    \end{align}
    \item Apply the oracle,
	\begin{align}
		\hat{U}_q \ket{\vec x} = i^{q(\vec x)} \ket{\vec x},
	\end{align}
	for bit-string \mbox{${\vec x}\in \{0,1\}^{n}$}, composed of a product of CZ gates\index{Controlled-Z (CZ) gates} and phase-shifts\index{Phase!Shift gates},
	\begin{align}
\hat{S}_i = \begin{pmatrix}
    1 & 0\\
    0 & -i
    \end{pmatrix}.
\end{align}
	\item Apply another $n$-qubit Hadamard transform,
    \begin{align}
    \hat{H}^{\otimes n}.
    \end{align}
    \item The output state is,
    \begin{align}
    \ket{\psi_\mathrm{out}} &= \hat{H}^{\otimes n}\cdot\hat{U}_q\cdot\hat{H}^{\otimes n}\ket{0}^{\otimes n}\nonumber\\
    &= \frac{1}{2^n}\sum_{\vec x,\vec z \in \{ 0,1\}^n} i^{q(\vec x)} (-1)^{\vec z^T \vec x}\ket{\vec z}.
    \end{align}
    \item Measure all qubits in the computational basis, yielding measurement outcomes \mbox{${\vec z}\in \{0,1\}^{n}$}, with probabilities,
    \begin{align}
   	 P(\vec z) = |\bra{\vec z}\hat{H}^{\otimes n}\cdot\hat{U}_q\cdot\hat{H}^{\otimes n}\ket{0}^{\otimes n}|^2.	
    \end{align}
	\item Repeat protocol \mbox{$O(\mathrm{poly}(n))$} times.
	\item $\Box$
\end{enumerate}
\begin{align}
\Qcircuit @C=1em @R=1.6em {
    \lstick{\ket{0}} & \gate{{H}} & \multigate{2}{{U}_q} & \gate{{H}} & \meter \\
    \lstick{\vdots} & \gate{{H}} & \ghost{{U}_q} & \gate{{H}} & \meter \\
    \lstick{\ket{0}} & \gate{{H}} & \ghost{{U}_q} & \gate{{H}} & \meter \\
} \nonumber
\end{align}
}
\end{mdframed}
\captionspacealg \caption{Quantum computing using shallow circuits, where circuit depth scales as \mbox{$O(\log d)$}.} \label{alg:shallow_circs}
\end{table}

Here a quantum circuit $\hat{U}_q$ acts on the computational basis states \mbox{$\vec x \in \{ 0,1\}^n$} via a generalised controlled-phase operation,
\begin{align}
\hat{U}_q \ket{\vec x} = i^{q(\vec x)} \ket{\vec x},
\end{align}
for some oracle function $q(\vec x)$. The solution is obtained by measuring the state $\ket{\Psi_q}$ in the computational basis. Here,
\begin{align}
\ket{\Psi_q} &= \hat{H}^{\otimes n }\cdot\hat{U}_q\cdot\hat{H}^{\otimes n} \ket{0}^{\otimes n} \nonumber \\
&= \frac{1}{2^n}\sum_{\vec x,\vec z \in \{ 0,1\}^n} i^{q(\vec x)} (-1)^{\vec z^T \vec x}\ket{\vec z}.
\end{align}

The circuit $\hat{U}_q$ can be decomposed into a product of CZ gates and phase-shift gates, which all commute, thereby allowing the shallow circuit-depth.

\begin{figure}[!htbp]
\includegraphics[clip=true, width=0.35\textwidth]{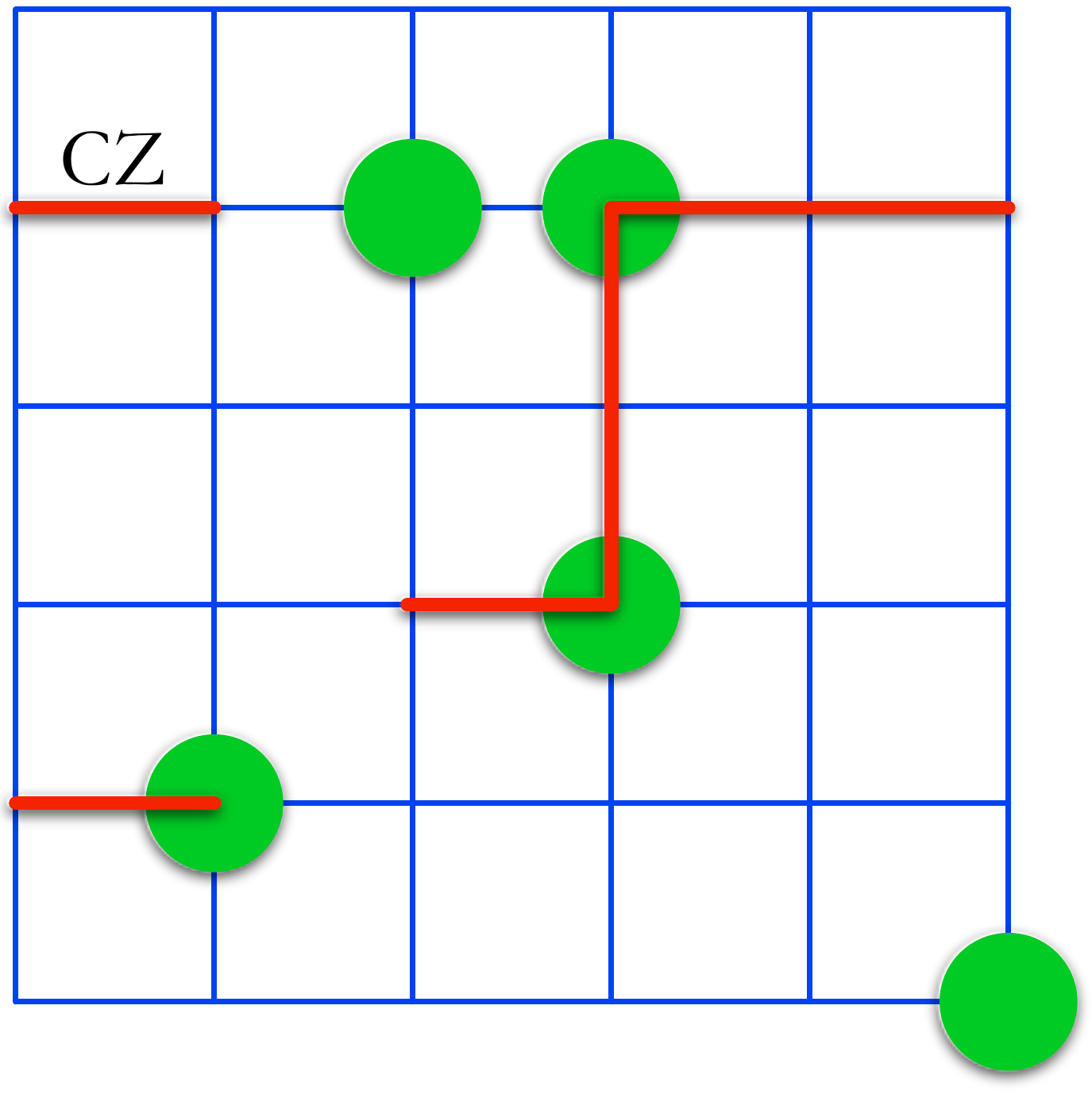}
\captionspacefig \caption{\label{fig:2DHLFgrid} Implementation of the HLF problem on a 2D grid. The qubits are represented as green circles (some omitted for clarity), and CZ gates are drawn in red.  To implement the circuit $\hat U_Q$, the qubits at the vertex of the graph $G(A)$ embedded into a 2D grid. Then $\hat U_q$ can be decomposed into a product of nearest-neighbour CZ gates and $S$.}
\end{figure}

If we lay out the circuit in 2D (hence the 2D HLF problem), it only requires classically-controlled Clifford gates\index{Clifford gates} between nearest neighbour qubits on a 2D grid. An example is shown in Fig.~\ref{fig:2DHLFgrid}. An instance of the HLF problem can be described by a graph $G(A)$ with $n$ vertices such that the off-diagonal part of A is the adjacency matrix of $G(A)$. Here $A_{i,j}=0$ unless $i,j$ are nearest neighbour vertices of the grid or $i = j$. We place qubits at each vertex and every edge corresponds to an input bit. The HLF problem is solved by applying nearest-neighbour CZ gates and $S_i$ phase gates at the correct site.

Now, comparing the quantum algorithm to a classical one, it has been proven that any probabilistic classical circuit with bounded input that solves the 2D HLF problem with high probability, $\epsilon \leq 1/8$, must have depth increasing logarithmically with input size \cite{bib:bravyi2018quantum}. The speed-up here is provided by quantum non-locality\index{Quantum non-locality}. 

The fact that a problem can be solved with constant depth quantum circuits, but not classical ones shows that there is a clear exponential computational separation between them. 

\latinquote{Sic semper tyrannis.}

%
% Physical Architectures For Quantum Computing
%

\section{Physical architectures for quantum computing} \label{sec:archs_QC} \index{Physical architectures}

\dropcap{T}{he} models for quantum computation introduced in the previous section are abstractions of algorithms in terms of elementary operations. But elementary operations must ultimately be physically realised. There are countless physical architectures for realising quantum computations, far too many to describe here, each with their own advantages and disadvantages, and it is far from clear which physical architecture(s) will ultimately win the quantum race.

Here we will summarise some of the physical architectures most applicable to networking. Since we reasonably anticipate that future quantum networking will be optically mediated, we focus on pure-optical and hybrid-optical architectures, on the basis that these will naturally lend themselves to optical interfacing.

%
% Universal Linear Optics
%

\subsection{Universal linear optics} \label{sec:KLM_univ} \index{Universal linear optics quantum computation}\index{Knill-Laflamme-Milburn (KLM)}

With single-photon encoding of qubits in the quantum network, the obvious architecture to implement quantum computation is linear optics quantum computing (LOQC) \cite{bib:KLM01} (KLM), since the states being processed by the computer are of the same form as the states traversing the network. See \cite{bib:Kok05, bib:KokLovettBook} for excellent introductions to this what has become a very broad and exciting field.

LOQC allows universal quantum computing to be implemented using single-photon polarisation or dual-rail encoding, with only linear optics interactions, i.e beamsplitter/phase-shifter networks \cite{bib:Reck94}, with the addition of quantum memory, and fast-feedforward, whereby some photons are measured, and the remaining part of the optical circuit is dynamically reconfigured based on the measurement outcomes. The former is readily available technology today, and elementary demonstrations have been performed \cite{bib:OBrien03, bib:carolan2015universal}, but the latter two have proven to be somewhat more challenging.

\begin{figure}[!htbp]
\includegraphics[clip=true, width=0.42\textwidth]{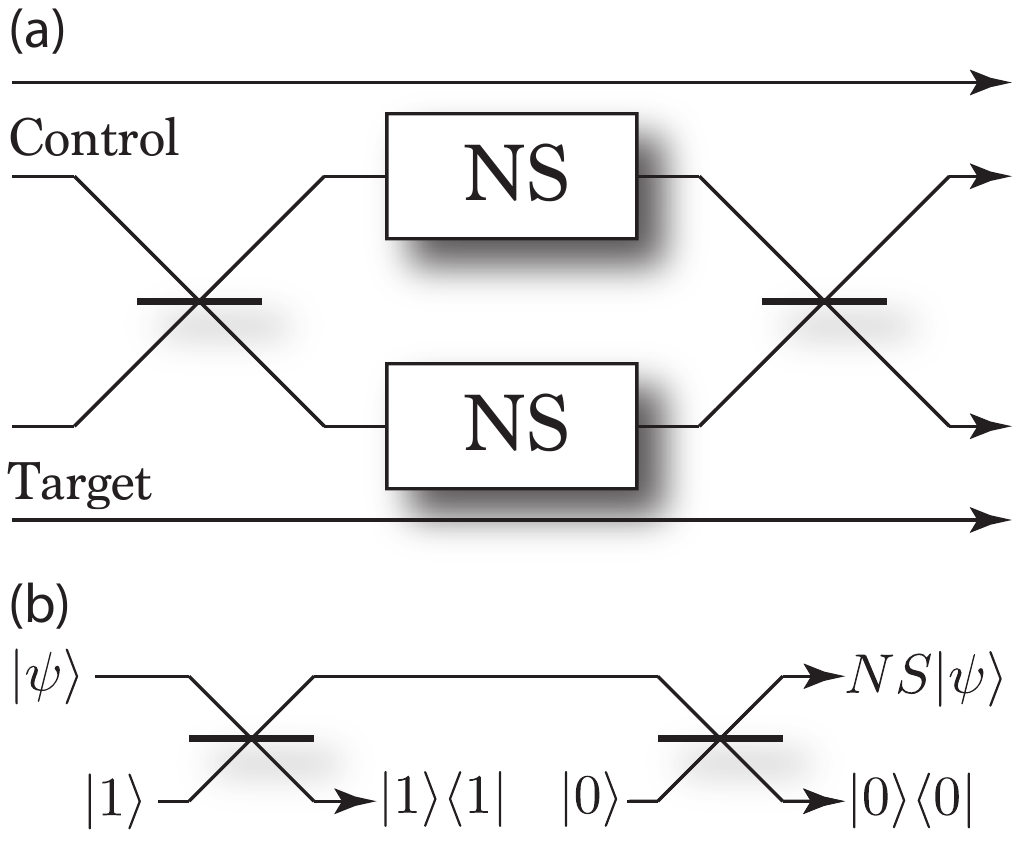}
\captionspacefig \caption{(a) A KLM CZ gate, employing dual-rail encoding, constructed from two non-linear sign-shift (NS) gates, which apply a $\pi$ phase-shift to only $\ket{2}$ terms in the photon-number basis. (b) Construction of the non-deterministic linear optics NS gate. Two ancillary states -- one $\ket{1}$ and one $\ket{0}$ -- are employed, and two photo-detectors post-select upon detecting $\ket{1}\bra{1}$ and $\ket{0}\bra{0}$ respectively. The beamsplitter reflectivities in (a) are 50:50, and in (b) chosen such that the amplitudes obey Eq.~(\ref{eq:NS_trans}).}. \label{fig:KLM_gate} \index{Knill-Laflamme-Milburn (KLM)}\index{Non-linear!Sign-shift (NS) gate}
\end{figure}

Originally it was believed that universal optical quantum computation, specifically the implementation of 2-qubit entangling gates (such as CNOT or CZ gates), would require extremely (and unrealistically) strong optical non-linearities that implement a non-linear sign-shift (NS) gate,
\begin{align} \label{eq:NS_trans}\index{Non-linear!Sign-shift (NS) gate}
NS: \alpha\ket{0}+\beta\ket{1}+\gamma\ket{2}\to\alpha\ket{0}+\beta\ket{1}-\gamma\ket{2},
\end{align}
in the photon-number basis, up to normalisation (which is determined by the post-selection success probability). That is, it applies a $\pi$ phase-shift to only the $\ket{2}$ component of a photon-number superposition. The breakthrough result by KLM demonstrated that this is in fact not the case at all. Instead, the NS gate can be implemented non-deterministically using post-selected linear optics. Two such NS gates allow the construction of a single CZ gate. The construction of the KLM NS and CZ gates are shown in Figs.~\ref{fig:KLM_gate} \& \ref{fig:KLM_explain}. Equivalently, a CNOT gate may be trivially constructed via conjugation by Hadamard gates, based on the identity \mbox{$\hat{H}\hat{Z}\hat{H}=\hat{X}$}.

\if 1\doublecol
	\begin{figure}[!htbp]
	\includegraphics[clip=true, width=0.43\textwidth]{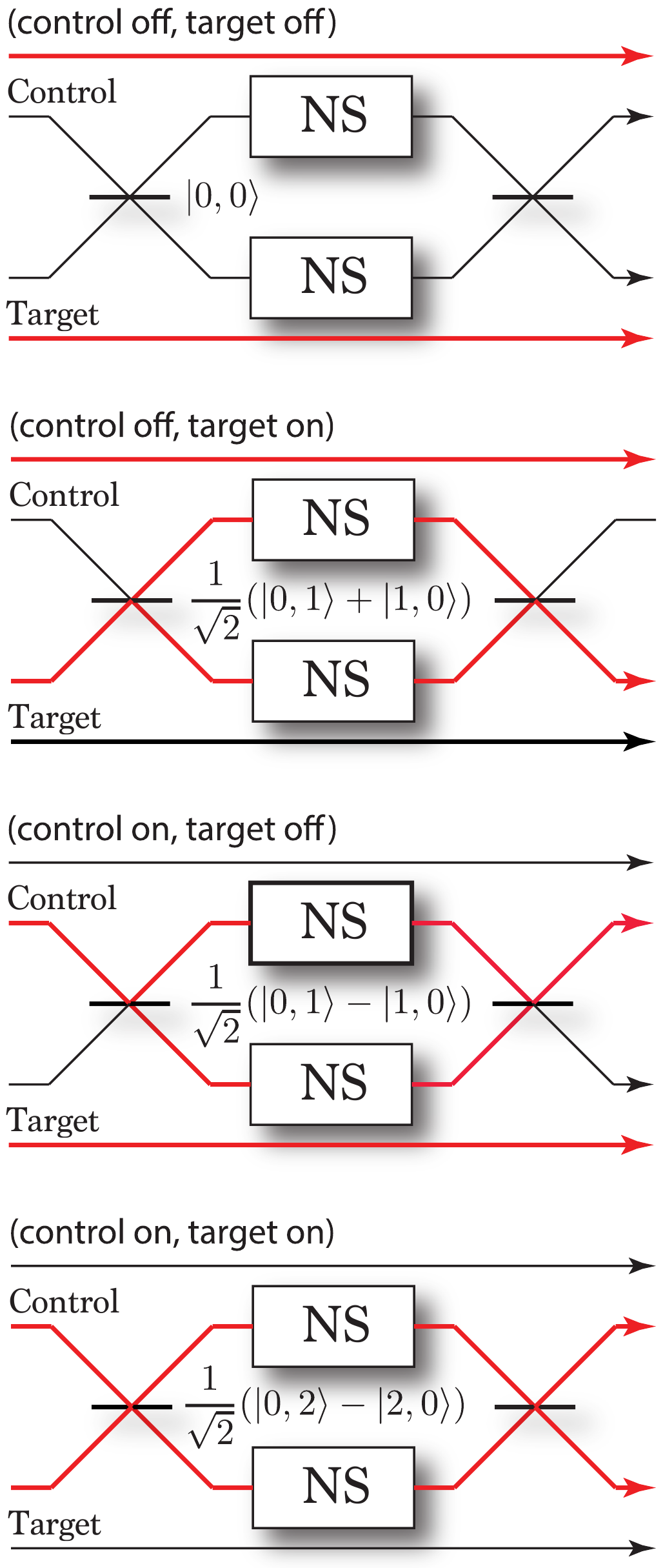}
	\captionspacefig \caption{Evolution of the four logical basis states through the KLM CZ gate. The NS gates do nothing in the first three cases, since they are operating only on vacuum and single-photon terms, which are left unchanged by the NS gate. In the last case, where both control and target are on, HOM interference results in photon bunching after the first beamsplitter, thereby creating two-photon terms. These terms inherit the $\pi$ phase-shift from the NS gate transformation, after which the final beamsplitter reverses the HOM photon bunching, yielding the same logical basis state with an acquired $\pi$ phase-shift.} \label{fig:KLM_explain} \index{Knill-Laflamme-Milburn (KLM)}\index{Controlled-Z (CZ) gates}
	\end{figure}
\else
	\begin{figure*}[!htbp]
	\includegraphics[clip=true, width=0.9\textwidth]{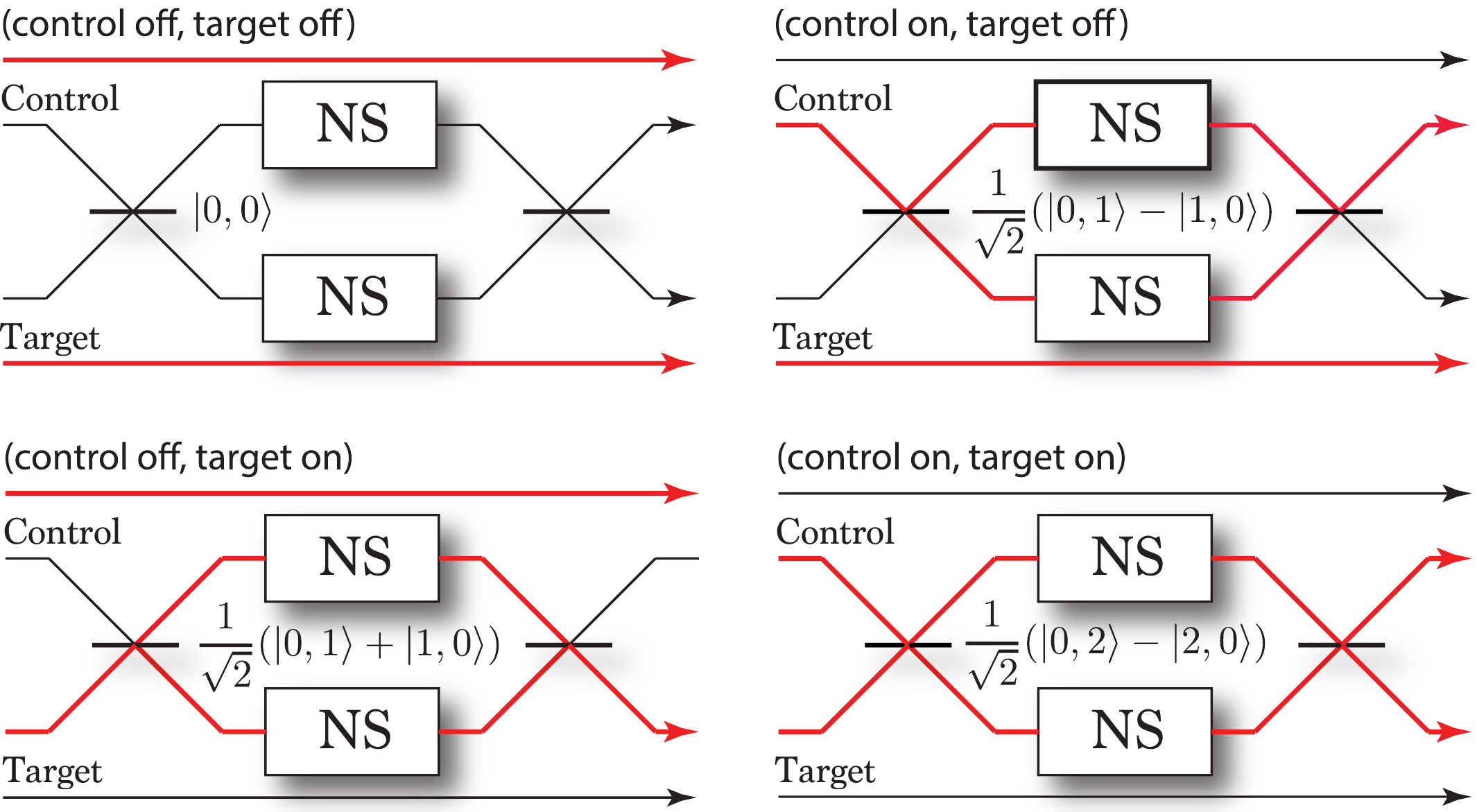}
	\captionspacefig \caption{Evolution of the four logical basis states through the KLM CZ gate. The NS gates do nothing in the first three cases, since they are operating only on vacuum and single-photon terms, which are left unchanged by the NS gate. In the last case, where both control and target are on, HOM interference results in photon bunching after the first beamsplitter, thereby creating two-photon terms. These terms inherit the $\pi$ phase-shift from the NS gate transformation, after which the final beamsplitter reverses the HOM photon bunching, yielding the same logical basis state with an acquired $\pi$ phase-shift.} \label{fig:KLM_explain} \index{Knill-Laflamme-Milburn (KLM)}\index{Controlled-Z (CZ) gates}
	\end{figure*}
\fi

Clearly this non-determinism is of immediate concern, since concatenating multiple gates would have exponentially decreasing success probability, making the protocol inefficient -- if the probability of a single gate succeeding is $p$, and we require that a circuit comprising $n$ of them all succeed, the success probability is clearly $p^n$.

The first key observation then is that gate teleportation can be used to shift this non-determinism to a resource state preparation stage, as described in detail in Sec.~\ref{sec:teleport_gate}. However, this is not the end of the story, since gate teleportation requires Bell state projections, which are themselves non-deterministic using purely linear optics (either using PBSs or CNOT gates).

\latinquote{Ignotum per ignotius}.

\begin{figure}[!htbp]
\includegraphics[clip=true, width=0.28\textwidth]{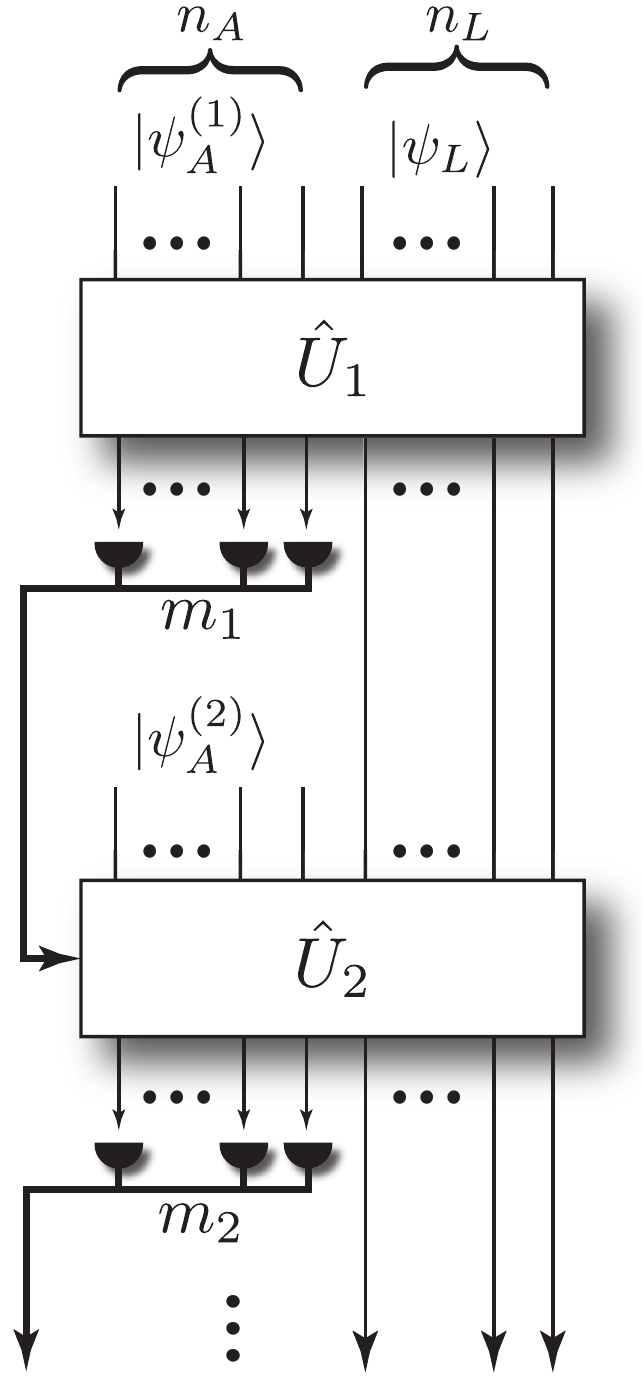}
\captionspacefig \caption{KLM architecture for universal LOQC. $n_L$ optical modes are associated with logical qubits in the state $\ket{\psi_L}$, with the remaining $n_A$ modes acting as ancillary states, $\ket{\psi_A}$. A round of passive linear optics is applied, $\hat{U}_1$. Then the ancillary modes are measured, yielding some set of measurement outcomes $m_1$. These are classically processed to determine what the next round of passive linear optics, $\hat{U}_2$, ought to be. This repeats some polynomial number of times, from which an arbitrary quantum computation can be implemented. The \textsc{BosonSampling} and quantum walk models are equivalent to taking just the first stage of this protocol: one round of input state, passive linear optics, and measurement.} \label{fig:KLM_protocol}\index{Knill-Laflamme-Milburn (KLM)}
\end{figure}

The final insight provided by KLM is that by concatenating these non-deterministic CNOT gates, we can inductively build up higher-level CNOT gates with ever increasing success probabilities, asymptoting to unity with high- (but polynomial-) depth concatenation. By combining these key insights, KLM were able to show that near-deterministic CNOT gates can be constructed using an efficient (polynomial) resource overhead, thereby enabling efficient universal quantum computation\footnote{Note that all single-qubit gates are trivially and deterministically implemented using wave-plates or beamsplitters, for polarisation or dual-rail encoding respectively. Thus, we need only concern ourselves with the challenges associated with implementing 2-qubit entangling gates.}. A sketch of the general KLM formalism is shown in Fig.~\ref{fig:KLM_protocol}.

Evolution via linear optics implements transformations of the form of Eq.~(\ref{eq:LO_unitary_map}), and may be implemented using the experimental architectures described in Sec.~\ref{sec:LO_ev_archs} and Fig.~\ref{fig:LO_archs}.

The measurements are implemented simply by number-resolved photo-detectors, implementing measurement projectors of the form \mbox{$\hat\Pi_n=\ket{n}\bra{n}$}, for the measurement outcome of $n$ photons (Sec.~\ref{sec:photo_detection}).

Since the original presentation of a universal LOQC gate set by KLM, numerous alternate implementations have been presented and experimentally demonstrated, with various pros and cons \cite{bib:Ralph01, bib:Pittman01, bib:Ralph02, bib:Knill02, bib:Pittman03, bib:MorYoran06}.

Significant progress is being made on reconfigurable, integrated LOQC devices \cite{bib:carolan2015universal}, but switching times remain orders of magnitude slower than that required for fast-feedforward. The resource overhead associated with overcoming the non-determinism of entangling gates is substantial in the original KLM proposal. But despite being improved upon by cluster state approaches, to be discussed next (Sec.~\ref{sec:CS_LO}), resource scaling remains daunting. It therefore seems most likely that certain elements from LOQC might be combined into hybrid architectures, to be discussed in detail in Sec.~\ref{sec:hybrid}.

%
% Cluster State Linear Optics
%

\subsection{Cluster state linear optics} \label{sec:CS_LO} \index{Cluster states}

Although the original KLM scheme is universal, and `efficient'\footnote{From a purely computer scientist's definition of `efficient = polynomial'\index{Efficiency}.}, resource usage can be reduced by orders of magnitude by combining concepts from LOQC with the cluster state formalism (Sec.~\ref{sec:CSQC}) or related concepts \cite{bib:YoranReznik03, bib:Nielsen04, bib:BrowneRudolph05, bib:GilchristHayes05, bib:Lim05, bib:LimBarrett05}.

Specifically, instead of using our non-deterministic KLM CZ gates within the circuit model formalism, they could be employed for the preparation of cluster states, since after all a CZ gate directly creates an edge in a cluster state graph.

We now review approaches for cluster state-based LOQC using non-deterministic entangling gates. A further discussion of this topic continues in Sec.~\ref{sec:module}, where we introduce modularised quantum computing from a cluster state perspective also using non-deterministic gates. We recommend beginning this topic here, and then skipping ahead to Sec.~\ref{sec:module} for continued discussion if interested.

%
% Fusion Gates
%

\subsubsection{Fusion gates}\label{sec:fusion_gates}\index{Fusion!Gates}

As introduced in Sec.~\ref{sec:CSQC}, a cluster state may be defined by the action of CZ gates upon a graph of qubits initialised into the $\ket{+}$ state. As we saw in the previous section, implementing these CZ gates is troublesome using linear optics, as it is non-deterministic and carries the burden of a large resource overhead. Nonetheless, it was shown early on \cite{bib:Nielsen04, bib:BrowneRudolph05} that by combining non-deterministic CZ gates with the cluster state formalism yields LOQC protocols far more efficient than the original KLM protocol for LOQC.

It was then noted \cite{browne2005resource} that CZ gates aren't required at all for the preparation of optical cluster states. Instead, parity measurements (Sec.~\ref{sec:bell_proj})\index{Bell!Measurements} operating in a rotated basis may be used to fuse smaller cluster states into larger ones, albeit acting destructively on two of the qubits, and also being non-deterministic, with a success probability of $1/2$. These gates have become known as \textit{fusion gates}, of which there are two types:
\begin{itemize}
	\item Type-I: destroy only a single photon, but require efficient number-resolved detection.
	\item Type-II: destroy two photons, but only require on/off detectors, since the gate succeeds upon coincidence events only and preserves photon-number.
\end{itemize}
Both types of gates have several highly favourable characteristics:
\begin{itemize}
	\item Unlike the KLM CZ gate, only HOM stability is required (Sec.~\ref{sec:opt_stab}). At no stage in the cluster state preparation procedure is any interferometric (i.e wavelength-scale) stability required.
	\item Gate failure is heralded by measurement of the wrong photon-number.
	\item Gate failure measures the respective qubits in the computational basis, thereby simply removing those qubits from the cluster state graph, whilst preserving the remainder of the state, which can be `recycled'\index{Cluster states!Recycling} for reuse.
\end{itemize}

\begin{figure}[!htbp]
\includegraphics[clip=true, width=0.4\textwidth]{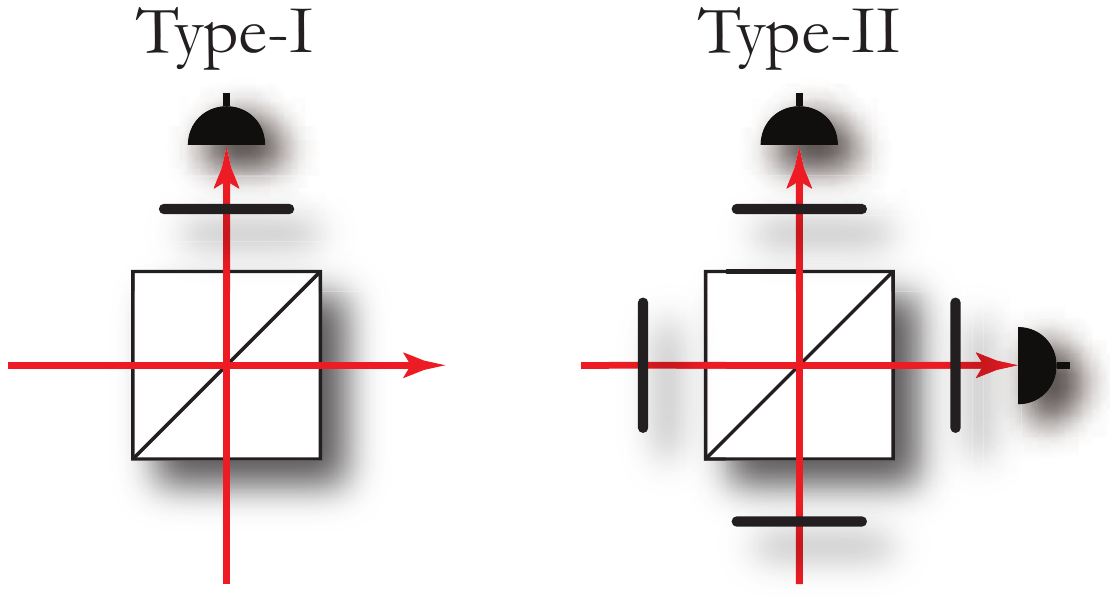}
\captionspacefig \caption{Linear optics cluster state fusion gates for polarisation-encoded photons. Both type-I and type-II gates employ a single polarising beamsplitter to mediate the entangling measurement. The black bars represent Hadamard gates in the polarisation basis (waveplates\index{Waveplates}). The type-I gate only measures a single photon, the other freely exiting the gate, which forms a part of the final cluster state. The type-II gate consumes two photons. Because the type-I gate does not measure in coincidence, as per the type-II gate, it requires number-resolved photo-detection, whereas bucket detectors suffice for type-II.} \label{fig:fusion_gates}
\end{figure}

The explicit construction of the linear optics fusion gates is shown in Fig.~\ref{fig:fusion_gates}.

%
% Fusion Strategies
%

\subsubsection{Fusion strategies} \index{Fusion!Strategies}

If a large cluster state has $n$ edges in its graph, single-shot state preparation will succeed with probability $p^n$ if individual gates succeed with probability $p$, implying that on average $1/p^n$ attempts will need to be made until success. Clearly this exponentiality doesn't lend itself to efficient implementation.

Thankfully, cluster states needn't be prepared in a single shot, since individual gate failures do not destroy the entire graph, but rather only cause localised damage to the graph in the vicinity of the gate.

Despite their non-determinism, numerous authors have examined approaches for efficiently preparing arbitrarily large cluster states using these destructive, non-deterministic gates \cite{bib:Nielsen04, bib:Kieling07, bib:RohdeBarrett07}. We refer to these schemes as `fusion strategies' -- simple algorithms for how to arrange qubits geometrically and the order in which to attempt bonding them.

These principles can be extended beyond LO to other schemes where entangling gates are inherently non-deterministic or sometimes fail in a heralded manner, e.g hybrid architectures (Sec.~\ref{sec:hybrid}), where a beamsplitter mediates entanglement via which-path erasure, but only successfully projects onto an entangled state with probability 1/2.

The key feature of all these fusion strategies is to employ `micro-clusters'\index{Micro-cluster states} as a primitive resource, which enable multiple bonding attempts between them via redundant vertices. We will now outline several of these schemes.

%
% Linear Clusters
%

\paragraph{Linear clusters}\index{Linear cluster states}

We begin with discussion of linear clusters as these are a particularly useful primitive resource for more advanced strategies. We briefly sketch out the formalism introduced by \cite{bib:RohdeBarrett07} for linear state preparation, which is applicable to a number of different variants of entangling gates.

A key observation was that although numerous strategies yield efficient state preparation, exact efficiencies are highly dependent on the ordering of bonding operations -- which clusters do we choose to bond together first?

Consider a non-deterministic KLM-type CZ gate\footnote{In reality, no one would use KLM-type gates for preparing cluster states, owing to their complexity compared to fusion gates. Rather, we use this gate for illustrative purposes, since its operation upon success and failure are very simple for exposition.}, which upon failure destroys its two input photons by measuring them in the computational $\hat{Z}$-basis, and leaves the number of qubits unchanged upon success and bonds them together.

To analyse the operation of such a non-deterministic protocol, we begin by defining a vector $\vec{n}_t$ at time $t$, which stores purely classical information. Specifically, the vector tells us how many clusters of every length we have stored in memory (except single-qubit clusters, which we assume `come for free'\footnote{As in beer.\index{Beer}}). For example, the second element of the vector tells us how many 3-qubit linear clusters we have in our possession.

\begin{figure}[!htbp]
	\includegraphics[clip=true, width=0.475\textwidth]{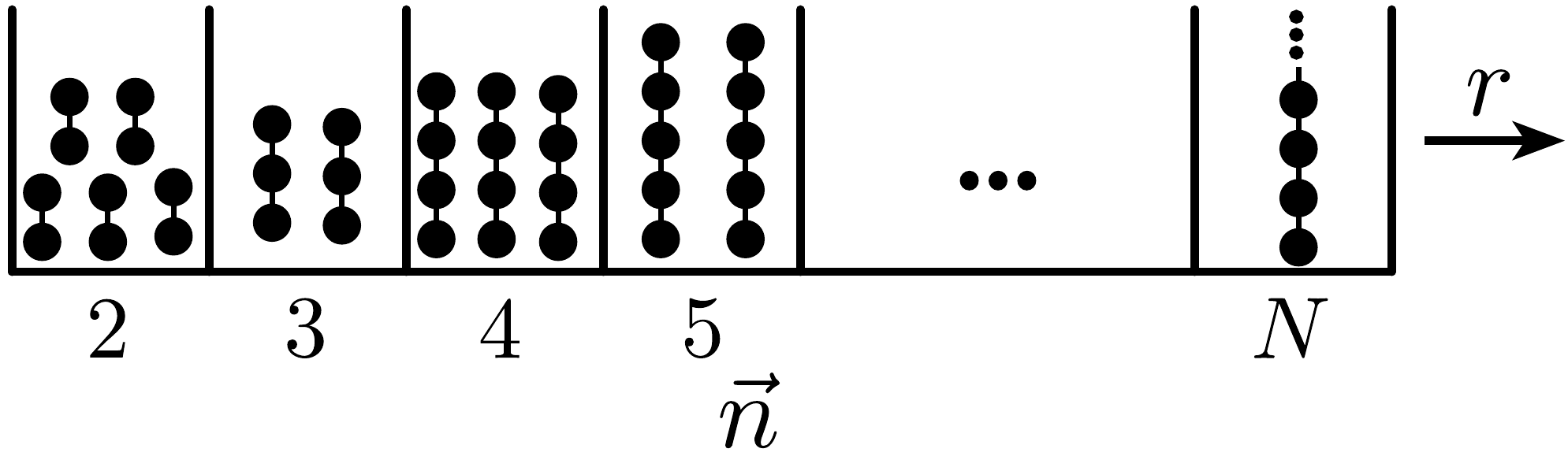}
	\captionspacefig \caption{Protocol for preparing linear cluster states using non-deterministic entangling gates and quantum memory. Each `bucket' holds a resource of micro-clusters of a given length, represented by the vector $\vec n$. Beginning with a resource of single qubits we repeatedly attempt bonding operations between micro-clusters in the buckets, according to a fusion strategy, $\mathcal{S}$. Proceeding as a biased random walk, the contents of the buckets shuffle around until ultimately (hopefully) clusters of the target length $N$ are prepared and steadily flow out as output at rate $r$ clusters per update operation. We assume a free supply of single qubits as a resource.}\label{fig:linear_cs_strategy}
\end{figure}

We then define a strategy, $\mathcal{S}$\index{Fusion!Strategies}, for choosing clusters we have stored and bonding them together to form larger clusters. The strategy acts on our cluster vector and updates it accordingly,
\begin{align}\index{Update rules}
\vec{n}_{t+1} = \mathcal{S}(\vec{n}_t).
\end{align}
That is, the length vector can be thought of as a series of `buckets' containing clusters of different lengths, and the strategy simply probabilistically shuffles the contents of the buckets around each time it is applied. The process can be thought of as a random walk\index{Random!Walks}, guided by a probabilistic update rule. Fig.~\ref{fig:linear_cs_strategy} outlines the protocol.

The strategy description, $\mathcal{S}$, is also responsible for taking care of updating the elements of $\vec{n}$ according to an update rule, which dictates how many photons are lost or gained upon success or failure of the non-deterministic gate. For example, the CZ gate we have employed here for our toy model destroys two qubits upon failure, but upon success creates a cluster of length given by the sum of the lengths of the clusters acted upon by the gate.

We are then interested in the \textit{rate}\index{Cluster states!Preparation rate}, $r$, at which large clusters are output from the protocol. This is simply extracted by defining a single parameter which counts the number of clusters in $\vec{n}$ above some predetermined target length, and normalises it by the total time taken to reach that point. The rate parameter converges asymptotically for long runtimes. Formally, the rate of preparation is given by,
\begin{align}
r = \lim_{t\to\infty} \frac{N_t}{t},
\end{align}
where $N_t$ is the total number of clusters of length greater than the target length at time $t$. The preparation rate is bounded by \mbox{$0\leq r\leq1$}. If the rate $r$ converges to a positive, finite value in the limit of large $t$, this implies state preparation proceeds in linear time and is therefore efficient.

This completes the theoretical analysis for different strategies and gate types, allowing us to explore different approaches tailored to different physical systems and their varying gate implementations.

One of the key outcomes was that a \textsc{Balanced} strategy\index{Balanced!Strategy} is optimal in terms of preparation rate. This is simply a strategy which preferentially always bonds clusters of equal length, beginning with the largest ones available. Asymmetric strategies, which bond clusters of differing lengths were found to be far less efficient.

\begin{figure}[!htbp]
\includegraphics[clip=true, width=0.475\textwidth]{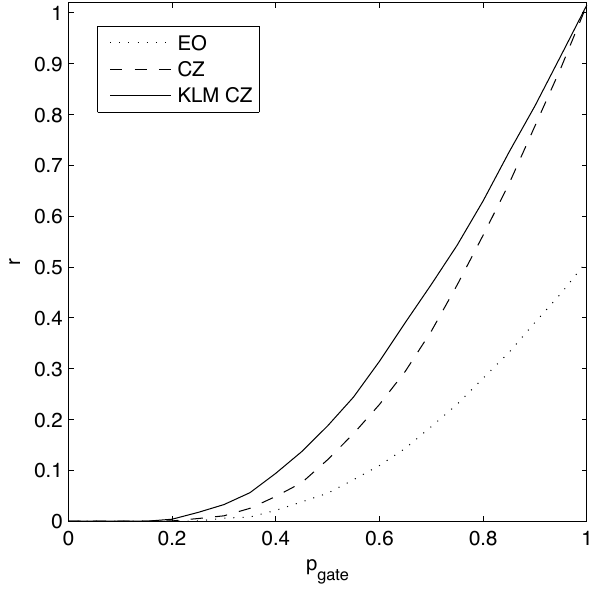}
\captionspacefig \caption{Linear micro-cluster preparation rates for three different types of entangling gates as described in \cite{bib:RohdeBarrett07}, against gate success probability, $p_\mathrm{gate}$. Here a \textsc{Balanced} fusion strategy is employed, whereby we only attempt to join clusters of equal length, always prioritising the largest ones available. This strategy was empirically found to perform better than any asymmetric strategies.}\label{fig:linear_cluster_state_r}
\end{figure}

Example simulated state preparation rate results are shown in Fig.~\ref{fig:linear_cluster_state_r} for various types of entangling gates and gate success probabilities.

%
% Lattice Clusters
%

\paragraph{Lattice clusters}\index{Lattice!Cluster states}

As we learnt from Sec.~\ref{sec:CSQC}, linear cluster states are not universal for quantum computation. What is required is lattices, where the rows and columns respectively map to logical qubits and time in the circuit model. There are numerous approaches one could employ to assemble such lattice clusters using non-deterministic gates, however the easiest to treat for illustrative purposes is to take a resource of linear clusters, prepared as described earlier, and weld them together according to some algorithm, enabling more complex two-dimensional topologies.

The central strategy is similar as for linear clusters -- we construct recyclable micro-clusters, which enable multiple bonding attempts, since gate failures only cause localised damage. The key difference now is that these redundant vertices must emanate in multiple directions so as to allow the more complex 2D topology.

Fig.~\ref{fig:micro_clusters} illustrates several topologies for micro-clusters, beginning with the linear micro-cluster that we employed previously for preparing 1D clusters, and two variations of micro-clusters that can be employed for 2D state preparation.

\if 1\doublecol
\begin{figure}[!htbp]
\includegraphics[clip=true, width=0.2\textwidth]{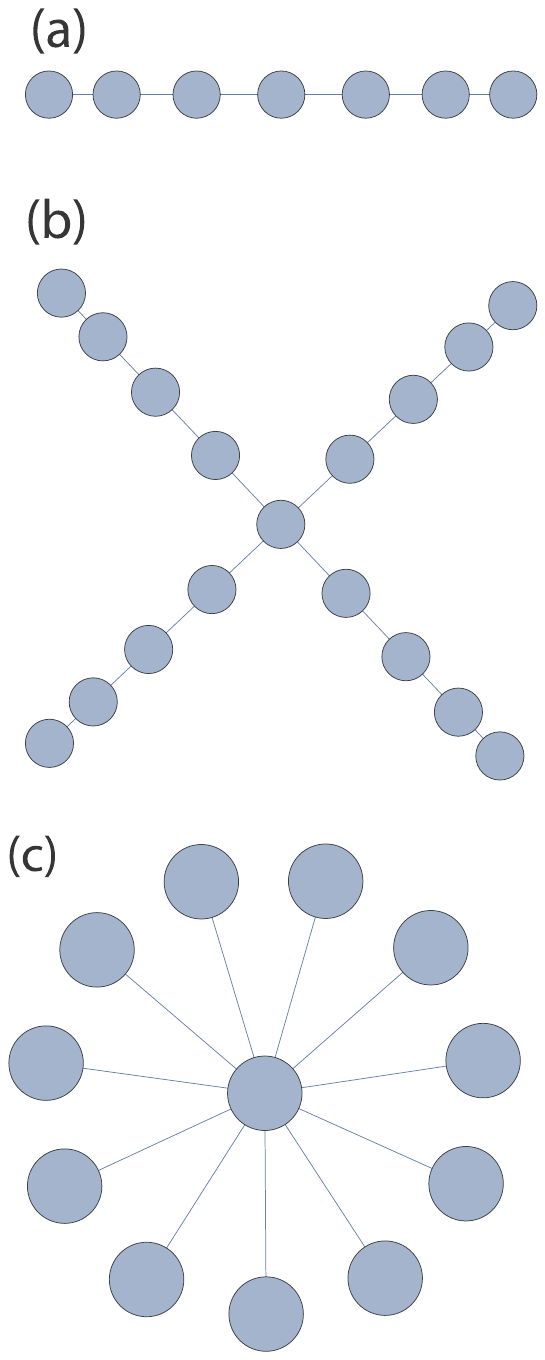}
\captionspacefig \caption{(a) Linear, (b) plus ($+$), and (c) star micro-cluster states.}\label{fig:micro_clusters}\index{Linear micro-cluster states}\index{Plus micro-cluster states}\index{Star micro-cluster states}\index{Micro-cluster states}
\end{figure}
\else
\begin{figure*}[!htbp]
\includegraphics[clip=true, width=0.7\textwidth]{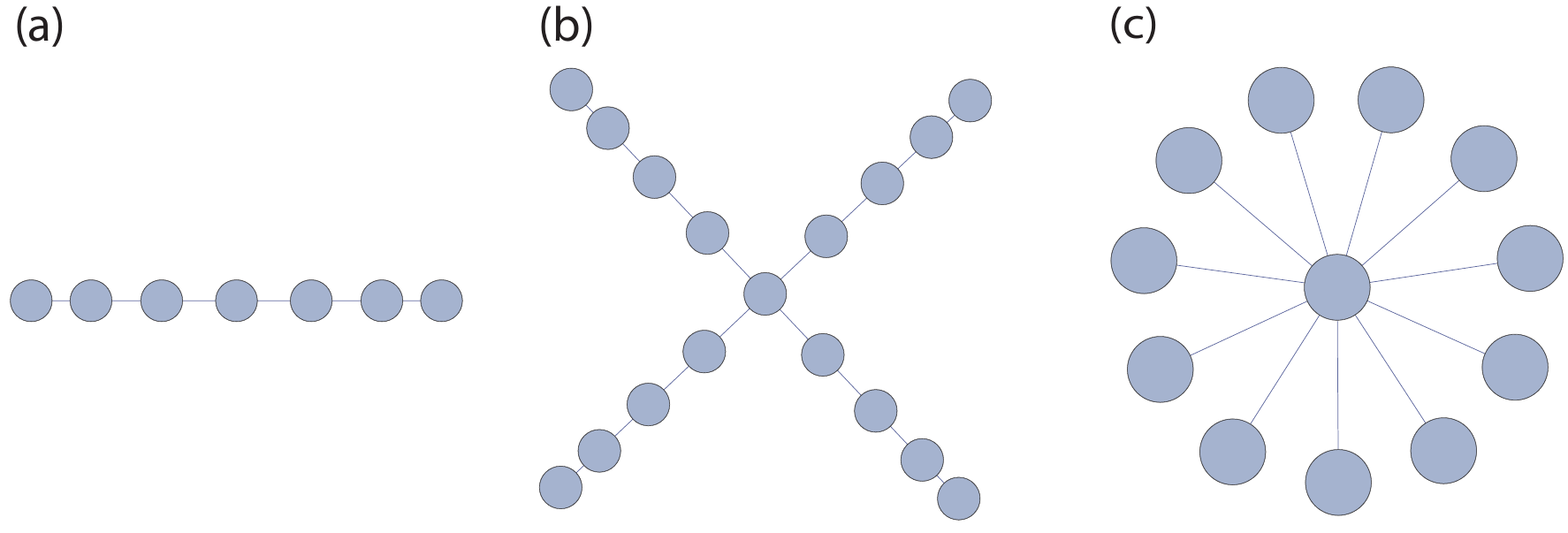}
\captionspacefig \caption{(a) Linear, (b) plus ($+$), and (c) star micro-cluster states.}\label{fig:micro_clusters}\index{Linear micro-cluster states}\index{Plus micro-cluster states}\index{Star micro-cluster states}\index{Micro-cluster states}
\end{figure*}
\fi

The $+$-cluster\index{Plus micro-cluster states} simply comprises four linear clusters emanating in the four directions, welded together at a central vertex. It's self-evident how this is subsequently applied to 2D state preparation -- we lay out the $+$-clusters in a grid, and attempt nearest neighbour bonding in each direction for every neighbouring pair of micro-clusters.

The star-cluster\index{Star micro-cluster states} similarly allows multiple bonding attempts in each direction. But now the dangling bonds are not uniquely associated with a particular direction, and may therefore be utilised when bonding to a neighbouring micro-cluster in any direction. This implies a modest efficiency improvement, since leftover vertices in any given direction needn't be wasted upon a successful bond in that direction. However, these micro-clusters are not as efficient to prepare as the $+$-clusters, since they do not straightforwardly arise from two fused linear clusters, which are highly efficient to prepare. Rather they must be prepared via a sequence of repeated successful bonding operations to the central node, where a single gate failure destroys the entire state.

In addition to an efficiency improvement in terms of the number of required physical qubits, minimising the number of redundant qubits that must be removed via $\hat{Y}$ measurements upon completion of the bonding strategy has another key benefit -- error accumulation \cite{bib:RohdeRalphMunro07}. Whenever a cluster state qubit is measured, the action of any error process that acted on that qubit will be teleported to its neighbour(s). For example, if we measure the first qubit in a linear cluster, which was previously acted upon by a depolarising channel\index{Depolarising channel}, the depolarisation process will be teleported to the neighbouring second qubit in the cluster. Thus, with high levels of redundancy, although this increases our chances of successfully joining two micro-clusters, it similarly increases the accumulation of errors. There is therefore a direct tradeoff between two undesirable error mechanisms -- gate failure, and logical errors. This tradeoff must be carefully managed in a real-world implementation.

\begin{figure}[!htbp]
\includegraphics[clip=true, width=0.475\textwidth]{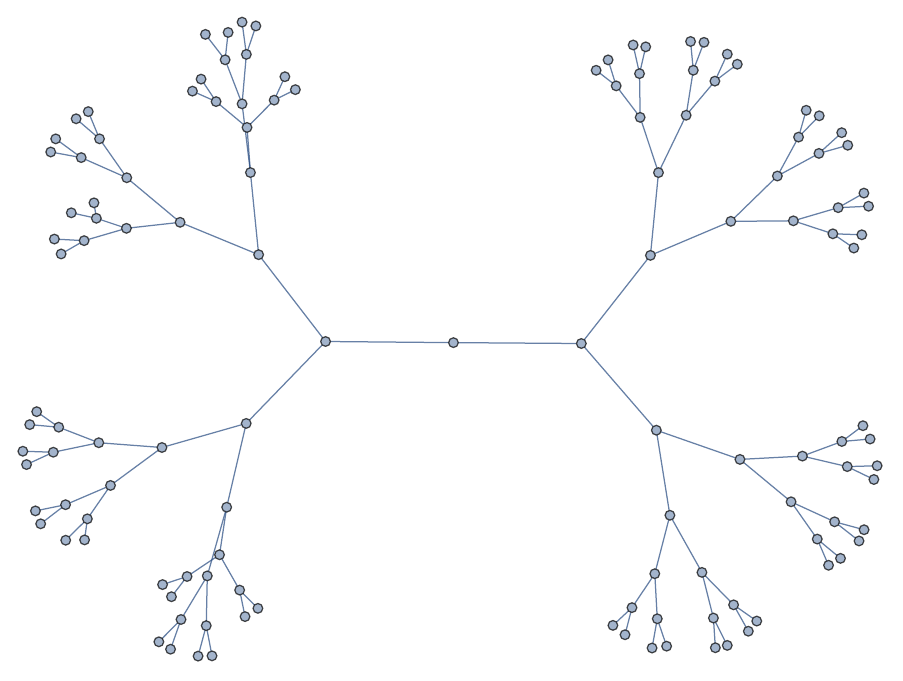}
\captionspacefig \caption{Snowflake micro-cluster comprising a binary tree structure, with depth \mbox{$d=7$}. Multiple copies of this micro-cluster can be placed side-by-side and fused together via attempting to bond the most outward available leaf qubits from neighbouring clusters. The tree structure allows excess qubits to be `pruned' via their trunks rather than leaves, bypassing the need to measure out every single leftover qubit, as is the case, for example, for $+$-clusters. This reduces the number of required pruning measurements from linear to logarithmic, similarly reducing the accumulation of errors associated with pruned qubits.} \label{fig:snowflake_graph}\index{Snowflake micro-cluster states}
\end{figure}

Having made this observation about error accumulation, is there a topology that is optimal? Yes there is -- the so-called snowflake cluster, shown in Fig.~\ref{fig:snowflake_graph}. He we take the $+$-cluster topology and replace the linear clusters emanating in each direction with binary tree graphs of some depth, $d$. This variation of micro-clusters has been studied in great detail both in optical and non-optical contexts \cite{PhysRevLett.104.050501}.

The endpoints (leaves) of each tree now provide the bonding opportunities for joining two neighbouring micro-clusters. There are $O(2^d)$ such opportunities. The bonding attempts proceed as expected, always exploiting the trees' outermost leaves.

Now the key feature is that when two sub-trees are successfully bonded via their leaves we do not need to measure out \textit{all} the leftover redundant vertices to reduce the graph to the desired residual topology. Instead we can `prune' away entire sub-trees by performing $\hat{Z}$ measurements at the base of their trunks. All vertices above the trunk will thereby be detached from the graph and needn't all be individually measured. Correspondingly, any error processes that had acted on the pruned vertices will not be teleported onto the main cluster, only the measurements acting on the trunks will contribute.

\begin{figure}[!htbp]
\includegraphics[clip=true, width=0.35\textwidth]{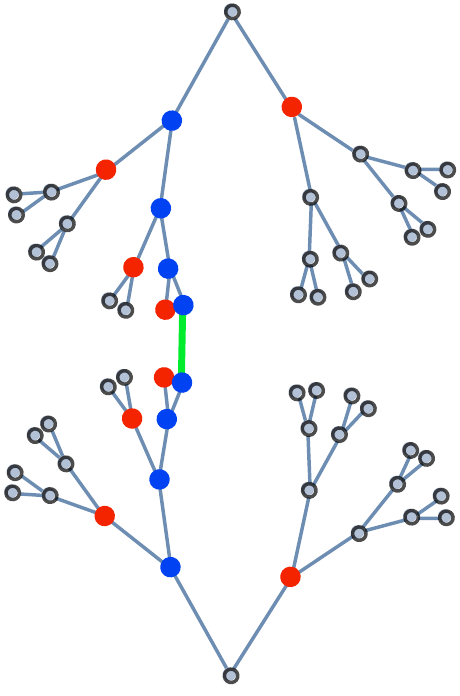}
\captionspacefig \caption{Pruning snowflake micro-clusters\index{Snowflake micro-cluster states} upon successful fusion of their outer leaves. Green indicates where the successful fusion operation took place. To remove all redundant nodes we measure the qubits marked in blue in the $\hat{Y}$-basis and the ones marked in red in the $\hat{Z}$ basis. This will discard all the other qubits marked in grey, modulo the two root qubits at the far top and bottom, which are left with a direct link between them. The total number of measured qubits scales logarithmically with the number of leaves.}\label{fig:snowflake_pruning}	
\end{figure}

Formally, for a linear subgraph of length $n$, there will be $O(n)$ leftover redundant qubits on average, which must \textit{all} be measured out using $\hat{Y}$ measurements. Thus, the residual state will have accumulated the action of $O(n)$ independent error processes. On the other hand, for a snowflake subgraph, the trees' depth scales as \mbox{$d=O(\log n)$}, and therefore at most \mbox{$O(\log n)$} qubits must be measured to prune away unwanted branches. Fig.~\ref{fig:snowflake_pruning} presents an example of how the pruning process works.

Keeping in mind that for an error process with error rate $p$, the net probability of an error occurring for $m$ independent channels is \mbox{$1-p^m$}, thus reducing $m$ from linear to logarithmic is highly favourable in terms of the accumulation of errors.

%
% On-Demand Cluster State Preparation
%

\paragraph{On-demand cluster state preparation}\index{On-demand!Cluster state preparation}

A beautiful feature of the cluster state model is that the entire cluster needn't be prepared in its entirety for computation to proceed. Instead the state can be grown via the fusion of additional qubits on-demand as the computation proceeds. This arises simply because all the entanglement in the graph is nearest neighbour only, i.e very short-range. So long as a gate failure doesn't lay its fingers on the leftmost column of qubits in the cluster we are in business. This means fewer quantum memories are required, which are very challenging optically. In non-optical, specifically matter qubit systems, this additionally means that physical qubits can be reused on-the-fly.

\begin{figure}[!htbp]
\includegraphics[clip=true, width=0.475\textwidth]{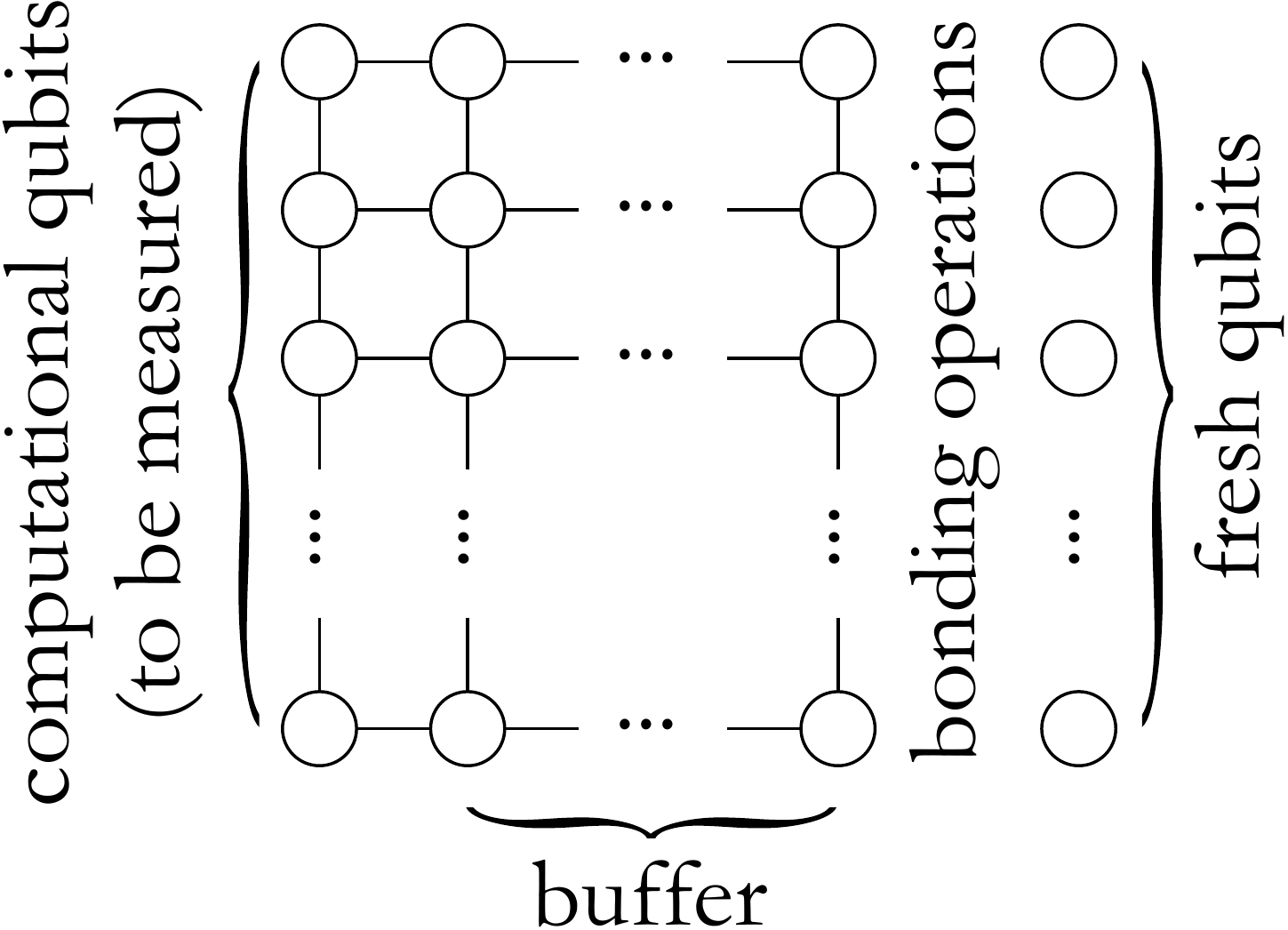}
\captionspacefig \caption{On-demand cluster state preparation. Every time a column of computational qubits are measured away from the lefthand side, thereby evolving the computation by a single step, we dynamically bond on a fresh column of qubits to the righthand side. The buffer in between provides the redundancy necessary when using non-deterministic gates.}\label{fig:on_demand_cs}
\end{figure}

The computation therefore proceeds as alternating applications of:
\begin{enumerate}
	\item Measure the leftmost column of physical qubits to evolve the computation by a single step.
	\item Bond on a new column of qubits to the rightmost column.
\end{enumerate}
The qubits in between the left and rightmost columns act as a buffer to give us some leeway when bonding operations fail. The architecture is shown in Fig.~\ref{fig:on_demand_cs}.

In Ref.~\cite{PhysRevA.73.012304}, the authors performed an analysis of this approach in the optical context and found that high-depth MBQC can be efficiently implemented using non-deterministic entangling operations, with significantly reduced quantum memory requirements compared to full in-advance state preparation.

In addition to technologically simplifying the architecture by reducing the number of required quantum memories, physical qubits are in existence within the computation for substantially reduced periods of time, since they are only prepared on-demand. This correspondingly reduces error rates.

%
% Weak Cross-Kerr Non-Linearities
%

\subsection{Weak cross-Kerr non-linearities} \index{Weak cross-Kerr non-linearity quantum computation}

In Sec.~\ref{sec:KLM_univ} we showed that with strong non-linearities\index{Non-linearities} at our disposal, scalable photonic quantum computing is possible. However, the interaction strengths of such materials available in the lab today are minuscule compared to the full $\pi$ phase-shift required for NS and CNOT gate constructions. In LOQC this problem is circumvented using measurement-induced non-linearities, which while being sufficiently strong, are also necessarily non-deterministic, mandating substantial error correction resource overheads to accommodate gate failure.

More recently, as an alternative to using post-selection to simulate strong optical non-linearities, it was shown that by introducing strong coherent states, the strength of the non-linear interaction can be effectively amplified arbitrarily, allowing even very weak non-linearities to be employed for deterministic entangling gate operations \cite{bib:Munro05}, compensated for using strong coherent states. Thankfully, strong coherent states are an easy resource to come by nowadays!

In this hybrid linear/non-linear optical architecture, a coherent state\index{Coherent states} is used as a `qubus' (quantum bus)\index{Qubus}, which is entangled with photonic qubits via weak non-linear interactions, thereby mediating long-range entangling operations. This can provide the sufficient entangling power needed to enable scalable, universal quantum computation.

Let us describe the operation of a parity measurement\index{Bell!Measurements} (i.e Bell analyser) device using weak non-linearities. This gate could subsequently be employed as a fusion gate\index{Fusion!Gates} (Sec.~\ref{sec:fusion_gates}) for building cluster states, and is therefore a resource for universal quantum computation. Simple extensions of this design idea easily extend to other non-trivial operations, such as CNOT and CZ gates or QND measurements\index{Quantum non-demolition measurements (QND)}. 

The key ingredient here is the cross-Kerr interaction\index{Cross-Kerr interaction}, which obeys the Hamiltonian,
\begin{align}
\hat{H}_\mathrm{ck} = \hbar \chi \hat{n}_a \hat{n}_b,	
\end{align}
where $\hat{n}_a$ and $\hat{n}_b$ are the photon-number operators\index{Photon-number!Operators} for modes $a$ and $b$ respectively, and $\chi$ is the interaction strength\index{Interaction!Strength}, which is typically very small in the lab. This Hamiltonian generates the unitary transformation,
\begin{align}
\hat{U}_\mathrm{ck} = e^{i\theta\hat{n}_a\hat{n}_b}.
\end{align}
The magnitude of the induced phase-shift is now proportional to photon-number and,  
\begin{align}
\theta = \chi t,	
\end{align}
where $t$ is the interaction time\index{Interaction!Time}. For strongly entangling gates we need \mbox{$\theta \approx \pi$}.

Applying this operation between a coherent state and a photon-number state implements the two-mode transformation, 
\begin{align}
\hat{U}_\mathrm{ck} \ket\alpha \ket{n} = \ket{\alpha e^{i\theta n}} \ket{n}.	
\end{align}
Note that the phase-shift accumulated by the coherent state is proportional to the photon-number in the other mode, thereby entangling the two modes via their shared dependence on $n$, effectively a photon-number-controlled phase-shift operation.

\begin{figure*}[!htpb]
	\includegraphics[clip=true, width=\textwidth]{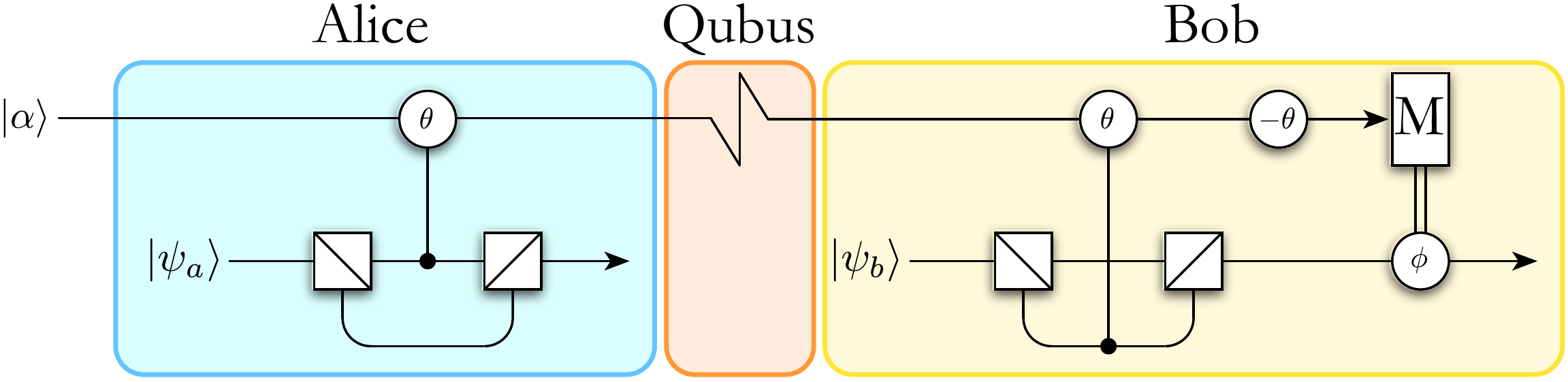}
\captionspacefig \caption{Construction of a two-mode, polarisation-encoded, photonic parity gate using an ancillary qubus coherent state $\ket\alpha$, mediating entangling photonic interactions between $\ket{\psi_a}$ and $\ket{\psi_b}$ via weak cross-Kerr non-linearities (denoted by the controlled-$\theta$ gates). The polarising beamsplitters\index{Polarising beamsplitters} are used to switch between polarisation and dual-rail encoding. The qubus is an optical channel, potentially across long distances over the quantum internet for performing distributed gates. The measurement $M$ is a homodyne measurement\index{Homodyne detection}, which feedforwards an $X$-quadrature measurement outcome to a local phase correction. This gate could be used for preparing distributed photonic cluster states, a resource for universal quantum computation.}\label{fig:weak_ck_parity}	
\end{figure*}

Consider the parity measurement circuit shown in Fig.~\ref{fig:weak_ck_parity}. Let us begin with two polarisation-encoded photonic qubits, which might be separated over long distances, and a coherent state in the shared qubus mode,
\begin{align}
\ket{\psi_\mathrm{in}} &= (\alpha_{HH} \ket{H}\ket{H} + \alpha_{HV} \ket{H}\ket{V} \nonumber\\
&+\alpha_{VH} \ket{V}\ket{H} + \alpha_{VV} \ket{V}\ket{V})\ket\alpha.
\end{align}
Applying the cross-Kerr interactions leaves us in the state,
\begin{align}
\hat{U}_\mathrm{gate} \ket{\psi_\mathrm{in}} &= [\alpha_{HH} \ket{H}\ket{H}] + \alpha_{VV} \ket{V}\ket{V}]\ket\alpha \nonumber\\
	&+ \alpha_{HV} \ket{H}\ket{V}\ket{\alpha e^{i\theta}}\nonumber\\
	&+ \alpha_{VH} \ket{V}\ket{H}\ket{\alpha e^{-i\theta}}.
\end{align}
Now the qubus state is in some superposition of $\ket\alpha$, $\ket{\alpha e^{i\theta}}$ and $\ket{\alpha e^{-i\theta}}$. The key observation is that these three coherent basis states become highly distinguishable for large $|\alpha|$ with non-zero $\theta$. Specifically, a homodyne measurement on the qubus that projects onto \mbox{$x=0$}, approximately leaves us in the state,
\begin{align}
	\ket{\psi_\mathrm{out}} = \alpha_{HH}\ket{H}\ket{H} + \alpha_{VV}\ket{V}\ket{V},
\end{align}
which corresponds to a maximally-entangling parity or Bell projection\index{Bell!Measurements}.

Clearly, since \mbox{$\braket{\alpha|\alpha e^{\pm i\theta}}\neq 0$}, there is some probability of error, associated with confusing the coherent basis states and hence their associated photonic qubit states. However, we asymptote towards perfect behaviour in the limit of large coherent qubus amplitudes, since,
\begin{align}	
\lim_{|\alpha|\to\infty}\braket{\alpha|\alpha e^{\pm i \theta}} = 0\,\,\forall\,\theta\neq 0.
\end{align}

This type of non-local gate lends itself very naturally to distributed, network-based implementation, where the quantum internet is employed to mediate the qubus, potentially over long distances. Note that unlike polarising beamsplitter-based fusion gates, this gate is non-destructive, and does not require measuring any photonic qubits, only the qubus.

The downside of this protocol, and other qubus-based protocols based on the same idea, is its sensitivity to loss. This is because the qubus is effectively in a cat state (Sec.~\ref{sec:cat_enc}), whose sensitivity to decoherence increases rapidly with the coherent amplitude $|\alpha|$. This will effectively place hard limits on how remote Alice and Bob can be in a distributed setting, depending on the loss characteristics of the quantum channel shared between them. It also presents the engineer with a direct tradeoff between decoherence of the qubus (undesirable) and distinguishability of the qubus basis states (desirable), both of which increase with $|\alpha|$.

%
% Passive Linear Optics
%

\subsection{Passive linear optics} \label{sec:passive_LO} \index{Passive linear optics quantum computation}

While the KLM protocol (and subsequent improvements, e.g using cluster states) are universal for quantum computing, some of the key technological requirements are very challenging, and unlikely to be achieved in the short-term. However, simplified yet non-universal models for optical quantum computing can abandon some of the more challenging requirements, nonetheless implementing a restricted set of post-classical quantum computations. In particular, we consider protocols requiring only photon-number state preparation, passive linear optics evolution [as per Eq.~(\ref{eq:LO_unitary_map})], and photo-detection.

Optically, the two main contenders for this are multi-photon quantum walks \cite{bib:Aharonov93, bib:Aharonov01, bib:Kempe03, bib:Childs09, bib:Salvador12, bib:RohdeMultiWalk11} and \textsc{BosonSampling} \cite{bib:AaronsonArkhipov10, bib:RohdeIntroBS15}, both closely related in that they require only passive linear optics and single-photon states, whilst mitigating the need for active switching, quantum memory and dynamic fast-feedforward. Since, evidence has been presented that similar passive linear optics protocols may implement computationally hard problems using states of light other than photon-number states \cite{bib:RandBS, bib:RohdePhotAdd15, bib:RohdeDisp15, bib:RohdeCat15}.

These protocols involve nothing more than evolving multiple single-photon states through beamsplitter networks and measuring the output photo-statistics. This is equivalent to just taking the first stage of the KLM protocol shown in Fig.~\ref{fig:KLM_protocol}.

Both quantum walks and \textsc{BosonSampling} have been subject to extensive experimental investigation in recent years \cite{bib:PeruzzoQW, bib:Broome10, bib:Schreiber11b, bib:Owens11, bib:RohdeQWExp12, bib:Broome2012, bib:Crespi3, bib:Tillmann4}.

Because these models are entirely passive, they can be made cloud-based very trivially: Alice prepares her permutation of single photons as the input state, sends it to Bob over the quantum network, who applies the passive operations before returning the state to Alice. In this case, no intermediate client/server interaction is required. Alternately, she could classically communicate a bit-string to Bob indicating the input photon-number configuration, in case she is unable to prepare it herself.

The \textsc{BosonSampling} and quantum walk models are based on single-photon encoding. However, passive linear optics could also be applied to other states of light. In particular, passive linear optics acting upon multi-mode coherent states implements the \textit{classical} computation of matrix multiplication.

%
% Boson-Sampling
%

\subsubsection{\textsc{BosonSampling}} \label{sec:boson_sampling} \index{Boson-sampling}

\textsc{BosonSampling} is the problem of sampling the output photon-number statistics of a linear optics interferometer fed with single-photon inputs. While not universal for quantum computing (in fact no one has any idea what to use it for at all!), there is strong evidence that it is a classically hard problem \cite{bib:AaronsonArkhipov10, bib:RohdeIntroBS15}.

The computational hardness of \textsc{BosonSampling} relates to the fact that the amplitudes in the output superpositions are proportional to matrix permanents, which are known to be \#\textbf{P}-hard\index{\#P} in general. This is believed to be a classically hard complexity class, even harder than \textbf{NP}-complete\index{NP \& NP-complete} in the complexity hierarchy, requiring exponential classical time to evaluate (see Fig.~\ref{fig:complexity_classes} for the believed complexity relationships). This yields computationally complex sampling problems.

%
% The Boson-Sampling Model
%

\paragraph{The \textsc{BosonSampling} model} \index{Boson-sampling!Model}

For an $m$-mode interferometer, and input state,
\begin{align}
\ket\psi_\mathrm{in} = \ket{T_1,\dots,T_m},
\end{align}
where there are $T_i$ photons in the $i$th input mode, the output superposition takes the form,
\begin{align}
\ket\psi_\mathrm{out} = \sum_S \gamma_{S,T} \ket{S_1,\dots,S_m},
\end{align}
where $S$ sums over all possible photon-number configurations at the output, of which there are,
\begin{align}
|S| = \binom{n+m-1}{n},
\end{align}
where there are $n$ photons in total in $m$ modes. It is assumed that,
\begin{align}
m=O(n^2),
\end{align}
which, for large $m$, puts us into the anti-bunched (i.e binary photon-number) regime with high probability\footnote{That is, we are unlikely to observe more than a single photon in any given output mode, placing us into a binary photo-detection regime. This condition has become known as the `bosonic birthday paradox' \cite{arkhipov2012bosonic}.}\index{Bosonic birthday paradox}, rendering non-number-resolved photo-detectors sufficient for physical implementation. However, this `no-collision' subspace remains exponentially large,
\begin{align}
|S_\mathrm{no\,collision}| = \binom{m}{n}.
\end{align}

The amplitudes $\gamma_{S,T}$ are given by,\index{Configuration amplitudes}
\begin{align}\label{eq:BS_perms}
	\gamma_{S,T} = \frac{\mathrm{Per}(U_{S,T})}{\sqrt{S_1!\dots S_m! T_1!\dots T_m!}},
\end{align}
and the associated configuration probabilities by,
\begin{align}
	P_{S,T} &= |\gamma_{S,T}|^2 \nonumber \\
	&= \frac{|\mathrm{Per}(U_{S,T})|^2}{S_1!\dots S_m! T_1!\dots T_m!}
\end{align}
where $\mathrm{Per}(\cdot)$ denotes the matrix permanent, and $U_{S,T}$ is a sub-matrix of $U$ -- the transfer matrix\index{Transfer matrices} associated with the particular input-to-output sample configuration -- obtained by taking $S_i$ copies of the $i$th row, and $T_j$ copies of the $j$th column of the linear optics unitary matrix $U$. For the purposes of the original complexity proof, the unitary is chosen randomly from the Haar-measure\footnote{The Haar-measure generalises the notion of a uniform distribution to higher-dimensional topologies than the real numbers, in this instance to the $\mathrm{SU}(n)$ group.}\index{Haar measure}, although it remains an open question as to what is the full class of unitaries that yield computationally hard problems.

\paragraph{The relationship to matrix permanents}\index{Permanents}

The observation that output probability amplitudes are related to matrix permanents as per Eq.~(\ref{eq:BS_perms}) is the most important one, as this is ultimately responsible for the computational hardness of the \textsc{BosonSampling} problem, since permanents are in general \textbf{\#P}-hard\index{\#P} problems, a classically inefficient complexity class.

The permanent of a square matrix is defined as\index{Permanents},
\begin{align}\label{eq:permanent}
\mathrm{Per}(A) = \sum_{\sigma\in S_n} \prod_{i=1}^n A_{i,\sigma_i},
\end{align}
which sums over $n!$ terms (super-exponential), where $S_n$ is the symmetric group, the group of permutations on $n$ elements, of which there are $n!$. Note the similarity with the definition for the matrix determinant\index{Determinants}, defines identically, but with the addition of an alternating $\pm$-sign in the terms. Despite this similarity, determinants reside in \textbf{P}, with an efficient classical algorithm. For this reason, Fermionic sampling\index{Fermionic sampling} yields an easy computational problem, since Fermionic sampling differs only in replacing the permanent with the determinant. The best-known classical algorithm for evaluating permanents by \cite{bib:RyserAlg} has exponential runtime,
\begin{align}
	O(2^n n^2).
\end{align}

\begin{figure}[!htbp]
\includegraphics[clip=true, width=0.475\textwidth]{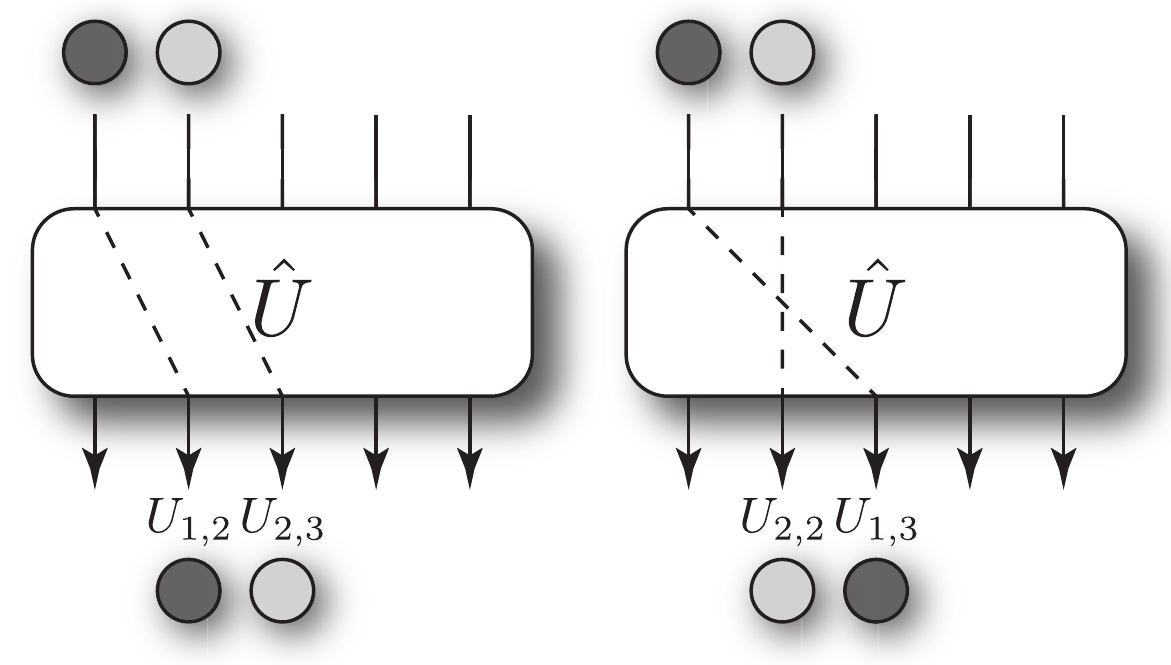}
\captionspacefig \caption{A linear optics interferometer $\hat{U}$, fed with 2 single-photon inputs, one in each of the first two modes. To calculate the output amplitude of one photon in each of the modes 2 and 3, we sum the amplitudes of all possible paths consistent with that output. In this example there are only two such paths -- either both photons pass straight through, or they swap positions. This summation yields a \mbox{$2\times 2$} matrix permanent, given by Eq.~(\ref{eq:BS_2_ph_comb}).}\label{fig:BS_2_comb}	
\end{figure}

To see how matrix permanents naturally arise in this setting, it is easiest to explain by example. In Fig.~\ref{fig:BS_2_comb} we illustrate a simple interferometer, fed with two photons. We wish to calculate the output amplitude of measuring a photon in each of the modes 2 and 3, given photons input at modes 1 and 2. To evaluate this amplitude we simply need to add up the amplitudes of all possible paths yielding the desired outcome. In this simple example this sum-of-paths\index{Sum-of-paths} is given by,
\begin{align} \label{eq:BS_2_ph_comb}
\gamma_{\{2,3\}} &= \underbrace{U_{1,2}U_{2,3}}_{\mathrm{don't\ swap}} + \underbrace{U_{1,3}U_{2,2}}_{\mathrm{swap}} \nonumber \\
&= \mathrm{Per}\begin{pmatrix}
   U_{1,2} & U_{2,2} \\
   U_{1,3} & U_{2,3}
  \end{pmatrix},
\end{align}
from which it is immediately clear that the amplitude is given by the sum of \mbox{$2!=2$} paths\footnote{The number of paths scales as $n!$ in general, which corresponds to the $n!$ order of the symmetric group, $S_n$, in the definition of the permanent from Eq.~(\ref{eq:permanent}).}, the permanent of the \mbox{$2\times 2$} matrix obtained from taking the columns (rows) of $\hat{U}$ where a photon is present at the respective input (output) mode.

\begin{figure}[!htbp]
\if 1\doublecol
	\includegraphics[clip=true, width=0.475\textwidth]{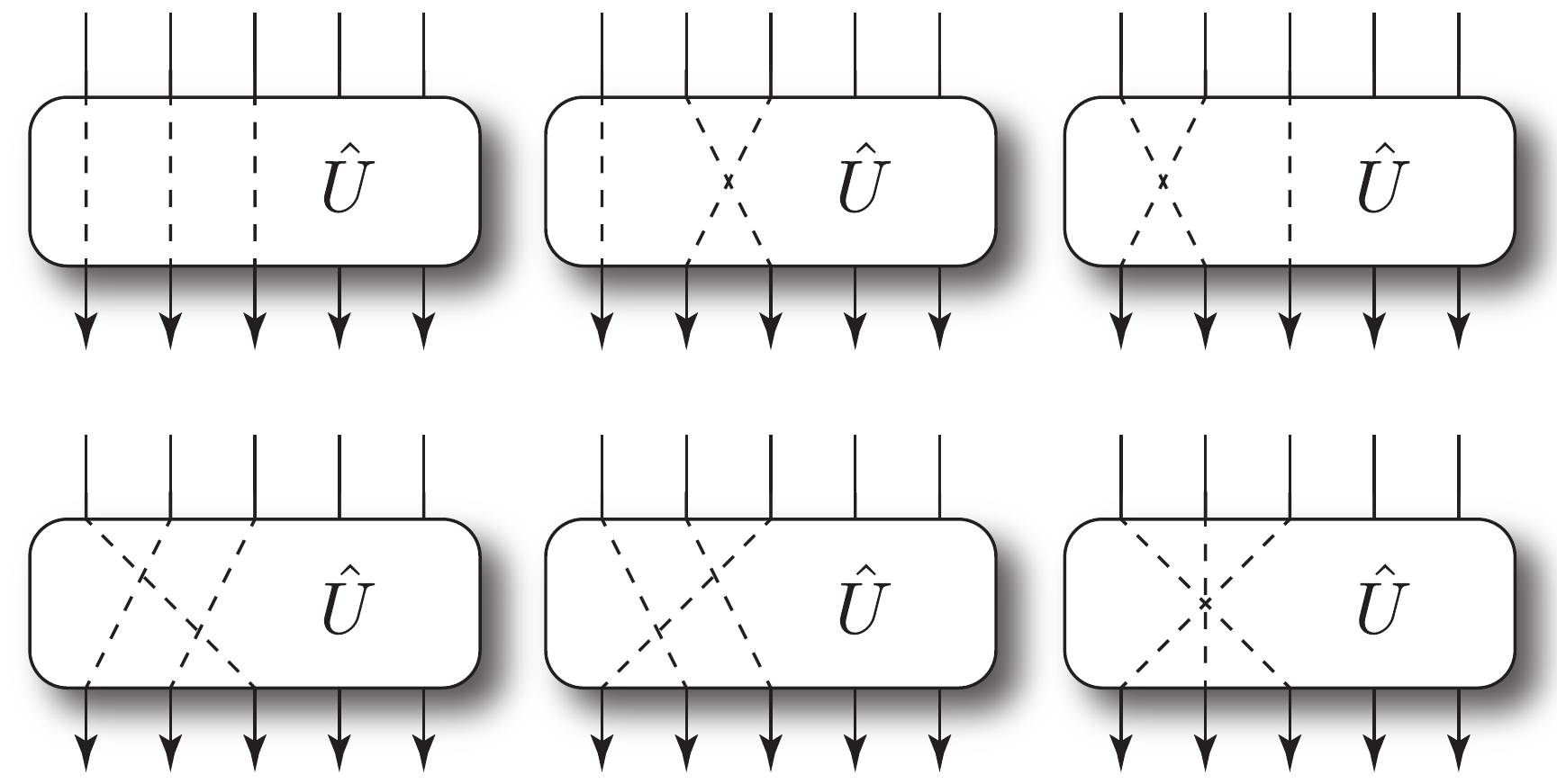}
\else
	\includegraphics[clip=true, width=0.6\textwidth]{BS_3_photon_combinatorics}
\fi
\captionspacefig \caption{A linear optics interferometer $\hat{U}$, fed with 3 single-photon inputs, in modes 1, 2 and 3. To calculate the output amplitude of one photon in each of the modes 1, 2 and 3, we sum the amplitudes of all possible paths consistent with that output. This summation yields a \mbox{$3\times 3$} matrix permanent, given by Eq.~(\ref{eq:BS_3_ph_comb}).}\label{fig:BS_3_comb}	
\end{figure}

In Fig.~\ref{fig:BS_3_comb} we present to next most sophisticated example of an interferometer fed by 3 photons, for which the sum-of-paths\index{Sum-of-paths} has \mbox{$3!=6$} terms, given by,
\begin{align} \label{eq:BS_3_ph_comb}
\gamma_{\{1,2,3\}} &= U_{1,1}U_{2,2}U_{3,3} + U_{1,1}U_{3,2}U_{2,3} \nonumber \\
&+ U_{2,1}U_{1,2}U_{3,3} + U_{2,1}U_{3,2}U_{1,3} \nonumber \\
&+ U_{3,1}U_{1,2}U_{2,3} + U_{3,1}U_{2,2}U_{1,3}
\nonumber \\
&= \mathrm{Per} \begin{pmatrix}
   U_{1,1} & U_{2,1} & U_{3,1} \\
   U_{1,2} & U_{2,2} & U_{3,2} \\
   U_{1,3} & U_{2,3} & U_{3,3}
  \end{pmatrix},
\end{align}
and it is now clear upon inspection that the amplitude is given by a \mbox{$3\times 3$} matrix permanent.

\paragraph{Problem description}\index{Boson-sampling!Problem description}

\begin{figure}[!htbp]
\includegraphics[clip=true, width=0.35\textwidth]{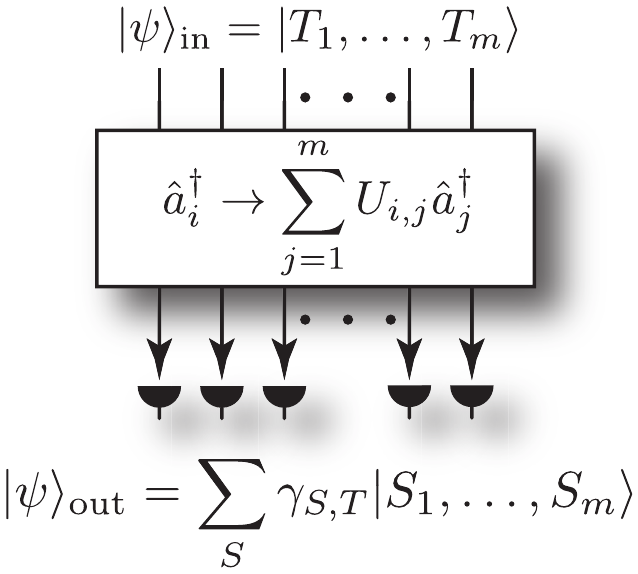}
\captionspacefig \caption{The \textsc{BosonSampling} model for non-universal linear optics quantum computing. $S$ (output) and $T$ (input) are represented in the photon-number basis. After application of the Haar-random linear optics unitary to the input multi-mode Fock state, the output superposition is sampled with coincidence photo-detection.} \label{fig:bs_model}
\end{figure}

The computational problem is simply to sample this probability distribution $P_{S,T}$, which the linear optics network can implement efficiently, but it is believed a classical computer cannot. The full model is shown in Fig.~\ref{fig:bs_model}.

For comparison, the equivalent classical protocol using distinguishable photons that evolve independently through the network would be described by,
\begin{align}
	P_{S,T} = \frac{\mathrm{Per}(|U_{S,T}|^2)}{S_1!\dots S_m! T_1!\dots T_m!},
\end{align}
which yields a classically efficient sampling problem. Thus, for \textsc{BosonSampling} the permanents are of complex-valued matrices, whereas for the equivalent classical problem the permanents are of positive real-valued matrices.

Very importantly, note that \textsc{BosonSampling} does \textit{not} let us efficiently \textit{calculate} matrix permanents. Rather, it samples across a distribution of an exponential number of permanents. This is because, for an exponentially large sample space, with only a polynomial number of measurement trials, we are unlikely to gain more than binary accuracy about individual amplitudes, which is insufficient for determining any particular permanent. It appears that God knows how to efficiently solve matrix permanents, but conspires against us such that we remain ignorant of them. \latinquote{Deus magnus est.}

The size of a boson-sampler required to exhibit post-classicality is under active debate, as has undergone much historical revision. But some recent estimates suggest that as many assssss \mbox{$n=50$} photons in \mbox{$m=2,500$} modes might be a rough guide for such a threshold \cite{neville2017no}. Needless to say, this is already an extremely challenging technological goal, suggesting that although the \textsc{BosonSampling} problem is far simpler than universal LOQC, it is far from simple.

\textsc{BosonSampling} in the presence of various error models, such as loss, source non-determinism and mode-mismatch, has been extensively investigated \cite{bib:RohdeErrBS12, bib:RohdeSPDC13, bib:ScottLost16, bib:RohdeArbSpec15, bib:RandBS}. 

%
% Multiplexed Boson-Sampling
%

\paragraph{Multiplexed \textsc{BosonSampling}} \index{Boson-sampling!Multiplexed}

As discussed in Sec.~\ref{sec:single_phot_src}, SPDC is the most common present-day implementation of single-photon sources. However, despite their ready availability, they suffer from non-determinism, with single-photon heralding probability given by Eq.~(\ref{eq:SPDC_p_prep}). To improve upon this, multiplexed sources can be employed \cite{bib:RohdeSPDC13}, improving effective single-photon preparation probabilities asymptotically to unity, as given by Eq.~(\ref{eq:SPDC_multiplex}).

However, rather than employing a multiplexed SPDC source in place of each of the required $n$ single-photons, we can instead employ a larger multiplexer that routes \mbox{$N\gg n$} sources to $n$ modes, which is far more efficient than $n$ independent multiplexed single-photon sources.

\begin{figure}[!htbp]
\includegraphics[clip=true, width=0.475\textwidth]{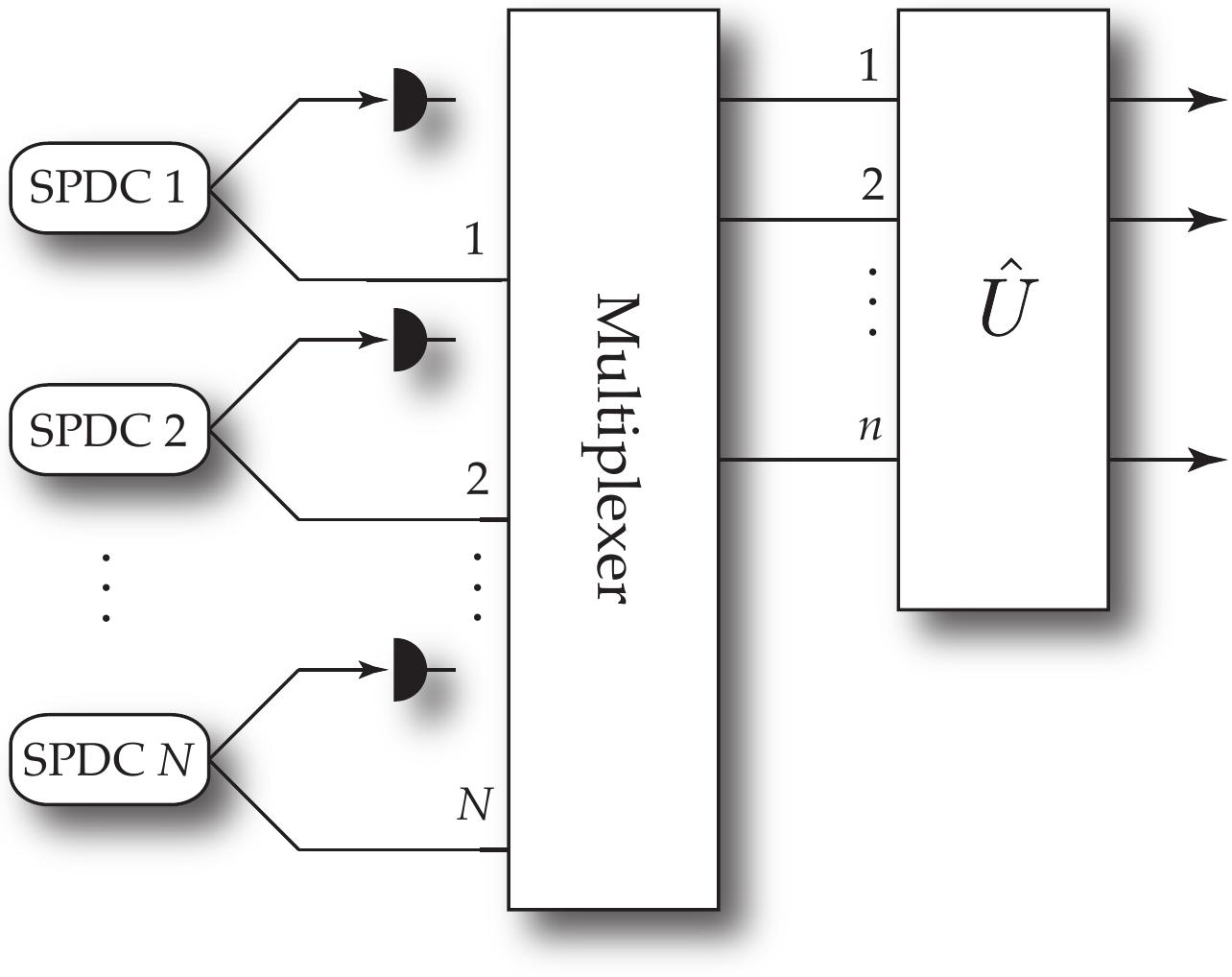}
\captionspacefig \caption{Model for multiplexed \textsc{BosonSampling}. We operate \mbox{$N\gg n$} SPDC sources in parallel, which are multiplexed to the first $n$ modes of the interferometer $\hat{U}$. With sufficiently large $N$ it becomes asymptotically certain that at least $n$ single-photons will be heralded, thereby successfully preparing the desired \textsc{BosonSampling} input state.} \label{fig:multiplexed_bs}\index{Multiplexed!Boson-sampling}
\end{figure}

The model is shown in Fig.~\ref{fig:multiplexed_bs}. We begin by operating $N$ SPDC sources in parallel. Clearly if $N$ is sufficiently large with respect to $n$, it becomes asymptotically certain that at least $n$ photons will be heralded. When this occurs, the successfully prepared $n$ photons -- in whatever configuration they happen to occur -- are routed to the first $n$ modes of the \textsc{BosonSampling} interferometer $\hat{U}$ by the multiplexer (which is classically controlled by the SPDC heralding outcomes), and the protocol proceeds as usual.

\begin{figure}[!htbp]
\includegraphics[clip=true, width=0.475\textwidth]{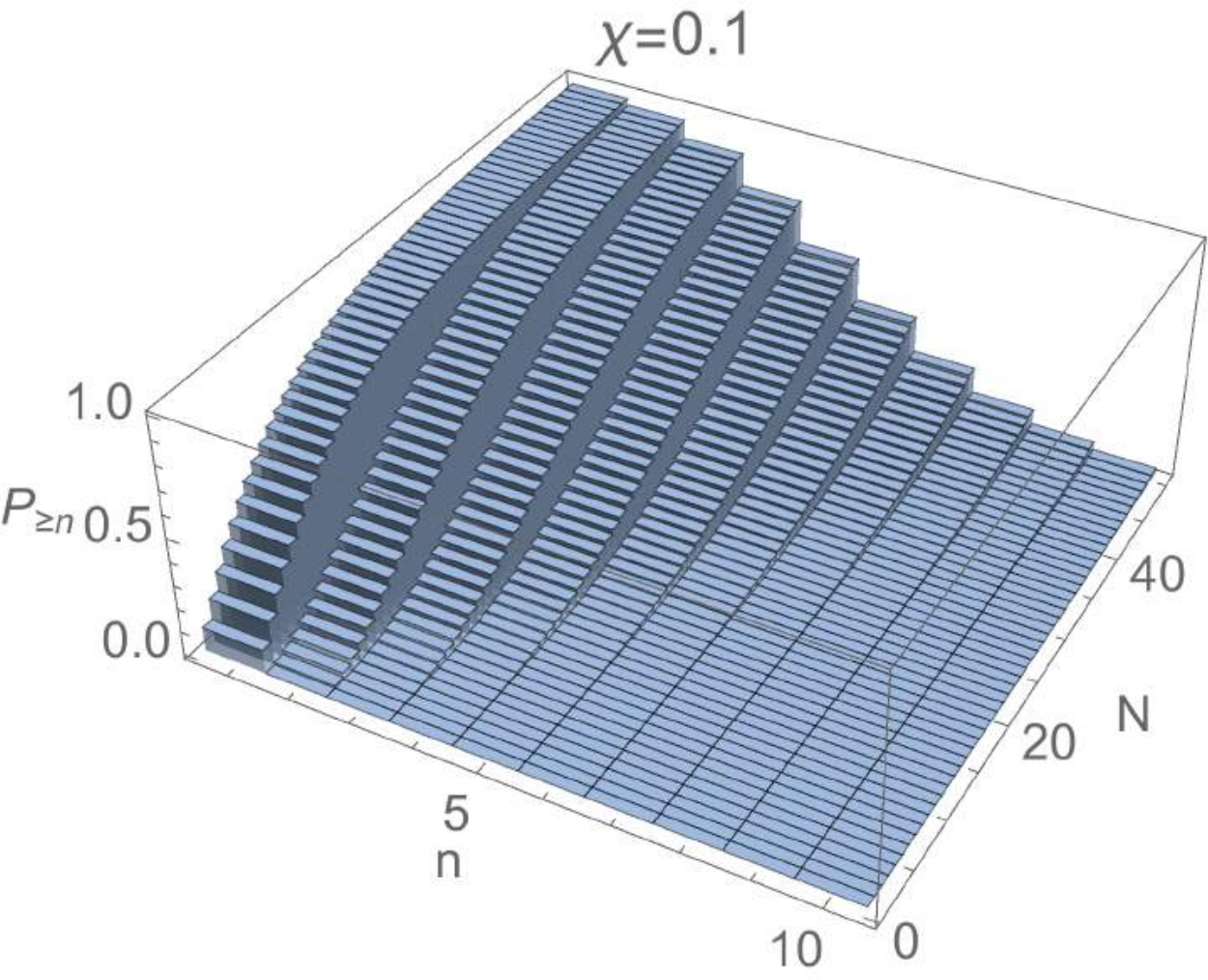}
\captionspacefig \caption{Probability of successfully preparing at least $n$ photons for \textsc{BosonSampling} from a multiplexed source comprising a bank of $N$ SPDC sources in parallel. For sufficiently large $N$, we prepare the desired $n$ photons with probability asymptoting to unity.} \label{fig:multiplex_bs_res}\index{Multiplexed!Boson-sampling}
\end{figure}

Specifically, the probability of at least $n$ successful single-photon heralding events occurring is,
\begin{align}
P_{\geq n} = \sum_{i=n}^\infty \binom{N}{i} 	{P_\mathrm{herald}}^i (1-P_\mathrm{herald})^{N-i},
\end{align}
where,
\begin{align}
	P_\mathrm{herald} = \chi^2(1-\chi^2),
\end{align}
is the probability of a single SPDC source heralding the preparation of a single-photon. This quantity asymptotes to unity for \mbox{$N\gg n$}, as shown in Fig.~\ref{fig:multiplex_bs_res}.

However, although this procedure works in-principle, it comes at the expense of a large number of sources, $N$, and more challengingly, fast-feedforward. Keep in mind that if we were able to perform complex fast-feedforward, we might be able to do much more (and far more interesting things) than just \textsc{BosonSampling} in the first place (i.e universal LOQC)!

%
% Scattershot Boson-Sampling
%

\paragraph{Scattershot \textsc{BosonSampling}} \index{Boson-sampling!Scattershot}

\begin{figure}[!htbp]
\includegraphics[clip=true, width=0.35\textwidth]{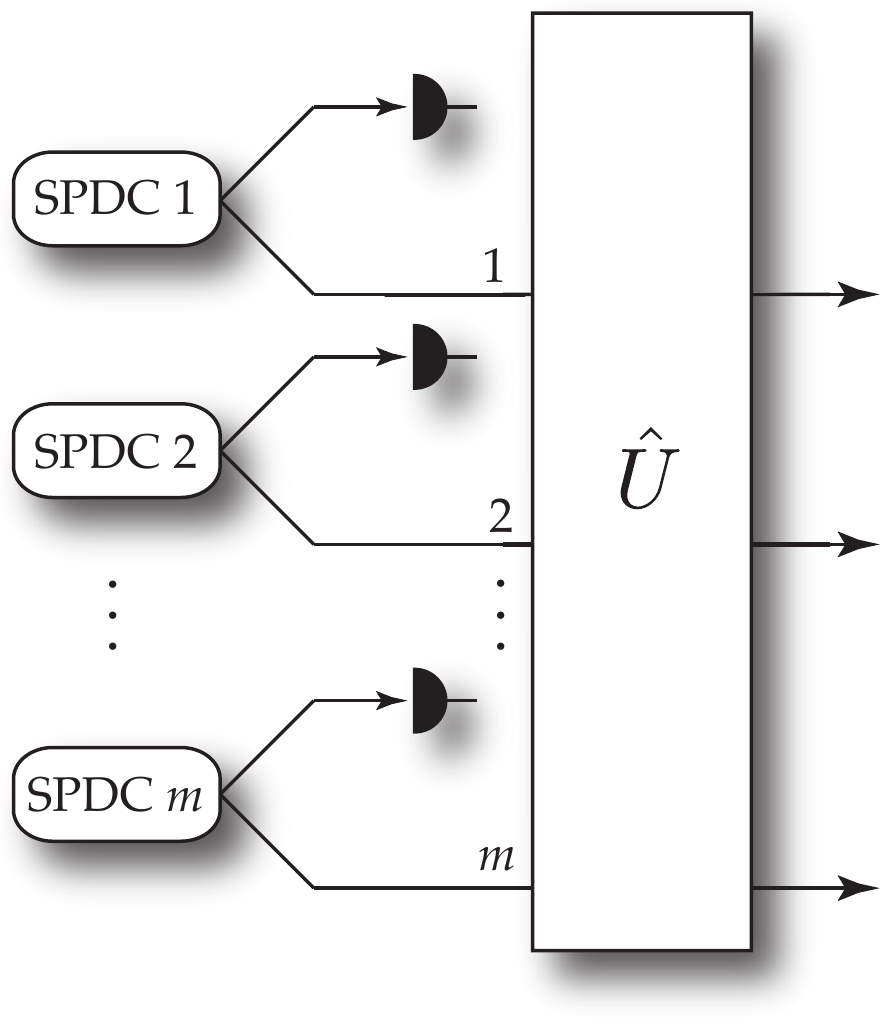}
\captionspacefig \caption{Model for `scattershot' \textsc{BosonSampling}. An SPDC source is inputted into all $m$ input modes. We post-select upon detecting a total of $n$ photons in the heralding modes, irrespective of their configuration, yielding an $n$-photon instance of \textsc{BosonSampling} with randomised input configuration. Unlike multiplexed architectures, the scheme remains entirely passive, without requiring adaptive fast-feedforward.} \label{fig:scattershot_model}
\end{figure}

A variation on SPDC-based \textsc{BosonSampling}, known as `scattershot'\index{Scattershot boson-sampling} \textsc{BosonSampling}, has been presented \cite{bib:RandBS}, which obviates the difficultly of fast multiplexing in the approach described previously. Here, rather than inputting an SPDC source into the first $n$ of the $m$ modes, we input a source into \textit{every} mode, i.e $m$ sources in total. We then accept all events with $n$ heralding successes in total, irrespective of the configuration in which they occur. This has the effect of implementing $n$-photon \textsc{BosonSampling} with an additional layer of randomisation on the input modes (i.e a randomisation in the input configuration, $T$, which is ordinarily fixed). However, since the algorithm is already randomised, this additional layer of randomisation does not undermine the complexity proofs, which hold as is. The scattershot model is shown in Fig.~\ref{fig:scattershot_model}.

\begin{figure}[!htbp]
\includegraphics[clip=true, width=0.475\textwidth]{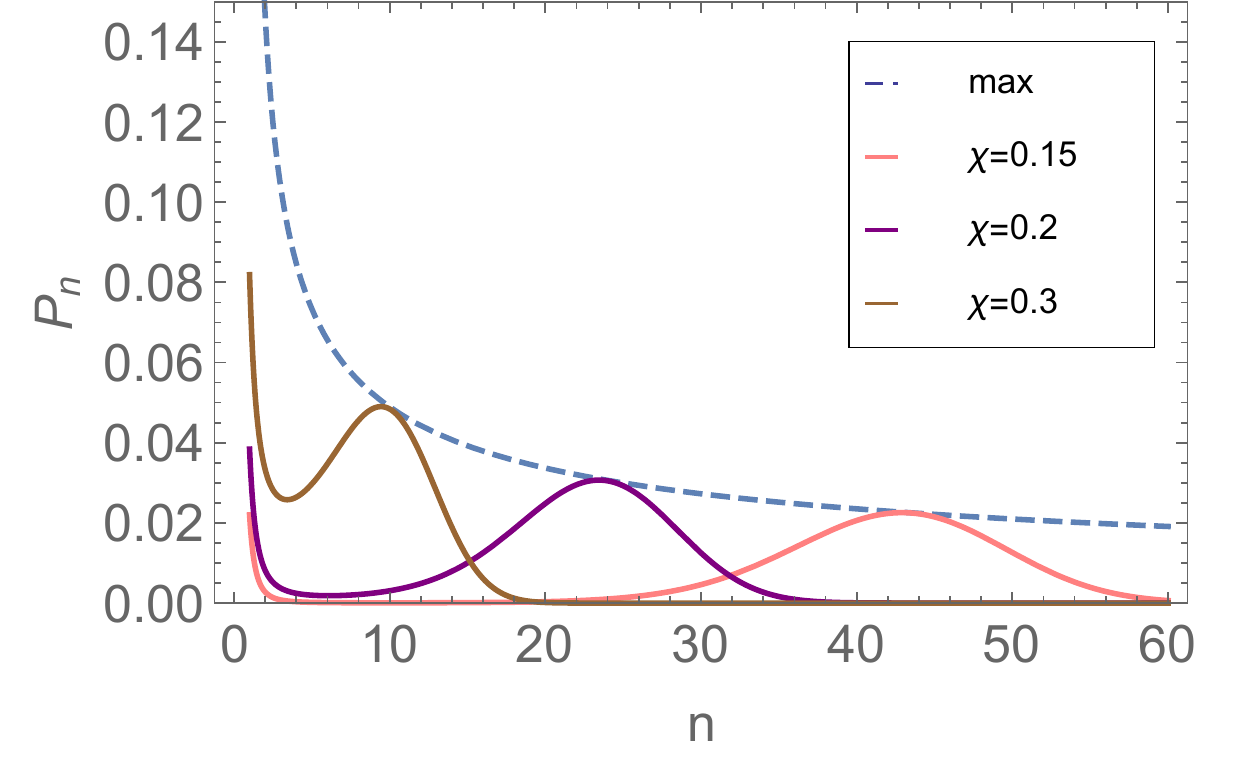}
\captionspacefig \caption{Probability of successfully implementing an instance of $n$-photon \textsc{BosonSampling} using the scattershot technique, whereby all input modes are fed with an SPDC source, and all $n$-photon heralding events are accepted, irrespective of their configuration.} \label{fig:scattershot_probs}
\end{figure}

By keeping all configurations of $n$ photons, rather than just the \mbox{$\ket{T}=\ket{1}^{\otimes n} \ket{0}^{\otimes (m-n)}$} case, we effectively boost the $n$-photon heralding probability from,
\begin{align}
	P_n = \chi^{2n}(1-\chi^2)^n,	
\end{align}
to,
\begin{align}
	P_n = \binom{n^2}{n}\chi^{2n}(1-\chi^2)^{n^2},	
\end{align}
exhibiting a binomial enhancement in $n$-photon events, yielding a significant improvement in count-rates. For a given desired photon-number $n$, choosing the value for the squeezing parameter, $\chi$, which maximises $P_n$, we obtain the optimised success probability,
\begin{align}
	P_n^{(\mathrm{opt})} \approx \frac{1}{e\sqrt{2\pi(n-1)}},
\end{align}
which exhibits only polynomial scaling against photon-number $n$, and is therefore scalable. This is shown in Fig.~\ref{fig:scattershot_probs}. This is in stark contrast to conventional \textsc{BosonSampling}, where the success probability decays exponentially with photon-number, and is therefore inefficient. Importantly, unlike the multiplexed approach, this efficiency improvement does not require any active elements, remaining in the true spirit of \textsc{BosonSampling}.

%
% Coherent States
%

\subsubsection{Coherent states} \label{sec:coherent_state_QC} \index{Coherent states!Computation}

A linear optics network acting on a tensor product of coherent state inputs implements simple matrix multiplication\index{Matrix!Multiplication} on the vector of coherent amplitudes. Specifically, Eq.~(\ref{eq:LO_unitary_map}) implies that for the input multi-mode coherent state,
\begin{align}
\ket{\psi}_\mathrm{in} = \ket{\vec\alpha} = \ket{\alpha_1,\dots,\alpha_m},
\end{align}
where $\alpha_i$ is the coherent amplitude of the $i$th mode, the linear map now takes the form,
\begin{align}
\beta_i = \sum_{j=1}^m U_{i,j} \alpha_j,
\end{align}
where the output state is the separable multi-mode coherent state,
\begin{align}
\ket{\psi}_\mathrm{out} = \ket{\vec\beta} = \ket{\beta_1,\dots,\beta_m}.
\end{align}
Equivalently, this could simply be expressed as the matrix equation,
\begin{align}\label{eq:coherent_state_LO_map}
\vec{\beta} = U\cdot\vec{\alpha}.
\end{align}

Of course this is not strictly a \textit{quantum} computation, since:
\begin{itemize}
\item It can be efficiently classically computed using $O(m^2)$ operations\footnote{Using the na\"ive element-wise approach, which can be further improved upon using more sophisticated contemporary algorithms.}, thus residing in \textbf{P}\index{P}.
\item Coherent states are considered classical states (i.e approximated by laser light\index{Lasers!Light}) with strictly positive Wigner and $P$-functions\index{Wigner function}\index{P-function}.
\item There is no entanglement between modes.
\item The algorithm offers no quantum (exponential) speedup.
\end{itemize}

Despite offering no direct quantum advantage, we introduce this model for restricted computation, since it lends itself very elegantly to a form of homomorphic encryption, to be described in detail in Sec.~\ref{sec:homo_coherent_state}.

The applications for matrix multiplication needn't be stated, as it forms such a ubiquitous elementary primitive throughout linear algebra and in solving systems of differential equations, with applications too many to count.

%
% Other Linear Optics Sampling Problems
%

\subsubsection{Other linear optics sampling problems} \label{sec:other_LO_samp_probs} \index{Linear optics!Sampling problems}

Beyond photonic \textsc{BosonSampling}, much investigation has explored the computational hardness of other types of linear optics sampling problems, using states beyond just single photons.

%
% Hard Problems
%

\paragraph{Hard problems}\index{Hard linear optics sampling problems}

In addition to photonic \textsc{BosonSampling}, several authors have presented strong evidence that other classes of quantum states of light exist, which yield computationally complex sampling problems under the action of linear optics. Most notably, such evidence has been provided for the following:
\begin{itemize}
\item \cite{bib:RandBS} considered two-mode squeezed vacuum (or SPDC) states, a type of Gaussian state with strictly positive Wigner function. This is the same as the scattershot model presented in Sec.~\ref{sec:boson_sampling}.\index{Two-mode squeezed vacuum states}\index{Gaussian states}
	\begin{align}
		\ket\psi_\mathrm{in} = \sqrt{1-\chi^2}\sum_{n=0}^\infty \chi^n\ket{n,n}.
	\end{align}
\item \cite{bib:RohdeDisp15} considered photon-added coherent states and displaced single-photon states.\index{Photon-added coherent states}\index{Displaced single-photon states}
	\begin{align}
		\ket\psi_\mathrm{in} &\propto \hat{a}^\dag\ket\alpha,\nonumber\\
		\ket\psi_\mathrm{in} &\propto \hat{D}(\alpha)\ket{1}.
	\end{align}
\item \cite{bib:RohdePhotAdd15} considered photon-added or -subtracted squeezed vacuum states.\index{Photon-added squeezed vacuum states}\index{Photon-subtracted squeezed vacuum states}
	\begin{align}
		\ket\psi_\mathrm{in} &\propto \hat{a}^\dag\hat{S}(\chi)\ket{0},\nonumber\\
		\ket\psi_\mathrm{in} &\propto \hat{a}\hat{S}(\chi)\ket{0}.
	\end{align}
\item \cite{bib:RohdeCat15} considered `cat' states -- superpositions of coherent states.\index{Cat states}
	\begin{align}
		\ket\psi_\mathrm{in} \propto \ket\alpha \pm \ket{-\alpha}.
	\end{align}
\end{itemize}

Preparation of all of these classes of quantum states of light present their own technological challenges, some very daunting, and all much harder to prepare than single-photons. Thus, the ability to outsource their preparation would be a useful application for the quantum cloud.

%
% Easy Problems
%

\paragraph{Easy problems}\label{sec:easy_LO_probs}\index{Easy linear optics sampling problems}

On the other hand, some classes of optical states are known to be efficiently classically simulable under linear optics evolution and photo-detection. This includes coherent states, thermal states, or any state with strictly positive $P$-function\footnote{A strictly positive $P$-function implies that the state can be considered a purely classical mixture of coherent states (each of which are classically efficient to simulate), according to some classical probability distribution.} (Sec.~\ref{sec:exotic}) \cite{bib:SalehQOCCC15, bib:SalehEffSim16}. Furthermore, Gaussian states evolved via linear optics and measured using Gaussian measurements have been shown to be computationally easy to simulate \cite{bib:Bartlett02, bib:Bartlett02b}.

While such negative results might be somewhat depressing, it is extremely insightful to understand these regimes, in the interest of avoiding investing excruciating effort into trying to instead fruitlessly prove that they are hard.

%\comment{Cite Saleh/Carlton paper on most optical states generating entanglement. Is it exponential entanglement? If so, is this a necessary but not sufficient condition for computational complexity.}

The simplest example of an easy such problem is coherent state linear optics\index{Coherent states!Linear optics}, as discussed in Sec.~\ref{sec:coherent_state_QC}. Taking this notion further, recall from Sec.~\ref{sec:exotic} that one of the phase-space\index{Phase!Space} representations for generic optical states is the $P$-function\index{P-function}, which represents a density operator as a sum over coherent states,
\begin{align}
\hat\rho = \int\!\!\!\int P(\alpha) \ket{\alpha}\bra{\alpha} d^2\alpha.
\end{align}
Here $P(\alpha)$ is a quasi-probability function\index{Quasi-probability functions}. Importantly, iff the $P$-function is strictly non-negative, \mbox{$P(\alpha)\geq 0 \,\,\forall \,\alpha$}, the optical state may be trivially interpreted as a purely classical mixture of coherent states ($P(\alpha)$ would be a delta function for pure coherent states). If, on the other hand, the $P$-function exhibits negativity for any $\alpha$, this interpretation breaks down and is indicative of the state exhibiting non-classical\index{Non-classical states} behaviour.

Alg.~\ref{alg:positive_P_sim} describes an efficient classical algorithm for simulating the output photo-statistics of a linear optics sampler fed with strictly non-negative $P$-function input states \cite{PhysRevLett.114.060501}.

\begin{table}[!htbp]
\begin{mdframed}[innertopmargin=3pt, innerbottommargin=3pt, nobreak]
\texttt{
function SimulatePositiveP($\vec{P}$):
\begin{enumerate}
	\item for(m$\in$modes) \{
	\setlength{\itemindent}{0.2in}
	\item Randomly choose a sample $\alpha_m$ from probability distribution function $P_m(\alpha)$\index{Probability distribution function}.
	\setlength{\itemindent}{0in}
		\item \}
		\item Evolve the set of input coherent state samples through the linear optics network,
		\begin{align}
		\vec\beta = U\cdot\vec\alpha.
		\end{align}
	\item for(m$\in$modes) \{
	\setlength{\itemindent}{0.2in}
	\item The probability of measuring $n$ photons in the $m$th mode is given by the distribution,
	\begin{align}
	D_{m,n} &= |\braket{\beta_m|n}|^2 \nonumber \\
	&= e^{-|\beta_m|^2} \frac{{\beta_m}^{2n}}{n!}.
	\end{align}
	\item Choose $n$ from this distribution.
	\setlength{\itemindent}{0in}
	\item \}
	\item return($\vec{n}$).
    \item $\Box$
\end{enumerate}}
\end{mdframed}
\captionspacealg \caption{Efficient classical algorithm for simulating any linear optics sampling problem whose input states have strictly non-negative $P$-functions, with output measured via photon-counting. $\vec{P}$ is the vector of $P$-functions for all modes.} \label{alg:positive_P_sim}
\end{table}

%
% Quantum Walks
%

\subsubsection{Quantum walks} \label{sec:QW} \index{Quantum walks}

Photonic quantum walks (QWs) are the other main contender for implementing restricted quantum computation, without requiring the full spectrum of challenging LOQC operations. The resource requirements are the same as for \textsc{BosonSampling}, the difference being that now instead of choosing a Haar-random unitary matrix for the interferometer, we choose one which encodes a graph. The photons are now referred to as `walkers', and they evolve by following edges within the graph, `hopping' between neighbouring vertices.

With only a single walker (photon), nothing computationally complex can occur in the system, since a single photon evolving under passive linear optics can be efficiently classically simulated\footnote{Note that the literature has described QW schemes, both discrete-time \cite{lovett2010universal} and continuous-time \cite{bib:Childs09}, that are universal for quantum computation. However, such universal schemes require an exponential number of vertices in the underlying graph, which clearly does not lend itself to efficient optical representation.}. However, once multiple walkers are introduced we have a system with almost identical features to \textsc{BosonSampling}, differing only in the structure of the linear optics unitary.

There are two predominant varieties of quantum walks: discrete- \cite{lovett2010universal} and continuous-time \cite{bib:Childs2009}, which we will now introduce. Algorithms have been described for both the discrete- and continuous-time QW models.

%
% Continuous-Time Quantum Walks
%

\paragraph{Continuous-time quantum walks}\index{Continuous-time quantum walks}

In the continuous-time QW model, a Hamiltonian, $\hat{H}_\mathrm{QW}$, encoding the (Hermitian) adjacency matrix of the QW's graph evolves the walker(s), generating a unitary evolution of the form,
\begin{align}\index{Quantum walks!Hamiltonian}
\hat{U}_\mathrm{QW}(t) = e^{-i\hat{H}_\mathrm{QW}t},
\end{align}
where \mbox{$t\in \mathbb{R}_+$}.

This model lends itself readily to optical wave-guide\index{Waveguides} implementation, where evanescent coupling between neighbouring wave-guides is inherently a continuous-time process. Fig.~\ref{fig:LO_archs}(c) illustrates an example implementation of a linear optics, continuous-time quantum walk on a line in an integrated wave-guide device.

In the context of linear optics, the evolution is best described using the coupled oscillator Hamiltonian\index{Coupled oscillator Hamiltonian},
\begin{align}\label{eq:coupled_osc_ham}
	\hat{H}_\mathrm{QW} = \sum_{i,j=1}^m c_{i,j} \hat{a}^\dag_i\hat{a}_j,
\end{align}
where $\hat{a}^\dag_i$ ($\hat{a}_i$) is the photonic creation (annihilation) operator for the $i$th of the $m$ modes, and the Hermitian matrix $c_{i,j}$ encodes the coupling strength between the $i$th and $j$th modes, which could correspond identically to the QW's graph adjacency matrix.

%
% Discrete-Time Quantum Walks
%

\paragraph{Discrete-time quantum walks}\index{Discrete-time quantum walks}

In the discrete-time QW model, each walker has access to an ancillary `coin' Hilbert space, which is used to record the direction of the walker through the graph. At each discrete time-step the coin is used to update the position (vertex) of the walker, before applying a unitary `coin' operator to the coin Hilbert space. The addition of the coin space is necessary to enable such quantum walks to reside on arbitrary graph topologies, whilst retaining unitarity in their evolution. 

\begin{figure}[!htbp]
\includegraphics[clip=true, width=0.475\textwidth]{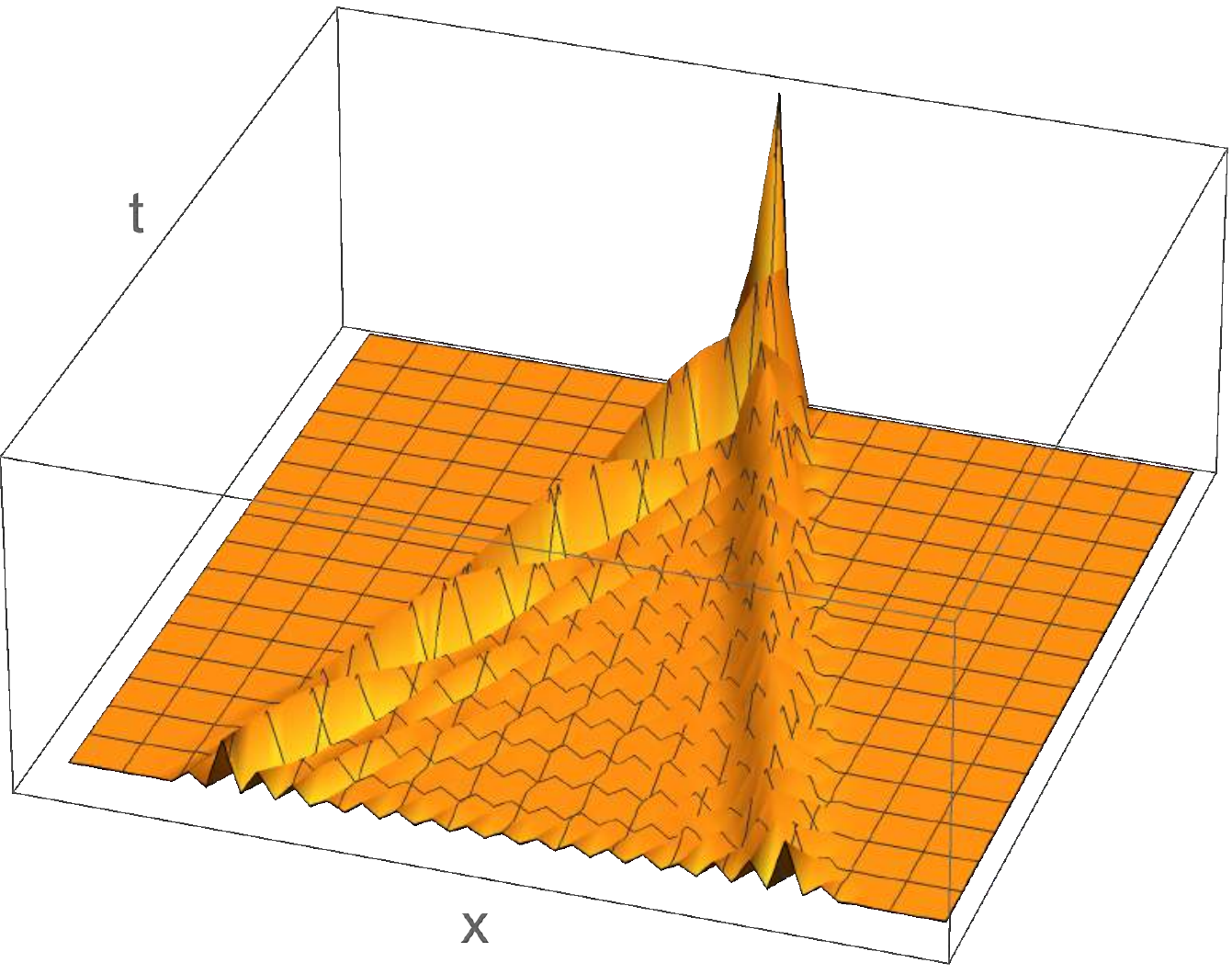}
\captionspacefig \caption{Evolution on a 1D quantum walk on a line with a Hadamard coin operator. The walker begins localised at the origin and then spreads out as a superposition over the position space ($x$) with time ($t$). A key feature of this distribution is that its variance grows quadratically with time, compared with linear growth for the equivalent classical random walk. This enhanced spreading forms the basis of the quantum walk search algorithm, with quadratic enhancement compared to a classical search \cite{childs2004spatial}.} \label{fig:QW_ev}
\end{figure}

We will briefly summarise the discrete-time QW model, as it most readily lends itself to linear optics implementation, and illustrate the parallels with \textsc{BosonSampling}. First, let us consider the standard simple example scenario of a single quantum walker, on a linear graph topology, using a Hadamard coin\index{Hadamard!Gate},
\begin{align}
\hat{C} = \hat{H} = \frac{1}{\sqrt{2}}\begin{pmatrix}
1 & 1 \\
1 & -1
\end{pmatrix},
\end{align}
which has been experimentally demonstrated using both bulk-optics \cite{bib:Broome10} and time-bin encoding \cite{bib:Schreiber10, bib:RohdeQWExp12}. The walker is defined by two Hilbert spaces: the position $x$, and the coin {$c=\pm 1$}, where $+1$ ($-1$) indicates that the walker is moving to the right (left). The basis states are then $\ket{x,c}$, and the state of the walker takes the form,
\begin{align}
\ket\psi = \sum_{x,c} \lambda_{x,c} \ket{x,c}.
\end{align}
The evolution of the walk is given by the coin and step operators,
\begin{align} \index{Coins!Operators}\index{Step operators}
\hat{C}\ket{x,\pm 1} &\to \frac{1}{\sqrt{2}}(\ket{x,+1}\pm \ket{x,-1}), \nonumber \\
\hat{S}\ket{x,c} &\to \ket{x+c,c}.
\end{align}
The Hadamard coin operator could be replaced with any arbitrary $\mathrm{SU}(2)$ matrix. The total time-evolution of the walk is then given by,
\begin{align}
\ket{\psi(t)} = (\hat{S}\hat{C})^t\ket{\psi(0)},
\end{align}
where \mbox{$t\in\mathbb{Z}_+$}. Upon measurement, the probability of the walker being at position $x$ is simply given by summing the probabilities over the coins at a given position,
\begin{align}
P(x) = |\lambda_{x,-1}|^2 + |\lambda_{x,+1}|^2.
\end{align}
An example of this kind of quantum walk is shown in Fig.~\ref{fig:QW_ev}.

The single-walker walk on a linear graph is not of computational interest as it can be efficiently classically simulated. However, the formalism is easily logically generalised to multiple walkers on arbitrary graph topologies. We will illustrate this using the formalism of \cite{bib:RohdeMultiWalk11}, where \textit{walker operators}\index{Walker operators}, rather than walker basis states are evolved under time-evolution (i.e we operate in the Heisenberg picture rather than the Schr{\" o}dinger picture). In an optical context, walker operators are identically photonic creation operators. The walker operators are of the form $\hat{w}(x,c)^\dag$, where $x$ denotes the vertex number currently occupied by the walker, and $c$ denotes the previous vertex occupied by the walker. The single-walker basis states are then of the form $\hat{w}(x,c)^\dag\ket{0}$, where $\ket{0}$ is the vacuum state containing no excitations. Notice the parallels to the previous example of a linear walk, where the coin degree of freedom specifies the direction the walker is following, which effectively acts as memory of the previous position.

\begin{figure}[!htbp]
\includegraphics[clip=true, width=0.2\textwidth]{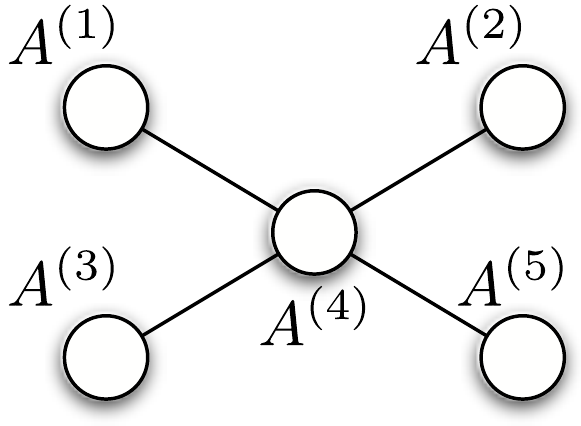}
\captionspacefig \caption{A simple example of a quantum walk on an irregular graph structure. Associated with each vertex $x$ is an $\mathrm{SU}(|n_x|)$ coin operator $A^{(x)}$, where $|n_x|$ is the number of neighbours to $x$.} \label{fig:QW_arbitrary_graph}\index{Quantum walk!Graphs}
\end{figure}

The coin and step operators now take the form,
\begin{align}
\hat{C}: \,\,\, &\hat{w}(x,c)^\dag \to \sum_{j\in n_x}A_{c,j}^{(x)} \hat{w}(x,j)^\dag, \nonumber \\
\hat{S}: \,\,\, &\hat{w}(x,j)^\dag \to \hat{w}(j,x)^\dag.
\end{align}
Here $n_x$ denotes the set of vertices neighbouring $x$. The coin operators $A^{(x)}$ are \mbox{$\mathrm{SU}(|n_x|)$} unitary matrices representing the weights of edges within this neighbourhood. The step operator, on the other hand, is simply a permutation. A simple example is shown in Fig.~\ref{fig:QW_arbitrary_graph}.

The total time evolution is defined analogously to before,
\begin{align}
\hat{U}_\mathrm{QW}(t) = (\hat{S}\hat{C})^t,
\end{align}
where \mbox{$t\in \mathbb{Z}_+$}.

With this formalism, multiple walkers are easily accommodated for simply with the addition of extra walker operators. Specifically, the $n$-walker basis states are of the form,
\begin{align}
\ket{\vec{x},\vec{c}} \propto \prod_{i=1}^n \hat{w}(x_i,c_i)^\dag \ket{0},
\end{align}
where we have ignored the normalisation factor, which is a function of the number of walkers in each basis state. Any graph topology can be represented, subject to the constraint that all $A^{(x)}$ are unitary. This implies that every vertex must have as many incoming as outgoing edges, which could be either directed or undirected, subject to this constraint.

Now the probabilities of measuring the walkers in different position configurations will be related to matrix permanents, in a similar manner to \textsc{BosonSampling}. But now the permanents will be of matrices that are functions of the set of $A^{(x)}$ matrices characterising the graph, rather than a Haar-random matrix.

\begin{figure}[!htbp]
\includegraphics[clip=true, width=0.475\textwidth]{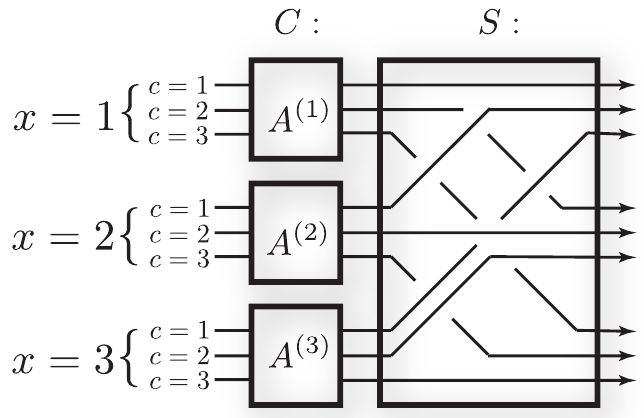}
\captionspacefig \caption{Linear optics decomposition for a single step of an arbitrary 3-vertex discrete-time quantum walk. The coin operators, $A^{(x)}$, may be arbitrary $\mathrm{SU}(3)$ matrices, whereas the step operator is simply a permutation of the optical modes. $O(|V|^2)$ optical modes are required, which scales efficiently.} \label{fig:QW_LO_representation}
\end{figure}

It was shown by \cite{bib:RohdeMultiWalk11} that any such walk can be efficiently represented using a linear optics decomposition comprising at most $O(|V|^2)$ optical modes. Such a decomposition for \mbox{$|V|=3$} is shown in Fig.~\ref{fig:QW_LO_representation}.

Because of the graph structure of the walk, it lends itself naturally to distributed implementation. We might imagine that different subgraphs -- or `widgets' \cite{bib:Lovett10, bib:Childs09} -- implement different computational primitives or subroutines. These widgets might be proprietary or expensive to implement, and are therefore best outsourced over a network.

%
% Continuous-Variables
%

\subsection{Continuous-variables} \label{sec:CV_QC} \index{Continuous-variables!Quantum computation}

\sectionby{Zixin Huang}\index{Zixin Huang}

Until now we have focussed on optical systems where quantum information is encoded into discrete variables\index{Discrete-variables}, such as photon-number or polarisation. However, quantum states of light can also be considered in terms of continuous-variables (CVs)\index{Continuous-variables} in phase-space\index{Phase!Space}.

In this picture, using squeezed states\index{Squeezed states} as a resource, qubits can be closely approximated by vacuum states squeezed in orthogonal directions, where the closeness of the approximation is determined by the squeezing parameter\footnote{In the limit of infinite squeezing the two orthogonally squeezed states becomes orthogonal, enabling the encoding of a genuine qubit.}. Squeezed state encoding of quantum information was introduced in Sec.~\ref{sec:squeezed_enc}.

A universal gate set can be constructed, enabling universal quantum computation to be implemented using such an encoding. All the necessary elements may be readily implemented using present-day quantum optics technology and numerous CV quantum protocols have been demonstrated in recent years \cite{bib:RevModPhys.77.513}. Most notably, very large-scale CV cluster states have been experimentally prepared in the laboratory \cite{bib:yoshikawa2016invited}.

%
% Encoding Quantum Information Using Squeezed States
%

\subsubsection{Encoding quantum information using squeezed states}\index{Continuous-variables!Encoding}

As discussed in Sec.~\ref{sec:squeezed_enc}, position and momentum eigenstates are orthogonal and may therefore be employed to encode a single qubit. However, these eigenstates have infinite energy. That is, they are infinitely squeezed in phase-space\index{Infinite squeezing}. But position and momentum eigenstates can be closely approximated using finite, but strong squeezing, where there is a direct tradeoff between energy and the quality of the qubit approximation. The squeezed states\index{Squeezed states} in the two quadratures\index{Quadratures} will now no longer be orthogonal, but will have a small overlap that asymptotes to zero as squeezing is increased. Thus, squeezed states can be used to approximate qubits using non-orthogonal basis states. Squeezed states may be prepared directly using a spontaneous parametric down-conversion (SPDC)\index{Spontaneous parametric down-conversion (SPDC)} process (Sec.~\ref{sec:single_phot_src}) \cite{bib:PhysRevLett.75.4337, bib:o2009photonic}. CV encoding using squeezed states is illustrated in phase-space in Fig.~\ref{fig:squeezed_state_encoding}.

Mathematically, there are two operators of interest, the single- and two-mode squeezing operators\index{Squeezing!Operators},
\begin{align}\label{eq:sq_op}
\hat{S}_\mathrm{single}(\xi) &= \exp\left[ \frac{1}{2}(\xi^*\hat{a}^2 - \xi{\hat{a}^{\dag 2}})\right],\nonumber\\
\hat{S}_\mathrm{two}(\xi) &= \exp\left[ \xi\hat{a}_1^\dag\hat{a}_2^\dag + \xi^*\hat{a}_1\hat{a}_2 \right],
\end{align}
where \mbox{$\xi\in \mathbb{C}$} is the squeezing parameter\index{Squeezing!Parameter}. Both of these may be realised physically using non-linear crystals\index{Non-linear!Crystals} with second order non-linearities\index{Second order non-linearities}, readily available in present-day labs.

%
% Phase-Shifters
%

\subsubsection{Phase-shifters}\index{Phase!Shifts}\index{Phase!Space!Rotations}

A phase-shifter\index{Phase!Shifts}, implementing the unitary operation,
\begin{align}
\hat{R}(\theta) = e^{i\theta \hat{n}},
\end{align}
where $\hat{n}$ is the photon-number operator\index{Photon-number!Operators}, rotates a state in phase-space by angle $\theta$ about the origin, implementing the transformation between the position and momentum operators,
\begin{align}\label{eq:xp_theta}
\begin{pmatrix}
\hat x_{\theta}\\
\hat p_{\theta}
\end{pmatrix}
=
\begin{pmatrix}
\cos\theta & \sin\theta \\
-\sin\theta & \cos\theta
\end{pmatrix}
\begin{pmatrix}
\hat x\\
\hat p
\end{pmatrix}.
\end{align}
It is evident upon inspection that this can be thought of as a single-qubit rotation in position/momentum space.

%
% Bell pairs
%

\subsubsection{Bell pairs}\index{Bell!States}\label{sec:CV_bell_pairs}\index{Continuous-variables!Bell states}

\begin{figure}[!htbp]
\includegraphics[clip=true, width=0.475\textwidth]{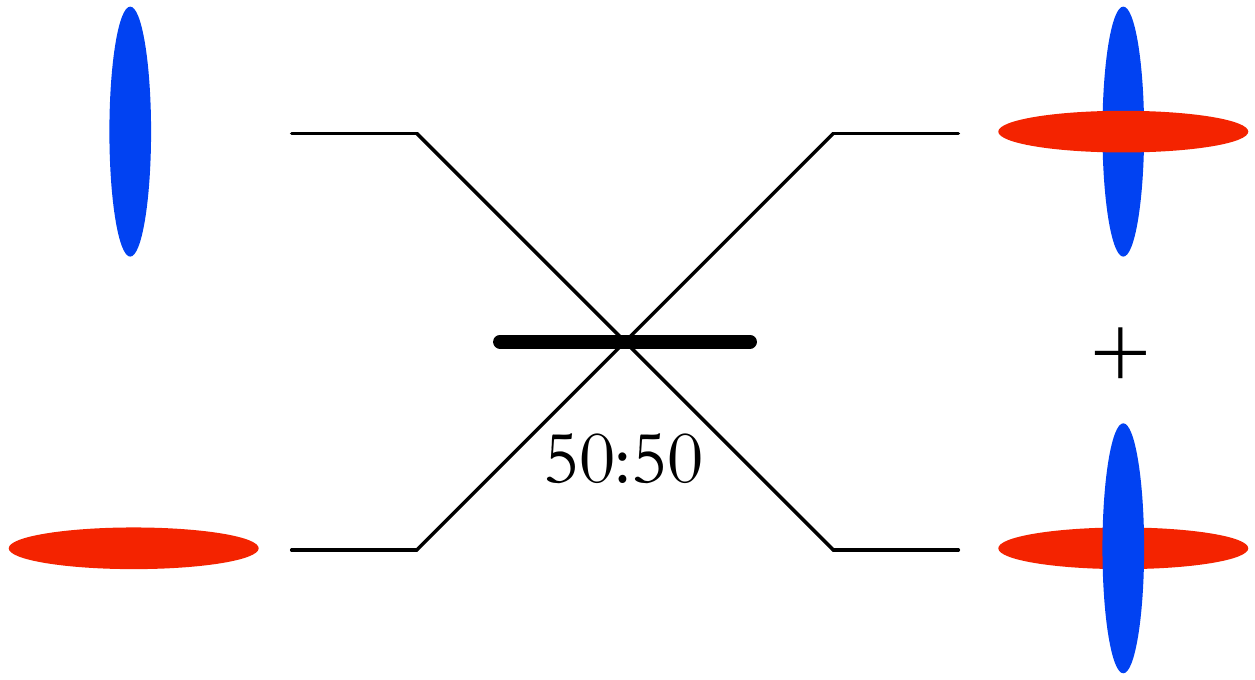}
\captionspacefig \caption{Preparation of an (approximate) CV Bell pair using a 50:50 beamsplitter to entangle two states, squeezed in orthogonal directions in phase-space.}\label{fig:CV_bell_pair}	
\end{figure}

In the squeezed state basis one can prepare Bell pairs in an analogous manner to doing so using single-photon encoding. Namely, mixing two orthogonally squeezed states (squeezed and anti-squeezed) on a 50:50 beamsplitter generates Bell-type entanglement. This is shown in Fig.~\ref{fig:CV_bell_pair}.

%
% Measurement
%

\subsubsection{Measurement}\index{Continuous-variables!Measurement}

Measurement of CV states may be performed using homodyne detection\index{Homodyne detection} (Sec.~\ref{sec:homodyne}), which performs a projection along an arbitrary axis in phase-space, allowing $\hat{x}$ and $\hat{p}$, or any linear combination of the two, to be directly sampled, i.e referring to Eq.~(\ref{eq:xp_theta}), we can access the observables $\hat{x}_\theta$ and $\hat{p}_\theta$ for any $\theta$.

Using a 50:50 beamsplitter to implement the reverse of Bell pair preparation, one can construct a Bell analyser for performing projections in the Bell basis\index{Bell!Measurements}, gifting us a highly-cherished entangling operation, useful for all manner of protocols (Part.~\ref{part:protocols}). The measurement is deterministic, and requires only homodyne detections.

% \comment{Is the Bell projection deterministic, and, can it distinguish all four?}
% \comment{The measurement is indeed deterministic. In CV there is a continuous family of Bell states, instead of four: the two-mode squeezed-vacuum with different phases.}

%
% Logical Operations
%

\subsubsection{Logical operations}\index{Continuous-variables!Logical operations}

In the position/momentum picture, the logical generalisations of the single-qubit Pauli $\hat{X}$ and $\hat{Z}$ gates may be thought of as displacements (Sec.~\ref{sec:non_lin_opt}) in the real and imaginary directions in phase-space \cite{bib:KokLovettBook},
\begin{align}
\hat{X}(s) \equiv \hat{D}(s)	, \, s\in\mathbb{R},\nonumber\\
\hat{Z}(t) \equiv \hat{D}(it), \, t\in\mathbb{R},
\end{align}
where $\hat{D}(\alpha)$ is the phase-space displacement operator\index{Displacement operator} from Eq.~(\ref{eq:disp_op}). These have the logical action,
\begin{align}
	\hat{X}(s)\ket{x} &= \ket{x+s},\nonumber \\
	\hat{Z}(t)\ket{p} &= \ket{p+t}. 
\end{align}

The logical generalisation of the CZ gate is,
\begin{align}
\hat{U}_\mathrm{CZ} = e^{\frac{i}{2} \hat x_1 \hat x_2},
\end{align}
which transforms two-mode quadrature eigenstates as,
\begin{align}
\hat{U}_\mathrm{CZ} \ket{s}_1 \ket{t}_2 = e^{\frac{i}{2} s_1 t_2} \ket{s}_1\ket{t}_2.
\end{align}
Intuitively, it is evident upon inspection of the form of the generalised CZ gate that it leaves logical basis state amplitudes unchanged, but adds state-dependent phases to them. Qualitatively, this is exactly what a regular CZ gate does in the space of two qubits, except that the basis is discrete rather than continuous.

A full set of circuit model CV gates, universal for CV quantum computation is summarised in Tab.~\ref{tab:CV_gates} \cite{bib:RevModPhys.84.621}.

\startnormtable
\begin{table*}[!htbp]
\begin{tabular}{ |c|c|c| } 
 \hline
 Qubit model gate &  CV equivalent & Implementation \\ 
  \hline\hline
 Pauli $X$ & $\hat{X}(s) = \exp[-i s \hat p]$ & Displacement \\ 
 Pauli $Z$ & $\hat{Z}(t) = \exp[i t \hat x]$ & Displacement \\ 
 Phase gate & $\hat{P}(\eta) = \exp[i \eta \hat x^2]$ & Single-mode squeezer \& quadrature rotation \\
Hadamard   & $\hat{F}=\exp[i \frac{\pi}{8}(\hat p^2+\hat x^2)]$ & Phase-shift \\
CZ		   & $\hat{U}_\mathrm{CZ}= \exp[\frac{i}{2}\hat x_1 \hat x_2]$ & Two beamsplitters \& two squeezers \\
CNOT 	   & $\hat{U}_\mathrm{CNOT} = \exp[-2i\hat x_1 \hat p_2]$ & CZ \& Hadamards \\
Non-linear phase gate &  $\hat{U}_\mathrm{NL}=\exp[i t\hat{x}^n],\,n\geq 3$       &  Probabilistic measurements \\
\hline
\end{tabular}
\captionspacetab \caption{Logical generalisations of a universal gate set to the CV model for quantum computing using squeezed state encoding.\label{tab:CV_gates}}
\end{table*}
\startalgtable

%
% Cluster States
%

\subsubsection{Cluster states}\index{Continuous-variables!Cluster states}

The CV model for quantum computation can be shown to be universal by operating within the cluster state model. As discussed in Sec.~\ref{sec:CSQC}, the basic primitive from which the universality of cluster states arises is the single-qubit teleporter (Fig.~\ref{fig:single_qubit_teleporter}), which enables arbitrary quantum information to be teleported through the substrate state, accumulating the action of single- and 2-qubit quantum gates in the process, thereby building up a large, arbitrary quantum computation as measurements proceed. Thus, to demonstrate the universality of the CV model we must first demonstrate an analogous circuit for single-mode CV state teleportation.

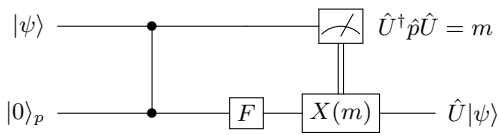
\begin{figure}[!htbp]
	\begin{align}
		\Qcircuit @C=.7em @R=.4em @! {
		\lstick{\ket{\psi}} & \ctrl{1} & \qw & \meter & \hat{U}^\dag\hat{p}\hat{U} = m \\
		\lstick{\ket{0}_p} & \ctrl{0} & \gate{F} & \gate{X(m)} \cwx & \rstick{\hat{U}\ket\psi} \qw \\
		} \nonumber
	\end{align}
	\captionspacefig \caption{The single-mode teleporter using CV state encoding, structurally almost identical to an ordinary single-qubit teleporter, just substituting the operations with their respective CV generalisations. The circuit teleports the input state $\ket\psi$ from the first mode to the second mode, accumulating the action of the single-mode gate $\hat{U}$ applied to the first mode prior to measurement.} \label{fig:CV_teleporter}\index{Continuous-variables!Teleporter}
\end{figure}

An optical circuit for a single-mode teleporter is shown in Fig.~\ref{fig:CV_teleporter}. Evidently, it is structurally almost identical to the standard single-qubit teleporter, just swapping out some operations for their direct CV equivalents.

This circuit has the desired property that, with classical feedforward and local corrections, we can teleport an arbitrary CV qubit state from one mode to another using the CV generalisation of the CZ gate as a resource, and accumulate single-mode unitaries in the process.

Having demonstrated an equivalent single-mode teleporter, we have all we need to construct arbitrary measurement-based quantum computations using the generalised CZ gates as the primitive entangling operation\index{Entangling operations} for preparation of the substrate graph state.

The addition of any non-Gaussian projective measurement allows universal quantum computation using CV cluster states. A class of such gates is the non-linear phase gate\index{Non-linear!Phase gate},
\begin{align}
	\hat{U}_\mathrm{NL}=\exp(it\hat x^{n}),
\end{align}
where \mbox{$n\geq 3$} (the cubic phase gate\index{Cubic phase gate}).

In order to perform universal quantum computation, one needs to implement Hamiltonians of arbitrary degree. In the Heisenberg picture, any Gaussian operation is at most quadratic in the Hamiltonian, and the commutators are also at most quadratic. If one can implement the cubic phase gate, the commutators will now be of degree 3 or 4, and by induction one can construct a Hamiltonian of any degree \cite{kok2010introduction}.

% Fault-tolerance

\subsubsection{Fault-tolerance}\index{Continuous-variables!Fault-tolerance}

As with any architecture for quantum computation, we must give consideration to the inevitable presence of noise corrupting our computation. To overcome this problem, and enable scalable quantum computation, we must demonstrate the capacity for the architecture to be made fault-tolerant.

It was shown by \cite{bib:PhysRevLett.100.030503} that the CV cluster state model can be made fault-tolerant using quantum error correcting codes, sufficient to enable arbitrary scalability.

Having demonstrated fault-tolerance, it can be concluded that CV optical quantum computation is viable, and scalable, with efficient error correction resource overheads.

%
% Hybrid Light-Matter Architectures
%

\subsection{Hybrid light-matter architectures} \label{sec:hybrid} \index{Hybrid!Architectures}

It is unlikely that future, large-scale quantum computers will be purely optical. Some other technologies have a more favourable outlook in terms of scalability. Nonetheless, when it comes to networking quantum computers, optics is the natural approach, motivating investigation into hybrid architectures, where qubits are represented using some non-optical system, but entangling operations (EOs)\index{Entangling operations} between them are mediated by optical states and linear optics \cite{bib:Duan06, bib:Beugnon06}.

\begin{figure}[!htbp]
\includegraphics[clip=true, width=0.475\textwidth]{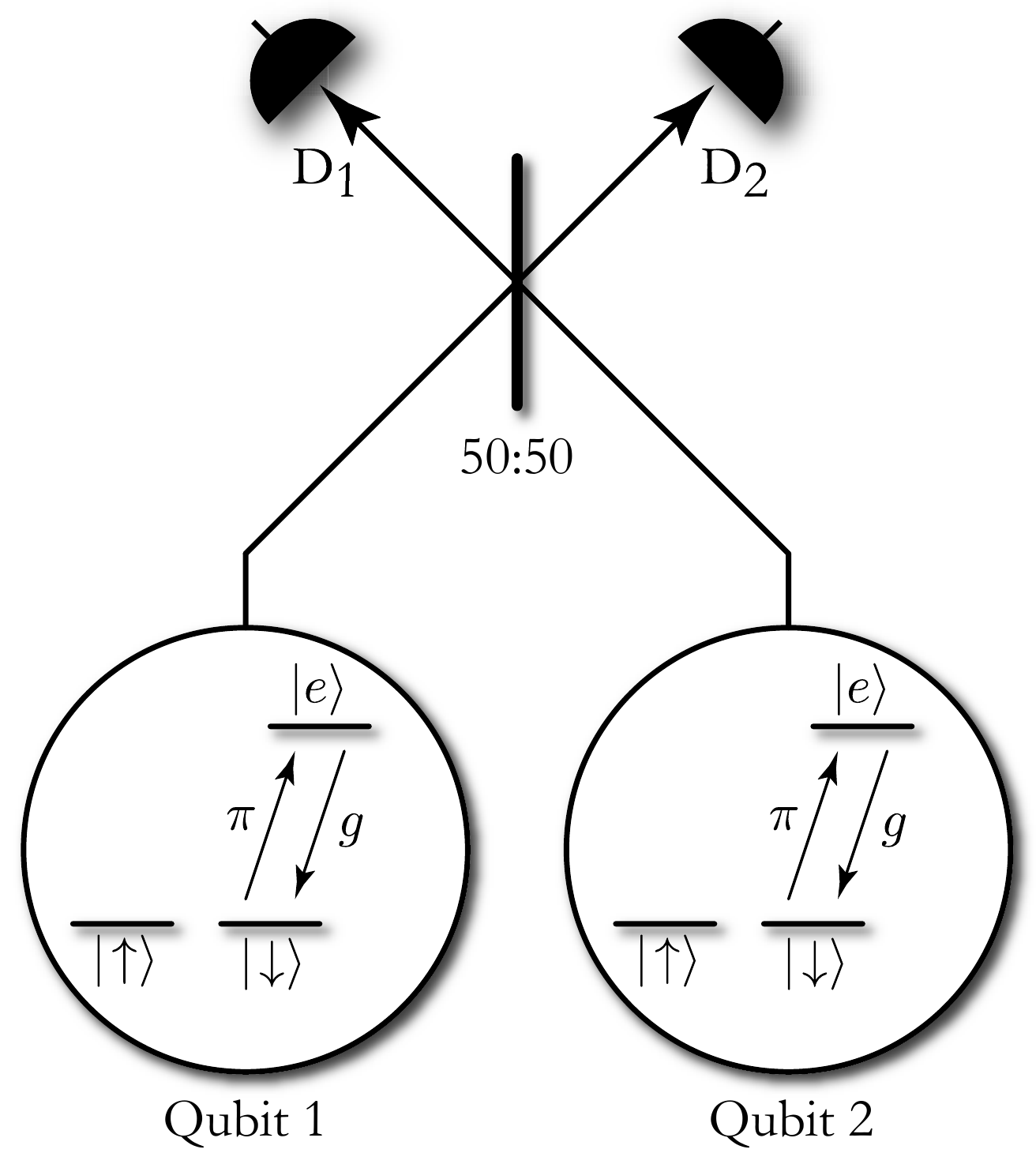}
\captionspacefig \caption{Two atomic systems in $\lambda$-configurations, each coupled with an optical mode. An EO between them is mediated via linear optics which-path erasure. Each system contains two degenerate ground states, which jointly encode a qubit (\mbox{$\ket{0}\equiv\ket{\!\uparrow}$}, \mbox{$\ket{1}\equiv\ket{\!\downarrow}$}), and an additional excited state ($\ket{e}$), which only couples to the $\ket{\!\downarrow}$ state. A $\pi$-pulse excites the electron from the $\ket{\!\downarrow}$ to the $\ket{e}$ state, after which emission of a photon is associated with a coherent relaxation back to $\ket{\!\downarrow}$. If the two optical modes are interfered on a 50:50 beamsplitter, and a single photon is detected between the two photo-detectors, $D_1$ and $D_2$, the two emission processes become indistinguishable, and which-path erasure entangles the two qubits by projecting them onto a maximally entangled Bell pair. More complicated networks based on this EO allow the preparation of cluster states, enabling universal quantum computation. In a quantum networking context, the matter qubits could be held by a client, and the optical interferometry implementing the computation outsourced to the cloud, i.e the PBS, $D_1$ and $D_2$ are implemented in the cloud. This would also facilitate the preparation of shared entangled states, where different clients possess parts of an entangled state, potentially physically separated over long distances.} \label{fig:barrett_kok}
\end{figure}

The natural example is matter qubits which couple to single-photon states, whereupon which-path erasure between coupled optical modes teleports entanglement onto the physical matter qubits\index{Entanglement!Teleportation}.\index{Which-path erasure} Similarly, measurement of the matter qubits may be performed by stimulating the emission\index{Stimulated emission} of photons from them. This idea has been applied to $\lambda$-configuration atomic qubits\index{Atomic!Qubits}\index{$\lambda$-configuration systems} \cite{bib:BarrettKok05}, shown in Fig.~\ref{fig:barrett_kok}, and atomic ensemble qubits \cite{bib:RohdeAtEns10} (Sec.~\ref{sec:atomic_ens}). The protocol is described in Alg.~\ref{alg:which_path}.

In principle, this technique could be applied to any physical system comprising natural or engineered $\lambda$-configured energy levels, which couple to accessible optical modes. This light-matter coupling\index{Light-matter!Coupling} may require carefully constructed optical cavities\index{Optical!Cavities}, as is the case for single atoms. Alternately, atomic ensemble qubits inherently undergo collective enhancement\index{Collective enhancement} in their light-matter coupling, mitigating the need for optical cavities.

Optically-mediated atomic ensemble architectures are particularly attractive, as discussed in Sec.~\ref{sec:atomic_ens}\index{Atomic!Ensembles}, owing to their long coherence lifetimes\index{Coherence!Time}, room temperature operation\index{Room temperature operation}, strong light-matter coupling, and robustness against qubit loss\index{Qubit loss}.

A novel `double heralding'\index{Double heralding} technique, introduced in \cite{bib:BarrettKok05}, allows photon loss\index{Photon loss} to be overcome during which-path erasure. Similarly, quantum states of light can be coupled to two-level quantum systems using Hamiltonians of the form shown in Eq.~(\ref{eq:two_level_hamil}). The preparation of long-distance entanglement between atomic systems has been demonstrated \cite{bib:Matsukevich05, bib:Matsukevich05b}

The attractive feature of this type of approach is that the actual entanglement is generated using all-optical operations, despite the underlying logical qubits being stationary and potentially physically separated a long distance apart, mitigating the need for direct matter-matter interactions, and enabling distributed computation. Optical interfacing is discussed in Sec.~\ref{sec:opt_inter}. This allows the EOs to be performed remotely in the cloud, without physically moving the stationary qubits. Such hybrid systems present an interesting platform for cloud quantum computing -- despite the qubits being stationary, we are able to outsource the interactions between them to distant servers or even satellites.

Importantly, the beamsplitter mediating the which-path erasure EO is based upon HOM interference, and therefore does not require interferometric stability, making the outsourcing process relatively robust and suitable for long-range operation.

This protocol can be regarded as a variation on the entanglement swapping protocol (Sec.~\ref{sec:swapping}), whereby entanglement between matter qubits and optical modes is swapped onto entanglement between the distinct matter qubits.

Alternately, if there is no direct line of quantum communication between two qubits, an EO can be performed by directly employing the same idea in reverse. We imagine that a third-party, such as a satellite, acts as a server for entangled Bell pairs. Two parties receive one qubit each from the pair. Then they perform an EO between their halves of the Bell pair and their local qubits. With appropriate local corrections, mediated by only cheap classical communication, this teleports the action of an EO onto the two qubits, creating a link between them.

Expanding upon this idea, we can envisage distributed models for quantum computation, where the qubits needn't even be of the same physical medium. We could, for example, entangle quantum dot qubits, atomic qubits, and atomic ensemble qubits with one another by coupling them to optical modes and performing which-path erasure between them. This enables distributed quantum computation between hosts possessing quantum infrastructure comprising different physical mediums (provided the photons emitted by those systems may be made indistinguishable, such that HOM interference is possible).

\begin{table}[!htbp]
\begin{mdframed}[innertopmargin=3pt, innerbottommargin=3pt, nobreak]
\texttt{
function WhichPathErasure():
\begin{enumerate}
\item Alice and Bob each prepare an equal superposition of the two logical basis states,
\begin{align}
\ket\psi_\mathrm{in} = &\frac{1}{2}(\ket{\!\uparrow}_{A_1}+\ket{\!\downarrow}_{A_1})\ket{0}_{A_2}\nonumber \\
&\cdot (\ket{\!\uparrow}_{B_1}+\ket{\!\downarrow}_{B_1})\ket{0}_{B_2},
\end{align}
where $A_1/B_1$ denote the matter qubits, and $A_2/B_2$ denote their coupled optical modes.
\item Apply a $\pi$-pulse to each qubit, inducing a \mbox{$\ket{\!\downarrow}\to\ket{e}$} transition,
\begin{align}
\ket\psi_\pi = \hat{U}_\pi\ket\psi_\mathrm{in} = &\frac{1}{2}(\ket{\!\uparrow}_{A_1}+\ket{e}_{A_1})\ket{0}_{A_2}\nonumber \\
&\cdot (\ket{\!\uparrow}_{B_1}+\ket{e}_{B_1})\ket{0}_{B_2}.
\end{align}
\item Wait for a coherent relaxation, inducing the transition \mbox{$\ket{e}\to\ket{\!\downarrow}\hat{a}^\dag$}, which emits a single photon,
\begin{align}
\ket\psi_\mathrm{relax} = \hat{U}_\mathrm{relax}\ket\psi_\pi = &\frac{1}{2}(\ket{\!\uparrow}_{A_1}+\ket{\!\downarrow}_{A_1}\hat{a}^\dag_{A_2})\ket{0}_{A_2}\nonumber \\
&\cdot (\ket{\!\uparrow}_{B_1}+\ket{\!\downarrow}_{B_1}\hat{a}^\dag_{B_2})\ket{0}_{B_2}.
\end{align}
\item Apply a 50:50 beamsplitter between the two optical modes,
\begin{align}
\ket\psi_\mathrm{BS} = \hat{U}_\mathrm{BS} \ket\psi_\mathrm{relax} = &\frac{1}{2}(\ket{\!\uparrow}_{A_1}+\ket{\!\downarrow}_{A_1}[\hat{a}^\dag_{A_2}+\hat{a}^\dag_{B_2}])\nonumber \\
&\cdot (\ket{\!\uparrow}_{B_1}+\ket{\!\downarrow}_{B_1}[\hat{a}^\dag_{A_2}-\hat{a}^\dag_{B_2}])\nonumber \\
&\cdot \ket{0}_{A_2}\ket{0}_{B_2}.
\end{align}
\item Conditional upon detecting exactly one photon between the output optical modes, we obtain,
\begin{align}
\ket\psi_\mathrm{out}^{1,0} = \bra{1,0}_{A_2,B_2} \ket\psi_\mathrm{BS} = \frac{1}{2} (\ket{\!\uparrow,\downarrow}_{A_1,B_1} + \ket{\!\downarrow,\uparrow}_{A_1,B_1}), \nonumber \\
\ket\psi_\mathrm{out}^{0,1} = \bra{0,1}_{A_2,B_2} \ket\psi_\mathrm{BS} = \frac{1}{2} (\ket{\!\uparrow,\downarrow}_{A_1,B_1} - \ket{\!\downarrow,\uparrow}_{A_1,B_1}),
\end{align}
which is a Bell pair between the matter qubits.
    \item $\Box$
\end{enumerate}}
\end{mdframed}
\captionspacealg \caption{Using which-path erasure to entangle two $\lambda$-configuration matter qubits via post-selected linear optics. Note that the two matter qubits could in principle be arbitrarily physically separated. Only the emitted photons need be brought together locally for the implementation of a beamsplitter operation. This lends such entanglement generation protocols to distributed implementation.} \label{alg:which_path}
\end{table}

%
% Atomic Ensembles
%

%\subsection{Atomic ensembles}

%\comment{To do}

%
% Ion Traps
%

%\subsection{Ion traps}\index{Ion traps}

%\comment{To do. Talk about optical interfacing.}

%
% Artificial Atoms
%

\subsection{Superconducting circuits}\index{Superconductors!Circuits}\label{sec:artificial_atoms}

\sectionby{Chandrashekar Radhakrishnan}\index{Chandrashekar Radhakrishnan}

\begin{figure}[!htbp]
\includegraphics[clip=true, width=0.475\textwidth]{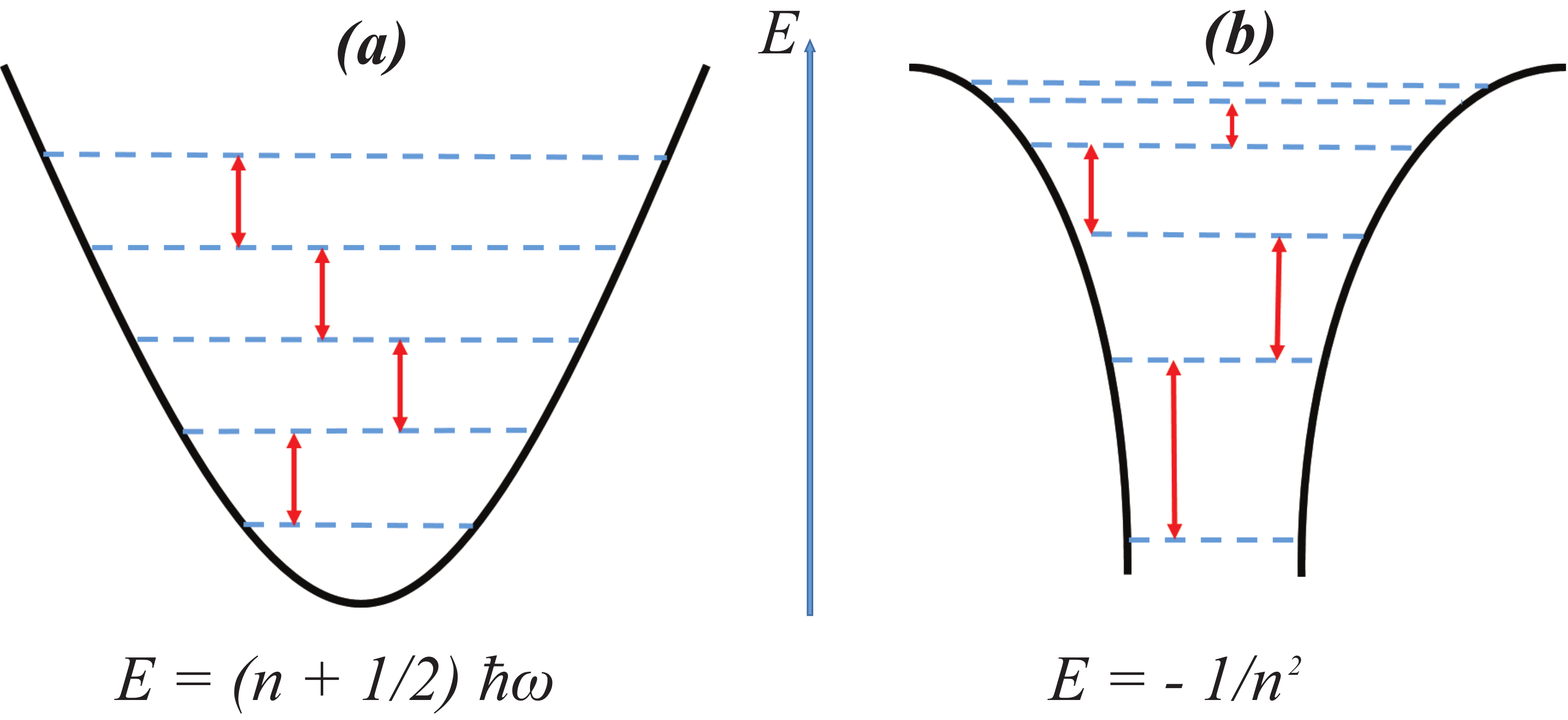}
\captionspacefig \caption{The energy levels of: (a) a quantum oscillator; and, (b) an atomic system. The quantum oscillator exhibits equidistant separation between energy levels, whereas for the atomic system the energy levels are non-uniform.}\label{fig:artificial_atom_energy_levels}\index{Energy levels}
\end{figure}

Qubits may be engineered by considering the lowest two energy levels of a quantum system. Based on the spacing between their energy levels, such quantum systems are classified into two categories:
\begin{itemize}
	\item Quantum harmonic oscillator type\index{Quantum harmonic oscillators}: have equal spacing between energy levels, given by,
	\begin{align}
	E_{n} &= \hbar \omega \left(n+\frac{1}{2}\right),
	\end{align}
	where \mbox{$n\in\mathbb{Z}^+$} denotes the discrete energy level, $\omega$ is optical frequency, and $E_0$ is the lowest-lying ground state energy\index{Ground states}.
	\item Atomic type\index{Artificial atoms}: have unequal spacing between energy levels, given by,
	\begin{align}
	E_{n} &= -\frac{E_{0}}{n^{2}}.
	\end{align}
\end{itemize}
We illustrate these two cases in Fig.~\ref{fig:artificial_atom_energy_levels}, with their corresponding energy level diagrams.

To construct a qubit we should be able to use external fields\index{Control fields} to control and selectively drive transitions between only two energy levels in the system. Such a procedure is easy to achieve in atomic systems, but it is not possible to address only two levels in a quantum oscillator due to the harmonicity\index{Harmonicity} (equal energy spacing) between energy levels. On the other hand, it's hard to work with individual natural atoms, mainly because of their size, which makes their individual isolation and control very challenging. To overcome this problem, we need to develop new quantum devices with anharmonic\index{Anharmonicity} energy spectra\index{Energy spectra}. Such devices are referred to as \textit{artificial atoms}\index{Artificial atoms}, due to their similarity to natural atoms in the anharmonicity of their energy level spectrum.

One of the most widely used types of artificial atom are superconducting qubits \cite{bib:martinis1985energy, bib:shnirman1997quantum, bib:averin1998adiabatic, bib:devoret2004superconducting, bib:makhlin2001quantum}\index{Superconductors!Qubits}, a class of non-linear quantum circuits\index{Non-linear!Quantum circuits}. An LC oscillator\index{LC!Oscillator} composed of an inductor $L$\index{Inductors} and capacitance $C$\index{Capacitors} is a typical example of a linear quantum circuit\index{Linear quantum circuits} with equal spacing. By introducing a Josephson junction\index{Josephson!Junction} into the linear quantum circuit we can make it non-linear, with anharmonic energy spectrum.

\begin{figure}[!htbp]
\includegraphics[clip=true, width=0.475\textwidth]{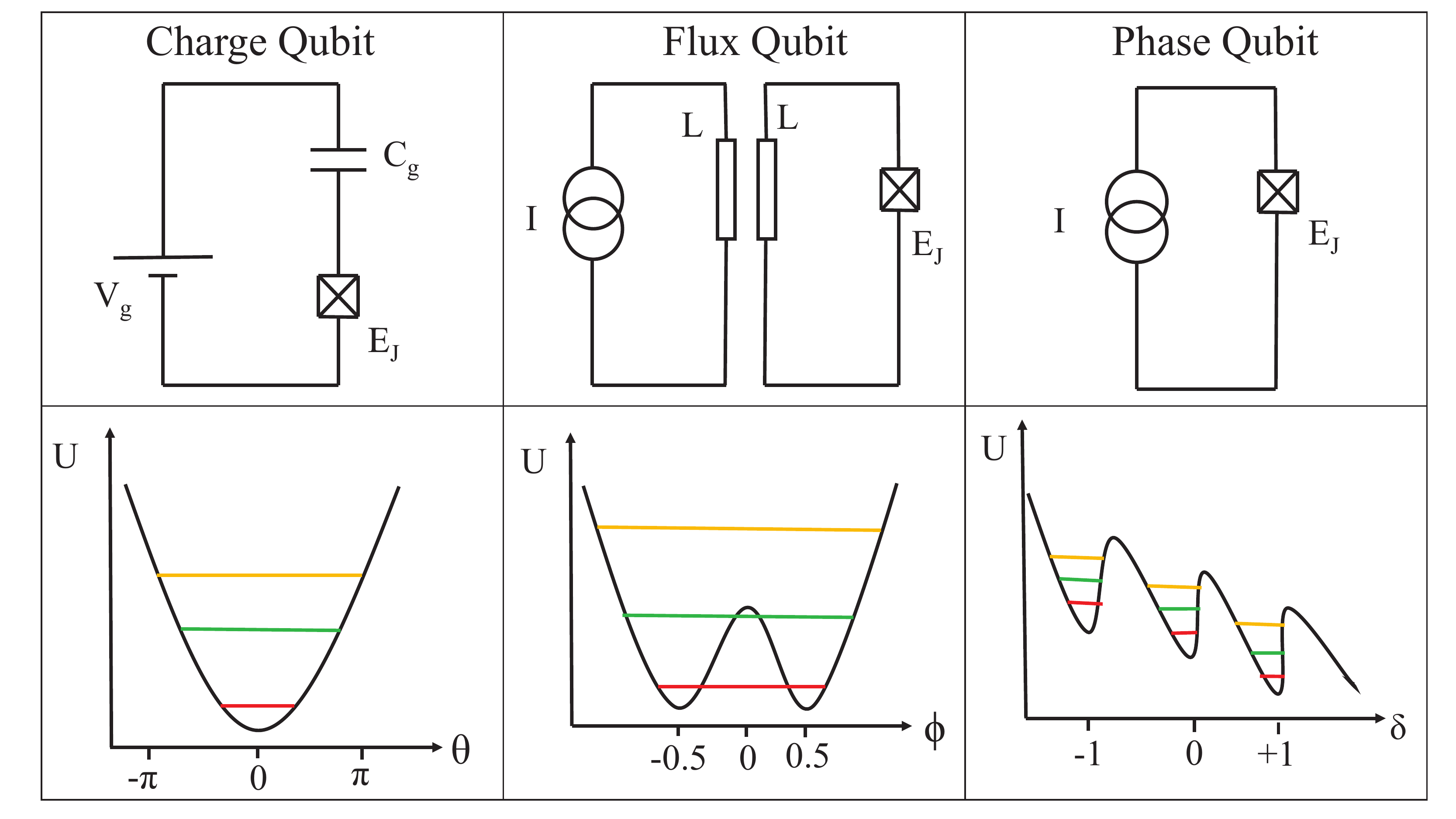}
\captionspacefig \caption{Simplified circuits for the different kinds of superconducting qubits, namely the charge qubit, flux qubit and phase qubit. Below each circuit are their respective energy level diagrams.}\label{fig:superconductor_circuits}
\end{figure}

A Josephson junction \cite{bib:josephson1974the} comprises two bulk superconducting materials\index{Superconductors}, separated by a thin layer of insulating material\index{Insulators}. In the superconducting phase the superconductors contain Cooper-pairs\index{Cooper-pairs}, composed of paired electrons. These Cooper-pairs move from one superconducting layer to another through the insulating layer via quantum tunnelling\index{Tunnelling}. The quantum mechanical nature of Josephson junctions is determined by two important energy scales:
\begin{itemize}
\item Josephson coupling energy, $E_{J}$.\index{Josephson!Coupling energy}
\item Coulomb energy, $E_C$.\index{Coulomb!Energy}
\end{itemize}
The ratio between these two energy scales determines the energy spectrum of the superconducting qubit. This yields three distinct types of qubits:
\begin{itemize}
\item Voltage-driven charge qubits \cite{bib:bouchiat1998quantum, bib:nakamura1999coherent}.\index{Charge!Qubits}
	\item Flux-driven flux qubits \cite{bib:friedman2000quantum, bib:van2000quantum}.\index{Flux!Qubits}
	\item Current-driven phase qubits \cite{bib:martinis2002rabi}.\index{Phase!Qubits}
\end{itemize} 
These circuits and energy level diagrams for these are illustrated in Fig.~\ref{fig:superconductor_circuits}.

\subsubsection{Charge qubits}\index{Charge!Qubits}

A non-linear quantum circuit driven by voltage\index{Voltage} is referred to as a charge qubit, whose Hamiltonian takes the form,
\begin{align}
\hat{H} = \frac{\hat{q}^{2}}{2C} - E_{J} \cos \left( \frac{2e}{\hbar} \hat\phi \right).
\label{circuitHamiltonian}
\end{align}
Here $\hat{q}$ is the charge in the superconducting system and $\hat\phi$ is flux\index{Flux}. The total capacitance of the circuit is given by $C$, and $E_{J}$ is the Josephson energy\index{Josephson!Energy}. The Hamiltonian in Eq.~(\ref{circuitHamiltonian}) can be rewritten as,
\begin{align}\label{eq:charge_qubit_hamiltonian}
\hat{H} = 4E_C (\hat{n} - n_{g})^{2} - E_{J} \cos (\hat{\phi}) .
\end{align}
The variables $\hat{q}$ and $\hat{\phi}$ are canonically conjugate\index{Canonically conjugate} and satisfy the commutation relation,
\begin{align}
[ \hat{\phi},\hat{q} ] = i \hbar.
\end{align}
In a truncated charge basis the Hamiltonian is,
\begin{align}
\hat{H} &= 4 E_C \sum_{n = -N}^{N} (\hat{n} - n_{g})^{2} \ket{n}\bra{n}\nonumber\\
&- E_{J} \sum_{n = -N}^{N-1} \ket{n+1}\bra{n} + \ket{n}\bra{n+1}.
\end{align}

The energy eigenstates are the charge states\index{Charge!States} $\ket{n}$, hence these qubits are referred to as charge qubits. In general the charge qubit \cite{bib:bouchiat1998quantum, bib:nakamura1999coherent} is operated in the region,
\begin{align}
	\frac{E_J}{E_C} \approx 1.
\end{align}

Charge qubits are highly sensitive to noise except at particular working points referred to as `sweet spots'\index{Sweet spots}. But it is experimentally difficult to control the voltage and current such that the qubit is maintained at these desired working conditions.

To overcome this, a special design of charge qubit known as the \textit{transmission line shunted plasma oscillation qubit} or `transmon' \cite{bib:koch2007charge}\index{Transmon} with,
\begin{align}
\frac{E_J}{E_C} \gg 1,
\end{align}
was suggested. The transmon is highly robust against external noise compared to the charge qubit. But the energy levels become more and more harmonic\index{Harmonicity} (i.e equally spaced) as we move away from the region,
\begin{align}
	\frac{E_{J}}{E_{C}} \approx 1.
\end{align}
Thus the charge qubits are designed by giving consideration to the trade-off between robustness against external noise and the anharmonicity\index{Anharmonicity} between the levels. A 20-qubit prototype quantum computer developed by IBM\index{IBM} employs transmon-type superconducting qubits \cite{bib:gambetta2017building}. 

\subsubsection{Flux qubits}\index{Flux!Qubits}

The flux qubit is popularly known as the RF SQUID (Radio Frequency Superconducting QUantum Interference Device)\index{SQUIDs}, which uses an AC current\index{Alternating current}. This qubit can be considered as the magnetic analogue of the charge qubit. In a charge qubit the Josephson junction is driven by a capacitor, but in a flux qubit, a superconducting transformer\index{Superconductors!Transformers} circuit generates the flux which drives the circuit. The Hamiltonian of the circuit is,
\begin{align}
\hat{H} = \frac{\hat{q}^{2}}{2 C_{J}} + \frac{\hat\phi^{2}}{2 L} - E_{J} \cos \left( \frac{2e}{\hbar}(\hat\phi - \phi_\mathrm{ext}) \right).
\end{align}

Here we can observe that there are three energy scales namely,
\begin{align}
E_J,& \nonumber\\
E_{C} &= \frac{2e^{2}}{C},\nonumber\\
E_{L} &= \frac{{\phi_{0}}^{2}}{2L}.
\end{align}
The quantum properties of the qubits depend on the interplay between these parameters. The Cooper-pairs in a flux qubit are confined to a double well potential\index{Double well potential}. The variables $Q$ and the total magnetic flux $\Phi$\index{Magnetic flux} are the conjugate variables, satisfying the commutation relation,
\begin{align}
[\hat{Q},\hat\Phi] = i \hbar.
\end{align}

Flux qubits are very robust against charge noise \cite{bib:you2005fast}, and hence have very long decoherence times\index{Decoherence!Times}, making them one of the most attractive qubit candidates for the construction of quantum computers. The early quantum computing devices developed by D-Wave\index{D-Wave} employ flux qubits \cite{bib:harris2018phase}.

\subsubsection{Phase qubits}\index{Phase!Qubits}

Current-driven superconducting qubits are referred to as phase qubits \cite{bib:martinis2002rabi}. They are commonly known as DC SQUIDs, and operate in the regime of very high values of $E_{J}/E_{C}$.

The Hamiltonian of a phase qubit is,
\begin{align}
\hat{H} = E_{C} \hat{p}^{2} - I \phi_{0} \hat\delta - I_{0} \phi_{0} \cos \hat\delta,
\label{eq:phase_qubit_hamiltonian}
\end{align}
where $\hat\delta$ is the gauge invariant\index{Gauge invariance} phase-difference operator\index{Phase!Difference operator} and the charge on the capacitor is $2pe$. These operators are conjugate variables, satisfying the commutation relation,
\begin{align}
	[\hat\delta, \hat{p}] = i \hbar.
\end{align}

In the phase qubit, Cooper-pairs\index{Cooper-pairs} experience a washboard potential\index{Washboard potential}. Since their decoherence times\index{Decoherence!Times} are very small compared to flux and charge qubits, they are not that widely employed.

\subsubsection{Quantum gates}\index{Quantum gates}

To build useful quantum information processing devices, we require quantum gates to act upon our superconducting qubits. This is a field under active development \cite{bib:blais2004cavity, bib:chow2011simple, bib:chow2013microwave}. Below we provide a brief description of the operation of single- and 2-qubit quantum gates based on superconducting qubits.

\paragraph{Single-qubit gates}

A single superconducting qubit, which is coherently controlled using microwaves, can be used as a quantum gate. Let us consider a cavity with resonant frequency $\omega_{r}$\index{Resonant frequency} and drive frequency $\omega_{d}$\index{Drive frequency}, where the difference \mbox{$\Delta_{r} = \omega_{r} - \omega_{d}$} is the detuning\index{Detuning} between the cavity and the drive. When \mbox{$\omega_{d} \approx \omega_{r}$} one can read the state of a superconducting qubit using microwaves. But when \mbox{$\omega_{d} = \omega_{q} \ll \omega_{r}$} the microwave can be used to perform gate operations on the qubit without measuring its state.

A system comprising a superconducting qubit and a microwave can be described using the Jaynes-Cummings Hamiltonian\index{Jaynes-Cummings Hamiltonian},
\begin{align}
\hat{H} = \Delta _{r} \hat{a}^{\dag} \hat{a} - \frac{\Delta_{q}}{2} \hat\sigma_{z} + g (\hat{a}^{\dag} \hat\sigma_{-} + \hat{a} \hat\sigma_{+}) + \xi(t) (\hat{a}^{\dag} + \hat{a}),
\label{eq:driven_jc_hamiltonian}
\end{align}
where $\hat{a}^{\dag}$ ($\hat{a}$) is the creation (annihilation) operator corresponding to the microwave photon, and $\sigma_{+}$ ($\sigma_{-}$) is the spin raising (lowering) operator\index{Spin operators}. The factors \mbox{$\Delta_{r} = \omega_{r} - \omega_{d}$} and \mbox{$\Delta_{q} = \omega_{q} - \omega_{d}$} are the detuning parameters\index{Detuning}. The factor $g$ is the coupling between the microwave photon and the qubit, and $\xi(t)$ is the envelope\index{Envelope} of the microwave pulse. The effective Hamiltonian is,
\begin{align}
\hat{H}_\mathrm{eff} &= \left( \Delta_{r} + \frac{g^{2}}{\Delta} \hat\sigma_{z} \right) \hat{a}^{\dag} a - \frac{1}{2} \left(\Delta_{q} - \frac{g^{2}}{\Delta} \right) \hat\sigma_{z} \nonumber\\
&+ \xi(t) (\hat{a}^{\dag} + \hat{a}) - \frac{g \xi(t)}{\Delta} \hat\sigma_{x}.
\end{align}

To perform an $X$-gate, we choose a drive frequency\index{Drive frequency},
\begin{align}
\omega_{d} = \omega_{q} - \frac{g^{2}}{\Delta}(2 \bar{n} + 1),
\end{align}
which causes the $\sigma_{z}$ term to disappear, leaving us with a pure $\sigma_{x}$ rotation. Using a phase-shifted drive,
\begin{align}
H_{d}(t) = \xi(t) i (\hat{a}^{\dag} - \hat{a}),
\end{align}
one might obtain a pure $\sigma_{y}$ rotation, yielding a $Y$-gate. Finally we note that using a drive,
\begin{align}
	\omega_{d} = \omega_{q} - \frac{g^{2}}{\Delta}(2 \bar{n} + 1) - 2 \xi(t)\frac{g}{\Delta},
\end{align}
we may construct a Hadamard gate. 

\paragraph{2-qubit gates}

Quantum gates operating on two qubits can be realised in many different ways. But in terms of their construction and operation, they can be divided into two classes. In the first class of quantum gates the superconducting qubits can be tuned over a wide range of frequencies. A good example of this is the iSWAP gate\index{iSWAP gate}, in which two Cooper-pair\index{Cooper-pairs} boxes are coupled via a transmission line resonator\index{Transmission line resonator}. In the rotating frame of reference\index{Rotating frame}, the effective Hamiltonian of the system is,
\begin{align}
\hat{H}_\mathrm{eff} &= \frac{g^{2}}{\Delta} \left( \hat{a}^{\dag} \hat{a} + \frac{1}{2} \right) (\hat\sigma_{z,1} + \hat\sigma_{z,2}) \nonumber\\
&- \frac{g^{2}}{\Delta} (\hat\sigma_{+,1} \hat\sigma_{-,2} + \hat\sigma_{+,2} \hat\sigma_{-,1}).
\end{align}

The parameters of the two qubits can be adjusted by tuning their flux. The interaction between qubits can be turned on and off by tuning the qubits in and out of resonance with one another. The advantage of the first class of quantum gates is that they can be operated in a region where the frequency of the two qubits differ from one another and the interaction between them is very strong. But the disadvantage is that they are sensitive to flux noise, hence requiring extra flux bias lines for tuning them properly.

The second class of quantum gates is built up of superconducting qubits with fixed frequencies, driven by microwaves. The cross-resonance gate\index{Cross-resonance gate}, the bSWAP\index{bSWAP gate}, and the MAP gate\index{MAP gate} belong to this class. The effective Hamiltonian of the cross-resonance gate is,
\begin{align}
\hat{H}_\mathrm{eff} = - \left( \frac{\tilde{\omega}_{1} - \tilde{\omega}_2}{2} \right) \hat\sigma_{z,1} + \frac{\Omega(t)}{2} \left(\hat\sigma_{x,1} - \frac{J}{\Delta_{12}} \hat\sigma_{z,1} \hat\sigma_{x,2} \right),
\end{align}
where,
\begin{align}
	\tilde{\omega}_{1} &= \omega_{1} + \frac{J^{2}}{\Delta_{12}}, \nonumber\\
		\tilde{\omega}_{2} &= \omega_{2} - \frac{J^{2}}{\Delta_{12}}, \nonumber\\
\Delta &= \omega_{1} -\omega_{2}.		
\end{align}
The factors $\omega_{1}$ and $\omega_{2}$ are the frequencies of the first and second qubits, and $\Delta_{12}$ is the detuning\index{Detuning}. The first qubit is rotating with frequency $\frac{1}{2}(\tilde{\omega_{1}} - \tilde{\omega_{2}})$ around the $Z$-axis, with a little shift in the $X$-direction, yielding an $X$-gate. Similarly we can construct a microwave-activated CZ (MAP) gate\index{MAP gate} using two transmons. The system of two transmons is modelled using a system of two coupled Duffing oscillators\index{Duffing oscillators}. The effective Hamiltonian in the two qubit space reads,
\begin{align}
\hat{H}_\mathrm{eff} &= - \frac{1}{2} \left( \omega_{01} - \frac{\zeta}{2} \right) \hat\sigma_{z,1} - \frac{1}{2} \left( \omega_{10} - \frac{\zeta}{2} \right) \hat\sigma_{z,2} \nonumber\\
&+ \frac{\zeta}{4} \hat\sigma_{z,1} \hat\sigma_{z,2}. 
\end{align}

Through this Hamiltonian we can realise a CZ gate, with gate time $514$ns and high fidelity. The second class of quantum gates have a longer coherence time\index{Coherence!Time}, since the superconducting qubits can be parked at the sweet spots\index{Sweet spots} of coherence where the effects of noise on the qubits are less substantial. But, control of the qubits is much harder, since we need to maintain them with the same qubit parameters for an extended period of time.

\latinquote{Acta deos numquam mortalia fallunt.}

%
% Quality of Service
%

\section{Error correction}\index{Quantum error correction (QEC)}\index{Quality of service (QoS)}\label{sec:QOS_chap}

\famousquote{Beware of bugs in the above code. I have only proved it correct, not tried it.}{Donald Knuth}\\

\dropcap{A}{s} with classical networks, quality of service (QoS) is a major consideration in any quantum network, hence the interest in cost vector analysis and optimisation. If we are transmitting quantum states over a quantum channel, there will inevitably be deterioration in the form of decoherence, loss, and other undesirable effects we wish to mitigate. In this section we review some of the essential quantum error correction (QEC) techniques that can positively improve the quality of transmitted quantum states, which do not have direct classical analogues.

For some simple quantum communications protocols, simple error correction (or even no error correction) may suffice. A full-fledged distributed quantum computation on the other hand will require error rates within a fault-tolerance threshold\index{Fault-tolerance!Thresholds}. Different error correcting codes have different error correcting power, and different resource overheads, which must be taken into consideration.

\subsection{3-qubit code}\index{3-qubit code}

The 3-qubit bit-flip code is traditionally used as a basic introduction to the concept of quantum error correction. However, the 3-qubit code \textit{does not} represent a full quantum code. This is due to the fact that the code cannot simultaneously correct for both $X$-errors and $Z$-errors, which is  required for correcting errors for an arbitrary error mapping on a single qubit. This code is a quantum analogue of the classical repetition code.

The 3-qubit code encodes a single logical qubit into three physical qubits with the property that it can correct for a single $X$-error. The two logical basis states $\ket{0}_L$ and $\ket{1}_L$ are defined as,
\begin{align}
\ket{0}_L = \ket{000}, \quad \quad \ket{1}_L = \ket{111},
\end{align}
such that an arbitrary single qubit state $\ket{\psi} = \alpha\ket{0} + \beta\ket{1}$ is mapped to,
\begin{align}
\alpha\ket{0} + \beta\ket{1} &\rightarrow \alpha\ket{0}_L + \beta\ket{1}_L \\
&= \alpha\ket{000} + 
\beta\ket{111} = \ket{\psi}_L.
\end{align}

As three individual $X$-errors are required to switch between logical states, $\ket{0}_L \leftrightarrow \ket{1}_L$, any single $X$-error, for example if the encoded state is $\ket{0}_L$. Results in a erred state that is closer to $\ket{0}_L$ than $\ket{1}_L$. The distance between two codeword states, $d$, defines the number of errors that can be corrected, $t$, as, $t = \lfloor(d-1)/2\rfloor$. In this case, $d=3$, hence $t=1$.  

How do we detect and correct errors without directly measuring or obtaining information about the logical state, which would collapse the computational wave-function and destroy the encoded information?  Two additional ancilla qubits, which are used to extract \textit{syndrome} information (information 
regarding possible errors) are introduced.  

Assuming all gate operations are perfect and the only place where the qubits are susceptible to error is the region between encoding and correction, correction proceeds by introducing two ancilla qubits and performing a sequence of CNOT gates, which checks the parity of the three qubits.  

Tab.~\ref{tab:errors} summarises the state of the whole system, for each possible error, just prior to measurement.

\startnormtable
\begin{table}[htbp!]
\begin{center}
\begin{tabular}{|c|c|}
\hline
Error location & Final state, $\ket{\text{data}}\ket{\text{ancilla}}$ \\
\hline \hline
No error & $\alpha\ket{000}\ket{00} + \beta\ket{111}\ket{00}$ \\
Qubit 1 & $\alpha\ket{100}\ket{11} + \beta\ket{011}\ket{11}$ \\
Qubit 2 & $\alpha\ket{010}\ket{10} + \beta\ket{101}\ket{10}$ \\
Qubit 3 & $\alpha\ket{001}\ket{01} + \beta\ket{110}\ket{01}$ \\
\hline
\end{tabular}
\caption{Final state of the five qubit system prior to the syndrome measurement for no error or a single 
$X$ error on one of the qubits. The last two qubits represent the state of the ancilla. Note that each possible error will result in a unique measurement result (syndrome) of the ancilla qubits. This allows for a $X$ correction gate to be applied to the data block which is classically controlled from the syndrome result. At no point during correction do we learn anything about $\alpha$ or $\beta$.} 
\label{tab:errors}
\end{center}
\end{table} 

For each possible situation, either no error or a single bit-flip error, the ancilla qubits are flipped to a unique state based on the parity of the data block. These qubits are then measured to obtain the classical {\em syndrome} result. The result of the measurement will then dictate if an $X$ correction gate needs to be applied to a specific qubit, 
\begin{widetext}
\begin{align}
&\text{Ancilla measurement:} \quad \ket{00}, \quad \text{Collapsed state:} \quad \alpha\ket{000} + \beta\ket{111} \quad \therefore \text{Clean state} \\
&\text{Ancilla measurement:} \quad \ket{01}, \quad \text{Collapsed state:} \quad \alpha\ket{001} + \beta\ket{110} \quad \therefore \text{Bit-flip on qubit 3} \nonumber\\
&\text{Ancilla measurement:} \quad \ket{10}, \quad \text{Collapsed state:} \quad \alpha\ket{010} + \beta\ket{101} \quad \therefore \text{Bit-flip on qubit 2} \nonumber\\
&\text{Ancilla measurement:} \quad \ket{11}, \quad \text{Collapsed state:} \quad \alpha\ket{100} + \beta\ket{011} \quad \therefore \text{Bit-flip on qubit 1} \nonumber
\end{align}
\end{widetext}
Provided that only a single error has occurred, the data block is restored.  

This code will only work if a maximum of one error occurs. If two $X$ errors occur, then the syndrome result becomes ambiguous. For example, if an $X$ error occurs on both qubits one and two, then the syndrome result will be $\ket{01}$. This will cause us to mis-correct by applying an $X$ gate to qubit 3. Therefore, two errors will induce a logical bit flip and causes the code to fail. This flow is illustrated in Alg.~\ref{alg:three_QEC}.

%The first described QEC\index{Quantum error correction (QEC)} code was the 3-qubit code\index{3-qubit code} (or redundancy code) by \cite{bib:Shor95} for redundantly encoding a single logical qubit into three encoded qubits, allowing the detection and correction of at most a single bit-flip error between the encoded qubits. Measurement of the three ancillary qubits yields a \textit{syndrome}\index{Syndromes}, which identifies where the error took place, allowing it to be subsequently corrected with feedforward. This protocol is shown in Alg.~\ref{alg:three_QEC}.

%By switching into a Hadamard-rotated basis, the same code could equivalently correct against at most a single phase-flip error (since \mbox{$\hat{H}\hat{X}\hat{H}=\hat{Z}$}).

\startalgtable
\begin{table}[!htbp]
\begin{mdframed}[innertopmargin=3pt, innerbottommargin=3pt, nobreak]
\texttt{
function ThreeQubitCode($\ket\psi$):
\begin{enumerate}
\item Using two CNOT gates, redundantly encode the logical single-qubit state,
\begin{align}
\ket\psi=\alpha\ket{0}+\beta\ket{1},
\end{align}
into the 3-qubit state,
\begin{align}
\ket\psi_R &= \hat{\mathrm{CNOT}}_{1,2}\hat{\mathrm{CNOT}}_{1,3}\ket\psi\ket{00} \nonumber \\
&= \alpha\ket{000}+\beta\ket{111}.
\end{align}
\item Independently apply bit-flip channels $\mathcal{E}_X$ to each of the three encoded qubits.
\item If exactly one bit-flip operation was applied in total, the three possible erroneous encoded states are,
\begin{align}
\ket\psi_1 &= \hat{X}_1 \ket\psi_R = \alpha\ket{001}+\beta\ket{110}, \nonumber \\
\ket\psi_2 &= \hat{X}_2 \ket\psi_R = \alpha\ket{010}+\beta\ket{101}, \nonumber \\
\ket\psi_3 &= \hat{X}_3 \ket\psi_R = \alpha\ket{100}+\beta\ket{011}.
\end{align}
\item Determine the parity of each of the three pairs of encoded qubits.
\item Assuming at most a single bit-flip operation has occurred on the encoded state, the three parity outcomes uniquely determine which encoded qubit the bit-flip was applied to.
\item Apply classically-controlled bit-flip recovery operations, $\mathcal{R}$, to correct the encoded state, recovering $\ket\psi_R$.
\item Apply the inverse of the encoding operation to recover $\ket\psi$.
\item $\Box$
\end{enumerate}}
\begin{align}
\Qcircuit @C=1.3em @R=.6em {
  & \lstick{\ket{\psi}} & \ctrl{2} & \gate{\mathcal{E}_X}  & \qw & \qw              & \ctrl{3}  & \qw       & \ctrl{4} & \qw & \multigate{2}{\ \mathcal{R}\ } & \qw \\
  & \lstick{\ket{0}}    & \targ    & \gate{\mathcal{E}_X}  & \qw & \qw              & \ctrl{2}  & \ctrl{2}  & \qw      & \qw & \ghost{\ \mathcal{R}\ } \qw & \qw \\
  & \lstick{\ket{0}}    & \targ    & \gate{\mathcal{E}_X}  & \qw & \qw              & \qw       & \ctrl{2}  & \ctrl{3} & \qw & \ghost{\ \mathcal{R}\ } \qw & \qw \\
  &          &          &          & & \lstick{\ket{0}} & \targ \qw & \qw       & \qw      & \meter & \control \cw \cwx \\
  &          &          &          & & \lstick{\ket{0}} & \qw       & \targ \qw & \qw      & \meter & \control \cw \cwx \\
  &          &          &          & & \lstick{\ket{0}} & \qw       & \qw       & \targ    & \meter & \control \cw \cwx
} \nonumber
\end{align}
\end{mdframed}
\captionspacealg \caption{3-qubit code for protecting against at most a single logical bit-flip error. The doubly-controlled CNOT gates represent parity measurements, \mbox{$n_3=n_1\oplus n_2$}, where $n_i$ represents the value of the $i$th qubit. In a Hadamard-rotated basis, the same circuit may be employed to protect against a single phase-flip error. And by concatenating the two we obtain a 9-qubit code protecting against a single depolarising error (i.e joint bit-flip/phase-flip), which is a universal single-qubit error model.} \label{alg:three_QEC}
\end{table}

%
% 9-Qubit Code
%

\subsection{9-qubit code}\index{9-qubit code}

The nine qubit error correcting code was first developed by Shor~\cite{bib:S95} in 1995 and is based on the 3-qubit repetition code. The Shor code can correct a logical qubit from one discrete bit-flip, one discrete phase-flip or one of each on any of the nine physical qubits and is therefore sufficient to correct for any continuous linear combination of errors on a single qubit.  

The two logical basis states for the code are,
\begin{equation}
\begin{aligned}
\ket{0}_L = \frac{1}{\sqrt{8}}(\ket{000}+\ket{111})(\ket{000}+\ket{111})(\ket{000}+\ket{111}), \\
\ket{1}_L = \frac{1}{\sqrt{8}}(\ket{000}-\ket{111})(\ket{000}-\ket{111})(\ket{000}-\ket{111}). \\
\end{aligned}
\end{equation}

Correction for $X$-errors occurs by treating each block of three qubits in the code identically to the three qubit code. Phase errors ($Z$-errors) are corrected by examining the sign differences between the three blocks.  

\begin{table}[!htbp]
\begin{mdframed}[innertopmargin=3pt, innerbottommargin=3pt, nobreak]
\texttt{
function NineQubitCode($\ket\psi$):
\begin{enumerate}
\item Using eight CNOT gates and three Hadamards, redundantly encode the logical single-qubit state,
\begin{align}
\ket\psi=\alpha\ket{0}+\beta\ket{1},
\end{align}
into the 9-qubit state,
\begin{align}
& \ket\psi\ket{00000000} \rightarrow \ket\psi_R \\
&= \frac{\alpha}{\sqrt{8}}(\ket{000}+\ket{111})(\ket{000}+\ket{111})(\ket{000}+\ket{111}) \nonumber \\
& + \frac{\beta}{\sqrt{8}}(\ket{000}-\ket{111})(\ket{000}-\ket{111})(\ket{000}-\ket{111}) \nonumber
\end{align}
\item Independently apply both bit-flip channels $\mathcal{E}_X$ and phase-flip channels $\mathcal{E}_Z$
to each of the three encoded qubits.
\item The nine qubit code is degenerate, meaning that multiple single qubit errors have the same effect on the code-state.  For all possible single qubit phase errors , there are only three possible changes to the logic state.
\begin{align}
\ket\psi_1 &= \hat{Z}_{1,2,3} \ket\psi_R \\
&= \frac{\alpha}{\sqrt{8}}(\ket{000}-\ket{111})(\ket{000}+\ket{111})(\ket{000}+\ket{111}) \nonumber \\
& + \frac{\beta}{\sqrt{8}}(\ket{000}+\ket{111})(\ket{000}-\ket{111})(\ket{000}-\ket{111}) 
, \nonumber \\
\ket\psi_2 &= \hat{Z}_{4,5,6} \ket\psi_R \nonumber \\
&= \frac{\alpha}{\sqrt{8}}(\ket{000}+\ket{111})(\ket{000}-\ket{111})(\ket{000}+\ket{111}) \nonumber \\
& + \frac{\beta}{\sqrt{8}}(\ket{000}-\ket{111})(\ket{000}+\ket{111})(\ket{000}-\ket{111}) 
, \nonumber \\
\ket\psi_3 &= \hat{Z}_{7,8,9} \ket\psi_R \nonumber \\
&= \frac{\alpha}{\sqrt{8}}(\ket{000}+\ket{111})(\ket{000}+\ket{111})(\ket{000}-\ket{111}) \nonumber \\
& + \frac{\beta}{\sqrt{8}}(\ket{000}-\ket{111})(\ket{000}-\ket{111})(\ket{000}+\ket{111}).\nonumber
\end{align}
\item Correct all $X$-errors by applying the three qubit correction protocol to each of the three block.
\item To check for $Z$-errors, compare the $\pm$ signs between blocks by first coupling all qubits in 
blocks one and two via CNOT gates to one ancilla and all qubits in blocks two and three to a second ancilla.
\item To correct for $Z$-errors, apply a physical $Z$-gate to any of the physical qubits in a block identified by the 
ancilla measurements (for a degenerate code, the correction gate does not need to be identical to the original error).
\item $\Box$
\end{enumerate}}
\end{mdframed}
\captionspacealg \caption{9-qubit code for protecting against at most a single bit- and/or phase-flip error. The code is a concatenation of three blocks of three bit-flip codes. Bit-flips are corrected within blocks of three qubits, while phase-flips are corrected by comparing blocks.} \label{alg:nine_QEC}
\end{table}

The nine qubit code is effectively a concatenation of two bit-flip codes. By combining three bit-flip encoded qubits into a 
second level encoded qubit, the lower level codes can correct for a single bit-flip error in any single block, while the 
upper level code corrects for a single phase flip error in any single block. 

Note that a phase flip on {\em any} one qubit in a block of three has the same effect, this is why the 9-qubit 
code is referred to as a degenerate code.  In other error correcting codes, such as the 5- or 7-qubit codes~\cite{bib:S96,bib:LMPZ96}, there is a one-to-one 
mapping between correctable errors and unique states, in degenerate codes such as this, the mapping is not unique.  If we know which block the error occurs it does not matter which qubit we apply the correction operator.  

Even if a bit {\em and} phase error occurs on the same qubit (i.e. a $Y$-error), the $X$ correction circuit will detect and correct for 
bit flips while the $Z$ correction circuit will detect and correct for phase flips.  The $X$-error correction 
has the ability to correct for up to three individual bit flips (provided each bit flip occurs in a different block of three).  However, in general 
the 9-qubit code is only a single error correcting code as it cannot handle more than a single bit flip error if they occur in certain locations.

%By taking two instances of the 3-qubit code, one implementing bit-flip correction and the other implementing phase-flip correction, and concatenating them, we can define a 9-qubit code, which protects against at most a single bit-flip and a single phase-flip. Joint protection against both the bit-flip and phase-flip operators ($\hat{X}$ and $\hat{Z}$) in turn allows error correction against a single depolarising error, the most general type of logical error, which completely destroys a single qubit.

%This is the simplest construction of a code which protects against a single depolarising event. But it is not the most efficient, and numerous more resource-savvy codes for protecting against Pauli errors exist, such as Steane's 7-qubit code\index{Steane code} \cite{bib:SteaneCode}.

%
% Stabiliser Codes
%

\subsection{Stabiliser codes}\index{Stabiliser!Codes}\label{sec:stab_code}

So far we have presented error correcting codes from the perspective of their state representations and their preparation and correction circuits. This is a rather inefficient method for describing the codes as the state representations and circuits clearly differ from code to code. The majority of error correcting codes that are used within the literature are members of a class known as stabiliser codes. Stabiliser codes are very useful to work with. The general formalism applies broadly and there exists general rules to construct preparation circuits, correction circuits and fault-tolerant logical gate operations once the stabiliser structure of the code is specified.  

The stabiliser formalism, first introduced by  Gottesman \cite{bib:G97+}, uses essentially the Heisenberg representation for quantum mechanics which describes quantum states in terms of operators rather that the basis states themselves.  An arbitrary state $\ket{\psi}$ is defined to be stabilised by some operator, $K$, if it is a 
$+1$ eigenstate of $K$,
\begin{align}
K\ket{\psi} = \ket{\psi}.
\end{align}
For example, the single qubit state $\ket{0}$ is stabilised by the operator $K = \sigma_z$,
\begin{align}
\sigma_z\ket{0} = \ket{0}.
\end{align}
Defining multi-qubit states with respect to this formalism relies on the group structure of multi-qubit operators.  

Within the group of all possible, single qubit operators, there exists a subgroup, denoted the Pauli group, $\mathcal{P}$, which contains the following elements,
\begin{align}
\mathcal{P} = \{\pm 1,\pm i\} \times \{\sigma_I, \sigma_x, \sigma_y, \sigma_z\}.
\end{align}
It is easy to check that these matrices form a group under multiplication through the commutation and anti-commutation rules for the Pauli matrices $\sigma_i$,
\begin{align}
[\sigma_i,\sigma_j] &= 2i\epsilon_{ijk}\sigma_k,\nonumber\\
\{\sigma_i,\sigma_j\} &= 2\delta_{ij},
\end{align}
where,
\begin{align}
\epsilon_{ijk} = \Bigg \{
\begin{array}{l}
+1\text{ for } (i,j,k) \in \{(1,2,3), (2,3,1), (3,1,2)\}\\
-1 \text{ for } (i,j,k) \in \{(1,3,2), (3,2,1), (2,1,3)\}\\
0 \text{    for } i=j, j=k, \text{ or } k=i
\end{array}
\end{align}
and
\begin{align}
\delta_{ij} = \Bigg \{
\begin{array}{cr}
1\text{ for } i = j\\
0 \text{ for } i \neq j.
\end{array}
\end{align}
The Pauli group extends over $N$-qubits by simply taking the $N$ fold tensor product of $\mathcal{P}$, i.e.
\begin{align}
\mathcal{P}_N &= \mathcal{P}^{\otimes N}.
\end{align}
An $N$-qubit stabiliser state, $\ket{\psi}_N$ is then defined via an $N$-element Abelian subgroup, $\mathcal{G}$, of the $N$-qubit Pauli group, in which $\ket{\psi}_N$ is a $+1$ eigenstate of each element, 
\begin{align}
\mathcal{G} = \{\; G_i \;|\; G_i\ket{\psi} = \ket{\psi}, \; [G_i,G_j] = 0 \; \forall \; (i,j) \} \subset \mathcal{P}_N.
\label{eq:stabdef}
\end{align}
Given this definition, the state $\ket{\psi}_N$ can be equivalently defined either through the state vector representation {\em or} by specifying the stabiliser set, $\mathcal{G}$.

Many extremely useful multi-qubit states are stabiliser states, including two-qubit Bell states, Greenberger-Horne-Zeilinger (GHZ) states \cite{greenberger1989going}, cluster states \cite{briegel2001persistent} and codeword states for QEC. As an example, consider a three qubit GHZ state, defined as,
\begin{align}
\ket{\text{GHZ}}_3 = \frac{\ket{000} + \ket{111}}{\sqrt{2}}.
\end{align}
This state can be expressed via any three linearly independent elements of the $\ket{\text{GHZ}}_3$ stabiliser group for example,
\begin{align}
G_1 &= \sigma_x\otimes \sigma_x \otimes \sigma_x \equiv XXX, \nonumber\\
G_2 &= \sigma_z\otimes \sigma_z \otimes \sigma_I \equiv ZZI, \nonumber\\
G_3 &= \sigma_I \otimes \sigma_z \otimes \sigma_z \equiv IZZ,
\end{align}
where the right-hand side of each equation is the short-hand representation of stabilisers. Note that these three operators form an Abelian group as,
\begin{align}
[G_i,G_j]\ket{\psi} &= G_iG_j\ket{\psi} - G_jG_i\ket{\psi} \nonumber\\
&= \ket{\psi}-\ket{\psi} = 0, \quad \forall \quad [i,j,\ket{\psi}].
\end{align}
Similarly, the four orthogonal Bell states,
\begin{align}
\ket{\Phi^{\pm}} &= \frac{\ket{00} \pm \ket{11}}{\sqrt{2}}, \\
\ket{\Psi^{\pm}} &= \frac{\ket{01} \pm \ket{10}}{\sqrt{2}},
\end{align}
are stabilised by the operators, $G_1 = (-1)^aXX$, and $G_2 = (-1)^b ZZ$. Where $[a,b] \in \{0,1\}$ and each of the four Bell states correspond to the four unique pairs, $\{\Phi^+,\Psi^+,\Phi^-,\Psi^-\} = \{ [0,0],[0,1],[1,0],[1,1]\}$.  
 
%\subsection{QEC with stabiliser codes}
%\label{sec:sec:QEC2}

The use of the stabiliser formalism to describe quantum error correction codes is extremely useful since it allows for easy synthesis of correction circuits and also clearly shows how logical operations can be performed directly on encoded data. As an introduction we will focus on arguably the most well known quantum code, the 7-qubit Steane code \cite{bib:S96}.  

The 7-qubit code represents a full quantum code that encodes seven physical qubits into one logical qubit, with the ability to correct for a single $X$ and/or $Z$ error. The $\ket{0}_L$ and $\ket{1}_L$ basis states are defined as,
\begin{widetext}
\begin{align}
|0\rangle_L = \frac{1}{\sqrt{8}}(&|0000000\rangle + |1010101\rangle + |0110011\rangle + |1100110\rangle + 
|0001111\rangle + |1011010\rangle + |0111100\rangle + |1101001\rangle),\nonumber\\
|1\rangle_L = \frac{1}{\sqrt{8}}(&|1111111\rangle + |0101010\rangle + |1001100\rangle + |0011001\rangle + 
|1110000\rangle + |0100101\rangle + |1000011\rangle + |0010110\rangle).
\label{eq:log}
\end{align}
\end{widetext}
The stabiliser set for the 7-qubit code is fully specified by the six operators,
\begin{align}\label{eq:stab7}
K^1 &= IIIXXXX, \nonumber\\
K^2 &= XIXIXIX, \nonumber\\ 
K^3 &= IXXIIXX, \nonumber\\
K^4 &= IIIZZZZ, \nonumber\\
K^5 &= ZIZIZIZ, \nonumber\\
K^6 &= IZZIIZZ.
\end{align}

As the 7-qubit codeword states are specified by only six stabilisers, the code contains two basis states, which are the logical states.  With a final operator, $K^7 = ZZZZZZZ=Z^{\otimes 7}$ fixing the state to one of the codewords, $K^7\ket{0}_L = \ket{0}_L$ and $K^7\ket{1}_L = -\ket{1}_L$. The 7-qubit code is defined as a $[[n,k,d]] = [[7,1,3]]$ quantum code, where $n=7$ physical qubits encode $k=1$ logical qubit with a distance between basis states $d=3$, correcting $t = (d-1)/2 = 1$ error.  Notice that the stabiliser set separates into $X$ and $Z$ sectors which defines the code as a Calderbank-Shor-Steane (CSS) code. CSS codes are extremely useful since they allow for straightforward logical gate operations to be applied directly to the encoded data and are easy to derive from classical codes.

Although the 7-qubit code is the most well known stabiliser code, there are other stabiliser codes which encode multiple logical qubits and correct for more errors~\cite{bib:G97+}. The downside to these lager codes is that they require more physical qubits and more complicated error correction circuits. Tables \ref{tab:9qubit} and \ref{tab:5qubit} show the stabiliser structure of two other well known codes, the 9-qubit code \cite{bib:S95} which we have examined and the 5-qubit code~\cite{bib:LMPZ96} which represents the smallest possible quantum code that corrects for a single error.  

\startnormtable
\begin{table}[htbp!]
\begin{center}
\vspace*{4pt}   
\begin{tabular}{|c|c|c|c|c|c|c|c|c|c|}
\hline
$K^1$ & $Z$&$Z$&$I$&$I$&$I$&$I$&$I$&$I$&$I$ \\
$K^2$ & $Z$&$I$&$Z$&$I$&$I$&$I$&$I$&$I$&$I$ \\
$K^3$ & $I$&$I$&$I$&$Z$&$Z$&$I$&$I$&$I$&$I$ \\
$K^4$ & $I$&$I$&$I$&$Z$&$I$&$Z$&$I$&$I$&$I$ \\
$K^5$ & $I$&$I$&$I$&$I$&$I$&$I$&$Z$&$Z$&$I$ \\
$K^6$ & $I$&$I$&$I$&$I$&$I$&$I$&$Z$&$I$&$Z$ \\
$K^7$ & $X$&$X$&$X$&$X$&$X$&$X$&$I$&$I$&$I$ \\
$K^8$ & $X$&$X$&$X$&$I$&$I$&$I$&$X$&$X$&$X$ \\
\hline
\end{tabular}
\caption{The eight stabilisers for the 9-qubit Shor code, encoding nine physical qubits into one logical qubit to correct for a single $X$ and/or $Z$ error.} 
\label{tab:9qubit}
\end{center}
\end{table} 

\begin{table}[htbp!]
\begin{center}
\vspace*{4pt}   
\begin{tabular}{|c|c|c|c|c|c|}
\hline
$K^1$ & $X$&$Z$&$Z$&$X$&$I$ \\
$K^2$ & $I$&$X$&$Z$&$Z$&$X$ \\
$K^3$ & $X$&$I$&$X$&$Z$&$Z$ \\
$K^4$ & $Z$&$X$&$I$&$X$&$Z$ \\
\hline
\end{tabular}
\caption{The four stabilisers for the $[[5,1,3]]$ quantum code, encoding five physical qubits into one logical qubit to correct for a single $X$ and/or $Z$ error. Unlike the 7- and 9-qubit codes, the $[[5,1,3]]$ code is a non-CSS code, since the stabiliser set does not separate into $X$ and $Z$ sectors.} 
\label{tab:5qubit}
\end{center}
\end{table} 

%\subsubsection{Performing error correction}\label{sec:stabilizer_correction}

Error correction using stabiliser codes is a straightforward extension of state preparation.  Consider an arbitrary single qubit state that has been encoded, 
\begin{align}
\alpha\ket{0} + \beta\ket{1} \rightarrow \alpha\ket{0}_L + \beta\ket{1}_L = \ket{\psi}_L.  
\end{align}
Now assume that an error occurs on one (or multiple) qubits which is described via the operator $E$, where $E$ is a combination of $X$ and/or $Z$ errors over the $N$ physical qubits of the logical state. By definition of stabiliser codes, $K^i\ket{\psi}_L = \ket{\psi}_L$, $i \in [1,..,N-k]$, for a code encoding $k$ logical qubits. Hence the erred state, $E\ket{\psi}_L$, satisfies,
\begin{align}
K^iE\ket{\psi}_L = (-1)^m EK^i\ket{\psi}_L = (-1)^m E\ket{\psi}_L.
\end{align}
where $m$ is defined as, $m=0$, if $[E,K^i]=0$ and $m=1$, if $\{E,K^i\} = 0$. Therefore, if the error operator commutes with the stabiliser, the state remains a $+1$ eigenstate of $K^i$, if the error operator anti-commutes with the stabiliser then the logical state is flips to now be a $-1$ eigenstate of $K^i$.  

Each of the code stabilisers are sequentially measured. Since a error free state is already a $+1$ eigenstate of all the stabilisers, any error which anti-commutes with a stabiliser will flip the eigenstate and consequently the parity measurement will return a result of $\ket{1}$.

Taking the $[[7,1,3]]$ code as an example, if the error operator is $E = X_i$, where $i = (1,\dots,7)$, representing a bit-flip on any {\em one} of the 7 physical qubits, then regardless of the location, $E$ will anti-commute with a unique combination of $(K^4,K^5,K^6)$. Hence the classical results of measuring these three operators will indicate if and where a single $X$ error has occurred. Similarly, if $E=Z_i$, then the error operator will anti-commute with a unique combination of, $(K^1,K^2,K^3)$. Consequently, the first three stabilisers for the $[[7,1,3]]$ code correspond to $Z$ sector correction while the second three stabilisers correspond to $X$ sector correction. Note, that correction for Pauli $Y$ errors are also taken care of by correcting in the $X$ and $Z$ sector since a $Y$ error on a single qubit is equivalent to both an $X$ and $Z$ error on the same qubit, i.e. $Y = iXZ$.  

\subsection{Surface codes}\index{Surface codes}\label{sec:surface_codes}

It has been shown that fault-tolerance is possible within the cluster state model \cite{bib:NielsenDawson04, bib:Dawson06} using variations of conventional QEC codes. However, more importantly, from cluster states certain \textit{topological QEC codes} can be readily constructed. This implements a form of QEC-encoded measurement-based quantum computing protocol, where the computation proceeds in a measurement-based fashion, but is `natively' fault-tolerant.

These codes have been shown to have very favourable fault-tolerance thresholds in terms of both depolarising noise and loss \cite{bib:StaceBarrettDohertyLoss, bib:BarrettStaceFT}, as well as frugal resource overhead compared to traditional concatenated codes. Additionally, loss- and gate-failure-tolerant codes, uniquely applicable to the cluster state model, have been described, with very favourable loss thresholds \cite{bib:Varnava05, bib:RalphHayes05, bib:Duan05}. 

Importantly, topological codes do not require joint measurements across the entire graph state, instead requiring only operations localised to small regions within the graph. Thanks to this, computation using such topological codes can remain distributed\index{Distributed quantum computation} without requiring the entire state to be held locally by a particular host, or requiring full access to the entire state by any particular user. As with cluster states, a large graph can be stitched together from a patchwork of smaller neighbouring graphs.

The most common topological code, which we will use here as an example, is the toric code\index{Toric code}, which resides on a lattice graph over the surface of a torus\footnote{As with cluster states, this graph needn't (but could) correspond to a network graph.}. As with cluster states (Sec.~\ref{sec:CSQC}), the toric code is most easily visualised in the stabiliser formalism\index{Stabiliser!Formalism}, and the toric code is an example of a stabiliser code\index{Stabiliser!Codes} (Sec.~\ref{sec:stab_code}). Consider a rectangular sub-graph of the torus. We place a qubit on each edge (not vertex) of the graph. Now we define two sets of stabiliser operators: \textit{star} ($\hat{S}_+$) and \textit{plaquette} ($\hat{S}_\square$) operators,
\begin{align} \index{Surface codes!Stabilisers}\index{Star operator}\index{Plaquette operator}
	\hat{S}_+(v) &= \prod_{i\in e(v)} \hat{X}_i, \nonumber \\
	\hat{S}_\square(p) &= \prod_{i\in e(p)} \hat{Z}_i,
\end{align}
where $e(v)$ are the edges neighbouring vertex $v$, and $e(p)$ are the edges surrounding plaquette $p$. By definition, the toric code state, $\ket\psi_\mathrm{toric}$, satisfies the stabiliser relations,
\begin{align}
	\hat{S}_+(v) \ket\psi_\mathrm{toric} &= \ket\psi_\mathrm{toric} \,\forall\, v, \nonumber \\
	\hat{S}_\square(p) \ket\psi_\mathrm{toric} &= \ket\psi_\mathrm{toric} \,\forall\, p.
\end{align}
The stabilisers can be visualised graphically as per Fig.~\ref{fig:toric_code}.

\if 1\doublecol
	\begin{figure}[!htbp]
		\includegraphics[clip=true, width=0.475\textwidth]{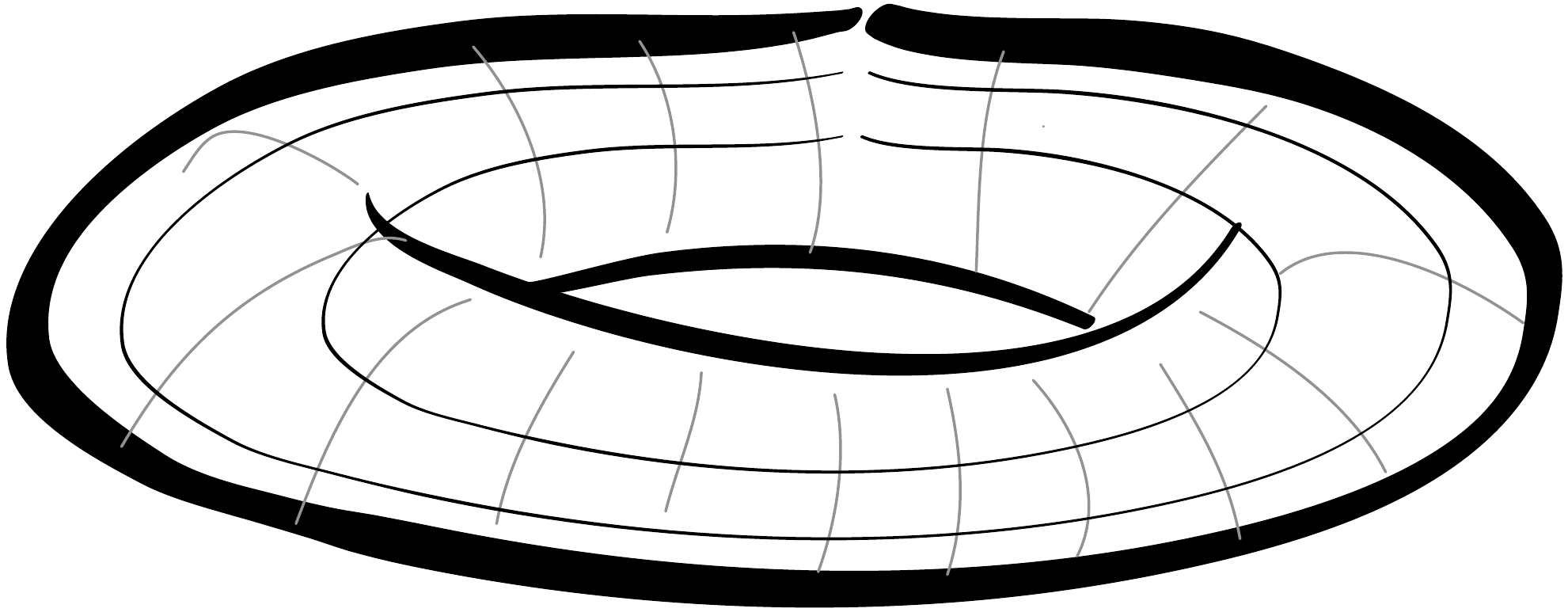}\\
		\includegraphics[clip=true, width=0.475\textwidth]{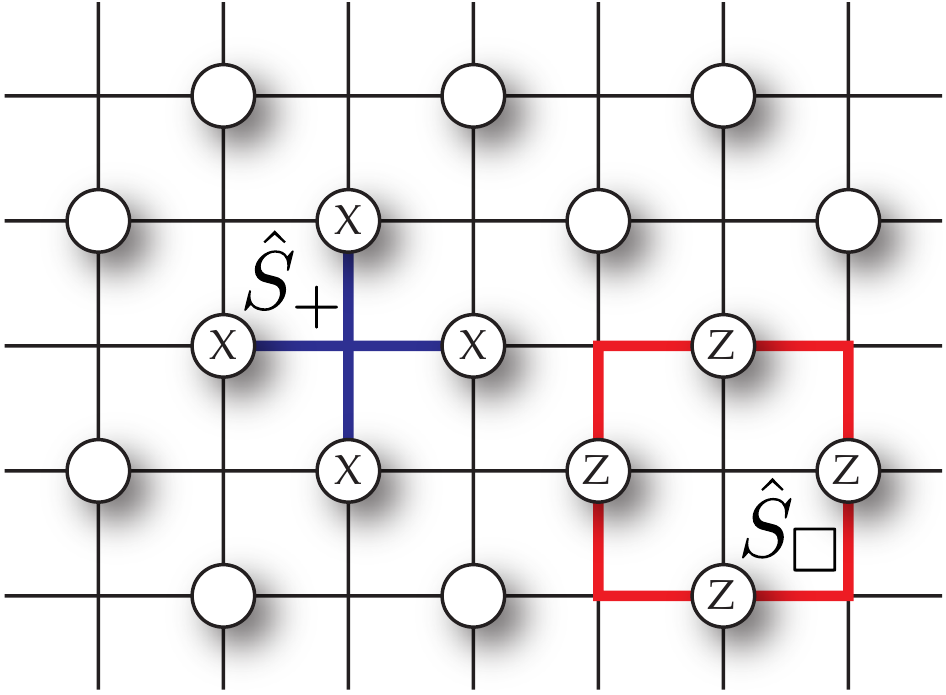}
		\captionspacefig \caption{Graph representation of the toric QEC code, and its associated stabilisers. The graph resides on a torus (top), while the physical qubits reside on the embedded lattice (bottom). The star and plaquette stabilisers across all vertices, jointly specify the state of the graph up to two missing degrees of freedom, which encode a single logical qubit. Thus, a logical qubit is encoded jointly across the entire graph, not at any specific vertex. Logical operations are performed via operations following topological paths through the lattice (not shown). The graph may be distributed across multiple hosts for distributed quantum computation.} \label{fig:toric_code}
	\end{figure}
\else
	\begin{figure*}[!htbp]
		\includegraphics[clip=true, width=0.475\textwidth]{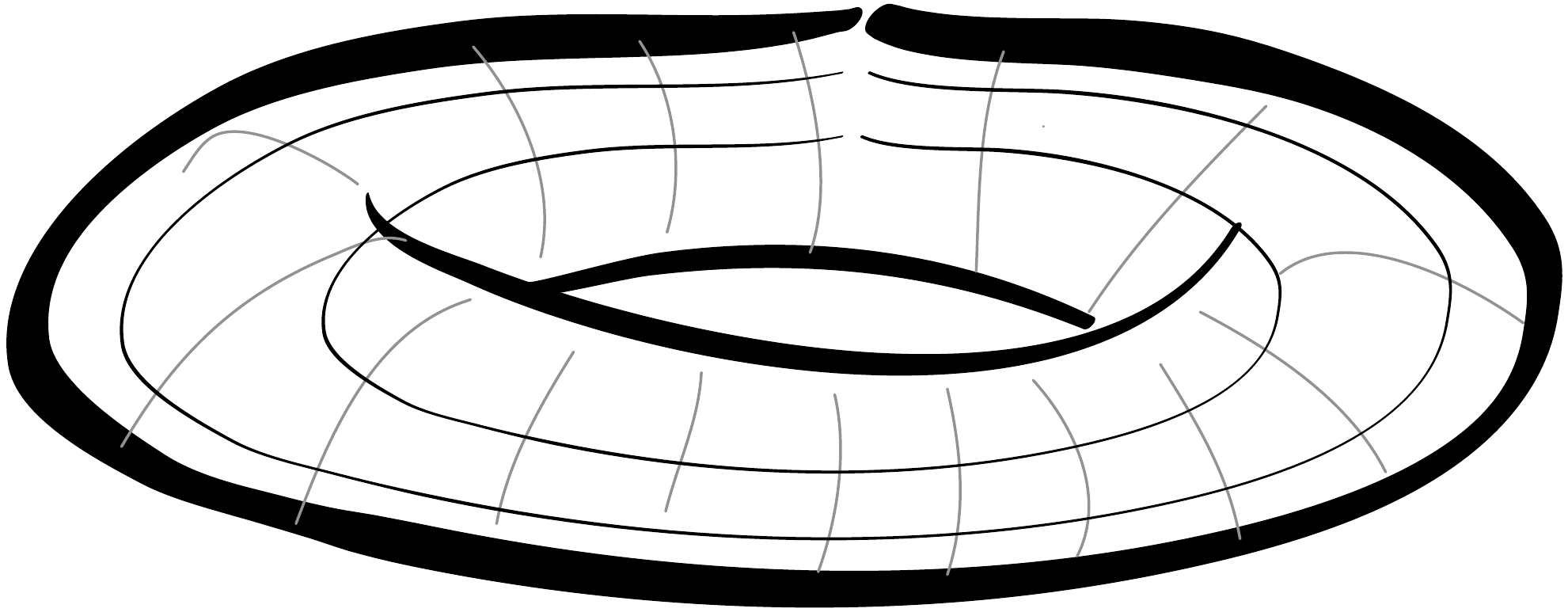}
		\includegraphics[clip=true, width=0.475\textwidth]{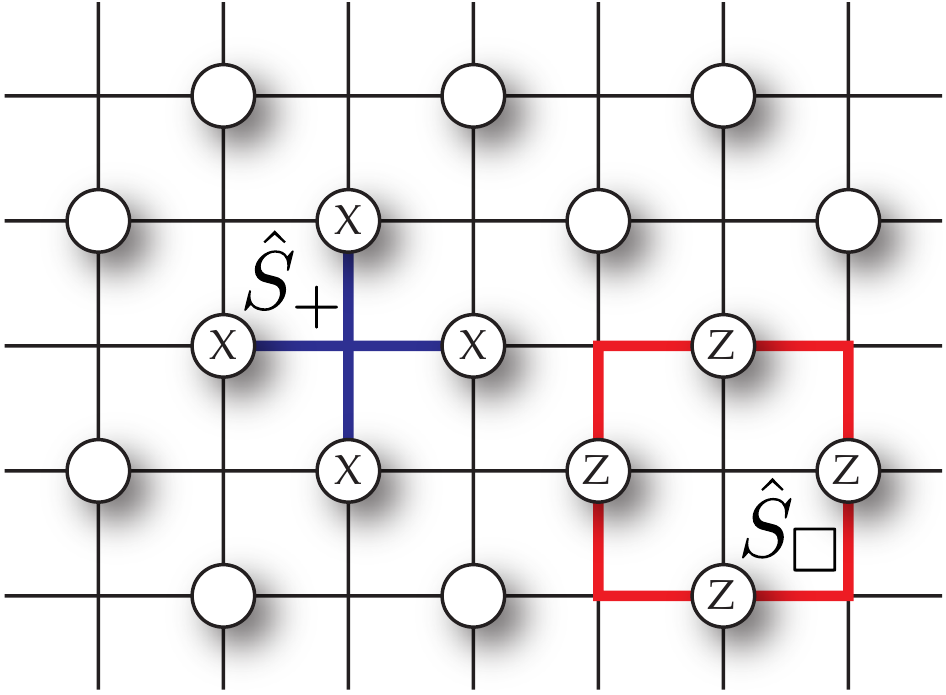}
		\captionspacefig \caption{Graph representation of the toric QEC code, and its associated stabilisers. The graph resides on a torus (left), while the physical qubits reside on the embedded lattice (right). The star and plaquette stabilisers across all vertices, jointly specify the state of the graph up to two missing degrees of freedom, which encode a single logical qubit. Thus, a logical qubit is encoded jointly across the entire graph, not at any specific vertex. Logical operations are performed via operations following topological paths through the lattice (not shown). The graph may be distributed across multiple hosts for distributed quantum computation.} \label{fig:toric_code}
	\end{figure*}
\fi

Unlike the cluster state stabilisers from Eq.~(\ref{eq:CS_stab}), these stabilisers are insufficient to fully characterise a unique quantum state. Rather, there are two unspecified degrees of freedom, which allows for a single qubit to be represented. Modifications of the topology, in the form of holes in the lattice (the genus of the topology), allow larger numbers of qubits to be encoded. Logical operations are implemented by performing sequences of local gates and measurements across topologies over the surface.

The important feature to note is that logical qubits encoded into the toric code do not reside locally at any of the physical qubits in the topology. Rather, they reside jointly across the entire graph, which, like cluster states, might be partitioned across multiple hosts, enabling distributed computation.

Having defined the toric code as such, QEC proceeds in a similar manner to any other stabiliser code -- we measure all the stabilisers, yielding a syndrome\index{Error!Syndrome}, from which we can determine geometrically where errors took place in the graph, which can subsequently be corrected (if below threshold).

The simplest example of error detection is the scenario where a single bit-flip ($\hat{X})$ error has occurred in the graph. Now exactly two plaquette stabilisers (the ones upon which the respective qubit acts) will yield the $-1$ measurement outcome, instead of the expected $+1$ outcome. These two stabilisers will necessarily be neighbouring ones, overlapping at the qubit where the error took place. Thus, using this geometric reasoning, we are able to identify the location of the single $\hat{X}$ error and subsequently correct it. On the other hand, if there were too many errors, it is possible they could conspire against us to create ambiguity in the geometric argument for the location of the errors. The QEC procedure is illustrated in Fig.~\ref{fig:toric_code}.

\begin{figure}[!htbp]
	\includegraphics[clip=true, width=0.3\textwidth]{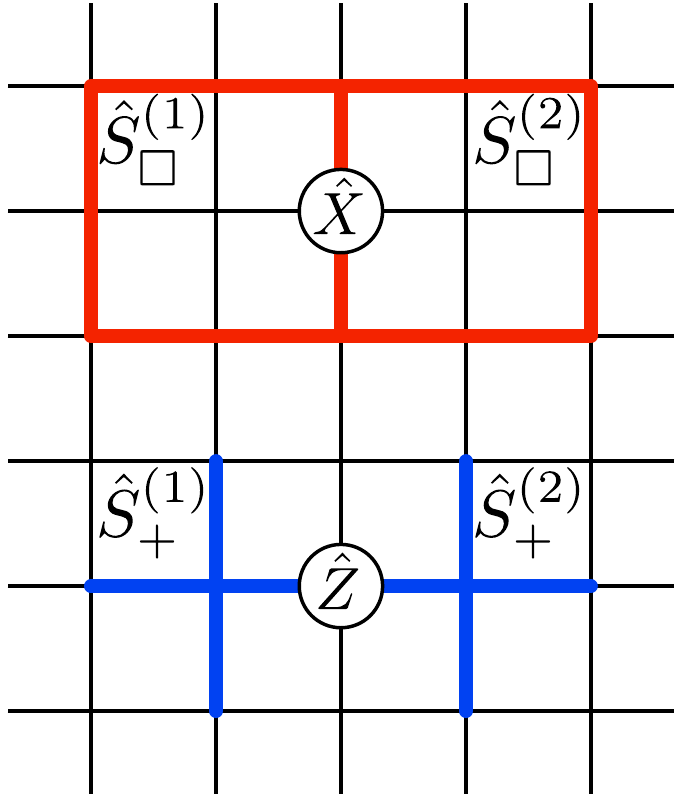}
	\captionspacefig \caption{QEC of a single bit-flip (top) and phase-flip (bottom) error via measuring two overlapping stabilisers. If there is at most one bit- or phase-flip error, as per this graphic, then upon measuring all the star and plaquette stabilisers, only those whose qubits are affected by the error channel will yield the $-1$ measurement outcome, the others all yielding $+1$. Such an error will affect exactly two stabilisers, whose qubit in common must have been the one affected by the respective error, which can subsequently be corrected since the type of the error and its location are known. Thus, in this example, all of the $\hat{S}_+^{(1)}$, $\hat{S}_+^{(2)}$, $\hat{S}_\square^{(1)}$ and $\hat{S}_\square^{(2)}$ stabilisers will yield $-1$ outcomes for the shown $\hat{X}$ and $\hat{Z}$ errors, whereas any of the remaining (not shown) $\hat{S}_k^{(j)}$ will yield $+1$ outcomes, which uniquely specifies where those errors took place.}\label{fig:toric_corr}	
\end{figure}

Importantly, the stabilisers are all defined over geometrically localised neighbourhood regions, and do not require long-range measurements, making this type of code suitable to distributed models for quantum computation, much like cluster states.

\begin{figure}[!htbp]
	\includegraphics[clip=true, width=0.475\textwidth]{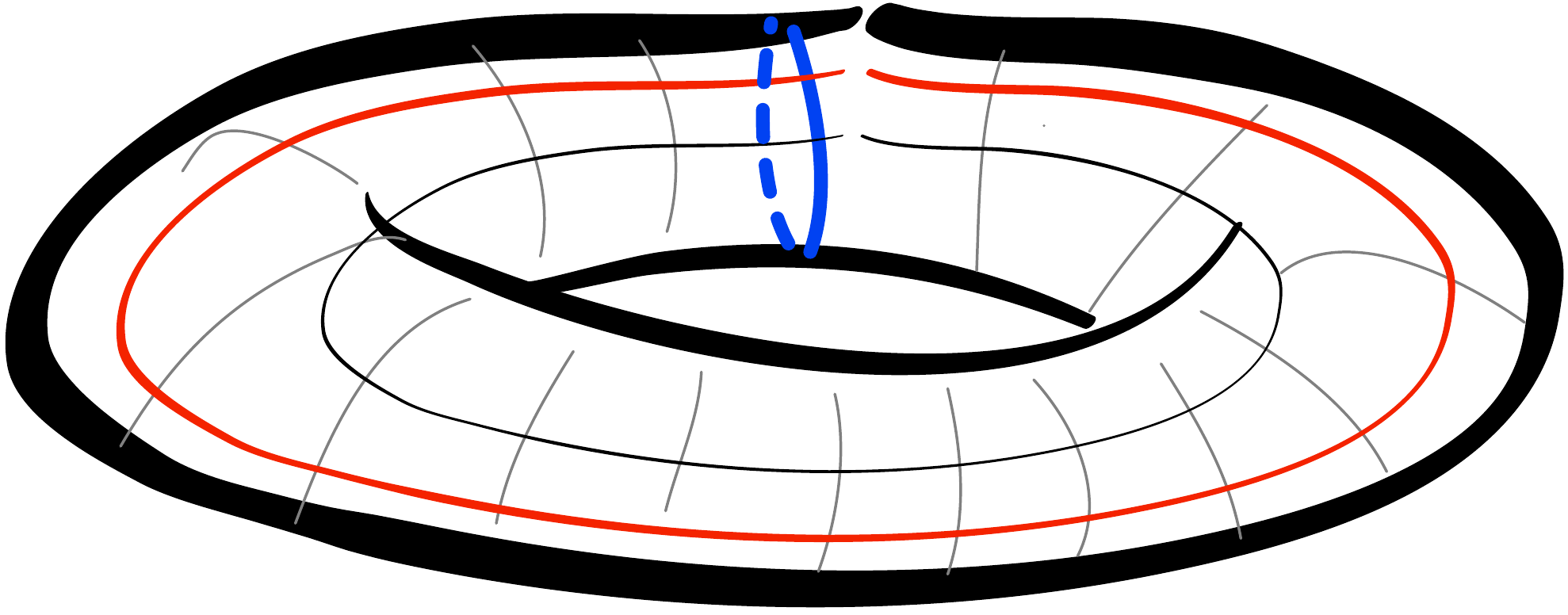}
	\captionspacefig \caption{Implementation of logical Pauli-$\hat{X}$ and -$\hat{Z}$ operations on a toric code qubit via the application of a chain of local Paulis. The chains must circumnavigate the torus around either axis. Any path topologically equivalent to such a circumnavigation of the torus is sufficient, enabling paths to bypass defects in the lattice.} \label{fig:toric_code_paulis}
\end{figure}

Logical operations similarly have geometric interpretations. Most simply, logical Pauli-$\hat{X}$ and -$\hat{Z}$ gates may be implemented by applying chains of local Pauli operations along topological paths, specifically closed loops around the two different axes of the torus (Fig.~\ref{fig:toric_code_paulis}. These topologies may be deformed, provided that they are topologically equivalent to the desired closed loops, hence the name `topological code'. With more qubits and more elaborate gates, similar topological definitions exist for implementing logical gates on encoded qubits.

%\comment{What about the actual computation? Can this still be distributed when we do the topological gates etc?}

%\comment{Figures for thresholds FT and QEC}

%\comment{Explain stabiliser measurement and how this yields error syndromes}

%\comment{Explain logical operations on code}

%\comment{Explain how number of logical qubits can be increased by changing topology}

%\comment{How to convert cluster states to topological codes, or any other way of preparing them. Add discussion of how to perform gates.}

%\comment{Figure for both examples of this -- under and over threshold!}

%\comment{Missing degree of freedom in stabilisers defines qubit. How does topology (genus) relate to number of logical qubits?}

%
% Unitary Error Averaging
%

\subsection{Unitary error averaging} \index{Quantum error correction (QEC)}\index{Unitary!Error averaging}\label{sec:error_averaging}

Aside from the \textit{active} QoS techniques discussed until now, there is also a recently described \textit{passive} technique that requires no feedforward to operate. This technique, called \textit{unitary error averaging} \cite{marshman2018passive,marshman2024unitary}, is formulated specifically in the linear optical context. It is an open question whether it generalises to other models, such as conventional quantum circuits.

Suppose we desire to implement some linear optics unitary network, $\hat{U}_\mathrm{target}$, but our fabrication techniques are imperfect and we instead implement a close approximation to it. How can we overcome this?

\begin{figure}[!htbp]
\includegraphics[clip=true, width=0.475\textwidth]{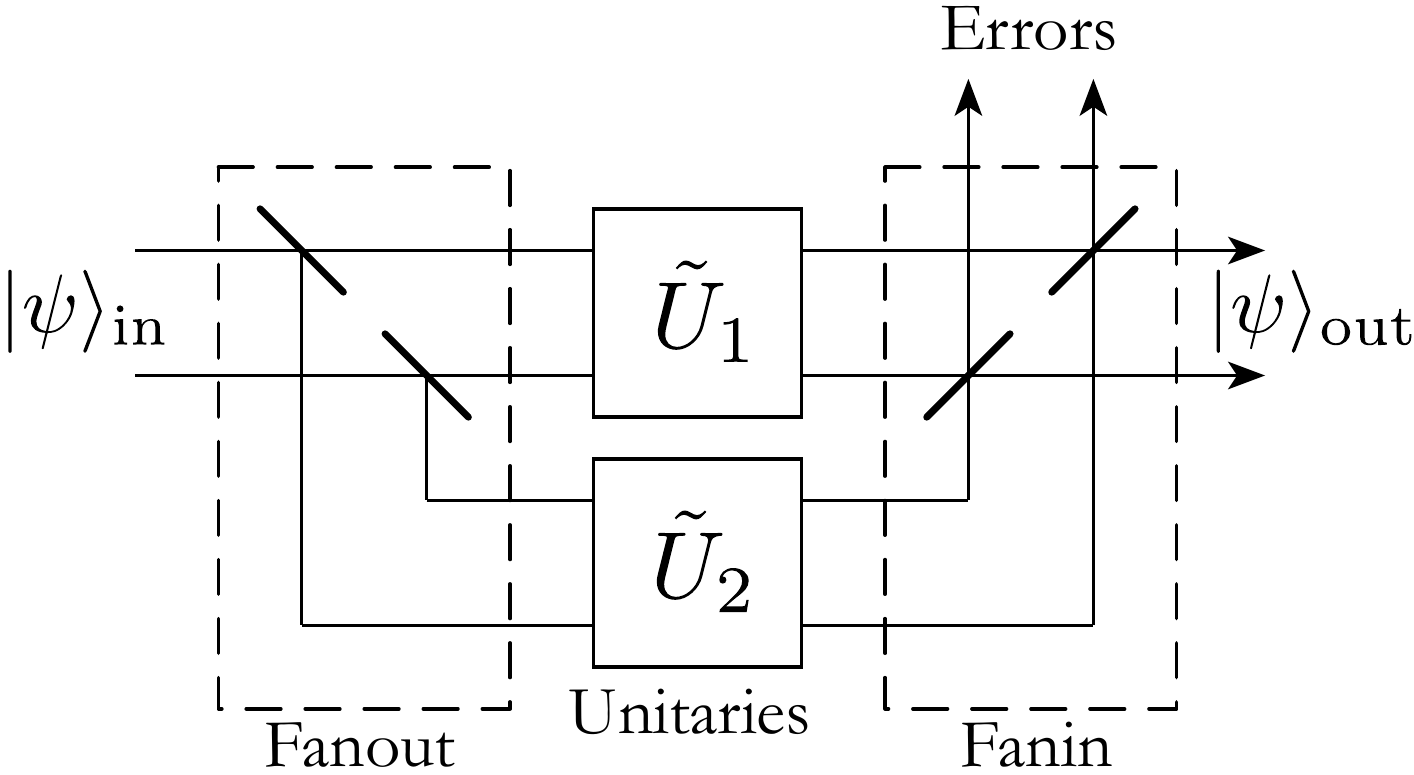}
\captionspacefig \caption{Protocol for error correction via unitary error averaging. Here our computation resides across \mbox{$N=2$} modes, and there are \mbox{$M=2$} redundant copies of the manufactured unitary, which straightforwardly generalises to arbitrary $M$ and $N$. A fanout operation splits the wave amplitudes from the set of $N$ input modes uniformly across $M$ sets of $N$ modes. Each set of $N$ evolves through an independent copy of the manufactured imperfect unitary, $\tilde{U}_i$. A reverse fanout operation recollects the bundles of modes. Upon post-selecting upon detecting all photons within the first $N$ modes, and none in the other ancillary modes, the state is projected into the subspace as if it had been acted upon by the arithmetic average of the imperfect unitaries. If photons are detected in the error modes, our logical state has effectively suffered photon loss and has been corrupted.} \label{fig:error_av_circuit}	
\end{figure}

Ref.~\cite{marshman2018passive} showed that by using an optical fanout\index{Fanout} operation, which splits a set of optical modes equally across multiple sets of modes in a type of redundant encoding\footnote{The optical fanout operation effectively maps single-photon states to W-states (Sec.~\ref{sec:W_state_prep})\index{W-states}.}, and passing each set through an independently manufactured copy of the imperfect unitary, $\tilde{U}_i$, upon performing a fan-in to recombine the modes, and post-selection on detecting all photons in the desired output modes, the output state will behave as if it had evolved through a network given by the arithmetic average of each of the imperfect copies of $\hat{U}_\mathrm{target}$,
\begin{align}
\hat{U}_\mathrm{av} = \frac{1}{M}\sum_{i=1}^M \tilde{U}_i.	
\end{align}
A schematic for the optical circuit is shown in Fig.~\ref{fig:error_av_circuit}.

Note that the protocol doesn't require that the $M$ instances of the unitary be held locally. They operate independently in parallel and could therefore be distributed, with the independent copies held by entirely different parties in distant locations (Sec.~\ref{sec:error_av_parallel}).\index{Distributed quantum computation}\index{Parallelisation}

The protocol is necessarily non-deterministic, since sometimes photons will be measured in the error modes, signalling that the output state has been projected onto an erroneous subspace. Thus, this is a heralded error correction scheme. When post-selection succeeds we have effectively evolved via the averaged operator, which statistically ought to more closely approximate $\hat{U}_\mathrm{target}$ than individual $\tilde{U}_i$.

%
% Qubit Loss Codes
%

\subsection{Qubit loss codes}\index{Qubit loss!Codes}

More severe than a simple, say, dephasing channel is qubit loss, whereby an entire physical qubit is lost to the environment, effectively tracing it out of the system. Depending on physical architecture, there are two ways in which qubit loss might manifest itself:
\begin{itemize}
\item Located errors\index{Located errors}: We know that a loss occurred and which physical qubit was affected.
\item Unlocated errors\index{Unlocated errors}: We do not know that a physical qubit was lost.
\end{itemize}

These manifestations arise naturally in different architectural contexts. For example, it is relatively easy to look at an atomic qubit to see whether it's still where it should be, whereas observing a photonic qubit typically requires destroying it, making the former vulnerable to located errors and the latter to unlocated errors.

It turns out that the distinction between these two modes of error can have a significant effect on the error correction thresholds of QECs. Specifically, a located error effectively gives us an additional classical bit of information diagnosing that an error took place, which we are not privy to for unlocated errors. This additional bit of information can be exploited by QEC codes to improve thresholds for located errors.

In \cite{bib:RohdeHaselgrove} a derivation of a generalisation of the quantum Hamming bound\index{Quantum Hamming bound} for non-degenerate\footnote{In a non-degenerate code, all correctable errors map the codeword to orthogonal states. For degenerate codes\index{Degenerate codes} multiple distinct errors can map the codeword to the same state. The Hamming bound exploits the non-degeneracy using a counting or `pigeonholing' argument to count correctable errors.} QEC codes\index{Non-degenerate codes} was presented, where the number of located and unlocated errors were treated as independent parameters. They found that for non-degenerate codes the Hamming bound allows exactly twice as many located as unlocated errors. Specifically, for $t_l$ located errors and $t_u$ unlocated errors in an $n$-qubit non-degenerate code encoding $k$ logical qubits, QEC is possible provided that the number of errors satisfied,
\begin{align}
\sum_{i=0}^{t_u + \lfloor t_l/2 \rfloor} \binom{n}{i}3^i 2^k \leq 2^n.
\end{align}

Subsequently, using surface codes \cite{bib:StaceBarrettDohertyLoss} arrived at the similar conclusion that QEC codes are more robust against located than unlocated errors.

%
% Gate Failure Codes
%

\subsection{Gate failure codes}\index{Gate failure codes}

Quantum gates, especially 2-qubit entangling ones in many architectures, can have some non-zero chance of failure. In linear optics this gate non-determinism is an inherent feature of entangling gates like the CNOT. In fact it is provable that such gates cannot be implemented deterministically. With a large number of such gates in a circuit, the probability of them all working drops exponentially with the number of non-deterministic gates, thereby undermining efficiency. Can we overcome this?

In Sec.~\ref{sec:CS_LO} we discuss techniques for performing scalable linear optics quantum computing with non-deterministic gates, within the cluster state formalism\index{Cluster states}. The idea is to construct micro-clusters\index{Micro-cluster states} with redundant dangling bonds, which facilitate multiple bonding attempts to merge smaller clusters into larger ones. Once all the entangling operations have been performed, all that remains is to perform a sequence of single-qubit unitaries and measurements, both of which are deterministic in-principle.

This same approach could be logically generalised to any physical architecture where entangling gates are non-deterministic, but single-qubit operations are deterministic. 

%
% Decoherence-Free Subspaces
%

\subsection{Decoherence-free subspaces}\index{Decoherence-free subspaces}

The QoS techniques discussed previously were based on the notion of performing measurements on quantum systems to project them into subspaces devoid of errors. For example, in the 3-qubit code, measurement of the syndrome qubits projects the encoded state into a subspace where there was either no error, or in which an error occurred whose location is known and may therefore be corrected.

An alternate approach is to encode quantum information into Hilbert spaces which are invariant under a given error model. Such spaces are referred to as \textit{decoherence-free subspaces} (DFSs). In this instance we assume the error model is known, for example a dephasing channel, such that we can choose the appropriate DFS.

To illustrate this idea we will consider encoding a single logical qubit into two physical qubits. The error model we will encode against is a collective $Z$-rotation, where the two physical qubits are subject to perfectly correlated $Z$ errors. This arises naturally in the context of, say, atomic qubits subject to the same external electromagnetic field, and therefore accumulate the associated phase errors in tandem.

A single-qubit $Z$-rotation of angle $\theta$ on the Bloch sphere\index{Bloch sphere} is given by,
\begin{align}
	\hat{Z}(\theta) = e^{i\frac{\theta}{2}\hat{Z}} = \begin{pmatrix}
  e^{i\frac{\theta}{2}} & 0 \\
  0 & e^{-i\frac{\theta}{2}}
\end{pmatrix}.
\end{align}
where $\hat{Z}$ is the usual Pauli phase-flip operator\index{Pauli!Operators} (Sec.~\ref{sec:circuit_model}). This operates on the physical basis states as,
\begin{align}
	\hat{Z}(\theta) \ket{0} &\to e^{i\frac{\theta}{2}}\ket{0}, \nonumber \\
	\hat{Z}(\theta) \ket{1} &\to e^{-i\frac{\theta}{2}}\ket{1}.
\end{align}

Now we employ the encoding for logical basis states,
\begin{align}
\ket{0}_L &\equiv \ket{0}_1\otimes\ket{1}_2,\nonumber \\
\ket{1}_L &\equiv \ket{1}_1\otimes\ket{0}_2,
\end{align}
Note that both logical basis states are invariant under a common $Z$-rotation (the other two physical basis states, $\ket{0}_1\otimes\ket{0}_2$ and $\ket{1}_1\otimes\ket{1}_2$ do not observe this property),
\begin{align}
	\hat{Z}_1(\theta)\hat{Z}_2(\theta)\ket{0}_L &= \ket{0}_L,\nonumber\\
	\hat{Z}_1(\theta)\hat{Z}_2(\theta)\ket{1}_L &= \ket{1}_L.
\end{align}
Thus, when acting on an arbitrary linear combination of these basis states (i.e a logical qubit),
\begin{align}
\ket\psi_L = \alpha \ket{0}_L + \beta\ket{1}_L,
\end{align}
via linearity the logical qubit must also be invariant under the collective error,
\begin{align}
	\hat{Z}_1(\theta)\hat{Z}_2(\theta)\ket\psi_L = \ket\psi_L.
\end{align}
This type of DFS encoding therefore protects a logical qubit against arbitrary, unknown but correlated $Z$-rotations.

The same principle can be logically extended to many other correlated error models. For example, operating in a rotated basis (under a Hadamard transform), one could similarly protect against correlated $X$-rotations using the encoding,
\begin{align}
\ket{0}_L &\equiv \ket{-}_1\otimes\ket{+}_2,\nonumber \\
\ket{1}_L &\equiv \ket{+}_1\otimes\ket{-}_2,
\end{align}
where \mbox{$\ket\pm=\frac{1}{\sqrt{2}}(\ket{0}\pm\ket{1})$}.

This idea, although very simple, is very powerful, since it is a completely passive form of error correction, requiring no syndrome measurements, feedforward or correction operations. It also arises quite naturally in some systems where undesired external fields act roughly uniformly across the physical qubits within a system.

%
% Dynamical Decoupling
%

\subsection{Dynamical decoupling}\index{Dynamical decoupling}

An alternate mechanism by which errors could be introduced into our system is via coupling to an external environment (for example via an electromagnetic field) introducing a persistent evolution of our qubits, which is slow-moving compared to the rate at which the implemented computation is evolving the system. We can model this as a joint system/environment Hamiltonian of the form,
\begin{align}\label{eq:dyn_dec_ham}
\hat{H}_\mathrm{total} = \lambda_\mathrm{comp}\hat{H}_\mathrm{comp} + \lambda_\mathrm{env}\hat{H}_\mathrm{env} + \lambda_\mathrm{int}\hat{H}_\mathrm{int},	
\end{align}
where the different components represent, in order, the Hamiltonians of the: total joint system; quantum computer (or system of interest); environment; interaction between system and environment. We are specifically operating in the regime where \mbox{$\lambda_\mathrm{int}\ll\lambda_\mathrm{comp}$}, such that the computation is the dominant term in the evolution and the environmental coupling can be treated as a small perturbation from the desired evolution.

The goal of dynamical decoupling is to minimise the influence of the system/environment interaction term, $\hat{H}_\mathrm{int}$, by manipulating the system in such a way that this term continuously cancels itself out over time.

Let us illustrate how this can be achieved using a simple example, whereby a single-qubit system couples with an environment which introduces a slow, unknown phase evolution. Discretising time, we can write this phase evolution as a Pauli $Z$-rotation on the Bloch sphere\index{Bloch sphere},
\begin{align}
	\hat{Z}(\theta) = e^{i\frac{\theta}{2}\hat{Z}} = \begin{pmatrix}
  e^{i\frac{\theta}{2}} & 0 \\
  0 & e^{-i\frac{\theta}{2}}
\end{pmatrix},
\end{align}
for some unknown, but small $\theta$. Next we observe that, quite obviously, a $Z$-rotation of angle $\theta$ can be trivially undone by applying another $Z$-rotation of angle $-\theta$, since,
\begin{align}
\hat{Z}(-\theta)\hat{Z}(\theta) = \hat\openone.	
\end{align}

Therefore, if the system/environment coupling introduced an evolution of $\hat{Z}(\theta)$ in the previous unit of time, our goal is to manipulate it into implementing $\hat{Z}(-\theta)$ during the next one. Alas, the environment is beyond our control and we cannot directly order it to reverse direction. We do, however, have complete control over our qubit system, so we will achieve the same outcome by flipping the direction of the Bloch sphere underneath the environments foot, allow it to take a step forward, before flipping it back.

How do we achieve this flip in the Bloch sphere? Simply by using the following identity from the algebra of the Pauli matrices\index{Pauli!Operators},
\begin{align}
\hat{X}\hat{Z}(\theta)\hat{X} = 	\hat{Z}(-\theta).
\end{align}
That is, applying a bit-flip to a qubit, followed by an arbitrary phase-rotation, followed by another bit-flip is equivalent to having taken the same phase-rotation in the reverse direction. Effectively we are tricking the environment into time-reversal!

Now if we proceed for two time-steps, once bit-flipped and another not, we have,
\begin{align}
\underbrace{\hat{Z}(\theta)}_{\mathrm{forward}}\cdot\underbrace{[\hat{X}\hat{Z}(\theta)\hat{X}]}_{\mathrm{reverse}} = \hat\openone,
\end{align}
and the unknown phase-rotation has been eliminated. The sequence is illustrated on the Bloch sphere\index{Bloch sphere} in Fig.~\ref{fig:bloch_sphere_dyn_dec}.

\if 1\doublecol
	\begin{figure}[!htbp]
	\includegraphics[clip=true, width=0.475\textwidth]{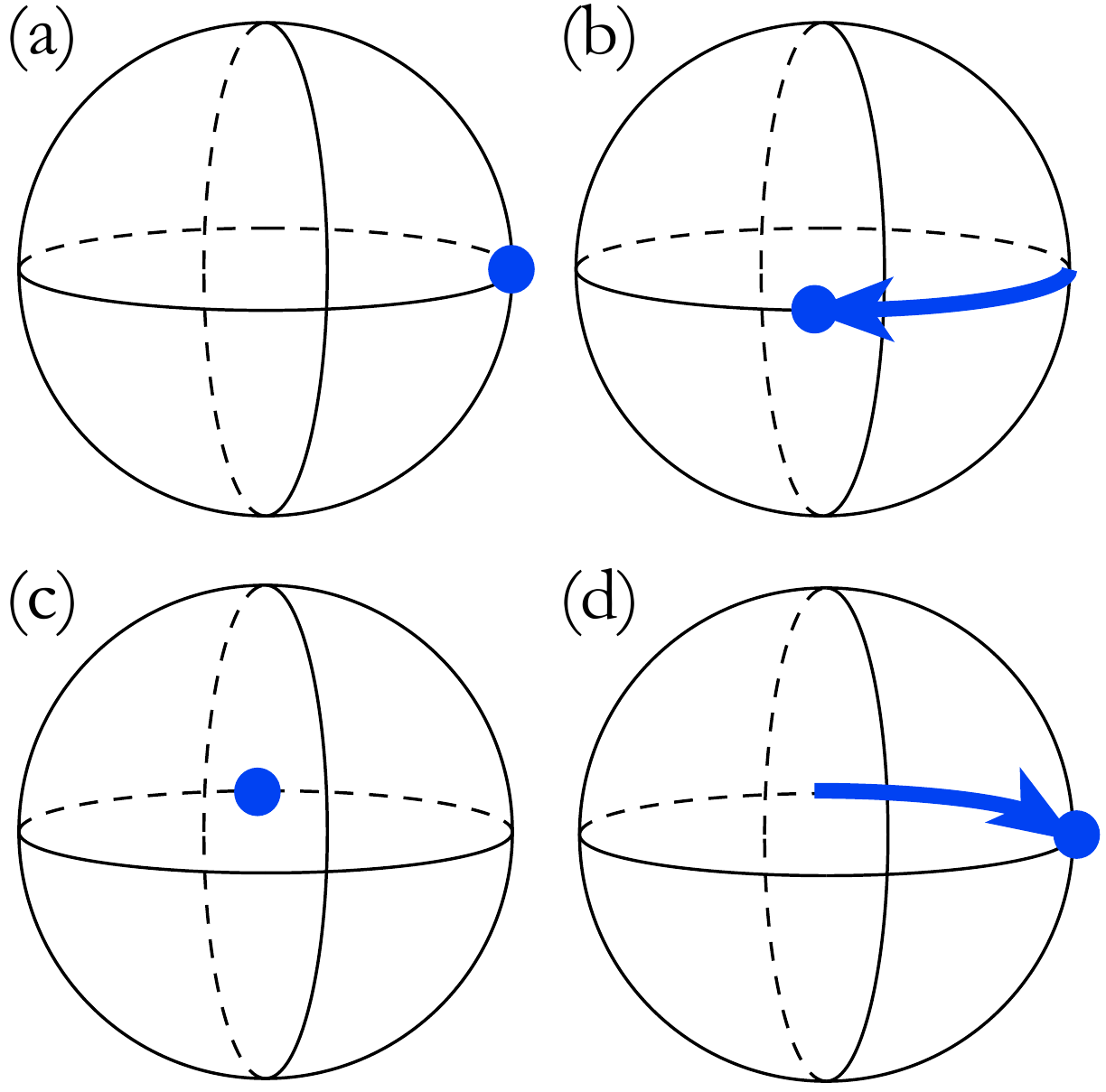}
	\captionspacefig \caption{Bloch sphere representation of a sequence to dynamically decouple an unknown $\hat{Z}$ rotation from a qubit. (a) Blue dot represents the initial state. (b) The state undergoes an unknown $\hat{Z}(\theta)$ rotation. (c) A bit-flip ($\hat{X}$) flips the Bloch sphere upside down. (d) Waiting for the same time as before such that the same unknown $\hat{Z}$-rotation takes place, the state ends up in its initial state, independent of the $\hat{Z}$-rotation angle $\theta$, but assuming it was the same on both occasions in (b) and (d).} \label{fig:bloch_sphere_dyn_dec}	
	\end{figure}
\else
	\begin{figure*}[!htbp]
	\includegraphics[clip=true, width=\textwidth]{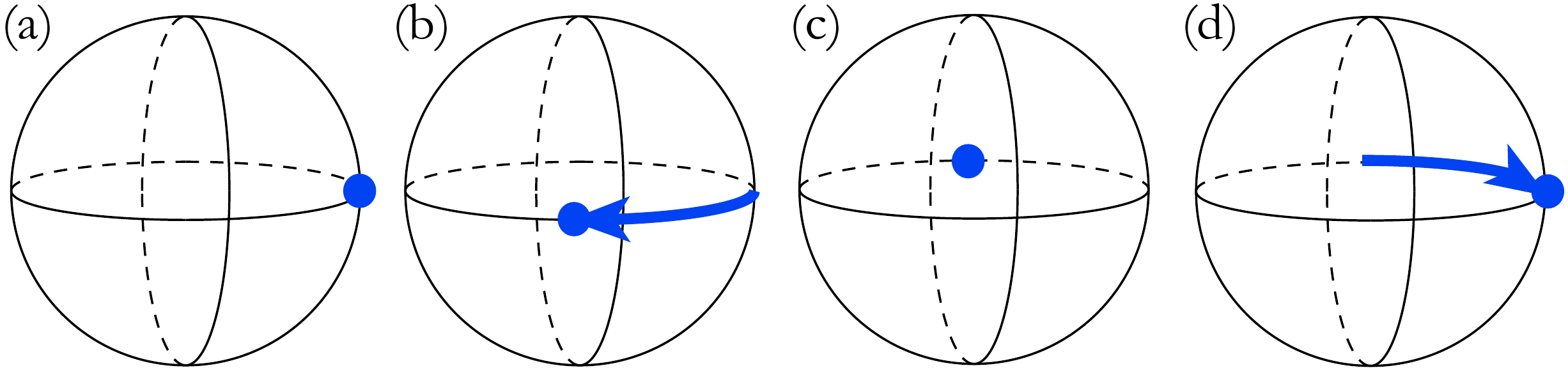}
	\captionspacefig \caption{Bloch sphere representation of a sequence to dynamically decouple an unknown $\hat{Z}$ rotation from a qubit. (a) Blue dot represents the initial state. (b) The state undergoes an unknown $\hat{Z}(\theta)$ rotation. (c) A bit-flip ($\hat{X}$) flips the Bloch sphere upside down. (d) Waiting for the same time as before such that the same unknown $\hat{Z}$-rotation takes place, the state ends up in its initial state, independent of the $\hat{Z}$-rotation angle $\theta$, but assuming it was the same on both occasions in (b) and (d).} \label{fig:bloch_sphere_dyn_dec}	
	\end{figure*}
\fi

The coupling to the external environment and the qubit state needn't be constant over time, but could be time-varying. In this instance, it is essential that the decoupling control sequence of bit-flip operations be much faster than the rate at which both the phase-rotations and computational operation are changing, such that consecutive unknown phase-rotations are approximately equal and thereby almost perfectly cancel. This rapid control sequence is sometimes referred to as `bang-bang' control\index{Bang-bang control}, since we are repeatedly implementing bit-flips at a fast rate. 

As described above, we have merely error corrected a quantum memory. Of course we wish to implement far more sophisticated evolutions. This requires breaking down the computational evolution (given by $\hat{H}_\mathrm{comp}$) into a large number of small, discrete steps. These are interspersed with our bang-bang control sequence so as to continuously remove any phase-errors accumulated during the course of the computation.

Dynamical decoupling extends to all manner of error models, beyond the simple $Z$-error model presented above. They are therefore a very powerful tool in error correction. However, unfortunately they are only naturally suited to continuous evolutions governed by Hamiltonians of the form shown in Eq.~(\ref{eq:dyn_dec_ham}), not to the more common discretised models such as the circuit and cluster state models.

%\comment{References}

%
% Continuous-Variables
%

\subsection{Continuous-variables}\index{Continuous-variables!Quantum error correction}

First CV QEC protocol is the direct analogue of the the qubit redundancy codes. However, the noise models do not correspond to what usually occur in Gaussian systems, namely, thermal noise\index{Thermal!Noise} and loss.

It has been proven that QEC of Gaussian noise on Gaussian states, using only Gaussian operations is impossible \cite{bib:PhysRevLett.102.120501}. It is closely related to entanglement distillation, discusses above. Several approaches have since been developed. 

This no-go theorem does not apply if the initial states are non-Gaussian. One can also encode qubits using non-Gaussian continuous-variable states \cite{bib:PhysRevA.64.012310}. Then error correction can be implemented using Gaussian operations, resulting in a Gaussian state. This protocol is known to be fault tolerant, however, the threshold requirements are quite stringent, such as needing the photon number in the squeezed state to be around 20 \cite{bib:PhysRevA.73.012325}.

Another approach is to error correct on Gaussian states using non-Gaussian operations. Such protocols have been proposed \cite{bib:PhysRevA.67.062320} and demonstrated \cite{bib:xiang2010heralded} for entanglement distillation. It has been shown that CV teleportation and distillation based on noiseless amplification can be combined to error correct Gaussian states against Gaussian noise \cite{bib:PhysRevA.84.022339}, implementable using linear optics and photon counting.
However, it is unclear whether such protocolsc can be made fault tolerant.

% \comment{To do -- can we be a bit more specific about the physical resource requirements and the specific noise models? Specify what Gaussian noise means.}

\latinquote{Ex luna scientia.}

\sketch{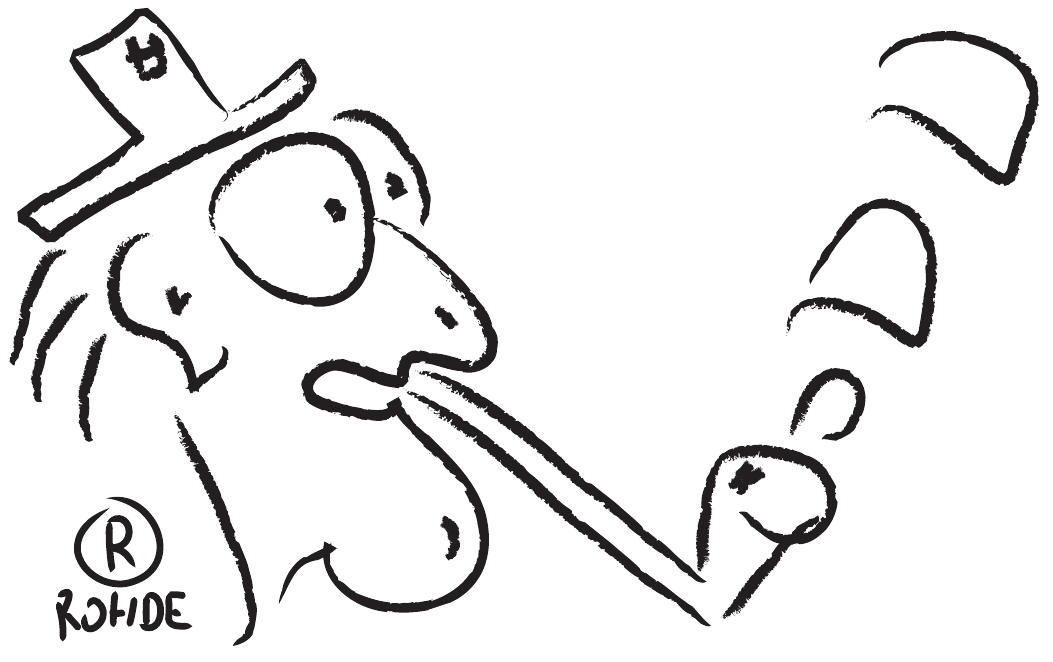}

\clearpage
% % %
% Cloud quantum computing
%

\part{Cloud quantum computing}\label{part:cloud_QC}\index{Cloud quantum computing}

\dropcap{F}{rom} the perspective of quantum computing, by far the most pressing goal for quantum networking is to facilitate \textit{cloud quantum computing}, whereby computations can be performed over a network via a client/server model. This will be of immense importance economically, allowing very expensive quantum computers to be accessible to end-users, who otherwise would have been priced out of the market. This economic model is critical to the early widespread adoption of quantum computation. Networking quantum computers is also of the immense importance to capitalise off the leverage associated with unifying quantum resources as opposed to utilising them in isolation (Sec.~\ref{sec:quant_ec_lev}).

There are several protocols necessary to facilitate cloud quantum computing. First of all, we must have a means by which to remotely process data prepared by a host on a server(s). At the most basic level, this simply involves communicating quantum and/or classical data from a client to a single server for processing, which returns quantum or classical information to the client -- \textit{outsourced quantum computation}. In the most general case, a computation may be processed by multiple servers, each responsible for a different part of the computation -- \textit{distributed quantum computation}.

Many real-world applications for quantum computing will involve sensitive data, in terms of both the information being processed and the algorithms being employed. This necessitates encryption protocols allowing computations to be performed securely over a network, such that intercept-resend attacks\index{Intercept-resend attacks} are unable to infer the client's data, and even the host itself is unable to do so -- \textit{homomorphic encryption} and \textit{blind quantum computing}. These form the basic building blocks from which a secure cloud-based model for quantum computing may be constructed, and economic models based on the outsourcing of computations may emerge.

The consumer of cloud quantum computing will of course need to be convinced that their data was processed faithfully, according to the desired algorithm. This requires \textit{verification protocols} to allow the server to prove to the client that their data was correctly and honestly processed. 

%
% The Quantum Cloud
%

\section{The Quantum Cloud\texttrademark} \label{sec:cloud} 
\index{Cloud quantum computing}

\dropcap{W}{e} begin by introducing the primitive building blocks for cloud quantum computing. These form the foundation for higher-level protocols to be discussed later in this part. 

%
% Outsourced Quantum Computation
%

\subsection{Outsourced quantum computation} \index{Outsourced!Quantum computation}

Most simply, an outsourced computation involves Alice preparing either a quantum or classical input state, which she would like processed on Bob's computer. Bob performs the computation and returns either a quantum or classical state to Alice.

The algorithm, which Bob implements, could either be stipulated by Alice, in which case she is purely licensing Bob's hardware, or by Bob, in which case she is licensing his hardware and software. In the case of classical input and classical output, such an outsourced computation is trivial from a networking perspective, requiring no usage of the quantum network whatsoever. In the case of quantum input and/or output data, the quantum network will be required.

Despite the model being very simple, there may still be stringent requirements on the costs in the network. When the result of the computation is returned to Alice, there may be fidelity requirements. An approximate solution to a problem, or a computation with any logical errors whatsoever, may be useless, particularly for algorithms, which are not efficiently verifiable. For example, if Alice is attempting to factorise a large number using Shor's algorithm\index{Shor's algorithm}, a number of incorrect digits may make the the correct solution effectively impossible to determine. Or if a large satisfiability problem is being solved, almost any classical bit-flip errors will invalidate the result, requiring additional computation by Alice to resolve (which may be exponentially complex to perform).

In the case of classical communication of input and output data, we can reasonably assume error-free communication, owing to its digital nature. However, in the case of quantum communication it is inevitable that at least some degree of noise will be present. Depending on the application, this may require the client and host to jointly implement a distributed implementation of QEC (Secs.~\ref{sec:QOS} \& \ref{sec:fault_tolerance}), whereby Alice and Bob communicate encoded states with one another, to which syndrome measurement and error correction are applied upon receipt. This will necessitate a limited amount of quantum processing to be directly available to Alice. In the case where she is completely starved of any quantum processing resources whatsoever, this may be a limiting factor. Otherwise, this type of cooperative QEC may be plausible.

%
% Distributed Quantum Computation
%

\subsection{Distributed quantum computation} \label{sec:dist_QC} \index{Distributed quantum computation}

The elementary model described above is very limited, as many realistic data processing applications will require multiple stages of computations to be performed, potentially by different hosts. For example, a client may need data processed using multiple proprietary algorithms owned by different hosts, and the processing will need to be distributed across the network \cite{bib:Cirac99}.

\subsubsection{In-parallel \& in-series computation}

\begin{figure}[!htbp]
\if 1\doublecol
	\includegraphics[clip=true, width=0.475\textwidth]{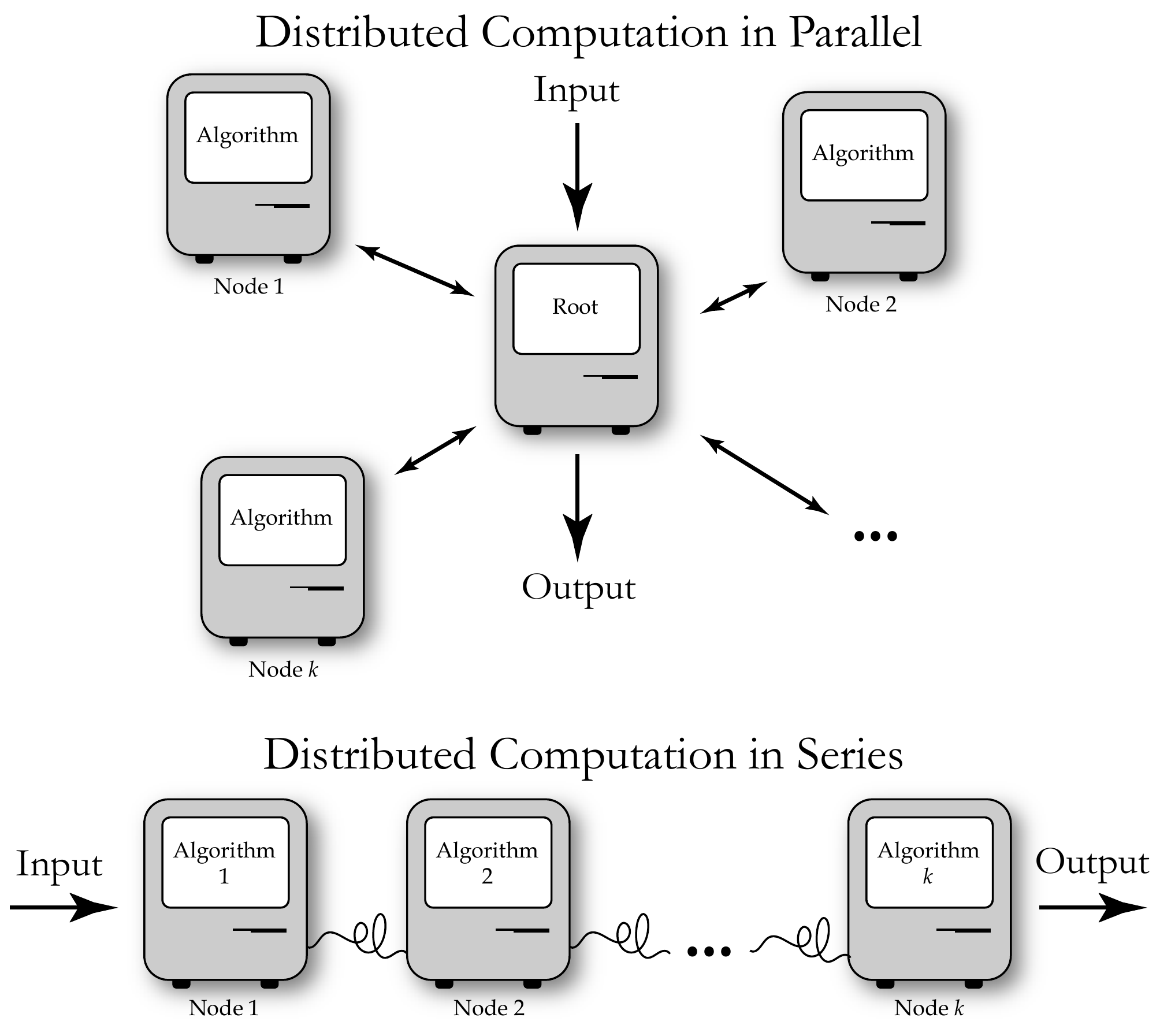}
\else
	\includegraphics[clip=true, width=0.6\textwidth]{distributed}
\fi
\captionspacefig \caption{Models for distributed computation in parallel and in series. In parallel, a root node oversees the total computation, delegating tasks to child nodes, which process data independently of one another before being merged. In series the nodes sequentially process data in a pipeline of algorithmic stages.} \label{fig:distributed}
\end{figure}

Classically, there are two main models for how a distributed computation may proceed -- in \textit{parallel}\index{Parallel!Computation}, or in \textit{series}\index{Series computation} -- whereby sub-algorithms are performed either side-by-side simultaneously, or one after another in a pipeline. The two models are illustrated in Fig.~\ref{fig:distributed}.

Classical parallel processing typically involves a root node\index{Root node}, which delegates tasks to be performed in parallel by a number of child nodes, and the results returned to the root node, which potentially applies an algorithm to merge the set of results, before returning a final result to the client. Classical models such as Google's \textsc{MapReduce} protocol \cite{bib:MapReduce}\index{MapReduce} are built on this idea.

In classical computing, parallel processing is widely employed to shorten algorithmic runtimes. However, the increase in clock-cycles scales only linearly with the number of nodes in the network: $k$-fold parallelisation yields an \mbox{$\sim k$}-fold speedup. For time-critical applications, such a linear improvement may already be highly beneficial, albeit costly.

The alternate scenario is in-series distributed computation, in which a computation proceeds through a pipeline of different stages, potentially performed by different hosts. This model allows a complex algorithm comprising smaller subroutines, each of which may be proprietary with different owners, to be delegated across the network. The different stages may communicate classical and/or quantum data. As with the simple single-host model, if the different stages of the processing pipeline are sharing quantum data, distributed QEC will generally be necessary to protect the computation in transit. This necessarily introduces an (efficient) overhead in the number of physical qubits being communicated across the network, introducing additional bandwidth costs, which must be accommodated for in networking strategies.

\subsubsection{Quantum enhancement}

The attractive feature of quantum computing is the potentially exponential improvement in algorithmic performance of certain tasks over their classical counterparts as the size of the computer grows. This exponential relationship implies that computation in general no longer has a simple linear tradeoff as the number of participating nodes increases. In Sec.~\ref{sec:comp_sc_func} we quantify this via so-called \textit{computational scaling functions}\index{Computational!Scaling functions} and study its economic implications in detail.

But not every effort at distributed quantum computation will automatically exhibit the holy grail of exponential speedup. The architecture and algorithm to which it is applied must be thoughtfully designed to fully exploit the computers' quantum power. A simple adaptation of in-series or in-parallel computation may not achieve this. Rather, we must cunningly exploit quantum entanglement between nodes to perform truly distributed computation, in the sense that no instance of an algorithm is uniquely associated with any given node, but is rather represented collectively across all of them.

Let us assume that we have such a carefully constructed distributed platform. Let $t_c$ be the time required by a classical algorithm to solve a given problem, and $t_q$ the time required to solve the same problem using a quantum algorithm. In the case of algorithms exhibiting exponential quantum speedup, we will have,
\begin{align}
t_c = O(\exp (t_q)).
\end{align}
If we now increase the quantum processing power (i.e number of nodes or qubits) $k$-fold, the equivalent classical processing time is (in the best case),
\begin{align}
t_c' &= O(\exp (t_q k)) \nonumber \\
&= O(\exp (t_q)^{k}) \nonumber \\
&= O({t_c}^{k}).
\end{align}
Thus, $k$-fold quantum enhancement corresponds to a $k$th-order exponential enhancement in the equivalent classical processing time, which clearly scales much more favourably than the linear $k$-fold enhancement offered by classical parallelisation.

For this scaling to be possible, we expect that nodes will need to communicate via quantum rather than purely classical channels, so as to preserve inter-node entanglement and mediate non-local gates across nodes.

% Quantum MapReduce

\subsubsection{Quantum MapReduce}\index{Quantum MapReduce}\label{sec:quant_map_reduce}

Designing native distributed algorithms is not trivial, and architectural constraints may physically limit the allowed set of inter-node operations available to us. Are there any simple constructions that allow us to achieve this? We will propose an approach to parallelised quantum computation based on a direct quantum adaptation of the classical \textsc{MapReduce} protocol.

\textsc{MapReduce}, originally developed by Google\index{Google} for large-scale parallel processing, is simply an elegant formalism for parallelising classical computations. There are three stages to the protocol:
\begin{enumerate}
	\item \textsc{Map}: a root node\index{Root node} generates $k$ instances of an algorithm, each with different input data (or a different random seed).
	\item \textsc{Execute}: each of the $k$ instances are executed independently on the $k$ nodes in parallel.
	\item \textsc{Reduce}: all outputs are returned to the root node, collated and combined together according to some algorithm, yielding the final output of the computation.
\end{enumerate}

Perhaps the simplest illustrative example is to consider the execution of a Monte Carlo simulation\index{Monte-Carlo simulations}. Here we wish to execute a large number of instances of the same problem, each with a different random seed, and average the results to yield a statistical outcome. Here the \textsc{Map} algorithm simply delegates out $k$ copies of the same algorithm, assigning each node a different random seed\index{Random!Seed}, and the \textsc{Reduce} algorithm needs only average their outputs. Note that the \textsc{Map} and \textsc{Reduce} algorithms are relatively simple, with the nodes operating in parallel doing all the hardcore number crunching.

\begin{figure}
\includegraphics[clip=true, width=0.475\textwidth]{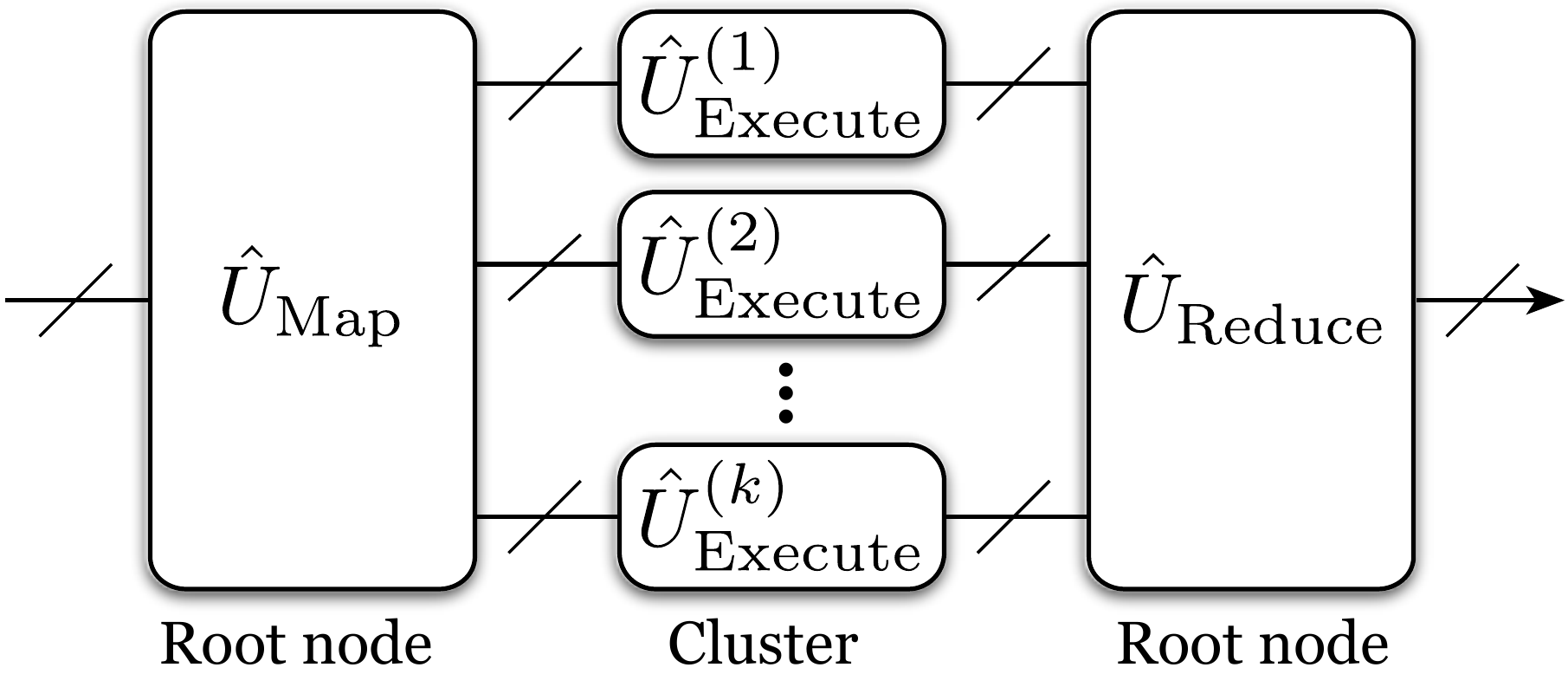}
\captionspacefig \caption{Structure of the quantum \textsc{MapReduce} protocol. All operations are unitary, and the \textsc{Map} and \textsc{Reduce} operations may be entangling in general. The \textsc{Execute} stage is separable into a tensor product of smaller \textsc{Execute} operations that are executed in parallel by the $k$ nodes.}\label{fig:quant_map_red}	
\end{figure}

Taking this model, one might intuitively follow a similar approach for quantum computation, where we simply replace all the operations with unitary processes, and replace the communication links with quantum channels. Now we have a model as shown in Fig.~\ref{fig:quant_map_red}.

The goal in this construction is to make the \textsc{Map} and \textsc{Reduce} operations be relatively very simple, e.g have low circuit depth, while the \textsc{Execute} operations are more challenging to implement. Note that the \textsc{Map} and \textsc{Reduce} operations are now unitary processes, rather than being, for example, simple classical dispatch and collate operations. This means that in general the \textsc{Map} operation will prepare entanglement between the \textsc{Execute} sub-computations, and \textsc{Reduce} might similarly implement non-separable entangling measurements to measure collective properties of the joint system.

This architecture is merely a direct mapping of classical \textsc{MapReduce} to the quantum setting. How might it be used? Consider quantum simulation\index{Quantum simulation}, where we aim to simulate a Hamiltonian of the form,
\begin{align}
\hat{H}_\mathrm{total} = \sum_i \hat{H}_i,	
\end{align}
where each of the $\hat{H}_i$ terms are local Hamiltonians acting on orthogonal Hilbert spaces. This implies that all terms commute,
\begin{align}
[\hat{H}_i,\hat{H}_j]=0,
\end{align}
and therefore the unitaries they generate,
\begin{align}
	\hat{U}_j=e^{-\frac{i\hat{H}_jt}{\hbar}},
\end{align}
have a separable tensor product structure,
\begin{align}
	\hat{U}_\mathrm{total}=\bigotimes_i \hat{U}_i.
\end{align}
This separability lends itself directly to the tensor product structure of the \textsc{Execute} unitaries. The \textsc{Map} operation could now be a stage for preparing entangled initial states (entangled across the different subsystems), and the \textsc{Reduce} operation might perform collective measurements or sampling.

% Distributed quantum search algorithm

\subsubsection{Distributed quantum search algorithm}\index{Distributed quantum search algorithm}

The Grover quantum search algorithm (Sec.~\ref{sec:quantum_search}) can be easily parallelised\index{Parallelisation} by partitioning the search space\index{Search space!Partitioning}, and allocating a different partition to each node.

Suppose we wish to search over the $N$-bit space $x$, to find a satisfying solution to some oracle\index{Oracles} function (e.g when solving an \textbf{NP}-complete problem\index{NP \& NP-complete}),
\begin{align}
x\,\, \mathrm{s.t.}\, f(x)=1.
\end{align}
Let there be $M$ nodes available for computation, where for simplicity we assume $M$ is a power of 2 (although the idea works generally for arbitrary $M$, albeit not as mathematically elegantly). We designate each of the nodes a $\log_2 M$-bit identification number\index{Identification numbers},
\begin{align}
	y=[0,M-1].
\end{align}
We now program each node to search over a smaller search space $x'$, which is \mbox{$N-\log_2 M$} bits in length, concatenated with the node's identification number to produce the full range of $x$. The input to each instance of the oracle is now,
\begin{align}
x = x'\frown y,
\end{align}
where `$\frown$' denotes binary string concatenation.

\begin{figure}[!htbp]
	\includegraphics[clip=true, width=0.475\textwidth]{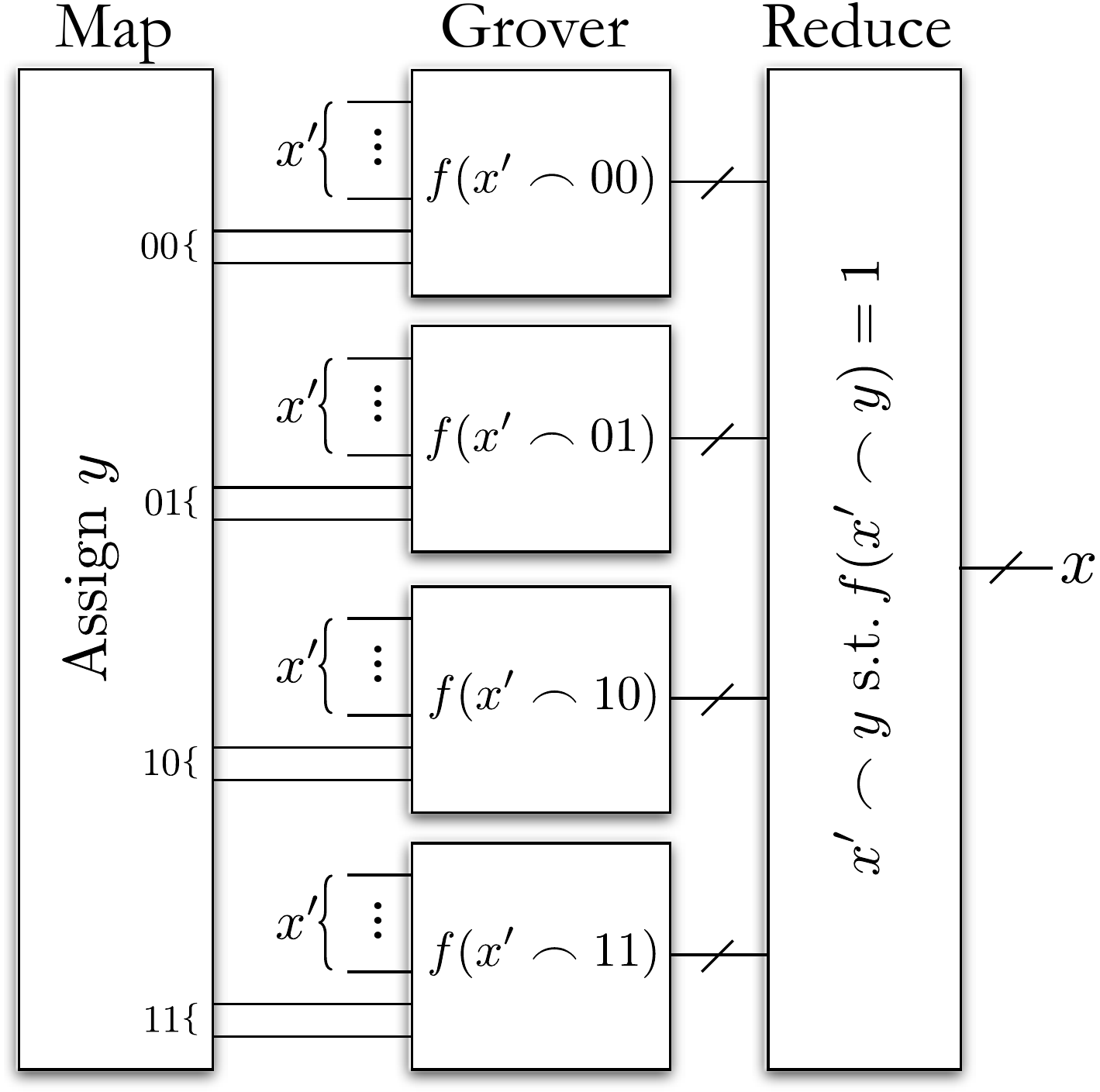}
	\captionspacefig \caption{Example of a quantum \textsc{MapReduce} protocol for implementing a distributed quantum search over four nodes. Each node performs a quantum search over the reduced space $x'$, concatenated with the identification number of the node, which recovers the full search-space $x$ across all the nodes collectively. This effectively partitions and allocates the search-space across the nodes, which implement their reduced searches in parallel. The net speedup provided by $M$ nodes operating in parallel scales as $O(\sqrt{M})$.}\label{fig:distributed_search}\index{Distributed quantum search algorithm}
\end{figure}

For example, with four nodes the 2-bit identification numbers are, 
\begin{align}
	y=\{00,01,10,11\}.
\end{align}
If the input search space is $N$-bits in length, then each of the nodes are assigned the search-space \mbox{$x'\frown y$}, where $x'$ is an \mbox{$N-2$}-bit number. Within each instance, the Grover search searches over only the reduced space $x'$, with $y$ a constant of the instance. Fig.~\ref{fig:distributed_search} illustrates the circuit schematic for the simple \mbox{$M=4$} example.

It can easily be seen that this approach is compatible with the general quantum \textsc{MapReduce} formalism (Sec.~\ref{sec:quant_map_reduce}), where the \textsc{Map} function assigns the partitions denoted by the node identification numbers $y$, the \textsc{Execute} functions implement the reduced searches associated with each instance, and the \textsc{Reduce} function collects satisfying arguments from the instances,
\begin{align}
	x'\,\, \mathrm{s.t.}\, f(x'\frown y)=1\,\,\forall\, y.
\end{align}

From the runtime of the Grover algorithm, it follows that the time required to solve the search problem on the initial full search-space is $O(\sqrt{2^N})$, whereas the time required in the parallelised implementation is only $O(\sqrt{2^{N-\log_2 M}})$. Thus, the need speedup is,
\begin{align}
	O\left(\frac{\sqrt{2^N}}{\sqrt{2^{N-\log_2 M}}}\right) = O(\sqrt{M}).
\end{align}

Evidently, the net computational speedup scales as a factor of the square root of the number of nodes in the parallelised implementation. Note that this approach does not exploit entanglement between nodes, and does not offer a `quantum' (i.e super-polynomial) speedup, since it's really just brute-force partitioning of a problem into smaller, quicker, bite-sized chunks that are attacked completely independently of one another, much like classical parallelisation.

\begin{figure}[!htpb]
	\includegraphics[clip=true, width=0.475\textwidth]{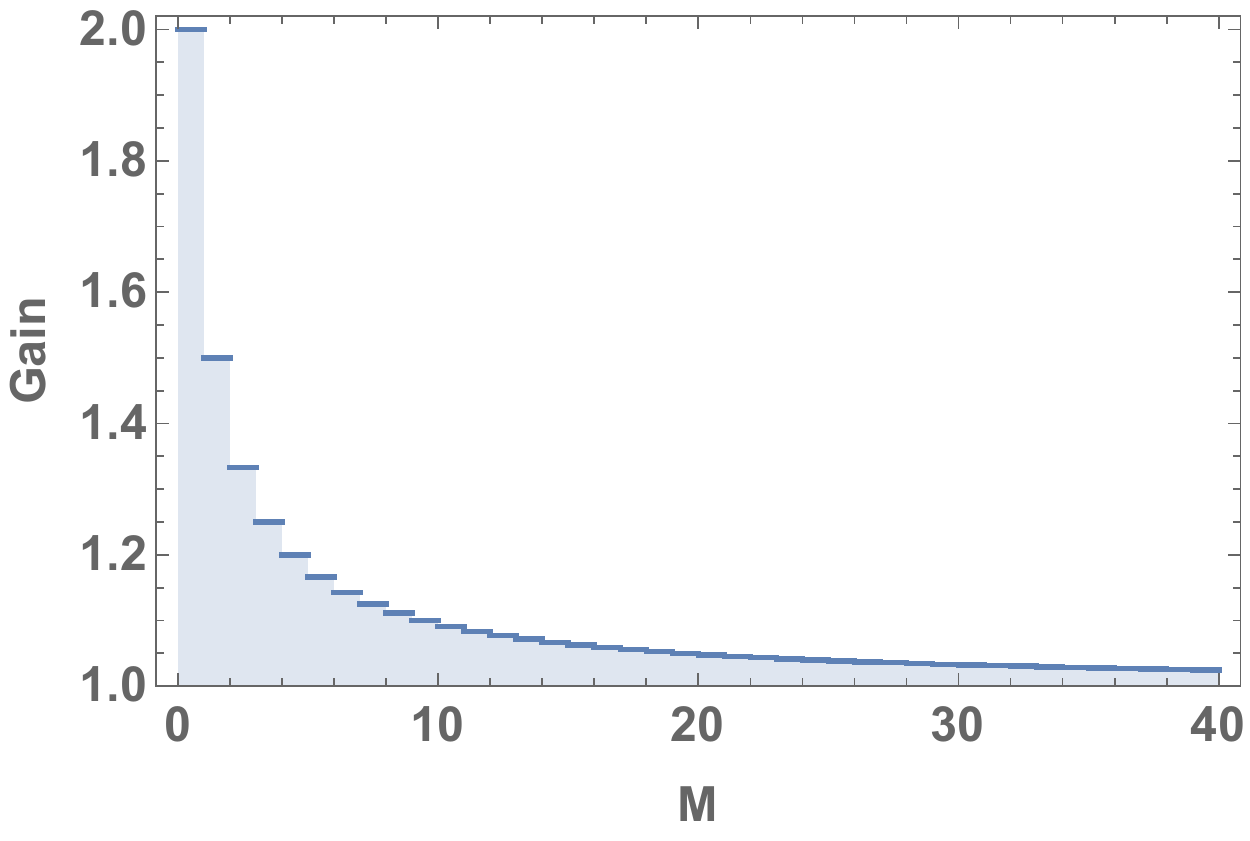}
	\captionspacefig \caption{Computational gain from adding a single extra node to a parallelised implementation of the quantum search algorithm. Asymptotically, the computational benefit vanishes.}\label{fig:dist_quant_search}
\end{figure}

To the contrary, unlike most quantum algorithms, whose power grows exponentially with the number of qubits (increasing returns\index{Increasing returns}), the distributed quantum search algorithm exhibits diminishing returns\index{Diminishing returns} with the degree of parallelisation -- the computational gain from adding one additional node to the network scales as,
\begin{align}
G=\sqrt{\frac{M+1}{M}},
\end{align}
shown in Fig.~\ref{fig:dist_quant_search}, which in the large $M$ limit asymptotes to,
\begin{align}
\lim_{M\to\infty} \sqrt{\frac{M+1}{M}} = 1.
\end{align}
That is, increasing the number of nodes from 1 to 2 has far greater net gain than increasing them from 100 to 101. In fact, asymptotically, the gain from adding an additional node vanishes in the limit of a high degree of parallelisation.

For this reason, parallelised implementation of a quantum search is not an example of a distributed quantum computation which achieves exponential gain with the addition of new nodes (i.e qubits). Rather, for this specific application it's far more optimal to consolidate quantum resources into a single larger instance of a quantum search algorithm than using the quantum \textsc{MapReduce}\index{Quantum MapReduce} architecture to parallelise it.

% Distributed unitary error averaging

\subsubsection{Distributed unitary error averaging}\index{Unitary!Error averaging}\index{Distributed unitary error averaging}\label{sec:error_av_parallel}

\begin{figure}[!htbp]
	\includegraphics[clip=true, width=0.475\textwidth]{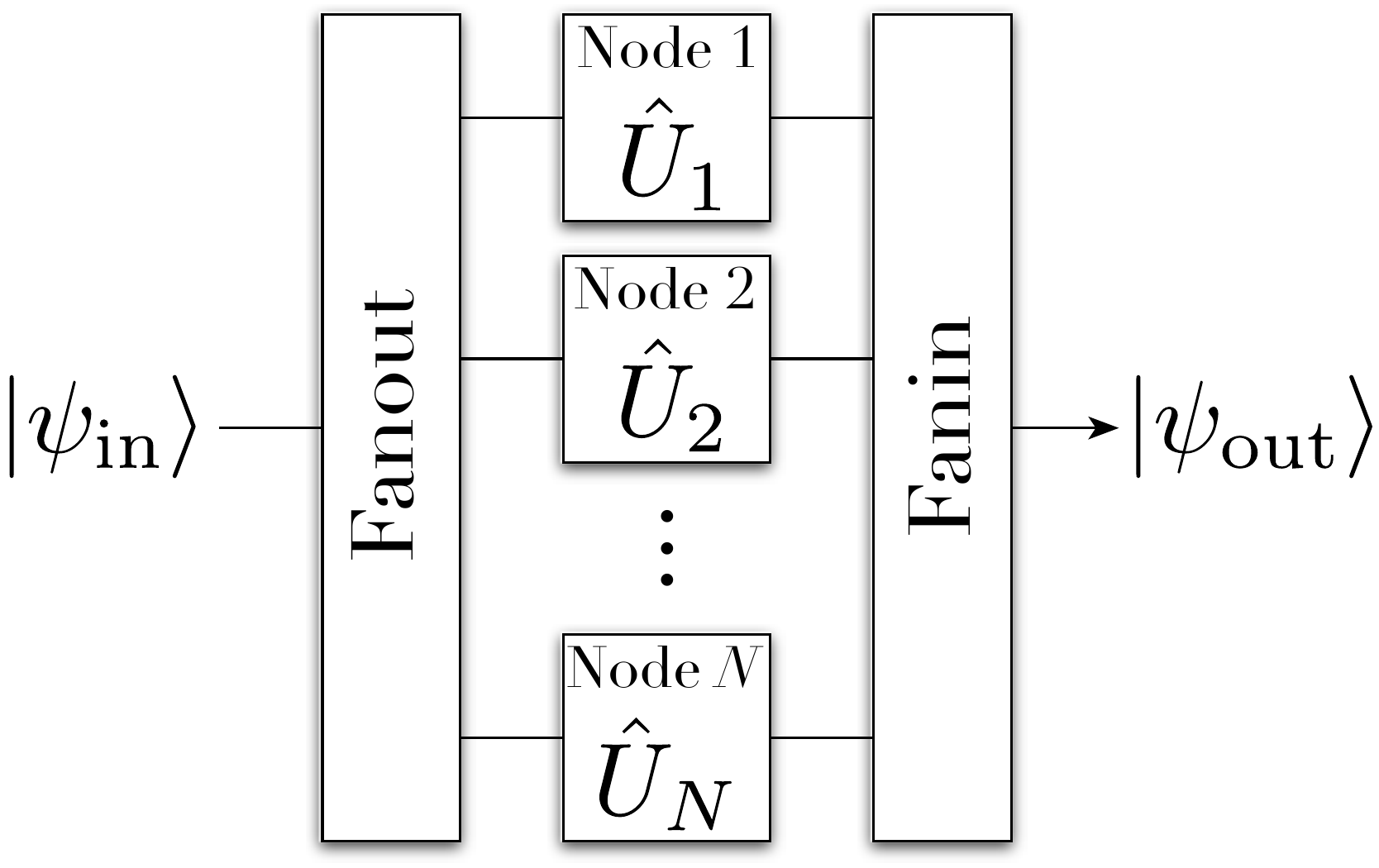}
	\captionspacefig \caption{Representing the unitary error averaging technique for error-correcting passive linear optics in parallelised form, consistent with the \textsc{Quantum MapReduce} structure.}\label{fig:error_av_map_reduce}
\end{figure}

In Sec.~\ref{sec:error_averaging} we introduced the unitary error averaging technique for minimising the errors associated with imperfect implementation of linear optics beamsplitter networks. This model is naturally of the form of \textsc{Quantum MapReduce}, where the \textsc{Map} and \textsc{Reduce} operations implement the fan-in and fan-out respectively, and the independent instances of the noisy unitary are executed in parallel on different nodes, as shown in Fig.~\ref{fig:error_av_map_reduce}.

Now the purpose of the parallelisation is not for computational gain, but rather for error minimisation. The more nodes involved in the parallelised execution, the smaller the final error rate.

\subsubsection{Delocalised computation}\index{Delocalised computation}

The cluster state (Sec.~\ref{sec:CSQC})\index{Cluster states}, topological code (Sec.~\ref{sec:surface_codes})\index{Topological!Codes} and quantum random walk (Sec.~\ref{sec:QW})\index{Quantum random walks} models for quantum computation may find themselves to be particularly well-suited to distributed implementation, since they naturally reside on graphs, whose nodes needn't be held locally by a single user, but could instead be shared across multiple hosts with the ability for graph nodes to intercommunicate. Then only classical communication is required to complete a computation and the quantum information is not localised to any particular node.

\begin{figure}[!htbp]
\includegraphics[clip=true, width=0.475\textwidth]{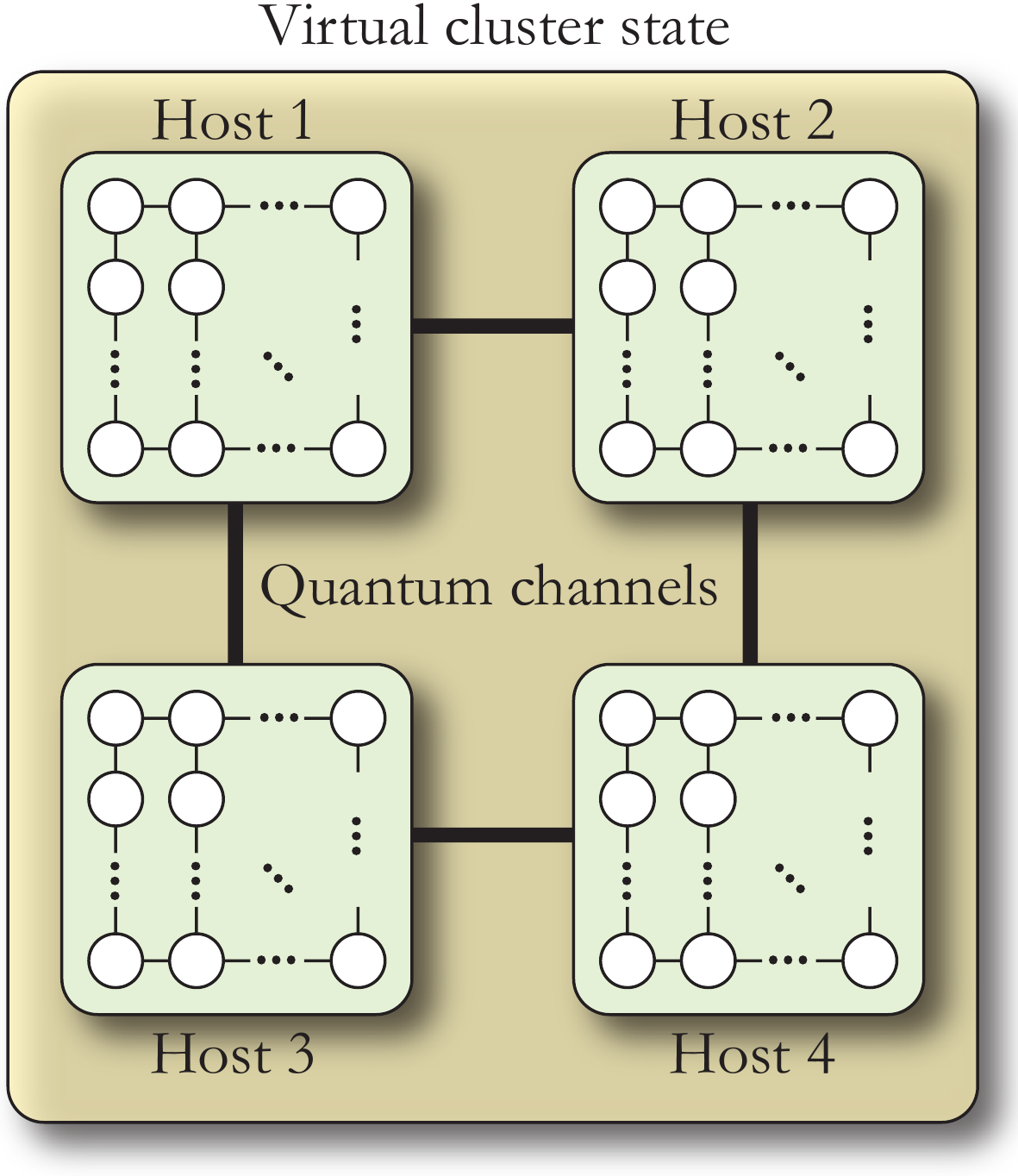} 
\captionspacefig \caption{Approach for constructing distributed cluster states (or topological codes, or other graph states) across multiple nodes. The quantum channels allow neighbouring clusters in the topology to be fused together, constructing a large virtual cluster state for distributed computation. The nodes could be arbitrarily separated with optically-mediated interconnects to enable fusing nodes together.} \label{fig:patchwork_cluster}
\end{figure}

Additionally, the entangling gates which build cluster states all commute and may be implemented simultaneously in parallel. This enables a distributed cluster state to be constructed in a `patchwork' fashion, as shown in Fig.~\ref{fig:patchwork_cluster}. Now the computation is truly distributed in the sense that the computation resides collectively across the distributed cluster state, held by any number of users. No instance of an algorithm can be uniquely associated with any given node.

This approach overlaps with the modularised approach for quantum computation discussed in the upcoming Sec.~\ref{sec:module}, the difference being that in distributed cluster states the goal is to delocalise computations due to resource constraints, whereas for modularised computation the motivation is largely economical, driven by economy of scale.

%
% Delegated Quantum Computation
%

\subsection{Delegated quantum computation} 
\index{Delegated!Quantum computation}

Taking the notions of outsourced and distributed quantum computation to the logical extreme, we can envisage the situation where Alice has no quantum resources whatsoever (state preparation, evolution or measurement), but knows exactly what the processing pipeline should entail, and who on the network has the different required quantum resources. We refer to this as \textit{delegated quantum computation}, where the entire processing pipeline is outsourced to a series of hosts.

To illustrate this, let us consider a simple example -- cat state quantum computation (Sec.~\ref{sec:cat_enc}). There are three main elements to the protocol:
\begin{enumerate}
\item Cat state preparation.
\item Post-selected linear optics with feedforward.
\item Continuous-variable measurement.
\end{enumerate}

\begin{figure}[!htbp]
\includegraphics[clip=true, width=0.475\textwidth]{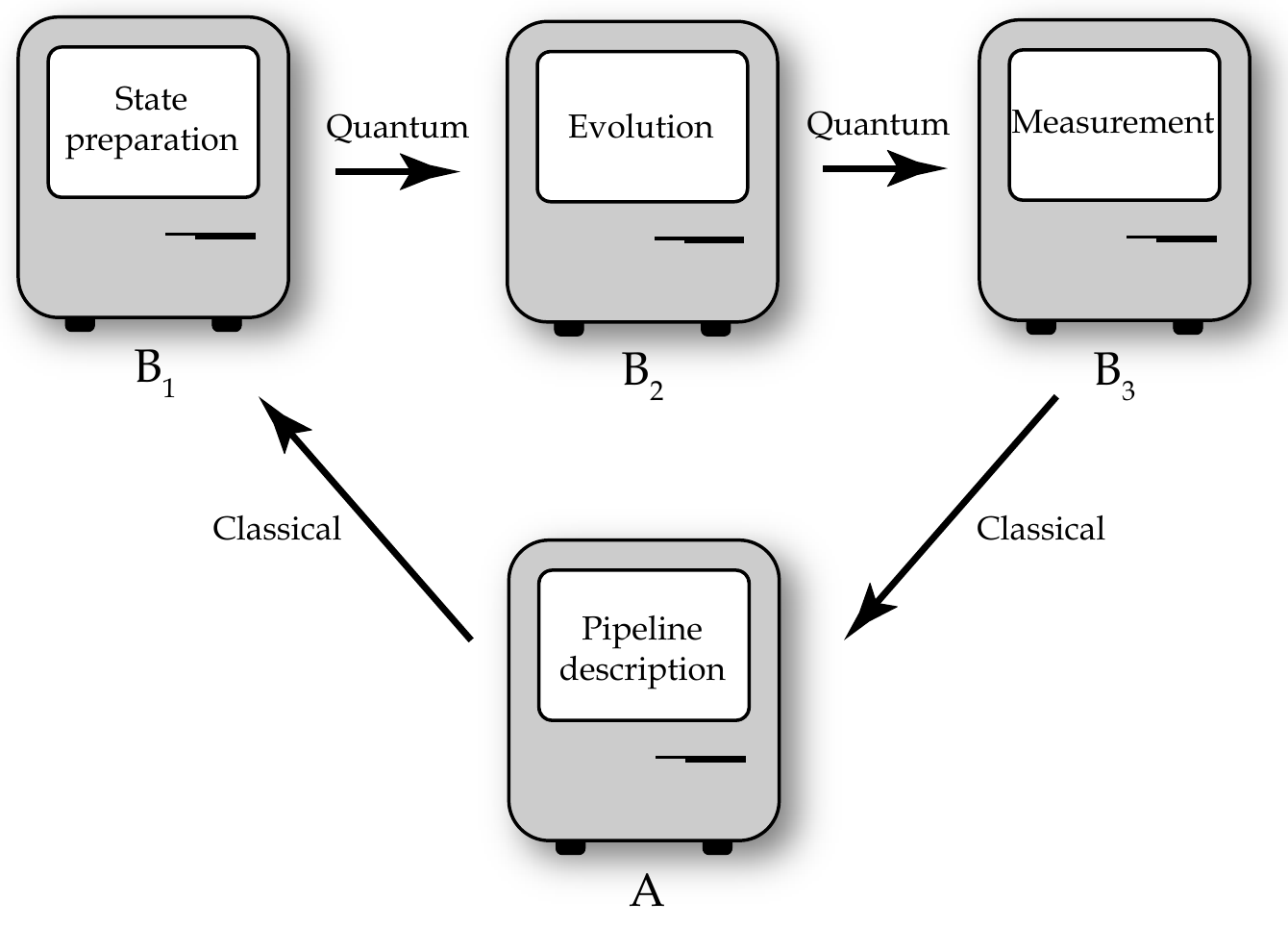}
\captionspacefig \caption{Delegated quantum computation, where each of the three computational stages (state preparation, evolution and measurement) are outsourced to the cloud without intermittent interaction with the client, $A$. $A$ provides only a classical description of the processing pipeline to be implemented, each stage of which is delegated to a server specialised in that particular task. Thus, the total processing pipeline takes the form \mbox{$A\to B_1\to B_2\to B_3\to A$}, where \mbox{$A\to B_1$} and \mbox{$B_3\to A$} are classical, and \mbox{$B_1\to B_2$} and \mbox{$B_2\to B_3$} are quantum channels.} \label{fig:delegated}
\end{figure}

Each of these stages present their own technological challenges, sufficiently challenging that one might wish to outsource all three stages. However, suppose there is no single host on the network with the ability to perform all three, but rather there are three hosts ($B_1$, $B_2$ and $B_3$), each specialising in just one of those tasks. In this instance, it would be most resource savvy for the network to implement the pipeline,
\begin{align}
	A\to B_1\to B_2\to B_3\to A,
\end{align}
without going back and forth to Alice after each step,
\begin{align}
	A\to B_1\to A\to B_2 \to A\to B_3\to A.
\end{align}
In fact, it may not even be technologically possible to implement back-and-forth to Alice if she has no capacity for handling quantum resources (i.e the \mbox{$A\leftrightarrow B$} stages are purely classical). An example of such a pipeline is shown in Fig.~\ref{fig:delegated}.

This can be achieved by adding a \textsc{Pipeline} field to the packet header prepared by Alice -- a FIFO queue describing the entire processing pipeline that Alice's packet (which initially contains only classical data) ought to follow through the network. Following completion of each stage of the pipeline we pop the stack and transmit the packet to the next specified host. Only at the very completion of the protocol is a packet (containing only classical data) returned to Alice.

Another good case study is quantum metrology using NOON states (Secs.~\ref{sec:NOON} \& \ref{sec:metrology}) for achieving Heisenberg limited precision. Preparing NOON states is extremely challenging, and additionally Alice may not possess the unknown phase to be measured, but rather wishes a NOON state, prepared by $B_1$, to be provided to a third-party, $B_2$, who applies the unknown phase, and passes the resulting state to $B_3$, who implements the required high-efficiency parity measurements required to complete the protocol. In this case, the pipeline would take the same form as above, again with no back-and-forth communication to Alice.

Such delegated protocols will be very useful in quantum networks, where different hosts specialise in different tasks (which may be the most economically efficient model), but poor old Alice specialises in none of them, despite knowing exactly what needs to be done. This would allow an aspiring undergraduate student, who is poor (aren't they all?), to sit in his bedroom at his classical PC, and implement entire distributed quantum information processing protocols in the cloud, with no quantum resources or interactions whatsoever.

%
% Modularised Quantum Computation
%

\subsection{Modularised quantum computation} \label{sec:module} \index{Modularised quantum computation}

How does one build a large-scale quantum computer, given the extremely daunting technological requirements and high costs? In any industry, economies of scale allow the mass production, and rapid reduction in price of technology. To achieve this, we must find a way to make quantum technologies commodity items, which avoid all the hassle of customised cutting-edge labs. What we really desire is production-line `Lego for Adults{\texttrademark}', allowing ad hoc connection of \textit{modules}, which implement small subsections of a larger computation \cite{bib:FowlerPrivate}.

\begin{figure}[!htbp]
	\includegraphics[clip=true, width=0.475\textwidth]{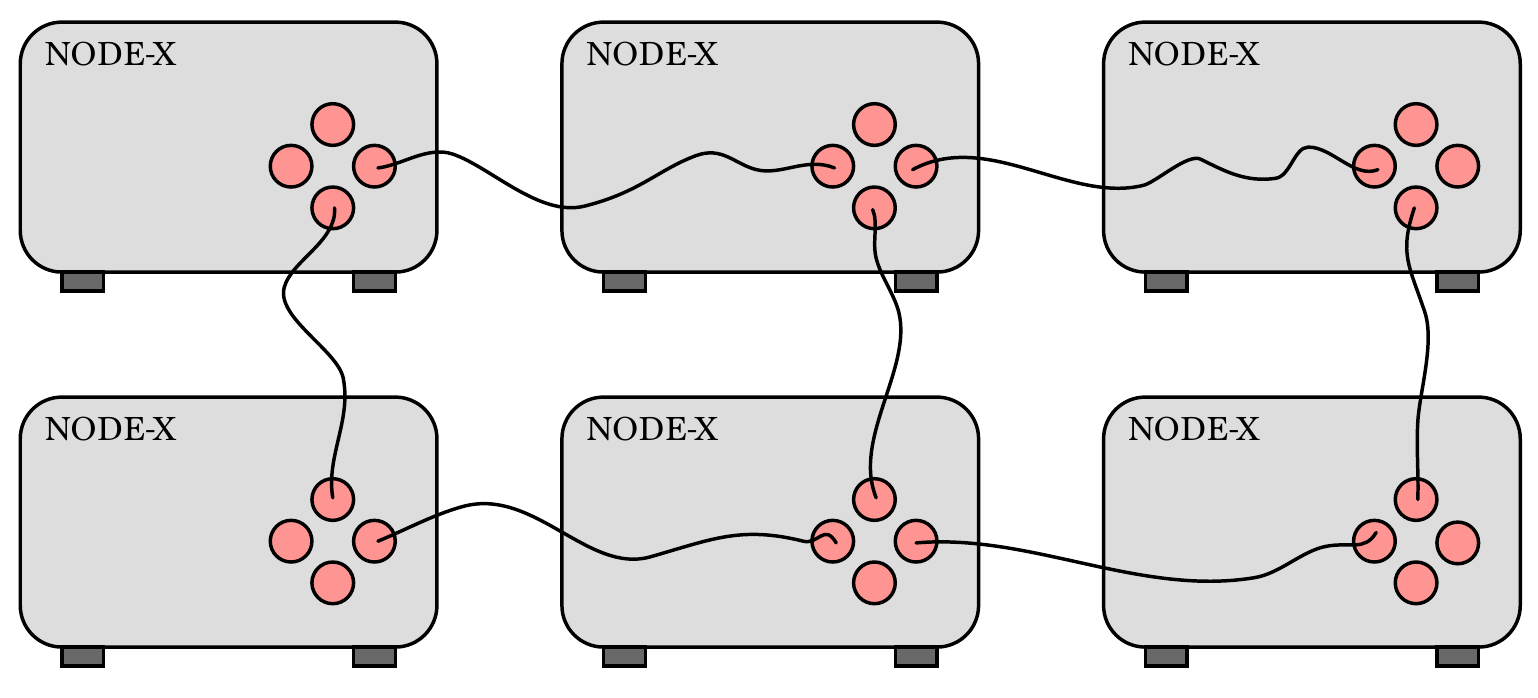}
	\captionspacefig \caption{A possible physical realisation of commercially produced quantum modules\index{Node-X}, forming a \mbox{$2\times 3$} patchwork of cluster states. Each hosts a relatively small number of qubits. The nodes each have four optical interconnects, which are used to connect the modules via optical fibre. Entangling operations performed on photons shared via the interconnects create inter-module entanglement links, yielding a distributed virtual quantum computer with far more qubits. The computation is truly distributed and cooperative, in the sense that the entire computation is non-local, instead being collectively distributed across all the nodes, which coordinate their local operations via only classical communication. An alternate implementation is to replace the inter-node quantum links with Bell pair distributers. Then entanglement swapping can be employed to swap the entanglement into a link between nodes.}\label{fig:modules_physical}
\end{figure}

We envisage that physically, a module is a black box with optical interconnects, that may be interconnected to form an arbitrary topology, yielding a physical platform as shown in Fig.~\ref{fig:modules_physical}. The user remains oblivious to the inner workings of the modules. The modules could all be identical, just patched together differently, paving the way for their mass production, and an associated quantum equivalent of Moore's Law, allowing them to become off-the-shelf commodity items over time. Then the cost of a quantum computer would simply scale linearly with its number of qubits.

The modules forming a particular computation could either be all owned by a single well-resourced operator, or alternately might be shared across multiple hosts, who network them remotely using EOs\index{Entangling operations} between emitted photons.

In Sec.~\ref{sec:dist_QC} we introduced the notion of distributed quantum computation. There the motivation was to enable a computation to be distributed across multiple servers, which either parallelise computation or process it as a pipeline in series.

An alternate direction, for economic reasons, is that it is unviable for a single server to host an entire computation. Rather, hosts will have limited capability, and performing large-scale computations will require employing a potentially large number of hosts cooperating and sharing resources with one another\footnote{Even some present-day massive-scale data processing and storage protocols are implemented virtually across multiple large-scale data-centres, which, for example, automatically handle geographically decentralised data redundancy and processing. Google and Amazon, for example, provide cloud services for this purpose, employed both internally, and licensed out to third parties, and the Apache Cassandra\index{Apache Cassandra project} project provides an open-source equivalent. The key is for the underlying protocol to abstract this away from the user, such that they interface with the data as though it were a local asset.}. This can be regarded as the most general incarnation of distributed computation.

This is not the same motivation as for in-series computation, where different servers in the pipeline have different proprietary algorithms as subroutines of a larger computation. And it also differs from in-parallel computation, where multiple servers implement the same algorithm on different data, which is subsequently merged by a root node, as per, for example, a \textsc{MapReduce}-style protocol.

Instead, the motivation is one of economics. First, individual servers will have finite resources, but there may be many of them, which can be networked to cooperatively implement a larger algorithm virtually. Second, because the modules in the architecture are identical and lend themselves to mass production, one can expect more favourable economics than that offered by a provider who sells full-fledged, customised quantum computers, which do not lend themselves to the same level of mass production.

The concept of this model is best explained using the optical cluster state formalism (Sec.~\ref{sec:CSQC}), which lends itself naturally to this approach. A rectangular lattice graph is sufficient for universal quantum computation, even if the cluster state graph is not local (but classical communication between nodes is allowed).

Let us first assume that we wish to construct a cluster state with $n_\mathrm{logical}$ logical qubits. We additionally allow each logical qubit to be the root node of a graph with a $+$-structure, where each branch comprises a chain of $n_\mathrm{ancilla}$ ancillary physical qubits. These are sometimes referred to as \textit{micro-clusters} \cite{bib:Nielsen04}\index{Micro-cluster states}. A single micro-cluster collectively forms a single \textit{module} in the topology. Our goal is to fuse modules via nearest neighbour entanglement to build up the desired distributed cluster state.

\if 1\doublecol
\begin{figure}[!htbp]
	\includegraphics[clip=true, width=0.4\textwidth]{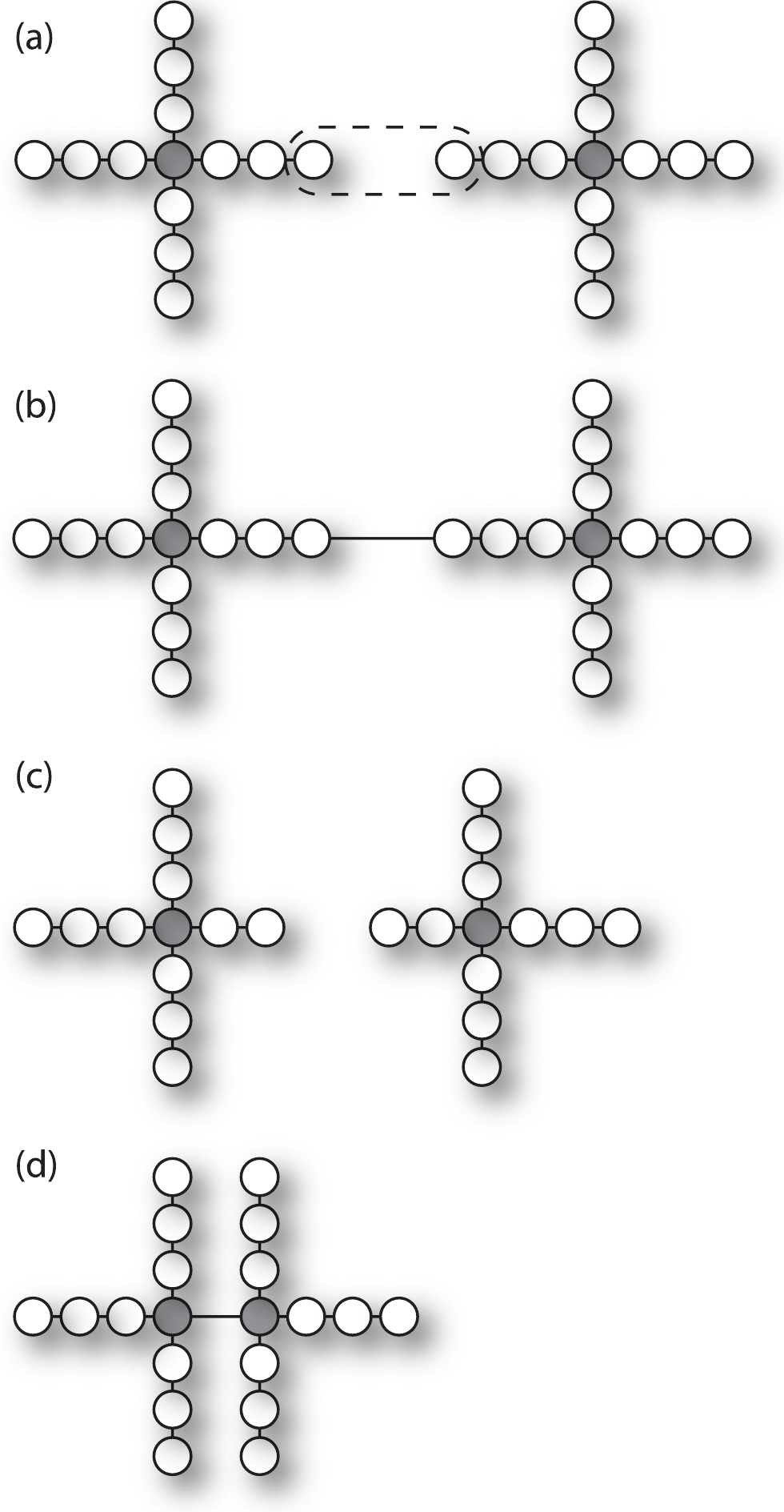}
	\captionspacefig \caption{Several cluster state identities for modularised quantum computation. (a) Two cluster states with a $+$-topology are fused together using an EO (dashed). (b) Upon success, an edge is created between the respective qubits. (c) Upon failure, both qubits are effectively measured in the $\hat{Z}$ basis, thereby removing them, and any associated edges, from the graph. (d) Following a successful EO, the unwanted ancillary qubits may be eliminated using measurements in the $\hat{Y}$ basis, creating edges between their neighbours. If the grey qubits represent the desired logical qubits, this can be used to remove the remainder of the branches emanating from them, thereby distilling the irregular graph down to a regular lattice.} \label{fig:plus_cluster_ident}
	\end{figure}
\else
	\begin{figure*}[!htbp]
	\includegraphics[clip=true, width=\textwidth]{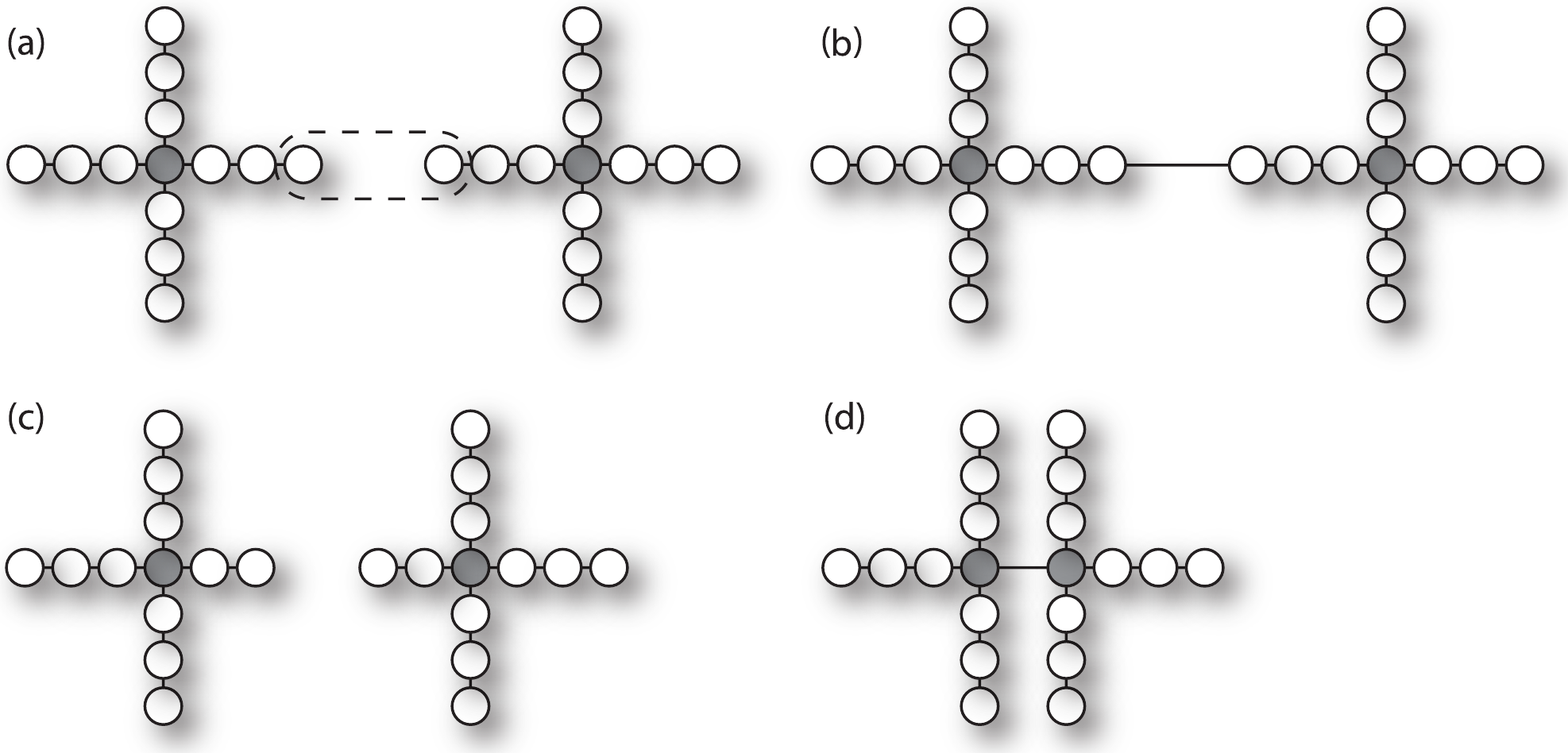}
	\captionspacefig \caption{Several cluster state identities for modularised quantum computation. (a) Two cluster states with a $+$-topology are fused together using an EO (dashed). (b) Upon success, an edge is created between the respective qubits. (c) Upon failure, both qubits are effectively measured in the $\hat{Z}$ basis, thereby removing them, and any associated edges, from the graph. (d) Following a successful EO, the unwanted ancillary qubits may be eliminated using measurements in the $\hat{Y}$ basis, creating edges between their neighbours. If the grey qubits represent the desired logical qubits, this can be used to remove the remainder of the branches emanating from them, thereby distilling the irregular graph down to a regular lattice.} \label{fig:plus_cluster_ident}
	\end{figure*}
\fi

We arrange the modules to internally represent a $+$-topology where each node has neighbouring branches in each of the up/down/left/right directions. But we imagine the situation whereby each logical qubit, along with its respective ancillary branches, is held by a different server. Thus, the final cluster state is truly decentralised across all the servers, and in general entire computations cannot be performed locally.

Using the ancillary states in the respective directions, we attempt to fuse neighbouring clusters using EOs, such as CZ gates (e.g a KLM CZ gate), linear optics \textit{fusion gates} (i.e rotated polarising beamsplitters followed by photo-detection, implementing which-path erasure\index{Which-path erasure}) \cite{bib:BrowneRudolph05}, or atoms with a $\lambda$-configuration coupled to photons \cite{bib:BarrettKok05}, which undergo which-path erasure (Sec.~\ref{sec:hybrid}). Importantly, using the fusion gate and which-path erasure approaches, only a single beamsplitter is required to perform the EO, which only necessitates high-visibility HOM interference, mitigating the need for far more challenging interferometric (MZ) stability (Sec.~\ref{sec:opt_stab}). This is delightful, as current leading quantum optics experiments routinely achieve HOM visibilities well in excess of 99\%.

An alternate fusion strategy is not to directly communicate qubits to be bonded, but instead rely off Bell pairs provided by a central authority. Each party then applies an EO between their half of the Bell pair and their target module qubit, which swaps the Bell pair entanglement onto the two respective module qubits (Sec.~\ref{sec:swapping}).

When an EO is successful, we have fused two modules together, albeit potentially with some leftover ancillary states between the logical qubits. When it fails, we have lost the respective ancillary states, and we attempt again using the next ancillary qubits in each of the the respective branches -- a kind of \textsc{Repeat Until Success} strategy. The bonding only fails if all $n_\mathrm{ancilla}$ EOs fail.

Note, however, that longer ancillary arms provide more opportunity for errors to accumulate \cite{bib:RohdeRalphMunro07}. Thus, despite its tolerance against gate failure, it is nonetheless highly desirable for EOs to be as deterministic as possible, so as to minimise the required number of ancillary qubits.

Upon successful bonding, any remaining ancillary qubits between the respective logical qubits are measured in the $\hat{Y}$ basis to remove them from the graph, whilst connecting their neighbours, leaving the two respective logical qubits as nearest neighbours in the graph. Now each module contains exactly one logical qubit, connected as desired to neighbouring modules. The relevant identities are shown in Fig.~\ref{fig:plus_cluster_ident}. Our goal is for the entire graph to have a lattice structure, once ancillary qubits have been measured out, as illustrated in Fig.~\ref{fig:module}.

\if 1\doublecol
\begin{figure}[!htbp]
\includegraphics[clip=true, width=0.35\textwidth]{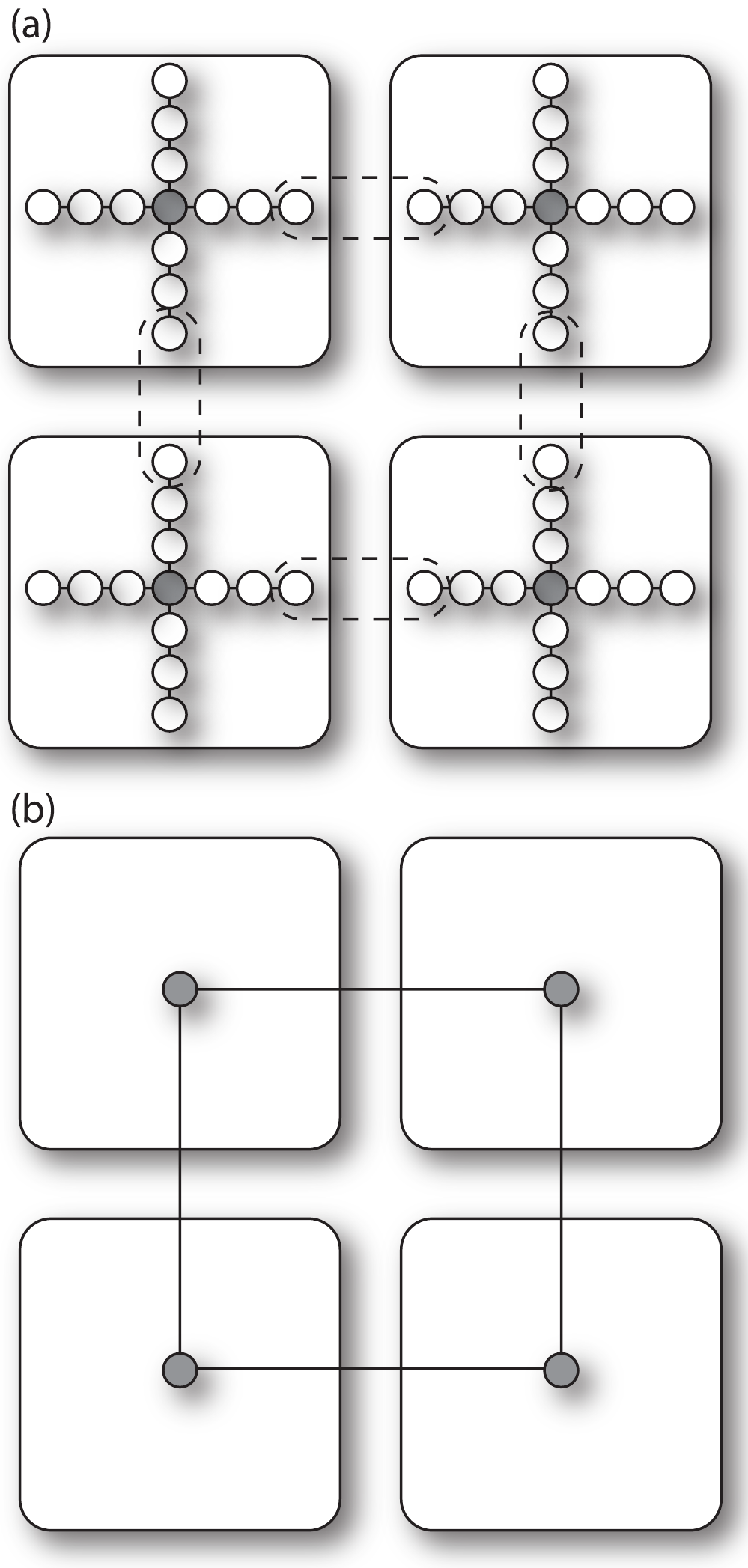}
\captionspacefig \caption{The modularised approach to scalable and economically efficient, distributed quantum computation using cluster states. The modules are all identical, and can be arbitrarily patched to one another, allowing the construction of arbitrary graph topologies. Because the modules are all identical, one might hope that mass production and economy of scale will drive down the cost of modules. We consider a simple \mbox{$2\times 2$} case where each module (rounded rectangles) comprises a single logical qubit (centre of each module in grey) and a number of ancillary qubits (white in each module), which facilitate bonding the logical qubits of nearest neighbours. The preparation of the modules is performed via nearest neighbour EOs (dashed ellipses), beginning at the end of branches, and working towards the root node upon each failure, until (hopefully) an EO is successful. (a) A \mbox{$2\times 2$} lattice of modules with their respective ancillary qubits. We attempt to bond the endpoints of chains using EOs. (b) Upon measuring the remaining ancillary qubits in the $\hat{Y}$ basis, only the logical qubits remain, with nearest neighbour bonds between adjacent modules, creating a distributed cluster state.} \label{fig:module}
\end{figure}
\else
\begin{figure*}[!htbp]
\includegraphics[clip=true, width=\textwidth]{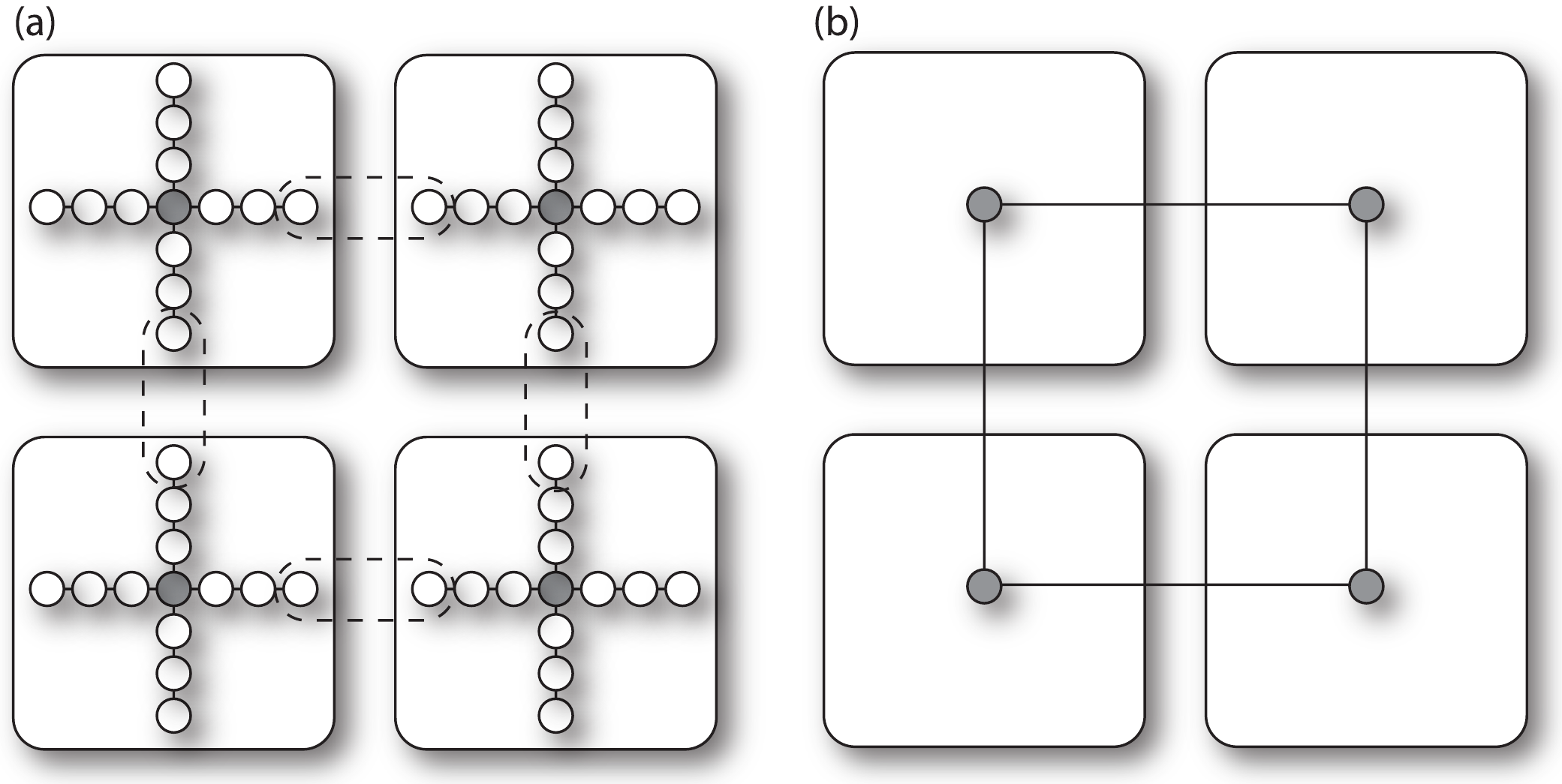}
\captionspacefig \caption{The modularised approach to scalable and economically efficient, distributed quantum computation using cluster states. The modules are all identical, and can be arbitrarily patched to one another, allowing the construction of arbitrary graph topologies. Because the modules are all identical, one might hope that mass production and economy of scale will drive down the cost of modules. We consider a simple \mbox{$2\times 2$} case where each module (rounded rectangles) comprises a single logical qubit (centre of each module in grey) and a number of ancillary qubits (white in each module), which facilitate bonding the logical qubits of nearest neighbours. The preparation of the modules is performed via nearest neighbour EOs (dashed ellipses), beginning at the end of branches, and working towards the root node upon each failure, until (hopefully) an EO is successful. (a) A \mbox{$2\times 2$} lattice of modules with their respective ancillary qubits. We attempt to bond the endpoints of chains using EOs. (b) Upon measuring the remaining ancillary qubits in the $\hat{Y}$ basis, only the logical qubits remain, with nearest neighbour bonds between adjacent modules, creating a distributed cluster state.} \label{fig:module}
\end{figure*}
\fi

This approach has been shown to be resource-efficient \cite{bib:YoranReznik03, bib:Nielsen04}. Let us perform a rudimentary analysis of the resource scaling of this type of approach. The probability of successfully creating an edge between two modules is,
\begin{align}
p_\mathrm{success} = 1 - {p_\mathrm{failure}}^{n_\mathrm{ancilla}},
\end{align}
where $p_\mathrm{success}$ is the probability of joining two modules, $p_\mathrm{failure}$ is the probability that a single EO fails, and $n_\mathrm{ancilla}$ is the number of ancillary qubits per chain. $p_\mathrm{success}$ can be made arbitrarily close to unity with sufficiently long ancillary chains, the required length of whom scales as,
\begin{align}
n_\mathrm{ancilla} = \frac{\log (1-p_\mathrm{success})}{\log (p_\mathrm{failure})}.
\end{align}

Now, for simplicity we will consider the preparation of linear cluster states, although these ideas can easily be extended to more complex topologies, such as 2D lattice graphs.

Let us assume we have a `primary' linear topology of modules, which we will incrementally attempt to `grow' by tacking on new modules to the end. When we do so, with probability $p_\mathrm{success}$ we grow the length of the primary by 1, otherwise we decrement it by 1. This proceeds as a random walk, with on average \mbox{$2p_\mathrm{success}-1$} new qubits added to the primary per time-step. Provided this number is positive, i.e \mbox{$p_\mathrm{success}>1/2$}, which can always be achieved with sufficient $n_\mathrm{ancilla}$, the length of the primary grows linearly over time, allowing efficient state preparation.

This is just a very primitive model for preparing linear cluster states, using an equally primitive \textsc{Incremental} strategy for constructing them using non-deterministic gates. As discussed in Sec.~\ref{sec:CSQC}, much further work has been performed on the resource scaling of efficiently preparing cluster states of different graph topologies using different non-deterministic bonding strategies.

\begin{figure}[!htbp]
\includegraphics[clip=true, width=0.3\textwidth]{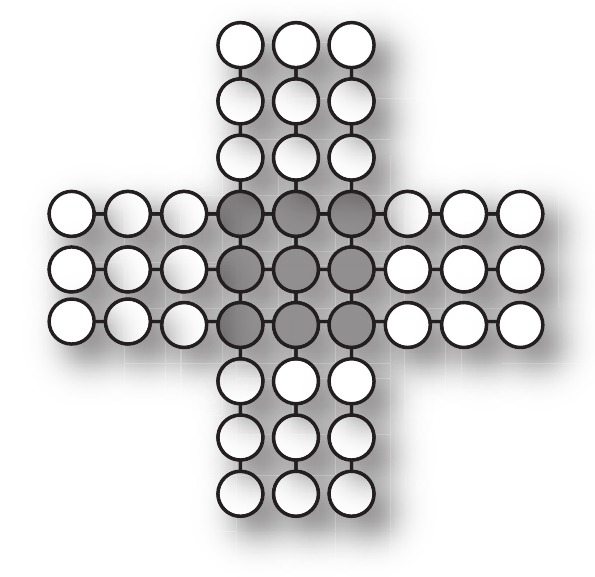}
\captionspacefig \caption{A larger cluster state module comprising a \mbox{$3\times 3$} lattice of logical qubits (grey), and dangling arms of ancillary qubits (white) in each direction for joining them to neighbouring modules. Fusing these modules enables the `patchwork' preparation of large, distributed lattices.} \label{fig:larger_module}
\end{figure}

Of course, we have used the most simple model for modules, where each accommodates a single logical qubit. In due course, we would expect commodity modules to become far more capable, and resource scaling to improve. We might envisage that each module houses a small square lattice of logical qubits, as shown in Fig.~\ref{fig:larger_module}, and the interconnects between them glue them together like a patchwork quilt.

%
% Fundamental Physics Experiments
%

\subsection{Outsourced quantum research} \index{Outsourced!Quantum research}\index{Fundamental physics experiments}

Thus far we have focussed on computation as the key utility for outsourced quantum technologies, and certainly this is likely to be the dominant driving force behind quantum outsourcing\index{Outsourced!Quantum technology}. But of course not everyone wants to only solve complex algorithmic problems. Others may wish to study quantum systems themselves from the perspective of basic science research\index{Basic science research}, or perform precise quantum metrology.

It is foreseeable that in the context of a true quantum internet, there will be a demand for not only the communication of bits and qubits, but more general `quantum assets'\index{Quantum assets} (Secs.~\ref{sec:introduction} \& \ref{part:quant_net}), involving all manner of state preparation, manipulation, evolution and measurement, potentially all performed by different interconnected parties, specialising in different aspects of quantum protocols. The demand for this will extend far beyond computation.

The availability of a globe-spanning quantum satellite network brings the opportunity for fundamental quantum mechanical experiments and unprecedented length scales and velocities in the future. Satellite-to-satellite\index{Satellites!Satellite-to-satellite communication} photon transfer can allow for ultra-long distance quantum communications that are not possible on Earth due to atmospheric loss. Another unique aspect of space is that satellites move at high velocities -- typically at $10^{-5} $ times the speed of light\index{Speed of light} for LEO satellites. The combination of both of these effects gives a unique opportunity for performing relativistic quantum information\index{Relativistic quantum information} experiments to test fundamental physics. 

We anticipate that some of the first experiments will be extensions of what are already performed on Earth. For example, one can perform increasingly long space-based Bell violation tests\index{Bell!Inequality} at unprecedented distances \cite{yin2017satellite}. Another possibility is to examine the speed of influence of entanglement \cite{bib:yin2013lower}. In space, such experiments could be extended much further, giving tighter bounds. There are demanding technical hurdles that must be overcome to succeed at such experiments, such as the necessity for synchronised clocks (Sec.~\ref{sec:clock_sync}).

 In addition to examining extensions of existing experiments, the high satellite velocities can be used to perform relativistic quantum information experiments, such as entanglement tests in the presence of special\index{Special relativity} and general relativity\index{General relativity}, Wheeler's delayed choice experiment\index{Wheeler's delayed choice experiment}, and enhanced quantum metrology \cite{bib:kaltenbaek2003proof, bib:scheidl2013quantum, bib:ahmadi2014relativistic}.

The QTCP protocol presented in Sec.~\ref{sec:QTCP} provides an extensible framework for facilitating these kinds of outsourced or delegated protocols\index{Outsourced!Protocols}\index{Delegated!Protocols} using generic quantum assets. Bear in mind that, as designed, the payload of QTCP packets could encapsulate all manner of optical states, or mediate long-distance interaction between them.

This model for quantum research could be invaluable to less-well-resourced researchers, for example in developing nations or not-so-well-funded universities, opening up a field of experimental research previously inaccessible to them. Indeed, some private and university sector operators are making elementary, remotely programmable quantum information processing protocols available over the internet, bringing this type of research within reach of researchers and even curious hobbyists around the globe.

While such early implementations fall far short of being truly reconfigurable, outsourced or delegated quantum protocols, applicable to a broad range of applications, they certainly already demonstrate the interest such models for outsourcing is generating within the research community, and the viability of further extending it.

Examples of how this type of model might be applied could include, but not be limited to research into:
\begin{itemize}
	\item Quantum information processing protocols, beyond only quantum computation, bits and qubits.
	\item Bose-Einstein condensates (BECs)\index{Bose-Einstein condensates (BECs)}.
	\item Light-matter interactions.\index{Light-matter!Interactions}
	\item Quantum thermodynamics and quantum statistical mechanics.\index{Quantum thermodynamics}\index{Quantum statistical mechanics}
	\item Quantum phase-transitions.\index{Quantum phase-transitions}
	\item Quantum optics, involving all manner of quantum states of light, beyond only those raised in Sec.~\ref{sec:opt_enc_of_qi}.\index{Quantum optics}
	\item Optical interferometry.\index{Optical!Interferometry}
	\item Providing a practical platform for university teaching and education.
\end{itemize}

In some instances, such outsourced quantum protocols might be applicable to encryption protocols, like those discussed in the next section (Sec.~\ref{sec:homo_blind}), enabling highly valuable secrecy for the experiments being conducted by researchers and their hard-earned results and ideas\footnote{Note, however, that the upcoming protocols are designed for application to particular optical states and protocols, and encryption schemes involving more generic quantum assets are likely to require some rethinking and adaptation (if possible at all, which isn't guaranteed!).}.

%
% The Globally Unified Quantum Cloud
%

\subsection{The globally unified quantum cloud}\index{Globally unified quantum cloud}\label{sec:glob_unif_quant_cloud}

In Sec.~\ref{sec:economics} we argue that in the quantum era it will be optimal to unify the world's quantum computers into a single virtual, distributed device, rather than utilising smaller individual quantum computers in isolation. This owes to the super-linear scaling in the power of a quantum computer against its number of constituent qubits, a phenomena unique to quantum computers with no classical parallel.

This economic imperative implies that the world's many clients of quantum computing will all be interacting will a single vendor -- the globally unified quantum cloud. This will create a competitive online marketplace for the licensing of timeshares in the utilisation of the unified device.

How this unified device will be managed, and by whom, is entirely open to speculation. Will a nation state or alliance of nation states monopolise it? Will a global consortium voluntarily emerge to manage the resources? Or will the whole thing be completely anarchic, potentially resulting in the fracturing of the unified device into several competing smaller ones? What policy and regulatory frameworks will emerge to oversee it? 

The answers to these questions are entirely uncertain. But what is certain is that there will be an extremely high level of unification of quantum resources via the quantum internet, massively enhancing its collective computational power.

The Quantum Cloud\texttrademark\, will be far more powerful than simply licensing compute-time from a single vendor. Its collective power will be far greater than the sum of its parts.

\latinquote{Timendi causa est nescire.}

\section{Encrypted cloud quantum computation} \label{sec:homo_blind} \index{Encrypted quantum computation}

\sectionby{Atul Mantri, Si-Hui Tan, Yingkai Ouyang \& Peter Rohde}\index{Atul Mantri}

\dropcap{E}{xtremely} important to many high-performance data-processing applications is security, as proprietary or sensitive data may be being dealt with. To address this, there are models for encrypted, outsourced computation that we focus on -- \textit{homomorphic encryption} \cite{bib:gentry2009fully, bib:van2010fully}\index{Homomorphic encryption} and \textit{blind quantum computation}\index{Blind quantum computation} \cite{bib:blind2, bib:broadbent2009universal, bib:barz2012demonstration, bib:PhysRevLett.108.200502, bib:Morimae3486, bib:Morimae5460}.

In both cases, Alice has secret data$^\copyright$, and wishes to not only ensure that an interceptor is unable to read it, but that even the server performing the computation isn't able to either -- she trusts no one. That is, she wishes the data to be processed in encrypted form, without first requiring decryption.

The difference between the two protocols lies in the treatment of algorithms:
\begin{itemize}
	\item Homomorphic encryption: Alice provides only the data, whereas Bob provides the processing and the algorithm it implements (which he would also like to keep to himself in general). When \textit{any} circuit is allowed, the protocol is said to be a \textit{fully homomorphic} encryption protocol (FHE). Otherwise, it is a \textit{somewhat-homomorphic} encryption protocol. Although homomorphic encryption protocols have been around for a few decades in the form of privacy homomorphisms \cite{bib:Rivest1978}, classical FHE has only been described very recently \cite{bib:gentry2009fully, bib:van2010fully}.
Data privacy is the usual focus, though privacy of the algorithm (circuit privacy) can also be discussed.
    
	\item Blind quantum computing: Alice provides both the algorithm \textit{and} the data, and wishes \textit{both} to remain secret to her. It is known that universal blind \textit{classical} computation is not possible, universal blind \textit{quantum} computation is.
\end{itemize}
Both of these seem like very challenging goals, yet significant developments have been made on both fronts in the quantum world, with efficient resource overheads associated with the encryption.

In the usual circuit model, blind quantum computation has been shown to be viable, and optimal bounds derived. Equivalently, such protocols have been described in the cluster state model (Sec.~\ref{sec:CSQC}). For universal computation, such protocols necessarily require classical interaction between the client and host. However, it was shown that in some restricted (i.e non-universal) models for optical quantum computation, specifically \textsc{BosonSampling}, quantum walks and coherent state passive linear optics, non-interactive, somewhat-homomorphic encryption may be implemented.

These encryption protocols induce a resource overhead in circuit size and number of qubits involved in the computation, with efficient scaling. 
They have differing amounts of information-theoretically secure (Sec.~\ref{sec:comp_vs_inf_th_sec}) data-hiding, enabling trustworthy outsourced processing of encrypted data, independent of the attack.

%
% Classical Computation 
%

\subsection{Classical homomorphic encryption} \index{Classical encrypted computation}
% \comment{A universal QC can implement any classical algorithm. So QCs with homo/BQC give us the means by which to perform encrypted classical computations, bypassing limitations imposed by purely classical protocols.}
% \comment{[Comment: The polynomial hierarchy is not contained in BQP. In fact, NP-complete problems are not contained in BQP.]}

% \subsubsection{Homomorphic encryption} \index{Homomorphic encryption}

To set the stage for our upcoming treatment of encrypted quantum computation protocols, we begin by reviewing recent developments in \textit{classical} homomorphic encryption, paying special interest to resource scaling and information-theoretic security.

% \comment{
The first FHE scheme was reported in Gentry's seminal paper \cite{bib:gentry2009fully}. He showed that if a homomorphic encryption scheme can evaluate its own decryption circuit, and also slightly augmented versions of it--a feature he calls {\it bootstrapping}, one can construct a FHE scheme from it. Then he constructed a somewhat homomorphic encryption protocol using ideal lattices, and via a clever transformation that decreases the complexity of its decryption circuit, showed that it is bootstrappable with respect to a universal set of gates. For a security parameter $\lambda$, Gentry's scheme has a $\widetilde{O}(\lambda^6)$ \footnote{The tilde in the big-O notation means that we are ignoring logarithmic factors.} bit bound on complexity for refreshing a ciphertext corresponding to a 1-bit plaintext \cite{bib:Gentrythesis}. This was subsequently reduced to $\widetilde{O}(\lambda^{3.5})$ \cite{bib:Damien2010}, $\widetilde{O}(\lambda)$ \cite{bib:Brakerski2011}, and ${\rm polylog (\lambda)}$ for any width-$\Omega(\lambda)$ circuit with $t$ gates \cite{bib:Craig2012}. 
% }

% \comment{
A homomorphic encryption scheme is made up of four algorithms: a key generation algorithm, \textsc{KeyGen}, an encryption algorithm, \textsc{Encrypt}, an evaluation algorithm, \textsc{Evaluate}, and a decryption algorithm, \textsc{Decrypt}. The four algorithms have the following inputs and outputs:
\begin{itemize}
\item \textsc{KeyGen}$(\lambda)$: Takes as input a security parameter $\lambda$, and outputs a public-key $pk$, and a secret-key $sk$.
\item \textsc{Encrypt}$(pk, \pi_i)$: Takes as input $pk$, and a plaintext $\pi_i$. It outputs a ciphertext $\psi_i$.
\item \textsc{Evaluate}$(pk, C, \Psi)$: Takes as input $pk$, a permitted circuit $C$, and $\Psi=(\psi_1,\ldots, \psi_t)$. It outputs a ciphertext $\psi$.
\item \textsc{Decrypt}$(sk,\psi)$: Takes as input $sk$, and $\psi$ and outputs $C(\pi_1,\ldots, \pi_t)$.
\end{itemize}
The computational complexity of all these algorithms must be polynomial in $\lambda$, and in the case of the evaluation algorithm, polynomial in the size of the evaluation circuit $C$. The condition that \textsc{Decrypt}$(sk,\psi)$ outputs $C(\pi_1,\ldots, \pi_t)$ is a condition known as correctness which we require of the homomorphic encryption scheme. Furthermore, we also require ciphertext size and decryption time to be upper bounded by a function of the security parameter $\lambda$, independently of $C$. This last condition is known as compactness, and is necessary to exclude trivial schemes such as that which decrypts the ciphertexts first, and then apply $C$.
% }

% \comment{
The specifics of these algorithm vary from scheme to scheme, and as is in the case of FHE, usually contains sub-algorithms within them. Making FHE practical is an active area of research. Much of the problem lies in the bootstrapping required in Gentry's scheme, and some of these efforts lies in reducing the overhead required in bootstrapping or removing the need for bootstrapping entirely.
% }

Although there exists a plethora of FHE schemes, the security of such schemes require us to assume that finding the solution to certain computational problems is very difficult. 
Two types of such problems are widely considered in lattice-based cryptography, and are (1) the Shortest Vector Problem (SVP), and (2) Learning with Errors (LWE) Problem. Gentry's original FHE was based on SVP, but over time, the schemes have evolved towards a LWE approach because of their lower overhead conceptually simplicity. An overview of advances, and applications of homomorphic encryption can be found in a recent review \cite{bib:Halevi2017}.

\subsection{Blind quantum computation} \label{sec:blind_qc} \index{Blind quantum computation}

Cluster states are highly entangled states that allow one to perform a quantum computation by (1) measuring the qubits individually, and (2) performing single qubit operations on the remaining qubits based on the obtained   measurement outcomes. When the cluster states are 
suitably chosen, it is possible to perform a universal quantum computation via such a measurement-based computation approach.

This measurement-based quantum computation (MBQC) approach, utilizing cluster states, forms the foundation for blind quantum computation protocols. 
In the context of blind quantum computation (BQC), the server generates the cluster state, but the client determines the measurement bases and interprets the results. 

The multitude of measurement bases available for the client to choose, combined with the probabilistic nature of quantum measurements, effectively hide the true computation from the server. The randomness in measurement outcomes allows the client to obfuscate the actual algorithm being performed, as the server cannot distinguish between intentional operations and random noise.
This ensures the privacy of both the quantum algorithm and the quantum data from the server. The terminology `blind' arises from the `blindness' of the server to the quantum algorithm and quantum data of the client.

The client retrieves the computed result by requiring the server to perform the requested measurements on the qubits, and send the measurement outcomes back to the client. 
The client then decrypts these outcomes using prior private information unbeknownst to the server, and obtains the 
some post-processing on the decrypted data to obtain the final computation result.

% When Alice the client has the limited quantum resources required to perform single-qubit measurements, and she knows the algorithm she wishes to implement, then by outsourcing just the cluster state preparation stage, whilst performing the single-qubit measurements herself, she can obtain \textit{perfect} secrecy of both her data and her algorithm, since no one else is involved in the processing stage.

% However, Alice may have access to no quantum resources whatsoever -- even single-qubit measurements -- requiring homomorphic encryption or blind quantum computing protocols that are native to the cluster state model. 
% Quantum computation can be discussed in both the circuit model and the cluster state model. For blind quantum computation, the cluster state model of computation is prevalently used, where a measurement-based quantum computation is performed on an initially prepared highly entangled quantum state. 

\subsection{Quantum homomorphic encryption} \index{Homomorphic encryption}

The key aspect in which quantum homomorphic encryption (QHE) differs from blind quantum computation is that the server and the client cannot communicate over multiple rounds. 
Namely, the client prepares the quantum data, encrypts the quantum data, and sends the quantum data to the server. 
The server then computes on the encrypted data without the client's help.
Upon completion of the computation, the server sends the computed data back to the client. 
The client is then no longer allowed to interact with the server, and decrypts the quantum data alone.

 There are two lines of research involving QHE. One considers QHE with computational hardness assumptions, and this type of security arises because classical homomorphic encryption protocols as a subroutines therein \cite{broadbent2015quantum,dulek2016quantum,mahadev2020classical}. Another line of research considers information-theoretic security requirements on the quantum data \cite{yu2014limitations,tan2016quantum,ouyang2018quantum,tan2018practical,ouyang2020homomorphic}, where one is not allowed to assume the computational hardness of any computational problem. 
 Such QHE schemes will surely be secure in a post-quantum world; the caveat of such schemes is that they might allow as rich a class of computations as QHE schemes based on computation-hardness assumptions, 
 because of the multitude of no-go results in quantum cryptography \cite{yu2014limitations,sikora,hu2023privacy}.
 In this chapter, we focus on QHE schemes with information-theoretic security.

\subsection{Passive optics} \index{Encrypted quantum computation!Passive optics}

The previously discussed schemes for encrypted universal quantum computation required a degree of client/server interaction via classical communication. But perhaps there are some restricted (i.e non-universal) models for optical quantum computation, which lend themselves to passive, non-interactive encryption? And perhaps these restrictions simplify the physical resource requirements for encryption?

Let us formalise some reasonable requirements for such a scheme. We will require that:
\begin{itemize}
\item Alice's encoding (state preparation) and decoding (measurement) operations are separable, single-mode operations (i.e she has no quantum power of entanglement at her disposal).
\item Bob's computation is non-interactive, requiring no input from Alice beyond her input state.
\item Bob's computation is passive, requiring no intermediate measurement and feedforward.
\item Other than this, there are no constraints on the structure of the encoding/decoding operations, or the optical quantum computation being implemented (e.g it could encompass more than just linear optics).
\end{itemize}

We can express these requirements very generally and elegantly in terms of a commutation relation between the encoding ($\hat{E}$), decoding ($\hat{D}$), and computational ($\hat{U}$) operations. Furthermore, for the protocol to hide information, the plaintext basis states must not be invariant under the encoding operations. This enforces the criteria,
\index{Criteria for encrypted passive optics}
\begin{definition}[Encrypted passive optics] \label{def:enc_pass}
Let \mbox{$k=\{k_1,\dots,k_m\}$} be a partition of the key $k$ into sub-keys $\{k_i\}$, one associated with each mode $i$. Let $\hat{E}_i(k_i)$ and $\hat{D}_i(\tilde k_i)$ be the encoding and decoding operations for the $i$th mode. $\tilde k$ is a potentially transformed version of $k$, to accommodate that the encryption and decryption keys may be asymmetric, in which case we require that $\tilde{k}$ be efficiently computable from $k$. Let $\hat{U}$ be the computation. Then, separability of the encoding and decoding operations requires the following commutation relation to hold,
\begin{align} \label{eq:gen_pass_hom}
\hat{U} \left[\bigotimes_{i=1}^m\hat{E}_i(k_i)\right] = \left[\bigotimes_{i=1}^m\hat{D}^\dag_i(\tilde k_i)\right] \hat{U}.
\end{align}
For the protocol to hide information, the plaintext basis states must not be invariant under the encoding operations,
\begin{align}
\left[\bigotimes_{i=1}^m\hat{E}_i(k_i)\right]\ket\psi_\mathrm{plaintext} \neq \ket\psi_\mathrm{plaintext}.
\end{align}
The state observed by Bob is the mixture of Alice's plaintext over the complete set of encoding operations, implementing a quantum process $\mathcal{E}$, with Kraus operators $\hat{E}(k)$,
\begin{align} \label{eq:mix_over_enc_ops}
\hat\rho_\mathrm{encoded} &= \mathcal{E}(\ket\psi_\mathrm{plaintext}\bra\psi_\mathrm{plaintext}) \nonumber \\
&= \sum_k \hat{E}(k)\ket\psi_\mathrm{plaintext}\bra\psi_\mathrm{plaintext} \hat{E}^\dag(k),
\end{align}
where,
\begin{align}
\hat{E}(k) = \bigotimes_{i=1}^m\hat{E}_i(k_i).
\end{align}
To minimise Bob's chances of guessing Alice's state, we would like to maximise the von Neuman entropy of Bob's state. For \mbox{$S(\hat\rho_\mathrm{encoded})=0$} we have no secrecy, whereas for maximal \mbox{$S(\hat\rho_\mathrm{encoded})$} we have maximal secrecy (for the given plaintext basis state).
\end{definition} 

\begin{figure}[!htbp]
\includegraphics[clip=true, width=0.425\textwidth]{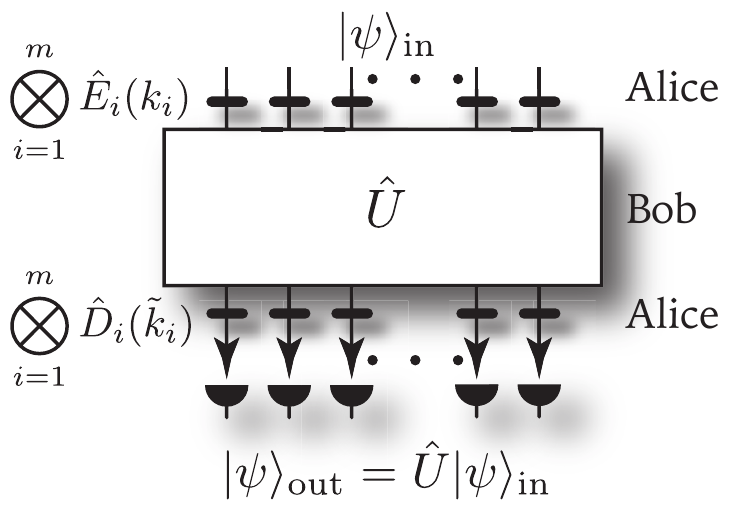}
\captionspacefig \caption{General structure for the relationship between the encoding ($\hat{E}$), decoding ($\hat{D}$), and computational ($\hat{U}$) operations in a passive, non-interactive, optical quantum computation, where Alice is restricted to non-entangling, single-mode encoding and decoding operations.} \label{fig:gen_pass_hom}
\end{figure}

Intuitively, this simply says that a tensor product of single-mode encoding operations commutes through the passive computation to yield a (potentially different) tensor product of single-mode decoding operations. This way, Alice's operations are all separable, requiring no entangling gates (after all, if she had access to entangling gates she might be able to do quantum computations herself!). This relationship can be illustrated as shown in Fig.~\ref{fig:gen_pass_hom}.

Importantly, note that devising a scheme satisfying this commutation relation does not automatically imply that it is secure -- it merely enforces the separability of Alice's encoding and decoding operations. A security proof is usually more technically challenging.

In the following sections we introduce non-interactive techniques for passive optical quantum computation based upon this general formalism. As encoding techniques compatible with the commutation relation from Eq.~(\ref{eq:gen_pass_hom}), we specifically introduce:
\begin{itemize}
\item \textit{Polarisation-key encoding} (Sec.~\ref{sec:phot_homo_enc}): a uniform random polarisation rotation is applied to each input mode, which we apply to photonic linear optics.\index{Polarisation!Key encoding}
\item \textit{Phase-key encoding} (Sec.~\ref{sec:homo_coherent_state}): a uniform random phase-shift is applied to each input mode, which we apply to the encryption of coherent states under evolution via linear optics and generalised non-linear phase-shift operations.\index{Phase!Key encoding}
\item \textit{Displacement-key encoding} (Sec.~\ref{sec:disp_key_enc}): an arbitrary configuration of random phase-space displacements is applied to the input modes, which in principle applies to any optical encoding.\index{Displacement-key encoding}
\end{itemize}
% However, we leave it as an open question for future work to fully characterise the set of compatible encoding, decoding and computational operations, and to evaluate their security for different choices of input states.

%
% Polarisation-Key Encoding
%

\subsubsection{Polarisation-key encoding} \label{sec:phot_homo_enc} \index{Polarisation!Key encoding}

It was recently shown that processing photonic states using passive linear optics -- i.e \textsc{BosonSampling} or quantum walks (Secs.~\ref{sec:boson_sampling} \& \ref{sec:QW}) -- may be trivially homomorphically encrypted with the addition of additional photons and randomised polarisation rotations on the inputs \cite{bib:RohdeQWEnc12}, so-called \textit{polarisation-key encoding}. This encryption does not require any client/server interaction, remaining completely passive, yet achieving near optimal secrecy, hiding $O(\log (m))$ bits of information in an $m$-mode interferometer. Furthermore, it does not impose an overhead in circuit complexity, only in the number of input photons.

\begin{figure}[!htbp]
\includegraphics[clip=true, width=0.425\textwidth]{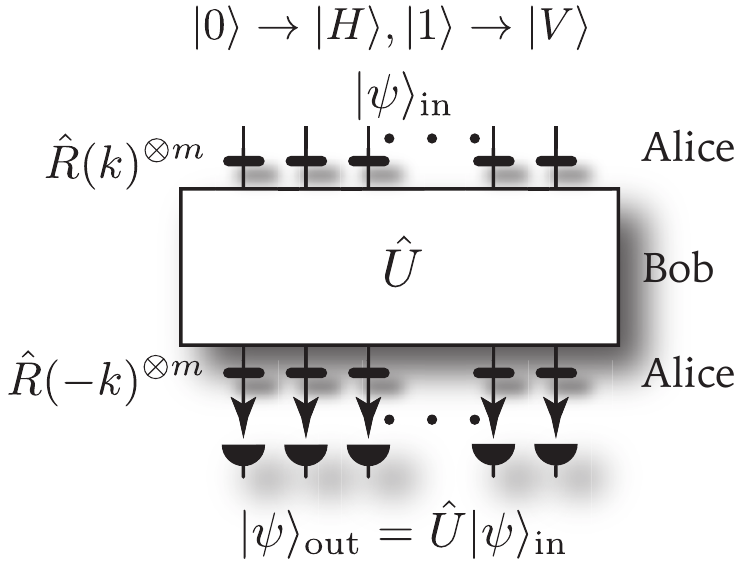}
\captionspacefig \caption{Protocol for implementing homomorphic encryption on photonic passive linear optics. Horizontal bars are wave-plates, implementing polarisation rotations $\hat{R}(\theta)$ -- the encryption and decryption operations performed by Alice. $\ket\psi_\mathrm{in}$ contains one photon per mode, polarisation-encoded such that vertically polarised photons belong to the desired computation, whilst the remaining horizontally polarised ones are dummies. The polarisation rotation angle, $k$, acts as Alice's private-key. The photo-detectors are polarisation-resolving, discarding all dummy horizontally polarised photons at the output. The algorithm is described in detail in Alg.~\ref{alg:homo_LO}.} \label{fig:BS_homo}
\end{figure}

For $m$ modes, the resource requirements are:
\begin{enumerate}
\item $m$ single-photons -- one per input mode.
\item $m$ classically controlled wave-plates, able to implement arbitrary polarisation rotations.
\item $m$ polarisation filters.
\item $m$ photo-detectors.
\item An \mbox{$m\times m$} linear optics network.
\end{enumerate}
The full protocol is described in Alg.~\ref{alg:homo_LO} and shown in Fig.~\ref{fig:BS_homo}.

\begin{table}[!htbp]
\begin{mdframed}[innertopmargin=3pt, innerbottommargin=3pt, nobreak]
\texttt{
function PolarisationKeyEncoding($S$,$k$):
\begin{enumerate}
    \item Alice meditates upon, but needn't actually prepare the state,
    \begin{align}
    \ket\psi_\mathrm{number} = \ket{S_1,\dots,S_m},    
    \end{align}
    where,
    \begin{align}
S_i\in\{0,1\},
    \end{align}
is the photon-number of the $i$th mode.
   \item Alice makes the substitutions from the photon-number basis into the polarisation basis, 
   \begin{align}
   \ket{0}&\to\ket{H}, \nonumber \\
   \ket{1}&\to\ket{V},
   \end{align}
   to obtain $\ket\psi_\mathrm{pol}$, containing $m$ photons in total, one per mode.
   \item Alice chooses a random private-key $k$ as a real number from the uniform distribution,
   \begin{align}
    k\in(0,2\pi).
    \end{align}
    \item Alice prepares the encoded state by applying the same polarisation rotation (using wave-plates), of angle $k$, to each mode,
   \begin{align}
   \ket\psi_\mathrm{enc} = \hat{R}(k)^{\otimes m}\ket\psi_\mathrm{pol},
   \end{align}
   where,
   \begin{align}
   \hat{R}(\theta) = \begin{pmatrix}
\cos\theta & -\sin\theta \\
\sin\theta & \cos\theta
\end{pmatrix}.
   \end{align}
    \item Alice sends $\ket\psi_\mathrm{enc}$ to Bob.
    \item Bob applies processing using his linear optics computer, to obtain,
    \begin{align}
    \ket\psi_\mathrm{enc\,comp} = \hat{U} \ket\psi_\mathrm{enc}.
    \end{align}
    \item Bob returns $\ket\psi_\mathrm{enc\,comp}$ to Alice.
    \item Alice applies the inverse of the encoding operation,
    \begin{align}
    \ket\psi_\mathrm{comp} = \hat{R}(-k)^{\otimes m}\ket\psi_\mathrm{enc\,comp}.
    \end{align}
    \item Alice applies polarisation filters to $\ket\psi_\mathrm{comp}$, discarding horizontally polarised photons.
    \item The remaining vertically polarised state is Alice's unencrypted output of the computation.
    \item $\Box$
\end{enumerate}}
\end{mdframed}
\captionspacealg \caption{Protocol for implementing homomorphic encryption on photonic passive linear optics, using polarisation-key encoding.} \label{alg:homo_LO}
\end{table}

The key idea here is that orthogonal polarisations do not interfere with one another under linear optics evolution. Thus, by inserting additional orthogonally polarised `dummy' photons, and applying uniform, random polarisation rotations, we can confuse any eavesdropper as to which photons belong to the computation, thereby hiding the secret data from them. Note that the encryption protocol does not affect the computation, since uniform polarisation rotations commute through linear optics circuits,
\begin{align} \label{eq:LO_key_commute}
\hat{R}(k)^{\otimes m} \hat{U} \hat{R}(-k)^{\otimes m} = \hat{R}(k)^{\otimes m} \hat{R}(-k)^{\otimes m} \hat{U} = \hat{U},
\end{align}
using the identity,
\begin{align}
\hat{R}(-k) = \hat{R}^\dag(k).	
\end{align}

Practically, $k$ could be chosen as some integer multiple of \mbox{$2\pi/d$}, where $d$ is the number of distinct keys, since an infinite precision key would be equivalent to an infinitely long key, were it represented as a bit-string. In this case, the information security of the protocol increases with $d$.

\cite{bib:RohdeQWEnc12} provided two relationships for the security of this protocol. First, the probability of Bob guessing Alice's input string approaches,
\begin{align}
P_\mathrm{guess} \leq \sqrt{\frac{8}{\pi m}},
\end{align}
for sufficiently large $m$ and $d$, which asymptotically (but unfortunately only polynomially\footnote{Note that an exponentially small bound is actually prohibited by no-go theorems for oblivious transfer and bit commitment \cite{bib:HKLo97, bib:SpekkensRudolphSecure}}) approaches 0.

Alternately, the mutual information between Alice and Bob, \mbox{$I(A;B)$}, can be upper-bounded using the Holevo quantity, $\chi$ \cite{holevo1973bounds}. That is, \mbox{$I(A;B)\leq\chi$}. The Holevo quantity is defined as,\index{Holevo quantity}
\begin{align}
\chi = S(\hat\rho) - \sum_i p_i S(\hat\rho_i),
\end{align}
where,
\begin{align}
\hat\rho = \sum_i p_i \hat\rho_i,
\end{align}
and $S(\cdot)$ denotes the von Neuman entropy\index{von Neuman entropy} (Sec.~\ref{sec:channel_cap}). Here $\hat\rho_i$ are the individual codewords, in our case the set of all polarisation-encoded basis states, and $p_i$ are their respective probabilities, which are uniform here.

The upper-bound stipulated by the Holevo quantity is an information-theoretic bound\index{Information-theoretic!Bound}, which holds under \textit{any} choice of measurement bases by Bob. Thus, it is impossible for Bob to extract more information about Alice's state than allowed by this bound.

For this protocol the Holevo quantity scales with the number of modes as,
\begin{align}
\chi(m) = m - \frac{1}{2}\log_2\left(\frac{\pi e m}{2}\right) + O\left(\frac{1}{m}\right),
\end{align}
for sufficiently large $d$. Since there are $m$ bits of information in Alice's input state, this implies that the protocol hides at least,
\begin{align}
\frac{1}{2}\log_2\left(\frac{\pi e m}{2}\right) + O\left(\frac{1}{m}\right),
\end{align}
bits of information from Bob.

Furthermore, because a single computation requires Alice to perform only a single call to Bob's algorithm, which we treat as a black box, Alice gains minimum knowledge about Bob's secret algorithm, which is optimal for Bob.

Note that while we have considered linear optics in the above discussion, we could in fact expand the list of ingredients available to the computation to include anything generated by a Hamiltonian that commutes with polarisation rotations,
\begin{align}
[\hat{R}(\theta),\hat{H}]=0.
\end{align}
This could include, for example, polarisation-independent non-linear operations.

% \comment{Discuss follow-up paper by Fitzsimons group on using other photonic degrees of freedom to enhance security.}

%
% Phase-Key Encoding
%

\subsubsection{Phase-key encoding} \label{sec:homo_coherent_state} \index{Phase!Key encoding}

As discussed in Sec.~\ref{sec:coherent_state_QC}, although not a \textit{quantum} computation, a system comprising multi-mode coherent state inputs, evolved via passive linear optics, implements simple matrix multiplication on the vector of input coherent state amplitudes,
\begin{align} \label{eq:betaUalpha}
\vec\beta = U\cdot\vec\alpha,
\end{align}
for input,
\begin{align}
\ket{\vec\alpha}=\ket{\alpha_1,\dots,\alpha_m},
\end{align}
and output,
\begin{align}
\ket{\vec\beta}=\ket{\beta_1,\dots,\beta_m}.
\end{align}
However, despite this being a classically efficient algorithm, it can experimentally be easily homomorphically encrypted with no computational resource overhead. This is in contrast to classical homomorphic encryption techniques, which incur a computational overhead.

The idea behind homomorphic encryption of coherent state linear optics is conceptually almost identical to the polarisation-space protocol for photonic linear optics (Sec.~\ref{sec:phot_homo_enc}). The key difference is that the random rotations are no longer applied in polarisation-space, but in phase-space as phase-rotations (\textit{phase-key encoding}). Specifically, the encryption/decryption operations are now given by the phase-shift operators, 
\begin{align}
\hat{R}(\phi) = \hat\Phi(\phi)=e^{-i\phi\hat{n}}.
\end{align}
Phase-shift operators acting on coherent states simply implement the transformation,
\begin{align}
\hat\Phi(\phi)\ket\alpha = \ket{e^{-i\phi}\alpha},
\end{align}
a simple rotation about the origin in phase-space.

Like polarisation rotations, uniform phase-shifts commute through linear optics networks, as per Eq.~(\ref{eq:LO_key_commute}), and thus the protocol has similar mathematical structure to the photonic case. Now the phase-shift angle, $\phi$, acts as Alice's private-key, which she applies uniformly to all input modes, applying inverse uniform phase-shifts after the computation to decrypt the state. The full algorithm is given in Alg.~\ref{alg:homo_coherent_LO}.

\begin{table}[!htbp]
\begin{mdframed}[innertopmargin=3pt, innerbottommargin=3pt, nobreak]
\texttt{
function PhaseKeyEncoding($\vec\alpha$,$k$):
\begin{enumerate}
    \item Alice prepares the input multi-mode coherent state,
    \begin{align}
    \ket\psi_\mathrm{in} = \ket{\vec\alpha} = \ket{\alpha_1,\dots,\alpha_m}.
    \end{align}
    \item Alice chooses a random private-key $k$ as a real number from the uniform distribution,
    \begin{align}
    k\in (0,2\pi).
    \end{align}
    \item Alice prepares the encoded state by applying the same phase-shift, of angle $k$, to each mode,
    \begin{align}
    \ket\psi_\mathrm{enc} = \hat\Phi(k)^{\otimes m}\ket\psi_\mathrm{in},
    \end{align}
    where,
    \begin{align}
    \hat\Phi(\phi) = e^{i\phi\hat{n}},
    \end{align}
    is the phase-shift operator.
    \item Alice sends $\ket\psi_\mathrm{enc}$ to Bob.
    \item Bob applies processing using his linear optics computer, to obtain,
    \begin{align}
    \ket\psi_\mathrm{enc\,comp} = \hat{U}\ket\psi_\mathrm{enc}.
    \end{align}
    \item Bob returns $\ket\psi_\mathrm{enc\,comp}$ to Alice.
    \item Alice applies the inverse of the encoding operation,
    \begin{align}
    \ket\psi_\mathrm{comp} = \hat\Phi(-k)^{\otimes m}\ket\psi_\mathrm{enc\,comp}.
    \end{align}
    \item The resulting state is,
    \begin{align}
    \ket\psi_\mathrm{comp} = \ket{\vec\beta} = \ket{\beta_1,\dots,\beta_m},
    \end{align}
    where,
    \begin{align}
    \vec\beta = U\cdot\vec\alpha.
    \end{align}
    \item $\Box$
\end{enumerate}}
\end{mdframed}
\captionspacealg \caption{Protocol for implementing homomorphic encryption on coherent state passive linear optics, using phase-key encoding.} \label{alg:homo_coherent_LO}
\end{table}

In fact, this encryption technique applies to more than just linear optics, but extends to also include generalised non-linear phase-shift operations, generated by Hamiltonians that are polynomials in the photon-number operators,
\begin{align}
\hat{H} = O(\mathrm{poly}(\hat{n}_1,\dots,\hat{n}_m)),
\end{align}
where $\hat{n}_i$ is the photon-number operator for the $i$th mode\index{Photon-number!Operators}. This observation follows trivially from the observation that the phase-shift encoding operations (which are generated by Hamiltonians linear in the photon-number operators) commute with any polynomial in the photon-number operators,
\begin{align}
[\hat{n}_i,\mathrm{poly}(\hat{n}_1,\dots,\hat{n}_m)] = 0\,\,\forall \, i.
\end{align}
This immediately significantly expands the class of operations available for the computation. In particular, while coherent states remain separable under linear optics evolution, the introduction of non-linear phase-shift operations enables quantum entanglement, and presumably a more powerful class of computations than simple matrix multiplication. It is unclear to us, however, exactly what this class of computations actually is.

Ref.~\cite{tan2018practical} evaluated the security of this protocol in the case where the basis states were restricted to the binary $\ket{\pm\alpha}$ states. Thus, each input mode encodes at most a single bit (zero bits for \mbox{$|\alpha|=0$}, approaching one bit for \mbox{$|\alpha|\to\infty$}). Rather than employing the mutual information, they resorted to the alternative approach of calculating the distinguishability of codewords under the trace distance (Sec.~\ref{sec:fid_metric}). This has a direct operational interpretation as the probability of Bob guessing Alice's state in the best case. If the basis codeword states observed by Bob are indistinguishable, they are effectively decorrelated from the plaintext basis states, preventing him from guessing Alice's plaintext state, whereas if they are distinguishable, he can.

Let $\vec{x}$ be the binary input string, and $\vec{0}$ be the special case of the all-zero string. Then, for unencrypted states, we have the trace distance,
\begin{align}
D_\mathrm{unenc} &= ||\hat\rho_{\vec{x}}-\hat\rho_{\vec{0}}||_\mathrm{tr} \nonumber \\
&= \sqrt{1-e^{-4\mathrm{wt}(\vec{x})|\alpha|^2}},
\end{align}
where $\mathrm{wt}(\vec{x})$ is the Hamming weight (number of 1s in the bit-string) of $\vec{x}$. On the other hand, for the encoded states, we have,
\begin{align}
D_\mathrm{enc} &= ||\mathcal{E}(\hat\rho_{\vec{x}})-\mathcal{E}(\hat\rho_{\vec{0}})||_\mathrm{tr} \\
&= \sum_{k=0}^{d-1} e^{-m|\alpha|^2}\frac{(m|\alpha|^2)^k}{k!}\sqrt{1-\left(\frac{m-2\mathrm{wt}(\vec{x})}{m}\right)^{2k}}, \nonumber
\end{align}
 where $\mathcal{E}$ denotes the mixture over encoding operations observed by Bob, as per Eq.~(\ref{eq:mix_over_enc_ops}), and there are $d$ distinct keys. 
% \comment{What's the limit as $d\to\infty$?}
 Since each term in the above sum is non-negative, the trace-distance $D_{\rm enc}$ is monotonically increasing as $d$ increases, which suggests that  (seems w (seems w

 We also define the ratio between these two distances,
\begin{align}
R=\frac{D_\mathrm{enc}}{D_\mathrm{unenc}	},
\end{align}
as an indicator of data-hiding. \mbox{$R<1$} is indicative that information is hidden from Bob.

These relationships are illustrated in Figs.~\ref{fig:homo_coh_st_tr} \& \ref{fig:homo_coh_st_ratio}\footnote{Note that the distance between any arbitrary pair of codewords may be obtained by replacing $\mathrm{wt}(\cdot)$ with their Hamming distance (in the simple case where one of the codewords is the $\vec{0}$ bit-string, the Hamming weight and Hamming distance are equivalent).}. Clearly, the encoded basis states exhibit greater indistinguishability than the unencoded ones, demonstrating that information is hidden from Bob. The trace distance does not have the elegant interpretation of `number of bits hidden' that the mutual information does. Rather, it gives us the probability of Bob correctly guessing Alice's state, demonstrating the degree of partial hiding of information.

\if 1\doublecol
\begin{figure}[!htbp]
\includegraphics[clip=true, width=0.475\textwidth]{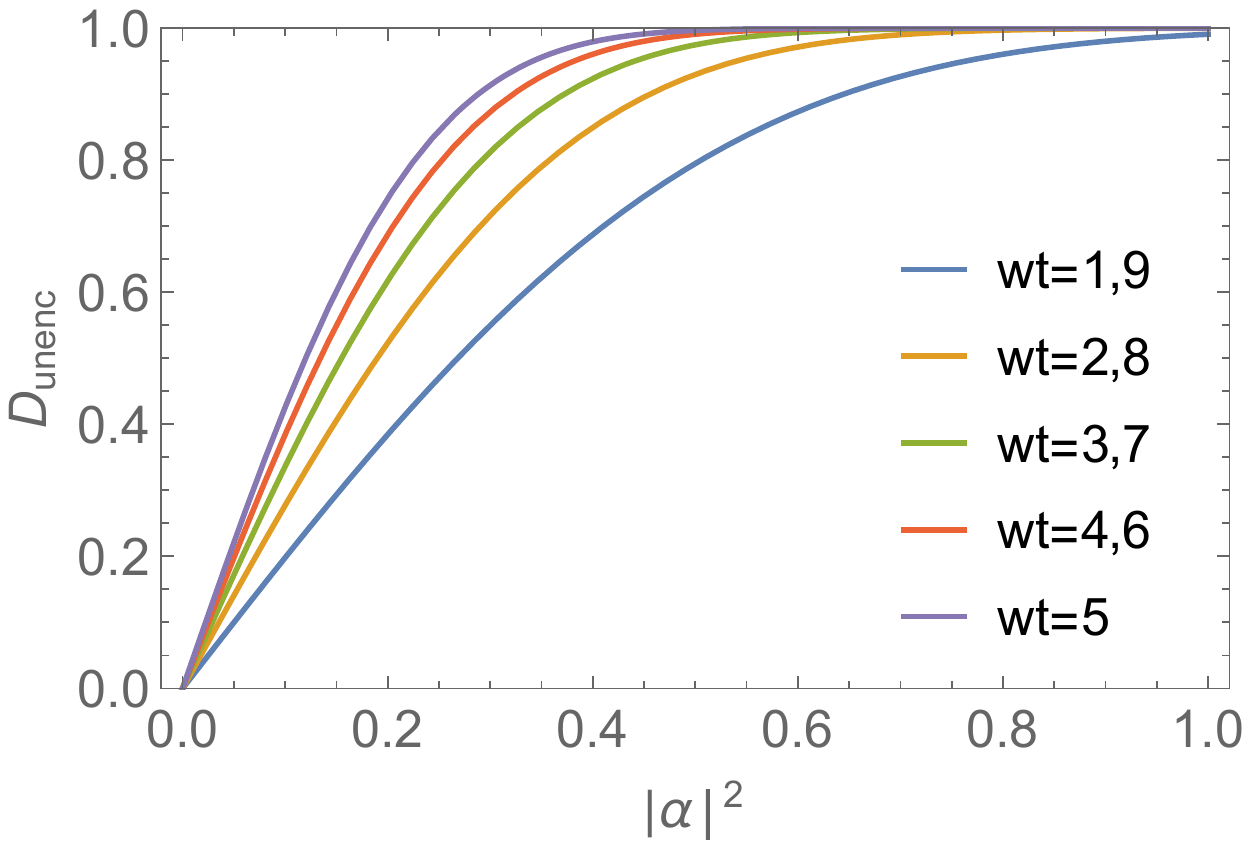}\\
\includegraphics[clip=true, width=0.475\textwidth]{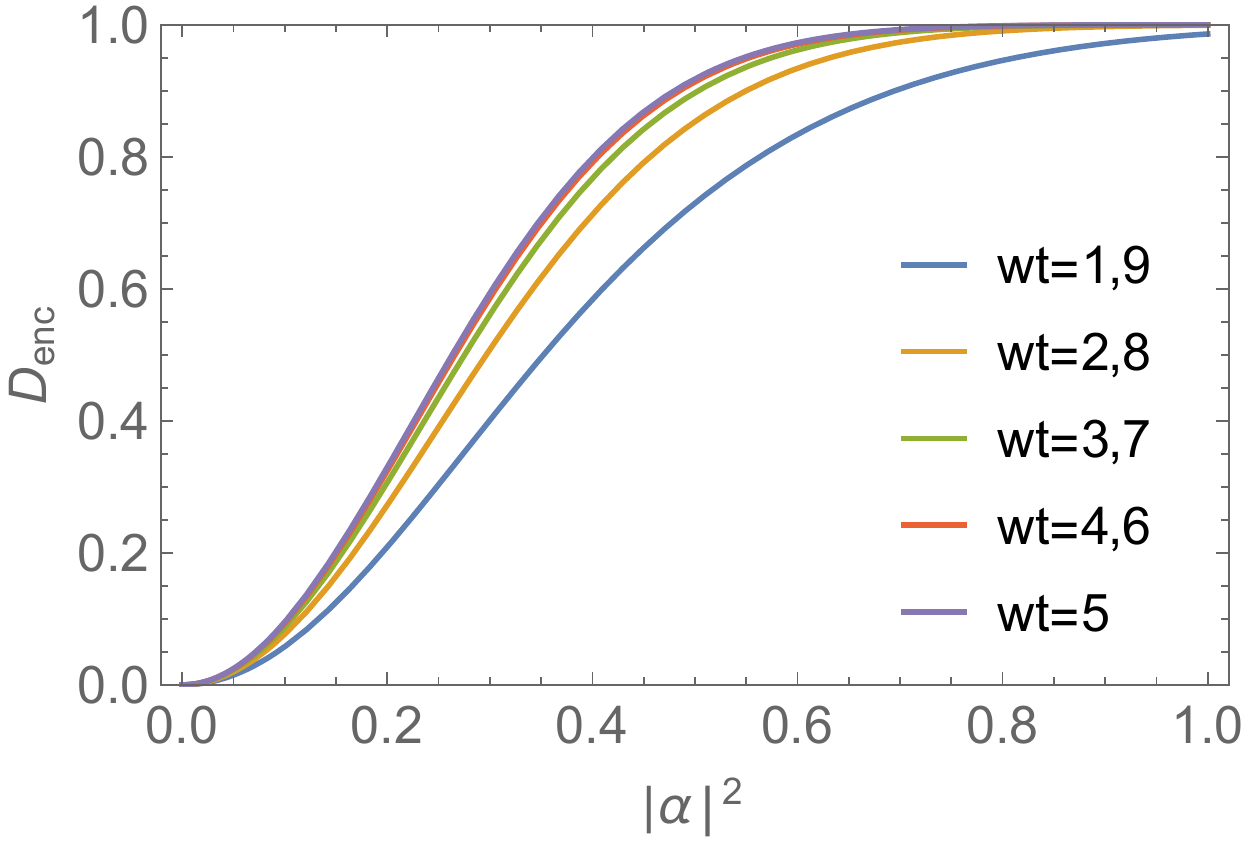}
\captionspacefig \caption{Trace distance between unencoded (top) and encoded (bottom) basis states for coherent state computation using phase-key encoding, with \mbox{$d=50$} keys and \mbox{$m=10$} modes. Each mode is inputted with one of two basis coherent states, \mbox{$\ket{\pm\alpha}$}. \mbox{$\mathrm{wt}(\vec{x})$} denotes the Hamming weight of bit-string $\vec{x}$. Two states are indistinguishable if their trace distance is 0, and distinguishable (orthogonal) if their trace distance is 1. Lower trace distance between encoded states implies a lower chance of Bob guessing Alice's plaintext input state.} \label{fig:homo_coh_st_tr}
\end{figure}
\else
\begin{figure*}[!htbp]
\includegraphics[clip=true, width=0.475\textwidth]{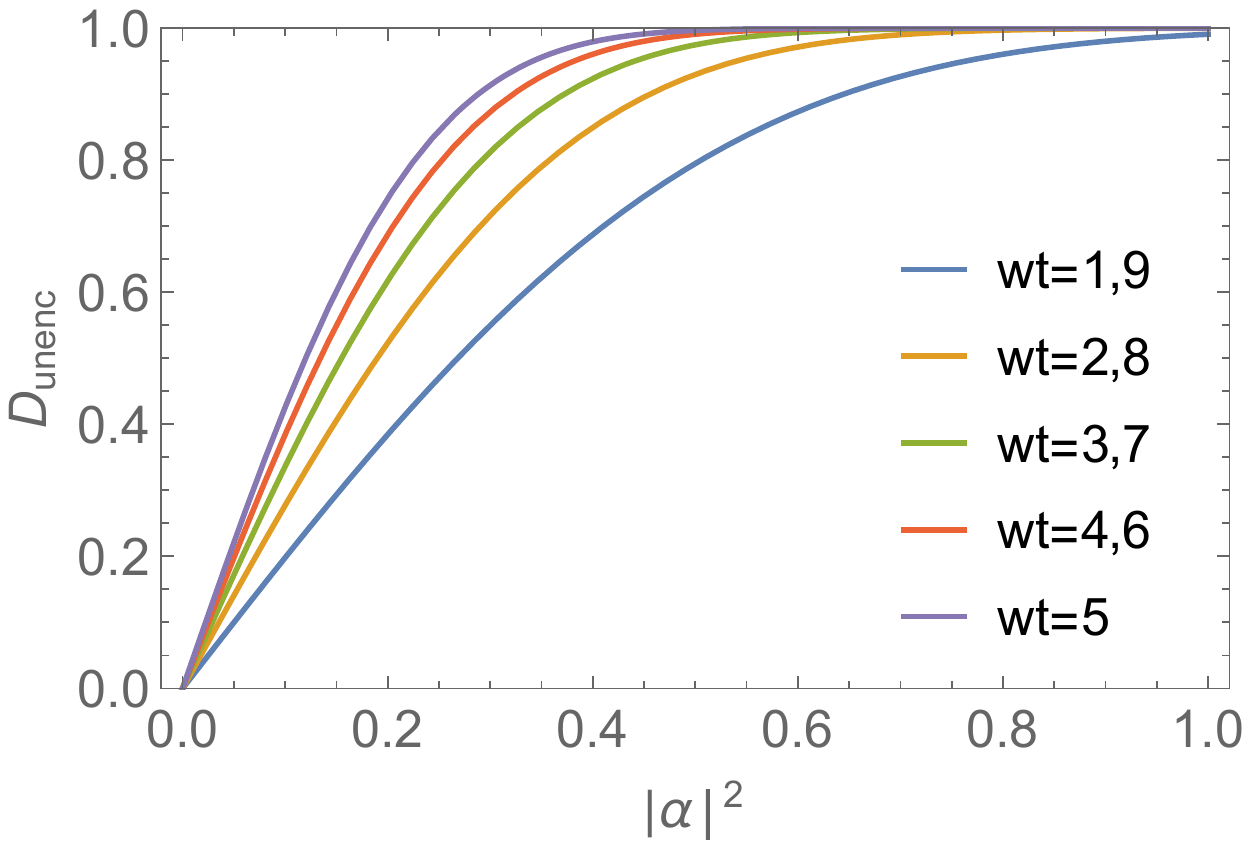}
\includegraphics[clip=true, width=0.475\textwidth]{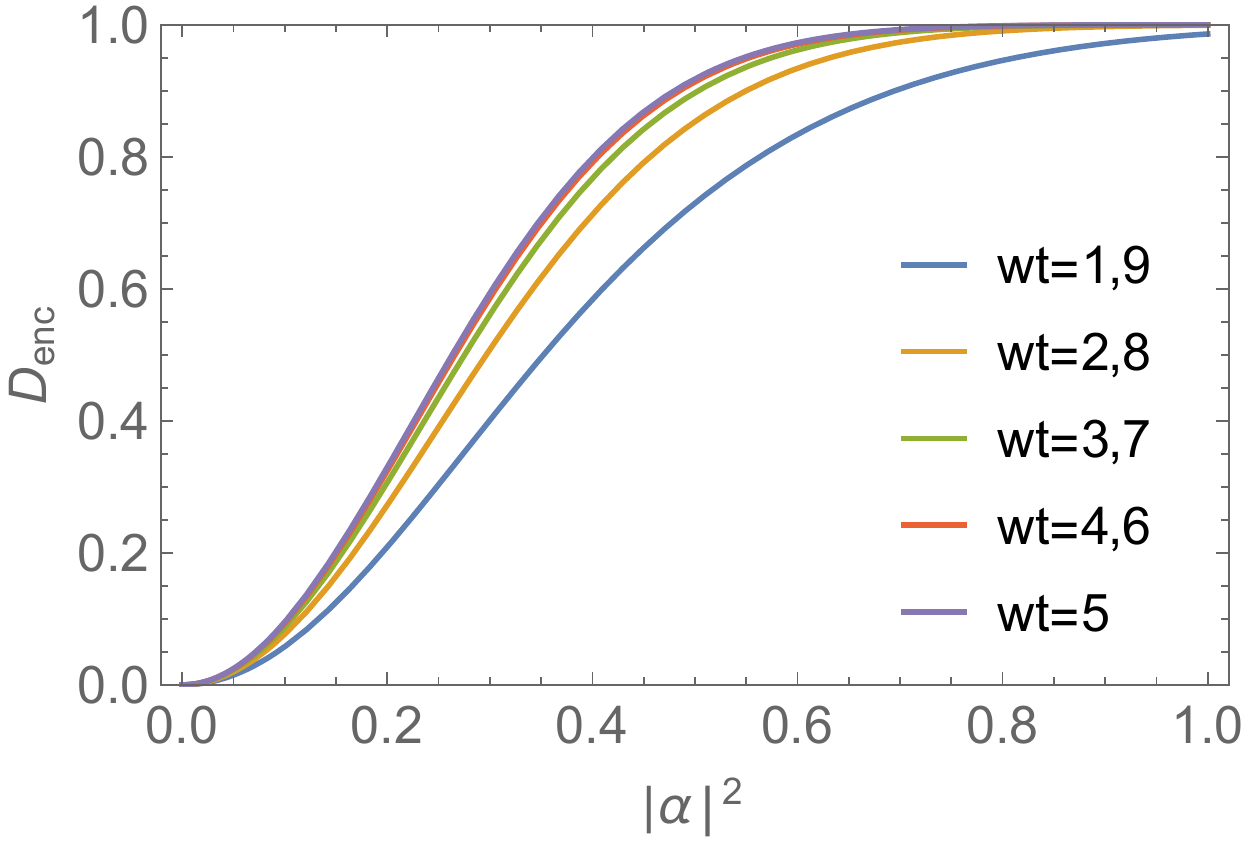}
\captionspacefig \caption{Trace distance between unencoded (top) and encoded (bottom) basis states for coherent state computation using phase-key encoding, with \mbox{$d=50$} keys and \mbox{$m=10$} modes. Each mode is inputted with one of two basis coherent states, \mbox{$\ket{\pm\alpha}$}. \mbox{$\mathrm{wt}(\vec{x})$} denotes the Hamming weight of bit-string $\vec{x}$. Two states are indistinguishable if their trace distance is 0, and distinguishable (orthogonal) if their trace distance is 1. Lower trace distance between encoded states implies a lower chance of Bob guessing Alice's plaintext input state.} \label{fig:homo_coh_st_tr}
\end{figure*}
\fi

\begin{figure}[!htbp]
\includegraphics[clip=true, width=0.475\textwidth]{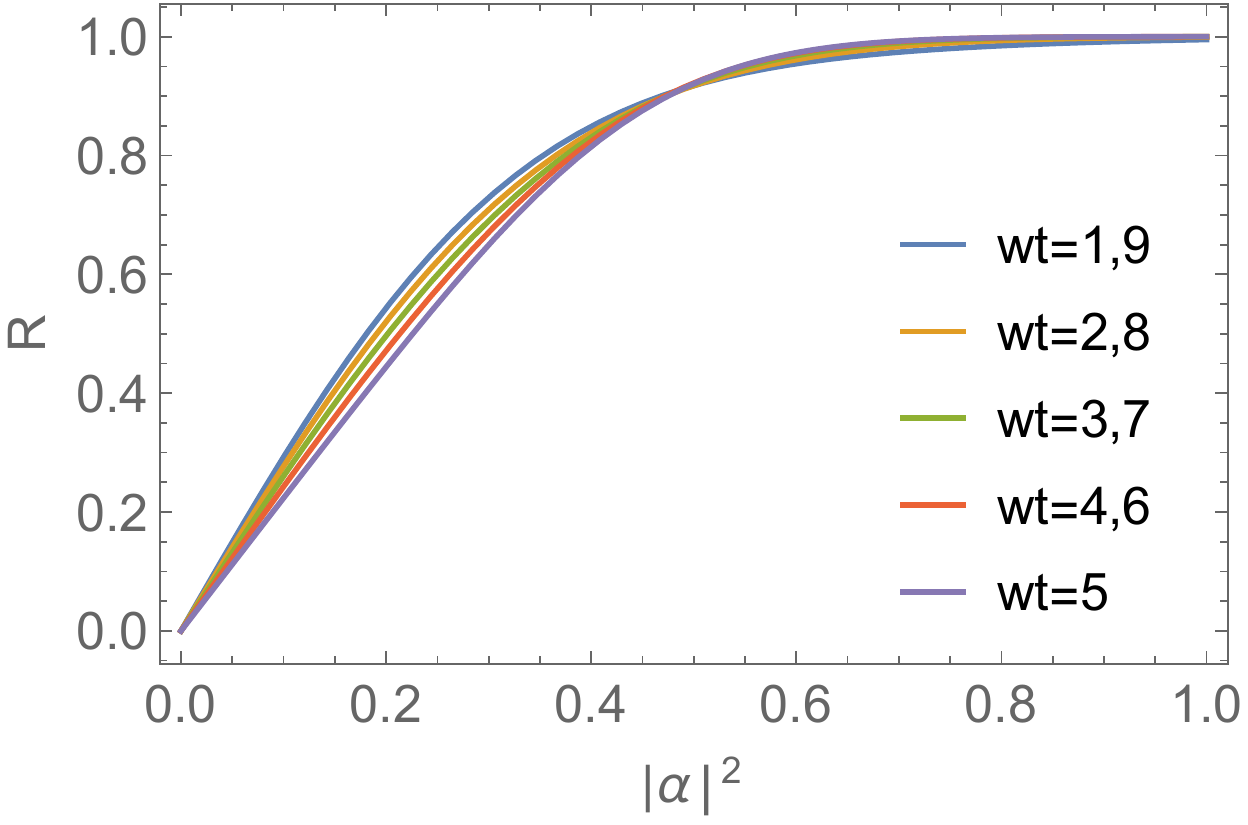}
\captionspacefig \caption{Ratio of the trace distance between unencoded and encoded basis states for coherent state computation using phase-key encoding, with \mbox{$d=50$} keys and \mbox{$m=10$} modes. \mbox{$R<1$} implies information hiding from Bob.} \label{fig:homo_coh_st_ratio}
\end{figure}

It seems plausible that this approach to homomorphic encryption in phase-space could be extended to other quantum states of light. However, as is typically the case, performing entropic security proofs is notoriously difficult, and this remains an open problem. It should be noted, however, that this approach will definitely \textit{not} work for any class of input states which are invariant under phase-shifts. This explicitly rules out employing this technique for, for example, photon-number states, which have no phase.

This protocol demonstrates that a simple optical system is able to homomorphically encrypt the classical computation of matrix multiplication, as well as the more general operations of non-linear phase-shifts, without incurring the computational resource overhead imposed by conventional classical homomorphic encryption techniques.

% \comment{Include figures for asymptotics against m and d.}

%
% Displacement-Key Encoding
%

\subsubsection{Displacement-key encoding} \label{sec:disp_key_enc} \index{Displacement-key encoding}

In the previous two sections we employed the encryption techniques of polarisation-key encoding and phase-key encoding. These are based on the observation that uniform polarisation- and phase-rotations commute through linear optics networks, as per Eq.~(\ref{eq:LO_key_commute}). Are there any other types of encoding operations that observe this property?

The other obvious candidate is \textit{displacement-key encoding}, whereby random displacements (not necessarily uniform across all modes) in phase-space (Sec.~\ref{sec:non_lin_opt}) form the encoding operations. Displacement operators exhibit the property that a tensor product of displacements commutes through linear optics circuits to yield a different combination of tensor products of displacements, where the displacement amplitudes obey the same relationship as for coherent states from Eq.~(\ref{eq:betaUalpha}). Specifically,
\begin{align}
\bigotimes_{i=1}^m \hat{D}_i(\alpha_i) \to \bigotimes_{j=1}^m \hat{D}_j(\beta_j),
\end{align}
where $\hat{D}_i(\alpha_i)$ is the displacement operator for the $i$th mode, with displacement amplitude $\alpha_i$, and the input and output displacement amplitudes are related according to,
\begin{align}
\vec\beta = \hat{U}\cdot\vec\alpha.
\end{align}

Based upon this observation, if the unitary $\hat{U}$ were known to Alice (i.e no secrecy for Bob's algorithm, unfortunately), she could efficiently encode and decode her state, since determining $\vec\beta$ from $\vec\alpha$ requires only classically-efficient matrix multiplication (residing in \textbf{P}\index{P}).

The algorithm for implementing displacement-key homomorphic encryption is shown in Alg.~\ref{alg:homo_disp_LO}.

\begin{table}[!htbp]
\begin{mdframed}[innertopmargin=3pt, innerbottommargin=3pt, nobreak]
\texttt{
function DisplacementKeyEncoding($\ket\psi$,$k$):
\begin{enumerate}
    \item Alice prepares the $m$-mode state $\ket\psi$.
    \item Alice chooses a set of independent complex displacement amplitudes as her private-key,
    \begin{align}
    k=\{\alpha_1,\dots,\alpha_m\}.
    \end{align}
    \item Alice applies the displacements to each mode, yielding her encrypted state,
    \begin{align}
    \ket\psi_\mathrm{enc} = \left[\bigotimes_{i=1}^m \hat{D}_i(\alpha_i)\right] \ket\psi,
    \end{align}
    where $\hat{D}_i(\alpha_i)$ is the displacement operator for the $i$th mode, with displacement amplitude $\alpha_i$.
    \item Alice sends the encrypted state to Bob.
    \item Bob applies the computation $\hat{U}$,
    \begin{align}
    \ket\psi_\mathrm{enc\, comp} = \hat{U}\ket\psi_\mathrm{enc}.
    \end{align}
    \item Bob returns the encrypted computed state to Alice.
    \item Alice calculates the inverse displacement amplitudes $\vec\beta$,
    \begin{align}
    \vec\beta = U\cdot\vec\alpha.
    \end{align}
    \item Alice applies the inverse displacements to each mode,
    \begin{align}
    \ket\psi_\mathrm{comp} &= \left[\bigotimes_{i=1}^m \hat{D}^\dag_i(\beta_i)\right] \ket\psi_\mathrm{enc\, comp} \nonumber \\
    &= \hat{U}\ket\psi,
    \end{align}
    \item $\ket\psi_\mathrm{comp}$ is the unencrypted computed output state.
    \item $\Box$
\end{enumerate}}
\end{mdframed}
\captionspacealg \caption{Protocol for implementing homomorphic encryption using displacement-key encoding.} \label{alg:homo_disp_LO}
\end{table}

Because performing the decoding operation requires solving the matrix multiplication problem to determine the decoding displacement amplitudes, this technique would obviously be inapplicable to encrypting, for example, coherent states, or other states which can be as efficiently classically simulated as matrix multiplication. Instead, it would only be relevant to linear optics sampling problems, which offer an exponential quantum speedup -- if performing the classical computation required for decryption is just as hard as performing the computation, Alice might as well do the computation herself!

A rigorous security proof of displacement-key encoding on a single mode was accomplished in Ref.~\cite{ouyang2020homomorphic}, which confirmed the intuitive arguments in Ref.~\cite{marshall2016continuous}.
In Ref.~\cite{ouyang2020homomorphic}, the authors evaluate an upper bound $\epsilon$ on the trace-distance between 
arbitrary states with at most $n$ photons, 
and which are randomly displaced according to a Gaussian distribution of standard deviation $\sigma$.
The upper bound on this trace distance is given by
\begin{align}
\epsilon = \frac{1}{\sigma^2} 
\left( 
\frac n 4
 + \frac 1 2
 \bigl(1 + \frac{ 1 }{2\sigma^2} \bigr)^n
 +4(n+1) \right) 
\end{align}
in \cite[Theorem 1]{ouyang2020homomorphic}.
For constant $n$ and increasing $\sigma$, this upper bound vanishes, 
and this confirms the indistinguishability of displacement-key encrypted states of bounded energy.
The trace-distance bound can also be easily extended to the multi-mode case by a simple application of a telescoping sum and a triangle inequality.
The proof here appeals to the continuous-variable representation of Fock states as Hermite functions in the quadrature basis.
The security analysis thus far applies to a much more general family of states as that considered earlier in this chapter on polarization-key encoding and phase-key encodings.

The trace-distance metric used in Ref.~\cite{ouyang2020homomorphic} has advantages over the mutual information that we have previously discussed. Namely, the trace distance directly quantifies the indistinguishability of quantum states
while the mutual information does not.

\subsection{Quantum homomorphic encryption based on codes}

Suppose that the only type of quantum computation that we wish to perform is a Clifford computation.
Such a type of unencrypted computation is deemed easy to perform in the usual quantum circuit model, where Clifford operations are free, and $T$ gates are costly. 

Now consider a qubit $\rho$ represented as a density matrix written in the Pauli basis,
 that is 
 $\rho = ( I  + r_X X + r_Y Y + r_Z Z)/2 $.
 Then we can consider an encoding that maps this qubit to an almost maximally mixed state on $n$ qubits, where $n$ is odd. 
 Namely, we consider the map $\mathcal R$ where
 $\mathcal R (\rho) = (  I^{\otimes n}  + r_X X^{\otimes n} + r_Y Y^{\otimes n} + r_Z Z^{\otimes n} )/2^n$.
 Such a map can be achieved with the help of $n-1$ ancilla qubits prepared in maximally mixed states, 
 and performing some CNOT gates according to the procedure described in \cite[Figure 2]{ouyang2018quantum}.
 In such an encoding, whenever $n$ modulo 4 is equals to 1, the transversal Clifford gates are equivalent to the transversal logical gates. 
 We may interpret $\mathcal R $ as a random encoding, where the quantum data is hardly protected; the loss of any one of the qubits leaves the remaining quantum state as a maximally mixed state on $n-1$ qubits.
 
 The quantum data encoded on $n$ qubits can be encrypted by a permutational-key encoding. For this, we can introduce another $n$ qubits as maximally mixed states, and apply a random permutation on the $2n$ qubits, where the distribution over which the permutation is chosen is uniformly random over the symmetric group of size $2n$. Intuitively, it is very hard to guess where the quantum data is after the random permutation, and hence different encrypted quantum states should not be very distinguishable. 
 This intuition can be made rigorous.

 Now given $p$ data qubits that we wish to encrypt.
 We map each data qubit into $2n$ qubits which we arrange row-wise. The permutation-key then acts on the columns, so that each encrypted data qubit uses the same permutation key.
 Then, the trace distance between any two such encrypted states is at most given by
 \begin{align}
 \epsilon \le  2^{p/2} \binom {2n}{n}^{-1/2}  ,
 \end{align}
 which for constant $p$ decreases exponentially with $n$.

 Quantum computation of Clifford gates can be done homomorphically, because the server would just apply the Cliffords uniformly across the columns. Since unitary operations leave maximally mixed states unchanged, it follows that even though the server applies unitaries on all columns, what effectively happens is that only on locations where the quantum information resides does the unitary apply non-trivially.  Decryption in this scheme is simple; one would just apply the unitary operations for the encryption and encoding in reverse.

 Such a scheme can not only support computations of Clifford circuits. Computation of a constant number of $T$-gates is also possible while retaining a exponentially good security measured in trace distance.  

  A variation of this scheme where there are no maximally mixed states in the columns is a Clifford computation on shared quantum secrets scheme \cite{ouyang2017computing}, which also allows a constant number of $T$-gates.

\latinquote{Audentes fortuna iuvat.}

%
% Verification Of Cloud Quantum Computing
%

\section{Verification of cloud quantum computing} \index{Verification}\label{sec:verification}

% \subsection{Randomised benchmarking}\label{sec:rand_bench}\index{Randomised benchmarking}

% \comment{Insert}

\subsection{Zero-knowledge proofs}\label{sec:ZKP}\index{Zero-knowledge proofs}

% \comment{Ryan's links for NEXP ZKPs.}

% \comment{General results.}

% \comment{Ryan review.}

A \textit{zero-knowledge proof} (ZKP) is an interactive protocol\index{Interactive protocols} between two parties -- a \textit{prover}\index{Prover} Peggy, and a \textit{verifier}\index{Verifier} Victor -- where Peggy wishes to efficiently prove to Victor that she knows the solution to a problem, without actually revealing it. Thus, a ZKP can serve as a signature that a problem has been faithfully solved, without disclosing the solution.

\begin{figure}[!htbp]
	\includegraphics[clip=true, width=0.475\textwidth]{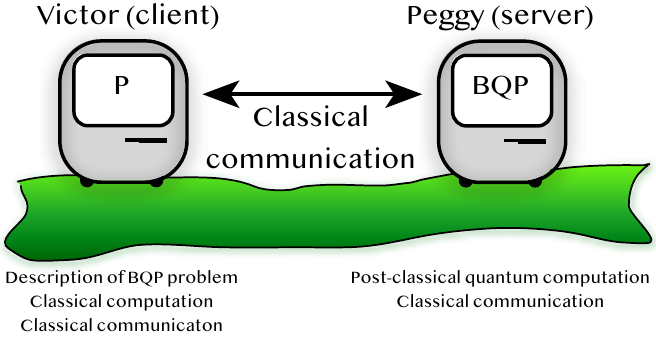}	
	\captionspacefig \caption{Problem description for using zero-knowledge proofs as a tool in the commercial outsourcing of quantum problems provided by a classical client (limited to both classical computation and communication) to a quantum-capable server. The proof provided by the server demonstrates that the client's computation has been faithfully executed, without disclosing the outcome before finalising the commercial transaction.}\label{fig:ZKP_problem_description}
\end{figure}

ZKPs are useful in a number of cryptographic applications, most notably in authentication\index{Authentication}. In the case of classical computing, a variety of free software packages are available for compiling ZKPs for generic code. However, ZKPs become far more valuable in the case of cloud quantum computing, where there is an inherent complexity asymmetry between the client Victor (classical resources only) and the server Peggy (full quantum resources), whereby efficient ZKPs become a useful transactional tool during computational outsourcing. In a commercial context, a client can convince themselves that a server has actually performed an outsourced computation before finalising the transaction. The problem description for this scenario is shown in Fig.~\ref{fig:ZKP_problem_description}.

A conceptually simple example for illustrating the operation of a ZKP protocol is the \textit{graph isomorphism problem}\index{Graph isomorphism}. Two graphs are \textit{isomorphic}, denoted \mbox{$G_1\sim G_2$}, if there exists a permutation\index{Permutation} \mbox{$\pi\in S_n$}\footnote{In group theory\index{Group theory}, $S_n$ denotes the symmetric group\index{Symmetric group}, the set of all permutations on $n$ elements, of which there are \mbox{$|S_n|=n!$} (the order of the group).} on their vertex labels that makes them equivalent, i.e \mbox{$G_1=\pi\cdot G_2$}\footnote{Here we have employed the operator notation that \mbox{$\pi\cdot G$} means `permutation $\pi$ applied to graph $G$'. Alternately, in matrix notation, where $\pi$ is a permutation matrix and $G$ is an adjacency matrix\index{Adjacency matrix}, this operation implies the matrix conjugation\index{Conjugation} \mbox{$\pi\cdot G\cdot\pi^\top$}.}. The graph isomorphism problem is to find $\pi$ for arbitrary $G_1$ and $G_2$.

This problem is clearly contained in \textbf{NP}, since permuting vertices in graphs and directly comparing them are both computationally straightforward, making verification of the problem a polynomial-time affair. However, it is believed that explicitly determining the respective permutation is difficult in general. This is intuitively unsurprising, since for a graph with $n$ vertices, there are $n!$ possible permutations to consider (which is super-exponential), and therefore a na{\" i}ve brute-force\index{Brute-force} approach would require $O(n!)$ comparisons in the worst-case\footnote{This is the worst-case scenario. In many special cases, knowledge of underlying graph structure can simplify this enormously, yielding classically efficient runtimes.}. This problem is not believed to be contained in either \textbf{P} or \textbf{NP}-complete, and therefore presumed to be \textbf{NP}-intermediate\index{\textbf{NP}-intermediate} (Sec.~\ref{sec:introduction}).

\if 1\doublecol
\begin{figure}[!htpb]
	\includegraphics[clip=true, width=0.475\textwidth]{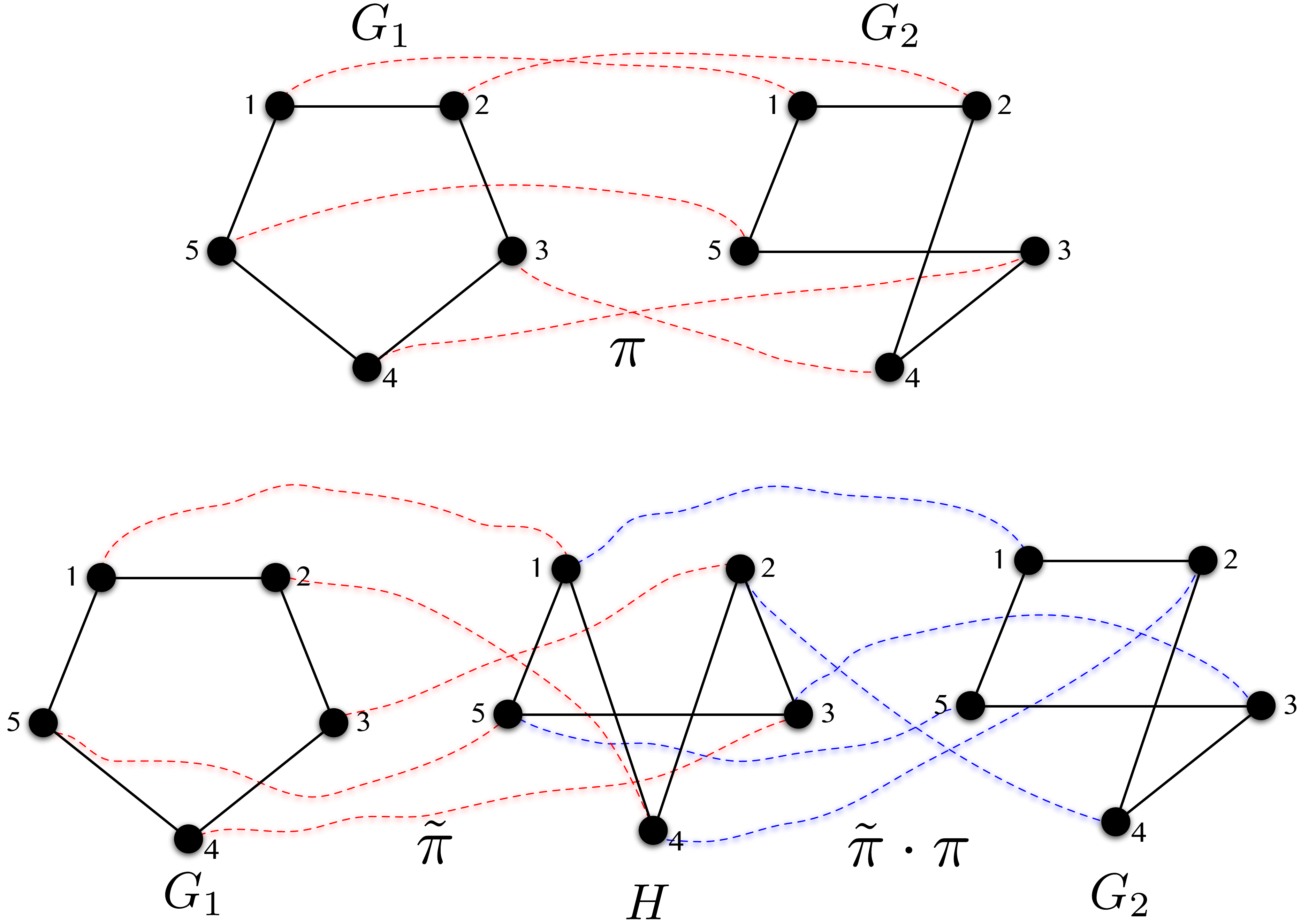}
\captionspacefig \caption{The isomorphisms \mbox{$G_1\sim G_2$} (top), and \mbox{$G_1\sim H\sim G_2$} (bottom). Coloured lines indicate the vertex relabelings associated with the respective isomorphisms. Knowing the latter two isomorphisms simultaneously implies knowledge of \mbox{$G_1\sim G_2$} via composition of the permutations. However, knowing only one of them does not, since $H$ is chosen randomly. By repeatedly proving knowledge of one of the isomorphisms with $H$ we achieve asymptotic certainty that the prover must have known \mbox{$G_1\sim G_2$}, without actually revealing the associated permutation.}\label{fig:ZKP_graph}
\end{figure}
\else
\begin{figure*}[!htpb]
	\includegraphics[clip=true, width=0.8\textwidth]{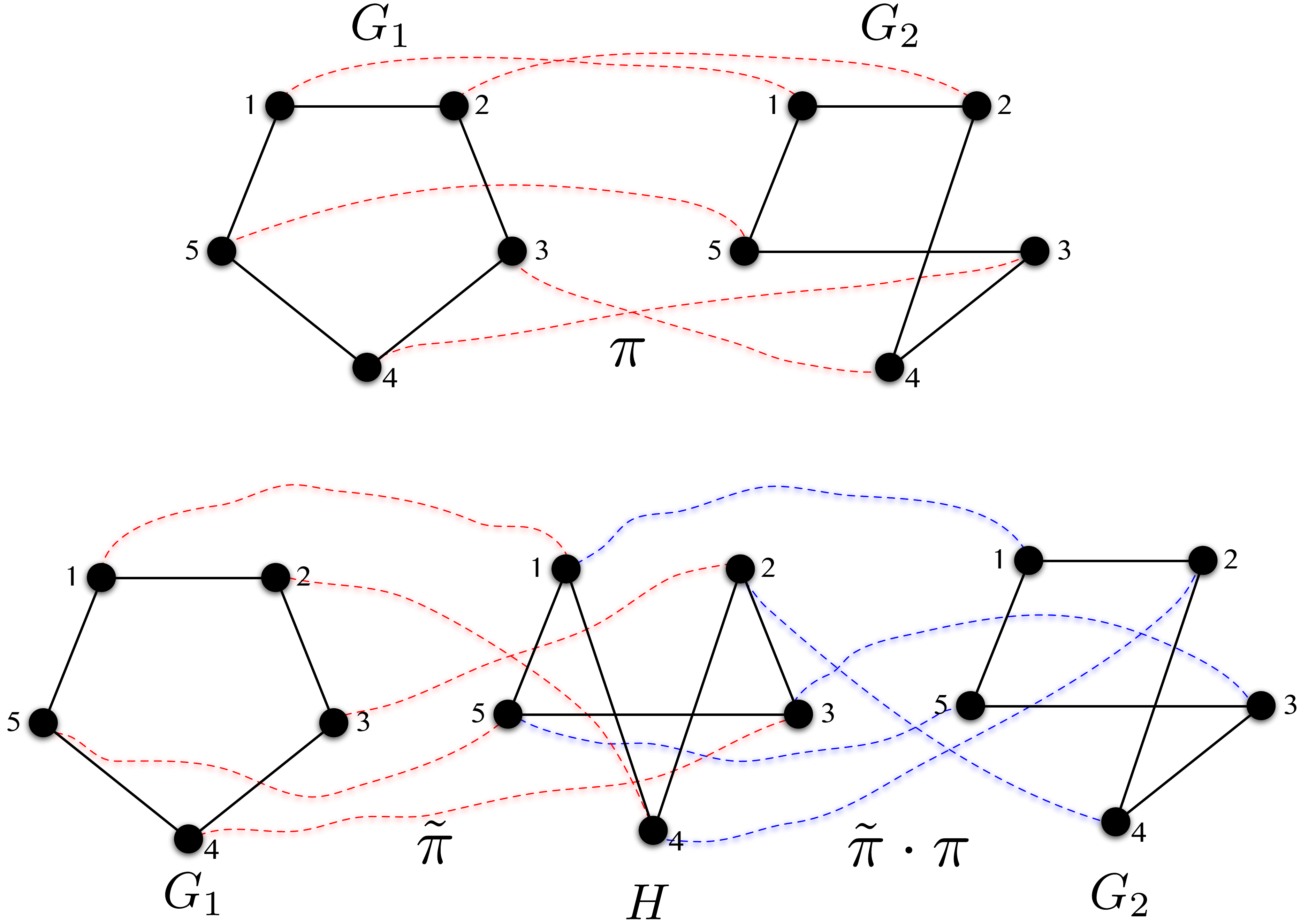}
\captionspacefig \caption{The isomorphisms \mbox{$G_1\sim G_2$} (top), and \mbox{$G_1\sim H\sim G_2$} (bottom). Coloured lines indicate the vertex relabelings associated with the respective isomorphisms. Knowing the latter two isomorphisms simultaneously implies knowledge of \mbox{$G_1\sim G_2$} via composition of the permutations. However, knowing only one of them does not, since $H$ is chosen randomly. By repeatedly proving knowledge of one of the isomorphisms with $H$ we achieve asymptotic certainty that the prover must have known \mbox{$G_1\sim G_2$}, without actually revealing the associated permutation.}\label{fig:ZKP_graph}
\end{figure*}
\fi

In the example isomorphism presented in Fig.~\ref{fig:ZKP_graph}(top), the vertex permutation mapping $G_1$ to $G_2$ could be expressed in vector form as,
\begin{align}
\pi = \begin{pmatrix}
	1 \\
	2 \\
	3 \\
	4 \\
	5
\end{pmatrix}_{G_1} \to \begin{pmatrix}
	1 \\
	2 \\
	4 \\
	3 \\
	5
\end{pmatrix}_{G_2},
\end{align}
where indices denote vertex labels, or equivalently via the permutation matrix,
\begin{align}
\pi = \begin{pmatrix}
	1 & 0 & 0 & 0 & 0 \\
	0 & 1 & 0 & 0 & 0 \\
	0 & 0 & 0 & 1 & 0 \\
	0 & 0 & 1 & 0 & 0 \\
	0 & 0 & 0 & 0 & 1
\end{pmatrix}.
\end{align}

\begin{table}[!htbp]
\begin{mdframed}[innertopmargin=3pt, innerbottommargin=3pt, nobreak]
\texttt{
function ZKP.GraphIsomorphism($G_1$, $G_2$):
\begin{enumerate}
	\item Graphs $G_1$ and $G_2$ are known to both verifier Victor, and prover Peggy.
	\item Peggy knows the permutation $\pi$ for the isomorphism \mbox{$G_1\sim G_2$},
	\begin{align}
		G_1 = \pi \cdot G_2.
	\end{align}
	\item Peggy wishes to prove to Victor that she knows $\pi$, without disclosing what it is.
	\item Peggy chooses another random permutation $\tilde\pi$, and constructs the new permuted graph $H$,
	\begin{align}
		H &= {\tilde\pi}\cdot G_1,\nonumber\\
		H &= {\tilde\pi}\cdot \pi \cdot G_2.
	\end{align}
	\item Peggy shares $H$ with Victor, randomly isomorphic to both $G_1$ and $G_2$.
	\item Victor randomly (\mbox{$p=1/2$}) asks Peggy to prove \textit{either} \mbox{$H\sim G_1$} or \mbox{$H\sim G_2$}.
	\item She accordingly reveals either $\tilde\pi$ or \mbox{$\tilde\pi\cdot\pi$} to Victor. He can now efficiently verify either \mbox{$H\sim G_1$} or \mbox{$H\sim G_2$} respectively, by performing the inverse permutation,
	\begin{align}
		G_1 &= {\tilde\pi}^{-1} \cdot H,\nonumber\\
		G_2 &= ({\tilde\pi}\cdot \pi)^{-1} \cdot H.
	\end{align}
	\item Victor is unable to determine $\pi$ from either scenario alone, but could were he to know \textit{both} isomorphisms simultaneously, since,
	\begin{align}
		\pi = {\tilde\pi}^{-1}\cdot (\tilde\pi\cdot\pi).
	\end{align}
	\item The above is repeated $n$ times. Each time, Peggy chooses a new random $\tilde\pi$.
	\item If Peggy does not actually know $\pi$, the probability of fraudulently passing this test $n$ times is,
	\begin{align}
		P_\text{deceive} = \frac{1}{2^n}	.
	\end{align}
	\item With confidence \mbox{$1-P_\text{deceive}$}, Victor knows that Peggy knows $\pi$, without knowing it himself.
	\item $\Box$
\end{enumerate}}
\end{mdframed}
\captionspacealg \caption{A zero-knowledge proof for the graph isomorphism problem. Victor (verifier) provides two graphs to Peggy (prover), who can demonstrate with asymptotic certainty that she knows their isomorphism, without disclosing the associated permutation relating them.} \label{alg:ZKP_graph}\index{Zero-knowledge proofs}\index{Graph isomorphism}
\end{table}

To provide a ZKP for this, Peggy's goal is to prove that she knows $\pi$, without explicitly revealing it. An efficient, randomised, classical ZKP protocol for achieving this is provided in Alg.~\ref{alg:ZKP_graph}, and a specific example illustrated in Fig.~\ref{fig:ZKP_graph}. The key underlying principle here is to obscure $\pi$ through randomisation, whereby instead of directly proving knowledge of $\pi$, it is implied via multiple proofs of isomorphisms with intermediate random graphs.

\latinquote{Semper fidelis.}

\sketch{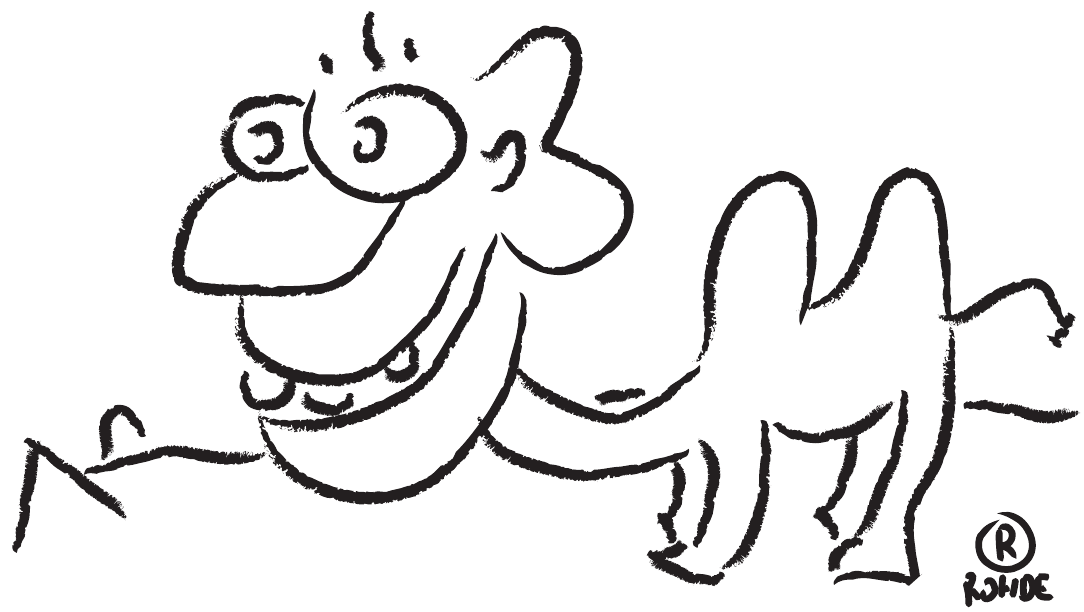}

\clearpage
%
% Economics & Politics
%

\part{Economics \& politics}\label{part:economics}\index{Economics}\index{Politics}

%
% Economics & Politics of the Quantum Internet
%

\famousquote{If you put the federal government in charge of the Sahara Desert, in five years there'd be a shortage of sand.}{Milton Friedman}
\newline

\famousquote{It is not from the benevolence of the butcher, the brewer, or the baker that we expect our dinner, but from their regard to their own interest.}{Adam Smith}
\newline

\dropcap{A}{ny} form of computation comes at an economic cost, but also brings with it a payoff. A key consideration in any model for computation is the tradeoff between the two. Because the computational power of quantum computers scales inherently differently than classical computers, we expect economic indicators to exhibit different scaling characteristics and dynamics also, thereby fundamentally altering the economic landscape of the post-quantum world.

We will now treat some of these economic issues in the context of a global network of unified quantum computing resources, which are then equitably time-shared\index{Time-sharing}. We argue in Sec.~\ref{sec:time_share} that this time-shared model for quantum computation is always more computationally efficient than having distinct quantum computers operating independently in parallel, owing to the super-linear scaling in their joint computational power. While this section provides mathematical details of various economic models, Secs.~\ref{sec:economics} \& \ref{sec:quant_coin_essay} provide a popular, high-level discussion surrounding these issues. Sec.~\ref{sec:summary_economic_models} summarises the various economic models we present in this part.

%
% Classical-Equivalent Computational Power & Computational Scaling Functions
%

\section{Classical-equivalent computational power \& computational scaling functions}\index{Classical-equivalent computational power}\index{Computational!Scaling functions}\label{sec:comp_sc_func}

\dropcap{L}{et} $t$ be the classical-equivalent runtime\index{Classical-equivalent computational power} of a quantum algorithm comprising $n$ qubits -- that is, how long would a given classical computer require to implement this $n$-qubit quantum computation? We define a \textit{computational scaling function}\index{Computational!Scaling functions} characterising this relationship,

\begin{definition}[Computational scaling functions] \label{def:scaling_func}\index{Computational!Scaling functions} 
The computational scaling function, $f_\mathrm{sc}$, relates the number of qubits held by a quantum computer, $n$, and the classical-equivalent runtime\index{Classical-equivalent computational power}, $t$, of the algorithm it implements,
\begin{align}
t = f_\mathrm{sc}(n),
\end{align}
	where $f_\mathrm{sc}$ is monotonically increasing, and depends heavily on both the algorithm being implemented, as well as the architecture of the computer, including the computational model and choice of fault-tolerance protocol.
\end{definition}

The exact form of the scaling function will be specific to the algorithm being deployed\footnote{For example, the \textit{circuit depth}\index{Circuit!Depth}, i.e number of gate applications in series, will heavily influence the number of classical steps required to simulate the circuit.}, and the computational model (e.g cluster states vs the circuit model, as well as choices in error correction, amongst other factors). Most notably, different quantum algorithms offer different scalings in their quantum speedup -- Grover's algorithm (Sec.~\ref{sec:quantum_search})\index{Grover's algorithm} offers only a quadratic quantum speedup, compared to the exponential speedup afforded by Shor's algorithm (Sec.~\ref{sec:shors_alg})\index{Shor's algorithm}. Thus, the computational scaling function depends on both the hardware and software, and may therefore differ between different users operating the same computer. We abstract this away and assume all these factors and resource overheads have been merged into the scaling function.

%
% Virtual Scaling Functions
%

\subsection{Virtual computational scaling functions}\index{Virtual computational scaling functions}

If a network of quantum computers were combined into a single, larger \textit{virtual quantum computer}\index{Virtual quantum computer} (Sec.~\ref{sec:GVQC}) using a distributed model for quantum computation (Sec.~\ref{sec:dist_QC}), we can define a computational scaling function relationship for the virtual device,

\begin{definition}[Virtual scaling function]\index{Virtual computational scaling functions}
The joint classical-equivalent runtime\index{Classical-equivalent computational power} of a distributed virtual quantum computation over a network is,
\begin{align}
t_\mathrm{joint} = f_\mathrm{sc}^\mathrm{virtual}(n_\mathrm{global}),
\end{align}
where,
\begin{align}
n_\mathrm{global} = \sum_{j\in\mathrm{nodes}} n_j,
\end{align}
is the total number of qubits in the network, with $j$ summing over all nodes in the network, each of which holds $n_j$ qubits. $f_\mathrm{sc}^\mathrm{virtual}$ is obtained from $f_\mathrm{sc}$ by factoring in network overheads and inefficiencies. With perfect network efficiency, \mbox{$f_\mathrm{sc}^\mathrm{virtual}=f_\mathrm{sc}$}.
\end{definition}

%
% Combined Computational Scaling Functions
%

\subsection{Combined computational scaling functions}\index{Combined computational scaling functions}\label{sec:comb_comp_sc_func}

Until now we have characterised the entire network by a single scaling function. Of course, the scaling functions observed by different market participants needn't all be the same, as they are functions of not only the hardware, but also the participants' different algorithmic applications (i.e software).

Consider taking a single unit of time (i.e we are ignoring cost discounting\index{Cost discounting} over multiple units of time) and dividing it amongst a number of nodes, $n_\mathrm{nodes}$, each with their own scaling function, $f_\mathrm{sc}^{(i)}$. The total classical-equivalent runtime of the computation is additive, given simply by a linear combination of the classical-equivalent processing times of the individual nodes. This yields the relationship for combining scaling functions,
\begin{definition}[Combined scaling functions]\index{Combined computational scaling functions}\label{def:comb_sc_func}
The effective combined computational scaling function, $f_\mathrm{sc}^\mathrm{(joint)}$, of a group of participants, each with their own scaling functions, $f_\mathrm{sc}^{(i)}$, is given by,
\begin{align}
	t_\mathrm{joint} &= \sum_{i=1}^{n_{\mathrm{nodes}}} \beta_i \cdot f_\mathrm{sc}^{(i)}(n_\mathrm{global}) \nonumber \\
	&= f_\mathrm{sc}^\mathrm{(joint)}(n_\mathrm{global}),
\end{align}
where $\beta_i$ characterise the share of processing time allocated to each node, and for normalisation,
\begin{align}
\sum_{i=1}^{n_\mathrm{nodes}} \beta_i = 1.
\end{align}
\end{definition}

Thus, the joint scaling function of the entire network is simply given by a linear combination (weighted average) of the scaling functions of the different market participants.

%
% Per-qubit computational power
%

\section{Per-qubit computational power}\label{sec:NPSF}\index{Per-qubit computational power}

\dropcap{O}{ne} parameter that appears ubiquitously in the upcoming economic models and warrants a definition of its own is the computational power of a quantum computer per qubit. This relates the power and size of the computer. We define this as the \textit{per-qubit computational power},

\begin{definition}[Per-qubit computational power]\label{def:NPSF}\index{Per-qubit computational power}
The per-qubit computational power is defined as the computational power per qubit,
\begin{align}
\chi_\mathrm{sc}(n) = \frac{f_\mathrm{sc}(n)}{n}.
\end{align}
\end{definition}

This parameter acts as an overall, network size-dependent price scaling factor on:
\begin{itemize}
\item Quantum computational leverage (Sec.~\ref{sec:quant_ec_lev}).
\item Cost of computation (Sec.~\ref{sec:cost_of_comp}).
\item Time-shared computational power (Sec.~\ref{sec:arb_free_time_share}).
\item Quantum computational leverage (Sec.~\ref{sec:quant_ec_lev}).
\item Forward contracts (Sec.~\ref{sec:for_contr}).
\end{itemize}

\begin{figure}[htb!]\index{Per-qubit computational power}
	\includegraphics[clip=true, width=0.475\textwidth]{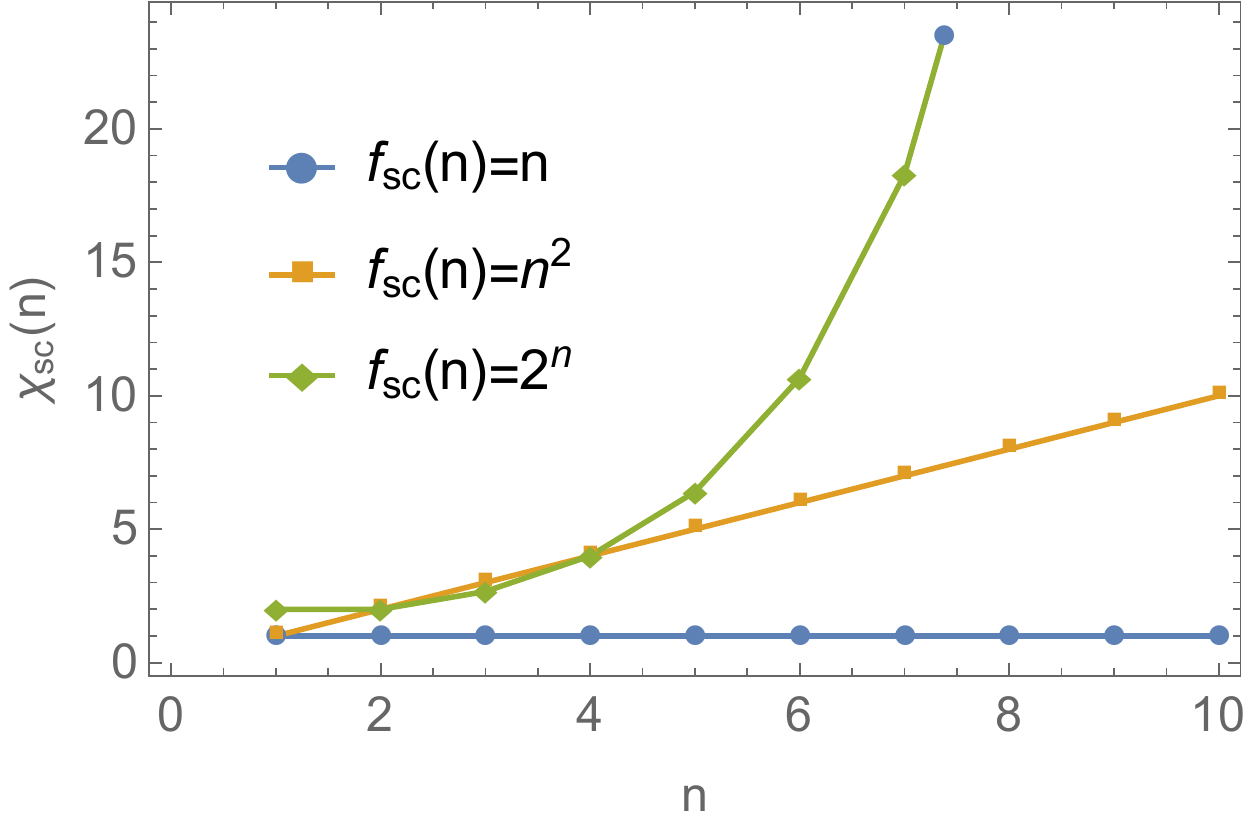}
	\captionspacefig \caption{Per-qubit computational power, $\chi_\mathrm{sc}$, as a function of several representative computational scaling functions, $f_\mathrm{sc}$, where $n$ is network size.} \label{fig:NPSF}
\end{figure}

This parameter lends itself to the elegant interpretation as a cost multiplier on qubit asset, dividend and derivative prices, which warrants investigation of its scaling characteristics, shown in Fig.~\ref{fig:NPSF}. 

Note that in the quantum context, the computational power per qubit is not intrinsic to the qubit itself, but depends on how many qubits it cooperates with, a phenomena which does not arise in the classical context.

The key observation is that this scaling factor is constant for classical computing, where the scaling function is linear, but monotonically increasing for any super-linear scaling function. For polynomial scaling functions, it has the effect of reducing the order of the polynomial by one. And for exponential scaling functions, it remains exponential.

%
% Time-Sharing
%

\section{Time-sharing}\label{sec:time_share}\index{Time-sharing}

\dropcap{S}{uppose} Alice and Bob both possessed expensive classical Cray\texttrademark\,supercomputers\index{Supercomputers}, both identical. They're both connected to the internet, so does it make sense to unify their computational resources over the network to construct a more powerful virtual machine\index{Virtual quantum computer}, which they subsequently time-share\index{Time-sharing} between themselves, or are they better off just using their own computers independently?

If there were an asymmetry in demand for computational resources, it would make perfect sense to unify computational resources, so as to mitigate wasting precious clock-cycles. However, if they were both heavy users, always consuming every last clock-cycle, it would make no difference: for a given computation, Alice and Bob could each be allocated half the processing time of the virtual supercomputer twice as powerful; or, each could exploit the full processing time of their half-as-fast computers. In either case, the dollar cost of the computation is the same. This simple observation follows trivially from the linear relationship between processing power and the number of CPUs in a classical computer.

More generally, in a networked environment where time-sharing\index{Time-sharing} of classical computational resources is applied equitably, proportionate to nodes' contribution to the network, the dollar cost per computation is (roughly, modulo parallelisation overheads) unaffected by the rest of the network. Instead, the motivation for networking computational resources is to improve efficiency by ensuring that clock-cycles are not wasted, but instead distributed according to demand by a scheduling algorithm, which could be market-driven, for example.

However, the computational power of a quantum computer generally doesn't scale linearly with its number of qubits, but super-linearly, often exponentially. This completely changes the economics, and market dynamics of networked quantum computers. Intuitively, we expect equitable time-sharing of unified quantum computational resources to offer more performance to all nodes than if they were to exclusively use their own resources in isolation. That is, the cost of a computation is reduced by resource-sharing, even after time-sharing.

For this reason, henceforth we will assume an environment in which owners of quantum hardware network and unify their computational power, sharing the virtual quantum computer's power between them.

In Sec.~\ref{sec:arb_free_time_share} we present an explicit model for equitable time-sharing, which is optimal from a market perspective.

%
% Economic Model Assumptions
%

\section{Economic model assumptions}

\dropcap{B}{efore} proceeding with explicit derivations of economic models, we state some assumptions about the dynamics of a marketplace in quantum assets. These assumptions are largely based on historical observations surrounding classical technologies that we might reasonably expect to also apply in the quantum era. However, given that the quantum marketplace is one that hasn't been explored in detail until now, it may be the case that some of these assumptions will require revision. Nonetheless, the general techniques we employ could readily be adapted to some relaxations and variations in these assumptions.

%
% Efficient Markets
%

\subsection{Efficient markets}\label{sec:eff_markets} \index{Efficient markets}

\famousquote{In a dream it's typical not to be rational.}{John Nash}
\newline

We make several assumptions about the efficiency of the quantum marketplace. These are largely based on the conventional efficient-market hypothesis (EMH)\index{Efficient-market hypothesis (EMH)}, readily taught in undergraduate ECON101 and subsequently summarily rejected upon entering ECON202. For ease of exposition, we will remain in the ECON101 classroom.

Some of these assumptions may reasonably turn out to be invalid, or require revision as we learn more about upcoming quantum technologies and the trajectories their marketplace will follow. However, for ease of exposition, and the purposes of presenting some initial rudimentary, \textit{qualitative} analyses and thought experiments, these assumptions simplify our derivations and act as a good starting point for future, more rigorous treatment (which we highly encourage!).

Given that the quantum marketplace doesn't actually exist yet, it isn't immediately clear which assumptions are likely to be valid or not, and future, more sophisticated models will inevitably need to make more appropriate assumptions. Certainly it's no secret that in conventional settings the EMH is flawed in many respects, and some of its idealised assumptions break down in reality.

\begin{postulate}[Efficient markets]\label{post:market_eff}\index{Efficient markets} We make the following efficiency assumptions on the dynamics of the quantum marketplace:
\begin{itemize}
	\item Qubits are a `scarce' resource: there is always positive, non-zero demand for them.\index{Scarcity}
	\item No wastage: quantum computational resources are always fully utilised, with no down-time.\index{Wastage}
	\item Transaction free: transaction costs are negligible, for both quantum assets and their derivatives.\index{Transaction cost}
	\item Negligible cost-of-carry: e.g storage and maintenance costs are negligible.\index{Cost of carry}
	\item High liquidity: it is always possible to execute transactions at market rates.\index{Liquidity}
	\item Perfect competition: there are no monopolies gouging prices, which are in equilibrium.\index{Perfect competition}
	\item Arbitrage-free: market rates for different assets and derivatives are perfectly consistent, with no opportunity for `free money' by trading on market discrepancies.\index{Arbitrage-free}
	\item Perfect information: all market participants have complete knowledge of all market variables, including one another.\index{Perfect information}
	\item Rational markets: all market participants act rationally\footnote{i.e with perfect economic self-interest \latinquote{Avaritia}.} upon available information.\index{Rational markets}
	\item Indefinite asset lifetime: there is no deterioration or death of quantum hardware over time.\index{Asset lifetime}
	\item There is a risk-free rate of return ($r_\mathrm{rf}$)\index{Risk-free!Rate of return}: the rate of growth exhibited by an investment into an optimal risk-free asset\footnote{Historically these risk-free assets are taken as being US government bonds\index{US government bonds}\index{Risk-free!Asset}, with the bond yield being the risk-free RoR.}
	\end{itemize}
\end{postulate}

%
% Central Mediating Authority
%

\subsection{Central mediating authority}\index{Central mediating authority}

In Sec.~\ref{sec:time_share} we argued that because of the super-linear scaling in the computational power of networked quantum computers, it will be most economically efficient to unify the world's entire collective quantum computational resources over the network and time-share their joint computational power. For this reason, we will assume that global quantum computing resources are unified, and time-shared equitably (as will be described in Sec.~\ref{sec:arb_free_time_share}), overseen by a trusted central authority, congruent with our efficient market assumptions (Sec.~\ref{sec:eff_markets}).

The role of the mediating authority is to perform process scheduling\index{Scheduling} -- equitably allocating algorithmic runtime on the virtual computer to the different network participants. This could be in the form of a state-backed authority, or open market-driven alliances. In any case, the job of the authority is a relatively straightforward one, and we will assume it induces negligible cost and computational overhead, remaining largely transparent to the end-user.

However, as discussed in Sec.~\ref{sec:GVQC}, it may be the case that competing strategic interests will drive a wedge between the quantum resources of competitors and adversaries, partitioning them into a set of smaller networks, divided across strategic boundaries. In this instance, the arguments presented in the upcoming sections will apply to these smaller, isolated networks individually.

%
% Network Growth
%

\subsection{Network growth} \index{Network!Growth}

We assume the number of qubits in the global network in the future is growing exponentially over time, i.e the rate of progress of quantum technology will observe a Moore's Law-like behaviour, as with the classical transistor.

This is a reasonable assumption based on the observation of this ubiquitous kind of behaviour in present-day technologies. Classical computing has been on a consistent exponential trajectory since the 1980's, and although it must eventually asymptote, it shows no sign of doing so in the immediate future. Quantum technologies sit at the entry point to this trajectory, and we expect it to continue for the medium-term. Thus, we let the number of qubits in the network be,
\begin{postulate}[Network growth]\label{post:net_growth}\index{Network!Growth}
The number of qubits in the global quantum internet is growing exponentially over time as,
\begin{align}
	N(t) = N_0 {\gamma_N}^{t},
\end{align}
where \mbox{$\gamma_N\geq 1$} characterises the rate of exponential growth in the number of qubits available to the quantum network.
\end{postulate}

The exact value of the growth rate, $\gamma_N$, is obviously unclear at such early stages in the development of the market and will ultimately be determined empirically. Although in the case of classical computing we have seen a very consistent doubling of computational power roughly every 18 months. This may very well be different for quantum technologies, owing to their fundamentally different engineering requirements (which are far more challenging in general).

%
% Hardware Cost
%

\subsection{Hardware cost} \index{Hardware cost}

Let the dollar cost of physical qubits follow Moore's Law-like dynamics, decreasing exponentially with time,
\begin{postulate}[Hardware cost]\label{post:hardware_cost}\index{Hardware cost}
The dollar-cost of a single physical qubit scales inverse exponentially against time as,
\begin{align}
	C(t) = C_0 {\gamma_C}^{-t},
\end{align}
where \mbox{$\gamma_C\geq 1$} characterises the decay rate.
\end{postulate}

This is consistent with the observed evolution of classical hardware since the beginning of the digital revolution, and it is reasonable to think that technological progress in the quantum era will follow a similar trajectory.

%
% Network Power
%

\section{Network power}\index{Network!Power}\label{sec:network_power}

\dropcap{F}{irst} and foremost, with a fully interconnected quantum computational network, what is the projection of its net computational power now and into the future? This is simply obtained via the joint computational scaling function applied to projected network size,

\begin{postulate}[Network power]\label{post:network_power}\index{Network!Power}
The combined computational power of the entire network, measured in classical-equivalent runtime (i.e FLOPs), is given by,
\begin{align}
P(t) &= f_\mathrm{sc}(n_\mathrm{global})\nonumber \\
&= f_\mathrm{sc}(N_0{\gamma_N}^t).
\end{align}
\end{postulate}

%
% Network Value
%

\section{Network value}\index{Network!Value}\label{sec:network_value}

\dropcap{T}{he} simplest economic metric one might define is the collective dollar value of the entire network. That is, the product of the number of qubits on the network and the dollar cost per physical qubit at a given time.

\begin{postulate}[Network value]\label{post:network_value}\index{Network!Value}
The dollar-value of the entire network is given by,
\begin{align}
	V(t) &= C(t) N(t) \nonumber \\
	&= C_0 N_0 \left(\frac{\gamma_N}{\gamma_C}\right)^t.
\end{align}
\end{postulate}

\begin{figure}[!htbp]
	\includegraphics[clip=true, width=0.4\textwidth]{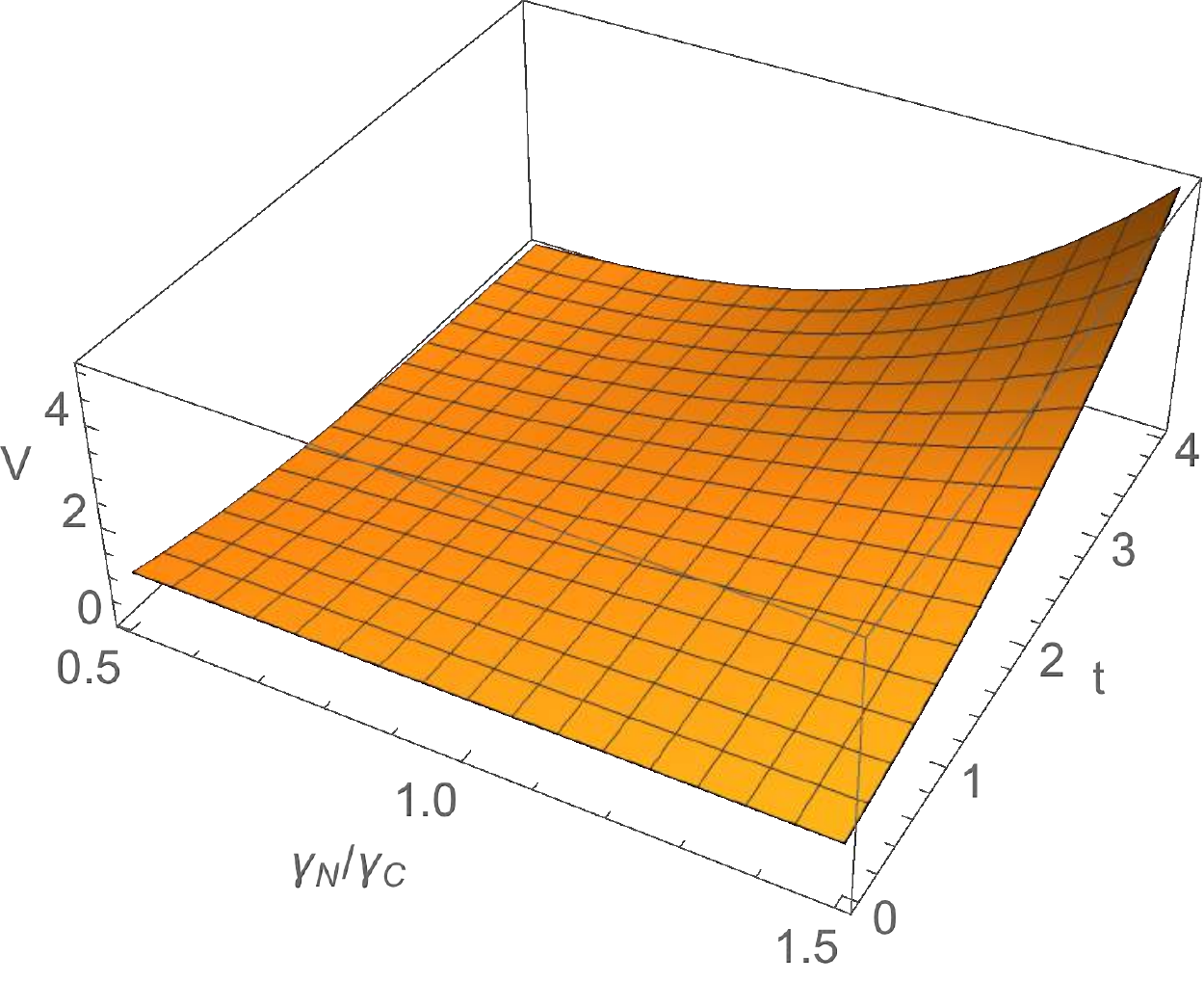}
	\captionspacefig \caption{Dollar value (in units of \mbox{$C_0N_0$}) of the network as a function of time, growth rate in the number of physical qubits, and rate of decay in the dollar value of physical qubits. When \mbox{$\gamma_N/\gamma_C=1$} the network's value remains constant over time. Above this the network's value appreciates exponentially, and below which it depreciates exponentially against time.} \label{fig:network_value}
\end{figure}

Note that the collective value of the network appreciates exponentially if the rate of network growth is greater than the rate of decay in the value of physical qubits, otherwise it depreciates. At \mbox{$\gamma_C=\gamma_N$} the network's dollar value remains constant over time, even if it continues expanding. This is shown in Fig.~\ref{fig:network_value}.

%
% Rate of Return
%

\section{Rate of return}\index{Rate of return}

\dropcap{T}{he} execution of computations typically has monetary value to the consumer. After all, they are paying hard-earned money for access to the technology!

Suppose the owners of the quantum hardware are not running computations themselves, but rather are collectively licensing out their joint compute-time to end-users. The hardware owners will of course be demanding a profit from their enterprise. The rate at which they earn back their investment into hardware via the licensing of compute-time, we will refer to as the rate of return (RoR), $\gamma_\mathrm{ror}$. We define this as,

\begin{postulate}[Rate of return]\label{post:RoR}\index{Rate of return}
The RoR is defined as,
\begin{align}
e^{\gamma_\mathrm{ror}(t)} = \frac{R(t)}{V(t)},
\end{align}
where $R(t)$ is the profit made by licensing out the network's joint compute-power for a single unit of time, given a present-day network value of $V(t)$.
\end{postulate}

A higher $\gamma_\mathrm{ror}$ implies a faster payback rate on hardware investment\footnote{We have parameterised the RoR as an exponential for convenience when performing derivations with compounding.}.

%
% Market Competitiveness
%

\section{Market competitiveness}\index{Market competitiveness}\index{Efficiency}

\dropcap{R}{ecall} the risk-free RoR is $r_\mathrm{rf}$\index{Risk-free!Rate of return}. The difference between the RoR on our investment into qubit assets and the risk-free rate effectively tells us our profitability relative to a baseline zero-risk asset. This difference in turn can be interpreted as an indicator of the competitiveness or efficiency of the market -- more efficient and competitive markets exhibit narrower profit windows. This is yields the figure of merit,

\begin{postulate}[Market competitiveness]\label{post:market_comp}\index{Market competitiveness}
The \textit{competitiveness} or \textit{efficiency} of the qubit market is given by the difference between the risk-free RoR and that of our physical qubits,
\begin{align}
\xi_\mathrm{comp} = \gamma_\mathrm{ror} - r_\mathrm{rf}.
\end{align}

There are three distinct regimes for market competitiveness:
\begin{itemize}
	\item \mbox{$\xi_\mathrm{comp}=0$}: The market exhibits perfect efficiency, since price competition\index{Price competition} is so strong that profit windows have narrowed to vanishing point. There is no profit incentive to buy into or sell qubits, since they have converged with the risk-free asset.
	\item \mbox{$\xi_\mathrm{comp}>0$}: The market is profit-making for qubit owners. There is a profit incentive to buy ownership of physical qubits and license them out on the time-share market.
	\item \mbox{$\xi_\mathrm{comp}<0$}: The market is loss-making for qubit owners. Purchasing of physical qubits is disincentivised, since it's more optimal to buy into the zero-risk asset than hold qubit assets.
\end{itemize}
\end{postulate}

The \mbox{$\xi_\textrm{comp}=0$} limit is really a fairly hypothetical regime which ought not to arise in real markets, which necessarily exhibit inefficiencies. However, highly-competitive real markets will asymptote to the efficient regime, \mbox{$\xi_\textrm{comp}\approx 0$}.

%
% Cost Of Computation
%

\section{Cost of computation}\label{sec:cost_of_comp} \index{Cost of computation}

As discussed in relation to combined computational scaling functions (Sec.~\ref{sec:comb_comp_sc_func})\index{Combined computational scaling functions}, different market participants will be executing different software applications on their share of the quantum computing resources, with differing QCLs\index{Quantum computational leverage}. Because the applications differ between users, as do their computational scaling functions ($f_\mathrm{sc}$), QCLs, so too does the monetary value of the computations they are performing. This yields the distinction between \textit{subjective}\index{Subjective value of computation} and \textit{objective}\index{Objective value of computation} value of computation:

\begin{itemize}
\item Subjective value of computation: the value to an end-user of a computation, measured in terms of their associated monetary profit from utilising its output, which is highly application-specific.
\item Objective value of computation: the cost of the physical hardware and infrastructure, which is not application-specific, but rather stipulated by technological and manufacturing progress.
\end{itemize}

This effectively implies that some users pay more for computation (in terms of return on investment\index{Return on investment (RoI)}) than others. While the objective cost of computation is conceptually simple to model (as performed in a rudimentary fashion in Sec.~\ref{sec:cost_of_comp}), the subjective cost is a highly non-trivial one. It will depend heavily on the scaling function of the algorithm run by a user, and of course the economic objectives of their computation -- a quantum simulation algorithm executed by an R\&D lab\index{R\&D} is likely to be of greater monetary value than an undergrad using his university's resources to execute the same task for completing an assignment!

\subsection{Objective value}\index{Objective value of computation}

\dropcap{I}{n} the same scenario as before, where compute-time is being licensed out to end-users, the hardware owner's return over a single unit of time equates to the cost of computation over that period.

Let $L(t)$ be the dollar-value of utilising the network's computing resources for a single unit of time. This is obtained as the return made on the value of the network per FLOP,
\begin{postulate}[Objective value of computation]\label{post:cost_comp}
The efficient-market dollar-value of a computation for a single unit of time at time $t$, per FLOP is,
\begin{align}\index{Cost of computation}
	L(t) &= \frac{e^{\gamma_\mathrm{ror}t} V(t)}{P(t)} \nonumber\\
	&= \frac{e^{\gamma_\mathrm{ror}t} C_0{\gamma_C}^{-t}}{\chi_\mathrm{sc}(N_0 {\gamma_N}^t)}.
\end{align}
\end{postulate}
which implies,
\begin{postulate}[Spot price of computation]\index{Spot price of computation} The present-day (\mbox{$t=0$}) spot price of a computation per FLOP is,
\begin{align}
L(0) = \frac{C_0}{\chi_\mathrm{sc}(N_0)}.	
\end{align}
\end{postulate}
That is, the value of computations simply approximates the return on initial hardware investment, scaled by its initial computational power, as is intuitively expected.

Note that if $f_\mathrm{sc}$ scales linearly, as per classical computation, we observe a regular exponential decay in the cost of computation, consistent with the classical Moore's Law. On the opposing extreme, for exponentially quantum-enhanced $f_\mathrm{sc}$, the cost of computation decreases super-exponentially with time, an economic behaviour unique to post-classical computation with no classical analogue.

The time-derivative of the cost of computation is strictly negative, assuming correctness of the growth and cost postulates (Post.~\ref{post:net_growth} \& \ref{post:cost_comp}),
\begin{align}
\frac{\partial L}{\partial t} \leq 0,	
\end{align}
which implies monotonic reduction in the cost of computation over time, unless network growth and cost completely freeze (\mbox{$\gamma_N=\gamma_C=0$}), in which case the cost of computation remains flat.

Examples of the temporal dynamics of the cost of computation are shown in Fig.~\ref{fig:econ_cost_of_comp} for representative computational scaling functions.

\if 1\doublecol
\begin{figure}[!htbp]
\includegraphics[clip=true, width=0.4\textwidth]{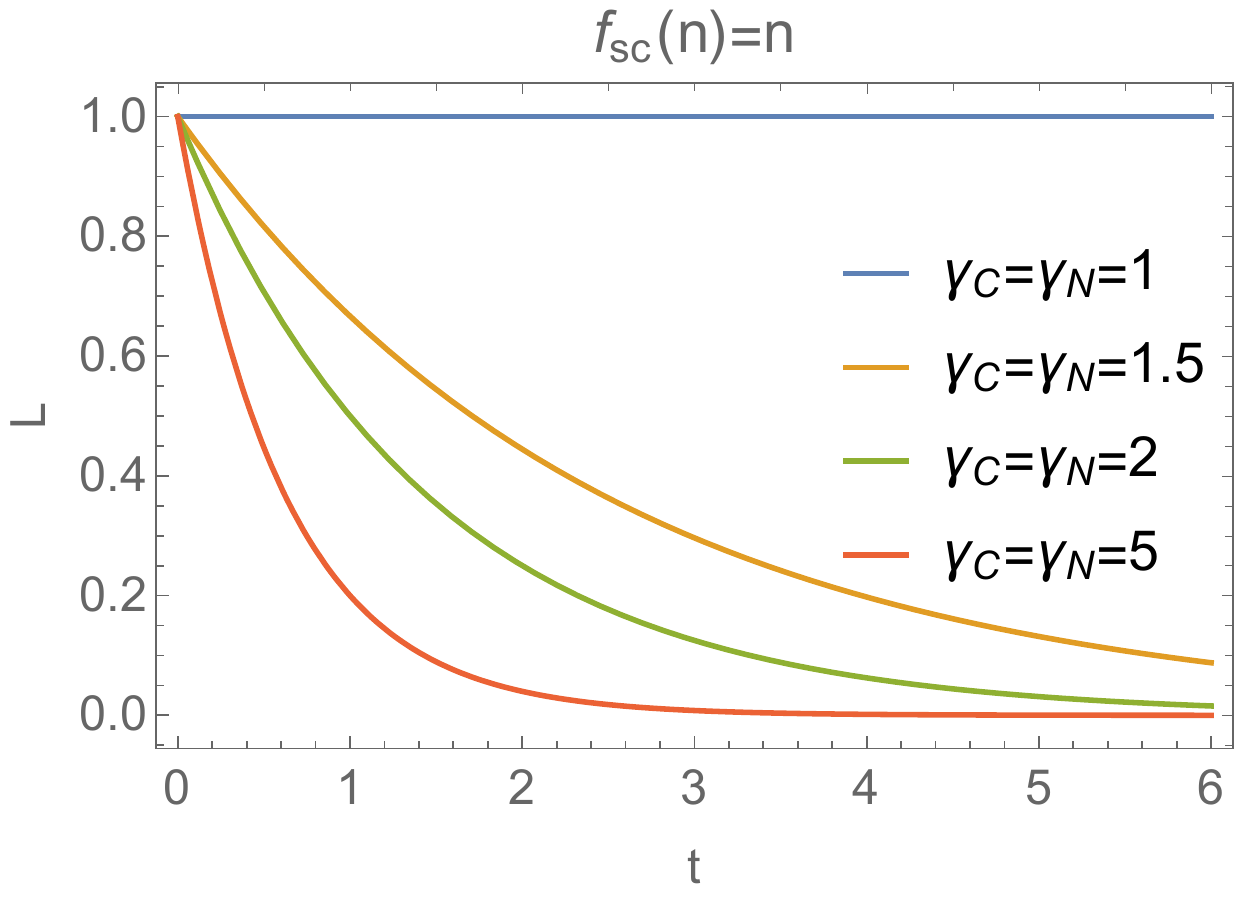}\\
\includegraphics[clip=true, width=0.4\textwidth]{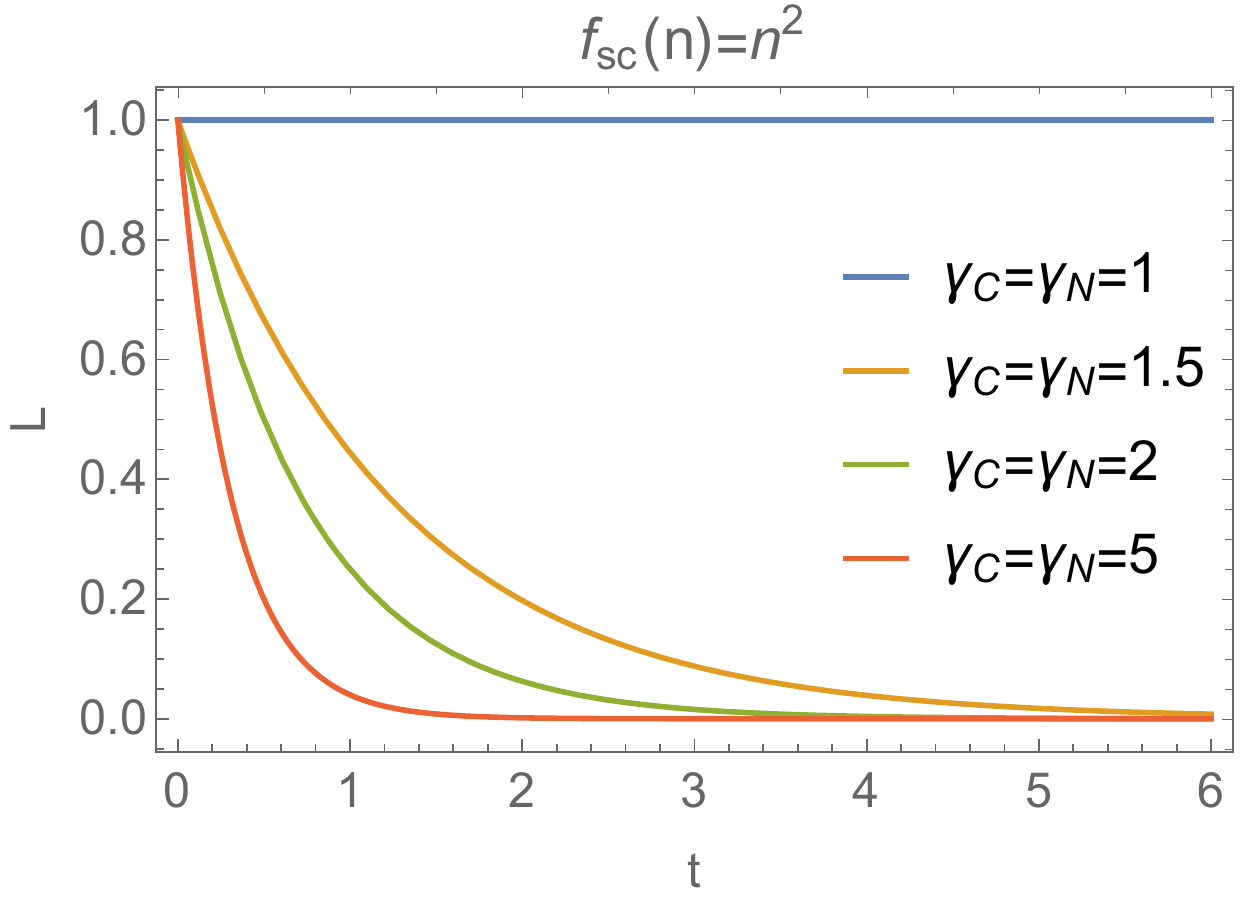}\\
\includegraphics[clip=true, width=0.4\textwidth]{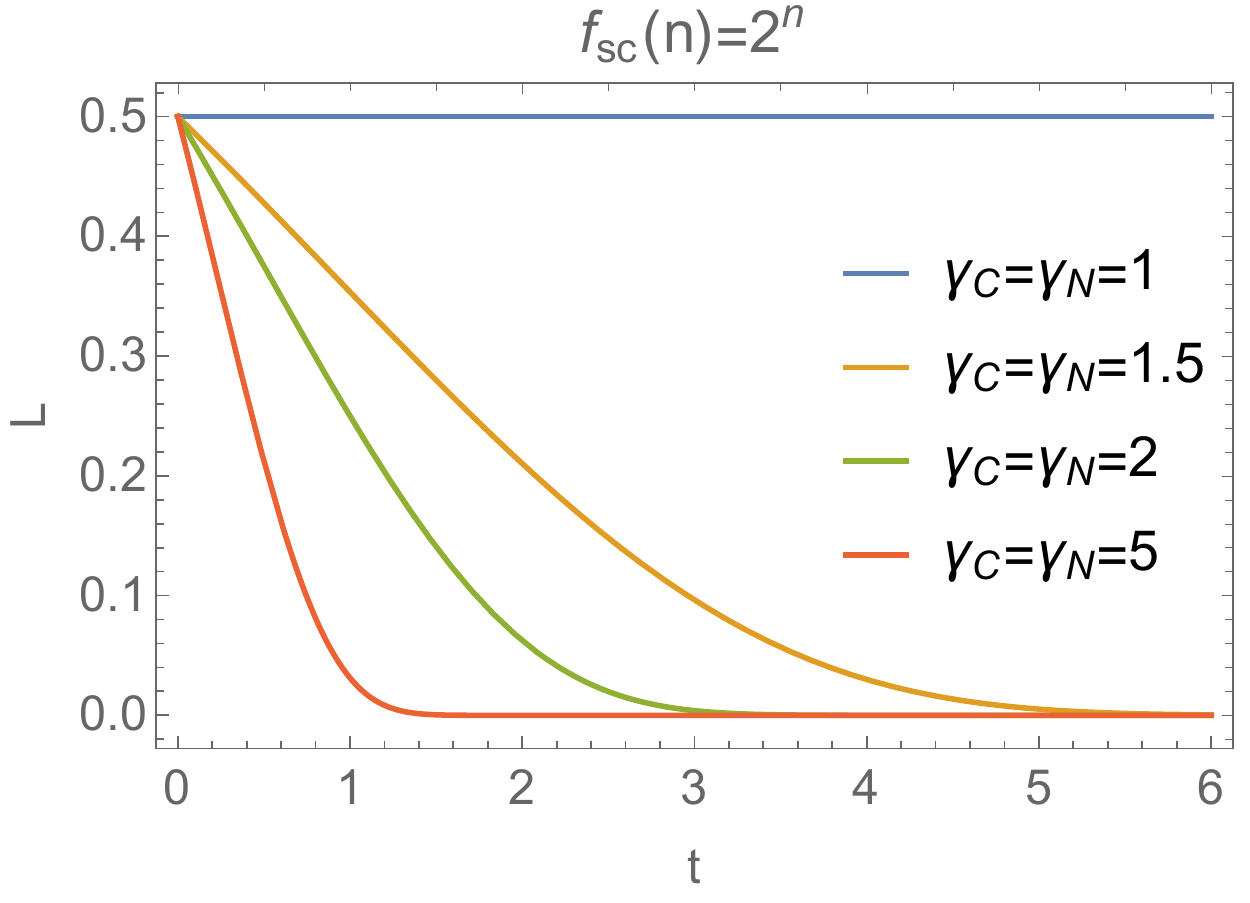}
\captionspacefig \caption{Examples of the temporal dynamics of the cost of computation for different scaling functions and exponential growth rates. Units are \mbox{$C_0=N_0=n=1$}, with RoR \mbox{$\gamma_\mathrm{ror}=0$}. A non-zero RoR would simply scale these figures by a constant factor of $e^{\gamma_\mathrm{ror}}$.}\label{fig:econ_cost_of_comp}
\end{figure}
\else
\begin{figure*}[!htbp]
\includegraphics[clip=true, width=0.325\textwidth]{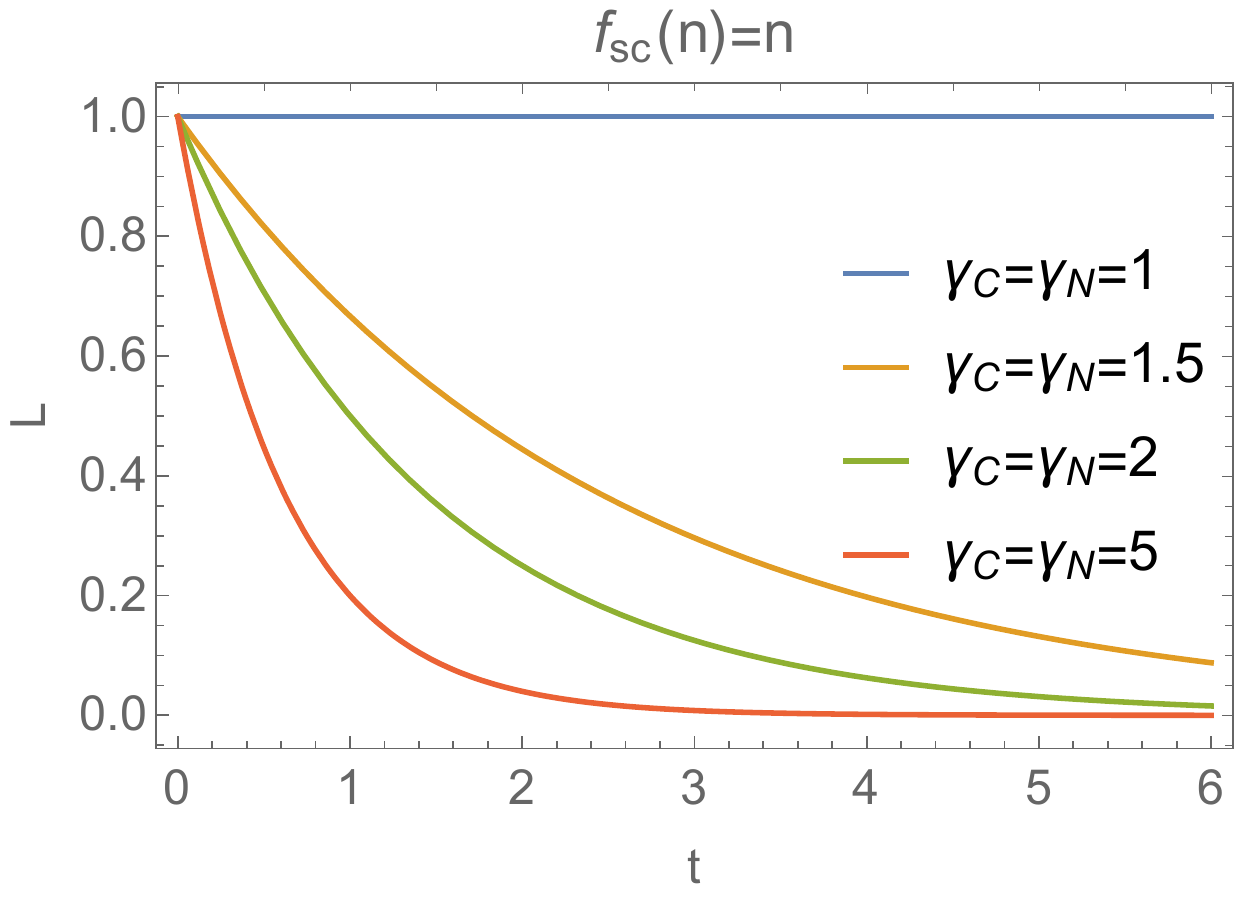}
\includegraphics[clip=true, width=0.325\textwidth]{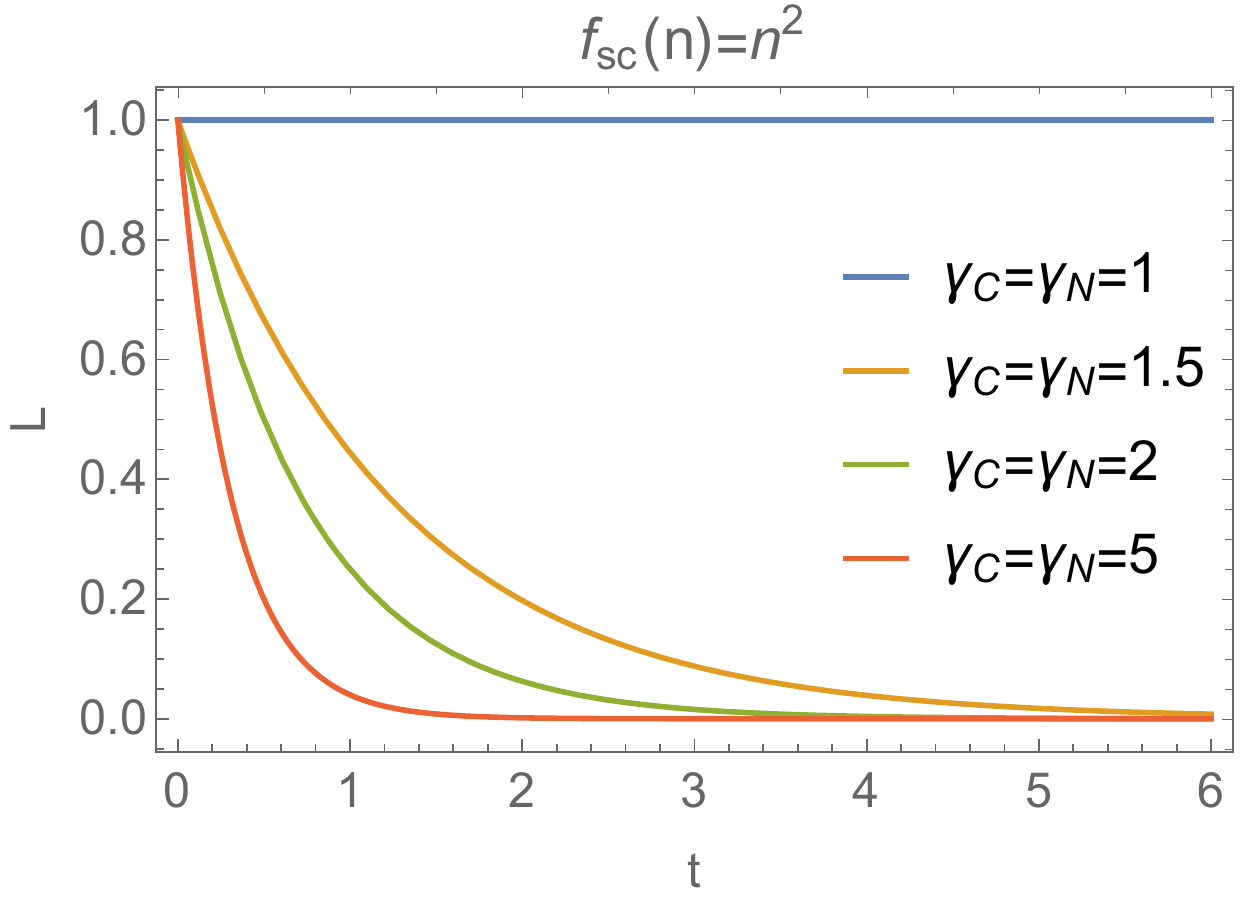}
\includegraphics[clip=true, width=0.325\textwidth]{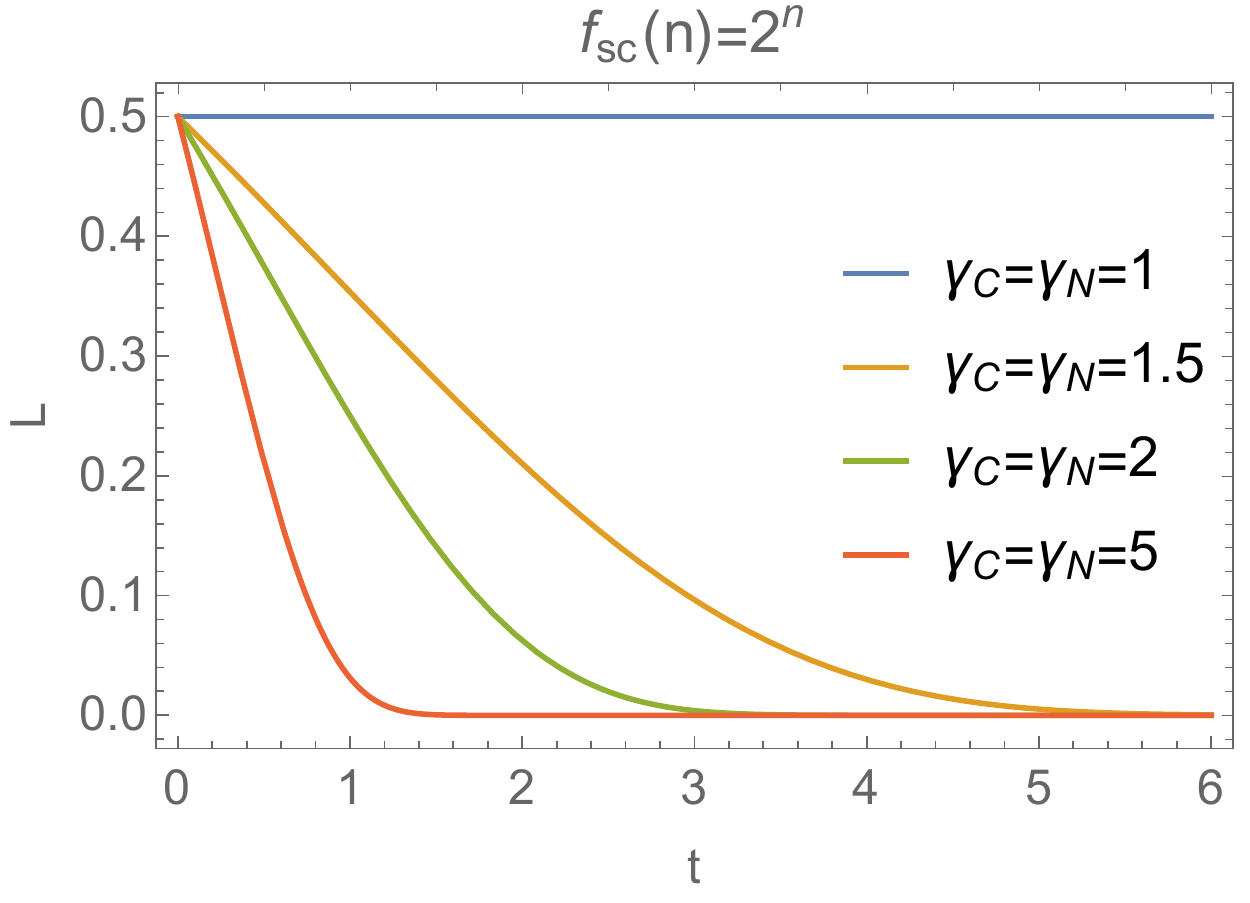}
\captionspacefig \caption{Examples of the temporal dynamics of the cost of computation for different scaling functions and exponential growth rates. Units are \mbox{$C_0=N_0=n=1$}, with RoR \mbox{$\gamma_\mathrm{ror}=0$}. A non-zero RoR would simply scale these figures by a constant factor of $e^{\gamma_\mathrm{ror}}$.}\label{fig:econ_cost_of_comp}
\end{figure*}
\fi

\subsection{Subjective value}\index{Subjective value of computation}

With access to $n$ qubits, let the subjective value extracted by user $i$ from its execution for a unit of time be characterised by $f_\mathrm{sub}^\mathrm{(joint)}(n)$. Then, the joint subjective value of computation (i.e total market subjective return) follows the same form as the joint computational scaling function given in Def.~\ref{def:comb_sc_func},

\begin{definition}[Subjective value of computation]\index{Subjective value of computation}\label{def:sub_val_comp}
The joint subjective value of a time-share in the global network, $f_\mathrm{sub}^\mathrm{(joint)}$, of a group of participants, each with their own subjective valuation functions, $f_\mathrm{sub}^{(i)}$,
\begin{align}
	f_\mathrm{sub}^\mathrm{(joint)} = \sum_{i=1}^{n_{\mathrm{nodes}}} \beta_i \cdot f_\mathrm{sub}^{(i)}(n_\mathrm{global}).
\end{align}
\end{definition}

%
% Arbitrage-Free Time-Sharing
%

\section{Arbitrage-free time-sharing model}\label{sec:arb_free_time_share} \index{Arbitrage-free!Time-sharing model}\index{Time-sharing}

\dropcap{I}{n} the context of our time-shared global network of unified quantum computers (Sec.~\ref{sec:time_share}), how do we fairly and equitably allocate time-shares between contributors? We now derive an elementary arbitrage-free model for equitable time-sharing in such a network.

Let,
\begin{align}
	0\leq r_n \leq 1,
\end{align}
be the proportion of compute-time allocated to a node in possession of $n$ qubits, in a global network of $n_\mathrm{global}$ qubits. Arbitrage in the value of physical qubits will enforce the linearity constraint,
\begin{align}
	r_{n_1+n_2} = r_{n_1} + r_{n_2}.
\end{align}
This constraint effectively mandates that `all qubits are created equal', and two qubits are twice as valuable as one \latinquote{Qubit aequalitatem}. Were, for example, a bundle of two qubits more expensive than two individual qubits purchased in isolation, a market participant could perform arbitrage and unfairly gain free compute-time by buying two qubits separately, unifying them, selling the bundle, buying them back individually, and repeating indefinitely until he seizes the entire network.

Additionally, we have assumed no compute-cycles are wasted -- compute-time is always fully utilised, as per Post.~\ref{post:market_eff}. Then it follows that the time-share of the combined resources of the entire network should be unity,
\begin{align}
	r_{n_\mathrm{global}}=1.
\end{align}
\mbox{$r_{n_\mathrm{global}}<1$} would imply inefficiency via wasted clock-cycles. Combining this with the linearity constraint implies the arbitrage-free time-sharing model,
\begin{definition}[Arbitrage-free time-sharing model] \label{def:arb_free_ts}\index{Arbitrage-free!Time-sharing model}
In an efficient market for unified quantum computing time-shares\index{Time-sharing}, a network participant in possession of $n$ of the entire $n_\mathrm{global}$ qubits in the network is entitled to the fraction of unified network compute time,
\begin{align}\index{Time-shared compute-time}
	r_n = \frac{n}{n_\mathrm{global}},
\end{align}
where,
\begin{align}
n_\mathrm{global} = \sum_{j\in\mathrm{nodes}} n_j,
\end{align}
is the total number of qubits in the network, and,
\begin{align}
0\leq r_n \leq 1.	
\end{align}
\mbox{$r_n=1$} iff the node has a complete monopoly over qubits, i.e \mbox{$n=n_\mathrm{global}$}.
\end{definition}

Based on this equitable model for time-sharing,
\begin{definition}[Time-shared computing power]\label{def:time_share_comp_power}\index{Time-shared computing power}
The computing power allocated to each user under the arbitrage-free time-sharing model is,
\begin{align}
	c_n &= r_n \cdot f_\mathrm{sc}(n_\mathrm{global}) \nonumber \\
	&= n \cdot \chi_\mathrm{sc}(n_\mathrm{global}).
\end{align}
\end{definition}

This model is intuitively unsurprising, since it is analogous to the case of classical computer clusters -- nodes receive a time-share proportional to the proportion of the hardware they are contributing to the network. However, it is important to point out that the arbitrage is taking place in the cost of physical qubits, but not in terms of the dollar value of their classical-equivalent processing power, since this is in general non-linearly related to the number of qubits. Arbitrage in computational power per se is complicated by the fact that it is a non-fungible asset that cannot be directly traded, or uniquely associated with a tangible, tradable asset -- its computational value is a function of other assets.

%
% Problem Size Scaling Functions
%

\section{Problem size scaling functions}\index{Problem size!Scaling functions}\label{sec:prob_sc_func}

\dropcap{T}{he} computational scaling function\index{Computational!Scaling functions} introduced previously expresses the power of a quantum computer in terms of its classical-equivalent runtime, or equivalently FLOPs\index{FLOPs}. However, this may not be the metric of interest when considering a computer's algorithmic power. In many situations, of far greater interest is the size of a problem\index{Problem size} instance that can be solved in a given timespan. For example, the FLOPs associated with solving an instance of a 3-\textsc{SAT} problem\index{3-SAT problem} grows exponentially with the number of clauses. When discussing the execution of this problem on a given computer, what we really want to know is how many clauses our device can cope with, rather than what the classical-equivalent runtime is.

This observation motivates us to re-parameterise the power of quantum computers in terms of the problem size of a given algorithm to be solved. Employing the same methodology as for computational scaling functions, we define the \textit{problem size scaling function}, which relates the size of an algorithmic problem to its classical equivalent runtime. Then equating the computational and problem size scaling function yields,

\begin{definition}[Problem size scaling function]\index{Problem size!Scaling functions}
The problem size scaling function relates the size of a problem instance ($s$), in some arbitrary metric, to its classical-equivalent runtime ($t$) under a time-shared network model,
\begin{align}
t = f_\mathrm{size}(s).
\end{align}
Equating this with the time-shared computational power yields,
\begin{align}
	n\cdot \chi_\mathrm{sc}(n_\mathrm{global}) = f_\mathrm{size}(s).
\end{align}
Isolating the problem size yields,
\begin{align}
s = f_\mathrm{size}^{-1}(n\cdot \chi_\mathrm{sc}(n_\mathrm{global})).
\end{align}
\end{definition}

We now consider several choices of scaling functions.

First let us consider the classical case of linear scaling functions (for both the computational and problem size scaling functions),
\begin{align}
	f_\mathrm{sc}(n) &= \alpha_\mathrm{sc} n,\nonumber\\
	f_\mathrm{size}(s) &= \alpha_\mathrm{size} s.
\end{align}
Solving for the problem size simply yields,
\begin{align}
s &= \frac{n}{\alpha_\mathrm{size}} \nonumber\\
&= O(1),
\end{align}
where $n$ is regarded as a constant, and $n_\mathrm{global}$ is a variable parameter of the network. That is, the problem sizes of solvable instances is independent of the size of the external network with whom we are time-sharing. This is to be expected, since these scaling functions are typical of classical computers.

For polynomial scaling functions,
\begin{align}
f_\mathrm{sc}(n) &= n^{p_\mathrm{sc}},\nonumber\\
f_\mathrm{size}(s) &= s^{p_\mathrm{size}}.
\end{align}
This yields problem size,
\begin{align}
	s &= (n \cdot {n_\mathrm{global}}^{p_\mathrm{sc}-1})^\frac{1}{p_\mathrm{size}} \nonumber\\
	&= O(\mathrm{poly}(n_\mathrm{global})),
\end{align}
demonstrating polynomial scaling in our solvable problem size against the size of the network.

For exponential scaling functions,
\begin{align}
f_\mathrm{sc}(n) &= e^{\alpha_\mathrm{sc}n},\nonumber\\
f_\mathrm{size}(s) &= e^{\alpha_\mathrm{size}s},
\end{align}
we obtain,
\begin{align}
s &= \log\left(n \frac{e^{\alpha_\mathrm{sc}n_\mathrm{global}}}{\alpha_\mathrm{sc}n_\mathrm{global}}\right) \nonumber\\
&= \alpha_\mathrm{sc}n_\mathrm{global} + \log(n)-\log(\alpha_\mathrm{sc}n_\mathrm{global}) \nonumber\\
&= O(n_\mathrm{global}),
\end{align}
demonstrating that the solvable problem size grows linearly with network size. That is to say, waiting for a doubling in the external network's size will also double the size of a \textbf{BQP}-complete\index{BQP} problem that can be solved in the same time.

%
% Quantum Computational Leverage
%

\section{Quantum computational leverage}\label{sec:quant_ec_lev}\index{Quantum computational leverage}

\dropcap{I}{n} Secs.~\ref{sec:dist_QC} \& \ref{sec:module} we introduced distributed and modularised quantum computation. Using this as a toy model, we will now investigate the market dynamics of uniting the quantum computational resources of multiple market participants, as per an equitable time-sharing model (Sec.~\ref{sec:time_share}). We envisage a model whereby network participants are contributing modules to the networked quantum computer, thereby unifying their computational power.

The $i$th node is contributing the fraction of the hardware $r_i$, and receives this same proportion of compute-time under the arbitrage-free time-sharing model (Def.~\ref{def:arb_free_ts}). This discounts his classical-equivalent processing time\index{Classical-equivalent computational power} to,
\begin{align}
\tau_i = t_\mathrm{joint} \cdot r_i.
\end{align}

We are now interested in quantifying how much better off individual contributors are under this model than they were individually. Let us define the \textit{quantum computational leverage}\index{Quantum computational leverage} (QCL) of a node's quantum computer to be the ratio between their unified time-shared and individual classical-equivalent processing times\index{Classical-equivalent computational power},
\begin{align}
\lambda_i = \frac{\tau_i}{t_i},
\end{align}
yielding the QCL formula,

\begin{definition}[Quantum computational leverage] \label{def:quant_econ_lev}\index{Quantum computational leverage!Formula}\index{Single-qubit quantum computational leverage}
For the $i$th node, and with scaling function $f_{sc}$, the QCL is defined as the ratio between the unified time-shared and individual classical-equivalent algorithmic runtimes,
\begin{align}
\lambda_i &= \frac{\tau_i}{t_i} \nonumber \\
&= \frac{n_i}{n_\mathrm{global}} \cdot \frac{f_{sc}(n_\mathrm{global})}{f_{sc}(n_i)} \nonumber \\
&= \frac{\chi_\mathrm{sc}(n_\mathrm{global})}{\chi_\mathrm{sc}(n_i)},\nonumber\\
\lambda_i^\mathrm{dB} &= 10\log_{10}(\lambda_i),
\end{align}
where,
\begin{align}
	n_\mathrm{global} = \sum_{j\in \mathrm{nodes}} n_j,
\end{align}
is the total number of qubits in the network. The logarithmic version of the representation in decibels is simply a convenience when dealing with exponential scaling functions.
\end{definition}

Effectively, the QCL tells us how much additional computational power we `get for free' by consolidating with the network.

It is extremely important to note that the QCL is asymmetric, in the sense that the leverage achieved by a given node is larger than the leverage achieved by the network, upon the user joining the network (assuming the remainder of the network comprises more qubits than the respective user).

More generally, smaller users achieve higher computational leverage from their investment into quantum hardware than larger users. Specifically,
\begin{align}
	\lambda_i<\lambda_j \,\,\mathrm{for}\,\,n_i>n_j.
\end{align}

For any super-linear scaling function we have \mbox{$\lambda_i > 1 \,\,\forall \, i$}, and for any linear scaling function we have \mbox{$\lambda_i = 1 \,\,\forall \, i$},
\begin{align}
	\lambda=1\,\,\forall\,\,f_\mathrm{sc}(n)=O(n), \nonumber \\
	\lambda>1\,\,\forall\,\,f_\mathrm{sc}(n)>O(n).	
\end{align}

For \mbox{$\lambda_i>1$} it is always computationally beneficial to all nodes to unify computational resources and time-share them equitably, as per the arbitrage-free time-sharing model. Similarly, the distributed network is better off accepting them into the network, albeit to a lesser extent for a large network.

This is in contrast to classical networks, where \mbox{$\lambda\approx 1$}, for any number of nodes in the network (i.e there is no leverage), and it makes no difference whether nodes unify resources or operate independently.

Finally, in the pathological case, where \mbox{$\lambda_i<1$}, nodes are better off working in isolation, a situation which would only naturally arise as a result of algorithmic inefficiencies in parallelisation or distribution.

\begin{definition}[Single-qubit QCL]\index{Single-qubit quantum computational leverage}
The single-qubit QCL is the leverage associated with adding a single qubit to the network, \mbox{$n=1$}, defined as,
\begin{align}
	\lambda_\mathrm{qubit} &= \frac{\chi_\mathrm{sc}(n_\mathrm{global})}{\chi_\mathrm{sc}(1)},\nonumber\\
	\lambda_\mathrm{qubit}^\mathrm{dB} &= 10\log_{10}(\lambda_\mathrm{qubit}).
\end{align}
\end{definition}

Using our postulate for network growth (Post.~\ref{post:net_growth}) yields the postulated time-dependent QCL,
\begin{postulate}[Time-dependent QCL]
The time-dependent QCL, based on the postulate of exponential network growth, is,
\begin{align}\index{Time-dependent quantum computational leverage}
\lambda_n(t) &= \frac{\chi_\mathrm{sc}(N_0{\gamma_N}^t)}{\chi_\mathrm{sc}(n)},\nonumber\\
\lambda_n^\mathrm{dB}(t) &= 10\log_{10}(\lambda_n(t)).
\end{align}
The initial (\mbox{$t=0$}) time-dependent QCL reduces to the standard QCL formula.
\end{postulate}
Note that for any super-linear scaling function, the time-dependent QCL grows exponentially over time, unlike the classical case where there is no leverage, which does not change over time (i.e \mbox{$\lambda_n(t)=1\,\,\forall\,n,t$}).

The leverage is not merely a function of the hardware, but also of the software applications running upon it, each of which associated with a unique scaling function. Furthermore, it is to be reasonably anticipated that the size of the quantum internet will increase monotonically over time, yielding ever increasing leverage on the initial hardware investment by network contributors.

\begin{figure}[!htbp]
\includegraphics[clip=true, width=0.475\textwidth]{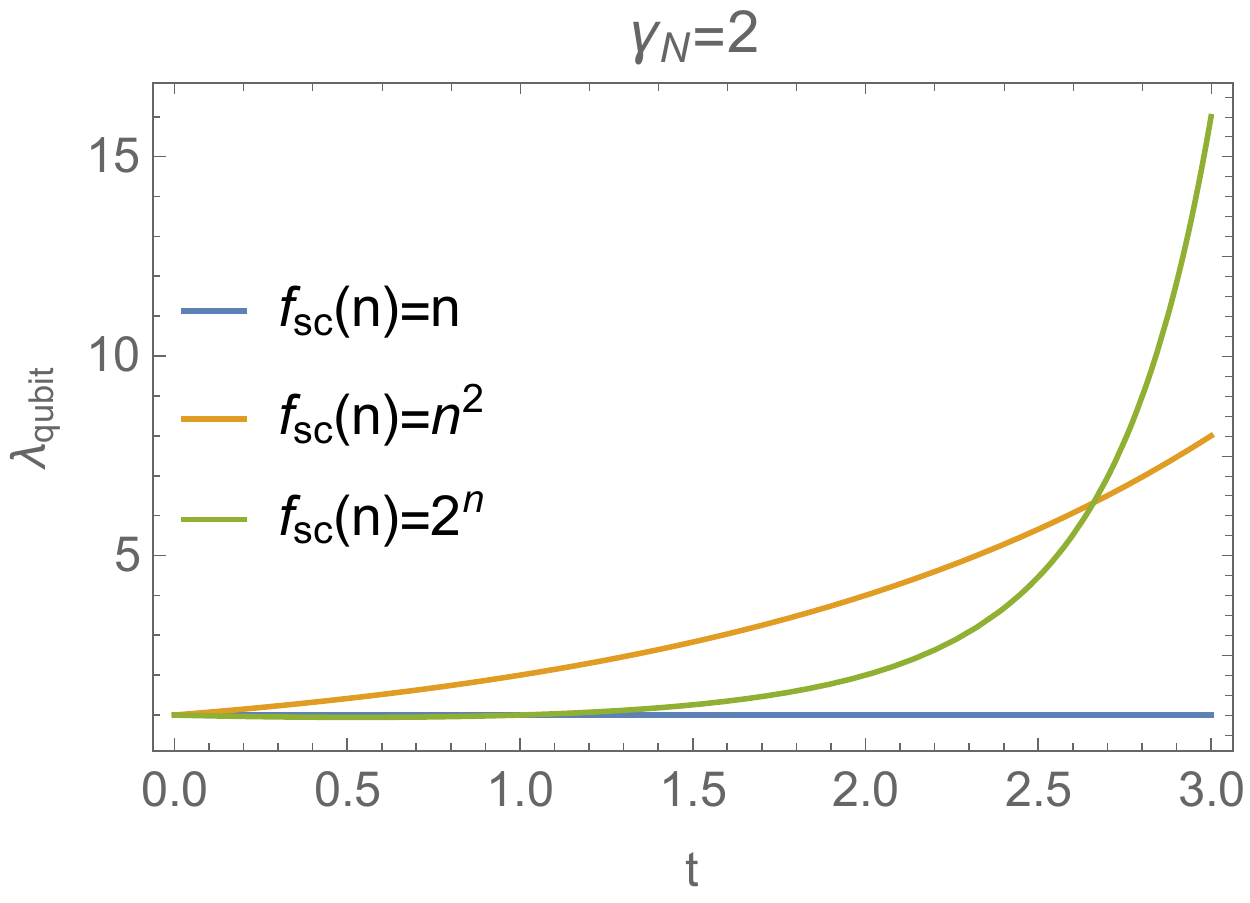}
\captionspacefig \caption{Time-dependent quantum computational leverage for a single qubit (\mbox{$n=1$}) with different computational scaling functions, under the assumption of exponential network growth in units of \mbox{$N_0=1$}.}\label{fig:time_dep_QCL}
\end{figure}

Examples of the temporal dynamics of the time-dependent single-qubit QCL are shown in Fig.~\ref{fig:time_dep_QCL}.

%
% Static Computational Return
%

\section{Static computational return}\index{Static computational return}\label{sec:static_comp_ret}

\dropcap{T}{he} computational leverage phenomena clearly implies that as the global quantum network expands over time, so too does the computational payback on investment into network expansion, or equivalently, the cost per unit of additional classical-equivalent processing time decreases.

Since exisiting network participants receive leverage upon \textit{other} participants joining the network, an investment into contributing modules has monotonically increasing computational return over time as the network expands, even if that participant ceases making further investment into the network. This is in contrast to classical networks, whereby the computational return upon an investment is fixed over time.

To formalise this, consider the case where a user purchases an initial $n$ qubits, while the global network expands over time as $N_t$. Then the classical-equivalent computational power of the user's fixed investment is,

\begin{definition}[Static computational return]\index{Static computational return} The static computational return is the classical-equivalent processing power of a user's time-share proportion (\mbox{$n/N_t$}), where the user has a fixed investment of $n$ qubits, whereas the network is allowed to expand over time arbitrarily as $N_t$ (e.g according to a quantum Moore's Law),
\begin{align}
	r_\mathrm{static}(t) &= \frac{n\cdot f_\mathrm{sc}(N_t)}{N_t}\nonumber\\
	&= n\cdot\chi_\mathrm{sc}(N_t),
\end{align}
which intuitively follows as the computational power per qubit\index{Per-qubit computational power} in the network, times the number of qubits in our possession.
\end{definition}

Graphical examples for this relationship are equivalent to those presented earlier in Fig.~\ref{fig:NPSF}. In particular, for linear classical scaling functions the return is constant, whereas for exponential quantum scaling functions the return is exponential in network size.

%
% Forward Contract Pricing Model
%

\section{Forward contract pricing model}\label{sec:for_contr}\index{Forward contract pricing model}

\dropcap{F}{orward} contracts are immensely useful in conventional markets, as a means by which to secure future use or ownership of an asset at predictable points in time. For example, farmers make heavy use of forward contracts to lock in sale of their produce before it has been harvested, such that the value is locked in in advance and the sale guaranteed, providing a very valuable hedging instrument\index{Hedging} for managing risk\index{Risk management}.

We envisage similar utility in the context of quantum computing. A company engaging in heavy use of computing power might have a need to perform certain computations at predictable points in the future. In this instance, forward contracts could be very helpful in reducing exposure to risk and guaranteeing access to the technology when needed, at a pre-agreed rate.

Now let us price forward contracts on units of computation, whereby we wish to pay today for the future use of a block of runtime on the global network.

The key observation is that a unit of computation (FLOP) does not carry over time. It must be utilised immediately and cannot be stored for future use. This simplifies the forward price of a unit of computation to simply be the future spot price, discounted by the risk-free RoR, yielding the forward contract pricing model for quantum computing time-shares,
\begin{definition}[Forward contract pricing model] \label{def:forward_cont}\index{Forward contract pricing model}
The efficient market price for a forward contract in a unit of network runtime at future time $T$ is,
\begin{align}
F(T) &= e^{-r_\mathrm{rf}T} L(T)\nonumber\\
&=\frac{e^{(\gamma_\mathrm{ror}-r_\mathrm{rf})T} C_0{\gamma_C}^{-T}}{\chi_\mathrm{sc}(N_0 {\gamma_N}^T)}.
\end{align}
\end{definition}

Note that in the limit of \mbox{$T\to 0$} this reduces to the spot price of the asset,
\begin{align}
	F(0)=L(0),
\end{align}
as expected.

\if 1\doublecol
\begin{figure}[!htbp]
\includegraphics[clip=true, width=0.4\textwidth]{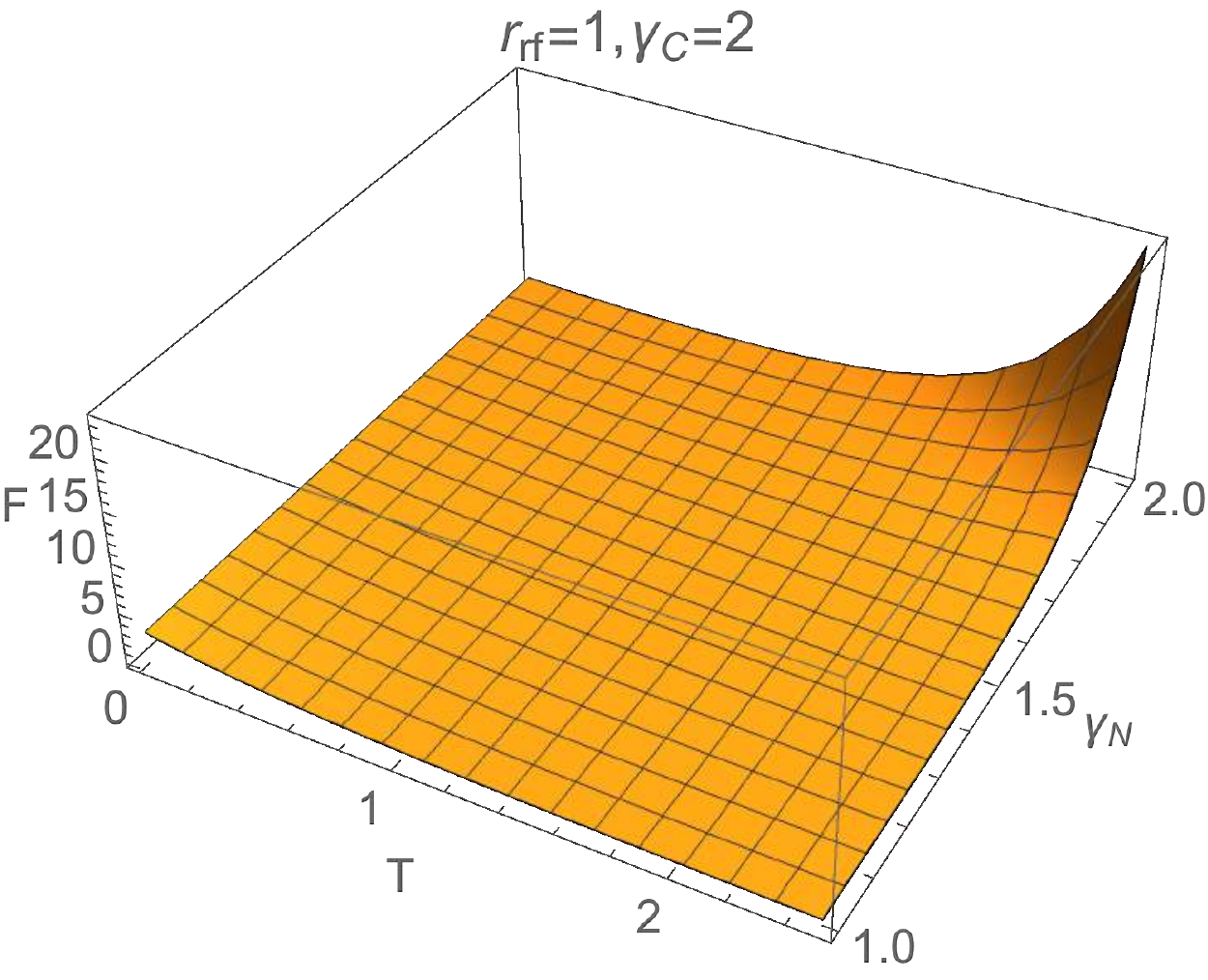}\\
\includegraphics[clip=true, width=0.4\textwidth]{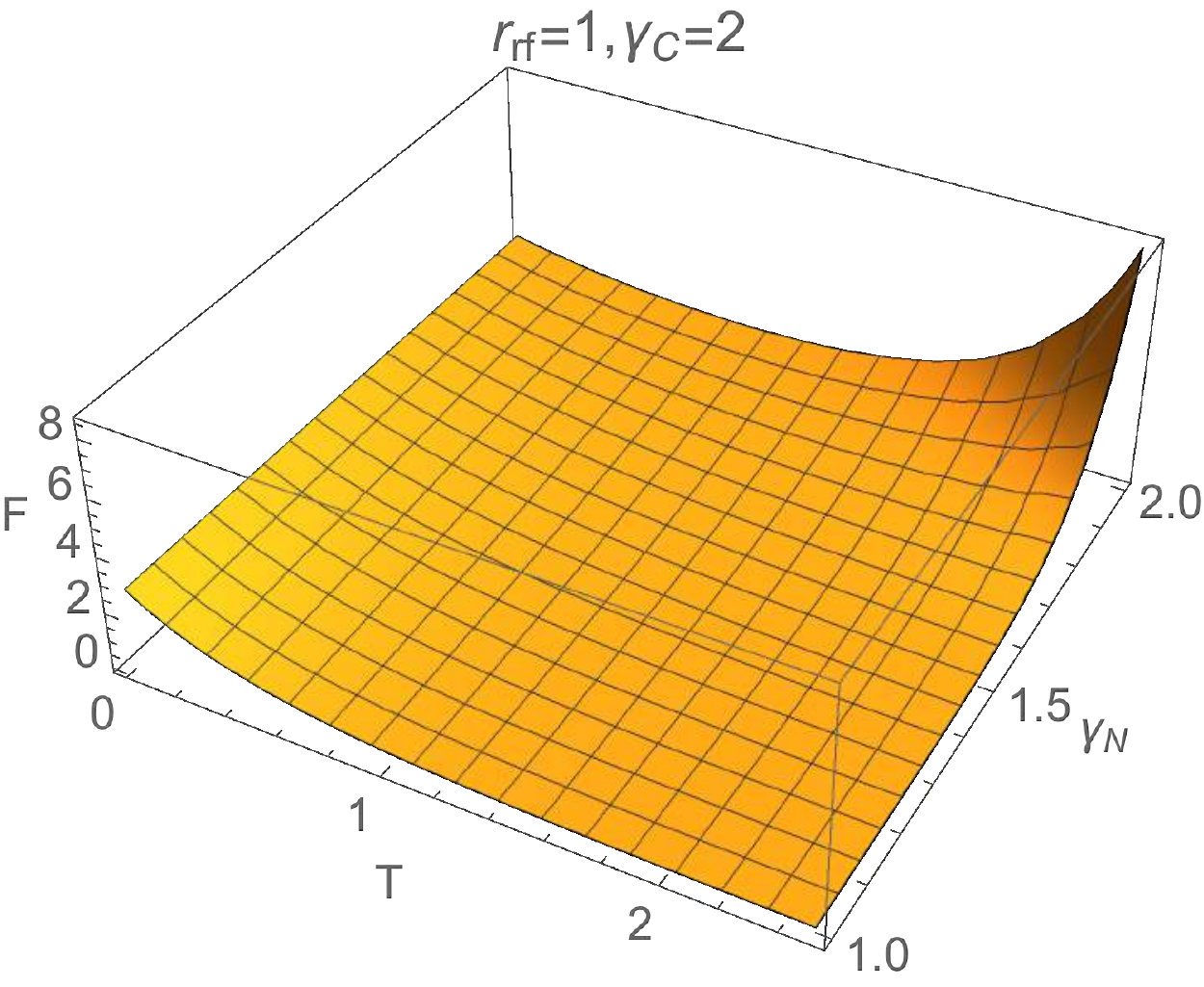}
\captionspacefig \caption{Forward price on a computation to be delivered at time $T$ in the future, in units \mbox{$C_0=N_0=1$}, where we are assuming an exponential scaling function, \mbox{$f_\mathrm{sc}(n)=e^n$}.}\label{fig:forward_cont_pricing_mod}
\end{figure}
\else
\begin{figure*}[!htbp]
\includegraphics[clip=true, width=0.475\textwidth]{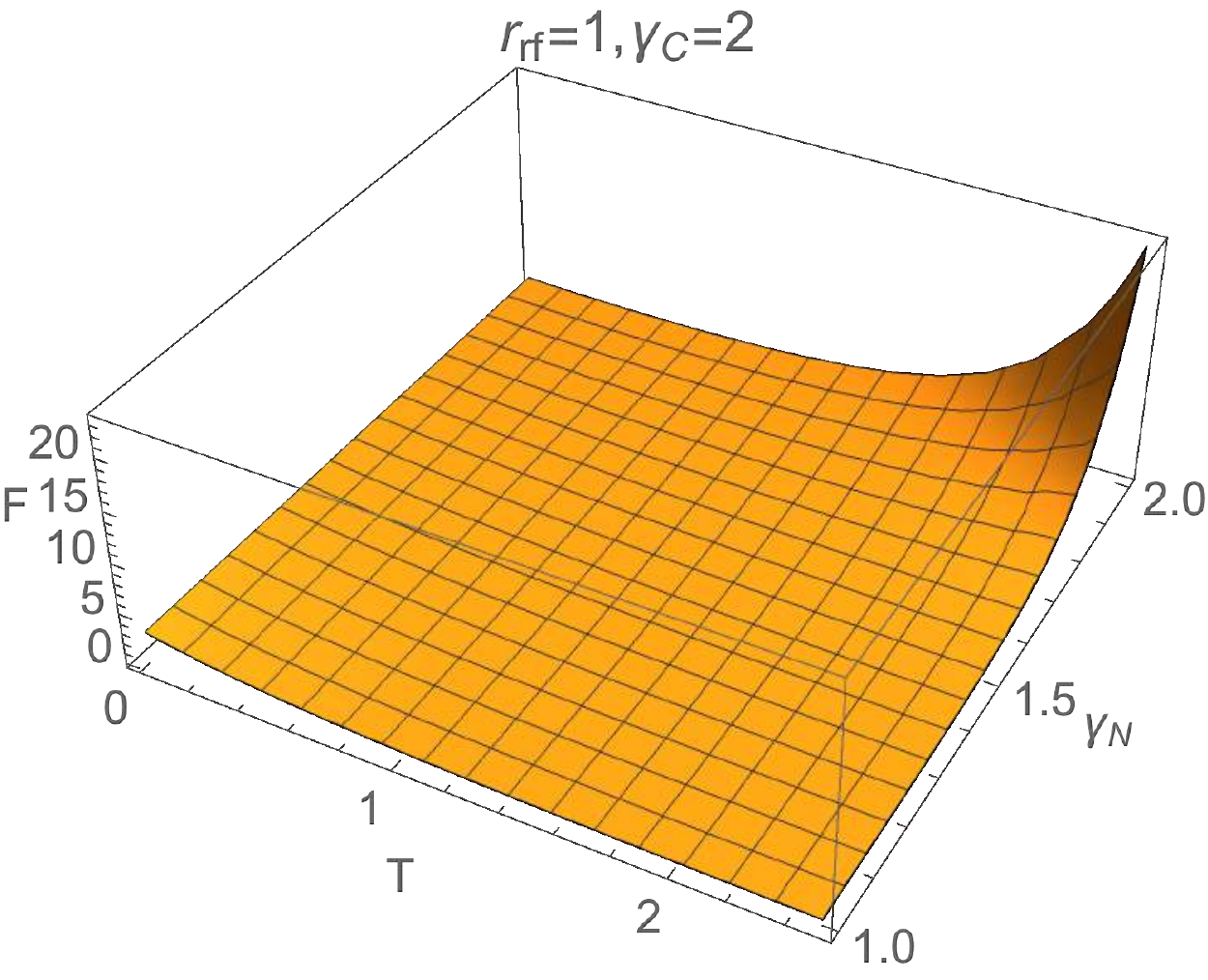}
\includegraphics[clip=true, width=0.475\textwidth]{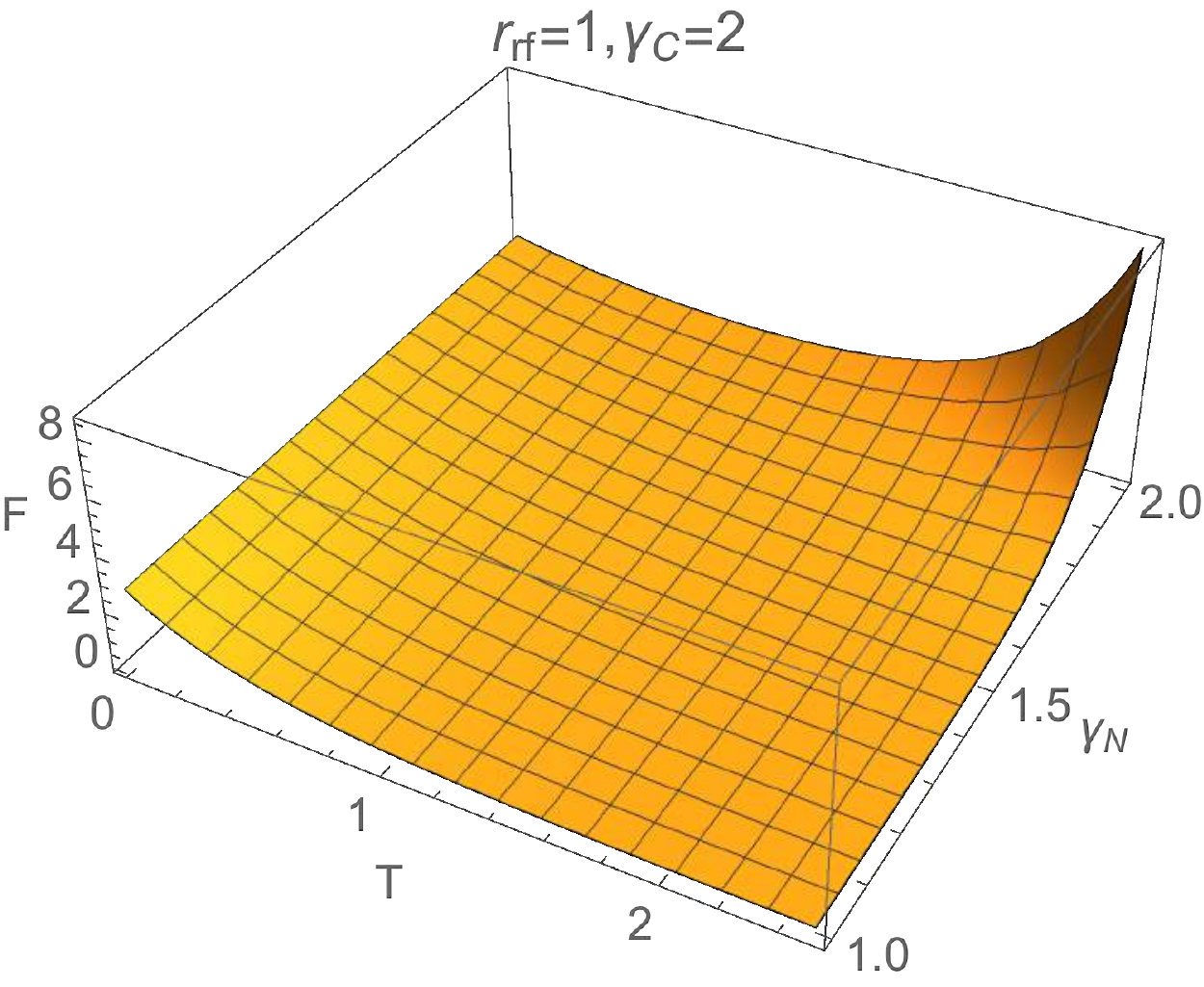}
\captionspacefig \caption{Forward price on a computation to be delivered at time $T$ in the future, in units \mbox{$C_0=N_0=1$}, where we are assuming an exponential scaling function, \mbox{$f_\mathrm{sc}(n)=e^n$}.}\label{fig:forward_cont_pricing_mod}
\end{figure*}
\fi

%
% Political Leverage
%

\section{Political leverage}\index{Political leverage}\label{sec:political_lev}

\dropcap{T}{he} asymmetry in computational leverage observed by parties of different sizes -- specifically, that parties possessing a smaller number of qubits observe greater leverage than those possessing a larger number of qubits -- inevitably will bring with it some power politics, with potentially interesting geo-political implications.

This asymmetry implies that in a globally unified network, were a large party to expel a small party from the network, it would be far more devastating to the computational power of the small party than the larger one. This suggests that inclusion in the global network could be a powerful tool of diplomacy\index{Diplomacy} in the quantum era, where threats of expulsion or resistance to inclusion is the modern day era of gunboat diplomacy\index{Gunboat diplomacy}.

To quantify this we introduce the \textit{political leverage} quantity -- the ratio between the computational leverages observed by two parties belonging to the same network. This directly quantifies the power asymmetry between them.

\begin{definition}[Political leverage]\index{Political leverage}
The political leverage is the ratio between the computational leverages of two parties residing on the same shared network,
\begin{align}
	\gamma_{A,B} &= \frac{\lambda_A}{\lambda_B}\nonumber\\
	&= \frac{\chi_\mathrm{sc}(n_B)}{\chi_\mathrm{sc}(n_A)},\nonumber\\
	\gamma_{A,B}^\mathrm{dB} &= 10\log_{10}(\gamma_{A,B}).
\end{align}
We have the trivial identity that the leverage of $A$ against $B$ is the inverse of the leverage of $B$ against $A$,
\begin{align}
	\gamma_{A,B} &= {\gamma_{B,A}}^{-1},\nonumber\\
	\gamma_{A,B}^\mathrm{dB} &= -\gamma_{B,A}^\mathrm{dB}.
\end{align}
\end{definition}
Note that when two parties are of equal size, there is no power asymmetry and \mbox{$\gamma_{A,B}=1$}. Otherwise, when $A$ and $B$ are unequal, then \mbox{$\gamma_{A,B}\neq 1$}, indicative of power asymmetry. With linear (classical) scaling functions, the political leverage is always unity, \mbox{$\gamma_{A,B}=1$}, regardless of any size asymmetry, whereas for super-linear scaling functions the political leverage diverges.

To the Machiavellian\index{Machiavelli} reader, this quantity can be thought of as answering the question `If I were to expel a party from the network, how much more would it hurt them than it would hurt me?'.

%
% QuantCoin - A Quantum Computation-Backed Cryptocurrency
%

\section{QuantCoin\texttrademark\,-- A quantum computation-backed cryptocurrency}\label{sec:quant_coin_technical}

\dropcap{A}{s} discussed in Sec.~\ref{sec:bitcoin_blockchain}, the Bitcoin\index{Bitcoin} mining\index{Bitcoin!Mining} process involves finding bit-strings that hash\index{Hash!Functions} under SHA256\index{SHA256} to a value within some relatively small range. This so-called `proof-of-work'\index{Proof-of-work} principle associates computational complexity with the mining process, and since the hashing functions are one-way functions, they must be evaluated via brute-force trial-and-error to find hits.

However, what a waste this is! Our proof-of-work is nothing more than hashing a huge number of random bit-strings, computations which are of no intrinsic value to anyone. The market value in turn has nothing to do with any inherent value earned during the mining process. Rather it is based purely on the psychology of scarcity\index{Scarcity}, since there is an upper bound on the number of Bitcoins that satisfy the legitimacy constraint.

What if we were to replace brute-force hashing of random data with computations of genuine monetary value? Then we would have a sounder currency, whose value derives from the monetary cost of executing useful computations. While it is not so easy to invent such a protocol for classical computation, the idea lends itself very naturally to quantum computation, owing to their ability to undergo encrypted computation\index{Encrypted quantum computation} and be subject to verification protocols\index{Verification!Protocols}.

Building upon some of the pricing models introduced earlier in this section, there are two main candidates for backing a cryptocurrency with quantum compute-time:
\begin{itemize}
	\item Spot market\index{Spot market}: we execute the computation immediately in exchange for a coin.
	\item Futures market\index{Futures market}: we own the right to utilise the computer at some fixed time in the future in exchange for a coin.
\end{itemize}
We consider the merits of both these candidates.

A popular-level essay on the future of quantum cryptocurrencies is presented in Sec.~\ref{sec:quant_coin_essay}.

%
% Spot Market Model
%

\subsection{Spot market model}\index{Spot market}

\begin{figure}[!htbp]
\includegraphics[clip=true, width=0.475\textwidth]{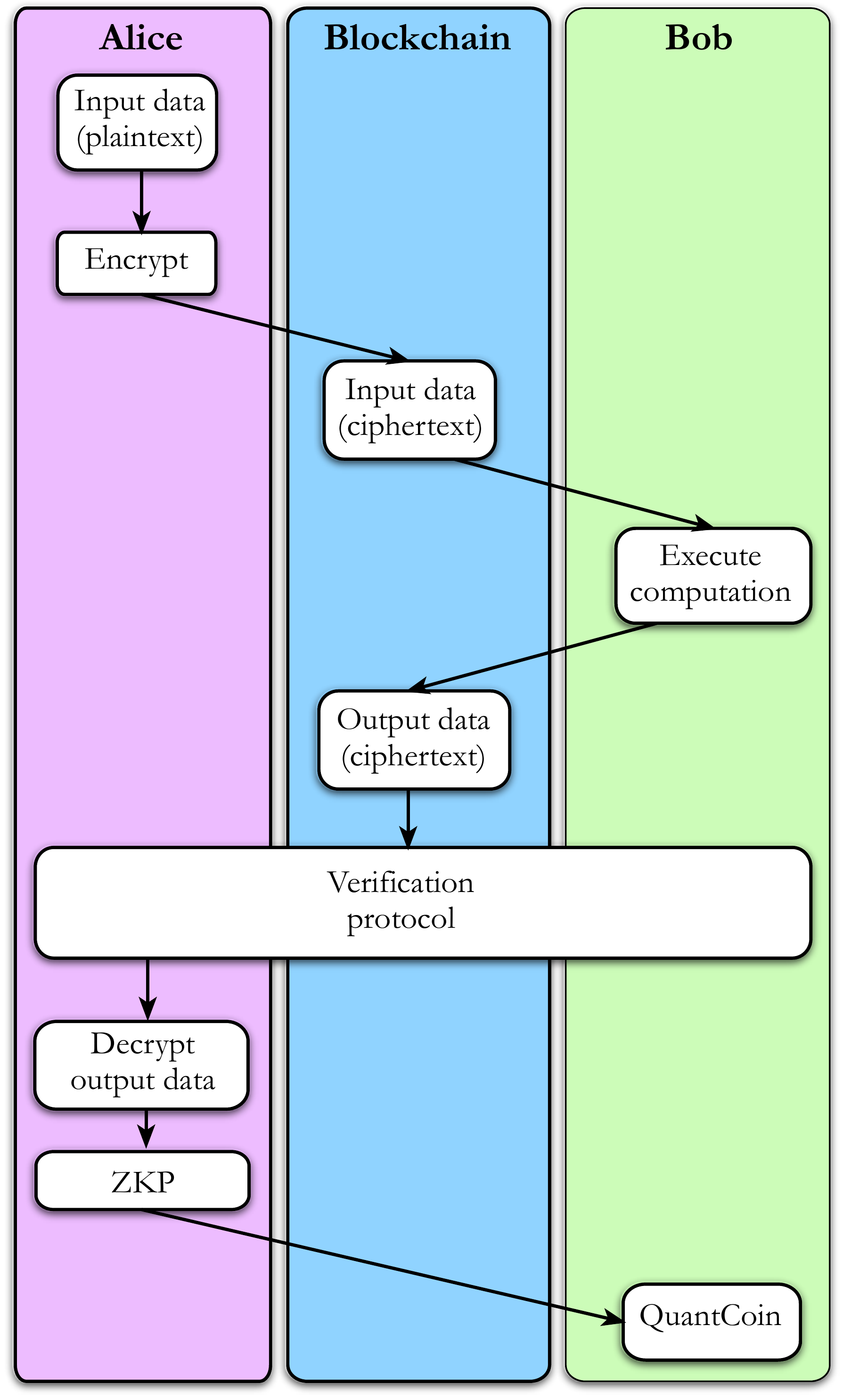}
\captionspacefig \caption{Flowchart for the QuantCoin\texttrademark\, protocol, introduced in Alg.~\ref{alg:quant_coin}.}\label{fig:quantcoin_protocol}	
\end{figure}

In Alg.~\ref{alg:quant_coin} we provide a very rough sketch for how a protocol based on the spot market might be constructed. A corresponding graphical flowchart is shown in Fig.~\ref{fig:quantcoin_protocol}. We present the ideas in a very high-level manner, abstracting away the physical implementation details of the computation, encryption, and verification protocols, instead envisaging that we can interface with them using a very high-level API\index{API}.

\begin{table}[!htbp]
\begin{mdframed}[innertopmargin=3pt, innerbottommargin=3pt, nobreak]
\texttt{
function QuantCoin($\hat{U}_\mathrm{comp}$, Blockchain, data):
\begin{enumerate}
	\item Alice homomorphically/blindly encrypts $data$,
	\begin{align}
		encryptedInput = homoEncrypt(data)
	\end{align}
	\item Alice commits the $encryptedInput$ to the public $Blockchain$,
	\begin{align}
		Alice.Blockchain.commit(encryptedInput)	
	\end{align}
	\item Bob processes the $encryptedInput$,
	\begin{align}
		encryptedOutput = \hat{U}_\mathrm{comp}(encryptedInput)
	\end{align}
	\item Bob commits encrypted output to the $Blockchain$,
	\begin{align}
		Bob.Blockchain.commit(encryptedOutput)	
	\end{align}
	\item Alice decrypts the output.
	\begin{align}
		output = homoDecrypt(encryptedOutput)
	\end{align}
	\item Alice and Bob execute a verification protocol,
	\begin{align}
		proof = verify(encryptedOutput)	
	\end{align}
	\item If successful, alice commits a zero-knowledge proof of the result to the $Blockchain$,
	\begin{align}
		Alice.Blockchain.commit(ZKP(input, output))
	\end{align}
	This signs off on the legitimacy of the mining entry previously committed by Bob.
	\item All users on the network can inspect the ZKP to validate the legitimacy of the execution, maintaining ignorance of the computational outcome.
	\item $\Box$
\end{enumerate}}
\end{mdframed}
\captionspacealg \caption{Sketch for how a quantum computation-backed cryptocurrency might be implemented. We have abstracted away the underlying Blockchain protocol, interfacing with it using a high-level API, since Blockchain technology is highly liable to evolve. We similarly call upon verification subroutines using a high-level implementation-independent API. The key technique to signing off on the legitimacy of a newly mined QuantCoin is for Alice to provide a zero-knowledge proof of the decrypted result (Sec.~\ref{sec:ZKP}), which maintains the secrecy of her data.} \label{alg:quant_coin}
\end{table}

It is evident from the flow of Alg.~\ref{alg:quant_coin} that the mining process now comprises solving an actual quantum computation of intrinsic value to Alice, since she is exchanging assets (e.g dollars or already-existing QuantCoins\texttrademark) in exchange for the computation. Completion of the computation followed by successful verification then further rewards Bob with a fresh QuantCoin\texttrademark\, courtesy of the distributed Blockchain algorithm.

The described protocol, in addition to mining a new coin, associated with the execution of a computation, acts as a currency converter for converting traditional assets (e.g dollars) into QuantCoins\texttrademark. This ability to currency convert is necessary, and completely differs from the original Bitcoin mining process, where coins are fabricated out of thin air by anyone and everyone, independent of their pre-existing monetary wealth -- Bitcoins are not created via conversion from any existing asset, they are an entirely new asset class of their own. The fact that the computation associated with each QuantCoin\texttrademark\, is of intrinsic value on the other hand, implies that Alice ought to be paying something for the service.

Note that we observe an expansion in the money supply with each successfully executed and verified computation -- one additional unit of QuantCoins\texttrademark\, is mined for every unit of computation implemented\footnote{To provide an analogy with the gold standard\index{Gold standard}, think of the physical quantum computer as the goldmine, and each unit of gold it produces as being a QuantCoin\texttrademark. The production of each unit of gold is associated with the utilisation of the mine for a particular amount of time, but the mine can in principle operate indefinitely, with no hard upper-bound on its total future gold yield.}. Unlike Bitcoin, there is no inherent theoretical upper limit on the number of coins that can exist. However the QuantCoin\texttrademark\, money supply\index{Money supply} will be limited for the practical reason that mining each one is associated with a monetary transaction between Alice and Bob, and Alice will eventually run out of assets to exchange for computations.

What relationships characterise the value of QuantCoins\texttrademark? First, in a perfectly efficient market (Sec.~\ref{sec:eff_markets}) we have,
\begin{align}\label{eq:value_quantcoins}
	P_\mathrm{coin} + P_\mathrm{reward} = P_\mathrm{exec},
\end{align}
where, 
$P_\mathrm{coin}$ is the dollar value of a QuantCoin\texttrademark\,, $P_\mathrm{reward}$ is the dollar value of the reward paid by Alice for execution, and $P_\mathrm{exec}$ is the dollar value of cost of execution for Bob.
 
Alternately, rather than Alice paying Bob's reward in dollars, she might pay for them in already-existing QuantCoins\texttrademark. Suppose Alice pays $\lambda$ QuantCoins\texttrademark\, as Bob's reward. Then Eq.~(\ref{eq:value_quantcoins}) reduces to,
\begin{align}
P_\mathrm{coin} =\frac{1}{\lambda+1}P_\mathrm{exec},
\end{align}
providing us with a simple financial model relating the market price of QuantCoins\texttrademark\, and the monetary cost of execution of computations.

Note that with exception to the scenario where Alice buys into QuantCoins\texttrademark\, using dollar currency (or any non-electronic asset that cannot be committed to the Blockchain), the entire protocol is self-enforcing via programmed Blockchain transactions. Bob doesn't get paid his newly earned and freshly printed QuantCoin\texttrademark\, until the verification of the computation has completed and the proof committed to the Blockchain. He therefore cannot get paid until he has executed the computation he promised to, and proven to the network that he actually did.

The main security risk is that of Bob taking Alice's dollars and running, upon receiving the upfront reward in dollars, which necessarily don't reside on the Blockchain since they are not crypto-assets. This could be addressed by introducing trusted third-party escrow agents into the protocol, as method currently used in some online dark markets\index{Dark markets}. 

However, if the upfront payment were being made in pre-existing QuantCoins\texttrademark\,, the Blockchain might be programmed to not release the reward until completion of the final verification stage of the protocol -- effectively an escrow programmed directly into the Blockchain for self-execution. Such self-executing smart-contracts are already a feature in some cryptocurrencies such as Ethereum\index{Ethereum}.

%
% Futures Market Model
%

\subsection{Futures market model}\index{Futures market}

As described above, the cryptocurrency is effectively backed by the spot market in computation -- we exchange currency for the execution of computations \textit{immediately}. However, one might also envisage more complex cryptocurrencies being backed by the futures market in the licensing of quantum compute-time down the line.

Intuitively, one would expect such a form of cryptocurrency to be sounder than the one backed by the spot market. This is because our spot market-derived coins, once mined are not guaranteed to be convertible to anything of value, including computations. Recall that the execution of the computation takes place immediately when the coin is created.

On the other hand, a QuantCoin\texttrademark\, backed by a guarantee to access quantum compute-time at a designated point in the future necessarily has value, so long as the demand for compute-time does, and maintains value until the contract matures and converts into compute-time at which point it becomes worthless.

We leave explicit construction of a futures-based QuantCoin\texttrademark\, model as an exercise for the interested reader, primarily because we haven't thought it through properly\index{Laziness}.

%
% Economic properties of the qubit marketplace
%

\section{Economic properties of the qubit marketplace}\index{Economics!Properties}\label{sec:econ_prop}

\famousquote{Economics is probably the weirdest academic discipline I've come across. I find myself constantly in a superposition of fascination and annoyance with how the field currently stands. How can smart people have come up with a collection of ideas that are simultaneously brilliant and ridiculous, insightful and delusional, pragmatic and useless?}{Andrew Ringsmuth}

\sectionby{Scott Harrison}\index{Scott Harrison}

\dropcap{T}{he} development and implementation of the quantum internet will give rise to a new tradable commodity -- the qubit. The pricing mechanisms associated with a qubit market were explained earlier in this part. Here we provide a broader discussion on the economic properties of such a marketplace. The are two areas of particular interest we will focus on:
\begin{enumerate}
	\item The responsiveness of the qubit market to price fluctuations, measured by elasticity.\index{Elasticity}
	\item The implications of qubit market properties for broader society in terms of pricing and taxation.\index{Taxation}
\end{enumerate}

\subsection{The concept of elasticity}\index{Elasticity}

Elasticity as a concept is measured through percentage changes. Starting with the demand for qubits as an example, the \textit{elasticity of demand}\index{Elasticity}, $E_d$, is the percentage change in the quantity demanded of a good, divided by the percentage change in the price of a good. Mathematically, the elasticity of demand is represented as,
\begin{align}\index{Elasticity!Formula}
E_d = \frac{\% \Delta Q_d}{\% \Delta P}.	
\end{align}

From this relationship, inferences about the underlying commodity can be made, summarised as follows:
\begin{itemize}
	\item \mbox{$|E_d|>1$}: the percentage change in quantity is greater than the percentage change in price, and is therefore \textit{elastic}. This indicates that demand for the asset in the market is responsive to small price changes.
	\item \mbox{$|E_d|<1$}: there is a proportionally larger change in price, for a smaller shift in demand. This indicates that demand is less responsive to price changes, and considered \textit{inelastic}.
	\item \mbox{$|E_d|=1$}: we have \textit{unit elasticity}, where the percentage change in quantity demanded is equal to the percentage change in market price.
\end{itemize}

Elasticity of demand is just one context where the concept of elasticity can be applied. Other contexts include: the elasticity of supply, measuring the supply side responsiveness to changes in price; income elasticity, which captures how the quantity of goods in the market change relative to changes in the income of consumers; and cross-price elasticity, which compares the percentage change in quantity of one good, relative to the percentage change in price of another good. An example that we will come back to is how changes in the price of quantum computing may have an effect on the quantity demanded of high-performance classical computing (i.e conventional supercomputing).

\subsection{Elasticity of the qubit market}\index{Elasticity!Qubits}

A number of factors will affect the elasticity of demand and supply in the qubit market. The most significant factor affecting the demand for qubits is the availability of substitutes\index{Substitution}. Given that quantum processing can efficiently solve unique problems that classical computing cannot, there are no close substitutes for qubits -- a transistor is no substitute for a qubit! Consequently, elasticity of demand will be relatively inelastic: the quantity of qubits demanded will be relatively unresponsive to price fluctuations and changes, since there are no viable alternatives to substitute with. 

The supply side of the equation is also initially going to be highly inelastic. However, as time progresses and technological advancements and enhancements increase the computational power of the quantum internet, the supply of qubits will become increasingly responsive to price fluctuations. However, the extent to how elastic the supply becomes over time is also going to be affected by potential for excess capacity given the exponential trajectory of the manufacture of qubits as new and improved fabrication technologies emerge -- the quantum Moore's Law\index{Quantum Moore's Law}.

%
% Economic Implications
%

\section{Economic implications}

\dropcap{O}{ur} analysis thus far has been very theoretical. But our observations have very tangible implications in the real-world. This has implications for governments, regulatory authorities, fiscal and technology policy, national security, and any end users of the quantum cloud.

%
% The Price To Pay For Isolationism
%

\subsection{The price to pay for isolationism}\index{Isolationism}

In many traditional sectors of the economy there is an economic incentive to directly compete against other market participants. However in the quantum era the incentive is for owners of quantum computing hardware to cooperate and contribute their resources to the quantum internet rather than go it alone, as a direct consequence of super-linear leverage.

Only those hardware owners who unite with the global network will benefit from its leverage and remain competitive. Those who choose not to participate in the global network will be priced out of the market via exponentially higher cost per FLOP (assuming all other costs are equal).

This effectively taxes the cost of computation for those who fail to unify their assets with the network. And it is in the direct economic self-interest of all market participants to contribute their resources to the time-shared quantum cloud.

%
% Taxation
%

\subsection{Taxation}\label{sec:taxation}\index{Taxation}

Any asset, dividend, derivative or other financial instrument will inevitably be subject to taxation. Any form of taxation has multiplier effects as the cost markup is repeatedly handed from one market participant to the next, influencing the chain of supply and demand along the way. However, this multiplier and other economic consequences are highly dependent on the asset undergoing transaction -- the economic implications of personal income tax are quite different to those of capital gains tax!

\subsubsection{Computational perspective}

We now consider the effect of taxation on quantum resources, specifically in the form of a \textit{qubit tax}\index{Taxation} -- a sales tax on the purchase of physical qubits. Although this model of taxation is unlikely to be implemented as we describe, it serves as an insightful test-bed for thought experiments into the qualitative implications of taxing quantum assets.

Imagine that consumers have an amount of capital available for the purchase of qubits. Let $\gamma_T$ be the rate of taxation (\mbox{$\gamma_T=1$} represents no taxation, \mbox{$\gamma_T>1$} represents positive taxation, and \mbox{$\gamma_T<1$} represents subsidisation). Then the cost of physical qubits is marked up by $\gamma_T$, reducing the number of qubits that can be afforded by the consumers to (assuming fixed capital available for purchasing),
\begin{align}
	N_\mathrm{tax} = \frac{N_\mathrm{no\,tax}}{\gamma_T}.
\end{align}

We now wish to understand how this taxation influences the computational power of the network. We define the \textit{tax performance multiplier}\index{Taxation!Performance multiplier},
\begin{definition}[Tax performance multiplier]
The \textit{tax performance multiplier}\index{Taxation!Performance multiplier}, is the ratio between computational scaling functions with and without qubit taxation,
\begin{align}
M(N_\mathrm{tax}) &= \frac{f_\mathrm{sc}(N_\mathrm{tax})}{f_\mathrm{sc}(N_\mathrm{no\,tax})} \nonumber \\
&= \frac{f_\mathrm{sc}(N_\mathrm{tax})}{f_\mathrm{sc}(N_\mathrm{tax} \gamma_T)},\nonumber\\
M^\mathrm{dB}(N_\mathrm{tax}) &= 10\log_{10}(M(N_\mathrm{tax})),
\end{align}
where the consumers have purchased $N_\mathrm{tax}$ qubits, after taxation, at a markup rate of $\gamma_T$.
\end{definition}
The tax performance multiplier effectively gives us a factor by which computational power is depreciated under taxation. We can accomodate for other models of taxation and regulation by choosing an appropriate relationship between $N_\mathrm{tax}$, $N_\mathrm{no\,tax}$, and the taxation and regulatory framework.

Using our illustrative examples of computational scaling functions (linear, polynomial and exponential), the respective tax performance multipliers are given by,
\begin{align}
M_\mathrm{linear}(N_\mathrm{tax}) &= \frac{1}{\gamma_T}, \nonumber \\
M_\mathrm{poly}(N_\mathrm{tax}) &= \frac{1}{{\gamma_T}^p}, \nonumber \\
M_\mathrm{exp} (N_\mathrm{tax}) &= e^{N_\mathrm{tax}(1-\gamma_T)}.
\end{align}

This demonstrates that the computational power of classical networks is simply inversely proportional to the rate of taxation, i.e a linear tax performance multiplier, as we intuitively expect. And for quadratic scaling functions the dependence is inverse quadratic in the taxation rate. In both cases the multiplier is a constant factor, independent of the network size. However, for exponential scaling functions we observe an exponential dependence on both the rate of taxation and the size of the network, shown in Fig.~\ref{fig:tax_exp}. Note that for large networks, executing computations with exponential scaling functions, there is enormous sensitivity to variations in tax rates, yielding very high leverage in computational return by tax rates.

This implies that as the quantum network expands over time, its joint processing power decreases exponentially with the rate of taxation, yielding an ever-decreasing performance multiplier. In Sec.~\ref{sec:economics} we discuss some of the implications of this uniquely quantum phenomena.

However, taxation could also be negative, in the form of subsidisation. In Fig.~\ref{fig:tax_exp} we focus on the region surrounding neutral taxation, showing small degrees of taxation and subsidisation on either side. Evidently, even small degrees of subsidisation have a very strong impact on the performance multiplier (more pronounced than the same rate of positive taxation!). This makes subsidisation of qubit expansion highly tempting.

\if 1\doublecol
\begin{figure}[!htbp]
\includegraphics[clip=true, width=0.475\textwidth]{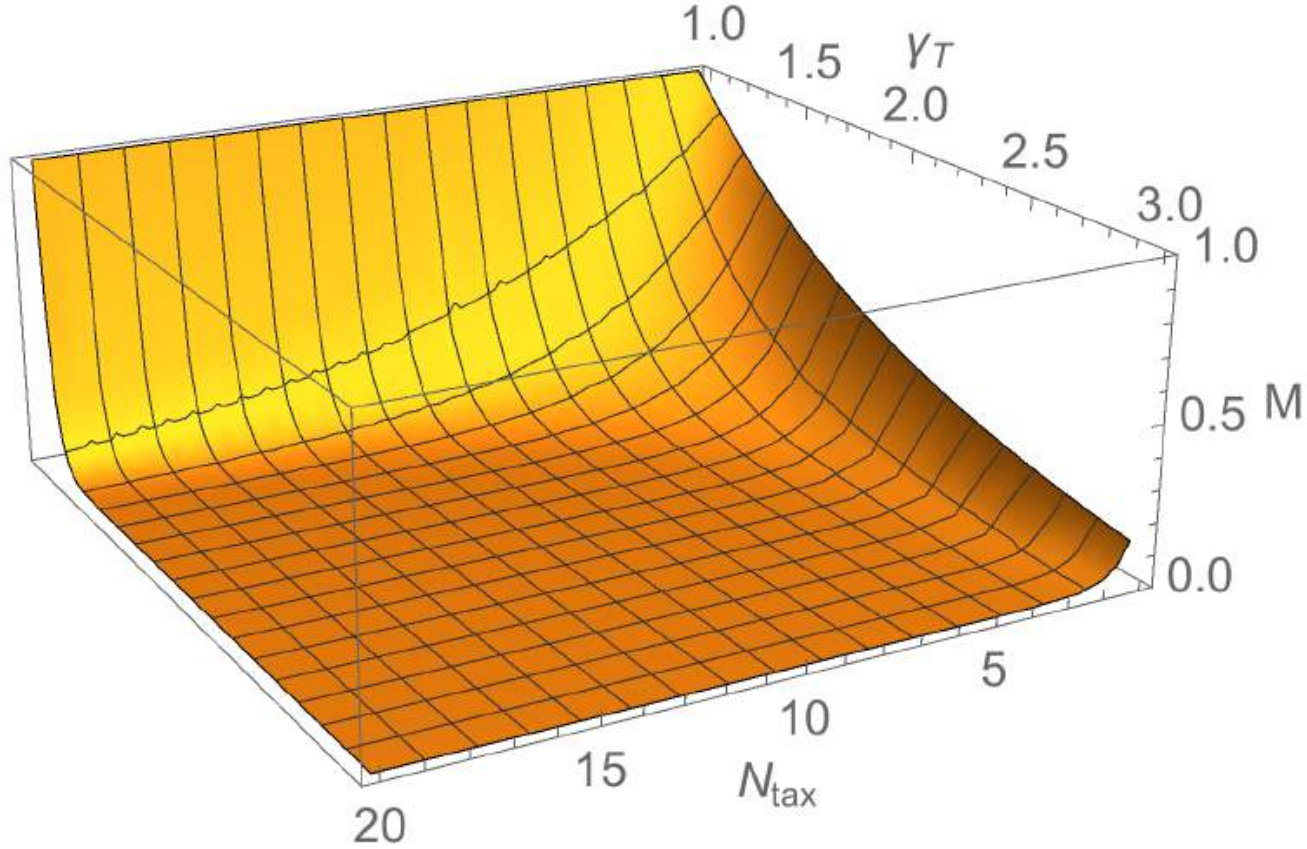}\\
\includegraphics[clip=true, width=0.475\textwidth]{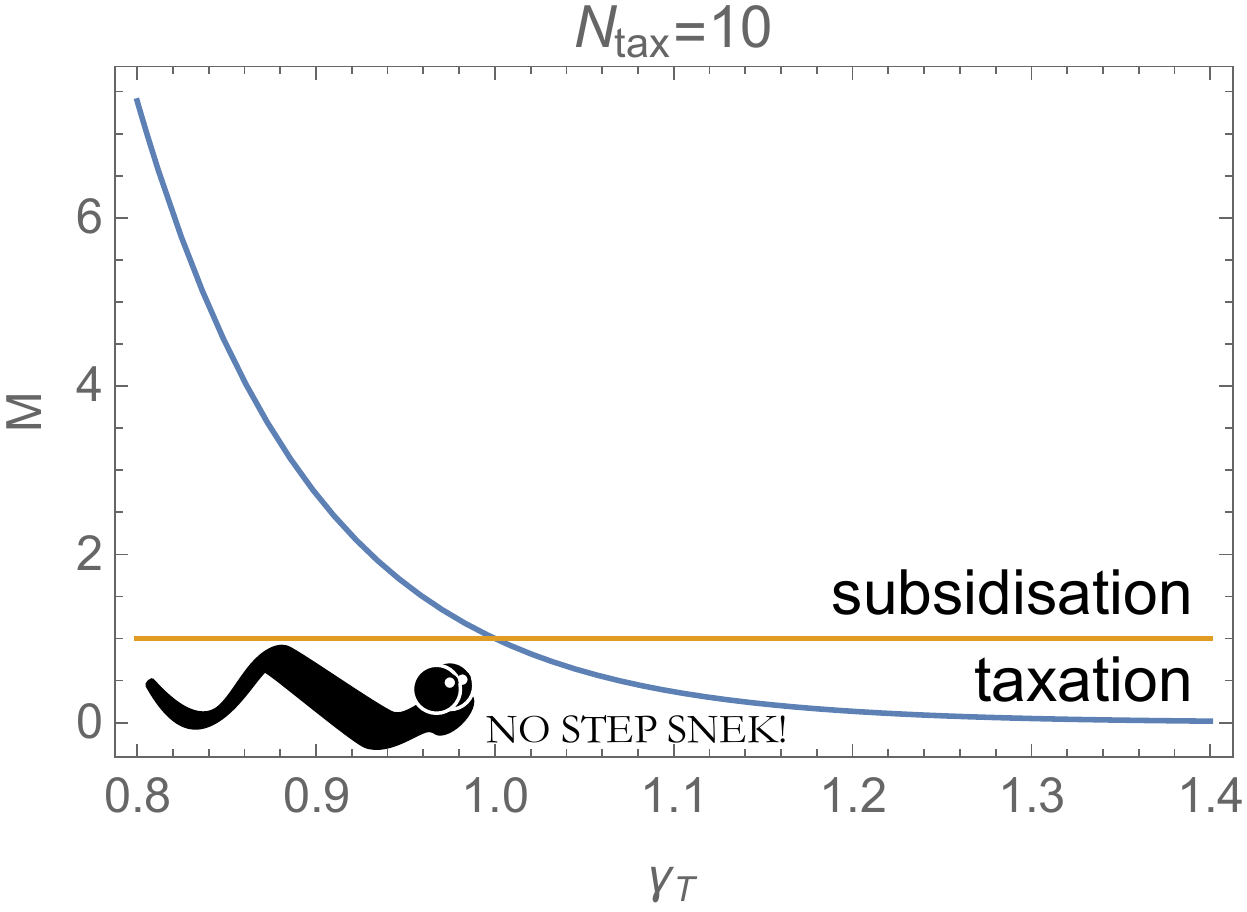}
\captionspacefig \caption{(top) Relationship between the tax performance multiplier, (positive) tax rate, and network size, assuming an exponential computational scaling function, in the regime of positive taxation, \mbox{$\gamma_T>1$}. (bottom) For \mbox{$N_\mathrm{tax}=10$}, a zoom into the region around neutral taxation, where \mbox{$\gamma_T\approx 1$}, showing slight degrees of both taxation (\mbox{$\gamma_T>1$}) and subsidisation (\mbox{$\gamma_T<1$}). Neutral taxation, \mbox{$\gamma_T=1$}, is shown in orange.}\label{fig:tax_exp}\index{Tea party}\index{Don't tread on me}
\end{figure}
\else
\begin{figure*}[!htbp]
\includegraphics[clip=true, width=0.475\textwidth]{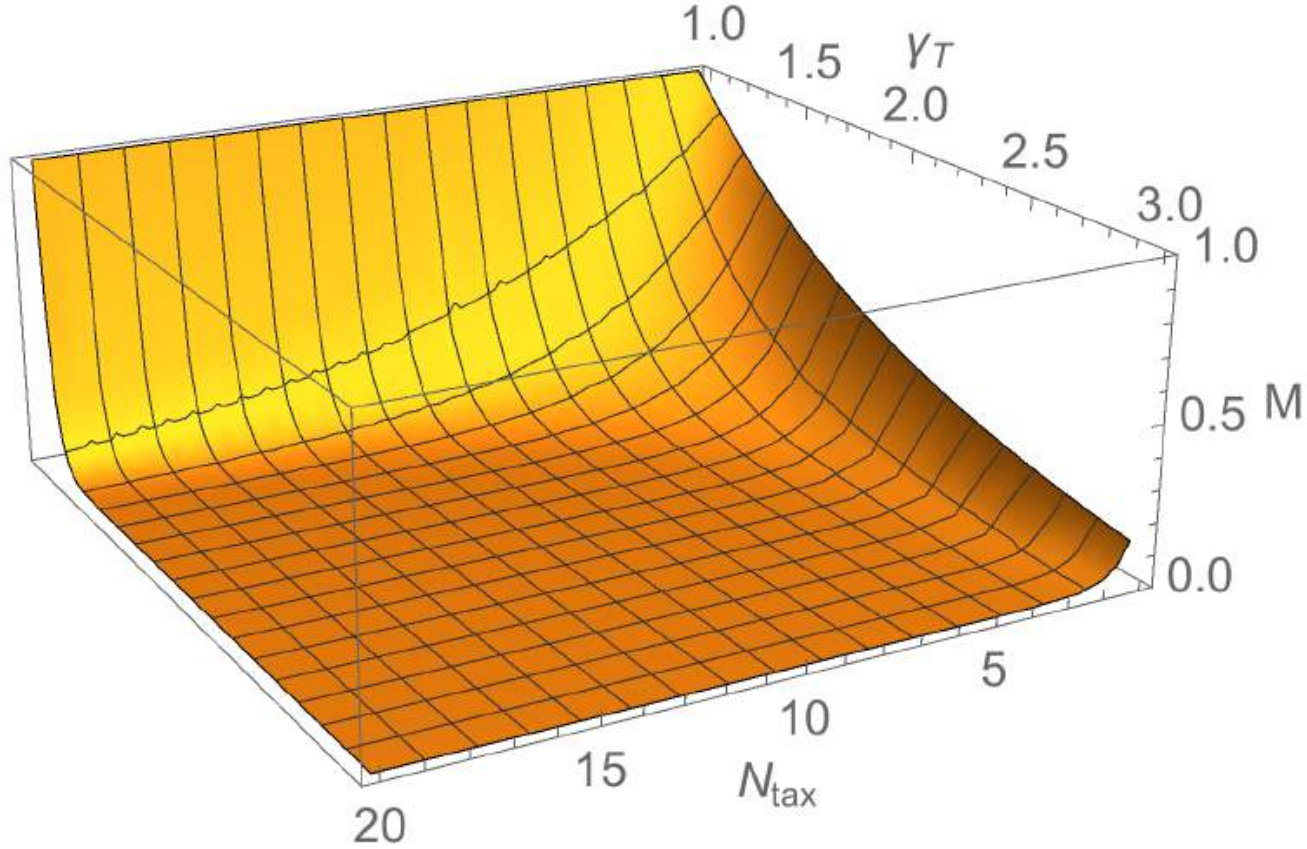}
\includegraphics[clip=true, width=0.475\textwidth]{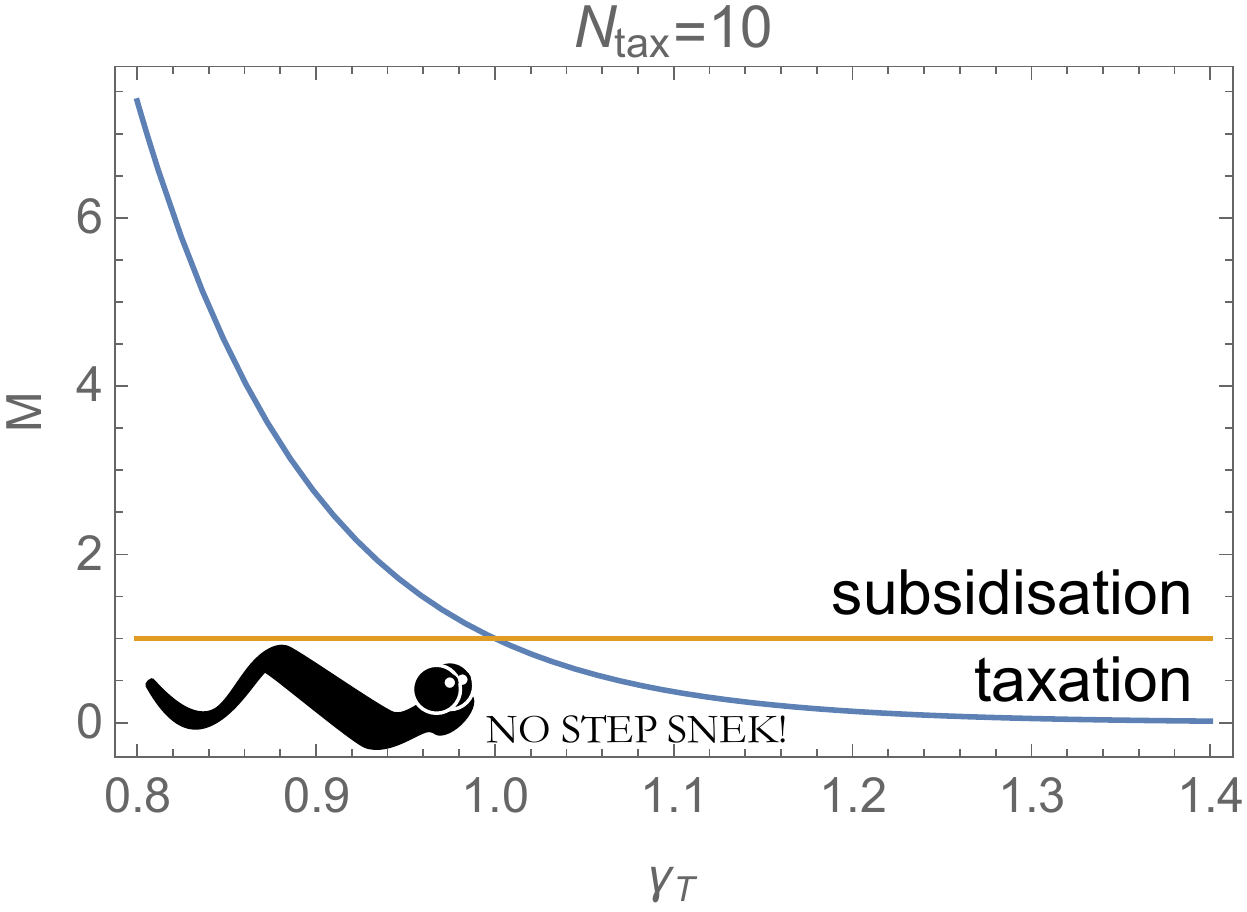}
\captionspacefig \caption{(left) Relationship between the tax performance multiplier, (positive) tax rate, and network size, assuming an exponential computational scaling function, in the regime of positive taxation, \mbox{$\gamma_T>1$}. (right) For \mbox{$N_\mathrm{tax}=10$}, a zoom into the region around neutral taxation, where \mbox{$\gamma_T\approx 1$}, showing slight degrees of both taxation (\mbox{$\gamma_T>1$}) and subsidisation (\mbox{$\gamma_T<1$}). Neutral taxation, \mbox{$\gamma_T=1$}, is shown in orange.}\label{fig:tax_exp}\index{Tea party}\index{Don't tread on me}
\end{figure*}
\fi

\subsubsection{Policy perspective}\index{Policy}\label{sec:policy}

\sectionby{Scott Harrison}\index{Scott Harrison}

Irrespective of the magnitude of change, the elasticities, especially on the demand side, indicate that the qubit would be ripe for the application of a consumer-driven tax. Graphically the imposition of a tax within a perfectly competitive market would take on the form of Fig.~\ref{fig:supply_demand}(a)\index{Supply \& demand curves}.

\if 1\doublecol
\begin{figure}[!htbp]
\includegraphics[clip=true, width=0.45\textwidth]{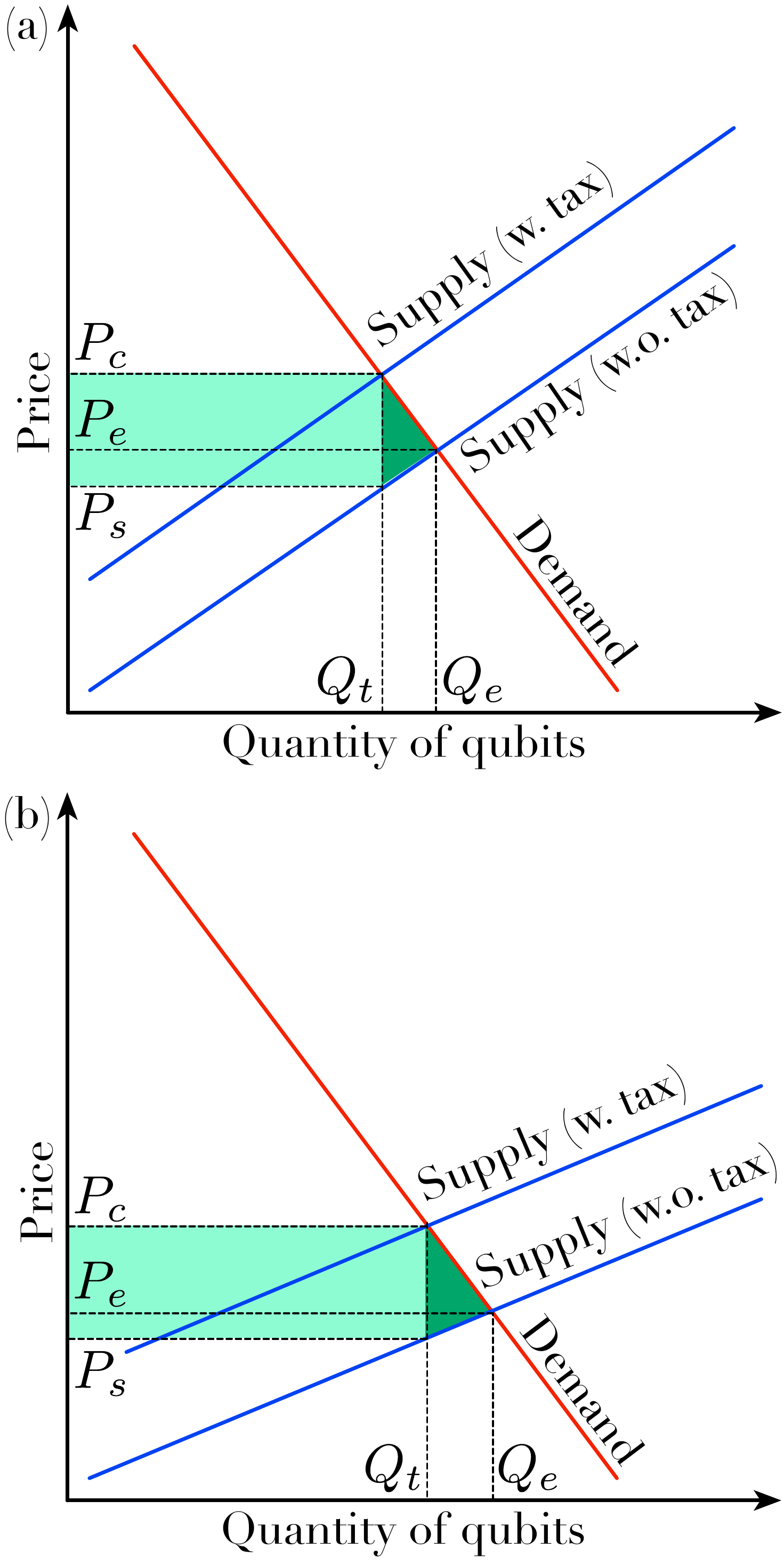}
\captionspacefig \caption{Hypothetical supply/demand curves\index{Supply \& demand curves}, showing the impact of taxation on price and supply, for both inelastic (a) and elastic (b) market dynamics. $Q_e$ and $P_e$ are the efficient market quantity and price of the asset respectively. $P_s$ is the price faced by suppliers, while $P_c$ is the consumer price under taxation. $Q_t$ is the quantity in a taxed environment. The net tax revenue collected is shaded in light teal, and the loss of market efficiency through the imposition of taxation is shaded in dark teal.}\index{Supply \& demand curves}\label{fig:supply_demand}	
\end{figure}
\else
\begin{figure*}[!htbp]
\includegraphics[clip=true, width=0.8\textwidth]{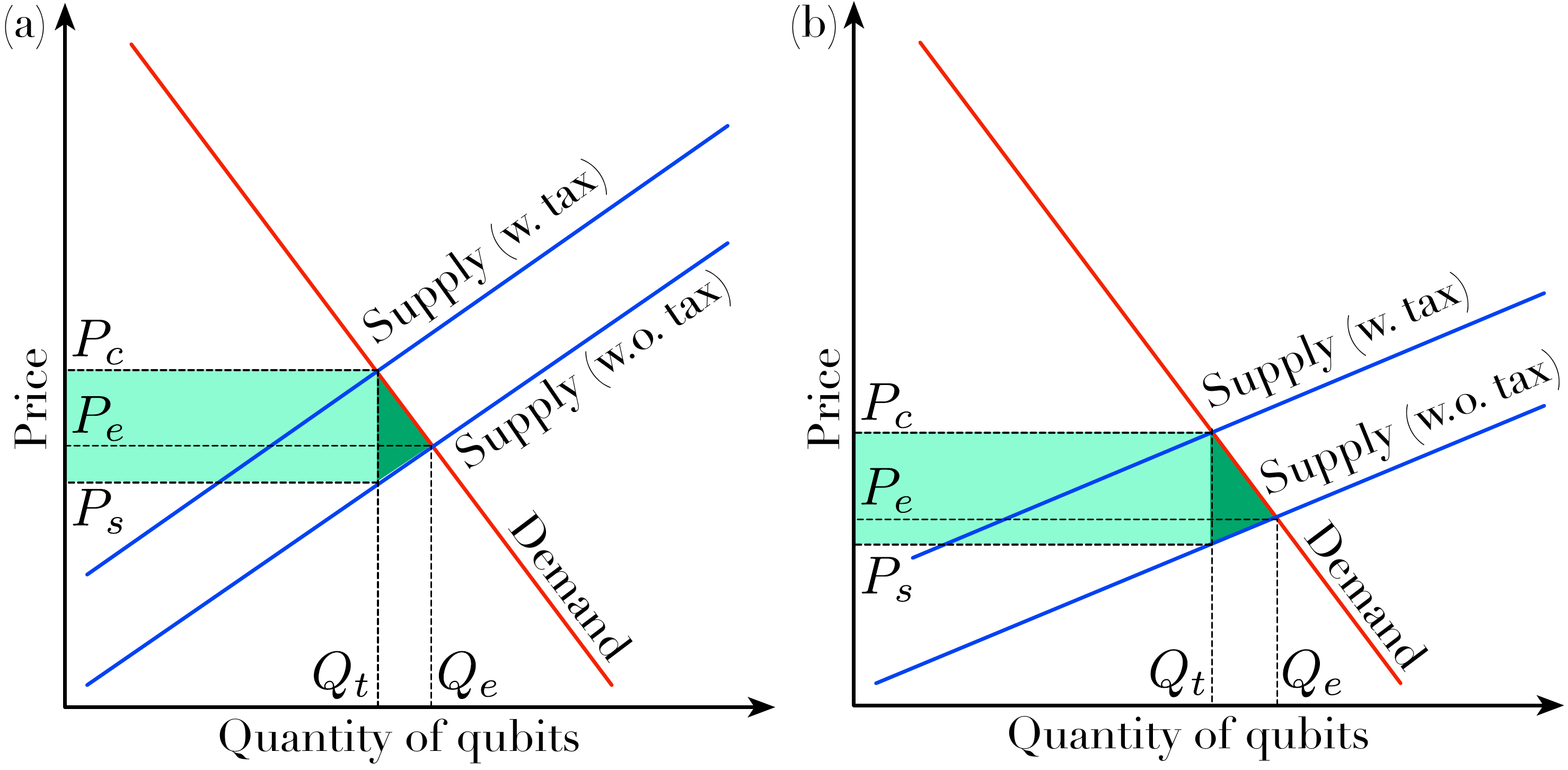}
\captionspacefig \caption{Hypothetical supply/demand curves, showing the impact of taxation on price and supply, for both inelastic (a) and elastic (b) market dynamics. $Q_e$ and $P_e$ are the efficient market quantity and price of the asset respectively. $P_s$ is the price faced by suppliers, while $P_c$ is the consumer price under taxation. $Q_t$ is the quantity in a taxed environment. The net tax revenue collected is shaded in light teal, and the loss of market efficiency through the imposition of taxation is shaded in dark teal.}\index{Supply \& demand curves}\label{fig:supply_demand}	
\end{figure*}
\fi

The graph indicates a downward sloping demand line, indicating that the lower the price, the higher the demand. The upward sloping supply curve reflects financial incentive, the higher the price, the greater the incentive there is for firms to increase the quantity of qubits available to the market. Correlating this to elasticities, a steeper demand or supply curve indicates a higher degree of inelasticity. As such, in Fig.~\ref{fig:supply_demand}(a)\index{Supply \& demand curves}, the slope of both the demand and supply curves are relatively inelastic (compared to a $45^\circ$ reference line). From this, the imposition of a consumer-based tax can be shown through the vertical shift of the supply curve, with the magnitude of the shift, \mbox{$P_c - P_s$}, indicating the per-qubit value of the tax. Other important observations from the graph are the tax revenue\index{Taxation!Revenue} collected (shaded in light teal), the loss of market efficiency\index{Market efficiency} through the imposition of a tax (shaded in dark teal), and a relatively small reduction in the quantity of qubits offered on the market from the efficient quantity, $Q_e$, to the quantity with the consumer tax, $Q_t$\index{Taxation}. 

An important implication of the imposition of the tax is the share of the taxation burden\index{Taxation}. The change in price from the efficient price, $P_e$, to the new market price faced by consumers, $P_c$, is somewhat equivalent to the shift from $P_e$ to the price point that the suppliers of qubits will receive, $P_s$. This means that the tax burden is likely to be equally shared between producers and consumers, which in the long term could act as a disincentive for increased production.

Fig.~\ref{fig:supply_demand}(b), however, shows a longer-term view of the qubit market, where the supply of qubits has become more elastic in nature. That is, the quantity supplied to the market becomes more sensitive to price fluctuations. This change in elasticity results in a shift in the tax burden, with a greater proportion of the tax now being paid by consumers, with only a small shift from the efficient price, $P_e$, to the price suppliers will face, $P_s$. 

From a policy implications perspective, this means that governments wanting to cash in on the new technology need to be cautious with the imposition of taxes relative to market maturity. Imposing a significant tax early on may only act as a disincentive to the development of the industry. However, once the market matures further, the imposition of a qubit tax would make strong economic sense, as there will be minimal loss of market efficiency. Importantly, such a tax on computational power alone could serve to be a relatively stable revenue generation tool for governments. 

%
% The Quantum Stockmarket
%

\subsection{The quantum stock market}\index{Quantum stock market}

In light of the distinction between subjective and objective value of computation, the question  is how to reconcile this distinction in value, given the diversity of applications in the quantum marketplace. This will supersede the na\"ive models for cost of computation presented earlier, which were based entirely on objective value. Of course, subjective value is what people are actually willing to pay for in the real-world!

This will give rise to a marketplace for tradable units of quantum computation, where the underlying asset is time-shares in the global network. We refer to this as the \textit{quantum stock market} -- a marketplace subject to ordinary supply and demand, economic, and of course psychological pressures. In a scenario where a large number of users are executing computations with high return (think the R\&D lab), asset values will be traded up. Contrarily, in a scenario of low-return computations (think our poor undergrad), they will be traded down. These market forces will be highly time-dependent, varying against many other factors in the economy, such as the emergence of new applications for quantum computation -- the discovery of an important new algorithm could spontaneously distort the market leading to major corrections.

The relative market value of computation will subsequently drive the direction of investment into quantum hardware, with carry-over effects on future market prices. If investment stagnates, so too will growth in computational dividends, driving up market rates by limiting supply (assuming positive growth in demand). This will, after market adjustment, drive investment back into the system to satisfy increasing demand. Thus, despite the present uncertainty into the future dynamics of the quantum stock market, we expect this positive feedback loop to ensure consistent, ongoing investment into the quantum network, and at least some marginal degree of price stability.

What is likely to arise is that most owners of quantum hardware will not be consumers, but rather investors, potentially highly speculative ones, who float their resources on the quantum stock market, betting on changes in demand for computation and their associated subjective cost. This trading could involve transactions in the direct underlying asset, future contracts (Sec.~\ref{sec:for_contr}) for locking in required computational power at future points in time, or more complex derivatives. For example, an investor anticipating a surge in high-value computations is likely to invest more heavily into hardware with the expectation of an uptrend in market rates of their licensing. And their return on investment\index{Return on investment (RoI)} will reflect these market dynamics.

As all markets for tradable assets do, sophisticated derivative markets will inevitably emerge, whereby people can speculate on or hedge against market dynamics, taking long, short, or more complex market positions, potentially in a highly-leveraged manner. As discussed in Sec.~\ref{sec:for_contr}, derivatives such as future contracts can be extremely helpful in enabling consumers to lock in future prices, creating a stable and predictable business climate. Similarly, other derivatives will enable market participants to hedge\index{Hedging} other quantum-related investments. For example, suppose an investor held a stake in an R\&D lab, highly reliant on quantum computing resources. By taking a leveraged long position on the market value for computation, he may limit losses on his R\&D investment associated with the higher price (and hence lower profit) they will be paying for computation. No doubt, market manipulation and all the usual nonsense and shenanigans\index{Shenanigans} will ensue.

%
% Geographic Localisation
%

\subsection{Geographic localisation}\index{Geographic!Localisation}\label{sec:geo_loc}

Because of the resource overheads associated with performing computations in a distributed manner, e.g via the resource costs associated with long-range repeater networks\index{Quantum repeater networks}, there is an economic imperative to localise quantum infrastructure, so as to mitigate this -- there is a clear economic benefit associated with housing qubits in close geographic proximity such that no long-distance quantum channels are required.

However, it's undesirable to \textit{entirely} centralise infrastructure of \textit{any} type, for two primary reasons:
\begin{itemize}
	\item Geostrategic competition\index{Geostrategic politics}: competing nation states or enterprises may not want essential infrastructure to be located entirely offshore, placing them at the mercy of their strategic competitors.
	\item Geographical redundancy\index{Geographic!Redundancy}: to eliminate single points of failure\index{Single points of failure} (SPOF), which undermine network robustness, it's desirable for infrastructure to be geographically decentralised. In present-day large-scale distributed classical platforms, geographical redundancy is a key consideration. Even though it would be most efficient if all data were completely centralised, obviating communications overheads, it would be catastrophic if a single earthquake (or war!) could decimate the entire system. For this reason, it is desirable to distribute failure modes.
\end{itemize}

Thus, we can reasonably anticipate that the quantum internet will not evolve like the classical `internet of things' (IoT)\index{Internet of things (IoT)}, whereby a massive number of ultra-small computational resources are scattered across the globe and networked. Rather, a relatively small number of central `hubs'\index{Hubs} are likely to emerge, which centralise enormous computational power, interconnected via the quantum internet to form the globally unified quantum cloud (Sec.~\ref{sec:glob_unif_quant_cloud}).

Much like the classical internet, it's to be expected the network that will emerge will exhibit a very hierarchical structure, following a Pareto distribution\index{Power!Law} in hub-size.

%
% Game theory of the qubit
%

\section{Game theory of the qubit marketplace}\index{Game theory}\label{sec:game_theory}

\sectionby{Scott Harrison \& Peter Rohde}\index{Scott Harrison}

\famousquote{People are always selling the idea that people with mental illness are suffering. I think madness can be an escape. If things are not so good, you maybe want to imagine something better.}{John Nash}
\newline

\startnormtable

\dropcap{E}{arlier} in this part, we established how a qubit market can function, and how the pricing mechanisms of various derivatives may work. One of the more interesting dimensions to the development of the quantum internet, is understanding \textit{how} the cooperation\index{Cooperation} between different suppliers will occur. Importantly, the expected high cost of  quantum hardware means that there may be a limited number of competing vendors. For profit maximisation\index{Profit maximisation} to occur, the most likely outcome is cooperation between them. The main question is, what can be learnt from applying game theoretic techniques to the strategies available to quantum computing vendors? And importantly, what are the implications associated with supply-side shifts, such as the imposition of taxation\index{Taxation}.

To analyse the decision making options for qubit suppliers, game theory is an analytical tool to understand `games' between players, where the outcome of the game is dependent on the various strategies employed by the players. The most well-known of these games is the prisoner's dilemma\index{Prisoner's dilemma} \cite{bib:Poundstone93}, describing the potential risks and rewards for two prisoners who are being independently questioned about a crime. 

Games can be classified based on their dimension, including the number of agents, the symmetry of the utility payoff, and whether they are cooperative. The Prisoner's dilemma is an example of a \textit{two-person, non-zero sum, non-cooperative game}\index{Two-person, non-zero sum, non-cooperative games} \cite{bib:Bacharach76}. More detailed examples of game theory have explored many of the base assumptions of this scenario, such as what if the prisoners are able to cooperate from the outset? How does this then result in maximising utility, and is cooperation always the best answer?

In the case of the quantum internet, it should be clear from the outset that there is a strong benefit associated with cooperation between vendors. Cooperative games form an important subset of the game theory domain, and are the most applicable to quantum computing, where `cooperation' translates to the unification of quantum computing resources into a larger distributed virtual quantum computer. As indicated previously, there will be exponential enhancements in computing power associated with unification, and as such, any qubit supplier will ultimately be able to produce excess computational power through networking and cooperating with others to exploit the computational leverage phenomenon (Sec.~\ref{sec:quant_ec_lev}). This idea is at the centre of the analysis when applying game theory to the decision making of suppliers.

\subsection{Key concepts}

For the uninitiated to economic analysis, particularly game theory, a few key concepts need to be established. This chapter by no means tries to cover these concepts in complete detail. For more detailed information we suggest referring to \cite{bib:Sugden04, bib:Bacharach76, bib:Straffin93}. Furthermore, the analysis at this point is only descriptive in nature, as a means of establishing the space where new research can be developed. Further more rigorous investigation is encouraged. The essential concepts that we rely on taken from game theory are:

\begin{itemize}
	\item \textit{Utility}\index{Utility}: this can be generally defined as `the ability to assign a number (utility) to each alternative so that, for any two alternatives, one is preferred to the other if and only if the utility of the first is greater than the utility of the second' \cite{bib:Fishburn70}. In this regard, utility is often seen as a representation of the overall benefit associated with a decision or preference. As utility is unobservable and may be defined essentially arbitrarily, it can become subjective in nature. However, the key is not whether the values assigned to any one preference are subjective, but rather whether the assigned values associated with competing decisions are comparable, so that they can be quantitatively ranked. Thus, utility is a tool for the comparison of benefits associated with decision outcomes.

	\item \textit{Utility payoff matrix}\index{Utility!Payoff matrices}: Utility values assigned to any possible decision can be formulated into a matrix, which collates all the possible decisions from a given decision-maker's perspective. In a game with two participants, this would result in a 2-dimensional matrix. With $n$ decision makers, there would be an $n$-dimensional matrix representation. The matrix also forms the basis for the graphical analysis undertaken throughout this section for the 2-player scenarios.

	\item \textit{Negotiation set}\index{Negotiation sets}: this defines a space of possible preferences. The negotiation set was first introduced in the seminal work of \cite{bib:NeumanMorgenstern44}. At its most basic, the negotiation set forms a set of bounds for the payoff matrix. This limits possible solutions for a game to strategies that would actually see an improvement in the individual's utility above the base alternative of making no decision at all. This is also referred to as the \textit{status quo}\index{Status quo}.
\end{itemize}

\subsection{Strategies}\index{Strategies}

To best develop an understanding of how a quantum internet game will be played out, we will begin by analysing the utility payoff space for classical computing. Currently we can easily define three key strategies for two market participants ($X$ and $Y$), who both act as both suppliers and consumers of computational resources. The three strategies we compare are:

\begin{itemize}
	\item \textsc{Isolation}\index{Isolation strategies}: $X$ and $Y$ build their own systems in isolation, which they utilise independently. There is no cooperation or networking of computational resources between them. This can be considered the status quo ($S$) for the players, as it represents the autarky position and defines the von Neuman \& Morgenstern utility space\index{Utility!Space}.  This is represented in Fig.~\ref{fig:game_theory_1}. Importantly, any changes to $X$'s computational capacity has no impact on $Y$'s. So $X$ can take any position along the horizontal axis, and there is no change in $Y$'s utility, and vice versa. The resulting \textit{utility point}\index{Utility!Point} is described here as the intersecting minimum of these options, represented by the point \mbox{$(S_x, S_y)$}.

\begin{figure}[!htbp]
\includegraphics[clip=true, width=0.475\textwidth]{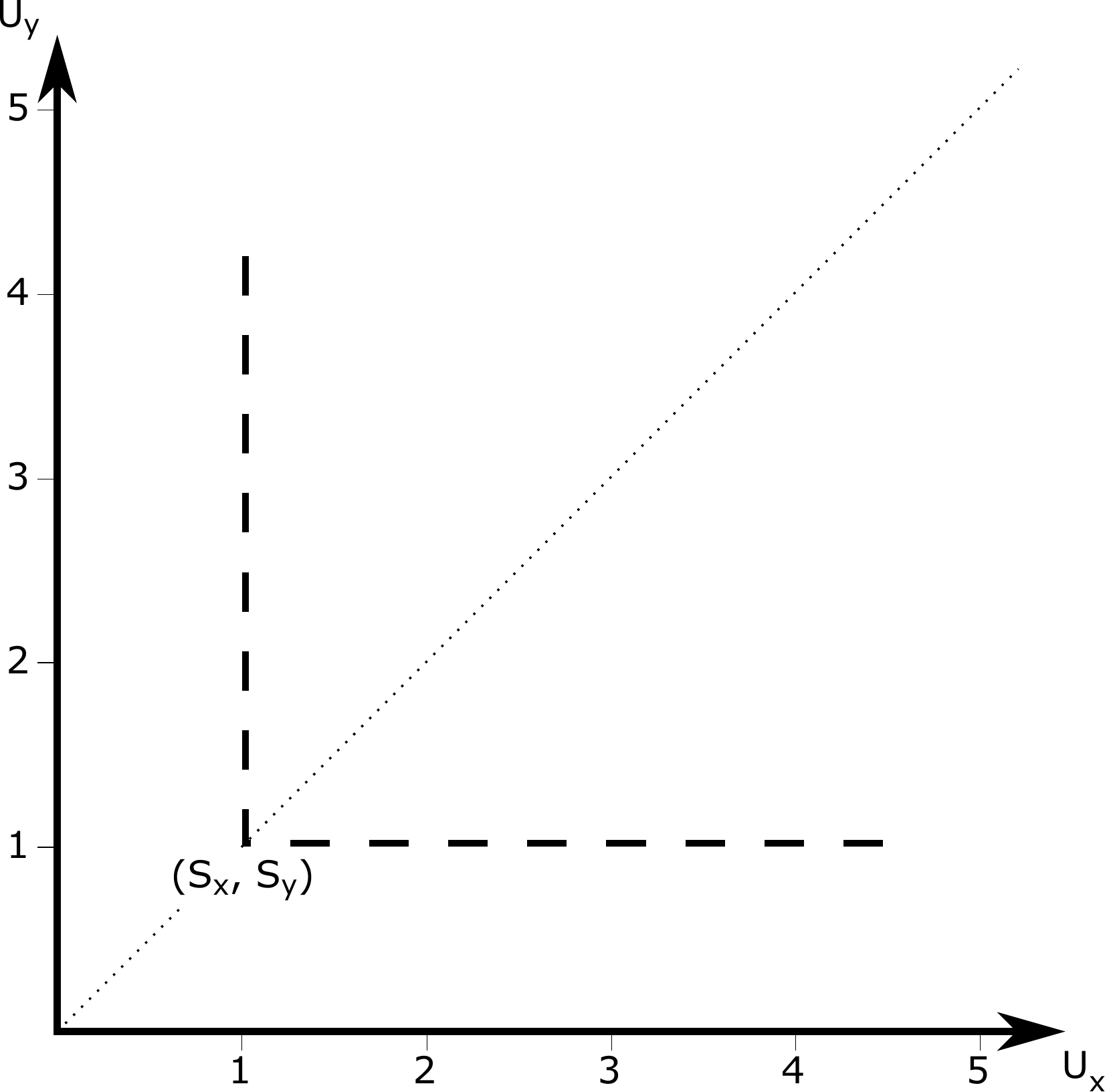}
\captionspacefig \caption{Utility space for two players with access to classical computing. The bold dashed lines show the minimum utility payoff for $X$ and $Y$.}\label{fig:game_theory_1}
\end{figure}

\item \textsc{License}\index{License strategies}: Either $X$ or $Y$ build their own systems, but then licence unused compute-cycles to the other. This means that the other player may still have access to their required net computational resources, but essentially outsources the setup costs and ongoing infrastructure maintenance. This improves the utilisation of the system for the player who licenses out, resulting in increased efficiency, profitability and subsequently utility (under the conditions of increasing economies of scale and assuming homogenous system requirements).

\item \textsc{Unify}\index{Unify strategies}: The two players consolidate their computational resources into a distributed cloud computing environment (Sec.~\ref{sec:dist_QC}), where the limitations of a single system are lifted, allowing for even better resource-sharing and improvement beyond what a single system solution can provide. In this scenario, both providers will be able to collaborate such that they can meet their individual computational needs, without having to fully build independent systems as before. Importantly though, while there may be a small loss in utility for one of the parties compared to the \textsc{License} strategy, unification allows an overall higher level of joint utility to be achieved, creating a Nash equilibrium\index{Nash equilibrium} at this point, $C$.
\end{itemize}

\subsection{Utility payoff behaviour}

In Tab.~\ref{tab:game_theory_1} we present an example utility payoff matrix (numbers chosen arbitrarily) between two players engaging in the above three strategies with classical computing resources.

\begin{table}[!htbp]
\resizebox{0.475\textwidth}{!}{\begin{tabular}{|c|c|c|c|c|}
\hline
Player               & \multicolumn{4}{c|}{$X$}                                            \\ \hline
\multirow{4}{*}{$Y$} & Strategy         & $X\rightarrow Y$ & $X+Y$       & $X\leftarrow Y$ \\ \cline{2-5} 
                     & $X\rightarrow Y$ & $(3,1.5)$        & $(1,1)$     & $(1,1)$         \\ \cline{2-5} 
                     & $X+Y$            & $(1,1)$          & $(2.5,2.5)$ & $(1,1)$         \\ \cline{2-5} 
                     & $X\leftarrow Y$  & $(1,1)$          & $(1,1)$     & $(1.5,3)$       \\ \hline
\end{tabular}}
\captionspacetab \caption{Example of a payoff matrix for two classical computing vendors. `\mbox{$\leftarrow/\rightarrow$}' indicates the \textsc{License} (from/to) strategy, `+' indicates the \textsc{Unify} strategy, and the off-diagonal combinations are the status quo \textsc{Isolation} strategy. $(X,Y)$ denotes the utility to players $X$ and $Y$ respectively. Note that there is some loss in net utility using \textsc{License}, owing to inefficiency\index{Inefficiency} through incurred transaction cost overheads\index{Transaction costs}.}\label{tab:game_theory_1}
\end{table}

\begin{figure}[!htbp]
\includegraphics[clip=true, width=0.475\textwidth]{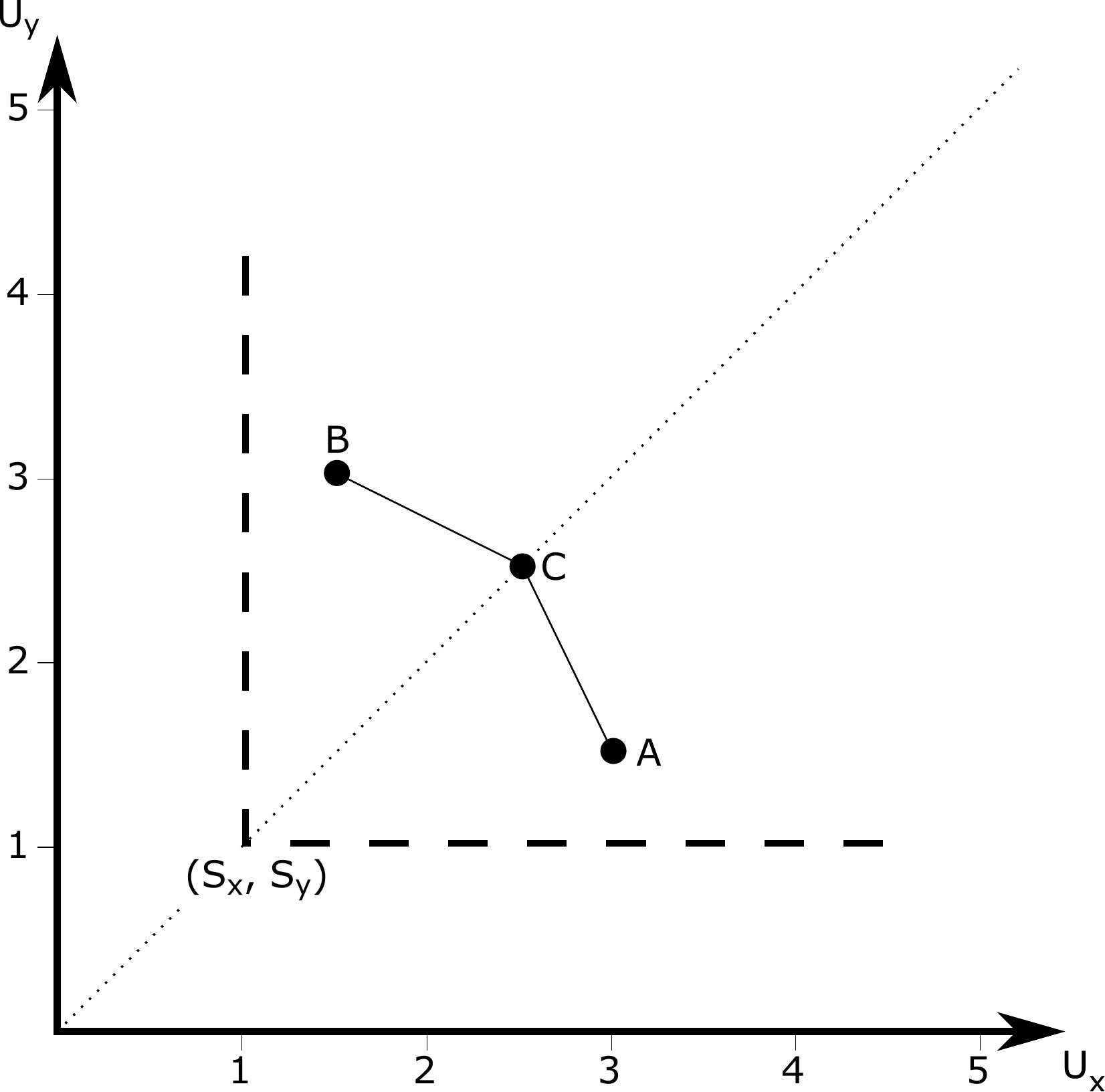}
\captionspacefig \caption{Utility payoff between two players in the classical computing environment. $A$ indicates the utility combination where $X$ builds the system and licenses out time to $Y$, and $B$ represents the reverse. $C$ is the point where both $X$ and $Y$ cooperate through distributed cloud computing, maximising the overall utility for both players. $A$ and $B$ derive identical utility values from the strategies, yielding symmetry about the diagonal axis. The lines $AC$ and $BC$ indicate possible solutions where mixed strategies \index{Mixed strategies} are employed, combining components of \textsc{License} and \textsc{Unify}. $C$ is also the Nash equilibrium\index{Nash equilibrium}, where cooperative bargaining would result in the best outcome overall.}\label{fig:game_theory_2}
\end{figure}

It's important to note that these values are totally arbitrary in nature, and do not have a `real-life' interpretation, other than understanding the preferencing of the described strategies. Translating the utility payoff matrix into a graphical representation yields Fig.~\ref{fig:game_theory_2}.

Now, quantum computing uses similar strategies for possible solutions, with one key difference. Using classical computing, the relationship for the \textsc{Unify} strategy was described as additive in nature, where the computational resources of both players are accumulated additively when unified into a larger virtual computer. In the corresponding quantum \textsc{Unify} strategy, this effect is enhanced by their super-linear computational leverage. For the sake of illustration, we will now assume this becomes multiplicative. Multiplicativity in computational power approximates behaviour under a \textsc{Unify} strategy when dealing with exponential scaling functions.

\begin{figure}[!htbp]
\includegraphics[clip=true, width=0.4\textwidth]{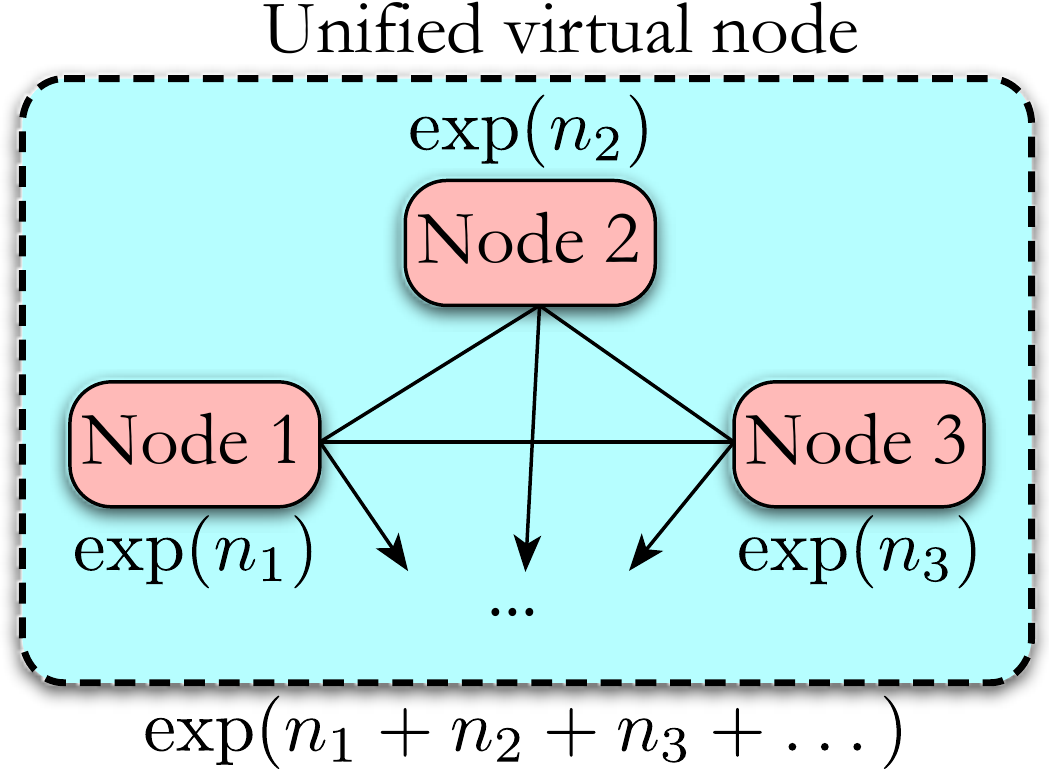}
\captionspacefig \caption{A distributed quantum computation across a number of nodes, each with $n_i$ qubits. The computational scaling function is chosen to be exponential in form,  yielding a classical-equivalent computational power of \mbox{$\mathrm{exp}(n_i)$} for each node. However, the joint computational power of the network is given by \mbox{$\mathrm{exp}(n_1+n_2+\dots)$}, which is exponentially greater than the sum of the individual computational powers, \mbox{$\mathrm{exp}(n_1)+\mathrm{exp}(n_2)+\dots$}, in general.}\label{fig:exp_coll_enh}
\end{figure}

This is easily intuitively seen as follows. Let the computational scaling function be an arbitrary exponential,\index{Multiplicativity in computational power}
\begin{align}
f_\mathrm{sc}(n) = O(\mathrm{exp}(n)).	
\end{align}
Then it immediately follows that the scaling function obeys the identity,
\begin{align}\label{eq:prod_comp_sc}
	f_\mathrm{sc}\left(\sum_i n_i \right) = \prod_i f_\mathrm{sc}(n_i),
\end{align}
yielding the multiplicative behaviour, shown diagrammatically in the context of a distributed quantum computation in Fig.~\ref{fig:exp_coll_enh}. Note that in the classical case the product in Eq.~(\ref{eq:prod_comp_sc}) would be become a sum, which is exponentially smaller in generally,
\begin{align}
	\prod_i f_\mathrm{sc}(n_i) \geq \sum_i f_\mathrm{sc}(n_i).
\end{align}

As such, should $X$ and $Y$ build identical quantum computers, the hypothesised payoff matrix would become as shown in Tab.~\ref{tab:game_theory_2}.

\begin{table}[!htbp]
\resizebox{0.475\textwidth}{!}{\begin{tabular}{|c|c|c|c|c|}
\hline
Player               & \multicolumn{4}{c|}{$X$}                                            \\ \hline
\multirow{4}{*}{$Y$} & Strategy         & $X\rightarrow Y$ & $X+Y$       & $X\leftarrow Y$ \\ \cline{2-5} 
                     & $X\rightarrow Y$ & $(3,1.5)$        & $(1,1)$     & $(1,1)$         \\ \cline{2-5} 
                     & $X+Y$            & $(1,1)$          & $(5,5)$ & $(1,1)$         \\ \cline{2-5} 
                     & $X\leftarrow Y$  & $(1,1)$          & $(1,1)$     & $(1.5,3)$       \\ \hline
\end{tabular}}
\captionspacetab \caption{Example utility payoff matrix for two players with quantum computing resources. Note the enhancement in the diagonal $X+Y$ matrix element, compared to the classical case.}\label{tab:game_theory_2}
\end{table}

\begin{figure}[!htbp]
\includegraphics[clip=true, width=0.475\textwidth]{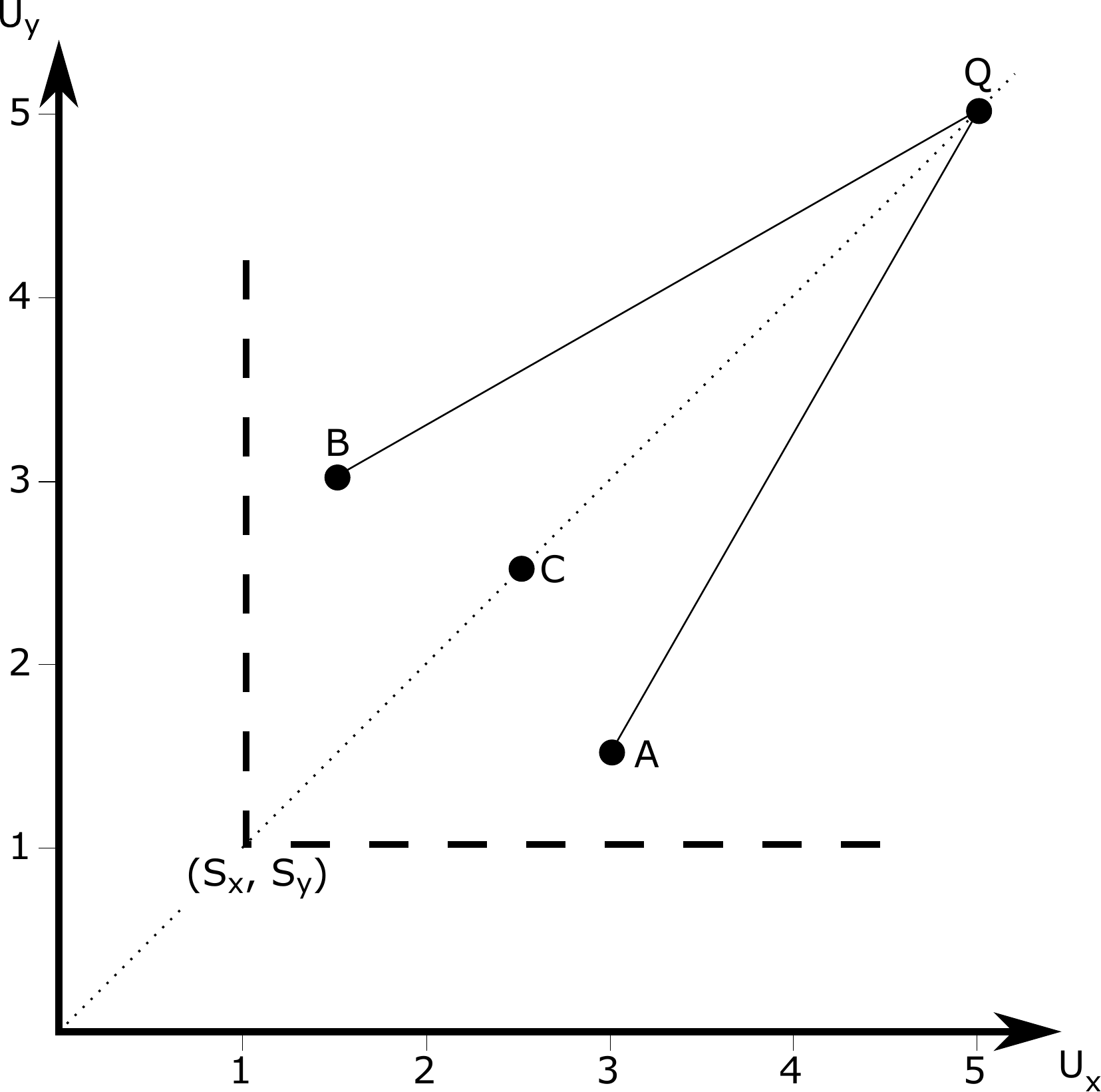}
\captionspacefig \caption{Tradeoff in utility between two players in the scenario of quantum computing. $Q$ now represents the \textsc{Unify} strategy, where there is a quantum enhancement in the utility payoff for $X$ and $Y$. Note that $Q$ offers strictly greater utility than $C$, the corresponding classical point. Any mixed strategy\index{Mixed strategy} combining \textsc{License} and \textsc{Unify} will result in a non-linear transition between points $A/B$ and $Q$.}\label{fig:game_theory_3}
\end{figure}

In this scenario, the \textsc{Isolation} and \textsc{License} strategies are assumed to yield the same utility payoffs as in the classical case. The only difference arises when $X$ and $Y$ \textsc{Unify}. The effect of quantum computational enhancement is to therefore amplify the cooperative elements in the utility payoff matrix, potentially by very large factors. This has the generic effect that in the quantum realm, cooperation is more highly incentivised than in the classical one. The resulting graphical representation of this payoff matrix is shown in Fig.~\ref{fig:game_theory_3}.

\subsection{Cooperative payoff enhancement}\index{Cooperative payoff enhancement}\label{sec:coop_enh}

An individual user, $i$, of a quantum computer operating on their own, observes computational power characterised by the appropriate computational scaling function acting on the number qubits in their possession, $f_\mathrm{sc}^{(i)}(n)$. This stipulates the utility payoff for that individual, allowing for trivial  construction (and efficient mathematical representation) of multi-player payoff matrices simply by characterising single-player payoffs independently.

Once cooperative strategies are introduced, the associated cooperative payoff matrix elements are transformed appropriately, according to the reallocation of resources and how they are collectively utilised.

We refer to this phenomena as \textit{cooperative payoff enhancement}, one which, depending on what cooperative techniques are employed, drastically alters the economic landscape, and its associated game-theoretic analysis and outcomes.

We will present an elementary analysis of this concept for the three different cooperation strategies introduced earlier, initially in the standard 2-player context, subsequently generalised to an arbitrary multi-player environment.

\subsubsection{\textsc{Isolation} strategies}\index{Isolation strategies}

When implementing quantum computations characterised by completely general computational scaling functions\index{Computational scaling functions}, $f^{(X,Y)}_\mathrm{sc}$ (which in general can be distinct for the two players, $X$ and $Y$), when using an \textsc{Isolation} strategy, the respective payoff matrix elements are simply given by,
\begin{align}\label{eq:util_payoff_iso}
\textsc{Isolation} = \begin{pmatrix}
 	f^{(X)}_\mathrm{sc}(n_X)\\
 	f^{(Y)}_\mathrm{sc}(n_Y)
 \end{pmatrix},
\end{align}
where payoff matrix elements are assumed to be in units of classical-equivalent processing time\footnote{There's nothing unique or special about using classical-equivalent computational power as our utility measure. Any other measure of `payoff' could be equally well justified, depending on circumstance. For example, one could instead represent utility in terms of the monetary value of computational power. In that case, we simply need to transform the payoff matrix elements using the cost of computation\index{Cost of computation} identity presented in Post.~\ref{post:cost_comp}.} (i.e FLOPs), and resource-sharing is based on the methodology for arbitrage-free time-share allocation presented in Sec.~\ref{sec:arb_free_time_share}.

\subsubsection{\textsc{License} strategies}\index{License strategies}

Elements associated with \textsc{License} strategies are transformed as,
\begin{widetext}
\begin{align}
\begin{pmatrix}
   f^{(X)}_\mathrm{sc}(n_X) \\
   f^{(Y)}_\mathrm{sc}(n_Y)
\end{pmatrix} \underset{\textsc{License}}{\longrightarrow} \begin{pmatrix}
   r_{X\to X} \cdot f^{(X)}_\mathrm{sc}(n_X) + r_{Y\to X} \cdot f^{(X)}_\mathrm{sc}(n_Y) \\
   r_{Y\to Y} \cdot f^{(Y)}_\mathrm{sc}(n_Y) + r_{X\to Y} \cdot f^{(Y)}_\mathrm{sc}(n_X)
\end{pmatrix},
\end{align}
\end{widetext}
where \mbox{$0\leq r_{i\to j}\leq 1$} denotes the proportion of $i$'s compute-time licensed to $j$.

This is easily logically generalised to an arbitrary multi-player setting, in which case the transformation becomes,
\begin{align}
	f_\mathrm{sc}^{(i)}(n_i) \underset{\textsc{License}}{\longrightarrow} \sum_{j=1}^N r_{j\to i} f_\mathrm{sc}^{(i)}(n_j),
\end{align}
where there are $N$ players, all engaging with one another using licensing only. The $r_{i\to j}$ parameters are normalised for all users such that,
\begin{align}
\sum_{j=1}^N r_{i\to j} \leq 1\,\,\forall\,i.
\end{align}

With equality, this normalisation implies perfect licensing efficiency (i.e no overheads) and no wasted clock-cycles (full utilisation). Inequality implies either inefficiency or under-utilisation. Since under this strategy net computational power is conserved (at best), it might appear mindless to employ it at all, given that there is no net gain. Whilst this is true, there may be ulterior motives for employing it. For example, it might be employed for the purposes of load balancing\index{Load balancing} across a distributed architecture, or implementing arbitrage\index{Arbitrage} between inconsistent market pricing of computational power between nodes.

Note that when \mbox{$r_{i\to j}=\delta_{i,j}$} (i.e \mbox{$r=I_N$} is the \mbox{$N\times N$} identity matrix, and there is no inter-player licensing) the \textsc{License} strategy simply reduces back to the \textsc{Isolation} strategy.

\subsubsection{\textsc{Unify} strategies}\index{Unify strategies}

Elements associated with \textsc{Unify} strategies will undergo the quantum utility payoff enhancement,
\begin{align}\label{eq:util_payoff_enh_1}
\begin{pmatrix}
	f^{(X)}_\mathrm{sc}(n_X)\\
	f^{(Y)}_\mathrm{sc}(n_Y)
\end{pmatrix}
 \underset{\textsc{Unify}}{\longrightarrow} \begin{pmatrix}
 	n_X \cdot \chi^{(X)}_\mathrm{sc}(n_X+n_Y)\\
 	n_Y \cdot \chi^{(Y)}_\mathrm{sc}(n_X+n_Y)
 \end{pmatrix},
\end{align}
from the definition for time-shared compute power given in Def.~\ref{def:time_share_comp_power}.

As before, we can logically generalise the payoff enhancement of a generalised \textsc{Unify} strategy to the multi-player scenario as,
\begin{align}
	f_\mathrm{sc}^{(i)}(n_i) \underset{\textsc{Unify}}{\longrightarrow} n_i \cdot \chi_\mathrm{sc}^{(i)}(n_\mathrm{global}).
\end{align}

Thus, it is evident that the enhancement in \textsc{Unify} strategies is highly dependent on:
\begin{itemize}
	\item The total number of qubits held between the players,
		\begin{align}
			n_\mathrm{global} = \sum_{j=1}^N n_j.
		\end{align}
	\item The proportion of the qubits held by each player,
		\begin{align}
			r_i = \frac{n_i}{\sum_{j=1}^N n_j}.
		\end{align}
	\item The respective algorithms to which the computational resources are being applied by each player, which influence the player-specific subjective scaling functions, $f^{(i)}_\mathrm{sc}$, independently.
\end{itemize}

The final point is particularly noteworthy, since it implies that optimal game-theoretic outcomes are not objective, but subjective, and highly dependent on how players are employing their computational resources, which may be highly distinct and change dynamically over time. If one player is employing an exponential scaling function, whereas the other is only employing a polynomial one, this could completely distort the utility payoff dynamics of the game in favour of the player who would otherwise have been weaker under symmetric scaling functions. This in turn could completely alter the landscape of how users choose strategies to play optimally.

\subsubsection{Strategic implications}\index{Strategy}

\famousquote{In savage countries they eat one another, in civilised ones they deceive one another; and that is what people call the way of the world!}{Arthur Schopenhauer}
\newline

It is clear that the \textsc{Unify} strategy, in which distinct quantum computing nodes are merged via the network into a larger distributed quantum computer, works to the (potentially exponential) benefit of all contributing parties. This distorts game-theoretic analysis of network participants compared to classical computing.

On one hand, the guaranteed mutual benefit of all players directly enhances their individual compute power. In a compute-centric world, where computation equates to productivity, this directly works in the self-interest of all.

However, taking a more strategic long-term perspective, despite self-enhancement, the associated enhancement of competitors may eventuate in outcomes that work against self-interest to a sufficient extent that it outweighs this benefit. Thus, cooperative enhancement, in an appropriate strategic context, could equally be tantamount to `adversarial enhancement'\index{Adversarial enhancement}, and be considered an overwhelming motivate to avoid cooperation with certain players.

Some non-technical discussion on the implications of these observations is presented in several of the essays in Part.~\ref{part:essays}.

\subsubsection{Inefficient markets}\index{Market inefficiency}

The utility payoff enhancement characterised by these transformations is based on the simplest of toy models, where unification is assumed to be perfectly efficient -- there are no overheads (e.g transaction or communication costs) associated with cooperation, nor are there any externalities, such as taxes or regulations. In reality, these assumptions are of course completely unrealistic. Thankfully, such secondary effects can be relatively easily incorporated into the model by modulating them with additional layers of transformations capturing these features.

For example, consider the unification of computational resources between two players residing in different jurisdictions, which levy import/export tariffs\index{Tariffs} against one another, an externality introducing inefficiency into cooperative strategies. When expressed in terms of the monetary cost of computation, this would effectively modulate the payoffs of \textsc{Unify} matrix elements by a tariff-dependent function.

\subsection{Mixed strategies}

\begin{figure}[!htbp]
\includegraphics[clip=true, width=0.475\textwidth]{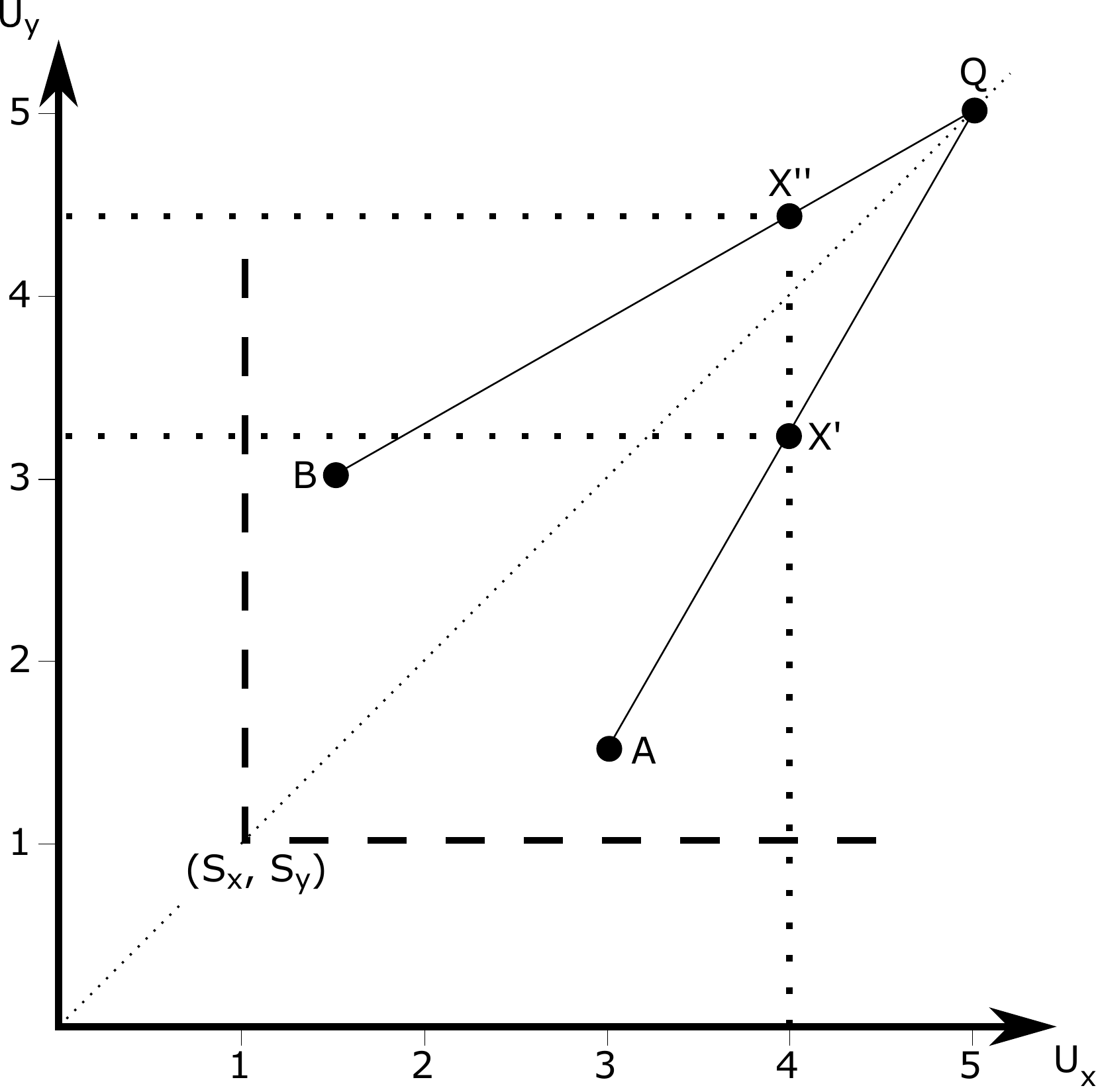}
\captionspacefig \caption{Utility payoff for two players with quantum computers, employing mixed strategies. $X'$ and $X''$ denote two unique solutions for $Y$ having a utility of 4. From $Y$'s perspective it makes no difference which strategy is chosen. However, $X$ can locally optimise their utility by choosing $X''$ over $X'$.}\label{fig:game_theory_4}
\end{figure}

How do mixed strategies\index{Mixed strategies} affect cooperation? To perform this analysis, assume that $X$ has chosen a mixed strategy comprising a combination of \textsc{License} and \textsc{Unify}. In the example shown in Fig.~\ref{fig:game_theory_4}, this results in a utility value of 4 for $X$, represented by the vertical dotted line.

\begin{figure}[!htbp]
\includegraphics[clip=true, width=0.475\textwidth]{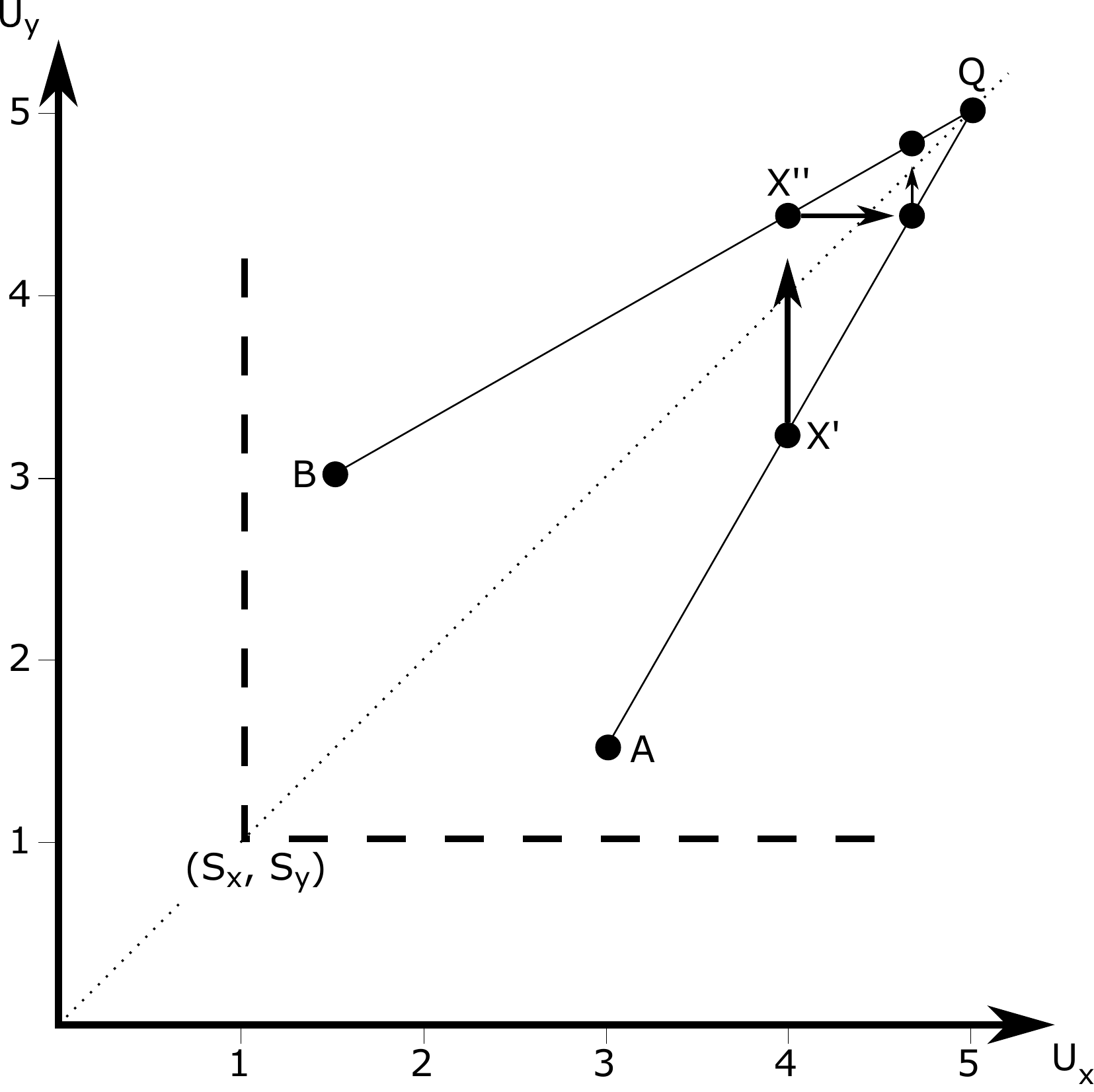}
\captionspacefig \caption{Adaptive strategy revision by $X$ and $Y$, yielding incremental local utility improvements without affecting the other, progressively marches us towards the optimal point $Q$ asymptotically.}\label{fig:game_theory_5}
\end{figure}

This results in two possible solutions, $X'$ and $X''$, that intersect the lines $AQ$ and $BQ$ respectively. $X'$ represents a point where more compute-time is allocated to the \textsc{License} strategy, and less to \textsc{Unify}, while $X''$ is the converse. $X$ is indifferent to both solutions, as they result in the same utility, but for $Y$, there is a clear choice. $X''$ will result in higher utility, placing an emphasis on cooperation. Once $Y$ has chosen their strategy, $X$ can further revise their original choice too, optimising their utility without undermining $Y$'s. This results in adaptive strategy revision, that asymptotes towards the optimal solution $Q$, as shown in Fig.~\ref{fig:game_theory_5}.

The conclusion from this is that while mixed strategies\index{Mixed strategies} comprising elements of \textsc{License} and \textsc{Unify} may be advantageous, there will always be an underlying pressure towards cooperation.

\subsection{Taxation}

What happens to the utility payoff matrix when taxation is imposed?\index{Taxation} The assumption that $X$ and $Y$ are operating in similar regulatory environments can be relaxed. What happens when $X$ is in a more heavily taxed environment than $Y$, or something impacts the utility achieved by the players in an asymmetric manner?

Economically, taxation operates by transferring utility from the supplier of a good to the government. This will result in some reduction in supply. However, in the \textsc{Unify} strategy, even small reductions in supply may be exponentially multiplied. The result is that all collaborative strategies will yield less utility for $X$, as it both reduces supply and transfers utility to government. The impact however is also felt by $Y$, as the overall joint computational power of the cloud is reduced. This results in a payoff matrix as shown in Tab.~\ref{tab:game_theory_3}.

\begin{table}[!htbp]
\resizebox{0.475\textwidth}{!}{\begin{tabular}{|c|c|c|c|c|}
\hline
Player               & \multicolumn{4}{c|}{$X$}                                            \\ \hline
\multirow{4}{*}{$Y$} & Strategy         & $X\rightarrow Y$ & $X+Y$       & $X\leftarrow Y$ \\ \cline{2-5} 
                     & $X\rightarrow Y$ & $(2,1.25)$        & $(1,1)$     & $(1,1)$         \\ \cline{2-5} 
                     & $X+Y$            & $(1,1)$          & $(3,4.5)$ & $(1,1)$         \\ \cline{2-5} 
                     & $X\leftarrow Y$  & $(1,1)$          & $(1,1)$     & $(1.5,3)$       \\ \hline
\end{tabular}}
\captionspacetab \caption{Example utility payoff matrix in the presence of taxation on quantum computers. The effect is a net depreciation in achievable utility.}\label{tab:game_theory_3}
\end{table}

When $X$ licenses compute-time to $Y$, there is a reduction in utility derived by $X$. The small reduction in supply will also mean that there is a reduction in the available compute-time available to be licensed to $Y$. For the \textsc{License} strategy, this means there would be a shift in utility from $(3,1.5)$ in Tab.~\ref{tab:game_theory_2} to $(2, 1.25)$ in Tab.~\ref{tab:game_theory_3}. Graphically, this is shown as the inward shift from $A$ to $A'$ in Fig.~\ref{fig:game_theory_4}. We also assume here that the imposition of the tax itself doesn't completely abolish any utility gains from the \textsc{License} strategy over the status quo.

\begin{figure}[!htbp]
\includegraphics[clip=true, width=0.475\textwidth]{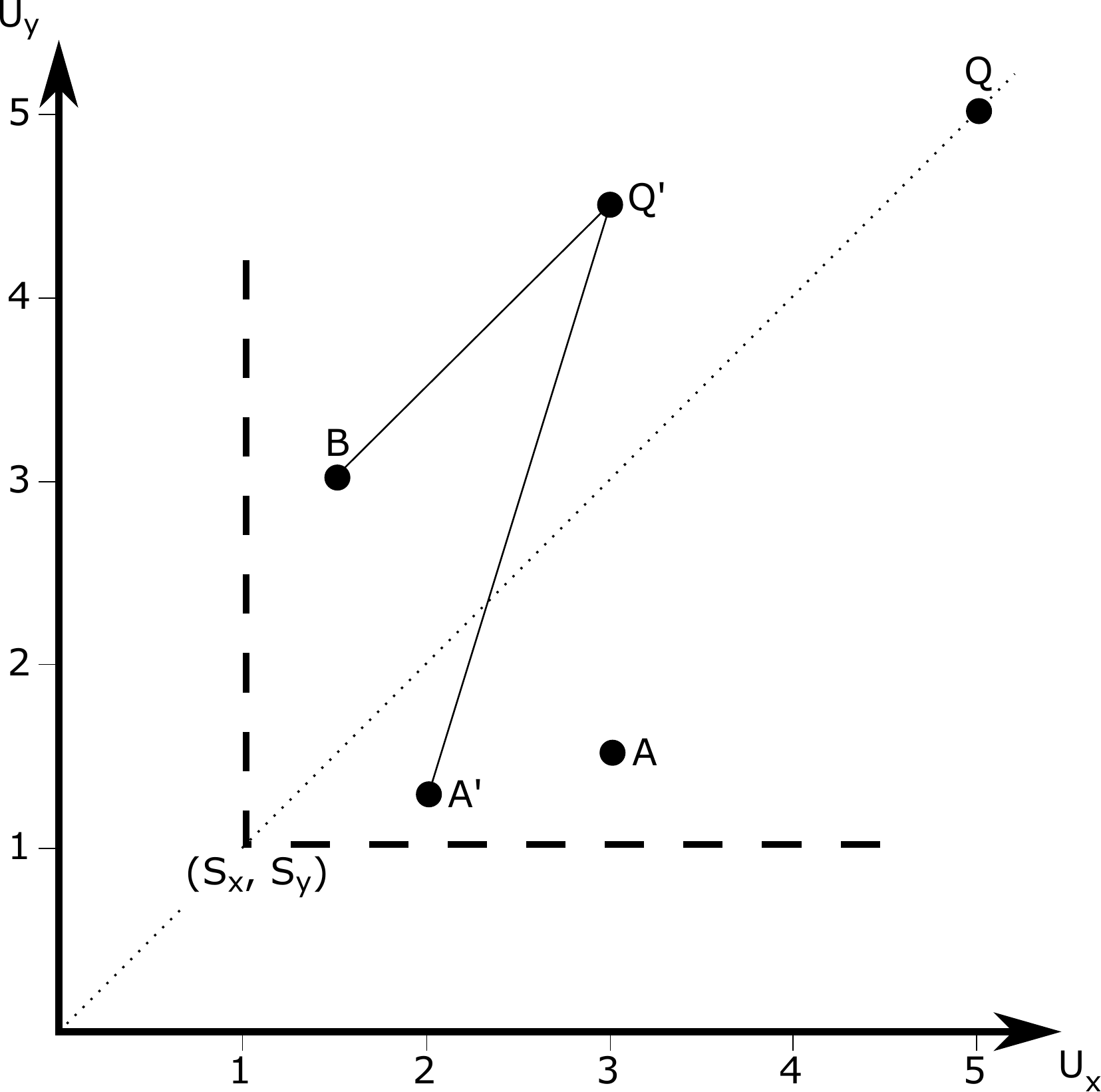}
\captionspacefig \caption{Utility payoff for two players with quantum computers, employing a \textsc{Unify} strategy, under the imposition of taxation.}\label{fig:game_theory_6}
\end{figure}

For the \textsc{Unify} strategy, there will be a depreciation in the final utility payoff. As such, the utility payoff shows a new combination of $(3,4.5)$, an asymmetric reduction from the previous position of $(5,5)$. Graphically, this is shown in Fig.~\ref{fig:game_theory_6} as a shift from $Q$ to $Q'$.

Two things are clear from both the matrix and its resulting graphical representation. The imposition of taxation affects both the \textsc{License} and \textsc{Unify} strategies, but unequally. For $X$, the imposition of taxation results in a greater loss in net utility with the \textsc{Unify} strategy. This would ultimately undermine the likelihood for $X$ to choose this strategy. Secondly, the rate at which utility is gained through a mixed strategy\index{Mixed strategies} (moving from $A'$ to $Q'$) is decreased. This implies that, with a \textsc{Unify} strategy, $X$ would be less motivated to cooperate with $Y$ than $Y$ would be to cooperate with $X$. $Q'$ would still lead to the maximum utility payoff, but the negotiation process to achieve it will be more difficult owing to the utility asymmetry. Thus, despite both players having identical systems, the regulatory asymmetry will strongly impact the likelihood of cooperation. This has important implications for future quantum policy-making.

\subsection{Resource asymmetry}\index{Resource asymmetry}

What happens with collaboration between players who have different systems? If this assumption is relaxed, what comes of the resulting utility payoff matrix and choice of strategies? The analysis for this assumption had already been presented in the previous discussion. The effective impact of taxation is that it will limit the amount of computational power delivered to market and capture governmental revenue from the sale of compute-time. This is analogous to what happens if $X$ was to build a less powerful system than $Y$. Through collaboration, they could still attain point $Q'$, shown in Fig.~\ref{fig:game_theory_4}, but this time the asymmetry arises as a design consequence rather than a regulatory imposition. Despite $X$ now not contributing as much to the overall computational power of the quantum cloud, choosing the maximising strategy of offering all available computing power to the cloud does maximise individual utility. Furthermore, the subjective nature of the utility derivation means there doesn't have to be a change from the described payoff matrix shown in Fig.~\ref{fig:game_theory_3}. Despite having a less powerful system, the proportional allocation of financial rewards relative to the contribution, means both $X$ and $Y$ could maintain a symmetric utility payoff outcome, like in Fig.~\ref{fig:game_theory_3}. In summary, the 2-player game shows that cooperation will result in the better outcome for both players.

\latinquote{Ceretis paribus.}

\subsection{Multi-player games}

What happens when the analysis moves beyond just two players? As established previously, the cost and scale limitations associated with quantum computing means there are likely to be limited vendors contributing to the quantum internet. However, it's also likely there will be more than just two. Formulating the $n$-person cooperative game opens up a plethora of possibilities that go beyond the scope of this introductory discussion, but one key takeaway point from the previous analysis is that in general there is a strong motivation for computational cooperation in the quantum world. This then introduces two possible scenarios:

\begin{itemize}
	\item There is complete cooperation between all suppliers, i.e the global virtual quantum computer discussed in Sec.~\ref{sec:glob_unif_quant_cloud}.
	\item Competing cartels\index{Cartels} develop, where for external reasons (e.g political, geographical, ideological, strategic), there is a benefit in cooperating with a limited number of players, and acting as a single supplier in direct competition with other cartels.
\end{itemize}
x
The first scenario is by far the most attractive, and in a world where free negotiations may take place, this is clearly the option resulting in both attaining the greatest computational power for the quantum cloud, and also the greatest utility maximisation for all participants involved. The end result will be similar to that shown in Fig.~\ref{fig:game_theory_3}. This implies that the quantum internet market structure would become like a natural monopoly, with diminishing long-run average costs as supply is increased. It would also mean that there is a strong case for some government involvement to prevent profit maximisation at the expense of efficiency.

However, history has shown us that another plausible outcome is the second scenario. While there are motivations to collaborate, there are also motivations to compete. Given the almost certain involvement of government intervention in the supply of quantum processing, the formulation of regional cartels due to external factors may also be a likely outcome. In such an environment, the cartels will internally operate as multi-player cooperative games, and externally towards other conglomerates, operate as separate multi-cartel competitive games \cite{bib:Bacharach76}. This result may still result in efficient Pareto optimal\index{Pareto optimal} outcomes at a market level, but will always fall short relative to a model of complete cooperation. This scenario will naturally end with an oligopolistic market structure. 

\subsection{Conclusions}

In summary, a game theory approach to understanding the quantum internet shows that there are strong motivations for quantum computing vendors to cooperate in order to globally maximise net utility. Furthermore, there is a strong potential to affect the possibility of cooperation through market distorting effects such as via the imposition of taxation or regulation. Finally, the two most likely market structures that will develop under the quantum internet are either a natural monopoly\index{Monopoly}, where some form of regulation will be required to ensure economically efficient production, or an oligopoly\index{Oligopoly}, where a few cartels will compete with each other to maximise their productive output. Either way, the quantum marketplace is an extremely interesting and largely unexplored avenue for future research, a new interdisciplinary field sitting at the intersection between economics and quantum information theory. It is also one of great relevance and importance for when this technology becomes a reality.

\latinquote{Natura nihil frustra facit.}

%
% Summary of Economic Models
%

\section{Summary of economic models}\index{Summary of economic models}\label{sec:summary_economic_models}

\dropcap{I}{n} Tab.~\ref{tab:summary_ec_models} we summarise the economic models and parameters we developed, and applied them to several illustrative scaling functions of particular interest: linear (i.e classical computing), polynomial, and exponential (i.e best-case quantum computing).

\latinquote{Errare humanum est.}

\startnormtable
\renewcommand{\arraystretch}{0.5}

\begin{table*}[!htbp]
{\footnotesize
\begin{tabular}{|m{0.21\linewidth}|m{0.21\linewidth}|m{0.15\linewidth}|m{0.155\linewidth}|m{0.225\linewidth}|}
	\hline
	\[\mathrm{Model}\] & \[\mathrm{General\, form}\] & \[f_\mathrm{sc}(n)=n\] \begin{center}(classical)\end{center} & \[f_\mathrm{sc}(n)=n^p\] \begin{center}(intermediate)\end{center} & \[f_\mathrm{sc}(n)=e^n\] \begin{center}(full quantum)\end{center}\\
	\hline \hline
	\begin{flushleft}Per-qubit computational power (Sec.~\ref{sec:NPSF})\end{flushleft} & \[\chi_\mathrm{sc}(n)=\frac{f_\mathrm{sc}(n)}{n}\] & \[1\] & \[n^{p-1}\] & \[\frac{e^n}{n}\]\\
	\hline
	\begin{flushleft}Network power (Sec.~\ref{sec:network_power})\end{flushleft} & \[P(t)=f_\mathrm{sc}(N_0{\gamma_N}^t)\] & \[N_0{\gamma_N}^t\] & \[\left(N_0{\gamma_N}^t\right)^p\] & \[ e^{N_0{\gamma_N}^t}\] \\
	\hline
	\begin{flushleft}Network value (Sec.~\ref{sec:network_value})\end{flushleft} & \[V(t)=C_0 N_0 \left(\frac{\gamma_N}{\gamma_C}\right)^t\] & \[C_0 N_0 \left(\frac{\gamma_N}{\gamma_C}\right)^t\] & \[C_0 N_0 \left(\frac{\gamma_N}{\gamma_C}\right)^t\] & \[C_0 N_0 \left(\frac{\gamma_N}{\gamma_C}\right)^t\] \\
	\hline
	\begin{flushleft}Spot price of computation (Sec.~\ref{sec:cost_of_comp})\end{flushleft} & \[L(0)=\frac{e^{\gamma_\mathrm{ror}} C_0}{\chi_\mathrm{sc}(N_0)}\] & \[e^{\gamma_\mathrm{ror}} C_0\] &  \[\frac{e^{\gamma_\mathrm{ror}}C_0}{{N_0}^{p-1}}\] & \[\frac{e^{\gamma_\mathrm{ror}}N_0C_0}{e^{N_0}}\] \\
	\hline
	\begin{flushleft}Future cost of computation (Sec.~\ref{sec:cost_of_comp})\end{flushleft} & \[L(t)=\frac{e^{\gamma_\mathrm{ror}} C_0{\gamma_C}^{-t}}{\chi_\mathrm{sc}(N_0 {\gamma_N}^t)}
\] & \[e^{\gamma_\mathrm{ror}} C_0{\gamma_C}^{-t} \] & \[ \frac{e^{\gamma_\mathrm{ror}} C_0{\gamma_C}^{-t}}{(N_0 {\gamma_N}^t)^{p-1}}
\] & \[ \frac{e^{\gamma_\mathrm{ror}} C_0N_0 \left(\frac{\gamma_N}{\gamma_C}\right)^t}{e^{N_0 {\gamma_N}^t}}\] \\
	\hline
	\begin{flushleft}Time-share computational power (Sec.~\ref{sec:arb_free_time_share}); Cooperative payoff enhancement (Sec.~\ref{sec:coop_enh})\end{flushleft} & \[c_n=n \cdot \chi_\mathrm{sc}(n_\mathrm{global})
\] & \[n\] & \[n\cdot{n_\mathrm{global}}^{p-1}\] & \[\frac{n e^{n_\mathrm{global}}}{n_\mathrm{global}}\]\\
	\hline
	\begin{flushleft}Problem size scaling function (Sec.~\ref{sec:prob_sc_func})\end{flushleft} & \[s = f_\mathrm{size}^{-1}(n\cdot \chi_\mathrm{sc}(n_\mathrm{global}))\] & \[\frac{n}{\alpha_\mathrm{size}}\] & \[(n \cdot {n_\mathrm{global}}^{p_\mathrm{sc}-1})^\frac{1}{p_\mathrm{size}}\] & \[\alpha_\mathrm{sc}n_\mathrm{global} + \log(n)\]\[-\log(\alpha_\mathrm{sc}n_\mathrm{global})\] \\
	\hline
	\begin{flushleft}Quantum computational leverage (Sec.~\ref{sec:quant_ec_lev})\end{flushleft} & \[\lambda_n=\frac{\chi_\mathrm{sc}(n_\mathrm{global})}{\chi_\mathrm{sc}(n)}\] & \[1\] & \[\left(\frac{n_\mathrm{global}}{n}\right)^{p-1}\] & \[\frac{n e^{n_\mathrm{global}}}{n_\mathrm{global}e^n}\]\\
	\hline
	\begin{flushleft}Single-qubit leverage (Sec.~\ref{sec:quant_ec_lev})\end{flushleft} & \[\lambda_\mathrm{qubit}=\frac{\chi_\mathrm{sc}(n_\mathrm{global})}{\chi_\mathrm{sc}(1)}\] & \[1\] & \[{n_\mathrm{global}}^{p-1}\] & \[\frac{e^{n_\mathrm{global}-1}}{n_\mathrm{global}}\]\\
	\hline
	\begin{flushleft}Time-dependent leverage (Sec.~\ref{sec:quant_ec_lev})\end{flushleft} & \[\lambda_n(t)=\frac{\chi_\mathrm{sc}(N_0{\gamma_N}^t)}{\chi_\mathrm{sc}(n)}\] &  \[1\] & \[\left(\frac{N_0{\gamma_N}^t}{n}\right)^{p-1}\] & \[\frac{n e^{N_0{\gamma_N}^t-n}}{N_0{\gamma_N}^t}\]\\
	\hline
	\begin{flushleft}Static computational return (Sec.~\ref{sec:static_comp_ret})\end{flushleft} & \[r_\mathrm{static}(t) = n\cdot\chi_\mathrm{sc}(N_t)\] & \[n\] & \[n{N_t}^{p-1}\] & \[\frac{n e^{N_t}}{N_t}\] \\
	\hline
	\begin{flushleft}Forward contract price (Sec.~\ref{sec:for_contr})\end{flushleft} & \[F(T)=\frac{e^{\gamma_\mathrm{ror}-r_\mathrm{rf}T} C_0{\gamma_C}^{-T}}{\chi_\mathrm{sc}(N_0 {\gamma_N}^T)}\]
 & \[e^{\gamma_\mathrm{ror}-r_\mathrm{rf}T} C_0{\gamma_C}^{-T}\] & \[\frac{e^{\gamma_\mathrm{ror}-r_\mathrm{rf}T} C_0{\gamma_C}^{-T}}{(N_0 {\gamma_N}^T)^{p-1}}\] & \[\frac{e^{\gamma_\mathrm{ror}-r_\mathrm{rf}T} C_0N_0\left(\frac{\gamma_N}{\gamma_C}\right)^T}{e^{N_0 {\gamma_N}^T}}\] \\
	\hline
	\begin{flushleft}Tax performance multiplier (Sec.~\ref{sec:taxation})\end{flushleft} & \[M(N_\mathrm{tax})=\frac{f_\mathrm{sc}(N_\mathrm{tax})}{f_\mathrm{sc}(N_\mathrm{tax} \gamma_T)}\] & \[\frac{1}{\gamma_T}\] & \[\frac{1}{{\gamma_T}^p}\] & \[e^{N_\mathrm{tax}(1-\gamma_T)}\]\\
	\hline
	\begin{flushleft}Political leverage (Sec.~\ref{sec:political_lev})\end{flushleft} & \[\gamma_{A,B}=\frac{\chi_\mathrm{sc}(n_B)}{\chi_\mathrm{sc}(n_A)}\] & \[1\] & \[\left(\frac{n_B}{n_A}\right)^{p-1}\] & \[\frac{e^{n_B}n_A}{e^{n_A}n_B}\] \\
	\hline
	\end{tabular}}
\captionspacetab \caption{Summary of the dynamics of various economic models under several computational scaling functions ($f_\mathrm{sc}$) of interest, where there are $n$ qubits held by the respective node and \mbox{$n_\mathrm{global}=\sum_{j\in \mathrm{nodes}} n_j$} qubits in the global network. $N_0$ is the initial number of qubits, undergoing growth rate $\gamma_N$. $C_0$ is the initial monetary cost per qubit, undergoing decay rate $\gamma_C$. $\gamma_\mathrm{ror}$ is the rate of return on the licensing of compute-time on qubits, and $r_\mathrm{rf}$ is the risk-free rate of return (typically the yield on US government bonds). $\gamma_T$ is the rate of taxation, and $N_\mathrm{tax}$ is the number of qubits that would exist on the network in the absence of taxation.} \label{tab:summary_ec_models}
\end{table*}

\renewcommand{\arraystretch}{1}
\startalgtable

\clearpage

\sketch{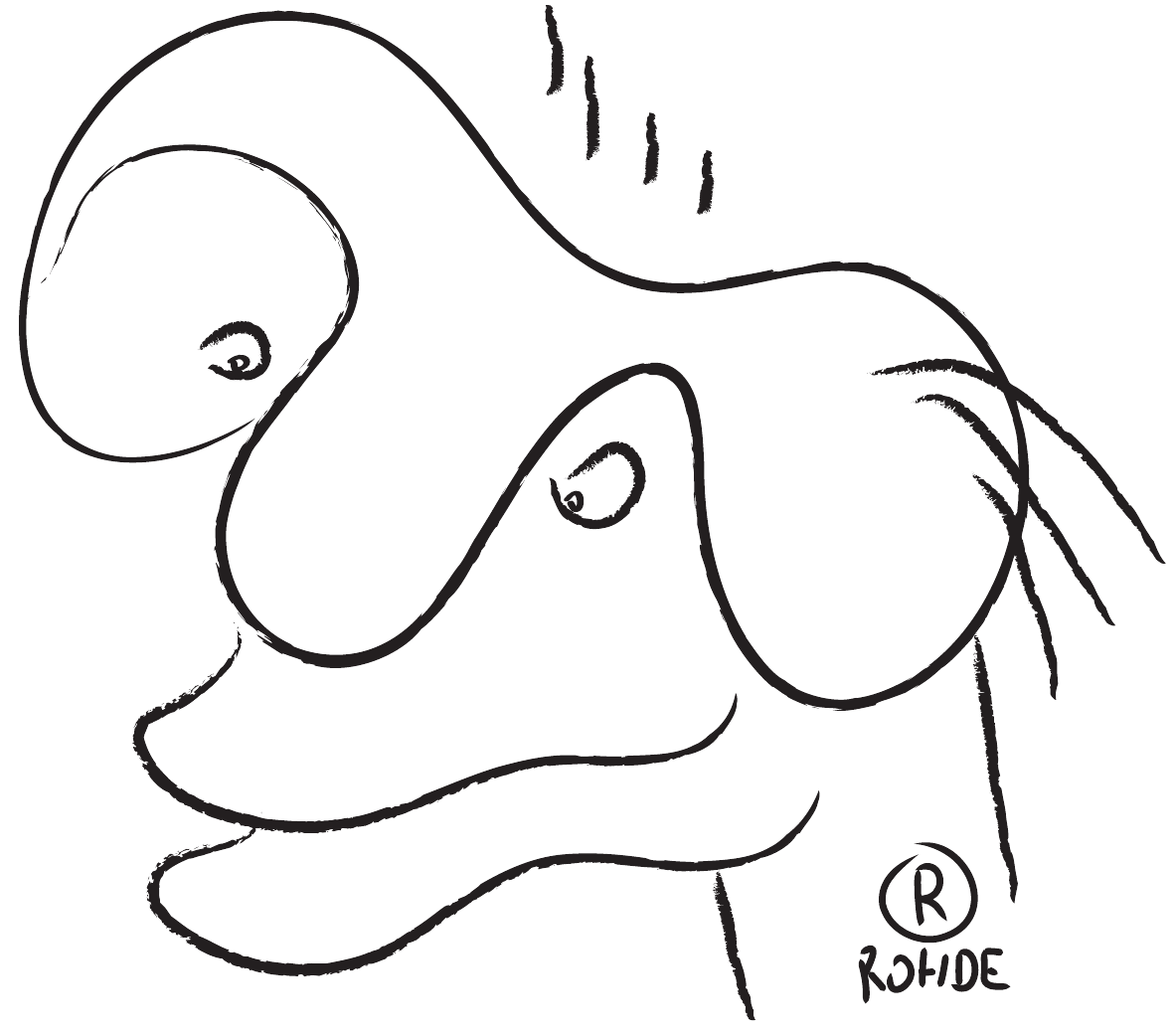}

\clearpage

%
% State of the art
%

\part{History \& state-of-the-art}\label{part:SotA}\index{State-of-the-art}\index{History}

%
% State-of-the-Art
%

\sectionby{He-Liang Huang \& Zu-en Su}\index{He-Liang Huang}\index{Zu-en Su}

\famousquote{We are not makers of history. We are made by history.}{Martin Luther King, Jr}
\newline

\startnormtable

\dropcap{W}{hen} will we have the quantum internet? This is a question with as many answers as there are people one asks. And the developments of the different components of such an internet will be staggered and under continual development -- a quantum internet with the capacity for long-distance QKD will likely arrive far sooner than one supporting fully distributed, blind quantum computation!

In this section we present a brief history of recent major developments and the state-of-the-art of some of the most important quantum technologies and protocols relevant to the quantum internet, with a view to understand trends in the development of the field, so as to position us to gauge the rate of progress and make educated predictions about future developments -- the timing of future developments is likely to have big strategic implications, and it's important we are able to anticipate technological transformations as radical as the quantum ones, so that we may be prepared for their fallout and consequences.

With this we aim to shed light on what quantum networking services could be readily implemented today, using present-day technology, and what is likely to be viable in the near and more distant future.

We will very succinctly summarise developments as timelines, without much elaboration on specific details. We provide detailed referencing such that the interested reader can follow up on specifics at their leisure. Effectively this section acts as a handbook for looking up important development events, milestones, and their references. We also provide links to many of the important review papers in the field, which are perhaps the best starting point for developing a deeper understanding as an outsider from the field.

The reader disinterested in a history lesson might comfortably skip this part, without serious risk of missing out on explanations of important topics (no new conceptual material is introduced in this part).

The tables summarising chronological timelines of major developments, and their referencing, are presented in bulk at the end of this part in Sec.~\ref{sec:timeline_tables}.

%
% Quantum Teleportation & Entanglement Distribution
%

\section{Quantum teleportation \& entanglement distribution} \index{Quantum state teleportation} \index{Quantum gate teleportation} \index{Entanglement!Distribution}

\dropcap{Q}{uantum} state teleportation (Sec.~\ref{sec:teleport}) has attracted broad experimental interest and been subject to widespread demonstration across many physical architectures, becoming one of the most well-investigated protocols in experimental quantum information science. 

Tab.~\ref{tab:state_tomo} summarises some of the notable recent developments in quantum teleportation and the entanglement distribution upon which it relies.

%
% Entanglement Swapping & Quantum Repeaters
%

\section{Entanglement swapping \& quantum repeaters} \index{Entanglement!Swapping} \index{Quantum repeater networks}
	\index{Entanglement!Distribution}

\dropcap{C}{losely} related to entanglement distribution is entanglement swapping (Sec.~\ref{sec:swapping}), where the goal is to entangle two remote parties, each of whom have one half of two distinct Bell pairs. As discussed in Sec.~\ref{sec:ent_ultimate}, entanglement swapping will be of fundamental importance in networks treating entanglement distribution as the most fundamental elementary resource, making it greatly applicable to quantum networking. 

Tabs.~\ref{tab:entanglement_swap1} \& \ref{tab:entanglement_swap2} summarise major recent developments in entanglement swapping and quantum repeater networks (in which entanglement swapping is also a basic primitive).

%
% Quantum Key Distribution
%

\section{Quantum key distribution} \index{Quantum key distribution (QKD)}\label{bib:QKD_state_of_art}

\dropcap{W}{ith} the development of quantum technology, QKD will gradually become increasingly practical, and economically accessible to consumers. Bennett (of BB84 protocol fame), first demonstrated a QKD protocol on an optical platform over a distance of just 30cm \cite{bib:JC_5_3}. Since then, experiments have developed rapidly from indoors to outdoors, over ever-increasing distances, and even recently between Earth-based ground stations and satellites in low-Earth orbit. Indeed, commercial QKD units are even available as off-the-shelf commodity products today, making QKD perhaps the first commercialised quantum technology.

Tab.~\ref{tab:QKD_table} summarises some major milestones in the development and deployment of QKD.

%
% Entanglement Purification
%

\section{Entanglement purification} \index{Entanglement!Purification}

\dropcap{E}{ntanglement} purification has been very successful in optical systems. This forms an essential primitive in long-range quantum repeater networks, where entanglement must be recovered from states that have degraded during transmission over long distances.

 Tab.~\ref{tab:ent_pur} summarises some of the  major recent developments in entanglement purification.

%
% State Preparation
%

\section{State preparation} \index{State preparation}

\dropcap{A}{ny} quantum protocol necessarily involves state preparation to create the inputs for the protocol. Different types of states come with their own unique challenges, and also their own unique utility.

In the context of the quantum internet, the most important class of quantum states is of course the optical ones, since light will almost inevitably be the primary information carrier in such a future network.

We now summarise some of the most common (and useful) optical states for quantum information processing applications.

%
% Coherent States
%

\subsection{Coherent states} \index{Coherent states!Preparation}

Coherent states are well approximated by laser\index{Lasers} or maser\index{Masers} light. Nowadays thousands of types of lasers are known with different power, temporal, spatial and spectral parameters, and the technology is already very mature and commercially available. Here we only introduce some of the basic concepts of lasers related to quantum networking.

A laser can be classified as operating in either a continuous\index{Continuous wave lasers} or pulsed\index{Pulsed lasers} regime. Most laser diodes\index{Lasers!Diodes} used in communication systems are continuous. But they can also be externally carved at some rate by modulators to create pulsed light. Usually, pulsed lasers are created by the technique of Q-switching\index{Q-switching} or mode-locking\index{Mode-locking}.

Different applications require lasers with different output power. Typical output powers of single-mode laser diodes are some tens of milliwatts, up to at most a few hundred milliwatts. However, multiple transverse mode diode lasers\index{Transverse mode diode lasers} can reach up to some tens of Watts, and can be used as pump sources for high-quality and high-power single-mode solid-state lasers. Such single-mode diode-pumped solid-state lasers can be further mode-locked to output femtosecond pulses, reaching as far as an average power of tens of Watts. Pulsed lasers can also be characterised by the peak power of each pulse. The peak power\index{Peak power} of a pulsed laser is many orders of magnitude greater than its average power\index{Average power}.

Tab.~\ref{tab:coherent_states} enumerates two types of lasers commonly used in quantum communication applications.

%
% Single-Photons
%

\subsection{Single-photons} \index{Photons!Preparation}

A highly attenuated laser can be used as a good approximation to a single-photon source when no more than one single-photon is used in an interferometric experiment, such as QKD. Otherwise, heralded or deterministic single-photon sources are required. This is because while the photo-statistics of a single-photon mimic those of a laser, the photo-statistics of multi-photon states do not, irrespective of the laser's power -- coherent light and multi-photon states obey fundamentally different rules of statistics.

Tab.~\ref{tab:single_photon_state} summarises some of the notable developments. A good review of solid-state single-photon emitters is \cite{bib:aharonovich2016solid}. Different sources are compared in the review by \cite{bib:eisaman2011}.

%
% Entangled States Based on Non-Linear Optics
%

\subsection{Entangled states based on non-linear optics} \index{Non-linear!Optics}

Non-linear optics is one of the most convenient and cheapest methods to produce entangled states. Tab.~\ref{tab:entangled_states} summarises some of the major developments. Further material can be found in the review papers \cite{bib:pan2012multiphoton, bib:ralph2009bright}.

%
% Non-Optical Systems
%

\section{Non-optical systems}\index{Non-optical systems}\index{Matter qubits}

\dropcap{A}{lthough} light is likely to be the information \textit{carrier} of the future quantum internet, it may very well not win the race of becoming the preferred means for computation or storage. Here, there are countless other possibilities to consider, of which we now outline a few.

%
% Atomic Ensembles
%

\subsection{Atomic ensembles} \index{Atomic!Ensembles}

A variety of techniques have been developed to create squeezing\index{Squeezing!Atomic ensembles} and entanglement in atomic ensembles. The main methods exploit atom-light interactions in cold gases, or interactions between particles such as atom-atom collisions in Bose-Einstein condensates\index{Bose-Einstein condensates (BECs)}, or combined electrostatic and ion-light interactions in ion chains. Atom-light interactions currently represent the most mature technology, and exhibit the highest squeezing of 20.1dB via an optical-cavity-based measurement \cite{bib:hosten2016measurement}.

Recent progress is summarised in Tab.~\ref{tab:atomic_ensembles}.

%
% Single Atoms
%

\subsection{Single atoms} \index{Single atoms}

Many experimental methods have been developed for measuring and manipulating individual quantum systems, including single atoms in a cavity, trapped ions\index{Trapped ions}, and neutral atoms\index{Neutral atoms} in an optical lattice\index{Optical!Lattice}. 

Tab.~\ref{tab:single_atoms} summarises some of the notable developments. More material can be found in the excellent review papers \cite{bib:blatt2008entangled, bib:haroche2006exploring, bib:leibfried2003quantum}.

%
% Quantum Dots
%

\subsection{Quantum dots} \index{Quantum dots!Sources}

In quantum dots, the recombination of an electron-hole pair\index{Electron-hole pair}, known as an \textit{exciton}, emits a single photon. Photon sources based on this concept are rapidly gaining popularity in present-day labs, and a summary of developments in this area is provided in Tab.~\ref{tab:quantum_dots}.

%
% Nitrogren-Vacancy Centres
%

\subsection{Nitrogen-vacancy centres} \index{Nitrogen-vacancy (NV) centres}

A nitrogen-vacancy (NV) centre in diamond refers to a nitrogen (N) atom replacing a carbon atom in the crystal lattice, neighbouring a single vacancy (V) defect \cite{bib:doherty2013nitrogen}. In such centres, both electrons and nuclear spins can exhibit long coherence times ($>$1ms for the electron spin, and $>$1s for the nuclear spin) even at room temperature \cite{bib:balasubramanian2009ultralong, bib:neumann2010quantum, bib:maurer2012room}.

Tab.~\ref{tab:NV_centres} summarises some of the recent developments in NV centre technology. Further material can be found in the review papers \cite{bib:doherty2013nitrogen, bib:atature2018material, bib:awschalom2018quantum}.

%
% Superconducting Rings
%

\subsection{Superconducting rings} \index{Superconductors!Rings}

Superconductors have become a promising candidate for processing quantum information, and much experimental progress has been made in recent years.

Tab.~\ref{tab:superconducting} summarises some of the recent achievements in superconducting quantum technology. The review paper \cite{bib:xiang2013hybrid} provides excellent follow-up material for further details.

%
% Measurement
%

\section{Measurement} \index{Measurement}

\dropcap{M}{any} quantum information protocols have been developing as two versions in parallel -- a discrete-variable\index{Discrete-variables} version, and a continuous-variable\index{Continuous-variables} version.

Optical discrete-variable systems are typically measured using photo-detectors, whereas continuous-variable schemes are typically measured using homodyning.

%
% Photo-Detection
%

\subsection{Photo-detection} \index{Photo-detection}

There are a variety of single-photon detectors, detailed descriptions of which can be found in several excellent review papers \cite{bib:eisaman2011, bib:hadfield2009}.

Tab.~\ref{tab:photodetection} summarises single-photon detectors that have good spectral response in the near-infrared regime.

%
% Homodyning
%

\subsection{Homodyning}\index{Homodyne detection}

Tab.~\ref{tab:homodyning} summarises some of the notable developments in homodyne detection, the most fundamental approach to quantum measurement in CV optical systems.

%
% Evolution of Optical States
%

\section{Evolution of optical states} \index{Evolution of optical states} \label{sec:LO_evolution}

\dropcap{Q}{uantum} states can be encoded in various degrees of freedom of a single photon. Tab.~\ref{tab:evolutionofstates} summarises some of the major developments in the controlled evolution of optical states.

There are two primary types of optical circuits: optical waveguides\index{Optical!Waveguides}, and fibre-loops\index{Fibre-loops}.

Optical waveguides can be miniaturised and integrated onto a small chip, which is alignment-free, compact, and phase-stable\index{Phase!Stability}.

Fibre-loop schemes are extremely resource-frugal -- resource requirements are essentially constant, and do not grow with the complexity of the underlying interferometer being implemented. 

Some of these techniques are summarised in Tab.~\ref{tab:waveguide_fibre}.

%
% Quantum Memory
%

\section{Quantum memory} \index{Quantum memory}

\dropcap{N}{umerous} excellent review articles exist \cite{bib:lvovsky2009optical, bib:simon2010quantum, bib:sangouard2011quantum, bib:bussieres2013prospective, bib:reiserer2015cavity}, providing detailed explanations of quantum memory schemes. Quantum memory is so ubiquitous and has such widespread utility that it has become a major research area in its own right.

The simplest approach to storing light is using an optical delay line\index{Delay line}, such as an optical fibre. Aside from loss, these also suffer that the storage time is fixed by the delay length, making on-demand output\index{On-demand!Output} extremely challenging without first having access to high-speed ($\sim$GHz) optical switches\index{Optical!Switches}.

Alternatively, light can be stored in a high-Q cavity\index{Cavities}. The light cycles back and forth between the reflecting boundaries, allowing it to be injected into and retrieved from the cavity \cite{bib:pittman2002single, bib:pittman2002cyclical, bib:leung2006quantum, bib:maitre1997quantum, bib:tanabe2007trapping, bib:tanabe2009dynamic}. Unfortunately, the storage of light in cavities suffers from a tradeoff between short cycle time\index{Cycle time} and long storage time\index{Storage time}, which limits efficiency. Therefore, whereas optical delay lines and nano-cavities could be appropriate for obtaining on-demand single photons from heralded sources \cite{bib:saglamyurek2015quantum, bib:jin2015telecom}, they may not be suitable for quantum memory or quantum repeaters.

Tab.~\ref{tab:memory} summarises some of the notable developments in quantum memories.

%
% Quantum Computation
%

\section{Quantum computation}\index{Quantum computing}

\dropcap{T}{he} Holy Grail\index{Holy Grail} of scalable, universal quantum computation -- perhaps the most exciting of future quantum developments -- has been developing rapidly across various physical platforms simultaneously in parallel, including: trapped atoms/ions\index{Trapped atoms}; nuclear magnetic resonance (NMR)\index{Nuclear magnetic resonance (NMR)}; photons; superconductors; spins in silicon\index{Spins in silicon}; and many, many more. An excellent point of reference is the review \cite{bib:ladd2010quantum}.

Tab.~\ref{tab:quantumcomputer} summarises some of current records for the number of qubits and attained fidelities in different physical systems for implementing quantum computation. Most noteworthy, IBM\index{IBM} and Google\index{Google} recently announced their record-breaking 50- and 72-qubit superconducting quantum processors respectively, a key step towards achieving the coveted goal of quantum supremacy\index{Quantum supremacy} \cite{bib:savage2018quantum, bib:neill2018blueprint, bib:harrow2017quantum}.

\latinquote{Om mani padme hum.}

%
% Tables
%

\section{Timelines \& references}\index{Timelines}\label{sec:timeline_tables}

\dropcap{T}{he} historical development timelines of the technologies referred to earlier in this part, and their references, now follow in table form.

\begin{table*}[!htbp]
\begin{tabular}{|p{0.755\linewidth}|p{0.22\linewidth}|}
	\hline
	\textbf{Summary} & \textbf{References \& years} \\
	\hline \hline
	A technique for the generation of high-intensity polarisation-entangled photon-pairs. For the partial Bell state projection a 50:50 beamsplitter was employed. & \cite{bib:PhysRevLett.75.4337, bib:Euro_25_559} \\
	\hline
	The first experimental demonstration of photonic quantum teleportation. Two photon-pairs were prepared by double-pumping a single non-linear beta-barium borate (BBO) crystal: one pair employed as the entanglement source; the other to prepare the state to teleport. The partial Bell state measurement was implemented using which-path erasure at a beamsplitter, with an efficiency of 25\%. & \cite{bib:Boumeester97} \\
	\hline
	Experimental demonstration of quantum state teleportation between two labs, separated by 55m, but connected by a 2km length of fibre, with photons at telecommunication wavelengths. This arouses the exciting prospect that future quantum networks might be able to piggyback off existing telecom infrastructure, which would be a boon to the quantum industry. & \cite{bib:Nat_421_509} \\
	\hline
	Quantum state teleportation was demonstrated between photonic and atomic qubits, a first step towards hybrid architectures, and an essential ingredient in interfacing optical and non-optical systems. & \cite{bib:Chen08} \\
	\hline
	Quantum teleportation over a 16km long, noisy, free-space channel between distant ground stations was demonstrated. This distance is of especial interest as it is significantly longer than the effective thickness of the atmosphere, equivalent to 5-10km of ground atmosphere. This is an exciting benchmark as it suggests that free-space ground-to-satellite teleportation may be viable. & \cite{bib:Nat_Phot_4_376, bib:PRL_94_150501} \\
	\hline
	The teleportation distance in free-space was extended to 97km over Qinghai Lake, and 143km between the two Canary Islands of La Palma and Tenerife. These overcame the daunting challenges associated with source targeting and tracking, for long-distance, free-space quantum teleportation, and paved the way for future satellite-based quantum teleportation. & \cite{bib:Nat_488_185, bib:Nat_489_269} \\
	\hline
	Accompanying the breakthrough of superconducting single-photon detectors with near-unit efficiency, 3-fold photo-detection for quantum teleportation was greatly enhanced by more than two orders of magnitude at telecom wavelengths, and the teleportation distance in optical fibre lengthened to 100km. & \cite{bib:Optica_2_832} \\
	\hline
	Quantum teleportation over fibre networks in Hefei and Calgary were demonstrated, with lengths of dozens of kilometres. & \cite{bib:sun2016quantum, bib:Nat_Phot_10_676} \\
	\hline
	The first quantum satellite for entanglement distribution was launched in China. In addition to teleportation, this could facilitate intercontinental QKD. The team is aiming to achieve quantum teleportation between ground stations and satellite, and even between pairs of distant ground stations, separated by over 1000km, using shared entanglement provided by the satellite. & \cite{liao2017satellite} \\
	\hline
	Using 5-photon entanglement, open-destination teleportation was implemented, whereby an unknown quantum state was teleported onto a superposition of 4 destination photons, which could be read out at any location -- a type of `broadcasting'. & \cite{bib:zhao2004experimental} \\
	\hline
	The state of a two-photon composite system was demonstrated -- a breakthrough in the teleportation of a single particle onto a complex system comprising multiple particles. & \cite{bib:Nat_Phys_2_678} \\
	\hline
	Quantum teleportation over multiple degrees of freedom of a single optical mode was demonstrated. & \cite{bib:Nat_518_516} \\
	\hline
	Teleportation of CV optical states was demonstrated. The advantage of this teleportation protocol was that it could be deterministic in principle,  overcoming the non-determinism inherent to partial Bell state projections using a beamsplitter. & \cite{bib:Science_282_706} \\
	\hline
	The deterministic teleportation of photonic qubits was demonstrated using hybrid techniques. & \cite{bib:Nat_500_315} \\
	\hline
	Quantum teleportation has also attracted great interest in other physical architectures. Demonstrations have been performed in various physical systems, including atoms, ions, electrons, and superconducting circuits. & \cite{bib:Nat_Phys_9_400, bib:Nat_429_734, bib:Nat_429_737, bib:Science_345_532, bib:Nat_500_319} \\
	\hline
Hybrid schemes combining different physical systems have been demonstrated, such as light-to-matter teleportation. Hybrid technologies are expected to play an important role in future quantum networks, where the underlying physical architecture for (say) a quantum computation is non-optical, but optics mediates the communication of quantum information. & \cite{bib:Kimble2008,gao2013quantum} \\
	\hline
\end{tabular}
\captionspacetab \caption{Developments in experimental quantum state teleportation and entanglement distribution.} \label{tab:state_tomo}
\end{table*}

\begin{table*}[!htbp]
	\begin{tabular}{|p{0.755\linewidth}|p{0.22\linewidth}|}
		\hline
		\textbf{Summary} & \textbf{References \& years} \\
		\hline \hline
		The first experimental demonstration of entanglement swapping. By pumping a BBO crystal in a double-pass configuration, two pairs of polarisation-entangled photons were generated to demonstrate the scheme. A visibility of \mbox{$0.65$} was observed, which clearly surpasses the classical limit of \mbox{$0.5$}. This was later improved in 2001 to a visibility of \mbox{$0.84$}, which violates the Bell inequality (for which the threshold is $0.71$).  & \cite{bib:PRL_80_3891, bib:jennewein2001experimental} \\
		\hline
		Aside from `event-ready' mode, a delayed-choice mode of operation for entanglement swapping was proposed, where entanglement is produced a posteriori, after the entangled particles have been measured and may no longer even exist.& \cite{bib:peres2000delayed}\\
		\hline
		Delayed-choice entanglement swapping was designed and realised. This was performed by adding two 10m optical fibre delays of about 50ns for both outputs of the Bell state measurement. & \cite{bib:PRL_88_017903}\\
		\hline
		A delayed-choice entanglement swapping experiment with vacuum-one-photon quantum states was realised. 
		%However, none of these demonstrations were active, random or delayed choice, which are required to ensure that photons cannot know in advance the basis choices for future measurements.
		& \cite{bib:PRA_66_024309}\\
		\hline
		By designing a special interferometer to realise active switching between Bell state measurement and separable state measurement, an entanglement swapping experiment with active delayed-choice was demonstrated. 
		%Subsequently, experimental demonstrations of entanglement swapping have evolved to become more complex and rigorous, finding uses in more sophisticated networking protocols.
		&\cite{bib:Nat_Phys_8_479}\\
		\hline 
		By using three pairs of polarisation-entangled photons and conducting two Bell state measurements, multistage entanglement swapping was realised.&\cite{bib:goebel08}\\
		\hline 
		Multi-particle entanglement swapping using a three-photon GHZ state was demonstrated.&\cite{bib:PRL_103_020501}\\
		\hline
		Entanglement swapping between photons that never coexisted was demonstrated. In their experiment, entangled photons are not only separated spatially, but also temporally.&\cite{bib:PRL_110_210403}\\
		\hline
		Hybrid entanglement swapping between discrete-variables and continuous-variables optical systems was realised experimentally.&\cite{bib:takeda2015entanglement}\\
		\hline
		%To develop a practical quantum network, entanglement swapping between independent entangled photon sources is a important. In the past two decades, entanglement swapping has been demonstrated in a large number of experiments across many physical architectures. However, in most experiments, entangled photons are generated by using the same laser, and therefore do not meet the requirements for independence. 
		Entanglement swapping based on independent entangled photon sources has been experimentally verified.
		%, but the distinguishability caused by photon propagation in the channel is still a great obstacle to realising entanglement swapping using independent sources under realistic conditions.
		& \cite{bib:PRL_96_110501, bib:Nat_Phys_3_692, bib:PRA_79_040302}\\
		\hline
		Entanglement swapping using independent entangled photon sources separated by 1.3km in a real-world environment was achieved. &\cite{bib:hensen2015loophole}\\
		\hline
		%However, the wavelength of the photons used in this experiment was 637nm (transmission loss \mbox{$\sim 15$dB/km}), which is not conducive to achieving long-distance entanglement swapping since it is far greater than the transmission loss of communication-band photons in fibre (\mbox{$\sim 0.2$dB/km}).
		 Entanglement swapping over 100km optical fibre with independent entangled photon-pair sources.&\cite{bib:sun2017entanglement}\\
		\hline
		%Entanglement swapping can also be directly used for QKD. Alice and Bob each have an entangled photon source, and one photon of each Bell pair is sent to a third-party measurement node, Eve. Similar to measurement-device-independent (MDI) QKD, the security of the generated private key does not depend on Eve's faithful execution of the operation. That is, Eve can be an untrusted third-party. This MDI property also reflects the physical beauty of quantum teleportation. Bell state measurements do not reveal any information about the quantum state, but can be used to restore the transmitted quantum state. On the other hand, quantum entanglement occurs between the remaining photons in the Bell pair of Alice and Bob.
		  %An entangled photon source can be considered as a basis-independent light source for QKD. Thus, the QKD realised by entanglement swapping has the characteristics of MDI and light source independence is suggested.&\cite{bib:PRL_90_057902, bib:NJP_10_2008}\\
		%\hline
		%An important application for entanglement swapping is that we can entangle spatially separated and independent matter qubits by coupling them with photons, upon which entanglement swapping is subsequently applied. This is an extremely important technique for hybrid quantum networks (Sec.~\ref{sec:hybrid}), where optical interactions mediate entanglement swapping between non-optical qubits.
		Starting with two entangled atom-photon pairs, we can project the two atomic qubits into a maximally entangled state by performing a Bell state measurement on the two photons.&\cite{bib:blinov2004observation, bib:PRL_96_030404}\\
		\hline
		Two trapped atomic ions separated 1m apart were entangled using entanglement swapping, exploiting interference between photons emitted by the ions. The fidelity of the states of the entangled ions was $0.63(3)$. In subsequent experiments, the ion-ion entanglement fidelity was improved to $0.81$.&\cite{bib:Nature_449_68,bib:PRL_100_150404}\\
		\hline
		Two atomic ensembles were entangled, each originally with a single emitted photon, by performing a joint Bell state measurement on the two single photons after they had passed through a 300m fibre-based communication channel.&\cite{bib:Nature_454_1098}\\\hline
		Robust entanglement (estimated state fidelity of $0.92\pm0.03$) between the two distant spins by entanglement swapping was generated. Such a high fidelity is sufficient to successfully perform loophole-free Bell inequality tests.&\cite{bib:hensen2015loophole}\\\hline
		%Entanglement swapping is a core element of quantum repeaters, which is of great significance to realising long-distance quantum communication. 
		%At present, the maximum transmission distance that can be achieved by QKD is 400km.&\cite{bib:arxiv_1606.06821}\\\hline
		Quantum repeaters, combining entanglement swapping and quantum memory, which provides a potential solution to this problem, was proposed in 1998 by Briegel \textit{et al.} And the first proposed practical quantum repeater architecture was proposed by Duan, Lukin, Cirac \& Zoller (DLCZ) using atomic ensembles and linear optics.&\cite{bib:BDCZ98, bib:Duan01}\\\hline
		To increase the repeater count-rate, various protocols  have been proposed.&\cite{bib:RMP_83_33, bib:PRA_79_042340, bib:PRA_92_012307, bib:PRA_81_052311, bib:PRA_81_052329, bib:NP_6_777, bib:MKLLJ14}\\\hline
	\end{tabular}
		\captionspacetab \caption{Developments in entanglement swapping and quantum repeaters.} \label{tab:entanglement_swap1}
\end{table*}

\begin{table*}[!htbp]
	\begin{tabular}{|p{0.755\linewidth}|p{0.22\linewidth}|}
		\hline
		\textbf{Summary} & \textbf{References \& years} \\
		\hline \hline
		The concept of all-photonic quantum repeaters, based on flying qubits, which entirely mitigate the need for a matter quantum memory, was introduced. &\cite{bib:azuma2015all}\\\hline
		Experimental demonstration of elementary segments of quantum repeaters were achieved.&\cite{bib:Sc_316_1316, bib:Nature_454_1098}.\\\hline
		In order to develop practical quantum repeaters, there are many experimental techniques that must be developed, such as multiplexing, and techniques based on non-degenerate photon-pair sources and quantum frequency conversion. &\cite{bib:PRA_76_050301, bib:PRA_82_010304, bib:PRL_113_053603, bib:PRL_98_060502,bib:Nat_469_508, bib:Nat_469_512, bib:PRL_112_040504, bib:PRA_92_012329,bib:NP_6_894, bib:NC_5_3376}\\\hline
		Quantum repeater techniques based on other physical systems have also been developed. &\cite{bib:NP_11_37, bib:Sc_337_72, bib:N_484_195, bib:bernien2013heralded}\\\hline
		%In general, to enable scaling up to repeaters with several links, many techniques need to be considerably improved and simplified, and it appears there is still a long way to go before building a first practical, long-distance quantum repeater.
	\end{tabular}
		\captionspacetab \caption{(continued) Developments in entanglement swapping and quantum repeaters.} \label{tab:entanglement_swap2}
\end{table*}

\begin{table*}[!htbp]
\begin{tabular}{|p{0.755\linewidth}|p{0.22\linewidth}|}
	\hline
	\textbf{Summary} & \textbf{References \& years} \\	\hline \hline
	Quantum cryptography using polarised photons in optical fibre over more than 1km. & \cite{bib:EL_23_383} \\
	\hline
	QKD experiment over 10km using phase-encoding. & \cite{bib:EL_29_634} \\
	\hline
	Outdoor experiment over 67km using a plug-and-play system to automatically maintain stabilisation. & \cite{bib:Arx0203118} \\
	\hline
	QKD based on decoy-states over more than 100km, marking the beginning of long-distance QKD. & \cite{bib:PRL_98_010505, bib:PRL_98_010504, bib:rosenberg2007long} \\
	\hline
	Decoy-state QKD over a 200km optical fibre cable through photon polarisation with a final key rate of 15Hz. & \cite{bib:OptExp_18_8587} \\
	\hline
	First realisation of a differential phase-shift (DPS) QKD protocol over a 42.1dB lossy channel and 200km of optical dispersion-shifted fibre. & \cite{bib:NP_1_343} \\
	\hline
	The DPS protocol over 50dB channel loss and 260km of optical fibre using superconducting detectors was realised. This is the first implementation of QKD over more than 50dB channel loss.&\cite{bib:OL_37_1008}\\
	\hline
	The coherent one way (COW) protocol QKD system with a maximum range of 250km at 42.6dB channel loss using ultra-low-loss fibre, with secret bit-rates up to 15Hz was realised. & \cite{bib:NJP_11_075003}\\
	\hline
	%Apart from using the QKD scheme based on state preparation and measurement, schemes based on entanglement distribution, mainly the E91  and BBM92 protocols, have been demonstrated, which are also undergoing extensive experimental investigation.&\cite{bib:PRL_67_661,bib:PRL_68_557}\\\hline
	Zeilinger's group distributed entangled single-photons over a free-space quantum channel, demonstrating the viability of free-space quantum communication. & \cite{bib:OE_13_202}\\
	\hline
	A complete experimental implementation of a QKD protocol over a free-space link using polarisation-entangled photon pairs. & \cite{bib:APL_89_101122}\\
	\hline
	The BBM92 QKD protocol based on polarisation encoding over 144km was realised. & \cite{bib:NP_3_481}\\
	\hline
	%The experiments listed above indicate that QKD protocols based on free-space entanglement distribution have the advantage of being less affected by decoherence, which lay a solid foundation for global and satellite-to-ground quantum communication.
	%Fibre loss increases exponentially with distance. However, the loss of free-space transmission increases very little with distance,  mainly related to the thickness of the atmosphere. Therefore, it is a perfectly reasonable solution to construct the global quantum internet based on satellite communication.
	To verify the feasibility of a quantum channel between space and Earth, a European Union group successfully received weak light pulses emitted from a ground station and reflected by a retroreflecting mirror on a low-Earth orbit satellite with orbital altitude of 1485km. & \cite{bib:NJP_10_033038}\\
	\hline
	In the context of rapidly moving platforms, QKD over 20km from an airplane to the ground was realised. & \cite{bib:NP_7_382}\\
	\hline
	Quantum communication with a hot-air balloon floating platform was accomplished successfully. & \cite{bib:NP_7_387}\\
	\hline
	%The experiments on aeroplanes and hot-air balloon systems demonstrate the feasibility of quantum communication in the condition of rapid motion, vibration, and random movement of satellites. At present, many countries including America, Canada, the European Union, China and Japan pay great attention to and support for accelerating the development of satellite-to-ground quantum communication. 
	The first quantum satellite was launched in China, and opened a platform for satellite-to-ground quantum communication at an intercontinental level.  & \cite{bib:gibney2016one, bib:liao2017satellite, yin2017satellite}\\
	\hline
%In addition to the ongoing expansion in distance, QKD is also being developed for P2P communication with quantum networks, which may be multi-user and of various and diverse topological structures. There is much competition and cooperation in this area.
The network of the American Defence Advanced Research Projects Agency (DARPA) connected the three nodes -- Harvard University, Boston University, and the BBN company -- in 2005, later increasing this to 10 nodes. & \cite{bib:QCC_2006_83}\\
\hline
%Since 2006, the EU has established a `SECOQ' network, combining the efforts of 41 research and industrial organisations from 12 countries, including the UK, France, Germany, and Austria.
Since 2006, the EU has established a `SECOQ' network, combining the efforts of 41 research and industrial organisations from 12 countries, including the UK, France, Germany, and Austria. A typical network employing a trusted repeater paradigm, with 6 nodes and 8 links was demonstrated in Vienna. & \cite{bib:NJP_11_075001}\\
\hline
National Institute of Communication Technology, together with Nippon Telegraph \& Telephone Corporation (NTT), Nippon Electric Company, Mitsubishi Electric Corporation, Toshiba European company, Switzerland IDQ Company and an Austrian team constructed a Tokyo QKD network in a metropolitan area, demonstrating the world's first secure TV conferencing over a distance of 45km. & \cite{bib:OExp_19_10387}\\
\hline
% The maximum distance is 9km, and the P2P bit-rate can reach 65kHz using superconducting detectors over 45km. \comment{How is the max distance both 45km and 9km???}
%In China, quantum networks are also developing rapidly.
 A 3-node network with a chained architecture, which demonstrated a cryptographically secure real-time voice call, was designed and constructed. & \cite{bib:OpEx17_6540}\\
 \hline
 A metropolitan all-pass and intercity quantum communication network in field fibre for 4 nodes was designed. Any 2 nodes can be connected in the network, enabling QKD between arbitrary pairs of users. &\cite{bib:OpEx_18_27217}\\
 \hline
 A QKD network with wavelength division multiplexers, realising 4 and 5 nodes with a star topology was constructed. &\cite{bib:PLA_372_3957,bib:OL_35_2454}\\
 \hline
%In 2012, a Chinese team constructed the largest metropolitan area quantum network in Hefei, linking 46 nodes to allow real-time voice communications, text messages and file transfers. A more than 2,000km quantum communication channel used by government bodies and banks under construction in Beijing and Shanghai will soon be fully operational. With the help of the new satellite, scientists will be able to test QKD, and other entanglement-based protocols, between the satellite and ground stations, and conduct secure quantum communications between Beijing and Xinjiang's Urumqi.
%With the distance and network coverage of quantum communication gradually increasing, the security of QKD systems draws more and more attention. Since 2012, the MDI QKD protocol has attracted much concern, because of its safety and practicability. 
The MDI QKD protocol in the laboratory over more than 80km of spooled fibre with time-bin encoding was demonstrated. They also tried outdoor experiments over 18.6km. &\cite{bib:PRL_111_130501}\\
\hline
A Brazilian group demonstrated the MDI QKD protocol using a polarisation encoding scheme. &\cite{bib:PRA_88_052303}\\
\hline
% However, these two demonstrations did not really distribute random key bits between two parties, and thus were not full MDI QKD demonstrations. Additionally, their system can be attacked by PNS or USD \comment{Define these acronyms!} sources and cannot generate secure key-bits in principle.
A full demonstration of time-bin phase-encoding MDI QKD over a 50km fibre link was reported. &\cite{bib:PRL_111_130502}\\
\hline
Polarisation-encoded MDI QKD with commercial off-the-shelf devices over 10km, with a secure key-rate of 0.0047Hz was implemented. &\cite{bib:PRL_112_190503}\\
\hline
The Chinese group continues to upgrade the performance of MDI QKD systems, making them viable over distances of up to 200km and 400km. &\cite{bib:PRL_113_190501, bib:yin2016measurement}
\\
\hline
A QKD system over 421km is achieved, by using a 3-state time-bin protocol combined with a one-decoy approach. & \cite{bib:boaron2018secure}
\\
\hline
\end{tabular}
\captionspacetab \caption{Developments in experimental QKD.} \label{tab:QKD_table}
\end{table*}

\begin{table*}[!htbp]
\begin{tabular}{|p{0.755\linewidth}|p{0.22\linewidth}|}
	\hline
	\textbf{Summary} & \textbf{References \& years} \\	\hline \hline
	An experimentally viable purification scheme, requiring only PBSs and post-selection, was proposed and demonstrated. The scheme was demonstrated again, using mixed states with fidelity $0.75$ ($0.8$), which they were able to purify to $0.92$ ($0.94$), a major improvement. & \cite{bib:Pan01, bib:Pan03} \\
	\hline
	A Bell experiment was performed using purified states. A state initially failing a Bell test successfully passed the test following entanglement purification. Unfortunately, the theoretical efficiency of the purification scheme is only $1/4$. & \cite{bib:PRL_94_040504, bib:Pan01} \\
	\hline
\end{tabular}
\captionspacetab \caption{Developments in experimental entanglement purification.} \label{tab:ent_pur}
\end{table*}

\begin{table*}[!htbp]
	\begin{tabular}{|p{0.755\linewidth}|p{0.22\linewidth}|}
		\hline
		\textbf{Summary} & \textbf{References \& years} \\
		\hline \hline
		 As one kind of single-longitudinal-mode laser, distributed feedback (DFB) lasers have a stable wavelength since they depend on etched gratings, which can only be tuned slightly with temperature. Thus, they are widely used in optical communication applications, such as QKD or dense wavelength division multiplexing, where it is desired to have a tuneable signal and extreme narrow linewidth. & \cite{bib:sun2016quantum, bib:liao2017long} \\
		\hline
		 In recent years, several QKD schemes have been designed and implemented using multi-longitudinal-mode Fabry-Perot (FP) lasers, mainly because they are significantly cheaper than DFB lasers. &  \cite{bib:choi2011quantum,  bib:wang2015experimental} \\
        \hline
	\end{tabular}
	\captionspacetab \caption{Two types of laser commonly used in quantum communication experiments.} \label{tab:coherent_states}
\end{table*}

\begin{table*}[!htbp]
	\begin{tabular}{|p{0.755\linewidth}|p{0.22\linewidth}|}
		\hline
	\textbf{Summary} & \textbf{References \& years} \\	\hline \hline
		Single photons can be easily created in pairs via SPDC. A state-of-the-art SPDC source at a wavelength of 788nm was reported simultaneously with a high brightness of $\sim$12 MHz/W, a collection efficiency of $\sim$70\% and an indistinguishability of $\sim 91\%$. & \cite{bib:wang2016experimental} \\
		\hline
		An SPDC source at a wavelength of around 1550nm was reported with simultaneously 97\% heralding efficiency and 96\% indistinguishability between independent single photons. & \cite{bib:zhong201812} \\
		\hline
		Heralded sources were integrated via four-wave mixing (FWM) in waveguides or optical fibres. Despite being a probabilistic process, SPDC or FWM enjoys a warm popularity in the labs of quantum optics since they have been technically mature and very cheap. That is why so far almost all the quantum information experiments in quantum networks were based on SPDC or FWM sources. & \cite{bib:silverstone2014, bib:spring2017chip, bib:goldschmidt2008, bib:smith2009} \\
		\hline
		Truly deterministic single-photon sources will be indispensable for future large-scale quantum networks. One of the most promising single-photon sources is based on quantum dots, which have been developed simultaneously exhibiting a high purity of 99.1\%, high indistinguishability of 98.5\%, and high extraction efficiency of 66\%. & \cite{bib:he2013on, bib:wei2014de, bib:ding2016on, bib:somaschi2016, bib:wang2016near, bib:loredo2016} \\
		\hline
	\end{tabular}
	\captionspacetab \caption{Some of the notable developments in single-photon state preparation.} \label{tab:single_photon_state}
\end{table*}

\begin{table*}[!htbp]
	\begin{tabular}{|p{0.755\linewidth}|p{0.22\linewidth}|}
		\hline
	\textbf{Summary} & \textbf{References \& years} \\	\hline \hline
        In 1999, the world's first multi-particle entanglement -- 3-photon GHZ entanglement -- was generated. After that, Pan and his colleagues broke the records continuously, realising 4-, 5-, 6-, 8-, 10-, and 12-photon entanglement, and 10- and 18-qubit hyper-entanglement\index{Hyper-entanglement} with multiple degrees of freedom. & \cite{bib:bouwmeester1999observation, bib:wang201818, bib:zhong201812}  \\
		\hline
		8-photon optical cluster states have been realised, and used to demonstrate topological error correction against a single error on any qubit. &  \cite{bib:yao2012experimental}  \\
		\hline
		5-photon NOON states have been realised by mixing quantum and classical light. & \cite{bib:afek2010high} \\
		\hline
		6-photon Holland-Burnett states\index{Holland-Burnett states} have been realised both at visible wavelengths and at telecom wavelengths. & \cite{bib:xiang2012optimal,  bib:jin2016detection} \\
		\hline
		Entangled photon-pairs were created with dimension up to \mbox{$15\times 15$} on a programmable silicon integrated circuit. & \cite{bib:wang2018multidimensional} \\
		\hline
		Zeilinger and his colleagues created orbital angular momentum (OAM) entangled photon-pairs differing by 600 in their quantum number. Four yeas later, they prepared \mbox{$100\times 100$}-dimensional OAM entangled photon pairs. & \cite{bib:fickler2012quantum} \\
		\hline
		Furusawa \textit{et al.} deterministically generate and fully characterise a continuous-variable cluster state containing more than 10,000 entangled modes. Three years later, they improved the system to more than one million modes. & \cite{bib:yokoyama2013ultra, bib:yoshikawa2016invited} \\
		\hline
        Small amplitude cat states were generated via photon subtraction from a squeezed vacuum state. & \cite{bib:neergaard2006generation,  bib:ourjoumtsev2006generating, bib:wakui2007photon} \\
        \hline
        Large amplitude cat states (i.e \mbox{${\left| \alpha  \right|^2} > 2.3$}) were prepared by reflecting half of a photon-number state into a momentum quadrature homodyne detector. & \cite{bib:ourjoumtsev2007generation, bib:takahashi2008generation} \\
		\hline
		Remote entanglement of two independent cat states was created by interfering small fractions of each pulse. & \cite{bib:ourjoumtsev2009preparation} \\
		\hline
	\end{tabular}
	\captionspacetab \caption{Developments in entangled state preparation, based on non-linear optics.} \label{tab:entangled_states}
\end{table*}

\begin{table*}[!htbp]
	\begin{tabular}{|p{0.755\linewidth}|p{0.22\linewidth}|}
		\hline
	\textbf{Summary} & \textbf{References \& years} \\	\hline \hline
		Atom-light interaction currently represents the most mature method to create atomic squeezing, and owns the highest squeezing of 20.1dB via an optical-cavity-based measurement. & \cite{bib:hosten2016measurement} \\
		\hline
		The detection of a single photon prepares almost 3000 atoms into an entangled Dicke state\index{Dicke state}. The state was reconstructed with a negative-valued Wigner function\index{Wigner function} -- an important hallmark of non-classicality. &  \cite{bib:mcconnell2015entanglement} \\
		\hline
		In 2010, two ensembles were entangled by storage of two entangled light fields; and ten years later, four quantum memories were entangled via spin waves. & \cite{bib:lukin2000entanglement, bib:choi2010entanglement} \\
		\hline
		Bose-Einstein condensates were used to create large ensembles of up to ${10^4}$ pair-correlated atoms with an interferometric sensitivity $-1.61$dB beyond the shot-noise limit. & \cite{bib:lucke2011twin} \\
		\hline
	\end{tabular}
	\captionspacetab \caption{Notable developments in atomic ensembles.} \label{tab:atomic_ensembles}
\end{table*}

\begin{table*}[!htbp]
	\begin{tabular}{|p{0.755\linewidth}|p{0.22\linewidth}|}
		\hline
	\textbf{Summary} & \textbf{References \& years} \\	\hline \hline
		In 2000, Haroche and his colleagues generated entanglement between two atoms and a single-photon cavity mode. Later, entanglement was demonstrated between two photons sequentially emitted by the same single atom in a cavity. An entanglement fidelity of 86\% between the two photons was obtained, and importantly, the entanglement generation efficiency was 15\%, a drastic improvement compared to free-space experiments with single atoms. & \cite{bib:rauschenbeutel2000step, bib:wilk2007single, bib:blinov2004observation} \\
		\hline
		In the atom-cavity system, one of the largest Schr{\"o}dinger cat states was created, with a spin of 25 on a Rydberg atom. & \cite{bib:facon2016sensitive} \\
		\hline
		In the recent year, Blatt \textit{et al.} reported scalable and deterministic generation of multi-ion entanglement, including 8-qubit W-states and 14-qubit GHZ states. & \cite{bib:haffner2005scalable, bib:monz2011}\\
		\hline
		Arbitrary coherent addressing of individual neutral atoms in a 3D optical lattice of up to 50 qubits was achieved. & \cite{bib:wang2015coherent} \\
		\hline		
		Several groups reported generation and detection of two neural-atom entanglement in optical lattices with a single-site-resolved technique. & \cite{bib:kaufman2015entangling, bib:islam2015measuring, bib:dai2016generation}\\
		\hline
	\end{tabular}
	\captionspacetab \caption{Notable developments in single-atom technology.} \label{tab:single_atoms}
\end{table*}

\begin{table*}[!htbp]
	\begin{tabular}{|p{0.755\linewidth}|p{0.22\linewidth}|}
		\hline
	\textbf{Summary} & \textbf{References \& years} \\	\hline \hline
		Spontaneous emission from a trion (the simplest charged excitons) state prepares a single electron spin in a quantum dot, while spin-photon entanglement was generated from such decay. &  \cite{bib:de2012quantum, bib:gao2012observation} \\
		\hline
	    Two-photon polarisation-entangled states can be generated from biexciton-exciton cascade radiative decay in a single quantum dot. The entangled photon pairs simultaneously exhibit a high purity of 0.004, high fidelity of 0.81, high two-photon interference visibility of 0.86, and on-demand generation efficiency of 0.86. & \cite{bib:muller2014demand} \\
		\hline
		A family of pyramidal site-controlled quantum dots was used to construct an array of entangled photon emitters with fidelities as high as 0.72. & \cite{bib:juska2013towards, bib:mohan2010polarization} \\
		\hline
		A cluster state of five sequentially-entangled  photons was generated by periodic timed excitation of a matter qubit. In each period, an entangled photon is added to the cluster state formed by the matter qubit and the previously emitted photons. & \cite{bib:schwartz2016deterministic} \\
		\hline
	\end{tabular}
	\captionspacetab \caption{Major developments with quantum dots.} \label{tab:quantum_dots}
\end{table*}

\begin{table*}[!htbp]
	\begin{tabular}{|p{0.755\linewidth}|p{0.22\linewidth}|}
		\hline
	\textbf{Summary} & \textbf{References \& years} \\	\hline \hline
		For NV centres, both electron and nuclear spins were observed with long coherence times ($>$1ms for the electron spin and $>$1s for the nuclear spin) even at room temperature. &  \cite{bib:balasubramanian2009ultralong, bib:neumann2010quantum, bib:maurer2012room} \\
		\hline
		Electron spin lifetime limited by phononic vacuum modes was observed. Negatively charged NV centres were proved to have exceptionally long longitudinal relaxation times, with $T_1$-times \index{T$_1$-time} of up to 8 hours. & \cite{bib:astner2018solid} \\
		\hline
		Bipartite entanglement was created between two nuclear spins coupled to a single NV centre in diamond, and tripartite entanglement was generated via the electron spin of the NV centre itself as the third qubit. & \cite{bib:neumann2008multipartite} \\
		\hline
		Quantum entanglement was created between a polarised optical photon and an NV centre qubit. Because NV centres couple to both optical and microwave fields, they can be used as a quantum interface between optical and solid-state systems.  & \cite{bib:togan2010quantum}\\
		\hline
		Room-temperature entanglement was created between two single electron spins over some 10nm distance in diamond. In the same year, heralded entanglement was created between two solid-state qubits located in independent low-temperature setups separated by 3m. & \cite{bib:dolde2013room, bib:bernien2013heralded} \\
		\hline
	\end{tabular}
	\captionspacetab \caption{Advances in nitrogen-vacancy (NV) centre technology.} \label{tab:NV_centres}
\end{table*}

\begin{table*}[!htbp]
	\begin{tabular}{|p{0.755\linewidth}|p{0.22\linewidth}|}
		\hline
	\textbf{Summary} & \textbf{References \& years} \\	\hline \hline
		There are three basic types of superconducting qubits -- charge, flux and phase. In 2007, a charge-insensitive qubit, called a transmon, was designed. &  \cite{bib:koch2007charge} \\
		\hline
		 Transmon qubits in a waveguide cavity, called circuit QED, achieved coherence times on the order of 0.1ms. & \cite{bib:paik2011observation, bib:rigetti2012superconducting} \\
		\hline
		In 2013, Martinis and his colleagues designed a cross-shaped transmon qubit, called an Xmon, which balances coherence, connectivity and fast control. In 2014, they constructed a 5-qubit GHZ state using five Xmons arranged in a linear array. In 2015, they reported the protection of states from environmental bit-flip errors in a 9-Xmon-qubit linear array. &  \cite{bib:barends2013coherent, bib:barends2014superconducting, bib:kelly2015state} \\
		\hline
		Pan \textit{et al.} reported the preparation and verification of genuine 10- and 12-qubit entanglement in a superconducting processor. & \cite{bib:gong2018genuine, bib:song201710} \\		
		\hline
		An 100-photon Schr{\"o}dinger cat state was created mapping from an arbitrary transmon qubit state, and extended to a superposition of up to four coherent states. & \cite{bib:vlastakis2013deterministically} \\
		\hline
		In 2016, the lifetime of microwave photons was extended to 0.3ms with error correction in superconducting circuits. &  \cite{bib:ofek2016extending} \\
		\hline
	\end{tabular}
	\captionspacetab \caption{Developments in superconducting rings for quantum information processing.} \label{tab:superconducting}
\end{table*}

\begin{table*}[!htbp]
	\begin{tabular}{|p{0.755\linewidth}|p{0.22\linewidth}|}
		\hline
	\textbf{Summary} & \textbf{References \& years} \\	\hline \hline
		\textit{Single photon avalanche diodes (SPAD) --}
		SPADs are the most compact and common single-photon detectors in the world. Commercially, silicon SPADs are available with detection efficiency around 60\% at 780nm, maximum count rates of 25MHz and dark-count rates as low as 25Hz. Several remarks are:
		
		\begin{itemize}
			
			\item Higher detection efficiency is possible. However, it is often at the cost of a lower maximum count rate (typical dead-time is 1$\mu$s, leading to 1MHz count-rates) and a larger dark-noise.
			
			\item Telecom band SPADs are usually based on InGaAs/InP, which suffer from low efficiencies of around 25\%. Another more efficient method is to convert the telecom photons into near infrared ones by up-conversion, and then detect them with a silicon SPAD. The detection efficiency can be improved to 30$\sim$40\%.
			
			\item SPADs are also commercially available in a multi-channel array with single power supply and individual inputs/outputs. Note that they are different from the multi-element SPAD arrays, which have either only one input or one output, and can be used to achieve photon-number resolution or high-speed single-photon detection.
			
		\end{itemize} &  \cite{bib:shentu2013ultralow} \\
		\hline
		\textit{Superconducting nanowire single photon detectors (SNSPD) --} 
		These are more efficient, but larger and more expensive than SPADs. Up until now, the highest detection efficiency of SNSPDs is 93\% at 1550nm, demonstrated in 2014. SNSPDs are commercially available with peak detection efficiencies higher than 80\% around 800nm, maximum dark-count rates of $100\sim 300$Hz, and a dead-time of $10\sim 70$ns. Similar performance is expected for optimised SNSPD at other wavelengths ($780\sim 1550$nm). However, SNSPDs must be operated at cryogenic temperatures of a few Kelvin to maintain their superconducting state. This is the main reason why SNSPDs are much more cumbersome and expensive than SPADs. &  \cite{bib:marsili2013} \\
		\hline
		\textit{Transition edge sensors (TES) --}
		These exhibit the highest detection efficiency, reported as high as 98\% (95\%) at wavelengths around 850nm (1556nm). Moreover, they are photon-number-resolving, owing to the linear response between resistance and temperature. However, the typical thermal recovery times of TESs are a fraction of a microsecond, limiting their applicability in high-speed quantum networks. Besides, TESs usually operate at ultra-low temperatures of 100mK, making it challenging to migrate the technique from lab to market. & \cite{bib:fukuda2011, bib:lita2008} \\
		\hline
	\end{tabular}
	\captionspacetab \caption{Some state-of-the-art single-photon detectors.}\label{tab:photodetection}
\end{table*}

\begin{table*}[!htbp]
	\begin{tabular}{|p{0.755\linewidth}|p{0.22\linewidth}|}
		\hline
	\textbf{Summary} & \textbf{References \& years} \\	\hline \hline
		The first time-domain homodyne detection was performed below 1KHz and achieved a shot-noise of 9dB. & \cite{bib:Smithey1993} \\
		\hline
		Several groups have constructed high speed and pulse-resolved homodyne detectors in the near infrared and telecom regimes. State-of-the-art homodyne detectors have achieved a shot-noise to electronic-noise ratio of 7.5$\sim$14dB, and a bandwidth of $\sim$100-300MHz, which enables quantum protocols with repetition rates ranging from tens of MHz to 100MHz. & \cite{bib:zavatta2002time, bib:okubo2008pulse, bib:kumar2012versatile, bib:chi2011balanced, bib:duan2013} \\
		\hline
	\end{tabular}
	\captionspacetab \caption{Some state-of-the-art homodyne detectors.} \label{tab:homodyning}
\end{table*}

\begin{table*}[!htbp]
	\begin{tabular}{|p{0.755\linewidth}|p{0.22\linewidth}|}
		\hline
	\textbf{Summary} & \textbf{References \& years} \\	\hline \hline
		For polarisation encoding, single-photon states can be easily manipulated by using a series of wave plates. CNOT gates with probability 1/9 have been reported in the coincidence basis. &  \cite{bib:Brien2003demonstration, bib:Kiesel2005, bib:Langford2005,  bib:Okamoto2005} \\
		\hline
		For path encoding, high-fidelity CNOT gate was reported with integrated optical waveguides. & \cite{bib:politi2008silica} \\
		\hline
		For time-bin encoding, heralded controlled-phase gates was reported. & \cite{bib:Humphreys2013} \\
		\hline
		Assisted by two independent photons, CNOT gates have been realised in a nondestructive manner. & \cite{bib:Bao2007Optical, bib:Zhao2005Experimental} \\
		\hline
		By using multi-level quantum systems to encode information, probabilistic 3-qubit and 4-qubit entangling gates were reported. & \cite{bib:lanyon2009simplifying, bib:starek2016} \\
		\hline
		Ultra-low loss multi-mode optical bulk circuits have been assembled and used for boson-sampling machines beating early classical computers. & \cite{bib:wang2017high, bib:wang2018toward} \\
		\hline
	\end{tabular}
	\captionspacetab \caption{Some of the important developments in the evolution of optical states.} \label{tab:evolutionofstates}
\end{table*}

\begin{table*}[!htbp]
	\begin{tabular}{|p{0.755\linewidth}|p{0.22\linewidth}|}
		\hline
	\textbf{Summary} & \textbf{References \& years} \\	\hline \hline
		Silica-on-Silicon (SoS) planar light-wave circuit (PLC). For the state-of-the-art technique, SoS-PLCs can achieve ultra-low propagation loss ($<$0.01 dB/cm) and coupling loss ($<$0.1 dB/facet). Nowadays, small-scale quantum circuits have reached coupling loss $\sim$0.4 dB/facet and each directional coupler (DC) loss $\sim$0.1 dB. &  \cite{bib:hibino2003silica, bib:carolan2015universal} \\
		\hline
		Fused silica waveguides written by femtosecond laser direct written (FLDW) technology. Unlike PLC fabrication, the FLDW is a powerful and flexible technique for 3D rapid fabrication, requiring no masking procedure or cleanroom environment. FLDW waveguides exhibit losses on the order of 0.1 dB/cm. & \cite{bib:sakuma2003ultra} \\
		\hline
		Telecom silicon waveguides have higher refractive index, thus allowing a dramatic reduction in the size. For example, a $2 \times 2$ coupler in silicon (27$\mu$m) is 40 times shorter than the one in silica (1.1 mm). The best propagation (coupling) losses can reach 0.1 dB/cm (0.5 dB/facet). &  \cite{bib:bonneau2012quantum, bib:lee2000, bib:almeida2003, bib:mcnab2003} \\
		\hline
		A programmable eight-mode photonic time-bin circuit has been constructed to implement three and four photons boson-sampling. The controlled switch was realised using a bulk electro-optic modulator (EOM) with a transmission of 97.3\%, and a single-loop efficiency was achieved with 83\%.  & \cite{bib:he2017time} \\
		\hline
	\end{tabular}
	\captionspacetab \caption{Some state-of-the-art programmable optical circuits.} \label{tab:waveguide_fibre}
\end{table*}

\begin{table*}[!htbp]
	\begin{tabular}{|p{0.755\linewidth}|p{0.22\linewidth}|}
		\hline
	\textbf{Summary} & \textbf{References \& years} \\
	\hline \hline
		One of the simplest way to storing light is to use an optical fibre. However, the storage time is hard to adjust and limited to tens of microseconds due to high transmission loss of ultra-long fibre.  & \cite{bib:landry2007quantum, bib:lvovsky2009optical} \\
		\hline
		An erbium-doped optical fibre was successfully used to store telecom-wavelength single photons and entangled photons. & \cite{bib:saglamyurek2015quantum, bib:jin2015telecom} \\
		\hline
		An ultra-small high-Q photonic-crystal nano-cavity was used to trapping and delaying photons for 1.45ns. & \cite{bib:tanabe2007trapping} \\
		\hline
		A quantum light-matter memory has been realised with simultaneously a readout efficiency of 76\% and a lifetime of 0.22s, which can support a sub-Hz entanglement distribution of up to 1,000km, for the first time going beyond the maximally achievable quantum communication distance using direct transmission. &  \cite{bib:yang2016efficient}\\
		\hline
		Electromagnetically induced transparency (EIT) in cold atomic ensembles has been demonstrated to store entangled photon pairs and squeezed vacuum states. & \cite{bib:Choi2008mapping, bib:appel2008quantum, bib:honda2008storage} \\
		\hline
		EIT was used to store a qubit with storage times reaching the millisecond range. & \cite{bib:lettner2011remote,  bib:riedl2012bose, bib:xu2013long} \\
		\hline
		EIT was used to store single-photon-level quantum images for 100 nanoseconds. & \cite{bib:ding2013single} \\
		\hline
        Nuclear spins of ${}^{31}$P impurities in an sample of ${}^{28}$Si were reported with a coherence time of as long as 192s at a temperature of 1.7K. & \cite{bib:steger2012quantum} \\
        \hline
        Ionised donors in silicon-28 was observed with a qubit storage time exceeding 39 minutes at room temperature. & \cite{bib:saeedi2013room} \\
        \hline
        Optically addressable nuclear spin in a solid have been demonstrated to achieve with remarkably long coherence times, and with a current record of six hours. & \cite{bib:zhong2015optically} \\
        \hline
	\end{tabular}
	\captionspacetab \caption{Some of the notable developments in quantum memory.} \label{tab:memory}
\end{table*}

\begin{table*}[!htbp]
	\begin{tabular}{|p{0.755\linewidth}|p{0.22\linewidth}|}
		\hline
			\textbf{Summary} & \textbf{References \& years} \\	
			\hline \hline
		Linear optics: The first demonstration of an entangling, photonic, 2-qubit CNOT gate using post-selected linear optics. & \cite{bib:OBrien03}\\
		\hline
		Trapped atoms/ions: A register of 20 individually controlled qubits was generated and characterised; Single-qubit gates with an error per gate of $3.8 \times {10^{ - 5}}$, and deterministic 2-qubit gate with a gate error of $8 \times {10^{ - 4}}$; 14-qubit entanglement with a 2-qubit fidelity of 98.6\%. & \cite{bib:friis2018observation, bib:Gaebler2016, bib:monz2011} \\
		\hline
		Nuclear magnetic resonance: 12-qubit system with the simulated fidelity over 99.7\% on each qubit; The average fidelity of 87.5\% on a 7-qubit entangling gate. & \cite{bib:Negrevergne2006, bib:lu2017enhancing, bib:Lu2015} \\
		\hline
		Linear optics: A fully programmable 2-qubit linear optics quantum processor with an average quantum process fidelity of 93.2\%. & \cite{bib:qiang2018large} \\
		\hline
		Boson-sampling: Several five-photon boson-sampling machines were built and beat early classical computers. & \cite{bib:zhong201812, bib:wang2018toward, bib:wang2017high} \\
		\hline
		Coherent Ising machine: An optical processing approach was used to model and optimise 2000-node optimisation problems based on a network of coupled optical pulses in a ring fibre. & \cite{bib:mcmahon2016fully, bib:inagaki2016coherent} \\
		\hline
		Superconductors: 9-qubit system with an average single-qubit gate fidelity of 99.92\% and a 2-qubit gate fidelity of up to 99.4\%. & \cite{bib:kelly2015state, bib:barends2014superconducting} \\
		\hline
		Spins in Silicon: A programmable 2-qubit quantum processor in silicon with state fidelities of 85$\sim$89\%; 2-qubit CNOT gate; An addressable single-qubit with a control fidelity of 99.6\%. & \cite{bib:watson2018programmable, bib:veldhorst2015two, bib:veldhorst2014addressable}. \\
		\hline
		Nitrogen-vacancy centres: A programmable 2-qubit solid-state quantum processor at room temperature. & \cite{bib:wu2018programmable} \\
		\hline
		Adiabatic quantum computation: A 2048-qubit D-Wave quantum processor was used to predict phase-transitions and topological phenomena. & \cite{bib:harris2018phase, bib:king2018observation} \\
		\hline
	\end{tabular}
	\captionspacetab \caption{Major milestones in the development of quantum computation.} \label{tab:quantumcomputer}
\end{table*}

\startalgtable

\sketch{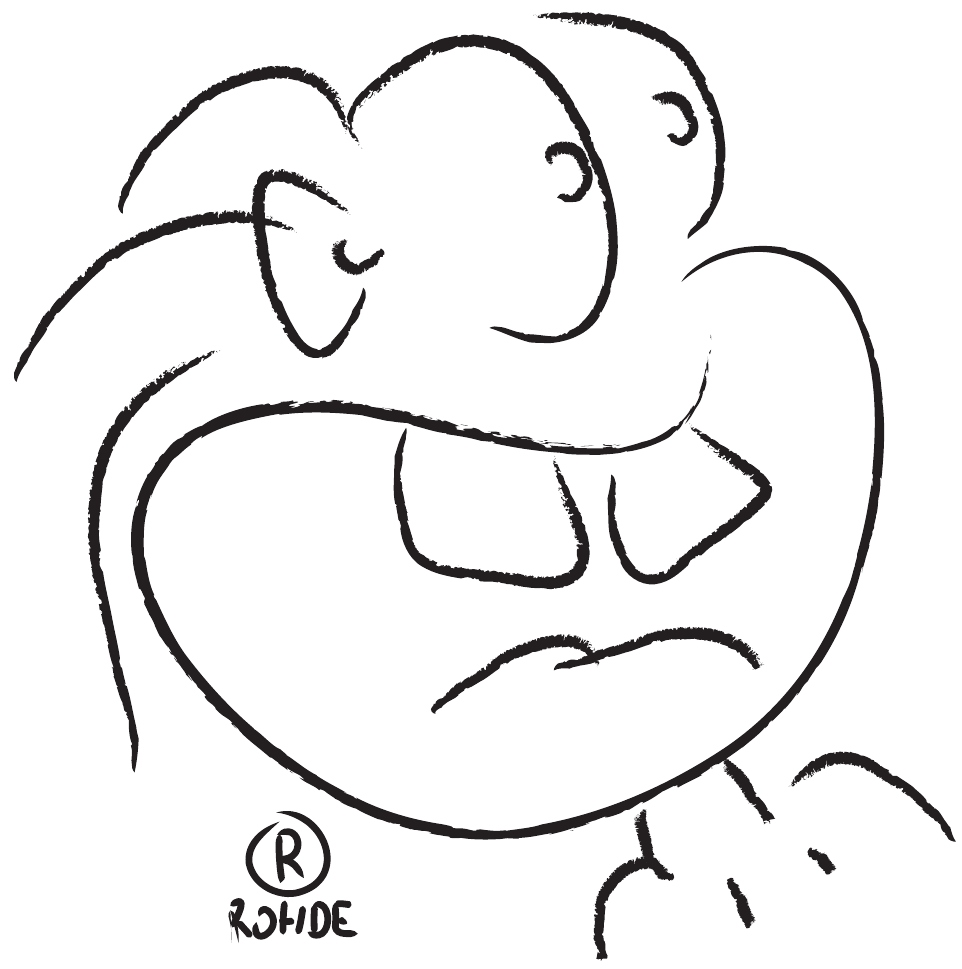}

\clearpage

%
% To the future & beyond
%

\part{To the future \& beyond}\label{part:future_beyond}

%
% To The Future & Beyond
%

\famousquote{The most exciting phrase to hear in science, the one that heralds new discoveries, is not `Eureka!', but `That's funny\ldots'}{Isaac Asimov}
\newline

\dropcap{I}{n} this part we highlight some of -- in the minds of the authors -- the most exciting and transformative applications for the future quantum internet and its monstrous unified quantum computational power. None of the technologies discussed here have presently been realised on any practical scale, but all have revolutionary potential to transform the global economy, how we interact with the world, and how we live our lives.

However, these predictions on the future of world-changing applications for quantum technology are merely speculation. We believe it is far more likely that the most impactful and reverberating applications for quantum technology likely haven't even been invented yet.

Currently, there is much ongoing research being conducted into new quantum algorithms and protocols, with major developments being made at a rapid pace. Most likely, mankind's quantum future will be far more exciting than our speculations here, much as speculation surrounding early digital computers never anticipated the digital technological revolution and its impact on every aspect of our modern lives.

\latinquote{Excelsior.}

%
% Space-Based Quantum Networks
%

\section{Space-based quantum networks}\label{sec:quant_space_race}\index{Space race}\index{Satellites}

\sectionby{Tim Byrnes}\index{Tim Byrnes}

\famousquote{The Earth is the cradle of humanity, but mankind cannot stay in the cradle forever.}{Konstantin Tsiolkovsky}
\newline

\dropcap{O}{ne} of the major hurdles that must be overcome before quantum networks achieve widespread commercial use is to span large distances between nodes distributing entanglement as a resource. The key technological and commercial centres of the world are decentralised, spread across the globe, hence connections over intercontinental distances are essential, with nodes separated potentially by thousands of kilometres. This distance is one of the major challenges facing the creation of large-scale quantum networks.

The most convenient way of transmitting quantum information is using optical fibres\index{Optical!Fibres}. However, it is well-known that due to the exponential scaling of loss with distance, the upper limit is of the order of several hundred kilometres. This shortcoming has inspired intensive research into extending these distances using quantum repeaters, discussed in detail in Sec.~\ref{sec:rep_net}\index{Quantum repeater networks}.

An alternative to fibre optic communication is using terrestrial free-space links\index{Terrestrial free-space communication}. Particular wavelength regimes exhibit low absorption\index{Absorption}, allowing the propagation of photons across long distances. Impressive experimental demonstrations of teleportation with free-space entanglement \cite{bib:NP_3_481, bib:Nat_489_269, bib:yin2013lower} over 100km have been achieved.

However, extending to larger distances has been problematic. Photon loss in free-space links is affected by weather\index{Weather conditions} conditions and other atmospheric effects\index{Atmospheric effects}\index{Photon loss}, such as pollution\index{Pollution}, which degrade visibility. The Earth's curvature\index{Earth curvature} provides an absolute upper limit, depending on the elevation of the source and receiver. For example, for an observer on a 30m tower, the horizon is at a distance of 20km, a relatively short distance. The observatory used in the experiments performed by the experiments of \cite{bib:NP_3_481, bib:Nat_489_269} were at an elevation of 2393m, allowing transmission over a distance of 143km.

\begin{figure}[!htbp]
\if 1\doublecol
	\includegraphics[clip=true, width=0.475\textwidth]{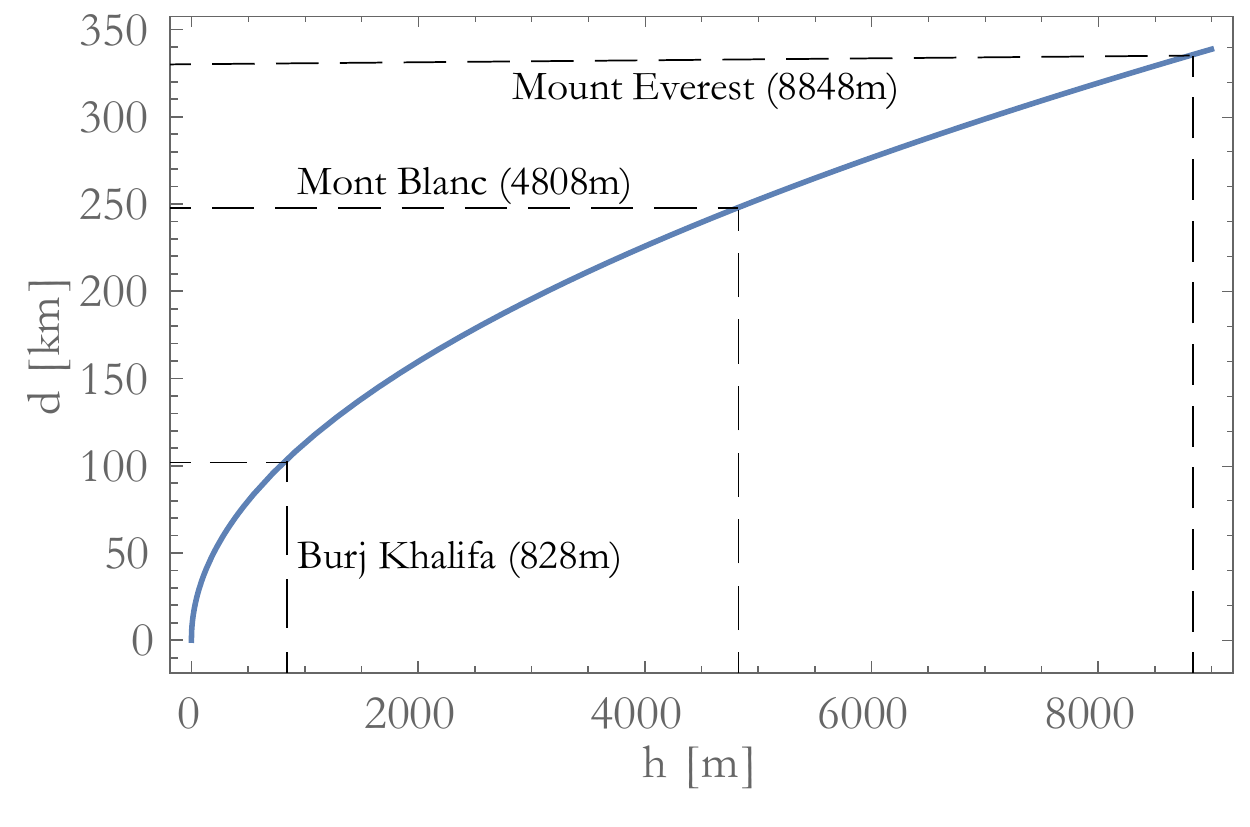}
\else
	\includegraphics[clip=true, width=0.6\textwidth]{distance_to_horizon}
\fi
\captionspacefig \caption{Distance to the horizon $d$ (in kilometres), at altitude $h$ (in meters), and several illustrative examples with their elevations. The range for $h$ spans from sea level to roughly the height of Mount Everest, thereby covering the full range of altitudes possible for ground-based network nodes. However, using more practically realistic bounds on altitude, say on the order of 5,000-6,000m (already very optimistic, and only achievable in some parts of the globe), line-of-sight distance to the horizon will be limited to be on the order of 200-250km. This limitation necessitates the use of repeater networks to enable the efficient approaches for constructing quantum communication channels that bypass this constraint imposed by the Earth's curvature.\index{Mount Everest}\index{Mont Blanc}\index{Burj Khalifa}} \label{fig:dist_hor}
\end{figure}

In general, the distance to the horizon $d$ (in kilometres)\index{Distance to horizon} on the Earth's surface at an altitude of $h$\index{Ground stations!Altitude} (in metres) is given by the approximation,
\begin{align}
d \approx 3.57\sqrt{h}.
\end{align}
This relationship is shown in Fig.~\ref{fig:dist_hor}, covering the range of altitudes from sea level\index{Sea level} to the summit of Mount Everest (8,848m)\index{Mount Everest}. It is evident that line-of-sight\index{Line-of-sight} channels across the Earth's surface are inherently limited by the curvature of the Earth to $\sim$250km, using altitudes that are practically realistic (i.e no ground stations on the summit of Everest thank you very much!).

One attractive possibility to overcome these issues is using space-based quantum communication. In this scenario, satellites orbiting the Earth would act as nodes of the quantum network, which could store, send and receive quantum information. At altitudes where satellites orbit the Earth, photon loss due to scattering\index{Photon loss}\index{Scattering} is negligible, and photons can propagate across extremely long distances unhindered -- despite various shortcomings of photonic quantum information, a key benefit is their robustness and preservation of coherence through empty space over very long distances. The main loss in this case is caused by diffraction\index{Diffraction}, due to the finite diameter of the receiver and transmitter hardware. For example, using reasonable diameters it is in the region of 40-80dB loss for low Earth orbit (LEO)\index{Low Earth orbit} satellites \cite{bib:aspelmeyer2003long, bib:liao2016ground}.

Repeater network protocols (Sec.~\ref{sec:rep_net}) are completely compatible in-principle with the space-based networks discussed here. In the context of a space-based network the key limitation we face is that the opposing hemisphere of the Earth is beyond direct line-of-sight\index{Line-of-sight}, prohibiting a direct P2P\index{Point-to-point (P2P)} communications channel. Instead, the long-distance channel must be broken down into shorter segments, within line-of-sight of one another, enabling, for example, entanglement to be incrementally swapped\index{Entanglement!Swapping} across neighbouring satellites and ultimately around the globe.

Involving ground-based quantum network nodes is not prohibitive as 80\% of the atmosphere by mass is located within the first 12km in altitude, yielding low effective thickness of the atmosphere ($\sim10$km)\index{Effective atmospheric thickness}. Despite this apparently low exposure to attenuation, the loss is nonetheless substantial, on the order of $\sim$40dB (depending on optical frequency) when the satellite is directly overhead. This significantly reduces effective count rates, a problem that does not represent any kind of engineering limitation, but rather an inherent physical one that cannot foreseeably be mitigated (it seems implausible that much can possibly be done to circumvent the atmosphere in satellite-to-ground communication!). This effective thickness of course changes with the azimuth of the satellite in the sky. Near the horizon the line-of-sight\index{Line-of-sight} between satellite and ground-station is far greater than when the satellite is directly overhead. The light trajectory beyond this effective thickness regime effectively passes through vacuum with attenuation\index{Photon loss} limited only by diffraction\index{Diffraction}. Thus, when directly overhead, a satellite-to-ground\index{Satellites!Satellite-to-ground communication} channel is more favourable in terms of signal attenuation than long-distance ground-to-ground\index{Ground-to-ground communication} channels over free-space\index{Free-space}.

Ground-to-satellite\index{Ground-to-satellite communication} and satellite-to-ground\index{Satellites!Satellite-to-ground communication} quantum communication have already been demonstrated, as will be discussed in detail in the next section. Such a satellite-based quantum communication system is naturally suited to various tasks, in particular QKD\index{Quantum key distribution (QKD)}, which does not require quantum memories\index{Quantum memory} or impose onerous timing constraints. However, with the addition of quantum memories the capabilities of such a quantum network could be greatly enhanced, enabling many of the applications and protocols described earlier (Part.~\ref{part:protocols}) at a global scale.

Naturally, such a space-based quantum network brings with it enormous engineering challenges that must be overcome prior to implementation. Creating even a short-distance quantum network is currently technologically challenging, let alone one that is loaded onto a satellite transmitting photons across distances potentially of the order of the diameter of the Earth. The cooperation of space agencies is necessary even for initial experiments.

As we describe more in the next section, several recent positive results have made the technology an immediate possibility in realising such a global level quantum network. In this section, we summarise the current technological state of the art internationally, and discuss some of the remaining major challenges.  

There are two main ingredients for communication in a global satellite-based quantum internet: satellite-to-satellite\index{Satellites!Satellite-to-satellite communication}, and ground-to-satellite\index{Ground-to-satellite communication}, shown in Fig.~\ref{fig:space_1}. Each bring with them their own engineering challenges.

\if 1\doublecol
\begin{figure}[!htbp]
\includegraphics[clip=true, width=0.475\textwidth]{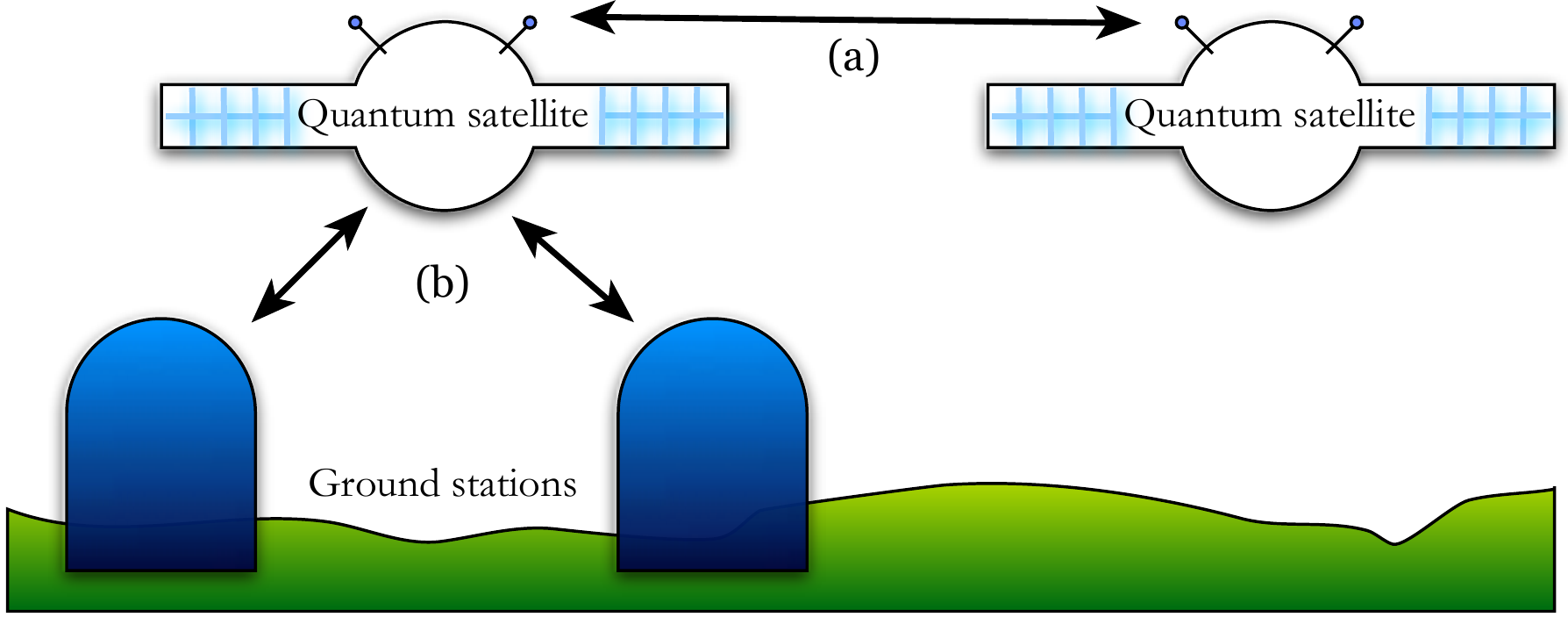}
\captionspacefig \caption{Various possibilities for space-based quantum communication. (a) Satellite-to-satellite quantum communication \cite{bib:byrnes2017lorentz}. (b) Ground-to-satellite quantum communication \cite{bib:armengol08}.}
\label{fig:space_1}
\end{figure}
\else
\begin{figure*}[!htbp]
\includegraphics[clip=true, width=\textwidth]{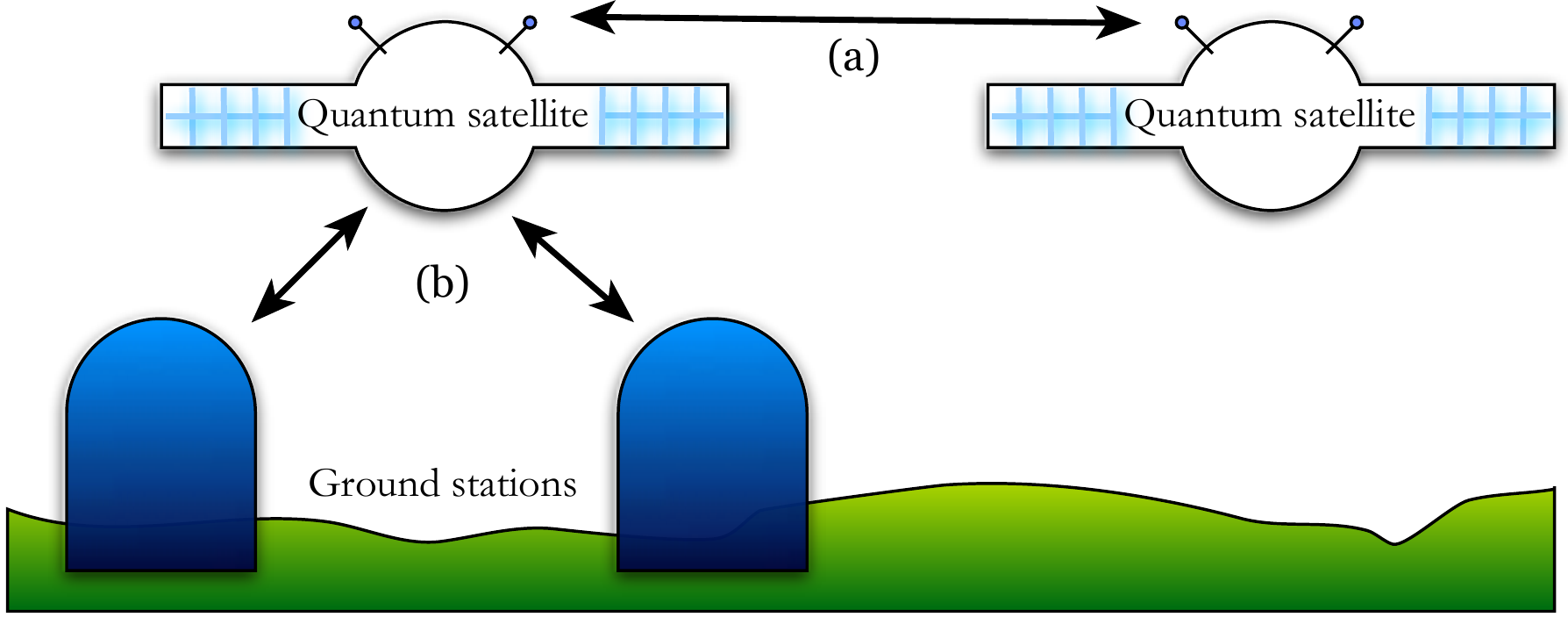}
\captionspacefig \caption{Various possibilities for space-based quantum communication. (a) Satellite-to-satellite quantum communication \cite{bib:byrnes2017lorentz}. (b) Ground-to-satellite quantum communication \cite{bib:armengol08}.}
\label{fig:space_1}
\end{figure*}
\fi 

%
% International efforts for space-based quantum communication
%

\subsection{International efforts}\index{International efforts}

We now briefly summarise some of the major international developments in the demonstration of satellite-based quantum communication. Some of these developments are extremely recent at the time of writing this work. It is a certainty that major developments will follow in the near future, with increasingly capable satellites, enabling the demonstration of ever more sophisticated quantum protocols.

\begin{figure*}[!htbp]
\includegraphics[clip=true, width=\textwidth]{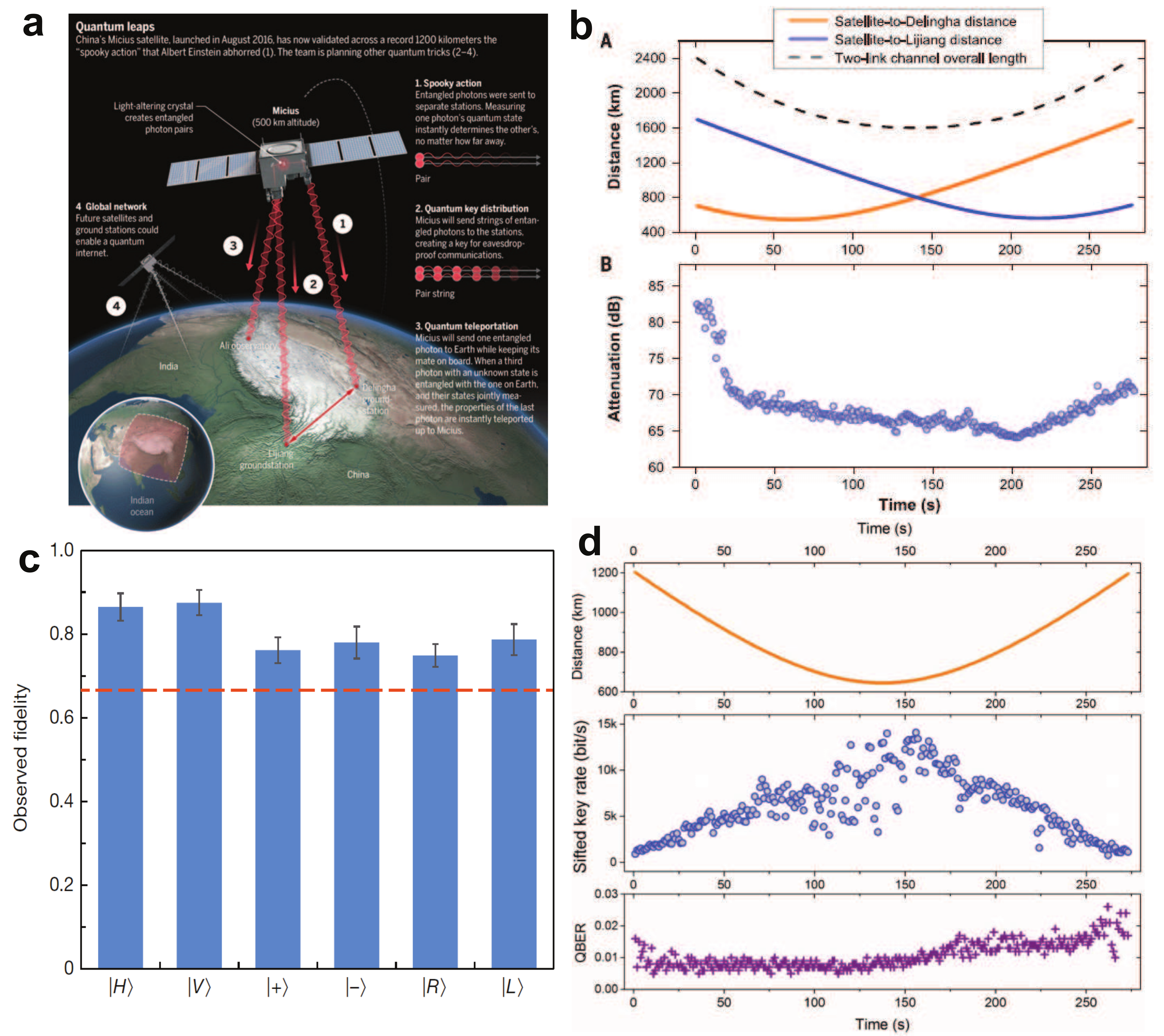}
\captionspacefig \caption{The Chinese Micius quantum communications satellite. (a) Schematic of the satellite and ground stations used to observe the entangled photons \cite{bib:popkin17}.
 % \comment{[TB: Took image from Science News, may need permission]. }
 Picture is watermarked}. (b) Attenuation during entanglement distribution from \cite{yin2017satellite}. (c) Fidelities achieved for teleportation of various states as marked from \cite{bib:ren2017ground}.
\label{fig:space_4}
\end{figure*}

%
% China: QUESS
%

\subsubsection{China: QUESS}\index{QUESS}

In 2016, the group at the University of Science \& Technology China (USTC), led by Jian-Wei Pan\index{Jian-Wei Pan}, launched QUESS (Quantum Science Experiment Satellite), a quantum communication satellite \cite{liao2017satellite, bib:xin11}. The main feature of the satellite is an ultra-bright source of polarisation entangled photons at a wavelength of 810nm, generated by SPDC\index{Spontaneous parametric down-conversion (SPDC)}. The source is capable of emitting $5.9 \times 10^6$ photon-pairs per second with a fidelity of $\sim 0.91$.

The other important piece of technology is the acquisition, pointing, and tracking technology\index{Acquisition}\index{Pointing}\index{Tracking}, which allows the ground stations to track the location of the satellite as it moves across the sky (which is very rapid for LEO), and vice versa. This is achieved using lasers at separate wavelengths, and allows the photon transmission and the ground-based receivers to be pointing to each other to within $\sim 1.2$mrad \cite{yin2017satellite}. The major aims of the project are to:
\begin{itemize}
\item Perform QKD\index{Quantum key distribution (QKD)} between Xinglong and Urumqi (a ground distance of 2,500km).
\item Test Bell's inequality\index{Bell!Inequality} over a distance of 1,200km.
\item Perform quantum state teleportation\index{Quantum state teleportation} between the satellite and Ali in Tibet (see Fig.~\ref{fig:space_4}).
\item (Ultimate goal) Perform QKD between Beijing and Vienna (a distance of 7,500km that extends well beyond the Earth's curvature).
\end{itemize}

At the time of writing, three main results have been reported. The first is an entanglement distribution experiment where a Bell violation was observed between ground stations at Delingha and Lijiang (separated by 1,203km) and Delingha and Urumqi (separated by 1,120km) \cite{yin2017satellite}. The satellite is in LEO at an elevation of $\sim500$km, and is in view from the observatories for a narrow duration of $\sim275$s.

The attenuation of the photons during the downlink transmission is shown in Fig.~\ref{fig:space_4}(b). Taking into account photon transmission distances, the attenuation rates are far better than the best performance of optical fibres at 0.16dB/km \cite{bib:yin2013lower} and even theoretical loss limits. The Bell violation recorded in this experiment was at the level of $2.37\pm 0.09>2$, and the non-locality\index{Non-locality} of the entanglement was confirmed.

The second experiment performed ground-to-satellite\index{Ground-to-satellite communication} QST\index{Quantum state teleportation} \cite{bib:ren2017ground}. Here, a single polarisation-encoded photon defines the qubit to be teleported, and the satellite acts as the receiver. The entangled photons are generated on the ground, at the observatory in Ngari, Tibet. As with the first experiment described above, the satellite is in LEO and sweeps across the sky, creating a narrow time-window during which the teleportation must be achieved. The longest distance for successful teleportation was $\sim 1400$km, when the satellite emerges from the horizon. When directly overhead, the satellite is at a distance of $\sim 500$km. As Fig.~\ref{fig:space_4}(c) shows, the fidelity of the types of states teleported are all above the theoretical classical bound of $2/3$, with an average of $0.80 \pm 0.01$.  

A third experiment performed QKD in a satellite-to-ground\index{Satellites!Satellite-to-ground communication} configuration, where the ground station was at Xinglong \cite{bib:liao2017satellite}. The protocol that was used was the decoy-state BB84\index{Decoy states}\index{BB84 protocol} protocol, which is robust against a photon-number splitting attack\index{Photon-number-splitting attacks}. A key-rate of between $1-12$kbit/s was achieved while the satellite was visible, depending upon the relative location in the sky, shown in Fig.~\ref{fig:space_4}(d).

%
% Japan: SOCRATES
%

\subsubsection{Japan: SOCRATES}\index{Space Optical Communications Research Advanced Technology Satellite (SOCRATES)}

SOCRATES (Space Optical Communications Research Advanced Technology Satellite) is a micro-satellite\index{Micro-satellites} developed by NICT (National Institute of Information \& Communications Technology\index{National Institute of Information \& Communications Technology (NICT)}) \cite{bib:horiuchi2015view, bib:toyoshima2015current, bib:takenaka2017}. This is a 48kg satellite of volume $50^3$cm$^3$ that demonstrates multi-purpose (i.e both classical and quantum) optical communication. Its primary mission is to demonstrate laser communication in space. As one of its subgoals, the on-board equipment is adapted to perform QKD experiments\index{Quantum key distribution (QKD)}. To this end, the satellite is equipped with a photon source, where the polarisation can be classically switched between non-orthogonal angles, necessary for BB84\index{BB84 protocol} for example.

Initial results indicated polarisation preservation of photons emitted from the satellite, a preliminary step towards performing protocols such as BB84\index{BB84 protocol} \cite{bib:carrasco2016leo}. More recent results showed quantum-limited satellite-to-ground\index{Satellites!Satellite-to-ground communication} communication was possible, attaining a quantum bit error rate below 5\% \cite{bib:takenaka2017} during the closest approach of the satellite to the ground at a distance of 744km. The downlink was established over distances of over 1,000km, although for obvious reasons the error rate was higher for longer distances.

%
% Singapore: Cubesat
%

\subsubsection{Singapore: Cubesat}\index{Cubesat}

The group at the National University of Singapore (NUS)\index{National University of Singapore (NUS)} also has taken the approach of using nano-satellites\index{Nano-satellites}, with the capability of generating correlated photon-pairs on-board \cite{bib:tang2016generation}. The nano-satellite weighs just 1.65kg and has all the components required for creating and detecting SPDC photon pairs. The nano-satellite approach greatly reduces the cost of space-launch, and the launch itself was performed by the Indian space agency. As photon generation and detection are both on the same satellite, no quantum communications channel (to other satellites or Earth) was established in the experiment, although it demonstrated in-principle the space-worthiness of the basic hardware components. 

%
% Europe: Retroreflector Satellite
%

\subsubsection{Europe: Retroreflector satellite}\index{Retroreflector satellite}

European groups used laser-ranging satellites fitted with corner-cube\index{Corner-cube} retroreflectors\index{Retroreflector satellite} to demonstrate quantum communication through the atmosphere \cite{bib:NJP_10_033038, bib:vallone15}. A ground-based photon source transmitted photons towards the satellite, which reflected them back down again to an Earth-based observatory. This confirmed the ability of photons to travel through the atmosphere into space and back, with a bit error ratio at the level of 5\%. Although no optical sources or detectors were placed on the satellite, there has been long-standing active interest in a European space-based quantum communication project, led in particular by the group of Anton Zeilinger at the University of Vienna\index{University of Vienna} \cite{bib:armengol08}. The aims of such projects are to perform space-based QKD\index{Quantum key distribution (QKD)}, with terrestrial free-space experiments \cite{bib:NP_3_481, bib:Nat_489_269} a part of the overall research effort. 

%
% Canada: QEYSSat
%

\subsubsection{Canada: QEYSSat}\index{QEYSSat}

The QEYSSat (Quantum EncrYption \& Science Satellite) proposes a micro-satellite\index{Micro-satellites} that incorporates a quantum receiver \cite{bib:jennewein2014qeyssat}. The satellite acts as a trusted node\index{Trusted nodes}, to which photons would be transmitted from Earth-based sources. This could be employed to perform QKD between two locations on Earth separated by large distances. No photon sources or quantum memories would be included on the QEYSSat. 

\begin{figure*}[!htbp]
\includegraphics[clip=true, width=\textwidth]{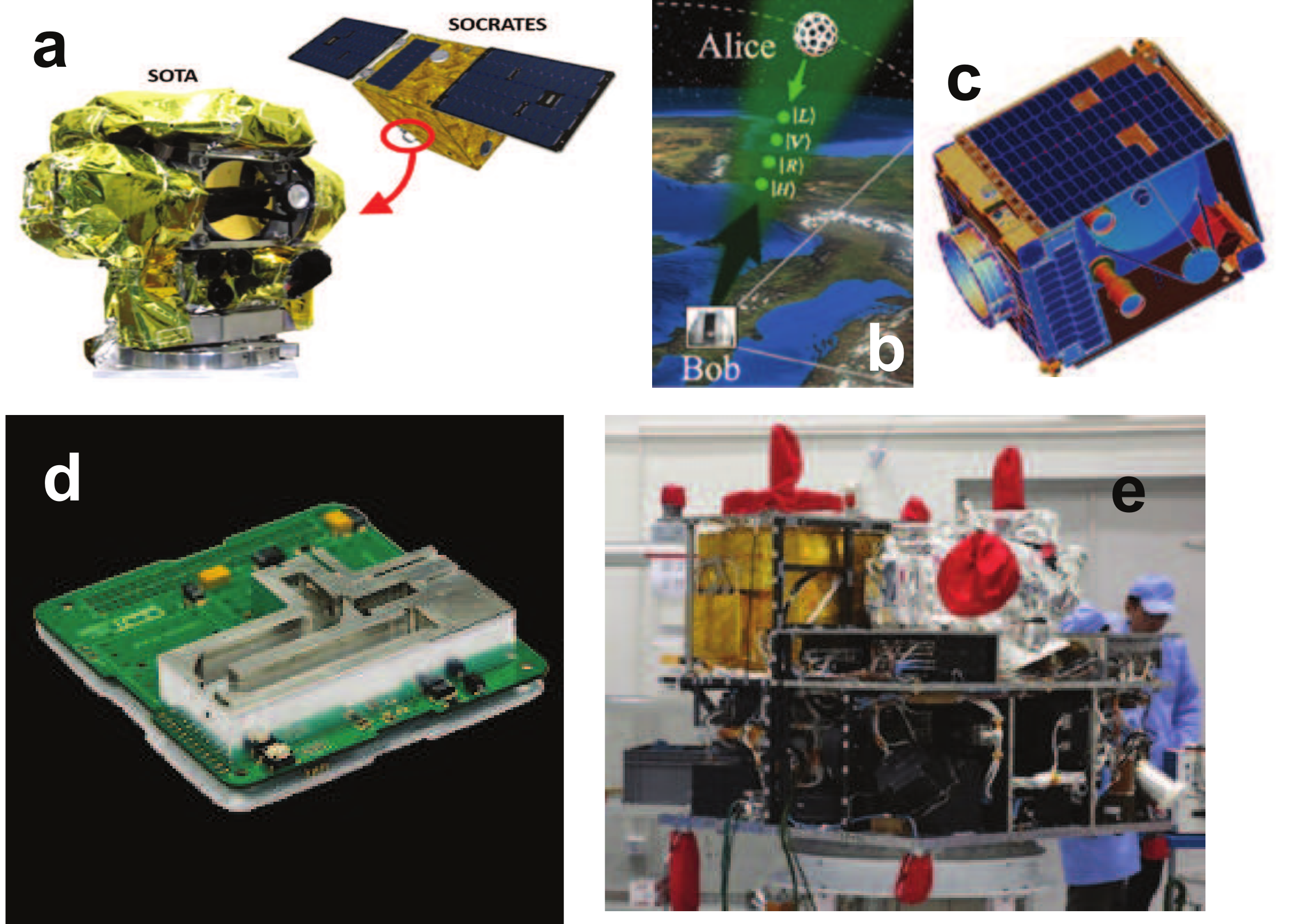}
\captionspacefig \caption{Satellites employing quantum technologies from various groups across the world: Japan \cite{bib:horiuchi2015view}, Italy \cite{bib:vallone15}, Canada \cite{bib:jennewein2014qeyssat}, and Singapore \cite{bib:tang2016generation}}
\label{fig:space_2}
\end{figure*}

%
% Future challenges
%

\subsection{Future challenges}

While early demonstrations of satellite-based quantum protocols have been successfully performed, future more sophisticated protocols will require more capable satellite hardware, which will inevitably begin to emerge in the immediate-term.

We identify two of the most pressing challenges facing a space-based quantum internet as:
\begin{itemize}
\item Coverage\index{Coverage}: how much of the surface of the globe will always be within line-of-sight of at least one satellite in the constellation network. Visibility of at least one satellite from any point on Earth will in-principle allow universal access to the quantum network, in the same way that the GPS constellation serves all points on the Earth's surface.
\item Precision\index{Precision}: many optical quantum protocols require strict synchronisation of photons undergoing interference. This becomes extremely challenging when those photons are originating from different satellite nodes, which must then be temporally synchronised on the scale of the photons' temporal wave-packets (i.e temporal mode-matching, discussed in Sec.~\ref{sec:MM_error}). Complicating things further, at satellite velocities relativistic effects become relevant to synchronisation. This is significant enough that the existing GPS\index{Global positioning system (GPS)} satellite positioning network must compensate for relativistic effects to maintain positioning accuracy.
\end{itemize}

%
% Coverage
%

\subsubsection{Coverage}\index{Coverage}

\begin{figure}[!htbp]
	\includegraphics[clip=true, width=0.475\textwidth]{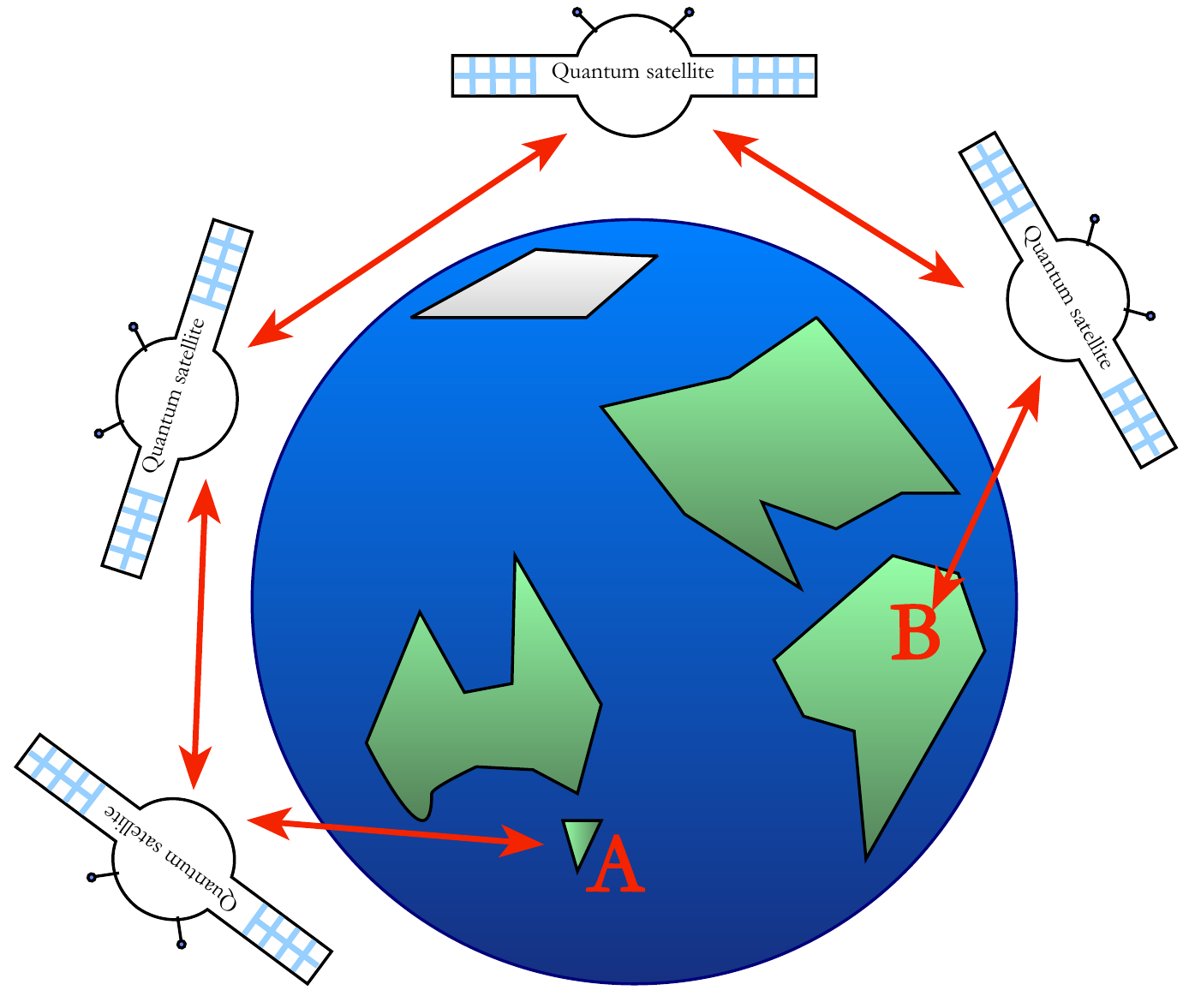}
	\captionspacefig \caption{Satellite constellation network for bypassing line-of-sight constraints in long-distance quantum communication. Neighbouring satellites connect via line-of-sight in space, and there are a sufficient number of satellites to enable global coverage such that all points on the Earth's surface are visible to at least one satellite at any given time, giving them indirect quantum access to any other point on the globe. In this simple example, entanglement swapping between satellites shares a Bell pair between Columbia and Tasmania, enabling the teleportation of certain Columbian products\index{Benzoylmethylecgonine} from Bogot{\'a} to the dance parties featuring at \href{https://soundcloud.com/peter-rohde}{DJ Rohde's} holiday home.\index{DJ Rohde}\index{Pablo Escobar}} \label{fig:sat_constellation}\index{Satellites!Constellations}
\end{figure}

Currently there are only two satellites with the capability of sending and/or receiving photons through space: the Chinese QUESS\index{QUESS}, and the Japanese SOCRATES\index{Space Optical Communications Research Advanced Technology Satellite (SOCRATES)} satellites. In order to achieve world-wide coverage multiple communicating satellites will be necessary such that there be a constellation of satellites\index{Satellites!Constellations} that are visible at all times, anywhere on Earth, in a similar manner to the GPS satellite network\index{Global positioning system (GPS)}, thereby bypassing line-of-sight constraints, as per Fig.~\ref{fig:sat_constellation}.

In the case of the QUESS satellite, the orbit is such that it passes over the observatories once per day, with a visibility window of $\sim 275$s. The coverage of a LEO satellite of elevation 500km has a maximum radius of $\sim 2,400$km from the point where the satellite is directly overhead, a limitation imposed by the curvature of the Earth. In practice this will be reduced, due to engineering imperfections and atmospheric influences, to distances on the order of $\sim 1,000$km (see Fig.~\ref{fig:space_4}). By placing satellites into higher orbits, coverage could be greatly increased, but this will introduce additional engineering challenges in establishing the more sophisticated satellite links, and effects such as dispersion which equates to loss. To this end, some of the main technologies that will need to be developed further are:
\begin{itemize}
\item Larger-sized telescopes with bigger apertures to increase collection efficiencies.
\item Better tracking systems\index{Tracking}, to maximise the overlap between detection apertures and photon spot-size.
\item Wave-front correction through adaptive optics \cite{bib:liao2017satellite}.
\end{itemize}

Currently, communications are performed at night due to the laser wavelengths that are employed, but plans for using telecommunication wavelengths pave the way for daytime quantum communication. For obvious reasons, this will be an essential ingredient in a global network of satellites forming a constellation capable of communication between opposing points on the Earth where night and day are reversed.

From the point of view of security, we anticipate that initial quantum networks for QKD will be based on trusted nodes\index{Trusted nodes}. Due to the inherent practical difficulty of tampering with space-based trusted nodes, it is likely this will offer a sufficient level of practical security for most realistic applications. Thus, using a constellation of satellites employing QKD can achieve long-distance cryptography by simply relaying the information, either by multiple satellite-to-ground\index{Satellites!Satellite-to-ground communication} or satellite-to-satellite\index{Satellites!Satellite-to-satellite communication} configurations. For ultimate security, not relying on the use of trusted nodes, one would employ alternative protocols that do not require line-of-sight quantum communication, such as the E91 protocol \cite{bib:PRL_67_661}\index{E91 protocol}. As this requires entanglement distribution, storage, and purification\index{Entanglement!Purification}, this would most likely be a second-generation technology after the trusted node QKD network is fully established and successfully demonstrated. 

%
% High-precision applications
%

\subsubsection{High-precision applications}\index{Precision}\index{High-precision applications}

For applications such as quantum clock synchronisation\index{Quantum clock synchronisation} (Sec.~\ref{sec:clock_sync}) and experiments testing fundamental physics at large length-scales and velocities, it is likely that extremely high fidelities of quantum operations will be required. For example, for clock synchronisation, even current atomic clocks\index{Atomic!Clocks} used as national standards have an accuracy at the level of 1 part in $10^{-14}$. Thus, such high-precision experiments will almost certainly follow the less demanding QKD experiments.

As the precision of technology improves, other sources of error may appear, which may require correction. It is well-known from existing GPS satellites\index{Global positioning system (GPS)}, that it is crucial to account for relativistic effects\index{Relativistic effects}, due to time-dilation\index{Time-dilation} and the gravitational red-shift\index{Gravitational red-shift}. Not accounting for these effects would seriously compromise the GPS system, with inaccuracies equating to an error on the order of 10km/day. This is due to the high velocities of LEO\index{Low Earth orbit} satellites, traveling at speeds of $\beta = 10^{-5}$ times the speed of light. Such relativistic effects can affect entanglement in the presence of diffracting photons \cite{bib:gingrich03}\index{Diffraction}. Even for single-photon transmission, polarisation-encoded photons can give rise to effects at the order of $\beta$ \cite{bib:byrnes2017lorentz}. Relativistically invariant entanglement distribution protocols\index{Relativistically invariant entanglement distribution} have been proposed to avoid such effects, which would otherwise require correction \cite{bib:yurtsever02, bib:li2003relativistic, bib:byrnes2017lorentz}.

\latinquote{Volens et potens.}

%
% Quantum Genetic Medicine
%

\section{Quantum genetic medicine}\index{Genetics!Medicine}\label{sec:genetic_medicine}

\famousquote{I have all these great genes, but they're recessive. That's the problem here.}{Bill Watterson}
\newline

\famousquote{The laws of genetics apply even if you refuse to learn them.}{Allison Plowden}
\newline

\dropcap{U}{ntil} now, medicine\index{Medicine} has largely relied on one-size-fits-all diagnosis and treatment options. Needless to say, everyone's different, manifest in their unique genotype\index{Genotype}. However, with access to individuals' unique genetic makeup, the next generation of medicine will become highly personalised, catering for individual genetic differences. People with different genetic predispositions, mutations, or traits will be able to have treatment options tailored to them.\index{Genetics!Mutations} Different cancer types, which are genetically distinct from one another, may be genetically targeted.\index{Cancer}

With the ability to sequence individuals' genomes extremely cheaply, we will open up entirely new medical possibilities for genetically-personalised treatments - the era of \textit{genetic medicine}, the next revolution in medicine.

Doing so will require highly complex processing of massive amounts of genetic and drug information, with demanding computational resource requirements. We foresee quantum computing as playing a central role in the processing pipeline of genetic drug development, some initial ideas for which we sketch here.

%
% The Human Genome Project
%

\subsection{The human genome project}\index{Genome}\index{Human genome project}

The first complete mapping of the human genome was a major scientific achievement \cite{10002010map}, opening up entirely new avenues for medical research\index{Medical!Research} that were previously never possible. It was, however, an extraordinarily expensive undertaking, costing on the order of \$1b.

Since then, genetic sequencing\index{Genetics!Sequencing} tools have undergone a massive technological transformation and are now available as commodity hardware at price-points accessible to any well-resourced bioscience lab. This now enables sequencing the human genome orders of magnitude cheaper than the first attempt. It is now possible to map a genome for $\sim$\$1,000's, and less sophisticated consumer-grade handheld devices are even becoming available.

Following its own technological Moore's Law\index{Moore's Law}, one can reasonably anticipate this process becoming sufficiently cheap that in the near future it will become economically viable (and desirable) for every individual to have their own personal genome fully sequenced and available for medical use.

\latinquote{Benedictus benedicat}.

%
% Short-Read Sequencing
%

\subsubsection{Short-read sequencing}

The major technological transition that has enabled this rapid progress is the adoption of next-generation sequencers (NGS)\index{Next-generation!Sequencers} based on \textit{short-read} technology\index{Short-reads}. Using this process, a DNA\index{DNA} sequence is not mapped exhaustively from beginning to end, but rather is chemically deconstructed into an enormous number of \textit{short-reads} - small genetic segments, each typically on the order of $\sim$50 base-pairs\index{Base-pairs} in length. Having prepared an enormous pool of such short-reads, they are then sequenced in parallel\index{Parallel!Sequencing}, yielding a large database of $N$ short strings, corresponding to small segments of the larger genome. The task then is to reconstruct a complete genome from this data.

Typically the end result is not a complete human genome from start to finish, but rather segments that capture regions of interest in the respective genome. In particular, certain genes or single-nucleotide polymorphisms (SNPs)\index{Single-nucleotide polymorphisms (SNPs)}\footnote{SNPs are positions in a genome whose nucleotides may differ between individuals within a species. Of course, the vast majority of the genetic makeup of individuals within a species is identical. Thus, the SNPs are sufficient to characterise the genetic differences between individuals and identify their genotype.} may be targeted as regions of interest during sequencing.

To achieve this, there are two approaches that are most commonly employed - \textit{de novo} assembly, and \textit{mapping}.

%
% De Novo Sequencing
%

\subsubsection{\textit{De novo} sequencing}\index{De novo sequencing}

In \textit{de novo} sequencing we treat the short-read data as a jigsaw puzzle that must be reassembled. We define an overlap threshold, $n$\index{Overlap threshold}, and upon comparing every string against every other, look at whether their ends overlap consistently by at least $n$ base-pairs. When a match is found, the respective short-reads are merged into a larger \textit{contig}\index{Contigs}. This process continues until no further sufficiently-overlapping strings are found. At this point we should, at least in principle, have a fully reassembled genome, or at least very large contigs belonging to it (assuming a sufficiently large pool of short-reads to begin with).

Open-source software for implementing \textit{de novo} assembly of short-read data is available. The well-known \textit{Velvet}\index{Velvet} package \cite{paszkiewicz2010novo}\index{Velvet} does this graph-theoretically using de Bruijn graph\index{de Bruijn graphs} \cite{compeau2011apply} representations for contigs/reads, with very efficient computational resource requirements.

The number of comparisons between short-reads scales only as $O(N^2)$, and employing hash table\index{Hash!Tables} representations for short-reads, the lookups for each individual comparison can be efficiently implemented in $O(1)$ time. The confidence that two contigs/reads actually overlap increases exponentially with $n$, but with increasing $n$ comes a reduction in the number of matches that will be identified - the tradeoff between \textit{sensitivity}\index{Sensitivity} and \textit{specificity}\index{Specificity} in contig reconstruction.

The \textit{de novo} approach is extremely powerful, as it does not require any genomic reference\index{Genome!Reference}. Rather, we can reconstruct a genome \textit{ab initio}, with sufficient read data.

\begin{figure}[!htbp]
\if 1\doublecol
	\includegraphics[clip=true, width=0.475\textwidth]{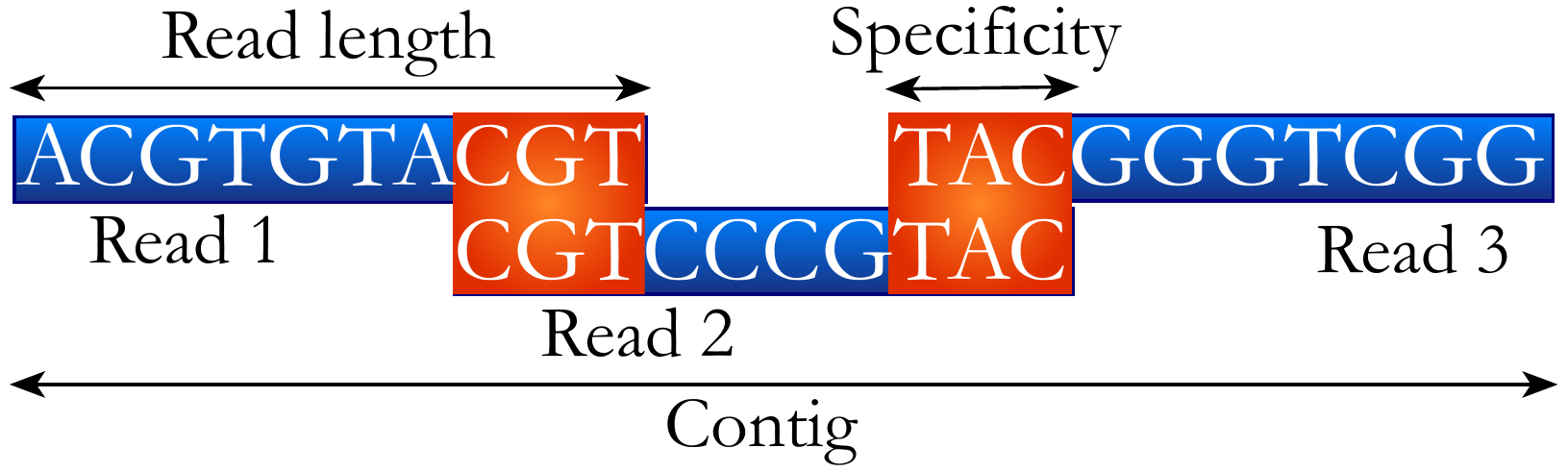}
\else
	\includegraphics[clip=true, width=0.7\textwidth]{genetic_sequencing_de_novo}
\fi
	\captionspacefig \caption{Three short-reads, each of 10 base-pairs in length (blue), which we wish to assemble into the largest possible contig\index{Contigs} belonging to the underlying genome. The specificity\index{Specificity} defines the number of endpoint base-pairs that must overlap for them to be considered a match (orange). In this case the specificity is 3, and 3 short-reads have been assembled into a single 24 base-pair contig. The greater the specificity the more accurate will be the matching, but the lower the probability of finding matches. A specificity too low will create false matches, whereas a specificity too high makes finding sufficiently overlapping short-reads unlikely. This makes optimum choice of specificity an important consideration, creating a direct tradeoff between reconstructed contig length and accuracy. This sequencing technique is best applied when there is no known reference genome for the data under analysis. For the technique to work effectively there must be a sufficient quantity of short-read data that with high probability overlaps between reads will exist. If the data is too sparse few overlaps are likely to be found. Thus, average reconstructed contig lengths grow with the amount of short-read data available to the engine.} \label{fig:gen_seq_de_novo}
\end{figure}

Fig.~\ref{fig:gen_seq_de_novo} provides a graphic example of \textit{de novo} sequencing.

%
% Mapping
%

\subsubsection{Mapping}\index{Mapping}

The failing of \textit{de novo} sequencing is that the pool of short-read data must be sufficiently large that all the pieces in the jigsaw puzzle are present, such that there are no gaps between neighbouring pieces, whereby contig `islands'\index{Contigs!Islands} remain isolated from other contigs. However, resources are of course always finite, as too is the number of short-reads available for reconstruction.

The other approach, \textit{mapping}, is to use an existing genome as a reference\index{Genome!Reference}. Rather than piece short-reads together, we compare them against this reference to infer their relative locations in the genome. When doing so, we allow some error threshold\index{Error!Threshold} in the matching. Specifically, we require the Hamming distance\index{Hamming distance} between a short-read and an equally-long segment of the reference be below some threshold. The flexibility offered by this threshold accommodates for mutational differences between the reference and the short-reads\index{Genetics!Mutations} (i.e the SNPs), which is ultimately what we wish to characterise.

To illustrate the power of this approach, consider the following. Only a minuscule fraction of the genome differs from one individual to the next - almost all of it is identical. Thus, mapping allows us to identify the mutational differences between individuals. Importantly, unlike \textit{de novo} sequencing, short-reads can be mapped to the reference even with incomplete data - contig `islands'\index{Contigs!Islands} needn't remain so. Non-overlapping short-reads can still be successfully mapped, making this approach applicable (but potentially incomplete) even with datasets insufficiently large for \textit{de novo} sequencing to be effective.

In the case of cancer diagnosis\index{Cancer}, for example, a cancer cell's genetic makeup is virtually identical to that of its host, modulo some small number of mutations that characterise the cancer. Using the mapping approach we can quickly identify these mutations, hence understanding the genetics of the underlying cancer, potentially opening opportunities for genetically-targeted treatment.

As with \textit{de novo} sequencing, open-source software for efficiently implementing mapping using commodity hardware is available.
 % \cite{Bfast etc}.

\begin{figure}[!htbp]
\if 1\doublecol
	\includegraphics[clip=true, width=0.475\textwidth]{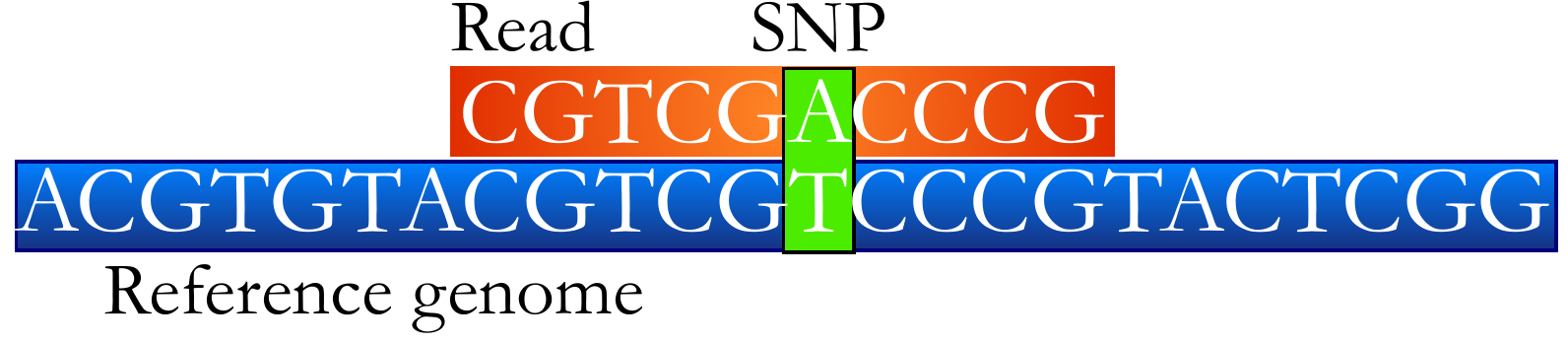}
\else
	\includegraphics[clip=true, width=0.7\textwidth]{genetic_sequencing_mapping}
\fi
	\captionspacefig \caption{In genome mapping we begin with a known reference genome (blue) to act as a comparison for short-reads (orange). The short-reads are slid along the reference until a location with sufficient overlap is found, at which point we record a match. By allowing some degree of single nucleotide mismatches (green) between the reference and short-read, we are able to characterise the SNPs in the short-read data. This sequencing technique is appropriate when we have a known genome for the species to which the short-read data belongs, but we wish to identify minor genetic mutations between the samples. For example, mapping short-reads belonging to human cancer cells to a human reference genome characterises the genetic mutations associated with that particular cancer. The technique works with any amount of short-read data, even just a single read, provided the reference is known.} \label{fig:gen_seq_mapping}
\end{figure}

Fig.~\ref{fig:gen_seq_mapping} provides a graphic example of mapping in the presence of a SNP.

%
% Genetic Medicine
%

\subsection{Genetic medicine}\index{Genetics!Medicine}

At its simplest, the design of genetically-tailored medical treatments will require understanding the interactions between drug compounds\index{Drug!Compounds} and genetic processes at the molecular level. For example, we might desire that a drug modifies gene expression\index{Gene expression} for specific genes, or the transcription\index{Transcription} of DNA\index{DNA} to specific proteins\index{Proteins}. In the case of cancer treatment, we may wish to inhibit the function of biochemical processes\index{Biochemical processes} reliant on genes exhibiting particular cancerous mutations\index{Genetics!Mutations}, whilst not affecting healthy, unmutated ones.

The drug design process for conventional medicines is an incredibly tedious one, requiring massive latitudinal and longitudinal studies on the effects of drugs on physiological symptoms. It is not uncommon for the entire drug development process, from design, to testing, to approval, to cost on the order of $\sim$\$100m's for a single drug. However, in the case of individually tailored medicine such studies are clearly not possible. This will require replacing human studies with accurate biomolecular-level simulations of these chemical processes. At this scale, interactions are inevitably quantum mechanical in nature, which must be accommodated for in simulations.

The key goal is to properly simulate the interaction of candidate drug compounds\index{Drug!Compounds}, ligands\index{Ligands}, or functional groups\index{Functional groups} with biochemical processes at the molecular level. Classical techniques for such simulations include estimating electron densities using density functional theory\index{Electron density}\index{Density functional theory} (the potential energy surface)\index{Electron density} of compounds, from which useful properties such as bonding affinities\index{Bonding affinity} may be estimated.

%
% Quantum Chemistry
%

\subsection{Quantum chemistry}\index{Quantum chemistry}

Despite the existence of classical estimation techniques, molecular-level interactions are necessarily quantum mechanical in nature. For this reason, classical simulation techniques are limited to approximations that ignore realistic and important quantum effects, since this would require exponential classical computational resources in general (i.e in general, simulating quantum systems is \textbf{BQP}-complete).

Using standard quantum chemistry techniques, one can construct accurate Hamiltonians\index{Molecular Hamiltonians}\index{Hamiltonians} describing the evolution of molecular systems at the quantum level. Efficiently implementing such simulations can then be performed on quantum computers using standard, efficient Hamiltonian simulation quantum algorithms (Sec.~\ref{sec:quantum_sim_alg})\index{Quantum simulation} for simulating the Schr\"odinger equation\index{Schr{\" o}dinger!Equation},
\begin{align}
	\hat{H}\ket{\psi(t)} = i\hbar \frac{\partial}{\partial t}\ket{\psi(t)},
\end{align}
where $\hat{H}$ is the system's Hamiltonian\index{Hamiltonians}, describing the time-dependent evolution of the state as,
\begin{align}
	\ket{\psi(t)} = e^{-\frac{i\hat{H}t}{\hbar}} \ket{\psi(0)}.
\end{align}

These Hamiltonians can be constructed \textit{ab initio}, combining relevant one- and two-body interactions\index{One-body interactions}\index{Two-body interactions}, such as Coulomb potential\index{Coulomb!Potential}, kinetic energy\index{Kinetic energy}, spin- and spin-orbit\index{Spin coupling}\index{Spin-orbit coupling} coupling, and magnetic and electric dipole terms appropriately. A net interaction Hamiltonian may be obtained by summing over the different one- ($i$) and two-body (\mbox{$j,k$}) terms for all particles comprising the system,
\begin{align}
	\hat{H}_\mathrm{int} = \sum_i \hat{H}_i + \sum_{j>k} \hat{H}_{j,k}.
\end{align}
This structure is of the form of Eq.~(\ref{eq:Ham_sim_Ham}) and thus lends itself to the quantum Hamiltonian simulation algorithm\index{Quantum simulation} (Sec.~\ref{sec:quantum_sim_alg}).

The dominant one-body interactions are kinetic energy\index{Kinetic energy!Hamiltonian} terms, of the form,
\begin{align}
\hat{H}_i^{(\mathrm{kinetic})} &= \frac{\hat{p}^2}{2m} \nonumber \\
&= -\frac{\hbar^2}{2m_i}\nabla^2_{\vec{r}_i},
\end{align}
where $\hat{p}$ is the momentum operator, $m_i$ is mass, $\vec{r}_i$ is the position vector, and,
\begin{align}
\nabla^2 = \frac{\partial^2}{\partial x^2} + \frac{\partial^2}{\partial y^2} + \frac{\partial^2}{\partial z^2}.
\end{align}
The dominant two-body terms are Coulomb (electrostatic) interactions\index{Coulomb!Interaction},
\begin{align}
\hat{H}_{j,k}^{(\mathrm{Coulomb})} = \frac{q_j q_k}{4\pi\epsilon_0|\vec{r}_j-\vec{r}_k|},
\end{align}
where $q_j$ denotes charge.

%
% Quantum Drug Trial Simulation
%

\subsection{Quantum drug trial simulations}\index{Quantum drug trial simulations}

These approaches borrowed from quantum chemistry allow simulation of the interaction between drug compounds and molecular biochemical interactions. The next step is to perform this process against a huge library of candidate drug compounds\index{Drug!Compounds!Libraries}, ligands or functional groups, from which, we hope, some candidates will exhibit desired characteristics, such as strong bonding affinities\index{Bonding affinity}.

To achieve this, let us construct a classical algorithm for exhaustively constructing and enumerating organic drug molecules or functional groups, up to some cutoff size\index{Cutoff size}. The length of this list grows exponentially with the cutoff size. Next we implement this algorithm unitarily (which is always possible, writing the classical circuit as a reversible one), and construct it as a quantum oracle\index{Oracles} such that every input state (which acts as a pointer\index{Pointers} to an element in the list) yields as output a qubit representation\index{Qubit molecule representation} of the corresponding drug compound. Specifically, the oracle implements a transformation of the form,
\begin{align}
\hat{U}_\mathrm{oracle}\ket{i}_\mathrm{pointer}\ket{0}_\mathrm{drug}^{\otimes n} \to \ket{i}_\mathrm{pointer}\ket{\psi_i}_\mathrm{drug}^{(n)},
\end{align}
where $i$ is the index (pointer) to the drug element in the enumeration, and $n$ is the number of qubits encoding the representation of the drug molecule\footnote{This may be implemented unitarily, which is easy to see via its preservation of orthonormality.} $\ket{\psi_i}_\mathrm{drug}^{(n)}$.

This oracle may be constructed by writing a classical algorithm that generates a search tree\index{Search tree} for exhaustively enumerating physically allowed drug compounds via brute-force. Writing this as a reversible quantum circuit then yields the oracle, where the output pointer register is entangled with the output drug description register, uniquely mapping them to one another.

Having constructed this oracle, we feed the output qubit molecule description into the Hamiltonian simulation algorithm\index{Quantum simulation}, with exponential speedup, yielding as output a `score' characterising some desired property of the interaction, such as bonding affinity\index{Bonding affinity}. This implements the simulate-and-score operator\index{Simulate-and-score operator},
\begin{align}
\hat{U}_\mathrm{sim} \ket{\psi_i}_\mathrm{drug}\ket{0}_\mathrm{score} \to \ket{\psi_i}_\mathrm{drug}\ket{s_i}_\mathrm{score}	
\end{align}

The joint oracle$\to$simulation algorithm is finally embedded within a quantum search subroutine\index{Grover's algorithm}, which searches over the input space enumerating the set of drug candidates for scores above some desired threshold. This yields a quadratic enhancement in the number of drug candidates that can be simulated in parallel. Successful search results, where a compound achieves the threshold score, are then tagged for further investigation as treatment drug candidates.

Combining the Grover search and quantum simulation subroutines, the total time-complexity of the entire (unencrypted) algorithmic pipeline is,
\begin{align}
	O\left(\frac{N\sqrt{M}t^2}{\epsilon}\right),
\end{align}
where $t$ is simulation time, $\epsilon$ is the simulation accuracy, $N$ is the number of local terms in the global Hamiltonian, and $M$ is the number of drug compound candidates in the algorithmic drug search space.

\begin{figure}[!htbp]
\if 1\doublecol
\includegraphics[clip=true, width=0.475\textwidth]{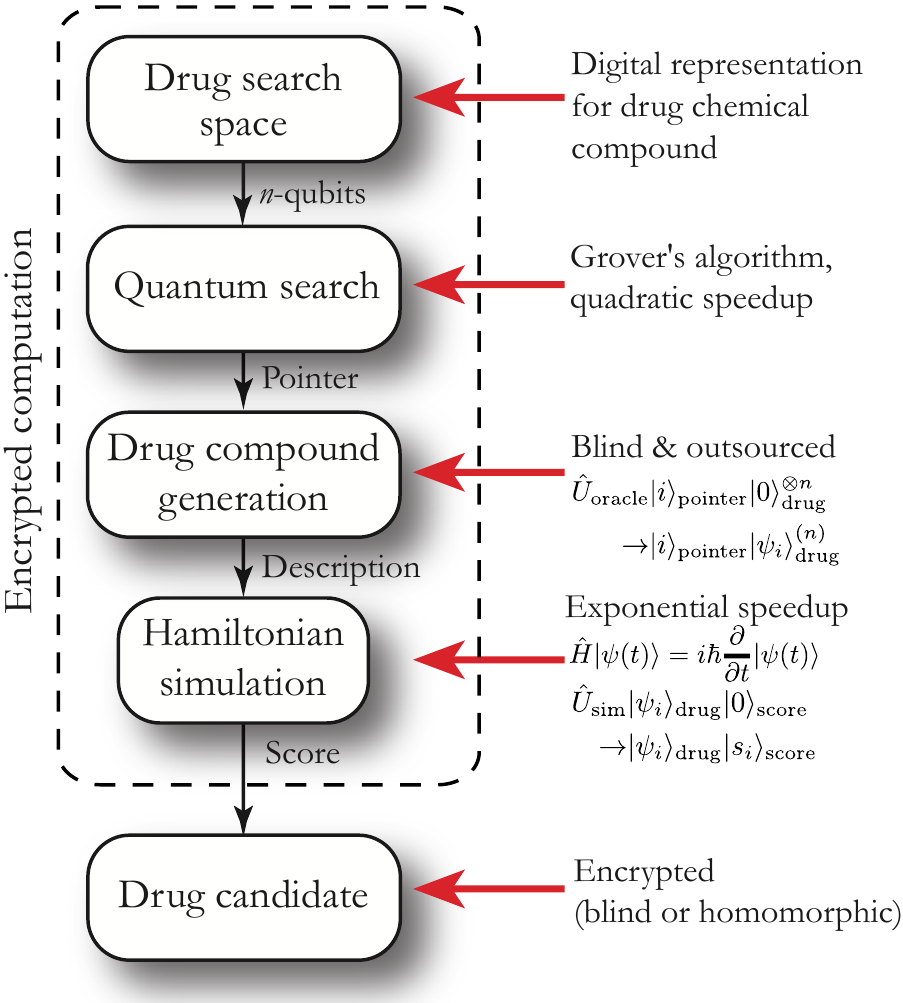}
\else
\includegraphics[clip=true, width=0.7\textwidth]{genetic_medicine_pipeline}
\fi
	\captionspacefig \caption{Pipeline for quantum-enhanced genetic drug development. A quantum search algorithm searches over algorithmically-generated drug compounds or functional groups, for each performing Hamiltonian simulation, to determine if its score is above a threshold. The quantum search subroutine yields a quadratic enhancement in the number of drugs that can be simulated in parallel, and the Hamiltonian simulation subroutine yields an exponential enhancement over classical simulation techniques.} \label{fig:genetic_med_pipe}
\end{figure}

The complete algorithmic pipeline is shown in Fig.~\ref{fig:genetic_med_pipe}.

%
% Cyborgs - The Era of Genetic Editing
%

\subsection{Cyborgs - The era of genetic editing}\index{Cyborgs}\index{Genetics!Editing}

In very recent years, new technologies have emerged for performing highly targeted gene editing, allowing genetic manipulations to be performed on living organisms. Using CRISPR\index{CRISPR} technology, for example, one can enable and disable entire functional genes, or perform even more intricate manipulations, right down to the single nucleotide level.

This exquisite control over the genomes of living organisms, something that would never have been dreamed possible only years ago, opens up our ability to enhance or suppress genetic characteristics of species, potentially facilitating everything from the development of highly improved agricultural crops, all the way to more ethically questionable motives such as genetic enhancement of the human race.

Using the genetic medicine pipeline discussed previously one might envisage even more ambitious goals in the era of genetic editing. Rather than test algorithmically-generated databases of drug compounds for their genetic effects, one might instead trial genetic variations for their outcomes at a molecular biological level. This might provide guidance and insight into genomic variations worth experimenting with for desired practical outcomes.

%
% In the Cloud
%

\subsection{In the cloud}\index{Cloud quantum computing}

The quantum genetic medicine processing pipeline lends itself well to several important cloud-based protocols. The pipeline comprises several distinct subroutines\index{Subroutines}, which might be outsourced or distributed independently of one another\index{Outsourced!Quantum computation}\index{Distributed quantum computation}.

An R\&D organisation might specialise in the algorithmic generation\index{Algorithmic generation} (or hard-coding\index{Hard-coding}) of candidate drug compounds\index{Compounds}, which they would like to retain as a trade secret\index{Trade secrets}, but desire to license\index{Outsourced!Quantum computation} to third-party drug designers\index{Drug!Design}, who wish to employ them in their computational pipeline\index{Quantum drug trial simulations}. The quantum search\index{Grover's algorithm} and Hamiltonian simulation\index{Quantum simulation} subroutines might similarly be outsourced or distributed over the cloud to vendors specialising in the implementation of these particular subroutines.

In an era where the genetic composition of every individual is fully characterised, data security will be of utmost importance - medical confidentiality\index{Medical!Confidentiality} will be lifted to an entirely new plane when people's genes are at stake, as the nefarious uses (and misuses\index{Nefarious uses}) for obtaining other people's genomes are immense. This provides a perfect example for the value of encrypted quantum computation\index{Encrypted quantum computation} (Sec.~\ref{sec:homo_blind}). If a medical lab is outsourcing some aspects of a computation involving clients' genetics, it is paramount that this be obscured from third-parties performing the computations in the interest of medical confidentiality. Computational efficiency aside, this consideration on its own already justifies implementing such simulation pipelines quantum mechanically, as encrypted classical computation is effectively unviable in general.
\latinquote{Annus mirabilis.}

%
% The QuantumMind - Quantum Machine Learning
%

\section{Quantum machine learning}\label{sec:quantum_mind}\index{Quantum machine learning}\index{Machine learning}

\famousquote{If computers get too powerful, we can organise them into committees. That'll do them in.}{Unknown}
\newline

\famousquote{My unshakeable belief in the stupidity of the animal man has never disappointed me and has often helped me throughout my life.}{Arthur Schopenhauer}

\sectionby{Nana Liu}\index{Nana Liu}
  
% \comment{Tidying, editing, proofing, structure. Section titles.}\\

\dropcap{I}{n} recent years, the implementation of artificial intelligence via machine learning techniques has become a forefront area of research, transforming many software industries and those that depend upon them. An obvious question to ask is to what extent machine learning techniques translate across to the quantum computing environment, and whether upon doing so there is an associated quantum enhancement. While this field of \textit{quantum machine learning} (QML) is far too extensive to summarise here, we will instead present a brief overview of how machine learning might apply in a networked quantum environment.

There are two distinct avenues to consider here. First, how does machine learning influence the operation of quantum networks? Second, how do quantum networks affect the implementation of machine learning? We will outline some of the key considerations here from both these opposing angles of consideration.

\subsection{Overview}

Today, classical machine learning is affecting the understanding and better regulation of the classical networking infrastructure underpinning the modern internet. This includes network pattern recognition, security and fault management, routing and traffic management, resource management, and distributed computation. In these applications, it may not solely be processing power that is of importance, but additionally reliability and security may also be paramount.

The large intersection between machine learning and network systems is perhaps unsurprising. Firstly, machine learning relies on access to data, and in many real-world scenarios, data naturally emerges from distributed sources. Secondly, especially for complex systems like large networks, the information to process is complex, containing many uncertainties, and subject to errors. Exactly solvable models in these regimes are rare, a scenario where estimation techniques based on machine learning are often helpful.

In the coming quantum era we can envision three distinct ways where quantum resources might be introduced: quantum communication; quantum processing at individual nodes; and, data that is inherently quantum in nature. To begin, we first make a classification of the four foreseeable network types (summarised in Tab.~\ref{tab:CommTable}). These can be classified as:

\begin{itemize}
\item CC: classical data and processing over a classical network (e.g the present-day internet).
\item CQ: classical data and processing over a quantum network.
\item QC: quantum data and processing over a classical network.
\item QQ: quantum data and processing over a quantum network -- a fully-quantum internet.
\end{itemize}

\startnormtable
\begin{table*}[htbp!]
\begin{tabular}{|l|l|l|l|}
\hline
Data/Network & Classical & Quantum \\
\hline
\hline
Classical & Current internet (CC) & Classical data in quantum network (CQ) \\
Quantum & Quantum data in classical network (QC) & Fully quantum internet (QQ)\\
\hline
\end{tabular}
\captionspacetab \caption{\label{tab:CommTable}Classical and quantum data in a network.}
\end{table*}

Now the inevitable question arises: how do these different scenarios relate to machine learning?

This is not yet an active research area in its own right. However, there are some preliminary toolkits that are starting to be developed in the new field of QML. By first summarising the intersection between the classical internet and classical machine learning, we gain insight into the kinds of tools required to start examining their quantum counterparts. For instance, we will see how aspects of quantum processing of quantum data might be aided by machine learning, how machine learning may be enhanced by quantum resources, and how machine learning may be implemented in these distributed quantum settings. A summary is presented in Tab.~\ref{tab:CommTable2}. 

\begin{table*}[htbp!]
\begin{tabular}{|l|l|l|l|}
\hline
Network  & Classical data & Quantum resources  & Quantum data \\
concerns & & and classical data & \\
\hline
\hline
Individual & Classical   & Quantum-enhanced  & Quantum learning  \\ 
computing & machine learning & machine learning & \\
\hline 
Security and faults & Machine learning for anomalies & Adversarial quantum & Anomaly detection and  \\
& and faults: detection and prediction; & machine learning  & change point detection \\
& Adversarial machine learning &  & for quantum data \\
\hline 
Routing and traffic & Machine learning for  & Open problems & Open problems \\
& traffic prediction, classification, &   &  \\
& congestion control and routing & & \\
\hline 
Distributed  & Distributed machine learning & Distributed quantum & Distributed quantum learning \\
computing & & machine learning & \\
\hline 
 Communication & Data compression & Open problems & Data compression and \\
 & and machine learning & & quantum machine learning \\
\hline 
\end{tabular}
\captionspacetab \caption{\label{tab:CommTable2} Classical and quantum machine learning applications in classical and quantum networks. Almost all of the categories here are very new and open to exploration in the quantum domain.}
\end{table*}
\startalgtable

\subsection{Classical machine learning in classical networks}\index{Classical machine learning!Classical networks}

To create and maintain efficient classical networks, one requires efficient and reliable processing at individual nodes, security, efficient routing and data transmission, efficient use of resources, and a means for distributed processing \cite{bib:boutaba2018comprehensive, bib:wang2018machine}. We now present a brief overview of how machine learning techniques apply in these areas. 

\subsubsection{Machine learning basics}

Machine learning algorithms allow us to make predictions about a current or future dataset without requiring explicit instruction on how to do so. Since the aim is directed more towards \textit{prediction}\index{Prediction} than purely \textit{estimation}\index{Estimation}, it differs from the field of statistical estimation\index{Statistical estimation}, although they share many techniques.

There are three main paradigms for machine learning:
\begin{itemize}
	\item Supervised learning\index{Supervised learning}: relies upon training data\index{Training data} from which inferences and predictions about new test data can be extracted.
	\item Unsupervised learning\index{Unsupervised learning}: makes inferences from the data at hand without training.
	\item Reinforcement learning\index{Reinforcement learning}: operates using a different framework, and aims to find the best action to take to maximise a given reward in a particular environment.
\end{itemize}

Machine learning is used regularly for data collection, feature engineering, and model learning. There are many excellent introductory texts on this topic \cite{bib:bishop2006pattern, bib:shalev2014understanding, bib:trevor2009elements, bib:marsland2011machine, bib:flach2012machine}.

\subsubsection{Security \& fault management}\index{Security}\index{Fault management}

There are two primary ways in which machine learning applies to managing security and faults in networks. The first is \textit{using} machine learning techniques to predict and detect security breaches and faults in the network, including anomaly detection. The second is in studying the security vulnerabilities of machine learning algorithms themselves, as the presence of adversaries is natural in real-world networks -- so-called `adversarial machine learning'\index{Adversarial machine learning}.

\paragraph{Anomaly detection \& fault management}\index{Anomaly detection}\index{Fault management}

When there are security breaches in a classical network, one desires the ability to predict and detect them, as well as a method for making protocols more robust against them. Machine learning is often used in anomaly and intrusion detection. These algorithms seek out unusual data or changes within it.

Broadly, there are three classes of anomalies: point, contextual, and collective, which refer respectively to single datum anomalies, unusual data with respect to a specified context, and clusters of data which suggest unusual behaviour. Both supervised and unsupervised algorithms are employed in these settings \cite{bib:thottan2003anomaly, bib:ahmed2007machine}. One of the prime challenges here is determining the presence of anomalies when limited data is available, and the associated rates of false identifications.

Fault management in a network is also extremely relevant, especially for complex networks more exposed to errors. We desire the prediction, detection and localisation of faults. Most applicable machine learning methods here employ supervised algorithms. However, the paucity of real training data (as opposed to synthetic data generated via simulation) means that algorithms might be poorly trained, especially in newly established networks \cite{bib:hood1997proactive, bib:kogeda2006prediction, bib:snow2005assessing}. This is to be reasonably anticipated with the deployment of a future quantum internet. To accommodate for this, new methods have arisen where unsupervised machine learning techniques are used instead to detect changes in the network rather than relying on labelled fault data \cite{bib:hajji2005statistical}.

In particular, to identify and localise unusual network behaviour, either due to natural faults or adversaries, network anomaly detection methods can be employed \cite{bib:ahmed2007machine, bib:fraley2017promise, bib:joseph2013machine}. Since results can be sensitive to the employed training data, it is important to examine which datasets are most appropriate for a given application \cite{bib:yavanoglu2017review}. Particularly, there have been many proposals for utilising anomaly detection in network intrusion. However, this approach has been criticised for its use in real-world scenarios, where it's often difficult to distinguish anomalies related to intrusions from those attributed to other factors, and the complexity of real-world networks may make it too difficult to define what even constitutes a `normal' signal \cite{bib:sommer2010outside}.

\paragraph{Adversarial machine learning}\index{Adversarial machine learning}

Machine learning algorithms themselves exhibit security vulnerabilities \cite{bib:huang2011adversarial}. There are two main types of attacks to which they are vulnerable:
\begin{itemize}
\item Evasion\index{Evasion}: directed at the test data.
\item Poisoning\index{Poisoning}: directed at the training data and machine learning models.
\end{itemize}

In real-world scenarios, data often originates from different sources, making adversarial attacks more likely. It has been discovered that many machine learning algorithms are in fact vulnerable to adversarial attacks, the first discovered in \cite{bib:szegedy2013intriguing}. A large proportion of the literature focuses on the details of specific algorithms: the detection of adversaries; their different methods of attack; and, the particular defences against them \cite{bib:kurakin2018adversarial}. However, recently, more foundational work has emerged, explaining the origins of this vulnerability as arising from the high dimensionality of the underlying data \cite{bib:goodfellow2014explaining, bib:gilmer2018adversarial, bib:mahloujifar2018curse}.

\subsubsection{Traffic management \& routing }\index{Traffic management}\index{Routing}\index{Congestion control}

The effective operation of large-scale networks requires automated management protocols. This includes efficient means for traffic prediction, traffic classification, routing, and congestion control\index{Congestion control}. Machine learning algorithms have been developed for all of these.

Predicting network traffic is becoming increasingly important, especially in diverse and complex networks. This is commonly addressed using time-series forecasting\index{Time-series forecasting} (TSF) methods. This can make use of either statistical analysis techniques, or supervised machine learning methods \cite{bib:bermolen2009support, bib:chabaa2010identification, bib:cortez2006internet}. Non-TSF methods also exist \cite{bib:chen2016predicting, bib:li2016inter}.

The most commonly used technique for traffic classification is the so-called flow feature-based technique\index{Flow feature}. This takes into account information about unidirectional packet transmissions. Here, supervised machine learning techniques have been found to be accurate. However, unsupervised techniques have been found to be more robust. Their joint application is a very powerful tool \cite{bib:erman2007offline, bib:zhang2015robust}.

Machine learning is most applicable to dynamic routing problems, requiring rapid updating of optimal routes. Since such settings require frequent reevaluation, reinforcement learning algorithms are most appropriate. In particular, Q-learning\index{Q-learning} has performed well in various networks \cite{bib:wang2006adaptive, bib:forster2007froms, arroyo2007q}.

Network congestion control is important to ensure stability and the minimisation of packet loss. Well-known congestion control methods like queue management already exist. However, machine learning can be used to enhance the effectiveness of congestion control in various scenarios, especially for packet-based TCP/IP networks \cite{bib:liu2002end, bib:barman2004model, bib:el2005improving}. 

\subsubsection{Distributed machine learning}\index{Distributed machine learning}

Distributed machine learning is simply the fusion of distributed computation with machine learning, where the learning algorithm is distributed across a network. This becomes highly relevant in several notable scenarios:
\begin{itemize}
\item Training and/or testing data originates from different sources. This is the naturally distributed setting.
\item There is too much data to store locally on a single device.
\item When fault-tolerance becomes important (e.g for high-value data), decentralised storage provides enhanced data integrity.
\end{itemize}

The toolbox and infrastructure for distributed machine learning is rapidly developing, and there are many known algorithms \cite{bib:peteiro2013survey, bib:florian2013}. Existing platforms catering for distributed machine learning include MLbase\index{MLbase} \cite{bib:MLbase}, Hadoop\index{Hadoop} \cite{bib:white2012hadoop}, and Spark\index{Spark} \cite{bib:shanahan2015large}.

Caution is required, however, as there are cases when one \textit{shouldn't} employ distributed machine learning, such as when:
\begin{itemize}
\item Communication and synchronisation between distributed parties presents a bottleneck for computation.
\item Developing and executing distributed software is too complicated.
\item One can run the same algorithm on a multi-core machine. This is possible with smart data-sampling, offline schemes, and efficient parallel codes.
\end{itemize}

\subsection{Machine learning on classical data with quantum resources}

There are at least three broad ways in which we can employ quantum resources for classical data over a network, specifically they:
\begin{itemize}
\item Enhance data-processing at individual nodes.
\item Improve security.
\item Enhance communication.
\end{itemize}

In a classical network, the first question is whether or not quantum resources can assist in any of the relevant algorithms. These belong to the class of quantum-enhanced machine learning algorithms.

In a quantum network with only classical data, communication complexity\index{Communication complexity} improvements are possible \cite{bib:brassard2003quantum}. It's unclear whether machine learning has utility in this setting, although there are some promising hints in this direction \cite{bib:kane2017communication, bib:balcan2012distributed, bib:conitzer2004communication}.

\subsubsection{Quantum machine learning overview}\index{Quantum machine learning}

Quantum(-enhanced) machine learning (QML) algorithms are quantum algorithms performing machine learning tasks, exhibiting super-classical enhancements. They have so far mostly concentrated on quantum speed-ups with respect to the dimensionality of the underlying data.

\paragraph{Fully-quantum algorithms}

The first of these algorithms relied on fully quantum devices, maintaining coherence throughout computation, requiring full fault-tolerance\index{Fault-tolerance}. For supervised learning algorithms claiming exponential quantum enhancement \cite{bib:biamonte2017quantum, bib:ciliberto2018quantum}, the HHL algorithm \cite{bib:harrow2009quantum}\index{HHL algorithm} for matrix inversion\index{Matrix inversion} is often employed. However, HHL exhibits a number of shortcomings, making it impractical for near-term quantum devices:
\begin{itemize}
\item The ability to efficiently encode classical data into quantum states and memory \cite{bib:aaronson2015read}.
\item Effective quantum state read-out \cite{bib:aaronson2015read}.
\item They generally require high circuit-depth.
\item There are restrictions on the sparsity and conditioning of the matrices to which the algorithm is applied.
\end{itemize}

Although subsequent developments have attempted to circumvent sparsity restrictions, and rather focus on low-rank matrices (e.g quantum principal component analysis for low-rank matrices \cite{bib:lloyd2014quantum}), recent work on quantum-inspired classical algorithms\index{Quantum-inspired classical algorithms} has demonstrated that previously undiscovered, efficient classical algorithms can exist \cite{bib:tang2018quantum, bib:gilyen2018quantum, bib:chia2018quantum}. In fact, classical sampling methods \cite{bib:tang2018quantum} for quantum-inspired machine learning algorithms suggest that classical methods for linear algebra problems in low-dimensions are likely to have efficient classical algorithms. Although these classical sampling methods are not yet more practical than existing classical sampling methods, they are still more practical than their quantum counterparts.

Another set of approaches, relying on amplitude amplification\index{Amplitude amplification} and Grover's search algorithm\index{Grover's algorithm}, can provide up to quadratic runtime enhancement. These include quantum algorithms for reinforcement learning \cite{bib:dunjko2016quantum}, and training of quantum perceptrons \cite{bib:kapoor2016quantum}\index{Perceptrons}. While theoretically appealing as long-term objectives, viable near-term proposals are absent.

\paragraph{Hybrid algorithms}\index{Hybrid algorithms}

To find algorithms practically viable in the the near future, research is increasingly devoting its attention to hybrid classical-quantum algorithms. These algorithms, which include variational methods for optimisation \cite{bib:moll2018quantum}, exhibit low circuit-depth, where the optimisation process is performed iteratively and classically. There are roughly two varieties: one that attempts to enhance classical algorithms with classical input data; and another where the quantum advantage lies in efficient quantum state preparation, thus relying on quantum input data. Prominent examples of the former include quantum approximate optimisation algorithms\index{Quantum approximate optimisation algorithms} (QAOA) \cite{bib:farhi2014quantum, bib:farhi2016quantum}, and the latter includes variational quantum eigensolvers (VQE) \cite{mcclean2016theory, bib:kandala2017hardware}, which we return to in the subsequent section.

Both QAOA and VQE can be considered as belonging to the same broader framework, and their optimisation component (which may be considered only as a component, not the entirety of machine learning) is performed classically. One begins with an ansatz quantum state. A unitary operation with classically-tuneable parameters is then applied to this state, and an observable whose expectation value represents the problem's cost function\index{Cost function} is subsequently measured. The classical parameters of the unitary are then iteratively adjusted until cost function minimum is reached (i.e a Hamiltonian ground state), for instance using the classical gradient-descent algorithm\index{Gradient-descent algorithm}.

In QAOA, the respective ground state encodes the classical solution to a classical optimisation problem, like \textsc{MaxCut}\index{MaxCut}, exhibiting efficient polynomial runtime. Thus, it is not a quantum-enhanced algorithm for a classical machine learning problem, but rather exploits a classical machine learning algorithm. It remains to be seen if optimisation problems more directly relevant to networking applications can be solved in this way.

Alternate frameworks have been developed to find quantum-enhanced algorithms that not only take advantage of classical optimisation algorithms, but also enhance classical machine learning algorithms. These new proposals include quantum circuit learning \cite{bib:mitarai2018quantum}\index{Quantum circuit learning}, quantum generalisations of neural networks \cite{wan2017quantum}\index{Quantum neural networks}, and Born machines \cite{bib:cheng2018information, bib:benedetti2018generative}\index{Born machines}. Theoretical demonstration of quantum enhancement in such settings remains an important open problem.

\subsubsection{Security \& other applications}\index{Security}

\paragraph{Anomaly detection}\index{Anomaly detection}

The chief machine learning method for detecting and averting faults and security breaches in classical networks is in anomaly detection. However, for anomalies in classical data, it appears unlikely that currently available QML algorithms can enhance detection speed or reliability. One of the primary reasons is the necessity for encoding classical data into quantum states, which can be very costly \cite{bib:aaronson2015read}. Thus, even if there are QML algorithms for anomaly detection in the computational stage of processing, state preparation and readout overheads may be prohibitive. However, this is no longer the case if we instead begin with quantum data, to be discussed in Sec.~\ref{sec:ml_quantum_data}.

\paragraph{Adversarial quantum machine learning}\index{Adversarial machine learning!Quantum}

Just as classical machine learning algorithms are vulnerable to attacks, so is QML. This is a very new field, known as \textit{adversarial QML}. As with adversarial machine learning, the aim is to find more robust QML algorithms, and some robust algorithms have indeed been proposed \cite{bib:wiebe2018hardening}. In addition to finding more robust algorithms, it's also important to understand the respective limitations on robustness, currently an open problem. A recent result suggests that the same quantum resource requirements may be necessary for detecting adversaries in higher dimensions as compared to quantum tomography \cite{paris2004quantum}. Thus, it remains unclear what the total resource cost of QML is in the presence of adversaries. However, there is the tantalising yet unexplored prospect that quantum resources may enhance the security of machine learning algorithms, in a similar way that information-theoretic security is afforded by quantum cryptographic protocols.

\paragraph{Other applications}

Whether or not there exist helpful QML techniques for traffic and routing management is currently very unclear, and may even appear unlikely. There may be some quantum-enhancements for machine learning algorithms applicable to traffic and routing management. However, the obstacle of efficient quantum encoding/decoding of classical data remains. The no-cloning theorem\index{No-cloning theorem} forbids state replication, and in general the overheads associated with encoding classical information into quantum states are very high \cite{bib:giovannetti2008quantum, bib:giovannetti2008architectures}, potentially outweighing any computational gain.

\subsubsection{Distributed quantum machine learning}\index{Distributed machine learning!Quantum}

The motives for considering distributed QML are similar to those for distributed classical machine learning. Suppose one wishes to perform distributed machine learning, either because the given data is naturally distributed or there is limited processing power on any given device. Then there are existing protocols for implementing general distributed quantum algorithms that might be helpful in delegating QML algorithms \cite{bib:beals2013efficient}.

Secure delegated quantum computational protocols \cite{bib:joe} can also be modified and applied to QML \cite{bib:sheng2017distributed, bib:bang2015protocol}. However, the same problem with state preparation could persist, for the server rather than the client. Alternatively, hybrid classical-quantum algorithms for distributed QML have been devised \cite{bib:yoo2014quantum}. Here, the quantum state preparation assumptions can be obviated by using a hybrid gate that takes in classical input data and implements classically-controlled unitary evolution.

\subsection{Machine learning with quantum data} \label{sec:ml_quantum_data}

Suppose our data is inherently quantum, in the form of quantum states or channels -- \textit{quantum data}\index{Quantum data}. We might additionally face restrictions in the number of copies we have access to, imposed by the no-cloning theorem\index{No-cloning theorem}.

In these cases, it has been found that classical machine learning methods may be helpful over traditional methods in dealing with quantum data. Another approach is to use quantum protocols to directly process quantum data. Learning protocols in the latter case belong to the field of quantum learning.

It is possible to process quantum data over both classical and quantum networks. Techniques from classical machine learning for quantum data may assist in the communication of quantum data over classical networks, while quantum learning protocols may be more appropriate over quantum networks. This is an exciting new research direction, as it is presently unclear whether such methods find utility.

\subsubsection{Classical machine learning for quantum data}

\paragraph{Tomography}\index{Tomography}

For classical processing of quantum data over a classical network, the first step is to find its classical description. The canonical methods for this are quantum state tomography\index{Quantum state tomography} and quantum process tomography\index{Quantum process tomography}. However, tomography is in general extremely resource intensive. Recent work has provided efficient methods for state tomography using classical machine learning techniques \cite{bib:Torlai2017, bib:Han2017}.

\paragraph{Separability}\index{Separability}

While tomography provides complete classical descriptions for quantum data, sometimes it may be sufficient to first classify data in terms of quantum characteristics. For instance, methods for classifying quantum states directly in terms of separability have been devised using classical machine learning \cite{bib:Ma2017, bib:Su2017, bib:Gao2018}. Here there are empirical demonstrations of some advantage compared to the CHSH inequality\index{CHSH inequality}. However, accumulating sufficient training data may still remain problematic for higher-dimensional states.

\paragraph{Automated experiment design}\index{Automated experiment design}

In a future quantum internet, it is desirable to find optimal methods for generating inter-node entanglement. It's also desirable for this process to be automated. Recently, such automated methods based on classical reinforcement learning \cite{bib:alexey} have been proposed to experimentally create a variety of entangled states, providing an exciting starting point for automated design in future quantum internet protocols.

\paragraph{Variational quantum eigensolvers}\index{Variational quantum eigensolvers}

We saw that variational quantum eigensolvers (VQE) rely on classical optimisation. When applied to quantum data, they have found success mostly in quantum chemistry \cite{mcclean2016theory, bib:moll2018quantum}\index{Quantum chemistry}. In the context of quantum networks, the most promising developments are perhaps in its applicability to quantum data compression\index{Quantum data compression} \cite{romero2017quantum}, which may improve quantum data communication.

\subsubsection{Quantum learning protocols}

\paragraph{Template matching}\index{Template matching}

The first quantum algorithms for processing quantum data most relevant to machine learning were quantum template-matching algorithms \cite{bib:sasaki1, bib:sasaki2}. These are classification algorithms\index{Classification algorithms}, where each class is represented by a quantum state: a `template'. The task is to find the class to which a given test quantum state belongs, where this state is not identical to any of the template states. It is unclear whether quantum template matching is directly applicable to quantum networks. However, the ideas introduced provide the key foundations for supervised learning of quantum data, which can be used in the quantum counterparts to supervised algorithms in traffic prediction, classification, and anomaly detection.

\paragraph{Learning quantum processes}

Suppose we want to transmit just enough information about a quantum process over a quantum network in order for the other parties to replicate it. For quantum data, we don't have access to the classical description a priori. Instead, we are only allowed to query the process a finite number of times. For a unitary operation, this problem is addressed in \cite{bisio2010optimal}, in a problem called the \textit{quantum learning of unitary operations}. A very interesting observation here is that the optimal strategy is semi-classical rather than fully-quantum, meaning it's sufficient for the classical data encoding the estimation of the unknown unitary to be stored. It remains an open question as to whether this extends to more general quantum processes.

\paragraph{Security}\index{Security}

In a future quantum network communicating quantum data, it becomes important to detect unusual behaviour in the incoming data-stream. These may present the first signs of a security breach or fault in the network. For dynamic time-series data, this is addressed by change point detection\index{Change point detection}. This has been extended to the quantum domain \cite{bib:gael1, bib:gael2}, where the optimal methods for detecting changes in quantum data are found using methods from state discrimination. For static data, anomaly detection methods based on machine learning become more appropriate as the definition of unusual behaviour is based on a priori training data. Classical anomaly detection algorithms have been applied to quantum data for the purpose of error detection \cite{bib:sara}, in the case where the classical description for quantum data is known. However, for cases where this classical description is unknown (as expected over a quantum internet), it is instead far more efficient to directly apply quantum algorithms. Examples of this include several quantum algorithms for anomaly detection \cite{bib:liu2018quantum}.

\latinquote{Grandescunt aucta labore.}

%
% Optimising the World
%

\section{Optimising the world}\label{sec:optimising_the_world}\index{Optimisation}

\famousquote{Premature optimisation is the root of all evil.}{Donald Knuth}
\newline

\dropcap{E}{very} aspect of our modern world is heavily optimised to improve efficiency, affordability, profitability, and every other kind of desirability. Even long before the digital revolution, which enabled the widespread implementation of modern optimisation-theoretic techniques, humans have always striven to make life as easy as possible -- the least effort for the greatest possible reward. Our modern application of optimisation theory is merely a logical extension of this to the many other facets of the 21st century economy that now exist.

\subsection{Optimisation in everyday life}

Broadly speaking, optimisation problems can all in some sense be thought of as \textit{resource allocation} problems\index{Resource!Allocation} -- given finite resources, that we would like to utilise most efficiently and for the greatest reward, what is the best way to allocate and distribute them within a complex system?

To lay the context for the remainder of the section, and illustrate the importance and transformative potential for improved optimisation approaches, we summarise several familiar everyday applications for optimisation theory that we all rely on, usually unwittingly, executed somewhere in the background up in the cloud, the end user oblivious to the gargantuan calculations being performed behind the scenes merely to update a barely noticeable on-screen logo. There are countless more examples than we can't possibly have the time and space to comprehensively summarise here.

The examples we present are all examples of \textit{satisfiability problems}\index{Satisfiability problems} -- the problem of trying to simultaneously satisfy a large number of potentially competing constraints in a system, with the best possible outcome.

It is well known that the most complex such satisfiability problems are \textbf{NP}-complete\index{NP \& NP-complete} in general, with no known efficient classical solutions (in fact even very trivial constructions of satisfiability problems are already \textbf{NP}-complete, e.g see Fig.~\ref{fig:3SAT}).

Here are a mere few notable everyday applications for optimisation theory.

\subsubsection{Traffic networks}\index{Traffic networks}

Given our highly interconnected road networks, what is the best route to take to get home, given the  complex and often unpredictable dynamics of road traffic? And how does that answer change when potentially millions of travellers are simultaneously competing to minimise their individual travel times?

To implement a straightforward \textsc{Greedy} strategy, a simple application of Dijkstra's shortest path algorithm (Sec.~\ref{sec:shortest_path}) will yield the optimal route for an individual user -- and very quickly too since it has only $O(n^2)$ complexity (i.e a \textbf{P} algorithm)!

However, in such multi-user systems, with so many users, \textsc{Greedy}\index{Greedy strategy} strategies, based on optimising users individually, are highly sub-optimal, and globally optimised algorithms must be pursued. For example, if there are a thousand drivers competing to get from Town Hall to Redfern, it makes no sense to optimise them individually, since then they will all be identically directed down George St, which will of course quickly saturate and congest. Obviously load-balancing across the various routes benefits everyone collectively \textit{and} individually.

The complexities of a global multi-user optimisation are intuitively evident -- with countless competing interests, finding a set of routes that minimises net transit time yields an enormous space of possibilities and variations to consider, far beyond individual optimisations performed independently. Alas, Dijkstra's tempting algorithm is not very useful, and we must employ harder algorithms, many of which are \textbf{NP}-complete. Some of these were introduced in the context of network packet routing algorithms in Sec.~\ref{sec:route_strats}, a conceptually identical problem to road traffic routing.

\begin{figure*}[!htbp]
	\includegraphics[clip=true, width=0.6\textwidth]{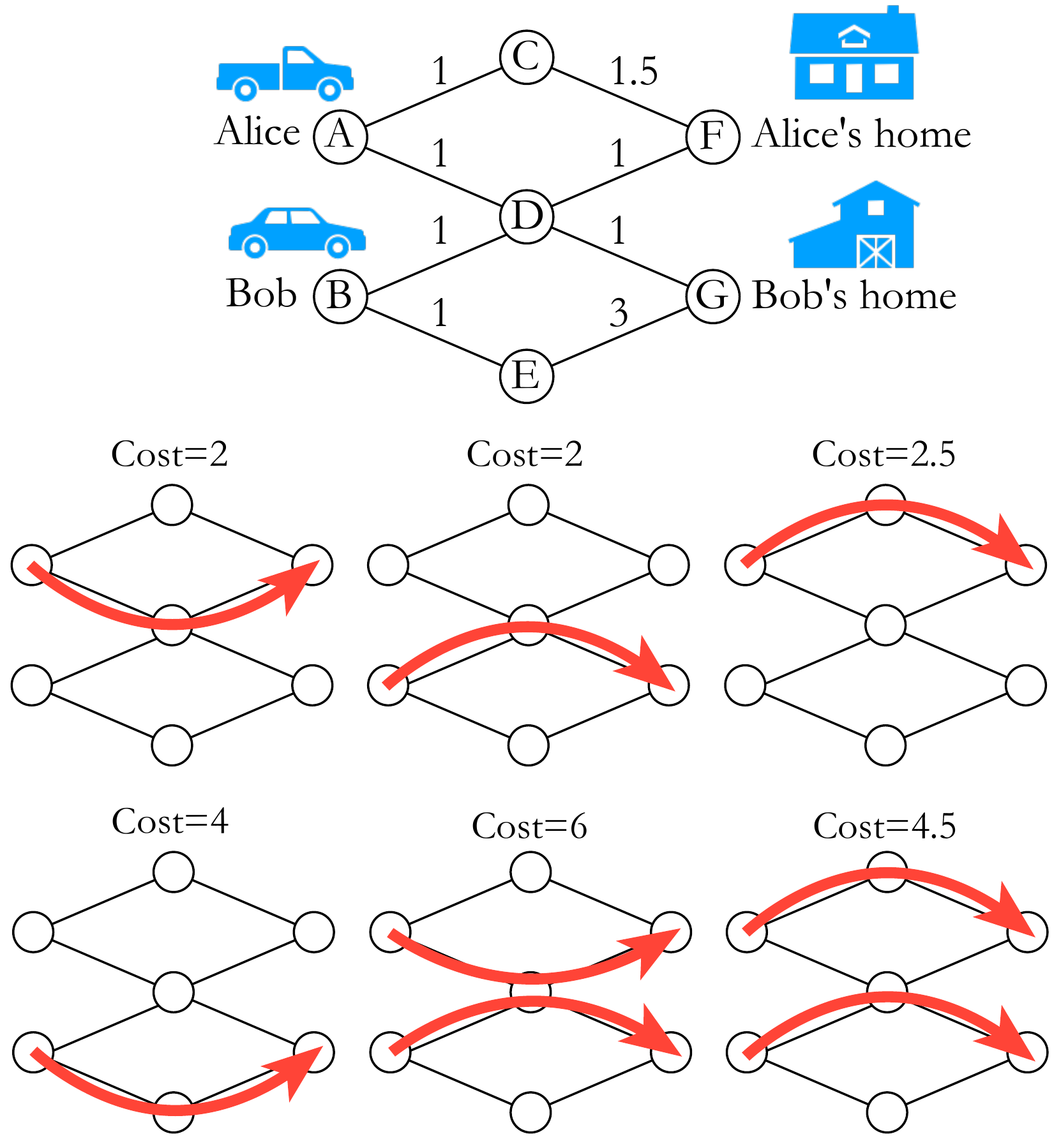}
	\captionspacefig \caption{A simple road network, where vertices represent intersections and edges represent roads. Edge weights indicate their associated travel time. The goal is for Alice and Bob to both drive home optimally, i.e minimising net commute time. Let us assume that the intersections do not delay single cars, but if they become congested (i.e more than one car simultaneously) they induce a time delay of 1 on all drivers present at the intersection. If Bob were to stay at work, and only Alice was attempting to drive home, her optimal route would be \mbox{$A\to D\to F$}, with a commute time of 2. Conversely, if Alice stayed at work and Bob was driving home, his optimal route would be \mbox{$B\to D\to G$}, also with a commute time of 2. However, if both Alice and Bob were driving home at the same time there would be congestion at intersection $D$, and their combined travel time would be 6, not the intended 4. Thus, by optimising their routes independently of one another their competition for intersection $D$ would penalise both of them by 1. On the other hand, a global optimisation would show that a more optimal routing would be for Alice to take \mbox{$A\to C\to F$} (time 2.5), with Bob remaining on \mbox{$B\to D\to G$} (time 2). This would penalise Alice by only 0.5 for taking the longer road, but both Alice and Bob would no longer be penalised at the congested intersection, reducing their collective transit time to 4.5, a saving of 1.5. Note that Alice and Bob are not only better off collectively (i.e their joint travel time), but also individually. That is, Alice making the sacrifice of taking the longer route home actually benefits her individually, with an individual saving of 0.5. In this trivial example there is only one point of conflict in the network. In general, as the number of users scales up, the distinct ways in which combinations of conflicts could emerge grows exponentially.}\label{fig:traffic_opt}
\end{figure*}

Fig.~\ref{fig:traffic_opt} provides a simple example illustrating how individual local optimisations can yield suboptimal routing, with a global optimisation yielding a better overall outcome, both collectively and individually.

\subsubsection{Public transport scheduling}\index{Public transport scheduling}

Given a fixed network of train lines, and a fixed number of trains, but the ability to schedule them freely, how does one schedule a roster that minimises average waiting times?

This might seem straightforward in simple test-cases. As before, optimising a single route, or several independent routes is trivial. But once complex, conflicting interdependencies are in place, conflict resolution becomes impossible to eliminate. Scheduling the Epping train to leave 5 minutes earlier will allow its passengers to catch the 5pm Hornsby train. But by leaving those few minutes earlier, passengers arriving from the Blue Mountains train will miss their connection and have to wait for the next one.

We are once again in a situation where we are overwhelmed with exponentially growing combinatorics to try and minimise the countless possible schedule conflicts that can occur. It is obvious upon inspection that finding a global optimum to this problem will not be possible via independent local optimisations, which do not accommodate for competing interdependencies. 

The importance of this problem is obvious. Minimising the resources required to operate public transport effectively could save enormous amounts of money in state budgets. Not to mention, passenger waiting time is of value too. If a million passengers lose just a few minutes of productivity per day to increased commute times, this can amount to billions of dollars a year in lost productivity to the broader economy.

\subsubsection{Economics}\index{Economics}

Governments inevitably have finite budgets\footnote{Except in Venezuela.}, and must therefore allocate tax revenue optimally. Public spending is associated with many of the same combinatorial difficulties as the previous examples.

The countless sectors of the economy are all interdependent with one another, and raise their own demands (constraints), which must be mutually satisfied. A minor misallocation of funding to one project could have flow-on effects to other interdependent programs, yielding another resource allocation nightmare.

Of course, these ideas don't just apply to government budgets, but also budgets at smaller scales, say within a large corporation with many distinct spending programs.

\subsubsection{Supply chain networks}\index{Supply chain networks}

The modern economic infrastructure for the production and distribution of goods is built upon supply chains -- complex, interdependent networks of the exchange of resources between entities as goods pass through their many stages of production before reaching the consumer. Each exchange of resources will typically be subject to its own constraints, such as strict time of arrival demands, or age for perishable goods. Effectively this yields a massive instance of a complex satisfiability problem, where all, or as many as possible of the demands of the units in the chain must be simultaneously met. Conceivably, a single unsatisfied demand may break the functionality of the entire supply chain! 

Bearing in mind that modern supply chains may be dealing with billions of dollars in resources at any given time, even minor improvements to their optimisation could be of enormous monetary value.

\subsection{Classical optimisation techniques}\index{Classical optimisation techniques}

Classical techniques for attacking optimisation problems, such as the ones presented above, typically come in the following flavours:
\begin{itemize}
	\item Heuristics\index{Heuristics}: efficient algorithms that find suboptimal, but satisfactory \textit{approximations} to the solution to a problem. These algorithms are fast and can be efficiently implemented classically, but can have highly variable accuracy in the solution they provide.
	\item Brute-force\index{Brute-force}: use raw computational power to exhaustively work through all the combinatorics of a problem. This is guaranteed to find the optimal solution, but is typically extremely slow and limited to small instances of the problem.
	\item Efficient classical algorithms: in the most ideal scenario, we may not need either of the above, since efficient classical optimisation algorithms may exist. Perhaps the best-known example of this is \textit{linear programming}\index{Linear programming}, whereby the relationships between entities in a system are defined via linear transformations. This class of problems is classically efficient to solve exactly. Clearly this won't solve \textbf{NP}-complete optimisation problems, but there are nonetheless many problems of interest in this category.
\end{itemize}

These techniques present us with a tradeoff between the optimality of a solution and the computational resources required to obtain it. Can quantum technology improve this tradeoff?

\subsection{Quantum enhancement}

We now provide a non-comprehensive overview of some of the better-known quantum optimisation algorithms. The speedups offered by these algorithms vary enormously, from simple quadratic enhancements all the way to exponential enhancement.

\subsubsection{Satisfiability problems}

%Many readers will have heard of the \textit{travelling salesman problem}\index{Travelling salesman problem}, the task of finding the shortest route through a weighted graph that traverses every vertex. This task is known to be \textbf{NP}-complete.\index{NP}

%Many other algorithms are also known to be \textbf{NP}-complete, a number of which that are relevant to networking are discussed in detail in Sec.~\ref{sec:graph_theory}, summarised in Tab.~\ref{tab:net_alg_sum}.

Alas, satisfiability and \textbf{NP}-complete problems are not believed to be efficiently solvable on quantum computers. This rules out a quantum future where the world's resources are perfectly allocated and optimised.

\begin{figure}[!htbp]
\includegraphics[clip=true, width=0.3\textwidth]{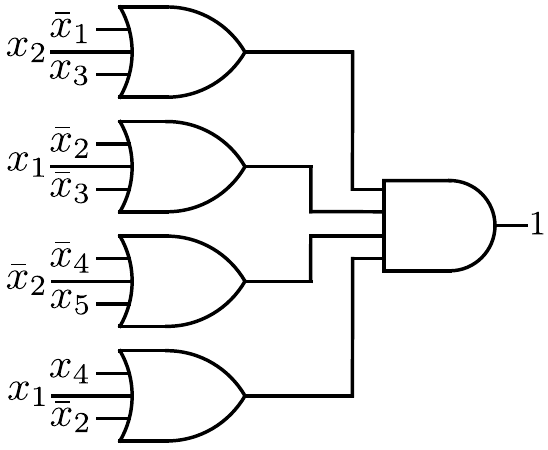}
\captionspacefig \caption{Digital circuit for an instance of the 3-\textsc{SAT} problem, with 4 clauses acting on input variables $\{x_i\}$. Each of the OR gates is input with some combination of 3 of the input bits or their compliments (i.e with a NOT gate). Each of these is referred to as a `clause', of which there are 4 in this example, but could be any number in general. The final AND gate requires that all clauses be simultaneously satisfied in order to yield a final output of `1'. The goal of the problem is to find an input bit-string $x$ that yields an output of `1'. In general, this may require exhaustively searching over the entire space of input states via brute-force, which exhibits time-complexity exponential in the length of the bit-string and associated number of clauses. This problem is proven to be \textbf{NP}-complete. Note that the similarly-defined 2-\textsc{SAT}\index{2-SAT problem} problem (i.e clauses each contain 2 input bits) is \textbf{P}, making 3-\textsc{SAT} the simplest model to consider in the study of \textbf{NP}-complete problems. For this reason 3-\textsc{SAT} is often used as the computational model when studying \textbf{NP}-complete problems.} \label{fig:3SAT}	
\end{figure}

However, such problems can be \textit{quadratically} enhanced in runtime by cunningly employing Grover's search algorithm (Sec.~\ref{sec:quantum_search}). To see this, note that all \textbf{NP}-complete problems can be efficiently mapped to one another with polynomial resource overhead (see Fig.~\ref{fig:complexity_classes}). Thus, we can restrict ourselves to considering the satisfiability problems\index{Satisfiability problems} discussed above -- the archetypal \textbf{NP}-complete problems. An example of the 3-\textsc{SAT} problem\index{3-SAT problem}, which is \textbf{NP}-complete, is shown in Fig.~\ref{fig:3SAT}.

By defining an oracle\index{Oracles} that implements a polynomial-time algorithm on $n$ qubits, a Grover search over the input space of $O(2^n)$ configurations will determine the satisfying input to the oracle for a given desired output, which acts as the tagged element within the search algorithm. The Grover search yields a quadratic speedup for this search compared to a brute-force classical search, therefore requiring only $O(2^{n/2})$ oracle calls. While this is short of the exponential speedup one might hope for, a quadratic speedup can nonetheless be very significant for large problem instances, where even minor improvements could be of enormous value.

\subsubsection{Graph colouring}\index{Graphs!Colouring}

One particularly useful approach to representing various optimisation problems is to relate them to the \textit{graph colouring problem}. This problem is known to be \textbf{NP}-complete, and when operating on a graph with $n$ vertices has $O(2^nn)$ classical runtime using the best known algorithm. Since this approach is \textbf{NP}-complete in general, and can therefore be mapped to a satisfiability problem, the quadratic enhancement offered by Grover is directly applicable.

\begin{figure}[!htbp]
\includegraphics[clip=true, width=0.35\textwidth]{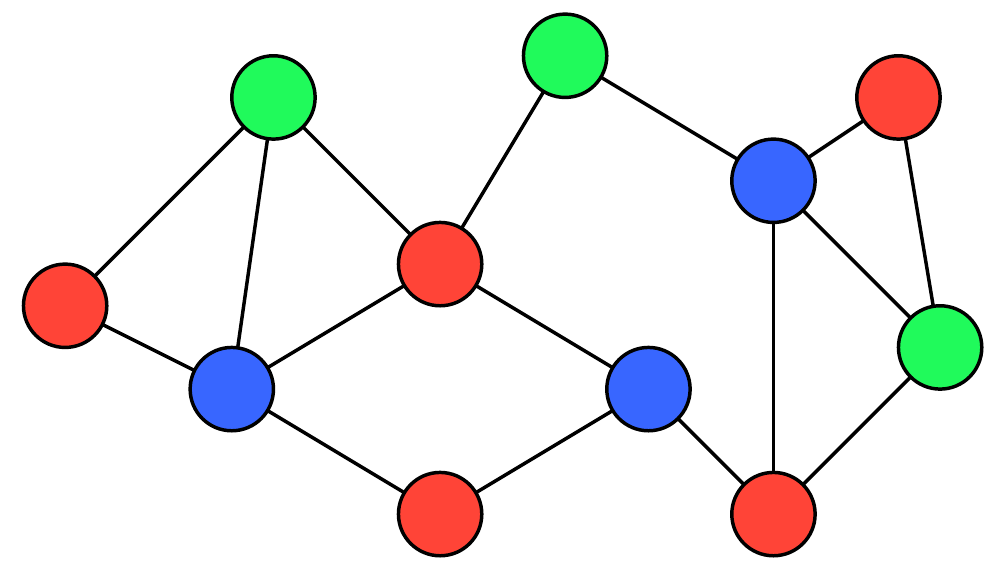}\\
\includegraphics[clip=true, width=0.35\textwidth]{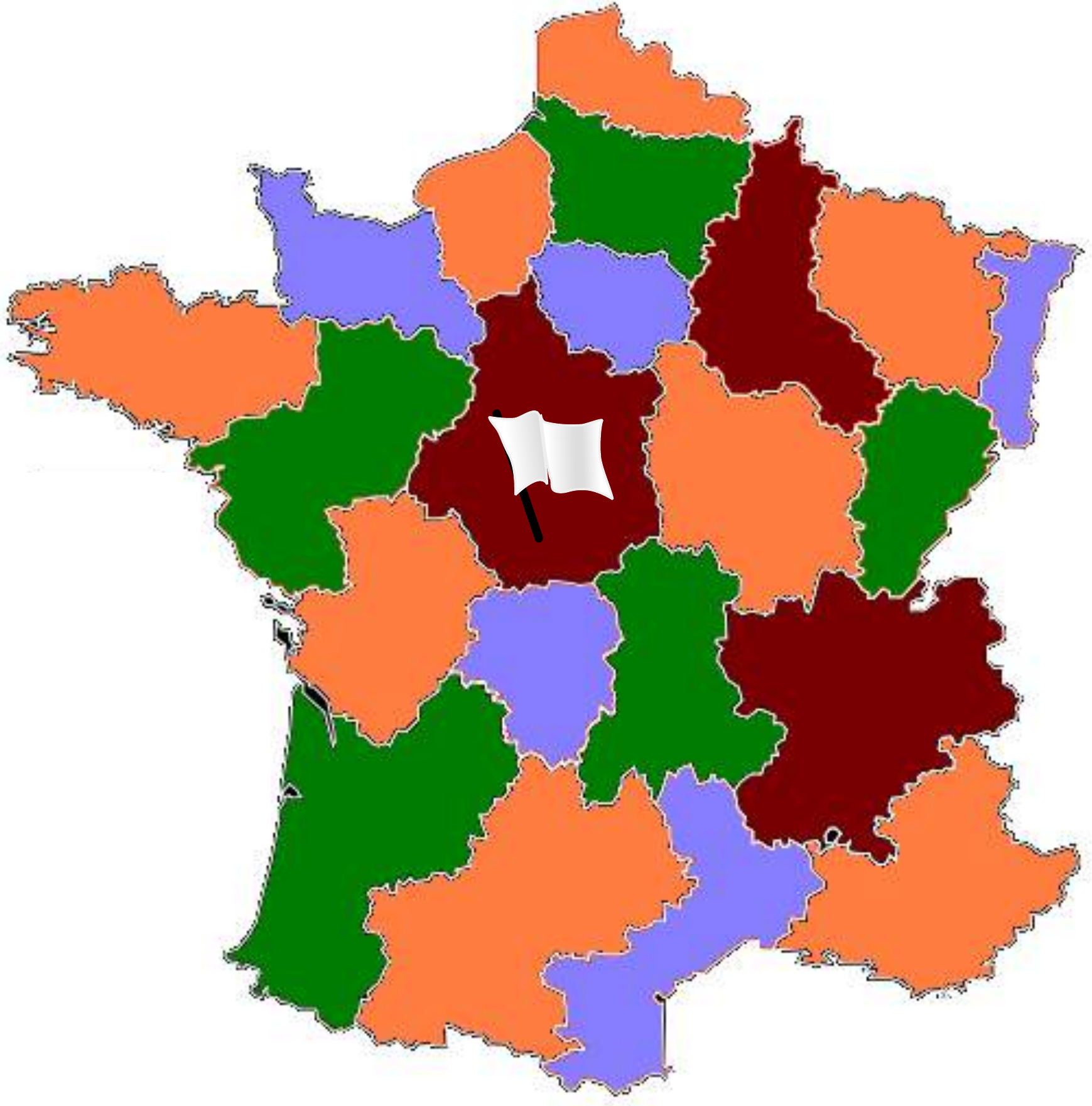}
\captionspacefig \caption{(top) Example of a graph with 3-colouring, such that no two vertices connected by an edge share the same colour. For this particular graph, 3-colouring is optimal (in the sense that the number of colours in minimised). It can easily be seen upon inspection that no 2-colouring is possible for this graph. Thus, this graph's chromatic number is \mbox{$\chi(G)=3$}\index{Chromatic number}. (bottom) A 4-colouring of the states of France\footnote{\famousquote{I say to you that you are better than a Frenchman. I would lay even money that you who are reading this are more than five feet seven in height, and weigh eleven stone; while a Frenchman is five feet four and does not weigh nine. The Frenchman has after his soup a dish of vegetables, where you have one of meat. You are a different and superior animal -- a French-beating animal (the history of hundreds of years has shown you to be so).}{William Thackeray}}\index{Elderberries}. Any planar map such as this has a chromatic number of at most 4 -- the `four colour theorem'\index{Four colour theorem}.}\index{Graphs!Colouring}\index{Surrender}\label{fig:graph_colouring}	
\end{figure}

The goal of the graph colouring problem is to take a graph $G$, and a set of $k$ colours, and assign a colour to each vertex such that no two vertices connected via an edge share the same colour. If such an assignment of colouring exists, we say the graph is $k$-colourable. The \textit{chromatic number}\index{Chromatic number} of $G$, denoted \mbox{$\chi(G)$}, is the smallest number $k$ such that the graph is $k$-colourable. An example is shown in Fig.~\ref{fig:graph_colouring}(top).

This problem relates closely to the \textit{map colouring problem}\index{Map colouring}, where the goal is to assign colours to states within a map, such that no two neighbouring states share the same colour, e.g Fig.~\ref{fig:graph_colouring}(bottom). The map colouring problem is a subset of the graph colouring problem, since a map is merely a planar graph. As opposed to graph colouring, where the chromatic number could be arbitrarily high, for map colouring the chromatic number is provably at most 4 -- the `four colour theorem'\index{Four colour theorem}.

How does this seemingly abstract mathematical problem relate to optimisation? Here are two simple examples illustrating the usefulness of the graph colouring problem in optimisation theory.

\paragraph{Time-scheduling}\index{Time-scheduling}

Our goal is to schedule the execution of a large number of tasks, so as to minimise total execution time. However, some tasks may conflict with others and cannot be executed simultaneously -- for example, they compete for access to a shared resource, which can only serve one task at a time.

Let the vertices in the graph represent tasks, and colours represent distinct points in time. If two tasks are in competition with one another and cannot be executed simultaneously, we designate an edge between them. Thus, the set of edges in this `conflict graph'\index{Conflict graph} fully characterises all conflicts in resource allocation between tasks.

Now finding a valid scheduling of tasks amounts to solving the graph colouring problem on the conflict graph, where the colouring specifies the temporal order in which tasks ought to be executed. The graph's chromatic number, $\chi(G)$, now tells us the minimum time in which all tasks can be executed subject to their constraints -- the optimised scheduling.

\paragraph{Broadcast networks}\index{Broadcast networks}

Consider a country comprising different regions, where each region is served by its own local radio station, operating at its own carrier frequency. But the radio signal within a region is strong enough to penetrate its immediate neighbouring regions also. For this reason, the frequency at which a regional station transmits should never be the same as that of its neighbours, so as to avoid signal interference.

In this incarnation of the map colouring problem, a valid colouring specifies a set of transmission frequencies that avoid interregional signal interference, and the chromatic number specifies the smallest possible radio spectrum\index{Radio spectrum} that must be allocated\footnote{Spectrum is a scarce resource\index{Scarcity}, in high demand, and worth a lot of money!}.

The same principles could apply to all manner of broadcast networks, such as the allocation of spectrum in cellular phone networks\index{Cellular phone networks}, where now map regions represent cellular transmission towers and their respective coverage.

\subsubsection{Non-satisfiability-based optimisation algorithms}

The optimisation problems discussed until now have been satisfiability problems residing in \textbf{NP}-complete, limited to only quadratic quantum enhancement owing to their utilisation of Grover's algorithm. Are there any other optimisation problems to which quantum computers lend themselves, potentially with the cherished exponential enhancement. There are, although the examples are far less ubiquitous than the plethora of satisfiability-based optimisation problems.

\subsubsection{Curve fitting}\index{Curve fitting}

A quantum-enhanced algorithm for performing least squares curve fitting\index{Least squares fitting} was described by \cite{harrow2009quantum,wiebe2012quantum}, potentially offering exponential enhancement, depending on the sparsity of the data.

As input it takes a set of data-points, \mbox{$(x_i,y_i)$}, and functions, $f_j$, outputting a set of coefficients, $\lambda_k$, defining a linear combination of those functions that fits the data, as well as a parameter characterising the quality of the fit, $E$. That is, we find the $\lambda$ that defines the fitting function\index{Fitting function},
\begin{align}
f_{\vec\lambda(x)} = \sum_k \lambda_k f_k(x),
\end{align}
such that the quality estimator\index{Quality estimator} is minimised according to a sum-of-squares\index{Sum-of-squares} metric,
\begin{align}
E = \sum_i |f_{\vec\lambda}(x_i)-y_i|^2.	
\end{align}

This can be thought of as an optimisation problem in the sense that our goal is to find a linear combination of functions that maximises the fit quality. This problem finds widespread use across many fields, particularly in statistics where it is ubiquitous.

\subsubsection{Semidefinite programming}

Semidefinite programming\index{Semidefinite programming} is a technique for finding solutions to constrained linear systems\index{Linear systems}. Although the technique is already classically efficient, running in polynomial time, a quantum-enhanced version has been described \cite{brandao2017quantum}, which promises exponential speedup in certain parameter regimes.

\subsection{An example of a poorly constructed joke}\index{Jokes}

\famousquote{How many apples grow on a tree? All of them.}{Dad}
\newline

A number of years ago there was a popular quantum computing joke circulating the internet. It went along the lines of:
\\
\\
\textit{Q:\,\,What happens when two quantum computers play chess?}\\
\textit{A:\,\,White e3, black resigns.}
\\

This joke is of course based on the assumption that quantum computers can efficiently and optimally solve the chess problem, foreseeing the entire future of a game, and determine the perfect optimal next move. Thus, after making the optimal first move `e3', black concedes that he cannot win.

However this is a poorly-designed joke, since it has been proven that chess is an \textbf{EXP}-complete problem\footnote{The complexity of chess is defined in the context where the game is generalised appropriately to an \mbox{$n\times n$} board, since complexity classes are all about computational scaling against problem size, and obviously any \textit{fixed} size instance of a game can be solved in $O(1)$ time (albeit with perhaps a very large constant!).}, believed to lie well outside of \textbf{BQP}, that which is efficiently solvable on quantum computers. Therefore this is a bad joke and it isn't funny.

However, although quantum computers are unlikely to be able to efficiently play chess \textit{optimally}, there may still be some room for improvement in \textit{approximating} optimal moves, to be discussed next in Sec.~\ref{sec:approx_optim}.

\subsection{Approximate optimisation}\label{sec:approx_optim}\index{Approximate optimisation}

Some problems in even harder classes than \textbf{NP}-complete can in some instances be \textit{approximated} using the same approach as for \textbf{NP}-complete problems. The key to solving such problems is to define an oracle\index{Oracles} that attributes a \textit{score}\index{Score} to a given input, rather than a yes/no answer to perfect satisfiability, and answers `yes' or `no' depending on whether that score is above some defined threshold for approximation. As an illustrative example, consider the optimisation of, say, a complex traffic network, where the goal is to maximise flow through the network. Then we might define our score to be some flow metric for the network's graph.

We then apply a Grover search repeatedly, each time incrementing this threshold until the algorithm outputs `no'. Then we know that the last input yielding the `yes' outcome had the highest score. The reason this approach is \textit{approximate} rather than \textit{exact} is that defining such a score-oracle mightn't be always efficiently implemented, or maybe it mightn't make sense at all to define score measures for a given problem. Alternately, maybe a particular problem inherently requires perfect satisfiability, and any imperfect approximation is insufficient -- if even a single constraint isn't satisfied, the system is broken.

An intuitive example of a problem taken from a far harder complexity class than \textbf{NP}-complete, that can be approximately optimised using the Grover approach is the game of chess\index{Chess}. Generalisation of the rules of chess to an \mbox{$n\times n$} board yields a game proven to reside in \textbf{EXP}-complete\index{EXP} -- the class of all problems requiring exponential runtime on a classical computer.

\begin{figure*}[!htbp]
	\includegraphics[clip=true, width=0.7\textwidth]{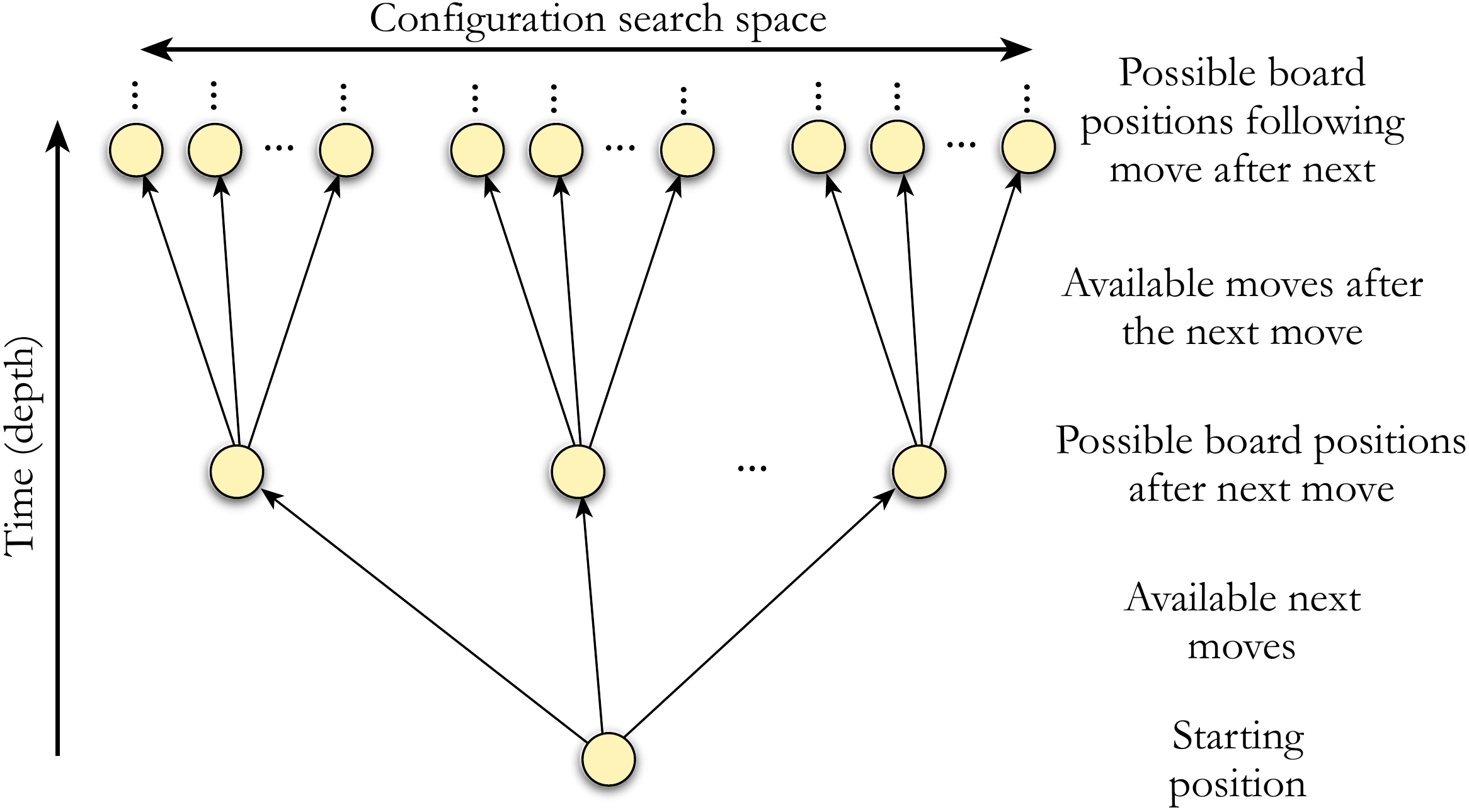}
%\fi
	\captionspacefig \caption{A search tree\index{Search tree}\index{Search space} is an outwardly-directed tree graph in which each vertex represents a state of the system (i.e `board position' in chess\index{Chess}). Edges represent actions (or `moves') available from the position from which the edge emanates. The root node represents the current position, and the depth of a vertex (i.e its distance from the root node) represents its future time. A path through the tree therefore represents a sequence of moves up to some point in future time. If the tree had maximum depth, a route (from root to leaf) would represent an entire single game, and the complete set of routes would represent all possible ways in which the game can be played. A complete analysis of a complete search tree with maximum depth would therefore allow us to exhaustively analyse all possible ways in which a game could be played to completion, enabling optimal gameplay. However, since the number of vertices/edges in a tree graph grows exponentially with depth, \mbox{$|V|=O(\exp(d))$}, the configuration search space in a game grows exponentially with how many moves into the future we wish to explore, thereby strongly limiting tree depth in practical implementations.}\label{fig:search_tree}
%\if 1\doublecol
%\end{figure*}
%\else
\end{figure*}

How do current classical computers play chess? They construct a \textit{search tree}\index{Search tree} (see Fig.~\ref{fig:search_tree}), which is explored exhaustively up to a given depth (bounded by computational resources), and assign a score to each board position, indicating its relative strength. The branch in the tree with the highest scores determines the next move.

%\if 1\doublecol
%	\includegraphics[clip=true, width=0.475\textwidth]{search_tree}
%\else
%\fi

This approach lends itself ideally to approximation via Grover. We take the same classical board-position-scoring algorithm, and implement it as an oracle. Grover is then able to search through a quadratically larger search tree, querying the oracle for scores above threshold.

Indeed this intuitive example is a very powerful one -- since all \textbf{EXP}-complete problems are by definition equivalent, and \textbf{EXP} contains many other lesser (but nonetheless hard) classes, the approximate optimisation technique immediately captures a vast array of interesting problems.

\subsection{How beneficial is quantum-enhanced optimisation?}

The utilisation of Grover's\index{Grover's algorithm} algorithm as the basis for improving optimisation problems may seem a little depressing, given that quantum search algorithms only confer a modest quadratic enhancement. By comparison, so many other quantum algorithms offer exponential speedup! How significant is such a quadratic enhancement actually?

In the context of a search tree\index{Search tree}, the size of the search space\index{Search space} (i.e number of leaves on the tree) is \mbox{$s=b^d$},
where $b$ is the tree's branching parameter (assumed uniform here for simplicity) and $d$ is tree depth. A quadratic enhancement in the searchable search space to \mbox{$s'=s^2$} effectively doubles the search tree depth that can be accommodated by a given number of oracle calls to \mbox{$d'=2d$}.

Needless to say, doubling one's foresight in planning for the future is highly advantageous and could lead to huge improvements in competitive advantage and efficiency in resource allocation. Doubling the number of moves a competitive chess\index{Chess} player could look ahead would result in their complete and utter dominance of the game!

Despite all this optimism, there remains one major concern, one shared by any quantum algorithm offering only polynomial enhancement. In any real-world quantum computer fault-tolerance\index{Fault-tolerance} is necessarily required to enable the algorithm to be executed faithfully in the inevitable presence of noise. However, fault-tolerant codes necessarily induce a time and space overhead (i.e in terms of the number of physical qubits, and algorithmic runtime). If this overhead is too large it could well overpower the   enhancement offered by the underlying algorithm. For this reason, real-world implementations of these types of algorithms will require very careful consideration of the construction of their associated error correction circuitry, such that there is still some net tangible algorithmic gain at the end. The quadratic enhancement in algorithmic runtime is actually a bound based on the ideal-case where there are no errors. In a fault-tolerant construction the actual achievable enhancement will necessarily be less than this. Specifically, for a given runtime, if the enhancement in the searchable search space is a polynomial of order $p$, this equates to an effective increase in search tree depth by a factor of $p$. For a useful quantum enhancement following fault-tolerance overhead we obviously want \mbox{$p>1$}.

Note that for the quantum search method for enhancing \textbf{NP}-complete problems, a quadratic improvement is known to be optimal. Therefore, strictly \mbox{$p\leq 2$}.

\subsection{Implications for the future}

While the dream-goals of perfectly optimised resource allocation and perfect economic efficiency are fantasy, we should not underestimate the impact that even modest enhancements in the efficiency of large-scale economic systems could have. While quantum computers will never perfectly solve humanity's immense resource allocation problems, even the more modest improvements they offer could be of immense value to the world economy, with enormous implications monetarily, socially, and environmentally. Indeed, optimisation problems are so ubiquitous that pursuing quantum-enhanced optimisation will be of central importance in the quantum era.

\latinquote{Ceteris paribus.}

\sketch{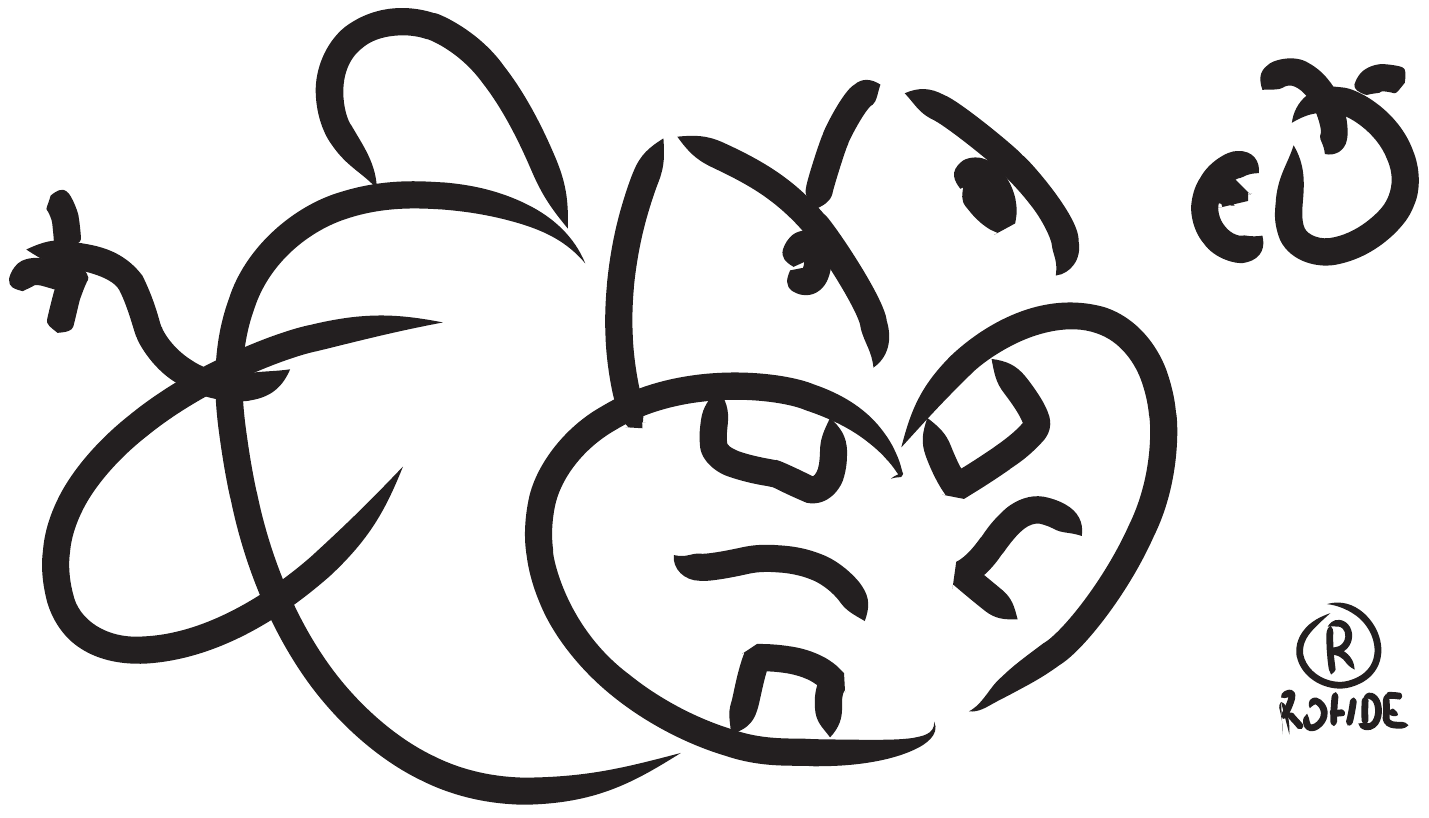}

%
% Roadmap - The way forward
%

%
% Roadmap - The way forward
%

\part{Roadmap -- The way forward}\index{Roadmap}\label{part:roadmap}

\famousquote{You got to go down a lot of wrong roads to find the right one.}{Bob Parsons}
\newline

\famousquote{The further a society drifts from the truth, the more it will hate those that speak it.}{George Orwell}

\section{The tech-web for the quantum future}

\begin{figure*}
	\includegraphics[clip=true, angle=-90, width=\textwidth]{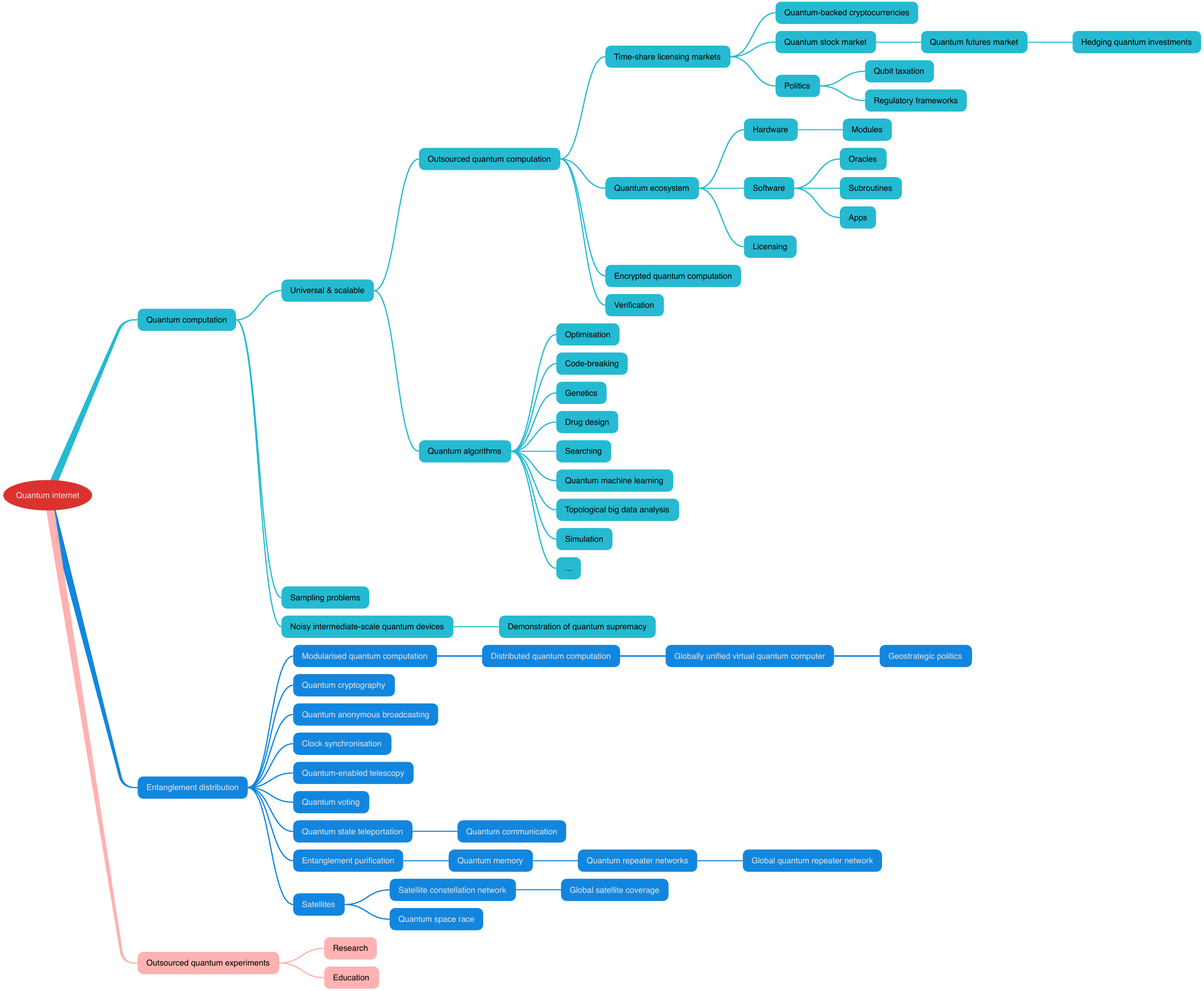}
	\caption{Relationships and dependencies in the development and deployment of the major concepts in the quantum internet -- the quantum \textit{tech-web}\index{Tech-web} -- showing the logical flow of developments that will need to take place to reach various technological milestones and destinations. The progression is highly non-linear, mandating many research programs operating in parallel to prevent technology bottlenecks.}\label{fig:roadmap}
\end{figure*}

\dropcap{W}{e} have in this work presented a vast number of ideas, protocols, applications and implications for future quantum technology, mediated by the quantum internet. The relationship between all these concepts is highly complex, with a large number of interdependencies. In Fig.~\ref{fig:roadmap} we express this as a mind-map timeline\index{Mind-map}\index{Timelines}, capturing these relationships to show the logical flow of developments that will need to take place to reach various technological milestones and destinations.

Note that the evolution of future quantum technologies proceeds as a \textit{tech-web}\index{Tech-web} rather than as a strictly linear progression. This necessitates that many quantum technology research projects will need to execute in parallel to achieve multiple desired endpoints in a timely manner without bottlenecks.

\latinquote{Aut inveniam viam aut faciam.}

\sketch{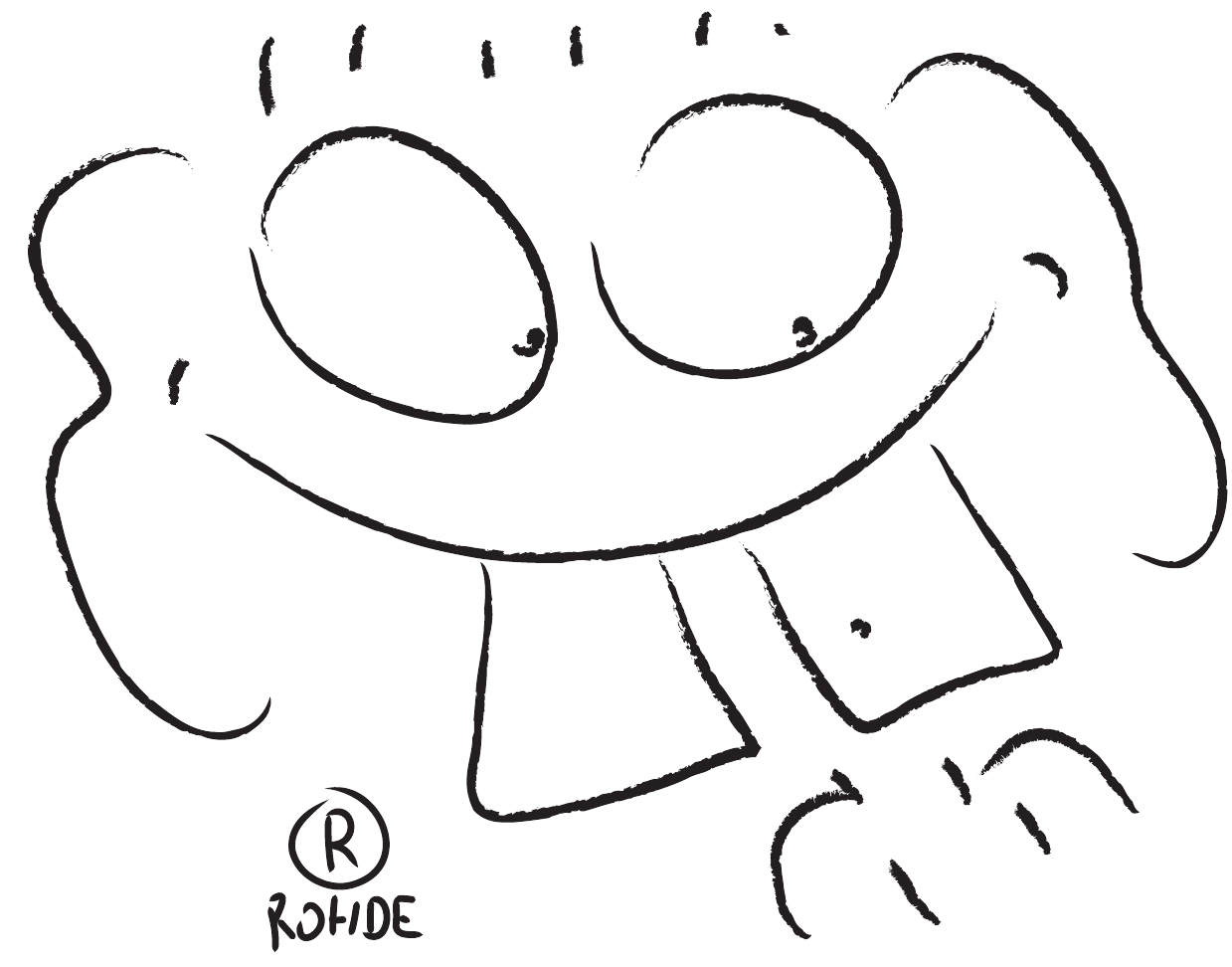}

\clearpage
%
% Essays - A New Frontier
%

\part{Essays -- A new frontier}\label{part:essays}\index{Essays}\index{A new frontier}

%
% A New Frontier - Towards the Singularity
%

\famousquote{Only the very weak-minded refuse to be influenced by literature and poetry.}{Cassandra Clare}
\newline

\famousquote{There is no harm in doubt and skepticism, for it is through these that new discoveries are made.}{Richard Feynman}
\newline

\dropcap{I}{n} this part we provide a non-technical outlook on the future quantum internet and its implications, for the benefit of the technically disinterested reader, who merely wishes to grasp some of the `big issues'. This section is in the form of a collection of short essays, requiring little or no technical background knowledge in quantum computation, quantum mechanics, or mathematics. 

We acknowledge that while parts of these essays are certainly highly plausible, if not certain, others are highly speculative, but nonetheless based on believable although somewhat futuristic (perhaps even bordering science fiction) reasoning. We can't predict the future. But at the very least we hope to stimulate the exchange of ideas, and promote their exploration and development. After all, the great ideas of the future always begin speculatively! We encourage the reader to critically question the ideas presented in these essays, and put forth their own thoughts and predictions for what the quantum future may bring and the implications it will have for humanity.

%
% The Era of Quantum Supremacy
%

\section{The era of quantum supremacy} \label{sec:era_quant} \index{Quantum supremacy}\index{SJW}

\famousquote{No matter how tiny you look, you can lead huge men if you have what the huge men don't have.}{Michael Bassey Johnson}
\newline

\dropcap{A}{} pertinent question to ask is `What is the timescale for useful quantum technologies? When will they be viable?'. The correct answer is likely very soon.

\begin{figure}[!htbp]
\if 1\doublecol
	\includegraphics[clip=true, width=0.475\textwidth]{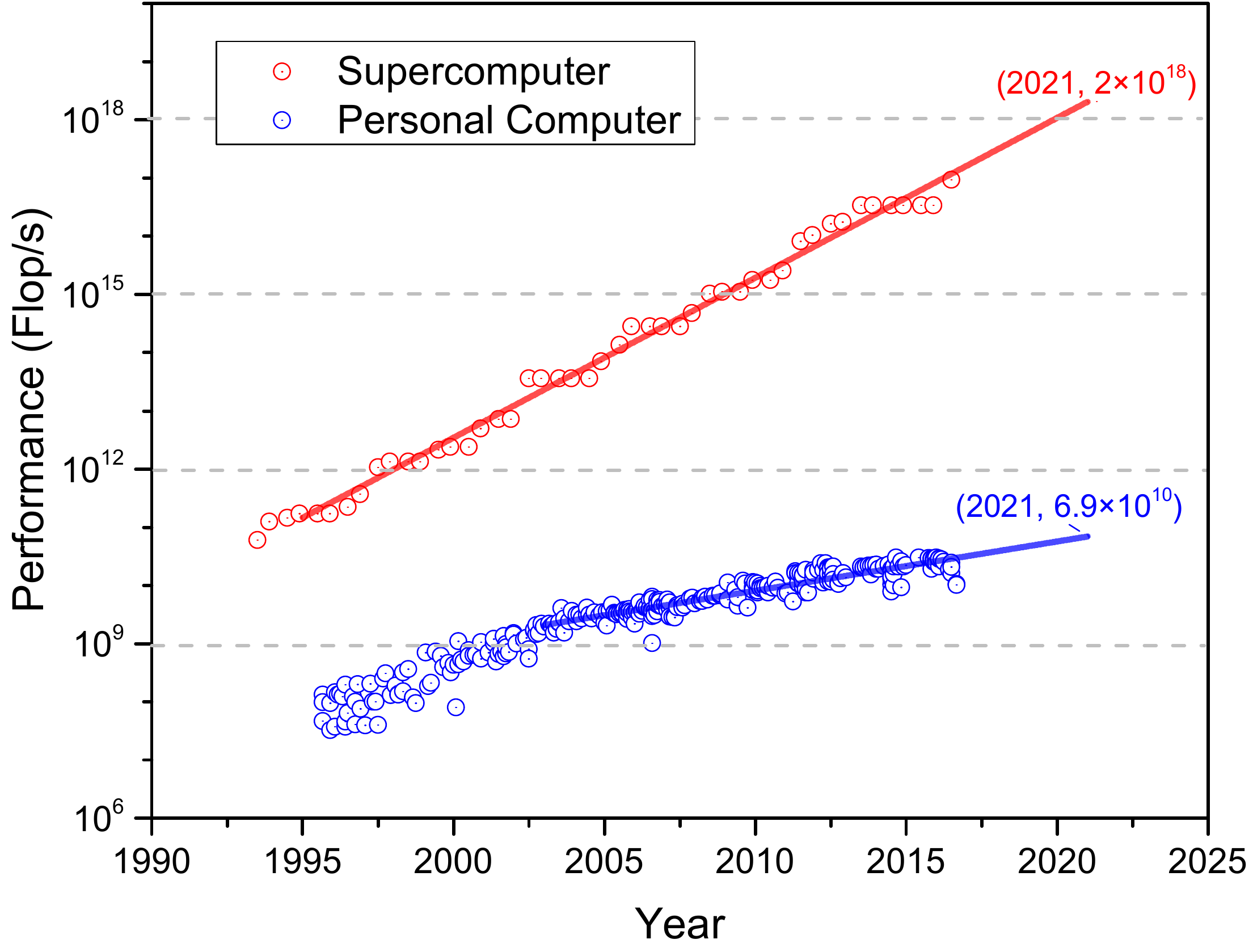}
\else
	\includegraphics[clip=true, width=0.7\textwidth]{moores_law}
\fi
\captionspacefig \caption{Historical trends in classical computing power for both PCs and top-end supercomputers, with an extrapolation 5 years into the future. The close fit to consistent exponential growth in performance over time is apparent from the logarithmic scale.} \label{fig:moores_law}
\end{figure}

From the perspective of classical computing, Moore's Law (observation!)\index{Moore's Law} for the exponential growth trend in classical computing power has proven to be a very accurate one. In Fig.~\ref{fig:moores_law} we illustrate the historical evolution in classical computing power, and extrapolate 5 years into the future.\index{Moore's Law}

To put this into context, current day microprocessors contain on the order of billions of single transistors. Current day  experimental quantum computers, on the other hand, contain fewer than 100 qubits. We sit at the mere very beginning of Gordon Moore's adventure through the quantum era.

While the power of classical computers scales at most linearly with the number of transistors, the classical-equivalent power\index{Classical-equivalent computational power} of quantum computers scales exponentially with the number of qubits (in the best-case scenario). The classical Moore's Law is close to saturation -- we simply can't make transistors too much smaller than they already are\footnote{Current transistor feature sizes are on the order of several hundred atoms. Under a Moore's Law prediction, we are likely to hit fundamental physical barriers in transistor size within a decade. Presumably, we can't make a transistor smaller than an atom!}! We therefore envisage a new Quantum Moore's Law\index{Quantum Moore's Law}, which follows a far more impressive trajectory than its classical counterpart. The point of critical mass in quantum computing will take place when the classical and Quantum Moore's Law extrapolations intersect, signalling the commencement of the \textit{post-classical era}\index{Post-classical era} of \textit{quantum supremacy}\index{Quantum supremacy}. Estimating this is more challenging than it sounds, since although the classical Moore's Law is extremely well established with an excellent fit to an exponential trajectory, there aren't yet enough data-points to make a confident prediction about a Quantum Moore's Law, to what trajectory it best fits, and at what rate it progresses, not to mention unforeseeable black swans.

Aside from quantum computing, theoretically unbreakable quantum crypto-systems, in the form of quantum key distribution (QKD), are already technologically viable\index{Quantum key distribution (QKD)}, and are in fact commercially available off-the-shelf today, as end-to-end units connectable via fibre-optics. Recently, satellite-based QKD was demonstrated, enabling direct intercontinental QKD over thousands of kilometres. Although only a single such satellite has been demonstrated, its success implies that constellations of interconnected such satellites are inevitable in the near future, enabling point-to-point QKD between any two points on Earth. It is likely the next space-race will be the one for quantum supremacy.

As the era of post-classical quantum computation edges closer, the importance of QKD networks will intensify, and along with it the demand for quantum networking infrastructure.

It is clear that humanity already sits at the precipice of harnessing quantum technologies, and must act quickly to enable them to be fully exploited as they emerge in the near future.

\latinquote{Ad astra per alas porci.}

%
% The Global Virtual Quantum Computer
%

\section{The global virtual quantum computer} \label{sec:GVQC} \index{Virtual quantum computer}

\famousquote{No generation has had the opportunity, as we now have, to build a global economy that leaves no-one behind. It is a wonderful opportunity, but also a profound responsibility.}{Bill Clinton}
\newline

\dropcap{F}{rom} the uniquely quantum phenomena that computational power can scale exponentially with the size of a quantum computer, as opposed to the linear relationship observed in classical computing, emerges an entirely new paradigm for future supercomputing. Rather than different quantum hardware vendors competing to have the biggest and best computers, using them independently in isolation, they are incentivised to unite their resources over the network and leverage (`piggyback') off one another, forming a larger and exponentially more powerful \textit{virtual quantum computer}, which could then be time-shared\index{Time-sharing} between them, to the benefit of all parties. The key observation is that \textit{all} contributing users to the network gain leverage from other users unifying their assets, irrespective of their size. In fact, this computational leverage\index{Computational!Leverage} is greater for smaller contributors than larger ones, making the benefits of this phenomenon disproportionately benefit the less-well-resourced. \famousquote{The only thing that will redeem mankind is cooperation.}{Bertrand Russell}

Users who make an initial fixed investment into quantum computing infrastructure, which they contribute to the network, but are then unable or unwilling to finance further expansion of, will nonetheless observe exponential growth in their computing power over time. That is, the computational dividend yielded by a fixed investment increases exponentially over time. This creates a very powerful economic model for investment into computational infrastructure with no classical parallel, which could be particularly valuable in developing nations or less-wealthy enterprises.

It follows that in the interests of economic efficiency, market forces will ensure that future quantum computers will \textit{all} be networked into a single \textit{global virtual quantum computer}\index{Virtual quantum computer}, providing exponentially greater computational power to all users than what they could have afforded on their own.

Vendors of quantum compute-time who do not unite with the global network will quickly be priced out of the market, owing to their reduced leverage, rendering the relative cost of their computations exponentially higher than vendors on the unified network. \famousquote{United we stand, divided we fall.}{Matthew Walker}

This might have very interesting implications for strategic adversaries -- government or private sector -- competing for computational supremacy, but nonetheless individually benefitting from jointly uniting their competing quantum resources. Bear in mind that using encrypted quantum computation all parties could maintain secrecy in their operations. Despite this secrecy, will the KGB\index{KGB} and NSA\index{NSA} really cooperate, to the benefit of both, or will the asymmetry in the computational leverage incentivise them to not unify resources and instead construct independent infrastructure?

The leverage asymmetry will be a key consideration in answering this question, since although both parties benefit on an absolute basis from unification, on a relative basis the weaker party achieves the higher computational leverage. For this reason, it is plausible the global virtual quantum computer will fracture\index{Fracturing}, dissolving into independent smaller virtual quantum computers, divided across geostrategic or competitive boundaries, with the stronger parties seceding from the union -- even though they would individually benefit computationally from unification, they may not wish the weaker ones to piggyback off them, achieving greater leverage than themselves\footnote{Insert jokes about Greece\index{Greece} and Germany\index{Germany} here --- \textit{Im Wandel der Zeiten -- Eine Geschichte der Zivilisation.}}.

\latinquote{Kyrie eleison.}

% \comment{Discuss adversarial enhancement, or in geo-strategy section.}

%
% The Economics of the Quantum Internet
%

\section{The economics of the quantum internet} \label{sec:economics} \index{Economics}

\famousquote{Underlying most arguments against the free market is a lack of belief in freedom itself.}{Milton Friedman}
\newline

\famousquote{Either we believe in free speech for those we despise or we don't believe in it at all.}{Noam Chomsky}
\newline

\dropcap{Q}{uantum} computers are highly likely to, at least initially, be extremely expensive, and affordable outright by few. Client/server economic models based on outsourcing\index{Outsourced!Quantum computation} of computations to servers on a network, will be essential to making quantum computing widely accessible. The protocols we have presented here pave the way for this type of economic model to emerge. It is paramount that the types of technologies introduced here be fully developed in time for the deployment of useful quantum computing hardware, such that they can be fully commercialised from day one of their availability, enabling widespread adoption, enhanced economy of scale, and rapid proliferation.

A key question regarding the economics of the quantum internet is the extent to which it will be able to piggyback off existing optical communications infrastructure, given that networking will almost inevitably be optically mediated. We have an existing intercontinental fibre-optic backbone, as well as sophisticated satellite networks. To what extent will this existing infrastructure (or future telecom/satellite infrastructure) be able to be exploited so as to avoid having to rebuild the entire future quantum internet infrastructure from scratch? This is a question worth billions of dollars. We also need to factor in that given the massive driving force behind telecom technology, its cost is following a Moore's Law-like\index{Moore's Law} trajectory of its own, and what costs a billion dollars today might cost a million dollars in a decade's time. In light of this, telecom wavelength quantum optics is being hotly pursued.

Technology should benefit humanity, not only an elite few \latinquote{Homo sum humani a me nihil alienum puto}. In light of this, who exactly will benefit from the quantum internet? Its beauty is that it doesn't create a system of winners and losers. Rather, it establishes a technological infrastructure from which all can benefit, rich or poor. Well-resourced operators who can afford quantum computers, for example, will benefit from being able to license out compute time on their computers, ensuring no wasted clock-cycles and maximising efficiency. The less-well-resourced will benefit in that they will have a means by which to access the extraordinary power of quantum computing on a licensed basis, facilitating access to infrastructure by those who otherwise would have been priced out of the market. This is essentially the same model as what is employed by some present-day supercomputer operators, enabling small players access to supercomputing infrastructure. The quantum internet is critical to achieving the same goal in the quantum era. This could have transformative effects on the developing world in particular. And many emerging industries, for whom access to quantum computation will be critical, but who cannot afford them, will benefit immensely from the client/server model.

Already today, even before the advent of useful post-classical quantum computers, we are seeing the emergence of the outsourced model for computation\index{Outsourced!Quantum computation}. IBM recently made an elementary 16-qubit quantum computer freely available for use via the cloud. Interested users can log in online, upload a circuit description for a quantum protocol, and have it executed remotely, with the results relayed back in real-time. Although still very primitive, this simple development already makes experimentation with elementary quantum protocols accessible to the poor layman, undergrad, or PhD student in a developing country, people who just a few years ago would never have dreamt of being able to run their own quantum information processing experiments! This effectively opens up research opportunities to people who otherwise would have been priced out of the market entirely, unable to compete with established, well-resourced labs. Evidently, the market already recognises the importance of outsourced models for quantum computation. We encourage the impatiently curious reader to log onto the `IBM Quantum Experience'\index{IBM!Quantum Experience} (\url{http://www.research.ibm.com/quantum/}) and take a shot at designing a 16-qubit quantum protocol, without even needing to be in the same country as the quantum computer.

The quantum internet will facilitate the communication and trade of quantum assets\index{Quantum assets} beyond just quantum computation and cryptography. There are many uses for various hard-to-prepare quantum states, for example in metrology, lithography, or research, where outsourcing complicated state preparation would be valuable. Alternately, performing some quantum measurements can be technologically challenging, and the ability to delegate them to someone better-equipped would be desirable. The quantum internet goes beyond just quantum computing. Rather, it extends to a full range of quantum resources and protocols.\index{State preparation}\index{Measurement}

To commodify quantum computing, if constructing large-scale quantum computers were a simple matter of plug-and-play, where QuantumLego\texttrademark\,building blocks\index{QuantumLego\texttrademark} are available off-the-shelf and straightforward to assemble even for monkeys, mass production would rapidly force down prices. By arbitrarily interconnecting these boxes, large-scale quantum computers could be scaled up with demand, with a trajectory following a new Quantum Moore's Law\index{Quantum Moore's Law}, with potentially super-exponential computational return.

We envisage that each of these commodity items is a black box, within which a relatively small number of qubits are held captive. Then, to build a larger quantum computer, we don't need to upgrade our boxes. Rather, we simply purchase more boxes to interconnect over the network -- modularised quantum computation\index{Modularised quantum computation}. This notion is tailored to graph states in particular -- because a graph state can be realised by nearest neighbour interactions alone, and since all preparation stages commute with one another, they naturally lend themselves to modularised, distributed preparation.\index{Modularised quantum computation}

Such an approach lends itself naturally to distributed computation, where modules may be shared across multiple users, with the economic benefit of maximising resource utilisation, and the practical benefit of the end-user effectively having a much larger quantum computer at their disposal.\index{Distributed quantum computation}

By having a standardised architecture for optically interconnecting modules, we also somewhat `future-proof' our hardware investment -- if interfacing modules is standardised, existing hardware can be fully compatible with newer, more capable module versions. We might envision the emergence of open standards on optical interconnects and fusion protocols\index{Open standards}.

On the other hand, if quantum computers were only ever sold as specialised, room-sized, all-in-one solutions (think D-Wave\texttrademark)\index{D-Wave}, such mass production would not experience the driving force of commodified, off-the-shelf building blocks, each of which is cheap, yet frugal in its computational power alone.

Essential to existing financial markets are pricing models for physical assets. Furthermore, derivative markets increase trading liquidity, market efficiency, enhance price discovery, and importantly, allow risk management via hedging and the ability to lock in future prices\index{Risk management}\index{Hedging}\index{Future contracts}. This is invaluable to traders of conventional commodities, and it is to be expected that it will be equally valuable to consumers of quantum resources. We have made initial steps in deriving pricing models for quantum assets and derivatives, which although they may require revision in the future real-world quantum marketplace, provide an initial qualitative understanding of quantum market dynamics.

Networked quantum computing will present new challenges for policy-makers\index{Policy-making}, whose fiscal policies strive to maximise economic efficiency and optimise resource allocation. Devising policies of taxation and a regulatory framework in the quantum era will require careful deliberation.

It is evident that taxation\index{Taxation} of qubits has far deeper economic implications than the taxation of other typical financial assets or classical technologies, owing to their exponential scaling characteristics. Generally speaking, taxation of an asset disincentivises its growth. But if the computational return on quantum assets grows exponentially with network size, so too will sensitivity to taxes that stifle it. This will require extremely prudent consideration when designing fiscal policies in the quantum era, so as to avoid exponential suppression of quantum-related economic activity.

Conversely, the exponential dependence on the rate of taxation could be exploited for leverage via subsidisation\index{Subsidisation}. It may be economically beneficial to subsidise quantum infrastructure, reaping its exponential payback, via taxation of other economic sectors, less sensitive to taxation.

The future quantum economy might be made more efficient by artificially transferring capital from low-multiplier sectors to high-multiplier quantum technologies. Or maybe the market will do this on its own accord\footnote{Have faith in the invisible hand.\index{Adam Smith}}? This is a uniquely quantum consideration that never previously applied to conventional supercomputing\index{Supercomputers}. The onset of the quantum era may redefine our entire economic mindset and fiscal policy-making, to adapt to the unique economic idiosyncrasies of this emerging technology.

\latinquote{Deus ex machina.}

%
% QuantCoin
%

%
% The Quantum Future of Cryptocurrencies
%

\section{The quantum future of cryptocurrencies}\index{Computation-backed currency}\index{Cryptocurrencies}\index{QuantCoin\texttrademark}\label{sec:quant_coin_essay}

\famousquote{Bitcoin is the most stellar and most useful system of mutual trust ever devised.}{Santosh Kalwar}
\newline

\famousquote{By 2030, some form of Crypto will become the global reserve currency but it will not be based on what exists today. Existing cryptos need to transform or will disappear. Also around 2030 or so, the first Nobel Prize in Economics will be awarded to a Cryptoeconomist.}{Tom Golway}
\newline

\dropcap{T}{he} advent of cryptocurrencies\index{Cryptocurrencies} (Sec.~\ref{sec:bitcoin_blockchain}) places the death of fiat currency\index{Fiat currency} firmly on the horizon \latinquote{Deo gratias}. Central banks around the world have been consistently inflating and devaluing national currencies, destroying their integrity through loose print-on-demand monetary policies to finance ever-increasing debt. National currencies and currency unions sit at the brink of crumbling. Can we opt out? Is there an alternative? Let us discuss an alternative!

What makes a sound currency\index{Sound currency}? First of all, it must exhibit scarcity\index{Scarcity} and be difficult or impossible to counterfeit\index{Counterfeit} -- it should not be possible to forge unlimited quantities out of thin air, a quality most certainly not inherent to the fiat currencies maintained by today's central banks. Second, its abundance and demand should exhibit relative stability and predictability over time, so as to create a stable money supply\index{Money supply} and inflationary/deflationary\index{Inflation}\index{Deflation} rate.

For these reasons, gold\index{Gold standard} for thousands of years was almost universally accepted as the legally recognised form of tender, since it is naturally scarce and much work must be invested into its production. For the same reasons, emerging cryptocurrencies like BitCoin\index{Bitcoin} have become widely adopted and even the norm in contemporary hyper-inflating\index{Hyper-inflation} economies like Venezuela where fiat currency has lost all integrity, as the cryptocurrencies exhibit these desired traits, immune to government. But rather than scarcity of a natural resource, we are dealing with artificial scarcity of bit-strings, cryptographically enforced to satisfy certain mathematical properties and constraints that cannot be easily counterfeited.

We propose that units of quantum computation meet these criteria very well. The only way to forge new computations is via investment into infrastructure, which has direct monetary cost and cannot be mitigated. Recent history has shown us that Moore's Law\index{Moore's Law} has made the growth in classical computing power highly predictable and relatively stable over time, and it is to be expected that a quantum Moore's Law\index{Quantum Moore's Law} will hold.

Time-shares in unified computing power over the quantum network, via licensing out qubits from hardware owners, would provide all these essential desired qualities for a sound currency. It can be envisaged that forward contracts in compute-time (Sec.~\ref{sec:for_contr}) would act as a good basis for backing a currency. Since these are nothing but simple forward contracts, they lend themselves to highly fluid, low-overhead trading on international markets.

Existing Blockchain-based cryptocurrencies like Bitcoin\index{Blockchain}\index{Bitcoin} (Sec.~\ref{sec:bitcoin_blockchain}) are actually examples of computation-backed currencies, where the mining process requires brute-force computation of a large number of SHA256\index{SHA256} hash functions\index{Hash!Functions}, to discover hashes satisfying certain constraints. Unfortunately, however, in the case of Bitcoin these computations are perfectly wasted, since they are not solving any problems of merit. Rather miners are made to perform them purely for the sake of imposing artificial scarcity via `proof-of-work'\index{Proof-of-work}\footnote{This proof-of-work notion was originally borrowed from the Hashcash\index{Hashcash} protocol for preventing email spamming.}.

QuantCoin\texttrademark\,\index{QuantCoin\texttrademark} (Sec.~\ref{sec:quant_coin_technical}) on the other hand backs the currency with real-world computations of value, as determined by market participants at the time, a far better utilisation of computational power, with far greater confidence in its objective monetary value. Such a currency is no longer backed purely by the psychology of scarcity\index{Scarcity}, but also the economic value of executing useful quantum algorithms on real-world data. Thanks to homomorphic encryption\index{Homomorphic encryption} and blind quantum computing\index{Blind quantum computation}, users' data may be protected from eavesdropping end-to-end during computation, whilst still allowing the computation to be associated with a unit of cryptocurrency.

Such currencies could be either commissioned and backed by nation states, or operate entirely in the private sector, leading us on a path to free banking\index{Free banking}, devoid of nation-backed currencies altogether.

Because future contracts have predetermined times until maturity, they also serve the very helpful role of being hedging\index{Hedging} instruments, an important tool for end-users of computation who may wish to lock in prices in advance to mitigate exposure to market risk.

Were a computation-backed currency to emerge, it would immediately further incentivise investment into expansion of quantum computational hardware, as it would be directly convertible to currency with zero overhead. The implications for compute-intensive industries would be immense, as there would be negligible transaction costs associated with the purchase of computation -- since contracts in computation \textit{are} the accepted currency -- thereby driving forward investment into the next technological revolution.

Consider the time-share future contract model\index{Time-sharing} as a basis for a currency. Unlike fiat currency, this monetary system would not be inflationary since the commodity backing the currency is one which cannot be easily counterfeited -- the only way to make more currency is to provide more genuine, functional, online qubits, which increases the money supply over time in tandem with the underlying asset backing it. This would in effect be a full-reserve banking system\index{Full-reserve banking}, where the direct one-to-one convertibility between currency (forward contracts on computation) and its backing asset (time-shared access to physical qubits) eliminates the money multiplier\index{Money multiplier}, a system essentially immune to bank runs\index{Bank runs}.

Because the currency is forward contracts in computing time-shares, not ownership of the physical underlying qubits, the qubits needn't change hands upon being utilised in monetary transactions. The currency could reside entirely on a distributed digital ledger\index{Distributed ledger} recording transactions in the future contracts, independent of trading in physical qubits, who owns them, or where they reside.

In a strategically fractured\index{Fracturing} world, where multiple, independent quantum internets may exist in isolation to one another, partitioned along geostrategic boundaries, each with their own local QuantCoin\texttrademark\,currencies, there would be an enormous monetary incentive to breaking down trade barriers and globalising the network by unifying smaller ones. This is contrary to nationalised fiat currencies, where there is little incentive for, yet much to lose by unifying currencies. Greed on behalf of those owning QuantCoins\texttrademark\,would therefore directly incentivise harmony and integration amongst all the world's leading players in the technological realm. Well-financed market participants would have much to lose from fracturing of the network. This could make QuantCoin\texttrademark\,a major driver towards international peace and prosperity in the quantum world of tomorrow.

Importantly, a computation-backed currency would be largely immune to political interference. Politicians would have zero ability to directly manipulate the money supply, short of suicidally self-destructive policies like shutting down or curtailing the quantum internet.

Having a sound monetary system, immunised against political interference, and incentivised to integrate, will play an important role in constraining the power of government and spreading economic liberty across the globe.

\latinquote{\href{https://youtu.be/lhyaiOZhpSg}{End the Fed!} Libertas justitia veritas. Libertas perfundet omnia luce.}\index{Ron Paul}

%
% Security Implications of the Global Quantum Internet
%

%
% Security Implications of the Global Quantum Internet
%

\section{Security implications of the global quantum internet} \label{sec:sec_imp} \index{Security!Implications}

\famousquote{Truth is treason in the empire of lies.}{George Orwell}
\newline

\famousquote{I don't even know why any of us are here. This is the worst job I've ever had.}{John Kelly}
\newline

\dropcap{W}{ith} any new technology comes ethical considerations. Who will have access to it, and how do they plan to use it? For this reason, many developed nations have export bans or restrictions in place on `dual-use' technologies -- those which have clearly legitimate and morally justifiable uses, but also nefarious ones by competitors and criminals. Nuclear technology is the obvious archetype. Quantum technologies (in particular quantum computing and quantum cryptography) are particularly vulnerable to dual-use, and for this reason are becoming subject to dual-use technology legislation, such as export controls, in some nations. In Australia, for example, legislation is being introduced criminalising the transfer of knowledge on certain quantum and cryptographic technologies to foreign nationals of certain target countries.

With a global QKD infrastructure\index{Quantum key distribution (QKD)} in place, any person on Earth with access would have uncrackable encryption at their fingertips. Whilst this might be welcomed by the populace of a despotic regime (or the libertarians in a democratic one), it would clearly be unwelcome for that level of secrecy and protection to be awarded to the regime itself. Similarly, criminal and terrorist organisations would be immune to government surveillance. With widespread global adoption of QKD technology, the signals intelligence agencies of nation states would become at least partially obsolete, leaving the NSA\index{National Security Agency (NSA)} and its Five Eyes\texttrademark\,partners furious\index{Five Eyes}.

Quantum computing also has dual-use potential. In fact, given their ability to compromise some existing cryptographic protocols, it appears highly likely that the first useful, post-classical quantum computers will find their way into the hands of national SIGINT agencies. Of course, it doesn't take much imagination to see that many other quantum algorithms could be employed for sinister purposes. For this reason we are likely to see export limitations placed on quantum computer technology in the future.

Much as the internet has eliminated national electronic borders, a quantum internet employed for distributed or outsourced computation, would make quantum computer technology available to hackers, criminals, terrorists, and strategically competing nations. And a distributed model for computation as unregulated as the classical internet would make it near impossible to prevent.

Many falsely argue that once quantum computers become available, capable of cracking current classical cryptographic codes, the world will have transitioned to quantum-proof QKD as a replacement encryption standard, and therefore that the security implications of quantum computing will not be relevant. In terms of an individual citizen's private online banking, this is largely true -- who wants to read a 10 year old bank transaction? However, when it comes to national security things aren't quite so rosy. This is because major national security agencies like the NSA of the United States systematically vacuum up astronomical amounts of internet traffic and store it for future reference, knowing that one day they may be able to crack it. Thus, if the KGB had at some point in the distant past electronically communicated sensitive information that was intercepted by the NSA but unable to be cracked at the time, when sufficiently sophisticated quantum computers become available, those messages may simply be pulled from the NSA's treasure trove and trivially cracked.

Bear in mind that as recently as the late 70's the Data Encryption Standard (DES)\index{Data Encryption Standard (DES)} was a US government-approved encryption standard. However, its mere 56-bit key-length is no match for a universal quantum computer. Therefore anything stored using this encryption standard in the past had better contain information that is irrelevant by the time the quantum computers arrive, since no doubt the NSA will immediately put them to good use cracking their entire historical catalogue of stored encrypted messages.

Combined with encrypted quantum computing protocols\index{Encrypted quantum computation}, no one would even know what they were up to when using this awesome computing power, and what they learn, they could keep to themselves.

\begin{figure}[!htbp]
\if 1\doublecol
	\includegraphics[clip=true, width=0.475\textwidth]{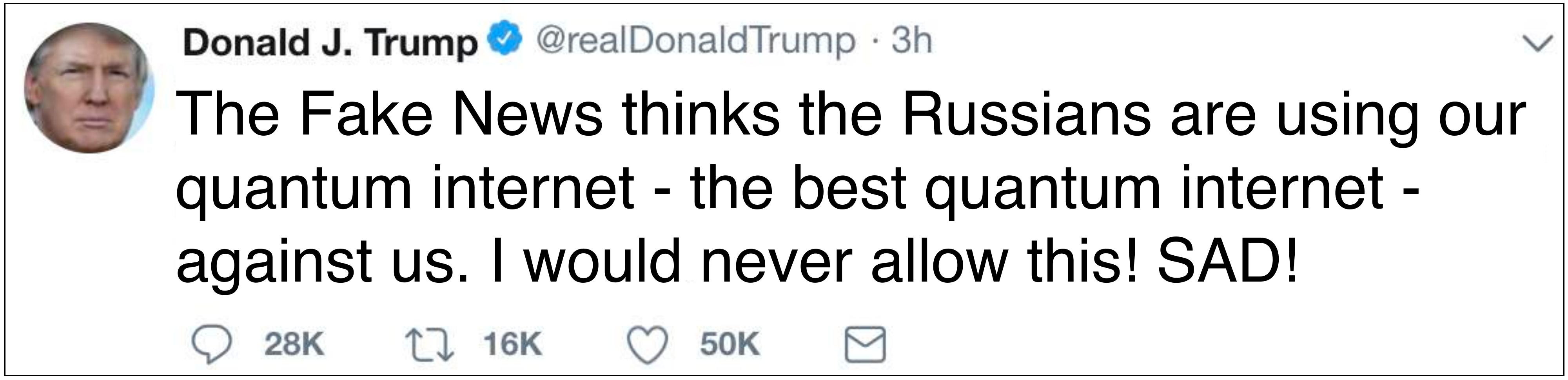}
\else
	\includegraphics[clip=true, width=0.6\textwidth]{trump_tweet}
\fi
\captionspacefig \caption{POTUS responds to Alg.~\ref{alg:russian}.}\label{fig:trump_tweet}\index{Trump}\index{Sad}\index{Covfefe}
\end{figure}

Alg.~\ref{alg:russian} describes a typical protocol for a particularly nefarious application for this, with the end result shown in Fig.~\ref{fig:trump_tweet}.

\begin{table}[!htbp]
\begin{mdframed}[innertopmargin=3pt, innerbottommargin=3pt, nobreak]
\texttt{
function MakeAmericaGreatAgain(Putin):
\begin{enumerate}
    \item A Russian bedroom hacker with no direct access to quantum technology, delegates a factorisation algorithm to the cloud using homomorphic encryption.
    \item The computation is physically executed on a server in the United States.
    \item The result is returned to our Russian comrade.
    \item The Russian uses the obtained private RSA key to hack Hillary's emails.
    \item The emails are strategically leaked during the next Presidential election.
    \item This swings the election in favour of Trump\texttrademark.
    \item The NSA and FBI have no clue who was behind it, since it was homomorphically encrypted.
    \item They blame Edward Snowden.
    \item Fox News\texttrademark\,calls for his execution.
    \item They'd kick themselves if they found out the algorithm was actually executed on US soil.
    \item Unless the NSA switches off the entire quantum internet, they can't prevent it from happening again in subsequent elections.
    \item return(America is Great Again\texttrademark).
    \item $\Box$
\end{enumerate}}
\end{mdframed}
\captionspacealg \caption{A typical example of a nefarious use for cloud quantum computation. See Fig.~\ref{fig:trump_tweet} for the outcome.} \label{alg:russian}\index{Trump}\index{Covfefe}
\end{table}

These are all legitimate concerns. But they are very much the same ones that detractors expressed about the classical internet and strong encryption. Nonetheless, it can be said that encryption and the internet have on balance been overwhelmingly beneficial to mankind, enabling unprecedented rates of technological and economic progress. Any attempts to eliminate or undermine them could be economically catastrophic.

We take the view that the same ethical stance ought to be applied in the quantum era. While quantum technologies clearly have dual-use potential, the magnitude of the implications they will have for scientific and technological progress overwhelms the discussed proliferation issues. No doubt, politicians will nonetheless attempt to regulate and restrict the quantum internet -- that's what governments like to do. But this will inevitably fail for the same underlying reasons that it failed for the classical internet -- no tech-savvy Chinese person can actually say they are hindered by the Great Firewall of China\texttrademark.

% \comment{Talk about QKD, post-classical crypto, halving private-key lengths. hacking stored public-key encrypted data from past is a security threat even once we have transitioned to post-classical crypto. NSA probably has mass storage of collected, but as yet uncracked data. Now they can work back through it.}

\latinquote{Annus horribilis.}

%
% Geostrategic Quantum Politics
%

\section{Geostrategic quantum politics}\index{Geostrategic politics}

\famousquote{The people can always be brought to the bidding of the leaders. That is easy. All you have to do is tell them they are being attacked and denounce the pacifists for lack of patriotism and exposing the country to danger. It works the same way in any country.}{Hermann G{\" o}ring}
\newline

\famousquote{What the United States fears the most is taking casualties. The loss of one super carrier would cost the US the lives of 5000 service men and women. Sinking two would double that toll. We'll see how frightened America is.}{Admiral Lou Yuan}
\newline

\dropcap{C}{omputation} is a commodity -- perhaps the most valuable of all in the 21st century economy -- and with any valuable, sought after commodity comes geostrategic powerplay. World powers fight wars, apply sanctions and use political leverage against one another to secure access to traditional commodities essential to economic progress and competitive advantage. It is to be expected that computation will be no different.

In conventional international relations\index{International relations}, political leverage between conflicting parties is achieved through alliances, shared common interests, threats of military action, and even more sinister possibilities. How will this differ in the quantum era?

The central point to note is the computational leverage phenomena associated with the quantum internet -- unification of resources is better for all. However, it is important to be cognisant that the leverage gained by parties unifying their resources with the cloud is asymmetric, biased in favour of (or against in an adversarial context) the weaker parties. That is, despite the fact that all players benefit from unification, smaller players relatively have more to gain. While this asymmetric computational leverage may seem favourable for the weaker parties, it also places them in a compromised situation whereby the threat of a major player expelling the smaller one from the network\footnote{Quantum internexit.} creates asymmetric political leverage in the opposite direction. A major player will have relatively little to lose under the expulsion of a smaller player. But the smaller player could suffer immensely in the relative power of their computational assets.

This observation leads to the foreseeable possibility that future trade-wars may be for computational power, with stronger parties exploiting their huge leverage over weaker parties for geopolitical objectives. Sanctions and political punishment in the quantum era may very well employ computational isolation of nation states or organisations.

It is foreseeable that the future quantum internet may become fractured\index{Fracturing} along geostrategic boundaries, with players (particularly stronger ones) unwilling to provide computational leverage to strategic competitors, even though on an absolute scale they would themselves benefit, since the leverage the competitor gains may compromise their own position, for example in cryptographic applications.

A further consideration is that the unification of quantum resources may very well require some form of central authority or marketplace to mediate the distribution and allocation of resources globally. Who will fill this role, and what strategic significance it will have is hard to predict. Certainly in the case of the United Nations, the Security Council, comprising a handful of self-declared world leaders, has immense geopolitical clout, with substantial power to influence international relations across the globe. Will the United Nations, under the supervision of the Security Council or some other politicised mediating authority, oversee the international quantum marketplace, or will some self-regulating, laissez-faire, libertarian utopia emerge under the guidance of the invisible hand. 

\latinquote{Magnus est mundus.}

%
% The Quantum Space Race
%

%
% The Quantum Space Race
%

\section{The quantum space race}\index{Space race}\label{sec:quant_space_race_essay}

\famousquote{We choose to go to the moon in this decade and do the other things, not because they are easy, but because they are hard, because that goal will serve to organise and measure the best of our energies and skills, because that challenge is one that we are willing to accept, one we are unwilling to postpone, and one which we intend to win, and the others, too.}{John F. Kennedy}
\newline

\dropcap{A}{t} the time of writing this book the world's first quantum-capable satellite\index{Quantum satellite} was very recently launched into low-Earth orbit by Chinese scientists \cite{yin2017satellite}. The key capability of the satellite was to distribute entangled pairs of photons between ground stations thousands of kilometres apart. Using these entangled pairs, quantum key distribution\index{Quantum key distribution (QKD)} was demonstrated, allowing theoretically unbreakable cryptography between the ground stations that no eavesdropper could compromise, guaranteed by the laws of physics.

However, entanglement distribution has many additional applications that are perhaps even more exciting than cryptography, most notably distributed quantum computation\index{Distributed quantum computation}, enabling the world's future quantum computers to be networked into a virtual device with exponentially greater power than the sum of the parts.

\begin{figure}[!htbp]
\includegraphics[clip=true, width=0.4\textwidth]{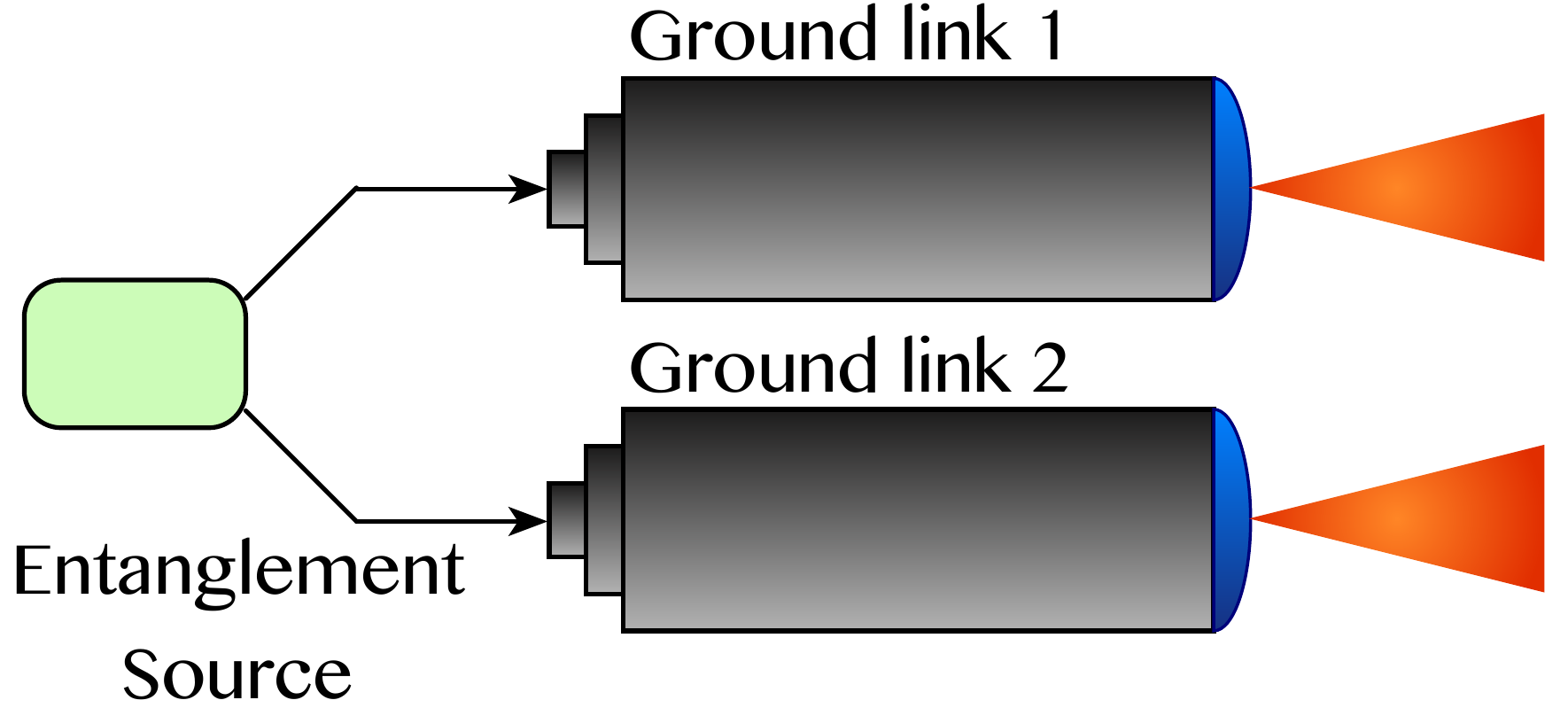}
\captionspacefig \caption{First-generation satellite\index{First-generation!Quantum satellites} for entanglement distribution. The on-board entanglement source couples to two telescopes, which lock onto independent ground stations using laser tracking\index{Lasers!Tracking}.}\label{fig:first_gen_sat}	
\end{figure}

The first-generation satellite\index{First-generation!Quantum satellites} that was recently developed merely contained an entanglement source, and two satellite-to-ground optical links via telescopes armed with laser tracking (Fig.~\ref{fig:first_gen_sat})\index{Lasers!Tracking}. However, this prototype is strictly restricted to distributing entanglement between two ground stations, both simultaneously in line-of-sight of the satellite.

\begin{figure}[!htbp]
\includegraphics[clip=true, width=0.475\textwidth]{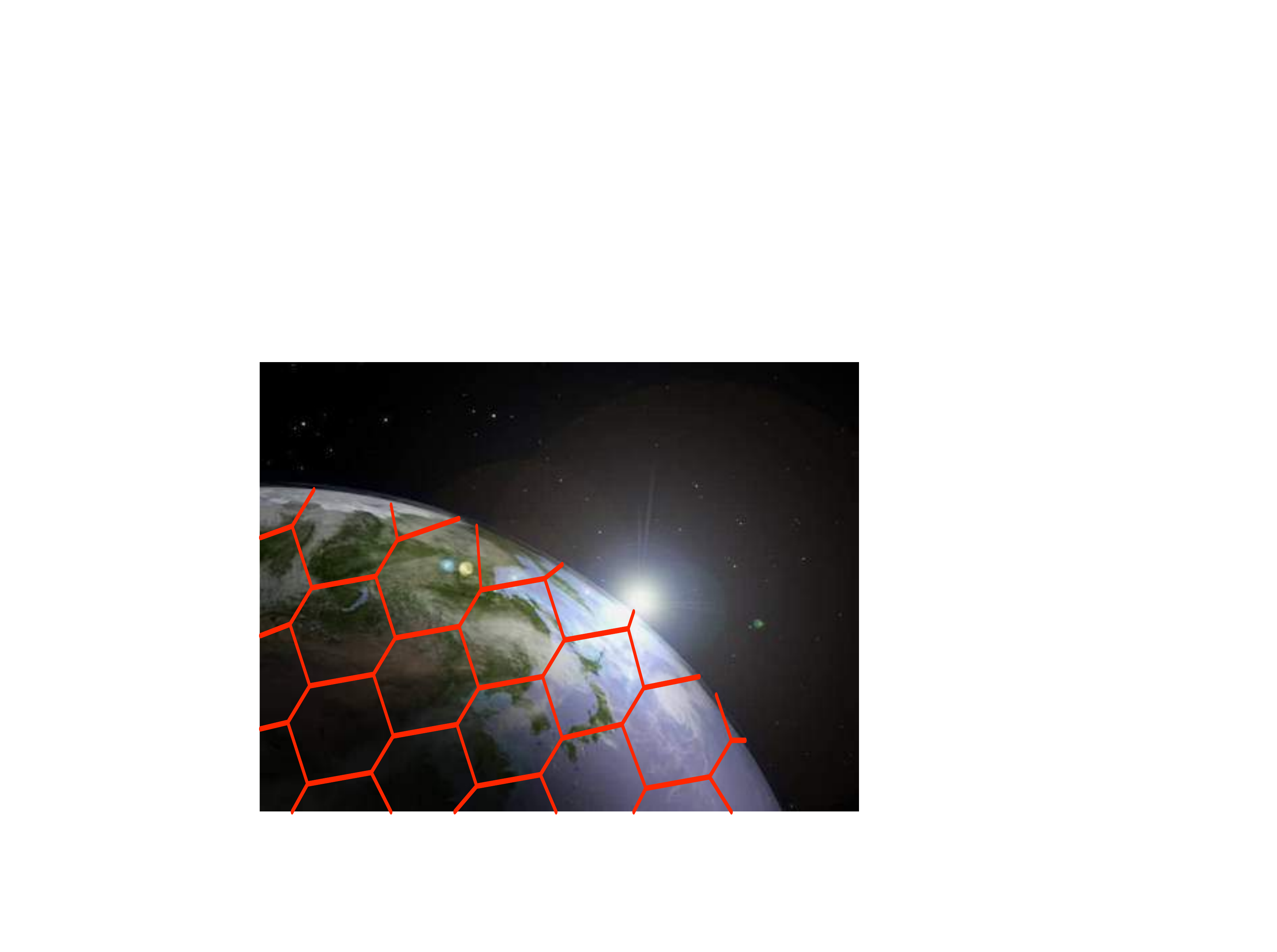}
\captionspacefig \caption{A honeycomb lattice\index{Honeycomb lattice} is the lowest order two-dimensional lattice that could be employed to construct a satellite constellation\index{Constellation network} network covering the Earth. Edges represent quantum communications channels, and their intersections are where the satellites reside. Such a network will require next-generation satellites\index{Next-generation!Quantum satellites} with satellite-to-satellite links.}\label{fig:sat_honeycomb}	
\end{figure}

To facilitate a true global network, next-generation satellites\index{Next-generation!Quantum satellites} will need to form a constellation\index{Constellation network} sufficiently dense that every point on the Earth's surface is always within line of sight of at least one satellite (Fig.~\ref{fig:sat_honeycomb}).

\begin{figure}[!htbp]
\if 1\doublecol
\includegraphics[clip=true, width=0.475\textwidth]{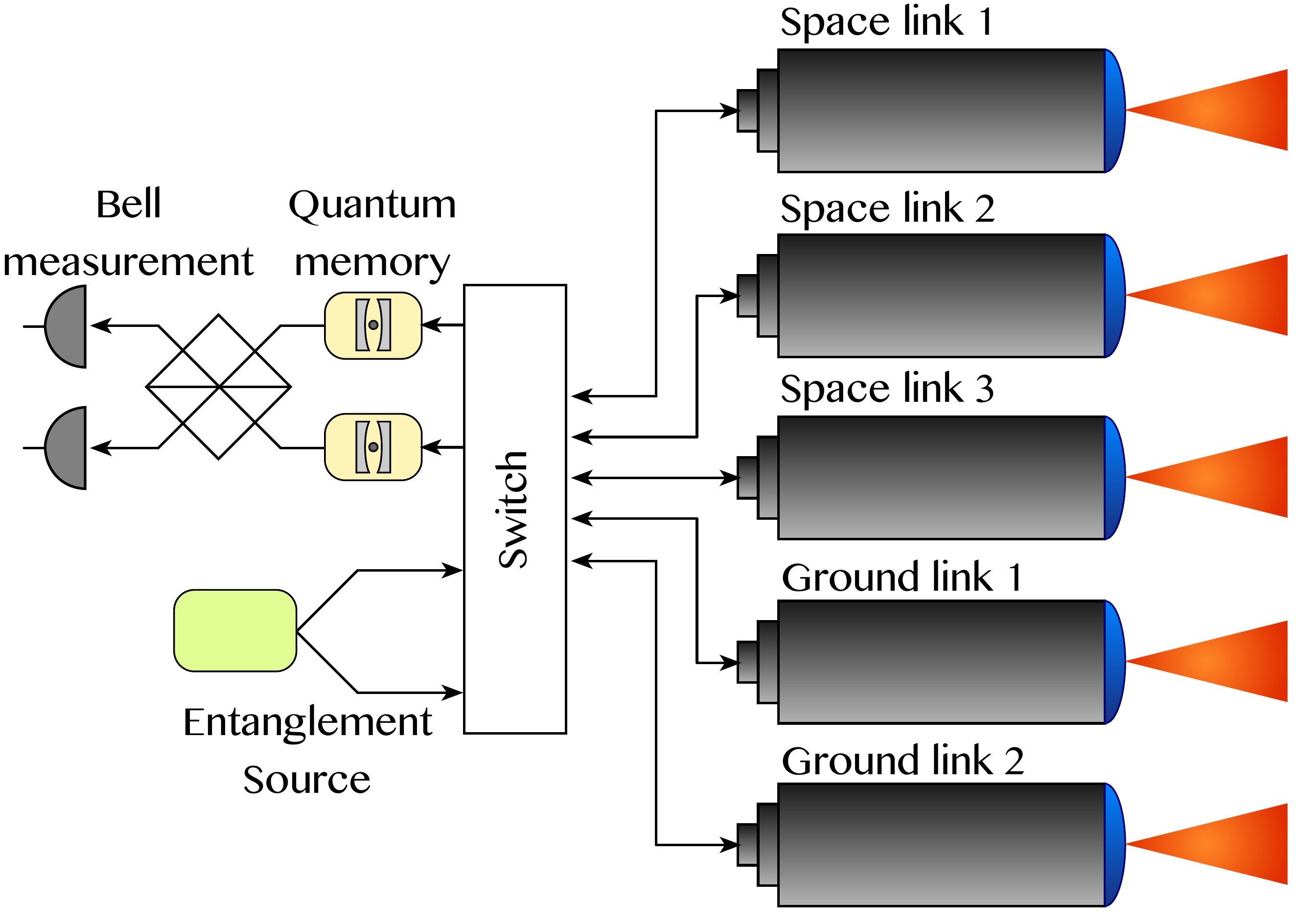}
\else
\includegraphics[clip=true, width=0.65\textwidth]{next_gen_satellite}
\fi
\captionspacefig \caption{A basic layout for how next-generation quantum satellites might be constructed. Each satellite is capable of both entangled state preparation, as well as entangling measurements. There are three space links for communicating with neighbouring satellites so as to enable a honeycomb lattice configuration, as well as two ground links, as per the first-generation satellite. The switch at the centre must be universal to enable arbitrary pairs of telescopes to couple with either the entanglement source or entangling measurement. The quantum memories\index{Quantum memory} prior to the entangling measurement facilitate synchronising distinct photons with different arrival times such that they can interfere.}\label{fig:next_gen_sat}	
\end{figure}

To enable a constellation, the satellites will need satellite-to-satellite links in addition to the satellite-to-ground links, such that they can relay the entanglement around the curvature of the Earth to overcome line-of-sight limitations. Additionally, they will need to do more than just prepare entangled states, but also perform entangling measurements, such that they can be configured as a quantum repeater network\index{Quantum repeater networks}. A concept model for a next-generation satellite with these essential capabilities is shown in Fig.~\ref{fig:next_gen_sat}.

While the next-generation satellite may appear only incrementally more complex than the first-generation one, it is in fact far more technologically challenging. The main obstacle is that when performing entangling measurements, photons must arrive at the detector simultaneously. Obviously this is hard to enforce in space over long distances on fast-moving objects. Therefore quantum memories\index{Quantum memory} will be required, such that the first of two arriving photons is held in memory until the second one arrives, at which point it is read out from memory and the two photons are jointly measured. Unfortunately, such quantum memories are still very much in their infancy, and not reliable enough or of sufficient quality that they are ready for prime-time applications like a global space-based repeater network. It is unclear how far-off these technologies are, despite being under intense investigation.

A global constellation network\index{Constellation network} may require hundreds or thousands of individual satellites. The key to deploying such a network will be via economies of scale\index{Economies of scale}. We must design a single standardised satellite (for example along the lines of that shown in Fig.~\ref{fig:next_gen_sat}), rather than a variety of more specialised models, make it as minimalistic as possible, and then mass produce them on a large scale. With this approach we can hope for economical deployment of a true space-based point-to-point\index{Point-to-point (P2P)!Network} global network.

The Chinese have successfully launched and demonstrated the first quantum satellite. This marks the beginning of the quantum space race\index{Space race}. Who will respond? For he who achieves a global network first will wield a huge competitive technological advantage in the upcoming era of the quantum internet and all its foreseeable and unforeseeable applications.

\latinquote{Carpe futurum.}

%
% The Quantum Mind
%

%\input{Sections/essay_the_quantum_mind}

%
% NISQ
%

\section{The near future: Noisy intermediate-scale quantum technology (NISQ)}\index{Noisy intermediate-scale quantum technology (NISQ)}\label{sec:NISQ}

\sectionby{Zixin Huang}\index{Zixin Huang}

\famousquote{The future belongs to those who believe in the beauty of their dreams.}{Eleanor Roosevelt}
\newline

\dropcap{I}{n} the near- to medium-term we are unlikely to make sufficient technological advances to realise fully scalable, fault-tolerant, universal quantum computation. But that doesn't mean we will have no quantum capabilities at all! Noisy intermediate-scale quantum technology (NISQ) refers to quantum processors which may be available in the next few years, with around 50 to a few hundred qubits. These are going to be noisy and will not have full quantum error-correcting capabilities. They are likely to be special-purpose devices targeted at specific applications, possibly yielding only approximate answers owing to the absence of fault-tolerance \cite{bib:preskill2018quantum}.

Although fully universal, fault-tolerant quantum computers are still somewhat distant, with advances in quantum control, we are now in the position to explore a new frontier of physics, where we have quantum entanglement as part of our computational toolbox. We may not yet have \textit{all} quantum capabilities, but we at least have some!

Scalable quantum computers, unlike classical ones, will be able to efficiently simulate any process that physically occurs in nature, enabling us to study the properties of complex molecules and new materials. This confidence is based on quantum complexity arguments\index{Computational!Complexity}, and our eventual capabilities to perform quantum error correction (which is admittedly very challenging and a potentially long-term vision). Both are based on quantum entanglement, a type of correlation between systems uniquely quantum mechanical, with no classical analogue. We have strong evidence that quantum computers have capabilities beyond classical computation. To illustrate this, consider the following:

\begin{itemize}
\item Quantum complexity: we have strong reason to believe that some tasks efficient on quantum computers may be computationally difficult classically. The best-known example is Shor's algorithm \cite{bib:ShorFactor}, allowing us to factorise large numbers exponentially  faster than using the best classical methods. Whilst we do not have a proof that an efficient classical algorithm doesn't exist, the brightest of mathematicians have been trying to find one for decades to no avail. Integer factorisation has significant implications for cryptography, where the security of some codes is underpinned by the believed computational complexity of this particular problem. 
\item Complexity theory arguments: computer scientists have shown that quantum states which can be easily prepared with a quantum computer have super-classical properties. For example, given single photons input into a multi-mode interferometer, it's hard for a classical computer to sample the probability distribution at the output, the so-called \textsc{BosonSampling} problem\index{Boson-sampling}. On the other hand, a quantum computer can trivially implement this experiment.
\item No known classical algorithm can efficiently simulate a universal, fault-tolerant quantum computer, or simulate general quantum systems.
\end{itemize}

As we see, there is a clear distinction between what is hard classically and quantum mechanically. Intense research efforts are being dedicated to understanding which problems exactly are hard for a classical computer but easy for a quantum one.

The huge obstacle that lies between us and building a scalable quantum computer is the need to keep the system isolated from the environment to minimise noise (environmental noise is the arch-enemy of quantum computation!), at the same time being able to control it with extraordinary precision. Eventually, we expect to be able to protect quantum systems using quantum error correction. However, in order to perform quantum error correction, we currently believe that perhaps $10^3$-$10^4$ physical qubits will be required to encode each logical qubit (depending on the physical architecture and its associated error rates). This adds huge overheads to the number of physical qubits needing to be individually prepared, manipulated and measured, all with extremely high fidelity. Therefore, reliable fault-tolerant quantum computers with quantum error correction are not likely going to be available in the near future.

In terms of the number of qubits, 50 is a significant number because it approximates the number of qubits we can still simulate by brute-force\index{Brute-force} with our most powerful existing classical computers \cite{bib:boixo2018characterizing} -- a benchmark for the meaning of the term \textit{quantum supremacy}\index{Quantum supremacy}. The main question is: when will quantum computers be able to solve useful problems faster than classical ones? This leads us onto several potential uses for limited quantum computation in the NISQ era:

\subsection{Quantum optimisers}\index{Quantum optimisation}

For many problems, there is a big gap between the approximation achieved by classical algorithms and the barrier of exact-case \textbf{NP}-hardness. We do not expect quantum computers to efficiently solve worst-case \textbf{NP}-hard problems, however, quantum devices may be able to find better \textit{approximate} solutions to such problems, or at least find such approximations more quickly. The vision for using NISQ to solve optimisation problems is a hybrid quantum-classical algorithm\index{Hybrid!Algorithms}. In this scheme we use the quantum device to produce and manipulate an $n$-qubit state, measure the qubits, then process the measurement outcomes classically. This then is utilised as feedback for the next round of quantum state preparation and evolution. The cycle is repeated until convergence is obtained to a quantum state from which the approximate answer can be extracted. Two such algorithms are known as \textit{quantum approximate optimisation algorithms} \cite{bib:farhi2014quantum}\index{Quantum approximate optimisation algorithms}, and \textit{variational quantum eigensolvers} \cite{bib:mcclean2016theory}\index{Variational quantum eigensolvers}.

\subsection{Quantum machine learning}\index{Quantum machine learning}

\famousquote{Many do not lose their mind because they do not have one.}{Arthur Schopenhauer}
\newline

Much of the quantum machine learning (QML) literature builds on algorithms which speed up problems in linear algebra \cite{bib:biamonte2017quantum}. One of the potentials for QML rests upon QRAM -- quantum random-access memory\index{Quantum random-access memory (QRAM)}. For classical data processing, by using QRAM we may be able to represent a large amount of classical data, $N$-bits, using only $O(\log N)$ qubits, an exponential improvement in resource efficiency. However, the bottleneck may be in the encoding/decoding of the QRAM, which may seemingly mitigate potential gains, owing to the fact that measurements yield only one element at a time, not the full exponentially-large structure. QML may find applications in a more natural setting where both the input an output are quantum states, for example, to control a quantum system, or in learning probability distributions where entanglement plays an important role.

\subsection{Quantum semidefinite programming}\index{Quantum semidefinite programming}

Semidefinite programming is the task of optimising a linear function, given some matrix inequality constraints. Classically, the problem can be solved in time polynomial in matrix size, and the number of constraints.

A quantum algorithm has been shown to find an approximate solution to this problem with an exponential speedup \cite{bib:brandao2017quantum, bib:brandao2017exponential}. In this algorithm, the initial state is a thermal state\index{Thermal!States} that is a function of the input matrices for the semidefinite program. The success of the implementation depends on whether the particular thermal state can be efficiently prepared. The output is a quantum state, which approximates the optimal matrix. The quantum state can be measured to extract (via sampling) features of this matrix. 

The crucial feature in the quantum algorithm is the preparation of a thermal state of non-zero temperature, suggesting the algorithm may be intrinsically robust against thermal noise -- this would be a fantastic trait to exhibit in the NISQ era of no fault-tolerance. It's therefore entirely possible that a quantum solver for semidefinite programs might be achievable with near-term NISQ technology.

\subsection{Quantum dynamics}\index{Quantum dynamics}

As was stressed previously, quantum computers are very well suited to studying highly entangled, multi-particle systems. It's the natural platform to simulate entangled states, where quantum computers appear to have a clear intrinsic advantage over classical ones.
 
With a universal quantum computer, we anticipate that studying quantum chemistry\index{Quantum chemistry} (especially noisy quantum chemistry) will be enabled. Ideally, if the noise model in the quantum computer can be cleverly mapped to be isomorphic to the noise present in the physical system being simulated, then noise becomes a feature not a bug! This could be used in the design of new pharmaceuticals\index{Drug!Design}, for example, as well as catalysts for improving the efficiency of nitrogen fixation\index{Nitrogen fixation} or carbon capture\index{Carbon capture}. We may be able to find new materials with better resistive properties, leading to more efficient transmission of electricity. However, these promises may not be fulfilled with NISQ, because algorithms to accurately simulate large molecules and materials may not succeed without quantum error correction.

We do know that classical computers are particularly inefficient at simulating quantum dynamics, i.e how highly entangled quantum states will evolve over time. Here quantum computers have a particularly obvious advantage, and one example would be quantum chaos\index{Quantum chaos}. In these systems entanglement spreads very rapidly. Insights might be gained using noisy devices on the order of only 100's of qubits, a perfect regime for the NISQ era.

We've barely had a glimpse of the promises of NISQ. But it's clear that although near-term devices will be limited, they may nonetheless open up exciting new prospects and computational applications, beyond the capabilities of present-day classical machinery.

\latinquote{Carpe futurum.}

%
% Future of QKD
%

\section{The future of quantum cryptography}\label{sec:essay_future_QKD}\index{Quantum key distribution (QKD)}\index{Quantum cryptography}
 
\sectionby{Zixin Huang}\index{Zixin Huang}
 
\famousquote{How long do you want these messages to remain secret?\ldots I want them to remain secret for as long as men are capable of evil.}{Neal Stephenson}
\newline
 
\dropcap{Q}{uantum} cryptography is the first field in quantum information task to reach commercialisation. At its early stages, quantum cryptography was almost synonymous with quantum key distribution (QKD), but has since branched and became one of the fastest growing areas in quantum information.
The purpose of QKD is to distribute a secret-key between two trusted parties who share a quantum channel, as well as a classical channel for authentication. Unlike current cryptography systems, which are secure based on the presumed limitations of an adversary's computer (\textit{computational security}\index{Computational!Security}), the security of QKD is based on the laws of quantum mechanics, providing guaranteed security unless our understanding of quantum physics is inherently wrong (\textit{information theoretic security}\index{Information-theoretic!Security}). In this section we discuss some of the challenges in QKD, as well as other aspects of quantum cryptography beyond traditional QKD.

The typical setting of QKD is as follows. There are two trusted parties who want to establish a secret-key, Alice and Bob. They share two channels: a quantum channel, which allows them to send quantum states (encoded in photons or other states of light) to one another; and, a classical channel, with which they can send classical messages. The communication over the classical channel is assumed to be public and completely insecure, and the eavesdropper, Eve, has full anonymous access to it. However, Eve cannot modify messages shared over the classical channel.

The quantum channel is subject to possible manipulation by Eve. The task of Alice and Bob is thus to guarantee security against an adversarial eavesdropper. The typical protocol assumes that Alice and Bob do not share any secret to begin with. The origin of the security of QKD springs from the fundamentals of quantum mechanics, that is, any act of measurement by an observer on a quantum state necessarily induces a change in the state -- measurement collapse\index{Measurement!Collapse}. This means that in combination with classical communication, actions of an eavesdropper cannot go undetected, ruling out intercept-resend attacks\index{Intercept-resend attacks} by Eve.

The ultimate goal of a QKD network is long distance secure quantum communication with imperfect sources.

Despite the significant advances in both the theoretical and experimental development of QKD, a number of challenges remain for it to be widely adopted in securing everyday communications \cite{bib:RevModPhys.81.1301, bib:diamanti2016practical}. Experimentally, much effort is being invested into improving the performance of QKD systems. On the theoretical side, showing the security of a QKD system with finite key-size is also a challenge, because information-theoretic security is achieved only when immunity against the most general (coherent) attack is proven \cite{bib:diamanti2016practical}.

\subsection{Performance}

Some of the criteria for assessing the performance of a QKD scheme include key-rate\index{Key-rate}, range\index{Range}, cost and robustness\index{Robustness}.

\subsubsection{Key-rate}\index{Key-rate}

Currently, a strong disparity exists between classical and quantum key distribution rates. Classical optical communication delivers on the order of $\sim$100Gbits/s per wavelength (for a frequency-multiplexed implementation\index{Multiplexing}), whereas communication rates only on the order of $\sim$Mbit/s are achievable using current QKD implementations.

The obtained key-rate depends on the performance of the detector used for measurement. For QKD based on single-photon detection techniques, to achieve a high bit-rate, one requires true single-photon states, in combination with detectors with high efficiency and short dead-time\index{Dead-time}, both of which effectively induce loss, mandating more trials. Current developments are promising, with a reported quantum efficiency of $93\%$ at telecom wavelengths \cite{bib:marsili2013detecting}.

For continuous-variable\index{Continuous-variables} QKD systems, increasing the bandwidth of the homodyne/heterodyne detectors\index{Homodyne detection}\index{Heterodyne detectors} whilst keeping the electronic noise low is essential.

\subsubsection{Range}\index{Range}

Extending the range\index{Range} of QKD systems is a major challenge and driving factor for QKD in terms of future network applications. Two approaches are being pursued -- free-space\index{Free-space} and quantum repeaters\index{Quantum repeaters}. A quantum repeater, similar to its classical analogue, is a device that can extend the range of quantum communication between sender and receiver. However, one cannot amplify the signal that contains the quantum information, owing to the no-cloning theorem\index{No-cloning theorem}, which prohibits making copies of unknown quantum states. The fact that an intercept-resend attack\index{Intercept-resend attacks} by Eve must disrupt the state of the system is the basis for the security of QKD -- one of the major limitations imposed by quantum mechanics works to our advantage!

A quantum repeater effectively needs to restore the quantum information without measuring it directly, and is extremely technologically challenging. Over optical fibre networks, the standard loss for 1550nm wavelength light is 0.2dB/km. Over long enough distances, this unavoidable loss will reduce the key-rate to a level of little practical relevance, therefore a ground-based solution would be to divide the entire channel into segments, where two partners exchange pairs of entangled photons and store it in a quantum memory\index{Quantum memory} \cite{bib:BDCZ98, bib:dur98}.

The second is to use free-space\index{Free-space} quantum communication techniques via satellite links. Satellite QKD is achievable with present-day technology. Here satellites are used as intermediate trusted nodes\index{Trusted nodes} for communication between locations on the ground. Direct links can be established between ground stations and the satellite, thus enabling communication between parties separated by long distances, potentially relaying across a satellite constellation network to overcome line-of-sight limitations from the Earth's curvature\index{Line-of-sight}\index{Earth curvature}. Satellite QKD suffers comparatively very low loss between satellites in orbit, but the satellite-to-ground links\index{Satellites!Satellite-to-ground communication}, which cannot be avoided at the endpoints, suffer around 40dB loss when propagating through the effective atmospheric thickness\index{Effective atmospheric thickness} of $\sim$10km when the satellite is directly overhead (and worse for satellites with lower azimuth). The atmospheric loss is a major hurdle, since distribution of a Bell pair between two ground stations effectively incurs 80dB inefficiency, meaning that only 1 in every 100,000,000 Bell pairs are successfully distributed \latinquote{Stupor}. Nonetheless, in China, satellite QKD over 1200km has been demonstrated \cite{bib:liao2017satellite}, sufficient for sharing a secret-key\index{Private-key} for private-key cryptography\index{Private-key!Cryptography} with guaranteed key secrecy.

\subsubsection{Cost \& robustness}
 
For QKD systems to be consumer-friendly, low cost and robustness are crucial features. Preferably QKD systems should make use of existing data fibre-optic infrastructure, since the use of dark fibres are not only expensive, but often unavailable \cite{bib:diamanti2016practical}, and there is a big economic incentive to reuse existing infrastructure rather than rebuild it from scratch. Single-photon detectors at room temperatures are also desirable, because this can remove the requirement for cryogenic cooling\index{Cryogenic cooling}, hence reducing power consumption and making consumer systems far more practical.

Integrated photonic platforms are being explored to reduce cost, since miniaturisation\index{Miniaturisation} can lead to light-weight, low-cost QKD modules that can be mass-manufactured, essential for economies of scale\index{Economies of scale}. 

Currently, two platforms are being explored: silicon\index{Silicon} \cite{bib:lim2014review}, and indium phosphide\index{Indium phosphide} \cite{bib:smit2014introduction}. A reconfigurable QKD system employing an In-P transmitter and silicon detectors has been demonstrated in the laboratory \cite{bib:sibson2017chip}.

\subsection{New protocols}

In parallel to hardware development, research efforts are being directed towards finding new QKD protocols which can outperform existing ones. Two of these are high-dimensional (HD) QKD\index{High-dimensional quantum key distribution} and the Round-Robin differential phase-shift protocol (RR-DPS)\index{Round-Robin!Differential phase-shift protocol}.

HD QKD aims at encoding more than one bit in each detected photon, which can increase the information capacity when the photon rate is limited. Security proofs against collective attacks are being developed, and an experiment has demonstrated an information capacity 6.9 bits per coincidence rate at 2.7Mbit/s over 20km \cite{bib:zhong2015photon}.

The RR-DPS protocol \cite{bib:sasaki2014practical} removes the need to monitor signal disturbance. In a conventional QKD protocol, the noise parameter needs to be estimated; and if high precision is required, the portion of the signal that is sacrificed increases, thus decreasing the efficiency of the protocol \cite{bib:cai2009finite, bib:hayashi2014security}. This protocol has a high tolerance to the qubit error rate ($<50\%$) \cite{bib:xu2015discrete}, and makes it attractive for implementation when high systematic errors are unavoidable.   

However, currently, neither of the protocols out-compete the more mature decoy-state BB84\index{BB84 protocol}\index{Decoy states}.

\subsection{Challenges in security}

Although QKD protocols are provably information-theoretically secure, physical implementations often contain imperfections which are not considered in the theoretical model -- no experiment ever perfectly matches its design! Attacks can be designed to exploit such imperfections, on either the source or the detector side.

Tab.~\ref{tab:attacks}, taken from \cite{bib:lo2014secure}, summarises some attacks against certain commercial and research systems.

\startnormtable
\begin{table*}[!htbp]
\begin{tabular}{|c|c|c|c|} 
 \hline
 Attack &  Targeted component & Tested system & References\\ 
  \hline
  \hline
Time shift
        & Detector & Commercial & \cite{bib:qi2005time, bib:PhysRevA.78.042333, bib:PhysRevA.74.022313}\\
Time information & Detector & Research & \cite{bib:lamas2007breaking} \\
Detector control & Detector  &   Commercial & \cite{bib:lydersen2010hacking, bib:yuan2010avoiding}\\
Detector control  & Detector  & Research & \cite{bib:gerhardt2011full} \\
Detector dead-time      & Detector  & Research   & \cite{bib:weier2011quantum}      \\
Channel calibration    & Detector  &  Commercial  & \cite{bib:jain2011device}      \\
Phase remapping  &  Phase modulator & Commercial & \cite{bib:xu2010experimental} \\
Phase information & Source & Research & \cite{bib:tang2013source}          \\
Device calibration  & Local oscillator & Research & \cite{bib:jouguet2013preventing} \\
                \hline
\end{tabular}
\captionspacetab \caption{\label{tab:attacks} Summary of various attacks against some commercial and 
research QKD systems.}
\end{table*}
\startalgtable

To regain security, a number of solutions have been proposed:

\subsubsection{QKD with imperfect sources}

The source is typically less vulnerable to attacks because Alice can prepare her quantum states in a protected environment, and we expect that she can characterise her source. Therefore, flaws in state preparation can be easily incorporated into the security proof\index{Security!Proofs}.

Loss-tolerant\index{Loss!Tolerance} protocols have been proposed \cite{bib:PhysRevA.90.052314}, and further developed by \cite{bib:PhysRevA.92.032305}, where decoy state\index{Decoy states} QKD with tight finite-key security has been employed. A wide range of imperfections with the laser source have been taken into account \cite{bib:mizutani2015finite}, including intensity fluctuations. A security proof\index{Security!Proofs} has shown that perfect phase randomisation is also not necessary \cite{bib:cao2015discrete}.

This provides strong evidence that secure quantum communication with imperfect sources is feasible \cite{bib:diamanti2016practical}. Intuitively, QKD with imperfect sources is viable because by assuming that states prepared by Alice are qubits, Eve cannot unambiguously discriminate Alice's states \cite{bib:diamanti2016practical} -- quantum measurement collapses quantum states\index{Measurement!Collapse}. 

\subsubsection{Measurement-device-independent QKD}\index{Measurement-device-independent quantum key distribution}

To prove security\index{Security!Proofs} for Bob's measurement device is more problematic, since Eve has complete access to the quantum channel and she can send any signal. 

One candidate for a long-term solution to side-channel attacks\index{Side-channel attacks}\footnote{A side-channel attack is one which exploits knowledge of the imperfect implementation of a system (e.g details of source or detector characteristics) to compromise security, rather than a weakness in the theoretical model underpinning it (normally approached using cryptanalysis\index{Cryptanalysis}).} is device-independent (DI) QKD \cite{bib:PhysRevLett.98.230501}\index{Device-independent quantum key distribution}. This relies on the violation of a Bell inequality\index{Bell!Inequality} \cite{bib:hensen2015loophole}, and the security can be proven without knowledge of the implementation. However, the expected secure key-rate is low even over short distances. A more practical approach is measurement-device-independent (MDI) QKD \cite{bib:PhysRevLett.108.130503}\index{Measurement-device-independent quantum key distribution}, which is immune to side-channel attacks\index{Side-channel attacks} against the measurement device. Here the device is treated as a black box\index{Black box}, and can be untrusted. However, an important assumption for MDI QKD is that Eve cannot interfere with the state preparation process, which is practically reasonable. 

Another candidate is detector-device-independent (DDI) QKD \cite{bib:lim2014detector, bib:PhysRevA.92.022337}\index{Detector-device-independent quantum key distribution}, which has been designed to take advantage of both the strong security of MDI-QKD, with the efficiency of conventional QKD. However, DDI-QKD has been shown to be vulnerable to certain attacks \cite{bib:PhysRevLett.117.250505}. 

The MDI-QKD protocol has been extended to the continuous variable\index{Continuous-variables} framework. However, this system requires homodyne detectors\index{Homodyne detection} with efficiency $>85\%$, and a reliable phase reference\index{Phase!Reference} between Alice and Bob.

We have discussed some significant remaining challenges in QKD. These range from theoretical security proofs to hardware developments. Advances in QKD will not only enable point-to-point quantum communication\index{Point-to-point (P2P)!Communication}, but have implications for a range of network applications, such as quantum secret sharing\index{Quantum secret sharing} \cite{bib:cleve1999share, bib:PhysRevA.61.042311, bib:PhysRevA.71.044301}, blind quantum computing\index{Blind quantum computation} \cite{bib:broadbent2009universal, bib:barz2012demonstration}, quantum anonymous broadcasting\index{Quantum anonymous broadcasting} \cite{bib:christandl2005quantum}, and many more.

As remarked in \cite{bib:diamanti2016practical}, \textit{``Determining the exact power and limitations of quantum communication is the subject of intense research efforts worldwide. The formidable developments that can be expected in the next few years will mark important milestones towards the quantum internet of the future.''}

% 
% Quantum Ecosystem
%

%
% The Quantum Ecosystem
%

\section{The quantum ecosystem}\index{Ecosystems}

\famousquote{The most dangerous worldview is the worldview of those who have not viewed the world.}{Alexander von Humboldt}
\newline

\begin{figure*}[!htpb]
\includegraphics[clip=true, width=\textwidth]{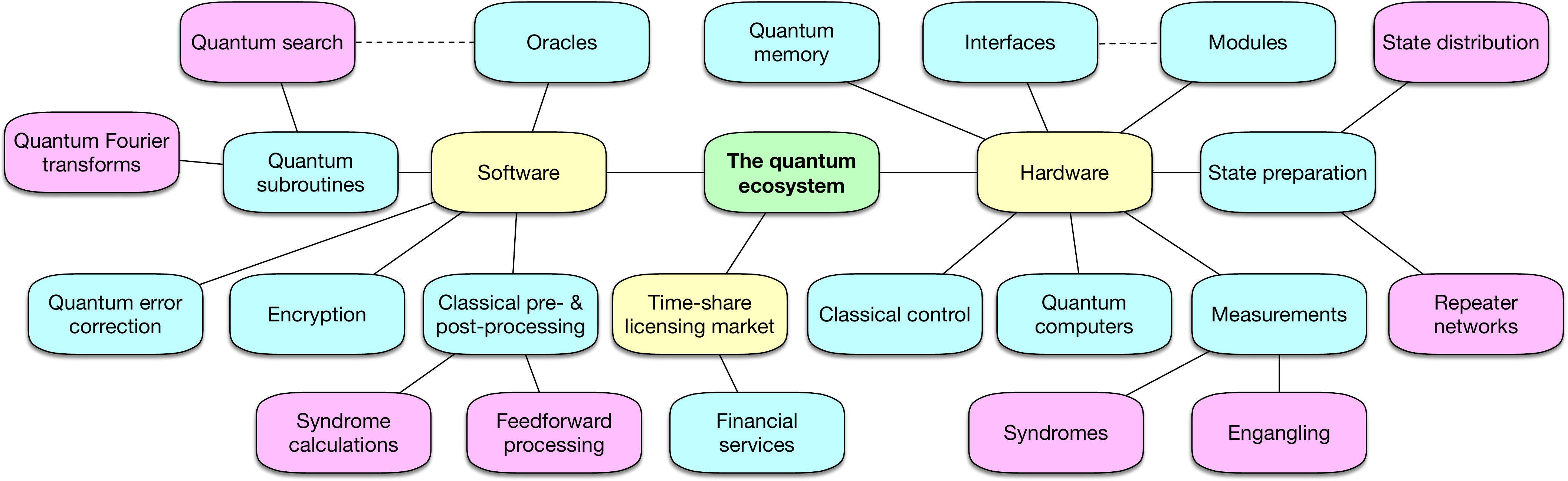}
\captionspacefig \caption{Map of just a few of the elements of the quantum ecosystem that are likely to arise with the advent of the quantum internet. The distinct units could become areas of specialisation for quantum vendors, which might be licensed out or sold to customers as discrete units, or as complete integrated processing pipelines, all outsourced and distributed over the quantum network.}\label{fig:ecosystem}\index{Ecosystems}	
\end{figure*}

\dropcap{A}{ssociated} with any new computer platform comes a hardware/software \textit{ecosystem}\index{Ecosystems} that evolves around it. If we consider the release of the original iPhone\index{iPhone} and its iOS\index{iOS} operating system, it wasn't just the product itself that was revolutionary, but the third-party software industry that emerged surrounding it, and it wasn't until this software ecosystem became established on the App Store\index{App Store} that the product realised its full potential and became truly transformative.

From the hardware perspective, it wasn't until interfacing standards such as USB\index{USB} and Wi-Fi\index{Wi-Fi} emerged, allowing the plethora of competing hardware products to arbitrarily interconnect and interface with one another, that the hardware realised its full potential.

In the quantum era we anticipate the same phenomena to arise. What will this \textit{quantum ecosystem} look like? Here are some of the elements that vendors might specialise in, providing compatible components for the quantum ecosystem:

\begin{itemize}
\item Quantum operations: vendors selling the capacity for non-trivial state preparation (e.g Bell, NOON\index{NOON states} and GHZ\index{GHZ states} states), or measurements (e.g complex entangling syndrome measurements\index{Syndromes!Measurements}).

\item Software subroutines and libraries:\index{Subroutines}\index{Libraries} much like classical code, many quantum computations (and other quantum protocols) can be decomposed into pipelines of subroutines. There are many quantum operations that arise repeatedly (such as quantum Fourier transforms and syndrome calculations), which vendors might specialise in for outsourcing.

\item Oracles:\index{Oracles} as an essential quantum software building block, oracles will become a fundamental unit for outsourcing. These oracles will store hard-corded or algorithmically-generated databases, or mathematical functions. For example, for use in genetic medicine (Sec.~\ref{sec:genetic_medicine}), such databases could algorithmically generate tables of candidate drug compounds, or they could implement mathematical functions whose input space is to be searched over when quantum-enhancing the solving of \textbf{NP}-complete\index{NP \& NP-complete} problems.

\item Interfacing:\index{Interfacing} \textit{de facto} standards will emerge for interconnecting quantum hardware units. Most notably, standards for optical interconnects will arise.

\item Modularisation:\index{Modularisation} arbitrarily-interconnectable units will develop, allowing quantum hardware to be constructed in an ad hoc, Lego-like\index{Lego} manner. These modules could implement small elements of a larger quantum computation, such as housing a small part of a larger graph state (Sec.~\ref{sec:module}), communications building blocks (such as transmitters or receivers of Bell pairs), or algorithmic building blocks such as quantum Fourier transforms (Sec.~\ref{sec:QFT_alg}).

\item Classical pre-, post- or intermediate-processing:\index{Classical processing} quantum computation, and other quantum protocols, typically require some degree of classical pre- or post-processing. These classical operations can be highly non-trivial. For example, a novel topological quantum error correcting code\index{Quantum error correction (QEC)} (Sec.~\ref{sec:surface_codes}) might require complex encoding, decoding and feedforward operations. Determining and implementing these operations may require complicated optimisation protocols. These might be outsourced to a specialised provider.

\item Classical control:\index{Classical control} many quantum protocols require intermediate classical control, i.e feedforward\index{Feedforward}. For example, in a quantum repeater network we must control the order of entangling operations and track the `Pauli frame'\index{Pauli!Reference frame}, a tally of the corrections accumulated by the final entangled Bell pair.

\item Quantum memory:\index{Quantum memory} storing qubits with long decoherence lifetimes is extremely challenging using today's technology, and it is foreseeable that vendors might specialise in this particular operation, especially once error correction is built into the memory. 

\item Quantum error correction:\index{Quantum error correction (QEC)} any given quantum error correcting code follows a well-defined recipe for encoding, correction, and decoding. Thus, it might become an example of a subroutine specialised in by a dedicated quantum error correction vendor, and licensed out as a building block for embedding into larger, fault-tolerant\index{Fault-tolerance} protocols.

\item Time-share licensing market:\index{Time-share licensing market} a market will emerge for the trade and allocation of time-shares on the global virtual quantum computer (Sec.~\ref{sec:glob_unif_quant_cloud}). Associated financial services industries\index{Financial services industries} will emerge around this marketplace, including secondary markets\index{Secondary markets}, derivative markets\index{Derivative markets}, managed funds in quantum infrastructure\index{Managed funds}, and IPO markets\index{Initial public offering (IPO) markets}.
\end{itemize}

A map of just a few of the potential major hardware and software elements to emerge in the quantum ecosystem is presented in Fig.~\ref{fig:ecosystem}.

\latinquote{Disiecti membra poetae.}

%
% The Quantum Singularity
%

%\input{Sections/essay_the_quantum_singularity}

%\latinquote{Non ducor, duco.}

\sketch{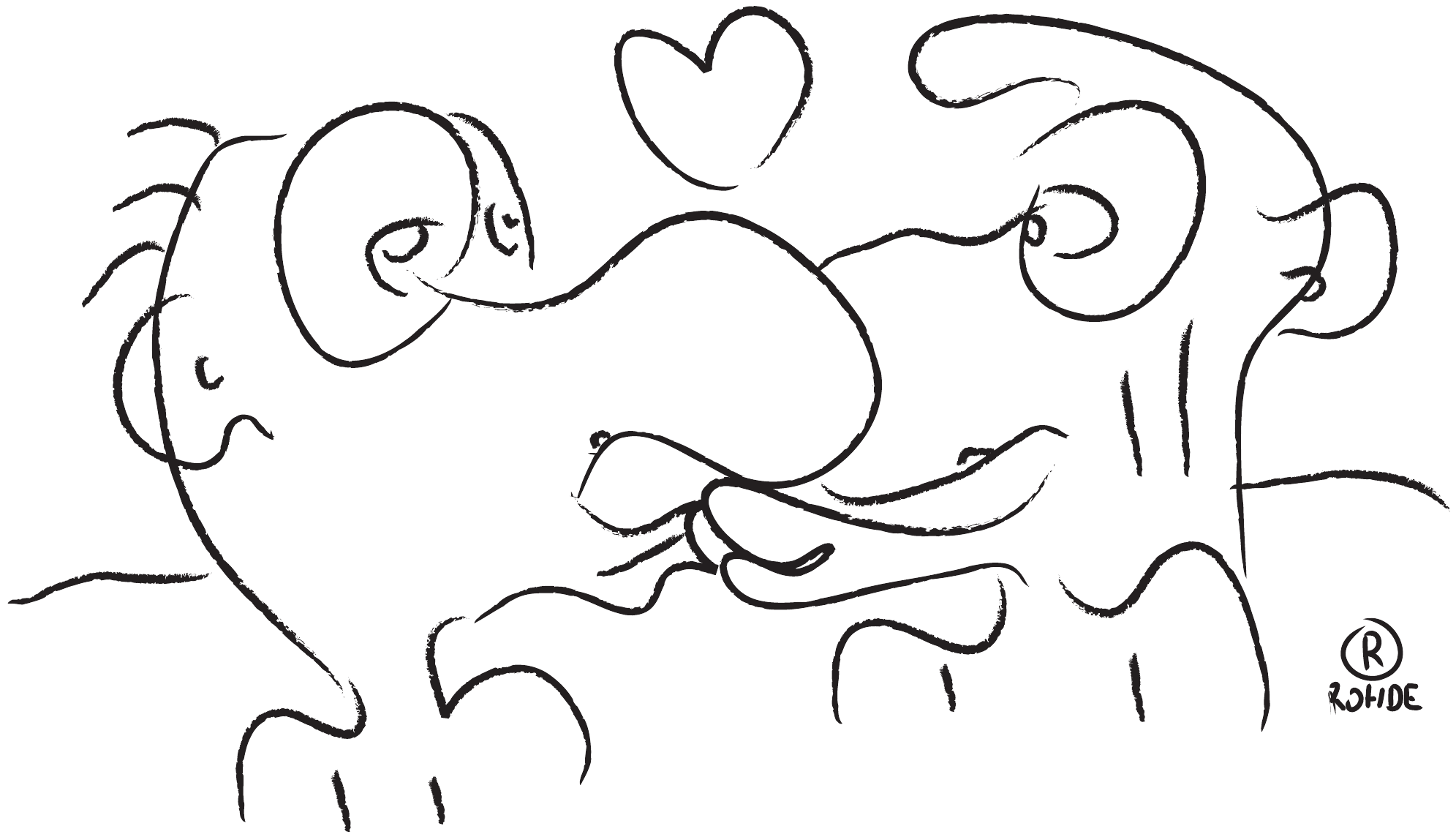}

%
% The end
%

\part{The end}\label{part:the_end}

%
% Conclusion - The Vision of the Quantum Internet
%

\famousquote{When something is important enough, you do it even if the odds are not in your favour.}{Elon Musk}
\newline

\famousquote{Be nice to nerds. Chances are you'll end up working for one.}{Bill Gates}
\newline

\famousquote{We are just an advanced breed of monkeys on a minor planet of a very average star. But we can understand the Universe. That makes us something very special.}{Stephen Hawking}
\newline

\section{Conclusion -- The vision of the quantum internet} \label{sec:vision_quant} \index{Vision of the quantum internet}\index{Conclusion}

\famousquote{We will either go down as the world's greatest statesmen, or its greatest villains}{Hermann G{\" o}ring}
\newline

\dropcap{Q}{uantum} technologies, particularly quantum computing, will truly revolutionise countless industries. With early demonstrations of key quantum technologies -- such as QKD, long distance quantum teleportation, and quantum computing -- becoming a reality, it is of utmost importance that networking protocols be pursued now.

We have presented an early formulation and analysis of quantum networking protocols with the vision of enabling a future quantum internet, where quantum resources can be shared and communicated in much the same way as is presently done with digital assets. Whilst it's hard to foresee exactly how future quantum networks will be implemented, as there are many unknowns, many of the central ideas presented here will be applicable across architectures and implementations on an ad hoc basis.

There are a number of schools of thought one might subscribe to when quantum networking. One might demand perfect data integrity and best-case network performance. But that would come at the expense of necessitating an all-powerful central authority to oversee all communications, ensuring that scheduling was absolutely perfect -- a potentially very challenging optimisation problem. Or one might tolerate lost data packets or suboptimal performance, at the expense of limiting applicability, but with the benefit of improved flexibility and reconfigurability. Or maybe some arbitrary compromise between different metrics and attributes is best. These are open questions that needn't have concrete, one-size-fits-all answers. They certainly needn't be answered right now.

The QTCP framework we presented is sufficiently flexible and extensible that these questions can be answered and enforced independently by different subnets, depending on their individual characteristics and requirements, in much the same way that every organisation connected to the classical internet today is free to structure their own LAN as they please, enforcing their own internal network policies.

The quantum internet will allow quantum computation to become distributed, not just outsourced. In the same way that many present-day classical algorithms are heavily parallelised and distributed across large clusters, CUDA cores\index{CUDA}, or even across the internet itself (e.g the SETI project\index{SETI project}), quantum networks will allow the distribution of quantum computation across many nodes, either in parallel, in series, or in a modularised fashion. This will be pivotal to achieving scalability. Keeping in mind that the classical-equivalent power of a quantum computer may grow exponentially with the number of qubits, it is highly desirable to squeeze out every last available qubit for our computations -- every qubit is worth more than the last!

Combined with recent advances in homomorphic encryption and blind quantum computation, commercial models for the distribution of quantum computation will emerge, allowing computational power to be outsourced, with both client and server confident in the security of their data and proprietary algorithms. This is a notion that is challenging on classical computers, but will be of utmost importance in quantum computing, where it is expected sensitive or valuable data and algorithms will often be at stake.

From the security perspective, the global quantum internet will enable an international QKD communications network with perfect secrecy, guaranteed to be information-theoretically secure by the laws of physics. This will be of immense economic and strategic benefit to commercial enterprises, governments, and individuals. Classical cryptography is already a multi-billion dollar industry worldwide. Quantum cryptography will supersede it, and be of especial importance in the era of quantum computers, which compromise some essential classical cryptographic protocols, such as RSA, which forms the basis of most current internet encryption, digital signatures, and the Blockchain/Bitcoin protocols. Not only is quantum cryptography being pursued optically, but even credit cards with embedded quantum circuitry are being actively developed to prevent fraud. Inevitably, this will require the communication between bank automats and servers to be mediated by a quantum network.

Already, off-the-shelf QKD systems are available as commodity items, from vendors such as MagiQ\index{MagiQ} and ID Quantique\index{ID Quantique}, which may be simply interconnected via an optical fibre link, thereby implementing end-to-end quantum cryptography in a modularised fashion. This is of a similar flavour to, and first technological step towards, modularised quantum computing, which would greatly enhance the economic viability and scalability of general purpose quantum computing by paving the way for the mass production of elementary interconnectable modules as commodity items.

We have focussed our attention thus far on the application of quantum networking to quantum information processing applications, such as quantum computing and quantum cryptography. However, with plug-and-play quantum resources available over a network, one might envisage far greater applicability than just these.

Of particular interest are the implications of quantum networking to basic science research. Presently, experimental quantum physics research is limited to well-resourced labs with access to state of the art equipment. With the ability to license these assets over a network, and dynamically interconnect them on an ad hoc basis, the ability to construct all manner of quantum experiments could be extended to all. An undergraduate laboratory would now have the ability to approach a host to politely borrow their state engineering technologies, send it to another with the ability to perform some evolution to that system, and to yet another to perform measurement and analysis of the output -- all from an undergrad lab equipped with nothing more than desktop PCs. This has broad implications for basic science research, opening it up to aspiring researchers across the globe, regardless of their direct access to cutting-edge tools. This will greatly expand the intellectual base for conducting quantum experimentation to the entire global scientific community, decimating the scientific monopolies controlled by a handful of world-leading, highly-resourced experimental teams.

The reality is that we are only just beginning to understand the full potential for quantum technologies, and as we learn more we will inevitably find new uses for networking them. The full potential of digital electronics was never fully realised (or anticipated) until the emergence of the internet. It is to be expected the same will hold in the quantum era, an era only in its inception.

Large-scale quantum computing may still seem a formidable, and somewhat long-term challenge. But it isn't likely to remain so. Once we have mastered the technological art of preparing qubits and implementing high-fidelity entangling operations between them, it's just a matter of sitting back and watching Gordon Moore\index{Moore's Law} perform his witchcraft, and scalability of quantum technology, and its rapid market-driven reduction in cost, will quickly ensue. The quantum internet will drive this rapid development by expanding both the supply and demand for access to this technology, and through unification of computational resources allow them to massively enhance their collective computational power, beyond their individual capabilities.

It is essential for the adoption and development of quantum technology, that quantum networking infrastructure be sufficiently well developed that it is ready to be deployed the minute the first useful, post-classical hardware becomes available. The proliferation of the defining technology of the 21st century depends upon it.

\begin{figure*}[!htbp]
	\includegraphics[clip=true, width=\textwidth]{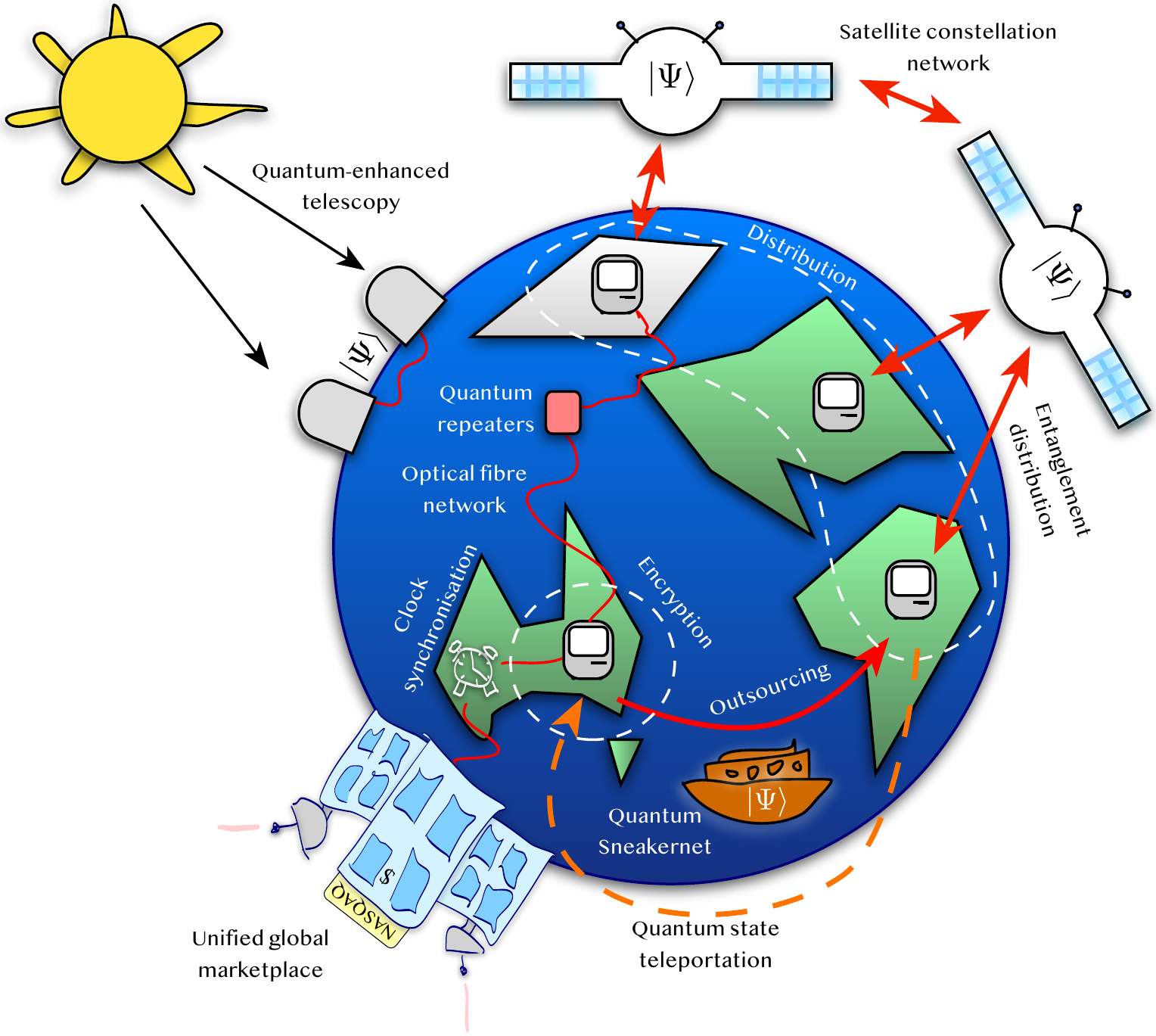}
	\captionspacefig \caption{Overview of some of the essential services integrated into a future globally-unified quantum internet ecosystem.}\label{fig:integrated_services_overview}
\end{figure*}

%
% How to learn more & get your hands dirty
%

\section{How to learn more \& get your hands dirty}\index{Learn more}

\dropcap{F}{or} the reader interested in delving deeper into quantum technology with a hands-on approach, there are a number of freely available software tools for simulating various quantum technologies, including quantum computing and quantum networks. We highlight some of the major ones below, all of which are freely available:

\begin{itemize}
	\item SimulaQron\index{SimulaQron} (\url{http://www.simulaqron.org}): Developed by QuTech Delft \cite{bib:AxelDahlbergQron}, a simulator for designing software for the quantum internet. The package includes its own quantum internet programming language for deploying networked quantum algorithms. The institute runs competitions for designing the best quantum internet app.
	\item Google Cirq\index{Cirq} (\url{https://github.com/quantumlib/Cirq}): Google's Python-based SDK for designing quantum software, specifically targeted at development for Google's own in-house quantum computing hardware platform.
	\item Microsoft Quantum Development Kit and the Q\# language\index{Q\#}\index{Microsoft Quantum Development Kit} (\url{https://www.microsoft.com/en-au/quantum/development-kit}): a full SDK for writing code executable on quantum computers. The Q\# language is used to describe quantum operations, which classical code written in C\#\index{C\#} can interface with and control. The platform is extremely versatile and allows highly complex quantum operations to be coded in a high-level, platform-independent manner. The platform includes a simulator for executing code for which the hardware doesn't yet exist.
	\item IBM Quantum Experience\index{IBM!Quantum Experience} (\url{http://www.research.ibm.com/quantum/}): IBM's cloud-based platform for remotely executing simple quantum protocols, as written and uploaded to the cloud by the user. The platform includes QISKit\index{QISKit} (\url{https://qiskit.org}), a Python-based SDK for executing quantum algorithms. It is based on the Open Quantum Assembly Language (OpenQASM) language\index{OpenQASM}.
	\item Rigetti Quantum Cloud Services (QCS)\index{Rigetti Quantum Cloud Services} (\url{https://www.rigetti.com/qcs}): Rigetti's cloud platform for accessing and deploying algorithms to their hardware over the cloud. Development is powered by the Forest SDK\index{Forest SDK}, their own in-house development platform.
	\item Keep looking, as the list is rapidly growing with new participants entering the market regularly\ldots
\end{itemize}

\latinquote{Opus dei.}

\clearpage

%
% Acknowledgments
%

\section*{Acknowledgments}

\begin{itemize}
\item We acknowledge Chris Ferrie, Joseph Clark, Klaus Rohde, Eric Cavalcanti, Charlotte Bridault, Marco Tomamichel, Ryan Mann, Jonathan Olson, Samuel Marks, Nicholas Umashev, Elija Perrier, Kate Hrayssi \& Alexis Shaw for helpful discussions.
\item We thank the authors of QCircuit (\url{http://physics.unm.edu/CQuIC/Qcircuit/}) for providing this very helpful package.
\item Peter P. Rohde is funded by an Australian Research Council Future Fellowship (project FT160100397).
\item ZH is supported by an ARC DECRA Fellowship (DE230100144) ``Quantum-enabled super-resolution imaging”.
\item Y.O. acknowledges support from EPSRC (Grant No. EP/W028115/1).
\item T.B. is supported by the National Natural Science Foundation of China (62071301); NYU-ECNU Institute of Physics at NYU Shanghai; the Joint Physics Research Institute Challenge Grant; the Science and Technology Commission of Shanghai Municipality (19XD1423000,22ZR1444600); the NYU Shanghai Boost Fund; the China Foreign Experts Program (G2021013002L); the NYU Shanghai Major-Grants Seed Fund; Tamkeen under the NYU Abu Dhabi Research Institute grant CG008; and the SMEC Scientific Research Innovation Project (2023ZKZD55).
\item H.-L. H. acknowledges support from the National Natural Science Foundation of China (Grant No. 12274464), and Natural Science Foundation of Henan (Grant No. 242300421049).
\item Nana Liu acknowledges funding from the Science and Technology Program of Shanghai, China (21JC1402900), NSFC grants No.12471411 and No. 12341104, the Shanghai Jiao Tong University 2030 Initiative, the Fundamental Research Funds for the Central Universities, Shanghai Pujiang Talent Grant (no. 20PJ1408400) and the NSFC International Young Scientists Project (no. 12050410230), the Shanghai Municipal Science and Technology Major Project (2021SHZDZX0102) and the Natural Science Foundation of Shanghai grant 21ZR1431000. 
\item B.Q.B. acknowledges support from the Australian Research Council Centre of Excellence for Quantum Computation and Communication
Technology (Project No. CE170100012).
\item WJM is  is supported by the MEXT Quantum Leap Flagship Program (MEXT Q-LEAP) under Grant No. JPMXS0118069605 and the JSPS KAKENHI Grant No. 21H04880.
\item Jonathan P. Dowling would like to acknowledge support from the US Air Force Office of Scientific Research, the Army Research Office, the National Science Foundation, and the Northrop-Grumman Corporation.
\item We thank almost unethical quantities of 1,3,7-Trimethylxanthine\index{1,3,7-Trimethylxanthine}\index{Caffeine} for aiding in the motivation for this work.

\end{itemize}

\clearpage

%
% List of Abbreviations
%

\nomenclature{\textbf{CDS}}{Credit default swap}
\nomenclature{\textbf{ZKP}}{Zero-knowledge proof}
\nomenclature{\textbf{HHL}}{Harrow-Hassidim-Lloyd}
\nomenclature{\textbf{TSF}}{Time-series forecasting}
\nomenclature{\textbf{QML}}{Quantum machine learning}
\nomenclature{\textbf{QAOA}}{Quantum approximate optimisation algorithms}
\nomenclature{\textbf{VQE}}{Variational quantum eigensolvers}
\nomenclature{\textbf{DoS}}{Denial of service}
\nomenclature{\textbf{QuSTICK}}{Quantum memory STICK}
\nomenclature{\textbf{PPQ}}{Price-per-qubit}
\nomenclature{\textbf{LHC}}{Large Hadron Collider}
\nomenclature{\textbf{QMU}}{Quantum memory unit}
\nomenclature{\textbf{GCHQ}}{Government Communications Headquarters}
\nomenclature{\textbf{COW}}{Coherent one way protocol}
\nomenclature{\textbf{DPS}}{Differential phase-shift}
\nomenclature{\textbf{IoT}}{Internet of things}
\nomenclature{\textbf{AQC}}{Adiabatic quantum computation}
\nomenclature{\textbf{Qubus}}{Quantum bus}
\nomenclature{\textbf{IPO}}{Initial public offering}
\nomenclature{\textbf{USB}}{Universal serial bus}
\nomenclature{\textbf{HLF}}{Hidden linear function}
\nomenclature{\textbf{ISP}}{Internet service provider}
\nomenclature{\textbf{SPOF}}{Single point of failure}
\nomenclature{\textbf{AC}}{Alternating current}
\nomenclature{\textbf{DC}}{Direct current}
\nomenclature{\textbf{RF}}{Radio frequency}
\nomenclature{\textbf{Transmon}}{Transmission line shunted plasma oscillation qubit}
\nomenclature{\textbf{QED}}{Quantum electrodynamics}
\nomenclature{\textbf{SQUID}}{Superconducting quantum interference device}
\nomenclature{\textbf{PNS}}{Photon-number-splitting}
\nomenclature{\textbf{DI}}{Device independent}
\nomenclature{\textbf{RR-DPS}}{Round-Robin differential phase-shift protocol}
\nomenclature{\textbf{QRAM}}{Quantum random-access memory}
\nomenclature{\textbf{NISQ}}{Noisy intermediate-scale quantum technology}
\nomenclature{\textbf{DDI}}{Detector device independent}
\nomenclature{\textbf{MDI}}{Measurement device independent}
\nomenclature{\textbf{HD}}{High-dimensional}
\nomenclature{\textbf{RR-DPS}}{Round-Robin differential phase-shift protocol}
\nomenclature{\textbf{SIGINT}}{Signals intelligence}
\nomenclature{\textbf{QCS}}{Rigetti Quantum Cloud Services}
\nomenclature{\textbf{OpenQASM}}{Open quantum assembly language}
\nomenclature{\textbf{SDK}}{Software development kit}
\nomenclature{\textbf{TEM}}{Transverse electro-magnetic}
\nomenclature{\textbf{VoIP}}{Voice over IP}
\nomenclature{\textbf{CRISPR}}{Clustered regularly interspaced short palindromic repeats}
\nomenclature{\textbf{CCD}}{Charge-coupled device}
\nomenclature{\textbf{API}}{Application programming interface}
\nomenclature{\textbf{RoR}}{Rate of return}
\nomenclature{\textbf{ASIC}}{Application-specific integrated circuit}
\nomenclature{\textbf{FPGA}}{Field-programmable gate array}
\nomenclature{\textbf{XOR}}{Exclusive-OR}
\nomenclature{\textbf{QFT}}{Quantum Fourier transform}
\nomenclature{\textbf{DFT}}{Discrete Fourier transform}
\nomenclature{\textbf{IQP}}{Instantaneous quantum protocol}
\nomenclature{\textbf{SHA}}{Secure Hash Algorithm}
\nomenclature{\textbf{PGP}}{Pretty Good Privacy}
\nomenclature{\textbf{KPA}}{Known plaintext attack}
\nomenclature{\textbf{CPA}}{Chosen plaintext attack}
\nomenclature{\textbf{AES}}{Advanced encryption standard}
\nomenclature{\textbf{GNFS}}{General number field sieve}
\nomenclature{\textbf{DES}}{Data encryption standard}
\nomenclature{\textbf{SNP}}{Single-nucleotide polymorphism}
\nomenclature{\textbf{NMR}}{Nuclear magnetic resonance}
\nomenclature{\textbf{QML}}{Quantum machine learning}
\nomenclature{\textbf{TDA}}{Topological data analysis}
\nomenclature{\textbf{CUDA}}{Compute unified device architecture}
\nomenclature{\textbf{SPDC}}{Spontaneous parametric down-conversion}
\nomenclature{\textbf{CV}}{Continuous-variable}
\nomenclature{\textbf{DV}}{Discrete-variable}
\nomenclature{\textbf{QKD}}{Quantum key distribution}
\nomenclature{\textbf{TCP}}{Transmission control protocol}
\nomenclature{\textbf{IP}}{Internet protocol}
\nomenclature{\textbf{QTCP}}{Quantum transmission control protocol}
\nomenclature{\textbf{BFS}}{Breadth-first-search}
\nomenclature{\textbf{DFS}}{Depth-first-search, decoherence-free subspace}
\nomenclature{\textbf{P2P}}{Point-to-point}
\nomenclature{\textbf{CNOT}}{Controlled-NOT}
\nomenclature{\textbf{CZ}}{Controlled-phase}
\nomenclature{\textbf{BGP}}{Border gateway protocol}
\nomenclature{\textbf{BQP}}{Bounded-error quantum polynomial-time}
\nomenclature{\textbf{NP}}{Non-deterministic polynomial-time}
\nomenclature{\textbf{P}}{Polynomial-time}
\nomenclature{\textbf{BPP}}{Bounded-error probabilisitic polynomial-time}
\nomenclature{\textbf{APD}}{Avalanche photo-diode}
\nomenclature{\textbf{QST}}{Quantum state tomography}
\nomenclature{\textbf{QPT}}{Quantum process tomography}
\nomenclature{\textbf{CDMA}}{Code division multiple access}
\nomenclature{\textbf{EGP}}{Exterior gateway protocol}
\nomenclature{\textbf{GHZ}}{Greenberger-Horne-Zeilinger}
\nomenclature{\textbf{MZ}}{Mach-Zehnder}
\nomenclature{\textbf{HOM}}{Hong-Ou-Mandel}
\nomenclature{\textbf{NV}}{Nitrogen-vacancy}
\nomenclature{\textbf{NS}}{Non-linear sign-shift}
\nomenclature{\textbf{QEC}}{Quantum error correction}
\nomenclature{\textbf{UDP}}{User datagram protocol}
\nomenclature{\textbf{QoS}}{Quality of service}
\nomenclature{\textbf{LAN}}{Local area network}
\nomenclature{\textbf{LEO}}{Low Earth orbit}
\nomenclature{\textbf{CP-map}}{Completely positive map}
\nomenclature{\textbf{MST}}{Minimum spanning tree}
\nomenclature{\textbf{VRP}}{Vehicle routing problem}
\nomenclature{\textbf{VRSP}}{Vehicle rescheduling problem}
\nomenclature{\textbf{QED}}{Quantum electrodynamics}
\nomenclature{\textbf{LOQC}}{Linear optics quantum computing}
\nomenclature{\textbf{KLM}}{Knill-Laflamme-Milburn}
\nomenclature{\textbf{BB84}}{Bennett-Brassard 1984}
\nomenclature{\textbf{SNL}}{Shot-noise limit}
\nomenclature{\textbf{HL}}{Heisenberg limit}
\nomenclature{\textbf{E91}}{Ekert 1991}
\nomenclature{\textbf{NOON}}{\mbox{$\frac{1}{\sqrt{2}} (\ket{N,0}+\ket{0,N})$}}
\nomenclature{\textbf{QW}}{Quantum walk}
\nomenclature{\textbf{HSW}}{Holevo-Schumacher-Westmoreland}
\nomenclature{\textbf{MORDOR}}{Motes-Olson-Rabeaux-Olson-Dowling-Rohde}
\nomenclature{\textbf{EO}}{Entangling operation}
\nomenclature{\textbf{EPR}}{Einstein-Podolsky-Rosen state}
\nomenclature{\textbf{BBO}}{Beta-barium borate}
\nomenclature{\textbf{PBS}}{Polarising beamsplitter}
\nomenclature{\textbf{BS}}{Beamsplitter}
\nomenclature{\textbf{QIP}}{Quantum information processing}
\nomenclature{\textbf{MBQC}}{Measurement-based quantum computing}
\nomenclature{\textbf{AOM}}{Acousto-optic modulator}
\nomenclature{\textbf{EOM}}{Electro-optic modulator}
\nomenclature{\textbf{RoI}}{Return on investment}
\nomenclature{\textbf{BEC}}{Bose-Einstein condensate}
\nomenclature{\textbf{FIFO}}{First in, first out}
\nomenclature{\textbf{RSA}}{Rivest-Shamir-Adleman}
\nomenclature{\textbf{FLOP}}{Floating point operation}
\nomenclature{\textbf{QCL}}{Quantum computational leverage}
\nomenclature{\textbf{DFB}}{Distributed feedback}
\nomenclature{\textbf{FB}}{Fabry-Perot}
\nomenclature{\textbf{DWDM}}{Dense wavelength division multiplexing}
\nomenclature{\textbf{FWM}}{Four-wave mixing}
\nomenclature{\textbf{SETI}}{Search for extra-terrestrial intelligence}
\nomenclature{\textbf{EMH}}{Efficient-market hypothesis}
\nomenclature{\textbf{WP}}{Wave-plate}
\nomenclature{\textbf{GPS}}{Global positioning system}
\nomenclature{\textbf{QUESS}}{Quantum Science Experiment Satellite}
\nomenclature{\textbf{SOCRATES}}{Space Optical Communications Research Advanced Technology Satellite}
\nomenclature{\textbf{QEYSSat}}{Quantum EncrYption and Science Satellite}
\nomenclature{\textbf{QAB}}{Quantum anonymous broadcasting}
\nomenclature{\textbf{ECT}}{Extended Church-Turing}
\nomenclature{\textbf{QND}}{Quantum non-demolition}
\nomenclature{\textbf{EAT}}{Effective atmospheric thickness}
\nomenclature{\textbf{POVM}}{Positive operator-valued measure}
\nomenclature{\textbf{R\&D}}{Research \& development}
\nomenclature{\textbf{NGS}}{Next-generation sequencers}
\nomenclature{\textbf{NICT}}{National Institute of Information \& Communications Technology}

\printnomenclature

% \bibliography{quantum_internet, nana_ML, sneakernet, additional_bib, BQB_bibs}

\begin{thebibliography}{901}%
\makeatletter
\providecommand \@ifxundefined [1]{%
 \@ifx{#1\undefined}
}%
\providecommand \@ifnum [1]{%
 \ifnum #1\expandafter \@firstoftwo
 \else \expandafter \@secondoftwo
 \fi
}%
\providecommand \@ifx [1]{%
 \ifx #1\expandafter \@firstoftwo
 \else \expandafter \@secondoftwo
 \fi
}%
\providecommand \natexlab [1]{#1}%
\providecommand \enquote  [1]{``#1''}%
\providecommand \bibnamefont  [1]{#1}%
\providecommand \bibfnamefont [1]{#1}%
\providecommand \citenamefont [1]{#1}%
\providecommand \href@noop [0]{\@secondoftwo}%
\providecommand \href [0]{\begingroup \@sanitize@url \@href}%
\providecommand \@href[1]{\@@startlink{#1}\@@href}%
\providecommand \@@href[1]{\endgroup#1\@@endlink}%
\providecommand \@sanitize@url [0]{\catcode `\\12\catcode `\$12\catcode
  `\&12\catcode `\#12\catcode `\^12\catcode `\_12\catcode `\%12\relax}%
\providecommand \@@startlink[1]{}%
\providecommand \@@endlink[0]{}%
\providecommand \url  [0]{\begingroup\@sanitize@url \@url }%
\providecommand \@url [1]{\endgroup\@href {#1}{\urlprefix }}%
\providecommand \urlprefix  [0]{URL }%
\providecommand \Eprint [0]{\href }%
\providecommand \doibase [0]{https://doi.org/}%
\providecommand \selectlanguage [0]{\@gobble}%
\providecommand \bibinfo  [0]{\@secondoftwo}%
\providecommand \bibfield  [0]{\@secondoftwo}%
\providecommand \translation [1]{[#1]}%
\providecommand \BibitemOpen [0]{}%
\providecommand \bibitemStop [0]{}%
\providecommand \bibitemNoStop [0]{.\EOS\space}%
\providecommand \EOS [0]{\spacefactor3000\relax}%
\providecommand \BibitemShut  [1]{\csname bibitem#1\endcsname}%
\let\auto@bib@innerbib\@empty
%</preamble>
\bibitem [{\citenamefont {Aaronson}(2015)}]{bib:aaronson2015read}%
  \BibitemOpen
  \bibfield  {author} {\bibinfo {author} {\bibnamefont {Aaronson},
  \bibfnamefont {Scott}}} (\bibinfo {year} {2015}),\ \bibfield  {title}
  {\enquote {\bibinfo {title} {Read the fine print},}\ }\href
  {https://doi.org/10.1038/nphys3272} {\bibfield  {journal} {\bibinfo
  {journal} {Nature Physics}\ }\textbf {\bibinfo {volume} {11}},\ \bibinfo
  {pages} {291}}\BibitemShut {NoStop}%
\bibitem [{\citenamefont {Aaronson}\ and\ \citenamefont
  {Arkhipov}(2011)}]{bib:AaronsonArkhipov10}%
  \BibitemOpen
  \bibfield  {author} {\bibinfo {author} {\bibnamefont {Aaronson},
  \bibfnamefont {Scott}}, and\ \bibinfo {author} {\bibfnamefont {Alex}\
  \bibnamefont {Arkhipov}}} (\bibinfo {year} {2011}),\ \bibfield  {title}
  {\enquote {\bibinfo {title} {The computational complexity of linear
  optics},}\ }\href {https://doi.org/10.1364/qim.2014.qth1a.2} {\bibinfo
  {journal} {Proceedings of ACM STOC (New York)}\ ,\ \bibinfo {pages}
  {333}}\BibitemShut {NoStop}%
\bibitem [{\citenamefont {Aaronson}\ and\ \citenamefont
  {Brod}(2016)}]{bib:ScottLost16}%
  \BibitemOpen
\bibfield  {journal} {  }\bibfield  {author} {\bibinfo {author} {\bibnamefont
  {Aaronson}, \bibfnamefont {Scott}}, and\ \bibinfo {author} {\bibfnamefont
  {Daniel~J.}\ \bibnamefont {Brod}}} (\bibinfo {year} {2016}),\ \bibfield
  {title} {\enquote {\bibinfo {title} {Bosonsampling with lost photons},}\
  }\href {https://doi.org/10.1103/physreva.93.012335} {\bibfield  {journal}
  {\bibinfo  {journal} {Physical Review A}\ }\textbf {\bibinfo {volume} {93}},\
  \bibinfo {pages} {012335}},\ \Eprint
  {https://arxiv.org/abs/arXiv:1510.05245v2} {arXiv:1510.05245v2} \BibitemShut
  {NoStop}%
\bibitem [{\citenamefont {Achilles}\ \emph {et~al.}(2004)\citenamefont
  {Achilles}, \citenamefont {Silberhorn}, \citenamefont {Sliwa}, \citenamefont
  {Banaszek}, \citenamefont {Walmsley}, \citenamefont {Fitch}, \citenamefont
  {Jacobs}, \citenamefont {Pittman},\ and\ \citenamefont
  {Franson}}]{bib:Achilles04}%
  \BibitemOpen
  \bibfield  {author} {\bibinfo {author} {\bibnamefont {Achilles},
  \bibfnamefont {Daryl}}, \bibinfo {author} {\bibfnamefont {Christine}\
  \bibnamefont {Silberhorn}}, \bibinfo {author} {\bibfnamefont {Cezary}\
  \bibnamefont {Sliwa}}, \bibinfo {author} {\bibfnamefont {Konrad}\
  \bibnamefont {Banaszek}}, \bibinfo {author} {\bibfnamefont {Ian~A.}\
  \bibnamefont {Walmsley}}, \bibinfo {author} {\bibfnamefont {Michael~J.}\
  \bibnamefont {Fitch}}, \bibinfo {author} {\bibfnamefont {Bryan~C.}\
  \bibnamefont {Jacobs}}, \bibinfo {author} {\bibfnamefont {Todd~B.}\
  \bibnamefont {Pittman}}, and\ \bibinfo {author} {\bibfnamefont {James~D.}\
  \bibnamefont {Franson}}} (\bibinfo {year} {2004}),\ \bibfield  {title}
  {\enquote {\bibinfo {title} {Photon number resolving detection using
  time-multiplexing},}\ }\href {https://doi.org/10.1080/09500340408235288}
  {\bibfield  {journal} {\bibinfo  {journal} {Journal of Modern Optics}\
  }\textbf {\bibinfo {volume} {51}},\ \bibinfo {pages} {1499}},\ \Eprint
  {https://arxiv.org/abs/arXiv:quant-ph/0310183v1} {arXiv:quant-ph/0310183v1}
  \BibitemShut {NoStop}%
\bibitem [{\citenamefont {Ac\'{\i}n}\ \emph {et~al.}(2007)\citenamefont
  {Ac\'{\i}n}, \citenamefont {Brunner}, \citenamefont {Gisin}, \citenamefont
  {Massar}, \citenamefont {Pironio},\ and\ \citenamefont
  {Scarani}}]{bib:PhysRevLett.98.230501}%
  \BibitemOpen
  \bibfield  {author} {\bibinfo {author} {\bibnamefont {Ac\'{\i}n},
  \bibfnamefont {Antonio}}, \bibinfo {author} {\bibfnamefont {Nicolas}\
  \bibnamefont {Brunner}}, \bibinfo {author} {\bibfnamefont {Nicolas}\
  \bibnamefont {Gisin}}, \bibinfo {author} {\bibfnamefont {Serge}\ \bibnamefont
  {Massar}}, \bibinfo {author} {\bibfnamefont {Stefano}\ \bibnamefont
  {Pironio}}, and\ \bibinfo {author} {\bibfnamefont {Valerio}\ \bibnamefont
  {Scarani}}} (\bibinfo {year} {2007}),\ \bibfield  {title} {\enquote {\bibinfo
  {title} {Device-independent security of quantum cryptography against
  collective attacks},}\ }\href {https://doi.org/10.1103/physrevlett.98.230501}
  {\bibfield  {journal} {\bibinfo  {journal} {Physical Review Letters}\
  }\textbf {\bibinfo {volume} {98}},\ \bibinfo {pages} {230501}},\ \Eprint
  {https://arxiv.org/abs/arXiv:quant-ph/0702152v2} {arXiv:quant-ph/0702152v2}
  \BibitemShut {NoStop}%
\bibitem [{\citenamefont {Afek}\ \emph {et~al.}(2010)\citenamefont {Afek},
  \citenamefont {Ambar},\ and\ \citenamefont {Silberberg}}]{bib:afek2010high}%
  \BibitemOpen
  \bibfield  {author} {\bibinfo {author} {\bibnamefont {Afek}, \bibfnamefont
  {Itai}}, \bibinfo {author} {\bibfnamefont {Oron}\ \bibnamefont {Ambar}}, and\
  \bibinfo {author} {\bibfnamefont {Yaron}\ \bibnamefont {Silberberg}}}
  (\bibinfo {year} {2010}),\ \bibfield  {title} {\enquote {\bibinfo {title}
  {High-noon states by mixing quantum and classical light},}\ }\href
  {https://doi.org/10.1126/science.1188172} {\bibfield  {journal} {\bibinfo
  {journal} {Science}\ }\textbf {\bibinfo {volume} {328}},\ \bibinfo {pages}
  {879}}\BibitemShut {NoStop}%
\bibitem [{\citenamefont {Aggarwal}\ \emph {et~al.}(2017)\citenamefont
  {Aggarwal}, \citenamefont {Brennen}, \citenamefont {Lee}, \citenamefont
  {Santha},\ and\ \citenamefont {Tomamichel}}]{bib:TomamichelBlockchain}%
  \BibitemOpen
  \bibfield  {author} {\bibinfo {author} {\bibnamefont {Aggarwal},
  \bibfnamefont {Divesh}}, \bibinfo {author} {\bibfnamefont {Gavin~K.}\
  \bibnamefont {Brennen}}, \bibinfo {author} {\bibfnamefont {Troy}\
  \bibnamefont {Lee}}, \bibinfo {author} {\bibfnamefont {Miklos}\ \bibnamefont
  {Santha}}, and\ \bibinfo {author} {\bibfnamefont {Marco}\ \bibnamefont
  {Tomamichel}}} (\bibinfo {year} {2017}),\ \bibfield  {title} {\enquote
  {\bibinfo {title} {Quantum attacks on bitcoin, and how to protect against
  them},}\ }\href {https://doi.org/10.5195/ledger.2018.127} {\
  10.5195/ledger.2018.127},\ \Eprint {https://arxiv.org/abs/arXiv:1710.10377}
  {arXiv:1710.10377} \BibitemShut {NoStop}%
\bibitem [{\citenamefont {Aharonov}\ \emph {et~al.}(2001)\citenamefont
  {Aharonov}, \citenamefont {Ambainis}, \citenamefont {Kempe},\ and\
  \citenamefont {Vazirani}}]{bib:Aharonov01}%
  \BibitemOpen
  \bibfield  {author} {\bibinfo {author} {\bibnamefont {Aharonov},
  \bibfnamefont {D}}, \bibinfo {author} {\bibfnamefont {A.}~\bibnamefont
  {Ambainis}}, \bibinfo {author} {\bibfnamefont {J.}~\bibnamefont {Kempe}},
  and\ \bibinfo {author} {\bibfnamefont {U.}~\bibnamefont {Vazirani}}}
  (\bibinfo {year} {2001}),\ \href@noop {} {\bibinfo  {journal} {in Proceedings
  of the 33rd Annual ACM Symposium on Theory of Computing STOC '01}\ ,\
  \bibinfo {pages} {50}}\BibitemShut {NoStop}%
\bibitem [{\citenamefont {Aharonov}\ and\ \citenamefont
  {Ben-Or}(1997)}]{bib:AB97}%
  \BibitemOpen
\bibfield  {journal} {  }\bibfield  {author} {\bibinfo {author} {\bibnamefont
  {Aharonov}, \bibfnamefont {D}}, and\ \bibinfo {author} {\bibfnamefont
  {M.}~\bibnamefont {Ben-Or}}} (\bibinfo {year} {1997}),\ \bibfield  {title}
  {\enquote {\bibinfo {title} {{Fault-tolerant Quantum Computation with
  constant error}},}\ }\href@noop {} {\bibinfo  {journal} {Proceedings of 29th
  Annual ACM Symposium on Theory of Computing}\ ,\ \bibinfo {pages}
  {46}}\BibitemShut {NoStop}%
\bibitem [{\citenamefont {Aharonov}\ \emph {et~al.}(2008)\citenamefont
  {Aharonov}, \citenamefont {Van~Dam}, \citenamefont {Kempe}, \citenamefont
  {Landau}, \citenamefont {Lloyd},\ and\ \citenamefont
  {Regev}}]{bib:aharonov2008adiabatic}%
  \BibitemOpen
\bibfield  {journal} {  }\bibfield  {author} {\bibinfo {author} {\bibnamefont
  {Aharonov}, \bibfnamefont {Dorit}}, \bibinfo {author} {\bibfnamefont {Wim}\
  \bibnamefont {Van~Dam}}, \bibinfo {author} {\bibfnamefont {Julia}\
  \bibnamefont {Kempe}}, \bibinfo {author} {\bibfnamefont {Zeph}\ \bibnamefont
  {Landau}}, \bibinfo {author} {\bibfnamefont {Seth}\ \bibnamefont {Lloyd}},
  and\ \bibinfo {author} {\bibfnamefont {Oded}\ \bibnamefont {Regev}}}
  (\bibinfo {year} {2008}),\ \bibfield  {title} {\enquote {\bibinfo {title}
  {Adiabatic quantum computation is equivalent to standard quantum
  computation},}\ }\href@noop {} {\bibfield  {journal} {\bibinfo  {journal}
  {SIAM review}\ }\textbf {\bibinfo {volume} {50}}~(\bibinfo {number} {4}),\
  \bibinfo {pages} {755--787}}\BibitemShut {NoStop}%
\bibitem [{\citenamefont {Aharonov}\ \emph {et~al.}(1993)\citenamefont
  {Aharonov}, \citenamefont {Davidovich},\ and\ \citenamefont
  {Zagury}}]{bib:Aharonov93}%
  \BibitemOpen
  \bibfield  {author} {\bibinfo {author} {\bibnamefont {Aharonov},
  \bibfnamefont {Y}}, \bibinfo {author} {\bibfnamefont {L.}~\bibnamefont
  {Davidovich}}, and\ \bibinfo {author} {\bibfnamefont {N.}~\bibnamefont
  {Zagury}}} (\bibinfo {year} {1993}),\ \href@noop {} {\bibfield  {journal}
  {\bibinfo  {journal} {Physical Review A}\ }\textbf {\bibinfo {volume} {48}},\
  \bibinfo {pages} {1687}}\BibitemShut {NoStop}%
\bibitem [{\citenamefont {Aharonovich}\ \emph {et~al.}(2016)\citenamefont
  {Aharonovich}, \citenamefont {Englund},\ and\ \citenamefont
  {Toth}}]{bib:aharonovich2016solid}%
  \BibitemOpen
  \bibfield  {author} {\bibinfo {author} {\bibnamefont {Aharonovich},
  \bibfnamefont {Igor}}, \bibinfo {author} {\bibfnamefont {Dirk}\ \bibnamefont
  {Englund}}, and\ \bibinfo {author} {\bibfnamefont {Milos}\ \bibnamefont
  {Toth}}} (\bibinfo {year} {2016}),\ \bibfield  {title} {\enquote {\bibinfo
  {title} {Solid-state single-photon emitters},}\ }\href
  {https://doi.org/10.1038/nphoton.2016.186} {\bibfield  {journal} {\bibinfo
  {journal} {Nature Photonics}\ }\textbf {\bibinfo {volume} {10}},\ \bibinfo
  {pages} {631}}\BibitemShut {NoStop}%
\bibitem [{\citenamefont {Aharonovich}\ \emph {et~al.}(2011)\citenamefont
  {Aharonovich}, \citenamefont {Greentree},\ and\ \citenamefont
  {Prawer}}]{SD-Aharonovich:2011aa}%
  \BibitemOpen
  \bibfield  {author} {\bibinfo {author} {\bibnamefont {Aharonovich},
  \bibfnamefont {Igor}}, \bibinfo {author} {\bibfnamefont {Andrew~D.}\
  \bibnamefont {Greentree}}, and\ \bibinfo {author} {\bibfnamefont {Steven}\
  \bibnamefont {Prawer}}} (\bibinfo {year} {2011}),\ \bibfield  {title}
  {\enquote {\bibinfo {title} {Diamond photonics},}\ }\href
  {https://doi.org/10.1038/nphoton.2011.54} {\bibfield  {journal} {\bibinfo
  {journal} {Nature Photonics}\ }\textbf {\bibinfo {volume} {5}},\ \bibinfo
  {pages} {397}}\BibitemShut {NoStop}%
\bibitem [{\citenamefont {Ahmadi}\ \emph {et~al.}(2014)\citenamefont {Ahmadi},
  \citenamefont {Bruschi}, \citenamefont {Sab{\'\i}n}, \citenamefont {Adesso},\
  and\ \citenamefont {Fuentes}}]{bib:ahmadi2014relativistic}%
  \BibitemOpen
  \bibfield  {author} {\bibinfo {author} {\bibnamefont {Ahmadi}, \bibfnamefont
  {Mehdi}}, \bibinfo {author} {\bibfnamefont {David~Edward}\ \bibnamefont
  {Bruschi}}, \bibinfo {author} {\bibfnamefont {Carlos}\ \bibnamefont
  {Sab{\'\i}n}}, \bibinfo {author} {\bibfnamefont {Gerardo}\ \bibnamefont
  {Adesso}}, and\ \bibinfo {author} {\bibfnamefont {Ivette}\ \bibnamefont
  {Fuentes}}} (\bibinfo {year} {2014}),\ \bibfield  {title} {\enquote {\bibinfo
  {title} {Relativistic quantum metrology: Exploiting relativity to improve
  quantum measurement technologies},}\ }\href
  {https://doi.org/10.1038/srep04996} {\bibfield  {journal} {\bibinfo
  {journal} {Scientific Reports}\ }\textbf {\bibinfo {volume} {4}},\ \bibinfo
  {pages} {4996}},\ \Eprint {https://arxiv.org/abs/arXiv:1307.7082v2}
  {arXiv:1307.7082v2} \BibitemShut {NoStop}%
\bibitem [{\citenamefont {Ahmed}\ \emph {et~al.}(2007)\citenamefont {Ahmed},
  \citenamefont {Oreshkin},\ and\ \citenamefont
  {Coates}}]{bib:ahmed2007machine}%
  \BibitemOpen
  \bibfield  {author} {\bibinfo {author} {\bibnamefont {Ahmed}, \bibfnamefont
  {Tarem}}, \bibinfo {author} {\bibfnamefont {Boris}\ \bibnamefont {Oreshkin}},
  and\ \bibinfo {author} {\bibfnamefont {Mark}\ \bibnamefont {Coates}}}
  (\bibinfo {year} {2007}),\ \bibfield  {title} {\enquote {\bibinfo {title}
  {Machine learning approaches to network anomaly detection},}\ }in\ \href@noop
  {} {\emph {\bibinfo {booktitle} {Proceedings of the 2nd USENIX workshop on
  Tackling computer systems problems with machine learning techniques}}},\
  p.~\bibinfo {pages} {1}\BibitemShut {NoStop}%
\bibitem [{\citenamefont {Ahuja}\ \emph {et~al.}(1995)\citenamefont {Ahuja},
  \citenamefont {Magnanti},\ and\ \citenamefont {Orlin}}]{ahuja1995network}%
  \BibitemOpen
  \bibfield  {author} {\bibinfo {author} {\bibnamefont {Ahuja}, \bibfnamefont
  {Ravindra~K}}, \bibinfo {author} {\bibfnamefont {Thomas~L}\ \bibnamefont
  {Magnanti}}, and\ \bibinfo {author} {\bibfnamefont {James~B}\ \bibnamefont
  {Orlin}}} (\bibinfo {year} {1995}),\ \href@noop {} {\emph {\bibinfo {title}
  {Network flows: theory, algorithms and applications}}}\ (\bibinfo
  {publisher} {Prentice hall})\BibitemShut {NoStop}%
\bibitem [{\citenamefont {Aichele}\ \emph {et~al.}(2002)\citenamefont
  {Aichele}, \citenamefont {Lvovsky},\ and\ \citenamefont
  {Schiller}}]{bib:Aichele02}%
  \BibitemOpen
  \bibfield  {author} {\bibinfo {author} {\bibnamefont {Aichele}, \bibfnamefont
  {T}}, \bibinfo {author} {\bibfnamefont {A.~I.}\ \bibnamefont {Lvovsky}}, and\
  \bibinfo {author} {\bibfnamefont {S.}~\bibnamefont {Schiller}}} (\bibinfo
  {year} {2002}),\ \bibfield  {title} {\enquote {\bibinfo {title} {Optical mode
  characterization of single photons prepared by means of conditional
  measurements on a biphoton state},}\ }\href
  {https://doi.org/10.1140/epjd/e20020028} {\bibfield  {journal} {\bibinfo
  {journal} {European Physics Journal D}\ }\textbf {\bibinfo {volume} {18}},\
  \bibinfo {pages} {237}}\BibitemShut {NoStop}%
\bibitem [{\citenamefont {Albash}\ and\ \citenamefont
  {Lidar}(2018)}]{bib:RevModPhys.90.015002}%
  \BibitemOpen
  \bibfield  {author} {\bibinfo {author} {\bibnamefont {Albash}, \bibfnamefont
  {Tameem}}, and\ \bibinfo {author} {\bibfnamefont {Daniel~A.}\ \bibnamefont
  {Lidar}}} (\bibinfo {year} {2018}),\ \bibfield  {title} {\enquote {\bibinfo
  {title} {Adiabatic quantum computation},}\ }\href
  {https://doi.org/10.1103/RevModPhys.90.015002} {\bibfield  {journal}
  {\bibinfo  {journal} {Reviews in Modern Physics}\ }\textbf {\bibinfo {volume}
  {90}},\ \bibinfo {pages} {015002}},\ \Eprint
  {https://arxiv.org/abs/arXiv:1611.04471v2} {arXiv:1611.04471v2} \BibitemShut
  {NoStop}%
\bibitem [{\citenamefont {Albert}\ and\ \citenamefont
  {Barabasi}(2002)}]{bib:BAfitness}%
  \BibitemOpen
  \bibfield  {author} {\bibinfo {author} {\bibnamefont {Albert}, \bibfnamefont
  {Reka}}, and\ \bibinfo {author} {\bibfnamefont {Albert-Laszlo}\ \bibnamefont
  {Barabasi}}} (\bibinfo {year} {2002}),\ \bibfield  {title} {\enquote
  {\bibinfo {title} {Statistical mechanics of complex networks},}\ }\href
  {https://doi.org/10.1103/RevModPhys.74.47} {\bibfield  {journal} {\bibinfo
  {journal} {Reviews of Modern Physics}\ }\textbf {\bibinfo {volume} {74}},\
  \bibinfo {pages} {47}},\ \Eprint
  {https://arxiv.org/abs/arXiv:cond-mat/0106096} {arXiv:cond-mat/0106096}
  \BibitemShut {NoStop}%
\bibitem [{\citenamefont {Albrecht}\ \emph {et~al.}(2014)\citenamefont
  {Albrecht}, \citenamefont {Farrera}, \citenamefont {Fernandez-Gonzalvo},
  \citenamefont {Cristiani},\ and\ \citenamefont
  {de~Riedmatten}}]{bib:NC_5_3376}%
  \BibitemOpen
  \bibfield  {author} {\bibinfo {author} {\bibnamefont {Albrecht},
  \bibfnamefont {Boris}}, \bibinfo {author} {\bibfnamefont {Pau}\ \bibnamefont
  {Farrera}}, \bibinfo {author} {\bibfnamefont {Xavier}\ \bibnamefont
  {Fernandez-Gonzalvo}}, \bibinfo {author} {\bibfnamefont {Matteo}\
  \bibnamefont {Cristiani}}, and\ \bibinfo {author} {\bibfnamefont {Hugues}\
  \bibnamefont {de~Riedmatten}}} (\bibinfo {year} {2014}),\ \bibfield  {title}
  {\enquote {\bibinfo {title} {A waveguide frequency converter connecting
  rubidium-based quantum memories to the telecom c-band},}\ }\href
  {https://doi.org/10.1038/ncomms4376} {\bibfield  {journal} {\bibinfo
  {journal} {Nature Communications}\ }\textbf {\bibinfo {volume} {5}},\
  \bibinfo {pages} {3376}},\ \Eprint {https://arxiv.org/abs/arXiv:1402.2866v1}
  {arXiv:1402.2866v1} \BibitemShut {NoStop}%
\bibitem [{\citenamefont {Almeida}\ \emph {et~al.}(2003)\citenamefont
  {Almeida}, \citenamefont {Panepucci},\ and\ \citenamefont
  {Lipson}}]{bib:almeida2003}%
  \BibitemOpen
  \bibfield  {author} {\bibinfo {author} {\bibnamefont {Almeida}, \bibfnamefont
  {Vilson~R}}, \bibinfo {author} {\bibfnamefont {Roberto~R}\ \bibnamefont
  {Panepucci}}, and\ \bibinfo {author} {\bibfnamefont {Michal}\ \bibnamefont
  {Lipson}}} (\bibinfo {year} {2003}),\ \bibfield  {title} {\enquote {\bibinfo
  {title} {Nanotaper for compact mode conversion},}\ }\href
  {https://doi.org/10.1364/ol.28.001302} {\bibfield  {journal} {\bibinfo
  {journal} {Optics Letters}\ }\textbf {\bibinfo {volume} {28}},\ \bibinfo
  {pages} {1302}}\BibitemShut {NoStop}%
\bibitem [{\citenamefont {Amos}\ \emph {et~al.}(2020)\citenamefont {Amos},
  \citenamefont {Georgiou}, \citenamefont {Kiayias},\ and\ \citenamefont
  {Zhandry}}]{bib:RMAZ20}%
  \BibitemOpen
  \bibfield  {author} {\bibinfo {author} {\bibnamefont {Amos}, \bibfnamefont
  {Ryan}}, \bibinfo {author} {\bibfnamefont {Marios}\ \bibnamefont {Georgiou}},
  \bibinfo {author} {\bibfnamefont {Aggelos}\ \bibnamefont {Kiayias}}, and\
  \bibinfo {author} {\bibfnamefont {Mark}\ \bibnamefont {Zhandry}}} (\bibinfo
  {year} {2020}),\ \bibfield  {title} {\enquote {\bibinfo {title} {One-shot
  signatures and applications to hybrid quantum/classical authentication},}\
  }in\ \href {https://doi.org/10.1145/3357713.3384304} {\emph {\bibinfo
  {booktitle} {Proceedings of the 52nd Annual ACM SIGACT Symposium on Theory of
  Computing}}},\ \bibinfo {series and number} {STOC 2020}\ (\bibinfo
  {publisher} {Association for Computing Machinery},\ \bibinfo {address} {New
  York, NY, USA})\ p.\ \bibinfo {pages} {255–268}\BibitemShut {NoStop}%
\bibitem [{\citenamefont {Andr\'e~Chailloux}\ and\ \citenamefont
  {Sikora}(2013)}]{sikora}%
  \BibitemOpen
  \bibfield  {author} {\bibinfo {author} {\bibnamefont {Andr\'e~Chailloux},
  \bibfnamefont {Iordanis~Kerenidis}}, and\ \bibinfo {author} {\bibfnamefont
  {Jamie}\ \bibnamefont {Sikora}}} (\bibinfo {year} {2013}),\ \bibfield
  {title} {\enquote {\bibinfo {title} {Lower bounds for quantum oblivious
  transfer},}\ }\href@noop {} {\bibfield  {journal} {\bibinfo  {journal}
  {Quantum Info. Comput.}\ }\textbf {\bibinfo {volume} {13}}}\BibitemShut
  {NoStop}%
\bibitem [{\citenamefont {Appel}\ \emph {et~al.}(2008)\citenamefont {Appel},
  \citenamefont {Figueroa}, \citenamefont {Korystov}, \citenamefont {Lobino},\
  and\ \citenamefont {Lvovsky}}]{bib:appel2008quantum}%
  \BibitemOpen
  \bibfield  {author} {\bibinfo {author} {\bibnamefont {Appel}, \bibfnamefont
  {J{\"u}rgen}}, \bibinfo {author} {\bibfnamefont {Eden}\ \bibnamefont
  {Figueroa}}, \bibinfo {author} {\bibfnamefont {Dmitry}\ \bibnamefont
  {Korystov}}, \bibinfo {author} {\bibfnamefont {Mirko}\ \bibnamefont
  {Lobino}}, and\ \bibinfo {author} {\bibfnamefont {AI}~\bibnamefont
  {Lvovsky}}} (\bibinfo {year} {2008}),\ \bibfield  {title} {\enquote {\bibinfo
  {title} {Quantum memory for squeezed light},}\ }\href
  {https://doi.org/10.1103/physrevlett.100.093602} {\bibfield  {journal}
  {\bibinfo  {journal} {Physical Review Letters}\ }\textbf {\bibinfo {volume}
  {100}},\ \bibinfo {pages} {093602}},\ \Eprint
  {https://arxiv.org/abs/arXiv:0709.2258v4} {arXiv:0709.2258v4} \BibitemShut
  {NoStop}%
\bibitem [{\citenamefont {Arkhipov}\ and\ \citenamefont
  {Kuperberg}(2012)}]{arkhipov2012bosonic}%
  \BibitemOpen
  \bibfield  {author} {\bibinfo {author} {\bibnamefont {Arkhipov},
  \bibfnamefont {Alex}}, and\ \bibinfo {author} {\bibfnamefont {Greg}\
  \bibnamefont {Kuperberg}}} (\bibinfo {year} {2012}),\ \bibfield  {title}
  {\enquote {\bibinfo {title} {The bosonic birthday paradox},}\ }\href@noop {}
  {\bibfield  {journal} {\bibinfo  {journal} {Geometry \& Topology Monographs}\
  }\textbf {\bibinfo {volume} {18}}~(\bibinfo {number} {1}),\ \bibinfo {pages}
  {10--2140}}\BibitemShut {NoStop}%
\bibitem [{\citenamefont {Armengol}\ \emph {et~al.}(2008)\citenamefont
  {Armengol}, \citenamefont {Furch}, \citenamefont {de~Matos}, \citenamefont
  {Minster}, \citenamefont {Cacciapuoti}, \citenamefont {Pfennigbauer},
  \citenamefont {Aspelmeyer}, \citenamefont {Jennewein}, \citenamefont {Ursin},
  \citenamefont {Schmitt-Manderbach} \emph {et~al.}}]{bib:armengol08}%
  \BibitemOpen
  \bibfield  {author} {\bibinfo {author} {\bibnamefont {Armengol},
  \bibfnamefont {Josep Maria~Perdigues}}, \bibinfo {author} {\bibfnamefont
  {Bernhard}\ \bibnamefont {Furch}}, \bibinfo {author} {\bibfnamefont
  {Clovis~Jacinto}\ \bibnamefont {de~Matos}}, \bibinfo {author} {\bibfnamefont
  {Olivier}\ \bibnamefont {Minster}}, \bibinfo {author} {\bibfnamefont {Luigi}\
  \bibnamefont {Cacciapuoti}}, \bibinfo {author} {\bibfnamefont {Martin}\
  \bibnamefont {Pfennigbauer}}, \bibinfo {author} {\bibfnamefont {Markus}\
  \bibnamefont {Aspelmeyer}}, \bibinfo {author} {\bibfnamefont {Thomas}\
  \bibnamefont {Jennewein}}, \bibinfo {author} {\bibfnamefont {Rupert}\
  \bibnamefont {Ursin}}, \bibinfo {author} {\bibfnamefont {Tobias}\
  \bibnamefont {Schmitt-Manderbach}},  \emph {et~al.}} (\bibinfo {year}
  {2008}),\ \bibfield  {title} {\enquote {\bibinfo {title} {Quantum
  communications at esa: Towards a space experiment on the iss},}\ }\href
  {https://doi.org/10.1016/j.actaastro.2007.12.039} {\bibfield  {journal}
  {\bibinfo  {journal} {Acta Astronautica}\ }\textbf {\bibinfo {volume} {63}},\
  \bibinfo {pages} {165}}\BibitemShut {NoStop}%
\bibitem [{\citenamefont {Arrighi}\ and\ \citenamefont
  {Salvail}(2006)}]{bib:blind2}%
  \BibitemOpen
  \bibfield  {author} {\bibinfo {author} {\bibnamefont {Arrighi}, \bibfnamefont
  {Pablo}}, and\ \bibinfo {author} {\bibfnamefont {Louis}\ \bibnamefont
  {Salvail}}} (\bibinfo {year} {2006}),\ \bibfield  {title} {\enquote {\bibinfo
  {title} {Blind quantum computation},}\ }\href
  {https://doi.org/10.1142/s0219749906002171} {\bibfield  {journal} {\bibinfo
  {journal} {International Journal of Quantum Information}\ }\textbf {\bibinfo
  {volume} {4}},\ \bibinfo {pages} {883}}\BibitemShut {NoStop}%
\bibitem [{\citenamefont {Arroyo-Valles}\ \emph {et~al.}(2007)\citenamefont
  {Arroyo-Valles}, \citenamefont {Alaiz-Rodriguez}, \citenamefont
  {Guerrero-Curieses},\ and\ \citenamefont {Cid-Sueiro}}]{arroyo2007q}%
  \BibitemOpen
  \bibfield  {author} {\bibinfo {author} {\bibnamefont {Arroyo-Valles},
  \bibfnamefont {Rocio}}, \bibinfo {author} {\bibfnamefont {Rocio}\
  \bibnamefont {Alaiz-Rodriguez}}, \bibinfo {author} {\bibfnamefont {Alicia}\
  \bibnamefont {Guerrero-Curieses}}, and\ \bibinfo {author} {\bibfnamefont
  {Jes{\'u}s}\ \bibnamefont {Cid-Sueiro}}} (\bibinfo {year} {2007}),\ \bibfield
   {title} {\enquote {\bibinfo {title} {Q-probabilistic routing in wireless
  sensor networks},}\ }in\ \href@noop {} {\emph {\bibinfo {booktitle} {2007 3rd
  International Conference on Intelligent Sensors, Sensor Networks and
  Information}}}\ (\bibinfo {organization} {IEEE})\ pp.\ \bibinfo {pages}
  {1--6}\BibitemShut {NoStop}%
\bibitem [{\citenamefont {Asavanant}\ \emph {et~al.}(2021)\citenamefont
  {Asavanant}, \citenamefont {Charoensombutamon}, \citenamefont {Yokoyama},
  \citenamefont {Ebihara}, \citenamefont {Nakamura}, \citenamefont {Alexander},
  \citenamefont {Endo}, \citenamefont {Yoshikawa}, \citenamefont {Menicucci},
  \citenamefont {Yonezawa},\ and\ \citenamefont
  {Furusawa}}]{asavanant2021logicalgates}%
  \BibitemOpen
  \bibfield  {author} {\bibinfo {author} {\bibnamefont {Asavanant},
  \bibfnamefont {Warit}}, \bibinfo {author} {\bibfnamefont {Baramee}\
  \bibnamefont {Charoensombutamon}}, \bibinfo {author} {\bibfnamefont {Shota}\
  \bibnamefont {Yokoyama}}, \bibinfo {author} {\bibfnamefont {Takeru}\
  \bibnamefont {Ebihara}}, \bibinfo {author} {\bibfnamefont {Tomohiro}\
  \bibnamefont {Nakamura}}, \bibinfo {author} {\bibfnamefont {Rafael~N.}\
  \bibnamefont {Alexander}}, \bibinfo {author} {\bibfnamefont {Mamoru}\
  \bibnamefont {Endo}}, \bibinfo {author} {\bibfnamefont {Jun-ichi}\
  \bibnamefont {Yoshikawa}}, \bibinfo {author} {\bibfnamefont {Nicolas~C.}\
  \bibnamefont {Menicucci}}, \bibinfo {author} {\bibfnamefont {Hidehiro}\
  \bibnamefont {Yonezawa}}, and\ \bibinfo {author} {\bibfnamefont {Akira}\
  \bibnamefont {Furusawa}}} (\bibinfo {year} {2021}),\ \bibfield  {title}
  {\enquote {\bibinfo {title} {Time-domain-multiplexed measurement-based
  quantum operations with 25-mhz clock frequency},}\ }\href
  {https://doi.org/10.1103/PhysRevApplied.16.034005} {\bibfield  {journal}
  {\bibinfo  {journal} {Phys. Rev. Appl.}\ }\textbf {\bibinfo {volume} {16}},\
  \bibinfo {pages} {034005}}\BibitemShut {NoStop}%
\bibitem [{\citenamefont {Aschauer}(2004)}]{bib:Aschauer2004}%
  \BibitemOpen
  \bibfield  {author} {\bibinfo {author} {\bibnamefont {Aschauer},
  \bibfnamefont {H}}} (\bibinfo {year} {2004}),\ \href@noop {} {Ph.D. thesis}\
  (\bibinfo  {school} {Ludwig Maximilians Universitat, Munchen})\BibitemShut
  {NoStop}%
\bibitem [{\citenamefont {Aspelmeyer}\ \emph {et~al.}(2003)\citenamefont
  {Aspelmeyer}, \citenamefont {Jennewein}, \citenamefont {Pfennigbauer},
  \citenamefont {Leeb},\ and\ \citenamefont
  {Zeilinger}}]{bib:aspelmeyer2003long}%
  \BibitemOpen
  \bibfield  {author} {\bibinfo {author} {\bibnamefont {Aspelmeyer},
  \bibfnamefont {Markus}}, \bibinfo {author} {\bibfnamefont {Thomas}\
  \bibnamefont {Jennewein}}, \bibinfo {author} {\bibfnamefont {Martin}\
  \bibnamefont {Pfennigbauer}}, \bibinfo {author} {\bibfnamefont {Walter~R}\
  \bibnamefont {Leeb}}, and\ \bibinfo {author} {\bibfnamefont {Anton}\
  \bibnamefont {Zeilinger}}} (\bibinfo {year} {2003}),\ \bibfield  {title}
  {\enquote {\bibinfo {title} {Long-distance quantum communication with
  entangled photons using satellites},}\ }\href
  {https://doi.org/10.1109/jstqe.2003.820918} {\bibfield  {journal} {\bibinfo
  {journal} {IEEE Journal of Selected Topics in Quantum Electronics}\ }\textbf
  {\bibinfo {volume} {9}},\ \bibinfo {pages} {1541}},\ \Eprint
  {https://arxiv.org/abs/arXiv:quant-ph/0305105v1} {arXiv:quant-ph/0305105v1}
  \BibitemShut {NoStop}%
\bibitem [{\citenamefont {Astner}\ \emph
  {et~al.}(2018{\natexlab{a}})\citenamefont {Astner}, \citenamefont {Gugler},
  \citenamefont {Angerer}, \citenamefont {Wald}, \citenamefont {Putz},
  \citenamefont {Mauser}, \citenamefont {Trupke}, \citenamefont {Sumiya},
  \citenamefont {Onoda}, \citenamefont {Isoya}, \citenamefont {Schmiedmayer},
  \citenamefont {Mohn},\ and\ \citenamefont {Majer}}]{SD-Astner:2018aa}%
  \BibitemOpen
  \bibfield  {author} {\bibinfo {author} {\bibnamefont {Astner}, \bibfnamefont
  {T}}, \bibinfo {author} {\bibfnamefont {J.}~\bibnamefont {Gugler}}, \bibinfo
  {author} {\bibfnamefont {A.}~\bibnamefont {Angerer}}, \bibinfo {author}
  {\bibfnamefont {S.}~\bibnamefont {Wald}}, \bibinfo {author} {\bibfnamefont
  {S.}~\bibnamefont {Putz}}, \bibinfo {author} {\bibfnamefont {N.~J.}\
  \bibnamefont {Mauser}}, \bibinfo {author} {\bibfnamefont {M.}~\bibnamefont
  {Trupke}}, \bibinfo {author} {\bibfnamefont {H.}~\bibnamefont {Sumiya}},
  \bibinfo {author} {\bibfnamefont {S.}~\bibnamefont {Onoda}}, \bibinfo
  {author} {\bibfnamefont {J.}~\bibnamefont {Isoya}}, \bibinfo {author}
  {\bibfnamefont {J.}~\bibnamefont {Schmiedmayer}}, \bibinfo {author}
  {\bibfnamefont {P.}~\bibnamefont {Mohn}}, and\ \bibinfo {author}
  {\bibfnamefont {J.}~\bibnamefont {Majer}}} (\bibinfo {year}
  {2018}{\natexlab{a}}),\ \bibfield  {title} {\enquote {\bibinfo {title}
  {Solid-state electron spin lifetime limited by phononic vacuum modes},}\
  }\href {https://doi.org/10.1038/s41563-017-0008-y} {\bibfield  {journal}
  {\bibinfo  {journal} {Nature Materials}\ }\textbf {\bibinfo {volume} {17}},\
  \bibinfo {pages} {313}}\BibitemShut {NoStop}%
\bibitem [{\citenamefont {Astner}\ \emph
  {et~al.}(2018{\natexlab{b}})\citenamefont {Astner}, \citenamefont {Gugler},
  \citenamefont {Angerer}, \citenamefont {Wald}, \citenamefont {Putz},
  \citenamefont {Mauser}, \citenamefont {Trupke}, \citenamefont {Sumiya},
  \citenamefont {Onoda}, \citenamefont {Isoya} \emph
  {et~al.}}]{bib:astner2018solid}%
  \BibitemOpen
  \bibfield  {author} {\bibinfo {author} {\bibnamefont {Astner}, \bibfnamefont
  {Thomas}}, \bibinfo {author} {\bibfnamefont {Johannes}\ \bibnamefont
  {Gugler}}, \bibinfo {author} {\bibfnamefont {Andreas}\ \bibnamefont
  {Angerer}}, \bibinfo {author} {\bibfnamefont {Sebastian}\ \bibnamefont
  {Wald}}, \bibinfo {author} {\bibfnamefont {Stefan}\ \bibnamefont {Putz}},
  \bibinfo {author} {\bibfnamefont {Norbert~J}\ \bibnamefont {Mauser}},
  \bibinfo {author} {\bibfnamefont {Michael}\ \bibnamefont {Trupke}}, \bibinfo
  {author} {\bibfnamefont {Hitoshi}\ \bibnamefont {Sumiya}}, \bibinfo {author}
  {\bibfnamefont {Shinobu}\ \bibnamefont {Onoda}}, \bibinfo {author}
  {\bibfnamefont {Junichi}\ \bibnamefont {Isoya}},  \emph {et~al.}} (\bibinfo
  {year} {2018}{\natexlab{b}}),\ \bibfield  {title} {\enquote {\bibinfo {title}
  {Solid-state electron spin lifetime limited by phononic vacuum modes},}\
  }\href {https://doi.org/10.1038/s41563-017-0008-y} {\bibfield  {journal}
  {\bibinfo  {journal} {Nature Materials}\ }\textbf {\bibinfo {volume} {17}},\
  \bibinfo {pages} {313}},\ \Eprint {https://arxiv.org/abs/arXiv:1706.09798v1}
  {arXiv:1706.09798v1} \BibitemShut {NoStop}%
\bibitem [{\citenamefont {Atat{\"u}re}\ \emph {et~al.}(2018)\citenamefont
  {Atat{\"u}re}, \citenamefont {Englund}, \citenamefont {Vamivakas},
  \citenamefont {Lee},\ and\ \citenamefont
  {Wrachtrup}}]{bib:atature2018material}%
  \BibitemOpen
  \bibfield  {author} {\bibinfo {author} {\bibnamefont {Atat{\"u}re},
  \bibfnamefont {Mete}}, \bibinfo {author} {\bibfnamefont {Dirk}\ \bibnamefont
  {Englund}}, \bibinfo {author} {\bibfnamefont {Nick}\ \bibnamefont
  {Vamivakas}}, \bibinfo {author} {\bibfnamefont {Sang-Yun}\ \bibnamefont
  {Lee}}, and\ \bibinfo {author} {\bibfnamefont {Joerg}\ \bibnamefont
  {Wrachtrup}}} (\bibinfo {year} {2018}),\ \bibfield  {title} {\enquote
  {\bibinfo {title} {Material platforms for spin-based photonic quantum
  technologies},}\ }\href {https://doi.org/10.1038/s41578-018-0008-9}
  {\bibfield  {journal} {\bibinfo  {journal} {Nature Reviews Materials}\
  }\textbf {\bibinfo {volume} {3}},\ \bibinfo {pages} {38}}\BibitemShut
  {NoStop}%
\bibitem [{\citenamefont {Averin}(1998)}]{bib:averin1998adiabatic}%
  \BibitemOpen
  \bibfield  {author} {\bibinfo {author} {\bibnamefont {Averin}, \bibfnamefont
  {DV}}} (\bibinfo {year} {1998}),\ \bibfield  {title} {\enquote {\bibinfo
  {title} {Adiabatic quantum computation with cooper pairs},}\ }\href
  {https://doi.org/10.1016/s0038-1098(97)10001-1} {\bibfield  {journal}
  {\bibinfo  {journal} {Solid State Communications}\ }\textbf {\bibinfo
  {volume} {105}},\ \bibinfo {pages} {659}},\ \Eprint
  {https://arxiv.org/abs/arXiv:quant-ph/9706026v1} {arXiv:quant-ph/9706026v1}
  \BibitemShut {NoStop}%
\bibitem [{\citenamefont {Avizienis}(1987)}]{bib:A87}%
  \BibitemOpen
  \bibfield  {author} {\bibinfo {author} {\bibnamefont {Avizienis},
  \bibfnamefont {A}}} (\bibinfo {year} {1987}),\ \href@noop {} {\emph {\bibinfo
  {title} {{The Evolution of Fault-Tolerant Computing}}}}\ (\bibinfo
  {publisher} {Springer -Verlag, New York})\BibitemShut {NoStop}%
\bibitem [{\citenamefont {Awschalom}\ \emph {et~al.}(2018)\citenamefont
  {Awschalom}, \citenamefont {Hanson}, \citenamefont {Wrachtrup},\ and\
  \citenamefont {Zhou}}]{bib:awschalom2018quantum}%
  \BibitemOpen
  \bibfield  {author} {\bibinfo {author} {\bibnamefont {Awschalom},
  \bibfnamefont {David~D}}, \bibinfo {author} {\bibfnamefont {Ronald}\
  \bibnamefont {Hanson}}, \bibinfo {author} {\bibfnamefont {J{\"o}rg}\
  \bibnamefont {Wrachtrup}}, and\ \bibinfo {author} {\bibfnamefont {Brian~B}\
  \bibnamefont {Zhou}}} (\bibinfo {year} {2018}),\ \bibfield  {title} {\enquote
  {\bibinfo {title} {Quantum technologies with optically interfaced solid-state
  spins},}\ }\href {https://doi.org/10.1038/s41566-018-0232-2} {\bibfield
  {journal} {\bibinfo  {journal} {Nature Photonics}\ }\textbf {\bibinfo
  {volume} {12}},\ \bibinfo {pages} {516}}\BibitemShut {NoStop}%
\bibitem [{\citenamefont {Azuma}\ \emph
  {et~al.}(2015{\natexlab{a}})\citenamefont {Azuma}, \citenamefont {Tamaki},\
  and\ \citenamefont {Lo}}]{bib:ATL13}%
  \BibitemOpen
  \bibfield  {author} {\bibinfo {author} {\bibnamefont {Azuma}, \bibfnamefont
  {K}}, \bibinfo {author} {\bibfnamefont {K.}~\bibnamefont {Tamaki}}, and\
  \bibinfo {author} {\bibfnamefont {H.~K.}\ \bibnamefont {Lo}}} (\bibinfo
  {year} {2015}{\natexlab{a}}),\ \bibfield  {title} {\enquote {\bibinfo {title}
  {All photonic quantum repeaters},}\ }\href
  {https://doi.org/10.1038/ncomms7787} {\bibfield  {journal} {\bibinfo
  {journal} {Nature Communications}\ }\textbf {\bibinfo {volume} {6}},\
  \bibinfo {pages} {6787}},\ \Eprint {https://arxiv.org/abs/arXiv:1309.7207v1}
  {arXiv:1309.7207v1} \BibitemShut {NoStop}%
\bibitem [{\citenamefont {Azuma}\ \emph
  {et~al.}(2015{\natexlab{b}})\citenamefont {Azuma}, \citenamefont {Tamaki},\
  and\ \citenamefont {Lo}}]{SD-Azuma:2015aa}%
  \BibitemOpen
  \bibfield  {author} {\bibinfo {author} {\bibnamefont {Azuma}, \bibfnamefont
  {Koji}}, \bibinfo {author} {\bibfnamefont {Kiyoshi}\ \bibnamefont {Tamaki}},
  and\ \bibinfo {author} {\bibfnamefont {Hoi-Kwong}\ \bibnamefont {Lo}}}
  (\bibinfo {year} {2015}{\natexlab{b}}),\ \bibfield  {title} {\enquote
  {\bibinfo {title} {All-photonic quantum repeaters},}\ }\href
  {https://doi.org/10.1038/ncomms7787} {\bibfield  {journal} {\bibinfo
  {journal} {Nature Communications}\ }\textbf {\bibinfo {volume} {6}},\
  \bibinfo {pages} {6787}},\ \Eprint {https://arxiv.org/abs/arXiv:1309.7207v1}
  {arXiv:1309.7207v1} \BibitemShut {NoStop}%
\bibitem [{\citenamefont {Azuma}\ \emph
  {et~al.}(2015{\natexlab{c}})\citenamefont {Azuma}, \citenamefont {Tamaki},\
  and\ \citenamefont {Lo}}]{bib:azuma2015all}%
  \BibitemOpen
  \bibfield  {author} {\bibinfo {author} {\bibnamefont {Azuma}, \bibfnamefont
  {Koji}}, \bibinfo {author} {\bibfnamefont {Kiyoshi}\ \bibnamefont {Tamaki}},
  and\ \bibinfo {author} {\bibfnamefont {Hoi-Kwong}\ \bibnamefont {Lo}}}
  (\bibinfo {year} {2015}{\natexlab{c}}),\ \bibfield  {title} {\enquote
  {\bibinfo {title} {All-photonic quantum repeaters},}\ }\href
  {https://doi.org/10.1038/ncomms7787} {\bibfield  {journal} {\bibinfo
  {journal} {Nature Communications}\ }\textbf {\bibinfo {volume} {6}},\
  \bibinfo {pages} {6787}},\ \Eprint {https://arxiv.org/abs/arXiv:1309.7207v1}
  {arXiv:1309.7207v1} \BibitemShut {NoStop}%
\bibitem [{\citenamefont {Bacharach}(1976)}]{bib:Bacharach76}%
  \BibitemOpen
  \bibfield  {author} {\bibinfo {author} {\bibnamefont {Bacharach},
  \bibfnamefont {M}}} (\bibinfo {year} {1976}),\ \href@noop {} {\emph {\bibinfo
  {title} {Economics and the Theory of Games}}}\ (\bibinfo  {publisher}
  {Macmillan})\BibitemShut {NoStop}%
\bibitem [{\citenamefont {Balasubramanian}\ \emph {et~al.}(2009)\citenamefont
  {Balasubramanian}, \citenamefont {Neumann}, \citenamefont {Twitchen},
  \citenamefont {Markham}, \citenamefont {Kolesov}, \citenamefont {Mizuochi},
  \citenamefont {Isoya}, \citenamefont {Achard}, \citenamefont {Beck},
  \citenamefont {Tissler} \emph {et~al.}}]{bib:balasubramanian2009ultralong}%
  \BibitemOpen
  \bibfield  {author} {\bibinfo {author} {\bibnamefont {Balasubramanian},
  \bibfnamefont {Gopalakrishnan}}, \bibinfo {author} {\bibfnamefont {Philipp}\
  \bibnamefont {Neumann}}, \bibinfo {author} {\bibfnamefont {Daniel}\
  \bibnamefont {Twitchen}}, \bibinfo {author} {\bibfnamefont {Matthew}\
  \bibnamefont {Markham}}, \bibinfo {author} {\bibfnamefont {Roman}\
  \bibnamefont {Kolesov}}, \bibinfo {author} {\bibfnamefont {Norikazu}\
  \bibnamefont {Mizuochi}}, \bibinfo {author} {\bibfnamefont {Junichi}\
  \bibnamefont {Isoya}}, \bibinfo {author} {\bibfnamefont {Jocelyn}\
  \bibnamefont {Achard}}, \bibinfo {author} {\bibfnamefont {Johannes}\
  \bibnamefont {Beck}}, \bibinfo {author} {\bibfnamefont {Julia}\ \bibnamefont
  {Tissler}},  \emph {et~al.}} (\bibinfo {year} {2009}),\ \bibfield  {title}
  {\enquote {\bibinfo {title} {Ultralong spin coherence time in isotopically
  engineered diamond},}\ }\href {https://doi.org/10.1038/nmat2420} {\bibfield
  {journal} {\bibinfo  {journal} {Nature Materials}\ }\textbf {\bibinfo
  {volume} {8}},\ \bibinfo {pages} {383}}\BibitemShut {NoStop}%
\bibitem [{\citenamefont {Balcan}\ \emph {et~al.}(2012)\citenamefont {Balcan},
  \citenamefont {Blum}, \citenamefont {Fine},\ and\ \citenamefont
  {Mansour}}]{bib:balcan2012distributed}%
  \BibitemOpen
  \bibfield  {author} {\bibinfo {author} {\bibnamefont {Balcan}, \bibfnamefont
  {Maria~Florina}}, \bibinfo {author} {\bibfnamefont {Avrim}\ \bibnamefont
  {Blum}}, \bibinfo {author} {\bibfnamefont {Shai}\ \bibnamefont {Fine}}, and\
  \bibinfo {author} {\bibfnamefont {Yishay}\ \bibnamefont {Mansour}}} (\bibinfo
  {year} {2012}),\ \bibfield  {title} {\enquote {\bibinfo {title} {Distributed
  learning, communication complexity and privacy},}\ }in\ \href@noop {} {\emph
  {\bibinfo {booktitle} {Conference on Learning Theory}}},\ p.~\bibinfo {pages}
  {26},\ \Eprint {https://arxiv.org/abs/arXiv:1204.3514v3} {arXiv:1204.3514v3}
  \BibitemShut {NoStop}%
\bibitem [{\citenamefont {Balcan}\ \emph {et~al.}(2013)\citenamefont {Balcan},
  \citenamefont {Ehrlich},\ and\ \citenamefont {Liang}}]{bib:florian2013}%
  \BibitemOpen
  \bibfield  {author} {\bibinfo {author} {\bibnamefont {Balcan}, \bibfnamefont
  {Maria-Florina~F}}, \bibinfo {author} {\bibfnamefont {Steven}\ \bibnamefont
  {Ehrlich}}, and\ \bibinfo {author} {\bibfnamefont {Yingyu}\ \bibnamefont
  {Liang}}} (\bibinfo {year} {2013}),\ \bibfield  {title} {\enquote {\bibinfo
  {title} {Distributed k-means and k-median clustering on general
  topologies},}\ }\href@noop {} {\bibfield  {journal} {\bibinfo  {journal}
  {Advances in Neural Information Processing Systems}\ ,\ \bibinfo {pages}
  {1995}}}\Eprint {https://arxiv.org/abs/arXiv:1306.0604v3} {arXiv:1306.0604v3}
  \BibitemShut {NoStop}%
\bibitem [{\citenamefont {Balensiefer}\ \emph {et~al.}(2005)\citenamefont
  {Balensiefer}, \citenamefont {Kregor-Stickles},\ and\ \citenamefont
  {Oskin}}]{bib:BSO05}%
  \BibitemOpen
  \bibfield  {author} {\bibinfo {author} {\bibnamefont {Balensiefer},
  \bibfnamefont {S}}, \bibinfo {author} {\bibfnamefont {L.}~\bibnamefont
  {Kregor-Stickles}}, and\ \bibinfo {author} {\bibfnamefont {M.}~\bibnamefont
  {Oskin}}} (\bibinfo {year} {2005}),\ \bibfield  {title} {\enquote {\bibinfo
  {title} {{An Evaluation Framework and Instruction Set Architecture for
  Ion-Trap based Quantum Micro-architectures}},}\ }\href@noop {} {\bibfield
  {journal} {\bibinfo  {journal} {SIGARCH Comput. Archit. News}\ }\textbf
  {\bibinfo {volume} {33}}~(\bibinfo {number} {2}),\ \bibinfo {pages}
  {186}}\BibitemShut {NoStop}%
\bibitem [{\citenamefont {Banaszek}\ and\ \citenamefont
  {Walmsley}(2003)}]{bib:Banaszek03}%
  \BibitemOpen
  \bibfield  {author} {\bibinfo {author} {\bibnamefont {Banaszek},
  \bibfnamefont {K}}, and\ \bibinfo {author} {\bibfnamefont {I.}~\bibnamefont
  {Walmsley}}} (\bibinfo {year} {2003}),\ \bibfield  {title} {\enquote
  {\bibinfo {title} {Photon counting with loop detector},}\ }\href
  {https://doi.org/10.1364/ol.28.000052} {\bibfield  {journal} {\bibinfo
  {journal} {Optics Letters}\ }\textbf {\bibinfo {volume} {28}},\ \bibinfo
  {pages} {52}},\ \Eprint {https://arxiv.org/abs/arXiv:quant-ph/0206162v1}
  {arXiv:quant-ph/0206162v1} \BibitemShut {NoStop}%
\bibitem [{\citenamefont {Bang}\ \emph {et~al.}(2015)\citenamefont {Bang},
  \citenamefont {Lee},\ and\ \citenamefont {Jeong}}]{bib:bang2015protocol}%
  \BibitemOpen
  \bibfield  {author} {\bibinfo {author} {\bibnamefont {Bang}, \bibfnamefont
  {Jeongho}}, \bibinfo {author} {\bibfnamefont {Seung-Woo}\ \bibnamefont
  {Lee}}, and\ \bibinfo {author} {\bibfnamefont {Hyunseok}\ \bibnamefont
  {Jeong}}} (\bibinfo {year} {2015}),\ \bibfield  {title} {\enquote {\bibinfo
  {title} {Protocol for secure quantum machine learning at a distant place},}\
  }\href {https://doi.org/10.1007/s11128-015-1089-7} {\bibfield  {journal}
  {\bibinfo  {journal} {Quantum Information Processing}\ }\textbf {\bibinfo
  {volume} {14}},\ \bibinfo {pages} {3933}},\ \Eprint
  {https://arxiv.org/abs/arXiv:1504.04929v2} {arXiv:1504.04929v2} \BibitemShut
  {NoStop}%
\bibitem [{\citenamefont {Bao}\ \emph {et~al.}(2007)\citenamefont {Bao},
  \citenamefont {Chen}, \citenamefont {Zhang}, \citenamefont {Yang},
  \citenamefont {Zhang}, \citenamefont {Yang},\ and\ \citenamefont
  {Pan}}]{bib:Bao2007Optical}%
  \BibitemOpen
  \bibfield  {author} {\bibinfo {author} {\bibnamefont {Bao}, \bibfnamefont
  {Xiao-Hui}}, \bibinfo {author} {\bibfnamefont {Teng-Yun}\ \bibnamefont
  {Chen}}, \bibinfo {author} {\bibfnamefont {Qiang}\ \bibnamefont {Zhang}},
  \bibinfo {author} {\bibfnamefont {Jian}\ \bibnamefont {Yang}}, \bibinfo
  {author} {\bibfnamefont {Han}\ \bibnamefont {Zhang}}, \bibinfo {author}
  {\bibfnamefont {Tao}\ \bibnamefont {Yang}}, and\ \bibinfo {author}
  {\bibfnamefont {Jian-Wei}\ \bibnamefont {Pan}}} (\bibinfo {year} {2007}),\
  \bibfield  {title} {\enquote {\bibinfo {title} {Optical nondestructive
  controlled-not gate without using entangled photons},}\ }\href
  {https://doi.org/10.1103/PhysRevLett.98.170502} {\bibfield  {journal}
  {\bibinfo  {journal} {Physical Review Letters}\ }\textbf {\bibinfo {volume}
  {98}},\ \bibinfo {pages} {170502}},\ \Eprint
  {https://arxiv.org/abs/arXiv:quant-ph/0610182v3} {arXiv:quant-ph/0610182v3}
  \BibitemShut {NoStop}%
\bibitem [{\citenamefont {Baragiola}\ \emph {et~al.}(2012)\citenamefont
  {Baragiola}, \citenamefont {Cook}, \citenamefont {Bra\ifmmode~\acute{n}\else
  \'{n}\fi{}czyk},\ and\ \citenamefont {Combes}}]{baragiola2012photons}%
  \BibitemOpen
  \bibfield  {author} {\bibinfo {author} {\bibnamefont {Baragiola},
  \bibfnamefont {Ben~Q}}, \bibinfo {author} {\bibfnamefont {Robert~L.}\
  \bibnamefont {Cook}}, \bibinfo {author} {\bibfnamefont {Agata~M.}\
  \bibnamefont {Bra\ifmmode~\acute{n}\else \'{n}\fi{}czyk}}, and\ \bibinfo
  {author} {\bibfnamefont {Joshua}\ \bibnamefont {Combes}}} (\bibinfo {year}
  {2012}),\ \bibfield  {title} {\enquote {\bibinfo {title} {$n$-photon wave
  packets interacting with an arbitrary quantum system},}\ }\href
  {https://doi.org/10.1103/PhysRevA.86.013811} {\bibfield  {journal} {\bibinfo
  {journal} {Phys. Rev. A}\ }\textbf {\bibinfo {volume} {86}},\ \bibinfo
  {pages} {013811}}\BibitemShut {NoStop}%
\bibitem [{\citenamefont {Baragiola}\ \emph {et~al.}(2019)\citenamefont
  {Baragiola}, \citenamefont {Pantaleoni}, \citenamefont {Alexander},
  \citenamefont {Karanjai},\ and\ \citenamefont
  {Menicucci}}]{baragiola2010allgaussian}%
  \BibitemOpen
  \bibfield  {author} {\bibinfo {author} {\bibnamefont {Baragiola},
  \bibfnamefont {Ben~Q}}, \bibinfo {author} {\bibfnamefont {Giacomo}\
  \bibnamefont {Pantaleoni}}, \bibinfo {author} {\bibfnamefont {Rafael~N.}\
  \bibnamefont {Alexander}}, \bibinfo {author} {\bibfnamefont {Angela}\
  \bibnamefont {Karanjai}}, and\ \bibinfo {author} {\bibfnamefont {Nicolas~C.}\
  \bibnamefont {Menicucci}}} (\bibinfo {year} {2019}),\ \bibfield  {title}
  {\enquote {\bibinfo {title} {All-gaussian universality and fault tolerance
  with the gottesman-kitaev-preskill code},}\ }\href
  {https://doi.org/10.1103/PhysRevLett.123.200502} {\bibfield  {journal}
  {\bibinfo  {journal} {Phys. Rev. Lett.}\ }\textbf {\bibinfo {volume} {123}},\
  \bibinfo {pages} {200502}}\BibitemShut {NoStop}%
\bibitem [{\citenamefont {Barends}\ \emph {et~al.}(2013)\citenamefont
  {Barends}, \citenamefont {Kelly}, \citenamefont {Megrant}, \citenamefont
  {Sank}, \citenamefont {Jeffrey}, \citenamefont {Chen}, \citenamefont {Yin},
  \citenamefont {Chiaro}, \citenamefont {Mutus}, \citenamefont {Neill} \emph
  {et~al.}}]{bib:barends2013coherent}%
  \BibitemOpen
  \bibfield  {author} {\bibinfo {author} {\bibnamefont {Barends}, \bibfnamefont
  {R}}, \bibinfo {author} {\bibfnamefont {J}~\bibnamefont {Kelly}}, \bibinfo
  {author} {\bibfnamefont {A}~\bibnamefont {Megrant}}, \bibinfo {author}
  {\bibfnamefont {D}~\bibnamefont {Sank}}, \bibinfo {author} {\bibfnamefont
  {E}~\bibnamefont {Jeffrey}}, \bibinfo {author} {\bibfnamefont
  {Yu}~\bibnamefont {Chen}}, \bibinfo {author} {\bibfnamefont {Y}~\bibnamefont
  {Yin}}, \bibinfo {author} {\bibfnamefont {B}~\bibnamefont {Chiaro}}, \bibinfo
  {author} {\bibfnamefont {J}~\bibnamefont {Mutus}}, \bibinfo {author}
  {\bibfnamefont {C}~\bibnamefont {Neill}},  \emph {et~al.}} (\bibinfo {year}
  {2013}),\ \bibfield  {title} {\enquote {\bibinfo {title} {Coherent josephson
  qubit suitable for scalable quantum integrated circuits},}\ }\href
  {https://doi.org/10.1103/physrevlett.111.080502} {\bibfield  {journal}
  {\bibinfo  {journal} {Physical Review Letters}\ }\textbf {\bibinfo {volume}
  {111}},\ \bibinfo {pages} {080502}},\ \Eprint
  {https://arxiv.org/abs/arXiv:1304.2322v1} {arXiv:1304.2322v1} \BibitemShut
  {NoStop}%
\bibitem [{\citenamefont {Barends}\ \emph {et~al.}(2014)\citenamefont
  {Barends}, \citenamefont {Kelly}, \citenamefont {Megrant}, \citenamefont
  {Veitia}, \citenamefont {Sank}, \citenamefont {Jeffrey}, \citenamefont
  {White}, \citenamefont {Mutus}, \citenamefont {Fowler}, \citenamefont
  {Campbell} \emph {et~al.}}]{bib:barends2014superconducting}%
  \BibitemOpen
  \bibfield  {author} {\bibinfo {author} {\bibnamefont {Barends}, \bibfnamefont
  {R}}, \bibinfo {author} {\bibfnamefont {J}~\bibnamefont {Kelly}}, \bibinfo
  {author} {\bibfnamefont {A}~\bibnamefont {Megrant}}, \bibinfo {author}
  {\bibfnamefont {A}~\bibnamefont {Veitia}}, \bibinfo {author} {\bibfnamefont
  {D}~\bibnamefont {Sank}}, \bibinfo {author} {\bibfnamefont {E}~\bibnamefont
  {Jeffrey}}, \bibinfo {author} {\bibfnamefont {TC}~\bibnamefont {White}},
  \bibinfo {author} {\bibfnamefont {J}~\bibnamefont {Mutus}}, \bibinfo {author}
  {\bibfnamefont {AG}~\bibnamefont {Fowler}}, \bibinfo {author} {\bibfnamefont
  {B}~\bibnamefont {Campbell}},  \emph {et~al.}} (\bibinfo {year} {2014}),\
  \bibfield  {title} {\enquote {\bibinfo {title} {Superconducting quantum
  circuits at the surface code threshold for fault tolerance},}\ }\href
  {https://doi.org/10.1038/nature13171} {\bibfield  {journal} {\bibinfo
  {journal} {Nature}\ }\textbf {\bibinfo {volume} {508}},\ \bibinfo {pages}
  {500}}\BibitemShut {NoStop}%
\bibitem [{\citenamefont {Barman}\ and\ \citenamefont
  {Matta}(2004)}]{bib:barman2004model}%
  \BibitemOpen
  \bibfield  {author} {\bibinfo {author} {\bibnamefont {Barman}, \bibfnamefont
  {Dhiman}}, and\ \bibinfo {author} {\bibfnamefont {Ibrahim}\ \bibnamefont
  {Matta}}} (\bibinfo {year} {2004}),\ \bibfield  {title} {\enquote {\bibinfo
  {title} {Model-based loss inference by tcp over heterogeneous networks},}\
  }in\ \href@noop {} {\emph {\bibinfo {booktitle} {Proceedings of WiOpt}}},\
  Vol.~\bibinfo {volume} {4}\BibitemShut {NoStop}%
\bibitem [{\citenamefont {Barnett}(2009)}]{barnett2009quantum}%
  \BibitemOpen
  \bibfield  {author} {\bibinfo {author} {\bibnamefont {Barnett}, \bibfnamefont
  {Stephen}}} (\bibinfo {year} {2009}),\ \href@noop {} {\emph {\bibinfo {title}
  {Quantum information}}},\ Vol.~\bibinfo {volume} {16}\ (\bibinfo  {publisher}
  {Oxford University Press})\BibitemShut {NoStop}%
\bibitem [{\citenamefont {Barrett}\ \emph {et~al.}(2004)\citenamefont
  {Barrett}, \citenamefont {Chiaverini}, \citenamefont {Schaetz}, \citenamefont
  {Britton}, \citenamefont {Itano}, \citenamefont {Jost}, \citenamefont
  {Knill}, \citenamefont {Langer}, \citenamefont {Leibfried}, \citenamefont
  {Ozeri} \emph {et~al.}}]{bib:Nat_429_737}%
  \BibitemOpen
  \bibfield  {author} {\bibinfo {author} {\bibnamefont {Barrett}, \bibfnamefont
  {MD}}, \bibinfo {author} {\bibfnamefont {J}~\bibnamefont {Chiaverini}},
  \bibinfo {author} {\bibfnamefont {T}~\bibnamefont {Schaetz}}, \bibinfo
  {author} {\bibfnamefont {J}~\bibnamefont {Britton}}, \bibinfo {author}
  {\bibfnamefont {WM}~\bibnamefont {Itano}}, \bibinfo {author} {\bibfnamefont
  {JD}~\bibnamefont {Jost}}, \bibinfo {author} {\bibfnamefont {E}~\bibnamefont
  {Knill}}, \bibinfo {author} {\bibfnamefont {C}~\bibnamefont {Langer}},
  \bibinfo {author} {\bibfnamefont {D}~\bibnamefont {Leibfried}}, \bibinfo
  {author} {\bibfnamefont {R}~\bibnamefont {Ozeri}},  \emph {et~al.}} (\bibinfo
  {year} {2004}),\ \bibfield  {title} {\enquote {\bibinfo {title}
  {Deterministic quantum teleportation of atomic qubits},}\ }\href
  {https://doi.org/10.1038/nature02608} {\bibfield  {journal} {\bibinfo
  {journal} {Nature}\ }\textbf {\bibinfo {volume} {429}},\ \bibinfo {pages}
  {737}}\BibitemShut {NoStop}%
\bibitem [{\citenamefont {Barrett}\ and\ \citenamefont
  {Kok}(2005)}]{bib:BarrettKok05}%
  \BibitemOpen
  \bibfield  {author} {\bibinfo {author} {\bibnamefont {Barrett}, \bibfnamefont
  {Sean~D}}, and\ \bibinfo {author} {\bibfnamefont {Pieter}\ \bibnamefont
  {Kok}}} (\bibinfo {year} {2005}),\ \bibfield  {title} {\enquote {\bibinfo
  {title} {Efficienct high-fidelity quantum computation using matter qubits and
  linear optics},}\ }\href {https://doi.org/10.1103/physreva.71.060310}
  {\bibfield  {journal} {\bibinfo  {journal} {Physical Review A}\ }\textbf
  {\bibinfo {volume} {71}},\ \bibinfo {pages} {060310(R)}},\ \Eprint
  {https://arxiv.org/abs/arXiv:quant-ph/0408040v2} {arXiv:quant-ph/0408040v2}
  \BibitemShut {NoStop}%
\bibitem [{\citenamefont {Barrett}\ \emph {et~al.}(2010)\citenamefont
  {Barrett}, \citenamefont {Rohde},\ and\ \citenamefont
  {Stace}}]{bib:RohdeAtEns10}%
  \BibitemOpen
  \bibfield  {author} {\bibinfo {author} {\bibnamefont {Barrett}, \bibfnamefont
  {Sean~D}}, \bibinfo {author} {\bibfnamefont {Peter~P.}\ \bibnamefont
  {Rohde}}, and\ \bibinfo {author} {\bibfnamefont {Thomas~M.}\ \bibnamefont
  {Stace}}} (\bibinfo {year} {2010}),\ \bibfield  {title} {\enquote {\bibinfo
  {title} {Scalable quantum computing with atomic ensembles},}\ }\href
  {https://doi.org/10.1088/1367-2630/12/9/093032} {\bibfield  {journal}
  {\bibinfo  {journal} {New Journal of Physics}\ }\textbf {\bibinfo {volume}
  {12}},\ \bibinfo {pages} {093032}}\BibitemShut {NoStop}%
\bibitem [{\citenamefont {Barrett}\ and\ \citenamefont
  {Stace}(2010)}]{bib:BarrettStaceFT}%
  \BibitemOpen
  \bibfield  {author} {\bibinfo {author} {\bibnamefont {Barrett}, \bibfnamefont
  {Sean~D}}, and\ \bibinfo {author} {\bibfnamefont {Thomas~M.}\ \bibnamefont
  {Stace}}} (\bibinfo {year} {2010}),\ \bibfield  {title} {\enquote {\bibinfo
  {title} {Fault tolerant quantum computation with very high threshold for loss
  errors},}\ }\href {https://doi.org/10.1103/physrevlett.105.200502} {\bibfield
   {journal} {\bibinfo  {journal} {Physical Review Letters}\ }\textbf {\bibinfo
  {volume} {105}},\ \bibinfo {pages} {200502}},\ \Eprint
  {https://arxiv.org/abs/arXiv:1005.2456v2} {arXiv:1005.2456v2} \BibitemShut
  {NoStop}%
\bibitem [{\citenamefont {Bartlett}\ and\ \citenamefont
  {Sanders}(2002)}]{bib:Bartlett02}%
  \BibitemOpen
  \bibfield  {author} {\bibinfo {author} {\bibnamefont {Bartlett},
  \bibfnamefont {Stephen~D}}, and\ \bibinfo {author} {\bibfnamefont {Barry~C.}\
  \bibnamefont {Sanders}}} (\bibinfo {year} {2002}),\ \bibfield  {title}
  {\enquote {\bibinfo {title} {Efficient classical simulation of optical
  quantum information circuits},}\ }\href
  {https://doi.org/10.1103/physrevlett.89.207903} {\bibfield  {journal}
  {\bibinfo  {journal} {Physical Review Letters}\ }\textbf {\bibinfo {volume}
  {89}},\ \bibinfo {pages} {207903}}\BibitemShut {NoStop}%
\bibitem [{\citenamefont {Bartlett}\ \emph {et~al.}(2002)\citenamefont
  {Bartlett}, \citenamefont {Sanders}, \citenamefont {Braunstein},\ and\
  \citenamefont {Nemoto}}]{bib:Bartlett02b}%
  \BibitemOpen
  \bibfield  {author} {\bibinfo {author} {\bibnamefont {Bartlett},
  \bibfnamefont {Stephen~D}}, \bibinfo {author} {\bibfnamefont {Barry~C.}\
  \bibnamefont {Sanders}}, \bibinfo {author} {\bibfnamefont {Samuel~L.}\
  \bibnamefont {Braunstein}}, and\ \bibinfo {author} {\bibfnamefont {Kae}\
  \bibnamefont {Nemoto}}} (\bibinfo {year} {2002}),\ \bibfield  {title}
  {\enquote {\bibinfo {title} {Efficient classical simulation of continuous
  variable quantum information processes},}\ }\href
  {https://doi.org/10.1007/978-94-015-1258-9_6} {\bibfield  {journal} {\bibinfo
   {journal} {Physical Review Letters}\ }\textbf {\bibinfo {volume} {88}},\
  \bibinfo {pages} {097904}}\BibitemShut {NoStop}%
\bibitem [{\citenamefont {Barz}\ \emph {et~al.}(2012)\citenamefont {Barz},
  \citenamefont {Kashefi}, \citenamefont {Broadbent}, \citenamefont
  {Fitzsimons}, \citenamefont {Zeilinger},\ and\ \citenamefont
  {Walther}}]{bib:barz2012demonstration}%
  \BibitemOpen
  \bibfield  {author} {\bibinfo {author} {\bibnamefont {Barz}, \bibfnamefont
  {Stefanie}}, \bibinfo {author} {\bibfnamefont {Elham}\ \bibnamefont
  {Kashefi}}, \bibinfo {author} {\bibfnamefont {Anne}\ \bibnamefont
  {Broadbent}}, \bibinfo {author} {\bibfnamefont {Joseph~F}\ \bibnamefont
  {Fitzsimons}}, \bibinfo {author} {\bibfnamefont {Anton}\ \bibnamefont
  {Zeilinger}}, and\ \bibinfo {author} {\bibfnamefont {Philip}\ \bibnamefont
  {Walther}}} (\bibinfo {year} {2012}),\ \bibfield  {title} {\enquote {\bibinfo
  {title} {Demonstration of blind quantum computing},}\ }\href
  {https://doi.org/10.1126/science.1214707} {\bibfield  {journal} {\bibinfo
  {journal} {Science}\ }\textbf {\bibinfo {volume} {335}},\ \bibinfo {pages}
  {303}},\ \Eprint {https://arxiv.org/abs/arXiv:1110.1381v1}
  {arXiv:1110.1381v1} \BibitemShut {NoStop}%
\bibitem [{\citenamefont {Barzanjeh}\ \emph {et~al.}(2011)\citenamefont
  {Barzanjeh}, \citenamefont {Vitali}, \citenamefont {Tombesi},\ and\
  \citenamefont {Milburn}}]{bib:barzanjeh2011entangling}%
  \BibitemOpen
  \bibfield  {author} {\bibinfo {author} {\bibnamefont {Barzanjeh},
  \bibfnamefont {Sh}}, \bibinfo {author} {\bibfnamefont {D}~\bibnamefont
  {Vitali}}, \bibinfo {author} {\bibfnamefont {P}~\bibnamefont {Tombesi}}, and\
  \bibinfo {author} {\bibfnamefont {GJ}~\bibnamefont {Milburn}}} (\bibinfo
  {year} {2011}),\ \bibfield  {title} {\enquote {\bibinfo {title} {Entangling
  optical and microwave cavity modes by means of a nanomechanical resonator},}\
  }\href {https://doi.org/10.1103/physreva.84.042342} {\bibfield  {journal}
  {\bibinfo  {journal} {Physical Review A}\ }\textbf {\bibinfo {volume} {84}},\
  \bibinfo {pages} {042342}},\ \Eprint
  {https://arxiv.org/abs/arXiv:1107.4152v3} {arXiv:1107.4152v3} \BibitemShut
  {NoStop}%
\bibitem [{\citenamefont {Beals}\ \emph {et~al.}(2013)\citenamefont {Beals},
  \citenamefont {Brierley}, \citenamefont {Gray}, \citenamefont {Harrow},
  \citenamefont {Kutin}, \citenamefont {Linden}, \citenamefont {Shepherd},\
  and\ \citenamefont {Stather}}]{bib:beals2013efficient}%
  \BibitemOpen
  \bibfield  {author} {\bibinfo {author} {\bibnamefont {Beals}, \bibfnamefont
  {Robert}}, \bibinfo {author} {\bibfnamefont {Stephen}\ \bibnamefont
  {Brierley}}, \bibinfo {author} {\bibfnamefont {Oliver}\ \bibnamefont {Gray}},
  \bibinfo {author} {\bibfnamefont {Aram~W}\ \bibnamefont {Harrow}}, \bibinfo
  {author} {\bibfnamefont {Samuel}\ \bibnamefont {Kutin}}, \bibinfo {author}
  {\bibfnamefont {Noah}\ \bibnamefont {Linden}}, \bibinfo {author}
  {\bibfnamefont {Dan}\ \bibnamefont {Shepherd}}, and\ \bibinfo {author}
  {\bibfnamefont {Mark}\ \bibnamefont {Stather}}} (\bibinfo {year} {2013}),\
  \bibfield  {title} {\enquote {\bibinfo {title} {Efficient distributed quantum
  computing},}\ }\href {https://doi.org/10.1098/rspa.2012.0686} {\bibfield
  {journal} {\bibinfo  {journal} {Proceedings of the Royal Society A}\ }\textbf
  {\bibinfo {volume} {469}},\ \bibinfo {pages} {20120686}},\ \Eprint
  {https://arxiv.org/abs/arXiv:1207.2307v2} {arXiv:1207.2307v2} \BibitemShut
  {NoStop}%
\bibitem [{\citenamefont {Bellman}(1958)}]{bellman1958routing}%
  \BibitemOpen
  \bibfield  {author} {\bibinfo {author} {\bibnamefont {Bellman}, \bibfnamefont
  {Richard}}} (\bibinfo {year} {1958}),\ \bibfield  {title} {\enquote {\bibinfo
  {title} {On a routing problem},}\ }\href@noop {} {\bibfield  {journal}
  {\bibinfo  {journal} {Quarterly of applied mathematics}\ }\textbf {\bibinfo
  {volume} {16}}~(\bibinfo {number} {1}),\ \bibinfo {pages}
  {87--90}}\BibitemShut {NoStop}%
\bibitem [{\citenamefont {Ben-Av}\ and\ \citenamefont
  {Exman}(2011)}]{bib:ben2011optimized}%
  \BibitemOpen
  \bibfield  {author} {\bibinfo {author} {\bibnamefont {Ben-Av}, \bibfnamefont
  {Radel}}, and\ \bibinfo {author} {\bibfnamefont {Iaakov}\ \bibnamefont
  {Exman}}} (\bibinfo {year} {2011}),\ \bibfield  {title} {\enquote {\bibinfo
  {title} {Optimized multiparty quantum clock synchronization},}\ }\href
  {https://doi.org/10.1103/physreva.84.014301} {\bibfield  {journal} {\bibinfo
  {journal} {Physical Review A}\ }\textbf {\bibinfo {volume} {84}},\ \bibinfo
  {pages} {014301}},\ \Eprint {https://arxiv.org/abs/arXiv:1105.0186v1}
  {arXiv:1105.0186v1} \BibitemShut {NoStop}%
\bibitem [{\citenamefont {Benedetti}\ \emph {et~al.}(2018)\citenamefont
  {Benedetti}, \citenamefont {Garcia-Pintos}, \citenamefont {Perdomo},
  \citenamefont {Leyton-Ortega}, \citenamefont {Nam},\ and\ \citenamefont
  {Perdomo-Ortiz}}]{bib:benedetti2018generative}%
  \BibitemOpen
  \bibfield  {author} {\bibinfo {author} {\bibnamefont {Benedetti},
  \bibfnamefont {Marcello}}, \bibinfo {author} {\bibfnamefont {Delfina}\
  \bibnamefont {Garcia-Pintos}}, \bibinfo {author} {\bibfnamefont {Oscar}\
  \bibnamefont {Perdomo}}, \bibinfo {author} {\bibfnamefont {Vicente}\
  \bibnamefont {Leyton-Ortega}}, \bibinfo {author} {\bibfnamefont {Yunseong}\
  \bibnamefont {Nam}}, and\ \bibinfo {author} {\bibfnamefont {Alejandro}\
  \bibnamefont {Perdomo-Ortiz}}} (\bibinfo {year} {2018}),\ \bibfield  {title}
  {\enquote {\bibinfo {title} {A generative modeling approach for benchmarking
  and training shallow quantum circuits},}\ }\href@noop {} {\ }\Eprint
  {https://arxiv.org/abs/arXiv:1801.07686} {arXiv:1801.07686} \BibitemShut
  {NoStop}%
\bibitem [{\citenamefont {Bengtsson}\ and\ \citenamefont
  {{\.Z}yczkowski}(2017)}]{bib:bengtsson2017geometry}%
  \BibitemOpen
  \bibfield  {author} {\bibinfo {author} {\bibnamefont {Bengtsson},
  \bibfnamefont {Ingemar}}, and\ \bibinfo {author} {\bibfnamefont {Karol}\
  \bibnamefont {{\.Z}yczkowski}}} (\bibinfo {year} {2017}),\ \href@noop {}
  {\emph {\bibinfo {title} {Geometry of quantum states: an introduction to
  quantum entanglement}}}\ (\bibinfo  {publisher} {Cambridge university
  press})\BibitemShut {NoStop}%
\bibitem [{\citenamefont {Benjamin}\ \emph {et~al.}(2005)\citenamefont
  {Benjamin}, \citenamefont {Eisert},\ and\ \citenamefont
  {Stace}}]{bib:BenjaminEisert05}%
  \BibitemOpen
  \bibfield  {author} {\bibinfo {author} {\bibnamefont {Benjamin},
  \bibfnamefont {S~C}}, \bibinfo {author} {\bibfnamefont {J.}~\bibnamefont
  {Eisert}}, and\ \bibinfo {author} {\bibfnamefont {T.~M.}\ \bibnamefont
  {Stace}}} (\bibinfo {year} {2005}),\ \bibfield  {title} {\enquote {\bibinfo
  {title} {Optical generation of matter qubit graph states},}\ }\href
  {https://doi.org/10.1088/1367-2630/7/1/194} {\bibfield  {journal} {\bibinfo
  {journal} {New Journal of Physics}\ }\textbf {\bibinfo {volume} {7}},\
  \bibinfo {pages} {194}}\BibitemShut {NoStop}%
\bibitem [{\citenamefont {Bennett}\ and\ \citenamefont
  {Brassard}(1984)}]{bib:BennetBrassard84}%
  \BibitemOpen
  \bibfield  {author} {\bibinfo {author} {\bibnamefont {Bennett}, \bibfnamefont
  {C~H}}, and\ \bibinfo {author} {\bibfnamefont {G.}~\bibnamefont {Brassard}}}
  (\bibinfo {year} {1984}),\ \bibfield  {title} {\enquote {\bibinfo {title}
  {Quantum cryptography: public-key distribution and coin tossing},}\ }\href
  {https://doi.org/10.1016/j.tcs.2014.05.025} {\bibinfo  {journal} {IEEE
  International Conference on Computers, Systems \& Signal Processing}\ ,\
  \bibinfo {pages} {175}}\BibitemShut {NoStop}%
\bibitem [{\citenamefont {Bennett}\ \emph
  {et~al.}(1993{\natexlab{a}})\citenamefont {Bennett}, \citenamefont
  {Brassard}, \citenamefont {Crepeau}, \citenamefont {Jozsa}, \citenamefont
  {Peres},\ and\ \citenamefont {Wootters}}]{bib:bennett93}%
  \BibitemOpen
\bibfield  {journal} {  }\bibfield  {author} {\bibinfo {author} {\bibnamefont
  {Bennett}, \bibfnamefont {C~H}}, \bibinfo {author} {\bibfnamefont
  {G.}~\bibnamefont {Brassard}}, \bibinfo {author} {\bibfnamefont
  {C.}~\bibnamefont {Crepeau}}, \bibinfo {author} {\bibfnamefont
  {R.}~\bibnamefont {Jozsa}}, \bibinfo {author} {\bibfnamefont
  {A.}~\bibnamefont {Peres}}, and\ \bibinfo {author} {\bibfnamefont {W.K.}\
  \bibnamefont {Wootters}}} (\bibinfo {year} {1993}{\natexlab{a}}),\ \bibfield
  {title} {\enquote {\bibinfo {title} {Teleporting an unknown quantum state via
  dual classical and einstein-podolsky-rosen channels},}\ }\href
  {https://doi.org/10.1103/physrevlett.70.1895} {\bibfield  {journal} {\bibinfo
   {journal} {Physical Review Letters}\ }\textbf {\bibinfo {volume} {70}},\
  \bibinfo {pages} {1895}}\BibitemShut {NoStop}%
\bibitem [{\citenamefont {Bennett}\ \emph
  {et~al.}(1996{\natexlab{a}})\citenamefont {Bennett}, \citenamefont
  {Brassard}, \citenamefont {Popescu}, \citenamefont {Schumacher},
  \citenamefont {Smolin},\ and\ \citenamefont {Wootters}}]{bib:Bennett96}%
  \BibitemOpen
  \bibfield  {author} {\bibinfo {author} {\bibnamefont {Bennett}, \bibfnamefont
  {C~H}}, \bibinfo {author} {\bibfnamefont {G.}~\bibnamefont {Brassard}},
  \bibinfo {author} {\bibfnamefont {S.}~\bibnamefont {Popescu}}, \bibinfo
  {author} {\bibfnamefont {B.}~\bibnamefont {Schumacher}}, \bibinfo {author}
  {\bibfnamefont {J.}~\bibnamefont {Smolin}}, and\ \bibinfo {author}
  {\bibfnamefont {W.~K.}\ \bibnamefont {Wootters}}} (\bibinfo {year}
  {1996}{\natexlab{a}}),\ \bibfield  {title} {\enquote {\bibinfo {title}
  {Purification of noisy entanglement and faithful teleportation via noisy
  channels},}\ }\href {https://doi.org/10.1103/physrevlett.76.722} {\bibfield
  {journal} {\bibinfo  {journal} {Physical Review Letters}\ }\textbf {\bibinfo
  {volume} {76}},\ \bibinfo {pages} {722}},\ \Eprint
  {https://arxiv.org/abs/arXiv:quant-ph/9511027v2} {arXiv:quant-ph/9511027v2}
  \BibitemShut {NoStop}%
\bibitem [{\citenamefont {Bennett}\ and\ \citenamefont
  {DiVincenzo}(2000)}]{bib:Bennett00}%
  \BibitemOpen
  \bibfield  {author} {\bibinfo {author} {\bibnamefont {Bennett}, \bibfnamefont
  {C~H}}, and\ \bibinfo {author} {\bibfnamefont {D.~P.}\ \bibnamefont
  {DiVincenzo}}} (\bibinfo {year} {2000}),\ \bibfield  {title} {\enquote
  {\bibinfo {title} {Quantum information and computation},}\ }\href
  {https://doi.org/10.1038/35005001} {\bibfield  {journal} {\bibinfo  {journal}
  {Nature}\ }\textbf {\bibinfo {volume} {404}},\ \bibinfo {pages}
  {247}}\BibitemShut {NoStop}%
\bibitem [{\citenamefont {Bennett}\ \emph
  {et~al.}(1996{\natexlab{b}})\citenamefont {Bennett}, \citenamefont
  {DiVincenzo}, \citenamefont {Smolin},\ and\ \citenamefont
  {Wootters}}]{bib:Bennettr1996a}%
  \BibitemOpen
  \bibfield  {author} {\bibinfo {author} {\bibnamefont {Bennett}, \bibfnamefont
  {C~H}}, \bibinfo {author} {\bibfnamefont {D.~P.}\ \bibnamefont {DiVincenzo}},
  \bibinfo {author} {\bibfnamefont {J.~A.}\ \bibnamefont {Smolin}}, and\
  \bibinfo {author} {\bibfnamefont {W.~K.}\ \bibnamefont {Wootters}}} (\bibinfo
  {year} {1996}{\natexlab{b}}),\ \bibfield  {title} {\enquote {\bibinfo {title}
  {Mixed state entanglement and quantum error correction},}\ }\href
  {https://doi.org/10.1103/physreva.54.3824} {\bibfield  {journal} {\bibinfo
  {journal} {Physical Review A}\ }\textbf {\bibinfo {volume} {54}},\ \bibinfo
  {pages} {3824}},\ \Eprint {https://arxiv.org/abs/arXiv:quant-ph/9604024v2}
  {arXiv:quant-ph/9604024v2} \BibitemShut {NoStop}%
\bibitem [{\citenamefont {Bennett}(1992)}]{bib:PhysRevLett.68.3121}%
  \BibitemOpen
  \bibfield  {author} {\bibinfo {author} {\bibnamefont {Bennett}, \bibfnamefont
  {Charles~H}}} (\bibinfo {year} {1992}),\ \bibfield  {title} {\enquote
  {\bibinfo {title} {Quantum cryptography using any two nonorthogonal
  states},}\ }\href {https://doi.org/10.1103/physrevlett.68.3121} {\bibfield
  {journal} {\bibinfo  {journal} {Physical Review Letters}\ }\textbf {\bibinfo
  {volume} {68}},\ \bibinfo {pages} {3121}}\BibitemShut {NoStop}%
\bibitem [{\citenamefont {Bennett}\ \emph {et~al.}(1997)\citenamefont
  {Bennett}, \citenamefont {Bernstein}, \citenamefont {Brassard},\ and\
  \citenamefont {Vazirani}}]{bennett1997strengths}%
  \BibitemOpen
  \bibfield  {author} {\bibinfo {author} {\bibnamefont {Bennett}, \bibfnamefont
  {Charles~H}}, \bibinfo {author} {\bibfnamefont {Ethan}\ \bibnamefont
  {Bernstein}}, \bibinfo {author} {\bibfnamefont {Gilles}\ \bibnamefont
  {Brassard}}, and\ \bibinfo {author} {\bibfnamefont {Umesh}\ \bibnamefont
  {Vazirani}}} (\bibinfo {year} {1997}),\ \bibfield  {title} {\enquote
  {\bibinfo {title} {Strengths and weaknesses of quantum computing},}\
  }\href@noop {} {\bibfield  {journal} {\bibinfo  {journal} {SIAM journal on
  Computing}\ }\textbf {\bibinfo {volume} {26}}~(\bibinfo {number} {5}),\
  \bibinfo {pages} {1510--1523}}\BibitemShut {NoStop}%
\bibitem [{\citenamefont {Bennett}\ \emph
  {et~al.}(1996{\natexlab{c}})\citenamefont {Bennett}, \citenamefont
  {Bernstein}, \citenamefont {Popescu},\ and\ \citenamefont
  {Schumacher}}]{bib:PRA_53_2046}%
  \BibitemOpen
  \bibfield  {author} {\bibinfo {author} {\bibnamefont {Bennett}, \bibfnamefont
  {Charles~H}}, \bibinfo {author} {\bibfnamefont {Herbert~J}\ \bibnamefont
  {Bernstein}}, \bibinfo {author} {\bibfnamefont {Sandu}\ \bibnamefont
  {Popescu}}, and\ \bibinfo {author} {\bibfnamefont {Benjamin}\ \bibnamefont
  {Schumacher}}} (\bibinfo {year} {1996}{\natexlab{c}}),\ \bibfield  {title}
  {\enquote {\bibinfo {title} {Concentrating partial entanglement by local
  operations},}\ }\href {https://doi.org/10.1103/physreva.53.2046} {\bibfield
  {journal} {\bibinfo  {journal} {Physical Review A}\ }\textbf {\bibinfo
  {volume} {53}},\ \bibinfo {pages} {2046}},\ \Eprint
  {https://arxiv.org/abs/arXiv:quant-ph/9511030v1} {arXiv:quant-ph/9511030v1}
  \BibitemShut {NoStop}%
\bibitem [{\citenamefont {Bennett}\ \emph {et~al.}(1992)\citenamefont
  {Bennett}, \citenamefont {Bessette}, \citenamefont {Brassard}, \citenamefont
  {Salvail},\ and\ \citenamefont {Smolin}}]{bib:JC_5_3}%
  \BibitemOpen
  \bibfield  {author} {\bibinfo {author} {\bibnamefont {Bennett}, \bibfnamefont
  {Charles~H}}, \bibinfo {author} {\bibfnamefont {Fran{\c{c}}ois}\ \bibnamefont
  {Bessette}}, \bibinfo {author} {\bibfnamefont {Gilles}\ \bibnamefont
  {Brassard}}, \bibinfo {author} {\bibfnamefont {Louis}\ \bibnamefont
  {Salvail}}, and\ \bibinfo {author} {\bibfnamefont {John}\ \bibnamefont
  {Smolin}}} (\bibinfo {year} {1992}),\ \bibfield  {title} {\enquote {\bibinfo
  {title} {Experimental quantum cryptography},}\ }\href
  {https://doi.org/10.1007/bf00191318} {\bibfield  {journal} {\bibinfo
  {journal} {Journal of Cryptography}\ }\textbf {\bibinfo {volume} {5}},\
  \bibinfo {pages} {3}}\BibitemShut {NoStop}%
\bibitem [{\citenamefont {Bennett}\ \emph
  {et~al.}(1993{\natexlab{b}})\citenamefont {Bennett}, \citenamefont
  {Brassard}, \citenamefont {Cr\'epeau}, \citenamefont {Jozsa}, \citenamefont
  {Peres},\ and\ \citenamefont {Wootters}}]{bib:PRL_70_1895}%
  \BibitemOpen
  \bibfield  {author} {\bibinfo {author} {\bibnamefont {Bennett}, \bibfnamefont
  {Charles~H}}, \bibinfo {author} {\bibfnamefont {Gilles}\ \bibnamefont
  {Brassard}}, \bibinfo {author} {\bibfnamefont {Claude}\ \bibnamefont
  {Cr\'epeau}}, \bibinfo {author} {\bibfnamefont {Richard}\ \bibnamefont
  {Jozsa}}, \bibinfo {author} {\bibfnamefont {Asher}\ \bibnamefont {Peres}},
  and\ \bibinfo {author} {\bibfnamefont {William~K.}\ \bibnamefont {Wootters}}}
  (\bibinfo {year} {1993}{\natexlab{b}}),\ \bibfield  {title} {\enquote
  {\bibinfo {title} {Teleporting an unknown quantum state via dual classical
  and einstein-podolsky-rosen channels},}\ }\href
  {https://doi.org/10.1103/physrevlett.70.1895} {\bibfield  {journal} {\bibinfo
   {journal} {Physical Review Letters}\ }\textbf {\bibinfo {volume} {70}},\
  \bibinfo {pages} {1895}}\BibitemShut {NoStop}%
\bibitem [{\citenamefont {Bennett}\ \emph
  {et~al.}(1996{\natexlab{d}})\citenamefont {Bennett}, \citenamefont
  {DiVincenzo}, \citenamefont {Smolin},\ and\ \citenamefont
  {Wootters}}]{bib:PRA_54_3824}%
  \BibitemOpen
  \bibfield  {author} {\bibinfo {author} {\bibnamefont {Bennett}, \bibfnamefont
  {Charles~H}}, \bibinfo {author} {\bibfnamefont {David~P}\ \bibnamefont
  {DiVincenzo}}, \bibinfo {author} {\bibfnamefont {John~A}\ \bibnamefont
  {Smolin}}, and\ \bibinfo {author} {\bibfnamefont {William~K}\ \bibnamefont
  {Wootters}}} (\bibinfo {year} {1996}{\natexlab{d}}),\ \bibfield  {title}
  {\enquote {\bibinfo {title} {Mixed-state entanglement and quantum error
  correction},}\ }\href {https://doi.org/10.1103/physreva.54.3824} {\bibfield
  {journal} {\bibinfo  {journal} {Physical Review A}\ }\textbf {\bibinfo
  {volume} {54}},\ \bibinfo {pages} {3824}},\ \Eprint
  {https://arxiv.org/abs/arXiv:quant-ph/9604024v2} {arXiv:quant-ph/9604024v2}
  \BibitemShut {NoStop}%
\bibitem [{\citenamefont {Bennett}\ \emph {et~al.}(1999)\citenamefont
  {Bennett}, \citenamefont {Shor}, \citenamefont {Smolin},\ and\ \citenamefont
  {Thapliyal}}]{bib:PhysRevLett.83.3081}%
  \BibitemOpen
  \bibfield  {author} {\bibinfo {author} {\bibnamefont {Bennett}, \bibfnamefont
  {Charles~H}}, \bibinfo {author} {\bibfnamefont {Peter~W.}\ \bibnamefont
  {Shor}}, \bibinfo {author} {\bibfnamefont {John~A.}\ \bibnamefont {Smolin}},
  and\ \bibinfo {author} {\bibfnamefont {Ashish~V.}\ \bibnamefont {Thapliyal}}}
  (\bibinfo {year} {1999}),\ \bibfield  {title} {\enquote {\bibinfo {title}
  {Entanglement-assisted classical capacity of noisy quantum channels},}\
  }\href {https://doi.org/10.1103/PhysRevLett.83.3081} {\bibfield  {journal}
  {\bibinfo  {journal} {Phys. Rev. Lett.}\ }\textbf {\bibinfo {volume} {83}},\
  \bibinfo {pages} {3081--3084}}\BibitemShut {NoStop}%
\bibitem [{\citenamefont {Bermolen}\ and\ \citenamefont
  {Rossi}(2009)}]{bib:bermolen2009support}%
  \BibitemOpen
  \bibfield  {author} {\bibinfo {author} {\bibnamefont {Bermolen},
  \bibfnamefont {Paola}}, and\ \bibinfo {author} {\bibfnamefont {Dario}\
  \bibnamefont {Rossi}}} (\bibinfo {year} {2009}),\ \bibfield  {title}
  {\enquote {\bibinfo {title} {Support vector regression for link load
  prediction},}\ }\href {https://doi.org/10.1016/j.comnet.2008.09.018}
  {\bibfield  {journal} {\bibinfo  {journal} {Computer Networks}\ }\textbf
  {\bibinfo {volume} {53}}~(\bibinfo {number} {2}),\ \bibinfo {pages}
  {191}}\BibitemShut {NoStop}%
\bibitem [{\citenamefont {Berndt}\ \emph {et~al.}(2000)\citenamefont {Berndt},
  \citenamefont {Dulberger},\ and\ \citenamefont {Rappaport}}]{SD-Berndt2000}%
  \BibitemOpen
  \bibfield  {author} {\bibinfo {author} {\bibnamefont {Berndt}, \bibfnamefont
  {E~R}}, \bibinfo {author} {\bibfnamefont {E.~R.}\ \bibnamefont {Dulberger}},
  and\ \bibinfo {author} {\bibfnamefont {N.~J.}\ \bibnamefont {Rappaport}}}
  (\bibinfo {year} {2000}),\ \bibfield  {title} {\enquote {\bibinfo {title}
  {Price and quality of desktop and mobile personal computers: A quarter
  century of history},}\ }\href
  {https://conference.nber.org/confer/2000/si2000/berndt.pdf} {\bibfield
  {journal} {\bibinfo  {journal} {Papers and Proceedings of the Hundred
  Thirteenth Annual Meeting of the American Economic Association}\ }\textbf
  {\bibinfo {volume} {91}},\ \bibinfo {pages} {268}}\BibitemShut {NoStop}%
\bibitem [{\citenamefont {Bernien}\ \emph {et~al.}(2013)\citenamefont
  {Bernien}, \citenamefont {Hensen}, \citenamefont {Pfaff}, \citenamefont
  {Koolstra}, \citenamefont {Blok}, \citenamefont {Robledo}, \citenamefont
  {Taminiau}, \citenamefont {Markham}, \citenamefont {Twitchen}, \citenamefont
  {Childress} \emph {et~al.}}]{bib:bernien2013heralded}%
  \BibitemOpen
  \bibfield  {author} {\bibinfo {author} {\bibnamefont {Bernien}, \bibfnamefont
  {Hannes}}, \bibinfo {author} {\bibfnamefont {Bas}\ \bibnamefont {Hensen}},
  \bibinfo {author} {\bibfnamefont {Wolfgang}\ \bibnamefont {Pfaff}}, \bibinfo
  {author} {\bibfnamefont {Gerwin}\ \bibnamefont {Koolstra}}, \bibinfo {author}
  {\bibfnamefont {MS}~\bibnamefont {Blok}}, \bibinfo {author} {\bibfnamefont
  {Lucio}\ \bibnamefont {Robledo}}, \bibinfo {author} {\bibfnamefont
  {TH}~\bibnamefont {Taminiau}}, \bibinfo {author} {\bibfnamefont {Matthew}\
  \bibnamefont {Markham}}, \bibinfo {author} {\bibfnamefont {DJ}~\bibnamefont
  {Twitchen}}, \bibinfo {author} {\bibfnamefont {Lilian}\ \bibnamefont
  {Childress}},  \emph {et~al.}} (\bibinfo {year} {2013}),\ \bibfield  {title}
  {\enquote {\bibinfo {title} {Heralded entanglement between solid-state qubits
  separated by three metres},}\ }\href {https://doi.org/10.1038/nature12016}
  {\bibfield  {journal} {\bibinfo  {journal} {Nature}\ }\textbf {\bibinfo
  {volume} {497}},\ \bibinfo {pages} {86}}\BibitemShut {NoStop}%
\bibitem [{\citenamefont {Bernstein}\ and\ \citenamefont
  {Vazirani}(1997)}]{bib:bernstein1997quantum}%
  \BibitemOpen
  \bibfield  {author} {\bibinfo {author} {\bibnamefont {Bernstein},
  \bibfnamefont {Ethan}}, and\ \bibinfo {author} {\bibfnamefont {Umesh}\
  \bibnamefont {Vazirani}}} (\bibinfo {year} {1997}),\ \bibfield  {title}
  {\enquote {\bibinfo {title} {Quantum complexity theory},}\ }\href
  {https://doi.org/10.1137/S0097539796300921} {\bibfield  {journal} {\bibinfo
  {journal} {SIAM Journal on Computing}\ }\textbf {\bibinfo {volume} {26}},\
  \bibinfo {pages} {1411}}\BibitemShut {NoStop}%
\bibitem [{\citenamefont {Berry}(2014)}]{bib:BerryLinear}%
  \BibitemOpen
  \bibfield  {author} {\bibinfo {author} {\bibnamefont {Berry}, \bibfnamefont
  {Dominic~W}}} (\bibinfo {year} {2014}),\ \bibfield  {title} {\enquote
  {\bibinfo {title} {High-order quantum algorithm for solving linear
  differential equations},}\ }\href
  {https://doi.org/10.1088/1751-8113/47/10/105301} {\bibfield  {journal}
  {\bibinfo  {journal} {Journal of Physics A: Mathematics \& Theoretical}\
  }\textbf {\bibinfo {volume} {47}},\ \bibinfo {pages} {105301}}\BibitemShut
  {NoStop}%
\bibitem [{\citenamefont {Beugnon}\ \emph {et~al.}(2006)\citenamefont
  {Beugnon}, \citenamefont {Jones}, \citenamefont {Dingjan}, \citenamefont
  {Darquie}, \citenamefont {Messin}, \citenamefont {Browaeys},\ and\
  \citenamefont {Grangier}}]{bib:Beugnon06}%
  \BibitemOpen
  \bibfield  {author} {\bibinfo {author} {\bibnamefont {Beugnon}, \bibfnamefont
  {J}}, \bibinfo {author} {\bibfnamefont {M.~P.~A.}\ \bibnamefont {Jones}},
  \bibinfo {author} {\bibfnamefont {J.}~\bibnamefont {Dingjan}}, \bibinfo
  {author} {\bibfnamefont {B.}~\bibnamefont {Darquie}}, \bibinfo {author}
  {\bibfnamefont {G.}~\bibnamefont {Messin}}, \bibinfo {author} {\bibfnamefont
  {A.}~\bibnamefont {Browaeys}}, and\ \bibinfo {author} {\bibfnamefont
  {P.}~\bibnamefont {Grangier}}} (\bibinfo {year} {2006}),\ \bibfield  {title}
  {\enquote {\bibinfo {title} {Quantum interference between two single photons
  emitted by independently trapped atoms},}\ }\href
  {https://doi.org/10.1038/nature04628} {\bibfield  {journal} {\bibinfo
  {journal} {Nature}\ }\textbf {\bibinfo {volume} {440}},\ \bibinfo {pages}
  {779}},\ \Eprint {https://arxiv.org/abs/arXiv:quant-ph/0610149v1}
  {arXiv:quant-ph/0610149v1} \BibitemShut {NoStop}%
\bibitem [{\citenamefont {Biamonte}\ \emph {et~al.}(2017)\citenamefont
  {Biamonte}, \citenamefont {Wittek}, \citenamefont {Pancotti}, \citenamefont
  {Rebentrost}, \citenamefont {Wiebe},\ and\ \citenamefont
  {Lloyd}}]{bib:biamonte2017quantum}%
  \BibitemOpen
  \bibfield  {author} {\bibinfo {author} {\bibnamefont {Biamonte},
  \bibfnamefont {Jacob}}, \bibinfo {author} {\bibfnamefont {Peter}\
  \bibnamefont {Wittek}}, \bibinfo {author} {\bibfnamefont {Nicola}\
  \bibnamefont {Pancotti}}, \bibinfo {author} {\bibfnamefont {Patrick}\
  \bibnamefont {Rebentrost}}, \bibinfo {author} {\bibfnamefont {Nathan}\
  \bibnamefont {Wiebe}}, and\ \bibinfo {author} {\bibfnamefont {Seth}\
  \bibnamefont {Lloyd}}} (\bibinfo {year} {2017}),\ \bibfield  {title}
  {\enquote {\bibinfo {title} {Quantum machine learning},}\ }\href
  {https://doi.org/10.1038/nature23474} {\bibfield  {journal} {\bibinfo
  {journal} {Nature}\ }\textbf {\bibinfo {volume} {549}},\ \bibinfo {pages}
  {195}},\ \Eprint {https://arxiv.org/abs/arXiv:1611.09347v2}
  {arXiv:1611.09347v2} \BibitemShut {NoStop}%
\bibitem [{\citenamefont {Bishop}(2006)}]{bib:bishop2006pattern}%
  \BibitemOpen
  \bibfield  {author} {\bibinfo {author} {\bibnamefont {Bishop}, \bibfnamefont
  {Christopher~M}}} (\bibinfo {year} {2006}),\ \href
  {https://doi.org/10.1198/tech.2007.s518} {\emph {\bibinfo {title} {Pattern
  recognition and machine learning}}}\ (\bibinfo  {publisher}
  {Springer})\BibitemShut {NoStop}%
\bibitem [{\citenamefont {Bisio}\ \emph {et~al.}(2010)\citenamefont {Bisio},
  \citenamefont {Chiribella}, \citenamefont {D’Ariano}, \citenamefont
  {Facchini},\ and\ \citenamefont {Perinotti}}]{bisio2010optimal}%
  \BibitemOpen
  \bibfield  {author} {\bibinfo {author} {\bibnamefont {Bisio}, \bibfnamefont
  {Alessandro}}, \bibinfo {author} {\bibfnamefont {Giulio}\ \bibnamefont
  {Chiribella}}, \bibinfo {author} {\bibfnamefont {Giacomo~Mauro}\ \bibnamefont
  {D’Ariano}}, \bibinfo {author} {\bibfnamefont {Stefano}\ \bibnamefont
  {Facchini}}, and\ \bibinfo {author} {\bibfnamefont {Paolo}\ \bibnamefont
  {Perinotti}}} (\bibinfo {year} {2010}),\ \bibfield  {title} {\enquote
  {\bibinfo {title} {Optimal quantum learning of a unitary transformation},}\
  }\href@noop {} {\bibfield  {journal} {\bibinfo  {journal} {Physical Review
  A—Atomic, Molecular, and Optical Physics}\ }\textbf {\bibinfo {volume}
  {81}}~(\bibinfo {number} {3}),\ \bibinfo {pages} {032324}}\BibitemShut
  {NoStop}%
\bibitem [{\citenamefont {Blais}\ \emph {et~al.}(2004)\citenamefont {Blais},
  \citenamefont {Huang}, \citenamefont {Wallraff}, \citenamefont {Girvin},\
  and\ \citenamefont {Schoelkopf}}]{bib:blais2004cavity}%
  \BibitemOpen
  \bibfield  {author} {\bibinfo {author} {\bibnamefont {Blais}, \bibfnamefont
  {Alexandre}}, \bibinfo {author} {\bibfnamefont {Ren-Shou}\ \bibnamefont
  {Huang}}, \bibinfo {author} {\bibfnamefont {Andreas}\ \bibnamefont
  {Wallraff}}, \bibinfo {author} {\bibfnamefont {Steven~M}\ \bibnamefont
  {Girvin}}, and\ \bibinfo {author} {\bibfnamefont {R~Jun}\ \bibnamefont
  {Schoelkopf}}} (\bibinfo {year} {2004}),\ \bibfield  {title} {\enquote
  {\bibinfo {title} {Cavity quantum electrodynamics for superconducting
  electrical circuits: An architecture for quantum computation},}\ }\href
  {https://doi.org/10.1103/physreva.69.062320} {\bibfield  {journal} {\bibinfo
  {journal} {Physical Review A}\ }\textbf {\bibinfo {volume} {69}},\ \bibinfo
  {pages} {062320}},\ \Eprint {https://arxiv.org/abs/arXiv:cond-mat/0402216v1}
  {arXiv:cond-mat/0402216v1} \BibitemShut {NoStop}%
\bibitem [{\citenamefont {Blatt}\ and\ \citenamefont
  {Wineland}(2008)}]{bib:blatt2008entangled}%
  \BibitemOpen
  \bibfield  {author} {\bibinfo {author} {\bibnamefont {Blatt}, \bibfnamefont
  {Rainer}}, and\ \bibinfo {author} {\bibfnamefont {David}\ \bibnamefont
  {Wineland}}} (\bibinfo {year} {2008}),\ \bibfield  {title} {\enquote
  {\bibinfo {title} {Entangled states of trapped atomic ions},}\ }\href
  {https://doi.org/10.1038/nature07125} {\bibfield  {journal} {\bibinfo
  {journal} {Nature}\ }\textbf {\bibinfo {volume} {453}},\ \bibinfo {pages}
  {1008}}\BibitemShut {NoStop}%
\bibitem [{\citenamefont {Blinov}\ \emph {et~al.}(2004)\citenamefont {Blinov},
  \citenamefont {Moehring}, \citenamefont {Duan},\ and\ \citenamefont
  {Monroe}}]{bib:blinov2004observation}%
  \BibitemOpen
  \bibfield  {author} {\bibinfo {author} {\bibnamefont {Blinov}, \bibfnamefont
  {BB}}, \bibinfo {author} {\bibfnamefont {DL}~\bibnamefont {Moehring}},
  \bibinfo {author} {\bibfnamefont {L-M}\ \bibnamefont {Duan}}, and\ \bibinfo
  {author} {\bibfnamefont {Chris}\ \bibnamefont {Monroe}}} (\bibinfo {year}
  {2004}),\ \bibfield  {title} {\enquote {\bibinfo {title} {Observation of
  entanglement between a single trapped atom and a single photon},}\ }\href
  {https://doi.org/10.1038/nature02377} {\bibfield  {journal} {\bibinfo
  {journal} {Nature}\ }\textbf {\bibinfo {volume} {428}},\ \bibinfo {pages}
  {153}}\BibitemShut {NoStop}%
\bibitem [{\citenamefont {Blum}\ \emph {et~al.}(2015)\citenamefont {Blum},
  \citenamefont {O'Brien}, \citenamefont {Lauk}, \citenamefont {Bushev},
  \citenamefont {Fleischhauer},\ and\ \citenamefont
  {Morigi}}]{bib:blum2015interfacing}%
  \BibitemOpen
  \bibfield  {author} {\bibinfo {author} {\bibnamefont {Blum}, \bibfnamefont
  {Susanne}}, \bibinfo {author} {\bibfnamefont {Christopher}\ \bibnamefont
  {O'Brien}}, \bibinfo {author} {\bibfnamefont {Nikolai}\ \bibnamefont {Lauk}},
  \bibinfo {author} {\bibfnamefont {Pavel}\ \bibnamefont {Bushev}}, \bibinfo
  {author} {\bibfnamefont {Michael}\ \bibnamefont {Fleischhauer}}, and\
  \bibinfo {author} {\bibfnamefont {Giovanna}\ \bibnamefont {Morigi}}}
  (\bibinfo {year} {2015}),\ \bibfield  {title} {\enquote {\bibinfo {title}
  {Interfacing microwave qubits and optical photons via spin ensembles},}\
  }\href {https://doi.org/10.1103/physreva.91.033834} {\bibfield  {journal}
  {\bibinfo  {journal} {Physical Review A}\ }\textbf {\bibinfo {volume} {91}},\
  \bibinfo {pages} {033834}},\ \Eprint
  {https://arxiv.org/abs/arXiv:1501.05860v1} {arXiv:1501.05860v1} \BibitemShut
  {NoStop}%
\bibitem [{\citenamefont {Boaron}\ \emph {et~al.}(2018)\citenamefont {Boaron},
  \citenamefont {Boso}, \citenamefont {Rusca}, \citenamefont {Vulliez},
  \citenamefont {Autebert}, \citenamefont {Caloz}, \citenamefont {Perrenoud},
  \citenamefont {Gras}, \citenamefont {Bussi{\`e}res}, \citenamefont {Li} \emph
  {et~al.}}]{bib:boaron2018secure}%
  \BibitemOpen
  \bibfield  {author} {\bibinfo {author} {\bibnamefont {Boaron}, \bibfnamefont
  {Alberto}}, \bibinfo {author} {\bibfnamefont {Gianluca}\ \bibnamefont
  {Boso}}, \bibinfo {author} {\bibfnamefont {Davide}\ \bibnamefont {Rusca}},
  \bibinfo {author} {\bibfnamefont {C{\'e}dric}\ \bibnamefont {Vulliez}},
  \bibinfo {author} {\bibfnamefont {Claire}\ \bibnamefont {Autebert}}, \bibinfo
  {author} {\bibfnamefont {Misael}\ \bibnamefont {Caloz}}, \bibinfo {author}
  {\bibfnamefont {Matthieu}\ \bibnamefont {Perrenoud}}, \bibinfo {author}
  {\bibfnamefont {Ga{\"e}tan}\ \bibnamefont {Gras}}, \bibinfo {author}
  {\bibfnamefont {F{\'e}lix}\ \bibnamefont {Bussi{\`e}res}}, \bibinfo {author}
  {\bibfnamefont {Ming-Jun}\ \bibnamefont {Li}},  \emph {et~al.}} (\bibinfo
  {year} {2018}),\ \bibfield  {title} {\enquote {\bibinfo {title} {Secure
  quantum key distribution over 421 km of optical fiber},}\ }\href
  {https://doi.org/10.1103/physrevlett.121.190502} {\bibfield  {journal}
  {\bibinfo  {journal} {Physical Review Letters}\ }\textbf {\bibinfo {volume}
  {121}},\ \bibinfo {pages} {190502}},\ \Eprint
  {https://arxiv.org/abs/arXiv:1807.03222v1} {arXiv:1807.03222v1} \BibitemShut
  {NoStop}%
\bibitem [{\citenamefont {Bochmann}\ \emph {et~al.}(2013)\citenamefont
  {Bochmann}, \citenamefont {Vainsencher}, \citenamefont {Awschalom},\ and\
  \citenamefont {Cleland}}]{bib:bochmann2013nanomechanical}%
  \BibitemOpen
  \bibfield  {author} {\bibinfo {author} {\bibnamefont {Bochmann},
  \bibfnamefont {Joerg}}, \bibinfo {author} {\bibfnamefont {Amit}\ \bibnamefont
  {Vainsencher}}, \bibinfo {author} {\bibfnamefont {David~D}\ \bibnamefont
  {Awschalom}}, and\ \bibinfo {author} {\bibfnamefont {Andrew~N}\ \bibnamefont
  {Cleland}}} (\bibinfo {year} {2013}),\ \bibfield  {title} {\enquote {\bibinfo
  {title} {Nanomechanical coupling between microwave and optical photons},}\
  }\href {https://doi.org/10.1038/nphys2748} {\bibfield  {journal} {\bibinfo
  {journal} {Nature Physics}\ }\textbf {\bibinfo {volume} {9}},\ \bibinfo
  {pages} {712}}\BibitemShut {NoStop}%
\bibitem [{\citenamefont {Bohm}\ and\ \citenamefont
  {Bub}(1966)}]{bohm1966proposed}%
  \BibitemOpen
  \bibfield  {author} {\bibinfo {author} {\bibnamefont {Bohm}, \bibfnamefont
  {David}}, and\ \bibinfo {author} {\bibfnamefont {Jeffrey}\ \bibnamefont
  {Bub}}} (\bibinfo {year} {1966}),\ \bibfield  {title} {\enquote {\bibinfo
  {title} {A proposed solution of the measurement problem in quantum mechanics
  by a hidden variable theory},}\ }\href@noop {} {\bibfield  {journal}
  {\bibinfo  {journal} {Reviews of Modern Physics}\ }\textbf {\bibinfo {volume}
  {38}}~(\bibinfo {number} {3}),\ \bibinfo {pages} {453}}\BibitemShut {NoStop}%
\bibitem [{\citenamefont {Boixo}\ \emph {et~al.}(2018)\citenamefont {Boixo},
  \citenamefont {Isakov}, \citenamefont {Smelyanskiy}, \citenamefont {Babbush},
  \citenamefont {Ding}, \citenamefont {Jiang}, \citenamefont {Bremner},
  \citenamefont {Martinis},\ and\ \citenamefont
  {Neven}}]{bib:boixo2018characterizing}%
  \BibitemOpen
  \bibfield  {author} {\bibinfo {author} {\bibnamefont {Boixo}, \bibfnamefont
  {Sergio}}, \bibinfo {author} {\bibfnamefont {Sergei~V}\ \bibnamefont
  {Isakov}}, \bibinfo {author} {\bibfnamefont {Vadim~N}\ \bibnamefont
  {Smelyanskiy}}, \bibinfo {author} {\bibfnamefont {Ryan}\ \bibnamefont
  {Babbush}}, \bibinfo {author} {\bibfnamefont {Nan}\ \bibnamefont {Ding}},
  \bibinfo {author} {\bibfnamefont {Zhang}\ \bibnamefont {Jiang}}, \bibinfo
  {author} {\bibfnamefont {Michael~J}\ \bibnamefont {Bremner}}, \bibinfo
  {author} {\bibfnamefont {John~M}\ \bibnamefont {Martinis}}, and\ \bibinfo
  {author} {\bibfnamefont {Hartmut}\ \bibnamefont {Neven}}} (\bibinfo {year}
  {2018}),\ \bibfield  {title} {\enquote {\bibinfo {title} {Characterizing
  quantum supremacy in near-term devices},}\ }\href
  {https://doi.org/10.1038/s41567-018-0124-x} {\bibfield  {journal} {\bibinfo
  {journal} {Nature Physics}\ }\textbf {\bibinfo {volume} {14}},\ \bibinfo
  {pages} {595}}\BibitemShut {NoStop}%
\bibitem [{\citenamefont {Bomb{\'\i}n}(2015)}]{SD-Bombin:2015aa}%
  \BibitemOpen
  \bibfield  {author} {\bibinfo {author} {\bibnamefont {Bomb{\'\i}n},
  \bibfnamefont {H{\'e}ctor}}} (\bibinfo {year} {2015}),\ \bibfield  {title}
  {\enquote {\bibinfo {title} {Gauge color codes: optimal transversal gates and
  gauge fixing in topological stabilizer codes},}\ }\href
  {https://doi.org/10.1088/1367-2630/17/8/083002} {\bibfield  {journal}
  {\bibinfo  {journal} {New Journal of Physics}\ }\textbf {\bibinfo {volume}
  {17}},\ \bibinfo {pages} {083002}}\BibitemShut {NoStop}%
\bibitem [{\citenamefont {Bonneau}\ \emph {et~al.}(2012)\citenamefont
  {Bonneau}, \citenamefont {Engin}, \citenamefont {Ohira}, \citenamefont
  {Suzuki}, \citenamefont {Yoshida}, \citenamefont {Iizuka}, \citenamefont
  {Ezaki}, \citenamefont {Natarajan}, \citenamefont {Tanner}, \citenamefont
  {Hadfield} \emph {et~al.}}]{bib:bonneau2012quantum}%
  \BibitemOpen
  \bibfield  {author} {\bibinfo {author} {\bibnamefont {Bonneau}, \bibfnamefont
  {Damien}}, \bibinfo {author} {\bibfnamefont {Erman}\ \bibnamefont {Engin}},
  \bibinfo {author} {\bibfnamefont {Kazuya}\ \bibnamefont {Ohira}}, \bibinfo
  {author} {\bibfnamefont {Nob}\ \bibnamefont {Suzuki}}, \bibinfo {author}
  {\bibfnamefont {Haruhiko}\ \bibnamefont {Yoshida}}, \bibinfo {author}
  {\bibfnamefont {Norio}\ \bibnamefont {Iizuka}}, \bibinfo {author}
  {\bibfnamefont {Mizunori}\ \bibnamefont {Ezaki}}, \bibinfo {author}
  {\bibfnamefont {Chandra~M}\ \bibnamefont {Natarajan}}, \bibinfo {author}
  {\bibfnamefont {Michael~G}\ \bibnamefont {Tanner}}, \bibinfo {author}
  {\bibfnamefont {Robert~H}\ \bibnamefont {Hadfield}},  \emph {et~al.}}
  (\bibinfo {year} {2012}),\ \bibfield  {title} {\enquote {\bibinfo {title}
  {Quantum interference and manipulation of entanglement in silicon wire
  waveguide quantum circuits},}\ }\href
  {https://doi.org/10.1088/1367-2630/14/4/045003} {\bibfield  {journal}
  {\bibinfo  {journal} {New Journal of Physics}\ }\textbf {\bibinfo {volume}
  {14}},\ \bibinfo {pages} {045003}}\BibitemShut {NoStop}%
\bibitem [{\citenamefont {Borregaard}\ \emph {et~al.}(2015)\citenamefont
  {Borregaard}, \citenamefont {K{\'o}m{\'a}r}, \citenamefont {Kessler},
  \citenamefont {Lukin},\ and\ \citenamefont
  {S{\o}rensen}}]{bib:PRA_92_012307}%
  \BibitemOpen
  \bibfield  {author} {\bibinfo {author} {\bibnamefont {Borregaard},
  \bibfnamefont {Johannes}}, \bibinfo {author} {\bibfnamefont {Peter}\
  \bibnamefont {K{\'o}m{\'a}r}}, \bibinfo {author} {\bibfnamefont {Eric~M}\
  \bibnamefont {Kessler}}, \bibinfo {author} {\bibfnamefont {Mikhail~D}\
  \bibnamefont {Lukin}}, and\ \bibinfo {author} {\bibfnamefont
  {Anders~S{\o}ndberg}\ \bibnamefont {S{\o}rensen}}} (\bibinfo {year} {2015}),\
  \bibfield  {title} {\enquote {\bibinfo {title} {Long-distance entanglement
  distribution using individual atoms in optical cavities},}\ }\href
  {https://doi.org/10.1103/physreva.92.012307} {\bibfield  {journal} {\bibinfo
  {journal} {Physical Review A}\ }\textbf {\bibinfo {volume} {92}},\ \bibinfo
  {pages} {012307}},\ \Eprint {https://arxiv.org/abs/arXiv:1504.03703v3}
  {arXiv:1504.03703v3} \BibitemShut {NoStop}%
\bibitem [{\citenamefont {Boruvka}(1926)}]{bib:Boruvka26}%
  \BibitemOpen
  \bibfield  {author} {\bibinfo {author} {\bibnamefont {Boruvka}, \bibfnamefont
  {Otakar}}} (\bibinfo {year} {1926}),\ \bibfield  {title} {\enquote {\bibinfo
  {title} {About a certain minimal problem},}\ }\href@noop {} {\bibfield
  {journal} {\bibinfo  {journal} {O Prace mor. prirodoved. spol. v Brne III}\
  }\textbf {\bibinfo {volume} {3}},\ \bibinfo {pages} {37}}\BibitemShut
  {NoStop}%
\bibitem [{\citenamefont {Bouchiat}\ \emph {et~al.}(1998)\citenamefont
  {Bouchiat}, \citenamefont {Vion}, \citenamefont {Joyez}, \citenamefont
  {Esteve},\ and\ \citenamefont {Devoret}}]{bib:bouchiat1998quantum}%
  \BibitemOpen
  \bibfield  {author} {\bibinfo {author} {\bibnamefont {Bouchiat},
  \bibfnamefont {Vincent}}, \bibinfo {author} {\bibfnamefont {D}~\bibnamefont
  {Vion}}, \bibinfo {author} {\bibfnamefont {Ph}~\bibnamefont {Joyez}},
  \bibinfo {author} {\bibfnamefont {D}~\bibnamefont {Esteve}}, and\ \bibinfo
  {author} {\bibfnamefont {MH}~\bibnamefont {Devoret}}} (\bibinfo {year}
  {1998}),\ \bibfield  {title} {\enquote {\bibinfo {title} {Quantum coherence
  with a single cooper pair},}\ }\href
  {https://doi.org/10.1238/physica.topical.076a00165} {\bibfield  {journal}
  {\bibinfo  {journal} {Physica Scripta}\ }\textbf {\bibinfo {volume} {1998}},\
  \bibinfo {pages} {165}}\BibitemShut {NoStop}%
\bibitem [{\citenamefont {Boumeester}\ \emph {et~al.}(1997)\citenamefont
  {Boumeester}, \citenamefont {Pan}, \citenamefont {Mattle}, \citenamefont
  {Eibl}, \citenamefont {Weinfurter},\ and\ \citenamefont
  {Zeilinger}}]{bib:Boumeester97}%
  \BibitemOpen
  \bibfield  {author} {\bibinfo {author} {\bibnamefont {Boumeester},
  \bibfnamefont {Dik}}, \bibinfo {author} {\bibfnamefont {Jian-Wei}\
  \bibnamefont {Pan}}, \bibinfo {author} {\bibfnamefont {Klaus}\ \bibnamefont
  {Mattle}}, \bibinfo {author} {\bibfnamefont {Manfred}\ \bibnamefont {Eibl}},
  \bibinfo {author} {\bibfnamefont {Harald}\ \bibnamefont {Weinfurter}}, and\
  \bibinfo {author} {\bibfnamefont {Anton}\ \bibnamefont {Zeilinger}}}
  (\bibinfo {year} {1997}),\ \bibfield  {title} {\enquote {\bibinfo {title}
  {Experimental quantum teleportation},}\ }\href
  {https://doi.org/10.1038/37539} {\bibfield  {journal} {\bibinfo  {journal}
  {Nature}\ }\textbf {\bibinfo {volume} {390}},\ \bibinfo {pages}
  {575}}\BibitemShut {NoStop}%
\bibitem [{\citenamefont {Bourassa}\ \emph {et~al.}(2021)\citenamefont
  {Bourassa}, \citenamefont {Alexander}, \citenamefont {Vasmer}, \citenamefont
  {Patil}, \citenamefont {Tzitrin}, \citenamefont {Matsuura}, \citenamefont
  {Su}, \citenamefont {Baragiola}, \citenamefont {Guha}, \citenamefont
  {Dauphinais}, \citenamefont {Sabapathy}, \citenamefont {Menicucci},\ and\
  \citenamefont {Dhand}}]{bourassa2021blueprint}%
  \BibitemOpen
  \bibfield  {author} {\bibinfo {author} {\bibnamefont {Bourassa},
  \bibfnamefont {J~Eli}}, \bibinfo {author} {\bibfnamefont {Rafael~N.}\
  \bibnamefont {Alexander}}, \bibinfo {author} {\bibfnamefont {Michael}\
  \bibnamefont {Vasmer}}, \bibinfo {author} {\bibfnamefont {Ashlesha}\
  \bibnamefont {Patil}}, \bibinfo {author} {\bibfnamefont {Ilan}\ \bibnamefont
  {Tzitrin}}, \bibinfo {author} {\bibfnamefont {Takaya}\ \bibnamefont
  {Matsuura}}, \bibinfo {author} {\bibfnamefont {Daiqin}\ \bibnamefont {Su}},
  \bibinfo {author} {\bibfnamefont {Ben~Q.}\ \bibnamefont {Baragiola}},
  \bibinfo {author} {\bibfnamefont {Saikat}\ \bibnamefont {Guha}}, \bibinfo
  {author} {\bibfnamefont {Guillaume}\ \bibnamefont {Dauphinais}}, \bibinfo
  {author} {\bibfnamefont {Krishna~K.}\ \bibnamefont {Sabapathy}}, \bibinfo
  {author} {\bibfnamefont {Nicolas~C.}\ \bibnamefont {Menicucci}}, and\
  \bibinfo {author} {\bibfnamefont {Ish}\ \bibnamefont {Dhand}}} (\bibinfo
  {year} {2021}),\ \bibfield  {title} {\enquote {\bibinfo {title} {Blueprint
  for a {S}calable {P}hotonic {F}ault-{T}olerant {Q}uantum {C}omputer},}\
  }\href {https://doi.org/10.22331/q-2021-02-04-392} {\bibfield  {journal}
  {\bibinfo  {journal} {Quantum}\ }\textbf {\bibinfo {volume} {5}},\ \bibinfo
  {pages} {392}}\BibitemShut {NoStop}%
\bibitem [{\citenamefont {Boutaba}\ \emph {et~al.}(2018)\citenamefont
  {Boutaba}, \citenamefont {Salahuddin}, \citenamefont {Limam}, \citenamefont
  {Ayoubi}, \citenamefont {Shahriar}, \citenamefont {Estrada-Solano},\ and\
  \citenamefont {Caicedo}}]{bib:boutaba2018comprehensive}%
  \BibitemOpen
  \bibfield  {author} {\bibinfo {author} {\bibnamefont {Boutaba}, \bibfnamefont
  {Raouf}}, \bibinfo {author} {\bibfnamefont {Mohammad~A.}\ \bibnamefont
  {Salahuddin}}, \bibinfo {author} {\bibfnamefont {Noura}\ \bibnamefont
  {Limam}}, \bibinfo {author} {\bibfnamefont {Sara}\ \bibnamefont {Ayoubi}},
  \bibinfo {author} {\bibfnamefont {Nashid}\ \bibnamefont {Shahriar}}, \bibinfo
  {author} {\bibfnamefont {Felipe}\ \bibnamefont {Estrada-Solano}}, and\
  \bibinfo {author} {\bibfnamefont {Oscar~M.}\ \bibnamefont {Caicedo}}}
  (\bibinfo {year} {2018}),\ \bibfield  {title} {\enquote {\bibinfo {title} {A
  comprehensive survey on machine learning for networking: evolution,
  applications and research opportunities},}\ }\href
  {https://doi.org/10.1186/s13174-018-0087-2} {\bibfield  {journal} {\bibinfo
  {journal} {Journal of Internet Services and Applications}\ }\textbf {\bibinfo
  {volume} {9}},\ \bibinfo {pages} {16}}\BibitemShut {NoStop}%
\bibitem [{\citenamefont {Bouwmeester}\ \emph {et~al.}(1999)\citenamefont
  {Bouwmeester}, \citenamefont {Pan}, \citenamefont {Daniell}, \citenamefont
  {Weinfurter},\ and\ \citenamefont
  {Zeilinger}}]{bib:bouwmeester1999observation}%
  \BibitemOpen
  \bibfield  {author} {\bibinfo {author} {\bibnamefont {Bouwmeester},
  \bibfnamefont {Dik}}, \bibinfo {author} {\bibfnamefont {Jian-Wei}\
  \bibnamefont {Pan}}, \bibinfo {author} {\bibfnamefont {Matthew}\ \bibnamefont
  {Daniell}}, \bibinfo {author} {\bibfnamefont {Harald}\ \bibnamefont
  {Weinfurter}}, and\ \bibinfo {author} {\bibfnamefont {Anton}\ \bibnamefont
  {Zeilinger}}} (\bibinfo {year} {1999}),\ \bibfield  {title} {\enquote
  {\bibinfo {title} {Observation of three-photon greenberger-horne-zeilinger
  entanglement},}\ }\href {https://doi.org/10.1103/physrevlett.82.1345}
  {\bibfield  {journal} {\bibinfo  {journal} {Physical Review Letters}\
  }\textbf {\bibinfo {volume} {82}},\ \bibinfo {pages} {1345}},\ \Eprint
  {https://arxiv.org/abs/arXiv:quant-ph/9810035v1} {arXiv:quant-ph/9810035v1}
  \BibitemShut {NoStop}%
\bibitem [{\citenamefont {Brady}\ \emph {et~al.}(2024)\citenamefont {Brady},
  \citenamefont {Eickbusch}, \citenamefont {Singh}, \citenamefont {Wu},\ and\
  \citenamefont {Zhuang}}]{brady2024GKPreview}%
  \BibitemOpen
  \bibfield  {author} {\bibinfo {author} {\bibnamefont {Brady}, \bibfnamefont
  {Anthony~J}}, \bibinfo {author} {\bibfnamefont {Alec}\ \bibnamefont
  {Eickbusch}}, \bibinfo {author} {\bibfnamefont {Shraddha}\ \bibnamefont
  {Singh}}, \bibinfo {author} {\bibfnamefont {Jing}\ \bibnamefont {Wu}}, and\
  \bibinfo {author} {\bibfnamefont {Quntao}\ \bibnamefont {Zhuang}}} (\bibinfo
  {year} {2024}),\ \bibfield  {title} {\enquote {\bibinfo {title} {Advances in
  bosonic quantum error correction with gottesman--kitaev--preskill codes:
  Theory, engineering and applications},}\ }\href
  {https://doi.org/https://doi.org/10.1016/j.pquantelec.2023.100496} {\bibfield
   {journal} {\bibinfo  {journal} {Progress in Quantum Electronics}\ }\textbf
  {\bibinfo {volume} {93}},\ \bibinfo {pages} {100496}}\BibitemShut {NoStop}%
\bibitem [{\citenamefont {Brakerski}\ \emph {et~al.}(2011)\citenamefont
  {Brakerski}, \citenamefont {Gentry},\ and\ \citenamefont
  {Vaikuntanathan}}]{bib:Brakerski2011}%
  \BibitemOpen
  \bibfield  {author} {\bibinfo {author} {\bibnamefont {Brakerski},
  \bibfnamefont {Zvika}}, \bibinfo {author} {\bibfnamefont {Craig}\
  \bibnamefont {Gentry}}, and\ \bibinfo {author} {\bibfnamefont {Vinod}\
  \bibnamefont {Vaikuntanathan}}} (\bibinfo {year} {2011}),\ \bibfield  {title}
  {\enquote {\bibinfo {title} {Fully homomorphic encryption without
  bootstrapping},}\ }\href@noop {} {\ }\Eprint
  {https://arxiv.org/abs/https://eprint.iacr.org/2011/277}
  {https://eprint.iacr.org/2011/277} \BibitemShut {NoStop}%
\bibitem [{\citenamefont {Brandao}\ \emph {et~al.}(2017)\citenamefont
  {Brandao}, \citenamefont {Kalev}, \citenamefont {Li}, \citenamefont {Lin},
  \citenamefont {Svore},\ and\ \citenamefont
  {Wu}}]{bib:brandao2017exponential}%
  \BibitemOpen
  \bibfield  {author} {\bibinfo {author} {\bibnamefont {Brandao}, \bibfnamefont
  {Fernando~GSL}}, \bibinfo {author} {\bibfnamefont {Amir}\ \bibnamefont
  {Kalev}}, \bibinfo {author} {\bibfnamefont {Tongyang}\ \bibnamefont {Li}},
  \bibinfo {author} {\bibfnamefont {Cedric Yen-Yu}\ \bibnamefont {Lin}},
  \bibinfo {author} {\bibfnamefont {Krysta~M}\ \bibnamefont {Svore}}, and\
  \bibinfo {author} {\bibfnamefont {Xiaodi}\ \bibnamefont {Wu}}} (\bibinfo
  {year} {2017}),\ \bibfield  {title} {\enquote {\bibinfo {title} {Exponential
  quantum speed-ups for semidefinite programming with applications to quantum
  learning},}\ }\href@noop {} {\ }\Eprint
  {https://arxiv.org/abs/arXiv:1710.02581} {arXiv:1710.02581} \BibitemShut
  {NoStop}%
\bibitem [{\citenamefont {Brand{\~a}o}\ \emph {et~al.}(2017)\citenamefont
  {Brand{\~a}o}, \citenamefont {Kalev}, \citenamefont {Li}, \citenamefont
  {Lin}, \citenamefont {Svore},\ and\ \citenamefont {Wu}}]{brandao2017quantum}%
  \BibitemOpen
  \bibfield  {author} {\bibinfo {author} {\bibnamefont {Brand{\~a}o},
  \bibfnamefont {Fernando~GSL}}, \bibinfo {author} {\bibfnamefont {Amir}\
  \bibnamefont {Kalev}}, \bibinfo {author} {\bibfnamefont {Tongyang}\
  \bibnamefont {Li}}, \bibinfo {author} {\bibfnamefont {Cedric Yen-Yu}\
  \bibnamefont {Lin}}, \bibinfo {author} {\bibfnamefont {Krysta~M}\
  \bibnamefont {Svore}}, and\ \bibinfo {author} {\bibfnamefont {Xiaodi}\
  \bibnamefont {Wu}}} (\bibinfo {year} {2017}),\ \bibfield  {title} {\enquote
  {\bibinfo {title} {Quantum sdp solvers: Large speed-ups, optimality, and
  applications to quantum learning},}\ }\href@noop {} {\bibinfo  {journal}
  {arXiv:1710.02581}\ }\BibitemShut {NoStop}%
\bibitem [{\citenamefont {Brandao}\ and\ \citenamefont
  {Svore}(2017)}]{bib:brandao2017quantum}%
  \BibitemOpen
\bibfield  {journal} {  }\bibfield  {author} {\bibinfo {author} {\bibnamefont
  {Brandao}, \bibfnamefont {Fernando~GSL}}, and\ \bibinfo {author}
  {\bibfnamefont {Krysta~M}\ \bibnamefont {Svore}}} (\bibinfo {year} {2017}),\
  \bibfield  {title} {\enquote {\bibinfo {title} {Quantum speed-ups for solving
  semidefinite programs},}\ }in\ \href {https://doi.org/10.1109/focs.2017.45}
  {\emph {\bibinfo {booktitle} {Symposium on Foundations of Computer Science
  (FOCS)}}},\ Vol.~\bibinfo {volume} {58},\ p.\ \bibinfo {pages}
  {415}\BibitemShut {NoStop}%
\bibitem [{\citenamefont {Branning}\ \emph {et~al.}(2000)\citenamefont
  {Branning}, \citenamefont {Grice}, \citenamefont {Erdmann},\ and\
  \citenamefont {Walmsley}}]{bib:Branning00}%
  \BibitemOpen
  \bibfield  {author} {\bibinfo {author} {\bibnamefont {Branning},
  \bibfnamefont {David}}, \bibinfo {author} {\bibfnamefont {Warren}\
  \bibnamefont {Grice}}, \bibinfo {author} {\bibfnamefont {Reinhard}\
  \bibnamefont {Erdmann}}, and\ \bibinfo {author} {\bibfnamefont {I.~A.}\
  \bibnamefont {Walmsley}}} (\bibinfo {year} {2000}),\ \bibfield  {title}
  {\enquote {\bibinfo {title} {Interferometric technique for engineering
  indistinguishability and entanglement of photon pairs},}\ }\href
  {https://doi.org/10.1103/physreva.62.013814} {\bibfield  {journal} {\bibinfo
  {journal} {Physical Review A}\ }\textbf {\bibinfo {volume} {62}},\ \bibinfo
  {pages} {013814}}\BibitemShut {NoStop}%
\bibitem [{\citenamefont {Brassard}(2003)}]{bib:brassard2003quantum}%
  \BibitemOpen
  \bibfield  {author} {\bibinfo {author} {\bibnamefont {Brassard},
  \bibfnamefont {Gilles}}} (\bibinfo {year} {2003}),\ \bibfield  {title}
  {\enquote {\bibinfo {title} {Quantum communication complexity},}\ }\href@noop
  {} {\bibfield  {journal} {\bibinfo  {journal} {Foundations of Physics}\
  }\textbf {\bibinfo {volume} {33}},\ \bibinfo {pages} {1593}}\BibitemShut
  {NoStop}%
\bibitem [{\citenamefont {Brattke}\ \emph {et~al.}(2001)\citenamefont
  {Brattke}, \citenamefont {Varcoe},\ and\ \citenamefont
  {Walther}}]{bib:Brattke01}%
  \BibitemOpen
  \bibfield  {author} {\bibinfo {author} {\bibnamefont {Brattke}, \bibfnamefont
  {S}}, \bibinfo {author} {\bibfnamefont {B.~T.~H.}\ \bibnamefont {Varcoe}},
  and\ \bibinfo {author} {\bibfnamefont {H.}~\bibnamefont {Walther}}} (\bibinfo
  {year} {2001}),\ \bibfield  {title} {\enquote {\bibinfo {title} {Generation
  of photon number states on demand via cavity quantum electrodynamics},}\
  }\href {https://doi.org/10.1103/physrevlett.86.3534} {\bibfield  {journal}
  {\bibinfo  {journal} {Physical Review Letters}\ }\textbf {\bibinfo {volume}
  {86}},\ \bibinfo {pages} {3534}}\BibitemShut {NoStop}%
\bibitem [{\citenamefont {Bratzik}\ \emph {et~al.}(2013)\citenamefont
  {Bratzik}, \citenamefont {Abruzzo}, \citenamefont {Kampermann},\ and\
  \citenamefont {Brub}}]{bib:braztzik2013}%
  \BibitemOpen
  \bibfield  {author} {\bibinfo {author} {\bibnamefont {Bratzik}, \bibfnamefont
  {Sylvia}}, \bibinfo {author} {\bibfnamefont {Silvestre}\ \bibnamefont
  {Abruzzo}}, \bibinfo {author} {\bibfnamefont {Hermann}\ \bibnamefont
  {Kampermann}}, and\ \bibinfo {author} {\bibfnamefont {Dagmar}\ \bibnamefont
  {Brub}}} (\bibinfo {year} {2013}),\ \bibfield  {title} {\enquote {\bibinfo
  {title} {Quantum repeaters and quantum key distribution: The impact of
  entanglement distillation on the secret-key rate},}\ }\href
  {https://doi.org/10.1103/physreva.87.062335} {\bibfield  {journal} {\bibinfo
  {journal} {Physical Review A}\ }\textbf {\bibinfo {volume} {86}},\ \bibinfo
  {pages} {062335}},\ \Eprint {https://arxiv.org/abs/arXiv:1303.3456v1}
  {arXiv:1303.3456v1} \BibitemShut {NoStop}%
\bibitem [{\citenamefont {Braunstein}\ and\ \citenamefont
  {Mann}(1995)}]{bib:BraunsteinMann95}%
  \BibitemOpen
  \bibfield  {author} {\bibinfo {author} {\bibnamefont {Braunstein},
  \bibfnamefont {S~L}}, and\ \bibinfo {author} {\bibfnamefont {A.}~\bibnamefont
  {Mann}}} (\bibinfo {year} {1995}),\ \bibfield  {title} {\enquote {\bibinfo
  {title} {Measurement of the bell operator and quantum teleportation},}\
  }\href {https://doi.org/10.1103/physreva.51.r1727} {\bibfield  {journal}
  {\bibinfo  {journal} {Physical Review A}\ }\textbf {\bibinfo {volume} {51}},\
  \bibinfo {pages} {R1727}}\BibitemShut {NoStop}%
\bibitem [{\citenamefont {Braunstein}\ and\ \citenamefont {van
  Loock}(2005)}]{bib:RevModPhys.77.513}%
  \BibitemOpen
  \bibfield  {author} {\bibinfo {author} {\bibnamefont {Braunstein},
  \bibfnamefont {Samuel~L}}, and\ \bibinfo {author} {\bibfnamefont {Peter}\
  \bibnamefont {van Loock}}} (\bibinfo {year} {2005}),\ \bibfield  {title}
  {\enquote {\bibinfo {title} {Quantum information with continuous
  variables},}\ }\href {https://doi.org/10.1103/revmodphys.77.513} {\bibfield
  {journal} {\bibinfo  {journal} {Reviews in Modern Physics}\ }\textbf
  {\bibinfo {volume} {77}},\ \bibinfo {pages} {513}},\ \Eprint
  {https://arxiv.org/abs/arXiv:quant-ph/0410100v1} {arXiv:quant-ph/0410100v1}
  \BibitemShut {NoStop}%
\bibitem [{\citenamefont {Bravyi}\ \emph {et~al.}(2018)\citenamefont {Bravyi},
  \citenamefont {Gosset},\ and\ \citenamefont
  {Koenig}}]{bib:bravyi2018quantum}%
  \BibitemOpen
  \bibfield  {author} {\bibinfo {author} {\bibnamefont {Bravyi}, \bibfnamefont
  {Sergey}}, \bibinfo {author} {\bibfnamefont {David}\ \bibnamefont {Gosset}},
  and\ \bibinfo {author} {\bibfnamefont {Robert}\ \bibnamefont {Koenig}}}
  (\bibinfo {year} {2018}),\ \bibfield  {title} {\enquote {\bibinfo {title}
  {Quantum advantage with shallow circuits},}\ }\href
  {https://doi.org/10.1126/science.aar3106} {\bibfield  {journal} {\bibinfo
  {journal} {Science}\ }\textbf {\bibinfo {volume} {362}},\ \bibinfo {pages}
  {308}},\ \Eprint {https://arxiv.org/abs/arXiv:1704.00690} {arXiv:1704.00690}
  \BibitemShut {NoStop}%
\bibitem [{\citenamefont {Brennen}\ \emph
  {et~al.}(2015{\natexlab{a}})\citenamefont {Brennen}, \citenamefont {Rohde},
  \citenamefont {Sanders},\ and\ \citenamefont {Singh}}]{bib:RohdeWavelet15}%
  \BibitemOpen
  \bibfield  {author} {\bibinfo {author} {\bibnamefont {Brennen}, \bibfnamefont
  {Gavin~K}}, \bibinfo {author} {\bibfnamefont {Peter}\ \bibnamefont {Rohde}},
  \bibinfo {author} {\bibfnamefont {Barry~C.}\ \bibnamefont {Sanders}}, and\
  \bibinfo {author} {\bibfnamefont {Sukhwinder}\ \bibnamefont {Singh}}}
  (\bibinfo {year} {2015}{\natexlab{a}}),\ \bibfield  {title} {\enquote
  {\bibinfo {title} {Multi-scale quantum simulation of quantum field theory
  using wavelets},}\ }\href {https://doi.org/10.1103/physreva.92.032315}
  {\bibfield  {journal} {\bibinfo  {journal} {Physical Review A}\ }\textbf
  {\bibinfo {volume} {92}},\ \bibinfo {pages} {032315}},\ \Eprint
  {https://arxiv.org/abs/arXiv:1412.0750v1} {arXiv:1412.0750v1} \BibitemShut
  {NoStop}%
\bibitem [{\citenamefont {Brennen}\ \emph
  {et~al.}(2015{\natexlab{b}})\citenamefont {Brennen}, \citenamefont {Rohde},
  \citenamefont {Sanders},\ and\ \citenamefont
  {Singh}}]{brennen2015multiscale}%
  \BibitemOpen
  \bibfield  {author} {\bibinfo {author} {\bibnamefont {Brennen}, \bibfnamefont
  {Gavin~K}}, \bibinfo {author} {\bibfnamefont {Peter}\ \bibnamefont {Rohde}},
  \bibinfo {author} {\bibfnamefont {Barry~C}\ \bibnamefont {Sanders}}, and\
  \bibinfo {author} {\bibfnamefont {Sukhwinder}\ \bibnamefont {Singh}}}
  (\bibinfo {year} {2015}{\natexlab{b}}),\ \bibfield  {title} {\enquote
  {\bibinfo {title} {Multiscale quantum simulation of quantum field theory
  using wavelets},}\ }\href@noop {} {\bibfield  {journal} {\bibinfo  {journal}
  {Physical Review A}\ }\textbf {\bibinfo {volume} {92}}~(\bibinfo {number}
  {3}),\ \bibinfo {pages} {032315}}\BibitemShut {NoStop}%
\bibitem [{\citenamefont {Breuckmann}\ \emph {et~al.}(2017)\citenamefont
  {Breuckmann}, \citenamefont {Vuillot}, \citenamefont {Campbell},
  \citenamefont {Krishna},\ and\ \citenamefont
  {Terhal}}]{SD-Breuckmann:2017aa}%
  \BibitemOpen
  \bibfield  {author} {\bibinfo {author} {\bibnamefont {Breuckmann},
  \bibfnamefont {Nikolas~P}}, \bibinfo {author} {\bibfnamefont {Christophe}\
  \bibnamefont {Vuillot}}, \bibinfo {author} {\bibfnamefont {Earl}\
  \bibnamefont {Campbell}}, \bibinfo {author} {\bibfnamefont {Anirudh}\
  \bibnamefont {Krishna}}, and\ \bibinfo {author} {\bibfnamefont {Barbara~M}\
  \bibnamefont {Terhal}}} (\bibinfo {year} {2017}),\ \bibfield  {title}
  {\enquote {\bibinfo {title} {Hyperbolic and semi-hyperbolic surface codes for
  quantum storage},}\ }\href {https://doi.org/10.1088/2058-9565/aa7d3b}
  {\bibfield  {journal} {\bibinfo  {journal} {Quantum Science and Technology}\
  }\textbf {\bibinfo {volume} {2}},\ \bibinfo {pages} {035007}},\ \Eprint
  {https://arxiv.org/abs/arXiv:1703.00590v2} {arXiv:1703.00590v2} \BibitemShut
  {NoStop}%
\bibitem [{\citenamefont {Briegel}\ \emph {et~al.}(1998)\citenamefont
  {Briegel}, \citenamefont {D{\"u}r}, \citenamefont {Cirac},\ and\
  \citenamefont {Zoller}}]{bib:BDCZ98}%
  \BibitemOpen
  \bibfield  {author} {\bibinfo {author} {\bibnamefont {Briegel}, \bibfnamefont
  {H~J}}, \bibinfo {author} {\bibfnamefont {W.}~\bibnamefont {D{\"u}r}},
  \bibinfo {author} {\bibfnamefont {J.I.}\ \bibnamefont {Cirac}}, and\ \bibinfo
  {author} {\bibfnamefont {P}~\bibnamefont {Zoller}}} (\bibinfo {year}
  {1998}),\ \bibfield  {title} {\enquote {\bibinfo {title} {Quantum repeaters:
  The role of imperfect local operations in quantum communication},}\ }\href
  {https://doi.org/10.1103/physrevlett.81.5932} {\bibfield  {journal} {\bibinfo
   {journal} {Physical Review Letters}\ }\textbf {\bibinfo {volume} {81}},\
  \bibinfo {pages} {5932}}\BibitemShut {NoStop}%
\bibitem [{\citenamefont {Briegel}\ and\ \citenamefont
  {Raussendorf}(2001)}]{briegel2001persistent}%
  \BibitemOpen
  \bibfield  {author} {\bibinfo {author} {\bibnamefont {Briegel}, \bibfnamefont
  {Hans~J}}, and\ \bibinfo {author} {\bibfnamefont {Robert}\ \bibnamefont
  {Raussendorf}}} (\bibinfo {year} {2001}),\ \bibfield  {title} {\enquote
  {\bibinfo {title} {Persistent entanglement in arrays of interacting
  particles},}\ }\href@noop {} {\bibfield  {journal} {\bibinfo  {journal}
  {Physical Review Letters}\ }\textbf {\bibinfo {volume} {86}}~(\bibinfo
  {number} {5}),\ \bibinfo {pages} {910}}\BibitemShut {NoStop}%
\bibitem [{\citenamefont {Broadbent}\ \emph {et~al.}(2009)\citenamefont
  {Broadbent}, \citenamefont {Fitzsimons},\ and\ \citenamefont
  {Kashefi}}]{bib:broadbent2009universal}%
  \BibitemOpen
  \bibfield  {author} {\bibinfo {author} {\bibnamefont {Broadbent},
  \bibfnamefont {Anne}}, \bibinfo {author} {\bibfnamefont {Joseph}\
  \bibnamefont {Fitzsimons}}, and\ \bibinfo {author} {\bibfnamefont {Elham}\
  \bibnamefont {Kashefi}}} (\bibinfo {year} {2009}),\ \bibfield  {title}
  {\enquote {\bibinfo {title} {Universal blind quantum computation},}\ }in\
  \href {https://doi.org/10.1109/focs.2009.36} {\emph {\bibinfo {booktitle}
  {IEEE Symposium on Foundations of Computer Science (FOCS)}}},\ Vol.~\bibinfo
  {volume} {50},\ p.\ \bibinfo {pages} {517},\ \Eprint
  {https://arxiv.org/abs/arXiv:0807.4154v3} {arXiv:0807.4154v3} \BibitemShut
  {NoStop}%
\bibitem [{\citenamefont {Broadbent}\ and\ \citenamefont
  {Jeffery}(2015)}]{broadbent2015quantum}%
  \BibitemOpen
  \bibfield  {author} {\bibinfo {author} {\bibnamefont {Broadbent},
  \bibfnamefont {Anne}}, and\ \bibinfo {author} {\bibfnamefont {Stacey}\
  \bibnamefont {Jeffery}}} (\bibinfo {year} {2015}),\ \bibfield  {title}
  {\enquote {\bibinfo {title} {Quantum homomorphic encryption for circuits of
  low t-gate complexity},}\ }in\ \href@noop {} {\emph {\bibinfo {booktitle}
  {Annual Cryptology Conference}}}\ (\bibinfo {organization} {Springer})\ pp.\
  \bibinfo {pages} {609--629}\BibitemShut {NoStop}%
\bibitem [{\citenamefont {Broadbent}\ and\ \citenamefont
  {Tapp}(2007)}]{broadbent2007information}%
  \BibitemOpen
  \bibfield  {author} {\bibinfo {author} {\bibnamefont {Broadbent},
  \bibfnamefont {Anne}}, and\ \bibinfo {author} {\bibfnamefont {Alain}\
  \bibnamefont {Tapp}}} (\bibinfo {year} {2007}),\ \bibfield  {title} {\enquote
  {\bibinfo {title} {Information-theoretic security without an honest
  majority},}\ }in\ \href@noop {} {\emph {\bibinfo {booktitle} {Advances in
  Cryptology-ASIACRYPT 2007: 13th International Conference on the Theory and
  Application of Cryptology and Information Security, Kuching, Malaysia,
  December 2-6, 2007. Proceedings 13}}}\ (\bibinfo {organization} {Springer})\
  pp.\ \bibinfo {pages} {410--426}\BibitemShut {NoStop}%
\bibitem [{\citenamefont {Broome}\ \emph {et~al.}(2010)\citenamefont {Broome},
  \citenamefont {Fedrizzi}, \citenamefont {Lanyon}, \citenamefont {Kassal},
  \citenamefont {Aspuru-Guzik},\ and\ \citenamefont {White}}]{bib:Broome10}%
  \BibitemOpen
  \bibfield  {author} {\bibinfo {author} {\bibnamefont {Broome}, \bibfnamefont
  {M~A}}, \bibinfo {author} {\bibfnamefont {A.}~\bibnamefont {Fedrizzi}},
  \bibinfo {author} {\bibfnamefont {B.~P.}\ \bibnamefont {Lanyon}}, \bibinfo
  {author} {\bibfnamefont {I.}~\bibnamefont {Kassal}}, \bibinfo {author}
  {\bibfnamefont {A.}~\bibnamefont {Aspuru-Guzik}}, and\ \bibinfo {author}
  {\bibfnamefont {A.~G.}\ \bibnamefont {White}}} (\bibinfo {year} {2010}),\
  \bibfield  {title} {\enquote {\bibinfo {title} {Discrete single-photon
  quantum walks with tunable decoherence},}\ }\href
  {https://doi.org/10.1103/physrevlett.104.153602} {\bibfield  {journal}
  {\bibinfo  {journal} {Physical Review Letters}\ }\textbf {\bibinfo {volume}
  {104}},\ \bibinfo {pages} {153602}},\ \Eprint
  {https://arxiv.org/abs/arXiv:1002.4923v2} {arXiv:1002.4923v2} \BibitemShut
  {NoStop}%
\bibitem [{\citenamefont {Broome}\ \emph {et~al.}(2013)\citenamefont {Broome},
  \citenamefont {Fedrizzi}, \citenamefont {Rahimi-Keshari}, \citenamefont
  {Dove}, \citenamefont {Aaronson}, \citenamefont {Ralph},\ and\ \citenamefont
  {White}}]{bib:Broome2012}%
  \BibitemOpen
  \bibfield  {author} {\bibinfo {author} {\bibnamefont {Broome}, \bibfnamefont
  {Matthew~A}}, \bibinfo {author} {\bibfnamefont {Alessandro}\ \bibnamefont
  {Fedrizzi}}, \bibinfo {author} {\bibfnamefont {Saleh}\ \bibnamefont
  {Rahimi-Keshari}}, \bibinfo {author} {\bibfnamefont {Justin}\ \bibnamefont
  {Dove}}, \bibinfo {author} {\bibfnamefont {Scott}\ \bibnamefont {Aaronson}},
  \bibinfo {author} {\bibfnamefont {Timothy~C.}\ \bibnamefont {Ralph}}, and\
  \bibinfo {author} {\bibfnamefont {Andrew~G.}\ \bibnamefont {White}}}
  (\bibinfo {year} {2013}),\ \bibfield  {title} {\enquote {\bibinfo {title}
  {Photonic boson sampling in a tunable circuit},}\ }\href@noop {} {\bibfield
  {journal} {\bibinfo  {journal} {Science}\ }\textbf {\bibinfo {volume}
  {339}},\ \bibinfo {pages} {6121}},\ \Eprint
  {https://arxiv.org/abs/arXiv:1212.2234v3} {arXiv:1212.2234v3} \BibitemShut
  {NoStop}%
\bibitem [{\citenamefont {Brown}(2013)}]{brown2013defects}%
  \BibitemOpen
  \bibfield  {author} {\bibinfo {author} {\bibnamefont {Brown}, \bibfnamefont
  {Benjamin~James}}} (\bibinfo {year} {2013}),\ \emph {\bibinfo {title}
  {Defects in topologically ordered lattice models}},\ \href@noop {} {Ph.D.
  thesis}\ (\bibinfo  {school} {Imperial College London})\BibitemShut {NoStop}%
\bibitem [{\citenamefont {Browne}\ \emph {et~al.}(2003)\citenamefont {Browne},
  \citenamefont {Eisert}, \citenamefont {Scheel},\ and\ \citenamefont
  {Plenio}}]{bib:PhysRevA.67.062320}%
  \BibitemOpen
  \bibfield  {author} {\bibinfo {author} {\bibnamefont {Browne}, \bibfnamefont
  {Daniel~E}}, \bibinfo {author} {\bibfnamefont {Jens}\ \bibnamefont {Eisert}},
  \bibinfo {author} {\bibfnamefont {Stefan}\ \bibnamefont {Scheel}}, and\
  \bibinfo {author} {\bibfnamefont {Martin~B.}\ \bibnamefont {Plenio}}}
  (\bibinfo {year} {2003}),\ \bibfield  {title} {\enquote {\bibinfo {title}
  {Driving non-gaussian to gaussian states with linear optics},}\ }\href
  {https://doi.org/10.1103/PhysRevA.67.062320} {\bibfield  {journal} {\bibinfo
  {journal} {Physical Review A}\ }\textbf {\bibinfo {volume} {67}},\ \bibinfo
  {pages} {062320}}\BibitemShut {NoStop}%
\bibitem [{\citenamefont {Browne}\ and\ \citenamefont
  {Rudolph}(2005{\natexlab{a}})}]{browne2005resource}%
  \BibitemOpen
  \bibfield  {author} {\bibinfo {author} {\bibnamefont {Browne}, \bibfnamefont
  {Daniel~E}}, and\ \bibinfo {author} {\bibfnamefont {Terry}\ \bibnamefont
  {Rudolph}}} (\bibinfo {year} {2005}{\natexlab{a}}),\ \bibfield  {title}
  {\enquote {\bibinfo {title} {Resource-efficient linear optical quantum
  computation},}\ }\href@noop {} {\bibfield  {journal} {\bibinfo  {journal}
  {Physical Review Letters}\ }\textbf {\bibinfo {volume} {95}}~(\bibinfo
  {number} {1}),\ \bibinfo {pages} {010501}}\BibitemShut {NoStop}%
\bibitem [{\citenamefont {Browne}\ and\ \citenamefont
  {Rudolph}(2005{\natexlab{b}})}]{bib:BrowneRudolph05}%
  \BibitemOpen
  \bibfield  {author} {\bibinfo {author} {\bibnamefont {Browne}, \bibfnamefont
  {Daniel~E}}, and\ \bibinfo {author} {\bibfnamefont {Terry}\ \bibnamefont
  {Rudolph}}} (\bibinfo {year} {2005}{\natexlab{b}}),\ \bibfield  {title}
  {\enquote {\bibinfo {title} {Resource-efficient linear optics quantum
  computation},}\ }\href {https://doi.org/10.1103/physrevlett.95.010501}
  {\bibfield  {journal} {\bibinfo  {journal} {Physical Review Letters}\
  }\textbf {\bibinfo {volume} {95}},\ \bibinfo {pages} {010501}},\ \Eprint
  {https://arxiv.org/abs/arXiv:quant-ph/0405157v2} {arXiv:quant-ph/0405157v2}
  \BibitemShut {NoStop}%
\bibitem [{\citenamefont {Brunel}\ \emph {et~al.}(1999)\citenamefont {Brunel},
  \citenamefont {Lounis}, \citenamefont {Tamarat},\ and\ \citenamefont
  {Orrit}}]{bib:Brunel99}%
  \BibitemOpen
  \bibfield  {author} {\bibinfo {author} {\bibnamefont {Brunel}, \bibfnamefont
  {C}}, \bibinfo {author} {\bibfnamefont {B.}~\bibnamefont {Lounis}}, \bibinfo
  {author} {\bibfnamefont {P.}~\bibnamefont {Tamarat}}, and\ \bibinfo {author}
  {\bibfnamefont {M.}~\bibnamefont {Orrit}}} (\bibinfo {year} {1999}),\
  \bibfield  {title} {\enquote {\bibinfo {title} {Triggered source of single
  photons based on controlled single molecule fluorescence},}\ }\href
  {https://doi.org/10.1103/physrevlett.83.2722} {\bibfield  {journal} {\bibinfo
   {journal} {Physical Review Letters}\ }\textbf {\bibinfo {volume} {83}},\
  \bibinfo {pages} {2722}}\BibitemShut {NoStop}%
\bibitem [{\citenamefont {Budinger}\ \emph {et~al.}(2024)\citenamefont
  {Budinger}, \citenamefont {Furusawa},\ and\ \citenamefont {van
  Loock}}]{budinger2024phasegate}%
  \BibitemOpen
  \bibfield  {author} {\bibinfo {author} {\bibnamefont {Budinger},
  \bibfnamefont {Niklas}}, \bibinfo {author} {\bibfnamefont {Akira}\
  \bibnamefont {Furusawa}}, and\ \bibinfo {author} {\bibfnamefont {Peter}\
  \bibnamefont {van Loock}}} (\bibinfo {year} {2024}),\ \bibfield  {title}
  {\enquote {\bibinfo {title} {All-optical quantum computing using cubic phase
  gates},}\ }\href {https://doi.org/10.1103/PhysRevResearch.6.023332}
  {\bibfield  {journal} {\bibinfo  {journal} {Phys. Rev. Res.}\ }\textbf
  {\bibinfo {volume} {6}},\ \bibinfo {pages} {023332}}\BibitemShut {NoStop}%
\bibitem [{\citenamefont {Bussi{\`e}res}\ \emph {et~al.}(2013)\citenamefont
  {Bussi{\`e}res}, \citenamefont {Sangouard}, \citenamefont {Afzelius},
  \citenamefont {de~Riedmatten}, \citenamefont {Simon},\ and\ \citenamefont
  {Tittel}}]{bib:bussieres2013prospective}%
  \BibitemOpen
  \bibfield  {author} {\bibinfo {author} {\bibnamefont {Bussi{\`e}res},
  \bibfnamefont {F{\'e}lix}}, \bibinfo {author} {\bibfnamefont {Nicolas}\
  \bibnamefont {Sangouard}}, \bibinfo {author} {\bibfnamefont {Mikael}\
  \bibnamefont {Afzelius}}, \bibinfo {author} {\bibfnamefont {Hugues}\
  \bibnamefont {de~Riedmatten}}, \bibinfo {author} {\bibfnamefont {Christoph}\
  \bibnamefont {Simon}}, and\ \bibinfo {author} {\bibfnamefont {Wolfgang}\
  \bibnamefont {Tittel}}} (\bibinfo {year} {2013}),\ \bibfield  {title}
  {\enquote {\bibinfo {title} {Prospective applications of optical quantum
  memories},}\ }\href {https://doi.org/10.1080/09500340.2013.856482} {\bibfield
   {journal} {\bibinfo  {journal} {Journal of Modern Optics}\ }\textbf
  {\bibinfo {volume} {60}},\ \bibinfo {pages} {1519}},\ \Eprint
  {https://arxiv.org/abs/arXiv:1306.6904v1} {arXiv:1306.6904v1} \BibitemShut
  {NoStop}%
\bibitem [{\citenamefont {Byrnes}\ \emph {et~al.}(2017)\citenamefont {Byrnes},
  \citenamefont {Ilyas}, \citenamefont {Tessler}, \citenamefont {Takeoka},
  \citenamefont {Jambulingam},\ and\ \citenamefont
  {Dowling}}]{bib:byrnes2017lorentz}%
  \BibitemOpen
  \bibfield  {author} {\bibinfo {author} {\bibnamefont {Byrnes}, \bibfnamefont
  {Tim}}, \bibinfo {author} {\bibfnamefont {Batyr}\ \bibnamefont {Ilyas}},
  \bibinfo {author} {\bibfnamefont {Louis}\ \bibnamefont {Tessler}}, \bibinfo
  {author} {\bibfnamefont {Masahiro}\ \bibnamefont {Takeoka}}, \bibinfo
  {author} {\bibfnamefont {Segar}\ \bibnamefont {Jambulingam}}, and\ \bibinfo
  {author} {\bibfnamefont {Jonathan~P}\ \bibnamefont {Dowling}}} (\bibinfo
  {year} {2017}),\ \bibfield  {title} {\enquote {\bibinfo {title} {Lorentz
  invariant entanglement distribution for the space-based quantum network},}\
  }\href@noop {} {\ }\Eprint {https://arxiv.org/abs/arXiv:1704.04774}
  {arXiv:1704.04774} \BibitemShut {NoStop}%
\bibitem [{\citenamefont {Cable}\ and\ \citenamefont
  {Dowling}(2007)}]{bib:Cable07}%
  \BibitemOpen
  \bibfield  {author} {\bibinfo {author} {\bibnamefont {Cable}, \bibfnamefont
  {H}}, and\ \bibinfo {author} {\bibfnamefont {J.~P.}\ \bibnamefont {Dowling}}}
  (\bibinfo {year} {2007}),\ \bibfield  {title} {\enquote {\bibinfo {title}
  {Efficient generation of large number-path entanglement using only linear
  optics and feed-forward},}\ }\href
  {https://doi.org/10.1103/physrevlett.99.163604} {\bibfield  {journal}
  {\bibinfo  {journal} {Physical Review Letters}\ }\textbf {\bibinfo {volume}
  {99}},\ \bibinfo {pages} {163604}}\BibitemShut {NoStop}%
\bibitem [{\citenamefont {Cahill}\ and\ \citenamefont
  {Glauber}(1969)}]{bib:CahillGlauber69}%
  \BibitemOpen
  \bibfield  {author} {\bibinfo {author} {\bibnamefont {Cahill}, \bibfnamefont
  {K~E}}, and\ \bibinfo {author} {\bibfnamefont {R.~J.}\ \bibnamefont
  {Glauber}}} (\bibinfo {year} {1969}),\ \bibfield  {title} {\enquote {\bibinfo
  {title} {Density operators and quasiprobability distributions},}\ }\href
  {https://doi.org/10.1103/physrev.177.1882} {\bibfield  {journal} {\bibinfo
  {journal} {Physical Review}\ }\textbf {\bibinfo {volume} {177}},\ \bibinfo
  {pages} {177}}\BibitemShut {NoStop}%
\bibitem [{\citenamefont {Cai}\ and\ \citenamefont
  {Scarani}(2009)}]{bib:cai2009finite}%
  \BibitemOpen
  \bibfield  {author} {\bibinfo {author} {\bibnamefont {Cai}, \bibfnamefont
  {Raymond~YQ}}, and\ \bibinfo {author} {\bibfnamefont {Valerio}\ \bibnamefont
  {Scarani}}} (\bibinfo {year} {2009}),\ \bibfield  {title} {\enquote {\bibinfo
  {title} {Finite-key analysis for practical implementations of quantum key
  distribution},}\ }\href {https://doi.org/10.1088/1367-2630/11/10/109801}
  {\bibfield  {journal} {\bibinfo  {journal} {New Journal of Physics}\ }\textbf
  {\bibinfo {volume} {11}},\ \bibinfo {pages} {045024}}\BibitemShut {NoStop}%
\bibitem [{\citenamefont {Calderbank}\ and\ \citenamefont
  {Shor}(1996)}]{bib:CalderbankShor96}%
  \BibitemOpen
  \bibfield  {author} {\bibinfo {author} {\bibnamefont {Calderbank},
  \bibfnamefont {A~R}}, and\ \bibinfo {author} {\bibfnamefont {Peter~W.}\
  \bibnamefont {Shor}}} (\bibinfo {year} {1996}),\ \bibfield  {title} {\enquote
  {\bibinfo {title} {Good quantum error-correcting codes exist},}\ }\href
  {https://doi.org/10.1103/physreva.54.1098} {\bibfield  {journal} {\bibinfo
  {journal} {Physical Review A}\ }\textbf {\bibinfo {volume} {54}},\ \bibinfo
  {pages} {1098}},\ \Eprint {https://arxiv.org/abs/arXiv:quant-ph/9512032v2}
  {arXiv:quant-ph/9512032v2} \BibitemShut {NoStop}%
\bibitem [{\citenamefont {Campagne-Ibarcq}\ \emph {et~al.}(2020)\citenamefont
  {Campagne-Ibarcq}, \citenamefont {Eickbusch}, \citenamefont {Touzard},
  \citenamefont {Zalys-Geller}, \citenamefont {Frattini}, \citenamefont
  {Sivak}, \citenamefont {Reinhold}, \citenamefont {Puri}, \citenamefont
  {Shankar}, \citenamefont {Schoelkopf}, \citenamefont {Frunzio}, \citenamefont
  {Mirrahimi},\ and\ \citenamefont {Devoret}}]{yale2020makingGKP}%
  \BibitemOpen
  \bibfield  {author} {\bibinfo {author} {\bibnamefont {Campagne-Ibarcq},
  \bibfnamefont {P}}, \bibinfo {author} {\bibfnamefont {A.}~\bibnamefont
  {Eickbusch}}, \bibinfo {author} {\bibfnamefont {S.}~\bibnamefont {Touzard}},
  \bibinfo {author} {\bibfnamefont {E.}~\bibnamefont {Zalys-Geller}}, \bibinfo
  {author} {\bibfnamefont {N.~E.}\ \bibnamefont {Frattini}}, \bibinfo {author}
  {\bibfnamefont {V.~V.}\ \bibnamefont {Sivak}}, \bibinfo {author}
  {\bibfnamefont {P.}~\bibnamefont {Reinhold}}, \bibinfo {author}
  {\bibfnamefont {S.}~\bibnamefont {Puri}}, \bibinfo {author} {\bibfnamefont
  {S.}~\bibnamefont {Shankar}}, \bibinfo {author} {\bibfnamefont {R.~J.}\
  \bibnamefont {Schoelkopf}}, \bibinfo {author} {\bibfnamefont
  {L.}~\bibnamefont {Frunzio}}, \bibinfo {author} {\bibfnamefont
  {M.}~\bibnamefont {Mirrahimi}}, and\ \bibinfo {author} {\bibfnamefont
  {M.~H.}\ \bibnamefont {Devoret}}} (\bibinfo {year} {2020}),\ \bibfield
  {title} {\enquote {\bibinfo {title} {Quantum error correction of a qubit
  encoded in grid states of an oscillator},}\ }\href
  {https://doi.org/10.1038/s41586-020-2603-3} {\bibfield  {journal} {\bibinfo
  {journal} {Nature}\ }\textbf {\bibinfo {volume} {584}}~(\bibinfo {number}
  {7821}),\ \bibinfo {pages} {368--372}}\BibitemShut {NoStop}%
\bibitem [{\citenamefont {Campbell}\ \emph {et~al.}(2007)\citenamefont
  {Campbell}, \citenamefont {Fitzsimons}, \citenamefont {Benjamin},\ and\
  \citenamefont {Kok}}]{bib:Campbell07}%
  \BibitemOpen
  \bibfield  {author} {\bibinfo {author} {\bibnamefont {Campbell},
  \bibfnamefont {Earl~T}}, \bibinfo {author} {\bibfnamefont {Joseph}\
  \bibnamefont {Fitzsimons}}, \bibinfo {author} {\bibfnamefont {Simon~C.}\
  \bibnamefont {Benjamin}}, and\ \bibinfo {author} {\bibfnamefont {Pieter}\
  \bibnamefont {Kok}}} (\bibinfo {year} {2007}),\ \bibfield  {title} {\enquote
  {\bibinfo {title} {Efficient growth of complex graph states via imperfect
  path erasure},}\ }\href {https://doi.org/10.1088/1367-2630/9/6/196}
  {\bibfield  {journal} {\bibinfo  {journal} {New Journal of Physics}\ }\textbf
  {\bibinfo {volume} {9}},\ \bibinfo {pages} {196}}\BibitemShut {NoStop}%
\bibitem [{\citenamefont {Cao}\ \emph {et~al.}(2015)\citenamefont {Cao},
  \citenamefont {Zhang}, \citenamefont {Lo},\ and\ \citenamefont
  {Ma}}]{bib:cao2015discrete}%
  \BibitemOpen
  \bibfield  {author} {\bibinfo {author} {\bibnamefont {Cao}, \bibfnamefont
  {Zhu}}, \bibinfo {author} {\bibfnamefont {Zhen}\ \bibnamefont {Zhang}},
  \bibinfo {author} {\bibfnamefont {Hoi-Kwong}\ \bibnamefont {Lo}}, and\
  \bibinfo {author} {\bibfnamefont {Xiongfeng}\ \bibnamefont {Ma}}} (\bibinfo
  {year} {2015}),\ \bibfield  {title} {\enquote {\bibinfo {title}
  {Discrete-phase-randomized coherent state source and its application in
  quantum key distribution},}\ }\href
  {https://doi.org/10.1088/1367-2630/17/5/053014} {\bibfield  {journal}
  {\bibinfo  {journal} {New Journal of Physics}\ }\textbf {\bibinfo {volume}
  {17}},\ \bibinfo {pages} {053014}}\BibitemShut {NoStop}%
\bibitem [{\citenamefont {Carolan}\ \emph {et~al.}(2015)\citenamefont
  {Carolan}, \citenamefont {Harrold}, \citenamefont {Sparrow}, \citenamefont
  {Mart{\'\i}n-L{\'o}pez}, \citenamefont {Russell}, \citenamefont
  {Silverstone}, \citenamefont {Shadbolt}, \citenamefont {Matsuda},
  \citenamefont {Oguma}, \citenamefont {Itoh} \emph
  {et~al.}}]{bib:carolan2015universal}%
  \BibitemOpen
  \bibfield  {author} {\bibinfo {author} {\bibnamefont {Carolan}, \bibfnamefont
  {Jacques}}, \bibinfo {author} {\bibfnamefont {Christopher}\ \bibnamefont
  {Harrold}}, \bibinfo {author} {\bibfnamefont {Chris}\ \bibnamefont
  {Sparrow}}, \bibinfo {author} {\bibfnamefont {Enrique}\ \bibnamefont
  {Mart{\'\i}n-L{\'o}pez}}, \bibinfo {author} {\bibfnamefont {Nicholas~J}\
  \bibnamefont {Russell}}, \bibinfo {author} {\bibfnamefont {Joshua~W}\
  \bibnamefont {Silverstone}}, \bibinfo {author} {\bibfnamefont {Peter~J}\
  \bibnamefont {Shadbolt}}, \bibinfo {author} {\bibfnamefont {Nobuyuki}\
  \bibnamefont {Matsuda}}, \bibinfo {author} {\bibfnamefont {Manabu}\
  \bibnamefont {Oguma}}, \bibinfo {author} {\bibfnamefont {Mikitaka}\
  \bibnamefont {Itoh}},  \emph {et~al.}} (\bibinfo {year} {2015}),\ \bibfield
  {title} {\enquote {\bibinfo {title} {Universal linear optics},}\ }\href
  {https://doi.org/10.1126/science.aab3642} {\bibfield  {journal} {\bibinfo
  {journal} {Science}\ }\textbf {\bibinfo {volume} {349}},\ \bibinfo {pages}
  {711}},\ \Eprint {https://arxiv.org/abs/arXiv:1505.01182v1}
  {arXiv:1505.01182v1} \BibitemShut {NoStop}%
\bibitem [{\citenamefont {Carrasco-Casado}\ \emph {et~al.}(2016)\citenamefont
  {Carrasco-Casado}, \citenamefont {Kunimori}, \citenamefont {Takenaka},
  \citenamefont {Kubo-Oka}, \citenamefont {Akioka}, \citenamefont {Fuse},
  \citenamefont {Koyama}, \citenamefont {Kolev}, \citenamefont {Munemasa},\
  and\ \citenamefont {Toyoshima}}]{bib:carrasco2016leo}%
  \BibitemOpen
  \bibfield  {author} {\bibinfo {author} {\bibnamefont {Carrasco-Casado},
  \bibfnamefont {Alberto}}, \bibinfo {author} {\bibfnamefont {Hiroo}\
  \bibnamefont {Kunimori}}, \bibinfo {author} {\bibfnamefont {Hideki}\
  \bibnamefont {Takenaka}}, \bibinfo {author} {\bibfnamefont {Toshihiro}\
  \bibnamefont {Kubo-Oka}}, \bibinfo {author} {\bibfnamefont {Maki}\
  \bibnamefont {Akioka}}, \bibinfo {author} {\bibfnamefont {Tetsuharu}\
  \bibnamefont {Fuse}}, \bibinfo {author} {\bibfnamefont {Yoshisada}\
  \bibnamefont {Koyama}}, \bibinfo {author} {\bibfnamefont {Dimitar}\
  \bibnamefont {Kolev}}, \bibinfo {author} {\bibfnamefont {Yasushi}\
  \bibnamefont {Munemasa}}, and\ \bibinfo {author} {\bibfnamefont {Morio}\
  \bibnamefont {Toyoshima}}} (\bibinfo {year} {2016}),\ \bibfield  {title}
  {\enquote {\bibinfo {title} {Leo-to-ground polarization measurements aiming
  for space qkd using small optical transponder (sota)},}\ }\href
  {https://doi.org/10.1364/oe.24.012254} {\bibfield  {journal} {\bibinfo
  {journal} {Optics Express}\ }\textbf {\bibinfo {volume} {24}},\ \bibinfo
  {pages} {12254}}\BibitemShut {NoStop}%
\bibitem [{\citenamefont {Chabaa}\ \emph {et~al.}(2010)\citenamefont {Chabaa},
  \citenamefont {Zeroual},\ and\ \citenamefont
  {Antari}}]{bib:chabaa2010identification}%
  \BibitemOpen
  \bibfield  {author} {\bibinfo {author} {\bibnamefont {Chabaa}, \bibfnamefont
  {Samira}}, \bibinfo {author} {\bibfnamefont {Abdelouhab}\ \bibnamefont
  {Zeroual}}, and\ \bibinfo {author} {\bibfnamefont {Jilali}\ \bibnamefont
  {Antari}}} (\bibinfo {year} {2010}),\ \bibfield  {title} {\enquote {\bibinfo
  {title} {Identification and prediction of internet traffic using artificial
  neural networks},}\ }\href {https://doi.org/10.4236/jilsa.2010.23018}
  {\bibfield  {journal} {\bibinfo  {journal} {Journal of Intelligent Learning
  Systems and Applications}\ }\textbf {\bibinfo {volume} {2}},\ \bibinfo
  {pages} {147}}\BibitemShut {NoStop}%
\bibitem [{\citenamefont {Chen}\ \emph {et~al.}(2009)\citenamefont {Chen},
  \citenamefont {Liang}, \citenamefont {Liu}, \citenamefont {Cai},
  \citenamefont {Ju}, \citenamefont {Liu}, \citenamefont {Wang}, \citenamefont
  {Yin}, \citenamefont {Chen}, \citenamefont {Chen} \emph
  {et~al.}}]{bib:OpEx17_6540}%
  \BibitemOpen
  \bibfield  {author} {\bibinfo {author} {\bibnamefont {Chen}, \bibfnamefont
  {Teng-Yun}}, \bibinfo {author} {\bibfnamefont {Hao}\ \bibnamefont {Liang}},
  \bibinfo {author} {\bibfnamefont {Yang}\ \bibnamefont {Liu}}, \bibinfo
  {author} {\bibfnamefont {Wen-Qi}\ \bibnamefont {Cai}}, \bibinfo {author}
  {\bibfnamefont {Lei}\ \bibnamefont {Ju}}, \bibinfo {author} {\bibfnamefont
  {Wei-Yue}\ \bibnamefont {Liu}}, \bibinfo {author} {\bibfnamefont {Jian}\
  \bibnamefont {Wang}}, \bibinfo {author} {\bibfnamefont {Hao}\ \bibnamefont
  {Yin}}, \bibinfo {author} {\bibfnamefont {Kai}\ \bibnamefont {Chen}},
  \bibinfo {author} {\bibfnamefont {Zeng-Bing}\ \bibnamefont {Chen}},  \emph
  {et~al.}} (\bibinfo {year} {2009}),\ \bibfield  {title} {\enquote {\bibinfo
  {title} {Field test of a practical secure communication network with
  decoy-state quantum cryptography},}\ }\href
  {https://doi.org/10.1364/oe.17.006540} {\bibfield  {journal} {\bibinfo
  {journal} {Optics Express}\ }\textbf {\bibinfo {volume} {17}},\ \bibinfo
  {pages} {6540}},\ \Eprint {https://arxiv.org/abs/arXiv:0810.1264v4}
  {arXiv:0810.1264v4} \BibitemShut {NoStop}%
\bibitem [{\citenamefont {Chen}\ \emph {et~al.}(2010)\citenamefont {Chen},
  \citenamefont {Wang}, \citenamefont {Liang}, \citenamefont {Liu},
  \citenamefont {Liu}, \citenamefont {Jiang}, \citenamefont {Wang},
  \citenamefont {Wan}, \citenamefont {Cai}, \citenamefont {Ju} \emph
  {et~al.}}]{bib:OpEx_18_27217}%
  \BibitemOpen
  \bibfield  {author} {\bibinfo {author} {\bibnamefont {Chen}, \bibfnamefont
  {Teng-Yun}}, \bibinfo {author} {\bibfnamefont {Jian}\ \bibnamefont {Wang}},
  \bibinfo {author} {\bibfnamefont {Hao}\ \bibnamefont {Liang}}, \bibinfo
  {author} {\bibfnamefont {Wei-Yue}\ \bibnamefont {Liu}}, \bibinfo {author}
  {\bibfnamefont {Yang}\ \bibnamefont {Liu}}, \bibinfo {author} {\bibfnamefont
  {Xiao}\ \bibnamefont {Jiang}}, \bibinfo {author} {\bibfnamefont {Yuan}\
  \bibnamefont {Wang}}, \bibinfo {author} {\bibfnamefont {Xu}~\bibnamefont
  {Wan}}, \bibinfo {author} {\bibfnamefont {Wen-Qi}\ \bibnamefont {Cai}},
  \bibinfo {author} {\bibfnamefont {Lei}\ \bibnamefont {Ju}},  \emph {et~al.}}
  (\bibinfo {year} {2010}),\ \bibfield  {title} {\enquote {\bibinfo {title}
  {Metropolitan all-pass and inter-city quantum communication network},}\
  }\href {https://doi.org/10.1364/oe.18.027217} {\bibfield  {journal} {\bibinfo
   {journal} {Optics Express}\ }\textbf {\bibinfo {volume} {18}},\ \bibinfo
  {pages} {27217}},\ \Eprint {https://arxiv.org/abs/arXiv:1008.1508v2}
  {arXiv:1008.1508v2} \BibitemShut {NoStop}%
\bibitem [{\citenamefont {Chen}\ \emph {et~al.}(2008)\citenamefont {Chen},
  \citenamefont {Chen}, \citenamefont {Yuan}, \citenamefont {Zhao},
  \citenamefont {Chuu}, \citenamefont {Schmiedmayer},\ and\ \citenamefont
  {Pan}}]{bib:Chen08}%
  \BibitemOpen
  \bibfield  {author} {\bibinfo {author} {\bibnamefont {Chen}, \bibfnamefont
  {Yu-Ao}}, \bibinfo {author} {\bibfnamefont {Shuai}\ \bibnamefont {Chen}},
  \bibinfo {author} {\bibfnamefont {Zhen-Sheng}\ \bibnamefont {Yuan}}, \bibinfo
  {author} {\bibfnamefont {Bo}~\bibnamefont {Zhao}}, \bibinfo {author}
  {\bibfnamefont {Chih-Sung}\ \bibnamefont {Chuu}}, \bibinfo {author}
  {\bibfnamefont {J{\" o}rg}\ \bibnamefont {Schmiedmayer}}, and\ \bibinfo
  {author} {\bibfnamefont {Jian-Wei}\ \bibnamefont {Pan}}} (\bibinfo {year}
  {2008}),\ \bibfield  {title} {\enquote {\bibinfo {title} {Memory-built-in
  quantum teleportation with photonic and atomic qubits},}\ }\href
  {https://doi.org/10.1038/nphys832} {\bibfield  {journal} {\bibinfo  {journal}
  {Nature Physics}\ }\textbf {\bibinfo {volume} {4}},\ \bibinfo {pages}
  {103}},\ \Eprint {https://arxiv.org/abs/arXiv:0705.1256v2}
  {arXiv:0705.1256v2} \BibitemShut {NoStop}%
\bibitem [{\citenamefont {Chen}\ \emph {et~al.}(2016)\citenamefont {Chen},
  \citenamefont {Wen}, \citenamefont {Geng} \emph
  {et~al.}}]{bib:chen2016predicting}%
  \BibitemOpen
  \bibfield  {author} {\bibinfo {author} {\bibnamefont {Chen}, \bibfnamefont
  {Zhitang}}, \bibinfo {author} {\bibfnamefont {Jiayao}\ \bibnamefont {Wen}},
  \bibinfo {author} {\bibfnamefont {Yanhui}\ \bibnamefont {Geng}},  \emph
  {et~al.}} (\bibinfo {year} {2016}),\ \bibfield  {title} {\enquote {\bibinfo
  {title} {Predicting future traffic using hidden markov models},}\ }in\ \href
  {https://doi.org/10.1109/icnp.2016.7785328} {\emph {\bibinfo {booktitle}
  {IEEE 24th International Conference on Network Protocols (ICNP)}}},\
  p.~\bibinfo {pages} {1}\BibitemShut {NoStop}%
\bibitem [{\citenamefont {Cheng}\ \emph {et~al.}(2018)\citenamefont {Cheng},
  \citenamefont {Chen},\ and\ \citenamefont {Wang}}]{bib:cheng2018information}%
  \BibitemOpen
  \bibfield  {author} {\bibinfo {author} {\bibnamefont {Cheng}, \bibfnamefont
  {Song}}, \bibinfo {author} {\bibfnamefont {Jing}\ \bibnamefont {Chen}}, and\
  \bibinfo {author} {\bibfnamefont {Lei}\ \bibnamefont {Wang}}} (\bibinfo
  {year} {2018}),\ \bibfield  {title} {\enquote {\bibinfo {title} {Information
  perspective to probabilistic modeling: Boltzmann machines versus born
  machines},}\ }\href {https://doi.org/10.3390/e20080583} {\bibfield  {journal}
  {\bibinfo  {journal} {Entropy}\ }\textbf {\bibinfo {volume} {20}},\ \bibinfo
  {pages} {583}},\ \Eprint {https://arxiv.org/abs/arXiv:1712.04144v1}
  {arXiv:1712.04144v1} \BibitemShut {NoStop}%
\bibitem [{\citenamefont {Chi}\ \emph {et~al.}(2011)\citenamefont {Chi},
  \citenamefont {Qi}, \citenamefont {Zhu}, \citenamefont {Qian}, \citenamefont
  {Lo}, \citenamefont {Youn}, \citenamefont {Lvovsky},\ and\ \citenamefont
  {Tian}}]{bib:chi2011balanced}%
  \BibitemOpen
  \bibfield  {author} {\bibinfo {author} {\bibnamefont {Chi}, \bibfnamefont
  {Yue-Meng}}, \bibinfo {author} {\bibfnamefont {Bing}\ \bibnamefont {Qi}},
  \bibinfo {author} {\bibfnamefont {Wen}\ \bibnamefont {Zhu}}, \bibinfo
  {author} {\bibfnamefont {Li}~\bibnamefont {Qian}}, \bibinfo {author}
  {\bibfnamefont {Hoi-Kwong}\ \bibnamefont {Lo}}, \bibinfo {author}
  {\bibfnamefont {Sun-Hyun}\ \bibnamefont {Youn}}, \bibinfo {author}
  {\bibfnamefont {AI}~\bibnamefont {Lvovsky}}, and\ \bibinfo {author}
  {\bibfnamefont {Liang}\ \bibnamefont {Tian}}} (\bibinfo {year} {2011}),\
  \bibfield  {title} {\enquote {\bibinfo {title} {A balanced homodyne detector
  for high-rate gaussian-modulated coherent-state quantum key distribution},}\
  }\href {https://doi.org/10.1088/1367-2630/13/1/013003} {\bibfield  {journal}
  {\bibinfo  {journal} {New Journal of Physics}\ }\textbf {\bibinfo {volume}
  {13}},\ \bibinfo {pages} {013003}}\BibitemShut {NoStop}%
\bibitem [{\citenamefont {Chia}\ \emph {et~al.}(2018)\citenamefont {Chia},
  \citenamefont {Lin},\ and\ \citenamefont {Wang}}]{bib:chia2018quantum}%
  \BibitemOpen
  \bibfield  {author} {\bibinfo {author} {\bibnamefont {Chia}, \bibfnamefont
  {Nai-Hui}}, \bibinfo {author} {\bibfnamefont {Han-Hsuan}\ \bibnamefont
  {Lin}}, and\ \bibinfo {author} {\bibfnamefont {Chunhao}\ \bibnamefont
  {Wang}}} (\bibinfo {year} {2018}),\ \bibfield  {title} {\enquote {\bibinfo
  {title} {Quantum-inspired sublinear classical algorithms for solving low-rank
  linear systems},}\ }\href@noop {} {\ }\Eprint
  {https://arxiv.org/abs/arXiv:1811.04852} {arXiv:1811.04852} \BibitemShut
  {NoStop}%
\bibitem [{\citenamefont {Childress}\ \emph {et~al.}(2006)\citenamefont
  {Childress}, \citenamefont {Taylor}, \citenamefont {S{\o}rensen},\ and\
  \citenamefont {Lukin}}]{bib:childress06}%
  \BibitemOpen
  \bibfield  {author} {\bibinfo {author} {\bibnamefont {Childress},
  \bibfnamefont {L}}, \bibinfo {author} {\bibfnamefont {J.~M.}\ \bibnamefont
  {Taylor}}, \bibinfo {author} {\bibfnamefont {A.~S.}\ \bibnamefont
  {S{\o}rensen}}, and\ \bibinfo {author} {\bibfnamefont {M.~D.}\ \bibnamefont
  {Lukin}}} (\bibinfo {year} {2006}),\ \bibfield  {title} {\enquote {\bibinfo
  {title} {Fault-tolerant quantum communication based on solid-state photon
  emitters},}\ }\href {https://doi.org/10.1103/physrevlett.96.070504}
  {\bibfield  {journal} {\bibinfo  {journal} {Physical Review Letters}\
  }\textbf {\bibinfo {volume} {96}},\ \bibinfo {pages} {070504}}\BibitemShut
  {NoStop}%
\bibitem [{\citenamefont {Childs}(2009{\natexlab{a}})}]{bib:Childs09}%
  \BibitemOpen
  \bibfield  {author} {\bibinfo {author} {\bibnamefont {Childs}, \bibfnamefont
  {Andrew~M}}} (\bibinfo {year} {2009}{\natexlab{a}}),\ \bibfield  {title}
  {\enquote {\bibinfo {title} {Universal computation by quantum walk},}\ }\href
  {https://doi.org/10.1103/physrevlett.102.180501} {\bibfield  {journal}
  {\bibinfo  {journal} {Physical Review Letters}\ }\textbf {\bibinfo {volume}
  {102}},\ \bibinfo {pages} {180501}},\ \Eprint
  {https://arxiv.org/abs/arXiv:0806.1972v1} {arXiv:0806.1972v1} \BibitemShut
  {NoStop}%
\bibitem [{\citenamefont {Childs}(2009{\natexlab{b}})}]{bib:Childs2009}%
  \BibitemOpen
  \bibfield  {author} {\bibinfo {author} {\bibnamefont {Childs}, \bibfnamefont
  {Andrew~M}}} (\bibinfo {year} {2009}{\natexlab{b}}),\ \bibfield  {title}
  {\enquote {\bibinfo {title} {Universal computation by quantum walk},}\ }\href
  {https://doi.org/10.1103/physrevlett.102.180501} {\bibfield  {journal}
  {\bibinfo  {journal} {Physical Review Letters}\ }\textbf {\bibinfo {volume}
  {102}},\ \bibinfo {pages} {180501}},\ \Eprint
  {https://arxiv.org/abs/arXiv:0806.1972v1} {arXiv:0806.1972v1} \BibitemShut
  {NoStop}%
\bibitem [{\citenamefont {Childs}\ \emph {et~al.}(2003)\citenamefont {Childs},
  \citenamefont {Cleve}, \citenamefont {Deotto}, \citenamefont {Farhi},
  \citenamefont {Gutmann},\ and\ \citenamefont
  {Spielman}}]{bib:childs2003exponential}%
  \BibitemOpen
  \bibfield  {author} {\bibinfo {author} {\bibnamefont {Childs}, \bibfnamefont
  {Andrew~M}}, \bibinfo {author} {\bibfnamefont {Richard}\ \bibnamefont
  {Cleve}}, \bibinfo {author} {\bibfnamefont {Enrico}\ \bibnamefont {Deotto}},
  \bibinfo {author} {\bibfnamefont {Edward}\ \bibnamefont {Farhi}}, \bibinfo
  {author} {\bibfnamefont {Sam}\ \bibnamefont {Gutmann}}, and\ \bibinfo
  {author} {\bibfnamefont {Daniel~A}\ \bibnamefont {Spielman}}} (\bibinfo
  {year} {2003}),\ \bibfield  {title} {\enquote {\bibinfo {title} {Exponential
  algorithmic speedup by a quantum walk},}\ }in\ \href
  {https://doi.org/10.1145/780551.780552} {\emph {\bibinfo {booktitle}
  {Proceedings of the thirty-fifth annual ACM symposium on Theory of
  computing}}},\ p.~\bibinfo {pages} {59},\ \Eprint
  {https://arxiv.org/abs/arXiv:quant-ph/0209131v2} {arXiv:quant-ph/0209131v2}
  \BibitemShut {NoStop}%
\bibitem [{\citenamefont {Childs}\ and\ \citenamefont
  {Goldstone}(2004)}]{childs2004spatial}%
  \BibitemOpen
  \bibfield  {author} {\bibinfo {author} {\bibnamefont {Childs}, \bibfnamefont
  {Andrew~M}}, and\ \bibinfo {author} {\bibfnamefont {Jeffrey}\ \bibnamefont
  {Goldstone}}} (\bibinfo {year} {2004}),\ \bibfield  {title} {\enquote
  {\bibinfo {title} {Spatial search by quantum walk},}\ }\href@noop {}
  {\bibfield  {journal} {\bibinfo  {journal} {Physical Review A—Atomic,
  Molecular, and Optical Physics}\ }\textbf {\bibinfo {volume} {70}}~(\bibinfo
  {number} {2}),\ \bibinfo {pages} {022314}}\BibitemShut {NoStop}%
\bibitem [{\citenamefont {Choi}\ \emph {et~al.}(2011)\citenamefont {Choi},
  \citenamefont {Young},\ and\ \citenamefont {Townsend}}]{bib:choi2011quantum}%
  \BibitemOpen
  \bibfield  {author} {\bibinfo {author} {\bibnamefont {Choi}, \bibfnamefont
  {Iris}}, \bibinfo {author} {\bibfnamefont {Robert~J}\ \bibnamefont {Young}},
  and\ \bibinfo {author} {\bibfnamefont {Paul~D}\ \bibnamefont {Townsend}}}
  (\bibinfo {year} {2011}),\ \bibfield  {title} {\enquote {\bibinfo {title}
  {Quantum information to the home},}\ }\href
  {https://doi.org/10.1364/ecoc.2011.we.10.p1.78} {\bibfield  {journal}
  {\bibinfo  {journal} {New Journal of Physics}\ }\textbf {\bibinfo {volume}
  {13}},\ \bibinfo {pages} {063039}}\BibitemShut {NoStop}%
\bibitem [{\citenamefont {Choi}\ \emph {et~al.}(2010)\citenamefont {Choi},
  \citenamefont {Goban}, \citenamefont {Papp}, \citenamefont {Van~Enk},\ and\
  \citenamefont {Kimble}}]{bib:choi2010entanglement}%
  \BibitemOpen
  \bibfield  {author} {\bibinfo {author} {\bibnamefont {Choi}, \bibfnamefont
  {KS}}, \bibinfo {author} {\bibfnamefont {A}~\bibnamefont {Goban}}, \bibinfo
  {author} {\bibfnamefont {SB}~\bibnamefont {Papp}}, \bibinfo {author}
  {\bibfnamefont {SJ}~\bibnamefont {Van~Enk}}, and\ \bibinfo {author}
  {\bibfnamefont {HJ}~\bibnamefont {Kimble}}} (\bibinfo {year} {2010}),\
  \bibfield  {title} {\enquote {\bibinfo {title} {Entanglement of spin waves
  among four quantum memories},}\ }\href {https://doi.org/10.1038/nature09568}
  {\bibfield  {journal} {\bibinfo  {journal} {Nature}\ }\textbf {\bibinfo
  {volume} {468}},\ \bibinfo {pages} {412}},\ \Eprint
  {https://arxiv.org/abs/arXiv:1007.1664v1} {arXiv:1007.1664v1} \BibitemShut
  {NoStop}%
\bibitem [{\citenamefont {Choi}\ \emph {et~al.}(2008)\citenamefont {Choi},
  \citenamefont {Deng}, \citenamefont {Laurat},\ and\ \citenamefont
  {Kimble}}]{bib:Choi2008mapping}%
  \BibitemOpen
  \bibfield  {author} {\bibinfo {author} {\bibnamefont {Choi}, \bibfnamefont
  {Kyung~Soo}}, \bibinfo {author} {\bibfnamefont {Hui}\ \bibnamefont {Deng}},
  \bibinfo {author} {\bibfnamefont {Julien}\ \bibnamefont {Laurat}}, and\
  \bibinfo {author} {\bibfnamefont {HJ}~\bibnamefont {Kimble}}} (\bibinfo
  {year} {2008}),\ \bibfield  {title} {\enquote {\bibinfo {title} {Mapping
  photonic entanglement into and out of a quantum memory},}\ }\href
  {https://doi.org/10.1364/fio.2008.ftuc3} {\bibfield  {journal} {\bibinfo
  {journal} {Nature}\ }\textbf {\bibinfo {volume} {452}},\ \bibinfo {pages}
  {67}}\BibitemShut {NoStop}%
\bibitem [{\citenamefont {Chou}\ \emph {et~al.}(2005)\citenamefont {Chou},
  \citenamefont {de~Riedmatten}, \citenamefont {Felinto}, \citenamefont
  {Polyakov}, \citenamefont {van Enk},\ and\ \citenamefont
  {Kimble}}]{bib:Chou05}%
  \BibitemOpen
  \bibfield  {author} {\bibinfo {author} {\bibnamefont {Chou}, \bibfnamefont
  {C~W}}, \bibinfo {author} {\bibfnamefont {H.}~\bibnamefont {de~Riedmatten}},
  \bibinfo {author} {\bibfnamefont {D.}~\bibnamefont {Felinto}}, \bibinfo
  {author} {\bibfnamefont {S.~V.}\ \bibnamefont {Polyakov}}, \bibinfo {author}
  {\bibfnamefont {S.~J.}\ \bibnamefont {van Enk}}, and\ \bibinfo {author}
  {\bibfnamefont {H.~J.}\ \bibnamefont {Kimble}}} (\bibinfo {year} {2005}),\
  \bibfield  {title} {\enquote {\bibinfo {title} {Measurement-induced
  entanglement for excitation stored in remote atomic ensembles},}\ }\href
  {https://doi.org/10.1038/nature04353} {\bibfield  {journal} {\bibinfo
  {journal} {Nature}\ }\textbf {\bibinfo {volume} {438}},\ \bibinfo {pages}
  {828}},\ \Eprint {https://arxiv.org/abs/arXiv:quant-ph/0510055v1}
  {arXiv:quant-ph/0510055v1} \BibitemShut {NoStop}%
\bibitem [{\citenamefont {Chou}\ \emph {et~al.}(2007)\citenamefont {Chou},
  \citenamefont {Laurat}, \citenamefont {Deng}, \citenamefont {Choi},
  \citenamefont {De~Riedmatten}, \citenamefont {Felinto},\ and\ \citenamefont
  {Kimble}}]{bib:Sc_316_1316}%
  \BibitemOpen
  \bibfield  {author} {\bibinfo {author} {\bibnamefont {Chou}, \bibfnamefont
  {Chin-Wen}}, \bibinfo {author} {\bibfnamefont {Julien}\ \bibnamefont
  {Laurat}}, \bibinfo {author} {\bibfnamefont {Hui}\ \bibnamefont {Deng}},
  \bibinfo {author} {\bibfnamefont {Kyung~Soo}\ \bibnamefont {Choi}}, \bibinfo
  {author} {\bibfnamefont {Hugues}\ \bibnamefont {De~Riedmatten}}, \bibinfo
  {author} {\bibfnamefont {Daniel}\ \bibnamefont {Felinto}}, and\ \bibinfo
  {author} {\bibfnamefont {H~Jeff}\ \bibnamefont {Kimble}}} (\bibinfo {year}
  {2007}),\ \bibfield  {title} {\enquote {\bibinfo {title} {Functional quantum
  nodes for entanglement distribution over scalable quantum networks},}\ }\href
  {https://doi.org/10.1126/science.1140300} {\bibfield  {journal} {\bibinfo
  {journal} {Science}\ }\textbf {\bibinfo {volume} {316}},\ \bibinfo {pages}
  {1316}},\ \Eprint {https://arxiv.org/abs/arXiv:quant-ph/0702057v2}
  {arXiv:quant-ph/0702057v2} \BibitemShut {NoStop}%
\bibitem [{\citenamefont {Chow}\ \emph {et~al.}(2011)\citenamefont {Chow},
  \citenamefont {C{\'o}rcoles}, \citenamefont {Gambetta}, \citenamefont
  {Rigetti}, \citenamefont {Johnson}, \citenamefont {Smolin}, \citenamefont
  {Rozen}, \citenamefont {Keefe}, \citenamefont {Rothwell}, \citenamefont
  {Ketchen} \emph {et~al.}}]{bib:chow2011simple}%
  \BibitemOpen
  \bibfield  {author} {\bibinfo {author} {\bibnamefont {Chow}, \bibfnamefont
  {Jerry~M}}, \bibinfo {author} {\bibfnamefont {AD}~\bibnamefont
  {C{\'o}rcoles}}, \bibinfo {author} {\bibfnamefont {Jay~M}\ \bibnamefont
  {Gambetta}}, \bibinfo {author} {\bibfnamefont {Chad}\ \bibnamefont
  {Rigetti}}, \bibinfo {author} {\bibfnamefont {BR}~\bibnamefont {Johnson}},
  \bibinfo {author} {\bibfnamefont {John~A}\ \bibnamefont {Smolin}}, \bibinfo
  {author} {\bibfnamefont {JR}~\bibnamefont {Rozen}}, \bibinfo {author}
  {\bibfnamefont {George~A}\ \bibnamefont {Keefe}}, \bibinfo {author}
  {\bibfnamefont {Mary~B}\ \bibnamefont {Rothwell}}, \bibinfo {author}
  {\bibfnamefont {Mark~B}\ \bibnamefont {Ketchen}},  \emph {et~al.}} (\bibinfo
  {year} {2011}),\ \bibfield  {title} {\enquote {\bibinfo {title} {Simple
  all-microwave entangling gate for fixed-frequency superconducting qubits},}\
  }\href {https://doi.org/10.1103/physrevlett.107.080502} {\bibfield  {journal}
  {\bibinfo  {journal} {Physical Review Letters}\ }\textbf {\bibinfo {volume}
  {107}},\ \bibinfo {pages} {080502}},\ \Eprint
  {https://arxiv.org/abs/arXiv:1106.0553v1} {arXiv:1106.0553v1} \BibitemShut
  {NoStop}%
\bibitem [{\citenamefont {Chow}\ \emph {et~al.}(2013)\citenamefont {Chow},
  \citenamefont {Gambetta}, \citenamefont {Cross}, \citenamefont {Merkel},
  \citenamefont {Rigetti},\ and\ \citenamefont
  {Steffen}}]{bib:chow2013microwave}%
  \BibitemOpen
  \bibfield  {author} {\bibinfo {author} {\bibnamefont {Chow}, \bibfnamefont
  {Jerry~M}}, \bibinfo {author} {\bibfnamefont {Jay~M}\ \bibnamefont
  {Gambetta}}, \bibinfo {author} {\bibfnamefont {Andrew~W}\ \bibnamefont
  {Cross}}, \bibinfo {author} {\bibfnamefont {Seth~T}\ \bibnamefont {Merkel}},
  \bibinfo {author} {\bibfnamefont {Chad}\ \bibnamefont {Rigetti}}, and\
  \bibinfo {author} {\bibfnamefont {M}~\bibnamefont {Steffen}}} (\bibinfo
  {year} {2013}),\ \bibfield  {title} {\enquote {\bibinfo {title}
  {Microwave-activated conditional-phase gate for superconducting qubits},}\
  }\href {https://doi.org/10.1088/1367-2630/15/11/115012} {\bibfield  {journal}
  {\bibinfo  {journal} {New Journal of Physics}\ }\textbf {\bibinfo {volume}
  {15}},\ \bibinfo {pages} {115012}},\ \Eprint
  {https://arxiv.org/abs/arXiv:1307.2594v1} {arXiv:1307.2594v1} \BibitemShut
  {NoStop}%
\bibitem [{\citenamefont {Christandl}\ and\ \citenamefont
  {Wehner}(2005{\natexlab{a}})}]{christandl2005quantum}%
  \BibitemOpen
  \bibfield  {author} {\bibinfo {author} {\bibnamefont {Christandl},
  \bibfnamefont {Matthias}}, and\ \bibinfo {author} {\bibfnamefont {Stephanie}\
  \bibnamefont {Wehner}}} (\bibinfo {year} {2005}{\natexlab{a}}),\ \bibfield
  {title} {\enquote {\bibinfo {title} {Quantum anonymous transmissions},}\ }in\
  \href@noop {} {\emph {\bibinfo {booktitle} {International conference on the
  theory and application of cryptology and information security}}}\ (\bibinfo
  {organization} {Springer})\ pp.\ \bibinfo {pages} {217--235}\BibitemShut
  {NoStop}%
\bibitem [{\citenamefont {Christandl}\ and\ \citenamefont
  {Wehner}(2005{\natexlab{b}})}]{bib:christandl2005quantum}%
  \BibitemOpen
  \bibfield  {author} {\bibinfo {author} {\bibnamefont {Christandl},
  \bibfnamefont {Matthias}}, and\ \bibinfo {author} {\bibfnamefont {Stephanie}\
  \bibnamefont {Wehner}}} (\bibinfo {year} {2005}{\natexlab{b}}),\ \bibfield
  {title} {\enquote {\bibinfo {title} {Quantum anonymous transmissions},}\ }in\
  \href {https://doi.org/10.1007/11593447_12} {\emph {\bibinfo {booktitle}
  {International Conference on the Theory and Application of Cryptology and
  Information Security}}},\ p.\ \bibinfo {pages} {217},\ \Eprint
  {https://arxiv.org/abs/arXiv:quant-ph/0409201v2} {arXiv:quant-ph/0409201v2}
  \BibitemShut {NoStop}%
\bibitem [{\citenamefont {Chuang}\ and\ \citenamefont
  {Nielsen}(1997)}]{bib:ChuangNielsen97}%
  \BibitemOpen
  \bibfield  {author} {\bibinfo {author} {\bibnamefont {Chuang}, \bibfnamefont
  {I~L}}, and\ \bibinfo {author} {\bibfnamefont {M.~A.}\ \bibnamefont
  {Nielsen}}} (\bibinfo {year} {1997}),\ \bibfield  {title} {\enquote {\bibinfo
  {title} {Prescription for experimental determination of the dynamics of a
  quantum black box},}\ }\href {https://doi.org/10.1080/095003497152609}
  {\bibfield  {journal} {\bibinfo  {journal} {Journal of Modern Optics}\
  }\textbf {\bibinfo {volume} {44}},\ \bibinfo {pages} {2455}}\BibitemShut
  {NoStop}%
\bibitem [{\citenamefont {Chuang}(2000)}]{bib:chuang2000quantum}%
  \BibitemOpen
  \bibfield  {author} {\bibinfo {author} {\bibnamefont {Chuang}, \bibfnamefont
  {Isaac~L}}} (\bibinfo {year} {2000}),\ \bibfield  {title} {\enquote {\bibinfo
  {title} {Quantum algorithm for distributed clock synchronization},}\ }\href
  {https://doi.org/10.1103/physrevlett.85.2006} {\bibfield  {journal} {\bibinfo
   {journal} {Physical Review Letters}\ }\textbf {\bibinfo {volume} {85}},\
  \bibinfo {pages} {2006}},\ \Eprint
  {https://arxiv.org/abs/arXiv:quant-ph/0005092v1} {arXiv:quant-ph/0005092v1}
  \BibitemShut {NoStop}%
\bibitem [{\citenamefont {Chui}(2016)}]{chui2016introduction}%
  \BibitemOpen
  \bibfield  {author} {\bibinfo {author} {\bibnamefont {Chui}, \bibfnamefont
  {Charles~K}}} (\bibinfo {year} {2016}),\ \href@noop {} {\emph {\bibinfo
  {title} {An introduction to wavelets}}}\ (\bibinfo  {publisher}
  {Elsevier})\BibitemShut {NoStop}%
\bibitem [{\citenamefont {Ciliberto}\ \emph {et~al.}(2018)\citenamefont
  {Ciliberto}, \citenamefont {Herbster}, \citenamefont {Ialongo}, \citenamefont
  {Pontil}, \citenamefont {Rocchetto}, \citenamefont {Severini},\ and\
  \citenamefont {Wossnig}}]{bib:ciliberto2018quantum}%
  \BibitemOpen
  \bibfield  {author} {\bibinfo {author} {\bibnamefont {Ciliberto},
  \bibfnamefont {Carlo}}, \bibinfo {author} {\bibfnamefont {Mark}\ \bibnamefont
  {Herbster}}, \bibinfo {author} {\bibfnamefont {Alessandro~Davide}\
  \bibnamefont {Ialongo}}, \bibinfo {author} {\bibfnamefont {Massimiliano}\
  \bibnamefont {Pontil}}, \bibinfo {author} {\bibfnamefont {Andrea}\
  \bibnamefont {Rocchetto}}, \bibinfo {author} {\bibfnamefont {Simone}\
  \bibnamefont {Severini}}, and\ \bibinfo {author} {\bibfnamefont {Leonard}\
  \bibnamefont {Wossnig}}} (\bibinfo {year} {2018}),\ \bibfield  {title}
  {\enquote {\bibinfo {title} {Quantum machine learning: a classical
  perspective},}\ }\href {https://doi.org/10.1098/rspa.2017.0551} {\bibfield
  {journal} {\bibinfo  {journal} {Proceedings of the Royal Society A:
  Mathematical, Physical and Engineering Sciences}\ }\textbf {\bibinfo {volume}
  {474}},\ \bibinfo {pages} {20170551}},\ \Eprint
  {https://arxiv.org/abs/arXiv:1707.08561v3} {arXiv:1707.08561v3} \BibitemShut
  {NoStop}%
\bibitem [{\citenamefont {Cirac}\ \emph {et~al.}(1999)\citenamefont {Cirac},
  \citenamefont {Ekert}, \citenamefont {Huelga},\ and\ \citenamefont
  {Macchiavello}}]{bib:Cirac99}%
  \BibitemOpen
  \bibfield  {author} {\bibinfo {author} {\bibnamefont {Cirac}, \bibfnamefont
  {J~I}}, \bibinfo {author} {\bibfnamefont {A.~K.}\ \bibnamefont {Ekert}},
  \bibinfo {author} {\bibfnamefont {S.~F.}\ \bibnamefont {Huelga}}, and\
  \bibinfo {author} {\bibfnamefont {C.}~\bibnamefont {Macchiavello}}} (\bibinfo
  {year} {1999}),\ \bibfield  {title} {\enquote {\bibinfo {title} {Distributed
  quantum computation over noisy channels},}\ }\href
  {https://doi.org/10.1103/physreva.59.4249} {\bibfield  {journal} {\bibinfo
  {journal} {Physical Review A}\ }\textbf {\bibinfo {volume} {59}},\ \bibinfo
  {pages} {1999}},\ \Eprint {https://arxiv.org/abs/arXiv:quant-ph/9803017v2}
  {arXiv:quant-ph/9803017v2} \BibitemShut {NoStop}%
\bibitem [{\citenamefont {Clausen}\ \emph {et~al.}(2011)\citenamefont
  {Clausen}, \citenamefont {Usmani}, \citenamefont {Bussi{\`e}res},
  \citenamefont {Sangouard}, \citenamefont {Afzelius}, \citenamefont
  {de~Riedmatten},\ and\ \citenamefont {Gisin}}]{bib:Nat_469_508}%
  \BibitemOpen
  \bibfield  {author} {\bibinfo {author} {\bibnamefont {Clausen}, \bibfnamefont
  {Christoph}}, \bibinfo {author} {\bibfnamefont {Imam}\ \bibnamefont
  {Usmani}}, \bibinfo {author} {\bibfnamefont {F{\'e}lix}\ \bibnamefont
  {Bussi{\`e}res}}, \bibinfo {author} {\bibfnamefont {Nicolas}\ \bibnamefont
  {Sangouard}}, \bibinfo {author} {\bibfnamefont {Mikael}\ \bibnamefont
  {Afzelius}}, \bibinfo {author} {\bibfnamefont {Hugues}\ \bibnamefont
  {de~Riedmatten}}, and\ \bibinfo {author} {\bibfnamefont {Nicolas}\
  \bibnamefont {Gisin}}} (\bibinfo {year} {2011}),\ \bibfield  {title}
  {\enquote {\bibinfo {title} {Quantum storage of photonic entanglement in a
  crystal},}\ }\href {https://doi.org/10.1038/nature09662} {\bibfield
  {journal} {\bibinfo  {journal} {Nature}\ }\textbf {\bibinfo {volume} {469}},\
  \bibinfo {pages} {508}},\ \Eprint {https://arxiv.org/abs/arXiv:1009.0489v3}
  {arXiv:1009.0489v3} \BibitemShut {NoStop}%
\bibitem [{\citenamefont {Clauser}\ \emph {et~al.}(1969)\citenamefont
  {Clauser}, \citenamefont {Horne}, \citenamefont {Shimony},\ and\
  \citenamefont {Holt}}]{SD-Clauser:1969aa}%
  \BibitemOpen
  \bibfield  {author} {\bibinfo {author} {\bibnamefont {Clauser}, \bibfnamefont
  {John~F}}, \bibinfo {author} {\bibfnamefont {Michael~A.}\ \bibnamefont
  {Horne}}, \bibinfo {author} {\bibfnamefont {Abner}\ \bibnamefont {Shimony}},
  and\ \bibinfo {author} {\bibfnamefont {Richard~A.}\ \bibnamefont {Holt}}}
  (\bibinfo {year} {1969}),\ \bibfield  {title} {\enquote {\bibinfo {title}
  {Proposed experiment to test local hidden-variable theories},}\ }\href
  {https://doi.org/10.1103/PhysRevLett.23.880} {\bibfield  {journal} {\bibinfo
  {journal} {Physical Review Letters}\ }\textbf {\bibinfo {volume} {23}},\
  \bibinfo {pages} {880}}\BibitemShut {NoStop}%
\bibitem [{\citenamefont {Cleve}\ \emph {et~al.}(1999)\citenamefont {Cleve},
  \citenamefont {Gottesman},\ and\ \citenamefont {Lo}}]{bib:cleve1999share}%
  \BibitemOpen
  \bibfield  {author} {\bibinfo {author} {\bibnamefont {Cleve}, \bibfnamefont
  {Richard}}, \bibinfo {author} {\bibfnamefont {Daniel}\ \bibnamefont
  {Gottesman}}, and\ \bibinfo {author} {\bibfnamefont {Hoi-Kwong}\ \bibnamefont
  {Lo}}} (\bibinfo {year} {1999}),\ \bibfield  {title} {\enquote {\bibinfo
  {title} {How to share a quantum secret},}\ }\href
  {https://doi.org/10.1103/physrevlett.83.648} {\bibfield  {journal} {\bibinfo
  {journal} {Physical Review Letters}\ }\textbf {\bibinfo {volume} {83}},\
  \bibinfo {pages} {648}},\ \Eprint
  {https://arxiv.org/abs/arXiv:quant-ph/9901025v1} {arXiv:quant-ph/9901025v1}
  \BibitemShut {NoStop}%
\bibitem [{\citenamefont {Cohen}\ and\ \citenamefont
  {Havlin}(2003)}]{bib:PhysRevLett.90.058701}%
  \BibitemOpen
  \bibfield  {author} {\bibinfo {author} {\bibnamefont {Cohen}, \bibfnamefont
  {Reuven}}, and\ \bibinfo {author} {\bibfnamefont {Shlomo}\ \bibnamefont
  {Havlin}}} (\bibinfo {year} {2003}),\ \bibfield  {title} {\enquote {\bibinfo
  {title} {Scale-free networks are ultrasmall},}\ }\href
  {https://doi.org/10.1103/physrevlett.90.058701} {\bibfield  {journal}
  {\bibinfo  {journal} {Physical Review Letters}\ }\textbf {\bibinfo {volume}
  {90}},\ \bibinfo {pages} {058701}},\ \Eprint
  {https://arxiv.org/abs/arXiv:cond-mat/0205476v2} {arXiv:cond-mat/0205476v2}
  \BibitemShut {NoStop}%
\bibitem [{\citenamefont {Cohen-Tannoudji}\ \emph {et~al.}(1992)\citenamefont
  {Cohen-Tannoudji}, \citenamefont {Dupont-Roc},\ and\ \citenamefont
  {Grynberg}}]{bib:Cohen-Tannoudji92}%
  \BibitemOpen
  \bibfield  {author} {\bibinfo {author} {\bibnamefont {Cohen-Tannoudji},
  \bibfnamefont {Claude}}, \bibinfo {author} {\bibfnamefont {Jacques}\
  \bibnamefont {Dupont-Roc}}, and\ \bibinfo {author} {\bibfnamefont {Gilbert}\
  \bibnamefont {Grynberg}}} (\bibinfo {year} {1992}),\ \href
  {https://doi.org/10.1063/1.2809840} {\emph {\bibinfo {title} {Atom-Photon
  Interactions: Basic Processes and Applications}}},\ \bibinfo {edition} {1st}\
  ed.\ (\bibinfo  {publisher} {Wiley-Interscience})\BibitemShut {NoStop}%
\bibitem [{\citenamefont {Collins}\ \emph {et~al.}(2007)\citenamefont
  {Collins}, \citenamefont {Jenkins}, \citenamefont {Kuzmich},\ and\
  \citenamefont {Kennedy}}]{bib:PRL_98_060502}%
  \BibitemOpen
  \bibfield  {author} {\bibinfo {author} {\bibnamefont {Collins}, \bibfnamefont
  {OA}}, \bibinfo {author} {\bibfnamefont {SD}~\bibnamefont {Jenkins}},
  \bibinfo {author} {\bibfnamefont {A}~\bibnamefont {Kuzmich}}, and\ \bibinfo
  {author} {\bibfnamefont {TAB}\ \bibnamefont {Kennedy}}} (\bibinfo {year}
  {2007}),\ \bibfield  {title} {\enquote {\bibinfo {title} {Multiplexed
  memory-insensitive quantum repeaters},}\ }\href
  {https://doi.org/10.1103/physrevlett.98.060502} {\bibfield  {journal}
  {\bibinfo  {journal} {Physical Review Letters}\ }\textbf {\bibinfo {volume}
  {98}},\ \bibinfo {pages} {060502}},\ \Eprint
  {https://arxiv.org/abs/arXiv:quant-ph/0610036v2} {arXiv:quant-ph/0610036v2}
  \BibitemShut {NoStop}%
\bibitem [{\citenamefont {Compeau}\ \emph {et~al.}(2011)\citenamefont
  {Compeau}, \citenamefont {Pevzner},\ and\ \citenamefont
  {Tesler}}]{compeau2011apply}%
  \BibitemOpen
  \bibfield  {author} {\bibinfo {author} {\bibnamefont {Compeau}, \bibfnamefont
  {Phillip~EC}}, \bibinfo {author} {\bibfnamefont {Pavel~A}\ \bibnamefont
  {Pevzner}}, and\ \bibinfo {author} {\bibfnamefont {Glenn}\ \bibnamefont
  {Tesler}}} (\bibinfo {year} {2011}),\ \bibfield  {title} {\enquote {\bibinfo
  {title} {How to apply de bruijn graphs to genome assembly},}\ }\href@noop {}
  {\bibfield  {journal} {\bibinfo  {journal} {Nature biotechnology}\ }\textbf
  {\bibinfo {volume} {29}}~(\bibinfo {number} {11}),\ \bibinfo {pages}
  {987--991}}\BibitemShut {NoStop}%
\bibitem [{\citenamefont {Conitzer}\ and\ \citenamefont
  {Sandholm}(2004)}]{bib:conitzer2004communication}%
  \BibitemOpen
  \bibfield  {author} {\bibinfo {author} {\bibnamefont {Conitzer},
  \bibfnamefont {Vincent}}, and\ \bibinfo {author} {\bibfnamefont {Tuomas}\
  \bibnamefont {Sandholm}}} (\bibinfo {year} {2004}),\ \bibfield  {title}
  {\enquote {\bibinfo {title} {Communication complexity as a lower bound for
  learning in games},}\ }in\ \href {https://doi.org/10.1145/1015330.1015351}
  {\emph {\bibinfo {booktitle} {ACM Proceedings of the twenty-first
  international conference on Machine learning}}},\ p.~\bibinfo {pages}
  {24}\BibitemShut {NoStop}%
\bibitem [{\citenamefont {Consortium}\ \emph {et~al.}(2010)\citenamefont
  {Consortium} \emph {et~al.}}]{10002010map}%
  \BibitemOpen
  \bibfield  {author} {\bibinfo {author} {\bibnamefont {Consortium},
  \bibfnamefont {1000 Genomes~Project}},  \emph {et~al.}} (\bibinfo {year}
  {2010}),\ \bibfield  {title} {\enquote {\bibinfo {title} {A map of human
  genome variation from population scale sequencing},}\ }\href@noop {}
  {\bibfield  {journal} {\bibinfo  {journal} {Nature}\ }\textbf {\bibinfo
  {volume} {467}}~(\bibinfo {number} {7319}),\ \bibinfo {pages}
  {1061}}\BibitemShut {NoStop}%
\bibitem [{\citenamefont {Cormen}\ \emph {et~al.}(2009)\citenamefont {Cormen},
  \citenamefont {Leiserson}, \citenamefont {Rivest},\ and\ \citenamefont
  {Stein}}]{bib:RivestAlgBook}%
  \BibitemOpen
  \bibfield  {author} {\bibinfo {author} {\bibnamefont {Cormen}, \bibfnamefont
  {Thomas~H}}, \bibinfo {author} {\bibfnamefont {Charles~E.}\ \bibnamefont
  {Leiserson}}, \bibinfo {author} {\bibfnamefont {Ronald~L.}\ \bibnamefont
  {Rivest}}, and\ \bibinfo {author} {\bibfnamefont {Clifford}\ \bibnamefont
  {Stein}}} (\bibinfo {year} {2009}),\ \href@noop {} {\emph {\bibinfo {title}
  {Introduction to Algorithms}}}\ (\bibinfo  {publisher} {MIT
  Press})\BibitemShut {NoStop}%
\bibitem [{\citenamefont {Cortez}\ \emph {et~al.}(2006)\citenamefont {Cortez},
  \citenamefont {Rio}, \citenamefont {Rocha},\ and\ \citenamefont
  {Sousa}}]{bib:cortez2006internet}%
  \BibitemOpen
  \bibfield  {author} {\bibinfo {author} {\bibnamefont {Cortez}, \bibfnamefont
  {Paulo}}, \bibinfo {author} {\bibfnamefont {Miguel}\ \bibnamefont {Rio}},
  \bibinfo {author} {\bibfnamefont {Miguel}\ \bibnamefont {Rocha}}, and\
  \bibinfo {author} {\bibfnamefont {Pedro}\ \bibnamefont {Sousa}}} (\bibinfo
  {year} {2006}),\ \bibfield  {title} {\enquote {\bibinfo {title} {Internet
  traffic forecasting using neural networks},}\ }in\ \href
  {https://doi.org/10.1109/ijcnn.2006.247142} {\emph {\bibinfo {booktitle}
  {IEEE International Joint Conference on Neural Networks (IJCNN'06)}}},\ p.\
  \bibinfo {pages} {2635}\BibitemShut {NoStop}%
\bibitem [{\citenamefont {Craig}\ and\ \citenamefont
  {Martin}(2021)}]{Gidney:2021}%
  \BibitemOpen
  \bibfield  {author} {\bibinfo {author} {\bibnamefont {Craig}, \bibfnamefont
  {Gidney}}, and\ \bibinfo {author} {\bibfnamefont {Ekerå}\ \bibnamefont
  {Martin}}} (\bibinfo {year} {2021}),\ \bibfield  {title} {\enquote {\bibinfo
  {title} {How to factor 2048 bit rsa integers in 8 hours using 20 million
  noisy qubits},}\ }\href {https://doi.org/10.22331/q-2021-04-15-433}
  {\bibfield  {journal} {\bibinfo  {journal} {Quantum}\ }\textbf {\bibinfo
  {volume} {5}},\ \bibinfo {pages} {433}}\BibitemShut {NoStop}%
\bibitem [{\citenamefont {Crespi}\ \emph {et~al.}(2012)\citenamefont {Crespi},
  \citenamefont {Osellame}, \citenamefont {Ramponi}, \citenamefont {Brod},
  \citenamefont {Galvao}, \citenamefont {Spagnolo}, \citenamefont {Vitelli},
  \citenamefont {Maiorino}, \citenamefont {Mataloni},\ and\ \citenamefont
  {Sciarrino}}]{bib:Crespi3}%
  \BibitemOpen
  \bibfield  {author} {\bibinfo {author} {\bibnamefont {Crespi}, \bibfnamefont
  {A}}, \bibinfo {author} {\bibfnamefont {R.}~\bibnamefont {Osellame}},
  \bibinfo {author} {\bibfnamefont {R.}~\bibnamefont {Ramponi}}, \bibinfo
  {author} {\bibfnamefont {D.~J.}\ \bibnamefont {Brod}}, \bibinfo {author}
  {\bibfnamefont {E.~F.}\ \bibnamefont {Galvao}}, \bibinfo {author}
  {\bibfnamefont {N.}~\bibnamefont {Spagnolo}}, \bibinfo {author}
  {\bibfnamefont {C.}~\bibnamefont {Vitelli}}, \bibinfo {author} {\bibfnamefont
  {E.}~\bibnamefont {Maiorino}}, \bibinfo {author} {\bibfnamefont
  {P.}~\bibnamefont {Mataloni}}, and\ \bibinfo {author} {\bibfnamefont
  {F.}~\bibnamefont {Sciarrino}}} (\bibinfo {year} {2012}),\ \bibfield  {title}
  {\enquote {\bibinfo {title} {Experimental boson sampling in arbitrary
  integrated photonic circuits},}\ }\href@noop {} {\bibfield  {journal}
  {\bibinfo  {journal} {Nature Photonics}\ }\textbf {\bibinfo {volume} {7}},\
  \bibinfo {pages} {545}},\ \Eprint {https://arxiv.org/abs/arXiv:1212.2783v1}
  {arXiv:1212.2783v1} \BibitemShut {NoStop}%
\bibitem [{\citenamefont {Curty}\ and\ \citenamefont
  {Santos}(2001)}]{SD-Curty:2001aa}%
  \BibitemOpen
  \bibfield  {author} {\bibinfo {author} {\bibnamefont {Curty}, \bibfnamefont
  {Marcos}}, and\ \bibinfo {author} {\bibfnamefont {David~J.}\ \bibnamefont
  {Santos}}} (\bibinfo {year} {2001}),\ \bibfield  {title} {\enquote {\bibinfo
  {title} {Quantum authentication of classical messages},}\ }\href
  {https://doi.org/10.1103/PhysRevA.64.062309} {\bibfield  {journal} {\bibinfo
  {journal} {Physical Review A}\ }\textbf {\bibinfo {volume} {64}},\ \bibinfo
  {pages} {062309}},\ \Eprint {https://arxiv.org/abs/arXiv:quant-ph/0103122v2}
  {arXiv:quant-ph/0103122v2} \BibitemShut {NoStop}%
\bibitem [{\citenamefont {Dahlberg}\ and\ \citenamefont
  {Wehner}(2018)}]{bib:AxelDahlbergQron}%
  \BibitemOpen
  \bibfield  {author} {\bibinfo {author} {\bibnamefont {Dahlberg},
  \bibfnamefont {Axel}}, and\ \bibinfo {author} {\bibfnamefont {Stephanie}\
  \bibnamefont {Wehner}}} (\bibinfo {year} {2018}),\ \bibfield  {title}
  {\enquote {\bibinfo {title} {Simulaqron - a simulator for developing quantum
  internet software},}\ }\href {https://doi.org/10.1088/2058-9565/aad56e}
  {\bibfield  {journal} {\bibinfo  {journal} {Quantum Science and Technology}\
  }\textbf {\bibinfo {volume} {4}},\ \bibinfo {pages} {015001}}\BibitemShut
  {NoStop}%
\bibitem [{\citenamefont {Dai}\ \emph {et~al.}(2016)\citenamefont {Dai},
  \citenamefont {Yang}, \citenamefont {Reingruber}, \citenamefont {Xu},
  \citenamefont {Jiang}, \citenamefont {Chen}, \citenamefont {Yuan},\ and\
  \citenamefont {Pan}}]{bib:dai2016generation}%
  \BibitemOpen
  \bibfield  {author} {\bibinfo {author} {\bibnamefont {Dai}, \bibfnamefont
  {Han-Ning}}, \bibinfo {author} {\bibfnamefont {Bing}\ \bibnamefont {Yang}},
  \bibinfo {author} {\bibfnamefont {Andreas}\ \bibnamefont {Reingruber}},
  \bibinfo {author} {\bibfnamefont {Xiao-Fan}\ \bibnamefont {Xu}}, \bibinfo
  {author} {\bibfnamefont {Xiao}\ \bibnamefont {Jiang}}, \bibinfo {author}
  {\bibfnamefont {Yu-Ao}\ \bibnamefont {Chen}}, \bibinfo {author}
  {\bibfnamefont {Zhen-Sheng}\ \bibnamefont {Yuan}}, and\ \bibinfo {author}
  {\bibfnamefont {Jian-Wei}\ \bibnamefont {Pan}}} (\bibinfo {year} {2016}),\
  \bibfield  {title} {\enquote {\bibinfo {title} {Generation and detection of
  atomic spin entanglement in optical lattices},}\ }\href
  {https://doi.org/10.1038/nphys3705} {\bibfield  {journal} {\bibinfo
  {journal} {Nature Physics}\ }\textbf {\bibinfo {volume} {12}},\ \bibinfo
  {pages} {783}},\ \Eprint {https://arxiv.org/abs/arXiv:1507.05937v1}
  {arXiv:1507.05937v1} \BibitemShut {NoStop}%
\bibitem [{\citenamefont {Dawson}\ \emph {et~al.}(2006)\citenamefont {Dawson},
  \citenamefont {Haselgrove},\ and\ \citenamefont {Nielsen}}]{bib:Dawson06}%
  \BibitemOpen
  \bibfield  {author} {\bibinfo {author} {\bibnamefont {Dawson}, \bibfnamefont
  {Christopher~M}}, \bibinfo {author} {\bibfnamefont {Henry~L.}\ \bibnamefont
  {Haselgrove}}, and\ \bibinfo {author} {\bibfnamefont {Michael~A.}\
  \bibnamefont {Nielsen}}} (\bibinfo {year} {2006}),\ \bibfield  {title}
  {\enquote {\bibinfo {title} {Noise thresholds for optical cluster-state
  quantum computation},}\ }\href {https://doi.org/10.1103/physreva.73.052306}
  {\bibfield  {journal} {\bibinfo  {journal} {Physical Review A}\ }\textbf
  {\bibinfo {volume} {73}},\ \bibinfo {pages} {052306}},\ \Eprint
  {https://arxiv.org/abs/arXiv:quant-ph/0601066v3} {arXiv:quant-ph/0601066v3}
  \BibitemShut {NoStop}%
\bibitem [{\citenamefont {De~Greve}\ \emph {et~al.}(2012)\citenamefont
  {De~Greve}, \citenamefont {Yu}, \citenamefont {McMahon}, \citenamefont
  {Pelc}, \citenamefont {Natarajan}, \citenamefont {Kim}, \citenamefont {Abe},
  \citenamefont {Maier}, \citenamefont {Schneider}, \citenamefont {Kamp} \emph
  {et~al.}}]{bib:de2012quantum}%
  \BibitemOpen
  \bibfield  {author} {\bibinfo {author} {\bibnamefont {De~Greve},
  \bibfnamefont {Kristiaan}}, \bibinfo {author} {\bibfnamefont {Leo}\
  \bibnamefont {Yu}}, \bibinfo {author} {\bibfnamefont {Peter~L}\ \bibnamefont
  {McMahon}}, \bibinfo {author} {\bibfnamefont {Jason~S}\ \bibnamefont {Pelc}},
  \bibinfo {author} {\bibfnamefont {Chandra~M}\ \bibnamefont {Natarajan}},
  \bibinfo {author} {\bibfnamefont {Na~Young}\ \bibnamefont {Kim}}, \bibinfo
  {author} {\bibfnamefont {Eisuke}\ \bibnamefont {Abe}}, \bibinfo {author}
  {\bibfnamefont {Sebastian}\ \bibnamefont {Maier}}, \bibinfo {author}
  {\bibfnamefont {Christian}\ \bibnamefont {Schneider}}, \bibinfo {author}
  {\bibfnamefont {Martin}\ \bibnamefont {Kamp}},  \emph {et~al.}} (\bibinfo
  {year} {2012}),\ \bibfield  {title} {\enquote {\bibinfo {title} {Quantum-dot
  spin-photon entanglement via frequency downconversion to telecom
  wavelength},}\ }\href {https://doi.org/10.1038/nature11577} {\bibfield
  {journal} {\bibinfo  {journal} {Nature}\ }\textbf {\bibinfo {volume} {491}},\
  \bibinfo {pages} {421}}\BibitemShut {NoStop}%
\bibitem [{\citenamefont {Dean}\ and\ \citenamefont
  {Ghemawat}(2008)}]{bib:MapReduce}%
  \BibitemOpen
  \bibfield  {author} {\bibinfo {author} {\bibnamefont {Dean}, \bibfnamefont
  {Jeffrey}}, and\ \bibinfo {author} {\bibfnamefont {Sanjay}\ \bibnamefont
  {Ghemawat}}} (\bibinfo {year} {2008}),\ \bibfield  {title} {\enquote
  {\bibinfo {title} {Mapreduce: simplified data processing on large
  clusters},}\ }\href {https://doi.org/10.1145/1327452.1327492} {\bibfield
  {journal} {\bibinfo  {journal} {Communications of the ACM}\ }\textbf
  {\bibinfo {volume} {51}},\ \bibinfo {pages} {107}}\BibitemShut {NoStop}%
\bibitem [{\citenamefont {Demkowicz-Dobrza{\'n}ski}\ \emph
  {et~al.}(2015)\citenamefont {Demkowicz-Dobrza{\'n}ski}, \citenamefont
  {Jarzyna},\ and\ \citenamefont {Ko{\l}ody{\'n}ski}}]{demkowicz2015quantum}%
  \BibitemOpen
  \bibfield  {author} {\bibinfo {author} {\bibnamefont
  {Demkowicz-Dobrza{\'n}ski}, \bibfnamefont {Rafal}}, \bibinfo {author}
  {\bibfnamefont {Marcin}\ \bibnamefont {Jarzyna}}, and\ \bibinfo {author}
  {\bibfnamefont {Jan}\ \bibnamefont {Ko{\l}ody{\'n}ski}}} (\bibinfo {year}
  {2015}),\ \bibfield  {title} {\enquote {\bibinfo {title} {Quantum limits in
  optical interferometry},}\ }\href@noop {} {\bibfield  {journal} {\bibinfo
  {journal} {Progress in Optics}\ }\textbf {\bibinfo {volume} {60}},\ \bibinfo
  {pages} {345--435}}\BibitemShut {NoStop}%
\bibitem [{\citenamefont {Deutsch}\ \emph {et~al.}(1996)\citenamefont
  {Deutsch}, \citenamefont {Ekert}, \citenamefont {Jozsa}, \citenamefont
  {Macchiavello}, \citenamefont {Popescu},\ and\ \citenamefont
  {Sanpera}}]{bib:Deutsch96}%
  \BibitemOpen
  \bibfield  {author} {\bibinfo {author} {\bibnamefont {Deutsch}, \bibfnamefont
  {D}}, \bibinfo {author} {\bibfnamefont {A.}~\bibnamefont {Ekert}}, \bibinfo
  {author} {\bibfnamefont {R.}~\bibnamefont {Jozsa}}, \bibinfo {author}
  {\bibfnamefont {C.}~\bibnamefont {Macchiavello}}, \bibinfo {author}
  {\bibfnamefont {S.}~\bibnamefont {Popescu}}, and\ \bibinfo {author}
  {\bibfnamefont {A.}~\bibnamefont {Sanpera}}} (\bibinfo {year} {1996}),\
  \bibfield  {title} {\enquote {\bibinfo {title} {Quantum privacy amplification
  and the security of quantum cryptography over noisy channels},}\ }\href
  {https://doi.org/10.1103/physrevlett.77.2818} {\bibfield  {journal} {\bibinfo
   {journal} {Physical Review Letters}\ }\textbf {\bibinfo {volume} {77}},\
  \bibinfo {pages} {2818}},\ \Eprint
  {https://arxiv.org/abs/arXiv:quant-ph/9604039v1} {arXiv:quant-ph/9604039v1}
  \BibitemShut {NoStop}%
\bibitem [{\citenamefont {Deutsch}(1985)}]{bib:Deutsch85}%
  \BibitemOpen
  \bibfield  {author} {\bibinfo {author} {\bibnamefont {Deutsch}, \bibfnamefont
  {David}}} (\bibinfo {year} {1985}),\ \bibfield  {title} {\enquote {\bibinfo
  {title} {Quantum theory, the church-turing principle and the universal
  quantum computer},}\ }\href {https://doi.org/10.1098/rspa.1985.0070}
  {\bibfield  {journal} {\bibinfo  {journal} {Proceedings of the Royal Society
  of London A}\ }\textbf {\bibinfo {volume} {400}},\ \bibinfo {pages}
  {97}}\BibitemShut {NoStop}%
\bibitem [{\citenamefont {Deutsch}\ and\ \citenamefont
  {Jozsa}(1992)}]{bib:DeutschJozsa92}%
  \BibitemOpen
  \bibfield  {author} {\bibinfo {author} {\bibnamefont {Deutsch}, \bibfnamefont
  {David}}, and\ \bibinfo {author} {\bibfnamefont {Richard}\ \bibnamefont
  {Jozsa}}} (\bibinfo {year} {1992}),\ \bibfield  {title} {\enquote {\bibinfo
  {title} {Rapid solution of problems by quantum computation},}\ }\href
  {https://doi.org/10.1098/rspa.1992.0167} {\bibfield  {journal} {\bibinfo
  {journal} {Proceedings of the Royal Society of London A}\ }\textbf {\bibinfo
  {volume} {439}},\ \bibinfo {pages} {553}}\BibitemShut {NoStop}%
\bibitem [{\citenamefont {Devitt}\ \emph {et~al.}(2016)\citenamefont {Devitt},
  \citenamefont {Greentree}, \citenamefont {Stephens},\ and\ \citenamefont
  {Van~Meter}}]{SD-Devitt:2016aa}%
  \BibitemOpen
  \bibfield  {author} {\bibinfo {author} {\bibnamefont {Devitt}, \bibfnamefont
  {Simon~J}}, \bibinfo {author} {\bibfnamefont {Andrew~D.}\ \bibnamefont
  {Greentree}}, \bibinfo {author} {\bibfnamefont {Ashley~M.}\ \bibnamefont
  {Stephens}}, and\ \bibinfo {author} {\bibfnamefont {Rodney}\ \bibnamefont
  {Van~Meter}}} (\bibinfo {year} {2016}),\ \bibfield  {title} {\enquote
  {\bibinfo {title} {High-speed quantum networking by ship},}\ }\href
  {https://doi.org/10.1038/srep36163} {\bibfield  {journal} {\bibinfo
  {journal} {Scientific Reports}\ }\textbf {\bibinfo {volume} {6}},\ \bibinfo
  {pages} {36163}}\BibitemShut {NoStop}%
\bibitem [{\citenamefont {Devitt}\ \emph
  {et~al.}(2013{\natexlab{a}})\citenamefont {Devitt}, \citenamefont {Munro},\
  and\ \citenamefont {Nemoto}}]{bib:devitt2013}%
  \BibitemOpen
  \bibfield  {author} {\bibinfo {author} {\bibnamefont {Devitt}, \bibfnamefont
  {Simon~J}}, \bibinfo {author} {\bibfnamefont {William~J.}\ \bibnamefont
  {Munro}}, and\ \bibinfo {author} {\bibfnamefont {Kae}\ \bibnamefont
  {Nemoto}}} (\bibinfo {year} {2013}{\natexlab{a}}),\ \bibfield  {title}
  {\enquote {\bibinfo {title} {Quantum error correction for beginners},}\
  }\href {https://doi.org/10.1088/0034-4885/76/7/076001} {\bibfield  {journal}
  {\bibinfo  {journal} {Reports on Progress in Physics}\ }\textbf {\bibinfo
  {volume} {76}},\ \bibinfo {pages} {076001}}\BibitemShut {NoStop}%
\bibitem [{\citenamefont {Devitt}\ \emph
  {et~al.}(2013{\natexlab{b}})\citenamefont {Devitt}, \citenamefont {Munro},\
  and\ \citenamefont {Nemoto}}]{SD-Devitt:2013aa}%
  \BibitemOpen
  \bibfield  {author} {\bibinfo {author} {\bibnamefont {Devitt}, \bibfnamefont
  {Simon~J}}, \bibinfo {author} {\bibfnamefont {William~J}\ \bibnamefont
  {Munro}}, and\ \bibinfo {author} {\bibfnamefont {Kae}\ \bibnamefont
  {Nemoto}}} (\bibinfo {year} {2013}{\natexlab{b}}),\ \bibfield  {title}
  {\enquote {\bibinfo {title} {Quantum error correction for beginners},}\
  }\href {https://doi.org/10.1088/0034-4885/76/7/076001} {\bibfield  {journal}
  {\bibinfo  {journal} {Reports on Progress in Physics}\ }\textbf {\bibinfo
  {volume} {76}},\ \bibinfo {pages} {076001}},\ \Eprint
  {https://arxiv.org/abs/arXiv:0905.2794v4} {arXiv:0905.2794v4} \BibitemShut
  {NoStop}%
\bibitem [{\citenamefont {Devitt}\ \emph {et~al.}(2011)\citenamefont {Devitt},
  \citenamefont {Munro},\ and\ \citenamefont {Nemoto}}]{SD-Devitt2011}%
  \BibitemOpen
  \bibfield  {author} {\bibinfo {author} {\bibnamefont {Devitt}, \bibfnamefont
  {SJ}}, \bibinfo {author} {\bibfnamefont {W.J.}\ \bibnamefont {Munro}}, and\
  \bibinfo {author} {\bibfnamefont {K.}~\bibnamefont {Nemoto}}} (\bibinfo
  {year} {2011}),\ \bibfield  {title} {\enquote {\bibinfo {title} {{High
  Performance Quantum Computing}},}\ }\href
  {https://doi.org/10.2201/niipi.2011.8.6} {\bibfield  {journal} {\bibinfo
  {journal} {Progress in Informatics}\ }\textbf {\bibinfo {volume} {8}},\
  \bibinfo {pages} {49}}\BibitemShut {NoStop}%
\bibitem [{\citenamefont {Devoret}\ \emph {et~al.}(2004)\citenamefont
  {Devoret}, \citenamefont {Wallraff},\ and\ \citenamefont
  {Martinis}}]{bib:devoret2004superconducting}%
  \BibitemOpen
  \bibfield  {author} {\bibinfo {author} {\bibnamefont {Devoret}, \bibfnamefont
  {Michel~H}}, \bibinfo {author} {\bibfnamefont {Andreas}\ \bibnamefont
  {Wallraff}}, and\ \bibinfo {author} {\bibfnamefont {John~M}\ \bibnamefont
  {Martinis}}} (\bibinfo {year} {2004}),\ \bibfield  {title} {\enquote
  {\bibinfo {title} {Superconducting qubits: A short review},}\ }\href@noop {}
  {\ }\Eprint {https://arxiv.org/abs/arXiv:cond-mat/0411174}
  {arXiv:cond-mat/0411174} \BibitemShut {NoStop}%
\bibitem [{\citenamefont {Diamanti}\ \emph {et~al.}(2016)\citenamefont
  {Diamanti}, \citenamefont {Lo}, \citenamefont {Qi},\ and\ \citenamefont
  {Yuan}}]{bib:diamanti2016practical}%
  \BibitemOpen
  \bibfield  {author} {\bibinfo {author} {\bibnamefont {Diamanti},
  \bibfnamefont {Eleni}}, \bibinfo {author} {\bibfnamefont {Hoi-Kwong}\
  \bibnamefont {Lo}}, \bibinfo {author} {\bibfnamefont {Bing}\ \bibnamefont
  {Qi}}, and\ \bibinfo {author} {\bibfnamefont {Zhiliang}\ \bibnamefont
  {Yuan}}} (\bibinfo {year} {2016}),\ \bibfield  {title} {\enquote {\bibinfo
  {title} {Practical challenges in quantum key distribution},}\ }\href
  {https://doi.org/10.1038/npjqi.2016.25} {\bibfield  {journal} {\bibinfo
  {journal} {NPJ Quantum Information}\ }\textbf {\bibinfo {volume} {2}},\
  \bibinfo {pages} {16025}},\ \Eprint
  {https://arxiv.org/abs/arXiv:1606.05853v1} {arXiv:1606.05853v1} \BibitemShut
  {NoStop}%
\bibitem [{\citenamefont {Didier}\ \emph {et~al.}(2014)\citenamefont {Didier},
  \citenamefont {Pugnetti}, \citenamefont {Blanter},\ and\ \citenamefont
  {Fazio}}]{bib:didier2014quantum}%
  \BibitemOpen
  \bibfield  {author} {\bibinfo {author} {\bibnamefont {Didier}, \bibfnamefont
  {Nicolas}}, \bibinfo {author} {\bibfnamefont {Stefano}\ \bibnamefont
  {Pugnetti}}, \bibinfo {author} {\bibfnamefont {Yaroslav~M}\ \bibnamefont
  {Blanter}}, and\ \bibinfo {author} {\bibfnamefont {Rosario}\ \bibnamefont
  {Fazio}}} (\bibinfo {year} {2014}),\ \bibfield  {title} {\enquote {\bibinfo
  {title} {Quantum transducer in circuit optomechanics},}\ }\href
  {https://doi.org/10.1016/j.ssc.2014.02.029} {\bibfield  {journal} {\bibinfo
  {journal} {Solid State Communications}\ }\textbf {\bibinfo {volume} {198}},\
  \bibinfo {pages} {61}},\ \Eprint {https://arxiv.org/abs/arXiv:1201.6293v1}
  {arXiv:1201.6293v1} \BibitemShut {NoStop}%
\bibitem [{\citenamefont {Diffie}\ and\ \citenamefont
  {Hellman}(2022)}]{diffie2022new}%
  \BibitemOpen
  \bibfield  {author} {\bibinfo {author} {\bibnamefont {Diffie}, \bibfnamefont
  {Whitfield}}, and\ \bibinfo {author} {\bibfnamefont {Martin~E}\ \bibnamefont
  {Hellman}}} (\bibinfo {year} {2022}),\ \bibfield  {title} {\enquote {\bibinfo
  {title} {New directions in cryptography},}\ }in\ \href@noop {} {\emph
  {\bibinfo {booktitle} {Democratizing Cryptography: The Work of Whitfield
  Diffie and Martin Hellman}}},\ pp.\ \bibinfo {pages} {365--390}\BibitemShut
  {NoStop}%
\bibitem [{\citenamefont {Dijkstra}(1959)}]{bib:Dijkstra59}%
  \BibitemOpen
  \bibfield  {author} {\bibinfo {author} {\bibnamefont {Dijkstra},
  \bibfnamefont {E~W}}} (\bibinfo {year} {1959}),\ \bibfield  {title} {\enquote
  {\bibinfo {title} {A note on two problems in connection with graphs},}\
  }\href {https://doi.org/10.1007/bf01386390} {\bibfield  {journal} {\bibinfo
  {journal} {Numerische Mathematik}\ }\textbf {\bibinfo {volume} {1}},\
  \bibinfo {pages} {269}}\BibitemShut {NoStop}%
\bibitem [{\citenamefont {Ding}\ \emph {et~al.}(2013)\citenamefont {Ding},
  \citenamefont {Zhou}, \citenamefont {Shi},\ and\ \citenamefont
  {Guo}}]{bib:ding2013single}%
  \BibitemOpen
  \bibfield  {author} {\bibinfo {author} {\bibnamefont {Ding}, \bibfnamefont
  {Dong-Sheng}}, \bibinfo {author} {\bibfnamefont {Zhi-Yuan}\ \bibnamefont
  {Zhou}}, \bibinfo {author} {\bibfnamefont {Bao-Sen}\ \bibnamefont {Shi}},
  and\ \bibinfo {author} {\bibfnamefont {Guang-Can}\ \bibnamefont {Guo}}}
  (\bibinfo {year} {2013}),\ \bibfield  {title} {\enquote {\bibinfo {title}
  {Single-photon-level quantum image memory based on cold atomic ensembles},}\
  }\href {https://doi.org/10.1038/ncomms3527} {\bibfield  {journal} {\bibinfo
  {journal} {Nature Communications}\ }\textbf {\bibinfo {volume} {4}},\
  \bibinfo {pages} {2527}},\ \Eprint {https://arxiv.org/abs/arXiv:1305.2675v3}
  {arXiv:1305.2675v3} \BibitemShut {NoStop}%
\bibitem [{\citenamefont {Ding}\ \emph {et~al.}(2016)\citenamefont {Ding},
  \citenamefont {He}, \citenamefont {Duan}, \citenamefont {Gregersen},
  \citenamefont {Chen}, \citenamefont {Unsleber}, \citenamefont {Maier},
  \citenamefont {Schneider}, \citenamefont {Kamp}, \citenamefont {H{\"o}fling}
  \emph {et~al.}}]{bib:ding2016on}%
  \BibitemOpen
  \bibfield  {author} {\bibinfo {author} {\bibnamefont {Ding}, \bibfnamefont
  {Xing}}, \bibinfo {author} {\bibfnamefont {Yu}~\bibnamefont {He}}, \bibinfo
  {author} {\bibfnamefont {Z-C}\ \bibnamefont {Duan}}, \bibinfo {author}
  {\bibfnamefont {Niels}\ \bibnamefont {Gregersen}}, \bibinfo {author}
  {\bibfnamefont {M-C}\ \bibnamefont {Chen}}, \bibinfo {author} {\bibfnamefont
  {S}~\bibnamefont {Unsleber}}, \bibinfo {author} {\bibfnamefont {Sebastian}\
  \bibnamefont {Maier}}, \bibinfo {author} {\bibfnamefont {Christian}\
  \bibnamefont {Schneider}}, \bibinfo {author} {\bibfnamefont {Martin}\
  \bibnamefont {Kamp}}, \bibinfo {author} {\bibfnamefont {Sven}\ \bibnamefont
  {H{\"o}fling}},  \emph {et~al.}} (\bibinfo {year} {2016}),\ \bibfield
  {title} {\enquote {\bibinfo {title} {On-demand single photons with high
  extraction efficiency and near-unity indistinguishability from a resonantly
  driven quantum dot in a micropillar},}\ }\href
  {https://doi.org/10.1103/physrevlett.116.020401} {\bibfield  {journal}
  {\bibinfo  {journal} {Physical Review Letters}\ }\textbf {\bibinfo {volume}
  {116}},\ \bibinfo {pages} {020401}},\ \Eprint
  {https://arxiv.org/abs/arXiv:1601.00284v2} {arXiv:1601.00284v2} \BibitemShut
  {NoStop}%
\bibitem [{\citenamefont {DiVincenzo}\ \emph {et~al.}(2004)\citenamefont
  {DiVincenzo}, \citenamefont {Horodecki}, \citenamefont {Leung}, \citenamefont
  {Smolin},\ and\ \citenamefont {Terhal}}]{DiVin}%
  \BibitemOpen
  \bibfield  {author} {\bibinfo {author} {\bibnamefont {DiVincenzo},
  \bibfnamefont {David~P}}, \bibinfo {author} {\bibfnamefont {Micha\l{}}\
  \bibnamefont {Horodecki}}, \bibinfo {author} {\bibfnamefont {Debbie~W.}\
  \bibnamefont {Leung}}, \bibinfo {author} {\bibfnamefont {John~A.}\
  \bibnamefont {Smolin}}, and\ \bibinfo {author} {\bibfnamefont {Barbara~M.}\
  \bibnamefont {Terhal}}} (\bibinfo {year} {2004}),\ \bibfield  {title}
  {\enquote {\bibinfo {title} {Locking classical correlations in quantum
  states},}\ }\href {https://doi.org/10.1103/PhysRevLett.92.067902} {\bibfield
  {journal} {\bibinfo  {journal} {Phys. Rev. Lett.}\ }\textbf {\bibinfo
  {volume} {92}},\ \bibinfo {pages} {067902}}\BibitemShut {NoStop}%
\bibitem [{\citenamefont {DiVincenzo}\ \emph {et~al.}(1998)\citenamefont
  {DiVincenzo}, \citenamefont {Shor},\ and\ \citenamefont
  {Smolin}}]{bib:PhysRevA.57.830}%
  \BibitemOpen
  \bibfield  {author} {\bibinfo {author} {\bibnamefont {DiVincenzo},
  \bibfnamefont {David~P}}, \bibinfo {author} {\bibfnamefont {Peter~W.}\
  \bibnamefont {Shor}}, and\ \bibinfo {author} {\bibfnamefont {John~A.}\
  \bibnamefont {Smolin}}} (\bibinfo {year} {1998}),\ \bibfield  {title}
  {\enquote {\bibinfo {title} {Quantum-channel capacity of very noisy
  channels},}\ }\href {https://doi.org/10.1103/PhysRevA.57.830} {\bibfield
  {journal} {\bibinfo  {journal} {Phys. Rev. A}\ }\textbf {\bibinfo {volume}
  {57}},\ \bibinfo {pages} {830--839}}\BibitemShut {NoStop}%
\bibitem [{\citenamefont {Doherty}\ \emph {et~al.}(2013)\citenamefont
  {Doherty}, \citenamefont {Manson}, \citenamefont {Delaney}, \citenamefont
  {Jelezko}, \citenamefont {Wrachtrup},\ and\ \citenamefont
  {Hollenberg}}]{bib:doherty2013nitrogen}%
  \BibitemOpen
  \bibfield  {author} {\bibinfo {author} {\bibnamefont {Doherty}, \bibfnamefont
  {Marcus~W}}, \bibinfo {author} {\bibfnamefont {Neil~B}\ \bibnamefont
  {Manson}}, \bibinfo {author} {\bibfnamefont {Paul}\ \bibnamefont {Delaney}},
  \bibinfo {author} {\bibfnamefont {Fedor}\ \bibnamefont {Jelezko}}, \bibinfo
  {author} {\bibfnamefont {J{\"o}rg}\ \bibnamefont {Wrachtrup}}, and\ \bibinfo
  {author} {\bibfnamefont {Lloyd~CL}\ \bibnamefont {Hollenberg}}} (\bibinfo
  {year} {2013}),\ \bibfield  {title} {\enquote {\bibinfo {title} {The
  nitrogen-vacancy colour centre in diamond},}\ }\href
  {https://doi.org/10.1016/j.physrep.2013.02.001} {\bibfield  {journal}
  {\bibinfo  {journal} {Physics Reports}\ }\textbf {\bibinfo {volume} {528}},\
  \bibinfo {pages} {1}},\ \Eprint {https://arxiv.org/abs/arXiv:1302.3288v1}
  {arXiv:1302.3288v1} \BibitemShut {NoStop}%
\bibitem [{\citenamefont {Dolde}\ \emph {et~al.}(2013)\citenamefont {Dolde},
  \citenamefont {Jakobi}, \citenamefont {Naydenov}, \citenamefont {Zhao},
  \citenamefont {Pezzagna}, \citenamefont {Trautmann}, \citenamefont {Meijer},
  \citenamefont {Neumann}, \citenamefont {Jelezko},\ and\ \citenamefont
  {Wrachtrup}}]{bib:dolde2013room}%
  \BibitemOpen
  \bibfield  {author} {\bibinfo {author} {\bibnamefont {Dolde}, \bibfnamefont
  {Florian}}, \bibinfo {author} {\bibfnamefont {Ingmar}\ \bibnamefont
  {Jakobi}}, \bibinfo {author} {\bibfnamefont {Boris}\ \bibnamefont
  {Naydenov}}, \bibinfo {author} {\bibfnamefont {Nan}\ \bibnamefont {Zhao}},
  \bibinfo {author} {\bibfnamefont {Sebastien}\ \bibnamefont {Pezzagna}},
  \bibinfo {author} {\bibfnamefont {Christina}\ \bibnamefont {Trautmann}},
  \bibinfo {author} {\bibfnamefont {Jan}\ \bibnamefont {Meijer}}, \bibinfo
  {author} {\bibfnamefont {Philipp}\ \bibnamefont {Neumann}}, \bibinfo {author}
  {\bibfnamefont {Fedor}\ \bibnamefont {Jelezko}}, and\ \bibinfo {author}
  {\bibfnamefont {J{\"o}rg}\ \bibnamefont {Wrachtrup}}} (\bibinfo {year}
  {2013}),\ \bibfield  {title} {\enquote {\bibinfo {title} {Room-temperature
  entanglement between single defect spins in diamond},}\ }\href
  {https://doi.org/10.1038/nphys2545} {\bibfield  {journal} {\bibinfo
  {journal} {Nature Physics}\ }\textbf {\bibinfo {volume} {9}},\ \bibinfo
  {pages} {139}}\BibitemShut {NoStop}%
\bibitem [{\citenamefont {Dowling}(2008)}]{bib:Dowling08}%
  \BibitemOpen
  \bibfield  {author} {\bibinfo {author} {\bibnamefont {Dowling}, \bibfnamefont
  {Jonathan~P}}} (\bibinfo {year} {2008}),\ \bibfield  {title} {\enquote
  {\bibinfo {title} {Quantum optical metrology - the lowdown on high-noon
  states},}\ }\href@noop {} {\bibfield  {journal} {\bibinfo  {journal}
  {Contemporary Physics}\ }\textbf {\bibinfo {volume} {49}},\ \bibinfo {pages}
  {125}}\BibitemShut {NoStop}%
\bibitem [{\citenamefont {Dowling}(2020)}]{dowling2020schrodinger}%
  \BibitemOpen
  \bibfield  {author} {\bibinfo {author} {\bibnamefont {Dowling}, \bibfnamefont
  {Jonathan~P}}} (\bibinfo {year} {2020}),\ \href@noop {} {\emph {\bibinfo
  {title} {Schr{\"o}dinger’s Web: Race to Build the Quantum Internet}}}\
  (\bibinfo  {publisher} {CRC Press})\BibitemShut {NoStop}%
\bibitem [{\citenamefont {Duan}\ \emph {et~al.}(2013)\citenamefont {Duan},
  \citenamefont {Jian}, \citenamefont {Chao}, \citenamefont {Peng},\ and\
  \citenamefont {Gui-Hua}}]{bib:duan2013}%
  \BibitemOpen
  \bibfield  {author} {\bibinfo {author} {\bibnamefont {Duan}, \bibfnamefont
  {Huang}}, \bibinfo {author} {\bibfnamefont {Fang}\ \bibnamefont {Jian}},
  \bibinfo {author} {\bibfnamefont {Wang}\ \bibnamefont {Chao}}, \bibinfo
  {author} {\bibfnamefont {Huang}\ \bibnamefont {Peng}}, and\ \bibinfo {author}
  {\bibfnamefont {Zeng}\ \bibnamefont {Gui-Hua}}} (\bibinfo {year} {2013}),\
  \bibfield  {title} {\enquote {\bibinfo {title} {A 300-mhz bandwidth balanced
  homodyne detector for continuous variable quantum key distribution},}\ }\href
  {https://doi.org/10.1088/0256-307x/30/11/114209} {\bibfield  {journal}
  {\bibinfo  {journal} {Chinese Physics Letters}\ }\textbf {\bibinfo {volume}
  {30}},\ \bibinfo {pages} {114209}}\BibitemShut {NoStop}%
\bibitem [{\citenamefont {Duan}(2002)}]{bib:Duan02}%
  \BibitemOpen
  \bibfield  {author} {\bibinfo {author} {\bibnamefont {Duan}, \bibfnamefont
  {L-M}}} (\bibinfo {year} {2002}),\ \bibfield  {title} {\enquote {\bibinfo
  {title} {Entangling many atomic ensembles through laser manipulation},}\
  }\href {https://doi.org/10.1103/physrevlett.88.170402} {\bibfield  {journal}
  {\bibinfo  {journal} {Physical Review Letters}\ }\textbf {\bibinfo {volume}
  {88}},\ \bibinfo {pages} {170402}},\ \Eprint
  {https://arxiv.org/abs/arXiv:quant-ph/0201128v2} {arXiv:quant-ph/0201128v2}
  \BibitemShut {NoStop}%
\bibitem [{\citenamefont {Duan}\ \emph {et~al.}(2000)\citenamefont {Duan},
  \citenamefont {Giedke}, \citenamefont {Cirac},\ and\ \citenamefont
  {Zoller}}]{bib:Duan00}%
  \BibitemOpen
  \bibfield  {author} {\bibinfo {author} {\bibnamefont {Duan}, \bibfnamefont
  {L-M}}, \bibinfo {author} {\bibfnamefont {G.}~\bibnamefont {Giedke}},
  \bibinfo {author} {\bibfnamefont {J.~I.}\ \bibnamefont {Cirac}}, and\
  \bibinfo {author} {\bibfnamefont {P.}~\bibnamefont {Zoller}}} (\bibinfo
  {year} {2000}),\ \bibfield  {title} {\enquote {\bibinfo {title} {Entanglement
  purification of gaussian continuous variable quantum states},}\ }\href
  {https://doi.org/10.1103/physrevlett.84.4002} {\bibfield  {journal} {\bibinfo
   {journal} {Physical Review Letters}\ }\textbf {\bibinfo {volume} {84}},\
  \bibinfo {pages} {4002}},\ \Eprint
  {https://arxiv.org/abs/arXiv:quant-ph/9912017v2} {arXiv:quant-ph/9912017v2}
  \BibitemShut {NoStop}%
\bibitem [{\citenamefont {Duan}\ \emph
  {et~al.}(2001{\natexlab{a}})\citenamefont {Duan}, \citenamefont {Lukin},
  \citenamefont {Cirac},\ and\ \citenamefont {Zoller}}]{bib:Duan01}%
  \BibitemOpen
  \bibfield  {author} {\bibinfo {author} {\bibnamefont {Duan}, \bibfnamefont
  {L-M}}, \bibinfo {author} {\bibfnamefont {M.~D.}\ \bibnamefont {Lukin}},
  \bibinfo {author} {\bibfnamefont {J.~I.}\ \bibnamefont {Cirac}}, and\
  \bibinfo {author} {\bibfnamefont {P.}~\bibnamefont {Zoller}}} (\bibinfo
  {year} {2001}{\natexlab{a}}),\ \bibfield  {title} {\enquote {\bibinfo {title}
  {Long-distance quantum communication with atomic ensembles and linear
  optics},}\ }\href {https://doi.org/10.1038/35106500} {\bibfield  {journal}
  {\bibinfo  {journal} {Nature}\ }\textbf {\bibinfo {volume} {414}},\ \bibinfo
  {pages} {413}},\ \Eprint {https://arxiv.org/abs/arXiv:quant-ph/0105105v1}
  {arXiv:quant-ph/0105105v1} \BibitemShut {NoStop}%
\bibitem [{\citenamefont {Duan}\ \emph
  {et~al.}(2001{\natexlab{b}})\citenamefont {Duan}, \citenamefont {Lukin},
  \citenamefont {Cirac},\ and\ \citenamefont {Zoller}}]{bib:DLCZ}%
  \BibitemOpen
  \bibfield  {author} {\bibinfo {author} {\bibnamefont {Duan}, \bibfnamefont
  {L-M}}, \bibinfo {author} {\bibfnamefont {MD}~\bibnamefont {Lukin}}, \bibinfo
  {author} {\bibfnamefont {J~Ignacio}\ \bibnamefont {Cirac}}, and\ \bibinfo
  {author} {\bibfnamefont {Peter}\ \bibnamefont {Zoller}}} (\bibinfo {year}
  {2001}{\natexlab{b}}),\ \bibfield  {title} {\enquote {\bibinfo {title}
  {Long-distance quantum communication with atomic ensembles and linear
  optics},}\ }\href {https://doi.org/10.1038/35106500} {\bibfield  {journal}
  {\bibinfo  {journal} {Nature}\ }\textbf {\bibinfo {volume} {414}},\ \bibinfo
  {pages} {413}},\ \Eprint {https://arxiv.org/abs/arXiv:quant-ph/0105105v1}
  {arXiv:quant-ph/0105105v1} \BibitemShut {NoStop}%
\bibitem [{\citenamefont {Duan}\ \emph {et~al.}(2006)\citenamefont {Duan},
  \citenamefont {Madsen}, \citenamefont {Moehring}, \citenamefont {Maunz},
  \citenamefont {Jr.},\ and\ \citenamefont {Monroe}}]{bib:Duan06}%
  \BibitemOpen
  \bibfield  {author} {\bibinfo {author} {\bibnamefont {Duan}, \bibfnamefont
  {L-M}}, \bibinfo {author} {\bibfnamefont {M.~J.}\ \bibnamefont {Madsen}},
  \bibinfo {author} {\bibfnamefont {D.~L.}\ \bibnamefont {Moehring}}, \bibinfo
  {author} {\bibfnamefont {P.}~\bibnamefont {Maunz}}, \bibinfo {author}
  {\bibfnamefont {R.~N.~Kohn}\ \bibnamefont {Jr.}}, and\ \bibinfo {author}
  {\bibfnamefont {C.}~\bibnamefont {Monroe}}} (\bibinfo {year} {2006}),\
  \bibfield  {title} {\enquote {\bibinfo {title} {Probabilistic quantum gates
  between remote atoms through interference of optical frequency qubits},}\
  }\href {https://doi.org/10.1103/physreva.73.062324} {\bibfield  {journal}
  {\bibinfo  {journal} {Physical Review A}\ }\textbf {\bibinfo {volume} {73}},\
  \bibinfo {pages} {062324}},\ \Eprint
  {https://arxiv.org/abs/arXiv:quant-ph/0603285v1} {arXiv:quant-ph/0603285v1}
  \BibitemShut {NoStop}%
\bibitem [{\citenamefont {Duan}\ and\ \citenamefont
  {Raussendorf}(2005)}]{bib:Duan05}%
  \BibitemOpen
  \bibfield  {author} {\bibinfo {author} {\bibnamefont {Duan}, \bibfnamefont
  {L-M}}, and\ \bibinfo {author} {\bibfnamefont {R.}~\bibnamefont
  {Raussendorf}}} (\bibinfo {year} {2005}),\ \bibfield  {title} {\enquote
  {\bibinfo {title} {Efficient quantum computation with probabilistic quantum
  gates},}\ }\href {https://doi.org/10.1103/physrevlett.95.080503} {\bibfield
  {journal} {\bibinfo  {journal} {Physical Review Letters}\ }\textbf {\bibinfo
  {volume} {95}},\ \bibinfo {pages} {080503}},\ \Eprint
  {https://arxiv.org/abs/arXiv:quant-ph/0502120v1} {arXiv:quant-ph/0502120v1}
  \BibitemShut {NoStop}%
\bibitem [{\citenamefont {Duivenvoorden}\ \emph {et~al.}(2017)\citenamefont
  {Duivenvoorden}, \citenamefont {Terhal},\ and\ \citenamefont
  {Weigand}}]{terhal2017sensor}%
  \BibitemOpen
  \bibfield  {author} {\bibinfo {author} {\bibnamefont {Duivenvoorden},
  \bibfnamefont {Kasper}}, \bibinfo {author} {\bibfnamefont {Barbara~M.}\
  \bibnamefont {Terhal}}, and\ \bibinfo {author} {\bibfnamefont {Daniel}\
  \bibnamefont {Weigand}}} (\bibinfo {year} {2017}),\ \bibfield  {title}
  {\enquote {\bibinfo {title} {Single-mode displacement sensor},}\ }\href
  {https://doi.org/10.1103/PhysRevA.95.012305} {\bibfield  {journal} {\bibinfo
  {journal} {Phys. Rev. A}\ }\textbf {\bibinfo {volume} {95}},\ \bibinfo
  {pages} {012305}}\BibitemShut {NoStop}%
\bibitem [{\citenamefont {Dulek}\ \emph {et~al.}(2016)\citenamefont {Dulek},
  \citenamefont {Schaffner},\ and\ \citenamefont
  {Speelman}}]{dulek2016quantum}%
  \BibitemOpen
  \bibfield  {author} {\bibinfo {author} {\bibnamefont {Dulek}, \bibfnamefont
  {Yfke}}, \bibinfo {author} {\bibfnamefont {Christian}\ \bibnamefont
  {Schaffner}}, and\ \bibinfo {author} {\bibfnamefont {Florian}\ \bibnamefont
  {Speelman}}} (\bibinfo {year} {2016}),\ \bibfield  {title} {\enquote
  {\bibinfo {title} {Quantum homomorphic encryption for polynomial-sized
  circuits},}\ }in\ \href@noop {} {\emph {\bibinfo {booktitle} {Advances in
  Cryptology--CRYPTO 2016: 36th Annual International Cryptology Conference,
  Santa Barbara, CA, USA, August 14-18, 2016, Proceedings, Part III 36}}}\
  (\bibinfo {organization} {Springer})\ pp.\ \bibinfo {pages}
  {3--32}\BibitemShut {NoStop}%
\bibitem [{\citenamefont {Dunjko}\ \emph {et~al.}(2012)\citenamefont {Dunjko},
  \citenamefont {Kashefi},\ and\ \citenamefont
  {Leverrier}}]{bib:PhysRevLett.108.200502}%
  \BibitemOpen
  \bibfield  {author} {\bibinfo {author} {\bibnamefont {Dunjko}, \bibfnamefont
  {Vedran}}, \bibinfo {author} {\bibfnamefont {Elham}\ \bibnamefont {Kashefi}},
  and\ \bibinfo {author} {\bibfnamefont {Anthony}\ \bibnamefont {Leverrier}}}
  (\bibinfo {year} {2012}),\ \bibfield  {title} {\enquote {\bibinfo {title}
  {Blind quantum computing with weak coherent pulses},}\ }\href
  {https://doi.org/10.1103/physrevlett.108.200502} {\bibfield  {journal}
  {\bibinfo  {journal} {Physical Review Letters}\ }\textbf {\bibinfo {volume}
  {108}},\ \bibinfo {pages} {200502}}\BibitemShut {NoStop}%
\bibitem [{\citenamefont {Dunjko}\ \emph {et~al.}(2016)\citenamefont {Dunjko},
  \citenamefont {Taylor},\ and\ \citenamefont
  {Briegel}}]{bib:dunjko2016quantum}%
  \BibitemOpen
  \bibfield  {author} {\bibinfo {author} {\bibnamefont {Dunjko}, \bibfnamefont
  {Vedran}}, \bibinfo {author} {\bibfnamefont {Jacob~M.}\ \bibnamefont
  {Taylor}}, and\ \bibinfo {author} {\bibfnamefont {Hans~J.}\ \bibnamefont
  {Briegel}}} (\bibinfo {year} {2016}),\ \bibfield  {title} {\enquote {\bibinfo
  {title} {Quantum-enhanced machine learning},}\ }\href
  {https://doi.org/10.1103/physrevlett.117.130501} {\bibfield  {journal}
  {\bibinfo  {journal} {Physical Review Letters}\ }\textbf {\bibinfo {volume}
  {117}},\ \bibinfo {pages} {130501}},\ \Eprint
  {https://arxiv.org/abs/arXiv:1610.08251v1} {arXiv:1610.08251v1} \BibitemShut
  {NoStop}%
\bibitem [{\citenamefont {Dunjko}\ \emph {et~al.}(2014)\citenamefont {Dunjko},
  \citenamefont {Wallden},\ and\ \citenamefont
  {Andersson}}]{bib:Dunjko2014prl}%
  \BibitemOpen
  \bibfield  {author} {\bibinfo {author} {\bibnamefont {Dunjko}, \bibfnamefont
  {Vedran}}, \bibinfo {author} {\bibfnamefont {Petros}\ \bibnamefont
  {Wallden}}, and\ \bibinfo {author} {\bibfnamefont {Erika}\ \bibnamefont
  {Andersson}}} (\bibinfo {year} {2014}),\ \bibfield  {title} {\enquote
  {\bibinfo {title} {Quantum digital signatures without quantum memory},}\
  }\href {https://doi.org/10.1103/PhysRevLett.112.040502} {\bibfield  {journal}
  {\bibinfo  {journal} {Phys. Rev. Lett.}\ }\textbf {\bibinfo {volume} {112}},\
  \bibinfo {pages} {040502}}\BibitemShut {NoStop}%
\bibitem [{\citenamefont {D{\"u}r}\ and\ \citenamefont
  {Briegel}(2007)}]{bib:dur07}%
  \BibitemOpen
  \bibfield  {author} {\bibinfo {author} {\bibnamefont {D{\"u}r}, \bibfnamefont
  {W}}, and\ \bibinfo {author} {\bibfnamefont {H.~J.}\ \bibnamefont {Briegel}}}
  (\bibinfo {year} {2007}),\ \bibfield  {title} {\enquote {\bibinfo {title}
  {Entanglement purification and quantum error correction},}\ }\href
  {https://doi.org/10.1088/0034-4885/70/8/r03} {\bibfield  {journal} {\bibinfo
  {journal} {Reports on Progress in Physics}\ }\textbf {\bibinfo {volume}
  {70}},\ \bibinfo {pages} {1381}}\BibitemShut {NoStop}%
\bibitem [{\citenamefont {D{\"u}r}\ \emph {et~al.}(1999)\citenamefont
  {D{\"u}r}, \citenamefont {Briegel}, \citenamefont {Cirac},\ and\
  \citenamefont {Zoller}}]{bib:dur98}%
  \BibitemOpen
  \bibfield  {author} {\bibinfo {author} {\bibnamefont {D{\"u}r}, \bibfnamefont
  {W}}, \bibinfo {author} {\bibfnamefont {H.~J.}\ \bibnamefont {Briegel}},
  \bibinfo {author} {\bibfnamefont {J.~I.}\ \bibnamefont {Cirac}}, and\
  \bibinfo {author} {\bibfnamefont {P.}~\bibnamefont {Zoller}}} (\bibinfo
  {year} {1999}),\ \bibfield  {title} {\enquote {\bibinfo {title} {Quantum
  repeaters based on entanglement purification},}\ }\href
  {https://doi.org/10.1103/physreva.59.169} {\bibfield  {journal} {\bibinfo
  {journal} {Physical Review A}\ }\textbf {\bibinfo {volume} {59}},\ \bibinfo
  {pages} {169}},\ \Eprint {https://arxiv.org/abs/arXiv:quant-ph/9808065v1}
  {arXiv:quant-ph/9808065v1} \BibitemShut {NoStop}%
\bibitem [{\citenamefont {Einstein}\ \emph {et~al.}(1935)\citenamefont
  {Einstein}, \citenamefont {Podolsky},\ and\ \citenamefont
  {Rosen}}]{bib:EPR35}%
  \BibitemOpen
  \bibfield  {author} {\bibinfo {author} {\bibnamefont {Einstein},
  \bibfnamefont {A}}, \bibinfo {author} {\bibfnamefont {B.}~\bibnamefont
  {Podolsky}}, and\ \bibinfo {author} {\bibfnamefont {N.}~\bibnamefont
  {Rosen}}} (\bibinfo {year} {1935}),\ \bibfield  {title} {\enquote {\bibinfo
  {title} {Can quantum-mechanical description of physical reality be considered
  complete?}}\ }\href {https://doi.org/10.1103/physrev.47.777} {\bibfield
  {journal} {\bibinfo  {journal} {Physical Review}\ }\textbf {\bibinfo {volume}
  {47}},\ \bibinfo {pages} {777}}\BibitemShut {NoStop}%
\bibitem [{\citenamefont {Eisaman}\ \emph {et~al.}(2011)\citenamefont
  {Eisaman}, \citenamefont {Fan}, \citenamefont {Migdall},\ and\ \citenamefont
  {Polyakov}}]{bib:eisaman2011}%
  \BibitemOpen
  \bibfield  {author} {\bibinfo {author} {\bibnamefont {Eisaman}, \bibfnamefont
  {MD}}, \bibinfo {author} {\bibfnamefont {J}~\bibnamefont {Fan}}, \bibinfo
  {author} {\bibfnamefont {A}~\bibnamefont {Migdall}}, and\ \bibinfo {author}
  {\bibfnamefont {SV}~\bibnamefont {Polyakov}}} (\bibinfo {year} {2011}),\
  \bibfield  {title} {\enquote {\bibinfo {title} {Invited review article:
  Single-photon sources and detectors},}\ }\href
  {https://doi.org/10.1063/1.3610677} {\bibfield  {journal} {\bibinfo
  {journal} {Review of Scientific Instruments}\ }\textbf {\bibinfo {volume}
  {82}},\ \bibinfo {pages} {071101}}\BibitemShut {NoStop}%
\bibitem [{\citenamefont {Ekert}(1991)}]{bib:PRL_67_661}%
  \BibitemOpen
  \bibfield  {author} {\bibinfo {author} {\bibnamefont {Ekert}, \bibfnamefont
  {Artur~K}}} (\bibinfo {year} {1991}),\ \bibfield  {title} {\enquote {\bibinfo
  {title} {Quantum cryptography based on bell's theorem},}\ }\href
  {https://doi.org/10.1103/physrevlett.67.661} {\bibfield  {journal} {\bibinfo
  {journal} {Physical Review Letters}\ }\textbf {\bibinfo {volume} {67}},\
  \bibinfo {pages} {661}}\BibitemShut {NoStop}%
\bibitem [{\citenamefont {El~Khayat}\ \emph {et~al.}(2005)\citenamefont
  {El~Khayat}, \citenamefont {Geurts},\ and\ \citenamefont
  {Leduc}}]{bib:el2005improving}%
  \BibitemOpen
  \bibfield  {author} {\bibinfo {author} {\bibnamefont {El~Khayat},
  \bibfnamefont {Ibtissam}}, \bibinfo {author} {\bibfnamefont {Pierre}\
  \bibnamefont {Geurts}}, and\ \bibinfo {author} {\bibfnamefont {Guy}\
  \bibnamefont {Leduc}}} (\bibinfo {year} {2005}),\ \bibfield  {title}
  {\enquote {\bibinfo {title} {Improving tcp in wireless networks with an
  adaptive machine-learnt classifier of packet loss causes},}\ }in\ \href
  {https://doi.org/10.1007/11422778_44} {\emph {\bibinfo {booktitle}
  {International Conference on Research in Networking}}},\ p.\ \bibinfo {pages}
  {549}\BibitemShut {NoStop}%
\bibitem [{\citenamefont {Elliott}(2006)}]{bib:QCC_2006_83}%
  \BibitemOpen
  \bibfield  {author} {\bibinfo {author} {\bibnamefont {Elliott}, \bibfnamefont
  {Chip}}} (\bibinfo {year} {2006}),\ \bibfield  {title} {\enquote {\bibinfo
  {title} {The darpa quantum network},}\ }\href@noop {} {\bibfield  {journal}
  {\bibinfo  {journal} {Quantum Communication \& Cryptography}\ ,\ \bibinfo
  {pages} {83}}}\Eprint {https://arxiv.org/abs/arXiv:quant-ph/0412029v1}
  {arXiv:quant-ph/0412029v1} \BibitemShut {NoStop}%
\bibitem [{\citenamefont {Enk}\ \emph {et~al.}(1998)\citenamefont {Enk},
  \citenamefont {Cirac},\ and\ \citenamefont {Zoller}}]{bib:enk98}%
  \BibitemOpen
  \bibfield  {author} {\bibinfo {author} {\bibnamefont {Enk}, \bibfnamefont
  {S}}, \bibinfo {author} {\bibfnamefont {J.~I.}\ \bibnamefont {Cirac}}, and\
  \bibinfo {author} {\bibfnamefont {P.}~\bibnamefont {Zoller}}} (\bibinfo
  {year} {1998}),\ \bibfield  {title} {\enquote {\bibinfo {title} {Photonic
  channels for quantum communication},}\ }\href
  {https://doi.org/10.1126/science.279.5348.205} {\bibfield  {journal}
  {\bibinfo  {journal} {Science}\ }\textbf {\bibinfo {volume} {279}},\ \bibinfo
  {pages} {205}}\BibitemShut {NoStop}%
\bibitem [{\citenamefont {Erman}\ \emph {et~al.}(2007)\citenamefont {Erman},
  \citenamefont {Mahanti}, \citenamefont {Arlitt}, \citenamefont {Cohen},\ and\
  \citenamefont {Williamson}}]{bib:erman2007offline}%
  \BibitemOpen
  \bibfield  {author} {\bibinfo {author} {\bibnamefont {Erman}, \bibfnamefont
  {Jeffrey}}, \bibinfo {author} {\bibfnamefont {Anirban}\ \bibnamefont
  {Mahanti}}, \bibinfo {author} {\bibfnamefont {Martin}\ \bibnamefont
  {Arlitt}}, \bibinfo {author} {\bibfnamefont {Ira}\ \bibnamefont {Cohen}},
  and\ \bibinfo {author} {\bibfnamefont {Carey}\ \bibnamefont {Williamson}}}
  (\bibinfo {year} {2007}),\ \bibfield  {title} {\enquote {\bibinfo {title}
  {Offline/realtime traffic classification using semi-supervised learning},}\
  }\href {https://doi.org/10.1016/j.peva.2007.06.014} {\bibfield  {journal}
  {\bibinfo  {journal} {Performance Evaluation}\ }\textbf {\bibinfo {volume}
  {64}},\ \bibinfo {pages} {1194}}\BibitemShut {NoStop}%
\bibitem [{\citenamefont {Esmaeil~Zadeh}\ \emph {et~al.}(2021)\citenamefont
  {Esmaeil~Zadeh}, \citenamefont {Chang}, \citenamefont {Los}, \citenamefont
  {Gyger}, \citenamefont {Elshaari}, \citenamefont {Steinhauer}, \citenamefont
  {Dorenbos},\ and\ \citenamefont {Zwiller}}]{zadeh2021SNSPDreview}%
  \BibitemOpen
  \bibfield  {author} {\bibinfo {author} {\bibnamefont {Esmaeil~Zadeh},
  \bibfnamefont {Iman}}, \bibinfo {author} {\bibfnamefont {J.}~\bibnamefont
  {Chang}}, \bibinfo {author} {\bibfnamefont {Johannes W.~N.}\ \bibnamefont
  {Los}}, \bibinfo {author} {\bibfnamefont {Samuel}\ \bibnamefont {Gyger}},
  \bibinfo {author} {\bibfnamefont {Ali~W.}\ \bibnamefont {Elshaari}}, \bibinfo
  {author} {\bibfnamefont {Stephan}\ \bibnamefont {Steinhauer}}, \bibinfo
  {author} {\bibfnamefont {Sander~N.}\ \bibnamefont {Dorenbos}}, and\ \bibinfo
  {author} {\bibfnamefont {Val}\ \bibnamefont {Zwiller}}} (\bibinfo {year}
  {2021}),\ \bibfield  {title} {\enquote {\bibinfo {title} {Superconducting
  nanowire single-photon detectors: A perspective on evolution,
  state-of-the-art, future developments, and applications},}\ }\href
  {https://doi.org/10.1063/5.0045990} {\bibfield  {journal} {\bibinfo
  {journal} {Applied Physics Letters}\ }\textbf {\bibinfo {volume}
  {118}}~(\bibinfo {number} {19}),\ \bibinfo {pages} {190502}},\ \Eprint
  {https://arxiv.org/abs/https://pubs.aip.org/aip/apl/article-pdf/doi/10.1063/5.0045990/20021815/190502\_1\_5.0045990.pdf}
  {https://pubs.aip.org/aip/apl/article-pdf/doi/10.1063/5.0045990/20021815/190502\_1\_5.0045990.pdf}
  \BibitemShut {NoStop}%
\bibitem [{\citenamefont {Facon}\ \emph {et~al.}(2016)\citenamefont {Facon},
  \citenamefont {Dietsche}, \citenamefont {Grosso}, \citenamefont {Haroche},
  \citenamefont {Raimond}, \citenamefont {Brune},\ and\ \citenamefont
  {Gleyzes}}]{bib:facon2016sensitive}%
  \BibitemOpen
  \bibfield  {author} {\bibinfo {author} {\bibnamefont {Facon}, \bibfnamefont
  {Adrien}}, \bibinfo {author} {\bibfnamefont {Eva-Katharina}\ \bibnamefont
  {Dietsche}}, \bibinfo {author} {\bibfnamefont {Dorian}\ \bibnamefont
  {Grosso}}, \bibinfo {author} {\bibfnamefont {Serge}\ \bibnamefont {Haroche}},
  \bibinfo {author} {\bibfnamefont {Jean-Michel}\ \bibnamefont {Raimond}},
  \bibinfo {author} {\bibfnamefont {Michel}\ \bibnamefont {Brune}}, and\
  \bibinfo {author} {\bibfnamefont {S{\'e}bastien}\ \bibnamefont {Gleyzes}}}
  (\bibinfo {year} {2016}),\ \bibfield  {title} {\enquote {\bibinfo {title} {A
  sensitive electrometer based on a rydberg atom in a schr{\"o}dinger-cat
  state},}\ }\href {https://doi.org/10.1038/nature18327} {\bibfield  {journal}
  {\bibinfo  {journal} {Nature}\ }\textbf {\bibinfo {volume} {535}},\ \bibinfo
  {pages} {262}}\BibitemShut {NoStop}%
\bibitem [{\citenamefont {Farhi}\ \emph {et~al.}(2014)\citenamefont {Farhi},
  \citenamefont {Goldstone},\ and\ \citenamefont
  {Gutmann}}]{bib:farhi2014quantum}%
  \BibitemOpen
  \bibfield  {author} {\bibinfo {author} {\bibnamefont {Farhi}, \bibfnamefont
  {Edward}}, \bibinfo {author} {\bibfnamefont {Jeffrey}\ \bibnamefont
  {Goldstone}}, and\ \bibinfo {author} {\bibfnamefont {Sam}\ \bibnamefont
  {Gutmann}}} (\bibinfo {year} {2014}),\ \bibfield  {title} {\enquote {\bibinfo
  {title} {A quantum approximate optimization algorithm},}\ }\href@noop {} {\
  }\Eprint {https://arxiv.org/abs/arXiv:1411.4028} {arXiv:1411.4028}
  \BibitemShut {NoStop}%
\bibitem [{\citenamefont {Farhi}\ and\ \citenamefont
  {Harrow}(2016)}]{bib:farhi2016quantum}%
  \BibitemOpen
  \bibfield  {author} {\bibinfo {author} {\bibnamefont {Farhi}, \bibfnamefont
  {Edward}}, and\ \bibinfo {author} {\bibfnamefont {Aram~W.}\ \bibnamefont
  {Harrow}}} (\bibinfo {year} {2016}),\ \bibfield  {title} {\enquote {\bibinfo
  {title} {Quantum supremacy through the quantum approximate optimization
  algorithm},}\ }\href@noop {} {\ }\Eprint
  {https://arxiv.org/abs/arXiv:1602.07674} {arXiv:1602.07674} \BibitemShut
  {NoStop}%
\bibitem [{\citenamefont {Fawzi}\ \emph {et~al.}(2018)\citenamefont {Fawzi},
  \citenamefont {Grospellier},\ and\ \citenamefont
  {Leverrier}}]{SD-Fawzi:2018aa}%
  \BibitemOpen
  \bibfield  {author} {\bibinfo {author} {\bibnamefont {Fawzi}, \bibfnamefont
  {O}}, \bibinfo {author} {\bibfnamefont {A.}~\bibnamefont {Grospellier}}, and\
  \bibinfo {author} {\bibfnamefont {A.}~\bibnamefont {Leverrier}}} (\bibinfo
  {year} {2018}),\ \bibfield  {title} {\enquote {\bibinfo {title} {Constant
  overhead quantum fault-tolerance with quantum expander codes},}\ }\href
  {https://doi.org/10.1109/FOCS.2018.00076} {\bibinfo  {journal} {IEEE 59th
  Annual Symposium on Foundations of Computer Science (FOCS)}\ ,\ \bibinfo
  {pages} {743}}\BibitemShut {NoStop}%
\bibitem [{\citenamefont {Feynman}(1985)}]{bib:Feynman85}%
  \BibitemOpen
\bibfield  {journal} {  }\bibfield  {author} {\bibinfo {author} {\bibnamefont
  {Feynman}, \bibfnamefont {Richard~P}}} (\bibinfo {year} {1985}),\ \bibfield
  {title} {\enquote {\bibinfo {title} {Quantum mechanical computers},}\ }\href
  {https://doi.org/10.1515/9781400886975-036} {\bibfield  {journal} {\bibinfo
  {journal} {Foundations of Physics}\ }\textbf {\bibinfo {volume} {16}},\
  \bibinfo {pages} {507}}\BibitemShut {NoStop}%
\bibitem [{\citenamefont {Fickler}\ \emph {et~al.}(2012)\citenamefont
  {Fickler}, \citenamefont {Lapkiewicz}, \citenamefont {Plick}, \citenamefont
  {Krenn}, \citenamefont {Schaeff}, \citenamefont {Ramelow},\ and\
  \citenamefont {Zeilinger}}]{bib:fickler2012quantum}%
  \BibitemOpen
  \bibfield  {author} {\bibinfo {author} {\bibnamefont {Fickler}, \bibfnamefont
  {Robert}}, \bibinfo {author} {\bibfnamefont {Radek}\ \bibnamefont
  {Lapkiewicz}}, \bibinfo {author} {\bibfnamefont {William~N}\ \bibnamefont
  {Plick}}, \bibinfo {author} {\bibfnamefont {Mario}\ \bibnamefont {Krenn}},
  \bibinfo {author} {\bibfnamefont {Christoph}\ \bibnamefont {Schaeff}},
  \bibinfo {author} {\bibfnamefont {Sven}\ \bibnamefont {Ramelow}}, and\
  \bibinfo {author} {\bibfnamefont {Anton}\ \bibnamefont {Zeilinger}}}
  (\bibinfo {year} {2012}),\ \bibfield  {title} {\enquote {\bibinfo {title}
  {Quantum entanglement of high angular momenta},}\ }\href
  {https://doi.org/10.1126/science.1227193} {\bibfield  {journal} {\bibinfo
  {journal} {Science}\ }\textbf {\bibinfo {volume} {338}},\ \bibinfo {pages}
  {640}},\ \Eprint {https://arxiv.org/abs/arXiv:1207.2376v3}
  {arXiv:1207.2376v3} \BibitemShut {NoStop}%
\bibitem [{\citenamefont {Fishburn}(1970)}]{bib:Fishburn70}%
  \BibitemOpen
  \bibfield  {author} {\bibinfo {author} {\bibnamefont {Fishburn},
  \bibfnamefont {P~C}}} (\bibinfo {year} {1970}),\ \href@noop {} {\emph
  {\bibinfo {title} {Utility Theory for Decision Making}}}\ (\bibinfo
  {publisher} {John Wiley \& Sons, New York})\BibitemShut {NoStop}%
\bibitem [{\citenamefont {Fitch}\ \emph {et~al.}(2003)\citenamefont {Fitch},
  \citenamefont {Jacobs}, \citenamefont {Pittman},\ and\ \citenamefont
  {Franson}}]{bib:Fitch03}%
  \BibitemOpen
  \bibfield  {author} {\bibinfo {author} {\bibnamefont {Fitch}, \bibfnamefont
  {M~J}}, \bibinfo {author} {\bibfnamefont {B.~C.}\ \bibnamefont {Jacobs}},
  \bibinfo {author} {\bibfnamefont {T.~B.}\ \bibnamefont {Pittman}}, and\
  \bibinfo {author} {\bibfnamefont {J.~D.}\ \bibnamefont {Franson}}} (\bibinfo
  {year} {2003}),\ \bibfield  {title} {\enquote {\bibinfo {title} {Photon
  number resolution using time-multiplexed single-photon detectors},}\ }\href
  {https://doi.org/10.1103/physreva.68.043814} {\bibfield  {journal} {\bibinfo
  {journal} {Physical Review A}\ }\textbf {\bibinfo {volume} {68}},\ \bibinfo
  {pages} {043814}},\ \Eprint {https://arxiv.org/abs/arXiv:quant-ph/0305193v3}
  {arXiv:quant-ph/0305193v3} \BibitemShut {NoStop}%
\bibitem [{\citenamefont {Fitzsimons}(2017)}]{SD-Fitzsimons:2017aa}%
  \BibitemOpen
  \bibfield  {author} {\bibinfo {author} {\bibnamefont {Fitzsimons},
  \bibfnamefont {Joseph~F}}} (\bibinfo {year} {2017}),\ \bibfield  {title}
  {\enquote {\bibinfo {title} {Private quantum computation: an introduction to
  blind quantum computing and related protocols},}\ }\href
  {https://doi.org/10.1038/s41534-017-0025-3} {\bibfield  {journal} {\bibinfo
  {journal} {NPJ Quantum Information}\ }\textbf {\bibinfo {volume} {3}},\
  \bibinfo {pages} {23}}\BibitemShut {NoStop}%
\bibitem [{\citenamefont {Fitzsimons}\ and\ \citenamefont
  {Kashefi}(2017)}]{bib:joe}%
  \BibitemOpen
  \bibfield  {author} {\bibinfo {author} {\bibnamefont {Fitzsimons},
  \bibfnamefont {Joseph~F}}, and\ \bibinfo {author} {\bibfnamefont {Elham}\
  \bibnamefont {Kashefi}}} (\bibinfo {year} {2017}),\ \bibfield  {title}
  {\enquote {\bibinfo {title} {Unconditionally verifiable blind quantum
  computation},}\ }\href {https://doi.org/10.1103/physreva.96.012303}
  {\bibfield  {journal} {\bibinfo  {journal} {Physical Review A}\ }\textbf
  {\bibinfo {volume} {96}},\ \bibinfo {pages} {012303}}\BibitemShut {NoStop}%
\bibitem [{\citenamefont {Flach}(2012)}]{bib:flach2012machine}%
  \BibitemOpen
  \bibfield  {author} {\bibinfo {author} {\bibnamefont {Flach}, \bibfnamefont
  {Peter}}} (\bibinfo {year} {2012}),\ \href@noop {} {\emph {\bibinfo {title}
  {Machine learning: the art and science of algorithms that make sense of
  data}}}\ (\bibinfo  {publisher} {Cambridge University Press})\BibitemShut
  {NoStop}%
\bibitem [{\citenamefont {Fl{\"u}hmann}\ \emph {et~al.}(2019)\citenamefont
  {Fl{\"u}hmann}, \citenamefont {Nguyen}, \citenamefont {Marinelli},
  \citenamefont {Negnevitsky}, \citenamefont {Mehta},\ and\ \citenamefont
  {Home}}]{fluhmann2019makingGKP}%
  \BibitemOpen
  \bibfield  {author} {\bibinfo {author} {\bibnamefont {Fl{\"u}hmann},
  \bibfnamefont {C}}, \bibinfo {author} {\bibfnamefont {T.~L.}\ \bibnamefont
  {Nguyen}}, \bibinfo {author} {\bibfnamefont {M.}~\bibnamefont {Marinelli}},
  \bibinfo {author} {\bibfnamefont {V.}~\bibnamefont {Negnevitsky}}, \bibinfo
  {author} {\bibfnamefont {K.}~\bibnamefont {Mehta}}, and\ \bibinfo {author}
  {\bibfnamefont {J.~P.}\ \bibnamefont {Home}}} (\bibinfo {year} {2019}),\
  \bibfield  {title} {\enquote {\bibinfo {title} {Encoding a qubit in a
  trapped-ion mechanical oscillator},}\ }\href
  {https://doi.org/10.1038/s41586-019-0960-6} {\bibfield  {journal} {\bibinfo
  {journal} {Nature}\ }\textbf {\bibinfo {volume} {566}}~(\bibinfo {number}
  {7745}),\ \bibinfo {pages} {513--517}}\BibitemShut {NoStop}%
\bibitem [{\citenamefont {Ford}(1956)}]{ford1956network}%
  \BibitemOpen
  \bibfield  {author} {\bibinfo {author} {\bibnamefont {Ford}, \bibfnamefont
  {Lester~Randolph}}} (\bibinfo {year} {1956}),\ \bibfield  {title} {\enquote
  {\bibinfo {title} {Network flow theory},}\ }\href@noop {} {\bibinfo
  {journal} {Rand Corporation Paper, Santa Monica, 1956}\ }\BibitemShut
  {NoStop}%
\bibitem [{\citenamefont {Ford}\ and\ \citenamefont
  {Fulkerson}(1956)}]{ford1956maximal}%
  \BibitemOpen
\bibfield  {journal} {  }\bibfield  {author} {\bibinfo {author} {\bibnamefont
  {Ford}, \bibfnamefont {Lester~Randolph}}, and\ \bibinfo {author}
  {\bibfnamefont {Delbert~R}\ \bibnamefont {Fulkerson}}} (\bibinfo {year}
  {1956}),\ \bibfield  {title} {\enquote {\bibinfo {title} {Maximal flow
  through a network},}\ }\href@noop {} {\bibfield  {journal} {\bibinfo
  {journal} {Canadian journal of Mathematics}\ }\textbf {\bibinfo {volume}
  {8}},\ \bibinfo {pages} {399--404}}\BibitemShut {NoStop}%
\bibitem [{\citenamefont {Forster}\ and\ \citenamefont
  {Murphy}(2007)}]{bib:forster2007froms}%
  \BibitemOpen
  \bibfield  {author} {\bibinfo {author} {\bibnamefont {Forster}, \bibfnamefont
  {Anna}}, and\ \bibinfo {author} {\bibfnamefont {Amy~L.}\ \bibnamefont
  {Murphy}}} (\bibinfo {year} {2007}),\ \bibfield  {title} {\enquote {\bibinfo
  {title} {Froms: Feedback routing for optimizing multiple sinks in wsn with
  reinforcement learning},}\ }in\ \href
  {https://doi.org/10.1109/issnip.2007.4496872} {\emph {\bibinfo {booktitle}
  {IEEE 3rd International Conference on Intelligent Sensors, Sensor Networks
  and Information (ISSNIP)}}},\ p.\ \bibinfo {pages} {371}\BibitemShut
  {NoStop}%
\bibitem [{\citenamefont {Fowler}\ \emph
  {et~al.}(2010{\natexlab{a}})\citenamefont {Fowler}, \citenamefont {D.~S},
  \citenamefont {Hill}, \citenamefont {Ladd}, \citenamefont {Meter},\ and\
  \citenamefont {Hollenberg}}]{bib:Fowler10}%
  \BibitemOpen
  \bibfield  {author} {\bibinfo {author} {\bibnamefont {Fowler}, \bibfnamefont
  {A~G}}, \bibinfo {author} {\bibfnamefont {Wang}\ \bibnamefont {D.~S}},
  \bibinfo {author} {\bibfnamefont {C.~D.}\ \bibnamefont {Hill}}, \bibinfo
  {author} {\bibfnamefont {T.~D.}\ \bibnamefont {Ladd}}, \bibinfo {author}
  {\bibfnamefont {R.~Van}\ \bibnamefont {Meter}}, and\ \bibinfo {author}
  {\bibfnamefont {L.~C.~L.}\ \bibnamefont {Hollenberg}}} (\bibinfo {year}
  {2010}{\natexlab{a}}),\ \bibfield  {title} {\enquote {\bibinfo {title}
  {Surface code quantum communication},}\ }\href
  {https://doi.org/10.1103/physrevlett.104.180503} {\bibfield  {journal}
  {\bibinfo  {journal} {Physical Review Letters}\ }\textbf {\bibinfo {volume}
  {104}},\ \bibinfo {pages} {180503}},\ \Eprint
  {https://arxiv.org/abs/arXiv:0910.4074v3} {arXiv:0910.4074v3} \BibitemShut
  {NoStop}%
\bibitem [{\citenamefont {Fowler}\ \emph
  {et~al.}(2012{\natexlab{a}})\citenamefont {Fowler}, \citenamefont
  {Mariantoni}, \citenamefont {Martinis},\ and\ \citenamefont
  {Cleland}}]{bib:FMMC12}%
  \BibitemOpen
  \bibfield  {author} {\bibinfo {author} {\bibnamefont {Fowler}, \bibfnamefont
  {AG}}, \bibinfo {author} {\bibfnamefont {M.}~\bibnamefont {Mariantoni}},
  \bibinfo {author} {\bibfnamefont {J.M.}\ \bibnamefont {Martinis}}, and\
  \bibinfo {author} {\bibfnamefont {A.N.}\ \bibnamefont {Cleland}}} (\bibinfo
  {year} {2012}{\natexlab{a}}),\ \bibfield  {title} {\enquote {\bibinfo {title}
  {{Surface codes: Towards practical large-scale quantum computation}},}\
  }\href@noop {} {\bibfield  {journal} {\bibinfo  {journal} {Phys. Rev. A.}\
  }\textbf {\bibinfo {volume} {86}},\ \bibinfo {pages} {032324}}\BibitemShut
  {NoStop}%
\bibitem [{\citenamefont {Fowler}(2016)}]{bib:FowlerPrivate}%
  \BibitemOpen
  \bibfield  {author} {\bibinfo {author} {\bibnamefont {Fowler}, \bibfnamefont
  {Austin}}} (\bibinfo {year} {2016}),\ \href@noop {} {\bibinfo  {journal}
  {Private communication}\ }\BibitemShut {NoStop}%
\bibitem [{\citenamefont {Fowler}\ \emph
  {et~al.}(2012{\natexlab{b}})\citenamefont {Fowler}, \citenamefont
  {Mariantoni}, \citenamefont {Martinis},\ and\ \citenamefont
  {Cleland}}]{SD-Fowler:2012aa}%
  \BibitemOpen
\bibfield  {journal} {  }\bibfield  {author} {\bibinfo {author} {\bibnamefont
  {Fowler}, \bibfnamefont {Austin~G}}, \bibinfo {author} {\bibfnamefont
  {Matteo}\ \bibnamefont {Mariantoni}}, \bibinfo {author} {\bibfnamefont
  {John~M.}\ \bibnamefont {Martinis}}, and\ \bibinfo {author} {\bibfnamefont
  {Andrew~N.}\ \bibnamefont {Cleland}}} (\bibinfo {year}
  {2012}{\natexlab{b}}),\ \bibfield  {title} {\enquote {\bibinfo {title}
  {Surface codes: Towards practical large-scale quantum computation},}\ }\href
  {https://doi.org/10.1103/PhysRevA.86.032324} {\bibfield  {journal} {\bibinfo
  {journal} {Physical Review A}\ }\textbf {\bibinfo {volume} {86}},\ \bibinfo
  {pages} {032324}},\ \Eprint {https://arxiv.org/abs/arXiv:1208.0928v2}
  {arXiv:1208.0928v2} \BibitemShut {NoStop}%
\bibitem [{\citenamefont {Fowler}\ \emph
  {et~al.}(2010{\natexlab{b}})\citenamefont {Fowler}, \citenamefont {Wang},
  \citenamefont {Hill}, \citenamefont {Ladd}, \citenamefont {Van~Meter},\ and\
  \citenamefont {Hollenberg}}]{SD-Fowler:2010aa}%
  \BibitemOpen
  \bibfield  {author} {\bibinfo {author} {\bibnamefont {Fowler}, \bibfnamefont
  {Austin~G}}, \bibinfo {author} {\bibfnamefont {David~S.}\ \bibnamefont
  {Wang}}, \bibinfo {author} {\bibfnamefont {Charles~D.}\ \bibnamefont {Hill}},
  \bibinfo {author} {\bibfnamefont {Thaddeus~D.}\ \bibnamefont {Ladd}},
  \bibinfo {author} {\bibfnamefont {Rodney}\ \bibnamefont {Van~Meter}}, and\
  \bibinfo {author} {\bibfnamefont {Lloyd C.~L.}\ \bibnamefont {Hollenberg}}}
  (\bibinfo {year} {2010}{\natexlab{b}}),\ \bibfield  {title} {\enquote
  {\bibinfo {title} {Surface code quantum communication},}\ }\href
  {https://doi.org/10.1103/PhysRevLett.104.180503} {\bibfield  {journal}
  {\bibinfo  {journal} {Physical Review Letters}\ }\textbf {\bibinfo {volume}
  {104}},\ \bibinfo {pages} {180503}},\ \Eprint
  {https://arxiv.org/abs/arXiv:0910.4074v3} {arXiv:0910.4074v3} \BibitemShut
  {NoStop}%
\bibitem [{\citenamefont {Fraley}\ and\ \citenamefont
  {Cannady}(2017)}]{bib:fraley2017promise}%
  \BibitemOpen
  \bibfield  {author} {\bibinfo {author} {\bibnamefont {Fraley}, \bibfnamefont
  {James~B}}, and\ \bibinfo {author} {\bibfnamefont {James}\ \bibnamefont
  {Cannady}}} (\bibinfo {year} {2017}),\ \bibfield  {title} {\enquote {\bibinfo
  {title} {The promise of machine learning in cybersecurity},}\ }in\ \href
  {https://doi.org/10.1109/secon.2017.7925283} {\emph {\bibinfo {booktitle}
  {IEEE SoutheastCon}}},\ p.~\bibinfo {pages} {1}\BibitemShut {NoStop}%
\bibitem [{\citenamefont {Fredman}\ and\ \citenamefont
  {Tarjan}(1984)}]{bib:FredmanLawrence84}%
  \BibitemOpen
  \bibfield  {author} {\bibinfo {author} {\bibnamefont {Fredman}, \bibfnamefont
  {Michael~Lawrence}}, and\ \bibinfo {author} {\bibfnamefont {Robert~E.}\
  \bibnamefont {Tarjan}}} (\bibinfo {year} {1984}),\ \bibfield  {title}
  {\enquote {\bibinfo {title} {Fibonacci heaps and their uses in improved
  network optimization algorithms},}\ }\bibfield  {booktitle} {\emph {\bibinfo
  {booktitle} {IEEE 25th Annual Symposium on Foundations of Computer
  Science}},\ }\href {https://doi.org/10.1109/SFCS.1984.715934} {\ \textbf
  {\bibinfo {volume} {346}},\ \bibinfo {pages} {338}}\BibitemShut {NoStop}%
\bibitem [{\citenamefont {Friedman}\ \emph {et~al.}(2000)\citenamefont
  {Friedman}, \citenamefont {Patel}, \citenamefont {Chen}, \citenamefont
  {Tolpygo},\ and\ \citenamefont {Lukens}}]{bib:friedman2000quantum}%
  \BibitemOpen
  \bibfield  {author} {\bibinfo {author} {\bibnamefont {Friedman},
  \bibfnamefont {Jonathan~R}}, \bibinfo {author} {\bibfnamefont {Vijay}\
  \bibnamefont {Patel}}, \bibinfo {author} {\bibfnamefont {Wei}\ \bibnamefont
  {Chen}}, \bibinfo {author} {\bibfnamefont {SK}~\bibnamefont {Tolpygo}}, and\
  \bibinfo {author} {\bibfnamefont {James~E}\ \bibnamefont {Lukens}}} (\bibinfo
  {year} {2000}),\ \bibfield  {title} {\enquote {\bibinfo {title} {Quantum
  superposition of distinct macroscopic states},}\ }\href
  {https://doi.org/10.1038/35017505} {\bibfield  {journal} {\bibinfo  {journal}
  {Nature}\ }\textbf {\bibinfo {volume} {406}},\ \bibinfo {pages}
  {43}}\BibitemShut {NoStop}%
\bibitem [{\citenamefont {Friis}\ \emph {et~al.}(2018)\citenamefont {Friis},
  \citenamefont {Marty}, \citenamefont {Maier}, \citenamefont {Hempel},
  \citenamefont {Holz{\"a}pfel}, \citenamefont {Jurcevic}, \citenamefont
  {Plenio}, \citenamefont {Huber}, \citenamefont {Roos}, \citenamefont {Blatt}
  \emph {et~al.}}]{bib:friis2018observation}%
  \BibitemOpen
  \bibfield  {author} {\bibinfo {author} {\bibnamefont {Friis}, \bibfnamefont
  {Nicolai}}, \bibinfo {author} {\bibfnamefont {Oliver}\ \bibnamefont {Marty}},
  \bibinfo {author} {\bibfnamefont {Christine}\ \bibnamefont {Maier}}, \bibinfo
  {author} {\bibfnamefont {Cornelius}\ \bibnamefont {Hempel}}, \bibinfo
  {author} {\bibfnamefont {Milan}\ \bibnamefont {Holz{\"a}pfel}}, \bibinfo
  {author} {\bibfnamefont {Petar}\ \bibnamefont {Jurcevic}}, \bibinfo {author}
  {\bibfnamefont {Martin~B}\ \bibnamefont {Plenio}}, \bibinfo {author}
  {\bibfnamefont {Marcus}\ \bibnamefont {Huber}}, \bibinfo {author}
  {\bibfnamefont {Christian}\ \bibnamefont {Roos}}, \bibinfo {author}
  {\bibfnamefont {Rainer}\ \bibnamefont {Blatt}},  \emph {et~al.}} (\bibinfo
  {year} {2018}),\ \bibfield  {title} {\enquote {\bibinfo {title} {Observation
  of entangled states of a fully controlled 20-qubit system},}\ }\href
  {https://doi.org/10.1103/physrevx.8.021012} {\bibfield  {journal} {\bibinfo
  {journal} {Physical Review X}\ }\textbf {\bibinfo {volume} {8}},\ \bibinfo
  {pages} {021012}},\ \Eprint {https://arxiv.org/abs/arXiv:1711.11092v3}
  {arXiv:1711.11092v3} \BibitemShut {NoStop}%
\bibitem [{\citenamefont {Fukuda}\ \emph {et~al.}(2011)\citenamefont {Fukuda},
  \citenamefont {Fujii}, \citenamefont {Numata}, \citenamefont {Amemiya},
  \citenamefont {Yoshizawa}, \citenamefont {Tsuchida}, \citenamefont {Fujino},
  \citenamefont {Ishii}, \citenamefont {Itatani}, \citenamefont {Inoue} \emph
  {et~al.}}]{bib:fukuda2011}%
  \BibitemOpen
  \bibfield  {author} {\bibinfo {author} {\bibnamefont {Fukuda}, \bibfnamefont
  {Daiji}}, \bibinfo {author} {\bibfnamefont {Go}~\bibnamefont {Fujii}},
  \bibinfo {author} {\bibfnamefont {Takayuki}\ \bibnamefont {Numata}}, \bibinfo
  {author} {\bibfnamefont {Kuniaki}\ \bibnamefont {Amemiya}}, \bibinfo {author}
  {\bibfnamefont {Akio}\ \bibnamefont {Yoshizawa}}, \bibinfo {author}
  {\bibfnamefont {Hidemi}\ \bibnamefont {Tsuchida}}, \bibinfo {author}
  {\bibfnamefont {Hidetoshi}\ \bibnamefont {Fujino}}, \bibinfo {author}
  {\bibfnamefont {Hiroyuki}\ \bibnamefont {Ishii}}, \bibinfo {author}
  {\bibfnamefont {Taro}\ \bibnamefont {Itatani}}, \bibinfo {author}
  {\bibfnamefont {Shuichiro}\ \bibnamefont {Inoue}},  \emph {et~al.}} (\bibinfo
  {year} {2011}),\ \bibfield  {title} {\enquote {\bibinfo {title}
  {Titanium-based transition-edge photon number resolving detector with 98\%
  detection efficiency with index-matched small-gap fiber coupling},}\ }\href
  {https://doi.org/10.1364/oe.19.000870} {\bibfield  {journal} {\bibinfo
  {journal} {Optics Express}\ }\textbf {\bibinfo {volume} {19}},\ \bibinfo
  {pages} {870}}\BibitemShut {NoStop}%
\bibitem [{\citenamefont {Fukui}\ \emph {et~al.}(2021)\citenamefont {Fukui},
  \citenamefont {Alexander},\ and\ \citenamefont {van
  Loock}}]{fukui2021GKPcomm}%
  \BibitemOpen
  \bibfield  {author} {\bibinfo {author} {\bibnamefont {Fukui}, \bibfnamefont
  {Kosuke}}, \bibinfo {author} {\bibfnamefont {Rafael~N.}\ \bibnamefont
  {Alexander}}, and\ \bibinfo {author} {\bibfnamefont {Peter}\ \bibnamefont
  {van Loock}}} (\bibinfo {year} {2021}),\ \bibfield  {title} {\enquote
  {\bibinfo {title} {All-optical long-distance quantum communication with
  gottesman-kitaev-preskill qubits},}\ }\href
  {https://doi.org/10.1103/PhysRevResearch.3.033118} {\bibfield  {journal}
  {\bibinfo  {journal} {Phys. Rev. Res.}\ }\textbf {\bibinfo {volume} {3}},\
  \bibinfo {pages} {033118}}\BibitemShut {NoStop}%
\bibitem [{\citenamefont {Fukui}\ \emph {et~al.}(2018)\citenamefont {Fukui},
  \citenamefont {Tomita}, \citenamefont {Okamoto},\ and\ \citenamefont
  {Fujii}}]{fukui2018analog}%
  \BibitemOpen
  \bibfield  {author} {\bibinfo {author} {\bibnamefont {Fukui}, \bibfnamefont
  {Kosuke}}, \bibinfo {author} {\bibfnamefont {Akihisa}\ \bibnamefont
  {Tomita}}, \bibinfo {author} {\bibfnamefont {Atsushi}\ \bibnamefont
  {Okamoto}}, and\ \bibinfo {author} {\bibfnamefont {Keisuke}\ \bibnamefont
  {Fujii}}} (\bibinfo {year} {2018}),\ \bibfield  {title} {\enquote {\bibinfo
  {title} {High-threshold fault-tolerant quantum computation with analog
  quantum error correction},}\ }\href
  {https://doi.org/10.1103/PhysRevX.8.021054} {\bibfield  {journal} {\bibinfo
  {journal} {Phys. Rev. X}\ }\textbf {\bibinfo {volume} {8}},\ \bibinfo {pages}
  {021054}}\BibitemShut {NoStop}%
\bibitem [{\citenamefont {Furusawa}\ \emph
  {et~al.}(1998{\natexlab{a}})\citenamefont {Furusawa}, \citenamefont
  {S{\o}rensen}, \citenamefont {Braunstein}, \citenamefont {Fuchs},
  \citenamefont {Kimble},\ and\ \citenamefont
  {Polzik}}]{furusawa1998teleportation}%
  \BibitemOpen
  \bibfield  {author} {\bibinfo {author} {\bibnamefont {Furusawa},
  \bibfnamefont {A}}, \bibinfo {author} {\bibfnamefont {J.~L.}\ \bibnamefont
  {S{\o}rensen}}, \bibinfo {author} {\bibfnamefont {S.~L.}\ \bibnamefont
  {Braunstein}}, \bibinfo {author} {\bibfnamefont {C.~A.}\ \bibnamefont
  {Fuchs}}, \bibinfo {author} {\bibfnamefont {H.~J.}\ \bibnamefont {Kimble}},
  and\ \bibinfo {author} {\bibfnamefont {E.~S.}\ \bibnamefont {Polzik}}}
  (\bibinfo {year} {1998}{\natexlab{a}}),\ \bibfield  {title} {\enquote
  {\bibinfo {title} {Unconditional quantum teleportation},}\ }\href
  {https://doi.org/10.1126/science.282.5389.706} {\bibfield  {journal}
  {\bibinfo  {journal} {Science}\ }\textbf {\bibinfo {volume} {282}}~(\bibinfo
  {number} {5389}),\ \bibinfo {pages} {706--709}},\ \Eprint
  {https://arxiv.org/abs/https://www.science.org/doi/pdf/10.1126/science.282.5389.706}
  {https://www.science.org/doi/pdf/10.1126/science.282.5389.706} \BibitemShut
  {NoStop}%
\bibitem [{\citenamefont {Furusawa}\ \emph
  {et~al.}(1998{\natexlab{b}})\citenamefont {Furusawa}, \citenamefont
  {S{\o}rensen}, \citenamefont {Braunstein}, \citenamefont {Fuchs},
  \citenamefont {Kimble},\ and\ \citenamefont {Polzik}}]{bib:Science_282_706}%
  \BibitemOpen
  \bibfield  {author} {\bibinfo {author} {\bibnamefont {Furusawa},
  \bibfnamefont {Akira}}, \bibinfo {author} {\bibfnamefont {Jens~Lykke}\
  \bibnamefont {S{\o}rensen}}, \bibinfo {author} {\bibfnamefont {Samuel~L}\
  \bibnamefont {Braunstein}}, \bibinfo {author} {\bibfnamefont {Christopher~A}\
  \bibnamefont {Fuchs}}, \bibinfo {author} {\bibfnamefont {H~Jeff}\
  \bibnamefont {Kimble}}, and\ \bibinfo {author} {\bibfnamefont {Eugene~S}\
  \bibnamefont {Polzik}}} (\bibinfo {year} {1998}{\natexlab{b}}),\ \bibfield
  {title} {\enquote {\bibinfo {title} {Unconditional quantum teleportation},}\
  }\href {https://doi.org/10.1126/science.282.5389.706} {\bibfield  {journal}
  {\bibinfo  {journal} {Science}\ }\textbf {\bibinfo {volume} {282}},\ \bibinfo
  {pages} {706}}\BibitemShut {NoStop}%
\bibitem [{\citenamefont {G\'{a}cs}(1983)}]{bib:G83}%
  \BibitemOpen
  \bibfield  {author} {\bibinfo {author} {\bibnamefont {G\'{a}cs},
  \bibfnamefont {P}}} (\bibinfo {year} {1983}),\ \bibfield  {title} {\enquote
  {\bibinfo {title} {{Reliable computation with cellular automata}},}\
  }\href@noop {} {\bibfield  {journal} {\bibinfo  {journal} {Proc. ACM Symp.
  Th. Comput.}\ }\textbf {\bibinfo {volume} {15}},\ \bibinfo {pages}
  {32}}\BibitemShut {NoStop}%
\bibitem [{\citenamefont {Gaebler}\ \emph {et~al.}(2016)\citenamefont
  {Gaebler}, \citenamefont {Tan}, \citenamefont {Lin}, \citenamefont {Wan},
  \citenamefont {Bowler}, \citenamefont {Keith}, \citenamefont {Glancy},
  \citenamefont {Coakley}, \citenamefont {Knill}, \citenamefont {Leibfried},\
  and\ \citenamefont {Wineland}}]{bib:Gaebler2016}%
  \BibitemOpen
  \bibfield  {author} {\bibinfo {author} {\bibnamefont {Gaebler}, \bibfnamefont
  {J~P}}, \bibinfo {author} {\bibfnamefont {T.~R.}\ \bibnamefont {Tan}},
  \bibinfo {author} {\bibfnamefont {Y.}~\bibnamefont {Lin}}, \bibinfo {author}
  {\bibfnamefont {Y.}~\bibnamefont {Wan}}, \bibinfo {author} {\bibfnamefont
  {R.}~\bibnamefont {Bowler}}, \bibinfo {author} {\bibfnamefont {A.~C.}\
  \bibnamefont {Keith}}, \bibinfo {author} {\bibfnamefont {S.}~\bibnamefont
  {Glancy}}, \bibinfo {author} {\bibfnamefont {K.}~\bibnamefont {Coakley}},
  \bibinfo {author} {\bibfnamefont {E.}~\bibnamefont {Knill}}, \bibinfo
  {author} {\bibfnamefont {D.}~\bibnamefont {Leibfried}}, and\ \bibinfo
  {author} {\bibfnamefont {D.~J.}\ \bibnamefont {Wineland}}} (\bibinfo {year}
  {2016}),\ \bibfield  {title} {\enquote {\bibinfo {title} {High-fidelity
  universal gate set for ${^{9}\mathrm{Be}}^{+}$ ion qubits},}\ }\href
  {https://doi.org/10.1103/PhysRevLett.117.060505} {\bibfield  {journal}
  {\bibinfo  {journal} {Physical Review Letters}\ }\textbf {\bibinfo {volume}
  {117}},\ \bibinfo {pages} {060505}}\BibitemShut {NoStop}%
\bibitem [{\citenamefont {Gambetta}\ \emph {et~al.}(2017)\citenamefont
  {Gambetta}, \citenamefont {Chow},\ and\ \citenamefont
  {Steffen}}]{bib:gambetta2017building}%
  \BibitemOpen
  \bibfield  {author} {\bibinfo {author} {\bibnamefont {Gambetta},
  \bibfnamefont {Jay~M}}, \bibinfo {author} {\bibfnamefont {Jerry~M}\
  \bibnamefont {Chow}}, and\ \bibinfo {author} {\bibfnamefont {Matthias}\
  \bibnamefont {Steffen}}} (\bibinfo {year} {2017}),\ \bibfield  {title}
  {\enquote {\bibinfo {title} {Building logical qubits in a superconducting
  quantum computing system},}\ }\href
  {https://doi.org/10.1038/s41534-016-0004-0} {\bibfield  {journal} {\bibinfo
  {journal} {NPJ Quantum Information}\ }\textbf {\bibinfo {volume} {3}},\
  \bibinfo {pages} {2}},\ \Eprint {https://arxiv.org/abs/arXiv:1510.04375v1}
  {arXiv:1510.04375v1} \BibitemShut {NoStop}%
\bibitem [{\citenamefont {Gao}\ \emph {et~al.}(2018)\citenamefont {Gao},
  \citenamefont {Qiao}, \citenamefont {Jiao}, \citenamefont {Ma}, \citenamefont
  {Hu}, \citenamefont {Ren}, \citenamefont {Yang}, \citenamefont {Tang},
  \citenamefont {Yung},\ and\ \citenamefont {Jin}}]{bib:Gao2018}%
  \BibitemOpen
  \bibfield  {author} {\bibinfo {author} {\bibnamefont {Gao}, \bibfnamefont
  {Jun}}, \bibinfo {author} {\bibfnamefont {Lu-Feng}\ \bibnamefont {Qiao}},
  \bibinfo {author} {\bibfnamefont {Zhi-Qiang}\ \bibnamefont {Jiao}}, \bibinfo
  {author} {\bibfnamefont {Yue-Chi}\ \bibnamefont {Ma}}, \bibinfo {author}
  {\bibfnamefont {Cheng-Qiu}\ \bibnamefont {Hu}}, \bibinfo {author}
  {\bibfnamefont {Ruo-Jing}\ \bibnamefont {Ren}}, \bibinfo {author}
  {\bibfnamefont {Ai-Lin}\ \bibnamefont {Yang}}, \bibinfo {author}
  {\bibfnamefont {Hao}\ \bibnamefont {Tang}}, \bibinfo {author} {\bibfnamefont
  {Man-Hong}\ \bibnamefont {Yung}}, and\ \bibinfo {author} {\bibfnamefont
  {Xian-Min}\ \bibnamefont {Jin}}} (\bibinfo {year} {2018}),\ \bibfield
  {title} {\enquote {\bibinfo {title} {Experimental machine learning of quantum
  states},}\ }\href {https://doi.org/10.1103/physrevlett.120.240501} {\bibfield
   {journal} {\bibinfo  {journal} {Physical Review Letters}\ }\textbf {\bibinfo
  {volume} {120}},\ 10.1103/physrevlett.120.240501}\BibitemShut {NoStop}%
\bibitem [{\citenamefont {Gao}\ \emph {et~al.}(2013)\citenamefont {Gao},
  \citenamefont {Fallahi}, \citenamefont {Togan}, \citenamefont {Delteil},
  \citenamefont {Chin}, \citenamefont {Miguel-Sanchez},\ and\ \citenamefont
  {Imamo{\u{g}}lu}}]{gao2013quantum}%
  \BibitemOpen
  \bibfield  {author} {\bibinfo {author} {\bibnamefont {Gao}, \bibfnamefont
  {WB}}, \bibinfo {author} {\bibfnamefont {Parisa}\ \bibnamefont {Fallahi}},
  \bibinfo {author} {\bibfnamefont {Emre}\ \bibnamefont {Togan}}, \bibinfo
  {author} {\bibfnamefont {Aymeric}\ \bibnamefont {Delteil}}, \bibinfo {author}
  {\bibfnamefont {YS}~\bibnamefont {Chin}}, \bibinfo {author} {\bibfnamefont
  {Javier}\ \bibnamefont {Miguel-Sanchez}}, and\ \bibinfo {author}
  {\bibfnamefont {A}~\bibnamefont {Imamo{\u{g}}lu}}} (\bibinfo {year} {2013}),\
  \bibfield  {title} {\enquote {\bibinfo {title} {Quantum teleportation from a
  propagating photon to a solid-state spin qubit},}\ }\href@noop {} {\bibfield
  {journal} {\bibinfo  {journal} {Nature communications}\ }\textbf {\bibinfo
  {volume} {4}}~(\bibinfo {number} {1}),\ \bibinfo {pages} {2744}}\BibitemShut
  {NoStop}%
\bibitem [{\citenamefont {Gao}\ \emph {et~al.}(2012)\citenamefont {Gao},
  \citenamefont {Fallahi}, \citenamefont {Togan}, \citenamefont
  {Miguel-S{\'a}nchez},\ and\ \citenamefont
  {Imamoglu}}]{bib:gao2012observation}%
  \BibitemOpen
  \bibfield  {author} {\bibinfo {author} {\bibnamefont {Gao}, \bibfnamefont
  {WB}}, \bibinfo {author} {\bibfnamefont {Parisa}\ \bibnamefont {Fallahi}},
  \bibinfo {author} {\bibfnamefont {Emre}\ \bibnamefont {Togan}}, \bibinfo
  {author} {\bibfnamefont {Javier}\ \bibnamefont {Miguel-S{\'a}nchez}}, and\
  \bibinfo {author} {\bibfnamefont {Atac}\ \bibnamefont {Imamoglu}}} (\bibinfo
  {year} {2012}),\ \bibfield  {title} {\enquote {\bibinfo {title} {Observation
  of entanglement between a quantum dot spin and a single photon},}\ }\href
  {https://doi.org/10.1038/nature11573} {\bibfield  {journal} {\bibinfo
  {journal} {Nature}\ }\textbf {\bibinfo {volume} {491}},\ \bibinfo {pages}
  {426}}\BibitemShut {NoStop}%
\bibitem [{\citenamefont {Gard}\ \emph {et~al.}(2015)\citenamefont {Gard},
  \citenamefont {Motes}, \citenamefont {Olson}, \citenamefont {Rohde},\ and\
  \citenamefont {Dowling}}]{bib:RohdeIntroBS15}%
  \BibitemOpen
  \bibfield  {author} {\bibinfo {author} {\bibnamefont {Gard}, \bibfnamefont
  {Bryan~T}}, \bibinfo {author} {\bibfnamefont {Keith~R.}\ \bibnamefont
  {Motes}}, \bibinfo {author} {\bibfnamefont {Jonathan~P.}\ \bibnamefont
  {Olson}}, \bibinfo {author} {\bibfnamefont {Peter~P.}\ \bibnamefont {Rohde}},
  and\ \bibinfo {author} {\bibfnamefont {Jonathan~P.}\ \bibnamefont {Dowling}}}
  (\bibinfo {year} {2015}),\ \enquote {\bibinfo {title} {An introduction to
  boson-sampling},}\ in\ \href {https://doi.org/10.1142/9789814678704_0008}
  {\emph {\bibinfo {booktitle} {From Atomic to Mesoscale: The Role of Quantum
  Coherence in Systems of Various Complexities}}}\ (\bibinfo  {publisher}
  {World Scientific Publishing})\ p.\ \bibinfo {pages} {Chapter 8},\ \Eprint
  {https://arxiv.org/abs/arXiv:1406.6767v1} {arXiv:1406.6767v1} \BibitemShut
  {NoStop}%
\bibitem [{\citenamefont {Garnerone}\ \emph {et~al.}(2012)\citenamefont
  {Garnerone}, \citenamefont {Zanardi},\ and\ \citenamefont
  {Lidar}}]{bib:PhysRevLett.108.230506}%
  \BibitemOpen
  \bibfield  {author} {\bibinfo {author} {\bibnamefont {Garnerone},
  \bibfnamefont {Silvano}}, \bibinfo {author} {\bibfnamefont {Paolo}\
  \bibnamefont {Zanardi}}, and\ \bibinfo {author} {\bibfnamefont {Daniel~A.}\
  \bibnamefont {Lidar}}} (\bibinfo {year} {2012}),\ \bibfield  {title}
  {\enquote {\bibinfo {title} {Adiabatic quantum algorithm for search engine
  ranking},}\ }\href {https://doi.org/10.1103/PhysRevLett.108.230506}
  {\bibfield  {journal} {\bibinfo  {journal} {Physical Review Letters}\
  }\textbf {\bibinfo {volume} {108}},\ \bibinfo {pages} {230506}},\ \Eprint
  {https://arxiv.org/abs/arXiv:1109.6546v5} {arXiv:1109.6546v5} \BibitemShut
  {NoStop}%
\bibitem [{\citenamefont {Gehrz}\ \emph {et~al.}(2007)\citenamefont {Gehrz},
  \citenamefont {Roellig}, \citenamefont {Werner}, \citenamefont {Fazio},
  \citenamefont {Houck}, \citenamefont {Low}, \citenamefont {Rieke},
  \citenamefont {Soifer}, \citenamefont {Levine},\ and\ \citenamefont
  {Romana}}]{SD-Gehrz2007}%
  \BibitemOpen
  \bibfield  {author} {\bibinfo {author} {\bibnamefont {Gehrz}, \bibfnamefont
  {RD}}, \bibinfo {author} {\bibfnamefont {T.L.}\ \bibnamefont {Roellig}},
  \bibinfo {author} {\bibfnamefont {M.W.}\ \bibnamefont {Werner}}, \bibinfo
  {author} {\bibfnamefont {G.G.}\ \bibnamefont {Fazio}}, \bibinfo {author}
  {\bibfnamefont {J.R.}\ \bibnamefont {Houck}}, \bibinfo {author}
  {\bibfnamefont {F.J.}\ \bibnamefont {Low}}, \bibinfo {author} {\bibfnamefont
  {G.H.}\ \bibnamefont {Rieke}}, \bibinfo {author} {\bibfnamefont {B.T.}\
  \bibnamefont {Soifer}}, \bibinfo {author} {\bibfnamefont {D.A.}\ \bibnamefont
  {Levine}}, and\ \bibinfo {author} {\bibfnamefont {E.A.}\ \bibnamefont
  {Romana}}} (\bibinfo {year} {2007}),\ \bibfield  {title} {\enquote {\bibinfo
  {title} {{The NASA Spitzer Space Telescope}},}\ }\href
  {https://doi.org/10.1063/1.2431313} {\bibfield  {journal} {\bibinfo
  {journal} {Review of Scientific Instruments}\ }\textbf {\bibinfo {volume}
  {78}},\ \bibinfo {pages} {011302}}\BibitemShut {NoStop}%
\bibitem [{\citenamefont {Gentry}(2009{\natexlab{a}})}]{bib:gentry2009fully}%
  \BibitemOpen
  \bibfield  {author} {\bibinfo {author} {\bibnamefont {Gentry}, \bibfnamefont
  {C}}} (\bibinfo {year} {2009}{\natexlab{a}}),\ \bibfield  {title} {\enquote
  {\bibinfo {title} {Fully homomorphic encryption using ideal lattices},}\ }in\
  \href {https://doi.org/10.1145/1536414.1536440} {\emph {\bibinfo {booktitle}
  {ACM symposium on theory of computing}}},\ Vol.~\bibinfo {volume} {41},\ p.\
  \bibinfo {pages} {169}\BibitemShut {NoStop}%
\bibitem [{\citenamefont {Gentry}(2009{\natexlab{b}})}]{bib:Gentrythesis}%
  \BibitemOpen
  \bibfield  {author} {\bibinfo {author} {\bibnamefont {Gentry}, \bibfnamefont
  {Craig}}} (\bibinfo {year} {2009}{\natexlab{b}}),\ \emph {\bibinfo {title} {A
  fully homomorphic encryption scheme}},\ \href
  {http://crypto.stanford.edu/craig} {Ph.D. thesis}\ (\bibinfo  {school}
  {Stanford University})\BibitemShut {NoStop}%
\bibitem [{\citenamefont {Gentry}\ \emph {et~al.}(2012)\citenamefont {Gentry},
  \citenamefont {Halevi},\ and\ \citenamefont {Smart}}]{bib:Craig2012}%
  \BibitemOpen
  \bibfield  {author} {\bibinfo {author} {\bibnamefont {Gentry}, \bibfnamefont
  {Craig}}, \bibinfo {author} {\bibfnamefont {Shai}\ \bibnamefont {Halevi}},
  and\ \bibinfo {author} {\bibfnamefont {Nigel~P.}\ \bibnamefont {Smart}}}
  (\bibinfo {year} {2012}),\ \bibfield  {title} {\enquote {\bibinfo {title}
  {Fully homomorphic encryption with polylog overhead},}\ }in\ \href
  {https://doi.org/10.1007/978-3-642-29011-4_28} {\emph {\bibinfo {booktitle}
  {Advances in Cryptology -- EUROCRYPT}}},\ \bibinfo {editor} {edited by\
  \bibinfo {editor} {\bibfnamefont {David}\ \bibnamefont {Pointcheval}}\ and\
  \bibinfo {editor} {\bibfnamefont {Thomas}\ \bibnamefont {Johansson}}},\ p.\
  \bibinfo {pages} {465}\BibitemShut {NoStop}%
\bibitem [{\citenamefont {Gerhardt}\ \emph {et~al.}(2011)\citenamefont
  {Gerhardt}, \citenamefont {Liu}, \citenamefont {Lamas-Linares}, \citenamefont
  {Skaar}, \citenamefont {Kurtsiefer},\ and\ \citenamefont
  {Makarov}}]{bib:gerhardt2011full}%
  \BibitemOpen
  \bibfield  {author} {\bibinfo {author} {\bibnamefont {Gerhardt},
  \bibfnamefont {Ilja}}, \bibinfo {author} {\bibfnamefont {Qin}\ \bibnamefont
  {Liu}}, \bibinfo {author} {\bibfnamefont {Antia}\ \bibnamefont
  {Lamas-Linares}}, \bibinfo {author} {\bibfnamefont {Johannes}\ \bibnamefont
  {Skaar}}, \bibinfo {author} {\bibfnamefont {Christian}\ \bibnamefont
  {Kurtsiefer}}, and\ \bibinfo {author} {\bibfnamefont {Vadim}\ \bibnamefont
  {Makarov}}} (\bibinfo {year} {2011}),\ \bibfield  {title} {\enquote {\bibinfo
  {title} {Full-field implementation of a perfect eavesdropper on a quantum
  cryptography system},}\ }\href {https://doi.org/10.1038/ncomms1348}
  {\bibfield  {journal} {\bibinfo  {journal} {Nature Communications}\ }\textbf
  {\bibinfo {volume} {2}},\ \bibinfo {pages} {349}},\ \Eprint
  {https://arxiv.org/abs/arXiv:1011.0105v2} {arXiv:1011.0105v2} \BibitemShut
  {NoStop}%
\bibitem [{\citenamefont {Gerry}\ and\ \citenamefont
  {Knight}(2005)}]{bib:GerryKnight05}%
  \BibitemOpen
  \bibfield  {author} {\bibinfo {author} {\bibnamefont {Gerry}, \bibfnamefont
  {Christopher~C}}, and\ \bibinfo {author} {\bibfnamefont {Peter~L.}\
  \bibnamefont {Knight}}} (\bibinfo {year} {2005}),\ \href@noop {} {\emph
  {\bibinfo {title} {Introductory quantum optics}}}\ (\bibinfo  {publisher}
  {Cambridge University Press})\BibitemShut {NoStop}%
\bibitem [{\citenamefont {Gibney}(2016)}]{bib:gibney2016one}%
  \BibitemOpen
  \bibfield  {author} {\bibinfo {author} {\bibnamefont {Gibney}, \bibfnamefont
  {Elizabeth}}} (\bibinfo {year} {2016}),\ \bibfield  {title} {\enquote
  {\bibinfo {title} {One giant step for quantum internet},}\ }\href@noop {}
  {\bibfield  {journal} {\bibinfo  {journal} {Nature}\ }\textbf {\bibinfo
  {volume} {535}},\ \bibinfo {pages} {478}}\BibitemShut {NoStop}%
\bibitem [{\citenamefont {Gilchrist}\ \emph {et~al.}(2007)\citenamefont
  {Gilchrist}, \citenamefont {Hayes},\ and\ \citenamefont
  {Ralph}}]{bib:GilchristHayes05}%
  \BibitemOpen
  \bibfield  {author} {\bibinfo {author} {\bibnamefont {Gilchrist},
  \bibfnamefont {A}}, \bibinfo {author} {\bibfnamefont {A.~J.~F.}\ \bibnamefont
  {Hayes}}, and\ \bibinfo {author} {\bibfnamefont {T.~C.}\ \bibnamefont
  {Ralph}}} (\bibinfo {year} {2007}),\ \bibfield  {title} {\enquote {\bibinfo
  {title} {Efficient parity encoded optical quantum computing},}\ }\href
  {https://doi.org/10.1103/physreva.75.052328} {\bibfield  {journal} {\bibinfo
  {journal} {Physical Review A}\ }\textbf {\bibinfo {volume} {75}},\ \bibinfo
  {pages} {052328}},\ \Eprint {https://arxiv.org/abs/arXiv:quant-ph/0505125v1}
  {arXiv:quant-ph/0505125v1} \BibitemShut {NoStop}%
\bibitem [{\citenamefont {Gilchrist}\ \emph {et~al.}(2004)\citenamefont
  {Gilchrist}, \citenamefont {Nemoto}, \citenamefont {Munro}, \citenamefont
  {Ralph}, \citenamefont {Glancy}, \citenamefont {Braunstein},\ and\
  \citenamefont {Milburn}}]{bib:Gilchrist04}%
  \BibitemOpen
  \bibfield  {author} {\bibinfo {author} {\bibnamefont {Gilchrist},
  \bibfnamefont {A}}, \bibinfo {author} {\bibfnamefont {Kae}\ \bibnamefont
  {Nemoto}}, \bibinfo {author} {\bibfnamefont {W.~J.}\ \bibnamefont {Munro}},
  \bibinfo {author} {\bibfnamefont {T.~C.}\ \bibnamefont {Ralph}}, \bibinfo
  {author} {\bibfnamefont {S.}~\bibnamefont {Glancy}}, \bibinfo {author}
  {\bibfnamefont {Samuel~L.}\ \bibnamefont {Braunstein}}, and\ \bibinfo
  {author} {\bibfnamefont {G.~J.}\ \bibnamefont {Milburn}}} (\bibinfo {year}
  {2004}),\ \bibfield  {title} {\enquote {\bibinfo {title} {Schr{\"o}dinger
  cats and their power for quantum information processing},}\ }\href
  {https://doi.org/10.1088/1464-4266/6/8/032} {\bibfield  {journal} {\bibinfo
  {journal} {Journal of Optics B}\ }\textbf {\bibinfo {volume} {6}},\ \bibinfo
  {pages} {S828}}\BibitemShut {NoStop}%
\bibitem [{\citenamefont {Gilchrist}\ \emph {et~al.}(2005)\citenamefont
  {Gilchrist}, \citenamefont {K.Langford},\ and\ \citenamefont
  {Nielsen}}]{bib:Gilchrist05}%
  \BibitemOpen
  \bibfield  {author} {\bibinfo {author} {\bibnamefont {Gilchrist},
  \bibfnamefont {Alexei}}, \bibinfo {author} {\bibfnamefont {Nathan}\
  \bibnamefont {K.Langford}}, and\ \bibinfo {author} {\bibfnamefont
  {Michael~A.}\ \bibnamefont {Nielsen}}} (\bibinfo {year} {2005}),\ \bibfield
  {title} {\enquote {\bibinfo {title} {Distance measures to compare real and
  ideal quantum processes},}\ }\href
  {https://doi.org/10.1103/physreva.71.062310} {\bibfield  {journal} {\bibinfo
  {journal} {Physical Review A}\ }\textbf {\bibinfo {volume} {71}},\ \bibinfo
  {pages} {062310}},\ \Eprint {https://arxiv.org/abs/arXiv:quant-ph/0408063v2}
  {arXiv:quant-ph/0408063v2} \BibitemShut {NoStop}%
\bibitem [{\citenamefont {Gilmer}\ \emph {et~al.}(2018)\citenamefont {Gilmer},
  \citenamefont {Metz}, \citenamefont {Faghri}, \citenamefont {Schoenholz},
  \citenamefont {Raghu}, \citenamefont {Wattenberg},\ and\ \citenamefont
  {Goodfellow}}]{bib:gilmer2018adversarial}%
  \BibitemOpen
  \bibfield  {author} {\bibinfo {author} {\bibnamefont {Gilmer}, \bibfnamefont
  {Justin}}, \bibinfo {author} {\bibfnamefont {Luke}\ \bibnamefont {Metz}},
  \bibinfo {author} {\bibfnamefont {Fartash}\ \bibnamefont {Faghri}}, \bibinfo
  {author} {\bibfnamefont {Samuel~S.}\ \bibnamefont {Schoenholz}}, \bibinfo
  {author} {\bibfnamefont {Maithra}\ \bibnamefont {Raghu}}, \bibinfo {author}
  {\bibfnamefont {Martin}\ \bibnamefont {Wattenberg}}, and\ \bibinfo {author}
  {\bibfnamefont {Ian}\ \bibnamefont {Goodfellow}}} (\bibinfo {year} {2018}),\
  \bibfield  {title} {\enquote {\bibinfo {title} {Adversarial spheres},}\
  }\href@noop {} {\ }\Eprint {https://arxiv.org/abs/arXiv:1801.02774}
  {arXiv:1801.02774} \BibitemShut {NoStop}%
\bibitem [{\citenamefont {Gily{\'e}n}\ \emph {et~al.}(2018)\citenamefont
  {Gily{\'e}n}, \citenamefont {Lloyd},\ and\ \citenamefont
  {Tang}}]{bib:gilyen2018quantum}%
  \BibitemOpen
  \bibfield  {author} {\bibinfo {author} {\bibnamefont {Gily{\'e}n},
  \bibfnamefont {Andr{\'a}s}}, \bibinfo {author} {\bibfnamefont {Seth}\
  \bibnamefont {Lloyd}}, and\ \bibinfo {author} {\bibfnamefont {Ewin}\
  \bibnamefont {Tang}}} (\bibinfo {year} {2018}),\ \bibfield  {title} {\enquote
  {\bibinfo {title} {Quantum-inspired low-rank stochastic regression with
  logarithmic dependence on the dimension},}\ }\href@noop {} {\ }\Eprint
  {https://arxiv.org/abs/arXiv:1811.04909} {arXiv:1811.04909} \BibitemShut
  {NoStop}%
\bibitem [{\citenamefont {Gimeno-Segovia}\ \emph
  {et~al.}(2015{\natexlab{a}})\citenamefont {Gimeno-Segovia}, \citenamefont
  {Shadbolt}, \citenamefont {Browne},\ and\ \citenamefont
  {Rudolph}}]{SD-Gimeno-Segovia:2015aa}%
  \BibitemOpen
  \bibfield  {author} {\bibinfo {author} {\bibnamefont {Gimeno-Segovia},
  \bibfnamefont {Mercedes}}, \bibinfo {author} {\bibfnamefont {Pete}\
  \bibnamefont {Shadbolt}}, \bibinfo {author} {\bibfnamefont {Dan~E.}\
  \bibnamefont {Browne}}, and\ \bibinfo {author} {\bibfnamefont {Terry}\
  \bibnamefont {Rudolph}}} (\bibinfo {year} {2015}{\natexlab{a}}),\ \bibfield
  {title} {\enquote {\bibinfo {title} {From three-photon
  greenberger-horne-zeilinger states to ballistic universal quantum
  computation},}\ }\href {https://doi.org/10.1103/PhysRevLett.115.020502}
  {\bibfield  {journal} {\bibinfo  {journal} {Physical Review Letters}\
  }\textbf {\bibinfo {volume} {115}},\ \bibinfo {pages} {020502}}\BibitemShut
  {NoStop}%
\bibitem [{\citenamefont {Gimeno-Segovia}\ \emph
  {et~al.}(2015{\natexlab{b}})\citenamefont {Gimeno-Segovia}, \citenamefont
  {Shadbolt}, \citenamefont {Browne},\ and\ \citenamefont
  {Rudolph}}]{bib:Gimeno-Segovia:2015aa}%
  \BibitemOpen
  \bibfield  {author} {\bibinfo {author} {\bibnamefont {Gimeno-Segovia},
  \bibfnamefont {Mercedes}}, \bibinfo {author} {\bibfnamefont {Pete}\
  \bibnamefont {Shadbolt}}, \bibinfo {author} {\bibfnamefont {Dan~E.}\
  \bibnamefont {Browne}}, and\ \bibinfo {author} {\bibfnamefont {Terry}\
  \bibnamefont {Rudolph}}} (\bibinfo {year} {2015}{\natexlab{b}}),\ \bibfield
  {title} {\enquote {\bibinfo {title} {From three-photon
  greenberger-horne-zeilinger states to ballistic universal quantum
  computation},}\ }\href {https://doi.org/10.1103/PhysRevLett.115.020502}
  {\bibfield  {journal} {\bibinfo  {journal} {Physical Review Letters}\
  }\textbf {\bibinfo {volume} {115}}~(\bibinfo {number} {2}),\ \bibinfo {pages}
  {020502--}}\BibitemShut {NoStop}%
\bibitem [{\citenamefont {Ginestra}\ and\ \citenamefont
  {Barabasi}(2001)}]{bib:BBfitness}%
  \BibitemOpen
  \bibfield  {author} {\bibinfo {author} {\bibnamefont {Ginestra},
  \bibfnamefont {Bianconi}}, and\ \bibinfo {author} {\bibfnamefont {A.~L.}\
  \bibnamefont {Barabasi}}} (\bibinfo {year} {2001}),\ \bibfield  {title}
  {\enquote {\bibinfo {title} {Competition and multiscaling in evolving
  networks},}\ }\href {https://doi.org/10.1209/epl/i2001-00260-6} {\bibfield
  {journal} {\bibinfo  {journal} {Europhysics Letters}\ }\textbf {\bibinfo
  {volume} {54}},\ \bibinfo {pages} {436}},\ \Eprint
  {https://arxiv.org/abs/arXiv:cond-mat/0011029} {arXiv:cond-mat/0011029}
  \BibitemShut {NoStop}%
\bibitem [{\citenamefont {Gingrich}\ \emph {et~al.}(2003)\citenamefont
  {Gingrich}, \citenamefont {Bergou},\ and\ \citenamefont
  {Adami}}]{bib:gingrich03}%
  \BibitemOpen
  \bibfield  {author} {\bibinfo {author} {\bibnamefont {Gingrich},
  \bibfnamefont {Robert~M}}, \bibinfo {author} {\bibfnamefont {Attila.~J.}\
  \bibnamefont {Bergou}}, and\ \bibinfo {author} {\bibfnamefont {Christoph}\
  \bibnamefont {Adami}}} (\bibinfo {year} {2003}),\ \bibfield  {title}
  {\enquote {\bibinfo {title} {Entangled light in moving frames},}\ }\href
  {https://doi.org/10.1364/fio.2003.waa7} {\bibfield  {journal} {\bibinfo
  {journal} {Physical Review A}\ }\textbf {\bibinfo {volume} {68}},\ \bibinfo
  {pages} {042102}}\BibitemShut {NoStop}%
\bibitem [{\citenamefont {Giovannetti}\ \emph
  {et~al.}(2008{\natexlab{a}})\citenamefont {Giovannetti}, \citenamefont
  {Lloyd},\ and\ \citenamefont {Maccone}}]{bib:giovannetti2008architectures}%
  \BibitemOpen
  \bibfield  {author} {\bibinfo {author} {\bibnamefont {Giovannetti},
  \bibfnamefont {Vittorio}}, \bibinfo {author} {\bibfnamefont {Seth}\
  \bibnamefont {Lloyd}}, and\ \bibinfo {author} {\bibfnamefont {Lorenzo}\
  \bibnamefont {Maccone}}} (\bibinfo {year} {2008}{\natexlab{a}}),\ \bibfield
  {title} {\enquote {\bibinfo {title} {Architectures for a quantum random
  access memory},}\ }\href {https://doi.org/10.1103/physreva.78.052310}
  {\bibfield  {journal} {\bibinfo  {journal} {Physical Review A}\ }\textbf
  {\bibinfo {volume} {78}},\ \bibinfo {pages} {052310}}\BibitemShut {NoStop}%
\bibitem [{\citenamefont {Giovannetti}\ \emph
  {et~al.}(2008{\natexlab{b}})\citenamefont {Giovannetti}, \citenamefont
  {Lloyd},\ and\ \citenamefont {Maccone}}]{bib:giovannetti2008quantum}%
  \BibitemOpen
  \bibfield  {author} {\bibinfo {author} {\bibnamefont {Giovannetti},
  \bibfnamefont {Vittorio}}, \bibinfo {author} {\bibfnamefont {Seth}\
  \bibnamefont {Lloyd}}, and\ \bibinfo {author} {\bibfnamefont {Lorenzo}\
  \bibnamefont {Maccone}}} (\bibinfo {year} {2008}{\natexlab{b}}),\ \bibfield
  {title} {\enquote {\bibinfo {title} {Quantum random access memory},}\ }\href
  {https://doi.org/10.1103/physrevlett.100.160501} {\bibfield  {journal}
  {\bibinfo  {journal} {Physical Review Letters}\ }\textbf {\bibinfo {volume}
  {100}},\ \bibinfo {pages} {160501}},\ \Eprint
  {https://arxiv.org/abs/arXiv:0708.1879v2} {arXiv:0708.1879v2} \BibitemShut
  {NoStop}%
\bibitem [{\citenamefont {Gisin}\ and\ \citenamefont
  {Thew}(2007)}]{bib:Gisin2007}%
  \BibitemOpen
  \bibfield  {author} {\bibinfo {author} {\bibnamefont {Gisin}, \bibfnamefont
  {N}}, and\ \bibinfo {author} {\bibfnamefont {R.}~\bibnamefont {Thew}}}
  (\bibinfo {year} {2007}),\ \bibfield  {title} {\enquote {\bibinfo {title}
  {Quantum communication},}\ }\href {https://doi.org/10.1038/nphoton.2007.22}
  {\bibfield  {journal} {\bibinfo  {journal} {Nature Photonics}\ }\textbf
  {\bibinfo {volume} {1}},\ \bibinfo {pages} {165}},\ \Eprint
  {https://arxiv.org/abs/arXiv:quant-ph/0703255v1} {arXiv:quant-ph/0703255v1}
  \BibitemShut {NoStop}%
\bibitem [{\citenamefont {Gisin}\ \emph {et~al.}(2002)\citenamefont {Gisin},
  \citenamefont {Ribordy}, \citenamefont {Tittel},\ and\ \citenamefont
  {Zbinden}}]{bib:Gisin02}%
  \BibitemOpen
  \bibfield  {author} {\bibinfo {author} {\bibnamefont {Gisin}, \bibfnamefont
  {Nicolas}}, \bibinfo {author} {\bibfnamefont {Gr\'egoire}\ \bibnamefont
  {Ribordy}}, \bibinfo {author} {\bibfnamefont {Wolfgang}\ \bibnamefont
  {Tittel}}, and\ \bibinfo {author} {\bibfnamefont {Hugo}\ \bibnamefont
  {Zbinden}}} (\bibinfo {year} {2002}),\ \bibfield  {title} {\enquote {\bibinfo
  {title} {Quantum cryptography},}\ }\href
  {https://doi.org/10.1103/revmodphys.74.145} {\bibfield  {journal} {\bibinfo
  {journal} {Reviews in Modern Physics}\ }\textbf {\bibinfo {volume} {74}},\
  \bibinfo {pages} {145}},\ \Eprint
  {https://arxiv.org/abs/arXiv:quant-ph/0101098v2} {arXiv:quant-ph/0101098v2}
  \BibitemShut {NoStop}%
\bibitem [{\citenamefont {Glancy}\ and\ \citenamefont
  {Knill}(2006)}]{bib:PhysRevA.73.012325}%
  \BibitemOpen
  \bibfield  {author} {\bibinfo {author} {\bibnamefont {Glancy}, \bibfnamefont
  {S}}, and\ \bibinfo {author} {\bibfnamefont {E.}~\bibnamefont {Knill}}}
  (\bibinfo {year} {2006}),\ \bibfield  {title} {\enquote {\bibinfo {title}
  {Error analysis for encoding a qubit in an oscillator},}\ }\href
  {https://doi.org/10.1103/PhysRevA.73.012325} {\bibfield  {journal} {\bibinfo
  {journal} {Physical Review A}\ }\textbf {\bibinfo {volume} {73}},\ \bibinfo
  {pages} {012325}}\BibitemShut {NoStop}%
\bibitem [{\citenamefont {Goebel}\ \emph {et~al.}(2008)\citenamefont {Goebel},
  \citenamefont {Wagenknecht}, \citenamefont {Zhang}, \citenamefont {Chen},
  \citenamefont {Chen}, \citenamefont {Schmiedmayer},\ and\ \citenamefont
  {Pan}}]{bib:goebel08}%
  \BibitemOpen
  \bibfield  {author} {\bibinfo {author} {\bibnamefont {Goebel}, \bibfnamefont
  {A~M}}, \bibinfo {author} {\bibfnamefont {G.}~\bibnamefont {Wagenknecht}},
  \bibinfo {author} {\bibfnamefont {Q.}~\bibnamefont {Zhang}}, \bibinfo
  {author} {\bibfnamefont {Y.}~\bibnamefont {Chen}}, \bibinfo {author}
  {\bibfnamefont {K.}~\bibnamefont {Chen}}, \bibinfo {author} {\bibfnamefont
  {J.}~\bibnamefont {Schmiedmayer}}, and\ \bibinfo {author} {\bibfnamefont
  {J.~W.}\ \bibnamefont {Pan}}} (\bibinfo {year} {2008}),\ \bibfield  {title}
  {\enquote {\bibinfo {title} {Multistage entanglement swapping},}\ }\href
  {https://doi.org/10.1103/physrevlett.101.080403} {\bibfield  {journal}
  {\bibinfo  {journal} {Physical Review Letters}\ }\textbf {\bibinfo {volume}
  {101}},\ \bibinfo {pages} {080403}},\ \Eprint
  {https://arxiv.org/abs/arXiv:0808.2972v1} {arXiv:0808.2972v1} \BibitemShut
  {NoStop}%
\bibitem [{\citenamefont {Goldberg}\ \emph {et~al.}(1989)\citenamefont
  {Goldberg}, \citenamefont {Tardos},\ and\ \citenamefont
  {Tarjan}}]{goldberg1989network}%
  \BibitemOpen
  \bibfield  {author} {\bibinfo {author} {\bibnamefont {Goldberg},
  \bibfnamefont {Andrew~V}}, \bibinfo {author} {\bibfnamefont {{\'E}va}\
  \bibnamefont {Tardos}}, and\ \bibinfo {author} {\bibfnamefont {Robert}\
  \bibnamefont {Tarjan}}} (\bibinfo {year} {1989}),\ \href@noop {} {\emph
  {\bibinfo {title} {Network flow algorithm}}},\ \bibinfo {type} {Tech. Rep.}\
  (\bibinfo  {institution} {Cornell University Operations Research and
  Industrial Engineering})\BibitemShut {NoStop}%
\bibitem [{\citenamefont {Goldschmidt}\ \emph {et~al.}(2008)\citenamefont
  {Goldschmidt}, \citenamefont {Eisaman}, \citenamefont {Fan}, \citenamefont
  {Polyakov},\ and\ \citenamefont {Migdall}}]{bib:goldschmidt2008}%
  \BibitemOpen
  \bibfield  {author} {\bibinfo {author} {\bibnamefont {Goldschmidt},
  \bibfnamefont {Elizabeth~A}}, \bibinfo {author} {\bibfnamefont {Matthew~D}\
  \bibnamefont {Eisaman}}, \bibinfo {author} {\bibfnamefont {Jingyun}\
  \bibnamefont {Fan}}, \bibinfo {author} {\bibfnamefont {Sergey~V}\
  \bibnamefont {Polyakov}}, and\ \bibinfo {author} {\bibfnamefont {Alan}\
  \bibnamefont {Migdall}}} (\bibinfo {year} {2008}),\ \bibfield  {title}
  {\enquote {\bibinfo {title} {Spectrally bright and broad fiber-based heralded
  single-photon source},}\ }\href {https://doi.org/10.1364/icqi.2008.qwc4}
  {\bibfield  {journal} {\bibinfo  {journal} {Physical Review A}\ }\textbf
  {\bibinfo {volume} {78}},\ \bibinfo {pages} {013844}}\BibitemShut {NoStop}%
\bibitem [{\citenamefont {Gong}\ \emph {et~al.}(2018)\citenamefont {Gong},
  \citenamefont {Chen}, \citenamefont {Zheng}, \citenamefont {Wang},
  \citenamefont {Zha}, \citenamefont {Deng}, \citenamefont {Yan}, \citenamefont
  {Rong}, \citenamefont {Wu}, \citenamefont {Li} \emph
  {et~al.}}]{bib:gong2018genuine}%
  \BibitemOpen
  \bibfield  {author} {\bibinfo {author} {\bibnamefont {Gong}, \bibfnamefont
  {Ming}}, \bibinfo {author} {\bibfnamefont {Ming-Cheng}\ \bibnamefont {Chen}},
  \bibinfo {author} {\bibfnamefont {Yarui}\ \bibnamefont {Zheng}}, \bibinfo
  {author} {\bibfnamefont {Shiyu}\ \bibnamefont {Wang}}, \bibinfo {author}
  {\bibfnamefont {Chen}\ \bibnamefont {Zha}}, \bibinfo {author} {\bibfnamefont
  {Hui}\ \bibnamefont {Deng}}, \bibinfo {author} {\bibfnamefont {Zhiguang}\
  \bibnamefont {Yan}}, \bibinfo {author} {\bibfnamefont {Hao}\ \bibnamefont
  {Rong}}, \bibinfo {author} {\bibfnamefont {Yulin}\ \bibnamefont {Wu}},
  \bibinfo {author} {\bibfnamefont {Shaowei}\ \bibnamefont {Li}},  \emph
  {et~al.}} (\bibinfo {year} {2018}),\ \bibfield  {title} {\enquote {\bibinfo
  {title} {Genuine 12-qubit entanglement on a superconducting quantum
  processor},}\ }\href@noop {} {\ }\Eprint
  {https://arxiv.org/abs/arXiv:1811.02292} {arXiv:1811.02292} \BibitemShut
  {NoStop}%
\bibitem [{\citenamefont {Gonz\'alez}\ \emph {et~al.}(2015)\citenamefont
  {Gonz\'alez}, \citenamefont {Reb\'on}, \citenamefont {Ferreira~da Silva},
  \citenamefont {Figueroa}, \citenamefont {Saavedra}, \citenamefont {Curty},
  \citenamefont {Lima}, \citenamefont {Xavier},\ and\ \citenamefont
  {Nogueira}}]{bib:PhysRevA.92.022337}%
  \BibitemOpen
  \bibfield  {author} {\bibinfo {author} {\bibnamefont {Gonz\'alez},
  \bibfnamefont {P}}, \bibinfo {author} {\bibfnamefont {L.}~\bibnamefont
  {Reb\'on}}, \bibinfo {author} {\bibfnamefont {T.}~\bibnamefont {Ferreira~da
  Silva}}, \bibinfo {author} {\bibfnamefont {M.}~\bibnamefont {Figueroa}},
  \bibinfo {author} {\bibfnamefont {C.}~\bibnamefont {Saavedra}}, \bibinfo
  {author} {\bibfnamefont {M.}~\bibnamefont {Curty}}, \bibinfo {author}
  {\bibfnamefont {G.}~\bibnamefont {Lima}}, \bibinfo {author} {\bibfnamefont
  {G.~B.}\ \bibnamefont {Xavier}}, and\ \bibinfo {author} {\bibfnamefont
  {W.~A.~T.}\ \bibnamefont {Nogueira}}} (\bibinfo {year} {2015}),\ \bibfield
  {title} {\enquote {\bibinfo {title} {Quantum key distribution with untrusted
  detectors},}\ }\href {https://doi.org/10.1103/physreva.92.022337} {\bibfield
  {journal} {\bibinfo  {journal} {Physical Review A}\ }\textbf {\bibinfo
  {volume} {92}},\ \bibinfo {pages} {022337}},\ \Eprint
  {https://arxiv.org/abs/arXiv:1410.1422v3} {arXiv:1410.1422v3} \BibitemShut
  {NoStop}%
\bibitem [{\citenamefont {Goodfellow}\ \emph {et~al.}(2014)\citenamefont
  {Goodfellow}, \citenamefont {Shlens},\ and\ \citenamefont
  {Szegedy}}]{bib:goodfellow2014explaining}%
  \BibitemOpen
  \bibfield  {author} {\bibinfo {author} {\bibnamefont {Goodfellow},
  \bibfnamefont {Ian~J}}, \bibinfo {author} {\bibfnamefont {Jonathon}\
  \bibnamefont {Shlens}}, and\ \bibinfo {author} {\bibfnamefont {Christian}\
  \bibnamefont {Szegedy}}} (\bibinfo {year} {2014}),\ \bibfield  {title}
  {\enquote {\bibinfo {title} {Explaining and harnessing adversarial
  examples},}\ }\href@noop {} {\ }\Eprint
  {https://arxiv.org/abs/arXiv:1412.6572} {arXiv:1412.6572} \BibitemShut
  {NoStop}%
\bibitem [{\citenamefont {Gottesman}(1997)}]{bib:G97+}%
  \BibitemOpen
  \bibfield  {author} {\bibinfo {author} {\bibnamefont {Gottesman},
  \bibfnamefont {D}}} (\bibinfo {year} {1997}),\ \bibfield  {title} {\enquote
  {\bibinfo {title} {{PhD Thesis (Caltech)}},}\ }\href@noop {} {\bibinfo
  {journal} {quant-ph/9705052}\ }\BibitemShut {NoStop}%
\bibitem [{\citenamefont {Gottesman}\ and\ \citenamefont
  {Chuang}(1999)}]{bib:GottesmanChuang99}%
  \BibitemOpen
\bibfield  {journal} {  }\bibfield  {author} {\bibinfo {author} {\bibnamefont
  {Gottesman}, \bibfnamefont {D}}, and\ \bibinfo {author} {\bibfnamefont
  {I.~L.}\ \bibnamefont {Chuang}}} (\bibinfo {year} {1999}),\ \bibfield
  {title} {\enquote {\bibinfo {title} {Demonstrating the viability of universal
  quantum computation using teleportation and single-qubit operations},}\
  }\href {https://doi.org/10.1038/46503} {\bibfield  {journal} {\bibinfo
  {journal} {Nature}\ }\textbf {\bibinfo {volume} {402}},\ \bibinfo {pages}
  {390}}\BibitemShut {NoStop}%
\bibitem [{\citenamefont {Gottesman}(2000)}]{bib:PhysRevA.61.042311}%
  \BibitemOpen
  \bibfield  {author} {\bibinfo {author} {\bibnamefont {Gottesman},
  \bibfnamefont {Daniel}}} (\bibinfo {year} {2000}),\ \bibfield  {title}
  {\enquote {\bibinfo {title} {Theory of quantum secret sharing},}\ }\href
  {https://doi.org/10.1103/physreva.61.042311} {\bibfield  {journal} {\bibinfo
  {journal} {Physical Review A}\ }\textbf {\bibinfo {volume} {61}},\ \bibinfo
  {pages} {042311}}\BibitemShut {NoStop}%
\bibitem [{\citenamefont {Gottesman}\ and\ \citenamefont
  {Chuang}(2001)}]{gottesman2001quantum}%
  \BibitemOpen
  \bibfield  {author} {\bibinfo {author} {\bibnamefont {Gottesman},
  \bibfnamefont {Daniel}}, and\ \bibinfo {author} {\bibfnamefont {Isaac}\
  \bibnamefont {Chuang}}} (\bibinfo {year} {2001}),\ \bibfield  {title}
  {\enquote {\bibinfo {title} {Quantum digital signatures},}\ }\href@noop {}
  {\bibinfo  {journal} {arXiv preprint quant-ph/0105032}\ }\BibitemShut
  {NoStop}%
\bibitem [{\citenamefont {Gottesman}\ \emph {et~al.}(2012)\citenamefont
  {Gottesman}, \citenamefont {Jennewein},\ and\ \citenamefont
  {Croke}}]{bib:PhysRevLett.109.070503}%
  \BibitemOpen
\bibfield  {journal} {  }\bibfield  {author} {\bibinfo {author} {\bibnamefont
  {Gottesman}, \bibfnamefont {Daniel}}, \bibinfo {author} {\bibfnamefont
  {Thomas}\ \bibnamefont {Jennewein}}, and\ \bibinfo {author} {\bibfnamefont
  {Sarah}\ \bibnamefont {Croke}}} (\bibinfo {year} {2012}),\ \bibfield  {title}
  {\enquote {\bibinfo {title} {Longer-baseline telescopes using quantum
  repeaters},}\ }\href {https://doi.org/10.1103/physrevlett.109.070503}
  {\bibfield  {journal} {\bibinfo  {journal} {Physical Review Letters}\
  }\textbf {\bibinfo {volume} {109}},\ \bibinfo {pages} {070503}},\ \Eprint
  {https://arxiv.org/abs/arXiv:1107.2939v2} {arXiv:1107.2939v2} \BibitemShut
  {NoStop}%
\bibitem [{\citenamefont {Gottesman}\ \emph
  {et~al.}(2001{\natexlab{a}})\citenamefont {Gottesman}, \citenamefont
  {Kitaev},\ and\ \citenamefont {Preskill}}]{gottesman2001encoding}%
  \BibitemOpen
  \bibfield  {author} {\bibinfo {author} {\bibnamefont {Gottesman},
  \bibfnamefont {Daniel}}, \bibinfo {author} {\bibfnamefont {Alexei}\
  \bibnamefont {Kitaev}}, and\ \bibinfo {author} {\bibfnamefont {John}\
  \bibnamefont {Preskill}}} (\bibinfo {year} {2001}{\natexlab{a}}),\ \bibfield
  {title} {\enquote {\bibinfo {title} {Encoding a qubit in an oscillator},}\
  }\href {https://doi.org/10.1103/PhysRevA.64.012310} {\bibfield  {journal}
  {\bibinfo  {journal} {Phys. Rev. A}\ }\textbf {\bibinfo {volume} {64}},\
  \bibinfo {pages} {012310}}\BibitemShut {NoStop}%
\bibitem [{\citenamefont {Gottesman}\ \emph
  {et~al.}(2001{\natexlab{b}})\citenamefont {Gottesman}, \citenamefont
  {Kitaev},\ and\ \citenamefont {Preskill}}]{bib:PhysRevA.64.012310}%
  \BibitemOpen
  \bibfield  {author} {\bibinfo {author} {\bibnamefont {Gottesman},
  \bibfnamefont {Daniel}}, \bibinfo {author} {\bibfnamefont {Alexei}\
  \bibnamefont {Kitaev}}, and\ \bibinfo {author} {\bibfnamefont {John}\
  \bibnamefont {Preskill}}} (\bibinfo {year} {2001}{\natexlab{b}}),\ \bibfield
  {title} {\enquote {\bibinfo {title} {Encoding a qubit in an oscillator},}\
  }\href {https://doi.org/10.1103/PhysRevA.64.012310} {\bibfield  {journal}
  {\bibinfo  {journal} {Physical Review A}\ }\textbf {\bibinfo {volume} {64}},\
  \bibinfo {pages} {012310}}\BibitemShut {NoStop}%
\bibitem [{\citenamefont {Greenberger}\ \emph
  {et~al.}(1989{\natexlab{a}})\citenamefont {Greenberger}, \citenamefont
  {Horne},\ and\ \citenamefont {(ed. M.~Kafatos)}}]{bib:GHZ89}%
  \BibitemOpen
  \bibfield  {author} {\bibinfo {author} {\bibnamefont {Greenberger},
  \bibfnamefont {Daniel~M}}, \bibinfo {author} {\bibfnamefont {Michael~A.}\
  \bibnamefont {Horne}}, and\ \bibinfo {author} {\bibfnamefont
  {Anton~Zeilinger}\ \bibnamefont {(ed. M.~Kafatos)}}} (\bibinfo {year}
  {1989}{\natexlab{a}}),\ \enquote {\bibinfo {title} {Going beyond bell's
  theorem},}\ in\ \href {https://doi.org/10.1007/978-94-017-0849-4_10} {\emph
  {\bibinfo {booktitle} {Bell's Theorem, Quantum Theory, and Conceptions of the
  Universe}}}\ (\bibinfo  {publisher} {Kluwer Academic, Dordrecht, The
  Netherlands})\ p.~\bibinfo {pages} {73}\BibitemShut {NoStop}%
\bibitem [{\citenamefont {Greenberger}\ \emph
  {et~al.}(1989{\natexlab{b}})\citenamefont {Greenberger}, \citenamefont
  {Horne},\ and\ \citenamefont {Zeilinger}}]{greenberger1989going}%
  \BibitemOpen
  \bibfield  {author} {\bibinfo {author} {\bibnamefont {Greenberger},
  \bibfnamefont {Daniel~M}}, \bibinfo {author} {\bibfnamefont {Michael~A}\
  \bibnamefont {Horne}}, and\ \bibinfo {author} {\bibfnamefont {Anton}\
  \bibnamefont {Zeilinger}}} (\bibinfo {year} {1989}{\natexlab{b}}),\ \bibfield
   {title} {\enquote {\bibinfo {title} {Going beyond bell’s theorem},}\ }in\
  \href@noop {} {\emph {\bibinfo {booktitle} {Bell’s theorem, quantum theory
  and conceptions of the universe}}}\ (\bibinfo  {publisher} {Springer})\ pp.\
  \bibinfo {pages} {69--72}\BibitemShut {NoStop}%
\bibitem [{\citenamefont {Grimsmo}\ \emph {et~al.}(2020)\citenamefont
  {Grimsmo}, \citenamefont {Combes},\ and\ \citenamefont
  {Baragiola}}]{Grimsmo2020rotation}%
  \BibitemOpen
  \bibfield  {author} {\bibinfo {author} {\bibnamefont {Grimsmo}, \bibfnamefont
  {Arne~L}}, \bibinfo {author} {\bibfnamefont {Joshua}\ \bibnamefont {Combes}},
  and\ \bibinfo {author} {\bibfnamefont {Ben~Q.}\ \bibnamefont {Baragiola}}}
  (\bibinfo {year} {2020}),\ \bibfield  {title} {\enquote {\bibinfo {title}
  {Quantum computing with rotation-symmetric bosonic codes},}\ }\href
  {https://doi.org/10.1103/PhysRevX.10.011058} {\bibfield  {journal} {\bibinfo
  {journal} {Phys. Rev. X}\ }\textbf {\bibinfo {volume} {10}},\ \bibinfo
  {pages} {011058}}\BibitemShut {NoStop}%
\bibitem [{\citenamefont {Gross}\ \emph {et~al.}(2006)\citenamefont {Gross},
  \citenamefont {Kieling},\ and\ \citenamefont {Eisert}}]{bib:Gross06}%
  \BibitemOpen
  \bibfield  {author} {\bibinfo {author} {\bibnamefont {Gross}, \bibfnamefont
  {D}}, \bibinfo {author} {\bibfnamefont {K.}~\bibnamefont {Kieling}}, and\
  \bibinfo {author} {\bibfnamefont {J.}~\bibnamefont {Eisert}}} (\bibinfo
  {year} {2006}),\ \bibfield  {title} {\enquote {\bibinfo {title} {Potential
  and limits to cluster state quantum computing using probabilistic gates},}\
  }\href {https://doi.org/10.1103/physreva.74.042343} {\bibfield  {journal}
  {\bibinfo  {journal} {Physical Review A}\ }\textbf {\bibinfo {volume} {74}},\
  \bibinfo {pages} {042343}},\ \Eprint
  {https://arxiv.org/abs/arXiv:quant-ph/0605014v2} {arXiv:quant-ph/0605014v2}
  \BibitemShut {NoStop}%
\bibitem [{\citenamefont {Gross}\ \emph {et~al.}(2010)\citenamefont {Gross},
  \citenamefont {Liu}, \citenamefont {Flammia}, \citenamefont {Becker},\ and\
  \citenamefont {Eisert}}]{gross2010quantum}%
  \BibitemOpen
  \bibfield  {author} {\bibinfo {author} {\bibnamefont {Gross}, \bibfnamefont
  {David}}, \bibinfo {author} {\bibfnamefont {Yi-Kai}\ \bibnamefont {Liu}},
  \bibinfo {author} {\bibfnamefont {Steven~T}\ \bibnamefont {Flammia}},
  \bibinfo {author} {\bibfnamefont {Stephen}\ \bibnamefont {Becker}}, and\
  \bibinfo {author} {\bibfnamefont {Jens}\ \bibnamefont {Eisert}}} (\bibinfo
  {year} {2010}),\ \bibfield  {title} {\enquote {\bibinfo {title} {Quantum
  state tomography via compressed sensing},}\ }\href@noop {} {\bibfield
  {journal} {\bibinfo  {journal} {Physical review letters}\ }\textbf {\bibinfo
  {volume} {105}}~(\bibinfo {number} {15}),\ \bibinfo {pages}
  {150401}}\BibitemShut {NoStop}%
\bibitem [{\citenamefont {Grover}(1996{\natexlab{a}})}]{bib:Grover96}%
  \BibitemOpen
  \bibfield  {author} {\bibinfo {author} {\bibnamefont {Grover}, \bibfnamefont
  {L~K}}} (\bibinfo {year} {1996}{\natexlab{a}}),\ \bibfield  {title} {\enquote
  {\bibinfo {title} {A fast quantum mechanical algorithm for database
  search},}\ }in\ \href {https://doi.org/10.1145/237814.237866} {\emph
  {\bibinfo {booktitle} {Proceedings of the 28th annual ACM symposium on theory
  of computing}}},\ p.\ \bibinfo {pages} {212}\BibitemShut {NoStop}%
\bibitem [{\citenamefont {Grover}(1996{\natexlab{b}})}]{grover1996fast}%
  \BibitemOpen
  \bibfield  {author} {\bibinfo {author} {\bibnamefont {Grover}, \bibfnamefont
  {Lov~K}}} (\bibinfo {year} {1996}{\natexlab{b}}),\ \bibfield  {title}
  {\enquote {\bibinfo {title} {A fast quantum mechanical algorithm for database
  search},}\ }in\ \href@noop {} {\emph {\bibinfo {booktitle} {Proceedings of
  the twenty-eighth annual ACM symposium on Theory of computing}}},\ pp.\
  \bibinfo {pages} {212--219}\BibitemShut {NoStop}%
\bibitem [{\citenamefont {Guha}\ \emph {et~al.}(2014)\citenamefont {Guha},
  \citenamefont {Hayden}, \citenamefont {Krovi}, \citenamefont {Lloyd},
  \citenamefont {Lupo}, \citenamefont {Shapiro}, \citenamefont {Takeoka},\ and\
  \citenamefont {Wilde}}]{bib:guha2014enigma}%
  \BibitemOpen
  \bibfield  {author} {\bibinfo {author} {\bibnamefont {Guha}, \bibfnamefont
  {Saikat}}, \bibinfo {author} {\bibfnamefont {Patrick}\ \bibnamefont
  {Hayden}}, \bibinfo {author} {\bibfnamefont {Hari}\ \bibnamefont {Krovi}},
  \bibinfo {author} {\bibfnamefont {Seth}\ \bibnamefont {Lloyd}}, \bibinfo
  {author} {\bibfnamefont {Cosmo}\ \bibnamefont {Lupo}}, \bibinfo {author}
  {\bibfnamefont {Jeffrey~H.}\ \bibnamefont {Shapiro}}, \bibinfo {author}
  {\bibfnamefont {Masahiro}\ \bibnamefont {Takeoka}}, and\ \bibinfo {author}
  {\bibfnamefont {Mark~M.}\ \bibnamefont {Wilde}}} (\bibinfo {year} {2014}),\
  \bibfield  {title} {\enquote {\bibinfo {title} {Quantum enigma machines and
  the locking capacity of a quantum channel},}\ }\href
  {https://doi.org/10.1103/PhysRevX.4.011016} {\bibfield  {journal} {\bibinfo
  {journal} {Phys. Rev. X}\ }\textbf {\bibinfo {volume} {4}},\ \bibinfo {pages}
  {011016}}\BibitemShut {NoStop}%
\bibitem [{\citenamefont {{Gyongyosi}}\ \emph {et~al.}(2018)\citenamefont
  {{Gyongyosi}}, \citenamefont {{Imre}},\ and\ \citenamefont
  {{Nguyen}}}]{bib:8242350}%
  \BibitemOpen
  \bibfield  {author} {\bibinfo {author} {\bibnamefont {{Gyongyosi}},
  \bibfnamefont {L}}, \bibinfo {author} {\bibfnamefont {S.}~\bibnamefont
  {{Imre}}}, and\ \bibinfo {author} {\bibfnamefont {H.~V.}\ \bibnamefont
  {{Nguyen}}}} (\bibinfo {year} {2018}),\ \bibfield  {title} {\enquote
  {\bibinfo {title} {A survey on quantum channel capacities},}\ }\href
  {https://doi.org/10.1109/COMST.2017.2786748} {\bibfield  {journal} {\bibinfo
  {journal} {IEEE Communications Surveys Tutorials}\ }\textbf {\bibinfo
  {volume} {20}}~(\bibinfo {number} {2}),\ \bibinfo {pages}
  {1149--1205}}\BibitemShut {NoStop}%
\bibitem [{\citenamefont {Hadfield}(2009)}]{bib:hadfield2009}%
  \BibitemOpen
  \bibfield  {author} {\bibinfo {author} {\bibnamefont {Hadfield},
  \bibfnamefont {Robert~H}}} (\bibinfo {year} {2009}),\ \bibfield  {title}
  {\enquote {\bibinfo {title} {Single-photon detectors for optical quantum
  information applications},}\ }\href
  {https://doi.org/10.1038/nphoton.2009.230} {\bibfield  {journal} {\bibinfo
  {journal} {Nature Photonics}\ }\textbf {\bibinfo {volume} {3}},\ \bibinfo
  {pages} {696}}\BibitemShut {NoStop}%
\bibitem [{\citenamefont {H{\"a}ffner}\ \emph {et~al.}(2005)\citenamefont
  {H{\"a}ffner}, \citenamefont {H{\"a}nsel}, \citenamefont {Roos},
  \citenamefont {Benhelm}, \citenamefont {Chwalla}, \citenamefont {K{\"o}rber},
  \citenamefont {Rapol}, \citenamefont {Riebe}, \citenamefont {Schmidt},
  \citenamefont {Becher} \emph {et~al.}}]{bib:haffner2005scalable}%
  \BibitemOpen
  \bibfield  {author} {\bibinfo {author} {\bibnamefont {H{\"a}ffner},
  \bibfnamefont {H}}, \bibinfo {author} {\bibfnamefont {W}~\bibnamefont
  {H{\"a}nsel}}, \bibinfo {author} {\bibfnamefont {CF}~\bibnamefont {Roos}},
  \bibinfo {author} {\bibfnamefont {J}~\bibnamefont {Benhelm}}, \bibinfo
  {author} {\bibfnamefont {M}~\bibnamefont {Chwalla}}, \bibinfo {author}
  {\bibfnamefont {T}~\bibnamefont {K{\"o}rber}}, \bibinfo {author}
  {\bibfnamefont {UD}~\bibnamefont {Rapol}}, \bibinfo {author} {\bibfnamefont
  {M}~\bibnamefont {Riebe}}, \bibinfo {author} {\bibfnamefont {PO}~\bibnamefont
  {Schmidt}}, \bibinfo {author} {\bibfnamefont {C}~\bibnamefont {Becher}},
  \emph {et~al.}} (\bibinfo {year} {2005}),\ \bibfield  {title} {\enquote
  {\bibinfo {title} {Scalable multiparticle entanglement of trapped ions},}\
  }\href {https://doi.org/10.1038/nature04279} {\bibfield  {journal} {\bibinfo
  {journal} {Nature}\ }\textbf {\bibinfo {volume} {438}},\ \bibinfo {pages}
  {643}},\ \Eprint {https://arxiv.org/abs/arXiv:quant-ph/0603217v1}
  {arXiv:quant-ph/0603217v1} \BibitemShut {NoStop}%
\bibitem [{\citenamefont {Hajji}(2005)}]{bib:hajji2005statistical}%
  \BibitemOpen
  \bibfield  {author} {\bibinfo {author} {\bibnamefont {Hajji}, \bibfnamefont
  {Hassan}}} (\bibinfo {year} {2005}),\ \bibfield  {title} {\enquote {\bibinfo
  {title} {Statistical analysis of network traffic for adaptive faults
  detection},}\ }\href {https://doi.org/10.1109/tnn.2005.853414} {\bibfield
  {journal} {\bibinfo  {journal} {IEEE Transactions on Neural Networks}\
  }\textbf {\bibinfo {volume} {16}},\ \bibinfo {pages} {1053}}\BibitemShut
  {NoStop}%
\bibitem [{\citenamefont {Halder}\ \emph {et~al.}(2007)\citenamefont {Halder},
  \citenamefont {Beveratos}, \citenamefont {Gisin}, \citenamefont {Scarani},
  \citenamefont {Simon},\ and\ \citenamefont {Zbinden}}]{bib:Nat_Phys_3_692}%
  \BibitemOpen
  \bibfield  {author} {\bibinfo {author} {\bibnamefont {Halder}, \bibfnamefont
  {Matth{\"a}us}}, \bibinfo {author} {\bibfnamefont {Alexios}\ \bibnamefont
  {Beveratos}}, \bibinfo {author} {\bibfnamefont {Nicolas}\ \bibnamefont
  {Gisin}}, \bibinfo {author} {\bibfnamefont {Valerio}\ \bibnamefont
  {Scarani}}, \bibinfo {author} {\bibfnamefont {Christoph}\ \bibnamefont
  {Simon}}, and\ \bibinfo {author} {\bibfnamefont {Hugo}\ \bibnamefont
  {Zbinden}}} (\bibinfo {year} {2007}),\ \bibfield  {title} {\enquote {\bibinfo
  {title} {Entangling independent photons by time measurement},}\ }\href
  {https://doi.org/10.1038/nphys700} {\bibfield  {journal} {\bibinfo  {journal}
  {Nature Physics}\ }\textbf {\bibinfo {volume} {3}},\ \bibinfo {pages}
  {692}},\ \Eprint {https://arxiv.org/abs/arXiv:0704.0758v1}
  {arXiv:0704.0758v1} \BibitemShut {NoStop}%
\bibitem [{\citenamefont {Halevi}(2017)}]{bib:Halevi2017}%
  \BibitemOpen
  \bibfield  {author} {\bibinfo {author} {\bibnamefont {Halevi}, \bibfnamefont
  {Shai}}} (\bibinfo {year} {2017}),\ \enquote {\bibinfo {title} {Homomorphic
  encryption},}\ in\ \href {https://doi.org/10.1007/978-3-319-57048-8_5} {\emph
  {\bibinfo {booktitle} {Tutorials on the Foundations of Cryptography:
  Dedicated to Oded Goldreich}}},\ \bibinfo {editor} {edited by\ \bibinfo
  {editor} {\bibfnamefont {Yehuda}\ \bibnamefont {Lindell}}}\ (\bibinfo
  {publisher} {Springer International Publishing},\ \bibinfo {address} {Cham})\
  p.\ \bibinfo {pages} {219}\BibitemShut {NoStop}%
\bibitem [{\citenamefont {Han}\ \emph {et~al.}(2010)\citenamefont {Han},
  \citenamefont {He}, \citenamefont {Heshami}, \citenamefont {Li},\ and\
  \citenamefont {Simon}}]{bib:PRA_81_052311}%
  \BibitemOpen
  \bibfield  {author} {\bibinfo {author} {\bibnamefont {Han}, \bibfnamefont
  {Yang}}, \bibinfo {author} {\bibfnamefont {Bing}\ \bibnamefont {He}},
  \bibinfo {author} {\bibfnamefont {Khabat}\ \bibnamefont {Heshami}}, \bibinfo
  {author} {\bibfnamefont {Cheng-Zu}\ \bibnamefont {Li}}, and\ \bibinfo
  {author} {\bibfnamefont {Christoph}\ \bibnamefont {Simon}}} (\bibinfo {year}
  {2010}),\ \bibfield  {title} {\enquote {\bibinfo {title} {Quantum repeaters
  based on rydberg-blockade-coupled atomic ensembles},}\ }\href
  {https://doi.org/10.1103/physreva.81.052311} {\bibfield  {journal} {\bibinfo
  {journal} {Physical Review A}\ }\textbf {\bibinfo {volume} {81}},\ \bibinfo
  {pages} {052311}},\ \Eprint {https://arxiv.org/abs/arXiv:1003.2353v1}
  {arXiv:1003.2353v1} \BibitemShut {NoStop}%
\bibitem [{\citenamefont {Hara}\ \emph {et~al.}(2014)\citenamefont {Hara},
  \citenamefont {Ono}, \citenamefont {Okamoto}, \citenamefont {Washio},\ and\
  \citenamefont {Takeuchi}}]{bib:sara}%
  \BibitemOpen
  \bibfield  {author} {\bibinfo {author} {\bibnamefont {Hara}, \bibfnamefont
  {Satoshi}}, \bibinfo {author} {\bibfnamefont {Takafumi}\ \bibnamefont {Ono}},
  \bibinfo {author} {\bibfnamefont {Ryo}\ \bibnamefont {Okamoto}}, \bibinfo
  {author} {\bibfnamefont {Takashi}\ \bibnamefont {Washio}}, and\ \bibinfo
  {author} {\bibfnamefont {Shigeki}\ \bibnamefont {Takeuchi}}} (\bibinfo {year}
  {2014}),\ \bibfield  {title} {\enquote {\bibinfo {title} {Anomaly detection
  in reconstructed quantum states using a machine-learning technique},}\ }\href
  {https://doi.org/10.1103/physreva.89.022104} {\bibfield  {journal} {\bibinfo
  {journal} {Physical Review A}\ }\textbf {\bibinfo {volume} {89}},\ \bibinfo
  {pages} {022104}},\ \Eprint {https://arxiv.org/abs/arXiv:1401.4785v1}
  {arXiv:1401.4785v1} \BibitemShut {NoStop}%
\bibitem [{\citenamefont {Haroche}\ and\ \citenamefont
  {Raimond}(2006)}]{bib:haroche2006exploring}%
  \BibitemOpen
  \bibfield  {author} {\bibinfo {author} {\bibnamefont {Haroche}, \bibfnamefont
  {Serge}}, and\ \bibinfo {author} {\bibfnamefont {Jean-Michel}\ \bibnamefont
  {Raimond}}} (\bibinfo {year} {2006}),\ \href@noop {} {\emph {\bibinfo {title}
  {Exploring the quantum: atoms, cavities, and photons}}}\ (\bibinfo
  {publisher} {Oxford University Press})\BibitemShut {NoStop}%
\bibitem [{\citenamefont {Harris}\ \emph {et~al.}(2018)\citenamefont {Harris},
  \citenamefont {Sato}, \citenamefont {Berkley}, \citenamefont {Reis},
  \citenamefont {Altomare}, \citenamefont {Amin}, \citenamefont {Boothby},
  \citenamefont {Bunyk}, \citenamefont {Deng}, \citenamefont {Enderud} \emph
  {et~al.}}]{bib:harris2018phase}%
  \BibitemOpen
  \bibfield  {author} {\bibinfo {author} {\bibnamefont {Harris}, \bibfnamefont
  {R}}, \bibinfo {author} {\bibfnamefont {Y}~\bibnamefont {Sato}}, \bibinfo
  {author} {\bibfnamefont {AJ}~\bibnamefont {Berkley}}, \bibinfo {author}
  {\bibfnamefont {M}~\bibnamefont {Reis}}, \bibinfo {author} {\bibfnamefont
  {F}~\bibnamefont {Altomare}}, \bibinfo {author} {\bibfnamefont
  {MH}~\bibnamefont {Amin}}, \bibinfo {author} {\bibfnamefont {K}~\bibnamefont
  {Boothby}}, \bibinfo {author} {\bibfnamefont {P}~\bibnamefont {Bunyk}},
  \bibinfo {author} {\bibfnamefont {C}~\bibnamefont {Deng}}, \bibinfo {author}
  {\bibfnamefont {C}~\bibnamefont {Enderud}},  \emph {et~al.}} (\bibinfo {year}
  {2018}),\ \bibfield  {title} {\enquote {\bibinfo {title} {Phase transitions
  in a programmable quantum spin glass simulator},}\ }\href
  {https://doi.org/10.1126/science.aat2025} {\bibfield  {journal} {\bibinfo
  {journal} {Science}\ }\textbf {\bibinfo {volume} {361}},\ \bibinfo {pages}
  {162}}\BibitemShut {NoStop}%
\bibitem [{\citenamefont {Harrow}\ \emph
  {et~al.}(2009{\natexlab{a}})\citenamefont {Harrow}, \citenamefont
  {Hassidim},\ and\ \citenamefont {Lloyd}}]{bib:harrow2009quantum}%
  \BibitemOpen
  \bibfield  {author} {\bibinfo {author} {\bibnamefont {Harrow}, \bibfnamefont
  {Aram~W}}, \bibinfo {author} {\bibfnamefont {Avinatan}\ \bibnamefont
  {Hassidim}}, and\ \bibinfo {author} {\bibfnamefont {Seth}\ \bibnamefont
  {Lloyd}}} (\bibinfo {year} {2009}{\natexlab{a}}),\ \bibfield  {title}
  {\enquote {\bibinfo {title} {Quantum algorithm for linear systems of
  equations},}\ }\href {https://doi.org/10.1103/physrevlett.103.150502}
  {\bibfield  {journal} {\bibinfo  {journal} {Physical Review Letters}\
  }\textbf {\bibinfo {volume} {103}},\ \bibinfo {pages} {150502}}\BibitemShut
  {NoStop}%
\bibitem [{\citenamefont {Harrow}\ \emph
  {et~al.}(2009{\natexlab{b}})\citenamefont {Harrow}, \citenamefont
  {Hassidim},\ and\ \citenamefont {Lloyd}}]{harrow2009quantum}%
  \BibitemOpen
  \bibfield  {author} {\bibinfo {author} {\bibnamefont {Harrow}, \bibfnamefont
  {Aram~W}}, \bibinfo {author} {\bibfnamefont {Avinatan}\ \bibnamefont
  {Hassidim}}, and\ \bibinfo {author} {\bibfnamefont {Seth}\ \bibnamefont
  {Lloyd}}} (\bibinfo {year} {2009}{\natexlab{b}}),\ \bibfield  {title}
  {\enquote {\bibinfo {title} {Quantum algorithm for linear systems of
  equations},}\ }\href@noop {} {\bibfield  {journal} {\bibinfo  {journal}
  {Physical review letters}\ }\textbf {\bibinfo {volume} {103}}~(\bibinfo
  {number} {15}),\ \bibinfo {pages} {150502}}\BibitemShut {NoStop}%
\bibitem [{\citenamefont {Harrow}\ and\ \citenamefont
  {Montanaro}(2017)}]{bib:harrow2017quantum}%
  \BibitemOpen
  \bibfield  {author} {\bibinfo {author} {\bibnamefont {Harrow}, \bibfnamefont
  {Aram~W}}, and\ \bibinfo {author} {\bibfnamefont {Ashley}\ \bibnamefont
  {Montanaro}}} (\bibinfo {year} {2017}),\ \bibfield  {title} {\enquote
  {\bibinfo {title} {Quantum computational supremacy},}\ }\href
  {https://doi.org/10.1038/nature23458} {\bibfield  {journal} {\bibinfo
  {journal} {Nature}\ }\textbf {\bibinfo {volume} {549}},\ \bibinfo {pages}
  {203}},\ \Eprint {https://arxiv.org/abs/arXiv:1809.07442v1}
  {arXiv:1809.07442v1} \BibitemShut {NoStop}%
\bibitem [{\citenamefont {Hart}\ \emph {et~al.}(1968)\citenamefont {Hart},
  \citenamefont {Nilsson},\ and\ \citenamefont {Raphael}}]{bib:Astar}%
  \BibitemOpen
  \bibfield  {author} {\bibinfo {author} {\bibnamefont {Hart}, \bibfnamefont
  {Peter~E}}, \bibinfo {author} {\bibfnamefont {Nils~J.}\ \bibnamefont
  {Nilsson}}, and\ \bibinfo {author} {\bibfnamefont {Bertram}\ \bibnamefont
  {Raphael}}} (\bibinfo {year} {1968}),\ \bibfield  {title} {\enquote {\bibinfo
  {title} {A formal basis for the heuristic determination of minimum cost
  paths},}\ }\href {https://doi.org/10.1109/tssc.1968.300136} {\bibfield
  {journal} {\bibinfo  {journal} {IEEE Transactions on Systems, Man, and
  Cybernetics.}\ }\textbf {\bibinfo {volume} {4}},\ \bibinfo {pages}
  {100}}\BibitemShut {NoStop}%
\bibitem [{\citenamefont {Haselgrove}\ and\ \citenamefont
  {Rohde}(2008)}]{bib:RohdeHaselgrove}%
  \BibitemOpen
  \bibfield  {author} {\bibinfo {author} {\bibnamefont {Haselgrove},
  \bibfnamefont {Henry~L}}, and\ \bibinfo {author} {\bibfnamefont {Peter~P.}\
  \bibnamefont {Rohde}}} (\bibinfo {year} {2008}),\ \bibfield  {title}
  {\enquote {\bibinfo {title} {Trade-off between the tolerance of located and
  unlocated errors in nondegenerate quantum error-correcting codes},}\
  }\href@noop {} {\bibfield  {journal} {\bibinfo  {journal} {Quantum
  Information \& Computation}\ }\textbf {\bibinfo {volume} {8}},\ \bibinfo
  {pages} {0399}},\ \Eprint {https://arxiv.org/abs/arXiv:quant-ph/0605183}
  {arXiv:quant-ph/0605183} \BibitemShut {NoStop}%
\bibitem [{\citenamefont {Hastie}\ \emph {et~al.}(2009)\citenamefont {Hastie},
  \citenamefont {Tibshirani},\ and\ \citenamefont
  {Friedman}}]{bib:trevor2009elements}%
  \BibitemOpen
  \bibfield  {author} {\bibinfo {author} {\bibnamefont {Hastie}, \bibfnamefont
  {Trevor}}, \bibinfo {author} {\bibfnamefont {Robert}\ \bibnamefont
  {Tibshirani}}, and\ \bibinfo {author} {\bibfnamefont {J.~H.}\ \bibnamefont
  {Friedman}}} (\bibinfo {year} {2009}),\ \href@noop {} {\enquote {\bibinfo
  {title} {The elements of statistical learning: data mining, inference, and
  prediction},}\ }\BibitemShut {NoStop}%
\bibitem [{\citenamefont {Hayashi}\ and\ \citenamefont
  {Nakayama}(2014)}]{bib:hayashi2014security}%
  \BibitemOpen
  \bibfield  {author} {\bibinfo {author} {\bibnamefont {Hayashi}, \bibfnamefont
  {Masahito}}, and\ \bibinfo {author} {\bibfnamefont {Ryota}\ \bibnamefont
  {Nakayama}}} (\bibinfo {year} {2014}),\ \bibfield  {title} {\enquote
  {\bibinfo {title} {Security analysis of the decoy method with the
  bennett--brassard 1984 protocol for finite key lengths},}\ }\href
  {https://doi.org/10.1088/1367-2630/16/6/063009} {\bibfield  {journal}
  {\bibinfo  {journal} {New Journal of Physics}\ }\textbf {\bibinfo {volume}
  {16}},\ \bibinfo {pages} {063009}}\BibitemShut {NoStop}%
\bibitem [{\citenamefont {He}\ \emph {et~al.}(2017{\natexlab{a}})\citenamefont
  {He}, \citenamefont {Ding}, \citenamefont {Su}, \citenamefont {Huang},
  \citenamefont {Qin}, \citenamefont {Wang}, \citenamefont {Unsleber},
  \citenamefont {Chen}, \citenamefont {Wang}, \citenamefont {He} \emph
  {et~al.}}]{he2017time}%
  \BibitemOpen
  \bibfield  {author} {\bibinfo {author} {\bibnamefont {He}, \bibfnamefont
  {Yu}}, \bibinfo {author} {\bibfnamefont {X}~\bibnamefont {Ding}}, \bibinfo
  {author} {\bibfnamefont {Z-E}\ \bibnamefont {Su}}, \bibinfo {author}
  {\bibfnamefont {H-L}\ \bibnamefont {Huang}}, \bibinfo {author} {\bibfnamefont
  {J}~\bibnamefont {Qin}}, \bibinfo {author} {\bibfnamefont {C}~\bibnamefont
  {Wang}}, \bibinfo {author} {\bibfnamefont {S}~\bibnamefont {Unsleber}},
  \bibinfo {author} {\bibfnamefont {C}~\bibnamefont {Chen}}, \bibinfo {author}
  {\bibfnamefont {H}~\bibnamefont {Wang}}, \bibinfo {author} {\bibfnamefont
  {Y-M}\ \bibnamefont {He}},  \emph {et~al.}} (\bibinfo {year}
  {2017}{\natexlab{a}}),\ \bibfield  {title} {\enquote {\bibinfo {title}
  {Time-bin-encoded boson sampling with a single-photon device},}\ }\href@noop
  {} {\bibfield  {journal} {\bibinfo  {journal} {Physical review letters}\
  }\textbf {\bibinfo {volume} {118}}~(\bibinfo {number} {19}),\ \bibinfo
  {pages} {190501}}\BibitemShut {NoStop}%
\bibitem [{\citenamefont {He}\ \emph {et~al.}(2017{\natexlab{b}})\citenamefont
  {He}, \citenamefont {Ding}, \citenamefont {Su}, \citenamefont {Huang},
  \citenamefont {Qin}, \citenamefont {Wang}, \citenamefont {Unsleber},
  \citenamefont {Chen}, \citenamefont {Wang}, \citenamefont {He} \emph
  {et~al.}}]{bib:he2017time}%
  \BibitemOpen
  \bibfield  {author} {\bibinfo {author} {\bibnamefont {He}, \bibfnamefont
  {Yu}}, \bibinfo {author} {\bibfnamefont {X}~\bibnamefont {Ding}}, \bibinfo
  {author} {\bibfnamefont {Z-E}\ \bibnamefont {Su}}, \bibinfo {author}
  {\bibfnamefont {H-L}\ \bibnamefont {Huang}}, \bibinfo {author} {\bibfnamefont
  {J}~\bibnamefont {Qin}}, \bibinfo {author} {\bibfnamefont {C}~\bibnamefont
  {Wang}}, \bibinfo {author} {\bibfnamefont {S}~\bibnamefont {Unsleber}},
  \bibinfo {author} {\bibfnamefont {C}~\bibnamefont {Chen}}, \bibinfo {author}
  {\bibfnamefont {H}~\bibnamefont {Wang}}, \bibinfo {author} {\bibfnamefont
  {Y-M}\ \bibnamefont {He}},  \emph {et~al.}} (\bibinfo {year}
  {2017}{\natexlab{b}}),\ \bibfield  {title} {\enquote {\bibinfo {title}
  {Time-bin-encoded boson sampling with a single-photon device},}\ }\href
  {https://doi.org/10.1103/physrevlett.118.190501} {\bibfield  {journal}
  {\bibinfo  {journal} {Physical Review Letters}\ }\textbf {\bibinfo {volume}
  {118}},\ \bibinfo {pages} {190501}}\BibitemShut {NoStop}%
\bibitem [{\citenamefont {He}\ \emph {et~al.}(2013)\citenamefont {He},
  \citenamefont {He}, \citenamefont {Wei}, \citenamefont {Wu}, \citenamefont
  {Atat{\"u}re}, \citenamefont {Schneider}, \citenamefont {H{\"o}fling},
  \citenamefont {Kamp}, \citenamefont {Lu},\ and\ \citenamefont
  {Pan}}]{bib:he2013on}%
  \BibitemOpen
  \bibfield  {author} {\bibinfo {author} {\bibnamefont {He}, \bibfnamefont
  {Yu-Ming}}, \bibinfo {author} {\bibfnamefont {Yu}~\bibnamefont {He}},
  \bibinfo {author} {\bibfnamefont {Yu-Jia}\ \bibnamefont {Wei}}, \bibinfo
  {author} {\bibfnamefont {Dian}\ \bibnamefont {Wu}}, \bibinfo {author}
  {\bibfnamefont {Mete}\ \bibnamefont {Atat{\"u}re}}, \bibinfo {author}
  {\bibfnamefont {Christian}\ \bibnamefont {Schneider}}, \bibinfo {author}
  {\bibfnamefont {Sven}\ \bibnamefont {H{\"o}fling}}, \bibinfo {author}
  {\bibfnamefont {Martin}\ \bibnamefont {Kamp}}, \bibinfo {author}
  {\bibfnamefont {Chao-Yang}\ \bibnamefont {Lu}}, and\ \bibinfo {author}
  {\bibfnamefont {Jian-Wei}\ \bibnamefont {Pan}}} (\bibinfo {year} {2013}),\
  \bibfield  {title} {\enquote {\bibinfo {title} {On-demand semiconductor
  single-photon source with near-unity indistinguishability},}\ }\href
  {https://doi.org/10.1038/nnano.2012.262} {\bibfield  {journal} {\bibinfo
  {journal} {Nature Nanotechnology}\ }\textbf {\bibinfo {volume} {8}},\
  \bibinfo {pages} {213}},\ \Eprint {https://arxiv.org/abs/arXiv:1303.4058v1}
  {arXiv:1303.4058v1} \BibitemShut {NoStop}%
\bibitem [{\citenamefont {Hensen}\ \emph {et~al.}(2015)\citenamefont {Hensen},
  \citenamefont {Bernien}, \citenamefont {Dr{\'e}au}, \citenamefont {Reiserer},
  \citenamefont {Kalb}, \citenamefont {Blok}, \citenamefont {Ruitenberg},
  \citenamefont {Vermeulen}, \citenamefont {Schouten}, \citenamefont
  {Abell{\'a}n} \emph {et~al.}}]{bib:hensen2015loophole}%
  \BibitemOpen
  \bibfield  {author} {\bibinfo {author} {\bibnamefont {Hensen}, \bibfnamefont
  {Bas}}, \bibinfo {author} {\bibfnamefont {Hannes}\ \bibnamefont {Bernien}},
  \bibinfo {author} {\bibfnamefont {Ana{\"\i}s~E}\ \bibnamefont {Dr{\'e}au}},
  \bibinfo {author} {\bibfnamefont {Andreas}\ \bibnamefont {Reiserer}},
  \bibinfo {author} {\bibfnamefont {Norbert}\ \bibnamefont {Kalb}}, \bibinfo
  {author} {\bibfnamefont {Machiel~S}\ \bibnamefont {Blok}}, \bibinfo {author}
  {\bibfnamefont {Just}\ \bibnamefont {Ruitenberg}}, \bibinfo {author}
  {\bibfnamefont {Raymond~FL}\ \bibnamefont {Vermeulen}}, \bibinfo {author}
  {\bibfnamefont {Raymond~N}\ \bibnamefont {Schouten}}, \bibinfo {author}
  {\bibfnamefont {Carlos}\ \bibnamefont {Abell{\'a}n}},  \emph {et~al.}}
  (\bibinfo {year} {2015}),\ \bibfield  {title} {\enquote {\bibinfo {title}
  {Loophole-free bell inequality violation using electron spins separated by
  1.3 kilometres},}\ }\href {https://doi.org/10.1038/nature15759} {\bibfield
  {journal} {\bibinfo  {journal} {Nature}\ }\textbf {\bibinfo {volume} {526}},\
  \bibinfo {pages} {682}}\BibitemShut {NoStop}%
\bibitem [{\citenamefont {Hibino}(2003)}]{bib:hibino2003silica}%
  \BibitemOpen
  \bibfield  {author} {\bibinfo {author} {\bibnamefont {Hibino}, \bibfnamefont
  {Yoshinori}}} (\bibinfo {year} {2003}),\ \bibfield  {title} {\enquote
  {\bibinfo {title} {Silica-based planar lightwave circuits and their
  applications},}\ }\href {https://doi.org/10.1557/mrs2003.102} {\bibfield
  {journal} {\bibinfo  {journal} {MRS Bulletin}\ }\textbf {\bibinfo {volume}
  {28}},\ \bibinfo {pages} {365}}\BibitemShut {NoStop}%
\bibitem [{\citenamefont {Hill}\ \emph
  {et~al.}(2015{\natexlab{a}})\citenamefont {Hill}, \citenamefont {Peretz},
  \citenamefont {Hile}, \citenamefont {House}, \citenamefont {Fuechsle},
  \citenamefont {Rogge}, \citenamefont {Simmons},\ and\ \citenamefont
  {Hollenberg}}]{SD-Hill:2015aa}%
  \BibitemOpen
  \bibfield  {author} {\bibinfo {author} {\bibnamefont {Hill}, \bibfnamefont
  {Charles~D}}, \bibinfo {author} {\bibfnamefont {Eldad}\ \bibnamefont
  {Peretz}}, \bibinfo {author} {\bibfnamefont {Samuel~J.}\ \bibnamefont
  {Hile}}, \bibinfo {author} {\bibfnamefont {Matthew~G.}\ \bibnamefont
  {House}}, \bibinfo {author} {\bibfnamefont {Martin}\ \bibnamefont
  {Fuechsle}}, \bibinfo {author} {\bibfnamefont {Sven}\ \bibnamefont {Rogge}},
  \bibinfo {author} {\bibfnamefont {Michelle~Y.}\ \bibnamefont {Simmons}}, and\
  \bibinfo {author} {\bibfnamefont {Lloyd C.~L.}\ \bibnamefont {Hollenberg}}}
  (\bibinfo {year} {2015}{\natexlab{a}}),\ \bibfield  {title} {\enquote
  {\bibinfo {title} {A surface code quantum computer in silicon},}\ }\href
  {https://doi.org/10.1126/sciadv.1500707} {\bibfield  {journal} {\bibinfo
  {journal} {Science Advances}\ }\textbf {\bibinfo {volume} {1}},\
  10.1126/sciadv.1500707}\BibitemShut {NoStop}%
\bibitem [{\citenamefont {Hill}\ \emph
  {et~al.}(2015{\natexlab{b}})\citenamefont {Hill}, \citenamefont {Peretz},
  \citenamefont {Hile}, \citenamefont {House}, \citenamefont {Fuechsle},
  \citenamefont {Rogge}, \citenamefont {Simmons},\ and\ \citenamefont
  {Hollenberg}}]{bib:Hill:2015aa}%
  \BibitemOpen
  \bibfield  {author} {\bibinfo {author} {\bibnamefont {Hill}, \bibfnamefont
  {Charles~D}}, \bibinfo {author} {\bibfnamefont {Eldad}\ \bibnamefont
  {Peretz}}, \bibinfo {author} {\bibfnamefont {Samuel~J.}\ \bibnamefont
  {Hile}}, \bibinfo {author} {\bibfnamefont {Matthew~G.}\ \bibnamefont
  {House}}, \bibinfo {author} {\bibfnamefont {Martin}\ \bibnamefont
  {Fuechsle}}, \bibinfo {author} {\bibfnamefont {Sven}\ \bibnamefont {Rogge}},
  \bibinfo {author} {\bibfnamefont {Michelle~Y.}\ \bibnamefont {Simmons}}, and\
  \bibinfo {author} {\bibfnamefont {Lloyd C.~L.}\ \bibnamefont {Hollenberg}}}
  (\bibinfo {year} {2015}{\natexlab{b}}),\ \bibfield  {title} {\enquote
  {\bibinfo {title} {A surface code quantum computer in silicon},}\ }\href
  {http://advances.sciencemag.org/content/1/9/e1500707.abstract} {\bibfield
  {journal} {\bibinfo  {journal} {Science Advances}\ }\textbf {\bibinfo
  {volume} {1}}~(\bibinfo {number} {9})}\BibitemShut {NoStop}%
\bibitem [{\citenamefont {Hofmann}\ \emph {et~al.}(2012)\citenamefont
  {Hofmann}, \citenamefont {Krug}, \citenamefont {Ortegel}, \citenamefont
  {G{\'e}rard}, \citenamefont {Weber}, \citenamefont {Rosenfeld},\ and\
  \citenamefont {Weinfurter}}]{bib:Sc_337_72}%
  \BibitemOpen
  \bibfield  {author} {\bibinfo {author} {\bibnamefont {Hofmann}, \bibfnamefont
  {Julian}}, \bibinfo {author} {\bibfnamefont {Michael}\ \bibnamefont {Krug}},
  \bibinfo {author} {\bibfnamefont {Norbert}\ \bibnamefont {Ortegel}}, \bibinfo
  {author} {\bibfnamefont {Lea}\ \bibnamefont {G{\'e}rard}}, \bibinfo {author}
  {\bibfnamefont {Markus}\ \bibnamefont {Weber}}, \bibinfo {author}
  {\bibfnamefont {Wenjamin}\ \bibnamefont {Rosenfeld}}, and\ \bibinfo {author}
  {\bibfnamefont {Harald}\ \bibnamefont {Weinfurter}}} (\bibinfo {year}
  {2012}),\ \bibfield  {title} {\enquote {\bibinfo {title} {Heralded
  entanglement between widely separated atoms},}\ }\href
  {https://doi.org/10.1126/science.1221856} {\bibfield  {journal} {\bibinfo
  {journal} {Science}\ }\textbf {\bibinfo {volume} {337}},\ \bibinfo {pages}
  {72}}\BibitemShut {NoStop}%
\bibitem [{\citenamefont {Holevo}(1998)}]{bib:holevo1998capacity}%
  \BibitemOpen
  \bibfield  {author} {\bibinfo {author} {\bibnamefont {Holevo}, \bibfnamefont
  {Alexander~S}}} (\bibinfo {year} {1998}),\ \bibfield  {title} {\enquote
  {\bibinfo {title} {The capacity of the quantum channel with general signal
  states},}\ }\href {https://doi.org/10.1109/18.651037} {\bibfield  {journal}
  {\bibinfo  {journal} {IEEE Transactions on Information Theory}\ }\textbf
  {\bibinfo {volume} {44}},\ \bibinfo {pages} {269}}\BibitemShut {NoStop}%
\bibitem [{\citenamefont {Holevo}(1973)}]{holevo1973bounds}%
  \BibitemOpen
  \bibfield  {author} {\bibinfo {author} {\bibnamefont {Holevo}, \bibfnamefont
  {Alexander~Semenovich}}} (\bibinfo {year} {1973}),\ \bibfield  {title}
  {\enquote {\bibinfo {title} {Bounds for the quantity of information
  transmitted by a quantum communication channel},}\ }\href@noop {} {\bibfield
  {journal} {\bibinfo  {journal} {Problemy Peredachi Informatsii}\ }\textbf
  {\bibinfo {volume} {9}}~(\bibinfo {number} {3}),\ \bibinfo {pages}
  {3--11}}\BibitemShut {NoStop}%
\bibitem [{\citenamefont {Honda}\ \emph {et~al.}(2008)\citenamefont {Honda},
  \citenamefont {Akamatsu}, \citenamefont {Arikawa}, \citenamefont {Yokoi},
  \citenamefont {Akiba}, \citenamefont {Nagatsuka}, \citenamefont {Tanimura},
  \citenamefont {Furusawa},\ and\ \citenamefont
  {Kozuma}}]{bib:honda2008storage}%
  \BibitemOpen
  \bibfield  {author} {\bibinfo {author} {\bibnamefont {Honda}, \bibfnamefont
  {Kazuhito}}, \bibinfo {author} {\bibfnamefont {Daisuke}\ \bibnamefont
  {Akamatsu}}, \bibinfo {author} {\bibfnamefont {Manabu}\ \bibnamefont
  {Arikawa}}, \bibinfo {author} {\bibfnamefont {Yoshihiko}\ \bibnamefont
  {Yokoi}}, \bibinfo {author} {\bibfnamefont {Keiichirou}\ \bibnamefont
  {Akiba}}, \bibinfo {author} {\bibfnamefont {Satoshi}\ \bibnamefont
  {Nagatsuka}}, \bibinfo {author} {\bibfnamefont {Takahito}\ \bibnamefont
  {Tanimura}}, \bibinfo {author} {\bibfnamefont {Akira}\ \bibnamefont
  {Furusawa}}, and\ \bibinfo {author} {\bibfnamefont {Mikio}\ \bibnamefont
  {Kozuma}}} (\bibinfo {year} {2008}),\ \bibfield  {title} {\enquote {\bibinfo
  {title} {Storage and retrieval of a squeezed vacuum},}\ }\href
  {https://doi.org/10.1103/physrevlett.100.093601} {\bibfield  {journal}
  {\bibinfo  {journal} {Physical Review Letters}\ }\textbf {\bibinfo {volume}
  {100}},\ \bibinfo {pages} {093601}},\ \Eprint
  {https://arxiv.org/abs/arXiv:0709.1785v2} {arXiv:0709.1785v2} \BibitemShut
  {NoStop}%
\bibitem [{\citenamefont {Hong}\ \emph {et~al.}(1987)\citenamefont {Hong},
  \citenamefont {Ou},\ and\ \citenamefont {Mandel}}]{bib:HOM87}%
  \BibitemOpen
  \bibfield  {author} {\bibinfo {author} {\bibnamefont {Hong}, \bibfnamefont
  {C~K}}, \bibinfo {author} {\bibfnamefont {Z.~Y.}\ \bibnamefont {Ou}}, and\
  \bibinfo {author} {\bibfnamefont {L.}~\bibnamefont {Mandel}}} (\bibinfo
  {year} {1987}),\ \bibfield  {title} {\enquote {\bibinfo {title} {Measurement
  of sub-picosecond time intervals between two photons by interference},}\
  }\href {https://doi.org/10.1103/physrevlett.59.2044} {\bibfield  {journal}
  {\bibinfo  {journal} {Physical Review Letters}\ }\textbf {\bibinfo {volume}
  {59}},\ \bibinfo {pages} {2044}}\BibitemShut {NoStop}%
\bibitem [{\citenamefont {Hood}\ and\ \citenamefont
  {Ji}(1997)}]{bib:hood1997proactive}%
  \BibitemOpen
  \bibfield  {author} {\bibinfo {author} {\bibnamefont {Hood}, \bibfnamefont
  {Cynthia~S}}, and\ \bibinfo {author} {\bibfnamefont {Chuanyi}\ \bibnamefont
  {Ji}}} (\bibinfo {year} {1997}),\ \bibfield  {title} {\enquote {\bibinfo
  {title} {Proactive network-fault detection [telecommunications]},}\ }\href
  {https://doi.org/10.1109/24.664004} {\bibfield  {journal} {\bibinfo
  {journal} {IEEE Transactions on reliability}\ }\textbf {\bibinfo {volume}
  {46}},\ \bibinfo {pages} {333}}\BibitemShut {NoStop}%
\bibitem [{\citenamefont {Horiuchi}(2015)}]{bib:horiuchi2015view}%
  \BibitemOpen
  \bibfield  {author} {\bibinfo {author} {\bibnamefont {Horiuchi},
  \bibfnamefont {Noriaki}}} (\bibinfo {year} {2015}),\ \bibfield  {title}
  {\enquote {\bibinfo {title} {View from... qcmc 2014: Expanding ambitions},}\
  }\href@noop {} {\bibfield  {journal} {\bibinfo  {journal} {Nature Photonics}\
  }\textbf {\bibinfo {volume} {9}},\ \bibinfo {pages} {13}}\BibitemShut
  {NoStop}%
\bibitem [{\citenamefont {Horodecki}\ \emph {et~al.}(2001)\citenamefont
  {Horodecki}, \citenamefont {Horodecki},\ and\ \citenamefont
  {Horodecki}}]{bib:horodecki2001separability}%
  \BibitemOpen
  \bibfield  {author} {\bibinfo {author} {\bibnamefont {Horodecki},
  \bibfnamefont {Micha{\l}}}, \bibinfo {author} {\bibfnamefont {Pawe{\l}}\
  \bibnamefont {Horodecki}}, and\ \bibinfo {author} {\bibfnamefont {Ryszard}\
  \bibnamefont {Horodecki}}} (\bibinfo {year} {2001}),\ \bibfield  {title}
  {\enquote {\bibinfo {title} {Separability of n-particle mixed states:
  necessary and sufficient conditions in terms of linear maps},}\ }\href@noop
  {} {\bibfield  {journal} {\bibinfo  {journal} {Physics Letters A}\ }\textbf
  {\bibinfo {volume} {283}}~(\bibinfo {number} {1-2}),\ \bibinfo {pages}
  {1--7}}\BibitemShut {NoStop}%
\bibitem [{\citenamefont {Horsman}\ \emph {et~al.}(2012)\citenamefont
  {Horsman}, \citenamefont {Fowler}, \citenamefont {Devitt},\ and\
  \citenamefont {van Meter}}]{SD-Horsman:2012aa}%
  \BibitemOpen
  \bibfield  {author} {\bibinfo {author} {\bibnamefont {Horsman}, \bibfnamefont
  {Clare}}, \bibinfo {author} {\bibfnamefont {Austin~G.}\ \bibnamefont
  {Fowler}}, \bibinfo {author} {\bibfnamefont {Simon}\ \bibnamefont {Devitt}},
  and\ \bibinfo {author} {\bibfnamefont {Rodney}\ \bibnamefont {van Meter}}}
  (\bibinfo {year} {2012}),\ \bibfield  {title} {\enquote {\bibinfo {title}
  {Surface code quantum computing by lattice surgery},}\ }\href
  {https://doi.org/10.1088/1367-2630/14/12/123011} {\bibfield  {journal}
  {\bibinfo  {journal} {New Journal of Physics}\ }\textbf {\bibinfo {volume}
  {14}},\ \bibinfo {pages} {123011}}\BibitemShut {NoStop}%
\bibitem [{\citenamefont {Hosten}\ \emph {et~al.}(2016)\citenamefont {Hosten},
  \citenamefont {Engelsen}, \citenamefont {Krishnakumar},\ and\ \citenamefont
  {Kasevich}}]{bib:hosten2016measurement}%
  \BibitemOpen
  \bibfield  {author} {\bibinfo {author} {\bibnamefont {Hosten}, \bibfnamefont
  {Onur}}, \bibinfo {author} {\bibfnamefont {Nils~J}\ \bibnamefont {Engelsen}},
  \bibinfo {author} {\bibfnamefont {Rajiv}\ \bibnamefont {Krishnakumar}}, and\
  \bibinfo {author} {\bibfnamefont {Mark~A}\ \bibnamefont {Kasevich}}}
  (\bibinfo {year} {2016}),\ \bibfield  {title} {\enquote {\bibinfo {title}
  {Measurement noise 100 times lower than the quantum-projection limit using
  entangled atoms},}\ }\href {https://doi.org/10.1038/nature16176} {\bibfield
  {journal} {\bibinfo  {journal} {Nature}\ }\textbf {\bibinfo {volume} {529}},\
  \bibinfo {pages} {505}}\BibitemShut {NoStop}%
\bibitem [{\citenamefont {Hu}\ \emph {et~al.}(2023)\citenamefont {Hu},
  \citenamefont {Ouyang},\ and\ \citenamefont {Tomamichel}}]{hu2023privacy}%
  \BibitemOpen
  \bibfield  {author} {\bibinfo {author} {\bibnamefont {Hu}, \bibfnamefont
  {Yanglin}}, \bibinfo {author} {\bibfnamefont {Yingkai}\ \bibnamefont
  {Ouyang}}, and\ \bibinfo {author} {\bibfnamefont {Marco}\ \bibnamefont
  {Tomamichel}}} (\bibinfo {year} {2023}),\ \bibfield  {title} {\enquote
  {\bibinfo {title} {Privacy and correctness trade-offs for
  information-theoretically secure quantum homomorphic encryption},}\
  }\href@noop {} {\bibfield  {journal} {\bibinfo  {journal} {Quantum}\ }\textbf
  {\bibinfo {volume} {7}},\ \bibinfo {pages} {976}}\BibitemShut {NoStop}%
\bibitem [{\citenamefont {Huang}\ \emph {et~al.}(2018)\citenamefont {Huang},
  \citenamefont {Wang}, \citenamefont {Rohde}, \citenamefont {Luo},
  \citenamefont {Zhao}, \citenamefont {Liu}, \citenamefont {Li}, \citenamefont
  {Liu}, \citenamefont {Lu},\ and\ \citenamefont
  {Pan}}]{huang2018demonstration}%
  \BibitemOpen
  \bibfield  {author} {\bibinfo {author} {\bibnamefont {Huang}, \bibfnamefont
  {He-Liang}}, \bibinfo {author} {\bibfnamefont {Xi-Lin}\ \bibnamefont {Wang}},
  \bibinfo {author} {\bibfnamefont {Peter~P}\ \bibnamefont {Rohde}}, \bibinfo
  {author} {\bibfnamefont {Yi-Han}\ \bibnamefont {Luo}}, \bibinfo {author}
  {\bibfnamefont {You-Wei}\ \bibnamefont {Zhao}}, \bibinfo {author}
  {\bibfnamefont {Chang}\ \bibnamefont {Liu}}, \bibinfo {author} {\bibfnamefont
  {Li}~\bibnamefont {Li}}, \bibinfo {author} {\bibfnamefont {Nai-Le}\
  \bibnamefont {Liu}}, \bibinfo {author} {\bibfnamefont {Chao-Yang}\
  \bibnamefont {Lu}}, and\ \bibinfo {author} {\bibfnamefont {Jian-Wei}\
  \bibnamefont {Pan}}} (\bibinfo {year} {2018}),\ \bibfield  {title} {\enquote
  {\bibinfo {title} {Demonstration of topological data analysis on a quantum
  processor},}\ }\href@noop {} {\bibfield  {journal} {\bibinfo  {journal}
  {Optica}\ }\textbf {\bibinfo {volume} {5}}~(\bibinfo {number} {2}),\ \bibinfo
  {pages} {193--198}}\BibitemShut {NoStop}%
\bibitem [{\citenamefont {Huang}\ \emph {et~al.}(2011)\citenamefont {Huang},
  \citenamefont {Joseph}, \citenamefont {Nelson}, \citenamefont {Rubinstein},\
  and\ \citenamefont {Tygar}}]{bib:huang2011adversarial}%
  \BibitemOpen
  \bibfield  {author} {\bibinfo {author} {\bibnamefont {Huang}, \bibfnamefont
  {Ling}}, \bibinfo {author} {\bibfnamefont {Anthony~D.}\ \bibnamefont
  {Joseph}}, \bibinfo {author} {\bibfnamefont {Blaine}\ \bibnamefont {Nelson}},
  \bibinfo {author} {\bibfnamefont {Benjamin}\ \bibnamefont {Rubinstein}}, and\
  \bibinfo {author} {\bibfnamefont {JD}~\bibnamefont {Tygar}}} (\bibinfo {year}
  {2011}),\ \bibfield  {title} {\enquote {\bibinfo {title} {Adversarial machine
  learning},}\ }in\ \href {https://doi.org/10.1145/2046684.2046692} {\emph
  {\bibinfo {booktitle} {Proceedings of the 4th ACM workshop on Security and
  artificial intelligence}}},\ p.~\bibinfo {pages} {43}\BibitemShut {NoStop}%
\bibitem [{\citenamefont {Huang}\ \emph {et~al.}(2024)\citenamefont {Huang},
  \citenamefont {Baragiola}, \citenamefont {Menicucci},\ and\ \citenamefont
  {Wilde}}]{PhysRevA.109.052434}%
  \BibitemOpen
  \bibfield  {author} {\bibinfo {author} {\bibnamefont {Huang}, \bibfnamefont
  {Zixin}}, \bibinfo {author} {\bibfnamefont {Ben~Q.}\ \bibnamefont
  {Baragiola}}, \bibinfo {author} {\bibfnamefont {Nicolas~C.}\ \bibnamefont
  {Menicucci}}, and\ \bibinfo {author} {\bibfnamefont {Mark~M.}\ \bibnamefont
  {Wilde}}} (\bibinfo {year} {2024}),\ \bibfield  {title} {\enquote {\bibinfo
  {title} {Limited quantum advantage for stellar interferometry via
  continuous-variable teleportation},}\ }\href
  {https://doi.org/10.1103/PhysRevA.109.052434} {\bibfield  {journal} {\bibinfo
   {journal} {Phys. Rev. A}\ }\textbf {\bibinfo {volume} {109}},\ \bibinfo
  {pages} {052434}}\BibitemShut {NoStop}%
\bibitem [{\citenamefont {Huang}\ \emph
  {et~al.}(2022{\natexlab{a}})\citenamefont {Huang}, \citenamefont {Brennen},\
  and\ \citenamefont {Ouyang}}]{PhysRevLett.129.210502}%
  \BibitemOpen
  \bibfield  {author} {\bibinfo {author} {\bibnamefont {Huang}, \bibfnamefont
  {Zixin}}, \bibinfo {author} {\bibfnamefont {Gavin~K.}\ \bibnamefont
  {Brennen}}, and\ \bibinfo {author} {\bibfnamefont {Yingkai}\ \bibnamefont
  {Ouyang}}} (\bibinfo {year} {2022}{\natexlab{a}}),\ \bibfield  {title}
  {\enquote {\bibinfo {title} {Imaging stars with quantum error correction},}\
  }\href {https://doi.org/10.1103/PhysRevLett.129.210502} {\bibfield  {journal}
  {\bibinfo  {journal} {Phys. Rev. Lett.}\ }\textbf {\bibinfo {volume} {129}},\
  \bibinfo {pages} {210502}}\BibitemShut {NoStop}%
\bibitem [{\citenamefont {Huang}\ \emph
  {et~al.}(2022{\natexlab{b}})\citenamefont {Huang}, \citenamefont {Joshi},
  \citenamefont {Aktas}, \citenamefont {Lupo}, \citenamefont {Quintavalle},
  \citenamefont {Venkatachalam}, \citenamefont {Wengerowsky}, \citenamefont
  {Lon{\v{c}}ari{\'c}}, \citenamefont {Neumann}, \citenamefont {Liu} \emph
  {et~al.}}]{huang2022experimental}%
  \BibitemOpen
  \bibfield  {author} {\bibinfo {author} {\bibnamefont {Huang}, \bibfnamefont
  {Zixin}}, \bibinfo {author} {\bibfnamefont {Siddarth~Koduru}\ \bibnamefont
  {Joshi}}, \bibinfo {author} {\bibfnamefont {Djeylan}\ \bibnamefont {Aktas}},
  \bibinfo {author} {\bibfnamefont {Cosmo}\ \bibnamefont {Lupo}}, \bibinfo
  {author} {\bibfnamefont {Armanda~O}\ \bibnamefont {Quintavalle}}, \bibinfo
  {author} {\bibfnamefont {Natarajan}\ \bibnamefont {Venkatachalam}}, \bibinfo
  {author} {\bibfnamefont {S{\"o}ren}\ \bibnamefont {Wengerowsky}}, \bibinfo
  {author} {\bibfnamefont {Martin}\ \bibnamefont {Lon{\v{c}}ari{\'c}}},
  \bibinfo {author} {\bibfnamefont {Sebastian~Philipp}\ \bibnamefont
  {Neumann}}, \bibinfo {author} {\bibfnamefont {Bo}~\bibnamefont {Liu}},  \emph
  {et~al.}} (\bibinfo {year} {2022}{\natexlab{b}}),\ \bibfield  {title}
  {\enquote {\bibinfo {title} {Experimental implementation of secure anonymous
  protocols on an eight-user quantum key distribution network},}\ }\href@noop
  {} {\bibfield  {journal} {\bibinfo  {journal} {npj quantum information}\
  }\textbf {\bibinfo {volume} {8}}~(\bibinfo {number} {1}),\ \bibinfo {pages}
  {25}}\BibitemShut {NoStop}%
\bibitem [{\citenamefont {Huang}\ \emph {et~al.}(2019)\citenamefont {Huang},
  \citenamefont {Rohde}, \citenamefont {Berry}, \citenamefont {Kok},
  \citenamefont {Dowling},\ and\ \citenamefont {Lupo}}]{huang2019boson}%
  \BibitemOpen
  \bibfield  {author} {\bibinfo {author} {\bibnamefont {Huang}, \bibfnamefont
  {Zixin}}, \bibinfo {author} {\bibfnamefont {Peter~P}\ \bibnamefont {Rohde}},
  \bibinfo {author} {\bibfnamefont {Dominic~W}\ \bibnamefont {Berry}}, \bibinfo
  {author} {\bibfnamefont {Pieter}\ \bibnamefont {Kok}}, \bibinfo {author}
  {\bibfnamefont {Jonathan~P}\ \bibnamefont {Dowling}}, and\ \bibinfo {author}
  {\bibfnamefont {Cosmo}\ \bibnamefont {Lupo}}} (\bibinfo {year} {2019}),\
  \bibfield  {title} {\enquote {\bibinfo {title} {Boson sampling private-key
  quantum cryptography},}\ }\href@noop {} {\bibinfo  {journal} {arXiv preprint
  arXiv:1905.03013}\ }\BibitemShut {NoStop}%
\bibitem [{\citenamefont {Hucul}\ \emph {et~al.}(2015)\citenamefont {Hucul},
  \citenamefont {Inlek}, \citenamefont {Vittorini}, \citenamefont {Crocker},
  \citenamefont {Debnath}, \citenamefont {Clark},\ and\ \citenamefont
  {Monroe}}]{bib:NP_11_37}%
  \BibitemOpen
\bibfield  {journal} {  }\bibfield  {author} {\bibinfo {author} {\bibnamefont
  {Hucul}, \bibfnamefont {D}}, \bibinfo {author} {\bibfnamefont
  {IV}~\bibnamefont {Inlek}}, \bibinfo {author} {\bibfnamefont {G}~\bibnamefont
  {Vittorini}}, \bibinfo {author} {\bibfnamefont {C}~\bibnamefont {Crocker}},
  \bibinfo {author} {\bibfnamefont {S}~\bibnamefont {Debnath}}, \bibinfo
  {author} {\bibfnamefont {SM}~\bibnamefont {Clark}}, and\ \bibinfo {author}
  {\bibfnamefont {Cl}~\bibnamefont {Monroe}}} (\bibinfo {year} {2015}),\
  \bibfield  {title} {\enquote {\bibinfo {title} {Modular entanglement of
  atomic qubits using photons and phonons},}\ }\href@noop {} {\bibfield
  {journal} {\bibinfo  {journal} {Nature Physics}\ }\textbf {\bibinfo {volume}
  {11}},\ \bibinfo {pages} {37}}\BibitemShut {NoStop}%
\bibitem [{\citenamefont {Humphreys}\ \emph {et~al.}(2013)\citenamefont
  {Humphreys}, \citenamefont {Metcalf}, \citenamefont {Spring}, \citenamefont
  {Moore}, \citenamefont {Jin}, \citenamefont {Barbieri}, \citenamefont
  {Kolthammer},\ and\ \citenamefont {Walmsley}}]{bib:Humphreys2013}%
  \BibitemOpen
  \bibfield  {author} {\bibinfo {author} {\bibnamefont {Humphreys},
  \bibfnamefont {Peter~C}}, \bibinfo {author} {\bibfnamefont {Benjamin~J.}\
  \bibnamefont {Metcalf}}, \bibinfo {author} {\bibfnamefont {Justin~B.}\
  \bibnamefont {Spring}}, \bibinfo {author} {\bibfnamefont {Merritt}\
  \bibnamefont {Moore}}, \bibinfo {author} {\bibfnamefont {Xian-Min}\
  \bibnamefont {Jin}}, \bibinfo {author} {\bibfnamefont {Marco}\ \bibnamefont
  {Barbieri}}, \bibinfo {author} {\bibfnamefont {W.~Steven}\ \bibnamefont
  {Kolthammer}}, and\ \bibinfo {author} {\bibfnamefont {Ian~A.}\ \bibnamefont
  {Walmsley}}} (\bibinfo {year} {2013}),\ \bibfield  {title} {\enquote
  {\bibinfo {title} {Linear optical quantum computing in a single spatial
  mode},}\ }\href {https://doi.org/10.1103/PhysRevLett.111.150501} {\bibfield
  {journal} {\bibinfo  {journal} {Physical Review Letters}\ }\textbf {\bibinfo
  {volume} {111}},\ \bibinfo {pages} {150501}},\ \Eprint
  {https://arxiv.org/abs/arXiv:1305.3592v2} {arXiv:1305.3592v2} \BibitemShut
  {NoStop}%
\bibitem [{\citenamefont {Hwang}(2003)}]{bib:PhysRevLett.91.057901}%
  \BibitemOpen
  \bibfield  {author} {\bibinfo {author} {\bibnamefont {Hwang}, \bibfnamefont
  {Won-Young}}} (\bibinfo {year} {2003}),\ \bibfield  {title} {\enquote
  {\bibinfo {title} {Quantum key distribution with high loss: Toward global
  secure communication},}\ }\href
  {https://doi.org/10.1103/physrevlett.91.057901} {\bibfield  {journal}
  {\bibinfo  {journal} {Physical Review Letters}\ }\textbf {\bibinfo {volume}
  {91}},\ \bibinfo {pages} {057901}},\ \Eprint
  {https://arxiv.org/abs/arXiv:quant-ph/0211153v5} {arXiv:quant-ph/0211153v5}
  \BibitemShut {NoStop}%
\bibitem [{\citenamefont {Imamo{\u{g}}lu}(2009)}]{bib:imamouglu2009cavity}%
  \BibitemOpen
  \bibfield  {author} {\bibinfo {author} {\bibnamefont {Imamo{\u{g}}lu},
  \bibfnamefont {Atac}}} (\bibinfo {year} {2009}),\ \bibfield  {title}
  {\enquote {\bibinfo {title} {Cavity qed based on collective magnetic dipole
  coupling: spin ensembles as hybrid two-level systems},}\ }\href
  {https://doi.org/10.1103/physrevlett.102.083602} {\bibfield  {journal}
  {\bibinfo  {journal} {Physical Review letters}\ }\textbf {\bibinfo {volume}
  {102}},\ \bibinfo {pages} {083602}},\ \Eprint
  {https://arxiv.org/abs/arXiv:0809.2909v1} {arXiv:0809.2909v1} \BibitemShut
  {NoStop}%
\bibitem [{\citenamefont {Inagaki}\ \emph {et~al.}(2016)\citenamefont
  {Inagaki}, \citenamefont {Haribara}, \citenamefont {Igarashi}, \citenamefont
  {Sonobe}, \citenamefont {Tamate}, \citenamefont {Honjo}, \citenamefont
  {Marandi}, \citenamefont {McMahon}, \citenamefont {Umeki}, \citenamefont
  {Enbutsu} \emph {et~al.}}]{bib:inagaki2016coherent}%
  \BibitemOpen
  \bibfield  {author} {\bibinfo {author} {\bibnamefont {Inagaki}, \bibfnamefont
  {Takahiro}}, \bibinfo {author} {\bibfnamefont {Yoshitaka}\ \bibnamefont
  {Haribara}}, \bibinfo {author} {\bibfnamefont {Koji}\ \bibnamefont
  {Igarashi}}, \bibinfo {author} {\bibfnamefont {Tomohiro}\ \bibnamefont
  {Sonobe}}, \bibinfo {author} {\bibfnamefont {Shuhei}\ \bibnamefont {Tamate}},
  \bibinfo {author} {\bibfnamefont {Toshimori}\ \bibnamefont {Honjo}}, \bibinfo
  {author} {\bibfnamefont {Alireza}\ \bibnamefont {Marandi}}, \bibinfo {author}
  {\bibfnamefont {Peter~L}\ \bibnamefont {McMahon}}, \bibinfo {author}
  {\bibfnamefont {Takeshi}\ \bibnamefont {Umeki}}, \bibinfo {author}
  {\bibfnamefont {Koji}\ \bibnamefont {Enbutsu}},  \emph {et~al.}} (\bibinfo
  {year} {2016}),\ \bibfield  {title} {\enquote {\bibinfo {title} {A coherent
  ising machine for 2000-node optimization problems},}\ }\href
  {https://doi.org/10.1126/science.aah4243} {\bibinfo  {journal} {Science}\ ,\
  \bibinfo {pages} {4243}}\BibitemShut {NoStop}%
\bibitem [{\citenamefont {Irwin}\ and\ \citenamefont
  {Hilton}(2005)}]{Irwin2005TES}%
  \BibitemOpen
\bibfield  {journal} {  }\bibfield  {author} {\bibinfo {author} {\bibnamefont
  {Irwin}, \bibfnamefont {KD}}, and\ \bibinfo {author} {\bibfnamefont {G.C.}\
  \bibnamefont {Hilton}}} (\bibinfo {year} {2005}),\ \enquote {\bibinfo {title}
  {Transition-edge sensors},}\ in\ \href {https://doi.org/10.1007/10933596_3}
  {\emph {\bibinfo {booktitle} {Cryogenic Particle Detection}}},\ \bibinfo
  {editor} {edited by\ \bibinfo {editor} {\bibfnamefont {Christian}\
  \bibnamefont {Enss}}}\ (\bibinfo  {publisher} {Springer Berlin Heidelberg},\
  \bibinfo {address} {Berlin, Heidelberg})\ pp.\ \bibinfo {pages}
  {63--150}\BibitemShut {NoStop}%
\bibitem [{\citenamefont {Islam}\ \emph {et~al.}(2015)\citenamefont {Islam},
  \citenamefont {Ma}, \citenamefont {Preiss}, \citenamefont {Tai},
  \citenamefont {Lukin}, \citenamefont {Rispoli},\ and\ \citenamefont
  {Greiner}}]{bib:islam2015measuring}%
  \BibitemOpen
  \bibfield  {author} {\bibinfo {author} {\bibnamefont {Islam}, \bibfnamefont
  {Rajibul}}, \bibinfo {author} {\bibfnamefont {Ruichao}\ \bibnamefont {Ma}},
  \bibinfo {author} {\bibfnamefont {Philipp~M}\ \bibnamefont {Preiss}},
  \bibinfo {author} {\bibfnamefont {M~Eric}\ \bibnamefont {Tai}}, \bibinfo
  {author} {\bibfnamefont {Alexander}\ \bibnamefont {Lukin}}, \bibinfo {author}
  {\bibfnamefont {Matthew}\ \bibnamefont {Rispoli}}, and\ \bibinfo {author}
  {\bibfnamefont {Markus}\ \bibnamefont {Greiner}}} (\bibinfo {year} {2015}),\
  \bibfield  {title} {\enquote {\bibinfo {title} {Measuring entanglement
  entropy in a quantum many-body system},}\ }\href
  {https://doi.org/10.1038/nature15750} {\bibfield  {journal} {\bibinfo
  {journal} {Nature}\ }\textbf {\bibinfo {volume} {528}},\ \bibinfo {pages}
  {77}}\BibitemShut {NoStop}%
\bibitem [{\citenamefont {Jain}\ \emph {et~al.}(2016)\citenamefont {Jain},
  \citenamefont {Stiller}, \citenamefont {Khan}, \citenamefont {Elser},
  \citenamefont {Marquardt},\ and\ \citenamefont
  {Leuchs}}]{bib:jain2016attacks}%
  \BibitemOpen
  \bibfield  {author} {\bibinfo {author} {\bibnamefont {Jain}, \bibfnamefont
  {Nitin}}, \bibinfo {author} {\bibfnamefont {Birgit}\ \bibnamefont {Stiller}},
  \bibinfo {author} {\bibfnamefont {Imran}\ \bibnamefont {Khan}}, \bibinfo
  {author} {\bibfnamefont {Dominique}\ \bibnamefont {Elser}}, \bibinfo {author}
  {\bibfnamefont {Christoph}\ \bibnamefont {Marquardt}}, and\ \bibinfo {author}
  {\bibfnamefont {Gerd}\ \bibnamefont {Leuchs}}} (\bibinfo {year} {2016}),\
  \bibfield  {title} {\enquote {\bibinfo {title} {Attacks on practical quantum
  key distribution systems (and how to prevent them)},}\ }\href
  {https://doi.org/10.1080/00107514.2016.1148333} {\bibfield  {journal}
  {\bibinfo  {journal} {Contemporary Physics}\ }\textbf {\bibinfo {volume}
  {57}},\ \bibinfo {pages} {366}},\ \Eprint
  {https://arxiv.org/abs/arXiv:1512.07990v2} {arXiv:1512.07990v2} \BibitemShut
  {NoStop}%
\bibitem [{\citenamefont {Jain}\ \emph {et~al.}(2011)\citenamefont {Jain},
  \citenamefont {Wittmann}, \citenamefont {Lydersen}, \citenamefont {Wiechers},
  \citenamefont {Elser}, \citenamefont {Marquardt}, \citenamefont {Makarov},\
  and\ \citenamefont {Leuchs}}]{bib:jain2011device}%
  \BibitemOpen
  \bibfield  {author} {\bibinfo {author} {\bibnamefont {Jain}, \bibfnamefont
  {Nitin}}, \bibinfo {author} {\bibfnamefont {Christoffer}\ \bibnamefont
  {Wittmann}}, \bibinfo {author} {\bibfnamefont {Lars}\ \bibnamefont
  {Lydersen}}, \bibinfo {author} {\bibfnamefont {Carlos}\ \bibnamefont
  {Wiechers}}, \bibinfo {author} {\bibfnamefont {Dominique}\ \bibnamefont
  {Elser}}, \bibinfo {author} {\bibfnamefont {Christoph}\ \bibnamefont
  {Marquardt}}, \bibinfo {author} {\bibfnamefont {Vadim}\ \bibnamefont
  {Makarov}}, and\ \bibinfo {author} {\bibfnamefont {Gerd}\ \bibnamefont
  {Leuchs}}} (\bibinfo {year} {2011}),\ \bibfield  {title} {\enquote {\bibinfo
  {title} {Device calibration impacts security of quantum key distribution},}\
  }\href {https://doi.org/10.1103/physrevlett.107.110501} {\bibfield  {journal}
  {\bibinfo  {journal} {Physical Review Letters}\ }\textbf {\bibinfo {volume}
  {107}},\ \bibinfo {pages} {110501}},\ \Eprint
  {https://arxiv.org/abs/arXiv:1103.2327v4} {arXiv:1103.2327v4} \BibitemShut
  {NoStop}%
\bibitem [{\citenamefont {Jansen}\ \emph {et~al.}(2007)\citenamefont {Jansen},
  \citenamefont {Ruskai},\ and\ \citenamefont {Seiler}}]{bib:jansen2007bounds}%
  \BibitemOpen
  \bibfield  {author} {\bibinfo {author} {\bibnamefont {Jansen}, \bibfnamefont
  {Sabine}}, \bibinfo {author} {\bibfnamefont {Mary-Beth}\ \bibnamefont
  {Ruskai}}, and\ \bibinfo {author} {\bibfnamefont {Ruedi}\ \bibnamefont
  {Seiler}}} (\bibinfo {year} {2007}),\ \bibfield  {title} {\enquote {\bibinfo
  {title} {Bounds for the adiabatic approximation with applications to quantum
  computation},}\ }\href@noop {} {\bibfield  {journal} {\bibinfo  {journal}
  {Journal of Mathematical Physics}\ }\textbf {\bibinfo {volume}
  {48}}~(\bibinfo {number} {10}),\ \bibinfo {pages} {102111}}\BibitemShut
  {NoStop}%
\bibitem [{\citenamefont {Jennewein}\ \emph {et~al.}(2014)\citenamefont
  {Jennewein}, \citenamefont {Bourgoin}, \citenamefont {Choi}, \citenamefont
  {D'Souza}, \citenamefont {Hudson}, \citenamefont {Laflamme}, \citenamefont
  {Higgins}, \citenamefont {Holloway}, \citenamefont {Meyer-Scott},
  \citenamefont {Erven} \emph {et~al.}}]{bib:jennewein2014qeyssat}%
  \BibitemOpen
  \bibfield  {author} {\bibinfo {author} {\bibnamefont {Jennewein},
  \bibfnamefont {T}}, \bibinfo {author} {\bibfnamefont {JP}~\bibnamefont
  {Bourgoin}}, \bibinfo {author} {\bibfnamefont {E}~\bibnamefont {Choi}},
  \bibinfo {author} {\bibfnamefont {I}~\bibnamefont {D'Souza}}, \bibinfo
  {author} {\bibfnamefont {D}~\bibnamefont {Hudson}}, \bibinfo {author}
  {\bibfnamefont {R}~\bibnamefont {Laflamme}}, \bibinfo {author} {\bibfnamefont
  {B}~\bibnamefont {Higgins}}, \bibinfo {author} {\bibfnamefont
  {C}~\bibnamefont {Holloway}}, \bibinfo {author} {\bibfnamefont
  {E}~\bibnamefont {Meyer-Scott}}, \bibinfo {author} {\bibfnamefont
  {C}~\bibnamefont {Erven}},  \emph {et~al.}} (\bibinfo {year} {2014}),\
  \bibfield  {title} {\enquote {\bibinfo {title} {Qeyssat: a mission proposal
  for a quantum receiver in space},}\ }\href
  {https://doi.org/10.1117/12.2041693} {\ 10.1117/12.2041693}\BibitemShut
  {NoStop}%
\bibitem [{\citenamefont {Jennewein}\ \emph
  {et~al.}(2001{\natexlab{a}})\citenamefont {Jennewein}, \citenamefont {Weihs},
  \citenamefont {Pan},\ and\ \citenamefont
  {Zeilinger}}]{bib:jennewein2001experimental}%
  \BibitemOpen
  \bibfield  {author} {\bibinfo {author} {\bibnamefont {Jennewein},
  \bibfnamefont {Thomas}}, \bibinfo {author} {\bibfnamefont {Gregor}\
  \bibnamefont {Weihs}}, \bibinfo {author} {\bibfnamefont {Jian-Wei}\
  \bibnamefont {Pan}}, and\ \bibinfo {author} {\bibfnamefont {Anton}\
  \bibnamefont {Zeilinger}}} (\bibinfo {year} {2001}{\natexlab{a}}),\ \bibfield
   {title} {\enquote {\bibinfo {title} {Experimental nonlocality proof of
  quantum teleportation and entanglement swapping},}\ }\href
  {https://doi.org/10.1103/physrevlett.88.017903} {\bibfield  {journal}
  {\bibinfo  {journal} {Physical Review Letters}\ }\textbf {\bibinfo {volume}
  {88}},\ \bibinfo {pages} {017903}},\ \Eprint
  {https://arxiv.org/abs/arXiv:quant-ph/0201134v1} {arXiv:quant-ph/0201134v1}
  \BibitemShut {NoStop}%
\bibitem [{\citenamefont {Jennewein}\ \emph
  {et~al.}(2001{\natexlab{b}})\citenamefont {Jennewein}, \citenamefont {Weihs},
  \citenamefont {Pan},\ and\ \citenamefont {Zeilinger}}]{bib:PRL_88_017903}%
  \BibitemOpen
  \bibfield  {author} {\bibinfo {author} {\bibnamefont {Jennewein},
  \bibfnamefont {Thomas}}, \bibinfo {author} {\bibfnamefont {Gregor}\
  \bibnamefont {Weihs}}, \bibinfo {author} {\bibfnamefont {Jian-Wei}\
  \bibnamefont {Pan}}, and\ \bibinfo {author} {\bibfnamefont {Anton}\
  \bibnamefont {Zeilinger}}} (\bibinfo {year} {2001}{\natexlab{b}}),\ \bibfield
   {title} {\enquote {\bibinfo {title} {Experimental nonlocality proof of
  quantum teleportation and entanglement swapping},}\ }\href
  {https://doi.org/10.1103/physrevlett.88.017903} {\bibfield  {journal}
  {\bibinfo  {journal} {Physical Review Letters}\ }\textbf {\bibinfo {volume}
  {88}},\ \bibinfo {pages} {017903}},\ \Eprint
  {https://arxiv.org/abs/arXiv:quant-ph/0201134v1} {arXiv:quant-ph/0201134v1}
  \BibitemShut {NoStop}%
\bibitem [{\citenamefont {Jeong}\ and\ \citenamefont
  {Ralph}(2007)}]{bib:JeongRalph05}%
  \BibitemOpen
  \bibfield  {author} {\bibinfo {author} {\bibnamefont {Jeong}, \bibfnamefont
  {H}}, and\ \bibinfo {author} {\bibfnamefont {T.~C.}\ \bibnamefont {Ralph}}}
  (\bibinfo {year} {2007}),\ \enquote {\bibinfo {title} {Schrodinger cat states
  for quantum information processing},}\ in\ \href
  {https://doi.org/10.1142/9781860948169_0009} {\emph {\bibinfo {booktitle}
  {Quantum Information with Continuous Variables of Atoms and Light}}}\
  (\bibinfo  {publisher} {Imperial College Press})\ \Eprint
  {https://arxiv.org/abs/arXiv:quant-ph/0509137} {arXiv:quant-ph/0509137}
  \BibitemShut {NoStop}%
\bibitem [{\citenamefont {Jiang}\ \emph {et~al.}(2009)\citenamefont {Jiang},
  \citenamefont {Taylor}, \citenamefont {Nemoto}, \citenamefont {Munro},
  \citenamefont {Meter},\ and\ \citenamefont {Lukin}}]{bib:jiang09}%
  \BibitemOpen
  \bibfield  {author} {\bibinfo {author} {\bibnamefont {Jiang}, \bibfnamefont
  {L}}, \bibinfo {author} {\bibfnamefont {J.~M.}\ \bibnamefont {Taylor}},
  \bibinfo {author} {\bibfnamefont {K.}~\bibnamefont {Nemoto}}, \bibinfo
  {author} {\bibfnamefont {W.~J}\ \bibnamefont {Munro}}, \bibinfo {author}
  {\bibfnamefont {R.~Van}\ \bibnamefont {Meter}}, and\ \bibinfo {author}
  {\bibfnamefont {M.~D.}\ \bibnamefont {Lukin}}} (\bibinfo {year} {2009}),\
  \bibfield  {title} {\enquote {\bibinfo {title} {Quantum repeater with
  encoding},}\ }\href {https://doi.org/10.1103/physreva.79.032325} {\bibfield
  {journal} {\bibinfo  {journal} {Physical Review A}\ }\textbf {\bibinfo
  {volume} {79}},\ \bibinfo {pages} {032325}},\ \Eprint
  {https://arxiv.org/abs/arXiv:0809.3629v2} {arXiv:0809.3629v2} \BibitemShut
  {NoStop}%
\bibitem [{\citenamefont {Jin}\ \emph {et~al.}(2015{\natexlab{a}})\citenamefont
  {Jin}, \citenamefont {Puigibert}, \citenamefont {Giner}, \citenamefont
  {Slater}, \citenamefont {Lamont}, \citenamefont {Verma}, \citenamefont
  {Shaw}, \citenamefont {Marsili}, \citenamefont {Nam}, \citenamefont {Oblak}
  \emph {et~al.}}]{bib:PRA_92_012329}%
  \BibitemOpen
  \bibfield  {author} {\bibinfo {author} {\bibnamefont {Jin}, \bibfnamefont
  {Jeongwan}}, \bibinfo {author} {\bibfnamefont {M~Grimau}\ \bibnamefont
  {Puigibert}}, \bibinfo {author} {\bibfnamefont {Lambert}\ \bibnamefont
  {Giner}}, \bibinfo {author} {\bibfnamefont {Joshua~A}\ \bibnamefont
  {Slater}}, \bibinfo {author} {\bibfnamefont {Michael~RE}\ \bibnamefont
  {Lamont}}, \bibinfo {author} {\bibfnamefont {Varun~B}\ \bibnamefont {Verma}},
  \bibinfo {author} {\bibfnamefont {MD}~\bibnamefont {Shaw}}, \bibinfo {author}
  {\bibfnamefont {Francesco}\ \bibnamefont {Marsili}}, \bibinfo {author}
  {\bibfnamefont {Sae~Woo}\ \bibnamefont {Nam}}, \bibinfo {author}
  {\bibfnamefont {Daniel}\ \bibnamefont {Oblak}},  \emph {et~al.}} (\bibinfo
  {year} {2015}{\natexlab{a}}),\ \bibfield  {title} {\enquote {\bibinfo {title}
  {Entanglement swapping with quantum-memory-compatible photons},}\ }\href
  {https://doi.org/10.1103/physreva.92.012329} {\bibfield  {journal} {\bibinfo
  {journal} {Physical Review A}\ }\textbf {\bibinfo {volume} {92}},\ \bibinfo
  {pages} {012329}},\ \Eprint {https://arxiv.org/abs/arXiv:1506.03704v1}
  {arXiv:1506.03704v1} \BibitemShut {NoStop}%
\bibitem [{\citenamefont {Jin}\ \emph {et~al.}(2015{\natexlab{b}})\citenamefont
  {Jin}, \citenamefont {Saglamyurek}, \citenamefont {Verma}, \citenamefont
  {Marsili}, \citenamefont {Nam}, \citenamefont {Oblak}, \citenamefont {Tittel}
  \emph {et~al.}}]{bib:jin2015telecom}%
  \BibitemOpen
  \bibfield  {author} {\bibinfo {author} {\bibnamefont {Jin}, \bibfnamefont
  {Jeongwan}}, \bibinfo {author} {\bibfnamefont {Erhan}\ \bibnamefont
  {Saglamyurek}}, \bibinfo {author} {\bibfnamefont {Varun}\ \bibnamefont
  {Verma}}, \bibinfo {author} {\bibfnamefont {Francesco}\ \bibnamefont
  {Marsili}}, \bibinfo {author} {\bibfnamefont {Sae~Woo}\ \bibnamefont {Nam}},
  \bibinfo {author} {\bibfnamefont {Daniel}\ \bibnamefont {Oblak}}, \bibinfo
  {author} {\bibfnamefont {Wolfgang}\ \bibnamefont {Tittel}},  \emph {et~al.}}
  (\bibinfo {year} {2015}{\natexlab{b}}),\ \bibfield  {title} {\enquote
  {\bibinfo {title} {Telecom-wavelength atomic quantum memory in optical fiber
  for heralded polarization qubits},}\ }\href
  {https://doi.org/10.1103/physrevlett.115.140501} {\bibfield  {journal}
  {\bibinfo  {journal} {Physical Review Letters}\ }\textbf {\bibinfo {volume}
  {115}},\ \bibinfo {pages} {140501}},\ \Eprint
  {https://arxiv.org/abs/arXiv:1506.04431v1} {arXiv:1506.04431v1} \BibitemShut
  {NoStop}%
\bibitem [{\citenamefont {Jin}\ \emph {et~al.}(2016)\citenamefont {Jin},
  \citenamefont {Fujiwara}, \citenamefont {Shimizu}, \citenamefont {Collins},
  \citenamefont {Buller}, \citenamefont {Yamashita}, \citenamefont {Miki},
  \citenamefont {Terai}, \citenamefont {Takeoka},\ and\ \citenamefont
  {Sasaki}}]{bib:jin2016detection}%
  \BibitemOpen
  \bibfield  {author} {\bibinfo {author} {\bibnamefont {Jin}, \bibfnamefont
  {Rui-Bo}}, \bibinfo {author} {\bibfnamefont {Mikio}\ \bibnamefont
  {Fujiwara}}, \bibinfo {author} {\bibfnamefont {Ryosuke}\ \bibnamefont
  {Shimizu}}, \bibinfo {author} {\bibfnamefont {Robert~J}\ \bibnamefont
  {Collins}}, \bibinfo {author} {\bibfnamefont {Gerald~S}\ \bibnamefont
  {Buller}}, \bibinfo {author} {\bibfnamefont {Taro}\ \bibnamefont
  {Yamashita}}, \bibinfo {author} {\bibfnamefont {Shigehito}\ \bibnamefont
  {Miki}}, \bibinfo {author} {\bibfnamefont {Hirotaka}\ \bibnamefont {Terai}},
  \bibinfo {author} {\bibfnamefont {Masahiro}\ \bibnamefont {Takeoka}}, and\
  \bibinfo {author} {\bibfnamefont {Masahide}\ \bibnamefont {Sasaki}}}
  (\bibinfo {year} {2016}),\ \bibfield  {title} {\enquote {\bibinfo {title}
  {Detection-dependent six-photon noon state interference},}\ }\href@noop {}
  {\bibfield  {journal} {\bibinfo  {journal} {Scientific Reports}\ }\textbf
  {\bibinfo {volume} {6}},\ \bibinfo {pages} {36914}},\ \Eprint
  {https://arxiv.org/abs/arXiv:1607.00926v1} {arXiv:1607.00926v1} \BibitemShut
  {NoStop}%
\bibitem [{\citenamefont {Jin}\ \emph {et~al.}(2010)\citenamefont {Jin},
  \citenamefont {Ren}, \citenamefont {Yang}, \citenamefont {Yi}, \citenamefont
  {Zhou}, \citenamefont {Xu}, \citenamefont {Wang}, \citenamefont {Yang},
  \citenamefont {Hu}, \citenamefont {Jiang} \emph
  {et~al.}}]{bib:Nat_Phot_4_376}%
  \BibitemOpen
  \bibfield  {author} {\bibinfo {author} {\bibnamefont {Jin}, \bibfnamefont
  {Xian-Min}}, \bibinfo {author} {\bibfnamefont {Ji-Gang}\ \bibnamefont {Ren}},
  \bibinfo {author} {\bibfnamefont {Bin}\ \bibnamefont {Yang}}, \bibinfo
  {author} {\bibfnamefont {Zhen-Huan}\ \bibnamefont {Yi}}, \bibinfo {author}
  {\bibfnamefont {Fei}\ \bibnamefont {Zhou}}, \bibinfo {author} {\bibfnamefont
  {Xiao-Fan}\ \bibnamefont {Xu}}, \bibinfo {author} {\bibfnamefont {Shao-Kai}\
  \bibnamefont {Wang}}, \bibinfo {author} {\bibfnamefont {Dong}\ \bibnamefont
  {Yang}}, \bibinfo {author} {\bibfnamefont {Yuan-Feng}\ \bibnamefont {Hu}},
  \bibinfo {author} {\bibfnamefont {Shuo}\ \bibnamefont {Jiang}},  \emph
  {et~al.}} (\bibinfo {year} {2010}),\ \bibfield  {title} {\enquote {\bibinfo
  {title} {Experimental free-space quantum teleportation},}\ }\href
  {https://doi.org/10.1038/nphoton.2010.87} {\bibfield  {journal} {\bibinfo
  {journal} {Nature Photonics}\ }\textbf {\bibinfo {volume} {4}},\ \bibinfo
  {pages} {376}}\BibitemShut {NoStop}%
\bibitem [{\citenamefont {Jones}\ \emph
  {et~al.}(2012{\natexlab{a}})\citenamefont {Jones}, \citenamefont {Van~Meter},
  \citenamefont {Fowler}, \citenamefont {McMahon}, \citenamefont {Kim},
  \citenamefont {Ladd},\ and\ \citenamefont {Yamamoto}}]{SD-Jones:2012aa}%
  \BibitemOpen
  \bibfield  {author} {\bibinfo {author} {\bibnamefont {Jones}, \bibfnamefont
  {N~Cody}}, \bibinfo {author} {\bibfnamefont {Rodney}\ \bibnamefont
  {Van~Meter}}, \bibinfo {author} {\bibfnamefont {Austin~G.}\ \bibnamefont
  {Fowler}}, \bibinfo {author} {\bibfnamefont {Peter~L.}\ \bibnamefont
  {McMahon}}, \bibinfo {author} {\bibfnamefont {Jungsang}\ \bibnamefont {Kim}},
  \bibinfo {author} {\bibfnamefont {Thaddeus~D.}\ \bibnamefont {Ladd}}, and\
  \bibinfo {author} {\bibfnamefont {Yoshihisa}\ \bibnamefont {Yamamoto}}}
  (\bibinfo {year} {2012}{\natexlab{a}}),\ \bibfield  {title} {\enquote
  {\bibinfo {title} {Layered architecture for quantum computing},}\ }\href
  {https://doi.org/10.1103/PhysRevX.2.031007} {\bibfield  {journal} {\bibinfo
  {journal} {Physical Review X}\ }\textbf {\bibinfo {volume} {2}},\ \bibinfo
  {pages} {031007}},\ \Eprint {https://arxiv.org/abs/arXiv:1010.5022v3}
  {arXiv:1010.5022v3} \BibitemShut {NoStop}%
\bibitem [{\citenamefont {Jones}\ \emph
  {et~al.}(2012{\natexlab{b}})\citenamefont {Jones}, \citenamefont {Van~Meter},
  \citenamefont {Fowler}, \citenamefont {McMahon}, \citenamefont {Kim},
  \citenamefont {Ladd},\ and\ \citenamefont {Yamamoto}}]{Jones:2012aa}%
  \BibitemOpen
  \bibfield  {author} {\bibinfo {author} {\bibnamefont {Jones}, \bibfnamefont
  {N~Cody}}, \bibinfo {author} {\bibfnamefont {Rodney}\ \bibnamefont
  {Van~Meter}}, \bibinfo {author} {\bibfnamefont {Austin~G.}\ \bibnamefont
  {Fowler}}, \bibinfo {author} {\bibfnamefont {Peter~L.}\ \bibnamefont
  {McMahon}}, \bibinfo {author} {\bibfnamefont {Jungsang}\ \bibnamefont {Kim}},
  \bibinfo {author} {\bibfnamefont {Thaddeus~D.}\ \bibnamefont {Ladd}}, and\
  \bibinfo {author} {\bibfnamefont {Yoshihisa}\ \bibnamefont {Yamamoto}}}
  (\bibinfo {year} {2012}{\natexlab{b}}),\ \bibfield  {title} {\enquote
  {\bibinfo {title} {Layered architecture for quantum computing},}\ }\href
  {https://doi.org/10.1103/PhysRevX.2.031007} {\bibfield  {journal} {\bibinfo
  {journal} {Physical Review X}\ }\textbf {\bibinfo {volume} {2}}~(\bibinfo
  {number} {3}),\ \bibinfo {pages} {031007--}}\BibitemShut {NoStop}%
\bibitem [{\citenamefont {Jordan}\ \emph {et~al.}(2012)\citenamefont {Jordan},
  \citenamefont {Lee},\ and\ \citenamefont {Preskill}}]{bib:JLP}%
  \BibitemOpen
  \bibfield  {author} {\bibinfo {author} {\bibnamefont {Jordan}, \bibfnamefont
  {Stephen~P}}, \bibinfo {author} {\bibfnamefont {Keith S.~M.}\ \bibnamefont
  {Lee}}, and\ \bibinfo {author} {\bibfnamefont {John}\ \bibnamefont
  {Preskill}}} (\bibinfo {year} {2012}),\ \bibfield  {title} {\enquote
  {\bibinfo {title} {Quantum algorithms for quantum field theories},}\ }\href
  {https://doi.org/10.1126/science.1217069} {\bibfield  {journal} {\bibinfo
  {journal} {Science}\ }\textbf {\bibinfo {volume} {336}},\ \bibinfo {pages}
  {1130}},\ \Eprint {https://arxiv.org/abs/arXiv:1111.3633v2}
  {arXiv:1111.3633v2} \BibitemShut {NoStop}%
\bibitem [{\citenamefont {Joseph}\ \emph {et~al.}(2013)\citenamefont {Joseph},
  \citenamefont {Laskov}, \citenamefont {Roli}, \citenamefont {Tygar},\ and\
  \citenamefont {Nelson}}]{bib:joseph2013machine}%
  \BibitemOpen
  \bibfield  {author} {\bibinfo {author} {\bibnamefont {Joseph}, \bibfnamefont
  {Anthony~D}}, \bibinfo {author} {\bibfnamefont {Pavel}\ \bibnamefont
  {Laskov}}, \bibinfo {author} {\bibfnamefont {Fabio}\ \bibnamefont {Roli}},
  \bibinfo {author} {\bibfnamefont {J.~Doug}\ \bibnamefont {Tygar}}, and\
  \bibinfo {author} {\bibfnamefont {Blaine}\ \bibnamefont {Nelson}}} (\bibinfo
  {year} {2013}),\ \bibfield  {title} {\enquote {\bibinfo {title} {Machine
  learning methods for computer security (dagstuhl perspectives workshop
  12371)},}\ }in\ \href@noop {} {\emph {\bibinfo {booktitle} {Dagstuhl
  Manifestos}}},\ Vol.~\bibinfo {volume} {3}\BibitemShut {NoStop}%
\bibitem [{\citenamefont {Josephson}(1974)}]{bib:josephson1974the}%
  \BibitemOpen
  \bibfield  {author} {\bibinfo {author} {\bibnamefont {Josephson},
  \bibfnamefont {B~D}}} (\bibinfo {year} {1974}),\ \bibfield  {title} {\enquote
  {\bibinfo {title} {The discovery of tunnelling supercurrents},}\ }\href
  {https://doi.org/10.1051/epn/19740503001} {\bibfield  {journal} {\bibinfo
  {journal} {Reviews in Modern Physics}\ }\textbf {\bibinfo {volume} {46}},\
  \bibinfo {pages} {251}}\BibitemShut {NoStop}%
\bibitem [{\citenamefont {Jouguet}\ \emph {et~al.}(2013)\citenamefont
  {Jouguet}, \citenamefont {Kunz-Jacques},\ and\ \citenamefont
  {Diamanti}}]{bib:jouguet2013preventing}%
  \BibitemOpen
  \bibfield  {author} {\bibinfo {author} {\bibnamefont {Jouguet}, \bibfnamefont
  {Paul}}, \bibinfo {author} {\bibfnamefont {S{\'e}bastien}\ \bibnamefont
  {Kunz-Jacques}}, and\ \bibinfo {author} {\bibfnamefont {Eleni}\ \bibnamefont
  {Diamanti}}} (\bibinfo {year} {2013}),\ \bibfield  {title} {\enquote
  {\bibinfo {title} {Preventing calibration attacks on the local oscillator in
  continuous-variable quantum key distribution},}\ }\href
  {https://doi.org/10.1103/physreva.87.062313} {\bibfield  {journal} {\bibinfo
  {journal} {Physical Review A}\ }\textbf {\bibinfo {volume} {87}},\ \bibinfo
  {pages} {062313}},\ \Eprint {https://arxiv.org/abs/arXiv:1304.7024v2}
  {arXiv:1304.7024v2} \BibitemShut {NoStop}%
\bibitem [{\citenamefont {Jozsa}\ \emph {et~al.}(2000)\citenamefont {Jozsa},
  \citenamefont {Abrams}, \citenamefont {Dowling},\ and\ \citenamefont
  {Williams}}]{bib:jozsa00}%
  \BibitemOpen
  \bibfield  {author} {\bibinfo {author} {\bibnamefont {Jozsa}, \bibfnamefont
  {Richard}}, \bibinfo {author} {\bibfnamefont {Daniel~S.}\ \bibnamefont
  {Abrams}}, \bibinfo {author} {\bibfnamefont {Jonathan~P.}\ \bibnamefont
  {Dowling}}, and\ \bibinfo {author} {\bibfnamefont {Colin~P.}\ \bibnamefont
  {Williams}}} (\bibinfo {year} {2000}),\ \bibfield  {title} {\enquote
  {\bibinfo {title} {Quantum clock synchronization based on shared prior
  entanglement},}\ }\href {https://doi.org/10.1103/physrevlett.85.2010}
  {\bibfield  {journal} {\bibinfo  {journal} {Physical Review Letters}\
  }\textbf {\bibinfo {volume} {85}},\ \bibinfo {pages} {2010}},\ \Eprint
  {https://arxiv.org/abs/arXiv:quant-ph/0004105v3} {arXiv:quant-ph/0004105v3}
  \BibitemShut {NoStop}%
\bibitem [{\citenamefont {Juska}\ \emph {et~al.}(2013)\citenamefont {Juska},
  \citenamefont {Dimastrodonato}, \citenamefont {Mereni}, \citenamefont
  {Gocalinska},\ and\ \citenamefont {Pelucchi}}]{bib:juska2013towards}%
  \BibitemOpen
  \bibfield  {author} {\bibinfo {author} {\bibnamefont {Juska}, \bibfnamefont
  {Gediminas}}, \bibinfo {author} {\bibfnamefont {Valeria}\ \bibnamefont
  {Dimastrodonato}}, \bibinfo {author} {\bibfnamefont {Lorenzo~O}\ \bibnamefont
  {Mereni}}, \bibinfo {author} {\bibfnamefont {Agnieszka}\ \bibnamefont
  {Gocalinska}}, and\ \bibinfo {author} {\bibfnamefont {Emanuele}\ \bibnamefont
  {Pelucchi}}} (\bibinfo {year} {2013}),\ \bibfield  {title} {\enquote
  {\bibinfo {title} {Towards quantum-dot arrays of entangled photon
  emitters},}\ }\href {https://doi.org/10.1038/nphoton.2013.128} {\bibfield
  {journal} {\bibinfo  {journal} {Nature Photonics}\ }\textbf {\bibinfo
  {volume} {7}},\ \bibinfo {pages} {527}},\ \Eprint
  {https://arxiv.org/abs/arXiv:1402.1709v1} {arXiv:1402.1709v1} \BibitemShut
  {NoStop}%
\bibitem [{\citenamefont {Kaltenbaek}\ \emph {et~al.}(2004)\citenamefont
  {Kaltenbaek}, \citenamefont {Aspelmeyer}, \citenamefont {Jennewein},
  \citenamefont {Brukner}, \citenamefont {Zeilinger}, \citenamefont
  {Pfennigbauer},\ and\ \citenamefont {Leeb}}]{bib:kaltenbaek2003proof}%
  \BibitemOpen
  \bibfield  {author} {\bibinfo {author} {\bibnamefont {Kaltenbaek},
  \bibfnamefont {Rainer}}, \bibinfo {author} {\bibfnamefont {Markus}\
  \bibnamefont {Aspelmeyer}}, \bibinfo {author} {\bibfnamefont {Thomas}\
  \bibnamefont {Jennewein}}, \bibinfo {author} {\bibfnamefont {Caslav}\
  \bibnamefont {Brukner}}, \bibinfo {author} {\bibfnamefont {Anton}\
  \bibnamefont {Zeilinger}}, \bibinfo {author} {\bibfnamefont {Martin}\
  \bibnamefont {Pfennigbauer}}, and\ \bibinfo {author} {\bibfnamefont
  {Walter~R.}\ \bibnamefont {Leeb}}} (\bibinfo {year} {2004}),\ \bibfield
  {title} {\enquote {\bibinfo {title} {Proof-of-concept experiments for quantum
  physics in space},}\ }in\ \href {https://doi.org/10.1117/12.506979} {\emph
  {\bibinfo {booktitle} {Proc. SPIE, Quantum Communications and Quantum
  Imaging}}},\ Vol.\ \bibinfo {volume} {5161},\ p.~\bibinfo {pages} {17},\
  \Eprint {https://arxiv.org/abs/arXiv:quant-ph/0308174v1}
  {arXiv:quant-ph/0308174v1} \BibitemShut {NoStop}%
\bibitem [{\citenamefont {Kaltenbaek}\ \emph {et~al.}(2009)\citenamefont
  {Kaltenbaek}, \citenamefont {Prevedel}, \citenamefont {Aspelmeyer},\ and\
  \citenamefont {Zeilinger}}]{bib:PRA_79_040302}%
  \BibitemOpen
  \bibfield  {author} {\bibinfo {author} {\bibnamefont {Kaltenbaek},
  \bibfnamefont {Rainer}}, \bibinfo {author} {\bibfnamefont {Robert}\
  \bibnamefont {Prevedel}}, \bibinfo {author} {\bibfnamefont {Markus}\
  \bibnamefont {Aspelmeyer}}, and\ \bibinfo {author} {\bibfnamefont {Anton}\
  \bibnamefont {Zeilinger}}} (\bibinfo {year} {2009}),\ \bibfield  {title}
  {\enquote {\bibinfo {title} {High-fidelity entanglement swapping with fully
  independent sources},}\ }\href {https://doi.org/10.1103/physreva.79.040302}
  {\bibfield  {journal} {\bibinfo  {journal} {Physical Review A}\ }\textbf
  {\bibinfo {volume} {79}},\ \bibinfo {pages} {040302}},\ \Eprint
  {https://arxiv.org/abs/arXiv:0809.3991v3} {arXiv:0809.3991v3} \BibitemShut
  {NoStop}%
\bibitem [{\citenamefont {Kandala}\ \emph {et~al.}(2017)\citenamefont
  {Kandala}, \citenamefont {Mezzacapo}, \citenamefont {Temme}, \citenamefont
  {Takita}, \citenamefont {Brink}, \citenamefont {Chow},\ and\ \citenamefont
  {Gambetta}}]{bib:kandala2017hardware}%
  \BibitemOpen
  \bibfield  {author} {\bibinfo {author} {\bibnamefont {Kandala}, \bibfnamefont
  {Abhinav}}, \bibinfo {author} {\bibfnamefont {Antonio}\ \bibnamefont
  {Mezzacapo}}, \bibinfo {author} {\bibfnamefont {Kristan}\ \bibnamefont
  {Temme}}, \bibinfo {author} {\bibfnamefont {Maika}\ \bibnamefont {Takita}},
  \bibinfo {author} {\bibfnamefont {Markus}\ \bibnamefont {Brink}}, \bibinfo
  {author} {\bibfnamefont {Jerry~M.}\ \bibnamefont {Chow}}, and\ \bibinfo
  {author} {\bibfnamefont {Jay~M.}\ \bibnamefont {Gambetta}}} (\bibinfo {year}
  {2017}),\ \bibfield  {title} {\enquote {\bibinfo {title} {Hardware-efficient
  variational quantum eigensolver for small molecules and quantum magnets},}\
  }\href {https://doi.org/10.1038/nature23879} {\bibfield  {journal} {\bibinfo
  {journal} {Nature}\ }\textbf {\bibinfo {volume} {549}},\ \bibinfo {pages}
  {242}},\ \Eprint {https://arxiv.org/abs/arXiv:1704.05018v2}
  {arXiv:1704.05018v2} \BibitemShut {NoStop}%
\bibitem [{\citenamefont {Kane}\ \emph {et~al.}(2017)\citenamefont {Kane},
  \citenamefont {Livni}, \citenamefont {Moran},\ and\ \citenamefont
  {Yehudayoff}}]{bib:kane2017communication}%
  \BibitemOpen
  \bibfield  {author} {\bibinfo {author} {\bibnamefont {Kane}, \bibfnamefont
  {Daniel~M}}, \bibinfo {author} {\bibfnamefont {Roi}\ \bibnamefont {Livni}},
  \bibinfo {author} {\bibfnamefont {Shay}\ \bibnamefont {Moran}}, and\ \bibinfo
  {author} {\bibfnamefont {Amir}\ \bibnamefont {Yehudayoff}}} (\bibinfo {year}
  {2017}),\ \bibfield  {title} {\enquote {\bibinfo {title} {On communication
  complexity of classification problems},}\ }\href@noop {} {\ }\Eprint
  {https://arxiv.org/abs/arXiv:1711.05893} {arXiv:1711.05893} \BibitemShut
  {NoStop}%
\bibitem [{\citenamefont {Kapoor}\ \emph {et~al.}(2016)\citenamefont {Kapoor},
  \citenamefont {Wiebe},\ and\ \citenamefont {Svore}}]{bib:kapoor2016quantum}%
  \BibitemOpen
  \bibfield  {author} {\bibinfo {author} {\bibnamefont {Kapoor}, \bibfnamefont
  {Ashish}}, \bibinfo {author} {\bibfnamefont {Nathan}\ \bibnamefont {Wiebe}},
  and\ \bibinfo {author} {\bibfnamefont {Krysta}\ \bibnamefont {Svore}}}
  (\bibinfo {year} {2016}),\ \bibfield  {title} {\enquote {\bibinfo {title}
  {Quantum perceptron models},}\ }in\ \href@noop {} {\emph {\bibinfo
  {booktitle} {Advances in Neural Information Processing Systems}}},\ p.\
  \bibinfo {pages} {3999},\ \Eprint {https://arxiv.org/abs/arXiv:1602.04799v1}
  {arXiv:1602.04799v1} \BibitemShut {NoStop}%
\bibitem [{\citenamefont {Kaufman}\ \emph {et~al.}(2015)\citenamefont
  {Kaufman}, \citenamefont {Lester}, \citenamefont {Foss-Feig}, \citenamefont
  {Wall}, \citenamefont {Rey},\ and\ \citenamefont
  {Regal}}]{bib:kaufman2015entangling}%
  \BibitemOpen
  \bibfield  {author} {\bibinfo {author} {\bibnamefont {Kaufman}, \bibfnamefont
  {AM}}, \bibinfo {author} {\bibfnamefont {BJ}~\bibnamefont {Lester}}, \bibinfo
  {author} {\bibfnamefont {M}~\bibnamefont {Foss-Feig}}, \bibinfo {author}
  {\bibfnamefont {ML}~\bibnamefont {Wall}}, \bibinfo {author} {\bibfnamefont
  {AM}~\bibnamefont {Rey}}, and\ \bibinfo {author} {\bibfnamefont
  {CA}~\bibnamefont {Regal}}} (\bibinfo {year} {2015}),\ \bibfield  {title}
  {\enquote {\bibinfo {title} {Entangling two transportable neutral atoms via
  local spin exchange},}\ }\href {https://doi.org/10.1038/nature16073}
  {\bibfield  {journal} {\bibinfo  {journal} {Nature}\ }\textbf {\bibinfo
  {volume} {527}},\ \bibinfo {pages} {208}},\ \Eprint
  {https://arxiv.org/abs/arXiv:1507.05586v2} {arXiv:1507.05586v2} \BibitemShut
  {NoStop}%
\bibitem [{\citenamefont {Kelly}\ \emph {et~al.}(2015)\citenamefont {Kelly},
  \citenamefont {Barends}, \citenamefont {Fowler}, \citenamefont {Megrant},
  \citenamefont {Jeffrey}, \citenamefont {White}, \citenamefont {Sank},
  \citenamefont {Mutus}, \citenamefont {Campbell}, \citenamefont {Chen} \emph
  {et~al.}}]{bib:kelly2015state}%
  \BibitemOpen
  \bibfield  {author} {\bibinfo {author} {\bibnamefont {Kelly}, \bibfnamefont
  {J}}, \bibinfo {author} {\bibfnamefont {R}~\bibnamefont {Barends}}, \bibinfo
  {author} {\bibfnamefont {AG}~\bibnamefont {Fowler}}, \bibinfo {author}
  {\bibfnamefont {A}~\bibnamefont {Megrant}}, \bibinfo {author} {\bibfnamefont
  {E}~\bibnamefont {Jeffrey}}, \bibinfo {author} {\bibfnamefont
  {TC}~\bibnamefont {White}}, \bibinfo {author} {\bibfnamefont {D}~\bibnamefont
  {Sank}}, \bibinfo {author} {\bibfnamefont {JY}~\bibnamefont {Mutus}},
  \bibinfo {author} {\bibfnamefont {B}~\bibnamefont {Campbell}}, \bibinfo
  {author} {\bibfnamefont {Yu}~\bibnamefont {Chen}},  \emph {et~al.}} (\bibinfo
  {year} {2015}),\ \bibfield  {title} {\enquote {\bibinfo {title} {State
  preservation by repetitive error detection in a superconducting quantum
  circuit},}\ }\href {https://doi.org/10.1038/nature14270} {\bibfield
  {journal} {\bibinfo  {journal} {Nature}\ }\textbf {\bibinfo {volume} {519}},\
  \bibinfo {pages} {66}},\ \Eprint {https://arxiv.org/abs/arXiv:1411.7403v1}
  {arXiv:1411.7403v1} \BibitemShut {NoStop}%
\bibitem [{\citenamefont {Kempe}(2003)}]{bib:Kempe03}%
  \BibitemOpen
  \bibfield  {author} {\bibinfo {author} {\bibnamefont {Kempe}, \bibfnamefont
  {J}}} (\bibinfo {year} {2003}),\ \bibfield  {title} {\enquote {\bibinfo
  {title} {Quantum random walks - an introductory overview},}\ }\href
  {https://doi.org/10.1080/00107151031000110776} {\bibfield  {journal}
  {\bibinfo  {journal} {Contemporary Physics}\ }\textbf {\bibinfo {volume}
  {44}},\ \bibinfo {pages} {307}},\ \Eprint
  {https://arxiv.org/abs/arXiv:quant-ph/0303081v1} {arXiv:quant-ph/0303081v1}
  \BibitemShut {NoStop}%
\bibitem [{\citenamefont {Khabiboulline}\ \emph
  {et~al.}(2019{\natexlab{a}})\citenamefont {Khabiboulline}, \citenamefont
  {Borregaard}, \citenamefont {De~Greve},\ and\ \citenamefont
  {Lukin}}]{PhysRevA.100.022316}%
  \BibitemOpen
  \bibfield  {author} {\bibinfo {author} {\bibnamefont {Khabiboulline},
  \bibfnamefont {E~T}}, \bibinfo {author} {\bibfnamefont {J.}~\bibnamefont
  {Borregaard}}, \bibinfo {author} {\bibfnamefont {K.}~\bibnamefont
  {De~Greve}}, and\ \bibinfo {author} {\bibfnamefont {M.~D.}\ \bibnamefont
  {Lukin}}} (\bibinfo {year} {2019}{\natexlab{a}}),\ \bibfield  {title}
  {\enquote {\bibinfo {title} {Quantum-assisted telescope arrays},}\ }\href
  {https://doi.org/10.1103/PhysRevA.100.022316} {\bibfield  {journal} {\bibinfo
   {journal} {Phys. Rev. A}\ }\textbf {\bibinfo {volume} {100}},\ \bibinfo
  {pages} {022316}}\BibitemShut {NoStop}%
\bibitem [{\citenamefont {Khabiboulline}\ \emph
  {et~al.}(2019{\natexlab{b}})\citenamefont {Khabiboulline}, \citenamefont
  {Borregaard}, \citenamefont {De~Greve},\ and\ \citenamefont
  {Lukin}}]{khabiboulline2019optical}%
  \BibitemOpen
  \bibfield  {author} {\bibinfo {author} {\bibnamefont {Khabiboulline},
  \bibfnamefont {Emil~T}}, \bibinfo {author} {\bibfnamefont {Johannes}\
  \bibnamefont {Borregaard}}, \bibinfo {author} {\bibfnamefont {Kristiaan}\
  \bibnamefont {De~Greve}}, and\ \bibinfo {author} {\bibfnamefont {Mikhail~D}\
  \bibnamefont {Lukin}}} (\bibinfo {year} {2019}{\natexlab{b}}),\ \bibfield
  {title} {\enquote {\bibinfo {title} {Optical interferometry with quantum
  networks},}\ }\href@noop {} {\bibfield  {journal} {\bibinfo  {journal} {Phys.
  Rev. Lett.}\ }\textbf {\bibinfo {volume} {123}}~(\bibinfo {number} {7}),\
  \bibinfo {pages} {070504}}\BibitemShut {NoStop}%
\bibitem [{\citenamefont {Khatri}\ and\ \citenamefont
  {Wilde}(2020)}]{khatri2020principles}%
  \BibitemOpen
  \bibfield  {author} {\bibinfo {author} {\bibnamefont {Khatri}, \bibfnamefont
  {Sumeet}}, and\ \bibinfo {author} {\bibfnamefont {Mark~M}\ \bibnamefont
  {Wilde}}} (\bibinfo {year} {2020}),\ \bibfield  {title} {\enquote {\bibinfo
  {title} {Principles of quantum communication theory: A modern approach},}\
  }\href@noop {} {\bibinfo  {journal} {arXiv preprint arXiv:2011.04672}\
  }\BibitemShut {NoStop}%
\bibitem [{\citenamefont {Kieling}\ \emph
  {et~al.}(2006{\natexlab{a}})\citenamefont {Kieling}, \citenamefont {Gross},\
  and\ \citenamefont {Eisert}}]{bib:Kieling06}%
  \BibitemOpen
\bibfield  {journal} {  }\bibfield  {author} {\bibinfo {author} {\bibnamefont
  {Kieling}, \bibfnamefont {K}}, \bibinfo {author} {\bibfnamefont
  {D.}~\bibnamefont {Gross}}, and\ \bibinfo {author} {\bibfnamefont
  {J.}~\bibnamefont {Eisert}}} (\bibinfo {year} {2006}{\natexlab{a}}),\
  \bibfield  {title} {\enquote {\bibinfo {title} {Minimal resources for linear
  optical one-way computing},}\ }\href
  {https://doi.org/10.1364/josab.24.000184} {\bibfield  {journal} {\bibinfo
  {journal} {Journal of the Optical Society of America B}\ ,\ \bibinfo {pages}
  {184}}}\Eprint {https://arxiv.org/abs/arXiv:quant-ph/0601190v2}
  {arXiv:quant-ph/0601190v2} \BibitemShut {NoStop}%
\bibitem [{\citenamefont {Kieling}\ \emph {et~al.}(2007)\citenamefont
  {Kieling}, \citenamefont {Gross},\ and\ \citenamefont
  {Eisert}}]{bib:Kieling07}%
  \BibitemOpen
  \bibfield  {author} {\bibinfo {author} {\bibnamefont {Kieling}, \bibfnamefont
  {K}}, \bibinfo {author} {\bibfnamefont {D.}~\bibnamefont {Gross}}, and\
  \bibinfo {author} {\bibfnamefont {J.}~\bibnamefont {Eisert}}} (\bibinfo
  {year} {2007}),\ \bibfield  {title} {\enquote {\bibinfo {title} {Cluster
  state preparation using gates operating at arbitrary success
  probabilities},}\ }\href {https://doi.org/10.1088/1367-2630/9/6/200}
  {\bibfield  {journal} {\bibinfo  {journal} {New Journal of Physics}\ }\textbf
  {\bibinfo {volume} {9}},\ \bibinfo {pages} {200}}\BibitemShut {NoStop}%
\bibitem [{\citenamefont {Kieling}\ \emph
  {et~al.}(2006{\natexlab{b}})\citenamefont {Kieling}, \citenamefont
  {Rudolph},\ and\ \citenamefont {Eisert}}]{bib:KielingRudolphEisert06}%
  \BibitemOpen
  \bibfield  {author} {\bibinfo {author} {\bibnamefont {Kieling}, \bibfnamefont
  {K}}, \bibinfo {author} {\bibfnamefont {T.}~\bibnamefont {Rudolph}}, and\
  \bibinfo {author} {\bibfnamefont {J.}~\bibnamefont {Eisert}}} (\bibinfo
  {year} {2006}{\natexlab{b}}),\ \bibfield  {title} {\enquote {\bibinfo {title}
  {Percolation, renormalization, and quantum computing with non-deterministic
  gates},}\ }\href {https://doi.org/10.1103/physrevlett.99.130501} {\bibfield
  {journal} {\bibinfo  {journal} {Physical Review Letters}\ }\textbf {\bibinfo
  {volume} {99}},\ \bibinfo {pages} {130501}},\ \Eprint
  {https://arxiv.org/abs/arXiv:quant-ph/0611140v3} {arXiv:quant-ph/0611140v3}
  \BibitemShut {NoStop}%
\bibitem [{\citenamefont {Kiesel}\ \emph {et~al.}(2005)\citenamefont {Kiesel},
  \citenamefont {Schmid}, \citenamefont {Weber}, \citenamefont {Ursin},\ and\
  \citenamefont {Weinfurter}}]{bib:Kiesel2005}%
  \BibitemOpen
  \bibfield  {author} {\bibinfo {author} {\bibnamefont {Kiesel}, \bibfnamefont
  {Nikolai}}, \bibinfo {author} {\bibfnamefont {Christian}\ \bibnamefont
  {Schmid}}, \bibinfo {author} {\bibfnamefont {Ulrich}\ \bibnamefont {Weber}},
  \bibinfo {author} {\bibfnamefont {Rupert}\ \bibnamefont {Ursin}}, and\
  \bibinfo {author} {\bibfnamefont {Harald}\ \bibnamefont {Weinfurter}}}
  (\bibinfo {year} {2005}),\ \bibfield  {title} {\enquote {\bibinfo {title}
  {Linear optics controlled-phase gate made simple},}\ }\href
  {https://doi.org/10.1103/PhysRevLett.95.210505} {\bibfield  {journal}
  {\bibinfo  {journal} {Physical Review Letters}\ }\textbf {\bibinfo {volume}
  {95}},\ \bibinfo {pages} {210505}}\BibitemShut {NoStop}%
\bibitem [{\citenamefont {Kim}\ \emph {et~al.}(2021)\citenamefont {Kim},
  \citenamefont {Song}, \citenamefont {Kim},\ and\ \citenamefont
  {Park}}]{kim2021quantumdotdetectors}%
  \BibitemOpen
  \bibfield  {author} {\bibinfo {author} {\bibnamefont {Kim}, \bibfnamefont
  {Jaehyun}}, \bibinfo {author} {\bibfnamefont {Seungho}\ \bibnamefont {Song}},
  \bibinfo {author} {\bibfnamefont {Yong-Hoon}\ \bibnamefont {Kim}}, and\
  \bibinfo {author} {\bibfnamefont {Sung~Kyu}\ \bibnamefont {Park}}} (\bibinfo
  {year} {2021}),\ \bibfield  {title} {\enquote {\bibinfo {title} {Recent
  progress of quantum dot-based photonic devices and systems: A comprehensive
  review of materials, devices, and applications},}\ }\href
  {https://doi.org/https://doi.org/10.1002/sstr.202000024} {\bibfield
  {journal} {\bibinfo  {journal} {Small Structures}\ }\textbf {\bibinfo
  {volume} {2}}~(\bibinfo {number} {3}),\ \bibinfo {pages} {2000024}},\ \Eprint
  {https://arxiv.org/abs/https://onlinelibrary.wiley.com/doi/pdf/10.1002/sstr.202000024}
  {https://onlinelibrary.wiley.com/doi/pdf/10.1002/sstr.202000024} \BibitemShut
  {NoStop}%
\bibitem [{\citenamefont {Kimble}(2008{\natexlab{a}})}]{bib:Kimble2008}%
  \BibitemOpen
  \bibfield  {author} {\bibinfo {author} {\bibnamefont {Kimble}, \bibfnamefont
  {H~J}}} (\bibinfo {year} {2008}{\natexlab{a}}),\ \bibfield  {title} {\enquote
  {\bibinfo {title} {The quantum internet},}\ }\href
  {https://doi.org/10.1038/nature07127} {\bibfield  {journal} {\bibinfo
  {journal} {Nature}\ }\textbf {\bibinfo {volume} {453}},\ \bibinfo {pages}
  {1023}},\ \Eprint {https://arxiv.org/abs/arXiv:0806.4195v1}
  {arXiv:0806.4195v1} \BibitemShut {NoStop}%
\bibitem [{\citenamefont {Kimble}(2008{\natexlab{b}})}]{bib:kimble2008quantum}%
  \BibitemOpen
  \bibfield  {author} {\bibinfo {author} {\bibnamefont {Kimble}, \bibfnamefont
  {H~Jeff}}} (\bibinfo {year} {2008}{\natexlab{b}}),\ \bibfield  {title}
  {\enquote {\bibinfo {title} {The quantum internet},}\ }\href
  {https://doi.org/10.1038/nature07127} {\bibfield  {journal} {\bibinfo
  {journal} {Nature}\ }\textbf {\bibinfo {volume} {453}},\ \bibinfo {pages}
  {1023}},\ \Eprint {https://arxiv.org/abs/arXiv:0806.4195v1}
  {arXiv:0806.4195v1} \BibitemShut {NoStop}%
\bibitem [{\citenamefont {King}\ \emph {et~al.}(2018)\citenamefont {King},
  \citenamefont {Carrasquilla}, \citenamefont {Raymond}, \citenamefont
  {Ozfidan}, \citenamefont {Andriyash}, \citenamefont {Berkley}, \citenamefont
  {Reis}, \citenamefont {Lanting}, \citenamefont {Harris}, \citenamefont
  {Altomare} \emph {et~al.}}]{bib:king2018observation}%
  \BibitemOpen
  \bibfield  {author} {\bibinfo {author} {\bibnamefont {King}, \bibfnamefont
  {Andrew~D}}, \bibinfo {author} {\bibfnamefont {Juan}\ \bibnamefont
  {Carrasquilla}}, \bibinfo {author} {\bibfnamefont {Jack}\ \bibnamefont
  {Raymond}}, \bibinfo {author} {\bibfnamefont {Isil}\ \bibnamefont {Ozfidan}},
  \bibinfo {author} {\bibfnamefont {Evgeny}\ \bibnamefont {Andriyash}},
  \bibinfo {author} {\bibfnamefont {Andrew}\ \bibnamefont {Berkley}}, \bibinfo
  {author} {\bibfnamefont {Mauricio}\ \bibnamefont {Reis}}, \bibinfo {author}
  {\bibfnamefont {Trevor}\ \bibnamefont {Lanting}}, \bibinfo {author}
  {\bibfnamefont {Richard}\ \bibnamefont {Harris}}, \bibinfo {author}
  {\bibfnamefont {Fabio}\ \bibnamefont {Altomare}},  \emph {et~al.}} (\bibinfo
  {year} {2018}),\ \bibfield  {title} {\enquote {\bibinfo {title} {Observation
  of topological phenomena in a programmable lattice of 1,800 qubits},}\ }\href
  {https://doi.org/10.1038/s41586-018-0410-x} {\bibfield  {journal} {\bibinfo
  {journal} {Nature}\ }\textbf {\bibinfo {volume} {560}},\ \bibinfo {pages}
  {456}},\ \Eprint {https://arxiv.org/abs/arXiv:1803.02047v1}
  {arXiv:1803.02047v1} \BibitemShut {NoStop}%
\bibitem [{\citenamefont {Kiraz}\ \emph {et~al.}(2004)\citenamefont {Kiraz},
  \citenamefont {Atat{\"u}re},\ and\ \citenamefont {Imamo{\u
  g}lu}}]{bib:Kiraz04}%
  \BibitemOpen
  \bibfield  {author} {\bibinfo {author} {\bibnamefont {Kiraz}, \bibfnamefont
  {A}}, \bibinfo {author} {\bibfnamefont {M.}~\bibnamefont {Atat{\"u}re}}, and\
  \bibinfo {author} {\bibfnamefont {A.}~\bibnamefont {Imamo{\u g}lu}}}
  (\bibinfo {year} {2004}),\ \bibfield  {title} {\enquote {\bibinfo {title}
  {Quantum-dot single-photon sources: Prospects for applications in linear
  optics quantum-information processing},}\ }\href
  {https://doi.org/10.1103/physreva.69.032305} {\bibfield  {journal} {\bibinfo
  {journal} {Physical Review A}\ }\textbf {\bibinfo {volume} {69}},\ \bibinfo
  {pages} {032305}},\ \Eprint {https://arxiv.org/abs/arXiv:quant-ph/0308117v2}
  {arXiv:quant-ph/0308117v2} \BibitemShut {NoStop}%
\bibitem [{\citenamefont {Kitaev}(1995)}]{kitaev1995quantum}%
  \BibitemOpen
  \bibfield  {author} {\bibinfo {author} {\bibnamefont {Kitaev}, \bibfnamefont
  {A~Yu}}} (\bibinfo {year} {1995}),\ \bibfield  {title} {\enquote {\bibinfo
  {title} {Quantum measurements and the abelian stabilizer problem},}\
  }\href@noop {} {\bibinfo  {journal} {arXiv preprint quant-ph/9511026}\
  }\BibitemShut {NoStop}%
\bibitem [{\citenamefont {Kitaev}(1997)}]{bib:K97}%
  \BibitemOpen
\bibfield  {journal} {  }\bibfield  {author} {\bibinfo {author} {\bibnamefont
  {Kitaev}, \bibfnamefont {AY}}} (\bibinfo {year} {1997}),\ \bibfield  {title}
  {\enquote {\bibinfo {title} {{Quantum Computations: algorithms and error
  correction}},}\ }\href@noop {} {\bibfield  {journal} {\bibinfo  {journal}
  {Russ. Math. Serv.}\ }\textbf {\bibinfo {volume} {52}}~(\bibinfo {number}
  {6}),\ \bibinfo {pages} {1191}}\BibitemShut {NoStop}%
\bibitem [{\citenamefont {Knill}(2002)}]{bib:Knill02}%
  \BibitemOpen
  \bibfield  {author} {\bibinfo {author} {\bibnamefont {Knill}, \bibfnamefont
  {E}}} (\bibinfo {year} {2002}),\ \bibfield  {title} {\enquote {\bibinfo
  {title} {Quantum gates using linear optics and postselection},}\ }\href
  {https://doi.org/10.1103/physreva.66.052306} {\bibfield  {journal} {\bibinfo
  {journal} {Physical Review A}\ }\textbf {\bibinfo {volume} {66}},\ \bibinfo
  {pages} {052306}}\BibitemShut {NoStop}%
\bibitem [{\citenamefont {Knill}\ and\ \citenamefont
  {Laflamme}(1997)}]{bib:Knill97}%
  \BibitemOpen
  \bibfield  {author} {\bibinfo {author} {\bibnamefont {Knill}, \bibfnamefont
  {E}}, and\ \bibinfo {author} {\bibfnamefont {R.}~\bibnamefont {Laflamme}}}
  (\bibinfo {year} {1997}),\ \bibfield  {title} {\enquote {\bibinfo {title}
  {Theory of quantum error-correcting codes},}\ }\href
  {https://doi.org/10.1103/physreva.55.900} {\bibfield  {journal} {\bibinfo
  {journal} {Physical Review A}\ }\textbf {\bibinfo {volume} {55}},\ \bibinfo
  {pages} {900}}\BibitemShut {NoStop}%
\bibitem [{\citenamefont {Knill}\ \emph {et~al.}(2001)\citenamefont {Knill},
  \citenamefont {Laflamme},\ and\ \citenamefont {Milburn}}]{bib:KLM01}%
  \BibitemOpen
  \bibfield  {author} {\bibinfo {author} {\bibnamefont {Knill}, \bibfnamefont
  {E}}, \bibinfo {author} {\bibfnamefont {R.}~\bibnamefont {Laflamme}}, and\
  \bibinfo {author} {\bibfnamefont {G.}~\bibnamefont {Milburn}}} (\bibinfo
  {year} {2001}),\ \bibfield  {title} {\enquote {\bibinfo {title} {A scheme for
  efficient quantum computation with linear optics},}\ }\href
  {https://doi.org/10.1038/35051009} {\bibfield  {journal} {\bibinfo  {journal}
  {Nature}\ }\textbf {\bibinfo {volume} {409}},\ \bibinfo {pages}
  {46}}\BibitemShut {NoStop}%
\bibitem [{\citenamefont {Knill}(2005)}]{bib:Knill2005}%
  \BibitemOpen
  \bibfield  {author} {\bibinfo {author} {\bibnamefont {Knill}, \bibfnamefont
  {Emanuel}}} (\bibinfo {year} {2005}),\ \bibfield  {title} {\enquote {\bibinfo
  {title} {Quantum computing with realistically noisy devices},}\ }\href
  {https://doi.org/10.1038/nature03350} {\bibfield  {journal} {\bibinfo
  {journal} {Nature}\ }\textbf {\bibinfo {volume} {434}},\ \bibinfo {pages}
  {39}}\BibitemShut {NoStop}%
\bibitem [{\citenamefont {Koch}\ \emph {et~al.}(2007)\citenamefont {Koch},
  \citenamefont {Terri}, \citenamefont {Gambetta}, \citenamefont {Houck},
  \citenamefont {Schuster}, \citenamefont {Majer}, \citenamefont {Blais},
  \citenamefont {Devoret}, \citenamefont {Girvin},\ and\ \citenamefont
  {Schoelkopf}}]{bib:koch2007charge}%
  \BibitemOpen
  \bibfield  {author} {\bibinfo {author} {\bibnamefont {Koch}, \bibfnamefont
  {Jens}}, \bibinfo {author} {\bibfnamefont {M~Yu}\ \bibnamefont {Terri}},
  \bibinfo {author} {\bibfnamefont {Jay}\ \bibnamefont {Gambetta}}, \bibinfo
  {author} {\bibfnamefont {Andrew~A}\ \bibnamefont {Houck}}, \bibinfo {author}
  {\bibfnamefont {DI}~\bibnamefont {Schuster}}, \bibinfo {author}
  {\bibfnamefont {J}~\bibnamefont {Majer}}, \bibinfo {author} {\bibfnamefont
  {Alexandre}\ \bibnamefont {Blais}}, \bibinfo {author} {\bibfnamefont
  {Michel~H}\ \bibnamefont {Devoret}}, \bibinfo {author} {\bibfnamefont
  {Steven~M}\ \bibnamefont {Girvin}}, and\ \bibinfo {author} {\bibfnamefont
  {Robert~J}\ \bibnamefont {Schoelkopf}}} (\bibinfo {year} {2007}),\ \bibfield
  {title} {\enquote {\bibinfo {title} {Charge-insensitive qubit design derived
  from the cooper pair box},}\ }\href
  {https://doi.org/10.1103/physreva.76.042319} {\bibfield  {journal} {\bibinfo
  {journal} {Physical Review A}\ }\textbf {\bibinfo {volume} {76}},\ \bibinfo
  {pages} {042319}},\ \Eprint {https://arxiv.org/abs/arXiv:cond-mat/0703002v2}
  {arXiv:cond-mat/0703002v2} \BibitemShut {NoStop}%
\bibitem [{\citenamefont {Kogeda}\ and\ \citenamefont
  {Agbinya}(2006)}]{bib:kogeda2006prediction}%
  \BibitemOpen
  \bibfield  {author} {\bibinfo {author} {\bibnamefont {Kogeda}, \bibfnamefont
  {P}}, and\ \bibinfo {author} {\bibfnamefont {Johnson~I.}\ \bibnamefont
  {Agbinya}}} (\bibinfo {year} {2006}),\ \bibfield  {title} {\enquote {\bibinfo
  {title} {Prediction of faults in cellular networks using bayesian network
  model},}\ }in\ \href@noop {} {\emph {\bibinfo {booktitle} {International
  conference on Wireless Broadband and Ultra Wideband Communication}}}\
  (\bibinfo {organization} {UTS ePress})\BibitemShut {NoStop}%
\bibitem [{\citenamefont {Kok}\ \emph {et~al.}(2005)\citenamefont {Kok},
  \citenamefont {Munro}, \citenamefont {Nemoto}, \citenamefont {Ralph},
  \citenamefont {Dowling},\ and\ \citenamefont {Milburn}}]{bib:Kok05}%
  \BibitemOpen
  \bibfield  {author} {\bibinfo {author} {\bibnamefont {Kok}, \bibfnamefont
  {P}}, \bibinfo {author} {\bibfnamefont {W.~J.}\ \bibnamefont {Munro}},
  \bibinfo {author} {\bibfnamefont {K.}~\bibnamefont {Nemoto}}, \bibinfo
  {author} {\bibfnamefont {T.~C.}\ \bibnamefont {Ralph}}, \bibinfo {author}
  {\bibfnamefont {Jonathan~P.}\ \bibnamefont {Dowling}}, and\ \bibinfo {author}
  {\bibfnamefont {G.~J.}\ \bibnamefont {Milburn}}} (\bibinfo {year} {2005}),\
  \bibfield  {title} {\enquote {\bibinfo {title} {Linear optical quantum
  computing with photonic qubits},}\ }\href
  {https://doi.org/10.1103/revmodphys.79.135} {\bibfield  {journal} {\bibinfo
  {journal} {Reviews in Modern Physics}\ }\textbf {\bibinfo {volume} {79}},\
  \bibinfo {pages} {135}}\BibitemShut {NoStop}%
\bibitem [{\citenamefont {Kok}\ and\ \citenamefont
  {Lovett}(2010{\natexlab{a}})}]{bib:KokLovettBook}%
  \BibitemOpen
  \bibfield  {author} {\bibinfo {author} {\bibnamefont {Kok}, \bibfnamefont
  {Pieter}}, and\ \bibinfo {author} {\bibfnamefont {Brendon~W.}\ \bibnamefont
  {Lovett}}} (\bibinfo {year} {2010}{\natexlab{a}}),\ \href@noop {} {\emph
  {\bibinfo {title} {Introduction to Optical Quantum Information Processing}}}\
  (\bibinfo  {publisher} {Cambridge University Press, Cambridge})\BibitemShut
  {NoStop}%
\bibitem [{\citenamefont {Kok}\ and\ \citenamefont
  {Lovett}(2010{\natexlab{b}})}]{kok2010introduction}%
  \BibitemOpen
  \bibfield  {author} {\bibinfo {author} {\bibnamefont {Kok}, \bibfnamefont
  {Pieter}}, and\ \bibinfo {author} {\bibfnamefont {Brendon~W}\ \bibnamefont
  {Lovett}}} (\bibinfo {year} {2010}{\natexlab{b}}),\ \href@noop {} {\emph
  {\bibinfo {title} {Introduction to optical quantum information processing}}}\
  (\bibinfo  {publisher} {Cambridge university press})\BibitemShut {NoStop}%
\bibitem [{\citenamefont {Komar}\ \emph {et~al.}(2014)\citenamefont {Komar},
  \citenamefont {Kessler}, \citenamefont {Bishof}, \citenamefont {Sorensen},
  \citenamefont {Ye},\ and\ \citenamefont {Lukin}}]{bib:komar14}%
  \BibitemOpen
  \bibfield  {author} {\bibinfo {author} {\bibnamefont {Komar}, \bibfnamefont
  {P}}, \bibinfo {author} {\bibfnamefont {E.~M.}\ \bibnamefont {Kessler}},
  \bibinfo {author} {\bibfnamefont {L.}~\bibnamefont {Bishof}, \bibfnamefont
  {M.;~Jiang}}, \bibinfo {author} {\bibfnamefont {A.~S.}\ \bibnamefont
  {Sorensen}}, \bibinfo {author} {\bibfnamefont {J.}~\bibnamefont {Ye}}, and\
  \bibinfo {author} {\bibfnamefont {M.~D.}\ \bibnamefont {Lukin}}} (\bibinfo
  {year} {2014}),\ \bibfield  {title} {\enquote {\bibinfo {title} {A quantum
  network of clocks},}\ }\href {https://doi.org/10.1038/nphys3000} {\bibfield
  {journal} {\bibinfo  {journal} {Nature Physics}\ }\textbf {\bibinfo {volume}
  {10}},\ \bibinfo {pages} {582}},\ \Eprint
  {https://arxiv.org/abs/arXiv:1310.6045v1} {arXiv:1310.6045v1} \BibitemShut
  {NoStop}%
\bibitem [{\citenamefont {Kong}\ \emph {et~al.}(2017)\citenamefont {Kong},
  \citenamefont {Xin}, \citenamefont {Wei}, \citenamefont {Wang}, \citenamefont
  {Li},\ and\ \citenamefont {Long}}]{bib:kong2017implementation}%
  \BibitemOpen
  \bibfield  {author} {\bibinfo {author} {\bibnamefont {Kong}, \bibfnamefont
  {Xiangyu}}, \bibinfo {author} {\bibfnamefont {Tao}\ \bibnamefont {Xin}},
  \bibinfo {author} {\bibfnamefont {ShiJie}\ \bibnamefont {Wei}}, \bibinfo
  {author} {\bibfnamefont {Bixue}\ \bibnamefont {Wang}}, \bibinfo {author}
  {\bibfnamefont {Keren}\ \bibnamefont {Li}}, and\ \bibinfo {author}
  {\bibfnamefont {GuiLu}\ \bibnamefont {Long}}} (\bibinfo {year} {2017}),\
  \bibfield  {title} {\enquote {\bibinfo {title} {Implementation of multiparty
  quantum clock synchronization},}\ }\href@noop {} {\ }\Eprint
  {https://arxiv.org/abs/arXiv:1708.06050} {arXiv:1708.06050} \BibitemShut
  {NoStop}%
\bibitem [{\citenamefont {Konno}\ \emph {et~al.}(2024)\citenamefont {Konno},
  \citenamefont {Asavanant}, \citenamefont {Hanamura}, \citenamefont
  {Nagayoshi}, \citenamefont {Fukui}, \citenamefont {Sakaguchi}, \citenamefont
  {Ide}, \citenamefont {China}, \citenamefont {Yabuno}, \citenamefont {Miki},
  \citenamefont {Terai}, \citenamefont {Takase}, \citenamefont {Endo},
  \citenamefont {Marek}, \citenamefont {Filip}, \citenamefont {van Loock},\
  and\ \citenamefont {Furusawa}}]{furusawa2024opticalGKPprep}%
  \BibitemOpen
  \bibfield  {author} {\bibinfo {author} {\bibnamefont {Konno}, \bibfnamefont
  {Shunya}}, \bibinfo {author} {\bibfnamefont {Warit}\ \bibnamefont
  {Asavanant}}, \bibinfo {author} {\bibfnamefont {Fumiya}\ \bibnamefont
  {Hanamura}}, \bibinfo {author} {\bibfnamefont {Hironari}\ \bibnamefont
  {Nagayoshi}}, \bibinfo {author} {\bibfnamefont {Kosuke}\ \bibnamefont
  {Fukui}}, \bibinfo {author} {\bibfnamefont {Atsushi}\ \bibnamefont
  {Sakaguchi}}, \bibinfo {author} {\bibfnamefont {Ryuhoh}\ \bibnamefont {Ide}},
  \bibinfo {author} {\bibfnamefont {Fumihiro}\ \bibnamefont {China}}, \bibinfo
  {author} {\bibfnamefont {Masahiro}\ \bibnamefont {Yabuno}}, \bibinfo {author}
  {\bibfnamefont {Shigehito}\ \bibnamefont {Miki}}, \bibinfo {author}
  {\bibfnamefont {Hirotaka}\ \bibnamefont {Terai}}, \bibinfo {author}
  {\bibfnamefont {Kan}\ \bibnamefont {Takase}}, \bibinfo {author}
  {\bibfnamefont {Mamoru}\ \bibnamefont {Endo}}, \bibinfo {author}
  {\bibfnamefont {Petr}\ \bibnamefont {Marek}}, \bibinfo {author}
  {\bibfnamefont {Radim}\ \bibnamefont {Filip}}, \bibinfo {author}
  {\bibfnamefont {Peter}\ \bibnamefont {van Loock}}, and\ \bibinfo {author}
  {\bibfnamefont {Akira}\ \bibnamefont {Furusawa}}} (\bibinfo {year} {2024}),\
  \bibfield  {title} {\enquote {\bibinfo {title} {Logical states for
  fault-tolerant quantum computation with propagating light},}\ }\href
  {https://doi.org/10.1126/science.adk7560} {\bibfield  {journal} {\bibinfo
  {journal} {Science}\ }\textbf {\bibinfo {volume} {383}}~(\bibinfo {number}
  {6680}),\ \bibinfo {pages} {289--293}},\ \Eprint
  {https://arxiv.org/abs/https://www.science.org/doi/pdf/10.1126/science.adk7560}
  {https://www.science.org/doi/pdf/10.1126/science.adk7560} \BibitemShut
  {NoStop}%
\bibitem [{\citenamefont {Kraska}\ \emph {et~al.}(2013)\citenamefont {Kraska},
  \citenamefont {Talwalkar}, \citenamefont {Duchi}, \citenamefont {Griffith},
  \citenamefont {Franklin},\ and\ \citenamefont {Jordan}}]{bib:MLbase}%
  \BibitemOpen
  \bibfield  {author} {\bibinfo {author} {\bibnamefont {Kraska}, \bibfnamefont
  {Tim}}, \bibinfo {author} {\bibfnamefont {Ameet}\ \bibnamefont {Talwalkar}},
  \bibinfo {author} {\bibfnamefont {John~C.}\ \bibnamefont {Duchi}}, \bibinfo
  {author} {\bibfnamefont {Rean}\ \bibnamefont {Griffith}}, \bibinfo {author}
  {\bibfnamefont {Michael~J.}\ \bibnamefont {Franklin}}, and\ \bibinfo {author}
  {\bibfnamefont {Michael~I.}\ \bibnamefont {Jordan}}} (\bibinfo {year}
  {2013}),\ \bibfield  {title} {\enquote {\bibinfo {title} {Mlbase: A
  distributed machine-learning system},}\ }\href@noop {} {\bibfield  {journal}
  {\bibinfo  {journal} {Cidr}\ }\textbf {\bibinfo {volume} {1}},\ \bibinfo
  {pages} {2}}\BibitemShut {NoStop}%
\bibitem [{\citenamefont {Krauter}\ \emph {et~al.}(2013)\citenamefont
  {Krauter}, \citenamefont {Salart}, \citenamefont {Muschik}, \citenamefont
  {Petersen}, \citenamefont {Shen}, \citenamefont {Fernholz},\ and\
  \citenamefont {Polzik}}]{bib:Nat_Phys_9_400}%
  \BibitemOpen
  \bibfield  {author} {\bibinfo {author} {\bibnamefont {Krauter}, \bibfnamefont
  {H}}, \bibinfo {author} {\bibfnamefont {D}~\bibnamefont {Salart}}, \bibinfo
  {author} {\bibfnamefont {CA}~\bibnamefont {Muschik}}, \bibinfo {author}
  {\bibfnamefont {Jonas~Meyer}\ \bibnamefont {Petersen}}, \bibinfo {author}
  {\bibfnamefont {Heng}\ \bibnamefont {Shen}}, \bibinfo {author} {\bibfnamefont
  {Thomas}\ \bibnamefont {Fernholz}}, and\ \bibinfo {author} {\bibfnamefont
  {Eugene~Simon}\ \bibnamefont {Polzik}}} (\bibinfo {year} {2013}),\ \bibfield
  {title} {\enquote {\bibinfo {title} {Deterministic quantum teleportation
  between distant atomic objects},}\ }\href {https://doi.org/10.1038/nphys2631}
  {\bibfield  {journal} {\bibinfo  {journal} {Nature Physics}\ }\textbf
  {\bibinfo {volume} {9}},\ \bibinfo {pages} {400}},\ \Eprint
  {https://arxiv.org/abs/arXiv:1212.6746v2} {arXiv:1212.6746v2} \BibitemShut
  {NoStop}%
\bibitem [{\citenamefont {Kr{\v{c}}o}\ and\ \citenamefont
  {Paul}(2002)}]{bib:krvco2002quantum}%
  \BibitemOpen
  \bibfield  {author} {\bibinfo {author} {\bibnamefont {Kr{\v{c}}o},
  \bibfnamefont {Marko}}, and\ \bibinfo {author} {\bibfnamefont {Prabasaj}\
  \bibnamefont {Paul}}} (\bibinfo {year} {2002}),\ \bibfield  {title} {\enquote
  {\bibinfo {title} {Quantum clock synchronization: Multiparty protocol},}\
  }\href {https://doi.org/10.1103/physreva.66.024305} {\bibfield  {journal}
  {\bibinfo  {journal} {Physical Review A}\ }\textbf {\bibinfo {volume} {66}},\
  \bibinfo {pages} {024305}},\ \Eprint
  {https://arxiv.org/abs/arXiv:quant-ph/0112111v1} {arXiv:quant-ph/0112111v1}
  \BibitemShut {NoStop}%
\bibitem [{\citenamefont {Kumar}\ \emph {et~al.}(2012)\citenamefont {Kumar},
  \citenamefont {Barrios}, \citenamefont {MacRae}, \citenamefont {Cairns},
  \citenamefont {Huntington},\ and\ \citenamefont
  {Lvovsky}}]{bib:kumar2012versatile}%
  \BibitemOpen
  \bibfield  {author} {\bibinfo {author} {\bibnamefont {Kumar}, \bibfnamefont
  {Ranjeet}}, \bibinfo {author} {\bibfnamefont {Erick}\ \bibnamefont
  {Barrios}}, \bibinfo {author} {\bibfnamefont {Andrew}\ \bibnamefont
  {MacRae}}, \bibinfo {author} {\bibfnamefont {E}~\bibnamefont {Cairns}},
  \bibinfo {author} {\bibfnamefont {EH}~\bibnamefont {Huntington}}, and\
  \bibinfo {author} {\bibfnamefont {AI}~\bibnamefont {Lvovsky}}} (\bibinfo
  {year} {2012}),\ \bibfield  {title} {\enquote {\bibinfo {title} {Versatile
  wideband balanced detector for quantum optical homodyne tomography},}\ }\href
  {https://doi.org/10.1016/j.optcom.2012.07.103} {\bibfield  {journal}
  {\bibinfo  {journal} {Optics Communications}\ }\textbf {\bibinfo {volume}
  {285}},\ \bibinfo {pages} {5259}},\ \Eprint
  {https://arxiv.org/abs/arXiv:1111.4012v1} {arXiv:1111.4012v1} \BibitemShut
  {NoStop}%
\bibitem [{\citenamefont {Kurakin}\ \emph {et~al.}(2018)\citenamefont
  {Kurakin}, \citenamefont {Goodfellow}, \citenamefont {Bengio}, \citenamefont
  {Dong}, \citenamefont {Liao}, \citenamefont {Liang}, \citenamefont {Pang},
  \citenamefont {Zhu}, \citenamefont {Hu}, \citenamefont {Xie} \emph
  {et~al.}}]{bib:kurakin2018adversarial}%
  \BibitemOpen
  \bibfield  {author} {\bibinfo {author} {\bibnamefont {Kurakin}, \bibfnamefont
  {Alexey}}, \bibinfo {author} {\bibfnamefont {Ian}\ \bibnamefont
  {Goodfellow}}, \bibinfo {author} {\bibfnamefont {Samy}\ \bibnamefont
  {Bengio}}, \bibinfo {author} {\bibfnamefont {Yinpeng}\ \bibnamefont {Dong}},
  \bibinfo {author} {\bibfnamefont {Fangzhou}\ \bibnamefont {Liao}}, \bibinfo
  {author} {\bibfnamefont {Ming}\ \bibnamefont {Liang}}, \bibinfo {author}
  {\bibfnamefont {Tianyu}\ \bibnamefont {Pang}}, \bibinfo {author}
  {\bibfnamefont {Jun}\ \bibnamefont {Zhu}}, \bibinfo {author} {\bibfnamefont
  {Xiaolin}\ \bibnamefont {Hu}}, \bibinfo {author} {\bibfnamefont {Cihang}\
  \bibnamefont {Xie}},  \emph {et~al.}} (\bibinfo {year} {2018}),\ \bibfield
  {title} {\enquote {\bibinfo {title} {Adversarial attacks and defences
  competition},}\ }in\ \href {https://doi.org/10.1007/978-3-319-94042-7_11}
  {\emph {\bibinfo {booktitle} {The NIPS'17 Competition: Building Intelligent
  Systems}}}\ (\bibinfo  {publisher} {Springer})\ p.\ \bibinfo {pages} {195},\
  \Eprint {https://arxiv.org/abs/arXiv:1804.00097v1} {arXiv:1804.00097v1}
  \BibitemShut {NoStop}%
\bibitem [{\citenamefont {Kurpiers}\ \emph {et~al.}(2018)\citenamefont
  {Kurpiers}, \citenamefont {Magnard}, \citenamefont {Walter}, \citenamefont
  {Royer}, \citenamefont {Pechal}, \citenamefont {Heinsoo}, \citenamefont
  {Salath{\'e}}, \citenamefont {Akin}, \citenamefont {Storz}, \citenamefont
  {Besse}, \citenamefont {Gasparinetti}, \citenamefont {Blais},\ and\
  \citenamefont {Wallraff}}]{SD-Kurpiers:2018aa}%
  \BibitemOpen
  \bibfield  {author} {\bibinfo {author} {\bibnamefont {Kurpiers},
  \bibfnamefont {P}}, \bibinfo {author} {\bibfnamefont {P.}~\bibnamefont
  {Magnard}}, \bibinfo {author} {\bibfnamefont {T.}~\bibnamefont {Walter}},
  \bibinfo {author} {\bibfnamefont {B.}~\bibnamefont {Royer}}, \bibinfo
  {author} {\bibfnamefont {M.}~\bibnamefont {Pechal}}, \bibinfo {author}
  {\bibfnamefont {J.}~\bibnamefont {Heinsoo}}, \bibinfo {author} {\bibfnamefont
  {Y.}~\bibnamefont {Salath{\'e}}}, \bibinfo {author} {\bibfnamefont
  {A.}~\bibnamefont {Akin}}, \bibinfo {author} {\bibfnamefont {S.}~\bibnamefont
  {Storz}}, \bibinfo {author} {\bibfnamefont {J.~C.}\ \bibnamefont {Besse}},
  \bibinfo {author} {\bibfnamefont {S.}~\bibnamefont {Gasparinetti}}, \bibinfo
  {author} {\bibfnamefont {A.}~\bibnamefont {Blais}}, and\ \bibinfo {author}
  {\bibfnamefont {A.}~\bibnamefont {Wallraff}}} (\bibinfo {year} {2018}),\
  \bibfield  {title} {\enquote {\bibinfo {title} {Deterministic quantum state
  transfer and remote entanglement using microwave photons},}\ }\href
  {https://doi.org/10.1038/s41586-018-0195-y} {\bibfield  {journal} {\bibinfo
  {journal} {Nature}\ }\textbf {\bibinfo {volume} {558}},\ \bibinfo {pages}
  {264}}\BibitemShut {NoStop}%
\bibitem [{\citenamefont {Kurtsiefer}\ \emph {et~al.}(2001)\citenamefont
  {Kurtsiefer}, \citenamefont {Zarda}, \citenamefont {Mayer},\ and\
  \citenamefont {Weinfurter}}]{bib:kurtsiefer2001breakdown}%
  \BibitemOpen
  \bibfield  {author} {\bibinfo {author} {\bibnamefont {Kurtsiefer},
  \bibfnamefont {Christian}}, \bibinfo {author} {\bibfnamefont {Patrick}\
  \bibnamefont {Zarda}}, \bibinfo {author} {\bibfnamefont {Sonja}\ \bibnamefont
  {Mayer}}, and\ \bibinfo {author} {\bibfnamefont {Harald}\ \bibnamefont
  {Weinfurter}}} (\bibinfo {year} {2001}),\ \bibfield  {title} {\enquote
  {\bibinfo {title} {The breakdown flash of silicon avalanche photodiodes-back
  door for eavesdropper attacks?}}\ }\href
  {https://doi.org/10.1080/09500340110070235} {\bibfield  {journal} {\bibinfo
  {journal} {Journal of Modern Optics}\ }\textbf {\bibinfo {volume} {48}},\
  \bibinfo {pages} {2039}}\BibitemShut {NoStop}%
\bibitem [{\citenamefont {Kwiat}\ \emph {et~al.}(1995)\citenamefont {Kwiat},
  \citenamefont {Mattle}, \citenamefont {Weinfurter}, \citenamefont
  {Zeilinger}, \citenamefont {Sergienko},\ and\ \citenamefont
  {Shih}}]{bib:PhysRevLett.75.4337}%
  \BibitemOpen
  \bibfield  {author} {\bibinfo {author} {\bibnamefont {Kwiat}, \bibfnamefont
  {Paul~G}}, \bibinfo {author} {\bibfnamefont {Klaus}\ \bibnamefont {Mattle}},
  \bibinfo {author} {\bibfnamefont {Harald}\ \bibnamefont {Weinfurter}},
  \bibinfo {author} {\bibfnamefont {Anton}\ \bibnamefont {Zeilinger}}, \bibinfo
  {author} {\bibfnamefont {Alexander~V.}\ \bibnamefont {Sergienko}}, and\
  \bibinfo {author} {\bibfnamefont {Yanhua}\ \bibnamefont {Shih}}} (\bibinfo
  {year} {1995}),\ \bibfield  {title} {\enquote {\bibinfo {title} {New
  high-intensity source of polarization-entangled photon pairs},}\ }\href
  {https://doi.org/10.1103/physrevlett.75.4337} {\bibfield  {journal} {\bibinfo
   {journal} {Physical Review Letters}\ }\textbf {\bibinfo {volume} {75}},\
  \bibinfo {pages} {4337}}\BibitemShut {NoStop}%
\bibitem [{\citenamefont {Ladd}\ \emph {et~al.}(2010)\citenamefont {Ladd},
  \citenamefont {Jelezko}, \citenamefont {Laflamme}, \citenamefont {Nakamura},
  \citenamefont {Monroe},\ and\ \citenamefont {O'Brien}}]{bib:ladd2010quantum}%
  \BibitemOpen
  \bibfield  {author} {\bibinfo {author} {\bibnamefont {Ladd}, \bibfnamefont
  {Thaddeus~D}}, \bibinfo {author} {\bibfnamefont {Fedor}\ \bibnamefont
  {Jelezko}}, \bibinfo {author} {\bibfnamefont {Raymond}\ \bibnamefont
  {Laflamme}}, \bibinfo {author} {\bibfnamefont {Yasunobu}\ \bibnamefont
  {Nakamura}}, \bibinfo {author} {\bibfnamefont {Christopher}\ \bibnamefont
  {Monroe}}, and\ \bibinfo {author} {\bibfnamefont {Jeremy~L}\ \bibnamefont
  {O'Brien}}} (\bibinfo {year} {2010}),\ \bibfield  {title} {\enquote {\bibinfo
  {title} {Quantum computers},}\ }\href {https://doi.org/10.1038/nature08812}
  {\bibfield  {journal} {\bibinfo  {journal} {Nature}\ }\textbf {\bibinfo
  {volume} {464}},\ \bibinfo {pages} {45}},\ \Eprint
  {https://arxiv.org/abs/arXiv:1009.2267v1} {arXiv:1009.2267v1} \BibitemShut
  {NoStop}%
\bibitem [{\citenamefont {Laflamme}\ \emph {et~al.}(1996)\citenamefont
  {Laflamme}, \citenamefont {Miquel}, \citenamefont {Paz},\ and\ \citenamefont
  {Zurek}}]{bib:LMPZ96}%
  \BibitemOpen
  \bibfield  {author} {\bibinfo {author} {\bibnamefont {Laflamme},
  \bibfnamefont {R}}, \bibinfo {author} {\bibfnamefont {C.}~\bibnamefont
  {Miquel}}, \bibinfo {author} {\bibfnamefont {J.P.}\ \bibnamefont {Paz}}, and\
  \bibinfo {author} {\bibfnamefont {W.H.}\ \bibnamefont {Zurek}}} (\bibinfo
  {year} {1996}),\ \bibfield  {title} {\enquote {\bibinfo {title} {{Perfect
  Quantum Error Correcting Code}},}\ }\href@noop {} {\bibfield  {journal}
  {\bibinfo  {journal} {Phys. Rev. Lett.}\ }\textbf {\bibinfo {volume} {77}},\
  \bibinfo {pages} {198}}\BibitemShut {NoStop}%
\bibitem [{\citenamefont {Lamas-Linares}\ and\ \citenamefont
  {Kurtsiefer}(2007)}]{bib:lamas2007breaking}%
  \BibitemOpen
  \bibfield  {author} {\bibinfo {author} {\bibnamefont {Lamas-Linares},
  \bibfnamefont {Ant{\'\i}a}}, and\ \bibinfo {author} {\bibfnamefont
  {Christian}\ \bibnamefont {Kurtsiefer}}} (\bibinfo {year} {2007}),\ \bibfield
   {title} {\enquote {\bibinfo {title} {Breaking a quantum key distribution
  system through a timing side channel},}\ }\href
  {https://doi.org/10.1364/oe.15.009388} {\bibfield  {journal} {\bibinfo
  {journal} {Optics Express}\ }\textbf {\bibinfo {volume} {15}},\ \bibinfo
  {pages} {9388}},\ \Eprint {https://arxiv.org/abs/arXiv:0704.3297v2}
  {arXiv:0704.3297v2} \BibitemShut {NoStop}%
\bibitem [{\citenamefont {Lamport}(1979)}]{lamport1979constructing}%
  \BibitemOpen
  \bibfield  {author} {\bibinfo {author} {\bibnamefont {Lamport}, \bibfnamefont
  {Leslie}}} (\bibinfo {year} {1979}),\ \href@noop {} {\emph {\bibinfo {title}
  {Constructing digital signatures from a one-way function}}},\ \bibinfo {type}
  {Tech. Rep.}\ (\bibinfo  {institution} {Technical Report CSL-98, SRI
  International Palo Alto})\BibitemShut {NoStop}%
\bibitem [{\citenamefont {Landry}\ \emph {et~al.}(2007)\citenamefont {Landry},
  \citenamefont {van Houwelingen}, \citenamefont {Beveratos}, \citenamefont
  {Zbinden},\ and\ \citenamefont {Gisin}}]{bib:landry2007quantum}%
  \BibitemOpen
  \bibfield  {author} {\bibinfo {author} {\bibnamefont {Landry}, \bibfnamefont
  {Olivier}}, \bibinfo {author} {\bibfnamefont {Jeroen Anton~Willem}\
  \bibnamefont {van Houwelingen}}, \bibinfo {author} {\bibfnamefont {Alexios}\
  \bibnamefont {Beveratos}}, \bibinfo {author} {\bibfnamefont {Hugo}\
  \bibnamefont {Zbinden}}, and\ \bibinfo {author} {\bibfnamefont {Nicolas}\
  \bibnamefont {Gisin}}} (\bibinfo {year} {2007}),\ \bibfield  {title}
  {\enquote {\bibinfo {title} {Quantum teleportation over the swisscom
  telecommunication network},}\ }\href
  {https://doi.org/10.1364/josab.24.000398} {\bibfield  {journal} {\bibinfo
  {journal} {JOSA B}\ }\textbf {\bibinfo {volume} {24}},\ \bibinfo {pages}
  {398}},\ \Eprint {https://arxiv.org/abs/arXiv:quant-ph/0605010v2}
  {arXiv:quant-ph/0605010v2} \BibitemShut {NoStop}%
\bibitem [{\citenamefont {Langford}\ \emph {et~al.}(2005)\citenamefont
  {Langford}, \citenamefont {Weinhold}, \citenamefont {Prevedel}, \citenamefont
  {Resch}, \citenamefont {Gilchrist}, \citenamefont {O'Brien}, \citenamefont
  {Pryde},\ and\ \citenamefont {White}}]{bib:Langford2005}%
  \BibitemOpen
  \bibfield  {author} {\bibinfo {author} {\bibnamefont {Langford},
  \bibfnamefont {N~K}}, \bibinfo {author} {\bibfnamefont {T.~J.}\ \bibnamefont
  {Weinhold}}, \bibinfo {author} {\bibfnamefont {R.}~\bibnamefont {Prevedel}},
  \bibinfo {author} {\bibfnamefont {K.~J.}\ \bibnamefont {Resch}}, \bibinfo
  {author} {\bibfnamefont {A.}~\bibnamefont {Gilchrist}}, \bibinfo {author}
  {\bibfnamefont {J.~L.}\ \bibnamefont {O'Brien}}, \bibinfo {author}
  {\bibfnamefont {G.~J.}\ \bibnamefont {Pryde}}, and\ \bibinfo {author}
  {\bibfnamefont {A.~G.}\ \bibnamefont {White}}} (\bibinfo {year} {2005}),\
  \bibfield  {title} {\enquote {\bibinfo {title} {Demonstration of a simple
  entangling optical gate and its use in bell-state analysis},}\ }\href
  {https://doi.org/10.1103/PhysRevLett.95.210504} {\bibfield  {journal}
  {\bibinfo  {journal} {Physical Review Letters}\ }\textbf {\bibinfo {volume}
  {95}},\ \bibinfo {pages} {210504}},\ \Eprint
  {https://arxiv.org/abs/arXiv:quant-ph/0506262v2} {arXiv:quant-ph/0506262v2}
  \BibitemShut {NoStop}%
\bibitem [{\citenamefont {Lanyon}\ \emph {et~al.}(2009)\citenamefont {Lanyon},
  \citenamefont {Barbieri}, \citenamefont {Almeida}, \citenamefont {Jennewein},
  \citenamefont {Ralph}, \citenamefont {Resch}, \citenamefont {Pryde},
  \citenamefont {O'Brien}, \citenamefont {Gilchrist},\ and\ \citenamefont
  {White}}]{bib:lanyon2009simplifying}%
  \BibitemOpen
  \bibfield  {author} {\bibinfo {author} {\bibnamefont {Lanyon}, \bibfnamefont
  {Benjamin~P}}, \bibinfo {author} {\bibfnamefont {Marco}\ \bibnamefont
  {Barbieri}}, \bibinfo {author} {\bibfnamefont {Marcelo~P}\ \bibnamefont
  {Almeida}}, \bibinfo {author} {\bibfnamefont {Thomas}\ \bibnamefont
  {Jennewein}}, \bibinfo {author} {\bibfnamefont {Timothy~C}\ \bibnamefont
  {Ralph}}, \bibinfo {author} {\bibfnamefont {Kevin~J}\ \bibnamefont {Resch}},
  \bibinfo {author} {\bibfnamefont {Geoff~J}\ \bibnamefont {Pryde}}, \bibinfo
  {author} {\bibfnamefont {Jeremy~L}\ \bibnamefont {O'Brien}}, \bibinfo
  {author} {\bibfnamefont {Alexei}\ \bibnamefont {Gilchrist}}, and\ \bibinfo
  {author} {\bibfnamefont {Andrew~G}\ \bibnamefont {White}}} (\bibinfo {year}
  {2009}),\ \bibfield  {title} {\enquote {\bibinfo {title} {Simplifying quantum
  logic using higher-dimensional hilbert spaces},}\ }\href@noop {} {\bibfield
  {journal} {\bibinfo  {journal} {Nature Physics}\ }\textbf {\bibinfo {volume}
  {5}},\ \bibinfo {pages} {134}}\BibitemShut {NoStop}%
\bibitem [{\citenamefont {Laurat}\ \emph {et~al.}(2007)\citenamefont {Laurat},
  \citenamefont {Choi}, \citenamefont {Deng}, \citenamefont {Chou},\ and\
  \citenamefont {Kimble}}]{bib:LauratKimble07}%
  \BibitemOpen
  \bibfield  {author} {\bibinfo {author} {\bibnamefont {Laurat}, \bibfnamefont
  {J}}, \bibinfo {author} {\bibfnamefont {K.~S.}\ \bibnamefont {Choi}},
  \bibinfo {author} {\bibfnamefont {H.}~\bibnamefont {Deng}}, \bibinfo {author}
  {\bibfnamefont {C.~W.}\ \bibnamefont {Chou}}, and\ \bibinfo {author}
  {\bibfnamefont {H.~J.}\ \bibnamefont {Kimble}}} (\bibinfo {year} {2007}),\
  \bibfield  {title} {\enquote {\bibinfo {title} {Heralded entanglement between
  atomic ensembles: Preparation, decoherence, and scaling},}\ }\href
  {https://doi.org/10.1103/physrevlett.99.180504} {\bibfield  {journal}
  {\bibinfo  {journal} {Physical Review Letters}\ }\textbf {\bibinfo {volume}
  {99}},\ \bibinfo {pages} {180504}},\ \Eprint
  {https://arxiv.org/abs/arXiv:0706.0528v2} {arXiv:0706.0528v2} \BibitemShut
  {NoStop}%
\bibitem [{\citenamefont {Lee}\ \emph {et~al.}(2002)\citenamefont {Lee},
  \citenamefont {Kok}, \citenamefont {Cerf},\ and\ \citenamefont
  {Dowling}}]{bib:PhysRevA.65.030101}%
  \BibitemOpen
  \bibfield  {author} {\bibinfo {author} {\bibnamefont {Lee}, \bibfnamefont
  {Hwang}}, \bibinfo {author} {\bibfnamefont {Pieter}\ \bibnamefont {Kok}},
  \bibinfo {author} {\bibfnamefont {Nicolas~J.}\ \bibnamefont {Cerf}}, and\
  \bibinfo {author} {\bibfnamefont {Jonathan~P.}\ \bibnamefont {Dowling}}}
  (\bibinfo {year} {2002}),\ \bibfield  {title} {\enquote {\bibinfo {title}
  {Linear optics and projective measurements alone suffice to create
  large-photon-number path entanglement},}\ }\href
  {https://doi.org/10.1103/physreva.65.030101} {\bibfield  {journal} {\bibinfo
  {journal} {Physical Review A}\ }\textbf {\bibinfo {volume} {65}},\ \bibinfo
  {pages} {030101}},\ \Eprint {https://arxiv.org/abs/arXiv:quant-ph/0109080v2}
  {arXiv:quant-ph/0109080v2} \BibitemShut {NoStop}%
\bibitem [{\citenamefont {Lee}\ \emph {et~al.}(2000)\citenamefont {Lee},
  \citenamefont {Lim}, \citenamefont {Luan}, \citenamefont {Agarwal},
  \citenamefont {Foresi},\ and\ \citenamefont {Kimerling}}]{bib:lee2000}%
  \BibitemOpen
  \bibfield  {author} {\bibinfo {author} {\bibnamefont {Lee}, \bibfnamefont
  {Kevin~K}}, \bibinfo {author} {\bibfnamefont {Desmond~R}\ \bibnamefont
  {Lim}}, \bibinfo {author} {\bibfnamefont {Hsin-Chiao}\ \bibnamefont {Luan}},
  \bibinfo {author} {\bibfnamefont {Anuradha}\ \bibnamefont {Agarwal}},
  \bibinfo {author} {\bibfnamefont {James}\ \bibnamefont {Foresi}}, and\
  \bibinfo {author} {\bibfnamefont {Lionel~C}\ \bibnamefont {Kimerling}}}
  (\bibinfo {year} {2000}),\ \bibfield  {title} {\enquote {\bibinfo {title}
  {Effect of size and roughness on light transmission in a si/sio2 waveguide:
  Experiments and model},}\ }\href {https://doi.org/10.1063/1.1308532}
  {\bibfield  {journal} {\bibinfo  {journal} {Applied Physics Letters}\
  }\textbf {\bibinfo {volume} {77}},\ \bibinfo {pages} {1617}}\BibitemShut
  {NoStop}%
\bibitem [{\citenamefont {Leghtas}\ \emph {et~al.}(2013)\citenamefont
  {Leghtas}, \citenamefont {Kirchmair}, \citenamefont {Vlastakis},
  \citenamefont {Schoelkopf}, \citenamefont {Devoret},\ and\ \citenamefont
  {Mirrahimi}}]{zaki2103cats}%
  \BibitemOpen
  \bibfield  {author} {\bibinfo {author} {\bibnamefont {Leghtas}, \bibfnamefont
  {Zaki}}, \bibinfo {author} {\bibfnamefont {Gerhard}\ \bibnamefont
  {Kirchmair}}, \bibinfo {author} {\bibfnamefont {Brian}\ \bibnamefont
  {Vlastakis}}, \bibinfo {author} {\bibfnamefont {Robert~J.}\ \bibnamefont
  {Schoelkopf}}, \bibinfo {author} {\bibfnamefont {Michel~H.}\ \bibnamefont
  {Devoret}}, and\ \bibinfo {author} {\bibfnamefont {Mazyar}\ \bibnamefont
  {Mirrahimi}}} (\bibinfo {year} {2013}),\ \bibfield  {title} {\enquote
  {\bibinfo {title} {Hardware-efficient autonomous quantum memory
  protection},}\ }\href {https://doi.org/10.1103/PhysRevLett.111.120501}
  {\bibfield  {journal} {\bibinfo  {journal} {Phys. Rev. Lett.}\ }\textbf
  {\bibinfo {volume} {111}},\ \bibinfo {pages} {120501}}\BibitemShut {NoStop}%
\bibitem [{\citenamefont {Leibfried}\ \emph {et~al.}(2003)\citenamefont
  {Leibfried}, \citenamefont {Blatt}, \citenamefont {Monroe},\ and\
  \citenamefont {Wineland}}]{bib:leibfried2003quantum}%
  \BibitemOpen
  \bibfield  {author} {\bibinfo {author} {\bibnamefont {Leibfried},
  \bibfnamefont {D}}, \bibinfo {author} {\bibfnamefont {R}~\bibnamefont
  {Blatt}}, \bibinfo {author} {\bibfnamefont {C}~\bibnamefont {Monroe}}, and\
  \bibinfo {author} {\bibfnamefont {D}~\bibnamefont {Wineland}}} (\bibinfo
  {year} {2003}),\ \bibfield  {title} {\enquote {\bibinfo {title} {Quantum
  dynamics of single trapped ions},}\ }\href
  {https://doi.org/10.1103/revmodphys.75.281} {\bibfield  {journal} {\bibinfo
  {journal} {Reviews in Modern Physics}\ }\textbf {\bibinfo {volume} {75}},\
  \bibinfo {pages} {281}}\BibitemShut {NoStop}%
\bibitem [{\citenamefont {Lekitsch}\ \emph
  {et~al.}(2017{\natexlab{a}})\citenamefont {Lekitsch}, \citenamefont {Weidt},
  \citenamefont {Fowler}, \citenamefont {M{\o}lmer}, \citenamefont {Devitt},
  \citenamefont {Wunderlich},\ and\ \citenamefont
  {Hensinger}}]{SD-Lekitsch:2017aa}%
  \BibitemOpen
  \bibfield  {author} {\bibinfo {author} {\bibnamefont {Lekitsch},
  \bibfnamefont {Bjoern}}, \bibinfo {author} {\bibfnamefont {Sebastian}\
  \bibnamefont {Weidt}}, \bibinfo {author} {\bibfnamefont {Austin~G.}\
  \bibnamefont {Fowler}}, \bibinfo {author} {\bibfnamefont {Klaus}\
  \bibnamefont {M{\o}lmer}}, \bibinfo {author} {\bibfnamefont {Simon~J.}\
  \bibnamefont {Devitt}}, \bibinfo {author} {\bibfnamefont {Christof}\
  \bibnamefont {Wunderlich}}, and\ \bibinfo {author} {\bibfnamefont
  {Winfried~K.}\ \bibnamefont {Hensinger}}} (\bibinfo {year}
  {2017}{\natexlab{a}}),\ \bibfield  {title} {\enquote {\bibinfo {title}
  {Blueprint for a microwave trapped ion quantum computer},}\ }\href
  {https://doi.org/10.1126/sciadv.1601540} {\bibfield  {journal} {\bibinfo
  {journal} {Science Advances}\ }\textbf {\bibinfo {volume} {3}},\
  10.1126/sciadv.1601540},\ \Eprint {https://arxiv.org/abs/arXiv:1508.00420v3}
  {arXiv:1508.00420v3} \BibitemShut {NoStop}%
\bibitem [{\citenamefont {Lekitsch}\ \emph
  {et~al.}(2017{\natexlab{b}})\citenamefont {Lekitsch}, \citenamefont {Weidt},
  \citenamefont {Fowler}, \citenamefont {M{\o}lmer}, \citenamefont {Devitt},
  \citenamefont {Wunderlich},\ and\ \citenamefont
  {Hensinger}}]{Lekitsch:2017aa}%
  \BibitemOpen
  \bibfield  {author} {\bibinfo {author} {\bibnamefont {Lekitsch},
  \bibfnamefont {Bjoern}}, \bibinfo {author} {\bibfnamefont {Sebastian}\
  \bibnamefont {Weidt}}, \bibinfo {author} {\bibfnamefont {Austin~G.}\
  \bibnamefont {Fowler}}, \bibinfo {author} {\bibfnamefont {Klaus}\
  \bibnamefont {M{\o}lmer}}, \bibinfo {author} {\bibfnamefont {Simon~J.}\
  \bibnamefont {Devitt}}, \bibinfo {author} {\bibfnamefont {Christof}\
  \bibnamefont {Wunderlich}}, and\ \bibinfo {author} {\bibfnamefont
  {Winfried~K.}\ \bibnamefont {Hensinger}}} (\bibinfo {year}
  {2017}{\natexlab{b}}),\ \bibfield  {title} {\enquote {\bibinfo {title}
  {Blueprint for a microwave trapped ion quantum computer},}\ }\href
  {http://advances.sciencemag.org/content/3/2/e1601540.abstract} {\bibfield
  {journal} {\bibinfo  {journal} {Science Advances}\ }\textbf {\bibinfo
  {volume} {3}}~(\bibinfo {number} {2})}\BibitemShut {NoStop}%
\bibitem [{\citenamefont {Lettner}\ \emph {et~al.}(2011)\citenamefont
  {Lettner}, \citenamefont {M{\"u}cke}, \citenamefont {Riedl}, \citenamefont
  {Vo}, \citenamefont {Hahn}, \citenamefont {Baur}, \citenamefont {Bochmann},
  \citenamefont {Ritter}, \citenamefont {D{\"u}rr},\ and\ \citenamefont
  {Rempe}}]{bib:lettner2011remote}%
  \BibitemOpen
  \bibfield  {author} {\bibinfo {author} {\bibnamefont {Lettner}, \bibfnamefont
  {Matthias}}, \bibinfo {author} {\bibfnamefont {Martin}\ \bibnamefont
  {M{\"u}cke}}, \bibinfo {author} {\bibfnamefont {Stefan}\ \bibnamefont
  {Riedl}}, \bibinfo {author} {\bibfnamefont {Christoph}\ \bibnamefont {Vo}},
  \bibinfo {author} {\bibfnamefont {Carolin}\ \bibnamefont {Hahn}}, \bibinfo
  {author} {\bibfnamefont {Simon}\ \bibnamefont {Baur}}, \bibinfo {author}
  {\bibfnamefont {J{\"o}rg}\ \bibnamefont {Bochmann}}, \bibinfo {author}
  {\bibfnamefont {Stephan}\ \bibnamefont {Ritter}}, \bibinfo {author}
  {\bibfnamefont {Stephan}\ \bibnamefont {D{\"u}rr}}, and\ \bibinfo {author}
  {\bibfnamefont {Gerhard}\ \bibnamefont {Rempe}}} (\bibinfo {year} {2011}),\
  \bibfield  {title} {\enquote {\bibinfo {title} {Remote entanglement between a
  single atom and a bose-einstein condensate},}\ }\href
  {https://doi.org/10.1109/cleoe.2011.5943434} {\bibfield  {journal} {\bibinfo
  {journal} {Physical Review Letters}\ }\textbf {\bibinfo {volume} {106}},\
  \bibinfo {pages} {210503}}\BibitemShut {NoStop}%
\bibitem [{\citenamefont {Leung}\ and\ \citenamefont
  {Ralph}(2006)}]{bib:leung2006quantum}%
  \BibitemOpen
  \bibfield  {author} {\bibinfo {author} {\bibnamefont {Leung}, \bibfnamefont
  {Patrick~M}}, and\ \bibinfo {author} {\bibfnamefont {Timothy~C}\ \bibnamefont
  {Ralph}}} (\bibinfo {year} {2006}),\ \bibfield  {title} {\enquote {\bibinfo
  {title} {Quantum memory scheme based on optical fibers and cavities},}\
  }\href {https://doi.org/10.1103/physreva.74.022311} {\bibfield  {journal}
  {\bibinfo  {journal} {Physical Review A}\ }\textbf {\bibinfo {volume} {74}},\
  \bibinfo {pages} {022311}}\BibitemShut {NoStop}%
\bibitem [{\citenamefont {Li}\ and\ \citenamefont
  {Du}(2003)}]{bib:li2003relativistic}%
  \BibitemOpen
  \bibfield  {author} {\bibinfo {author} {\bibnamefont {Li}, \bibfnamefont
  {Hui}}, and\ \bibinfo {author} {\bibfnamefont {Jiangfeng}\ \bibnamefont
  {Du}}} (\bibinfo {year} {2003}),\ \bibfield  {title} {\enquote {\bibinfo
  {title} {Relativistic invariant quantum entanglement between the spins of
  moving bodies},}\ }\href {https://doi.org/10.1103/physreva.68.022108}
  {\bibfield  {journal} {\bibinfo  {journal} {Physical Review A}\ }\textbf
  {\bibinfo {volume} {68}},\ \bibinfo {pages} {022108}},\ \Eprint
  {https://arxiv.org/abs/arXiv:quant-ph/0211159v2} {arXiv:quant-ph/0211159v2}
  \BibitemShut {NoStop}%
\bibitem [{\citenamefont {Li}\ \emph {et~al.}(2009)\citenamefont {Li},
  \citenamefont {Winter}, \citenamefont {Zou},\ and\ \citenamefont
  {Guo}}]{bib:PhysRevLett.103.120501}%
  \BibitemOpen
  \bibfield  {author} {\bibinfo {author} {\bibnamefont {Li}, \bibfnamefont
  {Ke}}, \bibinfo {author} {\bibfnamefont {Andreas}\ \bibnamefont {Winter}},
  \bibinfo {author} {\bibfnamefont {XuBo}\ \bibnamefont {Zou}}, and\ \bibinfo
  {author} {\bibfnamefont {GuangCan}\ \bibnamefont {Guo}}} (\bibinfo {year}
  {2009}),\ \bibfield  {title} {\enquote {\bibinfo {title} {Private capacity of
  quantum channels is not additive},}\ }\href
  {https://doi.org/10.1103/PhysRevLett.103.120501} {\bibfield  {journal}
  {\bibinfo  {journal} {Phys. Rev. Lett.}\ }\textbf {\bibinfo {volume} {103}},\
  \bibinfo {pages} {120501}}\BibitemShut {NoStop}%
\bibitem [{\citenamefont {Li}\ \emph {et~al.}(2021)\citenamefont {Li},
  \citenamefont {Liu}, \citenamefont {Lü}, \citenamefont {Hu}, \citenamefont
  {Xu},\ and\ \citenamefont {Zhang}}]{LI20211}%
  \BibitemOpen
  \bibfield  {author} {\bibinfo {author} {\bibnamefont {Li}, \bibfnamefont
  {Ming}}, \bibinfo {author} {\bibfnamefont {Run-Ran}\ \bibnamefont {Liu}},
  \bibinfo {author} {\bibfnamefont {Linyuan}\ \bibnamefont {Lü}}, \bibinfo
  {author} {\bibfnamefont {Mao-Bin}\ \bibnamefont {Hu}}, \bibinfo {author}
  {\bibfnamefont {Shuqi}\ \bibnamefont {Xu}}, and\ \bibinfo {author}
  {\bibfnamefont {Yi-Cheng}\ \bibnamefont {Zhang}}} (\bibinfo {year} {2021}),\
  \bibfield  {title} {\enquote {\bibinfo {title} {Percolation on complex
  networks: Theory and application},}\ }\href
  {https://doi.org/https://doi.org/10.1016/j.physrep.2020.12.003} {\bibfield
  {journal} {\bibinfo  {journal} {Physics Reports}\ }\textbf {\bibinfo {volume}
  {907}},\ \bibinfo {pages} {1--68}},\ \bibinfo {note} {percolation on complex
  networks: Theory and application}\BibitemShut {NoStop}%
\bibitem [{\citenamefont {Li}\ \emph {et~al.}(2016)\citenamefont {Li},
  \citenamefont {Liu}, \citenamefont {Yang}, \citenamefont {Hu},\ and\
  \citenamefont {Xu}}]{bib:li2016inter}%
  \BibitemOpen
  \bibfield  {author} {\bibinfo {author} {\bibnamefont {Li}, \bibfnamefont
  {Yi}}, \bibinfo {author} {\bibfnamefont {Hong}\ \bibnamefont {Liu}}, \bibinfo
  {author} {\bibfnamefont {Wenjun}\ \bibnamefont {Yang}}, \bibinfo {author}
  {\bibfnamefont {Dianming}\ \bibnamefont {Hu}}, and\ \bibinfo {author}
  {\bibfnamefont {Wei}\ \bibnamefont {Xu}}} (\bibinfo {year} {2016}),\
  \bibfield  {title} {\enquote {\bibinfo {title} {Inter-data-center network
  traffic prediction with elephant flows},}\ }in\ \href
  {https://doi.org/10.1109/noms.2016.7502814} {\emph {\bibinfo {booktitle}
  {IEEE/IFIP Network Operations and Management Symposium (NOMS)}}},\ p.\
  \bibinfo {pages} {206}\BibitemShut {NoStop}%
\bibitem [{\citenamefont {Liao}\ \emph
  {et~al.}(2017{\natexlab{a}})\citenamefont {Liao}, \citenamefont {Cai},
  \citenamefont {Liu}, \citenamefont {Zhang}, \citenamefont {Li}, \citenamefont
  {Ren}, \citenamefont {Yin}, \citenamefont {Shen}, \citenamefont {Cao},
  \citenamefont {Li}, \citenamefont {Li}, \citenamefont {Chen}, \citenamefont
  {Sun}, \citenamefont {Jia}, \citenamefont {Wu}, \citenamefont {Jiang},
  \citenamefont {Wang}, \citenamefont {Huang}, \citenamefont {Wang},
  \citenamefont {Zhou}, \citenamefont {Deng}, \citenamefont {Xi}, \citenamefont
  {Ma}, \citenamefont {Hu}, \citenamefont {Zhang}, \citenamefont {Chen},
  \citenamefont {Liu}, \citenamefont {Wang}, \citenamefont {Zhu}, \citenamefont
  {Lu}, \citenamefont {Shu}, \citenamefont {Peng}, \citenamefont {Wang},\ and\
  \citenamefont {Pan}}]{SD-Liao:2017aa}%
  \BibitemOpen
  \bibfield  {author} {\bibinfo {author} {\bibnamefont {Liao}, \bibfnamefont
  {Sheng-Kai}}, \bibinfo {author} {\bibfnamefont {Wen-Qi}\ \bibnamefont {Cai}},
  \bibinfo {author} {\bibfnamefont {Wei-Yue}\ \bibnamefont {Liu}}, \bibinfo
  {author} {\bibfnamefont {Liang}\ \bibnamefont {Zhang}}, \bibinfo {author}
  {\bibfnamefont {Yang}\ \bibnamefont {Li}}, \bibinfo {author} {\bibfnamefont
  {Ji-Gang}\ \bibnamefont {Ren}}, \bibinfo {author} {\bibfnamefont {Juan}\
  \bibnamefont {Yin}}, \bibinfo {author} {\bibfnamefont {Qi}~\bibnamefont
  {Shen}}, \bibinfo {author} {\bibfnamefont {Yuan}\ \bibnamefont {Cao}},
  \bibinfo {author} {\bibfnamefont {Zheng-Ping}\ \bibnamefont {Li}}, \bibinfo
  {author} {\bibfnamefont {Feng-Zhi}\ \bibnamefont {Li}}, \bibinfo {author}
  {\bibfnamefont {Xia-Wei}\ \bibnamefont {Chen}}, \bibinfo {author}
  {\bibfnamefont {Li-Hua}\ \bibnamefont {Sun}}, \bibinfo {author}
  {\bibfnamefont {Jian-Jun}\ \bibnamefont {Jia}}, \bibinfo {author}
  {\bibfnamefont {Jin-Cai}\ \bibnamefont {Wu}}, \bibinfo {author}
  {\bibfnamefont {Xiao-Jun}\ \bibnamefont {Jiang}}, \bibinfo {author}
  {\bibfnamefont {Jian-Feng}\ \bibnamefont {Wang}}, \bibinfo {author}
  {\bibfnamefont {Yong-Mei}\ \bibnamefont {Huang}}, \bibinfo {author}
  {\bibfnamefont {Qiang}\ \bibnamefont {Wang}}, \bibinfo {author}
  {\bibfnamefont {Yi-Lin}\ \bibnamefont {Zhou}}, \bibinfo {author}
  {\bibfnamefont {Lei}\ \bibnamefont {Deng}}, \bibinfo {author} {\bibfnamefont
  {Tao}\ \bibnamefont {Xi}}, \bibinfo {author} {\bibfnamefont {Lu}~\bibnamefont
  {Ma}}, \bibinfo {author} {\bibfnamefont {Tai}\ \bibnamefont {Hu}}, \bibinfo
  {author} {\bibfnamefont {Qiang}\ \bibnamefont {Zhang}}, \bibinfo {author}
  {\bibfnamefont {Yu-Ao}\ \bibnamefont {Chen}}, \bibinfo {author}
  {\bibfnamefont {Nai-Le}\ \bibnamefont {Liu}}, \bibinfo {author}
  {\bibfnamefont {Xiang-Bin}\ \bibnamefont {Wang}}, \bibinfo {author}
  {\bibfnamefont {Zhen-Cai}\ \bibnamefont {Zhu}}, \bibinfo {author}
  {\bibfnamefont {Chao-Yang}\ \bibnamefont {Lu}}, \bibinfo {author}
  {\bibfnamefont {Rong}\ \bibnamefont {Shu}}, \bibinfo {author} {\bibfnamefont
  {Cheng-Zhi}\ \bibnamefont {Peng}}, \bibinfo {author} {\bibfnamefont
  {Jian-Yu}\ \bibnamefont {Wang}}, and\ \bibinfo {author} {\bibfnamefont
  {Jian-Wei}\ \bibnamefont {Pan}}} (\bibinfo {year} {2017}{\natexlab{a}}),\
  \bibfield  {title} {\enquote {\bibinfo {title} {Satellite-to-ground quantum
  key distribution},}\ }\href {https://doi.org/10.1038/nature23655} {\bibfield
  {journal} {\bibinfo  {journal} {Nature}\ }\textbf {\bibinfo {volume} {549}},\
  \bibinfo {pages} {43}},\ \Eprint {https://arxiv.org/abs/arXiv:1707.00542v1}
  {arXiv:1707.00542v1} \BibitemShut {NoStop}%
\bibitem [{\citenamefont {Liao}\ \emph
  {et~al.}(2017{\natexlab{b}})\citenamefont {Liao}, \citenamefont {Cai},
  \citenamefont {Liu}, \citenamefont {Zhang}, \citenamefont {Li}, \citenamefont
  {Ren}, \citenamefont {Yin}, \citenamefont {Shen}, \citenamefont {Cao},
  \citenamefont {Li} \emph {et~al.}}]{liao2017satellite}%
  \BibitemOpen
  \bibfield  {author} {\bibinfo {author} {\bibnamefont {Liao}, \bibfnamefont
  {Sheng-Kai}}, \bibinfo {author} {\bibfnamefont {Wen-Qi}\ \bibnamefont {Cai}},
  \bibinfo {author} {\bibfnamefont {Wei-Yue}\ \bibnamefont {Liu}}, \bibinfo
  {author} {\bibfnamefont {Liang}\ \bibnamefont {Zhang}}, \bibinfo {author}
  {\bibfnamefont {Yang}\ \bibnamefont {Li}}, \bibinfo {author} {\bibfnamefont
  {Ji-Gang}\ \bibnamefont {Ren}}, \bibinfo {author} {\bibfnamefont {Juan}\
  \bibnamefont {Yin}}, \bibinfo {author} {\bibfnamefont {Qi}~\bibnamefont
  {Shen}}, \bibinfo {author} {\bibfnamefont {Yuan}\ \bibnamefont {Cao}},
  \bibinfo {author} {\bibfnamefont {Zheng-Ping}\ \bibnamefont {Li}},  \emph
  {et~al.}} (\bibinfo {year} {2017}{\natexlab{b}}),\ \bibfield  {title}
  {\enquote {\bibinfo {title} {Satellite-to-ground quantum key distribution},}\
  }\href@noop {} {\bibfield  {journal} {\bibinfo  {journal} {Nature}\ }\textbf
  {\bibinfo {volume} {549}}~(\bibinfo {number} {7670}),\ \bibinfo {pages}
  {43--47}}\BibitemShut {NoStop}%
\bibitem [{\citenamefont {Liao}\ \emph
  {et~al.}(2017{\natexlab{c}})\citenamefont {Liao}, \citenamefont {Cai},
  \citenamefont {Liu}, \citenamefont {Zhang}, \citenamefont {Li}, \citenamefont
  {Ren}, \citenamefont {Yin}, \citenamefont {Shen}, \citenamefont {Cao},
  \citenamefont {Li} \emph {et~al.}}]{bib:liao2017satellite}%
  \BibitemOpen
  \bibfield  {author} {\bibinfo {author} {\bibnamefont {Liao}, \bibfnamefont
  {Sheng-Kai}}, \bibinfo {author} {\bibfnamefont {Wen-Qi}\ \bibnamefont {Cai}},
  \bibinfo {author} {\bibfnamefont {Wei-Yue}\ \bibnamefont {Liu}}, \bibinfo
  {author} {\bibfnamefont {Liang}\ \bibnamefont {Zhang}}, \bibinfo {author}
  {\bibfnamefont {Yang}\ \bibnamefont {Li}}, \bibinfo {author} {\bibfnamefont
  {Ji-Gang}\ \bibnamefont {Ren}}, \bibinfo {author} {\bibfnamefont {Juan}\
  \bibnamefont {Yin}}, \bibinfo {author} {\bibfnamefont {Qi}~\bibnamefont
  {Shen}}, \bibinfo {author} {\bibfnamefont {Yuan}\ \bibnamefont {Cao}},
  \bibinfo {author} {\bibfnamefont {Zheng-Ping}\ \bibnamefont {Li}},  \emph
  {et~al.}} (\bibinfo {year} {2017}{\natexlab{c}}),\ \bibfield  {title}
  {\enquote {\bibinfo {title} {Satellite-to-ground quantum key distribution},}\
  }\href {https://doi.org/10.1038/nature23655} {\bibfield  {journal} {\bibinfo
  {journal} {Nature}\ }\textbf {\bibinfo {volume} {549}},\ \bibinfo {pages}
  {43}},\ \Eprint {https://arxiv.org/abs/arXiv:1707.00542v1}
  {arXiv:1707.00542v1} \BibitemShut {NoStop}%
\bibitem [{\citenamefont {Liao}\ \emph {et~al.}(2016)\citenamefont {Liao},
  \citenamefont {Yong}, \citenamefont {Liu}, \citenamefont {Shentu},
  \citenamefont {Li}, \citenamefont {Lin}, \citenamefont {Dai}, \citenamefont
  {Zhao}, \citenamefont {Li}, \citenamefont {Guan} \emph
  {et~al.}}]{bib:liao2016ground}%
  \BibitemOpen
  \bibfield  {author} {\bibinfo {author} {\bibnamefont {Liao}, \bibfnamefont
  {Sheng-Kai}}, \bibinfo {author} {\bibfnamefont {Hai-Lin}\ \bibnamefont
  {Yong}}, \bibinfo {author} {\bibfnamefont {Chang}\ \bibnamefont {Liu}},
  \bibinfo {author} {\bibfnamefont {Guo-Liang}\ \bibnamefont {Shentu}},
  \bibinfo {author} {\bibfnamefont {Dong-Dong}\ \bibnamefont {Li}}, \bibinfo
  {author} {\bibfnamefont {Jin}\ \bibnamefont {Lin}}, \bibinfo {author}
  {\bibfnamefont {Hui}\ \bibnamefont {Dai}}, \bibinfo {author} {\bibfnamefont
  {Shuang-Qiang}\ \bibnamefont {Zhao}}, \bibinfo {author} {\bibfnamefont
  {Bo}~\bibnamefont {Li}}, \bibinfo {author} {\bibfnamefont {Jian-Yu}\
  \bibnamefont {Guan}},  \emph {et~al.}} (\bibinfo {year} {2016}),\ \bibfield
  {title} {\enquote {\bibinfo {title} {Ground test of satellite constellation
  based quantum communication},}\ }\href@noop {} {\bibfield  {journal}
  {\bibinfo  {journal} {Nature Photonics}\ }\textbf {\bibinfo {volume} {11}},\
  \bibinfo {pages} {509}},\ \Eprint {https://arxiv.org/abs/arXiv:1611.09982v1}
  {arXiv:1611.09982v1} \BibitemShut {NoStop}%
\bibitem [{\citenamefont {Liao}\ \emph
  {et~al.}(2017{\natexlab{d}})\citenamefont {Liao}, \citenamefont {Yong},
  \citenamefont {Liu}, \citenamefont {Shentu}, \citenamefont {Li},
  \citenamefont {Lin}, \citenamefont {Dai}, \citenamefont {Zhao}, \citenamefont
  {Li}, \citenamefont {Guan} \emph {et~al.}}]{bib:liao2017long}%
  \BibitemOpen
  \bibfield  {author} {\bibinfo {author} {\bibnamefont {Liao}, \bibfnamefont
  {Sheng-Kai}}, \bibinfo {author} {\bibfnamefont {Hai-Lin}\ \bibnamefont
  {Yong}}, \bibinfo {author} {\bibfnamefont {Chang}\ \bibnamefont {Liu}},
  \bibinfo {author} {\bibfnamefont {Guo-Liang}\ \bibnamefont {Shentu}},
  \bibinfo {author} {\bibfnamefont {Dong-Dong}\ \bibnamefont {Li}}, \bibinfo
  {author} {\bibfnamefont {Jin}\ \bibnamefont {Lin}}, \bibinfo {author}
  {\bibfnamefont {Hui}\ \bibnamefont {Dai}}, \bibinfo {author} {\bibfnamefont
  {Shuang-Qiang}\ \bibnamefont {Zhao}}, \bibinfo {author} {\bibfnamefont
  {Bo}~\bibnamefont {Li}}, \bibinfo {author} {\bibfnamefont {Jian-Yu}\
  \bibnamefont {Guan}},  \emph {et~al.}} (\bibinfo {year}
  {2017}{\natexlab{d}}),\ \bibfield  {title} {\enquote {\bibinfo {title}
  {Long-distance free-space quantum key distribution in daylight towards
  inter-satellite communication},}\ }\href
  {https://doi.org/10.1038/nphoton.2017.116} {\bibfield  {journal} {\bibinfo
  {journal} {Nature Photonics}\ }\textbf {\bibinfo {volume} {11}},\ \bibinfo
  {pages} {509}}\BibitemShut {NoStop}%
\bibitem [{\citenamefont {Lim}\ \emph {et~al.}(2014{\natexlab{a}})\citenamefont
  {Lim}, \citenamefont {Song}, \citenamefont {Fang}, \citenamefont {Li},
  \citenamefont {Tu}, \citenamefont {Duan}, \citenamefont {Chen}, \citenamefont
  {Tern},\ and\ \citenamefont {Liow}}]{bib:lim2014review}%
  \BibitemOpen
  \bibfield  {author} {\bibinfo {author} {\bibnamefont {Lim}, \bibfnamefont
  {Andy Eu-Jin}}, \bibinfo {author} {\bibfnamefont {Junfeng}\ \bibnamefont
  {Song}}, \bibinfo {author} {\bibfnamefont {Qing}\ \bibnamefont {Fang}},
  \bibinfo {author} {\bibfnamefont {Chao}\ \bibnamefont {Li}}, \bibinfo
  {author} {\bibfnamefont {Xiaoguang}\ \bibnamefont {Tu}}, \bibinfo {author}
  {\bibfnamefont {Ning}\ \bibnamefont {Duan}}, \bibinfo {author} {\bibfnamefont
  {Kok~Kiong}\ \bibnamefont {Chen}}, \bibinfo {author} {\bibfnamefont {Roger
  Poh-Cher}\ \bibnamefont {Tern}}, and\ \bibinfo {author} {\bibfnamefont
  {Tsung-Yang}\ \bibnamefont {Liow}}} (\bibinfo {year} {2014}{\natexlab{a}}),\
  \bibfield  {title} {\enquote {\bibinfo {title} {Review of silicon photonics
  foundry efforts},}\ }\href {https://doi.org/10.1109/jstqe.2013.2293274}
  {\bibfield  {journal} {\bibinfo  {journal} {IEEE Journal of Selected Topics
  in Quantum Electronics}\ }\textbf {\bibinfo {volume} {20}},\ \bibinfo {pages}
  {405}}\BibitemShut {NoStop}%
\bibitem [{\citenamefont {Lim}\ \emph {et~al.}(2014{\natexlab{b}})\citenamefont
  {Lim}, \citenamefont {Korzh}, \citenamefont {Martin}, \citenamefont
  {Bussieres}, \citenamefont {Thew},\ and\ \citenamefont
  {Zbinden}}]{bib:lim2014detector}%
  \BibitemOpen
  \bibfield  {author} {\bibinfo {author} {\bibnamefont {Lim}, \bibfnamefont
  {Charles Ci~Wen}}, \bibinfo {author} {\bibfnamefont {Boris}\ \bibnamefont
  {Korzh}}, \bibinfo {author} {\bibfnamefont {Anthony}\ \bibnamefont {Martin}},
  \bibinfo {author} {\bibfnamefont {F{\'e}lix}\ \bibnamefont {Bussieres}},
  \bibinfo {author} {\bibfnamefont {Rob}\ \bibnamefont {Thew}}, and\ \bibinfo
  {author} {\bibfnamefont {Hugo}\ \bibnamefont {Zbinden}}} (\bibinfo {year}
  {2014}{\natexlab{b}}),\ \bibfield  {title} {\enquote {\bibinfo {title}
  {Detector-device-independent quantum key distribution},}\ }\href
  {https://doi.org/10.1063/1.4903350} {\bibfield  {journal} {\bibinfo
  {journal} {Applied Physics Letters}\ }\textbf {\bibinfo {volume} {105}},\
  \bibinfo {pages} {221112}},\ \Eprint
  {https://arxiv.org/abs/arXiv:1410.1850v2} {arXiv:1410.1850v2} \BibitemShut
  {NoStop}%
\bibitem [{\citenamefont {Lim}\ \emph {et~al.}(2005{\natexlab{a}})\citenamefont
  {Lim}, \citenamefont {Barrett}, \citenamefont {Beige}, \citenamefont {Kok},\
  and\ \citenamefont {Kwek}}]{bib:LimBarrett05}%
  \BibitemOpen
  \bibfield  {author} {\bibinfo {author} {\bibnamefont {Lim}, \bibfnamefont
  {Yuan~Liang}}, \bibinfo {author} {\bibfnamefont {Sean~D.}\ \bibnamefont
  {Barrett}}, \bibinfo {author} {\bibfnamefont {Almut}\ \bibnamefont {Beige}},
  \bibinfo {author} {\bibfnamefont {Pieter}\ \bibnamefont {Kok}}, and\ \bibinfo
  {author} {\bibfnamefont {Leong~Chuan}\ \bibnamefont {Kwek}}} (\bibinfo {year}
  {2005}{\natexlab{a}}),\ \bibfield  {title} {\enquote {\bibinfo {title}
  {Repeat-until-success quantum computing using stationary and flying
  qubits},}\ }\href {https://doi.org/10.1103/physreva.73.012304} {\bibfield
  {journal} {\bibinfo  {journal} {Physical Review A}\ }\textbf {\bibinfo
  {volume} {73}},\ \bibinfo {pages} {012304}},\ \Eprint
  {https://arxiv.org/abs/arXiv:quant-ph/0508218v3} {arXiv:quant-ph/0508218v3}
  \BibitemShut {NoStop}%
\bibitem [{\citenamefont {Lim}\ \emph {et~al.}(2006)\citenamefont {Lim},
  \citenamefont {Barrett}, \citenamefont {Beige}, \citenamefont {Kok},\ and\
  \citenamefont {Kwek}}]{PhysRevA.73.012304}%
  \BibitemOpen
  \bibfield  {author} {\bibinfo {author} {\bibnamefont {Lim}, \bibfnamefont
  {Yuan~Liang}}, \bibinfo {author} {\bibfnamefont {Sean~D.}\ \bibnamefont
  {Barrett}}, \bibinfo {author} {\bibfnamefont {Almut}\ \bibnamefont {Beige}},
  \bibinfo {author} {\bibfnamefont {Pieter}\ \bibnamefont {Kok}}, and\ \bibinfo
  {author} {\bibfnamefont {Leong~Chuan}\ \bibnamefont {Kwek}}} (\bibinfo {year}
  {2006}),\ \bibfield  {title} {\enquote {\bibinfo {title}
  {Repeat-until-success quantum computing using stationary and flying
  qubits},}\ }\href {https://doi.org/10.1103/PhysRevA.73.012304} {\bibfield
  {journal} {\bibinfo  {journal} {Phys. Rev. A}\ }\textbf {\bibinfo {volume}
  {73}},\ \bibinfo {pages} {012304}}\BibitemShut {NoStop}%
\bibitem [{\citenamefont {Lim}\ \emph {et~al.}(2005{\natexlab{b}})\citenamefont
  {Lim}, \citenamefont {Beige},\ and\ \citenamefont {Kwek}}]{bib:Lim05}%
  \BibitemOpen
  \bibfield  {author} {\bibinfo {author} {\bibnamefont {Lim}, \bibfnamefont
  {Yuan~Liang}}, \bibinfo {author} {\bibfnamefont {Almut}\ \bibnamefont
  {Beige}}, and\ \bibinfo {author} {\bibfnamefont {Leong~Chuan}\ \bibnamefont
  {Kwek}}} (\bibinfo {year} {2005}{\natexlab{b}}),\ \bibfield  {title}
  {\enquote {\bibinfo {title} {Repeat-until-success linear optics distributed
  quantum computing},}\ }\href {https://doi.org/10.1103/physrevlett.95.030505}
  {\bibfield  {journal} {\bibinfo  {journal} {Physical Review Letters}\
  }\textbf {\bibinfo {volume} {95}},\ \bibinfo {pages} {030505}}\BibitemShut
  {NoStop}%
\bibitem [{\citenamefont {Lita}\ \emph {et~al.}(2008)\citenamefont {Lita},
  \citenamefont {Miller},\ and\ \citenamefont {Nam}}]{bib:lita2008}%
  \BibitemOpen
  \bibfield  {author} {\bibinfo {author} {\bibnamefont {Lita}, \bibfnamefont
  {Adriana~E}}, \bibinfo {author} {\bibfnamefont {Aaron~J}\ \bibnamefont
  {Miller}}, and\ \bibinfo {author} {\bibfnamefont {Sae~Woo}\ \bibnamefont
  {Nam}}} (\bibinfo {year} {2008}),\ \bibfield  {title} {\enquote {\bibinfo
  {title} {Counting near-infrared single-photons with 95\% efficiency},}\
  }\href {https://doi.org/10.1364/oe.16.003032} {\bibfield  {journal} {\bibinfo
   {journal} {Optics Express}\ }\textbf {\bibinfo {volume} {16}},\ \bibinfo
  {pages} {3032}}\BibitemShut {NoStop}%
\bibitem [{\citenamefont {Liu}\ \emph {et~al.}(2002)\citenamefont {Liu},
  \citenamefont {Matta},\ and\ \citenamefont {Crovella}}]{bib:liu2002end}%
  \BibitemOpen
  \bibfield  {author} {\bibinfo {author} {\bibnamefont {Liu}, \bibfnamefont
  {Jun}}, \bibinfo {author} {\bibfnamefont {Ibrahim}\ \bibnamefont {Matta}},
  and\ \bibinfo {author} {\bibfnamefont {Mark}\ \bibnamefont {Crovella}}}
  (\bibinfo {year} {2002}),\ \href@noop {} {\emph {\bibinfo {title} {End-to-end
  inference of loss nature in a hybrid wired/wireless environment}}},\ \bibinfo
  {type} {Tech. Rep.}\ (\bibinfo  {institution} {Boston University Computer
  Science Department})\BibitemShut {NoStop}%
\bibitem [{\citenamefont {Liu}\ and\ \citenamefont
  {Rebentrost}(2018)}]{bib:liu2018quantum}%
  \BibitemOpen
  \bibfield  {author} {\bibinfo {author} {\bibnamefont {Liu}, \bibfnamefont
  {Nana}}, and\ \bibinfo {author} {\bibfnamefont {Patrick}\ \bibnamefont
  {Rebentrost}}} (\bibinfo {year} {2018}),\ \bibfield  {title} {\enquote
  {\bibinfo {title} {Quantum machine learning for quantum anomaly detection},}\
  }\href {https://doi.org/10.1103/physreva.97.042315} {\bibfield  {journal}
  {\bibinfo  {journal} {Physical Review A}\ }\textbf {\bibinfo {volume} {97}},\
  \bibinfo {pages} {042315}},\ \Eprint
  {https://arxiv.org/abs/arXiv:1710.07405v1} {arXiv:1710.07405v1} \BibitemShut
  {NoStop}%
\bibitem [{\citenamefont {Liu}\ \emph {et~al.}(2010)\citenamefont {Liu},
  \citenamefont {Chen}, \citenamefont {Wang}, \citenamefont {Cai},
  \citenamefont {Wan}, \citenamefont {Chen}, \citenamefont {Wang},
  \citenamefont {Liu}, \citenamefont {Liang}, \citenamefont {Yang} \emph
  {et~al.}}]{bib:OptExp_18_8587}%
  \BibitemOpen
  \bibfield  {author} {\bibinfo {author} {\bibnamefont {Liu}, \bibfnamefont
  {Yang}}, \bibinfo {author} {\bibfnamefont {Teng-Yun}\ \bibnamefont {Chen}},
  \bibinfo {author} {\bibfnamefont {Jian}\ \bibnamefont {Wang}}, \bibinfo
  {author} {\bibfnamefont {Wen-Qi}\ \bibnamefont {Cai}}, \bibinfo {author}
  {\bibfnamefont {Xu}~\bibnamefont {Wan}}, \bibinfo {author} {\bibfnamefont
  {Luo-Kan}\ \bibnamefont {Chen}}, \bibinfo {author} {\bibfnamefont {Jin-Hong}\
  \bibnamefont {Wang}}, \bibinfo {author} {\bibfnamefont {Shu-Bin}\
  \bibnamefont {Liu}}, \bibinfo {author} {\bibfnamefont {Hao}\ \bibnamefont
  {Liang}}, \bibinfo {author} {\bibfnamefont {Lin}\ \bibnamefont {Yang}},
  \emph {et~al.}} (\bibinfo {year} {2010}),\ \bibfield  {title} {\enquote
  {\bibinfo {title} {Decoy-state quantum key distribution with polarized
  photons over 200 km},}\ }\href {https://doi.org/10.1364/oe.18.008587}
  {\bibfield  {journal} {\bibinfo  {journal} {Optics Expresss}\ }\textbf
  {\bibinfo {volume} {18}},\ \bibinfo {pages} {8587}}\BibitemShut {NoStop}%
\bibitem [{\citenamefont {Liu}\ \emph {et~al.}(2013)\citenamefont {Liu},
  \citenamefont {Chen}, \citenamefont {Wang}, \citenamefont {Liang},
  \citenamefont {Shentu}, \citenamefont {Wang}, \citenamefont {Cui},
  \citenamefont {Yin}, \citenamefont {Liu}, \citenamefont {Li} \emph
  {et~al.}}]{bib:PRL_111_130502}%
  \BibitemOpen
  \bibfield  {author} {\bibinfo {author} {\bibnamefont {Liu}, \bibfnamefont
  {Yang}}, \bibinfo {author} {\bibfnamefont {Teng-Yun}\ \bibnamefont {Chen}},
  \bibinfo {author} {\bibfnamefont {Liu-Jun}\ \bibnamefont {Wang}}, \bibinfo
  {author} {\bibfnamefont {Hao}\ \bibnamefont {Liang}}, \bibinfo {author}
  {\bibfnamefont {Guo-Liang}\ \bibnamefont {Shentu}}, \bibinfo {author}
  {\bibfnamefont {Jian}\ \bibnamefont {Wang}}, \bibinfo {author} {\bibfnamefont
  {Ke}~\bibnamefont {Cui}}, \bibinfo {author} {\bibfnamefont {Hua-Lei}\
  \bibnamefont {Yin}}, \bibinfo {author} {\bibfnamefont {Nai-Le}\ \bibnamefont
  {Liu}}, \bibinfo {author} {\bibfnamefont {Li}~\bibnamefont {Li}},  \emph
  {et~al.}} (\bibinfo {year} {2013}),\ \bibfield  {title} {\enquote {\bibinfo
  {title} {Experimental measurement-device-independent quantum key
  distribution},}\ }\href {https://doi.org/10.1103/physrevlett.111.130502}
  {\bibfield  {journal} {\bibinfo  {journal} {Physical Review Letters}\
  }\textbf {\bibinfo {volume} {111}},\ \bibinfo {pages} {130502}},\ \Eprint
  {https://arxiv.org/abs/arXiv:1209.6178v1} {arXiv:1209.6178v1} \BibitemShut
  {NoStop}%
\bibitem [{\citenamefont {Lloyd}(1996)}]{bib:lloyd1996universal}%
  \BibitemOpen
  \bibfield  {author} {\bibinfo {author} {\bibnamefont {Lloyd}, \bibfnamefont
  {Seth}}} (\bibinfo {year} {1996}),\ \bibfield  {title} {\enquote {\bibinfo
  {title} {Universal quantum simulators},}\ }\href
  {https://doi.org/10.1126/science.273.5278.1073} {\bibfield  {journal}
  {\bibinfo  {journal} {Science}\ }\textbf {\bibinfo {volume} {273}},\ \bibinfo
  {pages} {1073}}\BibitemShut {NoStop}%
\bibitem [{\citenamefont {Lloyd}(2013)}]{bib:LloydEnigma}%
  \BibitemOpen
  \bibfield  {author} {\bibinfo {author} {\bibnamefont {Lloyd}, \bibfnamefont
  {Seth}}} (\bibinfo {year} {2013}),\ \bibfield  {title} {\enquote {\bibinfo
  {title} {Quantum enigma machines},}\ }\href@noop {} {\ }\Eprint
  {https://arxiv.org/abs/arXiv:1307.0380} {arXiv:1307.0380} \BibitemShut
  {NoStop}%
\bibitem [{\citenamefont {Lloyd}\ \emph {et~al.}(2016)\citenamefont {Lloyd},
  \citenamefont {Garnerone},\ and\ \citenamefont
  {Zanardi}}]{bib:lloyd2016quantum}%
  \BibitemOpen
  \bibfield  {author} {\bibinfo {author} {\bibnamefont {Lloyd}, \bibfnamefont
  {Seth}}, \bibinfo {author} {\bibfnamefont {Silvano}\ \bibnamefont
  {Garnerone}}, and\ \bibinfo {author} {\bibfnamefont {Paolo}\ \bibnamefont
  {Zanardi}}} (\bibinfo {year} {2016}),\ \bibfield  {title} {\enquote {\bibinfo
  {title} {Quantum algorithms for topological and geometric analysis of
  data},}\ }\href {https://doi.org/10.1038/ncomms10138} {\bibfield  {journal}
  {\bibinfo  {journal} {Nature Communications}\ }\textbf {\bibinfo {volume}
  {7}},\ \bibinfo {pages} {10138}}\BibitemShut {NoStop}%
\bibitem [{\citenamefont {Lloyd}\ \emph {et~al.}(2013)\citenamefont {Lloyd},
  \citenamefont {Mohseni},\ and\ \citenamefont
  {Rebentrost}}]{bib:lloyd2013quantum}%
  \BibitemOpen
  \bibfield  {author} {\bibinfo {author} {\bibnamefont {Lloyd}, \bibfnamefont
  {Seth}}, \bibinfo {author} {\bibfnamefont {Masoud}\ \bibnamefont {Mohseni}},
  and\ \bibinfo {author} {\bibfnamefont {Patrick}\ \bibnamefont {Rebentrost}}}
  (\bibinfo {year} {2013}),\ \bibfield  {title} {\enquote {\bibinfo {title}
  {Quantum algorithms for supervised and unsupervised machine learning},}\
  }\href@noop {} {\ }\Eprint {https://arxiv.org/abs/arXiv:1307.0411}
  {arXiv:1307.0411} \BibitemShut {NoStop}%
\bibitem [{\citenamefont {Lloyd}\ \emph {et~al.}(2014)\citenamefont {Lloyd},
  \citenamefont {Mohseni},\ and\ \citenamefont
  {Rebentrost}}]{bib:lloyd2014quantum}%
  \BibitemOpen
  \bibfield  {author} {\bibinfo {author} {\bibnamefont {Lloyd}, \bibfnamefont
  {Seth}}, \bibinfo {author} {\bibfnamefont {Masoud}\ \bibnamefont {Mohseni}},
  and\ \bibinfo {author} {\bibfnamefont {Patrick}\ \bibnamefont {Rebentrost}}}
  (\bibinfo {year} {2014}),\ \bibfield  {title} {\enquote {\bibinfo {title}
  {Quantum principal component analysis},}\ }\href
  {https://doi.org/10.1038/nphys3029} {\bibfield  {journal} {\bibinfo
  {journal} {Nature Physics}\ }\textbf {\bibinfo {volume} {10}},\ \bibinfo
  {pages} {631}},\ \Eprint {https://arxiv.org/abs/arXiv:1307.0401v2}
  {arXiv:1307.0401v2} \BibitemShut {NoStop}%
\bibitem [{\citenamefont {Lo}(1997)}]{bib:HKLo97}%
  \BibitemOpen
  \bibfield  {author} {\bibinfo {author} {\bibnamefont {Lo}, \bibfnamefont
  {Hoi-Kwong}}} (\bibinfo {year} {1997}),\ \bibfield  {title} {\enquote
  {\bibinfo {title} {Insecurity of quantum secure computations},}\ }\href
  {https://doi.org/10.1103/physreva.56.1154} {\bibfield  {journal} {\bibinfo
  {journal} {Physical Review A}\ }\textbf {\bibinfo {volume} {56}},\ \bibinfo
  {pages} {1154}},\ \Eprint {https://arxiv.org/abs/arXiv:quant-ph/9611031v2}
  {arXiv:quant-ph/9611031v2} \BibitemShut {NoStop}%
\bibitem [{\citenamefont {Lo}\ \emph {et~al.}(2012)\citenamefont {Lo},
  \citenamefont {Curty},\ and\ \citenamefont
  {Qi}}]{bib:PhysRevLett.108.130503}%
  \BibitemOpen
  \bibfield  {author} {\bibinfo {author} {\bibnamefont {Lo}, \bibfnamefont
  {Hoi-Kwong}}, \bibinfo {author} {\bibfnamefont {Marcos}\ \bibnamefont
  {Curty}}, and\ \bibinfo {author} {\bibfnamefont {Bing}\ \bibnamefont {Qi}}}
  (\bibinfo {year} {2012}),\ \bibfield  {title} {\enquote {\bibinfo {title}
  {Measurement-device-independent quantum key distribution},}\ }\href
  {https://doi.org/10.1103/physrevlett.108.130503} {\bibfield  {journal}
  {\bibinfo  {journal} {Physical Review Letters}\ }\textbf {\bibinfo {volume}
  {108}},\ \bibinfo {pages} {130503}},\ \Eprint
  {https://arxiv.org/abs/arXiv:1109.1473v2} {arXiv:1109.1473v2} \BibitemShut
  {NoStop}%
\bibitem [{\citenamefont {Lo}\ \emph {et~al.}(2014)\citenamefont {Lo},
  \citenamefont {Curty},\ and\ \citenamefont {Tamaki}}]{bib:lo2014secure}%
  \BibitemOpen
  \bibfield  {author} {\bibinfo {author} {\bibnamefont {Lo}, \bibfnamefont
  {Hoi-Kwong}}, \bibinfo {author} {\bibfnamefont {Marcos}\ \bibnamefont
  {Curty}}, and\ \bibinfo {author} {\bibfnamefont {Kiyoshi}\ \bibnamefont
  {Tamaki}}} (\bibinfo {year} {2014}),\ \bibfield  {title} {\enquote {\bibinfo
  {title} {Secure quantum key distribution},}\ }\href
  {https://doi.org/10.1038/nphoton.2014.149} {\bibfield  {journal} {\bibinfo
  {journal} {Nature Photonics}\ }\textbf {\bibinfo {volume} {8}},\ \bibinfo
  {pages} {595}},\ \Eprint {https://arxiv.org/abs/arXiv:1505.05303v1}
  {arXiv:1505.05303v1} \BibitemShut {NoStop}%
\bibitem [{\citenamefont {Lo}\ \emph {et~al.}(2005)\citenamefont {Lo},
  \citenamefont {Ma},\ and\ \citenamefont {Chen}}]{bib:PhysRevLett.94.230504}%
  \BibitemOpen
  \bibfield  {author} {\bibinfo {author} {\bibnamefont {Lo}, \bibfnamefont
  {Hoi-Kwong}}, \bibinfo {author} {\bibfnamefont {Xiongfeng}\ \bibnamefont
  {Ma}}, and\ \bibinfo {author} {\bibfnamefont {Kai}\ \bibnamefont {Chen}}}
  (\bibinfo {year} {2005}),\ \bibfield  {title} {\enquote {\bibinfo {title}
  {Decoy state quantum key distribution},}\ }\href
  {https://doi.org/10.1103/physrevlett.94.230504} {\bibfield  {journal}
  {\bibinfo  {journal} {Physical Review Letters}\ }\textbf {\bibinfo {volume}
  {94}},\ \bibinfo {pages} {230504}},\ \Eprint
  {https://arxiv.org/abs/arXiv:quant-ph/0411004v4} {arXiv:quant-ph/0411004v4}
  \BibitemShut {NoStop}%
\bibitem [{\citenamefont {Loock}\ \emph {et~al.}(2006)\citenamefont {Loock},
  \citenamefont {Ladd}, \citenamefont {Sanaka}, \citenamefont {Yamaguchi},
  \citenamefont {Nemoto}, \citenamefont {Munro},\ and\ \citenamefont
  {Yamamoto}}]{bib:loock06}%
  \BibitemOpen
  \bibfield  {author} {\bibinfo {author} {\bibnamefont {Loock}, \bibfnamefont
  {P~Van}}, \bibinfo {author} {\bibfnamefont {T.~D.}\ \bibnamefont {Ladd}},
  \bibinfo {author} {\bibfnamefont {K.}~\bibnamefont {Sanaka}}, \bibinfo
  {author} {\bibfnamefont {F.}~\bibnamefont {Yamaguchi}}, \bibinfo {author}
  {\bibfnamefont {K.}~\bibnamefont {Nemoto}}, \bibinfo {author} {\bibfnamefont
  {W.~J.}\ \bibnamefont {Munro}}, and\ \bibinfo {author} {\bibfnamefont
  {Y.}~\bibnamefont {Yamamoto}}} (\bibinfo {year} {2006}),\ \bibfield  {title}
  {\enquote {\bibinfo {title} {Hybrid quantum repeater using bright coherent
  light},}\ }\href {https://doi.org/10.1103/physrevlett.96.240501} {\bibfield
  {journal} {\bibinfo  {journal} {Physical Review Letters}\ }\textbf {\bibinfo
  {volume} {96}},\ \bibinfo {pages} {240501}},\ \Eprint
  {https://arxiv.org/abs/arXiv:quant-ph/0510202v3} {arXiv:quant-ph/0510202v3}
  \BibitemShut {NoStop}%
\bibitem [{\citenamefont {Loredo}\ \emph {et~al.}(2016)\citenamefont {Loredo},
  \citenamefont {Zakaria}, \citenamefont {Somaschi}, \citenamefont {Anton},
  \citenamefont {de~Santis}, \citenamefont {Giesz}, \citenamefont {Grange},
  \citenamefont {Broome}, \citenamefont {Gazzano}, \citenamefont {Coppola}
  \emph {et~al.}}]{bib:loredo2016}%
  \BibitemOpen
  \bibfield  {author} {\bibinfo {author} {\bibnamefont {Loredo}, \bibfnamefont
  {Juan~C}}, \bibinfo {author} {\bibfnamefont {Nor~A}\ \bibnamefont {Zakaria}},
  \bibinfo {author} {\bibfnamefont {Niccolo}\ \bibnamefont {Somaschi}},
  \bibinfo {author} {\bibfnamefont {Carlos}\ \bibnamefont {Anton}}, \bibinfo
  {author} {\bibfnamefont {Lorenzo}\ \bibnamefont {de~Santis}}, \bibinfo
  {author} {\bibfnamefont {Valerian}\ \bibnamefont {Giesz}}, \bibinfo {author}
  {\bibfnamefont {Thomas}\ \bibnamefont {Grange}}, \bibinfo {author}
  {\bibfnamefont {Matthew~A}\ \bibnamefont {Broome}}, \bibinfo {author}
  {\bibfnamefont {Olivier}\ \bibnamefont {Gazzano}}, \bibinfo {author}
  {\bibfnamefont {Guillaume}\ \bibnamefont {Coppola}},  \emph {et~al.}}
  (\bibinfo {year} {2016}),\ \bibfield  {title} {\enquote {\bibinfo {title}
  {Scalable performance in solid-state single-photon sources},}\ }\href
  {https://doi.org/10.1364/optica.3.000433} {\bibfield  {journal} {\bibinfo
  {journal} {Optica}\ }\textbf {\bibinfo {volume} {3}},\ \bibinfo {pages}
  {433}},\ \Eprint {https://arxiv.org/abs/arXiv:1601.00654v2}
  {arXiv:1601.00654v2} \BibitemShut {NoStop}%
\bibitem [{\citenamefont {Lovett}\ \emph
  {et~al.}(2010{\natexlab{a}})\citenamefont {Lovett}, \citenamefont {Cooper},
  \citenamefont {Everitt}, \citenamefont {Trevers},\ and\ \citenamefont
  {Kendon}}]{lovett2010universal}%
  \BibitemOpen
  \bibfield  {author} {\bibinfo {author} {\bibnamefont {Lovett}, \bibfnamefont
  {Neil~B}}, \bibinfo {author} {\bibfnamefont {Sally}\ \bibnamefont {Cooper}},
  \bibinfo {author} {\bibfnamefont {Matthew}\ \bibnamefont {Everitt}}, \bibinfo
  {author} {\bibfnamefont {Matthew}\ \bibnamefont {Trevers}}, and\ \bibinfo
  {author} {\bibfnamefont {Viv}\ \bibnamefont {Kendon}}} (\bibinfo {year}
  {2010}{\natexlab{a}}),\ \bibfield  {title} {\enquote {\bibinfo {title}
  {Universal quantum computation using the discrete-time quantum walk},}\
  }\href@noop {} {\bibfield  {journal} {\bibinfo  {journal} {Physical Review
  A—Atomic, Molecular, and Optical Physics}\ }\textbf {\bibinfo {volume}
  {81}}~(\bibinfo {number} {4}),\ \bibinfo {pages} {042330}}\BibitemShut
  {NoStop}%
\bibitem [{\citenamefont {Lovett}\ \emph
  {et~al.}(2010{\natexlab{b}})\citenamefont {Lovett}, \citenamefont {Cooper},
  \citenamefont {Everitt}, \citenamefont {Trevers},\ and\ \citenamefont
  {Kendon}}]{bib:Lovett10}%
  \BibitemOpen
  \bibfield  {author} {\bibinfo {author} {\bibnamefont {Lovett}, \bibfnamefont
  {Neil~B}}, \bibinfo {author} {\bibfnamefont {Sally}\ \bibnamefont {Cooper}},
  \bibinfo {author} {\bibfnamefont {Matthew}\ \bibnamefont {Everitt}}, \bibinfo
  {author} {\bibfnamefont {Matthew}\ \bibnamefont {Trevers}}, and\ \bibinfo
  {author} {\bibfnamefont {Viv}\ \bibnamefont {Kendon}}} (\bibinfo {year}
  {2010}{\natexlab{b}}),\ \bibfield  {title} {\enquote {\bibinfo {title}
  {Universal quantum computation using the discrete time quantum walk},}\
  }\href {https://doi.org/10.1103/physreva.81.042330} {\bibfield  {journal}
  {\bibinfo  {journal} {Physical Review A}\ }\textbf {\bibinfo {volume} {81}},\
  \bibinfo {pages} {042330}},\ \Eprint
  {https://arxiv.org/abs/arXiv:0910.1024v3} {arXiv:0910.1024v3} \BibitemShut
  {NoStop}%
\bibitem [{\citenamefont {Lu}\ \emph {et~al.}(2009)\citenamefont {Lu},
  \citenamefont {Yang},\ and\ \citenamefont {Pan}}]{bib:PRL_103_020501}%
  \BibitemOpen
  \bibfield  {author} {\bibinfo {author} {\bibnamefont {Lu}, \bibfnamefont
  {Chao-Yang}}, \bibinfo {author} {\bibfnamefont {Tao}\ \bibnamefont {Yang}},
  and\ \bibinfo {author} {\bibfnamefont {Jian-Wei}\ \bibnamefont {Pan}}}
  (\bibinfo {year} {2009}),\ \bibfield  {title} {\enquote {\bibinfo {title}
  {Experimental multiparticle entanglement swapping for quantum networking},}\
  }\href {https://doi.org/10.1103/physrevlett.103.020501} {\bibfield  {journal}
  {\bibinfo  {journal} {Physical Review Letters}\ }\textbf {\bibinfo {volume}
  {103}},\ \bibinfo {pages} {020501}}\BibitemShut {NoStop}%
\bibitem [{\citenamefont {Lu}\ \emph {et~al.}(2015)\citenamefont {Lu},
  \citenamefont {Li}, \citenamefont {Trottier}, \citenamefont {Li},
  \citenamefont {Brodutch}, \citenamefont {Krismanich}, \citenamefont
  {Ghavami}, \citenamefont {Dmitrienko}, \citenamefont {Long}, \citenamefont
  {Baugh},\ and\ \citenamefont {Laflamme}}]{bib:Lu2015}%
  \BibitemOpen
  \bibfield  {author} {\bibinfo {author} {\bibnamefont {Lu}, \bibfnamefont
  {Dawei}}, \bibinfo {author} {\bibfnamefont {Hang}\ \bibnamefont {Li}},
  \bibinfo {author} {\bibfnamefont {Denis-Alexandre}\ \bibnamefont {Trottier}},
  \bibinfo {author} {\bibfnamefont {Jun}\ \bibnamefont {Li}}, \bibinfo {author}
  {\bibfnamefont {Aharon}\ \bibnamefont {Brodutch}}, \bibinfo {author}
  {\bibfnamefont {Anthony~P.}\ \bibnamefont {Krismanich}}, \bibinfo {author}
  {\bibfnamefont {Ahmad}\ \bibnamefont {Ghavami}}, \bibinfo {author}
  {\bibfnamefont {Gary~I.}\ \bibnamefont {Dmitrienko}}, \bibinfo {author}
  {\bibfnamefont {Guilu}\ \bibnamefont {Long}}, \bibinfo {author}
  {\bibfnamefont {Jonathan}\ \bibnamefont {Baugh}}, and\ \bibinfo {author}
  {\bibfnamefont {Raymond}\ \bibnamefont {Laflamme}}} (\bibinfo {year}
  {2015}),\ \bibfield  {title} {\enquote {\bibinfo {title} {Experimental
  estimation of average fidelity of a clifford gate on a 7-qubit quantum
  processor},}\ }\href {https://doi.org/10.1103/PhysRevLett.114.140505}
  {\bibfield  {journal} {\bibinfo  {journal} {Physical Review Letters}\
  }\textbf {\bibinfo {volume} {114}},\ \bibinfo {pages} {140505}},\ \Eprint
  {https://arxiv.org/abs/arXiv:1411.7993v1} {arXiv:1411.7993v1} \BibitemShut
  {NoStop}%
\bibitem [{\citenamefont {Lu}\ \emph {et~al.}(2017)\citenamefont {Lu},
  \citenamefont {Li}, \citenamefont {Li}, \citenamefont {Katiyar},
  \citenamefont {Park}, \citenamefont {Feng}, \citenamefont {Xin},
  \citenamefont {Li}, \citenamefont {Long}, \citenamefont {Brodutch} \emph
  {et~al.}}]{bib:lu2017enhancing}%
  \BibitemOpen
  \bibfield  {author} {\bibinfo {author} {\bibnamefont {Lu}, \bibfnamefont
  {Dawei}}, \bibinfo {author} {\bibfnamefont {Keren}\ \bibnamefont {Li}},
  \bibinfo {author} {\bibfnamefont {Jun}\ \bibnamefont {Li}}, \bibinfo {author}
  {\bibfnamefont {Hemant}\ \bibnamefont {Katiyar}}, \bibinfo {author}
  {\bibfnamefont {Annie~Jihyun}\ \bibnamefont {Park}}, \bibinfo {author}
  {\bibfnamefont {Guanru}\ \bibnamefont {Feng}}, \bibinfo {author}
  {\bibfnamefont {Tao}\ \bibnamefont {Xin}}, \bibinfo {author} {\bibfnamefont
  {Hang}\ \bibnamefont {Li}}, \bibinfo {author} {\bibfnamefont {Guilu}\
  \bibnamefont {Long}}, \bibinfo {author} {\bibfnamefont {Aharon}\ \bibnamefont
  {Brodutch}},  \emph {et~al.}} (\bibinfo {year} {2017}),\ \bibfield  {title}
  {\enquote {\bibinfo {title} {Enhancing quantum control by bootstrapping a
  quantum processor of 12 qubits},}\ }\href
  {https://doi.org/10.1038/s41534-017-0045-z} {\bibfield  {journal} {\bibinfo
  {journal} {NPJ Quantum Information}\ }\textbf {\bibinfo {volume} {3}},\
  \bibinfo {pages} {45}},\ \Eprint {https://arxiv.org/abs/arXiv:1701.01198v2}
  {arXiv:1701.01198v2} \BibitemShut {NoStop}%
\bibitem [{\citenamefont {Lu}\ \emph {et~al.}(2018)\citenamefont {Lu},
  \citenamefont {Huang}, \citenamefont {Li}, \citenamefont {Li}, \citenamefont
  {Chen}, \citenamefont {Lu}, \citenamefont {Ji}, \citenamefont {Shen},
  \citenamefont {Zhou},\ and\ \citenamefont {Zeng}}]{bib:Su2017}%
  \BibitemOpen
  \bibfield  {author} {\bibinfo {author} {\bibnamefont {Lu}, \bibfnamefont
  {Sirui}}, \bibinfo {author} {\bibfnamefont {Shilin}\ \bibnamefont {Huang}},
  \bibinfo {author} {\bibfnamefont {Keren}\ \bibnamefont {Li}}, \bibinfo
  {author} {\bibfnamefont {Jun}\ \bibnamefont {Li}}, \bibinfo {author}
  {\bibfnamefont {Jianxin}\ \bibnamefont {Chen}}, \bibinfo {author}
  {\bibfnamefont {Dawei}\ \bibnamefont {Lu}}, \bibinfo {author} {\bibfnamefont
  {Zhengfeng}\ \bibnamefont {Ji}}, \bibinfo {author} {\bibfnamefont
  {Yi}~\bibnamefont {Shen}}, \bibinfo {author} {\bibfnamefont {Duanlu}\
  \bibnamefont {Zhou}}, and\ \bibinfo {author} {\bibfnamefont {Bei}\
  \bibnamefont {Zeng}}} (\bibinfo {year} {2018}),\ \bibfield  {title} {\enquote
  {\bibinfo {title} {A separability-entanglement classifier via machine
  learning},}\ }\href {https://doi.org/10.1103/physreva.98.012315} {\bibfield
  {journal} {\bibinfo  {journal} {Physical Review A}\ }\textbf {\bibinfo
  {volume} {98}},\ \bibinfo {pages} {012315}},\ \Eprint
  {https://arxiv.org/abs/arXiv:1705.01523} {arXiv:1705.01523} \BibitemShut
  {NoStop}%
\bibitem [{\citenamefont {L{\"u}cke}\ \emph {et~al.}(2011)\citenamefont
  {L{\"u}cke}, \citenamefont {Scherer}, \citenamefont {Kruse}, \citenamefont
  {Pezz{\'e}}, \citenamefont {Deuretzbacher}, \citenamefont {Hyllus},
  \citenamefont {Peise}, \citenamefont {Ertmer}, \citenamefont {Arlt},
  \citenamefont {Santos} \emph {et~al.}}]{bib:lucke2011twin}%
  \BibitemOpen
  \bibfield  {author} {\bibinfo {author} {\bibnamefont {L{\"u}cke},
  \bibfnamefont {Bernd}}, \bibinfo {author} {\bibfnamefont {Manuel}\
  \bibnamefont {Scherer}}, \bibinfo {author} {\bibfnamefont {Jens}\
  \bibnamefont {Kruse}}, \bibinfo {author} {\bibfnamefont {Luca}\ \bibnamefont
  {Pezz{\'e}}}, \bibinfo {author} {\bibfnamefont {Frank}\ \bibnamefont
  {Deuretzbacher}}, \bibinfo {author} {\bibfnamefont {Phillip}\ \bibnamefont
  {Hyllus}}, \bibinfo {author} {\bibfnamefont {Jan}\ \bibnamefont {Peise}},
  \bibinfo {author} {\bibfnamefont {Wolfgang}\ \bibnamefont {Ertmer}}, \bibinfo
  {author} {\bibfnamefont {Jan}\ \bibnamefont {Arlt}}, \bibinfo {author}
  {\bibfnamefont {Luis}\ \bibnamefont {Santos}},  \emph {et~al.}} (\bibinfo
  {year} {2011}),\ \bibfield  {title} {\enquote {\bibinfo {title} {Twin matter
  waves for interferometry beyond the classical limit},}\ }\href
  {https://doi.org/10.1126/science.1208798} {\bibfield  {journal} {\bibinfo
  {journal} {Science}\ }\textbf {\bibinfo {volume} {334}},\ \bibinfo {pages}
  {773}},\ \Eprint {https://arxiv.org/abs/arXiv:1204.4102v1}
  {arXiv:1204.4102v1} \BibitemShut {NoStop}%
\bibitem [{\citenamefont {Ludlow}\ \emph {et~al.}(2015)\citenamefont {Ludlow},
  \citenamefont {Boyd}, \citenamefont {Ye}, \citenamefont {Peik},\ and\
  \citenamefont {Schmidt}}]{bib:ludlow2015optical}%
  \BibitemOpen
  \bibfield  {author} {\bibinfo {author} {\bibnamefont {Ludlow}, \bibfnamefont
  {Andrew~D}}, \bibinfo {author} {\bibfnamefont {Martin~M}\ \bibnamefont
  {Boyd}}, \bibinfo {author} {\bibfnamefont {Jun}\ \bibnamefont {Ye}}, \bibinfo
  {author} {\bibfnamefont {Ekkehard}\ \bibnamefont {Peik}}, and\ \bibinfo
  {author} {\bibfnamefont {Piet~O}\ \bibnamefont {Schmidt}}} (\bibinfo {year}
  {2015}),\ \bibfield  {title} {\enquote {\bibinfo {title} {Optical atomic
  clocks},}\ }\href {https://doi.org/10.1103/revmodphys.87.637} {\bibfield
  {journal} {\bibinfo  {journal} {Reviews in Modern Physics}\ }\textbf
  {\bibinfo {volume} {87}},\ \bibinfo {pages} {637}}\BibitemShut {NoStop}%
\bibitem [{\citenamefont {Lukin}\ \emph {et~al.}(2000)\citenamefont {Lukin},
  \citenamefont {Yelin},\ and\ \citenamefont
  {Fleischhauer}}]{bib:lukin2000entanglement}%
  \BibitemOpen
  \bibfield  {author} {\bibinfo {author} {\bibnamefont {Lukin}, \bibfnamefont
  {MD}}, \bibinfo {author} {\bibfnamefont {SF}~\bibnamefont {Yelin}}, and\
  \bibinfo {author} {\bibfnamefont {M}~\bibnamefont {Fleischhauer}}} (\bibinfo
  {year} {2000}),\ \bibfield  {title} {\enquote {\bibinfo {title} {Entanglement
  of atomic ensembles by trapping correlated photon states},}\ }\href
  {https://doi.org/10.1103/physrevlett.84.4232} {\bibfield  {journal} {\bibinfo
   {journal} {Physical Review Letters}\ }\textbf {\bibinfo {volume} {84}},\
  \bibinfo {pages} {4232}},\ \Eprint
  {https://arxiv.org/abs/arXiv:quant-ph/9912046v1} {arXiv:quant-ph/9912046v1}
  \BibitemShut {NoStop}%
\bibitem [{\citenamefont {Lund}\ \emph {et~al.}(2014)\citenamefont {Lund},
  \citenamefont {Laing}, \citenamefont {Rahimi-Keshari}, \citenamefont
  {Rudolph}, \citenamefont {O'Brien},\ and\ \citenamefont
  {Ralph}}]{bib:RandBS}%
  \BibitemOpen
  \bibfield  {author} {\bibinfo {author} {\bibnamefont {Lund}, \bibfnamefont
  {A~P}}, \bibinfo {author} {\bibfnamefont {A.}~\bibnamefont {Laing}}, \bibinfo
  {author} {\bibfnamefont {S.}~\bibnamefont {Rahimi-Keshari}}, \bibinfo
  {author} {\bibfnamefont {T.}~\bibnamefont {Rudolph}}, \bibinfo {author}
  {\bibfnamefont {J.~L}\ \bibnamefont {O'Brien}}, and\ \bibinfo {author}
  {\bibfnamefont {T.~C.}\ \bibnamefont {Ralph}}} (\bibinfo {year} {2014}),\
  \bibfield  {title} {\enquote {\bibinfo {title} {Boson sampling from gaussian
  states},}\ }\href {https://doi.org/10.1103/physrevlett.113.100502} {\bibfield
   {journal} {\bibinfo  {journal} {Physical Review Letters}\ }\textbf {\bibinfo
  {volume} {113}},\ \bibinfo {pages} {100502}},\ \Eprint
  {https://arxiv.org/abs/arXiv:1305.4346v3} {arXiv:1305.4346v3} \BibitemShut
  {NoStop}%
\bibitem [{\citenamefont {Lund}\ \emph {et~al.}(2008)\citenamefont {Lund},
  \citenamefont {Ralph},\ and\ \citenamefont
  {Haselgrove}}]{bib:PhysRevLett.100.030503}%
  \BibitemOpen
  \bibfield  {author} {\bibinfo {author} {\bibnamefont {Lund}, \bibfnamefont
  {A~P}}, \bibinfo {author} {\bibfnamefont {T.~C.}\ \bibnamefont {Ralph}}, and\
  \bibinfo {author} {\bibfnamefont {H.~L.}\ \bibnamefont {Haselgrove}}}
  (\bibinfo {year} {2008}),\ \bibfield  {title} {\enquote {\bibinfo {title}
  {Fault-tolerant linear optical quantum computing with small-amplitude
  coherent states},}\ }\href {https://doi.org/10.1103/PhysRevLett.100.030503}
  {\bibfield  {journal} {\bibinfo  {journal} {Phys. Rev. Lett.}\ }\textbf
  {\bibinfo {volume} {100}},\ \bibinfo {pages} {030503}}\BibitemShut {NoStop}%
\bibitem [{\citenamefont {Lund}\ \emph {et~al.}(2017)\citenamefont {Lund},
  \citenamefont {Bremner},\ and\ \citenamefont {Ralph}}]{lund2017quantum}%
  \BibitemOpen
  \bibfield  {author} {\bibinfo {author} {\bibnamefont {Lund}, \bibfnamefont
  {Austin~P}}, \bibinfo {author} {\bibfnamefont {Michael~J}\ \bibnamefont
  {Bremner}}, and\ \bibinfo {author} {\bibfnamefont {Timothy~C}\ \bibnamefont
  {Ralph}}} (\bibinfo {year} {2017}),\ \bibfield  {title} {\enquote {\bibinfo
  {title} {Quantum sampling problems, bosonsampling and quantum supremacy},}\
  }\href@noop {} {\bibfield  {journal} {\bibinfo  {journal} {npj Quantum
  Information}\ }\textbf {\bibinfo {volume} {3}}~(\bibinfo {number} {1}),\
  \bibinfo {pages} {15}}\BibitemShut {NoStop}%
\bibitem [{\citenamefont {Lupo}(2015)}]{lupo2015Entropy}%
  \BibitemOpen
  \bibfield  {author} {\bibinfo {author} {\bibnamefont {Lupo}, \bibfnamefont
  {Cosmo}}} (\bibinfo {year} {2015}),\ \bibfield  {title} {\enquote {\bibinfo
  {title} {Quantum data locking for secure communication against an
  eavesdropper with time-limited storage},}\ }\href@noop {} {\bibfield
  {journal} {\bibinfo  {journal} {Entropy}\ }\textbf {\bibinfo {volume}
  {17}}~(\bibinfo {number} {5}),\ \bibinfo {pages} {3194--3204}}\BibitemShut
  {NoStop}%
\bibitem [{\citenamefont {Lupo}\ and\ \citenamefont
  {Lloyd}(2015)}]{lupo2015NJP}%
  \BibitemOpen
  \bibfield  {author} {\bibinfo {author} {\bibnamefont {Lupo}, \bibfnamefont
  {Cosmo}}, and\ \bibinfo {author} {\bibfnamefont {Seth}\ \bibnamefont
  {Lloyd}}} (\bibinfo {year} {2015}),\ \bibfield  {title} {\enquote {\bibinfo
  {title} {Quantum data locking for high-rate private communication},}\
  }\href@noop {} {\bibfield  {journal} {\bibinfo  {journal} {New Journal of
  Physics}\ }\textbf {\bibinfo {volume} {17}}~(\bibinfo {number} {3}),\
  \bibinfo {pages} {033022}}\BibitemShut {NoStop}%
\bibitem [{\citenamefont {Lupo}\ \emph {et~al.}(2014)\citenamefont {Lupo},
  \citenamefont {Wilde},\ and\ \citenamefont {Lloyd}}]{lupo2014pra}%
  \BibitemOpen
  \bibfield  {author} {\bibinfo {author} {\bibnamefont {Lupo}, \bibfnamefont
  {Cosmo}}, \bibinfo {author} {\bibfnamefont {Mark~M.}\ \bibnamefont {Wilde}},
  and\ \bibinfo {author} {\bibfnamefont {Seth}\ \bibnamefont {Lloyd}}}
  (\bibinfo {year} {2014}),\ \bibfield  {title} {\enquote {\bibinfo {title}
  {Robust quantum data locking from phase modulation},}\ }\href
  {https://doi.org/10.1103/PhysRevA.90.022326} {\bibfield  {journal} {\bibinfo
  {journal} {Phys. Rev. A}\ }\textbf {\bibinfo {volume} {90}},\ \bibinfo
  {pages} {022326}}\BibitemShut {NoStop}%
\bibitem [{\citenamefont {Lvovsky}\ \emph {et~al.}(2009)\citenamefont
  {Lvovsky}, \citenamefont {Sanders},\ and\ \citenamefont
  {Tittel}}]{bib:lvovsky2009optical}%
  \BibitemOpen
  \bibfield  {author} {\bibinfo {author} {\bibnamefont {Lvovsky}, \bibfnamefont
  {Alexander~I}}, \bibinfo {author} {\bibfnamefont {Barry~C}\ \bibnamefont
  {Sanders}}, and\ \bibinfo {author} {\bibfnamefont {Wolfgang}\ \bibnamefont
  {Tittel}}} (\bibinfo {year} {2009}),\ \bibfield  {title} {\enquote {\bibinfo
  {title} {Optical quantum memory},}\ }\href
  {https://doi.org/10.1038/nphoton.2009.231} {\bibfield  {journal} {\bibinfo
  {journal} {Nature Photonics}\ }\textbf {\bibinfo {volume} {3}},\ \bibinfo
  {pages} {706}}\BibitemShut {NoStop}%
\bibitem [{\citenamefont {Lydersen}\ \emph {et~al.}(2010)\citenamefont
  {Lydersen}, \citenamefont {Wiechers}, \citenamefont {Wittmann}, \citenamefont
  {Elser}, \citenamefont {Skaar},\ and\ \citenamefont
  {Makarov}}]{bib:lydersen2010hacking}%
  \BibitemOpen
  \bibfield  {author} {\bibinfo {author} {\bibnamefont {Lydersen},
  \bibfnamefont {Lars}}, \bibinfo {author} {\bibfnamefont {Carlos}\
  \bibnamefont {Wiechers}}, \bibinfo {author} {\bibfnamefont {Christoffer}\
  \bibnamefont {Wittmann}}, \bibinfo {author} {\bibfnamefont {Dominique}\
  \bibnamefont {Elser}}, \bibinfo {author} {\bibfnamefont {Johannes}\
  \bibnamefont {Skaar}}, and\ \bibinfo {author} {\bibfnamefont {Vadim}\
  \bibnamefont {Makarov}}} (\bibinfo {year} {2010}),\ \bibfield  {title}
  {\enquote {\bibinfo {title} {Hacking commercial quantum cryptography systems
  by tailored bright illumination},}\ }\href
  {https://doi.org/10.1038/nphoton.2010.214} {\bibfield  {journal} {\bibinfo
  {journal} {Nature Photonics}\ }\textbf {\bibinfo {volume} {4}},\ \bibinfo
  {pages} {686}},\ \Eprint {https://arxiv.org/abs/arXiv:1008.4593v2}
  {arXiv:1008.4593v2} \BibitemShut {NoStop}%
\bibitem [{\citenamefont {Ma}\ \emph {et~al.}(2012{\natexlab{a}})\citenamefont
  {Ma}, \citenamefont {Herbst}, \citenamefont {Scheidl}, \citenamefont {Wang},
  \citenamefont {Kropatschek}, \citenamefont {Naylor}, \citenamefont
  {Wittmann}, \citenamefont {Mech}, \citenamefont {Kofler}, \citenamefont
  {Anisimova}, \citenamefont {Makarov}, \citenamefont {Jennewein},
  \citenamefont {Ursin},\ and\ \citenamefont {Zeilinger}}]{SD-Ma:2012aa}%
  \BibitemOpen
  \bibfield  {author} {\bibinfo {author} {\bibnamefont {Ma}, \bibfnamefont
  {Xiao-Song}}, \bibinfo {author} {\bibfnamefont {Thomas}\ \bibnamefont
  {Herbst}}, \bibinfo {author} {\bibfnamefont {Thomas}\ \bibnamefont
  {Scheidl}}, \bibinfo {author} {\bibfnamefont {Daqing}\ \bibnamefont {Wang}},
  \bibinfo {author} {\bibfnamefont {Sebastian}\ \bibnamefont {Kropatschek}},
  \bibinfo {author} {\bibfnamefont {William}\ \bibnamefont {Naylor}}, \bibinfo
  {author} {\bibfnamefont {Bernhard}\ \bibnamefont {Wittmann}}, \bibinfo
  {author} {\bibfnamefont {Alexandra}\ \bibnamefont {Mech}}, \bibinfo {author}
  {\bibfnamefont {Johannes}\ \bibnamefont {Kofler}}, \bibinfo {author}
  {\bibfnamefont {Elena}\ \bibnamefont {Anisimova}}, \bibinfo {author}
  {\bibfnamefont {Vadim}\ \bibnamefont {Makarov}}, \bibinfo {author}
  {\bibfnamefont {Thomas}\ \bibnamefont {Jennewein}}, \bibinfo {author}
  {\bibfnamefont {Rupert}\ \bibnamefont {Ursin}}, and\ \bibinfo {author}
  {\bibfnamefont {Anton}\ \bibnamefont {Zeilinger}}} (\bibinfo {year}
  {2012}{\natexlab{a}}),\ \bibfield  {title} {\enquote {\bibinfo {title}
  {Quantum teleportation over 143 kilometres using active feed-forward},}\
  }\href {https://doi.org/10.1038/nature11472} {\bibfield  {journal} {\bibinfo
  {journal} {Nature}\ }\textbf {\bibinfo {volume} {489}},\ \bibinfo {pages}
  {269}}\BibitemShut {NoStop}%
\bibitem [{\citenamefont {Ma}\ \emph {et~al.}(2012{\natexlab{b}})\citenamefont
  {Ma}, \citenamefont {Herbst}, \citenamefont {Scheidl}, \citenamefont {Wang},
  \citenamefont {Kropatschek}, \citenamefont {Naylor}, \citenamefont
  {Wittmann}, \citenamefont {Mech}, \citenamefont {Kofler}, \citenamefont
  {Anisimova} \emph {et~al.}}]{bib:Nat_489_269}%
  \BibitemOpen
  \bibfield  {author} {\bibinfo {author} {\bibnamefont {Ma}, \bibfnamefont
  {Xiao-Song}}, \bibinfo {author} {\bibfnamefont {Thomas}\ \bibnamefont
  {Herbst}}, \bibinfo {author} {\bibfnamefont {Thomas}\ \bibnamefont
  {Scheidl}}, \bibinfo {author} {\bibfnamefont {Daqing}\ \bibnamefont {Wang}},
  \bibinfo {author} {\bibfnamefont {Sebastian}\ \bibnamefont {Kropatschek}},
  \bibinfo {author} {\bibfnamefont {William}\ \bibnamefont {Naylor}}, \bibinfo
  {author} {\bibfnamefont {Bernhard}\ \bibnamefont {Wittmann}}, \bibinfo
  {author} {\bibfnamefont {Alexandra}\ \bibnamefont {Mech}}, \bibinfo {author}
  {\bibfnamefont {Johannes}\ \bibnamefont {Kofler}}, \bibinfo {author}
  {\bibfnamefont {Elena}\ \bibnamefont {Anisimova}},  \emph {et~al.}} (\bibinfo
  {year} {2012}{\natexlab{b}}),\ \bibfield  {title} {\enquote {\bibinfo {title}
  {Quantum teleportation over 143 kilometres using active feed-forward},}\
  }\href {https://doi.org/10.1038/nature11472} {\bibfield  {journal} {\bibinfo
  {journal} {Nature}\ }\textbf {\bibinfo {volume} {489}},\ \bibinfo {pages}
  {269}}\BibitemShut {NoStop}%
\bibitem [{\citenamefont {Ma}\ \emph {et~al.}(2012{\natexlab{c}})\citenamefont
  {Ma}, \citenamefont {Zotter}, \citenamefont {Kofler}, \citenamefont {Ursin},
  \citenamefont {Jennewein}, \citenamefont {Brukner},\ and\ \citenamefont
  {Zeilinger}}]{bib:Nat_Phys_8_479}%
  \BibitemOpen
  \bibfield  {author} {\bibinfo {author} {\bibnamefont {Ma}, \bibfnamefont
  {Xiao-song}}, \bibinfo {author} {\bibfnamefont {Stefan}\ \bibnamefont
  {Zotter}}, \bibinfo {author} {\bibfnamefont {Johannes}\ \bibnamefont
  {Kofler}}, \bibinfo {author} {\bibfnamefont {Rupert}\ \bibnamefont {Ursin}},
  \bibinfo {author} {\bibfnamefont {Thomas}\ \bibnamefont {Jennewein}},
  \bibinfo {author} {\bibfnamefont {{\v{C}}aslav}\ \bibnamefont {Brukner}},
  and\ \bibinfo {author} {\bibfnamefont {Anton}\ \bibnamefont {Zeilinger}}}
  (\bibinfo {year} {2012}{\natexlab{c}}),\ \bibfield  {title} {\enquote
  {\bibinfo {title} {Experimental delayed-choice entanglement swapping},}\
  }\href {https://doi.org/10.1038/nphys2294} {\bibfield  {journal} {\bibinfo
  {journal} {Nature Physics}\ }\textbf {\bibinfo {volume} {8}},\ \bibinfo
  {pages} {479}},\ \Eprint {https://arxiv.org/abs/arXiv:1203.4834v2}
  {arXiv:1203.4834v2} \BibitemShut {NoStop}%
\bibitem [{\citenamefont {Ma}\ and\ \citenamefont {Yung}(2017)}]{bib:Ma2017}%
  \BibitemOpen
  \bibfield  {author} {\bibinfo {author} {\bibnamefont {Ma}, \bibfnamefont
  {Y-C}}, and\ \bibinfo {author} {\bibfnamefont {M.-H.}\ \bibnamefont {Yung}}}
  (\bibinfo {year} {2017}),\ \bibfield  {title} {\enquote {\bibinfo {title}
  {Transforming bell's inequalities into state classifiers with machine
  learning},}\ }\href {https://doi.org/10.1038/s41534-018-0081-3} {\
  10.1038/s41534-018-0081-3},\ \Eprint {https://arxiv.org/abs/arXiv:1705.00813}
  {arXiv:1705.00813} \BibitemShut {NoStop}%
\bibitem [{\citenamefont {Mahadev}(2020)}]{mahadev2020classical}%
  \BibitemOpen
  \bibfield  {author} {\bibinfo {author} {\bibnamefont {Mahadev}, \bibfnamefont
  {Urmila}}} (\bibinfo {year} {2020}),\ \bibfield  {title} {\enquote {\bibinfo
  {title} {Classical homomorphic encryption for quantum circuits},}\
  }\href@noop {} {\bibfield  {journal} {\bibinfo  {journal} {SIAM Journal on
  Computing}\ }\textbf {\bibinfo {volume} {52}}~(\bibinfo {number} {6}),\
  \bibinfo {pages} {FOCS18--189}}\BibitemShut {NoStop}%
\bibitem [{\citenamefont {Mahloujifar}\ \emph {et~al.}(2018)\citenamefont
  {Mahloujifar}, \citenamefont {Diochnos},\ and\ \citenamefont
  {Mahmoody}}]{bib:mahloujifar2018curse}%
  \BibitemOpen
  \bibfield  {author} {\bibinfo {author} {\bibnamefont {Mahloujifar},
  \bibfnamefont {Saeed}}, \bibinfo {author} {\bibfnamefont {Dimitrios~I.}\
  \bibnamefont {Diochnos}}, and\ \bibinfo {author} {\bibfnamefont {Mohammad}\
  \bibnamefont {Mahmoody}}} (\bibinfo {year} {2018}),\ \bibfield  {title}
  {\enquote {\bibinfo {title} {The curse of concentration in robust learning:
  Evasion and poisoning attacks from concentration of measure},}\ }\href@noop
  {} {\ }\Eprint {https://arxiv.org/abs/arXiv:1809.03063} {arXiv:1809.03063}
  \BibitemShut {NoStop}%
\bibitem [{\citenamefont {Maitre}\ \emph {et~al.}(1997)\citenamefont {Maitre},
  \citenamefont {Hagley}, \citenamefont {Nogues}, \citenamefont {Wunderlich},
  \citenamefont {Goy}, \citenamefont {Brune}, \citenamefont {Raimond},\ and\
  \citenamefont {Haroche}}]{bib:maitre1997quantum}%
  \BibitemOpen
  \bibfield  {author} {\bibinfo {author} {\bibnamefont {Maitre}, \bibfnamefont
  {X}}, \bibinfo {author} {\bibfnamefont {E}~\bibnamefont {Hagley}}, \bibinfo
  {author} {\bibfnamefont {G}~\bibnamefont {Nogues}}, \bibinfo {author}
  {\bibfnamefont {C}~\bibnamefont {Wunderlich}}, \bibinfo {author}
  {\bibfnamefont {P}~\bibnamefont {Goy}}, \bibinfo {author} {\bibfnamefont
  {M}~\bibnamefont {Brune}}, \bibinfo {author} {\bibfnamefont {JM}~\bibnamefont
  {Raimond}}, and\ \bibinfo {author} {\bibfnamefont {S}~\bibnamefont
  {Haroche}}} (\bibinfo {year} {1997}),\ \bibfield  {title} {\enquote {\bibinfo
  {title} {Quantum memory with a single photon in a cavity},}\ }\href
  {https://doi.org/10.1103/physrevlett.79.769} {\bibfield  {journal} {\bibinfo
  {journal} {Physical Review Letters}\ }\textbf {\bibinfo {volume} {79}},\
  \bibinfo {pages} {769}}\BibitemShut {NoStop}%
\bibitem [{\citenamefont {Makarov}\ \emph {et~al.}(2006)\citenamefont
  {Makarov}, \citenamefont {Anisimov},\ and\ \citenamefont
  {Skaar}}]{bib:PhysRevA.74.022313}%
  \BibitemOpen
  \bibfield  {author} {\bibinfo {author} {\bibnamefont {Makarov}, \bibfnamefont
  {Vadim}}, \bibinfo {author} {\bibfnamefont {Andrey}\ \bibnamefont
  {Anisimov}}, and\ \bibinfo {author} {\bibfnamefont {Johannes}\ \bibnamefont
  {Skaar}}} (\bibinfo {year} {2006}),\ \bibfield  {title} {\enquote {\bibinfo
  {title} {Effects of detector efficiency mismatch on security of quantum
  cryptosystems},}\ }\href {https://doi.org/10.1103/physreva.74.022313}
  {\bibfield  {journal} {\bibinfo  {journal} {Physical Review A}\ }\textbf
  {\bibinfo {volume} {74}},\ \bibinfo {pages} {022313}},\ \Eprint
  {https://arxiv.org/abs/arXiv:quant-ph/0511032v3} {arXiv:quant-ph/0511032v3}
  \BibitemShut {NoStop}%
\bibitem [{\citenamefont {Makhlin}\ \emph {et~al.}(2001)\citenamefont
  {Makhlin}, \citenamefont {Sch{\"o}n},\ and\ \citenamefont
  {Shnirman}}]{bib:makhlin2001quantum}%
  \BibitemOpen
  \bibfield  {author} {\bibinfo {author} {\bibnamefont {Makhlin}, \bibfnamefont
  {Yuriy}}, \bibinfo {author} {\bibfnamefont {Gerd}\ \bibnamefont {Sch{\"o}n}},
  and\ \bibinfo {author} {\bibfnamefont {Alexander}\ \bibnamefont {Shnirman}}}
  (\bibinfo {year} {2001}),\ \bibfield  {title} {\enquote {\bibinfo {title}
  {Quantum-state engineering with josephson-junction devices},}\ }\href
  {https://doi.org/10.1103/revmodphys.73.357} {\bibfield  {journal} {\bibinfo
  {journal} {Reviews of Modern Physics}\ }\textbf {\bibinfo {volume} {73}},\
  \bibinfo {pages} {357}},\ \Eprint
  {https://arxiv.org/abs/arXiv:cond-mat/0011269v1} {arXiv:cond-mat/0011269v1}
  \BibitemShut {NoStop}%
\bibitem [{\citenamefont {Marchese}\ and\ \citenamefont
  {Kok}(2023)}]{PhysRevLett.130.160801}%
  \BibitemOpen
  \bibfield  {author} {\bibinfo {author} {\bibnamefont {Marchese},
  \bibfnamefont {Marta~Maria}}, and\ \bibinfo {author} {\bibfnamefont {Pieter}\
  \bibnamefont {Kok}}} (\bibinfo {year} {2023}),\ \bibfield  {title} {\enquote
  {\bibinfo {title} {Large baseline optical imaging assisted by single photons
  and linear quantum optics},}\ }\href
  {https://doi.org/10.1103/PhysRevLett.130.160801} {\bibfield  {journal}
  {\bibinfo  {journal} {Phys. Rev. Lett.}\ }\textbf {\bibinfo {volume} {130}},\
  \bibinfo {pages} {160801}}\BibitemShut {NoStop}%
\bibitem [{\citenamefont {Marcikic}\ \emph {et~al.}(2003)\citenamefont
  {Marcikic}, \citenamefont {De~Riedmatten}, \citenamefont {Tittel},
  \citenamefont {Zbinden},\ and\ \citenamefont {Gisin}}]{bib:Nat_421_509}%
  \BibitemOpen
  \bibfield  {author} {\bibinfo {author} {\bibnamefont {Marcikic},
  \bibfnamefont {Ivan}}, \bibinfo {author} {\bibfnamefont {Hugues}\
  \bibnamefont {De~Riedmatten}}, \bibinfo {author} {\bibfnamefont {Wolfgang}\
  \bibnamefont {Tittel}}, \bibinfo {author} {\bibfnamefont {Hugo}\ \bibnamefont
  {Zbinden}}, and\ \bibinfo {author} {\bibfnamefont {Nicolas}\ \bibnamefont
  {Gisin}}} (\bibinfo {year} {2003}),\ \bibfield  {title} {\enquote {\bibinfo
  {title} {Long-distance teleportation of qubits at telecommunication
  wavelengths},}\ }\href {https://doi.org/10.1038/nature01376} {\bibfield
  {journal} {\bibinfo  {journal} {Nature}\ }\textbf {\bibinfo {volume} {421}},\
  \bibinfo {pages} {509}}\BibitemShut {NoStop}%
\bibitem [{\citenamefont {Marcikic}\ \emph {et~al.}(2006)\citenamefont
  {Marcikic}, \citenamefont {Lamas-Linares},\ and\ \citenamefont
  {Kurtsiefer}}]{bib:APL_89_101122}%
  \BibitemOpen
  \bibfield  {author} {\bibinfo {author} {\bibnamefont {Marcikic},
  \bibfnamefont {Ivan}}, \bibinfo {author} {\bibfnamefont {Ant{\'\i}a}\
  \bibnamefont {Lamas-Linares}}, and\ \bibinfo {author} {\bibfnamefont
  {Christian}\ \bibnamefont {Kurtsiefer}}} (\bibinfo {year} {2006}),\ \bibfield
   {title} {\enquote {\bibinfo {title} {Free-space quantum key distribution
  with entangled photons},}\ }\href {https://doi.org/10.1063/1.2348775}
  {\bibfield  {journal} {\bibinfo  {journal} {Applied Physics Letters}\
  }\textbf {\bibinfo {volume} {89}},\ \bibinfo {pages} {101122}},\ \Eprint
  {https://arxiv.org/abs/arXiv:quant-ph/0606072v2} {arXiv:quant-ph/0606072v2}
  \BibitemShut {NoStop}%
\bibitem [{\citenamefont {Marshall}\ \emph {et~al.}(2016)\citenamefont
  {Marshall}, \citenamefont {Jacobsen}, \citenamefont {Sch{\"a}fermeier},
  \citenamefont {Gehring}, \citenamefont {Weedbrook},\ and\ \citenamefont
  {Andersen}}]{marshall2016continuous}%
  \BibitemOpen
  \bibfield  {author} {\bibinfo {author} {\bibnamefont {Marshall},
  \bibfnamefont {Kevin}}, \bibinfo {author} {\bibfnamefont {Christian~S}\
  \bibnamefont {Jacobsen}}, \bibinfo {author} {\bibfnamefont {Clemens}\
  \bibnamefont {Sch{\"a}fermeier}}, \bibinfo {author} {\bibfnamefont {Tobias}\
  \bibnamefont {Gehring}}, \bibinfo {author} {\bibfnamefont {Christian}\
  \bibnamefont {Weedbrook}}, and\ \bibinfo {author} {\bibfnamefont {Ulrik~L}\
  \bibnamefont {Andersen}}} (\bibinfo {year} {2016}),\ \bibfield  {title}
  {\enquote {\bibinfo {title} {Continuous-variable quantum computing on
  encrypted data},}\ }\href@noop {} {\bibfield  {journal} {\bibinfo  {journal}
  {Nature communications}\ }\textbf {\bibinfo {volume} {7}}~(\bibinfo {number}
  {1}),\ \bibinfo {pages} {13795}}\BibitemShut {NoStop}%
\bibitem [{\citenamefont {Marshman}\ \emph {et~al.}(2018)\citenamefont
  {Marshman}, \citenamefont {Lund}, \citenamefont {Rohde},\ and\ \citenamefont
  {Ralph}}]{marshman2018passive}%
  \BibitemOpen
  \bibfield  {author} {\bibinfo {author} {\bibnamefont {Marshman},
  \bibfnamefont {Ryan~J}}, \bibinfo {author} {\bibfnamefont {Austin~P}\
  \bibnamefont {Lund}}, \bibinfo {author} {\bibfnamefont {Peter~P}\
  \bibnamefont {Rohde}}, and\ \bibinfo {author} {\bibfnamefont {Timothy~C}\
  \bibnamefont {Ralph}}} (\bibinfo {year} {2018}),\ \bibfield  {title}
  {\enquote {\bibinfo {title} {Passive quantum error correction of linear
  optics networks through error averaging},}\ }\href@noop {} {\bibfield
  {journal} {\bibinfo  {journal} {Physical Review A}\ }\textbf {\bibinfo
  {volume} {97}}~(\bibinfo {number} {2}),\ \bibinfo {pages}
  {022324}}\BibitemShut {NoStop}%
\bibitem [{\citenamefont {Marshman}\ \emph {et~al.}(2024)\citenamefont
  {Marshman}, \citenamefont {Singh}, \citenamefont {Ralph},\ and\ \citenamefont
  {Lund}}]{marshman2024unitary}%
  \BibitemOpen
  \bibfield  {author} {\bibinfo {author} {\bibnamefont {Marshman},
  \bibfnamefont {Ryan~J}}, \bibinfo {author} {\bibfnamefont {Deepesh}\
  \bibnamefont {Singh}}, \bibinfo {author} {\bibfnamefont {Timothy~C}\
  \bibnamefont {Ralph}}, and\ \bibinfo {author} {\bibfnamefont {Austin~P}\
  \bibnamefont {Lund}}} (\bibinfo {year} {2024}),\ \bibfield  {title} {\enquote
  {\bibinfo {title} {Unitary averaging with fault and loss tolerance},}\
  }\href@noop {} {\bibfield  {journal} {\bibinfo  {journal} {Physical Review
  A}\ }\textbf {\bibinfo {volume} {109}}~(\bibinfo {number} {6}),\ \bibinfo
  {pages} {062436}}\BibitemShut {NoStop}%
\bibitem [{\citenamefont {Marsili}\ \emph
  {et~al.}(2013{\natexlab{a}})\citenamefont {Marsili}, \citenamefont {Verma},
  \citenamefont {Stern}, \citenamefont {Harrington}, \citenamefont {Lita},
  \citenamefont {Gerrits}, \citenamefont {Vayshenker}, \citenamefont {Baek},
  \citenamefont {Shaw}, \citenamefont {Mirin} \emph
  {et~al.}}]{bib:marsili2013}%
  \BibitemOpen
  \bibfield  {author} {\bibinfo {author} {\bibnamefont {Marsili}, \bibfnamefont
  {F}}, \bibinfo {author} {\bibfnamefont {Varun~B}\ \bibnamefont {Verma}},
  \bibinfo {author} {\bibfnamefont {Jeffrey~A}\ \bibnamefont {Stern}}, \bibinfo
  {author} {\bibfnamefont {S}~\bibnamefont {Harrington}}, \bibinfo {author}
  {\bibfnamefont {Adriana~E}\ \bibnamefont {Lita}}, \bibinfo {author}
  {\bibfnamefont {Thomas}\ \bibnamefont {Gerrits}}, \bibinfo {author}
  {\bibfnamefont {Igor}\ \bibnamefont {Vayshenker}}, \bibinfo {author}
  {\bibfnamefont {Burm}\ \bibnamefont {Baek}}, \bibinfo {author} {\bibfnamefont
  {Matthew~D}\ \bibnamefont {Shaw}}, \bibinfo {author} {\bibfnamefont
  {Richard~P}\ \bibnamefont {Mirin}},  \emph {et~al.}} (\bibinfo {year}
  {2013}{\natexlab{a}}),\ \bibfield  {title} {\enquote {\bibinfo {title}
  {Detecting single infrared photons with 93\% system efficiency},}\ }\href
  {https://doi.org/10.1038/nphoton.2013.13} {\bibfield  {journal} {\bibinfo
  {journal} {Nature Photonics}\ }\textbf {\bibinfo {volume} {7}},\ \bibinfo
  {pages} {210}},\ \Eprint {https://arxiv.org/abs/arXiv:1209.5774v1}
  {arXiv:1209.5774v1} \BibitemShut {NoStop}%
\bibitem [{\citenamefont {Marsili}\ \emph
  {et~al.}(2013{\natexlab{b}})\citenamefont {Marsili}, \citenamefont {Verma},
  \citenamefont {Stern}, \citenamefont {Harrington}, \citenamefont {Lita},
  \citenamefont {Gerrits}, \citenamefont {Vayshenker}, \citenamefont {Baek},
  \citenamefont {Shaw}, \citenamefont {Mirin} \emph
  {et~al.}}]{bib:marsili2013detecting}%
  \BibitemOpen
  \bibfield  {author} {\bibinfo {author} {\bibnamefont {Marsili}, \bibfnamefont
  {F}}, \bibinfo {author} {\bibfnamefont {Varun~B}\ \bibnamefont {Verma}},
  \bibinfo {author} {\bibfnamefont {Jeffrey~A}\ \bibnamefont {Stern}}, \bibinfo
  {author} {\bibfnamefont {S}~\bibnamefont {Harrington}}, \bibinfo {author}
  {\bibfnamefont {Adriana~E}\ \bibnamefont {Lita}}, \bibinfo {author}
  {\bibfnamefont {Thomas}\ \bibnamefont {Gerrits}}, \bibinfo {author}
  {\bibfnamefont {Igor}\ \bibnamefont {Vayshenker}}, \bibinfo {author}
  {\bibfnamefont {Burm}\ \bibnamefont {Baek}}, \bibinfo {author} {\bibfnamefont
  {Matthew~D}\ \bibnamefont {Shaw}}, \bibinfo {author} {\bibfnamefont
  {Richard~P}\ \bibnamefont {Mirin}},  \emph {et~al.}} (\bibinfo {year}
  {2013}{\natexlab{b}}),\ \bibfield  {title} {\enquote {\bibinfo {title}
  {Detecting single infrared photons with 93\% system efficiency},}\ }\href
  {https://doi.org/10.1038/nphoton.2013.13} {\bibfield  {journal} {\bibinfo
  {journal} {Nature Photonics}\ }\textbf {\bibinfo {volume} {7}},\ \bibinfo
  {pages} {210}},\ \Eprint {https://arxiv.org/abs/arXiv:1209.5774v1}
  {arXiv:1209.5774v1} \BibitemShut {NoStop}%
\bibitem [{\citenamefont {Marsland}(2011)}]{bib:marsland2011machine}%
  \BibitemOpen
  \bibfield  {author} {\bibinfo {author} {\bibnamefont {Marsland},
  \bibfnamefont {Stephen}}} (\bibinfo {year} {2011}),\ \href
  {https://doi.org/10.1201/9781420067194} {\emph {\bibinfo {title} {Machine
  learning: an algorithmic perspective}}}\ (\bibinfo  {publisher} {Chapman and
  Hall/CRC})\BibitemShut {NoStop}%
\bibitem [{\citenamefont {Martinis}\ \emph {et~al.}(1985)\citenamefont
  {Martinis}, \citenamefont {Devoret},\ and\ \citenamefont
  {Clarke}}]{bib:martinis1985energy}%
  \BibitemOpen
  \bibfield  {author} {\bibinfo {author} {\bibnamefont {Martinis},
  \bibfnamefont {John~M}}, \bibinfo {author} {\bibfnamefont {Michel~H}\
  \bibnamefont {Devoret}}, and\ \bibinfo {author} {\bibfnamefont {John}\
  \bibnamefont {Clarke}}} (\bibinfo {year} {1985}),\ \bibfield  {title}
  {\enquote {\bibinfo {title} {Energy-level quantization in the zero-voltage
  state of a current-biased josephson junction},}\ }\href
  {https://doi.org/10.1103/physrevlett.55.1543} {\bibfield  {journal} {\bibinfo
   {journal} {Physical Review Letters}\ }\textbf {\bibinfo {volume} {55}},\
  \bibinfo {pages} {1543}}\BibitemShut {NoStop}%
\bibitem [{\citenamefont {Martinis}\ \emph {et~al.}(2002)\citenamefont
  {Martinis}, \citenamefont {Nam}, \citenamefont {Aumentado},\ and\
  \citenamefont {Urbina}}]{bib:martinis2002rabi}%
  \BibitemOpen
  \bibfield  {author} {\bibinfo {author} {\bibnamefont {Martinis},
  \bibfnamefont {John~M}}, \bibinfo {author} {\bibfnamefont {S}~\bibnamefont
  {Nam}}, \bibinfo {author} {\bibfnamefont {J}~\bibnamefont {Aumentado}}, and\
  \bibinfo {author} {\bibfnamefont {C}~\bibnamefont {Urbina}}} (\bibinfo {year}
  {2002}),\ \bibfield  {title} {\enquote {\bibinfo {title} {Rabi oscillations
  in a large josephson-junction qubit},}\ }\href
  {https://doi.org/10.1103/physrevlett.89.117901} {\bibfield  {journal}
  {\bibinfo  {journal} {Physical Review Letters}\ }\textbf {\bibinfo {volume}
  {89}},\ \bibinfo {pages} {117901}}\BibitemShut {NoStop}%
\bibitem [{\citenamefont {Matsukevich}\ \emph
  {et~al.}(2005{\natexlab{a}})\citenamefont {Matsukevich}, \citenamefont
  {Chaneli\`ere}, \citenamefont {Bhattacharya}, \citenamefont {Lan},
  \citenamefont {Jenkins}, \citenamefont {Kennedy},\ and\ \citenamefont
  {Kuzmich}}]{bib:Matsukevich05}%
  \BibitemOpen
  \bibfield  {author} {\bibinfo {author} {\bibnamefont {Matsukevich},
  \bibfnamefont {D~N}}, \bibinfo {author} {\bibfnamefont {T.}~\bibnamefont
  {Chaneli\`ere}}, \bibinfo {author} {\bibfnamefont {M.}~\bibnamefont
  {Bhattacharya}}, \bibinfo {author} {\bibfnamefont {S.-Y.}\ \bibnamefont
  {Lan}}, \bibinfo {author} {\bibfnamefont {S.~D.}\ \bibnamefont {Jenkins}},
  \bibinfo {author} {\bibfnamefont {T.~A.~B.}\ \bibnamefont {Kennedy}}, and\
  \bibinfo {author} {\bibfnamefont {A.}~\bibnamefont {Kuzmich}}} (\bibinfo
  {year} {2005}{\natexlab{a}}),\ \bibfield  {title} {\enquote {\bibinfo {title}
  {Entanglement of a photon and a collective atomic excitation},}\ }\href
  {https://doi.org/10.1103/physrevlett.95.040405} {\bibfield  {journal}
  {\bibinfo  {journal} {Physical Review Letters}\ }\textbf {\bibinfo {volume}
  {95}},\ \bibinfo {pages} {040405}},\ \Eprint
  {https://arxiv.org/abs/arXiv:quant-ph/0507161v1} {arXiv:quant-ph/0507161v1}
  \BibitemShut {NoStop}%
\bibitem [{\citenamefont {Matsukevich}\ \emph
  {et~al.}(2005{\natexlab{b}})\citenamefont {Matsukevich}, \citenamefont
  {Chaneli\`ere}, \citenamefont {Jenkins}, \citenamefont {Lan}, \citenamefont
  {Kennedy},\ and\ \citenamefont {Kuzmich}}]{bib:Matsukevich05b}%
  \BibitemOpen
  \bibfield  {author} {\bibinfo {author} {\bibnamefont {Matsukevich},
  \bibfnamefont {D~N}}, \bibinfo {author} {\bibfnamefont {T.}~\bibnamefont
  {Chaneli\`ere}}, \bibinfo {author} {\bibfnamefont {S.~D.}\ \bibnamefont
  {Jenkins}}, \bibinfo {author} {\bibfnamefont {S.-Y.}\ \bibnamefont {Lan}},
  \bibinfo {author} {\bibfnamefont {T.~A.~B.}\ \bibnamefont {Kennedy}}, and\
  \bibinfo {author} {\bibfnamefont {A.}~\bibnamefont {Kuzmich}}} (\bibinfo
  {year} {2005}{\natexlab{b}}),\ \bibfield  {title} {\enquote {\bibinfo {title}
  {Entanglement of remote atomic qubits},}\ }\href
  {https://doi.org/10.1103/physrevlett.96.030405} {\bibfield  {journal}
  {\bibinfo  {journal} {Physical Review Letters}\ }\textbf {\bibinfo {volume}
  {96}},\ \bibinfo {pages} {030405}},\ \Eprint
  {https://arxiv.org/abs/arXiv:quant-ph/0511012v1} {arXiv:quant-ph/0511012v1}
  \BibitemShut {NoStop}%
\bibitem [{\citenamefont {Matsukevich}\ \emph {et~al.}(2008)\citenamefont
  {Matsukevich}, \citenamefont {Maunz}, \citenamefont {Moehring}, \citenamefont
  {Olmschenk},\ and\ \citenamefont {Monroe}}]{bib:PRL_100_150404}%
  \BibitemOpen
  \bibfield  {author} {\bibinfo {author} {\bibnamefont {Matsukevich},
  \bibfnamefont {DN}}, \bibinfo {author} {\bibfnamefont {Peter}\ \bibnamefont
  {Maunz}}, \bibinfo {author} {\bibfnamefont {DL}~\bibnamefont {Moehring}},
  \bibinfo {author} {\bibfnamefont {Steven}\ \bibnamefont {Olmschenk}}, and\
  \bibinfo {author} {\bibfnamefont {Chris}\ \bibnamefont {Monroe}}} (\bibinfo
  {year} {2008}),\ \bibfield  {title} {\enquote {\bibinfo {title} {Bell
  inequality violation with two remote atomic qubits},}\ }\href
  {https://doi.org/10.1103/physrevlett.100.150404} {\bibfield  {journal}
  {\bibinfo  {journal} {Physical Review Letters}\ }\textbf {\bibinfo {volume}
  {100}},\ \bibinfo {pages} {150404}},\ \Eprint
  {https://arxiv.org/abs/arXiv:0801.2184v1} {arXiv:0801.2184v1} \BibitemShut
  {NoStop}%
\bibitem [{\citenamefont {Matsuzaki}\ \emph {et~al.}(2010)\citenamefont
  {Matsuzaki}, \citenamefont {Benjamin},\ and\ \citenamefont
  {Fitzsimons}}]{PhysRevLett.104.050501}%
  \BibitemOpen
  \bibfield  {author} {\bibinfo {author} {\bibnamefont {Matsuzaki},
  \bibfnamefont {Yuichiro}}, \bibinfo {author} {\bibfnamefont {Simon~C.}\
  \bibnamefont {Benjamin}}, and\ \bibinfo {author} {\bibfnamefont {Joseph}\
  \bibnamefont {Fitzsimons}}} (\bibinfo {year} {2010}),\ \bibfield  {title}
  {\enquote {\bibinfo {title} {Probabilistic growth of large entangled states
  with low error accumulation},}\ }\href
  {https://doi.org/10.1103/PhysRevLett.104.050501} {\bibfield  {journal}
  {\bibinfo  {journal} {Phys. Rev. Lett.}\ }\textbf {\bibinfo {volume} {104}},\
  \bibinfo {pages} {050501}}\BibitemShut {NoStop}%
\bibitem [{\citenamefont {Maurer}\ \emph {et~al.}(2012)\citenamefont {Maurer},
  \citenamefont {Kucsko}, \citenamefont {Latta}, \citenamefont {Jiang},
  \citenamefont {Yao}, \citenamefont {Bennett}, \citenamefont {Pastawski},
  \citenamefont {Hunger}, \citenamefont {Chisholm}, \citenamefont {Markham}
  \emph {et~al.}}]{bib:maurer2012room}%
  \BibitemOpen
  \bibfield  {author} {\bibinfo {author} {\bibnamefont {Maurer}, \bibfnamefont
  {Peter~Christian}}, \bibinfo {author} {\bibfnamefont {Georg}\ \bibnamefont
  {Kucsko}}, \bibinfo {author} {\bibfnamefont {Christian}\ \bibnamefont
  {Latta}}, \bibinfo {author} {\bibfnamefont {Liang}\ \bibnamefont {Jiang}},
  \bibinfo {author} {\bibfnamefont {Norman~Ying}\ \bibnamefont {Yao}}, \bibinfo
  {author} {\bibfnamefont {Steven~D}\ \bibnamefont {Bennett}}, \bibinfo
  {author} {\bibfnamefont {Fernando}\ \bibnamefont {Pastawski}}, \bibinfo
  {author} {\bibfnamefont {David}\ \bibnamefont {Hunger}}, \bibinfo {author}
  {\bibfnamefont {Nicholas}\ \bibnamefont {Chisholm}}, \bibinfo {author}
  {\bibfnamefont {Matthew}\ \bibnamefont {Markham}},  \emph {et~al.}} (\bibinfo
  {year} {2012}),\ \bibfield  {title} {\enquote {\bibinfo {title}
  {Room-temperature quantum bit memory exceeding one second},}\ }\href
  {https://doi.org/10.1126/science.1220513} {\bibfield  {journal} {\bibinfo
  {journal} {Science}\ }\textbf {\bibinfo {volume} {336}},\ \bibinfo {pages}
  {1283}}\BibitemShut {NoStop}%
\bibitem [{\citenamefont {McClean}\ \emph
  {et~al.}(2016{\natexlab{a}})\citenamefont {McClean}, \citenamefont {Romero},
  \citenamefont {Babbush},\ and\ \citenamefont
  {Aspuru-Guzik}}]{mcclean2016theory}%
  \BibitemOpen
  \bibfield  {author} {\bibinfo {author} {\bibnamefont {McClean}, \bibfnamefont
  {Jarrod~R}}, \bibinfo {author} {\bibfnamefont {Jonathan}\ \bibnamefont
  {Romero}}, \bibinfo {author} {\bibfnamefont {Ryan}\ \bibnamefont {Babbush}},
  and\ \bibinfo {author} {\bibfnamefont {Al{\'a}n}\ \bibnamefont
  {Aspuru-Guzik}}} (\bibinfo {year} {2016}{\natexlab{a}}),\ \bibfield  {title}
  {\enquote {\bibinfo {title} {The theory of variational hybrid
  quantum-classical algorithms},}\ }\href@noop {} {\bibfield  {journal}
  {\bibinfo  {journal} {New Journal of Physics}\ }\textbf {\bibinfo {volume}
  {18}}~(\bibinfo {number} {2}),\ \bibinfo {pages} {023023}}\BibitemShut
  {NoStop}%
\bibitem [{\citenamefont {McClean}\ \emph
  {et~al.}(2016{\natexlab{b}})\citenamefont {McClean}, \citenamefont {Romero},
  \citenamefont {Babbush},\ and\ \citenamefont
  {Aspuru-Guzik}}]{bib:mcclean2016theory}%
  \BibitemOpen
  \bibfield  {author} {\bibinfo {author} {\bibnamefont {McClean}, \bibfnamefont
  {Jarrod~R}}, \bibinfo {author} {\bibfnamefont {Jonathan}\ \bibnamefont
  {Romero}}, \bibinfo {author} {\bibfnamefont {Ryan}\ \bibnamefont {Babbush}},
  and\ \bibinfo {author} {\bibfnamefont {Al{\'a}n}\ \bibnamefont
  {Aspuru-Guzik}}} (\bibinfo {year} {2016}{\natexlab{b}}),\ \bibfield  {title}
  {\enquote {\bibinfo {title} {The theory of variational hybrid
  quantum-classical algorithms},}\ }\href
  {https://doi.org/10.1088/1367-2630/18/2/023023} {\bibfield  {journal}
  {\bibinfo  {journal} {New Journal of Physics}\ }\textbf {\bibinfo {volume}
  {18}},\ \bibinfo {pages} {023023}}\BibitemShut {NoStop}%
\bibitem [{\citenamefont {McConnell}\ \emph {et~al.}(2015)\citenamefont
  {McConnell}, \citenamefont {Zhang}, \citenamefont {Hu}, \citenamefont
  {{\'C}uk},\ and\ \citenamefont
  {Vuleti{\'c}}}]{bib:mcconnell2015entanglement}%
  \BibitemOpen
  \bibfield  {author} {\bibinfo {author} {\bibnamefont {McConnell},
  \bibfnamefont {Robert}}, \bibinfo {author} {\bibfnamefont {Hao}\ \bibnamefont
  {Zhang}}, \bibinfo {author} {\bibfnamefont {Jiazhong}\ \bibnamefont {Hu}},
  \bibinfo {author} {\bibfnamefont {Senka}\ \bibnamefont {{\'C}uk}}, and\
  \bibinfo {author} {\bibfnamefont {Vladan}\ \bibnamefont {Vuleti{\'c}}}}
  (\bibinfo {year} {2015}),\ \bibfield  {title} {\enquote {\bibinfo {title}
  {Entanglement with negative wigner function of almost 3,000 atoms heralded by
  one photon},}\ }\href {https://doi.org/10.1038/nature14293} {\bibfield
  {journal} {\bibinfo  {journal} {Nature}\ }\textbf {\bibinfo {volume} {519}},\
  \bibinfo {pages} {439}}\BibitemShut {NoStop}%
\bibitem [{\citenamefont {McEliece}(1978)}]{mceliece1978public}%
  \BibitemOpen
  \bibfield  {author} {\bibinfo {author} {\bibnamefont {McEliece},
  \bibfnamefont {Robert~J}}} (\bibinfo {year} {1978}),\ \bibfield  {title}
  {\enquote {\bibinfo {title} {A public-key cryptosystem based on algebraic},}\
  }\href@noop {} {\bibfield  {journal} {\bibinfo  {journal} {Coding Thv}\
  }\textbf {\bibinfo {volume} {4244}},\ \bibinfo {pages}
  {114--116}}\BibitemShut {NoStop}%
\bibitem [{\citenamefont {McMahon}\ \emph {et~al.}(2016)\citenamefont
  {McMahon}, \citenamefont {Marandi}, \citenamefont {Haribara}, \citenamefont
  {Hamerly}, \citenamefont {Langrock}, \citenamefont {Tamate}, \citenamefont
  {Inagaki}, \citenamefont {Takesue}, \citenamefont {Utsunomiya}, \citenamefont
  {Aihara} \emph {et~al.}}]{bib:mcmahon2016fully}%
  \BibitemOpen
  \bibfield  {author} {\bibinfo {author} {\bibnamefont {McMahon}, \bibfnamefont
  {Peter~L}}, \bibinfo {author} {\bibfnamefont {Alireza}\ \bibnamefont
  {Marandi}}, \bibinfo {author} {\bibfnamefont {Yoshitaka}\ \bibnamefont
  {Haribara}}, \bibinfo {author} {\bibfnamefont {Ryan}\ \bibnamefont
  {Hamerly}}, \bibinfo {author} {\bibfnamefont {Carsten}\ \bibnamefont
  {Langrock}}, \bibinfo {author} {\bibfnamefont {Shuhei}\ \bibnamefont
  {Tamate}}, \bibinfo {author} {\bibfnamefont {Takahiro}\ \bibnamefont
  {Inagaki}}, \bibinfo {author} {\bibfnamefont {Hiroki}\ \bibnamefont
  {Takesue}}, \bibinfo {author} {\bibfnamefont {Shoko}\ \bibnamefont
  {Utsunomiya}}, \bibinfo {author} {\bibfnamefont {Kazuyuki}\ \bibnamefont
  {Aihara}},  \emph {et~al.}} (\bibinfo {year} {2016}),\ \bibfield  {title}
  {\enquote {\bibinfo {title} {A fully programmable 100-spin coherent ising
  machine with all-to-all connections},}\ }\href
  {https://doi.org/10.1126/science.aah5178} {\bibfield  {journal} {\bibinfo
  {journal} {Science}\ }\textbf {\bibinfo {volume} {354}},\ \bibinfo {pages}
  {614}}\BibitemShut {NoStop}%
\bibitem [{\citenamefont {McNab}\ \emph {et~al.}(2003)\citenamefont {McNab},
  \citenamefont {Moll},\ and\ \citenamefont {Vlasov}}]{bib:mcnab2003}%
  \BibitemOpen
  \bibfield  {author} {\bibinfo {author} {\bibnamefont {McNab}, \bibfnamefont
  {Sharee~J}}, \bibinfo {author} {\bibfnamefont {Nikolaj}\ \bibnamefont
  {Moll}}, and\ \bibinfo {author} {\bibfnamefont {Yurii~A}\ \bibnamefont
  {Vlasov}}} (\bibinfo {year} {2003}),\ \bibfield  {title} {\enquote {\bibinfo
  {title} {Ultra-low loss photonic integrated circuit with membrane-type
  photonic crystal waveguides},}\ }\href {https://doi.org/10.1364/oe.11.002927}
  {\bibfield  {journal} {\bibinfo  {journal} {Optics Express}\ }\textbf
  {\bibinfo {volume} {11}},\ \bibinfo {pages} {2927}}\BibitemShut {NoStop}%
\bibitem [{\citenamefont {Megidish}\ \emph {et~al.}(2013)\citenamefont
  {Megidish}, \citenamefont {Halevy}, \citenamefont {Shacham}, \citenamefont
  {Dvir}, \citenamefont {Dovrat},\ and\ \citenamefont
  {Eisenberg}}]{bib:PRL_110_210403}%
  \BibitemOpen
  \bibfield  {author} {\bibinfo {author} {\bibnamefont {Megidish},
  \bibfnamefont {E}}, \bibinfo {author} {\bibfnamefont {A}~\bibnamefont
  {Halevy}}, \bibinfo {author} {\bibfnamefont {T}~\bibnamefont {Shacham}},
  \bibinfo {author} {\bibfnamefont {T}~\bibnamefont {Dvir}}, \bibinfo {author}
  {\bibfnamefont {L}~\bibnamefont {Dovrat}}, and\ \bibinfo {author}
  {\bibfnamefont {HS}~\bibnamefont {Eisenberg}}} (\bibinfo {year} {2013}),\
  \bibfield  {title} {\enquote {\bibinfo {title} {Entanglement swapping between
  photons that have never coexisted},}\ }\href
  {https://doi.org/10.1103/physrevlett.110.210403} {\bibfield  {journal}
  {\bibinfo  {journal} {Physical Review Letters}\ }\textbf {\bibinfo {volume}
  {110}},\ \bibinfo {pages} {210403}}\BibitemShut {NoStop}%
\bibitem [{\citenamefont {Melnikov}\ \emph {et~al.}(2018)\citenamefont
  {Melnikov}, \citenamefont {Nautrup}, \citenamefont {Krenn}, \citenamefont
  {Dunjko}, \citenamefont {Tiersch}, \citenamefont {Zeilinger},\ and\
  \citenamefont {Briegel}}]{bib:alexey}%
  \BibitemOpen
  \bibfield  {author} {\bibinfo {author} {\bibnamefont {Melnikov},
  \bibfnamefont {Alexey~A}}, \bibinfo {author} {\bibfnamefont
  {Hendrik~Poulsen}\ \bibnamefont {Nautrup}}, \bibinfo {author} {\bibfnamefont
  {Mario}\ \bibnamefont {Krenn}}, \bibinfo {author} {\bibfnamefont {Vedran}\
  \bibnamefont {Dunjko}}, \bibinfo {author} {\bibfnamefont {Markus}\
  \bibnamefont {Tiersch}}, \bibinfo {author} {\bibfnamefont {Anton}\
  \bibnamefont {Zeilinger}}, and\ \bibinfo {author} {\bibfnamefont {Hans~J.}\
  \bibnamefont {Briegel}}} (\bibinfo {year} {2018}),\ \bibfield  {title}
  {\enquote {\bibinfo {title} {Active learning machine learns to create new
  quantum experiments},}\ }\href {https://doi.org/10.1073/pnas.1714936115}
  {\bibfield  {journal} {\bibinfo  {journal} {Proceedings of the National
  Academy of Sciences}\ }\textbf {\bibinfo {volume} {115}},\ \bibinfo {pages}
  {1221}},\ \Eprint {https://arxiv.org/abs/arXiv:1706.00868v3}
  {arXiv:1706.00868v3} \BibitemShut {NoStop}%
\bibitem [{\citenamefont {Menicucci}(2014)}]{menicucci2014fault}%
  \BibitemOpen
  \bibfield  {author} {\bibinfo {author} {\bibnamefont {Menicucci},
  \bibfnamefont {Nicolas~C}}} (\bibinfo {year} {2014}),\ \bibfield  {title}
  {\enquote {\bibinfo {title} {Fault-tolerant measurement-based quantum
  computing with continuous-variable cluster states},}\ }\href
  {https://doi.org/10.1103/PhysRevLett.112.120504} {\bibfield  {journal}
  {\bibinfo  {journal} {Phys. Rev. Lett.}\ }\textbf {\bibinfo {volume} {112}},\
  \bibinfo {pages} {120504}}\BibitemShut {NoStop}%
\bibitem [{\citenamefont {Menicucci}\ \emph {et~al.}(2018)\citenamefont
  {Menicucci}, \citenamefont {Baragiola}, \citenamefont {Demarie},\ and\
  \citenamefont {Brennen}}]{bib:MenicucciExpQAB}%
  \BibitemOpen
  \bibfield  {author} {\bibinfo {author} {\bibnamefont {Menicucci},
  \bibfnamefont {Nicolas~C}}, \bibinfo {author} {\bibfnamefont {Ben~Q.}\
  \bibnamefont {Baragiola}}, \bibinfo {author} {\bibfnamefont {Tommaso~F.}\
  \bibnamefont {Demarie}}, and\ \bibinfo {author} {\bibfnamefont {Gavin~K.}\
  \bibnamefont {Brennen}}} (\bibinfo {year} {2018}),\ \bibfield  {title}
  {\enquote {\bibinfo {title} {Anonymous broadcasting of classical information
  with a continuous-variable topological quantum code},}\ }\href
  {https://doi.org/10.1103/physreva.97.032345} {\bibfield  {journal} {\bibinfo
  {journal} {Physical Review A}\ }\textbf {\bibinfo {volume} {97}},\ \bibinfo
  {pages} {032345}},\ \Eprint {https://arxiv.org/abs/arXiv:1503.00717v4}
  {arXiv:1503.00717v4} \BibitemShut {NoStop}%
\bibitem [{\citenamefont {Menicucci}\ \emph {et~al.}(2006)\citenamefont
  {Menicucci}, \citenamefont {van Loock}, \citenamefont {Gu}, \citenamefont
  {Weedbrook}, \citenamefont {Ralph},\ and\ \citenamefont
  {Nielsen}}]{menicucci2006universal}%
  \BibitemOpen
  \bibfield  {author} {\bibinfo {author} {\bibnamefont {Menicucci},
  \bibfnamefont {Nicolas~C}}, \bibinfo {author} {\bibfnamefont {Peter}\
  \bibnamefont {van Loock}}, \bibinfo {author} {\bibfnamefont {Mile}\
  \bibnamefont {Gu}}, \bibinfo {author} {\bibfnamefont {Christian}\
  \bibnamefont {Weedbrook}}, \bibinfo {author} {\bibfnamefont {Timothy~C.}\
  \bibnamefont {Ralph}}, and\ \bibinfo {author} {\bibfnamefont {Michael~A.}\
  \bibnamefont {Nielsen}}} (\bibinfo {year} {2006}),\ \bibfield  {title}
  {\enquote {\bibinfo {title} {Universal quantum computation with
  continuous-variable cluster states},}\ }\href
  {https://doi.org/10.1103/PhysRevLett.97.110501} {\bibfield  {journal}
  {\bibinfo  {journal} {Phys. Rev. Lett.}\ }\textbf {\bibinfo {volume} {97}},\
  \bibinfo {pages} {110501}}\BibitemShut {NoStop}%
\bibitem [{\citenamefont {Mensen}\ \emph {et~al.}(2021)\citenamefont {Mensen},
  \citenamefont {Baragiola},\ and\ \citenamefont
  {Menicucci}}]{mensen2021phase}%
  \BibitemOpen
  \bibfield  {author} {\bibinfo {author} {\bibnamefont {Mensen}, \bibfnamefont
  {Lucas~J}}, \bibinfo {author} {\bibfnamefont {Ben~Q.}\ \bibnamefont
  {Baragiola}}, and\ \bibinfo {author} {\bibfnamefont {Nicolas~C.}\
  \bibnamefont {Menicucci}}} (\bibinfo {year} {2021}),\ \bibfield  {title}
  {\enquote {\bibinfo {title} {Phase-space methods for representing,
  manipulating, and correcting gottesman-kitaev-preskill qubits},}\ }\href
  {https://doi.org/10.1103/PhysRevA.104.022408} {\bibfield  {journal} {\bibinfo
   {journal} {Phys. Rev. A}\ }\textbf {\bibinfo {volume} {104}},\ \bibinfo
  {pages} {022408}}\BibitemShut {NoStop}%
\bibitem [{\citenamefont {Merkle}(1978)}]{SD-Merkle:1978:SCO:359460.359473}%
  \BibitemOpen
  \bibfield  {author} {\bibinfo {author} {\bibnamefont {Merkle}, \bibfnamefont
  {Ralph~C}}} (\bibinfo {year} {1978}),\ \bibfield  {title} {\enquote {\bibinfo
  {title} {Secure communications over insecure channels},}\ }\href
  {https://doi.org/10.1145/359460.359473} {\bibfield  {journal} {\bibinfo
  {journal} {Communications ACM}\ }\textbf {\bibinfo {volume} {21}},\ \bibinfo
  {pages} {294}}\BibitemShut {NoStop}%
\bibitem [{\citenamefont {van Meter}\ \emph {et~al.}(2007)\citenamefont {van
  Meter}, \citenamefont {Lougovski}, \citenamefont {Uskov}, \citenamefont
  {Kieling}, \citenamefont {Eisert},\ and\ \citenamefont
  {Dowling}}]{bib:PhysRevA.76.063808}%
  \BibitemOpen
  \bibfield  {author} {\bibinfo {author} {\bibnamefont {van Meter},
  \bibfnamefont {N~M}}, \bibinfo {author} {\bibfnamefont {P.}~\bibnamefont
  {Lougovski}}, \bibinfo {author} {\bibfnamefont {D.~B.}\ \bibnamefont
  {Uskov}}, \bibinfo {author} {\bibfnamefont {K.}~\bibnamefont {Kieling}},
  \bibinfo {author} {\bibfnamefont {J.}~\bibnamefont {Eisert}}, and\ \bibinfo
  {author} {\bibfnamefont {Jonathan~P.}\ \bibnamefont {Dowling}}} (\bibinfo
  {year} {2007}),\ \bibfield  {title} {\enquote {\bibinfo {title} {General
  linear-optical quantum state generation scheme: Applications to maximally
  path-entangled states},}\ }\href {https://doi.org/10.1103/physreva.76.063808}
  {\bibfield  {journal} {\bibinfo  {journal} {Physical Review A}\ }\textbf
  {\bibinfo {volume} {76}},\ \bibinfo {pages} {063808}},\ \Eprint
  {https://arxiv.org/abs/arXiv:quant-ph/0612154v2} {arXiv:quant-ph/0612154v2}
  \BibitemShut {NoStop}%
\bibitem [{\citenamefont {Metodiev}\ \emph {et~al.}(2004)\citenamefont
  {Metodiev}, \citenamefont {Cross}, \citenamefont {Thaker}, \citenamefont
  {Brown}, \citenamefont {Copsey}, \citenamefont {Chong},\ and\ \citenamefont
  {Chuang}}]{bib:MCTBCCC04}%
  \BibitemOpen
  \bibfield  {author} {\bibinfo {author} {\bibnamefont {Metodiev},
  \bibfnamefont {T}}, \bibinfo {author} {\bibfnamefont {A.}~\bibnamefont
  {Cross}}, \bibinfo {author} {\bibfnamefont {D.}~\bibnamefont {Thaker}},
  \bibinfo {author} {\bibfnamefont {K.}~\bibnamefont {Brown}}, \bibinfo
  {author} {\bibfnamefont {D.}~\bibnamefont {Copsey}}, \bibinfo {author}
  {\bibfnamefont {F.T.}\ \bibnamefont {Chong}}, and\ \bibinfo {author}
  {\bibfnamefont {I.L.}\ \bibnamefont {Chuang}}} (\bibinfo {year} {2004}),\
  \bibfield  {title} {\enquote {\bibinfo {title} {{Preliminary Results on
  Simulating a Scalable Fault-Tolerant Ion Trap system for quantum
  computation}},}\ }in\ \href@noop {} {\emph {\bibinfo {booktitle} {3rd
  Workshop on Non-Silicon Computing (NSC-3),
  online:www.csif.cs.ucdavis.edu/~metodiev/papers/NSC3-setso.pdf}}}\BibitemShut
  {NoStop}%
\bibitem [{\citenamefont {Milgram}(1967)}]{SD-Milgram1967}%
  \BibitemOpen
  \bibfield  {author} {\bibinfo {author} {\bibnamefont {Milgram}, \bibfnamefont
  {S}}} (\bibinfo {year} {1967}),\ \bibfield  {title} {\enquote {\bibinfo
  {title} {The small world problem},}\ }\href
  {https://doi.org/10.1037/e400002009-005} {\bibfield  {journal} {\bibinfo
  {journal} {Psychology Today}\ }\textbf {\bibinfo {volume} {2}},\ \bibinfo
  {pages} {60}}\BibitemShut {NoStop}%
\bibitem [{\citenamefont {Mitarai}\ \emph {et~al.}(2018)\citenamefont
  {Mitarai}, \citenamefont {Negoro}, \citenamefont {Kitagawa},\ and\
  \citenamefont {Fujii}}]{bib:mitarai2018quantum}%
  \BibitemOpen
  \bibfield  {author} {\bibinfo {author} {\bibnamefont {Mitarai}, \bibfnamefont
  {Kosuke}}, \bibinfo {author} {\bibfnamefont {Makoto}\ \bibnamefont {Negoro}},
  \bibinfo {author} {\bibfnamefont {Masahiro}\ \bibnamefont {Kitagawa}}, and\
  \bibinfo {author} {\bibfnamefont {Keisuke}\ \bibnamefont {Fujii}}} (\bibinfo
  {year} {2018}),\ \bibfield  {title} {\enquote {\bibinfo {title} {Quantum
  circuit learning},}\ }\href {https://doi.org/10.1103/physreva.98.032309}
  {\bibfield  {journal} {\bibinfo  {journal} {Physical Review A}\ }\textbf
  {\bibinfo {volume} {98}},\ \bibinfo {pages} {032309}},\ \Eprint
  {https://arxiv.org/abs/arXiv:1803.00745v2} {arXiv:1803.00745v2} \BibitemShut
  {NoStop}%
\bibitem [{\citenamefont {Mizutani}\ \emph {et~al.}(2015)\citenamefont
  {Mizutani}, \citenamefont {Curty}, \citenamefont {Lim}, \citenamefont
  {Imoto},\ and\ \citenamefont {Tamaki}}]{bib:mizutani2015finite}%
  \BibitemOpen
  \bibfield  {author} {\bibinfo {author} {\bibnamefont {Mizutani},
  \bibfnamefont {Akihiro}}, \bibinfo {author} {\bibfnamefont {Marcos}\
  \bibnamefont {Curty}}, \bibinfo {author} {\bibfnamefont {Charles Ci~Wen}\
  \bibnamefont {Lim}}, \bibinfo {author} {\bibfnamefont {Nobuyuki}\
  \bibnamefont {Imoto}}, and\ \bibinfo {author} {\bibfnamefont {Kiyoshi}\
  \bibnamefont {Tamaki}}} (\bibinfo {year} {2015}),\ \bibfield  {title}
  {\enquote {\bibinfo {title} {Finite-key security analysis of quantum key
  distribution with imperfect light sources},}\ }\href
  {https://doi.org/10.1088/1367-2630/17/9/093011} {\bibfield  {journal}
  {\bibinfo  {journal} {New Journal of Physics}\ }\textbf {\bibinfo {volume}
  {17}},\ \bibinfo {pages} {093011}}\BibitemShut {NoStop}%
\bibitem [{\citenamefont {Moehring}\ \emph {et~al.}(2007)\citenamefont
  {Moehring}, \citenamefont {Maunz}, \citenamefont {Olmschenk}, \citenamefont
  {Younge}, \citenamefont {Matsukevich}, \citenamefont {Duan},\ and\
  \citenamefont {Monroe}}]{bib:Nature_449_68}%
  \BibitemOpen
  \bibfield  {author} {\bibinfo {author} {\bibnamefont {Moehring},
  \bibfnamefont {DL}}, \bibinfo {author} {\bibfnamefont {P}~\bibnamefont
  {Maunz}}, \bibinfo {author} {\bibfnamefont {S}~\bibnamefont {Olmschenk}},
  \bibinfo {author} {\bibfnamefont {KC}~\bibnamefont {Younge}}, \bibinfo
  {author} {\bibfnamefont {DN}~\bibnamefont {Matsukevich}}, \bibinfo {author}
  {\bibfnamefont {L-M}\ \bibnamefont {Duan}}, and\ \bibinfo {author}
  {\bibfnamefont {C}~\bibnamefont {Monroe}}} (\bibinfo {year} {2007}),\
  \bibfield  {title} {\enquote {\bibinfo {title} {Entanglement of single-atom
  quantum bits at a distance},}\ }\href {https://doi.org/10.1038/nature06118}
  {\bibfield  {journal} {\bibinfo  {journal} {Nature}\ }\textbf {\bibinfo
  {volume} {449}},\ \bibinfo {pages} {68}}\BibitemShut {NoStop}%
\bibitem [{\citenamefont {Mohan}\ \emph {et~al.}(2010)\citenamefont {Mohan},
  \citenamefont {Felici}, \citenamefont {Gallo}, \citenamefont {Dwir},
  \citenamefont {Rudra}, \citenamefont {Faist},\ and\ \citenamefont
  {Kapon}}]{bib:mohan2010polarization}%
  \BibitemOpen
  \bibfield  {author} {\bibinfo {author} {\bibnamefont {Mohan}, \bibfnamefont
  {Arun}}, \bibinfo {author} {\bibfnamefont {Marco}\ \bibnamefont {Felici}},
  \bibinfo {author} {\bibfnamefont {Pascal}\ \bibnamefont {Gallo}}, \bibinfo
  {author} {\bibfnamefont {Benjamin}\ \bibnamefont {Dwir}}, \bibinfo {author}
  {\bibfnamefont {Alok}\ \bibnamefont {Rudra}}, \bibinfo {author}
  {\bibfnamefont {J{\'e}r{\^o}me}\ \bibnamefont {Faist}}, and\ \bibinfo
  {author} {\bibfnamefont {Elyahou}\ \bibnamefont {Kapon}}} (\bibinfo {year}
  {2010}),\ \bibfield  {title} {\enquote {\bibinfo {title}
  {Polarization-entangled photons produced with high-symmetry site-controlled
  quantum dots},}\ }\href {https://doi.org/10.1038/nphoton.2010.2} {\bibfield
  {journal} {\bibinfo  {journal} {Nature Photonics}\ }\textbf {\bibinfo
  {volume} {4}},\ \bibinfo {pages} {302}}\BibitemShut {NoStop}%
\bibitem [{\citenamefont {Moll}\ \emph {et~al.}(2018)\citenamefont {Moll},
  \citenamefont {Barkoutsos}, \citenamefont {Bishop}, \citenamefont {Chow},
  \citenamefont {Cross}, \citenamefont {Egger}, \citenamefont {Filipp},
  \citenamefont {Fuhrer}, \citenamefont {Gambetta}, \citenamefont {Ganzhorn}
  \emph {et~al.}}]{bib:moll2018quantum}%
  \BibitemOpen
  \bibfield  {author} {\bibinfo {author} {\bibnamefont {Moll}, \bibfnamefont
  {Nikolaj}}, \bibinfo {author} {\bibfnamefont {Panagiotis}\ \bibnamefont
  {Barkoutsos}}, \bibinfo {author} {\bibfnamefont {Lev~S.}\ \bibnamefont
  {Bishop}}, \bibinfo {author} {\bibfnamefont {Jerry~M.}\ \bibnamefont {Chow}},
  \bibinfo {author} {\bibfnamefont {Andrew}\ \bibnamefont {Cross}}, \bibinfo
  {author} {\bibfnamefont {Daniel~J.}\ \bibnamefont {Egger}}, \bibinfo {author}
  {\bibfnamefont {Stefan}\ \bibnamefont {Filipp}}, \bibinfo {author}
  {\bibfnamefont {Andreas}\ \bibnamefont {Fuhrer}}, \bibinfo {author}
  {\bibfnamefont {Jay~M.}\ \bibnamefont {Gambetta}}, \bibinfo {author}
  {\bibfnamefont {Marc}\ \bibnamefont {Ganzhorn}},  \emph {et~al.}} (\bibinfo
  {year} {2018}),\ \bibfield  {title} {\enquote {\bibinfo {title} {Quantum
  optimization using variational algorithms on near-term quantum devices},}\
  }\href {https://doi.org/10.1088/2058-9565/aab822} {\bibfield  {journal}
  {\bibinfo  {journal} {Quantum Science and Technology}\ }\textbf {\bibinfo
  {volume} {3}},\ \bibinfo {pages} {030503}},\ \Eprint
  {https://arxiv.org/abs/arXiv:1710.01022v2} {arXiv:1710.01022v2} \BibitemShut
  {NoStop}%
\bibitem [{\citenamefont {Monz}\ \emph {et~al.}(2011)\citenamefont {Monz},
  \citenamefont {Schindler}, \citenamefont {Barreiro}, \citenamefont {Chwalla},
  \citenamefont {Nigg}, \citenamefont {Coish}, \citenamefont {Harlander},
  \citenamefont {H{\"a}nsel}, \citenamefont {Hennrich},\ and\ \citenamefont
  {Blatt}}]{bib:monz2011}%
  \BibitemOpen
  \bibfield  {author} {\bibinfo {author} {\bibnamefont {Monz}, \bibfnamefont
  {Thomas}}, \bibinfo {author} {\bibfnamefont {Philipp}\ \bibnamefont
  {Schindler}}, \bibinfo {author} {\bibfnamefont {Julio~T}\ \bibnamefont
  {Barreiro}}, \bibinfo {author} {\bibfnamefont {Michael}\ \bibnamefont
  {Chwalla}}, \bibinfo {author} {\bibfnamefont {Daniel}\ \bibnamefont {Nigg}},
  \bibinfo {author} {\bibfnamefont {William~A}\ \bibnamefont {Coish}}, \bibinfo
  {author} {\bibfnamefont {Maximilian}\ \bibnamefont {Harlander}}, \bibinfo
  {author} {\bibfnamefont {Wolfgang}\ \bibnamefont {H{\"a}nsel}}, \bibinfo
  {author} {\bibfnamefont {Markus}\ \bibnamefont {Hennrich}}, and\ \bibinfo
  {author} {\bibfnamefont {Rainer}\ \bibnamefont {Blatt}}} (\bibinfo {year}
  {2011}),\ \bibfield  {title} {\enquote {\bibinfo {title} {14-qubit
  entanglement: Creation and coherence},}\ }\href
  {https://doi.org/10.1103/physrevlett.106.130506} {\bibfield  {journal}
  {\bibinfo  {journal} {Physical Review Letters}\ }\textbf {\bibinfo {volume}
  {106}},\ \bibinfo {pages} {130506}},\ \Eprint
  {https://arxiv.org/abs/arXiv:1009.6126v2} {arXiv:1009.6126v2} \BibitemShut
  {NoStop}%
\bibitem [{\citenamefont {Moore}(1959)}]{moore1959shortest}%
  \BibitemOpen
  \bibfield  {author} {\bibinfo {author} {\bibnamefont {Moore}, \bibfnamefont
  {Edward~F}}} (\bibinfo {year} {1959}),\ \bibfield  {title} {\enquote
  {\bibinfo {title} {The shortest path through a maze},}\ }in\ \href@noop {}
  {\emph {\bibinfo {booktitle} {Proc. of the International Symposium on the
  Theory of Switching}}}\ (\bibinfo {organization} {Harvard University Press})\
  pp.\ \bibinfo {pages} {285--292}\BibitemShut {NoStop}%
\bibitem [{\citenamefont {Mor}\ and\ \citenamefont
  {Yoran}(2006)}]{bib:MorYoran06}%
  \BibitemOpen
  \bibfield  {author} {\bibinfo {author} {\bibnamefont {Mor}, \bibfnamefont
  {Tal}}, and\ \bibinfo {author} {\bibfnamefont {Nadav}\ \bibnamefont {Yoran}}}
  (\bibinfo {year} {2006}),\ \bibfield  {title} {\enquote {\bibinfo {title}
  {Methods for scalable optical quantum computation},}\ }\href
  {https://doi.org/10.1103/physrevlett.97.090501} {\bibfield  {journal}
  {\bibinfo  {journal} {Physical Review Letters}\ }\textbf {\bibinfo {volume}
  {97}},\ \bibinfo {pages} {090501}},\ \Eprint
  {https://arxiv.org/abs/arXiv:quant-ph/0603118v1} {arXiv:quant-ph/0603118v1}
  \BibitemShut {NoStop}%
\bibitem [{\citenamefont {Morimae}\ \emph {et~al.}(2015)\citenamefont
  {Morimae}, \citenamefont {Dunjko},\ and\ \citenamefont
  {Kashefi}}]{bib:Morimae3486}%
  \BibitemOpen
  \bibfield  {author} {\bibinfo {author} {\bibnamefont {Morimae}, \bibfnamefont
  {Tomoyuki}}, \bibinfo {author} {\bibfnamefont {Vedran}\ \bibnamefont
  {Dunjko}}, and\ \bibinfo {author} {\bibfnamefont {Elham}\ \bibnamefont
  {Kashefi}}} (\bibinfo {year} {2015}),\ \bibfield  {title} {\enquote {\bibinfo
  {title} {Ground state blind quantum computation on aklt state},}\ }\href@noop
  {} {\bibfield  {journal} {\bibinfo  {journal} {Quantum Information and
  Computation}\ }\textbf {\bibinfo {volume} {15}},\ \bibinfo {pages} {0200}},\
  \Eprint {https://arxiv.org/abs/arXiv:1009.3486v2} {arXiv:1009.3486v2}
  \BibitemShut {NoStop}%
\bibitem [{\citenamefont {Morimae}\ and\ \citenamefont
  {Fujii}(2013)}]{bib:Morimae5460}%
  \BibitemOpen
  \bibfield  {author} {\bibinfo {author} {\bibnamefont {Morimae}, \bibfnamefont
  {Tomoyuki}}, and\ \bibinfo {author} {\bibfnamefont {Keisuke}\ \bibnamefont
  {Fujii}}} (\bibinfo {year} {2013}),\ \bibfield  {title} {\enquote {\bibinfo
  {title} {Blind topological measurement-based quantum computation},}\
  }\href@noop {} {\bibfield  {journal} {\bibinfo  {journal} {Physical Review
  A}\ }\textbf {\bibinfo {volume} {87}},\ \bibinfo {pages}
  {050301(R)}}\BibitemShut {NoStop}%
\bibitem [{\citenamefont {Motes}\ \emph {et~al.}(2017)\citenamefont {Motes},
  \citenamefont {Baragiola}, \citenamefont {Gilchrist},\ and\ \citenamefont
  {Menicucci}}]{motes2017encoding}%
  \BibitemOpen
  \bibfield  {author} {\bibinfo {author} {\bibnamefont {Motes}, \bibfnamefont
  {Keith~R}}, \bibinfo {author} {\bibfnamefont {Ben~Q.}\ \bibnamefont
  {Baragiola}}, \bibinfo {author} {\bibfnamefont {Alexei}\ \bibnamefont
  {Gilchrist}}, and\ \bibinfo {author} {\bibfnamefont {Nicolas~C.}\
  \bibnamefont {Menicucci}}} (\bibinfo {year} {2017}),\ \bibfield  {title}
  {\enquote {\bibinfo {title} {Encoding qubits into oscillators with atomic
  ensembles and squeezed light},}\ }\href
  {https://doi.org/10.1103/PhysRevA.95.053819} {\bibfield  {journal} {\bibinfo
  {journal} {Phys. Rev. A}\ }\textbf {\bibinfo {volume} {95}},\ \bibinfo
  {pages} {053819}}\BibitemShut {NoStop}%
\bibitem [{\citenamefont {Motes}\ \emph {et~al.}(2015)\citenamefont {Motes},
  \citenamefont {Dowling}, \citenamefont {Gilchrist},\ and\ \citenamefont
  {Rohde}}]{bib:RohdeLoop15}%
  \BibitemOpen
  \bibfield  {author} {\bibinfo {author} {\bibnamefont {Motes}, \bibfnamefont
  {Keith~R}}, \bibinfo {author} {\bibfnamefont {Jonathan~P.}\ \bibnamefont
  {Dowling}}, \bibinfo {author} {\bibfnamefont {Alexei}\ \bibnamefont
  {Gilchrist}}, and\ \bibinfo {author} {\bibfnamefont {Peter~P.}\ \bibnamefont
  {Rohde}}} (\bibinfo {year} {2015}),\ \bibfield  {title} {\enquote {\bibinfo
  {title} {Implementing scalable boson sampling with time-bin encoding:
  Analysis of loss, mode mismatch, and time jitter},}\ }\href@noop {}
  {\bibfield  {journal} {\bibinfo  {journal} {Physical Review A}\ }\textbf
  {\bibinfo {volume} {92}},\ \bibinfo {pages} {052319}},\ \Eprint
  {https://arxiv.org/abs/arXiv:1507.07185v1} {arXiv:1507.07185v1} \BibitemShut
  {NoStop}%
\bibitem [{\citenamefont {Motes}\ \emph {et~al.}(2013)\citenamefont {Motes},
  \citenamefont {Dowling},\ and\ \citenamefont {Rohde}}]{bib:RohdeSPDC13}%
  \BibitemOpen
  \bibfield  {author} {\bibinfo {author} {\bibnamefont {Motes}, \bibfnamefont
  {Keith~R}}, \bibinfo {author} {\bibfnamefont {Jonathan~P.}\ \bibnamefont
  {Dowling}}, and\ \bibinfo {author} {\bibfnamefont {Peter~P.}\ \bibnamefont
  {Rohde}}} (\bibinfo {year} {2013}),\ \bibfield  {title} {\enquote {\bibinfo
  {title} {Spontaneous parametric down-conversion photon sources are scalable
  in the asymptotic limit for boson-sampling},}\ }\href
  {https://doi.org/10.1103/physreva.88.063822} {\bibfield  {journal} {\bibinfo
  {journal} {Physical Review A}\ }\textbf {\bibinfo {volume} {88}},\ \bibinfo
  {pages} {063822}},\ \Eprint {https://arxiv.org/abs/arXiv:1307.8238v4}
  {arXiv:1307.8238v4} \BibitemShut {NoStop}%
\bibitem [{\citenamefont {Mukai}\ \emph {et~al.}(2020)\citenamefont {Mukai},
  \citenamefont {Sakata}, \citenamefont {Devitt}, \citenamefont {Wang},
  \citenamefont {Zhou}, \citenamefont {Nakajima},\ and\ \citenamefont
  {Tsai}}]{bib:Mukai_2020}%
  \BibitemOpen
  \bibfield  {author} {\bibinfo {author} {\bibnamefont {Mukai}, \bibfnamefont
  {Hiroto}}, \bibinfo {author} {\bibfnamefont {Keiichi}\ \bibnamefont
  {Sakata}}, \bibinfo {author} {\bibfnamefont {Simon~J}\ \bibnamefont
  {Devitt}}, \bibinfo {author} {\bibfnamefont {Rui}\ \bibnamefont {Wang}},
  \bibinfo {author} {\bibfnamefont {Yu}~\bibnamefont {Zhou}}, \bibinfo {author}
  {\bibfnamefont {Yukito}\ \bibnamefont {Nakajima}}, and\ \bibinfo {author}
  {\bibfnamefont {Jaw-Shen}\ \bibnamefont {Tsai}}} (\bibinfo {year} {2020}),\
  \bibfield  {title} {\enquote {\bibinfo {title} {Pseudo-2d superconducting
  quantum computing circuit for the surface code: proposal and preliminary
  tests},}\ }\href {https://doi.org/10.1088/1367-2630/ab7d7d} {\bibfield
  {journal} {\bibinfo  {journal} {New Journal of Physics}\ }\textbf {\bibinfo
  {volume} {22}}~(\bibinfo {number} {4}),\ \bibinfo {pages}
  {043013}}\BibitemShut {NoStop}%
\bibitem [{\citenamefont {Muller}\ \emph {et~al.}(1993)\citenamefont {Muller},
  \citenamefont {Breguet},\ and\ \citenamefont {Gisin}}]{bib:EL_23_383}%
  \BibitemOpen
  \bibfield  {author} {\bibinfo {author} {\bibnamefont {Muller}, \bibfnamefont
  {Antoine}}, \bibinfo {author} {\bibfnamefont {J}~\bibnamefont {Breguet}},
  and\ \bibinfo {author} {\bibfnamefont {N}~\bibnamefont {Gisin}}} (\bibinfo
  {year} {1993}),\ \bibfield  {title} {\enquote {\bibinfo {title} {Experimental
  demonstration of quantum cryptography using polarized photons in optical
  fibre over more than 1 km},}\ }\href
  {https://doi.org/10.1209/0295-5075/23/6/001} {\bibfield  {journal} {\bibinfo
  {journal} {Europhysics Letters}\ }\textbf {\bibinfo {volume} {23}},\ \bibinfo
  {pages} {383}}\BibitemShut {NoStop}%
\bibitem [{\citenamefont {M{\"u}ller}\ \emph {et~al.}(2014)\citenamefont
  {M{\"u}ller}, \citenamefont {Bounouar}, \citenamefont {J{\"o}ns},
  \citenamefont {Gl{\"a}ssl},\ and\ \citenamefont
  {Michler}}]{bib:muller2014demand}%
  \BibitemOpen
  \bibfield  {author} {\bibinfo {author} {\bibnamefont {M{\"u}ller},
  \bibfnamefont {Markus}}, \bibinfo {author} {\bibfnamefont {Samir}\
  \bibnamefont {Bounouar}}, \bibinfo {author} {\bibfnamefont {Klaus~D}\
  \bibnamefont {J{\"o}ns}}, \bibinfo {author} {\bibfnamefont {M}~\bibnamefont
  {Gl{\"a}ssl}}, and\ \bibinfo {author} {\bibfnamefont {P}~\bibnamefont
  {Michler}}} (\bibinfo {year} {2014}),\ \bibfield  {title} {\enquote {\bibinfo
  {title} {On-demand generation of indistinguishable polarization-entangled
  photon pairs},}\ }\href {https://doi.org/10.1038/nphoton.2013.377} {\bibfield
   {journal} {\bibinfo  {journal} {Nature Photonics}\ }\textbf {\bibinfo
  {volume} {8}},\ \bibinfo {pages} {224}},\ \Eprint
  {https://arxiv.org/abs/arXiv:1308.4257v2} {arXiv:1308.4257v2} \BibitemShut
  {NoStop}%
\bibitem [{\citenamefont {M{\"u}ller-Quade}\ and\ \citenamefont
  {Renner}(2009)}]{muller2009composability}%
  \BibitemOpen
  \bibfield  {author} {\bibinfo {author} {\bibnamefont {M{\"u}ller-Quade},
  \bibfnamefont {J{\"o}rn}}, and\ \bibinfo {author} {\bibfnamefont {Renato}\
  \bibnamefont {Renner}}} (\bibinfo {year} {2009}),\ \bibfield  {title}
  {\enquote {\bibinfo {title} {Composability in quantum cryptography},}\
  }\href@noop {} {\bibfield  {journal} {\bibinfo  {journal} {New Journal of
  Physics}\ }\textbf {\bibinfo {volume} {11}}~(\bibinfo {number} {8}),\
  \bibinfo {pages} {085006}}\BibitemShut {NoStop}%
\bibitem [{\citenamefont {Munro}\ \emph {et~al.}(2010)\citenamefont {Munro},
  \citenamefont {Harrison}, \citenamefont {Stephens}, \citenamefont {Devitt},\
  and\ \citenamefont {Nemoto}}]{bib:munro10}%
  \BibitemOpen
  \bibfield  {author} {\bibinfo {author} {\bibnamefont {Munro}, \bibfnamefont
  {W~J}}, \bibinfo {author} {\bibfnamefont {K.~A.}\ \bibnamefont {Harrison}},
  \bibinfo {author} {\bibfnamefont {A.~M.}\ \bibnamefont {Stephens}}, \bibinfo
  {author} {\bibfnamefont {S.~J.}\ \bibnamefont {Devitt}}, and\ \bibinfo
  {author} {\bibfnamefont {K.}~\bibnamefont {Nemoto}}} (\bibinfo {year}
  {2010}),\ \bibfield  {title} {\enquote {\bibinfo {title} {From quantum
  multiplexing to high-performance quantum networking},}\ }\href
  {https://doi.org/10.1038/nphoton.2010.213} {\bibfield  {journal} {\bibinfo
  {journal} {Nature Photonics}\ }\textbf {\bibinfo {volume} {4}},\ \bibinfo
  {pages} {792}}\BibitemShut {NoStop}%
\bibitem [{\citenamefont {Munro}\ \emph {et~al.}(2008)\citenamefont {Munro},
  \citenamefont {Meter}, \citenamefont {Louis},\ and\ \citenamefont
  {Nemoto}}]{bib:munro08}%
  \BibitemOpen
  \bibfield  {author} {\bibinfo {author} {\bibnamefont {Munro}, \bibfnamefont
  {W~J}}, \bibinfo {author} {\bibfnamefont {R.~Van}\ \bibnamefont {Meter}},
  \bibinfo {author} {\bibfnamefont {S.~G.~R.}\ \bibnamefont {Louis}}, and\
  \bibinfo {author} {\bibfnamefont {K.}~\bibnamefont {Nemoto}}} (\bibinfo
  {year} {2008}),\ \bibfield  {title} {\enquote {\bibinfo {title}
  {High-bandwidth hybrid quantum repeater},}\ }\href
  {https://doi.org/10.1103/physrevlett.101.040502} {\bibfield  {journal}
  {\bibinfo  {journal} {Physical Review Letters}\ }\textbf {\bibinfo {volume}
  {101}},\ \bibinfo {pages} {040502}}\BibitemShut {NoStop}%
\bibitem [{\citenamefont {Munro}\ \emph {et~al.}(2005)\citenamefont {Munro},
  \citenamefont {Nemoto},\ and\ \citenamefont {Spiller}}]{bib:Munro05}%
  \BibitemOpen
  \bibfield  {author} {\bibinfo {author} {\bibnamefont {Munro}, \bibfnamefont
  {W~J}}, \bibinfo {author} {\bibfnamefont {K.}~\bibnamefont {Nemoto}}, and\
  \bibinfo {author} {\bibfnamefont {T.~P.}\ \bibnamefont {Spiller}}} (\bibinfo
  {year} {2005}),\ \bibfield  {title} {\enquote {\bibinfo {title} {Weak
  non-linearities: a new route to optical quantum computation},}\ }\href
  {https://doi.org/10.1088/1367-2630/7/1/137} {\bibfield  {journal} {\bibinfo
  {journal} {New Journal of Physics}\ }\textbf {\bibinfo {volume} {7}},\
  \bibinfo {pages} {137}}\BibitemShut {NoStop}%
\bibitem [{\citenamefont {Munro}\ \emph
  {et~al.}(2012{\natexlab{a}})\citenamefont {Munro}, \citenamefont {Stephens},
  \citenamefont {Devitt}, \citenamefont {Harrison},\ and\ \citenamefont
  {Nemoto}}]{bib:munro12}%
  \BibitemOpen
  \bibfield  {author} {\bibinfo {author} {\bibnamefont {Munro}, \bibfnamefont
  {W~J}}, \bibinfo {author} {\bibfnamefont {A.~M.}\ \bibnamefont {Stephens}},
  \bibinfo {author} {\bibfnamefont {S.~J.}\ \bibnamefont {Devitt}}, \bibinfo
  {author} {\bibfnamefont {K.~A.}\ \bibnamefont {Harrison}}, and\ \bibinfo
  {author} {\bibfnamefont {K.}~\bibnamefont {Nemoto}}} (\bibinfo {year}
  {2012}{\natexlab{a}}),\ \bibfield  {title} {\enquote {\bibinfo {title}
  {Quantum communication without the necessity of quantum memories},}\ }\href
  {https://doi.org/10.1038/nphoton.2012.243} {\bibfield  {journal} {\bibinfo
  {journal} {Nature Photonics}\ }\textbf {\bibinfo {volume} {6}},\ \bibinfo
  {pages} {777}}\BibitemShut {NoStop}%
\bibitem [{\citenamefont {Munro}\ \emph
  {et~al.}(2012{\natexlab{b}})\citenamefont {Munro}, \citenamefont {Stephens},
  \citenamefont {Devitt}, \citenamefont {Harrison},\ and\ \citenamefont
  {Nemoto}}]{SD-Munro:2012aa}%
  \BibitemOpen
  \bibfield  {author} {\bibinfo {author} {\bibnamefont {Munro}, \bibfnamefont
  {W~J}}, \bibinfo {author} {\bibfnamefont {A.~M.}\ \bibnamefont {Stephens}},
  \bibinfo {author} {\bibfnamefont {S.~J.}\ \bibnamefont {Devitt}}, \bibinfo
  {author} {\bibfnamefont {K.~A.}\ \bibnamefont {Harrison}}, and\ \bibinfo
  {author} {\bibfnamefont {Kae}\ \bibnamefont {Nemoto}}} (\bibinfo {year}
  {2012}{\natexlab{b}}),\ \bibfield  {title} {\enquote {\bibinfo {title}
  {Quantum communication without the necessity of quantum memories},}\ }\href
  {https://doi.org/10.1038/nphoton.2012.243} {\bibfield  {journal} {\bibinfo
  {journal} {Nature Photonics}\ }\textbf {\bibinfo {volume} {6}},\ \bibinfo
  {pages} {777}}\BibitemShut {NoStop}%
\bibitem [{\citenamefont {Munro}\ \emph {et~al.}(2015)\citenamefont {Munro},
  \citenamefont {Azuma}, \citenamefont {Tamaki},\ and\ \citenamefont
  {Nemoto}}]{bib:WJM2015}%
  \BibitemOpen
  \bibfield  {author} {\bibinfo {author} {\bibnamefont {Munro}, \bibfnamefont
  {William~J}}, \bibinfo {author} {\bibfnamefont {Koji}\ \bibnamefont {Azuma}},
  \bibinfo {author} {\bibfnamefont {Kiyoshi}\ \bibnamefont {Tamaki}}, and\
  \bibinfo {author} {\bibfnamefont {Kae}\ \bibnamefont {Nemoto}}} (\bibinfo
  {year} {2015}),\ \bibfield  {title} {\enquote {\bibinfo {title} {Inside
  quantum repeaters},}\ }\href {https://doi.org/10.1109/jstqe.2015.2392076}
  {\bibfield  {journal} {\bibinfo  {journal} {IEEE Journal of Selected Topics
  in Quantum Electronics}\ }\textbf {\bibinfo {volume} {21}},\ \bibinfo {pages}
  {6400813}}\BibitemShut {NoStop}%
\bibitem [{\citenamefont {Munro}\ \emph
  {et~al.}(2012{\natexlab{c}})\citenamefont {Munro}, \citenamefont {Stephens},
  \citenamefont {Devitt}, \citenamefont {Harrison},\ and\ \citenamefont
  {Nemoto}}]{bib:NP_6_777}%
  \BibitemOpen
  \bibfield  {author} {\bibinfo {author} {\bibnamefont {Munro}, \bibfnamefont
  {WJ}}, \bibinfo {author} {\bibfnamefont {AM}~\bibnamefont {Stephens}},
  \bibinfo {author} {\bibfnamefont {SJ}~\bibnamefont {Devitt}}, \bibinfo
  {author} {\bibfnamefont {KA}~\bibnamefont {Harrison}}, and\ \bibinfo {author}
  {\bibfnamefont {Kae}\ \bibnamefont {Nemoto}}} (\bibinfo {year}
  {2012}{\natexlab{c}}),\ \bibfield  {title} {\enquote {\bibinfo {title}
  {Quantum communication without the necessity of quantum memories},}\ }\href
  {https://doi.org/10.1038/nphoton.2012.243} {\bibfield  {journal} {\bibinfo
  {journal} {Nature Photonics}\ }\textbf {\bibinfo {volume} {6}},\ \bibinfo
  {pages} {777}}\BibitemShut {NoStop}%
\bibitem [{\citenamefont {Muralidharan}\ \emph {et~al.}(2014)\citenamefont
  {Muralidharan}, \citenamefont {Kim}, \citenamefont {L{\"u}tkenhaus},
  \citenamefont {Lukin},\ and\ \citenamefont {Jiang}}]{bib:MKLLJ14}%
  \BibitemOpen
  \bibfield  {author} {\bibinfo {author} {\bibnamefont {Muralidharan},
  \bibfnamefont {S}}, \bibinfo {author} {\bibfnamefont {J.}~\bibnamefont
  {Kim}}, \bibinfo {author} {\bibfnamefont {N.}~\bibnamefont {L{\"u}tkenhaus}},
  \bibinfo {author} {\bibfnamefont {N.~M.~D.}\ \bibnamefont {Lukin}}, and\
  \bibinfo {author} {\bibfnamefont {L.}~\bibnamefont {Jiang}}} (\bibinfo {year}
  {2014}),\ \bibfield  {title} {\enquote {\bibinfo {title} {Ultrafast and
  fault-tolerant quantum communication across long distances},}\ }\href
  {https://doi.org/10.1103/physrevlett.112.250501} {\bibfield  {journal}
  {\bibinfo  {journal} {Physical Review Letters}\ }\textbf {\bibinfo {volume}
  {112}},\ \bibinfo {pages} {250501}},\ \Eprint
  {https://arxiv.org/abs/arXiv:1310.5291v2} {arXiv:1310.5291v2} \BibitemShut
  {NoStop}%
\bibitem [{\citenamefont {Muralidharan}\ \emph {et~al.}(2015)\citenamefont
  {Muralidharan}, \citenamefont {Li}, \citenamefont {Kim}, \citenamefont
  {Lutkenhaus}, \citenamefont {Lukin},\ and\ \citenamefont
  {Jiang}}]{bib:Muralidharan2016}%
  \BibitemOpen
  \bibfield  {author} {\bibinfo {author} {\bibnamefont {Muralidharan},
  \bibfnamefont {Sreraman}}, \bibinfo {author} {\bibfnamefont {Linshu}\
  \bibnamefont {Li}}, \bibinfo {author} {\bibfnamefont {Jungsang}\ \bibnamefont
  {Kim}}, \bibinfo {author} {\bibfnamefont {Norbert}\ \bibnamefont
  {Lutkenhaus}}, \bibinfo {author} {\bibfnamefont {Mikhail~D.}\ \bibnamefont
  {Lukin}}, and\ \bibinfo {author} {\bibfnamefont {Liang}\ \bibnamefont
  {Jiang}}} (\bibinfo {year} {2015}),\ \bibfield  {title} {\enquote {\bibinfo
  {title} {Optimal architectures for long distance quantum communication},}\
  }\href {https://doi.org/10.1038/srep20463} {\bibfield  {journal} {\bibinfo
  {journal} {Scientific Reports}\ }\textbf {\bibinfo {volume} {6}},\ \bibinfo
  {pages} {20463}}\BibitemShut {NoStop}%
\bibitem [{\citenamefont {Nagayama}\ \emph {et~al.}(2017)\citenamefont
  {Nagayama}, \citenamefont {Fowler}, \citenamefont {Horsman}, \citenamefont
  {Devitt},\ and\ \citenamefont {Meter}}]{SD-Nagayama:2017aa}%
  \BibitemOpen
  \bibfield  {author} {\bibinfo {author} {\bibnamefont {Nagayama},
  \bibfnamefont {Shota}}, \bibinfo {author} {\bibfnamefont {Austin~G}\
  \bibnamefont {Fowler}}, \bibinfo {author} {\bibfnamefont {Dominic}\
  \bibnamefont {Horsman}}, \bibinfo {author} {\bibfnamefont {Simon~J}\
  \bibnamefont {Devitt}}, and\ \bibinfo {author} {\bibfnamefont {Rodney~Van}\
  \bibnamefont {Meter}}} (\bibinfo {year} {2017}),\ \bibfield  {title}
  {\enquote {\bibinfo {title} {Surface code error correction on a defective
  lattice},}\ }\href {https://doi.org/10.1088/1367-2630/aa5918} {\bibfield
  {journal} {\bibinfo  {journal} {New Journal of Physics}\ }\textbf {\bibinfo
  {volume} {19}},\ \bibinfo {pages} {023050}},\ \Eprint
  {https://arxiv.org/abs/arXiv:1607.00627v2} {arXiv:1607.00627v2} \BibitemShut
  {NoStop}%
\bibitem [{\citenamefont {Nakamura}\ \emph {et~al.}(1999)\citenamefont
  {Nakamura}, \citenamefont {Pashkin},\ and\ \citenamefont
  {Tsai}}]{bib:nakamura1999coherent}%
  \BibitemOpen
  \bibfield  {author} {\bibinfo {author} {\bibnamefont {Nakamura},
  \bibfnamefont {Yasunobu}}, \bibinfo {author} {\bibfnamefont {Yu~A}\
  \bibnamefont {Pashkin}}, and\ \bibinfo {author} {\bibfnamefont
  {JS}~\bibnamefont {Tsai}}} (\bibinfo {year} {1999}),\ \bibfield  {title}
  {\enquote {\bibinfo {title} {Coherent control of macroscopic quantum states
  in a single-cooper-pair box},}\ }\href {https://doi.org/10.1038/19718}
  {\bibfield  {journal} {\bibinfo  {journal} {Nature}\ }\textbf {\bibinfo
  {volume} {398}},\ \bibinfo {pages} {786}},\ \Eprint
  {https://arxiv.org/abs/arXiv:cond-mat/9904003v1} {arXiv:cond-mat/9904003v1}
  \BibitemShut {NoStop}%
\bibitem [{\citenamefont {Nauerth}\ \emph {et~al.}(2013)\citenamefont
  {Nauerth}, \citenamefont {Moll}, \citenamefont {Rau}, \citenamefont {Fuchs},
  \citenamefont {Horwath}, \citenamefont {Frick},\ and\ \citenamefont
  {Weinfurter}}]{bib:NP_7_382}%
  \BibitemOpen
  \bibfield  {author} {\bibinfo {author} {\bibnamefont {Nauerth}, \bibfnamefont
  {Sebastian}}, \bibinfo {author} {\bibfnamefont {Florian}\ \bibnamefont
  {Moll}}, \bibinfo {author} {\bibfnamefont {Markus}\ \bibnamefont {Rau}},
  \bibinfo {author} {\bibfnamefont {Christian}\ \bibnamefont {Fuchs}}, \bibinfo
  {author} {\bibfnamefont {Joachim}\ \bibnamefont {Horwath}}, \bibinfo {author}
  {\bibfnamefont {Stefan}\ \bibnamefont {Frick}}, and\ \bibinfo {author}
  {\bibfnamefont {Harald}\ \bibnamefont {Weinfurter}}} (\bibinfo {year}
  {2013}),\ \bibfield  {title} {\enquote {\bibinfo {title} {Air-to-ground
  quantum communication},}\ }\href {https://doi.org/10.1038/nphoton.2013.46}
  {\bibfield  {journal} {\bibinfo  {journal} {Nature Photonics}\ }\textbf
  {\bibinfo {volume} {7}},\ \bibinfo {pages} {382}}\BibitemShut {NoStop}%
\bibitem [{\citenamefont {Neergaard-Nielsen}\ \emph {et~al.}(2006)\citenamefont
  {Neergaard-Nielsen}, \citenamefont {Nielsen}, \citenamefont {Hettich},
  \citenamefont {M{\o}lmer},\ and\ \citenamefont
  {Polzik}}]{bib:neergaard2006generation}%
  \BibitemOpen
  \bibfield  {author} {\bibinfo {author} {\bibnamefont {Neergaard-Nielsen},
  \bibfnamefont {Jonas~S}}, \bibinfo {author} {\bibfnamefont {B~Melholt}\
  \bibnamefont {Nielsen}}, \bibinfo {author} {\bibfnamefont {C}~\bibnamefont
  {Hettich}}, \bibinfo {author} {\bibfnamefont {Klaus}\ \bibnamefont
  {M{\o}lmer}}, and\ \bibinfo {author} {\bibfnamefont {Eugene~S}\ \bibnamefont
  {Polzik}}} (\bibinfo {year} {2006}),\ \bibfield  {title} {\enquote {\bibinfo
  {title} {Generation of a superposition of odd photon number states for
  quantum information networks},}\ }\href
  {https://doi.org/10.1103/physrevlett.97.083604} {\bibfield  {journal}
  {\bibinfo  {journal} {Physical Review Letters}\ }\textbf {\bibinfo {volume}
  {97}},\ \bibinfo {pages} {083604}},\ \Eprint
  {https://arxiv.org/abs/arXiv:quant-ph/0602198v2} {arXiv:quant-ph/0602198v2}
  \BibitemShut {NoStop}%
\bibitem [{\citenamefont {Negrevergne}\ \emph {et~al.}(2006)\citenamefont
  {Negrevergne}, \citenamefont {Mahesh}, \citenamefont {Ryan}, \citenamefont
  {Ditty}, \citenamefont {Cyr-Racine}, \citenamefont {Power}, \citenamefont
  {Boulant}, \citenamefont {Havel}, \citenamefont {Cory},\ and\ \citenamefont
  {Laflamme}}]{bib:Negrevergne2006}%
  \BibitemOpen
  \bibfield  {author} {\bibinfo {author} {\bibnamefont {Negrevergne},
  \bibfnamefont {C}}, \bibinfo {author} {\bibfnamefont {T.~S.}\ \bibnamefont
  {Mahesh}}, \bibinfo {author} {\bibfnamefont {C.~A.}\ \bibnamefont {Ryan}},
  \bibinfo {author} {\bibfnamefont {M.}~\bibnamefont {Ditty}}, \bibinfo
  {author} {\bibfnamefont {F.}~\bibnamefont {Cyr-Racine}}, \bibinfo {author}
  {\bibfnamefont {W.}~\bibnamefont {Power}}, \bibinfo {author} {\bibfnamefont
  {N.}~\bibnamefont {Boulant}}, \bibinfo {author} {\bibfnamefont
  {T.}~\bibnamefont {Havel}}, \bibinfo {author} {\bibfnamefont {D.~G.}\
  \bibnamefont {Cory}}, and\ \bibinfo {author} {\bibfnamefont {R.}~\bibnamefont
  {Laflamme}}} (\bibinfo {year} {2006}),\ \bibfield  {title} {\enquote
  {\bibinfo {title} {Benchmarking quantum control methods on a 12-qubit
  system},}\ }\href {https://doi.org/10.1103/PhysRevLett.96.170501} {\bibfield
  {journal} {\bibinfo  {journal} {Physical Review Letters}\ }\textbf {\bibinfo
  {volume} {96}},\ \bibinfo {pages} {170501}},\ \Eprint
  {https://arxiv.org/abs/arXiv:quant-ph/0603248v1} {arXiv:quant-ph/0603248v1}
  \BibitemShut {NoStop}%
\bibitem [{\citenamefont {Neill}\ \emph {et~al.}(2018)\citenamefont {Neill},
  \citenamefont {Roushan}, \citenamefont {Kechedzhi}, \citenamefont {Boixo},
  \citenamefont {Isakov}, \citenamefont {Smelyanskiy}, \citenamefont {Megrant},
  \citenamefont {Chiaro}, \citenamefont {Dunsworth}, \citenamefont {Arya} \emph
  {et~al.}}]{bib:neill2018blueprint}%
  \BibitemOpen
  \bibfield  {author} {\bibinfo {author} {\bibnamefont {Neill}, \bibfnamefont
  {C}}, \bibinfo {author} {\bibfnamefont {P}~\bibnamefont {Roushan}}, \bibinfo
  {author} {\bibfnamefont {K}~\bibnamefont {Kechedzhi}}, \bibinfo {author}
  {\bibfnamefont {S}~\bibnamefont {Boixo}}, \bibinfo {author} {\bibfnamefont
  {SV}~\bibnamefont {Isakov}}, \bibinfo {author} {\bibfnamefont
  {V}~\bibnamefont {Smelyanskiy}}, \bibinfo {author} {\bibfnamefont
  {A}~\bibnamefont {Megrant}}, \bibinfo {author} {\bibfnamefont
  {B}~\bibnamefont {Chiaro}}, \bibinfo {author} {\bibfnamefont {A}~\bibnamefont
  {Dunsworth}}, \bibinfo {author} {\bibfnamefont {K}~\bibnamefont {Arya}},
  \emph {et~al.}} (\bibinfo {year} {2018}),\ \bibfield  {title} {\enquote
  {\bibinfo {title} {A blueprint for demonstrating quantum supremacy with
  superconducting qubits},}\ }\href {https://doi.org/10.1126/science.aao4309}
  {\bibfield  {journal} {\bibinfo  {journal} {Science}\ }\textbf {\bibinfo
  {volume} {360}},\ \bibinfo {pages} {195}},\ \Eprint
  {https://arxiv.org/abs/arXiv:1709.06678v1} {arXiv:1709.06678v1} \BibitemShut
  {NoStop}%
\bibitem [{\citenamefont {Nemoto}\ \emph
  {et~al.}(2014{\natexlab{a}})\citenamefont {Nemoto}, \citenamefont {Trupke},
  \citenamefont {Devitt}, \citenamefont {Stephens}, \citenamefont
  {Scharfenberger}, \citenamefont {Buczak}, \citenamefont {N{\"o}bauer},
  \citenamefont {Everitt}, \citenamefont {Schmiedmayer},\ and\ \citenamefont
  {Munro}}]{SD-Nemoto:2014aa}%
  \BibitemOpen
  \bibfield  {author} {\bibinfo {author} {\bibnamefont {Nemoto}, \bibfnamefont
  {Kae}}, \bibinfo {author} {\bibfnamefont {Michael}\ \bibnamefont {Trupke}},
  \bibinfo {author} {\bibfnamefont {Simon~J.}\ \bibnamefont {Devitt}}, \bibinfo
  {author} {\bibfnamefont {Ashley~M.}\ \bibnamefont {Stephens}}, \bibinfo
  {author} {\bibfnamefont {Burkhard}\ \bibnamefont {Scharfenberger}}, \bibinfo
  {author} {\bibfnamefont {Kathrin}\ \bibnamefont {Buczak}}, \bibinfo {author}
  {\bibfnamefont {Tobias}\ \bibnamefont {N{\"o}bauer}}, \bibinfo {author}
  {\bibfnamefont {Mark~S.}\ \bibnamefont {Everitt}}, \bibinfo {author}
  {\bibfnamefont {J{\"o}rg}\ \bibnamefont {Schmiedmayer}}, and\ \bibinfo
  {author} {\bibfnamefont {William~J.}\ \bibnamefont {Munro}}} (\bibinfo {year}
  {2014}{\natexlab{a}}),\ \bibfield  {title} {\enquote {\bibinfo {title}
  {Photonic architecture for scalable quantum information processing in
  diamond},}\ }\href {https://doi.org/10.1103/PhysRevX.4.031022} {\bibfield
  {journal} {\bibinfo  {journal} {Physical Review X}\ }\textbf {\bibinfo
  {volume} {4}},\ \bibinfo {pages} {031022}},\ \Eprint
  {https://arxiv.org/abs/arXiv:1309.4277v1} {arXiv:1309.4277v1} \BibitemShut
  {NoStop}%
\bibitem [{\citenamefont {Nemoto}\ \emph
  {et~al.}(2014{\natexlab{b}})\citenamefont {Nemoto}, \citenamefont {Trupke},
  \citenamefont {Devitt}, \citenamefont {Stephens}, \citenamefont
  {Scharfenberger}, \citenamefont {Buczak}, \citenamefont {N{\"o}bauer},
  \citenamefont {Everitt}, \citenamefont {Schmiedmayer},\ and\ \citenamefont
  {Munro}}]{Nemoto:2014aa}%
  \BibitemOpen
  \bibfield  {author} {\bibinfo {author} {\bibnamefont {Nemoto}, \bibfnamefont
  {Kae}}, \bibinfo {author} {\bibfnamefont {Michael}\ \bibnamefont {Trupke}},
  \bibinfo {author} {\bibfnamefont {Simon~J.}\ \bibnamefont {Devitt}}, \bibinfo
  {author} {\bibfnamefont {Ashley~M.}\ \bibnamefont {Stephens}}, \bibinfo
  {author} {\bibfnamefont {Burkhard}\ \bibnamefont {Scharfenberger}}, \bibinfo
  {author} {\bibfnamefont {Kathrin}\ \bibnamefont {Buczak}}, \bibinfo {author}
  {\bibfnamefont {Tobias}\ \bibnamefont {N{\"o}bauer}}, \bibinfo {author}
  {\bibfnamefont {Mark~S.}\ \bibnamefont {Everitt}}, \bibinfo {author}
  {\bibfnamefont {J{\"o}rg}\ \bibnamefont {Schmiedmayer}}, and\ \bibinfo
  {author} {\bibfnamefont {William~J.}\ \bibnamefont {Munro}}} (\bibinfo {year}
  {2014}{\natexlab{b}}),\ \bibfield  {title} {\enquote {\bibinfo {title}
  {Photonic architecture for scalable quantum information processing in
  diamond},}\ }\href {https://doi.org/10.1103/PhysRevX.4.031022} {\bibfield
  {journal} {\bibinfo  {journal} {Physical Review X}\ }\textbf {\bibinfo
  {volume} {4}}~(\bibinfo {number} {3}),\ \bibinfo {pages}
  {031022--}}\BibitemShut {NoStop}%
\bibitem [{\citenamefont {Neumann}(1955)}]{bib:N55}%
  \BibitemOpen
  \bibfield  {author} {\bibinfo {author} {\bibnamefont {Neumann}, \bibfnamefont
  {J~Von}}} (\bibinfo {year} {1955}),\ \bibfield  {title} {\enquote {\bibinfo
  {title} {{Probabilistic logics and the synthesis of reliable organisms from
  unreliable components}},}\ }\href@noop {} {\bibfield  {journal} {\bibinfo
  {journal} {Automata Studies}\ }\textbf {\bibinfo {volume} {43}}}\BibitemShut
  {NoStop}%
\bibitem [{\citenamefont {von Neumann}\ and\ \citenamefont
  {Morgenstern}(2007)}]{bib:NeumanMorgenstern44}%
  \BibitemOpen
  \bibfield  {author} {\bibinfo {author} {\bibnamefont {von Neumann},
  \bibfnamefont {John}}, and\ \bibinfo {author} {\bibfnamefont {Oskar}\
  \bibnamefont {Morgenstern}}} (\bibinfo {year} {2007}),\ \href@noop {} {\emph
  {\bibinfo {title} {Theory of Games and Economic Behavior}}}\ (\bibinfo
  {publisher} {Princeton University Press})\BibitemShut {NoStop}%
\bibitem [{\citenamefont {Neumann}\ \emph {et~al.}(2010)\citenamefont
  {Neumann}, \citenamefont {Kolesov}, \citenamefont {Naydenov}, \citenamefont
  {Beck}, \citenamefont {Rempp}, \citenamefont {Steiner}, \citenamefont
  {Jacques}, \citenamefont {Balasubramanian}, \citenamefont {Markham},
  \citenamefont {Twitchen} \emph {et~al.}}]{bib:neumann2010quantum}%
  \BibitemOpen
  \bibfield  {author} {\bibinfo {author} {\bibnamefont {Neumann}, \bibfnamefont
  {P}}, \bibinfo {author} {\bibfnamefont {R}~\bibnamefont {Kolesov}}, \bibinfo
  {author} {\bibfnamefont {B}~\bibnamefont {Naydenov}}, \bibinfo {author}
  {\bibfnamefont {J}~\bibnamefont {Beck}}, \bibinfo {author} {\bibfnamefont
  {F}~\bibnamefont {Rempp}}, \bibinfo {author} {\bibfnamefont {M}~\bibnamefont
  {Steiner}}, \bibinfo {author} {\bibfnamefont {V}~\bibnamefont {Jacques}},
  \bibinfo {author} {\bibfnamefont {G}~\bibnamefont {Balasubramanian}},
  \bibinfo {author} {\bibfnamefont {ML}~\bibnamefont {Markham}}, \bibinfo
  {author} {\bibfnamefont {DJ}~\bibnamefont {Twitchen}},  \emph {et~al.}}
  (\bibinfo {year} {2010}),\ \bibfield  {title} {\enquote {\bibinfo {title}
  {Quantum register based on coupled electron spins in a room-temperature
  solid},}\ }\href {https://doi.org/10.1038/nphys1536} {\bibfield  {journal}
  {\bibinfo  {journal} {Nature Physics}\ }\textbf {\bibinfo {volume} {6}},\
  \bibinfo {pages} {249}}\BibitemShut {NoStop}%
\bibitem [{\citenamefont {Neumann}\ \emph {et~al.}(2008)\citenamefont
  {Neumann}, \citenamefont {Mizuochi}, \citenamefont {Rempp}, \citenamefont
  {Hemmer}, \citenamefont {Watanabe}, \citenamefont {Yamasaki}, \citenamefont
  {Jacques}, \citenamefont {Gaebel}, \citenamefont {Jelezko},\ and\
  \citenamefont {Wrachtrup}}]{bib:neumann2008multipartite}%
  \BibitemOpen
  \bibfield  {author} {\bibinfo {author} {\bibnamefont {Neumann}, \bibfnamefont
  {P}}, \bibinfo {author} {\bibfnamefont {N}~\bibnamefont {Mizuochi}}, \bibinfo
  {author} {\bibfnamefont {F}~\bibnamefont {Rempp}}, \bibinfo {author}
  {\bibfnamefont {Philip}\ \bibnamefont {Hemmer}}, \bibinfo {author}
  {\bibfnamefont {H}~\bibnamefont {Watanabe}}, \bibinfo {author} {\bibfnamefont
  {S}~\bibnamefont {Yamasaki}}, \bibinfo {author} {\bibfnamefont
  {V}~\bibnamefont {Jacques}}, \bibinfo {author} {\bibfnamefont {Torsten}\
  \bibnamefont {Gaebel}}, \bibinfo {author} {\bibfnamefont {F}~\bibnamefont
  {Jelezko}}, and\ \bibinfo {author} {\bibfnamefont {J}~\bibnamefont
  {Wrachtrup}}} (\bibinfo {year} {2008}),\ \bibfield  {title} {\enquote
  {\bibinfo {title} {Multipartite entanglement among single spins in
  diamond},}\ }\href {https://doi.org/10.1126/science.1157233} {\bibfield
  {journal} {\bibinfo  {journal} {Science}\ }\textbf {\bibinfo {volume}
  {320}},\ \bibinfo {pages} {1326}}\BibitemShut {NoStop}%
\bibitem [{\citenamefont {Neville}\ \emph {et~al.}(2017)\citenamefont
  {Neville}, \citenamefont {Sparrow}, \citenamefont {Clifford}, \citenamefont
  {Johnston}, \citenamefont {Birchall}, \citenamefont {Montanaro},\ and\
  \citenamefont {Laing}}]{neville2017no}%
  \BibitemOpen
  \bibfield  {author} {\bibinfo {author} {\bibnamefont {Neville}, \bibfnamefont
  {Alex}}, \bibinfo {author} {\bibfnamefont {Chris}\ \bibnamefont {Sparrow}},
  \bibinfo {author} {\bibfnamefont {Rapha{\"e}l}\ \bibnamefont {Clifford}},
  \bibinfo {author} {\bibfnamefont {Eric}\ \bibnamefont {Johnston}}, \bibinfo
  {author} {\bibfnamefont {Patrick~M}\ \bibnamefont {Birchall}}, \bibinfo
  {author} {\bibfnamefont {Ashley}\ \bibnamefont {Montanaro}}, and\ \bibinfo
  {author} {\bibfnamefont {Anthony}\ \bibnamefont {Laing}}} (\bibinfo {year}
  {2017}),\ \bibfield  {title} {\enquote {\bibinfo {title} {No imminent quantum
  supremacy by boson sampling},}\ }\href@noop {} {\bibinfo  {journal} {arXiv
  preprint arXiv:1705.00686}\ }\BibitemShut {NoStop}%
\bibitem [{\citenamefont {Nielsen}(2004)}]{bib:Nielsen04}%
  \BibitemOpen
\bibfield  {journal} {  }\bibfield  {author} {\bibinfo {author} {\bibnamefont
  {Nielsen}, \bibfnamefont {M~A}}} (\bibinfo {year} {2004}),\ \bibfield
  {title} {\enquote {\bibinfo {title} {Optical quantum computation using
  cluster states},}\ }\href {https://doi.org/10.1103/physrevlett.93.040503}
  {\bibfield  {journal} {\bibinfo  {journal} {Physical Review Letters}\
  }\textbf {\bibinfo {volume} {93}},\ \bibinfo {pages} {040503}},\ \Eprint
  {https://arxiv.org/abs/arXiv:quant-ph/0402005v1} {arXiv:quant-ph/0402005v1}
  \BibitemShut {NoStop}%
\bibitem [{\citenamefont {Nielsen}(2006)}]{bib:Nielsen06}%
  \BibitemOpen
  \bibfield  {author} {\bibinfo {author} {\bibnamefont {Nielsen}, \bibfnamefont
  {M~A}}} (\bibinfo {year} {2006}),\ \bibfield  {title} {\enquote {\bibinfo
  {title} {Cluster-state quantum computation},}\ }\href
  {https://doi.org/10.1016/s0034-4877(06)80014-5} {\bibfield  {journal}
  {\bibinfo  {journal} {Reviews in Mathematical Physics}\ }\textbf {\bibinfo
  {volume} {57}},\ \bibinfo {pages} {147}},\ \Eprint
  {https://arxiv.org/abs/arXiv:quant-ph/0504097v2} {arXiv:quant-ph/0504097v2}
  \BibitemShut {NoStop}%
\bibitem [{\citenamefont {Nielsen}\ and\ \citenamefont
  {Chuang}(2000)}]{bib:NielsenChuang00}%
  \BibitemOpen
  \bibfield  {author} {\bibinfo {author} {\bibnamefont {Nielsen}, \bibfnamefont
  {M~A}}, and\ \bibinfo {author} {\bibfnamefont {I.~L.}\ \bibnamefont
  {Chuang}}} (\bibinfo {year} {2000}),\ \href
  {https://doi.org/10.1017/cbo9780511976667} {\emph {\bibinfo {title} {Quantum
  Computation and Quantum Information}}}\ (\bibinfo  {publisher} {Cambridge
  University Press, Cambridge})\BibitemShut {NoStop}%
\bibitem [{\citenamefont {Nielsen}\ and\ \citenamefont
  {Dawson}(2005)}]{bib:NielsenDawson04}%
  \BibitemOpen
  \bibfield  {author} {\bibinfo {author} {\bibnamefont {Nielsen}, \bibfnamefont
  {M~A}}, and\ \bibinfo {author} {\bibfnamefont {C.~M.}\ \bibnamefont
  {Dawson}}} (\bibinfo {year} {2005}),\ \bibfield  {title} {\enquote {\bibinfo
  {title} {Fault-tolerant quantum computation with cluster states},}\ }\href
  {https://doi.org/10.1103/physreva.71.042323} {\bibfield  {journal} {\bibinfo
  {journal} {Physical Review A}\ }\textbf {\bibinfo {volume} {71}},\ \bibinfo
  {pages} {042323}},\ \Eprint {https://arxiv.org/abs/arXiv:quant-ph/0405134v2}
  {arXiv:quant-ph/0405134v2} \BibitemShut {NoStop}%
\bibitem [{\citenamefont {Niset}\ \emph {et~al.}(2009)\citenamefont {Niset},
  \citenamefont {Fiur\'a\ifmmode~\check{s}\else \v{s}\fi{}ek},\ and\
  \citenamefont {Cerf}}]{bib:PhysRevLett.102.120501}%
  \BibitemOpen
  \bibfield  {author} {\bibinfo {author} {\bibnamefont {Niset}, \bibfnamefont
  {Julien}}, \bibinfo {author} {\bibfnamefont {Jarom\'{\i}r}\ \bibnamefont
  {Fiur\'a\ifmmode~\check{s}\else \v{s}\fi{}ek}}, and\ \bibinfo {author}
  {\bibfnamefont {Nicolas~J.}\ \bibnamefont {Cerf}}} (\bibinfo {year} {2009}),\
  \bibfield  {title} {\enquote {\bibinfo {title} {No-go theorem for gaussian
  quantum error correction},}\ }\href
  {https://doi.org/10.1103/physrevlett.102.120501} {\bibfield  {journal}
  {\bibinfo  {journal} {Physical Review Letters}\ }\textbf {\bibinfo {volume}
  {102}},\ \bibinfo {pages} {120501}},\ \Eprint
  {https://arxiv.org/abs/arXiv:0811.3128v1} {arXiv:0811.3128v1} \BibitemShut
  {NoStop}%
\bibitem [{\citenamefont {Noh}\ \emph {et~al.}(2019)\citenamefont {Noh},
  \citenamefont {Albert},\ and\ \citenamefont {Jiang}}]{noh2019capacity}%
  \BibitemOpen
  \bibfield  {author} {\bibinfo {author} {\bibnamefont {Noh}, \bibfnamefont
  {Kyungjoo}}, \bibinfo {author} {\bibfnamefont {Victor~V.}\ \bibnamefont
  {Albert}}, and\ \bibinfo {author} {\bibfnamefont {Liang}\ \bibnamefont
  {Jiang}}} (\bibinfo {year} {2019}),\ \bibfield  {title} {\enquote {\bibinfo
  {title} {Quantum capacity bounds of gaussian thermal loss channels and
  achievable rates with gottesman-kitaev-preskill codes},}\ }\href
  {https://doi.org/10.1109/TIT.2018.2873764} {\bibfield  {journal} {\bibinfo
  {journal} {IEEE Transactions on Information Theory}\ }\textbf {\bibinfo
  {volume} {65}}~(\bibinfo {number} {4}),\ \bibinfo {pages}
  {2563--2582}}\BibitemShut {NoStop}%
\bibitem [{\citenamefont {O'Brien}\ \emph {et~al.}(2004)\citenamefont
  {O'Brien}, \citenamefont {Pryde}, \citenamefont {Gilchrist}, \citenamefont
  {James}, \citenamefont {Langford}, \citenamefont {Ralph},\ and\ \citenamefont
  {White}}]{bib:OBrien04}%
  \BibitemOpen
  \bibfield  {author} {\bibinfo {author} {\bibnamefont {O'Brien}, \bibfnamefont
  {J~L}}, \bibinfo {author} {\bibfnamefont {G.~J.}\ \bibnamefont {Pryde}},
  \bibinfo {author} {\bibfnamefont {A.}~\bibnamefont {Gilchrist}}, \bibinfo
  {author} {\bibfnamefont {D.~F.~V.}\ \bibnamefont {James}}, \bibinfo {author}
  {\bibfnamefont {N.~K.}\ \bibnamefont {Langford}}, \bibinfo {author}
  {\bibfnamefont {T.~C.}\ \bibnamefont {Ralph}}, and\ \bibinfo {author}
  {\bibfnamefont {A.~G.}\ \bibnamefont {White}}} (\bibinfo {year} {2004}),\
  \bibfield  {title} {\enquote {\bibinfo {title} {Quantum process tomography of
  a controlled-not gate},}\ }\href
  {https://doi.org/10.1103/physrevlett.93.080502} {\bibfield  {journal}
  {\bibinfo  {journal} {Physical Review Letters}\ }\textbf {\bibinfo {volume}
  {93}},\ \bibinfo {pages} {080502}},\ \Eprint
  {https://arxiv.org/abs/arXiv:quant-ph/0402166v2} {arXiv:quant-ph/0402166v2}
  \BibitemShut {NoStop}%
\bibitem [{\citenamefont {O'Brien}\ \emph
  {et~al.}(2003{\natexlab{a}})\citenamefont {O'Brien}, \citenamefont {Pryde},
  \citenamefont {White}, \citenamefont {Ralph},\ and\ \citenamefont
  {Branning}}]{bib:OBrien03}%
  \BibitemOpen
  \bibfield  {author} {\bibinfo {author} {\bibnamefont {O'Brien}, \bibfnamefont
  {J~L}}, \bibinfo {author} {\bibfnamefont {G.~J.}\ \bibnamefont {Pryde}},
  \bibinfo {author} {\bibfnamefont {A.~G.}\ \bibnamefont {White}}, \bibinfo
  {author} {\bibfnamefont {T.~C.}\ \bibnamefont {Ralph}}, and\ \bibinfo
  {author} {\bibfnamefont {D.}~\bibnamefont {Branning}}} (\bibinfo {year}
  {2003}{\natexlab{a}}),\ \bibfield  {title} {\enquote {\bibinfo {title}
  {Demonstration of an all-optical quantum controlled-not gate},}\ }\href@noop
  {} {\bibfield  {journal} {\bibinfo  {journal} {Nature}\ }\textbf {\bibinfo
  {volume} {426}},\ \bibinfo {pages} {264}}\BibitemShut {NoStop}%
\bibitem [{\citenamefont {O'Brien}\ \emph {et~al.}(2009)\citenamefont
  {O'Brien}, \citenamefont {Furusawa},\ and\ \citenamefont
  {Vu{\v{c}}kovi{\'c}}}]{bib:o2009photonic}%
  \BibitemOpen
  \bibfield  {author} {\bibinfo {author} {\bibnamefont {O'Brien}, \bibfnamefont
  {Jeremy~L}}, \bibinfo {author} {\bibfnamefont {Akira}\ \bibnamefont
  {Furusawa}}, and\ \bibinfo {author} {\bibfnamefont {Jelena}\ \bibnamefont
  {Vu{\v{c}}kovi{\'c}}}} (\bibinfo {year} {2009}),\ \bibfield  {title}
  {\enquote {\bibinfo {title} {Photonic quantum technologies},}\ }\href
  {https://doi.org/10.1038/nphoton.2009.229} {\bibfield  {journal} {\bibinfo
  {journal} {Nature Photonics}\ }\textbf {\bibinfo {volume} {3}},\ \bibinfo
  {pages} {687}},\ \Eprint {https://arxiv.org/abs/arXiv:1003.3928v1}
  {arXiv:1003.3928v1} \BibitemShut {NoStop}%
\bibitem [{\citenamefont {O'Brien}\ \emph
  {et~al.}(2003{\natexlab{b}})\citenamefont {O'Brien}, \citenamefont {Pryde},
  \citenamefont {White}, \citenamefont {Ralph},\ and\ \citenamefont
  {Branning}}]{bib:Brien2003demonstration}%
  \BibitemOpen
  \bibfield  {author} {\bibinfo {author} {\bibnamefont {O'Brien}, \bibfnamefont
  {Jeremy~L}}, \bibinfo {author} {\bibfnamefont {Geoffrey~J}\ \bibnamefont
  {Pryde}}, \bibinfo {author} {\bibfnamefont {Andrew~G}\ \bibnamefont {White}},
  \bibinfo {author} {\bibfnamefont {Timothy~C}\ \bibnamefont {Ralph}}, and\
  \bibinfo {author} {\bibfnamefont {David}\ \bibnamefont {Branning}}} (\bibinfo
  {year} {2003}{\natexlab{b}}),\ \bibfield  {title} {\enquote {\bibinfo {title}
  {Demonstration of an all-optical quantum controlled-not gate},}\ }\href
  {https://doi.org/10.1038/nature02054} {\bibfield  {journal} {\bibinfo
  {journal} {Nature}\ }\textbf {\bibinfo {volume} {426}},\ \bibinfo {pages}
  {264}},\ \Eprint {https://arxiv.org/abs/arXiv:quant-ph/0403062v1}
  {arXiv:quant-ph/0403062v1} \BibitemShut {NoStop}%
\bibitem [{\citenamefont {Ofek}\ \emph {et~al.}(2016)\citenamefont {Ofek},
  \citenamefont {Petrenko}, \citenamefont {Heeres}, \citenamefont {Reinhold},
  \citenamefont {Leghtas}, \citenamefont {Vlastakis}, \citenamefont {Liu},
  \citenamefont {Frunzio}, \citenamefont {Girvin}, \citenamefont {Jiang} \emph
  {et~al.}}]{bib:ofek2016extending}%
  \BibitemOpen
  \bibfield  {author} {\bibinfo {author} {\bibnamefont {Ofek}, \bibfnamefont
  {Nissim}}, \bibinfo {author} {\bibfnamefont {Andrei}\ \bibnamefont
  {Petrenko}}, \bibinfo {author} {\bibfnamefont {Reinier}\ \bibnamefont
  {Heeres}}, \bibinfo {author} {\bibfnamefont {Philip}\ \bibnamefont
  {Reinhold}}, \bibinfo {author} {\bibfnamefont {Zaki}\ \bibnamefont
  {Leghtas}}, \bibinfo {author} {\bibfnamefont {Brian}\ \bibnamefont
  {Vlastakis}}, \bibinfo {author} {\bibfnamefont {Yehan}\ \bibnamefont {Liu}},
  \bibinfo {author} {\bibfnamefont {Luigi}\ \bibnamefont {Frunzio}}, \bibinfo
  {author} {\bibfnamefont {SM}~\bibnamefont {Girvin}}, \bibinfo {author}
  {\bibfnamefont {L}~\bibnamefont {Jiang}},  \emph {et~al.}} (\bibinfo {year}
  {2016}),\ \bibfield  {title} {\enquote {\bibinfo {title} {Extending the
  lifetime of a quantum bit with error correction in superconducting
  circuits},}\ }\href {https://doi.org/10.1038/nature18949} {\bibfield
  {journal} {\bibinfo  {journal} {Nature}\ }\textbf {\bibinfo {volume} {536}},\
  \bibinfo {pages} {441}}\BibitemShut {NoStop}%
\bibitem [{\citenamefont {Okamoto}\ \emph {et~al.}(2005)\citenamefont
  {Okamoto}, \citenamefont {Hofmann}, \citenamefont {Takeuchi},\ and\
  \citenamefont {Sasaki}}]{bib:Okamoto2005}%
  \BibitemOpen
  \bibfield  {author} {\bibinfo {author} {\bibnamefont {Okamoto}, \bibfnamefont
  {Ryo}}, \bibinfo {author} {\bibfnamefont {Holger~F.}\ \bibnamefont
  {Hofmann}}, \bibinfo {author} {\bibfnamefont {Shigeki}\ \bibnamefont
  {Takeuchi}}, and\ \bibinfo {author} {\bibfnamefont {Keiji}\ \bibnamefont
  {Sasaki}}} (\bibinfo {year} {2005}),\ \bibfield  {title} {\enquote {\bibinfo
  {title} {Demonstration of an optical quantum controlled-not gate without path
  interference},}\ }\href {https://doi.org/10.1103/PhysRevLett.95.210506}
  {\bibfield  {journal} {\bibinfo  {journal} {Physical Review Letters}\
  }\textbf {\bibinfo {volume} {95}},\ \bibinfo {pages} {210506}},\ \Eprint
  {https://arxiv.org/abs/arXiv:quant-ph/0506263v1} {arXiv:quant-ph/0506263v1}
  \BibitemShut {NoStop}%
\bibitem [{\citenamefont {Okubo}\ \emph {et~al.}(2008)\citenamefont {Okubo},
  \citenamefont {Hirano}, \citenamefont {Zhang},\ and\ \citenamefont
  {Hirano}}]{bib:okubo2008pulse}%
  \BibitemOpen
  \bibfield  {author} {\bibinfo {author} {\bibnamefont {Okubo}, \bibfnamefont
  {Ryuhi}}, \bibinfo {author} {\bibfnamefont {Mayumi}\ \bibnamefont {Hirano}},
  \bibinfo {author} {\bibfnamefont {Yun}\ \bibnamefont {Zhang}}, and\ \bibinfo
  {author} {\bibfnamefont {Takuya}\ \bibnamefont {Hirano}}} (\bibinfo {year}
  {2008}),\ \bibfield  {title} {\enquote {\bibinfo {title} {Pulse-resolved
  measurement of quadrature phase amplitudes of squeezed pulse trains at a
  repetition rate of 76 mhz},}\ }\href {https://doi.org/10.1364/ol.33.001458}
  {\bibfield  {journal} {\bibinfo  {journal} {Optics Letters}\ }\textbf
  {\bibinfo {volume} {33}},\ \bibinfo {pages} {1458}}\BibitemShut {NoStop}%
\bibitem [{\citenamefont {Olson}\ \emph {et~al.}(2015)\citenamefont {Olson},
  \citenamefont {Seshadreesan}, \citenamefont {Motes}, \citenamefont {Rohde},\
  and\ \citenamefont {Dowling}}]{bib:RohdePhotAdd15}%
  \BibitemOpen
  \bibfield  {author} {\bibinfo {author} {\bibnamefont {Olson}, \bibfnamefont
  {Jonathan~P}}, \bibinfo {author} {\bibfnamefont {Kaushik~P.}\ \bibnamefont
  {Seshadreesan}}, \bibinfo {author} {\bibfnamefont {Keith~R.}\ \bibnamefont
  {Motes}}, \bibinfo {author} {\bibfnamefont {Peter~P.}\ \bibnamefont {Rohde}},
  and\ \bibinfo {author} {\bibfnamefont {Jonathan~P.}\ \bibnamefont {Dowling}}}
  (\bibinfo {year} {2015}),\ \bibfield  {title} {\enquote {\bibinfo {title}
  {Sampling arbitrary photon-added or photon-subtracted squeezed states is in
  the same complexity class as boson sampling},}\ }\href
  {https://doi.org/10.1103/physreva.91.022317} {\bibfield  {journal} {\bibinfo
  {journal} {Physical Review A}\ }\textbf {\bibinfo {volume} {91}},\ \bibinfo
  {pages} {022317}},\ \Eprint {https://arxiv.org/abs/arXiv:1406.7821v3}
  {arXiv:1406.7821v3} \BibitemShut {NoStop}%
\bibitem [{\citenamefont {Orlin}(1997)}]{orlin1997polynomial}%
  \BibitemOpen
  \bibfield  {author} {\bibinfo {author} {\bibnamefont {Orlin}, \bibfnamefont
  {James~B}}} (\bibinfo {year} {1997}),\ \bibfield  {title} {\enquote {\bibinfo
  {title} {A polynomial time primal network simplex algorithm for minimum cost
  flows},}\ }\href@noop {} {\bibfield  {journal} {\bibinfo  {journal}
  {Mathematical Programming}\ }\textbf {\bibinfo {volume} {78}},\ \bibinfo
  {pages} {109--]129}}\BibitemShut {NoStop}%
\bibitem [{\citenamefont {Oszmaniec}\ and\ \citenamefont
  {Brod}(2018)}]{oszmaniec2018classical}%
  \BibitemOpen
  \bibfield  {author} {\bibinfo {author} {\bibnamefont {Oszmaniec},
  \bibfnamefont {Micha{\l}}}, and\ \bibinfo {author} {\bibfnamefont {Daniel~J}\
  \bibnamefont {Brod}}} (\bibinfo {year} {2018}),\ \bibfield  {title} {\enquote
  {\bibinfo {title} {Classical simulation of photonic linear optics with lost
  particles},}\ }\href@noop {} {\bibfield  {journal} {\bibinfo  {journal} {New
  Journal of Physics}\ }\textbf {\bibinfo {volume} {20}}~(\bibinfo {number}
  {9}),\ \bibinfo {pages} {092002}}\BibitemShut {NoStop}%
\bibitem [{\citenamefont {Ourjoumtsev}\ \emph {et~al.}(2009)\citenamefont
  {Ourjoumtsev}, \citenamefont {Ferreyrol}, \citenamefont {Tualle-Brouri},\
  and\ \citenamefont {Grangier}}]{bib:ourjoumtsev2009preparation}%
  \BibitemOpen
  \bibfield  {author} {\bibinfo {author} {\bibnamefont {Ourjoumtsev},
  \bibfnamefont {Alexei}}, \bibinfo {author} {\bibfnamefont {Franck}\
  \bibnamefont {Ferreyrol}}, \bibinfo {author} {\bibfnamefont {Rosa}\
  \bibnamefont {Tualle-Brouri}}, and\ \bibinfo {author} {\bibfnamefont
  {Philippe}\ \bibnamefont {Grangier}}} (\bibinfo {year} {2009}),\ \bibfield
  {title} {\enquote {\bibinfo {title} {Preparation of non-local superpositions
  of quasi-classical light states},}\ }\href
  {https://doi.org/10.1038/nphys1199} {\bibfield  {journal} {\bibinfo
  {journal} {Nature Physics}\ }\textbf {\bibinfo {volume} {5}},\ \bibinfo
  {pages} {189}}\BibitemShut {NoStop}%
\bibitem [{\citenamefont {Ourjoumtsev}\ \emph {et~al.}(2007)\citenamefont
  {Ourjoumtsev}, \citenamefont {Jeong}, \citenamefont {Tualle-Brouri},\ and\
  \citenamefont {Grangier}}]{bib:ourjoumtsev2007generation}%
  \BibitemOpen
  \bibfield  {author} {\bibinfo {author} {\bibnamefont {Ourjoumtsev},
  \bibfnamefont {Alexei}}, \bibinfo {author} {\bibfnamefont {Hyunseok}\
  \bibnamefont {Jeong}}, \bibinfo {author} {\bibfnamefont {Rosa}\ \bibnamefont
  {Tualle-Brouri}}, and\ \bibinfo {author} {\bibfnamefont {Philippe}\
  \bibnamefont {Grangier}}} (\bibinfo {year} {2007}),\ \bibfield  {title}
  {\enquote {\bibinfo {title} {Generation of optical 'schr{\"o}dinger cats'
  from photon number states},}\ }\href {https://doi.org/10.1038/nature06054}
  {\bibfield  {journal} {\bibinfo  {journal} {Nature}\ }\textbf {\bibinfo
  {volume} {448}},\ \bibinfo {pages} {784}}\BibitemShut {NoStop}%
\bibitem [{\citenamefont {Ourjoumtsev}\ \emph {et~al.}(2006)\citenamefont
  {Ourjoumtsev}, \citenamefont {Tualle-Brouri}, \citenamefont {Laurat},\ and\
  \citenamefont {Grangier}}]{bib:ourjoumtsev2006generating}%
  \BibitemOpen
  \bibfield  {author} {\bibinfo {author} {\bibnamefont {Ourjoumtsev},
  \bibfnamefont {Alexei}}, \bibinfo {author} {\bibfnamefont {Rosa}\
  \bibnamefont {Tualle-Brouri}}, \bibinfo {author} {\bibfnamefont {Julien}\
  \bibnamefont {Laurat}}, and\ \bibinfo {author} {\bibfnamefont {Philippe}\
  \bibnamefont {Grangier}}} (\bibinfo {year} {2006}),\ \bibfield  {title}
  {\enquote {\bibinfo {title} {Generating optical schr{\"o}dinger kittens for
  quantum information processing},}\ }\href
  {https://doi.org/10.1126/science.1122858} {\bibfield  {journal} {\bibinfo
  {journal} {Science}\ }\textbf {\bibinfo {volume} {312}},\ \bibinfo {pages}
  {83}}\BibitemShut {NoStop}%
\bibitem [{\citenamefont {Ouyang}\ \emph {et~al.}(2018)\citenamefont {Ouyang},
  \citenamefont {Tan},\ and\ \citenamefont {Fitzsimons}}]{ouyang2018quantum}%
  \BibitemOpen
  \bibfield  {author} {\bibinfo {author} {\bibnamefont {Ouyang}, \bibfnamefont
  {Yingkai}}, \bibinfo {author} {\bibfnamefont {Si-Hui}\ \bibnamefont {Tan}},
  and\ \bibinfo {author} {\bibfnamefont {Joseph}\ \bibnamefont {Fitzsimons}}}
  (\bibinfo {year} {2018}),\ \bibfield  {title} {\enquote {\bibinfo {title}
  {Quantum homomorphic encryption from quantum codes},}\ }\href
  {https://doi.org/10.1103/PhysRevA.98.042334} {\bibfield  {journal} {\bibinfo
  {journal} {Physical Review A}\ }\textbf {\bibinfo {volume} {98}},\ \bibinfo
  {pages} {042334}}\BibitemShut {NoStop}%
\bibitem [{\citenamefont {Ouyang}\ \emph {et~al.}(2020)\citenamefont {Ouyang},
  \citenamefont {Tan}, \citenamefont {Fitzsimons},\ and\ \citenamefont
  {Rohde}}]{ouyang2020homomorphic}%
  \BibitemOpen
  \bibfield  {author} {\bibinfo {author} {\bibnamefont {Ouyang}, \bibfnamefont
  {Yingkai}}, \bibinfo {author} {\bibfnamefont {Si-Hui}\ \bibnamefont {Tan}},
  \bibinfo {author} {\bibfnamefont {Joseph}\ \bibnamefont {Fitzsimons}}, and\
  \bibinfo {author} {\bibfnamefont {Peter~P.}\ \bibnamefont {Rohde}}} (\bibinfo
  {year} {2020}),\ \bibfield  {title} {\enquote {\bibinfo {title} {Homomorphic
  encryption of linear optics quantum computation on almost arbitrary states of
  light with asymptotically perfect security},}\ }\href
  {https://doi.org/10.1103/PhysRevResearch.2.013332} {\bibfield  {journal}
  {\bibinfo  {journal} {Physical Review Research}\ }\textbf {\bibinfo {volume}
  {2}},\ \bibinfo {pages} {013332}}\BibitemShut {NoStop}%
\bibitem [{\citenamefont {Ouyang}\ \emph {et~al.}(2017)\citenamefont {Ouyang},
  \citenamefont {Tan}, \citenamefont {Zhao},\ and\ \citenamefont
  {Fitzsimons}}]{ouyang2017computing}%
  \BibitemOpen
  \bibfield  {author} {\bibinfo {author} {\bibnamefont {Ouyang}, \bibfnamefont
  {Yingkai}}, \bibinfo {author} {\bibfnamefont {Si-Hui}\ \bibnamefont {Tan}},
  \bibinfo {author} {\bibfnamefont {Liming}\ \bibnamefont {Zhao}}, and\
  \bibinfo {author} {\bibfnamefont {Joseph~F}\ \bibnamefont {Fitzsimons}}}
  (\bibinfo {year} {2017}),\ \bibfield  {title} {\enquote {\bibinfo {title}
  {Computing on quantum shared secrets},}\ }\href@noop {} {\bibfield  {journal}
  {\bibinfo  {journal} {Physical Review A}\ }\textbf {\bibinfo {volume}
  {96}}~(\bibinfo {number} {5}),\ \bibinfo {pages} {052333}}\BibitemShut
  {NoStop}%
\bibitem [{\citenamefont {Owens}\ \emph {et~al.}(2011)\citenamefont {Owens},
  \citenamefont {Broome}, \citenamefont {Biggerstaff}, \citenamefont {Goggin},
  \citenamefont {Fedrizzi}, \citenamefont {Linjordet}, \citenamefont {Ams},
  \citenamefont {Marshall}, \citenamefont {Twamley}, \citenamefont {Withford},\
  and\ \citenamefont {White}}]{bib:Owens11}%
  \BibitemOpen
  \bibfield  {author} {\bibinfo {author} {\bibnamefont {Owens}, \bibfnamefont
  {J~O}}, \bibinfo {author} {\bibfnamefont {M.~A.}\ \bibnamefont {Broome}},
  \bibinfo {author} {\bibfnamefont {D.~N.}\ \bibnamefont {Biggerstaff}},
  \bibinfo {author} {\bibfnamefont {M.~E.}\ \bibnamefont {Goggin}}, \bibinfo
  {author} {\bibfnamefont {A.}~\bibnamefont {Fedrizzi}}, \bibinfo {author}
  {\bibfnamefont {T.}~\bibnamefont {Linjordet}}, \bibinfo {author}
  {\bibfnamefont {M.}~\bibnamefont {Ams}}, \bibinfo {author} {\bibfnamefont
  {G.~D.}\ \bibnamefont {Marshall}}, \bibinfo {author} {\bibfnamefont
  {J.}~\bibnamefont {Twamley}}, \bibinfo {author} {\bibfnamefont {M.~J.}\
  \bibnamefont {Withford}}, and\ \bibinfo {author} {\bibfnamefont {A.~G.}\
  \bibnamefont {White}}} (\bibinfo {year} {2011}),\ \bibfield  {title}
  {\enquote {\bibinfo {title} {Two-photon quantum walks in an elliptical
  direct-write waveguide array},}\ }\href
  {https://doi.org/10.1088/1367-2630/13/7/075003} {\bibfield  {journal}
  {\bibinfo  {journal} {New Journal of Physics}\ }\textbf {\bibinfo {volume}
  {13}},\ \bibinfo {pages} {075003}}\BibitemShut {NoStop}%
\bibitem [{\citenamefont {Oxborrow}\ and\ \citenamefont
  {Sinclair}(2005)}]{bib:Oxborrow05}%
  \BibitemOpen
  \bibfield  {author} {\bibinfo {author} {\bibnamefont {Oxborrow},
  \bibfnamefont {M}}, and\ \bibinfo {author} {\bibfnamefont {A.~G.}\
  \bibnamefont {Sinclair}}} (\bibinfo {year} {2005}),\ \bibfield  {title}
  {\enquote {\bibinfo {title} {Single-photon sources},}\ }\href
  {https://doi.org/10.1080/00107510512331337936} {\bibfield  {journal}
  {\bibinfo  {journal} {Contemporary Physics}\ }\textbf {\bibinfo {volume}
  {46}},\ \bibinfo {pages} {173}}\BibitemShut {NoStop}%
\bibitem [{\citenamefont {Paik}\ \emph {et~al.}(2011)\citenamefont {Paik},
  \citenamefont {Schuster}, \citenamefont {Bishop}, \citenamefont {Kirchmair},
  \citenamefont {Catelani}, \citenamefont {Sears}, \citenamefont {Johnson},
  \citenamefont {Reagor}, \citenamefont {Frunzio}, \citenamefont {Glazman}
  \emph {et~al.}}]{bib:paik2011observation}%
  \BibitemOpen
  \bibfield  {author} {\bibinfo {author} {\bibnamefont {Paik}, \bibfnamefont
  {Hanhee}}, \bibinfo {author} {\bibfnamefont {DI}~\bibnamefont {Schuster}},
  \bibinfo {author} {\bibfnamefont {Lev~S}\ \bibnamefont {Bishop}}, \bibinfo
  {author} {\bibfnamefont {G}~\bibnamefont {Kirchmair}}, \bibinfo {author}
  {\bibfnamefont {G}~\bibnamefont {Catelani}}, \bibinfo {author} {\bibfnamefont
  {AP}~\bibnamefont {Sears}}, \bibinfo {author} {\bibfnamefont
  {BR}~\bibnamefont {Johnson}}, \bibinfo {author} {\bibfnamefont
  {MJ}~\bibnamefont {Reagor}}, \bibinfo {author} {\bibfnamefont
  {L}~\bibnamefont {Frunzio}}, \bibinfo {author} {\bibfnamefont
  {LI}~\bibnamefont {Glazman}},  \emph {et~al.}} (\bibinfo {year} {2011}),\
  \bibfield  {title} {\enquote {\bibinfo {title} {Observation of high coherence
  in josephson junction qubits measured in a three-dimensional circuit qed
  architecture},}\ }\href {https://doi.org/10.1103/physrevlett.107.240501}
  {\bibfield  {journal} {\bibinfo  {journal} {Physical Review Letters}\
  }\textbf {\bibinfo {volume} {107}},\ \bibinfo {pages} {240501}},\ \Eprint
  {https://arxiv.org/abs/arXiv:1105.4652v4} {arXiv:1105.4652v4} \BibitemShut
  {NoStop}%
\bibitem [{\citenamefont {Pan}\ \emph {et~al.}(2000)\citenamefont {Pan},
  \citenamefont {Bouwmeester}, \citenamefont {Daniell}, \citenamefont
  {Weinfurter},\ and\ \citenamefont {Zeilinger}}]{pan2000experimental}%
  \BibitemOpen
  \bibfield  {author} {\bibinfo {author} {\bibnamefont {Pan}, \bibfnamefont
  {Jian-Wei}}, \bibinfo {author} {\bibfnamefont {Dik}\ \bibnamefont
  {Bouwmeester}}, \bibinfo {author} {\bibfnamefont {Matthew}\ \bibnamefont
  {Daniell}}, \bibinfo {author} {\bibfnamefont {Harald}\ \bibnamefont
  {Weinfurter}}, and\ \bibinfo {author} {\bibfnamefont {Anton}\ \bibnamefont
  {Zeilinger}}} (\bibinfo {year} {2000}),\ \bibfield  {title} {\enquote
  {\bibinfo {title} {Experimental test of quantum nonlocality in three-photon
  greenberger--horne--zeilinger entanglement},}\ }\href@noop {} {\bibfield
  {journal} {\bibinfo  {journal} {Nature}\ }\textbf {\bibinfo {volume}
  {403}}~(\bibinfo {number} {6769}),\ \bibinfo {pages} {515--519}}\BibitemShut
  {NoStop}%
\bibitem [{\citenamefont {Pan}\ \emph {et~al.}(1998)\citenamefont {Pan},
  \citenamefont {Bouwmeester}, \citenamefont {Weinfurter},\ and\ \citenamefont
  {Zeilinger}}]{bib:PRL_80_3891}%
  \BibitemOpen
  \bibfield  {author} {\bibinfo {author} {\bibnamefont {Pan}, \bibfnamefont
  {Jian-Wei}}, \bibinfo {author} {\bibfnamefont {Dik}\ \bibnamefont
  {Bouwmeester}}, \bibinfo {author} {\bibfnamefont {Harald}\ \bibnamefont
  {Weinfurter}}, and\ \bibinfo {author} {\bibfnamefont {Anton}\ \bibnamefont
  {Zeilinger}}} (\bibinfo {year} {1998}),\ \bibfield  {title} {\enquote
  {\bibinfo {title} {Experimental entanglement swapping: Entangling photons
  that never interacted},}\ }\href
  {https://doi.org/10.1103/physrevlett.80.3891} {\bibfield  {journal} {\bibinfo
   {journal} {Physical Review Letters}\ }\textbf {\bibinfo {volume} {80}},\
  \bibinfo {pages} {3891}}\BibitemShut {NoStop}%
\bibitem [{\citenamefont {Pan}\ \emph {et~al.}(2012)\citenamefont {Pan},
  \citenamefont {Chen}, \citenamefont {Lu}, \citenamefont {Weinfurter},
  \citenamefont {Zeilinger},\ and\ \citenamefont {\ifmmode~\dot{Z}\else
  \.{Z}\fi{}ukowski}}]{bib:pan2012multiphoton}%
  \BibitemOpen
  \bibfield  {author} {\bibinfo {author} {\bibnamefont {Pan}, \bibfnamefont
  {Jian-Wei}}, \bibinfo {author} {\bibfnamefont {Zeng-Bing}\ \bibnamefont
  {Chen}}, \bibinfo {author} {\bibfnamefont {Chao-Yang}\ \bibnamefont {Lu}},
  \bibinfo {author} {\bibfnamefont {Harald}\ \bibnamefont {Weinfurter}},
  \bibinfo {author} {\bibfnamefont {Anton}\ \bibnamefont {Zeilinger}}, and\
  \bibinfo {author} {\bibfnamefont {Marek}\ \bibnamefont {\ifmmode~\dot{Z}\else
  \.{Z}\fi{}ukowski}}} (\bibinfo {year} {2012}),\ \bibfield  {title} {\enquote
  {\bibinfo {title} {Multiphoton entanglement and interferometry},}\ }\href
  {https://doi.org/10.1103/revmodphys.84.777} {\bibfield  {journal} {\bibinfo
  {journal} {Reviews in Modern Physics}\ }\textbf {\bibinfo {volume} {84}},\
  \bibinfo {pages} {777}},\ \Eprint {https://arxiv.org/abs/arXiv:0805.2853v2}
  {arXiv:0805.2853v2} \BibitemShut {NoStop}%
\bibitem [{\citenamefont {Pan}\ \emph {et~al.}(2003)\citenamefont {Pan},
  \citenamefont {Gasparoni}, \citenamefont {Ursin}, \citenamefont {Weihs},\
  and\ \citenamefont {Zeilinger}}]{bib:Pan03}%
  \BibitemOpen
  \bibfield  {author} {\bibinfo {author} {\bibnamefont {Pan}, \bibfnamefont
  {Jian-Wei}}, \bibinfo {author} {\bibfnamefont {Sara}\ \bibnamefont
  {Gasparoni}}, \bibinfo {author} {\bibfnamefont {Rupert}\ \bibnamefont
  {Ursin}}, \bibinfo {author} {\bibfnamefont {Gregor}\ \bibnamefont {Weihs}},
  and\ \bibinfo {author} {\bibfnamefont {Anton}\ \bibnamefont {Zeilinger}}}
  (\bibinfo {year} {2003}),\ \bibfield  {title} {\enquote {\bibinfo {title}
  {Experimental entanglement purification of arbitrary unknown states},}\
  }\href {https://doi.org/10.1038/nature01623} {\bibfield  {journal} {\bibinfo
  {journal} {Nature}\ }\textbf {\bibinfo {volume} {423}},\ \bibinfo {pages}
  {417}}\BibitemShut {NoStop}%
\bibitem [{\citenamefont {Pan}\ \emph {et~al.}(2001)\citenamefont {Pan},
  \citenamefont {Simon}, \citenamefont {Brukner},\ and\ \citenamefont
  {Zeilinger}}]{bib:Pan01}%
  \BibitemOpen
  \bibfield  {author} {\bibinfo {author} {\bibnamefont {Pan}, \bibfnamefont
  {Jian-Wei}}, \bibinfo {author} {\bibfnamefont {Christoph}\ \bibnamefont
  {Simon}}, \bibinfo {author} {\bibfnamefont {{\u C}aslav}\ \bibnamefont
  {Brukner}}, and\ \bibinfo {author} {\bibfnamefont {Anton}\ \bibnamefont
  {Zeilinger}}} (\bibinfo {year} {2001}),\ \bibfield  {title} {\enquote
  {\bibinfo {title} {Entanglement purification for quantum communication},}\
  }\href@noop {} {\bibfield  {journal} {\bibinfo  {journal} {Nature}\ }\textbf
  {\bibinfo {volume} {410}},\ \bibinfo {pages} {1067}}\BibitemShut {NoStop}%
\bibitem [{\citenamefont {Pantaleoni}\ \emph {et~al.}(2021)\citenamefont
  {Pantaleoni}, \citenamefont {Baragiola},\ and\ \citenamefont
  {Menicucci}}]{pantaleoni2021hidden}%
  \BibitemOpen
  \bibfield  {author} {\bibinfo {author} {\bibnamefont {Pantaleoni},
  \bibfnamefont {Giacomo}}, \bibinfo {author} {\bibfnamefont {Ben~Q.}\
  \bibnamefont {Baragiola}}, and\ \bibinfo {author} {\bibfnamefont
  {Nicolas~C.}\ \bibnamefont {Menicucci}}} (\bibinfo {year} {2021}),\ \bibfield
   {title} {\enquote {\bibinfo {title} {Hidden qubit cluster states},}\ }\href
  {https://doi.org/10.1103/PhysRevA.104.012431} {\bibfield  {journal} {\bibinfo
   {journal} {Phys. Rev. A}\ }\textbf {\bibinfo {volume} {104}},\ \bibinfo
  {pages} {012431}}\BibitemShut {NoStop}%
\bibitem [{\citenamefont {Paris}\ and\ \citenamefont
  {Rehacek}(2004)}]{paris2004quantum}%
  \BibitemOpen
  \bibfield  {author} {\bibinfo {author} {\bibnamefont {Paris}, \bibfnamefont
  {Matteo}}, and\ \bibinfo {author} {\bibfnamefont {Jaroslav}\ \bibnamefont
  {Rehacek}}} (\bibinfo {year} {2004}),\ \href@noop {} {\emph {\bibinfo {title}
  {Quantum state estimation}}},\ Vol.\ \bibinfo {volume} {649}\ (\bibinfo
  {publisher} {Springer Science \& Business Media})\BibitemShut {NoStop}%
\bibitem [{\citenamefont {Paszkiewicz}\ and\ \citenamefont
  {Studholme}(2010)}]{paszkiewicz2010novo}%
  \BibitemOpen
  \bibfield  {author} {\bibinfo {author} {\bibnamefont {Paszkiewicz},
  \bibfnamefont {Konrad}}, and\ \bibinfo {author} {\bibfnamefont {David~J}\
  \bibnamefont {Studholme}}} (\bibinfo {year} {2010}),\ \bibfield  {title}
  {\enquote {\bibinfo {title} {De novo assembly of short sequence reads},}\
  }\href@noop {} {\bibfield  {journal} {\bibinfo  {journal} {Briefings in
  bioinformatics}\ }\textbf {\bibinfo {volume} {11}}~(\bibinfo {number} {5}),\
  \bibinfo {pages} {457--472}}\BibitemShut {NoStop}%
\bibitem [{\citenamefont {Peev}\ \emph
  {et~al.}(2009{\natexlab{a}})\citenamefont {Peev}, \citenamefont {Pacher},
  \citenamefont {All{\'e}aume}, \citenamefont {Barreiro}, \citenamefont
  {Bouda}, \citenamefont {Boxleitner}, \citenamefont {Debuisschert},
  \citenamefont {Diamanti}, \citenamefont {Dianati}, \citenamefont {Dynes},
  \citenamefont {Fasel}, \citenamefont {Fossier}, \citenamefont {F{\"u}rst},
  \citenamefont {Gautier}, \citenamefont {Gay}, \citenamefont {Gisin},
  \citenamefont {Grangier}, \citenamefont {Happe}, \citenamefont {Hasani},
  \citenamefont {Hentschel}, \citenamefont {H{\"u}bel}, \citenamefont {Humer},
  \citenamefont {L{\"a}nger}, \citenamefont {Legr{\'e}}, \citenamefont
  {Lieger}, \citenamefont {Lodewyck}, \citenamefont {Lor{\"u}nser},
  \citenamefont {L{\"u}tkenhaus}, \citenamefont {Marhold}, \citenamefont
  {Matyus}, \citenamefont {Maurhart}, \citenamefont {Monat}, \citenamefont
  {Nauerth}, \citenamefont {Page}, \citenamefont {Poppe}, \citenamefont
  {Querasser}, \citenamefont {Ribordy}, \citenamefont {Robyr}, \citenamefont
  {Salvail}, \citenamefont {Sharpe}, \citenamefont {Shields}, \citenamefont
  {Stucki}, \citenamefont {Suda}, \citenamefont {Tamas}, \citenamefont
  {Themel}, \citenamefont {Thew}, \citenamefont {Thoma}, \citenamefont
  {Treiber}, \citenamefont {Trinkler}, \citenamefont {Tualle-Brouri},
  \citenamefont {Vannel}, \citenamefont {Walenta}, \citenamefont {Weier},
  \citenamefont {Weinfurter}, \citenamefont {Wimberger}, \citenamefont {Yuan},
  \citenamefont {Zbinden},\ and\ \citenamefont {Zeilinger}}]{SD-Peev:2009aa}%
  \BibitemOpen
  \bibfield  {author} {\bibinfo {author} {\bibnamefont {Peev}, \bibfnamefont
  {M}}, \bibinfo {author} {\bibfnamefont {C}~\bibnamefont {Pacher}}, \bibinfo
  {author} {\bibfnamefont {R}~\bibnamefont {All{\'e}aume}}, \bibinfo {author}
  {\bibfnamefont {C}~\bibnamefont {Barreiro}}, \bibinfo {author} {\bibfnamefont
  {J}~\bibnamefont {Bouda}}, \bibinfo {author} {\bibfnamefont {W}~\bibnamefont
  {Boxleitner}}, \bibinfo {author} {\bibfnamefont {T}~\bibnamefont
  {Debuisschert}}, \bibinfo {author} {\bibfnamefont {E}~\bibnamefont
  {Diamanti}}, \bibinfo {author} {\bibfnamefont {M}~\bibnamefont {Dianati}},
  \bibinfo {author} {\bibfnamefont {J~F}\ \bibnamefont {Dynes}}, \bibinfo
  {author} {\bibfnamefont {S}~\bibnamefont {Fasel}}, \bibinfo {author}
  {\bibfnamefont {S}~\bibnamefont {Fossier}}, \bibinfo {author} {\bibfnamefont
  {M}~\bibnamefont {F{\"u}rst}}, \bibinfo {author} {\bibfnamefont {J-D}\
  \bibnamefont {Gautier}}, \bibinfo {author} {\bibfnamefont {O}~\bibnamefont
  {Gay}}, \bibinfo {author} {\bibfnamefont {N}~\bibnamefont {Gisin}}, \bibinfo
  {author} {\bibfnamefont {P}~\bibnamefont {Grangier}}, \bibinfo {author}
  {\bibfnamefont {A}~\bibnamefont {Happe}}, \bibinfo {author} {\bibfnamefont
  {Y}~\bibnamefont {Hasani}}, \bibinfo {author} {\bibfnamefont {M}~\bibnamefont
  {Hentschel}}, \bibinfo {author} {\bibfnamefont {H}~\bibnamefont {H{\"u}bel}},
  \bibinfo {author} {\bibfnamefont {G}~\bibnamefont {Humer}}, \bibinfo {author}
  {\bibfnamefont {T}~\bibnamefont {L{\"a}nger}}, \bibinfo {author}
  {\bibfnamefont {M}~\bibnamefont {Legr{\'e}}}, \bibinfo {author}
  {\bibfnamefont {R}~\bibnamefont {Lieger}}, \bibinfo {author} {\bibfnamefont
  {J}~\bibnamefont {Lodewyck}}, \bibinfo {author} {\bibfnamefont
  {T}~\bibnamefont {Lor{\"u}nser}}, \bibinfo {author} {\bibfnamefont
  {N}~\bibnamefont {L{\"u}tkenhaus}}, \bibinfo {author} {\bibfnamefont
  {A}~\bibnamefont {Marhold}}, \bibinfo {author} {\bibfnamefont
  {T}~\bibnamefont {Matyus}}, \bibinfo {author} {\bibfnamefont {O}~\bibnamefont
  {Maurhart}}, \bibinfo {author} {\bibfnamefont {L}~\bibnamefont {Monat}},
  \bibinfo {author} {\bibfnamefont {S}~\bibnamefont {Nauerth}}, \bibinfo
  {author} {\bibfnamefont {J-B}\ \bibnamefont {Page}}, \bibinfo {author}
  {\bibfnamefont {A}~\bibnamefont {Poppe}}, \bibinfo {author} {\bibfnamefont
  {E}~\bibnamefont {Querasser}}, \bibinfo {author} {\bibfnamefont
  {G}~\bibnamefont {Ribordy}}, \bibinfo {author} {\bibfnamefont
  {S}~\bibnamefont {Robyr}}, \bibinfo {author} {\bibfnamefont {L}~\bibnamefont
  {Salvail}}, \bibinfo {author} {\bibfnamefont {A~W}\ \bibnamefont {Sharpe}},
  \bibinfo {author} {\bibfnamefont {A~J}\ \bibnamefont {Shields}}, \bibinfo
  {author} {\bibfnamefont {D}~\bibnamefont {Stucki}}, \bibinfo {author}
  {\bibfnamefont {M}~\bibnamefont {Suda}}, \bibinfo {author} {\bibfnamefont
  {C}~\bibnamefont {Tamas}}, \bibinfo {author} {\bibfnamefont {T}~\bibnamefont
  {Themel}}, \bibinfo {author} {\bibfnamefont {R~T}\ \bibnamefont {Thew}},
  \bibinfo {author} {\bibfnamefont {Y}~\bibnamefont {Thoma}}, \bibinfo {author}
  {\bibfnamefont {A}~\bibnamefont {Treiber}}, \bibinfo {author} {\bibfnamefont
  {P}~\bibnamefont {Trinkler}}, \bibinfo {author} {\bibfnamefont
  {R}~\bibnamefont {Tualle-Brouri}}, \bibinfo {author} {\bibfnamefont
  {F}~\bibnamefont {Vannel}}, \bibinfo {author} {\bibfnamefont {N}~\bibnamefont
  {Walenta}}, \bibinfo {author} {\bibfnamefont {H}~\bibnamefont {Weier}},
  \bibinfo {author} {\bibfnamefont {H}~\bibnamefont {Weinfurter}}, \bibinfo
  {author} {\bibfnamefont {I}~\bibnamefont {Wimberger}}, \bibinfo {author}
  {\bibfnamefont {Z~L}\ \bibnamefont {Yuan}}, \bibinfo {author} {\bibfnamefont
  {H}~\bibnamefont {Zbinden}}, and\ \bibinfo {author} {\bibfnamefont
  {A}~\bibnamefont {Zeilinger}}} (\bibinfo {year} {2009}{\natexlab{a}}),\
  \bibfield  {title} {\enquote {\bibinfo {title} {The secoqc quantum key
  distribution network in vienna},}\ }\href
  {https://doi.org/10.1088/1367-2630/11/7/075001} {\bibfield  {journal}
  {\bibinfo  {journal} {New Journal of Physics}\ }\textbf {\bibinfo {volume}
  {11}},\ \bibinfo {pages} {075001}}\BibitemShut {NoStop}%
\bibitem [{\citenamefont {Peev}\ \emph
  {et~al.}(2009{\natexlab{b}})\citenamefont {Peev}, \citenamefont {Pacher},
  \citenamefont {All{\'e}aume}, \citenamefont {Barreiro}, \citenamefont
  {Bouda}, \citenamefont {Boxleitner}, \citenamefont {Debuisschert},
  \citenamefont {Diamanti}, \citenamefont {Dianati}, \citenamefont {Dynes}
  \emph {et~al.}}]{bib:NJP_11_075001}%
  \BibitemOpen
  \bibfield  {author} {\bibinfo {author} {\bibnamefont {Peev}, \bibfnamefont
  {Momtchil}}, \bibinfo {author} {\bibfnamefont {Christoph}\ \bibnamefont
  {Pacher}}, \bibinfo {author} {\bibfnamefont {Romain}\ \bibnamefont
  {All{\'e}aume}}, \bibinfo {author} {\bibfnamefont {Claudio}\ \bibnamefont
  {Barreiro}}, \bibinfo {author} {\bibfnamefont {Jan}\ \bibnamefont {Bouda}},
  \bibinfo {author} {\bibfnamefont {W}~\bibnamefont {Boxleitner}}, \bibinfo
  {author} {\bibfnamefont {Thierry}\ \bibnamefont {Debuisschert}}, \bibinfo
  {author} {\bibfnamefont {Eleni}\ \bibnamefont {Diamanti}}, \bibinfo {author}
  {\bibfnamefont {M}~\bibnamefont {Dianati}}, \bibinfo {author} {\bibfnamefont
  {JF}~\bibnamefont {Dynes}},  \emph {et~al.}} (\bibinfo {year}
  {2009}{\natexlab{b}}),\ \bibfield  {title} {\enquote {\bibinfo {title} {The
  secoqc quantum key distribution network in vienna},}\ }\href
  {https://doi.org/10.1364/ofc.2009.othl2} {\bibfield  {journal} {\bibinfo
  {journal} {New Journal of Physics}\ }\textbf {\bibinfo {volume} {11}},\
  \bibinfo {pages} {075001}}\BibitemShut {NoStop}%
\bibitem [{\citenamefont {Peng}\ \emph {et~al.}(2005)\citenamefont {Peng},
  \citenamefont {Yang}, \citenamefont {Bao}, \citenamefont {Zhang},
  \citenamefont {Jin}, \citenamefont {Feng}, \citenamefont {Yang},
  \citenamefont {Yang}, \citenamefont {Yin}, \citenamefont {Zhang} \emph
  {et~al.}}]{bib:PRL_94_150501}%
  \BibitemOpen
  \bibfield  {author} {\bibinfo {author} {\bibnamefont {Peng}, \bibfnamefont
  {Cheng-Zhi}}, \bibinfo {author} {\bibfnamefont {Tao}\ \bibnamefont {Yang}},
  \bibinfo {author} {\bibfnamefont {Xiao-Hui}\ \bibnamefont {Bao}}, \bibinfo
  {author} {\bibfnamefont {Jun}\ \bibnamefont {Zhang}}, \bibinfo {author}
  {\bibfnamefont {Xian-Min}\ \bibnamefont {Jin}}, \bibinfo {author}
  {\bibfnamefont {Fa-Yong}\ \bibnamefont {Feng}}, \bibinfo {author}
  {\bibfnamefont {Bin}\ \bibnamefont {Yang}}, \bibinfo {author} {\bibfnamefont
  {Jian}\ \bibnamefont {Yang}}, \bibinfo {author} {\bibfnamefont {Juan}\
  \bibnamefont {Yin}}, \bibinfo {author} {\bibfnamefont {Qiang}\ \bibnamefont
  {Zhang}},  \emph {et~al.}} (\bibinfo {year} {2005}),\ \bibfield  {title}
  {\enquote {\bibinfo {title} {Experimental free-space distribution of
  entangled photon pairs over 13 km: towards satellite-based global quantum
  communication},}\ }\href {https://doi.org/10.1103/physrevlett.94.150501}
  {\bibfield  {journal} {\bibinfo  {journal} {Physical Review Letters}\
  }\textbf {\bibinfo {volume} {94}},\ \bibinfo {pages} {150501}}\BibitemShut
  {NoStop}%
\bibitem [{\citenamefont {Peng}\ \emph {et~al.}(2007)\citenamefont {Peng},
  \citenamefont {Zhang}, \citenamefont {Yang}, \citenamefont {Gao},
  \citenamefont {Ma}, \citenamefont {Yin}, \citenamefont {Zeng}, \citenamefont
  {Yang}, \citenamefont {Wang},\ and\ \citenamefont {Pan}}]{bib:PRL_98_010505}%
  \BibitemOpen
  \bibfield  {author} {\bibinfo {author} {\bibnamefont {Peng}, \bibfnamefont
  {Cheng-Zhi}}, \bibinfo {author} {\bibfnamefont {Jun}\ \bibnamefont {Zhang}},
  \bibinfo {author} {\bibfnamefont {Dong}\ \bibnamefont {Yang}}, \bibinfo
  {author} {\bibfnamefont {Wei-Bo}\ \bibnamefont {Gao}}, \bibinfo {author}
  {\bibfnamefont {Huai-Xin}\ \bibnamefont {Ma}}, \bibinfo {author}
  {\bibfnamefont {Hao}\ \bibnamefont {Yin}}, \bibinfo {author} {\bibfnamefont
  {He-Ping}\ \bibnamefont {Zeng}}, \bibinfo {author} {\bibfnamefont {Tao}\
  \bibnamefont {Yang}}, \bibinfo {author} {\bibfnamefont {Xiang-Bin}\
  \bibnamefont {Wang}}, and\ \bibinfo {author} {\bibfnamefont {Jian-Wei}\
  \bibnamefont {Pan}}} (\bibinfo {year} {2007}),\ \bibfield  {title} {\enquote
  {\bibinfo {title} {Experimental long-distance decoy-state quantum key
  distribution based on polarization encoding},}\ }\href
  {https://doi.org/10.1103/physrevlett.98.010505} {\bibfield  {journal}
  {\bibinfo  {journal} {Physical Review Letters}\ }\textbf {\bibinfo {volume}
  {98}},\ \bibinfo {pages} {010505}},\ \Eprint
  {https://arxiv.org/abs/arXiv:quant-ph/0607129v3} {arXiv:quant-ph/0607129v3}
  \BibitemShut {NoStop}%
\bibitem [{\citenamefont {Peres}(2000)}]{bib:peres2000delayed}%
  \BibitemOpen
  \bibfield  {author} {\bibinfo {author} {\bibnamefont {Peres}, \bibfnamefont
  {Asher}}} (\bibinfo {year} {2000}),\ \bibfield  {title} {\enquote {\bibinfo
  {title} {Delayed choice for entanglement swapping},}\ }\href
  {https://doi.org/10.1080/095003400148105} {\bibfield  {journal} {\bibinfo
  {journal} {Journal of Modern Optics}\ }\textbf {\bibinfo {volume} {47}},\
  \bibinfo {pages} {139}},\ \Eprint
  {https://arxiv.org/abs/arXiv:quant-ph/9904042v1} {arXiv:quant-ph/9904042v1}
  \BibitemShut {NoStop}%
\bibitem [{\citenamefont {Peruzzo}\ \emph {et~al.}(2010)\citenamefont
  {Peruzzo}, \citenamefont {Lobino}, \citenamefont {Matthews}, \citenamefont
  {Matsuda}, \citenamefont {Politi}, \citenamefont {Poulios}, \citenamefont
  {Zhou}, \citenamefont {Lahini}, \citenamefont {Ismail}, \citenamefont
  {WÃ¶rhoff}, \citenamefont {Bromberg}, \citenamefont {Silberberg},
  \citenamefont {Thompson},\ and\ \citenamefont {O'Brien}}]{bib:PeruzzoQW}%
  \BibitemOpen
  \bibfield  {author} {\bibinfo {author} {\bibnamefont {Peruzzo}, \bibfnamefont
  {Alberto}}, \bibinfo {author} {\bibfnamefont {Mirko}\ \bibnamefont {Lobino}},
  \bibinfo {author} {\bibfnamefont {Jonathan C.~F.}\ \bibnamefont {Matthews}},
  \bibinfo {author} {\bibfnamefont {Nobuyuki}\ \bibnamefont {Matsuda}},
  \bibinfo {author} {\bibfnamefont {Alberto}\ \bibnamefont {Politi}}, \bibinfo
  {author} {\bibfnamefont {Konstantinos}\ \bibnamefont {Poulios}}, \bibinfo
  {author} {\bibfnamefont {Xiao-Qi}\ \bibnamefont {Zhou}}, \bibinfo {author}
  {\bibfnamefont {Yoav}\ \bibnamefont {Lahini}}, \bibinfo {author}
  {\bibfnamefont {Nur}\ \bibnamefont {Ismail}}, \bibinfo {author}
  {\bibfnamefont {Kerstin}\ \bibnamefont {WÃ¶rhoff}}, \bibinfo {author}
  {\bibfnamefont {Yaron}\ \bibnamefont {Bromberg}}, \bibinfo {author}
  {\bibfnamefont {Yaron}\ \bibnamefont {Silberberg}}, \bibinfo {author}
  {\bibfnamefont {Mark~G.}\ \bibnamefont {Thompson}}, and\ \bibinfo {author}
  {\bibfnamefont {Jeremy~L.}\ \bibnamefont {O'Brien}}} (\bibinfo {year}
  {2010}),\ \bibfield  {title} {\enquote {\bibinfo {title} {Quantum walks of
  correlated photons},}\ }\href {https://doi.org/10.1126/science.1193515}
  {\bibfield  {journal} {\bibinfo  {journal} {Science}\ }\textbf {\bibinfo
  {volume} {329}},\ \bibinfo {pages} {1500}}\BibitemShut {NoStop}%
\bibitem [{\citenamefont {Peteiro-Barral}\ and\ \citenamefont
  {Guijarro-Berdi{\~n}as}(2013)}]{bib:peteiro2013survey}%
  \BibitemOpen
  \bibfield  {author} {\bibinfo {author} {\bibnamefont {Peteiro-Barral},
  \bibfnamefont {Diego}}, and\ \bibinfo {author} {\bibfnamefont {Bertha}\
  \bibnamefont {Guijarro-Berdi{\~n}as}}} (\bibinfo {year} {2013}),\ \bibfield
  {title} {\enquote {\bibinfo {title} {A survey of methods for distributed
  machine learning},}\ }\href@noop {} {\bibfield  {journal} {\bibinfo
  {journal} {Progress in Artificial Intelligence}\ }\textbf {\bibinfo {volume}
  {2}},\ \bibinfo {pages} {1}}\BibitemShut {NoStop}%
\bibitem [{\citenamefont {Pfaff}\ \emph {et~al.}(2014)\citenamefont {Pfaff},
  \citenamefont {Hensen}, \citenamefont {Bernien}, \citenamefont {van Dam},
  \citenamefont {Blok}, \citenamefont {Taminiau}, \citenamefont {Tiggelman},
  \citenamefont {Schouten}, \citenamefont {Markham}, \citenamefont {Twitchen}
  \emph {et~al.}}]{bib:Science_345_532}%
  \BibitemOpen
  \bibfield  {author} {\bibinfo {author} {\bibnamefont {Pfaff}, \bibfnamefont
  {Wolfgang}}, \bibinfo {author} {\bibfnamefont {BJ}~\bibnamefont {Hensen}},
  \bibinfo {author} {\bibfnamefont {Hannes}\ \bibnamefont {Bernien}}, \bibinfo
  {author} {\bibfnamefont {Suzanne~B}\ \bibnamefont {van Dam}}, \bibinfo
  {author} {\bibfnamefont {Machiel~S}\ \bibnamefont {Blok}}, \bibinfo {author}
  {\bibfnamefont {Tim~H}\ \bibnamefont {Taminiau}}, \bibinfo {author}
  {\bibfnamefont {Marijn~J}\ \bibnamefont {Tiggelman}}, \bibinfo {author}
  {\bibfnamefont {Raymond~N}\ \bibnamefont {Schouten}}, \bibinfo {author}
  {\bibfnamefont {Matthew}\ \bibnamefont {Markham}}, \bibinfo {author}
  {\bibfnamefont {Daniel~J}\ \bibnamefont {Twitchen}},  \emph {et~al.}}
  (\bibinfo {year} {2014}),\ \bibfield  {title} {\enquote {\bibinfo {title}
  {Unconditional quantum teleportation between distant solid-state quantum
  bits},}\ }\href {https://doi.org/10.1126/science.1253512} {\bibfield
  {journal} {\bibinfo  {journal} {Science}\ }\textbf {\bibinfo {volume}
  {345}},\ \bibinfo {pages} {532}}\BibitemShut {NoStop}%
\bibitem [{\citenamefont {Pirandola}\ \emph {et~al.}(2019)\citenamefont
  {Pirandola}, \citenamefont {Andersen}, \citenamefont {Banchi}, \citenamefont
  {Berta}, \citenamefont {Bunandar}, \citenamefont {Colbeck}, \citenamefont
  {Englund}, \citenamefont {Gehring}, \citenamefont {Lupo}, \citenamefont
  {Ottaviani} \emph {et~al.}}]{pirandola2019advances}%
  \BibitemOpen
  \bibfield  {author} {\bibinfo {author} {\bibnamefont {Pirandola},
  \bibfnamefont {S}}, \bibinfo {author} {\bibfnamefont {UL}~\bibnamefont
  {Andersen}}, \bibinfo {author} {\bibfnamefont {L}~\bibnamefont {Banchi}},
  \bibinfo {author} {\bibfnamefont {M}~\bibnamefont {Berta}}, \bibinfo {author}
  {\bibfnamefont {D}~\bibnamefont {Bunandar}}, \bibinfo {author} {\bibfnamefont
  {R}~\bibnamefont {Colbeck}}, \bibinfo {author} {\bibfnamefont
  {D}~\bibnamefont {Englund}}, \bibinfo {author} {\bibfnamefont
  {T}~\bibnamefont {Gehring}}, \bibinfo {author} {\bibfnamefont
  {C}~\bibnamefont {Lupo}}, \bibinfo {author} {\bibfnamefont {C}~\bibnamefont
  {Ottaviani}},  \emph {et~al.}} (\bibinfo {year} {2019}),\ \bibfield  {title}
  {\enquote {\bibinfo {title} {Advances in quantum cryptography},}\ }\href@noop
  {} {\bibinfo  {journal} {arXiv preprint arXiv:1906.01645}\ }\BibitemShut
  {NoStop}%
\bibitem [{\citenamefont {Pittman}\ \emph {et~al.}(2003)\citenamefont
  {Pittman}, \citenamefont {Fitch}, \citenamefont {Jacobs},\ and\ \citenamefont
  {Franson}}]{bib:Pittman03}%
  \BibitemOpen
\bibfield  {journal} {  }\bibfield  {author} {\bibinfo {author} {\bibnamefont
  {Pittman}, \bibfnamefont {T~B}}, \bibinfo {author} {\bibfnamefont {M.~J.}\
  \bibnamefont {Fitch}}, \bibinfo {author} {\bibfnamefont {B.~C.}\ \bibnamefont
  {Jacobs}}, and\ \bibinfo {author} {\bibfnamefont {J.~D.}\ \bibnamefont
  {Franson}}} (\bibinfo {year} {2003}),\ \bibfield  {title} {\enquote {\bibinfo
  {title} {Experimental controlled-not logic gate for single photons in the
  coincidence basis},}\ }\href {https://doi.org/10.1103/physreva.68.032316}
  {\bibfield  {journal} {\bibinfo  {journal} {Physical Review A}\ }\textbf
  {\bibinfo {volume} {68}},\ \bibinfo {pages} {032316}},\ \Eprint
  {https://arxiv.org/abs/arXiv:quant-ph/0303095v2} {arXiv:quant-ph/0303095v2}
  \BibitemShut {NoStop}%
\bibitem [{\citenamefont {Pittman}\ \emph {et~al.}(2001)\citenamefont
  {Pittman}, \citenamefont {Jacobs},\ and\ \citenamefont
  {Franson}}]{bib:Pittman01}%
  \BibitemOpen
  \bibfield  {author} {\bibinfo {author} {\bibnamefont {Pittman}, \bibfnamefont
  {T~B}}, \bibinfo {author} {\bibfnamefont {B.~C.}\ \bibnamefont {Jacobs}},
  and\ \bibinfo {author} {\bibfnamefont {J.~D.}\ \bibnamefont {Franson}}}
  (\bibinfo {year} {2001}),\ \bibfield  {title} {\enquote {\bibinfo {title}
  {Probabilistic quantum logic operations using polarizing beam splitters},}\
  }\href {https://doi.org/10.1103/physreva.64.062311} {\bibfield  {journal}
  {\bibinfo  {journal} {Physical Review A}\ }\textbf {\bibinfo {volume} {64}},\
  \bibinfo {pages} {062311}},\ \Eprint
  {https://arxiv.org/abs/arXiv:quant-ph/0107091v2} {arXiv:quant-ph/0107091v2}
  \BibitemShut {NoStop}%
\bibitem [{\citenamefont {Pittman}\ and\ \citenamefont
  {Franson}(2002)}]{bib:pittman2002cyclical}%
  \BibitemOpen
  \bibfield  {author} {\bibinfo {author} {\bibnamefont {Pittman}, \bibfnamefont
  {TB}}, and\ \bibinfo {author} {\bibfnamefont {JD}~\bibnamefont {Franson}}}
  (\bibinfo {year} {2002}),\ \bibfield  {title} {\enquote {\bibinfo {title}
  {Cyclical quantum memory for photonic qubits},}\ }\href
  {https://doi.org/10.1103/physreva.66.062302} {\bibfield  {journal} {\bibinfo
  {journal} {Physical Review A}\ }\textbf {\bibinfo {volume} {66}},\ \bibinfo
  {pages} {062302}},\ \Eprint {https://arxiv.org/abs/arXiv:quant-ph/0207041v1}
  {arXiv:quant-ph/0207041v1} \BibitemShut {NoStop}%
\bibitem [{\citenamefont {Pittman}\ \emph {et~al.}(2002)\citenamefont
  {Pittman}, \citenamefont {Jacobs},\ and\ \citenamefont
  {Franson}}]{bib:pittman2002single}%
  \BibitemOpen
  \bibfield  {author} {\bibinfo {author} {\bibnamefont {Pittman}, \bibfnamefont
  {TB}}, \bibinfo {author} {\bibfnamefont {BC}~\bibnamefont {Jacobs}}, and\
  \bibinfo {author} {\bibfnamefont {JD}~\bibnamefont {Franson}}} (\bibinfo
  {year} {2002}),\ \bibfield  {title} {\enquote {\bibinfo {title} {Single
  photons on pseudodemand from stored parametric down-conversion},}\ }\href
  {https://doi.org/10.1103/physreva.66.042303} {\bibfield  {journal} {\bibinfo
  {journal} {Physical Review A}\ }\textbf {\bibinfo {volume} {66}},\ \bibinfo
  {pages} {042303}},\ \Eprint {https://arxiv.org/abs/arXiv:quant-ph/0205103v1}
  {arXiv:quant-ph/0205103v1} \BibitemShut {NoStop}%
\bibitem [{\citenamefont {Politi}\ \emph {et~al.}(2008)\citenamefont {Politi},
  \citenamefont {Cryan}, \citenamefont {Rarity}, \citenamefont {Yu},\ and\
  \citenamefont {O'Brien}}]{bib:politi2008silica}%
  \BibitemOpen
  \bibfield  {author} {\bibinfo {author} {\bibnamefont {Politi}, \bibfnamefont
  {Alberto}}, \bibinfo {author} {\bibfnamefont {Martin~J}\ \bibnamefont
  {Cryan}}, \bibinfo {author} {\bibfnamefont {John~G}\ \bibnamefont {Rarity}},
  \bibinfo {author} {\bibfnamefont {Siyuan}\ \bibnamefont {Yu}}, and\ \bibinfo
  {author} {\bibfnamefont {Jeremy~L}\ \bibnamefont {O'Brien}}} (\bibinfo {year}
  {2008}),\ \bibfield  {title} {\enquote {\bibinfo {title} {Silica-on-silicon
  waveguide quantum circuits},}\ }\href
  {https://doi.org/10.1126/science.1155441} {\bibfield  {journal} {\bibinfo
  {journal} {Science}\ }\textbf {\bibinfo {volume} {320}},\ \bibinfo {pages}
  {646}},\ \Eprint {https://arxiv.org/abs/arXiv:0802.0136v1}
  {arXiv:0802.0136v1} \BibitemShut {NoStop}%
\bibitem [{\citenamefont {Popkin}(2017)}]{bib:popkin17}%
  \BibitemOpen
  \bibfield  {author} {\bibinfo {author} {\bibnamefont {Popkin}, \bibfnamefont
  {Gabriel}}} (\bibinfo {year} {2017}),\ \bibfield  {title} {\enquote {\bibinfo
  {title} {China's quantum satellite achieves `spooky action' at record
  distance},}\ }\href {https://doi.org/10.1126/science.aan6972} {\bibfield
  {journal} {\bibinfo  {journal} {Science News}\
  }10.1126/science.aan6972}\BibitemShut {NoStop}%
\bibitem [{\citenamefont {Poundstone}(1993)}]{bib:Poundstone93}%
  \BibitemOpen
  \bibfield  {author} {\bibinfo {author} {\bibnamefont {Poundstone},
  \bibfnamefont {W}}} (\bibinfo {year} {1993}),\ \href@noop {} {\emph {\bibinfo
  {title} {Prisoner's Dilemma/John von Neumann, Game Theory and the Puzzle of
  the Bomb}}}\ (\bibinfo  {publisher} {Anchor})\BibitemShut {NoStop}%
\bibitem [{\citenamefont {Preskill}(2000)}]{bib:preskill2000quantum}%
  \BibitemOpen
  \bibfield  {author} {\bibinfo {author} {\bibnamefont {Preskill},
  \bibfnamefont {John}}} (\bibinfo {year} {2000}),\ \bibfield  {title}
  {\enquote {\bibinfo {title} {Quantum clock synchronization and quantum error
  correction},}\ }\href@noop {} {\ }\Eprint
  {https://arxiv.org/abs/quant-ph/0010098} {quant-ph/0010098} \BibitemShut
  {NoStop}%
\bibitem [{\citenamefont
  {Preskill}(2018{\natexlab{a}})}]{SD-Preskill2018quantumcomputingin}%
  \BibitemOpen
  \bibfield  {author} {\bibinfo {author} {\bibnamefont {Preskill},
  \bibfnamefont {John}}} (\bibinfo {year} {2018}{\natexlab{a}}),\ \bibfield
  {title} {\enquote {\bibinfo {title} {Quantum computing in the {NISQ} era and
  beyond},}\ }\href {https://doi.org/10.22331/q-2018-08-06-79} {\bibfield
  {journal} {\bibinfo  {journal} {Quantum}\ }\textbf {\bibinfo {volume} {2}},\
  \bibinfo {pages} {79}}\BibitemShut {NoStop}%
\bibitem [{\citenamefont
  {Preskill}(2018{\natexlab{b}})}]{bib:preskill2018quantum}%
  \BibitemOpen
  \bibfield  {author} {\bibinfo {author} {\bibnamefont {Preskill},
  \bibfnamefont {John}}} (\bibinfo {year} {2018}{\natexlab{b}}),\ \bibfield
  {title} {\enquote {\bibinfo {title} {Quantum computing in the nisq era and
  beyond},}\ }\href {https://doi.org/10.22331/q-2018-08-06-79} {\
  10.22331/q-2018-08-06-79},\ \Eprint {https://arxiv.org/abs/arXiv:1801.00862}
  {arXiv:1801.00862} \BibitemShut {NoStop}%
\bibitem [{\citenamefont {Qi}\ \emph {et~al.}(2005)\citenamefont {Qi},
  \citenamefont {Fung}, \citenamefont {Lo},\ and\ \citenamefont
  {Ma}}]{bib:qi2005time}%
  \BibitemOpen
  \bibfield  {author} {\bibinfo {author} {\bibnamefont {Qi}, \bibfnamefont
  {Bing}}, \bibinfo {author} {\bibfnamefont {Chi-Hang~Fred}\ \bibnamefont
  {Fung}}, \bibinfo {author} {\bibfnamefont {Hoi-Kwong}\ \bibnamefont {Lo}},
  and\ \bibinfo {author} {\bibfnamefont {Xiongfeng}\ \bibnamefont {Ma}}}
  (\bibinfo {year} {2005}),\ \bibfield  {title} {\enquote {\bibinfo {title}
  {Time-shift attack in practical quantum cryptosystems},}\ }\href@noop {} {\
  }\Eprint {https://arxiv.org/abs/arXiv:quant-ph/0512080}
  {arXiv:quant-ph/0512080} \BibitemShut {NoStop}%
\bibitem [{\citenamefont {Qiang}\ \emph {et~al.}(2018)\citenamefont {Qiang},
  \citenamefont {Zhou}, \citenamefont {Wang}, \citenamefont {Wilkes},
  \citenamefont {Loke}, \citenamefont {O'Gara}, \citenamefont {Kling},
  \citenamefont {Marshall}, \citenamefont {Santagati}, \citenamefont {Ralph}
  \emph {et~al.}}]{bib:qiang2018large}%
  \BibitemOpen
  \bibfield  {author} {\bibinfo {author} {\bibnamefont {Qiang}, \bibfnamefont
  {Xiaogang}}, \bibinfo {author} {\bibfnamefont {Xiaoqi}\ \bibnamefont {Zhou}},
  \bibinfo {author} {\bibfnamefont {Jianwei}\ \bibnamefont {Wang}}, \bibinfo
  {author} {\bibfnamefont {Callum~M}\ \bibnamefont {Wilkes}}, \bibinfo {author}
  {\bibfnamefont {Thomas}\ \bibnamefont {Loke}}, \bibinfo {author}
  {\bibfnamefont {Sean}\ \bibnamefont {O'Gara}}, \bibinfo {author}
  {\bibfnamefont {Laurent}\ \bibnamefont {Kling}}, \bibinfo {author}
  {\bibfnamefont {Graham~D}\ \bibnamefont {Marshall}}, \bibinfo {author}
  {\bibfnamefont {Raffaele}\ \bibnamefont {Santagati}}, \bibinfo {author}
  {\bibfnamefont {Timothy~C}\ \bibnamefont {Ralph}},  \emph {et~al.}} (\bibinfo
  {year} {2018}),\ \bibfield  {title} {\enquote {\bibinfo {title} {Large-scale
  silicon quantum photonics implementing arbitrary two-qubit processing},}\
  }\href {https://doi.org/10.1038/s41566-018-0236-y} {\bibfield  {journal}
  {\bibinfo  {journal} {Nature Photonics}\ }\textbf {\bibinfo {volume} {12}},\
  \bibinfo {pages} {534}},\ \Eprint {https://arxiv.org/abs/arXiv:1809.09791v1}
  {arXiv:1809.09791v1} \BibitemShut {NoStop}%
\bibitem [{\citenamefont {Qin}\ \emph {et~al.}(2016)\citenamefont {Qin},
  \citenamefont {Kumar},\ and\ \citenamefont {All\'eaume}}]{bib:qin2016pra}%
  \BibitemOpen
  \bibfield  {author} {\bibinfo {author} {\bibnamefont {Qin}, \bibfnamefont
  {Hao}}, \bibinfo {author} {\bibfnamefont {Rupesh}\ \bibnamefont {Kumar}},
  and\ \bibinfo {author} {\bibfnamefont {Romain}\ \bibnamefont {All\'eaume}}}
  (\bibinfo {year} {2016}),\ \bibfield  {title} {\enquote {\bibinfo {title}
  {Quantum hacking: Saturation attack on practical continuous-variable quantum
  key distribution},}\ }\href {https://doi.org/10.1103/PhysRevA.94.012325}
  {\bibfield  {journal} {\bibinfo  {journal} {Phys. Rev. A}\ }\textbf {\bibinfo
  {volume} {94}},\ \bibinfo {pages} {012325}}\BibitemShut {NoStop}%
\bibitem [{\citenamefont {Quan}\ \emph {et~al.}(2016)\citenamefont {Quan},
  \citenamefont {Zhai}, \citenamefont {Wang}, \citenamefont {Hou},
  \citenamefont {Wang}, \citenamefont {Xiang}, \citenamefont {Liu},
  \citenamefont {Zhang},\ and\ \citenamefont
  {Dong}}]{bib:quan2016demonstration}%
  \BibitemOpen
  \bibfield  {author} {\bibinfo {author} {\bibnamefont {Quan}, \bibfnamefont
  {Runai}}, \bibinfo {author} {\bibfnamefont {Yiwei}\ \bibnamefont {Zhai}},
  \bibinfo {author} {\bibfnamefont {Mengmeng}\ \bibnamefont {Wang}}, \bibinfo
  {author} {\bibfnamefont {Feiyan}\ \bibnamefont {Hou}}, \bibinfo {author}
  {\bibfnamefont {Shaofeng}\ \bibnamefont {Wang}}, \bibinfo {author}
  {\bibfnamefont {Xiao}\ \bibnamefont {Xiang}}, \bibinfo {author}
  {\bibfnamefont {Tao}\ \bibnamefont {Liu}}, \bibinfo {author} {\bibfnamefont
  {Shougang}\ \bibnamefont {Zhang}}, and\ \bibinfo {author} {\bibfnamefont
  {Ruifang}\ \bibnamefont {Dong}}} (\bibinfo {year} {2016}),\ \bibfield
  {title} {\enquote {\bibinfo {title} {Demonstration of quantum synchronization
  based on second-order quantum coherence of entangled photons},}\ }\href
  {https://doi.org/10.1038/srep30453} {\bibfield  {journal} {\bibinfo
  {journal} {Scientific Reports}\ }\textbf {\bibinfo {volume} {6}},\
  10.1038/srep30453},\ \Eprint {https://arxiv.org/abs/arXiv:1602.06371v1}
  {arXiv:1602.06371v1} \BibitemShut {NoStop}%
\bibitem [{\citenamefont {Rabl}\ \emph {et~al.}(2010)\citenamefont {Rabl},
  \citenamefont {Kolkowitz}, \citenamefont {Koppens}, \citenamefont {Harris},
  \citenamefont {Zoller},\ and\ \citenamefont {Lukin}}]{bib:rabl2010quantum}%
  \BibitemOpen
  \bibfield  {author} {\bibinfo {author} {\bibnamefont {Rabl}, \bibfnamefont
  {Peter}}, \bibinfo {author} {\bibfnamefont {Shimon~Jacob}\ \bibnamefont
  {Kolkowitz}}, \bibinfo {author} {\bibfnamefont {FHL}\ \bibnamefont
  {Koppens}}, \bibinfo {author} {\bibfnamefont {JGE}\ \bibnamefont {Harris}},
  \bibinfo {author} {\bibfnamefont {P}~\bibnamefont {Zoller}}, and\ \bibinfo
  {author} {\bibfnamefont {Mikhail~D}\ \bibnamefont {Lukin}}} (\bibinfo {year}
  {2010}),\ \bibfield  {title} {\enquote {\bibinfo {title} {A quantum spin
  transducer based on nanoelectromechanical resonator arrays},}\ }\href
  {https://doi.org/10.1038/nphys1679} {\bibfield  {journal} {\bibinfo
  {journal} {Nature Physics}\ }\textbf {\bibinfo {volume} {6}},\ \bibinfo
  {pages} {602}},\ \Eprint {https://arxiv.org/abs/arXiv:0908.0316v1}
  {arXiv:0908.0316v1} \BibitemShut {NoStop}%
\bibitem [{\citenamefont {Radnaev}\ \emph {et~al.}(2010)\citenamefont
  {Radnaev}, \citenamefont {Dudin}, \citenamefont {Zhao}, \citenamefont {Jen},
  \citenamefont {Jenkins}, \citenamefont {Kuzmich},\ and\ \citenamefont
  {Kennedy}}]{bib:NP_6_894}%
  \BibitemOpen
  \bibfield  {author} {\bibinfo {author} {\bibnamefont {Radnaev}, \bibfnamefont
  {AG}}, \bibinfo {author} {\bibfnamefont {YO}~\bibnamefont {Dudin}}, \bibinfo
  {author} {\bibfnamefont {R}~\bibnamefont {Zhao}}, \bibinfo {author}
  {\bibfnamefont {HH}~\bibnamefont {Jen}}, \bibinfo {author} {\bibfnamefont
  {SD}~\bibnamefont {Jenkins}}, \bibinfo {author} {\bibfnamefont
  {A}~\bibnamefont {Kuzmich}}, and\ \bibinfo {author} {\bibfnamefont {TAB}\
  \bibnamefont {Kennedy}}} (\bibinfo {year} {2010}),\ \bibfield  {title}
  {\enquote {\bibinfo {title} {A quantum memory with telecom-wavelength
  conversion},}\ }\href {https://doi.org/10.1038/nphys1773} {\bibfield
  {journal} {\bibinfo  {journal} {Nature Physics}\ }\textbf {\bibinfo {volume}
  {6}},\ \bibinfo {pages} {894}}\BibitemShut {NoStop}%
\bibitem [{\citenamefont {Rahimi-Keshari}\ \emph
  {et~al.}(2015{\natexlab{a}})\citenamefont {Rahimi-Keshari}, \citenamefont
  {Lund},\ and\ \citenamefont {Ralph}}]{bib:SalehQOCCC15}%
  \BibitemOpen
  \bibfield  {author} {\bibinfo {author} {\bibnamefont {Rahimi-Keshari},
  \bibfnamefont {Saleh}}, \bibinfo {author} {\bibfnamefont {Austin~P.}\
  \bibnamefont {Lund}}, and\ \bibinfo {author} {\bibfnamefont {Timothy~C.}\
  \bibnamefont {Ralph}}} (\bibinfo {year} {2015}{\natexlab{a}}),\ \bibfield
  {title} {\enquote {\bibinfo {title} {What can quantum optics say about
  computational complexity theory?}}\ }\href
  {https://doi.org/10.1103/physrevlett.114.060501} {\bibfield  {journal}
  {\bibinfo  {journal} {Physical Review Letters}\ }\textbf {\bibinfo {volume}
  {114}},\ \bibinfo {pages} {060501}},\ \Eprint
  {https://arxiv.org/abs/arXiv:1408.3712v2} {arXiv:1408.3712v2} \BibitemShut
  {NoStop}%
\bibitem [{\citenamefont {Rahimi-Keshari}\ \emph
  {et~al.}(2015{\natexlab{b}})\citenamefont {Rahimi-Keshari}, \citenamefont
  {Lund},\ and\ \citenamefont {Ralph}}]{PhysRevLett.114.060501}%
  \BibitemOpen
  \bibfield  {author} {\bibinfo {author} {\bibnamefont {Rahimi-Keshari},
  \bibfnamefont {Saleh}}, \bibinfo {author} {\bibfnamefont {Austin~P.}\
  \bibnamefont {Lund}}, and\ \bibinfo {author} {\bibfnamefont {Timothy~C.}\
  \bibnamefont {Ralph}}} (\bibinfo {year} {2015}{\natexlab{b}}),\ \bibfield
  {title} {\enquote {\bibinfo {title} {What can quantum optics say about
  computational complexity theory?}}\ }\href
  {https://doi.org/10.1103/PhysRevLett.114.060501} {\bibfield  {journal}
  {\bibinfo  {journal} {Phys. Rev. Lett.}\ }\textbf {\bibinfo {volume} {114}},\
  \bibinfo {pages} {060501}}\BibitemShut {NoStop}%
\bibitem [{\citenamefont {Rahimi-Keshari}\ \emph {et~al.}(2016)\citenamefont
  {Rahimi-Keshari}, \citenamefont {Ralph},\ and\ \citenamefont
  {Caves}}]{bib:SalehEffSim16}%
  \BibitemOpen
  \bibfield  {author} {\bibinfo {author} {\bibnamefont {Rahimi-Keshari},
  \bibfnamefont {Saleh}}, \bibinfo {author} {\bibfnamefont {Timothy~C.}\
  \bibnamefont {Ralph}}, and\ \bibinfo {author} {\bibfnamefont {Carlton~M.}\
  \bibnamefont {Caves}}} (\bibinfo {year} {2016}),\ \bibfield  {title}
  {\enquote {\bibinfo {title} {Sufficient conditions for efficient classical
  simulation of quantum optics},}\ }\href
  {https://doi.org/10.1103/physrevx.6.021039} {\bibfield  {journal} {\bibinfo
  {journal} {Physical Review X}\ }\textbf {\bibinfo {volume} {6}},\ \bibinfo
  {pages} {021039}},\ \Eprint {https://arxiv.org/abs/arXiv:1511.06526v2}
  {arXiv:1511.06526v2} \BibitemShut {NoStop}%
\bibitem [{\citenamefont {Raimond}\ \emph {et~al.}(2001)\citenamefont
  {Raimond}, \citenamefont {Brune},\ and\ \citenamefont
  {Haroche}}]{bib:raimond2001manipulating}%
  \BibitemOpen
  \bibfield  {author} {\bibinfo {author} {\bibnamefont {Raimond}, \bibfnamefont
  {Jean-Michel}}, \bibinfo {author} {\bibfnamefont {M}~\bibnamefont {Brune}},
  and\ \bibinfo {author} {\bibfnamefont {Serge}\ \bibnamefont {Haroche}}}
  (\bibinfo {year} {2001}),\ \bibfield  {title} {\enquote {\bibinfo {title}
  {Manipulating quantum entanglement with atoms and photons in a cavity},}\
  }\href {https://doi.org/10.1103/revmodphys.73.565} {\bibfield  {journal}
  {\bibinfo  {journal} {Reviews in Modern Physics}\ }\textbf {\bibinfo {volume}
  {73}},\ \bibinfo {pages} {565}}\BibitemShut {NoStop}%
\bibitem [{\citenamefont {Raju~Valivarthi}\ and\ \citenamefont
  {Tittel}(2016)}]{bib:Nat_Phot_10_676}%
  \BibitemOpen
  \bibfield  {author} {\bibinfo {author} {\bibnamefont {Raju~Valivarthi},
  \bibfnamefont {Marcelli Grimau~Puigibert, Qiang Zhou Gabriel H Aguilar Varun
  B Verma Francesco Marsili Matthew D Shaw Sae Woo Nam Daniel~Oblak}}, and\
  \bibinfo {author} {\bibfnamefont {Wolfgang}\ \bibnamefont {Tittel}}}
  (\bibinfo {year} {2016}),\ \bibfield  {title} {\enquote {\bibinfo {title}
  {Quantum teleportation across a metropolitan fibre network},}\ }\href
  {https://doi.org/10.1038/nphoton.2016.180} {\bibfield  {journal} {\bibinfo
  {journal} {Nature Photonics}\ }\textbf {\bibinfo {volume} {10}},\ \bibinfo
  {pages} {676}},\ \Eprint {https://arxiv.org/abs/arXiv:1605.08814v1}
  {arXiv:1605.08814v1} \BibitemShut {NoStop}%
\bibitem [{\citenamefont {Ralph}(2011)}]{bib:PhysRevA.84.022339}%
  \BibitemOpen
  \bibfield  {author} {\bibinfo {author} {\bibnamefont {Ralph}, \bibfnamefont
  {T~C}}} (\bibinfo {year} {2011}),\ \bibfield  {title} {\enquote {\bibinfo
  {title} {Quantum error correction of continuous-variable states against
  gaussian noise},}\ }\href {https://doi.org/10.1103/PhysRevA.84.022339}
  {\bibfield  {journal} {\bibinfo  {journal} {Physical Review A}\ }\textbf
  {\bibinfo {volume} {84}},\ \bibinfo {pages} {022339}}\BibitemShut {NoStop}%
\bibitem [{\citenamefont {Ralph}\ \emph
  {et~al.}(2005{\natexlab{a}})\citenamefont {Ralph}, \citenamefont {Hayes},\
  and\ \citenamefont {Gilchrist}}]{bib:ralph05}%
  \BibitemOpen
  \bibfield  {author} {\bibinfo {author} {\bibnamefont {Ralph}, \bibfnamefont
  {T~C}}, \bibinfo {author} {\bibfnamefont {A.}~\bibnamefont {Hayes}}, and\
  \bibinfo {author} {\bibfnamefont {A.}~\bibnamefont {Gilchrist}}} (\bibinfo
  {year} {2005}{\natexlab{a}}),\ \bibfield  {title} {\enquote {\bibinfo {title}
  {Loss-tolerant optical qubits},}\ }\href
  {https://doi.org/10.1103/physrevlett.95.100501} {\bibfield  {journal}
  {\bibinfo  {journal} {Physical Review Letters}\ }\textbf {\bibinfo {volume}
  {95}},\ \bibinfo {pages} {100501}},\ \Eprint
  {https://arxiv.org/abs/arXiv:quant-ph/0501184v1} {arXiv:quant-ph/0501184v1}
  \BibitemShut {NoStop}%
\bibitem [{\citenamefont {Ralph}\ \emph {et~al.}(2002)\citenamefont {Ralph},
  \citenamefont {Langford}, \citenamefont {Bell},\ and\ \citenamefont
  {White}}]{bib:Ralph02}%
  \BibitemOpen
  \bibfield  {author} {\bibinfo {author} {\bibnamefont {Ralph}, \bibfnamefont
  {T~C}}, \bibinfo {author} {\bibfnamefont {N.~K.}\ \bibnamefont {Langford}},
  \bibinfo {author} {\bibfnamefont {T.~B.}\ \bibnamefont {Bell}}, and\ \bibinfo
  {author} {\bibfnamefont {A.~G.}\ \bibnamefont {White}}} (\bibinfo {year}
  {2002}),\ \bibfield  {title} {\enquote {\bibinfo {title} {Linear optical
  controlled-not gate in the coincidence basis},}\ }\href
  {https://doi.org/10.1103/physreva.65.062324} {\bibfield  {journal} {\bibinfo
  {journal} {Physical Review A}\ }\textbf {\bibinfo {volume} {65}},\ \bibinfo
  {pages} {062324}}\BibitemShut {NoStop}%
\bibitem [{\citenamefont {Ralph}\ \emph {et~al.}(2001)\citenamefont {Ralph},
  \citenamefont {White}, \citenamefont {Munro},\ and\ \citenamefont
  {Milburn}}]{bib:Ralph01}%
  \BibitemOpen
  \bibfield  {author} {\bibinfo {author} {\bibnamefont {Ralph}, \bibfnamefont
  {T~C}}, \bibinfo {author} {\bibfnamefont {A.~G.}\ \bibnamefont {White}},
  \bibinfo {author} {\bibfnamefont {W.~J.}\ \bibnamefont {Munro}}, and\
  \bibinfo {author} {\bibfnamefont {G.~J.}\ \bibnamefont {Milburn}}} (\bibinfo
  {year} {2001}),\ \bibfield  {title} {\enquote {\bibinfo {title} {Simple
  scheme for efficient linear optics quantum gates},}\ }\href
  {https://doi.org/10.1103/physreva.65.012314} {\bibfield  {journal} {\bibinfo
  {journal} {Physical Review A}\ }\textbf {\bibinfo {volume} {65}},\ \bibinfo
  {pages} {012314}},\ \Eprint {https://arxiv.org/abs/arXiv:quant-ph/0108049v3}
  {arXiv:quant-ph/0108049v3} \BibitemShut {NoStop}%
\bibitem [{\citenamefont {Ralph}\ \emph
  {et~al.}(2005{\natexlab{b}})\citenamefont {Ralph}, \citenamefont {Hayes},\
  and\ \citenamefont {Gilchrist}}]{bib:RalphHayes05}%
  \BibitemOpen
  \bibfield  {author} {\bibinfo {author} {\bibnamefont {Ralph}, \bibfnamefont
  {Timothy~C}}, \bibinfo {author} {\bibfnamefont {A.~J.~F.}\ \bibnamefont
  {Hayes}}, and\ \bibinfo {author} {\bibfnamefont {Alexei}\ \bibnamefont
  {Gilchrist}}} (\bibinfo {year} {2005}{\natexlab{b}}),\ \bibfield  {title}
  {\enquote {\bibinfo {title} {Loss-tolerant optical qubits},}\ }\href
  {https://doi.org/10.1103/physrevlett.95.100501} {\bibfield  {journal}
  {\bibinfo  {journal} {Physical Review Letters}\ }\textbf {\bibinfo {volume}
  {95}},\ \bibinfo {pages} {100501}},\ \Eprint
  {https://arxiv.org/abs/arXiv:quant-ph/0501184v1} {arXiv:quant-ph/0501184v1}
  \BibitemShut {NoStop}%
\bibitem [{\citenamefont {Ralph}\ and\ \citenamefont
  {Lam}(2009)}]{bib:ralph2009bright}%
  \BibitemOpen
  \bibfield  {author} {\bibinfo {author} {\bibnamefont {Ralph}, \bibfnamefont
  {Timothy~C}}, and\ \bibinfo {author} {\bibfnamefont {Ping~K}\ \bibnamefont
  {Lam}}} (\bibinfo {year} {2009}),\ \bibfield  {title} {\enquote {\bibinfo
  {title} {A bright future for quantum communications},}\ }\href
  {https://doi.org/10.1038/nphoton.2009.222} {\bibfield  {journal} {\bibinfo
  {journal} {Nature Photonics}\ }\textbf {\bibinfo {volume} {3}},\ \bibinfo
  {pages} {671}}\BibitemShut {NoStop}%
\bibitem [{\citenamefont {Ran{\v c}i{\'c}}\ \emph {et~al.}(2017)\citenamefont
  {Ran{\v c}i{\'c}}, \citenamefont {Hedges}, \citenamefont {Ahlefeldt},\ and\
  \citenamefont {Sellars}}]{SD-Rancic:2017aa}%
  \BibitemOpen
  \bibfield  {author} {\bibinfo {author} {\bibnamefont {Ran{\v c}i{\'c}},
  \bibfnamefont {Milo{\v s}}}, \bibinfo {author} {\bibfnamefont {Morgan~P.}\
  \bibnamefont {Hedges}}, \bibinfo {author} {\bibfnamefont {Rose~L.}\
  \bibnamefont {Ahlefeldt}}, and\ \bibinfo {author} {\bibfnamefont
  {Matthew~J.}\ \bibnamefont {Sellars}}} (\bibinfo {year} {2017}),\ \bibfield
  {title} {\enquote {\bibinfo {title} {Coherence time of over a second in a
  telecom-compatible quantum memory storage material},}\ }\href
  {https://doi.org/10.1038/nphys4254} {\bibfield  {journal} {\bibinfo
  {journal} {Nature Physics}\ }\textbf {\bibinfo {volume} {14}},\ \bibinfo
  {pages} {50}},\ \Eprint {https://arxiv.org/abs/arXiv:1611.04315v2}
  {arXiv:1611.04315v2} \BibitemShut {NoStop}%
\bibitem [{\citenamefont {Rauschenbeutel}\ \emph {et~al.}(2000)\citenamefont
  {Rauschenbeutel}, \citenamefont {Nogues}, \citenamefont {Osnaghi},
  \citenamefont {Bertet}, \citenamefont {Brune}, \citenamefont {Raimond},\ and\
  \citenamefont {Haroche}}]{bib:rauschenbeutel2000step}%
  \BibitemOpen
  \bibfield  {author} {\bibinfo {author} {\bibnamefont {Rauschenbeutel},
  \bibfnamefont {Arno}}, \bibinfo {author} {\bibfnamefont {Gilles}\
  \bibnamefont {Nogues}}, \bibinfo {author} {\bibfnamefont {Stefano}\
  \bibnamefont {Osnaghi}}, \bibinfo {author} {\bibfnamefont {Patrice}\
  \bibnamefont {Bertet}}, \bibinfo {author} {\bibfnamefont {Michel}\
  \bibnamefont {Brune}}, \bibinfo {author} {\bibfnamefont {Jean-Michel}\
  \bibnamefont {Raimond}}, and\ \bibinfo {author} {\bibfnamefont {Serge}\
  \bibnamefont {Haroche}}} (\bibinfo {year} {2000}),\ \bibfield  {title}
  {\enquote {\bibinfo {title} {Step-by-step engineered multiparticle
  entanglement},}\ }\href {https://doi.org/10.1126/science.288.5473.2024}
  {\bibfield  {journal} {\bibinfo  {journal} {Science}\ }\textbf {\bibinfo
  {volume} {288}},\ \bibinfo {pages} {2024}}\BibitemShut {NoStop}%
\bibitem [{\citenamefont {Raussendorf}\ and\ \citenamefont
  {Briegel}(2001)}]{bib:Raussendorf01}%
  \BibitemOpen
  \bibfield  {author} {\bibinfo {author} {\bibnamefont {Raussendorf},
  \bibfnamefont {R}}, and\ \bibinfo {author} {\bibfnamefont {H.~J.}\
  \bibnamefont {Briegel}}} (\bibinfo {year} {2001}),\ \bibfield  {title}
  {\enquote {\bibinfo {title} {A one-way quantum computer},}\ }\href
  {https://doi.org/10.1103/physrevlett.86.5188} {\bibfield  {journal} {\bibinfo
   {journal} {Physical Review Letters}\ }\textbf {\bibinfo {volume} {86}},\
  \bibinfo {pages} {5188}}\BibitemShut {NoStop}%
\bibitem [{\citenamefont {Raussendorf}\ \emph {et~al.}(2003)\citenamefont
  {Raussendorf}, \citenamefont {Browne},\ and\ \citenamefont
  {Briegel}}]{bib:Raussendorf03}%
  \BibitemOpen
  \bibfield  {author} {\bibinfo {author} {\bibnamefont {Raussendorf},
  \bibfnamefont {R}}, \bibinfo {author} {\bibfnamefont {D.~E.}\ \bibnamefont
  {Browne}}, and\ \bibinfo {author} {\bibfnamefont {H.~J.}\ \bibnamefont
  {Briegel}}} (\bibinfo {year} {2003}),\ \bibfield  {title} {\enquote {\bibinfo
  {title} {Measurement-based quantum computation on cluster states},}\ }\href
  {https://doi.org/10.1103/physreva.68.022312} {\bibfield  {journal} {\bibinfo
  {journal} {Physical Review A}\ }\textbf {\bibinfo {volume} {68}},\ \bibinfo
  {pages} {022312}},\ \Eprint {https://arxiv.org/abs/arXiv:quant-ph/0301052v2}
  {arXiv:quant-ph/0301052v2} \BibitemShut {NoStop}%
\bibitem [{\citenamefont {Rebentrost}\ \emph {et~al.}(2014)\citenamefont
  {Rebentrost}, \citenamefont {Mohseni},\ and\ \citenamefont
  {Lloyd}}]{bib:PhysRevLett.113.130503}%
  \BibitemOpen
  \bibfield  {author} {\bibinfo {author} {\bibnamefont {Rebentrost},
  \bibfnamefont {Patrick}}, \bibinfo {author} {\bibfnamefont {Masoud}\
  \bibnamefont {Mohseni}}, and\ \bibinfo {author} {\bibfnamefont {Seth}\
  \bibnamefont {Lloyd}}} (\bibinfo {year} {2014}),\ \bibfield  {title}
  {\enquote {\bibinfo {title} {Quantum support vector machine for big data
  classification},}\ }\href {https://doi.org/10.1103/physrevlett.113.130503}
  {\bibfield  {journal} {\bibinfo  {journal} {Physical Review Letters}\
  }\textbf {\bibinfo {volume} {113}},\ \bibinfo {pages} {130503}},\ \Eprint
  {https://arxiv.org/abs/arXiv:1307.0471v3} {arXiv:1307.0471v3} \BibitemShut
  {NoStop}%
\bibitem [{\citenamefont {Reck}\ \emph {et~al.}(1994)\citenamefont {Reck},
  \citenamefont {Zeilinger}, \citenamefont {Bernstein},\ and\ \citenamefont
  {Bertani}}]{bib:Reck94}%
  \BibitemOpen
  \bibfield  {author} {\bibinfo {author} {\bibnamefont {Reck}, \bibfnamefont
  {M}}, \bibinfo {author} {\bibfnamefont {A.}~\bibnamefont {Zeilinger}},
  \bibinfo {author} {\bibfnamefont {H.~J.}\ \bibnamefont {Bernstein}}, and\
  \bibinfo {author} {\bibfnamefont {P.}~\bibnamefont {Bertani}}} (\bibinfo
  {year} {1994}),\ \bibfield  {title} {\enquote {\bibinfo {title} {Experimental
  realization of any discrete unitary operator},}\ }\href
  {https://doi.org/10.1103/physrevlett.73.58} {\bibfield  {journal} {\bibinfo
  {journal} {Physical Review Letters}\ }\textbf {\bibinfo {volume} {73}},\
  \bibinfo {pages} {58}}\BibitemShut {NoStop}%
\bibitem [{\citenamefont {Reiher}\ \emph {et~al.}(2017)\citenamefont {Reiher},
  \citenamefont {Wiebe}, \citenamefont {Svore}, \citenamefont {Wecker},\ and\
  \citenamefont {Troyer}}]{SD-Reiher:2017aa}%
  \BibitemOpen
  \bibfield  {author} {\bibinfo {author} {\bibnamefont {Reiher}, \bibfnamefont
  {Markus}}, \bibinfo {author} {\bibfnamefont {Nathan}\ \bibnamefont {Wiebe}},
  \bibinfo {author} {\bibfnamefont {Krysta~M.}\ \bibnamefont {Svore}}, \bibinfo
  {author} {\bibfnamefont {Dave}\ \bibnamefont {Wecker}}, and\ \bibinfo
  {author} {\bibfnamefont {Matthias}\ \bibnamefont {Troyer}}} (\bibinfo {year}
  {2017}),\ \bibfield  {title} {\enquote {\bibinfo {title} {Elucidating
  reaction mechanisms on quantum computers},}\ }\href
  {https://doi.org/10.1073/pnas.1619152114} {\bibfield  {journal} {\bibinfo
  {journal} {Proceedings of the National Academy of Sciences}\
  }10.1073/pnas.1619152114},\ \Eprint
  {https://arxiv.org/abs/arXiv:1605.03590v2} {arXiv:1605.03590v2} \BibitemShut
  {NoStop}%
\bibitem [{\citenamefont {Reiserer}\ and\ \citenamefont
  {Rempe}(2015)}]{bib:reiserer2015cavity}%
  \BibitemOpen
  \bibfield  {author} {\bibinfo {author} {\bibnamefont {Reiserer},
  \bibfnamefont {Andreas}}, and\ \bibinfo {author} {\bibfnamefont {Gerhard}\
  \bibnamefont {Rempe}}} (\bibinfo {year} {2015}),\ \bibfield  {title}
  {\enquote {\bibinfo {title} {Cavity-based quantum networks with single atoms
  and optical photons},}\ }\href {https://doi.org/10.1103/revmodphys.87.1379}
  {\bibfield  {journal} {\bibinfo  {journal} {Reviews in Modern Physics}\
  }\textbf {\bibinfo {volume} {87}},\ \bibinfo {pages} {1379}},\ \Eprint
  {https://arxiv.org/abs/arXiv:1412.2889v2} {arXiv:1412.2889v2} \BibitemShut
  {NoStop}%
\bibitem [{\citenamefont {Ren}\ and\ \citenamefont
  {Hofmann}(2012)}]{bib:ren2012clock}%
  \BibitemOpen
  \bibfield  {author} {\bibinfo {author} {\bibnamefont {Ren}, \bibfnamefont
  {Changliang}}, and\ \bibinfo {author} {\bibfnamefont {Holger~F}\ \bibnamefont
  {Hofmann}}} (\bibinfo {year} {2012}),\ \bibfield  {title} {\enquote {\bibinfo
  {title} {Clock synchronization using maximal multipartite entanglement},}\
  }\href {https://doi.org/10.1103/physreva.86.014301} {\bibfield  {journal}
  {\bibinfo  {journal} {Physical Review A}\ }\textbf {\bibinfo {volume} {86}},\
  \bibinfo {pages} {014301}},\ \Eprint
  {https://arxiv.org/abs/arXiv:1203.4300v2} {arXiv:1203.4300v2} \BibitemShut
  {NoStop}%
\bibitem [{\citenamefont {Ren}\ \emph {et~al.}(2017)\citenamefont {Ren},
  \citenamefont {Xu}, \citenamefont {Yong}, \citenamefont {Zhang},
  \citenamefont {Liao}, \citenamefont {Yin}, \citenamefont {Liu}, \citenamefont
  {Cai}, \citenamefont {Yang}, \citenamefont {Li} \emph
  {et~al.}}]{bib:ren2017ground}%
  \BibitemOpen
  \bibfield  {author} {\bibinfo {author} {\bibnamefont {Ren}, \bibfnamefont
  {Ji-Gang}}, \bibinfo {author} {\bibfnamefont {Ping}\ \bibnamefont {Xu}},
  \bibinfo {author} {\bibfnamefont {Hai-Lin}\ \bibnamefont {Yong}}, \bibinfo
  {author} {\bibfnamefont {Liang}\ \bibnamefont {Zhang}}, \bibinfo {author}
  {\bibfnamefont {Sheng-Kai}\ \bibnamefont {Liao}}, \bibinfo {author}
  {\bibfnamefont {Juan}\ \bibnamefont {Yin}}, \bibinfo {author} {\bibfnamefont
  {Wei-Yue}\ \bibnamefont {Liu}}, \bibinfo {author} {\bibfnamefont {Wen-Qi}\
  \bibnamefont {Cai}}, \bibinfo {author} {\bibfnamefont {Meng}\ \bibnamefont
  {Yang}}, \bibinfo {author} {\bibfnamefont {Li}~\bibnamefont {Li}},  \emph
  {et~al.}} (\bibinfo {year} {2017}),\ \bibfield  {title} {\enquote {\bibinfo
  {title} {Ground-to-satellite quantum teleportation},}\ }\href
  {https://doi.org/10.1038/nature23675} {\bibfield  {journal} {\bibinfo
  {journal} {Nature}\ }\textbf {\bibinfo {volume} {549}},\ \bibinfo {pages}
  {70}},\ \Eprint {https://arxiv.org/abs/arXiv:1707.00934v1}
  {arXiv:1707.00934v1} \BibitemShut {NoStop}%
\bibitem [{\citenamefont {Resch}\ \emph {et~al.}(2005)\citenamefont {Resch},
  \citenamefont {Lindenthal}, \citenamefont {Blauensteiner}, \citenamefont
  {B{\"o}hm}, \citenamefont {Fedrizzi}, \citenamefont {Kurtsiefer},
  \citenamefont {Poppe}, \citenamefont {Schmitt-Manderbach}, \citenamefont
  {Taraba}, \citenamefont {Ursin} \emph {et~al.}}]{bib:OE_13_202}%
  \BibitemOpen
  \bibfield  {author} {\bibinfo {author} {\bibnamefont {Resch}, \bibfnamefont
  {KJ}}, \bibinfo {author} {\bibfnamefont {M}~\bibnamefont {Lindenthal}},
  \bibinfo {author} {\bibfnamefont {B}~\bibnamefont {Blauensteiner}}, \bibinfo
  {author} {\bibfnamefont {HR}~\bibnamefont {B{\"o}hm}}, \bibinfo {author}
  {\bibfnamefont {A}~\bibnamefont {Fedrizzi}}, \bibinfo {author} {\bibfnamefont
  {C}~\bibnamefont {Kurtsiefer}}, \bibinfo {author} {\bibfnamefont
  {A}~\bibnamefont {Poppe}}, \bibinfo {author} {\bibfnamefont {T}~\bibnamefont
  {Schmitt-Manderbach}}, \bibinfo {author} {\bibfnamefont {M}~\bibnamefont
  {Taraba}}, \bibinfo {author} {\bibfnamefont {R}~\bibnamefont {Ursin}},  \emph
  {et~al.}} (\bibinfo {year} {2005}),\ \bibfield  {title} {\enquote {\bibinfo
  {title} {Distributing entanglement and single photons through an intra-city,
  free-space quantum channel},}\ }\href
  {https://doi.org/10.1364/opex.13.000202} {\bibfield  {journal} {\bibinfo
  {journal} {Optics Express}\ }\textbf {\bibinfo {volume} {13}},\ \bibinfo
  {pages} {202}},\ \Eprint {https://arxiv.org/abs/arXiv:quant-ph/0501008v2}
  {arXiv:quant-ph/0501008v2} \BibitemShut {NoStop}%
\bibitem [{\citenamefont {Riebe}\ \emph {et~al.}(2004)\citenamefont {Riebe},
  \citenamefont {H{\"a}ffner}, \citenamefont {Roos}, \citenamefont
  {H{\"a}nsel}, \citenamefont {Benhelm}, \citenamefont {Lancaster},
  \citenamefont {K{\"o}rber}, \citenamefont {Becher}, \citenamefont
  {Schmidt-Kaler}, \citenamefont {James} \emph {et~al.}}]{bib:Nat_429_734}%
  \BibitemOpen
  \bibfield  {author} {\bibinfo {author} {\bibnamefont {Riebe}, \bibfnamefont
  {Mark}}, \bibinfo {author} {\bibfnamefont {H}~\bibnamefont {H{\"a}ffner}},
  \bibinfo {author} {\bibfnamefont {CF}~\bibnamefont {Roos}}, \bibinfo {author}
  {\bibfnamefont {W}~\bibnamefont {H{\"a}nsel}}, \bibinfo {author}
  {\bibfnamefont {J}~\bibnamefont {Benhelm}}, \bibinfo {author} {\bibfnamefont
  {GPT}\ \bibnamefont {Lancaster}}, \bibinfo {author} {\bibfnamefont
  {TW}~\bibnamefont {K{\"o}rber}}, \bibinfo {author} {\bibfnamefont
  {C}~\bibnamefont {Becher}}, \bibinfo {author} {\bibfnamefont {F}~\bibnamefont
  {Schmidt-Kaler}}, \bibinfo {author} {\bibfnamefont {DFV}\ \bibnamefont
  {James}},  \emph {et~al.}} (\bibinfo {year} {2004}),\ \bibfield  {title}
  {\enquote {\bibinfo {title} {Deterministic quantum teleportation with
  atoms},}\ }\href {https://doi.org/10.1038/nature02570} {\bibfield  {journal}
  {\bibinfo  {journal} {Nature}\ }\textbf {\bibinfo {volume} {429}},\ \bibinfo
  {pages} {734}}\BibitemShut {NoStop}%
\bibitem [{\citenamefont {Riedl}\ \emph {et~al.}(2012)\citenamefont {Riedl},
  \citenamefont {Lettner}, \citenamefont {Vo}, \citenamefont {Baur},
  \citenamefont {Rempe},\ and\ \citenamefont {D{\"u}rr}}]{bib:riedl2012bose}%
  \BibitemOpen
  \bibfield  {author} {\bibinfo {author} {\bibnamefont {Riedl}, \bibfnamefont
  {Stefan}}, \bibinfo {author} {\bibfnamefont {Matthias}\ \bibnamefont
  {Lettner}}, \bibinfo {author} {\bibfnamefont {Christoph}\ \bibnamefont {Vo}},
  \bibinfo {author} {\bibfnamefont {Simon}\ \bibnamefont {Baur}}, \bibinfo
  {author} {\bibfnamefont {Gerhard}\ \bibnamefont {Rempe}}, and\ \bibinfo
  {author} {\bibfnamefont {Stephan}\ \bibnamefont {D{\"u}rr}}} (\bibinfo {year}
  {2012}),\ \bibfield  {title} {\enquote {\bibinfo {title} {Bose-einstein
  condensate as a quantum memory for a photonic polarization qubit},}\ }\href
  {https://doi.org/10.1103/physreva.85.022318} {\bibfield  {journal} {\bibinfo
  {journal} {Physical Review A}\ }\textbf {\bibinfo {volume} {85}},\ \bibinfo
  {pages} {022318}},\ \Eprint {https://arxiv.org/abs/arXiv:1112.4733v2}
  {arXiv:1112.4733v2} \BibitemShut {NoStop}%
\bibitem [{\citenamefont {Riel{\"a}nder}\ \emph {et~al.}(2014)\citenamefont
  {Riel{\"a}nder}, \citenamefont {Kutluer}, \citenamefont {Ledingham},
  \citenamefont {G{\"u}ndo{\u{g}}an}, \citenamefont {Fekete}, \citenamefont
  {Mazzera},\ and\ \citenamefont {de~Riedmatten}}]{bib:PRL_112_040504}%
  \BibitemOpen
  \bibfield  {author} {\bibinfo {author} {\bibnamefont {Riel{\"a}nder},
  \bibfnamefont {Daniel}}, \bibinfo {author} {\bibfnamefont {Kutlu}\
  \bibnamefont {Kutluer}}, \bibinfo {author} {\bibfnamefont {Patrick~M}\
  \bibnamefont {Ledingham}}, \bibinfo {author} {\bibfnamefont {Mustafa}\
  \bibnamefont {G{\"u}ndo{\u{g}}an}}, \bibinfo {author} {\bibfnamefont {Julia}\
  \bibnamefont {Fekete}}, \bibinfo {author} {\bibfnamefont {Margherita}\
  \bibnamefont {Mazzera}}, and\ \bibinfo {author} {\bibfnamefont {Hugues}\
  \bibnamefont {de~Riedmatten}}} (\bibinfo {year} {2014}),\ \bibfield  {title}
  {\enquote {\bibinfo {title} {Quantum storage of heralded single photons in a
  praseodymium-doped crystal},}\ }\href
  {https://doi.org/10.1103/physrevlett.112.040504} {\bibfield  {journal}
  {\bibinfo  {journal} {Physical Review Letters}\ }\textbf {\bibinfo {volume}
  {112}},\ \bibinfo {pages} {040504}},\ \Eprint
  {https://arxiv.org/abs/arXiv:1310.8261v2} {arXiv:1310.8261v2} \BibitemShut
  {NoStop}%
\bibitem [{\citenamefont {Rigetti}\ \emph {et~al.}(2012)\citenamefont
  {Rigetti}, \citenamefont {Gambetta}, \citenamefont {Poletto}, \citenamefont
  {Plourde}, \citenamefont {Chow}, \citenamefont {C{\'o}rcoles}, \citenamefont
  {Smolin}, \citenamefont {Merkel}, \citenamefont {Rozen}, \citenamefont
  {Keefe} \emph {et~al.}}]{bib:rigetti2012superconducting}%
  \BibitemOpen
  \bibfield  {author} {\bibinfo {author} {\bibnamefont {Rigetti}, \bibfnamefont
  {Chad}}, \bibinfo {author} {\bibfnamefont {Jay~M}\ \bibnamefont {Gambetta}},
  \bibinfo {author} {\bibfnamefont {Stefano}\ \bibnamefont {Poletto}}, \bibinfo
  {author} {\bibfnamefont {BLT}\ \bibnamefont {Plourde}}, \bibinfo {author}
  {\bibfnamefont {Jerry~M}\ \bibnamefont {Chow}}, \bibinfo {author}
  {\bibfnamefont {AD}~\bibnamefont {C{\'o}rcoles}}, \bibinfo {author}
  {\bibfnamefont {John~A}\ \bibnamefont {Smolin}}, \bibinfo {author}
  {\bibfnamefont {Seth~T}\ \bibnamefont {Merkel}}, \bibinfo {author}
  {\bibfnamefont {JR}~\bibnamefont {Rozen}}, \bibinfo {author} {\bibfnamefont
  {George~A}\ \bibnamefont {Keefe}},  \emph {et~al.}} (\bibinfo {year}
  {2012}),\ \bibfield  {title} {\enquote {\bibinfo {title} {Superconducting
  qubit in a waveguide cavity with a coherence time approaching 0.1 ms},}\
  }\href {https://doi.org/10.1103/physrevb.86.100506} {\bibfield  {journal}
  {\bibinfo  {journal} {Physical Review B}\ }\textbf {\bibinfo {volume} {86}},\
  \bibinfo {pages} {100506}}\BibitemShut {NoStop}%
\bibitem [{\citenamefont {Ritter}\ \emph {et~al.}(2012)\citenamefont {Ritter},
  \citenamefont {N{\"o}lleke}, \citenamefont {Hahn}, \citenamefont {Reiserer},
  \citenamefont {Neuzner}, \citenamefont {Uphoff}, \citenamefont {M{\"u}cke},
  \citenamefont {Figueroa}, \citenamefont {Bochmann},\ and\ \citenamefont
  {Rempe}}]{bib:N_484_195}%
  \BibitemOpen
  \bibfield  {author} {\bibinfo {author} {\bibnamefont {Ritter}, \bibfnamefont
  {Stephan}}, \bibinfo {author} {\bibfnamefont {Christian}\ \bibnamefont
  {N{\"o}lleke}}, \bibinfo {author} {\bibfnamefont {Carolin}\ \bibnamefont
  {Hahn}}, \bibinfo {author} {\bibfnamefont {Andreas}\ \bibnamefont
  {Reiserer}}, \bibinfo {author} {\bibfnamefont {Andreas}\ \bibnamefont
  {Neuzner}}, \bibinfo {author} {\bibfnamefont {Manuel}\ \bibnamefont
  {Uphoff}}, \bibinfo {author} {\bibfnamefont {Martin}\ \bibnamefont
  {M{\"u}cke}}, \bibinfo {author} {\bibfnamefont {Eden}\ \bibnamefont
  {Figueroa}}, \bibinfo {author} {\bibfnamefont {Joerg}\ \bibnamefont
  {Bochmann}}, and\ \bibinfo {author} {\bibfnamefont {Gerhard}\ \bibnamefont
  {Rempe}}} (\bibinfo {year} {2012}),\ \bibfield  {title} {\enquote {\bibinfo
  {title} {An elementary quantum network of single atoms in optical
  cavities},}\ }\href {https://doi.org/10.1038/nature11023} {\bibfield
  {journal} {\bibinfo  {journal} {Nature}\ }\textbf {\bibinfo {volume} {484}},\
  \bibinfo {pages} {195}},\ \Eprint {https://arxiv.org/abs/arXiv:1202.5955v1}
  {arXiv:1202.5955v1} \BibitemShut {NoStop}%
\bibitem [{\citenamefont {Rivest}\ \emph
  {et~al.}(1978{\natexlab{a}})\citenamefont {Rivest}, \citenamefont {Shamir},\
  and\ \citenamefont {Adleman}}]{bib:RSA}%
  \BibitemOpen
  \bibfield  {author} {\bibinfo {author} {\bibnamefont {Rivest}, \bibfnamefont
  {R}}, \bibinfo {author} {\bibfnamefont {A.}~\bibnamefont {Shamir}}, and\
  \bibinfo {author} {\bibfnamefont {L.}~\bibnamefont {Adleman}}} (\bibinfo
  {year} {1978}{\natexlab{a}}),\ \bibfield  {title} {\enquote {\bibinfo {title}
  {A method for obtaining digital signatures and public-key cryptosystems.}}\
  }\href {https://doi.org/10.21236/ada606588} {\bibfield  {journal} {\bibinfo
  {journal} {Communications of the ACM}\ }\textbf {\bibinfo {volume} {21}},\
  \bibinfo {pages} {120}}\BibitemShut {NoStop}%
\bibitem [{\citenamefont {Rivest}\ \emph
  {et~al.}(1978{\natexlab{b}})\citenamefont {Rivest}, \citenamefont {Adleman},\
  and\ \citenamefont {Dertouzos}}]{bib:Rivest1978}%
  \BibitemOpen
  \bibfield  {author} {\bibinfo {author} {\bibnamefont {Rivest}, \bibfnamefont
  {R~L}}, \bibinfo {author} {\bibfnamefont {L}~\bibnamefont {Adleman}}, and\
  \bibinfo {author} {\bibfnamefont {M~L}\ \bibnamefont {Dertouzos}}} (\bibinfo
  {year} {1978}{\natexlab{b}}),\ \bibfield  {title} {\enquote {\bibinfo {title}
  {On data banks and privacy homomorphisms},}\ }\href@noop {} {\bibinfo
  {journal} {Foundations of Secure Computation, Academia Press}\ ,\ \bibinfo
  {pages} {169}}\BibitemShut {NoStop}%
\bibitem [{\citenamefont {Rohde}(2012)}]{bib:RohdeArbLow12}%
  \BibitemOpen
\bibfield  {journal} {  }\bibfield  {author} {\bibinfo {author} {\bibnamefont
  {Rohde}, \bibfnamefont {Peter~P}}} (\bibinfo {year} {2012}),\ \bibfield
  {title} {\enquote {\bibinfo {title} {Optical quantum computing with photons
  of arbitrarily low fidelity and purity},}\ }\href
  {https://doi.org/10.1103/physreva.86.052321} {\bibfield  {journal} {\bibinfo
  {journal} {Physical Review A}\ }\textbf {\bibinfo {volume} {86}},\ \bibinfo
  {pages} {052321}},\ \Eprint {https://arxiv.org/abs/arXiv:1208.2475v2}
  {arXiv:1208.2475v2} \BibitemShut {NoStop}%
\bibitem [{\citenamefont {Rohde}(2015{\natexlab{a}})}]{bib:RohdeArbSpec15}%
  \BibitemOpen
  \bibfield  {author} {\bibinfo {author} {\bibnamefont {Rohde}, \bibfnamefont
  {Peter~P}}} (\bibinfo {year} {2015}{\natexlab{a}}),\ \bibfield  {title}
  {\enquote {\bibinfo {title} {Boson-sampling with photons of arbitrary
  spectral structure},}\ }\href {https://doi.org/10.1103/physreva.91.012307}
  {\bibfield  {journal} {\bibinfo  {journal} {Physical Review A}\ }\textbf
  {\bibinfo {volume} {91}},\ \bibinfo {pages} {012307}},\ \Eprint
  {https://arxiv.org/abs/arXiv:1410.3979v1} {arXiv:1410.3979v1} \BibitemShut
  {NoStop}%
\bibitem [{\citenamefont {Rohde}(2015{\natexlab{b}})}]{bib:RohdeUnivLoop15}%
  \BibitemOpen
  \bibfield  {author} {\bibinfo {author} {\bibnamefont {Rohde}, \bibfnamefont
  {Peter~P}}} (\bibinfo {year} {2015}{\natexlab{b}}),\ \bibfield  {title}
  {\enquote {\bibinfo {title} {A simple scheme for universal linear optics
  quantum computing with constant experimental complexity using fiber-loops},}\
  }\href {https://doi.org/10.1103/physreva.91.012306} {\bibfield  {journal}
  {\bibinfo  {journal} {Physical Review A}\ }\textbf {\bibinfo {volume} {91}},\
  \bibinfo {pages} {012306}},\ \Eprint
  {https://arxiv.org/abs/arXiv:1410.0433v1} {arXiv:1410.0433v1} \BibitemShut
  {NoStop}%
\bibitem [{\citenamefont {Rohde}\ and\ \citenamefont
  {Barrett}(2007)}]{bib:RohdeBarrett07}%
  \BibitemOpen
  \bibfield  {author} {\bibinfo {author} {\bibnamefont {Rohde}, \bibfnamefont
  {Peter~P}}, and\ \bibinfo {author} {\bibfnamefont {Sean~D.}\ \bibnamefont
  {Barrett}}} (\bibinfo {year} {2007}),\ \bibfield  {title} {\enquote {\bibinfo
  {title} {Strategies for the preparation of large cluster states using
  non-deterministic gates},}\ }\href
  {https://doi.org/10.1088/1367-2630/9/6/198} {\bibfield  {journal} {\bibinfo
  {journal} {New Journal of Physics}\ }\textbf {\bibinfo {volume} {9}},\
  \bibinfo {pages} {198}}\BibitemShut {NoStop}%
\bibitem [{\citenamefont {Rohde}\ \emph {et~al.}(2012)\citenamefont {Rohde},
  \citenamefont {Fitzsimons},\ and\ \citenamefont
  {Gilchrist}}]{bib:RohdeQWEnc12}%
  \BibitemOpen
  \bibfield  {author} {\bibinfo {author} {\bibnamefont {Rohde}, \bibfnamefont
  {Peter~P}}, \bibinfo {author} {\bibfnamefont {Joseph~F.}\ \bibnamefont
  {Fitzsimons}}, and\ \bibinfo {author} {\bibfnamefont {Alexei}\ \bibnamefont
  {Gilchrist}}} (\bibinfo {year} {2012}),\ \bibfield  {title} {\enquote
  {\bibinfo {title} {Quantum walks with encrypted data},}\ }\href
  {https://doi.org/10.1103/physrevlett.109.150501} {\bibfield  {journal}
  {\bibinfo  {journal} {Physical Review Letters}\ }\textbf {\bibinfo {volume}
  {109}},\ \bibinfo {pages} {150501}},\ \Eprint
  {https://arxiv.org/abs/arXiv:1204.3370v1} {arXiv:1204.3370v1} \BibitemShut
  {NoStop}%
\bibitem [{\citenamefont {Rohde}\ \emph {et~al.}(2013)\citenamefont {Rohde},
  \citenamefont {Fitzsimons},\ and\ \citenamefont
  {Gilchrist}}]{bib:RohdeInfCap13}%
  \BibitemOpen
  \bibfield  {author} {\bibinfo {author} {\bibnamefont {Rohde}, \bibfnamefont
  {Peter~P}}, \bibinfo {author} {\bibfnamefont {Joseph~F.}\ \bibnamefont
  {Fitzsimons}}, and\ \bibinfo {author} {\bibfnamefont {Alexei}\ \bibnamefont
  {Gilchrist}}} (\bibinfo {year} {2013}),\ \bibfield  {title} {\enquote
  {\bibinfo {title} {The information capacity of a single photon},}\ }\href
  {https://doi.org/10.1103/physreva.88.022310} {\bibfield  {journal} {\bibinfo
  {journal} {Physical Review A}\ }\textbf {\bibinfo {volume} {88}},\ \bibinfo
  {pages} {022310}},\ \Eprint {https://arxiv.org/abs/arXiv:1211.1427v1}
  {arXiv:1211.1427v1} \BibitemShut {NoStop}%
\bibitem [{\citenamefont {Rohde}\ \emph
  {et~al.}(2015{\natexlab{a}})\citenamefont {Rohde}, \citenamefont {Helt},
  \citenamefont {Steel},\ and\ \citenamefont
  {Gilchrist}}]{bib:RohdeLoopMulti15}%
  \BibitemOpen
  \bibfield  {author} {\bibinfo {author} {\bibnamefont {Rohde}, \bibfnamefont
  {Peter~P}}, \bibinfo {author} {\bibfnamefont {L.~G.}\ \bibnamefont {Helt}},
  \bibinfo {author} {\bibfnamefont {M.~J.}\ \bibnamefont {Steel}}, and\
  \bibinfo {author} {\bibfnamefont {Alexei}\ \bibnamefont {Gilchrist}}}
  (\bibinfo {year} {2015}{\natexlab{a}}),\ \bibfield  {title} {\enquote
  {\bibinfo {title} {Multiplexed single-photon state preparation using a
  fibre-loop architecture},}\ }\href
  {https://doi.org/10.1103/physreva.92.053829} {\bibfield  {journal} {\bibinfo
  {journal} {Physical Review A}\ }\textbf {\bibinfo {volume} {92}},\ \bibinfo
  {pages} {053829}},\ \Eprint {https://arxiv.org/abs/arXiv:1503.03546v2}
  {arXiv:1503.03546v2} \BibitemShut {NoStop}%
\bibitem [{\citenamefont {Rohde}\ \emph
  {et~al.}(2007{\natexlab{a}})\citenamefont {Rohde}, \citenamefont {Mauerer},\
  and\ \citenamefont {Silberhorn}}]{bib:RohdeMauererSilberhorn07}%
  \BibitemOpen
  \bibfield  {author} {\bibinfo {author} {\bibnamefont {Rohde}, \bibfnamefont
  {Peter~P}}, \bibinfo {author} {\bibfnamefont {Wolfgang}\ \bibnamefont
  {Mauerer}}, and\ \bibinfo {author} {\bibfnamefont {Christine}\ \bibnamefont
  {Silberhorn}}} (\bibinfo {year} {2007}{\natexlab{a}}),\ \bibfield  {title}
  {\enquote {\bibinfo {title} {Spectral structure and decompositions of optical
  states, and their applications},}\ }\href
  {https://doi.org/10.1088/1367-2630/9/4/091} {\bibfield  {journal} {\bibinfo
  {journal} {New Journal of Physics}\ }\textbf {\bibinfo {volume} {9}},\
  \bibinfo {pages} {91}}\BibitemShut {NoStop}%
\bibitem [{\citenamefont {Rohde}\ \emph
  {et~al.}(2015{\natexlab{b}})\citenamefont {Rohde}, \citenamefont {Motes},
  \citenamefont {Knott}, \citenamefont {Fitzsimons}, \citenamefont {Munro},\
  and\ \citenamefont {Dowling}}]{bib:RohdeCat15}%
  \BibitemOpen
  \bibfield  {author} {\bibinfo {author} {\bibnamefont {Rohde}, \bibfnamefont
  {Peter~P}}, \bibinfo {author} {\bibfnamefont {Keith~R.}\ \bibnamefont
  {Motes}}, \bibinfo {author} {\bibfnamefont {Paul}\ \bibnamefont {Knott}},
  \bibinfo {author} {\bibfnamefont {Joseph}\ \bibnamefont {Fitzsimons}},
  \bibinfo {author} {\bibfnamefont {William}\ \bibnamefont {Munro}}, and\
  \bibinfo {author} {\bibfnamefont {Jonathan~P.}\ \bibnamefont {Dowling}}}
  (\bibinfo {year} {2015}{\natexlab{b}}),\ \bibfield  {title} {\enquote
  {\bibinfo {title} {Evidence for the conjecture that sampling generalized cat
  states with linear optics is hard},}\ }\href
  {https://doi.org/10.1103/physreva.91.012342} {\bibfield  {journal} {\bibinfo
  {journal} {Physical Review A}\ }\textbf {\bibinfo {volume} {91}},\ \bibinfo
  {pages} {012342}}\BibitemShut {NoStop}%
\bibitem [{\citenamefont {Rohde}\ \emph
  {et~al.}(2005{\natexlab{a}})\citenamefont {Rohde}, \citenamefont {Pryde},
  \citenamefont {O'Brien},\ and\ \citenamefont {Ralph}}]{bib:RohdeGateChar05}%
  \BibitemOpen
  \bibfield  {author} {\bibinfo {author} {\bibnamefont {Rohde}, \bibfnamefont
  {Peter~P}}, \bibinfo {author} {\bibfnamefont {G.~J.}\ \bibnamefont {Pryde}},
  \bibinfo {author} {\bibfnamefont {J.~L.}\ \bibnamefont {O'Brien}}, and\
  \bibinfo {author} {\bibfnamefont {Timothy~C.}\ \bibnamefont {Ralph}}}
  (\bibinfo {year} {2005}{\natexlab{a}}),\ \bibfield  {title} {\enquote
  {\bibinfo {title} {Quantum-gate characterization in an extended hilbert
  space},}\ }\href {https://doi.org/10.1103/physreva.72.032306} {\bibfield
  {journal} {\bibinfo  {journal} {Physical Review A}\ }\textbf {\bibinfo
  {volume} {72}},\ \bibinfo {pages} {032306}},\ \Eprint
  {https://arxiv.org/abs/arXiv:quant-ph/0411144v2} {arXiv:quant-ph/0411144v2}
  \BibitemShut {NoStop}%
\bibitem [{\citenamefont {Rohde}\ and\ \citenamefont
  {Ralph}(2005)}]{bib:RohdeFreqTemp05}%
  \BibitemOpen
  \bibfield  {author} {\bibinfo {author} {\bibnamefont {Rohde}, \bibfnamefont
  {Peter~P}}, and\ \bibinfo {author} {\bibfnamefont {Timothy~C.}\ \bibnamefont
  {Ralph}}} (\bibinfo {year} {2005}),\ \bibfield  {title} {\enquote {\bibinfo
  {title} {Frequency and temporal effects in linear optical quantum
  computing},}\ }\href {https://doi.org/10.1103/physreva.71.032320} {\bibfield
  {journal} {\bibinfo  {journal} {Physical Review A}\ }\textbf {\bibinfo
  {volume} {71}},\ \bibinfo {pages} {032320}},\ \Eprint
  {https://arxiv.org/abs/arXiv:quant-ph/0407002v3} {arXiv:quant-ph/0407002v3}
  \BibitemShut {NoStop}%
\bibitem [{\citenamefont {Rohde}\ and\ \citenamefont
  {Ralph}(2006)}]{bib:RohdeRalph06}%
  \BibitemOpen
  \bibfield  {author} {\bibinfo {author} {\bibnamefont {Rohde}, \bibfnamefont
  {Peter~P}}, and\ \bibinfo {author} {\bibfnamefont {Timothy~C.}\ \bibnamefont
  {Ralph}}} (\bibinfo {year} {2006}),\ \bibfield  {title} {\enquote {\bibinfo
  {title} {Error models for mode-mismatch in linear optics quantum
  computing},}\ }\href {https://doi.org/10.1103/physreva.73.062312} {\bibfield
  {journal} {\bibinfo  {journal} {Physical Review A}\ }\textbf {\bibinfo
  {volume} {73}},\ \bibinfo {pages} {062312}},\ \Eprint
  {https://arxiv.org/abs/arXiv:quant-ph/0602004v1} {arXiv:quant-ph/0602004v1}
  \BibitemShut {NoStop}%
\bibitem [{\citenamefont {Rohde}\ and\ \citenamefont
  {Ralph}(2011)}]{bib:RohdeTimeRes11}%
  \BibitemOpen
  \bibfield  {author} {\bibinfo {author} {\bibnamefont {Rohde}, \bibfnamefont
  {Peter~P}}, and\ \bibinfo {author} {\bibfnamefont {Timothy~C.}\ \bibnamefont
  {Ralph}}} (\bibinfo {year} {2011}),\ \bibfield  {title} {\enquote {\bibinfo
  {title} {Time-resolved detection and mode-mismatch in a linear optics quantum
  gate},}\ }\href {https://doi.org/10.1088/1367-2630/13/5/053036} {\bibfield
  {journal} {\bibinfo  {journal} {New Journal of Physics}\ }\textbf {\bibinfo
  {volume} {13}},\ \bibinfo {pages} {053036}}\BibitemShut {NoStop}%
\bibitem [{\citenamefont {Rohde}\ and\ \citenamefont
  {Ralph}(2012)}]{bib:RohdeErrBS12}%
  \BibitemOpen
  \bibfield  {author} {\bibinfo {author} {\bibnamefont {Rohde}, \bibfnamefont
  {Peter~P}}, and\ \bibinfo {author} {\bibfnamefont {Timothy~C.}\ \bibnamefont
  {Ralph}}} (\bibinfo {year} {2012}),\ \bibfield  {title} {\enquote {\bibinfo
  {title} {Error tolerance of the bosonsampling model for linear optics quantum
  computing},}\ }\href {https://doi.org/10.1103/physreva.85.022332} {\bibfield
  {journal} {\bibinfo  {journal} {Physical Review A}\ }\textbf {\bibinfo
  {volume} {85}},\ \bibinfo {pages} {022332}},\ \Eprint
  {https://arxiv.org/abs/arXiv:1111.2426v1} {arXiv:1111.2426v1} \BibitemShut
  {NoStop}%
\bibitem [{\citenamefont {Rohde}\ \emph {et~al.}(2006)\citenamefont {Rohde},
  \citenamefont {Ralph},\ and\ \citenamefont {Munro}}]{bib:RohdeOptEntPur06}%
  \BibitemOpen
  \bibfield  {author} {\bibinfo {author} {\bibnamefont {Rohde}, \bibfnamefont
  {Peter~P}}, \bibinfo {author} {\bibfnamefont {Timothy~C.}\ \bibnamefont
  {Ralph}}, and\ \bibinfo {author} {\bibfnamefont {William~J.}\ \bibnamefont
  {Munro}}} (\bibinfo {year} {2006}),\ \bibfield  {title} {\enquote {\bibinfo
  {title} {Practical limitations in optical entanglement purification},}\
  }\href {https://doi.org/10.1103/physreva.73.030301} {\bibfield  {journal}
  {\bibinfo  {journal} {Physical Review A}\ }\textbf {\bibinfo {volume} {73}},\
  \bibinfo {pages} {030301(R)}},\ \Eprint
  {https://arxiv.org/abs/arXiv:quant-ph/0511268v1} {arXiv:quant-ph/0511268v1}
  \BibitemShut {NoStop}%
\bibitem [{\citenamefont {Rohde}\ \emph
  {et~al.}(2007{\natexlab{b}})\citenamefont {Rohde}, \citenamefont {Ralph},\
  and\ \citenamefont {Munro}}]{bib:RohdeRalphMunro07}%
  \BibitemOpen
  \bibfield  {author} {\bibinfo {author} {\bibnamefont {Rohde}, \bibfnamefont
  {Peter~P}}, \bibinfo {author} {\bibfnamefont {Timothy~C.}\ \bibnamefont
  {Ralph}}, and\ \bibinfo {author} {\bibfnamefont {William~J.}\ \bibnamefont
  {Munro}}} (\bibinfo {year} {2007}{\natexlab{b}}),\ \bibfield  {title}
  {\enquote {\bibinfo {title} {Error tolerance and tradeoffs in loss- and
  failure-tolerant quantum computing schemes},}\ }\href
  {https://doi.org/10.1103/physreva.75.010302} {\bibfield  {journal} {\bibinfo
  {journal} {Physical Review A}\ }\textbf {\bibinfo {volume} {75}},\ \bibinfo
  {pages} {010302(R)}},\ \Eprint
  {https://arxiv.org/abs/arXiv:quant-ph/0603130v3} {arXiv:quant-ph/0603130v3}
  \BibitemShut {NoStop}%
\bibitem [{\citenamefont {Rohde}\ \emph
  {et~al.}(2005{\natexlab{b}})\citenamefont {Rohde}, \citenamefont {Ralph},\
  and\ \citenamefont {Nielsen}}]{bib:RohdeOptPhot05}%
  \BibitemOpen
  \bibfield  {author} {\bibinfo {author} {\bibnamefont {Rohde}, \bibfnamefont
  {Peter~P}}, \bibinfo {author} {\bibfnamefont {Timothy~C.}\ \bibnamefont
  {Ralph}}, and\ \bibinfo {author} {\bibfnamefont {Michael~A.}\ \bibnamefont
  {Nielsen}}} (\bibinfo {year} {2005}{\natexlab{b}}),\ \bibfield  {title}
  {\enquote {\bibinfo {title} {Optimal photons for quantum information
  processing},}\ }\href {https://doi.org/10.1103/physreva.72.052332} {\bibfield
   {journal} {\bibinfo  {journal} {Physical Review A}\ }\textbf {\bibinfo
  {volume} {72}},\ \bibinfo {pages} {052332}},\ \Eprint
  {https://arxiv.org/abs/arXiv:quant-ph/0505139v1} {arXiv:quant-ph/0505139v1}
  \BibitemShut {NoStop}%
\bibitem [{\citenamefont {Rohde}\ \emph {et~al.}(2011)\citenamefont {Rohde},
  \citenamefont {Schreiber}, \citenamefont {Stefanak}, \citenamefont {Jex},\
  and\ \citenamefont {Silberhorn}}]{bib:RohdeMultiWalk11}%
  \BibitemOpen
  \bibfield  {author} {\bibinfo {author} {\bibnamefont {Rohde}, \bibfnamefont
  {Peter~P}}, \bibinfo {author} {\bibfnamefont {Andreas}\ \bibnamefont
  {Schreiber}}, \bibinfo {author} {\bibfnamefont {Martin}\ \bibnamefont
  {Stefanak}}, \bibinfo {author} {\bibfnamefont {Igor}\ \bibnamefont {Jex}},
  and\ \bibinfo {author} {\bibfnamefont {Christine}\ \bibnamefont
  {Silberhorn}}} (\bibinfo {year} {2011}),\ \bibfield  {title} {\enquote
  {\bibinfo {title} {Multi-walker discrete time quantum walks on arbitrary
  graphs, their properties, and their photonic implementation},}\ }\href
  {https://doi.org/10.1088/1367-2630/13/1/013001} {\bibfield  {journal}
  {\bibinfo  {journal} {New Journal of Physics}\ }\textbf {\bibinfo {volume}
  {13}},\ \bibinfo {pages} {013001}}\BibitemShut {NoStop}%
\bibitem [{\citenamefont {Rohde}\ \emph
  {et~al.}(2007{\natexlab{c}})\citenamefont {Rohde}, \citenamefont {Webb},
  \citenamefont {Huntington},\ and\ \citenamefont
  {Ralph}}]{bib:RohdeCompDet07}%
  \BibitemOpen
  \bibfield  {author} {\bibinfo {author} {\bibnamefont {Rohde}, \bibfnamefont
  {Peter~P}}, \bibinfo {author} {\bibfnamefont {James~G.}\ \bibnamefont
  {Webb}}, \bibinfo {author} {\bibfnamefont {Elanor~H.}\ \bibnamefont
  {Huntington}}, and\ \bibinfo {author} {\bibfnamefont {Timothy~C.}\
  \bibnamefont {Ralph}}} (\bibinfo {year} {2007}{\natexlab{c}}),\ \bibfield
  {title} {\enquote {\bibinfo {title} {Comparison of architectures for
  approximating number-resolving photo-detection using non-number-resolving
  detectors},}\ }\href@noop {} {\bibfield  {journal} {\bibinfo  {journal} {New
  Journal of Physics}\ }\textbf {\bibinfo {volume} {9}},\ \bibinfo {pages}
  {233}},\ \Eprint {https://arxiv.org/abs/arXiv:0705.4003v1}
  {arXiv:0705.4003v1} \BibitemShut {NoStop}%
\bibitem [{\citenamefont {Roland}\ and\ \citenamefont
  {Cerf}(2002)}]{bib:PhysRevA.65.042308}%
  \BibitemOpen
  \bibfield  {author} {\bibinfo {author} {\bibnamefont {Roland}, \bibfnamefont
  {J\'er\'emie}}, and\ \bibinfo {author} {\bibfnamefont {Nicolas~J.}\
  \bibnamefont {Cerf}}} (\bibinfo {year} {2002}),\ \bibfield  {title} {\enquote
  {\bibinfo {title} {Quantum search by local adiabatic evolution},}\ }\href
  {https://doi.org/10.1103/PhysRevA.65.042308} {\bibfield  {journal} {\bibinfo
  {journal} {Physical Review A}\ }\textbf {\bibinfo {volume} {65}},\ \bibinfo
  {pages} {042308}},\ \Eprint {https://arxiv.org/abs/arXiv:quant-ph/0107015v1}
  {arXiv:quant-ph/0107015v1} \BibitemShut {NoStop}%
\bibitem [{\citenamefont {Romero}\ \emph {et~al.}(2017)\citenamefont {Romero},
  \citenamefont {Olson},\ and\ \citenamefont
  {Aspuru-Guzik}}]{romero2017quantum}%
  \BibitemOpen
  \bibfield  {author} {\bibinfo {author} {\bibnamefont {Romero}, \bibfnamefont
  {Jonathan}}, \bibinfo {author} {\bibfnamefont {Jonathan~P}\ \bibnamefont
  {Olson}}, and\ \bibinfo {author} {\bibfnamefont {Alan}\ \bibnamefont
  {Aspuru-Guzik}}} (\bibinfo {year} {2017}),\ \bibfield  {title} {\enquote
  {\bibinfo {title} {Quantum autoencoders for efficient compression of quantum
  data},}\ }\href@noop {} {\bibfield  {journal} {\bibinfo  {journal} {Quantum
  Science and Technology}\ }\textbf {\bibinfo {volume} {2}}~(\bibinfo {number}
  {4}),\ \bibinfo {pages} {045001}}\BibitemShut {NoStop}%
\bibitem [{\citenamefont {Rosenberg}\ \emph {et~al.}(2007)\citenamefont
  {Rosenberg}, \citenamefont {Harrington}, \citenamefont {Rice}, \citenamefont
  {Hiskett}, \citenamefont {Peterson}, \citenamefont {Hughes}, \citenamefont
  {Lita}, \citenamefont {Nam},\ and\ \citenamefont
  {Nordholt}}]{bib:rosenberg2007long}%
  \BibitemOpen
  \bibfield  {author} {\bibinfo {author} {\bibnamefont {Rosenberg},
  \bibfnamefont {Danna}}, \bibinfo {author} {\bibfnamefont {Jim~W}\
  \bibnamefont {Harrington}}, \bibinfo {author} {\bibfnamefont {Patrick~R}\
  \bibnamefont {Rice}}, \bibinfo {author} {\bibfnamefont {Philip~A}\
  \bibnamefont {Hiskett}}, \bibinfo {author} {\bibfnamefont {Charles~G}\
  \bibnamefont {Peterson}}, \bibinfo {author} {\bibfnamefont {Richard~J}\
  \bibnamefont {Hughes}}, \bibinfo {author} {\bibfnamefont {Adriana~E}\
  \bibnamefont {Lita}}, \bibinfo {author} {\bibfnamefont {Sae~Woo}\
  \bibnamefont {Nam}}, and\ \bibinfo {author} {\bibfnamefont {Jane~E}\
  \bibnamefont {Nordholt}}} (\bibinfo {year} {2007}),\ \bibfield  {title}
  {\enquote {\bibinfo {title} {Long-distance decoy-state quantum key
  distribution in optical fiber},}\ }\href
  {https://doi.org/10.1103/physrevlett.98.010503} {\bibfield  {journal}
  {\bibinfo  {journal} {Physical Review Letters}\ }\textbf {\bibinfo {volume}
  {98}},\ \bibinfo {pages} {010503}},\ \Eprint
  {https://arxiv.org/abs/arXiv:quant-ph/0607186v2} {arXiv:quant-ph/0607186v2}
  \BibitemShut {NoStop}%
\bibitem [{\citenamefont {Rubenok}\ \emph {et~al.}(2013)\citenamefont
  {Rubenok}, \citenamefont {Slater}, \citenamefont {Chan}, \citenamefont
  {Lucio-Martinez},\ and\ \citenamefont {Tittel}}]{bib:PRL_111_130501}%
  \BibitemOpen
  \bibfield  {author} {\bibinfo {author} {\bibnamefont {Rubenok}, \bibfnamefont
  {Allison}}, \bibinfo {author} {\bibfnamefont {Joshua~A}\ \bibnamefont
  {Slater}}, \bibinfo {author} {\bibfnamefont {Philip}\ \bibnamefont {Chan}},
  \bibinfo {author} {\bibfnamefont {Itzel}\ \bibnamefont {Lucio-Martinez}},
  and\ \bibinfo {author} {\bibfnamefont {Wolfgang}\ \bibnamefont {Tittel}}}
  (\bibinfo {year} {2013}),\ \bibfield  {title} {\enquote {\bibinfo {title}
  {Real-world two-photon interference and proof-of-principle quantum key
  distribution immune to detector attacks},}\ }\href
  {https://doi.org/10.1103/physrevlett.111.130501} {\bibfield  {journal}
  {\bibinfo  {journal} {Physical Review Letters}\ }\textbf {\bibinfo {volume}
  {111}},\ \bibinfo {pages} {130501}},\ \Eprint
  {https://arxiv.org/abs/arXiv:1304.2463v1} {arXiv:1304.2463v1} \BibitemShut
  {NoStop}%
\bibitem [{\citenamefont {Ryser}(1963)}]{bib:RyserAlg}%
  \BibitemOpen
  \bibfield  {author} {\bibinfo {author} {\bibnamefont {Ryser}, \bibfnamefont
  {Herbert~John}}} (\bibinfo {year} {1963}),\ \href@noop {} {\bibfield
  {journal} {\bibinfo  {journal} {Combinatorial Mathematics, Carus Mathematical
  Monograph}\ }\textbf {\bibinfo {volume} {14}}}\BibitemShut {NoStop}%
\bibitem [{\citenamefont {Saeedi}\ \emph {et~al.}(2013)\citenamefont {Saeedi},
  \citenamefont {Simmons}, \citenamefont {Salvail}, \citenamefont {Dluhy},
  \citenamefont {Riemann}, \citenamefont {Abrosimov}, \citenamefont {Becker},
  \citenamefont {Pohl}, \citenamefont {Morton},\ and\ \citenamefont
  {Thewalt}}]{bib:saeedi2013room}%
  \BibitemOpen
  \bibfield  {author} {\bibinfo {author} {\bibnamefont {Saeedi}, \bibfnamefont
  {Kamyar}}, \bibinfo {author} {\bibfnamefont {Stephanie}\ \bibnamefont
  {Simmons}}, \bibinfo {author} {\bibfnamefont {Jeff~Z}\ \bibnamefont
  {Salvail}}, \bibinfo {author} {\bibfnamefont {Phillip}\ \bibnamefont
  {Dluhy}}, \bibinfo {author} {\bibfnamefont {Helge}\ \bibnamefont {Riemann}},
  \bibinfo {author} {\bibfnamefont {Nikolai~V}\ \bibnamefont {Abrosimov}},
  \bibinfo {author} {\bibfnamefont {Peter}\ \bibnamefont {Becker}}, \bibinfo
  {author} {\bibfnamefont {Hans-Joachim}\ \bibnamefont {Pohl}}, \bibinfo
  {author} {\bibfnamefont {John~JL}\ \bibnamefont {Morton}}, and\ \bibinfo
  {author} {\bibfnamefont {Mike~LW}\ \bibnamefont {Thewalt}}} (\bibinfo {year}
  {2013}),\ \bibfield  {title} {\enquote {\bibinfo {title} {Room-temperature
  quantum bit storage exceeding 39 minutes using ionized donors in
  silicon-28},}\ }\href {https://doi.org/10.1126/science.1239584} {\bibfield
  {journal} {\bibinfo  {journal} {Science}\ }\textbf {\bibinfo {volume}
  {342}},\ \bibinfo {pages} {830}}\BibitemShut {NoStop}%
\bibitem [{\citenamefont {Saglamyurek}\ \emph {et~al.}(2015)\citenamefont
  {Saglamyurek}, \citenamefont {Jin}, \citenamefont {Verma}, \citenamefont
  {Shaw}, \citenamefont {Marsili}, \citenamefont {Nam}, \citenamefont {Oblak},\
  and\ \citenamefont {Tittel}}]{bib:saglamyurek2015quantum}%
  \BibitemOpen
  \bibfield  {author} {\bibinfo {author} {\bibnamefont {Saglamyurek},
  \bibfnamefont {Erhan}}, \bibinfo {author} {\bibfnamefont {Jeongwan}\
  \bibnamefont {Jin}}, \bibinfo {author} {\bibfnamefont {Varun~B}\ \bibnamefont
  {Verma}}, \bibinfo {author} {\bibfnamefont {Matthew~D}\ \bibnamefont {Shaw}},
  \bibinfo {author} {\bibfnamefont {Francesco}\ \bibnamefont {Marsili}},
  \bibinfo {author} {\bibfnamefont {Sae~Woo}\ \bibnamefont {Nam}}, \bibinfo
  {author} {\bibfnamefont {Daniel}\ \bibnamefont {Oblak}}, and\ \bibinfo
  {author} {\bibfnamefont {Wolfgang}\ \bibnamefont {Tittel}}} (\bibinfo {year}
  {2015}),\ \bibfield  {title} {\enquote {\bibinfo {title} {Quantum storage of
  entangled telecom-wavelength photons in an erbium-doped optical fibre},}\
  }\href {https://doi.org/10.1038/nphoton.2014.311} {\bibfield  {journal}
  {\bibinfo  {journal} {Nature Photonics}\ }\textbf {\bibinfo {volume} {9}},\
  \bibinfo {pages} {83}},\ \Eprint {https://arxiv.org/abs/arXiv:1409.0831v2}
  {arXiv:1409.0831v2} \BibitemShut {NoStop}%
\bibitem [{\citenamefont {Saglamyurek}\ \emph {et~al.}(2011)\citenamefont
  {Saglamyurek}, \citenamefont {Sinclair}, \citenamefont {Jin}, \citenamefont
  {Slater}, \citenamefont {Oblak}, \citenamefont {Bussieres}, \citenamefont
  {George}, \citenamefont {Ricken}, \citenamefont {Sohler},\ and\ \citenamefont
  {Tittel}}]{bib:Nat_469_512}%
  \BibitemOpen
  \bibfield  {author} {\bibinfo {author} {\bibnamefont {Saglamyurek},
  \bibfnamefont {Erhan}}, \bibinfo {author} {\bibfnamefont {Neil}\ \bibnamefont
  {Sinclair}}, \bibinfo {author} {\bibfnamefont {Jeongwan}\ \bibnamefont
  {Jin}}, \bibinfo {author} {\bibfnamefont {Joshua~A}\ \bibnamefont {Slater}},
  \bibinfo {author} {\bibfnamefont {Daniel}\ \bibnamefont {Oblak}}, \bibinfo
  {author} {\bibfnamefont {F{\'e}lix}\ \bibnamefont {Bussieres}}, \bibinfo
  {author} {\bibfnamefont {Mathew}\ \bibnamefont {George}}, \bibinfo {author}
  {\bibfnamefont {Raimund}\ \bibnamefont {Ricken}}, \bibinfo {author}
  {\bibfnamefont {Wolfgang}\ \bibnamefont {Sohler}}, and\ \bibinfo {author}
  {\bibfnamefont {Wolfgang}\ \bibnamefont {Tittel}}} (\bibinfo {year} {2011}),\
  \bibfield  {title} {\enquote {\bibinfo {title} {Broadband waveguide quantum
  memory for entangled photons},}\ }\href
  {https://doi.org/10.1109/phosst.2011.5999934} {\bibfield  {journal} {\bibinfo
   {journal} {Nature}\ }\textbf {\bibinfo {volume} {469}},\ \bibinfo {pages}
  {512}}\BibitemShut {NoStop}%
\bibitem [{\citenamefont {Sajeed}\ \emph {et~al.}(2016)\citenamefont {Sajeed},
  \citenamefont {Huang}, \citenamefont {Sun}, \citenamefont {Xu}, \citenamefont
  {Makarov},\ and\ \citenamefont {Curty}}]{bib:PhysRevLett.117.250505}%
  \BibitemOpen
  \bibfield  {author} {\bibinfo {author} {\bibnamefont {Sajeed}, \bibfnamefont
  {Shihan}}, \bibinfo {author} {\bibfnamefont {Anqi}\ \bibnamefont {Huang}},
  \bibinfo {author} {\bibfnamefont {Shihai}\ \bibnamefont {Sun}}, \bibinfo
  {author} {\bibfnamefont {Feihu}\ \bibnamefont {Xu}}, \bibinfo {author}
  {\bibfnamefont {Vadim}\ \bibnamefont {Makarov}}, and\ \bibinfo {author}
  {\bibfnamefont {Marcos}\ \bibnamefont {Curty}}} (\bibinfo {year} {2016}),\
  \bibfield  {title} {\enquote {\bibinfo {title} {Insecurity of
  detector-device-independent quantum key distribution},}\ }\href
  {https://doi.org/10.1103/physrevlett.117.250505} {\bibfield  {journal}
  {\bibinfo  {journal} {Physical Review Letters}\ }\textbf {\bibinfo {volume}
  {117}},\ \bibinfo {pages} {250505}},\ \Eprint
  {https://arxiv.org/abs/arXiv:1607.05814v1} {arXiv:1607.05814v1} \BibitemShut
  {NoStop}%
\bibitem [{\citenamefont {Sakuma}\ \emph {et~al.}(2003)\citenamefont {Sakuma},
  \citenamefont {Ishikawa}, \citenamefont {Shikata}, \citenamefont {Fukuda},\
  and\ \citenamefont {Hosoya}}]{bib:sakuma2003ultra}%
  \BibitemOpen
  \bibfield  {author} {\bibinfo {author} {\bibnamefont {Sakuma}, \bibfnamefont
  {Ken}}, \bibinfo {author} {\bibfnamefont {Shimon}\ \bibnamefont {Ishikawa}},
  \bibinfo {author} {\bibfnamefont {Tomoko}\ \bibnamefont {Shikata}}, \bibinfo
  {author} {\bibfnamefont {Takeshi}\ \bibnamefont {Fukuda}}, and\ \bibinfo
  {author} {\bibfnamefont {Hideyuki}\ \bibnamefont {Hosoya}}} (\bibinfo {year}
  {2003}),\ \bibfield  {title} {\enquote {\bibinfo {title} {Ultra low-loss
  waveguides of 0.12 db/cm directly written in pure silica glass by femtosecond
  laser pulses},}\ }in\ \href {https://doi.org/10.1109/ofc.2003.315926} {\emph
  {\bibinfo {booktitle} {Optical Fiber Communication Conference}}}\ (\bibinfo
  {organization} {Optical Society of America})\ p.\ \bibinfo {pages}
  {ThD2}\BibitemShut {NoStop}%
\bibitem [{\citenamefont {Sakurai}(1994)}]{bib:Sakurai94}%
  \BibitemOpen
  \bibfield  {author} {\bibinfo {author} {\bibnamefont {Sakurai}, \bibfnamefont
  {J~J}}} (\bibinfo {year} {1994}),\ \href
  {https://doi.org/10.1017/9781108499996} {\emph {\bibinfo {title} {Modern
  Quantum Mechanics}}}\ (\bibinfo  {publisher} {Addison-Wesley})\BibitemShut
  {NoStop}%
\bibitem [{\citenamefont {Sangouard}\ \emph {et~al.}(2009)\citenamefont
  {Sangouard}, \citenamefont {Dubessy},\ and\ \citenamefont
  {Simon}}]{bib:PRA_79_042340}%
  \BibitemOpen
  \bibfield  {author} {\bibinfo {author} {\bibnamefont {Sangouard},
  \bibfnamefont {Nicolas}}, \bibinfo {author} {\bibfnamefont {Romain}\
  \bibnamefont {Dubessy}}, and\ \bibinfo {author} {\bibfnamefont {Christoph}\
  \bibnamefont {Simon}}} (\bibinfo {year} {2009}),\ \bibfield  {title}
  {\enquote {\bibinfo {title} {Quantum repeaters based on single trapped
  ions},}\ }\href {https://doi.org/10.1103/physreva.79.042340} {\bibfield
  {journal} {\bibinfo  {journal} {Physical Review A}\ }\textbf {\bibinfo
  {volume} {79}},\ \bibinfo {pages} {042340}},\ \Eprint
  {https://arxiv.org/abs/arXiv:0902.3127v2} {arXiv:0902.3127v2} \BibitemShut
  {NoStop}%
\bibitem [{\citenamefont {Sangouard}\ \emph
  {et~al.}(2011{\natexlab{a}})\citenamefont {Sangouard}, \citenamefont {Simon},
  \citenamefont {De~Riedmatten},\ and\ \citenamefont
  {Gisin}}]{bib:sangouard2011quantum}%
  \BibitemOpen
  \bibfield  {author} {\bibinfo {author} {\bibnamefont {Sangouard},
  \bibfnamefont {Nicolas}}, \bibinfo {author} {\bibfnamefont {Christoph}\
  \bibnamefont {Simon}}, \bibinfo {author} {\bibfnamefont {Hugues}\
  \bibnamefont {De~Riedmatten}}, and\ \bibinfo {author} {\bibfnamefont
  {Nicolas}\ \bibnamefont {Gisin}}} (\bibinfo {year} {2011}{\natexlab{a}}),\
  \bibfield  {title} {\enquote {\bibinfo {title} {Quantum repeaters based on
  atomic ensembles and linear optics},}\ }\href
  {https://doi.org/10.1103/revmodphys.83.33} {\bibfield  {journal} {\bibinfo
  {journal} {Reviews in Modern Physics}\ }\textbf {\bibinfo {volume} {83}},\
  \bibinfo {pages} {33}}\BibitemShut {NoStop}%
\bibitem [{\citenamefont {Sangouard}\ \emph
  {et~al.}(2011{\natexlab{b}})\citenamefont {Sangouard}, \citenamefont {Simon},
  \citenamefont {De~Riedmatten},\ and\ \citenamefont {Gisin}}]{bib:RMP_83_33}%
  \BibitemOpen
  \bibfield  {author} {\bibinfo {author} {\bibnamefont {Sangouard},
  \bibfnamefont {Nicolas}}, \bibinfo {author} {\bibfnamefont {Christoph}\
  \bibnamefont {Simon}}, \bibinfo {author} {\bibfnamefont {Hugues}\
  \bibnamefont {De~Riedmatten}}, and\ \bibinfo {author} {\bibfnamefont
  {Nicolas}\ \bibnamefont {Gisin}}} (\bibinfo {year} {2011}{\natexlab{b}}),\
  \bibfield  {title} {\enquote {\bibinfo {title} {Quantum repeaters based on
  atomic ensembles and linear optics},}\ }\href
  {https://doi.org/10.1103/revmodphys.83.33} {\bibfield  {journal} {\bibinfo
  {journal} {Reviews in Modern Physics}\ }\textbf {\bibinfo {volume} {83}},\
  \bibinfo {pages} {33}}\BibitemShut {NoStop}%
\bibitem [{\citenamefont {Sangouard}\ \emph {et~al.}(2007)\citenamefont
  {Sangouard}, \citenamefont {Simon}, \citenamefont {Min{\'a}{\v{r}}},
  \citenamefont {Zbinden}, \citenamefont {De~Riedmatten},\ and\ \citenamefont
  {Gisin}}]{bib:PRA_76_050301}%
  \BibitemOpen
  \bibfield  {author} {\bibinfo {author} {\bibnamefont {Sangouard},
  \bibfnamefont {Nicolas}}, \bibinfo {author} {\bibfnamefont {Christoph}\
  \bibnamefont {Simon}}, \bibinfo {author} {\bibfnamefont {Ji{\v{r}}{\'\i}}\
  \bibnamefont {Min{\'a}{\v{r}}}}, \bibinfo {author} {\bibfnamefont {Hugo}\
  \bibnamefont {Zbinden}}, \bibinfo {author} {\bibfnamefont {Hugues}\
  \bibnamefont {De~Riedmatten}}, and\ \bibinfo {author} {\bibfnamefont
  {Nicolas}\ \bibnamefont {Gisin}}} (\bibinfo {year} {2007}),\ \bibfield
  {title} {\enquote {\bibinfo {title} {Long-distance entanglement distribution
  with single-photon sources},}\ }\href
  {https://doi.org/10.1103/physreva.76.050301} {\bibfield  {journal} {\bibinfo
  {journal} {Physical Review A}\ }\textbf {\bibinfo {volume} {76}},\ \bibinfo
  {pages} {050301}},\ \Eprint {https://arxiv.org/abs/arXiv:0706.1924v1}
  {arXiv:0706.1924v1} \BibitemShut {NoStop}%
\bibitem [{\citenamefont {Sangouard}\ \emph
  {et~al.}(2011{\natexlab{c}})\citenamefont {Sangouard}, \citenamefont {Simon},
  \citenamefont {de~Riedmatten},\ and\ \citenamefont
  {Gisin}}]{bib:sangouard11}%
  \BibitemOpen
  \bibfield  {author} {\bibinfo {author} {\bibnamefont {Sangouard},
  \bibfnamefont {Nicolas}}, \bibinfo {author} {\bibfnamefont {Christoph}\
  \bibnamefont {Simon}}, \bibinfo {author} {\bibfnamefont {Hugues}\
  \bibnamefont {de~Riedmatten}}, and\ \bibinfo {author} {\bibfnamefont
  {Nicolas}\ \bibnamefont {Gisin}}} (\bibinfo {year} {2011}{\natexlab{c}}),\
  \bibfield  {title} {\enquote {\bibinfo {title} {Quantum repeaters based on
  atomic ensembles and linear optics},}\ }\href
  {https://doi.org/10.1103/revmodphys.83.33} {\bibfield  {journal} {\bibinfo
  {journal} {Reviews in Modern Physics}\ }\textbf {\bibinfo {volume} {83}},\
  \bibinfo {pages} {33}}\BibitemShut {NoStop}%
\bibitem [{\citenamefont {Sangouard}\ \emph
  {et~al.}(2011{\natexlab{d}})\citenamefont {Sangouard}, \citenamefont {Simon},
  \citenamefont {de~Riedmatten},\ and\ \citenamefont
  {Gisin}}]{SD-Sangouard:2011aa}%
  \BibitemOpen
  \bibfield  {author} {\bibinfo {author} {\bibnamefont {Sangouard},
  \bibfnamefont {Nicolas}}, \bibinfo {author} {\bibfnamefont {Christoph}\
  \bibnamefont {Simon}}, \bibinfo {author} {\bibfnamefont {Hugues}\
  \bibnamefont {de~Riedmatten}}, and\ \bibinfo {author} {\bibfnamefont
  {Nicolas}\ \bibnamefont {Gisin}}} (\bibinfo {year} {2011}{\natexlab{d}}),\
  \bibfield  {title} {\enquote {\bibinfo {title} {Quantum repeaters based on
  atomic ensembles and linear optics},}\ }\href
  {https://doi.org/10.1103/RevModPhys.83.33} {\bibfield  {journal} {\bibinfo
  {journal} {Reviews of Modern Physics}\ }\textbf {\bibinfo {volume} {83}},\
  \bibinfo {pages} {33}}\BibitemShut {NoStop}%
\bibitem [{\citenamefont {Santori}\ \emph {et~al.}(2001)\citenamefont
  {Santori}, \citenamefont {Pelton}, \citenamefont {Solomon}, \citenamefont
  {Dale},\ and\ \citenamefont {Yamamoto}}]{bib:Santori01}%
  \BibitemOpen
  \bibfield  {author} {\bibinfo {author} {\bibnamefont {Santori}, \bibfnamefont
  {C}}, \bibinfo {author} {\bibfnamefont {M.}~\bibnamefont {Pelton}}, \bibinfo
  {author} {\bibfnamefont {G.}~\bibnamefont {Solomon}}, \bibinfo {author}
  {\bibfnamefont {Y.}~\bibnamefont {Dale}}, and\ \bibinfo {author}
  {\bibfnamefont {Y.}~\bibnamefont {Yamamoto}}} (\bibinfo {year} {2001}),\
  \bibfield  {title} {\enquote {\bibinfo {title} {Triggered single photons from
  a quantum dot},}\ }\href {https://doi.org/10.1103/physrevlett.86.1502}
  {\bibfield  {journal} {\bibinfo  {journal} {Physical Review Letters}\
  }\textbf {\bibinfo {volume} {86}},\ \bibinfo {pages} {1502}},\ \Eprint
  {https://arxiv.org/abs/arXiv:cond-mat/0012379v1} {arXiv:cond-mat/0012379v1}
  \BibitemShut {NoStop}%
\bibitem [{\citenamefont {Sarandy}\ and\ \citenamefont
  {Lidar}(2005)}]{bib:PhysRevLett.95.250503}%
  \BibitemOpen
  \bibfield  {author} {\bibinfo {author} {\bibnamefont {Sarandy}, \bibfnamefont
  {M~S}}, and\ \bibinfo {author} {\bibfnamefont {D.~A.}\ \bibnamefont {Lidar}}}
  (\bibinfo {year} {2005}),\ \bibfield  {title} {\enquote {\bibinfo {title}
  {Adiabatic quantum computation in open systems},}\ }\href
  {https://doi.org/10.1103/PhysRevLett.95.250503} {\bibfield  {journal}
  {\bibinfo  {journal} {Physical Review Letters}\ }\textbf {\bibinfo {volume}
  {95}},\ \bibinfo {pages} {250503}},\ \Eprint
  {https://arxiv.org/abs/arXiv:quant-ph/0502014v2} {arXiv:quant-ph/0502014v2}
  \BibitemShut {NoStop}%
\bibitem [{\citenamefont {Sasaki}\ and\ \citenamefont
  {Carlini}(2002)}]{bib:sasaki2}%
  \BibitemOpen
  \bibfield  {author} {\bibinfo {author} {\bibnamefont {Sasaki}, \bibfnamefont
  {Masahide}}, and\ \bibinfo {author} {\bibfnamefont {Alberto}\ \bibnamefont
  {Carlini}}} (\bibinfo {year} {2002}),\ \bibfield  {title} {\enquote {\bibinfo
  {title} {Quantum learning and universal quantum matching machine},}\ }\href
  {https://doi.org/10.1103/physreva.66.022303} {\bibfield  {journal} {\bibinfo
  {journal} {Physical Review A}\ }\textbf {\bibinfo {volume} {66}},\ \bibinfo
  {pages} {022303}},\ \Eprint {https://arxiv.org/abs/arXiv:quant-ph/0202173v1}
  {arXiv:quant-ph/0202173v1} \BibitemShut {NoStop}%
\bibitem [{\citenamefont {Sasaki}\ \emph {et~al.}(2001)\citenamefont {Sasaki},
  \citenamefont {Carlini},\ and\ \citenamefont {Jozsa}}]{bib:sasaki1}%
  \BibitemOpen
  \bibfield  {author} {\bibinfo {author} {\bibnamefont {Sasaki}, \bibfnamefont
  {Masahide}}, \bibinfo {author} {\bibfnamefont {Alberto}\ \bibnamefont
  {Carlini}}, and\ \bibinfo {author} {\bibfnamefont {Richard}\ \bibnamefont
  {Jozsa}}} (\bibinfo {year} {2001}),\ \bibfield  {title} {\enquote {\bibinfo
  {title} {Quantum template matching},}\ }\href
  {https://doi.org/10.1103/physreva.64.022317} {\bibfield  {journal} {\bibinfo
  {journal} {Physical Review A}\ }\textbf {\bibinfo {volume} {64}},\ \bibinfo
  {pages} {022317}}\BibitemShut {NoStop}%
\bibitem [{\citenamefont {Sasaki}\ \emph {et~al.}(2011)\citenamefont {Sasaki},
  \citenamefont {Fujiwara}, \citenamefont {Ishizuka}, \citenamefont {Klaus},
  \citenamefont {Wakui}, \citenamefont {Takeoka}, \citenamefont {Miki},
  \citenamefont {Yamashita}, \citenamefont {Wang}, \citenamefont {Tanaka} \emph
  {et~al.}}]{bib:OExp_19_10387}%
  \BibitemOpen
  \bibfield  {author} {\bibinfo {author} {\bibnamefont {Sasaki}, \bibfnamefont
  {Masahide}}, \bibinfo {author} {\bibfnamefont {M}~\bibnamefont {Fujiwara}},
  \bibinfo {author} {\bibfnamefont {H}~\bibnamefont {Ishizuka}}, \bibinfo
  {author} {\bibfnamefont {W}~\bibnamefont {Klaus}}, \bibinfo {author}
  {\bibfnamefont {K}~\bibnamefont {Wakui}}, \bibinfo {author} {\bibfnamefont
  {M}~\bibnamefont {Takeoka}}, \bibinfo {author} {\bibfnamefont
  {S}~\bibnamefont {Miki}}, \bibinfo {author} {\bibfnamefont {T}~\bibnamefont
  {Yamashita}}, \bibinfo {author} {\bibfnamefont {Z}~\bibnamefont {Wang}},
  \bibinfo {author} {\bibfnamefont {A}~\bibnamefont {Tanaka}},  \emph {et~al.}}
  (\bibinfo {year} {2011}),\ \bibfield  {title} {\enquote {\bibinfo {title}
  {Field test of quantum key distribution in the tokyo qkd network},}\ }\href
  {https://doi.org/10.1364/oe.19.010387} {\bibfield  {journal} {\bibinfo
  {journal} {Optics Express}\ }\textbf {\bibinfo {volume} {19}},\ \bibinfo
  {pages} {10387}},\ \Eprint {https://arxiv.org/abs/arXiv:1103.3566v1}
  {arXiv:1103.3566v1} \BibitemShut {NoStop}%
\bibitem [{\citenamefont {Sasaki}\ \emph {et~al.}(2014)\citenamefont {Sasaki},
  \citenamefont {Yamamoto},\ and\ \citenamefont
  {Koashi}}]{bib:sasaki2014practical}%
  \BibitemOpen
  \bibfield  {author} {\bibinfo {author} {\bibnamefont {Sasaki}, \bibfnamefont
  {Toshihiko}}, \bibinfo {author} {\bibfnamefont {Yoshihisa}\ \bibnamefont
  {Yamamoto}}, and\ \bibinfo {author} {\bibfnamefont {Masato}\ \bibnamefont
  {Koashi}}} (\bibinfo {year} {2014}),\ \bibfield  {title} {\enquote {\bibinfo
  {title} {Practical quantum key distribution protocol without monitoring
  signal disturbance},}\ }\href {https://doi.org/10.1038/nature13303}
  {\bibfield  {journal} {\bibinfo  {journal} {Nature}\ }\textbf {\bibinfo
  {volume} {509}},\ \bibinfo {pages} {475}}\BibitemShut {NoStop}%
\bibitem [{\citenamefont {Savage}(2018)}]{bib:savage2018quantum}%
  \BibitemOpen
  \bibfield  {author} {\bibinfo {author} {\bibnamefont {Savage}, \bibfnamefont
  {Neil}}} (\bibinfo {year} {2018}),\ \bibfield  {title} {\enquote {\bibinfo
  {title} {Quantum computers compete for supremacy},}\ }\href@noop {}
  {\bibfield  {journal} {\bibinfo  {journal} {Scientific American}\ }\textbf
  {\bibinfo {volume} {27}},\ \bibinfo {pages} {108}}\BibitemShut {NoStop}%
\bibitem [{\citenamefont {Scarani}\ \emph {et~al.}(2009)\citenamefont
  {Scarani}, \citenamefont {Bechmann-Pasquinucci}, \citenamefont {Cerf},
  \citenamefont {Du\ifmmode~\check{s}\else \v{s}\fi{}ek}, \citenamefont
  {L\"utkenhaus},\ and\ \citenamefont {Peev}}]{bib:RevModPhys.81.1301}%
  \BibitemOpen
  \bibfield  {author} {\bibinfo {author} {\bibnamefont {Scarani}, \bibfnamefont
  {Valerio}}, \bibinfo {author} {\bibfnamefont {Helle}\ \bibnamefont
  {Bechmann-Pasquinucci}}, \bibinfo {author} {\bibfnamefont {Nicolas~J.}\
  \bibnamefont {Cerf}}, \bibinfo {author} {\bibfnamefont {Miloslav}\
  \bibnamefont {Du\ifmmode~\check{s}\else \v{s}\fi{}ek}}, \bibinfo {author}
  {\bibfnamefont {Norbert}\ \bibnamefont {L\"utkenhaus}}, and\ \bibinfo
  {author} {\bibfnamefont {Momtchil}\ \bibnamefont {Peev}}} (\bibinfo {year}
  {2009}),\ \bibfield  {title} {\enquote {\bibinfo {title} {The security of
  practical quantum key distribution},}\ }\href
  {https://doi.org/10.1103/revmodphys.81.1301} {\bibfield  {journal} {\bibinfo
  {journal} {Reviews in Modern Physics}\ }\textbf {\bibinfo {volume} {81}},\
  \bibinfo {pages} {1301}},\ \Eprint {https://arxiv.org/abs/arXiv:0802.4155v3}
  {arXiv:0802.4155v3} \BibitemShut {NoStop}%
\bibitem [{\citenamefont {Scheidl}\ \emph {et~al.}(2013)\citenamefont
  {Scheidl}, \citenamefont {Wille},\ and\ \citenamefont
  {Ursin}}]{bib:scheidl2013quantum}%
  \BibitemOpen
  \bibfield  {author} {\bibinfo {author} {\bibnamefont {Scheidl}, \bibfnamefont
  {Thomas}}, \bibinfo {author} {\bibfnamefont {Eric}\ \bibnamefont {Wille}},
  and\ \bibinfo {author} {\bibfnamefont {Rupert}\ \bibnamefont {Ursin}}}
  (\bibinfo {year} {2013}),\ \bibfield  {title} {\enquote {\bibinfo {title}
  {Quantum optics experiments using the international space station: a
  proposal},}\ }\href {https://doi.org/10.1088/1367-2630/15/4/043008}
  {\bibfield  {journal} {\bibinfo  {journal} {New Journal of Physics}\ }\textbf
  {\bibinfo {volume} {15}},\ \bibinfo {pages} {043008}}\BibitemShut {NoStop}%
\bibitem [{\citenamefont {Schmitt-Manderbach}\ \emph
  {et~al.}(2007{\natexlab{a}})\citenamefont {Schmitt-Manderbach}, \citenamefont
  {Weier}, \citenamefont {F{\"u}rst}, \citenamefont {Ursin}, \citenamefont
  {Tiefenbacher}, \citenamefont {Scheidl}, \citenamefont {Perdigues},
  \citenamefont {Sodnik}, \citenamefont {Kurtsiefer}, \citenamefont {Rarity},
  \citenamefont {Zeilinger},\ and\ \citenamefont
  {Weinfurter}}]{SD-Schmitt-Manderbach:2007aa}%
  \BibitemOpen
  \bibfield  {author} {\bibinfo {author} {\bibnamefont {Schmitt-Manderbach},
  \bibfnamefont {Tobias}}, \bibinfo {author} {\bibfnamefont {Henning}\
  \bibnamefont {Weier}}, \bibinfo {author} {\bibfnamefont {Martin}\
  \bibnamefont {F{\"u}rst}}, \bibinfo {author} {\bibfnamefont {Rupert}\
  \bibnamefont {Ursin}}, \bibinfo {author} {\bibfnamefont {Felix}\ \bibnamefont
  {Tiefenbacher}}, \bibinfo {author} {\bibfnamefont {Thomas}\ \bibnamefont
  {Scheidl}}, \bibinfo {author} {\bibfnamefont {Josep}\ \bibnamefont
  {Perdigues}}, \bibinfo {author} {\bibfnamefont {Zoran}\ \bibnamefont
  {Sodnik}}, \bibinfo {author} {\bibfnamefont {Christian}\ \bibnamefont
  {Kurtsiefer}}, \bibinfo {author} {\bibfnamefont {John~G.}\ \bibnamefont
  {Rarity}}, \bibinfo {author} {\bibfnamefont {Anton}\ \bibnamefont
  {Zeilinger}}, and\ \bibinfo {author} {\bibfnamefont {Harald}\ \bibnamefont
  {Weinfurter}}} (\bibinfo {year} {2007}{\natexlab{a}}),\ \bibfield  {title}
  {\enquote {\bibinfo {title} {Experimental demonstration of free-space
  decoy-state quantum key distribution over 144 km},}\ }\href
  {https://doi.org/10.1103/PhysRevLett.98.010504} {\bibfield  {journal}
  {\bibinfo  {journal} {Physical Review Letters}\ }\textbf {\bibinfo {volume}
  {98}},\ \bibinfo {pages} {010504}}\BibitemShut {NoStop}%
\bibitem [{\citenamefont {Schmitt-Manderbach}\ \emph
  {et~al.}(2007{\natexlab{b}})\citenamefont {Schmitt-Manderbach}, \citenamefont
  {Weier}, \citenamefont {F{\"u}rst}, \citenamefont {Ursin}, \citenamefont
  {Tiefenbacher}, \citenamefont {Scheidl}, \citenamefont {Perdigues},
  \citenamefont {Sodnik}, \citenamefont {Kurtsiefer}, \citenamefont {Rarity}
  \emph {et~al.}}]{bib:PRL_98_010504}%
  \BibitemOpen
  \bibfield  {author} {\bibinfo {author} {\bibnamefont {Schmitt-Manderbach},
  \bibfnamefont {Tobias}}, \bibinfo {author} {\bibfnamefont {Henning}\
  \bibnamefont {Weier}}, \bibinfo {author} {\bibfnamefont {Martin}\
  \bibnamefont {F{\"u}rst}}, \bibinfo {author} {\bibfnamefont {Rupert}\
  \bibnamefont {Ursin}}, \bibinfo {author} {\bibfnamefont {Felix}\ \bibnamefont
  {Tiefenbacher}}, \bibinfo {author} {\bibfnamefont {Thomas}\ \bibnamefont
  {Scheidl}}, \bibinfo {author} {\bibfnamefont {Josep}\ \bibnamefont
  {Perdigues}}, \bibinfo {author} {\bibfnamefont {Zoran}\ \bibnamefont
  {Sodnik}}, \bibinfo {author} {\bibfnamefont {Christian}\ \bibnamefont
  {Kurtsiefer}}, \bibinfo {author} {\bibfnamefont {John~G}\ \bibnamefont
  {Rarity}},  \emph {et~al.}} (\bibinfo {year} {2007}{\natexlab{b}}),\
  \bibfield  {title} {\enquote {\bibinfo {title} {Experimental demonstration of
  free-space decoy-state quantum key distribution over 144 km},}\ }\href
  {https://doi.org/10.1103/physrevlett.98.010504} {\bibfield  {journal}
  {\bibinfo  {journal} {Physical Review Letters}\ }\textbf {\bibinfo {volume}
  {98}},\ \bibinfo {pages} {010504}}\BibitemShut {NoStop}%
\bibitem [{\citenamefont {Schneier}(1996)}]{bib:Schneier96}%
  \BibitemOpen
  \bibfield  {author} {\bibinfo {author} {\bibnamefont {Schneier},
  \bibfnamefont {Bruce}}} (\bibinfo {year} {1996}),\ \href@noop {} {\emph
  {\bibinfo {title} {Applied Cryptography}}}\ (\bibinfo  {publisher} {John
  Wiley \& Sons})\BibitemShut {NoStop}%
\bibitem [{\citenamefont {Schreiber}\ \emph {et~al.}(2011)\citenamefont
  {Schreiber}, \citenamefont {Cassemiro}, \citenamefont {Potocek},
  \citenamefont {Gabris}, \citenamefont {Jex},\ and\ \citenamefont
  {Silberhorn}}]{bib:Schreiber11b}%
  \BibitemOpen
  \bibfield  {author} {\bibinfo {author} {\bibnamefont {Schreiber},
  \bibfnamefont {A}}, \bibinfo {author} {\bibfnamefont {K.~N.}\ \bibnamefont
  {Cassemiro}}, \bibinfo {author} {\bibfnamefont {V.}~\bibnamefont {Potocek}},
  \bibinfo {author} {\bibfnamefont {A.}~\bibnamefont {Gabris}}, \bibinfo
  {author} {\bibfnamefont {I.}~\bibnamefont {Jex}}, and\ \bibinfo {author}
  {\bibfnamefont {Ch.}\ \bibnamefont {Silberhorn}}} (\bibinfo {year} {2011}),\
  \bibfield  {title} {\enquote {\bibinfo {title} {Decoherence and disorder in
  quantum walks: From ballistic spread to localization},}\ }\href
  {https://doi.org/10.1103/physrevlett.106.180403} {\bibfield  {journal}
  {\bibinfo  {journal} {Physical Review Letters}\ }\textbf {\bibinfo {volume}
  {106}},\ \bibinfo {pages} {180403}},\ \Eprint
  {https://arxiv.org/abs/arXiv:1101.2638v1} {arXiv:1101.2638v1} \BibitemShut
  {NoStop}%
\bibitem [{\citenamefont {Schreiber}\ \emph {et~al.}(2010)\citenamefont
  {Schreiber}, \citenamefont {Cassemiro}, \citenamefont {Poto{\u c}ek},
  \citenamefont {G{\' a}bris}, \citenamefont {Mosley}, \citenamefont
  {Andersson}, \citenamefont {Jex},\ and\ \citenamefont
  {Silberhorn}}]{bib:Schreiber10}%
  \BibitemOpen
  \bibfield  {author} {\bibinfo {author} {\bibnamefont {Schreiber},
  \bibfnamefont {A}}, \bibinfo {author} {\bibfnamefont {K.~N.}\ \bibnamefont
  {Cassemiro}}, \bibinfo {author} {\bibfnamefont {V.}~\bibnamefont {Poto{\u
  c}ek}}, \bibinfo {author} {\bibfnamefont {A.}~\bibnamefont {G{\' a}bris}},
  \bibinfo {author} {\bibfnamefont {P.~J.}\ \bibnamefont {Mosley}}, \bibinfo
  {author} {\bibfnamefont {E.}~\bibnamefont {Andersson}}, \bibinfo {author}
  {\bibfnamefont {I.}~\bibnamefont {Jex}}, and\ \bibinfo {author}
  {\bibfnamefont {Ch.}\ \bibnamefont {Silberhorn}}} (\bibinfo {year} {2010}),\
  \bibfield  {title} {\enquote {\bibinfo {title} {Photons walking the line: A
  quantum walk with adjustable coin operations},}\ }\href
  {https://doi.org/10.1103/physrevlett.104.050502} {\bibfield  {journal}
  {\bibinfo  {journal} {Physical Review Letters}\ }\textbf {\bibinfo {volume}
  {104}},\ \bibinfo {pages} {050502}},\ \Eprint
  {https://arxiv.org/abs/arXiv:0910.2197v2} {arXiv:0910.2197v2} \BibitemShut
  {NoStop}%
\bibitem [{\citenamefont {Schreiber}\ \emph {et~al.}(2012)\citenamefont
  {Schreiber}, \citenamefont {Gabris}, \citenamefont {Rohde}, \citenamefont
  {Laiho}, \citenamefont {Stefanak}, \citenamefont {Potocek}, \citenamefont
  {Hamilton}, \citenamefont {Jex},\ and\ \citenamefont
  {Silberhorn}}]{bib:RohdeQWExp12}%
  \BibitemOpen
  \bibfield  {author} {\bibinfo {author} {\bibnamefont {Schreiber},
  \bibfnamefont {Andreas}}, \bibinfo {author} {\bibfnamefont {Aurel}\
  \bibnamefont {Gabris}}, \bibinfo {author} {\bibfnamefont {Peter~P.}\
  \bibnamefont {Rohde}}, \bibinfo {author} {\bibfnamefont {Kaisa}\ \bibnamefont
  {Laiho}}, \bibinfo {author} {\bibfnamefont {Martin}\ \bibnamefont
  {Stefanak}}, \bibinfo {author} {\bibfnamefont {Vaclav}\ \bibnamefont
  {Potocek}}, \bibinfo {author} {\bibfnamefont {Craig}\ \bibnamefont
  {Hamilton}}, \bibinfo {author} {\bibfnamefont {Igor}\ \bibnamefont {Jex}},
  and\ \bibinfo {author} {\bibfnamefont {Christine}\ \bibnamefont
  {Silberhorn}}} (\bibinfo {year} {2012}),\ \bibfield  {title} {\enquote
  {\bibinfo {title} {A 2d quantum walk simulation of two-particle dynamics},}\
  }\href {https://doi.org/10.1126/science.1218448} {\bibfield  {journal}
  {\bibinfo  {journal} {Science}\ }\textbf {\bibinfo {volume} {336}},\ \bibinfo
  {pages} {55}},\ \Eprint {https://arxiv.org/abs/arXiv:1204.3555v1}
  {arXiv:1204.3555v1} \BibitemShut {NoStop}%
\bibitem [{\citenamefont {Schuetz}\ \emph {et~al.}(2015)\citenamefont
  {Schuetz}, \citenamefont {Kessler}, \citenamefont {Giedke}, \citenamefont
  {Vandersypen}, \citenamefont {Lukin},\ and\ \citenamefont
  {Cirac}}]{bib:schuetz2015universal}%
  \BibitemOpen
  \bibfield  {author} {\bibinfo {author} {\bibnamefont {Schuetz}, \bibfnamefont
  {MJA}}, \bibinfo {author} {\bibfnamefont {EM}~\bibnamefont {Kessler}},
  \bibinfo {author} {\bibfnamefont {G}~\bibnamefont {Giedke}}, \bibinfo
  {author} {\bibfnamefont {LMK}\ \bibnamefont {Vandersypen}}, \bibinfo {author}
  {\bibfnamefont {MD}~\bibnamefont {Lukin}}, and\ \bibinfo {author}
  {\bibfnamefont {JI}~\bibnamefont {Cirac}}} (\bibinfo {year} {2015}),\
  \bibfield  {title} {\enquote {\bibinfo {title} {Universal quantum transducers
  based on surface acoustic waves},}\ }\href
  {https://doi.org/10.1103/physrevx.5.031031} {\bibfield  {journal} {\bibinfo
  {journal} {Physical Review X}\ }\textbf {\bibinfo {volume} {5}},\ \bibinfo
  {pages} {031031}},\ \Eprint {https://arxiv.org/abs/arXiv:1504.05127v2}
  {arXiv:1504.05127v2} \BibitemShut {NoStop}%
\bibitem [{\citenamefont {Schumacher}\ and\ \citenamefont
  {Westmoreland}(1997)}]{bib:schumacher1997sending}%
  \BibitemOpen
  \bibfield  {author} {\bibinfo {author} {\bibnamefont {Schumacher},
  \bibfnamefont {Benjamin}}, and\ \bibinfo {author} {\bibfnamefont {Michael~D}\
  \bibnamefont {Westmoreland}}} (\bibinfo {year} {1997}),\ \bibfield  {title}
  {\enquote {\bibinfo {title} {Sending classical information via noisy quantum
  channels},}\ }\href {https://doi.org/10.1103/physreva.56.131} {\bibfield
  {journal} {\bibinfo  {journal} {Physical Review A}\ }\textbf {\bibinfo
  {volume} {56}},\ \bibinfo {pages} {131}}\BibitemShut {NoStop}%
\bibitem [{\citenamefont {Schwartz}\ \emph {et~al.}(2016)\citenamefont
  {Schwartz}, \citenamefont {Cogan}, \citenamefont {Schmidgall}, \citenamefont
  {Don}, \citenamefont {Gantz}, \citenamefont {Kenneth}, \citenamefont
  {Lindner},\ and\ \citenamefont {Gershoni}}]{bib:schwartz2016deterministic}%
  \BibitemOpen
  \bibfield  {author} {\bibinfo {author} {\bibnamefont {Schwartz},
  \bibfnamefont {Ido}}, \bibinfo {author} {\bibfnamefont {Dan}\ \bibnamefont
  {Cogan}}, \bibinfo {author} {\bibfnamefont {Emma~R}\ \bibnamefont
  {Schmidgall}}, \bibinfo {author} {\bibfnamefont {Yaroslav}\ \bibnamefont
  {Don}}, \bibinfo {author} {\bibfnamefont {Liron}\ \bibnamefont {Gantz}},
  \bibinfo {author} {\bibfnamefont {Oded}\ \bibnamefont {Kenneth}}, \bibinfo
  {author} {\bibfnamefont {Netanel~H}\ \bibnamefont {Lindner}}, and\ \bibinfo
  {author} {\bibfnamefont {David}\ \bibnamefont {Gershoni}}} (\bibinfo {year}
  {2016}),\ \bibfield  {title} {\enquote {\bibinfo {title} {Deterministic
  generation of a cluster state of entangled photons},}\ }\href
  {https://doi.org/10.1126/science.aah4758} {\bibfield  {journal} {\bibinfo
  {journal} {Science}\ }\textbf {\bibinfo {volume} {354}},\ \bibinfo {pages}
  {434}},\ \Eprint {https://arxiv.org/abs/arXiv:1606.07492v1}
  {arXiv:1606.07492v1} \BibitemShut {NoStop}%
\bibitem [{\citenamefont {Sciarrino}\ \emph {et~al.}(2002)\citenamefont
  {Sciarrino}, \citenamefont {Lombardi}, \citenamefont {Milani},\ and\
  \citenamefont {De~Martini}}]{bib:PRA_66_024309}%
  \BibitemOpen
  \bibfield  {author} {\bibinfo {author} {\bibnamefont {Sciarrino},
  \bibfnamefont {Fabio}}, \bibinfo {author} {\bibfnamefont {Egilberto}\
  \bibnamefont {Lombardi}}, \bibinfo {author} {\bibfnamefont {Giorgio}\
  \bibnamefont {Milani}}, and\ \bibinfo {author} {\bibfnamefont {Francesco}\
  \bibnamefont {De~Martini}}} (\bibinfo {year} {2002}),\ \bibfield  {title}
  {\enquote {\bibinfo {title} {Delayed-choice entanglement swapping with
  vacuum--one-photon quantum states},}\ }\href
  {https://doi.org/10.1103/physreva.66.024309} {\bibfield  {journal} {\bibinfo
  {journal} {Physical Review A}\ }\textbf {\bibinfo {volume} {66}},\ \bibinfo
  {pages} {024309}}\BibitemShut {NoStop}%
\bibitem [{\citenamefont {Sentís}\ \emph {et~al.}(2016)\citenamefont
  {Sentís}, \citenamefont {Bagan}, \citenamefont {Calsamiglia}, \citenamefont
  {Chiribella},\ and\ \citenamefont {Munoz-Tapia}}]{bib:gael1}%
  \BibitemOpen
  \bibfield  {author} {\bibinfo {author} {\bibnamefont {Sentís}, \bibfnamefont
  {Gael}}, \bibinfo {author} {\bibfnamefont {Emilio}\ \bibnamefont {Bagan}},
  \bibinfo {author} {\bibfnamefont {John}\ \bibnamefont {Calsamiglia}},
  \bibinfo {author} {\bibfnamefont {Giulio}\ \bibnamefont {Chiribella}}, and\
  \bibinfo {author} {\bibfnamefont {Ramon}\ \bibnamefont {Munoz-Tapia}}}
  (\bibinfo {year} {2016}),\ \bibfield  {title} {\enquote {\bibinfo {title}
  {Quantum change point},}\ }\href
  {https://doi.org/10.1103/physrevlett.117.150502} {\bibfield  {journal}
  {\bibinfo  {journal} {Physical Review Letters}\ }\textbf {\bibinfo {volume}
  {117}},\ \bibinfo {pages} {150502}},\ \Eprint
  {https://arxiv.org/abs/arXiv:1605.01916v3} {arXiv:1605.01916v3} \BibitemShut
  {NoStop}%
\bibitem [{\citenamefont {Serafini}(2023)}]{serafini2023quantum}%
  \BibitemOpen
  \bibfield  {author} {\bibinfo {author} {\bibnamefont {Serafini},
  \bibfnamefont {Alessio}}} (\bibinfo {year} {2023}),\ \href@noop {} {\emph
  {\bibinfo {title} {Quantum continuous variables: a primer of theoretical
  methods}}}\ (\bibinfo  {publisher} {CRC press})\BibitemShut {NoStop}%
\bibitem [{\citenamefont {Seshadreesan}\ \emph {et~al.}(2015)\citenamefont
  {Seshadreesan}, \citenamefont {Olson}, \citenamefont {Motes}, \citenamefont
  {Rohde},\ and\ \citenamefont {Dowling}}]{bib:RohdeDisp15}%
  \BibitemOpen
  \bibfield  {author} {\bibinfo {author} {\bibnamefont {Seshadreesan},
  \bibfnamefont {Kaushik~P}}, \bibinfo {author} {\bibfnamefont {Jonathan~P.}\
  \bibnamefont {Olson}}, \bibinfo {author} {\bibfnamefont {Keith~R.}\
  \bibnamefont {Motes}}, \bibinfo {author} {\bibfnamefont {Peter~P.}\
  \bibnamefont {Rohde}}, and\ \bibinfo {author} {\bibfnamefont {Jonathan~P.}\
  \bibnamefont {Dowling}}} (\bibinfo {year} {2015}),\ \bibfield  {title}
  {\enquote {\bibinfo {title} {Boson sampling with displaced single-photon fock
  states versus single-photon-added coherent states - the quantum-classical
  divide and computational-complexity transitions in linear optics},}\ }\href
  {https://doi.org/10.1103/physreva.91.022334} {\bibfield  {journal} {\bibinfo
  {journal} {Physical Review A}\ }\textbf {\bibinfo {volume} {91}},\ \bibinfo
  {pages} {022334}},\ \Eprint {https://arxiv.org/abs/arXiv:1402.0531v3}
  {arXiv:1402.0531v3} \BibitemShut {NoStop}%
\bibitem [{\citenamefont {Shalev-Shwartz}\ and\ \citenamefont
  {Ben-David}(2014)}]{bib:shalev2014understanding}%
  \BibitemOpen
  \bibfield  {author} {\bibinfo {author} {\bibnamefont {Shalev-Shwartz},
  \bibfnamefont {Shai}}, and\ \bibinfo {author} {\bibfnamefont {Shai}\
  \bibnamefont {Ben-David}}} (\bibinfo {year} {2014}),\ \href@noop {} {\emph
  {\bibinfo {title} {Understanding machine learning: From theory to
  algorithms}}}\ (\bibinfo  {publisher} {Cambridge University
  Press})\BibitemShut {NoStop}%
\bibitem [{\citenamefont {Shanahan}\ and\ \citenamefont
  {Dai}(2015)}]{bib:shanahan2015large}%
  \BibitemOpen
  \bibfield  {author} {\bibinfo {author} {\bibnamefont {Shanahan},
  \bibfnamefont {James~G}}, and\ \bibinfo {author} {\bibfnamefont {Laing}\
  \bibnamefont {Dai}}} (\bibinfo {year} {2015}),\ \bibfield  {title} {\enquote
  {\bibinfo {title} {Large scale distributed data science using apache
  spark},}\ }in\ \href {https://doi.org/10.1145/2783258.2789993} {\emph
  {\bibinfo {booktitle} {Proceedings of the 21th ACM SIGKDD international
  conference on knowledge discovery and data mining}}},\ p.\ \bibinfo {pages}
  {2323}\BibitemShut {NoStop}%
\bibitem [{\citenamefont {Shannon}(1949)}]{SD-Shannon:1949aa}%
  \BibitemOpen
  \bibfield  {author} {\bibinfo {author} {\bibnamefont {Shannon}, \bibfnamefont
  {C~E}}} (\bibinfo {year} {1949}),\ \bibfield  {title} {\enquote {\bibinfo
  {title} {Communication theory of secrecy systems},}\ }\href
  {https://doi.org/10.1002/j.1538-7305.1949.tb00928.x} {\bibfield  {journal}
  {\bibinfo  {journal} {The Bell System Technical Journal}\ }\textbf {\bibinfo
  {volume} {28}},\ \bibinfo {pages} {656}}\BibitemShut {NoStop}%
\bibitem [{\citenamefont {Sheng}\ and\ \citenamefont
  {Zhou}(2017)}]{bib:sheng2017distributed}%
  \BibitemOpen
  \bibfield  {author} {\bibinfo {author} {\bibnamefont {Sheng}, \bibfnamefont
  {Yu-Bo}}, and\ \bibinfo {author} {\bibfnamefont {Lan}\ \bibnamefont {Zhou}}}
  (\bibinfo {year} {2017}),\ \bibfield  {title} {\enquote {\bibinfo {title}
  {Distributed secure quantum machine learning},}\ }\href
  {https://doi.org/10.1016/j.scib.2017.06.007} {\bibfield  {journal} {\bibinfo
  {journal} {Science Bulletin}\ }\textbf {\bibinfo {volume} {62}},\ \bibinfo
  {pages} {1025}}\BibitemShut {NoStop}%
\bibitem [{\citenamefont {Shentu}\ \emph {et~al.}(2013)\citenamefont {Shentu},
  \citenamefont {Pelc}, \citenamefont {Wang}, \citenamefont {Sun},
  \citenamefont {Zheng}, \citenamefont {Fejer}, \citenamefont {Zhang},\ and\
  \citenamefont {Pan}}]{bib:shentu2013ultralow}%
  \BibitemOpen
  \bibfield  {author} {\bibinfo {author} {\bibnamefont {Shentu}, \bibfnamefont
  {Guo-Liang}}, \bibinfo {author} {\bibfnamefont {Jason~S}\ \bibnamefont
  {Pelc}}, \bibinfo {author} {\bibfnamefont {Xiao-Dong}\ \bibnamefont {Wang}},
  \bibinfo {author} {\bibfnamefont {Qi-Chao}\ \bibnamefont {Sun}}, \bibinfo
  {author} {\bibfnamefont {Ming-Yang}\ \bibnamefont {Zheng}}, \bibinfo {author}
  {\bibfnamefont {MM}~\bibnamefont {Fejer}}, \bibinfo {author} {\bibfnamefont
  {Qiang}\ \bibnamefont {Zhang}}, and\ \bibinfo {author} {\bibfnamefont
  {Jian-Wei}\ \bibnamefont {Pan}}} (\bibinfo {year} {2013}),\ \bibfield
  {title} {\enquote {\bibinfo {title} {Ultralow noise up-conversion detector
  and spectrometer for the telecom band},}\ }\href
  {https://doi.org/10.1364/oe.21.013986} {\bibfield  {journal} {\bibinfo
  {journal} {Optics Express}\ }\textbf {\bibinfo {volume} {21}},\ \bibinfo
  {pages} {13986}}\BibitemShut {NoStop}%
\bibitem [{\citenamefont {Shnirman}\ \emph {et~al.}(1997)\citenamefont
  {Shnirman}, \citenamefont {Sch{\"o}n},\ and\ \citenamefont
  {Hermon}}]{bib:shnirman1997quantum}%
  \BibitemOpen
  \bibfield  {author} {\bibinfo {author} {\bibnamefont {Shnirman},
  \bibfnamefont {Alexander}}, \bibinfo {author} {\bibfnamefont {Gerd}\
  \bibnamefont {Sch{\"o}n}}, and\ \bibinfo {author} {\bibfnamefont {Ziv}\
  \bibnamefont {Hermon}}} (\bibinfo {year} {1997}),\ \bibfield  {title}
  {\enquote {\bibinfo {title} {Quantum manipulations of small josephson
  junctions},}\ }\href {https://doi.org/10.1103/physrevlett.79.2371} {\bibfield
   {journal} {\bibinfo  {journal} {Physical Review Letters}\ }\textbf {\bibinfo
  {volume} {79}},\ \bibinfo {pages} {2371}},\ \Eprint
  {https://arxiv.org/abs/arXiv:cond-mat/9706016v2} {arXiv:cond-mat/9706016v2}
  \BibitemShut {NoStop}%
\bibitem [{\citenamefont {Shor}(1994)}]{bib:ShorFactor}%
  \BibitemOpen
  \bibfield  {author} {\bibinfo {author} {\bibnamefont {Shor}, \bibfnamefont
  {Peter~W}}} (\bibinfo {year} {1994}),\ \bibfield  {title} {\enquote {\bibinfo
  {title} {Algorithms for quantum computation: discrete logarithms and
  factoring},}\ }in\ \href@noop {} {\emph {\bibinfo {booktitle} {Symposium on
  the Foundations of Computer Science}}},\ Vol.~\bibinfo {volume} {35},\ p.\
  \bibinfo {pages} {124}\BibitemShut {NoStop}%
\bibitem [{\citenamefont {Shor}(1995{\natexlab{a}})}]{bib:Shor95}%
  \BibitemOpen
  \bibfield  {author} {\bibinfo {author} {\bibnamefont {Shor}, \bibfnamefont
  {Peter~W}}} (\bibinfo {year} {1995}{\natexlab{a}}),\ \bibfield  {title}
  {\enquote {\bibinfo {title} {Scheme for reducing decoherence in quantum
  computer memory},}\ }\href {https://doi.org/10.1103/physreva.52.r2493}
  {\bibfield  {journal} {\bibinfo  {journal} {Physical Review A}\ }\textbf
  {\bibinfo {volume} {52}},\ \bibinfo {pages} {R2493}}\BibitemShut {NoStop}%
\bibitem [{\citenamefont {Shor}(1995{\natexlab{b}})}]{bib:S95}%
  \BibitemOpen
  \bibfield  {author} {\bibinfo {author} {\bibnamefont {Shor}, \bibfnamefont
  {PW}}} (\bibinfo {year} {1995}{\natexlab{b}}),\ \bibfield  {title} {\enquote
  {\bibinfo {title} {{Scheme for reducing decoherence in quantum computer
  memory}},}\ }\href@noop {} {\bibfield  {journal} {\bibinfo  {journal} {Phys.
  Rev. A.}\ }\textbf {\bibinfo {volume} {52}},\ \bibinfo {pages}
  {R2493}}\BibitemShut {NoStop}%
\bibitem [{\citenamefont {Shumeiko}(2016)}]{bib:shumeiko2016quantum}%
  \BibitemOpen
  \bibfield  {author} {\bibinfo {author} {\bibnamefont {Shumeiko},
  \bibfnamefont {Vitaly~S}}} (\bibinfo {year} {2016}),\ \bibfield  {title}
  {\enquote {\bibinfo {title} {Quantum acousto-optic transducer for
  superconducting qubits},}\ }\href
  {https://doi.org/10.1103/physreva.93.023838} {\bibfield  {journal} {\bibinfo
  {journal} {Physical Review A}\ }\textbf {\bibinfo {volume} {93}},\ \bibinfo
  {pages} {023838}},\ \Eprint {https://arxiv.org/abs/arXiv:1511.03819v2}
  {arXiv:1511.03819v2} \BibitemShut {NoStop}%
\bibitem [{\citenamefont {Sibson}\ \emph {et~al.}(2017)\citenamefont {Sibson},
  \citenamefont {Erven}, \citenamefont {Godfrey}, \citenamefont {Miki},
  \citenamefont {Yamashita}, \citenamefont {Fujiwara}, \citenamefont {Sasaki},
  \citenamefont {Terai}, \citenamefont {Tanner}, \citenamefont {Natarajan}
  \emph {et~al.}}]{bib:sibson2017chip}%
  \BibitemOpen
  \bibfield  {author} {\bibinfo {author} {\bibnamefont {Sibson}, \bibfnamefont
  {Philip}}, \bibinfo {author} {\bibfnamefont {Chris}\ \bibnamefont {Erven}},
  \bibinfo {author} {\bibfnamefont {Mark}\ \bibnamefont {Godfrey}}, \bibinfo
  {author} {\bibfnamefont {Shigehito}\ \bibnamefont {Miki}}, \bibinfo {author}
  {\bibfnamefont {Taro}\ \bibnamefont {Yamashita}}, \bibinfo {author}
  {\bibfnamefont {Mikio}\ \bibnamefont {Fujiwara}}, \bibinfo {author}
  {\bibfnamefont {Masahide}\ \bibnamefont {Sasaki}}, \bibinfo {author}
  {\bibfnamefont {Hirotaka}\ \bibnamefont {Terai}}, \bibinfo {author}
  {\bibfnamefont {Michael~G}\ \bibnamefont {Tanner}}, \bibinfo {author}
  {\bibfnamefont {Chandra~M}\ \bibnamefont {Natarajan}},  \emph {et~al.}}
  (\bibinfo {year} {2017}),\ \bibfield  {title} {\enquote {\bibinfo {title}
  {Chip-based quantum key distribution},}\ }\href
  {https://doi.org/10.1038/ncomms13984} {\bibfield  {journal} {\bibinfo
  {journal} {Nature Communications}\ }\textbf {\bibinfo {volume} {8}},\
  \bibinfo {pages} {13984}}\BibitemShut {NoStop}%
\bibitem [{\citenamefont {Ferreira~da Silva}\ \emph {et~al.}(2013)\citenamefont
  {Ferreira~da Silva}, \citenamefont {Vitoreti}, \citenamefont {Xavier},
  \citenamefont {do~Amaral}, \citenamefont {Tempor\~ao},\ and\ \citenamefont
  {von~der Weid}}]{bib:PRA_88_052303}%
  \BibitemOpen
  \bibfield  {author} {\bibinfo {author} {\bibnamefont {Ferreira~da Silva},
  \bibfnamefont {T}}, \bibinfo {author} {\bibfnamefont {D.}~\bibnamefont
  {Vitoreti}}, \bibinfo {author} {\bibfnamefont {G.~B.}\ \bibnamefont
  {Xavier}}, \bibinfo {author} {\bibfnamefont {G.~C.}\ \bibnamefont
  {do~Amaral}}, \bibinfo {author} {\bibfnamefont {G.~P.}\ \bibnamefont
  {Tempor\~ao}}, and\ \bibinfo {author} {\bibfnamefont {J.~P.}\ \bibnamefont
  {von~der Weid}}} (\bibinfo {year} {2013}),\ \bibfield  {title} {\enquote
  {\bibinfo {title} {Proof-of-principle demonstration of
  measurement-device-independent quantum key distribution using polarization
  qubits},}\ }\href {https://doi.org/10.1103/physreva.88.052303} {\bibfield
  {journal} {\bibinfo  {journal} {Physical Review A}\ }\textbf {\bibinfo
  {volume} {88}},\ \bibinfo {pages} {052303}},\ \Eprint
  {https://arxiv.org/abs/arXiv:1207.6345v3} {arXiv:1207.6345v3} \BibitemShut
  {NoStop}%
\bibitem [{\citenamefont {Silverstone}\ \emph {et~al.}(2014)\citenamefont
  {Silverstone}, \citenamefont {Bonneau}, \citenamefont {Ohira}, \citenamefont
  {Suzuki}, \citenamefont {Yoshida}, \citenamefont {Iizuka}, \citenamefont
  {Ezaki}, \citenamefont {Natarajan}, \citenamefont {Tanner}, \citenamefont
  {Hadfield} \emph {et~al.}}]{bib:silverstone2014}%
  \BibitemOpen
  \bibfield  {author} {\bibinfo {author} {\bibnamefont {Silverstone},
  \bibfnamefont {Joshua~W}}, \bibinfo {author} {\bibfnamefont {Damien}\
  \bibnamefont {Bonneau}}, \bibinfo {author} {\bibfnamefont {Kazuya}\
  \bibnamefont {Ohira}}, \bibinfo {author} {\bibfnamefont {Nob}\ \bibnamefont
  {Suzuki}}, \bibinfo {author} {\bibfnamefont {Haruhiko}\ \bibnamefont
  {Yoshida}}, \bibinfo {author} {\bibfnamefont {Norio}\ \bibnamefont {Iizuka}},
  \bibinfo {author} {\bibfnamefont {Mizunori}\ \bibnamefont {Ezaki}}, \bibinfo
  {author} {\bibfnamefont {Chandra~M}\ \bibnamefont {Natarajan}}, \bibinfo
  {author} {\bibfnamefont {Michael~G}\ \bibnamefont {Tanner}}, \bibinfo
  {author} {\bibfnamefont {Robert~H}\ \bibnamefont {Hadfield}},  \emph
  {et~al.}} (\bibinfo {year} {2014}),\ \bibfield  {title} {\enquote {\bibinfo
  {title} {On-chip quantum interference between silicon photon-pair sources},}\
  }\href@noop {} {\bibfield  {journal} {\bibinfo  {journal} {Nature Photonics}\
  }\textbf {\bibinfo {volume} {8}},\ \bibinfo {pages} {104}},\ \Eprint
  {https://arxiv.org/abs/arXiv:1304.1490v3} {arXiv:1304.1490v3} \BibitemShut
  {NoStop}%
\bibitem [{\citenamefont {Simon}\ \emph
  {et~al.}(2010{\natexlab{a}})\citenamefont {Simon}, \citenamefont {Afzelius},
  \citenamefont {Appel}, \citenamefont {de~La~Giroday}, \citenamefont
  {Dewhurst}, \citenamefont {Gisin}, \citenamefont {Hu}, \citenamefont
  {Jelezko}, \citenamefont {Kr{\"o}ll}, \citenamefont {M{\"u}ller} \emph
  {et~al.}}]{bib:simon2010quantum}%
  \BibitemOpen
  \bibfield  {author} {\bibinfo {author} {\bibnamefont {Simon}, \bibfnamefont
  {Christoph}}, \bibinfo {author} {\bibfnamefont {Mikael}\ \bibnamefont
  {Afzelius}}, \bibinfo {author} {\bibfnamefont {J{\"u}rgen}\ \bibnamefont
  {Appel}}, \bibinfo {author} {\bibfnamefont {A~Boyer}\ \bibnamefont
  {de~La~Giroday}}, \bibinfo {author} {\bibfnamefont {SJ}~\bibnamefont
  {Dewhurst}}, \bibinfo {author} {\bibfnamefont {Nicolas}\ \bibnamefont
  {Gisin}}, \bibinfo {author} {\bibfnamefont {CY}~\bibnamefont {Hu}}, \bibinfo
  {author} {\bibfnamefont {F}~\bibnamefont {Jelezko}}, \bibinfo {author}
  {\bibfnamefont {Stefan}\ \bibnamefont {Kr{\"o}ll}}, \bibinfo {author}
  {\bibfnamefont {JH}~\bibnamefont {M{\"u}ller}},  \emph {et~al.}} (\bibinfo
  {year} {2010}{\natexlab{a}}),\ \bibfield  {title} {\enquote {\bibinfo {title}
  {Quantum memories},}\ }\href {https://doi.org/10.1140/epjd/e2010-00103-y}
  {\bibfield  {journal} {\bibinfo  {journal} {European Physics Journal D}\
  }\textbf {\bibinfo {volume} {58}},\ \bibinfo {pages} {1}}\BibitemShut
  {NoStop}%
\bibitem [{\citenamefont {Simon}\ \emph
  {et~al.}(2010{\natexlab{b}})\citenamefont {Simon}, \citenamefont
  {De~Riedmatten},\ and\ \citenamefont {Afzelius}}]{bib:PRA_82_010304}%
  \BibitemOpen
  \bibfield  {author} {\bibinfo {author} {\bibnamefont {Simon}, \bibfnamefont
  {Christoph}}, \bibinfo {author} {\bibfnamefont {Hugues}\ \bibnamefont
  {De~Riedmatten}}, and\ \bibinfo {author} {\bibfnamefont {Mikael}\
  \bibnamefont {Afzelius}}} (\bibinfo {year} {2010}{\natexlab{b}}),\ \bibfield
  {title} {\enquote {\bibinfo {title} {Temporally multiplexed quantum repeaters
  with atomic gases},}\ }\href {https://doi.org/10.1103/physreva.82.010304}
  {\bibfield  {journal} {\bibinfo  {journal} {Physical Review A}\ }\textbf
  {\bibinfo {volume} {82}},\ \bibinfo {pages} {010304}},\ \Eprint
  {https://arxiv.org/abs/arXiv:1007.5028v1} {arXiv:1007.5028v1} \BibitemShut
  {NoStop}%
\bibitem [{\citenamefont {Sinclair}\ \emph {et~al.}(2014)\citenamefont
  {Sinclair}, \citenamefont {Saglamyurek}, \citenamefont {Mallahzadeh},
  \citenamefont {Slater}, \citenamefont {George}, \citenamefont {Ricken},
  \citenamefont {Hedges}, \citenamefont {Oblak}, \citenamefont {Simon},
  \citenamefont {Sohler} \emph {et~al.}}]{bib:PRL_113_053603}%
  \BibitemOpen
  \bibfield  {author} {\bibinfo {author} {\bibnamefont {Sinclair},
  \bibfnamefont {Neil}}, \bibinfo {author} {\bibfnamefont {Erhan}\ \bibnamefont
  {Saglamyurek}}, \bibinfo {author} {\bibfnamefont {Hassan}\ \bibnamefont
  {Mallahzadeh}}, \bibinfo {author} {\bibfnamefont {Joshua~A}\ \bibnamefont
  {Slater}}, \bibinfo {author} {\bibfnamefont {Mathew}\ \bibnamefont {George}},
  \bibinfo {author} {\bibfnamefont {Raimund}\ \bibnamefont {Ricken}}, \bibinfo
  {author} {\bibfnamefont {Morgan~P}\ \bibnamefont {Hedges}}, \bibinfo {author}
  {\bibfnamefont {Daniel}\ \bibnamefont {Oblak}}, \bibinfo {author}
  {\bibfnamefont {Christoph}\ \bibnamefont {Simon}}, \bibinfo {author}
  {\bibfnamefont {Wolfgang}\ \bibnamefont {Sohler}},  \emph {et~al.}} (\bibinfo
  {year} {2014}),\ \bibfield  {title} {\enquote {\bibinfo {title} {Spectral
  multiplexing for scalable quantum photonics using an atomic frequency comb
  quantum memory and feed-forward control},}\ }\href
  {https://doi.org/10.1103/physrevlett.113.053603} {\bibfield  {journal}
  {\bibinfo  {journal} {Physical Review Letters}\ }\textbf {\bibinfo {volume}
  {113}},\ \bibinfo {pages} {053603}},\ \Eprint
  {https://arxiv.org/abs/arXiv:1309.3202v3} {arXiv:1309.3202v3} \BibitemShut
  {NoStop}%
\bibitem [{\citenamefont {Singh}\ \emph {et~al.}(2023)\citenamefont {Singh},
  \citenamefont {Muraleedharan}, \citenamefont {Fu}, \citenamefont {Cheng},
  \citenamefont {Newton}, \citenamefont {Rohde},\ and\ \citenamefont
  {Brennen}}]{singh2023proof}%
  \BibitemOpen
  \bibfield  {author} {\bibinfo {author} {\bibnamefont {Singh}, \bibfnamefont
  {Deepesh}}, \bibinfo {author} {\bibfnamefont {Gopikrishnan}\ \bibnamefont
  {Muraleedharan}}, \bibinfo {author} {\bibfnamefont {Boxiang}\ \bibnamefont
  {Fu}}, \bibinfo {author} {\bibfnamefont {Chen-Mou}\ \bibnamefont {Cheng}},
  \bibinfo {author} {\bibfnamefont {Nicolas~Roussy}\ \bibnamefont {Newton}},
  \bibinfo {author} {\bibfnamefont {Peter~P}\ \bibnamefont {Rohde}}, and\
  \bibinfo {author} {\bibfnamefont {Gavin~K}\ \bibnamefont {Brennen}}}
  (\bibinfo {year} {2023}),\ \bibfield  {title} {\enquote {\bibinfo {title}
  {Proof-of-work consensus by quantum sampling},}\ }\href@noop {} {\bibinfo
  {journal} {arXiv preprint arXiv:2305.19865}\ }\BibitemShut {NoStop}%
\bibitem [{\citenamefont {Smit}\ \emph {et~al.}(2014)\citenamefont {Smit},
  \citenamefont {Leijtens}, \citenamefont {Ambrosius}, \citenamefont {Bente},
  \citenamefont {Van~der Tol}, \citenamefont {Smalbrugge}, \citenamefont
  {De~Vries}, \citenamefont {Geluk}, \citenamefont {Bolk}, \citenamefont
  {Van~Veldhoven} \emph {et~al.}}]{bib:smit2014introduction}%
  \BibitemOpen
\bibfield  {journal} {  }\bibfield  {author} {\bibinfo {author} {\bibnamefont
  {Smit}, \bibfnamefont {Meint}}, \bibinfo {author} {\bibfnamefont {Xaveer}\
  \bibnamefont {Leijtens}}, \bibinfo {author} {\bibfnamefont {Huub}\
  \bibnamefont {Ambrosius}}, \bibinfo {author} {\bibfnamefont {Erwin}\
  \bibnamefont {Bente}}, \bibinfo {author} {\bibfnamefont {Jos}\ \bibnamefont
  {Van~der Tol}}, \bibinfo {author} {\bibfnamefont {Barry}\ \bibnamefont
  {Smalbrugge}}, \bibinfo {author} {\bibfnamefont {Tjibbe}\ \bibnamefont
  {De~Vries}}, \bibinfo {author} {\bibfnamefont {Erik-Jan}\ \bibnamefont
  {Geluk}}, \bibinfo {author} {\bibfnamefont {Jeroen}\ \bibnamefont {Bolk}},
  \bibinfo {author} {\bibfnamefont {Rene}\ \bibnamefont {Van~Veldhoven}},
  \emph {et~al.}} (\bibinfo {year} {2014}),\ \bibfield  {title} {\enquote
  {\bibinfo {title} {An introduction to inp-based generic integration
  technology},}\ }\href {https://doi.org/10.1088/0268-1242/29/8/083001}
  {\bibfield  {journal} {\bibinfo  {journal} {Semiconductor Science \&
  Technology}\ }\textbf {\bibinfo {volume} {29}},\ \bibinfo {pages}
  {083001}}\BibitemShut {NoStop}%
\bibitem [{\citenamefont {Smith}\ \emph {et~al.}(2009)\citenamefont {Smith},
  \citenamefont {Mahou}, \citenamefont {Cohen}, \citenamefont {Lundeen},\ and\
  \citenamefont {Walmsley}}]{bib:smith2009}%
  \BibitemOpen
  \bibfield  {author} {\bibinfo {author} {\bibnamefont {Smith}, \bibfnamefont
  {Brian~J}}, \bibinfo {author} {\bibfnamefont {P}~\bibnamefont {Mahou}},
  \bibinfo {author} {\bibfnamefont {Offir}\ \bibnamefont {Cohen}}, \bibinfo
  {author} {\bibfnamefont {JS}~\bibnamefont {Lundeen}}, and\ \bibinfo {author}
  {\bibfnamefont {IA}~\bibnamefont {Walmsley}}} (\bibinfo {year} {2009}),\
  \bibfield  {title} {\enquote {\bibinfo {title} {Photon pair generation in
  birefringent optical fibers},}\ }\href {https://doi.org/10.1364/oe.17.023589}
  {\bibfield  {journal} {\bibinfo  {journal} {Optics Express}\ }\textbf
  {\bibinfo {volume} {17}},\ \bibinfo {pages} {23589}},\ \Eprint
  {https://arxiv.org/abs/arXiv:0909.4319v2} {arXiv:0909.4319v2} \BibitemShut
  {NoStop}%
\bibitem [{\citenamefont {Smithey}\ \emph {et~al.}(1993)\citenamefont
  {Smithey}, \citenamefont {Beck}, \citenamefont {Raymer},\ and\ \citenamefont
  {Faridani}}]{bib:Smithey1993}%
  \BibitemOpen
  \bibfield  {author} {\bibinfo {author} {\bibnamefont {Smithey}, \bibfnamefont
  {D~T}}, \bibinfo {author} {\bibfnamefont {M.}~\bibnamefont {Beck}}, \bibinfo
  {author} {\bibfnamefont {M.~G.}\ \bibnamefont {Raymer}}, and\ \bibinfo
  {author} {\bibfnamefont {A.}~\bibnamefont {Faridani}}} (\bibinfo {year}
  {1993}),\ \bibfield  {title} {\enquote {\bibinfo {title} {Measurement of the
  wigner distribution and the density matrix of a light mode using optical
  homodyne tomography: Application to squeezed states and the vacuum},}\ }\href
  {https://doi.org/10.1103/PhysRevLett.70.1244} {\bibfield  {journal} {\bibinfo
   {journal} {Physical Review Letters}\ }\textbf {\bibinfo {volume} {70}},\
  \bibinfo {pages} {1244}}\BibitemShut {NoStop}%
\bibitem [{\citenamefont {Snow}\ \emph {et~al.}(2005)\citenamefont {Snow},
  \citenamefont {Rastogi},\ and\ \citenamefont
  {Weckman}}]{bib:snow2005assessing}%
  \BibitemOpen
  \bibfield  {author} {\bibinfo {author} {\bibnamefont {Snow}, \bibfnamefont
  {A}}, \bibinfo {author} {\bibfnamefont {P.}~\bibnamefont {Rastogi}}, and\
  \bibinfo {author} {\bibfnamefont {G.}~\bibnamefont {Weckman}}} (\bibinfo
  {year} {2005}),\ \bibfield  {title} {\enquote {\bibinfo {title} {Assessing
  dependability of wireless networks using neural networks},}\ }in\ \href
  {https://doi.org/10.1109/milcom.2005.1606090} {\emph {\bibinfo {booktitle}
  {IEEE Military Communications Conference (MILCOM)}}},\ p.\ \bibinfo {pages}
  {2809}\BibitemShut {NoStop}%
\bibitem [{\citenamefont {Somaschi}\ \emph {et~al.}(2016)\citenamefont
  {Somaschi}, \citenamefont {Giesz}, \citenamefont {De~Santis}, \citenamefont
  {Loredo}, \citenamefont {Almeida}, \citenamefont {Hornecker}, \citenamefont
  {Portalupi}, \citenamefont {Grange}, \citenamefont {Anton}, \citenamefont
  {Demory} \emph {et~al.}}]{bib:somaschi2016}%
  \BibitemOpen
  \bibfield  {author} {\bibinfo {author} {\bibnamefont {Somaschi},
  \bibfnamefont {N}}, \bibinfo {author} {\bibfnamefont {V}~\bibnamefont
  {Giesz}}, \bibinfo {author} {\bibfnamefont {L}~\bibnamefont {De~Santis}},
  \bibinfo {author} {\bibfnamefont {JC}~\bibnamefont {Loredo}}, \bibinfo
  {author} {\bibfnamefont {MP}~\bibnamefont {Almeida}}, \bibinfo {author}
  {\bibfnamefont {G}~\bibnamefont {Hornecker}}, \bibinfo {author}
  {\bibfnamefont {SL}~\bibnamefont {Portalupi}}, \bibinfo {author}
  {\bibfnamefont {T}~\bibnamefont {Grange}}, \bibinfo {author} {\bibfnamefont
  {C}~\bibnamefont {Anton}}, \bibinfo {author} {\bibfnamefont {J}~\bibnamefont
  {Demory}},  \emph {et~al.}} (\bibinfo {year} {2016}),\ \bibfield  {title}
  {\enquote {\bibinfo {title} {Near-optimal single-photon sources in the solid
  state},}\ }\href {https://doi.org/10.1038/nphoton.2016.23} {\bibfield
  {journal} {\bibinfo  {journal} {Nature Photonics}\ }\textbf {\bibinfo
  {volume} {10}},\ \bibinfo {pages} {340}},\ \Eprint
  {https://arxiv.org/abs/arXiv:1510.06499v2} {arXiv:1510.06499v2} \BibitemShut
  {NoStop}%
\bibitem [{\citenamefont {Sommer}\ and\ \citenamefont
  {Paxson}(2010)}]{bib:sommer2010outside}%
  \BibitemOpen
  \bibfield  {author} {\bibinfo {author} {\bibnamefont {Sommer}, \bibfnamefont
  {Robin}}, and\ \bibinfo {author} {\bibfnamefont {Vern}\ \bibnamefont
  {Paxson}}} (\bibinfo {year} {2010}),\ \bibfield  {title} {\enquote {\bibinfo
  {title} {Outside the closed world: On using machine learning for network
  intrusion detection},}\ }in\ \href {https://doi.org/10.1109/sp.2010.25}
  {\emph {\bibinfo {booktitle} {IEEE Symposium on Security and Privacy
  (SP)}}},\ p.\ \bibinfo {pages} {305}\BibitemShut {NoStop}%
\bibitem [{\citenamefont {Song}\ \emph {et~al.}(2017)\citenamefont {Song},
  \citenamefont {Xu}, \citenamefont {Liu}, \citenamefont {Yang}, \citenamefont
  {Zheng}, \citenamefont {Deng}, \citenamefont {Xie}, \citenamefont {Huang},
  \citenamefont {Guo}, \citenamefont {Zhang} \emph {et~al.}}]{bib:song201710}%
  \BibitemOpen
  \bibfield  {author} {\bibinfo {author} {\bibnamefont {Song}, \bibfnamefont
  {Chao}}, \bibinfo {author} {\bibfnamefont {Kai}\ \bibnamefont {Xu}}, \bibinfo
  {author} {\bibfnamefont {Wuxin}\ \bibnamefont {Liu}}, \bibinfo {author}
  {\bibfnamefont {Chui-ping}\ \bibnamefont {Yang}}, \bibinfo {author}
  {\bibfnamefont {Shi-Biao}\ \bibnamefont {Zheng}}, \bibinfo {author}
  {\bibfnamefont {Hui}\ \bibnamefont {Deng}}, \bibinfo {author} {\bibfnamefont
  {Qiwei}\ \bibnamefont {Xie}}, \bibinfo {author} {\bibfnamefont {Keqiang}\
  \bibnamefont {Huang}}, \bibinfo {author} {\bibfnamefont {Qiujiang}\
  \bibnamefont {Guo}}, \bibinfo {author} {\bibfnamefont {Libo}\ \bibnamefont
  {Zhang}},  \emph {et~al.}} (\bibinfo {year} {2017}),\ \bibfield  {title}
  {\enquote {\bibinfo {title} {10-qubit entanglement and parallel logic
  operations with a superconducting circuit},}\ }\href
  {https://doi.org/10.1103/physrevlett.119.180511} {\bibfield  {journal}
  {\bibinfo  {journal} {Physical Review Letters}\ }\textbf {\bibinfo {volume}
  {119}},\ \bibinfo {pages} {180511}},\ \Eprint
  {https://arxiv.org/abs/arXiv:1703.10302v2} {arXiv:1703.10302v2} \BibitemShut
  {NoStop}%
\bibitem [{\citenamefont {Spekkens}\ and\ \citenamefont
  {Rudolph}(2001)}]{bib:SpekkensRudolphSecure}%
  \BibitemOpen
  \bibfield  {author} {\bibinfo {author} {\bibnamefont {Spekkens},
  \bibfnamefont {R~W}}, and\ \bibinfo {author} {\bibfnamefont {T.}~\bibnamefont
  {Rudolph}}} (\bibinfo {year} {2001}),\ \bibfield  {title} {\enquote {\bibinfo
  {title} {Degrees of concealment and bindingness in quantum bit commitment
  protocols},}\ }\href {https://doi.org/10.1103/physreva.65.012310} {\bibfield
  {journal} {\bibinfo  {journal} {Physical Review A}\ }\textbf {\bibinfo
  {volume} {65}},\ \bibinfo {pages} {012310}},\ \Eprint
  {https://arxiv.org/abs/arXiv:quant-ph/0106019v2} {arXiv:quant-ph/0106019v2}
  \BibitemShut {NoStop}%
\bibitem [{\citenamefont {Spring}\ \emph {et~al.}(2017)\citenamefont {Spring},
  \citenamefont {Mennea}, \citenamefont {Metcalf}, \citenamefont {Humphreys},
  \citenamefont {Gates}, \citenamefont {Rogers}, \citenamefont {S{\"o}ller},
  \citenamefont {Smith}, \citenamefont {Kolthammer}, \citenamefont {Smith}
  \emph {et~al.}}]{bib:spring2017chip}%
  \BibitemOpen
  \bibfield  {author} {\bibinfo {author} {\bibnamefont {Spring}, \bibfnamefont
  {Justin~B}}, \bibinfo {author} {\bibfnamefont {Paolo~L}\ \bibnamefont
  {Mennea}}, \bibinfo {author} {\bibfnamefont {Benjamin~J}\ \bibnamefont
  {Metcalf}}, \bibinfo {author} {\bibfnamefont {Peter~C}\ \bibnamefont
  {Humphreys}}, \bibinfo {author} {\bibfnamefont {James~C}\ \bibnamefont
  {Gates}}, \bibinfo {author} {\bibfnamefont {Helen~L}\ \bibnamefont {Rogers}},
  \bibinfo {author} {\bibfnamefont {Christoph}\ \bibnamefont {S{\"o}ller}},
  \bibinfo {author} {\bibfnamefont {Brian~J}\ \bibnamefont {Smith}}, \bibinfo
  {author} {\bibfnamefont {W~Steven}\ \bibnamefont {Kolthammer}}, \bibinfo
  {author} {\bibfnamefont {Peter~GR}\ \bibnamefont {Smith}},  \emph {et~al.}}
  (\bibinfo {year} {2017}),\ \bibfield  {title} {\enquote {\bibinfo {title}
  {Chip-based array of near-identical, pure, heralded single-photon sources},}\
  }\href {https://doi.org/10.1364/optica.4.000090} {\bibfield  {journal}
  {\bibinfo  {journal} {Optica}\ ,\ \bibinfo {pages} {90}}}\Eprint
  {https://arxiv.org/abs/arXiv:1603.06984v1} {arXiv:1603.06984v1} \BibitemShut
  {NoStop}%
\bibitem [{\citenamefont {Stace}\ and\ \citenamefont
  {Barrett}(2010)}]{SD-Stace:2010aa}%
  \BibitemOpen
  \bibfield  {author} {\bibinfo {author} {\bibnamefont {Stace}, \bibfnamefont
  {Thomas~M}}, and\ \bibinfo {author} {\bibfnamefont {Sean~D.}\ \bibnamefont
  {Barrett}}} (\bibinfo {year} {2010}),\ \bibfield  {title} {\enquote {\bibinfo
  {title} {Error correction and degeneracy in surface codes suffering loss},}\
  }\href {https://doi.org/10.1103/PhysRevA.81.022317} {\bibfield  {journal}
  {\bibinfo  {journal} {Physical Review A}\ }\textbf {\bibinfo {volume} {81}},\
  \bibinfo {pages} {022317}},\ \Eprint
  {https://arxiv.org/abs/arXiv:0912.1159v1} {arXiv:0912.1159v1} \BibitemShut
  {NoStop}%
\bibitem [{\citenamefont {Stace}\ \emph {et~al.}(2009)\citenamefont {Stace},
  \citenamefont {Barrett},\ and\ \citenamefont
  {Doherty}}]{bib:StaceBarrettDohertyLoss}%
  \BibitemOpen
  \bibfield  {author} {\bibinfo {author} {\bibnamefont {Stace}, \bibfnamefont
  {Thomas~M}}, \bibinfo {author} {\bibfnamefont {Sean~D.}\ \bibnamefont
  {Barrett}}, and\ \bibinfo {author} {\bibfnamefont {Andrew~C.}\ \bibnamefont
  {Doherty}}} (\bibinfo {year} {2009}),\ \bibfield  {title} {\enquote {\bibinfo
  {title} {Thresholds for topological codes in the presence of loss},}\ }\href
  {https://doi.org/10.1103/physrevlett.102.200501} {\bibfield  {journal}
  {\bibinfo  {journal} {Physical Review Letters}\ }\textbf {\bibinfo {volume}
  {102}},\ \bibinfo {pages} {200501}},\ \Eprint
  {https://arxiv.org/abs/arXiv:0904.3556v1} {arXiv:0904.3556v1} \BibitemShut
  {NoStop}%
\bibitem [{\citenamefont {Stannigel}\ \emph {et~al.}(2010)\citenamefont
  {Stannigel}, \citenamefont {Rabl}, \citenamefont {S{\o}rensen}, \citenamefont
  {Zoller},\ and\ \citenamefont {Lukin}}]{bib:stannigel2010optomechanical}%
  \BibitemOpen
  \bibfield  {author} {\bibinfo {author} {\bibnamefont {Stannigel},
  \bibfnamefont {Kai}}, \bibinfo {author} {\bibfnamefont {Peter}\ \bibnamefont
  {Rabl}}, \bibinfo {author} {\bibfnamefont {Anders~S}\ \bibnamefont
  {S{\o}rensen}}, \bibinfo {author} {\bibfnamefont {Peter}\ \bibnamefont
  {Zoller}}, and\ \bibinfo {author} {\bibfnamefont {Mikhail~D}\ \bibnamefont
  {Lukin}}} (\bibinfo {year} {2010}),\ \bibfield  {title} {\enquote {\bibinfo
  {title} {Optomechanical transducers for long-distance quantum
  communication},}\ }\href {https://doi.org/10.1103/physrevlett.105.220501}
  {\bibfield  {journal} {\bibinfo  {journal} {Physical Review Letters}\
  }\textbf {\bibinfo {volume} {105}},\ \bibinfo {pages} {220501}},\ \Eprint
  {https://arxiv.org/abs/arXiv:1006.4361v2} {arXiv:1006.4361v2} \BibitemShut
  {NoStop}%
\bibitem [{\citenamefont {St{\'a}rek}\ \emph {et~al.}(2016)\citenamefont
  {St{\'a}rek}, \citenamefont {Mi{\v{c}}uda}, \citenamefont {Mikov{\'a}},
  \citenamefont {Straka}, \citenamefont {Du{\v{s}}ek}, \citenamefont
  {Je{\v{z}}ek},\ and\ \citenamefont {Fiur{\'a}{\v{s}}ek}}]{bib:starek2016}%
  \BibitemOpen
  \bibfield  {author} {\bibinfo {author} {\bibnamefont {St{\'a}rek},
  \bibfnamefont {R}}, \bibinfo {author} {\bibfnamefont {M}~\bibnamefont
  {Mi{\v{c}}uda}}, \bibinfo {author} {\bibfnamefont {M}~\bibnamefont
  {Mikov{\'a}}}, \bibinfo {author} {\bibfnamefont {I}~\bibnamefont {Straka}},
  \bibinfo {author} {\bibfnamefont {M}~\bibnamefont {Du{\v{s}}ek}}, \bibinfo
  {author} {\bibfnamefont {M}~\bibnamefont {Je{\v{z}}ek}}, and\ \bibinfo
  {author} {\bibfnamefont {J}~\bibnamefont {Fiur{\'a}{\v{s}}ek}}} (\bibinfo
  {year} {2016}),\ \bibfield  {title} {\enquote {\bibinfo {title} {Experimental
  investigation of a four-qubit linear-optical quantum logic circuit},}\ }\href
  {https://doi.org/10.1038/srep33475} {\bibfield  {journal} {\bibinfo
  {journal} {Scientific Reports}\ }\textbf {\bibinfo {volume} {6}},\ \bibinfo
  {pages} {33475}}\BibitemShut {NoStop}%
\bibitem [{\citenamefont {Statista}(2024)}]{sneakernet_web_3}%
  \BibitemOpen
  \bibfield  {author} {\bibinfo {author} {\bibnamefont {Statista},}} (\bibinfo
  {year} {2024}),\ \href
  {https://doi.org/http://www.statista.com/statistics/499431/global-ip-data-traffic-forecast}
  {\
  http://www.statista.com/statistics/499431/global-ip-data-traffic-forecast}\BibitemShut
  {NoStop}%
\bibitem [{\citenamefont {Steane}(1996)}]{bib:S96}%
  \BibitemOpen
  \bibfield  {author} {\bibinfo {author} {\bibnamefont {Steane}, \bibfnamefont
  {AM}}} (\bibinfo {year} {1996}),\ \bibfield  {title} {\enquote {\bibinfo
  {title} {{Multiple particle interference and quantum error correction}},}\
  }\href@noop {} {\bibfield  {journal} {\bibinfo  {journal} {Proc. Royal
  Society of London A.}\ }\textbf {\bibinfo {volume} {452}},\ \bibinfo {pages}
  {2551}}\BibitemShut {NoStop}%
\bibitem [{\citenamefont {Steffen}\ \emph {et~al.}(2013)\citenamefont
  {Steffen}, \citenamefont {Salathe}, \citenamefont {Oppliger}, \citenamefont
  {Kurpiers}, \citenamefont {Baur}, \citenamefont {Lang}, \citenamefont
  {Eichler}, \citenamefont {Puebla-Hellmann}, \citenamefont {Fedorov},\ and\
  \citenamefont {Wallraff}}]{bib:Nat_500_319}%
  \BibitemOpen
  \bibfield  {author} {\bibinfo {author} {\bibnamefont {Steffen}, \bibfnamefont
  {Lars}}, \bibinfo {author} {\bibfnamefont {Yves}\ \bibnamefont {Salathe}},
  \bibinfo {author} {\bibfnamefont {Markus}\ \bibnamefont {Oppliger}}, \bibinfo
  {author} {\bibfnamefont {Philipp}\ \bibnamefont {Kurpiers}}, \bibinfo
  {author} {\bibfnamefont {Matthias}\ \bibnamefont {Baur}}, \bibinfo {author}
  {\bibfnamefont {Christian}\ \bibnamefont {Lang}}, \bibinfo {author}
  {\bibfnamefont {Christopher}\ \bibnamefont {Eichler}}, \bibinfo {author}
  {\bibfnamefont {Gabriel}\ \bibnamefont {Puebla-Hellmann}}, \bibinfo {author}
  {\bibfnamefont {Arkady}\ \bibnamefont {Fedorov}}, and\ \bibinfo {author}
  {\bibfnamefont {Andreas}\ \bibnamefont {Wallraff}}} (\bibinfo {year}
  {2013}),\ \bibfield  {title} {\enquote {\bibinfo {title} {Deterministic
  quantum teleportation with feed-forward in a solid state system},}\ }\href
  {https://doi.org/10.1038/nature12422} {\bibfield  {journal} {\bibinfo
  {journal} {Nature}\ }\textbf {\bibinfo {volume} {500}},\ \bibinfo {pages}
  {319}}\BibitemShut {NoStop}%
\bibitem [{\citenamefont {Steger}\ \emph {et~al.}(2012)\citenamefont {Steger},
  \citenamefont {Saeedi}, \citenamefont {Thewalt}, \citenamefont {Morton},
  \citenamefont {Riemann}, \citenamefont {Abrosimov}, \citenamefont {Becker},\
  and\ \citenamefont {Pohl}}]{bib:steger2012quantum}%
  \BibitemOpen
  \bibfield  {author} {\bibinfo {author} {\bibnamefont {Steger}, \bibfnamefont
  {M}}, \bibinfo {author} {\bibfnamefont {K}~\bibnamefont {Saeedi}}, \bibinfo
  {author} {\bibfnamefont {MLW}\ \bibnamefont {Thewalt}}, \bibinfo {author}
  {\bibfnamefont {JJL}\ \bibnamefont {Morton}}, \bibinfo {author}
  {\bibfnamefont {H}~\bibnamefont {Riemann}}, \bibinfo {author} {\bibfnamefont
  {NV}~\bibnamefont {Abrosimov}}, \bibinfo {author} {\bibfnamefont
  {P}~\bibnamefont {Becker}}, and\ \bibinfo {author} {\bibfnamefont {H-J}\
  \bibnamefont {Pohl}}} (\bibinfo {year} {2012}),\ \bibfield  {title} {\enquote
  {\bibinfo {title} {Quantum information storage for over 180s using donor
  spins in a 28si `semiconductor vacuum'},}\ }\href
  {https://doi.org/10.1126/science.1217635} {\bibfield  {journal} {\bibinfo
  {journal} {Science}\ }\textbf {\bibinfo {volume} {336}},\ \bibinfo {pages}
  {1280}}\BibitemShut {NoStop}%
\bibitem [{\citenamefont {Stehl{\'e}}\ and\ \citenamefont
  {Steinfeld}(2010)}]{bib:Damien2010}%
  \BibitemOpen
  \bibfield  {author} {\bibinfo {author} {\bibnamefont {Stehl{\'e}},
  \bibfnamefont {Damien}}, and\ \bibinfo {author} {\bibfnamefont {Ron}\
  \bibnamefont {Steinfeld}}} (\bibinfo {year} {2010}),\ \bibfield  {title}
  {\enquote {\bibinfo {title} {Faster fully homomorphic encryption},}\ }in\
  \href {https://doi.org/10.1007/978-3-642-17373-8_22} {\emph {\bibinfo
  {booktitle} {Advances in Cryptology - ASIACRYPT}}},\ \bibinfo {editor}
  {edited by\ \bibinfo {editor} {\bibfnamefont {Masayuki}\ \bibnamefont
  {Abe}}},\ p.\ \bibinfo {pages} {377}\BibitemShut {NoStop}%
\bibitem [{\citenamefont {Stephens}\ \emph {et~al.}(2008)\citenamefont
  {Stephens}, \citenamefont {Fowler},\ and\ \citenamefont
  {Hollenberg}}]{bib:SFH07}%
  \BibitemOpen
  \bibfield  {author} {\bibinfo {author} {\bibnamefont {Stephens},
  \bibfnamefont {A}}, \bibinfo {author} {\bibfnamefont {A.G.}\ \bibnamefont
  {Fowler}}, and\ \bibinfo {author} {\bibfnamefont {L.C.L.}\ \bibnamefont
  {Hollenberg}}} (\bibinfo {year} {2008}),\ \bibfield  {title} {\enquote
  {\bibinfo {title} {{Universal Fault-Tolerant Computation on bilinear nearest
  neighbor arrays}},}\ }\href@noop {} {\bibfield  {journal} {\bibinfo
  {journal} {Quant. Inf. Comp.}\ }\textbf {\bibinfo {volume} {8}},\ \bibinfo
  {pages} {330}}\BibitemShut {NoStop}%
\bibitem [{\citenamefont {Stephens}(2014{\natexlab{a}})}]{bib:S14}%
  \BibitemOpen
  \bibfield  {author} {\bibinfo {author} {\bibnamefont {Stephens},
  \bibfnamefont {AM}}} (\bibinfo {year} {2014}{\natexlab{a}}),\ \bibfield
  {title} {\enquote {\bibinfo {title} {{Fault-tolerant thresholds for quantum
  error correction with the surface code}},}\ }\href@noop {} {\bibfield
  {journal} {\bibinfo  {journal} {Phys. Rev. A.}\ }\textbf {\bibinfo {volume}
  {89}},\ \bibinfo {pages} {022321}}\BibitemShut {NoStop}%
\bibitem [{\citenamefont {Stephens}(2014{\natexlab{b}})}]{SD-Stephens:2014aa}%
  \BibitemOpen
  \bibfield  {author} {\bibinfo {author} {\bibnamefont {Stephens},
  \bibfnamefont {Ashley~M}}} (\bibinfo {year} {2014}{\natexlab{b}}),\ \bibfield
   {title} {\enquote {\bibinfo {title} {Fault-tolerant thresholds for quantum
  error correction with the surface code},}\ }\href
  {https://doi.org/10.1103/PhysRevA.89.022321} {\bibfield  {journal} {\bibinfo
  {journal} {Physical Review A}\ }\textbf {\bibinfo {volume} {89}},\ \bibinfo
  {pages} {022321}},\ \Eprint {https://arxiv.org/abs/arXiv:1311.5003v2}
  {arXiv:1311.5003v2} \BibitemShut {NoStop}%
\bibitem [{\citenamefont {Stephens}\ \emph {et~al.}(2013)\citenamefont
  {Stephens}, \citenamefont {Huang}, \citenamefont {Nemoto},\ and\
  \citenamefont {Munro}}]{bib:Stephens2013}%
  \BibitemOpen
  \bibfield  {author} {\bibinfo {author} {\bibnamefont {Stephens},
  \bibfnamefont {Ashley~M}}, \bibinfo {author} {\bibfnamefont {Jingjing}\
  \bibnamefont {Huang}}, \bibinfo {author} {\bibfnamefont {Kae}\ \bibnamefont
  {Nemoto}}, and\ \bibinfo {author} {\bibfnamefont {William~J.}\ \bibnamefont
  {Munro}}} (\bibinfo {year} {2013}),\ \bibfield  {title} {\enquote {\bibinfo
  {title} {Hybrid-system approach to fault-tolerant quantum communication},}\
  }\href {https://doi.org/10.1103/physreva.87.052333} {\bibfield  {journal}
  {\bibinfo  {journal} {Physical Review A}\ }\textbf {\bibinfo {volume} {87}},\
  \bibinfo {pages} {052333}},\ \Eprint
  {https://arxiv.org/abs/arXiv:1209.3851v2} {arXiv:1209.3851v2} \BibitemShut
  {NoStop}%
\bibitem [{\citenamefont {Straffin}(1993)}]{bib:Straffin93}%
  \BibitemOpen
  \bibfield  {author} {\bibinfo {author} {\bibnamefont {Straffin},
  \bibfnamefont {P~D}}} (\bibinfo {year} {1993}),\ \bibfield  {title} {\enquote
  {\bibinfo {title} {Game theory and strategy},}\ }\href@noop {} {\bibfield
  {journal} {\bibinfo  {journal} {Mathematical Association of America}\
  }\textbf {\bibinfo {volume} {36}}}\BibitemShut {NoStop}%
\bibitem [{\citenamefont {Stucki}\ \emph {et~al.}(2002)\citenamefont {Stucki},
  \citenamefont {Gisin}, \citenamefont {Guinnard}, \citenamefont {Ribordy},\
  and\ \citenamefont {Zbinden}}]{bib:Arx0203118}%
  \BibitemOpen
  \bibfield  {author} {\bibinfo {author} {\bibnamefont {Stucki}, \bibfnamefont
  {D}}, \bibinfo {author} {\bibfnamefont {N}~\bibnamefont {Gisin}}, \bibinfo
  {author} {\bibfnamefont {O}~\bibnamefont {Guinnard}}, \bibinfo {author}
  {\bibfnamefont {G}~\bibnamefont {Ribordy}}, and\ \bibinfo {author}
  {\bibfnamefont {H}~\bibnamefont {Zbinden}}} (\bibinfo {year} {2002}),\
  \bibfield  {title} {\enquote {\bibinfo {title} {Quantum key distribution over
  67 km with a plug \& play system},}\ }\href
  {https://doi.org/10.1088/1367-2630/4/1/341} {\bibfield  {journal} {\bibinfo
  {journal} {New Journal of Physics}\ }\textbf {\bibinfo {volume} {4}},\
  \bibinfo {pages} {41}}\BibitemShut {NoStop}%
\bibitem [{\citenamefont {Stucki}\ \emph {et~al.}(2009)\citenamefont {Stucki},
  \citenamefont {Walenta}, \citenamefont {Vannel}, \citenamefont {Thew},
  \citenamefont {Gisin}, \citenamefont {Zbinden}, \citenamefont {Gray},
  \citenamefont {Towery},\ and\ \citenamefont {Ten}}]{bib:NJP_11_075003}%
  \BibitemOpen
  \bibfield  {author} {\bibinfo {author} {\bibnamefont {Stucki}, \bibfnamefont
  {Damien}}, \bibinfo {author} {\bibfnamefont {Nino}\ \bibnamefont {Walenta}},
  \bibinfo {author} {\bibfnamefont {Fabien}\ \bibnamefont {Vannel}}, \bibinfo
  {author} {\bibfnamefont {Robert~Thomas}\ \bibnamefont {Thew}}, \bibinfo
  {author} {\bibfnamefont {Nicolas}\ \bibnamefont {Gisin}}, \bibinfo {author}
  {\bibfnamefont {Hugo}\ \bibnamefont {Zbinden}}, \bibinfo {author}
  {\bibfnamefont {S}~\bibnamefont {Gray}}, \bibinfo {author} {\bibfnamefont
  {CR}~\bibnamefont {Towery}}, and\ \bibinfo {author} {\bibfnamefont
  {S}~\bibnamefont {Ten}}} (\bibinfo {year} {2009}),\ \bibfield  {title}
  {\enquote {\bibinfo {title} {High rate, long-distance quantum key
  distribution over 250 km of ultra low loss fibres},}\ }\href
  {https://doi.org/10.1088/1367-2630/11/7/075003} {\bibfield  {journal}
  {\bibinfo  {journal} {New Journal of Physics}\ }\textbf {\bibinfo {volume}
  {11}},\ \bibinfo {pages} {075003}}\BibitemShut {NoStop}%
\bibitem [{\citenamefont {Sugden}(2004)}]{bib:Sugden04}%
  \BibitemOpen
  \bibfield  {author} {\bibinfo {author} {\bibnamefont {Sugden}, \bibfnamefont
  {Robert}}} (\bibinfo {year} {2004}),\ \href
  {https://doi.org/10.1057/9780230536791} {\emph {\bibinfo {title} {The
  Economics of Rights, Co-operation and Welfare}}}\ (\bibinfo  {publisher}
  {Palgrave Macmillan})\BibitemShut {NoStop}%
\bibitem [{\citenamefont {Sun}\ \emph {et~al.}(2017)\citenamefont {Sun},
  \citenamefont {Jiang}, \citenamefont {Mao}, \citenamefont {You},
  \citenamefont {Zhang}, \citenamefont {Zhang}, \citenamefont {Jiang},
  \citenamefont {Chen}, \citenamefont {Li}, \citenamefont {Huang} \emph
  {et~al.}}]{bib:sun2017entanglement}%
  \BibitemOpen
  \bibfield  {author} {\bibinfo {author} {\bibnamefont {Sun}, \bibfnamefont
  {Qi-Chao}}, \bibinfo {author} {\bibfnamefont {Yang-Fan}\ \bibnamefont
  {Jiang}}, \bibinfo {author} {\bibfnamefont {Ya-Li}\ \bibnamefont {Mao}},
  \bibinfo {author} {\bibfnamefont {Li-Xing}\ \bibnamefont {You}}, \bibinfo
  {author} {\bibfnamefont {Wei}\ \bibnamefont {Zhang}}, \bibinfo {author}
  {\bibfnamefont {Wei-Jun}\ \bibnamefont {Zhang}}, \bibinfo {author}
  {\bibfnamefont {Xiao}\ \bibnamefont {Jiang}}, \bibinfo {author}
  {\bibfnamefont {Teng-Yun}\ \bibnamefont {Chen}}, \bibinfo {author}
  {\bibfnamefont {Hao}\ \bibnamefont {Li}}, \bibinfo {author} {\bibfnamefont
  {Yi-Dong}\ \bibnamefont {Huang}},  \emph {et~al.}} (\bibinfo {year} {2017}),\
  \bibfield  {title} {\enquote {\bibinfo {title} {Entanglement swapping over
  100 km optical fiber with independent entangled photon-pair sources},}\
  }\href {https://doi.org/10.1364/optica.4.001214} {\bibfield  {journal}
  {\bibinfo  {journal} {Optica}\ }\textbf {\bibinfo {volume} {4}},\ \bibinfo
  {pages} {1214}}\BibitemShut {NoStop}%
\bibitem [{\citenamefont {Sun}\ \emph {et~al.}(2016)\citenamefont {Sun},
  \citenamefont {Mao}, \citenamefont {Chen}, \citenamefont {Zhang},
  \citenamefont {Jiang}, \citenamefont {Zhang}, \citenamefont {Zhang},
  \citenamefont {Miki}, \citenamefont {Yamashita}, \citenamefont {Terai} \emph
  {et~al.}}]{bib:sun2016quantum}%
  \BibitemOpen
  \bibfield  {author} {\bibinfo {author} {\bibnamefont {Sun}, \bibfnamefont
  {Qi-Chao}}, \bibinfo {author} {\bibfnamefont {Ya-Li}\ \bibnamefont {Mao}},
  \bibinfo {author} {\bibfnamefont {Si-Jing}\ \bibnamefont {Chen}}, \bibinfo
  {author} {\bibfnamefont {Wei}\ \bibnamefont {Zhang}}, \bibinfo {author}
  {\bibfnamefont {Yang-Fan}\ \bibnamefont {Jiang}}, \bibinfo {author}
  {\bibfnamefont {Yan-Bao}\ \bibnamefont {Zhang}}, \bibinfo {author}
  {\bibfnamefont {Wei-Jun}\ \bibnamefont {Zhang}}, \bibinfo {author}
  {\bibfnamefont {Shigehito}\ \bibnamefont {Miki}}, \bibinfo {author}
  {\bibfnamefont {Taro}\ \bibnamefont {Yamashita}}, \bibinfo {author}
  {\bibfnamefont {Hirotaka}\ \bibnamefont {Terai}},  \emph {et~al.}} (\bibinfo
  {year} {2016}),\ \bibfield  {title} {\enquote {\bibinfo {title} {Quantum
  teleportation with independent sources and prior entanglement distribution
  over a network},}\ }\href {https://doi.org/10.1038/nphoton.2016.179}
  {\bibfield  {journal} {\bibinfo  {journal} {Nature Photonics}\ }\textbf
  {\bibinfo {volume} {10}},\ \bibinfo {pages} {671}}\BibitemShut {NoStop}%
\bibitem [{\citenamefont {Svore}\ \emph {et~al.}(2007)\citenamefont {Svore},
  \citenamefont {DiVincenzo},\ and\ \citenamefont {Terhal}}]{bib:SDT07}%
  \BibitemOpen
  \bibfield  {author} {\bibinfo {author} {\bibnamefont {Svore}, \bibfnamefont
  {KM}}, \bibinfo {author} {\bibfnamefont {D.P.}\ \bibnamefont {DiVincenzo}},
  and\ \bibinfo {author} {\bibfnamefont {B.M.}\ \bibnamefont {Terhal}}}
  (\bibinfo {year} {2007}),\ \bibfield  {title} {\enquote {\bibinfo {title}
  {{Noise Threshold for a Fault-Tolerant Two-Dimensional Lattice
  Architecture}},}\ }\href@noop {} {\bibfield  {journal} {\bibinfo  {journal}
  {Quant. Inf. Comp.}\ }\textbf {\bibinfo {volume} {7}},\ \bibinfo {pages}
  {297}}\BibitemShut {NoStop}%
\bibitem [{\citenamefont {Szegedy}\ \emph {et~al.}(2013)\citenamefont
  {Szegedy}, \citenamefont {Zaremba}, \citenamefont {Sutskever}, \citenamefont
  {Bruna}, \citenamefont {Erhan}, \citenamefont {Goodfellow},\ and\
  \citenamefont {Fergus}}]{bib:szegedy2013intriguing}%
  \BibitemOpen
  \bibfield  {author} {\bibinfo {author} {\bibnamefont {Szegedy}, \bibfnamefont
  {Christian}}, \bibinfo {author} {\bibfnamefont {Wojciech}\ \bibnamefont
  {Zaremba}}, \bibinfo {author} {\bibfnamefont {Ilya}\ \bibnamefont
  {Sutskever}}, \bibinfo {author} {\bibfnamefont {Joan}\ \bibnamefont {Bruna}},
  \bibinfo {author} {\bibfnamefont {Dumitru}\ \bibnamefont {Erhan}}, \bibinfo
  {author} {\bibfnamefont {Ian}\ \bibnamefont {Goodfellow}}, and\ \bibinfo
  {author} {\bibfnamefont {Rob}\ \bibnamefont {Fergus}}} (\bibinfo {year}
  {2013}),\ \bibfield  {title} {\enquote {\bibinfo {title} {Intriguing
  properties of neural networks},}\ }\href@noop {} {\ }\Eprint
  {https://arxiv.org/abs/arXiv:1312.6199} {arXiv:1312.6199} \BibitemShut
  {NoStop}%
\bibitem [{\citenamefont {Szkopek}\ \emph {et~al.}(2006)\citenamefont
  {Szkopek}, \citenamefont {Boykin}, \citenamefont {Fan}, \citenamefont
  {Roychowdhury}, \citenamefont {Yablonovitch}, \citenamefont {Simms},
  \citenamefont {Gyure},\ and\ \citenamefont {Fong}}]{bib:SBFRYSGF06}%
  \BibitemOpen
  \bibfield  {author} {\bibinfo {author} {\bibnamefont {Szkopek}, \bibfnamefont
  {T}}, \bibinfo {author} {\bibfnamefont {P.O.}\ \bibnamefont {Boykin}},
  \bibinfo {author} {\bibfnamefont {H.}~\bibnamefont {Fan}}, \bibinfo {author}
  {\bibfnamefont {V.P.}\ \bibnamefont {Roychowdhury}}, \bibinfo {author}
  {\bibfnamefont {E.}~\bibnamefont {Yablonovitch}}, \bibinfo {author}
  {\bibfnamefont {G.}~\bibnamefont {Simms}}, \bibinfo {author} {\bibfnamefont
  {M.}~\bibnamefont {Gyure}}, and\ \bibinfo {author} {\bibfnamefont
  {B.}~\bibnamefont {Fong}}} (\bibinfo {year} {2006}),\ \bibfield  {title}
  {\enquote {\bibinfo {title} {{Threshold Error Penalty for Fault-Tolerant
  Computation with Nearest Neighbour Communication}},}\ }\href@noop {}
  {\bibfield  {journal} {\bibinfo  {journal} {IEEE Trans. Nano.}\ }\textbf
  {\bibinfo {volume} {5}}~(\bibinfo {number} {1}),\ \bibinfo {pages}
  {42}}\BibitemShut {NoStop}%
\bibitem [{\citenamefont {Takahashi}\ \emph {et~al.}(2008)\citenamefont
  {Takahashi}, \citenamefont {Wakui}, \citenamefont {Suzuki}, \citenamefont
  {Takeoka}, \citenamefont {Hayasaka}, \citenamefont {Furusawa},\ and\
  \citenamefont {Sasaki}}]{bib:takahashi2008generation}%
  \BibitemOpen
  \bibfield  {author} {\bibinfo {author} {\bibnamefont {Takahashi},
  \bibfnamefont {Hiroki}}, \bibinfo {author} {\bibfnamefont {Kentaro}\
  \bibnamefont {Wakui}}, \bibinfo {author} {\bibfnamefont {Shigenari}\
  \bibnamefont {Suzuki}}, \bibinfo {author} {\bibfnamefont {Masahiro}\
  \bibnamefont {Takeoka}}, \bibinfo {author} {\bibfnamefont {Kazuhiro}\
  \bibnamefont {Hayasaka}}, \bibinfo {author} {\bibfnamefont {Akira}\
  \bibnamefont {Furusawa}}, and\ \bibinfo {author} {\bibfnamefont {Masahide}\
  \bibnamefont {Sasaki}}} (\bibinfo {year} {2008}),\ \bibfield  {title}
  {\enquote {\bibinfo {title} {Generation of large-amplitude coherent-state
  superposition via ancilla-assisted photon subtraction},}\ }\href
  {https://doi.org/10.1103/physrevlett.101.233605} {\bibfield  {journal}
  {\bibinfo  {journal} {Physical Review Letters}\ }\textbf {\bibinfo {volume}
  {101}},\ \bibinfo {pages} {233605}},\ \Eprint
  {https://arxiv.org/abs/arXiv:0806.2965v2} {arXiv:0806.2965v2} \BibitemShut
  {NoStop}%
\bibitem [{\citenamefont {Takase}\ \emph {et~al.}(2024)\citenamefont {Takase},
  \citenamefont {Hanamura}, \citenamefont {Nagayoshi}, \citenamefont
  {Bourassa}, \citenamefont {Alexander}, \citenamefont {Kawasaki},
  \citenamefont {Asavanant}, \citenamefont {Endo},\ and\ \citenamefont
  {Furusawa}}]{takase2024breeding}%
  \BibitemOpen
  \bibfield  {author} {\bibinfo {author} {\bibnamefont {Takase}, \bibfnamefont
  {Kan}}, \bibinfo {author} {\bibfnamefont {Fumiya}\ \bibnamefont {Hanamura}},
  \bibinfo {author} {\bibfnamefont {Hironari}\ \bibnamefont {Nagayoshi}},
  \bibinfo {author} {\bibfnamefont {J.~Eli}\ \bibnamefont {Bourassa}}, \bibinfo
  {author} {\bibfnamefont {Rafael~N.}\ \bibnamefont {Alexander}}, \bibinfo
  {author} {\bibfnamefont {Akito}\ \bibnamefont {Kawasaki}}, \bibinfo {author}
  {\bibfnamefont {Warit}\ \bibnamefont {Asavanant}}, \bibinfo {author}
  {\bibfnamefont {Mamoru}\ \bibnamefont {Endo}}, and\ \bibinfo {author}
  {\bibfnamefont {Akira}\ \bibnamefont {Furusawa}}} (\bibinfo {year} {2024}),\
  \bibfield  {title} {\enquote {\bibinfo {title} {Generation of flying logical
  qubits using generalized photon subtraction with adaptive gaussian
  operations},}\ }\href {https://doi.org/10.1103/PhysRevA.110.012436}
  {\bibfield  {journal} {\bibinfo  {journal} {Phys. Rev. A}\ }\textbf {\bibinfo
  {volume} {110}},\ \bibinfo {pages} {012436}}\BibitemShut {NoStop}%
\bibitem [{\citenamefont {Takeda}\ \emph {et~al.}(2015)\citenamefont {Takeda},
  \citenamefont {Fuwa}, \citenamefont {van Loock},\ and\ \citenamefont
  {Furusawa}}]{bib:takeda2015entanglement}%
  \BibitemOpen
  \bibfield  {author} {\bibinfo {author} {\bibnamefont {Takeda}, \bibfnamefont
  {Shuntaro}}, \bibinfo {author} {\bibfnamefont {Maria}\ \bibnamefont {Fuwa}},
  \bibinfo {author} {\bibfnamefont {Peter}\ \bibnamefont {van Loock}}, and\
  \bibinfo {author} {\bibfnamefont {Akira}\ \bibnamefont {Furusawa}}} (\bibinfo
  {year} {2015}),\ \bibfield  {title} {\enquote {\bibinfo {title} {Entanglement
  swapping between discrete and continuous variables},}\ }\href
  {https://doi.org/10.1103/physrevlett.114.100501} {\bibfield  {journal}
  {\bibinfo  {journal} {Physical Review Letters}\ }\textbf {\bibinfo {volume}
  {114}},\ \bibinfo {pages} {100501}},\ \Eprint
  {https://arxiv.org/abs/arXiv:1411.1310v2} {arXiv:1411.1310v2} \BibitemShut
  {NoStop}%
\bibitem [{\citenamefont {Takeda}\ \emph {et~al.}(2013)\citenamefont {Takeda},
  \citenamefont {Mizuta}, \citenamefont {Fuwa}, \citenamefont {van Loock},\
  and\ \citenamefont {Furusawa}}]{bib:Nat_500_315}%
  \BibitemOpen
  \bibfield  {author} {\bibinfo {author} {\bibnamefont {Takeda}, \bibfnamefont
  {Shuntaro}}, \bibinfo {author} {\bibfnamefont {Takahiro}\ \bibnamefont
  {Mizuta}}, \bibinfo {author} {\bibfnamefont {Maria}\ \bibnamefont {Fuwa}},
  \bibinfo {author} {\bibfnamefont {Peter}\ \bibnamefont {van Loock}}, and\
  \bibinfo {author} {\bibfnamefont {Akira}\ \bibnamefont {Furusawa}}} (\bibinfo
  {year} {2013}),\ \bibfield  {title} {\enquote {\bibinfo {title}
  {Deterministic quantum teleportation of photonic quantum bits by a hybrid
  technique},}\ }\href {https://doi.org/10.1038/nature12366} {\bibfield
  {journal} {\bibinfo  {journal} {Nature}\ }\textbf {\bibinfo {volume} {500}},\
  \bibinfo {pages} {315}},\ \Eprint {https://arxiv.org/abs/arXiv:1402.4895v1}
  {arXiv:1402.4895v1} \BibitemShut {NoStop}%
\bibitem [{\citenamefont {Takenaka}\ \emph
  {et~al.}(2017{\natexlab{a}})\citenamefont {Takenaka}, \citenamefont
  {Carrasco-Casado}, \citenamefont {Fujiwara}, \citenamefont {Kitamura},
  \citenamefont {Sasaki},\ and\ \citenamefont
  {Toyoshima}}]{SD-Takenaka:2017aa}%
  \BibitemOpen
  \bibfield  {author} {\bibinfo {author} {\bibnamefont {Takenaka},
  \bibfnamefont {Hideki}}, \bibinfo {author} {\bibfnamefont {Alberto}\
  \bibnamefont {Carrasco-Casado}}, \bibinfo {author} {\bibfnamefont {Mikio}\
  \bibnamefont {Fujiwara}}, \bibinfo {author} {\bibfnamefont {Mitsuo}\
  \bibnamefont {Kitamura}}, \bibinfo {author} {\bibfnamefont {Masahide}\
  \bibnamefont {Sasaki}}, and\ \bibinfo {author} {\bibfnamefont {Morio}\
  \bibnamefont {Toyoshima}}} (\bibinfo {year} {2017}{\natexlab{a}}),\ \bibfield
   {title} {\enquote {\bibinfo {title} {Satellite-to-ground quantum-limited
  communication using a 50-kg-class microsatellite},}\ }\href
  {https://doi.org/10.1038/nphoton.2017.107} {\bibfield  {journal} {\bibinfo
  {journal} {Nature Photonics}\ }\textbf {\bibinfo {volume} {11}},\ \bibinfo
  {pages} {502}}\BibitemShut {NoStop}%
\bibitem [{\citenamefont {Takenaka}\ \emph
  {et~al.}(2017{\natexlab{b}})\citenamefont {Takenaka}, \citenamefont
  {Carrasco-Casado}, \citenamefont {Fujiwara}, \citenamefont {Kitamura},
  \citenamefont {Sasaki},\ and\ \citenamefont {Toyoshima}}]{bib:takenaka2017}%
  \BibitemOpen
  \bibfield  {author} {\bibinfo {author} {\bibnamefont {Takenaka},
  \bibfnamefont {Hideki}}, \bibinfo {author} {\bibfnamefont {Alberto}\
  \bibnamefont {Carrasco-Casado}}, \bibinfo {author} {\bibfnamefont {Mikio}\
  \bibnamefont {Fujiwara}}, \bibinfo {author} {\bibfnamefont {Mitsuo}\
  \bibnamefont {Kitamura}}, \bibinfo {author} {\bibfnamefont {Masahide}\
  \bibnamefont {Sasaki}}, and\ \bibinfo {author} {\bibfnamefont {Morio}\
  \bibnamefont {Toyoshima}}} (\bibinfo {year} {2017}{\natexlab{b}}),\ \bibfield
   {title} {\enquote {\bibinfo {title} {Satellite-to-ground quantum-limited
  communication using a 50-kg-class microsatellite},}\ }\href
  {https://doi.org/10.1038/nphoton.2017.107} {\bibfield  {journal} {\bibinfo
  {journal} {Nature Photonics}\ }\textbf {\bibinfo {volume} {11}},\ \bibinfo
  {pages} {502}}\BibitemShut {NoStop}%
\bibitem [{\citenamefont {Takesue}\ \emph {et~al.}(2015)\citenamefont
  {Takesue}, \citenamefont {Dyer}, \citenamefont {Stevens}, \citenamefont
  {Verma}, \citenamefont {Mirin},\ and\ \citenamefont
  {Nam}}]{bib:Optica_2_832}%
  \BibitemOpen
  \bibfield  {author} {\bibinfo {author} {\bibnamefont {Takesue}, \bibfnamefont
  {Hiroki}}, \bibinfo {author} {\bibfnamefont {Shellee~D}\ \bibnamefont
  {Dyer}}, \bibinfo {author} {\bibfnamefont {Martin~J}\ \bibnamefont
  {Stevens}}, \bibinfo {author} {\bibfnamefont {Varun}\ \bibnamefont {Verma}},
  \bibinfo {author} {\bibfnamefont {Richard~P}\ \bibnamefont {Mirin}}, and\
  \bibinfo {author} {\bibfnamefont {Sae~Woo}\ \bibnamefont {Nam}}} (\bibinfo
  {year} {2015}),\ \bibfield  {title} {\enquote {\bibinfo {title} {Quantum
  teleportation over 100 km of fiber using highly efficient superconducting
  nanowire single-photon detectors},}\ }\href
  {https://doi.org/10.1364/optica.2.000832} {\bibfield  {journal} {\bibinfo
  {journal} {Optica}\ }\textbf {\bibinfo {volume} {2}},\ \bibinfo {pages}
  {832}},\ \Eprint {https://arxiv.org/abs/arXiv:1510.00476v1}
  {arXiv:1510.00476v1} \BibitemShut {NoStop}%
\bibitem [{\citenamefont {Takesue}\ \emph {et~al.}(2007)\citenamefont
  {Takesue}, \citenamefont {Nam}, \citenamefont {Zhang}, \citenamefont
  {Hadfield}, \citenamefont {Honjo}, \citenamefont {Tamaki},\ and\
  \citenamefont {Yamamoto}}]{bib:NP_1_343}%
  \BibitemOpen
  \bibfield  {author} {\bibinfo {author} {\bibnamefont {Takesue}, \bibfnamefont
  {Hiroki}}, \bibinfo {author} {\bibfnamefont {Sae~Woo}\ \bibnamefont {Nam}},
  \bibinfo {author} {\bibfnamefont {Qiang}\ \bibnamefont {Zhang}}, \bibinfo
  {author} {\bibfnamefont {Robert~H}\ \bibnamefont {Hadfield}}, \bibinfo
  {author} {\bibfnamefont {Toshimori}\ \bibnamefont {Honjo}}, \bibinfo {author}
  {\bibfnamefont {Kiyoshi}\ \bibnamefont {Tamaki}}, and\ \bibinfo {author}
  {\bibfnamefont {Yoshihisa}\ \bibnamefont {Yamamoto}}} (\bibinfo {year}
  {2007}),\ \bibfield  {title} {\enquote {\bibinfo {title} {Quantum key
  distribution over a 40-db channel loss using superconducting single-photon
  detectors},}\ }\href {https://doi.org/10.1038/nphoton.2007.75} {\bibfield
  {journal} {\bibinfo  {journal} {Nature Photonics}\ }\textbf {\bibinfo
  {volume} {1}},\ \bibinfo {pages} {343}},\ \Eprint
  {https://arxiv.org/abs/arXiv:0706.0397v1} {arXiv:0706.0397v1} \BibitemShut
  {NoStop}%
\bibitem [{\citenamefont {Tamaki}\ \emph {et~al.}(2014)\citenamefont {Tamaki},
  \citenamefont {Curty}, \citenamefont {Kato}, \citenamefont {Lo},\ and\
  \citenamefont {Azuma}}]{bib:PhysRevA.90.052314}%
  \BibitemOpen
  \bibfield  {author} {\bibinfo {author} {\bibnamefont {Tamaki}, \bibfnamefont
  {Kiyoshi}}, \bibinfo {author} {\bibfnamefont {Marcos}\ \bibnamefont {Curty}},
  \bibinfo {author} {\bibfnamefont {Go}~\bibnamefont {Kato}}, \bibinfo {author}
  {\bibfnamefont {Hoi-Kwong}\ \bibnamefont {Lo}}, and\ \bibinfo {author}
  {\bibfnamefont {Koji}\ \bibnamefont {Azuma}}} (\bibinfo {year} {2014}),\
  \bibfield  {title} {\enquote {\bibinfo {title} {Loss-tolerant quantum
  cryptography with imperfect sources},}\ }\href
  {https://doi.org/10.1103/physreva.90.052314} {\bibfield  {journal} {\bibinfo
  {journal} {Physical Review A}\ }\textbf {\bibinfo {volume} {90}},\ \bibinfo
  {pages} {052314}},\ \Eprint {https://arxiv.org/abs/arXiv:1312.3514v2}
  {arXiv:1312.3514v2} \BibitemShut {NoStop}%
\bibitem [{\citenamefont {Tan}\ \emph {et~al.}(2016)\citenamefont {Tan},
  \citenamefont {Kettlewell}, \citenamefont {Ouyang}, \citenamefont {Chen},\
  and\ \citenamefont {Fitzsimons}}]{tan2016quantum}%
  \BibitemOpen
  \bibfield  {author} {\bibinfo {author} {\bibnamefont {Tan}, \bibfnamefont
  {Si-Hui}}, \bibinfo {author} {\bibfnamefont {Joshua~A}\ \bibnamefont
  {Kettlewell}}, \bibinfo {author} {\bibfnamefont {Yingkai}\ \bibnamefont
  {Ouyang}}, \bibinfo {author} {\bibfnamefont {Lin}\ \bibnamefont {Chen}}, and\
  \bibinfo {author} {\bibfnamefont {Joseph~F}\ \bibnamefont {Fitzsimons}}}
  (\bibinfo {year} {2016}),\ \bibfield  {title} {\enquote {\bibinfo {title} {A
  quantum approach to homomorphic encryption},}\ }\href
  {https://doi.org/10.1038/srep33467} {\bibfield  {journal} {\bibinfo
  {journal} {Scientific Reports}\ }\textbf {\bibinfo {volume} {6}},\ \bibinfo
  {pages} {33467}}\BibitemShut {NoStop}%
\bibitem [{\citenamefont {Tan}\ \emph {et~al.}(2018)\citenamefont {Tan},
  \citenamefont {Ouyang},\ and\ \citenamefont {Rohde}}]{tan2018practical}%
  \BibitemOpen
  \bibfield  {author} {\bibinfo {author} {\bibnamefont {Tan}, \bibfnamefont
  {Si-Hui}}, \bibinfo {author} {\bibfnamefont {Yingkai}\ \bibnamefont
  {Ouyang}}, and\ \bibinfo {author} {\bibfnamefont {Peter~P}\ \bibnamefont
  {Rohde}}} (\bibinfo {year} {2018}),\ \bibfield  {title} {\enquote {\bibinfo
  {title} {Practical somewhat-secure quantum somewhat-homomorphic encryption
  with coherent states},}\ }\href {https://doi.org/10.1103/PhysRevA.97.042308}
  {\bibfield  {journal} {\bibinfo  {journal} {Physical Review A}\ }\textbf
  {\bibinfo {volume} {97}}~(\bibinfo {number} {4}),\ \bibinfo {pages}
  {042308}}\BibitemShut {NoStop}%
\bibitem [{\citenamefont {Tan}\ and\ \citenamefont
  {Rohde}(2018)}]{bib:TanRohdeRev}%
  \BibitemOpen
  \bibfield  {author} {\bibinfo {author} {\bibnamefont {Tan}, \bibfnamefont
  {Si-Hui}}, and\ \bibinfo {author} {\bibfnamefont {Peter~P.}\ \bibnamefont
  {Rohde}}} (\bibinfo {year} {2018}),\ \bibfield  {title} {\enquote {\bibinfo
  {title} {The resurgence of the linear optics quantum interferometer - recent
  advances \& applications},}\ }\href@noop {} {\ }\Eprint
  {https://arxiv.org/abs/arXiv:1805.11827} {arXiv:1805.11827} \BibitemShut
  {NoStop}%
\bibitem [{\citenamefont {Tanabe}\ \emph {et~al.}(2007)\citenamefont {Tanabe},
  \citenamefont {Notomi}, \citenamefont {Kuramochi}, \citenamefont {Shinya},\
  and\ \citenamefont {Taniyama}}]{bib:tanabe2007trapping}%
  \BibitemOpen
  \bibfield  {author} {\bibinfo {author} {\bibnamefont {Tanabe}, \bibfnamefont
  {Takasumi}}, \bibinfo {author} {\bibfnamefont {Masaya}\ \bibnamefont
  {Notomi}}, \bibinfo {author} {\bibfnamefont {Eiichi}\ \bibnamefont
  {Kuramochi}}, \bibinfo {author} {\bibfnamefont {Akihiko}\ \bibnamefont
  {Shinya}}, and\ \bibinfo {author} {\bibfnamefont {Hideaki}\ \bibnamefont
  {Taniyama}}} (\bibinfo {year} {2007}),\ \bibfield  {title} {\enquote
  {\bibinfo {title} {Trapping and delaying photons for one nanosecond in an
  ultrasmall high-q photonic-crystal nanocavity},}\ }\href
  {https://doi.org/10.1038/nphoton.2006.51} {\bibfield  {journal} {\bibinfo
  {journal} {Nature Photonics}\ }\textbf {\bibinfo {volume} {1}},\ \bibinfo
  {pages} {49}}\BibitemShut {NoStop}%
\bibitem [{\citenamefont {Tanabe}\ \emph {et~al.}(2009)\citenamefont {Tanabe},
  \citenamefont {Notomi}, \citenamefont {Taniyama},\ and\ \citenamefont
  {Kuramochi}}]{bib:tanabe2009dynamic}%
  \BibitemOpen
  \bibfield  {author} {\bibinfo {author} {\bibnamefont {Tanabe}, \bibfnamefont
  {Takasumi}}, \bibinfo {author} {\bibfnamefont {Masaya}\ \bibnamefont
  {Notomi}}, \bibinfo {author} {\bibfnamefont {Hideaki}\ \bibnamefont
  {Taniyama}}, and\ \bibinfo {author} {\bibfnamefont {Eiichi}\ \bibnamefont
  {Kuramochi}}} (\bibinfo {year} {2009}),\ \bibfield  {title} {\enquote
  {\bibinfo {title} {Dynamic release of trapped light from an ultrahigh-q
  nanocavity via adiabatic frequency tuning},}\ }\href
  {https://doi.org/10.1103/physrevlett.102.043907} {\bibfield  {journal}
  {\bibinfo  {journal} {Physical Review Letters}\ }\textbf {\bibinfo {volume}
  {102}},\ \bibinfo {pages} {043907}},\ \Eprint
  {https://arxiv.org/abs/arXiv:0812.4144v1} {arXiv:0812.4144v1} \BibitemShut
  {NoStop}%
\bibitem [{\citenamefont {Tanenbaum}(2002)}]{bib:TanenbaumNet}%
  \BibitemOpen
  \bibfield  {author} {\bibinfo {author} {\bibnamefont {Tanenbaum},
  \bibfnamefont {Andrew~S}}} (\bibinfo {year} {2002}),\ \href@noop {} {\emph
  {\bibinfo {title} {Computer networks}}}\ (\bibinfo  {publisher} {Prentice
  Hall})\BibitemShut {NoStop}%
\bibitem [{\citenamefont {Tang}(2018)}]{bib:tang2018quantum}%
  \BibitemOpen
  \bibfield  {author} {\bibinfo {author} {\bibnamefont {Tang}, \bibfnamefont
  {Ewin}}} (\bibinfo {year} {2018}),\ \bibfield  {title} {\enquote {\bibinfo
  {title} {Quantum-inspired classical algorithms for principal component
  analysis and supervised clustering},}\ }\href@noop {} {\ }\Eprint
  {https://arxiv.org/abs/arXiv:1811.00414} {arXiv:1811.00414} \BibitemShut
  {NoStop}%
\bibitem [{\citenamefont {Tang}\ \emph
  {et~al.}(2014{\natexlab{a}})\citenamefont {Tang}, \citenamefont {Yin},
  \citenamefont {Chen}, \citenamefont {Liu}, \citenamefont {Zhang},
  \citenamefont {Jiang}, \citenamefont {Zhang}, \citenamefont {Wang},
  \citenamefont {You}, \citenamefont {Guan} \emph
  {et~al.}}]{bib:PRL_113_190501}%
  \BibitemOpen
  \bibfield  {author} {\bibinfo {author} {\bibnamefont {Tang}, \bibfnamefont
  {Yan-Lin}}, \bibinfo {author} {\bibfnamefont {Hua-Lei}\ \bibnamefont {Yin}},
  \bibinfo {author} {\bibfnamefont {Si-Jing}\ \bibnamefont {Chen}}, \bibinfo
  {author} {\bibfnamefont {Yang}\ \bibnamefont {Liu}}, \bibinfo {author}
  {\bibfnamefont {Wei-Jun}\ \bibnamefont {Zhang}}, \bibinfo {author}
  {\bibfnamefont {Xiao}\ \bibnamefont {Jiang}}, \bibinfo {author}
  {\bibfnamefont {Lu}~\bibnamefont {Zhang}}, \bibinfo {author} {\bibfnamefont
  {Jian}\ \bibnamefont {Wang}}, \bibinfo {author} {\bibfnamefont {Li-Xing}\
  \bibnamefont {You}}, \bibinfo {author} {\bibfnamefont {Jian-Yu}\ \bibnamefont
  {Guan}},  \emph {et~al.}} (\bibinfo {year} {2014}{\natexlab{a}}),\ \bibfield
  {title} {\enquote {\bibinfo {title} {Measurement-device-independent quantum
  key distribution over 200 km},}\ }\href
  {https://doi.org/10.1103/physrevlett.113.190501} {\bibfield  {journal}
  {\bibinfo  {journal} {Physical Review Letters}\ }\textbf {\bibinfo {volume}
  {113}},\ \bibinfo {pages} {190501}},\ \Eprint
  {https://arxiv.org/abs/arXiv:1407.8012v1} {arXiv:1407.8012v1} \BibitemShut
  {NoStop}%
\bibitem [{\citenamefont {Tang}\ \emph {et~al.}(2013)\citenamefont {Tang},
  \citenamefont {Yin}, \citenamefont {Ma}, \citenamefont {Fung}, \citenamefont
  {Liu}, \citenamefont {Yong}, \citenamefont {Chen}, \citenamefont {Peng},
  \citenamefont {Chen},\ and\ \citenamefont {Pan}}]{bib:tang2013source}%
  \BibitemOpen
  \bibfield  {author} {\bibinfo {author} {\bibnamefont {Tang}, \bibfnamefont
  {Yan-Lin}}, \bibinfo {author} {\bibfnamefont {Hua-Lei}\ \bibnamefont {Yin}},
  \bibinfo {author} {\bibfnamefont {Xiongfeng}\ \bibnamefont {Ma}}, \bibinfo
  {author} {\bibfnamefont {Chi-Hang~Fred}\ \bibnamefont {Fung}}, \bibinfo
  {author} {\bibfnamefont {Yang}\ \bibnamefont {Liu}}, \bibinfo {author}
  {\bibfnamefont {Hai-Lin}\ \bibnamefont {Yong}}, \bibinfo {author}
  {\bibfnamefont {Teng-Yun}\ \bibnamefont {Chen}}, \bibinfo {author}
  {\bibfnamefont {Cheng-Zhi}\ \bibnamefont {Peng}}, \bibinfo {author}
  {\bibfnamefont {Zeng-Bing}\ \bibnamefont {Chen}}, and\ \bibinfo {author}
  {\bibfnamefont {Jian-Wei}\ \bibnamefont {Pan}}} (\bibinfo {year} {2013}),\
  \bibfield  {title} {\enquote {\bibinfo {title} {Source attack of decoy-state
  quantum key distribution using phase information},}\ }\href
  {https://doi.org/10.1103/physreva.88.022308} {\bibfield  {journal} {\bibinfo
  {journal} {Physical Review A}\ }\textbf {\bibinfo {volume} {88}},\ \bibinfo
  {pages} {022308}},\ \Eprint {https://arxiv.org/abs/arXiv:1304.2541v2}
  {arXiv:1304.2541v2} \BibitemShut {NoStop}%
\bibitem [{\citenamefont {Tang}\ \emph
  {et~al.}(2014{\natexlab{b}})\citenamefont {Tang}, \citenamefont {Liao},
  \citenamefont {Xu}, \citenamefont {Qi}, \citenamefont {Qian},\ and\
  \citenamefont {Lo}}]{bib:PRL_112_190503}%
  \BibitemOpen
  \bibfield  {author} {\bibinfo {author} {\bibnamefont {Tang}, \bibfnamefont
  {Zhiyuan}}, \bibinfo {author} {\bibfnamefont {Zhongfa}\ \bibnamefont {Liao}},
  \bibinfo {author} {\bibfnamefont {Feihu}\ \bibnamefont {Xu}}, \bibinfo
  {author} {\bibfnamefont {Bing}\ \bibnamefont {Qi}}, \bibinfo {author}
  {\bibfnamefont {Li}~\bibnamefont {Qian}}, and\ \bibinfo {author}
  {\bibfnamefont {Hoi-Kwong}\ \bibnamefont {Lo}}} (\bibinfo {year}
  {2014}{\natexlab{b}}),\ \bibfield  {title} {\enquote {\bibinfo {title}
  {Experimental demonstration of polarization encoding
  measurement-device-independent quantum key distribution},}\ }\href
  {https://doi.org/10.1103/physrevlett.112.190503} {\bibfield  {journal}
  {\bibinfo  {journal} {Physical Review Letters}\ }\textbf {\bibinfo {volume}
  {112}},\ \bibinfo {pages} {190503}},\ \Eprint
  {https://arxiv.org/abs/arXiv:1306.6134v2} {arXiv:1306.6134v2} \BibitemShut
  {NoStop}%
\bibitem [{\citenamefont {Tang}\ \emph
  {et~al.}(2016{\natexlab{a}})\citenamefont {Tang}, \citenamefont
  {Chandrasekara}, \citenamefont {Tan}, \citenamefont {Cheng}, \citenamefont
  {Sha}, \citenamefont {Hiang}, \citenamefont {Oi},\ and\ \citenamefont
  {Ling}}]{SD-Tang:2016aa}%
  \BibitemOpen
  \bibfield  {author} {\bibinfo {author} {\bibnamefont {Tang}, \bibfnamefont
  {Zhongkan}}, \bibinfo {author} {\bibfnamefont {Rakhitha}\ \bibnamefont
  {Chandrasekara}}, \bibinfo {author} {\bibfnamefont {Yue~Chuan}\ \bibnamefont
  {Tan}}, \bibinfo {author} {\bibfnamefont {Cliff}\ \bibnamefont {Cheng}},
  \bibinfo {author} {\bibfnamefont {Luo}\ \bibnamefont {Sha}}, \bibinfo
  {author} {\bibfnamefont {Goh~Cher}\ \bibnamefont {Hiang}}, \bibinfo {author}
  {\bibfnamefont {Daniel K.~L.}\ \bibnamefont {Oi}}, and\ \bibinfo {author}
  {\bibfnamefont {Alexander}\ \bibnamefont {Ling}}} (\bibinfo {year}
  {2016}{\natexlab{a}}),\ \bibfield  {title} {\enquote {\bibinfo {title}
  {Generation and analysis of correlated pairs of photons aboard a
  nanosatellite},}\ }\href {https://doi.org/10.1103/PhysRevApplied.5.054022}
  {\bibfield  {journal} {\bibinfo  {journal} {Physical Review Applied}\
  }\textbf {\bibinfo {volume} {5}},\ \bibinfo {pages} {054022}},\ \Eprint
  {https://arxiv.org/abs/arXiv:1603.06659v1} {arXiv:1603.06659v1} \BibitemShut
  {NoStop}%
\bibitem [{\citenamefont {Tang}\ \emph
  {et~al.}(2016{\natexlab{b}})\citenamefont {Tang}, \citenamefont
  {Chandrasekara}, \citenamefont {Tan}, \citenamefont {Cheng}, \citenamefont
  {Sha}, \citenamefont {Hiang}, \citenamefont {Oi},\ and\ \citenamefont
  {Ling}}]{bib:tang2016generation}%
  \BibitemOpen
  \bibfield  {author} {\bibinfo {author} {\bibnamefont {Tang}, \bibfnamefont
  {Zhongkan}}, \bibinfo {author} {\bibfnamefont {Rakhitha}\ \bibnamefont
  {Chandrasekara}}, \bibinfo {author} {\bibfnamefont {Yue~Chuan}\ \bibnamefont
  {Tan}}, \bibinfo {author} {\bibfnamefont {Cliff}\ \bibnamefont {Cheng}},
  \bibinfo {author} {\bibfnamefont {Luo}\ \bibnamefont {Sha}}, \bibinfo
  {author} {\bibfnamefont {Goh~Cher}\ \bibnamefont {Hiang}}, \bibinfo {author}
  {\bibfnamefont {Daniel~KL}\ \bibnamefont {Oi}}, and\ \bibinfo {author}
  {\bibfnamefont {Alexander}\ \bibnamefont {Ling}}} (\bibinfo {year}
  {2016}{\natexlab{b}}),\ \bibfield  {title} {\enquote {\bibinfo {title}
  {Generation and analysis of correlated pairs of photons aboard a
  nanosatellite},}\ }\href {https://doi.org/10.1103/physrevapplied.5.054022}
  {\bibfield  {journal} {\bibinfo  {journal} {Physical Review Applied}\
  }\textbf {\bibinfo {volume} {5}},\ \bibinfo {pages} {054022}},\ \Eprint
  {https://arxiv.org/abs/arXiv:1603.06659v1} {arXiv:1603.06659v1} \BibitemShut
  {NoStop}%
\bibitem [{\citenamefont {Tavakoli}\ \emph {et~al.}(2015)\citenamefont
  {Tavakoli}, \citenamefont {Cabello}, \citenamefont {{\.Z}ukowski},\ and\
  \citenamefont {Bourennane}}]{bib:tavakoli2015quantum}%
  \BibitemOpen
  \bibfield  {author} {\bibinfo {author} {\bibnamefont {Tavakoli},
  \bibfnamefont {Armin}}, \bibinfo {author} {\bibfnamefont {Ad{\'a}n}\
  \bibnamefont {Cabello}}, \bibinfo {author} {\bibfnamefont {Marek}\
  \bibnamefont {{\.Z}ukowski}}, and\ \bibinfo {author} {\bibfnamefont
  {Mohamed}\ \bibnamefont {Bourennane}}} (\bibinfo {year} {2015}),\ \bibfield
  {title} {\enquote {\bibinfo {title} {Quantum clock synchronization with a
  single qudit},}\ }\href {https://doi.org/10.1038/srep07982} {\bibfield
  {journal} {\bibinfo  {journal} {Scientific Reports}\ }\textbf {\bibinfo
  {volume} {5}},\ \bibinfo {pages} {7982}}\BibitemShut {NoStop}%
\bibitem [{\citenamefont {Terhal}(2015)}]{SD-Terhal:2015aa}%
  \BibitemOpen
  \bibfield  {author} {\bibinfo {author} {\bibnamefont {Terhal}, \bibfnamefont
  {Barbara~M}}} (\bibinfo {year} {2015}),\ \bibfield  {title} {\enquote
  {\bibinfo {title} {Quantum error correction for quantum memories},}\ }\href
  {https://doi.org/10.1103/RevModPhys.87.307} {\bibfield  {journal} {\bibinfo
  {journal} {Reviews of Modern Physics}\ }\textbf {\bibinfo {volume} {87}},\
  \bibinfo {pages} {307}},\ \Eprint {https://arxiv.org/abs/arXiv:1302.3428v7}
  {arXiv:1302.3428v7} \BibitemShut {NoStop}%
\bibitem [{\citenamefont {Thottan}\ and\ \citenamefont
  {Ji}(2003)}]{bib:thottan2003anomaly}%
  \BibitemOpen
  \bibfield  {author} {\bibinfo {author} {\bibnamefont {Thottan}, \bibfnamefont
  {Marina}}, and\ \bibinfo {author} {\bibfnamefont {Chuanyi}\ \bibnamefont
  {Ji}}} (\bibinfo {year} {2003}),\ \bibfield  {title} {\enquote {\bibinfo
  {title} {Anomaly detection in ip networks},}\ }\href
  {https://doi.org/10.1109/tsp.2003.814797} {\bibfield  {journal} {\bibinfo
  {journal} {IEEE Transactions on signal processing}\ }\textbf {\bibinfo
  {volume} {51}},\ \bibinfo {pages} {2191}}\BibitemShut {NoStop}%
\bibitem [{\citenamefont {Tillmann}\ \emph {et~al.}(2013)\citenamefont
  {Tillmann}, \citenamefont {Daki}, \citenamefont {Heilmann}, \citenamefont
  {Nolte}, \citenamefont {Szameit},\ and\ \citenamefont
  {Walther}}]{bib:Tillmann4}%
  \BibitemOpen
  \bibfield  {author} {\bibinfo {author} {\bibnamefont {Tillmann},
  \bibfnamefont {Max}}, \bibinfo {author} {\bibfnamefont {Borivoje}\
  \bibnamefont {Daki}}, \bibinfo {author} {\bibfnamefont {Ren{' e}}\
  \bibnamefont {Heilmann}}, \bibinfo {author} {\bibfnamefont {Stefan}\
  \bibnamefont {Nolte}}, \bibinfo {author} {\bibfnamefont {Alexander}\
  \bibnamefont {Szameit}}, and\ \bibinfo {author} {\bibfnamefont {Philip}\
  \bibnamefont {Walther}}} (\bibinfo {year} {2013}),\ \bibfield  {title}
  {\enquote {\bibinfo {title} {Experimental boson sampling},}\ }\href
  {https://doi.org/10.1038/nphoton.2013.102} {\bibfield  {journal} {\bibinfo
  {journal} {Nature Photonics}\ }\textbf {\bibinfo {volume} {7}},\ \bibinfo
  {pages} {540}},\ \Eprint {https://arxiv.org/abs/arXiv:1212.2240v1}
  {arXiv:1212.2240v1} \BibitemShut {NoStop}%
\bibitem [{\citenamefont {Togan}\ \emph {et~al.}(2010)\citenamefont {Togan},
  \citenamefont {Chu}, \citenamefont {Trifonov}, \citenamefont {Jiang},
  \citenamefont {Maze}, \citenamefont {Childress}, \citenamefont {Dutt},
  \citenamefont {S{\o}rensen}, \citenamefont {Hemmer}, \citenamefont {Zibrov}
  \emph {et~al.}}]{bib:togan2010quantum}%
  \BibitemOpen
  \bibfield  {author} {\bibinfo {author} {\bibnamefont {Togan}, \bibfnamefont
  {Emre}}, \bibinfo {author} {\bibfnamefont {Yiwen}\ \bibnamefont {Chu}},
  \bibinfo {author} {\bibfnamefont {AS}~\bibnamefont {Trifonov}}, \bibinfo
  {author} {\bibfnamefont {Liang}\ \bibnamefont {Jiang}}, \bibinfo {author}
  {\bibfnamefont {Jeronimo}\ \bibnamefont {Maze}}, \bibinfo {author}
  {\bibfnamefont {Lilian}\ \bibnamefont {Childress}}, \bibinfo {author}
  {\bibfnamefont {MV~Gurudev}\ \bibnamefont {Dutt}}, \bibinfo {author}
  {\bibfnamefont {Anders~S{\o}ndberg}\ \bibnamefont {S{\o}rensen}}, \bibinfo
  {author} {\bibfnamefont {PR}~\bibnamefont {Hemmer}}, \bibinfo {author}
  {\bibfnamefont {AS}~\bibnamefont {Zibrov}},  \emph {et~al.}} (\bibinfo {year}
  {2010}),\ \bibfield  {title} {\enquote {\bibinfo {title} {Quantum
  entanglement between an optical photon and a solid-state spin qubit},}\
  }\href {https://doi.org/10.1364/fio.2011.fthl4} {\bibfield  {journal}
  {\bibinfo  {journal} {Nature}\ }\textbf {\bibinfo {volume} {466}},\ \bibinfo
  {pages} {730}}\BibitemShut {NoStop}%
\bibitem [{\citenamefont {Torlai}\ \emph {et~al.}(2018)\citenamefont {Torlai},
  \citenamefont {Mazzola}, \citenamefont {Carrasquilla}, \citenamefont
  {Troyer}, \citenamefont {Melko},\ and\ \citenamefont
  {Carleo}}]{bib:Torlai2017}%
  \BibitemOpen
  \bibfield  {author} {\bibinfo {author} {\bibnamefont {Torlai}, \bibfnamefont
  {Giacomo}}, \bibinfo {author} {\bibfnamefont {Guglielmo}\ \bibnamefont
  {Mazzola}}, \bibinfo {author} {\bibfnamefont {Juan}\ \bibnamefont
  {Carrasquilla}}, \bibinfo {author} {\bibfnamefont {Matthias}\ \bibnamefont
  {Troyer}}, \bibinfo {author} {\bibfnamefont {Roger}\ \bibnamefont {Melko}},
  and\ \bibinfo {author} {\bibfnamefont {Giuseppe}\ \bibnamefont {Carleo}}}
  (\bibinfo {year} {2018}),\ \bibfield  {title} {\enquote {\bibinfo {title}
  {Neural-network quantum state tomography},}\ }\href
  {https://doi.org/10.1038/s41567-018-0048-5} {\bibfield  {journal} {\bibinfo
  {journal} {Nature Physics}\ }\textbf {\bibinfo {volume} {14}},\ \bibinfo
  {pages} {447}}\BibitemShut {NoStop}%
\bibitem [{\citenamefont {Tosi}\ \emph {et~al.}(2017)\citenamefont {Tosi},
  \citenamefont {Mohiyaddin}, \citenamefont {Schmitt}, \citenamefont {Tenberg},
  \citenamefont {Rahman}, \citenamefont {Klimeck},\ and\ \citenamefont
  {Morello}}]{SD-Tosi:2017aa}%
  \BibitemOpen
  \bibfield  {author} {\bibinfo {author} {\bibnamefont {Tosi}, \bibfnamefont
  {Guilherme}}, \bibinfo {author} {\bibfnamefont {Fahd~A.}\ \bibnamefont
  {Mohiyaddin}}, \bibinfo {author} {\bibfnamefont {Vivien}\ \bibnamefont
  {Schmitt}}, \bibinfo {author} {\bibfnamefont {Stefanie}\ \bibnamefont
  {Tenberg}}, \bibinfo {author} {\bibfnamefont {Rajib}\ \bibnamefont {Rahman}},
  \bibinfo {author} {\bibfnamefont {Gerhard}\ \bibnamefont {Klimeck}}, and\
  \bibinfo {author} {\bibfnamefont {Andrea}\ \bibnamefont {Morello}}} (\bibinfo
  {year} {2017}),\ \bibfield  {title} {\enquote {\bibinfo {title} {Silicon
  quantum processor with robust long-distance qubit couplings},}\ }\href
  {https://doi.org/10.1038/s41467-017-00378-x} {\bibfield  {journal} {\bibinfo
  {journal} {Nature Communications}\ }\textbf {\bibinfo {volume} {8}},\
  \bibinfo {pages} {450}}\BibitemShut {NoStop}%
\bibitem [{\citenamefont {Townsend}\ \emph {et~al.}(1993)\citenamefont
  {Townsend}, \citenamefont {Rarity},\ and\ \citenamefont
  {Tapster}}]{bib:EL_29_634}%
  \BibitemOpen
  \bibfield  {author} {\bibinfo {author} {\bibnamefont {Townsend},
  \bibfnamefont {Paul~D}}, \bibinfo {author} {\bibfnamefont {JG}~\bibnamefont
  {Rarity}}, and\ \bibinfo {author} {\bibfnamefont {PR}~\bibnamefont
  {Tapster}}} (\bibinfo {year} {1993}),\ \bibfield  {title} {\enquote {\bibinfo
  {title} {Single photon interference in 10 km long optical fibre
  interferometer},}\ }\href {https://doi.org/10.1049/el:19930424} {\bibfield
  {journal} {\bibinfo  {journal} {Electronics Letters}\ }\textbf {\bibinfo
  {volume} {29}},\ \bibinfo {pages} {634}}\BibitemShut {NoStop}%
\bibitem [{\citenamefont {Toyoshima}\ \emph {et~al.}(2015)\citenamefont
  {Toyoshima}, \citenamefont {Fuse}, \citenamefont {Kolev}, \citenamefont
  {Takenaka}, \citenamefont {Munemasa}, \citenamefont {Iwakiri}, \citenamefont
  {Suzuki}, \citenamefont {Koyama}, \citenamefont {Kubooka}, \citenamefont
  {Akioka} \emph {et~al.}}]{bib:toyoshima2015current}%
  \BibitemOpen
  \bibfield  {author} {\bibinfo {author} {\bibnamefont {Toyoshima},
  \bibfnamefont {Morio}}, \bibinfo {author} {\bibfnamefont {Tetsuharu}\
  \bibnamefont {Fuse}}, \bibinfo {author} {\bibfnamefont {Dimitar~R}\
  \bibnamefont {Kolev}}, \bibinfo {author} {\bibfnamefont {Hideki}\
  \bibnamefont {Takenaka}}, \bibinfo {author} {\bibfnamefont {Yasushi}\
  \bibnamefont {Munemasa}}, \bibinfo {author} {\bibfnamefont {Naohiko}\
  \bibnamefont {Iwakiri}}, \bibinfo {author} {\bibfnamefont {Kenji}\
  \bibnamefont {Suzuki}}, \bibinfo {author} {\bibfnamefont {Yoshisada}\
  \bibnamefont {Koyama}}, \bibinfo {author} {\bibfnamefont {Toshihiro}\
  \bibnamefont {Kubooka}}, \bibinfo {author} {\bibfnamefont {Maki}\
  \bibnamefont {Akioka}},  \emph {et~al.}} (\bibinfo {year} {2015}),\ \bibfield
   {title} {\enquote {\bibinfo {title} {Current status of research and
  development on space laser communications technologies and future plans in
  nict},}\ }in\ \href {https://doi.org/10.1109/icsos.2015.7425055} {\emph
  {\bibinfo {booktitle} {Space Optical Systems and Applications (ICSOS), 2015
  IEEE International Conference on}}},\ p.~\bibinfo {pages} {1}\BibitemShut
  {NoStop}%
\bibitem [{\citenamefont {Travaglione}\ and\ \citenamefont
  {Milburn}(2002)}]{traviglione2002GKPprep}%
  \BibitemOpen
  \bibfield  {author} {\bibinfo {author} {\bibnamefont {Travaglione},
  \bibfnamefont {B~C}}, and\ \bibinfo {author} {\bibfnamefont {G.~J.}\
  \bibnamefont {Milburn}}} (\bibinfo {year} {2002}),\ \bibfield  {title}
  {\enquote {\bibinfo {title} {Preparing encoded states in an oscillator},}\
  }\href {https://doi.org/10.1103/PhysRevA.66.052322} {\bibfield  {journal}
  {\bibinfo  {journal} {Phys. Rev. A}\ }\textbf {\bibinfo {volume} {66}},\
  \bibinfo {pages} {052322}}\BibitemShut {NoStop}%
\bibitem [{\citenamefont {Tuckett}\ \emph {et~al.}(2018)\citenamefont
  {Tuckett}, \citenamefont {Bartlett},\ and\ \citenamefont
  {Flammia}}]{SD-Tuckett:2018aa}%
  \BibitemOpen
  \bibfield  {author} {\bibinfo {author} {\bibnamefont {Tuckett}, \bibfnamefont
  {David~K}}, \bibinfo {author} {\bibfnamefont {Stephen~D.}\ \bibnamefont
  {Bartlett}}, and\ \bibinfo {author} {\bibfnamefont {Steven~T.}\ \bibnamefont
  {Flammia}}} (\bibinfo {year} {2018}),\ \bibfield  {title} {\enquote {\bibinfo
  {title} {Ultrahigh error threshold for surface codes with biased noise},}\
  }\href {https://doi.org/10.1103/PhysRevLett.120.050505} {\bibfield  {journal}
  {\bibinfo  {journal} {Physical Review Letters}\ }\textbf {\bibinfo {volume}
  {120}},\ \bibinfo {pages} {050505}},\ \Eprint
  {https://arxiv.org/abs/arXiv:1708.08474v3} {arXiv:1708.08474v3} \BibitemShut
  {NoStop}%
\bibitem [{\citenamefont {ucsusa.org}(2024)}]{sneakernet_web_1}%
  \BibitemOpen
  \bibfield  {author} {\bibinfo {author} {\bibnamefont {ucsusa.org},}}
  (\bibinfo {year} {2024}),\ \href
  {https://doi.org/https://www.ucsusa.org/resources/satellite-database} {\
  https://www.ucsusa.org/resources/satellite-database}\BibitemShut {NoStop}%
\bibitem [{\citenamefont {U'Ren}\ \emph {et~al.}(2003)\citenamefont {U'Ren},
  \citenamefont {Banaszek},\ and\ \citenamefont {Walmsley}}]{bib:URen03}%
  \BibitemOpen
  \bibfield  {author} {\bibinfo {author} {\bibnamefont {U'Ren}, \bibfnamefont
  {A~B}}, \bibinfo {author} {\bibfnamefont {K.}~\bibnamefont {Banaszek}}, and\
  \bibinfo {author} {\bibfnamefont {I.~A.}\ \bibnamefont {Walmsley}}} (\bibinfo
  {year} {2003}),\ \bibfield  {title} {\enquote {\bibinfo {title} {Photon
  engineering for quantum information processing},}\ }\href@noop {} {\bibfield
  {journal} {\bibinfo  {journal} {Quantum Information \& Computation}\ }\textbf
  {\bibinfo {volume} {3}},\ \bibinfo {pages} {480}},\ \Eprint
  {https://arxiv.org/abs/arXiv:quant-ph/0305192v1} {arXiv:quant-ph/0305192v1}
  \BibitemShut {NoStop}%
\bibitem [{\citenamefont {U'Ren}\ \emph {et~al.}(2005)\citenamefont {U'Ren},
  \citenamefont {Silberhorn}, \citenamefont {Banaszek}, \citenamefont
  {Walmsley}, \citenamefont {Erdman}, \citenamefont {Grice},\ and\
  \citenamefont {Raymer}}]{bib:URen05}%
  \BibitemOpen
  \bibfield  {author} {\bibinfo {author} {\bibnamefont {U'Ren}, \bibfnamefont
  {A~B}}, \bibinfo {author} {\bibfnamefont {C.}~\bibnamefont {Silberhorn}},
  \bibinfo {author} {\bibfnamefont {K.}~\bibnamefont {Banaszek}}, \bibinfo
  {author} {\bibfnamefont {I.~A.}\ \bibnamefont {Walmsley}}, \bibinfo {author}
  {\bibfnamefont {R.}~\bibnamefont {Erdman}}, \bibinfo {author} {\bibfnamefont
  {W.~P.}\ \bibnamefont {Grice}}, and\ \bibinfo {author} {\bibfnamefont
  {M.~G.}\ \bibnamefont {Raymer}}} (\bibinfo {year} {2005}),\ \bibfield
  {title} {\enquote {\bibinfo {title} {Generation of pure-state single-photon
  wavepackets by conditional preparation based on spontaneous parametric
  downconversion},}\ }\href@noop {} {\bibfield  {journal} {\bibinfo  {journal}
  {Laser Physics}\ }\textbf {\bibinfo {volume} {15}},\ \bibinfo {pages}
  {146}},\ \Eprint {https://arxiv.org/abs/arXiv:quant-ph/0611019v1}
  {arXiv:quant-ph/0611019v1} \BibitemShut {NoStop}%
\bibitem [{\citenamefont {Ursin}\ \emph
  {et~al.}(2007{\natexlab{a}})\citenamefont {Ursin}, \citenamefont
  {Tiefenbacher}, \citenamefont {Schmitt-Manderbach}, \citenamefont {Weier},
  \citenamefont {Scheidl}, \citenamefont {Lindenthal}, \citenamefont
  {Blauensteiner}, \citenamefont {Jennewein}, \citenamefont {Perdigues},
  \citenamefont {Trojek}, \citenamefont {{\"O}mer}, \citenamefont {F{\"u}rst},
  \citenamefont {Meyenburg}, \citenamefont {Rarity}, \citenamefont {Sodnik},
  \citenamefont {Barbieri}, \citenamefont {Weinfurter},\ and\ \citenamefont
  {Zeilinger}}]{SD-Ursin:2007aa}%
  \BibitemOpen
  \bibfield  {author} {\bibinfo {author} {\bibnamefont {Ursin}, \bibfnamefont
  {R}}, \bibinfo {author} {\bibfnamefont {F.}~\bibnamefont {Tiefenbacher}},
  \bibinfo {author} {\bibfnamefont {T.}~\bibnamefont {Schmitt-Manderbach}},
  \bibinfo {author} {\bibfnamefont {H.}~\bibnamefont {Weier}}, \bibinfo
  {author} {\bibfnamefont {T.}~\bibnamefont {Scheidl}}, \bibinfo {author}
  {\bibfnamefont {M.}~\bibnamefont {Lindenthal}}, \bibinfo {author}
  {\bibfnamefont {B.}~\bibnamefont {Blauensteiner}}, \bibinfo {author}
  {\bibfnamefont {T.}~\bibnamefont {Jennewein}}, \bibinfo {author}
  {\bibfnamefont {J.}~\bibnamefont {Perdigues}}, \bibinfo {author}
  {\bibfnamefont {P.}~\bibnamefont {Trojek}}, \bibinfo {author} {\bibfnamefont
  {B.}~\bibnamefont {{\"O}mer}}, \bibinfo {author} {\bibfnamefont
  {M.}~\bibnamefont {F{\"u}rst}}, \bibinfo {author} {\bibfnamefont
  {M.}~\bibnamefont {Meyenburg}}, \bibinfo {author} {\bibfnamefont
  {J.}~\bibnamefont {Rarity}}, \bibinfo {author} {\bibfnamefont
  {Z.}~\bibnamefont {Sodnik}}, \bibinfo {author} {\bibfnamefont
  {C.}~\bibnamefont {Barbieri}}, \bibinfo {author} {\bibfnamefont
  {H.}~\bibnamefont {Weinfurter}}, and\ \bibinfo {author} {\bibfnamefont
  {A.}~\bibnamefont {Zeilinger}}} (\bibinfo {year} {2007}{\natexlab{a}}),\
  \bibfield  {title} {\enquote {\bibinfo {title} {Entanglement-based quantum
  communication over 144 km},}\ }\href {https://doi.org/10.1038/nphys629}
  {\bibfield  {journal} {\bibinfo  {journal} {Nature Physics}\ }\textbf
  {\bibinfo {volume} {3}},\ \bibinfo {pages} {481}}\BibitemShut {NoStop}%
\bibitem [{\citenamefont {Ursin}\ \emph
  {et~al.}(2007{\natexlab{b}})\citenamefont {Ursin}, \citenamefont
  {Tiefenbacher}, \citenamefont {Schmitt-Manderbach}, \citenamefont {Weier},
  \citenamefont {Scheidl}, \citenamefont {Lindenthal}, \citenamefont
  {Blauensteiner}, \citenamefont {Jennewein}, \citenamefont {Perdigues},
  \citenamefont {Trojek} \emph {et~al.}}]{bib:NP_3_481}%
  \BibitemOpen
  \bibfield  {author} {\bibinfo {author} {\bibnamefont {Ursin}, \bibfnamefont
  {Rupert}}, \bibinfo {author} {\bibfnamefont {F}~\bibnamefont {Tiefenbacher}},
  \bibinfo {author} {\bibfnamefont {T}~\bibnamefont {Schmitt-Manderbach}},
  \bibinfo {author} {\bibfnamefont {H}~\bibnamefont {Weier}}, \bibinfo {author}
  {\bibfnamefont {Thomas}\ \bibnamefont {Scheidl}}, \bibinfo {author}
  {\bibfnamefont {M}~\bibnamefont {Lindenthal}}, \bibinfo {author}
  {\bibfnamefont {B}~\bibnamefont {Blauensteiner}}, \bibinfo {author}
  {\bibfnamefont {T}~\bibnamefont {Jennewein}}, \bibinfo {author}
  {\bibfnamefont {J}~\bibnamefont {Perdigues}}, \bibinfo {author}
  {\bibfnamefont {P}~\bibnamefont {Trojek}},  \emph {et~al.}} (\bibinfo {year}
  {2007}{\natexlab{b}}),\ \bibfield  {title} {\enquote {\bibinfo {title}
  {Entanglement-based quantum communication over 144 km},}\ }\href
  {https://doi.org/10.1038/nphys629} {\bibfield  {journal} {\bibinfo  {journal}
  {Nature Physics}\ }\textbf {\bibinfo {volume} {3}},\ \bibinfo {pages}
  {481}}\BibitemShut {NoStop}%
\bibitem [{\citenamefont {Vaccaro}\ \emph {et~al.}(2007)\citenamefont
  {Vaccaro}, \citenamefont {Spring},\ and\ \citenamefont
  {Chefles}}]{bib:VaccaroVoting}%
  \BibitemOpen
  \bibfield  {author} {\bibinfo {author} {\bibnamefont {Vaccaro}, \bibfnamefont
  {J~A}}, \bibinfo {author} {\bibfnamefont {Joseph}\ \bibnamefont {Spring}},
  and\ \bibinfo {author} {\bibfnamefont {Anthony}\ \bibnamefont {Chefles}}}
  (\bibinfo {year} {2007}),\ \bibfield  {title} {\enquote {\bibinfo {title}
  {Quantum protocols for anonymous voting and surveying},}\ }\href
  {https://doi.org/10.1103/physreva.75.012333} {\bibfield  {journal} {\bibinfo
  {journal} {Physical Review A}\ }\textbf {\bibinfo {volume} {75}},\ \bibinfo
  {pages} {012333}},\ \Eprint {https://arxiv.org/abs/arXiv:quant-ph/0504161v2}
  {arXiv:quant-ph/0504161v2} \BibitemShut {NoStop}%
\bibitem [{\citenamefont {Valahu}\ \emph {et~al.}(2024)\citenamefont {Valahu},
  \citenamefont {Stafford}, \citenamefont {Zixin~Huang}, \citenamefont
  {Chalermpusitarak}, \citenamefont {Menicucci},\ and\ \citenamefont
  {Joshua~Combes}}]{valahu2024sensing}%
  \BibitemOpen
  \bibfield  {author} {\bibinfo {author} {\bibnamefont {Valahu}, \bibfnamefont
  {Christophe~H}}, \bibinfo {author} {\bibfnamefont {Matthew~P.}\ \bibnamefont
  {Stafford}}, \bibinfo {author} {\bibfnamefont {Maverick J.~Millican}\
  \bibnamefont {Zixin~Huang}, \bibfnamefont {Vassili G.~Matsos}}, \bibinfo
  {author} {\bibfnamefont {Teerawat}\ \bibnamefont {Chalermpusitarak}},
  \bibinfo {author} {\bibfnamefont {Nicolas~C.}\ \bibnamefont {Menicucci}},
  and\ \bibinfo {author} {\bibfnamefont {Ting Rei~Tan}\ \bibnamefont
  {Joshua~Combes}, \bibfnamefont {Ben Q. Baragiola~and}}} (\bibinfo {year}
  {2024}),\ \href {https://arxiv.org/abs/2412.04865} {\enquote {\bibinfo
  {title} {Quantum-enhanced multi-parameter sensing in a single mode},}\
  }\BibitemShut {NoStop}%
\bibitem [{\citenamefont {Valencia}\ \emph {et~al.}(2004)\citenamefont
  {Valencia}, \citenamefont {Scarcelli},\ and\ \citenamefont
  {Shih}}]{bib:valencia2004distant}%
  \BibitemOpen
  \bibfield  {author} {\bibinfo {author} {\bibnamefont {Valencia},
  \bibfnamefont {Alejandra}}, \bibinfo {author} {\bibfnamefont {Giuliano}\
  \bibnamefont {Scarcelli}}, and\ \bibinfo {author} {\bibfnamefont {Yanhua}\
  \bibnamefont {Shih}}} (\bibinfo {year} {2004}),\ \bibfield  {title} {\enquote
  {\bibinfo {title} {Distant clock synchronization using entangled photon
  pairs},}\ }\href {https://doi.org/10.1063/1.1797561} {\bibfield  {journal}
  {\bibinfo  {journal} {Applied Physics Letters}\ }\textbf {\bibinfo {volume}
  {85}},\ \bibinfo {pages} {2655}},\ \Eprint
  {https://arxiv.org/abs/arXiv:quant-ph/0407204v1} {arXiv:quant-ph/0407204v1}
  \BibitemShut {NoStop}%
\bibitem [{\citenamefont {Vallone}\ \emph {et~al.}(2015)\citenamefont
  {Vallone}, \citenamefont {Bacco}, \citenamefont {Dequal}, \citenamefont
  {Gaiarin}, \citenamefont {Luceri}, \citenamefont {Bianco},\ and\
  \citenamefont {Villoresi}}]{bib:vallone15}%
  \BibitemOpen
  \bibfield  {author} {\bibinfo {author} {\bibnamefont {Vallone}, \bibfnamefont
  {Giuseppe}}, \bibinfo {author} {\bibfnamefont {Davide}\ \bibnamefont
  {Bacco}}, \bibinfo {author} {\bibfnamefont {Daniele}\ \bibnamefont {Dequal}},
  \bibinfo {author} {\bibfnamefont {Simone}\ \bibnamefont {Gaiarin}}, \bibinfo
  {author} {\bibfnamefont {Vincenza}\ \bibnamefont {Luceri}}, \bibinfo {author}
  {\bibfnamefont {Giuseppe}\ \bibnamefont {Bianco}}, and\ \bibinfo {author}
  {\bibfnamefont {Paolo}\ \bibnamefont {Villoresi}}} (\bibinfo {year} {2015}),\
  \bibfield  {title} {\enquote {\bibinfo {title} {Experimental satellite
  quantum communications},}\ }\href
  {https://doi.org/10.1103/physrevlett.115.040502} {\bibfield  {journal}
  {\bibinfo  {journal} {Physical Review Letters}\ }\textbf {\bibinfo {volume}
  {115}},\ \bibinfo {pages} {040502}},\ \Eprint
  {https://arxiv.org/abs/arXiv:1406.4051v1} {arXiv:1406.4051v1} \BibitemShut
  {NoStop}%
\bibitem [{\citenamefont {Van Der~Wal}\ \emph {et~al.}(2000)\citenamefont {Van
  Der~Wal}, \citenamefont {Ter~Haar}, \citenamefont {Wilhelm}, \citenamefont
  {Schouten}, \citenamefont {Harmans}, \citenamefont {Orlando}, \citenamefont
  {Lloyd},\ and\ \citenamefont {Mooij}}]{bib:van2000quantum}%
  \BibitemOpen
  \bibfield  {author} {\bibinfo {author} {\bibnamefont {Van Der~Wal},
  \bibfnamefont {Caspar~H}}, \bibinfo {author} {\bibfnamefont {ACJ}\
  \bibnamefont {Ter~Haar}}, \bibinfo {author} {\bibfnamefont {FK}~\bibnamefont
  {Wilhelm}}, \bibinfo {author} {\bibfnamefont {RN}~\bibnamefont {Schouten}},
  \bibinfo {author} {\bibfnamefont {CJPM}\ \bibnamefont {Harmans}}, \bibinfo
  {author} {\bibfnamefont {TP}~\bibnamefont {Orlando}}, \bibinfo {author}
  {\bibfnamefont {Seth}\ \bibnamefont {Lloyd}}, and\ \bibinfo {author}
  {\bibfnamefont {JE}~\bibnamefont {Mooij}}} (\bibinfo {year} {2000}),\
  \bibfield  {title} {\enquote {\bibinfo {title} {Quantum superposition of
  macroscopic persistent-current states},}\ }\href
  {https://doi.org/10.1126/science.290.5492.773} {\bibfield  {journal}
  {\bibinfo  {journal} {Science}\ }\textbf {\bibinfo {volume} {290}},\ \bibinfo
  {pages} {773}}\BibitemShut {NoStop}%
\bibitem [{\citenamefont {Van~Dijk}\ \emph {et~al.}(2010)\citenamefont
  {Van~Dijk}, \citenamefont {Gentry}, \citenamefont {Halevi},\ and\
  \citenamefont {Vaikuntanathan}}]{bib:van2010fully}%
  \BibitemOpen
  \bibfield  {author} {\bibinfo {author} {\bibnamefont {Van~Dijk},
  \bibfnamefont {M}}, \bibinfo {author} {\bibfnamefont {C.}~\bibnamefont
  {Gentry}}, \bibinfo {author} {\bibfnamefont {S.}~\bibnamefont {Halevi}}, and\
  \bibinfo {author} {\bibfnamefont {V.}~\bibnamefont {Vaikuntanathan}}}
  (\bibinfo {year} {2010}),\ \bibfield  {title} {\enquote {\bibinfo {title}
  {Fully homomorphic encryption over the integers},}\ }\href
  {https://doi.org/10.1007/978-3-642-13190-5_2} {\bibinfo  {journal} {Advances
  in Cryptology -- EUROCRYPT}\ ,\ \bibinfo {pages} {24}}\BibitemShut {NoStop}%
\bibitem [{\citenamefont {Van~Meter}(2014)}]{bib:van2014quantum}%
  \BibitemOpen
\bibfield  {journal} {  }\bibfield  {author} {\bibinfo {author} {\bibnamefont
  {Van~Meter}, \bibfnamefont {R}}} (\bibinfo {year} {2014}),\ \href
  {https://doi.org/10.1002/9781118648919} {\emph {\bibinfo {title} {Quantum
  Networking}}}\ (\bibinfo  {publisher} {Wiley})\BibitemShut {NoStop}%
\bibitem [{\citenamefont {Varnava}\ \emph {et~al.}(2006)\citenamefont
  {Varnava}, \citenamefont {Browne},\ and\ \citenamefont
  {Rudolph}}]{bib:Varnava05}%
  \BibitemOpen
  \bibfield  {author} {\bibinfo {author} {\bibnamefont {Varnava}, \bibfnamefont
  {Michael}}, \bibinfo {author} {\bibfnamefont {Daniel~E.}\ \bibnamefont
  {Browne}}, and\ \bibinfo {author} {\bibfnamefont {Terry}\ \bibnamefont
  {Rudolph}}} (\bibinfo {year} {2006}),\ \bibfield  {title} {\enquote {\bibinfo
  {title} {Loss tolerant one-way quantum computation - a horticultural
  approach},}\ }\href@noop {} {\bibfield  {journal} {\bibinfo  {journal}
  {Physical Review Letters}\ }\textbf {\bibinfo {volume} {97}},\ \bibinfo
  {pages} {120501}}\BibitemShut {NoStop}%
\bibitem [{\citenamefont {Vasconcelos}\ \emph {et~al.}(2010)\citenamefont
  {Vasconcelos}, \citenamefont {Sanz},\ and\ \citenamefont
  {Glancy}}]{glancy2010breeding}%
  \BibitemOpen
  \bibfield  {author} {\bibinfo {author} {\bibnamefont {Vasconcelos},
  \bibfnamefont {H~M}}, \bibinfo {author} {\bibfnamefont {L.}~\bibnamefont
  {Sanz}}, and\ \bibinfo {author} {\bibfnamefont {S.}~\bibnamefont {Glancy}}}
  (\bibinfo {year} {2010}),\ \bibfield  {title} {\enquote {\bibinfo {title}
  {All-optical generation of states for ``encoding a qubit in an
  oscillator''},}\ }\href {https://doi.org/10.1364/OL.35.003261} {\bibfield
  {journal} {\bibinfo  {journal} {Opt. Lett.}\ }\textbf {\bibinfo {volume}
  {35}}~(\bibinfo {number} {19}),\ \bibinfo {pages} {3261--3263}}\BibitemShut
  {NoStop}%
\bibitem [{\citenamefont {Veldhorst}\ \emph {et~al.}(2014)\citenamefont
  {Veldhorst}, \citenamefont {Hwang}, \citenamefont {Yang}, \citenamefont
  {Leenstra}, \citenamefont {De~Ronde}, \citenamefont {Dehollain},
  \citenamefont {Muhonen}, \citenamefont {Hudson}, \citenamefont {Itoh},
  \citenamefont {Morello} \emph {et~al.}}]{bib:veldhorst2014addressable}%
  \BibitemOpen
  \bibfield  {author} {\bibinfo {author} {\bibnamefont {Veldhorst},
  \bibfnamefont {M}}, \bibinfo {author} {\bibfnamefont {JCC}\ \bibnamefont
  {Hwang}}, \bibinfo {author} {\bibfnamefont {CH}~\bibnamefont {Yang}},
  \bibinfo {author} {\bibfnamefont {AW}~\bibnamefont {Leenstra}}, \bibinfo
  {author} {\bibfnamefont {B}~\bibnamefont {De~Ronde}}, \bibinfo {author}
  {\bibfnamefont {JP}~\bibnamefont {Dehollain}}, \bibinfo {author}
  {\bibfnamefont {JT}~\bibnamefont {Muhonen}}, \bibinfo {author} {\bibfnamefont
  {FE}~\bibnamefont {Hudson}}, \bibinfo {author} {\bibfnamefont {Kohei~M}\
  \bibnamefont {Itoh}}, \bibinfo {author} {\bibfnamefont {A}~\bibnamefont
  {Morello}},  \emph {et~al.}} (\bibinfo {year} {2014}),\ \bibfield  {title}
  {\enquote {\bibinfo {title} {An addressable quantum dot qubit with
  fault-tolerant control-fidelity},}\ }\href
  {https://doi.org/10.1038/nnano.2014.216} {\bibfield  {journal} {\bibinfo
  {journal} {Nature Nanotechnology}\ }\textbf {\bibinfo {volume} {9}},\
  \bibinfo {pages} {981}},\ \Eprint {https://arxiv.org/abs/arXiv:1407.1950v1}
  {arXiv:1407.1950v1} \BibitemShut {NoStop}%
\bibitem [{\citenamefont {Veldhorst}\ \emph {et~al.}(2015)\citenamefont
  {Veldhorst}, \citenamefont {Yang}, \citenamefont {Hwang}, \citenamefont
  {Huang}, \citenamefont {Dehollain}, \citenamefont {Muhonen}, \citenamefont
  {Simmons}, \citenamefont {Laucht}, \citenamefont {Hudson}, \citenamefont
  {Itoh} \emph {et~al.}}]{bib:veldhorst2015two}%
  \BibitemOpen
  \bibfield  {author} {\bibinfo {author} {\bibnamefont {Veldhorst},
  \bibfnamefont {M}}, \bibinfo {author} {\bibfnamefont {CH}~\bibnamefont
  {Yang}}, \bibinfo {author} {\bibfnamefont {JCC}\ \bibnamefont {Hwang}},
  \bibinfo {author} {\bibfnamefont {W}~\bibnamefont {Huang}}, \bibinfo {author}
  {\bibfnamefont {JP}~\bibnamefont {Dehollain}}, \bibinfo {author}
  {\bibfnamefont {JT}~\bibnamefont {Muhonen}}, \bibinfo {author} {\bibfnamefont
  {S}~\bibnamefont {Simmons}}, \bibinfo {author} {\bibfnamefont
  {A}~\bibnamefont {Laucht}}, \bibinfo {author} {\bibfnamefont
  {FE}~\bibnamefont {Hudson}}, \bibinfo {author} {\bibfnamefont {Kohei~M}\
  \bibnamefont {Itoh}},  \emph {et~al.}} (\bibinfo {year} {2015}),\ \bibfield
  {title} {\enquote {\bibinfo {title} {A two-qubit logic gate in silicon},}\
  }\href {https://doi.org/10.1038/nature15263} {\bibfield  {journal} {\bibinfo
  {journal} {Nature}\ }\textbf {\bibinfo {volume} {526}},\ \bibinfo {pages}
  {410}},\ \Eprint {https://arxiv.org/abs/arXiv:1411.5760v1}
  {arXiv:1411.5760v1} \BibitemShut {NoStop}%
\bibitem [{\citenamefont {Venegas-Andraca}(2012)}]{bib:Salvador12}%
  \BibitemOpen
  \bibfield  {author} {\bibinfo {author} {\bibnamefont {Venegas-Andraca},
  \bibfnamefont {Salvador~E}}} (\bibinfo {year} {2012}),\ \bibfield  {title}
  {\enquote {\bibinfo {title} {Quantum walks: a comprehensive review},}\ }\href
  {https://doi.org/10.1007/s11128-012-0432-5} {\bibfield  {journal} {\bibinfo
  {journal} {Quantum Information Processing}\ }\textbf {\bibinfo {volume}
  {11}},\ \bibinfo {pages} {1015}}\BibitemShut {NoStop}%
\bibitem [{\citenamefont {Villoresi}\ \emph {et~al.}(2008)\citenamefont
  {Villoresi}, \citenamefont {Jennewein}, \citenamefont {Tamburini},
  \citenamefont {Aspelmeyer}, \citenamefont {Bonato}, \citenamefont {Ursin},
  \citenamefont {Pernechele}, \citenamefont {Luceri}, \citenamefont {Bianco},
  \citenamefont {Zeilinger} \emph {et~al.}}]{bib:NJP_10_033038}%
  \BibitemOpen
  \bibfield  {author} {\bibinfo {author} {\bibnamefont {Villoresi},
  \bibfnamefont {Paolo}}, \bibinfo {author} {\bibfnamefont {Thomas}\
  \bibnamefont {Jennewein}}, \bibinfo {author} {\bibfnamefont {Fabrizio}\
  \bibnamefont {Tamburini}}, \bibinfo {author} {\bibfnamefont {Markus}\
  \bibnamefont {Aspelmeyer}}, \bibinfo {author} {\bibfnamefont {Cristian}\
  \bibnamefont {Bonato}}, \bibinfo {author} {\bibfnamefont {Rupert}\
  \bibnamefont {Ursin}}, \bibinfo {author} {\bibfnamefont {Claudio}\
  \bibnamefont {Pernechele}}, \bibinfo {author} {\bibfnamefont {Vincenza}\
  \bibnamefont {Luceri}}, \bibinfo {author} {\bibfnamefont {Giuseppe}\
  \bibnamefont {Bianco}}, \bibinfo {author} {\bibfnamefont {Anton}\
  \bibnamefont {Zeilinger}},  \emph {et~al.}} (\bibinfo {year} {2008}),\
  \bibfield  {title} {\enquote {\bibinfo {title} {Experimental verification of
  the feasibility of a quantum channel between space and earth},}\ }\href
  {https://doi.org/10.1088/1367-2630/10/3/033038} {\bibfield  {journal}
  {\bibinfo  {journal} {New Journal of Physics}\ }\textbf {\bibinfo {volume}
  {10}},\ \bibinfo {pages} {033038}}\BibitemShut {NoStop}%
\bibitem [{\citenamefont {Vinay}\ and\ \citenamefont
  {Kok}(2018)}]{bib:PhysRevA.97.042335}%
  \BibitemOpen
  \bibfield  {author} {\bibinfo {author} {\bibnamefont {Vinay}, \bibfnamefont
  {Scott~E}}, and\ \bibinfo {author} {\bibfnamefont {Pieter}\ \bibnamefont
  {Kok}}} (\bibinfo {year} {2018}),\ \bibfield  {title} {\enquote {\bibinfo
  {title} {Extended analysis of the trojan-horse attack in quantum key
  distribution},}\ }\href {https://doi.org/10.1103/PhysRevA.97.042335}
  {\bibfield  {journal} {\bibinfo  {journal} {Physical Review A}\ }\textbf
  {\bibinfo {volume} {97}},\ \bibinfo {pages} {042335}}\BibitemShut {NoStop}%
\bibitem [{\citenamefont {Vlastakis}\ \emph {et~al.}(2013)\citenamefont
  {Vlastakis}, \citenamefont {Kirchmair}, \citenamefont {Leghtas},
  \citenamefont {Nigg}, \citenamefont {Frunzio}, \citenamefont {Girvin},
  \citenamefont {Mirrahimi}, \citenamefont {Devoret},\ and\ \citenamefont
  {Schoelkopf}}]{bib:vlastakis2013deterministically}%
  \BibitemOpen
  \bibfield  {author} {\bibinfo {author} {\bibnamefont {Vlastakis},
  \bibfnamefont {Brian}}, \bibinfo {author} {\bibfnamefont {Gerhard}\
  \bibnamefont {Kirchmair}}, \bibinfo {author} {\bibfnamefont {Zaki}\
  \bibnamefont {Leghtas}}, \bibinfo {author} {\bibfnamefont {Simon~E}\
  \bibnamefont {Nigg}}, \bibinfo {author} {\bibfnamefont {Luigi}\ \bibnamefont
  {Frunzio}}, \bibinfo {author} {\bibfnamefont {Steven~M}\ \bibnamefont
  {Girvin}}, \bibinfo {author} {\bibfnamefont {Mazyar}\ \bibnamefont
  {Mirrahimi}}, \bibinfo {author} {\bibfnamefont {Michel~H}\ \bibnamefont
  {Devoret}}, and\ \bibinfo {author} {\bibfnamefont {Robert~J}\ \bibnamefont
  {Schoelkopf}}} (\bibinfo {year} {2013}),\ \bibfield  {title} {\enquote
  {\bibinfo {title} {Deterministically encoding quantum information using
  100-photon schr{\"o}dinger cat states},}\ }\href
  {https://doi.org/10.1126/science.1243289} {\bibfield  {journal} {\bibinfo
  {journal} {Science}\ }\textbf {\bibinfo {volume} {342}},\ \bibinfo {pages}
  {607}}\BibitemShut {NoStop}%
\bibitem [{\citenamefont {Volz}\ \emph {et~al.}(2006)\citenamefont {Volz},
  \citenamefont {Weber}, \citenamefont {Schlenk}, \citenamefont {Rosenfeld},
  \citenamefont {Vrana}, \citenamefont {Saucke}, \citenamefont {Kurtsiefer},\
  and\ \citenamefont {Weinfurter}}]{bib:PRL_96_030404}%
  \BibitemOpen
  \bibfield  {author} {\bibinfo {author} {\bibnamefont {Volz}, \bibfnamefont
  {J{\"u}rgen}}, \bibinfo {author} {\bibfnamefont {Markus}\ \bibnamefont
  {Weber}}, \bibinfo {author} {\bibfnamefont {Daniel}\ \bibnamefont {Schlenk}},
  \bibinfo {author} {\bibfnamefont {Wenjamin}\ \bibnamefont {Rosenfeld}},
  \bibinfo {author} {\bibfnamefont {Johannes}\ \bibnamefont {Vrana}}, \bibinfo
  {author} {\bibfnamefont {Karen}\ \bibnamefont {Saucke}}, \bibinfo {author}
  {\bibfnamefont {Christian}\ \bibnamefont {Kurtsiefer}}, and\ \bibinfo
  {author} {\bibfnamefont {Harald}\ \bibnamefont {Weinfurter}}} (\bibinfo
  {year} {2006}),\ \bibfield  {title} {\enquote {\bibinfo {title} {Observation
  of entanglement of a single photon with a trapped atom},}\ }\href
  {https://doi.org/10.1103/physrevlett.96.030404} {\bibfield  {journal}
  {\bibinfo  {journal} {Physical Review Letters}\ }\textbf {\bibinfo {volume}
  {96}},\ \bibinfo {pages} {030404}}\BibitemShut {NoStop}%
\bibitem [{\citenamefont {Wakui}\ \emph {et~al.}(2007)\citenamefont {Wakui},
  \citenamefont {Takahashi}, \citenamefont {Furusawa},\ and\ \citenamefont
  {Sasaki}}]{bib:wakui2007photon}%
  \BibitemOpen
  \bibfield  {author} {\bibinfo {author} {\bibnamefont {Wakui}, \bibfnamefont
  {Kentaro}}, \bibinfo {author} {\bibfnamefont {Hiroki}\ \bibnamefont
  {Takahashi}}, \bibinfo {author} {\bibfnamefont {Akira}\ \bibnamefont
  {Furusawa}}, and\ \bibinfo {author} {\bibfnamefont {Masahide}\ \bibnamefont
  {Sasaki}}} (\bibinfo {year} {2007}),\ \bibfield  {title} {\enquote {\bibinfo
  {title} {Photon subtracted squeezed states generated with periodically poled
  ktiopo 4},}\ }\href {https://doi.org/10.1364/oe.15.003568} {\bibfield
  {journal} {\bibinfo  {journal} {Optics Express}\ }\textbf {\bibinfo {volume}
  {15}},\ \bibinfo {pages} {3568}}\BibitemShut {NoStop}%
\bibitem [{\citenamefont {Wallden}\ \emph {et~al.}(2015)\citenamefont
  {Wallden}, \citenamefont {Dunjko}, \citenamefont {Kent},\ and\ \citenamefont
  {Andersson}}]{bib:Wallden2015pra}%
  \BibitemOpen
  \bibfield  {author} {\bibinfo {author} {\bibnamefont {Wallden}, \bibfnamefont
  {Petros}}, \bibinfo {author} {\bibfnamefont {Vedran}\ \bibnamefont {Dunjko}},
  \bibinfo {author} {\bibfnamefont {Adrian}\ \bibnamefont {Kent}}, and\
  \bibinfo {author} {\bibfnamefont {Erika}\ \bibnamefont {Andersson}}}
  (\bibinfo {year} {2015}),\ \bibfield  {title} {\enquote {\bibinfo {title}
  {Quantum digital signatures with quantum-key-distribution components},}\
  }\href {https://doi.org/10.1103/PhysRevA.91.042304} {\bibfield  {journal}
  {\bibinfo  {journal} {Phys. Rev. A}\ }\textbf {\bibinfo {volume} {91}},\
  \bibinfo {pages} {042304}}\BibitemShut {NoStop}%
\bibitem [{\citenamefont {Wallraff}\ \emph {et~al.}(2004)\citenamefont
  {Wallraff}, \citenamefont {Schuster}, \citenamefont {Blais}, \citenamefont
  {Frunzio}, \citenamefont {Huang}, \citenamefont {Majer}, \citenamefont
  {Kumar}, \citenamefont {Girvin},\ and\ \citenamefont
  {Schoelkopf}}]{bib:wallraff2004strong}%
  \BibitemOpen
  \bibfield  {author} {\bibinfo {author} {\bibnamefont {Wallraff},
  \bibfnamefont {Andreas}}, \bibinfo {author} {\bibfnamefont {David~I}\
  \bibnamefont {Schuster}}, \bibinfo {author} {\bibfnamefont {Alexandre}\
  \bibnamefont {Blais}}, \bibinfo {author} {\bibfnamefont {L}~\bibnamefont
  {Frunzio}}, \bibinfo {author} {\bibfnamefont {R-S}\ \bibnamefont {Huang}},
  \bibinfo {author} {\bibfnamefont {J}~\bibnamefont {Majer}}, \bibinfo {author}
  {\bibfnamefont {S}~\bibnamefont {Kumar}}, \bibinfo {author} {\bibfnamefont
  {Steven~M}\ \bibnamefont {Girvin}}, and\ \bibinfo {author} {\bibfnamefont
  {Robert~J}\ \bibnamefont {Schoelkopf}}} (\bibinfo {year} {2004}),\ \bibfield
  {title} {\enquote {\bibinfo {title} {Strong coupling of a single photon to a
  superconducting qubit using circuit quantum electrodynamics},}\ }\href
  {https://doi.org/10.1038/nature02851} {\bibfield  {journal} {\bibinfo
  {journal} {Nature}\ }\textbf {\bibinfo {volume} {431}},\ \bibinfo {pages}
  {162}}\BibitemShut {NoStop}%
\bibitem [{\citenamefont {Walshe}\ \emph {et~al.}(2023)\citenamefont {Walshe},
  \citenamefont {Alexander}, \citenamefont {Matsuura}, \citenamefont
  {Baragiola},\ and\ \citenamefont {Menicucci}}]{walshe2023equivalent}%
  \BibitemOpen
  \bibfield  {author} {\bibinfo {author} {\bibnamefont {Walshe}, \bibfnamefont
  {Blayney~W}}, \bibinfo {author} {\bibfnamefont {Rafael~N.}\ \bibnamefont
  {Alexander}}, \bibinfo {author} {\bibfnamefont {Takaya}\ \bibnamefont
  {Matsuura}}, \bibinfo {author} {\bibfnamefont {Ben~Q.}\ \bibnamefont
  {Baragiola}}, and\ \bibinfo {author} {\bibfnamefont {Nicolas~C.}\
  \bibnamefont {Menicucci}}} (\bibinfo {year} {2023}),\ \bibfield  {title}
  {\enquote {\bibinfo {title} {Equivalent noise properties of scalable
  continuous-variable cluster states},}\ }\href
  {https://doi.org/10.1103/PhysRevA.108.042602} {\bibfield  {journal} {\bibinfo
   {journal} {Phys. Rev. A}\ }\textbf {\bibinfo {volume} {108}},\ \bibinfo
  {pages} {042602}}\BibitemShut {NoStop}%
\bibitem [{\citenamefont {Walshe}\ \emph {et~al.}(2021)\citenamefont {Walshe},
  \citenamefont {Alexander}, \citenamefont {Menicucci},\ and\ \citenamefont
  {Baragiola}}]{walsh2021streamlined}%
  \BibitemOpen
  \bibfield  {author} {\bibinfo {author} {\bibnamefont {Walshe}, \bibfnamefont
  {Blayney~W}}, \bibinfo {author} {\bibfnamefont {Rafael~N.}\ \bibnamefont
  {Alexander}}, \bibinfo {author} {\bibfnamefont {Nicolas~C.}\ \bibnamefont
  {Menicucci}}, and\ \bibinfo {author} {\bibfnamefont {Ben~Q.}\ \bibnamefont
  {Baragiola}}} (\bibinfo {year} {2021}),\ \bibfield  {title} {\enquote
  {\bibinfo {title} {Streamlined quantum computing with macronode cluster
  states},}\ }\href {https://doi.org/10.1103/PhysRevA.104.062427} {\bibfield
  {journal} {\bibinfo  {journal} {Phys. Rev. A}\ }\textbf {\bibinfo {volume}
  {104}},\ \bibinfo {pages} {062427}}\BibitemShut {NoStop}%
\bibitem [{\citenamefont {Walshe}\ \emph {et~al.}(2020)\citenamefont {Walshe},
  \citenamefont {Baragiola}, \citenamefont {Alexander},\ and\ \citenamefont
  {Menicucci}}]{walshe2020gateteleportation}%
  \BibitemOpen
  \bibfield  {author} {\bibinfo {author} {\bibnamefont {Walshe}, \bibfnamefont
  {Blayney~W}}, \bibinfo {author} {\bibfnamefont {Ben~Q.}\ \bibnamefont
  {Baragiola}}, \bibinfo {author} {\bibfnamefont {Rafael~N.}\ \bibnamefont
  {Alexander}}, and\ \bibinfo {author} {\bibfnamefont {Nicolas~C.}\
  \bibnamefont {Menicucci}}} (\bibinfo {year} {2020}),\ \bibfield  {title}
  {\enquote {\bibinfo {title} {Continuous-variable gate teleportation and
  bosonic-code error correction},}\ }\href
  {https://doi.org/10.1103/PhysRevA.102.062411} {\bibfield  {journal} {\bibinfo
   {journal} {Phys. Rev. A}\ }\textbf {\bibinfo {volume} {102}},\ \bibinfo
  {pages} {062411}}\BibitemShut {NoStop}%
\bibitem [{\citenamefont {Walshe}\ \emph {et~al.}(2024)\citenamefont {Walshe},
  \citenamefont {Baragiola}, \citenamefont {Ferretti}, \citenamefont {Gefaell},
  \citenamefont {Vasmer}, \citenamefont {Weil}, \citenamefont {Matsuura},
  \citenamefont {Thomas~Jaeken}, \citenamefont {Zhihua~Han},\ and\
  \citenamefont {Ilan~Tzitrin}}]{walshe2024totl}%
  \BibitemOpen
  \bibfield  {author} {\bibinfo {author} {\bibnamefont {Walshe}, \bibfnamefont
  {Blayney~W}}, \bibinfo {author} {\bibfnamefont {Ben~Q.}\ \bibnamefont
  {Baragiola}}, \bibinfo {author} {\bibfnamefont {Hugo}\ \bibnamefont
  {Ferretti}}, \bibinfo {author} {\bibfnamefont {Jos{\'e}}\ \bibnamefont
  {Gefaell}}, \bibinfo {author} {\bibfnamefont {Michael}\ \bibnamefont
  {Vasmer}}, \bibinfo {author} {\bibfnamefont {Ryohei}\ \bibnamefont {Weil}},
  \bibinfo {author} {\bibfnamefont {Takaya}\ \bibnamefont {Matsuura}}, \bibinfo
  {author} {\bibfnamefont {Giacomo~Pantaleoni}\ \bibnamefont {Thomas~Jaeken}},
  \bibinfo {author} {\bibfnamefont {Nicolas C.~Menicucci}\ \bibnamefont
  {Zhihua~Han}}, and\ \bibinfo {author} {\bibfnamefont {Rafael N.~Alexander}\
  \bibnamefont {Ilan~Tzitrin}}} (\bibinfo {year} {2024}),\ \href
  {https://arxiv.org/abs/2408.04126} {\enquote {\bibinfo {title}
  {Linear-optical quantum computation with arbitrary error-correcting codes},}\
  }\BibitemShut {NoStop}%
\bibitem [{\citenamefont {Walther}\ \emph {et~al.}(2005)\citenamefont
  {Walther}, \citenamefont {Resch}, \citenamefont {Brukner}, \citenamefont
  {Steinberg}, \citenamefont {Pan},\ and\ \citenamefont
  {Zeilinger}}]{bib:PRL_94_040504}%
  \BibitemOpen
  \bibfield  {author} {\bibinfo {author} {\bibnamefont {Walther}, \bibfnamefont
  {P}}, \bibinfo {author} {\bibfnamefont {KJ}~\bibnamefont {Resch}}, \bibinfo
  {author} {\bibfnamefont {{\v{C}}}~\bibnamefont {Brukner}}, \bibinfo {author}
  {\bibfnamefont {AM}~\bibnamefont {Steinberg}}, \bibinfo {author}
  {\bibfnamefont {J-W}\ \bibnamefont {Pan}}, and\ \bibinfo {author}
  {\bibfnamefont {A}~\bibnamefont {Zeilinger}}} (\bibinfo {year} {2005}),\
  \bibfield  {title} {\enquote {\bibinfo {title} {Quantum nonlocality obtained
  from local states by entanglement purification},}\ }\href
  {https://doi.org/10.1103/physrevlett.94.040504} {\bibfield  {journal}
  {\bibinfo  {journal} {Physical Review Letters}\ }\textbf {\bibinfo {volume}
  {94}},\ \bibinfo {pages} {040504}},\ \Eprint
  {https://arxiv.org/abs/arXiv:quant-ph/0502026v1} {arXiv:quant-ph/0502026v1}
  \BibitemShut {NoStop}%
\bibitem [{\citenamefont {Wan}\ \emph {et~al.}(2017)\citenamefont {Wan},
  \citenamefont {Dahlsten}, \citenamefont {Kristj{\'a}nsson}, \citenamefont
  {Gardner},\ and\ \citenamefont {Kim}}]{wan2017quantum}%
  \BibitemOpen
  \bibfield  {author} {\bibinfo {author} {\bibnamefont {Wan}, \bibfnamefont
  {Kwok~Ho}}, \bibinfo {author} {\bibfnamefont {Oscar}\ \bibnamefont
  {Dahlsten}}, \bibinfo {author} {\bibfnamefont {Hl{\'e}r}\ \bibnamefont
  {Kristj{\'a}nsson}}, \bibinfo {author} {\bibfnamefont {Robert}\ \bibnamefont
  {Gardner}}, and\ \bibinfo {author} {\bibfnamefont {MS}~\bibnamefont {Kim}}}
  (\bibinfo {year} {2017}),\ \bibfield  {title} {\enquote {\bibinfo {title}
  {Quantum generalisation of feedforward neural networks},}\ }\href@noop {}
  {\bibfield  {journal} {\bibinfo  {journal} {npj Quantum information}\
  }\textbf {\bibinfo {volume} {3}}~(\bibinfo {number} {1}),\ \bibinfo {pages}
  {36}}\BibitemShut {NoStop}%
\bibitem [{\citenamefont {Wang}\ \emph
  {et~al.}(2011{\natexlab{a}})\citenamefont {Wang}, \citenamefont {Fowler},\
  and\ \citenamefont {Hollenberg}}]{SD-Wang:2011aa}%
  \BibitemOpen
  \bibfield  {author} {\bibinfo {author} {\bibnamefont {Wang}, \bibfnamefont
  {David~S}}, \bibinfo {author} {\bibfnamefont {Austin~G.}\ \bibnamefont
  {Fowler}}, and\ \bibinfo {author} {\bibfnamefont {Lloyd C.~L.}\ \bibnamefont
  {Hollenberg}}} (\bibinfo {year} {2011}{\natexlab{a}}),\ \bibfield  {title}
  {\enquote {\bibinfo {title} {Surface code quantum computing with error rates
  over 1{\%}},}\ }\href {https://doi.org/10.1103/PhysRevA.83.020302} {\bibfield
   {journal} {\bibinfo  {journal} {Physical Review A}\ }\textbf {\bibinfo
  {volume} {83}},\ \bibinfo {pages} {020302}}\BibitemShut {NoStop}%
\bibitem [{\citenamefont {Wang}\ \emph
  {et~al.}(2011{\natexlab{b}})\citenamefont {Wang}, \citenamefont {Fowler},\
  and\ \citenamefont {Hollenberg}}]{bib:WFH11}%
  \BibitemOpen
  \bibfield  {author} {\bibinfo {author} {\bibnamefont {Wang}, \bibfnamefont
  {DS}}, \bibinfo {author} {\bibfnamefont {A.G.}\ \bibnamefont {Fowler}}, and\
  \bibinfo {author} {\bibfnamefont {L.C.L.}\ \bibnamefont {Hollenberg}}}
  (\bibinfo {year} {2011}{\natexlab{b}}),\ \bibfield  {title} {\enquote
  {\bibinfo {title} {{Quantum computing with nearest neighbor interactions and
  error rates over 1\%}},}\ }\href@noop {} {\bibfield  {journal} {\bibinfo
  {journal} {Phys. Rev. A.}\ }\textbf {\bibinfo {volume} {83}},\ \bibinfo
  {pages} {020302(R)}}\BibitemShut {NoStop}%
\bibitem [{\citenamefont {Wang}\ \emph
  {et~al.}(2010{\natexlab{a}})\citenamefont {Wang}, \citenamefont {Fowler},
  \citenamefont {Stephens},\ and\ \citenamefont {Hollenberg}}]{bib:WFSH09}%
  \BibitemOpen
  \bibfield  {author} {\bibinfo {author} {\bibnamefont {Wang}, \bibfnamefont
  {DS}}, \bibinfo {author} {\bibfnamefont {A.G.}\ \bibnamefont {Fowler}},
  \bibinfo {author} {\bibfnamefont {A.M.}\ \bibnamefont {Stephens}}, and\
  \bibinfo {author} {\bibfnamefont {L.C.L.}\ \bibnamefont {Hollenberg}}}
  (\bibinfo {year} {2010}{\natexlab{a}}),\ \bibfield  {title} {\enquote
  {\bibinfo {title} {{Threshold Error rates for the toric and surface
  codes}},}\ }\href@noop {} {\bibfield  {journal} {\bibinfo  {journal} {Quant.
  Inf. Comp.}\ }\textbf {\bibinfo {volume} {10}},\ \bibinfo {pages}
  {456}}\BibitemShut {NoStop}%
\bibitem [{\citenamefont {Wang}\ \emph
  {et~al.}(2016{\natexlab{a}})\citenamefont {Wang}, \citenamefont {Duan},
  \citenamefont {Li}, \citenamefont {Chen}, \citenamefont {Li}, \citenamefont
  {He}, \citenamefont {Chen}, \citenamefont {He}, \citenamefont {Ding},
  \citenamefont {Peng} \emph {et~al.}}]{bib:wang2016near}%
  \BibitemOpen
  \bibfield  {author} {\bibinfo {author} {\bibnamefont {Wang}, \bibfnamefont
  {Hui}}, \bibinfo {author} {\bibfnamefont {Z-C}\ \bibnamefont {Duan}},
  \bibinfo {author} {\bibfnamefont {Y-H}\ \bibnamefont {Li}}, \bibinfo {author}
  {\bibfnamefont {Si}~\bibnamefont {Chen}}, \bibinfo {author} {\bibfnamefont
  {J-P}\ \bibnamefont {Li}}, \bibinfo {author} {\bibfnamefont {Y-M}\
  \bibnamefont {He}}, \bibinfo {author} {\bibfnamefont {M-C}\ \bibnamefont
  {Chen}}, \bibinfo {author} {\bibfnamefont {Yu}~\bibnamefont {He}}, \bibinfo
  {author} {\bibfnamefont {X}~\bibnamefont {Ding}}, \bibinfo {author}
  {\bibfnamefont {Cheng-Zhi}\ \bibnamefont {Peng}},  \emph {et~al.}} (\bibinfo
  {year} {2016}{\natexlab{a}}),\ \bibfield  {title} {\enquote {\bibinfo {title}
  {Near-transform-limited single photons from an efficient solid-state quantum
  emitter},}\ }\href {https://doi.org/10.1103/physrevlett.116.213601}
  {\bibfield  {journal} {\bibinfo  {journal} {Physical Review Letters}\
  }\textbf {\bibinfo {volume} {116}},\ \bibinfo {pages} {213601}},\ \Eprint
  {https://arxiv.org/abs/arXiv:1602.07386v2} {arXiv:1602.07386v2} \BibitemShut
  {NoStop}%
\bibitem [{\citenamefont {Wang}\ \emph
  {et~al.}(2017{\natexlab{a}})\citenamefont {Wang}, \citenamefont {He},
  \citenamefont {Li}, \citenamefont {Su}, \citenamefont {Li}, \citenamefont
  {Huang}, \citenamefont {Ding}, \citenamefont {Chen}, \citenamefont {Liu},
  \citenamefont {Qin} \emph {et~al.}}]{bib:wang2017high}%
  \BibitemOpen
  \bibfield  {author} {\bibinfo {author} {\bibnamefont {Wang}, \bibfnamefont
  {Hui}}, \bibinfo {author} {\bibfnamefont {Yu}~\bibnamefont {He}}, \bibinfo
  {author} {\bibfnamefont {Yu-Huai}\ \bibnamefont {Li}}, \bibinfo {author}
  {\bibfnamefont {Zu-En}\ \bibnamefont {Su}}, \bibinfo {author} {\bibfnamefont
  {Bo}~\bibnamefont {Li}}, \bibinfo {author} {\bibfnamefont {He-Liang}\
  \bibnamefont {Huang}}, \bibinfo {author} {\bibfnamefont {Xing}\ \bibnamefont
  {Ding}}, \bibinfo {author} {\bibfnamefont {Ming-Cheng}\ \bibnamefont {Chen}},
  \bibinfo {author} {\bibfnamefont {Chang}\ \bibnamefont {Liu}}, \bibinfo
  {author} {\bibfnamefont {Jian}\ \bibnamefont {Qin}},  \emph {et~al.}}
  (\bibinfo {year} {2017}{\natexlab{a}}),\ \bibfield  {title} {\enquote
  {\bibinfo {title} {High-efficiency multiphoton boson sampling},}\ }\href
  {https://doi.org/10.1038/nphoton.2017.63} {\bibfield  {journal} {\bibinfo
  {journal} {Nature Photonics}\ }\textbf {\bibinfo {volume} {11}},\ \bibinfo
  {pages} {36}}\BibitemShut {NoStop}%
\bibitem [{\citenamefont {Wang}\ \emph
  {et~al.}(2018{\natexlab{a}})\citenamefont {Wang}, \citenamefont {Li},
  \citenamefont {Jiang}, \citenamefont {He}, \citenamefont {Li}, \citenamefont
  {Ding}, \citenamefont {Chen}, \citenamefont {Qin}, \citenamefont {Peng},
  \citenamefont {Schneider} \emph {et~al.}}]{bib:wang2018toward}%
  \BibitemOpen
  \bibfield  {author} {\bibinfo {author} {\bibnamefont {Wang}, \bibfnamefont
  {Hui}}, \bibinfo {author} {\bibfnamefont {Wei}\ \bibnamefont {Li}}, \bibinfo
  {author} {\bibfnamefont {Xiao}\ \bibnamefont {Jiang}}, \bibinfo {author}
  {\bibfnamefont {Y-M}\ \bibnamefont {He}}, \bibinfo {author} {\bibfnamefont
  {Y-H}\ \bibnamefont {Li}}, \bibinfo {author} {\bibfnamefont {Xing}\
  \bibnamefont {Ding}}, \bibinfo {author} {\bibfnamefont {M-C}\ \bibnamefont
  {Chen}}, \bibinfo {author} {\bibfnamefont {Jian}\ \bibnamefont {Qin}},
  \bibinfo {author} {\bibfnamefont {C-Z}\ \bibnamefont {Peng}}, \bibinfo
  {author} {\bibfnamefont {Christian}\ \bibnamefont {Schneider}},  \emph
  {et~al.}} (\bibinfo {year} {2018}{\natexlab{a}}),\ \bibfield  {title}
  {\enquote {\bibinfo {title} {Toward scalable boson sampling with photon
  loss},}\ }\href {https://doi.org/10.1103/physrevlett.120.230502} {\bibfield
  {journal} {\bibinfo  {journal} {Physical Review Letters}\ }\textbf {\bibinfo
  {volume} {120}},\ \bibinfo {pages} {230502}},\ \Eprint
  {https://arxiv.org/abs/arXiv:1801.08282v1} {arXiv:1801.08282v1} \BibitemShut
  {NoStop}%
\bibitem [{\citenamefont {Wang}\ \emph {et~al.}(2013)\citenamefont {Wang},
  \citenamefont {Yang}, \citenamefont {Liao}, \citenamefont {Zhang},
  \citenamefont {Shen}, \citenamefont {Hu}, \citenamefont {Wu}, \citenamefont
  {Yang}, \citenamefont {Jiang}, \citenamefont {Tang} \emph
  {et~al.}}]{bib:NP_7_387}%
  \BibitemOpen
  \bibfield  {author} {\bibinfo {author} {\bibnamefont {Wang}, \bibfnamefont
  {Jian-Yu}}, \bibinfo {author} {\bibfnamefont {Bin}\ \bibnamefont {Yang}},
  \bibinfo {author} {\bibfnamefont {Sheng-Kai}\ \bibnamefont {Liao}}, \bibinfo
  {author} {\bibfnamefont {Liang}\ \bibnamefont {Zhang}}, \bibinfo {author}
  {\bibfnamefont {Qi}~\bibnamefont {Shen}}, \bibinfo {author} {\bibfnamefont
  {Xiao-Fang}\ \bibnamefont {Hu}}, \bibinfo {author} {\bibfnamefont {Jin-Cai}\
  \bibnamefont {Wu}}, \bibinfo {author} {\bibfnamefont {Shi-Ji}\ \bibnamefont
  {Yang}}, \bibinfo {author} {\bibfnamefont {Hao}\ \bibnamefont {Jiang}},
  \bibinfo {author} {\bibfnamefont {Yan-Lin}\ \bibnamefont {Tang}},  \emph
  {et~al.}} (\bibinfo {year} {2013}),\ \bibfield  {title} {\enquote {\bibinfo
  {title} {Direct and full-scale experimental verifications towards
  ground-satellite quantum key distribution},}\ }\href
  {https://doi.org/10.1038/nphoton.2013.89} {\bibfield  {journal} {\bibinfo
  {journal} {Nature Photonics}\ }\textbf {\bibinfo {volume} {7}},\ \bibinfo
  {pages} {387}},\ \Eprint {https://arxiv.org/abs/arXiv:1210.7556v2}
  {arXiv:1210.7556v2} \BibitemShut {NoStop}%
\bibitem [{\citenamefont {Wang}\ \emph
  {et~al.}(2018{\natexlab{b}})\citenamefont {Wang}, \citenamefont {Paesani},
  \citenamefont {Ding}, \citenamefont {Santagati}, \citenamefont {Skrzypczyk},
  \citenamefont {Salavrakos}, \citenamefont {Tura}, \citenamefont {Augusiak},
  \citenamefont {Man{\v{c}}inska}, \citenamefont {Bacco} \emph
  {et~al.}}]{bib:wang2018multidimensional}%
  \BibitemOpen
  \bibfield  {author} {\bibinfo {author} {\bibnamefont {Wang}, \bibfnamefont
  {Jianwei}}, \bibinfo {author} {\bibfnamefont {Stefano}\ \bibnamefont
  {Paesani}}, \bibinfo {author} {\bibfnamefont {Yunhong}\ \bibnamefont {Ding}},
  \bibinfo {author} {\bibfnamefont {Raffaele}\ \bibnamefont {Santagati}},
  \bibinfo {author} {\bibfnamefont {Paul}\ \bibnamefont {Skrzypczyk}}, \bibinfo
  {author} {\bibfnamefont {Alexia}\ \bibnamefont {Salavrakos}}, \bibinfo
  {author} {\bibfnamefont {Jordi}\ \bibnamefont {Tura}}, \bibinfo {author}
  {\bibfnamefont {Remigiusz}\ \bibnamefont {Augusiak}}, \bibinfo {author}
  {\bibfnamefont {Laura}\ \bibnamefont {Man{\v{c}}inska}}, \bibinfo {author}
  {\bibfnamefont {Davide}\ \bibnamefont {Bacco}},  \emph {et~al.}} (\bibinfo
  {year} {2018}{\natexlab{b}}),\ \bibfield  {title} {\enquote {\bibinfo {title}
  {Multidimensional quantum entanglement with large-scale integrated optics},}\
  }\href {https://doi.org/10.1126/science.aar7053} {\bibfield  {journal}
  {\bibinfo  {journal} {Science}\ ,\ \bibinfo {pages} {7053}}}\Eprint
  {https://arxiv.org/abs/arXiv:1803.04449v1} {arXiv:1803.04449v1} \BibitemShut
  {NoStop}%
\bibitem [{\citenamefont {Wang}\ \emph
  {et~al.}(2017{\natexlab{b}})\citenamefont {Wang}, \citenamefont {Han},
  \citenamefont {Wang}, \citenamefont {Li}, \citenamefont {Mu}, \citenamefont
  {Fan},\ and\ \citenamefont {Wang}}]{bib:Han2017}%
  \BibitemOpen
  \bibfield  {author} {\bibinfo {author} {\bibnamefont {Wang}, \bibfnamefont
  {Jun}}, \bibinfo {author} {\bibfnamefont {Zhao-Yu}\ \bibnamefont {Han}},
  \bibinfo {author} {\bibfnamefont {Song-Bo}\ \bibnamefont {Wang}}, \bibinfo
  {author} {\bibfnamefont {Zeyang}\ \bibnamefont {Li}}, \bibinfo {author}
  {\bibfnamefont {Liang-Zhu}\ \bibnamefont {Mu}}, \bibinfo {author}
  {\bibfnamefont {Heng}\ \bibnamefont {Fan}}, and\ \bibinfo {author}
  {\bibfnamefont {Lei}\ \bibnamefont {Wang}}} (\bibinfo {year}
  {2017}{\natexlab{b}}),\ \bibfield  {title} {\enquote {\bibinfo {title}
  {Efficient quantum tomography with fidelity estimation},}\ }\href@noop {} {\
  }\Eprint {https://arxiv.org/abs/arXiv:1712.03213} {arXiv:1712.03213}
  \BibitemShut {NoStop}%
\bibitem [{\citenamefont {Wang}\ \emph
  {et~al.}(2015{\natexlab{a}})\citenamefont {Wang}, \citenamefont {Chen},
  \citenamefont {Ju}, \citenamefont {Xu}, \citenamefont {Zhao}, \citenamefont
  {Chen}, \citenamefont {Chen}, \citenamefont {Chen},\ and\ \citenamefont
  {Pan}}]{bib:wang2015experimental}%
  \BibitemOpen
  \bibfield  {author} {\bibinfo {author} {\bibnamefont {Wang}, \bibfnamefont
  {Liu-Jun}}, \bibinfo {author} {\bibfnamefont {Luo-Kan}\ \bibnamefont {Chen}},
  \bibinfo {author} {\bibfnamefont {Lei}\ \bibnamefont {Ju}}, \bibinfo {author}
  {\bibfnamefont {Mu-Lan}\ \bibnamefont {Xu}}, \bibinfo {author} {\bibfnamefont
  {Yong}\ \bibnamefont {Zhao}}, \bibinfo {author} {\bibfnamefont {Kai}\
  \bibnamefont {Chen}}, \bibinfo {author} {\bibfnamefont {Zeng-Bing}\
  \bibnamefont {Chen}}, \bibinfo {author} {\bibfnamefont {Teng-Yun}\
  \bibnamefont {Chen}}, and\ \bibinfo {author} {\bibfnamefont {Jian-Wei}\
  \bibnamefont {Pan}}} (\bibinfo {year} {2015}{\natexlab{a}}),\ \bibfield
  {title} {\enquote {\bibinfo {title} {Experimental multiplexing of quantum key
  distribution with classical optical communication},}\ }\href
  {https://doi.org/10.1063/1.4913483} {\bibfield  {journal} {\bibinfo
  {journal} {Applied Physics Letters}\ }\textbf {\bibinfo {volume} {106}},\
  \bibinfo {pages} {081108}}\BibitemShut {NoStop}%
\bibitem [{\citenamefont {Wang}\ \emph
  {et~al.}(2018{\natexlab{c}})\citenamefont {Wang}, \citenamefont {Cui},
  \citenamefont {Wang}, \citenamefont {Xiao},\ and\ \citenamefont
  {Jiang}}]{bib:wang2018machine}%
  \BibitemOpen
  \bibfield  {author} {\bibinfo {author} {\bibnamefont {Wang}, \bibfnamefont
  {Mowei}}, \bibinfo {author} {\bibfnamefont {Yong}\ \bibnamefont {Cui}},
  \bibinfo {author} {\bibfnamefont {Xin}\ \bibnamefont {Wang}}, \bibinfo
  {author} {\bibfnamefont {Shihan}\ \bibnamefont {Xiao}}, and\ \bibinfo
  {author} {\bibfnamefont {Junchen}\ \bibnamefont {Jiang}}} (\bibinfo {year}
  {2018}{\natexlab{c}}),\ \bibfield  {title} {\enquote {\bibinfo {title}
  {Machine learning for networking: Workflow, advances and opportunities},}\
  }\href {https://doi.org/10.1109/mnet.2017.1700200} {\bibfield  {journal}
  {\bibinfo  {journal} {IEEE Network}\ }\textbf {\bibinfo {volume} {32}},\
  \bibinfo {pages} {92}},\ \Eprint {https://arxiv.org/abs/arXiv:1709.08339v2}
  {arXiv:1709.08339v2} \BibitemShut {NoStop}%
\bibitem [{\citenamefont {Wang}\ and\ \citenamefont
  {Wang}(2006)}]{bib:wang2006adaptive}%
  \BibitemOpen
  \bibfield  {author} {\bibinfo {author} {\bibnamefont {Wang}, \bibfnamefont
  {Ping}}, and\ \bibinfo {author} {\bibfnamefont {Ting}\ \bibnamefont {Wang}}}
  (\bibinfo {year} {2006}),\ \bibfield  {title} {\enquote {\bibinfo {title}
  {Adaptive routing for sensor networks using reinforcement learning},}\ }in\
  \href {https://doi.org/10.1109/cit.2006.34} {\emph {\bibinfo {booktitle} {The
  Sixth IEEE International Conference on Computer and Information Technology
  (CIT'06)}}},\ p.\ \bibinfo {pages} {219}\BibitemShut {NoStop}%
\bibitem [{\citenamefont {Wang}\ \emph {et~al.}(2012)\citenamefont {Wang},
  \citenamefont {Chen}, \citenamefont {Guo}, \citenamefont {Yin}, \citenamefont
  {Li}, \citenamefont {Zhou}, \citenamefont {Guo},\ and\ \citenamefont
  {Han}}]{bib:OL_37_1008}%
  \BibitemOpen
  \bibfield  {author} {\bibinfo {author} {\bibnamefont {Wang}, \bibfnamefont
  {Shuang}}, \bibinfo {author} {\bibfnamefont {Wei}\ \bibnamefont {Chen}},
  \bibinfo {author} {\bibfnamefont {Jun-Fu}\ \bibnamefont {Guo}}, \bibinfo
  {author} {\bibfnamefont {Zhen-Qiang}\ \bibnamefont {Yin}}, \bibinfo {author}
  {\bibfnamefont {Hong-Wei}\ \bibnamefont {Li}}, \bibinfo {author}
  {\bibfnamefont {Zheng}\ \bibnamefont {Zhou}}, \bibinfo {author}
  {\bibfnamefont {Guang-Can}\ \bibnamefont {Guo}}, and\ \bibinfo {author}
  {\bibfnamefont {Zheng-Fu}\ \bibnamefont {Han}}} (\bibinfo {year} {2012}),\
  \bibfield  {title} {\enquote {\bibinfo {title} {2 ghz clock quantum key
  distribution over 260 km of standard telecom fiber},}\ }\href
  {https://doi.org/10.1364/ol.37.001008} {\bibfield  {journal} {\bibinfo
  {journal} {Optics Letters}\ }\textbf {\bibinfo {volume} {37}},\ \bibinfo
  {pages} {1008}},\ \Eprint {https://arxiv.org/abs/arXiv:1203.4323v1}
  {arXiv:1203.4323v1} \BibitemShut {NoStop}%
\bibitem [{\citenamefont {Wang}\ \emph
  {et~al.}(2010{\natexlab{b}})\citenamefont {Wang}, \citenamefont {Chen},
  \citenamefont {Yin}, \citenamefont {Zhang}, \citenamefont {Zhang},
  \citenamefont {Li}, \citenamefont {Xu}, \citenamefont {Zhou}, \citenamefont
  {Yang}, \citenamefont {Huang} \emph {et~al.}}]{bib:OL_35_2454}%
  \BibitemOpen
  \bibfield  {author} {\bibinfo {author} {\bibnamefont {Wang}, \bibfnamefont
  {Shuang}}, \bibinfo {author} {\bibfnamefont {Wei}\ \bibnamefont {Chen}},
  \bibinfo {author} {\bibfnamefont {Zhen-Qiang}\ \bibnamefont {Yin}}, \bibinfo
  {author} {\bibfnamefont {Yang}\ \bibnamefont {Zhang}}, \bibinfo {author}
  {\bibfnamefont {Tao}\ \bibnamefont {Zhang}}, \bibinfo {author} {\bibfnamefont
  {Hong-Wei}\ \bibnamefont {Li}}, \bibinfo {author} {\bibfnamefont {Fang-Xing}\
  \bibnamefont {Xu}}, \bibinfo {author} {\bibfnamefont {Zheng}\ \bibnamefont
  {Zhou}}, \bibinfo {author} {\bibfnamefont {Yang}\ \bibnamefont {Yang}},
  \bibinfo {author} {\bibfnamefont {Da-Jun}\ \bibnamefont {Huang}},  \emph
  {et~al.}} (\bibinfo {year} {2010}{\natexlab{b}}),\ \bibfield  {title}
  {\enquote {\bibinfo {title} {Field test of wavelength-saving quantum key
  distribution network},}\ }\href@noop {} {\bibfield  {journal} {\bibinfo
  {journal} {Optics Letters}\ }\textbf {\bibinfo {volume} {35}},\ \bibinfo
  {pages} {2454}},\ \Eprint {https://arxiv.org/abs/arXiv:1203.4321v1}
  {arXiv:1203.4321v1} \BibitemShut {NoStop}%
\bibitem [{\citenamefont {Wang}\ \emph
  {et~al.}(2015{\natexlab{b}})\citenamefont {Wang}, \citenamefont {Cai},
  \citenamefont {Su}, \citenamefont {Chen}, \citenamefont {Wu}, \citenamefont
  {Li}, \citenamefont {Liu}, \citenamefont {Lu},\ and\ \citenamefont
  {Pan}}]{bib:Nat_518_516}%
  \BibitemOpen
  \bibfield  {author} {\bibinfo {author} {\bibnamefont {Wang}, \bibfnamefont
  {Xi-Lin}}, \bibinfo {author} {\bibfnamefont {Xin-Dong}\ \bibnamefont {Cai}},
  \bibinfo {author} {\bibfnamefont {Zu-En}\ \bibnamefont {Su}}, \bibinfo
  {author} {\bibfnamefont {Ming-Cheng}\ \bibnamefont {Chen}}, \bibinfo {author}
  {\bibfnamefont {Dian}\ \bibnamefont {Wu}}, \bibinfo {author} {\bibfnamefont
  {Li}~\bibnamefont {Li}}, \bibinfo {author} {\bibfnamefont {Nai-Le}\
  \bibnamefont {Liu}}, \bibinfo {author} {\bibfnamefont {Chao-Yang}\
  \bibnamefont {Lu}}, and\ \bibinfo {author} {\bibfnamefont {Jian-Wei}\
  \bibnamefont {Pan}}} (\bibinfo {year} {2015}{\natexlab{b}}),\ \bibfield
  {title} {\enquote {\bibinfo {title} {Quantum teleportation of multiple
  degrees of freedom of a single photon},}\ }\href
  {https://doi.org/10.1038/nature14246} {\bibfield  {journal} {\bibinfo
  {journal} {Nature}\ }\textbf {\bibinfo {volume} {518}},\ \bibinfo {pages}
  {516}}\BibitemShut {NoStop}%
\bibitem [{\citenamefont {Wang}\ \emph
  {et~al.}(2016{\natexlab{b}})\citenamefont {Wang}, \citenamefont {Chen},
  \citenamefont {Li}, \citenamefont {Huang}, \citenamefont {Liu}, \citenamefont
  {Chen}, \citenamefont {Luo}, \citenamefont {Su}, \citenamefont {Wu},
  \citenamefont {Li} \emph {et~al.}}]{bib:wang2016experimental}%
  \BibitemOpen
  \bibfield  {author} {\bibinfo {author} {\bibnamefont {Wang}, \bibfnamefont
  {Xi-Lin}}, \bibinfo {author} {\bibfnamefont {Luo-Kan}\ \bibnamefont {Chen}},
  \bibinfo {author} {\bibfnamefont {Wei}\ \bibnamefont {Li}}, \bibinfo {author}
  {\bibfnamefont {He-Liang}\ \bibnamefont {Huang}}, \bibinfo {author}
  {\bibfnamefont {Chang}\ \bibnamefont {Liu}}, \bibinfo {author} {\bibfnamefont
  {Chao}\ \bibnamefont {Chen}}, \bibinfo {author} {\bibfnamefont {Yi-Han}\
  \bibnamefont {Luo}}, \bibinfo {author} {\bibfnamefont {Zu-En}\ \bibnamefont
  {Su}}, \bibinfo {author} {\bibfnamefont {Dian}\ \bibnamefont {Wu}}, \bibinfo
  {author} {\bibfnamefont {Zheng-Da}\ \bibnamefont {Li}},  \emph {et~al.}}
  (\bibinfo {year} {2016}{\natexlab{b}}),\ \bibfield  {title} {\enquote
  {\bibinfo {title} {Experimental ten-photon entanglement},}\ }\href
  {https://doi.org/10.1103/physrevlett.117.210502} {\bibfield  {journal}
  {\bibinfo  {journal} {Physical Review Letters}\ }\textbf {\bibinfo {volume}
  {117}},\ \bibinfo {pages} {210502}},\ \Eprint
  {https://arxiv.org/abs/arXiv:1605.08547v3} {arXiv:1605.08547v3} \BibitemShut
  {NoStop}%
\bibitem [{\citenamefont {Wang}\ \emph
  {et~al.}(2018{\natexlab{d}})\citenamefont {Wang}, \citenamefont {Luo},
  \citenamefont {Huang}, \citenamefont {Chen}, \citenamefont {Su},
  \citenamefont {Liu}, \citenamefont {Chen}, \citenamefont {Li}, \citenamefont
  {Fang}, \citenamefont {Jiang} \emph {et~al.}}]{bib:wang201818}%
  \BibitemOpen
  \bibfield  {author} {\bibinfo {author} {\bibnamefont {Wang}, \bibfnamefont
  {Xi-Lin}}, \bibinfo {author} {\bibfnamefont {Yi-Han}\ \bibnamefont {Luo}},
  \bibinfo {author} {\bibfnamefont {He-Liang}\ \bibnamefont {Huang}}, \bibinfo
  {author} {\bibfnamefont {Ming-Cheng}\ \bibnamefont {Chen}}, \bibinfo {author}
  {\bibfnamefont {Zu-En}\ \bibnamefont {Su}}, \bibinfo {author} {\bibfnamefont
  {Chang}\ \bibnamefont {Liu}}, \bibinfo {author} {\bibfnamefont {Chao}\
  \bibnamefont {Chen}}, \bibinfo {author} {\bibfnamefont {Wei}\ \bibnamefont
  {Li}}, \bibinfo {author} {\bibfnamefont {Yu-Qiang}\ \bibnamefont {Fang}},
  \bibinfo {author} {\bibfnamefont {Xiao}\ \bibnamefont {Jiang}},  \emph
  {et~al.}} (\bibinfo {year} {2018}{\natexlab{d}}),\ \bibfield  {title}
  {\enquote {\bibinfo {title} {18-qubit entanglement with six photons' three
  degrees of freedom},}\ }\href
  {https://doi.org/10.1103/physrevlett.120.260502} {\bibfield  {journal}
  {\bibinfo  {journal} {Physical Review Letters}\ }\textbf {\bibinfo {volume}
  {120}},\ \bibinfo {pages} {260502}}\BibitemShut {NoStop}%
\bibitem [{\citenamefont {Wang}\ \emph
  {et~al.}(2015{\natexlab{c}})\citenamefont {Wang}, \citenamefont {Zhang},
  \citenamefont {Corcovilos}, \citenamefont {Kumar},\ and\ \citenamefont
  {Weiss}}]{bib:wang2015coherent}%
  \BibitemOpen
  \bibfield  {author} {\bibinfo {author} {\bibnamefont {Wang}, \bibfnamefont
  {Yang}}, \bibinfo {author} {\bibfnamefont {Xianli}\ \bibnamefont {Zhang}},
  \bibinfo {author} {\bibfnamefont {Theodore~A}\ \bibnamefont {Corcovilos}},
  \bibinfo {author} {\bibfnamefont {Aishwarya}\ \bibnamefont {Kumar}}, and\
  \bibinfo {author} {\bibfnamefont {David~S}\ \bibnamefont {Weiss}}} (\bibinfo
  {year} {2015}{\natexlab{c}}),\ \bibfield  {title} {\enquote {\bibinfo {title}
  {Coherent addressing of individual neutral atoms in a 3d optical lattice},}\
  }\href {https://doi.org/10.1103/physrevlett.115.043003} {\bibfield  {journal}
  {\bibinfo  {journal} {Physical Review Letters}\ }\textbf {\bibinfo {volume}
  {115}},\ \bibinfo {pages} {043003}},\ \Eprint
  {https://arxiv.org/abs/arXiv:1504.02117v1} {arXiv:1504.02117v1} \BibitemShut
  {NoStop}%
\bibitem [{\citenamefont {Wang}\ \emph {et~al.}(2023)\citenamefont {Wang},
  \citenamefont {Zhang},\ and\ \citenamefont {Lorenz}}]{wang2023astronomical}%
  \BibitemOpen
  \bibfield  {author} {\bibinfo {author} {\bibnamefont {Wang}, \bibfnamefont
  {Yunkai}}, \bibinfo {author} {\bibfnamefont {Yujie}\ \bibnamefont {Zhang}},
  and\ \bibinfo {author} {\bibfnamefont {Virginia~O}\ \bibnamefont {Lorenz}}}
  (\bibinfo {year} {2023}),\ \bibfield  {title} {\enquote {\bibinfo {title}
  {Astronomical interferometry using continuous variable quantum
  teleportation},}\ }\href@noop {} {\bibinfo  {journal} {arXiv:2308.12851}\
  }\BibitemShut {NoStop}%
\bibitem [{\citenamefont {Watson}\ \emph {et~al.}(2018)\citenamefont {Watson},
  \citenamefont {Philips}, \citenamefont {Kawakami}, \citenamefont {Ward},
  \citenamefont {Scarlino}, \citenamefont {Veldhorst}, \citenamefont {Savage},
  \citenamefont {Lagally}, \citenamefont {Friesen}, \citenamefont {Coppersmith}
  \emph {et~al.}}]{bib:watson2018programmable}%
  \BibitemOpen
\bibfield  {journal} {  }\bibfield  {author} {\bibinfo {author} {\bibnamefont
  {Watson}, \bibfnamefont {TF}}, \bibinfo {author} {\bibfnamefont {SGJ}\
  \bibnamefont {Philips}}, \bibinfo {author} {\bibfnamefont {Erika}\
  \bibnamefont {Kawakami}}, \bibinfo {author} {\bibfnamefont {DR}~\bibnamefont
  {Ward}}, \bibinfo {author} {\bibfnamefont {Pasquale}\ \bibnamefont
  {Scarlino}}, \bibinfo {author} {\bibfnamefont {Menno}\ \bibnamefont
  {Veldhorst}}, \bibinfo {author} {\bibfnamefont {DE}~\bibnamefont {Savage}},
  \bibinfo {author} {\bibfnamefont {MG}~\bibnamefont {Lagally}}, \bibinfo
  {author} {\bibfnamefont {Mark}\ \bibnamefont {Friesen}}, \bibinfo {author}
  {\bibfnamefont {SN}~\bibnamefont {Coppersmith}},  \emph {et~al.}} (\bibinfo
  {year} {2018}),\ \bibfield  {title} {\enquote {\bibinfo {title} {A
  programmable two-qubit quantum processor in silicon},}\ }\href
  {https://doi.org/10.1038/nature25766} {\bibfield  {journal} {\bibinfo
  {journal} {Nature}\ }\textbf {\bibinfo {volume} {555}},\ \bibinfo {pages}
  {633}},\ \Eprint {https://arxiv.org/abs/arXiv:1708.04214v2}
  {arXiv:1708.04214v2} \BibitemShut {NoStop}%
\bibitem [{\citenamefont {Weedbrook}\ \emph {et~al.}(2012)\citenamefont
  {Weedbrook}, \citenamefont {Pirandola}, \citenamefont {Garc\'{\i}a-Patr\'on},
  \citenamefont {Cerf}, \citenamefont {Ralph}, \citenamefont {Shapiro},\ and\
  \citenamefont {Lloyd}}]{bib:RevModPhys.84.621}%
  \BibitemOpen
  \bibfield  {author} {\bibinfo {author} {\bibnamefont {Weedbrook},
  \bibfnamefont {Christian}}, \bibinfo {author} {\bibfnamefont {Stefano}\
  \bibnamefont {Pirandola}}, \bibinfo {author} {\bibfnamefont {Ra\'ul}\
  \bibnamefont {Garc\'{\i}a-Patr\'on}}, \bibinfo {author} {\bibfnamefont
  {Nicolas~J.}\ \bibnamefont {Cerf}}, \bibinfo {author} {\bibfnamefont
  {Timothy~C.}\ \bibnamefont {Ralph}}, \bibinfo {author} {\bibfnamefont
  {Jeffrey~H.}\ \bibnamefont {Shapiro}}, and\ \bibinfo {author} {\bibfnamefont
  {Seth}\ \bibnamefont {Lloyd}}} (\bibinfo {year} {2012}),\ \bibfield  {title}
  {\enquote {\bibinfo {title} {Gaussian quantum information},}\ }\href
  {https://doi.org/10.1103/revmodphys.84.621} {\bibfield  {journal} {\bibinfo
  {journal} {Reviews in Modern Physics}\ }\textbf {\bibinfo {volume} {84}},\
  \bibinfo {pages} {621}},\ \Eprint {https://arxiv.org/abs/arXiv:1110.3234v1}
  {arXiv:1110.3234v1} \BibitemShut {NoStop}%
\bibitem [{\citenamefont {Wehner}\ \emph {et~al.}(2018)\citenamefont {Wehner},
  \citenamefont {Elkouss},\ and\ \citenamefont {Hanson}}]{SD-Wehner:2018aa}%
  \BibitemOpen
  \bibfield  {author} {\bibinfo {author} {\bibnamefont {Wehner}, \bibfnamefont
  {Stephanie}}, \bibinfo {author} {\bibfnamefont {David}\ \bibnamefont
  {Elkouss}}, and\ \bibinfo {author} {\bibfnamefont {Ronald}\ \bibnamefont
  {Hanson}}} (\bibinfo {year} {2018}),\ \bibfield  {title} {\enquote {\bibinfo
  {title} {Quantum internet: A vision for the road ahead},}\ }\href
  {https://doi.org/10.1126/science.aam9288} {\bibfield  {journal} {\bibinfo
  {journal} {Science}\ }\textbf {\bibinfo {volume} {362}},\ \bibinfo {pages}
  {eaam9288}}\BibitemShut {NoStop}%
\bibitem [{\citenamefont {Wei}\ \emph {et~al.}(2014)\citenamefont {Wei},
  \citenamefont {He}, \citenamefont {Chen}, \citenamefont {Hu}, \citenamefont
  {He}, \citenamefont {Wu}, \citenamefont {Schneider}, \citenamefont {Kamp},
  \citenamefont {H{\"o}fling}, \citenamefont {Lu} \emph
  {et~al.}}]{bib:wei2014de}%
  \BibitemOpen
  \bibfield  {author} {\bibinfo {author} {\bibnamefont {Wei}, \bibfnamefont
  {Yu-Jia}}, \bibinfo {author} {\bibfnamefont {Yu-Ming}\ \bibnamefont {He}},
  \bibinfo {author} {\bibfnamefont {Ming-Cheng}\ \bibnamefont {Chen}}, \bibinfo
  {author} {\bibfnamefont {Yi-Nan}\ \bibnamefont {Hu}}, \bibinfo {author}
  {\bibfnamefont {Yu}~\bibnamefont {He}}, \bibinfo {author} {\bibfnamefont
  {Dian}\ \bibnamefont {Wu}}, \bibinfo {author} {\bibfnamefont {Christian}\
  \bibnamefont {Schneider}}, \bibinfo {author} {\bibfnamefont {Martin}\
  \bibnamefont {Kamp}}, \bibinfo {author} {\bibfnamefont {Sven}\ \bibnamefont
  {H{\"o}fling}}, \bibinfo {author} {\bibfnamefont {Chao-Yang}\ \bibnamefont
  {Lu}},  \emph {et~al.}} (\bibinfo {year} {2014}),\ \bibfield  {title}
  {\enquote {\bibinfo {title} {Deterministic and robust generation of single
  photons from a single quantum dot with 99.5\% indistinguishability using
  adiabatic rapid passage},}\ }\href {https://doi.org/10.1021/nl503081n}
  {\bibfield  {journal} {\bibinfo  {journal} {Nanotechnology Letters}\ }\textbf
  {\bibinfo {volume} {14}},\ \bibinfo {pages} {6515}}\BibitemShut {NoStop}%
\bibitem [{\citenamefont {Weier}\ \emph {et~al.}(2011)\citenamefont {Weier},
  \citenamefont {Krauss}, \citenamefont {Rau}, \citenamefont {F{\"u}rst},
  \citenamefont {Nauerth},\ and\ \citenamefont
  {Weinfurter}}]{bib:weier2011quantum}%
  \BibitemOpen
  \bibfield  {author} {\bibinfo {author} {\bibnamefont {Weier}, \bibfnamefont
  {Henning}}, \bibinfo {author} {\bibfnamefont {Harald}\ \bibnamefont
  {Krauss}}, \bibinfo {author} {\bibfnamefont {Markus}\ \bibnamefont {Rau}},
  \bibinfo {author} {\bibfnamefont {Martin}\ \bibnamefont {F{\"u}rst}},
  \bibinfo {author} {\bibfnamefont {Sebastian}\ \bibnamefont {Nauerth}}, and\
  \bibinfo {author} {\bibfnamefont {Harald}\ \bibnamefont {Weinfurter}}}
  (\bibinfo {year} {2011}),\ \bibfield  {title} {\enquote {\bibinfo {title}
  {Quantum eavesdropping without interception: an attack exploiting the dead
  time of single-photon detectors},}\ }\href
  {https://doi.org/10.1088/1367-2630/13/7/073024} {\bibfield  {journal}
  {\bibinfo  {journal} {New Journal of Physics}\ }\textbf {\bibinfo {volume}
  {13}},\ \bibinfo {pages} {073024}}\BibitemShut {NoStop}%
\bibitem [{\citenamefont {Weinfurter}(1994)}]{bib:Euro_25_559}%
  \BibitemOpen
  \bibfield  {author} {\bibinfo {author} {\bibnamefont {Weinfurter},
  \bibfnamefont {Harald}}} (\bibinfo {year} {1994}),\ \bibfield  {title}
  {\enquote {\bibinfo {title} {Experimental bell state analysis},}\ }\href
  {https://doi.org/10.1209/0295-5075/25/8/001} {\bibfield  {journal} {\bibinfo
  {journal} {Europhysics Letters}\ }\textbf {\bibinfo {volume} {25}},\ \bibinfo
  {pages} {559}}\BibitemShut {NoStop}%
\bibitem [{\citenamefont {White}(2012)}]{bib:white2012hadoop}%
  \BibitemOpen
  \bibfield  {author} {\bibinfo {author} {\bibnamefont {White}, \bibfnamefont
  {Tom}}} (\bibinfo {year} {2012}),\ \href@noop {} {\emph {\bibinfo {title}
  {Hadoop: The definitive guide}}}\ (\bibinfo  {publisher}
  {O'Reilly})\BibitemShut {NoStop}%
\bibitem [{\citenamefont {Wiebe}\ \emph {et~al.}(2012)\citenamefont {Wiebe},
  \citenamefont {Braun},\ and\ \citenamefont {Lloyd}}]{wiebe2012quantum}%
  \BibitemOpen
  \bibfield  {author} {\bibinfo {author} {\bibnamefont {Wiebe}, \bibfnamefont
  {Nathan}}, \bibinfo {author} {\bibfnamefont {Daniel}\ \bibnamefont {Braun}},
  and\ \bibinfo {author} {\bibfnamefont {Seth}\ \bibnamefont {Lloyd}}}
  (\bibinfo {year} {2012}),\ \bibfield  {title} {\enquote {\bibinfo {title}
  {Quantum algorithm for data fitting},}\ }\href@noop {} {\bibfield  {journal}
  {\bibinfo  {journal} {Physical review letters}\ }\textbf {\bibinfo {volume}
  {109}}~(\bibinfo {number} {5}),\ \bibinfo {pages} {050505}}\BibitemShut
  {NoStop}%
\bibitem [{\citenamefont {Wiebe}\ and\ \citenamefont
  {Kumar}(2018)}]{bib:wiebe2018hardening}%
  \BibitemOpen
  \bibfield  {author} {\bibinfo {author} {\bibnamefont {Wiebe}, \bibfnamefont
  {Nathan}}, and\ \bibinfo {author} {\bibfnamefont {Ram Shankar~Siva}\
  \bibnamefont {Kumar}}} (\bibinfo {year} {2018}),\ \bibfield  {title}
  {\enquote {\bibinfo {title} {Hardening quantum machine learning against
  adversaries},}\ }\href {https://doi.org/10.1088/1367-2630/aae71a} {\bibfield
  {journal} {\bibinfo  {journal} {New Journal of Physics}\
  }10.1088/1367-2630/aae71a},\ \Eprint
  {https://arxiv.org/abs/arXiv:1711.06652v1} {arXiv:1711.06652v1} \BibitemShut
  {NoStop}%
\bibitem [{\citenamefont {Wikipedia}(2024)}]{sneakernet_web_2}%
  \BibitemOpen
  \bibfield  {author} {\bibinfo {author} {\bibnamefont {Wikipedia},}} (\bibinfo
  {year} {2024}),\ \bibfield  {title} {\enquote {\bibinfo {title} {List of
  international submarine communications cables. (september 11)},}\ }\href@noop
  {} {\ }\BibitemShut {NoStop}%
\bibitem [{\citenamefont {Wilde}(2013)}]{bib:wilde2013quantum}%
  \BibitemOpen
  \bibfield  {author} {\bibinfo {author} {\bibnamefont {Wilde}, \bibfnamefont
  {Mark~M}}} (\bibinfo {year} {2013}),\ \href@noop {} {\emph {\bibinfo {title}
  {Quantum information theory}}}\ (\bibinfo  {publisher} {Cambridge University
  Press})\BibitemShut {NoStop}%
\bibitem [{\citenamefont {Wilk}\ \emph {et~al.}(2007)\citenamefont {Wilk},
  \citenamefont {Webster}, \citenamefont {Kuhn},\ and\ \citenamefont
  {Rempe}}]{bib:wilk2007single}%
  \BibitemOpen
  \bibfield  {author} {\bibinfo {author} {\bibnamefont {Wilk}, \bibfnamefont
  {Tatjana}}, \bibinfo {author} {\bibfnamefont {Simon~C}\ \bibnamefont
  {Webster}}, \bibinfo {author} {\bibfnamefont {Axel}\ \bibnamefont {Kuhn}},
  and\ \bibinfo {author} {\bibfnamefont {Gerhard}\ \bibnamefont {Rempe}}}
  (\bibinfo {year} {2007}),\ \bibfield  {title} {\enquote {\bibinfo {title}
  {Single-atom single-photon quantum interface},}\ }\href
  {https://doi.org/10.1126/science.1143835} {\bibfield  {journal} {\bibinfo
  {journal} {Science}\ }\textbf {\bibinfo {volume} {317}},\ \bibinfo {pages}
  {488}}\BibitemShut {NoStop}%
\bibitem [{\citenamefont {Wu}\ \emph {et~al.}(2018)\citenamefont {Wu},
  \citenamefont {Wang}, \citenamefont {Qin}, \citenamefont {Rong},\ and\
  \citenamefont {Du}}]{bib:wu2018programmable}%
  \BibitemOpen
  \bibfield  {author} {\bibinfo {author} {\bibnamefont {Wu}, \bibfnamefont
  {Yang}}, \bibinfo {author} {\bibfnamefont {Ya}~\bibnamefont {Wang}}, \bibinfo
  {author} {\bibfnamefont {Xi}~\bibnamefont {Qin}}, \bibinfo {author}
  {\bibfnamefont {Xing}\ \bibnamefont {Rong}}, and\ \bibinfo {author}
  {\bibfnamefont {Jiangfeng}\ \bibnamefont {Du}}} (\bibinfo {year} {2018}),\
  \bibfield  {title} {\enquote {\bibinfo {title} {A programmable two-qubit
  solid-state quantum processor under ambient conditions},}\ }\href@noop {} {\
  }\Eprint {https://arxiv.org/abs/arXiv:1808.05733} {arXiv:1808.05733}
  \BibitemShut {NoStop}%
\bibitem [{\citenamefont {Xiang}\ \emph {et~al.}(2012)\citenamefont {Xiang},
  \citenamefont {Hofmann},\ and\ \citenamefont {Pryde}}]{bib:xiang2012optimal}%
  \BibitemOpen
  \bibfield  {author} {\bibinfo {author} {\bibnamefont {Xiang}, \bibfnamefont
  {G~Y}}, \bibinfo {author} {\bibfnamefont {H.~F.}\ \bibnamefont {Hofmann}},
  and\ \bibinfo {author} {\bibfnamefont {G.~J.}\ \bibnamefont {Pryde}}}
  (\bibinfo {year} {2012}),\ \bibfield  {title} {\enquote {\bibinfo {title}
  {Optimal multi-photon phase sensing with a single interference fringe},}\
  }\href {https://doi.org/10.1038/srep02684} {\bibfield  {journal} {\bibinfo
  {journal} {Scientific Reports}\ }\textbf {\bibinfo {volume} {3}},\ \bibinfo
  {pages} {2684}},\ \Eprint {https://arxiv.org/abs/arXiv:1307.1523v1}
  {arXiv:1307.1523v1} \BibitemShut {NoStop}%
\bibitem [{\citenamefont {Xiang}\ \emph {et~al.}(2010)\citenamefont {Xiang},
  \citenamefont {Ralph}, \citenamefont {Lund}, \citenamefont {Walk},\ and\
  \citenamefont {Pryde}}]{bib:xiang2010heralded}%
  \BibitemOpen
  \bibfield  {author} {\bibinfo {author} {\bibnamefont {Xiang}, \bibfnamefont
  {Guo-Yong}}, \bibinfo {author} {\bibfnamefont {TC}~\bibnamefont {Ralph}},
  \bibinfo {author} {\bibfnamefont {AP}~\bibnamefont {Lund}}, \bibinfo {author}
  {\bibfnamefont {N}~\bibnamefont {Walk}}, and\ \bibinfo {author}
  {\bibfnamefont {Geoff~J}\ \bibnamefont {Pryde}}} (\bibinfo {year} {2010}),\
  \bibfield  {title} {\enquote {\bibinfo {title} {Heralded noiseless linear
  amplification and distillation of entanglement},}\ }\href
  {https://doi.org/10.1038/nphoton.2010.35} {\bibfield  {journal} {\bibinfo
  {journal} {Nature Photonics}\ }\textbf {\bibinfo {volume} {4}},\ \bibinfo
  {pages} {316}}\BibitemShut {NoStop}%
\bibitem [{\citenamefont {Xiang}\ \emph {et~al.}(2013)\citenamefont {Xiang},
  \citenamefont {Ashhab}, \citenamefont {You},\ and\ \citenamefont
  {Nori}}]{bib:xiang2013hybrid}%
  \BibitemOpen
  \bibfield  {author} {\bibinfo {author} {\bibnamefont {Xiang}, \bibfnamefont
  {Ze-Liang}}, \bibinfo {author} {\bibfnamefont {Sahel}\ \bibnamefont
  {Ashhab}}, \bibinfo {author} {\bibfnamefont {JQ}~\bibnamefont {You}}, and\
  \bibinfo {author} {\bibfnamefont {Franco}\ \bibnamefont {Nori}}} (\bibinfo
  {year} {2013}),\ \bibfield  {title} {\enquote {\bibinfo {title} {Hybrid
  quantum circuits: Superconducting circuits interacting with other quantum
  systems},}\ }\href {https://doi.org/10.1103/revmodphys.85.623} {\bibfield
  {journal} {\bibinfo  {journal} {Reviews in Modern Physics}\ }\textbf
  {\bibinfo {volume} {85}},\ \bibinfo {pages} {623}},\ \Eprint
  {https://arxiv.org/abs/arXiv:1204.2137v5} {arXiv:1204.2137v5} \BibitemShut
  {NoStop}%
\bibitem [{\citenamefont {Xin}(2011)}]{bib:xin11}%
  \BibitemOpen
  \bibfield  {author} {\bibinfo {author} {\bibnamefont {Xin}, \bibfnamefont
  {Hao}}} (\bibinfo {year} {2011}),\ \bibfield  {title} {\enquote {\bibinfo
  {title} {Chinese academy takes space under its wing},}\ }\href
  {https://doi.org/10.1126/science.332.6032.904} {\bibfield  {journal}
  {\bibinfo  {journal} {Science}\ }\textbf {\bibinfo {volume} {332}},\ \bibinfo
  {pages} {904}}\BibitemShut {NoStop}%
\bibitem [{\citenamefont {Xu}\ \emph {et~al.}(2015{\natexlab{a}})\citenamefont
  {Xu}, \citenamefont {Curty}, \citenamefont {Qi}, \citenamefont {Qian},\ and\
  \citenamefont {Lo}}]{bib:xu2015discrete}%
  \BibitemOpen
  \bibfield  {author} {\bibinfo {author} {\bibnamefont {Xu}, \bibfnamefont
  {Feihu}}, \bibinfo {author} {\bibfnamefont {Marcos}\ \bibnamefont {Curty}},
  \bibinfo {author} {\bibfnamefont {Bing}\ \bibnamefont {Qi}}, \bibinfo
  {author} {\bibfnamefont {Li}~\bibnamefont {Qian}}, and\ \bibinfo {author}
  {\bibfnamefont {Hoi-Kwong}\ \bibnamefont {Lo}}} (\bibinfo {year}
  {2015}{\natexlab{a}}),\ \bibfield  {title} {\enquote {\bibinfo {title}
  {Discrete and continuous variables for measurement-device-independent quantum
  cryptography},}\ }\href {https://doi.org/10.1038/nphoton.2015.206} {\bibfield
   {journal} {\bibinfo  {journal} {Nature Photonics}\ }\textbf {\bibinfo
  {volume} {9}},\ \bibinfo {pages} {772}}\BibitemShut {NoStop}%
\bibitem [{\citenamefont {Xu}\ \emph {et~al.}(2010)\citenamefont {Xu},
  \citenamefont {Qi},\ and\ \citenamefont {Lo}}]{bib:xu2010experimental}%
  \BibitemOpen
  \bibfield  {author} {\bibinfo {author} {\bibnamefont {Xu}, \bibfnamefont
  {Feihu}}, \bibinfo {author} {\bibfnamefont {Bing}\ \bibnamefont {Qi}}, and\
  \bibinfo {author} {\bibfnamefont {Hoi-Kwong}\ \bibnamefont {Lo}}} (\bibinfo
  {year} {2010}),\ \bibfield  {title} {\enquote {\bibinfo {title} {Experimental
  demonstration of phase-remapping attack in a practical quantum key
  distribution system},}\ }\href
  {https://doi.org/10.1088/1367-2630/12/11/113026} {\bibfield  {journal}
  {\bibinfo  {journal} {New Journal of Physics}\ }\textbf {\bibinfo {volume}
  {12}},\ \bibinfo {pages} {113026}}\BibitemShut {NoStop}%
\bibitem [{\citenamefont {Xu}\ \emph {et~al.}(2015{\natexlab{b}})\citenamefont
  {Xu}, \citenamefont {Wei}, \citenamefont {Sajeed}, \citenamefont {Kaiser},
  \citenamefont {Sun}, \citenamefont {Tang}, \citenamefont {Qian},
  \citenamefont {Makarov},\ and\ \citenamefont {Lo}}]{bib:PhysRevA.92.032305}%
  \BibitemOpen
  \bibfield  {author} {\bibinfo {author} {\bibnamefont {Xu}, \bibfnamefont
  {Feihu}}, \bibinfo {author} {\bibfnamefont {Kejin}\ \bibnamefont {Wei}},
  \bibinfo {author} {\bibfnamefont {Shihan}\ \bibnamefont {Sajeed}}, \bibinfo
  {author} {\bibfnamefont {Sarah}\ \bibnamefont {Kaiser}}, \bibinfo {author}
  {\bibfnamefont {Shihai}\ \bibnamefont {Sun}}, \bibinfo {author}
  {\bibfnamefont {Zhiyuan}\ \bibnamefont {Tang}}, \bibinfo {author}
  {\bibfnamefont {Li}~\bibnamefont {Qian}}, \bibinfo {author} {\bibfnamefont
  {Vadim}\ \bibnamefont {Makarov}}, and\ \bibinfo {author} {\bibfnamefont
  {Hoi-Kwong}\ \bibnamefont {Lo}}} (\bibinfo {year} {2015}{\natexlab{b}}),\
  \bibfield  {title} {\enquote {\bibinfo {title} {Experimental quantum key
  distribution with source flaws},}\ }\href
  {https://doi.org/10.1103/physreva.92.032305} {\bibfield  {journal} {\bibinfo
  {journal} {Phys. Rev. A}\ }\textbf {\bibinfo {volume} {92}},\ \bibinfo
  {pages} {032305}},\ \Eprint {https://arxiv.org/abs/arXiv:1408.3667v2}
  {arXiv:1408.3667v2} \BibitemShut {NoStop}%
\bibitem [{\citenamefont {Xu}\ \emph {et~al.}(2013)\citenamefont {Xu},
  \citenamefont {Wu}, \citenamefont {Tian}, \citenamefont {Chen}, \citenamefont
  {Zhang}, \citenamefont {Yan}, \citenamefont {Li}, \citenamefont {Wang},
  \citenamefont {Xie},\ and\ \citenamefont {Peng}}]{bib:xu2013long}%
  \BibitemOpen
  \bibfield  {author} {\bibinfo {author} {\bibnamefont {Xu}, \bibfnamefont
  {Zhongxiao}}, \bibinfo {author} {\bibfnamefont {Yuelong}\ \bibnamefont {Wu}},
  \bibinfo {author} {\bibfnamefont {Long}\ \bibnamefont {Tian}}, \bibinfo
  {author} {\bibfnamefont {Lirong}\ \bibnamefont {Chen}}, \bibinfo {author}
  {\bibfnamefont {Zhiying}\ \bibnamefont {Zhang}}, \bibinfo {author}
  {\bibfnamefont {Zhihui}\ \bibnamefont {Yan}}, \bibinfo {author}
  {\bibfnamefont {Shujing}\ \bibnamefont {Li}}, \bibinfo {author}
  {\bibfnamefont {Hai}\ \bibnamefont {Wang}}, \bibinfo {author} {\bibfnamefont
  {Changde}\ \bibnamefont {Xie}}, and\ \bibinfo {author} {\bibfnamefont
  {Kunchi}\ \bibnamefont {Peng}}} (\bibinfo {year} {2013}),\ \bibfield  {title}
  {\enquote {\bibinfo {title} {Long lifetime and high-fidelity quantum memory
  of photonic polarization qubit by lifting zeeman degeneracy},}\ }\href
  {https://doi.org/10.1103/physrevlett.111.240503} {\bibfield  {journal}
  {\bibinfo  {journal} {Physical Review Letters}\ }\textbf {\bibinfo {volume}
  {111}},\ \bibinfo {pages} {240503}},\ \Eprint
  {https://arxiv.org/abs/arXiv:1306.4262v4} {arXiv:1306.4262v4} \BibitemShut
  {NoStop}%
\bibitem [{\citenamefont {Yang}\ \emph {et~al.}(2016)\citenamefont {Yang},
  \citenamefont {Wang}, \citenamefont {Bao},\ and\ \citenamefont
  {Pan}}]{bib:yang2016efficient}%
  \BibitemOpen
  \bibfield  {author} {\bibinfo {author} {\bibnamefont {Yang}, \bibfnamefont
  {Sheng-Jun}}, \bibinfo {author} {\bibfnamefont {Xu-Jie}\ \bibnamefont
  {Wang}}, \bibinfo {author} {\bibfnamefont {Xiao-Hui}\ \bibnamefont {Bao}},
  and\ \bibinfo {author} {\bibfnamefont {Jian-Wei}\ \bibnamefont {Pan}}}
  (\bibinfo {year} {2016}),\ \bibfield  {title} {\enquote {\bibinfo {title} {An
  efficient quantum light--matter interface with sub-second lifetime},}\ }\href
  {https://doi.org/10.1038/nphoton.2016.51} {\bibfield  {journal} {\bibinfo
  {journal} {Nature Photonics}\ }\textbf {\bibinfo {volume} {10}},\ \bibinfo
  {pages} {381}},\ \Eprint {https://arxiv.org/abs/arXiv:1511.00407v1}
  {arXiv:1511.00407v1} \BibitemShut {NoStop}%
\bibitem [{\citenamefont {Yang}\ \emph {et~al.}(2006)\citenamefont {Yang},
  \citenamefont {Zhang}, \citenamefont {Chen}, \citenamefont {Lu},
  \citenamefont {Yin}, \citenamefont {Pan}, \citenamefont {Wei}, \citenamefont
  {Tian},\ and\ \citenamefont {Zhang}}]{bib:PRL_96_110501}%
  \BibitemOpen
  \bibfield  {author} {\bibinfo {author} {\bibnamefont {Yang}, \bibfnamefont
  {Tao}}, \bibinfo {author} {\bibfnamefont {Qiang}\ \bibnamefont {Zhang}},
  \bibinfo {author} {\bibfnamefont {Teng-Yun}\ \bibnamefont {Chen}}, \bibinfo
  {author} {\bibfnamefont {Shan}\ \bibnamefont {Lu}}, \bibinfo {author}
  {\bibfnamefont {Juan}\ \bibnamefont {Yin}}, \bibinfo {author} {\bibfnamefont
  {Jian-Wei}\ \bibnamefont {Pan}}, \bibinfo {author} {\bibfnamefont {Zhi-Yi}\
  \bibnamefont {Wei}}, \bibinfo {author} {\bibfnamefont {Jing-Rong}\
  \bibnamefont {Tian}}, and\ \bibinfo {author} {\bibfnamefont {Jie}\
  \bibnamefont {Zhang}}} (\bibinfo {year} {2006}),\ \bibfield  {title}
  {\enquote {\bibinfo {title} {Experimental synchronization of independent
  entangled photon sources},}\ }\href
  {https://doi.org/10.1103/physrevlett.96.110501} {\bibfield  {journal}
  {\bibinfo  {journal} {Physical Review Letters}\ }\textbf {\bibinfo {volume}
  {96}},\ \bibinfo {pages} {110501}},\ \Eprint
  {https://arxiv.org/abs/arXiv:quant-ph/0502146v1} {arXiv:quant-ph/0502146v1}
  \BibitemShut {NoStop}%
\bibitem [{\citenamefont {Yao}\ \emph {et~al.}(2012)\citenamefont {Yao},
  \citenamefont {Wang}, \citenamefont {Chen}, \citenamefont {Gao},
  \citenamefont {Fowler}, \citenamefont {Raussendorf}, \citenamefont {Chen},
  \citenamefont {Liu}, \citenamefont {Lu}, \citenamefont {Deng} \emph
  {et~al.}}]{bib:yao2012experimental}%
  \BibitemOpen
  \bibfield  {author} {\bibinfo {author} {\bibnamefont {Yao}, \bibfnamefont
  {Xing-Can}}, \bibinfo {author} {\bibfnamefont {Tian-Xiong}\ \bibnamefont
  {Wang}}, \bibinfo {author} {\bibfnamefont {Hao-Ze}\ \bibnamefont {Chen}},
  \bibinfo {author} {\bibfnamefont {Wei-Bo}\ \bibnamefont {Gao}}, \bibinfo
  {author} {\bibfnamefont {Austin~G}\ \bibnamefont {Fowler}}, \bibinfo {author}
  {\bibfnamefont {Robert}\ \bibnamefont {Raussendorf}}, \bibinfo {author}
  {\bibfnamefont {Zeng-Bing}\ \bibnamefont {Chen}}, \bibinfo {author}
  {\bibfnamefont {Nai-Le}\ \bibnamefont {Liu}}, \bibinfo {author}
  {\bibfnamefont {Chao-Yang}\ \bibnamefont {Lu}}, \bibinfo {author}
  {\bibfnamefont {You-Jin}\ \bibnamefont {Deng}},  \emph {et~al.}} (\bibinfo
  {year} {2012}),\ \bibfield  {title} {\enquote {\bibinfo {title} {Experimental
  demonstration of topological error correction},}\ }\href
  {https://doi.org/10.1038/nature10770} {\bibfield  {journal} {\bibinfo
  {journal} {Nature}\ }\textbf {\bibinfo {volume} {482}},\ \bibinfo {pages}
  {489}},\ \Eprint {https://arxiv.org/abs/arXiv:1202.5459v1}
  {arXiv:1202.5459v1} \BibitemShut {NoStop}%
\bibitem [{\citenamefont {Yavanoglu}\ and\ \citenamefont
  {Aydos}(2017)}]{bib:yavanoglu2017review}%
  \BibitemOpen
  \bibfield  {author} {\bibinfo {author} {\bibnamefont {Yavanoglu},
  \bibfnamefont {Ozlem}}, and\ \bibinfo {author} {\bibfnamefont {Murat}\
  \bibnamefont {Aydos}}} (\bibinfo {year} {2017}),\ \bibfield  {title}
  {\enquote {\bibinfo {title} {A review on cyber security datasets for machine
  learning algorithms},}\ }in\ \href
  {https://doi.org/10.1109/bigdata.2017.8258167} {\emph {\bibinfo {booktitle}
  {IEEE International Conference on Big Data}}},\ p.\ \bibinfo {pages}
  {2186}\BibitemShut {NoStop}%
\bibitem [{\citenamefont {Yin}\ \emph {et~al.}(2016)\citenamefont {Yin},
  \citenamefont {Chen}, \citenamefont {Yu}, \citenamefont {Liu}, \citenamefont
  {You}, \citenamefont {Zhou}, \citenamefont {Chen}, \citenamefont {Mao},
  \citenamefont {Huang}, \citenamefont {Zhang} \emph
  {et~al.}}]{bib:yin2016measurement}%
  \BibitemOpen
  \bibfield  {author} {\bibinfo {author} {\bibnamefont {Yin}, \bibfnamefont
  {Hua-Lei}}, \bibinfo {author} {\bibfnamefont {Teng-Yun}\ \bibnamefont
  {Chen}}, \bibinfo {author} {\bibfnamefont {Zong-Wen}\ \bibnamefont {Yu}},
  \bibinfo {author} {\bibfnamefont {Hui}\ \bibnamefont {Liu}}, \bibinfo
  {author} {\bibfnamefont {Li-Xing}\ \bibnamefont {You}}, \bibinfo {author}
  {\bibfnamefont {Yi-Heng}\ \bibnamefont {Zhou}}, \bibinfo {author}
  {\bibfnamefont {Si-Jing}\ \bibnamefont {Chen}}, \bibinfo {author}
  {\bibfnamefont {Yingqiu}\ \bibnamefont {Mao}}, \bibinfo {author}
  {\bibfnamefont {Ming-Qi}\ \bibnamefont {Huang}}, \bibinfo {author}
  {\bibfnamefont {Wei-Jun}\ \bibnamefont {Zhang}},  \emph {et~al.}} (\bibinfo
  {year} {2016}),\ \bibfield  {title} {\enquote {\bibinfo {title}
  {Measurement-device-independent quantum key distribution over a 404 km
  optical fiber},}\ }\href {https://doi.org/10.1103/physrevlett.117.190501}
  {\bibfield  {journal} {\bibinfo  {journal} {Physical Review Letters}\
  }\textbf {\bibinfo {volume} {117}},\ \bibinfo {pages} {190501}}\BibitemShut
  {NoStop}%
\bibitem [{\citenamefont {Yin}\ \emph {et~al.}(2017{\natexlab{a}})\citenamefont
  {Yin}, \citenamefont {Cao}, \citenamefont {Li}, \citenamefont {Liao},
  \citenamefont {Zhang}, \citenamefont {Ren}, \citenamefont {Cai},
  \citenamefont {Liu}, \citenamefont {Li}, \citenamefont {Dai}, \citenamefont
  {Li}, \citenamefont {Lu}, \citenamefont {Gong}, \citenamefont {Xu},
  \citenamefont {Li}, \citenamefont {Li}, \citenamefont {Yin}, \citenamefont
  {Jiang}, \citenamefont {Li}, \citenamefont {Jia}, \citenamefont {Ren},
  \citenamefont {He}, \citenamefont {Zhou}, \citenamefont {Zhang},
  \citenamefont {Wang}, \citenamefont {Chang}, \citenamefont {Zhu},
  \citenamefont {Liu}, \citenamefont {Chen}, \citenamefont {Lu}, \citenamefont
  {Shu}, \citenamefont {Peng}, \citenamefont {Wang},\ and\ \citenamefont
  {Pan}}]{SD-Yin:2017aa}%
  \BibitemOpen
  \bibfield  {author} {\bibinfo {author} {\bibnamefont {Yin}, \bibfnamefont
  {Juan}}, \bibinfo {author} {\bibfnamefont {Yuan}\ \bibnamefont {Cao}},
  \bibinfo {author} {\bibfnamefont {Yu-Huai}\ \bibnamefont {Li}}, \bibinfo
  {author} {\bibfnamefont {Sheng-Kai}\ \bibnamefont {Liao}}, \bibinfo {author}
  {\bibfnamefont {Liang}\ \bibnamefont {Zhang}}, \bibinfo {author}
  {\bibfnamefont {Ji-Gang}\ \bibnamefont {Ren}}, \bibinfo {author}
  {\bibfnamefont {Wen-Qi}\ \bibnamefont {Cai}}, \bibinfo {author}
  {\bibfnamefont {Wei-Yue}\ \bibnamefont {Liu}}, \bibinfo {author}
  {\bibfnamefont {Bo}~\bibnamefont {Li}}, \bibinfo {author} {\bibfnamefont
  {Hui}\ \bibnamefont {Dai}}, \bibinfo {author} {\bibfnamefont {Guang-Bing}\
  \bibnamefont {Li}}, \bibinfo {author} {\bibfnamefont {Qi-Ming}\ \bibnamefont
  {Lu}}, \bibinfo {author} {\bibfnamefont {Yun-Hong}\ \bibnamefont {Gong}},
  \bibinfo {author} {\bibfnamefont {Yu}~\bibnamefont {Xu}}, \bibinfo {author}
  {\bibfnamefont {Shuang-Lin}\ \bibnamefont {Li}}, \bibinfo {author}
  {\bibfnamefont {Feng-Zhi}\ \bibnamefont {Li}}, \bibinfo {author}
  {\bibfnamefont {Ya-Yun}\ \bibnamefont {Yin}}, \bibinfo {author}
  {\bibfnamefont {Zi-Qing}\ \bibnamefont {Jiang}}, \bibinfo {author}
  {\bibfnamefont {Ming}\ \bibnamefont {Li}}, \bibinfo {author} {\bibfnamefont
  {Jian-Jun}\ \bibnamefont {Jia}}, \bibinfo {author} {\bibfnamefont
  {Ge}~\bibnamefont {Ren}}, \bibinfo {author} {\bibfnamefont {Dong}\
  \bibnamefont {He}}, \bibinfo {author} {\bibfnamefont {Yi-Lin}\ \bibnamefont
  {Zhou}}, \bibinfo {author} {\bibfnamefont {Xiao-Xiang}\ \bibnamefont
  {Zhang}}, \bibinfo {author} {\bibfnamefont {Na}~\bibnamefont {Wang}},
  \bibinfo {author} {\bibfnamefont {Xiang}\ \bibnamefont {Chang}}, \bibinfo
  {author} {\bibfnamefont {Zhen-Cai}\ \bibnamefont {Zhu}}, \bibinfo {author}
  {\bibfnamefont {Nai-Le}\ \bibnamefont {Liu}}, \bibinfo {author}
  {\bibfnamefont {Yu-Ao}\ \bibnamefont {Chen}}, \bibinfo {author}
  {\bibfnamefont {Chao-Yang}\ \bibnamefont {Lu}}, \bibinfo {author}
  {\bibfnamefont {Rong}\ \bibnamefont {Shu}}, \bibinfo {author} {\bibfnamefont
  {Cheng-Zhi}\ \bibnamefont {Peng}}, \bibinfo {author} {\bibfnamefont
  {Jian-Yu}\ \bibnamefont {Wang}}, and\ \bibinfo {author} {\bibfnamefont
  {Jian-Wei}\ \bibnamefont {Pan}}} (\bibinfo {year} {2017}{\natexlab{a}}),\
  \bibfield  {title} {\enquote {\bibinfo {title} {Satellite-based entanglement
  distribution over 1200 kilometers},}\ }\href
  {https://doi.org/10.1126/science.aan3211} {\bibfield  {journal} {\bibinfo
  {journal} {Science}\ }\textbf {\bibinfo {volume} {356}},\ \bibinfo {pages}
  {1140}}\BibitemShut {NoStop}%
\bibitem [{\citenamefont {Yin}\ \emph {et~al.}(2017{\natexlab{b}})\citenamefont
  {Yin}, \citenamefont {Cao}, \citenamefont {Li}, \citenamefont {Liao},
  \citenamefont {Zhang}, \citenamefont {Ren}, \citenamefont {Cai},
  \citenamefont {Liu}, \citenamefont {Li}, \citenamefont {Dai} \emph
  {et~al.}}]{yin2017satellite}%
  \BibitemOpen
  \bibfield  {author} {\bibinfo {author} {\bibnamefont {Yin}, \bibfnamefont
  {Juan}}, \bibinfo {author} {\bibfnamefont {Yuan}\ \bibnamefont {Cao}},
  \bibinfo {author} {\bibfnamefont {Yu-Huai}\ \bibnamefont {Li}}, \bibinfo
  {author} {\bibfnamefont {Sheng-Kai}\ \bibnamefont {Liao}}, \bibinfo {author}
  {\bibfnamefont {Liang}\ \bibnamefont {Zhang}}, \bibinfo {author}
  {\bibfnamefont {Ji-Gang}\ \bibnamefont {Ren}}, \bibinfo {author}
  {\bibfnamefont {Wen-Qi}\ \bibnamefont {Cai}}, \bibinfo {author}
  {\bibfnamefont {Wei-Yue}\ \bibnamefont {Liu}}, \bibinfo {author}
  {\bibfnamefont {Bo}~\bibnamefont {Li}}, \bibinfo {author} {\bibfnamefont
  {Hui}\ \bibnamefont {Dai}},  \emph {et~al.}} (\bibinfo {year}
  {2017}{\natexlab{b}}),\ \bibfield  {title} {\enquote {\bibinfo {title}
  {Satellite-based entanglement distribution over 1200 kilometers},}\ }\href
  {https://doi.org/10.1126/science.aan3211} {\bibfield  {journal} {\bibinfo
  {journal} {Science}\ }\textbf {\bibinfo {volume} {356}},\ \bibinfo {pages}
  {1140}}\BibitemShut {NoStop}%
\bibitem [{\citenamefont {Yin}\ \emph {et~al.}(2013)\citenamefont {Yin},
  \citenamefont {Cao}, \citenamefont {Yong}, \citenamefont {Ren}, \citenamefont
  {Liang}, \citenamefont {Liao}, \citenamefont {Zhou}, \citenamefont {Liu},
  \citenamefont {Wu}, \citenamefont {Pan} \emph {et~al.}}]{bib:yin2013lower}%
  \BibitemOpen
  \bibfield  {author} {\bibinfo {author} {\bibnamefont {Yin}, \bibfnamefont
  {Juan}}, \bibinfo {author} {\bibfnamefont {Yuan}\ \bibnamefont {Cao}},
  \bibinfo {author} {\bibfnamefont {Hai-Lin}\ \bibnamefont {Yong}}, \bibinfo
  {author} {\bibfnamefont {Ji-Gang}\ \bibnamefont {Ren}}, \bibinfo {author}
  {\bibfnamefont {Hao}\ \bibnamefont {Liang}}, \bibinfo {author} {\bibfnamefont
  {Sheng-Kai}\ \bibnamefont {Liao}}, \bibinfo {author} {\bibfnamefont {Fei}\
  \bibnamefont {Zhou}}, \bibinfo {author} {\bibfnamefont {Chang}\ \bibnamefont
  {Liu}}, \bibinfo {author} {\bibfnamefont {Yu-Ping}\ \bibnamefont {Wu}},
  \bibinfo {author} {\bibfnamefont {Ge-Sheng}\ \bibnamefont {Pan}},  \emph
  {et~al.}} (\bibinfo {year} {2013}),\ \bibfield  {title} {\enquote {\bibinfo
  {title} {Lower bound on the speed of nonlocal correlations without locality
  and measurement choice loopholes},}\ }\href
  {https://doi.org/10.1103/physrevlett.110.260407} {\bibfield  {journal}
  {\bibinfo  {journal} {Physical Review Letters}\ }\textbf {\bibinfo {volume}
  {110}},\ \bibinfo {pages} {260407}}\BibitemShut {NoStop}%
\bibitem [{\citenamefont {Yin}\ \emph {et~al.}(2012)\citenamefont {Yin},
  \citenamefont {Ren}, \citenamefont {Lu}, \citenamefont {Cao}, \citenamefont
  {Yong}, \citenamefont {Wu}, \citenamefont {Liu}, \citenamefont {Liao},
  \citenamefont {Zhou}, \citenamefont {Jiang} \emph
  {et~al.}}]{bib:Nat_488_185}%
  \BibitemOpen
  \bibfield  {author} {\bibinfo {author} {\bibnamefont {Yin}, \bibfnamefont
  {Juan}}, \bibinfo {author} {\bibfnamefont {Ji-Gang}\ \bibnamefont {Ren}},
  \bibinfo {author} {\bibfnamefont {He}~\bibnamefont {Lu}}, \bibinfo {author}
  {\bibfnamefont {Yuan}\ \bibnamefont {Cao}}, \bibinfo {author} {\bibfnamefont
  {Hai-Lin}\ \bibnamefont {Yong}}, \bibinfo {author} {\bibfnamefont {Yu-Ping}\
  \bibnamefont {Wu}}, \bibinfo {author} {\bibfnamefont {Chang}\ \bibnamefont
  {Liu}}, \bibinfo {author} {\bibfnamefont {Sheng-Kai}\ \bibnamefont {Liao}},
  \bibinfo {author} {\bibfnamefont {Fei}\ \bibnamefont {Zhou}}, \bibinfo
  {author} {\bibfnamefont {Yan}\ \bibnamefont {Jiang}},  \emph {et~al.}}
  (\bibinfo {year} {2012}),\ \bibfield  {title} {\enquote {\bibinfo {title}
  {Quantum teleportation and entanglement distribution over 100-kilometre
  free-space channels},}\ }\href {https://doi.org/10.1038/nature11332}
  {\bibfield  {journal} {\bibinfo  {journal} {Nature}\ }\textbf {\bibinfo
  {volume} {488}},\ \bibinfo {pages} {185}},\ \Eprint
  {https://arxiv.org/abs/arXiv:1205.2024v2} {arXiv:1205.2024v2} \BibitemShut
  {NoStop}%
\bibitem [{\citenamefont {Yokoyama}\ \emph {et~al.}(2013)\citenamefont
  {Yokoyama}, \citenamefont {Ukai}, \citenamefont {Armstrong}, \citenamefont
  {Sornphiphatphong}, \citenamefont {Kaji}, \citenamefont {Suzuki},
  \citenamefont {Yoshikawa}, \citenamefont {Yonezawa}, \citenamefont
  {Menicucci},\ and\ \citenamefont {Furusawa}}]{bib:yokoyama2013ultra}%
  \BibitemOpen
  \bibfield  {author} {\bibinfo {author} {\bibnamefont {Yokoyama},
  \bibfnamefont {Shota}}, \bibinfo {author} {\bibfnamefont {Ryuji}\
  \bibnamefont {Ukai}}, \bibinfo {author} {\bibfnamefont {Seiji~C}\
  \bibnamefont {Armstrong}}, \bibinfo {author} {\bibfnamefont {Chanond}\
  \bibnamefont {Sornphiphatphong}}, \bibinfo {author} {\bibfnamefont
  {Toshiyuki}\ \bibnamefont {Kaji}}, \bibinfo {author} {\bibfnamefont
  {Shigenari}\ \bibnamefont {Suzuki}}, \bibinfo {author} {\bibfnamefont
  {Jun-ichi}\ \bibnamefont {Yoshikawa}}, \bibinfo {author} {\bibfnamefont
  {Hidehiro}\ \bibnamefont {Yonezawa}}, \bibinfo {author} {\bibfnamefont
  {Nicolas~C}\ \bibnamefont {Menicucci}}, and\ \bibinfo {author} {\bibfnamefont
  {Akira}\ \bibnamefont {Furusawa}}} (\bibinfo {year} {2013}),\ \bibfield
  {title} {\enquote {\bibinfo {title} {Ultra-large-scale continuous-variable
  cluster states multiplexed in the time domain},}\ }\href
  {https://doi.org/10.1038/nphoton.2013.287} {\bibfield  {journal} {\bibinfo
  {journal} {Nature Photonics}\ }\textbf {\bibinfo {volume} {7}},\ \bibinfo
  {pages} {982}},\ \Eprint {https://arxiv.org/abs/arXiv:1306.3366v2}
  {arXiv:1306.3366v2} \BibitemShut {NoStop}%
\bibitem [{\citenamefont {Yoo}\ \emph {et~al.}(2014)\citenamefont {Yoo},
  \citenamefont {Bang}, \citenamefont {Lee},\ and\ \citenamefont
  {Lee}}]{bib:yoo2014quantum}%
  \BibitemOpen
  \bibfield  {author} {\bibinfo {author} {\bibnamefont {Yoo}, \bibfnamefont
  {Seokwon}}, \bibinfo {author} {\bibfnamefont {Jeongho}\ \bibnamefont {Bang}},
  \bibinfo {author} {\bibfnamefont {Changhyoup}\ \bibnamefont {Lee}}, and\
  \bibinfo {author} {\bibfnamefont {Jinhyoung}\ \bibnamefont {Lee}}} (\bibinfo
  {year} {2014}),\ \bibfield  {title} {\enquote {\bibinfo {title} {A quantum
  speedup in machine learning: finding an n-bit boolean function for a
  classification},}\ }\href {https://doi.org/10.1088/1367-2630/16/10/103014}
  {\bibfield  {journal} {\bibinfo  {journal} {New Journal of Physics}\ }\textbf
  {\bibinfo {volume} {16}},\ \bibinfo {pages} {103014}},\ \Eprint
  {https://arxiv.org/abs/arXiv:1303.6055v4} {arXiv:1303.6055v4} \BibitemShut
  {NoStop}%
\bibitem [{\citenamefont {Yoran}\ and\ \citenamefont
  {Reznik}(2003)}]{bib:YoranReznik03}%
  \BibitemOpen
  \bibfield  {author} {\bibinfo {author} {\bibnamefont {Yoran}, \bibfnamefont
  {N}}, and\ \bibinfo {author} {\bibfnamefont {B.}~\bibnamefont {Reznik}}}
  (\bibinfo {year} {2003}),\ \bibfield  {title} {\enquote {\bibinfo {title}
  {Deterministic linear optics quantum computation with single photon
  qubits},}\ }\href {https://doi.org/10.1103/physrevlett.91.037903} {\bibfield
  {journal} {\bibinfo  {journal} {Physical Review Letters}\ }\textbf {\bibinfo
  {volume} {91}},\ \bibinfo {pages} {037903}}\BibitemShut {NoStop}%
\bibitem [{\citenamefont {Yoshikawa}\ \emph {et~al.}(2016)\citenamefont
  {Yoshikawa}, \citenamefont {Yokoyama}, \citenamefont {Kaji}, \citenamefont
  {Sornphiphatphong}, \citenamefont {Shiozawa}, \citenamefont {Makino},\ and\
  \citenamefont {Furusawa}}]{bib:yoshikawa2016invited}%
  \BibitemOpen
  \bibfield  {author} {\bibinfo {author} {\bibnamefont {Yoshikawa},
  \bibfnamefont {Jun-ichi}}, \bibinfo {author} {\bibfnamefont {Shota}\
  \bibnamefont {Yokoyama}}, \bibinfo {author} {\bibfnamefont {Toshiyuki}\
  \bibnamefont {Kaji}}, \bibinfo {author} {\bibfnamefont {Chanond}\
  \bibnamefont {Sornphiphatphong}}, \bibinfo {author} {\bibfnamefont
  {Yu}~\bibnamefont {Shiozawa}}, \bibinfo {author} {\bibfnamefont {Kenzo}\
  \bibnamefont {Makino}}, and\ \bibinfo {author} {\bibfnamefont {Akira}\
  \bibnamefont {Furusawa}}} (\bibinfo {year} {2016}),\ \bibfield  {title}
  {\enquote {\bibinfo {title} {Invited article: Generation of one-million-mode
  continuous-variable cluster state by unlimited time-domain multiplexing},}\
  }\href {https://doi.org/10.1063/1.4962732} {\bibfield  {journal} {\bibinfo
  {journal} {APL Photonics}\ }\textbf {\bibinfo {volume} {1}},\ \bibinfo
  {pages} {060801}}\BibitemShut {NoStop}%
\bibitem [{\citenamefont {You}\ \emph {et~al.}(2005)\citenamefont {You},
  \citenamefont {Nakamura},\ and\ \citenamefont {Nori}}]{bib:you2005fast}%
  \BibitemOpen
  \bibfield  {author} {\bibinfo {author} {\bibnamefont {You}, \bibfnamefont
  {JQ}}, \bibinfo {author} {\bibfnamefont {Y}~\bibnamefont {Nakamura}}, and\
  \bibinfo {author} {\bibfnamefont {Franco}\ \bibnamefont {Nori}}} (\bibinfo
  {year} {2005}),\ \bibfield  {title} {\enquote {\bibinfo {title} {Fast two-bit
  operations in inductively coupled flux qubits},}\ }\href
  {https://doi.org/10.1103/physrevb.71.024532} {\bibfield  {journal} {\bibinfo
  {journal} {Physical Review B}\ }\textbf {\bibinfo {volume} {71}},\ \bibinfo
  {pages} {024532}},\ \Eprint {https://arxiv.org/abs/arXiv:cond-mat/0309491v3}
  {arXiv:cond-mat/0309491v3} \BibitemShut {NoStop}%
\bibitem [{\citenamefont {Yu}\ \emph {et~al.}(2014)\citenamefont {Yu},
  \citenamefont {P{\'e}rez-Delgado},\ and\ \citenamefont
  {Fitzsimons}}]{yu2014limitations}%
  \BibitemOpen
  \bibfield  {author} {\bibinfo {author} {\bibnamefont {Yu}, \bibfnamefont
  {Li}}, \bibinfo {author} {\bibfnamefont {Carlos~A}\ \bibnamefont
  {P{\'e}rez-Delgado}}, and\ \bibinfo {author} {\bibfnamefont {Joseph~F}\
  \bibnamefont {Fitzsimons}}} (\bibinfo {year} {2014}),\ \bibfield  {title}
  {\enquote {\bibinfo {title} {Limitations on information-theoretically-secure
  quantum homomorphic encryption},}\ }\href@noop {} {\bibfield  {journal}
  {\bibinfo  {journal} {Physical Review A}\ }\textbf {\bibinfo {volume}
  {90}}~(\bibinfo {number} {5}),\ \bibinfo {pages} {050303}}\BibitemShut
  {NoStop}%
\bibitem [{\citenamefont {Yu}\ \emph {et~al.}(2018)\citenamefont {Yu},
  \citenamefont {Huang}, \citenamefont {Tang}, \citenamefont {Jia},
  \citenamefont {Wang}, \citenamefont {Ke}, \citenamefont {Liu}, \citenamefont
  {Liu}, \citenamefont {Zhou}, \citenamefont {Cheng}, \citenamefont {Xu},
  \citenamefont {Wu}, \citenamefont {Zhao}, \citenamefont {Xiang},
  \citenamefont {Li}, \citenamefont {Guo}, \citenamefont {Sentís},\ and\
  \citenamefont {Muñoz-Tapia}}]{bib:gael2}%
  \BibitemOpen
  \bibfield  {author} {\bibinfo {author} {\bibnamefont {Yu}, \bibfnamefont
  {Shang}}, \bibinfo {author} {\bibfnamefont {Chang-Jiang}\ \bibnamefont
  {Huang}}, \bibinfo {author} {\bibfnamefont {Jian-Shun}\ \bibnamefont {Tang}},
  \bibinfo {author} {\bibfnamefont {Zhih-Ahn}\ \bibnamefont {Jia}}, \bibinfo
  {author} {\bibfnamefont {Yi-Tao}\ \bibnamefont {Wang}}, \bibinfo {author}
  {\bibfnamefont {Zhi-Jin}\ \bibnamefont {Ke}}, \bibinfo {author}
  {\bibfnamefont {Wei}\ \bibnamefont {Liu}}, \bibinfo {author} {\bibfnamefont
  {Xiao}\ \bibnamefont {Liu}}, \bibinfo {author} {\bibfnamefont {Zong-Quan}\
  \bibnamefont {Zhou}}, \bibinfo {author} {\bibfnamefont {Ze-Di}\ \bibnamefont
  {Cheng}}, \bibinfo {author} {\bibfnamefont {Jin-Shi}\ \bibnamefont {Xu}},
  \bibinfo {author} {\bibfnamefont {Yu-Chun}\ \bibnamefont {Wu}}, \bibinfo
  {author} {\bibfnamefont {Yuan-Yuan}\ \bibnamefont {Zhao}}, \bibinfo {author}
  {\bibfnamefont {Guo-Yong}\ \bibnamefont {Xiang}}, \bibinfo {author}
  {\bibfnamefont {Chuan-Feng}\ \bibnamefont {Li}}, \bibinfo {author}
  {\bibfnamefont {Guang-Can}\ \bibnamefont {Guo}}, \bibinfo {author}
  {\bibfnamefont {Gael}\ \bibnamefont {Sentís}}, and\ \bibinfo {author}
  {\bibfnamefont {Ramon}\ \bibnamefont {Muñoz-Tapia}}} (\bibinfo {year}
  {2018}),\ \bibfield  {title} {\enquote {\bibinfo {title} {Experimentally
  detecting a quantum change point via the bayesian inference},}\ }\href
  {https://doi.org/10.1103/physreva.98.040301} {\bibfield  {journal} {\bibinfo
  {journal} {Physical Review A}\ }\textbf {\bibinfo {volume} {98}}~(\bibinfo
  {number} {4}),\ \bibinfo {pages} {040301}},\ \Eprint
  {https://arxiv.org/abs/arXiv:1801.07508v1} {arXiv:1801.07508v1} \BibitemShut
  {NoStop}%
\bibitem [{\citenamefont {Yuan}\ \emph {et~al.}(2008)\citenamefont {Yuan},
  \citenamefont {Chen}, \citenamefont {Zhao}, \citenamefont {Chen},
  \citenamefont {Schmiedmayer},\ and\ \citenamefont
  {Pan}}]{bib:Nature_454_1098}%
  \BibitemOpen
  \bibfield  {author} {\bibinfo {author} {\bibnamefont {Yuan}, \bibfnamefont
  {Zhen-Sheng}}, \bibinfo {author} {\bibfnamefont {Yu-Ao}\ \bibnamefont
  {Chen}}, \bibinfo {author} {\bibfnamefont {Bo}~\bibnamefont {Zhao}}, \bibinfo
  {author} {\bibfnamefont {Shuai}\ \bibnamefont {Chen}}, \bibinfo {author}
  {\bibfnamefont {J{\"o}rg}\ \bibnamefont {Schmiedmayer}}, and\ \bibinfo
  {author} {\bibfnamefont {Jian-Wei}\ \bibnamefont {Pan}}} (\bibinfo {year}
  {2008}),\ \bibfield  {title} {\enquote {\bibinfo {title} {Experimental
  demonstration of a bdcz quantum repeater node},}\ }\href
  {https://doi.org/10.1038/nature07241} {\bibfield  {journal} {\bibinfo
  {journal} {Nature}\ }\textbf {\bibinfo {volume} {454}},\ \bibinfo {pages}
  {1098}},\ \Eprint {https://arxiv.org/abs/arXiv:0803.1810v2}
  {arXiv:0803.1810v2} \BibitemShut {NoStop}%
\bibitem [{\citenamefont {Yuan}\ \emph {et~al.}(2010)\citenamefont {Yuan},
  \citenamefont {Dynes},\ and\ \citenamefont {Shields}}]{bib:yuan2010avoiding}%
  \BibitemOpen
  \bibfield  {author} {\bibinfo {author} {\bibnamefont {Yuan}, \bibfnamefont
  {ZL}}, \bibinfo {author} {\bibfnamefont {JF}~\bibnamefont {Dynes}}, and\
  \bibinfo {author} {\bibfnamefont {AJ}~\bibnamefont {Shields}}} (\bibinfo
  {year} {2010}),\ \bibfield  {title} {\enquote {\bibinfo {title} {Avoiding the
  blinding attack in qkd},}\ }\href {https://doi.org/10.1038/nphoton.2010.269}
  {\bibfield  {journal} {\bibinfo  {journal} {Nature Photonics}\ }\textbf
  {\bibinfo {volume} {4}},\ \bibinfo {pages} {800}}\BibitemShut {NoStop}%
\bibitem [{\citenamefont {Yurtsever}\ and\ \citenamefont
  {Dowling}(2002)}]{bib:yurtsever02}%
  \BibitemOpen
  \bibfield  {author} {\bibinfo {author} {\bibnamefont {Yurtsever},
  \bibfnamefont {Ulvi}}, and\ \bibinfo {author} {\bibfnamefont {Jonathan~P.}\
  \bibnamefont {Dowling}}} (\bibinfo {year} {2002}),\ \bibfield  {title}
  {\enquote {\bibinfo {title} {Lorentz-invariant look at quantum
  clock-synchronization protocols based on distributed entanglement},}\ }\href
  {https://doi.org/10.1103/physreva.65.052317} {\bibfield  {journal} {\bibinfo
  {journal} {Physical Review A}\ }\textbf {\bibinfo {volume} {65}},\ \bibinfo
  {pages} {052317}},\ \Eprint {https://arxiv.org/abs/arXiv:quant-ph/0010097v1}
  {arXiv:quant-ph/0010097v1} \BibitemShut {NoStop}%
\bibitem [{\citenamefont {Zalka}(1999)}]{zalka1999grover}%
  \BibitemOpen
  \bibfield  {author} {\bibinfo {author} {\bibnamefont {Zalka}, \bibfnamefont
  {Christof}}} (\bibinfo {year} {1999}),\ \bibfield  {title} {\enquote
  {\bibinfo {title} {Grover’s quantum searching algorithm is optimal},}\
  }\href@noop {} {\bibfield  {journal} {\bibinfo  {journal} {Physical Review
  A}\ }\textbf {\bibinfo {volume} {60}}~(\bibinfo {number} {4}),\ \bibinfo
  {pages} {2746}}\BibitemShut {NoStop}%
\bibitem [{\citenamefont {Zavatta}\ \emph {et~al.}(2002)\citenamefont
  {Zavatta}, \citenamefont {Bellini}, \citenamefont {Ramazza}, \citenamefont
  {Marin},\ and\ \citenamefont {Arecchi}}]{bib:zavatta2002time}%
  \BibitemOpen
  \bibfield  {author} {\bibinfo {author} {\bibnamefont {Zavatta}, \bibfnamefont
  {Alessandro}}, \bibinfo {author} {\bibfnamefont {Marco}\ \bibnamefont
  {Bellini}}, \bibinfo {author} {\bibfnamefont {Pier~Luigi}\ \bibnamefont
  {Ramazza}}, \bibinfo {author} {\bibfnamefont {Francesco}\ \bibnamefont
  {Marin}}, and\ \bibinfo {author} {\bibfnamefont {Fortunato~Tito}\
  \bibnamefont {Arecchi}}} (\bibinfo {year} {2002}),\ \bibfield  {title}
  {\enquote {\bibinfo {title} {Time-domain analysis of quantum states of light:
  noise characterization and homodyne tomography},}\ }\href
  {https://doi.org/10.1364/josab.19.001189} {\bibfield  {journal} {\bibinfo
  {journal} {JOSA B}\ }\textbf {\bibinfo {volume} {19}},\ \bibinfo {pages}
  {1189}}\BibitemShut {NoStop}%
\bibitem [{\citenamefont {Zehnder}(1891)}]{bib:Zehnder1}%
  \BibitemOpen
  \bibfield  {author} {\bibinfo {author} {\bibnamefont {Zehnder}, \bibfnamefont
  {Ludwig}}} (\bibinfo {year} {1891}),\ \bibfield  {title} {\enquote {\bibinfo
  {title} {Ein neuer interferenzrefraktor},}\ }\href@noop {} {\bibfield
  {journal} {\bibinfo  {journal} {Zeitschrift f{\"u}r Instrumentenkunde}\
  }\textbf {\bibinfo {volume} {11}},\ \bibinfo {pages} {275}}\BibitemShut
  {NoStop}%
\bibitem [{\citenamefont {Zehnder}(1892)}]{bib:Zehnder2}%
  \BibitemOpen
  \bibfield  {author} {\bibinfo {author} {\bibnamefont {Zehnder}, \bibfnamefont
  {Ludwig}}} (\bibinfo {year} {1892}),\ \bibfield  {title} {\enquote {\bibinfo
  {title} {{\"U}ber einen interferenzrefraktor},}\ }\href@noop {} {\bibfield
  {journal} {\bibinfo  {journal} {Zeitschrift f{\"u}r Instrumentenkunde}\
  }\textbf {\bibinfo {volume} {12}},\ \bibinfo {pages} {89}}\BibitemShut
  {NoStop}%
\bibitem [{\citenamefont {Zhang}\ \emph {et~al.}(2004)\citenamefont {Zhang},
  \citenamefont {Long}, \citenamefont {Deng}, \citenamefont {Liu},\ and\
  \citenamefont {Lu}}]{bib:zhang2004nuclear}%
  \BibitemOpen
  \bibfield  {author} {\bibinfo {author} {\bibnamefont {Zhang}, \bibfnamefont
  {Jingfu}}, \bibinfo {author} {\bibfnamefont {Gui~Lu}\ \bibnamefont {Long}},
  \bibinfo {author} {\bibfnamefont {Zhiwei}\ \bibnamefont {Deng}}, \bibinfo
  {author} {\bibfnamefont {Wenzhang}\ \bibnamefont {Liu}}, and\ \bibinfo
  {author} {\bibfnamefont {Zhiheng}\ \bibnamefont {Lu}}} (\bibinfo {year}
  {2004}),\ \bibfield  {title} {\enquote {\bibinfo {title} {Nuclear magnetic
  resonance implementation of a quantum clock synchronization algorithm},}\
  }\href {https://doi.org/10.1103/physreva.70.062322} {\bibfield  {journal}
  {\bibinfo  {journal} {Physical Review A}\ }\textbf {\bibinfo {volume} {70}},\
  \bibinfo {pages} {062322}},\ \Eprint
  {https://arxiv.org/abs/arXiv:quant-ph/0406208v2} {arXiv:quant-ph/0406208v2}
  \BibitemShut {NoStop}%
\bibitem [{\citenamefont {Zhang}\ \emph {et~al.}(2015)\citenamefont {Zhang},
  \citenamefont {Chen}, \citenamefont {Xiang}, \citenamefont {Zhou},\ and\
  \citenamefont {Wu}}]{bib:zhang2015robust}%
  \BibitemOpen
  \bibfield  {author} {\bibinfo {author} {\bibnamefont {Zhang}, \bibfnamefont
  {Jun}}, \bibinfo {author} {\bibfnamefont {Xiao}\ \bibnamefont {Chen}},
  \bibinfo {author} {\bibfnamefont {Yang}\ \bibnamefont {Xiang}}, \bibinfo
  {author} {\bibfnamefont {Wanlei}\ \bibnamefont {Zhou}}, and\ \bibinfo
  {author} {\bibfnamefont {Jie}\ \bibnamefont {Wu}}} (\bibinfo {year} {2015}),\
  \bibfield  {title} {\enquote {\bibinfo {title} {Robust network traffic
  classification},}\ }\href {https://doi.org/10.1109/tnet.2014.2320577}
  {\bibfield  {journal} {\bibinfo  {journal} {IEEE/ACM Transactions on
  Networking (TON)}\ }\textbf {\bibinfo {volume} {23}},\ \bibinfo {pages}
  {1257}}\BibitemShut {NoStop}%
\bibitem [{\citenamefont {Zhang}\ \emph {et~al.}(2006)\citenamefont {Zhang},
  \citenamefont {Goebel}, \citenamefont {Wagenknecht}, \citenamefont {Chen},
  \citenamefont {Zhao}, \citenamefont {Yang}, \citenamefont {Mair},
  \citenamefont {Schmiedmayer},\ and\ \citenamefont
  {Pan}}]{bib:Nat_Phys_2_678}%
  \BibitemOpen
  \bibfield  {author} {\bibinfo {author} {\bibnamefont {Zhang}, \bibfnamefont
  {Qiang}}, \bibinfo {author} {\bibfnamefont {Alexander}\ \bibnamefont
  {Goebel}}, \bibinfo {author} {\bibfnamefont {Claudia}\ \bibnamefont
  {Wagenknecht}}, \bibinfo {author} {\bibfnamefont {Yu-Ao}\ \bibnamefont
  {Chen}}, \bibinfo {author} {\bibfnamefont {Bo}~\bibnamefont {Zhao}}, \bibinfo
  {author} {\bibfnamefont {Tao}\ \bibnamefont {Yang}}, \bibinfo {author}
  {\bibfnamefont {Alois}\ \bibnamefont {Mair}}, \bibinfo {author}
  {\bibfnamefont {J{\"o}rg}\ \bibnamefont {Schmiedmayer}}, and\ \bibinfo
  {author} {\bibfnamefont {Jian-Wei}\ \bibnamefont {Pan}}} (\bibinfo {year}
  {2006}),\ \bibfield  {title} {\enquote {\bibinfo {title} {Experimental
  quantum teleportation of a 2-qubit composite system},}\ }\href@noop {}
  {\bibfield  {journal} {\bibinfo  {journal} {Nature Physics}\ }\textbf
  {\bibinfo {volume} {2}},\ \bibinfo {pages} {678}}\BibitemShut {NoStop}%
\bibitem [{\citenamefont {Zhang}\ \emph {et~al.}(2008)\citenamefont {Zhang},
  \citenamefont {Mo}, \citenamefont {Han},\ and\ \citenamefont
  {Guo}}]{bib:PLA_372_3957}%
  \BibitemOpen
  \bibfield  {author} {\bibinfo {author} {\bibnamefont {Zhang}, \bibfnamefont
  {Tao}}, \bibinfo {author} {\bibfnamefont {Xiao-Fan}\ \bibnamefont {Mo}},
  \bibinfo {author} {\bibfnamefont {Zheng-Fu}\ \bibnamefont {Han}}, and\
  \bibinfo {author} {\bibfnamefont {Guang-Can}\ \bibnamefont {Guo}}} (\bibinfo
  {year} {2008}),\ \bibfield  {title} {\enquote {\bibinfo {title} {Extensible
  router for a quantum key distribution network},}\ }\href
  {https://doi.org/10.1016/j.physleta.2008.03.023} {\bibfield  {journal}
  {\bibinfo  {journal} {Physics Letters A}\ }\textbf {\bibinfo {volume}
  {372}},\ \bibinfo {pages} {3957}}\BibitemShut {NoStop}%
\bibitem [{\citenamefont {Zhang}\ \emph {et~al.}(2005)\citenamefont {Zhang},
  \citenamefont {Li},\ and\ \citenamefont {Man}}]{bib:PhysRevA.71.044301}%
  \BibitemOpen
  \bibfield  {author} {\bibinfo {author} {\bibnamefont {Zhang}, \bibfnamefont
  {Zhan-jun}}, \bibinfo {author} {\bibfnamefont {Yong}\ \bibnamefont {Li}},
  and\ \bibinfo {author} {\bibfnamefont {Zhong-xiao}\ \bibnamefont {Man}}}
  (\bibinfo {year} {2005}),\ \bibfield  {title} {\enquote {\bibinfo {title}
  {Multiparty quantum secret sharing},}\ }\href
  {https://doi.org/10.1103/physreva.71.044301} {\bibfield  {journal} {\bibinfo
  {journal} {Physical Review A}\ }\textbf {\bibinfo {volume} {71}},\ \bibinfo
  {pages} {044301}},\ \Eprint {https://arxiv.org/abs/arXiv:quant-ph/0412203v1}
  {arXiv:quant-ph/0412203v1} \BibitemShut {NoStop}%
\bibitem [{\citenamefont {Zhao}\ \emph {et~al.}(2010)\citenamefont {Zhao},
  \citenamefont {M{\"u}ller}, \citenamefont {Hammerer},\ and\ \citenamefont
  {Zoller}}]{bib:PRA_81_052329}%
  \BibitemOpen
  \bibfield  {author} {\bibinfo {author} {\bibnamefont {Zhao}, \bibfnamefont
  {Bo}}, \bibinfo {author} {\bibfnamefont {Markus}\ \bibnamefont {M{\"u}ller}},
  \bibinfo {author} {\bibfnamefont {Klemens}\ \bibnamefont {Hammerer}}, and\
  \bibinfo {author} {\bibfnamefont {Peter}\ \bibnamefont {Zoller}}} (\bibinfo
  {year} {2010}),\ \bibfield  {title} {\enquote {\bibinfo {title} {Efficient
  quantum repeater based on deterministic rydberg gates},}\ }\href
  {https://doi.org/10.1103/physreva.81.052329} {\bibfield  {journal} {\bibinfo
  {journal} {Physical Review A}\ }\textbf {\bibinfo {volume} {81}},\ \bibinfo
  {pages} {052329}},\ \Eprint {https://arxiv.org/abs/arXiv:1003.1911v1}
  {arXiv:1003.1911v1} \BibitemShut {NoStop}%
\bibitem [{\citenamefont {Zhao}\ \emph {et~al.}(2008)\citenamefont {Zhao},
  \citenamefont {Fung}, \citenamefont {Qi}, \citenamefont {Chen},\ and\
  \citenamefont {Lo}}]{bib:PhysRevA.78.042333}%
  \BibitemOpen
  \bibfield  {author} {\bibinfo {author} {\bibnamefont {Zhao}, \bibfnamefont
  {Yi}}, \bibinfo {author} {\bibfnamefont {Chi-Hang~Fred}\ \bibnamefont
  {Fung}}, \bibinfo {author} {\bibfnamefont {Bing}\ \bibnamefont {Qi}},
  \bibinfo {author} {\bibfnamefont {Christine}\ \bibnamefont {Chen}}, and\
  \bibinfo {author} {\bibfnamefont {Hoi-Kwong}\ \bibnamefont {Lo}}} (\bibinfo
  {year} {2008}),\ \bibfield  {title} {\enquote {\bibinfo {title} {Quantum
  hacking: Experimental demonstration of time-shift attack against practical
  quantum-key-distribution systems},}\ }\href
  {https://doi.org/10.1103/physreva.78.042333} {\bibfield  {journal} {\bibinfo
  {journal} {Physical Review A}\ }\textbf {\bibinfo {volume} {78}},\ \bibinfo
  {pages} {042333}},\ \Eprint {https://arxiv.org/abs/arXiv:0704.3253v3}
  {arXiv:0704.3253v3} \BibitemShut {NoStop}%
\bibitem [{\citenamefont {Zhao}\ \emph
  {et~al.}(2004{\natexlab{a}})\citenamefont {Zhao}, \citenamefont {Chen},
  \citenamefont {Zhang}, \citenamefont {Yang}, \citenamefont {Briegel},\ and\
  \citenamefont {Pan}}]{bib:PanOpenDest}%
  \BibitemOpen
  \bibfield  {author} {\bibinfo {author} {\bibnamefont {Zhao}, \bibfnamefont
  {Zhi}}, \bibinfo {author} {\bibfnamefont {Yu-Ao}\ \bibnamefont {Chen}},
  \bibinfo {author} {\bibfnamefont {An-Ning}\ \bibnamefont {Zhang}}, \bibinfo
  {author} {\bibfnamefont {Tao}\ \bibnamefont {Yang}}, \bibinfo {author}
  {\bibfnamefont {Hans~J.}\ \bibnamefont {Briegel}}, and\ \bibinfo {author}
  {\bibfnamefont {Jian-Wei}\ \bibnamefont {Pan}}} (\bibinfo {year}
  {2004}{\natexlab{a}}),\ \bibfield  {title} {\enquote {\bibinfo {title}
  {Experimental demonstration of five-photon entanglement and open-destination
  teleportation},}\ }\href {https://doi.org/10.1038/nature02643} {\bibfield
  {journal} {\bibinfo  {journal} {Nature}\ }\textbf {\bibinfo {volume} {430}},\
  \bibinfo {pages} {54}},\ \Eprint
  {https://arxiv.org/abs/arXiv:quant-ph/0402096} {arXiv:quant-ph/0402096}
  \BibitemShut {NoStop}%
\bibitem [{\citenamefont {Zhao}\ \emph
  {et~al.}(2004{\natexlab{b}})\citenamefont {Zhao}, \citenamefont {Chen},
  \citenamefont {Zhang}, \citenamefont {Yang}, \citenamefont {Briegel},\ and\
  \citenamefont {Pan}}]{bib:zhao2004experimental}%
  \BibitemOpen
  \bibfield  {author} {\bibinfo {author} {\bibnamefont {Zhao}, \bibfnamefont
  {Zhi}}, \bibinfo {author} {\bibfnamefont {Yu-Ao}\ \bibnamefont {Chen}},
  \bibinfo {author} {\bibfnamefont {An-Ning}\ \bibnamefont {Zhang}}, \bibinfo
  {author} {\bibfnamefont {Tao}\ \bibnamefont {Yang}}, \bibinfo {author}
  {\bibfnamefont {Hans~J}\ \bibnamefont {Briegel}}, and\ \bibinfo {author}
  {\bibfnamefont {Jian-Wei}\ \bibnamefont {Pan}}} (\bibinfo {year}
  {2004}{\natexlab{b}}),\ \bibfield  {title} {\enquote {\bibinfo {title}
  {Experimental demonstration of five-photon entanglement and open-destination
  teleportation},}\ }\href {https://doi.org/10.1038/nature02643} {\bibfield
  {journal} {\bibinfo  {journal} {Nature}\ }\textbf {\bibinfo {volume} {430}},\
  \bibinfo {pages} {54}},\ \Eprint
  {https://arxiv.org/abs/arXiv:quant-ph/0402096v1} {arXiv:quant-ph/0402096v1}
  \BibitemShut {NoStop}%
\bibitem [{\citenamefont {Zhao}\ \emph {et~al.}(2005)\citenamefont {Zhao},
  \citenamefont {Zhang}, \citenamefont {Chen}, \citenamefont {Zhang},
  \citenamefont {Du}, \citenamefont {Yang},\ and\ \citenamefont
  {Pan}}]{bib:Zhao2005Experimental}%
  \BibitemOpen
  \bibfield  {author} {\bibinfo {author} {\bibnamefont {Zhao}, \bibfnamefont
  {Zhi}}, \bibinfo {author} {\bibfnamefont {An-Ning}\ \bibnamefont {Zhang}},
  \bibinfo {author} {\bibfnamefont {Yu-Ao}\ \bibnamefont {Chen}}, \bibinfo
  {author} {\bibfnamefont {Han}\ \bibnamefont {Zhang}}, \bibinfo {author}
  {\bibfnamefont {Jiang-Feng}\ \bibnamefont {Du}}, \bibinfo {author}
  {\bibfnamefont {Tao}\ \bibnamefont {Yang}}, and\ \bibinfo {author}
  {\bibfnamefont {Jian-Wei}\ \bibnamefont {Pan}}} (\bibinfo {year} {2005}),\
  \bibfield  {title} {\enquote {\bibinfo {title} {Experimental demonstration of
  a nondestructive controlled-not quantum gate for two independent photon
  qubits},}\ }\href {https://doi.org/10.1103/PhysRevLett.94.030501} {\bibfield
  {journal} {\bibinfo  {journal} {Physical Review Letters}\ }\textbf {\bibinfo
  {volume} {94}},\ \bibinfo {pages} {030501}},\ \Eprint
  {https://arxiv.org/abs/arXiv:quant-ph/0404129v2} {arXiv:quant-ph/0404129v2}
  \BibitemShut {NoStop}%
\bibitem [{\citenamefont {Zhong}\ \emph {et~al.}(2018)\citenamefont {Zhong},
  \citenamefont {Li}, \citenamefont {Li}, \citenamefont {Peng}, \citenamefont
  {Su}, \citenamefont {Hu}, \citenamefont {He}, \citenamefont {Ding},
  \citenamefont {Zhang}, \citenamefont {Li} \emph {et~al.}}]{bib:zhong201812}%
  \BibitemOpen
  \bibfield  {author} {\bibinfo {author} {\bibnamefont {Zhong}, \bibfnamefont
  {Han-Sen}}, \bibinfo {author} {\bibfnamefont {Yuan}\ \bibnamefont {Li}},
  \bibinfo {author} {\bibfnamefont {Wei}\ \bibnamefont {Li}}, \bibinfo {author}
  {\bibfnamefont {Li-Chao}\ \bibnamefont {Peng}}, \bibinfo {author}
  {\bibfnamefont {Zu-En}\ \bibnamefont {Su}}, \bibinfo {author} {\bibfnamefont
  {Yi}~\bibnamefont {Hu}}, \bibinfo {author} {\bibfnamefont {Yu-Ming}\
  \bibnamefont {He}}, \bibinfo {author} {\bibfnamefont {Xing}\ \bibnamefont
  {Ding}}, \bibinfo {author} {\bibfnamefont {W-J}\ \bibnamefont {Zhang}},
  \bibinfo {author} {\bibfnamefont {Hao}\ \bibnamefont {Li}},  \emph {et~al.}}
  (\bibinfo {year} {2018}),\ \bibfield  {title} {\enquote {\bibinfo {title}
  {12-photon entanglement and scalable scattershot boson sampling with optimal
  entangled-photon pairs from parametric down-conversion},}\ }\href@noop {} {\
  }\Eprint {https://arxiv.org/abs/arXiv:1810.04823} {arXiv:1810.04823}
  \BibitemShut {NoStop}%
\bibitem [{\citenamefont {Zhong}\ \emph
  {et~al.}(2015{\natexlab{a}})\citenamefont {Zhong}, \citenamefont {Hedges},
  \citenamefont {Ahlefeldt}, \citenamefont {Bartholomew}, \citenamefont
  {Beavan}, \citenamefont {Wittig}, \citenamefont {Longdell},\ and\
  \citenamefont {Sellars}}]{SD-Zhong:2015aa}%
  \BibitemOpen
  \bibfield  {author} {\bibinfo {author} {\bibnamefont {Zhong}, \bibfnamefont
  {Manjin}}, \bibinfo {author} {\bibfnamefont {Morgan~P.}\ \bibnamefont
  {Hedges}}, \bibinfo {author} {\bibfnamefont {Rose~L.}\ \bibnamefont
  {Ahlefeldt}}, \bibinfo {author} {\bibfnamefont {John~G.}\ \bibnamefont
  {Bartholomew}}, \bibinfo {author} {\bibfnamefont {Sarah~E.}\ \bibnamefont
  {Beavan}}, \bibinfo {author} {\bibfnamefont {Sven~M.}\ \bibnamefont
  {Wittig}}, \bibinfo {author} {\bibfnamefont {Jevon~J.}\ \bibnamefont
  {Longdell}}, and\ \bibinfo {author} {\bibfnamefont {Matthew~J.}\ \bibnamefont
  {Sellars}}} (\bibinfo {year} {2015}{\natexlab{a}}),\ \bibfield  {title}
  {\enquote {\bibinfo {title} {Optically addressable nuclear spins in a solid
  with a six-hour coherence time},}\ }\href
  {https://doi.org/10.1038/nature14025} {\bibfield  {journal} {\bibinfo
  {journal} {Nature}\ }\textbf {\bibinfo {volume} {517}},\ \bibinfo {pages}
  {177}}\BibitemShut {NoStop}%
\bibitem [{\citenamefont {Zhong}\ \emph
  {et~al.}(2015{\natexlab{b}})\citenamefont {Zhong}, \citenamefont {Hedges},
  \citenamefont {Ahlefeldt}, \citenamefont {Bartholomew}, \citenamefont
  {Beavan}, \citenamefont {Wittig}, \citenamefont {Longdell},\ and\
  \citenamefont {Sellars}}]{bib:zhong2015optically}%
  \BibitemOpen
  \bibfield  {author} {\bibinfo {author} {\bibnamefont {Zhong}, \bibfnamefont
  {Manjin}}, \bibinfo {author} {\bibfnamefont {Morgan~P}\ \bibnamefont
  {Hedges}}, \bibinfo {author} {\bibfnamefont {Rose~L}\ \bibnamefont
  {Ahlefeldt}}, \bibinfo {author} {\bibfnamefont {John~G}\ \bibnamefont
  {Bartholomew}}, \bibinfo {author} {\bibfnamefont {Sarah~E}\ \bibnamefont
  {Beavan}}, \bibinfo {author} {\bibfnamefont {Sven~M}\ \bibnamefont {Wittig}},
  \bibinfo {author} {\bibfnamefont {Jevon~J}\ \bibnamefont {Longdell}}, and\
  \bibinfo {author} {\bibfnamefont {Matthew~J}\ \bibnamefont {Sellars}}}
  (\bibinfo {year} {2015}{\natexlab{b}}),\ \bibfield  {title} {\enquote
  {\bibinfo {title} {Optically addressable nuclear spins in a solid with a
  six-hour coherence time},}\ }\href {https://doi.org/10.1038/nature14025}
  {\bibfield  {journal} {\bibinfo  {journal} {Nature}\ }\textbf {\bibinfo
  {volume} {517}},\ \bibinfo {pages} {177}}\BibitemShut {NoStop}%
\bibitem [{\citenamefont {Zhong}\ \emph
  {et~al.}(2015{\natexlab{c}})\citenamefont {Zhong}, \citenamefont {Zhou},
  \citenamefont {Horansky}, \citenamefont {Lee}, \citenamefont {Verma},
  \citenamefont {Lita}, \citenamefont {Restelli}, \citenamefont {Bienfang},
  \citenamefont {Mirin}, \citenamefont {Gerrits} \emph
  {et~al.}}]{bib:zhong2015photon}%
  \BibitemOpen
  \bibfield  {author} {\bibinfo {author} {\bibnamefont {Zhong}, \bibfnamefont
  {Tian}}, \bibinfo {author} {\bibfnamefont {Hongchao}\ \bibnamefont {Zhou}},
  \bibinfo {author} {\bibfnamefont {Robert~D}\ \bibnamefont {Horansky}},
  \bibinfo {author} {\bibfnamefont {Catherine}\ \bibnamefont {Lee}}, \bibinfo
  {author} {\bibfnamefont {Varun~B}\ \bibnamefont {Verma}}, \bibinfo {author}
  {\bibfnamefont {Adriana~E}\ \bibnamefont {Lita}}, \bibinfo {author}
  {\bibfnamefont {Alessandro}\ \bibnamefont {Restelli}}, \bibinfo {author}
  {\bibfnamefont {Joshua~C}\ \bibnamefont {Bienfang}}, \bibinfo {author}
  {\bibfnamefont {Richard~P}\ \bibnamefont {Mirin}}, \bibinfo {author}
  {\bibfnamefont {Thomas}\ \bibnamefont {Gerrits}},  \emph {et~al.}} (\bibinfo
  {year} {2015}{\natexlab{c}}),\ \bibfield  {title} {\enquote {\bibinfo {title}
  {Photon-efficient quantum key distribution using time--energy entanglement
  with high-dimensional encoding},}\ }\href
  {https://doi.org/10.1088/1367-2630/17/2/022002} {\bibfield  {journal}
  {\bibinfo  {journal} {New Journal of Physics}\ }\textbf {\bibinfo {volume}
  {17}},\ \bibinfo {pages} {022002}}\BibitemShut {NoStop}%
\bibitem [{\citenamefont {Zukowski}\ \emph {et~al.}(1993)\citenamefont
  {Zukowski}, \citenamefont {Zeilinger}, \citenamefont {Horne},\ and\
  \citenamefont {Ekert}}]{bib:Zukowski93}%
  \BibitemOpen
  \bibfield  {author} {\bibinfo {author} {\bibnamefont {Zukowski},
  \bibfnamefont {M}}, \bibinfo {author} {\bibfnamefont {A.}~\bibnamefont
  {Zeilinger}}, \bibinfo {author} {\bibfnamefont {M.~A.}\ \bibnamefont
  {Horne}}, and\ \bibinfo {author} {\bibfnamefont {A.~K.}\ \bibnamefont
  {Ekert}}} (\bibinfo {year} {1993}),\ \bibfield  {title} {\enquote {\bibinfo
  {title} {Event-ready-detectors bell experiment via entanglement swapping},}\
  }\href {https://doi.org/10.1103/physrevlett.71.4287} {\bibfield  {journal}
  {\bibinfo  {journal} {Physical Review Letters}\ }\textbf {\bibinfo {volume}
  {71}},\ \bibinfo {pages} {4287}}\BibitemShut {NoStop}%
\bibitem [{\citenamefont {{\.Z}ukowski}\ \emph {et~al.}(1993)\citenamefont
  {{\.Z}ukowski}, \citenamefont {Zeilinger}, \citenamefont {Horne},\ and\
  \citenamefont {Ekert}}]{SD-Zukowski:1993aa}%
  \BibitemOpen
  \bibfield  {author} {\bibinfo {author} {\bibnamefont {{\.Z}ukowski},
  \bibfnamefont {M}}, \bibinfo {author} {\bibfnamefont {A.}~\bibnamefont
  {Zeilinger}}, \bibinfo {author} {\bibfnamefont {M.~A.}\ \bibnamefont
  {Horne}}, and\ \bibinfo {author} {\bibfnamefont {A.~K.}\ \bibnamefont
  {Ekert}}} (\bibinfo {year} {1993}),\ \bibfield  {title} {\enquote {\bibinfo
  {title} {``event-ready-detectors'' bell experiment via entanglement
  swapping},}\ }\href {https://doi.org/10.1103/PhysRevLett.71.4287} {\bibfield
  {journal} {\bibinfo  {journal} {Physical Review Letters}\ }\textbf {\bibinfo
  {volume} {71}},\ \bibinfo {pages} {4287}}\BibitemShut {NoStop}%
\bibitem [{\citenamefont {Zurek}(2001)}]{Zurek2001subplanck}%
  \BibitemOpen
  \bibfield  {author} {\bibinfo {author} {\bibnamefont {Zurek}, \bibfnamefont
  {Wojciech~Hubert}}} (\bibinfo {year} {2001}),\ \bibfield  {title} {\enquote
  {\bibinfo {title} {Sub-planck structure in phase space and its relevance for
  quantum decoherence},}\ }\href {https://doi.org/10.1038/35089017} {\bibfield
  {journal} {\bibinfo  {journal} {Nature}\ }\textbf {\bibinfo {volume}
  {412}}~(\bibinfo {number} {6848}),\ \bibinfo {pages} {712--717}}\BibitemShut
  {NoStop}%
\end{thebibliography}
%apsrmp4-2.bst 2018-12-27 (MD) hand-edited version of apsrmp4-1.bst
%Control: author (3) reversed first dotless

% \input{Sections/index}
% 
\clearpage
% % \input{Sections/about_the_authors}

% \clearpage
% 

%
 % Dedication
%

% \clearpage
\if 1\doublecol
	\onecolumngrid
\fi

\section{Dedications}%
\begin{center}
	\href{https://www.youtube.com/watch?v=XVWDQwsMnEE}{\includegraphics[clip=true, width=0.7\textwidth]{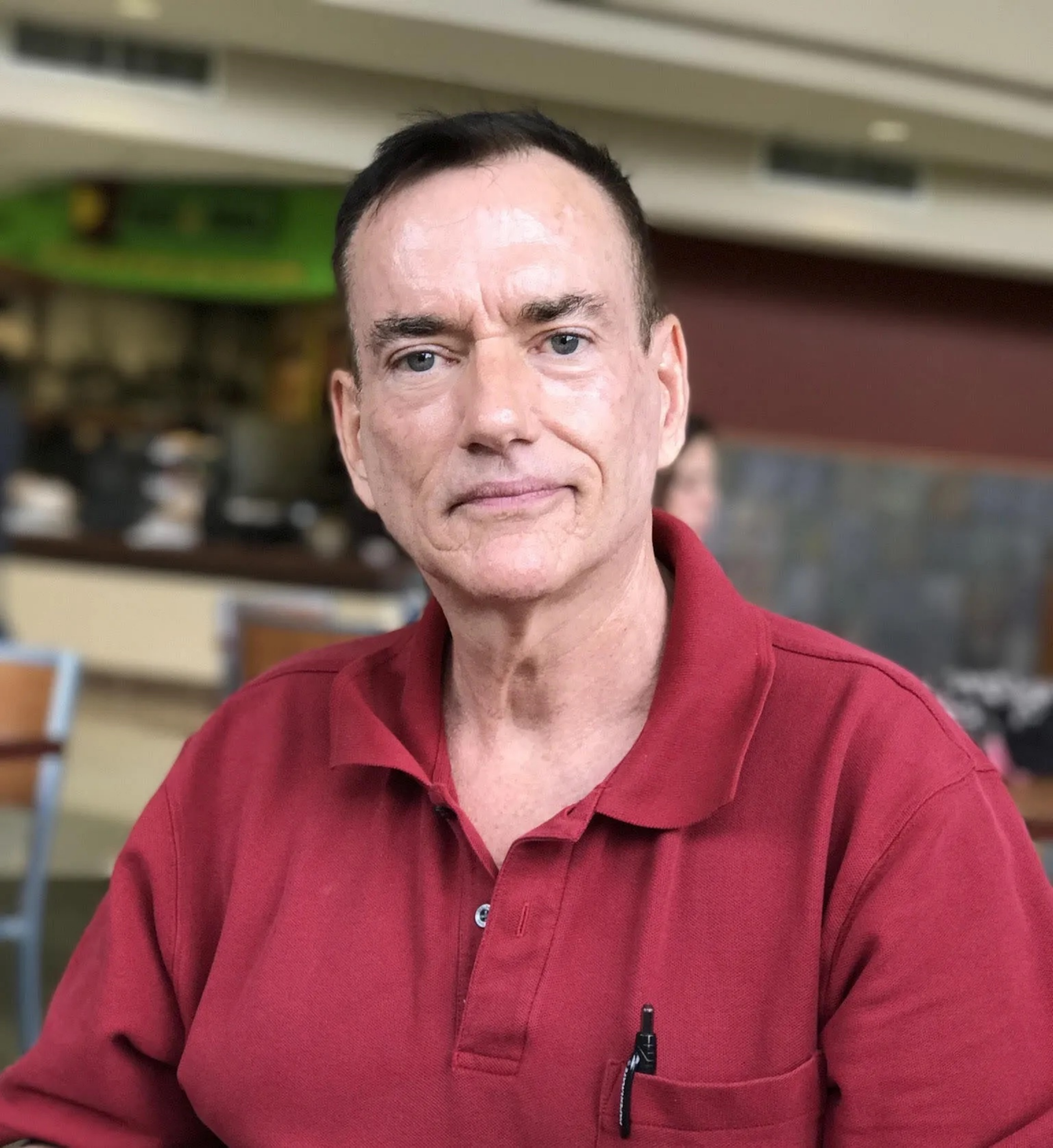}}\\
	{\famousquote{Don't drink and derive.}{Jonathan P. Dowling}}\index{Jon Dowling}\index{Latin}
\end{center}

\vspace*{100pt}

\sketch{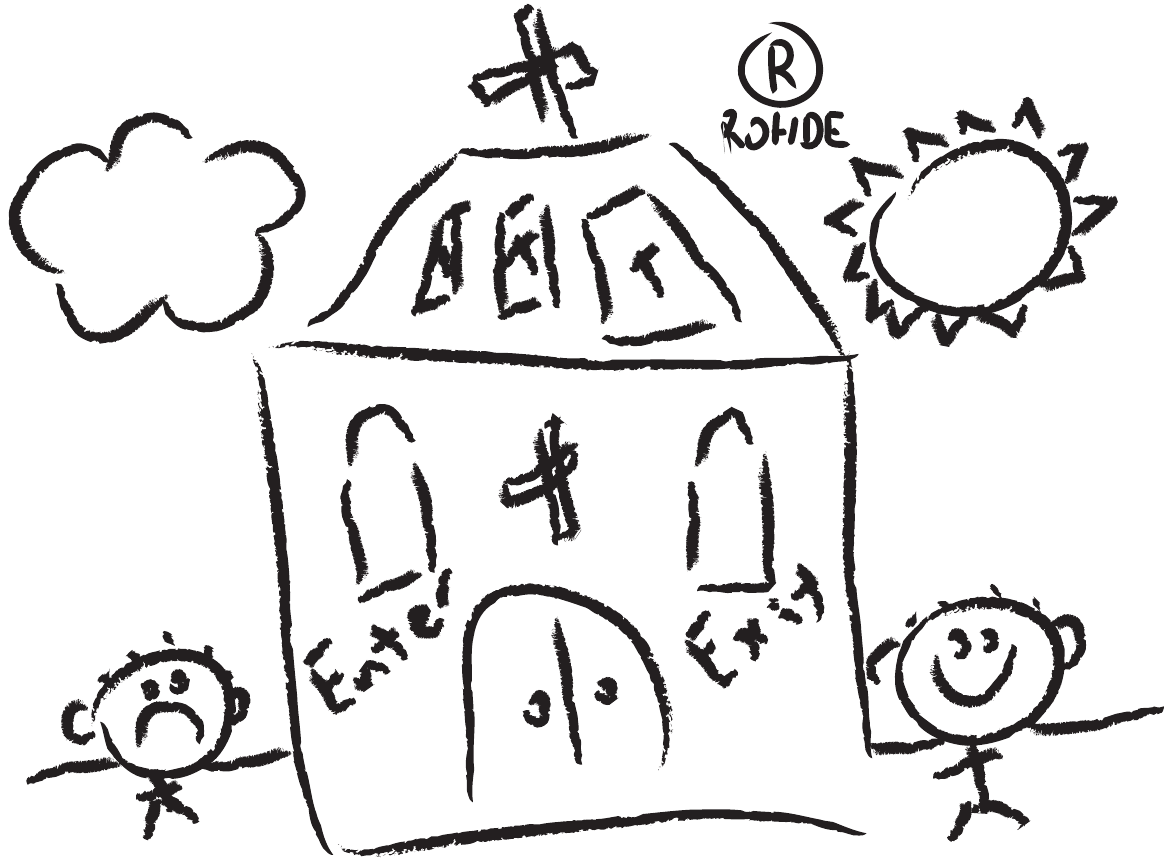}\famousquote{We all die. The goal isn't to live forever, the goal is to create something that will.}{Chuck Palahniuk} 

\vspace*{180pt}
\begin{center}
	\href{https://youtu.be/fuF-z-Aq37o}{\includegraphics[clip=true, width=0.75\textwidth]{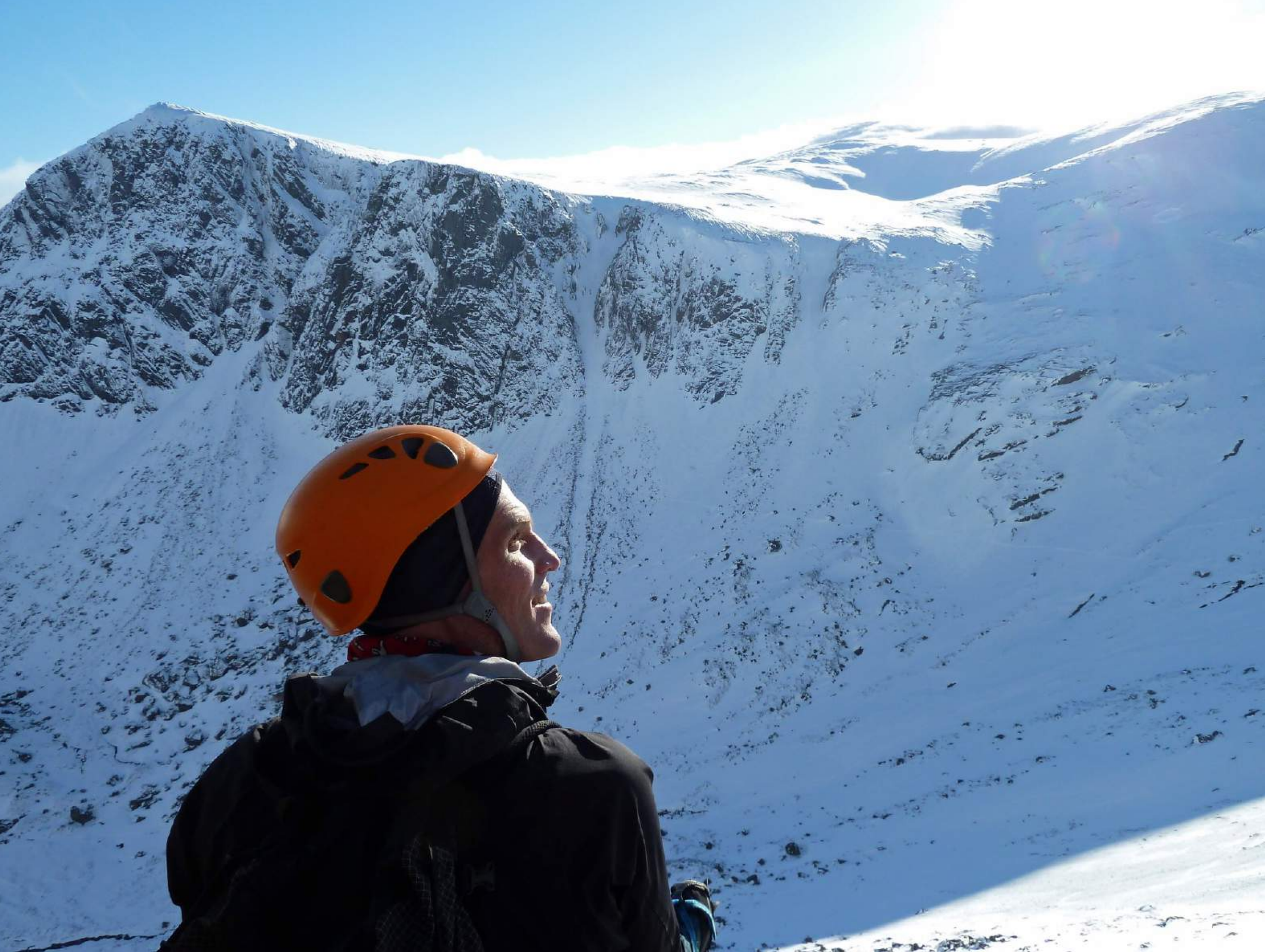}}\\
	\textit{To those who die doing what they love.\\ --- Finis vitae sed non amoris.}\index{Peter Metcalfe}\index{Latin}
\end{center}

% \newpage

\if 1\doublecol
	\twocolumngrid
\fi

% % \clearpage

\end{document}